# CEPC

# *Technical Design Report*

## Accelerator

The CEPC Study Group

# CEPC

# *Technical Design Report*

## Accelerator

The CEPC Study Group

# CEPC

# *Technical Design Report*

Waleed Abdallah[15], Tiago Carlos Adorno de Freitas[268], Konstantin Afanaciev[95], Shakeel Ahmad[153], Ijaz Ahmed[3], Xiaocong Ai[278], Abid Aleem[91], Wolfgang Altmannshofer[215], Fabio Alves[135], Weiming An[11], Rui An[56], Daniele Paolo Anderle[176], Stefan Antusch[211], Yasuo Arai[65], Andrej Arbuzov[113], Abdesslam Arhrib[206], Mustafa Ashry[15], Sha Bai[91], Yu Bai[177], Yang Bai[255], Vipul Bairathi[98], Csaba Balazs[132], Philip Bambade[121], Yong Ban[155], Triparno Bandyopadhyay[180], Shou-Shan Bao[167], Desmond P. Barber[233], Ayse Bat[5], Varvara Batozskaya[91,138], Subash Chandra Behera[74], Alexander Belyaev[244], Michele Bertucci[81], Xiao-Jun Bi[91], Yuanjie Bi[183], Tianjian Bian[24], Fabrizio Bianchi[87], Thomas Biekötter[114], Michela Biglietti[85], Shalva Bilanishvili[117], Deng Binglin[91], Denis Bodrov[175], Anton Bogomyagkov[13,150], Serge Bondarenko[113], Stewart Boogert[34], Maarten Boonekamp[104], Marcello Borri[199], Angelo Bosotti[81], Vincent Boudry[123], Mohammed Boukidi[14], Igor Boyko[113], Ivanka Bozovic[257], Giuseppe Bozzi[214], Jean-Claude Brient[123], Anastasiia Budzinskaya[90], Masroor Bukhari[107], Vladimir Bytev[113], Giacomo Cacciapaglia[103], Hua Cai[22], Wenyong Cai[270], Wujun Cai[157], Yijian Cai[156], Yizhou Cai[135], Yuchen Cai[91], Haiying Cai[116], Huacheng Cai[235], Lorenzo Calibbi[136], Junsong Cang[62], Guofu Cao[91], Jianshe Cao[91], Antoine Chance[104], Xuejun Chang[270], Yue Chang[136], Zhe Chang[91], Xinyuan Chang[91], Wei Chao[11], Auttakit Chatrabhuti[32], Yimin Che[135], Yuzhi Che[91], Bin Chen[91], Danping Chen[169], Fuqing Chen[91], Fusan Chen[91], Gang Chen[91], Guoming Chen[91], Hua-Xing Chen[177], Huirun Chen[135], Jinhui Chen[91], Ji-Yuan Chen[170], Kai Chen[17], Mali Chen[91], Mingjun Chen[91], Mingshui Chen[91], Ning Chen[136], Shanhong Chen[91], Shanzhen Chen[91], Shao-Long Chen[17], Shaomin Chen[197], Shiqiang Chen[135], Tianlu Chen[196], Wei Chen[183], Xiang Chen[170], Xiaoyu Chen[156], Xin Chen[135], Xun Chen[170], Xurong Chen[92], Ye Chen[91], Ying Chen[91], Yukai Chen[91], Zelin Chen[135], Zilin Chen[91], Gang Chen[146], Boping Chen[187], Chunhui Chen[102], Hok Chuen Cheng[188], Huajie Cheng[144], Shan Cheng[70], `Tongguang Cheng[6], Yunlong Chi[91], Pietro Chimenti[203], Wen Han Chiu[223], Guk Cho[272], Ming-Chung Chu[188], Xiaotong Chu[91], Ziliang Chu[91], Guglielmo Coloretti[253], Andreas Crivellin[253], Hanhua Cui[91], Xiaohao Cui[91], Zhaoyuan Cui[210], Brunella D'Anzi[208], Ling-Yun Dai[70], Xinchen Dai[155], Xuwen Dai[91], Antonio De Maria[135], Nicola De Filippis[77], Christophe De La Taille[152], Francesca De Mori[249], Chiara De Sio[213], Elisa Del Core[81], Shuangxue Deng[157], Wei-Tian Deng[68], Zhi Deng[197], Ziyan Deng[91], Bhupal Dev[260], Tang Dewen[243], Biagio Di Micco[79], Ran Ding[2], Siqin Ding[197], Yadong Ding[261], Haiyi Dong[91], Jianing Dong[167], Jing Dong[91], Lan Dong[91], Mingyi Dong[91], Xu Dong[45], Yipei Dong[156], Yubing Dong[91], Milos Dordevic[257], Marco Drewes[207], Mingxuan Du[135], Mingxuan Du[155], Qianqian Du[49], Xiaokang Du[94], Yanyan Du[269], Yong Du[170], Yunfei Du[91], Chun-Gui Duan[57], Zhe Duan[91], Yahor Dydyshka[113], Ulrik Egede[132], Walaa Elmetenawee[76], Yun Eo[272], Ka Yan Fan[191], Kuanjun Fan[68], Yunyun Fan[91], Bo Fang[263], Shuangshi Fang[91], Yuquan Fang[91], Ada Farilla[85], Riccardo Farinelli[78], Muhammad Farooq[250], Angeles Faus Golfe[121], Almaz Fazliakhmetov[90], Rujun Fei[157], Bo Feng[7], Chong Feng[37], Junhua Feng[183], Xu Feng[155], Zhuoran Feng[139],

Zhuoran Feng[139], Luis Roberto Flores Castillo[188], Etienne Forest[65], Andrew Fowlie[268], Harald Fox[125], Hai-Bing Fu[51], Jinyu Fu[91], Benjamin Fuks[122], Yoshihiro Funakoshi[65], Emidio Gabrielli[248], Nan Gan[91], Li Gang[2], Jie Gao[91], Meisen Gao[45], Wenbin Gao[91], Wenchun Gao[91], Yu Gao[91], Yuanning Gao[155], Zhanxiang Gao[135], Yanyan Gao[218], Kun Ge[9], Shao-Feng Ge[170], Zhenwu Ge[135], Li-Sheng Geng[6], Qinglin Geng[167], Chao-Qiang Geng[142], Swagata Ghosh[73], Antonio Gioiosa[232], Leonid Gladilin[130], Ti Gong[64], Stefania Gori[215], Quanbu Gou[91], Sebastian Grinstein[89], Chenxi Gu[123], Gerardo Guillermo[91], Joao Guimaraes da Costa[91], Dizhou Guo[91], Fangyi Guo[91,23], Jiacheng Guo[24], Jun Guo[170], Lei Guo[91,64], Lei Guo[31], Xia Guo[151], Xin-Heng Guo[11], Xinyang Guo[37], Yun Guo[49], Yunqiang Guo[157], Yuping Guo[45], Zhi-Hui Guo[57], Alejandro Gutiérrez-Rodríguez[200], Seungkyu Ha[272], Noman Habib[91], Jan Hajer[19], Francois Hammer[154], Chengcheng Han[183], Huayong Han[52], Jifeng Han[173], Liang Han[239], Liangliang Han[135], Ruixiong Han[91], Yang Han[183], Yezi Han[156], Yuanying Han[91], Tao Han[235], Jiankui Hao[155], Xiqing Hao[62], XiqingHao[62], Chuanqi He[176], Dayong He[91], Dongbing He[169], Guangyuan He[157], Hong-Jian He[170], Jibo He[217], Jun He[91], Longyan He[91], Xiang He[91], Xiao-Gang He[170], Zhenqiang He[91], Klaus Heinemann[233], Sven Heinemeyer[100], Yuekun Heng[91], María A. Hernández-Ruíz[200], Jiamin Hong[156], Yuenkeung Hor[183], George W.S. Hou[141], Xiantao Hou[91], Xiaonan Hou[91], Zhilong Hou[91], Suen Hou[93], Caishi Hu[157], Chen Hu[168], Dake Hu[157], Haiming Hu[91], Jiagen Hu[157], Jun Hu[91], Kun Hu[167], Shouyang Hu[24], Yongcai Hu[149], Yu Hu[91], Zhen Hu[197], Zhehao Hua[91], Jianfei Hua[197], Chao-Shang Huang[96], Fa Peng Huang[183], Guangshun Huang[239], Jinshu Huang[91], Ke Huang[156], Liangsheng Huang[91], Shuhui Huang[191], Xingtao Huang[167], Xu-Guang Huang[45], Yanping Huang[91], Yonggang Huang[22], Yongsheng Huang[183], Zimiao Huang[135], Chen Huanyuan[275], Changgi Huh[118], Jiaqi Hui[170], Lihua Huo[91], Talab Hussain[246], Kyuyeong Hwang[272], Ara Ioannisian[271], Munawar Iqbal[246], Paul Jackson[209], Shahriyar Jafarzade[106], Haeun Jang[272], Seoyun Jang[272], Daheng Ji[91], Qingping Ji[62], Quan Ji[91], Xiaolu Ji[91], Jingguang Jia[26], Jinsheng Jia[22], Xuewei Jia[91], Zihang Jia[135], Cailian Jiang[135], Han Ren Jiang[156], Houbing Jiang[263], Jun Jiang[167], Xiaowei Jiang[91], Xin Jiang[179], Xuhui Jiang[91], Yongcheng Jiang[91], Zhongjian Jiang[156], Cheng Jiang[218], Ruiqi Jiao[197], Dapeng Jin[91], Shan Jin[135], Song Jin[91], Yi Jin[224], Junji Jis[263], Sunghoon Jung[164], Goran Kacarevic[257], Eric Kajfasz[36], Lidia Kalinovskaya[113], Aleksei Kampf[40], Wen Kang[91], Xian-Wei Kang[11], Xiaolin Kang[28], Biswajit Karmakar[242], Zhiyong Ke[91], Rijeesh Keloth[258], Alamgir Khan[101], Hamzeh Khanpour[1,240], Khanchai Khosonthongkee[185], KhanchaiKhosonthongkee[185], Bobae Kim[118], Dongwoon Kim[272], Mi Ran Kim[184], Minsuk Kim[47], Sungwon Kim[272], On Kim[231], Michael Klasen[254], Sanghyun Ko[164], Ivan Koop[13,150], Vitaliy Kornienko[40], Bryan Kortman[139], Gennady Kozlov[113], Shiqing Kuang[67], Mukesh Kumar[247], Chia Ming Kuo[137], Tsz Hong Kwok[66], François Sylvain Ren Lagarde[170], Pei-Zhu Lai[137], Imad Laktineh[227], Xiaofei Lan[30], Zuxiu Lan[270], Lia Lavezzi[84], Justin Lee[8], Junghyun Lee[118], Sehwook Lee[118], Ge Lei[91], Roy Lemmon[199,226], Yongxiang Leng[179], Sze Ching Leung[235], Hai Tao Li[167], Bingzhi Li[276], Bo Li[91], Bo Li[269], Changhong Li[273], Chao Li[265], Cheng Li[183], Cheng Li[239], Chunhua Li[128], Cui Li[161], Dazhang Li[91], Dikai Li[172], Fei Li[91], Gang Li[91], Gang Li[91], Gang Li[161], Gaosong Li[91], Haibo Li[91], Haifeng Li[167], Hai-Jun Li[96], Haotian Li[135], Hengne Li[176], Honglei Li[224], Huijing Li[62], Jialin Li[170], Jingyi Li[91], Jinmian Li[173], Jun Li[156], Leyi Li[167], Liang Li[170], Ling Li[156], Mei Li[91], Meng Li[91], Minxian Li[91], Pei-Rong Li[126], Qiang Li[149], Shaopeng Li[91], Shenghe Li[21], Shu Li[170], Shuo Li[135], Teng Li[167],

Tiange Li[64], Tong Li[136], Weichang Li[169], Weidong Li[91], Wenjun Li[62], Xiaoling Li[167], Xiaomei Li[24], Xiaonan Li[91], Xiaoping Li[91], Xiaoting Li[91], Xin Li[92], Xinqiang Li[17], Xuekang Li[37], Yang Li[239], Yanwei Li[270], Yiming Li[91], Ying Li[269], Ying-Ying Li[239], Yonggang Li[157], Yonglin Li[135], Yufeng Li[91], Yuhui Li[91], Zhan Li[91], Zhao Li[91], Zhiji Li[157], Tong Li[119], Lingfeng Li[12], Fei Li[197], Jing Liang[91], Jinhan Liang[176], Zhijun Liang[91], Guangrui Liao[49], Hean Liao[264], Jiajun Liao[183], Libo Liao[183], Longzhou Liao[69], Yi Liao[176], Yipu Liao[91], Ayut Limphirat[185], AyutLimphirat[185], Tao Lin[91], Weiping Lin[173], Yufu Lin[49], Yugen Lin[91], Beijiang Liu[91], Bo Liu[91], Danning Liu[170], Dong Liu[167], Fu-Hu Liu[171], Hongbang Liu[50], Huangcheng Liu[157], Hui Liu[22], Huiling Liu[60], Jia Liu[91], Jia Liu[155], Jiaming Liu[91], Jianbei Liu[239], Jianyi Liu[197], Jingdong Liu[91], Jinhua Liu[156], Kai Liu[126], Kang Liu[170], Kun Liu[170], Mengyao Liu[167], Peng Liu[91], Pengcheng Liu[157], Qibin Liu[170], Shan Liu[9], Shidong Liu[161], Shuang Liu[197], Shubin Liu[239], Tao Liu[91], Tao Liu[190], Tong Liu[45], Wei Liu[134], Xiang Liu[126], Xiao-Hai Liu[195], Xiaohui Liu[11], Xiaoyu Liu[91], Xin Liu[108], Xinglin Liu[45], Xingquan Liu[173], Yang Liu[183], Yanlin Liu[167], Yao-Bei Liu[61], Yi Liu[20], Yiming Liu[10], Yong Liu[91], Yonglu Liu[143], Yu Liu[91], Yubin Liu[136], Yudong Liu[91], Yulong Liu[91], Zhaofeng Liu[91], Zhen Liu[91], Zhenchao Liu[266], Zhi Liu[91], Zhi-Feng Liu[277], Zhiqing Liu[167], Zhongfu Liu[37], Zuowei Liu[135], Mia Liu[159], Zhen Liu[230], Xiaoyang Liu[91], Xinchou Lou[91,245], Cai-Dian Lu[91], Jun-Xu Lu[6], Qiu Zhen Lu[156], Shang Lu[91], Shang Lu[91], Wenxi Lu[135], Xiaohan Lu[91], Yunpeng Lu[91], Zhiyong Lu[24], Xianguo Lu[251], Wei Lu[197], Bayarto Lubsandorzhiev[90], Sultim Lubsandorzhiev[90], Arslan Lukanov[90], Jinliang Luo[243], Tao Luo[45], xiaoan Luo[91], Xiaofeng Luo[17], Xiaolan Luo[91], Jindong Lv[135], Feng Lyu[91], Xiao-Rui Lyu[217], Kun-Feng Lyu[230], Ande Ma[91], Hong-Hao Ma[49], Jun-Li Ma[91], Kai Ma[166], Lishuang Ma[91], Na Ma[91], Renjie Ma[135], Weihu Ma[45], Xinpeng Ma[91], Yanling Ma[270], Yan-Qing Ma[155], Yongsheng Ma[91], Zhonghui Ma[91], Zhongjian Ma[91], Yang Ma[80], Mousam Maity[158], Lining Mao[170], Yanmin Mao[270], Yaxian Mao[17], Aurélien Martens[121], Caccia Massimo Luigi Maria[204], Shigeki Matsumoto[115], Bruce Mellado[247,105], Davide Meloni[236], Lingling Men[91], Cai Meng[91], Lingxin Meng[125], Zhenghui Mi[91], Yuhui Miao[135], Mauro Migliorati[237], Lei Ming[183], Vasiliki A. Mitsou[99], Laura Monaco[81], Arthur Moraes[268], Karabo Mosala[247], Ahmad Moursy[15], Lichao Mu[270], Zhihui Mu[91], Nickolai Muchnoi[13,150], Daniel Muenstermann[125], DanielMuenstermann[125], Pankaj Munbodh[215], William John Murray[251,163], Jérôme Nanni[123], Dmitry Nanzanov[90], Changshan Nie[270], Sergei Nikitin[13,150], Feipeng Ning[91], Guozhu Ning[58], Jia-Shu Niu[171], Juan–Juan Niu[49], Yan Niu[167], Edward Khomotso Nkadimeng[105], Kazuhito Ohmi[65], Katsunobu Oide[221,42,66], Hideki Okawa[91], Mohamed Ouchemhou[14], Qun Ouyang[91], Daniele Paesani[83], Carlo Pagani[81], Stathes Paganis[141], Collette Pakuza[42,234], Jiangyang Pan[157], Juntong Pan[91], Tong Pan[188], Xiang Pan[175], Papia Panda[222], Saraswati Pandey[4], Mila Pandurovic[257], Rocco Paparella[81], Roman Pasechnik[129], Emilie Passemar[205,75], Hua Pei[17], Xiaohua Peng[91], Xinye Peng[27], Yuemei Peng[91], Jialun Ping[133], Ronggang Ping[91], Souvik Priyam Adhya[71], Baohua Qi[91], Hang Qi[239], Huirong Qi[91], Ming Qi[135], Sen Qian[91], Zhuoni Qian[53], Congfeng Qiao[217], Guangyou Qin[17], Jiajia Qin[243], Laishun Qin[25], Liqing Qin[49], Qin Qin[68], Xiaoshuai Qin[167], Zhonghua Qin[91], Guofeng Qu[173], Antonio Racioppi[140], Michael Ramsey-Musolf[170], Shabbar Raza[43], Vladimir Rekovic[257], Jing Ren[54], Jürgen Reuter[39], Tania Robens[97], Giancarlo Rossi[238], Manqi Ruan[91], Manqi Ruan[91], Leonid Rumyantsev[40], Min Sang Ryu[118], Renat Sadykov[113], Minjing Sang[91], Juan José Sanz-Cillero[201], Miroslav Saur[155], Nishil Savla[44], Michael A. Schmidt[193],

Daniele Sertore[81], Ron Settles[131], Peng Sha[91], Ding-Yu Shao[45], Ligang Shao[91], Hua-Sheng Shao[122], Xin She[91], Chuang Shen[91], Hong-Fei Shen[91], Jian-Ming Shen[70], Peixun Shen[91], Qiuping Shen[170], Zhongtao Shen[239], Shuqi Sheng[91], Haoyu Shi[91], Hua Shi[91], Qi Shi[217], Shusu Shi[17], Xiaolei Shi[91], Xin Shi[91], Yukun Shi[239], Zhan Shi[37], Ian Shipsey[234], Gary Shiu[252], Chang Shu[91], Zong-Guo Si[167], Andrei Sidorenkov[90], Ivan Smiljanić[257], Aodong Song[167], Huayang Song[96], Jiaojiao Song[62], Jinxing Song[24], Siyuan Song[170], Weimin Song[109], Weizheng Song[91], Zhi Song[197], Shashwat Sourav[72], Paolo Spruzzola[81], Feng Su[160], Shengsen Su[91], Wei Su[183], Shufang Su[210], Yanfeng Sui[91], Zexuan Sui[54], Michael Sullivan[174], Baiyang Sun[135], Guoqiang Sun[270], Hao Sun[38], Hao-Kai Sun[91], Junfeng Sun[62], Liang Sun[263], Mengcheng Sun[156], Pengfei Sun[24], Sichun Sun[10], Xianjing Sun[91], Xiaohu Sun[155], Xilei Sun[91], Xingyang Sun[135], Xin-Yuan Sun[111], Yanjun Sun[147], Yongzhao Sun[91], Yue Sun[91], Zheng Sun[173], Zheng Sun[91], Narumon Suwonjandee[32], Elsayed Tag Eldin[46], Biao Tan[91], Bo Tang[157], Chuanxiang Tang[197], Gao Tang[25], Guangyi Tang[91], Jian Tang[183], Jingyu Tang[239], Liang Tang[57], Ying'Ao Tang[263], Junquan Tao[91], Abdel Nasser Tawfik[46], Geoffrey Taylor[192], Valery Telnov[13,150], Saike Tian[91], Riccardo Torre[82], Wladyslaw Henryk Trzaska[225], Dmitri Tsybychev[182], Yanjun Tu[191], Shengquan Tuo[256], Michael Tytgat[259], Ghalib Ul Islam[219], Nikita Ushakov[90], German Valencia[132], Jaap Velthuis[213], Alessandro Vicini[229], Trevor Vickey[241], Ivana Vidakovic[257], Henri Videau[123], Raymond Volkas[192], Dmitry Voronin[90], Natasa Vukasinovic[257], Xia Wan[165], Xuying Wan[156], Xiao Wang[132], Anqing Wang[167], Bin Wang[91], Chengtao Wang[91], Chuanye Wang[135], Ci Wang[54], Dayong Wang[155], Dou Wang[91], En Wang[278], Fei Wang[156], Fei Wang[278], Guanwen Wang[91], Guo-Li Wang[58], Haijing Wang[91], Haolin Wang[176], Jia Wang[149], Jian Wang[167], Jianchun Wang[91], Jianli Wang[91], Jiawei Wang[190], Jin Wang[91], Jin-Wei Wang[220], Joseph Wang[198], Kechen Wang[262], Lechun Wang[157], Lei Wang[91], Liguo Wang[37], Lijiao Wang[91], Lu Wang[270], Meng Wang[167], Na Wang[91], Pengcheng Wang[91], Qian Wang[176], Qun Wang[239], Shu Lin Wang[156], Shudong Wang[91], Taofeng Wang[6], Tianhong Wang[55], Tianyang Wang[274], Tong Wang[91], Wei Wang[91], Wei Wang[170], Xiaolong Wang[45], Xiaolong Wang[91], Xiaoning Wang[91], Xiao-Ping Wang[6], Xiongfei Wang[126], Xujian Wang[91], Yaping Wang[17], Yaqian Wang[58], Yi Wang[91], Yiao Wang[91], Yifang Wang[91], Yilun Wang[135], Yiwei Wang[91], You-Kai Wang[165], Yuanping Wang[60], Yuexin Wang[91,23], Yuhao Wang[135], Yu-Ming Wang[136], Yuting Wang[91], Zhen Wang[170], Zhigang Wang[91], Weiping Wang[112], Zeren Simon Wang[142,16], Biao Wang[178], Hui Wang[162], Lian-Tao Wang[216], Zihui Wang[145], Zirui Wang[228], Jia Wang[91], Tong Wang[91], Daihui Wei[49], Shujun Wei[91], Wei Wei[91], Xiaomin Wei[149], Yuanyuan Wei[91], Yingjie Wei[234], Liangjian Wen[91], Xuejun Wen[157], Yufeng Wen[111], Martin White[209], Peter Williams[181], Zef Wolffs[139], William John Womersley[218], Baona Wu[135], Bobing Wu[91], Guanjian Wu[91], Jinfei Wu[91,23], Lei Wu[133], Lina Wu[267], Linghui Wu[91], Minlin Wu[183], Peiwen Wu[177], Qi Wu[91], Qun Wu[167], Tianya Wu[91], Xiang Wu[157], Xiaohong Wu[41], Xing-Gang Wu[31], Xuehui Wu[91], Yaru Wu[91], Yongcheng Wu[133], Yuwen Wu[91], Zhi Wu[91], Xin Wu[221], Lei Xia[239], Ligang Xia[135], Shang Xia[91], Benhou Xiang[91], Dao Xiang[170], Zhiyu Xiang[91], Bo-Wen Xiao[189], Chu-Wen Xiao[49], Dong Xiao[126], Guangyan Xiao[135], Han Xiao[91], Meng Xiao[277], Ouzheng Xiao[91], Rui-Qing Xiao[170], Xiang Xiao[183], Yichen Xiao[157], Ying Xiao[37], Yu Xiao[270], Yunlong Xiao[45], Zhenjun Xiao[133], Hengyuan Xiao[197], Nian Xie[91], Yuehong Xie[17], Tianmu Xin[91], Ye Xing[29], Zhizhong Xing[91], Da Xu[91], Fang Xu[45], Fanrong Xu[110], Haisheng Xu[91], Haocheng Xu[91], Ji Xu[278], Miaofu Xu[91],

Qingjin Xu[91], Qingnian Xu[88], Wei Xu[91], Wei Xu[91], Weixi Xu[135], Xinping Xu[175], Zhen Xu[277], Zijun Xu[91], Zehua Xu[120], Yaoyuan Xu[127], Feifei Xue[149], Baojun Yan[91], Bin Yan[91], Fen Yan[91], Fucheng Yan[156], Jiaming Yan[156], Liang Yan[45], Luping Yan[91], Qi-Shu Yan[217], Wenbiao Yan[239], Yupeng Yan[185], Luping Yan[91], Haoyue Yan [91], Dong Yang[9], Fengying Yang[157], Guicheng Yang[156], Haijun Yang[170], Jin Min Yang[96], Jing Yang[270], Lan Yang[135], Li Yang[270], Li Lin Yang[277], Lili Yang[183], Litao Yang[197], Mei Yang[91], Qiaoli Yang[110], Tiansen Yang[151], Xiaochen Yang[91], Yingjun Yang[270], Yueling Yang[62], Zhengyong Yang[157], Zhenwei Yang[155], Youhua Yang[48], Xiancong Yang[197], De-Liang Yao[157], Shi Yao[156], Lei Ye[156], Lingxi Ye[91], Mei Ye[91], Rui Ye[91], Rui Ye[157], Yecheng Ye[135], Vitaly Yermolchyk[113], Kai Yi[133], Li Yi[167], Yang Yi[70], Di Yin[91], Peng-Fei Yin[91], Shenghua Yin[26], Ze Yin[270], Zhongbao Yin[17], Zhang Yinhong[91], Hwi Dong Yoo[272], Zhengyun You[183], Charles Young[174], Boxiang Yu[91], Chenghui Yu[91], Fusheng Yu[126], Jie-Sheng Yu[70], Jinqing Yu[70], Lingda Yu[91], Zhao-Huan Yu[183], Felix Yu[112], Bingrong Yu[35], Changzheng Yuan[91], Li Yuan[6], Xing-Bo Yuan[17], Youjin Yuan[92], Junhui Yue[91], Qian Yue[197], Baobiao Yue[194], Un Nisa Zaib[42], Riccardo Zanzottera[86], Hao Zeng[91], Ming Zeng[91], Jian Zhai[270], Jiyuan Zhai[91], Xin Zhe Zhai[91], Xi-Jie Zhan[58], Ben-Wei Zhang[17], Bolun Zhang[91], Di Zhang[63], Guangyi Zhang[62], Hao Zhang[91], Hong-Hao Zhang[183], Huaqiao Zhang[91], Hui Zhang[270], Jialiang Zhang[135], Jianyu Zhang[217], Jianzhong Zhang[54], Jiehao Zhang[91], Jielei Zhang[64], Jingru Zhang[91], Jinxian Zhang[91], Junsong Zhang[91], Junxing Zhang[37], Lei Zhang[91], Lei Zhang[133], Liang Zhang[167], Licheng Zhang[155], Liming Zhang[197], Linhao Zhang[239], Luyan Zhang[91], Mengchao Zhang[110], Rao Zhang[173], Shulei Zhang[70], Wan Zhang[91], Wenchao Zhang[165], Xiangzhen Zhang[91], Xiaomei Zhang[91], Xiaoming Zhang[17], Xiaoxu Zhang[135], Xiaoyu Zhang[133], Xuantong Zhang[91], Xueyao Zhang[167], Yang Zhang[91], Yang Zhang[278], Yanxi Zhang[155], Yao Zhang[91], Ying Zhang[91], Yixiang Zhang[135], Yizhou Zhang[91], Yongchao Zhang[177], Yu Zhang[59], Yuan Zhang[91], Yujie Zhang[6], Yulei Zhang[170], Yumei Zhang[183], Yunlong Zhang[239], Zhandong Zhang[91], Zhaoru Zhang[91], Zhen-Hua Zhang[243], Zhenyu Zhang[263], Zhichao Zhang[18], Zhi-Qing Zhang[63], Zhuo Zhang[91], Zhiqing Zhang[121], Cong Zhang[212], Tianliang Zhang[197], Luyan Zhang[91], Guang Zhao[91], Hongyun Zhao[148], Jie Zhao[91], Jingxia Zhao[91], Jingyi Zhao[91], Ling Zhao[91], Luyang Zhao[91], Mei Zhao[91], Minggang Zhao[136], Mingrui Zhao[24], Qiang Zhao[91], Ruiguang Zhao[149], Tongxian Zhao[91], Yaliang Zhao[91], Ying Zhao[91], Yue Zhao[170], Zhiyu Zhao[170], Zhuo Zhao[91], Alexey Zhemchugov[113], Hongjuan Zheng[91], Jinchao Zheng[135], Liang Zheng[28], Ran Zheng[149], shanxi zheng[151], Xu-Chang Zheng[31], Wang Zhile[55], Weicai Zhong[157], Yi-Ming Zhong[33], Chen Zhou[155], Daicui Zhou[17], Jianxin Zhou[91], Jing Zhou[24], Jing Zhou[91], Ning Zhou[170], Qi-Dong Zhou[167], Shiyu Zhou[197], Shun Zhou[91], Sihong Zhou[88], Xiang Zhou[263], Xingyu Zhou[128], Yang Zhou[91], Yong Zhou[135], Yu-Feng Zhou[96], Zusheng Zhou[91], Demin Zhou[65], Dechong Zhu[91], Hongbo Zhu[277], Huaxing Zhu[155], Jingya Zhu[64], Kai Zhu[91], Pengxuan Zhu[96], Ruilin Zhu[133], Xianglei Zhu[197], Yingshun Zhu[91], Yongfeng Zhu[91], Xiao Zhuang[24], Xuai Zhuang[91], Mikhail Zobov[124], Zhanguo Zong[65], Cong Zou[270], Hongying Zou[270], the CEPC Collaboration[279]

[1] AGH University of Science and Technology, Faculty of Physics and Applied Computer Science, Krakow
[2] Anhui University, Hefei, Anhui
[3] Applied Physics Department, Federal Urdu University of Arts, Science and Technology, Islamabad
[4] Banaras Hindu University, Institute of Science, Varanasi
[5] Bandirma Onyedi Eylul University, Bandırma
[6] Beihang University, Beijing
[7] Beijing Computational Science Research Center, Beijing
[8] Beijing Conveyi LTD, Beijing
[9] Beijing Glass Research Institute Co., Ltd, Beijing
[10] Beijing Institute of Technology, Beijing
[11] Beijing Normal University, Beijing
[12] Brown University, Providence, RI
[13] Budker Institute of Nuclear Physics, Novosibirsk
[14] Cadi Ayyad University, Laboratory of Fundamental and Applied Physics, Polydisciplinary Faculty, Marrakech
[15] Cairo University, Giza
[16] Center for Theory and Computation, National Tsing Hua University, Hsinchu
[17] Central China Normal University, Wuhan, Hubei
[18] Central South University, Changsha, Hunan
[19] Centro de Fisica Teorica de Particulas, Universidade Tecnica de Lisboa, Lisboa
[20] Chengdu KaitengSifang Digital Radio \& TV Equipment Co., Ltd, Chengdu, Sichuan
[21] China academy of engineering physics, Mianyang, Sichuan
[22] China building materials academy, Beijing
[23] China Center of Advanced Science and Technology, Beijing
[24] China Institute of Atomic Energy, Beijing
[25] China Jiliang University, Hangzhou, Zhejiang
[26] China Nuclear Instrument CO., LTD, Beijing
[27] China University of Geosciences, Beijing
[28] China University of Geosciences, Wuhan, Hubei
[29] China University of Mining and Technology, Xuzhou, Jiangsu
[30] China West Normal University, Nanchong, Sichuan
[31] Chongqing University, Department of Physics, Chongqing
[32] Chulalongkorn University, Bangkok
[33] City University of Hong Kong, Department of Physics, Hong Kong
[34] Cockcroft Institute, Daresbury
[35] Cornell University, Ithaca, NY
[36] CPPM (AMU/IPhU & CNRS/IN2P3), Marseille
[37] Dalian Minzu University, Dalian, Liaoning
[38] Dalian University of Technology, Department of Physics, Dalian
[39] Deutsches Elektronen-Synchrotron DESY, Hamburg
[40] Dzhelepov Laboratory of Nuclear Problems of Joint Institute for Nuclear Research (DLNP JINR), Dubna
[41] East China University Of Science And Technology, Shanghai
[42] European Organization for Nuclear Research (CERN), Geneva

[43] Federal Urdu University of Arts, Science and Technology, Department of Physics, Karachi
[44] Fergusson College, Pune
[45] Fudan University, Shanghai
[46] Future University in Egypt (FUE), New Cairo
[47] Gangneung-Wonju National University, Gangneung
[48] Genova University, Genova
[49] Guangxi Normal University, Guilin, Guangxi
[50] Guangxi University, Nanning, Guangxi
[51] Guizhou Minzu University, Department of Physics, Guiyang
[52] Guizhou University of Finance and Economics, Guiyang, Guizhou
[53] Hangzhou Normal University, Hangzhou, Zhejiang
[54] Harbin Engineering University, Harbin, Heilongjiang
[55] Harbin Institute of Technology, Harbin, Heilongjiang
[56] Harvard University, Cambridge, MA
[57] Hebei Normal University, Shijiazhuang, Hebei
[58] Hebei University, Baoding, Hebei
[59] Hefei University of Technology, School of Physics, Hefei, Anhui
[60] Henan Academy of Sciences, Zhengzhou, Henan
[61] Henan Institute of Science and Technology, Xinxiang, Henan
[62] Henan Normal University, Xinxiang, Henan
[63] Henan University of Technology, Zhengzhou, Henan
[64] Henan University, Kaifeng, Henan
[65] High Energy Accelerator Research Organization (KEK), Tsukuba
[66] Hong Kong university of Science and Technology, Hong Kong
[67] Huaibei Normal University, Huaibei, Anhui
[68] Huazhong University of Science and Technology, Wuhan, Hubei
[69] Hubei University of Automotive Technology, Shiyan
[70] Hunan University, Changsha, Hunan
[71] IFJ-PAN, Krakow
[72] Indian Institute of Science Education and Research, Bhopal
[73] Indian Institute of Technology Kharagpur, Kharagpur
[74] Indian Institute of Technology Madras (IN), Chennai
[75] Indiana University, Department of Physics, Bloomington, IN
[76] INFN - Sezione di Bari, Bari
[77] INFN - Sezione di Bari, University and Politecnico of Bari
[78] INFN - Sezione di Ferrara and University of Ferrara, Ferrara
[79] INFN - Sezione di Roma Tre and University of Roma Tre, Rome
[80] INFN Bologna, Bologna
[81] INFN Milano - LASA, Segrate Milano
[82] INFN, Sezione di Genova, Genova
[83] INFN-Laboratori Nazionali di Frascati, Frascati
[84] INFN-Sezione Di Torino, Torino
[85] INFN-Sezione di Roma Tre, Rome
[86] INFN-Università di Milano, Milano
[87] INFN-University of Torino, Torino
[88] Inner Mongolia University, Hohhot, Inner Mongolia
[89] Institut de Física d'Altes Energies (IFAE), Barcelona

[90] Institute for Nuclear Research of the Russian Academy of Sciences, Moscow
[91] Institute of High Energy Physics, Chinese Academy of Sciences, Beijing
[92] Institute of Modern Physics, Chinese Academy of Sciences, Lanzhou, Gansu
[93] Institute of Physics, Academia Sinica, Taipei
[94] Institute of Physics, Henan Academy of Sciences, Zhengzhou, Henan
[95] Institute of Power Engineering of National Academy of Sciences of Belarus, Minsk
[96] Institute of Theoretical Physics, Chinese Academy of Sciences, Beijing
[97] Institute Rudjer Boskovic, Zagreb
[98] Instituto de Alta Investigación, Universidad de Tarapacá, Arica
[99] Instituto de Física Corpuscular (IFIC), CSIC - Universidad de Valencia
[100] Instituto de Fisica Teorica UAM-CSIC, Madrid
[101] International Islamic University, Islamabad
[102] Iowa State University, Department of Physics and Astronomy, Ames
[103] IP2I de Lyon, Lyon
[104] IRFU, CEA, Universite Paris-Saclay, Paris
[105] iThemba Laboratory for Accelerator, Based Sciences, Somerset West
[106] Jan Kochanowski University of Kielce, Kielce
[107] Jazan University, Jazan
[108] Jiangsu Normal University, Department of Physics, Xuzhou
[109] Jilin University, Jilin, Changchun
[110] Jinan University, Guangzhou, Guangdong
[111] Jinggangshan University, Ji'an, Jiangxi
[112] Johannes Gutenberg University Mainz, Mainz
[113] Joint Institute for Nuclear Research, Dubna
[114] Karlsruhe Institute of Technology, Institute for Theoretical Physics, Karlsruhe
[115] Kavli Institute for the Physics and Mathematics of the Universe (KIPMU), University of Tokyo, Kashiwa, Chiba
[116] Korea University, Department of Physics, Seoul
[117] Kutaisi International University, Kutaisi
[118] Kyungpook National University, Daegu
[119] Laboratoire AstroParticule et Cosmologie, CNRS/IN2P3, Paris
[120] Laboratoire de Physique de Clermont (LPC), Aubière
[121] Laboratoire de Physique des 2 infinis Irène Joliot-Curie – IJCLab, Orsay
[122] Laboratoire de Physique Théorique et Hautes Energies (LPTHE), UMR 7589, Sorbonne Université et CNRS, Paris
[123] Laboratoire Leprince-Ringuet (LLR), CNRS, École Polytechnique, Institut Polytechnique de Paris, Palaiseau
[124] Laboratori Nazionali di Frascati, Frascati
[125] Lancaster University, Lancaster
[126] Lanzhou University, Lanzhou, Gansu
[127] Lawrence Berkeley National Laboratory, Berkeley, CA
[128] Liaoning Normal University, Dalian, Liaoning
[129] Lund university, Department of Physics, Lund
[130] M.V. Lomonosov Moscow State University, Skobeltsyn Institute of Nuclear Physics (SINP MSU), Moscow
[131] Max Planck Institute for Physics, Munich
[132] Monash University, Melbourne
[133] Nanjing Normal University, Nanjing, Jiangsu

[134] Nanjing University of Science and Technology, Nanjing, Jiangsu
[135] Nanjing University, Nanjing, Jiangsu
[136] Nankai University, Tianjin
[137] National Central University, Taoyuan
[138] National Centre for Nuclear Research, Warsaw
[139] National Institute for Subatomic Physics (NIKHEF), Amsterdam
[140] National Institute of Chemical Physics and Biophysics, Tallinn
[141] National Taiwan University, Department of Physics, Taipei
[142] National Tsing Hua University, Hsinchu
[143] National University of Defense Technology, Changsha, Hunan
[144] Naval University of Engineering, Wuhan, Hubei
[145] New York University, New York, NY
[146] Niels Bohr institute, Copenhagen
[147] Northwest Normal University, Lanzhou, Gansu
[148] Northwest Rare Metal Research Institute, Shizuishan, Ningxia
[149] Northwestern Polytechnical University, Xi'an, Shaanxi
[150] Novosibirsk State University, Novosibirsk
[151] Nuclear Industry Huzhou Survey Planning & Design Institute Co., Ltd. Huzhou, Hunan
[152] OMEGA CNRS/IN2P3, Palaiseau
[153] Pakistan Institute of Nuclear Science and Technology (PINSTECH), Islamabad
[154] Paris Observatory, Paris
[155] Peking University, Beijing
[156] PowerChina Huadong Engineering Corporation Limited, Hangzhou, Zhejiang
[157] PowerChina Zhongnan Engineering Corporation Limited, Changsha, Hunan
[158] Presidency University, Department of Physics, Kolkata
[159] Purdue University, West Lafayette, IN
[160] Qingdao Academy, Qingdao, Shandong
[161] Qufu Normal University, Qufu, Shandong
[162] Rutgers, The State University of New Jersey, New Brunswick, NJ
[163] Science &Technology Facilities Council (STFC/RAL), Oxfordshire
[164] Seoul National University, Seoul
[165] Shaanxi Normal University, Xi'an, Shaanxi
[166] Shaanxi University of Technology, Hanzhong, Zhejiang
[167] Shandong University, Jinan, Shandong
[168] Shanghai Institute of Ceramics, Chinese Academy of Sciences, Shanghai
[169] Shanghai Institute of Optics and Fine Mechanics, Chinese Academy of Sciences, Shanghai
[170] Shanghai Jiao Tong University, Shanghai
[171] Shanxi University, Taiyuan, Shanxi
[172] Shenzhen Technology University, Shenzhen, Guangdong
[173] Sichuan University, Chengdu, Sichuan
[174] SLAC National Accelerator Laboratory, Menlo Park, CA
[175] Soochow University, Suzhou, Jiangsu
[176] South China Normal University, Guangzhou, Guangdong
[177] Southeast University, Nanjing, Jiangsu
[178] Southern Methodist University, Department of Physics, Dallas
[179] Southwest Jiaotong University, Chengdu, Sichuan

[180] SRM Institute of Science and Technology, Department of Physics and Nanotechnology, Kattankulathur
[181] STFC Daresbury Laboratory & Cockcroft Institute, Daresbury
[182] Stony Brook University, Stonybrook, NY
[183] Sun Yat-Sen University, Guangzhou, Guangdong
[184] SungKyunKwan University, Suwon
[185] Suranaree University of Technology, Nakhon Ratchasima
[186] Syracuse University, Syracuse, NY
[187] Tel Aviv University, Tel Aviv
[188] The Chinese University of Hong Kong, Hong Kong
[189] The Chinese University of Hong Kong-Shenzhen, School of Science and Engineering, Shenzhen, Guangdong
[190] The Hong Kong University of Science and Technology, Hong Kong
[191] The University of Hong Kong, Hong Kong
[192] The University of Melbourne, Melbourne
[193] The University of New South Wales, Sydney
[194] The University of Wuppertal, Wuppertal
[195] Tianjin University, Tianjin
[196] Tibet University, Lhasa, Tibet
[197] Tsinghua University, Beijing
[198] Twofish Enterprises (Asia) Limited, Hong Kong
[199] UKRI-STFC Daresbury Laboratory, Warrington
[200] Universidad Autónoma de Zacatecas, Zacatecas
[201] Universidad Complutense de Madrid, Madrid
[202] Universidad de Oviedo and Instituto de Ciencias y Tecnologías Espaciales de Asturias (ICTEA), Oviedo
[203] Universidade Estadual de Londrina, Londrina
[204] Università dell'Insubria, Como
[205] Universitat de València, IFIC,Departament de Física Teòrica, Paterna
[206] Université Abdelmalek Essaadi, Faculty of sciences and techniques, Tangier
[207] Université Catholique de Louvain (UC Louvain), Belgium
[208] University and INFN Bari, Bari
[209] University of Adelaide, Adelaide
[210] University of Arizona, Tucson, AZ
[211] University of Basel, Basel
[212] University of Bonn, Physics Institute, Bonn
[213] University of Bristol, Bristol
[214] University of Cagliari and INFN, Cagliari
[215] University of California Santa Cruz, Santa Cruz, CA
[216] University of Chicago, Chicago, IL
[217] University of Chinese Academy of Sciences, Beijing
[218] University of Edinburgh, Edinburgh
[219] University of Education Lahore, Lahore
[220] University of Electronic Science and Technology of China, Chengdu, Sichuan
[221] University of Geneva, Geneva
[222] University of Hyderabad, Hyderabad
[223] University of Illinois, Urbana, IL
[224] University of Jinan, School of Physics and technology, Jinan, Shandong

[225] University of Jyvaskyla, Jyvaskyla
[226] University of Liverpool, Liverpool
[227] University of Lyon1, Lyon
[228] University of Michigan, College of literature, science and the arts, Ann Arbor, MI
[229] University of Milano, Milano
[230] University of Minnesota, Minneapolis, MN
[231] University of Mississippi, University, MS
[232] University of Molise and INFN Roma Tor Vergata, Campobasso
[233] University of New Mexico, Albuquerque, NM
[234] University of Oxford, Oxford
[235] University of Pittsburgh, Pittsburgh, PA
[236] University of Roma Tre, Rome
[237] University of Rome la Sapienza - Roma, Rome
[238] University of Rome Tor Vergata and INFN, Rome
[239] University of Science and Technology of China, Hefei, Anhui
[240] University of Science and Technology of Mazandaran, Department of Physics, Behshahr
[241] University of Sheffield, Sheffield
[242] University of Silesia, Katowice
[243] University of South China, Hengyang, Hunan
[244] University of Southampton, Southampton
[245] University of Texas at Dallas, Richardson, TX
[246] University of the Punjab, Centre for High Energy Physics, Lahore
[247] University of the Witwatersrand, Johannesburg
[248] University of Trieste, Department of Physics, Trieste
[249] University of Turin and INFN sez. di Torino, Turin
[250] University of Virginia, Charlottesville, VA
[251] University of Warwick, Coventry
[252] University of Wisconsin-Madison, Madison, WI
[253] University of Zurich, Zurich
[254] University of Muenster, Muenster
[255] University of Wisconsin–Madison, Department of Physics, Madison
[256] Vanderbilt University, Nashville, TN
[257] Vinca Institute of Nuclear Sciences, University of Belgrade, Belgrade
[258] Virginia Tech, Blacksburg, VA
[259] Vrije Universiteit Brussel (VUB), Brussels, Belgium
[260] Washington University, St. Louis, MO
[261] Wuhan Second Ship Design and Research Institute, Wuhan, Hubei
[262] Wuhan University of Technology, Wuhan, Hubei
[263] Wuhan University, Wuhan, Hubei
[264] Wuxi Toly Electric Works Co., Ltd., Wuxi, Jiangsu
[265] Xi‘an Superconducting Magnet Technology Co., Ltd, Xi'an, Shaanxi
[266] Xi’an Jiaotong University, Xi'an, Shaanxi
[267] Xi’an Technological University, Xi'an, Shanxi
[268] Xi'an Jiaotong-Liverpool University, Suzhou, Jiangsu
[269] Yantai University, Yantai, Shandong
[270] Yellow River Engineering Consulting Co., Ltd., Zhengzhou, Henan
[271] Yerevan Physics Institute, Institute for Theoretical Physics and Modeling, Yerevan

[272] Yonsei university, Seoul
[273] Yunnan University, Kunming, Yunnan
[274] Zhangjiang Laboratory, Shanghai
[275] Zhejiang Institute of Geosciences, Hangzhou, Zhejiang
[276] Zhejiang Lab, Hangzhou, Zhejiang
[277] Zhejiang University, Hangzhou, Zhejiang
[278] Zhengzhou University, Zhengzhou, Henan
[279] cepc-office@ihep.ac.cn

# Acknowledgements

The completion of the CEPC Accelerator Technical Design Report (TDR) owes its success to the diligent efforts of the CEPC accelerator research team, spearheaded by the Institute of High Energy Physics (IHEP) of the Chinese Academy of Sciences (CAS). Collaborations with both domestic and international institutes played pivotal roles by offering invaluable advice and support, contributing significantly to the TDR's creation.

This study received unwavering support from diverse funding sources, including the National Key Program for S&T Research and Development of the Ministry of Science and Technology (MOST), Yifang Wang's Science Studio of the Ten Thousand Talents Project, the CAS Key Foreign Cooperation Grant, the National Natural Science Foundation of China (NSFC), Beijing Municipal Science & Technology Commission, the CAS Focused Science Grant, the IHEP Innovation Grant, the CAS Lead Special Training Program, the CAS Center for Excellence in Particle Physics, the CAS International Partnership Program, and the CAS/SAFEA International Partnership Program for Creative Research Teams.

Over the past five years of prototypes R&D and physics design optimization, the CEPC International Accelerator Review Committee (IARC) and the International Advisory Committee (IAC) have been indispensable. Their generous support and insightful suggestions have significantly shaped the CEPC Accelerator TDR. Additionally, several ad hoc committees established during the TDR review process – the International Technical Review Committee, the Domestic Civil Engineering Cost Review Committee, the International Civil Engineering Cost Review Subpanel, and the International Cost Review Committee – played crucial roles in finalizing the TDR. This document stands as a testament to the collaborative efforts of hundreds of scientists and engineers from the host institute, IHEP, along with numerous domestic and international institutes.

# Contents

# Executive Summary

Following the landmark discovery of the Higgs boson in 2012 at CERN's Large Hadron Collider (LHC), the global High Energy Physics (HEP) community has reached a consensus on the importance of an $e^+e^-$ Higgs factory as the next future collider. This consensus is reflected in the strategic plans of various regions and countries worldwide.

In Europe, the 2020 Update of the European Strategy for Particle Physics concluded that an electron-positron Higgs factory is the highest priority for the next collider. Similarly, in the United States, using Higgs as a tool for discovery is one of the five science drivers identified by the P5 in 2014. A recent 2023 report from P5 recommended for the US to actively engage in feasibility and design studies for an off-shore Higgs factory. In Japan, the International Linear Collider (ILC) has adapted its baseline design to function as a dedicated Higgs factory by reducing the center-of-mass energy from 500 GeV to 250 GeV. Meanwhile, in China, the Circular Electron Positron Collider (CEPC) has been identified as the top future particle accelerator in the planning study conducted by the Chinese Academy of Sciences (CAS).

The CEPC was initially proposed during an ICFA workshop titled "Accelerators for a Higgs Factory: Linear vs Circular" in November 2012 at Fermilab, among many other proposals worldwide. A distinctive aspect of the CEPC proposal was its emphasis on the potential follow-up of a proton collider (SPPC) to be situated in the same tunnel, extending the energy frontier beyond the LHC. This visionary idea gained widespread recognition for charting a new course in high-energy physics. The CEPC-SPPC kick-off meeting convened in September 2013. The release of the Preliminary Conceptual Design Report (Pre-CDR or the White Report) in 2015 and the subsequent Conceptual Design Report (CDR or the Blue Report) in 2018 marked significant milestones in its development. This Technical Design Report (TDR or the Green Report) presents the comprehensive work carried out by hundreds of scientists and engineers involved in advancing this ambitious project.

The TDR confirms that the CEPC is fully prepared for implementation, with notable progress made in the physical design since the CDR. The collider's energy scale ranges from the Z-pole at 91 GeV to $W^+W^-$ at 160 GeV, ZH at 240 GeV, and up to $t\bar{t}$ at 360 GeV. The achieved luminosities at these energy scales are comparable to those of other Higgs factories, including the FCC-ee. Extensive research and development efforts have been completed for the technical systems of the CEPC, with all hardware prototypes meeting or exceeding design requirements. These accomplishments demonstrate the readiness of the CEPC for the construction phase, expected to begin around 2027-2028 during China's 15th Five-Year Plan, pending government approval.

The CEPC is a large-scale scientific project comprising four accelerators: a 30 GeV Linac, a 1.1 GeV Damping Ring, a Booster with an energy up to 180 GeV, and a Collider operating at four different energy modes (Z, W, H, and $t\bar{t}$). These machines are connected by ten transport lines. While the Linac and Damping Ring are constructed on the surface, the Booster and Collider are situated in an underground ring tunnel with a circumference of 100 km, which also reserves space for a future Super Proton Proton Collider (SPPC).

The Collider features a double-ring structure, with electron and positron beams circulating in opposite directions within separate beam pipes. They collide at two interaction points (IPs) where large detectors are being designed. The CEPC Booster, positioned atop the Collider in the same tunnel, functions as a synchrotron featuring a 30 GeV injection energy and an extraction energy equal to the beam collision energy. To maintain constant luminosity, top-up injection will be employed. The 30 GeV Linac, a 1.8 km injector to the Booster, accelerates both electrons and positrons using S-band and C-band RF systems. A 1.1 GeV Damping Ring reduces positron emittance for the Booster.

The tunnel, primarily consisting of hard rock, provides a stable foundation and allows for accommodating the future SPPC without removing the CEPC collider ring. This opens up exciting possibilities for ep and e-ion physics in addition to the CEPC's ee physics and the SPPC's pp and ion-ion physics. Furthermore, the Collider can operate as a synchrotron radiation (SR) light source, extending the usable SR spectrum into an unprecedented energy and brightness range. The design includes two gamma-ray beamlines, and the Linac can be transformed into a high-energy x-ray free electron laser (FEL) by adding an undulator.

Detailed investigations have been carried out for three among several potential sites – Qinhuangdao, Changsha, and Huzhou – all of which meet the general site requirements.

The baseline design of the CEPC involves three collision modes (H, Z, and W) with a synchrotron radiation power of 30 MW per beam. The operation plan follows a "10-2-1" scheme, dedicating 10 years as a Higgs factory, 2 years as a Z factory, and 1 year as a W factory. Two upgrade plans are also considered, aiming to increase beam power to 50 MW and energy for $t\bar{t}$ collision, allowing for an additional 5 years of operation.

Construction of the CEPC is expected to commence around 2027-2028 and is estimated to take approximately 8 years. Following commissioning, the physics program can begin as early as the mid-2030s. A detailed construction cost estimate, based on a Work Breakdown Structure (WBS), has been conducted, encompassing the accelerator complex, two detectors, gamma-ray beamlines, conventional facilities, commissioning, and contingency. The total cost is estimated at approximately RMB 36.4 billion (USD 5.2 billion). Various funding scenarios are under consideration, with contributions expected from the central government, local government, and the international community.

The CEPC represents a crucial component of the global strategic plan for High Energy Physics, fostering collaboration among scientists worldwide and driving advancements in our understanding of the fundamental nature of matter, energy, and the universe.

# 1 Introduction

## 1.1 A Brief History of Particle Colliders

The idea of colliding two particle beams to fully exploit the energy of accelerated particles was first proposed by Rolf Widerøe in 1943 [1], and a patent was awarded in 1953. Serious efforts to design and build storage ring-based colliders began in 1956 [2,3]. The first three colliders – AdA in Italy, CBX in the US, and VEP-1 in the then Soviet Union – came into operation in the 1960s. A number of other colliders followed [4,5]. They define the energy frontier in particle physics and are often termed "*discovery machines*" as many of the constituents in the Standard Model were discovered in colliders including the Higgs boson.

Unlike in a fixed-target experiment, in which the center of mass energy $E_{\mathrm{cm}}$ increases as the square root of the beam energy, in a collider $E_{\mathrm{cm}}$ increases linearly with the beam energy. This is a main advantage of colliders.

There are two categories of colliders: lepton colliders ($e^-e^-$, $e^+e^-$) and hadron colliders (pp, $p\bar{p}$, $e^-p$, $e^-$-ion, p-ion, ion-ion). Among the 30-plus colliders that have been built in the past six decades, a majority (80%) are $e^+e^-$ colliders (Table 1.1); the other 20% are hadron colliders (Table 1.2), including one electron-ion collider (EIC) that is currently under construction. There are also other types of colliders, e.g., $\mu^+\mu^-$ and $\gamma\gamma$, but those are still in the R&D stage and have not yet been built due to technical difficulties. While the basic beam physics is the same for all types of colliders, each type has its own specific challenging beam physics issues and employs different accelerator technologies.

The two most important parameters of a collider are the center-of-mass energy $E_{\mathrm{cm}}$ and the luminosity $L$. The latter determines the rate of collision events.

Table 1.1 lists two dozen lepton colliders that have been constructed and operated, along with their location, maximum beam energy and luminosity achieved, and operation period as a collider [6,7]. Table 1.2 lists seven hadron colliders.

A collider can be linear or circular (storage ring). Most colliders are storage rings: either single ring (for particle-antiparticle collisions only) or double ring (for collisions of any two types of particles). The highest energy of a circular $e^+e^-$ collider is limited by synchrotron radiation, which increases as the fourth power of the beam energy. To avoid this problem, linear $e^+e^-$ collider was proposed [8]. But linear colliders are technically more challenging. To date only one (SLC) has been built and operated.

Presently the LHC has recorded the highest collision energy (13.6 TeV), and the SuperKEKB the highest luminosity ($4.65 \times 10^{34}$ cm$^{-2}$s$^{-1}$).

**Table 1.1:** Lepton Colliders (all $e^+e^-$ unless marked as $e^-e^-$) [a] DR: Double ring, SR: Single ring, LC: Linear collider. [b] Highest achieved. [c] Princeton-Stanford Colliding Beam Experiment. [d] Collisions achieved when operated in Orsay. [e] Using a detector with no solenoid field. [f] Design goal.

| **Location** | **Name** (type[a]) | **Beam Energy**[b] (GeV) | **Luminosity**[b] ($cm^{-2}s^{-1}$) | **Operation Period** |
|---|---|---|---|---|
| Stanford/SLAC, USA | CBX[c] ($e^-e^-$ DR) | 0.5 | $2 \times 10^{28}$ | 1963-1968 |
| | SPEAR (SR) | 4 | $1.2 \times 10^{31}$ | 1972-1988 |
| | PEP (SR) | 15 | $6 \times 10^{31}$ | 1980-1990 |
| | SLC (LC) | 49 | $2.5 \times 10^{30}$ | 1989-1998 |
| | PEP-II (DR) | 9 ($e^-$) 3.1 ($e^+$) | $1.2 \times 10^{34}$ | 1998-2008 |
| Frascati, Italy | AdA (SR) | 0.25 | $5 \times 10^{25}$ [d] | 1961-1964 |
| | ADONE (SR) | 1.5 | $6 \times 10^{29}$ | 1969-1993 |
| | DAΦNE (SR) | 0.51 | $2.4 \times 10^{32}$<br>$4.5 \times 10^{32}$ [e] | 1999-present |
| BINP, Russia | VEP-1 ($e^-e^-$ DR) | 0.16 | $5 \times 10^{27}$ | 1964-1968 |
| | VEPP-2 (SR) | 0.67 | $4 \times 10^{28}$ | 1966-1970 |
| | VEPP-2M (SR) | 0.7 | $5 \times 10^{30}$ (@0.511) | 1974-2000 |
| | VEPP-3 (SR) | 1.55 | $2 \times 10^{27}$ | 1974-1975 |
| | VEPP-4M (SR) | 6 | $2 \times 10^{31}$ | 1984-present |
| | VEPP-2000 (SR) | 1 | $5 \times 10^{31}$ | 2010-present |
| Cambridge, USA | CEA Bypass (SR) | 3 | $8 \times 10^{27}$ | 1971-1973 |
| Orsay, France | ACO (SR) | 0.54 | $1 \times 10^{29}$ | 1965-1975 |
| | DCI (DR) | 1.8 | $1.7 \times 10^{30}$ | 1977-1985 |
| DESY, Germany | DORIS (SR) | 5.6 | $3.3 \times 10^{31}$ | 1973-1993 |
| | PETRA (SR) | 23.4 | $2.4 \times 10^{31}$ (@17.5) | 1978-1986 |
| CERN, Europe | LEP (SR) | 104.5 | $1 \times 10^{32}$ | 1989-2000 |
| Cornell, USA | CESR (SR) | 5.5 | $1.3 \times 10^{33}$ | 1979-2008 |
| KEK, Japan | TRISTAN (SR) | 32 | $4.1 \times 10^{31}$ | 1986-1995 |
| | KEKB (DR) | 8 ($e^-$) 3.5 ($e^+$) | $2.1 \times 10^{34}$ | 1998-2010 |
| | SuperKEKB (DR) | 7 ($e^-$) 4 ($e^+$) | $8 \times 10^{35}$ [f] | 2016-present |
| IHEP, China | BEPC (SR) | 2.4 | $1 \times 10^{31}$ (@1.84) | 1988-2004 |
| | BEPC II (DR) | 2.47 | $1 \times 10^{33}$ (@1.89) | 2009-present |

**Table 1.2:** Hadron Colliders [a] DR: Double ring, SR: Single ring. [b] Highest achieved. [c] Energy per nucleon. [d] Design goal.

| **Location** | **Name** (type[a]) | **Beam Energy**[b] (GeV) | **Luminosity**[b] ($cm^{-2}s^{-1}$) | **Operation Period** |
|---|---|---|---|---|
| CERN, Europe | ISR (pp DR) | 31.4 | $1.4 \times 10^{32}$ | 1971-1984 |
| | S$p\bar{p}$S ($p\bar{p}$ SR) | 315 | $6 \times 10^{30}$ | 1981-1991 |
| | LHC (pp, ii, pi DR) | 6800 (p)<br>2510/n (Pb) [c]<br>6500 (p) 2560/n (Pb) [c] | $2.1 \times 10^{34}$<br>$6.1 \times 10^{27}$<br>$9 \times 10^{29}$ | 2009-present |
| Fermilab, USA | Tevatron ($p\bar{p}$ SR) | 980 | $4.3 \times 10^{32}$ | 1987-2011 |
| DESY, Germany | HERA (ep DR) | 27.5 ($e^-$) 920 (p) | $5.3 \times 10^{31}$ | 1992-2007 |
| Brookhaven, USA | RHIC (pp, ii DR) | 250 (p)<br>100/n (Au) [c] | $2.5 \times 10^{32}$<br>$1.6 \times 10^{28}$ | 2000-present |
| | EIC (ep, ei DR) [d] | 18 ($e^-$) 275 (p)<br>18 ($e^-$) 110/n (Au) [c] | $1 \times 10^{34}$<br>$3 \times 10^{30}$ | (under construction) |

Several future colliders are currently under consideration. The highest priority on the list is a Higgs factory, which has $E_{cm}$ ~ 240 GeV, $L$ ~ $10^{34}$ $cm^{-2}s^{-1}$ and would be capable of generating millions of Higgs particles for physics study. It can be either linear such as ILC [9] and CLIC [10], or circular such as FCC-ee [11] and CEPC [12]. The ILC employs high gradient superconducting RF systems to accelerate electron and positron to collision energy, whereas the CLIC uses a novel technology, two-beam acceleration for that purpose. The design of the FCC-ee and the CEPC is similar: both would build a ~100 km round tunnel deep underground to accommodate an $e^+e^-$ collider. Later the same tunnel can be used for a future pp collider of $E_{cm}$ ~ 100 TeV, $L$ ~ $10^{35}$ $cm^{-2}s^{-1}$, e.g., FCC-hh [13] and SPPC [14]. A future hadron collider is not only more costly than a Higgs factory but also technically more challenging. Critical issues such as high field (16 Tesla or higher) superconducting magnets, synchrotron radiation problem in a cryogenic environment, and a sophisticated beam collimation system for quench protection, must be adequately addressed before the construction can begin.

A muon collider ($\mu^+\mu^-$) has several advantages over an $e^+e^-$ one, e.g., it has minimal synchrotron radiation or beamstrahlung problem, the two major limiting factors of $e^+e^-$ colliders. This is because the strength of these phenomena decreases as the fourth power of the particle mass, and the muon mass is ~207 times that of an electron. Therefore, a TeV muon collider ring is smaller than a GeV $e^+e^-$ collider and can be readily accommodated within the boundary of several exiting laboratories. The difficulty, however, is how to obtain small size intense muon beams required by a collider during muon's short lifetime (2.2 μs). Two approaches are currently under study: one is to strike a high-power (~MW) intense proton beam on a target (e.g., mercury) to generate intense muons and cool them; another is to use positrons to hit electrons on an internal target in a ring to generate small emittance muons and accumulate them [15].

A photon collider ($\gamma\gamma$) is based on inverse Compton scattering: when a low energy (~eV) photon hits a high energy (~MeV, GeV or TeV) electron, the photon will be back scattered and carry a portion of the electron energy. Two such photons can be made to collide generating new particles. For a long time, a photon collider was thought to be an add-on to a high-energy (~GeV) linear $e^+e^-$ collider [16]. Recently, however, a low-energy (~MeV) stand-alone photon collider has received considerable attention as it can be built quickly at a low cost and is able to explore valuable new physics [17].

## 1.2 Higgs Factory and CEPC

The Higgs boson occupies a unique and fundamental position among all fundamental particles. It endows all other particles with their mass, and the mass of the Higgs has a direct bearing on all the other particles that make up our reality. Although the Large Hadron Collider successfully discovered the Higgs boson, the generation of a vast number of other particles creates a high level of noise. To conduct a focused study of the Higgs boson within a relatively pristine environment, there is a pressing need to construct an electron-positron "Higgs factory."

Due to the Higgs boson's relatively low mass (125 GeV), the construction of an electron-positron collider with a circular design, known as the CEPC (Circular Electron Positron Collider), becomes feasible. Extensive research has indicated that a circumference of 100 km offers the best balance between cost and performance, minimizing the cost per particle produced for the intended objectives [18]. Furthermore, it allows for the future incorporation of a super proton-proton collider (SPPC) with a center-of-mass energy of 125 TeV within the same tunnel, alongside the CEPC.

Circular electron-positron colliders have a long history of development and operation, benefiting from well-established design theories, tools, and mature technologies. The accumulated experience in this field is extensive. The highest center-of-mass energy attained at CERN's collider LEP2 reached 209 GeV, while the highest luminosity was achieved at SuperKEKB with a value of $4.65 \times 10^{34}$ $cm^{-2}s^{-1}$.

The CEPC is ambitious in its pursuit of both high beam energy and high luminosity, which introduces new challenges in terms of physical design and the need for key technology research and development. Nonetheless, valuable lessons and insights can be drawn from the experience gained through previous machines, particularly LEP2, KEKB, and SuperKEKB.

The design of the CEPC has been the outcome of a decade-long research and development endeavor since its initial proposal in 2012. The core of the CEPC consists of the Collider and Booster rings, housed within a 100 km underground tunnel. The design incorporates mature technologies such as magnets, superconducting RF cavities, vacuum systems, cryogenic systems, instrumentation, and RF power sources, which have been thoroughly developed and proven.

Concurrently, extensive efforts have been dedicated to optimizing the physical design to achieve high luminosity. These endeavors have now reached a successful milestone with the publication of the Technical Design Report (TDR). The TDR demonstrates that the CEPC can attain the required luminosity to fulfill scientific objectives. Furthermore, it confirms that there are no major technological barriers that would hinder mass production, indicating that the project is technologically feasible.

Circular electron-positron colliders offer a significant advantage over linear colliders, particularly in terms of high luminosity, which refers to the rate of particle collisions achievable. The design of the CEPC targets a collision frequency of approximately 1.5 MHz for Higgs operation. To optimize cost, the CEPC utilizes high-Q superconducting radio-frequency cavities and high-efficiency klystrons. This approach allows the CEPC to achieve world-class levels of luminosity in Higgs, W, and Z operations.

Figure 1.1 illustrates a comparison of the CEPC's luminosity with other state-of-the-art electron-positron colliders proposed worldwide. The CEPC's design luminosity is comparable to that of the FCC-ee in the Higgs, Z, and W energy regions.

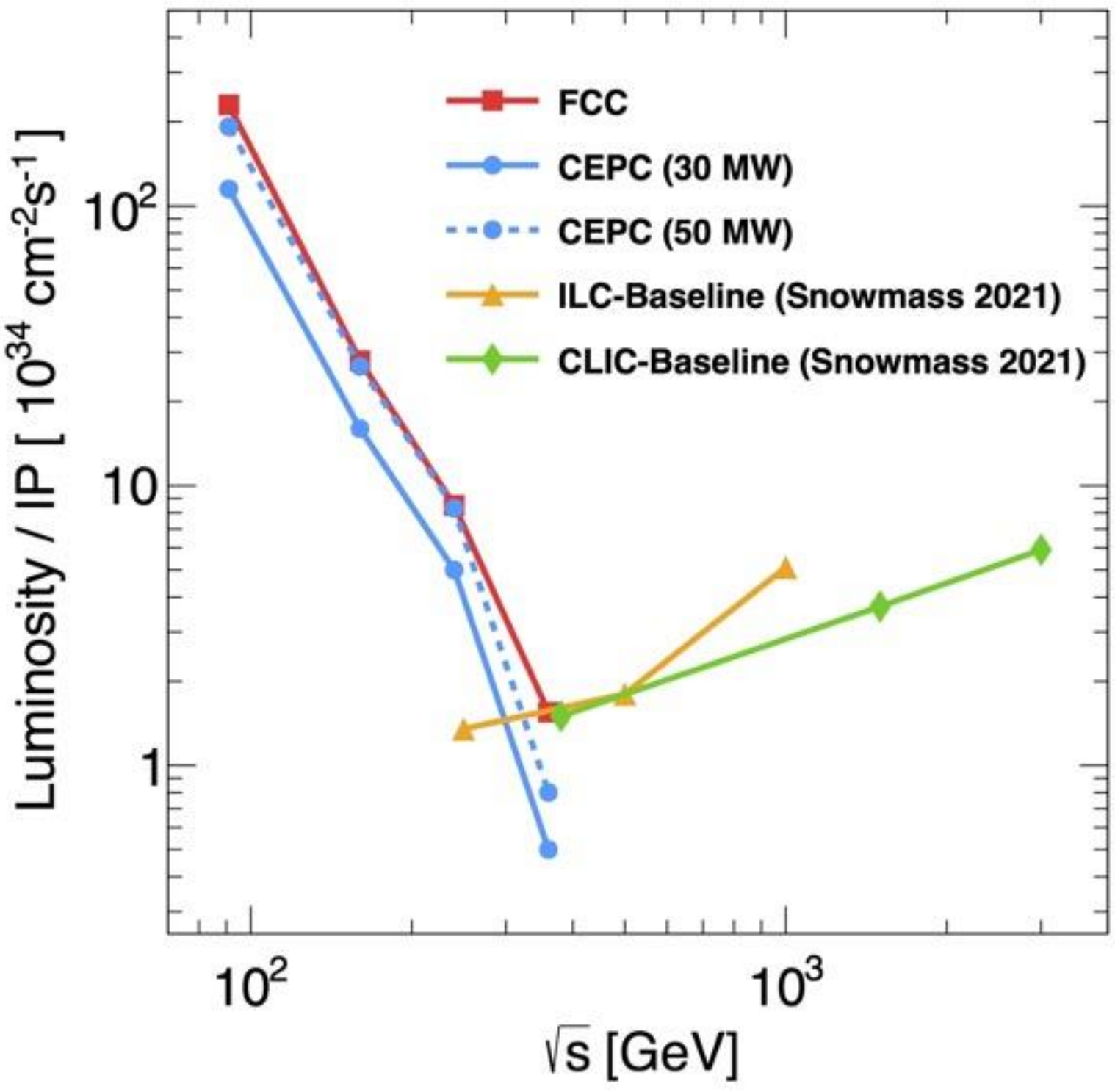

**Figure 1.1:** Comparison of the CEPC luminosity and those of other electron-positron colliders under consideration in the world HEP community.

The accelerator design has undergone extensive optimization to maximize its scientific potential while taking into account factors such as cost, technical feasibility, and operational efficiency. The resulting design successfully fulfills the required luminosity specifications, and the key design objectives are outlined in Tables 1.3 and 1.4.

The current baseline configuration of the CEPC features two interaction points (IPs) and a single beam radiation power of 30 MW (as indicated in Table 1.3). It is important to note that this baseline design can be enhanced to a high-power scheme of 50 MW through international collaboration (as presented in Table 1.4).

In the baseline design, the CEPC operates at three distinct center-of-mass energy levels to target specific particles and their associated phenomena. These energy levels are as follows: 240 GeV for Higgs boson studies, 160 GeV for W boson investigations, and 91 GeV for Z boson studies.

Additionally, to augment the research capabilities and delve deeper into the understanding of the top quark, an optional plan has been proposed to increase the center-of-mass energy to 360 GeV. This enhancement in energy levels presents researchers with a valuable opportunity to carry out more extensive investigations into the characteristics and behavior of the top quark, thereby pushing the boundaries of knowledge in this particular field of study.

**Table 1.3:** Primary CEPC design objectives (@ 30 MW)

| Parameter | Operation mode | | | |
|---|---|---|---|---|
| | H | Z | W | $t\bar{t}$ |
| Colliding particles | $e^{+}, e^{-}$ | | | |
| Center-of-mass energy (GeV) | 240 | 91 | 160 | 360 |
| Luminosity ($10^{34}$ cm$^{-2}$s$^{-1}$) | 5 | 115 | 16 | 0.5 |
| No. of interaction points | 2 | | | |

**Table 1.4:** Primary CEPC design objectives (@ 50 MW)

| Parameter | Operation mode | | | |
|---|---|---|---|---|
| | H | Z | W | $t\bar{t}$ |
| Colliding particles | $e^{+}, e^{-}$ | | | |
| Center-of-mass energy (GeV) | 240 | 91 | 160 | 360 |
| Luminosity ($10^{34}$ cm$^{-2}$s$^{-1}$) | 8.3 | 192 | 27 | 0.8 |
| No. of interaction points | 2 | | | |

## 1.3 CEPC and FCC-ee

The CEPC and FCC-ee share numerous similarities, such as being hosted in circular deep underground tunnels with large circumferences, utilizing separate rings for electron and positron beams, employing superconducting RF technology for particle acceleration, and incorporating a full energy booster and top-up injection. The luminosity at the H, W, and Z energy levels is comparable for both colliders. However, there are several differences between the two machines:

1. Schedule: The FCC-ee plans to commence operations in the late 2040s, whereas the CEPC aims to begin experiments and data collection in the middle 2030s.
2. Cost: Due to variations in purchasing power and labor costs, the construction cost of the CEPC is expected to be half that of the FCC-ee or even less.
3. Location: While the CEPC site has yet to be determined, it offers flexibility as it can be built in a green field at any chosen location. The FCC-ee, on the other hand, will be situated in the vicinity of the Geneva area and will have less flexibility in terms of location.
4. Operation plan: The CEPC plans to operate for 10 years as a Higgs factory, followed by 2 years as a Z factory, 1 year as a W factory, and an optional 5 years for $t\bar{t}$. In contrast, the FCC-ee will initially operate in Z mode for 4 years, then spend 1-2 years in W mode, 3 years in Higgs mode, and 5 years in $t\bar{t}$ mode.
5. Tunnel size: The CEPC tunnel is longer (100 km compared to 91 km for the FCC-ee) and larger (6 m × 5 m gate-shape or Φ6.5 m round shape compared to Φ5.5 m for the FCC-ee). This enables both the CEPC and a future pp collider SPPC to be accommodated in the same tunnel.

6. Booster location: In the CEPC, the Booster is positioned on top of the Collider and suspended from the ceiling, whereas in the FCC-ee, the Booster is parallel to the Collider.
7. RF frequency choice: The CEPC utilizes 650 MHz for the Collider and 1.3 GHz for the Booster. The former is also the RF frequency chosen by PIP-II, and the latter is consistent with the frequencies used by XFEL, LCLS II, SHINE, and ILC. This alignment opens possibilities for collaboration on RF and klystron technology between the CEPC and accelerator projects in other countries. The FCC-ee, on the other hand, adopts 400 MHz and 800 MHz, which have been widely employed in various CERN accelerators for decades.
8. Injector to the Booster: The CEPC employs a 30 GeV linac, while the FCC-ee uses a 6 GeV linac and a 20 GeV pre-booster combination.
9. Number of interaction points (IPs): The CEPC has two IPs, whereas the FCC-ee allows for either two or four IPs.
10. Power consumption and carbon footprint: The power consumption of the two facilities is comparable, with the disparity arising from varying assumptions regarding klystron and operational efficiency. However, a significant distinction exists in the carbon intensity of electricity production (measured by $CO_2$ emissions per generated MWh) between China and Switzerland/France due to China's prevalent use of coal as an energy source. Fortunately, China has taken a proactive stance to achieve a carbon peak by 2030 and carbon neutrality by 2060. Consequently, when the CEPC commences operations in the mid-2030s, it is anticipated that its carbon footprint will be considerably diminished.

## 1.4 Structure of the Report

This report consists of 12 chapters and 11 appendices.

Chapter 1 is an introduction,

Chapter 2 is the overview of the CEPC layout and the expected performance.

Chapter 3 presents the operation plan for Higgs, W, Z. The baseline power is 30 MW, and the 50 MW upgrade plan is introduced. The upgrade plan for the $t\bar{t}$ operation is also introduced. It also discussed the scenario to the future SPPC and various collision schemes such as electron-positron collider.

Chapter 4-7 comprise a major part of the report, providing detailed descriptions of the accelerator complex design and key technology research and development.

Chapter 4 specifically focuses on the heart of CEPC, the Collider, and covers all aspects of its physical design. This includes a discussion on lattice optimization for the arc-region, straight section, and interaction region (IR) in great detail. Dynamic aperture calculation with all kinds of errors is introduced, along with their mitigation measures. Important issues such as beam-beam effects and collective instabilities are also discussed. The comprehensive studies for the Machine-Detector-Interface (MDI) region constitute a major part of this chapter, with radiation background due to synchrotron radiation, beam losses, etc. calculated and corresponding collimator and shielding schemes analyzed. Along with theoretical studies, the chapter describes the fruitful research and development results of key technologies, such as superconducting RF cavities, high-efficiency klystrons, different types of dual-aperture magnets, key components for the vacuum system, electro-static deflector, magnet power source, and more.

Chapter 5 provides an overview of the studies conducted for the Booster. The chapter begins by introducing the lattice design and compares two schemes: the FODO and TME lattice. The decision to use the TME lattice is discussed, and the key technology breakthroughs, including the 1.3GHz Superconducting RF system and weak field dipoles, are presented in detail.

Chapter 6 focuses on the Linac injection process. The new baseline for injection energy is 30 GeV, which helps reduce the cost of the Booster dipole. The chapter demonstrates start-to-end simulations for both electron and positron beam lines and introduces the RF system, including pre-bunching modules, S-band accelerators, and C-band accelerators. The electron and positron source R&D status is also presented in this chapter, along with the Damping Ring design.

Chapter 7 discusses the communal facilities necessary for the successful operation of CEPC. Important sustaining systems, including the cryogenic system, survey and alignment, and machine protection system, are introduced.

Chapter 8 outlines the plans for upgrading CEPC to the super Proton-Proton collider, SPPC. This will have a center-of-mass energy of 125 TeV, and an injector chain, including a proton Linac and three synchrotrons, will increase the proton energy up to 2 TeV the injection energy. The crucial technology for SPPC will be superconducting high field magnets, and the pre-study results for the HTS magnet are presented.

Chapter 9 discusses the plans for the conventional facilities needed for CEPC, including the civil constructions, electricity power and transformer substation schemes, cooling water system, ventilation system, and other auxiliary systems required for building and maintaining the machine during its many years of operation.

Chapter 10 focuses on the important preparations prior to the CEPC construction, including environmental protection regulations and healthy and safety issues. It includes calculations indicating that the CEPC construction and operation will meet these regulations.

Chapter 11 discusses project planning, including project management, procurement and bidding regulations, financial models, and industrialization and mass production of components. This chapter also analyzes the social impact of the CEPC project. In addition to the significant technological advancements in accelerator design and cutting-edge technologies, the CEPC will have a significant impact on promoting frontier science and industrial development. The CEPC project is enormous and the city where it is located will attract thousands of top-level scientists, engineers, and supporting personnel, significantly boosting local economic and technological growth.

Chapter 12 presents essential aspects of project cost and implementation and operation schedules. A detailed estimate of the construction cost is based on a Work Breakdown Structure (WBS), and a comprehensive project timeline is presented.

Appendix 1 lists the detailed parameters for the accelerator, including the components in the Collider, the Booster, and the Linac. It also summarizes the parameters of the Damping Ring, the electron and positron sources, and the beam transport lines.

Appendix 2 provides a detailed list of all the components and their specifications.

Appendix 3 gives a breakdown of the electric power requirements and summarizes the power requirement.

Appendix 4 discusses the project risks and the corresponding mitigation measures.

Appendix 5 focuses on the application of CEPC in multidisciplinary research. With a beam energy of over 100 GeV, CEPC can generate intensive $\gamma$-rays with energies higher than 100 MeV. Moreover, the 30 GeV Linac can drive many scientific research projects in addition to serving as the CEPC injector, such as a free electron laser (FEL) and a photon collider.

Appendix 6 focuses on the frontier research for the plasma wakefield accelerator. Due to the prominent advantages of high acceleration gradient, the plasma accelerator is a research focus in recent years. The CEPC uses the plasma accelerator as a possible alternative to the Linac injector. However, the CEPC imposes more requirements, including the positron acceleration, cascaded acceleration with high energy transformation ratio, and large bunch charge. Therefore, the pre-studies of the plasma accelerator for the CEPC will promote revolutionary acceleration theory and technology development.

Appendix 7 briefly analyzes the potential possibilities of the CEPC for other collision modes, including e-p, e-A, and heavy ion collisions. The advantages and limitations of each mode are discussed, along with the potential physics goals that can be achieved.

Appendix 8 introduces the opportunities for polarization implementation in the CEPC. Different polarization schemes and their utilities are analyzed, and the potential physics benefits that can be achieved with polarized collisions are discussed.

Appendix 9 introduces the development status for the electronic database, named Deep-C, for the CEPC. The importance of data management in large-scale experiments is highlighted, and the design and functions of the Deep-C database are presented, including data storage, analysis, and sharing. The current status of the development and implementation of Deep-C is also discussed.

Appendix 10 presents an initial project construction schedule timeline, spanning a duration of 8 years.

Appendix 11 is a list of members from several advisory and review committees.

The CEPC project is a significant step forward for China in the field of particle physics. As the successor to the BEPC-II, the electron-positron circular collider currently operating at IHEP in Beijing, the CEPC represents a grand plan proposed, studied, and to be constructed by Chinese scientists in close collaboration with international partners. The consensus in the particle physics community is that an electron-positron collider is the ideal Higgs factory, and the CEPC will allow for research into the most important scientific questions in particle physics with unparalleled precision. Although the challenges are significant, the potential benefits are enormous. The CEPC project aligns with China's national priorities for basic science and will promote China's stature in the global particle physics research community.

The CEPC project has the potential to span several decades and educate generations of scientists and engineers. With its grand upgrade potential, it will be at the forefront of high-energy physics research for years to come. The collaboration of scientists, engineers, and theorists from around the world will conduct comprehensive research programs and explore the frontiers of science and technology. The CEPC will attract top-level talent and promote international cooperation, leading to a deeper understanding of the fundamental nature of the universe and taking mankind to new heights of knowledge.

# 2 Machine Layout and Performance

## 2.1 Machine Layout

The main component of the CEPC accelerator complex is the Collider ring, which has a circumference of 100 kilometers and is designed to optimize the performance-cost ratio in producing Higgs, Z, W, and $t\bar{t}$ particles. The long perimeter allows for potential future use of the CEPC tunnel for the Super Proton-Proton Collider (SPPC) if a breakthrough in high-field superconducting magnets is achieved in the next 20 years. The CEPC tunnel will be either a gate-shape, 6-meter wide and 5-meter high, or a round shape with a diameter of 6.5 meters, which can accommodate both CEPC and SPPC as shown in Fig. 2.1. Two excavation methods, TBM and drilling and blasting, will be considered. The SPPC ring will be placed at the outer side of the tunnel, occupying more space with superconducting high field magnets and cryogenic modules. The CEPC Booster and Collider rings will be located on the inner side of the tunnel, with the Booster hung from the ceiling and the Collider ring placed on the floor, separated by a 2.4-meter distance to avoid magnetic interference.

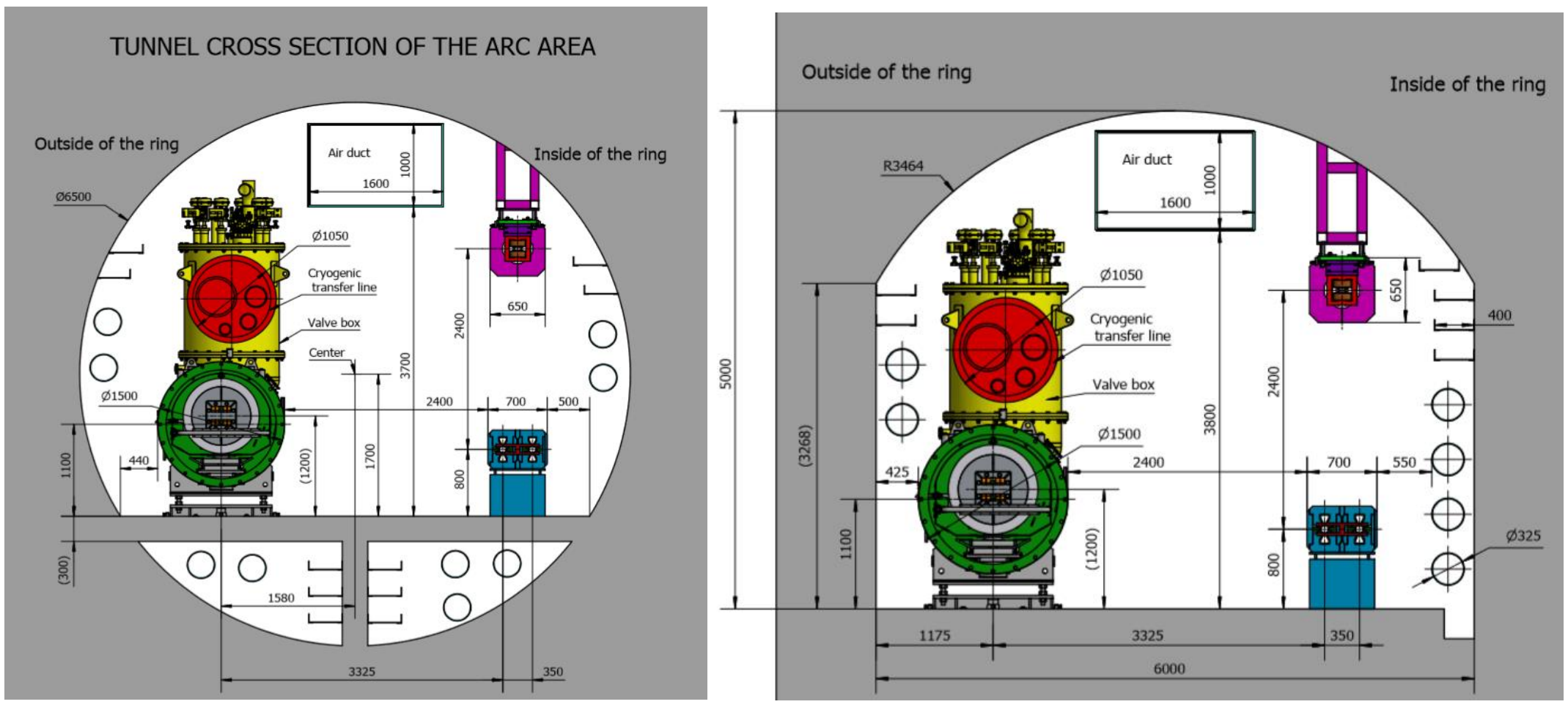

**Figure 2.1:** CEPC Tunnel cross section. Left: the round shape tunnel excavated with TBM; Right: the gate-shape tunnel exacavated with drilling and blasting.

The CEPC accelerator complex shown in Fig. 2.2 mainly consists of the Linac, Booster and Collider.

The Linac is built on the ground level. It raises the electron and positron beam energy up to 30 GeV. A damping ring is combined with the Linac to reduce the positron beam emittance.

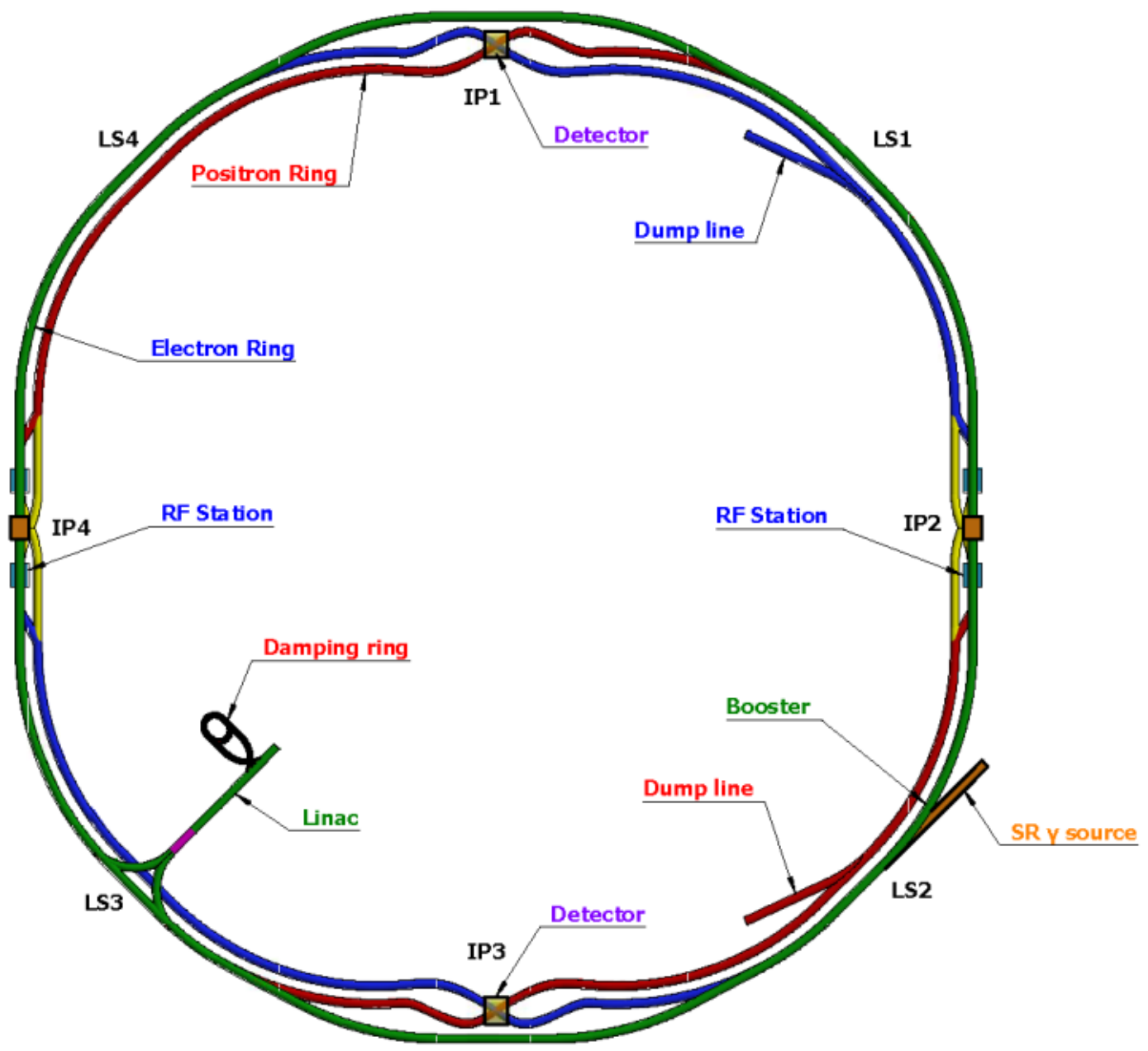

**Figurate 2.2:** Layout of the CEPC accelerator complex.

The Booster is located underground, approximately 100 meters below the surface. There are two transmission lines: one for the electron beam and one for the positron beam. These lines, which have a slope of 1:10, transport the beams from the surface down to the subterranean tunnel.

Four linear sections (LS1 – LS4) in the Collider and Booster are used for injection, extraction, and beam dump. Two sections use off-axis injection, which applies to Higgs, W, Z, and $t\bar{t}$ modes. The other two sections use on-axis injection, which is only used for the Higgs mode. This relaxed the requirement for the dynamic aperture of the collider.

Two long straight sections at IP2 and IP4 in the Booster are allocated for the 1.3 GHz superconducting RF stations, which use the same TESLA technology that is the foundation of the International Linear Collider. The beams can be accelerated to different energies, ranging from 30 GeV up to 180 GeV.

Another two long straight sections at IP1 and IP3 in the Booster are bypass beamlines in a separate tunnel, which is necessary for the Booster beam to avoid the CEPC detectors.

The Collider shares the same footprint with the Booster except at IP1 and IP3 as explained above. It has four linear sections for injection, extraction, and beam dump, two long straight sections (IP1 and IP3) for detectors, and two long straight sections (IP2 and IP4) for RF.

In the RF section, the 650 MHz superconducting RF (SRF) stations are used to compensate for the synchrotron radiation (SR) energy loss of the circulating beams. The SRF station is split into two parts in each of the two straight sections. During Higgs operation, when the SR energy loss is high, both parts of the SRF system are used, and the Collider turns into a partial-double ring configuration where each of the ringd is only half filled with bunches. This allows the shared-use mode, while still meeting the luminosity goals with optimal performance. However, during W and Z operations, where the energy loss is relatively low, the electron and positron beams each pass through only one of the two parts. This allows the Collider to operate in a full double-ring configuration, where the entire rings are filled with bunches, resulting in enhanced luminosity.

IP1 and IP3 in the Collider are the interaction regions (IRs) where electron and positron beams collide. In order to maximize luminosity, an advanced crab-waist scheme with a large Piwinski angle collision is implemented, with careful optimization of the lattice parameters for the quadrupoles and sextupoles. Figure 2.3 shows the layout of the central region of the IR, which includes superconducting final-focusing quadrupoles in cryogenic modules and a beryllium pipe where the electron-positron collision occurs. Surrounding the beryllium pipe and the final-focusing quadrupoles are various components of the detector and a superconducting solenoid magnet (not shown in the figure). This area is referred to as the Machine-Detector Interface (MDI), as both accelerator and detector components must share space. A more detailed layout drawing of the MDI can be found in Sec. 4.2.6. The diameter of the beryllium pipe is 20 mm and is narrower than the connecting Y-crotch chambers, where the electron and positron beams are separated and further transmitted to avoid excessive power of the Higher-Order Modes (HOMs). The crossing half-angle is 16.5 mrad to provide enough space for the superconducting quadrupole coils in a two-in-one type of configuration, saving space for a room temperature vacuum chamber. The detector includes a cylindrical drift chamber surrounded by an electromagnetic calorimeter, with the superconducting solenoid outside. Collimators and shielding masks are also provided to reduce background radiation and the effects of beam loss in the detector.

The remaining part of the Collider is the arc region, which is filled with various types of dual-aperture and single-aperture magnets. The distance between the electron and positron rings in this region is 0.35 m. Prototype measurements have shown that the field quality of the dual-aperture dipoles and quadrupoles meets the design requirements with this distance. Additionally, every two adjacent sextupoles (single aperture) make up one group and are powered by an independent power supply, allowing for flexible tuning capabilities.

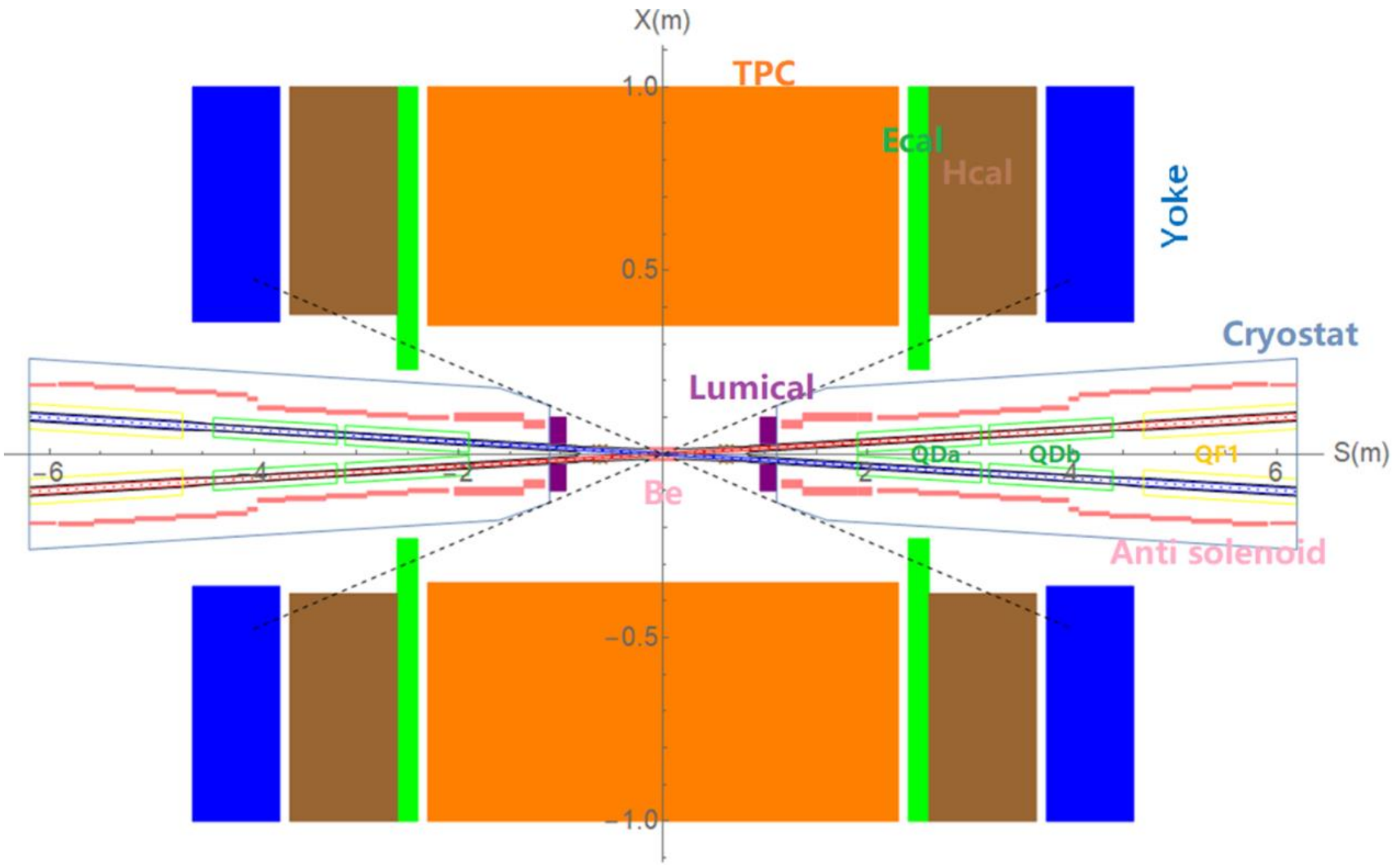

**Figure 2.3:** Layout of the center part of the Interaction Region and the Machine Detector Interface.

## 2.2 Machine Performance

The circumference of the CEPC is a critical and fundamental design parameter. A comprehensive study has been conducted to determine the best performance-cost ratio and the most promising future prospects. An established evaluation model was used to calculate the instant luminosity, construction and operation costs associated with different circumferences. Taking into consideration the overall project cost, the desired number of particles, and the corresponding cost per particle, it has been concluded that the optimal circumference for Higgs operation is 80 km. However, including other physical objectives, such as the Z-pole operation, the high-energy upgrade for the top factory, and the potential SPPC project, the optimal circumference is 100 km [1].

The operations cost is a significant consideration besides the construction cost. To reduce operational costs, the synchrotron-radiation power per beam is limited to 30 MW in the baseline design. However, if there is additional budget available, the CEPC can be upgraded to 50 MW for higher luminosity. The primary scientific goal is to build a Higgs factory, so the beam energy is set at 120 GeV. Nonetheless, the accelerator has a design compatible with operation at energies of 45.5 GeV and 80 GeV for the Z and W production, respectively. The CEPC has two Interaction Points (IPs), which have been optimized since the publication of the CDR to improve luminosity. At each IP, the instantaneous luminosity for Higgs operation is $5\times10^{34}$ $cm^{-2}s^{-1}$, for Z operation it is $115\times10^{34}$ $cm^{-2}s^{-1}$, and for W operation it is $16\times10^{34}$ $cm^{-2}s^{-1}$ in the 30 MW power operation mode. The corresponding beam currents are 16.7 mA for Higgs, 803.5 mA for Z, and 84.1 mA for W. If the power is upgraded to 50 MW, the instantaneous luminosity per IP increases to $8.3\times10^{34}$ $cm^{-2}s^{-1}$ for Higgs, $192\times10^{34}$ $cm^{-2}s^{-1}$ for Z, and $26.7\times10^{34}$ $cm^{-2}s^{-1}$ for W, with beam currents of 27.8 mA for Higgs, 1340.9 mA for Z, and 140.2 mA for W.

To ensure safe and stable operation of the CEPC, beam dynamics issues such as dynamic aperture with magnetic field and alignment errors, beam-beam effects, collective instabilities, beam lifetime, radiation background in the MDI region, and shielding schemes have been thoroughly studied in addition to lattice optimization for high luminosity. Furthermore, specific designs enable the energy of the operation to be switched without changing the hardware. Figure 2.4 illustrates that each RF station is divided into two sections. During Higgs operation, both sections are utilized by the electron and positron beams, resulting in a partial double-ring collider where the two beams share the RF straight section. However, during W/Z operations, only one of the two RF sections is used by each beam, resulting in a true double-ring collider. This flexible energy switching scheme supports physics research, particularly for precision measurement.

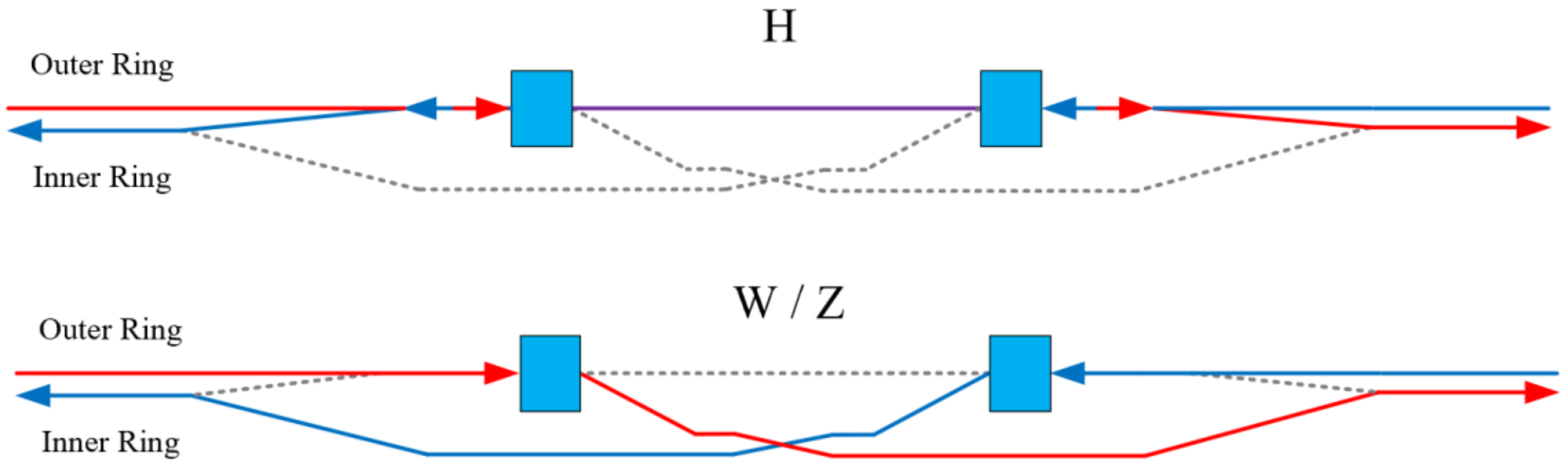

**Figure 2.4:** The switching operation scheme compatible with the H/W/Z modes: the RF system is divided into two sections. When running at a higher energy for Higgs mode, the electron and positron beams pass through both sections; when running at a lower energy for W and Z, both the electron and positron beams pass through only one RF section.

After sufficient data collection in the H/Z/W operation, the CEPC energy can be upgraded to 180 GeV for $t\bar{t}$ operation. This will require the installation of additional 650 MHz SRF cavities in the Collider RF stations and 1.3 GHz SRF cavities in the Booster RF stations. The $t\bar{t}$ luminosity per IP is expected to be $0.5\times10^{34}$ cm$^{-2}$s$^{-1}$ at 30 MW and $0.8\times10^{34}$ cm$^{-2}$s$^{-1}$ at 50 MW.

The baseline design for the detector solenoid is 3 T with a length of 7.6 m. However, the fringe field of the detector solenoid and anti-solenoids causes an x-y coupling effect. The coupling factor is 0.2% for Higgs and W and 0.34% for $t\bar{t}$. The most significant coupling effect and resulting vertical emittance growth occur during Z operation. Therefore, the solenoid field will be reduced to 2 T in Z operation when the coupling factor reduces to 0.5%.

In high-energy operation, the beamstrahlung effect is the dominant factor limiting the beam lifetime, with a lifetime of 40 minutes at the Higgs energy and 23 minutes at the $t\bar{t}$ energy. At lower energies, the Bhabha effect becomes dominant, with beam lifetimes of 80/60 minutes for Z/W operations, respectively. The energy acceptance for the dynamic aperture should be greater than 1.6% for Higgs, including errors. For other energies such as Z/W/$t\bar{t}$, the energy acceptance is 1.3%, 1.2%, and 2.0%, respectively.

In the Collider, the beam-stay-clear (BSC) region is defined as BSC(x) = ± ($18\sigma_x$ + 3 mm) and BSC(y) = ± ($22\sigma_y$ + 3 mm) for the horizontal and vertical planes, respectively. In the Booster, in the low-energy region close to the injection energies (30-80 GeV), it is defined as BSC(x,y) = $4\sigma_{x,y}$ + 5 mm. At higher energies, the BSCs are defined as: at 120 GeV, BSC(x) = $6\sigma_x$ + 3 mm; BSC(y) = $39\sigma_y$ + 3 mm; and at 180 GeV, BSC(x) = $6\sigma_x$ + 3 mm; BSC(y) = $50\sigma_y$ + 3 mm.

The Booster utilizes 1.3 GHz 9-cell superconducting RF cavities. During injection from the Linac at an energy of 30 GeV, the threshold for single-bunch current is 6.3 µA for Higgs and 5.8 µA for W/Z. At the extraction energy of 120 GeV, the threshold for single-bunch current is 70 µA, while it is 22.16 µA for the W mode at 80 GeV, and 9.57 µA for the Z mode at 45 GeV.

On-axis injection is used from the Linac to the Booster. For the injection from the Booster to the Collider for Z/W modes, off-axis injection is used. However, for the Higgs energy, the injection from the Booster to the Collider adopts an on-axis scheme, which requires a smaller dynamic aperture to ensure sufficient safety margins. Additionally, a special swap-out injection/extraction scheme is designed for Higgs top-up injection. In this scheme, bunches with sufficiently reduced charges are extracted from the Collider to the Booster and are merged with the existing bunches there to recover the bunch charge. The merged bunches are then injected back to the Collider using on-axis mode.

The baseline design for the CEPC includes 2 interaction points (IPs). However, a schematic design for 4 IPs was preliminarily studied and compared with the 2 IP scheme in terms of performance-cost ratio, assuming equal intervals between the IPs [1]. The number of RF stations is increased to 4 and placed in the middle of two adjacent IPs to ensure equal beam energy at each IP. The bunch pattern is also redesigned to be suitable for the 4 IPs, and the collision mode is changed to the two-by-two mode with the introduction of the two-bunch-train scheme.

Figures 2.5 and 2.6 compare the total cost (which is the sum of construction and operation costs in RMB) of the 2 IP and 4 IP schemes for Higgs operation at two different SR powers of 30 MW and 50 MW. As shown in Figure 2.5, for the power of 30 MW, the total cost linearly increases with the produced Higgs number, and the cost is equal for both schemes at 1.5 million Higgs production, with the increasing slope of 2 IPs being larger. For the power of 50 MW, as shown in Figure 2.6, the crossing point is at the Higgs production of 1.9 million. The scientific goal is to generate a few million Higgs, and therefore, the 2 IP and 4 IP schemes have similar performance in terms of economic efficiency. Considering other issues such as risk control and simpler configuration, it was decided to retain the 2 IP scheme as the baseline CEPC design.

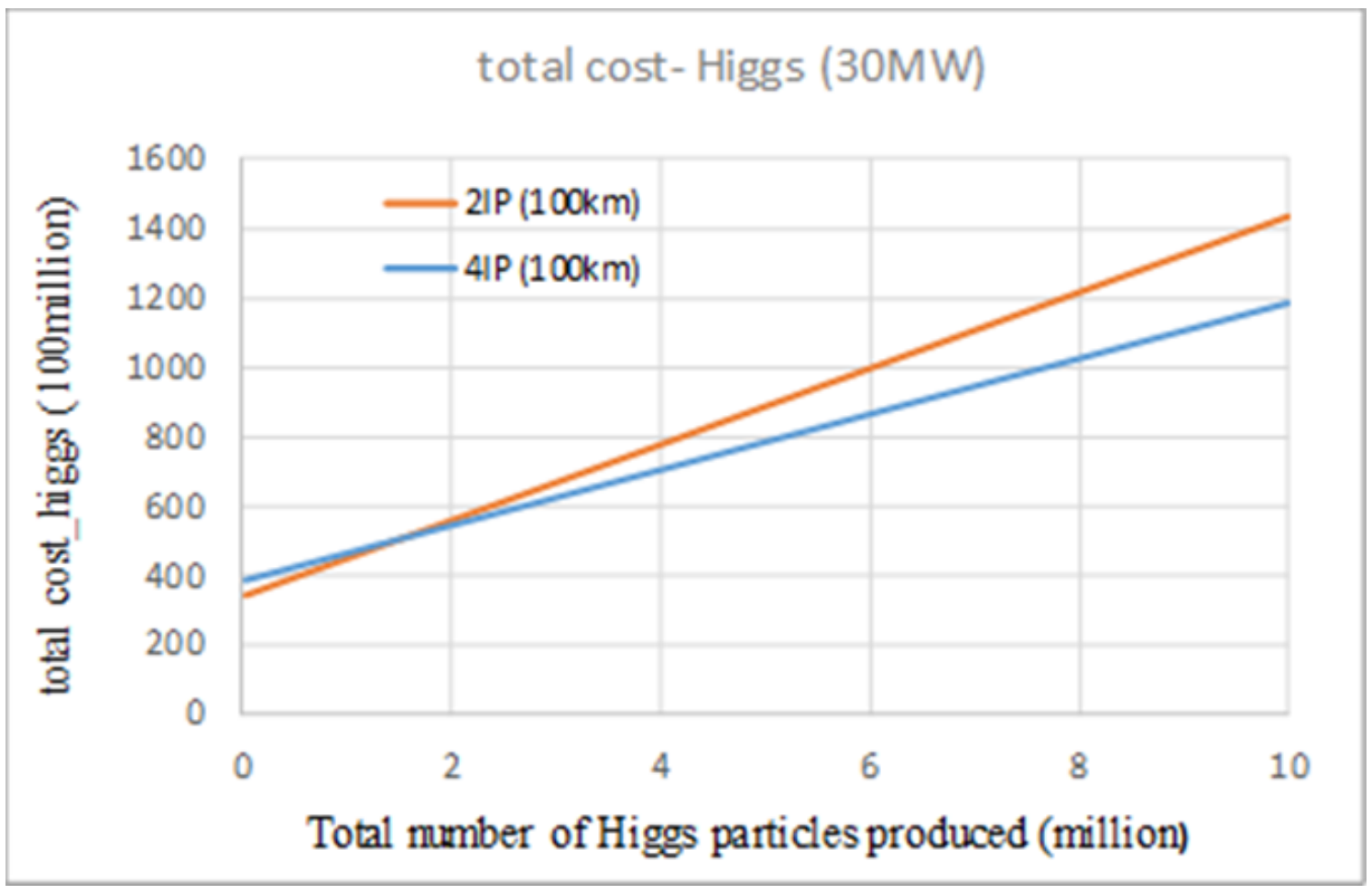

**Figure 2.5:** Comparison of total cost of a Higgs factory between 2 IPs and 4 IPs scheme with 30 MW SR power per beam.

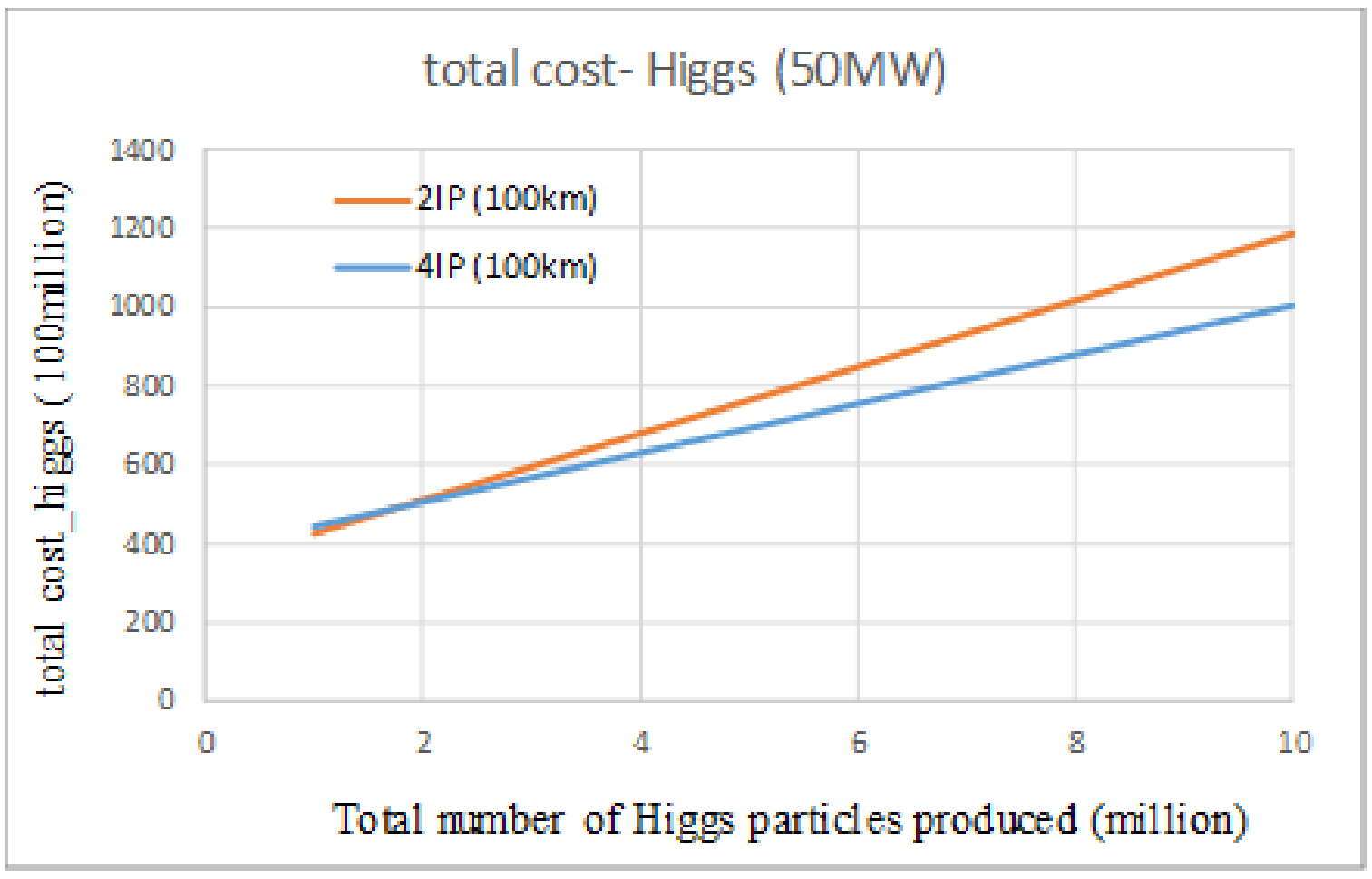

**Figure 2.6:** Comparison of total cost of a Higgs factory between 2 IPs and 4 IPs scheme with 50 MW SR power per beam.

# 3 Operation Scenarios

The CEPC has a flexible operation scenario, with the ability to switch between three energies: H ($e^+e^- \rightarrow ZH$), Z ($e^+e^- \rightarrow Z$) and W ($e^+e^- \rightarrow W^+W^-$). The primary goal is to operate as a Higgs factory, and therefore a "10-2-1" operation plan has been developed. The plan is to operate in Higgs mode for 10 years, generating approximately 2.6 million Higgs at 30 MW synchrotron radiation (SR) power per beam or 4.3 million Higgs at 50 MW. Afterward, the CEPC will operate as a Z factory for two years, generating about $2.5\times10^{12}$ Z particles at 30 MW, followed by one year of W operation, yielding 130 million $W^+W^-$ pairs at 30 MW.

Following the initial 13 years of operation for H/Z/W, a significant upgrade will take place to enable $t\bar{t}$ operation. The superconducting RF station will be separated into five parts. The middle part of the Collider will have 650 MHz 5-cell cavities added for $t\bar{t}$, with 2-cell cavities for H/W and 1-cell cavities for Z on the two sides. In addition, more 1.3 GHz 9-cell cavities will be added in the middle of the Booster for $t\bar{t}$. Using all cavities, a beam energy of 180 GeV can be achieved. Additionally, if necessary, it will still be possible to operate at lower energies of H/W or Z by bypassing parts of the cavities. Figure 3.1 illustrates the SRF operational scenario for $t\bar{t}$ Collider.

Once sufficient data has been collected for the H/Z/W and $t\bar{t}$ operations, it is expected that the superconducting technology for high-field magnets will have matured. At that point, the CEPC tunnel can be used to construct the Super Proton-Proton Collider (SPPC), while still reserving the CEPC for electron-positron collisions. This will greatly enhance the CEPC/SPPC capability to explore new particle physics through various collision modes.

To ensure that the particle-production estimates for the CEPC are realistic and based on proper models, the performance of four world-renowned lepton colliders has been investigated: LEP, KEKB, PEP-II, and BEPC-II.

LEP, including LEP1 and LEP2, operated from 1989 to 2000 for a total of 12 years. The operation time per year varied from 2,669 to 5,496 hours, with an average of 4,240 hours per year. Among the effective machine operation time, data collection time occupied 35% to 59%, with an average ratio of 41% [1].

KEKB operated from 1998 to 2010, with an average operation time of 5,060 hours per year and an average efficiency of 73% [2].

PEP-II operated from 1999 to 2008. The statistics for the last six years of operation, from 2003 to 2008, showed an average operation time of 5,750 hours per year and an average efficiency of 58% [3].

BEPC-II was designed, constructed, and operated by the Institute of High Energy Physics (IHEP), which also proposed the CEPC. BEPC-II has been running since 2008 and sustains high-energy physics research as well as multidisciplinary research as a synchrotron radiation (SR) facility. According to statistics from the three most recent years, 2020 to 2022, the operation time for each year was 7,296, 6,648, and 7,392 hours, respectively. The operation time in 2021 was short because a special insertion device was installed for the SR operation. The efficiency for high-energy physics operation was 78.7%, 69.1%, and 66.5%, respectively [4].

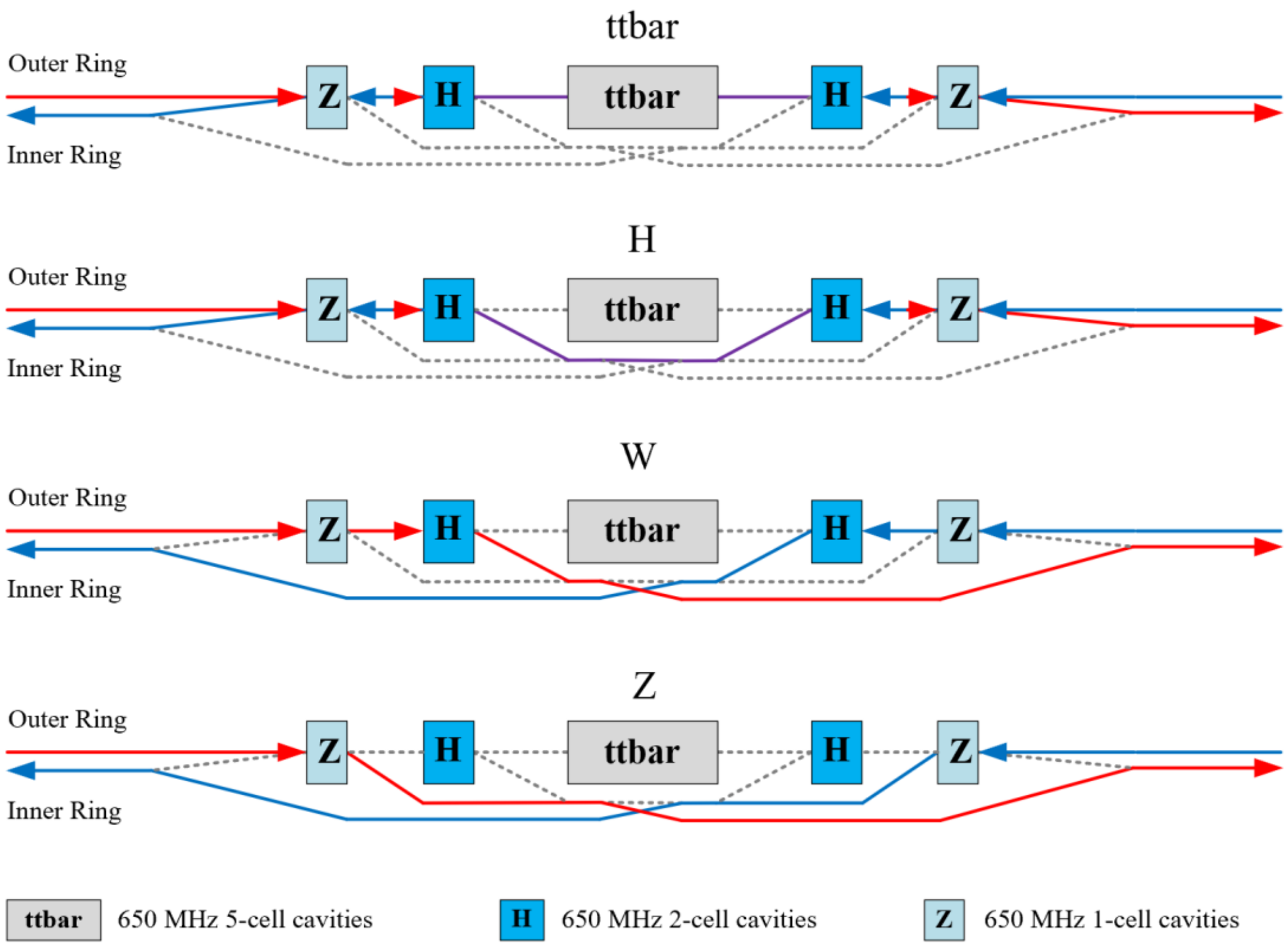

**Figure 3.1:** RF scheme for the $t\bar{t}$ Collider upgrade.

Based on the experience gained from previous lepton colliders, the operation plan for the CEPC will be as follows:

- The CEPC will be operational for a period of 8-10 months each year, with the total operation time ranging from 6,000 hours (equivalent to 250 days) to 7,500 hours (equivalent to 312 days).
- The efficiency for the collection of physics data is assumed to be 60% during the initial years of operation, gradually increasing to 75% as the operation becomes more mature.
- Consequently, the effective time available for data acquisition will be 3,600 hours based on the most conservative assumption (6,000 hours × 60%), and up to 5,600 hours in the best performing year (7,500 hours × 75%).

Tables 3.1 and 3.2 present the luminosity, integrated luminosity, operation years, and expected event statistics for the H/Z/W experiments using 30 MW and 50 MW SR power, respectively. It is important to note that the integrated luminosity and event numbers provided in these tables, as well as the subsequent discussions, are calculated using the most conservative assumption of 3,600 hours per year for data collection. However, if we were to consider a more optimistic assumption of 5,600 hours per year for data collection, these figures could potentially increase by up to 50%.

**Table 3.1:** CEPC operation plan (@ 30 MW)

| Particle | $E_{c.m.}$ (GeV) | $L$ per IP ($10^{34}$ cm$^{-2}$s$^{-1}$) | Integrated $L$ per year (ab$^{-1}$, 2 IPs) | Years | Total Integrated $L$ (ab$^{-1}$, 2 IPs) | Total no. of events |
|---|---|---|---|---|---|---|
| H | 240 | 5 | 1.3 | 10 | 13 | $2.6 \times 10^{6}$ |
| Z | 91 | 115* | 30 | 2 | 60 | $2.5 \times 10^{12}$ |
| W | 160 | 16 | 4,2 | 1 | 4.2 | $1.3 \times 10^{8}$ |
| $t\bar{t}$** | 360 | 0.5 | 0.13 | 5 | 0.65 | $0.4 \times 10^{6}$ |

* Detector solenoid field is 2 Tesla during Z operation.
** $t\bar{t}$ operation is optional.

**Table 3.2:** CEPC operation plan (@ 50 MW)

| Particle | $E_{c.m.}$ (GeV) | $L$ per IP ($10^{34}$ cm$^{-2}$s$^{-1}$) | Integrated $L$ per year (ab$^{-1}$, 2 IPs) | Years | Total Integrated $L$ (ab$^{-1}$, 2 IPs) | Total no. of events |
|---|---|---|---|---|---|---|
| H | 240 | 8.3 | 2.2 | 10 | 21.6 | $4.3 \times 10^{6}$ |
| Z | 91 | 192* | 50 | 2 | 100 | $4.1 \times 10^{12}$ |
| W | 160 | 26.7 | 6.9 | 1 | 6.9 | $2.1 \times 10^{8}$ |
| $t\bar{t}$** | 360 | 0.8 | 0.2 | 5 | 1.0 | $0.6 \times 10^{6}$ |

* Detector solenoid field is 2 Tesla during Z operation.
** $t\bar{t}$ operation is optional.

The 10-year Higgs operation with 30 MW SR power will yield 2.6 million Higgs bosons. If the SR power is increased to 50 MW, the CEPC can produce 4.3 million Higgs bosons. This will allow for precision measurements of Higgs coupling to the sub-percent level, improving the accuracy by an order of magnitude from what is achievable at the HL-LHC. The Higgs width can be determined independently of any specific model, and exotic Higgs decay branching ratios can be probed down to a level of $10^{-4}$, providing access to new physics such as Higgs portal dark matter.

The 2-year Z pole operation with 30 MW SR power will produce 2.5 trillion particles, a hundred thousand times more than the total generation produced during the entire LEP operation period. This will greatly benefit electroweak precision measurements and provide opportunities to explore new physics.

The 1-year W operation will allow for the measurement of the W width with an accuracy of about 3 MeV. 130 million $W^{+}W^{-}$ pairs will be produced, about 3200 times more than LEP.

Following the 13-year operation plan for the H/Z/W experiments, the CEPC has an optional proposal to upgrade its center-of-mass energy to 180 GeV and continue data collection specifically for $t\bar{t}$ (top quark-antiquark) events for an additional 5 years. According to Tables 3.1 and 3.2, this upgrade would result in the generation of 0.4 million $t\bar{t}$ events at 30 MW and 0.6 million $t\bar{t}$ events at 50 MW. With this extension, the CEPC's operation would extend until the mid-21st century. At that point, mature high-temperature superconducting (HTS) high-field magnet technology is expected to be available, enabling the construction of the Superconducting Proton-Proton Collider (SPPC).

The SPPC will share the same tunnel as the CEPC and will be positioned adjacent to the outer wall. The tunnel has sufficient space to accommodate both the SPPC and CEPC collider rings. The two SPPC detectors will be situated near the two CEPC RF straight sections at IP2 and IP4. Given the SPPC's location on the outer side and the CEPC on the inner side, it will be more convenient to establish the required SPPC bypass at the CEPC detector positions IP1 and IP3. As a result, there will be flexibility to operate the SPPC in pp or ion-ion mode alongside the CEPC, or vice versa, enabling the CEPC to operate in e+e– mode while the SPPC is in operation.

Moreover, this arrangement creates the potential for research in e-p and e-ion physics by simultaneously operating the CEPC and SPPC.

# 4 Collider

## 4.1 Main Parameters

The CEPC has two interaction points (IPs) for $e^+e^-$ collisions and is designed to operate in four energy modes ($t\bar{t}$, Higgs, W, and Z). A tentative "10-2-1-5" operation plan is to run the CEPC first as a Higgs factory for 10 years to produce three million or more Higgs particles, followed by 2 years of operation as a Super Z factory to create one trillion Z bosons, and then 1 year as a W factory to produce approximately 100 million W bosons. Finally, an upgrade will enable the CEPC to operate at the $t\bar{t}$ energy. The CEPC's circumference is approximately 100 km, which is compatible with the SPPC, a proton-proton collider designed to operate at 125 TeV center-of-mass energy using 20 Tesla superconducting dipole magnets.

The main parameters of the CEPC, as described in this TDR, are listed in Table 4.1.1. The luminosities of the CEPC are mainly limited by the synchrotron radiation power. The assumed limit is 30 MW per beam, with consideration of the grid distribution status, electricity capability in China, and the operation cost due to power consumption. However, this limitation can be increased to 50 MW with an upgrade. The luminosity at the Higgs mode is $5 \times 10^{34}$ cm$^{-2}$ s$^{-1}$ with 268 bunches, while at the W mode, it is $1.6 \times 10^{35}$ cm$^{-2}$ s$^{-1}$ with 1,297 bunches. At the Z pole, the luminosity is $1.2 \times 10^{36}$ cm$^{-2}$ s$^{-1}$ with 11,934 bunches and a 2 T detector solenoid. The bunch number at the Z pole is limited by the electron cloud instability of the positron beam. The minimum bunch separation for Z mode, taking into account electron cloud effects and timing structure [1], is 23 ns, leaving 18% beam gap for ion cleaning and kicker rise.

The luminosities listed in Table 4.1.1 include the bunch lengthening effect determined from the impedance of the whole ring and the beam-beam effect. The bunch lengthening is approximately 78% for Higgs, 250% for Z, 96% for W, and 32% for $t\bar{t}$.

At higher energy regions, the dominant constraint for beam lifetime is the beamstrahlung effect. It is 40 minutes at Higgs energy and 23 minutes at $t\bar{t}$ energy. The bhabha lifetime due to collision is 40 minutes for Higgs and 81 minutes for $t\bar{t}$. The dynamic aperture's energy acceptance needs to be larger than 1.6% for Higgs, taking into account error effects. The total energy spread including the beamstrahlung effect is about 170% of the natural energy spread at Higgs energy.

The vertical emittance growth for Z mode is the most serious issue among the four operating energies due to the fringe field of the detector solenoid and anti-solenoids. Therefore, a larger coupling factor of 0.5% is chosen at the Z pole, with a 2 T detector solenoid. For Higgs and W mode, the coupling factor is 0.2%, while it is 0.34% for $t\bar{t}$, which requires a 3 T detector solenoid.

**Table 4.1.1:** CEPC baseline parameters in TDR

| | **Higgs** | **Z** | **W** | $t\bar{t}$ |
|---|---|---|---|---|
| Number of IPs | 2 | | | |
| Circumference (km) | 100.0 | | | |
| SR power per beam (MW) | 30 | | | |
| Half crossing angle at IP (mrad) | 16.5 | | | |
| Bending radius (km) | 10.7 | | | |
| Energy (GeV) | 120 | 45.5 | 80 | 180 |
| Energy loss per turn (GeV) | 1.8 | 0.037 | 0.357 | 9.1 |
| Damping time $\tau_x/\tau_y/\tau_z$ (ms) | 44.6/44.6/22.3 | 816/816/408 | 150/150/75 | 13.2/13.2/6.6 |
| Piwinski angle | 4.88 | 24.23 | 5.98 | 1.23 |
| Bunch number | 268 | 11934 | 1297 | 35 |
| Bunch spacing (ns) | 591 (53% gap) | 23 (18% gap) | 257 | 4524 (53% gap) |
| Bunch population ($10^{11}$) | 1.3 | 1.4 | 1.35 | 2.0 |
| Beam current (mA) | 16.7 | 803.5 | 84.1 | 3.3 |
| Phase advance of arc FODO (°) | 90 | 60 | 60 | 90 |
| Momentum compaction ($10^{-5}$) | 0.71 | 1.43 | 1.43 | 0.71 |
| Beta functions at IP $\beta_x^*/\beta_y^*$ (m/mm) | 0.3/1 | 0.13/0.9 | 0.21/1 | 1.04/2.7 |
| Emittance $\varepsilon_x/\varepsilon_y$ (nm/pm) | 0.64/1.3 | 0.27/1.4 | 0.87/1.7 | 1.4/4.7 |
| Betatron tune $\nu_x/\nu_y$ | 445/445 | 317/317 | 317/317 | 445/445 |
| Beam size at IP $\sigma_x/\sigma_y$ (um/nm) | 14/36 | 6/35 | 13/42 | 39/113 |
| Bunch length (natural/total) (mm) | 2.3/4.1 | 2.5/8.7 | 2.5/4.9 | 2.2/2.9 |
| Energy spread (natural/total) (%) | 0.10/0.17 | 0.04/0.13 | 0.07/0.14 | 0.15/0.20 |
| Energy acceptance (DA/RF) (%) | 1.6/2.2 | 1.0/1.7 | 1.05/2.5 | 2.0/2.6 |
| Beam-beam parameters $\xi_x/\xi_y$ | 0.015/0.11 | 0.004/0.127 | 0.012/0.113 | 0.071/0.1 |
| RF voltage (GV) | 2.2 | 0.12 | 0.7 | 10 |
| RF frequency (MHz) | 650 | | | |
| Longitudinal tune $\nu_s$ | 0.049 | 0.035 | 0.062 | 0.078 |
| Beam lifetime (Bhabha/beamstrahlung) (min) | 40/40 | 90/2800 | 60/195 | 81/23 |
| Beam lifetime requirement (min) | 18 | 77 | 22 | 18 |
| Hourglass Factor | 0.9 | 0.97 | 0.9 | 0.89 |
| Luminosity per IP ($10^{34}$ cm$^{-2}$ s$^{-1}$) | 5.0 | 115 | 16 | 0.5 |

The CEPC upgrade parameters with 50 MW synchrotron radiation power at Higgs, W, Z and $t\bar{t}$ energy operations are shown in Table 4.1.2.

**Table 4.1.2:** CEPC main parameters with 50 MW upgrade

| | Higgs | Z | W | $t\bar{t}$ |
|---|---|---|---|---|
| Number of IPs | 2 | | | |
| Circumference (km) | 100.0 | | | |
| SR power per beam (MW) | 50 | | | |
| Half crossing angle at IP (mrad) | 16.5 | | | |
| Bending radius (km) | 10.7 | | | |
| Energy (GeV) | 120 | 45.5 | 80 | 180 |
| Energy loss per turn (GeV) | 1.8 | 0.037 | 0.357 | 9.1 |
| Damping time $\tau_x/\tau_y/\tau_z$ (ms) | 44.6/44.6/22.3 | 816/816/408 | 150/150/75 | 13.2/13.2/6.6 |
| Piwinski angle | 4.88 | 29.52 | 5.98 | 1.23 |
| Bunch number | 446 | 13104 | 2162 | 58 |
| Bunch spacing (ns) | 355 (53% gap) | 23 (10% gap) | 154 | 2714 (53% gap) |
| Bunch population ($10^{11}$) | 1.3 | 2.14 | 1.35 | 2.0 |
| Beam current (mA) | 27.8 | 1340.9 | 140.2 | 5.5 |
| Phase advance of arc FODO (°) | 90 | 60 | 60 | 90 |
| Momentum compaction ($10^{-5}$) | 0.71 | 1.43 | 1.43 | 0.71 |
| Beta functions at IP $\beta_x^*/\beta_y^*$ (m/mm) | 0.3/1 | 0.13/0.9 | 0.21/1 | 1.04/2.7 |
| Emittance $\varepsilon_x/\varepsilon_y$ (nm/pm) | 0.64/1.3 | 0.27/1.4 | 0.87/1.7 | 1.4/4.7 |
| Betatron tune $\nu_x/\nu_y$ | 445/445 | 317/317 | 317/317 | 445/445 |
| Beam size at IP $\sigma_x/\sigma_y$ (um/nm) | 14/36 | 6/35 | 13/42 | 39/113 |
| Bunch length (natural/total) (mm) | 2.3/4.1 | 2.7/10.6 | 2.5/4.9 | 2.2/2.9 |
| Energy spread (natural/total) (%) | 0.10/0.17 | 0.04/0.15 | 0.07/0.14 | 0.15/0.20 |
| Energy acceptance (DA/RF) (%) | 1.6/2.2 | 1.0/1.5 | 1.05/2.5 | 2.0/2.6 |
| Beam-beam parameters $\xi_x/\xi_y$ | 0.015/0.11 | 0.0045/0.13 | 0.012/0.113 | 0.071/0.1 |
| RF voltage (GV) | 2.2 | 0.1 | 0.7 | 10 |
| RF frequency (MHz) | 650 | | | |
| Longitudinal tune $\nu_s$ | 0.049 | 0.032 | 0.062 | 0.078 |
| Beam lifetime (Bhabha/beamstrahlung) (min) | 40/40 | 90/930 | 60/195 | 81/23 |
| Beam lifetime requirement (min) | 20 | 81 | 25 | 18 |
| Hourglass Factor | 0.9 | 0.97 | 0.9 | 0.89 |
| Luminosity per IP ($10^{34}$ cm$^{-2}$ s$^{-1}$) | 8.3 | 192 | 26.7 | 0.8 |

### 4.1.1 Constraints for Parameter Choices

The CEPC has several constraints on its main parameters, which must be taken into account when making choices about the collider's design.

1) Synchrotron radiation (SR) power:
   The maximum SR power per beam is currently limited to 30 MW in order to control the total AC power of the project. However, a future upgrade to 50 MW SR power/beam should be feasible.

2) Beam-beam effect:
   The beam-beam tune shift is a limit beyond which the beam emittance will blow up [2]. This is about 0.11 for Higgs, 0.12 for W, 0.13 for Z and 0.1 for $t\bar{t}$. The beam-beam limit at the W/Z is mainly determined by the coherent X-Z instability instead of the beamstrahlung lifetime as in the $t\bar{t}$/Higgs mode. A larger phase advance of the FODO cell (60°/60°) for the collider ring optics is chosen at the W/Z mode to supress the beam-beam instability when we take both the beam-beam effect and longitudinal impedance into account [3].
3) Beam lifetime due to beamstrahlung:
   Beam lifetime is mainly limited by the emission of single photons in the tail of the beamstrahlung spectra. At Higgs energy, the beam lifetime is about 40 minutes, and the requirement for energy acceptance of the dynamic aperture is 1.6% including lattice errors and beam-beam effects.
4) Additional energy spread due to beamstrahlung:
   To maintain the uniformity of beam energy, the energy spread induced by beamstrahlung must be controlled. This limit has been set to 70% of the natural energy spread for Higgs.
5) High-order mode (HOM) power per cavity:
   HOM power is a limit for the coaxial coupler, and the bunch charge and total beam current must be controlled; otherwise, the HOM power cannot be fully extracted, and the HOM couplers will be damaged. The HOM power of a 2-cell cavity has been constrained to be less than 2 kW. The single cell cavities will be added for the Z operation, and the 5-cell cavities are considered as the first choice for the $t\bar{t}$ upgrade.

### 4.1.2 Luminosity

The luminosity for a circular $e^+e^-$ collider is given by:

$$L_0[cm^{-2}s^{-1}] = 2.17 \times 10^{34}(1+r)\xi_y \frac{eE_0(GeV)N_bN_e}{T_0(s)\beta_y^*(cm)} F_h \tag{4.1.1}$$

where $N_e$ is the bunch population, $N_b$ the bunch number, $T_0$ the revolution time, $\beta_y$* the vertical betatron function at the interaction point, $r$ the IP beam size ratio $\sigma_y^*/\sigma_x^*$ and $\xi_y$ the vertical beam-beam tune shift. $F_h$ is the luminosity reduction factor due to the hourglass effect.

### 4.1.3 Crab Waist Scheme

The crab waist scheme [4] increases the luminosity by suppressing the hourglass effect and the vertical blow up of the colliding bunches in order to reach high luminosity. This scheme requires a large Piwinski angle, which is defined as

$$\Phi = \frac{\sigma_z^*}{\sigma_x^*} \tan\theta_h \gg 1 \tag{4.1.2}$$

where $\theta_h$ is the half crossing angle. The crab waist enhancement factor, which is related to the Piwinski angle and the $\beta_y$ at IP, is introduced to estimate the potential of the maximum allowed beam-beam tune shift. In order to increase the luminosity, the Piwinski angle should be larger than one. The crossing angle is 33 mrad for CEPC. The Piwinski angle is 4.9 for Higgs, 6.0 for W, 24.2 for Z and 1.2 for ttbar. Thus, the luminosity enhancement by the crab waist scheme is about 2.0 for Higgs, 3.0 for W, 5.5 for Z and 1.3 for $t\bar{t}$.

The strong $x$ and $y$ betatron resonances are supressed by a pair of sextupoles placed on both sides of the IP, in phase with the IP in the horizontal plane and π/2 phase shift in the vertical plane.

With a large Piwinski angle, the overlapping area of the colliding beams becomes much smaller than $\sigma_z$, and an effective bunch length using the real overlap has been defined in order to assess the hourglass effect [5].

### 4.1.4 **Beam-beam Tune Shift**

The beam-beam tune shifts with large crossing angle are estimated by [6]

$$\xi_x = \frac{r_e N_e \beta_x^*}{2\pi\gamma\sigma_x^*\sqrt{1+\Phi^2}\left(\sigma_x^*\sqrt{1+\Phi^2}+\sigma_y^*\right)}, \quad \xi_y = \frac{r_e N_e \beta_y^*}{2\pi\gamma\sigma_y^*\left(\sigma_x^*\sqrt{1+\Phi^2}+\sigma_y^*\right)}. \qquad (4.1.3)$$

The values of beam-beam tune shifts in table 4.1.1 are calculated by beam-beam simulations.

### 4.1.5 **RF Parameters**

The primary function of the RF system is to compensate for synchrotron radiation loss and provide the required bunch length and energy acceptance. The parameters of the RF system for the collider ring are presented in Table 4.1.3. The superconducting RF cavity is selected for its high CW gradient, high energy efficiency, and low impendence. The selection of a 650 MHz RF frequency strikes a balance between beam stability and cost.

**Table 4.1.3:** Main RF parameters of the Collider Ring

| | $t\bar{t}$ 30/50 MW | | Higgs 30/50 MW | W 30/50 MW | Z 30/50 MW |
|---|---|---|---|---|---|
| | New cavities | Higgs cavities | | | |
| Beam energy [GeV] | 180 | | 120 | 80 | 45.5 |
| SR power / beam [MW] | 30 / 50 | | 30 / 50 | 30 / 50 | 30 / 50 |
| RF voltage [GV] | 10 (6.1 + 3.9) | | 2.2 | 0.7 | 0.12 / 0.1 |
| RF frequency [MHz] | 650 | | 650 | 650 | 650 |
| Syn. phase from crest [deg] | 24.5 | | 35.1 | 59.3 | 68.3 |
| Beam current / beam [mA] | 3.4 / 5.6 | | 16.7 / 27.8 | 84 / 140 | 801 / 1345 |
| Bunch charge [nC] | 32 | | 21 | 21.6 | 22.4 / 34.2 |
| Bunch length [mm] | 2.9 | | 4.1 | 4.9 | 8.7 / 10.6 |
| Cavity number | 192 | 336 | 192 / 336 | 96 / 168 / ring | 30 / 50 / ring |
| Cell number / cavity | 5 | 2 | 2 | 2 | 1 |
| Operation gradient [MV/m] | 27.6 | 25.2 | 24.9 / 14.2 | 15.9 / 9.1 | 17.4 / 8.7 |
| $Q_0$ @ 2 K at operating gradient | 3E10 | 3E10 | 3E10 | 3E10 | 2E10 |
| HOM power / cavity [kW] | 0.4 / 0.66 | 0.16 / 0.26 | 0.4 / 0.67 | 0.93 / 1.54 | 2.9 / 6.2 |
| Input power / cavity [kW] | 188 / 315 | 71 / 118 | 313 / 298 | 313 / 298 | 1000 |
| Klystron number | 48 / 96 | 168 | 96 / 168 | 96 / 168 | 60 / 100 |
| Cryomodule number | 48 | 56 | 32 / 56 | 32 / 56 | 60 / 100 |

## 4.2 Collider Accelerator Physics

### 4.2.1 Optics

#### 4.2.1.1 *Optics Design*

The Circular Electron and Positron Collider (CEPC) [1] is a double ring collider with two interaction points. Fig. 4.2.1.1 shows the layout. The lattice of the collider ring was designed with the following requirements:

- Double ring, 100 km circumference, 2 interaction points (IPs).
- SR power limited to 30 MW per beam (upgradable to 50 MW).
- Operation in the Higgs mode is the top priority while remaining compatible with operating in the $t\bar{t}$/W/Z modes.
- Compatible with the Super Proton Proton Collider (SPPC) in a common tunnel.
- 2 interaction regions: crab-waist collisions [2], local chromaticity correction, asymmetric acceptance [3].
- 8 arc regions: dual-aperture dipole and quadrupole [4] magnets to reduce power, non-interleaved sextupoles.
- 2 RF regions: shared cavities for two beams to economize operating costs. Flexible switching between compatible modes is provided.
- 4 short straight sections provide on-axis injection for the Higgs mode and off-axis injection for the other modes and beam dumping.

A conceptual design has been made for the luminosity goal 3/10 × $10^{34}$ cm$^{-2}$s$^{-1}$ for Higgs/W at 30 MW and 32 × $10^{34}$ cm$^{-2}$s$^{-1}$ for Z at 16.5 MW [1].

Primarily to increase the luminosity at the Higgs and Z energies, a high luminosity scheme [5, 6] has been proposed. This is done mainly by squeezing the vertical beta function at the IP, and by increasing the beam current at the Z.

The beam parameters for the $t\bar{t}$/Higgs/W/Z/ modes are listed in Section 4.1. For Higgs operations, the vertical beta function at IP $\beta_y$* has been squeezed from 1.5 mm to 1.0 mm and the emittance reduced from 1.21 nm to 0.64 nm while the single bunch population $Ne$ is decreased from 15 × $10^{10}$ to 13 × $10^{10}$ to reduce the energy acceptance requirement.

For the Z running, the beam current has been almost doubled as the high-order mode problem can be solved by using high-current 1-cell cavities. Larger momentum compaction factor $\alpha_p$, emittance $\epsilon_x$ and longitudinal tune $Q_s$ are favoured to suppress the microwave and transfer-mode coupling instabilies [7] and to increase the stable tune area which shrinks when considering both the beam-beam effect and longitudinal impedance [8]. Thus, the phase advance of the arc cell was reduced from 90 degrees (for Higgs/$t\bar{t}$) to 60 degrees. For W running, the phase advance was reduced to 60 degrees as well to increase the stable tune area. For $t\bar{t}$ energy running, the luminosity has not been pushed to a high value [9]. An asymmetric acceptance is required in the lattice design due to the asymmetric energy distribution due to strong synchrotron radiation [3].

The spacing of 0.35 m between the two beams is chosen for design of the twin-aperture magnets. For the other regions, independent magnets are chosen for flexibility. Nearby magnets in the two rings have a longitudinal separation of 0.3 m to provide enough space for independent magnets.

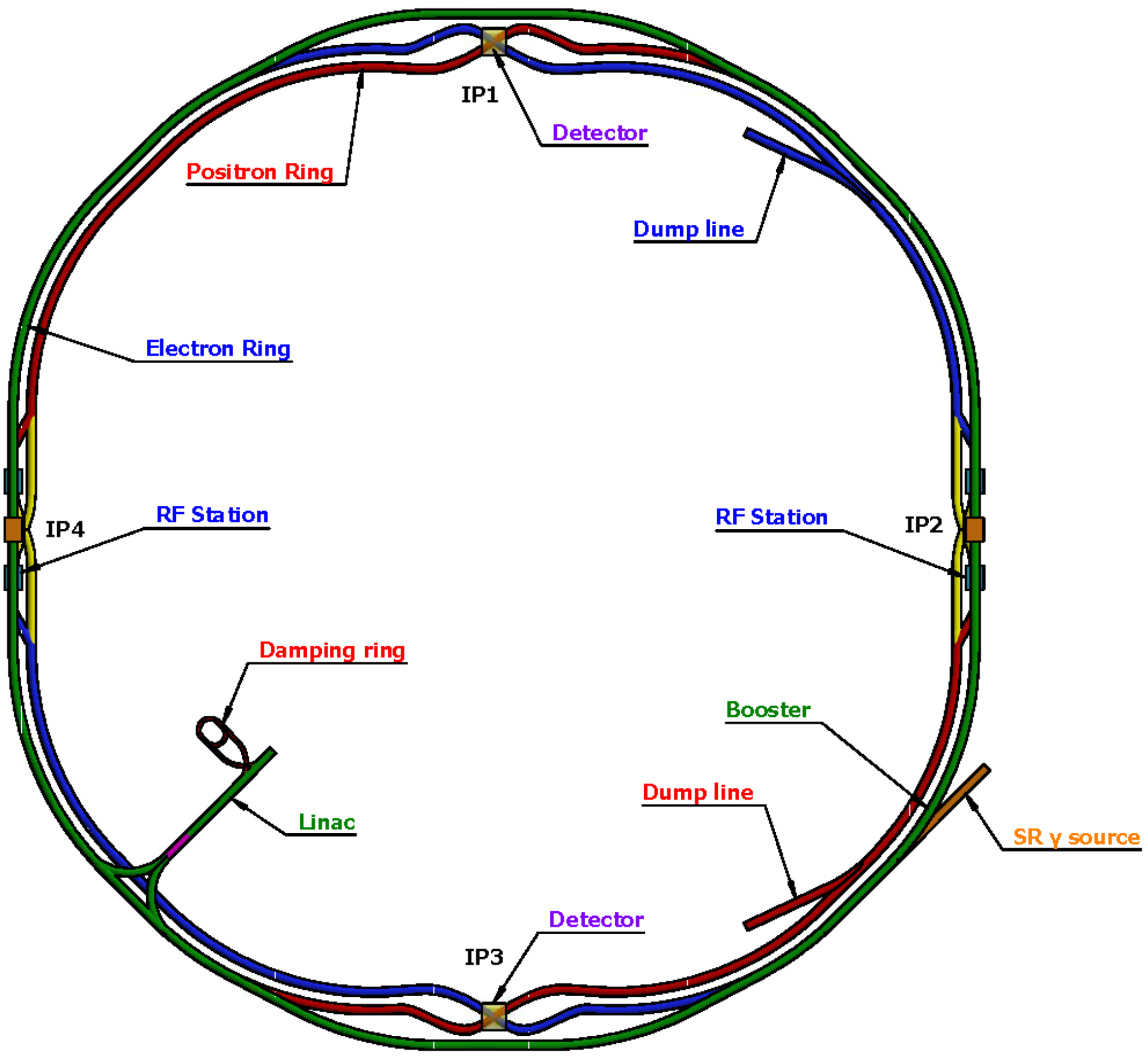

**Figure 4.2.1.1:** Layout of CEPC accelerator complex. The blue and brown circles indicate the two Collider rings, and the green circle is the Booster.

#### *4.2.1.1.1 Interaction Region*

The interaction region was designed to provide local chromaticity correction generated by the final doublet magnets and crab waist collision. It consists of modular sections including the final transformer (FT), chromaticity correction for the vertical plane (CCY), chromaticity correction for the horizontal plane (CCX), crab waist section (CW) and matching transformer (MT) [10-12].

The FT consists of two quadrupole doublets. The phase advance is π in the vertical plane and than π in the horizontal plane. At the end of FT is the first image point. The CCY actually consists of four FODO cells whose phase advances are π/2 in both planes and begin with a half defocusing quadrupole. Four identical dipoles are used to make dispersion bumps. A pair of sextupoles is placed at the two beta peaks to compensate the vertical chromaticity generated by the final defocusing quadrupole. The geometric sextupole aberrations are cancelled by the –I transformation between the paired sextupoles. At the end of CCY, there is the second image point which is identical to the first one. The CCX is similar to the CCY. It begins with a half focusing quadrupole. The MT also consists of two quadrupole doublets. With MT, the Twiss functions are matched to the arc section of the ring and make the total phase advance of FF to be 6π.

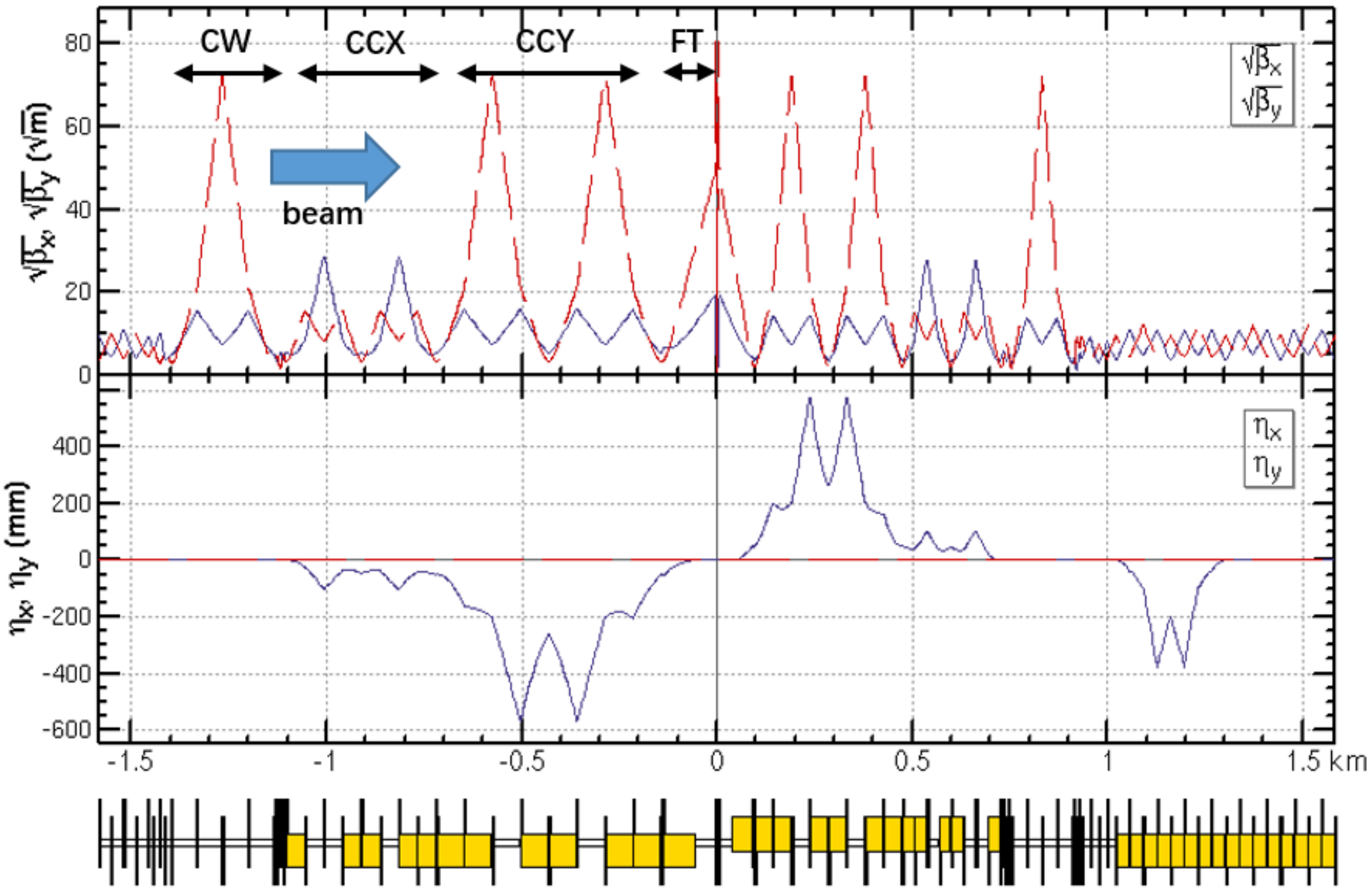

**Figure 4.2.1.2:** Optics of the interaction region for Higgs energy.

Figures 4.2.1.2 and 4.2.1.3 depict the lattice design and geometry of the interaction region surrounding IP3. The interaction point is positioned at the center. To facilitate gentler bends in the upstream section of the IP, an asymmetric lattice configuration is employed. Reverse bending is implemented in the final bends to prevent synchrotron radiation from reaching the IP chamber. The geometry retraction of the interaction region relative to the ring is devised to achieve smoother bends in the bypass of the SPPC, which is located on the outer side of the CEPC.

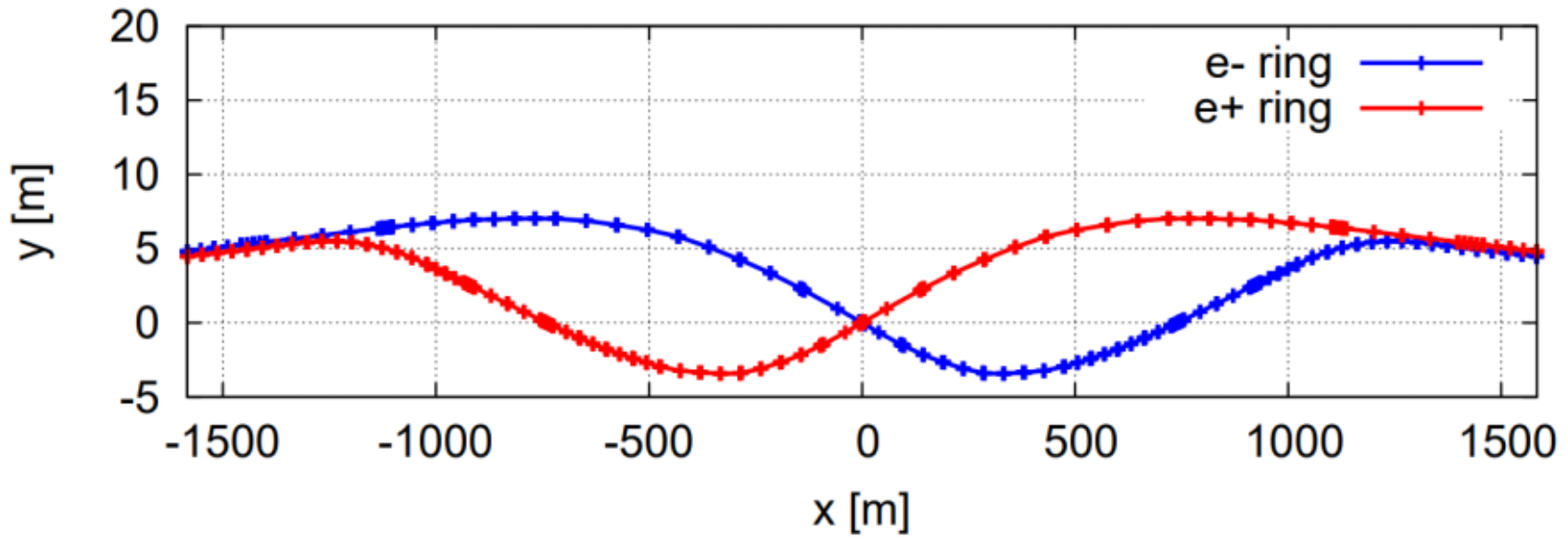

**Figure 4.2.1.3:** Geometry of the interaction region.

To make the lattice robust and provide a good starting point for dynamic aperture optimization, the length from the IP to the first quadrupole, $L^*$, is reduced from 2.2 m to 1.9 m [5,13]. The space in the final doublet cryo-module is used without changing the front-end position, i.e. not affecting the detector design. The dynamic aperture reduction due to synchrotron-radiation damping and fluctuation due to quadrupole is significant especially in the vertical plane [14]. The synchrotron radiation power of beam due to betatron motion in the quadrupole is proportional to $\Sigma(K_1 l)2\beta/l$, where $\beta$, $K_1$ and $l$ are the beta function at the quadrupoles, and the normalized strength and lengths of the quadrupoles respectively. The normalized strength and vertical beta function of the final quadrupole Q1 is largest in the ring. Thus, the contribution of Q1 is dominant. A longer Q1 will significantly decrease the power in the vertical plane and thus helps to increase the dynamic aperture [14,15]. The length of Q1 has been increased from 2 m to 2.5 m to ease the quadrupole radiation effect. To further ease the quadrupole radiation effect, Q1 was divided into two equal-length quadrupoles and the strength of the second one is lowered to accommodate the larger vertical beta function [16]. The comparison of the final doublet parameter for the Higgs energy is shown in Table 4.2.1.1.and illustrates the modifications in this TDR from the CDR.

**Table 4.2.1.1:** The comparison of the fin-l doublet parameter for the Higgs energy**.**

| | L* [m] | LQ1 [m] | LQ2 [m] | GQ1 [T/m] | GQ2 [T/m] | d [m] |
|---|---|---|---|---|---|---|
| CDR | 2.2 | 2.0 | 1.5 | -136 | 111 | 0.3 |
| TDR | 1.9 | 1.22/1.22 | 1.5 | -141/-85 | +95 | 0.3 |

The final quadrupoles are optimized at the Higgs energy. Thus, the strength of other modes doesn't exceed those in the Higgs mode. The layout of the final quadrupoles is shown in Fig. 4.2.1.4. The length and strength of the final quadrupoles for the compatible modes are listed in Table 4.2.1.2.

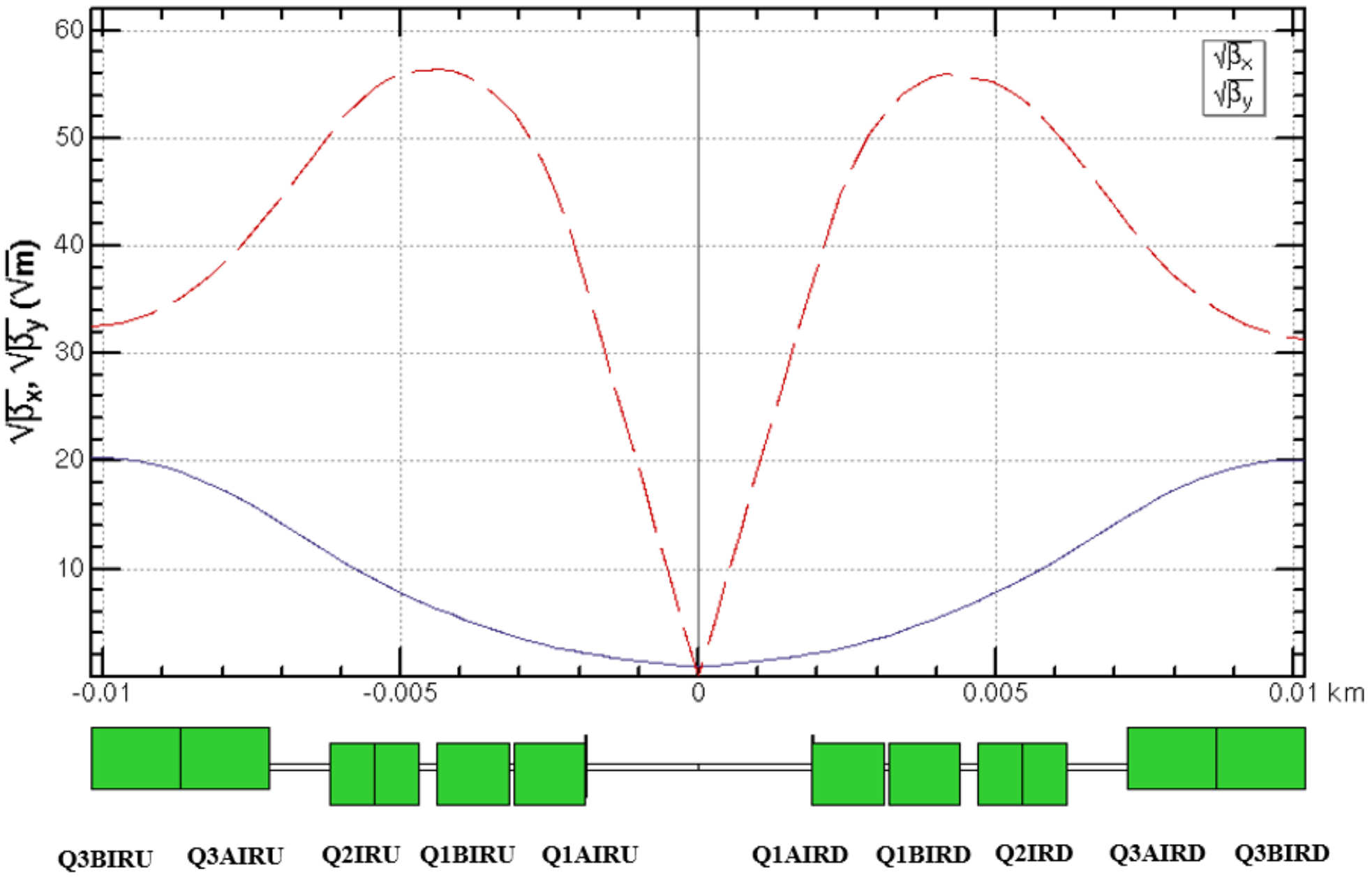

**Fig 4.2.1.4:** Layout of the final quadrupoles.

**Table 4.2.1.2:** The length and strength of the final quadrupoles for the compatible modes.

| Quadrupole | L [m] | Distance from IP [m] | Strength [T/m] | | | |
|---|---|---|---|---|---|---|
| | | | $t\bar{t}$ | Higgs | W | Z |
| Q1AIRU | 1.21 | 1.9 | -141 | -141 | -94 | -110 |
| Q1BIRU | 1.21 | 3.19 | -59 | -85 | -56 | +65 |
| Q2IRU | 1.5 | 4.7 | -51 | +95 | +63 | 0 |
| Q3IRU | 3 | 7.2 | +40 | 0 | 0 | 0 |
| Q1AIRD | 1.21 | -1.9 | -142 | -142 | -95 | -110 |
| Q1BIRD | 1.21 | -3.19 | -64 | -85 | -57 | +65 |
| Q2IRD | 1.5 | -4.7 | -47 | +96 | +64 | 0 |
| Q3IRD | 3 | -7.2 | +40 | 0 | 0 | 0 |

The chromaticity up to 3rd order is corrected with pairs of main sextupoles, phase tuning and additional sextupoles respectively. Most of the resonance driving terms (RDT) due to sextupoles in 3$^{rd}$ and 4$^{th}$ order Lie operators are almost cancelled [10,11]. The tune shift due to the finite length of the main sextupoles is corrected with additional weak sextupoles [17].

#### *4.2.1.1.2 Arc Region*

For the arc region, a FODO cell structure is chosen to provide a large filling factor. The 90/90 degrees phase advances and non-interleaved sextupole scheme are selected due to their property of aberration cancellation. Fig 4.2.1.5 shows the cell structure for the $t\bar{t}$/Higgs and W/Z modes.

A shorter cell length was used to squeeze the emittance from 1.2 nm to 0.64 nm. To increase the bend filling factor, the layout of the sextupoles in the two rings was changed from staggered to parallel and the left drifts are used for a longer bend. The length of the arc quadrupole is increased from 2 m to 3 m for optimization of the quadrupole radiation effect. The 2$^{nd}$-order chromaticity is a main aberration for optimization of momentum acceptance with a twice-repeated non-interleaved sextupole structure [3] which was used in the CDR. In the CDR lattice, the 2$^{nd}$-order chromaticity generated in the arc region was corrected with the IR knobs of phase advance or weak quadrupole at the first image point. However, the IR knobs generate distortions at the IP (beta, alpha and dispersion functions) especially in the horizontal plane. An 8-repeated sextupoles structure with 23 FODO cells generates much less 2$^{nd}$-order chromaticity for the horizontal plane [18] as the coupling term disappears. Fig 4.2.1.6 shows the comparison of the one –I sextupole and four –I sextupole schemes.

As shown in the Fig 4.2.1.7, the distribution of sextupoles for Higgs and $t\bar{t}$ mode allowed selection of –I sextupole pairs for W and Z mode. Totally, there are 6×6 = 36 cases for 23 cells. Fig. 4.2.1.8 shows the 2$^{nd}$ and 3$^{rd}$ order chromaticity for different cases. There are many more combinations if we choose different cases in each arc section (184 cells). Cancellation of main aberration has a period of 23×3 = 69 cells as shown in the Fig 4.2.1.9.

The nonlinearity cancellation structure with two families of sextupoles provided a good starting point for dynamic aperture optimization. To further optimize dynamic aperture and other aspects of beam performance, additional families were employed.

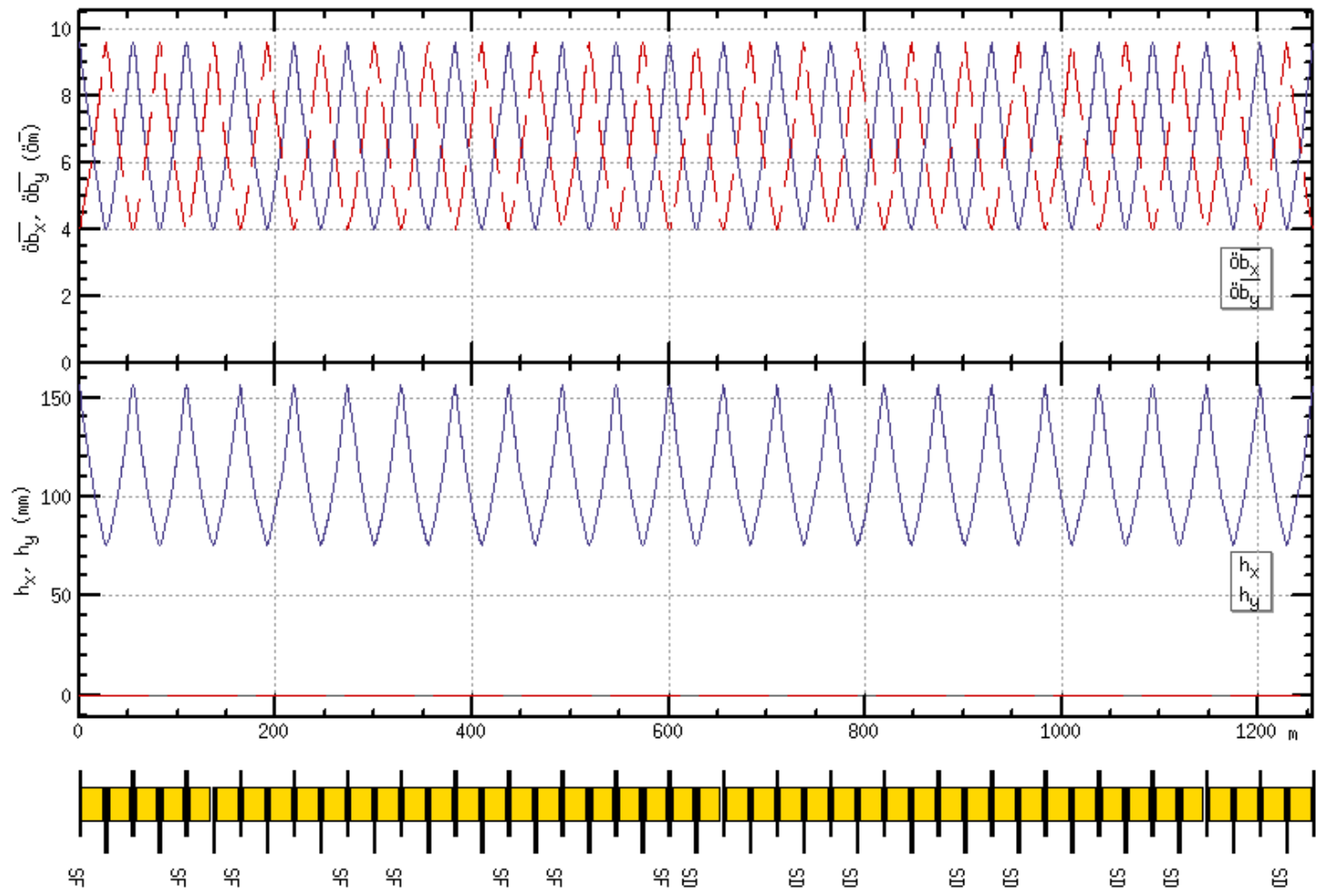

(a)

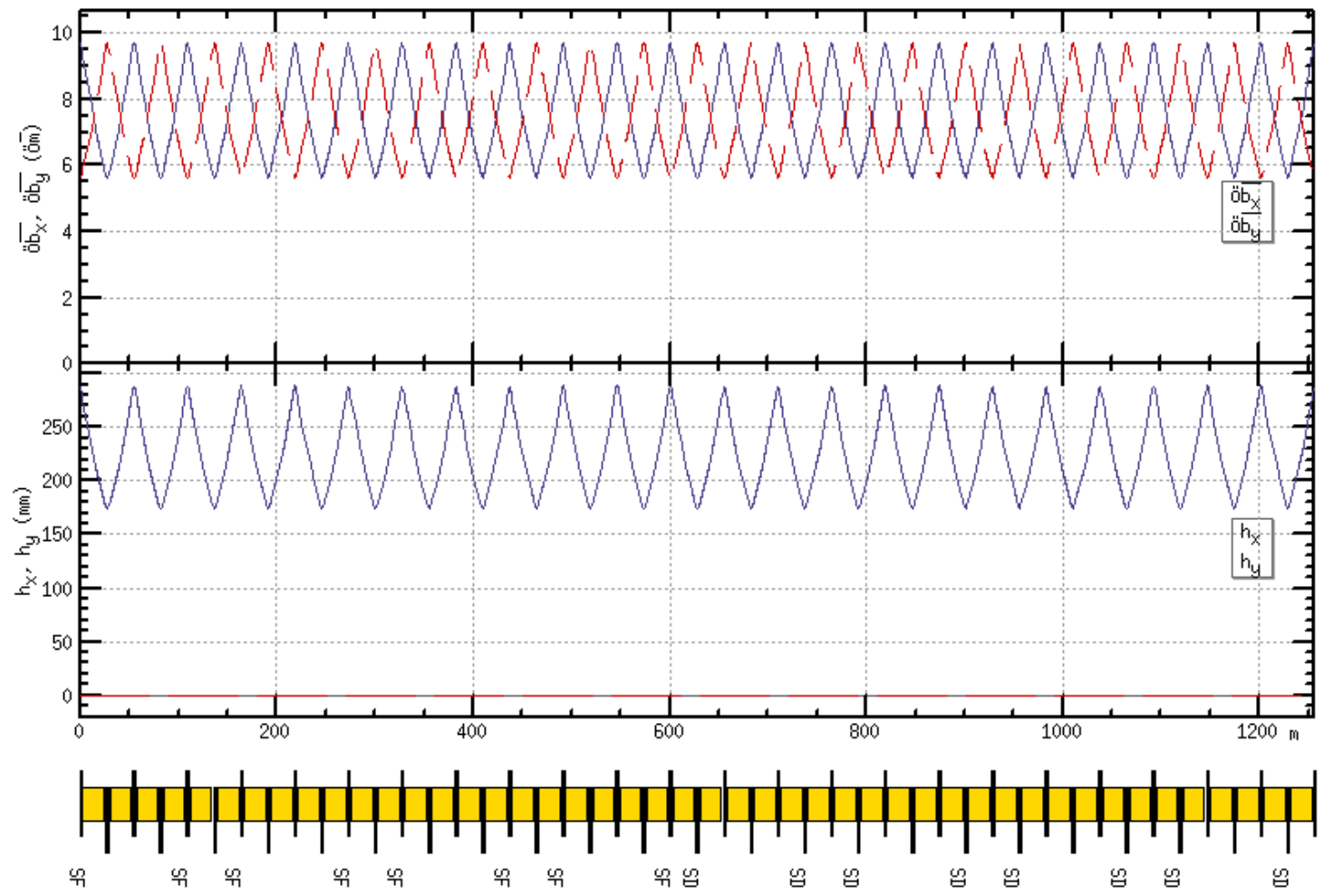

(b)

**Fig 4.2.1.5:** Optics functions of the (a) H/$t\bar{t}$ and (b) W/Z modes.

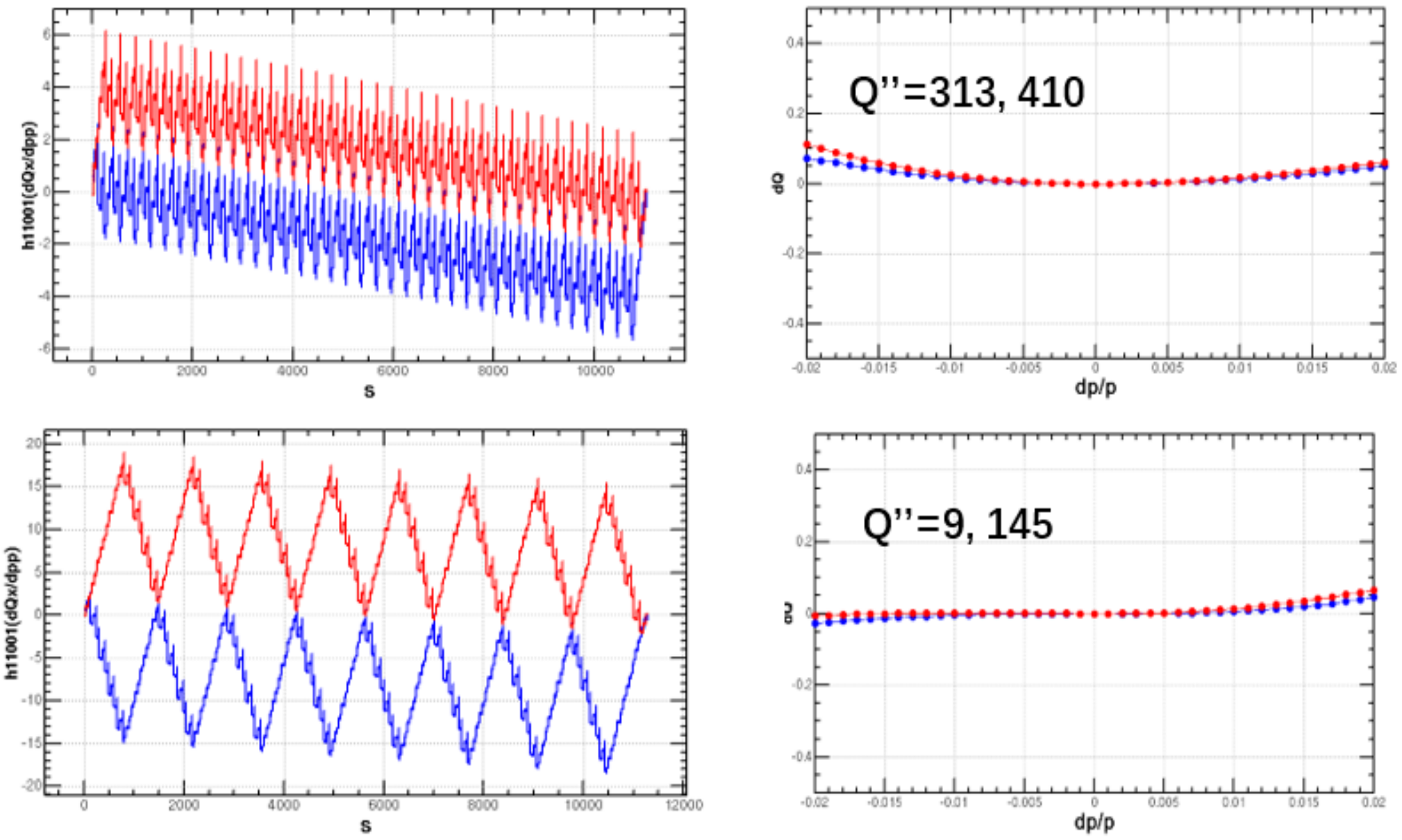

**Fig 4.2.1.6:** Comparison of the one –I sextupole and four –I sextupole schemes.

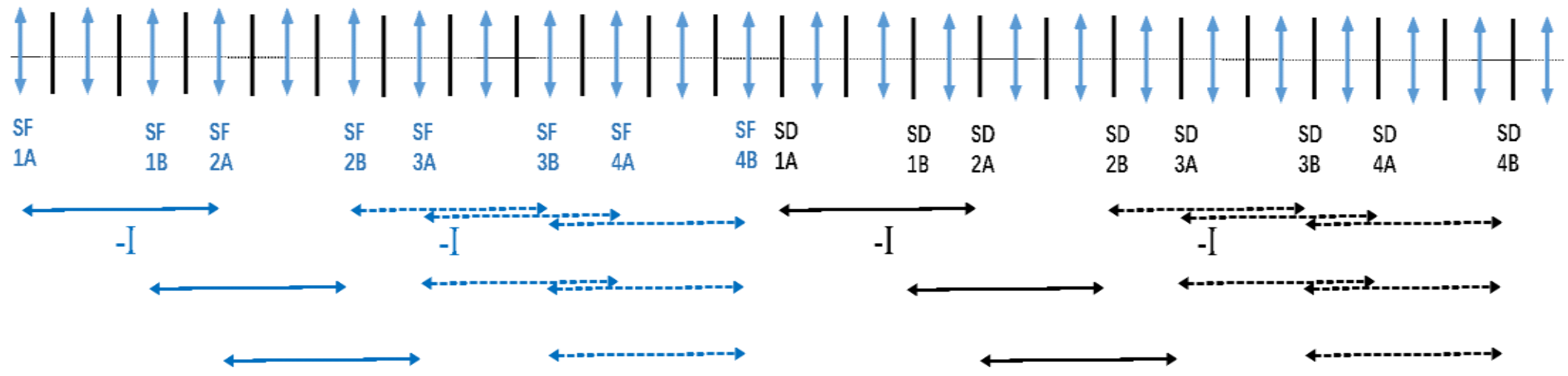

**Fig 4.2.1.7:** The sextupole distributions for the Higgs/$t\bar{t}$ modes and re-combinations for the W/Z modes in each period of the arc region.

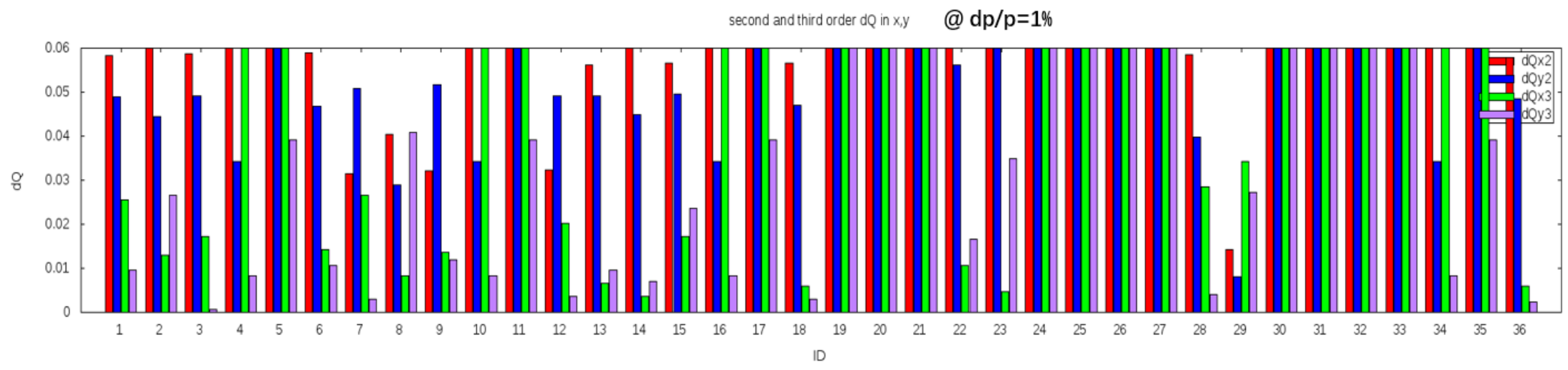

**Fig 4.2.1.8:** The 2$^{nd}$ and 3$^{rd}$ order dQ @dp/p=1% in the x and y plane.

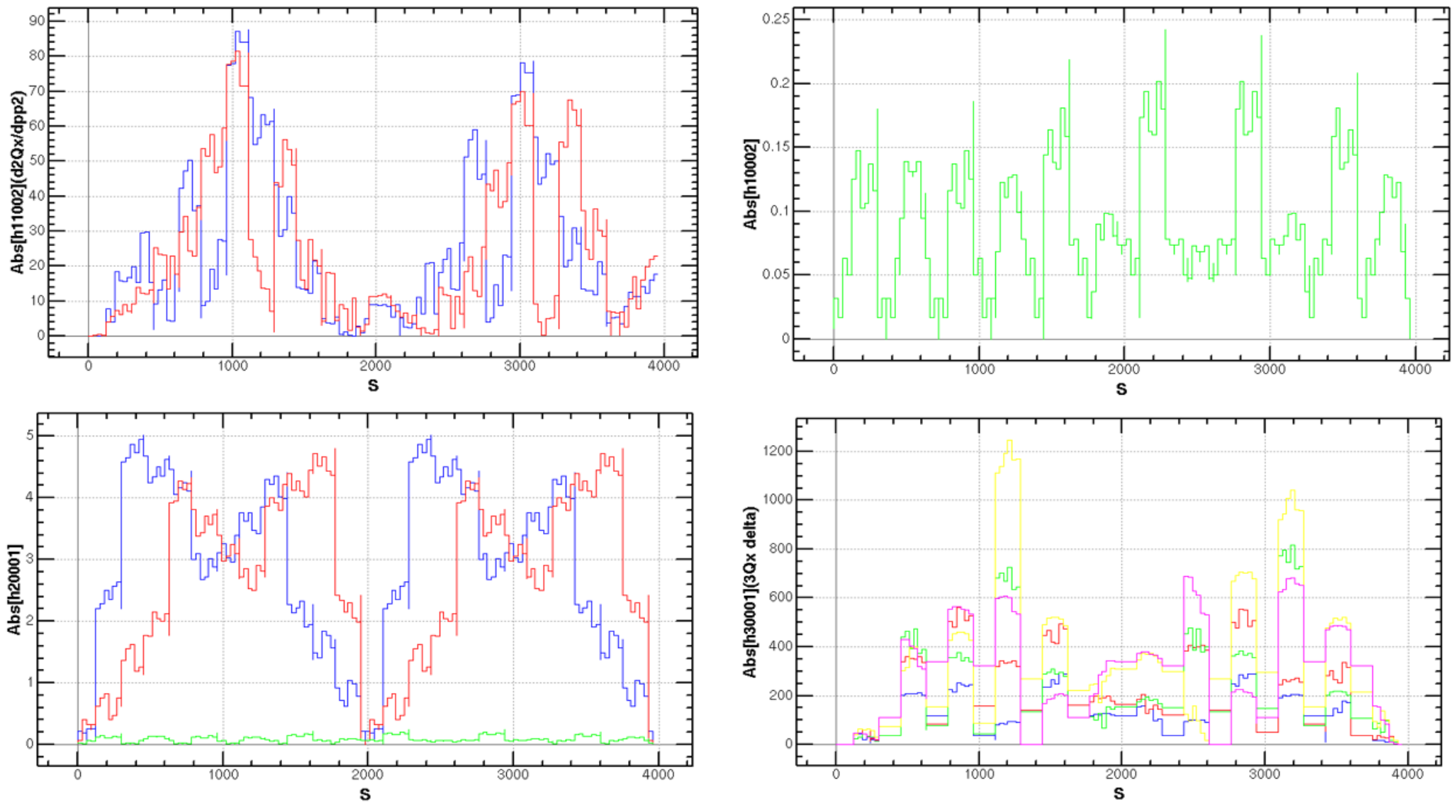

**Fig 4.2.1.9:** Cancellation of the main aberration with 69 cells.

The dispersion suppressor at the ends of the arc region was designed with a half-bending-angle FODO structure. The geometry was re-matched by adjusting the length of nearby drifts.

#### *4.2.1.1.3 RF Region*

A new RF staging was proposed for the high luminosity scheme [19]. The design philosophy is to maintain Higgs running as a first priority and at the same time have flexible switching to keep cost low at an early stage and obtain high luminosity for all modes. The three stages are shown in Figure 4.2.1.10.

In stage 1, the layout and parameters are the same as in the CDR except for a longer central part. For the Higgs mode, the 2-cell RF cavities are shared by the two rings. Thus, each beam will be only filled in half the ring. W and Z modes use the same cavities with Higgs mode to save costs. But lower energy leads to a lower total RF voltage; thus, fewer cavities are used in W and Z modes to lower the impedance. To split the difference of cavity numbers in W and Z modes, half the number of Higgs cavities used in W and Z modes. This provides high luminosity at Higgs and medium or low luminosity at Z due to beam current limit with 2-cell cavities.

An electrostatic separator is combined with a dipole magnet to avoid bending of the incoming beam [3]. The gradient of the electrostatic separator is 2.0 MV/m and its total length 50 m. The magnetic field of the dipole is 60 Gauss. After this combined component, there is a drift as long as 75 m to make the two beam separation distance as large as 10 cm at the entrance of the quadrupole. To limit the beta functions, two triplets are used. Then the beam is further separated with dipoles. The deviation of the outgoing beam is 0.35 m for Higgs mode and 1.0 m for W and Z mode to bypass the half of cryo-modules whose radius is around 0.75 m. In the straight section for cryo-modules, small average beta functions are favoured to reduce the multi-bunch instability caused by the RF cavities. Thus, phase advances of 90/90 degrees is chosen. A quadrupole distance of 13.7 m allows room for a cryo-module.

In stage 2, the high-current 1-cell Z cavities will be installed at a reserved outer side of the RF region. The low-current Higgs cavities are bypassed with the transfer line shown in green when running at the Z. It provides high luminosity at the Z. At this stage, the machine can switch to Higgs and W running when the beams go through both the 2-cell and 1-cell cavities.

In stage 3, the 5-cell low-current, high-gradient and high-Q cavities will be installed in the center for the $t\bar{t}$ running. At this stage, the machine can switch to W/Z running the same as with stage 2 and it is possible to switch to Higgs running by bypassing the $t\bar{t}$ cavities with the transfer line shown in black in the Fig. 4.2.1.10.

The synchrotron radiation resulting from the bypass can be effectively blocked within the vacuum chamber, with the exception of the transition from $t\bar{t}$ to Higgs in stage 3. The shielding scheme for this transition will be thoroughly examined and developed in the engineering design stage. Figures 4.2.1.11-13 illustrate the beam optics specific to each stage of the four modes, while Figures 4.2.1.14 provide a visual representation of the geometry employed in each stage of the four modes.

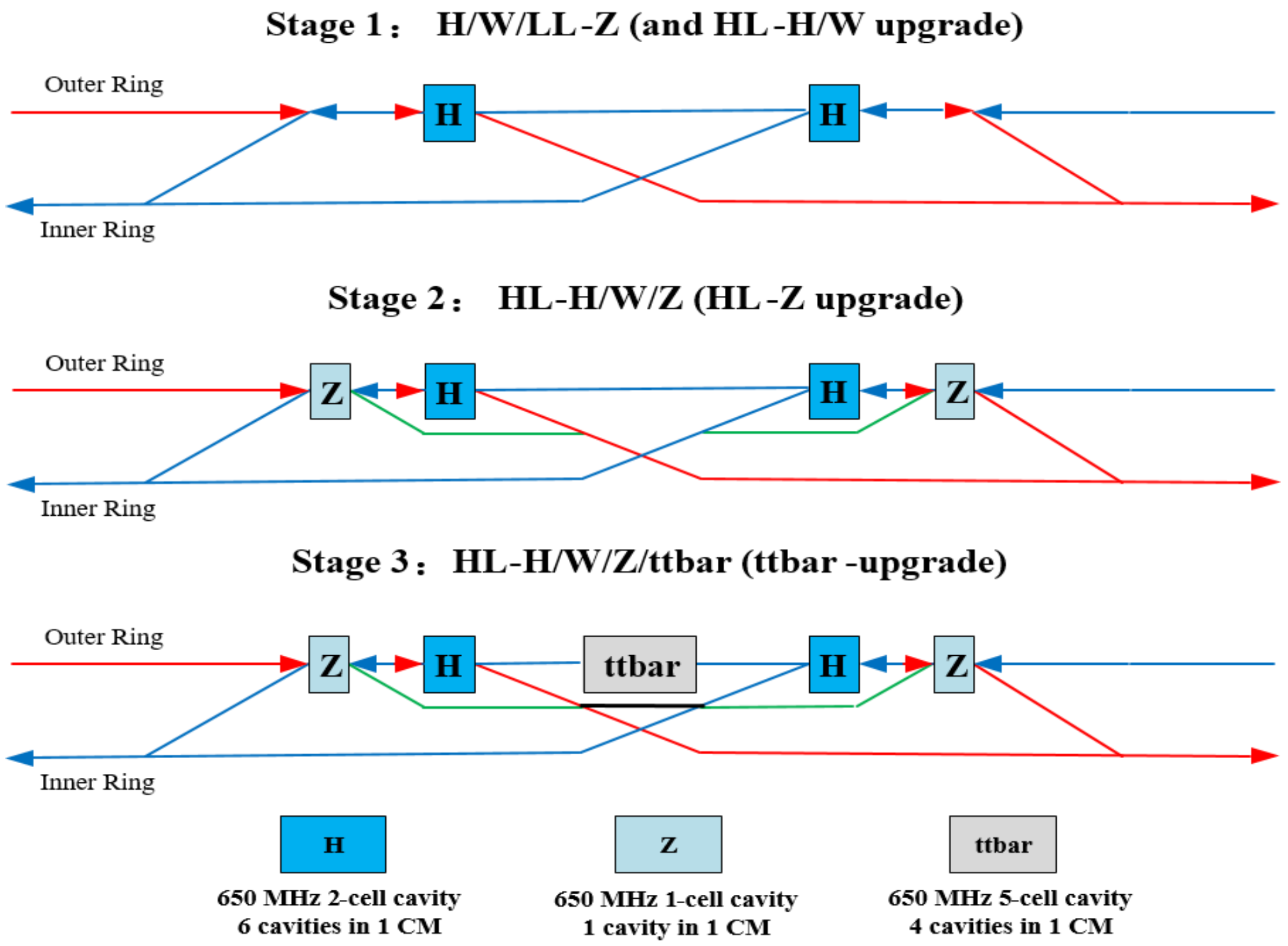

**Figure 4.2.1.10:** Layout of RF region for the 3 stages. In stage 1, the 2-cell H cavities are shared by two rings for the Higgs and W running; Half the number of H cavities are bypassed for W and Z running. In stage 2, the 1-cell Z cavities will be installed, and the low current H cavities are bypassed with the transfer line (shown in green) when running in Z mode; For Higgs and W running, the beams go through both the 2-cell and 1-cell cavities. In stage 3, the 5-cell $t\bar{t}$ cavities will be installed in the center for $t\bar{t}$ running; For W/Z running, the beams go through the same beamline as in stage 2 and is possible to switch to Higgs running by bypassing the $t\bar{t}$ cavities with a transfer line (in black colour).

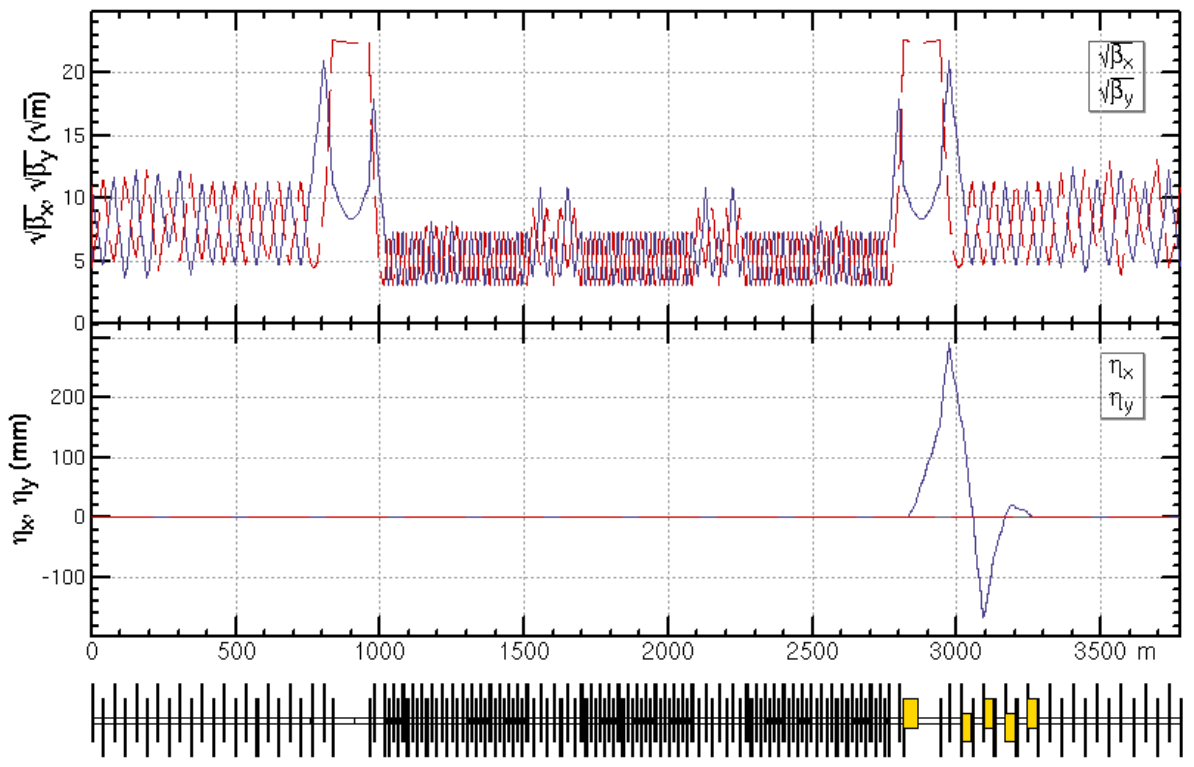

**Figure 4.2.1.11:** Optics of the RF region for Higgs at stage 1/2 and $t\bar{t}$ at stage 3.

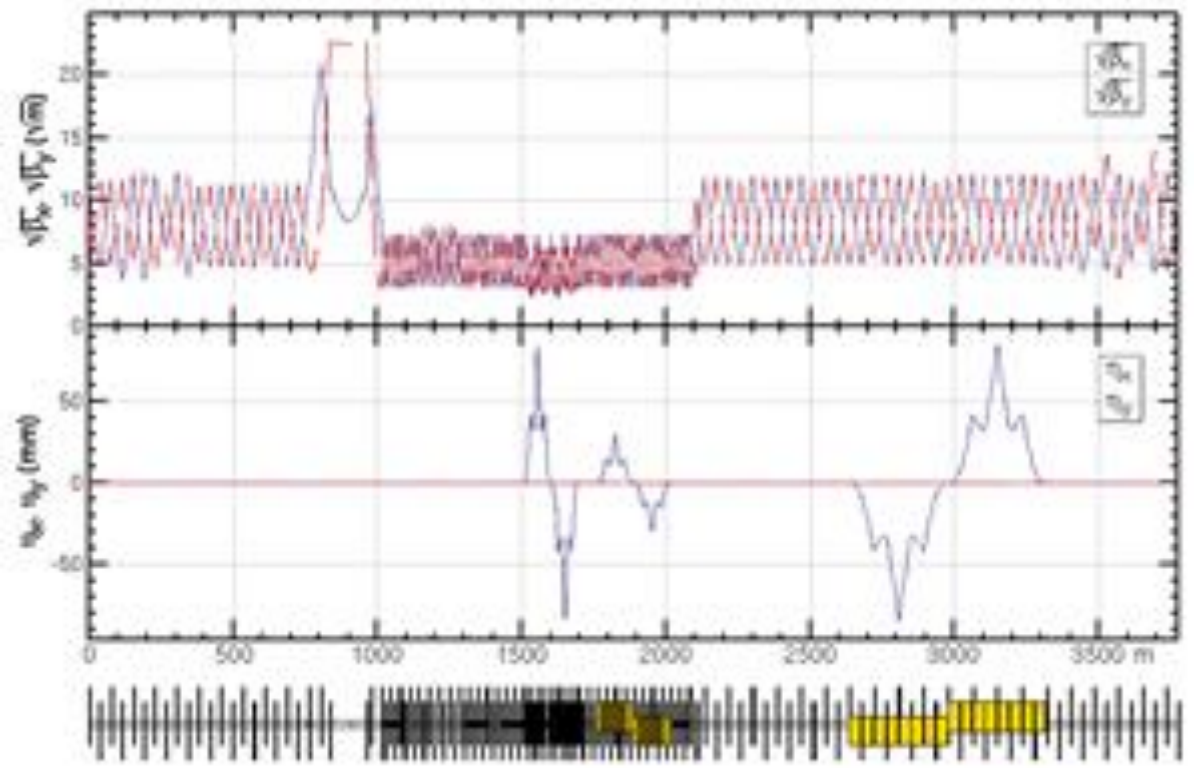

**Figure 4.2.1.12:** Optics of the RF region for W at stage 1/2/3 and Z at stage 1.

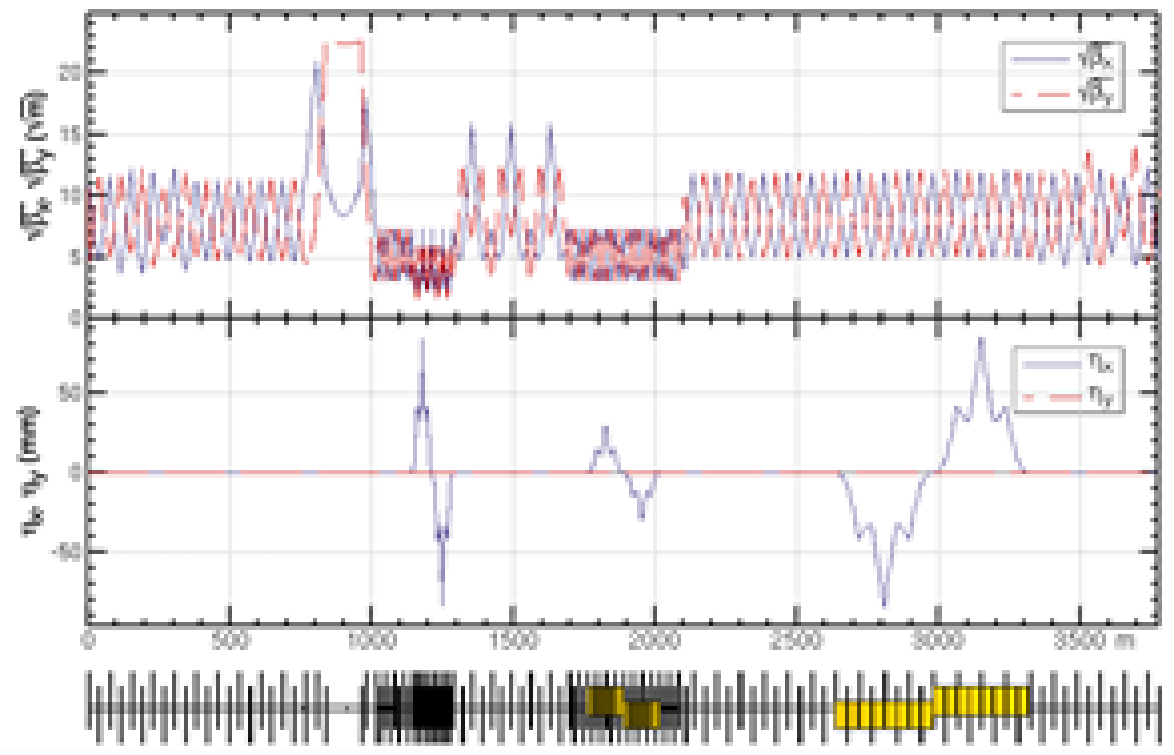

**Figure 4.2.1.13:** Optics of the RF region for Z at stage 2/3.

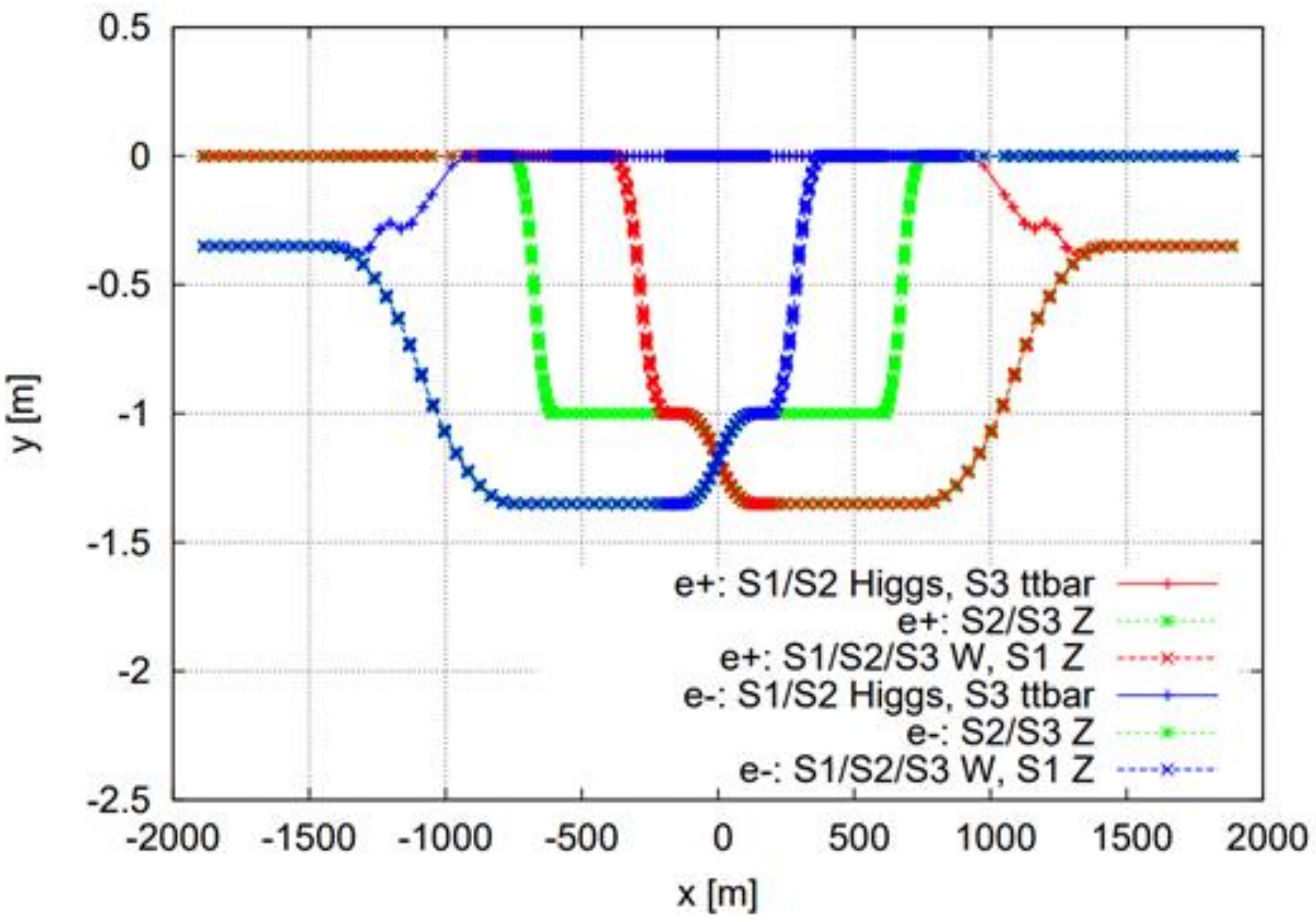

**Figure 4.2.1.14:** Geometry of the beam lines in the RF region for the three stages.

### *4.2.1.1.4 Straight Sections*

The functions in the straight section are phase-advance tuning and injection. For Higgs mode, the on-axis injection scheme is adopted to reduce the requirement from injection while in the off-axis injection scheme for $t\bar{t}$, W and Z modes. Independent magnets are used for the two rings and a longitudinal distance of 0.3 m between the two quadrupoles in the two rings allows a larger size of quadrupoles.

Fig. 4.2.1.15 shows the beam optics for the half ring of the CEPC collider ring. The lattice functions start from the interaction point and the center parts are RF regions.

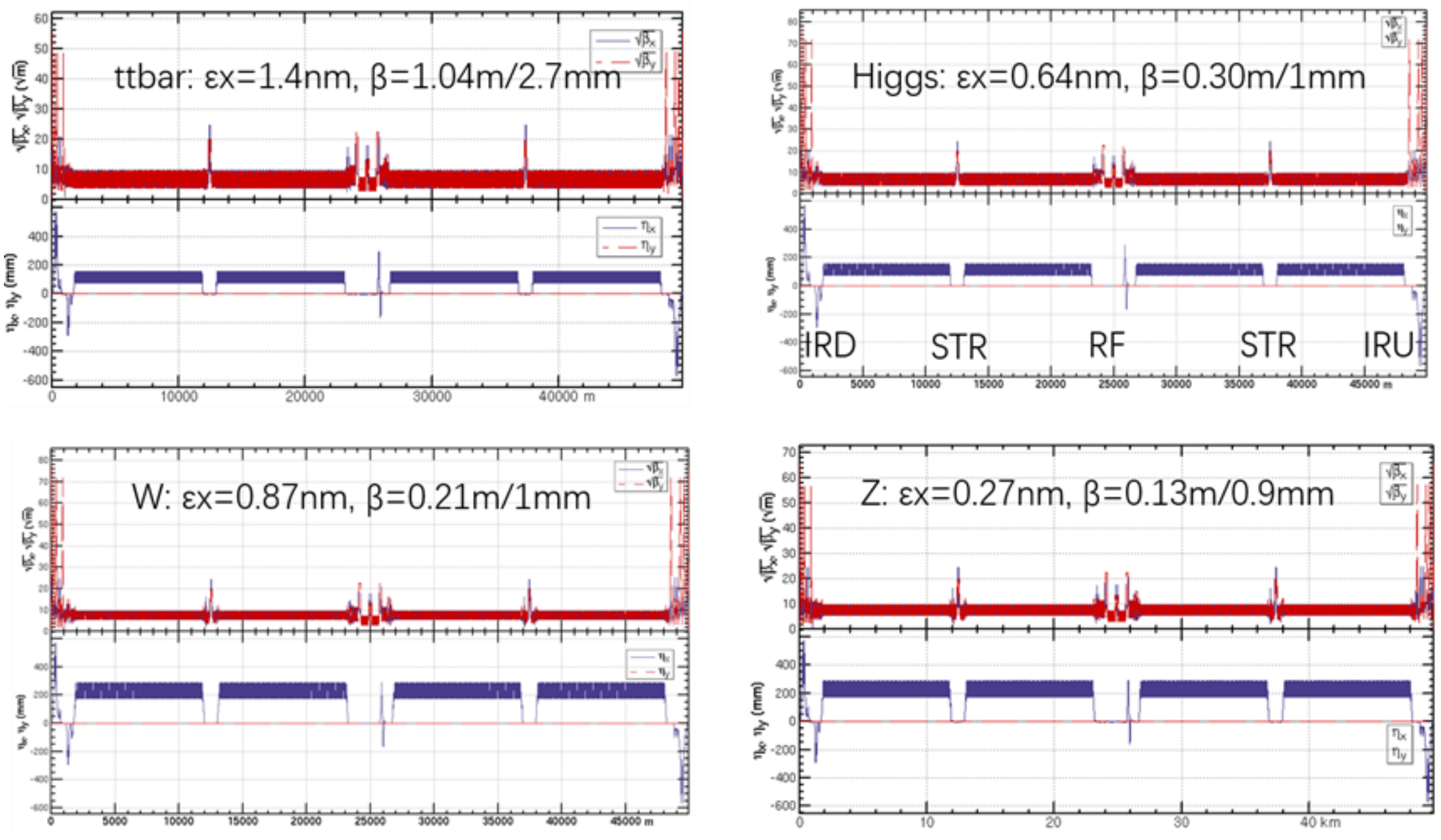

**Figure 4.2.1.15:** Beam optics for half ring of the CEPC collider ring. The lattice functions start from the interaction point and the center parts are RF regions.

#### *4.2.1.1.5 Energy Sawtooth*

At an energy as high as 120 GeV, synchrotron radiation has a pronounced effect on the beam behaviour [20]. With only two RF stations, the sawtooth orbit in the CEPC rings can be as large as around 1 mm for Higgs running. Beams with this off-center orbit will see extra fields in magnets, which will result in ~5% distortion of beam optics and DA reduction. The sawtooth effect is expected to be curable by tapering the magnet strengths corresponding to the beam energy in each magnet, and it's been proven that the beam optics and DA can be recovered with this tapering [1,3]. The sawtooth orbit becomes 1 μm after tapering. The orbit and optics before and after tapering the magnet strength in the CEPC collider ring at Higgs energy is shown in Fig. 4.2.1.16 (a) and (b). The maximum beam energy of the CEPC will be approximately 180 GeV, with a maximum energy deviation of about 1.3%, necessitating an adjustment range of magnet strength around 2.5%.

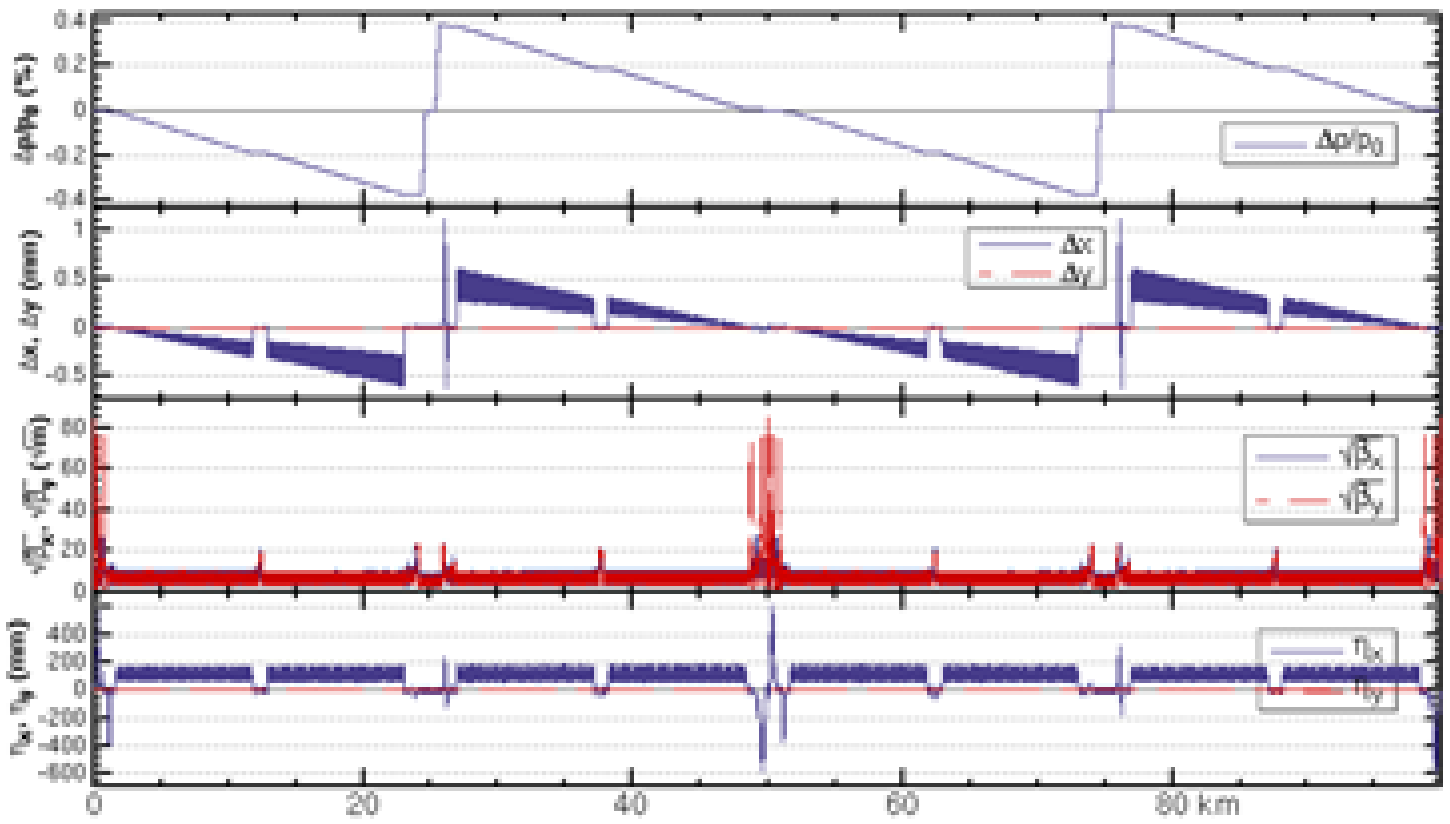

(a)

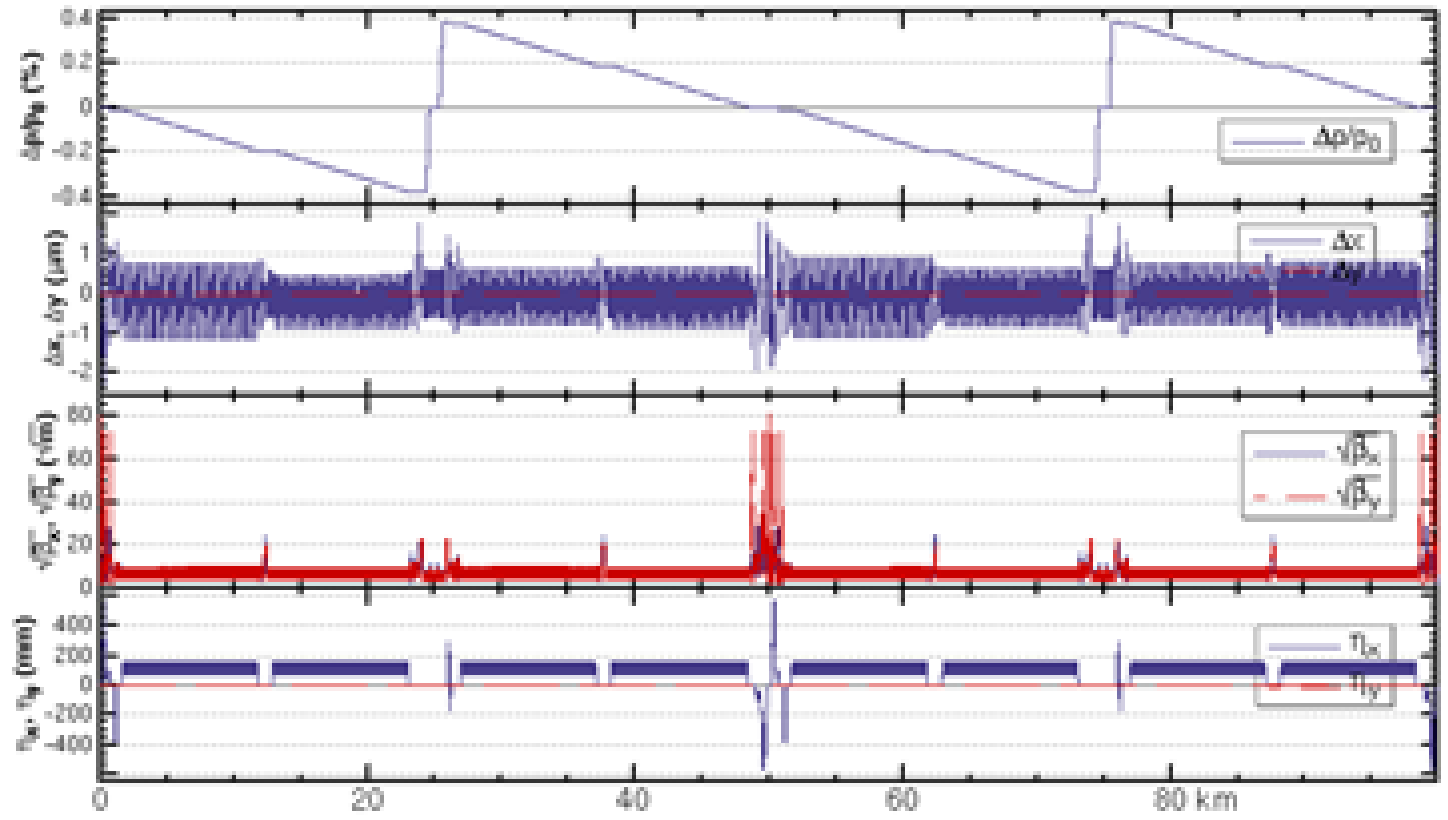

(b)

**Figure 4.2.1.16:** The orbit and optics before (a) and after (b) tapering the magnets strength in the CEPC collider ring at Higgs energy.

#### *4.2.1.1.6 Solenoid Effects*

Achieving high luminosity requires very low vertical emittance. Vertical emittance growth can result from transverse coupling in the arc. Additionally, the solenoid field in the interaction region may contribute to vertical emittance, as the solenoid field in the longitudinal direction ($B_z$) can directly excite transverse coupling, and the fringe field ($B_r$) combined with the crossing angle can induce vertical dispersion, particularly in the large beta region far from the IP.

The solenoid field distribution is depicted in Figure 4.3.4.19. To reduce vertical emittance, an anti-solenoid is positioned as close to the IP as feasible. This cancels out the $\int B_z\, dz$ between the IP and the faces of the final quadrupoles, minimizing transverse coupling. Subsequently, the $B_z$ field distribution at the anti-solenoid fringe is optimized for smoothness to reduce fringe field $B_r$. Lastly, beta function distortion is corrected using the final quadrupoles.

The same solenoid configuration will be used for $t\bar{t}$/Higgs/W/Z. The detector solenoid field is 3 Tesla for $t\bar{t}$/Higgs/W energy and 2 Tesla for the Z energy, primarily due to strict vertical emittance constraints. The optics near the IP with the solenoid field at Higgs energy is illustrated in Figure 4.2.1.17. As indicated in Table 4.2.1.3, the vertical emittance induced by the solenoid field complies with the machine parameter requirements. Furthermore, the impact of the solenoid field on the dynamic aperture is negligible.

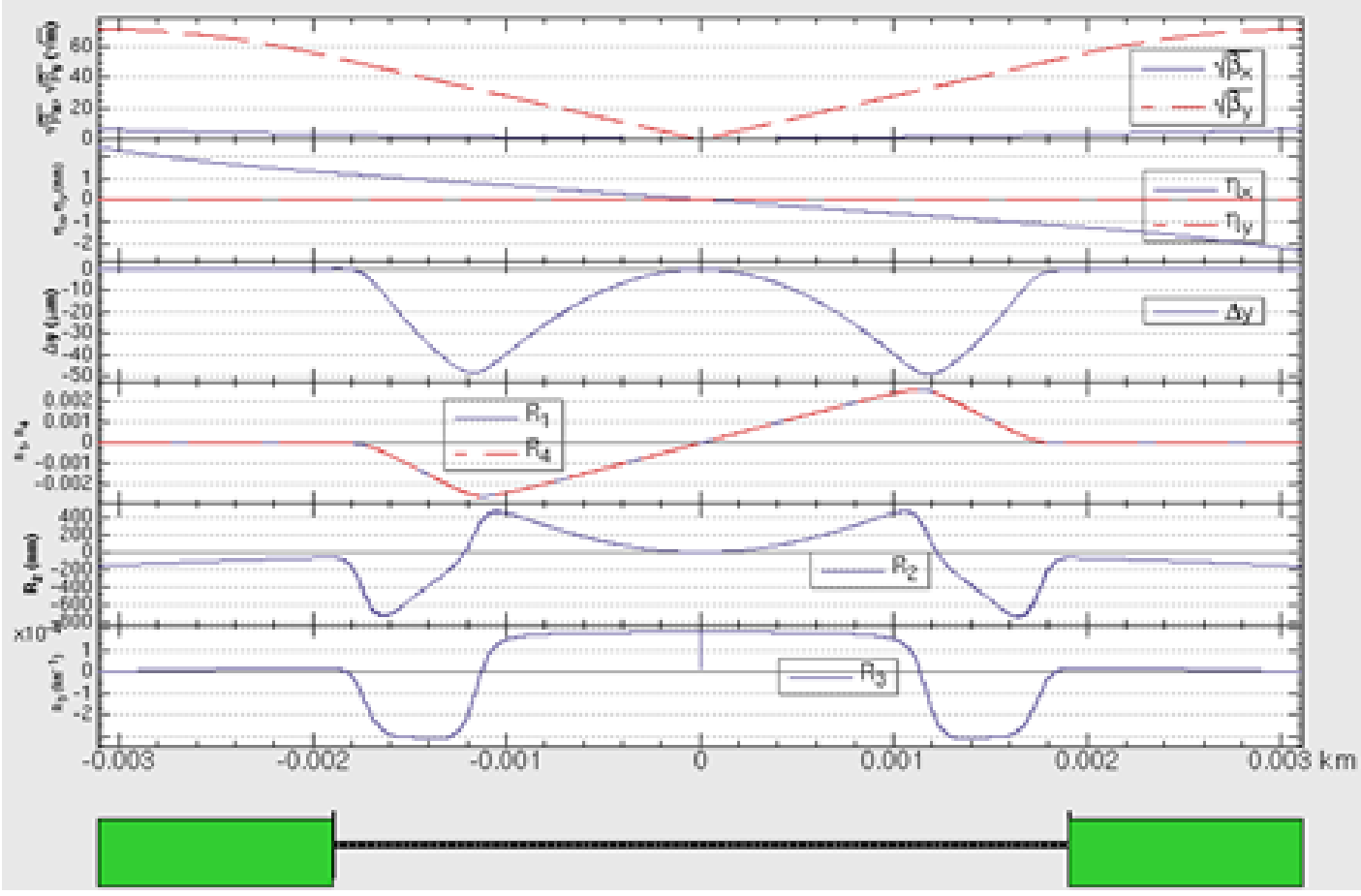

**Figure 4.2.1.17**: Optics near the IP with solenoid field at Higgs energy.

**Table 4.2.1.3:** Vertical emittance due to the solenoid field

| | **$t\bar{t}$**<br>**3 T** | **Higgs**<br>**3 T** | **W**<br>**3 T** | **Z**<br>**2 T** |
|---|---|---|---|---|
| Nominal $\varepsilon_y$ [pm] | 4.7 | 1.3 | 1.7 | 1.4 |
| Solenoid induced $\varepsilon_y$ [pm] | 0.04 | 0.36 | 1.20 | 0.95 |

### 4.2.1.2 *Dynamic Aperture Optimization*

For the CEPC, the dynamic aperture is significantly influenced by strong synchrotron radiation especially at Higgs and $t\bar{t}$ energies. Strong synchrotron radiation causes strong radiation damping which helps enlarge the dynamic aperture to some extent [3,14,21]. The effect, shown in Fig. 4.2.1.18, is clear, especially for large off-momentum particles. The radiation damping with dipole increases the dynamic aperture and the peaks for on-momentum particles are kept as small energy oscillations are generated with large transverse amplitude. The radiation damping with quadrupoles destroys the peak of on-momentum dynamic aperture as large energy oscillations are generated with transverse amplitude which is larger than the half-width of the dynamic aperture peak. For the vertical plane, the dynamic aperture is limited by the self-inducing parametric resonance [22].

Quantum fluctuations in synchrotron radiation are considered in SAD, where the random diffusion due to synchrotron radiation in particle tracking is implemented in each magnet. Quantum excitation with dipole fluctuate the dynamic aperture a bit around the result including damping. In the horizontal plane, the quantum excitation with quadrupole fluctuates the dynamic aperture a bit around the result including damping as a quite uniform distribution of the radiation power in the quadrupole. For the vertical plane, the dynamic-aperture fluctuation is large as the radiation power is dominant in the final quadrupoles [14]. The dynamic aperture for the same lattice with and without radiation fluctuations is shown in Fig. 4.2.1.19, which shows that the fluctuation mainly comes from the vertical direction. The difference mainly comes from radiation in the final-focus quadrupoles in the interaction region.

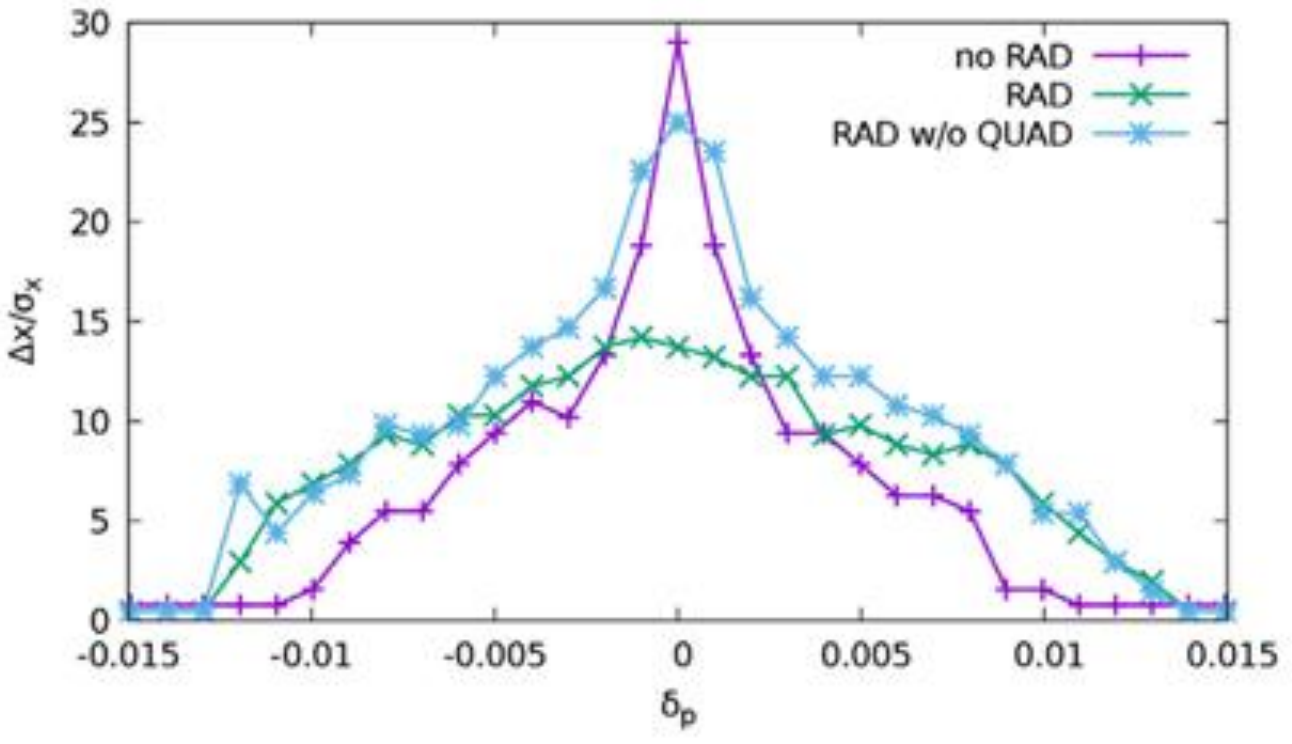

**Figure 4.2.1.18**: DA in horizontal direction with radiation damping on/off and without radiation damping from quadrupoles [21].

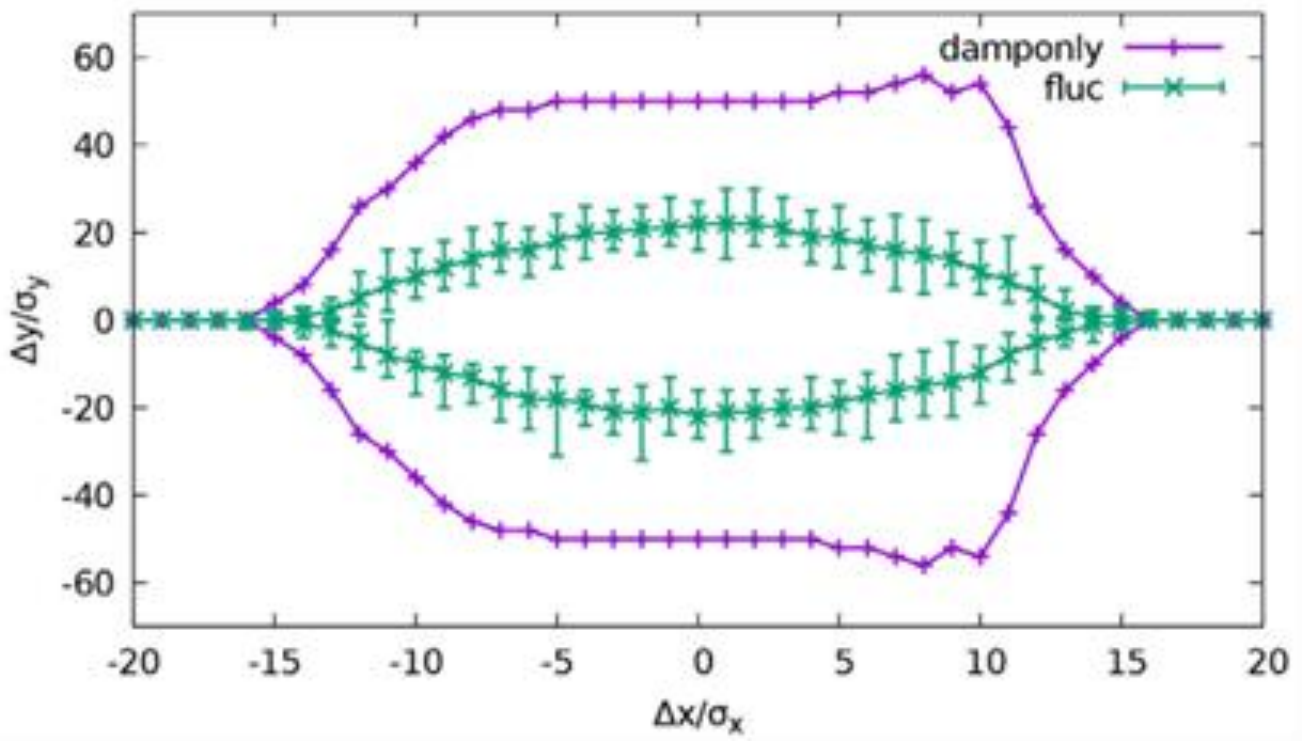

**Figure 4.2.1.19:** On-momentum DA in transverse space with radiation fluctuation and with radiation damping in each element [21].

A differential evolution algorithm-based optimization code has been developed, which is a multi-objective code called MODE [21,23]. It was shown that multi-objective optimization is an effective way to optimize the dynamic aperture for both on-momentum and off-momentum particles including the effects of synchrotron radiation in all the magnets. In order to reduce the random noise, the DA result is clipped to ensure the DA at large momentum deviation will be less than that at small deviation, as shown in Fig. 4.2.1.20.

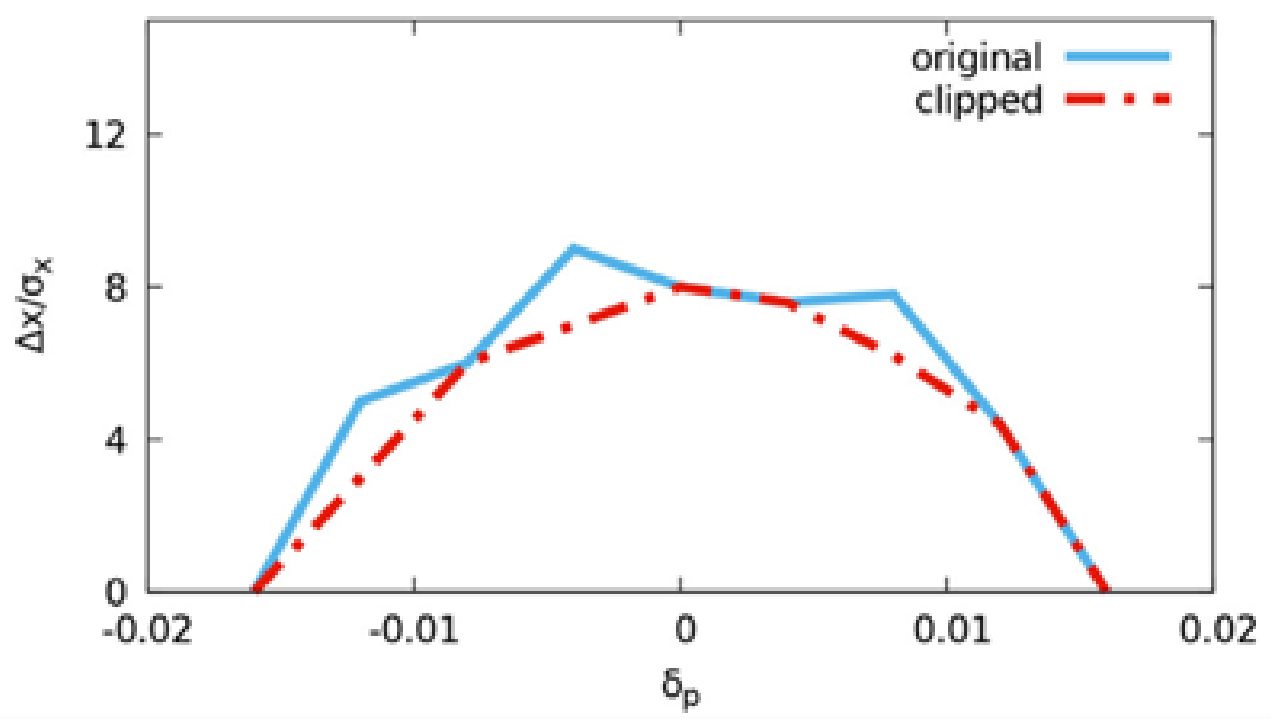

**Figure 4.2.1.20:** An example of DA clipping [21].

The tracking to get dynamic aperture without errors was done with turns for one transvers damping time for 4 initial phases. The effects of synchrotron motion, radiation loss in all magnets, magnet strength tapering with beam energy, crab-waist sextupoles, Maxwellian fringe fields, kinematic terms at the interaction points, finite length of sextupoles etc. are included in the tracking. The SAD code is used to do the optics calculation and dynamic-aperture tracking.

In CEPC, the dynamic aperture is limited by a combination of various complicated effects. To maximize the dynamic aperture, a large number of knobs is always necessary. For the Higgs and $t\bar{t}$ modes, the dynamic aperture was optimized with a total of 84 knobs including the strengths of families of 64 sextupoles in the arc region, 8 sextupoles, and 4 multipoles in the interaction region and 8 phase advances between different regions. For the Z and W modes, the dynamic aperture was optimized with a total of 116 knobs including the strength of families of 96 sextupoles in the arc region, 8 sextupoles, and 4 multipoles in the interaction region and 8 phase advances between different regions. The optimized DA at $t\bar{t}$/Higgs/W/Z/ energies is shown in Figs. 4.2.1.21. The requirements of dynamic aperture from injection and beam-beam effect to get efficient injection and adequate beam lifetime are listed in the Tab. 4.2.1.4. The DA meets the requirement of injection and colliding-beam lifetime.

**Table 4.2.1.4:** Requirements on dynamic aperture

| Operation mode | On-axis injection | Off-axis injection |
|---|---|---|
| $t\bar{t}$ | - | $11\sigma_x \times 16\sigma_y \times 2.0\%$ |
| Higgs | $8\sigma_x \times 20\sigma_y \times 1.6\%$ | $13.5\sigma_x \times 20\sigma_y \times 1.6\%$ |
| W | - | $8.5\sigma_x \times 20\sigma_y \times 1.05\%$ |
| Z | - | $11\sigma_x \times 23\sigma_y \times 1.0\%$ |

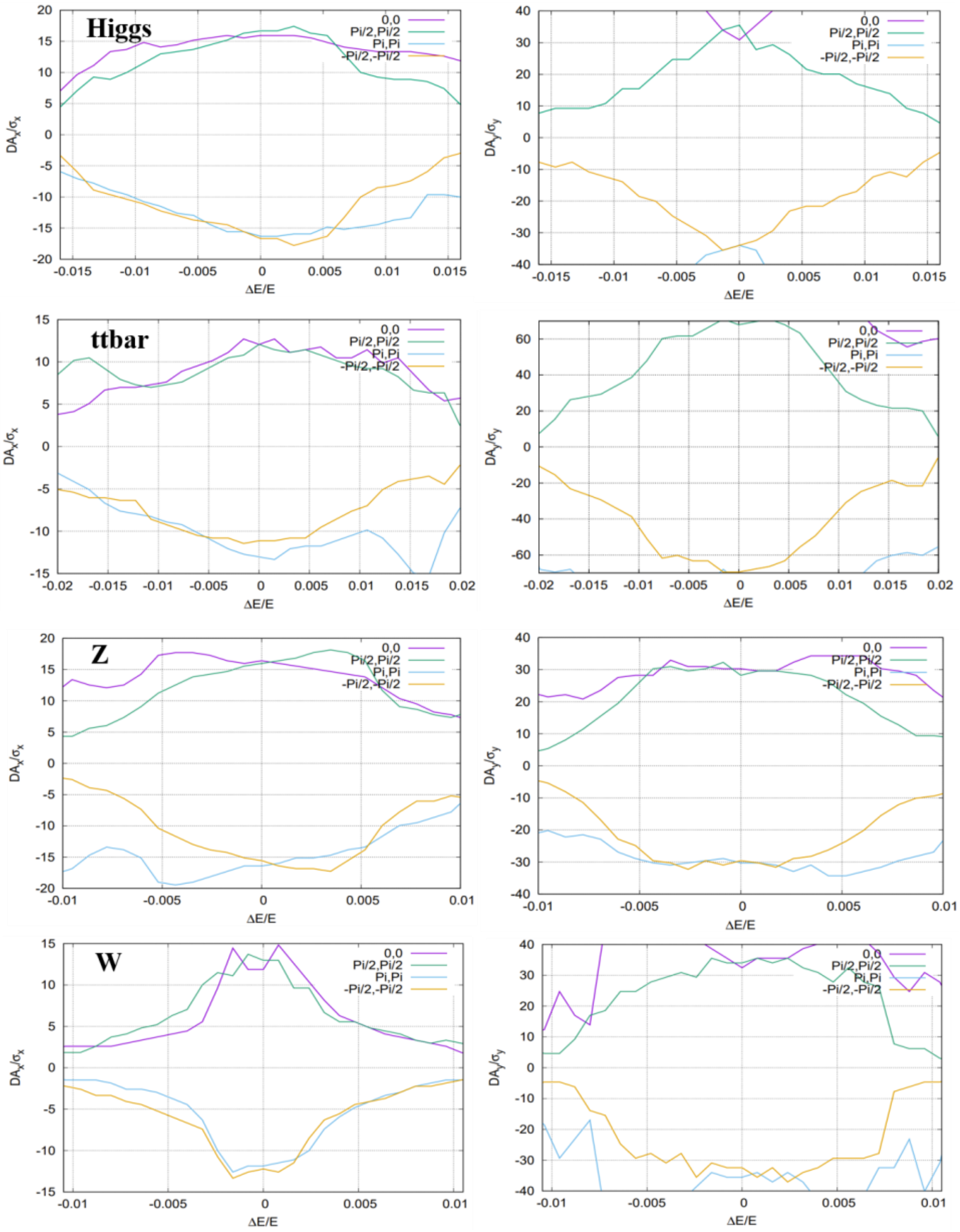

**Figure 4.2.1.21:** Dynamic aperture at IP with 90% survival of 100 samples for the four modes. The left and right figures show the normalized dynamic aperture in the horizontal and vertical directions respectively. The four lines in each figure show the dynamic aperture with different initial phases. The particles were tracked with turns for one transvers damping time.

Dynamic aperture is usually strongly correlated to the beam lifetime. A longer beam lifetime is expected with a larger dynamic aperture. However, it is found that sometimes larger dynamic aperture does not ensure longer lifetime [21, 24, 25]. The turns of tracking for dynamic aperture, about 150 for Higgs energy, is much less than that for lifetime which is about 50 000. Thus, in these simulations, the dynamic aperture is a result of short-term behaviour, but beam lifetime is a long-term result. Simulation showed that a shorter lifetime means that more large-amplitude particles exist in equilibrium distribution which is a non-Gaussian distribution. None of the particles is lost because the amplitude is far from the boundary of dynamic aperture [21].

An analysis based on the diffusion map is developed to describe the influence of combined effects where radiation fluctuation and beamstrahlung dominate [21]. The function in diffusion map based on tracking is defined as

$$f(ax, ay) = log_{10}\left(\sum_{turn} \sigma_a^2\right)$$

where $\sigma_a = \sqrt{\sigma_{ax}^2 + \sigma_{ay}^2 + \sigma_{az}^2}$, $\sigma_{ai}$ is represented as the RMS value of particles' amplitude ai $= \sqrt{2J_i/\varepsilon_i}$ (i=x,y,z). Fig. 4.2.1.22 show the diffusion maps of four cases with different configuration of sextupoles which have almost the same dynamic aperture but the case 3 and case 4 have better lifetime.

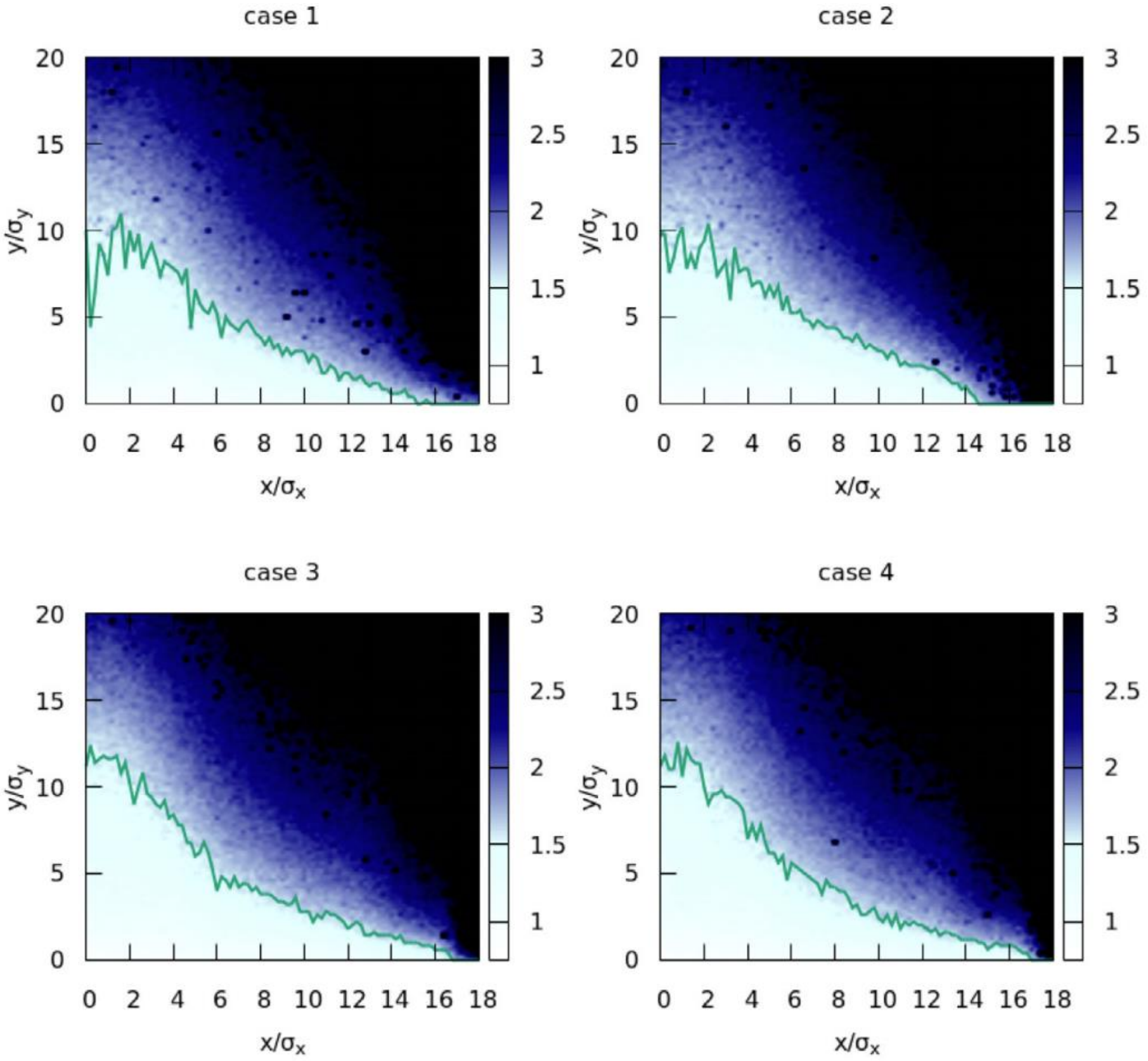

**Figure 4.2.1.22:** The diffusion maps of four cases with CEPC collider lattice at Higgs energy [21]. The green lines are corresponding to the value of the function $f$ = 1.52. The particles were tracked for only 25 turns at each initial amplitude.

This method presents good consistency with beam lifetime by tracking. Constraints of diffusion rate in amplitude space are used in optimization of dynamic aperture, solutions always present good results of beam lifetime. Another method [26] of using a beam with bigger beam size than the equilibrium and tracking with less turns than beam-beam simulation to optimize the lifetime itself was also applied as one of constraint. During the tracking of dynamic aperture and lifetime, the lattice nonlinearity, synchrotron radiation, weak-strong beam-beam interaction and beamstrahlung effects were included. Following optimization with 256/128/128 families of the arc sextupoles, the beam tail distribution and lifetime for the Higgs/$t\bar{t}$/W energy are depicted in Figure 4.2.1.23 and Table 4.2.1.5. The simulated beam lifetime, accounting for lattice, beam-beam, and beamstrahlung effects at the Higgs/$t\bar{t}$/W energy, meets the specified requirement. Investigations regarding beam lifetime at Z energy are currently in progress.

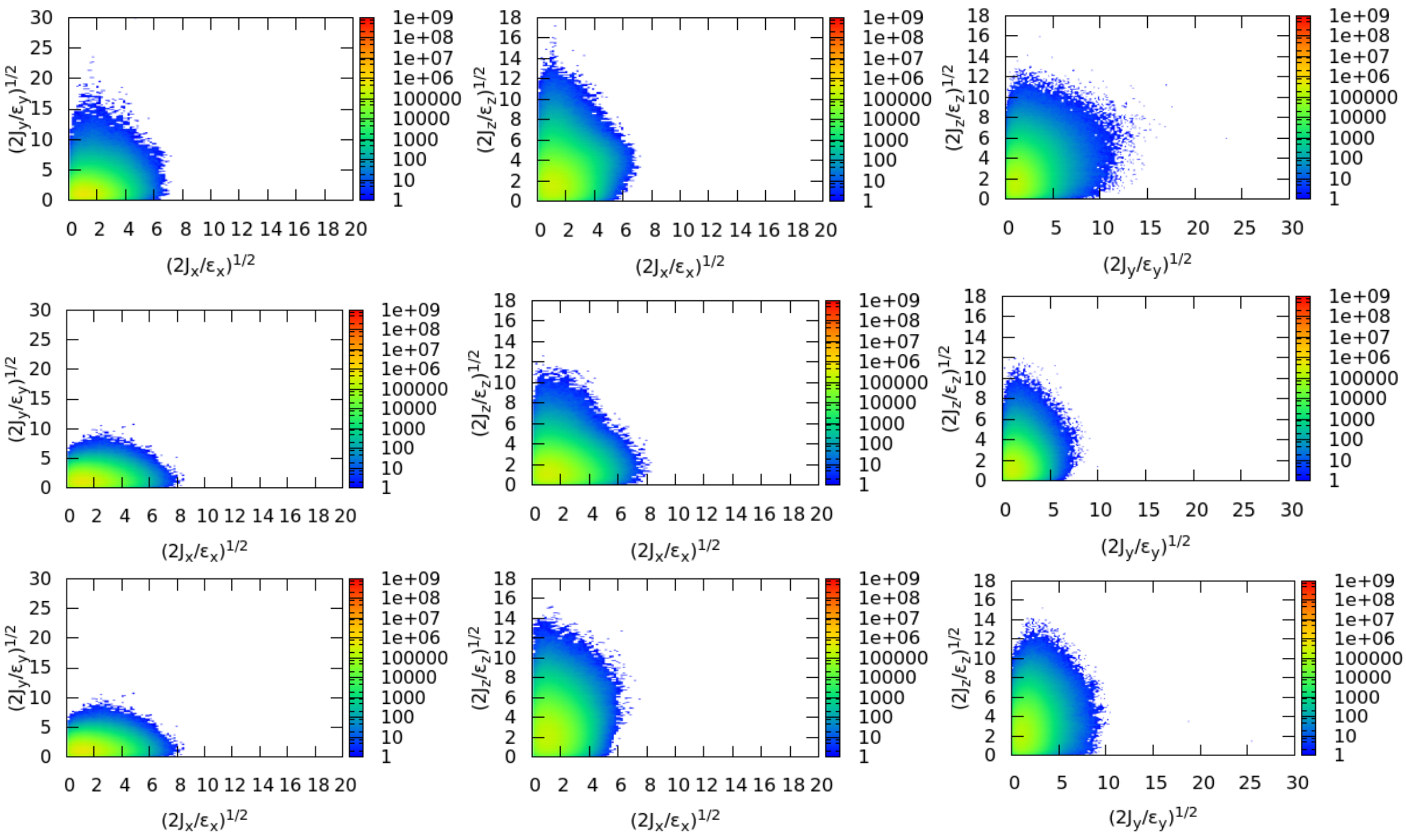

**Figure 4.2.1.23:** Beam tail distribution in x–y, x-z and y-z space for the tracking result at Higgs/ttbar/W (top/middle/bottom) energy with lattice, beam-beam interaction and beamstrahlung effects.

**Table 4.2.1.5:** Beam lifetime with lattice, beam-beam interaction and beamstrahlung effects

| | **Higgs** | **$t\bar{t}$** | **W** |
|---|---|---|---|
| Beam lifetime requirement for top-up injection [min] | 18 | 18 | 22 |
| Bhabha lifetime [min] | 40 | 80 | 60 |
| Required beam lifetime with lattice + beambeam + beamstrahlung [min] | 33 | 24 | 35 |
| Simulated beam lifetime with lattice + beambeam + beamstrahlung [min] | 139 | 278 | 139 |

### 4.2.1.3 *Performance with Errors*

The low emittance and small beta function of the lattices described above have an enhanced sensitivity to the misalignment of magnets and field errors, especially the misalignment of quadrupole magnets in the interaction region (IR). The imperfection of magnets will cause distortion of closed orbit and optics, and then decrease the dynamic aperture. A series of correction methods are performed to recover the design requirements of emittance and dynamic aperture.

#### *4.2.1.3.1 Error Assumptions*

In the correction study of the Higgs lattice in the CDR [1, 27-28], a couple of cases with different misalignment assumptions of magnets were studied. In particular, misalignment of quadrupoles in the interaction region were studied with 100 random lattice seeds. The closed orbit, dispersion, beta-beating, emittance and dynamical aperture after correction are comparable, while the correction time and iteration time are increased, and the convergence rate is decreased as the errors increase. The most imperfect error assumption was studied for these four lattices in the TDR. Table 4.2.1.6 and Table 4.2.1.7 list the misalignment errors with respect to the reference orbit and field errors of all magnets, respectively. All error sources follow a Gaussian distribution truncated at ±3$\sigma$.

**Table 4.2.1.6**: Misalignment RMS error requirements for the CEPC collider ring.

| Component | $\Delta$x (μm) | $\Delta$y (μm) | $\Delta\theta_z$ (μrad) |
|---|---|---|---|
| Dipole | 100 | 100 | 100 |
| Quadrupole | 100 | 100 | 100 |
| Sextupole | 100 | 100 | 100 |

**Table 4.2.1.7**: Field error requirements for the CEPC collider ring.

| Component | Field error (RMS) |
|---|---|
| Dipole | 0.01% |
| Quadrupole | 0.02% |
| Sextupole | 0.02% |

#### *4.2.1.3.2 Correction Scheme*

The correction algorithms are mainly based on SAD [29] and Matlab-based accelerator toolbox (AT) software [30,31]. Beam-position monitors (BPM) and correctors are arranged to correct the closed orbit. Two BPMs and a pair of correctors (one each for horizontal and vertical) are installed in each cell. For the cells accommodating sextupoles, horizontal and vertical correctors are produced by the sextupole trims. For other cells, individual horizontal and vertical corrector magnets are located close to the focusing and defocusing quadrupoles, respectively. Therefore, there are four BPMs and four corrector magnets for each plane per betatron period. Firstly, a closed-orbit distortion (COD) correction was performed with sextupoles off; then the sextupoles were turned on and the COD correction repeated. The precision of the beam-based alignment (BBA) for sextupoles is assumed to be 10 μm. The dispersion correction and beta-beating correction are also used for optics correction. Dispersion-free steering (DFS) [32] and linear optics from closed-orbits algorithm (LOCO) [33] are the corresponding methods. The coupling and vertical dispersion correction are used to decrease the vertical emittance. The above correction scheme is iterated until the emittance and tracking dynamic aperture satisfy the design requirements. To offer a broader insight into the effectiveness of error correction mechanisms, we use the correction of a lattice seed in the Higgs mode as an illustrative example.

##### 4.2.1.3.2.1 Closed-Orbit Correction

Closed-orbit distortion is caused by quadrupole misalignments and integral field errors in the dipoles. Lattices with errors have no closed orbit before correction. The orbit correction algorithm is based on the response matrix. By inverting the response matrix using singular value decomposition (SVD) [34], the steering strengths can be derived to correct the distorted orbits. Figure 4.2.1.24 shows the closed orbit after correction. The maximum and RMS closed orbit is about 0.1 mm and 40 μm for both horizontal and vertical orbit, respectively. Figure 4.2.1.25 shows the corresponding strength of horizontal and vertical correctors; the maximum and RMS strength of both horizontal correctors and vertical correctors are about $1.0 \times 10^{-4}$ and $7.7 \times 10^{-6}$, respectively.

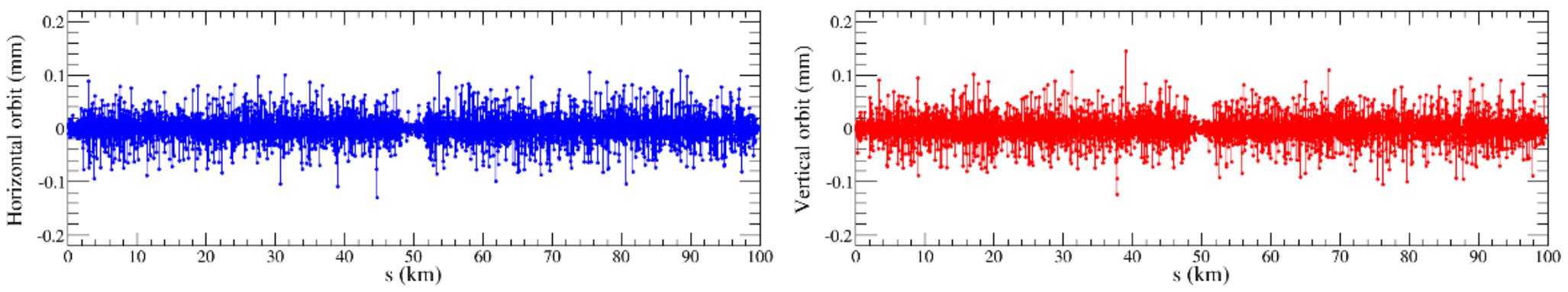

**Figure 4.2.1.24**: The closed orbit after COD correction for a lattice seed of the Higgs mode.

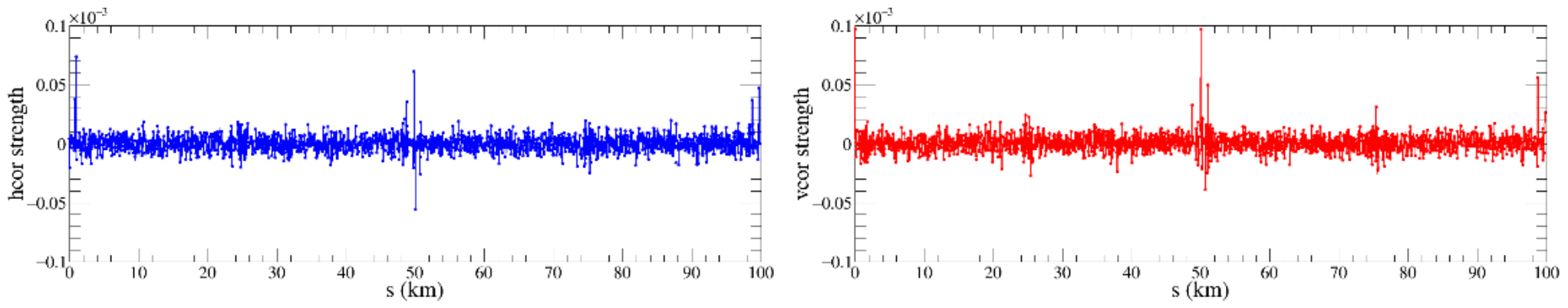

**Figure 4.2.1.25**: The strength of horizontal (left) and vertical (right) correctors for a lattice seed of the Higgs mode.

#### 4.2.1.3.2.2 Dispersion Correction

The principle of DFS is a simultaneous correction of the orbit and the dispersion by knob corrections, the strength of correctors is calculated by the following function:

$$\begin{pmatrix}(1-\alpha)\vec{\mu}\\ \alpha\vec{D}_{\mu}\end{pmatrix} + ((1-\alpha)A\alpha B)\vec{\theta} = 0,$$

where vectors $\vec{\mu}$, $\vec{D}_{\mu}$ and $\vec{\theta}$ are the orbit vector, dispersion vector and corrector strengths vector, respectively. $A$ and $B$ are the orbit response matrix and dispersion response matrix. The weight factor α is used to shift from a pure orbit (α=0) to a pure dispersion correction (α=1). In general, the optimum closed orbit and dispersion rms are not of the same magnitude and α must be adjusted for a given lattice. Figure 4.2.1.26 shows the dispersion correction result of one lattice seed. We can see that the RMS horizontal and vertical dispersion has been reduced from 16.4 mm to 0.4 mm and 11.1 mm to 0.4 mm, respectively.

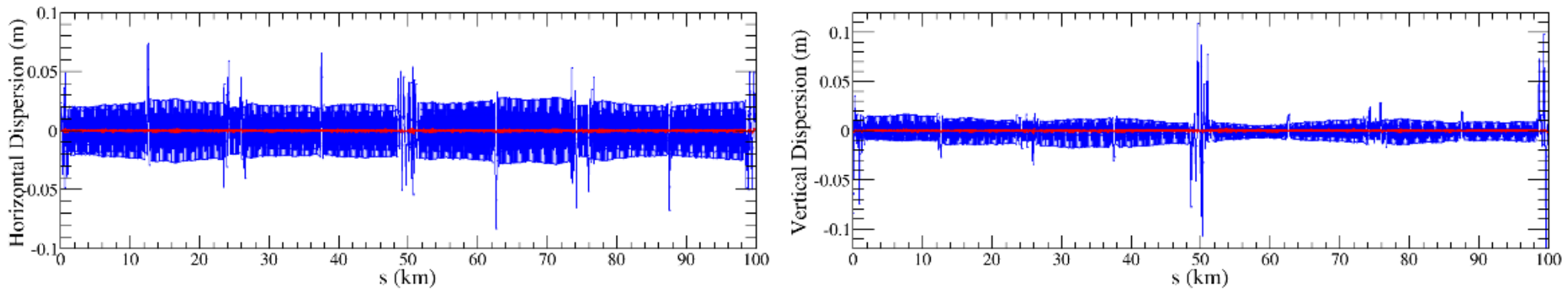

**Figure 4.2.1.26:** The horizontal dispersion (left) and vertical dispersion (right) for a lattice seed in the Higgs mode, where the blue curves are the dispersion distortion before dispersion correction and the red curves are the dispersion distortion after dispersion correction.

#### 4.2.1.3.2.3 Beta-Beating Correction

The beta-beating correction is performed for lattice seeds with dispersion correction. LOCO based on AT is applied to restore the optics. By fitting the response matrix, the quadrupole strengths that best reduce the beta-beatings and dispersion distortions are determined. Figure 4.2.1.27 shows the beta-beating correction result of one lattice seed. We find the RMS beta-beating distortions are reduced significantly.

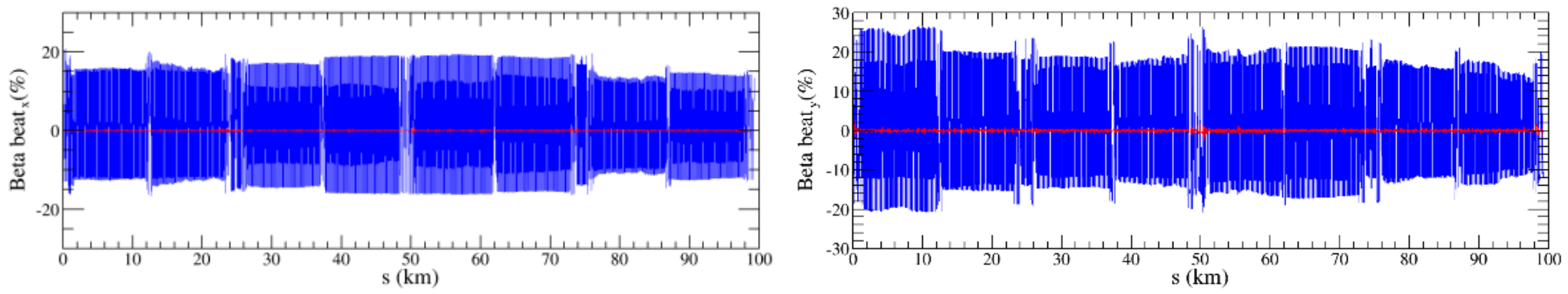

**Figure 4.2.1.27**: The horizontal (left) and vertical (right) relative beta-beating distortion for a lattice seed in the Higgs mode, where the blue curves are the relative beta-beating distortion before correction and the red curves are the relative beta-beating distortion after correction.

#### 4.2.1.3.2.4 Coupling Correction

The rolls (angular error in placement) of quadrupoles and the feed-down effect on sextupoles give rise to coupling. Skew coils on sextupoles and some independent skew quadrupoles are used to minimize the coupling response matrix by LOCO. The vertical orbit distortion due to a horizontal deflection at a BPM is

$$\frac{\Delta y_{cod}}{\Delta x_{cod}} = \bar{c}_{b,22}k_1 + \bar{c}_{b,12}k_2 + \bar{c}_{c,11}k_3 + \bar{c}_{c,12}k_4,$$

where $k_1$, $k_2, k_3$, and $k_4$ are the strength of skew quadrupoles, which relate only to the decoupled linear optics and $\bar{c}_{b,22}, \bar{c}_{b,12}, \bar{c}_{c,11}$, and $\bar{c}_{c,12}$ are the local coupling parameters. Figure 4.2.1.28 shows the calculated strength of skew quadrupoles. The emittance after correction is evaluated to be $\epsilon_x = 0.666$ nm and $\epsilon_y = 0.209$ pm, the emittance ratio $(\epsilon_y/\epsilon_x)$ is 0.03%, which satisfies the design requirement. Figure 4.2.1.29 shows the tracking dynamic aperture, which also satisfies the requirements of on-axis injection, which is $8\sigma_x \times 20\sigma_y \times 1.6\%$.

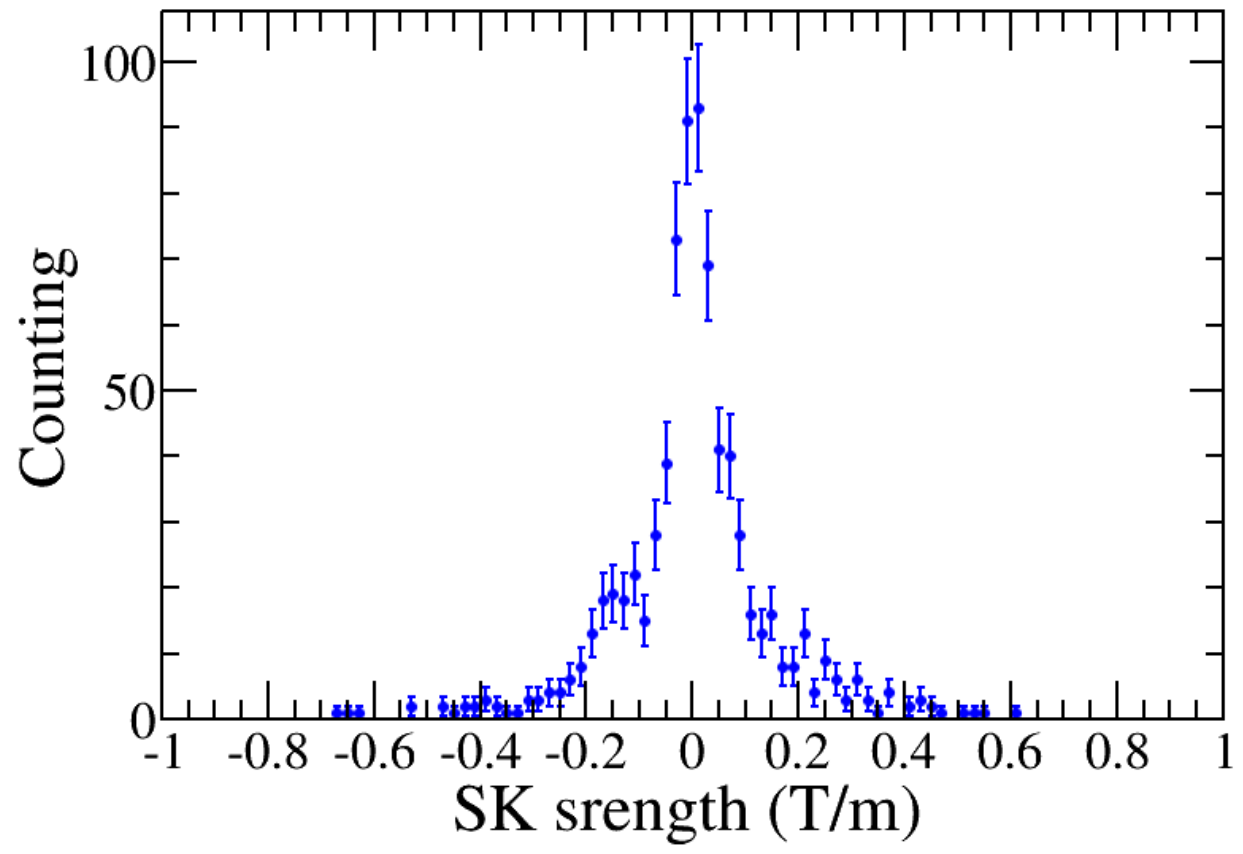

**Figure 4.2.1.28:** The strength of skew quadrupoles for a lattice seed in the Higgs mode.

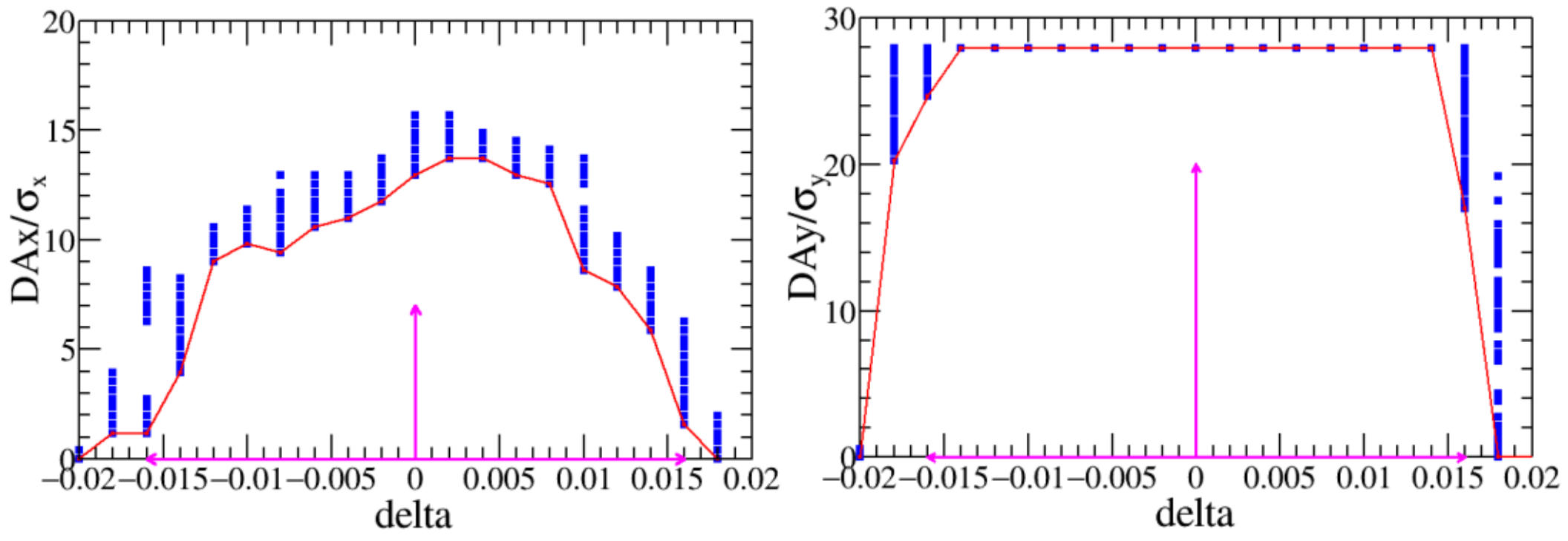

**Figure 4.2.1.29:** The dynamic aperture for a lattice seed in the Higgs mode.

#### *4.2.1.3.3 Performance after Correction*

To reduce statistical fluctuation, 100 random lattices seeds with errors listed in Table 4.2.1.6 and Table 4.2.1.7 were generated for error correction. Table 4.2.1.8 lists the correction performance for the high-luminosity lattice of four modes.

**Table 4.2.1.8:** Correction performance for the high-luminosity lattice of four modes.

| RMS | Higgs | Z | W | $t\bar{t}$ |
|---|---|---|---|---|
| Orbit (μm) | < 50 | < 50 | < 50 | < 50 |
| Dispersion (mm) | 1.8/0.9 | 2.8/1.4 | 2.7/1.8 | 0.6/0.3 |
| Beta-beating (%) | 1.0/2.8 | 2.0/3.0 | 0.5/2.5 | 1.1/1.2 |

##### 4.2.1.3.3.1 Higgs Mode

Figure 4.2.1.30 shows the RMS closed orbit after correction for all 100 lattice seeds of Higgs mode. To achieve the fully converged COD correction, the iteration times, the steps in response matrix calculation and the number of different error steps during the orbit correction were increased. The RMS closed orbits after correction were calculated to be lower than 40 μm.

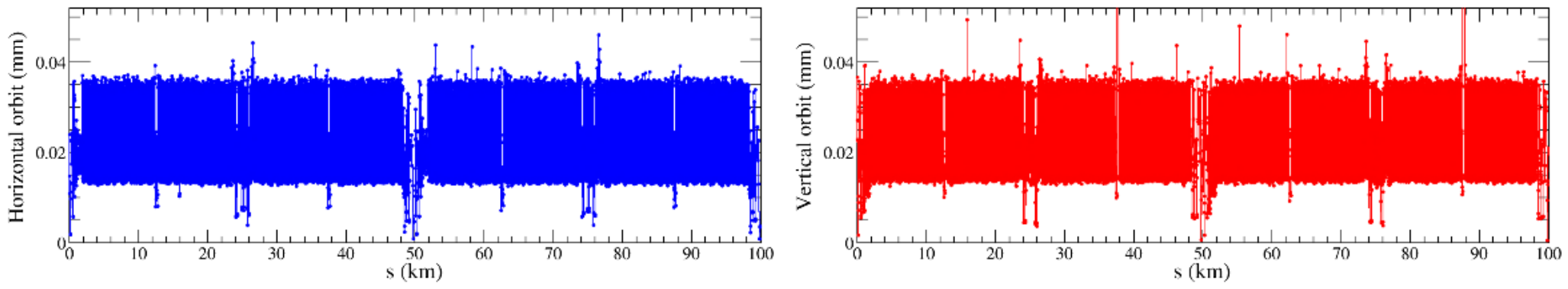

**Figure 4.2.1.30**: The RMS closed orbit after COD correction for 100 lattice seeds in the Higgs mode.

After passing through the orbit correction, the dispersion correction for these lattice seeds were performed. Figure 4.2.1.31 shows the RMS dispersion after correction for all lattice seeds; the RMS horizontal dispersion decreased from 23.1 mm to 1.8 mm, and the RMS vertical dispersion decreased from 31.9 mm to 0.9 mm.

The beta-beating correction was performed for lattice seeds after the above dispersion correction. Figure 4.2.1.32 shows the RMS relative beta-beating distortion after correction for all lattice seeds, where the RMS horizontal beta-beating decreased from 5.2% to 1.0% (about a factor of 5) and the RMS vertical beta-beating decreased from 83.2% to 2.8% (about a factor of 30).

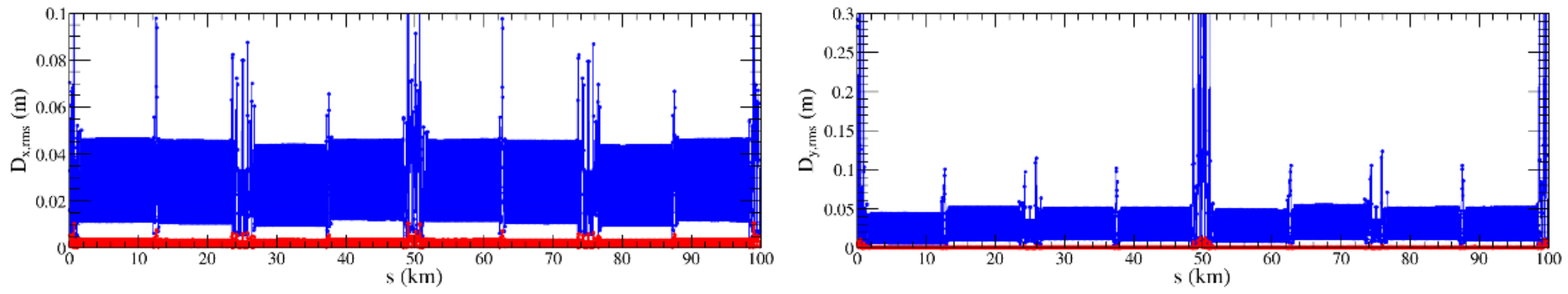

**Figure 4.2.1.31:** The RMS horizontal dispersion (left) and RMS vertical dispersion (right) of 100 lattice seeds in the Higgs mode, where the blue curves are the dispersion distortion before dispersion correction and the red curves the dispersion distortion after dispersion correction.

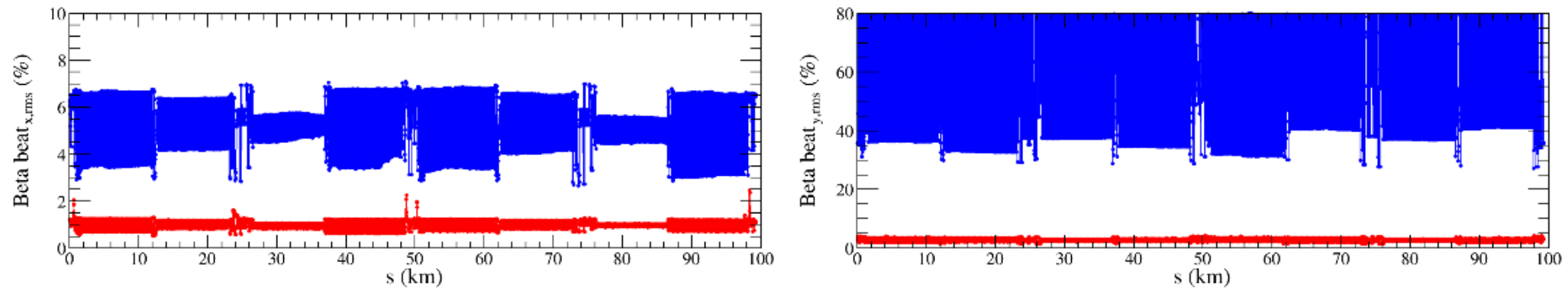

**Figure 4.2.1.32:** The RMS horizontal beat beating (left) and RMS vertical beta-beating (right) of 100 lattice seeds in the Higgs mode, where the blue curves are the beta-beating distortion before correction and the red curves the beta-beating distortion after correction.

#### 4.2.1.3.3.2 Z Mode

Figure 4.2.1.33 to Fig. 4.2.1.35 show the correction result of all 100 lattice seeds in Z mode. The RMS closed orbit after correction were calculated to be lower than 40 μm, which is comparable with that of the Higgs mode. The RMS horizontal and vertical dispersion after correction decreased from 25.4 mm to 2.8 mm, and from 30.5 mm to 1.4 mm, respectively. The RMS horizontal and vertical relative beta-beating after correction decreased from 7.5% to 2.0% (about 4 times) and from 148.8% to 3.0% (about 50 times), respectively.

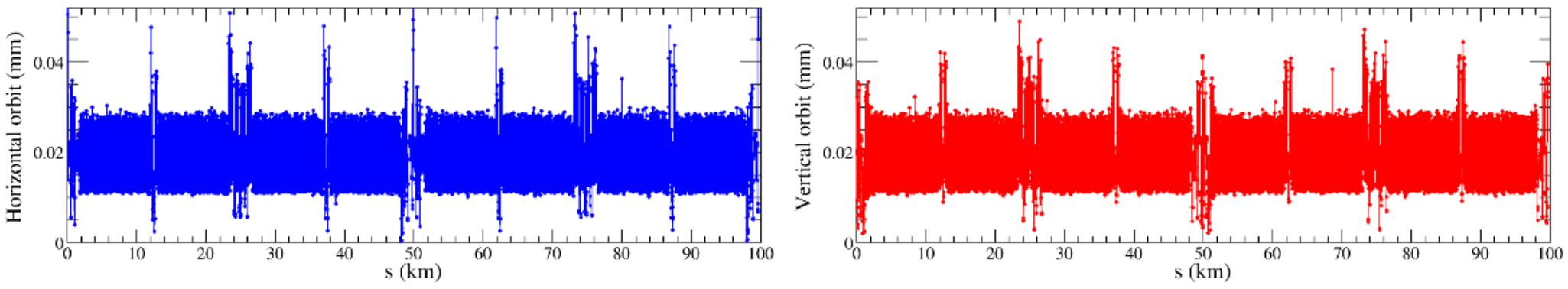

**Figure 4.2.1.33:** The RMS closed orbit after COD correction for 100 lattice seeds in the Z mode.

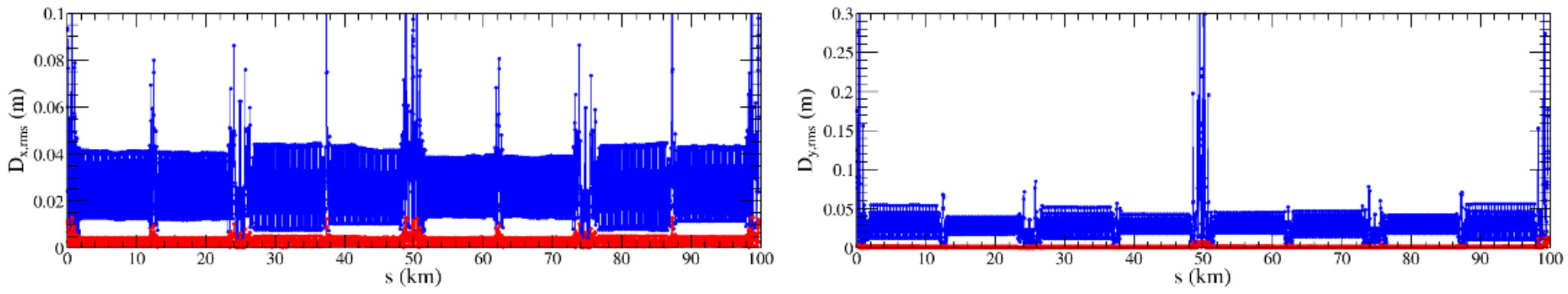

**Figure 4.2.1.34:** The RMS horizontal dispersion (left) and RMS vertical dispersion (right) of 100 Z lattice seeds, where the blue curves are the dispersion distortion before dispersion correction and the red curves the dispersion distortion after dispersion correction.

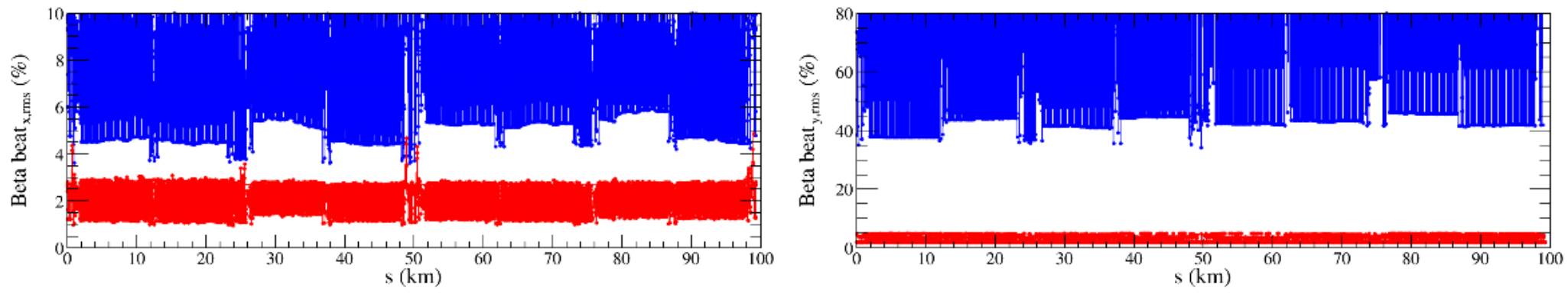

**Figure 4.2.1.35:** The (upper) RMS horizontal beta-beating and (lower) RMS vertical beta-beating of 100 Z lattice seeds, where the blue curves are the beta-beating before correction and the red curves are the beta-beating distortion after dispersion correction.

#### 4.2.1.3.3.3 W Mode

Figures 4.2.1.36 to Fig. 4.2.1.38 show the correction result of all 100 lattice seeds in W mode. The RMS closed orbit after correction is calculated to be lower than 40 μm. The RMS horizontal and vertical dispersion after correction decreased from 85.0 mm to 2.7 mm, and from 92.4 mm to 1.8 mm, respectively. The RMS horizontal and vertical relative beta-beating after correction decreased from 8.1% to 0.5% (about 16 times) and from 106.8% to 2.5% (about 43 times), respectively.

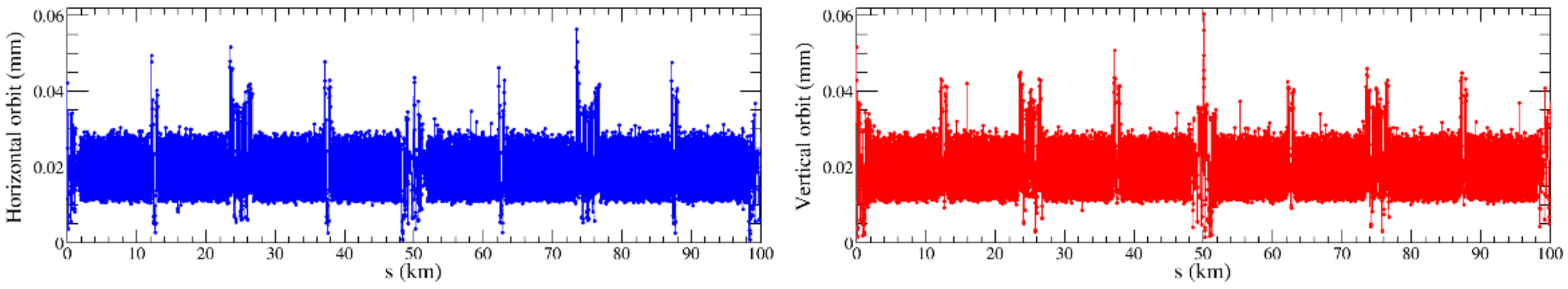

**Figure 4.2.1.36:** The RMS closed orbit after COD correction for 100 Z lattice seeds.

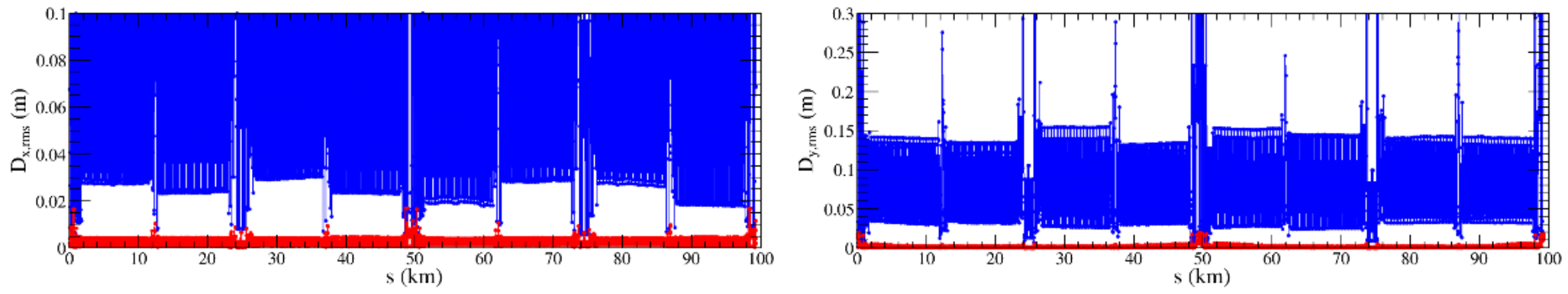

**Figure 4.2.1.37:** The RMS horizontal dispersion (left) and RMS vertical dispersion (right) for 100 W lattice seeds, where the blue curves are the dispersion distortion before dispersion correction and the red curves the dispersion distortion after dispersion correction.

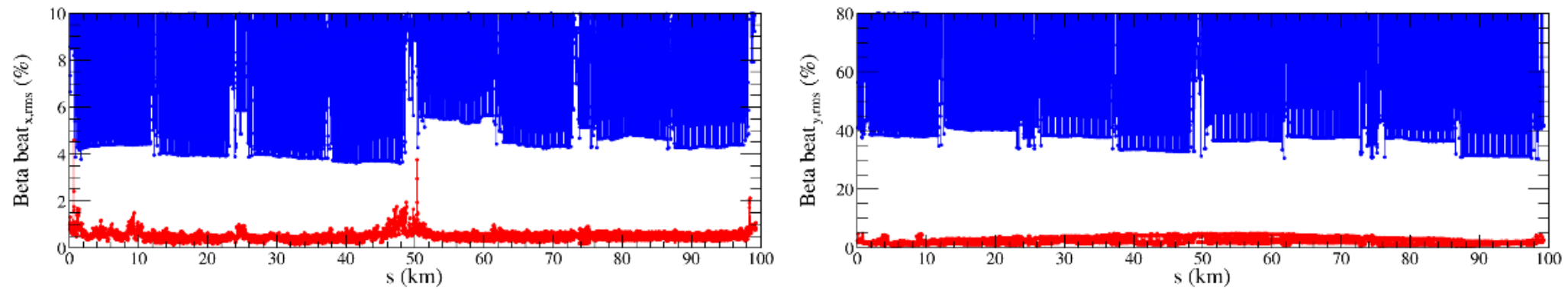

**Figure 4.2.1.38:** The RMS horizontal beta-beating (left) and RMS vertical beta-beating (right) for 100 W lattice seeds, where the blue curves are the beta-beating before correction and the red curves the beta-beating distortion after dispersion correction.

#### 4.2.1.3.3.4 $t\bar{t}$ Mode

Figure 4.2.1.39 to Fig. 4.2.1.41 show the correction result of all 100 lattice seeds in $t\bar{t}$ mode. The RMS closed orbit after correction were calculated to be lower than 40 µm. The RMS horizontal and vertical dispersions after correction were decreased from 85.0 mm to 2.7 mm, and from 8.8 mm to 0.6 mm, respectively. The RMS horizontal and vertical relative beta-beating after correction was decreased from 3.7% to 1.1% (about 3 times) and from 11.1% to 1.2% (about 10 times), respectively.

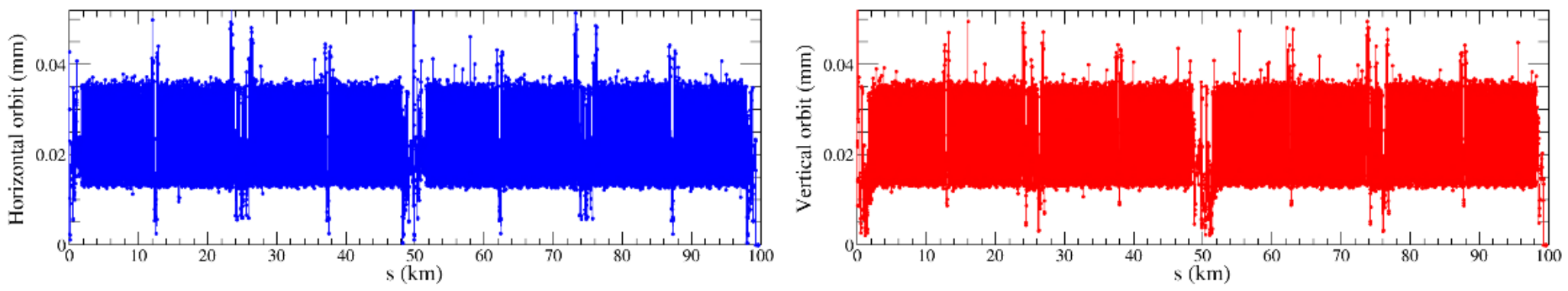

**Figure 4.2.1.39**: The RMS closed orbit after COD correction for 100 $t\bar{t}$ lattice seeds.

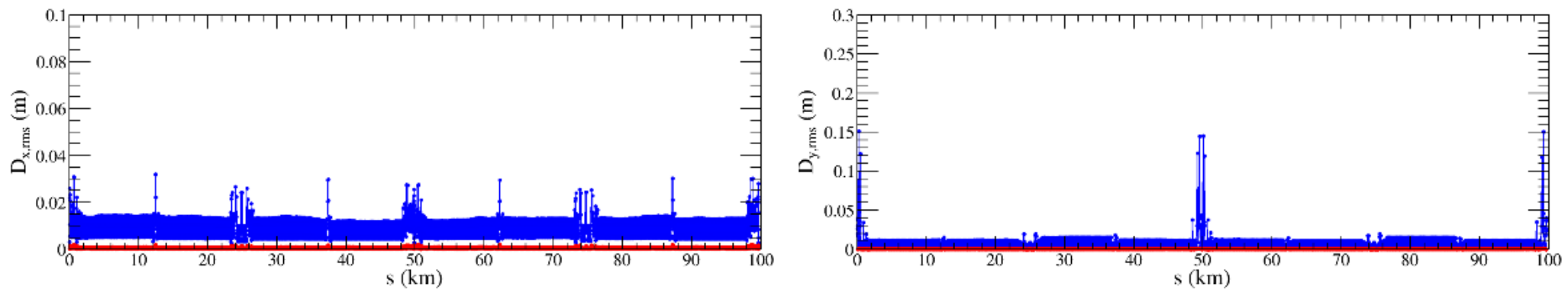

**Figure 4.2.1.40**: The RMS horizontal dispersion (left) and RMS vertical dispersion (right) for 100 $t\bar{t}$ lattice seeds, where the blue curves are the dispersion distortion before dispersion correction and the red curves the dispersion distortion after dispersion correction.

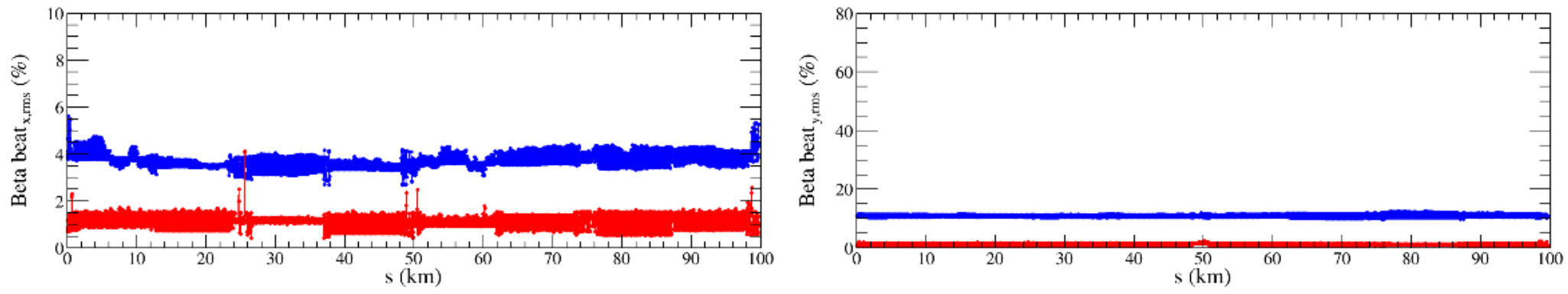

**Figure 4.2.1.41:** The RMS horizontal beta-beating (left) and (lower) RMS vertical beta-beating (right) for 100 $t\bar{t}$ lattice seeds, where the blue curves are the beta-beating before correction and the red curves the beta-beating distortion after dispersion correction.

#### 4.2.1.3.3.5 Dynamic Aperture Tracking

The dynamic aperture in Higgs mode is tracked over 145 turns, about one damping time. All lattice seeds in Higgs mode with the above error correction are used to track the dynamic aperture, as shown in Fig. 4.2.1.42. Figure 4.2.1.43 shows the dynamic aperture of 100 lattice seeds in Z mode with error correction, where each lattice seed is tracked over 2,518 turns. Figure 4.2.1.44 illustrates the dynamic aperture of 100 lattice seeds in W mode with error correction, with each lattice seed tracked over 896 turns. In Figure 4.2.1.45, we see the dynamic aperture of 100 lattice seeds in $t\bar{t}$ mode with error correction, each tracked over 40 turns. The blue lines in both figures represent the dynamic aperture of individual lattice seeds. It is noteworthy that over 90% of the lattice seeds, when subjected to global error corrections, meet the top-up injection requirement. To further optimize the dynamic aperture in the presence of errors, we are currently fine-tuning IP tuning, which includes local orbits, optics functions, and couplings. This process will be completed in the next step.

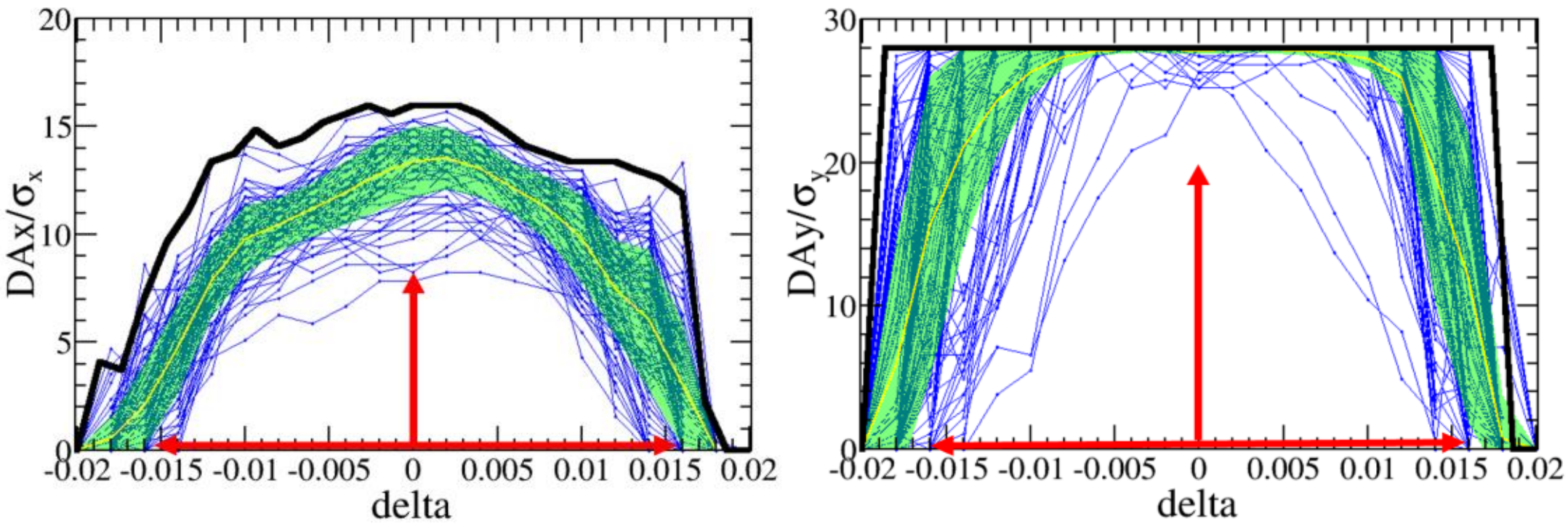

**Figure 4.2.1.42:** Dynamic aperture (DA) for the Higgs lattice with global error correction, where the yellow lines and green bands are the mean value and its corresponding statistical errors. The black line is the DA of bare lattice. The blue lines are the DA of each lattice seed. 90% lattice seeds with error correction fulfil the requirement of on axis top-up injection $8\sigma_x \times 20\sigma_y \times 1.6\%$.

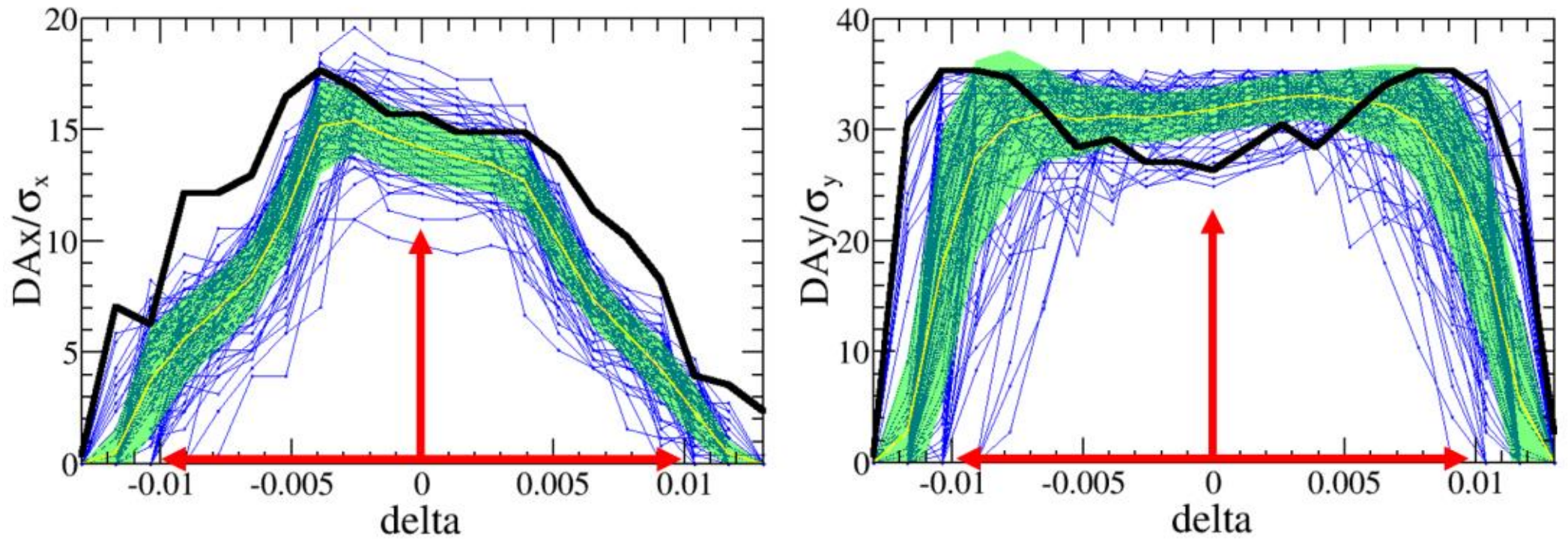

**Figure 4.2.1.43:** Dynamic aperture for the Z lattice with global error correction, where the yellow lines and green bands are the mean value and its corresponding statistical errors. The black line is the DA of bare lattice. The blue lines are the DA of each lattice seed. 94% lattice seeds with error correction fulfil the requirement of off axis top-up injection $11\sigma_x \times 23\sigma_y \times 1.0\%$

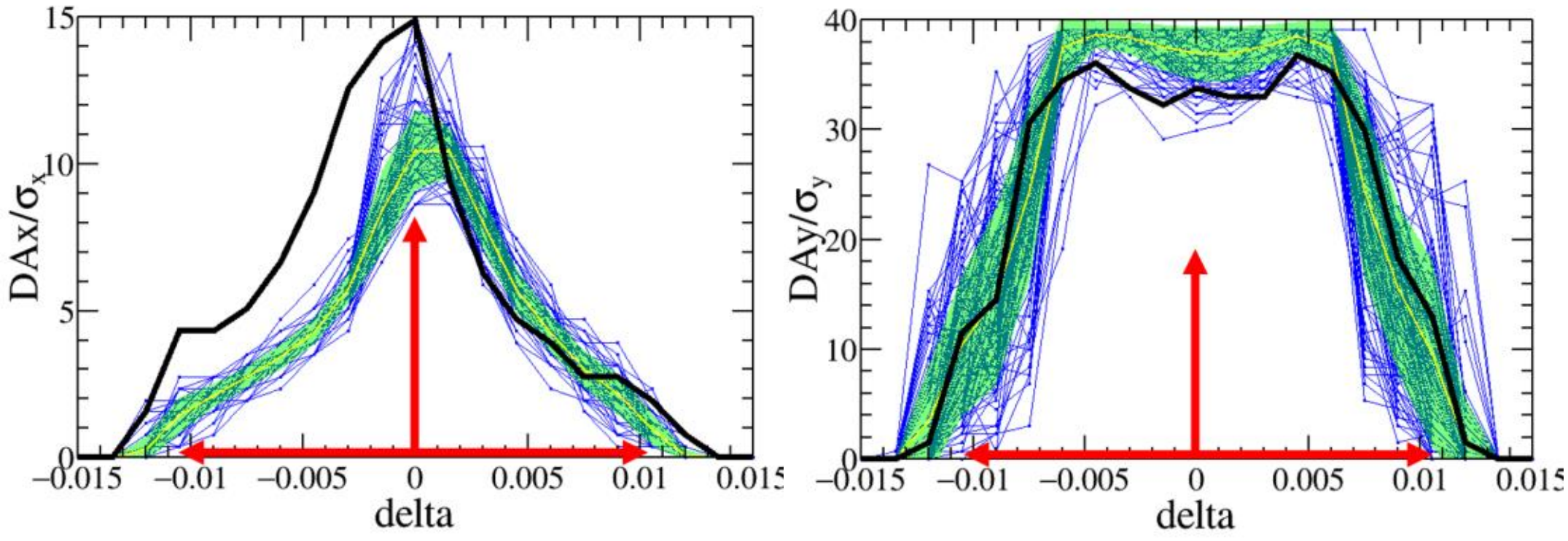

**Figure 4.2.1.44:** Dynamic aperture for the W lattice with global error correction, where the yellow lines and green bands are the mean value and its corresponding statistical errors. The black line is the DA of bare lattice. The blue lines are the DA of each lattice seed. 91% lattice seeds with error correction fulfil the requirement of off-axis top-up injection $8.5\sigma_x \times 20\sigma_y \times 1.05\%$.

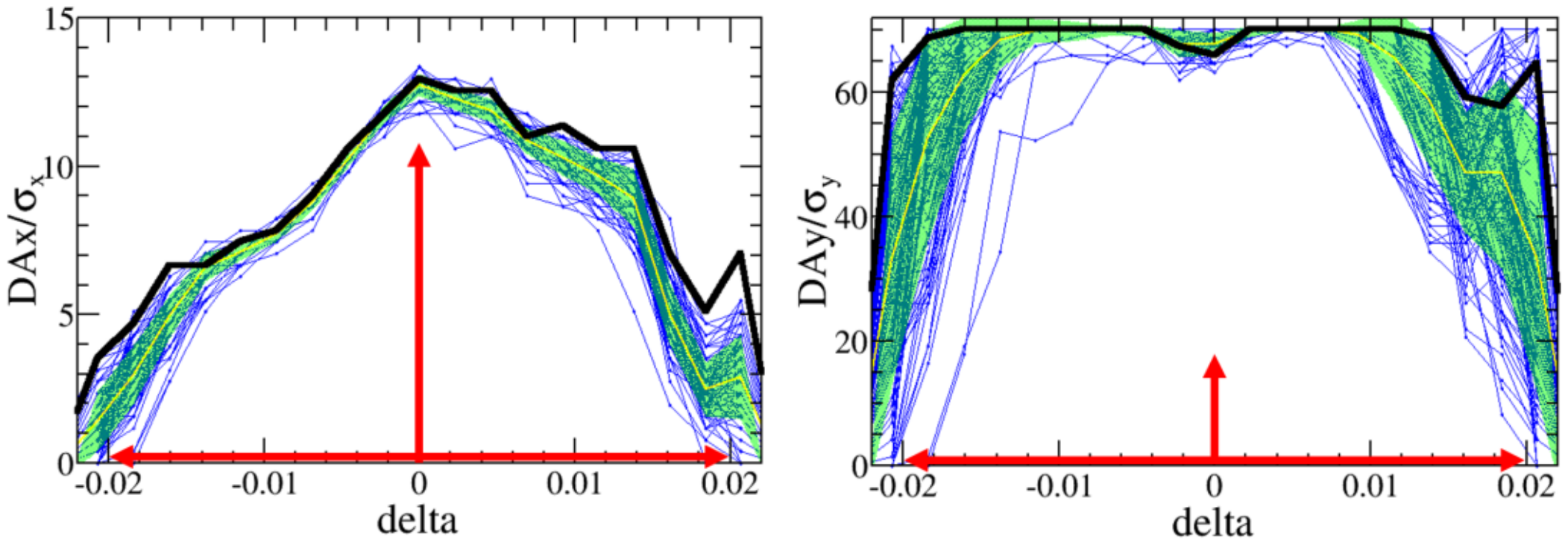

**Figure 4.2.1.45:** Dynamic aperture for the $t\bar{t}$ lattice with global error correction, where the yellow lines and green bands are the mean value and its corresponding statistical errors. The black line is the DA of bare lattice. The blue lines are the DA of each lattice seed. 90% lattice seeds with error correction fulfil the requirement of off-axis top-up injection $11\sigma_x \times 16\sigma_y \times 2.0\%$.

### 4.2.2 **Beam-beam Effects**

The beam-beam interaction in storage rings has been studied since the 1960s. A lot of fruitful work has been done since then in order to improve collider performance. Different techniques and collision schemes have been proposed to increase the luminosity. For the most recent generation of electron-positron factories, colliding multibunch beams was one of the main ingredients to increase their luminosity. A crossing angle between colliding bunches was necessary to alleviate the parasitic beam-beam interaction and operations with a horizontal crossing angle were successful at several lepton particle factories such as KEKB, DAΦNE and BEPCII. For these machines, the Piwinski angle $\Phi = (\sigma_z/\sigma_x)\tan\theta$ was chosen to be relatively modest, of the order of 0.5 (1.7 in DAΦNE only after 2007), to avoid an excessive geometric luminosity reduction and to diminish the strength of synchro-betatron resonances arising from the beam-beam interaction at a non-zero crossing angle.

The new crab-waist (CW) collision scheme was proposed in 2005. The large Piwinski angle allows the vertical beta function $\beta_y$ at the interaction point (IP) to be squeezed down to the scale $\sigma_x/\theta$ [1]. Dedicated CW sextupoles eliminate the beam-beam resonances arising (in collisions without CW) due to vertical motion modulation by the horizontal betatron oscillations [2,3]. For the first time, a combination of large Piwinski angle and crab-waist collisions has been successfully applied in DAΦNE [4] and recently exploited at SuperKEKB, giving very promising results [5]. In order to reach high luminosity, CEPC is expected to use the crab-waist collision scheme with a large Piwinski angle.

For parameters as extreme as those of CEPC, several new effects become important for the collider performance such as beamstrahlung [6], coherent X-Z instability [7], 3D flip-flop [8] and TMCI due to combined effect of beam-beam interaction and machine impedance [9]. Beamstrahlung is the synchrotron radiation induced by the beam-beam force. The X-Z instability, which appears in the correlated head-tail motion of two colliding beams, is a novel coherent beam-beam instability in collisions with a large crossing angle. The beam dynamics becomes essentially three dimensional and the longitudinal motion can no longer be considered as frozen. Moreover, interplay of different beam dynamics effects becomes very important for the choice of the collider parameters and their optimization.

In CEPC, beamstrahlung leads to a substantial energy-spread growth and bunch elongation in beam-beam collisions [10]. On the other hand, the bunch electromagnetic interaction with adjacent accelerator componants, described in terms of beam coupling impedance, also results in a notable bunch lengthening [10,11]. In addition, the impedance is responsible for synchrotron tune reduction and synchrotron-tune-spread increase [11]. In the conceptual design of CEPC [10] these effects, beamstrahlung and the impedance related effects, have been studied separately. Here more emphasis is placed on the combined effect including the effect of impedance and lattice nonlinearity.

#### 4.2.2.1 *Introduction of the Simulation and Analysis Method*

The simulation code IBB has been developed for the design and optimization of BEPCII [12] and has been extended to support large Piwinski-angle collisions, beamstrahlung effect, and multiple bunches/multiple IPs [13]. In the code, the one turn map is the following: (1) beam-beam interaction at IP, where beamstrahlung may be considered; (2) linear map including the effects of synchrotron radiation (damping and fluctuation) [14] during transport through the arc; (3) longitudinal and transverse wakefield of the entire ring.

Due to the hourglass effect, the geometric luminosity reduction and the transverse beam-beam blowup depend on the bunch length. In order to take this effect into account, the colliding bunches were sliced in the longitudinal direction. To speed up the convergence of the slice number, the potential interpolation method was used [15].

The Lorentz boost map [16] is used to consider the horizontal crossing angle. The bunch-slice number is usually about ten times the Piwinski angle. In order to calculate the beam-beam potential, the Poisson equation was solved directly by applying the particle-in-cell (PIC) method. The fast Fourier transform (FFT) method using a shifted integrated Green function [17] was adopted. This is very helpful to handle the separated slice-by-slice collision in large Piwinski-angle collisions. The strong-strong simulations of the beam-beam interaction with a large Piwinski angle are very time consuming [18], so the synchro-beam mapping method [19] to model the slice-slice collision was preferred, where a Gaussian approximation was used. In addition, the slice transverse-RMS size in each slice-slice collision was updated. The Gaussian strong-strong method is about one order of magnitude faster than the PIC strong-strong method. The energy of beamstrahlung photons emitted during collisions is much higher than that from the normal bending magnets. The random emitted photon energy is modelled using bending-magnet radiation with the Monte-Carlo method [20].

For the analysis study, the key is the model of beam-beam cross wake force [7,9]. In order to consider the influence of potential-well distortion due to longitudinal impedance. a method has been developed as follows [21]: 1) the dipole moment is azimuthally expanded in longitudinal phase; 2) in the arc section, the synchro-betatron motion could be described, where synchrotron tune is a function of $J$ considering potential-well distortion; 3) the beam-beam kick could also be modeled for the Fourie-series coefficient; 4) the revolution matrix could be constructed and used to analyze stability. Since the mode index $l$ is infinite, and the action $J$ is continuous, $l$ was truncated at $\pm l_{max}$, and $J$ discretized at sample points.

#### 4.2.2.2 *Coherent Instability*

The new-found X-Z instability would influence the beam-beam performance seriously. New machine parameters have been optimized iteratively. Here, some simulation results for the different modes ($t\bar{t}$/Higgs/W/Z) are presented in Fig. 4.2.2.1. With the new impedance budget, the machine parameters are well optimized, which should ensure a stable tune area large enough to suppress the X-Z instability.

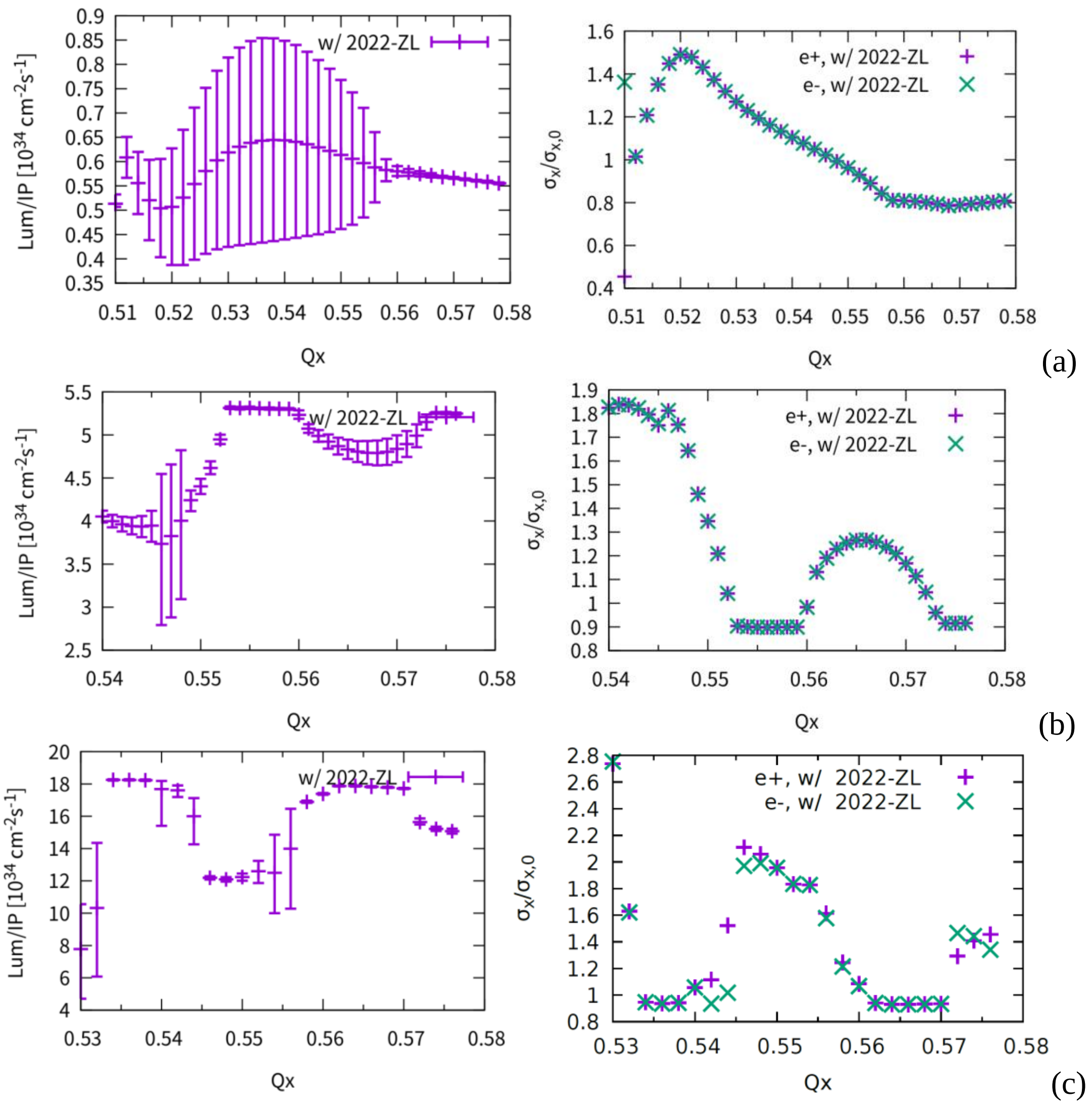

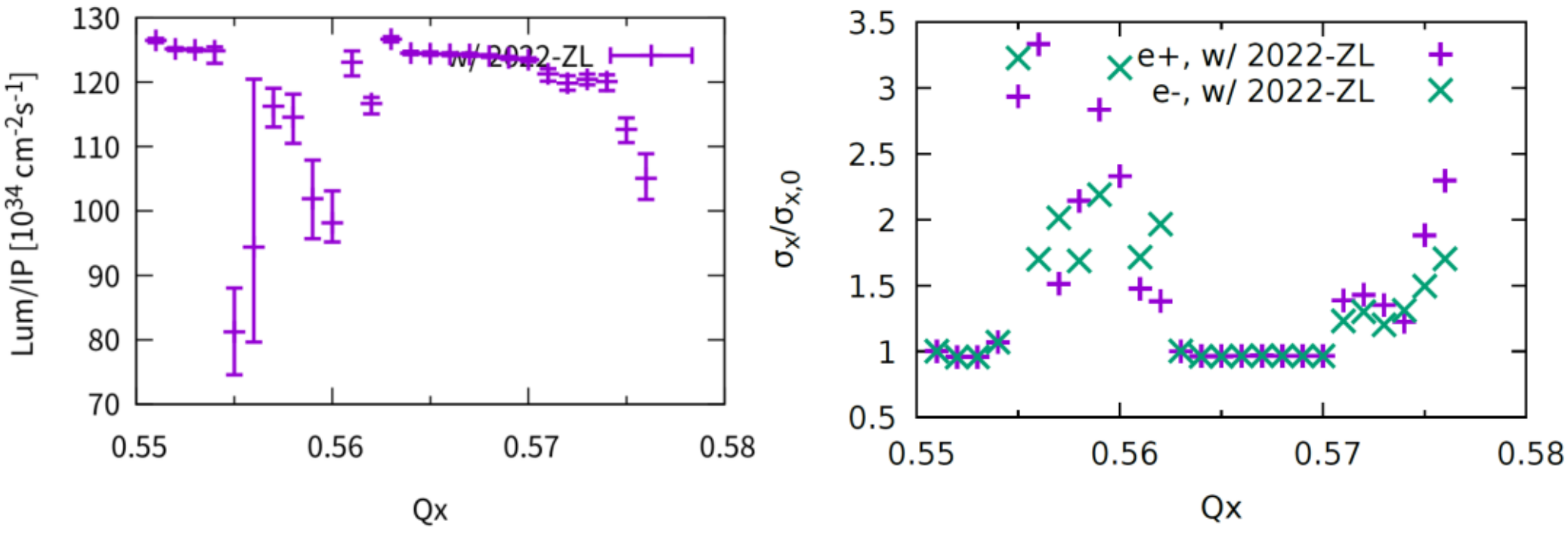

(d)

**Figure 4.2.2.1:** Luminosity and Horizontal beam-size blowup of $t\bar{t}$ (a) /Higgs (b) /W (c) /Z (d) mode versus horizontal tune

It is also a concern if the finite dispersion would enhance the X-Z instability since the longitudinal impedance is distributed along the ring. The longitudinal wakefield kick is placed at different betatron phase advance from IP, as shown in Fig. 4.2.2.2. No clear effect induced by the dispersion was found.

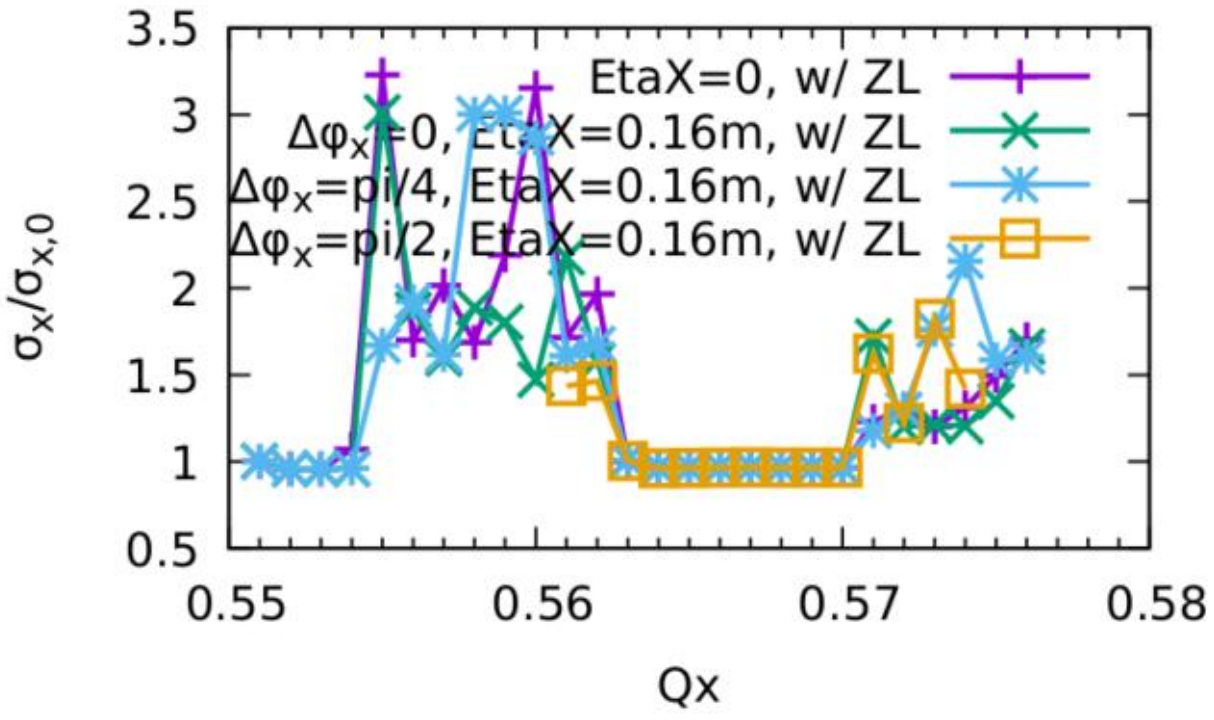

**Figure 4.2.2.2:** Effect of finite dispersion with longitudinal wakefields (ZL)

After longitudinal impedance is considered, it is natural to check if transverse impedance would induce some effect in the collision. Fig. 4.2.2.3 shows the transverse impedance effect in different modes. Collision stability is nearly not impacted for $t\bar{t}$/Higgs/W modes. However, Z mode is very different, almost no stable working points being found. The bunch population of transverse-mode-coupling-instbility (TMCI) threshold at Z mode is about $21\times10^{10}$ considering bunch lengthening by beamstrahlung but without collisions. On the other hand, collisions are stable without transverse impedance. The conclusion is that the instability is due to the combined effects of the beam-beam interaction and transverse impedance. More details show that the new instability is sigma-mode dominated. In fact, if only horizontal impedance is considered, the collision is stable. The main issue comes from vertical impedance.

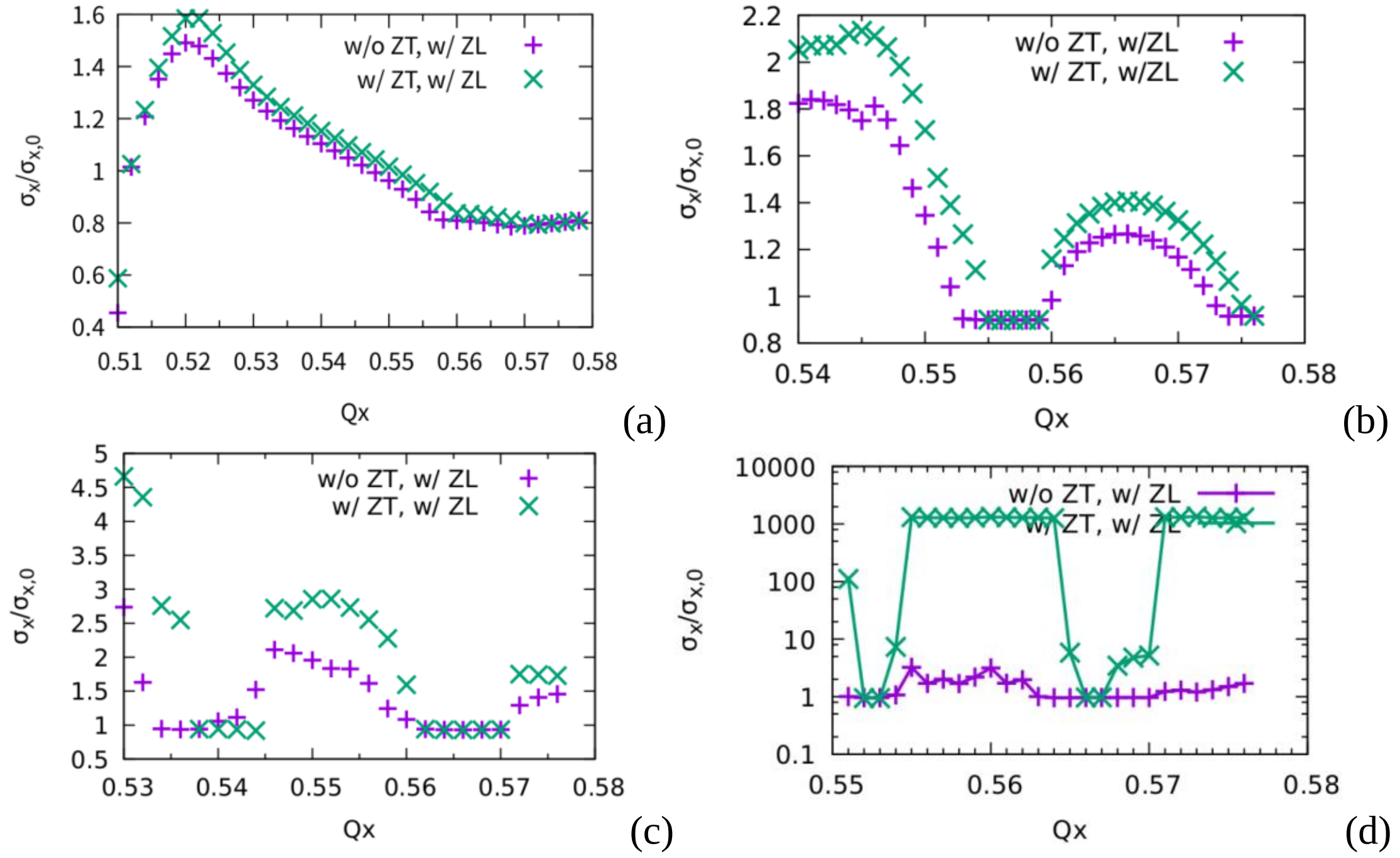

**Figure 4.2.2.3:** Beam-size blowup versus horizontal tune in different modes considering transverse impdedance: $t\bar{t}$ (a) /Higgs (b) /W (c) /Z (d)

Some analysis work has been done to understand the combined phenonomenon of transverse impedance and beam-beam interaction. The vertical instability is induced by coupling between the azimuthal 0 mode and −1 mode, which is shown in Fig. 4.2.2.4. The zero mode ($l$=0) tune would decrease with bunch population due to ring impedance, while −1 mode tune would increase with bunch population due to beam-beam interaction. The two modes would merge at a bunch population lower than the conventional TMCI threshold. In the specific case, the bunch population of instability threshold reduced from $21\times10^{10}$ (without beam-beam) to $11\times10^{10}$ (with beam-beam). The eigenvector of this specific TMCI mode obtained in analysis agrees well with that vertical centroid distribution in simulation [9]. Different from the X-Z instability, the vertical TMCI mode is not related with the locality of beam-beam interaction.

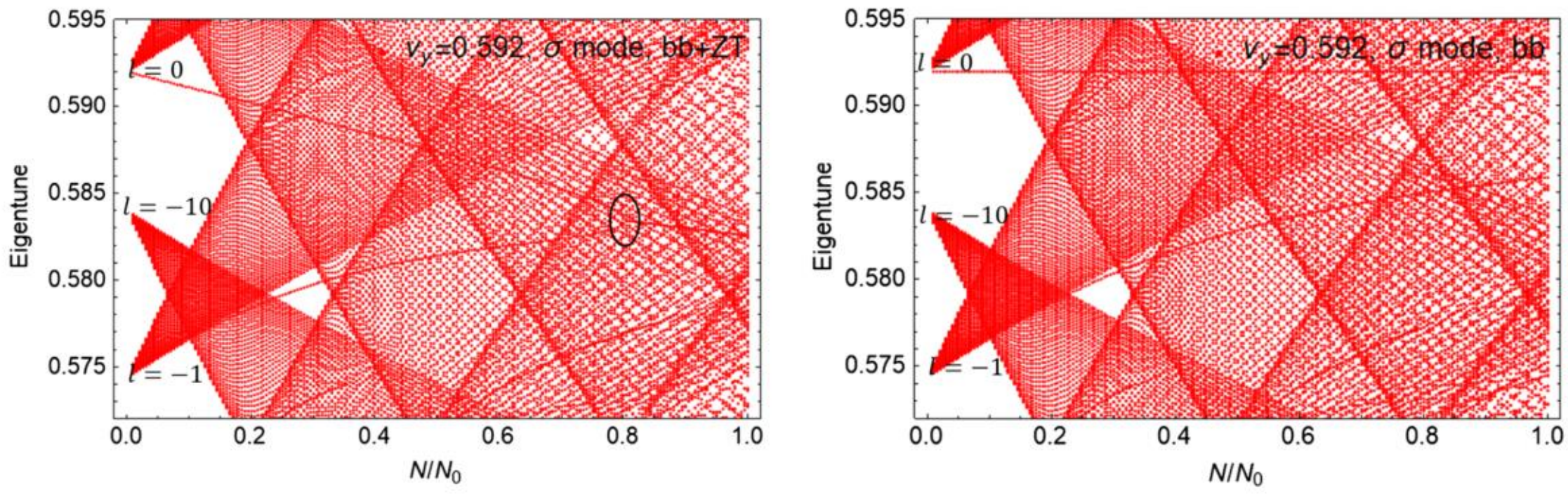

**Figure 4.2.2.4:** Mode analysis of vertical oscillation versus bunch current ($\nu_y$ = 0.592, Z mode). Ring impedance is considered in the left figure, while not considered in the right figure.

To mitigate the new instability, two methods were tried. One used finite vertical chromaticity; the other asymmetric tunes. Fig. 4.2.2.5 shows the simulation results with $Q_y$'= 0/5/10. Finite chromaticity could help suppress the instability, but a large tune chromaticity $Q_y$'~10 would be required for the mitigation. This could be a new constraint on the optics design and optimization. Fig. 4.2.2.6 shows the simulation results with asymmetric vertical tunes. When the tune gap between colliding bunches is greater than 0.01, it is very useful to suppress the sigma-mode beam-beam instability. With the help of asymmetric tunes, the chromaticity could be reduced to about 5, which ensure the stable collision.

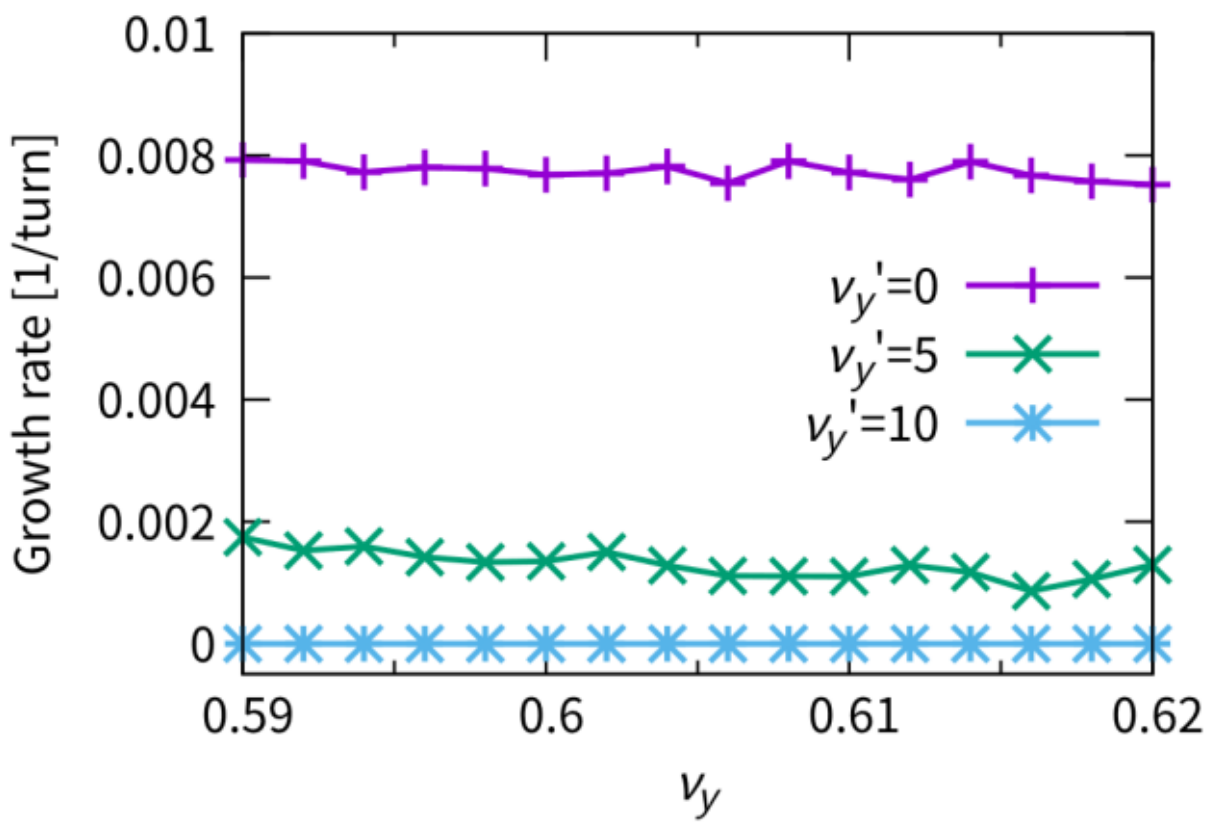

**Figure 4.2.2.5:** Effect of vertical tune chromaticity to suppress the instability considering transverse impedance (Z mode)

Concerns exist regarding a higher impedance budget potentially causing greater instability due to ongoing impedance evaluations. To perform a margin check, we raise the transverse impedance by a factor of 1.5. This adjustment results in a reduction of the TMCI threshold (without beam-beam) from $21\times10^{10}$ to $15\times10^{10}$ for Z mode, rendering the collision unstable at the design bunch population. Even a large chromaticity value (Qy' ~ 10) cannot mitigate this instability. However, given the expected strength of the transverse feedback system with a damping rate of approximately 0.1, we implement a pure resistive feedback in combination with significant chromaticity. This approach stabilizes collisions with bunch populations of both $14\times10^{10}$ and $21\times10^{10}$. Further in-depth research will be conducted in the future to analyze the combined effects of the feedback system, ring impedance, and beam-beam interactions.

Recent analysis and simulation studies also indicate that the TMCI instability could be mitigated by a significant hourglass effect [22]. The current value of $\beta_y^*\theta/\sigma_x$ is approximately 2.5 for Z mode, representing the ratio between $\beta_y^*$ and the effective interaction length (where θ is half the crossing angle). When $\beta_y^* < \sigma_x/\theta$, the instability disappears, possibly due to the influence of strong beam-beam nonlinearity forces.

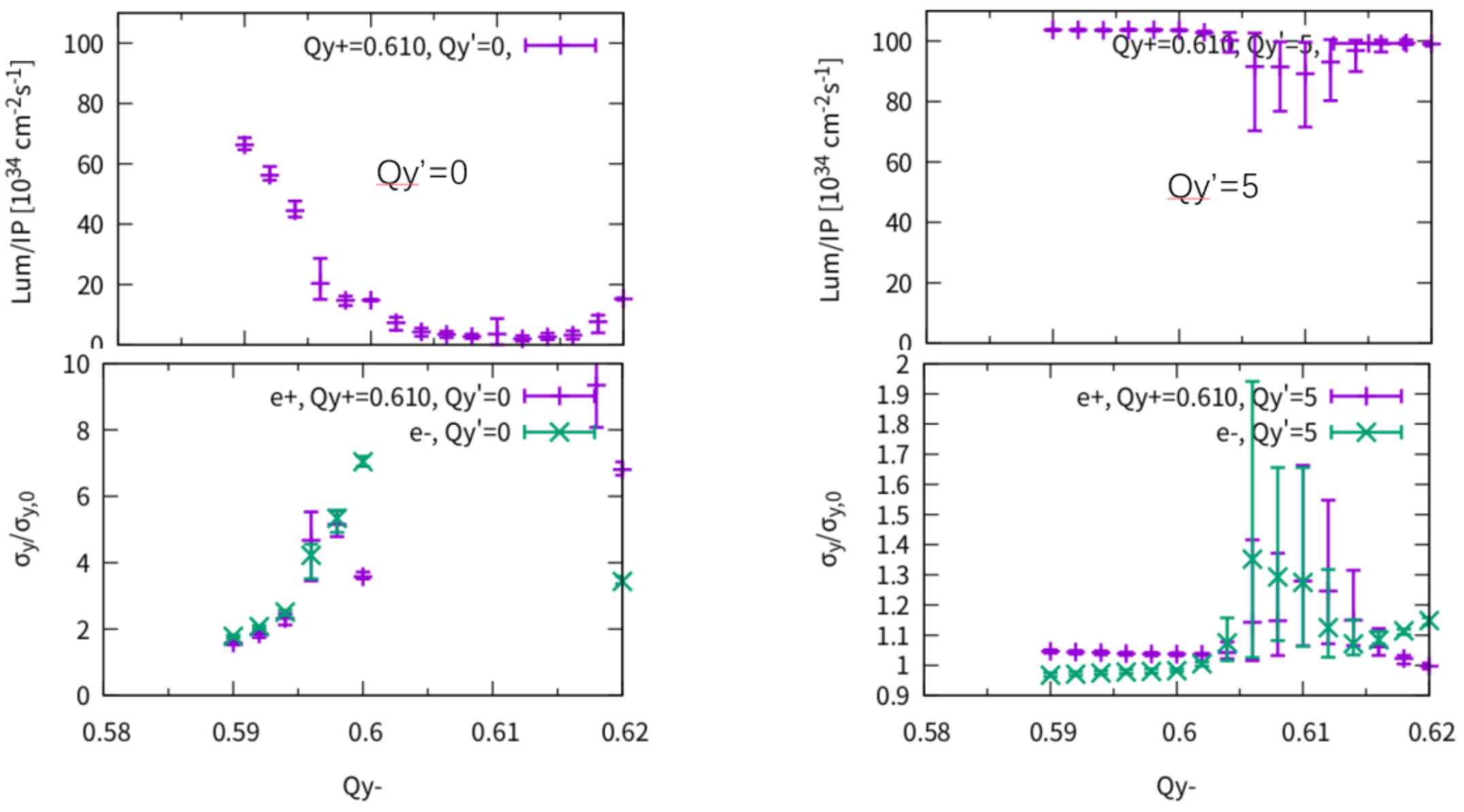

**Figure 4.2.2.6:** Luminosity with asymmetric vertical tunes. Two cases are shown (Qy'=0/5). One beam's tune is fixed at 0.610, while that of the other is varied.

### 4.2.2.3 *Incoherent Beam-beam Effect*

This section focuses on the beam lifetime as limited by the beam-beam interaction. One method is to consider only the beam-beam interaction and the related beamstrahlung effect, where IBB and a strong-strong model are used. The other method is to focus on the combined effect of lattice nonlinearity and the beam-beam interaction (beamstrahlung), where SAD and a weak-strong model are used.

Only the Higgs mode is considered, since it is most critical. Figure 4.2.2.7 shows the strong-strong simulation result obtained with IBB. The beamstrahlung lifetime is very sensitive to the momentum acceptance and the bunch population difference.

In the evaluation of dynamic aperture, typically only two times the damping time is tracked. It has been found that direct dynamic-aperture optimization could not ensure the survival of particles after more turns [23]. The combined effect of beam-beam nonlinearity, beamstrahlung, lattice nonlinearity and strong SR radiation especially in the IR magnets makes the physics very complicated. Tracking results from optimization solutions indicate that a larger dynamic aperture may not necessarily result in a longer lifetime. We have introduced a method known as diffusion-map analysis to quantify the impact of combined effects, particularly when radiation fluctuation and beamstrahlung are dominant [23]. This method demonstrates strong alignment with beam lifetime results obtained through tracking. Fig. 4.2.2.8 shows the beam equilibritum distribution in different cases w/ and w/o collision, before and after optimization. As an optimization example, only 32 arc sextupole families are used here.

To address the challenges posed by CEPC, we are in the process of developing advanced beam dynamics tracking code for more realistic simulations that account for factors such as lattice, impedance, strong-strong beam-beam interactions, and so on. Additionally, we plan to investigate the impact of machine errors on beam-beam performance and explore how they interact with the luminosity-tuning parameters for future compensation.

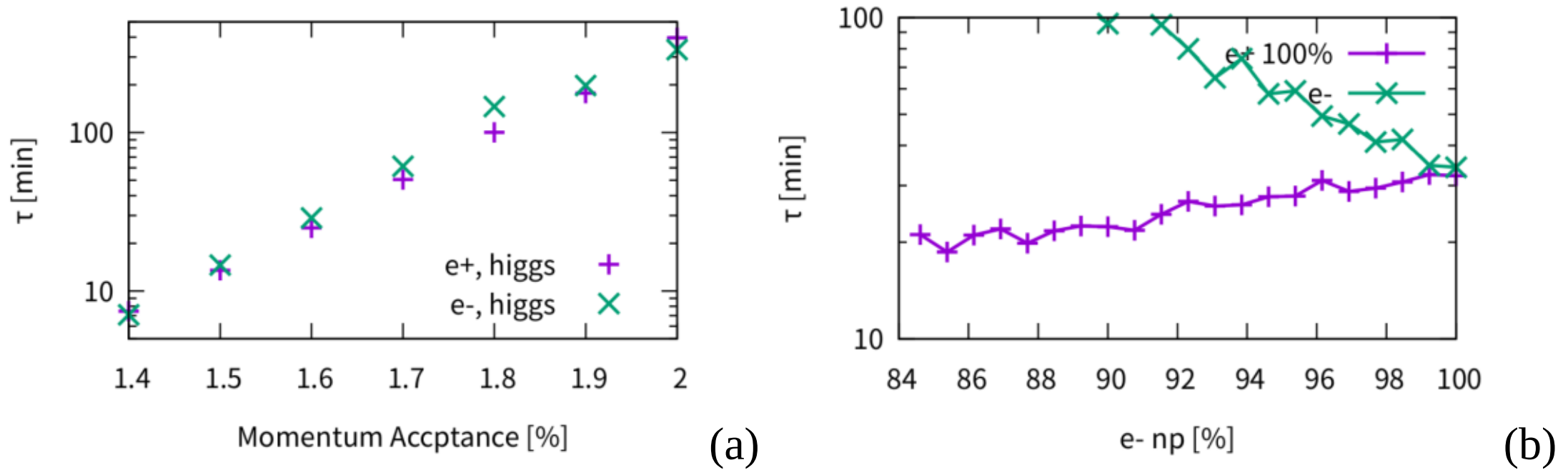

**Figure 4.2.2.7:** Beamstrahlung lifetime by strong-strong simulation with symmetric (a) and asymmetric (b) bunch population.

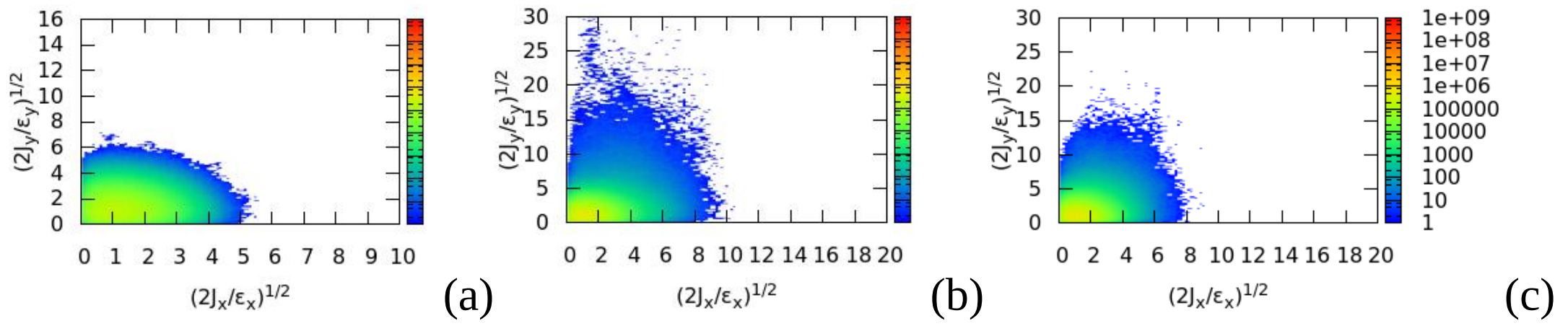

**Figure 4.2.2.8:** Equlibrium beam distribution. (a) w/o collision. (b) w/ beam-beam interaction and beamstrahlung effect, before sextupoles optimization. (c) w/ beam-beam interaction and beamstrahlung effect, after sextupoles optimization.

### 4.2.3 Beam Instability and Beam Lifetime

The interaction between the beam and the vacuum chamber surroundings can lead to collective instabilities, which can degrade the beam quality and cause beam losses, thereby limiting the performance of the machine. This section focuses on investigating the modeling of impedance and its impact on single bunch and multi-bunch collective instabilities. Furthermore, it also discusses the two-stream effects, such as electron cloud effects and beam ion instabilities. Various issues concerning beam lifetime will also be discussed. The key beam parameters utilized in the subsequent studies are presented in Table 4.2.3.1.

**Table 4.2.3.1:** Main beam parameters used in the beam instability studies.

| | **Higgs** | **Z** | **W** | $t\bar{t}$ |
|---|---|---|---|---|
| Circumference (km) | 100.0 | | | |
| Energy (GeV) | 120 | 45.5 | 80 | 180 |
| Damping time $\tau_x/\tau_y/\tau_z$(ms) | 44.6/44.6/22.3 | 850/850/425 | 156/156/78 | 13.2/13.2/6.6 |
| Bunch number | 268 | 11934 | 1297 | 35 |
| Bunch spacing (ns) | 591 (53% gap) | 23 (18% gap) | 257 | 4524 (53% gap) |
| Bunch population ($10^{11}$) | 1.3 | 1.4 | 1.35 | 2.0 |
| Beam current (mA) | 16.7 | 803.5 | 84.1 | 3.3 |
| Momentum compaction ($10^{-5}$) | 0.71 | 1.43 | 1.43 | 0.71 |
| Emittance $\varepsilon_x/\varepsilon_y$ (nm/pm) | 0.64/1.3 | 0.27/1.4 | 0.87/1.7 | 1.4/4.7 |
| Betatron tune $\nu_x/\nu_y$ | 445.10/445.22 | 317.10/317.22 | 317.10/317.22 | 445.10/445.22 |
| Bunch length (natural/total) (mm) | 2.3/4.1 | 2.5/8.7 | 2.5/4.9 | 2.2/2.9 |
| Energy spread (natural/total) (%) | 0.10/0.17 | 0.04/0.13 | 0.07/0.14 | 0.15/0.20 |
| RF frequency (MHz) | 650 | | | |
| Longitudinal tune $\nu_s$ | 0.049 | 0.035 | 0.062 | 0.078 |
| Luminosity per IP ($10^{34}$/cm$^2$/s) | 5.0 | 115 | 16 | 0.5 |

#### 4.2.3.1 *Impedance*

Extensive impedance modeling of the key vacuum components has been conducted since the initiation of the project [1-4]. In the Conceptual Design Report (CDR), the vacuum chamber originally possessed an elliptical cross-section. However, due to the significant incoherent tune shift resulting from its non-axial symmetry [5-6], it has been redesigned to adopt a circular shape. Consequently, the impedance model has been updated accordingly, incorporating additional contributors. A comprehensive impedance model, based on the round beam pipe, has been developed, considering various elements such as resistive walls, RF cavities, flanges, bellows, gate valves, vacuum pumps, Beam Position Monitors (BPMs), Interaction Region (IR) collimators, electro separators, Interaction Point (IP) chambers, and vacuum transitions.

Among the various impedance contributors, the resistive wall impedance, arising from the finite conductivity of the vacuum chambers, plays a significant role. The primary vacuum chamber is constructed from copper and features a 0.2 μm non-evaporable getter (NEG) coating on its inner surface. The theoretical formulas for multilayer cylindrical beam pipes have been employed to assess the longitudinal and transverse impedances [7]. The obtained results are depicted by the magenta curve in Figures 4.2.3.1 and 4.2.3.2. Additionally, the contribution of resistive wall impedance from the IP chamber and collimators in the interaction region has been computed. However, their impact on the total resistive wall impedance is minimal due to the relatively shorter length of these chambers compared to the main chamber.

The geometrical impedance resulting from the discontinuity of the vacuum chambers is individually calculated, assuming no crosstalk between adjacent components. The total impedance for the ring is obtained by summing up the individual contributions. To evaluate the geometrical impedance, a numerical code such as CST Particle Studio and ABCI is utilized. The longitudinal and transverse impedance arising from various elements are illustrated in Figures 4.2.3.1 and 4.2.3.2. To manage the impedance contributions, measures such as smooth transitions and RF shielding are implemented for components with aperture changes or cavity structures.

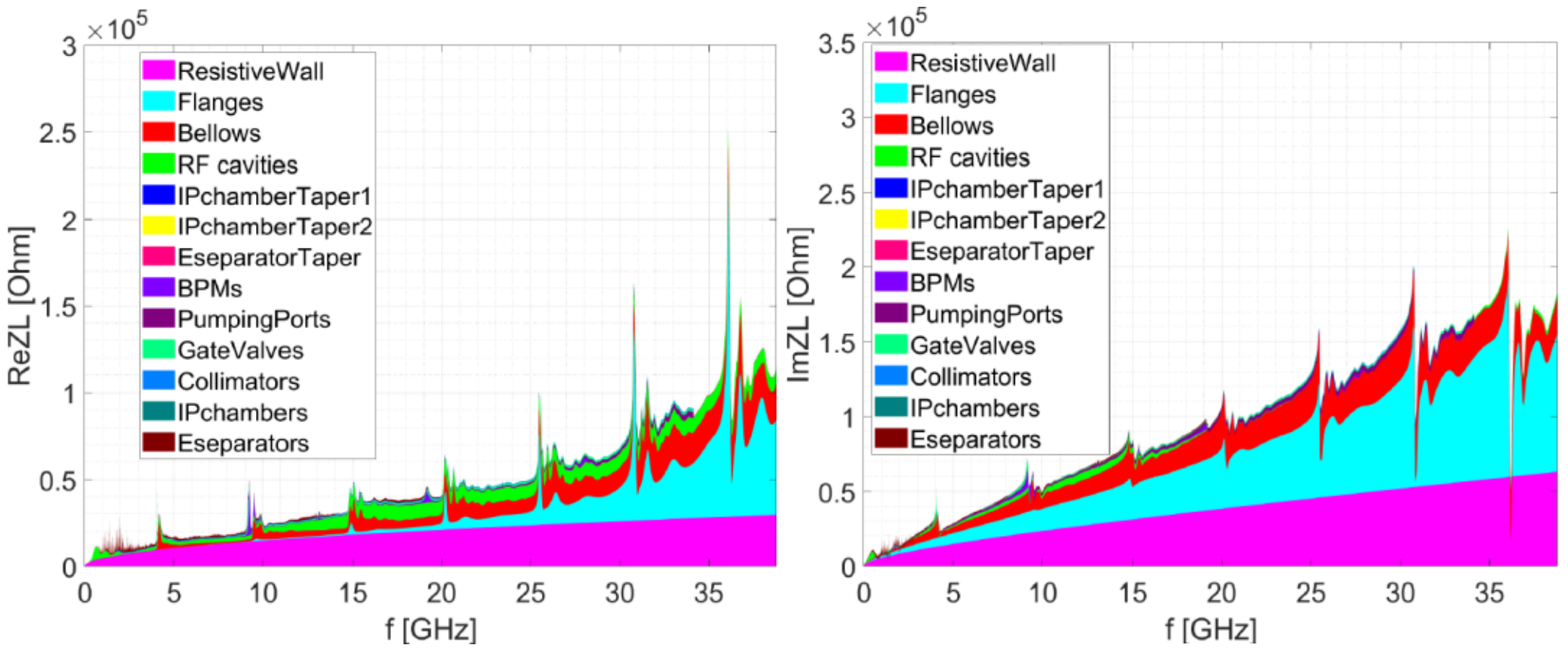

**Figure 4.2.3.1:** Real part (left) and imaginary part (right) of the longitudinal impedance.

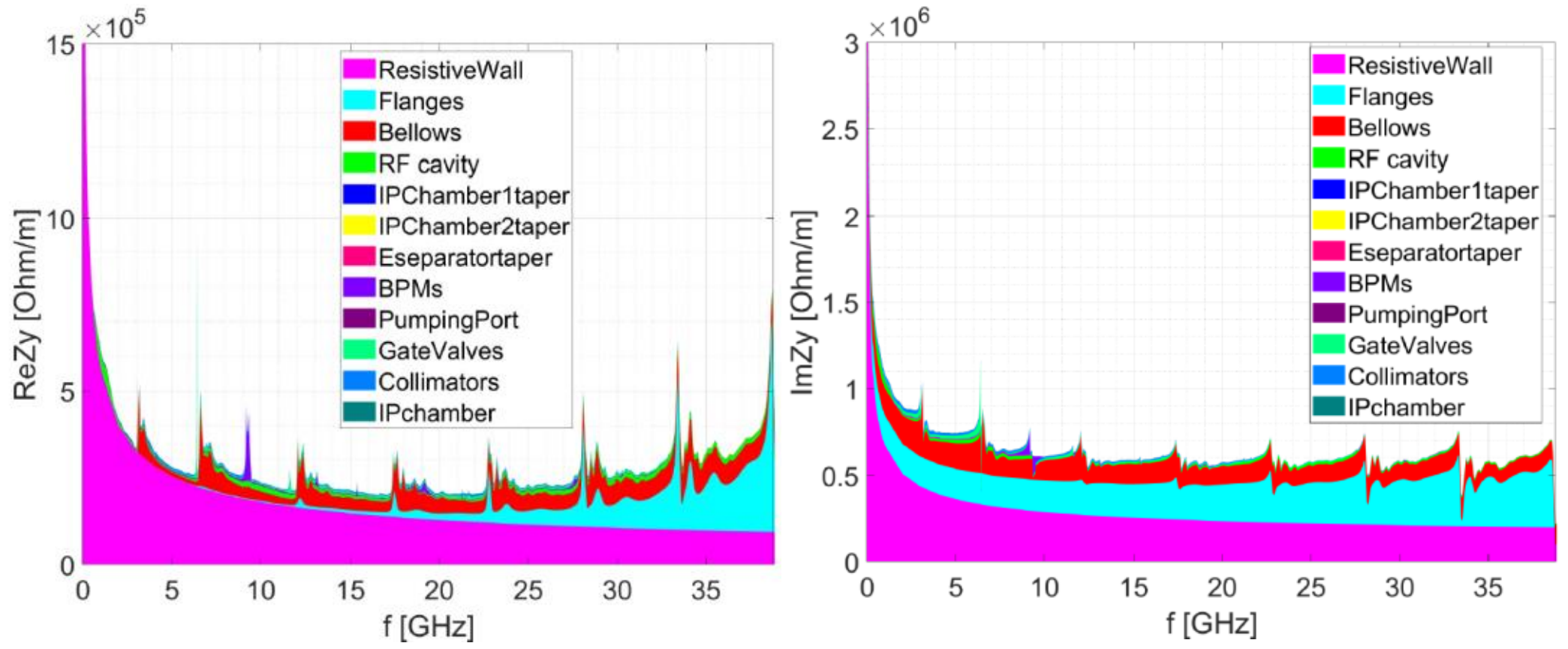

**Figure 4.2.3.2:** Real part (left) and imaginary part (right) of the transverse impedance.

Table 4.2.3.2 presents the elements included in the impedance model and their corresponding broadband effective impedances. The effective impedances consist of the longitudinal loss factor ($k_{loss}$), longitudinal effective impedance ($Z_{||}/n$), and transverse kick factor ($k_y$), as defined in [8]. The inductance ($L$) is obtained by dividing the longitudinal effective impedance ($Z_{||}/n$) by the angular revolution frequency. The evaluation of effective impedances assumes a Gaussian bunch with an rms bunch length of 3 mm.

The results indicate that both the longitudinal and transverse broadband impedances are primarily influenced by the resistive wall, as well as elements with significant quantities, such as the flanges and bellows. Meanwhile, the loss factor is predominantly contributed to by the resistive wall, RF cavities, and bellows.

**Table 4.2.3.2:** Summary of the broadband effective impedance budget with rms bunch length of 3 mm

| **Components** | $Z_{\|}/n$, **mΩ** | ***L*, nH** | $k_{loss}$, **V/pC** | $k_y$, **kV/pC/m** |
|---|---|---|---|---|
| Resistive wall | 6.2 | 329.2 | 363.7 | 11.3 |
| RF cavities | 0.5 | 24.8 | 101.2 | 0.5 |
| Flanges | 5.2 | 276.1 | 37.3 | 5.2 |
| BPMs | 0.04 | 2.0 | 9.5 | 0.2 |
| Bellows | 2.9 | 154.6 | 87.4 | 3.9 |
| Gate Valves | 0.2 | 11.4 | 14.5 | 0.4 |
| Pumping ports | 0.3 | 18.4 | 2.3 | 0.2 |
| Collimators | 0.04 | 1.7 | 23.4 | 0.6 |
| IP chambers | 0.004 | 0.2 | 0.3 | 0.05 |
| Electro-separators | -0.1 | -5.4 | 34.5 | 0.1 |
| Taper transitions | 0.04 | 2.3 | 2.5 | 0.09 |
| Total | 15.3 | 815.3 | 676.6 | 22.5 |

### 4.2.3.2 *Single-bunch Effects*

The impedance model developed earlier is utilized to estimate single bunch instabilities in both the longitudinal and transverse planes using conventional analytical theories. These estimations are performed for various operation scenarios, and the corresponding results are presented in Table 4.2.3.3.

The collective effects primarily impose constraints on the Z-pole operation mode due to its specific characteristics, including low beam energy, high beam current and bunch intensity, and slow synchrotron radiation damping. In the case of high-energy scenarios, the main concern lies in the longitudinal broadband impedance exceeding the Boussard-Keil-Schnell criterion [9-10]. It should be noted that this theoretical criterion generally underestimates the threshold for microwave instability. Nevertheless, surpassing this criterion can lead to perturbations in the longitudinal beam dynamics, potentially causing degradation in the beam-beam interaction. Section 4.2.2 provides more comprehensive studies considering both beam-beam and longitudinal impedance, which consistently demonstrate a reduction in the stable beam-beam tune area due to the presence of impedance [11-12].

For the Z-pole mode, both longitudinal and transverse impedances exceed the instability threshold. Consequently, particle tracking simulations are employed to investigate collective instability issues specifically related to high luminosity Z.

**Table 4.2.3.3:** Rough estimation of the single bunch instability threshold

| | **Higgs** | **Z** | **W** | $\boldsymbol{t\bar{t}}$ |
|---|---|---|---|---|
| Longitudinal threshold $Z_{\parallel}/n$, mΩ | 6.5 | 0.7 | 4.1 | 14.4 |
| Transverse threshold $k_y$, V/pC/m | 69.7 | 12.4 | 40.2 | 109.8 |

The microwave instability, although not resulting in beam losses, can have detrimental effects on the luminosity by distorting the beam distribution, altering the synchrotron tune, and increasing the beam energy spread. Figures 4.2.3.3 and 4.2.3.4 illustrate the variations of bunch length and beam energy spread as the single bunch population increases for the Higgs and Z modes, respectively.

For the Higgs mode, the microwave instability threshold is approximately 100 nC, which is well above the design bunch current. However, at the design bunch intensity, the impedance will cause the bunch length to increase by around 30%.

On the other hand, for the Z mode, the microwave instability threshold occurs at about half of the design bunch intensity. At the design bunch current, the rms bunch length and beam energy spread will increase by approximately 140% and 35% respectively.

Furthermore, the longitudinal impedance will induce incoherent synchrotron tune spread and bunch distortion, as depicted in Figure 4.2.3.5. These effects will also impact the transverse beam dynamics, as further discussed in the subsequent part of this section.

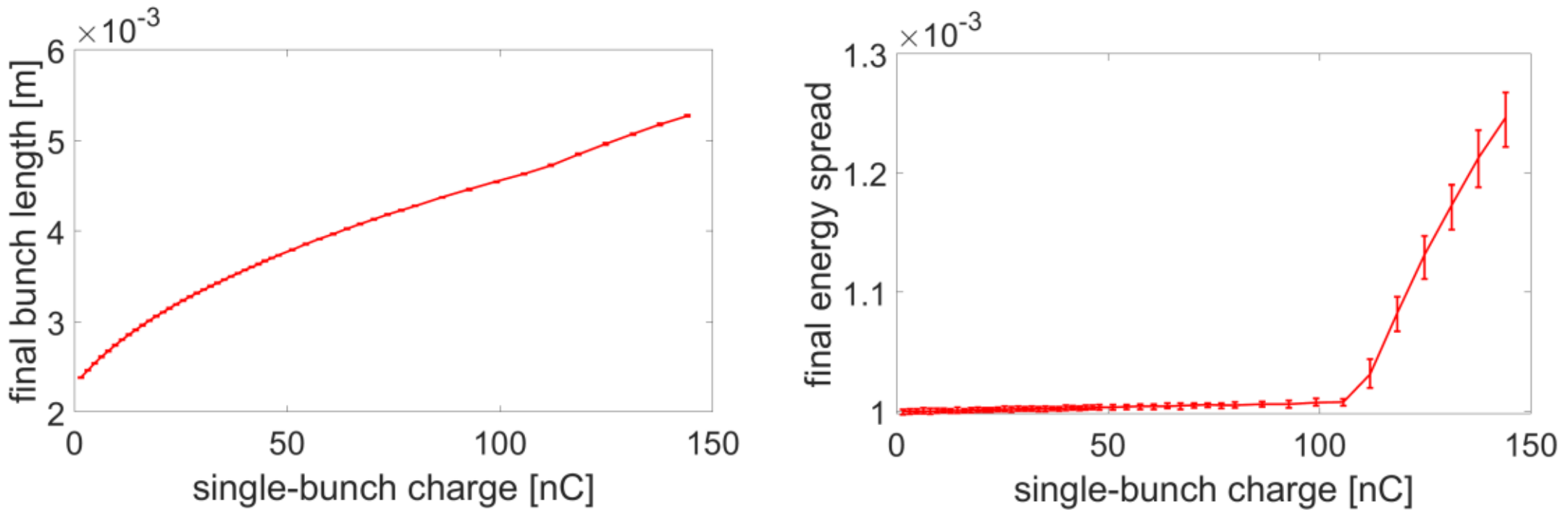

**Figure 4.2.3.3:** Variation of the bunch length (left) and beam energy spread (right) with bunch intensity for Higgs.

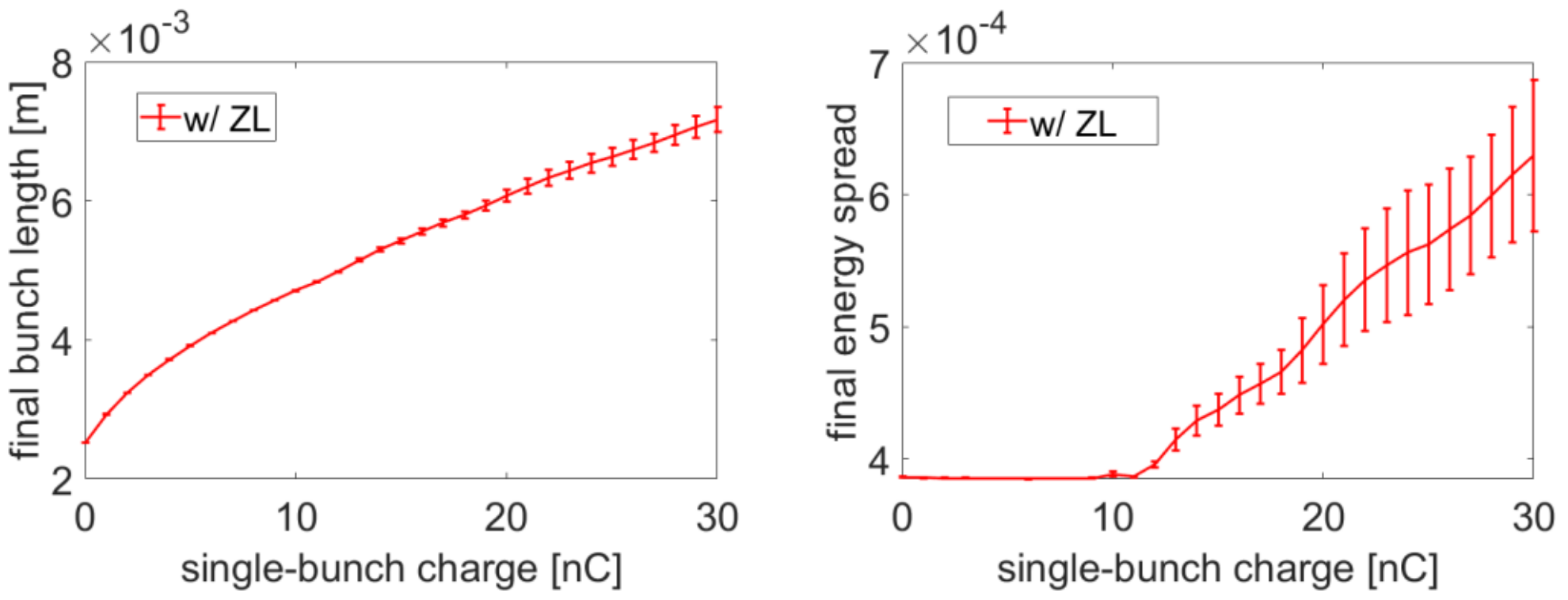

**Figure 4.2.3.4:** Variation of the bunch length (left) and beam energy spread (right) with bunch intensity for Z.

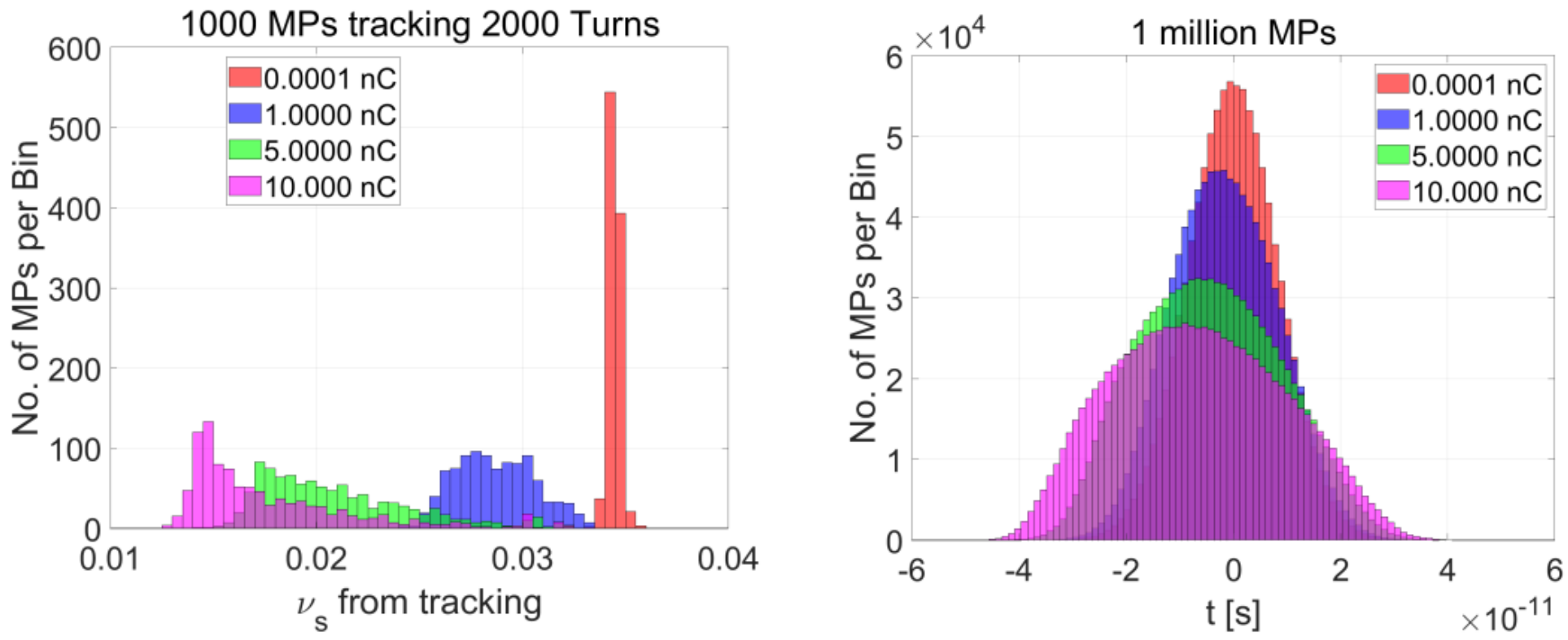

**Figure 4.2.3.5:** The incoherent synchrotron tune spread (left) and bunch distortion (right) at different bunch intensities for Z (MP stands for macroparticles).

Due to the necessity of maintaining stable beam-beam interaction, the beam will undergo collision injection, resulting in significant bunch lengthening through beamstrahlung. This elongation of the bunch helps mitigate the impact of longitudinal impedance on beam dynamics. However, it is still crucial to conduct a thorough investigation of the longitudinal impedance's effect on beam-beam interaction.

Detailed studies that consider both longitudinal impedance and beam-beam interaction reveal that the presence of longitudinal impedance contributes to increased instability in the beam-beam interaction, particularly due to X-Z coupling [13]. Specifically, the stable region of the working point contracts and shifts as a result of the longitudinal impedance. Therefore, a comprehensive understanding of the impact of impedance on beam-beam interaction is essential.

To mitigate these effects, rough estimations are performed by multiplying the real or imaginary part of the longitudinal impedance by a factor, providing a general idea of how to optimize the longitudinal impedance. The results demonstrate that increasing the real part of the impedance significantly reduces the size of the stable working point region, while the location of the stable region remains relatively unchanged. Conversely, increasing the imaginary part of the impedance shifts the position of the stable region, while the width of the stable region remains nearly unchanged. Hence, to optimize the impedance and mitigate its effects, controlling the real part of the longitudinal impedance is of greater importance.

In the transverse case, the presence of impedance can lead to fast instabilities through transverse mode coupling, often resulting in beam losses. Eigen mode analysis is conducted considering both with and without bunch lengthening caused by longitudinal impedance for Z, as illustrated in Figure 4.2.3.6. It is observed that considering only bunch lengthening extends the instability threshold to higher bunch intensities.

To study the transverse beam dynamics consistently, including the effects of longitudinal impedance, particle tracking simulations using the Elegant code [14-15] are performed. The simulation results, presented in Figure 4.2.3.7, align with the analytical estimations when longitudinal impedance is not taken into account. However, when longitudinal impedance is included, the variation of mode 0 agrees with the analytical estimations considering only bunch lengthening. Nonetheless, the presence of longitudinal impedance reduces the instability threshold significantly compared to the theoretical estimation shown in the right plot of Figure 4.2.3.6. This reduction is primarily due to the smaller incoherent synchrotron tune induced by the longitudinal impedance.

Furthermore, mode analysis is conducted, incorporating longitudinal perturbation resulting from beamstrahlung-induced bunch lengthening, using the method developed in Reference [12]. The results in Figure 4.2.3.8 demonstrate that in the absence of longitudinal impedance, the threshold is increased compared to Figure 4.2.3.6 due to beamstrahlung-induced bunch lengthening. However, with the inclusion of longitudinal impedance, higher-order modes shift towards mode 0 with a wider bandwidth, and the instability threshold is reduced compared to the case without longitudinal impedance. The threshold bunch intensity with impedance and beamstrahlung is approximately $19\times10^{10}$, as per the analytical investigation.

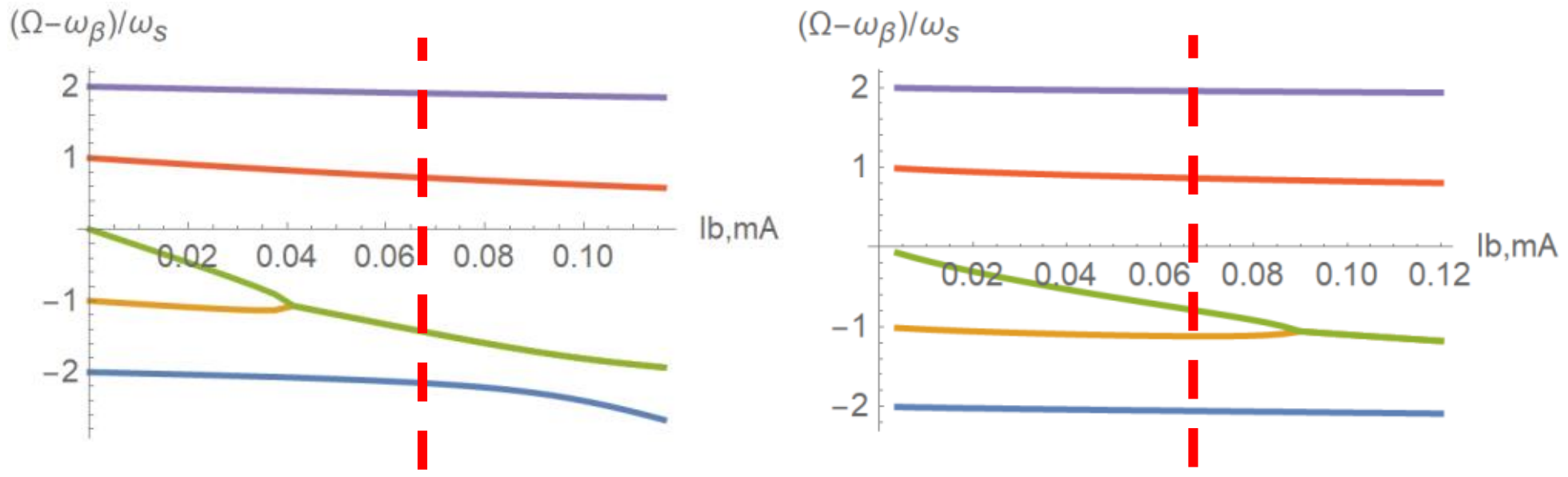

**Figure 4.2.3.6:** Dependence of the transverse modes with single bunch current in the case of without (left) and with (right) longitudinal impedance bunch lengthening for Z. The red dashed lines represent the design bunch intensity.

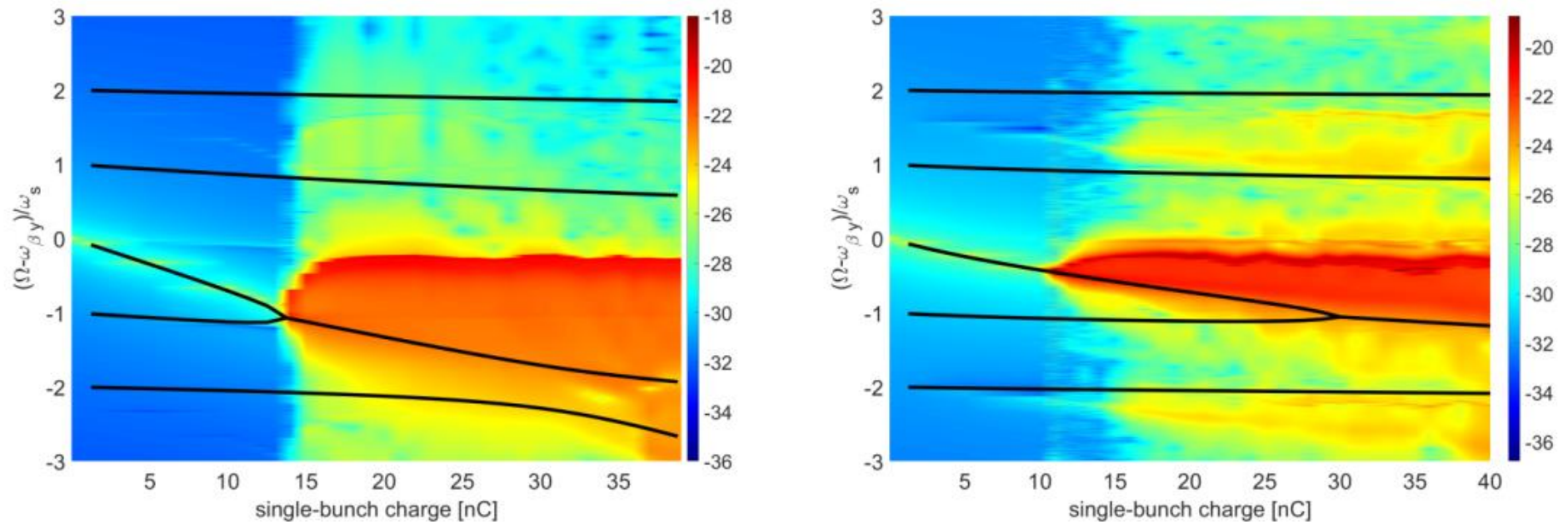

**Figure 4.2.3.7:** Simulation results of TMCI without longitudinal impedance (left) and with longitudinal impedance (right) for Z. Different color represents the amplitude of the transverse beam dynamics. The results are also compared with the theoretical estimations as shown by the black lines.

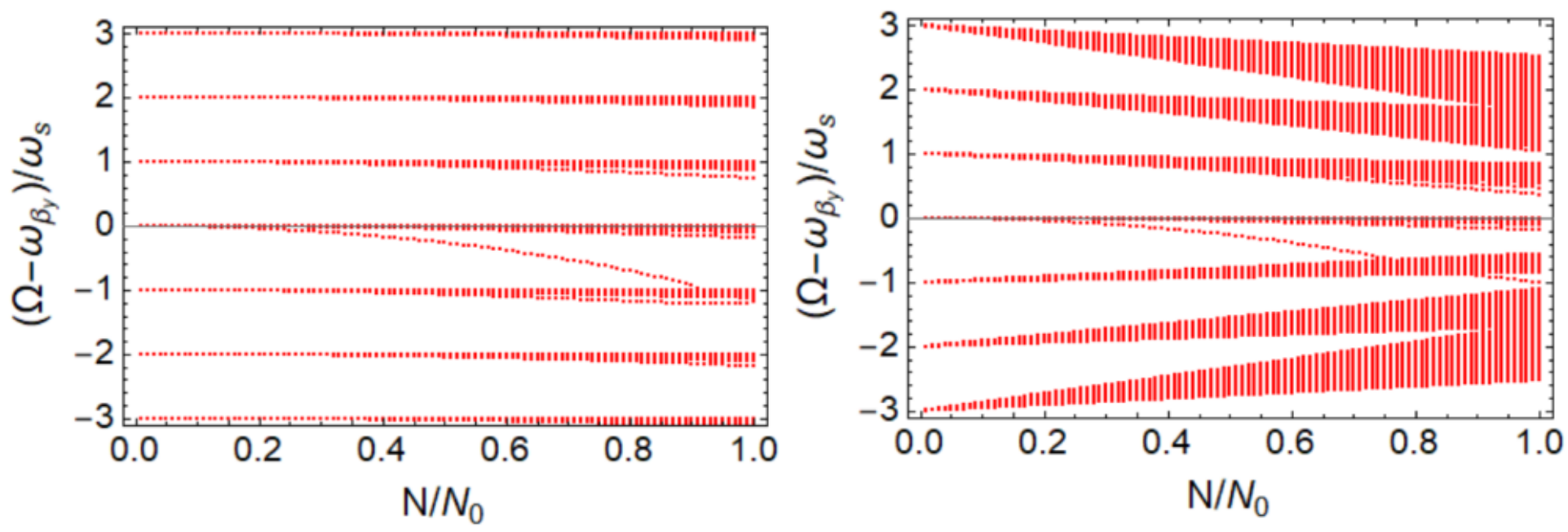

**Figure 4.2.3.8:** Transverse mode analysis including radial mode with lengthened bunch from beamstrahlung without (left) and with (right) longitudinal perturbation for Z ($N_0 = 2.49\times10^{11}$).

In the aforementioned single bunch instability studies, the impedance model from Section 4.2.3.1 was employed. However, it's important to note that the impedance model is subject to continuous refinement alongside hardware developments, potentially resulting in increased impedance contributions. Their impact on impedance and instabilities was estimated. Initially, the effective impedance was assessed by incorporating more intricate collimator designs, detailed vacuum chamber materials, and rough impedance models for the feedback kickers. The findings indicate a 3% increase in total longitudinal broadband impedance and a 10% increase in the loss factor. Consequently, the longitudinal beam dynamics remain relatively consistent with prior results. However, the transverse kick factor experiences an approximate 50% rise, with collimators contributing around 15% to this increase. As a linear scaling, the threshold bunch population for the single beam, as determined by TMCI, decreases from $6.3\times10^{10}$ to $3.2\times10^{10}$. Considering the stabilizing influence of the opposing colliding beam, the instability threshold reduces from $19\times10^{10}$ to $9.3\times10^{10}$. In this scenario, the combination of transverse bunch-by-bunch feedback and a positive chromaticity of approximately 10 is validated as an effective means to stabilize the beam [16]. While the required chromaticity for instability damping is considerable, its impact on single-particle beam dynamics is expected to be minor, given the natural chromaticity level is around 1000. Nevertheless, further detailed simulations will be conducted to validate this assertion.

Moreover, meticulous control of the impedance in the injection and extraction elements is imperative. The adoption of a low impedance design for machine protection elements, such as the nonlinear collimators proposed in SuperKEKB [17], will be contemplated. Anticipated is that the amalgamated effects of chromaticity and feedback will adequately suppress instability resulting from increased impedance, thereby avoiding any hindrance to machine performance in Z mode.

#### 4.2.3.3 ***Multi-bunch Effects***

In high-energy colliders with a large circumference, the dominant instability is often the transverse resistive wall instability. The growth time of the coupled bunch instability is primarily determined by the zero-frequency resonance of the transverse resistive wall impedance.

Analytical calculations indicate that for the Higgs and $t\bar{t}$ modes at the design bunch current, the growth times are significantly slower than synchrotron radiation damping, suggesting beam stability. However, for the W and Z operation modes, at beam energies corresponding to these particles, the instability growth times are 38 ms and 2 ms, respectively. These values are much faster than radiation damping, particularly for the Z operation mode.

Figure 4.2.3.9 illustrates the growth rate versus mode number for the Z mode. When considering only synchrotron radiation damping, thousands of modes could become unstable. The most critical mode occurs at a frequency of 2.338 kHz with a growth time of approximately 2 ms. The first five unstable modes have growth times of less than 5 ms, equivalent to around 15 revolution turns. Therefore, to mitigate this instability, an efficient bunch-by-bunch feedback damping system is essential. This system may involve multiple feedback systems distributed around the ring, as well as specialized single-mode damping designed to suppress the resistive wall instability.

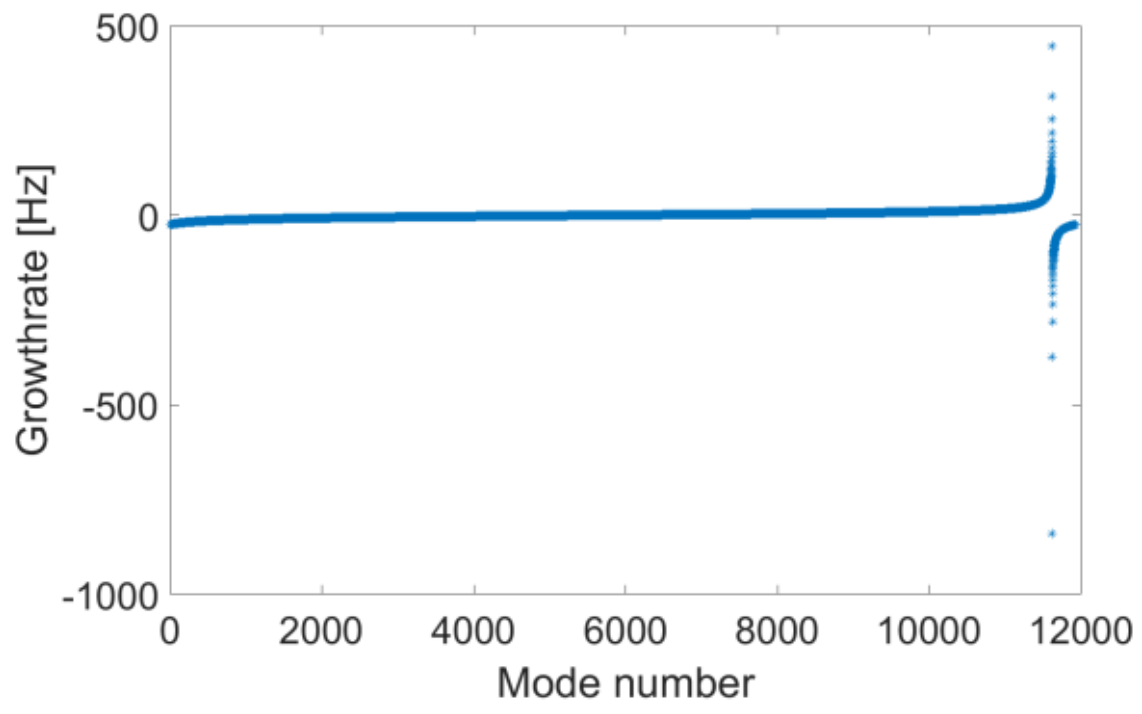

**Figure 4.2.3.9:** Resistive wall instability growth rate versus mode number for Z operation mode**.**

The influence of high-order modes (HOMs) originating from RF cavities on the coupled bunch instability is considered. Superconducting RF cavities operating at a frequency of 650 MHz are utilized.

For the Higgs, W, and low luminosity Z operation modes (with a total beam power of 10 W), 2-cell cavities are employed. The cutoff frequencies for the TM01 mode and TE11 mode, determined by the radius of the side beampipe, are approximately 1.35 GHz and 1.04 GHz, respectively. The longitudinal and transverse HOMs below the cutoff frequency are measured for the 2-cell cavities. The results, shown in Figure 4.2.3.10, are compared with the threshold determined by synchrotron radiation damping. With the 2-cell cavities, the impedance exceeds the threshold for the Z mode in both the longitudinal and transverse planes. In the resonant case, the instability has a growth rate of 114 ms and 19 ms in the longitudinal and transverse directions, respectively. Furthermore, in the W operation mode, the transverse instability is present when considering only synchrotron radiation damping. In the resonant case, the instability exhibits a growth rate of 105 ms. These instabilities can be further mitigated by the implementation of a bunch-by-bunch feedback system.

For the high luminosity Z mode (with a total beam power of 30 W or 50 W), the 1-cell cavities are utilized, which effectively damp the HOMs below the threshold determined by synchrotron radiation damping. Figure 4.2.3.11 depicts a comparison of the Higher Order Modes (HOMs) in the 1-cell cavities and their relation to the instability threshold for the high luminosity Z mode.

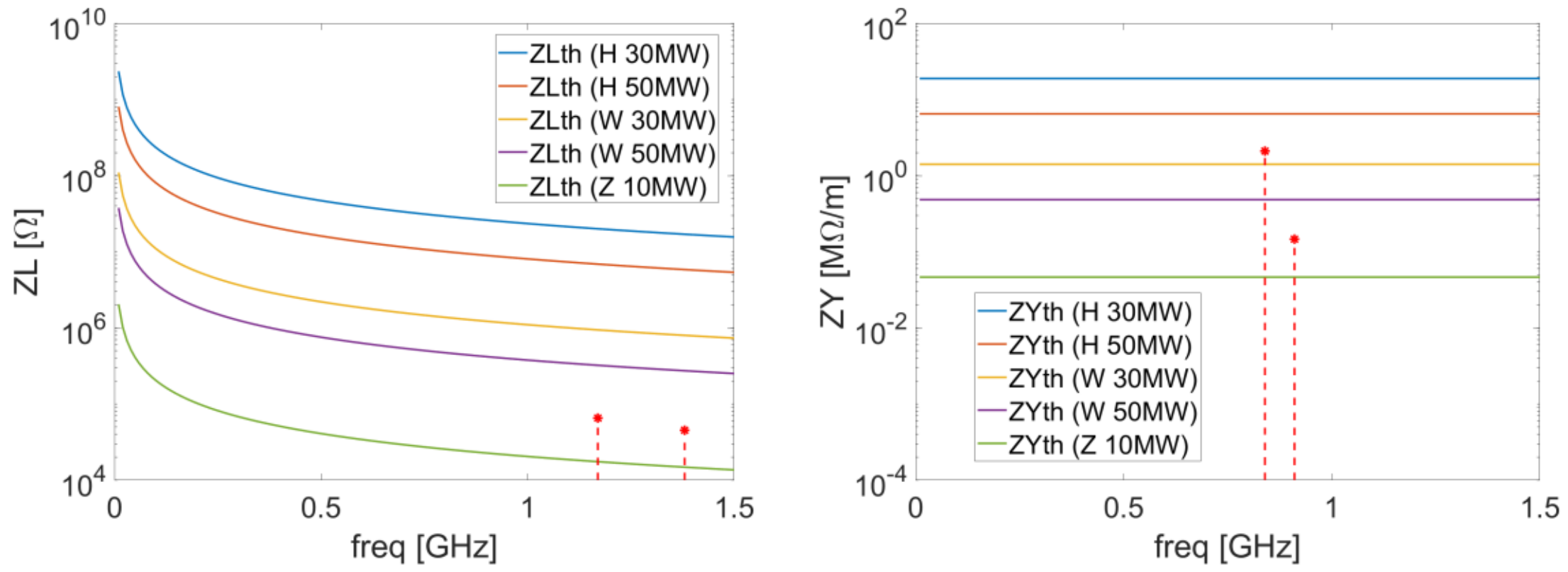

**Figure 4.2.3.10:** Impedance of the high order modes of the 2-cell RF cavity and compared with the threshold for different operation modes (left: longitudinal, right: transverse).

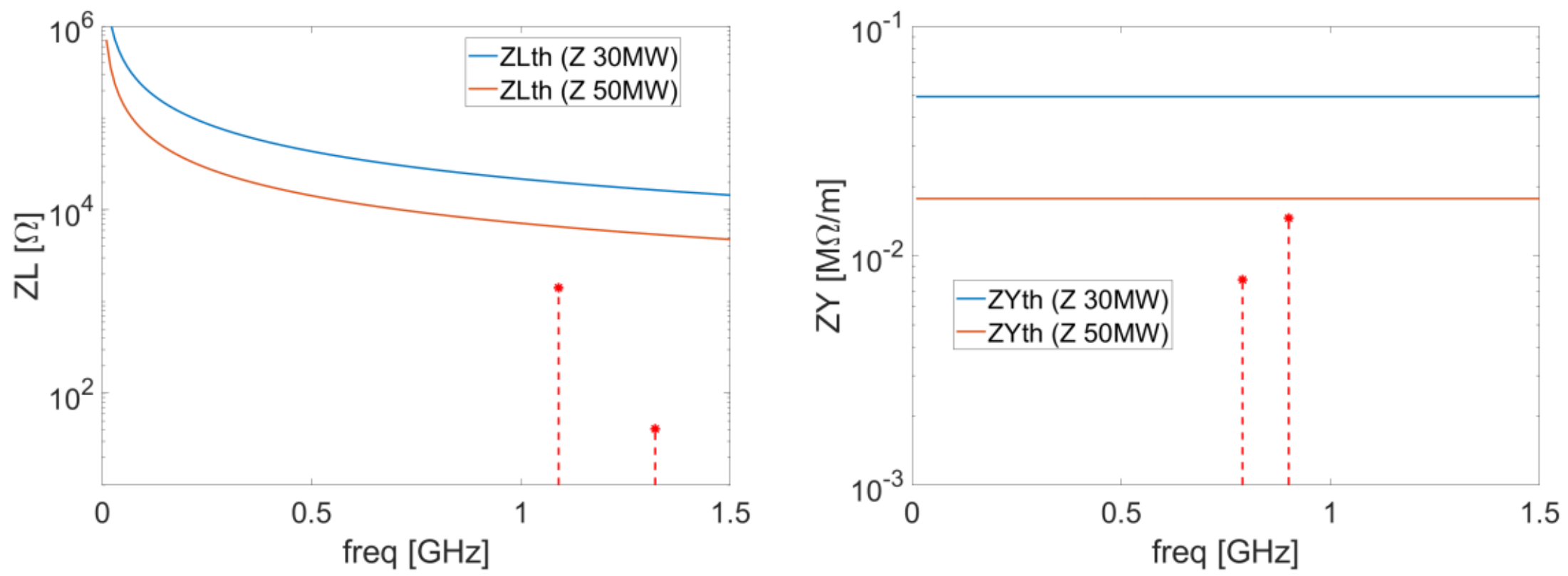

**Figure 4.2.3.11:** Impedance of the high order modes of the 1-cell RF cavity and compared with the threshold for different operation modes (left: longitudinal, right: transverse).

The narrowband impedances originating from components such as flanges, bellows, gate valves, and BPMs are effectively damped below the threshold determined by synchrotron radiation damping. This indicates that their impact on the coupled bunch instability is minimal. However, the narrowband impedance generated by the electro separators and IP chambers requires further investigation to assess its potential influence on the coupled bunch instability. HOM damping, such as using a ferrite damper, will be explored to address impedance issues.

### 4.2.3.4 *Electron Cloud Effect*

The build-up of accumulated photon electrons and secondary electrons has been one of the most serious restrictions on collider luminosity in PEP II, KEKB, LHC, and BEPC. The electron cloud (EC) can induce emittance blow-up and coupled bunch instabilities, and lead to luminosity degradation. For CEPC, the photon electrons and secondary electron emission will be the main contributions to the electron cloud. The necessary condition for electron amplification is that the average secondary electron emission yield (SEY) exceeds one.

Assuming uniform synchrotron radiation interaction with the vacuum chamber, the number of photons emitted by a positron per meter is given by [18]:

$$N_\gamma = \frac{5\pi}{\sqrt{3}}\frac{\alpha\gamma}{C} \tag{4.2.3.1}$$

where α = 1/137, γ represents relativistic energy, and C is the machine's circumference. Considering a circular cross-section for the beam pipe without an antechamber, a substantial value of 0.1 is assumed for the number of electrons produced when a photon strikes a wall. Consequently, for the Z mode, the number of electrons emitted by a positron per meter is 0.006 $m^{-1}$.

The simulated electron cloud accumulation in both the drift space and bending magnet for the Z mode is illustrated in Figure 4.2.3.12. In both cases, the maximum volume density of the electron cloud reaches approximately $1\times10^{11}$ $m^{-3}$.

Figure 4.2.3.13 displays the simulated electron cloud buildup in the drift space for the Higgs mode. The electron cloud is effectively damped during the bunch spacing, and the maximum volume density of the electron cloud in this case is around $2\times10^{9}$ $m^{-3}$.

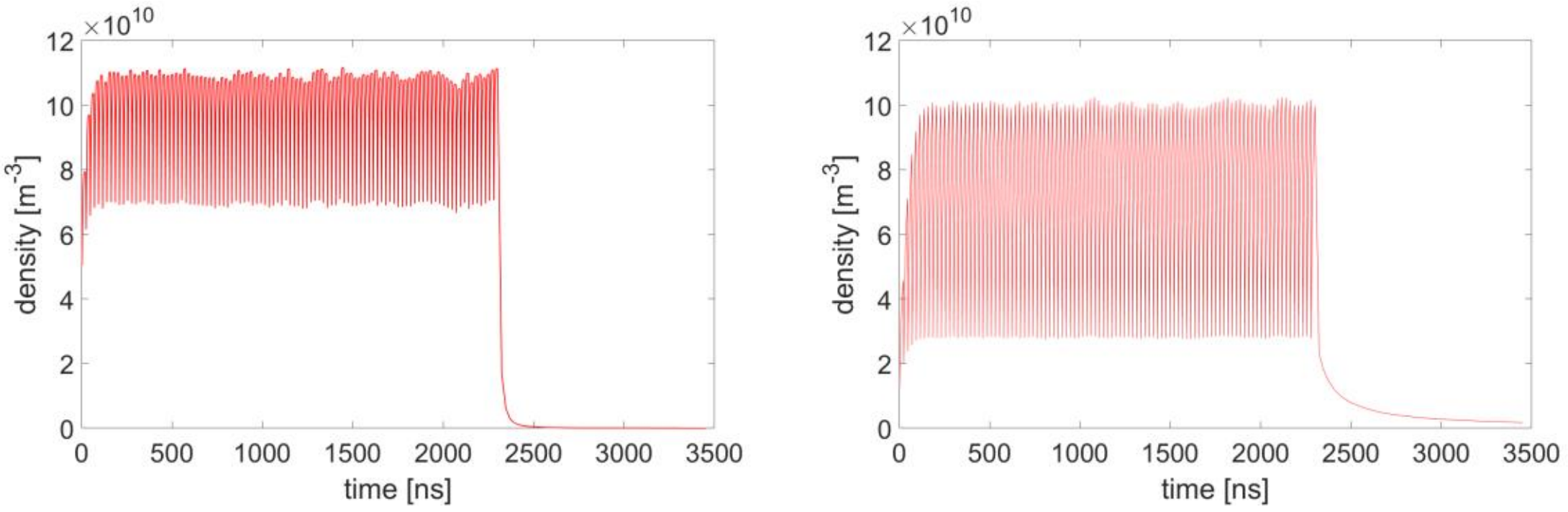

**Figure 4.2.3.12:** Electron cloud build-up (volume density) in the drift space (left plot) and in bending manget (right plot) in Z mode, where secondary emission yield is 1.2.

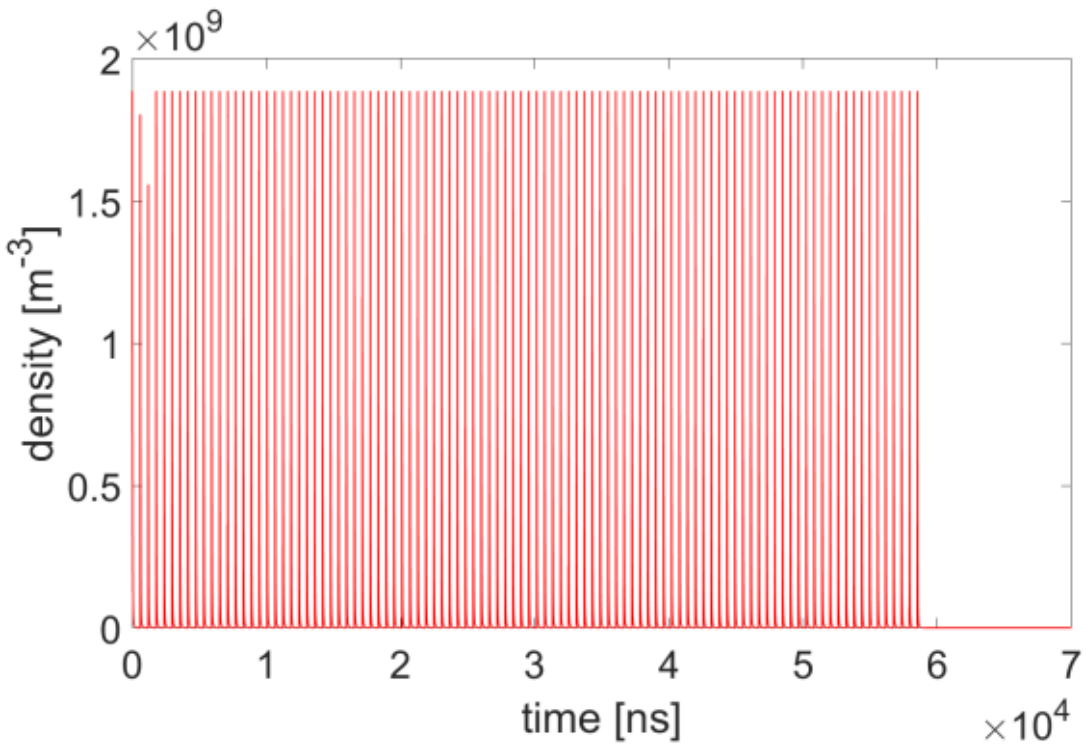

**Figure 4.2.3.13:** Electron cloud build-up (volume density) in the drift space (left plot) and in bending manget (right plot) in Higgs mode, where secondary emission yield is 1.2.

**Table 4.2.3.4:** Estimates on electron cloud instability for CEPC

| **Parameters** | **H** | **Z** |
|---|---|---|
| Bunch population ($10^{10}$) | 13 | 14 |
| Bunch number | 268 | 11934 |
| Emittance $x/y$ (nm) | 0.64/0.0013 | 0.27/0.0014 |
| Bunch spacing (ns) | 591 | 23 |
| Neutralization volume density ($10^{12}/m^3$) | 0.3 | 8.2 |
| Simulated electron cloud density ($10^{11}/m^3$) | 0.02 | 1.0 |
| Growth time (ms) | 687.4 | 4.2 |
| Threshold electron density ($10^{11}/m^3$) | 7.3 | 1.4 |

In the strong head-tail instability model, the threshold density of the electron cloud for single-bunch instability can be expressed as [19-20]

$$\rho_{e,th} = (2\gamma \nu_s \omega_e \sigma_z / c)/(\sqrt{3} KQr_e C\beta) \qquad (4.2.3.2)$$

where $K=\omega_e\sigma_z/c$, $Q$ depends on the nonlinear interaction, and $\omega_e$ the electron oscillation frequency. Assuming $Q=\pi/\sqrt{3}$, the threshold electron densities for various bunch spacings with SEY=1.2 are summarized in Table 4.2.3.4. In the case of Higgs, the threshold electron density significantly exceeds the simulated electron density, indicating that the electron cloud effect is expected to have minimal impact on the Higgs operation mode. For Z, the threshold electron density is approximately 40% higher than the simulated electron density.

Furthermore, electron cloud (EC) density was simulated for Z using various secondary electron yields (SEY) and compared with the threshold density, as illustrated in Fig. 4.2.3.14. It is observed that a SEY lower than 1.3 is adequate to mitigate the electron cloud instability in the Z pole operation mode. With the application of NEG coating to the inner surface of the beam pipe, it becomes feasible to maintain the maximum SEY below 1.1 after beam scrubbing. Further in-depth simulations will be conducted to assess the electron cloud's impact on beam dynamics.

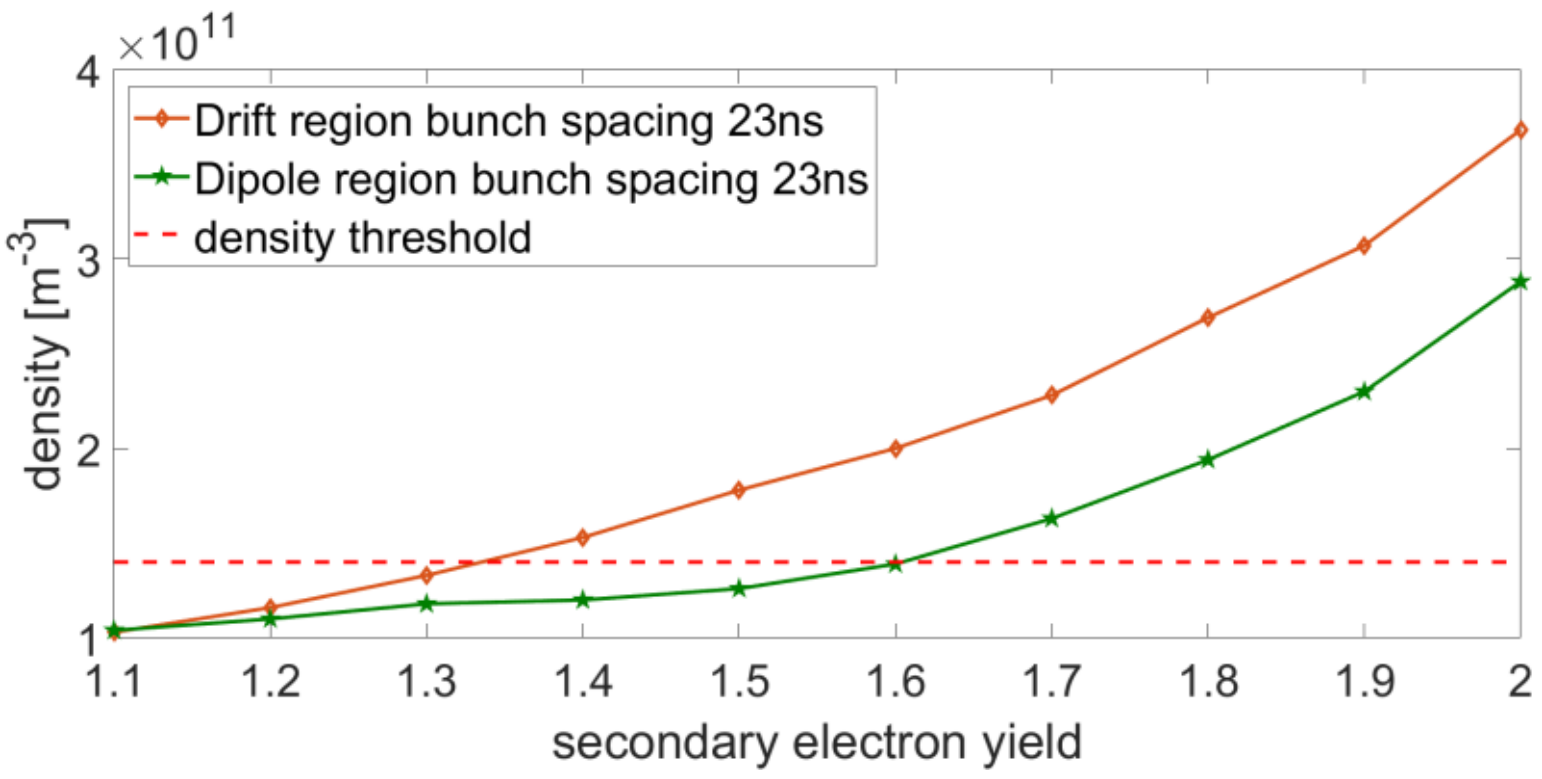

**Figure 4.2.3.14:** Electron cloud density for Z with different SEY and compared with the threshold density for single-bunch instability.

The electron cloud can cause coupled bunch instability by linking the oscillation between subsequent bunches. The growth rate for the coupled bunch instability is calculated as [20]:

$$1/\tau_{e,CB} = (2r_e n_b c^2)/(\gamma\omega_\beta ab L_{sep}) \tag{4.2.3.3}$$

Estimated growth times for simulated EC density at SEY=1.2 are listed in Table 4.2.3.4. A transverse feedback system with a damping time of a millisecond scale is required to suppress this instability.

### 4.2.3.5 *Beam-Ion Instability*

Trapped ions can contribute to various beam instabilities, including bunch centroid oscillation, beam size blow-up, and tune shift variation along the bunch train. However, they rarely directly cause beam losses. To optimize the beam operation for different energy scenarios, the evolution of beam ion density along the bunch train passage is calculated analytically, considering a vacuum chamber pressure of 1 nTorr and considering CO+ as the only ion species. The average ion density and the corresponding instability growth time are summarized in Table 4.2.3.5.

For the Higgs and $t\bar{t}$ processes, the synchrotron radiation is expected to effectively damp the instability. However, for the Z and W processes, the instability growth time is faster than the synchrotron radiation damping time. Therefore, the implementation of a transverse bunch-by-bunch feedback system is necessary to mitigate this effect.

The investigation focuses on the possibility of ion trapping and fast beam ion instability specifically for the Z process. The linear theory suggests that ions with a relative molecular mass larger than the critical mass $A_{x,y}$ can be trapped. Fig. 4.2.3.15 illustrates the critical mass distribution along the ring with a vacuum pressure of 1 nTorr and uniform bunch filling. The results indicate that only $CO_2^+$ ions will be trapped in locations around the interaction point with a large betatron function $\beta_y$. The trapped ion region constitutes less than 0.1% of the total area.

**Table 4.2.3.5:** Average ion density and instability growth time under design beam current and vacuum pressure of 1 nTorr.

| Parameters | Higgs | Z | W | $t\bar{t}$ |
|---|---|---|---|---|
| $L_{sep}\omega_{ion}/c_0$ | 6.2 | 0.7 | 2.4 | 14.7 |
| $\rho_{ion,\ ave}$ [$e^{11}$ $m^{-3}$] | 0.6 | 4.1 | 0.8 | 0.1 |
| $\tau_e$ [ms] | 7.9 | 0.1 | 1.9 | 105.4 |
| $\tau_H$ [ms] | 44.1 | 4.3 | 24.3 | 220.5 |
| $\Delta\nu_y$ | 0.0006 | 0.02 | 0.002 | 0.0001 |

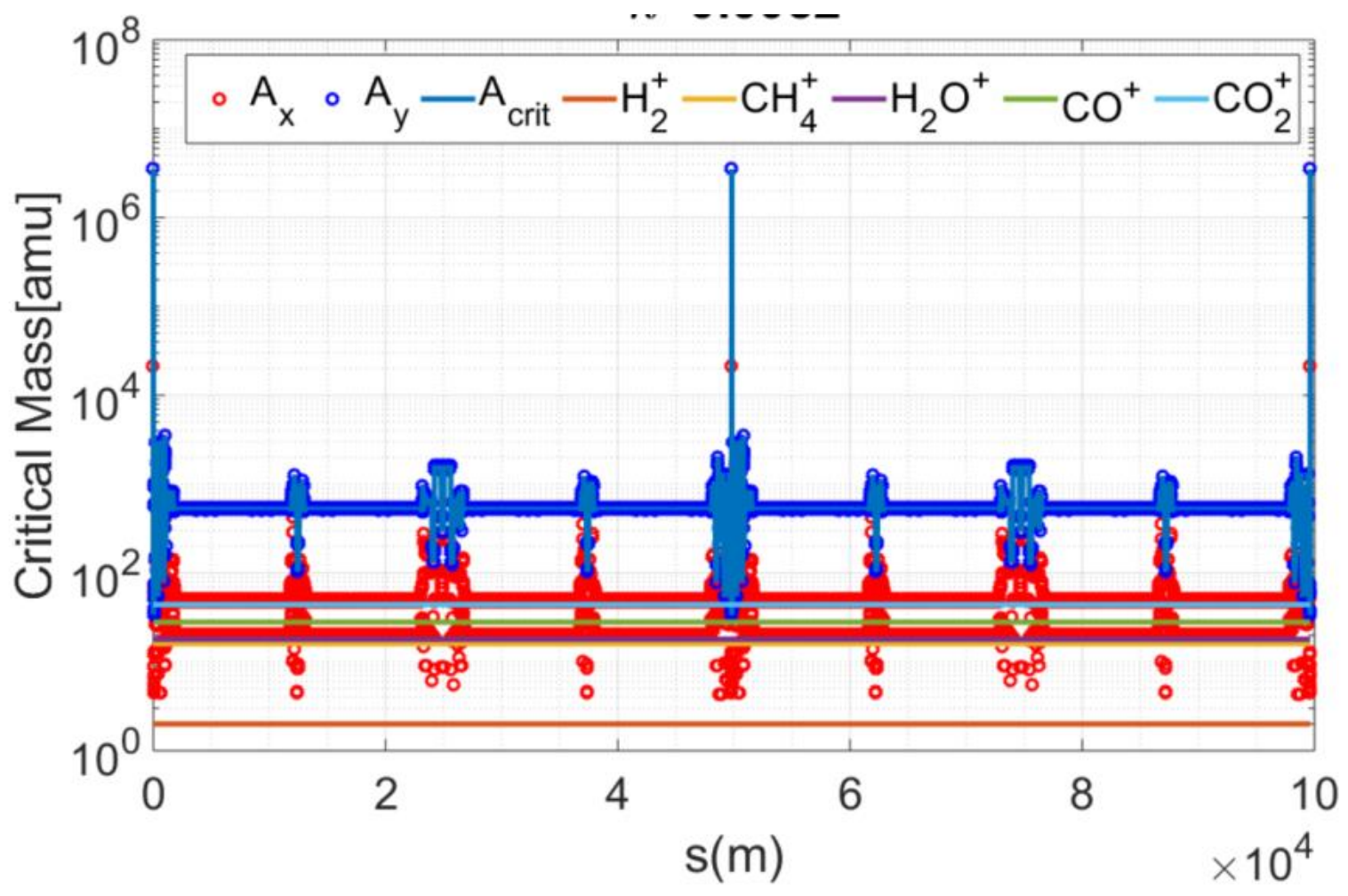

**Figure 4.2.3.15:** Relative molecular mass along the collider ring for Z mode.

In addition, particle tracking simulations are performed to study the ion build-up along the bunch train and the perturbation of transverse beam dynamics. Bunch centroid oscillations are obtained for both uniform and multi-train filling patterns with a bunch spacing of 23 ns and bunch intensity of $14\times10^{10}$. The evolution of the transverse oscillation amplitude for the uniform filling pattern is shown in Figures 4.2.3.16-19, indicating beam centroid oscillations exceeding 10% of the transverse beam size for most cases, with the amplitude dependent on the betatron functions. For the multi-train filling pattern, with 144 bunch trains each containing 83 bunches, Figure 4.2.3.20 shows a comparison of the beam centroid oscillation with the uniform filling pattern, revealing the effectiveness of multi-train filling in mitigating beam ion instability. There is no significant increase in the transverse beam size, although some minor oscillations are present, requiring the implementation of a bunch-by-bunch feedback system to mitigate the instability.

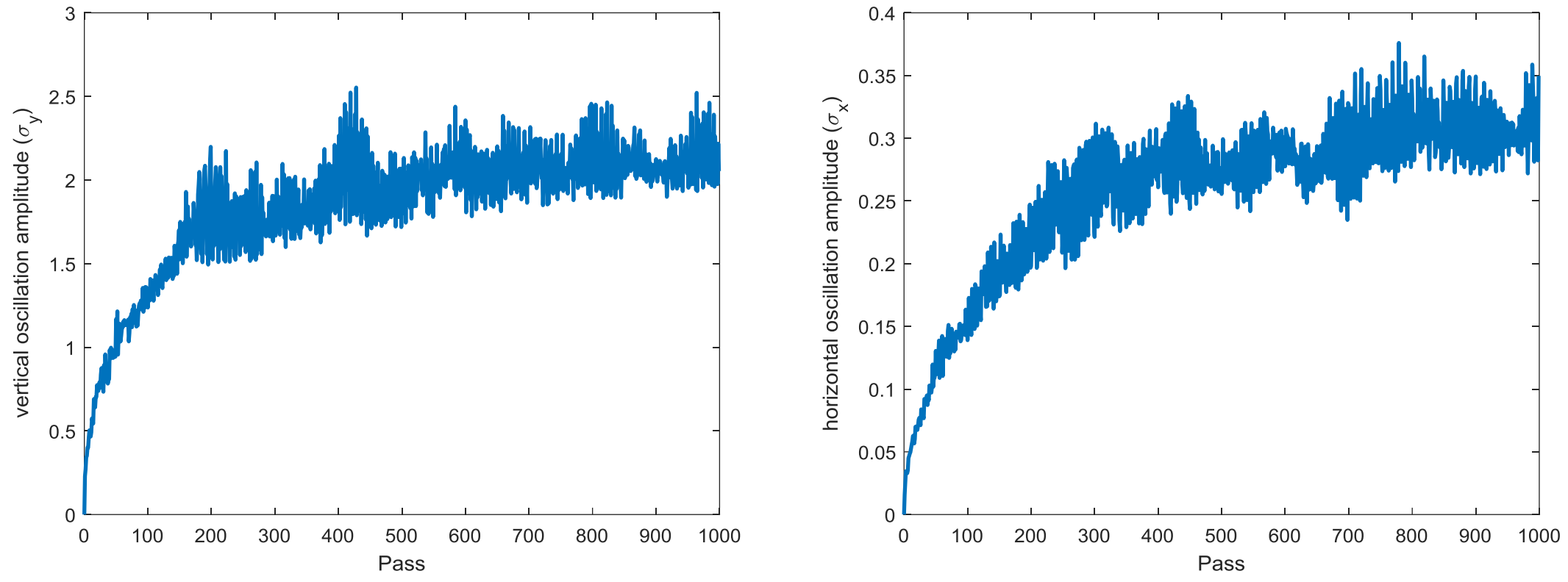

**Figure 4.2.3.16:** Beam centroid oscillation with uniform filling at betatron functions of $\beta_x/\beta_y$ = 92m/32m (left: vertical plane, right: horizontal plane).

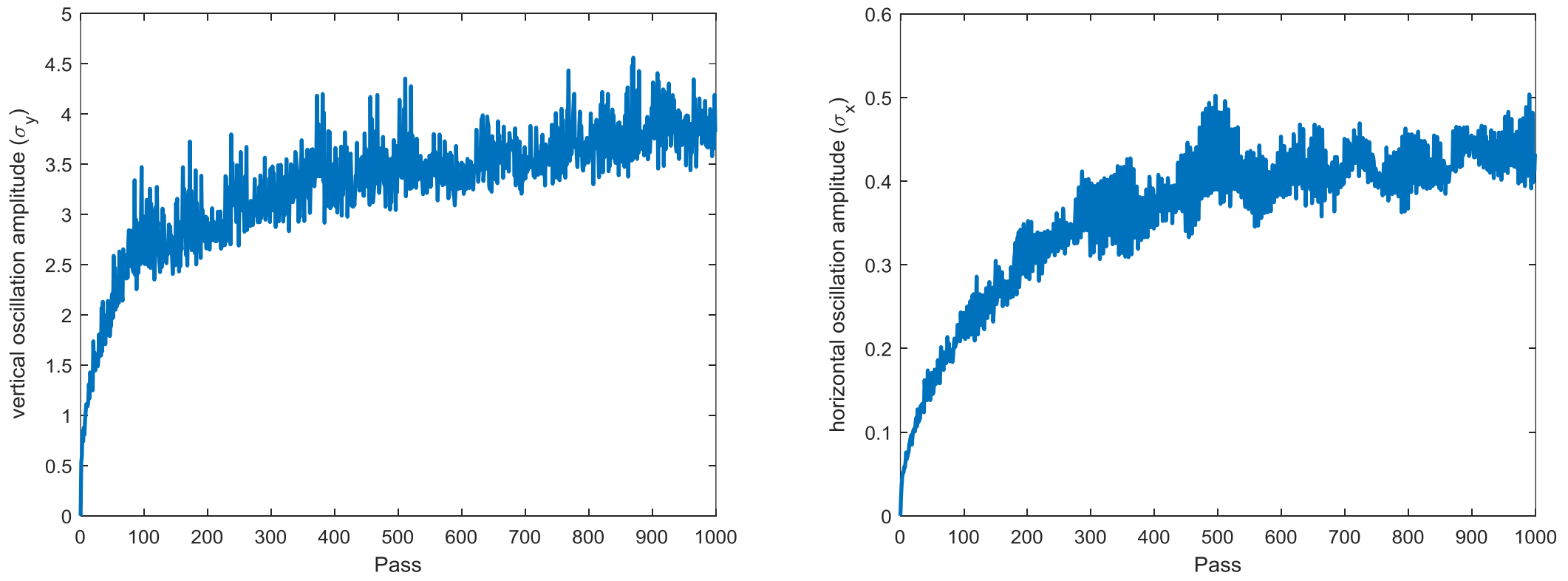

**Figure 4.2.3.17:** Beam centroid oscillation with uniform filling at betatron functions of $\beta_x/\beta_y$ = 69m/96m (left: vertical plane, right: horizontal plane).

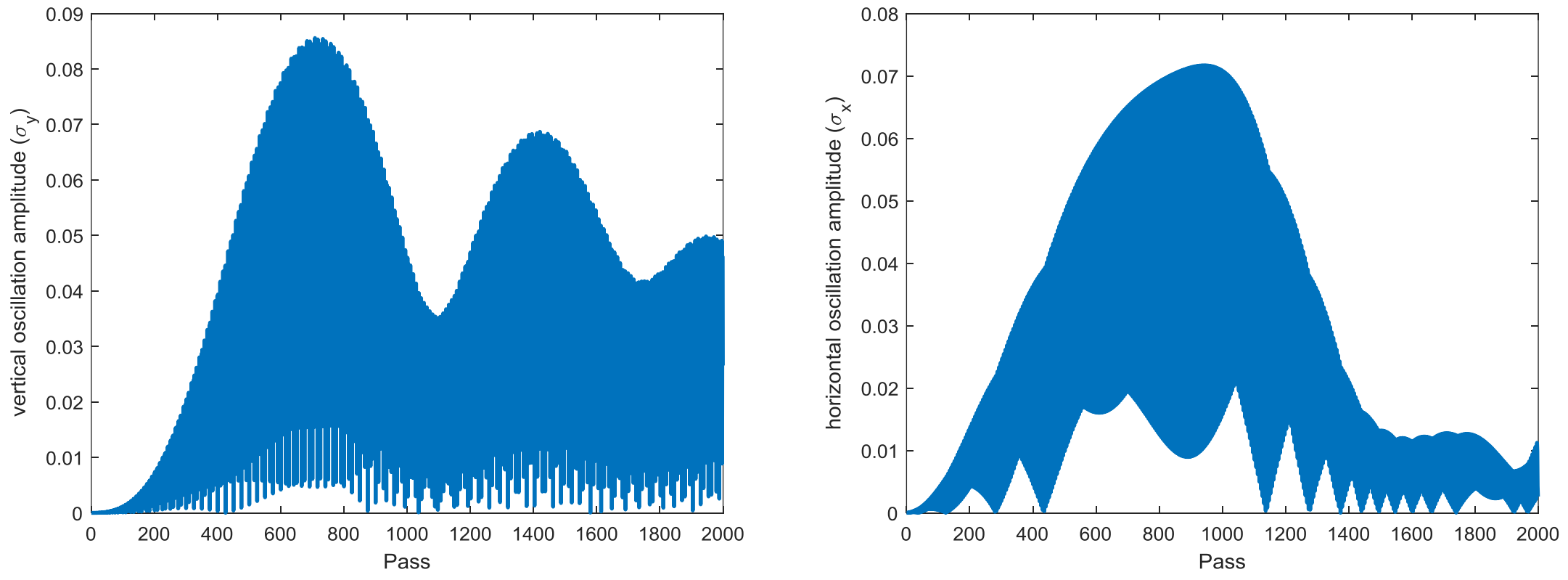

**Figure 4.2.3.18:** Beam centroid oscillation with uniform filling at betatron functions of $\beta_x/\beta_y$ = 0.13m/0.9mm (left: vertical plane, right: horizontal plane).

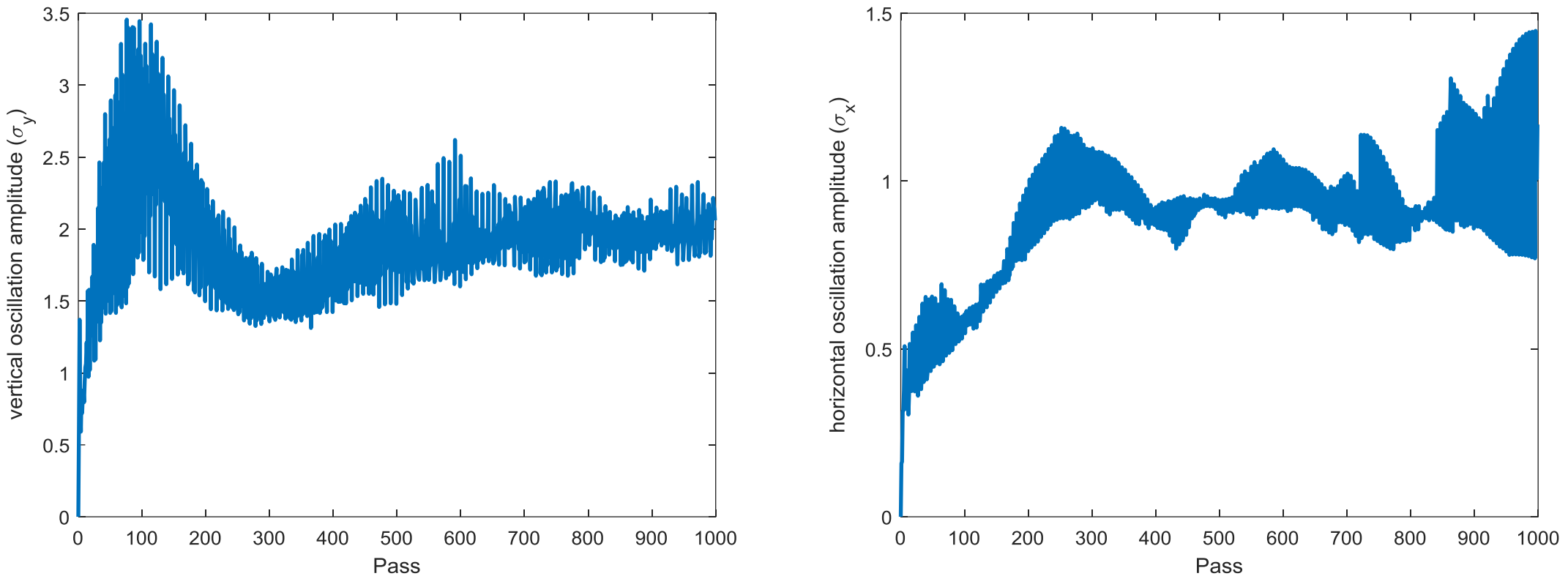

**Figure 4.2.3.19:** Beam centroid oscillation with uniform filling at betatron functions of $\beta_x/\beta_y$ = 151m/4.4km (left: vertical plane, right: horizontal plane).

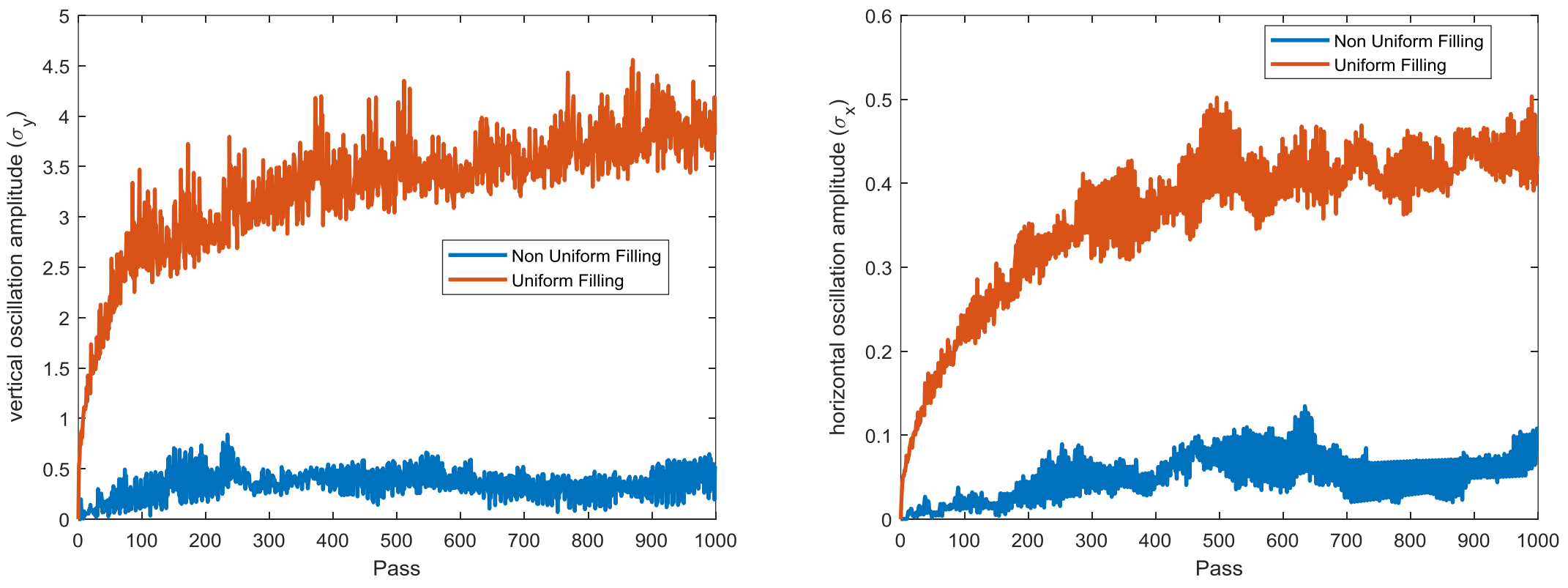

**Figure 4.2.3.20:** Comparison of beam centroid oscillation with uniform filling and multi-train filling pattern at betatron functions of $\beta_x/\beta_y$ = 69m/96m (left: vertical plane, right: horizontal plane).

### 4.2.3.6 ***Beam Lifetime***

#### *4.2.3.6.1 Beamstrahlung Lifetime*

If the strength of the beamstrahlung is such that particles have energy after collision outside of the ring's energy acceptance, they may exit the beam, hit the vacuum chamber walls, and reduce beam lifetime by emitting single photons in the tail of the beamstrahlung spectra [21]. To determine the beamstrahlung lifetime, the analytic formulas from [22] have been used. Beam-beam simulations have also been performed. The two approaches yield somewhat different results; the analytical calculation gives about 40 minutes for Higgs, while the beam-beam simulation gives 35 minutes. For $t\bar{t}$, the beamstrahlung lifetime is approximately 23 minutes according to the analytical estimation.

#### *4.2.3.6.2 Bhabha Lifetime*

The lifetime due to the radiative Bhabha scattering is the dominant item and can expressed by:

$$\tau_L = \frac{I}{eLn_{IP}\sigma_{ee}f_0} \tag{4.2.3.4}$$

where the cross section of the radiative Bhabha process can be calculated by:

$$\sigma_{ee} = \frac{16\alpha r_e^2}{3}\left(\left(\ln\frac{1}{\eta}+\eta-\frac{3}{8}\eta^2-\frac{5}{8}\right)\left[\ln\left(\sqrt{2}\frac{a}{\lambda_p}\right)+\frac{\gamma_E}{2}\right]+\frac{1}{4}\left(\frac{13}{3}\ln\frac{1}{\eta}+\frac{13\eta}{3}-\frac{3}{2}\eta^2-\frac{17}{6}\right)\right),$$

$$a = \sqrt{2}\frac{\sigma_x\sigma_y}{\sigma_x+\sigma_y}. \qquad (4.2.3.5)$$

where $\lambda_p$ and $\gamma_E$ denote the electron Compton wavelength and the Euler's constant, respectively. The simulation code BBBrem is also used to calculate the cross section, and the results are very close to the one obtained from the analytic formula (4.2.3.5). In the CEPC, it is $1.27\times10^{-25}$ cm$^2$ for Higgs, $1.36\times10^{-25}$ cm$^2$ for W, $1.34\times10^{-25}$ cm$^2$ for Z, and $1.32\times10^{-25}$ cm$^2$ for $t\bar{t}$. Therefore, the lifetime due to the radiative Bhabha scattering is about 40 minutes for H, 1.0 hours for W, 1.5 hours for Z, and 1.3 hours for $t\bar{t}$.

Equation (4.2.3.4) reveals that there is an inverse relationship between lifetime and luminosity. Therefore, it is necessary to find a balance between these two factors.

##### *4.2.3.6.3 Touschek Lifetime*

The Touchek lifetime is caused by single large-angle Coulomb scattering between particles in the beam, similar to the intra-beam scattering (IBS) process. This can result in random transformation of transverse momenta into longitudinal momenta, leading to a reduction of the beam lifetime if the energy deviation of particles exceeds the energy acceptance, which is limited by the longitudinal dynamic aperture in the case of CEPC. The Higgs has a Touchek lifetime of about 119 hours, W has 45 hours, Z has 10 hours, and $t\bar{t}$ has about 1,690 hours.

##### *4.2.3.6.4 Quantum Lifetime*

Lifetimes are reduced due to particle losses caused by finite transverse apertures and energy acceptance in Gaussian particle distributions, which can be calculated by:

$$\tau_q = \frac{1}{2}\tau_u\frac{e^{\xi}}{\xi}, \quad \text{with} \quad \xi = \frac{A_u^2}{2\sigma_u^2} \qquad (4.2.3.6)$$

where $u = x$, $y$ or $s$, $\sigma_u$ is the beam size in the three directions, and $A_u$ is the limiting half apertures that are determined by the transverse dynamic aperture or the energy acceptance of the dynamic aperture. Taking into account a more realistic particle distribution, the quantum lifetimes of the CEPC are determined through particle tracking. This evaluation includes the bare lattice, beam-beam effect, and beamstrahlung effect after multi-sextupole optimization. The calculated lifetimes are approximately 100 minutes for Higgs/W and 300 minutes for $t\bar{t}$.

##### *4.2.3.6.5 Vacuum Lifetime*

The beam chamber in a storage ring is not a perfect vacuum, and particles in the beam can scatter off residual gas atoms, resulting in two principal effects:

- Elastic Scattering: where the stored particle is transversely deflected and increases its betatron oscillation amplitude, potentially causing the particle to be lost at either the physical aperture or the dynamic aperture. Two types of elastic scattering processes occur: Rutherford scattering, in which the particle is scattered by the point-like Coulomb field of the residual gas atom's nucleus, and elastic scattering with electrons outside the nucleus of the residual gas.

- Inelastic Scattering: in addition to deflection, there is the possibility that a light quantum is emitted during the collision, changing the particle's energy, or the particle transfers energy to the residual gas atom. In both cases, the particle loses energy and can be lost at the RF acceptance limit or the off-momentum dynamic aperture limit. Two types of inelastic scattering processes occur: Bremsstrahlung scattering, in which the particle interacts with the nucleus of a residual gas atom and emits a photon, and inelastic scattering from an electron of the residual gas atom, in which the momentum transfer excites the atom.

4.2.3.6.5.1 Rutherford Scattering

Elastic scattering of beam particles off residual gas nuclei results in angular deflection, which can cause the betatron amplitude to exceed the transverse acceptance (physical or dynamic aperture) of the ring, leading to particle loss. The beam lifetime due to elastic scattering for given residual gas pressure $P$ can be calculated using the following formula [23]:

$$\frac{1}{\tau_{elastic1}} = \frac{2\pi r_0^2 c N_A}{R\gamma^2} \frac{P[Pa]}{T[K]} \oint \frac{\beta_{x,y}}{A^{min}[m \cdot rad]} \sum_i^n Z_i (Z_i + 1) N_i r_{pi} \tag{4.2.3.7}$$

where $N_A$ is the Avogadro's number, $R$ = 8.314 $J/(K{\cdot}\mathrm{mol})$ is the universal gas constant, $T(K)$ is the residual gas temperature, $P(Pa)$ is the pressure, $n$ is the residual gas partial components number, $r_{pi}$ is the partial fraction of the residual gas components, $Z_i$ is the atomic number, $N_i$ is the number of atoms per molecule, and $A_{min}$ is the transverse acceptance (the smaller one of the dynamic aperture and physical aperture).

4.2.3.6.5.2 Eleastic Scattering from an Electron of the Residual Gas Atom

During the scattering of the particle beam with electrons outside the nucleus of the residual gas, the incident high-energy particles can transfer some of their energy to the electrons. If the energy deviation of the particles exceeds the momentum acceptance, they will be lost. The lifetime due to this type of elastic scattering is given by:

$$\frac{1}{\tau_{elastic2}} = \frac{2\pi r_e^2 Z}{\gamma} \frac{1}{\delta_m} cn \tag{4.2.3.8}$$

where $n$ is the density of the residual gas, which is calculated by:

$$n[m^{-3}] = 3{,}217 \times 10^{22} P[Torr] \tag{4.2.3.9}$$

4.2.3.6.5.3 Bremsstrahlung Scattering

Bremsstrahlung is a process where an electron in the beam loses energy through interaction with the residual gas atoms and emits a photon. If the electron's energy deviation exceeds the limiting momentum half-aperture (either the RF bucket half-height or the dynamic aperture) of the ring, it will be lost. The lifetime due to bremsstrahlung is given by [23]:

$$\frac{1}{\tau_{inelastic1}} = \frac{4r_0^2 cN_A}{137R} L(\delta_m) \frac{P[Pa]}{T[K]} \sum_i^n \ln\frac{183}{Z_i^{1/3}} Z_i (Z_i + \xi_i) r_{pi} N_i \qquad (4.2.3.10)$$

with

$$\xi_i = \ln\left(1440 Z_i^{-2/3}\right) / \ln\left(183 Z_i^{-1/3}\right) \qquad (4.2.3.11)$$

$$L(\delta_m) = 4/3 \cdot \left(\ln(1/\delta_m) - 5/8\right) \qquad (4.2.3.12)$$

where $\delta_m$ is the momentum acceptance of the ring. Other variables are explained in the sections above.

4.2.3.6.5.4 Inelastic Scattering from an Electron of the Residual Gas Atom

In the process of inelastic scattering of beam particles with the electrons outside the nucleus of the residual gas, the shell electrons are excited and emit photons, leading to a reduction in beam lifetime, which is given by:

$$\frac{1}{\tau_{inelastic2}} = \frac{4r_e^2 Z}{137} \frac{4}{3} (ln\frac{2.5\gamma}{\delta_m} - 1.4)(ln\frac{1}{\delta_m} - \frac{5}{8}) \qquad (4.2.3.13)$$

### *4.2.3.6.6 Summary of Beam Lifetime*

The total lifetime can be calculated by:

$$\frac{1}{\tau_{total}} = \sum_i \frac{1}{\tau_i} \qquad (4.2.3.14)$$

where $i$ = 1, 2, 3, …, $\tau_i$ represents a specific type of lifetime. Table 4.2.3.6 summarizes the beam lifetime of the CEPC at the Higgs energy.

**Table 4.2.3.6:** Summary of the CEPC beam lifetime at the Higgs energy ($P_{SR}$ = 30 MW).

| | $t\bar{t}$ | **Higgs** | **W** | **Z** |
|---|---|---|---|---|
| Bhabha Lifetime (min) | 80 | 40 | 60 | 90 |
| Touschek Lifetime (hour) | 1404 | 119 | 30 | 7 |
| Vacuum Lifetime (hour) | 25 | 15 | 9 | 5 |
| Quantum Lifetime (min) (lattice+BB+BS)* | 278 | 139 | 139 | - |
| Total lifetime (min) | 62 | 30 | 38 | - |
| Lifetime requirement for top-up injection (min) | 18 | 18 | 22 | 77 |

* Simulated beam lifetime with bare lattice, beam-beam effect and beamstrahlung effect after multi-sextupole optimization.

The lifetimes for Higgs/W/$t\bar{t}$ all meet the specified requirements, but the lifetime for Z has not yet reached the required level and remains under investigation.

### 4.2.4 Synchrotron Radiation and Shield

#### 4.2.4.1 *Introduction*

Synchrotron radiation (SR) is a major source of unwanted energy loss and heat deposition in colliders [1]. It has a broad energy spectrum ranging from visible light to several MeVs [2-3], and is extremely powerful, often a thousand times greater than regular beam losses [4]. As a result, SR significantly contributes to high radiation dose rates in the tunnel air and nearby components, leading to numerous issues such as vacuum chamber and air heating, ozone and nitrogen-oxide formation, and radiation damage [5-6]. To mitigate SR-related damage, special vacuum-chamber designs or shielding are necessary [4].

#### 4.2.4.2 *Synchrotron Radiation from Bending Magnets*

The CEPC, with a circumference of 100 km, features bending arcs and straight sections. Dipole magnets for bending and quadrupole magnets for focusing are installed in the bending areas. As electrons and positrons pass through the dipole magnets, a significant amount of synchrotron radiation is emitted. To assess the impact of SR, parameters such as radiation power, critical energy, and the number of emitted photons must be considered. The total energy loss per unit length is given by:

$$\mathit{\Delta} E = 14.08\frac{E^4}{\rho^2}, \tag{4.2.4.1}$$

where $\mathit{\Delta} E$ is the energy loss per unit length in keV/m, $E$ is the energy of electrons and positrons in GeV, $\rho$ is the bending radius in meters. The total SR power per complete turn can be calculated by

$$P = 88.46\frac{E^4 I}{\rho}, \tag{4.2.4.2}$$

where I is the circulating particle current in mA. Combining the above equations, the total SR power emitted by the electrons or positrons per unit length is:

$$P = 14.08\frac{E^4 I}{\rho^2}. \tag{4.2.4.3}$$

The critical energy $E_c$ of the synchrotron spectrum, which characterizes the "hardness" of the radiation, is given by:

$$E_c = 2.218\frac{E^3}{\rho}, \tag{4.2.4.4}$$

The photon spectrum from a single radiating electron can be expressed as:

$$\frac{d^2 N}{d\varepsilon dt} = \left(\frac{2\alpha}{h\sqrt{3}}\right)\left(\frac{1}{\gamma^2}\right)\int_{\varepsilon/E_c}^{\infty} K_{5/3}(\eta) d\eta, \tag{4.2.4.5}$$

where $\alpha$ is the fine-structure constant, $h$ the Planck constant, $\gamma$ the ratio of total energy $E$ to the rest mass of electron and positron, and $K_{5/3}(\eta)$ is a modified Bessel function of order 5/3. A single electron radiates at a rate of:

$$\frac{d^2N}{d\varepsilon dt} = (\frac{5.32 \times 10^5}{E^2}) \int_{\varepsilon/E_c}^{\infty} K_{5/3}(\eta) d\eta. \quad (4.2.4.6)$$

As the distance *ds* travelled in a time interval of *dt* equals *cdt*, the loss along the orbit for a single electron is:

$$\frac{d^2N}{d\varepsilon ds} = (\frac{1.775 \times 10^{-3}}{E^2}) \int_{\varepsilon/E_c}^{\infty} K_{5/3}(\eta) d\eta. \quad (4.2.4.7)$$

Let $\frac{\varepsilon}{E_c} = r$. Then the total number of photons emitted by an electron per meter is:

$$\frac{dN}{ds} = \int_0^{\infty} E_c \left(\frac{d^2N}{d\varepsilon ds}\right) dr = 3.936 \left(\frac{E}{\rho}\right) \int_0^{\infty} \int_r^{\infty} K_{5/3}(\eta) d\eta dr = 19.4 \frac{E}{\rho}. \quad (4.2.4.8)$$

The synchrotron-radiation spectrum, *R*(*k*), in units of photons per MeV per electron per meter of bending length is also expressed as:

$$R(k) = \frac{aS(x)}{x}, \quad (4.2.4.9)$$

where $a$ is a parameter expressed by $3809.5/E^2$, and $E$ the beam energy in GeV. The universal function $S(x)$ can be calculated by the following formula:

$$S\left(\frac{\omega}{\omega_c}\right) = 0.4652 \frac{\omega}{\omega_c} \int_{\omega/\omega_c}^{\infty} K_{5/3}(\eta) d\eta, \quad (4.2.4.10)$$

where $\omega$ is the angular frequency of the photon in rad/s and $\omega_c$ is the angular frequency of the photon at critical energy in rad/s. Photons are emitted tangentially to the curved trajectory of the beam into a cone, of which the opening angle can be characterized by:

$$\varphi = \frac{1}{\gamma} = \frac{m_0 c^2}{E}. \quad (4.2.4.11)$$

The parameters of the SR for the CEPC collider dipoles at 50 MW are shown in Table 4.2.4.1. The critical energy and opening angle can be calculated using the above equations.

**Table 4.2.4.1** Parameters of Collider SR in 50MW case.

| Parameters | Symbols | Values | | | | Units |
|---|---|---|---|---|---|---|
| Beam energy | $E$ | 120 | 45.5 | 80 | 180 | GeV |
| Beam current | $I$ | 28.8 | 1340.9 | 140.2 | 5.5 | mA |
| Bending radius | $\rho$ | 10700 | | | | m |
| Power per unit length | $P$ | 744 | | | | W/m |
| Critical energy | $E_c$ | 357.5 | 19.5 | 105.9 | 1208.9 | keV |
| Bending angle | $\Theta$ | 2.659 | | | | mrad |
| Opening angle | $\Phi$ | 4.258 | 11.23 | 6.388 | 2.839 | μrad |

Synchrotron radiation possesses features that could significantly impact this study. Figure 4.2.4.1 displays the photon spectra of SR produced in the CEPC collider dipoles at different electron energies, namely, 45.5 GeV, 80 GeV, 120 GeV and 180 GeV. The abscissa denotes the energy, and the ordinate denotes the number of photons.

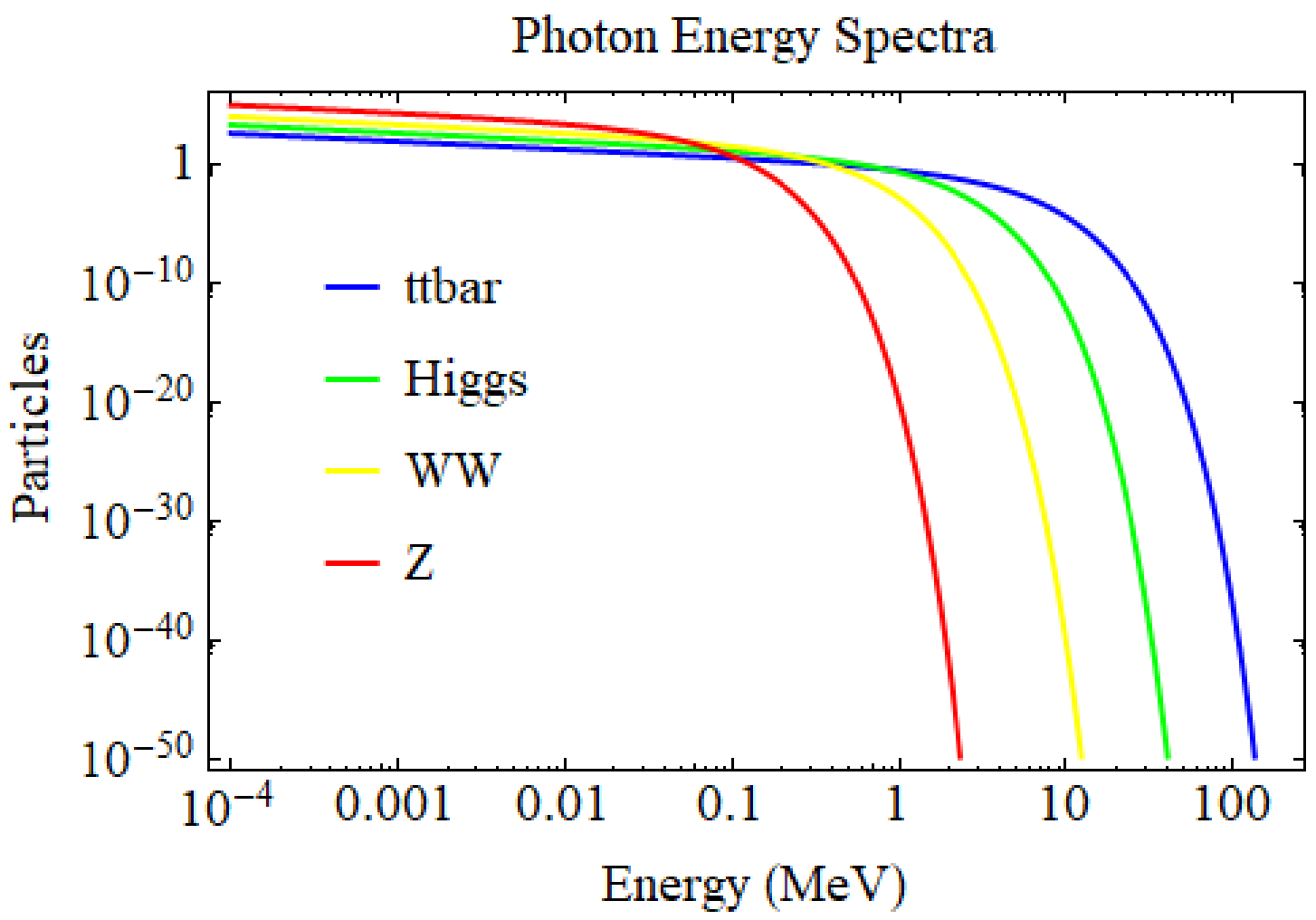

**Fig. 4.2.4.1** The photon spectrum of the SR emitted at different beam energy.

The total number of photons emitted per unit length calculated using Equation 4.2.4.8 is shown in Table 4.2.4.2

. **Table 4.2.4.2** Parameters of Collider SR

| Parameters | Values | | | | Units |
|---|---|---|---|---|---|
| Beam energy | 120 | 45.5 | 80 | 180 | GeV |
| Number of SR photons | $4.0 \times 10^{16}$ | $7.4 \times 10^{17}$ | $1.3 \times 10^{17}$ | $1.2 \times 10^{16}$ | $s^{-1}m^{-1}$ |
| Ave. energy of photons | 110.1 | 6.0 | 37.1 | 372.3 | keV |

### 4.2.4.3 *Monte Carlo Simulation*

In this section, we present Monte-Carlo simulation results, which involve repeated random sampling to derive numerical outcomes. These simulations essentially emulate "virtual experiments" performed via computer simulation, utilizing diverse algorithms within the Monte-Carlo particle-transport code FLUKA [7-9]. These simulations are facilitated through the user-friendly interface Flair [10], and they involve the generation of synchrotron radiation (SR) photons using FLUKA's built-in functions.

The CEPC CDR describes the energy deposition caused by SR and estimates the absorbed dose for the collider ring. To shield SR, a two-centimeter layer of lead is utilized, ensuring that the absorbed doses to the coil insulations and the tunnel decrease to a safe level. This section delves into an extended examination of synchrotron radiation shielding beyond what was covered in the CDR, and it provides an overview of the optimization process for lead.

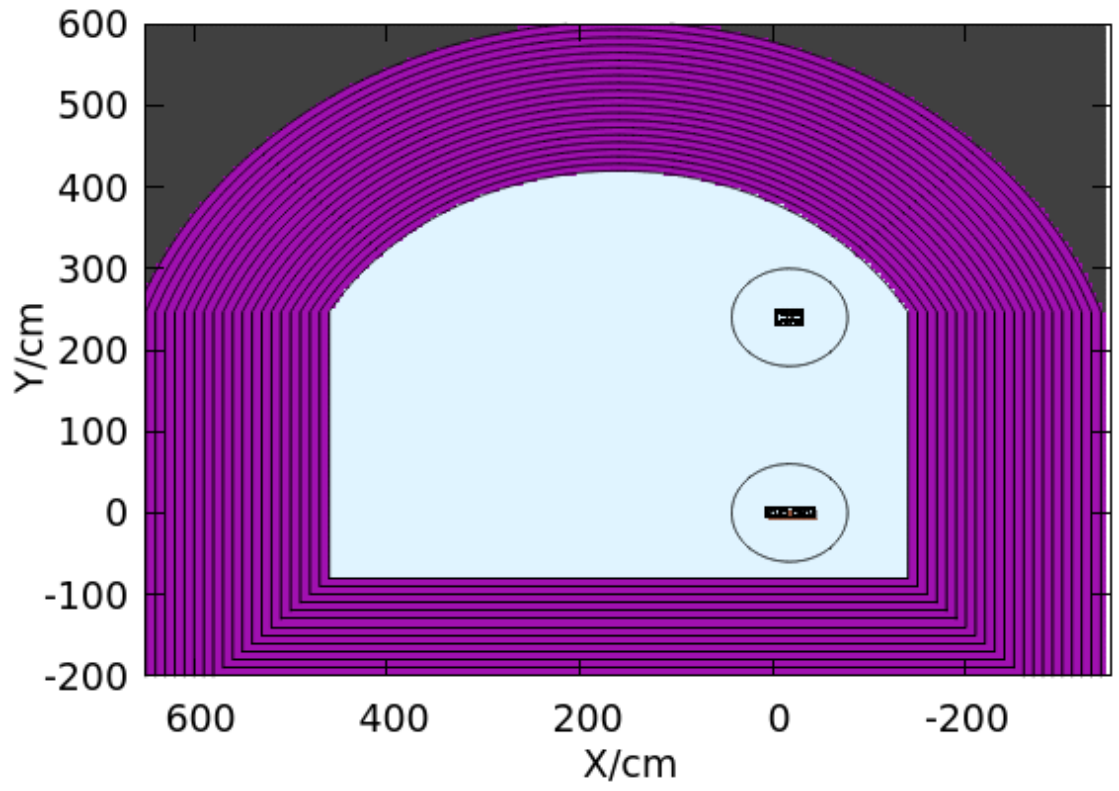

**Figure 4.2.4.2:** The cross section of the CEPC tunnel. The purple layers represent the rock wall. Inside the lower circle is the Collider ring, and inside the upper circle the Booster ring.

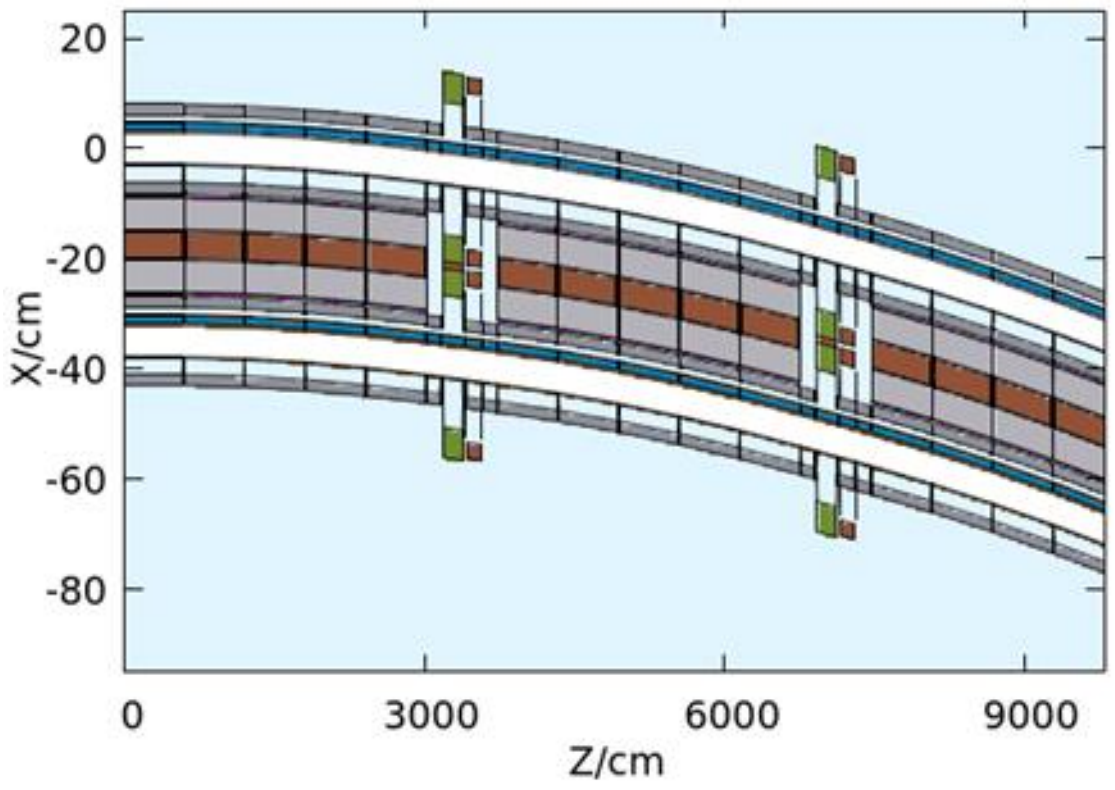

**Figure 4.2.4.3:** Top-down view of a 100-meter segment of the Collider.

The tunnel geometry has been updated to reflect the revised CEPC designs, as shown in Figure 4.2.4.2. The rock layers are depicted in purple, and the collider and booster rings are visible. The collider is located in the bottom circle, while the top circle contains the booster ring. Figure 4.2.4.3 provides a top-down view of a segment of the Collider, which consists of three dipoles, two quadrupoles, two sextupoles, and four drift chambers. Each dipole is segmented into five parts, with distances between segments and adjacent magnets on the order of 10 cm.

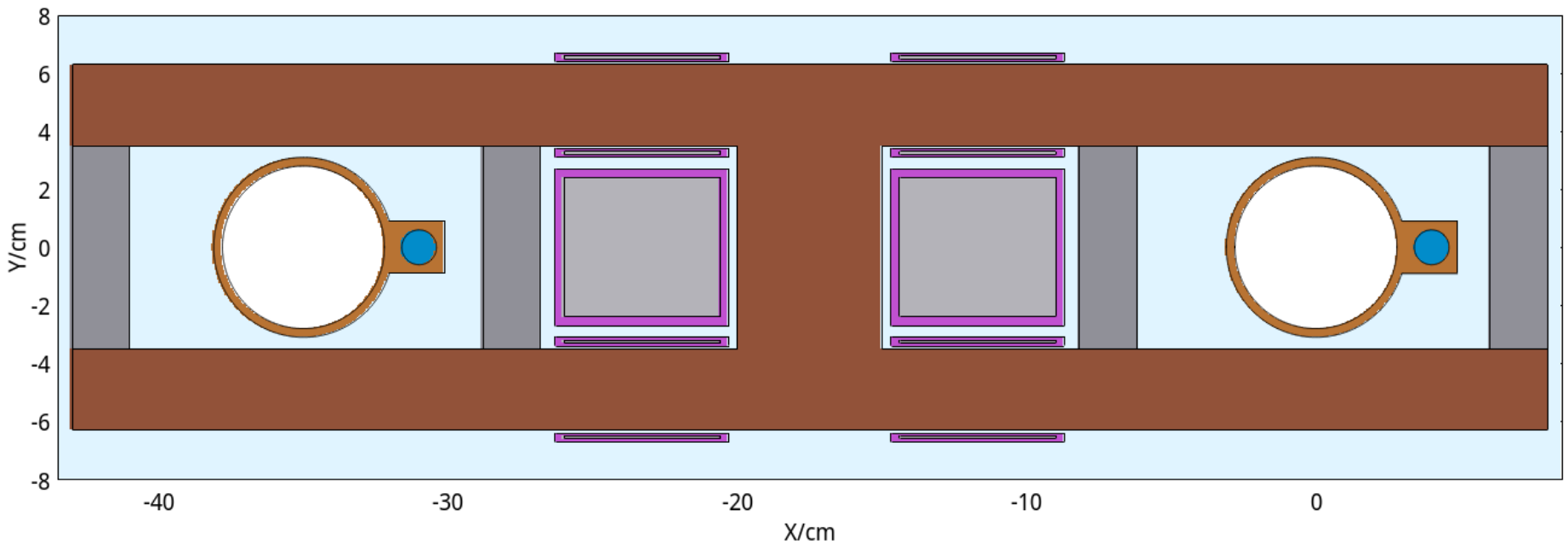

**Figure 4.2.4.4:** The X-Y cross section of the Collider dipole. The beam pipes are represented by light brown shapes, while cooling water is denoted by blue shapes. The magnet cores are depicted in dark brown, lead shielding is shown in dark grey, coils are presented in light gray, and insulations composed of epoxy resin are colored purple.

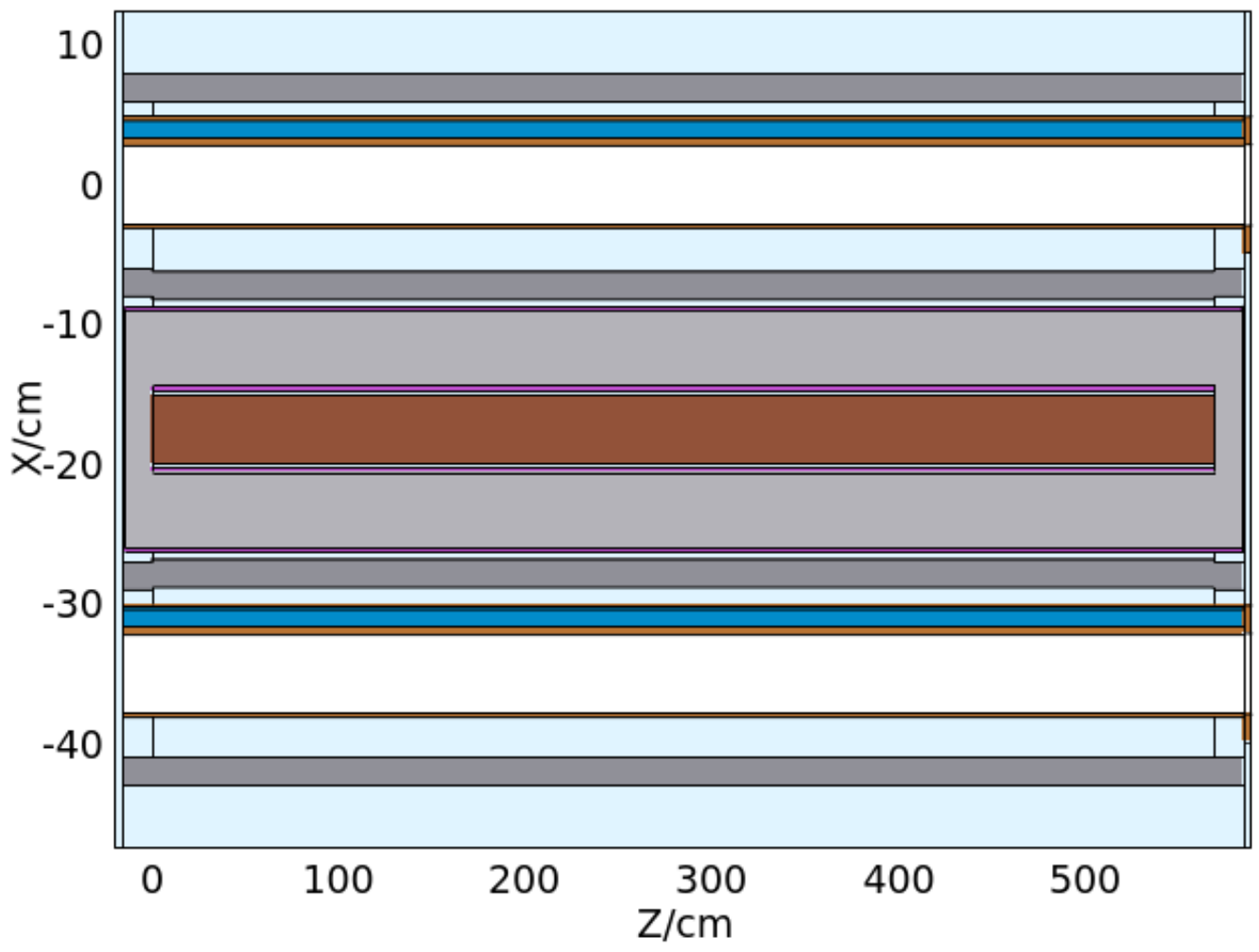

**Figure 4.2.4.5:** The X-Z cross section of the Collider dipole. Colors are explained in Fig. 4.2.4.4.

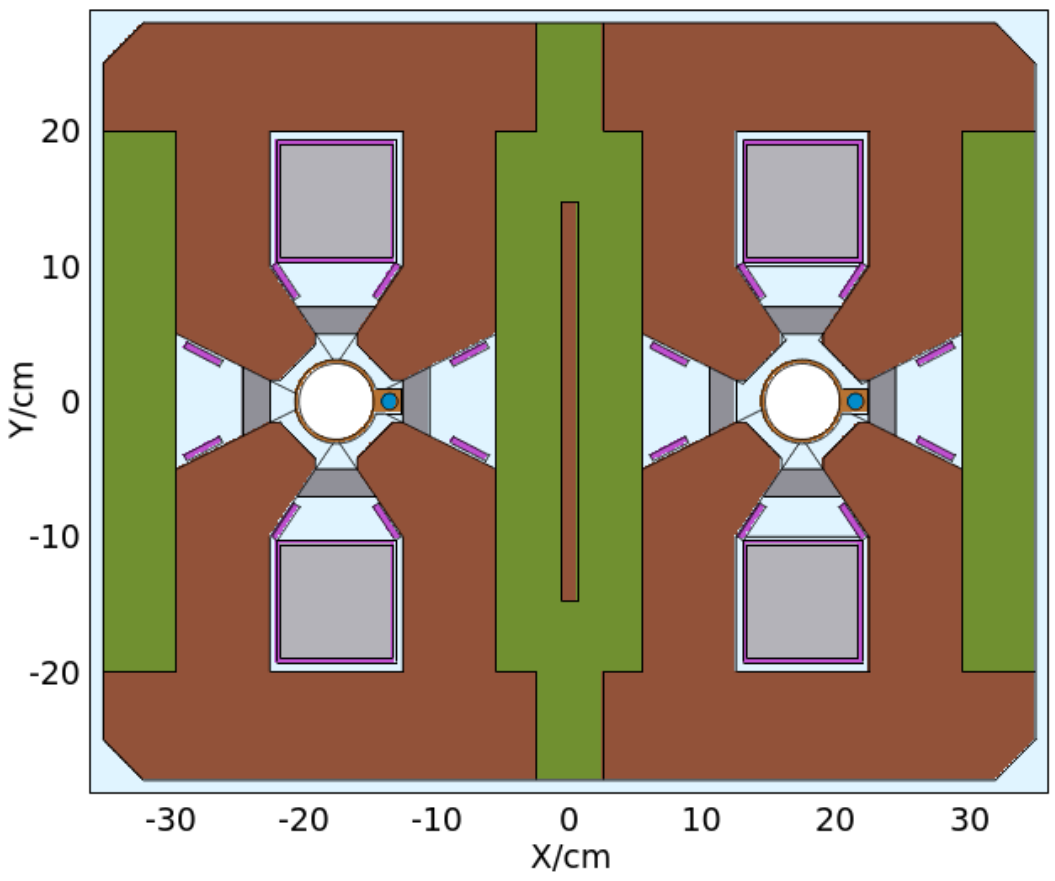

**Figure 4.2.4.6**: The X-Y cross section of the Collider quadrupole. Colors are the same as in Fig. 4.2.4.4. Green represents the stainless steel.

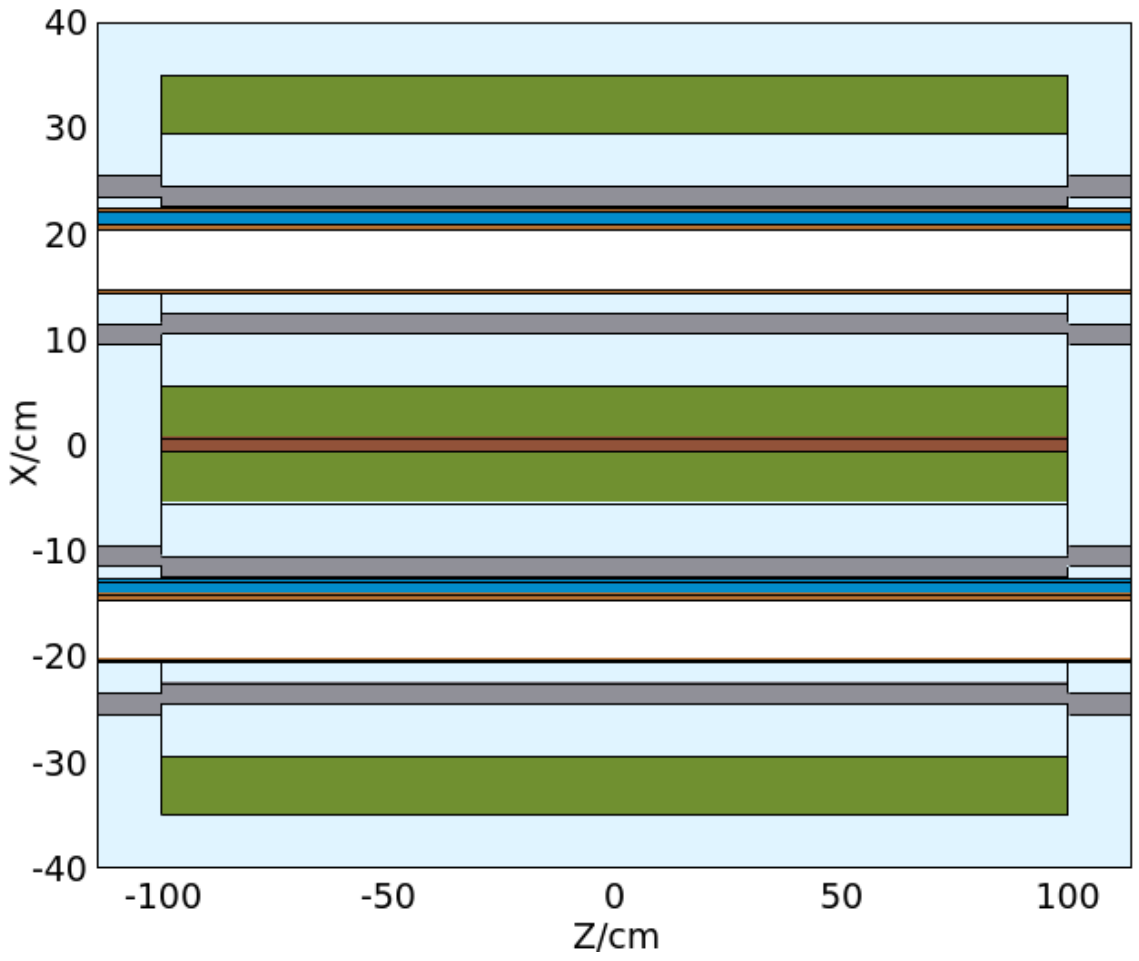

**Figure 4.2.4.7**: The X-Z cross section of the Collider quadrupole. Colors are explained in Fig. 4.2.4.6.

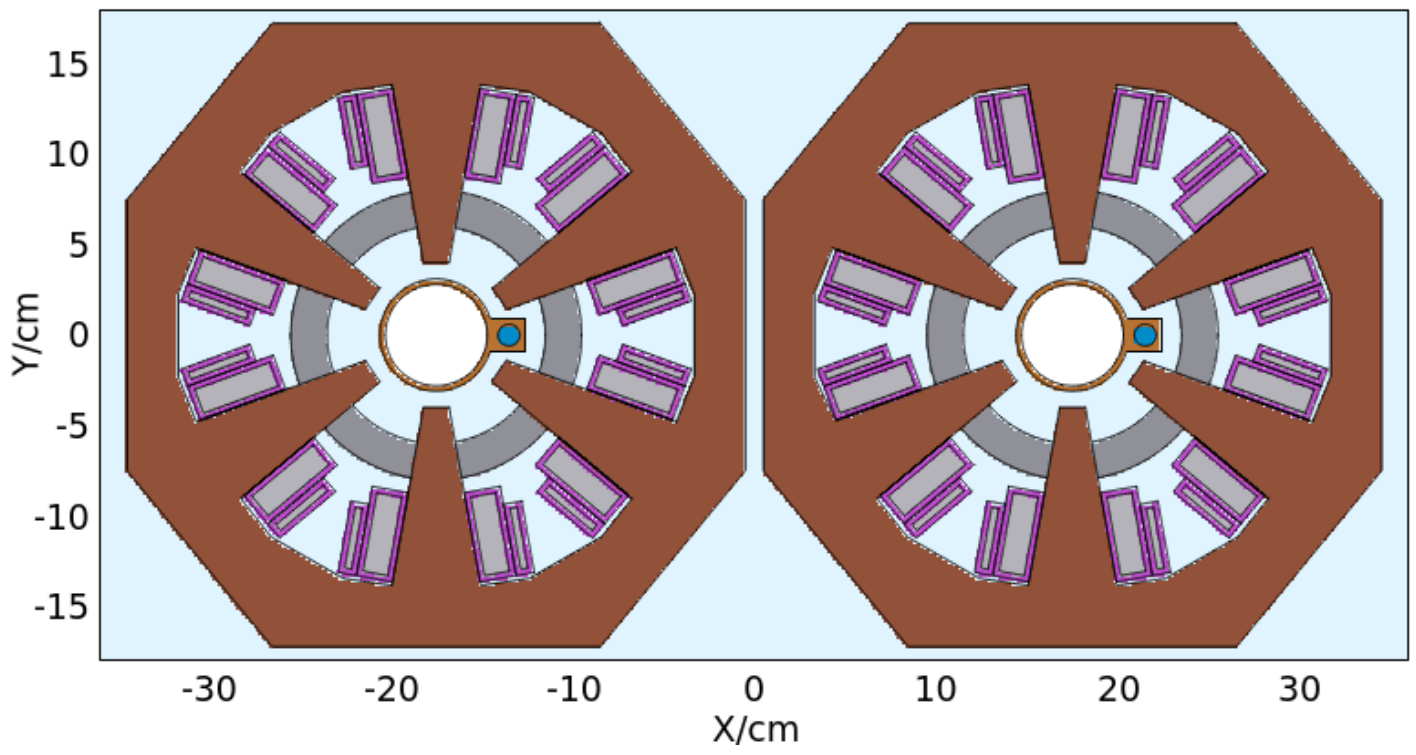

**Figure 4.2.4.8:** The X-Y cross section of Collider sextupole. Colors are explained in Fig. 4.2.4.4.

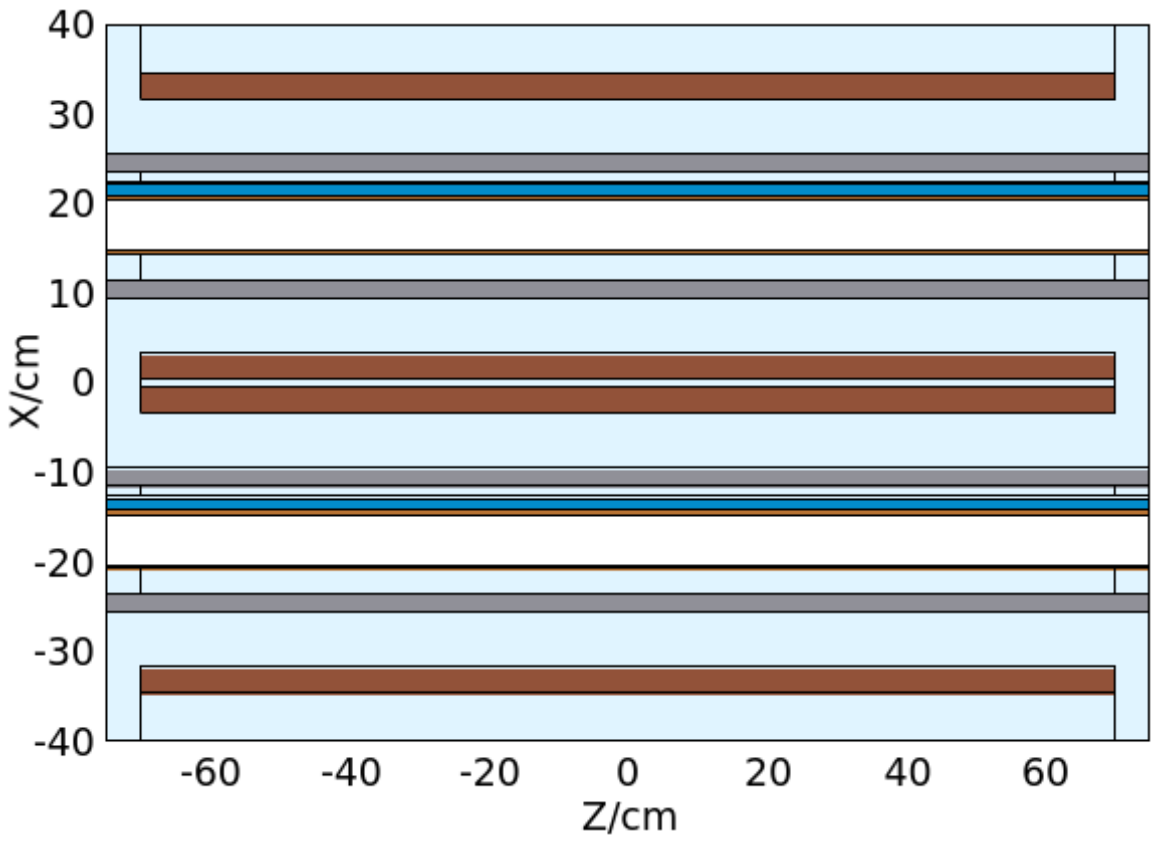

**Figure 4.2.4.9**: The X-Z cross section of Collider sextupole.

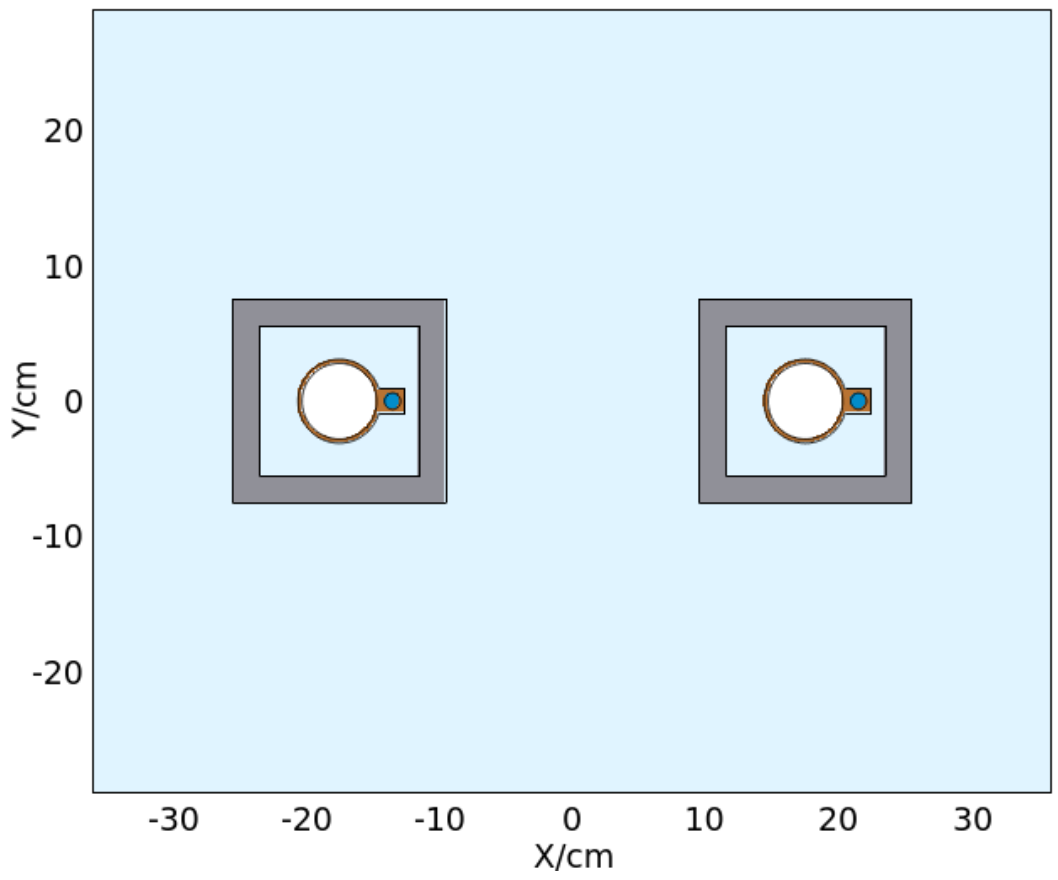

**Figure 4.2.4.10:** The X-Y cross section of the Collider vacuum chamber and lead shielding in the drift space. Colors are explained in Fig. 4.2.4.4.

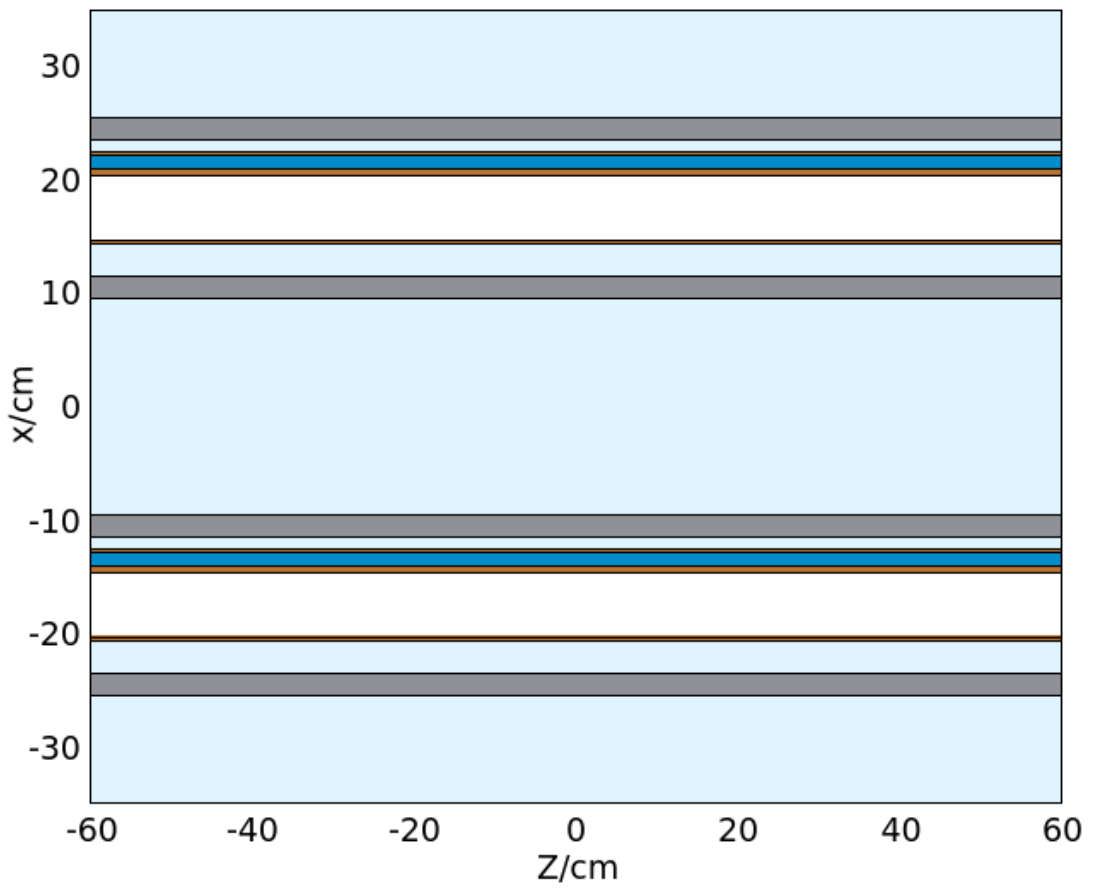

**Figure 4.2.4.11:** The X-Z cross section of the Collider vacuum chamber and lead shielding in the drift space.

The magnet geometries are depicted in Figures 4.2.4.4 - 4.2.4.11. The Collider beam pipes are composed of 3mm-thick copper. Each beam pipe is equipped with a cooling channel on the outside.

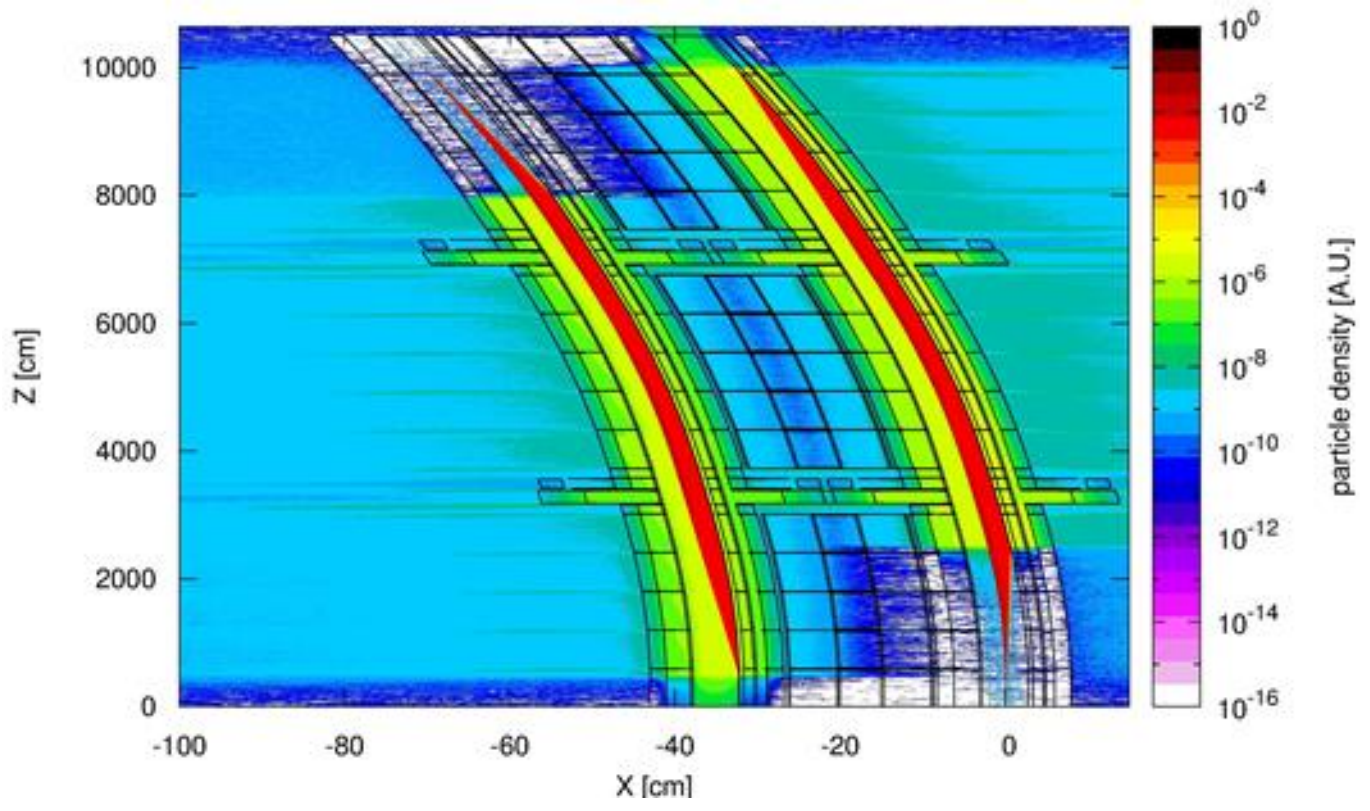

**Figure 4.2.4.12:** Particle fluence in the Collider plane.

### 4.2.4.4 *Absorbed Doses to Insulations*

The particle fluence of SR photons and secondary particles is shown in Figure 4.2.4.12. Red regions indicate higher concentration of SR photons. Photons generated at the entrance of the dipole hit the downstream beam pipe roughly 20 meters away, which can be calculated using the dipole's bending angle and the beam pipe radius. In addition, high fluence regions (sharp peaks) were found in the gaps between neighboring magnets due to the lead shielding not covering the beam pipes near the flange. An example of a flange between a drift chamber and quadrupole is shown in Figure 4.2.4.13, where there is a gap in the lead covering. The particle fluence near the flange is shown in Figure 4.2.4.14. Secondary particles escaping from this gap can affect the coils and other equipment. Detailed design of these areas will be completed in the engineering design phase.

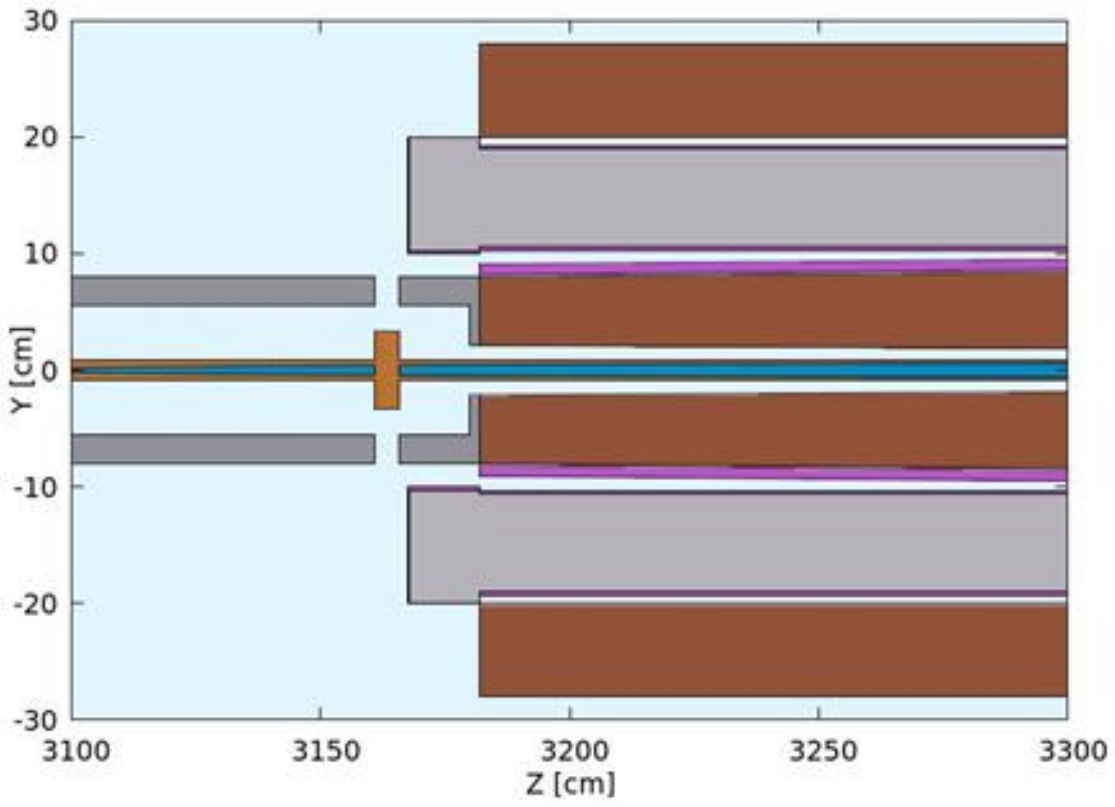

**Figure 4.2.4.13:** Side view of a flange between the drift chamber and the quadrupole.

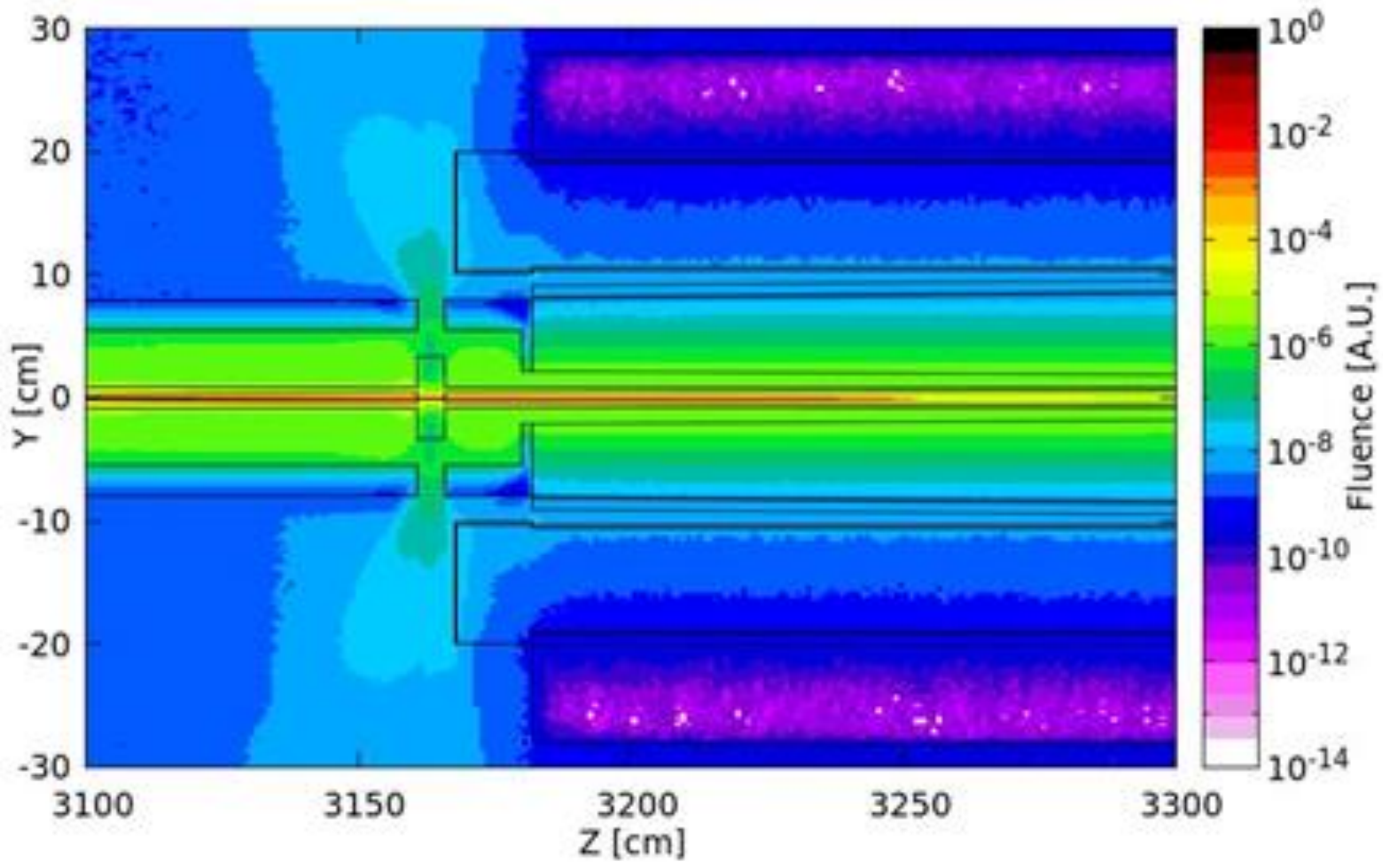

**Figure 4.2.4.14:** Fluence around a flange

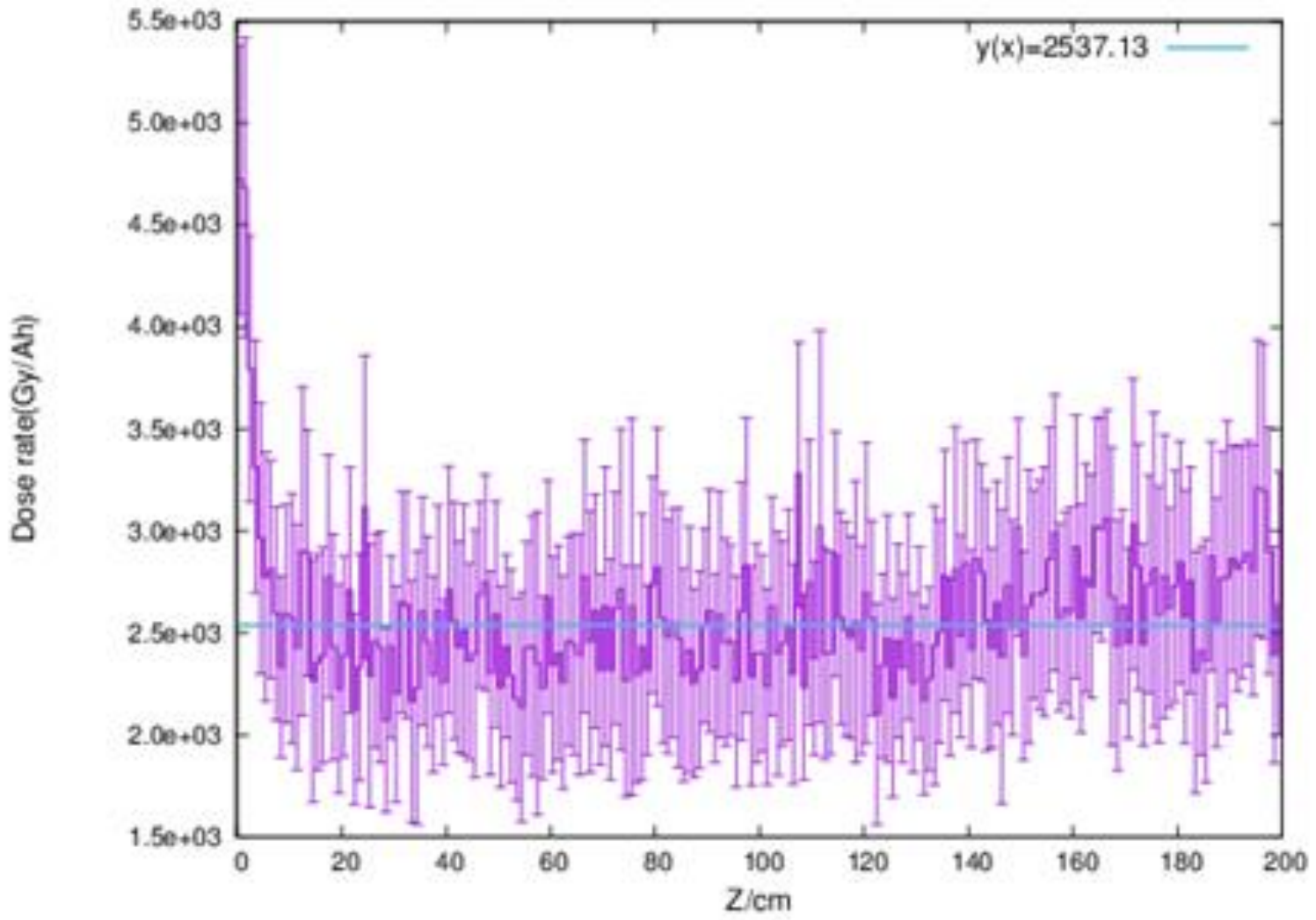

**Figure 4.2.4.15:** Absorbed doses in a quadrupole.

The coil insulation is particularly susceptible to radiation. Figure 4.2.4.15 illustrates the absorbed doses within the insulations of quadrupole coils. It shows higher doses on the left side, primarily attributed to the gap in the lead near the flange, as previously discussed. In these areas with elevated doses, the lead shielding will be designed in conjunction with the vacuum equipment and flange. In other magnet sections, the absorbed doses are relatively uniform, allowing for the use of the same lead thickness. Table 4.2.4.3 provides details on the absorbed doses within coil insulations in the dose-flat region for various operational modes and lead thicknesses.

**Table 4.2.4.3:** Absorbed dose rate in coils, unit: Gy/A-hour

| | | Higgs | Z | WW | $t\bar{t}$ |
|---|---|---|---|---|---|
| 2.5cm lead | Dipole | $(2.7 \pm 0.7) \times 10^3$ | 0 | $45 \pm 40$ | $(5.5 \pm 0.5) \times 10^4$ |
| | Quadrupole | $(2.7 \pm 0.6) \times 10^3$ | 0 | $22 \pm 12$ | $(3.2 \pm 0.5) \times 10^4$ |
| | Sextupole | $(4.2 \pm 0.7) \times 10^3$ | 0 | $27 \pm 13$ | $(5.2 \pm 0.5) \times 10^4$ |
| 2cm lead | Dipole | $(3.6 \pm 0.7) \times 10^3$ | 0 | $150 \pm 120$ | $(9.4 \pm 1.5) \times 10^4$ |
| | Quadrupole | $(3.0 \pm 0.2) \times 10^3$ | 0 | $17 \pm 7$ | $(5.7 \pm 0.4) \times 10^4$ |
| | Sextupole | $(5.4 \pm 0.8) \times 10^3$ | 0 | $87 \pm 60$ | $(9.3 \pm 1.3) \times 10^4$ |
| 1.5cm lead | Dipole | $(1.0 \pm 0.2) \times 10^4$ | 0 | $260 \pm 200$ | $(2.3 \pm 0.5) \times 10^5$ |
| | Quadrupole | $(6.1 \pm 0.5) \times 10^3$ | 0 | $220 \pm 50$ | $(1.0 \pm 0.2) \times 10^5$ |
| | Sextupole | $(1.8 \pm 0.3) \times 10^4$ | 0 | $650 \pm 200$ | $(3.0 \pm 0.5) \times 10^5$ |

Chapter 3 of this report outlines that the CEPC is planned to be in operation for 8 months annually, totaling 6,000 hours. This operational schedule is used to calculate the cumulative absorbed doses for magnet coil insulations, as illustrated in Figure 4.2.4.16, considering a 10-year Higgs operation, 2-year Z operation, and 1-year W operation. Figure 4.2.4.17 displays the absorbed doses when an additional 5-year $t\bar{t}$ operation is included. These plots also include the upper limit for absorbed dose in epoxy resin, which is measured at $2 \times 10^7$ Gy [11].

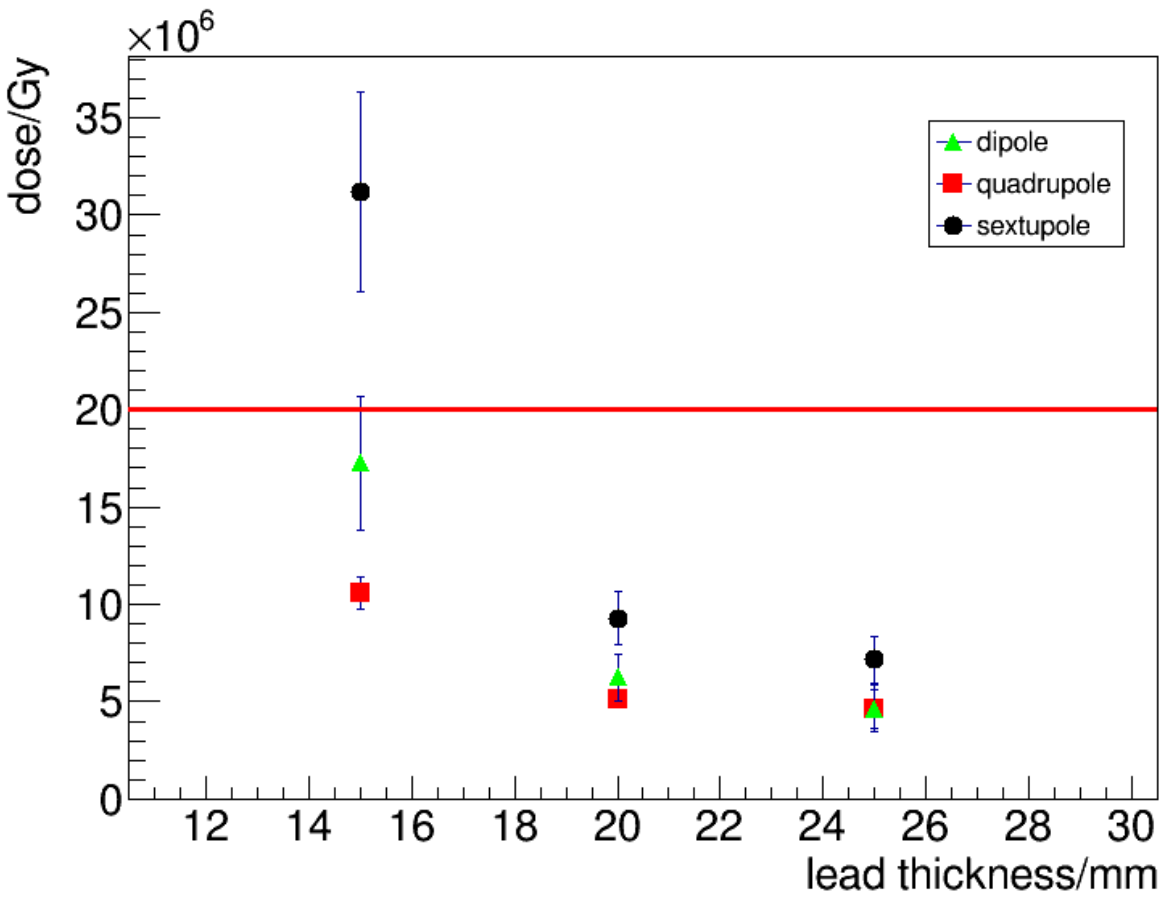

**Figure 4.2.4.16:** Absorbed doses after a 13-year operation.

Upon analysis, it is evident that for a 13-year operation schedule encompassing Higgs, Z, and W operations, the necessary lead thickness can be minimized to less than 2 cm. Specifically, for quadrupoles, the thickness can be further reduced to below 1.5 cm. However, if $t\bar{t}$ operation is incorporated into the schedule, the lead thickness must be increased. For an 18-year schedule, a lead thickness of 2.5 cm proves to be adequate.

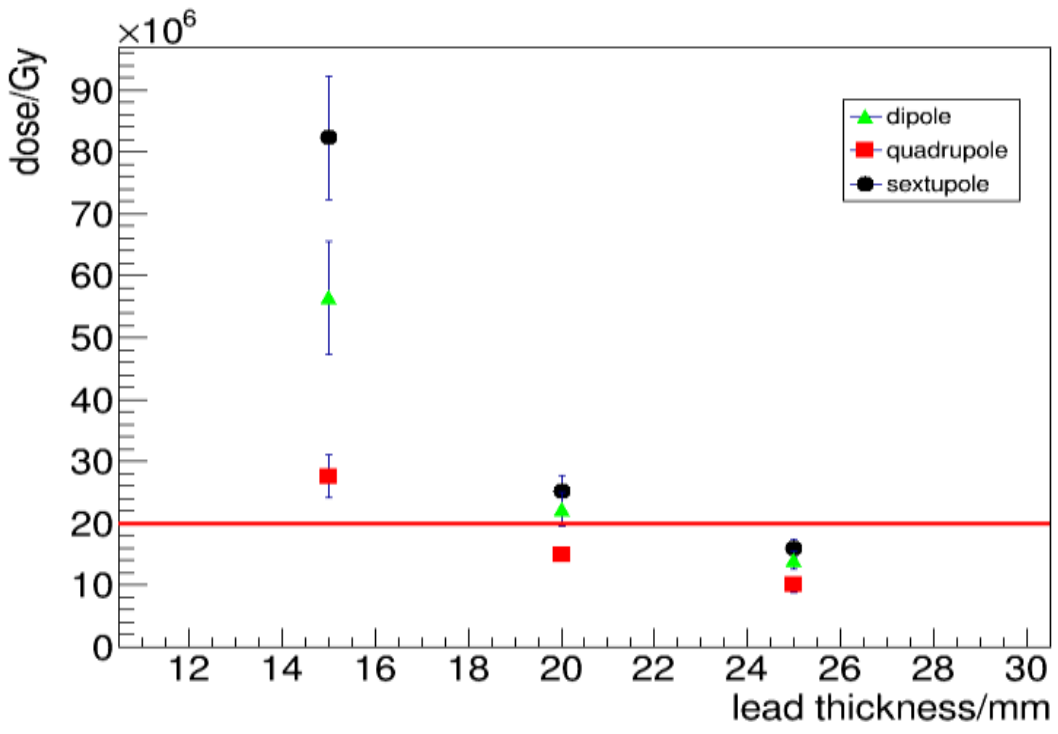

**Figure 4.2.4.17** Absorbed doses after a 18-year operation.

### 4.2.4.5 *Energy Deposition in Magnets and Tunnel*

Considering the synchrotron radiation power of 50 MW per beam, significant amounts of energy are deposited in both the magnets and the tunnel. In Table 4.2.4.4, the energy depositions specifically for the Higgs operation are provided, excluding the energy depositions in the Booster magnets, which can be found in Section 5.2.4 of the report. Furthermore, Figure 4.2.4.18 visually presents the distribution of absorbed doses in the tunnel during the Higgs operation, providing additional insights into the energy deposition patterns. It's important to note that this analysis disregards heat-conduction effects.

**Table 4.2.4.4** Energy deposition with total SR power in different parts of the Collider.

| | Collider beam-pipes | Air | Wall |
|---|---|---|---|
| Energy deposition [%] | 75.2 | 0.02 | 1.1 |
| | Collider dipole | Collider quadrupole | Collider sextupole |
| Energy deposition [%] | 8.6 | 1.0 | 0.6 |
| | Lead in dipole | Lead in quadrupole | Lead in sextupole |
| Energy deposition [%] | 6.2 | 1.4 | 1.4 |
| | Colling water around beam pipes | | |
| Energy deposition [%] | 0.2 | | |

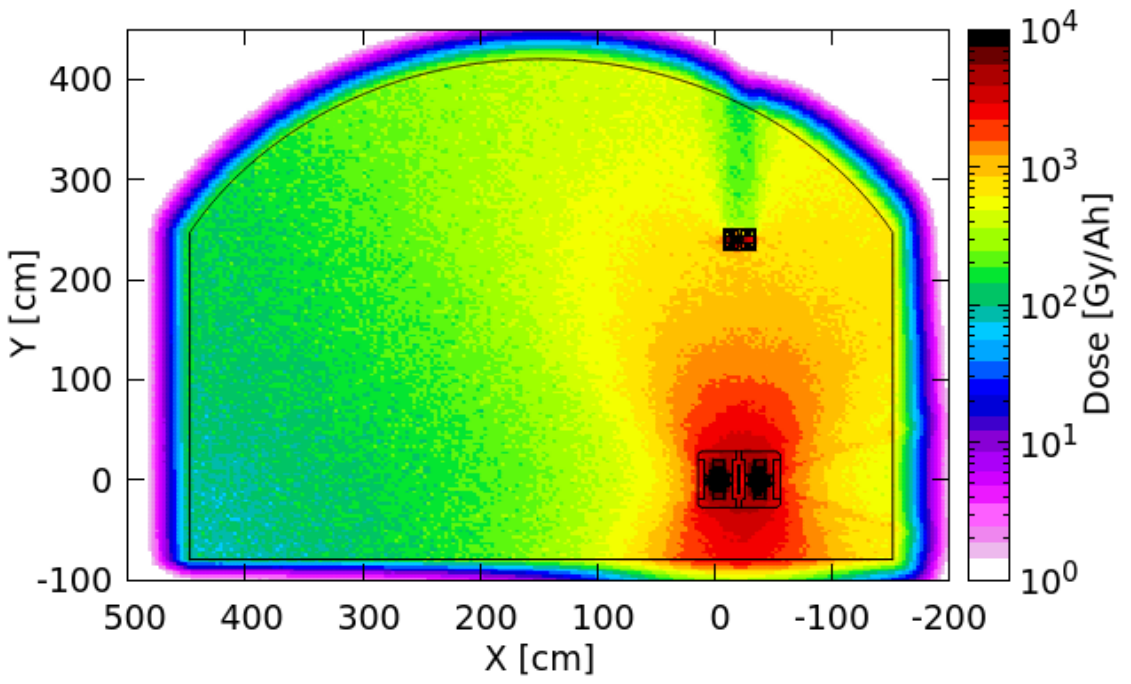

**Figure 4.2.4.18** Absorbed dose distribution in the tunnel.

### 4.2.5 Injection and Beam Dump

The CEPC accelerator complex comprises a double-ring Collider, a Booster, a Linac, and several transport lines. The injection system's function is to introduce electron or positron beams from the booster into the Collider ring to satisfy the CEPC Collider's operational requirements. Conversely, the extraction system draws the circulating beam out of the Collider ring and delivers it to the dumping system or re-injects it into the Booster. Owing to the continuous particle loss in the beams caused by the radiative Bhabha and beamstrahlung effect, the lifetime of the beam in the Collider ring may be only a few tens of minutes [1]. For the CEPC's operation, maximizing the integrated brightness requires top-up injection to become an unavoidable choice [2]. Moreover, to restart the operation in the case of total beam loss, the injection time from scratch should also be controlled within a reasonable range.

Figure 4.2.5.1 displays the geometry of the CEPC complex., with the Collider and Booster located in the same underground tunnel and the Linac built at the ground level. The electrons and positrons generated and accelerated to 30 GeV in the Linac are injected into the Booster ring and then accelerated to full energy before being injected into the Collider. Table 4.2.5.1 lists some essential parameters of the CEPC. The Collider ring is mainly injected using the traditional off-axis injection method, considering the injection process's robustness. Only at Higgs energy if the Collider's dynamic aperture does not satisfy the off-axis injection requirements, the on-axis injection design option will be added. Some components included in the Collider injection and extraction section are shown in Table 4.2.5.2.

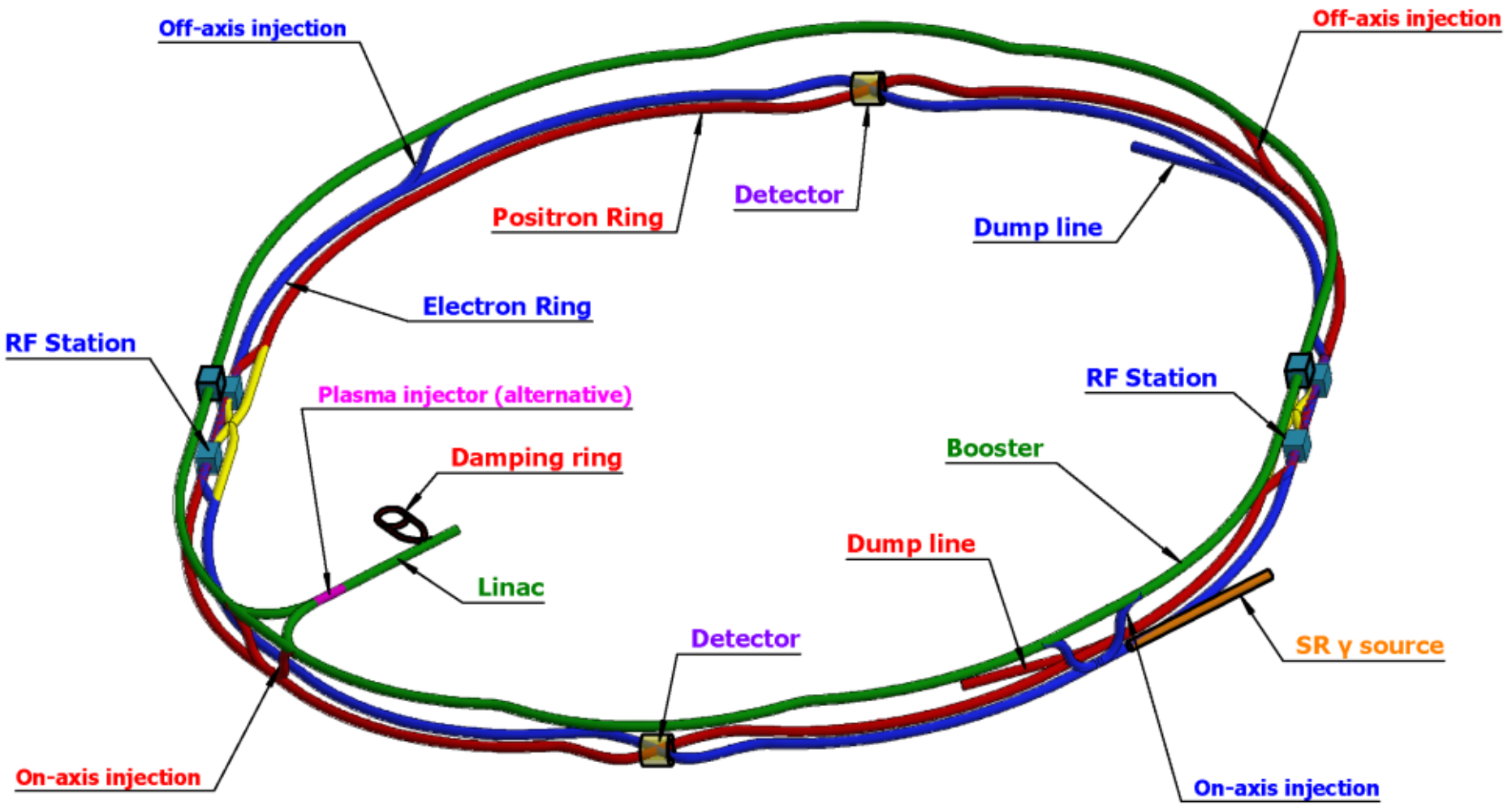

**Figure 4.2.5.1:** A schematic plot of the CEPC.

**Table 4.2.5.1:** Main parameters for injection system design.

| | **Higgs** | **W** | **Z** | $t\bar{t}$ |
|---|---|---|---|---|
| Bunch number in Collider | 268 | 1297 | 11934 | 35 |
| Bunch separation (ns) | 591 | 257 | 23 | 4524 |
| Bunch number in Booster | 268 | 1297 | 3978 | 37 |
| Lifetime (min) | 20 | 55 | 80 | 18 |
| Interval for top-up (s) | 37 | 100 | 146 | 33 |

**Table 4.2.5.2:** Some main components in the collider injection and extraction section.

| | **Numbers** | **Length (m)** | **Max Field (T)** |
|---|---|---|---|
| Kickers for off-axis injection | 8 | 1 | 0.06 |
| Septum for off-axis injection | 2 | 36 | 0.43 |
| Kickers for on-axis injection | 2 | 1 | 0.08 |
| Septum for on-axis injection | 2 | 30 | 0.47 |
| Kickers to the dump | 2 | 2 | 0.11 |
| Septum to the dump | 2 | 36 | 0.43 |
| Kickers to the booster | 2 | 1 | 0.08 |
| Septum to the booster | 2 | 30 | 0.47 |
| Dipoles | 24 | 5 | 1.5 |
| Quadrupoles | 40 | 2 | |
| Correctors | 30 | 0.3 | |
| BPMs | 32 | | |

### 4.2.5.1 *Off-axis Injection*

The off-axis injection for the CEPC Collider ring follows a traditional design that is similar to many existing accelerator machines. The Booster is located 2.4 meters above the Collider in the tunnel, and the injection transport line brings the beam down to the Collider ring with the smallest possible angle. A pulsed kicker bump is used to make the circulating beam as close as possible to the septum cutting plate and minimize residual oscillation. Figure 4.2.5.2 shows the layout of the injection area. Kickers generate a closed orbit, bringing the circulating beam close to the septum blade, and the septum is used to kick the injected beam into the collider plane. The injected beam and circulating beam are separated by the septum blade thickness (2 mm) and the stay clear region. Figure 4.2.5.3 shows the phase space and the relationship between the injection and circulating bunches. Residual oscillation, which results from the spacing between the injection beam and the circulating beam, propagates downstream until radiation damping suppresses it. To improve the injection efficiency, the off-axis injection design needs to optimize the parameters to minimize the injected beam trajectory oscillation.

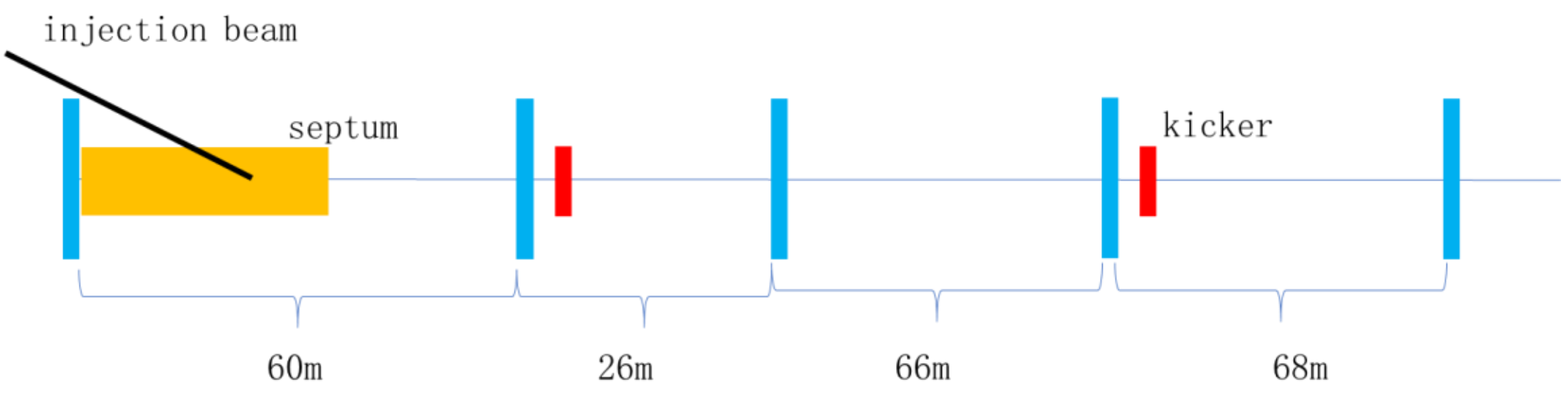

**Figure 4.2.5.2:** Schematic view of the injection region

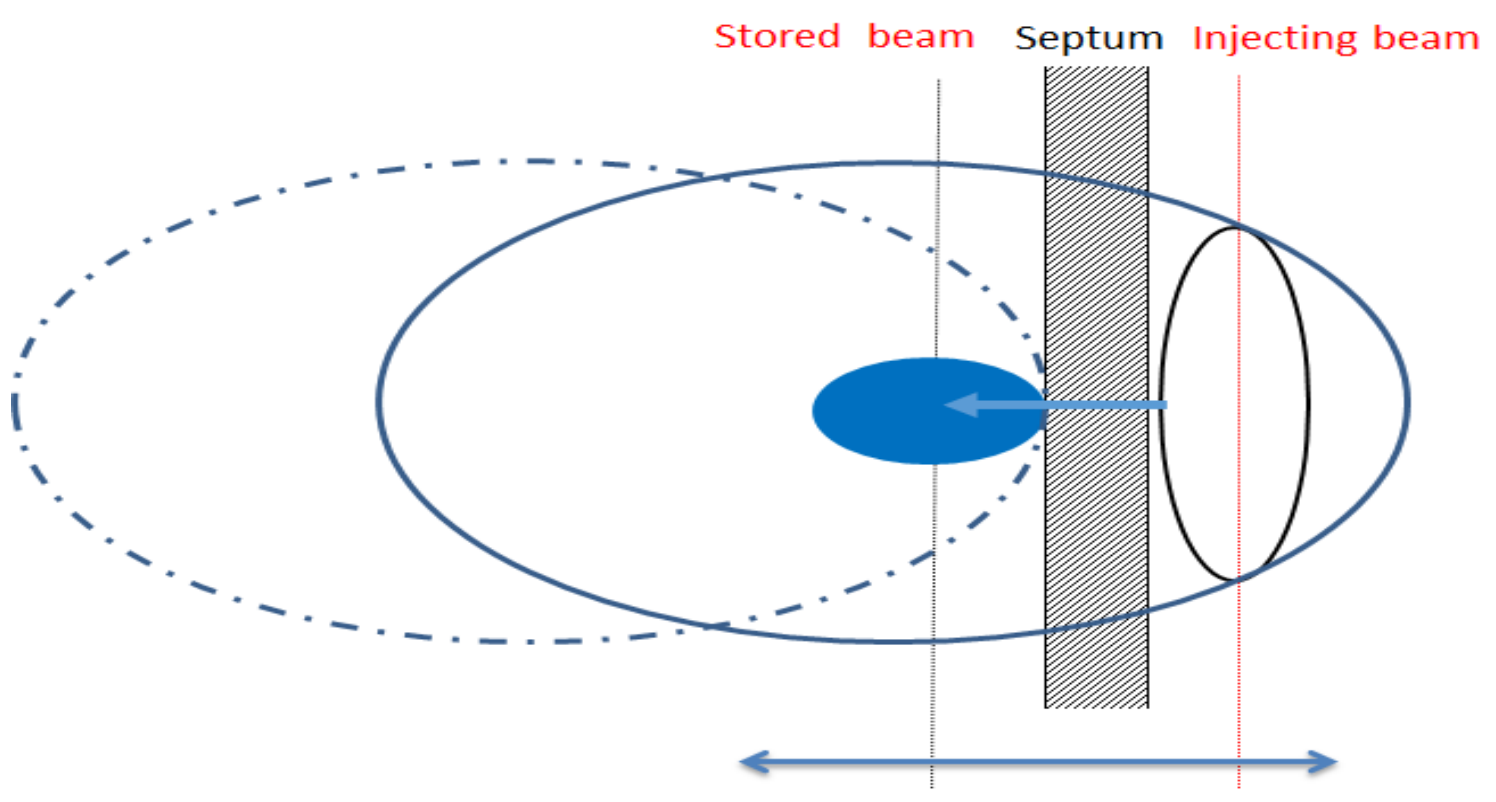

**Figure 4.2.5.3:** A plot of the phase space for CEPC off-axis injection

The CEPC Collider ring has eight arc areas, two IP sections, two RF sections, and four linear section areas, with two of the linear sections designated for off-axis injection of electrons and positrons, respectively. The Twiss parameters at the injection exit are optimized to reduce the need for dynamic aperture, and the horizontal beta function of the injection area is increased to 1800 m to improve the injection efficiency. Figure 4.2.5.4 shows the beta functions of the optics at the injection cell.

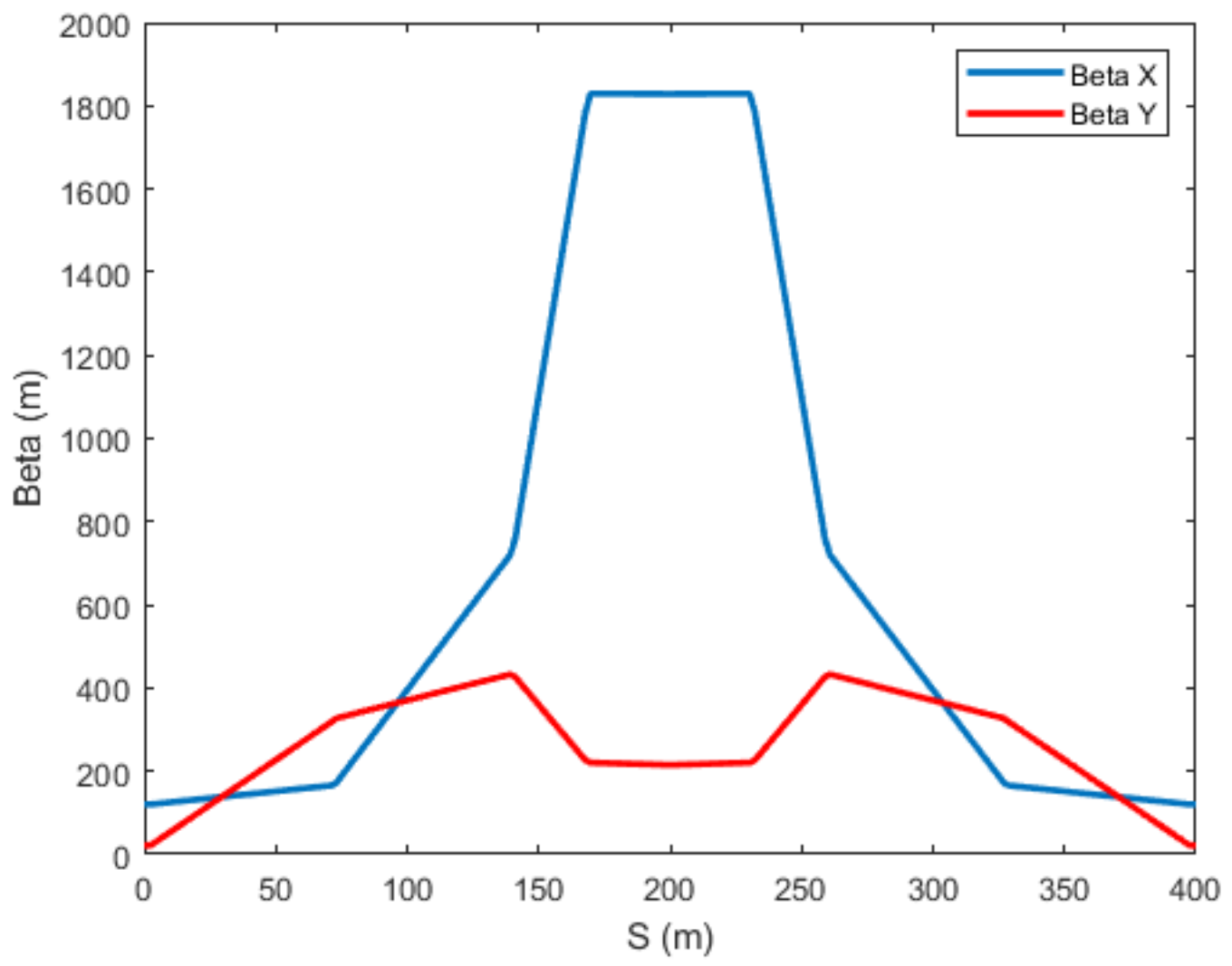

**Figure 4.2.5.4:** Beta functions at the off-axis injection cell

The standard top-up injection operation maintains the Collider beam current within ±3% of the nominal value. The bunch pattern in the Booster is the same as that in the collision for tt, Higgs, and W energies. However, for the Z energy, the Booster current threshold limits the number of bunches to 1/3 of that in the collision ring. Injection is performed bunch by bunch for tt, Higgs, and W energy, and train by train for Z energy. Figure 4.2.5.5 shows the time structure of the kicker magnetic field in the Collider ring under both injection modes.

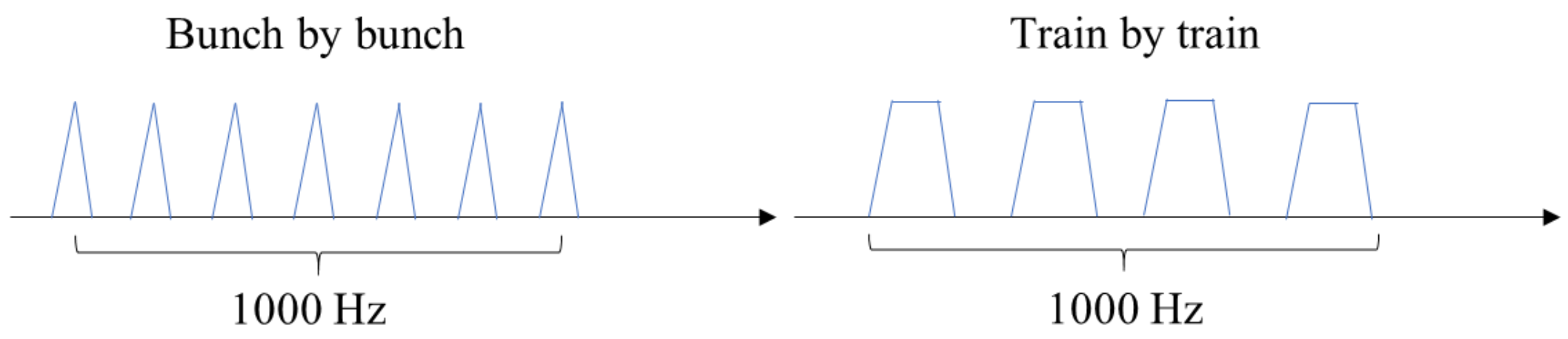

**Figure 4.2.5.5:** Magnetic field time structure of the kicker in the collider in the bunch-by-bunch injection mode and train-by-train injection mode, respectively.

The off-axis injection time required for CEPC is mainly determined by the injection time from the Linac to the Booster, the ramp speed of the Booster, and the extraction time from the Booster. Injection time varies based on the energy mode, with the Z-mode requiring three ramps to complete one injection owing to the reduced number of booster bunches. Table 4.2.5.3 shows the injection time required for each mode, and Figure 4.2.5.6 provides a specific injection time structure.

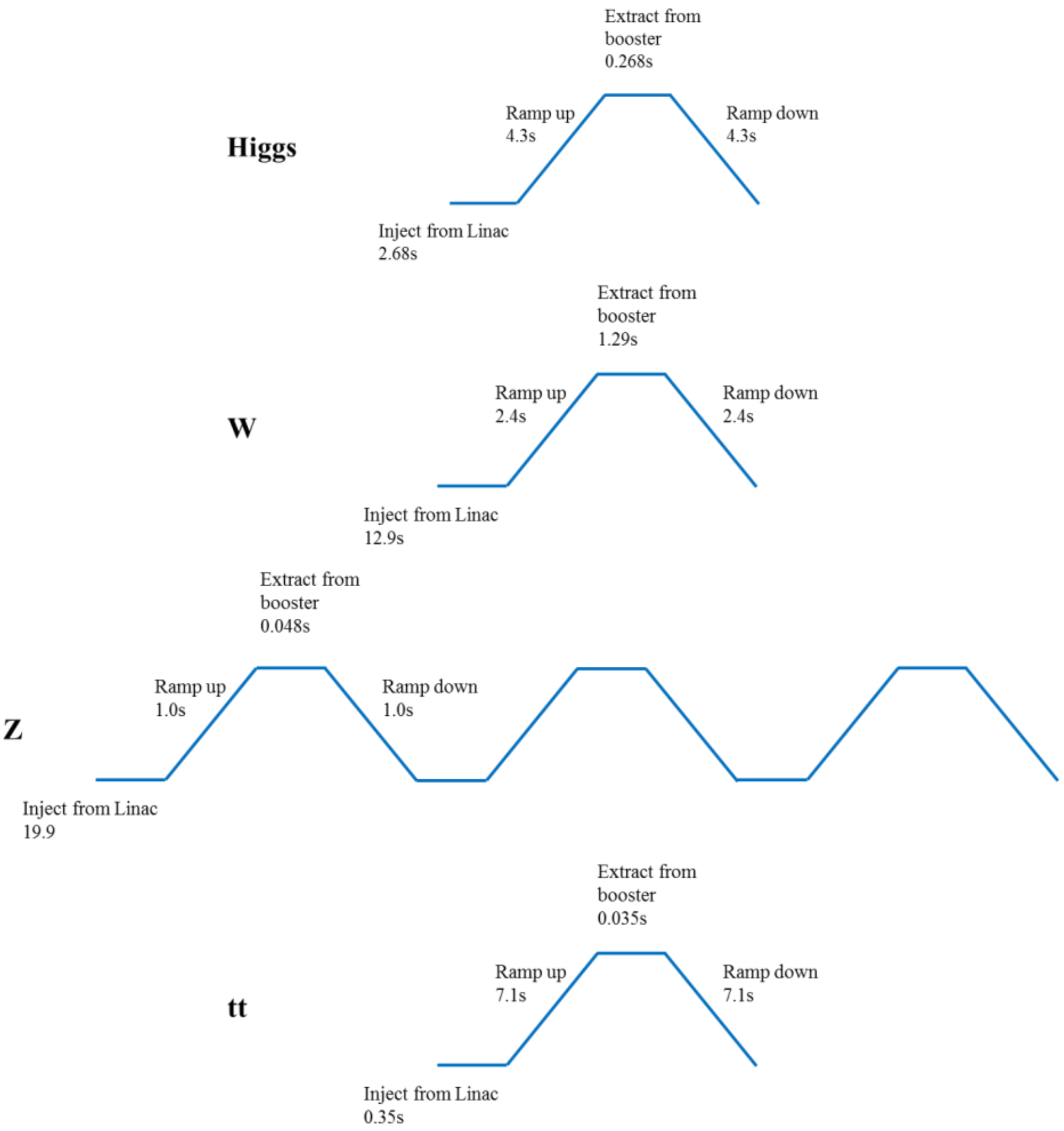

**Figure 4.2.5.6:** Injection time structure for different energy modes.

**Table 4.2.5.3:** Injection time for different energy modes

| | **Higgs** | **W** | **Z** | $t\bar{t}$ |
|---|---|---|---|---|
| Interval for top up (s) | 37 | 100 | 146 | 33 |
| Injection time for both e+/e- (s) | 24 | 40 | 139 | 30 |

### 4.2.5.2 *On-axis Injection*

The on-axis injection strategy for the CEPC involves using the Booster as an accumulator ring and injecting a large bunch from the Collider into the Booster. This approach enables off-axis injection and bunch mergence to be performed in the Booster, which has a sufficiently large dynamic aperture. The injection process is illustrated in Figure 4.2.5.7 and involves filling the booster with small bunches initially (with a bunch charge of approximately 3% of that in the collider), ramping it up to 120 GeV, and then injecting several circulating bunches of the Collider back into the Booster ring. The injected bunches merge with the small bunches in the Booster after four damping times, and the merged bunches are injected back into the same empty buckets until the Booster becomes empty.

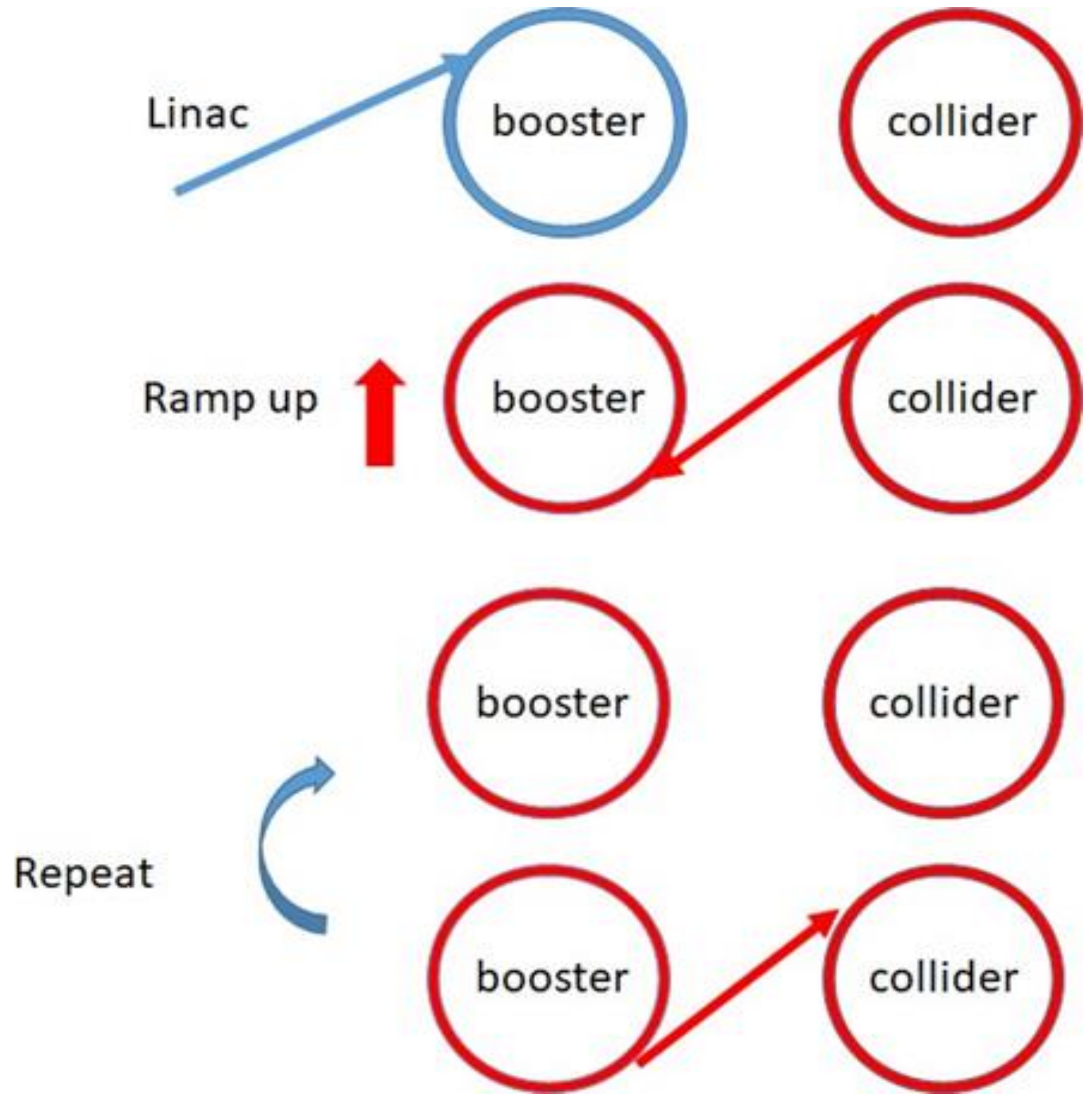

**Figure 4.2.5.7:** A sketch of the on-axis injection process.

The on-axis injection method reduces the required horizontal dynamic aperture in the Collider from 13 $\sigma_x$ to 8 $\sigma_x$, and the number of exchanged bunches is limited by the total current in the Booster. The Booster current threshold is 1 mA, and the time structure of the Booster is shown in Fig 4.2.5.8. Each on-axis injection takes approximately 35 s, which is less than the 47 s required by the beam lifetime in the Collider.

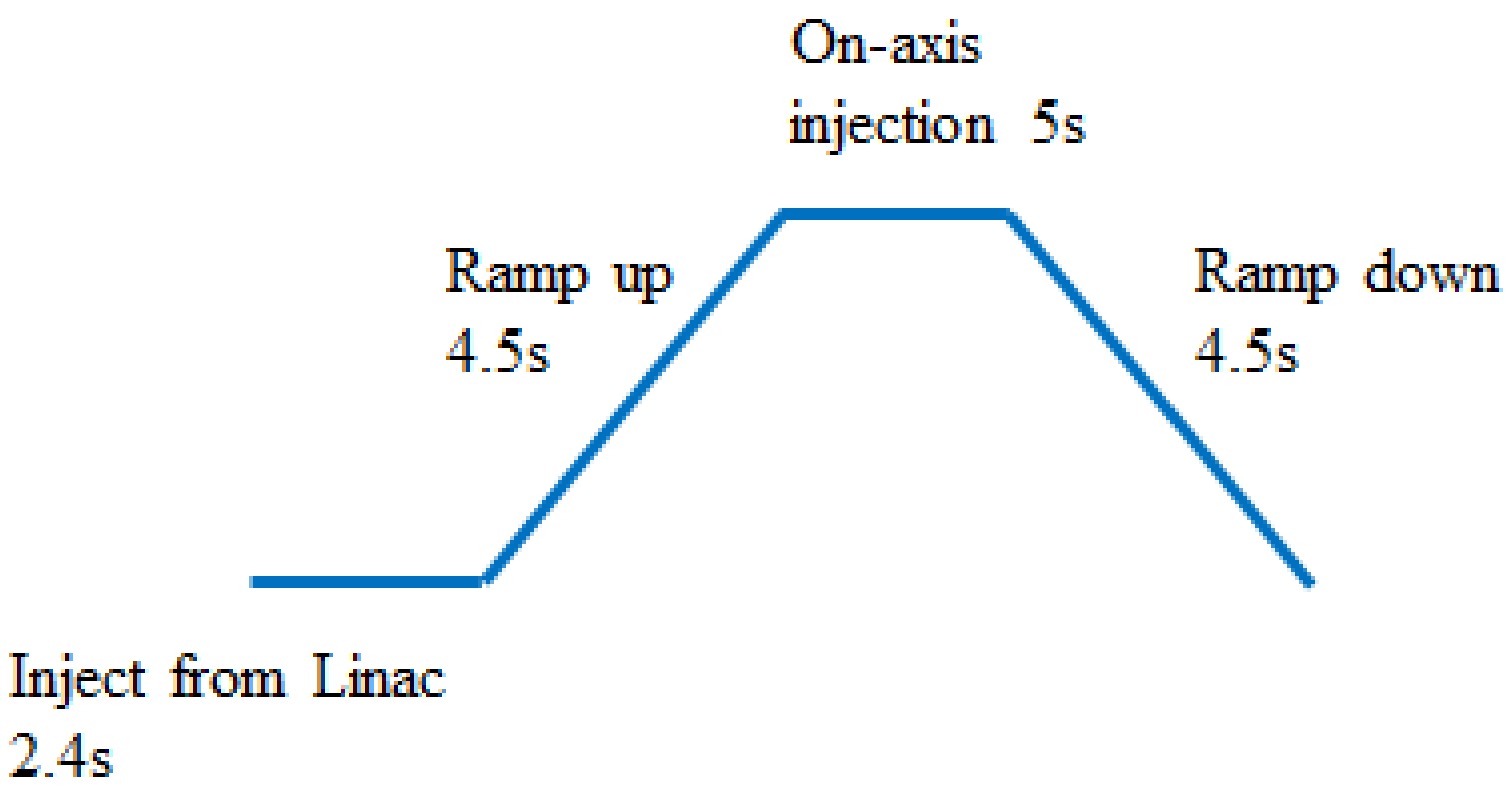

**Figure 4.2.5.8:** The injection time structure of the Higgs on-axis injection

### 4.2.5.3 *Beam Dump*

The Collider beam dump system consists of two components: an extraction line and an absorber core, both of which require appropriate shielding.

**Table 4.2.5.4:** Parameters for beam dump.

| | Higgs | Z | WW | $t\bar{t}$ |
|---|---|---|---|---|
| Beam energy [GeV] | 120 | 80 | 45.5 | 180 |
| Bunch population/$10^{10}$ | 13 | 21.4 | 13.5 | 20 |
| Number of bunches | 446 | 13104 | 2162 | 58 |
| Total energy [MJ] | 1.1 | 20 | 3.7 | 0.33 |

**Table 4.2.5.5:** Parameters of different magnets for extraction line

| | | Extraction kicker | Septum | Dilution kickers |
|---|---|---|---|---|
| Length [m] | | 2 | 36 | 10 |
| Magnetic induction [Gs] | Higgs | 740 | 2867 | 40 (Max.) |
| | Z | 280 | 1075 | |
| | WW | 493 | 1911 | |
| | $t\bar{t}$ | 1110 | 4300 | |

Figure 4.2.5.9 shows the design of the extraction line, which includes an extraction kicker, a septum, and two dilution kickers that sweep bunches horizontally and vertically. The dump core is located at the end of the extraction line, about 100 m away from the dilution kicker. Table 4.2.5.5 lists the parameters of magnets in the extraction line, including the different magnetic inductions required for the extraction kicker and septum for different operation modes. The length of the Collider is 100 km, meaning it takes 333 µs to dump beam in one turn. The waveforms of the voltage of dilution kickers during this time interval are shown in Figures 4.2.5.10 and 4.2.5.11. The shorter the distance from the dilution kickers to the dump core, the higher the magnetic field can be, resulting in lower costs. In the next engineering design phase, we will optimize the magnetic induction and drift distance, considering both magnet design and tunnel costs. The calculation of temperature rises will be discussed in more detail below.

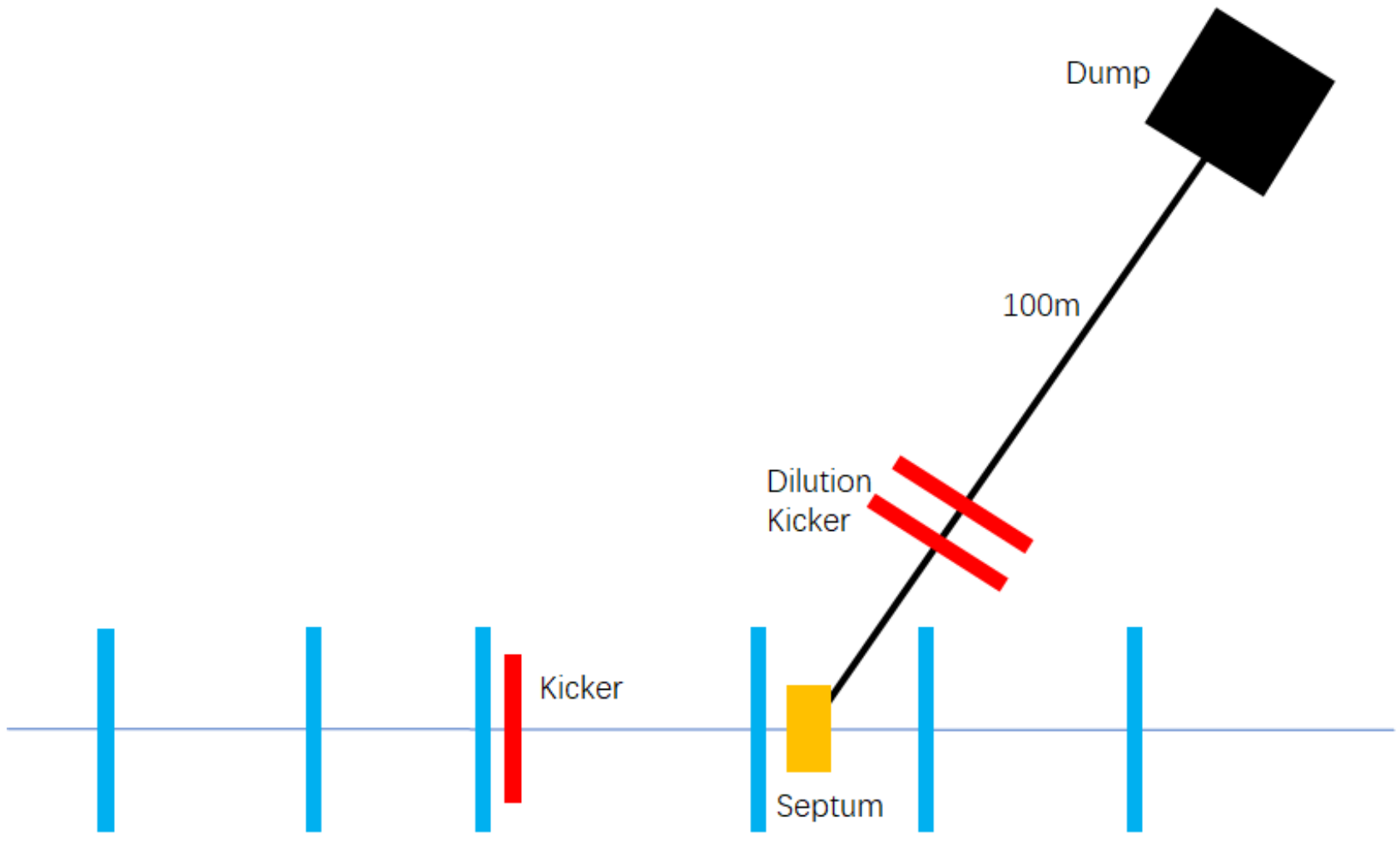

**Figure 4.2.5.9:** Extraction line for CEPC dumping system.

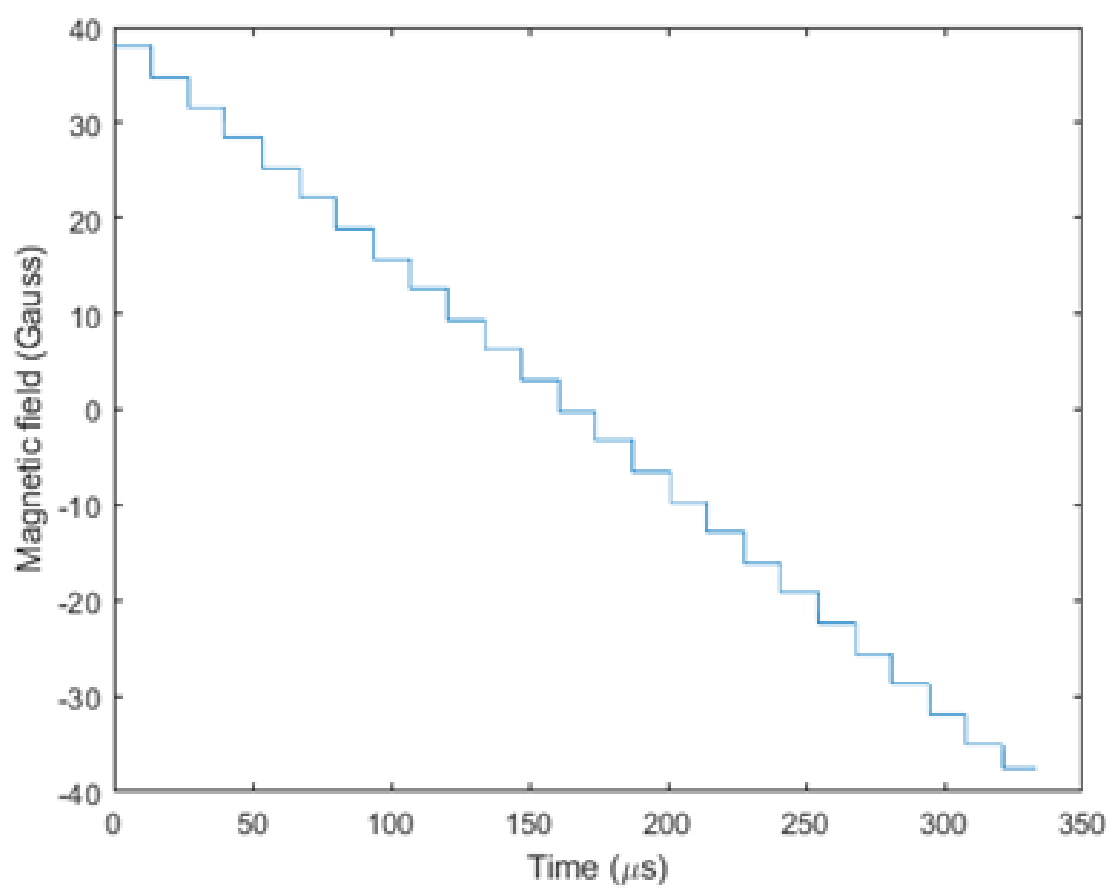

**Figure 4.2.5.10:** Field waveform of the horizontal dilution kicker.

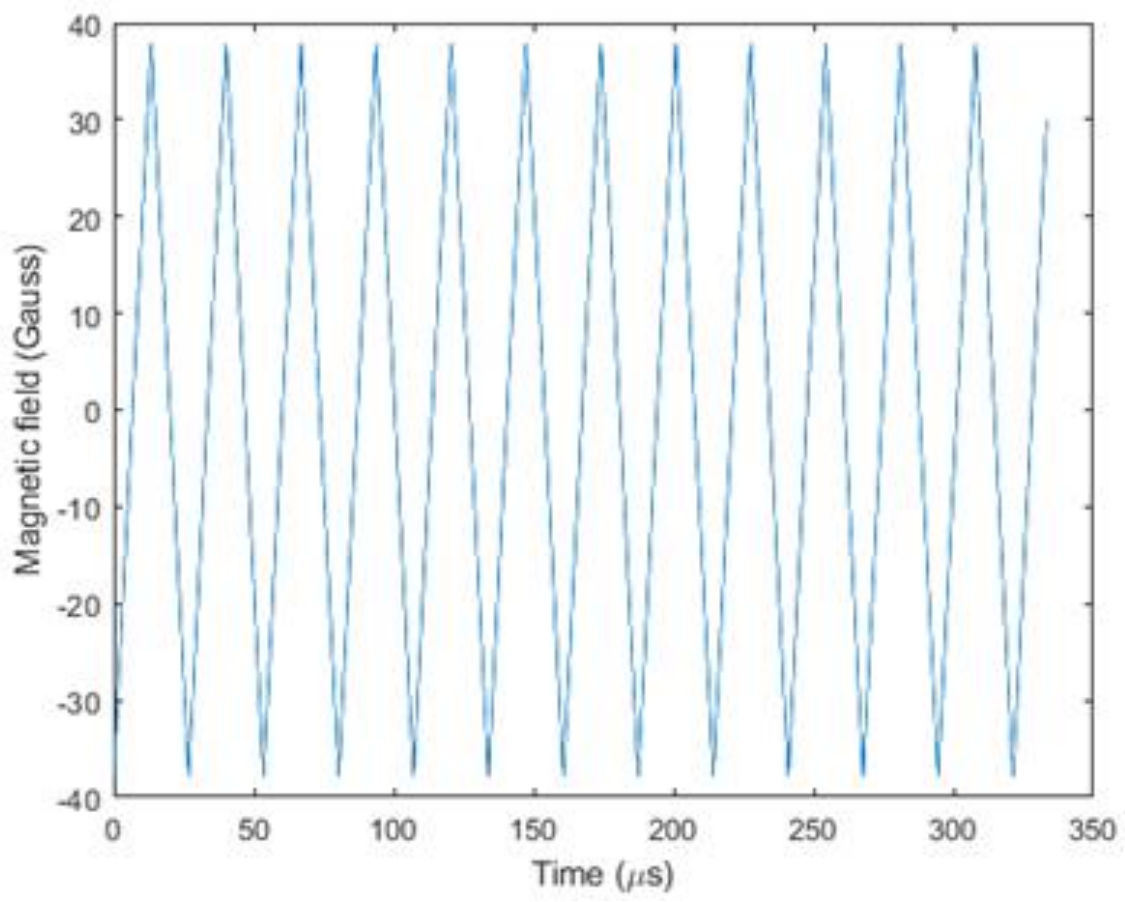

**Figure 4.2.5.11:** Field waveform of the vertical dilution kicker

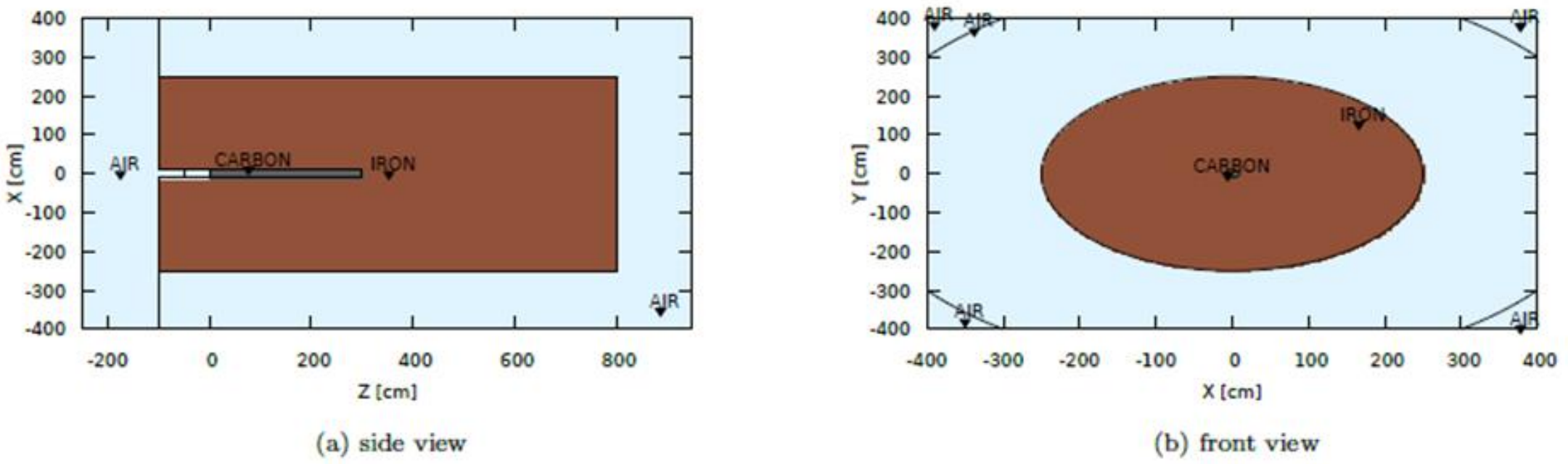

**Figure 4.2.5.12:** The setup of collider dump

Regarding the materials of the absorber, a combination of low-mass-density and higher-mass-density materials is ideal for reducing power dissipation per unit length while ensuring energy capture and compactness. To achieve this, we propose a structure with a low mass-density core embedded in a high mass-density shell. Graphite is chosen as the core material due to its higher melting point compared to aluminium, while iron is selected as the material for the dump shell around the core because of its similarity in stopping power to copper and its lower cost compared to other high-Z elements.

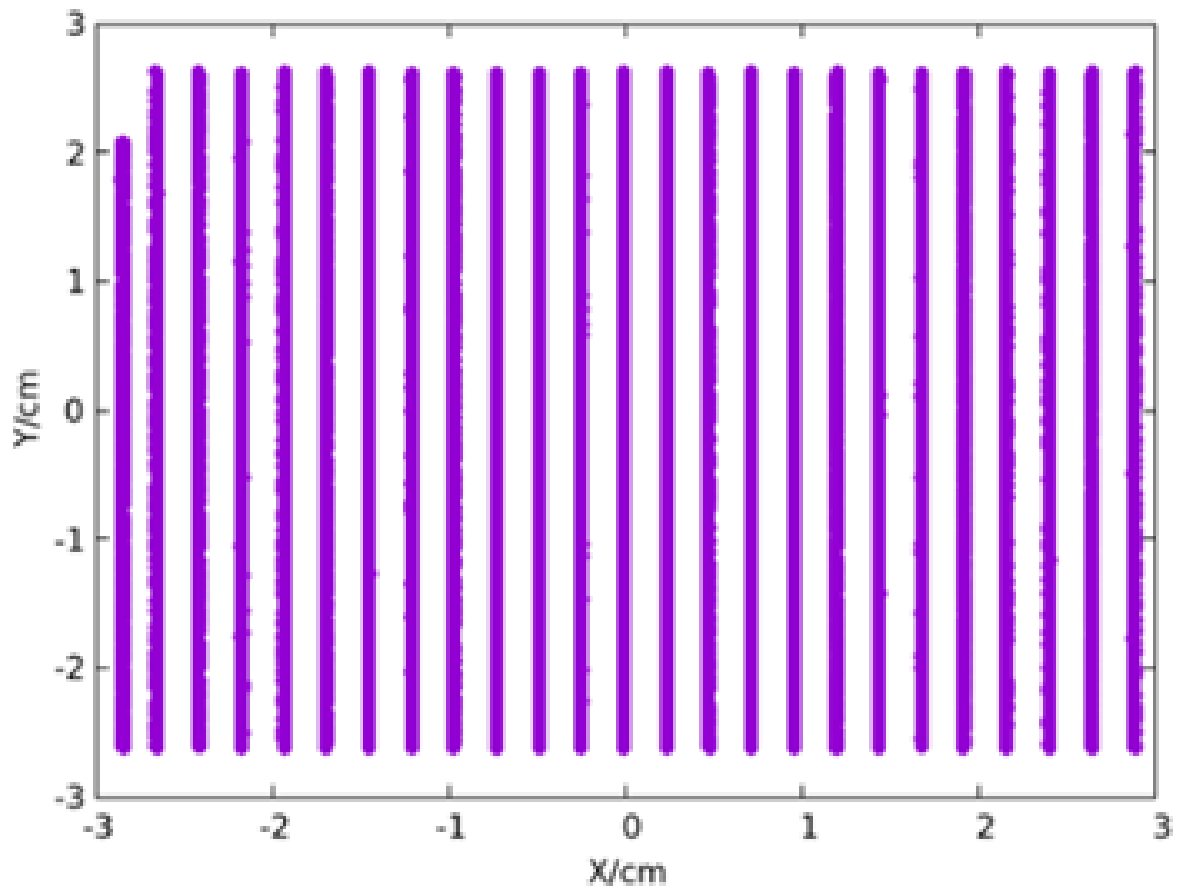

**Figure 4.2.5.13:** bunch distribution at Z mode

To evaluate the dump design, we used the Monte-Carlo particle transport code FLUKA [1-3] and the user-friendly interface Flair [4]. The dump geometry, as shown in Figure 4.2.5.12, consists of a graphite cylinder for the core and an iron shell. The radius of the graphite core is approximately 10 cm, which is the Moliere radius of graphite with the radius of the maximum sweeping areas roughly added [5-6]. The length of the graphite core and the size of the iron shell are optimized based on temperature rise and dose equivalent constraints, where the dose equivalent should not exceed 5.5 mSv/h at Z mode.

**Table 4.2.5.6:** Maximum temperature rise in the dump after dumping once.

| | Higgs | Z | WW | $t\bar{t}$ |
|---|---|---|---|---|
| Sweeping area [$cm^2$] | $1 \times 2$ | $6 \times 6$ | $3 \times 3$ | $0.8 \times 1.5$ |
| Min. bunch size [$mm^2$] | $9.2 \times 0.05$ | $8.5 \times 0.05$ | $9.2 \times 0.05$ | $9.7 \times 0.05$ |
| Max. temperature rise [°C] | 510 ± 15 | 2620 ± 15 | 1020 ± 30 | 194 ± 2 |

The MATLAB Accelerator Toolbox [7] can be used to simulate the bunch distributions at the entrance of the dump core during different operation modes, as shown in Figure 4.2.5.13. With these distributions as the source in FLUKA, we can calculate the deposited energy in the dump and the resulting temperature rises using the equation:

$$\Delta T = \frac{E_{dep}}{\rho \cdot C_{heat}}.$$

in which $\Delta T$ is the temperature rise, $E_{dep}$ is the deposited energy density, $\rho$ is the mass density and $C_{heat}$ is the specific heat capacity. The maximum temperature rises in the graphite core are listed in Table 4.2.5.6, which are lower than the melting point of graphite but high enough to cause a reaction with oxygen. To prevent fire and chemical reactions, an inert gas will be filled around the graphite core [8]. It should be noted that these temperature rises do not take heat conduction into account and represent the most critical scenario compared to the case where not all bunches are dumped. As the dumping frequency is assumed to be less than once per hour, the deposited heat must be conducted away during the dumping intervals. The need for a cooling system and its design will be addressed in future studies.

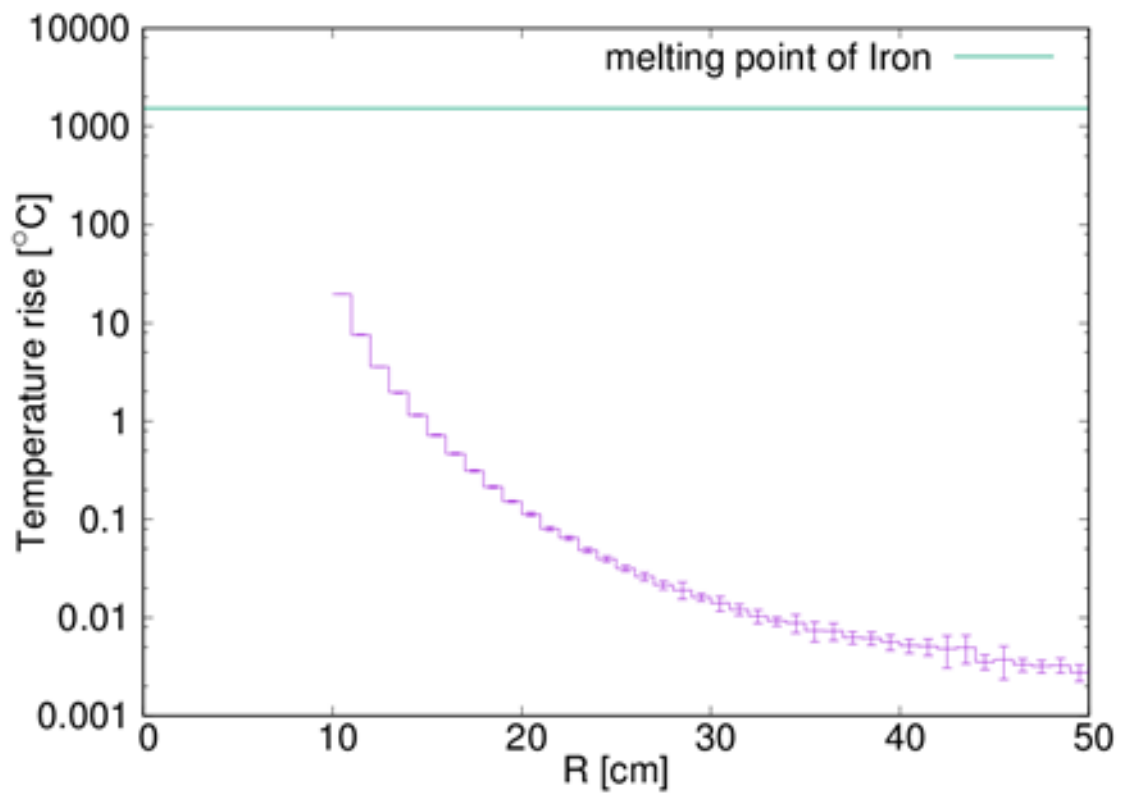

**Figure 4.2.5.14:** Temperature rise in iron: radial direction.

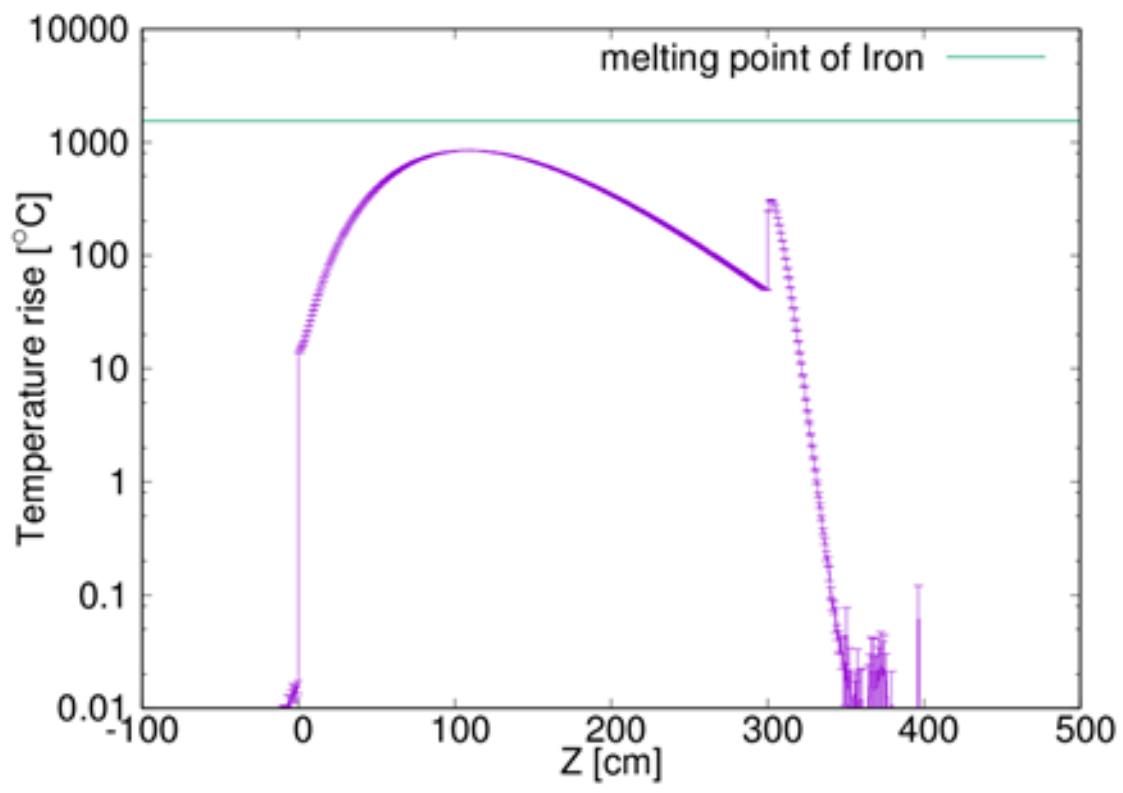

**Figure 4.2.5.15:** Temperature rise in iron: longitudinal direction.

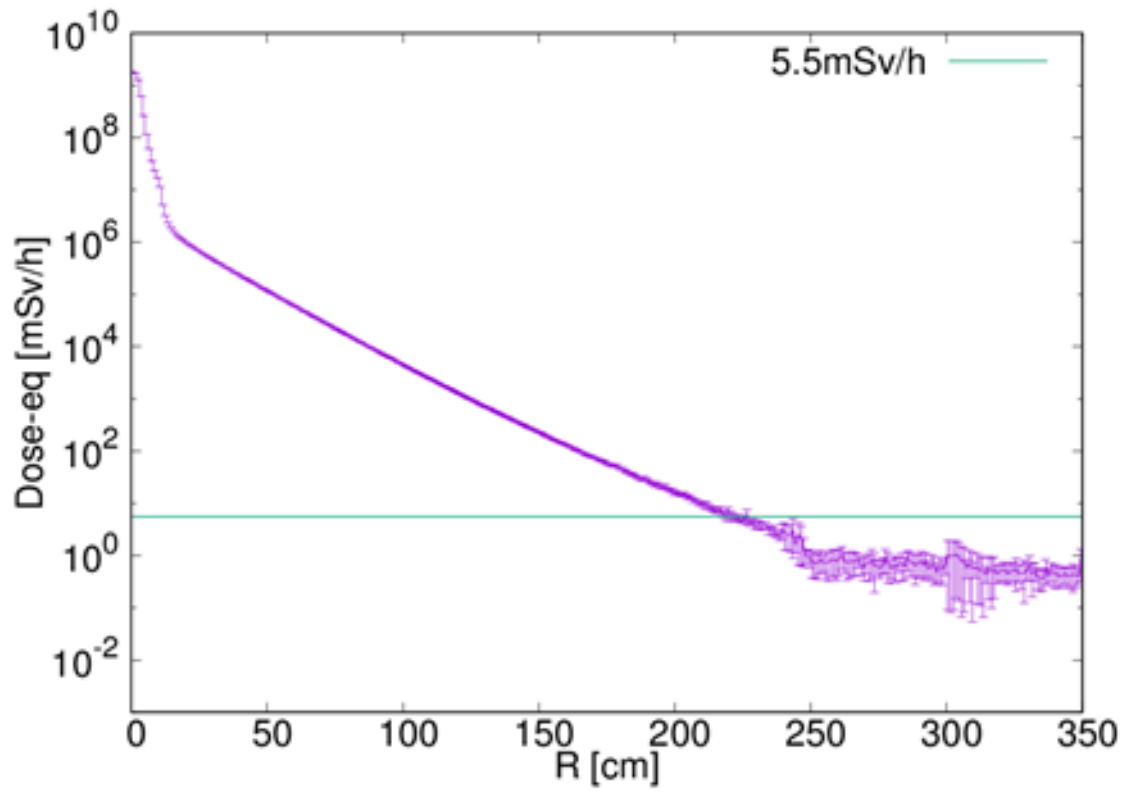

**Figure 4.2.5.16:** Dose equivalent in iron: radial direction.

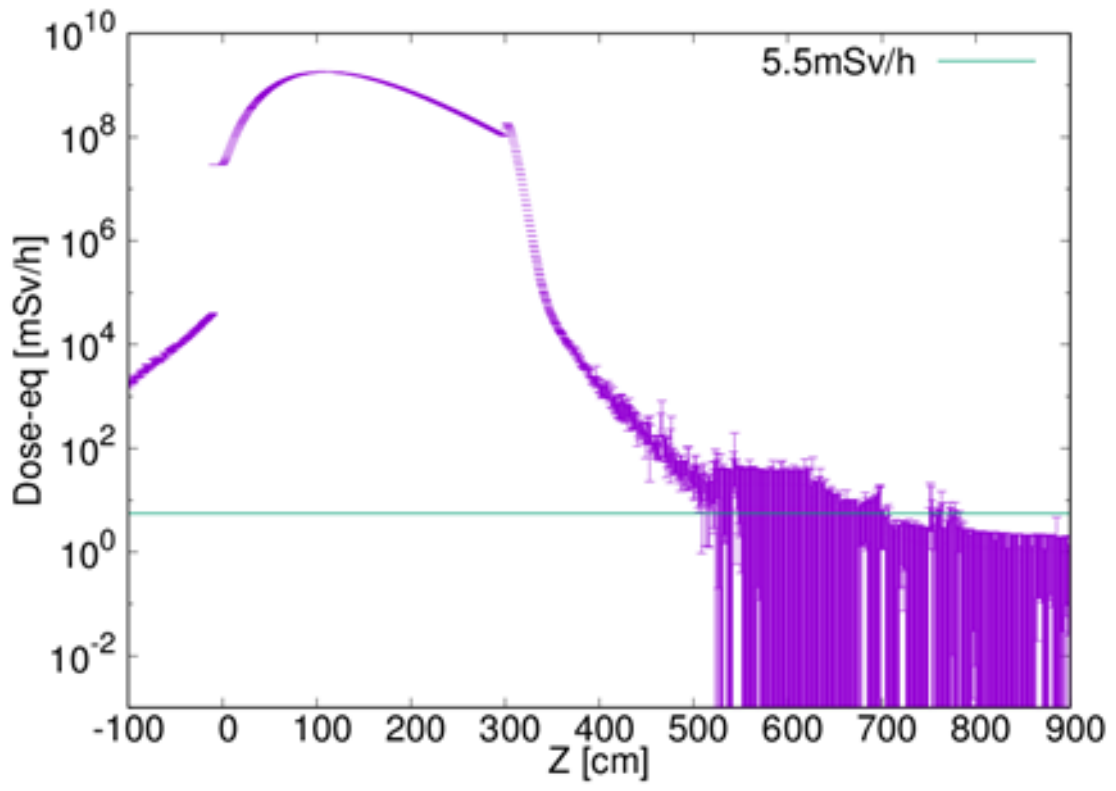

**Figure 4.2.5.17:** Dose equivalent in iron: longitudinal direction.

The temperature rises of the iron shell are presented in Figures 4.2.5.14 and 4.2.5.15. Figure 4.2.5.14 shows that the iron in the radial direction will not melt when the radius of the graphite core is 10 cm. Through simulations with varying lengths of graphite core, we determined that the iron shell will not melt if the length of the graphite core is longer than 3 m, as shown in Figure 4.2.5.15. Regarding the dose equivalent constraint, Figures 4.2.5.16 and 4.2.5.17 show the dose equivalent distributions. To limit the dose equivalent to less than 5.5 mSv/h, we determined that the radius of the iron shielding should be larger than 2.4 m, and the length of the iron shielding should be longer than 8 m.

For the Z operation, we study the thermal conductivity after all bunches have been dumped. The results show that one hour after the dumping process, the maximum temperature rise decreases to 40 degrees Celsius without cooling. This temperature is close to room temperature. Additionally, the dumping frequency is assumed to be lower than once per hour. Therefore, a water-cooling system is not required for the beam dump.

### 4.2.6 **Machine-Detector Interface**

#### 4.2.6.1 *Introduction and MDI Layout*

The machine-detector interface, shown in Figure 4.2.6.1, is approximately ±7 m in length in the Interaction Region (IR) and contains various elements, including the detector solenoid, luminosity calorimeter, interaction region beam pipe, beryllium pipe, cryostat, and bellows. To shield the detector, the cryostat contains final doublet superconducting magnets and an anti-solenoid covered by a thin layer of tungsten (~1 cm). The CEPC detector consists of a cylindrical drift chamber surrounded by an electromagnetic calorimeter embedded in a 3 Tesla (2 Tesla in Z) superconducting solenoid of length 7.3 m. The accelerator components inside the detector must not interfere with the detector devices, and after optimization, they occupy a conical space with an opening angle of 6.78 degrees. The crossing angle between electron and positron beams is 33 mrad in the horizontal plane, and the final focusing quadrupole is 1.9 m from the IP [1]. The MDI parameters are shown in Table 4.2.6.1.

**Table 4.2.6.1:** MDI Parameters

| | range | Peak field in coil | Central field gradient | Bending angle | length | Beam stay clear region | Minimal distance between two aperture | Inner diameter of beam pipe | Outer diameter of beam pipe | Critical energy (Horizontal) | Critical energy (Vertical) | SR power (Horizontal) | SR power (Vertical) |
|---|---|---|---|---|---|---|---|---|---|---|---|---|---|
| L* | 0~1.9m | | | | 1.9m | | | | | | | | |
| Crossing angle | 33mrad | | | | | | | | | | | | |
| MDI length | ±7m | | | | | | | | | | | | |
| Detector requirement of accelerator components in opening angle | 8.11° | | | | | | | | | | | | |
| QDa/QDb | | 3.5/2.7T | 142/85T/m | | 1.21m | 14.9/18.2mm | 62.71/105.28mm | 20/23mm | 26/29mm | 724.7/663.1keV | 396.3/263keV | 212.2/239.23W | 99.9/42.8W |
| QF1 | | 3.7T | 96.7T/m | | 1.5m | 24.48mm | 155.11mm | 32mm | 38mm | 675.2keV | 499.4keV | 472.9W | 135.1W |
| Lumical | 0.94~1.11m | | | | 0.16m | | | | | | | | |
| Anti-solenoid before QD0 | | 8.8T | | | 1.1m | | | | | | | | |
| Anti-solenoid QD0 | | 3T | | | 2.5m | | | | | | | | |
| Anti-solenoid QF1 | | 3T | | | 1.5m | | | | | | | | |
| Beryllium pipe | | | | | ±85mm | | | 20mm | | | | | |
| Last B upstream | 64.97~153.5m | | | 0.77mrad | 88.5m | | | | | 33.3keV | | | |
| First B downstream | 44.4~102m | | | 1.17mrad | 57.6m | | | | | 77.9keV | | | |
| Beampipe within QDa/QDb | | | | | 1.21m | | | | | | | 1.19/1.31W | |
| Beampipe within QF1 | | | | | 1.5m | | | | | | | 2.39W | |
| Beampipe between QD0/QF1 | | | | | 0.3m | | | | | | | 26.5W | |

The vacuum requirement for the interaction region is $3\times10^{-9}$ torr, the same as in the arc area. Two methods can be used to achieve these levels of vacuum: utilizing multiple vacuum pumps or coating the inner surface of the vacuum chamber with NEG. Due to limited space in the interaction region, coating with NEG is the preferred method. The inner diameter of the MDI vacuum chamber gradually reduces from 32 mm to 20 mm over a length of 6 meters, and materials such as CrZrCu (18150) or tungsten alloy will be used. The small aperture of the vacuum chamber and the requirement to inhibit secondary electron emission from the positron beam make NEG coating the ideal choice.

The luminosity calorimeter comprises three parts, located at 0.65~1.11 m. The physical and mechanical design of the IR beam pipe is presented in Sec. 4.2.6.2, and other design details can be found in Sec. 4.3.4 for the final doublet quadrupoles, Sec. 7.1.8 for the cryostat chamber in the MDI, and Sec. 4.3.10.5 for other mechanical designs of the MDI.

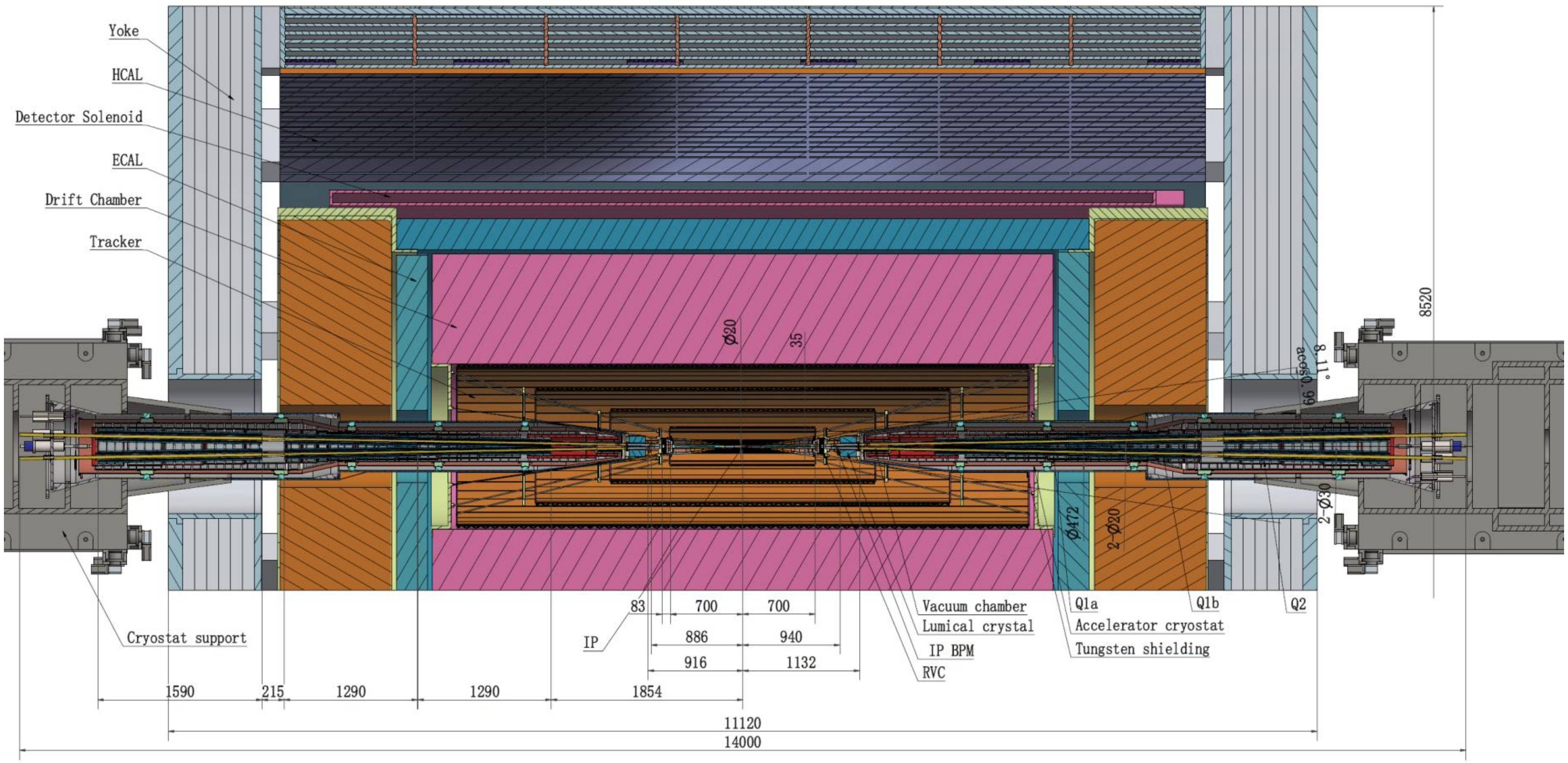

**Fig 4.2.6.1:** CEPC MDI layout.

*IR Superconducting magnet*

On each side of the IP, a final doublet of quadrupoles (Q1a/Q1b and Q2) is used to control the beta function at the IP. Both Q1a/Q1b and Q2 are double aperture superconducting coils. The beam-stay-clear (BSC) region in horizontal and vertical are defined as $BSC_x = \pm(18\sigma_x + 3mm)$ and $BSC_y = \pm(22\sigma_y + 3mm)$ to account for injection and beam tail effects after collision [2-4]. The final doublet quadrupoles are designed based on these parameters, which are listed in tables 4.2.6.2, 4.2.6.3, and 4.2.6.4.

**Table 4.2.6.2:** Q1a design parameters

| **QD0** | **Horizontal BSC** **$2 \times (18\sigma_x + 3)$** | **Vertical BSC** **$2 \times (22\sigma_y + 3)$** | **e+e- beam center distance** |
|---|---|---|---|
| Entrance | 9.2 mm | 12.69 mm | 62.71 mm |
| Middle | 10.41 mm | 14.33 mm | 82.51 mm |
| Exit | 12.24 mm | 14.92 mm | 102.64 mm |
| Good field region | Horizontal 12.24 mm; Vertical 14.92 mm | | |
| Effective length | 1.21 m | | |
| Distance from IP | 1.9 m | | |
| Gradient | 142 T/m | | |

**Table 4.2.6.3:** Q1b design parameters

| **QD0** | **Horizontal BSC** **$2 \times (18\sigma_x + 3)$** | **Vertical BSC** **$2 \times (22\sigma_y + 3)$** | **e+e- beam center distance** |
|---|---|---|---|
| Entrance | 12.53 mm | 14.92 mm | 105.28 mm |
| Middle | 14.97 mm | 14.58 mm | 125.08 mm |
| Exit | 18.17 mm | 13.57 mm | 145.21 mm |
| Good field region | Horizontal 18.17 mm Vertical 14.92 mm | | |
| Effective length | 1.21 m | | |
| Distance from IP | 3.19 m | | |
| Gradient | 85.4 T/m | | |

**Table 4.2.6.4:** Q2 design parameters

| **QF1** | **Horizontal BSC** **$2 \times (18\sigma_x + 3)$** | **Vertical BSC** **$2 \times (22\sigma_y + 3)$** | **e+e- beam center distance** |
|---|---|---|---|
| Entrance | 19.95 mm | 12.92 mm | 155.11 mm |
| Middle | 23.38 mm | 11.72 mm | 179.87 mm |
| Exit | 24.48 mm | 11.31 mm | 204.62 mm |
| Good field region | Horizontal 24.48 mm; Vertical 12.92 mm | | |
| Effective length | 1.5 m | | |
| Distance from IP | 4.7 m | | |
| Gradient | 96.7 T/m | | |

Design of the final doublet quadrupoles is shown in Fig 4.2.6.2.

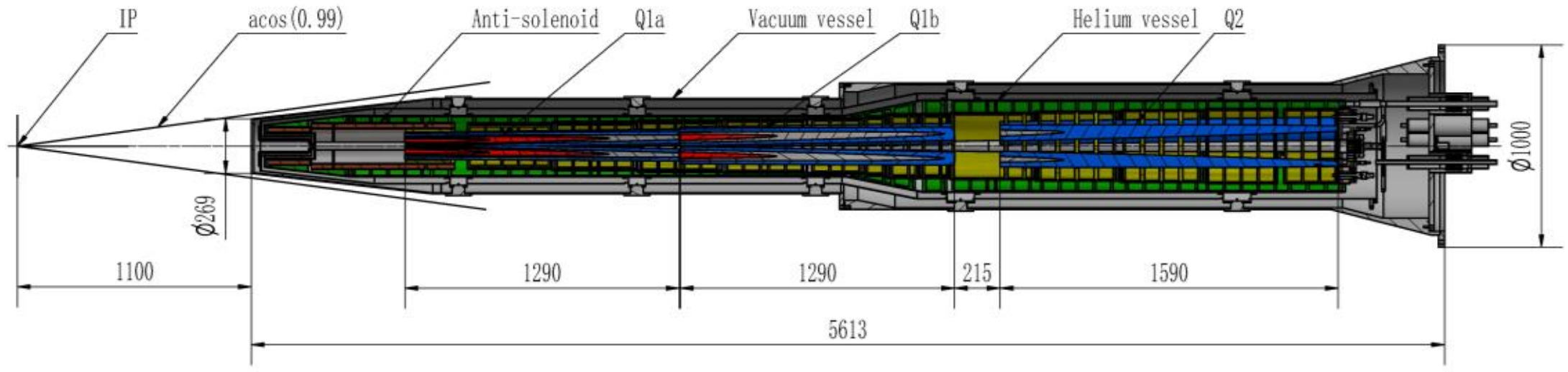

**Fig 4.2.6.2:** Design of the final doublet quadrupoles

*IP BPM*

Beam Position Monitors (BPMs) positioned around the Interaction Point (IP) can function as reliable detectors for determining beam arrival time, in addition to measuring beam position. They serve as the final and most dependable means of monitoring the beam state at the collision point.

To ensure temporal and spatial coincidence between the electron and position, the signal from the IP BPM is utilized. A positional scan is conducted by adjusting the angle ($x$, $y$, $x'$, $y'$) of one beam (either positron or electron) while keeping the other beam unchanged. The process is then repeated by exchanging the beams, allowing for a comprehensive scan of the beam position. Through several iterations, the parameters of the electron and position can be adjusted to reach an ideal state.

In order for the electron and position to arrive at the collision point simultaneously, the RF phase of the electrons is modified. This adjustment ensures that the time difference between the BPM signal of the position and the electron in a shared pipe is equal to the theoretical value $\Delta t = 2 \times L/c$, where $c$ represents the speed of light and $L$ is the distance between the BPM and the collision point. By maintaining this time difference, synchronization of the electron and position arrival times is achieved, resulting in improved accuracy during the collision process.

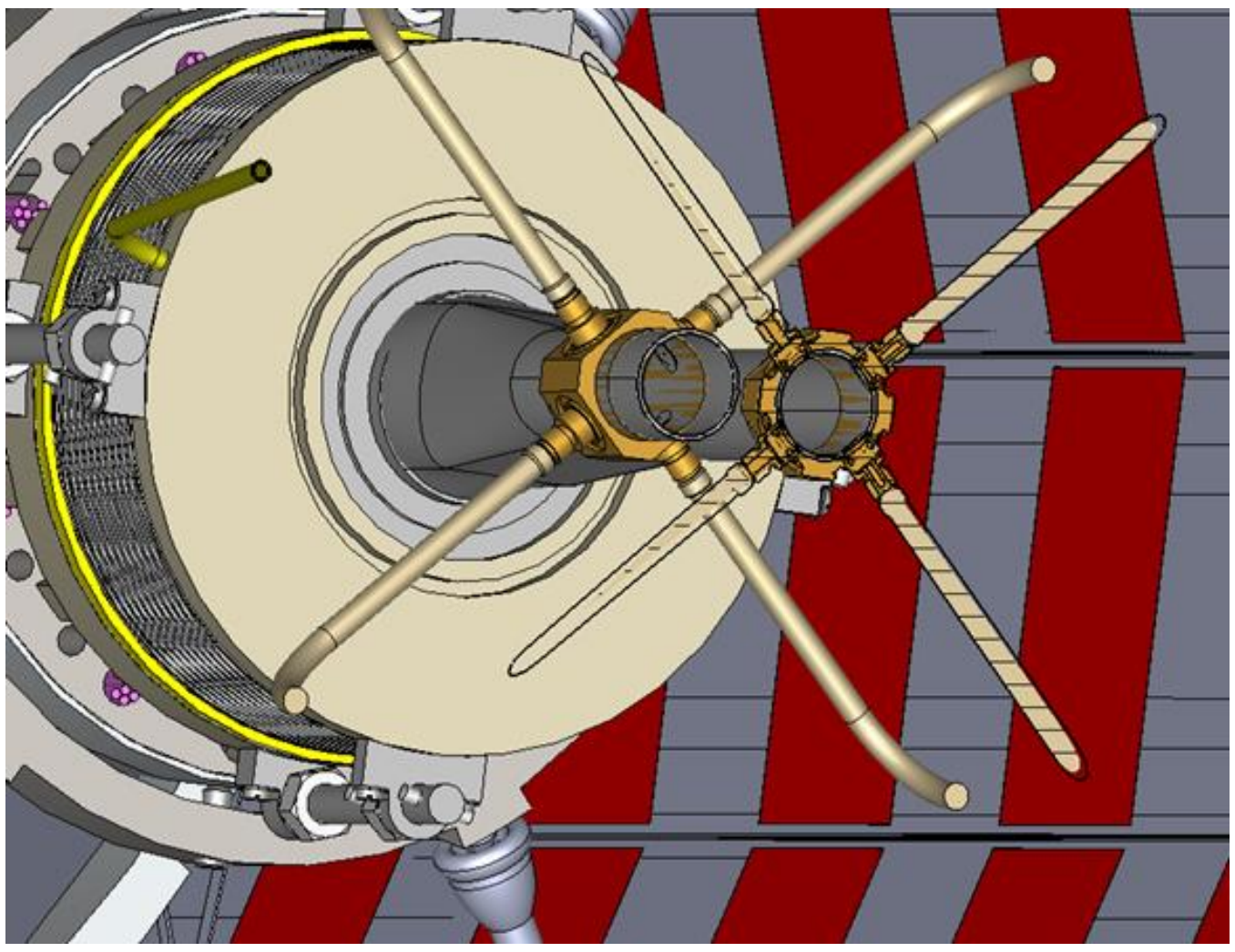

**Fig 4.2.6.3:** 4-button electrodes BPM in the IR.

*Luminosity Calorimeter*

The CEPC physics program focuses on precise measurements of cross sections in the Higgs and EM sectors of the Standard Model. The integrated luminosity of $e^+e^-$ collisions needs to be determined with a precision of $10^{-3}$ and $10^{-4}$ at the center-of-energies corresponding to the Higgs mass and the $Z^0$ pole, respectively. To achieve this, the Luminosity Calorimeter (LumiCal) is positioned in the Machine Detector Interface (MDI) to detect Bhabha events, which result from the elastic scattering of $e^+e^-$ collisions in the low theta region. The Bhabha cross section, described by the QED theory up to near $10^{-4}$, serves as the best reference for luminosity measurements.

The LumiCal faces challenges that need to be addressed. It must achieve geometric coverage with a cross section larger than the $Z^0$ production (41 nb for $Z^0 \rightarrow qq$). Additionally, precise measurement of the scattered electron's theta angle is necessary. The $e^+e^-$ collisions at the IP occur with the z position distributed within the spread of the electron bunch. Electrons that traverse beam-pipe materials are deviated due to multiple scattering. Furthermore, about 1% of the total cross section comprises radiative Bhabha events where electrons are accompanied by photons.

The technological options for designing LumiCal are being studied, focusing on precise detection of the polar angle and energy of electrons and photons originating from the IP. The MDI includes the design of the beam-pipe structure for LumiCal. It incorporates tracking-type position detectors and a $2X_0$ crystal-array stacked within the volume of the Vertex detector. To minimize deviation of the electron impact position, the area with beam-pipe materials is kept minimal.

Furthermore, the LumiCal includes position detectors and crystal arrays on the quadruple magnet situated behind the beam-pipe flanges and bellows. These are used for measuring the energy of the electron shower.

The beam-pipe in the MDI is designed as a racetrack shape, consisting of two merged tubes with a diameter of 20 mm. This configuration limits the LumiCal's coverage of the polar angle (theta) to the range of 25-80 mrad. The Bhabha cross section is twice that of $Z^0 \rightarrow qq$ production, which requires a tolerance of less than 1 μrad on the lower theta edge to achieve a precision of $10^{-4}$ on integrated luminosity.

For the construction and monitoring of LumiCal, mean errors of less than 1 μm are necessary in terms of the beam center, and a deviation of 30 μm in the z-axis to the $e^+e^-$ interaction point.

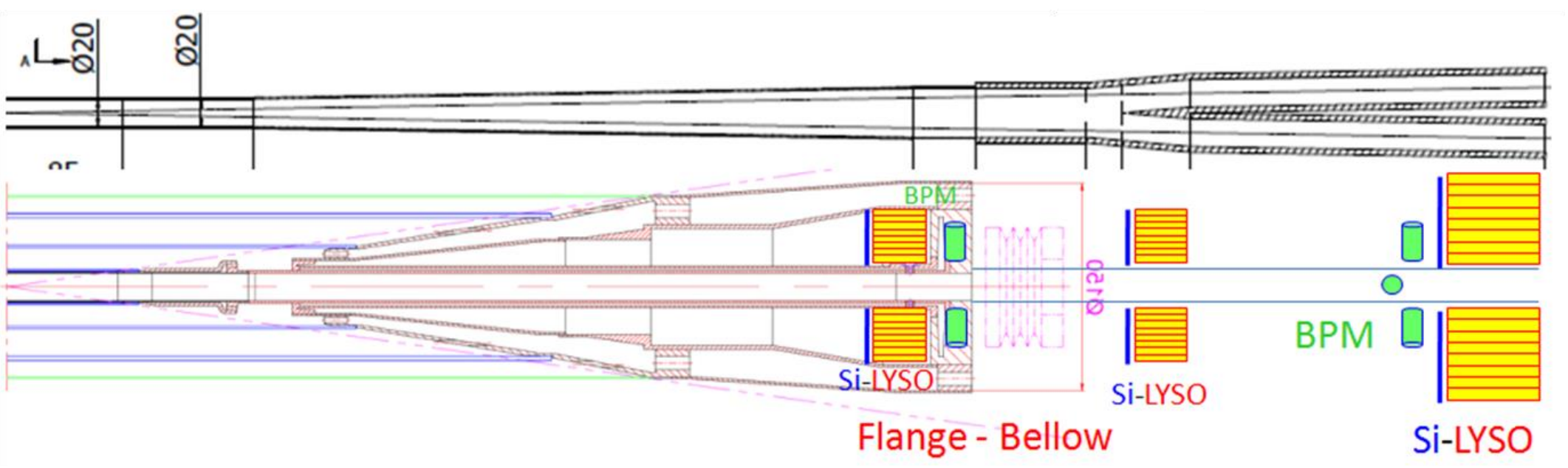

**Fig 4.2.6.4:** LumiCal is illustrated with Silicon and Crystal detectors, in ssemblies before and after the beam-pipe flange and bellow, for electron impact position and energy measurements.

## 4.2.6.2 *Interacton Region Beam Pipe Structure*

### *4.2.6.2.1 Introduction*

The IR vacuum chamber must have a large beam stay clear (BSC) region to reduce detector background and radiation dose due to beam loss. To maintain precise shaping, the chambers must be manufactured using digital-controlled machines and carefully welded to prevent deformation.

In the current design, the inner diameter of the beryllium pipe is 20 mm, chosen to address mechanical assembly and beam background issues. The beryllium pipe has a longitudinal length of 85 mm. To address bremsstrahlung incoherent pairs, the beam pipe shape between 180-655 mm is conical. A bellows is installed in the crotch region, located approximately 0.7 m from the IP, which has a crotch point located at 805 mm from the IP with slope. A racetrack shape beam pipe is used between 805-855 mm from the IP, with an inner diameter of 39 mm for a single pipe and 20 mm for double pipes, chosen to control the heating problem of HOM. Finally, a room temperature beam pipe is utilized for the beam pipe within the final doublet quadrupoles.

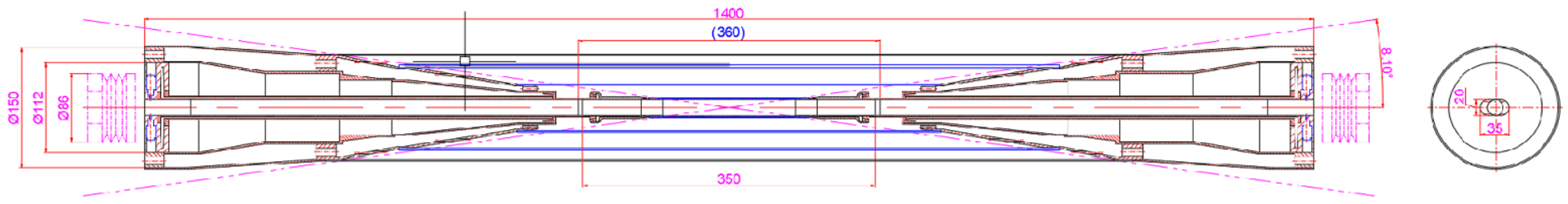

**Fig 4.2.6.5:** CEPC IR central beam pipe mechanical design.

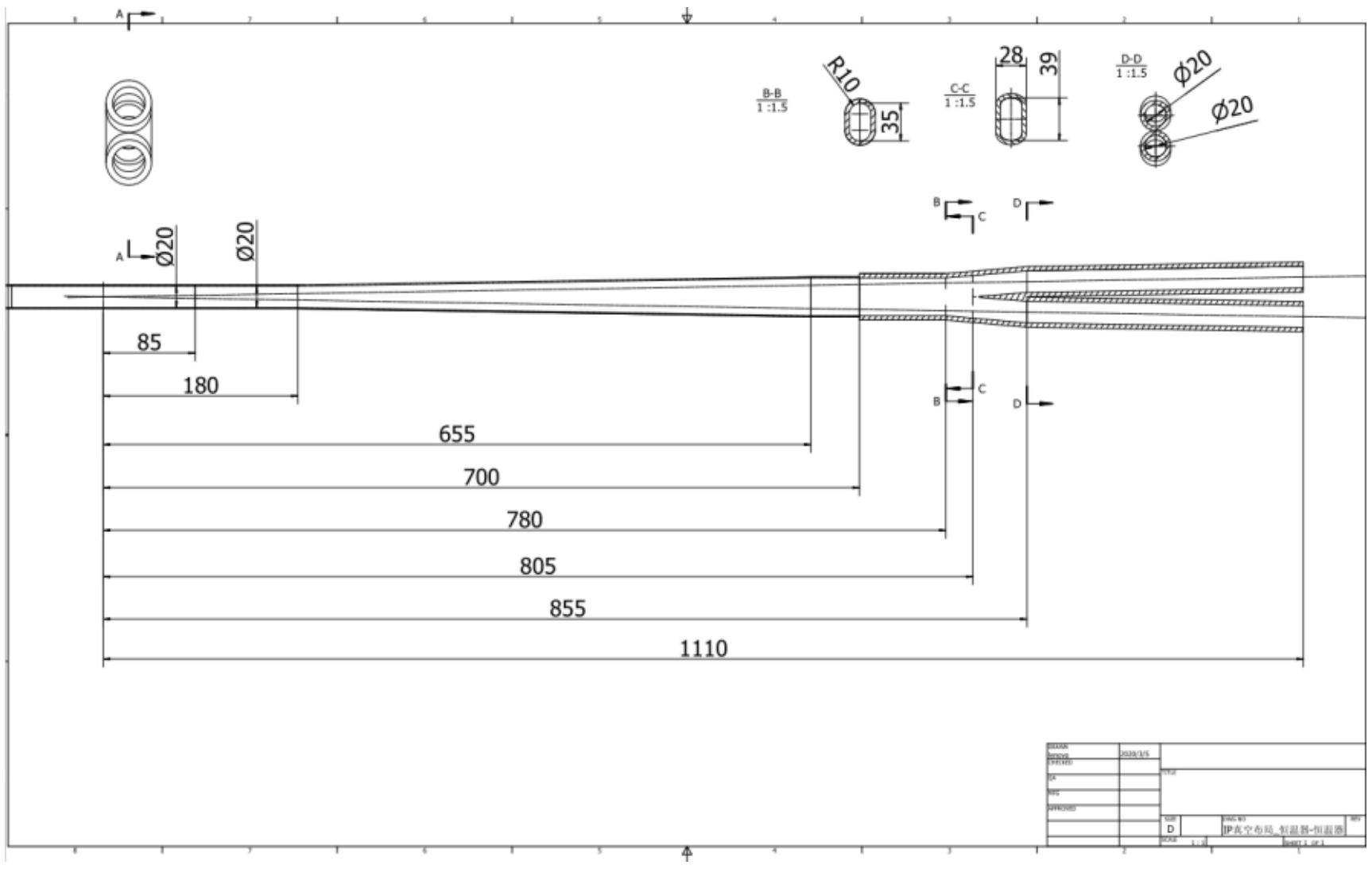

**Fig 4.2.6.6:** CEPC IR beam pipe mechanical design.

**Table 4.2.6.5:** CEPC IR central beam pipe design.

| From IP(mm) | Shape | Inner diameter(mm) | Material | Marker |
|---|---|---|---|---|
| 0-85 | Circular | 20 | Be | |
| 85-150 | Circular | 20 | Al | |
| 180-655 | Cone | 20~35 | Al | Taper: 1:70 |
| 655-700 | Circular | 35 | Al | |
| 700-780 | Circular | 35 | Cu | |
| 780-805 | Cone | 35~39 | Cu | |
| 805-855 | Race-track | 39~20 double pipe | Cu | |

#### *4.2.6.2.2 Mechanical Design*

*General introduction*

The central beam pipe in the CEPC detector runs through the center and is connected to the accelerator vacuum tube at the front and rear ends. It operates in an ultra-high vacuum environment and houses the positron-electron collision center. The beam pipe has an inner diameter of 20 mm and a total length of 1400 mm.

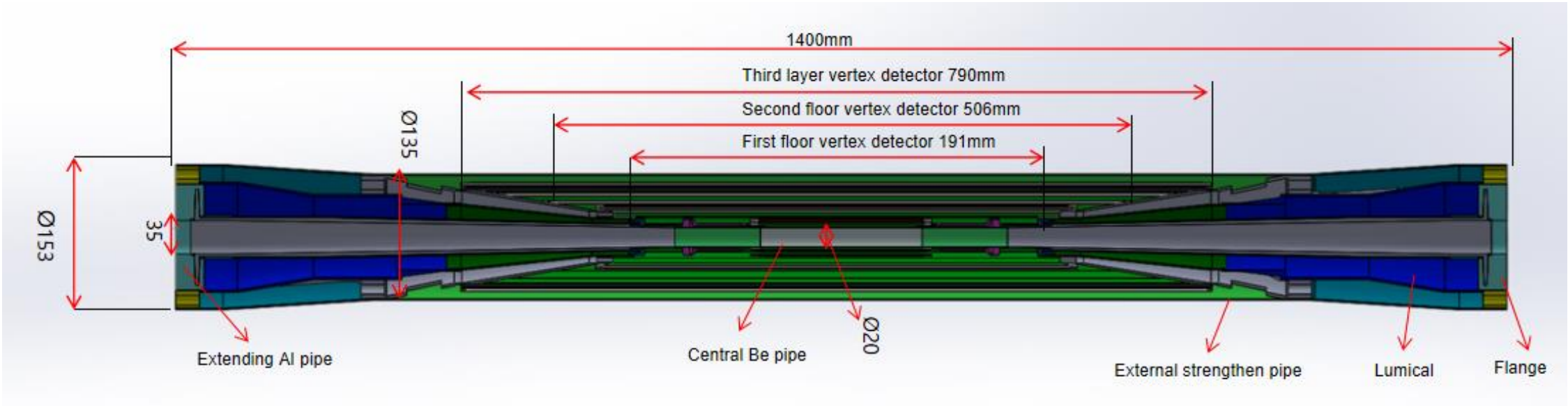

**Fig 4.2.6.7:** Overall layout of the beam pipe

*Central beryllium pipe*

The central beam pipe is designed with a double-layer beryllium pipe, consisting of an inner layer with a thickness of 0.2 mm and an outer layer with a thickness of 0.15 mm. A 0.35 mm cooling gap is incorporated between the two layers, which is filled with cooling oil. The oil enters through the left oil cooling amplification chamber, passes through the gap, and is discharged through the right oil cooling amplification chamber. This cooling mechanism effectively removes the high-order film heat from the inner wall of the beam pipe.

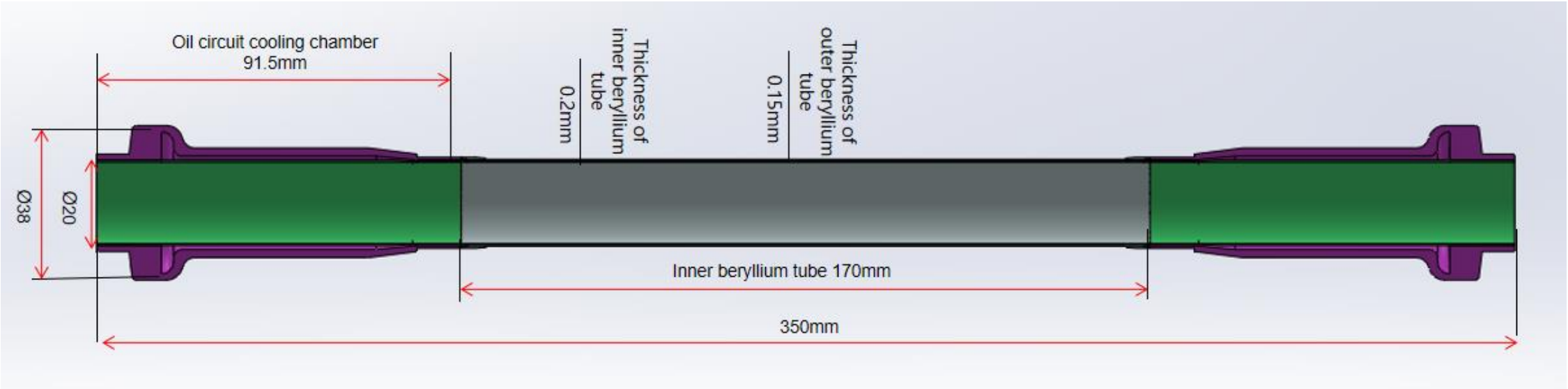

**Fig 4.2.6.8:** Central beryllium pipe

*<u>Structure of air-cooled amplification chamber</u>*

According to the physical requirements, a three-layer vertex detector will be attached to the beryllium pipe. The vertex detector is a thin strip structure and will be supported by two aluminum cylinders on both sides of the beam pipe. The aluminum support cylinders are divided into three steps to install the three-layer vertex detector. As the vertex detector chip and electronics generate heat when working, a cooling structure is necessary to maintain their normal working temperature. The designed cooling system utilizes air cooling to dissipate the heat from the detector. Air ducts are set on the aluminum support cylinders at both ends. Low-temperature compressed air is led from the left pipe to the left air-cooling amplification chamber, and then blown to the vertex detector through the air duct. The detector's heat is taken away through low-temperature air cooling and finally discharged from the right air-cooling channel.

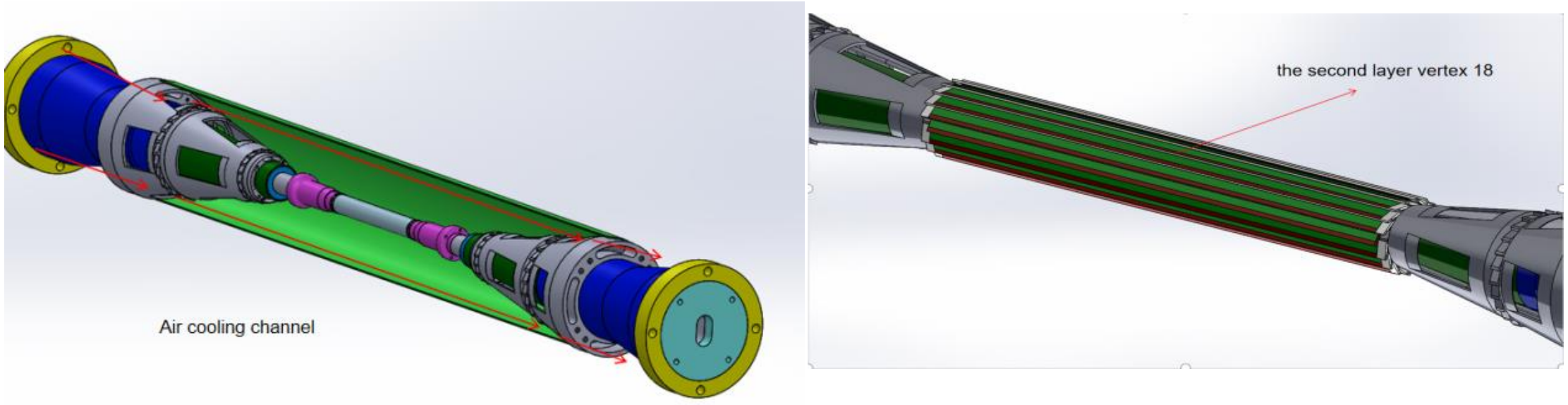

**Fig 4.2.6.9:** Vertex detector distribution and air-cooling channel

*<u>Epitaxial aluminum tube structure</u>*

Two epitaxial aluminum tubes are situated on either side of the beam tube, with two groups of vertex detectors arranged according to the physical design. The vertex detector is a long, thin strip structure that is fastened to the epitaxial aluminum tube using screws. The tube features a double-layer structure, with a partition dividing the internal gap into upper and lower layers. The cooling water channel is formed by the inlet of water from the lower layer and the outlet of water from the upper layer, which takes away the high order film heat of the inner wall and the heat generated by the vertex detector chip.

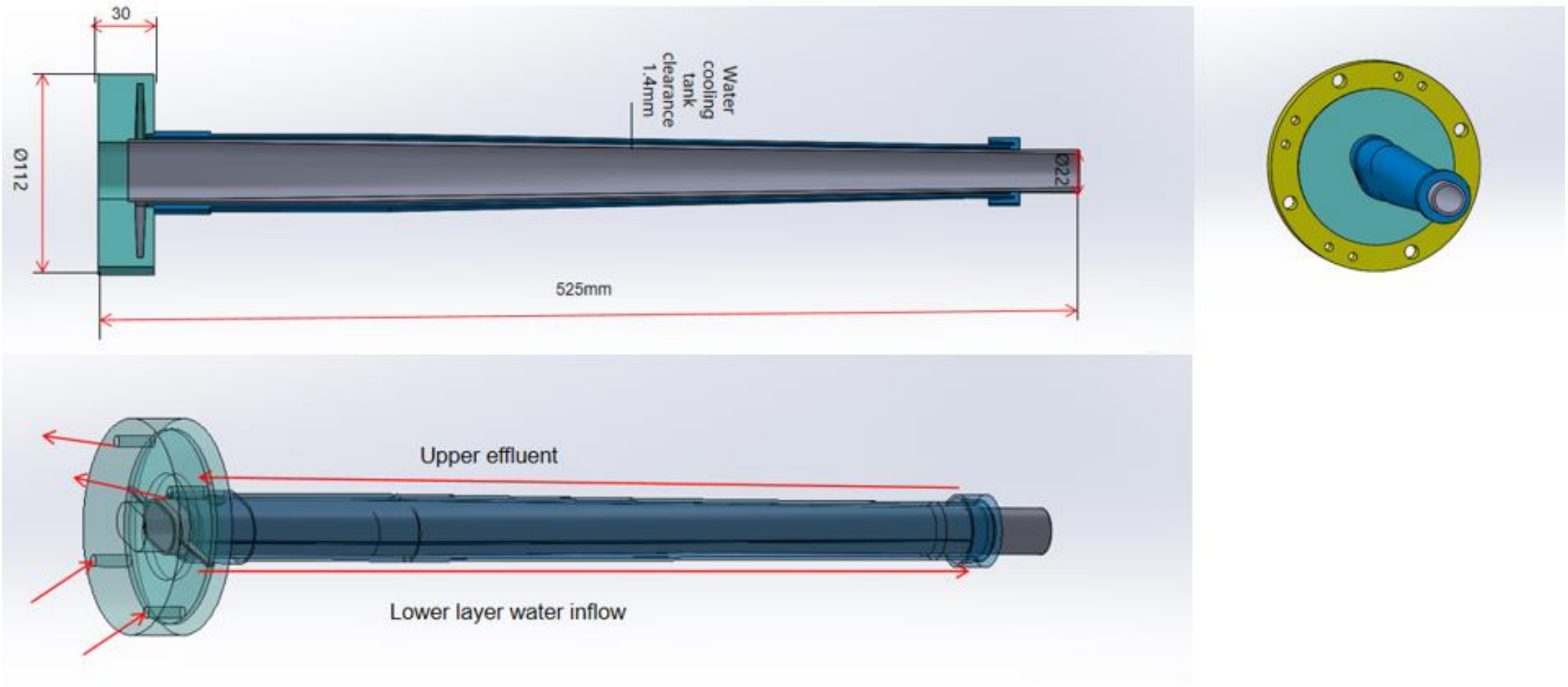

**Fig 4.2.6.10:** Epitaxial aluminum tube structure

*End flange structure*

In an ultra-high vacuum environment, vacuum flanges are required at both ends of the beam pipe to connect to the front and rear accelerator vacuum tubes via flanges. These flanges also provide outlet space for oil cooling pipes, air cooling pipes, detector cables, and other components within the beam pipe.

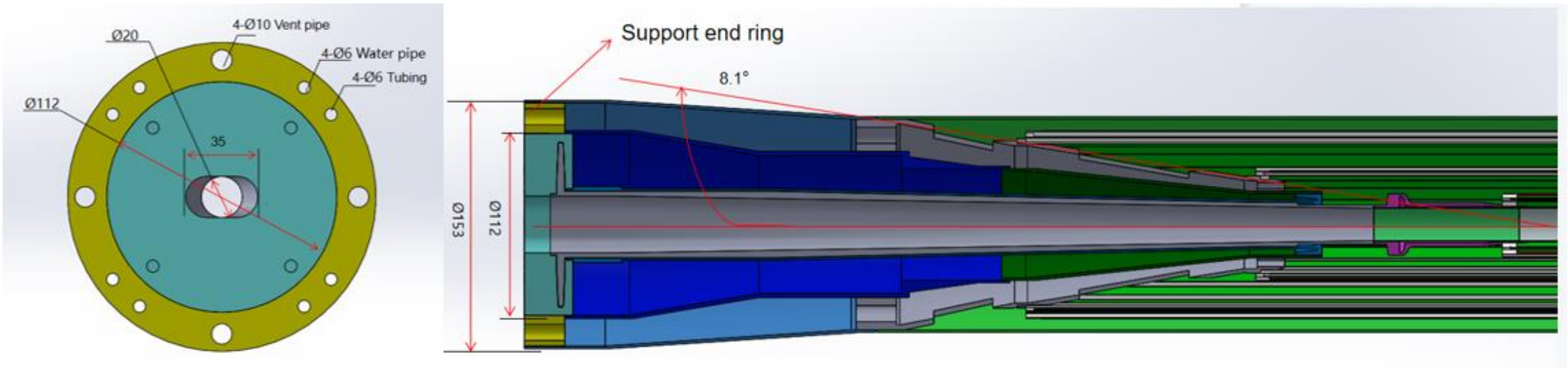

**Fig 4.2.6.11:** End flange structure

*External stiffening tube structure*

During the installation of the beam pipe, it needs to be pushed into the center of the CEPC detector using a cantilever. Therefore, the overall structure of the beam pipe needs to provide sufficient strength and stiffness for the cantilever installation. To achieve this, a carbon fiber sleeve will be added to the outside of the beam pipe. The use of carbon fiber material, which is lightweight and high in strength, will improve the stiffness and strength of the beam pipe and provide protection for the internal detector.

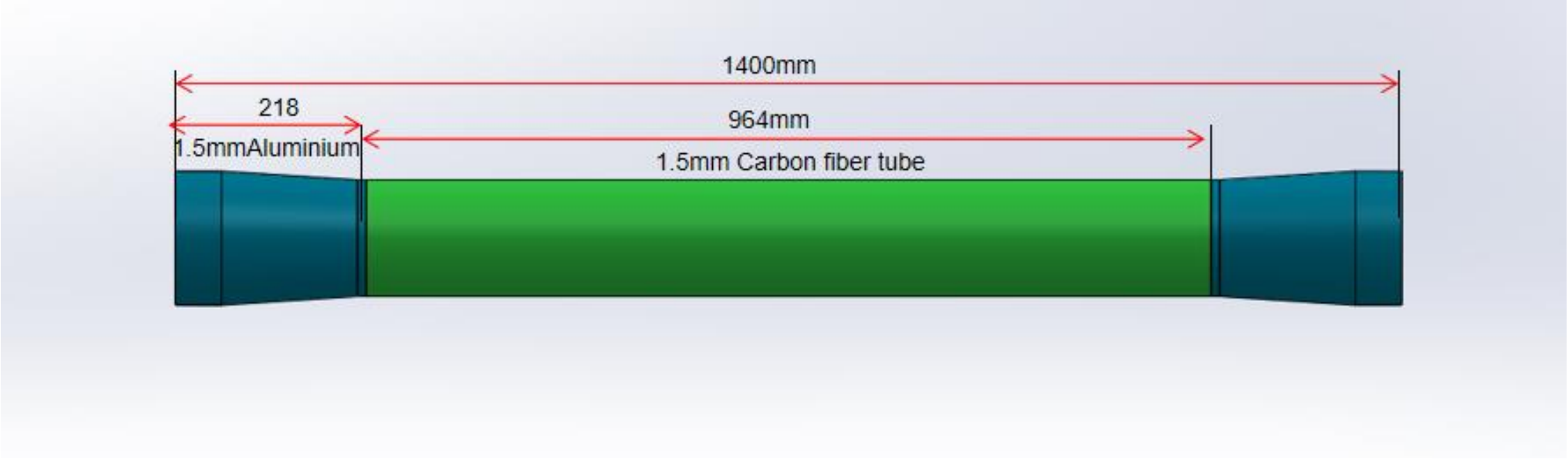

**Fig 4.2.6.12:** Externally stiffening tube structure

*Analysis of overall strength of beam pipe*

Based on the overall strength analysis results of the beam pipe, it was found that when one end of the beam pipe is supported while the other end is floating, during the simulated cantilever installation the maximum sink at the end was 1.2 mm and the maximum stress at the junction of the central beryllium pipe and the epitaxial aluminum pipe was 34.6 MPa. However, when both ends of the beam pipe are supported, the maximum sink at the center during the simulated installation is 0.0025 mm, and the maximum stress is 2.7 MPa, which meets the required use requirements and ensures the overall structural safety.

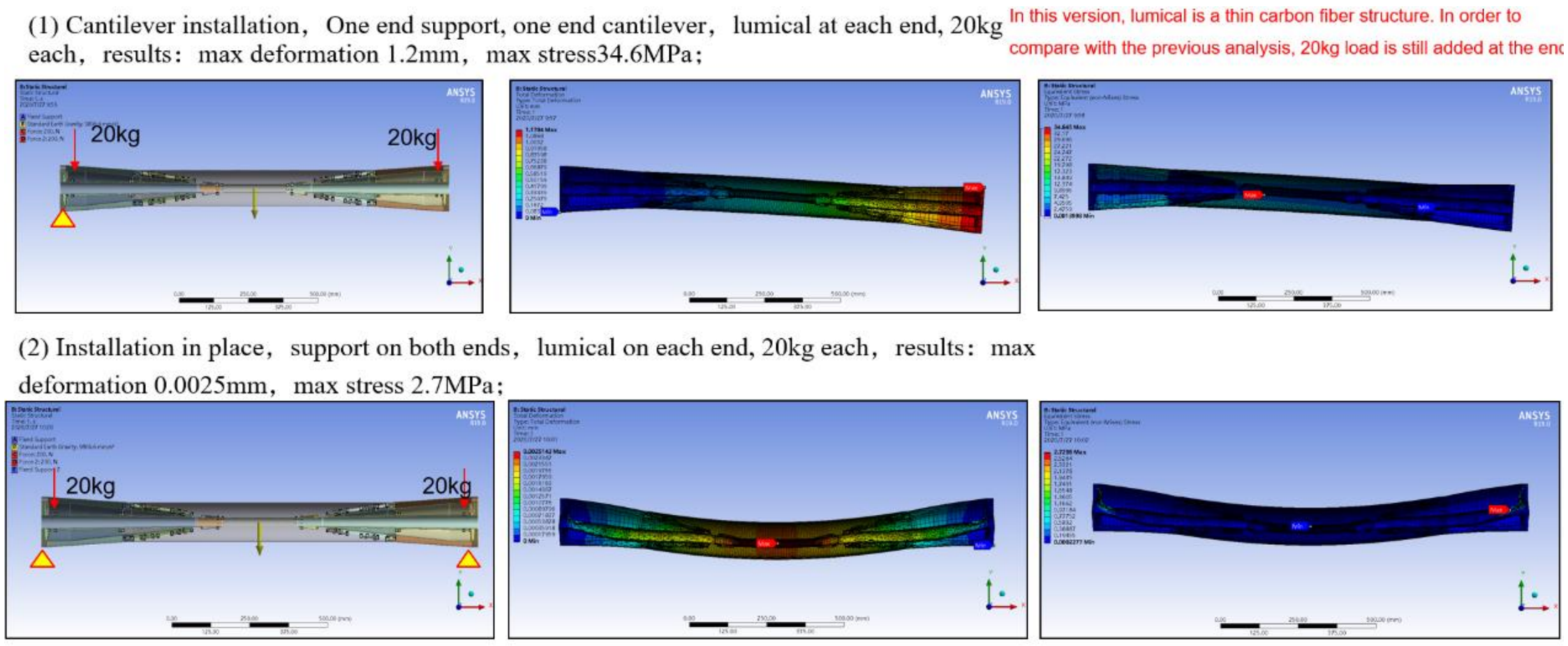

**Fig 4.2.6.13:** Beam tube strength analysis

### *4.2.6.2.3 Thermal Analysis*

#### 4.2.6.2.3.1 Impedance and HOM Heating

The vacuum chamber in the interaction region of CEPC consists of a central pipe region shared by the positron and electron beams, a Y-crotch, and a separated beam pipe. The shape and aperture of the IP chamber must meet the following conditions: (1) avoid direct hitting of synchrotron radiation (SR) on the middle beryllium pipe from the final magnets and minimize reflected SR into the IR pipe, and (2) eliminate the high-order mode (HOM) created by the Y-crotch in the IP chamber.

To suppress impedance and HOM trapping in this region, the central aperture size is 20 mm, the same as the separated beam pipe. The round central beryllium pipe is connected to the Y-crotch by a racetrack taper as a transition. With the Y-crotch in the IR region, all the direct SR from the last bend and 85% of the reflected photon can be stopped. The schematic for the IP chamber is shown in Figure 4.2.6.15. The impedance can be estimated with CST, and it is clear that the longitudinal impedance in the lower frequency region is at a lower level, as shown in Figure 4.2.6.14. There is only one single narrow impedance with a frequency of 8.1 GHz that can produce HOM heating in the IR. For the IP chamber, the wake for the most part is resistance and also has some inductance, $L$ ~ 0.21nH, $k_l$ = 0.0185V/pC.

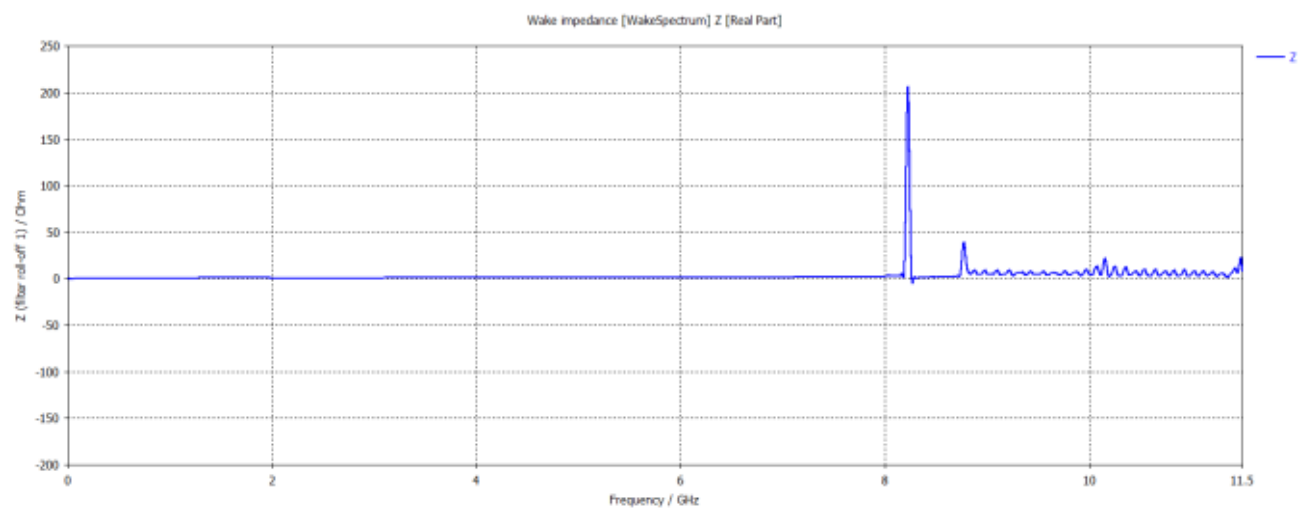

**Fig 4.2.6.14:** Longitudinal impedance distribution

Due to the Y-crotch in the IP chamber, some high-order electromagnetic fields may be trapped in the IR region. The trapped HOM loss factor of the IP chamber was calculated with CST for the current IR structure. The trapped mode loss factor in the IR region is about 0.0323 V/pC (for a bunch length of 5 mm, including the Ohmic loss by materials), corresponding to the power of 24 W, 117.1 W, and 1160 W for the Higgs, W, and Z modes, respectively.

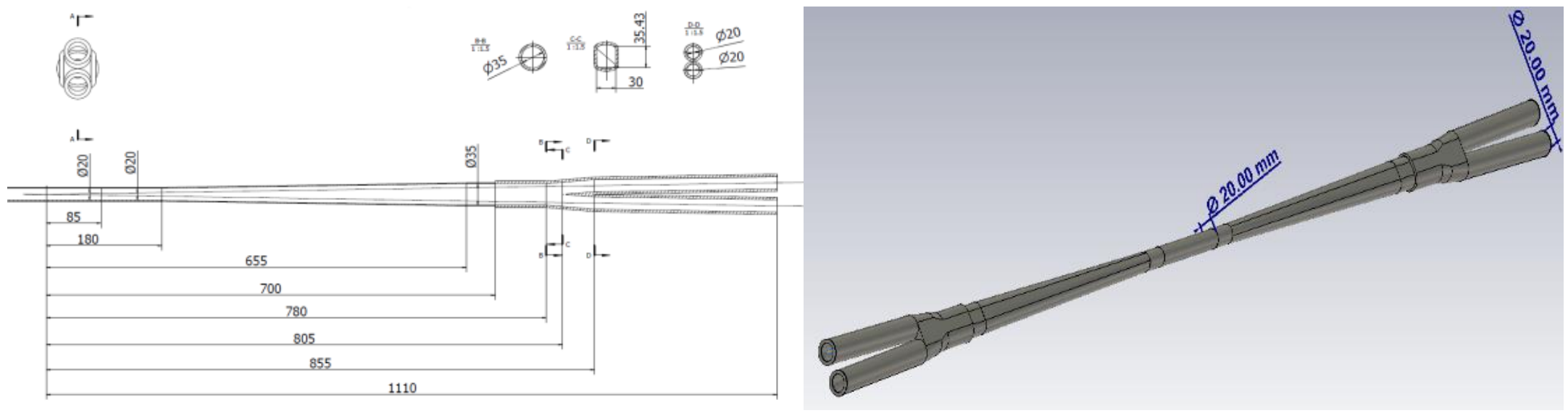

**Fig 4.2.6.15:** The schematic of IP chamber

The power distribution in different regions of the IP chamber is crucial for the cooling system. The trapped HOM, which originates from the Y-crotch, will transfer heating power and deposit it in different parts of the IP chamber. Table 4.2.6.5 lists the power disposition and power density, indicating that the most power deposition position is on the beryllium pipe.

**Talbe 4.2.6.6:** Power distribution in different region of the IP chamber

| Position | Start-end (mm) | Material | Length (mm) | Higgs (W / W/cm$^2$) | W (W / W/cm$^2$) | Z (W / W/cm$^2$) |
|---|---|---|---|---|---|---|
| Be pipe | 0-85 | Be | 85 | 1.13 / 0.021 | 5.587 / 0.105 | 55.295 / 1.35 |
| Be pipe transition | 85-180 | Al | 95 | 0.61 / 0.01 | 2.950 / 0.049 | 29.280 / 0.491 |
| Transition pipe | 180-655 | Al | 475 | 6.99 / 0.017 | 34.48 / 0.085 | 341.562 / 0.83 |
| Transition | 655-700 | Al | 45 | 0.62 / 0.015 | 2.95 / 0.071 | 29.28 / 0.701 |
| RVC bellows | 700-780 | Cu | 80 | 0.52 / 0.007 | 2.532 / 0.034 | 25.002 / 0.337 |
| Transition on Y-crotch | 780-805 | Cu | 25 | 0.16 / 0.007 | 0.785 / 0.032 | 7.822 / 0.316 |
| Y- crotch | 805-855 | Cu | 50 | 0.33 / 0.005 | 1.572 / 0.024 | 15.626 / 0.241 |
| Quadrupole pipe | 855-1100 | Cu | 245 | 1.58 / 0.005 | 7.735 / 0.024 | 75.594 / 0.24 |
| Total | 0-1100 | - | 1100 | 12.0 / 0.011 | 58.594 / 0.056 | 580.46 / 0.56 |

#### 4.2.6.2.3.2 Synchrotron Radiation

*SR in normal condition*

To allow for softer bends in the upstream of the interaction point (IP), an asymmetric lattice is used. The last bending magnet before the IP has a reverse bending direction to prevent synchrotron radiation from hitting the central IP beam pipe. As a result, under normal conditions, there are no synchrotron radiation photons hitting the beam pipe generated from the last bending magnet in the upstream of the IP.

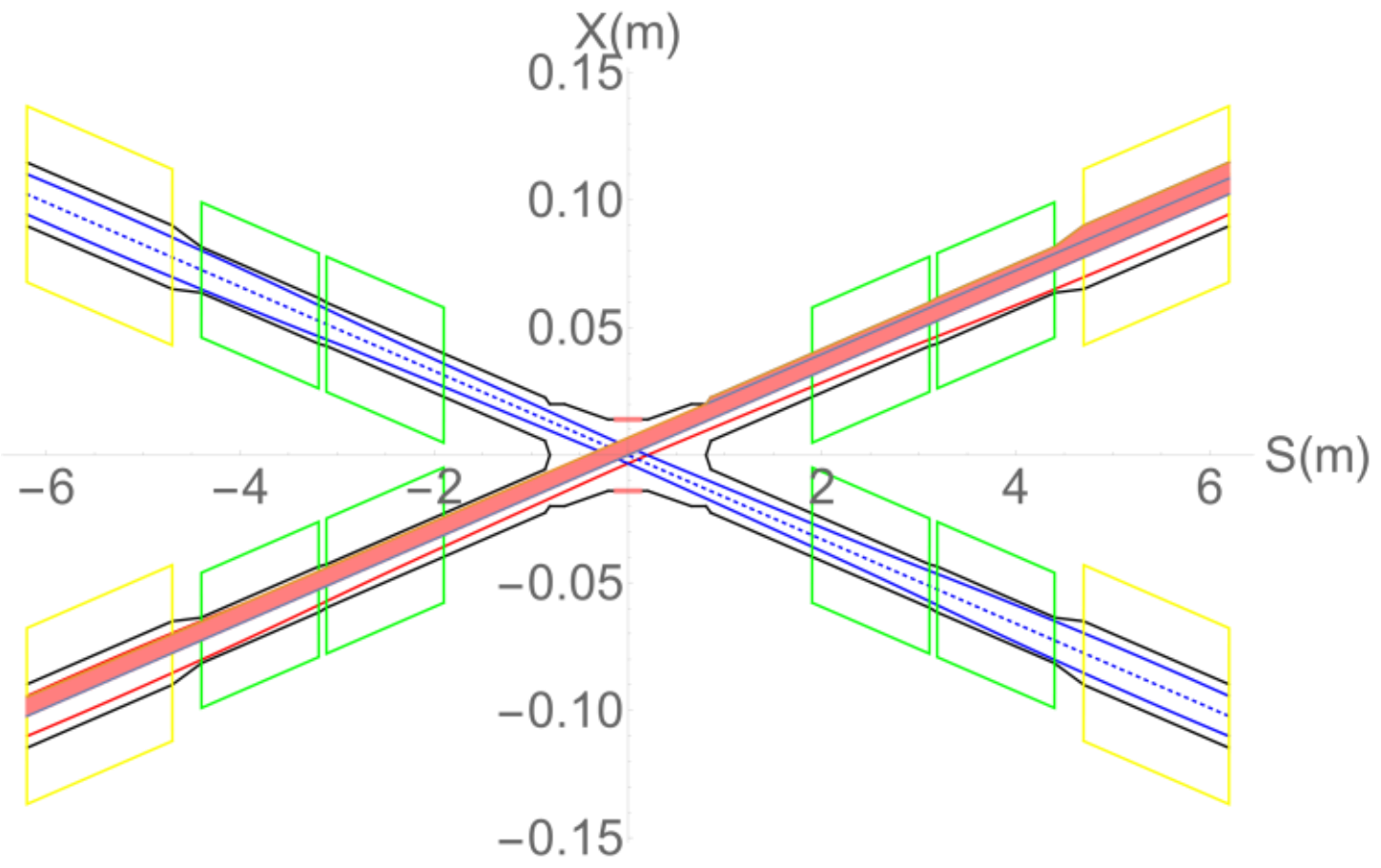

**Fig 4.2.6.16:** SR from last bending magnet in upstream of IP in normal conditions

Both a "room temperature" beam pipe and conduction cooled superconducting magnet are necessary. A single-layer beam pipe with water cooling has been implemented for the IR beam pipe of the accelerator part beyond 700 mm away from the IP. The synchrotron radiation heat load distribution for this implementation is shown in Table 4.2.6.7.

**Table 4.2.6.7:** SR heat load distribution from last bending magnet in upstream of IP

| Region | SR heat load | SR average power density |
|---|---|---|
| 0~780mm | 0 | 0 |
| 780mm~805mm | 23.04W | 256W/cm$^2$ |
| 805mm~855mm | 53.39W | 296.6 W/cm$^2$ |
| 855mm~1.9m (QDa entrance) | 4.32W | 1.15 W/cm$^2$ |
| QDa | 3.28W | 0.75 W/cm$^2$ |
| QDa ~ QDb | 22.92W | 79.58 W/cm$^2$ |
| QDb | 3.96W | 0.91 W/cm$^2$ |
| QDb~QF1 | 71.04W | 65.8 W/cm$^2$ |
| QF1 | 7.26W | 1.34 W/cm$^2$ |

However, some secondaries generated within the QD beam pipe will hit the detector beam pipe, including the beryllium part. Therefore, mitigation methods must be studied. Synchrotron radiation (SR) photons generated from the FD magnets will hit downstream of the IR beam pipe. The once-scattered photons will not go into the detector beam pipe but will go even farther away from the IP region.

*SR in extreme condition*

In extreme situations, such as power lost in a magnet, the entire ring orbit will immediately experience significant distortion, and the beam may be lost if the distortion exceeds certain limits. The beam will be stopped within 0.5 ms in extreme cases, which can happen at least 10 times per day. However, the deviation of the beam orbit will not affect detector operation since the high background section will be removed during data analysis.

In abnormal conditions, synchrotron radiation photons will hit the bellows and beryllium pipe in the IR. Althugh there is no cooling at the bellows, heat load is not a concern as it is a transient effect. However, the detector background and radiation dose should be considered under abnormal conditions.

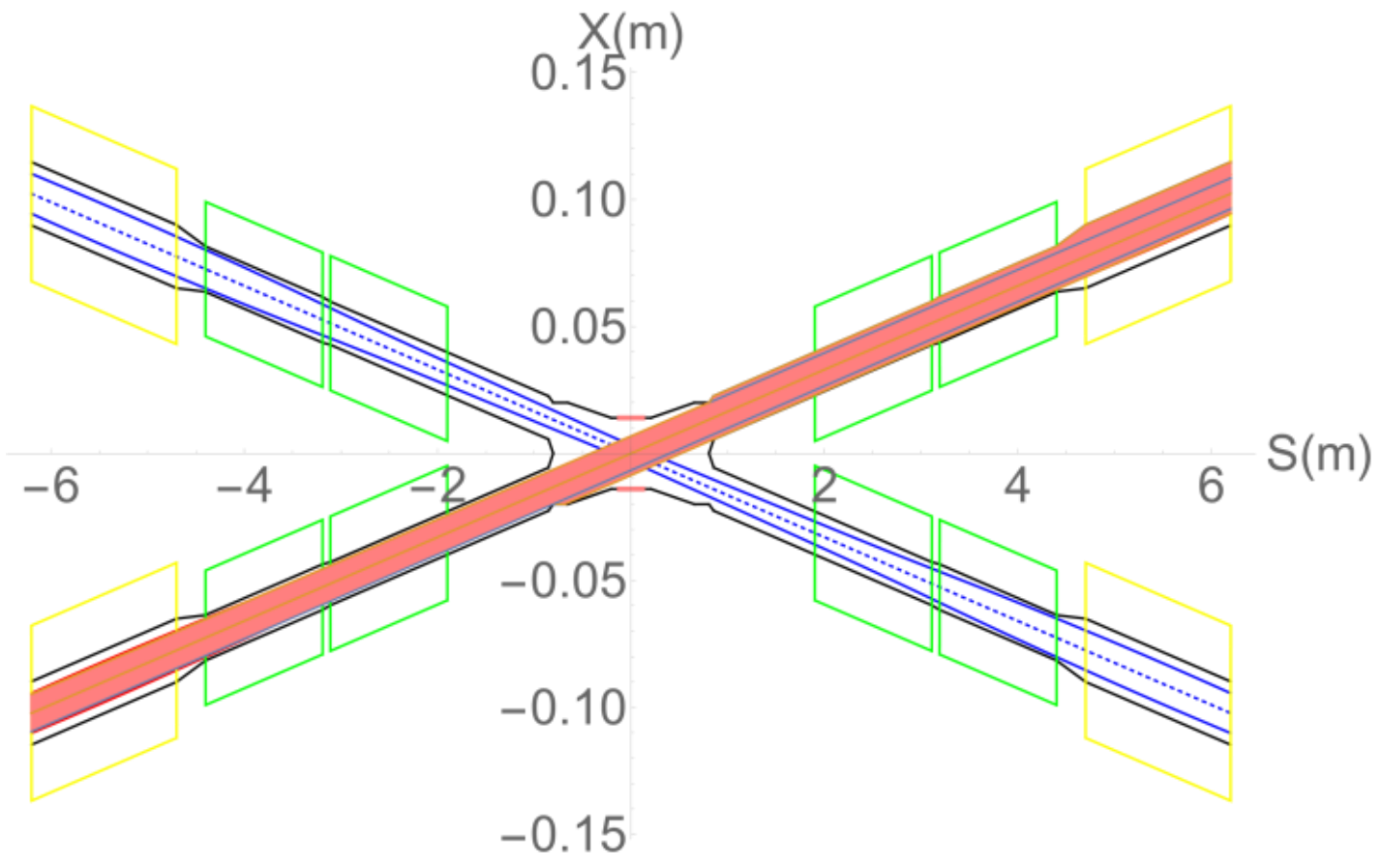

**Fig 4.2.6.17:** SR from upstream of IP in extreme conditions

*SR from solenoid combined field*

Since the solenoids used in the accelerator can couple the horizontal and vertical trajectories, any deviation in the horizontal trajectory will also affect the vertical trajectory, as shown in figure 4.2.6.18.

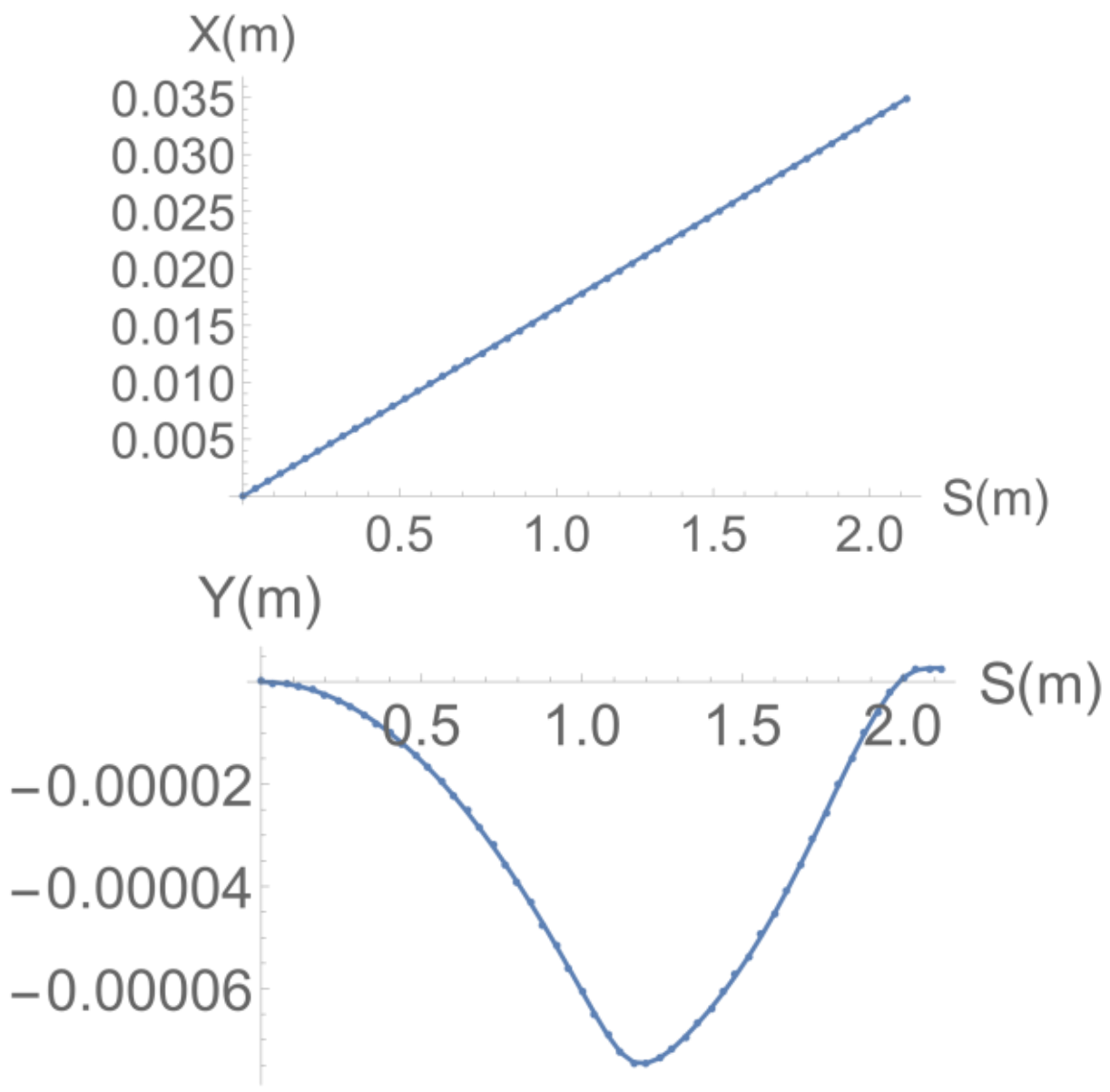

**Fig 4.2.6.18:** Horizontal and vertical trajectory due to solenoid combined field.

In addition, synchrotron radiation is generated tangentially along the vertical trajectory in the solenoid combined field, as seen in figure 4.2.6.19. As the combined solenoid field strength is quite high, with a maximum of about 5.81 T, the transverse magnetic field component is also significant. Therefore, synchrotron radiation generated from the vertical trajectory in the solenoid combined field must be considered.

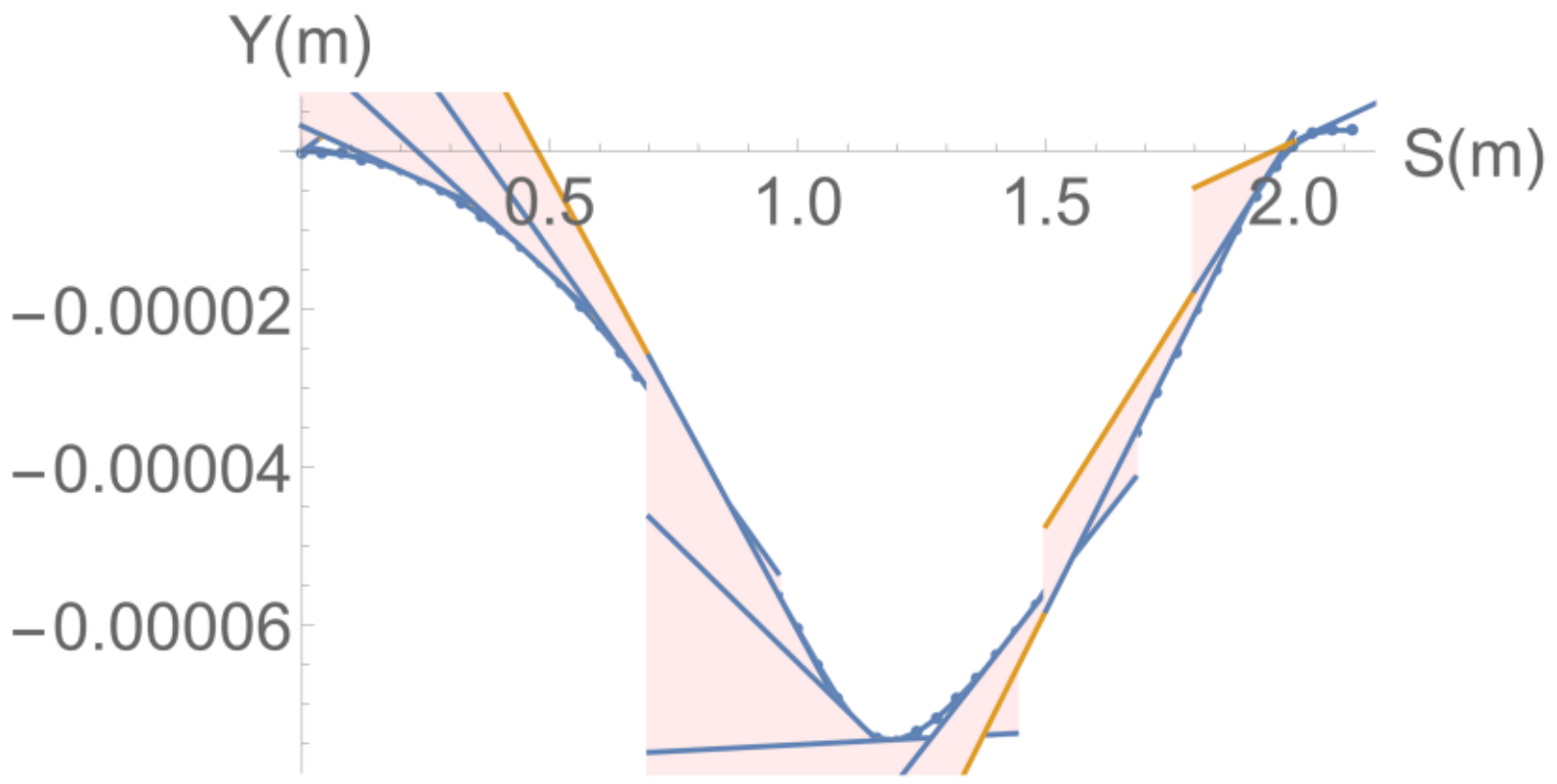

**Fig 4.2.6.19:** Synchrotron radiation fan from vertical trajectory due to solenoid combined field.

The critical energy distribution of the synchrotron radiation generated from the solenoid combined field effect along the longitudinal direction is presented in figure 4.2.6.20, where the maximum SR critical energy is 918 keV.

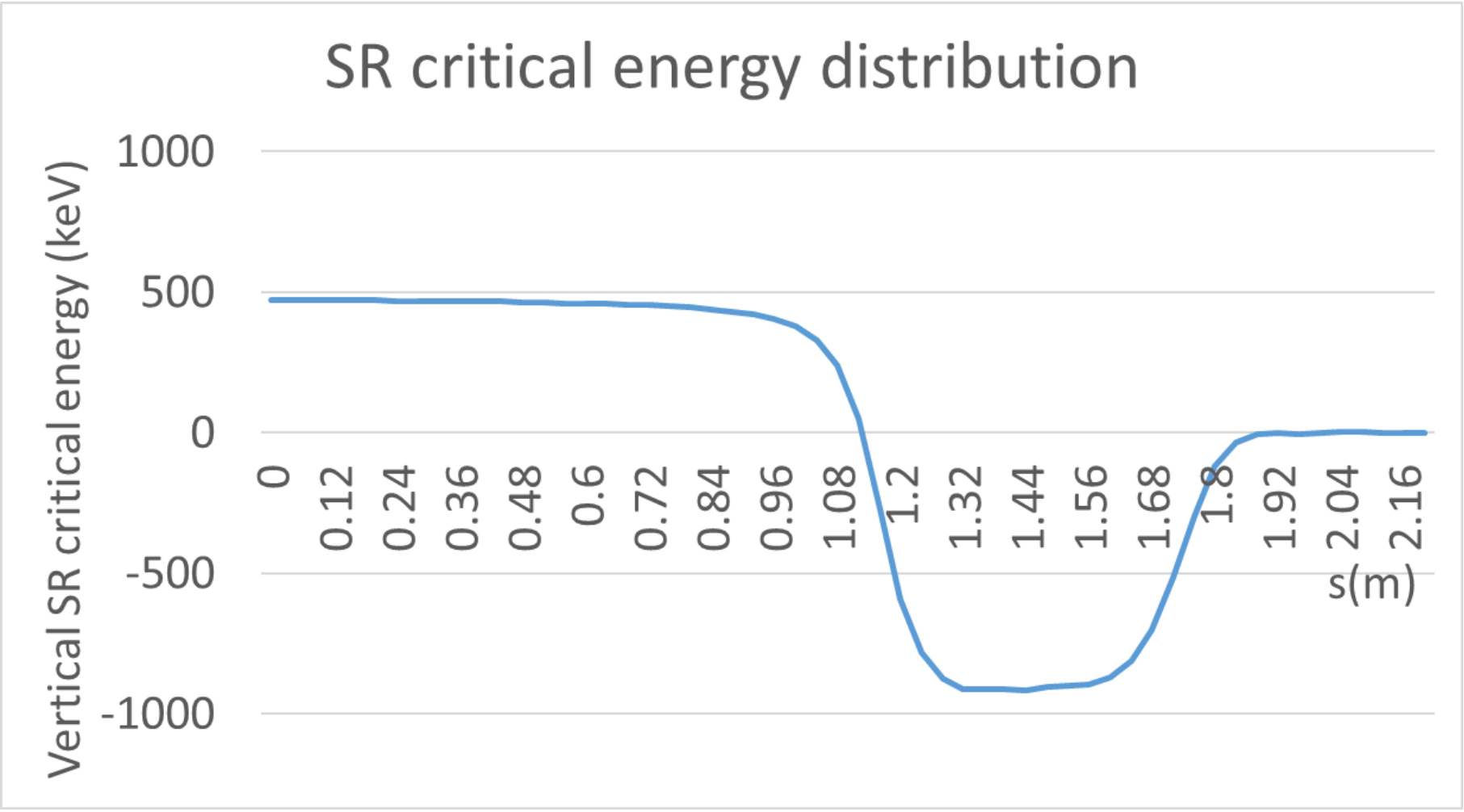

**Fig 4.2.6.20:** Synchrotron radiation critical energy distribution due to solenoid combined field.

Furthermore, the power distribution of the synchrotron radiation generated from the solenoid combined field is shown in figure 4.2.6.21, where the maximum SR power is 40.6 W.

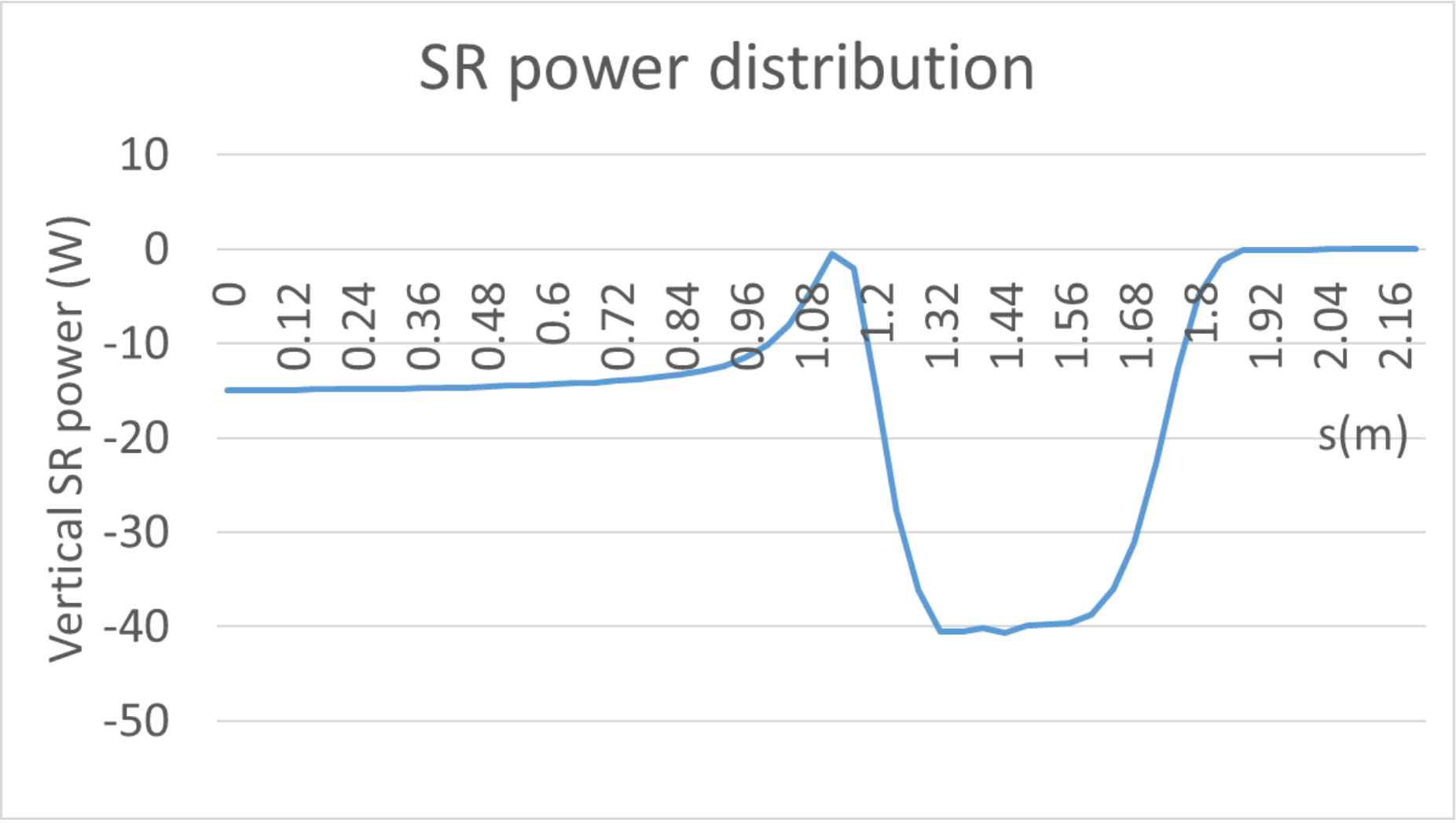

**Fig 4.2.6.21:** Synchrotron radiation power distribution due to solenoid combined field.

The synchrotron radiation fan due to the solenoid combined field is focused in a very narrow angle from -116 µrad to 131 µrad. Although the synchrotron radiation will not hit the detector's beryllium pipe and will not cause background to the detector, it will hit the beam pipe located about 213.5 m downstream from the IP. Therefore, a water-cooling structure is required.

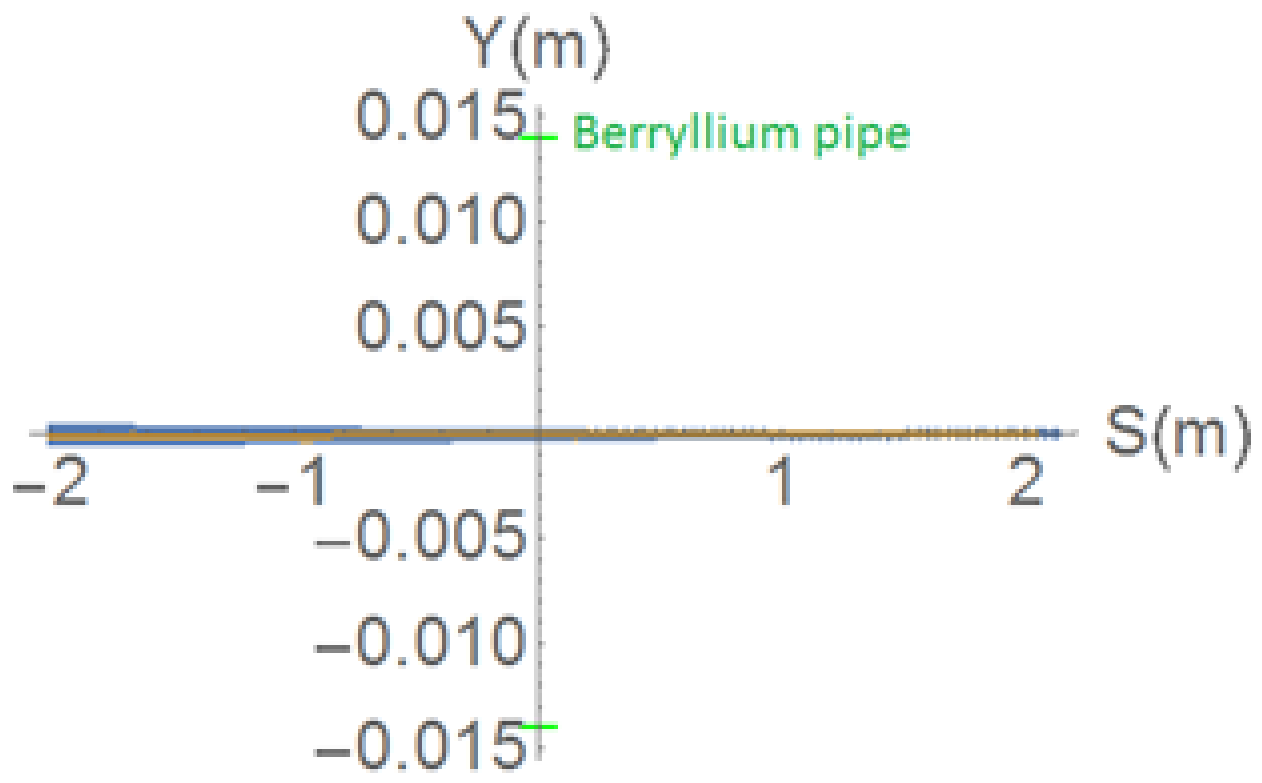

**Fig 4.2.6.22:** SR from solenoid combined field.

4.2.6.2.3.3 Beam Loss Background

The CEPC beam particles may experience abrupt energy loss through scattering processes such as Radiative Bhabha scattering, Beamstrahlung, beam-gas scattering, and beam-thermal photon scattering. With consideration for beam-beam effects and errors, the CEPC energy acceptance is approximately 1.5% based on the optimized lattice's off-momentum dynamic aperture. If a particle's energy loss exceeds 1.5% of the beam energy, it will be lost from the beam and may collide with the vacuum chamber. If this happens near the IR, it will cause a heat load on the beam pipe, potentially leading to superconducting magnet quenching. Table 4.2.6.8 shows the heat load distribution from beam loss background.

**Table 4.2.6.8:** Heat load distribution from beam loss backgrounds.

| Region | RBB | Beamstrahlung | Beam-Gas | BTH |
|---|---|---|---|---|
| Berryllium pipe | 6.7mW | 0 | 0 | 0 |
| Detector beam pipe | 0.024W | 0 | 4.8μW | 1.2μW |
| Accelerator beam pipe before Q1a | 0.17w | 0 | 4.2μW | 1.2μW |
| Q1a~Q1b | 2.13w | 3.8μW | 5.9μW | 1.8μW |
| Q1b~Q2 | 0.01w | 3.8μW | 0.5μW | 0.6μW |
| Q2 | 0.26mW | 0 | 3.7μW | 0.66μW |

The heat load in the IR from beam loss backgrounds is relatively small compared to the heat load generated from synchrotron radiation and higher-order modes (HOM).

4.2.6.2.3.4 Beam Pipe Thermal Analysis

Heat with a uniform distribution will be generated on the inner walls of the central beryllium and epitaxial aluminum pipes due to high order film effects. To mitigate this, oil cooling is utilized for the central beryllium pipe, while water cooling is used for the epitaxial aluminum pipe. Based on calculations, the highest temperature occurs on the inner wall of the amplification chamber pipe at the outlet of the oil cooling system. In the Z mode, the highest temperature reaches 48.1℃, while in the High Luminosity Z mode, it reaches 81.2 ℃. The temperature of other parts must meet the engineering requirements.

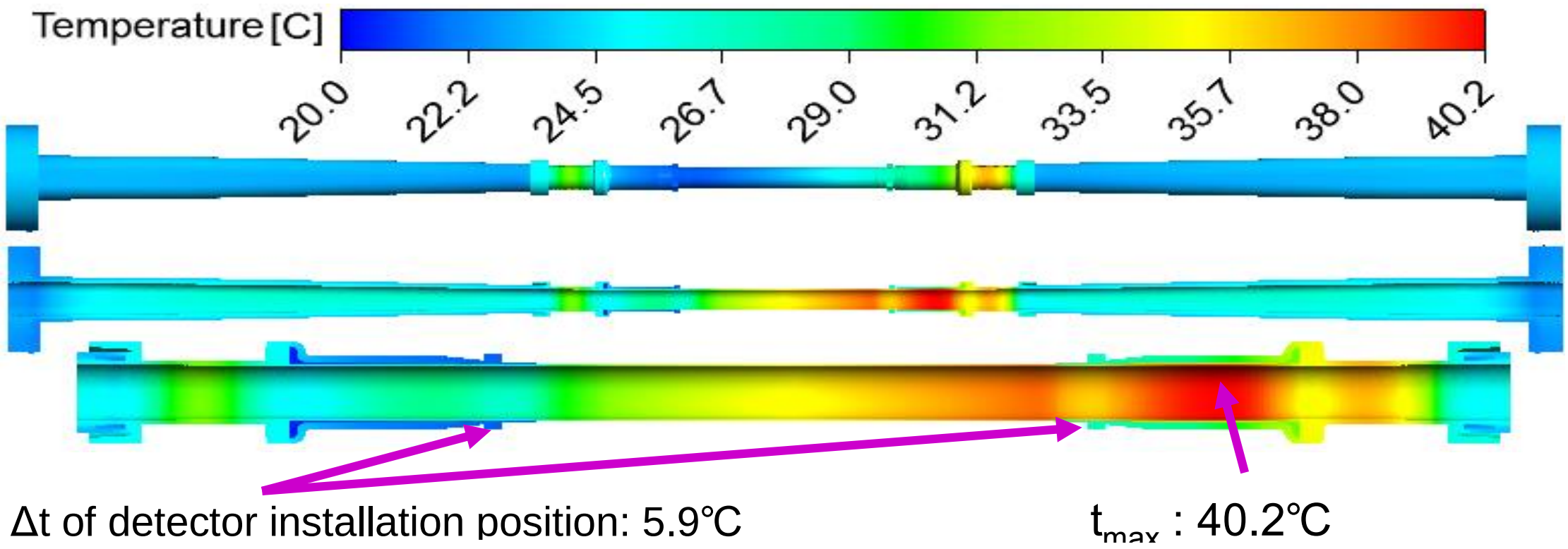

**Fig 4.2.6.23:** Cooling calculation of central beryllium pipe

### 4.2.6.3 *Background Estimation and Mitigation*

#### *4.2.6.3.1 Introduction*

Beam particles may lose energy through various scattering processes such as radiative Bhabha, beamstrahlung, beam-gas, or beam-thermal photon scattering. The energy acceptance of the CEPC, after optimizing the lattice and considering beam-beam effects and errors, is 1.3%, 1.2%, 1.6%, and 2.3% at a center-of-mass energy of 90 GeV, 180 GeV, 240 GeV, and 360 GeV, respectively. If the energy loss exceeds the energy acceptance, particles will be lost from the beam and may collide with the vacuum chamber or detectors in the IR. In addition, synchrotron radiation and pair-production processes can affect the IR components.

#### 4.2.6.3.2 *Photon Background: SR, Mask and Pair Production*

Photons originating from the Synchrotron Radiation and Pair Production processes represent the most important radiation backgrounds at the CEPC.

4.2.6.3.2.1 Synchrotron Radiation

Synchrotron radiation (SR) is a significant background at circular colliders, including the CEPC. SR photons at the CEPC will range from a few keV to several MeV. While the central beryllium beam pipe in the interaction region is designed to avoid direct hitting of SR photons, secondary photons scattered by elements in the interaction region must be carefully considered. Simulation of the SR photons using BDSim [5] and Geant4 [6] shows that approximately 40,000 photons could hit the beryllium beam pipe per bunch crossing, posing a potential threat to the detector, particularly the first layer of the Vertex detector, as illustrated in Figure 4.2.6.24.

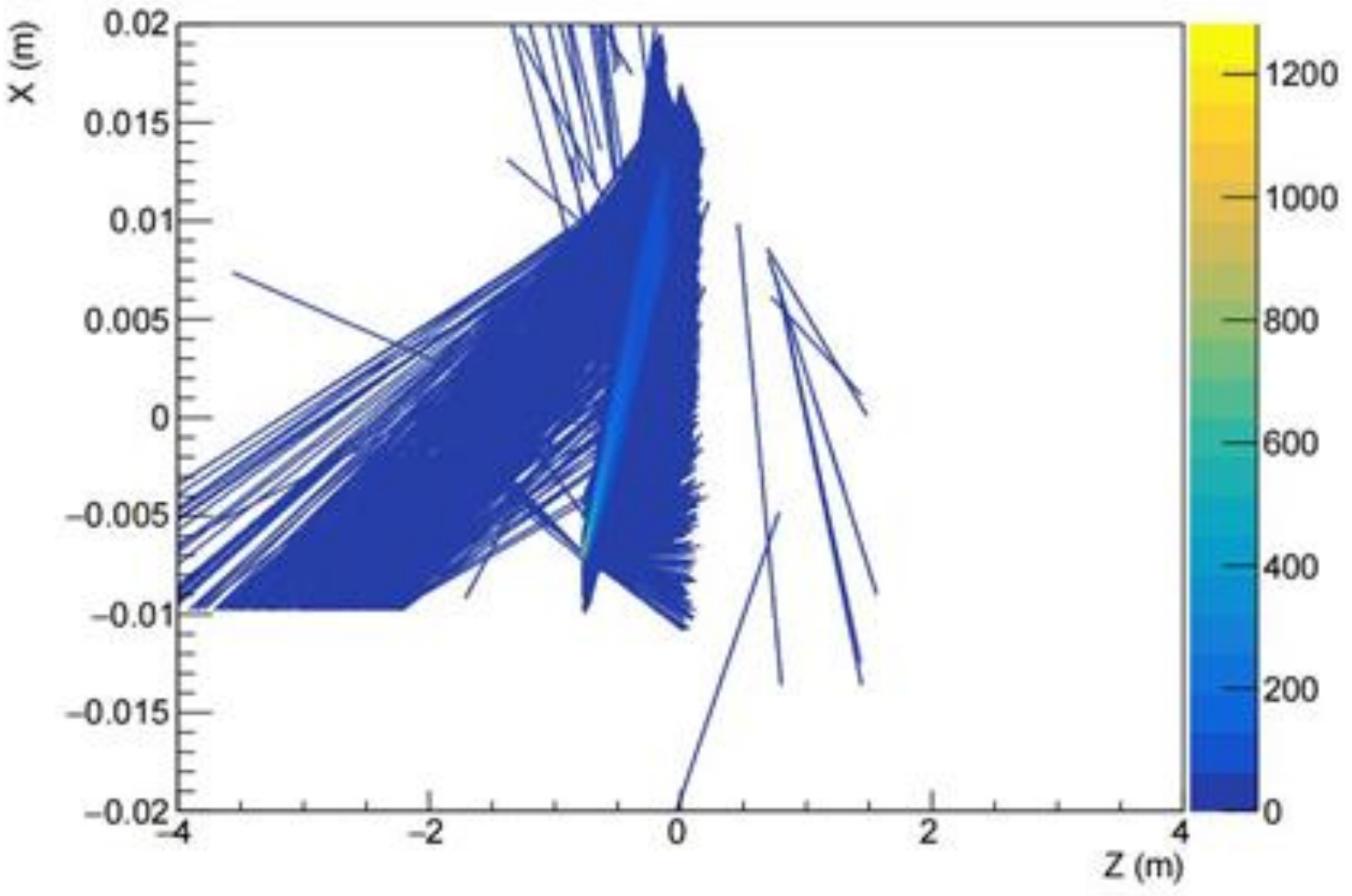

**Fig 4.2.6.24:** Distribution of the energy and generation position of the secondary SR photons hitting the beryllium pipe.

To mitigate this problem, high-Z material masks have been introduced to reduce the number of SR photons that could hit the beryllium beam pipe or the detector. Tape-shaped designs were used to reduce impedance and high-order mode heat load. The masks' size, material, position, and shape have been studied, and copper masks have been found to be effective. The primary mask, located at -1.9 m, the exit of the upstream QDa, has a height of 4 mm and a taper of 1:5, with additional masks at the entrance of every quadrupole. Additionally, the inner wall of the beryllium beam pipe will be coated with a thin layer of gold (~5 µm) to further suppress SR photons' penetration.

4.2.6.3.2.2 Pair Production

The pinch effect in beam-beam interactions can result in the emission of beamstrahlung photons. This can lead to the production of electron-positron pairs either coherently or incoherently through the interaction of real and/or virtual photons. While coherent pair production was found to be negligible, the focus was on incoherent pair production. It should be noted that most of the beamstrahlung photons, electrons, and positrons are produced in the forward direction and exit the interaction region without hitting the beam pipe. However, there is still a small fraction of particles with sufficient energy and a large polar angle that can hit the beam pipe.

To simulate the pair production process, GUINEA-PIG++ [7] was used. Each colliding bunch was divided into longitudinal slices, which were then cut into transverse cells. The beam particles were replaced with macro particles, and their charge was distributed onto the grid. The interaction of the two bunches was simulated by a grid passing through another grid of the opposite bunch. Electron pairs were generated and tracked with the deflection of the electromagnetic field of the beam until the two grids separated completely. The GUINEA-PIG++ code has been updated to incorporate the external magnetic field for circular colliders. The pair production distribution is shown in Figure 4.2.6.25.

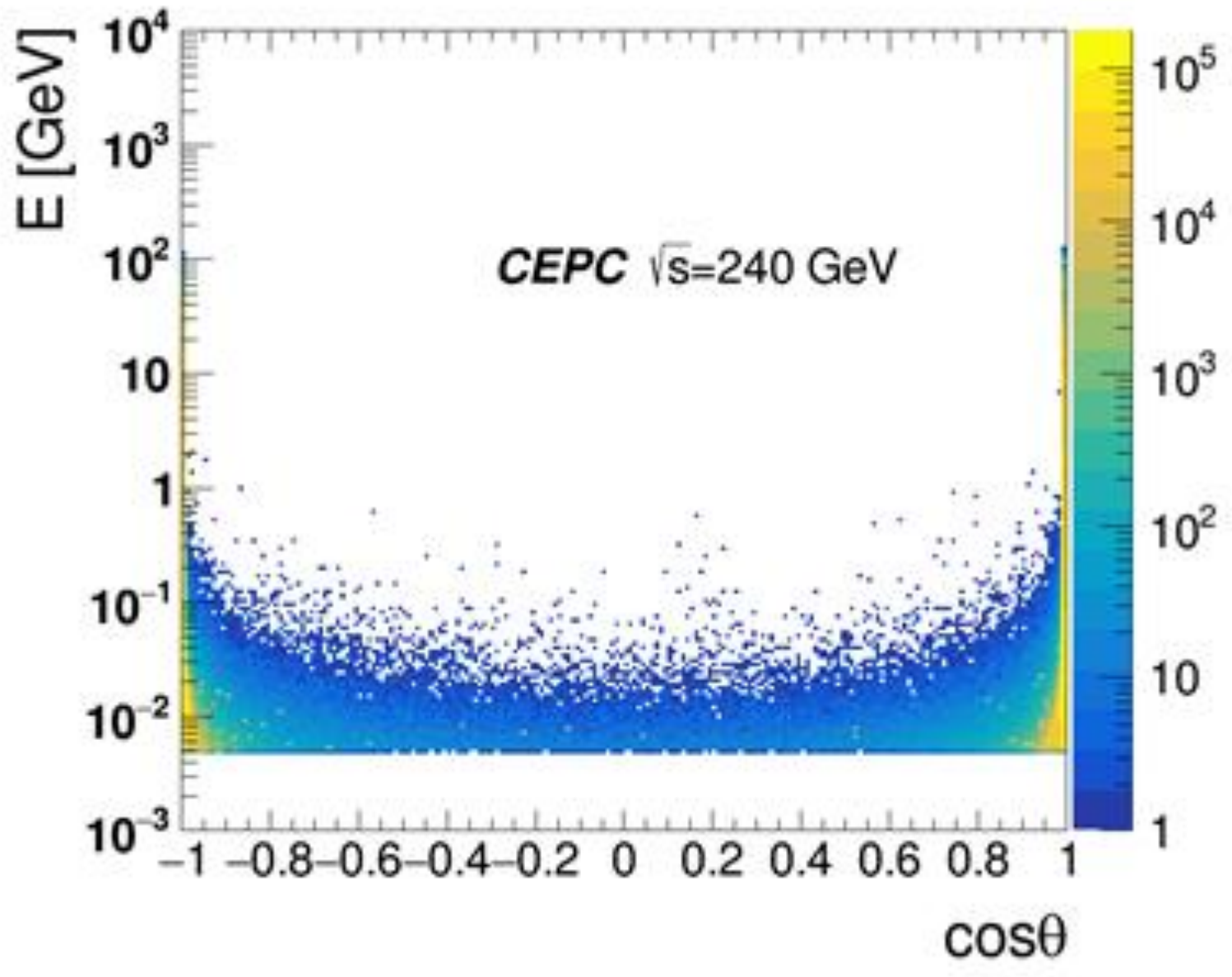

**Fig 4.2.6.25:** Distribution of pair production

#### *4.2.6.3.3 Beam Loss Background, Collimator and Shielding*

Beam loss mechanisms and the associated beam lifetimes are listed in Table 4.2.6.9.

**Table 4.2.6.9** Beam lifetime of CEPC

| Beam Loss Backgrounds | Beam Lifetime in Higgs Mode |
|---|---|
| Quantum effect | >1000 h |
| Touschek Scattering | 119 h |
| Beam-Gas Coulomb | >400 h |
| Beam-Gas Bremsstrahlung | 10 h |
| Beam Thermal Photon | 50.66 h |
| Radiative Bhabha | 40 min |
| Beamstrahlung | 40 min |

Touschek scattering and beam-gas Coulomb scattering can be ignored due to their long lifetimes. However, beamstrahlung [8], radiative Bhabha scattering [9], beam-thermal photon scattering [10], and beam gas bremsstrahlung, especially beamstrahlung and radiative Bhabha scattering, must be carefully analyzed and collimated due to their shorter lifetimes. To address this, we have implemented several collimators listed in Table 4.2.6.10, while adopting new parameters.

**Table 4.2.6.10:** Collimator design

| **Name** | **Position** | **Distance to IP (m)** | **Beta function (m)** | **Horizontal Dispersion (m)** | **Phase** | **BSC/2 (m)** | **Range of half width allowed (mm)** |
|---|---|---|---|---|---|---|---|
| APTX1 | D1I.785 | 40412 | 90 | 0.15 | 183.29 | 0.00732 | 1.44~7.32 |
| APTX2 | D1I.788 | 40467 | 90 | 0.15 | 183.54 | 0.00732 | 1.44~7,32 |
| APTY1 | D1I.791 | 40518 | 18.6 | 0.15 | 183.73 | 0.00324 | 0.03~3.24 |
| APTY2 | D1I.794 | 40572 | 18.6 | 0.15 | 183.98 | 0.00324 | 0.03~3.24 |
| APTX3 | D1O.10 | 1721 | 90 | 0.15 | 7.00 | 0.00732 | 1.44~7,32 |
| APTX4 | D1O.14 | 1776 | 90 | 0.15 | 7.25 | 0.00732 | 1.44~7.32 |
| APTY3 | D1O.5 | 1639 | 90 | 0.04 | 6.47 | 0.00406 | 0.065~4 |
| APTY4 | D1O.8 | 1694 | 90 | 0.07 | 6.72 | 0.00406 | 0.065~4 |
| APTX5 | DMBV01IR U0 | 56.3 | 152.4 | 0 | 444.8 | 0.0086 | 1.8~8.6 |

The collimators implemented per IP per ring are listed in Table 4.2.6.10. In simulations, all the horizontal collimator apertures are set to 8 mm (~17$\sigma_x$), while all the vertical collimator apertures are set to 6 mm (~125$\sigma_y$) in order to limit their contributions to impedance.

The collimators successfully shield the beamstrahlung backgrounds from the IR and nearly all multi-turn losses. Figures 4.2.6.26 and 4.2.6.27 show the loss rates and power distribution in the IR of CEPC, respectively, with $\sqrt{s}$ = 240 GeV. The upstream loss rate and deposited power are slightly smaller than the downstream. However, deposited power on both the upstream and downstream could harm the magnet components and coil of the SCQs. To address this issue, a tungsten alloy-based beam pipe is being considered for use in this region. In addition, a 1 cm thick tungsten shielding will be placed outside the cryostat chamber to further shield the detector.

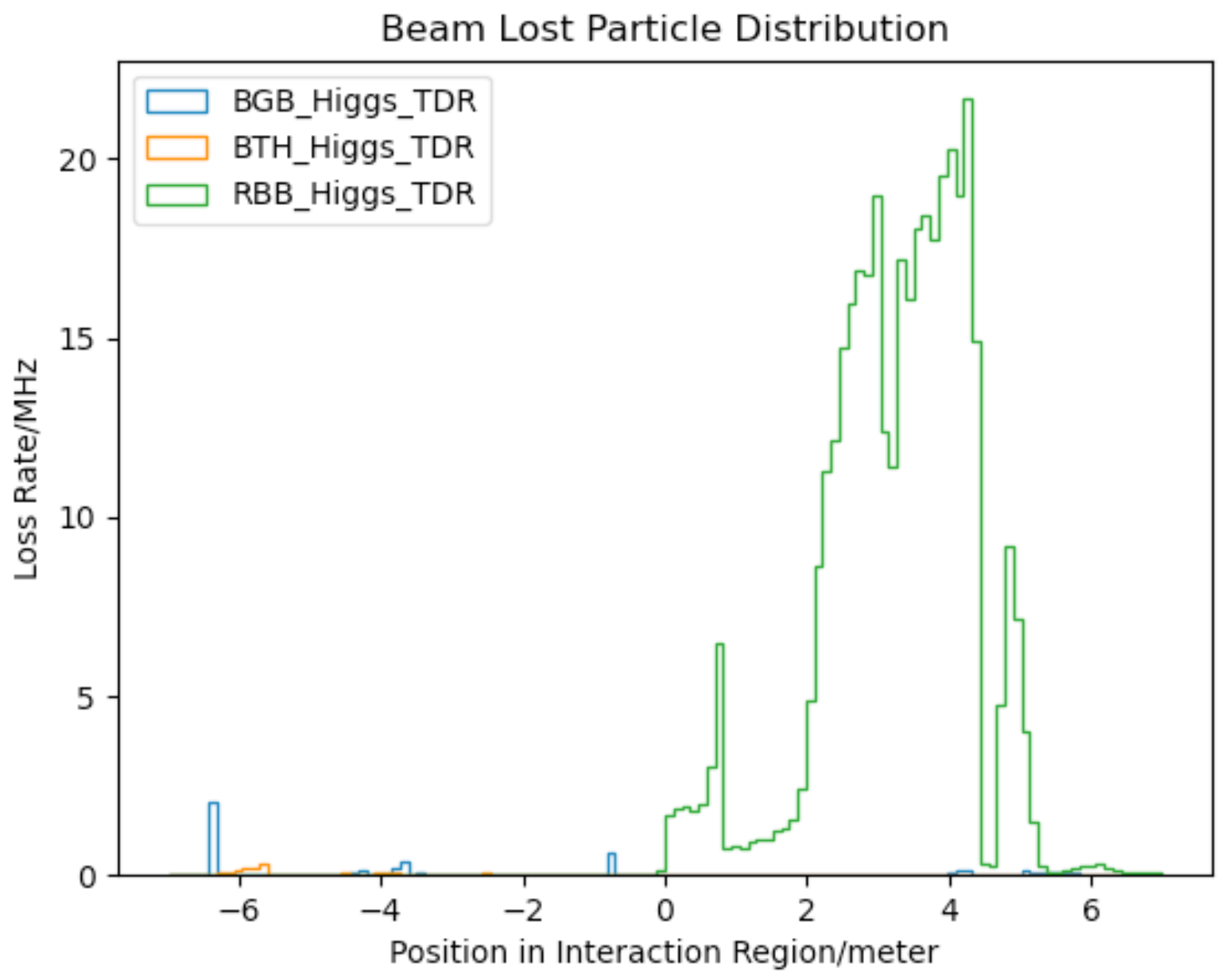

**Fig 4.2.6.26:** Loss rate in the IR, Higgs Mode

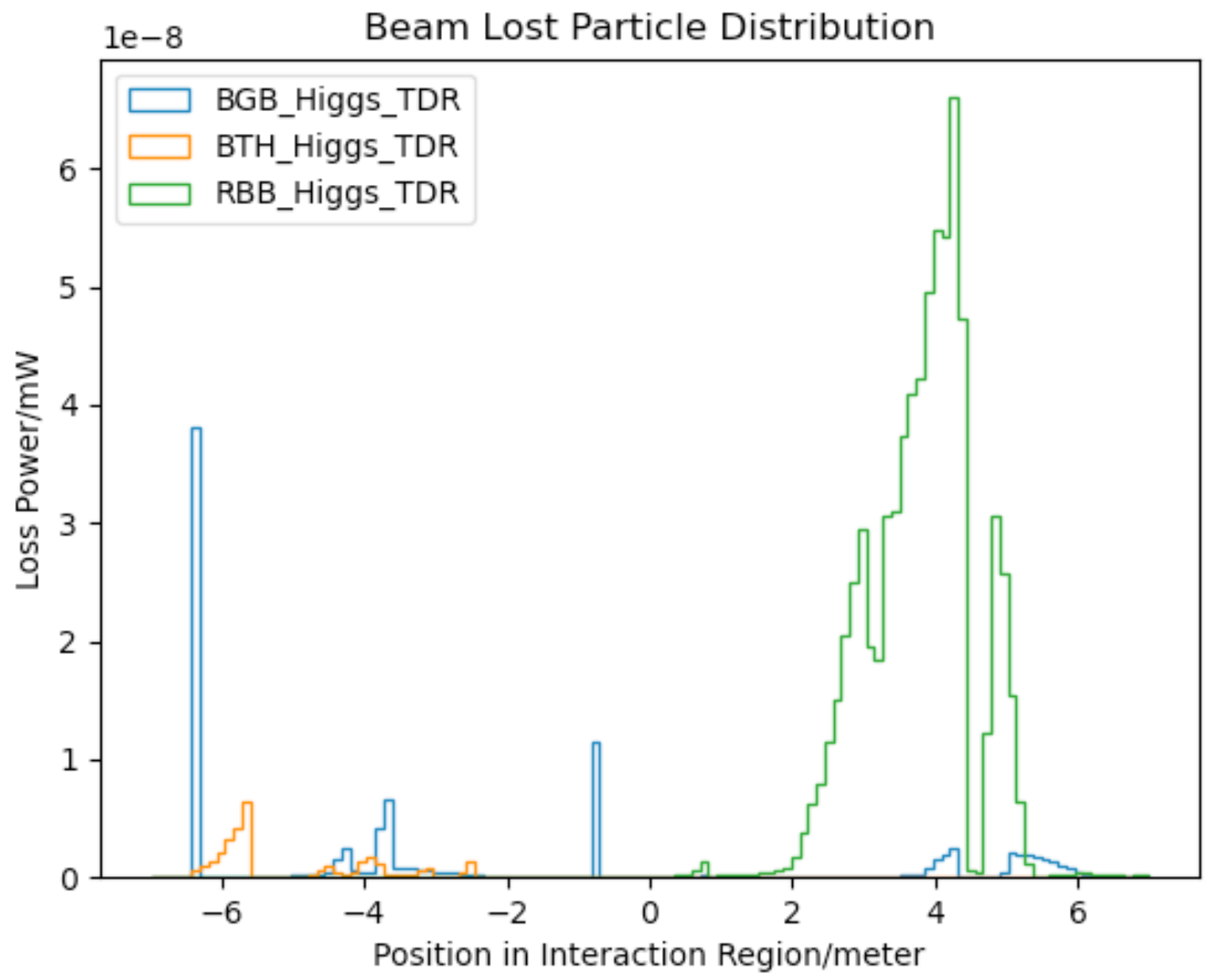

**Fig 4.2.6.27:** Loss power in the IR, Higgs Mode

#### *4.2.6.3.4 Injection Background*

When a charge is injected into a circulating beam bunch, it perturbs the injected bunch, leading to a higher background rate in the detector for a few milliseconds after the injection [11]. There are two injection modes: full injection to an empty ring and top-up injection. For the first mode, since the detector high voltage is off and detector measurement is off, background should not be considered. However, the potential beam-loss-induced background effects from continuous top-up injection in the IR need to be analyzed. Table 4.2.6.11 below shows the injection parameters for CEPC.

**Table 4.2.6.11:** CPEC injection parameters at different operation mode.

| **Parametrs** | $t\bar{t}$ | **Higgs** | **W** | **Z** |
|---|---|---|---|---|
| Bunch number | 37 | 240 | 1230 | 3840 |
| Bunch charge (nC) | 0.96 | 0.7 | 0.73 | 0.8 |
| Beam current (mA) | 0.11 | 0.51 | 2.69 | 9.2 |
| Beta function (x/y) | 200/55 | 200/55 | 200/55 | 200/55 |
| Emittance (nm) | 2.83 | 1.26 | 0.56 | 0.19 |
| Bunch length (mm) | 2.0 | 2.0 | 1.7 | 0.96 |

A simplified model is used to analyze the effects of continuous top-up injection on potential beam-loss-induced backgrounds in the IR, with a focus on radiative Bhabha scatte.

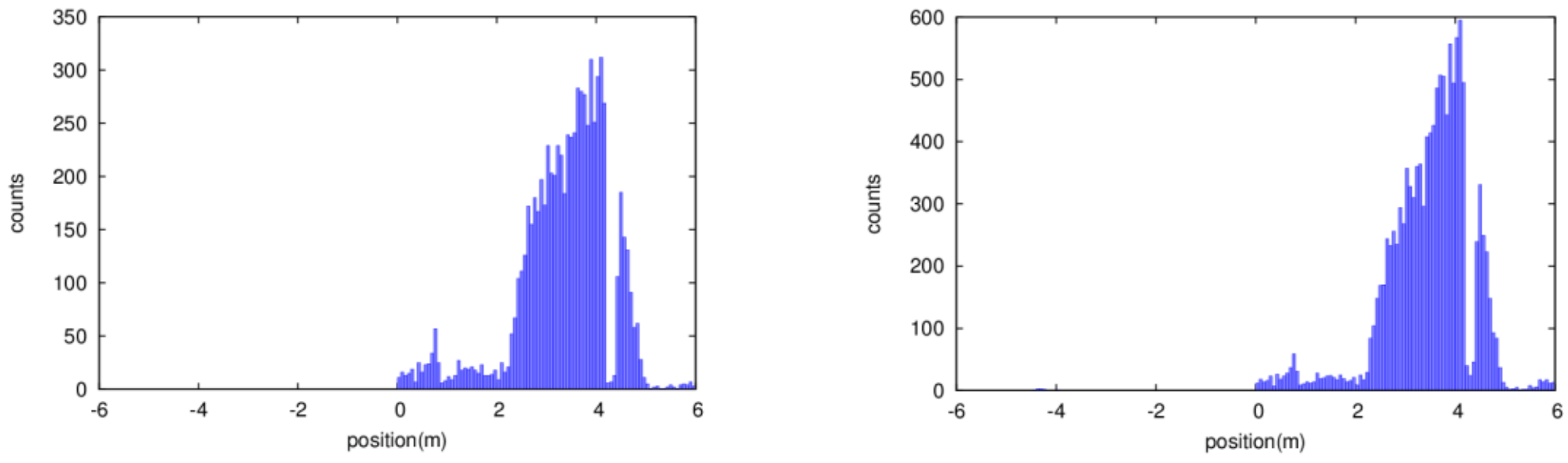

**Fig 4.2.6.28**. Beam loss background in ±6m around IP from circular beam (left) and injection beam(right)

Before the injection beam is damped by the radiation damping and/or transverse-feedback system, the presence of beam tails from kicker errors and imperfectly corrected X-Y coupling after the injection point should be considered.

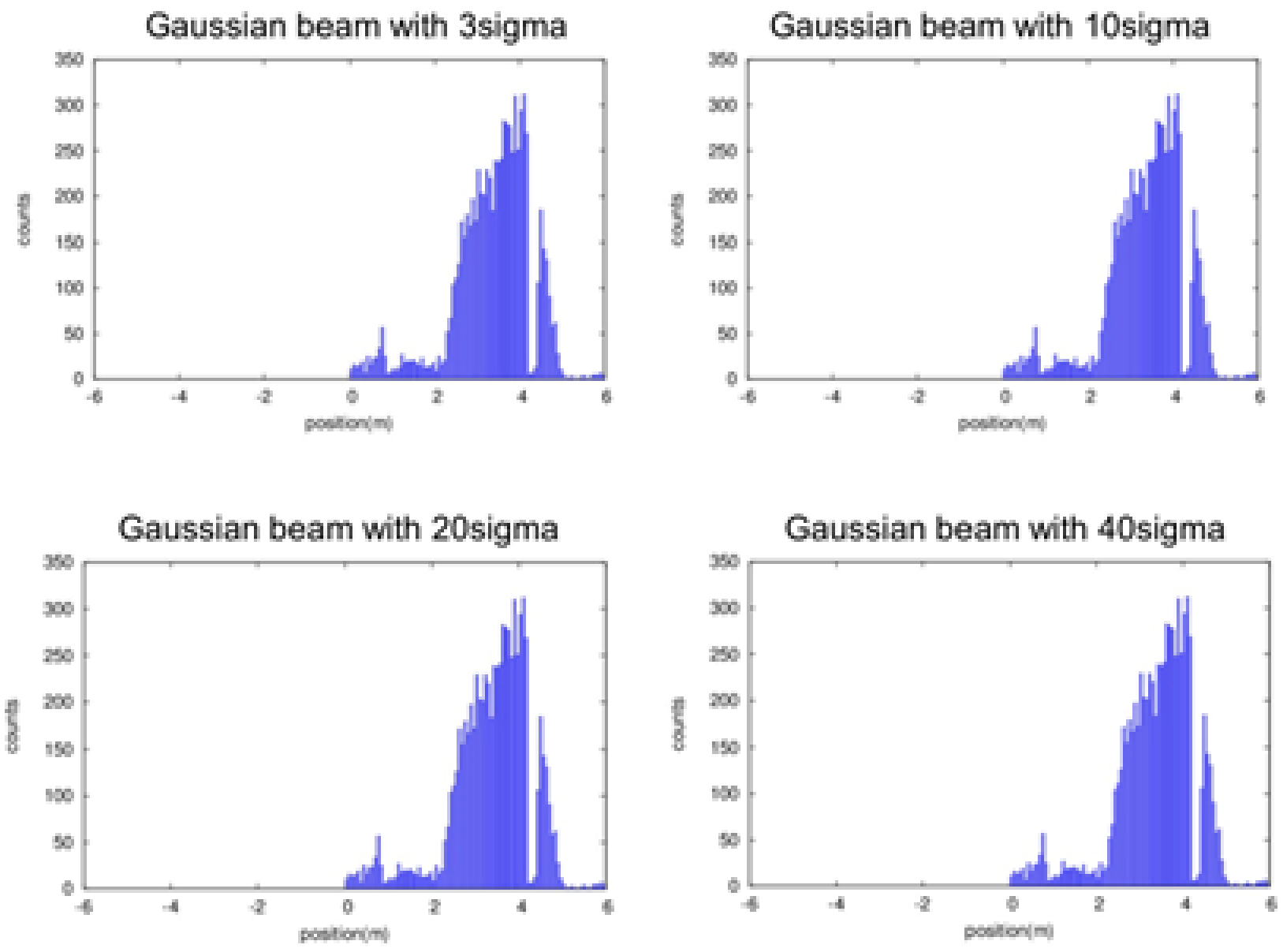

**Fig 4.2.6.29:** Beam loss background in ±6m around IP with the presence of beam tails

Tolerances such as too large emittances and imperfect beams from the booster must also be taken into account.

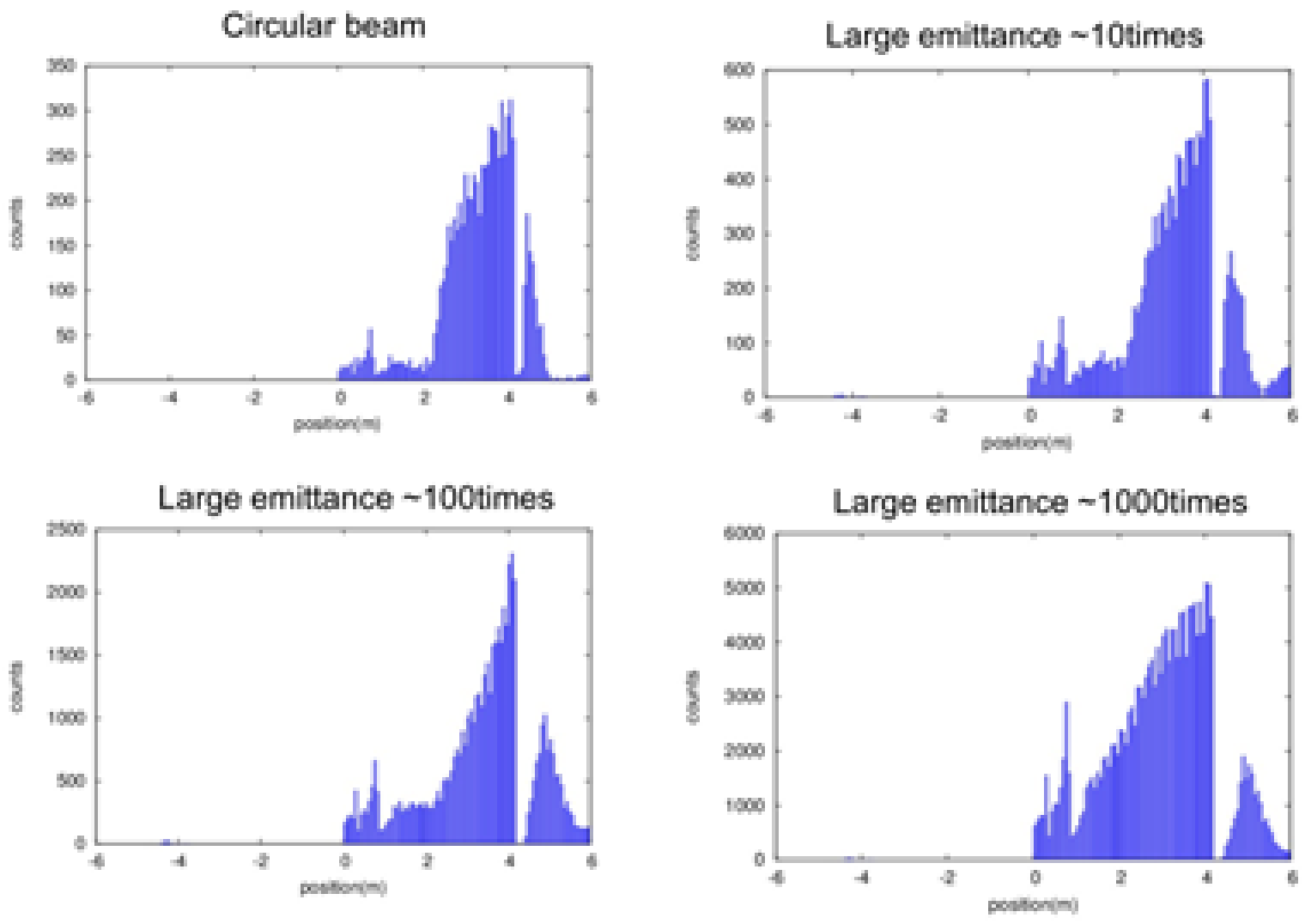

**Fig 4.2.6.30:** Beam loss background in ±6m around IP with large emittances to imperfect beams from the booster

The existing collimation system in the upstream of the IP can cope well with almost no beam loss background. However, the beam loss background in the downstream of the IP can significantly increase and may damage the outer layer or endcap detector due to radiation background. It may also damage the superconducting magnet coils and cause quench. Adding tungsten shielding may be a better choice, and the tight space in the IP region requires a tungsten-alloy beam pipe under design in the engineering stage.

The beam distribution in the booster and injected into the collider is typically a Gaussian distribution but may become non-Gaussian due to interaction effects such as beam-gas scattering. The non-Gaussian distribution is evaluated by introducing a uniform and a double-Gaussian beam distribution, and the beam halo occupancy in the whole beam distribution may be much larger than in the Gaussian distribution. The results shown in Figure 4.2.6.29 indicates that there is no significant increase in the beam loss background, and the existing collimation system can cope well.

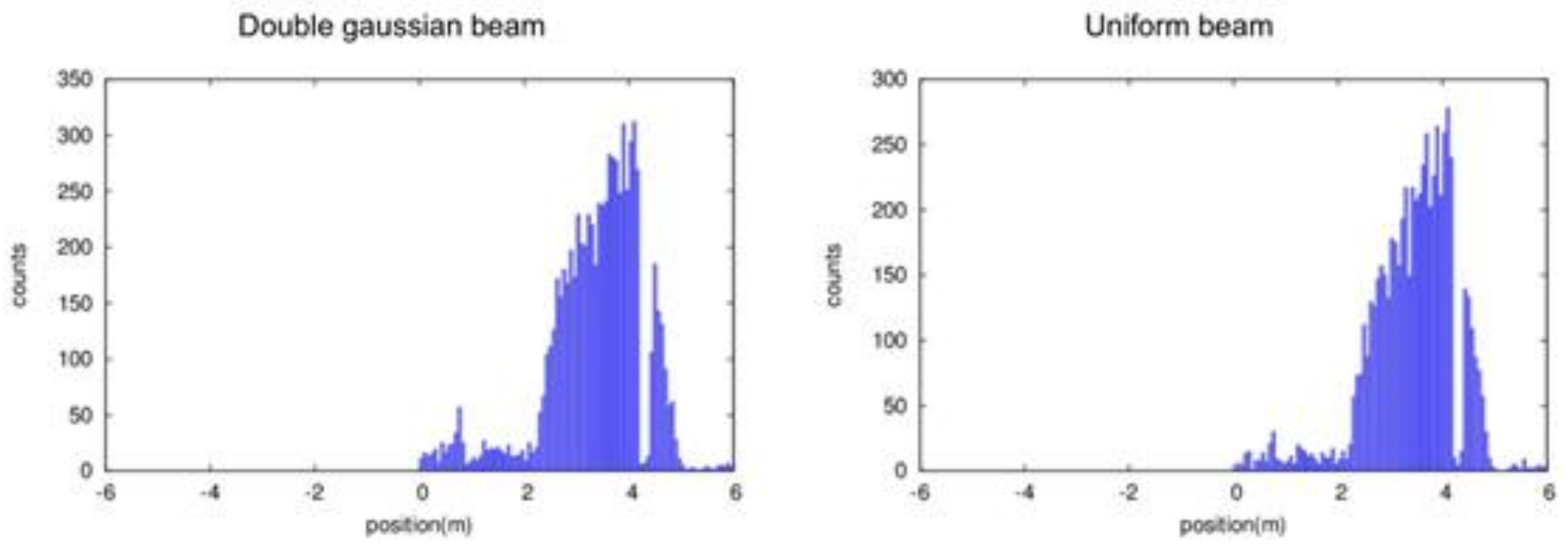

**Fig 4.2.6.31:** Beam loss background in ± 6m around IP with non-Gaussian injection beam

## 4.3 Collider Technical Systems

### 4.3.1 Superconducting RF System

#### 4.3.1.1 *Introduction*

CEPC is designed to operate in four different modes (Higgs, W, Z-pole, and $t\bar{t}$) with a wide range of beam parameters. The Collider beam energy ranges from 45.5 to 180 GeV with a beam current of 5.6 mA to 1.4 A and beam synchrotron radiation (SR) power of 30 to 50 MW. The Booster beam energy ranges from 30 GeV to the top-off injection energy of each mode, with the beam current ranging from 0.1 to 30 mA.

The superconducting RF (SRF) system for both the Collider and the Booster is optimized for the Higgs mode of 30 MW SR power per beam of the Collider, and upgradable to higher power and/or energy by adding cavities and/or RF power sources. Mode switching is required between the operation modes, especially Z, W, and Higgs.

There are two RF sections located at two long straight sections. The Collider 650 MHz cryomodules will be mounted on the tunnel floor, and the Booster 1.3 GHz cryomodules will be hung from the ceiling at a different beamline height and in between Collider cryomodule sections. The Collider cavities operate in CW mode, the Booster cavities operate in fast voltage ramp mode.

Considering the high RF voltage of $t\bar{t}$ and Higgs and the low RF voltage of W and Z, the Collider double-ring is designed to have shared cavities for $t\bar{t}$ and Higgs operation and separate cavities for W and Z operation. This common cavity scheme will reduce the total number of cavities and cryomodules as well as the cryogenics by half.

Due to the wide range of machine parameters in terms of RF voltage, beam current, and cavity impedance, etc., it is impossible to have a single common SRF system for the highest possible luminosity in each mode ($t\bar{t}$, Higgs, W, and Z) with up to 50 MW SR power per beam. A staged SRF complex with bypass lines for both Collider and Booster is inevitable. To enable flexibility and mode switching, the lower current cavities should be placed in the center of an RF section, and the higher current cavities on each side. The RF cavity layout is "Z–Higgs(W)–$t\bar{t}$–Higgs(W)–Z" in an RF section for both Collider and Booster (Figure 4.3.1.1). In this configuration, the higher current beam can bypass the lower current cavities without causing HOM power and beam instability issues, and the lower current beam can go through the cavities of the two rings for the Collider and use the RF voltage of the higher current cavities.

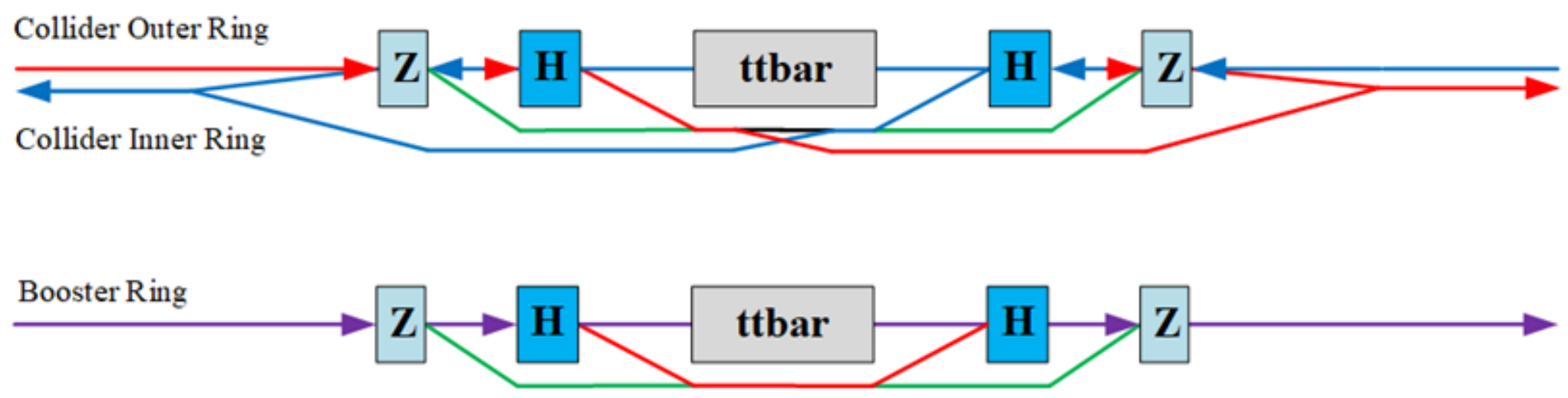

**Figure 4.3.1.1:** Cavity bypass scheme for CEPC mode switching.

The electron or positron beams will go through all the RF cavities for $t\bar{t}$ and Higgs operation. Less than half of the buckets will be filled to avoid collisions in the RF section. When operating at W and low-luminosity Z, part of the Higgs cavities will be used, and the electron or positron beams will go through only half of the cavities in an RF section. The W and Z bunches will be quasi-uniformly distributed in the two rings.

#### 4.3.1.2 *Collider RF Layout and Parameters*

The Collider Higgs mode will require 30 MW SR power per beam and will use a 650 MHz RF system with 192 2-cell cavities. Each of the 11 m-long Collider cryomodules will contain six 650 MHz 2-cell cavities. Each cavity will have two detachable coaxial HOM couplers mounted on the cavity beam pipe with a HOM power-handling capacity of 1 kW. Additionally, each cryomodule will have two beamline HOM absorbers at room temperature outside the vacuum vessel with a HOM power-handling capacity of 5 kW each.

To upgrade to the Higgs 50 MW SR power per beam, the input coupler power per cavity will be kept at a 300 kW level and additional 144 2-cell cavities will be added, resulting in a total of 336 2-cell cavities for 50 MW Higgs operation. During 30 MW and 50 MW modes in W running, only half of the Higgs cavities per ring will be utilized; Higgs 2-cell cavities are generally compatible with W-mode operation.

The HOM power limit per cavity and the fast-growing coupled-bunch instabilities (CBI) driven by both fundamental modes (FM) and higher-order modes (HOM) of the RF cavities will determine to a large extent the highest beam current and luminosity obtainable in the Z mode.

Using Higgs 2-cell cavities, Z mode can only operate at low luminosity and low power (up to about 10 MW SR power per beam). The low luminosity Z mode will use 48 cavities per ring and park the other 48 cavities in the ring. However, the beam will go through all the 96 cavities and interact with the cavity impedance. Therefore, the unused cavities will be kept at 2 K to extract the HOM power. They will be symmetrically detuned to cancel FM impedance and adjust the variable input coupler to high loaded Q with a narrow bandwidth to avoid FM CBI.

For high luminosity Z running, i.e., 30 and 50 MW SR per beam, 30 and 50 KEKB/BEPCII-type high current SRF cryomodules for each ring will be chosen as the baseline due to their excellent HOM damping. The high-current beam will bypass the Higgs/$t\bar{t}$ cavities. The cryomodule will have only one 1-cell cavity inside with HOM absorbers at room temperature. The cavity number is a balance between the input power and the stored energy per cavity for the sake of FM CBI and transient beam loading. Fewer cavities and cell numbers are preferred to have high stored energy and low impedance. The counter-phasing operation method could be used to further increase the cavity stored energy. A multi-1-cell-cavity cryomodule with deep HOM damping (LHC type) will be developed as an alternative.

During $t\bar{t}$ 30 MW mode, 240 5-cell 650 MHz cavities will be added in addition to the 336 Higgs 2-cell cavities. Each cryomodule will contain four 5-cell cavities. For $t\bar{t}$ 50 MW running, the klystron numbers will be doubled. It is possible to switch between $t\bar{t}$ and Higgs/W/Z modes as shown in Figure 4.3.1.1 with careful design to avoid SR light from hitting the cavities.

The main parameters of CEPC SRF system is listed in Table 4.3.1.1.

**Table 4.3.1.1:** CEPC Collider RF Parameters

| | $t\bar{t}$ 30/50 MW | | Higgs 30/50 MW | W 30/50 MW | Z 30/50 MW |
|---|---|---|---|---|---|
| | New cavities | Higgs cavities | | | |
| Beam energy [GeV] | 180 | | 120 | 80 | 45.5 |
| SR power / beam [MW] | 30 / 50 | | 30 / 50 | 30 / 50 | 30 / 50 |
| Luminosity / IP [$10^{34}$ $cm^{-2}s^{-1}$] | 0.5 / 0.8 | | 5 / 8.3 | 16 / 26.7 | 115 / 192 |
| SR loss / turn [GV] | 9.1 | | 1.8 | 0.357 | 0.037 |
| RF voltage [GV] | 10 (6.1 + 3.9) | | 2.2 | 0.7 | 0.12 / 0.1 |
| Syn. phase from crest [deg] | 24.5 | | 35.1 | 59.3 | 68.3 |
| Beam current / ring [mA] | 3.4 / 5.6 | | 16.7 / 27.8 | 84 / 140 | 801 / 1345 |
| Bunch charge [nC] | 32 | | 21 | 21.6 | 22.4 / 34.2 |
| Bunch number / beam | 35 / 58 | | 268 / 446 | 1297 / 2162 | 11934 / 13104 |
| Bunch length [mm] | 2.9 | | 4.1 | 4.9 | 8.7 / 10.6 |
| Syn. oscillation period [ms] | 4.4 | | 6.6 | 5.5 | 10.5 |
| Longitudinal damping time [ms] | 6.6 | | 22.2 | 74.7 | 410.2 |
| 650 MHz cavity number | 192 | 336 | 192 / 336 | 96 / 168 / ring | 30 / 50 / ring |
| Cell number / cavity | 5 | 2 | 2 | 2 | 1 |
| Cavity effective length [m] | 1.15 | 0.46 | 0.46 | 0.46 | 0.23 |
| R/Q [Ω] | 532.5 | 213 | 213 | 213 | 106.5 |
| Cavity operation gradient [MV/m] | 27.6 | 25.2 | 24.9 / 14.2 | 15.9 / 9.1 | 17.4 / 8.7 |
| $Q_0$ @ 2 K at operating gradient | 3E10 | 3E10 | 3E10 | 3E10 | 2E10 |
| HOM power / cavity [kW] | 0.4 / 0.66 | 0.16 / 0.26 | 0.4 / 0.67 | 0.93 / 1.54 | 2.9 / 6.2 |
| Input power / cavity [kW] | 188 / 315 | 71 / 118 | 313 / 298 | 313 / 298 | 1000 |
| Optimal $Q_L$ | 1E7 / 6E6 | 9E6 / 5.4E6 | 2.0E6 / 6.8E5 | 8E5 / 2.7E5 | 1.5E5 / 3.8E4 |
| Cavity bandwidth [kHz] | 0.06 / 0.11 | 0.07 / 0.12 | 0.3 / 1.0 | 0.8 / 2.4 | 4.3 / 17.3 |
| Optimal detuning [kHz] | 0.01 / 0.02 | 0.02 / 0.03 | 0.1 / 0.2 | 0.7 / 2 | 6.7 / 21.7 |
| Cavity stored energy [J] | 464 | 155 | 151 / 49 | 61 / 20 | 37 / 9 |
| Cavity time constant [μs] | 4928 / 2946 | 4367 / 2635 | 966 / 331 | 391 / 134 | 74 / 18 |
| Cavity number / klystron | 4 / 2 | 2 | 2 | 2 | 1 |
| Klystron max output power [kW] | 800 | 800 | 800 | 800 | 1200 |
| Klystron number | 48 / 96 | 168 | 96 / 168 | 96 / 168 | 60 / 100 |
| Cavity number / cryomodule | 4 | 6 | 6 | 6 | 1 |
| Cryomodule number | 48 | 56 | 32 / 56 | 32 / 56 | 60 / 100 |
| Total cavity wall loss @ 2 K [kW] | 12.1 | 7.1 | 3.9 / 2.3 | 1.6 / 0.9 | 0.45 / 0.2 |
| RF section length [m] (Bypass lines not included) | 768 | 896 | 512 / 896 | 256 / 448 / ring | 160 / 256 / ring |

### 4.3.1.3 ***Beam-Cavity Interaction***

#### *4.3.1.3.1 Transient Beam Loading*

Transient beam loading is a major concern in large rings with uneven fill patterns. Although uneven fills with beam gaps are necessary for beam abort and ion-clearing, they can result in a bunch phase shift, smaller energy acceptance, and possible lifetime and luminosity degradation. In CEPC Z mode, even a 1% beam gap to mitigate ion-trapping and fast beam-ion instability (FBII) can create a significant bunch phase shift. To minimize these effects, the fill pattern should consist of many small gaps and short bunch trains, rather than one long gap.

In the on-axis injection process of the Higgs mode, transient beam loading can cause a significant phase shift between the colliding bunches in the two beams, negatively impacting the luminosity of the machine. However, by carefully choosing which bunches to extract, it is possible to mitigate the phase mismatching and recover a major portion of the integrated luminosity. With the help of a strong direct feedback loop, it is possible to further reduce the mismatching without putting too much stress on the RF power generator. The simulation results from the time-domain tracking code are presented in Figure 4.3.1.2.

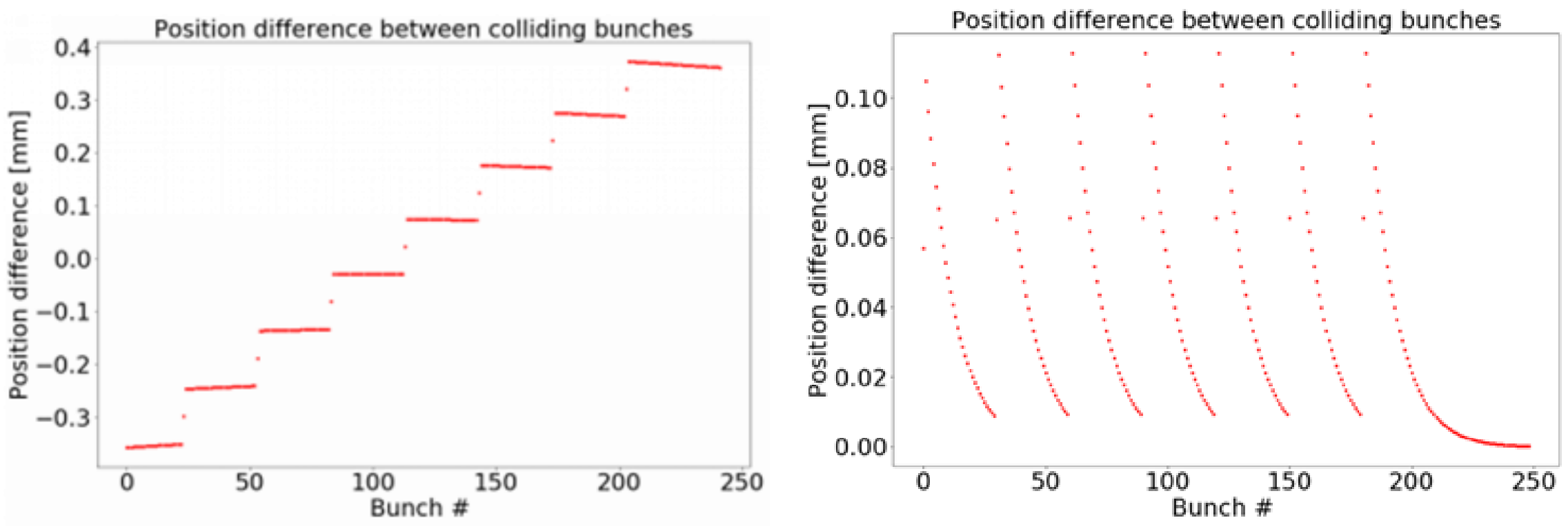

**Figure 4.3.1.2:** Position differences between the colliding bunches at IP. Left: Equal spacing extraction from one beam, no feedback. Right: Equal spacing extraction with direct feedback (gain = 100).

#### *4.3.1.3.2 Cavity Fundamental-Mode Instability*

The frequency detuning of the cavities in CEPC's W and Z modes is significant compared to the revolution frequency in the Collider ring due to the low cavity voltage and relatively high beam loading. This, combined with a low matched quality factor, $Q_L$, can result in more than 10 unstable coupled-bunch modes in some operation modes [1-3]. To reduce the growth rate of these dangerous modes to a level (100 ms) at which the longitudinal bunch-by-bunch feedback system to work, a direct feedback loop and one-turn-delay feedback with a comb filter are required to reduce the effective impedance seen by the beam. A loop delay of around 300 ns is assumed, which is achievable.

For W mode, a loop with a gain of 10 is sufficient to reduce the growth rates from over 100 Hz to nearly 10 Hz, which is below the radiation damping rate. For the low-luminosity Z mode, with symmetrically parked cavities and direct feedback loop gain of 11, the growth rates can be reduced to a level that can be damped by the radiation and bunch-by-bunch feedback system together. For high-luminosity Z mode, dedicated 1-cell cavities are used to reduce the total R/Q. Figure 4.3.1.3 shows the impedance spectrum of the 30 MW and 50 MW Z mode. To bring down the growth rates of the coupled bunch modes to 13 Hz, direct feedback with gains of 11 and 9 for the 30 MW and 50 MW Z mode, respectively, must be applied.

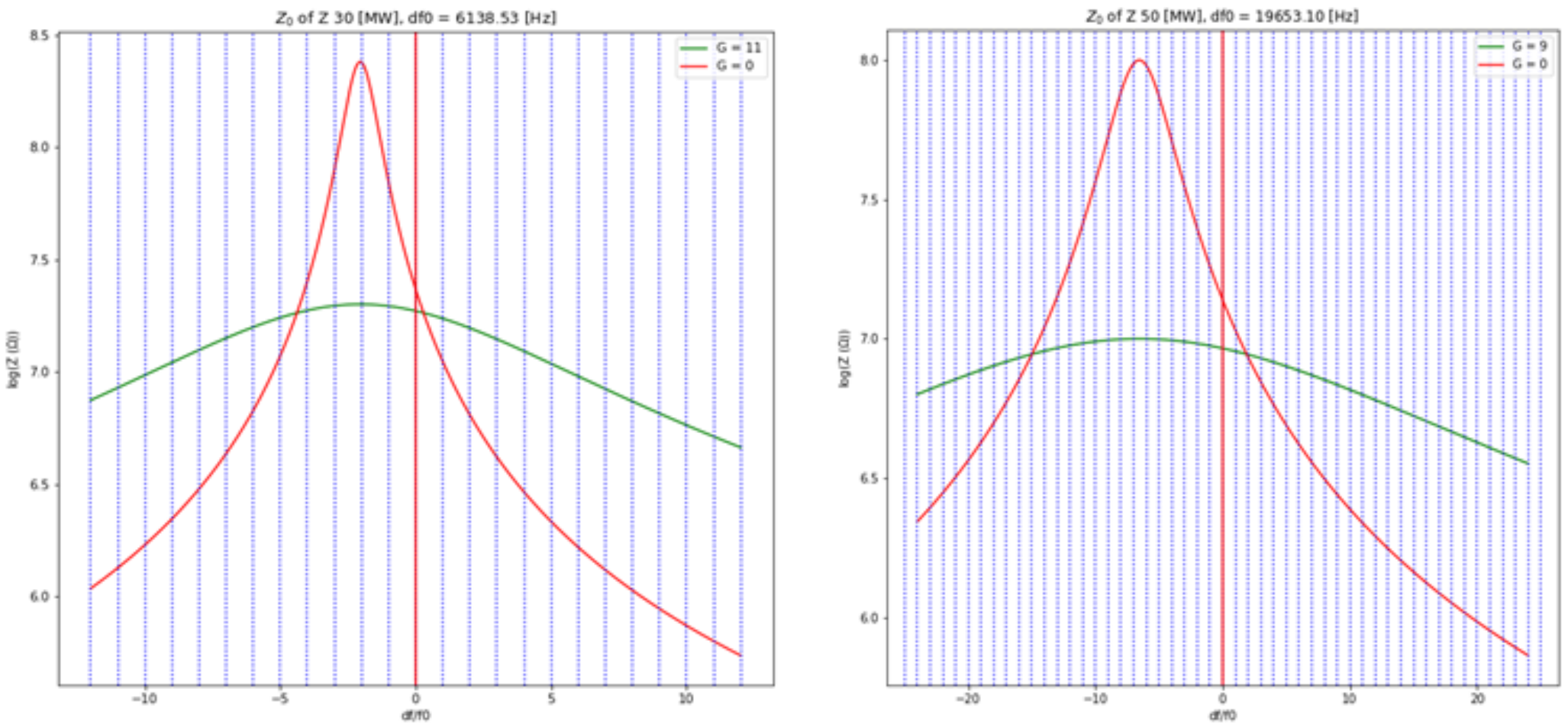

**Figure 4.3.1.3:** Impedance spectrum of 30 MW Z mode and 50 MW Z mode.

**Table 4.3.1.2:** Growth rates (Hz) of the first 15 FM CBI modes in 30/50 MW Z mode

| Mode Index | 30 MW, G=0 | 30 MW, G=11 | 50 MW, G=0 | 50 MW, G=9 |
|---|---|---|---|---|
| -1 | 128.30 | 4.82 | 30.77 | 1.38 |
| -2 | 533.36 | 9.02 | 68.11 | 2.75 |
| -3 | 184.36 | 11.92 | 118.38 | 4.06 |
| -4 | 53.94 | 13.38 | 191.05 | 5.27 |
| -5 | 22.94 | 13.59 | 291.00 | 6.38 |
| -6 | 12.00 | 12.93 | 388.09 | 7.36 |
| -7 | 7.11 | 11.75 | 400.55 | 8.21 |
| -8 | 4.58 | 10.37 | 316.83 | 8.92 |
| -9 | 3.13 | 8.99 | 218.89 | 9.50 |
| -10 | 2.24 | 7.70 | 147.65 | 9.93 |
| -11 | 1.66 | 6.56 | 101.81 | 10.23 |
| -12 | 1.26 | 5.58 | 72.58 | 10.40 |
| -13 | 0.99 | 4.75 | 53.45 | 10.46 |
| -14 | 0.78 | 4.06 | 40.52 | 10.41 |
| -15 | 0.63 | 3.49 | 31.48 | 10.27 |

#### *4.3.1.3.3 Cavity HOM CBI*

Table 4.3.1.3 provides the threshold of the external quality factor of the high R/Q higher-order modes (HOMs). With the damping performance of the HOM coupler discussed in section 4.3.1.6, most of the modes of $t\bar{t}$, Higgs, and W are stable, except for the TE111 mode of W. To improve the HOM coupler performance, three times deeper damping is required. However, the HOM frequency spread of cavities can help to relax the requirements.

For the 10 MW Z mode, a 100 ms longitudinal and 1 ms transverse bunch-by-bunch feedback system is necessary and sufficient to mitigate the instability. For the 50 MW Z mode, the external quality factors (Qe) of KEKB/BEPC-II type cavity modes can be reduced to the level of 100 to meet the damping requirement.

**Table 4.3.1.3:** Damping requirements of prominent HOMs of the Collider cavity

| Modes | $t\bar{t}$ 50 MW | | Higgs 50 MW | W 50 MW | Z 10 MW | Z 50 MW |
|---|---|---|---|---|---|---|
| | New cavities | Higgs cavities | | | | |
| | 192 5-cell | 336 2-cell | 336 2-cel | 168 2-cell | 96 2-cell | 50 1-cell |
| TM011 | 5.4E+06 | 4.1E+06 | 1.1E+05 | 1.0E+04 | 5.6E+02 | 4.8E+02 |
| TM020 | 7.0E+06 | 1.7E+08 | 4.6E+06 | 4.3E+05 | 2.4E+04 | 1.4E+04 |
| TE111 | 3.4E+05 | 5.8E+05 | 2.3E+04 | 3.6E+03 | 3.4E+02 | 2.3E+02 |
| TM110 | 4.2E+05 | 3.9E+05 | 1.5E+04 | 2.4E+03 | 2.3E+02 | 1.3E+02 |

### 4.3.1.4 *Cavity*

The 650 MHz 2-cell cavity is constructed using bulk niobium and is designed to operate at 2 K. The cavity has been designed to meet the following specifications: a vertical acceptance test with $Q_0 > 5\times10^{10}$ at 30 MV/m, a horizontal acceptance test with $Q_0 > 4\times10^{10}$ at 28 MV/m, and a normal operation gradient with $Q_0 > 3\times10^{10}$ at 25 MV/m for long-term operation.

Table 4.3.1.4 provides the key RF parameters for the main cavity [4]. The mechanical structure of the cavity has been optimized to reduce pressure sensitivity (*df/dp*) and mechanical stress, with the helium vessel being carefully designed to minimize these effects. The length of the cavity beam pipes, HOM coupler ports, and input coupler port should be sufficient to ensure minimal power dissipation on the gaskets and flange surfaces but should not be so long that they exceed the Nb critical temperature. To avoid additional dissipation at different joints, special gapless gaskets will be utilized. In future designs, cooling of cavity ports by an extended helium vessel should be considered, especially for the power coupler.

**Table 4.3.1.4:** Main parameters of the CEPC 650 MHz 2-cell cavity

| Parameter | Unit | Value |
|---|---|---|
| Cavity effective length | m | 0.46 |
| Cavity iris diameter | mm | 150 |
| Beam tube diameter | mm | 160 |
| Cell-to-cell coupling | - | 2.7 % |
| R/Q | Ω | 211 |
| Geometry factor | Ω | 279 |
| $E_{peak}/E_{acc}$ | - | 2.35 |
| $B_{peak}/E_{acc}$ | mT/(MV/m) | 4.2 |

Three prototypes of the 650 MHz 2-cell cavities were manufactured and subsequently treated with BCP since the EP tool was unavailable at the time. The cavity has been designed with two detachable double-notch higher-order-mode (HOM) couplers, as shown in Figure 4.3.1.4. The results of the vertical tests are presented in Figure 4.3.1.5 [5, 6]. Two of the cavities have already been installed in a test cryomodule.

EP and mid-T treatments were then carried out on several 650 MHz single-cell cavities to further improve their gradient and Q [7, 8]. The results of the vertical tests are shown in Figure 4.3.1.6. EP-treated cavities can meet the operational specifications and can attain very high gradients of up to 42 MV/m. Mid-T-treated cavities with high Q can satisfy the vertical test acceptance criteria. EP and mid-T treatments will be applied to 2-cell cavities, and the cavities will be installed in the test cryomodule to evaluate the module's performance.

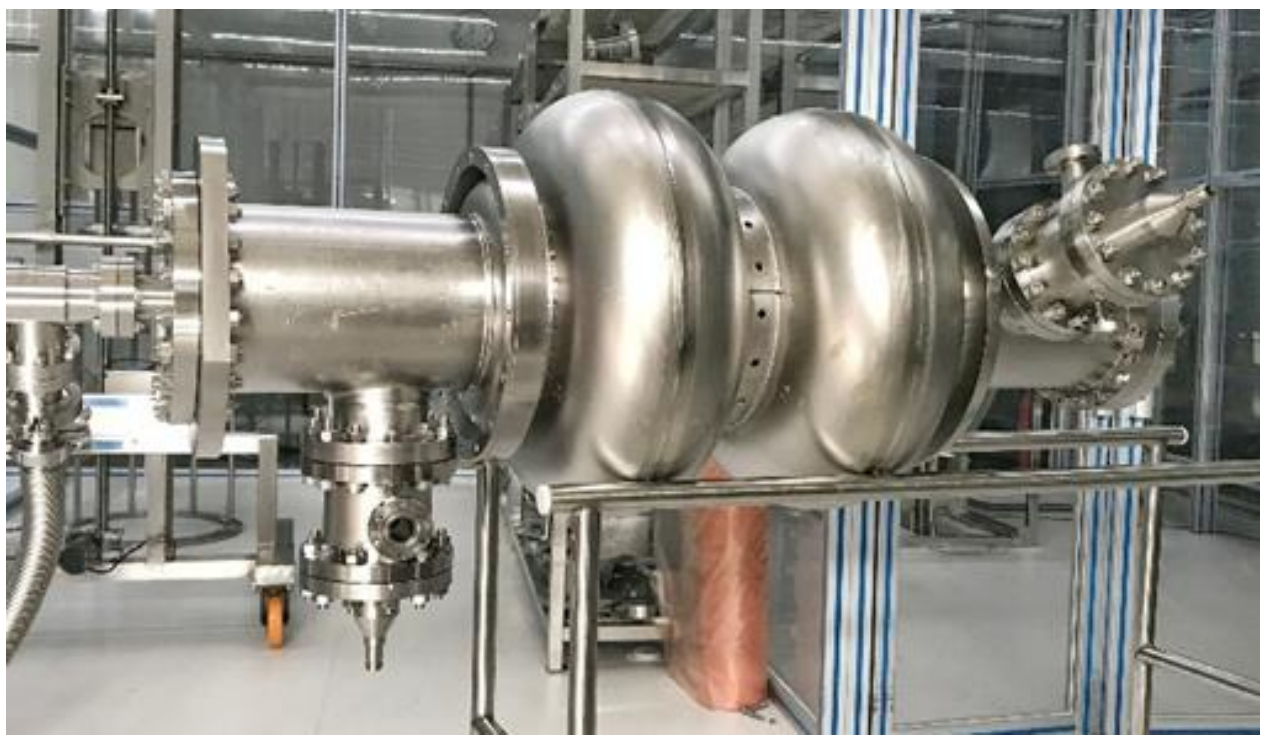

**Figure 4.3.1.4:** The 650 MHz 2-cell cavity with detachable HOM couplers

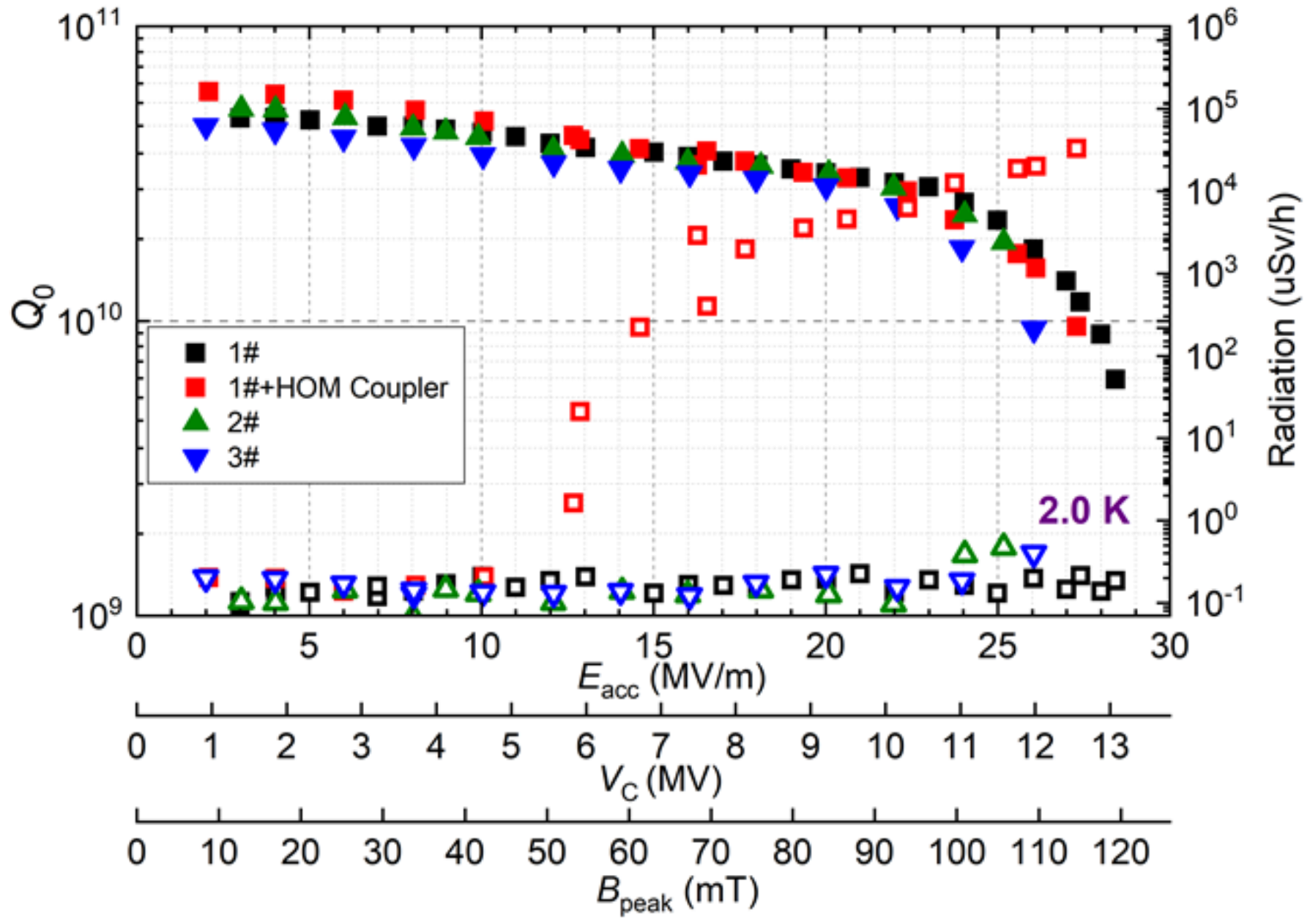

**Figure 4.3.1.5:** Vertical test results of BCP 650 MHz 2-cell cavities

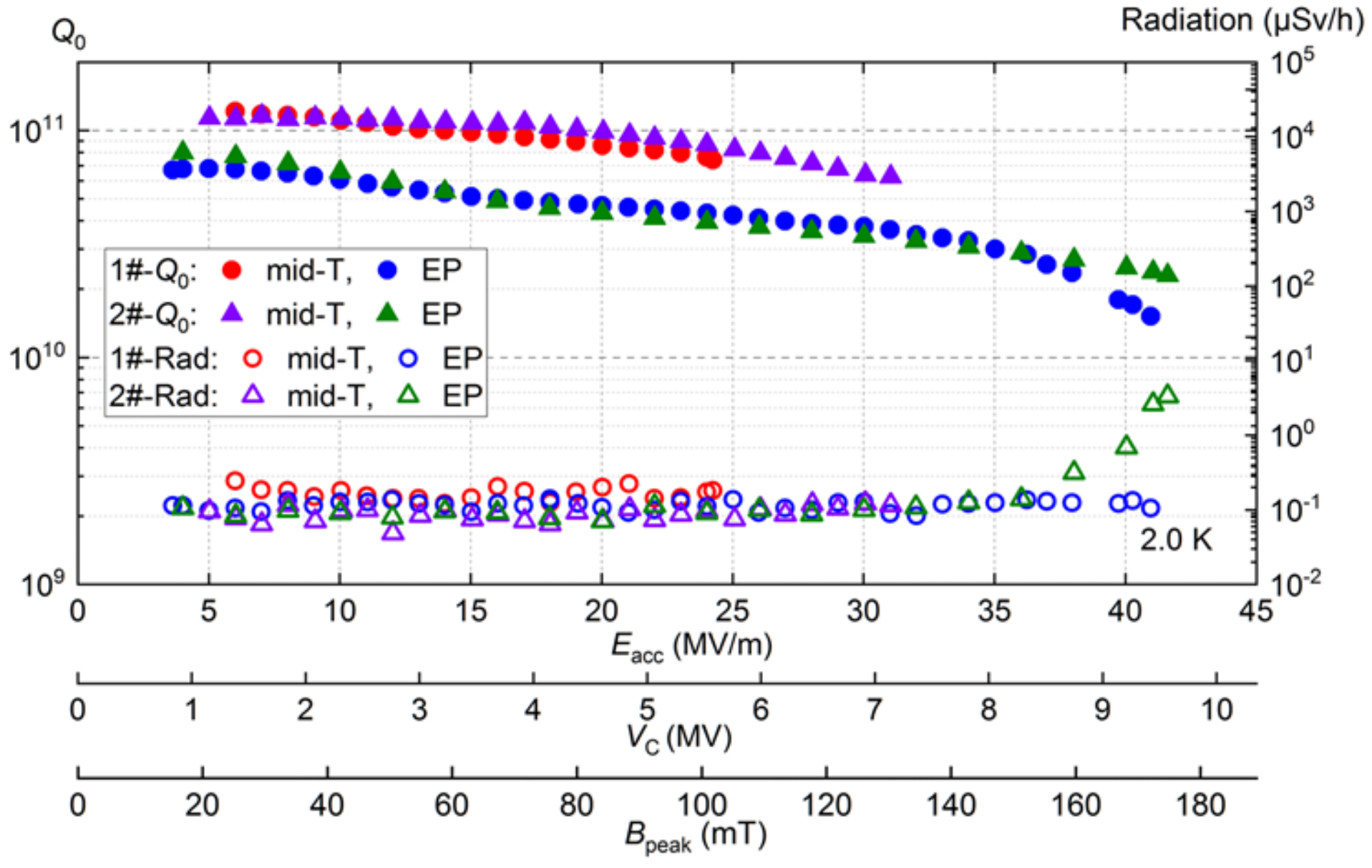

**Figure 4.3.1.6:** Vertical test results of EP and Mid-T 650 MHz single-cell cavities

### 4.3.1.5 *Power Coupler*

To address the issue of RF mismatch at different cavity voltages and beam currents for various operation modes, the power coupler needs to have a wide range of variable coupling to avoid excess power. Additionally, small number of cavities for lower cost and impedance, as well as shared cavities that double the beam current, require the power coupler to have high power capacity while maintaining low heat load. Therefore, the power coupler is a crucial component for the stable operation of the CEPC Collider RF system and its power consumption. The 1 MW input power per cavity for the high-luminosity Z mode presents a significant challenge for both the power coupler and the klystron. As an alternative solution, the power could be provided by two 600 kW klystrons through two power couplers.

The power coupler for Higgs operation faces several challenges, including variable coupling with a wide range, high power-handling capability (CW 300 kW), single window coupler and cavity clean assembly, very small heat load, simple structure for low cost and high yield, and high reliability. The inner conductor of the variable coupler is water-cooled and has bellows. The outer conductor is He-gas-cooled.

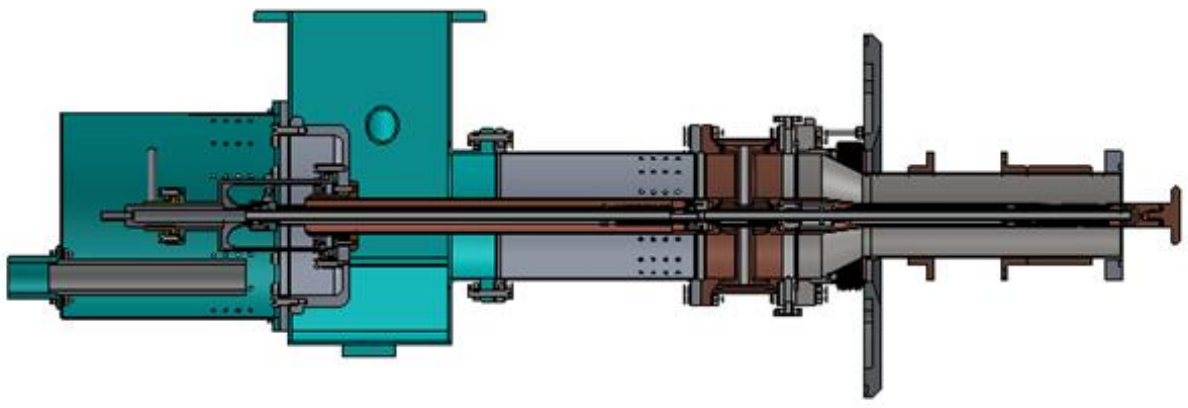

**Figure 4.3.1.7:** 650 MHz variable coupler design

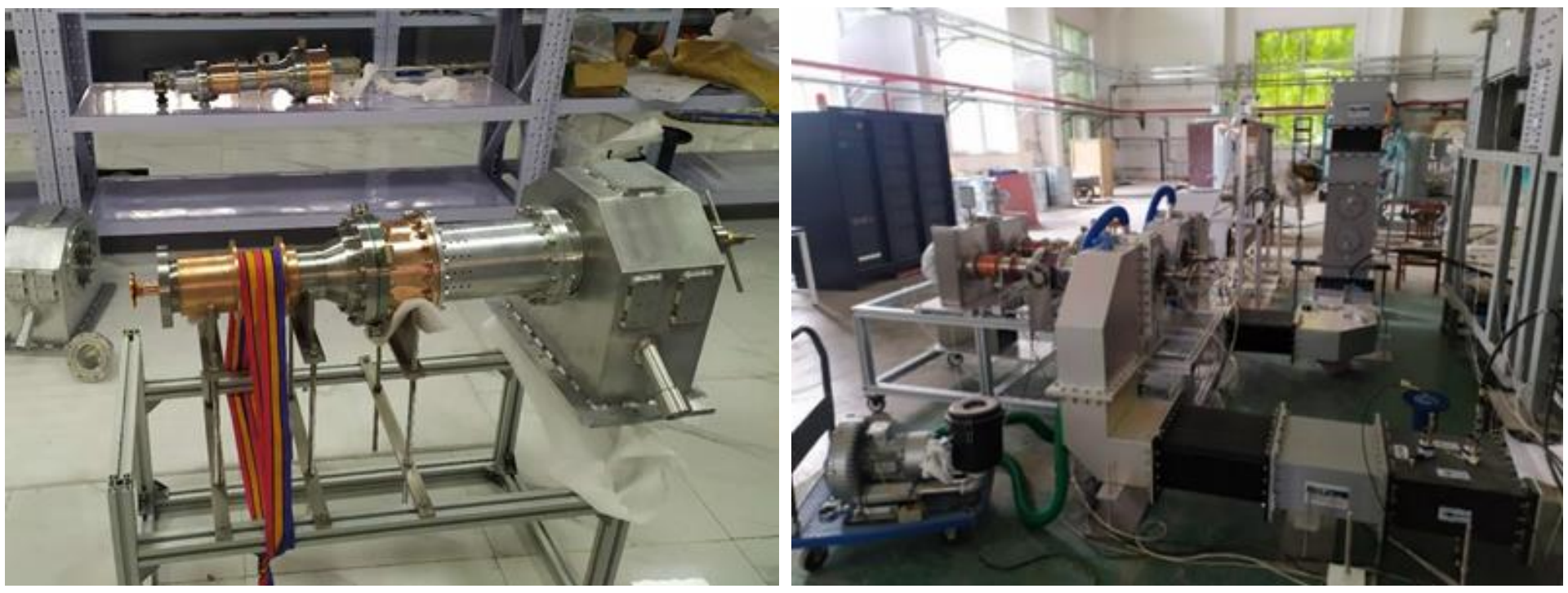

**Figure 4.3.1.8:** 650 MHz variable coupler and the high-power conditioning test

Two prototypes of 650 MHz variable couplers have been manufactured and tested with continuous wave (CW) travelling wave power of 150 kW (which is the limit for the solid-state amplifiers) and with a standing wave power of 100 kW, which corresponds to a travelling wave power of 400 kW at the window.

The power coupler is designed to operate with 300 kW of CW travelling wave power for the Higgs mode and has a heat load limit of 0.6 W/3.8 W/15 W (dynamic) and 0.05 W/0.5 W/1 W (static) at temperatures of 2 K/5 K/80 K.

#### 4.3.1.6 *HOM Damper*

To prevent cryogenic loss and beam instabilities, higher-order-modes (HOMs) that are excited by the intense beam bunches must be damped. The 650 MHz 2-cell cavities have beam pipes with cut-off frequencies of 1.471 GHz (TM01) and 1.126 GHz (TE11). The HOM coupler, which is mounted on the beam pipe, couples all the HOM power below the cut-off frequency. The propagating modes are then absorbed by two HOM absorbers at room temperature outside the cryomodule. However, some modes far above the cut-off frequency may become trapped among cavities in the cryomodule due to the large frequency spread, which is another concern related to the HOMs..

To damp different polarized HOMs, two HOM couplers per cavity with a 110-degree angle between them are used [9, 10]. The HOM coupler is designed to transfer up to 1 kW HOM power (expected in the worst case) and operates within the range of 780 MHz to 1471 MHz. To prevent the leakage of fundamental power from the HOM coupler, the external quality factor ($Q_e$) for the fundamental mode is designed to be larger than $10^{12}$. With a good control of fabrication and assembly tolerance, the double-notch filter does not need tuning because of its wide bandwidth (~ 100 MHz). To extract the HOM power, a specially designed rigid coaxial line will be used.

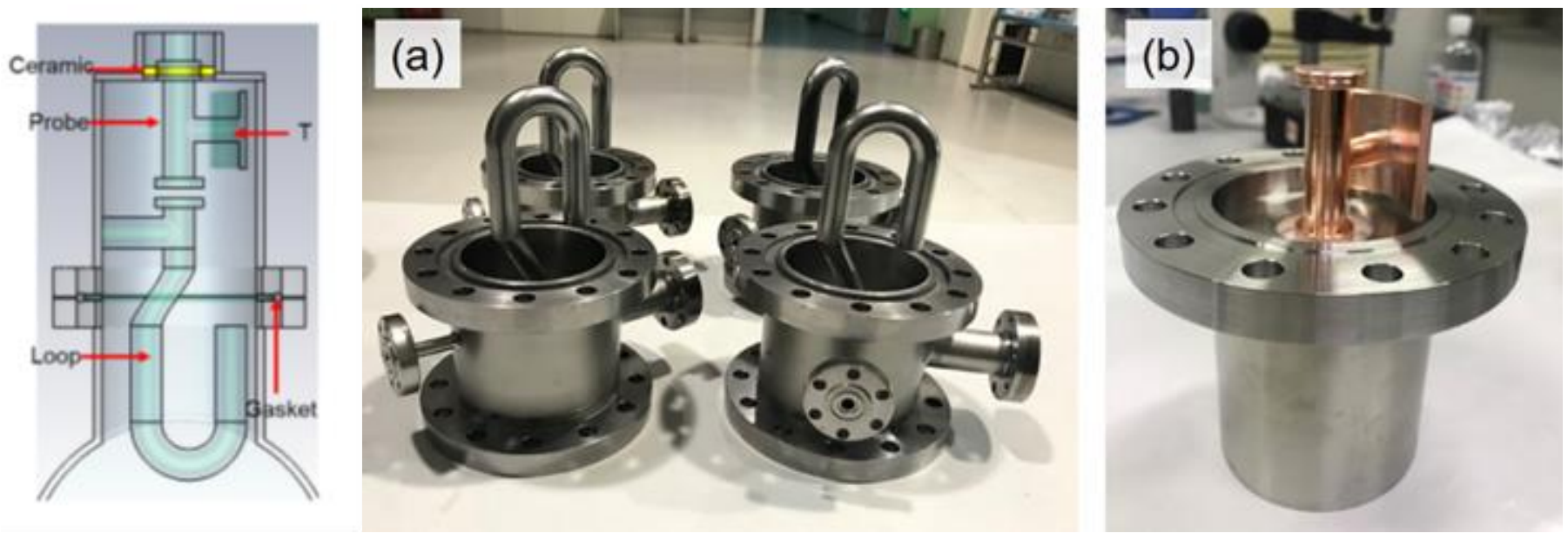

**Figure 4.3.1.9:** Double notch HOM coupler with two parts: (a) niobium hook part (b) copper probe part

The external quality factor $Q_e$ of the four HOM couplers (HOM1 to HOM4) for the fundamental mode was measured before, during, and after the vertical test. During the test in the cryomodule, the fundamental mode $Q_e$ at 2 K was measured as HOM1 $6.8\times10^{13}$, HOM2 $1.9\times10^{13}$, and HOM3 $2.1\times10^{12}$, demonstrating excellent filter performance without requiring tuning.

Figure 4.3.1.10 summarizes the measured $Q_e$ of monopole and dipole modes below 2 GHz, along with the simulated results. The damping results at 2 K were found to be within acceptable limits for the $Q_e$ of the HOMs for Higgs.

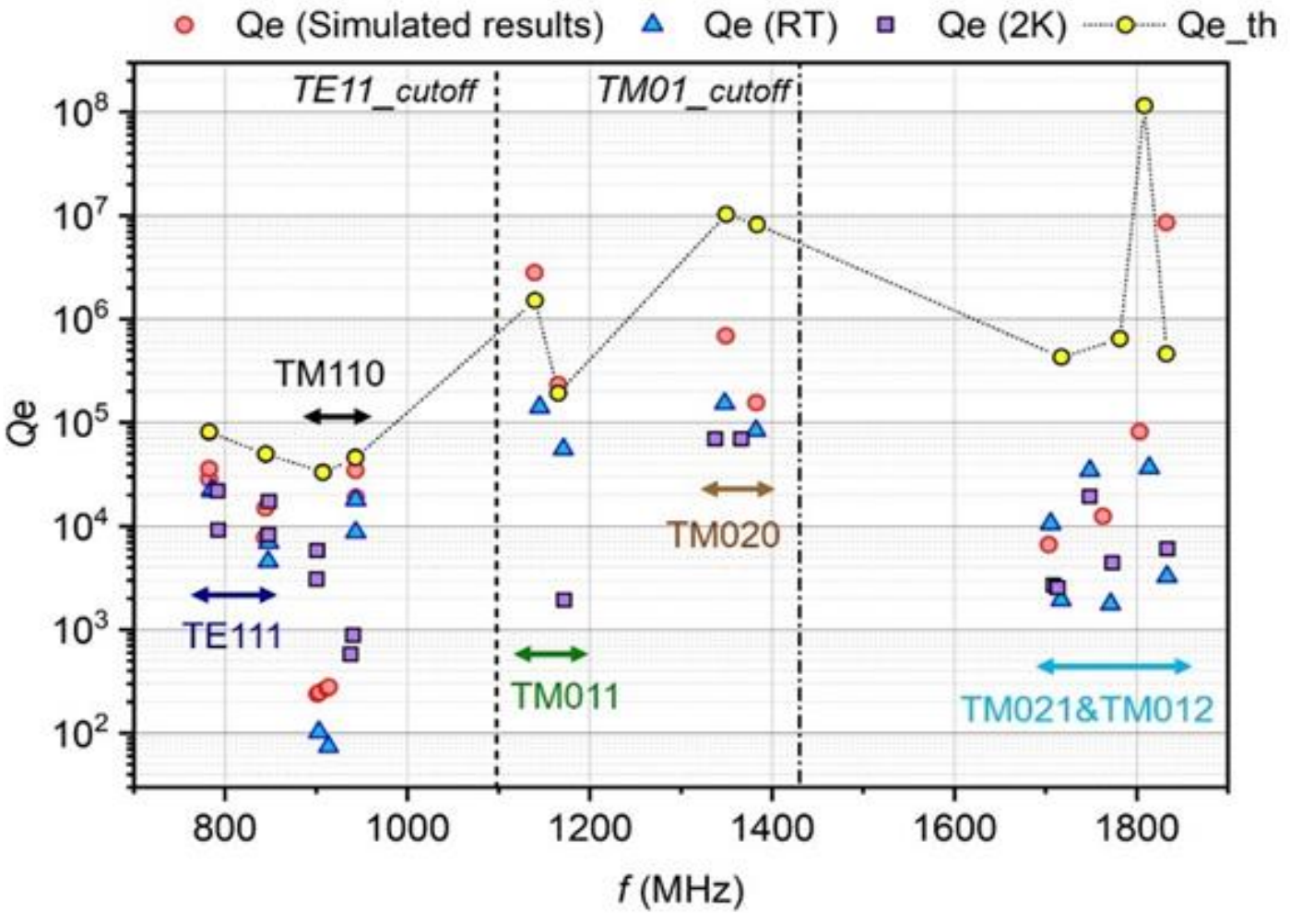

**Figure 4.3.1.10:** Measured $Q_e$ of HOMs below 2 GHz under room temperature and 2 K compared with simulated results. ''RT'' stands for room temperature. ''$Q_e$_th'' stands for the acceptable limits for the $Q_e$ of the various modes.

The HOM absorber is primarily utilized to suppress the HOM power above 1.4 GHz. Because of the short bunch length and wide HOM frequency range, a composite of SiC and AlN is selected as the absorbing material for the cavity HOM. A high-power test of the HOM absorber, with a power of 5 kW, demonstrates high absorbing efficiency and satisfies the specifications of the CEPC.

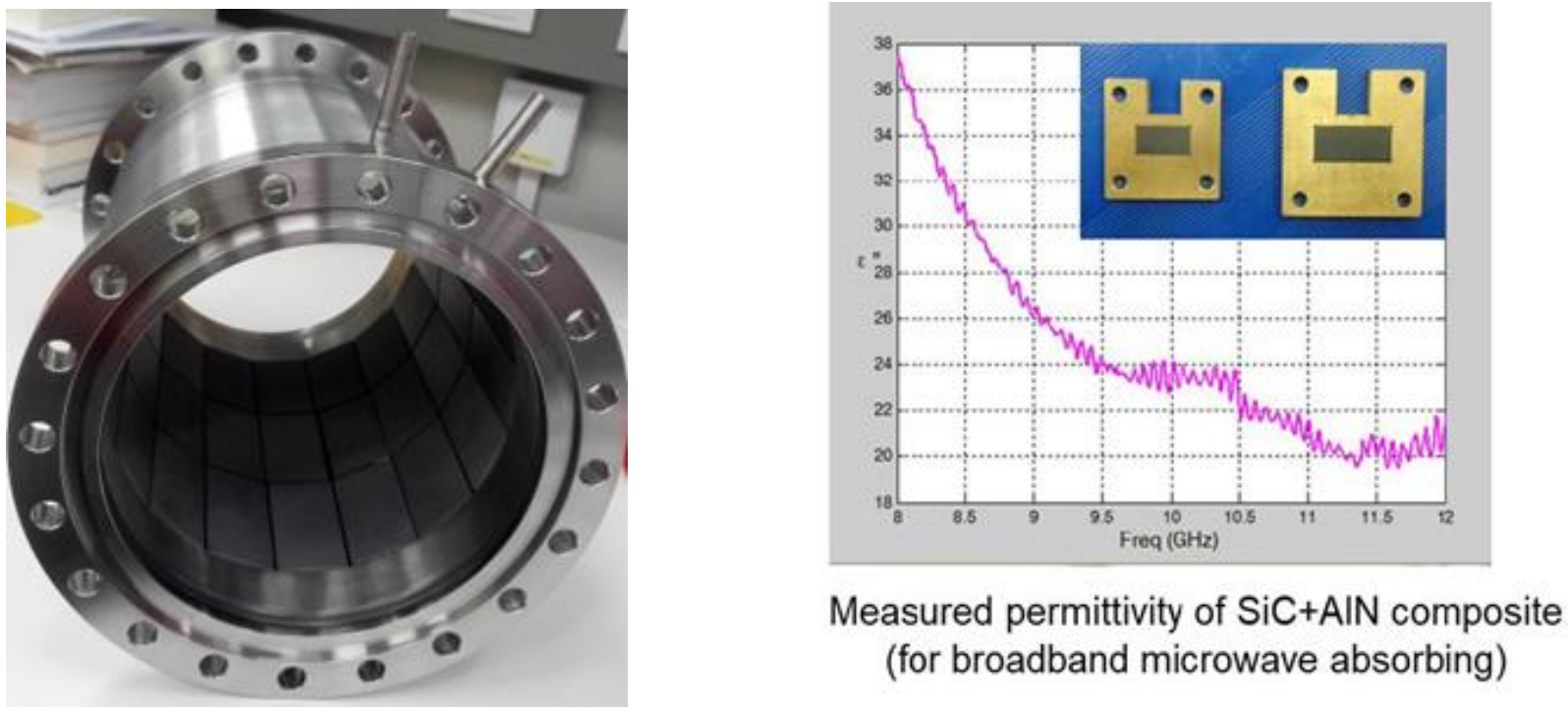

**Figure 4.3.1.11:** HOM absorber and the material permittivity measurement

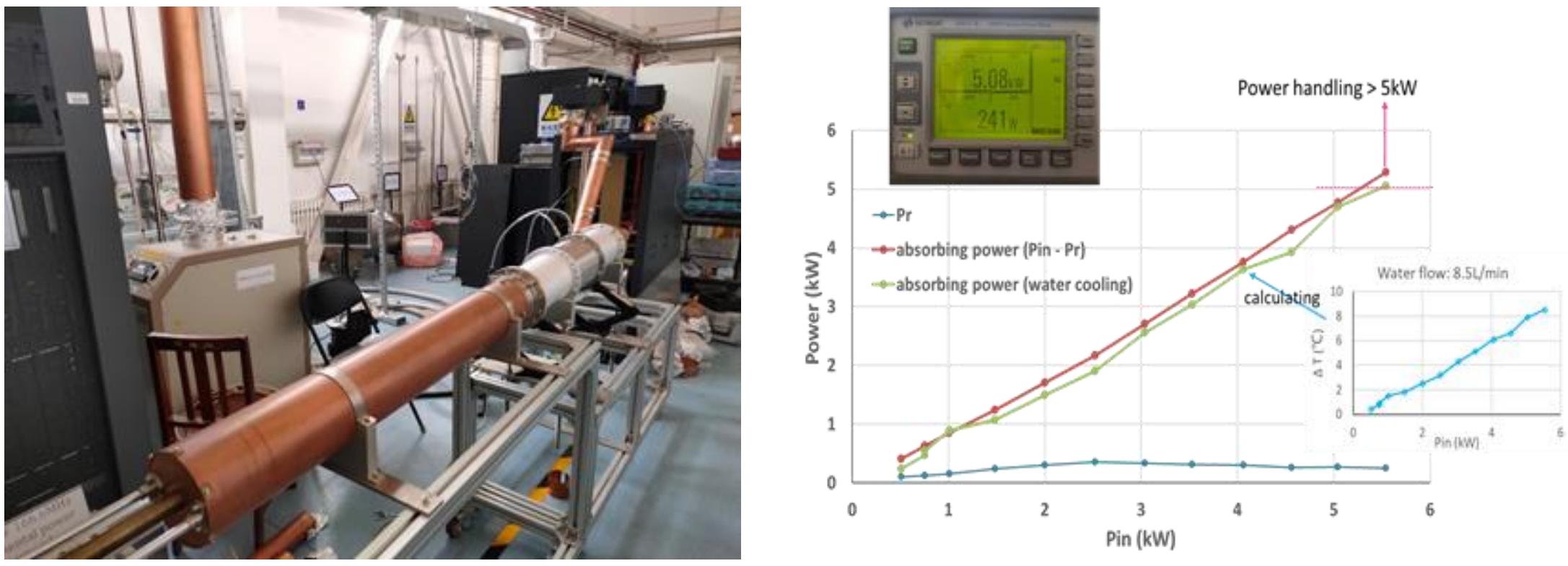

**Figure 4.3.1.12:** High power test of the HOM absorber

### 4.3.1.7 *Cryomodule*

The 650 MHz cavity cryomodule is designed to house six 2-cell 650 MHz superconducting cavities, operating at a superfluid helium environment of 2 K, for the Higgs mode. To assess the viability and effectiveness of the cryomodule design, a test cryomodule (TCM) was created. The TCM is equipped with two 650 MHz 2-cell cavities, two high-power couplers, and three higher-order-mode couplers. Figure 4.3.1.13 shows the structure of the TCM, which has been assembled, installed in the tunnel, and connected with the 2 K distribution valve (Figure 4.3.1.14).

To tune the cavity, a double-lever tuner is utilized. The actuation system, comprising a low-temperature stepper motor and two piezoelectric ceramic actuators with large tuning stroke, is connected to a secondary arm that is hinged at one end and pushes the flange on the cavity end plate through two rods in the middle.

To address the large beam pipe of the 650 MHz cavity and the north-to-south orientation of the TCM, a global warm magnetic shield inside the cryomodule vacuum vessel and a local cold magnetic shield around the cavity helium vessel are utilized. The remnant magnetic field at the cavity wall is below 2 mG at 2 K, which is adequet for high Q operation.

During the first high-power horizontal test at 2 K, the cavities achieved 8 MV/m with $Q_0 = 2\times10^{10}$. However, the gradient limit is thought to be due to inadequate helium gas cooling of the power-coupler outer conductor. Further optimization of the helium gas cooling will increase the cavity performance in the module. A beam test of the cryomodule with a DC photo-cathode gun is planned.

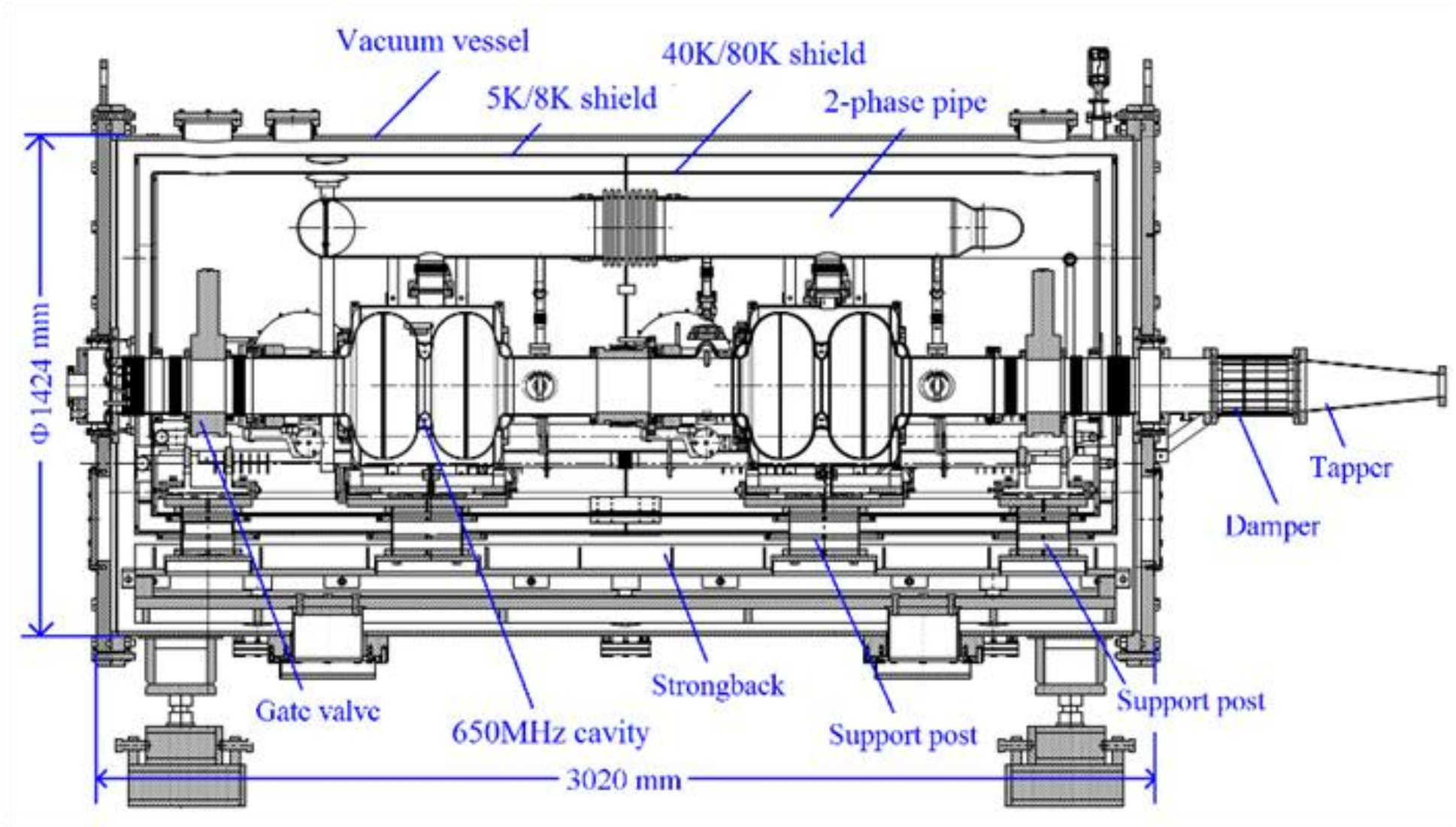

a) Longitudinal section

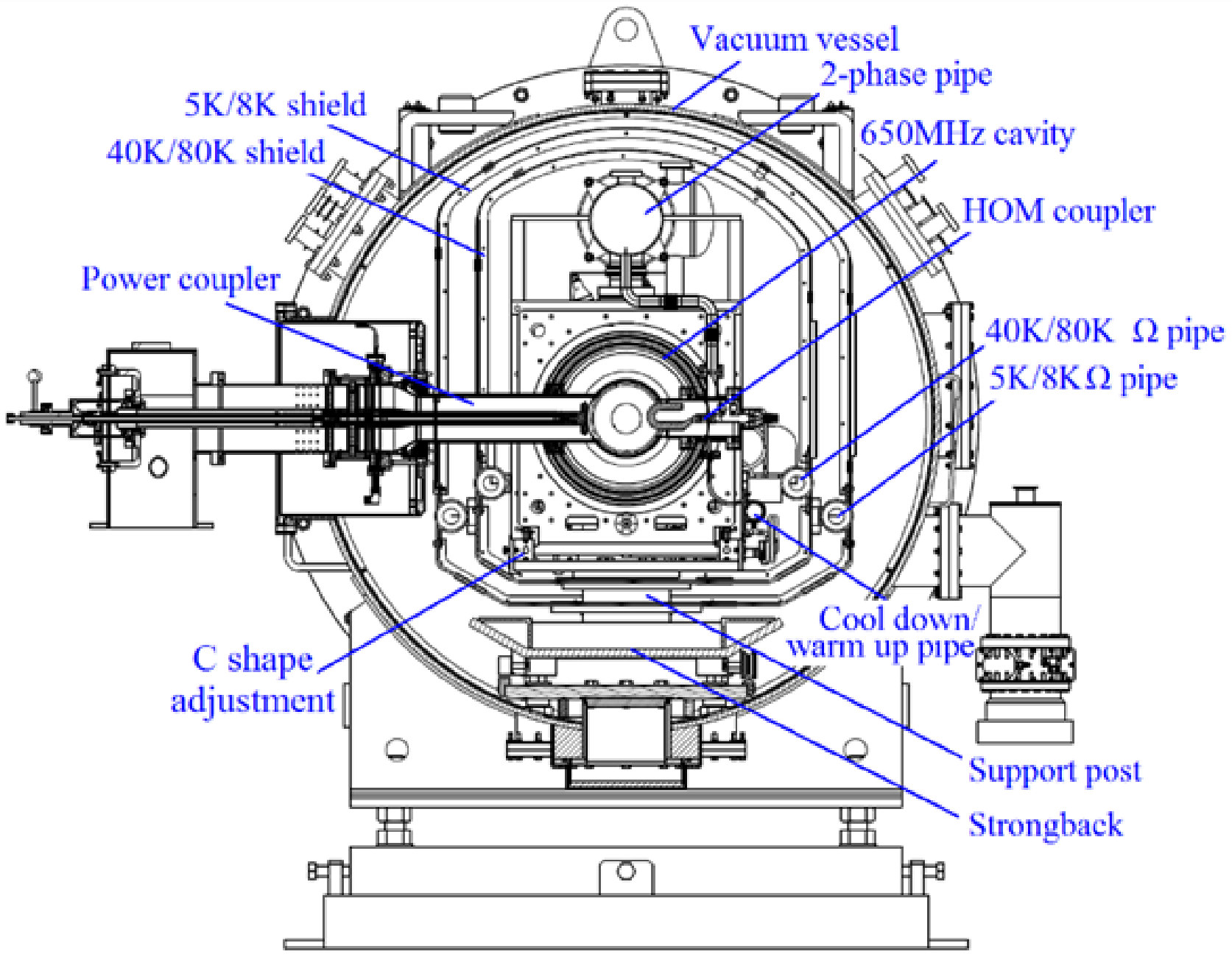

b) Transverse section

**Figure 4.3.1.13:** Structure of the 650 MHz test cryomodule.

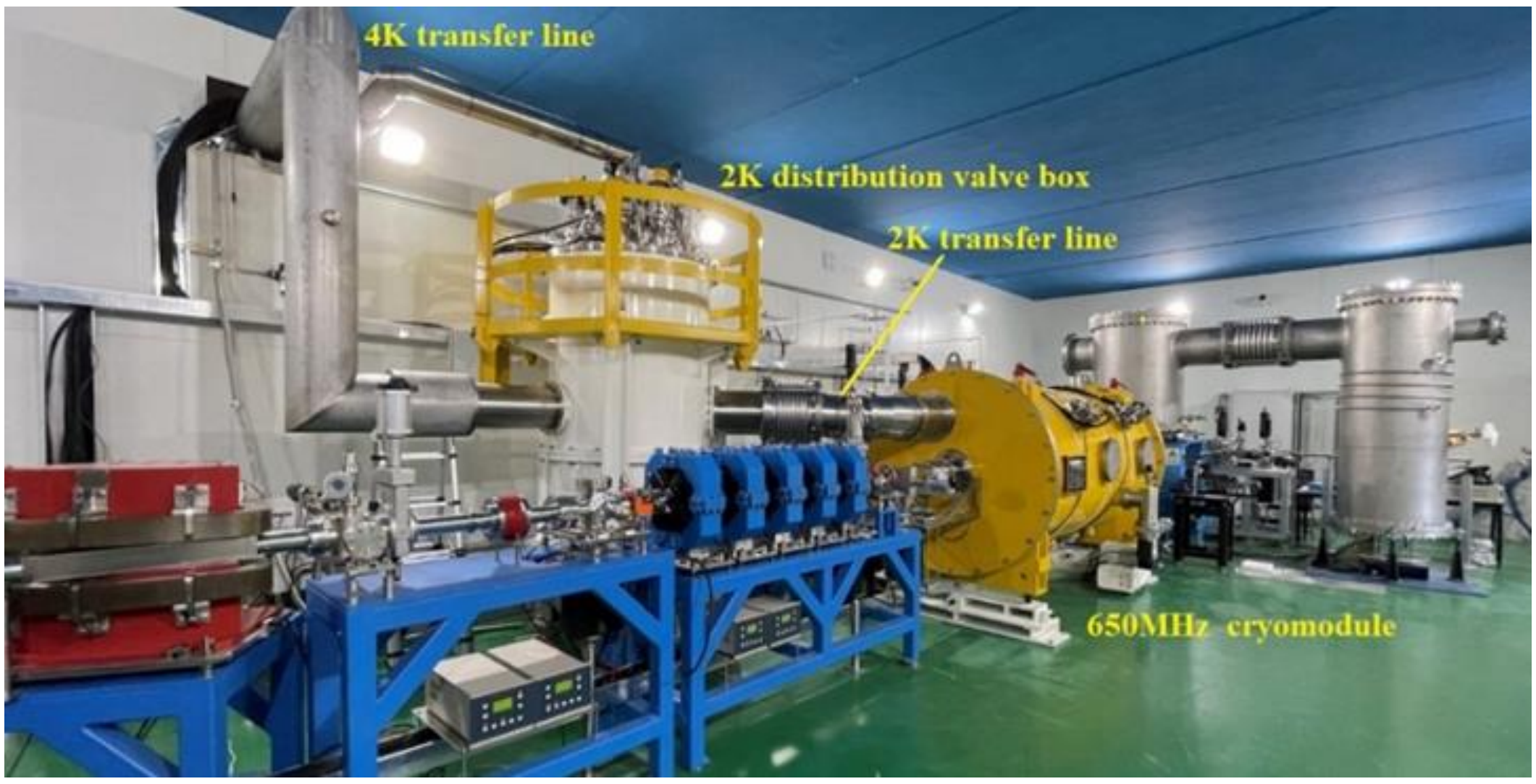

**Figure 4.3.1.14:** 650 MHz test cryomodule and 2 K valve box.

## 4.3.2 RF Power Source

### 4.3.2.1 *Introduction*

The RF power source must remain stable to ensure that any fluctuations in the supplied RF power do not have a significant impact on the Collider beam quality. The RF power source gives energy to the electrons, compensating for their energy loss from synchrotron radiation and from interactions with the beam chamber impedance. Additionally, the RF power source must provide energy to the beam when ramping up to higher energy levels, as well as powering the capture and focus of the electrons into bunches. The beam and RF stations are two dynamic systems that strongly interact, which makes stability considerations for the combined system complex.

The CEPC synchrotron radiation power is expected to be 60 MW (positron beam 30 MW and electron beam 30 MW), making efficiency from direct current (DC) to beam power a crucial consideration in the cost of implementing the project. A high-power klystron is an attractive choice for an RF power source due to its potential for higher efficiency than a solid-state amplifier and its stable operation compared to an inductive output tube (IOT). For optimal performance, a single klystron with a specified saturation power of 600 to 700 kW can power two cavities. This approach considers the klystron's operation in a linear region and takes into account the losses of a circulator and waveguide. For maximum reliability, a single 800 kW klystron amplifier can drive two collider cavities through a magic tee, two rated circulators, and the loads, while also allowing for power redundancy and considering the klystron's operation lifetime.

### 4.3.2.2 *Klystron*

The acquisition of a high-efficiency RF power source for CEPC is a critical concern. Klystron efficiency is well-known to be strongly dependent on the beam perveance of the tube. For pulsed klystrons with high perveance, the efficiency typically ranges between 40% and 45%, while for klystrons that operate in CW or long-pulsed mode, the perveance is relatively low, and the efficiency can reach 65%. Multi-beam klystrons are preferred for their high efficiency of over 65% [1]. In a recent theoretical calculation [2], a 90% RF power-conversion efficiency was achieved. Considering this recent high-efficiency approach, our design goal is to achieve approximately 80%. It should be noted that as the efficiency increases, there is a greater likelihood of unstable oscillations due to backscattering of electrons generated at the output cavity gap. Another issue to consider is how to increase the efficiency at the operating point (the linear region of the power transfer curve). Additionally, it is necessary to take into account the power consumption of the focusing electromagnet. The design parameters for a 650 MHz klystron for the case of conventional single-beam, high-efficiency, and MBK approaches are listed in Table 4.3.2.1.

**Table 4.3.2.2:** CEPC Klystron Key Design Parameters

| Parameters | Units | Values |
|---|---|---|
| Centre frequency | MHz | 650±0.5 |
| Output power | kW | 800 |
| Efficiency (Goal) | % | 80 |

After decades of development, several bunching methods have been proposed to improve the efficiency of klystrons after decades of development. The second-harmonic bunching method has been successfully applied in most high-power ultra-high-frequency (UHF) CW klystrons [3]. Other methods such as the bunching-alignment-collection (BAC) [3], core stabilization method (CSM) [4], and core oscillation method (COM) [5] are being validated in the laboratory.

The Institute of High Energy Physics (IHEP) in China is currently developing a high-efficiency klystron with a frequency of 650 MHz/800 kW. This is the first time such a klystron is being designed and manufactured in China. The goal is to achieve an efficiency of around 80%. Over the last five years, several prototypes of the klystron have been produced to reach this objective.

The first klystron prototype used the traditional bunching method to achieve an efficiency of 65%. However, the design efficiency goal is more than 80%, and to reach this target, several other bunching methods, including high-voltage klystron and multi-beam klystron, were tested. The first prototype has been designed, fabricated, and tested at high power.

In pulse mode, we achieved 804 kW peak power with 65.3% efficiency, and in CW mode, we obtained approximately 700 kW average power [6]. Figure 4.3.2.1 shows the first prototype on the test stand along with the test results.

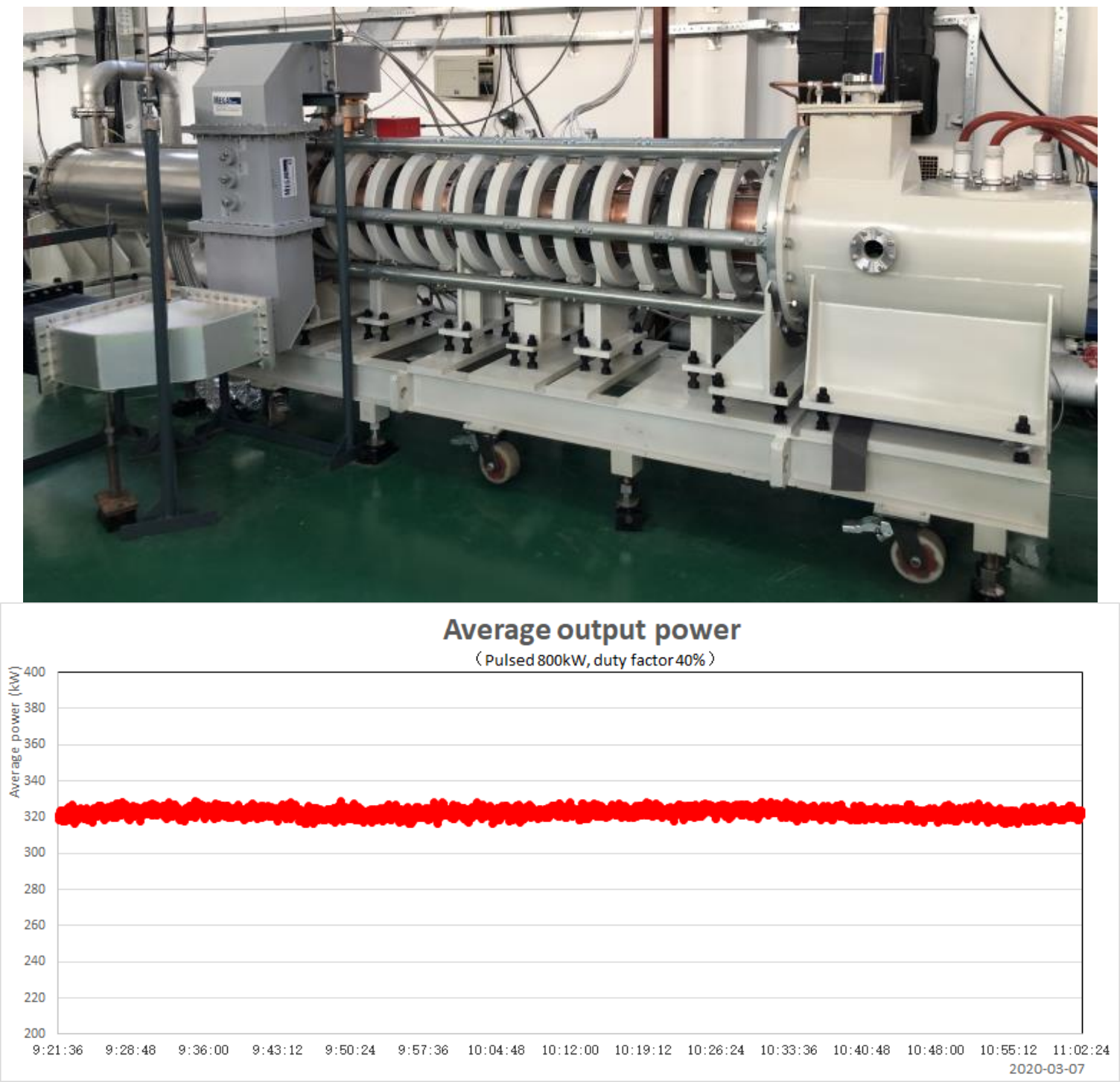

**Figure 4.3.2.1:** First prototype on the test stand (top) and test results (bottom).

Acquiring a high-efficiency RF power source for the CEPC project is a crucial aspect. Two schemes have been proposed for the development of high-efficiency klystrons: high-voltage klystrons and multi-beam klystrons [7]. The design parameters for both schemes are provided in Table 4.3.2.2.

**Table 4.3.2.2:** Klystron design parameters

| Parameters | Scheme 1 | Scheme 2 |
|---|---|---|
| Operation frequency (MHz) | 650 | 650 |
| Beam voltage (kV) | 110 | 54 |
| Current (A) | 9.1 | 2.51 |
| Beam number | 1 | 8 |
| Beam pervence | 0.25 | 0.2 |
| Efficiency (%) | ~80 | > 80 |

Scheme 1 has achieved efficiencies of approximately 85.6%, 81.4%, and 78.2% using AJDISK (1D), EMSYS (2D), and CST (3D) based on second- and third-harmonic cavities. An ongoing design with a significantly higher klystron gun voltage (120 kV) is expected to yield even higher efficiency. Figure 4.3.2.2 displays the design results. In July 2022, the first stage of the high-power test for the second prototype was completed, achieving 70.5% efficiency at 630 kW CW power. Test results and a photo of the second prototype on the test stand are presented in Figure 4.3.2.3.

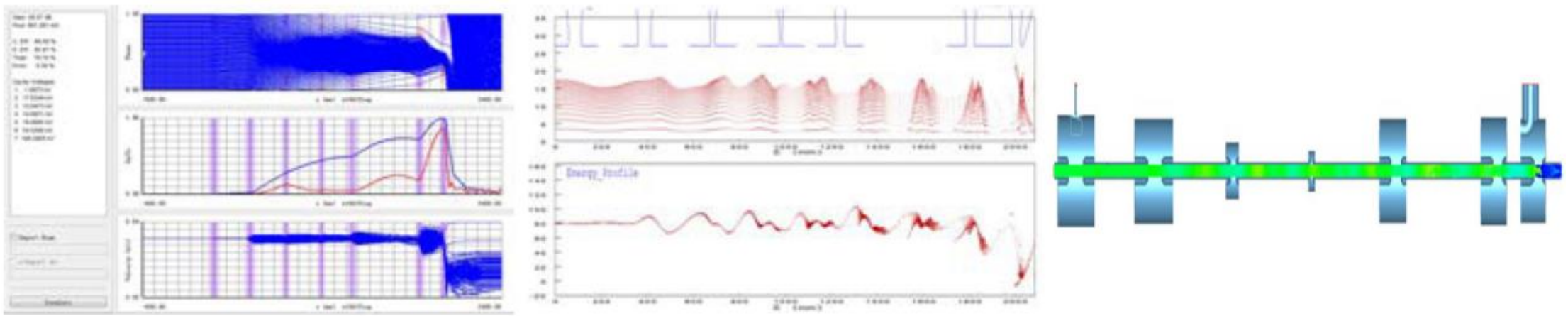

**Figure 4.3.2.2**: High efficiency klystron simulation results.

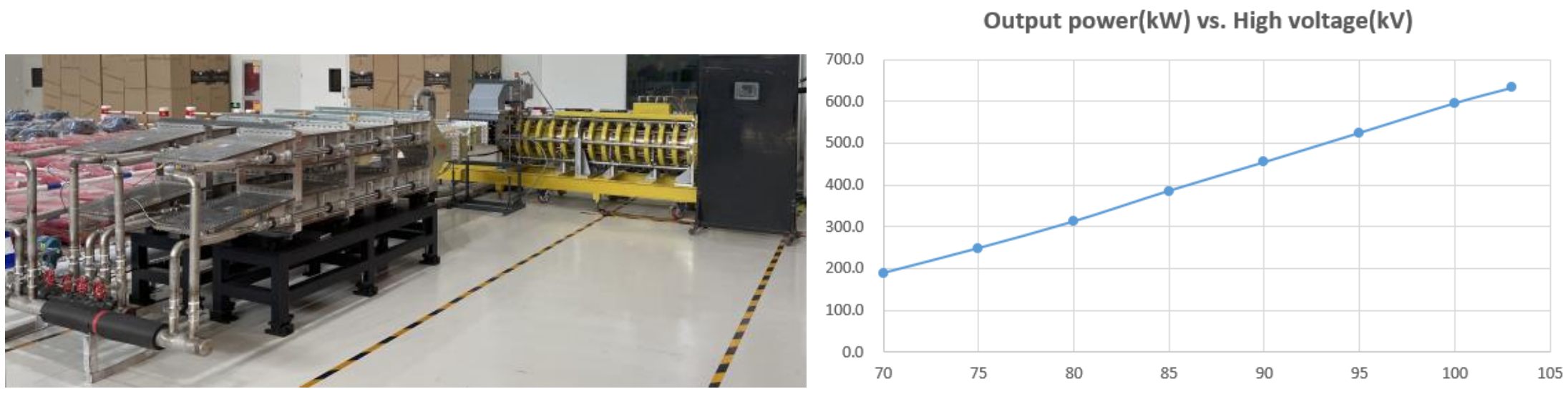

**Figure 4.3.2.3**: Second klystron prototype (left) and first stage test result (right).

To achieve high efficiency, the beam perveance of a multi-beam klystron (MBK) must be selected with care. To simplify this process, we use a methodology [8-9] that transforms coaxial multi-beam cavity parameters to single beamlet cavity parameters. This allows us to extend 1D RF circuit simulation results to MBK, which are considered roughly equivalent. Once the main parameters of the MBK are chosen, we carry out 3D PIC simulations to analyze its performance. Figure 4.3.2.4 depicts the simulation of the MBK model and bunched beams in CST [10], a 3D PIC code. The MBK model consists of seven cavities, including one second-harmonic cavity and one third-harmonic cavity. The two penultimate cavities gradually bunch particles before they enter the output cavity. We are currently manufacturing the third prototype (Multi-beam klystron), which is scheduled for testing in the middle of next year.

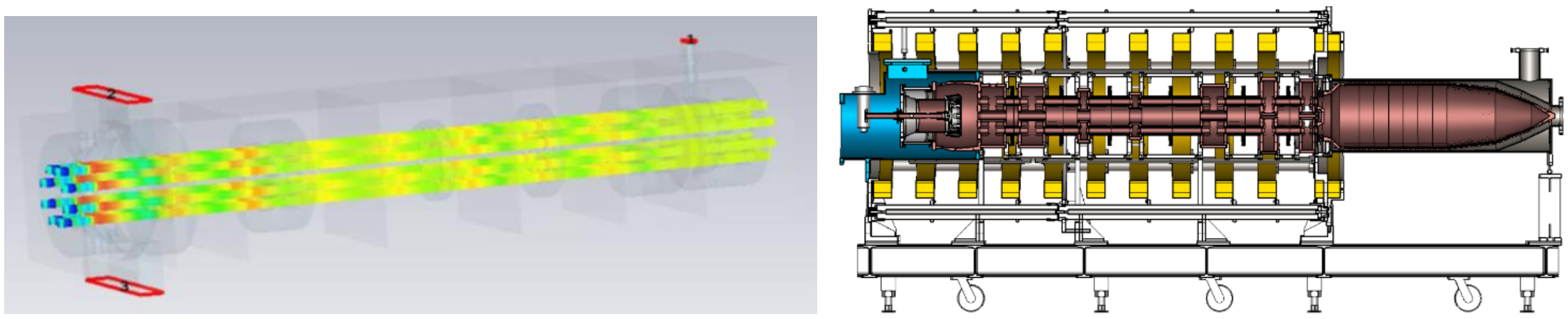

**Figure 4.3.2.4**: CST simulation result (left) and mechanical model (right).

### 4.3.2.3 *PSM Power Supply*

The high-voltage power supply for superconducting RF power sources is a power conversion device based on power electronics converters. It converts AC power from the distribution network into RF modulated high-voltage for the speed control tube. This tube is a high-power device, and its key indicators, such as high-voltage ripple, long-term stability, and control accuracy, are crucial factors affecting the amplitude and phase stability of RF power. Conventional power-conversion devices use pulse step modulation (PSM) high-voltage power supplies, with gas pedal field voltage levels ranging from 20 kV to 200 kV and power ranging from 500 kW to 10 MW. However, due to the early development and application of power electronics, these devices are limited in terms of high-voltage ripple and long-term stability. The current performance of mature products generally results in high-voltage ripple and long-term stability in the order of 1%. They have a power factor of 0.9 and total efficiency of 90%. As a result, the RF power performance is not optimal, which limits the beam quality of the particle gas pedal.

In contrast, a continuous wave superconducting RF system requires the RF power to be gradually increased from narrow pulse to continuous wave during high-power conditioning and beam macro-pulse experiments. Conventional high-voltage power supplies always maintain high-power continuous-wave outputs, and the RF-power pulse width is varied only by adjusting the excitation. This process results in a large amount of power being wasted. With the rapid development of semiconductor technology and power electronics technology and the application of advanced control strategies, there are many possibilities for higher performance and more efficient use of energy for power conversion devices in future advanced accelerators.

Figure 4.3.2.5 shows the PSM high-voltage power-supply topology, which utilizes a simple, reliable, and redundant design. It comprises of a multi-winding phase-shifting isolation transformer, a unit module, filtering and firing-suppression circuits, and a cascade of unit modules to form a high-voltage output. The operating mode waveforms of the prototype design are illustrated in Figure 4.3.2.6. To meet the high-energy and energy-saving requirements of CEPC, we propose improving the traditional PSM high-voltage topology by incorporating advanced automatic-control-algorithm strategies. This will not only enhance the long-term stability of the high-voltage power supply in continuous wave mode but also reduce the output ripple. Additionally, it will allow the regulation mode of duty cycle to be realized. We have taken into account modulation control, load-firing energy limitations, control accuracy, and firing-energy limitations, among other factors. Table 4.3.2.3 gives the main parameters.

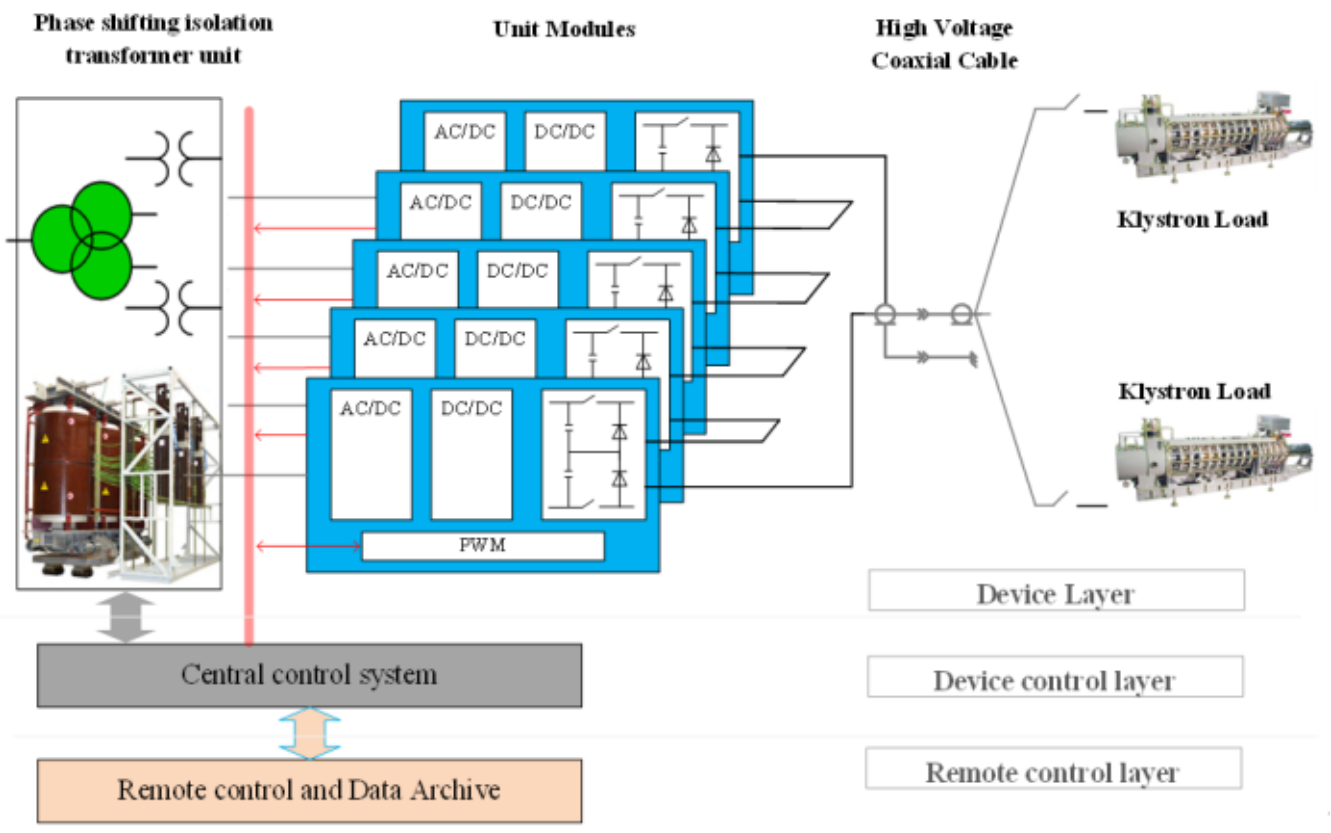

**Figure 4.3.2.5**: High-voltage power-supply topology

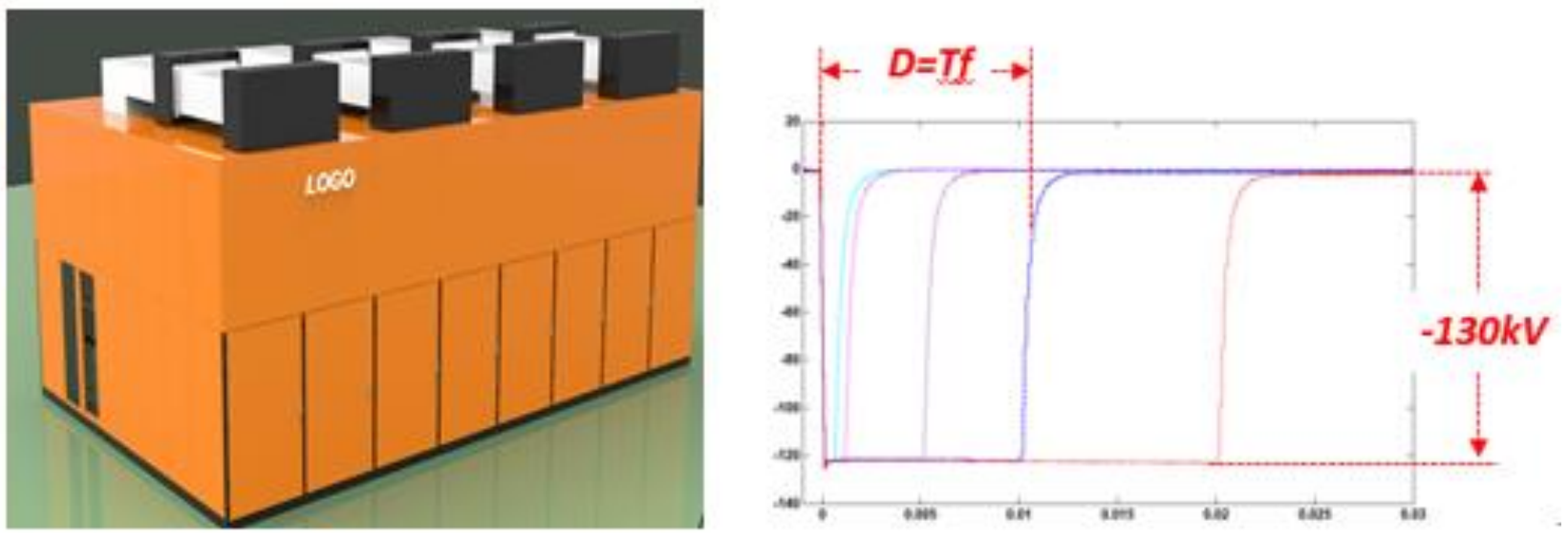

**Figure 4.3.2.6**: Design prototypes and pulse modes

**Table 4.3.2.3:** PSM main working parameters

| **Operation mode** | **1** | **2** | **3** | **4** | **5** |
|---|---|---|---|---|---|
| Cathode voltage (kV) | 81.5 | 100 | 110 | 120 | 130 |
| Beam current (A) | 15.1 | 10.6 | 9.3 | 8.3 | 7.6 |
| Maximum parameter | 130 kV DC, 16A | | | | |
| Operation Mode | Pulse or DC | | | | |
| Pulse width | 100 μs～DC | | | | |
| Repetition frequency | 0.1～100Hz | | | | |
| Pulse front | ≤ 200μs | | | | |
| Pulse trailing edge | ≤ 200μs | | | | |
| Pulse top fluctuation | ≤ 1% | | | | |
| Protection time | ≤ 5μs | | | | |
| Transferring energy | 10 J | | | | |
| Duty cycle | 0～100% | | | | |
| Power factor | ≥ 95% | | | | |

### 4.3.2.4 *RF-Power Transfer System*

The SRF system of the Collider is composed of 240 2-cell cavities, requiring a minimum of 300 kW of power transmission to meet the total power demands for radiated, HOM, and reflected power [11]. Table 4.3.2.4 outlines the specific RF power demands.

**Table 4.3.2.4:** SRF system parameters

| Parameters | Higgs |
|---|---|
| Operation frequency (MHz) | 650 +/-0.5 |
| Cavity number | 240 |
| Coupler input power (kW) | 300 |

The 240 cavities will be divided into 2 sections, with each section containing 120 cavities. To drive 2 cavities with equal power division, a single 800 kW klystron amplifier will be selected. This choice is technically justified by better control of microphonic noise and minimum perturbation in the event of a klystron trip [12].

The system accounts for the linear operation of the klystron and transmission losses. The RF power generated by the klystron will be transmitted through a WR1500 aluminum waveguide. Figure 4.3.2.7 and Figure 4.3.2.8 depict the placement of the klystron and waveguide in the RF Transmission System (RFTS), respectively.

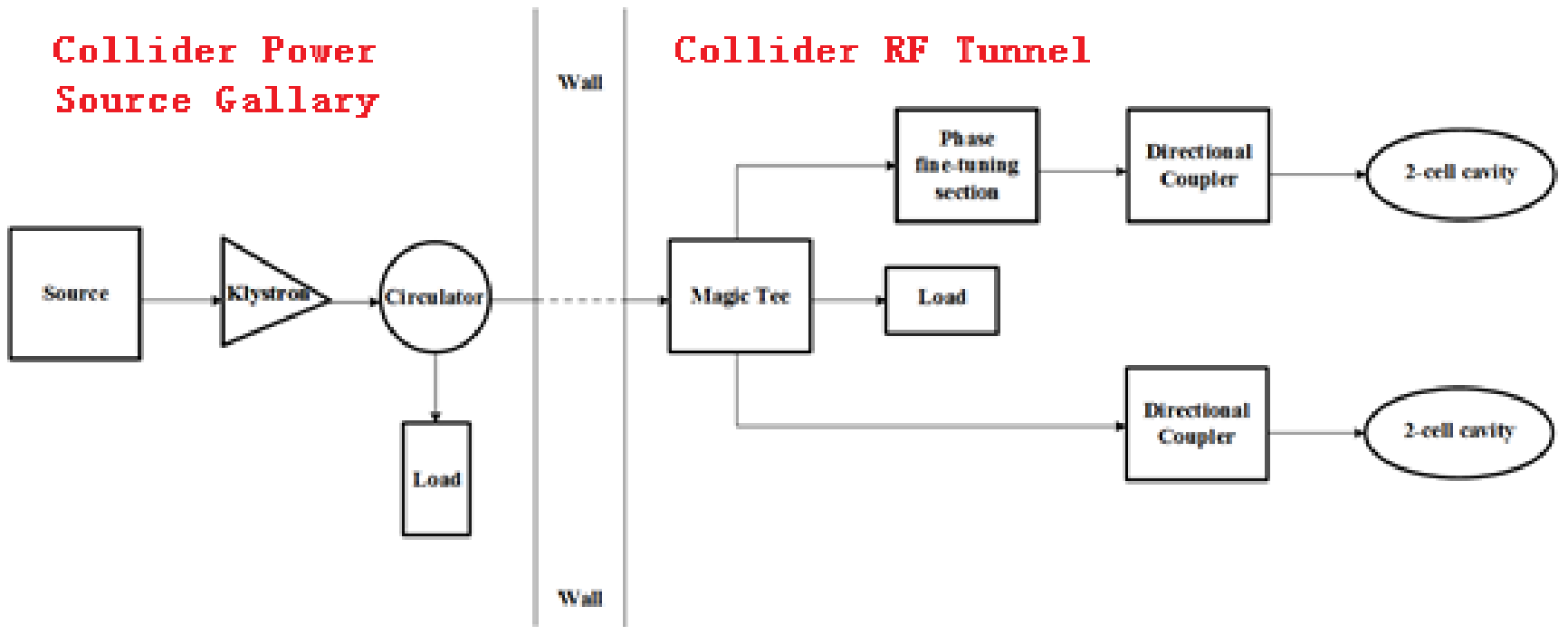

**Figure 4.3.2.7:** Schematic of the Collider RFTS

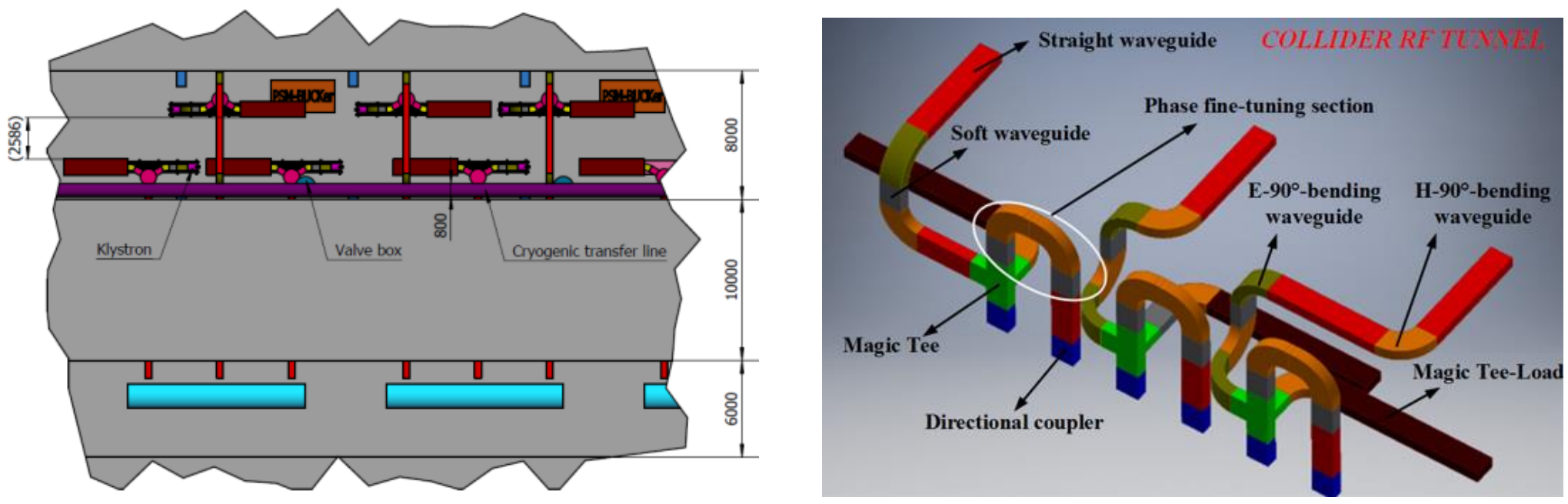

**Figure 4.3.2.8:** Schematic of the Collider RFTS

To protect the klystron, a CW 800 kW circulator will be installed in each feed line to block power that is either reflected or discharged from the cavity. The circulator is a three-port device, with power from the klystron fed into the first port and transmitted to the second port. Any power reflected to the second port will be transmitted to the third port, which is connected to a ferrite load. This setup protects the klystron, particularly its fatal output ceramic window. The klystron, circulator, and ferrite load can now be manufactured in-house after years of hard work, with high power almost qualified and further improvements in the efficiency of the klystron anticipated.

The power from the klystron is divided equally using a Magic Tee, and a ferrite load is used on the fourth port to absorb any power, usually reflected power from downstream.

The directional coupler [13-14] provides both forward and backward (reflected) pick-up signals, with directivity better than 30 dB. These signals are used for monitoring, close-loop control, and interlock protection during operation.

Flexible waveguides [15] are adopted for both installation connections and a phase fine-tuning section. They eliminate installation errors, reduce vibration transmission, and offset the influence of thermal expansion and contraction in the installation connection. In the phase fine-tuning section, they are used to compensate for any phase differences introduced in the system by slightly changing the length so that the phase of the two signals divided by the Magic Tee accurately matches the design. Alternatively, a waveguide phase shifter could be used for one of the divided powers, but it would significantly increase the cost, which is a considerable factor for such a large collider.

The klystron, circulator, and ferrite load will be placed in the auxiliary tunnel, while the Magic Tee and phase fine-tuning section will be placed in the main tunnel with the cavities. The PSM power supply will be located on the surface with the transformer and rectifiers.

The total length of the waveguide transmission path is approximately 20 meters, and the standard transmission loss of a WR1500 aluminum waveguide at 650 MHz is 0.0048 dB/m. Therefore, the total loss for a 20-meter waveguide is 0.096 dB, which means 97.8% of the klystron output power can be transmitted to the cavities (800×97.8% = 782.4 kW, >600 kW). This meets the requirements in Table 4.3.2.4. Even with a klystron output power of 720 kW (which is 90% of the maximum output power of 800 kW), the corresponding power delivered to the cavities will still be 704.16 kW, which is more than enough compared to the required 600 kW.

#### 4.3.2.5 ***Low Level RF System***

The LLRF system ensures the stability of the superconducting cavities and high-power sources while monitoring all signals in the High-level RF system and interacting with other systems. Each of the 120 sets of LLRF hardware is installed beside each klystron, as two 650MHz 2-cell cavities are powered by a single klystron in the collider ring, resulting in a total of 240 cavities and 120 klystrons. During operation, 18 channels of signals should be sampled synchronously, and two pickup signals from two cavities should be sampled and vector-summed to calculate the phase errors of the cavities [16]. Functions of the LLRF include:

- regulation of phase, amplitude, and frequency using feedback and feedforward techniques
- compensation for beam loading and stability of beam-cavity coupling
- suppression of microphonics and disturbances
- detection of quench and protection of the cavities
- monitoring of the forward, reverse, and pickup power of all signals on the high-power source chain
- calculation of the vector sum of double cavities driven by one klystron
- detection of reverse power and protection of the klystron/SSA
- interface to/from interlock/BI/timing system
- remote diagnosis and control of LLRF infrastructure
- smart and automated RF-system operation
- high reliability and long-term stability
- ease of maintenance for operators feedforward

**Table 4.3.2.5:** Requirements for LLRF

| | Phase stabilization | Amplitude stabilization | Run mode |
|---|---|---|---|
| Booster | 0.1deg (rms) | 0.1% (rms) | CW |
| Collider | 0.1deg (rms) | 0.1% (rms) | CW |
| Linac | 0.2deg (rms) | 0.2% (rms) | Pulsed |

Maintenance and reliability are crucial for a large project like CEPC, as controller stations are installed kilometers apart from each other. On-site maintenance during every RF system trip is not feasible, so the LLRF system must be remotely accessible and controllable directly into the rack/crate and even at the electronics board level through Ethernet. This requirement calls for remote management and automated operation of the LLRF hardware. The LLRF system should monitor all the RF system signals simultaneously, such as 18 ADC channels using a 16-bit, 125 MHz sampling clock, which requires a minimum data rate of 4.5 GBytes.

The PCI-Express bus-based MicroTCA standard will serve as the hardware platform of the LLRF system since VME/CPCI/PXI bus cannot meet the requirements. The LLRF layout for the high-power source and cavities is illustrated in Figure 4.3.2.9. Directly sampling 20 650 MHz signals is possible using a 130 MHz ADC clock, with the benefit of simplifying the system by totally canceling LO signal generation and down-converters. A single LLRF system for two cavities implements 2 tuners and 4 piezo controllers.

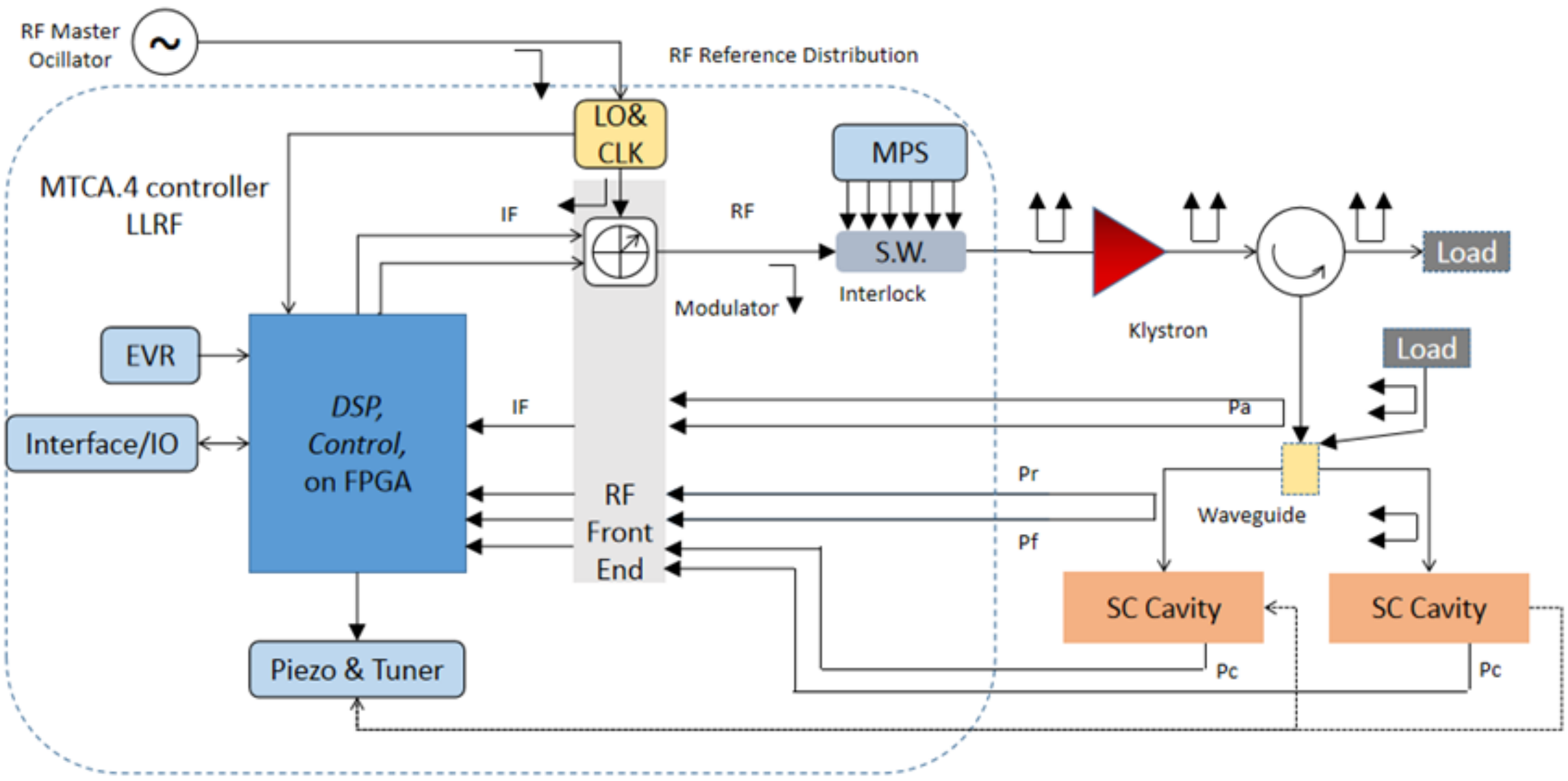

**Figure 4.3.2.9:** Scheme of collider RF control system

For each control system, one LLRF controller digitizes two pickup signals from the superconducting cavities probe, with another two LLRF DAQ modules sampling all the signals in the high-power RF and reference signals. The Tuner&Piezo module receives cavity frequency detuning errors and adjusts the tuners and piezos to keep the cavities resonant. An embedded or standalone timing receiver recovers the event timing signals from the timing system and distributes to LLRF controller boards. The local protection interface module reacts between LLRF fault detection and machine protection system. Two power supplies and two CPUs run on hot-swap mode. The MicroTCA MCH (MicroTCA Carrier Hub) with a PCIE switch and remote management function is necessary for data exchange and easy maintenance.

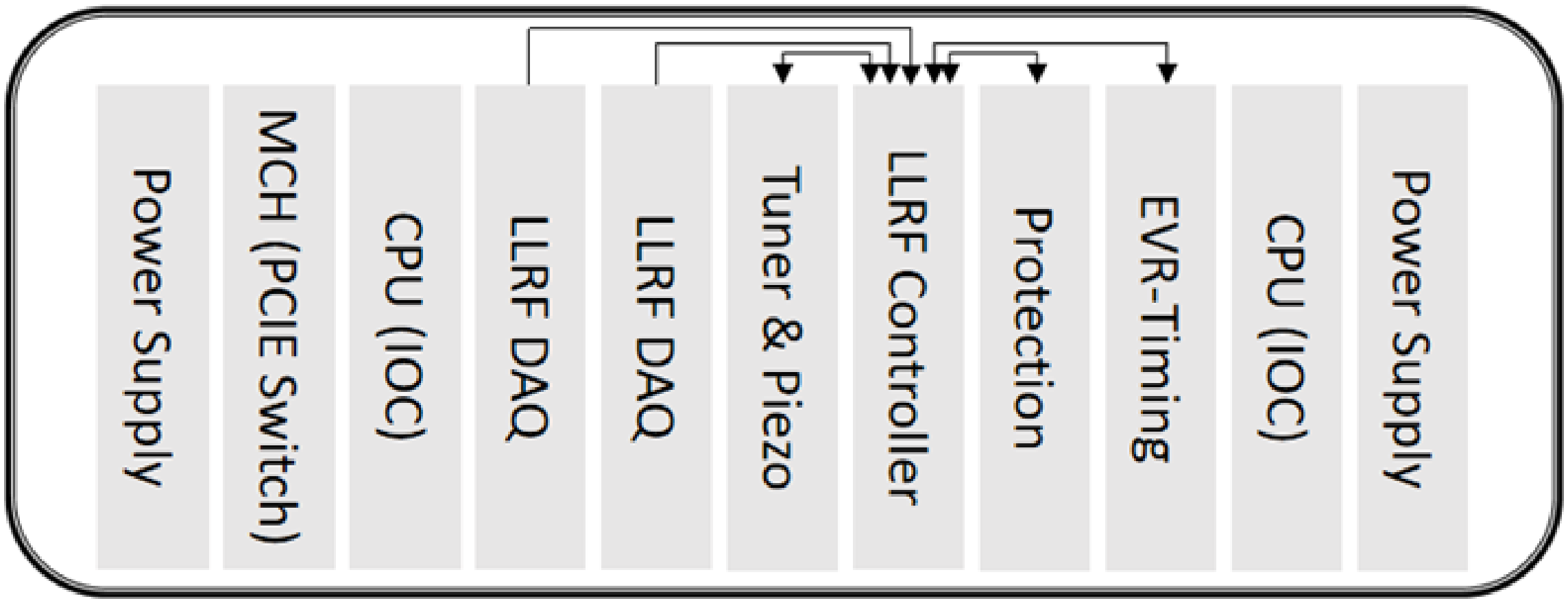

**Figure 4.3.2.10**: Functionality in one RF control chassis

Status of the LLRF development：

A two-cell 650 MHz superconducting cavity cryo-module has been successfully installed and tested on the PAPS (Plateform of Advanced Proton Source) beam test-bench platform. The LLRF system has been developed for the high-power RF and cavity system, and the base firmware, including digital algorithms in FPGA and software on the CPU, has been validated and can be used on CEPC with minor re-configuration. The piezo and tuner controls have been integrated into the LLRF controller board and a PLC module, respectively. The long-term stability and phase noise of the system have been demonstrated to meet the requirements. The timing fan-out module has been designed and implemented on the PAPS/HEPS Linac and BECP-II Linac machines. A universal DAQ and control module developed by IHEP has been tested and its firmware build is in progress for CSNS-II and BEPCII-Upgrade projects.

To focus on the development of major components, the piezo and tuner controls will be upgraded to a single standalone module, and the klystron and cavity protection module will be redesigned for universal purposes. Although the MicroTCA standard power supply, crate, CPU, and MCH are mostly commercially available, domestic development is required. Therefore, a research and industry alliance should be established to coordinate future control system development.

Feedback control of longitudinal instability:

Z mode has lower RF voltage, significantly more bunches, higher beam current, and a narrower optimal detuning bandwidth compared to the Higgs and W modes. This can lead to significant beam-cavity stability challenges, particularly with fundamental-mode instability. The cavity's narrow-band impedance can generate a long-range wake field and increase the longitudinal oscillation amplitude of the bunches, resulting in bunch loss and reduced machine luminosity.

Based on the influence analysis of fundamental-mode-induced longitudinal instability, 9 unstable modes are identified for the 30 MW operation mode, and 27 for the 50 MW operation mode, with significant growth rates. The LLRF system is essential for controlling the cavity field and ensuring beam stability. The direct RF feedback and digital comb-filter control algorithm have been tested on the 650 MHz dummy cavity [17]. The loop delay, consisting of the klystron, waveguide, digital processor signal cable, totals within 3 μs. Simulation results of direct RF feedback and digital comb-filter implemented in the controller are displayed in Figure 4.3.2.11, with the left side showing Z(30 MW) and the right side showing Z(50 MW) modes with appropriate feedback gain. Here, fundamental impedance represents the beam-cavity coupling. When only direct RF feedback is used, not all oscillation modes are suppressed. However, when additional one-turn feedback (comb-filter) is applied alongside direct RF feedback, all unstable modes are effectively controlled as per requirements.

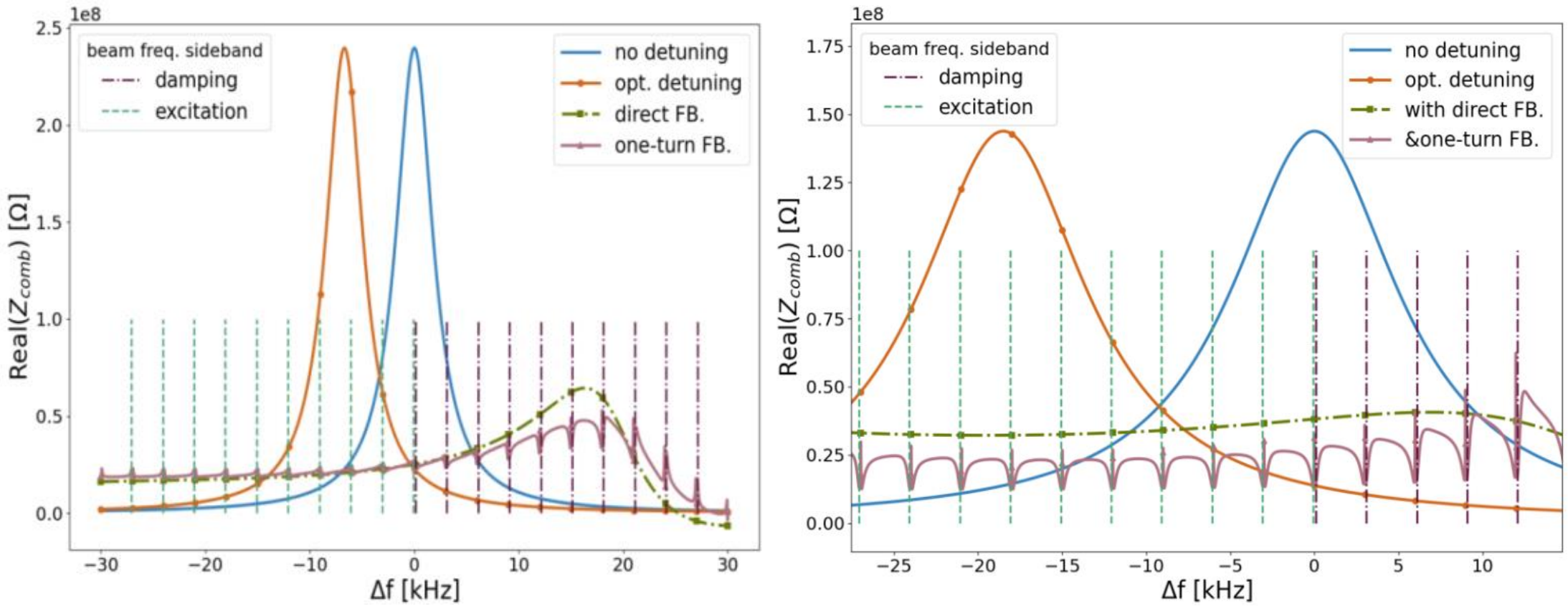

**Figure 4.3.2.11**: Fundamental impedance changes with additional one-turn delay feedback. Gain = 0.5 in Z(30 MW) mode (left) and Gain = 1.4 in Z(50 MW) mode (right).

The direct RF feedback and comb-filter feedback can be made equivalent using a PID controller in the modern digital LLRF system. Simulation results indicate that the control performance is directly linked to the closed-loop gain, where the gain of direct feedback corresponds to the proportional factor (P) of the PID controller.

<u>Beam commissioning and automation of the LLRF for the machine:</u>

The LLRF system will be installed alongside the klystron station to ensure simultaneous operation with the high-power source. Special attention will be given to increasing the system's reliability through automation. Automation software, including frequency tuning, phase-shifter tuning, conditioning, abnormal handling and trip recovery, and optimization of system parameters, has already been developed for existing machines like HEPS and C-ADS.

To achieve optimal performance during machine and beam commissioning, the LLRF system will require measurement of system parameters, calibration of klystron gain, and detection of beam-related detuning, microphonics, and mechanical electronic coupling.

#### 4.3.2.6 *Phase Reference System*

The Phase Reference System [18-20] offers a stable, low-noise, and coherent phase reference for the entire accelerator complex, both in the short and long term. The Low-Level Radio Frequency (LLRF) system stabilizes the amplitude and phase of the accelerating field, meeting the requirements of beam physics. The LLRF system also controls the amplitude stabilization of multiple cavities individually. However, the phase reference of the LLRF system must be correlated and relative to ensure optimal beam quality. Therefore, it is essential to derive or synchronize the phase reference for each cavity/LLRF with a single source. The system is designed to be highly reliable, easily extendable, and simple to maintain. It also features real-time self-status monitoring, such as detecting slow phase drift across multiple reference paths.

The phase-reference system of CEPC consists of three parts:

1) The master oscillator provides a frequency standard of 650 MHz for the Collider, 1.3 GHz for the Booster, and 2860 MHz, 572 MHz, and 143 MHz for the Injector. It serves as the drive and reference for all the accelerating tubes, superconducting RF cavities, LLRF system, and beam diagnosis system.
2) The system also provides a stable reference transfer to remote stations such as RF/LLRF, timing system, beam instrumentation system, detector electronics, and plasma accelerator laser, maintaining phase and frequency stability (or coherence) along the accelerator complex during long-term operation.
3) Signals are distributed locally to multiple endpoints.

A compact 2-layer phase reference system has been designed, consisting of long-distance distribution and short-distance local routing. Each long-distance distribution path carries only one reference signal, which is then distributed locally to the desired stations. The layout of the phase-reference system is shown in Figure 4.3.2.12.

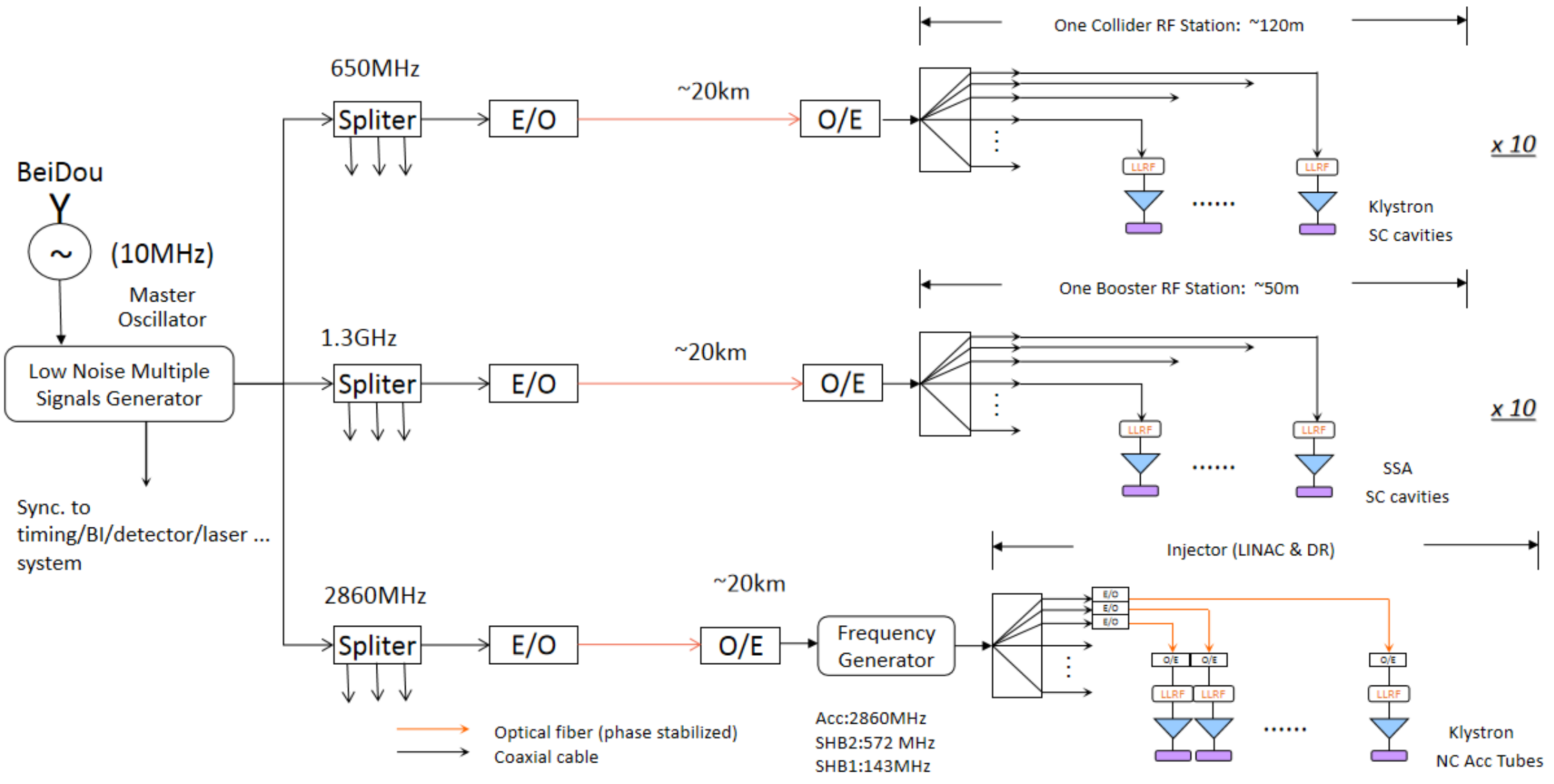

**Figure 4.3.2.12**: Layout of the phase reference system

a) The phase/frequency standard source:
The phase drift of the standard source Master Oscillator (MO) is a common factor affecting all downstream reference phases. Additionally, the phase noise of the MO introduces timing jitter to the reference signal in the time domain. Long-term operation is affected by the frequency stability of the MO. Therefore, the standard source must have low phase noise and high frequency stability, with an atomic clock or crystal oscillator that has a phase noise of –150 dBc/Hz @10 kHz and a frequency stability of $10^{-11}$. The Low Noise Multiple Signals Generator (LNMSG) produces three main frequencies for the collider: 650 MHz, 1.3 GHz, and 2.830 GHz, delivered to the collider, booster, and Linac respectively. Each frequency can be tuned within ±1 MHz for operation. The LNMSG is locked with a 10 MHz Rb clock and BeiDou Navigation Satellite System to improve long-term frequency stability.

b) Distribution of the RF reference signals with several frequencies:
To overcome attenuation, temperature drift, mechanical vibration and electromagnetic-radiation issues when transporting an RF signal through km-long coaxial cables, phase-stabilized optical fiber is selected as the RF transport medium. However, the length variation of optical fiber caused by temperature/humidity change also results in phase drift of signal transmitting. Therefore, the length of optical fibers must be stabilized using measure-and-feedback techniques, and the phase variation can be monitored and recorded simultaneously. A long-time stability to within 50 fs rms (0.05° for 2860 MHz) after 1 km has been achieved based on tested results. As illustrated in Figure 4.3.2.13, each RF station will receive one phase-reference signal. To further reduce phase drift during long-distance transfer, domestic-made phase-stabilized optical fiber with a measured delay coefficient of around 7 ps/km/K, five times lower than ordinary signal-mode optical fiber, will be selected. The phase variation caused by 4 K average temperature fluctuation would be within 560 ps, which can be compensated for by an electronic phase shifter or delay line.

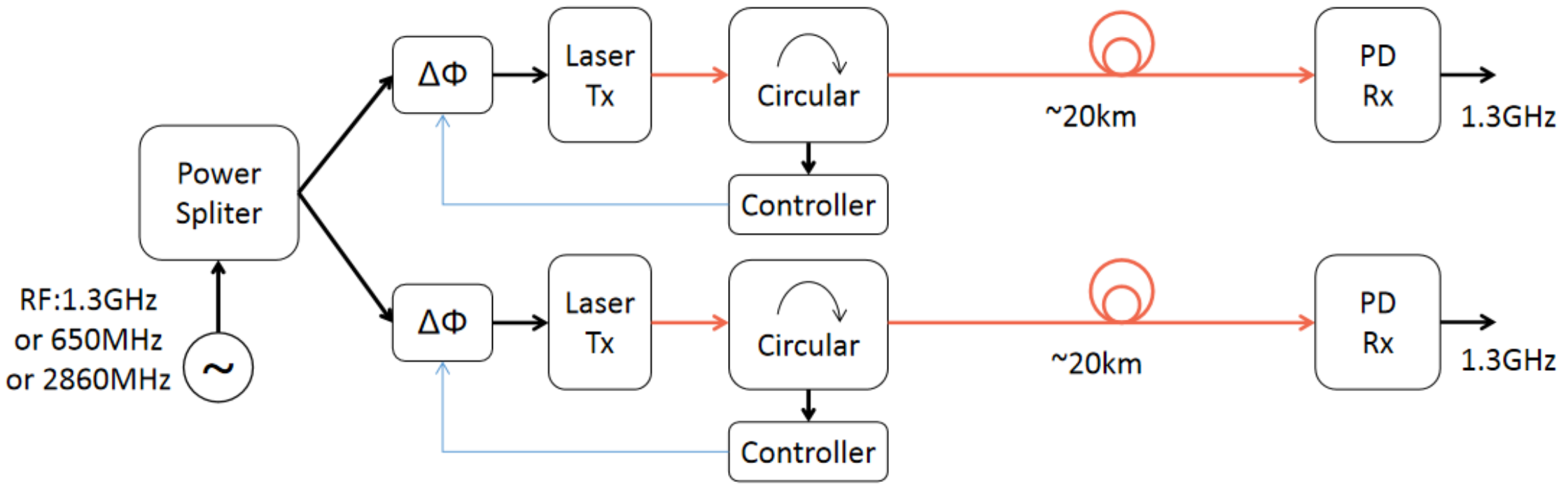

**Figure 4.3.2.13**: Principle of stable RF transfer

c) Receiver of the RF reference signals and short-distance routing:
The collider and the booster contain 10 RF stations (including 4 half-stations at detector points). Each RF station is less than 120 m so within one RF station the phase drift is small, and the reference signal can be distributed through coaxial cables with careful environmental control. There is only one operation frequency each for the collider (650 MHz) and the booster (1.3 GHz).
The injector Linac is more than 500m long and the phase stability requirement is higher than the rings. It will also adopt the optical fiber to transfer the phase reference signal. The injector Linac contains S-band accelerating tubes and Sub-Harmonic Buncher (SHB).

### 4.3.3 **Magnets**

#### 4.3.3.1 ***Overview of the Collider Magnets***

The CEPC Collider employs a double-ring lattice structure, and its magnet system consists of over 17,000 magnets, including 3,170 dipoles, 4,140 quadrupoles, 3,176 sextupoles, and 7,088 correctors. Most of these magnets are conventional, with some exceptions for quadrupoles near the IR region, which are superconducting. These magnets cover over 90% of the circumference of the 100 km ring [1]. To optimize power consumption, manufacturing, and operation costs of the magnets, a significant effort has been made, similar to that in the LEP and FCC-ee [2-4].

Special technologies have been implemented to reduce magnet costs, including core steel dilution for dipoles and using aluminum coils instead of copper. To decrease power consumption, the current density of the coils has been designed to a relatively low value. Furthermore, the magnets are designed to operate at low current and high voltage, which reduces power loss in cables. The dual-aperture magnet design for the main dipoles and quadrupoles can reduce power loss by approximately 50%. In this design, the $e^+$ and $e^-$ beam centers are separated by 350 mm.

Table 4.3.3.1 to Table 4.3.3.3 list the parameters of the main magnets of the CEPC Collider.

**Table 4.3.3.1:** The magnets requirement of CEPC collider.

| Name | Quantity | Gap [mm] | Bmax [Gs] | $L_{eff}$ [m] | GFR (H×V) [mm] | Quality [$10^{-4}$] |
|---|---|---|---|---|---|---|
| B0A | 1024 | 66 | 597 | 19.743 | 11.6×4 | 3 |
| B0B | 1920 | 66 | 597 | 23.142 | 11.6×4 | 3 |

**Table 4.3.3.2:** The quadrupole magnets requirement of CEPC collider.

| Name | Quantity | Aperture [mm] | $G_{max}$ [T/m] | $L_{eff}$ [m] | GFR (R) [mm] | Quality [$10^{-4}$] |
|---|---|---|---|---|---|---|
| Q | 3008 | 72 | 10.6 | 3 | 11.6 | 3 |
| QH | 8 | 72 | 10.6 | 1.5 | 11.6 | 3 |

**Table 4.3.3.3:** The main sextupole magnet requirements of CEPC collider.

| Name | Quantity | Aperture [mm] | $G_{max}$ [T/m$^2$] | $L_{eff}$ [m] | GFR (R) [mm] | Quality [$10^{-4}$] |
|---|---|---|---|---|---|---|
| SFI | 512 | 76 | 825 | 1.4 | 11.6 | 3 |
| SDI | 1024 | 76 | –847 | 1.4 | 11.6 | 3 |
| SFO | 512 | 76 | 825 | 1.4 | 11.6 | 3 |
| SDO | 1024 | 76 | –847 | 1.4 | 11.6 | 3 |

The magnets in the CEPC are designed to cover four different field strengths or gradients to accommodate the four different beam energies ranging from 45.5 GeV to 180 GeV, corresponding to the Z, W, Higgs, and ttbar modes. Due to the large beam energy involved, synchrotron radiation is a significant issue, and water cooling in the vacuum chamber is utilized to absorb the power. Lead blocks are also inserted between the coil and vacuum chamber for protection.

This section focuses on the design and study of the main magnets, which are simulated using OPERA 2D/3D [5]. Additionally, two magnet prototypes have been built to verify the design.

#### 4.3.3.2 *Dual-Aperture Dipole Magnets*

As the circumference of the Collider is large (100 km), the dipole magnets have very low field strength, primarily 153 Gs at 45.5 GeV and only 600 Gs at the maximum energy of 180 GeV. Out of the 3,170 dipole magnets, 2,944 are primary series dual-aperture magnets. The I-shaped yoke consists of two plate poles and a narrow center yoke to enhance the field strength in the steel, particularly at low energy levels, and improve the magnetic performance of the magnet. The dual-aperture structure provides identical fields in both apertures. Due to their length, the bending magnets are divided into several sections, sharing a common coil for convenient production and transportation. The cross-section of the dual-aperture dipole magnets is depicted in Fig. 4.3.3.1.

The dipoles are excited by two turns of aluminum busbars, connected in series in the tunnel to form a single powering circuit. An 8 mm cooling hole is used for the excitation bars.

The shielding blocks made of 30 mm thick lead will be added between the coil and the vacuum chamber to protect against radiation from high-energy $e^+$/ $e^-$ beams. These blocks are placed on both sides of the vacuum chamber. To account for the beam energy sawtooth effect, trim coils are used for both apertures to adjust the field independently by ±1.5%. The trim coils are wound on the upper and lower yoke of the dipole. According to simulation results, the trim coils in one aperture do not appear to have a significant effect on the field in the other aperture [7].

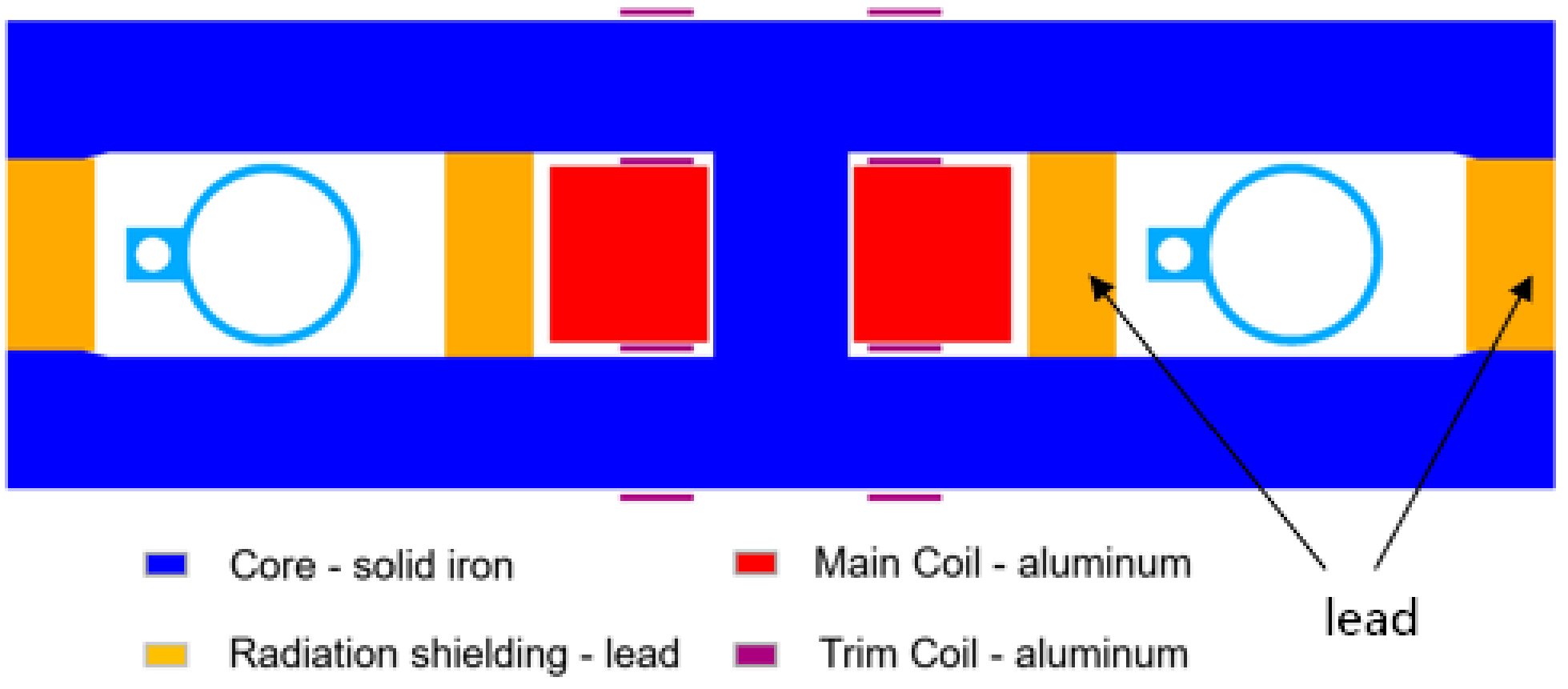

**Fig.4.3.3.1:** Cross section of dual-aperture dipole magnet

In the R&D stage, a prototype of a dual-aperture dipole magnet with sextupole component was built [7-8] and is as shown in Fig. 4.3.3.2. The dipole field is the same in the two apertures and the sextupole gradients have opposite polarities.

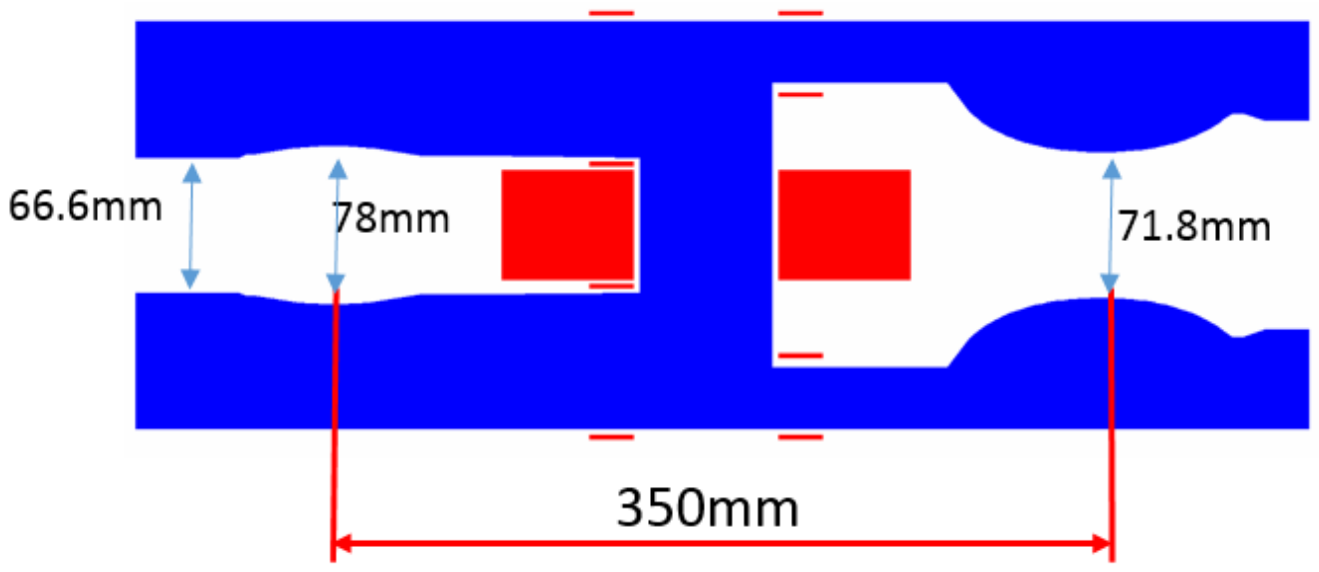

**Fig.4.3.3.2:** Cross section of dual-aperture dipole magnet with sextupole component

The length of the prototype magnet is only 1 m, so an end chamfer was used to reduce the end effect and achieve good integral field homogeneity as shown in Fig. 4.3.3.3. The fields are very low in the yoke and are independent in the two apertures. When the trim coils are excited with an adjustment capability of ±1.5%, the multipole fields in the two apertures are almost the same as the original one, which are shown in Table 4.3.3.4.

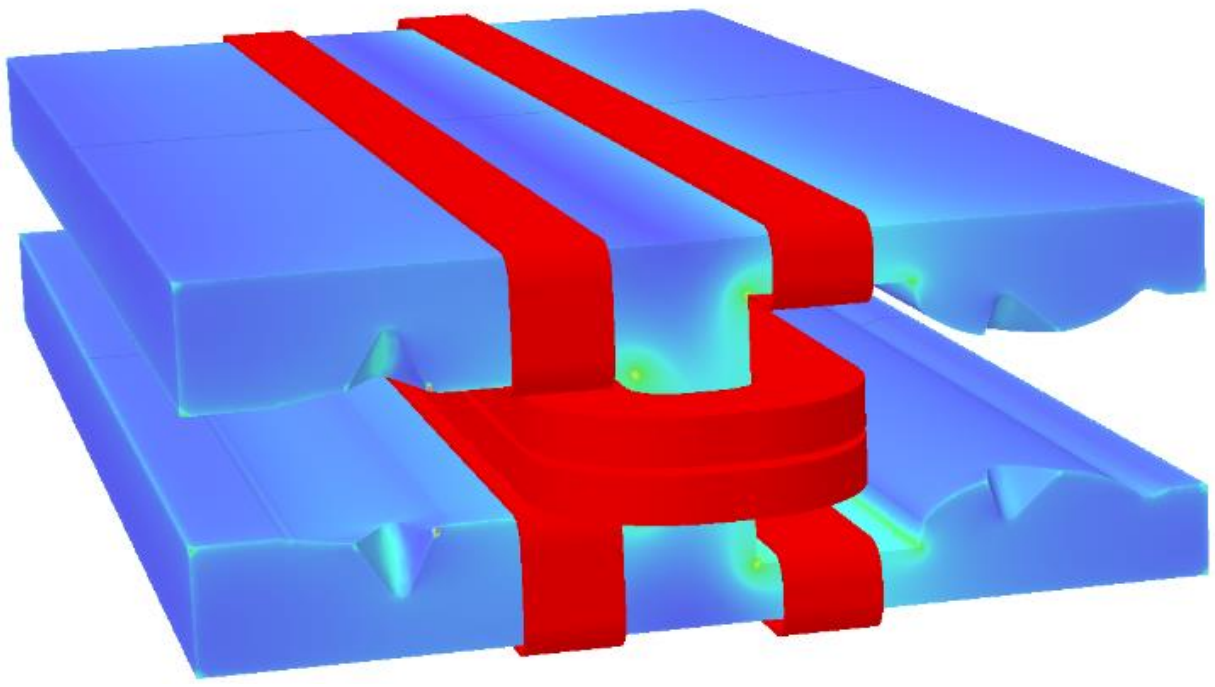

**Fig.4.3.3.3:** 3D field distribution in twin aperture dipole-sextupole prototype magnet

**Table 4.3.3.4:** 3D Calculated integrated field harmonics in the two apertures (units)

| n | bn_left | bn_left+1.5% | bn_left-1.5% | bn_right |
|---|---|---|---|---|
| 1 | 10000 | 10000 | 10000 | 10000 |
| 2 | -2.72 | -2.51 | -1.83 | 0.23 |
| 3 | 113.60 | 113.80 | 113.85 | -155.236 |
| 4 | 2.28 | 2.35 | 2.38 | -0.44 |
| 5 | -3.11 | -3.09 | -3.08 | 2.89 |

The twin aperture magnet is illustrated in Fig. 4.3.3.4. Its yoke is made of DT4 iron and divided into three parts: upper and lower half poles and a thin center yoke. The pole profiles are machined by a CNC machine. The common coil is composed of four-turn aluminum bus bars and is air-cooled. The enameled copper wires of the trim coils are wound on the yoke. Supporters at the C opening can reduce the pole deformation caused by its weight.

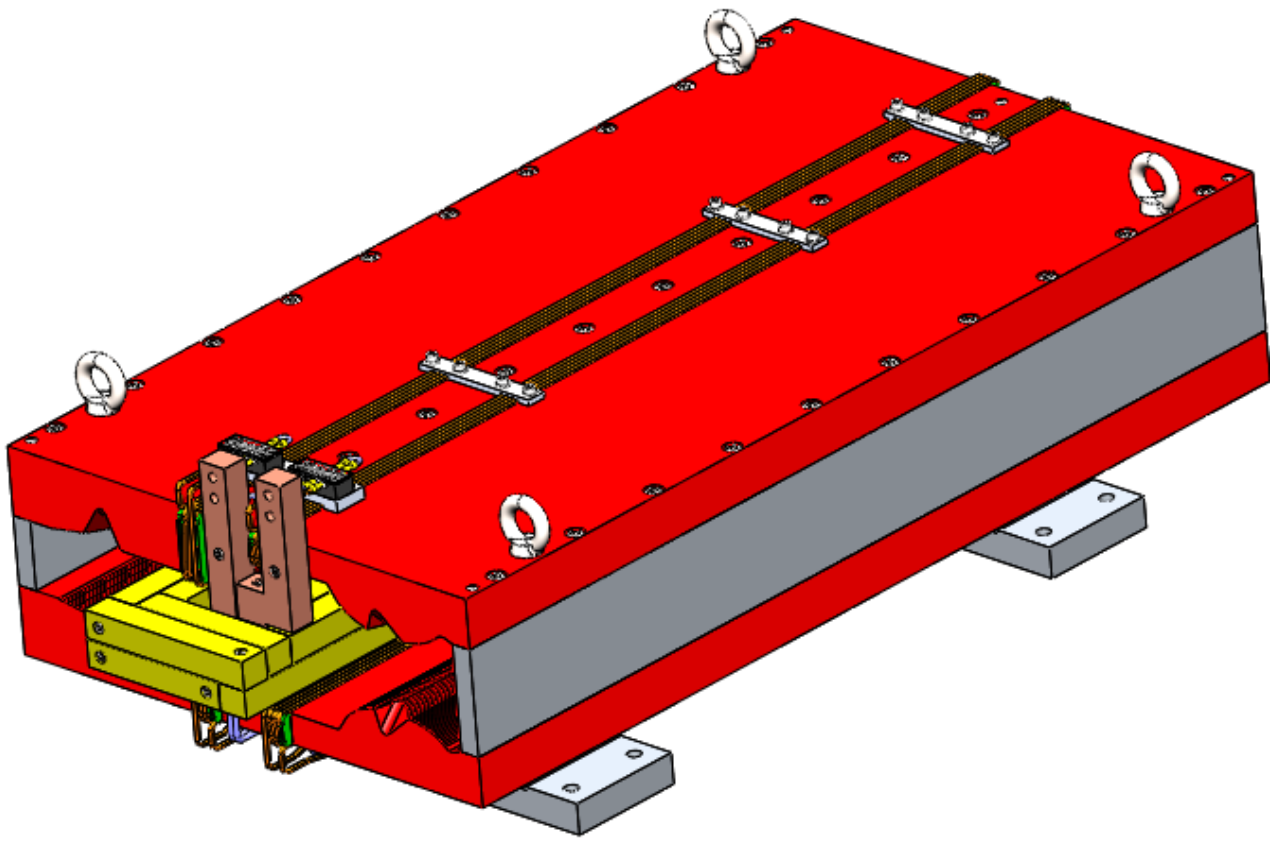

**Fig.4.3.3.4:** 3D sketch of dual-aperture dipole magnet prototype

Magnetic field mapping has been provided by a Hall probe measuring system and a stretched wire system. The measured center field difference is less than 0.3% in the two apertures. However, there is a slight difference in the transfer function of the two apertures, as shown in Fig. 4.3.3.5. This discrepancy may be due to the effect of the different residual field at the two apertures. The measured transverse field distribution at 45.5 GeV (Fig. 4.3.3.6) is consistent with OPERA 3D simulations. The multipole field coefficients were obtained from the integral field distribution by polynomial fitting. The b3 component is 113.24 and -157.23 units in the two apertures, respectively, and the other higher-order harmonics are below 5 units, which satisfies the design requirements.

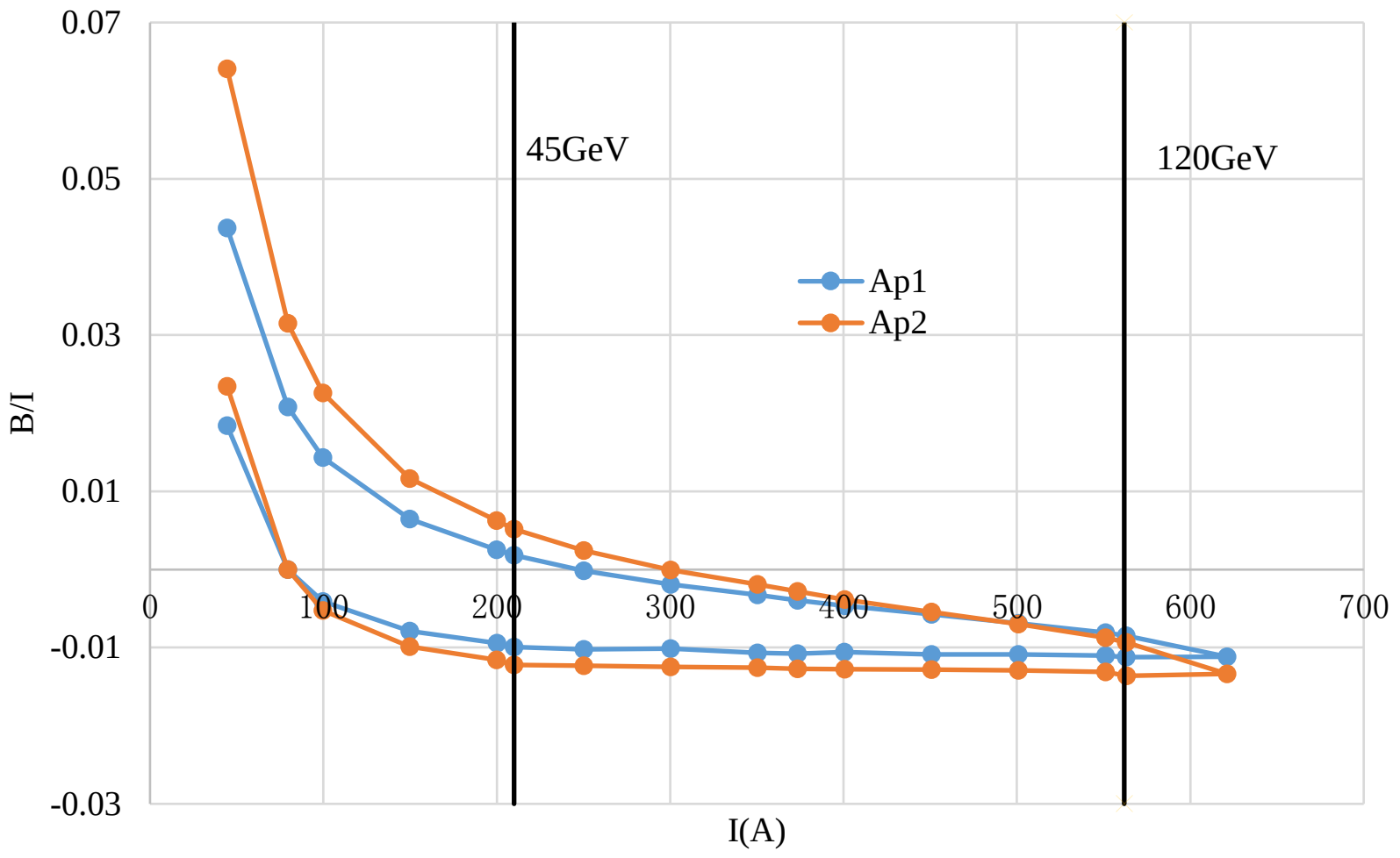

**Fig. 4.3.3.5:** Transfer function in the two apertures

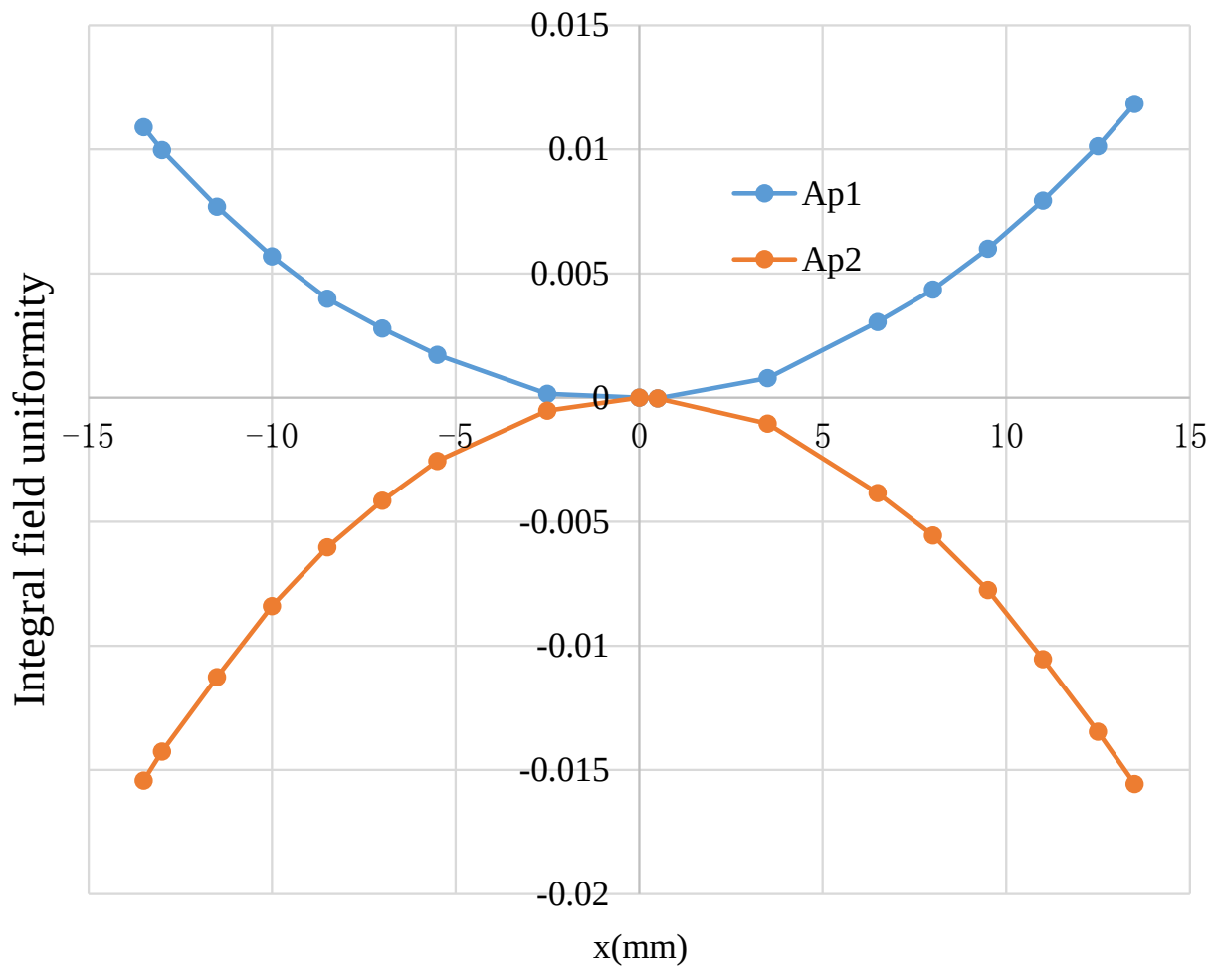

**Fig. 4.3.3.6:** Transverse distribution of the integral field in two apertures @45.5 GeV

A full-scale model of a 5.7 m-long dual-aperture dipole has been developed, which has a simple solid DT4 core pole shape. The 2D cross section is displayed in Fig. 4.3.3.7, and the mechanical design is illustrated in Fig. 4.3.3.8.

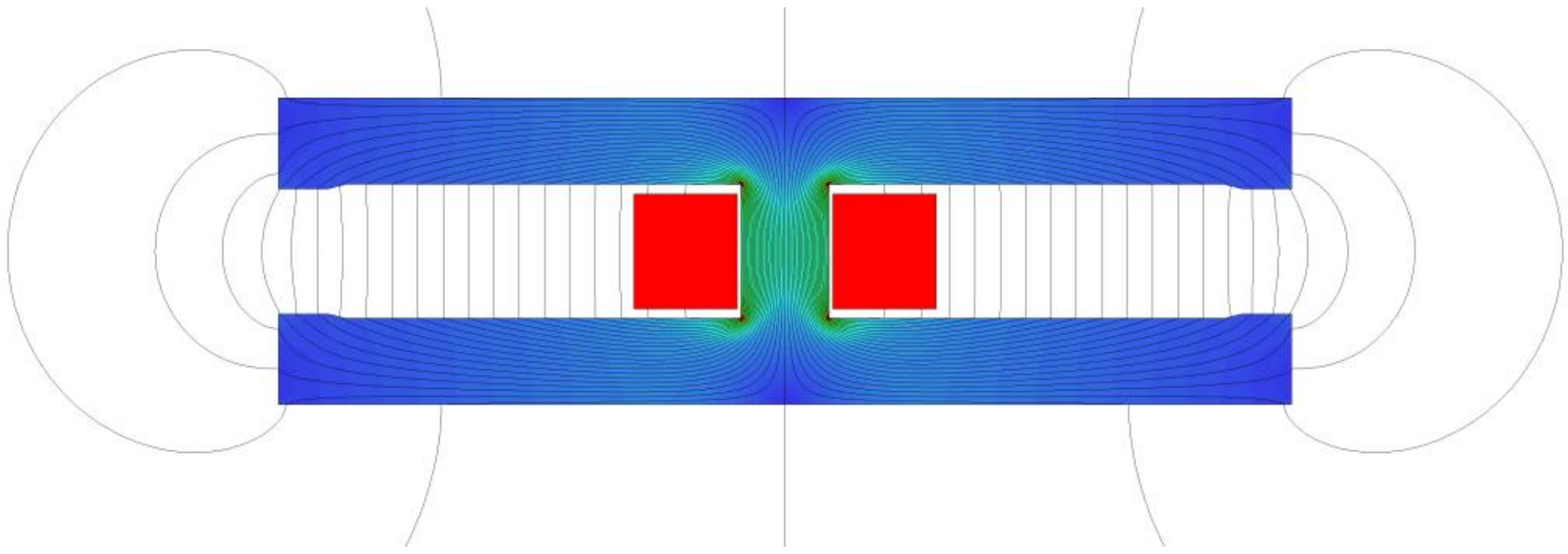

**Fig.4.3.3.7:** Bmod and flux in the 2D cross section of dual-aperture dipole magnet

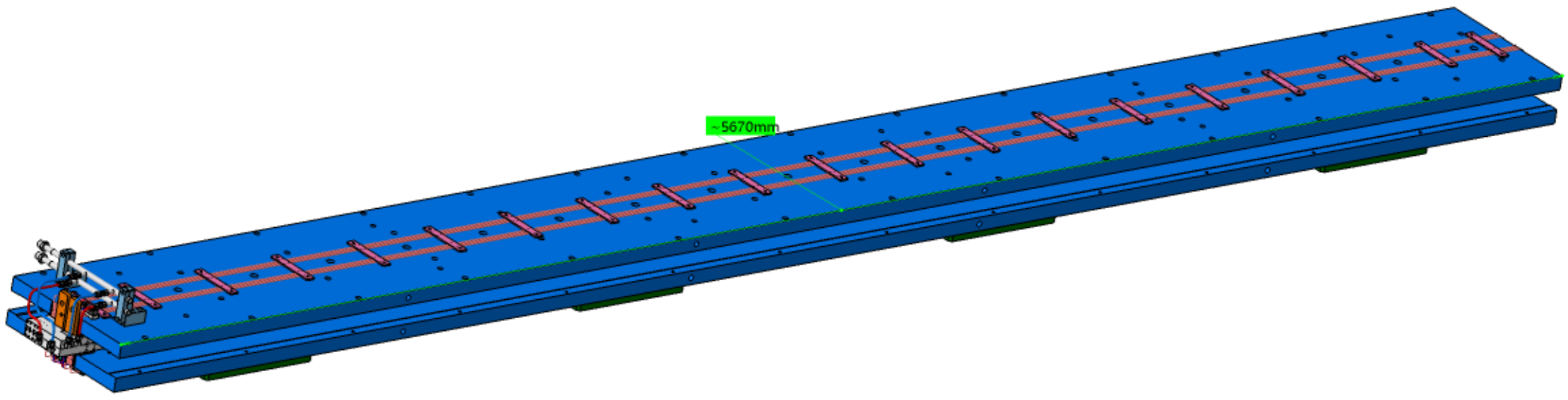

**Fig.4.3.3.8:** 3D drawing of the full-scale model.

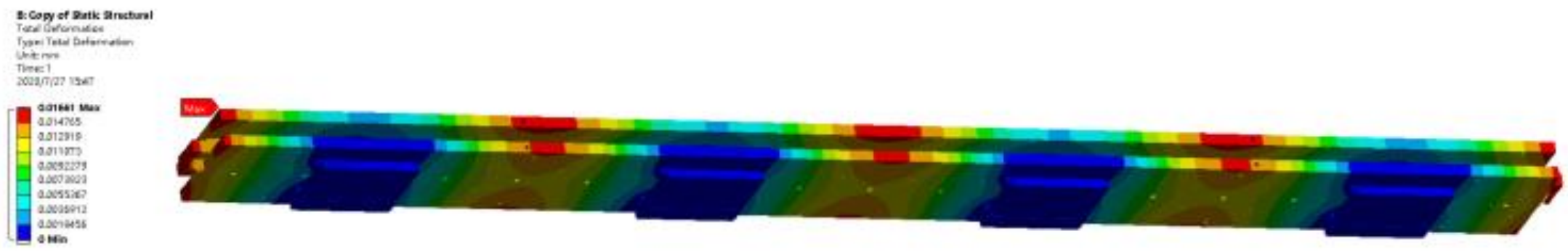

**Fig.4.3.3.9:** Deformation of the DAD

The ANSYS mechanical simulation indicates that the maximum deformation is 0.017 mm at the edge area, as depicted in Fig. 4.3.3.9. The main excitation busbars are made of pure aluminum, with a cross section of 60 mm by 27 mm and a cooling hole of 5 mm in diameter. The dipole magnet has one coil with two turns, assembled by several blocks of aluminum busbars. Currently, this prototype is in production, with the iron being finished and preassembled, as shown in Fig. 4.3.3.10. After the final assembly, the magnet is depicted in Fig. 4.3.3.11. Mechanical tolerances have been checked, and the gap height's tolerance is within ±0.05 mm. This magnet prototype is undergoing measurement.

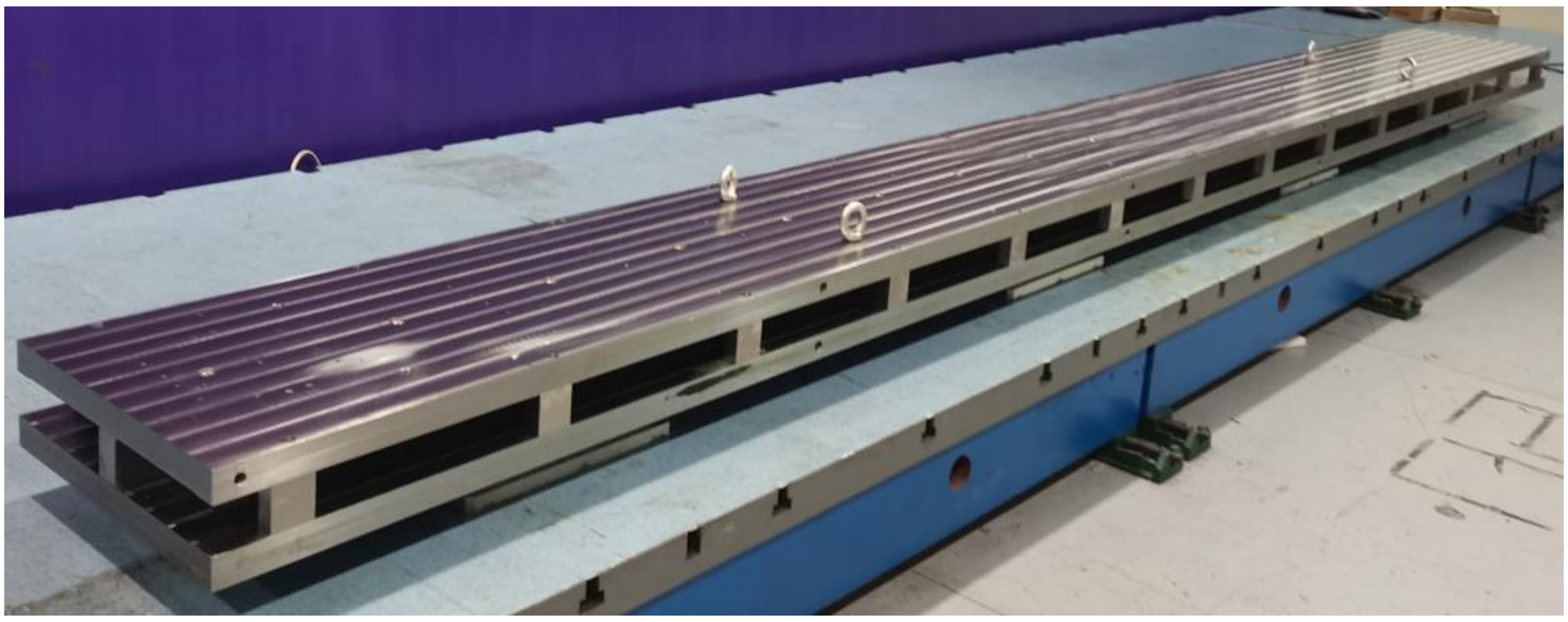

**Fig.4.3.3.10:** Iron of 5.7 m-long dipole magnet

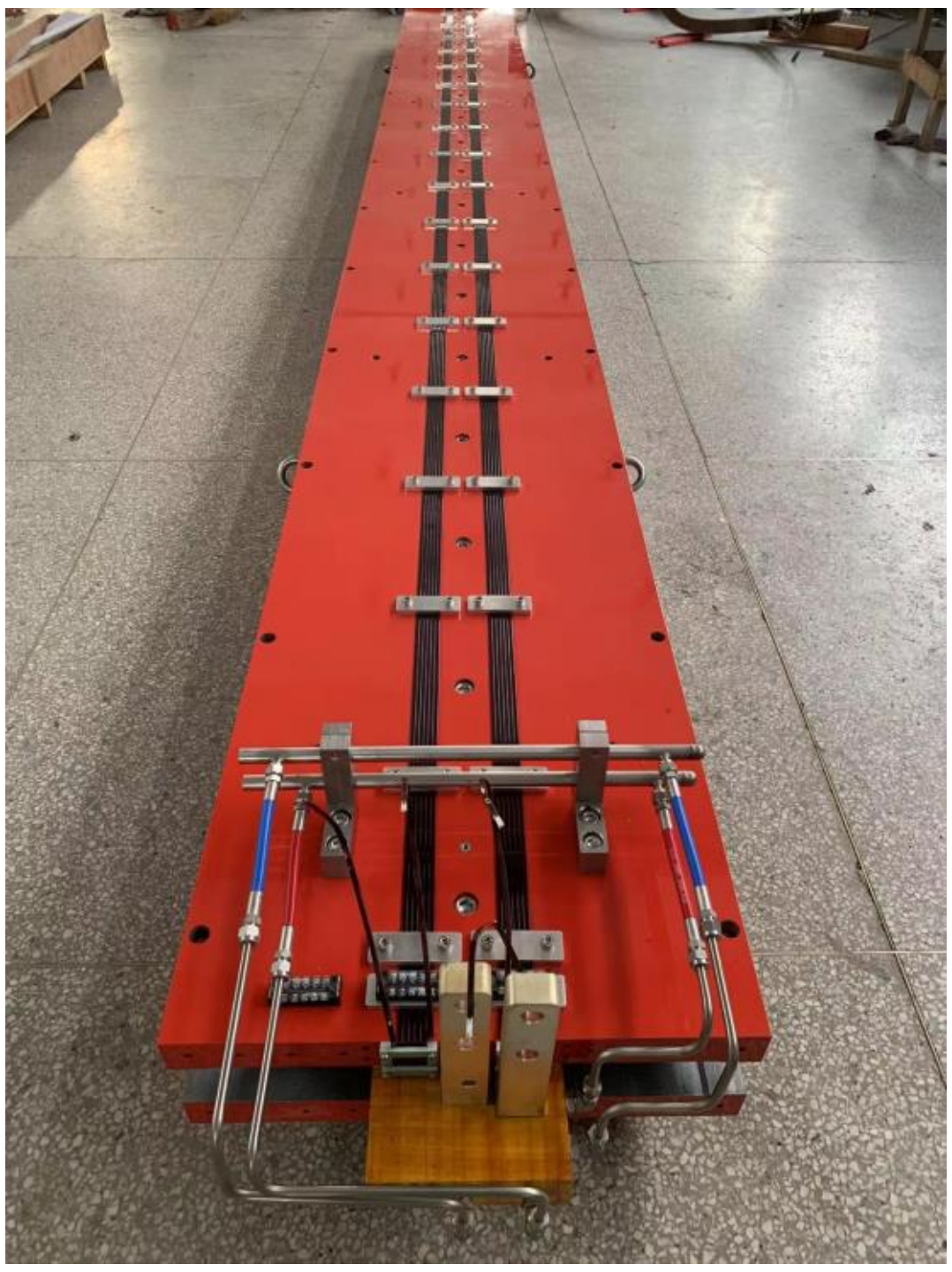

**Fig.4.3.3.11:** The 5.7 m long dual-aperture dipole magnet prototype.

Magnetic measurements for the 5.7 m prototype employ a rotating coil measurement system, as depicted in Fig. 4.3.3.12, and are subject to absolute calibration using a Hall probe. The powering cycle spans from 0 to 1,623 A and back to 0, repeating three times. The integral transfer function curves are illustrated in Fig. 4.3.3.13. Notably, the disparity in integral field between the two apertures remains consistently below 0.1% across all energy levels.

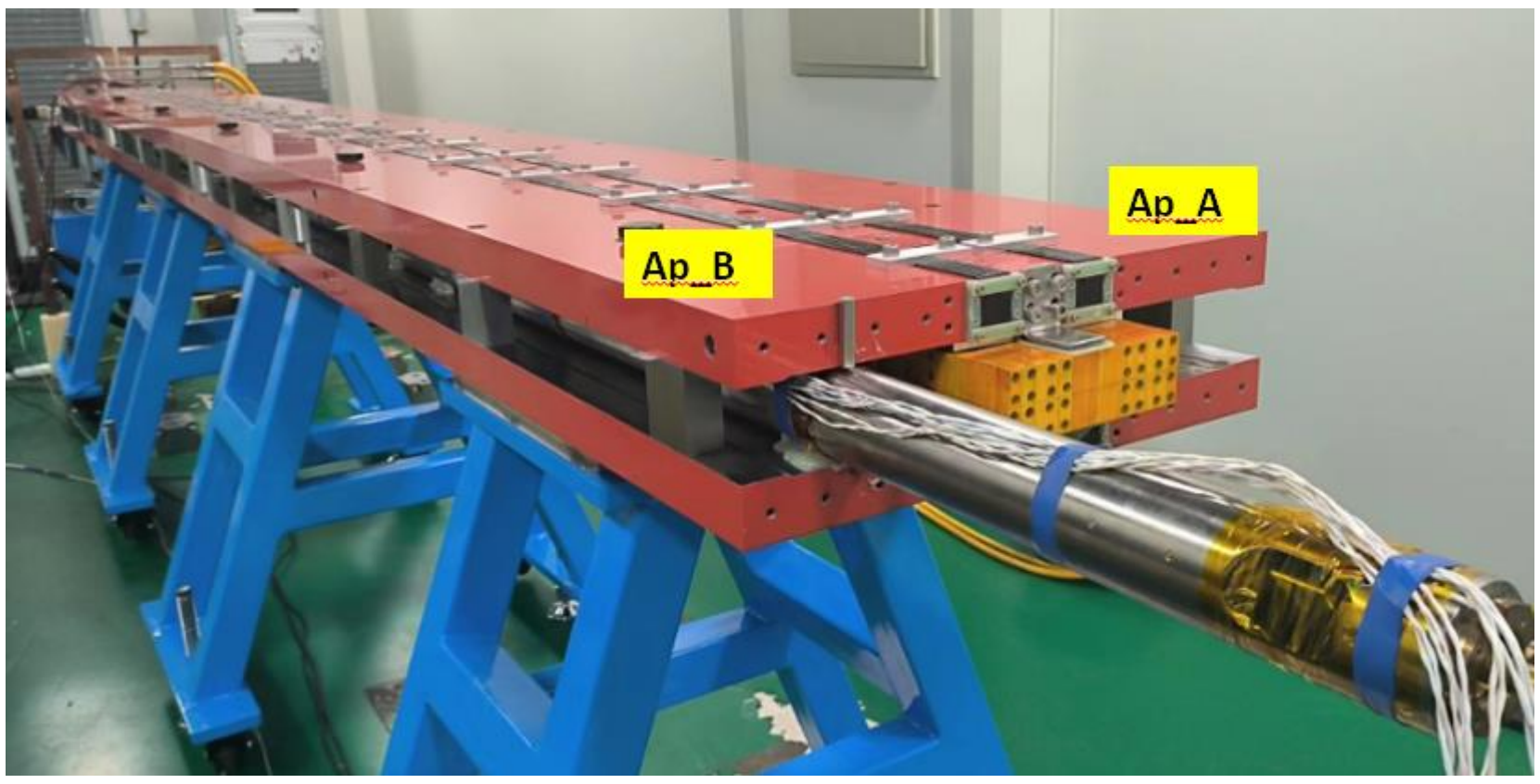

**Fig.4.3.3.12:** Long dual-aperture dipole magnet prototype on a rotating coil measurement system

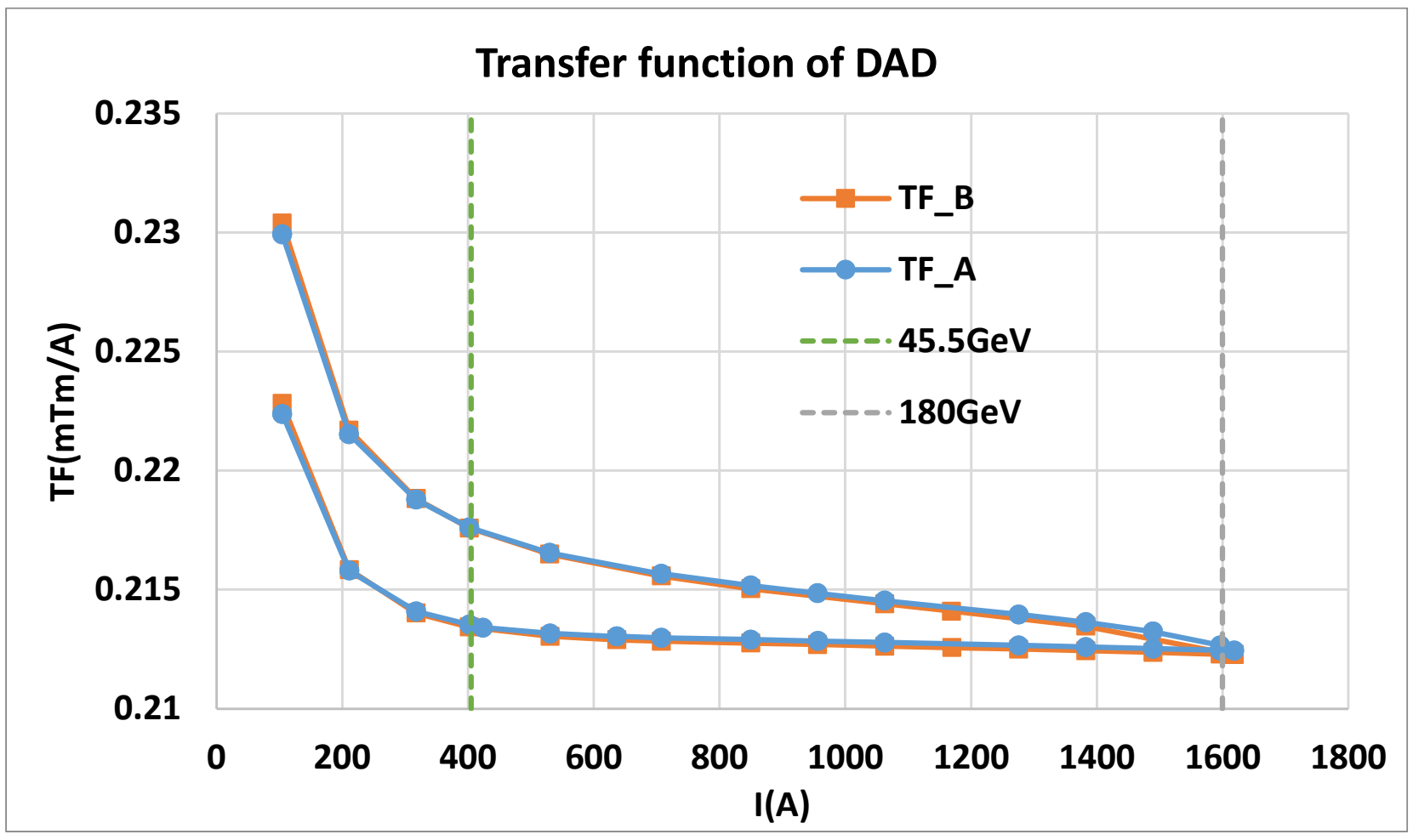

**Fig.4.3.3.13:** Integrated transfer function of the two apetures of the prototype dipole.

The magnetic field uniformity remains consistent across four energy levels and within two apertures. In the case of this prototype, it is anticipated that higher harmonics will not exceed $5\times10^{-4}$. Table 4.3.3.5 presents the measured harmonics within the two apertures, with the sextupole component being the most significant.

**Table 4.3.3.5:** High harmonics in the two apertuers @120GeV

| n | bn_A | bn_B |
|---|---|---|
| 2 | 0 | 0 |
| 3 | 3.92 | 3.88 |
| 4 | 1.03 | -1.22 |
| 5 | 0.47 | 0.54 |

The full length of the dipole magnet is approximately 20 meters, with the iron divided into four cores, each measuring about 5 meters in length. A schematic diagram of the iron and coils is presented in Figure 4.3.3.14. Aluminum busbars, each 11 meters long, are used to accommodate pairs of dipole cores. The interconnection between the cores will be welded in place in the tunnel, and similar connections are made for the trim coils using connectors and cables.

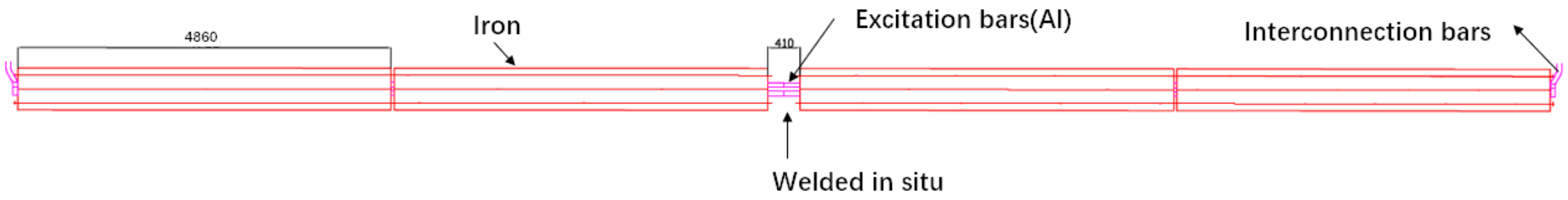

**Fig.4.3.3.14:** Four-core dipole with aluminium busbar coil.

The batch dipoles have similar parameters and designs, with the basic parameters listed in Table 4.3.3.6.

**Table 4.3.3.6:** Main parameters of the dual-aperture dipole magnet

| Magnet name | B0A | B0B |
|---|---|---|
| Quantity | 1024 | 1920 |
| Magnetic length[m] | 19.743 | 23.142 |
| Aperture [mm] | 66 | 66 |
| Field strength [Gs] @45.5 GeV | 150.9 | 150.9 |
| Field strength [Gs] @180 GeV | 597 | 597 |
| Good field region [mm] | ±11.6 | ±11.6 |
| Field errors [$\times10^{-4}$] | 3 | 3 |
| Ampere turns [At] | 3200 | 3200 |
| Turns per pole | 2 | 2 |
| Current[A] @180GeV | 1600 | 1600 |
| Size of conductor [mm×mm] | 60×27-Al | 60×27-Al |
| Max current density [A/mm$^2$] | 1.0 | 1.0 |
| Resistance of the coil [mΩ] | 1.55 | 1.79 |
| Power loss (W) @180GeV | 4.0 | 4.6 |
| Core height [mm] | 160 | 160 |
| Core width [mm] | 530 | 530 |
| Core length [m] | 4.86 m×4 | 5.72 m×4 |

### 4.3.3.3 *Dual-Aperture Quadrupole Magnets*

The number of quadrupoles in the design is 3008. A dual-aperture configuration is utilized to reduce power consumption by 50%. The two apertures of the dual-aperture quadrupole (DAQ) have opposite polarities and are separated by 350 mm. Two racetrack main coils are shared between the apertures, and trim coils in each aperture have ±1.5% adjustment capability to correct for beam sawtooth energy variations. Fig. 4.3.3.15 depicts the cross section of the DAQ, which includes trim coils wound on the yoke and lead blocks for radiation protection.

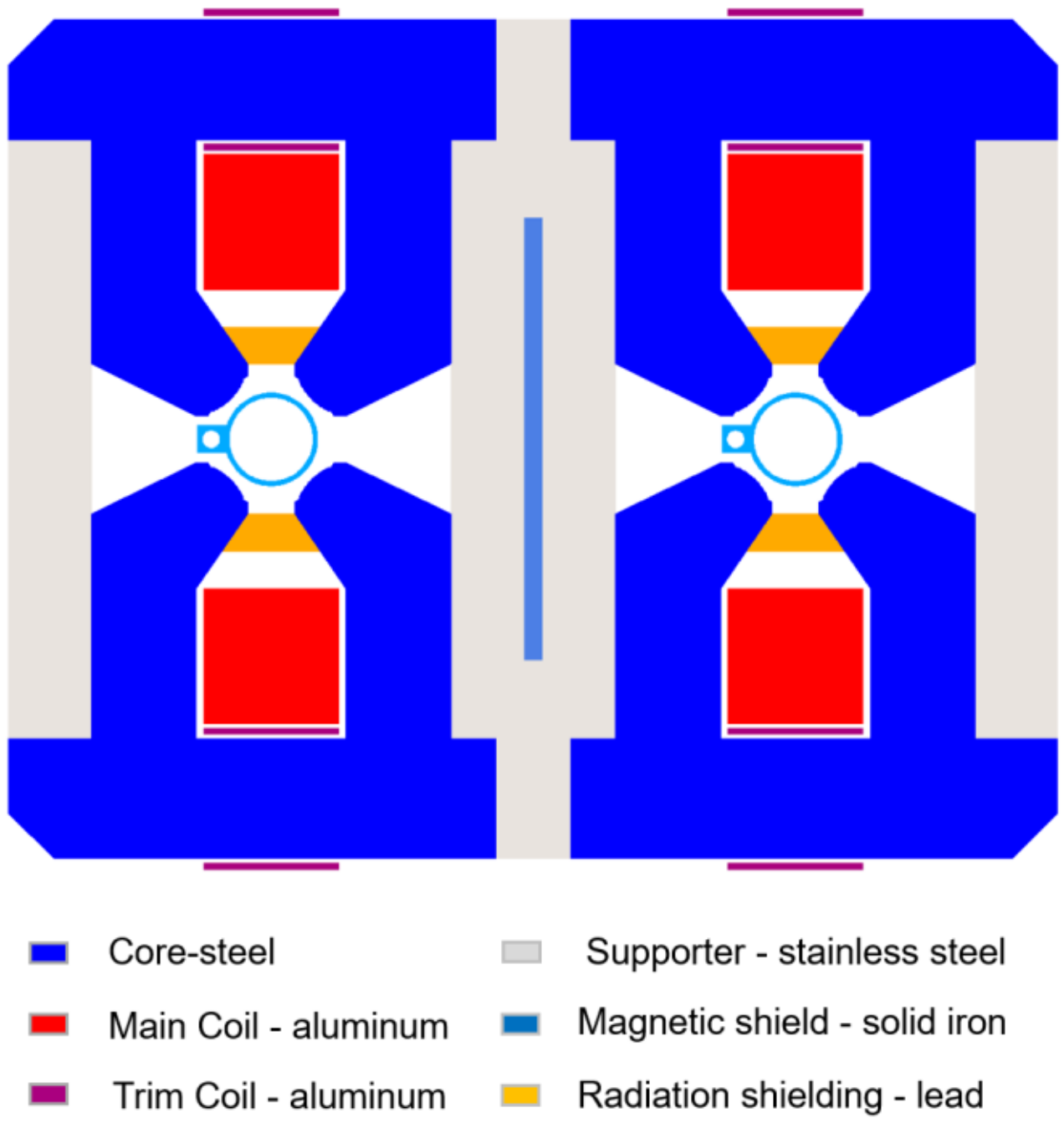

**Fig.4.3.3.15:** Cross section of dual-aperture quadrupole

Simulations reveal a cross-talk effect between the two apertures. The iron is initially mechanically decoupled into separate cores with a stainless-steel spacer, and the profile of the poles is optimized. However, field coupling introduces non-systematic harmonics, even with a 50 mm gap between the yokes. To address these non-systematic harmonics, a pure iron shielding plate is inserted between the two apertures. The thickness of the shielding plate significantly affects the non-systematic harmonics in both apertures, as illustrated in Fig. 4.3.3.16.

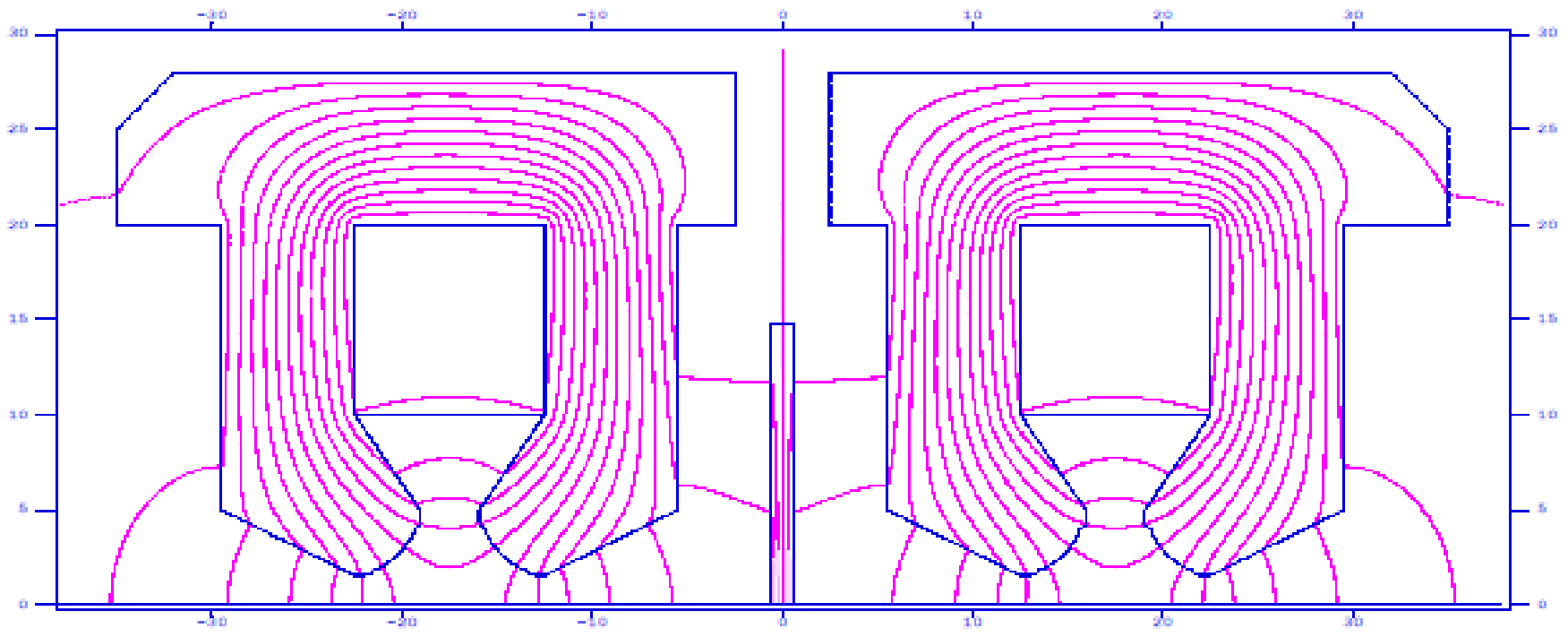

**Fig.4.3.3.16:** The magnetic flux in the dual-aperture quadrupole

A 1-meter-long prototype of a quadrupole was developed and measured using a Hall probe measurement system (Fig. 4.3.3.17). The magnet's iron core is made of 0.5 mm thick laminated low carbon silicon steel sheets, while a hollow aluminum conductor was used to reduce the price and weight. The whole magnet will be assembled with many blocks, and each piece of the pole will be compressed by a longitudinal rod, behaving like a prestressed beam.

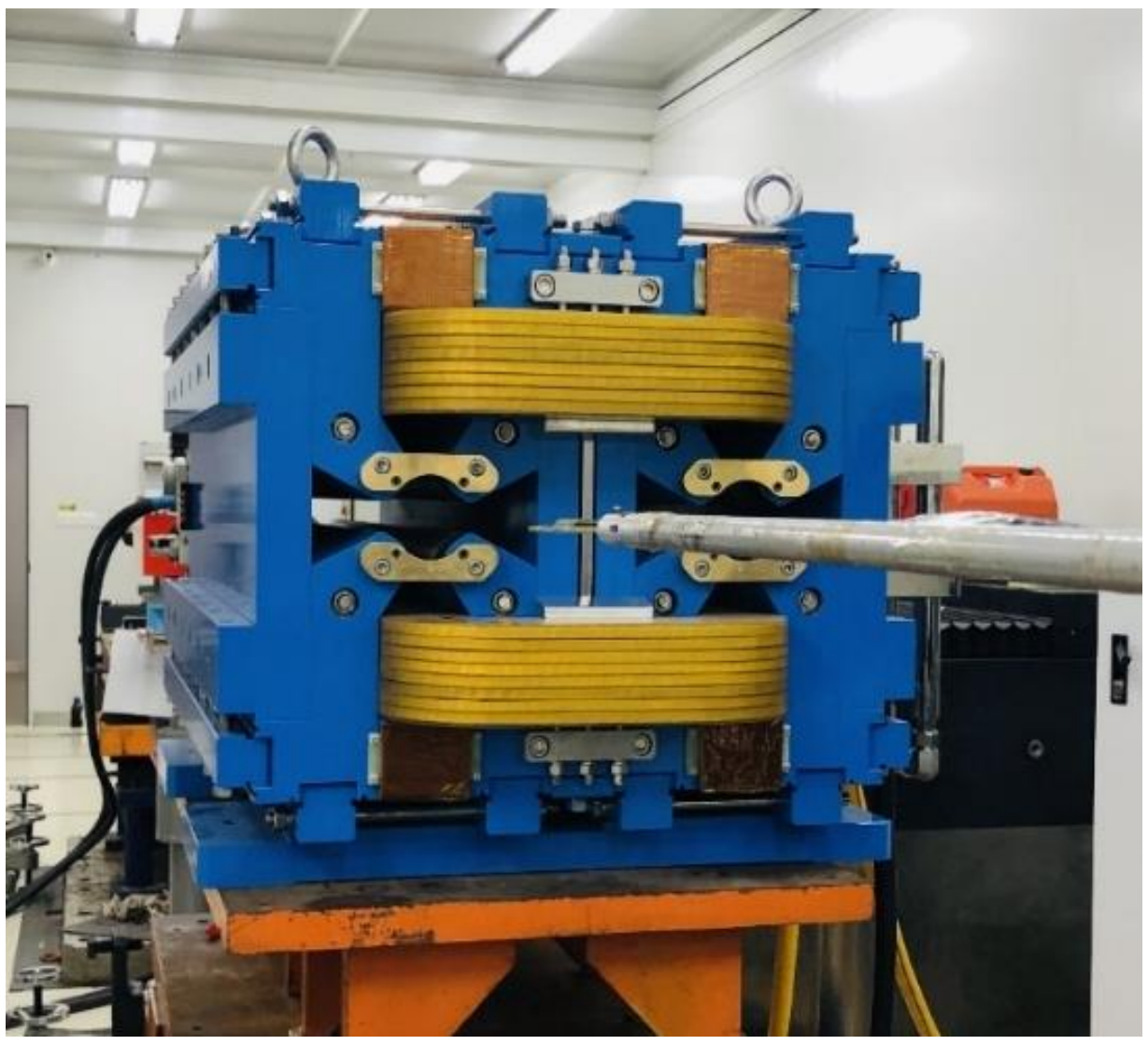

**Fig.4.3.3.17:** The DAQ prototype on a Hall probe measurement system

After the prototype's fabrication, it was measured using a Hall probe measurement system, which revealed a large magnetic shift in the X-axis during current ramping due to the center DT4 sheet's saturation, as shown in Fig. 4.3.3.18.

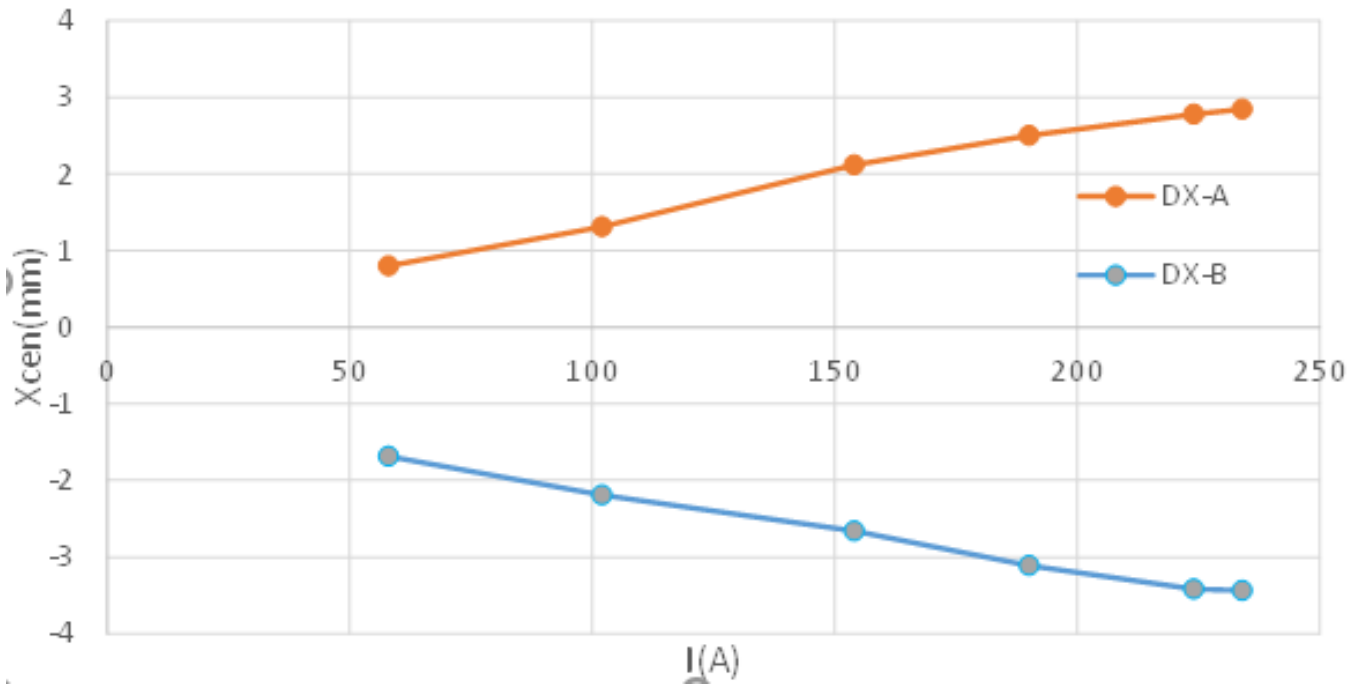

**Fig.4.3.3.18:** Magnetic center X0 shift with current

To address this cross-talk effect, a modification was made. In Fig. 4.3.3.19, a center shim was used to balance the magnetic flux symmetrically in one single aperture, based on the prototype.

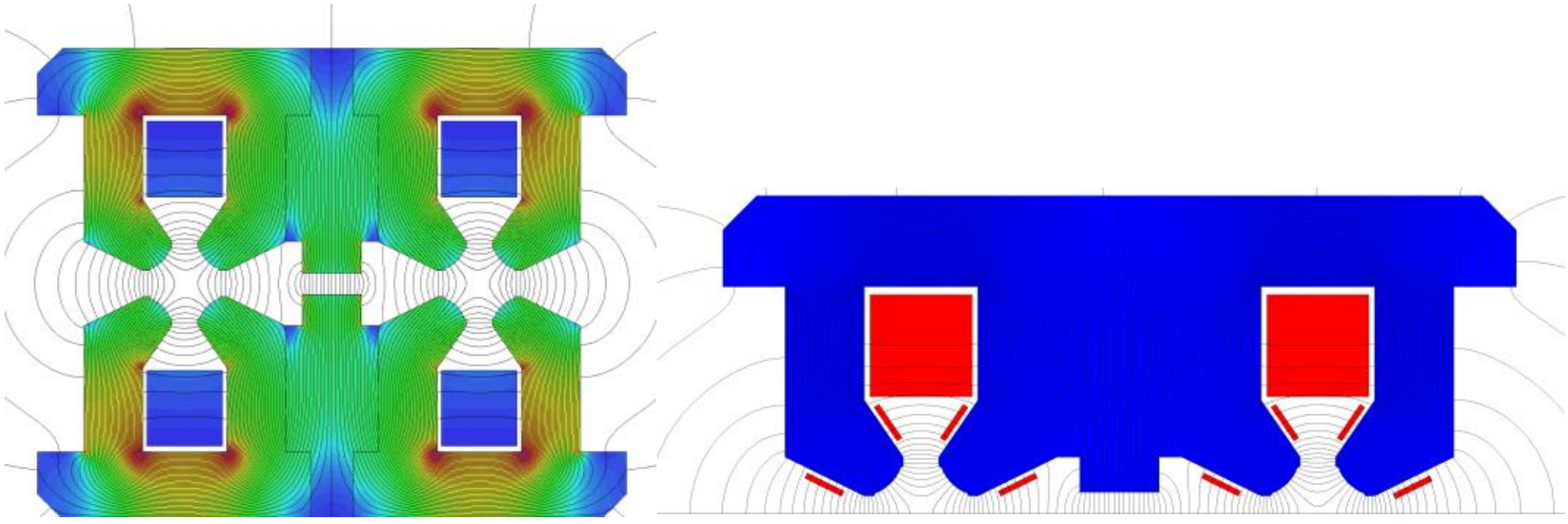

**Fig. 4.3.3.19:** The center shim and the layout of the trim coils

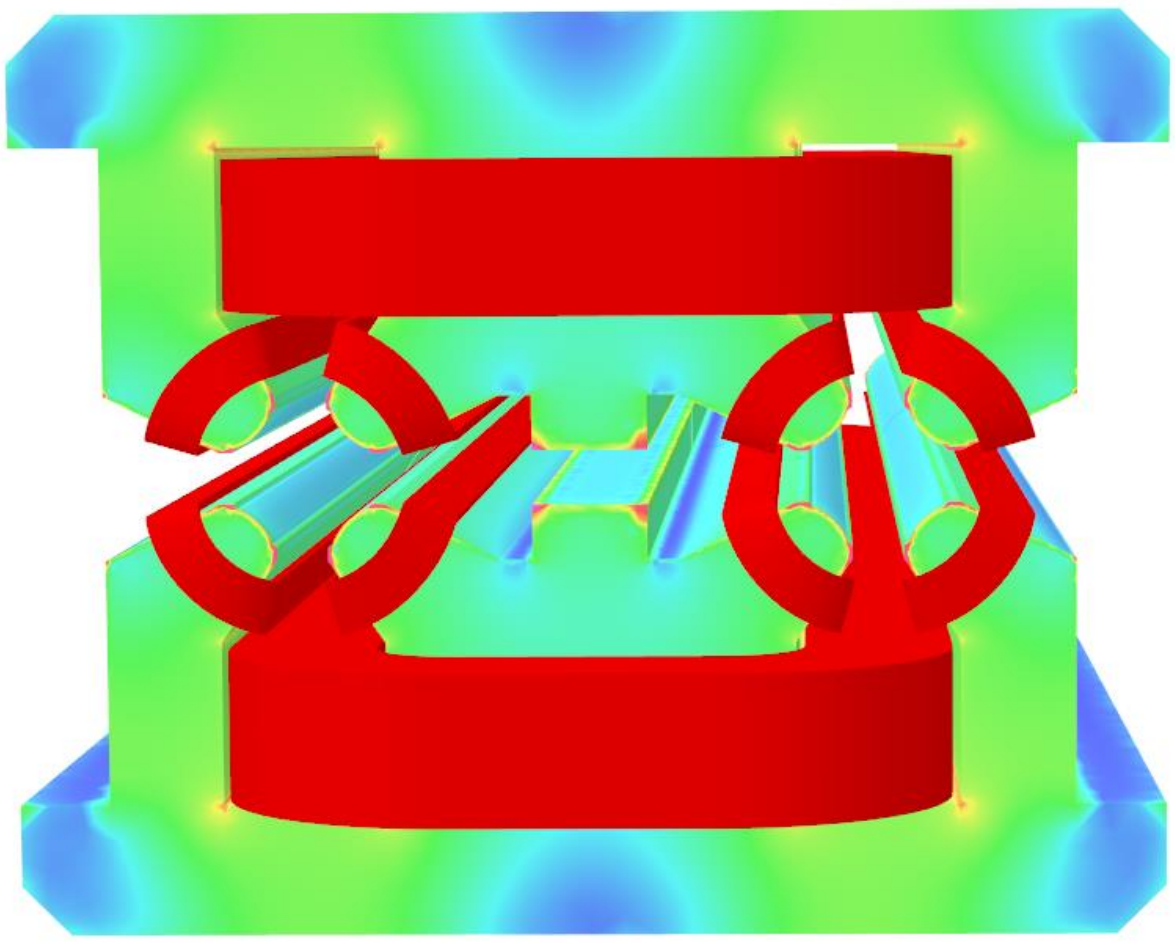

**Fig. 4.3.3.20:** 3D simulation of the modified DAQ

The trim coil is wound on the pole instead of the yoke, providing 1.5% adjustability without affecting the field harmonics. A 3D simulation is shown in Fig.4.3.3.20.

After modification, the DAQ prototype is measured using a rotating coil measurement system [9-10], as shown in Fig.4.3.3.21. The relative function in both apertures is shown in Fig.4.3.3.22, indicating a small difference between the two apertures and no iron saturation.

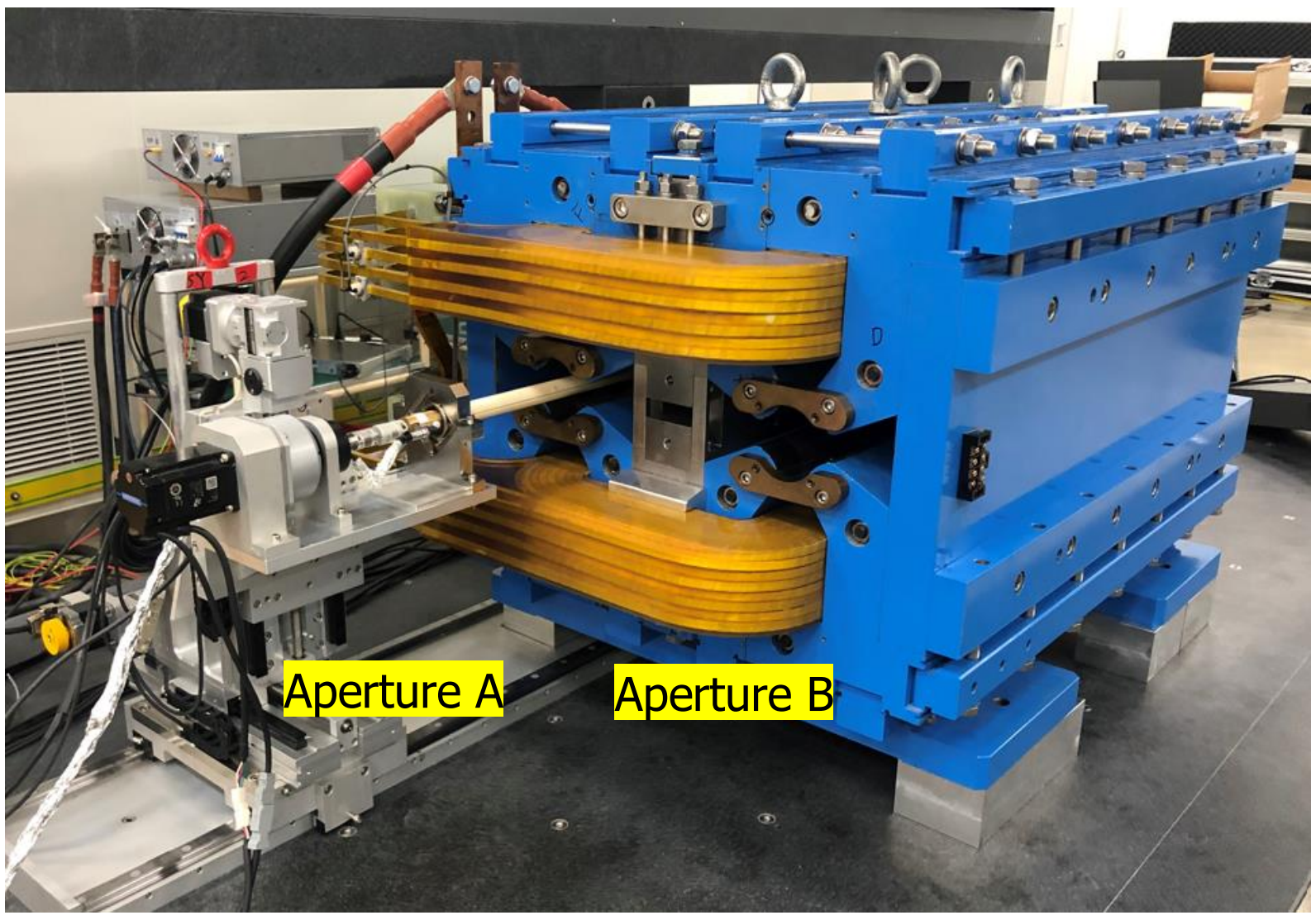

**Fig. 4.3.3.21:** DAQ prototype on a rotating coil measurement system

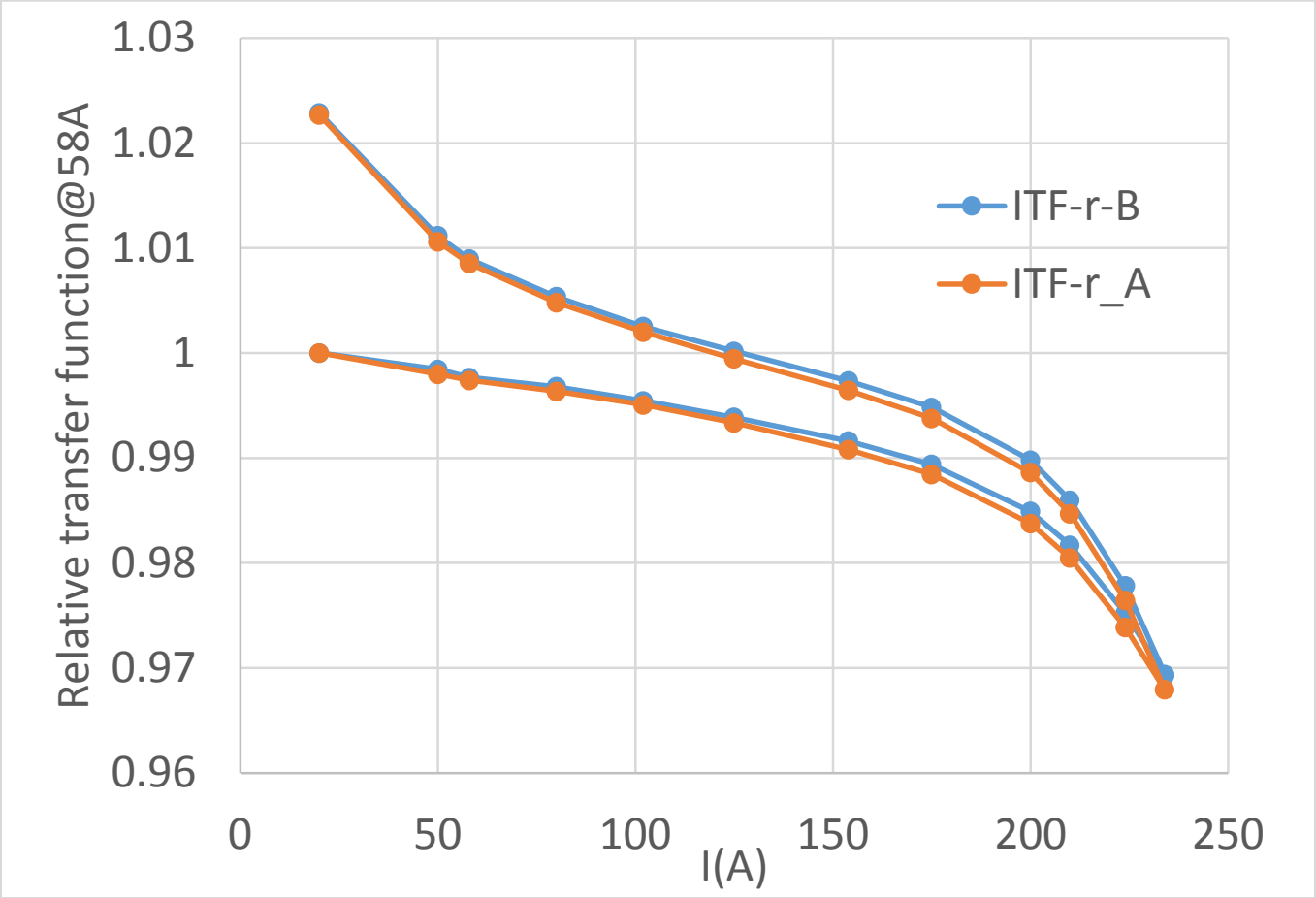

**Fig.4.3.3.22:** Relative transfer function of DAQ

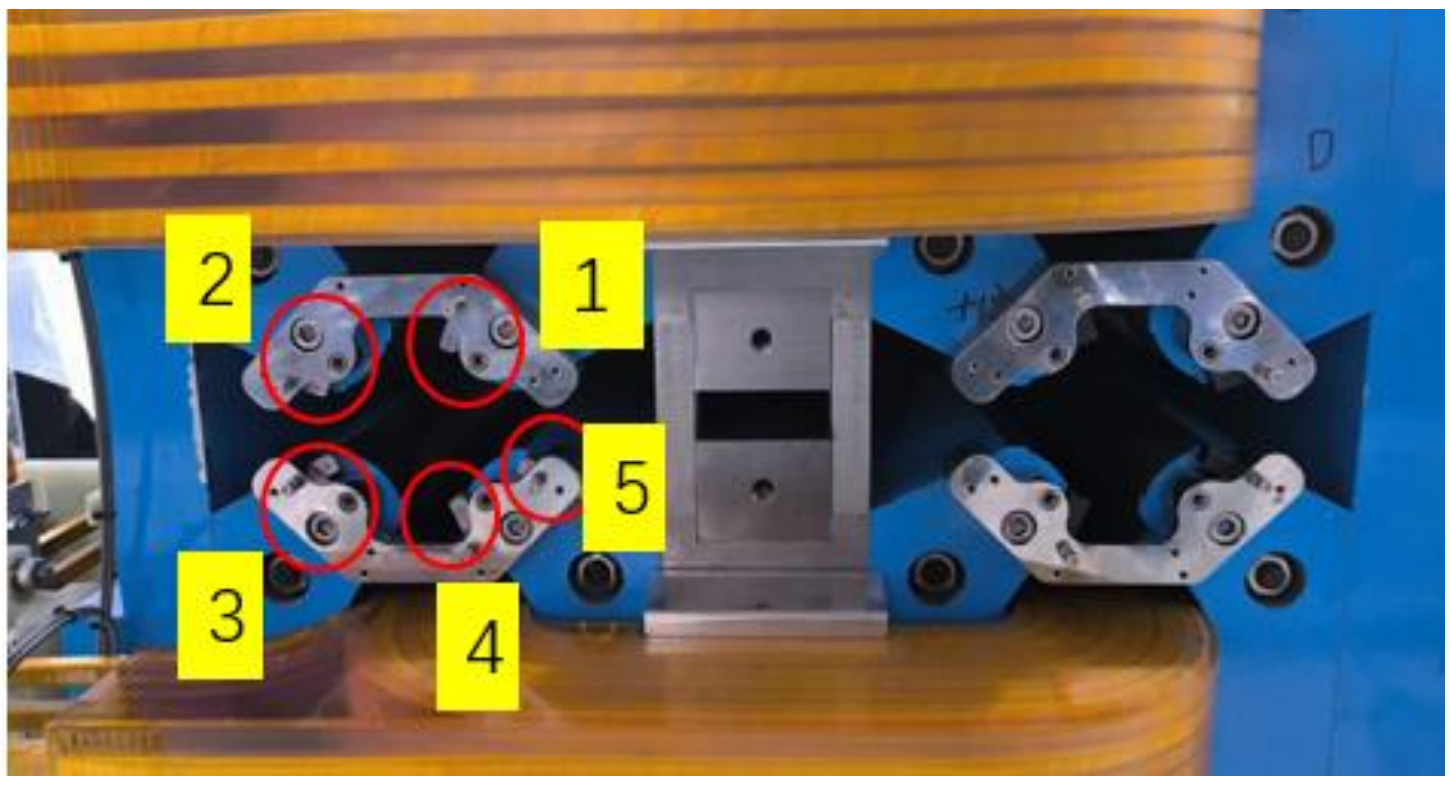

**Fig.4.3.3.23:** The magic fingers in the two apertures

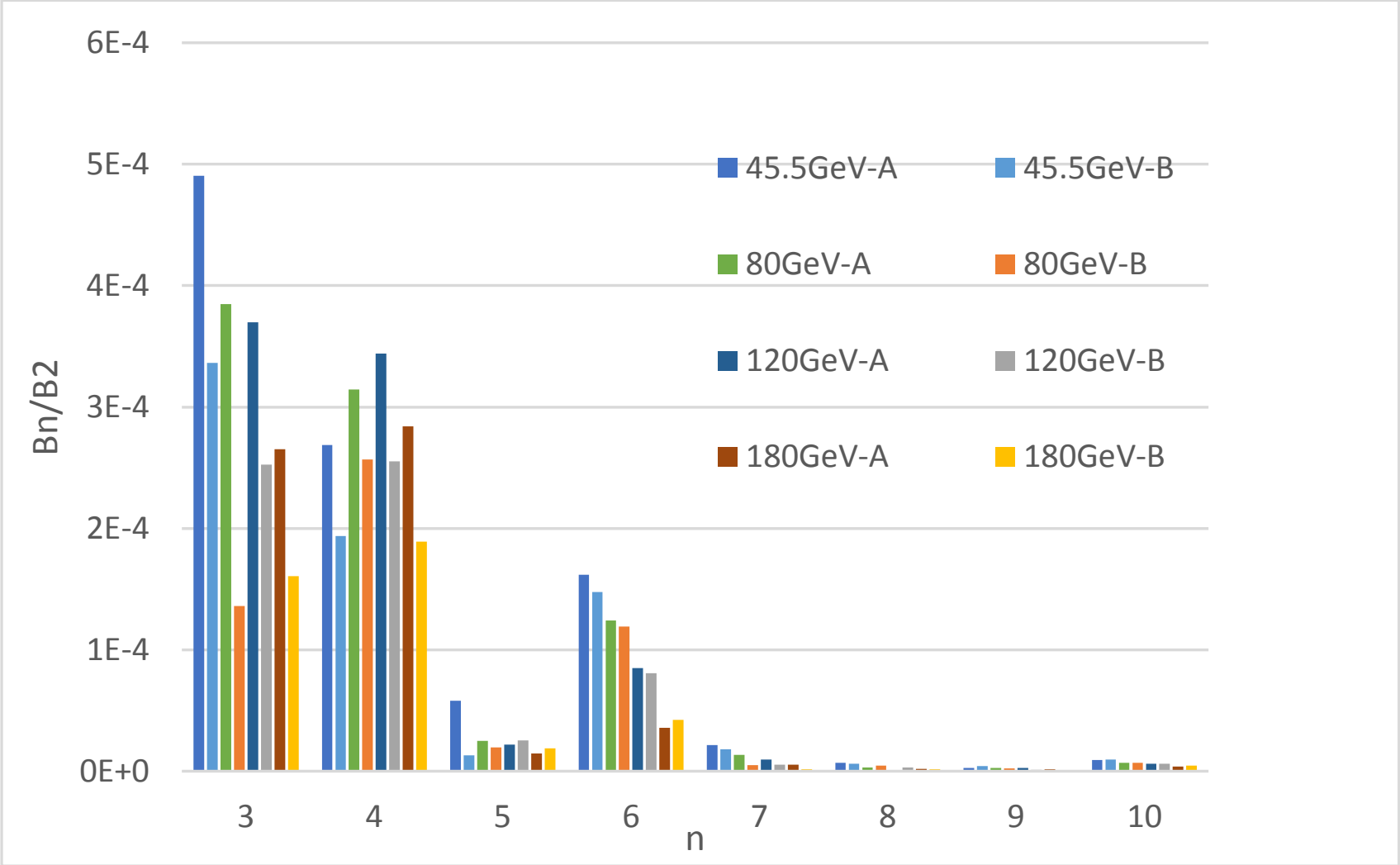

**Fig.4.3.3.24:** Higher harmonics in the two apertures at four energies

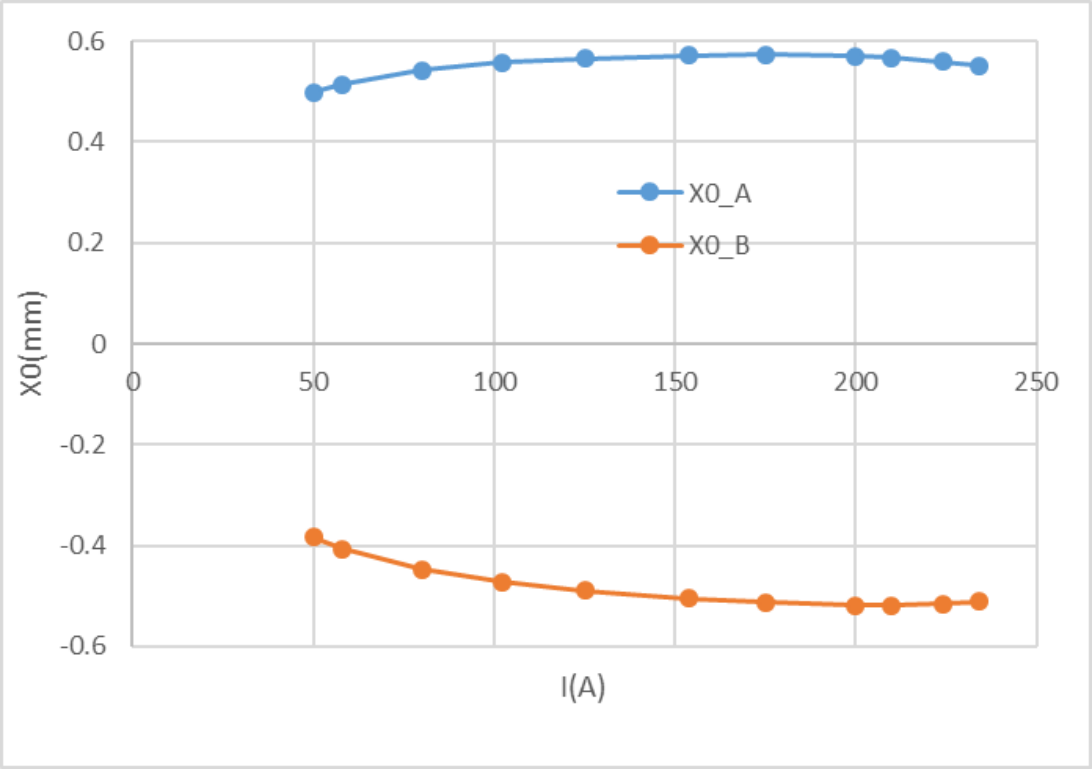

**Fig.4.3.3.25:** Magnetic center in X0 shift in the energy range.

High harmonics are measured, and all except the sextupole component are within 1 unit due to exceeding tolerances or compensation for cross talk. The sextupole component is around 9 units, but after compensation using the magic fingers as shown in Fig.4.3.3.23, it is reduced to within 5 units. Multipole components at four energies in both apertures are plotted in Fig.4.3.3.24. Magnetic center shift in X0 is reduced from 2 mm to 0.1 mm, as shown in Fig. 4.3.3.25, indicating the effectiveness of the center shim. Table 4.3.3.7 lists the main parameters. These results will inform further optimization for batch production.

**Table 4.3.3.7:** Main parameters of the dual-aperture quadrupole magnets

| Magnet name | Q |
|---|---|
| Quantity | 3008 |
| Magnetic length[m] | 3 |
| Aperture [mm] | 72 |
| Field gradient [T/m] @45.5 GeV | 1.9 |
| Field strength [Gs] @180 GeV | 10.6 |
| Radius for good field region [mm] | 11.6 |
| Field errors [$\times 10^{-4}$] | 3 |
| Ampere turns [At] | 11369 |
| Turns per magnet | 144 |
| Current[A]@180GeV | 158 |
| Size of conductor [mm×mm] | 11×11, d7, r1.5-Al |
| Max current density [A/mm$^2$] | 2 |
| Resistance of the coil [mΩ] | 377 |
| Power loss (kW) @180GeV | 9.4 |
| Water drop (Kg/cm$^2$) | 6 |
| Number of water loops | 8 |
| Velocity [m/s] | 1.41 |
| Temperature rise [°C] | 5 |
| Core height [mm] | 560 |
| Core width [mm] | 700 |
| Core length [m] | 3 |
| Magnet weight [ton] | 6 |

### 4.3.3.4 *Sextupole Magnets*

The sextupoles consist of two individual parallel magnets instead of a dual-aperture component. As the required strength for the D and F sextupoles is similar, they can be designed as a common type. However, due to the proximity of the $e^+$ and $e^-$ beams in the presence of dual-aperture dipoles and quadrupoles, the size of the sextupoles is limited, and the space between neighboring sextupoles is restricted. Figure 4.3.3.26 illustrates the cross sections and positions of two neighboring sextupoles in the two rings.

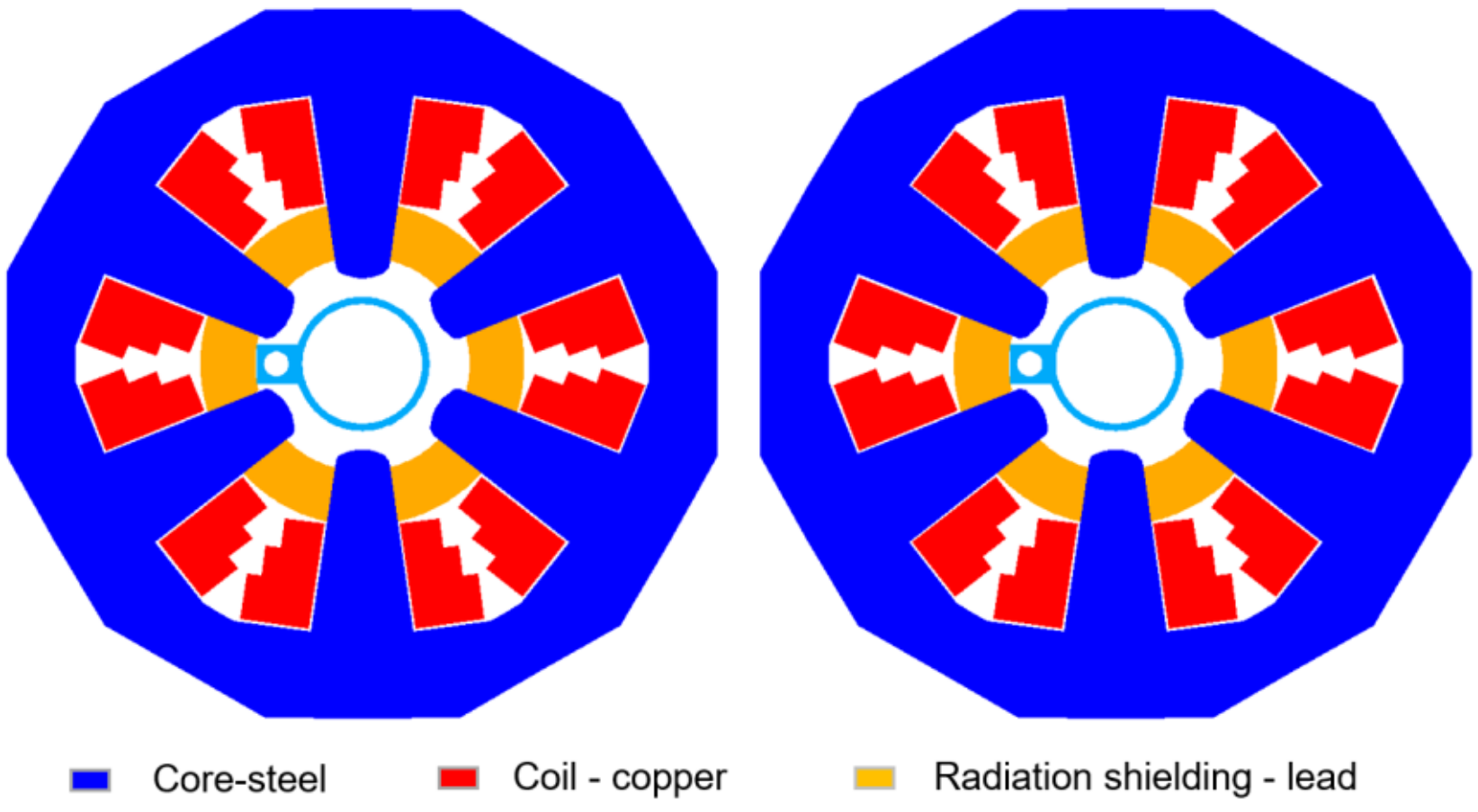

**Fig.4.3.3.26:** Cross section of two neighboring sextupoles in two rings

Despite the proximity of the two sextupoles, the field interference between them is negligible. The pole surface profile is optimized to compensate for the harmonics, as shown in Fig. 4.3.3.27 for the magnetic strength in the 3D model and Fig. 4.3.3.28 for the 3D mechanical design of the sextupoles. To cope with the narrow space and high gradient, copper conductor is used, resulting in high current density and power. The design parameters are listed in Table 4.3.3.8.

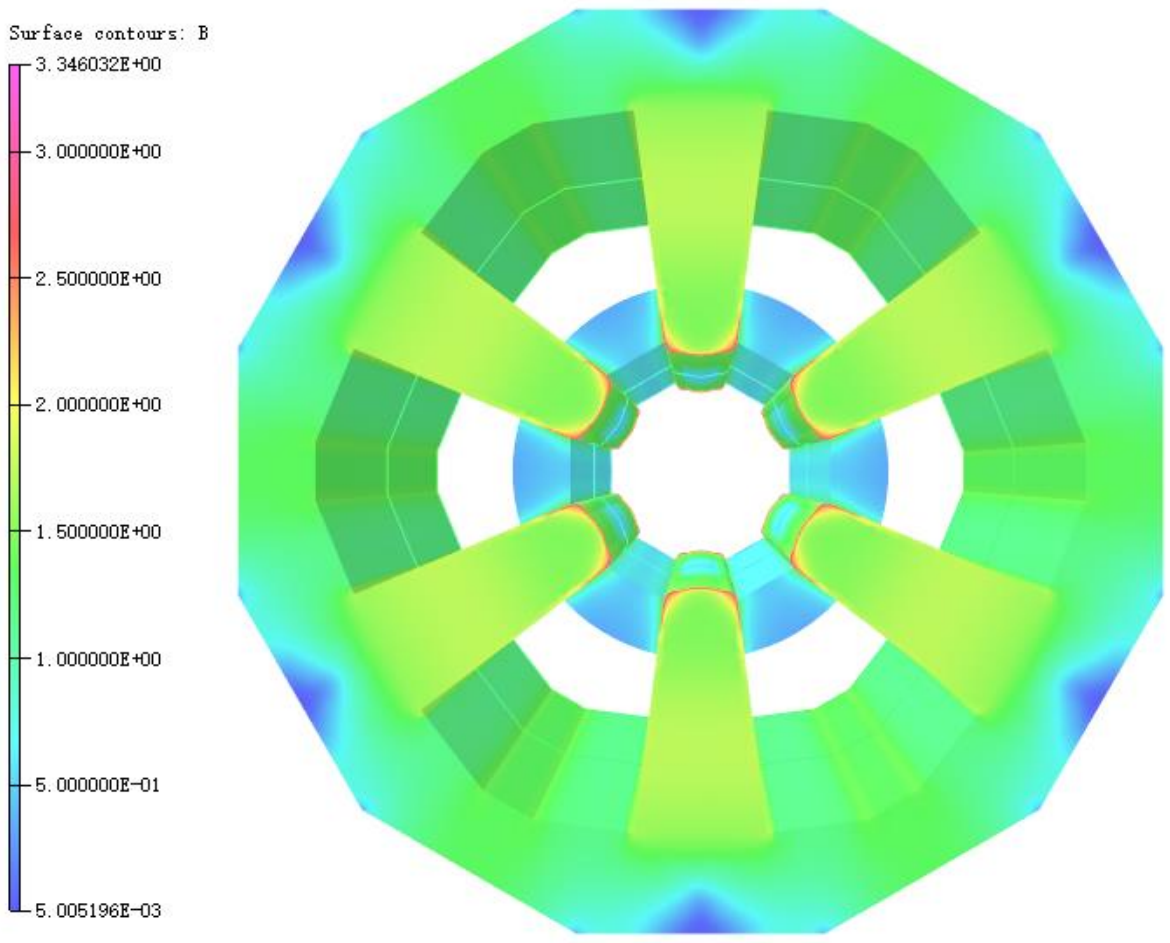

**Fig. 4.3.3.27:** Bmod distribution in the sextupole iron.

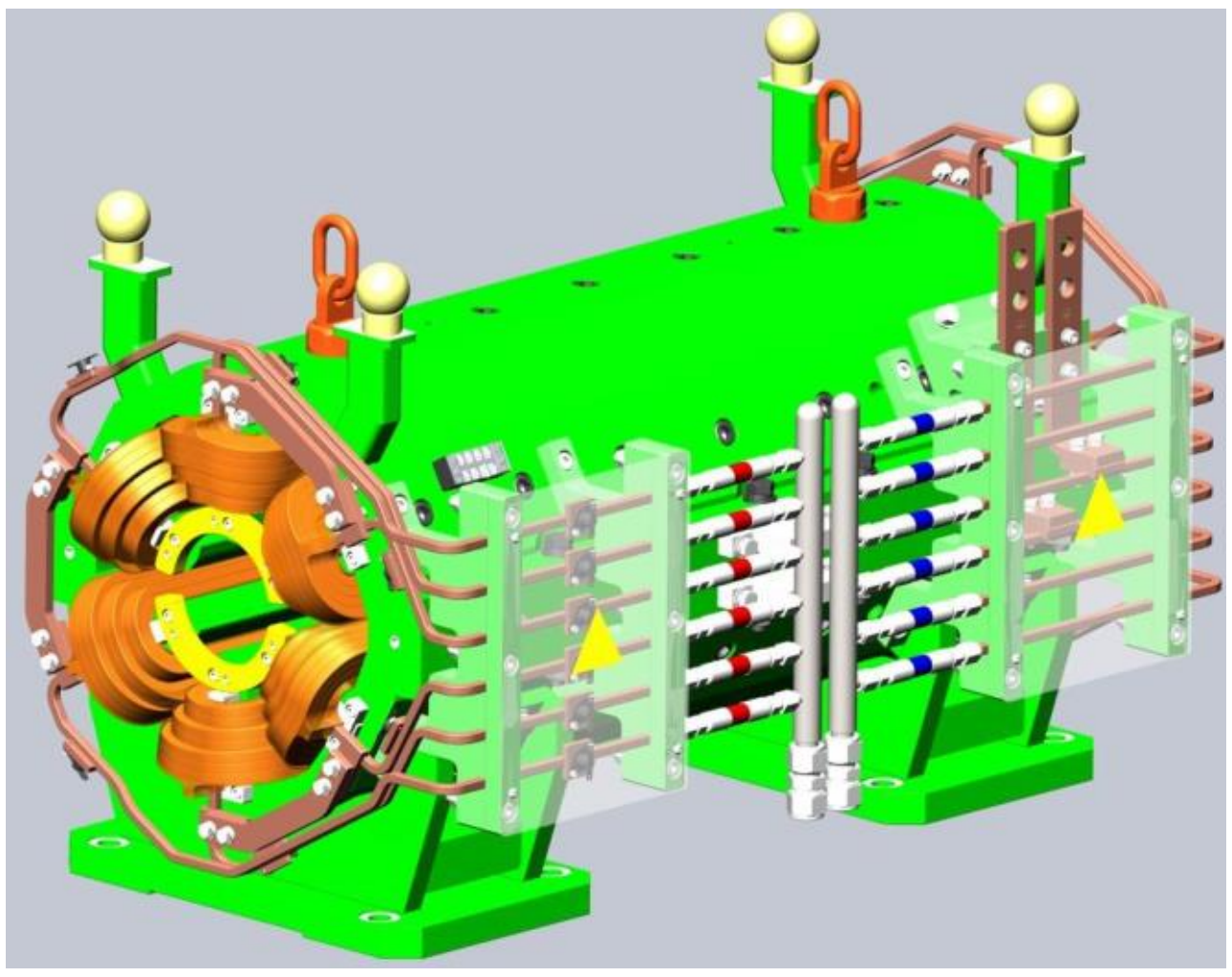

**Fig.4.3.3.28:** 3D mechanical design of sextupole.

**Table 4.3.3.8:** Main parameters of the sextupoles

| Magnet name | SFI/SFO | SDI/SDO |
|---|---|---|
| Quantity | 512/512 | 1024/1024 |
| Magnetic length[m] | 1.4 | 1.4 |
| Aperture [mm] | 76 | 76 |
| Field gradient [T/m$^2$] @45.5 GeV | 208.5 | -214. |
| Field strength [T/m$^2$] @180 GeV | 825 | -847 |
| Radius for good field region [mm] | 11.6 | 11.6 |
| Field errors [$\times 10^{-4}$] | 3 | 3 |
| Ampere turns [At] | 6087 | 6248 |
| Turns per pole | 25 | 25 |
| Current[A]@180GeV | 244 | 250 |
| Size of conductor [mm*mm] | 7×7, d4, r1-Cu | 7×7, d4, r1- Cu |
| Max current density [A/mm^2] | 6.68 | 6.86 |
| Resistance of the coil [mΩ] | 0.24 | 0.24 |
| Power loss/magnet (kW) @180GeV | 14.26 | 15.03 |
| Average Power Loss/magnet (kW) | 6.28 | 6.61 |
| Core height [mm] | 320 | 320 |
| Core width [mm] | 320 | 320 |
| Core length [m] | 1.4 | 1.4 |

#### 4.3.3.5 *Corrector Magnets*

Two types of orbit correction dipoles are needed to meet the beam optics requirements in both the arcs and the acceleration regions. The horizontal and vertical correctors share the same magnetic circuit, field strength, and length, and have similar physical and mechanical designs. Table 4.3.3.9 lists the key characteristics of these correction dipole magnets.

**Table 4.3.3.9:** Main requirements of the correctors

| Name | Quantity | Aperture [mm] | $B_{max}$ [T] | $L_{eff}$ [m] | GFR (H×V) [mm] | Quality [$10^{-4}$] |
|---|---|---|---|---|---|---|
| CH | 3544 | 66 | 0.0225 | 0.875 | 11.6×4 | 10 |
| CV | 3544 | 66 | 0.0225 | 0.875 | 11.6×4 | 10 |

The CH and CV correctors are used for closed-orbit corrections in the horizontal and vertical directions, respectively. They have the same cross section, with the CV being a 90° rotation of the CH. The correctors use a window frame design due to the low magnetic field strength, with the yoke made of solid iron and the coils wound around it with solid enamelled wires. The conductor has a rectangular cross section of 9.3 mm by 4.3 mm. The 2D cross sections of the correctors are shown in Fig.4.3.3.29, and the basic design parameters are listed in Table 4.3.3.10.

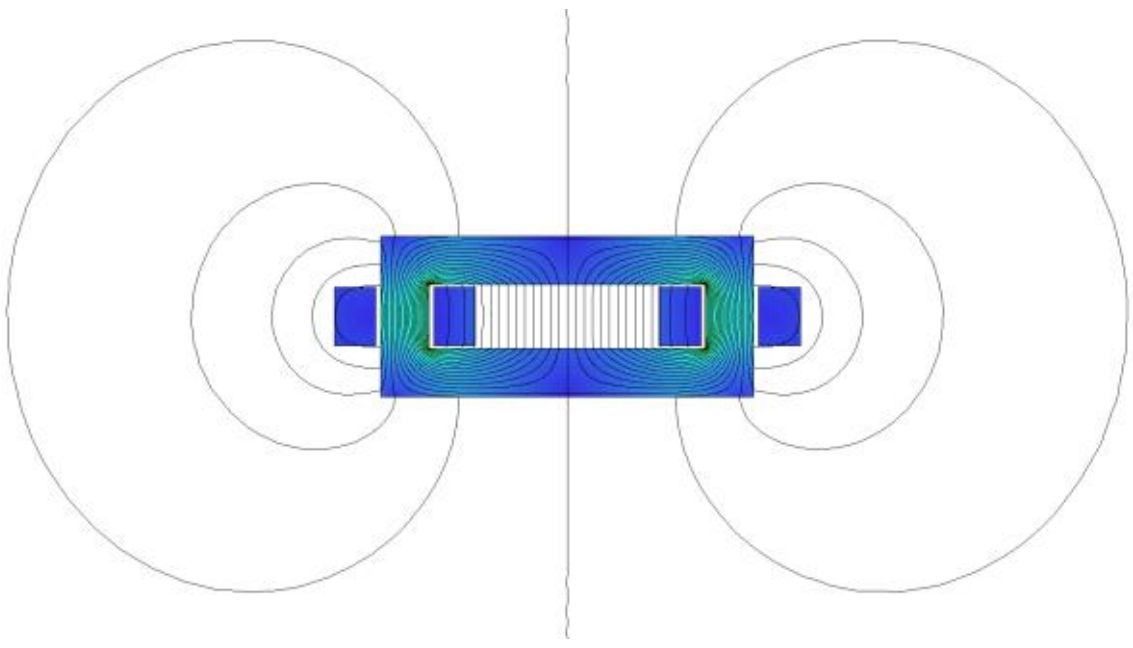

**Figure 4.3.3.29:** Magnetic flux distribution in CH correctors

**Table 4.3.310:** Main parameters of CH and CV

| Magnet name | CH | CV |
|---|---|---|
| Quantity | 3544 | 3544 |
| Magnetic length[m] | 0.875 | 0.875 |
| Aperture [mm] | 66 | 66 |
| Field strength [Gs] @45.5 GeV | 56.8 | 56.8 |
| Field strength [Gs] @180 GeV | 225 | 225 |
| Good field region [mm] | 11.6 | 11.6 |
| Field errors [$\times 10^{-4}$] | 10 | 10 |
| Ampere turns [At] | 1205 | 1205 |
| Turns per pole | 36 | 36 |
| Current[A]@180GeV | 17 | 17 |
| Size of conductor [mm*mm] | 9×4- Al | 9×4-Al |
| Max current density [A/mm^2] | 0.48 | 0.48 |
| Resistance of the coil [Ω] | 0.13 | 0.13 |
| Power loss/magnet (W) @180GeV | 37 | 37 |
| Core height [mm] | 0.17 | 0.3 |
| Core width [mm] | 0.3 | 0.17 |
| Core length [m] | 0.81 | 0.81 |

#### 4.3.3.6 *Field Measurements*

Magnetic field measurement is crucial for testing the quality of magnets, providing quality control for magnet production, and important parameters for accelerator operation. Four types of magnets need to be measured for the collider: dipoles, quadrupoles, sextupoles, and correctors. Dipole and corrector magnets can be measured as one type, and quadrupole and sextupole magnets are measured as another type based on the similarity of measurement contents.

Four parameters must be considered for dipole and corrector magnets:

1. Central field B (x=0, y=0, z=0) and its uniformity distribution
2. Integral field B (x, y=0, z) and its uniformity distribution
3. Effective length ($L_{eff}$)
4. Excitation curve

For the prototype magnet, it is important to study all four aspects in detail. However, for batch magnets, items 2 and 4 are sufficient to reflect the overall quality of magnets and meet the needs of accelerator operation, making them the key parameters for batch magnet detection.

For Quadrupole and Sextupole magnets, attention should be given to the following four parameters:

1. Gradient integral field (G×L) and excitation curve
2. Central gradient field ($G_0$) and Effective length ($L_{eff}$)
3. Harmonics
4. Magnetic center offset (dx, dy)

For prototype magnets, these four aspects require detailed study. For batch magnets, items 1, 3, and 4 reflect the overall quality of magnets and the needs of accelerator operation, making them the key parameters for batch magnet production.

The magnetic field measurement system is crucial for detecting magnetic field quality and is one of the essential pieces of equipment for accelerator construction. Various types of magnetic field measurement systems have been developed to meet different measurement needs, including the Hall probe, rotating coil, stretched wire, and flip coil systems. However, due to the large number of CEPC magnets, the long mechanical length, and the weak magnetic field of some magnets, accurately measuring the magnetic field quickly and efficiently is a significant challenge. To address this challenge, two types of 6.5-meter-long magnetic measurement system prototypes, using rotating coil and Hall probe systems, respectively, are under development (shown in Fig.4.3.3.30 and Fig.4.3.3.31).

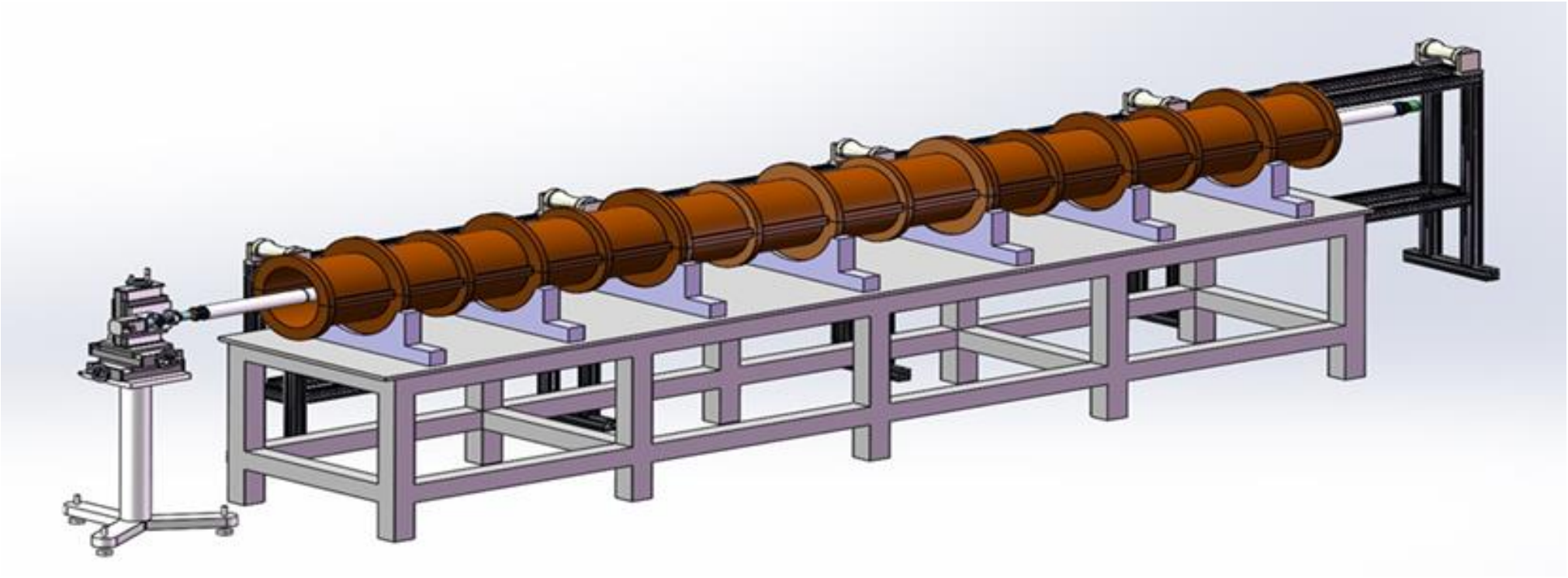

**Figure 4.3.3.30:** Schematic diagram of rotating coil in 5.1m CT prototype magnet

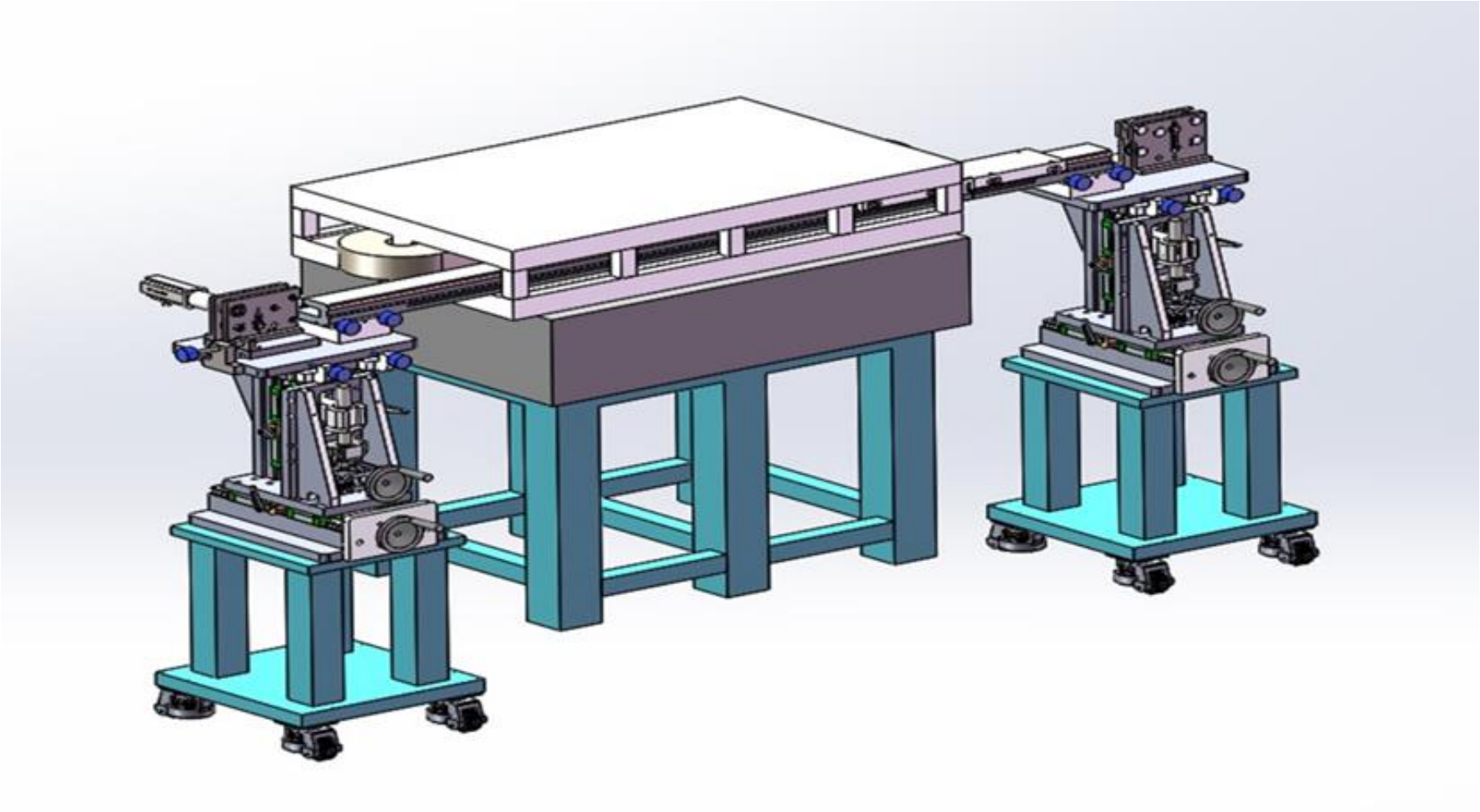

**Figure 4.3.3.31**：Schematic diagram of Hall probe measurement system (the figure shows a 1.5m long model)

The Hall probe measurement system provides detailed local magnetic field information, but it takes a long time to complete a measurement. In contrast, the rotating coil measurement system can efficiently and quickly perform integral magnetic field measurements but cannot measure local magnetic fields. By combining the strengths of the two measurement methods, they can complement each other and meet the requirements of magnet quality inspection and accelerator operation. This combination is the most efficient for CEPC accelerator construction.

To complete all the magnet measurement tasks of the CEPC collider within three years, the following equipment and environment will be built:

1. Test site: a 2000 square-meter area equipped with a crane
2. Magnet power supply: 39 sets
3. Deionized cooling water
4. Collimating equipment: 6 sets
5. Rotating coil system: 37 sets
6. Hall probe system: 2 sets

Measuring the enormous number of magnets in the CEPC project poses a significant challenge. To prepare for the CEPC project's construction, ongoing research and development efforts focus on new signal acquisition equipment, automatic transportation of long magnets, automatic alignment, rapid connection of cooling water and cables, new measurement principles, and technologies. The ultimate goal is to establish an automatic, efficient, and precise magnetic field measurement laboratory capable of undertaking the CEPC batch magnet measurement task.

### 4.3.4 **Superconducting Magnets in the Interaction Region**

#### 4.3.4.1 *Introduction*

To achieve high luminosity in the CEPC, final focus superconducting magnets are necessary on both sides of the interaction point (IP) in the interaction region (IR). The quadrupole magnet QD0 in the CDR [1] is now divided into two quadrupole magnets Q1a and Q1b, and the requirements for the final focus quadrupoles are based on L* of 1.9 m and a beam crossing angle of 33 mrad [2].

Figure 4.3.4.1 illustrates that compact high-gradient quadrupoles Q1a, Q1b, and Q2 must be present on both sides of the IP. These dual-aperture quadrupoles operate entirely within the solenoid field of the detector magnet, which has a central field strength of 3.0 T. To reduce the impact of the longitudinal solenoid field on the particle beams, anti-solenoids must be placed before Q1a as well as outside Q1a, Q1b, and Q2. The magnetic field direction of these anti-solenoids is opposite to that of the detector solenoid, and the total integral longitudinal field produced by the detector solenoid and anti-solenoid coils is zero. Additionally, it is necessary to maintain a total solenoid field inside the Q1a, Q1b, and Q2 magnet aperture that is as close to zero as possible [1,3].

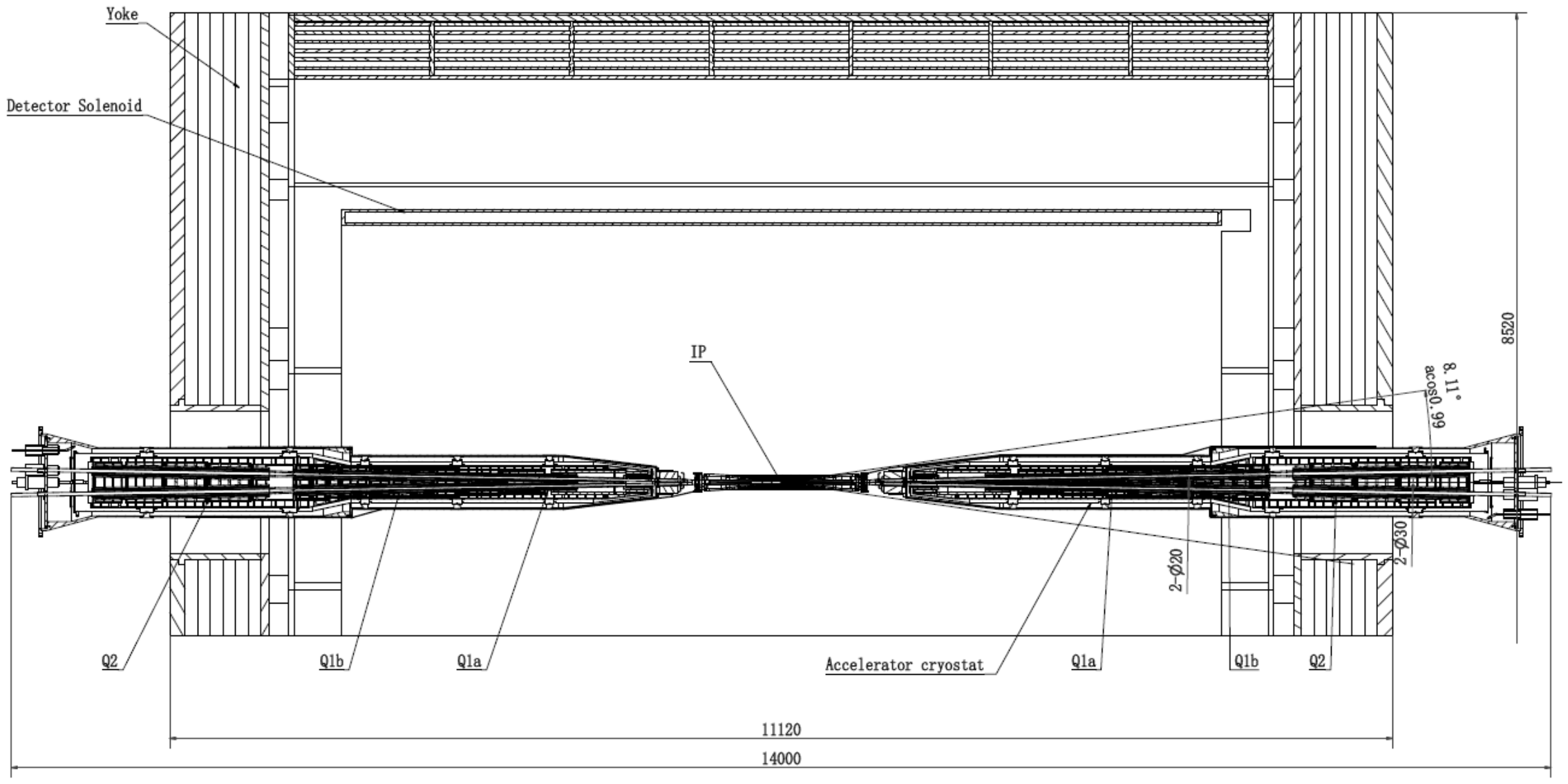

**Figure 4.3.4.1:** Layout of CEPC interaction region SC magnets.

The Machine Detector Interface (MDI) layout dictates that accelerator devices can only commence after 1.1 m from the IP along the longitudinal axis. This constraint limits the space available for the anti-solenoid before Q1a. Moreover, the angle of the accelerator magnet observed from the IP point must be small and conform to detector requirements. Given the high field strength of twin aperture quadrupole magnets, the anti-solenoid's high central field, and the limited space, interaction region quadrupole magnets and anti-solenoids will employ superconducting technology.

### 4.3.4.2 *Superconducting Quadrupole Magnet Q1a*

#### *4.3.4.2.1 Overall Design*

The dual-aperture superconducting quadrupole magnet Q1a is positioned 1.9 m away from the interaction point (IP), with design requirements listed in Table 4.3.4.1. With a crossing angle of 33 mrad between the two aperture centerlines, the distance between their centers varies longitudinally, with a minimum of only 62.71 mm, providing limited radial space. The gradient of Q1a must be 142.3 T/m, with a magnetic length of 1.21 m and field harmonics in the good field region less than $5\times10^{-4}$. Despite significant field crosstalk between the two apertures, the dipole field at the center of each aperture must be less than 3 mT.

**Table 4.3.4.1:** Design requirements of the dual aperture superconducting quadrupole magnet Q1a

| Parameter | Value | Unit |
|---|---|---|
| Field gradient | 142.3 | T/m |
| Magnetic length | 1210 | mm |
| Reference radius | 7.46 | mm |
| Minimum distance between two aperture centerlines | 62.71 | mm |
| High order field harmonics | $\leq 5\times10^{-4}$ | |
| Dipole field at the center of each aperture | $\leq 3$ | mT |

#### 4.3.4.2.2 *Field Calculation without Iron Yoke*

The layout of the CEPC interaction region places a combined superconducting magnet on a cantilever support, which extends deep into the detector. This fixing method imposes stringent weight requirements on the superconducting magnets. Ideally, the iron yoke should be removed, leaving only the coils to minimize the weight of the superconducting magnet. Two design schemes are being studied: one involves a pure coil magnet structure without an iron yoke, while the other includes an iron yoke in the magnet structure.

##### *4.3.4.2.2.1 Cos2θ Quadrupole Coil Structure without Iron Yoke*

The field polarity, gradient, and quality requirements are identical in both apertures of Q1a, and the inner radius of the quadrupole coil is chosen to be 20 mm, taking into account the beam pipe and helium vessel size.

The cos2θ quadrupole coil structure employs Rutherford cable made of either 0.5 mm HTS or LTS strand. The 2D model is established and magnetic field calculations are performed using ROXIE [4]. Magnet Q1a is designed with a two-layer cos2θ quadrupole coil, with each layer comprising two blocks separated by a wedge. The Rutherford cable, twisted by 10 strands, has a trapezoidal angle of 2.1 degrees.

Figure 4.3.4.2 shows the two-dimensional simulation model of the single-aperture Q1a. Due to its small inner diameter of only 40 mm, the quadrupole coil winding process is challenging. In the cross-sectional design, particular attention is given to ensuring that the coil pole width is wider than 10 mm. This ensures that the minimum bending radius of the Rutherford cable at the coil end is larger than 5 mm, which is deemed a safe value that almost does not degrade the critical current of the cable. The coil has an inner radius of 20 mm, outer radius of 25.65 mm, and a distance of 0.35 mm between the two layers. The Rutherford cable's design current is 2650 A, and the coil's peak field is 3.572 T. All the computed field harmonics are less than 1 unit ($1\times10^{-4}$) at reference radius of 7.46 mm, which satisfies the design requirements.

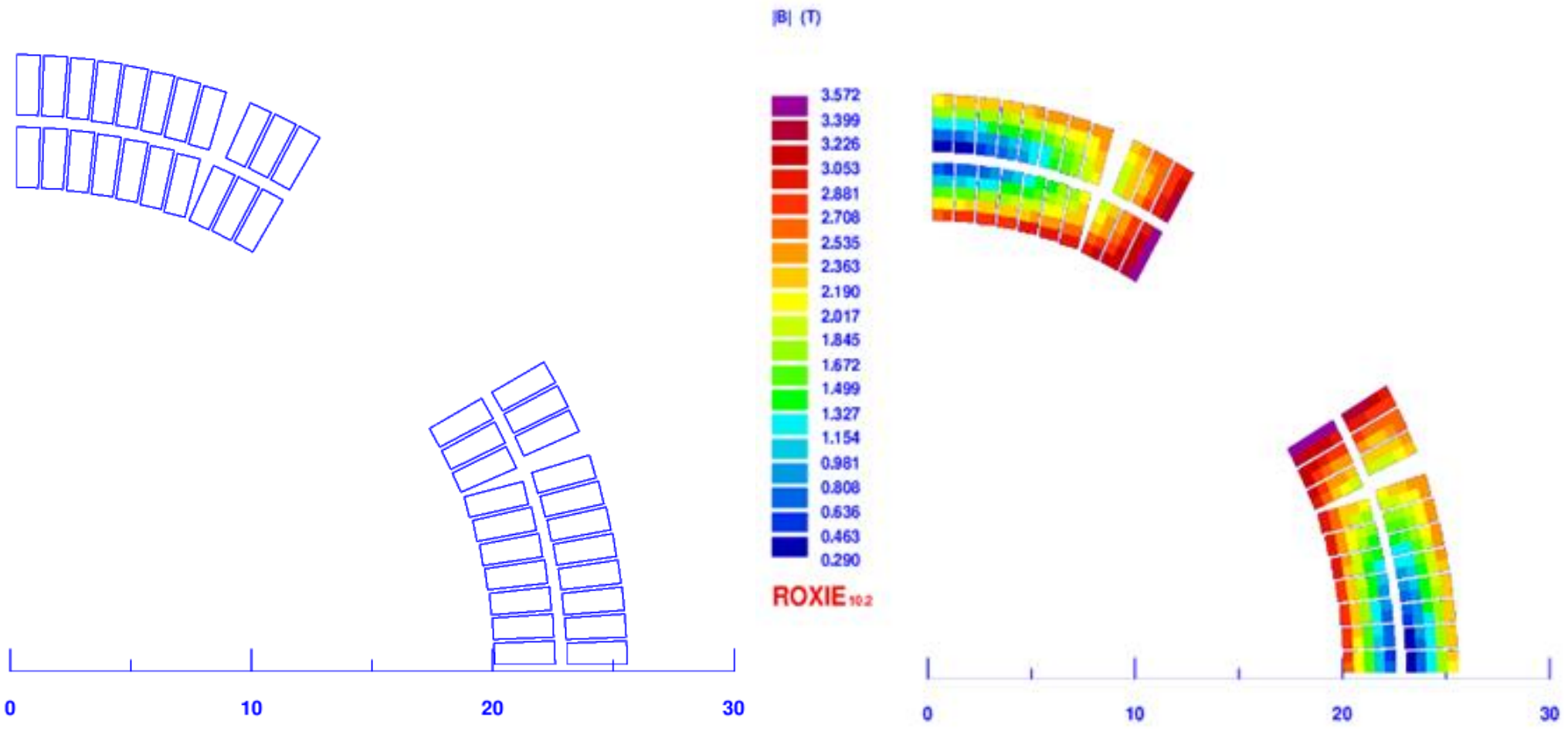

**Figure 4.3.4.2:** Layout of 2D cos2θ quadrupole coils

##### *4.3.4.2.2.2 CCT Quadrupole Coil Structure without Iron Yoke*

The CCT quadrupole coil is wound using 10 HTS or LTS strands, arranged in 2×5 formation, with a single NbTi strand diameter of 0.5 mm. To reduce calculation time, the model is simplified, using one rectangular conductor instead of 10 strands. The coil has a cross-sectional size of 1 mm × 2.5 mm, and the electrical characteristic parameter is the average effect of 10 strands on the cross-section [6]. The first layer coil has an inner radius of 22 mm, while the second layer coil has an inner radius of 25.5 mm. To accommodate the narrow space of the interaction region, the tilt angle is set at 15 degrees, reducing the magnet's longitudinal length and making conductor transition smoother. The single-aperture magnet comprises two layers of coils, and its OPERA coil simulation results are presented in Figure 4.3.4.3 [7]. The design current in each strand is 472.5 A, and the coil's peak field is 4.25 T.

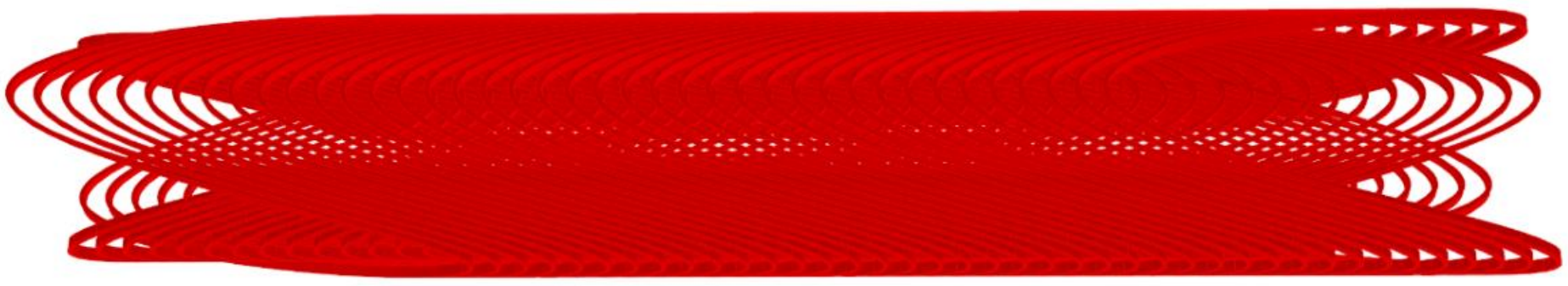

**Figure 4.3.4.3:** CCT quadrupole coil model

*4.3.4.2.2.3 Serpentine Quadrupole Coil Structure without Iron Yoke*

The third option is to use serpentine coil [8]. Round wire with a diameter of 0.5 mm is used in the Serpentine Quadrupole Coil for Q1a. The wire can be either HTS Bi-2212 or LTS NbTi. The coil has a total of 8 layers, and the inner radius is 20 mm. There are 136 turns of conductors in each pole, and the excitation current is 480 A. Figure 4.3.4.4 shows the 2D simulation model of Q1a with the Serpentine Quadrupole Coil. Each multipole field in a single aperture is smaller than $1\times10^{-4}$.

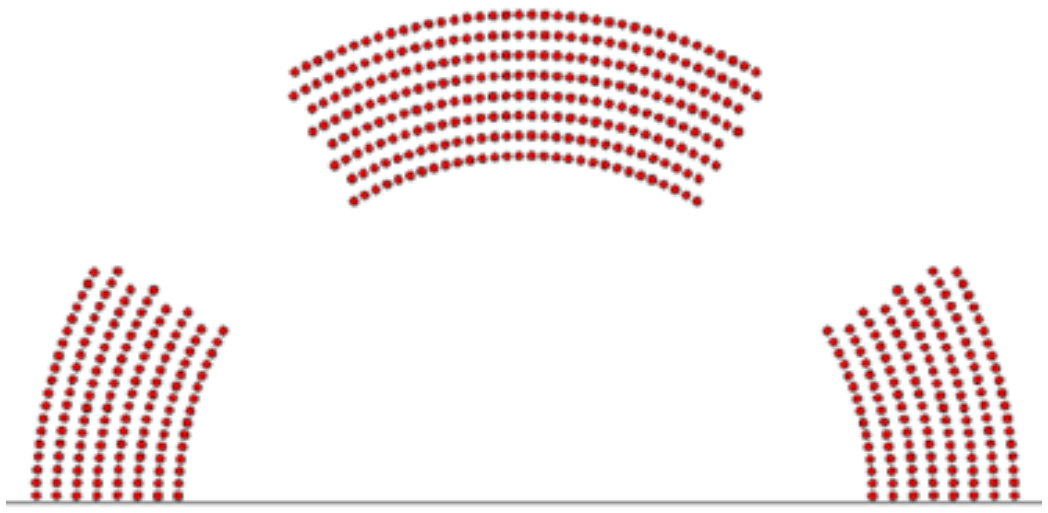

**Figure 4.3.4.4:** Serpentine quadrupole coil model

*4.3.4.2.3 Crosstalk between Two Aperture Quadrupole Coils*

In Figure 4.3.4.5, two single-aperture quadrupole coils of Q1a are shown in 3D, distributed at an angle of 33 mrad and 1.9 meters away from the IP. The field crosstalk between the two apertures creates a dipole field that reaches about 100 mT, which is far greater than the acceptable limit of 3 mT. Figure 4.3.4.6 shows the dipole field at the center of the aperture along the longitudinal direction. This crosstalk issue is not limited to cos2θ quadrupole coils, but also occurs in CCT and serpentine quadrupole coils.

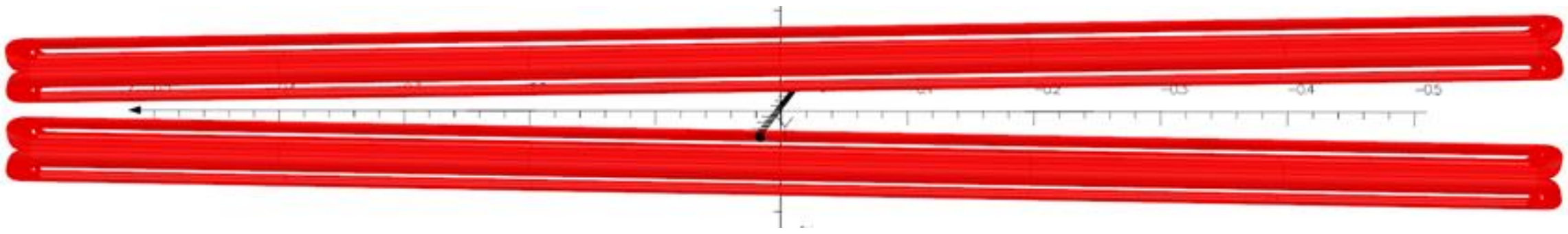

**Figure 4.3.4.5:** 3D simulation model of dual-aperture quadrupole Q1a with no iron yoke (dimensions in m).

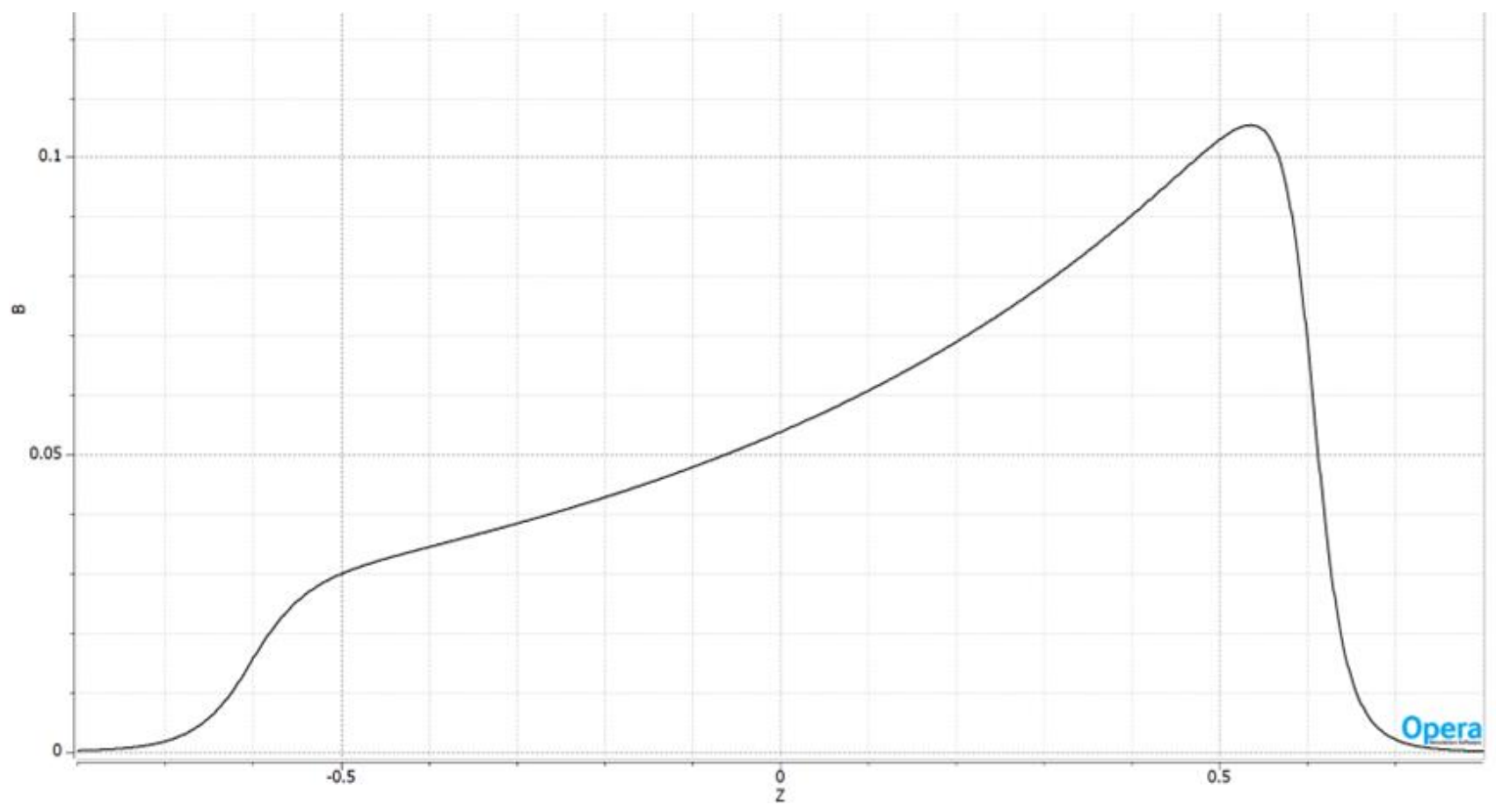

**Figure 4.3.4.6:** Dipole field at the center of the aperture in dual-aperture Q1a with no iron yoke (dimensions in m).

While all three types of single-aperture quadrupole coil structures meet the physical design requirements individually, the dipole field generated by crosstalk between the two apertures cannot be resolved when combined. Therefore, in our baseline design, an iron yoke is added outside the coil to enhance the field gradient, reduce the coil excitation current, and shield the field crosstalk between the two apertures.

*4.3.4.2.4 Field Calculation with Iron Yoke*

*4.3.4.2.4.1 Cos2θ Quadrupole Coil Structure with Iron Yoke*

In our baseline scheme, we utilize Cos2θ quadrupole coils with an iron yoke. The iron yoke, made of FeCoV, is placed outside the coil, with an inner radius of 30.5 mm and an outer radius of 44 mm. By incorporating the iron yoke, the excitation current in the strand reduces to 202 A. Figure 4.3.4.7 and Figure 4.3.4.8 depict the 2D and 3D field calculation models, respectively, for single aperture Q1a with an iron yoke.

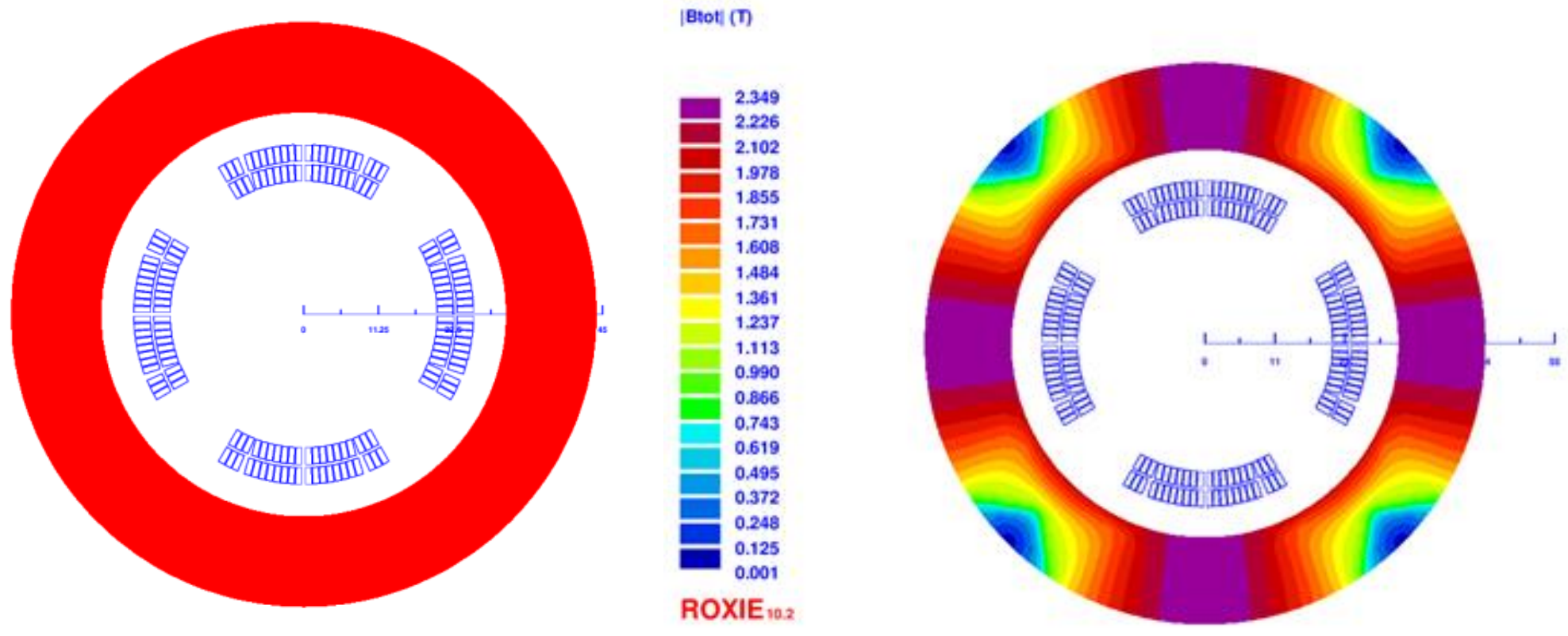

**Figure 4.3.4.7:** 2D simulation result of cos2θ quadrupole coil with iron yoke

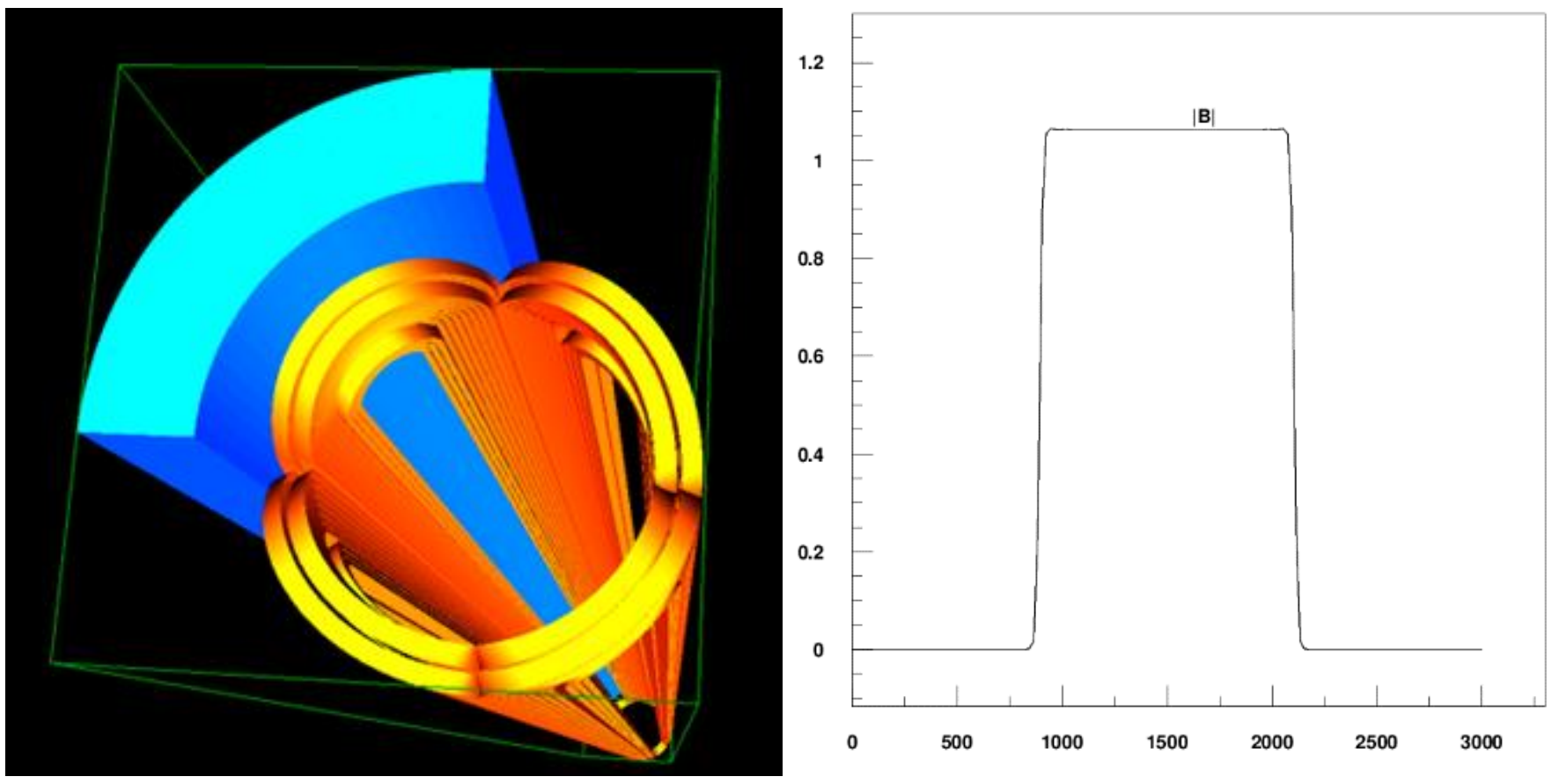

**Figure 4.3.4.8:** 3D simulation of single aperture cos2θ quadrupole coil with iron yoke

The three-dimensional field simulation model of Q1a is based on the two-dimensional model, where the quadrupole coil is stretched along the Z-axis to form a complete magnet. To cover the entire quadrupole coil, the length of the iron yoke is selected to be 1240 mm. The coil end is divided into conductor groups, and the detailed end shape of the coil is optimized to ensure that each integrated multipole field content meets the design requirements.

*4.3.4.2.4.2 CCT Quadrupole Coil Structure with Iron Yoke*

The 3D simulation model of the Q1a quadrupole magnet in the CCT option is shown in Figure 4.3.4.9. The iron yoke has an inner radius of 30.5 mm and an outer radius of 44 mm. The excitation current in the strand decreases to 324 A.

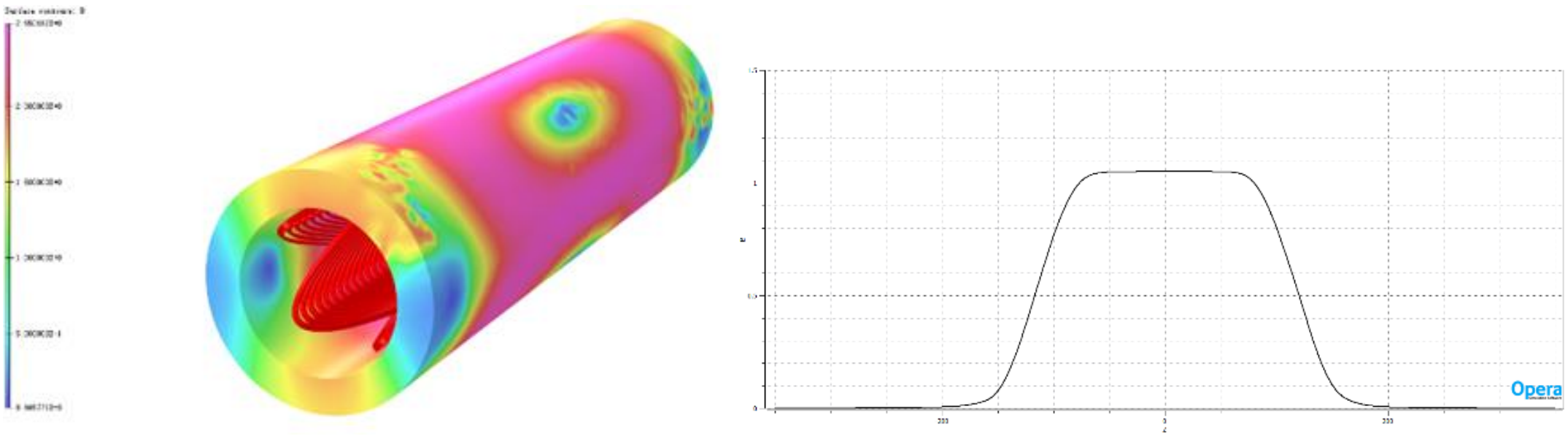

**Figure 4.3.4.9:** Simulation model of Q1a quadrupole magnet in the CCT option

*4.3.4.2.4.3 Serpentine Quadrupole Coil Structure with Iron Yoke*

Figure 4.3.4.10 shows the 2D simulation model of Q1a quadrupole magnet with iron yoke in the serpentine coil option. The inner and outer radius of the iron yoke are 30.5 mm and 44 mm, respectively, and the excitation current in the strand decreases to 334 A.

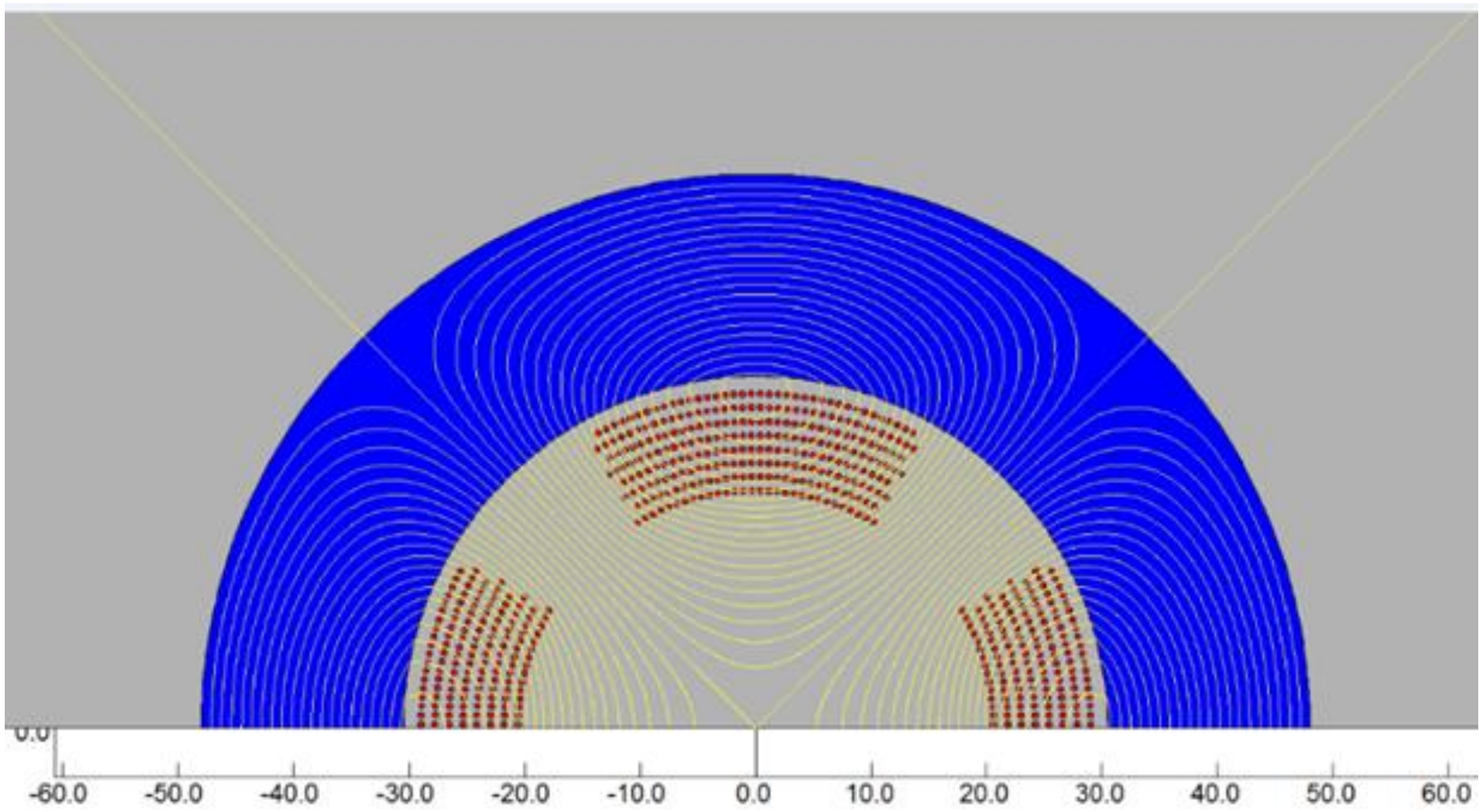

**Figure 4.3.4.10:** Simulation model of Q1a quadrupole magnet in the serpentine coil option

Table 4.3.4.2 and Table 4.3.4.3 present a parameter comparison of the three design coil structures of Q1a with the iron-free option and iron option, respectively, while maintaining the same design requirements. The strands employed in the coil design have a uniform diameter of 0.5 mm. The critical current-carrying characteristics of NbTi strands at 4.2 K are as follows: 1,642 A/mm² @ 3T, 1,389 A/mm² @ 4T, and 1,173 A/mm² @ 5T. In comparison, the current-carrying properties of Bi-2212 strands at 4.2 K are: 2,100 A/mm² @ 3T, 1,980 A/mm² @ 4T, and 1,810 A/mm² @ 5T [9].

**Table 4.3.4.2:** Comparison of Q1a parameters of three kinds of coil structures (iron-free)

| Coil type | Cos2θ coil | CCT coil | Serpentine coil |
|---|---|---|---|
| Excitation current in strand (A) | 265 | 472.5 | 480 |
| Current density Je on wire (A/mm$^2$) | 1350 | 2406 | 2445 |
| Peak field in coil (T) | 3.6 | 4.3 | 4.2 |

**Table 4.3.4.3:** Comparison of Q1a parameters of three kinds of coil structures (with iron)

| Coil type | Cos2θ coil | CCT coil | Serpentine coil |
|---|---|---|---|
| Excitation current in strand (A) | 202 | 324 | 334 |
| Current density Je on wire (A/mm$^2$) | 1029 | 1650 | 1701 |
| Peak field in coil (T) | 3.5 | 3.8 | 3.8 |
| Loadline (NbTi) | 79% | - | - |
| Loadline (Bi-2212) | 60% | 88% | 89% |

After considering the physical requirements of Q1a, all three coil structures were designed using 0.5mm diameter strands. Among the three designs, the cos2θ quadrupole coil has the smallest current in the strand and the smallest peak field, meeting the gradient requirements. Therefore, the cos2θ coil is selected as the baseline design for Q1a, while the CCT and Serpentine coils are considered as alternative designs.

#### *4.3.4.2.5 3D Field Simulation of Double Aperture Quadrupole Q1a*

The two apertures of Q1a magnet must meet the same polarity, magnetic field gradient, and field quality requirements. However, there is not enough space to place two single apertures side by side, so a compact dual-aperture magnet design is adopted. As shown in Figure 4.3.4.11, the two single apertures intersect in the middle part, and the iron yoke made of FeCoV is shared by the two apertures. At the end closer to the IP, the maximum dipole field at the center of the aperture is 2 mT, which meets the design requirements.

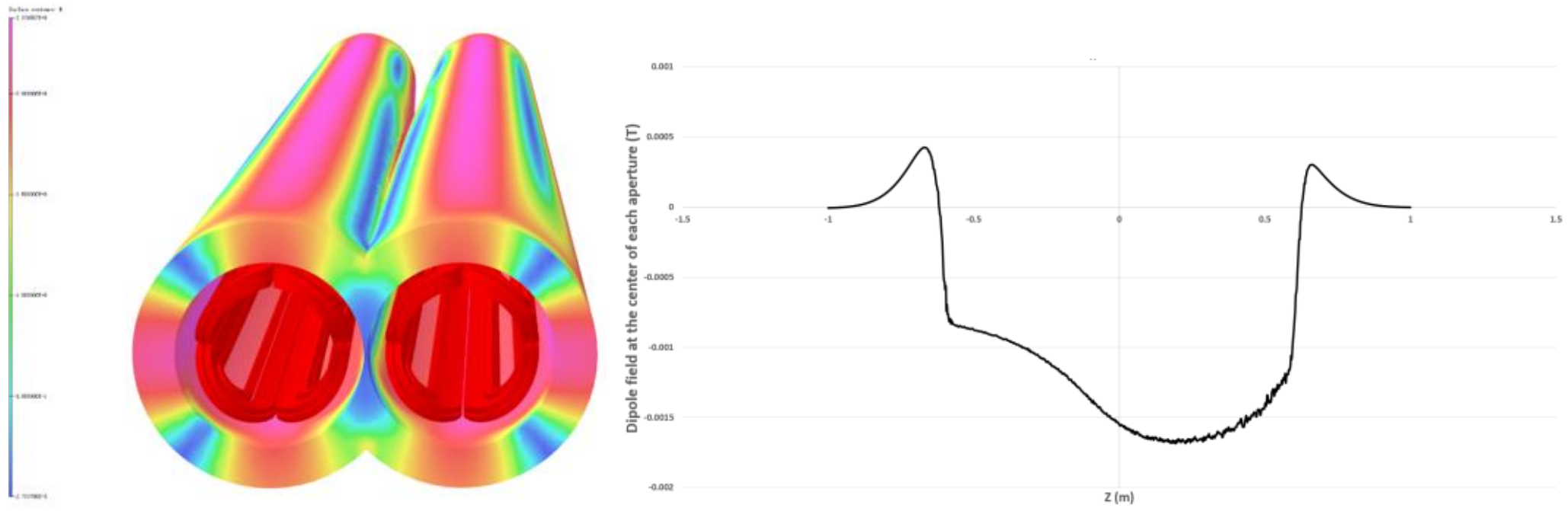

(a) (b)

**Figure 4.3.4.11:** Field simulation of dual-aperture quadrupole magnet Q1a: (a) 3D moldel, (b) field distribution along the magnet axis.

All calculated integrated field harmonics are smaller than 1 unit. Table 4.3.4.4 lists the design parameters of the dua aperture Q1a magnet. The working temperature of Q1 magnet is 4.2 K.

**Table 4.3.4.4:** Design parameters of the cos2θ Q1a quadrupole magnet with iron yoke

| Parameter | Q1a dual aperture |
|---|---|
| Field gradient (T/m) | 142.41 |
| Magnetic length (m) | 1211.80 |
| Coil turns per pole | 21 |
| Excitation current (A ) | 2020 |
| Coil layers | 2 |
| Conductor (HTS or LTS) | Rutherford Cable, width 2.5 mm, mid thickness 0.93 mm, keystone angle 2.1 deg, Cu:Sc = 1.3, 10 strands |
| Maximum dipole field at the center of each aperture (mT) | 2.497 |
| Stored energy (KJ) (dual aperture) | 11.5 |
| Inductance (mH ) | 5.64 |
| Peak field in coil (T) | 3.42 |
| Load line | 78.79% (NbTi) |
| Integrated field harmonics | $b_6 = -0.61$<br>$b_{10} = -0.24$ |
| Coil inner diameter (mm) | 40 |
| Coil outer diameter (mm) | 51.3 |
| Yoke outer diameter (mm) | 88 |
| X direction Lorentz force/octant (kN) | 62.33 |
| Y direction Lorentz force/octant (kN) | –58.59 |
| Net weight (kg) | 93 |

### 4.3.4.3 *Superconducting Quadrupole Magnet Q1b*

#### *4.3.4.3.1 Overall Design*

The dual-aperture superconducting quadrupole magnet Q1b was relocated to a position 3.19 m away from the IP, with a center-to-center distance between the two apertures of only 105.28 mm. Q1b requires a gradient of 85.4 T/m and a magnetic length of 1.21 m, with magnetic field harmonics in the good field region of less than $5\times10^{-4}$. Due to the small aperture separation distance, the field cross talk between the two apertures in Q1b is significant, and the dipole field at the center of each aperture must be less than 3 mT. Table 4.3.4.5 summarizes the design requirements for the double-aperture superconducting quadrupole magnet Q1b.

**Table 4.3.4.5:** Design requirements of the dual aperture superconducting quadrupole magnet Q1b

| Parameter | Value | Unit |
|---|---|---|
| Field gradient | 85.4 | T/m |
| Magnetic length | 1210 | mm |
| Reference radius | 9.085 | mm |
| Minimum distance between two aperture centerlines | 105.28 | mm |
| High order field harmonics | $\leq 5\times10^{-4}$ | |
| Dipole field at the center of each aperture | ≤ 3 | mT |

#### *4.3.4.3.2 2D Field Calculation with Iron Yoke*

The cos2θ quadrupole coil structure for magnet Q1b utilizes a Rutherford cable made of 0.5 mm strand, which can be either HTS Bi-2212 or LTS NbTi. The 2D model is established and magnetic field calculations are performed using ROXIE. The design of the magnet is based on a two-layer cos2θ quadrupole coil, with the first layer consisting of a single block and the second layer consisting of two blocks separated by a wedge. The Rutherford cable is twisted with a trapezoidal angle of 1.9 degrees and has 12 strands.

Figure 4.3.4.12 shows the two-dimensional simulation model of the single-aperture Q1b magnet. The inner and outer radius of the coil are 26 mm and 32.15 mm, and the distance between the two layers is 0.35 mm. The design current in the Rutherford cable is 1590 A, and the peak field in the coil is 2.675 T. The iron yoke made of FeCoV is added outside the collar to enhance the field gradient, reduce the coil excitation current, and shield the field crosstalk. The inner radius and outer radius of the iron yoke are 39 mm and 51.7 mm.

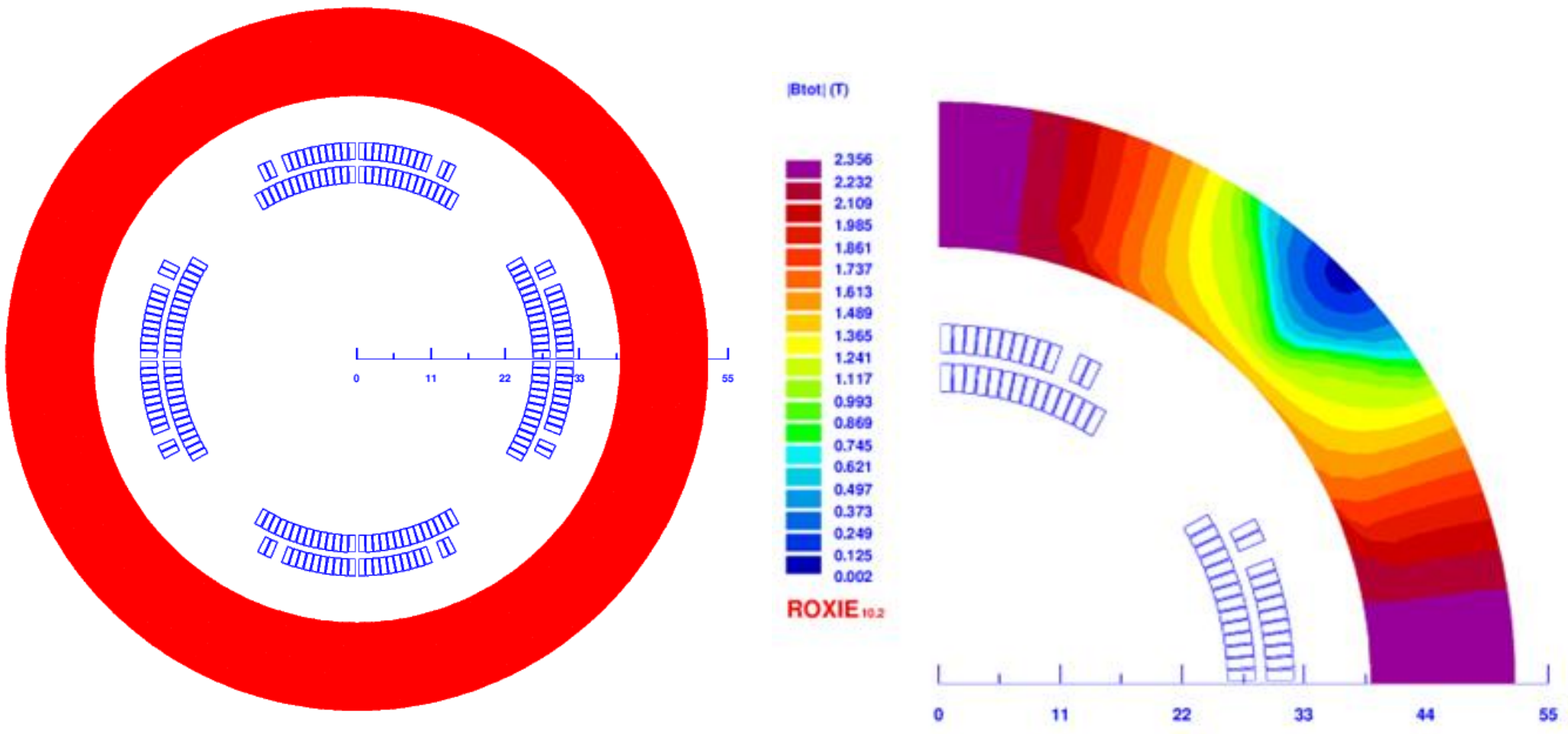

**Figure 4.3.4.12:** 2D field calculation of quadrupole magnet Q1b

Figure 4.3.4.13 shows that the two apertures of the Q1b magnet are designed according to the same polarity, magnetic field gradient, and field quality requirements in each aperture. The minimum distance between the two apertures of Q1b is 105.28 mm, while the outer radius of the iron yoke is 51.7 mm, so Q1b two apertures does not need to share the iron yoke like Q1a. At the end of Q1b near Q1a, the maximum value of dipole magnetic field at the center of each aperture is 2.3 mT.

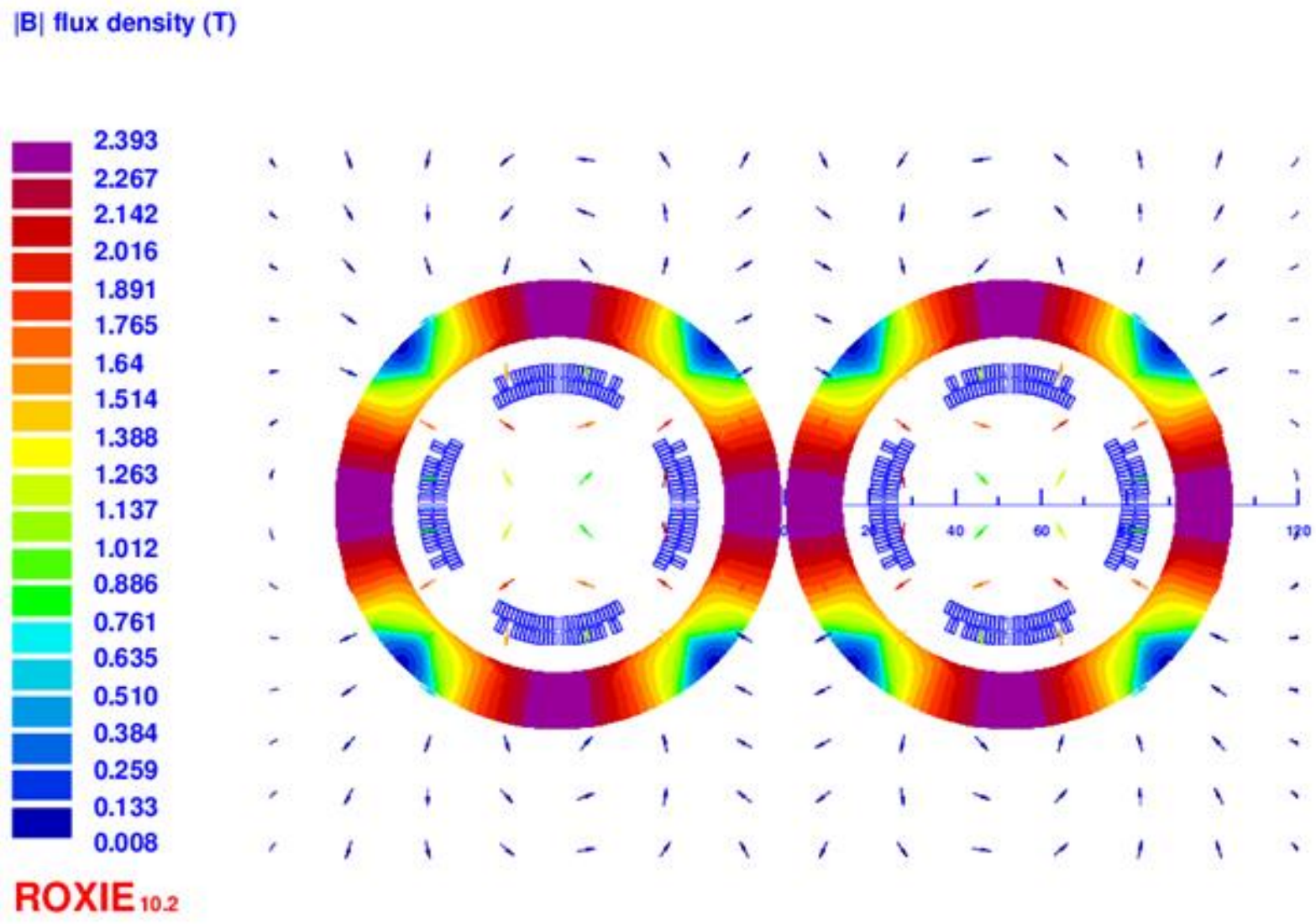

**Figure 4.3.4.13:** 2D Layout of dual aperture quadrupole magnet Q1b

### *4.3.4.3.3 3D Field Calculation*

The 3D model in Figure 4.3.4.14 shows that one block of the 2D model was divided into two blocks in the 3D model to create a smoother transition at the end of the coil layer. By adjusting the longitudinal length of the four blocks, the integrated field harmonics in the 3D model are less than 1 unit. The lengths of the four block straight lines are 604 mm, 592 mm, 604 mm, and 594.27 mm, respectively. The iron yoke has a length of 1240 mm, ensuring complete coverage of the coil. All calculated integrated field harmonics are smaller than 1 unit. Table 4.3.4.6 lists the design parameters of quadrupole magnet Q1b. The working temperature of Q1b magnet is also 4.2 K.

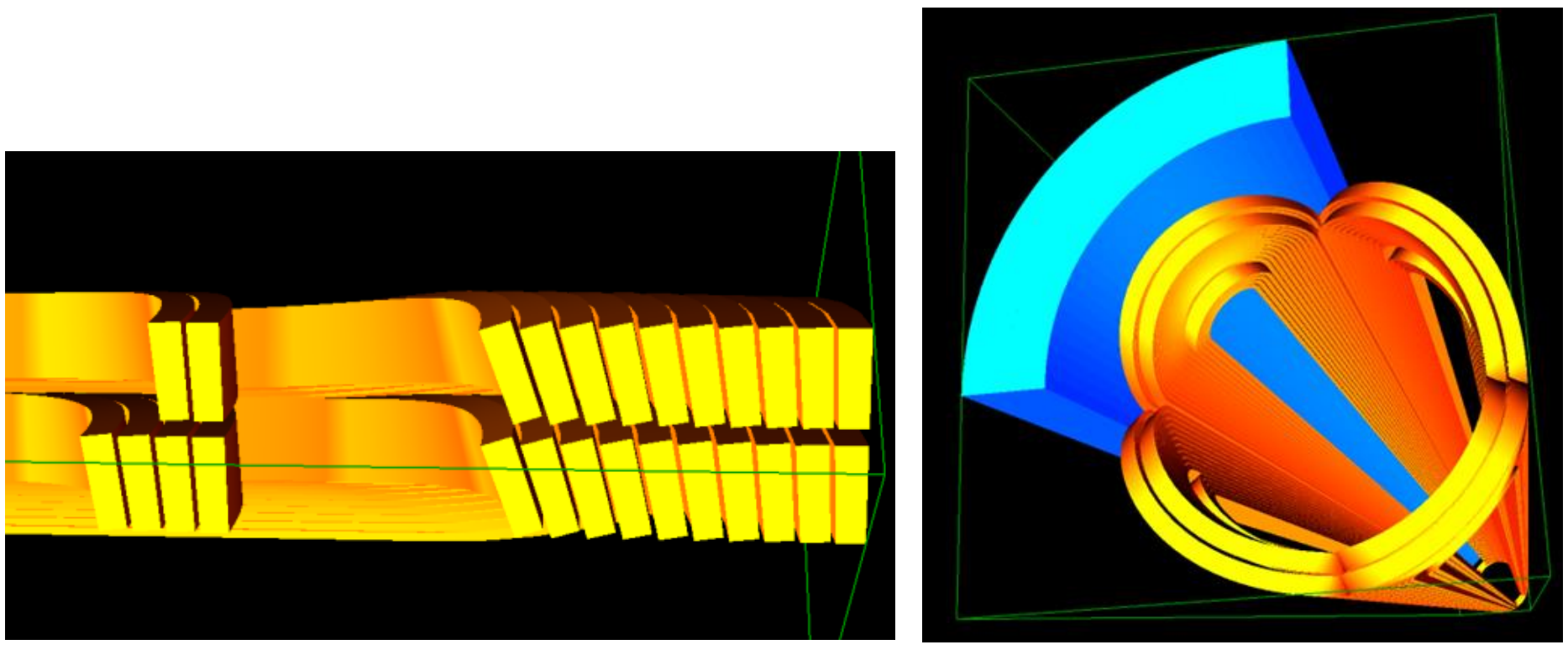

**Figure 4.3.4.14:** 3D model of single aperture quadrupole magnet Q1b

**Table 4.3.4.6:** Design parameters of dual aperture quadrupole coils Q1b with iron yoke

| Parameter | Q1b dual aperture |
|---|---|
| Field gradient (T/m) | 85.5 |
| Magnetic length (m) | 1211.84 |
| Coil turns per pole | 26 |
| Excitation current (A ) | 1590 |
| Coil layers | 2 |
| Conductor (HTS or LTS) | Rutherford Cable, width 3 mm, mid thickness 0.93 mm, keystone angle 1.9 deg, Cu:Sc = 1.3, 12 strands |
| Maximum dipole field at the center of each aperture (mT) | 2.30 |
| Stored energy (KJ) (double aperture) | 11.03 |
| Inductance (mH ) | 8.75 |
| Peak field in coil (T) | 2.68 |
| Load line | 55.96% (NbTi) |
| Integrated field harmonics | $b_6$ = 0.25<br>$b_{10}$ = –0.14 |
| Coil inner diameter (mm) | 52 |
| Coil outer diameter (mm) | 64.3 |
| Yoke outer diameter (mm) | 103.4 |
| X direction Lorentz force/octant (kN) | 45.86 |
| Y direction Lorentz force/octant (kN) | –44.69 |
| Net weight (kg) | 124 |

### 4.3.4.4 ***Superconducting Quadrupole Magnet Q2***

#### *4.3.4.4.1 Overall Design*

The dual-aperture superconducting quadrupole magnet Q2 has been relocated to a position 4.7 m away from the IP. The center lines of the two apertures must maintain a minimum distance of 155.11 mm. For superconducting magnet Q2, a gradient of 96.7 T/m and a magnetic length of 1.5 m are necessary. The magnetic field harmonics in the good field region must not exceed $5\times10^{-4}$. Additionally, taking into account the field crosstalk of the two apertures, the dipole field at the center of each aperture should be no more than 3 mT. The requirements for the design of the dual-aperture superconducting quadrupole magnet Q2 can be found in Table 4.3.4.7.

**Table 4.3.4.7:** Design requirements of the dual aperture superconducting quadrupole magnet Q2

| Parameter | Value | Unit |
|---|---|---|
| Field gradient | 96.7 | T/m |
| Magnetic length | 1500 | mm |
| Reference radius | 12.24 | mm |
| Minimum distance between two aperture centerlines | 155.11 | mm |
| High order field harmonics | $\leq 5\times10^{-4}$ | |
| Dipole field at the center of each aperture | $\leq 3$ | mT |

### *4.3.4.4.2 2D Field Calculation with Iron Yoke*

In the cos2θ quadrupole coil structure, either HTS or LTS Rutherford cable made of 0.5 mm strand is utilized. The 2D model is established and the magnetic field calculation is performed. Magnet Q2 is designed with a double layer cos2θ quadrupole coil. Each layer consists of only one block, and the Rutherford cable with a trapezoidal angle of 1.9 degrees is twisted by 12 NbTi strands. The two-dimensional simulation model of the single-aperture model Q2 is presented in Figure 4.3.4.15, with inner and outer coil radii of 31 mm and 37.65 mm, and a distance of 0.35 mm between the two layers. The Rutherford cable is designed to handle a current of 1925 A, with a peak field in coil of 3.656 T. An iron yoke made of FeCoV is incorporated outside the collar to enhance field gradient, decrease the coil excitation current, and shield field crosstalk, with inner and outer radii of 44 mm and 63.2 mm.

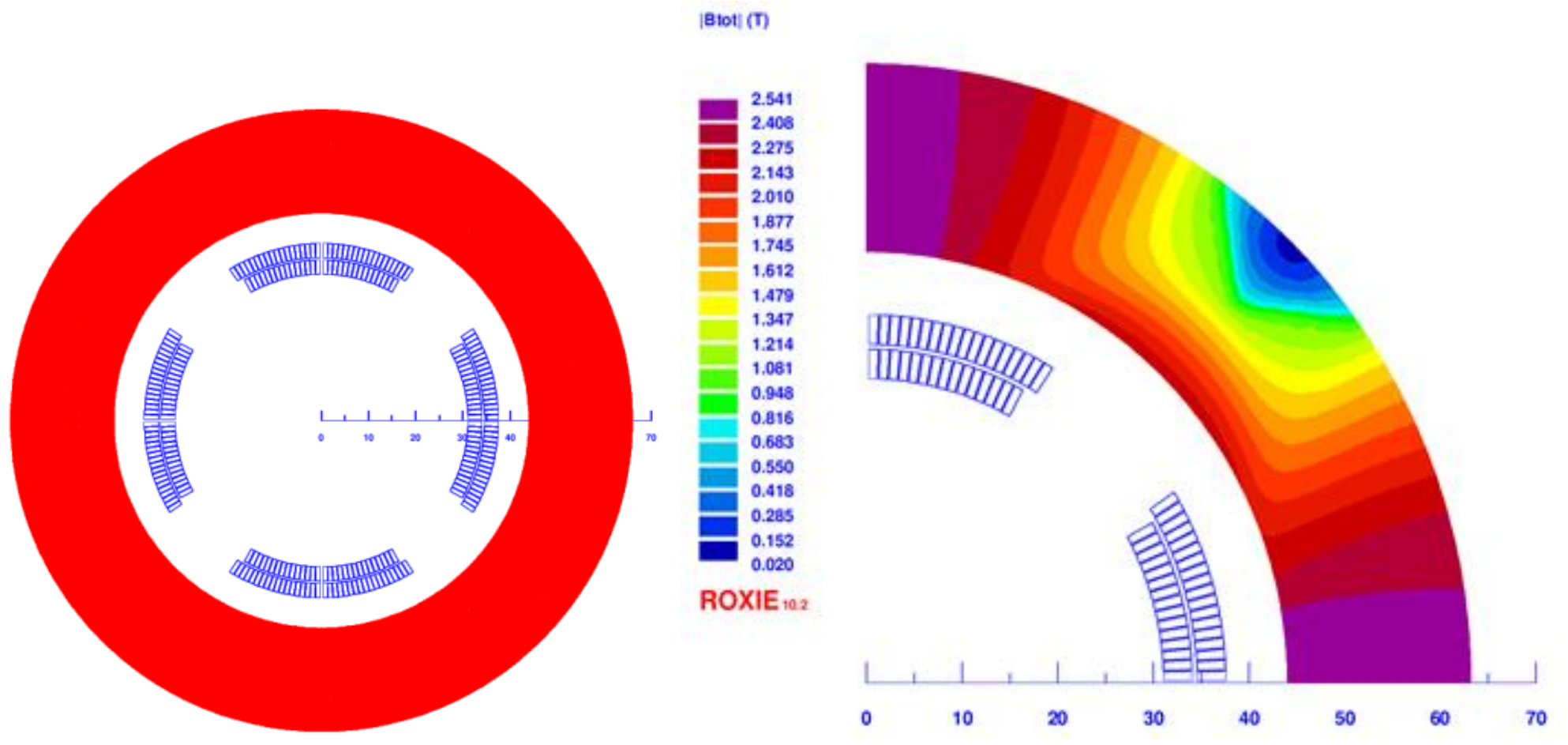

**Figure 4.3.4.15:** Layout and field calculation of quadrupole magnet Q2 with iron yoke

As illustrated in Figure 4.3.4.16, the two apertures of the Q2 magnet adhere to the same polarity, magnetic field gradient, and field quality requirements for each aperture. The minimum distance between the two apertures of the superconducting magnet Q2 is 155.11 mm, and the outer radius of the iron yoke is 63.2 mm, therefore, Q2 two apertures are not required to share the iron yoke like Q1a. The maximum value of dipole magnetic field at the center of each aperture at the end of Q2, near Q1b, is 2.5 mT.

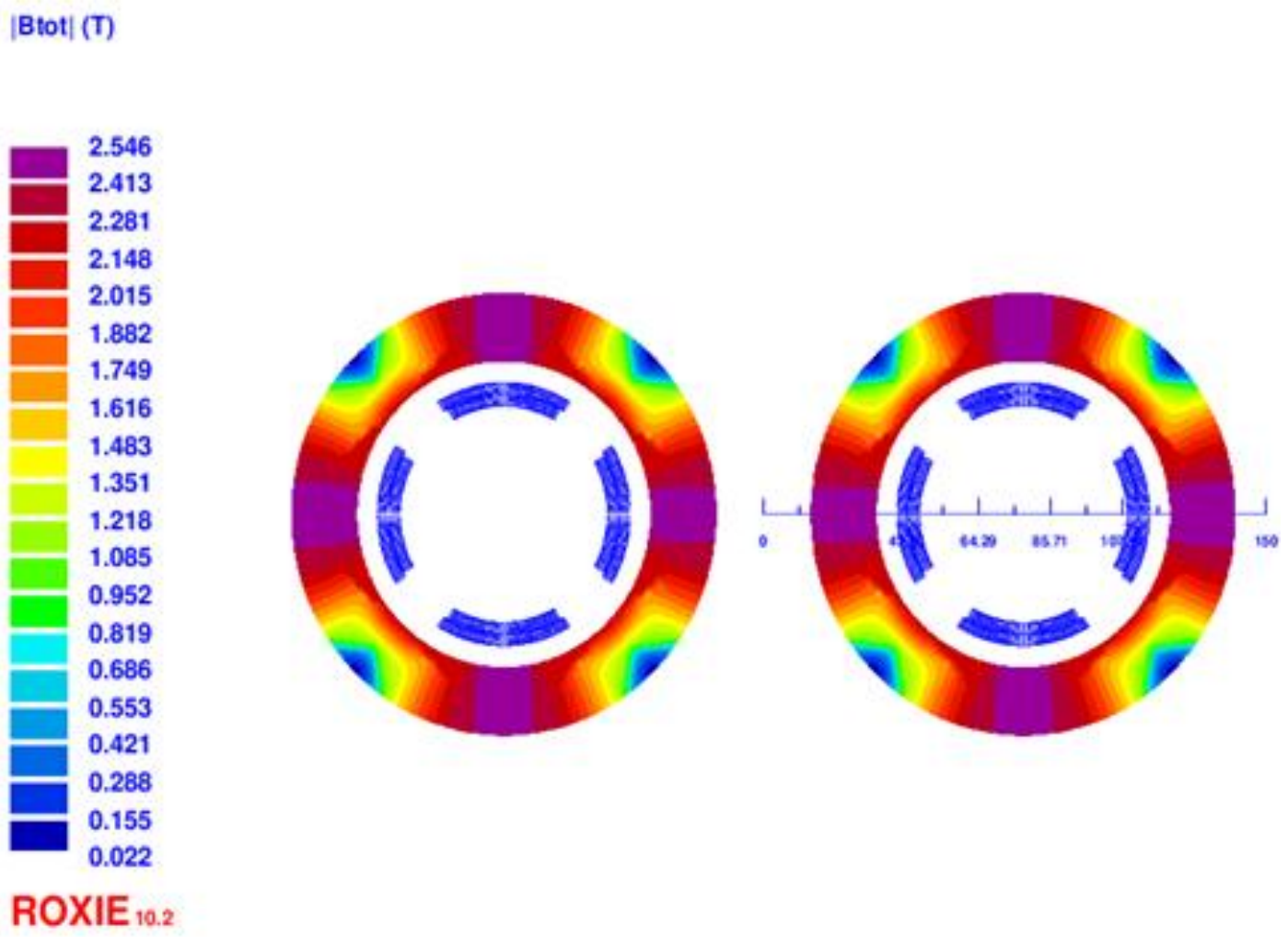

**Figure 4.3.4.16:** Layout of quadrupole magnet Q2 with iron yoke

*4.3.4.4.3 3D Field Calculation*

In the three-dimensional model shown in Figure 4.3.4.17, one block from the two-dimensional model was divided into two blocks to make the end transition of the coil layer smoother. By adjusting the longitudinal length of the four blocks, the three-dimensional integrated field harmonics were reduced to less than 1 unit. The lengths of the four block straight lines are 739 mm, 765 mm, 739 mm, and 765 mm respectively. The length of the iron yoke is 1540 mm, ensuring complete coverage of the coil.

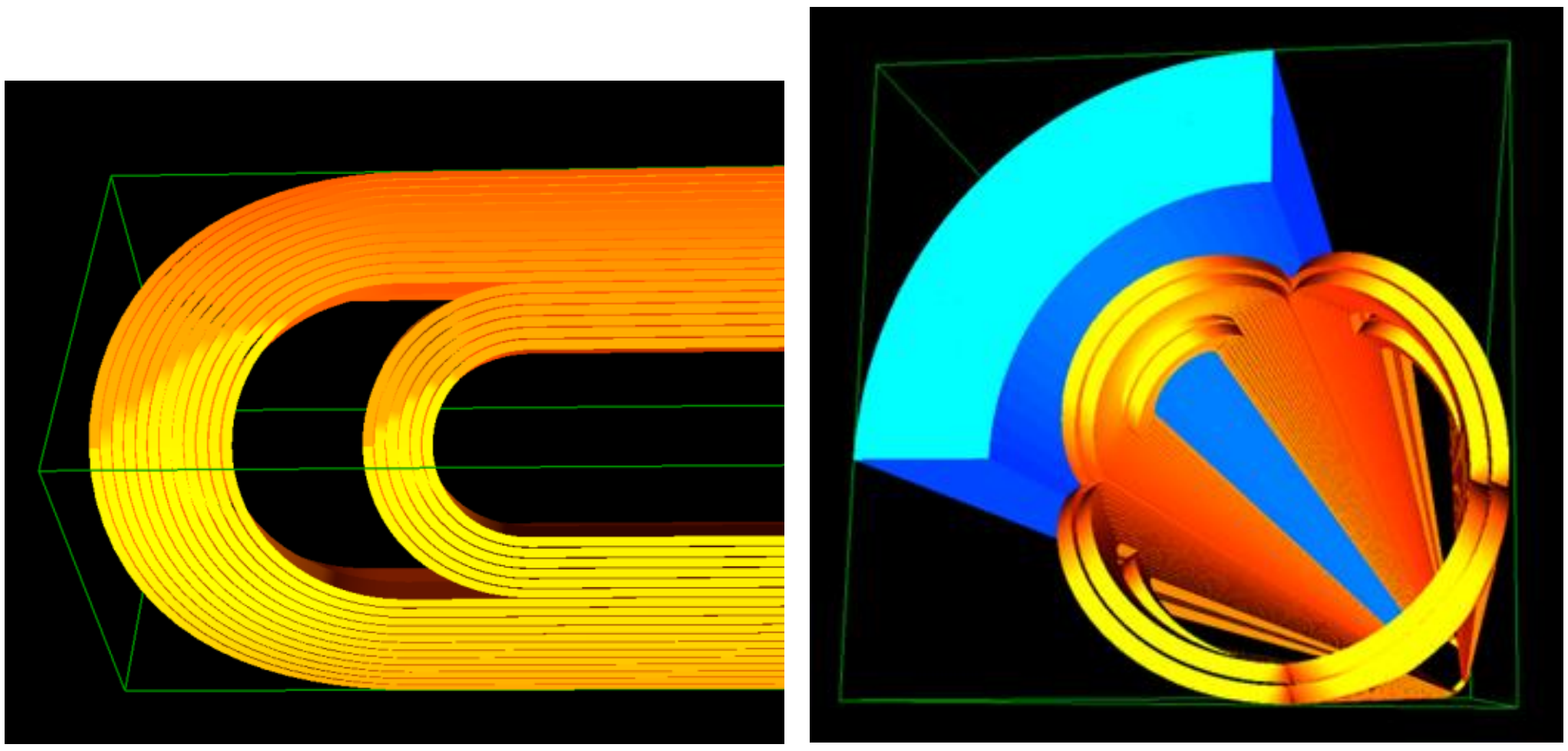

**Figure 4.3.4.17:** 3D field simulation of quadrupole magnet Q2 with iron yoke

All calculated integrated field harmonics are smaller than 1 unit at reference radius of 12.24 mm, indicating that the 3D magnetic field performance meets the requirement. Table 4.3.4.8 summarizes the design parameters of the Q2 quadrupole magnet.

**Table 4.3.4.8:** Design parameters of dual aperture quadrupole magnet Q2 with iron yoke

| Parameter | Q2 dual aperture |
|---|---|
| Field gradient (T/m) | 97.7 |
| Magnetic length (m) | 1502.08 |
| Coil turns per pole | 33 |
| Excitation current (A ) | 1925 |
| Coil layers | 2 |
| Conductor (HTS or LTS) | Rutherford Cable, width 3 mm, mid thickness 0.93 mm, keystone angle 1.9 deg, Cu:Sc = 1.3, 12 strands |
| Maximum dipole field at the center of each aperture (mT) | 2.54 |
| Stored energy (KJ) (double aperture) | 33.28 |
| Inductance (mH ) | 18.19 |
| Peak field in coil (T) | 3.66 |
| Load line | 72.05% (NbTi) |
| Integrated field harmonics | $b_6 = -0.52$<br>$b_{10} = -0.49$ |
| Coil inner diameter (mm) | 62 |
| Coil outer diameter (mm) | 75.30 |
| Yoke outer diameter (mm) | 126.4 |
| X direction Lorentz force/octant (kN) | 126.94 |
| Y direction Lorentz force/octant (kN) | –112.68 |
| Net weight (kg) | 235 |

### 4.3.4.5 *Superconducting Anti-Solenoid*

#### *4.3.4.5.1 Overall Design*

Below are the summarized requirements for the anti-solenoid:

a) The total longitudinal field integral generated by the anti-solenoid and detector solenoid coils should be zero.
b) The longitudinal solenoid field inside Q1a, Q1b, and Q2 should be limited to a few hundred Gauss at each longitudinal position.
c) The solenoid field distribution along the longitudinal direction should meet the vertical emittance requirement of the beam optics.
d) The angle of the anti-solenoid visible at the collision point should comply with the detector requirements.

#### *4.3.4.5.2 Magnetic Field Analysis*

The anti-solenoid design in this study closely resembles that of CDR [10], with a similar coil layout. It has been meticulously designed to meet the specified requirements. The anti-solenoid will be constructed using rectangular HTS or LTS conductor. To account for the slowly decreasing magnetic field of the detector solenoid along the longitudinal direction, and to minimize magnet size, energy, and cost, the anti-solenoid will be divided into 30 sections of varying inner coil diameters. Although these sections are connected in series, the current in select sections of the anti-solenoid can be adjusted with auxiliary power supplies as needed.

The anti-solenoid flux lines are calculated using an axial-symmetric model in OPERA-2D, as shown in Figure 4.3.4.18.

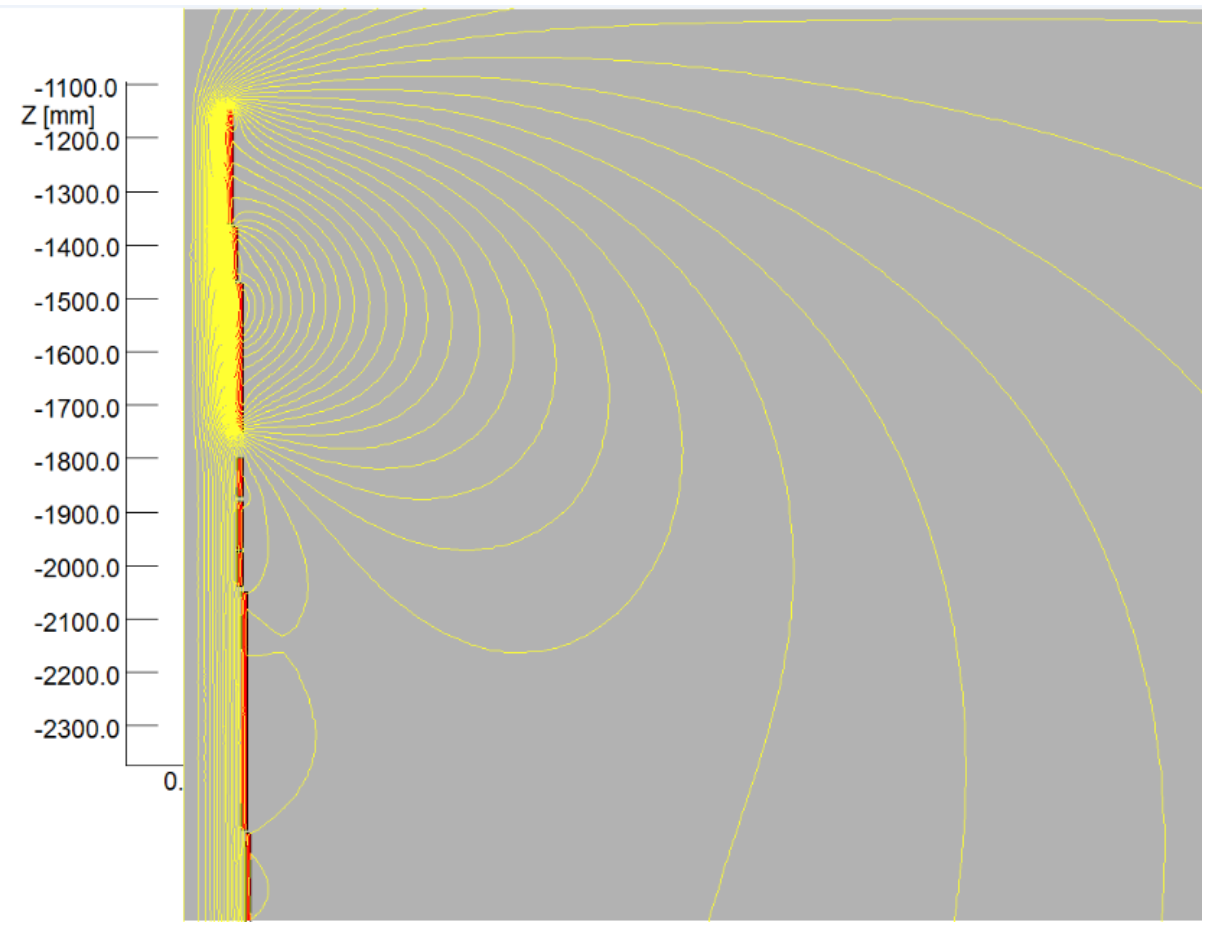

**Figure 4.3.4.18:** Simulation of anti-solenoid flux lines

The anti-solenoid has an outer coil diameter of 380 mm and is excited with a current of 1300 A. The working temperature of the anti-solenoid is 4.2 K, and the stored energy is 660 KJ. The weight of the anti-solenoid on each side of the IP point is 302 kg. Combined field simulation model of detector solenoid and anti-solenoid is presented in Figure 4.3.4.19. Figure 4.3.4.20 shows the distribution of the combined field generated by the detector solenoid and anti-solenoid along the longitudinal direction.

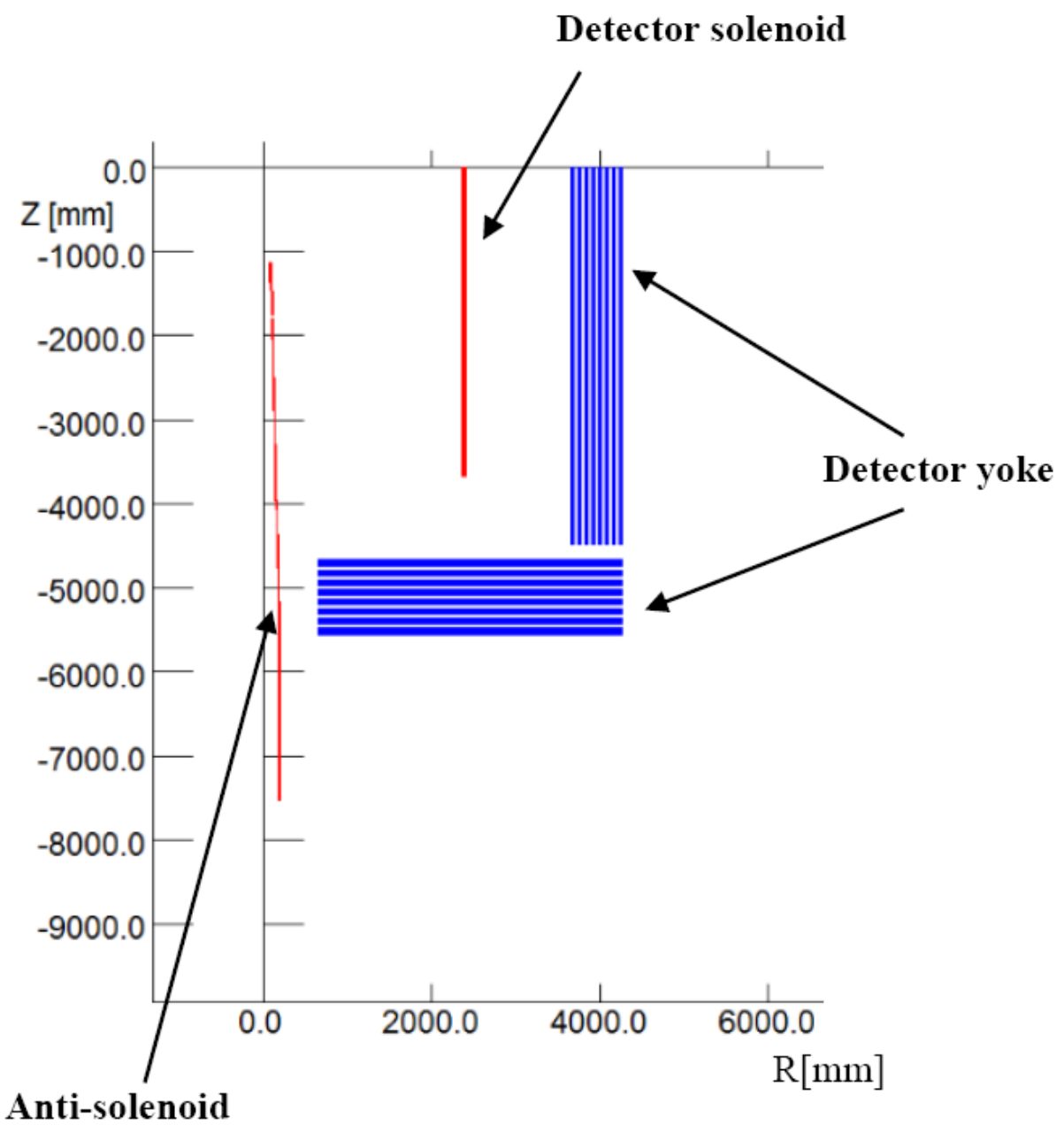

**Figure 4.3.4.19:** One half of the simulation model of the detector solenoid and anti-solenoid. The horizontal axis is the radial direction, and the vertical axis the longitudinal direction; (0, 0) is the interaction point; dimensions in mm.

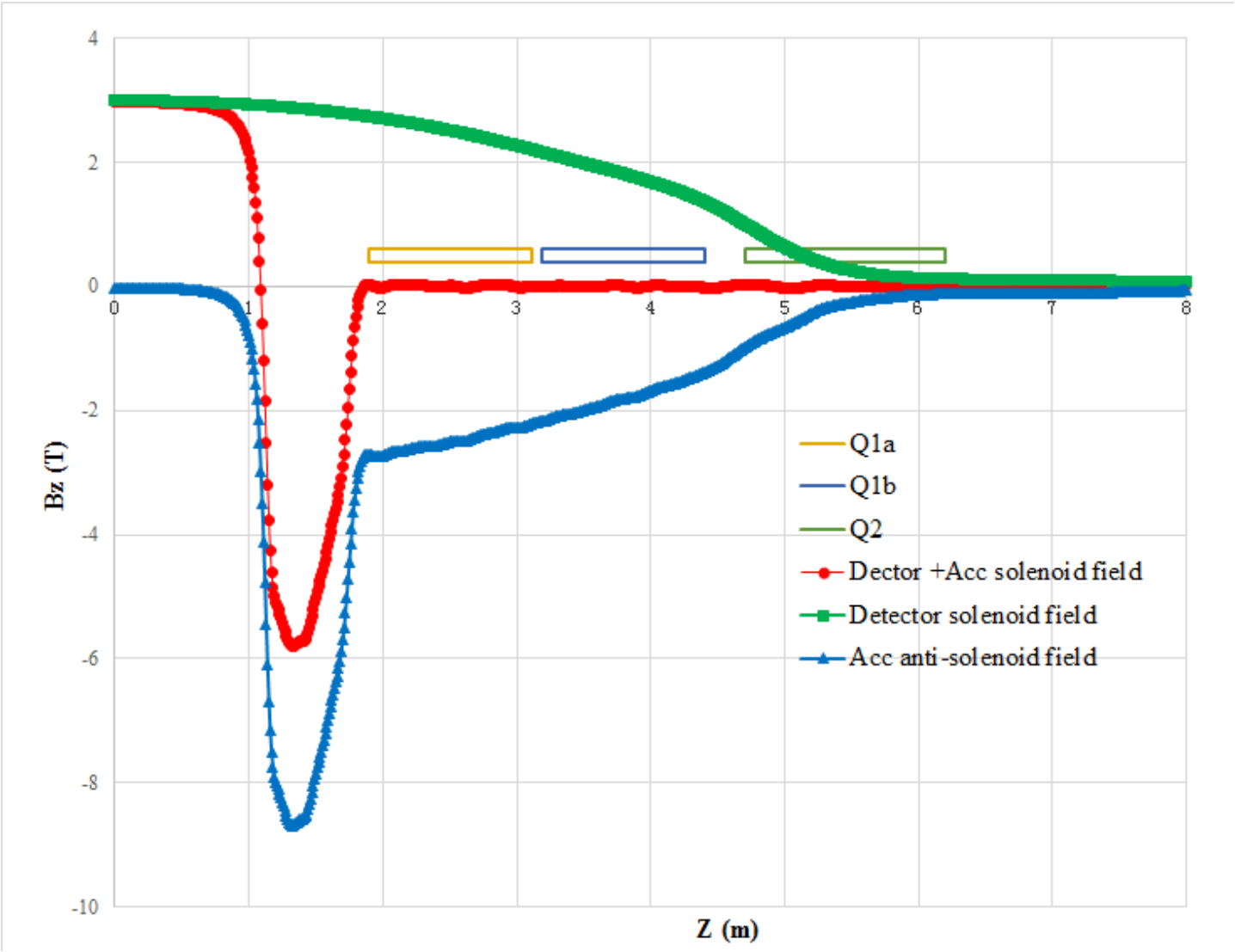

**Figure 4.3.4.20:** The distribution of the combined field of the detector solenoid and anti-solenoid along the longitudinal direction

The strongest central field is in the first section of the anti-solenoid with a peak value of 8.8 T. So YBCO or Nb3Sn conductors will be utilized for the first section of the anti-solenoid, whereas NbTi conductor will be employed for the remaining anti-solenoid sections with lower magnetic fields. The solenoid field inside each quadrupole magnet along the longitudinal position is less than 300 Gs. The combined field distribution of the anti-solenoid and detector solenoid magnet meets the design requirements.

#### 4.3.4.6 *R&D of 0.5 m Single Aperture Short Model Quadrupole Magnet*

In the R&D process, the initial phase involves researching and mastering the primary key technologies of superconducting quadrupole magnets. This is accomplished by creating a single-aperture quadrupole model magnet with a length of 0.5 meters using NbTi conductor, in collaboration with Hefei KEYE Company. The primary objective of this initial R&D phase is to confirm whether a high magnetic field gradient can be attained.

The single-aperture short model quadrupole magnet's overall design features a self-supporting structure consisting of a coil, collar, and iron yoke. In the physical design, a double-layer Cos2θ coil configuration is employed for the superconducting quadrupole magnet. To mimic the actual coil winding process, we construct the coil turn by turn, transitioning from the inner block to the outer block at the coil's end during field simulation. Figure 4.3.4.21 displays a complete coil and the cross-section of the short model magnet.

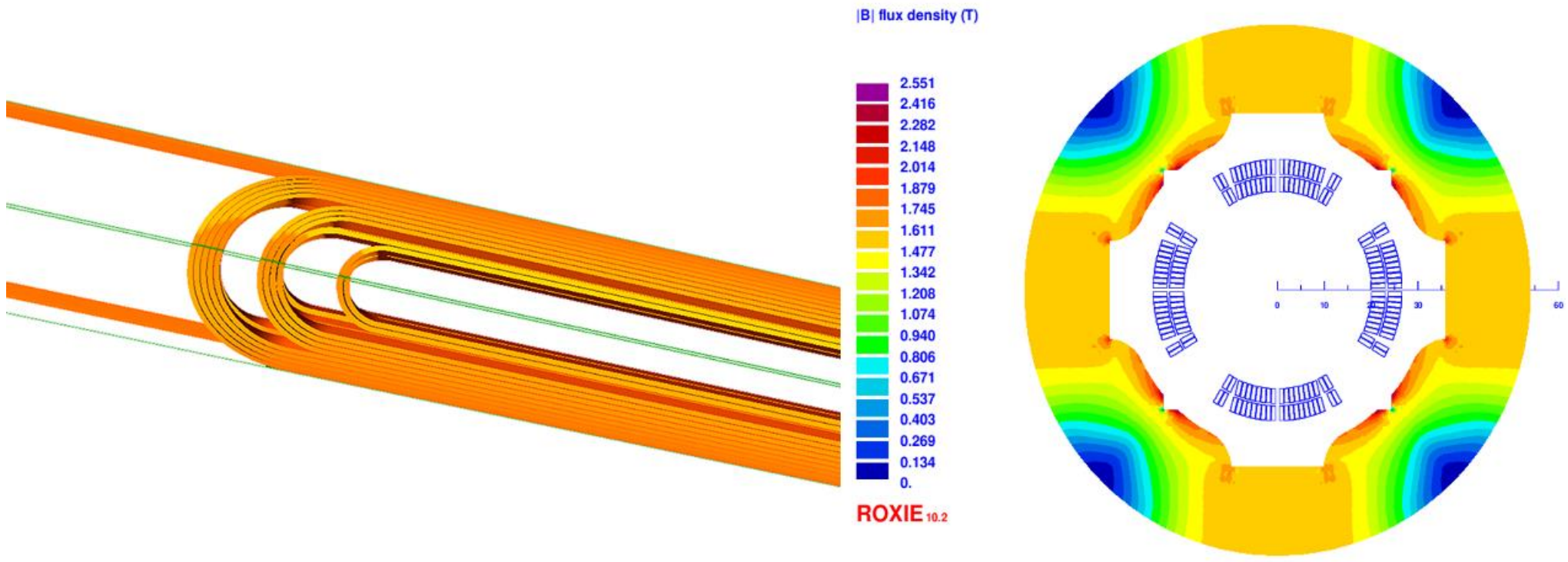

**Figure 4.3.4.21:** Physical design of the 0.5m single aperture short model quadrupole magnet.

The manufacturing of the 0.5-meter single-aperture short model quadrupole was finished in August 2022, as depicted in Figure 4.3.4.22.

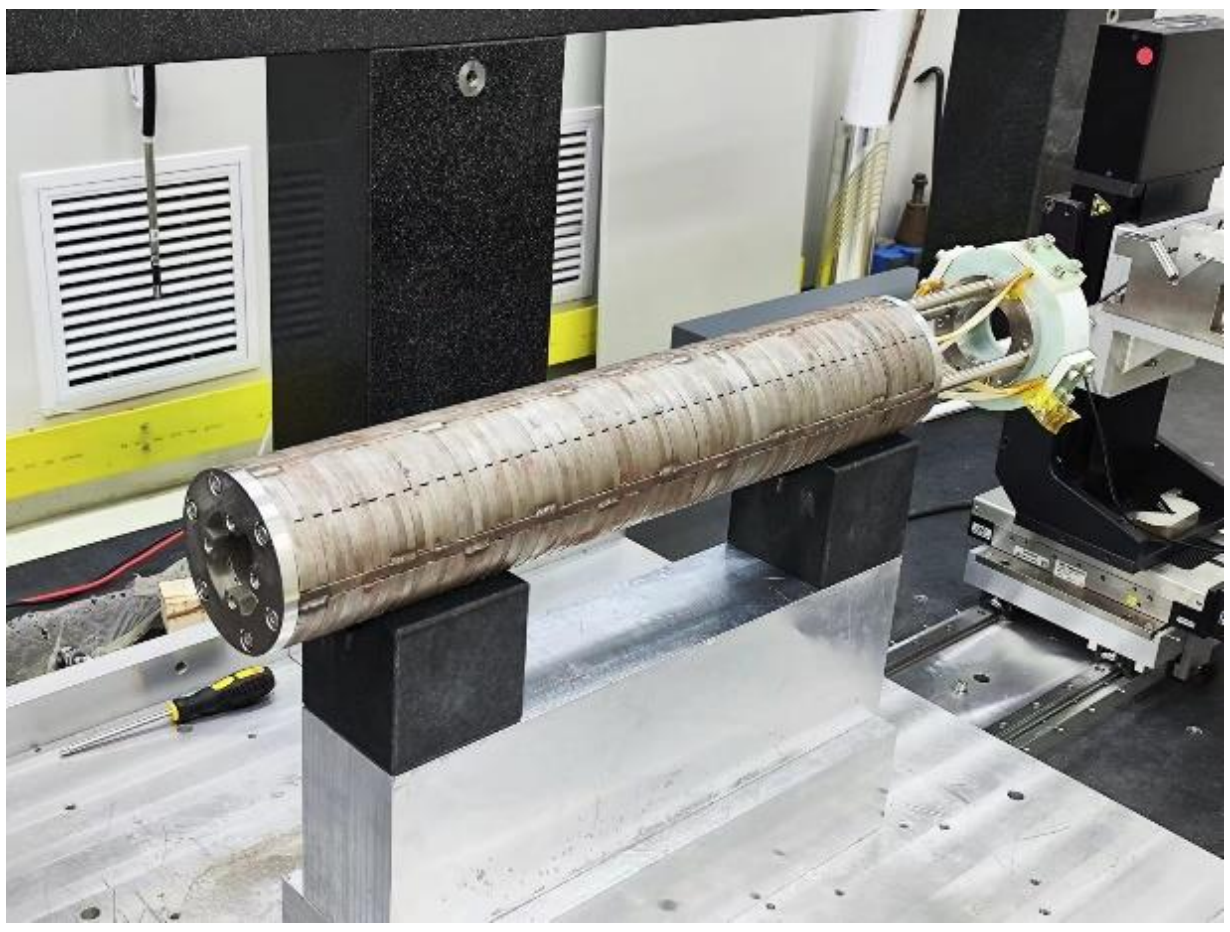

**Figure 4.3.4.22:** A 0.5m single aperture short model quadrupole magnet.

Room temperature rotating coil magnetic field measurements were conducted, revealing that the measured sextupole and octupole field components exceed $1\times10^{-3}$. This indicates the necessity for enhancements in magnet fabrication and assembly accuracy.

At IHEP, a cryogenic excitation test of the 0.5-meter single-aperture short model quadrupole magnet at 4.2 K was carried out within a vertical Dewar. The purpose of this test was to confirm the attainability of a high magnetic field gradient.

During the cryogenic excitation process, the magnet successfully achieved the design current of 2,115A (136 T/m). Subsequently, the excitation current was incrementally increased to assess the safety margin of the superconducting magnet. Ultimately, the current reached 2,500A, which aligns with the maximum design current of the Dewar current lead. The results of the cryogenic excitation test are illustrated in Figure 4.3.4.23.

The theoretical magnetic field gradient for the 0.5-meter single-aperture short model quadrupole at 2,500A is calculated to be 160.6 T/m, surpassing the magnetic field gradient established during the CEPC technical design stage (TDR). Importantly, the short model quadrupole magnet remained quench-free throughout the entire cryogenic excitation process.

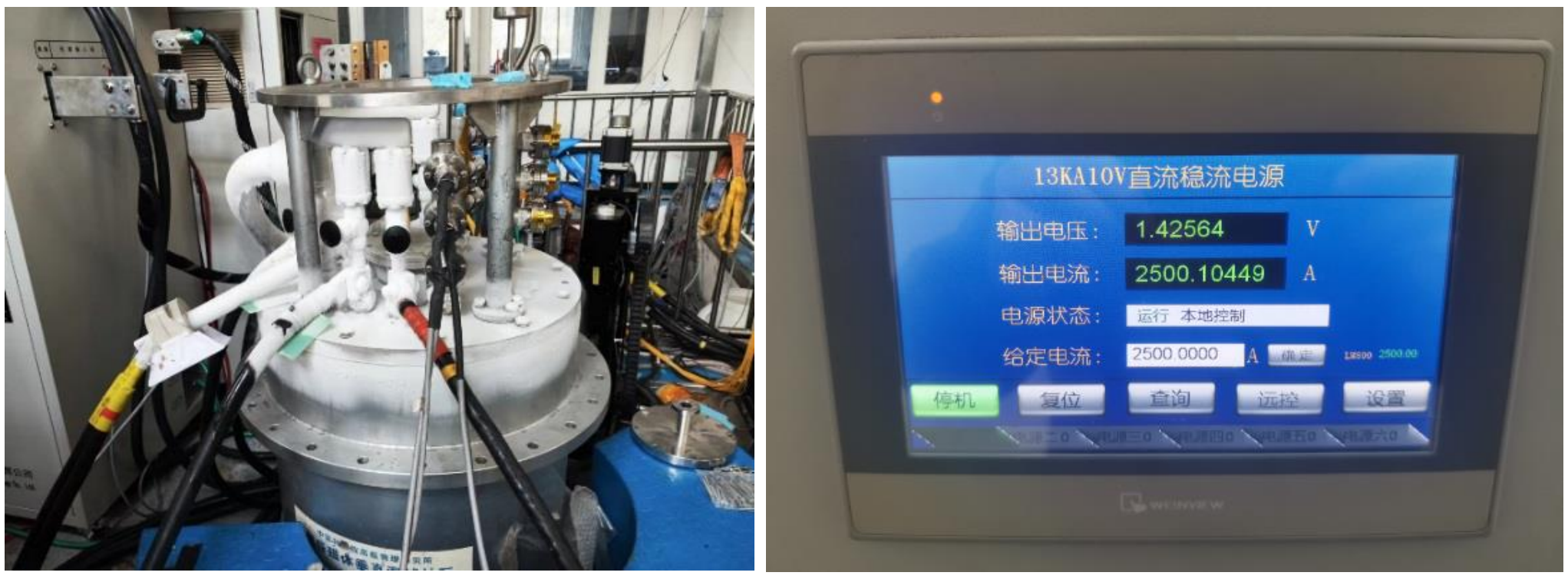

**Figure 4.3.4.23:** Cryogenic test of the 0.5m single aperture short model quadrupole magnet.

The successful cryogenic excitation of the CEPC 0.5-meter single-aperture short model quadrupole magnet demonstrates the mastery of key technologies related to design, manufacturing, assembly, and cryogenic vertical testing of short superconducting quadrupole magnets utilizing LTS Cos2θ coils. This achievement confirms the attainability of the high field gradient of 142.3 T/m for quadrupole magnets as specified in the CEPC TDR stage.

### 4.3.5 Magnet Power Supplies

#### 4.3.5.1 *Introduction*

The primary function of the collider magnet power supply is to provide the necessary excitation current to the magnet. This is achieved by setting the current of the power supply based on the design requirements of accelerator physics during the operation of the collider. This process is carried out through the central control system to create the working mode required for the operation of the accelerator.

The main power supply for the storage ring's magnets is unipolar, with differences only in power output rating and stability requirements. The power supply is utilized to provide long-term, stable DC excitation current for the magnets in order to achieve the design goals of accelerator physics.

To achieve the full ring and local correction of the beam orbit, the correction magnet power supply in the collider is bipolar.

Given the massive size of CEPC, which spans approximately 100 kilometers in length, comprising of 8 curved arc sections and 8 straight sections, a significant number of power supplies are required to provide excitation current for various types of magnets. All power supplies are rated for 120 GeV operation plus a 10% safety margin in both current and voltage.

The latest accelerator physics design requires approximately 17,500 magnets to be powered, with a total power consumption of around 30 MW. As a result, reducing unnecessary power consumption, mainly the power loss on the load cable, is a critical aspect of power system design.

Power supply design experience has shown that cable loss can be reduced by decreasing magnet excitation current, but this necessitates an increase in the power supply output voltage rating. Therefore, to reduce the complexity of power supply design, the power supply output voltage rating is limited to DC1000 V.

To save cables, a single power supply or two power supplies will be used for the same type of magnets in one arc, which will be installed in surface buildings located close to the access shafts. Power supplies for single magnet loads will be installed in auxiliary tunnels around the main tunnel. There are a total of 96 auxiliary tunnels distributed uniformly around the ring, with all eight surface buildings located close to the access shafts, as depicted in Figure 4.3.5.1.

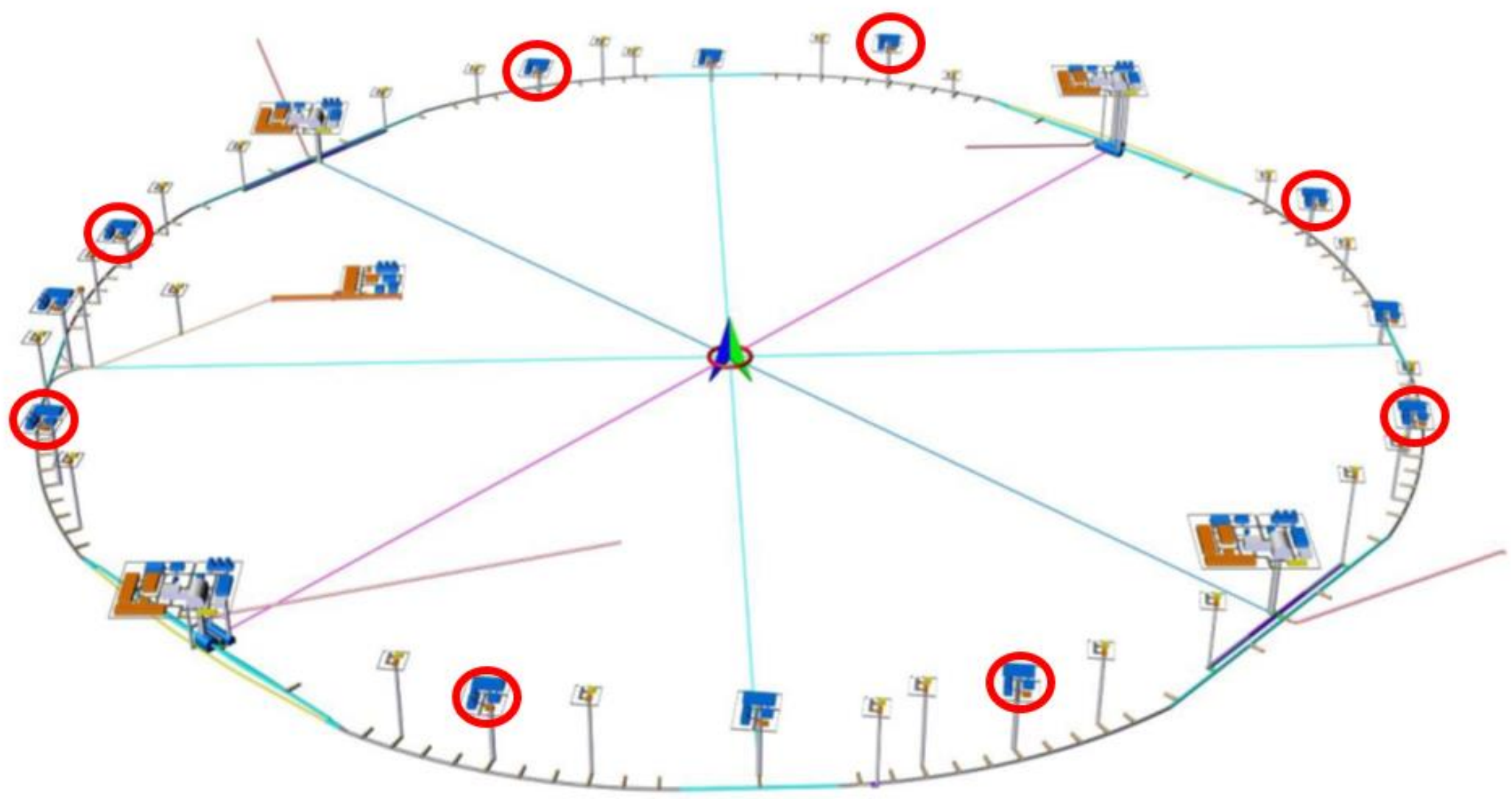

**Figure 4.3.5.1:** Layout of surface and underground CEPC structures

### 4.3.5.2 *Types of Power Supplies*

Table 4.3.5.1a-d presents the specifications of the main magnet and correction magnet power supplies for the collider.

**Table 4.3.5.1a:** Power supply requirements for dipole magnet

| Collider | Quantity | Stability / 8 hrs | Output Rating |
|---|---|---|---|
| D-aperture B0 | 16 | 10 ppm | 1300A/500V |
| D-aperture B1 | 16 | 10 ppm | 150A/50V |
| BMV01IRD | 4 | 10 ppm | 90A/40V |
| BMVIRD | 4 | 10 ppm | 250A/90V |
| BMHIRD | 4 | 10 ppm | 160A/50V |
| BGM1 | 4 | 10 ppm | 150A/60V |
| BGM2 | 4 | 10 ppm | 80A/60V |
| BMV01IRU | 4 | 10 ppm | 40A/50V |
| BMVIRU | 4 | 10 ppm | 160A/70V |
| BMHIRU | 4 | 10 ppm | 70A/50V |
| BRF0 | 2 | 10 ppm | 90A/50V |
| BRF | 4 | 10 ppm | 210A/60V |
| **Totle** | **70** | | |

**Table 4.3.5.1b:** Power supply requirements for quadrupole magnet

| Collider | Quantity | Stability / 8 hrs | Output Rating |
|---|---|---|---|
| D-aperture Q | 320 | 10 ppm | 120A/450V |
| D-aperture QH | 8 | 10 ppm | 120A/50V |
| Q3IRD | 4 | 10 ppm | 160A/50V |
| Q4IRD | 4 | 10 ppm | 170A/50V |
| Q5IRD | 4 | 10 ppm | 40A/30V |
| QDVHIRD | 4 | 10 ppm | 120A/180V |
| QFVIRD | 4 | 10 ppm | 120A/150V |
| QFHHIRD | 4 | 10 ppm | 160A/240V |
| QDHIRD | 4 | 10 ppm | 160A/190V |
| QCMIRD | 4 | 10 ppm | 140A/510V |
| QDCIRD | 4 | 10 ppm | 100A/120V |
| QFCIRD | 4 | 10 ppm | 90A/100V |
| QMIRD | 4 | 10 ppm | 120A/810V |
| QMIRD | 4 | 10 ppm | 110A/150V |
| QMIRD | 4 | 10 ppm | 100A/120V |
| QDGMH | 4 | 10 ppm | 110A/440V |
| QFGM | 4 | 10 ppm | 110A/350V |
| Q3IRUJ2 | 4 | 10 ppm | 90A/40V |
| Q4IRUJ2 | 4 | 10 ppm | 90A/40V |
| QFVIRU | 4 | 10 ppm | 90A/110V |
| QDVHIRU | 4 | 10 ppm | 80A/130V |
| QDHIRU | 4 | 10 ppm | 120A/140V |
| QFHHIRU | 4 | 10 ppm | 80A/170V |
| QCMIRU | 4 | 10 ppm | 100A/450V |
| QDCIRU | 4 | 10 ppm | 70A/90V |
| QFCIRU | 4 | 10 ppm | 70A/80V |
| QMIRU | 4 | 10 ppm | 100A/140V |
| QMIRU | 4 | 10 ppm | 100A/140V |
| QMIRU | 4 | 10 ppm | 70A/110V |
| QMIRU | 4 | 10 ppm | 70A/110V |
| QSEP | 20 | 10 ppm | 140A/60V |
| QRF | 12 | 10 ppm | 70A/410V |
| QFSTRH | 8 | 10 ppm | 50A/30V |
| QFSTR | 12 | 10 ppm | 70A/60V |
| QSEP | 12 | 10 ppm | 90A/40V |
| QSTRH | 4 | 10 ppm | 80A/40V |
| QSTR | 16 | 10 ppm | 100A/50V |
| QAI | 40 | 10 ppm | 100A/60V |
| QHAI | 4 | 10 ppm | 60A/40V |
| QHAO | 4 | 10 ppm | 60A/40V |
| QAO | 44 | 10 ppm | 80A/60V |
| QSTR | 32 | 10 ppm | 60A/50V |
| QSTRH | 4 | 10 ppm | 60A/40V |
| QSEP | 12 | 10 ppm | 90A/50V |
| QDI | 48 | 10 ppm | 70A/60V |
| QSTRHI | 8 | 10 ppm | 70A/40V |
| QINJI | 32 | 10 ppm | 90A/30V |
| QFI | 44 | 10 ppm | 70A/60V |
| QDO | 48 | 10 ppm | 70A/60V |
| QFSTRHO | 8 | 10 ppm | 70A/40V |
| QINJO | 32 | 10 ppm | 70A/40V |
| QFO | 44 | 10 ppm | 70A/60V |
| **Totle** | **928** | | |

**Table 4.3.5.1c:** Power supply requirements for sextupole magnet

| Collider | Quantity | Stability / 8 hrs | Output Rating |
|---|---|---|---|
| SFI | 256 | 20 ppm | 200A/110V |
| SDI | 256 | 20 ppm | 200A/200V |
| SFO | 256 | 20 ppm | 200A/110V |
| SDO | 256 | 20 ppm | 200A/200V |
| VSCIRD | 16 | 20 ppm | 100A/30V |
| HSCIRD | 16 | 20 ppm | 150A/30V |
| SC0IRD | 4 | 20 ppm | 150A/50V |
| VSCIRU | 16 | 20 ppm | 100A/30V |
| HSCIRU | 16 | 20 ppm | 120A/30V |
| SC0IRU | 4 | 20 ppm | 150A/50V |
| **Totle** | **1096** | | |

**Table 4.3.5.1d:** Power supply requirements for corrector

| Collider | Quantity | Stability / 8 hrs | Output Rating |
|---|---|---|---|
| CH | 3544 | 100 ppm | ±20A/±20V |
| CV | 3544 | 100 ppm | ±20A/±20V |
| B0A | 2048 | 100 ppm | ±6A/±12V |
| B0B | 3840 | 100 ppm | ±6A/±12V |
| Q | 6016 | 100 ppm | ±15A/±10V |
| **Totle** | **18992** | | |

1) Dipole Magnet Power Supplies:
The collider is equipped with a total of 3,170 dipole magnets, with 3,008 dual-aperture dipoles located in the arc sections and 162 single-aperture dipoles distributed mainly in the IR sections and RF zones. To minimize power loss, the dual-aperture dipoles within a half-arc are connected in series and powered by a single power supply. As a result, 16 dipole power supplies, each with a power rating of 0.6 MW (including an allowance for cable losses), will be installed in surface buildings. The single-aperture dipole magnets, on the other hand, will be powered independently and installed in an auxiliary underground tunnel, as close as possible to the magnet load.

2) Quadrupole Magnet Power Supplies:
   The Collider is equipped with 3016 dual-aperture quadrupoles and 1116 single-aperture quadrupoles. The dual-aperture quadrupoles are arranged into 320 families, each comprising 9 or 10 series-connected magnets that are powered by an individual power supply. The single-aperture quadrupoles are powered independently. To accommodate the power supplies, all of them will be installed in the auxiliary stub tunnel that runs parallel to the main tunnel.

3) Sextupole Magnet Power Supplies:
   There are 3072 sextupoles in the arc region of the Collider. To meet the accelerator physics requirements, every two adjacent magnets are connected in series and powered by one power supply. The remaining 104 magnets in the RF region (32 of them are superconducting magnets) are powered independently. All the power supplies for the sextupoles and RF region magnets will be installed in the auxiliary tunnel.

4) Corrector Power Supplies:
   The total number of correction CH and CV magnets is approximately 7088. In addition, each of all dual aperture magnets (B0 and Q) has 2 trim coils. All correction magnets and trim coils are powered independently.

Since magnet parameters may undergo future changes, we have not currently optimized the parameters of the magnet system to minimize the diversity of required power supplies. The requested power supply stability ranges from 10 ppm to 100 ppm peak to peak, based on calculations by the Beam Dynamic Group.

#### 4.3.5.3 *Design of the Power Supply System*

The power supply system design is guided by the following principles:

1) Meet the requirements of accelerator physics design to ensure beam position and collider operating point stability.
2) Maintain effective communication with the magnet system to select optimal electrical parameters.
3) Use switching power supply as the primary topology mode to reduce volume, improve efficiency, and facilitate digital control.
4) Employ modular design to achieve high current and high voltage output.
5) Utilize high precision and digital current closed-loop control in the power supply.
6) Maintain a cable current density of less than 2 A/mm$^2$.
7) Use water cooling for power supplies with power greater than 1 kW and forced air cooling for others.
8) Attain a power factor of cosø = 0.95 in the main network.
9) Achieve power supply efficiency of $\eta = 0.9$.

The CEPC collider power supply will utilize digital control technology, enabling adjustments and optimization of power control loop parameters through software. This provides convenience for power supply commissioning and maintenance. The power control system utilizes a professional signal processing chip, reducing the impact of ambient temperature on power supply output performance and improving reliability. As intelligent technology continues to develop, intelligent monitoring of the power supply can be achieved without the need for hardware changes. Figure 4.3.5.2 (a) and (b) illustrate two main digital power supply architectures for the CEPC collider power supply.

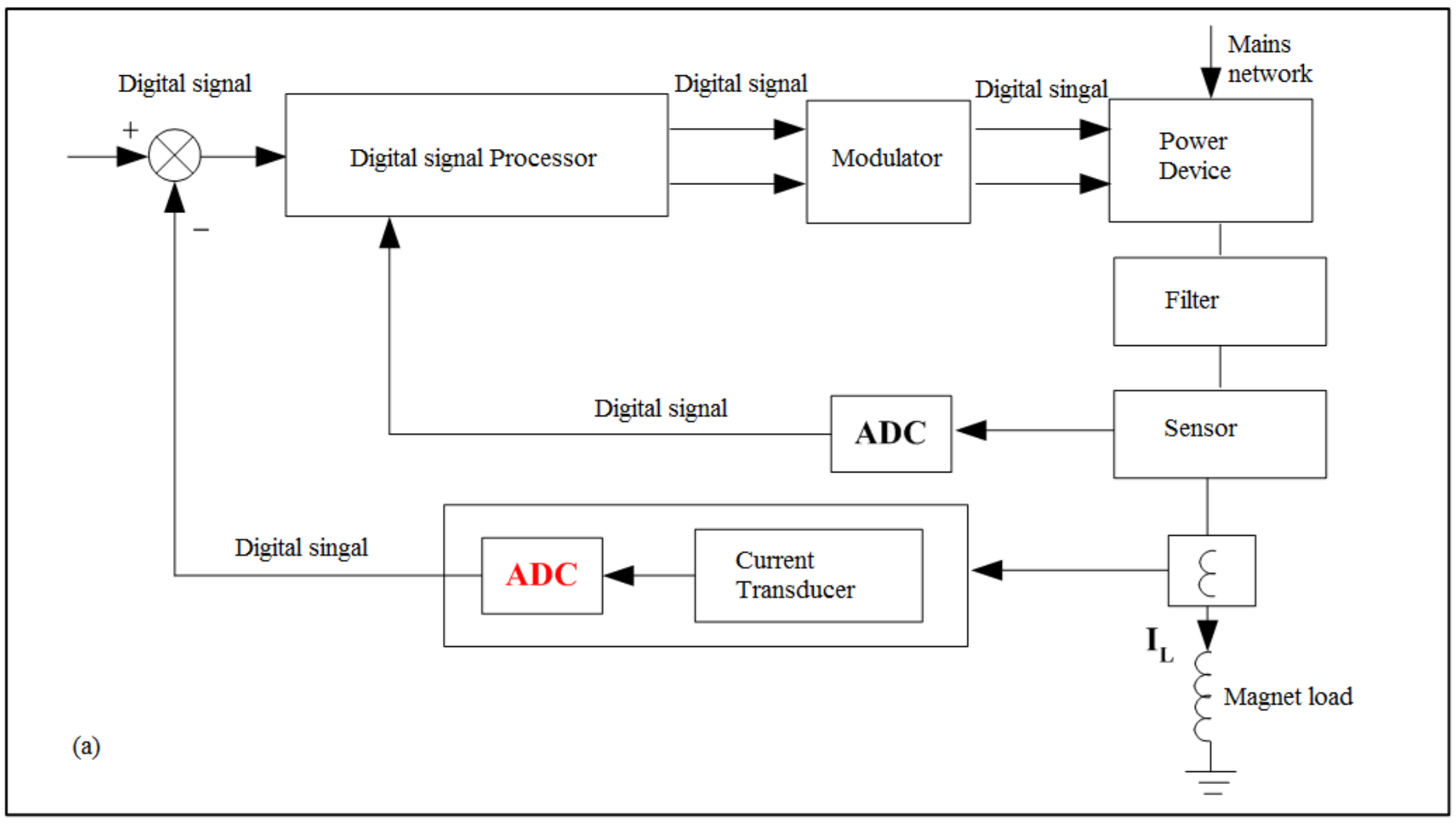

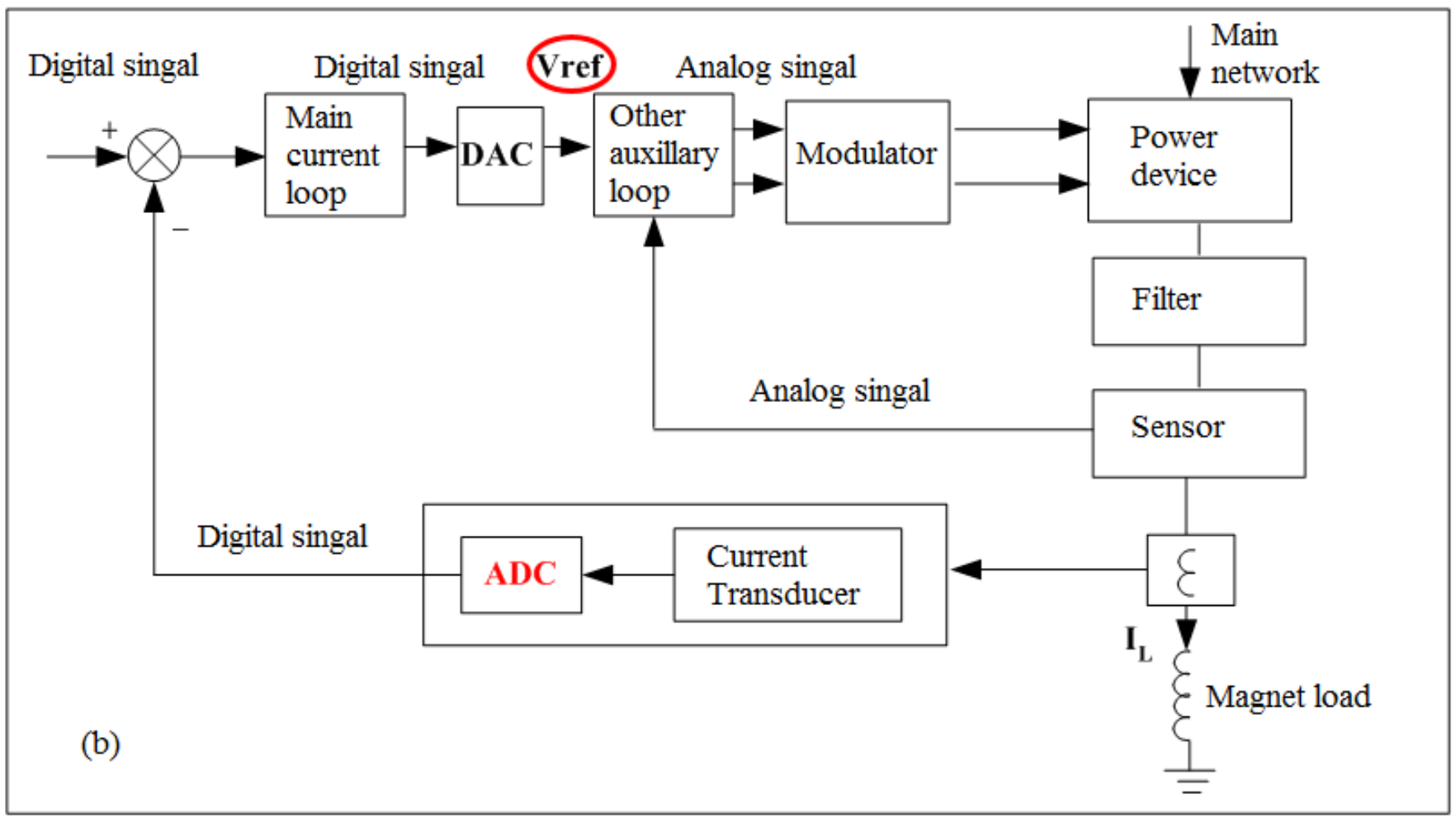

**Figure 4.3.5.2**: Top: Fully digital + switch topology. Bottom: Digital current loop + arbitrary topology.

The all-digital switching power supply operates by utilizing the digital controller of the power supply to achieve digital adjustment and control of all control circuits. It generates ultra-precise pulse width modulation (PWM) signals via the hardware with a minimum adjustment of less than 150 ps. As the minimum frequency of the PWM signal is 20 kHz (50 μs), it satisfies the stability requirements of the power supply in CEPC, with an accuracy of 3 PPM (three parts per million). This architecture provides the simplest closed-loop control structure for the digital switching power supply.

The digital controller employs the digital current loop and arbitrary topology mode to achieve high-precision current closed-loop control. The output of the current loop serves as a reference for the analog voltage loop after being passed through digital-to-analog conversion (DAC). This enables the power part, functioning as a voltage source, to operate in arbitrary topology mode. This control mode leverages the benefits of digital control while simultaneously surmounting the limitations of digital PWM control precision and dependence on topological structure.

Based on the prototype development results, the preferred choice for correcting magnet power supply is the structure depicted in Figure 4.3.5.2 (a), as it simplifies circuit design. Meanwhile, for other high-stability DC stable current power supplies, the preferred choice is the structure shown in Figure 4.3.5.2 (b). The digital controller completes various current indicators, while the analog voltage loop suppresses voltage ripple. Additionally, the design of multi-module series-parallel current balancing and voltage balancing is executed through the analog control circuit. This approach enhances the flexibility and selectivity of power supply design.

The main topology of the main magnet power supply operating in a single quadrant is depicted in Figure 4.3.5.3. The circuit employs a DC source plus chopper approach, with the DC source utilizing a rectifier transformer and a diode full bridge rectifier structure.

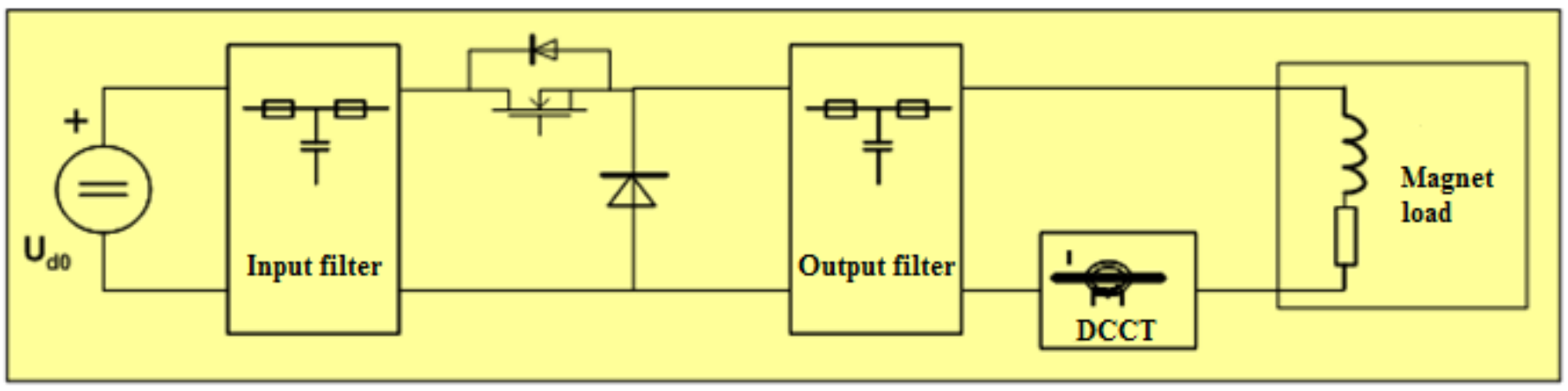

**Figure 4.3.5.3:** Schematic diagram of the single quadrant power supply topology.

Figure 4.3.5.4 shows the primary topology of the main magnet power supply working in two quadrants, which uses the DC source plus H bridge scheme. The DC source adopts the rectifier transformer and diode full bridge rectifier structure.

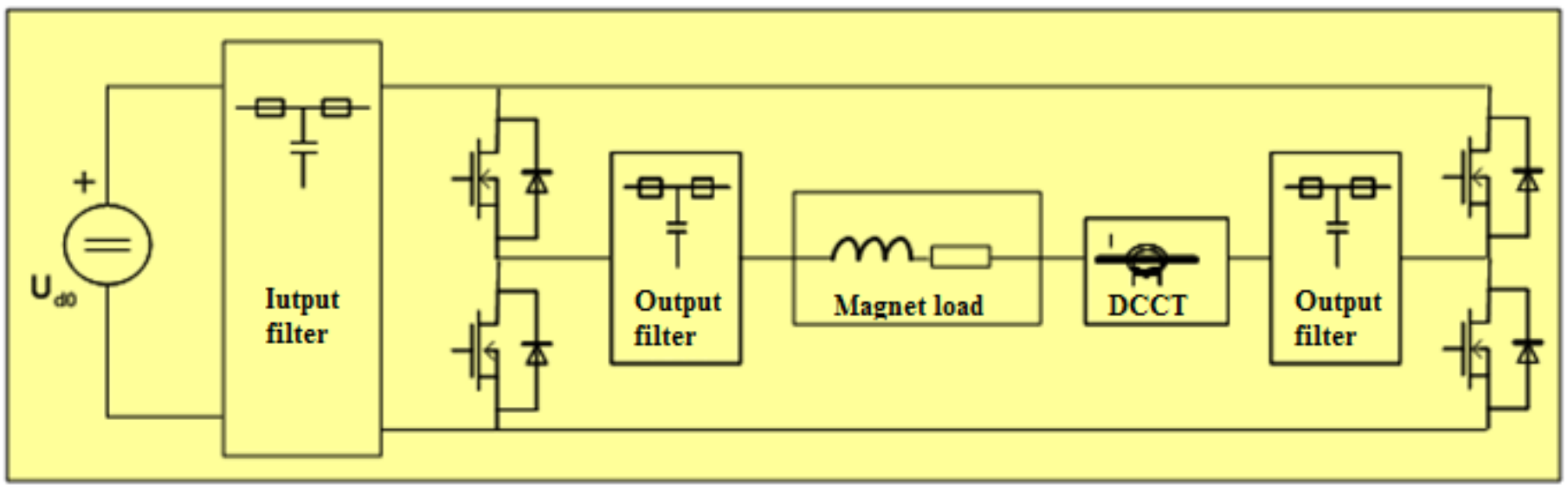

**Figure 4.3.5.4:** Schematic diagram of the two-quadrant power supply topology.

4.3.5.4 ***Design of Digital Power Supply Cntrol Module***

The digital control module is the key component that enables high-precision control of the power supply. It serves as the executive element for achieving digital control, and its control algorithms provide diversity and robustness.

To meet the control precision requirements of CEPC's power supply system, the second generation of the digital power supply control module (DPSCM-II) will be utilized, which was developed for HEPS. DPSCM-II maintains the overall architecture of the first generation DPSCM but employs a field-programmable gate array (FPGA) as the data processing core and implements digital control of the power supply through a system-on-a-programmable-chip (SOPC) approach. Figure 4.3.5-5 depicts the power supply structure embedded in DPSCM-II.

The DPSCM-II hardware consists of several components, including the main board (DPSCM_MB), the high-precision ADC control board (DPSCM_AD), the digital-to-analog conversion between the digital current closed-loop and other analog control loops of the power supply (DAC DPSCM_DA), and the power supply monitoring interface circuit (DSPCM_MDA) composed of a multi-channel DAC. The main board also includes several interface circuits with other systems, such as the optical fiber interface for remote control, the optical fiber interface with the timing system, the control interface for displaying power supply parameters, the PWM synchronization signal with other power supplies, and the human-machine interface for local debugging. Figure 4.3.5.6 displays photographs of the primary hardware components of DPSCM-II.

The main board is capable of generating ultra-high precision PWM signals with a control step size of less than 150 ps, allowing for full digital control of the power supply. Only the current outer loop needs to be realized through digital closed-loop control using DSPCM_DA to accommodate various topologies. To ensure high stability, DPSCM_AD will be thermostatically controlled.

Conventional analog power supply control systems rely on simulator components, making them highly sensitive to temperature changes in the environment. As a result, high-precision stable current power supplies require strict ambient temperature control. By adopting digital control, the power supply's closed-loop control system uses digital signal processing, reducing the impact of temperature on performance.

In high-precision magnet power supplies, the output of the current sampling device (DCCT) is an analog voltage (current) signal that must be converted to digital before participating in the digital closed-loop feedback adjustment of the power supply. Therefore, ADC is the most temperature-dependent component in the entire power control system. To address this, a high-precision ADC application circuit with local constant temperature control will be designed and integrated into the power digital control system. This circuit not only provides current sampling for digital closed-loop feedback but also minimizes the power supply's ambient temperature requirements.

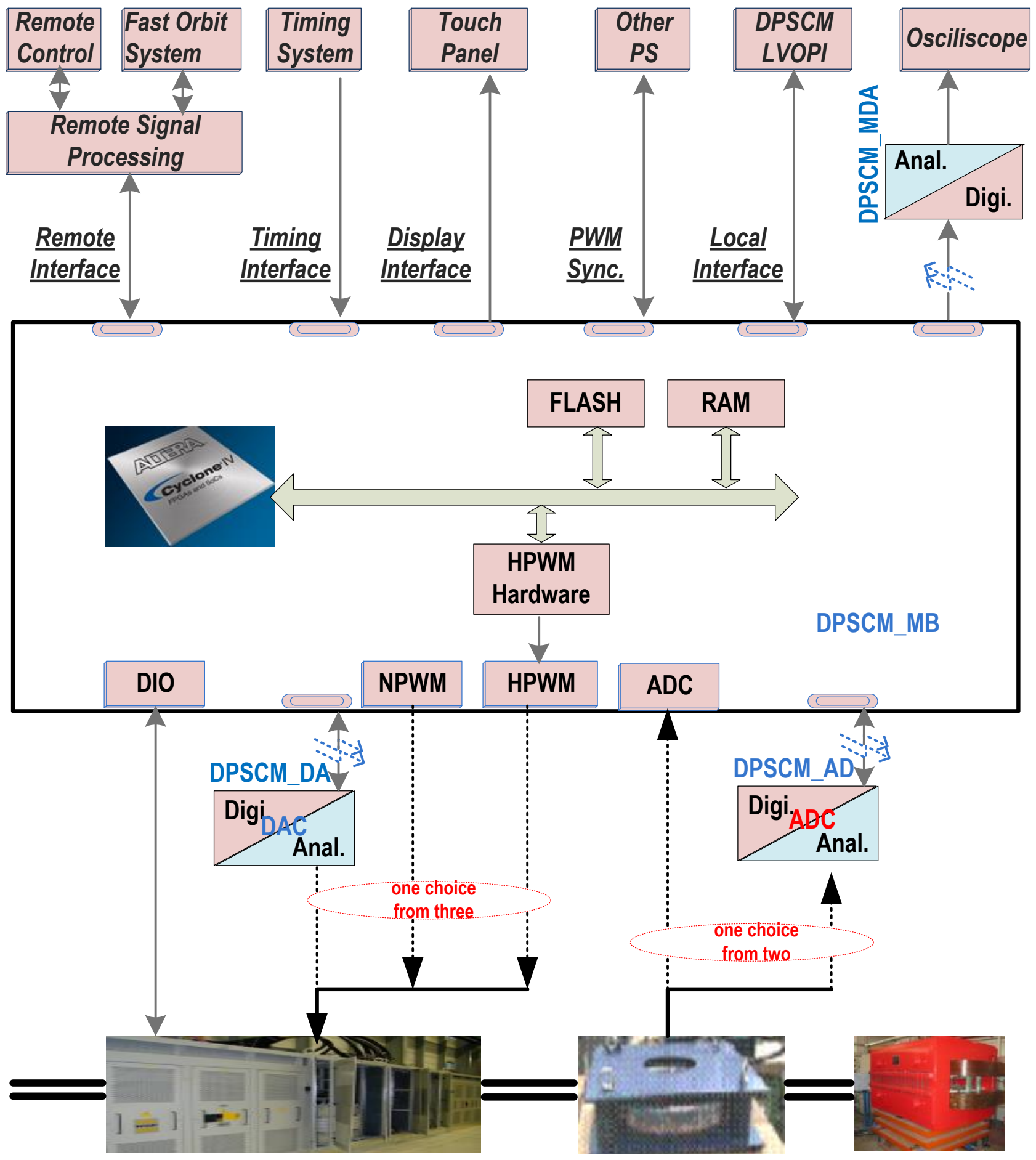

**Figure 4.3.5.5:** Digital power block diagram embedded in DPSCM-II.

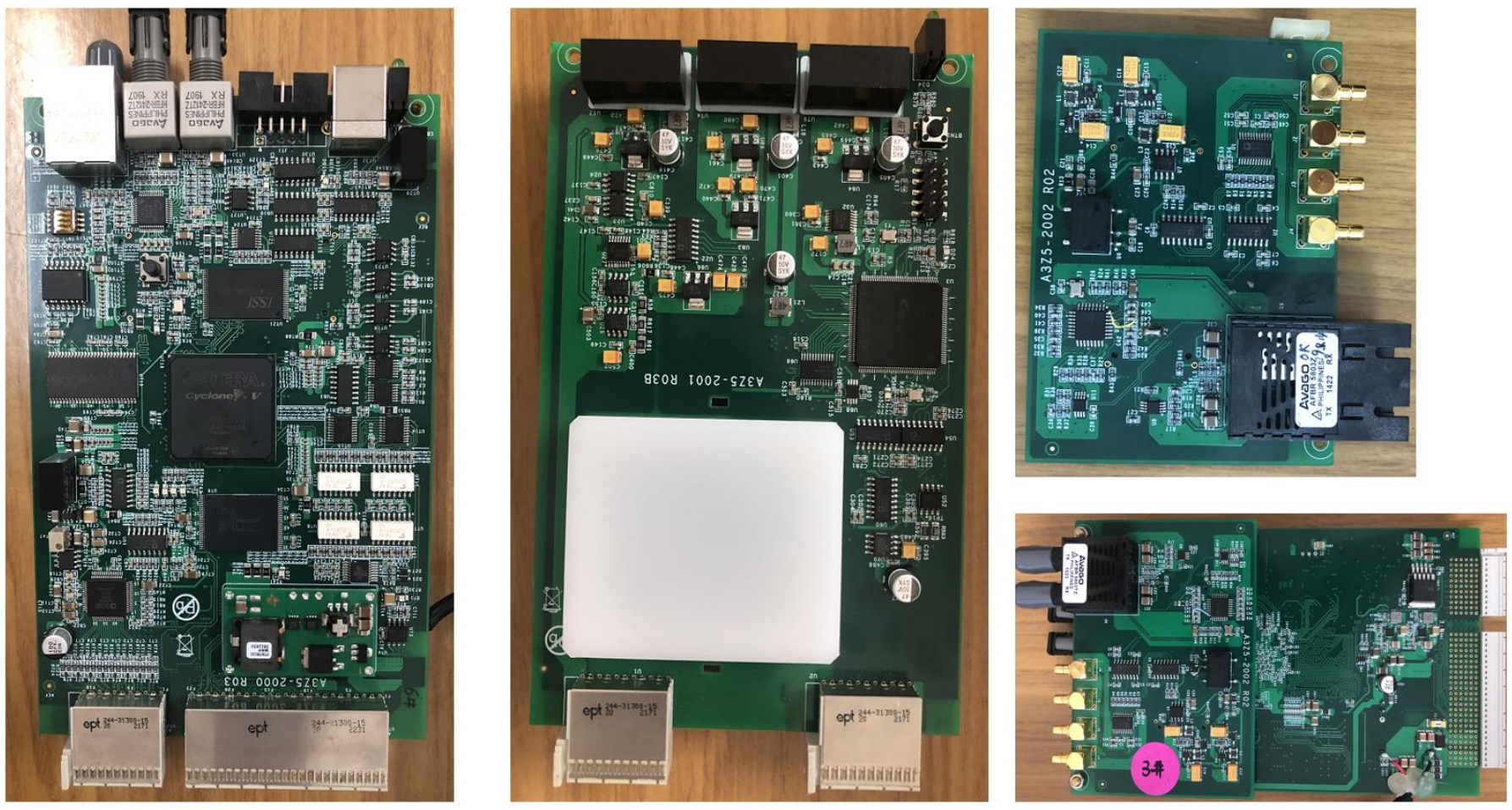

**Figure 4.3.5.6:** The main hardware of DPSCM-II.

### 4.3.5.5 *Topology of Power Supplies*

The power supply design utilizes a DC source and chopper mode. The DC source supplies a stable DC voltage and multiple 12/24 pulse rectifiers or PWM rectifiers are used to reduce harmonic current. A step-down or step-up chopper is implemented to control input power fluctuations and achieve output current control.

The modular design of the high-power supply allows for both high voltage and large current output through series and parallel modules, while also improving maintainability. Figure 4.3.5.7 depicts the block diagram of a high-power supply. The power supply topology utilizes staggered parallel technology in the power output circuit. The front DC source of the power supply comprises two 12-phase diode rectifier circuits connected in series, forming an equivalent 24-phase power rectifier circuit. In the output section of the power supply, six groups of BUCK circuits are utilized in parallel to provide a stable current output. The switching frequency of the BUCK circuit is set at 40 kHz.

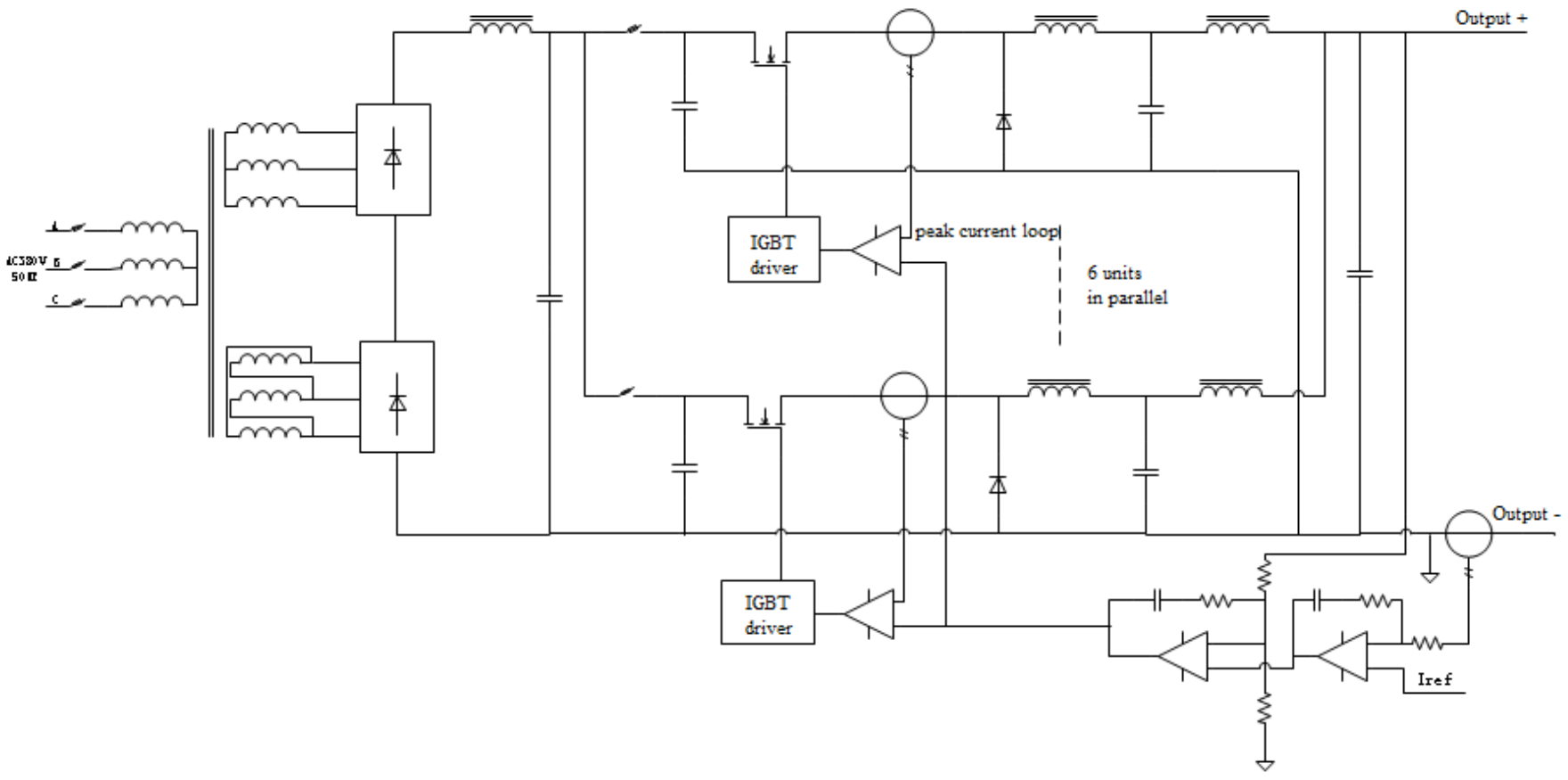

**Figure 4.3.5.7:** High power supply structure diagram.

This design has been implemented for a high-power supply utilized in BEPCII, capable of delivering an output of 1700A/110V, as illustrated in Figure 4.3.5.8. To enhance the output current stability of the power supply, a precision sampling system with temperature control has been incorporated. This system effectively mitigates temperature drift in the precision ADC, ensuring accurate and reliable measurements. The temperature control system maintains a temperature range within ±0.1℃.

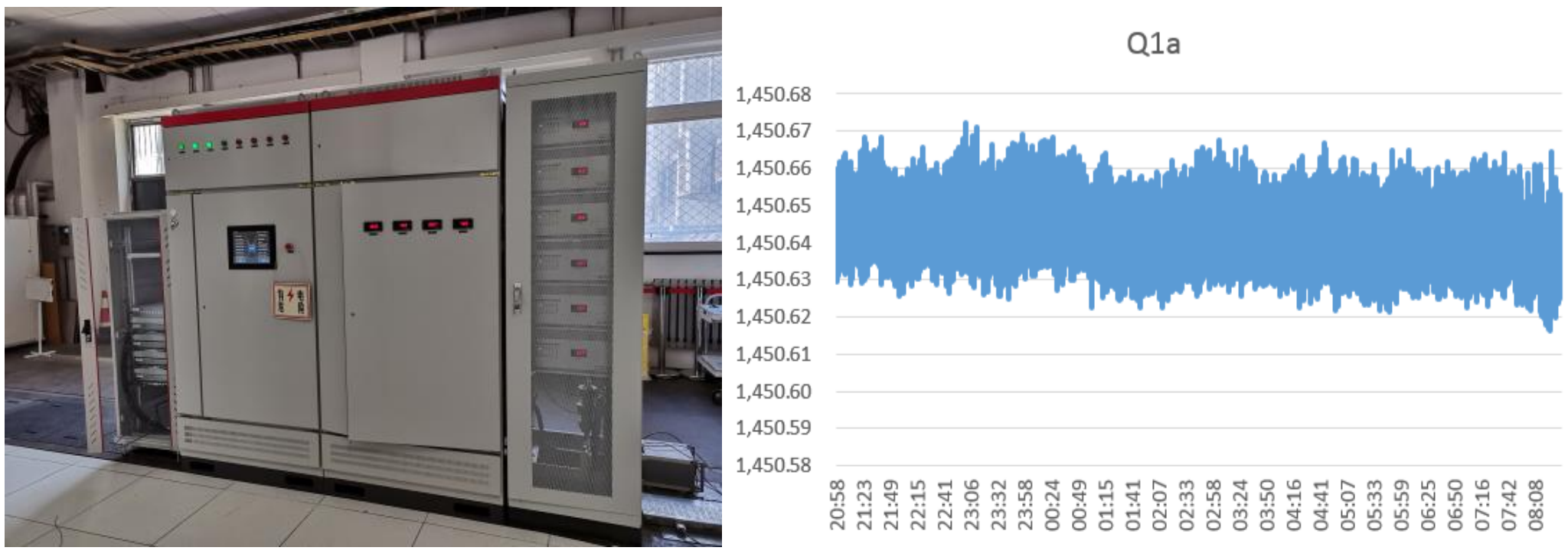

**Figure 4.3.5.8:** 1700A/110V High-power supply for BEPC II and the test curve of output current stability.

For the middle power supply, Figure 4.3.5.9 displays the block diagram of two modules in parallel with 300A/50V output. The equivalent switching frequency can be increased through the multiplex processing of PWM signal after series and parallel pulse width adjustment, resulting in improved power response speed, simplified output filter design, and reduced switching loss. The modular structure enhances production technology and maintainability.

The module is designed using PWM control to achieve constant frequency regulation, which optimizes filter design. Based on zero conversion converter technology, a "soft switching" PWM DC/DC full-bridge converter controlled by phase shift is designed. It uses the leakage inductance of the high-frequency transformer or primary-side inductance in series and the parasitic capacitance of the switching tube to realize zero-voltage switching of the switching tube. Combining the advantages of resonant and PWM power supplies, this converter is especially suitable for middle-power DC power supplies.

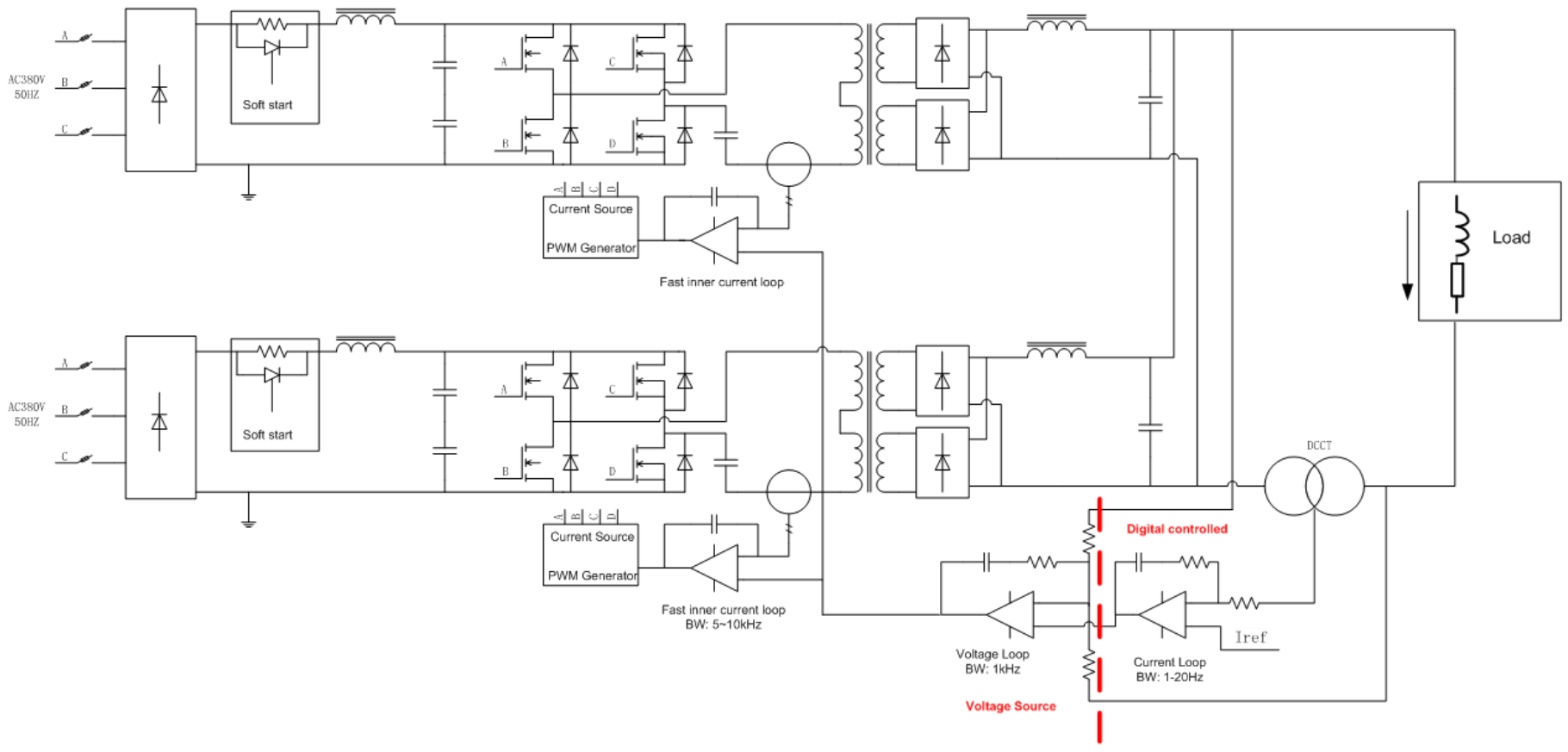

**Figure 4.3.5.9:** Structure diagram of double - module phase - shifted zero - voltage switching full - bridge converter.

The digital controller ensures stability and accuracy of the output current through analog sampling, AD conversion algorithms, and control parameters. To minimize voltage ripple, it is advisable to employ a three-loop control structure. The outer loop consists of a digital current loop, the middle loop functions as a voltage loop, and the inner loop acts as a peak current loop for the primary side of the high-frequency transformer. This configuration enhances the power supply's dynamic response and effectively mitigates grid fluctuations.

Figure 4.3.5.10 showcases the prototype of the power supply utilized in HEPS, along with the test results demonstrating its current stability.

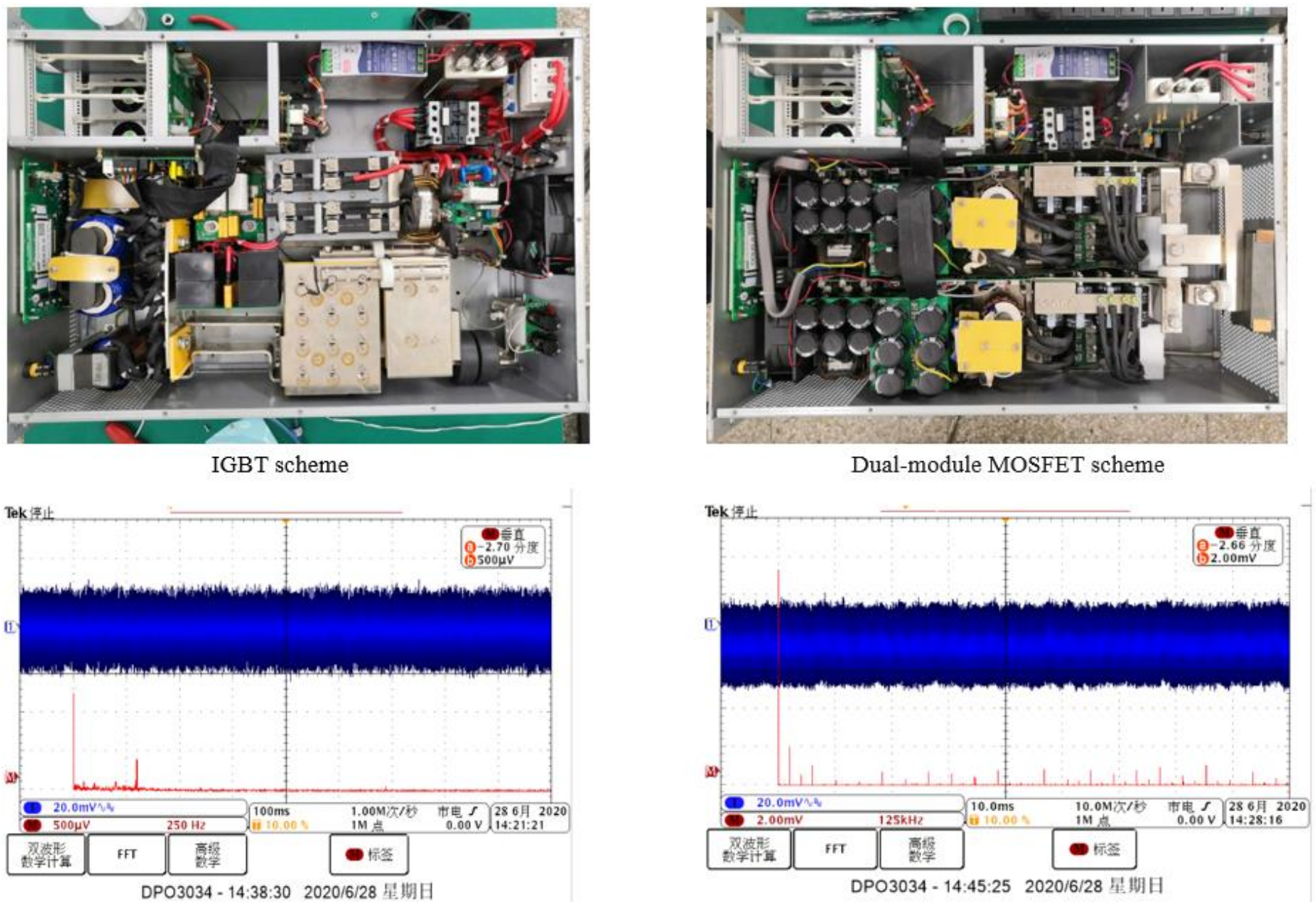

**Figure 4.3.5.10:** Prototype of power supply for the HEPS and the test curve of output current stability.

The CEPC correction magnet power supply uses a two-stage control topology. The first stage is a DC voltage regulator that ensures a stable output DC voltage. The second stage is a bidirectional H-type high-frequency inverter bridge structure (as show in Figure 4.3.5.11) that has two functions: a) the output current can be positive or negative; and b) the current feedback mode ensures stable output current. The circuit topology consists of four high-frequency power switching tubes that form a full-bridge chopper circuit, and two diagonal bridge arm switching tubes that complement each other to switch on and off. This topology guarantees accurate zero-point output and smooth positive and negative current commutation.

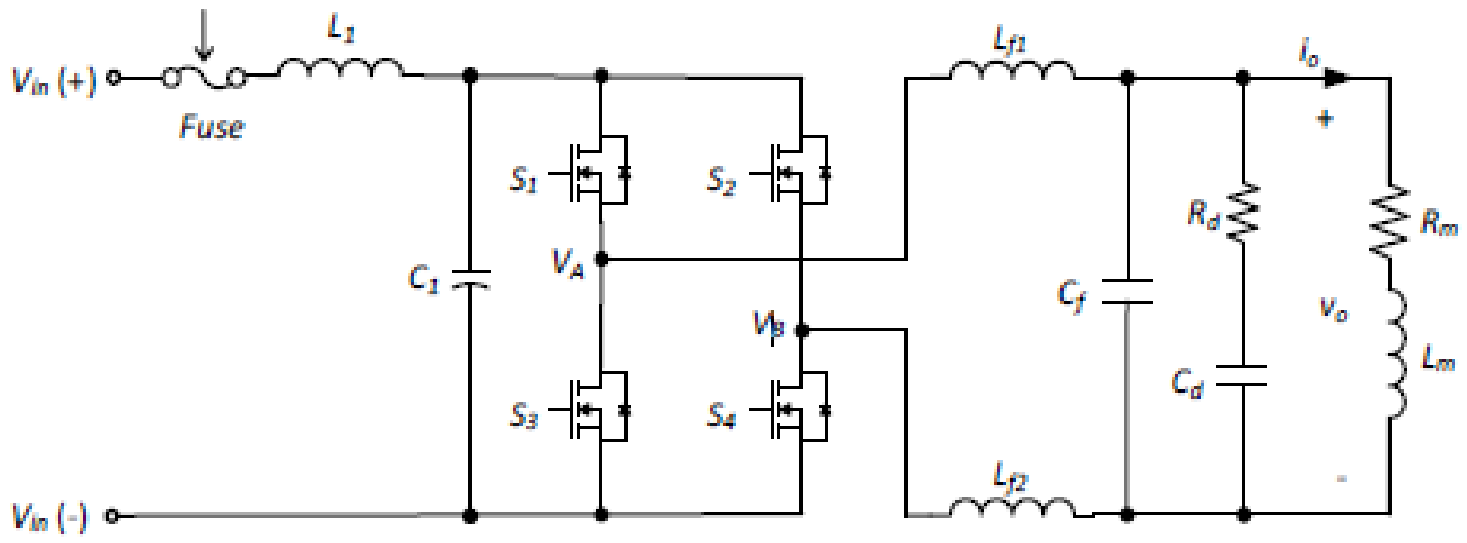

**Figure 4.3.5.11:** Topology of low-power supply for corrector magnet.

For a significant quantity of low-power supplies with outputs below 100 W, it is recommended to utilize the framework depicted in Figure 4.3.5.12. This approach involves multiple power supplies sharing a single DC source and a single power controller. By implementing this configuration, the overall size, number of controllers, and cost of the power system are substantially reduced.

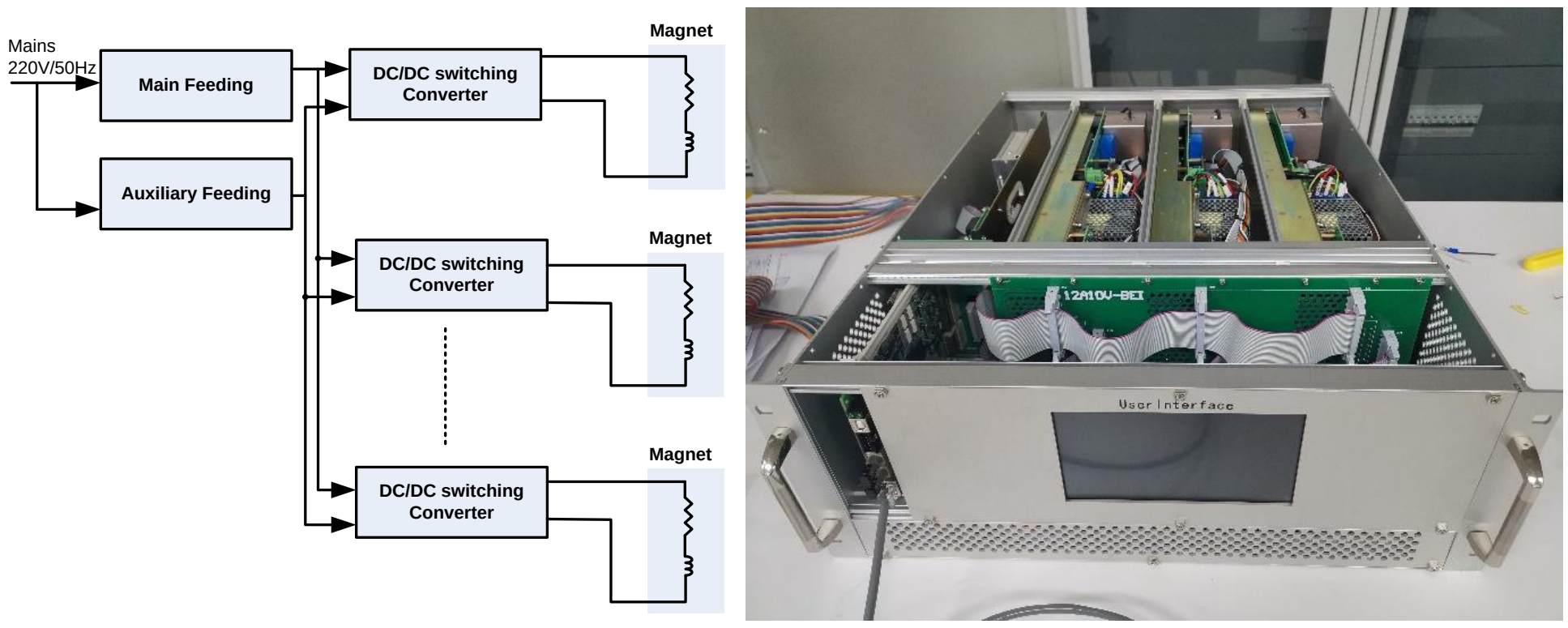

**Figure 4.3.5.11:** Low-power supply structure diagram for corrector magnet and a prototype for BEPC II.

Furthermore, each of the dual aperture magnets (B0 and Q) is equipped with two trim coils, which require independent power sources. These trim coil power supplies should be located in close proximity to the magnets to minimize cable loss and need to be either radiation-tolerant or shielded. The development of such power supplies is currently in progress.

### 4.3.5.6 *Power Supply Interlocks*

All power supplies are equipped with dedicated protective interlocks to safeguard against damage caused by fluctuations in cooling conditions, AC power interference, and deviations from setpoint ranges. Additionally, a safety interlock, controlled by a remote personal safety system, is in place to prevent the power system from activation in cases where the machine safety system deems it necessary.

## 4.3.6 **Vacuum System**

### 4.3.6.1 *Introduction*

In a collider, the beam lifetime and stability are critical factors that determine the quality of the experimental data. When the circulating particles interact with the residual gas molecules, they experience energy loss and get scattered, leading to beam losses and increased background in the detector.

One significant challenge in maintaining the beam quality is the intense synchrotron radiation emitted by the circulating particles in a forward-directed narrow cone. This radiation generates a high-energy photon flux, leading to strong outgassing from the vacuum chamber and dynamic pressure increase, limiting the beam lifetime and causing increased background in the experiments. To counter this issue, the pumping system must maintain the operating pressure even under a large dynamic photodesorption gas load.

Overall, achieving and maintaining stable beam conditions in the collider requires a robust pumping system and careful monitoring of the vacuum conditions to minimize beam losses and background levels.

To estimate beam-gas lifetime in a storage ring, it is crucial to know the residual gas composition. Typically, desorbed hydrogen ($H_2$) makes up over 90%, while the concentration of carbon monoxide (CO) and carbon dioxide ($CO_2$) combined is below 10%, after NEG coating. Heavy molecules like argon (Ar) are particularly problematic, whereas light molecules like $H_2$ are less critical. Table 4.3.6.1 outlines the basic requirements for an ultra-high vacuum system.

**Table 4.3.6.1:** Basic requirements for the ultra-high vacuum system

| Modes | E (Gev) | Beam gas scattering lifetime (hours) | Vacuum requirement (Torr) |
|---|---|---|---|
| Higgs | 120 | 10 | $2 \times 10^{-9}$ |
| W | 80 | 5 | $1.5 \times 10^{-9}$ |
| Z | 45.5 | 3 | $8 \times 10^{-10}$ |
| tt | 180 | 15 | $1 \times 10^{-8}$ |

In a collider, several key factors contribute to its successful operation:

- Good beam lifetime must be achieved soon after the initial startup with a stored beam;
- The vacuum system must be capable of quick recovery after sections are exposed to air for maintenance or repairs;
- The chamber wall must be as smooth as possible to minimize electromagnetic fields induced by the beam;
- Very low pressure must be attained in the interaction regions to minimize detector backgrounds from beam-gas scattering, ideally $3\times10^{-10}$ Torr or lower outside of the Q1 magnet;
- Sufficient cooling is necessary to safely dissipate the heat load associated with both synchrotron radiation and higher-order-mode (HOM) losses.

Table 4.3.6.2 provides a comparison of the vacuum system parameters of various colliders with double rings for electrons and positrons that have been constructed to date.

**Table 4.3.6.2:** Comparison of vacuum-related parameters in several storage rings

| Parameter | PEP II | | KEKB | | LEP2 | CEPC | |
|---|---|---|---|---|---|---|---|
| | $e^+$ | $e^-$ | $e^+$ | $e^-$ | $e^+$ $e^-$ | $e^+$ | $e^-$ |
| Energy [GeV] | 3.11 | 9.00 | 3.5 | 8.0 | 96 | 45.5~180 | |
| Beam current [A] | 2.14 | 0.95 | 2.6 | 1.1 | 2×0.007 | 0.0033~1.39 | |
| Circumference [m] | 2199.32 | | 3016.26 | | 26700 | 100000 | |
| Bending radius [m] | 13.75 | 165 | 16.31 | 104.46 | 3096.18 | 10700 | |
| Arc beam pipe material | Extruded aluminum, TiN coating | Extruded copper | Extruded copper | | Extruded aluminum, Lead shielding | Extruded copper, NEG coating | |
| Arc beam pipe shape (mm×mm) | Ellipse with antechamber 95×55 | Octagon 90×50 | Circle 94 | Racetrack 104×50 | Ellipse 131×70 | Circle 56 | |
| Pump type in arcs | TSP, IP | IP | NEGs, IP | NEGs, IP | NEGs, TSP, IP | NEGs, IP | |

### 4.3.6.2 ***Synchrotron Radiation Power and Gas Load***

The design of the vacuum system must take into account two issues related to synchrotron radiation. The first is heating of the vacuum chamber walls from the high thermal flux, while the second is the strong gas desorption resulting from both photon-desorption and thermal desorption. When a beam starts circulating, the dynamic pressure induced by synchrotron radiation can increase by several orders of magnitude. In this section, we will examine the effects of these issues and evaluate their impact on the system.

#### *4.3.6.2.1 Synchrotron Radiation Power*

To estimate the heat load, the well-known expression for the synchrotron radiation power (in kW) emitted by an electron beam in uniform circular motion is applicable [1]:

$$P_{SR} = \frac{88.5E^4 I}{\rho} \tag{4.3.6.1}$$

where $E$ is the beam energy (in GeV), $I$ is the total beam current (in A), and ρ is the bending radius of the dipole (in meters). The linear power density (in kW/m) along the circumference is given by:

$$P_L = \frac{P_{SR}}{2\pi\rho} = \frac{88.5E^4 I}{2\pi\rho^2} \tag{4.3.6.2}$$

For CEPC, with $E$ = 120 GeV, $I$ = 0.0174 A, and $\rho$ = 10700 m, Eqs. (4.3.6.1) and (4.3.6.2) imply that the total synchrotron radiation power $P_{SR}$ is 29.8 MW, and the linear power density $P_L$ is 444 W/m.

*4.3.6.2.2 Gas Load*

The gas load in the vacuum chamber arises from two processes: thermal outgassing and synchrotron-radiation-induced photodesorption. To estimate the desorption rate, the approach developed by Grobner et al. [2] was used. The effective gas load due to photodesorption can be calculated as:

$$Q_{gas} = 24.2EI\eta \quad \text{[Torr·L/s]}, \qquad (4.3.6.3)$$

where $E$ is the beam energy in GeV, $I$ the beam current in A, and $\eta$ the photodesorption coefficient in molecules/photon. The photodesorption coefficient $\eta$ is a property of the chamber and depends on several factors:

- the chamber material;
- the fabrication and preparation of the chamber material;
- the amount of prior exposure to radiation;
- the photon angle of incidence;
- the photon energy.

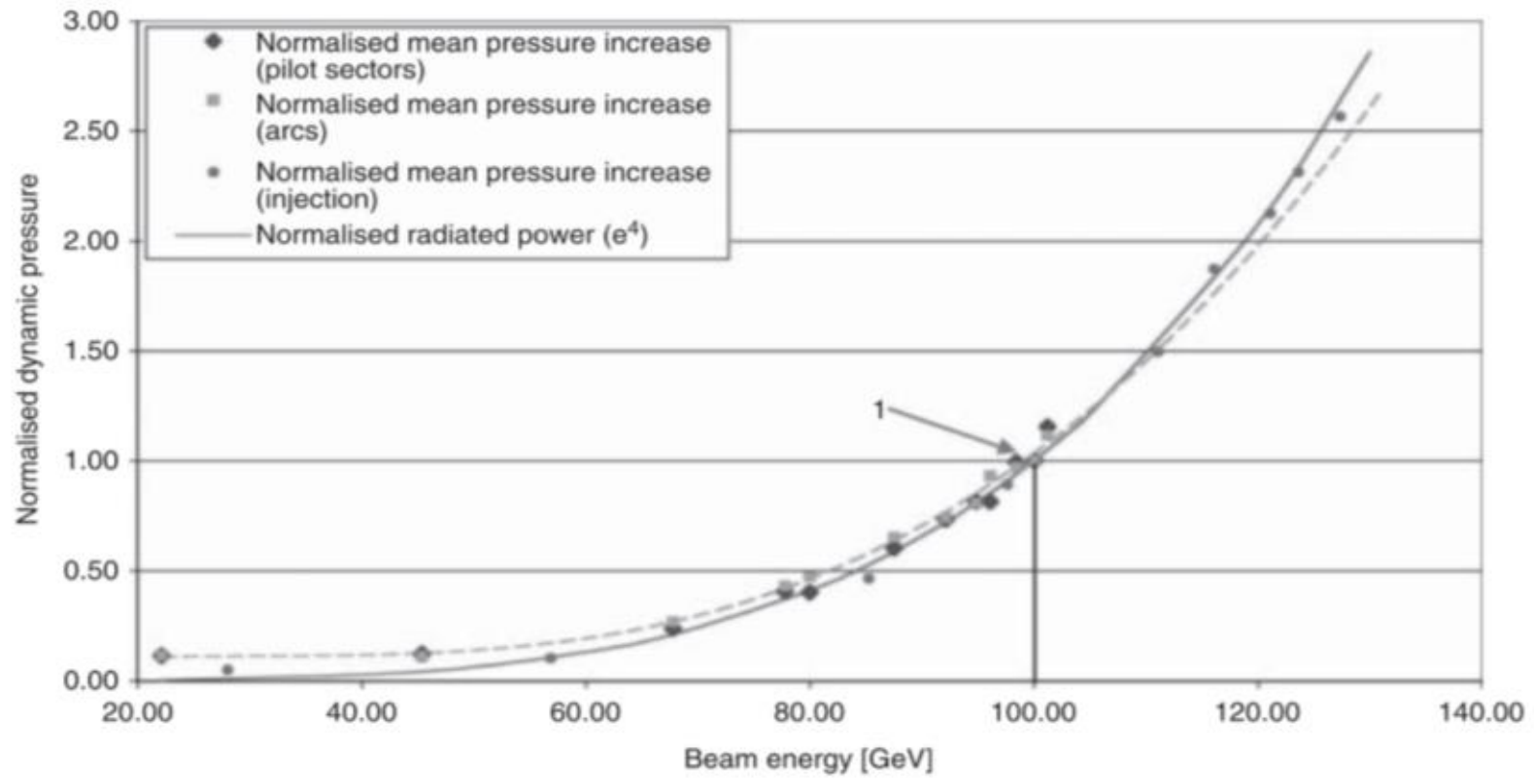

**Figure 4.3.6.1:** Normalised pressure increase in LEP as a function of beam energy [2].

**Table 4.3.6.3:**The η under different beam energy and beam density

| Modes | E (Gev) | η (molecules/photon) |
|---|---|---|
| Higgs | 120 | 2.00E-05 |
| W | 80 | 1.00E-05 |
| Z | 45.5 | 2.00E-06 |
| tt | 180 | 1.35E-03 |

The vacuum increases exponentially as the beam energy increases, as shown in Figure 4.3.6.1. Experimental measurements indicate that a copper chamber may eventually develop an effective $\eta \approx 10^{-6}$ [1]. Due to the very high energy of the beam, η is given a conservative value, as shown in Table 4.3.6.3. For a vacuum chamber with a desorption coefficient of $\eta = 2\times10^{-5}$ in the Higgs mode, for example, the dynamic gas load is

$$Q_{gas} = 4.84 \times 10^{-4} EI \quad \text{[Torr·L/s]} \qquad (4.3.6.4)$$

and the linear gas load (average gas load per meter) is:

$$Q_L = \frac{Q_{gas}}{2\pi\rho} \text{ [Torr·L/s·m].} \qquad (4.3.6.5)$$

The total dynamic gas load of $Q_{gas}$ = 9.7×10$^{-4}$ Torr·L/s is obtained, along with a linear SR gas load of $Q_{LSR}$ = 1.44×10$^{-8}$ Torr·L/s/m. Assuming a thermal outgassing rate of 1×10$^{-12}$ Torr·L/s·cm$^2$ for the vacuum chambers and a circular cross section of D = 56 mm, the linear thermal gas load $Q_{LT}$ is estimated to be 1.76×10$^{-9}$ Torr·L/s/m. Thus, the total linear gas load is 1.62×10$^{-8}$ Torr·L/s/m. Table 4.3.6.4 shows the gas load under different beam energy and beam density.

**Table 4.3.6.4:** The gas load under different beam energy and beam density

| The total synchrotron radiation power $P_{SR}$ = 30 MW | | | | | |
|---|---|---|---|---|---|
| Mode | E | I | PSD | $Q_{gas}$ | $Q_{LSR}$ |
| | Gev | A | molecules/photon | Torr·L/s | Torr·L/s·m |
| Higgs | 120 | 0.0167 | 2.00E-05 | 9.70E-04 | 1.44E-08 |
| W | 80 | 0.084 | 1.00E-05 | 1.63E-03 | 2.42E-08 |
| Z | 45.5 | 0.803 | 2.00E-06 | 1.77E-03 | 2.63E-08 |
| tt | 180 | 0.0033 | 1.35E-03 | 1.94E-02 | 2.89E-07 |
| The total synchrotron radiation power $P_{SR}$ = 50 MW | | | | | |
| Higgs | 120 | 0.029 | 2.00E-05 | 1.68E-03 | 2.51E-08 |
| W | 80 | 0.147 | 1.00E-05 | 2.85E-03 | 4.24E-08 |
| Z | 45.5 | 1.39 | 2.00E-06 | 3.06E-03 | 4.56E-08 |
| tt | 180 | 0.0054 | 1.35E-03 | 3.18E-02 | 4.73E-07 |

### 4.3.6.3 *Vacuum System Components*

The NEG coating is used to suppress the e-cloud of the positron ring, which can cause beam instabilities, heat loads, and pressure increases in the vacuum system. It also provides distributed pumping speed for both the positron and electron rings simultaneously. The thickness of the NEG coating should be lower than 200 nm due to the impedance of the beam. The length of the vacuum chambers varies from 3.6 m to 11.3 m, depending on the layout of magnets, with the dipole magnet vacuum chamber being the longest and the quadrupole magnet being the shortest. The vacuum chamber extension is significant because the NEG coating needs to be activated by baking, which requires a temperature of 180℃ and may increase to 250℃ after several activations. To absorb the extension and misalignment of vacuum chambers, an RF shielding bellows is employed. The 100 km circumference of the ring is subdivided into 520 sectors using all-metal gate valves. The main devices of the collider ring are shown in Table 4.3.6.5.

**Table 4.3.6.5:** The main devices of the collider vacuum system (per ring)

| Device | Specification | Quantity |
|---|---|---|
| Vacuum chamber | D56/OFHC | 94.4 km |
| NEG coating | 200 nm | Both $e^+$/ $e^-$ ring |
| Sputtering ion pump | 50-100 L/s | 6667 |
| All metal gate valve | DN63/spring fingers type | 520 |
| RF shielding bellow | DN63 | 103840 |
| Vacuum Gauge | CCG | 2160 |
| RGA | Mass 100 | 520 |
| Baking | 220 V | Both $e^+$/ $e^-$ ring |

### 4.3.6.4 ***Vacuum Chamber***

#### *4.3.6.4.1 Vacuum Chamber Material*

The vacuum chamber needs to be water-cooled, and of high electrical and thermal conductivity one (made of either aluminum or copper) due to the power deposited by synchrotron radiation. In LEP, extruded aluminum chambers were used, which were water-cooled and covered with lead cladding to protect other components from radiation damage [3].

Copper is the preferred material for the vacuum chamber due to its naturally lower molecular yield, lower electrical resistance, and smaller radiation length, which provides more efficiency in preventing photons from escaping through the vacuum chamber walls and damaging the magnets and other components. Additionally, copper's excellent thermal conductivity makes it the preferred material for the vacuum chamber walls in the arcs, which are subjected to very high thermal loads. Vacuum chambers in the straight sections will be fabricated from stainless steel.

Copper has been extensively used for B-factory vacuum chambers [4], and research has shown that its initial molecular yields are nearly 1-2 orders of magnitude lower than aluminum [5-6]. PSD (Photon Stimulated Desorption) tests on copper at DCI have demonstrated that a photodesorption coefficient of $10^{-6}$ can be achieved in a reasonable amount of time at high current [1]. This low photodesorption coefficient allows for the design of a vacuum chamber with a conventional elliptical or circular shape, instead of being forced to adopt an antechamber design that is more difficult and expensive to fabricate. While copper may initially seem more expensive, the relative simplicity of its shape, the reduction in the amount of pumping needed, and the shortening of the vacuum system commissioning time offset this apparent cost disadvantage.

To absorb active gases such as $H_2$, CO, $CO_2$, and reduce secondary electron yields, NEG coating will be applied to both the electron and positron rings, thereby avoiding e-cloud instability in the positron ring. Radiation shielding will be accomplished by placing lead blocks inside the magnets. Unlike in the CDR, copper vacuum chambers will also be used for the electron rings.

#### *4.3.6.4.2 Vacuum Chamber Shape*

To eliminate quadrupolar wakes, the elliptical vacuum chamber in the Collider ring, which is 75 mm wide by 56 mm high in the CDR, will be replaced by circular chambers with a diameter of 56 mm (as shown in Figure 4.3.6.2). The dipole chamber in the Collider ring has a maximum length of 11.3 m and a wall thickness of 3 mm. A cooling channel attached to the outer wall of the beam duct is used to carry away the heat generated by synchrotron radiation hitting the chamber wall. The beam duct of the collider ring will be extruded from full lengths of UNS C10100, a high-purity, oxygen-free, high-conductivity and high mechanical strength alloy of copper, while the cooling channel will be fabricated from USN C10300, an oxygen-free copper alloy. The vacuum flanges will be made of stainless steel.

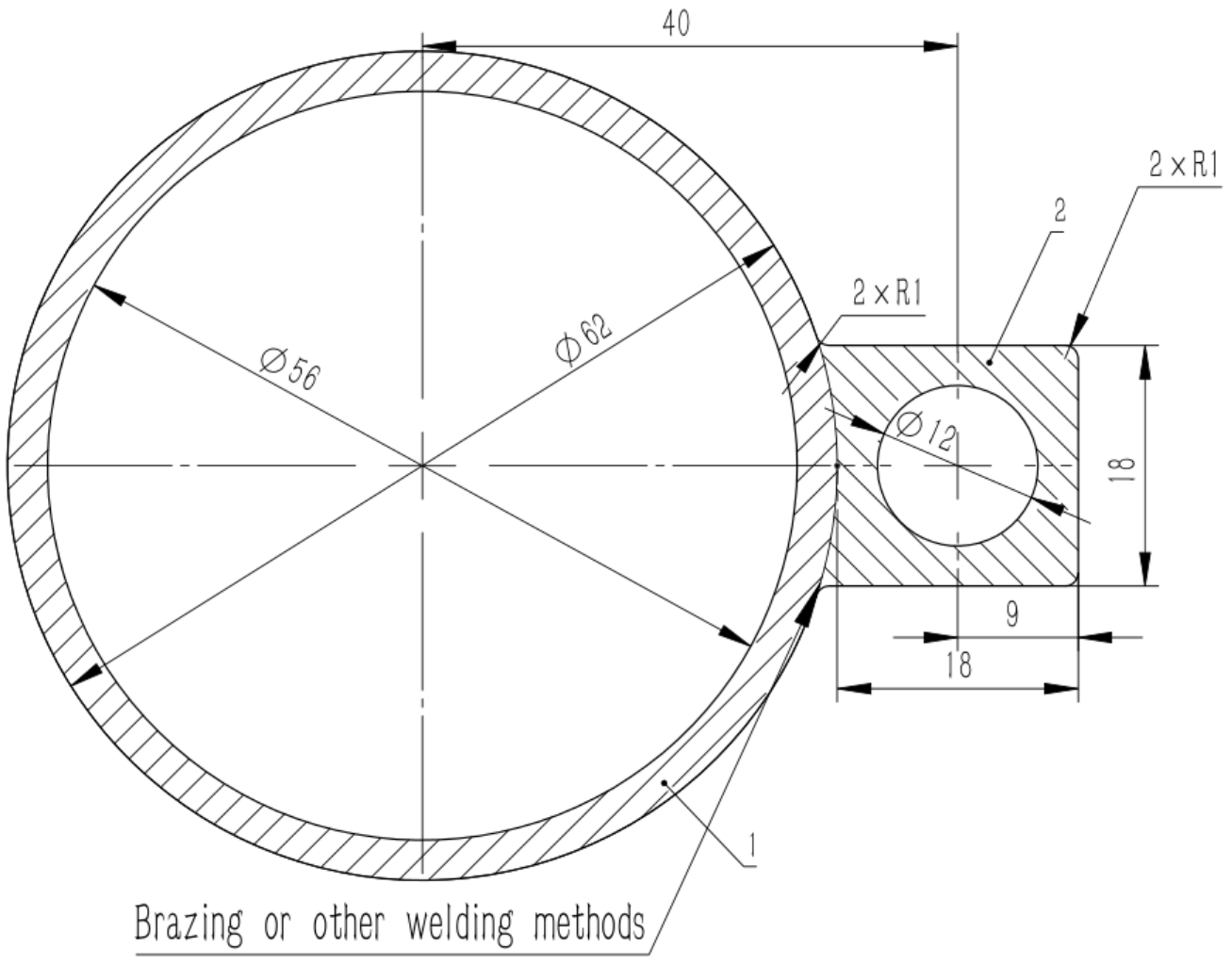

**Figure 4.3.6.2:** Copper dipole vacuum chamber

One of the primary challenges in designing the vacuum chamber is to effectively manage the high thermal synchrotron radiation power that is incident on the vacuum chamber wall. The linear power density of the synchrotron radiation reaches up to 444 W/m. Finite-element analysis of a dipole chamber that is exposed to this power has shown that the highest temperature reaches 46.3℃ when a convective heat transfer coefficient of $1\times10^{-3}$ $W/mm^2\cdot$℃ is selected. The maximum stress that is experienced is 6.3 MPa, and the total deformation is 0.1 mm per meter, which are within the safety limits. The results of the finite-element analysis are depicted in Figure 4.3.6.3.

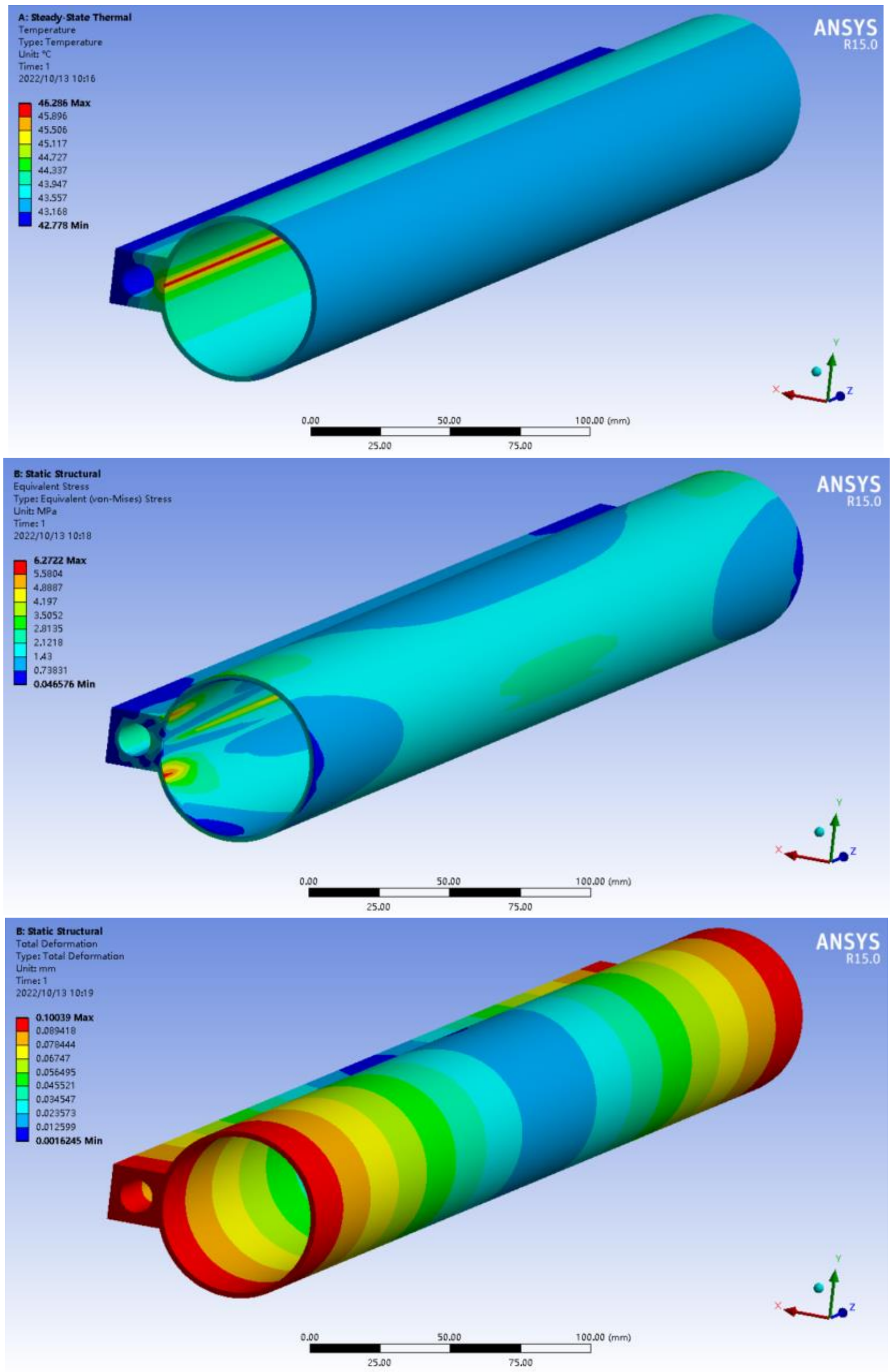

**Figure 4.3.6.3**: Finite-element analysis of a copper dipole vacuum chamber: Top – Temperature increase due to sychrotron radiation; Middle – Stress distribution resulting from nonuniform temperature distribution; Bottom – Deformation analysis.

The vacuum chamber is fabricated using an extruded copper chamber and a cooling channel with two conflate-type end flanges. The copper chamber and cooling channel are both drawn to their final shape to produce a minimum half-hard temper. The components are then thoroughly cleaned and joined together by electron-beam welding. The sub-assembly is then stretch-formed to its correct radius, and the ends are machined and cleaned. Finally, the end flanges are TIG-brazed onto the ends of the chamber to complete the assembly [1]. The use of a one-piece chamber extrusion eliminates all longitudinal vacuum welds, resulting in a more accurate and dependable chamber.

The synchrotron radiation continuously deposits 5 W of heat on the stainless-steel flange. Thermal analysis indicates that the maximum temperature with 5 W of uniformly distributed power in the flange is 58 °C. When the powers of 2.5 W and 5 W are concentrated in a small block with a diameter of 1.2 mm, which is employed to protect the downstream devices, the maximum temperatures are 56 °C and 90 °C, respectively. These temperatures are within the safe range for stainless steel. Using the coefficient of expansion for stainless steel, which is $1.7 \times 10^{-5}$ m/(m·K), the temperature difference of 90 °C and 56 °C for a flange pair will result in an expansion of $5.78 \times 10^{-4}$ m/m. Assuming the flange has a diameter of 10 cm, the expansion will be $5.78 \times 10^{-5}$ m or 57.8 microns, which is still within the tight tolerance for vacuum.

To eliminate the potential NEG coating poisoning caused by a significant accumulation of carbon and oxygen in the oxide layer on the inner surface of the chamber, a two-step process is employed. First, ammonium persulphate is used as an etchant to remove a 50 μm thick layer, ensuring the effective removal of the oxide layer within an approximate time of 30 minutes. Subsequently, a 5-minute chromic acid passivation is conducted to create a new carbon and oxygen-free passivation layer.

*4.3.6.4.3 Vacuum Chamber prototype Manufacture*

The Cu beam pipe and water-cooling channel are manufactured separately via extrusion, and then joined together by brazing. Flanges made of stainless steel are used, and a rotatable flange is located at one end of the vacuum chamber, as shown in Figure 4.3.6.4. The flanges are welded to the beam pipe using high-temperature brazing solder, while low-temperature brazing solder is used to join the beam pipe to the water-cooling channel. Due to the difficulty in finding a 6-meter-long high-temperature vacuum furnace, low-temperature brazing solder is used for this joint.

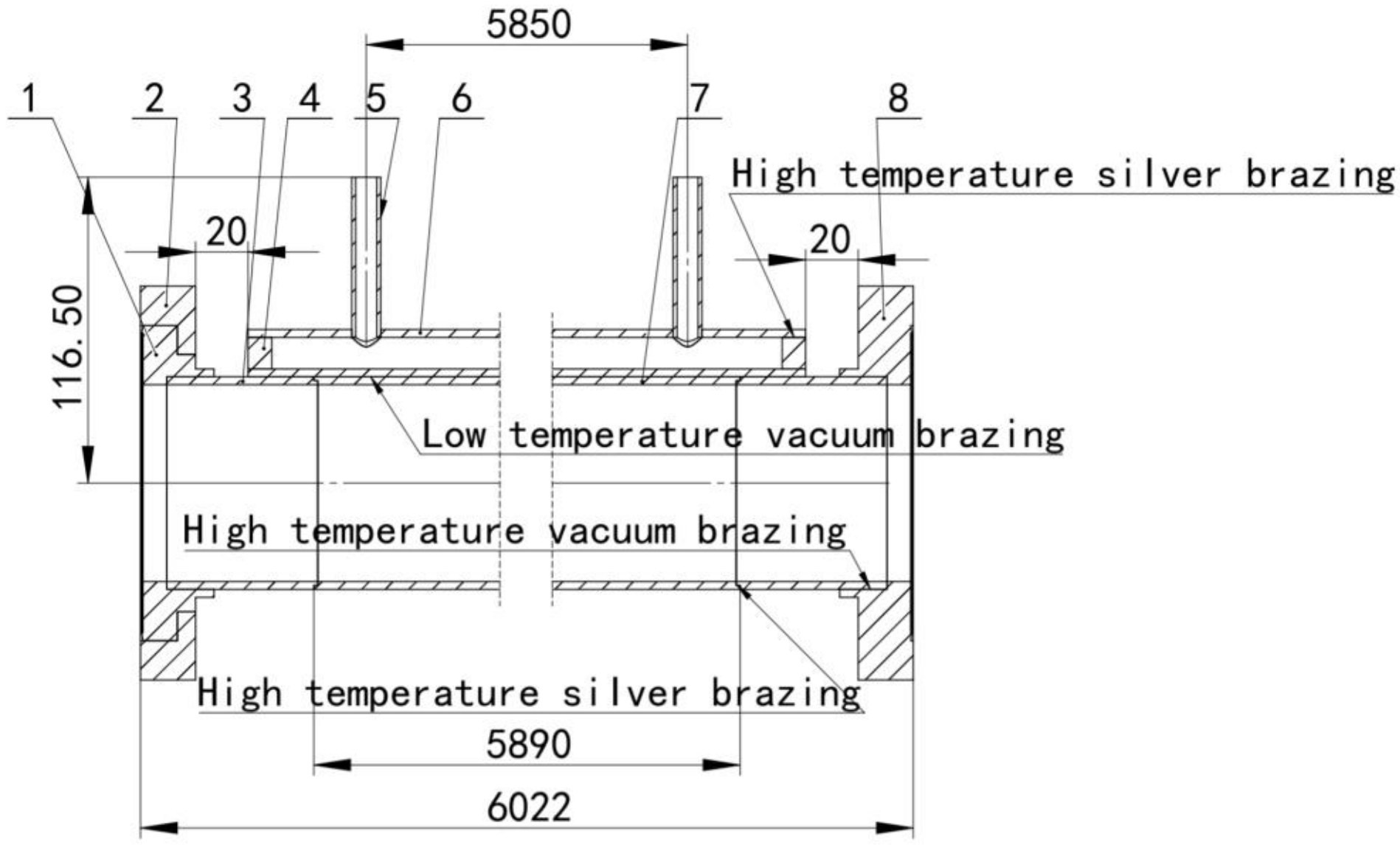

**Figure 4.3.6.4:** Diagram of a copper vacuum chamber

A 6-meter-long vacuum furnace has been fabricated specifically for low-temperature brazing solder to weld the water-cooling channels of the Cu chambers. The welding seams are checked using wire-electrode cutting, and the joints are found to be smooth with good contact. Prototypes of the copper and aluminum vacuum chambers, each with a length of 6 meters, have been successfully fabricated and tested. The resulting chambers meet all engineering requirements, as demonstrated in Figure 4.3.6.5.

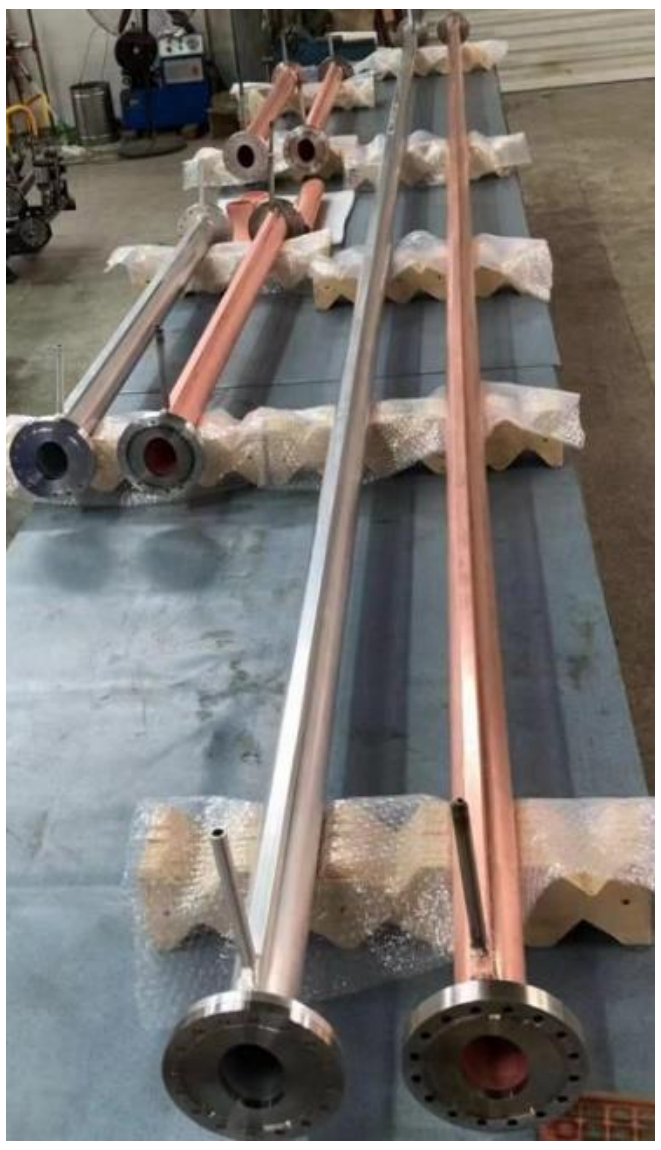

**Figure 4.3.6.5:** Prototypes of vacuum chamber

### 4.3.6.5 *Bellows Module with RF Shielding*

The main purpose of the bellows module is to accommodate thermal expansion of the chambers and to allow for lateral, longitudinal, and angular offsets resulting from tolerances and alignment. This is achieved while maintaining a uniform chamber cross-section to reduce the impedance experienced by the beam. **Error! Reference source not ound.** depicts the RF shielding bellows module, which is designed to fulfill these requirements.

The typical RF shield is composed of many narrow Be-Cu fingers that slide along the inside of the beam passage as the bellows compresses. One of the primary concerns with this finger-type RF shield is the strength of the contact force. It is crucial that each contact finger maintains an appropriate contact force with the beam tube to ensure adequate electrical contact with the high-frequency current. A higher contact force generally results in better electrical contact, but it also increases the risk of abrasion and dust generation during mechanical flexing. Maintaining a minimum contact force is essential to prevent excess heating and arcing at the contact point. The leakage of HOM RF from the gaps between the contact fingers into the inside of the bellows is another significant issue that must be addressed.

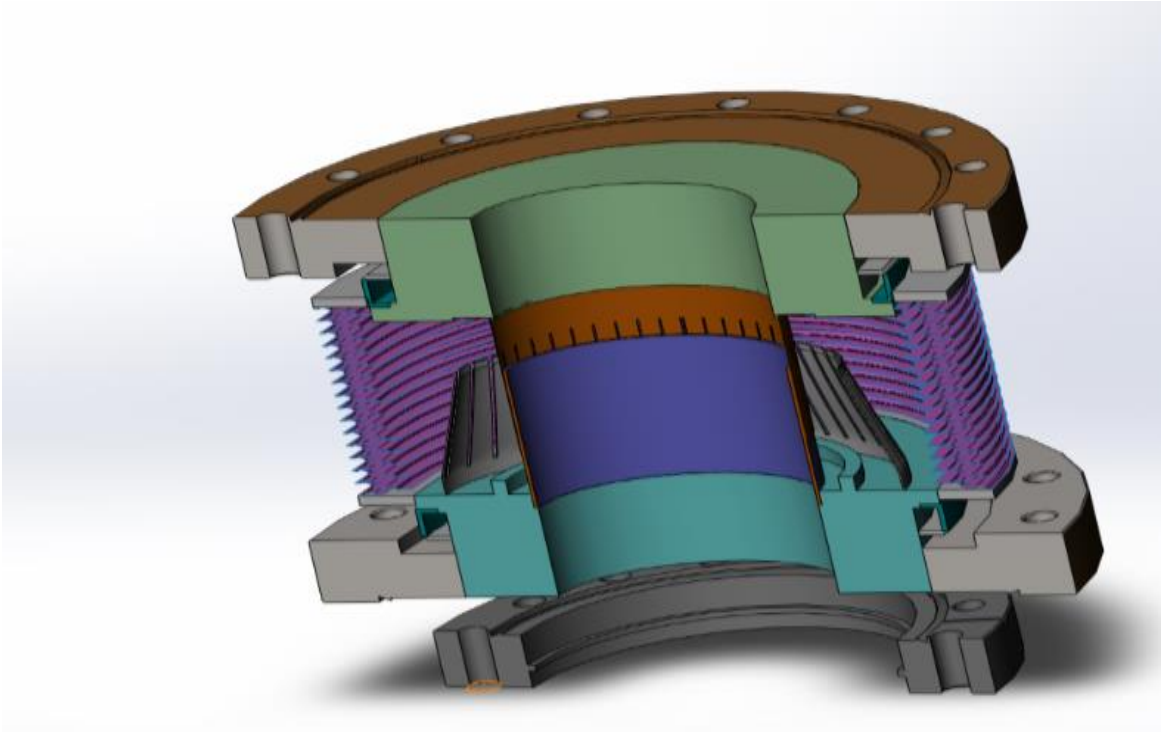

**Figure 4.3.6.6**: RF shielding bellows module.

Each finger of the RF shield is designed to maintain a relatively high contact pressure of 125±25 g/finger. The length of the slit between the fingers is 20 mm, and the RF shield can accommodate a maximum expansion of 10 mm and a contraction of 20 mm, allowing for a 2 mm offset. The step at the contact point is limited to less than 1 mm. The cooling water channel is responsible for managing synchrotron radiation power, Joule loss, HOM heat load on the inner surface, and any leaked HOM power inside the bellows. The RF fingers are protected by a mask located in the upstream vacuum chamber to shield them from synchrotron radiation irradiation.

#### *4.3.6.5.1 Key Parameters*

Contact force: As previously mentioned, the contact force should be carefully controlled within a reasonable range. Based on the experience gained from BEPC II, this indicator is set at 125 ± 25 g.

Radial offset: The purpose of this indicator is to compensate for any position errors between the elements on the two sides of the corrugated pipe. The value is determined to be 2 mm.

Tension and compression: The corrugated pipe can be stretched up to 5 mm and compressed by up to 40 mm.

#### *4.3.6.5.2 Structural Design and Processing Technology*

As shown in Figure 4.3.6.7, the RF shield structure consists of three parts: the spring-finger, contact finger, and inner tube.

The spring-finger is the core of the entire shield structure, made of a 0.4 mm-thick nickel-alloy plate. The end of the finger bends outward, and there is a raised structure with a larger curvature in the contact part to ensure point contact and prevent heat caused by unreliable contact. To ensure good electrical contact, the surface of the spring-finger is plated with silver.

The contact-finger is made of beryllium copper and has a width of 4.5 mm to 5.5 mm. The width of the gap between fingers is 0.5 mm, which is used to remove the gas between the corrugated pipe and the shield structure.

The inner tube is made of stainless steel and features a protruding part on the lateral side of its end that contacts with the contact-finger to avoid heat caused by unreliable contact.

To ensure the uniformity and consistency of the spring-finger, a specially designed forming mold and welding fixture were utilized. The spring-finger was formed and then welded with the fixing-ring using the welding fixture. After that, it was further welded and fixed with the endplate. Two prototype RF shield bellows were manufactured, one of which is depicted in Figure 4.3.6.8.

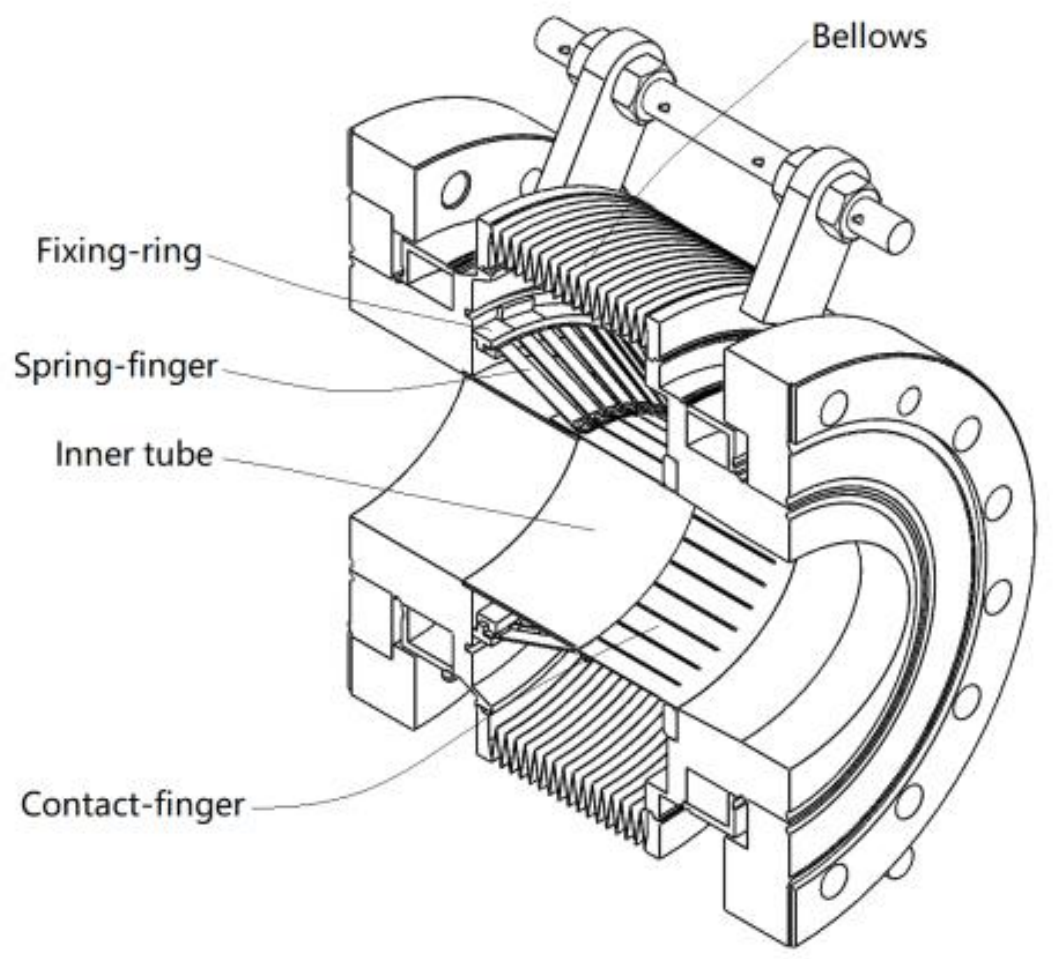

**Figure 4.3.6.7:** Schematic diagram of the structure of the RF shield bellows.

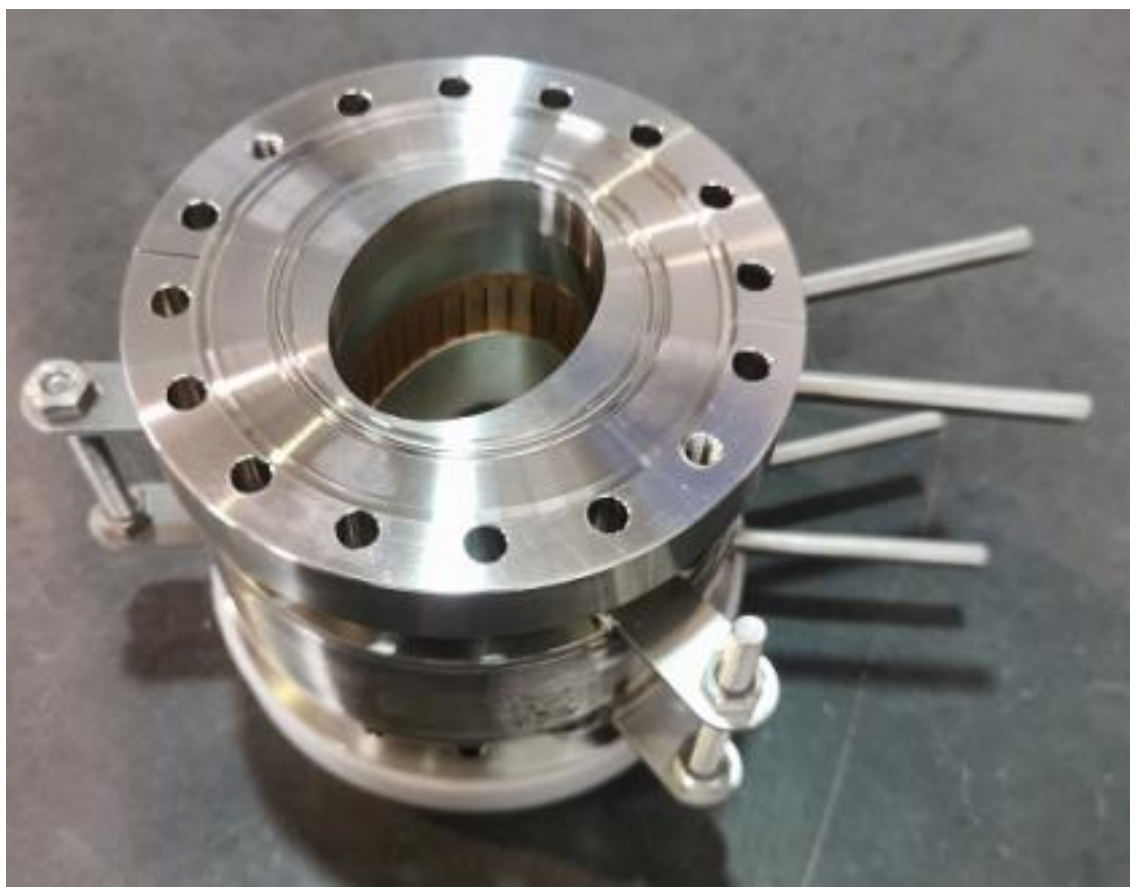

**Figure 4.3.6.8:** Prototype of RF shield bellows

#### *4.3.6.5.3 Testing*

To ensure the accuracy and consistency of the contact force, a specialized testing device was designed and used during the prototype development process, as shown in Figure 4.3.6.9. A polyimide tape with a thickness of 0.1 was used to insulate the beryllium copper and inner tube, with the overall thickness being the same as that of the actual contact-finger. One end of the indicator light (or multimeter) was connected to the spring-finger, and the other end was connected to the beryllium copper. The force gauge was connected to a spring with a proper elastic coefficient, with the other end of the spring being connected to a soft coil that got close to the contact point between the spring-finger and beryllium copper, hooking the spring-finger in place.

During testing, if the applied tensile force is smaller than or equal to the contact force, the relative position of the spring-finger and beryllium copper will not change. The tensile force will change into the tensile deformation of the spring, and the indicator light will remain in the same state.

When the tensile force exceeds the contact force, the spring-finger is pulled up, the beryllium copper has an open circuit, and the indicator light turns off. At this point, the reading of the force gauge should be equal to the maximum tension the shield can withstand.

The method described above was used to test the contact force of the prototype, and the results are presented in Figure 4.3.6.10. The measured contact force was found to be in the range of 120 g to 130 g, which falls within the acceptable range for the project.

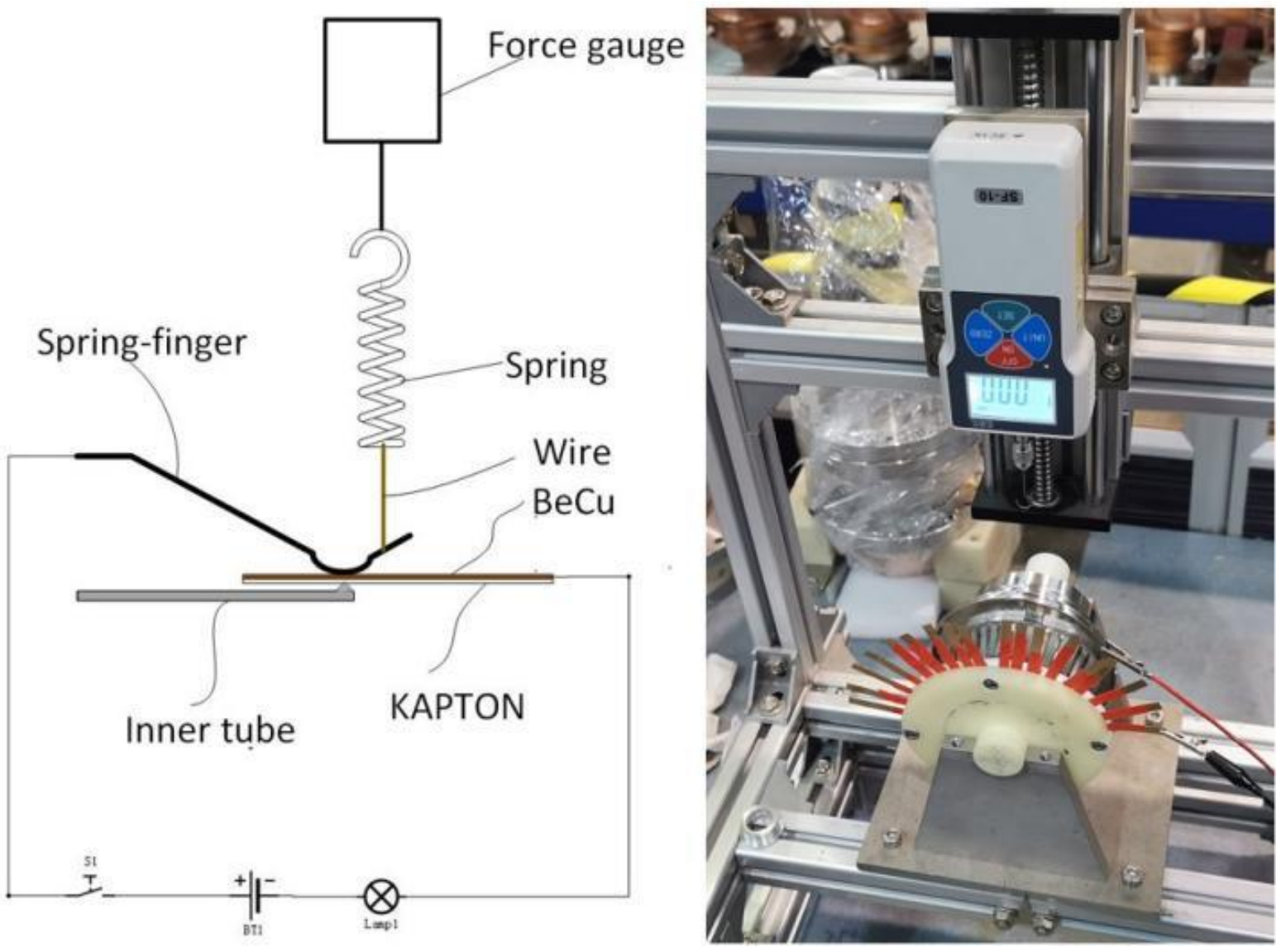

**Figure 4.3.6.9:** Schematic diagram of the testing device of contact force and its physical photograph.

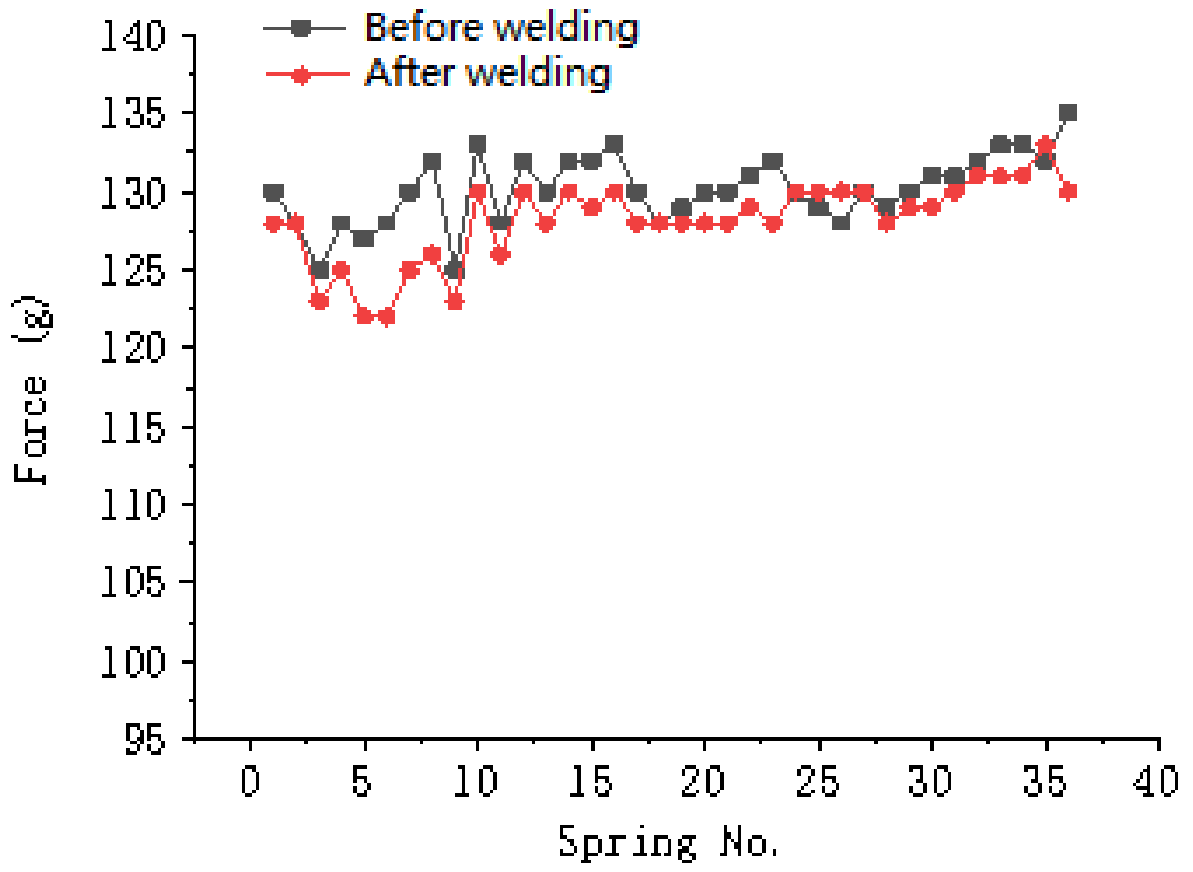

**Figure 4.3.6.10:** Contact force test results of RF shield bellows prototype.

### 4.3.6.6 *Pumping System*

The ring of 100 km circumference will be divided into 520 sectors using all-metal gate valves. These valves will enable pump down from atmospheric pressure, leak detection, bake-out, and vacuum interlock protection to be carried out in manageable sections. Roughing down to approximately $10^{-7}$ Torr will be accomplished using an oil-free turbomolecular pump group. Non-evaporable getter (NEG)-coated copper chambers will be used for main pumping in both the electron and positron rings. Sputter ion pumps will be employed to maintain pressure and pump off $CH_4$ and noble gases that cannot be pumped off by the NEG pump. Depending on the available space, NEG pumps, sublimation pumps, and sputter ion pumps will be used in the pumping system for the interaction regions where the detectors are located.

##### *4.3.6.6.1 NEG Coating*

The NEG coating is a deposition of a titanium, zirconium, vanadium alloy on the inner surface of the chamber, typically achieved through sputtering. In the straight sections of the LHC, approximately 6 km of vacuum chambers were coated with NEG [7]. The utilization of NEG coatings has been widespread in the fourth generation of light sources to meet the stringent vacuum requirements, primarily due to the low conductance of the vacuum chambers.

The presence of NEG coating improves the vacuum in the chamber through two mechanisms: reduced desorption yield and direct pumping by the NEG alloy coating [5]. However, when the NEG-coated surface is exposed to air, it becomes saturated and loses its pumping performance. Activation of the NEG coating involves heating, which facilitates the diffusion of gas atoms from the saturated surface layer into the bulk of the NEG coating, thus producing a fresh and reactive NEG surface. The activation process typically requires relatively low temperatures, such as 200°C for 24 hours [8]. In cases where aluminum chambers are used, which cannot withstand high-temperature bake-out, even lower activation temperatures (e.g., 160°C for 48 hours) have been successfully applied.

A DC magnetron sputtering facility has been established at IHEP for the production of NEG coatings, as depicted in Figure 4.3.6.11. Prior to the coating process, thorough cleaning of the NEG-coated chamber is conducted to prevent any significant contamination or surface defects that may adversely affect the quality of the film.

For achieving a uniform thickness distribution, each dipole chamber is equipped with three cathodes made of twisted wires of titanium (Ti), zirconium (Zr), and vanadium (V), each having a diameter of 1 mm. To maintain the proximity of the cathode wires to the chamber's axis, several ceramic spacers are strategically placed along the chamber's length, along with two adapters at the ends.

To create the necessary magnetic field, a solenoid with dimensions of 1500 mm in length and 280 mm in diameter is externally mounted on DT4. The ion pump is utilized to attain ultra-high vacuum (UHV), and the NEG bulk pump is employed to evacuate residual gases such as CO and $H_2O$ to achieve UHV conditions.

The chambers are initially evacuated using a turbomolecular pump group, reaching a range of $10^{-9}$ mbar. They are then subjected to a 48-hour bake-out process at a temperature of 200℃, followed by a helium leak test to ensure their integrity before the coating process. A Residual Gas Analyzer (RGA) is utilized for monitoring residual gases during the coating process.

During the coating process, krypton gas is used as the working gas, set at approximately 0.01 mbar. The chamber temperature is maintained at around 120℃ to facilitate the sputtering process.

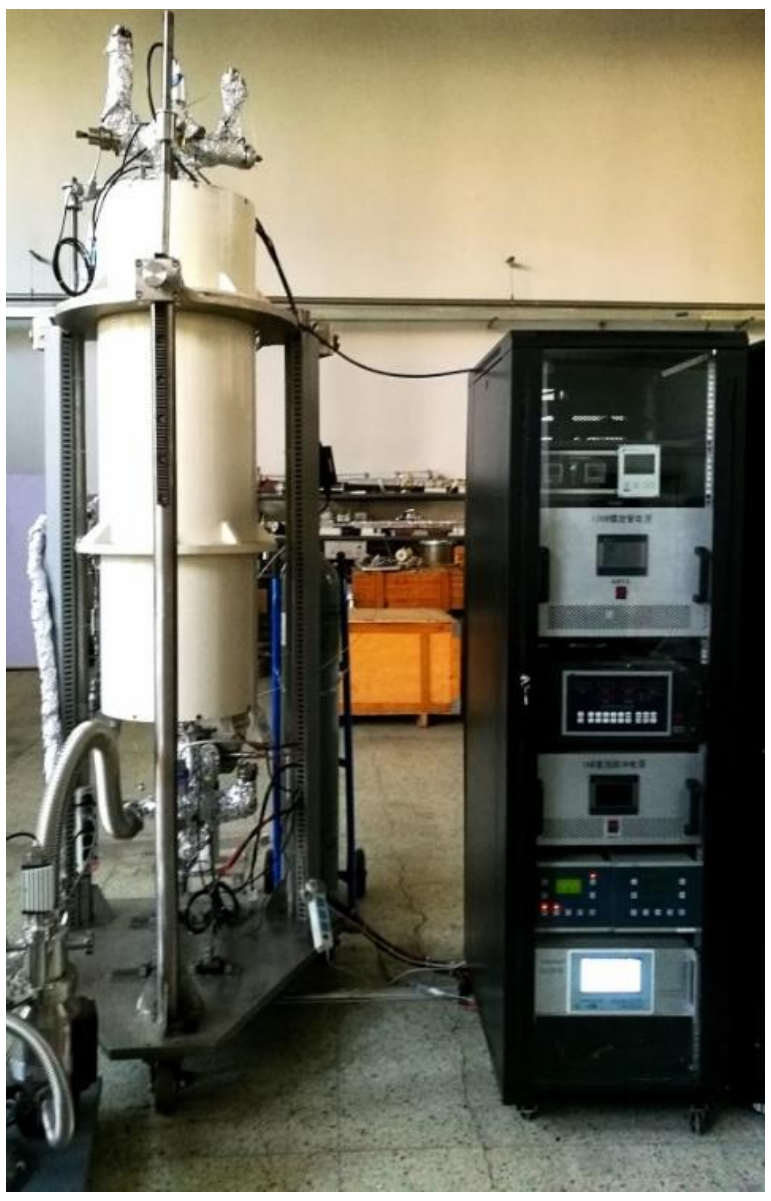

**Figure 4.3.6.11:** Prototype of NEG coating facility.

Due to impedance limitations, it is crucial to adhere to a maximum thickness of 200 nm for the TiZrV NEG coating. However, this constraint imposes a limitation on the lifetime of the coating, allowing for fewer than 10 reactivation cycles.

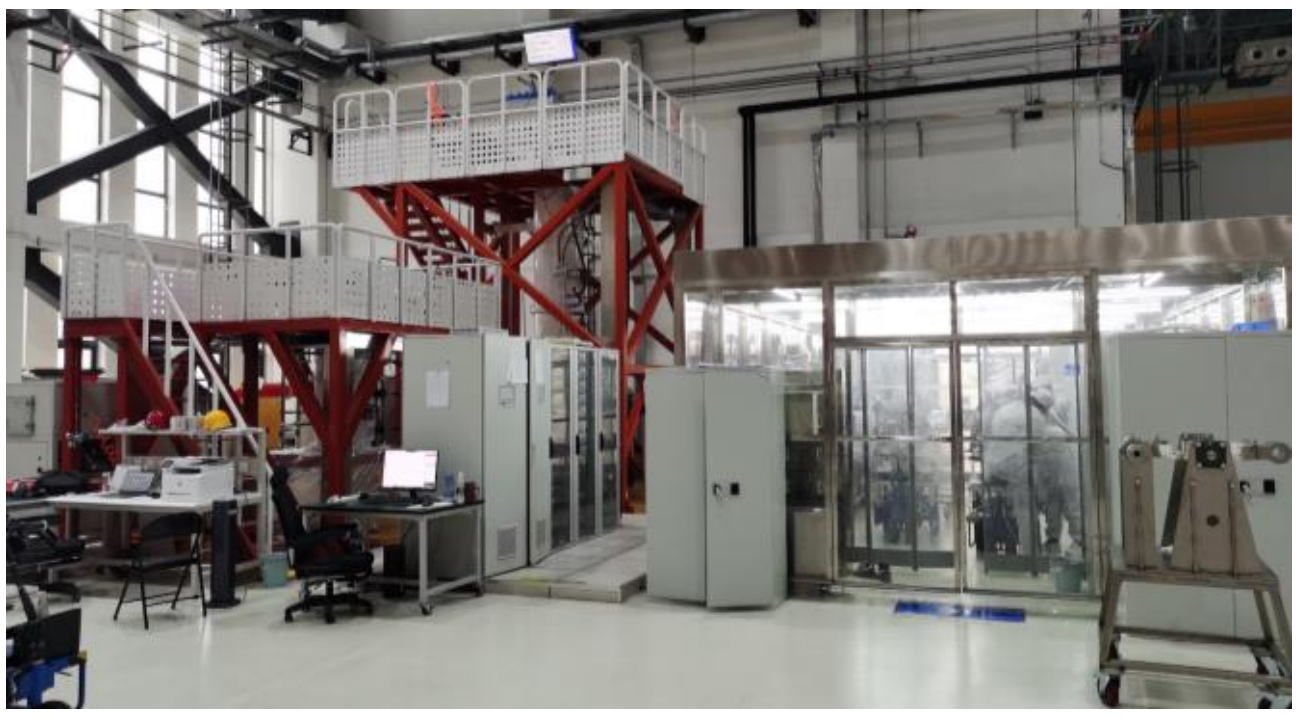

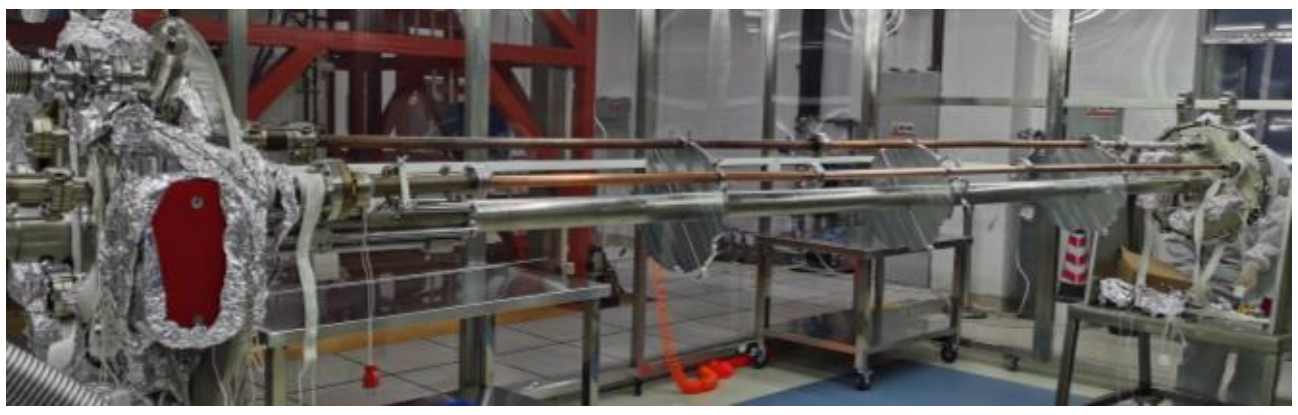

**Figure 4.3.6.12:** Top: Large-scale NEG coating production facilities. Bottom: Installation of vacuum chambers in both parallel and series configurations for mass coating.

Massive production of NEG coating poses a substantial challenge due to the intricate process and the large quantity of vacuum chambers involved. As a preliminary estimate, to complete approximately 190 kilometers of vacuum chambers for the CEPC Collider within a 5-year timeframe, a facility with dimensions of 25 meters in height and an area of 7000 $m^2$ would require the installation of 25 coating facilities. Figure 4.3.6.12 provides an illustration of a large-scale production facility from HEPS, which is presently under construction in Huairou, Beijing.

#### 4.3.6.6.1.1 Pumping Properties Evaluation

Given that the residual gases in the accelerator primarily consist of $H_2$ and CO (constituting approximately 99%) and a smaller portion of Ar and $CH_4$, the pumping properties of $H_2$ and CO are of utmost importance when considering the NEG coating. To evaluate the performance of the NEG coating for accelerator applications, it is crucial to characterize its pumping speed and absorbing capacity.

The measurement of pumping speed and absorbing capacity is carried out using the transmission factor method, which was first introduced in NEG coating pipes by C. Benvenuti in 1999 [9]. The schematic diagram of the pumping speed measurement setup is depicted in Figure 4.3.6.13. Following each activation cycle of the NEG coating, test gases are injected into the vacuum chamber using variable leak valves. The gases then pass through an orifice into the NEG coated pipe, where a significant portion of the gases is absorbed by the NEG coating. Two Residual Gas Analyzers (RGAs) are positioned at the ends of the pipe to monitor the pressure levels. By measuring the pressures P1 and P2 and calculating the ratio P2/P1, the sticking factor or pumping speed can be determined using molflow simulation.

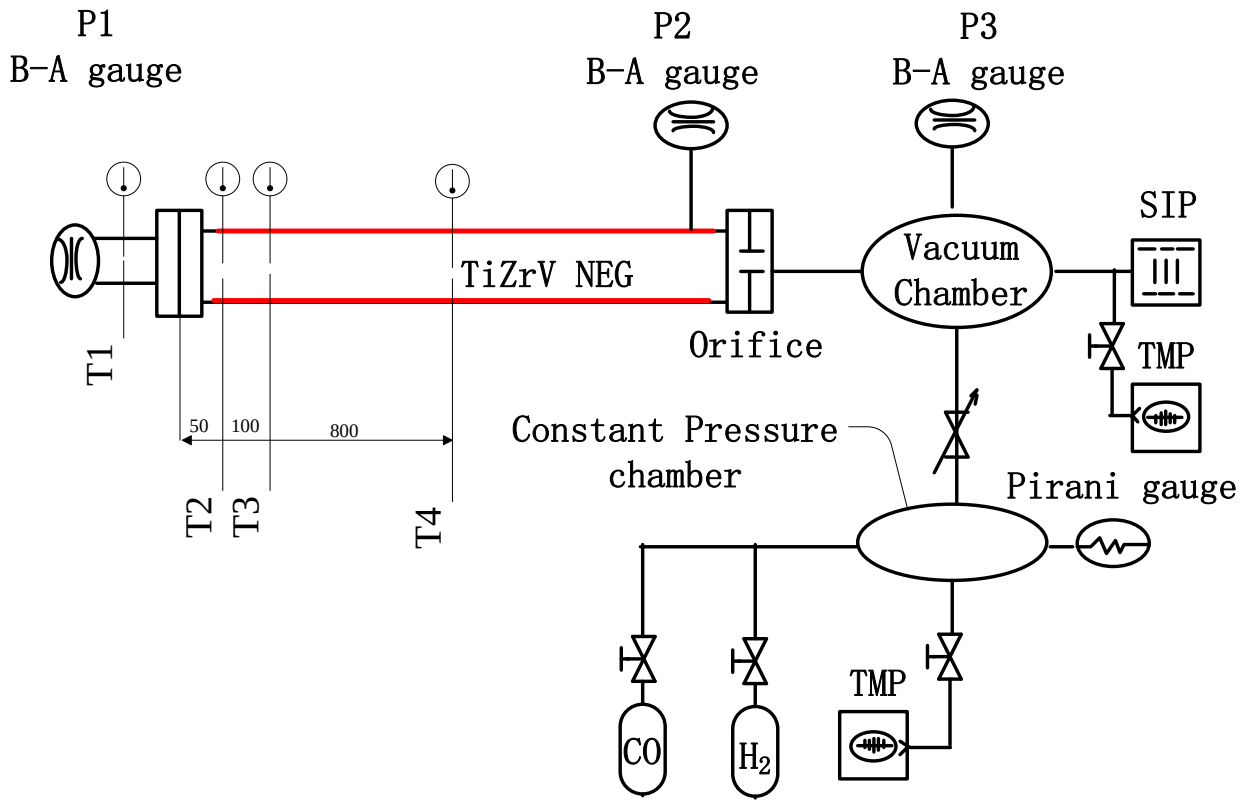

**Figure 4.3.6.13:** Schematic diagram of the pumping performance test facility.

The pumping speed of CO by the NEG coating is approximately 6-8 times higher than that of $H_2$, while the capacity of CO is much smaller than that of $H_2$. The pumping speed of $H_2$ and the capacity of CO are therefore considered to be more critical factors. Test results for the pumping speed of $H_2$ at different activation temperatures are presented in Figure 4.3.6.14 and capacity of CO are presented in Figure 4.3.6.15. The NEG coatings of TiZrV and TiZrVHf can be effectively activated at 160℃. The pumping speed of $H_2$ increases as the activation temperature rises from 160℃ to 250℃ but decreases when the activation temperature exceeds 250℃.

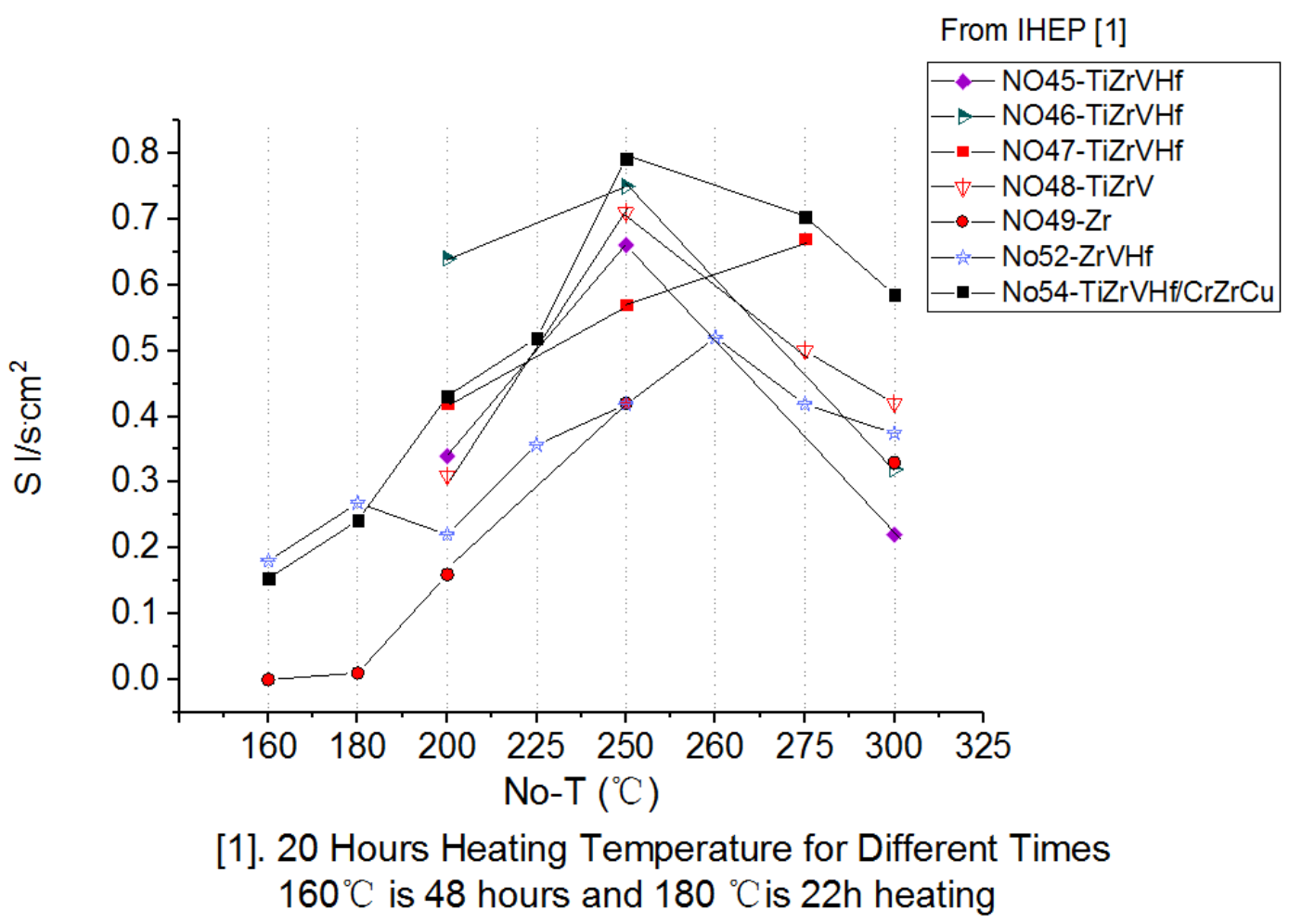

**Figure 4.3.6.14:** Pumping speed of NEG coating under different activation temperature.

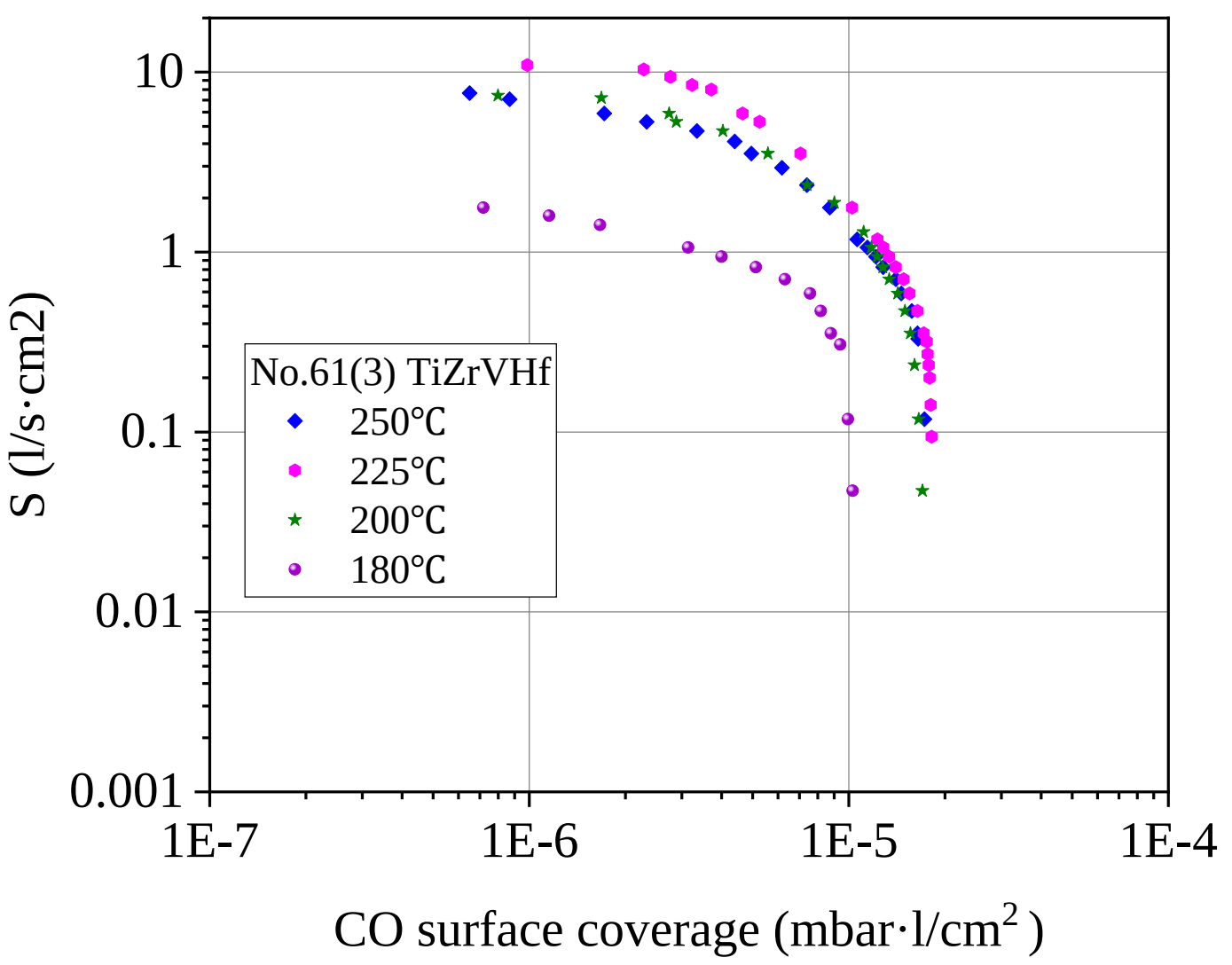

**Figure 4.3.6.15:** CO capacity of NEG coating under different activation temperature.

When all the absorption sites of the NEG coating are filled, its ability to pump gases diminishes. Molecules that contain oxygen atoms, such as $H_2O$, CO, $O_2$, etc., chemically adhere to the surface of the NEG coating, resulting in the formation of metallic oxides ($M_xO_y$). During the activation process, the metal oxides gradually decrease, and oxygen atoms diffuse into the bulk of the coating due to concentration gradients. This activation mechanism imposes limitations on the number of reactivation cycles that can be performed. Figure 4.3.6.16 presents the results regarding the lifetime of a 1 μm thick NEG coating, highlighting its finite lifespan.

The pumping performance of the NEG coating is inversely related to the duration of exposure to air. This suggests that chambers coated with a getter should be adequately preserved before being utilized as a "pump" [10].

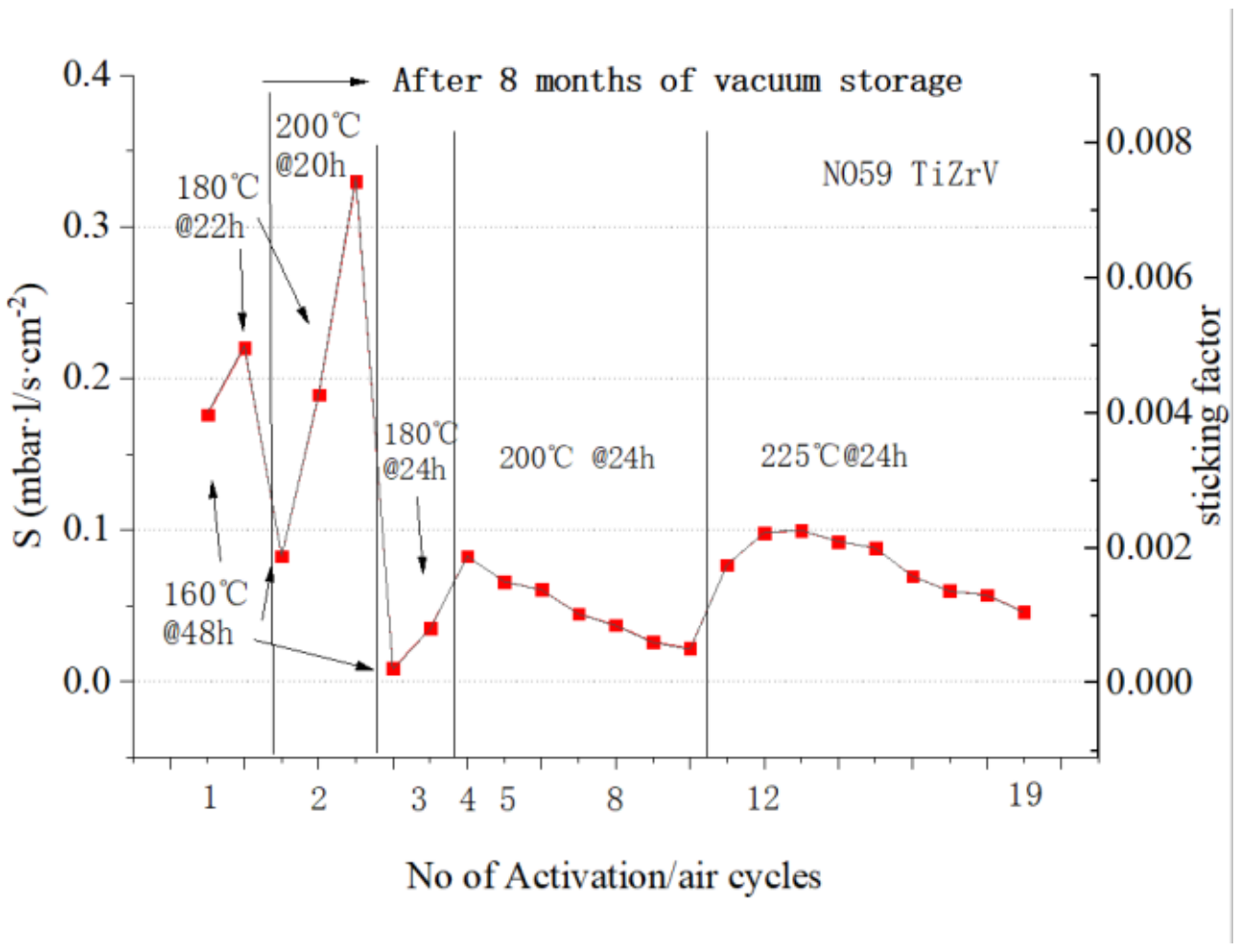

**Figure 4.3.6.16:** The lifetime of NEG coating of 1 µm thickness.

#### 4.3.6.6.1.2 Secondary Electron Yield (SEY) of NEG Coating

The presence of an electron cloud, generated by secondary electrons, can lead to beam instability within a positron storage ring. Copper, for instance, has a Secondary Electron Yield (SEY) of approximately 1.5 [2], making it prone to electron cloud formation. However, in applications like the CEPC positron ring, materials with an SEY below 1.1 are required to mitigate electron cloud effects. NEG coating serves as a promising solution to suppress the electron cloud and provide distributed pumping in the Collider rings.

To investigate the SEY properties of NEG coating, a test facility was established at the China Spallation Neutron Source (CSNS), which is a department of IHEP [11]. The SEY curve was measured for TiZrV and TiZrVHf NEG coatings baked at different temperatures. Figure 4.3.6.17 illustrates the relationship between SEY and primary electron energy for these coatings. The results indicate that as the activation temperature increases, the SEY decreases.

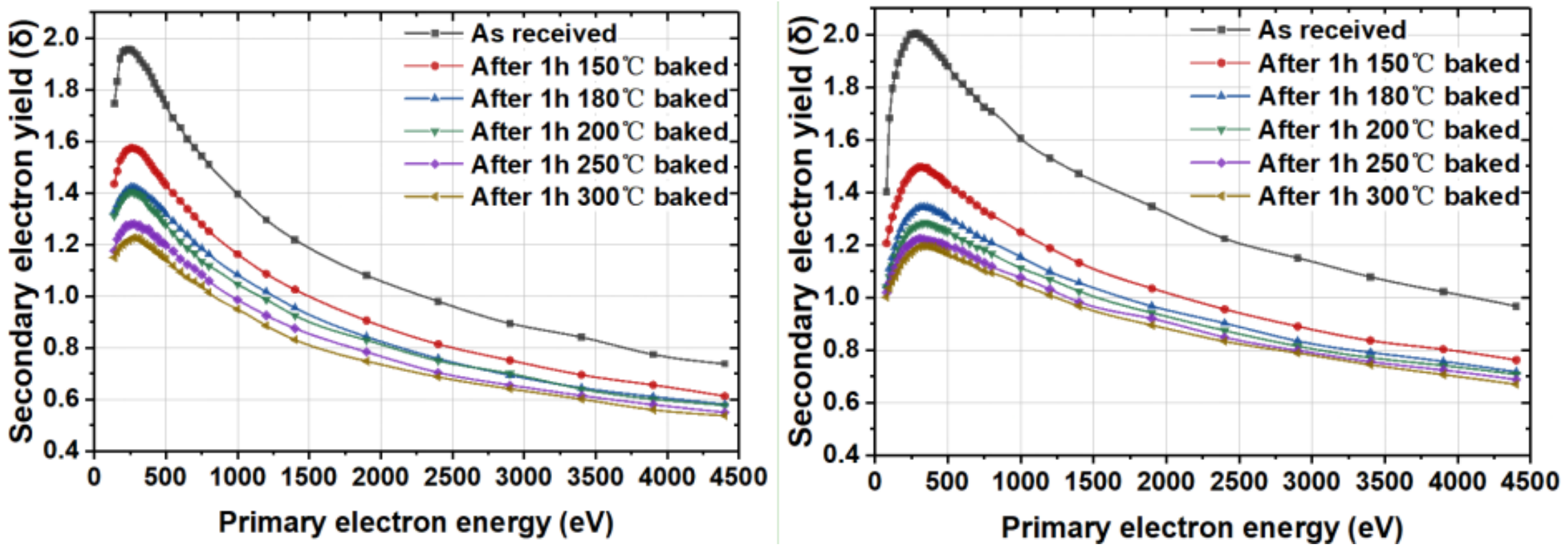

**Figure 4.3.6.17:** SEY versus primary electron energy of the TiZrV (left) TiZrVHf (right) NEG coating.

#### 4.3.6.6.1.3 Thickness Distribution

The thickness distribution of the NEG coating along the axis of the vacuum chamber depends on factors such as the uniformity of plasma discharge pressure, magnetic field, and cathode position. Extensive experiments conducted in HEPS have demonstrated that achieving a distribution within ±30% error is relatively straightforward. However, due to the lower electrical conductivity of the NEG coating compared to copper, the thickness of the coating is limited by the requirement of machine impedance. Simulation results analyzing the vacuum chamber impedance of the NEG coating indicate that an average thickness exceeding 200 nm should be avoided. Such a small thickness would negatively impact the reactivation lifetime of the coating.

##### *4.3.6.6.2 Sputter Ion Pumps*

Sputter ion pumps will be necessary to pump Ar, He, and $CH_4$, which cannot be absorbed by the NEG coating. The number of sputter ion pumps needed will depend on the amount of inert gases, like $CH_4$, Ar, in the residual gas during the beam cleaning period. To ensure efficient pumping, a sputter ion pump will be installed at intervals of 11 to 22 meters, with the option to double the number if needed. These pumps will be turned on only after the NEG coating has been fully activated and the pressure has reached $10^{-7}$ Torr or lower. This will allow multiple pumps to be connected in parallel to a common power supply.

Sputter ion pumps are known for their high reliability, long life, lack of moving parts, and high radiation resistance. In addition, the ion pump current is directly proportional to the vacuum pressure and can be used to provide a detailed pressure profile around the ring. To protect the ion pumps from damage, their power supplies are designed to trip if the ion current exceeds a pre-set value. The leakage current of the pumps is less than 10% of the current drawn at $1\times10^{-9}$ Torr, making them suitable for use as a pressure monitor. Ion pump currents can be stored in a databank, which allows operators to quickly identify any issues that arise.

#### 4.3.6.7 ***Vacuum Measurement and Control***

Due to the large size of CEPC, it is not feasible to install vacuum gauges at short intervals throughout the entire ring. Instead, only specific sections such as the injection regions, RF cavities, and interaction regions will be equipped with cold cathode gauges and residual gas analyzers. For the rest of the ring, continuous monitoring of the current of the sputter ion pumps should provide sufficient pressure measurements down to $10^{-9}$ Torr. If needed, mobile diagnostic equipment can be brought to specific locations during pump down, leak detection, and bake-out when the machine is accessible.

The vacuum system control will be integrated into the overall computer control system, including the monitoring of sputter ion pumps, vacuum gauges, sector valves, and water cooling of vacuum chambers. Critical interlocks, such as those for sector valves, RF cavities, and water cooling, will be hard-wired for reliability. Local and temporary controls will be managed using mobile terminals.

To ensure safety in the high radiation environment of the tunnel, all vacuum electronic devices will be located in the service buildings.

#### 4.3.6.8 ***Vacuum System Preparation and Conditioning***

The Linac, transport line, and Booster vacuum systems will undergo conditioning tailored to the beam current, as they exhibit a low gas load. Conversely, the Collider vacuum system will undergo a baking process to activate the NEG coating. As each section of the vacuum system is constructed, it will be pumped down, ion pumps and vacuum gauges turned on, and the NEG coating can be activated. By adopting this strategy, significant time savings can be achieved once the entire accelerator system is fully assembled.

In the initial stages of beam conditioning, the photo-stimulated gas load tends to be notably high. However, as the conditioning process advances, the gas load gradually diminishes. This reduction in gas load arises from the decline in photo-stimulated desorption (PSD) within the vacuum chambers and the removal of certain residual gases and contaminants.

Drawing from the valuable insights garnered from SuperKEKB, it has been observed that PSD decreases significantly after accumulating approximately 1000 Ampere-hours (Ah) of beam current [12].

### 4.3.7 **Instrumentation and Feedback**

#### 4.3.7.1 *Introduction*

The beam instrumentation system is composed of multiple beam monitors and signal processing electronics. Its purpose is to provide accurate and comprehensive information, enabling accelerator physicists and machine operators to enhance injection efficiency, optimize lattice parameters, monitor beam behavior, and increase luminosity.

Efficient commissioning of the CEPC relies heavily on its instrumentation. Due to the ring's large size, unique challenges arise. Copper cables are impractical for transmitting signals over long distances, so analog signals should be digitized in the tunnel, and data should be transmitted using optical fibers from electronics near the monitors to local stations in an auxiliary tunnel. Positrons and electrons pass through the same monitors and are differentiated by polarity. In summary, our design philosophy is:

- Satisfying long-term stable operation requirements.
- Appropriate precision and speed for parameter measurements.
- A large dynamic range under different conditions.
- Minimizing coupling impedance.

To save costs, it is advisable to utilize in-house developed components wherever possible. Our primary objectives are to quickly and accurately monitor beam status, efficiently measure and control bunch current, and remedy beam instabilities. The measurement of beam orbit is particularly critical, particularly in the interaction region. By understanding beam position, offset, and crossing angle, we can optimize luminosity. Several subsystems are employed to accomplish this, including BPMs for beam position, the DCCT for average beam current measurement, the tune measurement system, and the synchrotron light-based measurement system for beam profile monitoring and bunch length measurement. These systems are detailed in Table 4.3.7.1.

**Table 4.3.7.1:** Main technical parameters of the Collider Beam Instrumentation Systems

| **Sub-systems** | | **Method** | **Parameter** | **Amounts** |
|---|---|---|---|---|
| Beam position monitor | Closed orbit | Button electrode BPM | Measurement area (x´y)：±20mm×±10mm<br>Resolution：<0.6um<br>Measurement time of COD：< 4 s | 3544 |
| | Bunch by bunch | Button electrode BPM | Measurement area (x´y)：±40mm×±20mm<br>Resolution：0.1mm | |
| Bunch current | | BCM | Measurement range：10mA / per bunch<br>Relatively precision：1/4095 | 2 |
| Average current | | DCCT | Dynamic measurement range：0.0~1.5A<br>Linearity：0.1 %<br>Zero drift: <0.05mA | 2 |
| Beam size | | Double slit interferometer<br>X ray pin hole | Resolution:0.2 μm | 4 |
| Bunch length | | Streak camera<br>Two photon intensity interferometer | Resolution:1ps@10ps | 2 |
| Tune measurement | | Frequency sweeping method | Resolution:0.001 | 2 |
| | | DDD | Resolution:0.001 | |
| Beam loss monitor | | PIN-diode | Dynamic range:120 dB<br>Maximum counting rates≥10 MHz | 5800 |
| Feedback system | | TFB | Damping time<=1ms | 4 |
| | | LFB | Damping time<=12ms | 4 |

### 4.3.7.2 *Beam Position Measurement*

BPMs are one of the most critical instruments in CEPC as they measure many machine parameters, including:

- Beam position and stabilization.
- Orbit concepts such as beam-based alignment, corrections, and "golden orbit."
- Optics functions including beta-function, dispersion, tune, coupling, chromaticity and RF frequency.
- Stabilizing the beam through feedback and addressing instabilities.

With one Beam Position Monitor (BPM) located near each quadrupole, there will be a total of 3544 BPMs, including some additional ones at specific locations. To accommodate these BPMs and other instruments, we will establish 50 local stations in an auxiliary tunnel, with each station controlling approximately 70 BPMs. The front-end and digital electronics of the BPMs will be located in the main tunnel. Due to the limited space there, it is optimal to position the BPMs underneath the magnet girder.

The digital electronics will transmit data via optical fiber to local stations, where a multiplex system will enable communication with other local stations and the central control room. At the special regions near the IP or the local chromaticity correction sextupoles, high-resolution BPMs are necessary. Moreover, we will consider implementing non-linear mapping for the off-axis position of the crossed beams, an orbit slow feedback system, and an IP point orbit feedback system based on the BPM signals.

#### *4.3.7.2.1 Mechanical Construction*

The following are the design criteria:

- Compact dresign to save space.
- Skewed sensor positions to avoid direct impact from synchrotron radiation.
- Minimum RF loading and beam coupling to the higher order modes.
- High precision for asy interchangeability.
- Flanges for easy replacement in the event of a leak.
- Corrosion resistance and ability to withstand baking up to 300°C.

A capacitive monitor with a button-like electrode, commonly used in other electron machines, is the preferred solution. To achieve the necessary geometrical accuracy, a machined block will be used to mount the four buttons.

#### *4.3.7.2.2 Feedthrough Design*

To avoid reflections and withstand high temperatures up to 200℃, a matched feedthrough must be designed. Figure 4.3.7.1(a) illustrates the characteristic impedance results of a feedthrough measured by a time-domain reflectometer (TDR) with a Tektronix DSA8200 digital serial analyzer equipped with an 80E04 sampling module. A smooth transition structure with a 50 Ω impedance can minimize reflection. As shown in Figure 4.3.7.1(b), a safe distance between the welding point and sealed dielectric is required. The button radius ($r_b$), button height ($t_b$), and button gap ($g_b$) are three important parameters for the feedthrough, and their effects are shown in Table 4.3.7.2. The basic structure of the feedthrough comprises four parts: the button, the pin, the support dielectric material (which also seals the vacuum), and the housing, as shown in Figure 4.3.7.1(b). Further design details can be found in reference [1].

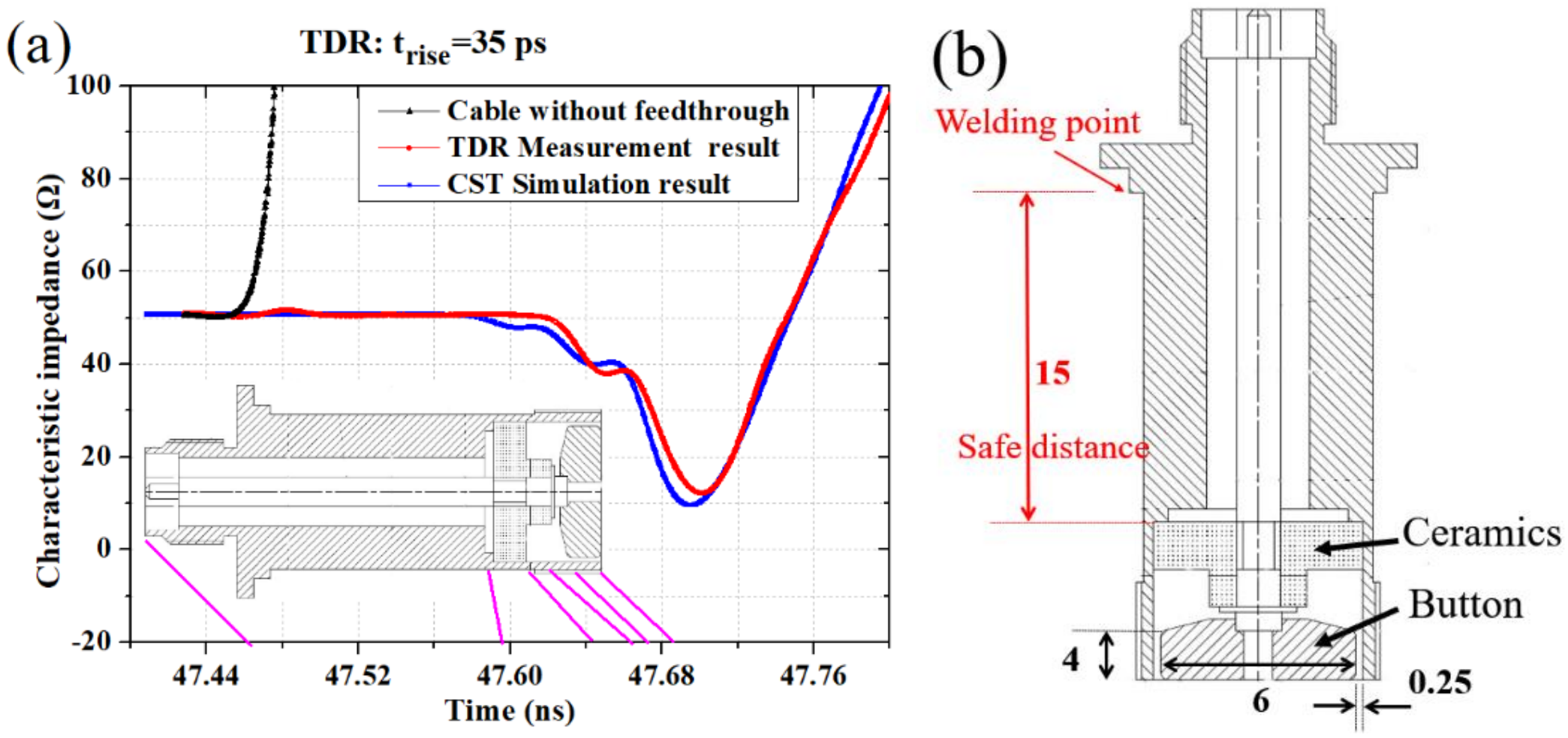

**Figure 4.3.7.1:** (a) The impedance of a feedthrough, (b) the schematic of the feedthrough with a safe distance of 15 mm.

**Table. 4.3.7.2:** The effects of button design parameter change (increase)

| Parameter | Feature | Advantage | Disadvantage |
|---|---|---|---|
| $r_b$ | Larger area and $C_b$ | Higher signal level | Lower resolution |
| $t_b$ | Larger volume | Lower wake impedance | Lower resolution, |
| $g_b$ | More trapped modes | Higher resolution | Higher wake impedance |

*4.3.7.2.3 Button BPM Design*

The BPM transfer impedance $Z_t(\omega)$ relates the voltage response signal $V(\omega)$ to the bunched beam current signal $I(\omega)$ via Ohm's law:

$$V(\omega)=Z_t(\omega)\times I(\omega) \tag{4.3.7.1}$$

$Zt(\omega)$ is a crucial parameter of a BPM pickup, which is determined by the geometry and shape of the BPM [2-3]. It represents a function of the angular frequency $\omega$ and exhibits a frequency characteristic similar to that of a high-pass filter. For broadband BPM pickups such as button-style electrodes, the behavior of the beam position is independent of frequency. In practical applications, $Z_t$ is typically on the order of ~1 Ω at frequencies above 1 GHz, which indirectly determines the resolution potential of a button BPM when connected to suitable read-out electronics. The characteristics of a BPM can be described using three parameters, namely, the transfer impedance $Z_t$, the longitudinal beam-coupling impedance $Z_l$, and the signal power $\langle P\rangle$ output from a button BPM electrode [4].

$$Z_t=\frac{r_b{}^2R_0\omega}{2bc\left(1+R_0{}^2C_b{}^2\omega^2\right)}\left(R_0C_b\omega+j\right) \tag{4.3.7.2}$$

$$Z_l=\frac{r_b{}^4\omega}{4b^2c^2C_b\left(1+R_0{}^2C_b{}^2\omega^2\right)}\left(R_0C_b\omega+j\right) \tag{4.3.7.3}$$

$$\langle P\rangle = \frac{I^2 r_b^{\ 4} \omega^2 R_0}{8b^2 c^2\left(1+R_0^{\ 2} C_b^{\ 2} \omega^2\right)} \tag{4.3.7.4}$$

in which $C_b$ represents the button capacitance to the ground, while $b$ represents the distance between the beam and the BPM button. For a circular beam pipe, $b$ is the beam pipe radius with the beam in the center. $R_0$ refers to the load impedance of the BPM electronics, which is 50 Ω for the CEPC. The speed of light is represented by $c$, and the beam current is represented by $I$.

Assuming $r_b$ = 3 mm, $t_b$ = 4 mm, $g_b$ = 0.25 mm, $C_b$ = 3.0 pF, $I$ = 0.2 A, $b$ = 28 mm, and $R_0$ = 50 Ω, the results shown in Figure 4.3.7.2 indicate that the transfer impedance at 650 MHz is 0.09 Ω. Additionally, the signal power is –24.5 dBm and –10.6 dBm for current levels of 200 mA and 1000 mA, respectively.

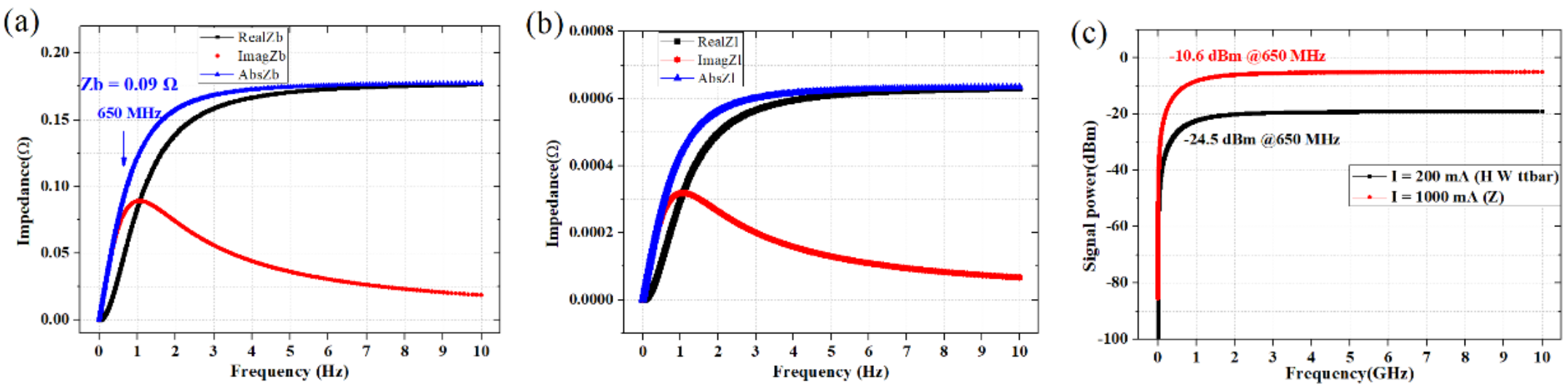

**Figure 4.3.7.2:** (a) Transfer impedance $Z_t$ , (b) the longitudinal beam-coupling impedance $Z_l$ , (c) the signal power results calculated by Eqs. 4.3.7.2-4.3.7.4.

To investigate the response of a pick-up electrode to a beam, CST particle studio wake-field simulations were conducted on the beam parameters and vacuum pipe [5]. Figure 4.3.7.3 (a) displays the CST model for button BPMs, where $r_b$ = 3 mm, $t_b$ = 4 mm, $g_b$ = 0.25 mm, and $b$ = 28 mm. The bunch current, pickup signal, and longitudinal potential computed by CST are depicted in Figures 4.3.7.3 (b-d). The simulation parameters are provided in Table 4.3.7.3. For the simulation, a bunch charge and length of 1 nC and 4.1 mm were chosen for Higgs or other modes. The transfer impedance and signal level, calculated by CST and Eqs. 4.3.7.2-4.3.7.4, are summarized in Table 4.3.7.4.

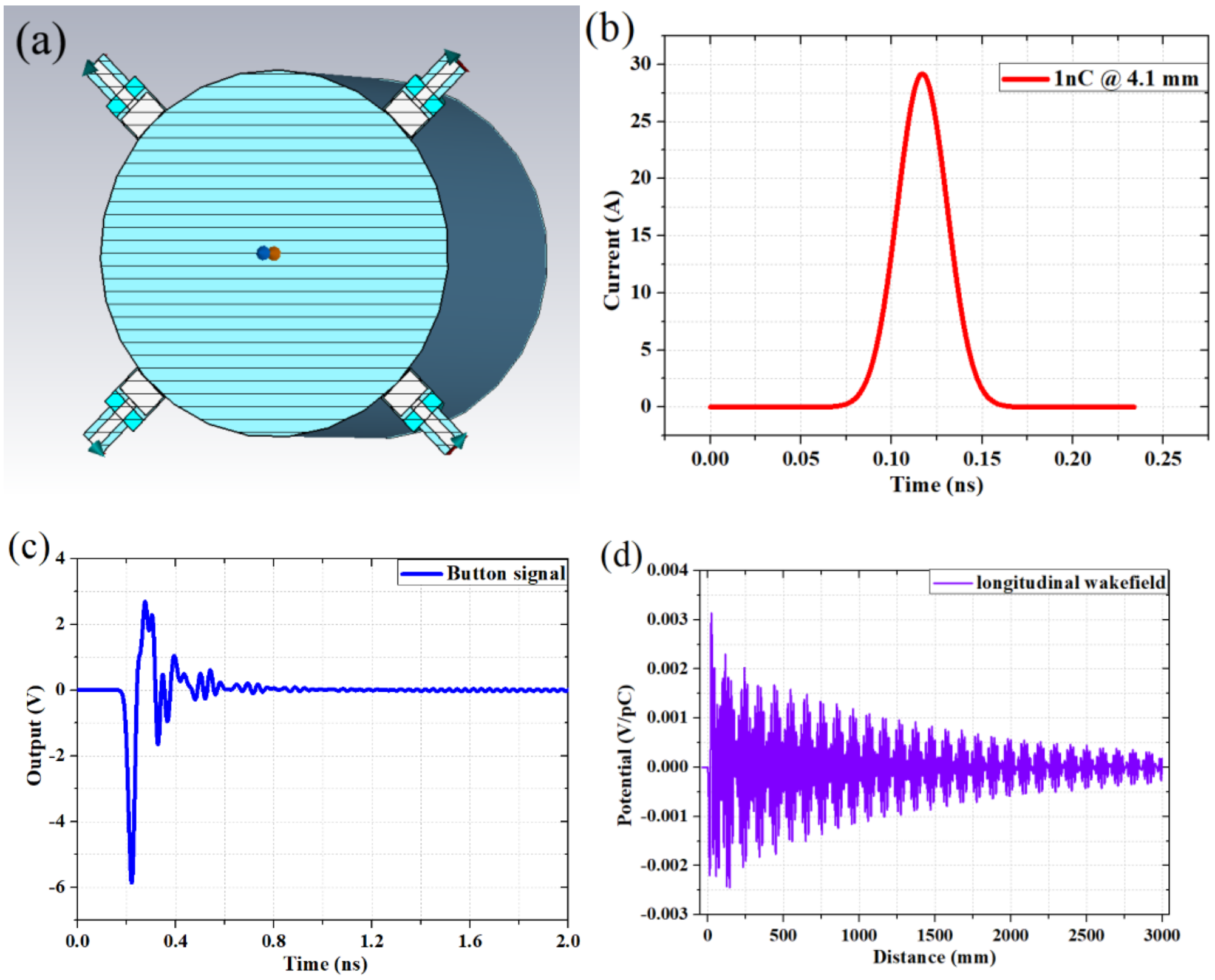

**Figure 4.3.7.3**: (a) CST model, (b) the bunch current where charge is 1 nC (position) and RMS length is 4.1 mm, (c) the signal of a pickup (electron), (d) the longitudinal potential calculated by CST.

**Table 4.3.7.3:** Main parameters used for CST simulation and analytical calculations.

| | | |
|---|---|---|
| Bunch | Length | 4.1 mm |
| | Charge | 1 nC |
| Mechanical | $r_b$ | 3 mm |
| | $t_b$ | 4 mm |
| | $g_b$ | 0.25 mm |
| | $b$ | 28 mm |

**Table 4.3.7.4:** CST simulation and analytical calculation results for the CEPC BPMs

| | CST results | Analytical results |
|---|---|---|
| Transfer impedance (650 MHz) | 0.09 Ω | 0.09 Ω |
| Signal power (200 mA) | –20.46 dBm | –24.5 dBm |

Given the cable attenuation coefficient of 1 dB/10m, the signal power may not be sufficient for achieving high resolution in the BPM readout electronics. As demonstrated previously, the most effective approach for boosting the signal is to increase the radius of the button [6]. In order to increase rb from 2 mm to 8 mm, we utilized CST to calculate the transfer impedance and signal power, and the results are presented in Figure 4.3.7.4 using Eqs. 4.3.7.2-4.3.7.4. Additionally, we calculated the capacitance of the feedthrough by applying Eq. 4.3.7.5.

$$C_{\text{coax}} = \frac{2\pi\varepsilon l_{\text{coax}}}{\ln(\frac{R_{\text{out}}}{R_{\text{in}}})} \tag{4.3.7.5}$$

where $R_{out}$ is the inner radius of the outer conductor, $R_{in}$ is the outer radius of the inner conductor, $\varepsilon$ is the dielectric constant of the material located between the outer and inner conductors, and $l_{coax}$ is the length of the coaxial structure.

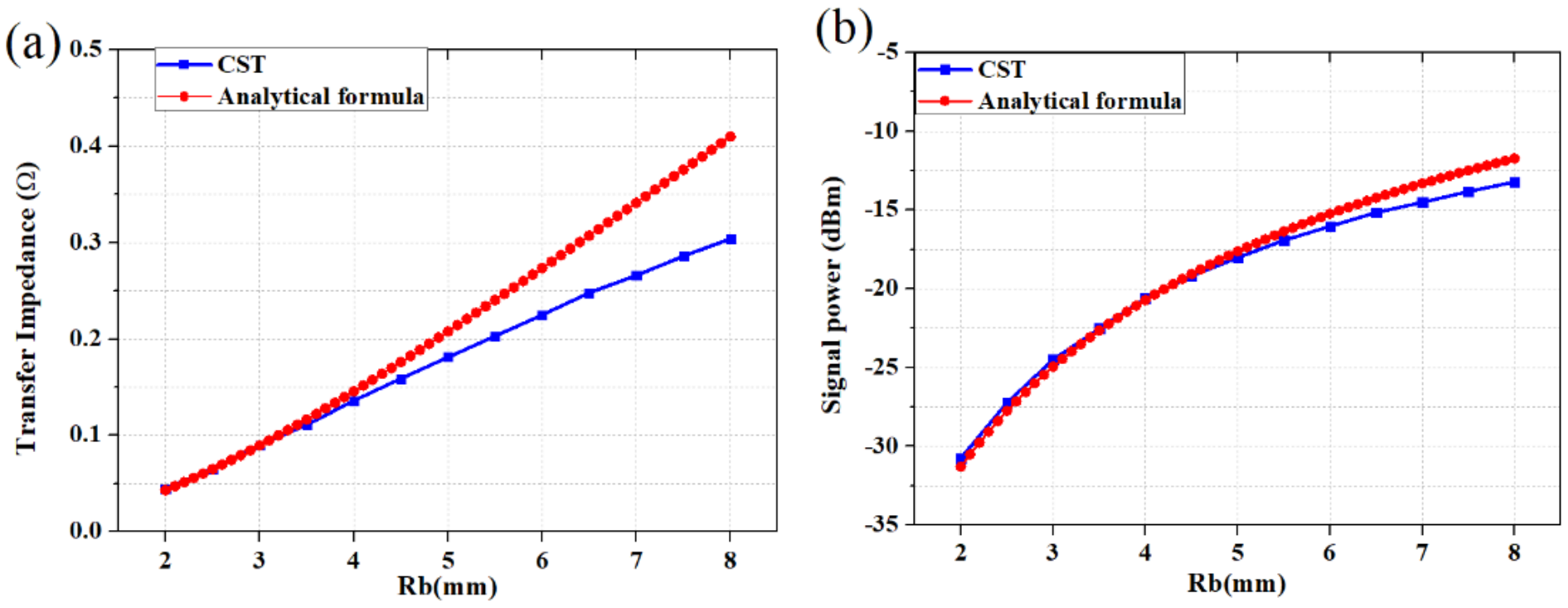

Figure 4.3.7.4: (a) Transfer impedance, (b) the signal power calculated by CST and Eqs. 4.3.7.2-4.3.7.4, with the parameters $I$ = 0.2 A, $t_b$ = 4 mm, $g_b$ = 0.25 mm, $f_0$ = 650 MHz.

The position of the beam ($x$, $y$) can be estimated by using the following polynomial ($X_{raw}$, $Y_{raw}$):

$$X_{raw} = \frac{V_a+V_d\text{-}V_b\text{-}V_c}{V_a+V_b+V_c+V_d}, \quad Y_{raw} = \frac{V_a+V_b\text{-}V_c\text{-}V_d}{V_a+V_b+V_c+V_d} \tag{4.3.7.6}$$

$$x=\sum_{i=0}^{n}\sum_{j=0}^{i} A_{i\text{-}j,j} X_{raw}^{\ i\text{-}j}\ Y_{raw}^{\ j},\ y=\sum_{i=0}^{n}\sum_{j=0}^{i} B_{i\text{-}j,j} X_{raw}^{\ i\text{-}j}\ Y_{raw}^{\ j} \tag{4.3.7.7}$$

The order of the polynomial used in the equation is denoted by $n$, which is typically chosen based on the required calibration accuracy and falls within the range of 1-5. Figure 4.3.7.5 (a) displays the simulation mapping within a range of ±15 mm × ±15 mm. The errors associated with high-order fitting are significantly smaller than those of the linear fit due to the non-linear response of the BPM when the beam is off-center. Figure 4.3.7.5 (b) illustrates the error distribution for $n$ = 5, while Table 4.3.7.5 outlines the fitting coefficients for $n$ = 1, 2, and 3. Both $A_{1,0}$ and $B_{0,1}$ are equal to 18.36 mm. Table 4.3.7.6 presents the signal of CEPC in various modes, with the differing power levels primarily arising from the time structure of the bunch, as depicted in Figure 4.3.7.6.

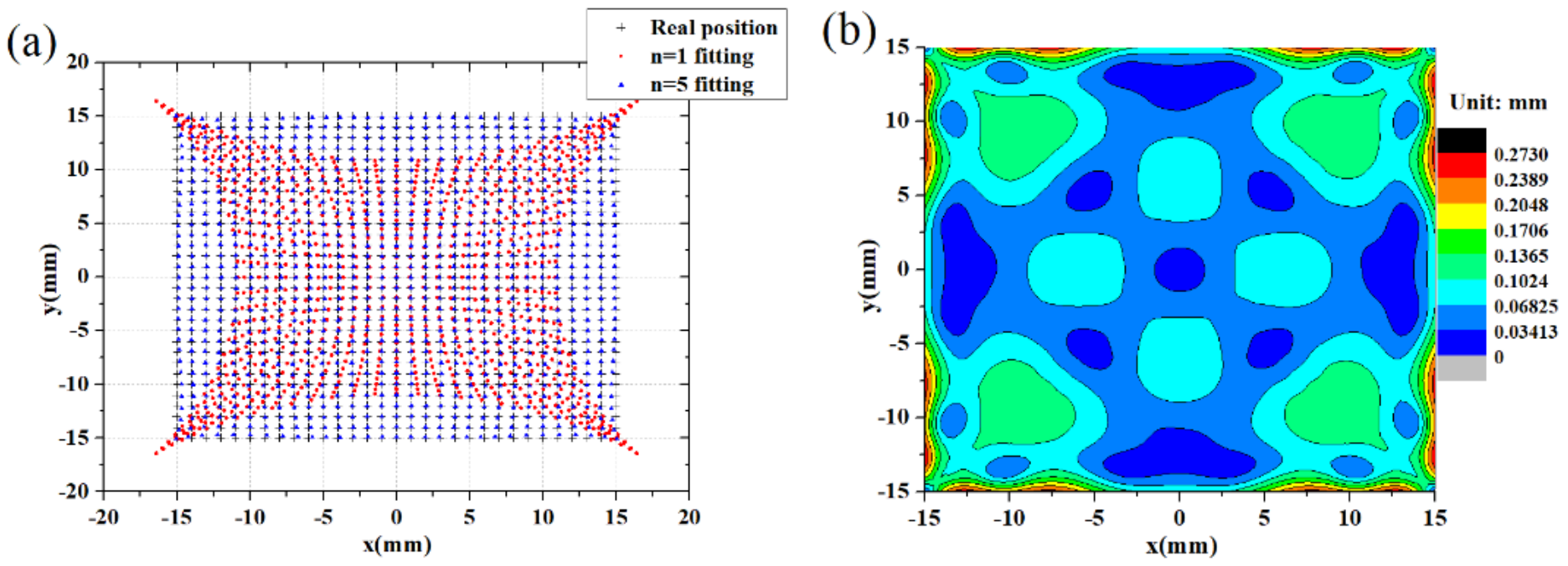

**Figure 4.3.7.5:** (a) Mapping plot in a range ±15 mm × ±15 mm with a step of 1 mm. (b) Fitting error distribution with high order fit (*n* = 5).

**Table 4.3.7.5:** The fitting coefficients by Least Squares method for n =1, 2, 3

| A(n=1) | B(n=1) | A(n=2) | B(n=1) | A(n=3) | B(n=3) |
|---|---|---|---|---|---|
| -0.00030 | 0.00030 | -0.00042 | 0.00042 | -0.00046 | 0.000453 |
| 18.36346 | 0.00023 | 18.36346 | 0.00023 | 18.32287 | 0.000352 |
| 0.00024 | 18.36346 | 0.00024 | 18.36346 | 0.00033 | 18.32289 |
| 0 | 0 | 0.00133 | 0.000842 | -0.00082 | -0.00100 |
| 0 | 0 | -0.00021 | 0.000195 | -0.00040 | 0.000375 |
| 0 | 0 | -0.00081 | -0.00134 | 0.00106 | 0.0008122 |
| 0 | 0 | 0 | 0 | 19.97593 | 0.0002306 |
| 0 | 0 | 0 | 0 | -0.00067 | -22.76355 |
| 0 | 0 | 0 | 0 | -22.76354 | -0.00072 |
| 0 | 0 | 0 | 0 | 0.00027 | 19.97590 |

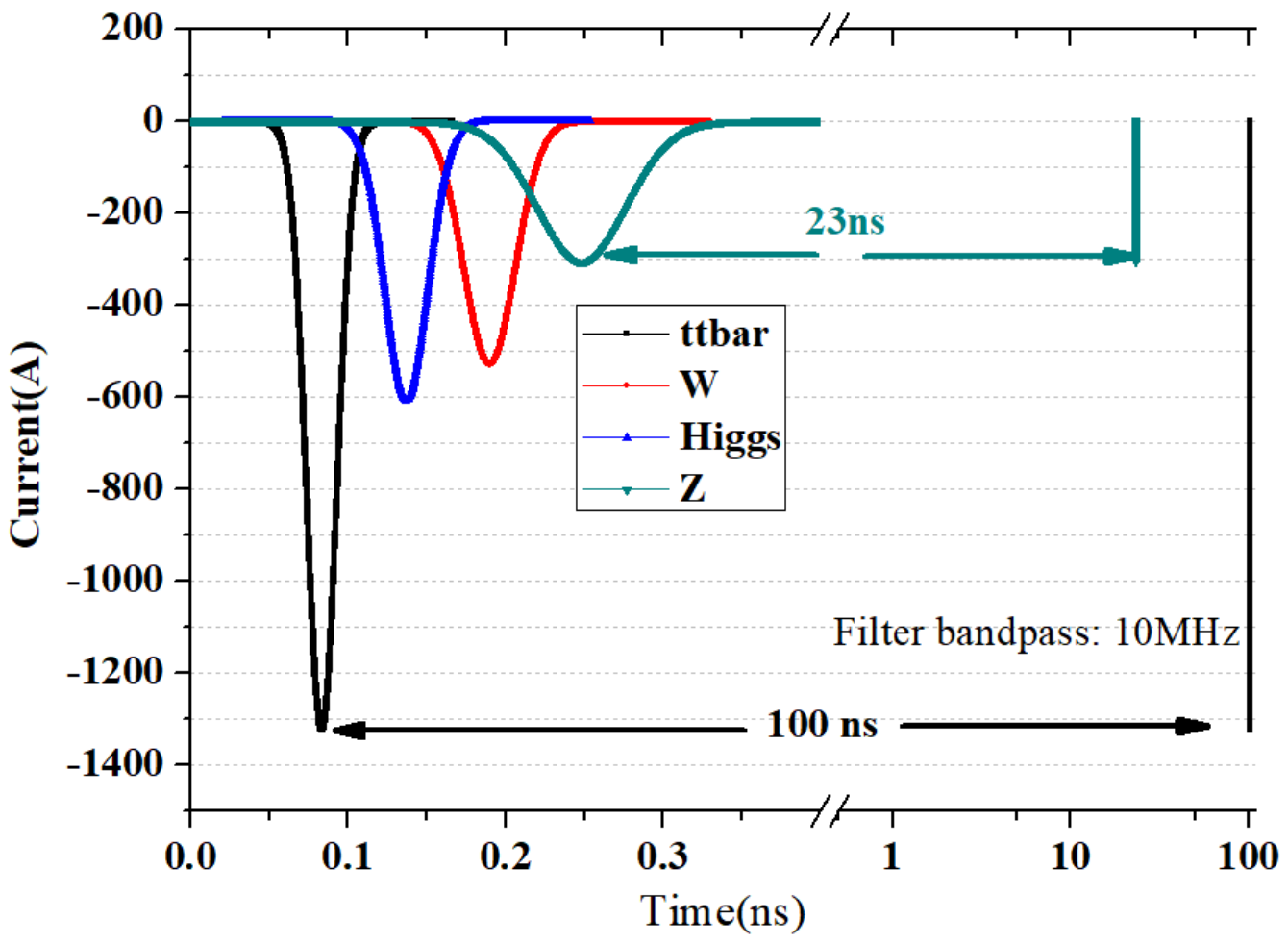

**Figure 4.3.7.6:** Bunch current of the CEPC

**Table 4.3.7.6:** Parameters used for BPM CST simulation.

| | Bunch charge (nC) | Bunch length (mm) | Peak current (A) | Average current (A) | 650MHz (dBm) |
|---|---|---|---|---|---|
| Higgs | 20.8 | 4.1 | 607 | 0.208 | –24.4 |
| Z | 22.4 | 8.7 | 308 | 0.974 | –10.9 |
| W | 21.6 | 4.9 | 527 | 0.216 | –23.7 |
| $t\bar{t}$ | 32.0 | 2.9 | 1315 | 0.320 | –20.2 |

### *4.3.7.2.4 Convention*

The CEPC Collider RF BPM pickups and associated cables are labeled TI, TO, BI, BO. The associated signal names are A, B, C, and D. The beam is assumed to flow into page in a clockwise direction.

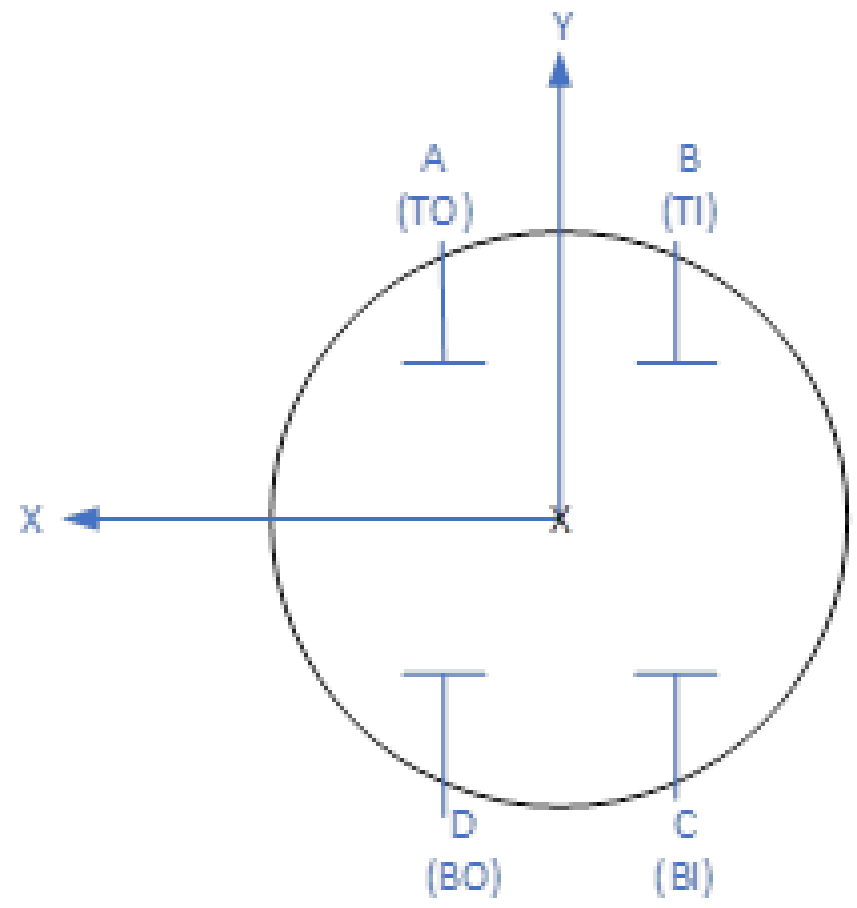

**Figure 4.3.7.7:** BPM pickups and direction

Signal processing conventions are based on the following formulae:

$$X = K_x \left[\frac{(A + D) - (B + C)}{A + B + C + D}\right] - X_{offset} \qquad (4.3.7.8)$$

$$Y = K_y \left[\frac{(A + B) - (C + D)}{A + B + C + D}\right] - Y_{offset} \qquad (4.3.7.9)$$

$$Q = K_X \left[\frac{(A + C) - (B + D)}{A + B + C + D}\right] - Q_{offset} \qquad (4.3.7.10)$$

### *4.3.7.2.5 System Archetecture*

The hardware system can be categorized into two main parts: the analog front-end (AFE) for radio frequency signal conditioning and the digital signal processing part (DFE). The RF signal conditioning part consists of the RF signal conditioning module, ADC analog-to-digital conversion module, clock processing module, and pilot signal generation module. On the other hand, the digital signal processing part includes the BPM signal acquisition module, signal conditioning module, data acquisition and transfer module, EPICS IOC module, and CSS control module. You can see the hardware architecture diagram of the digital BPM signal processor system in Figure 4.3.7.8.

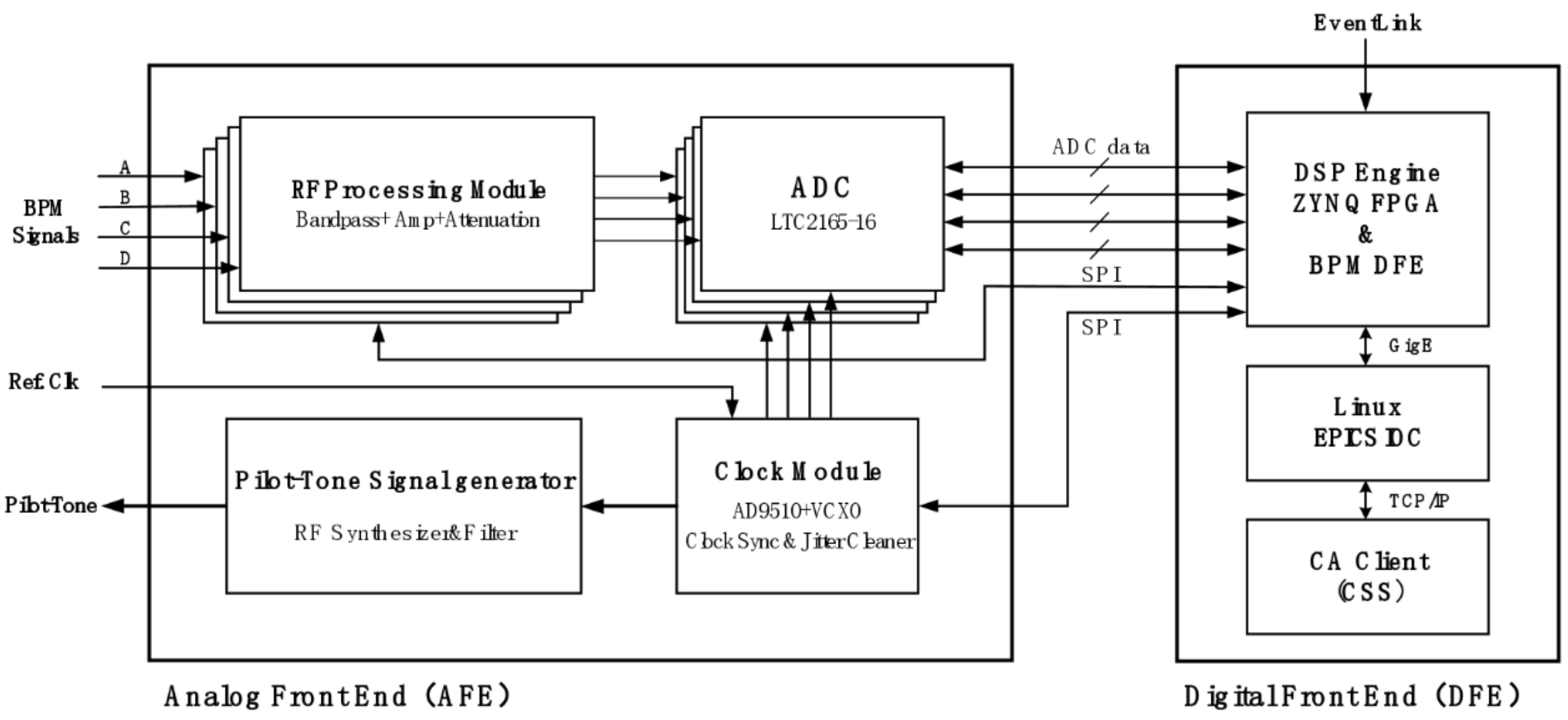

**Figure 4.3.7.8:** Hardware architecture of digital BPM signal processor system

The CEPC RF BPM utilizes a bandpass sampling architecture. Initially, the RF signal undergoes a bandpass filtering process, which centers on the RF harmonic frequency. This procedure is aimed at expanding the 20 ps bunch length while minimizing the amplitude. A bandpass filter topology utilizing RF SAW has been chosen, offering a passband 3 dB width of approximately 20 MHz. In addition, the injection system will utilize an in-band pilot tone, positioned between the 3rd and 4th revolution line above the central frequency line.

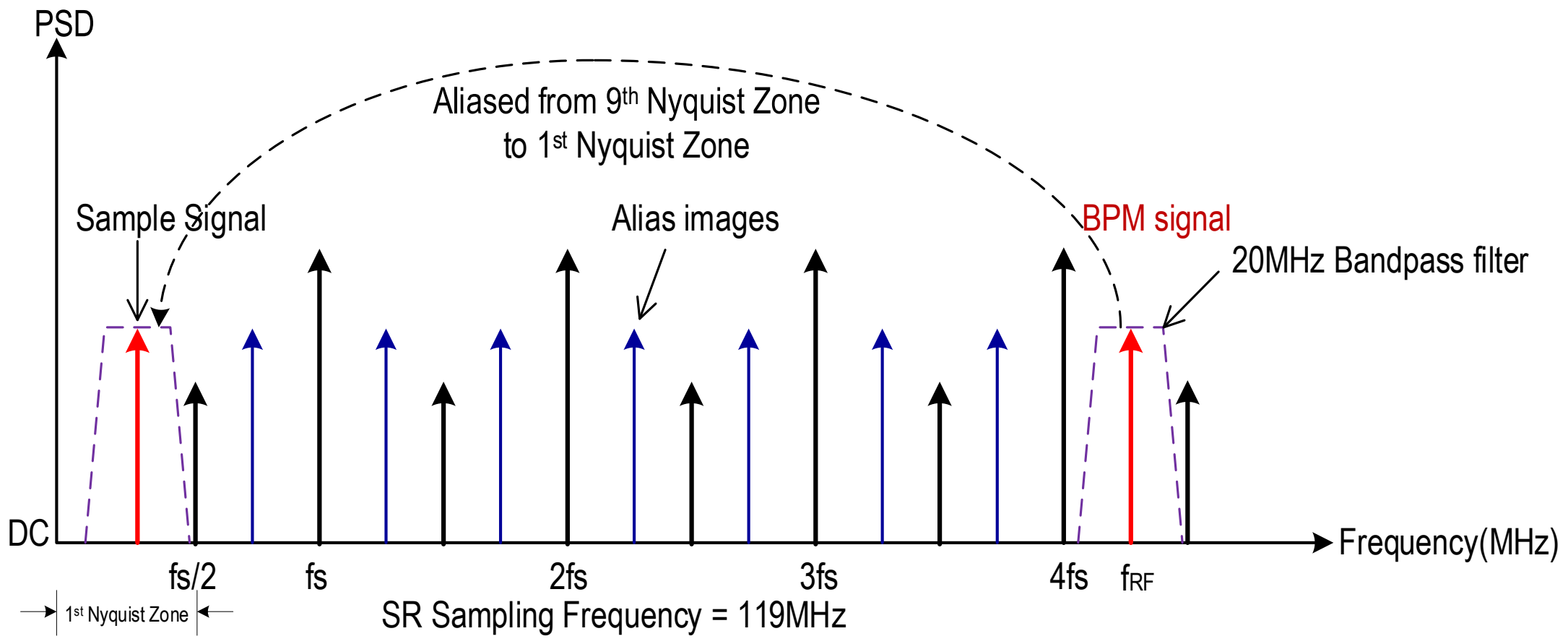

**Figure 4.3.7.9:** Bandpass Sampling

To maintain uniformity between the Booster and SR, the ADC sampling frequency has been set as the 310th harmonic of the storage ring. Given that the Booster's circumference is 1/5th that of the storage ring, the sampling harmonic aligns with the 62nd harmonic of the Booster revolution frequency. Consequently, the ADC generates 62 samples per revolution.

The ADC clock is synchronized to the Booster machine clock via an external machine clock reference obtained from the timing system. A PLL is used to multiply the Booster machine clock by 62, thereby achieving a sampling frequency of 119 MHz. Additionally, the sub-sampled 499.8 MHz RF fundamental signal is converted to a digital IF frequency by adjusting the frequency according to the formula $|F_{rf} - 4{\cdot}F_s| = 23.8$ MHz, in which $F_{rf}$ is the RF frequency and $F_s$ the sampling frequency.

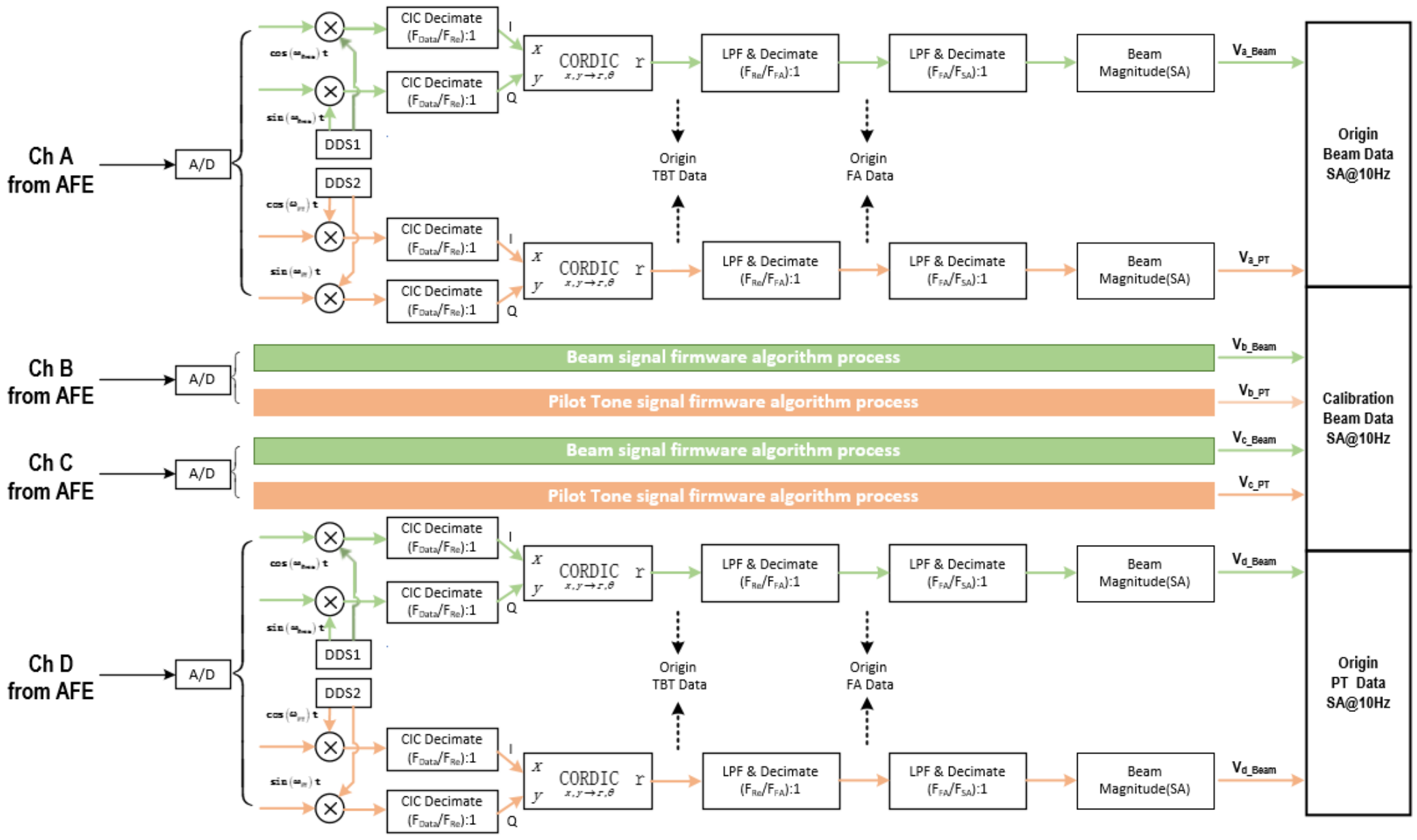

**Figure 4.3.7.10:** Signal Processing Architecture

The diagram presented above provides a clear illustration of the signal processing involved in the RF BPM at a high level. Specifically, the AFE is responsible for providing 4 channels of 16-bit 119 Msps data to the DFE.

The DFE then performs two essential functions: it first corrects for systematic gain errors, which are derived from bench-level testing and calibration. This correction is followed by a dynamic calibration block, which maintains long-term stability using a pilot-tone integrated on approximately a seconds time-scale.

The integral result of the pilot-tone is used to dynamically adjust a digital multiplier, which in turn, normalizes the four channels. This process ensures long-term stability of the system.

Once the dynamic calibration output is obtained, it is fed into a band-pass filter (BPF) that helps to suppress any dynamic low-frequency and DC components generated by the ADC, as well as any spurious or harmonic high-frequency content.

Following the BPF is a digital down-converter, which is utilized to translate the digital intermediate frequency (IF) to baseband. After this, the signal is passed through a 200 KHz Finite Impulse Response (FIR) filter, which helps to suppress any out-of-band energy from the Time-base Trigger (TbT).

Finally, the position of the signal is calculated using the standard difference/sum technique in the TbT processor.

### *4.3.7.2.6 Passive Pilot Tone Combiner*

In the tunnel, a passive combiner module is installed to inject the pilot tone into the signal path. The combiner design includes subminiature, stripline directional couplers, and a 4-way power splitter used to distribute the pilot-tone evenly across all four channels.

To ensure proper functioning of the module, it will be isolated from the system ground by a RO4350B plate placed between the module and the girder. Additionally, the screws used to fasten the module to the girder will be isolated from the electronics ground within the module.

Furthermore, signal ground will also be isolated using a coupling capacitor, which allows for the AC return path of RF currents but inhibits the flow of low-frequency noise energy. The signal path will also be isolated using RF coupling capacitors to ensure proper transmission of the signal.

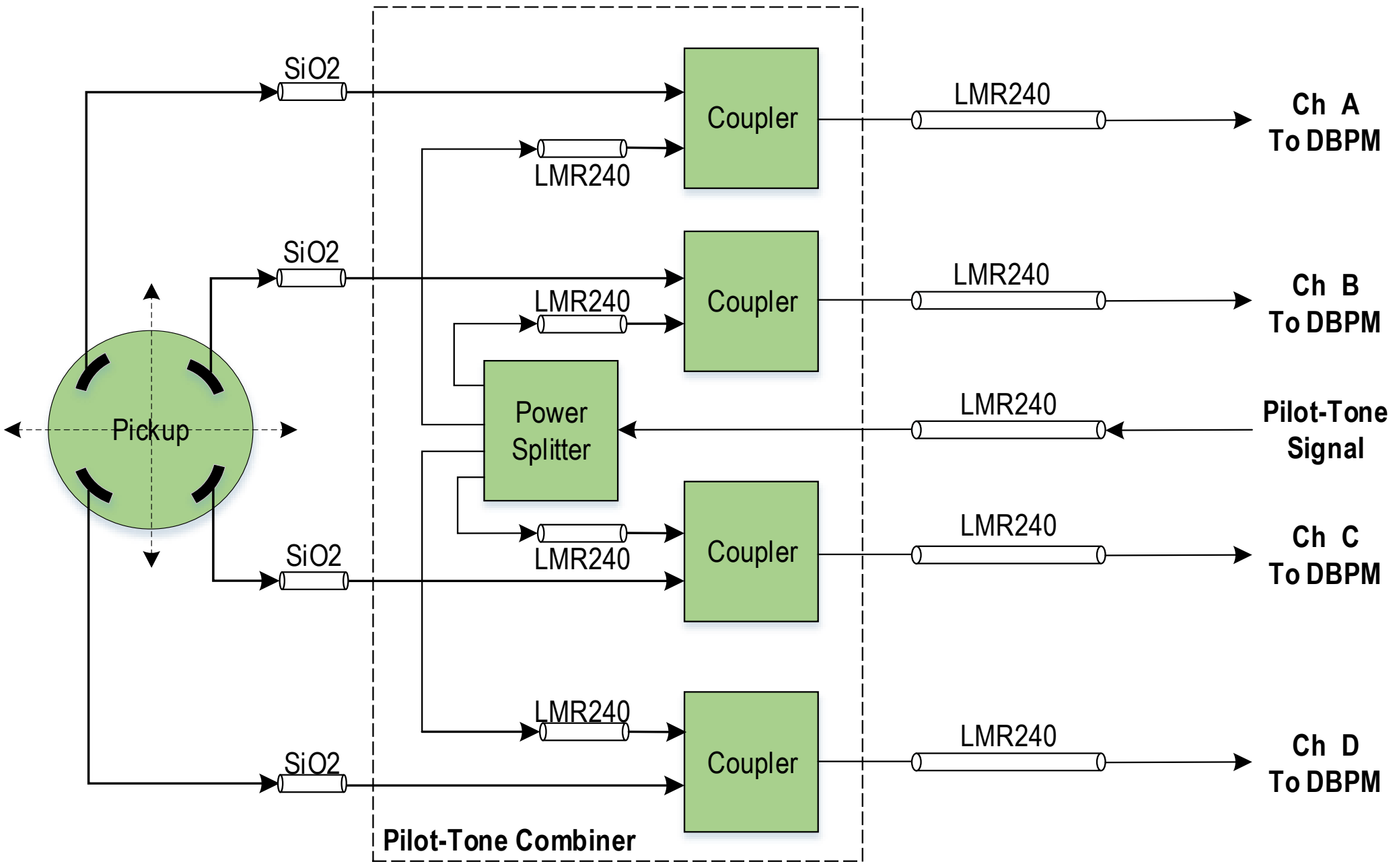

**Figure 4.3.7.11:** Passive Pilot Tone Combiner

### *4.3.7.2.7 AFE Functional Description*

The figure below illustrates the AFE receiver architecture.

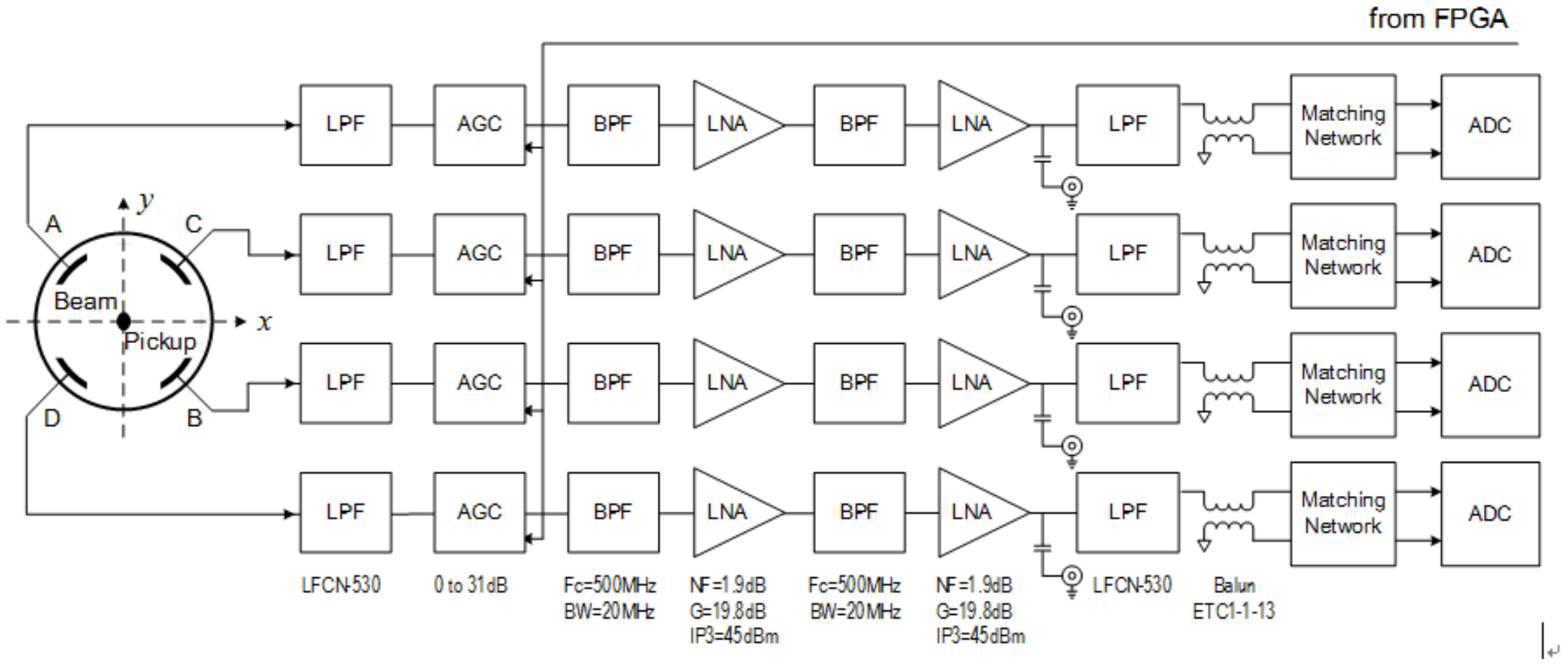

**Figure 4.3.7.12:** AFE receiver architecture

To ensure that out-of-band energy is attenuated, and peak voltage amplitude is reduced, an input low-pass filter (LPF) with a cutoff frequency of 530 MHz is utilized. Following this is a digital attenuator that allows for attenuation ranging from 0 dB to 31.25 dB in 0.25 dB steps.

Next in the signal path is a surface acoustic wave (SAW) band-pass filter (BPF) with a 20 MHz bandwidth. Both amplifiers in the receiver utilize the SOT-89 standard package style, providing flexibility in terms of amplifier selection.

The first amplifier is chosen for its high gain and low noise figure, maximizing the low-level signal performance observed in single-bunch and low current fills. The second BPF is identical to the first and serves to reduce spurious energy and filter noise prior to sub-sampling.

Finally, the second amplifier is chosen for its very high IP3, allowing for maximum linear dynamic performance of the ADC.

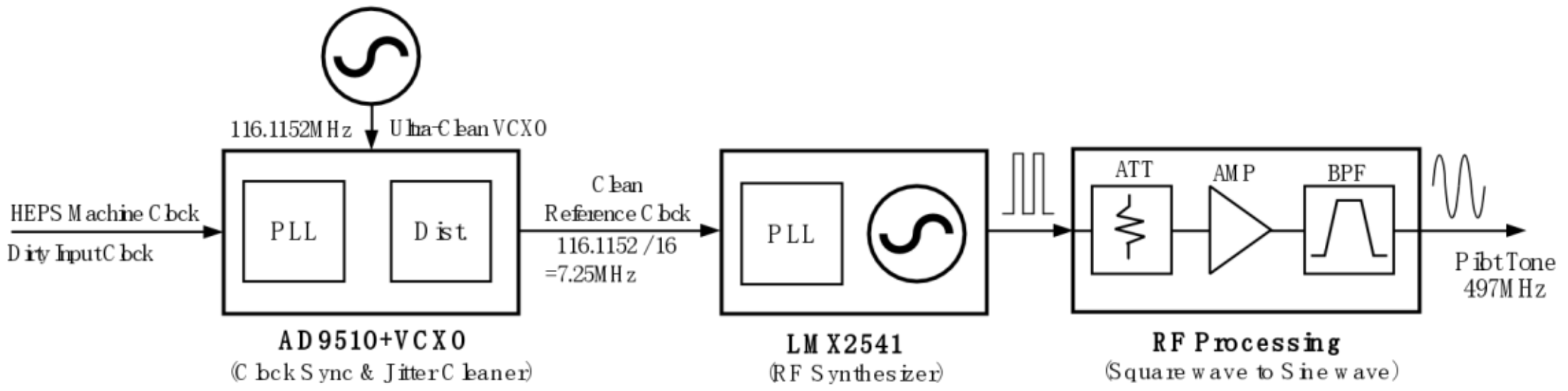

**Figure 4.3.7.13:** AFE Pilot tone synthesizer

Figure 4.3.7.13 depicts the AFE ADC Clock synthesizer and pilot-tone synthesizer. The ADC clock synthesizer is phase-locked to an external machine clock reference, which is buffered and sent to the DFE to provide a time reference for TbT processing.

The AD9510 generates four phase-matched ADC clocks that are also phase-locked to the machine clock. A phase noise test port is provided primarily for analysis during development, and there is a provision for an external clock that is also used during development. However, both the external clock and phase noise test port connectors are local to the board and are not brought out to the rear panel.

The pilot-tone synthesizer, on the other hand, generates a continuous wave (CW) signal that is also phase-locked to the machine clock.

*4.3.7.2.8 The Digital Front and DFE Hardware*

The digital section of the beam position monitor electronics is commonly known as the DFE, and it is responsible for performing all of the digital signal processing of the button signals, as well as communicating the results with the control system.

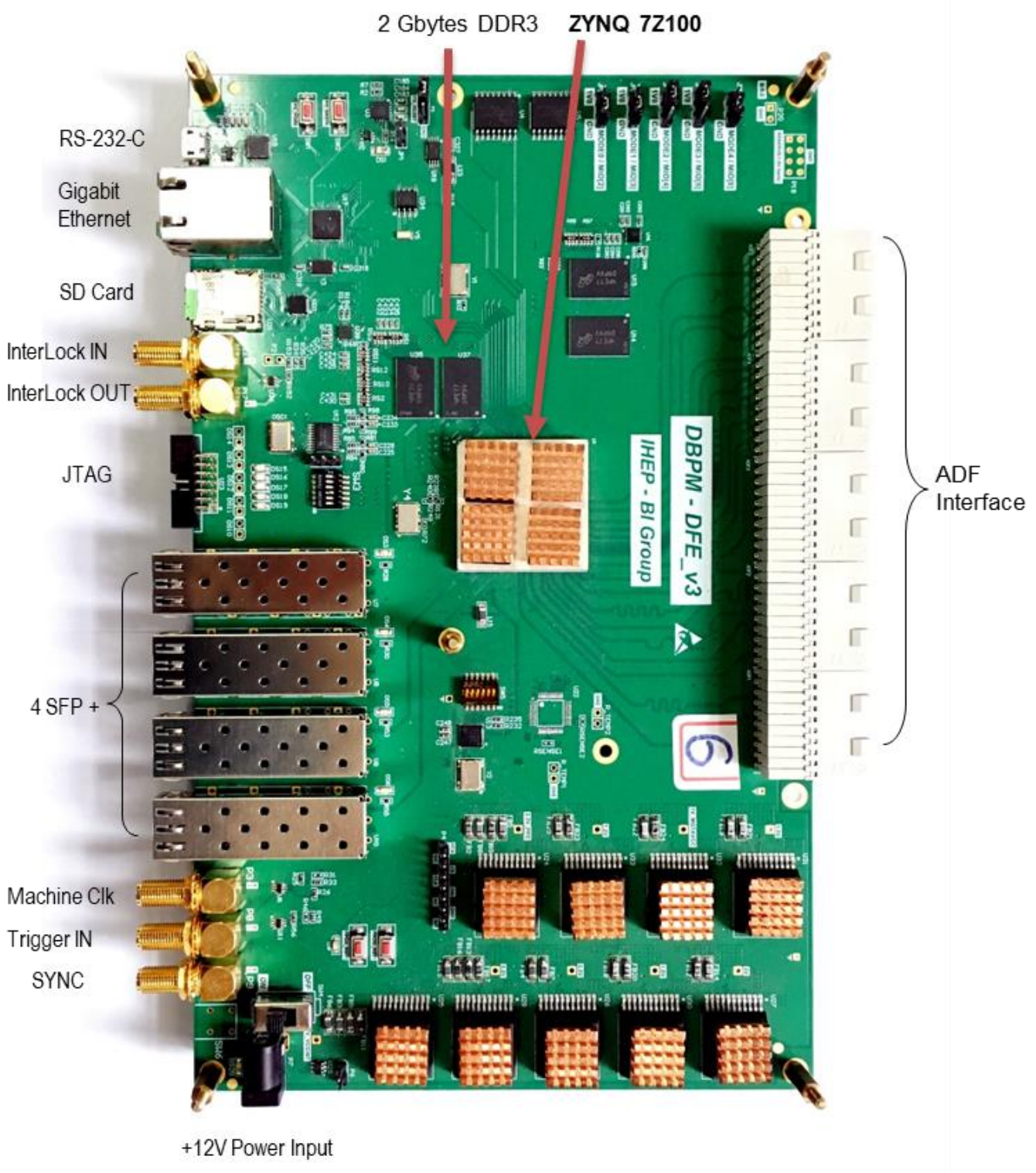

**Figure 4.3.7.14:** The picture of digital front end electronics PCB

The core component of the DFE is the Xilinx ZYNQ 7100 Field Programmable Gate Array (FPGA), which plays a crucial role in the digital signal processing chain, as will be discussed in the following section. The FPGA also includes a soft-core Debian processor that manages all on-board functionality and communication with the control system [7].

The board is equipped with a 256 MB SO-DIMM DDR3 memory module, which serves as both the program memory space for the PS and a sizable buffer area for raw ADC data and turn-by-turn data. This memory module can achieve a total throughput of 4 GB/sec, allowing it to store continuous bursts of raw ADC data and turn-by-turn data, as well as provide program and data memory for the PS. For slow controls, communication with the control system is facilitated through a gigabit Ethernet interface, and a TCPIP stack is implemented on the PS processor to ensure reliable communication.

This interface provides position data at a rate of 10 Hz, and also allows for on-demand requests and configuration settings. In addition, the DFE provides 6 SFP modules for other communication needs, supporting data rates of up to 6 Gbit/sec. For fast orbit feedback, low latency and high determinism are crucial. Therefore, the 10 kHz updated data bypasses the Microblaze and connects directly from the FPGA fabric to the high-speed SERDES section of the FPGA. This data is transmitted to dedicated cell-controllers, which can process the data and provide deterministic, low-latency updates to the fast magnet power supplies. For non-volatile memory, the DFE is equipped with a single 256 Mbit FLASH memory, which stores the FPGA bitstream as well as program data for the Microblaze.

Upon power-up, the FPGA configures itself using the program stored in the FLASH memory. Following this, when the Microblaze comes out of reset, a small boot-loader program is executed from internal memory. This program retrieves the final program from FLASH and transfers it to the DDR memory for execution. The interface between the DFE and AFE is established through five high-speed differential Tyco Hm-Zd connectors, which utilize LVDS signaling levels. Of these connectors, four are dedicated to the four 16-bit ADCs running at 119 MHz, while the fifth connector facilitates configuration and status signals exchange between the two boards.

#### *4.3.7.2.9 Data Flow*

The data flow path block diagram inside the FPGA is depicted in Figure 4.3.7.15, which illustrates the main blocks comprising the FPGA: the digital signal processing block, the MicroBlaze Processor with its related peripherals, and the cell controller interface.

The four ADC channels are inputted to the FPGA, which processes them in the digital signal processing (DSP) block, as described in the next section. At the output of the DSP block, four data streams are available to the user: Raw ADC data, Turn-by-Turn (TbT) data, Fast Average (FA) data, and Slow Average (SA) data. Upon trigger, up to 1 million points of Raw ADC data, TbT data, and FA data can be stored in the DDR3 Memory. The SA data can be continuously streamed to the user.

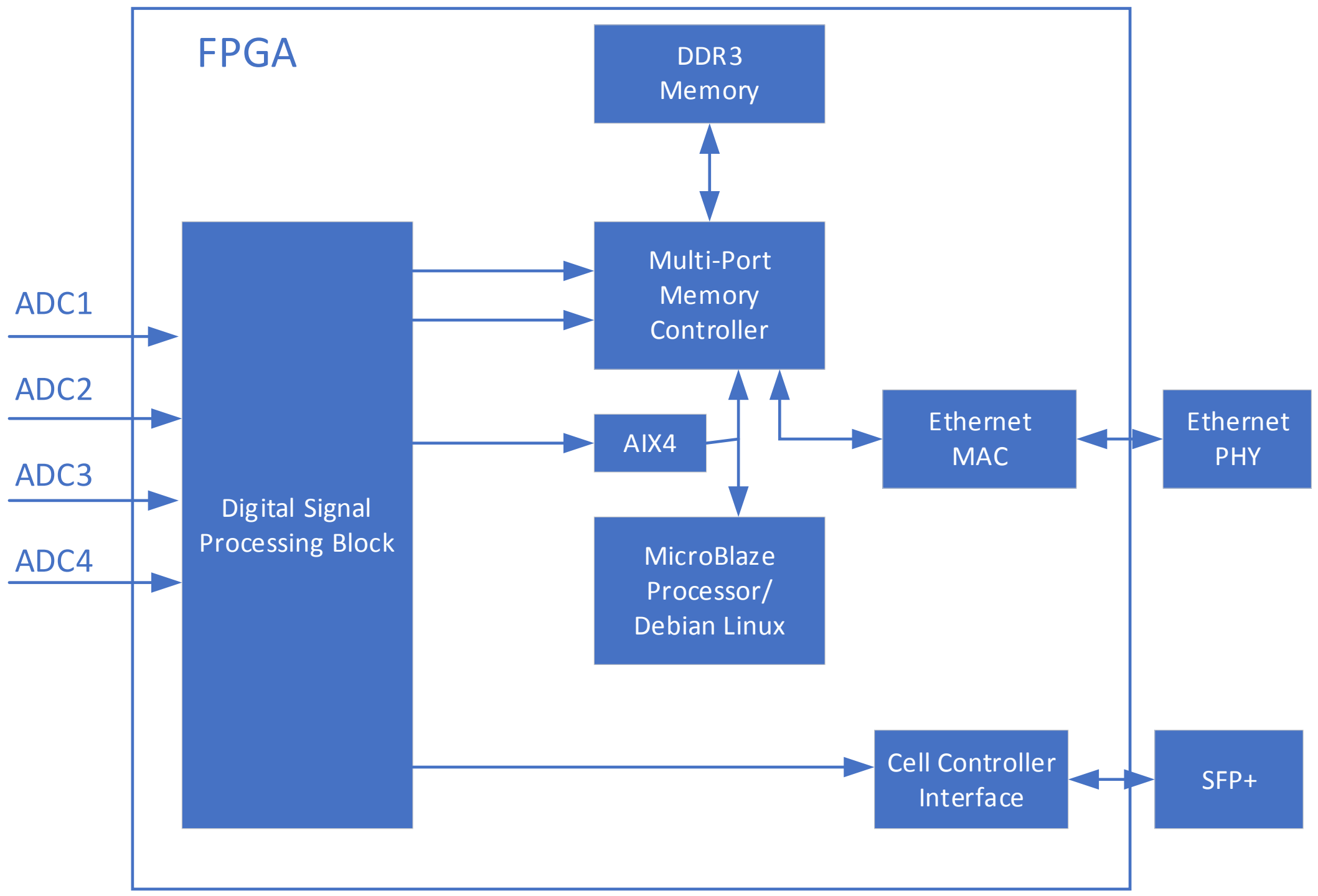

**Figure 4.3.7.15:** DFE High-Level Diagram

#### *4.3.7.2.10 FPGA Fixed-Point Digital Signal Processing*

The FPGA is the primary signal processing element where all digital signal processing occurs. The chosen FPGA is the ZYNQ 7100 which contains hard multipliers to facilitate complex filtering. To implement the signal processing, the MATLAB/Simulink environment and the Xilinx System Generator toolbox are utilized. This approach enables the modeling and simulation of the system at a higher level than HDL code.

The top-level block diagram of the signal processing chain is illustrated in Figure 4.3.7.16. The digital signal processing chain consists of four identical channels, each containing a digital down-converter followed by a programmable length averager that sums the magnitude outputs over a single turn. The position is then calculated, followed by additional filtering and decimation. The table above shows the data streams available to the user.

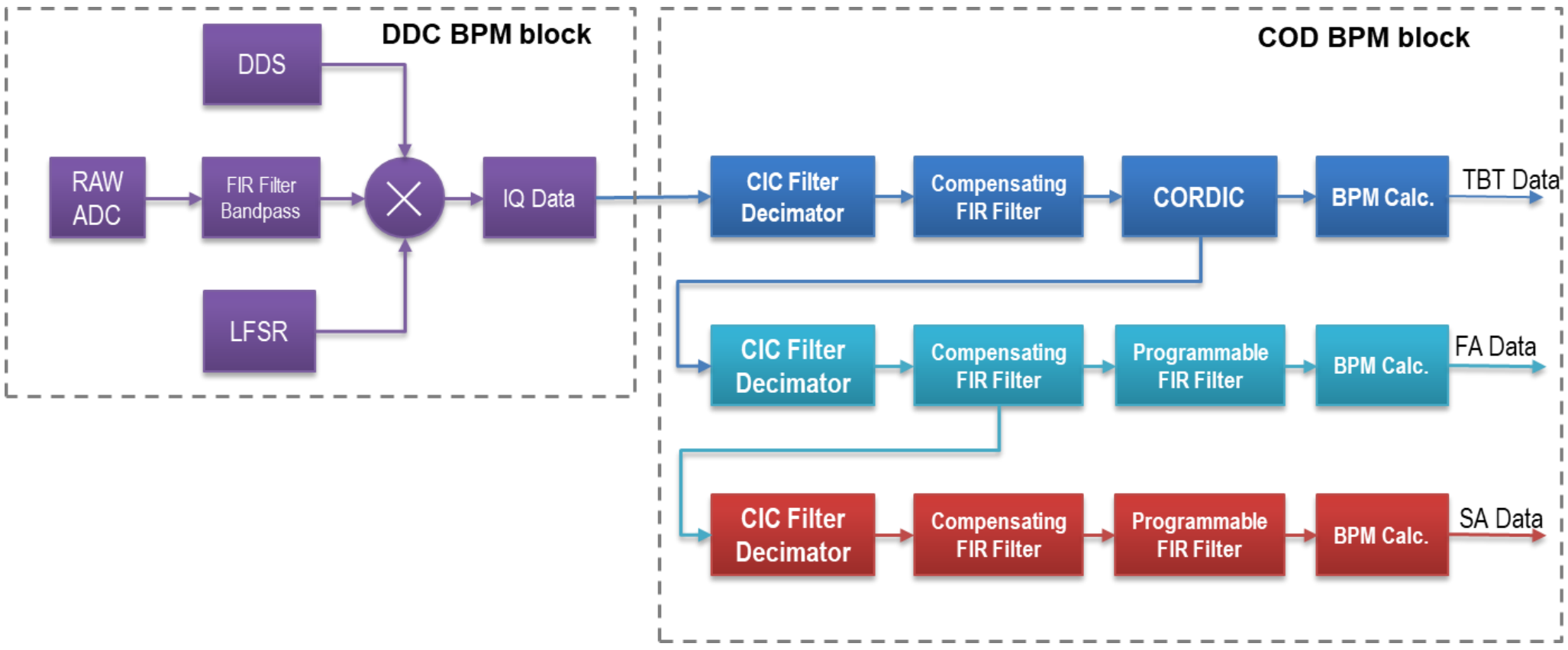

**Figure 4.3.7.16:** FPGA DSP Block Diagram

#### *4.3.7.2.11 BPM Data for Fast Orbit Feedback System*

Due to the 100 km circumference of the CEPC ring, implementing a global beam orbit feedback system would be time-consuming due to the extensive BPM data and lengthy control logic processing. This would significantly reduce the orbit feedback system's bandwidth and hinder high-frequency noise suppression. To address this, a local fast beam orbit feedback system is essential. For the fast orbit feedback control system near the Interaction Point (IP), a high feedback frequency is crucial, with the primary goal of enhancing CEPC's collider luminosity.

In CEPC, the local fast orbit feedback system is constructed using global orbit measurements. It utilizes local BPM data and a dedicated orbit feedback algorithm to achieve swift and localized orbit adjustments. This approach enhances the feedback frequency, effectively suppresses high-frequency orbit noise, and improves beam orbit stability. The orbit control at the IP in CEPC similarly relies on local, fast orbit feedback to enhance collider luminosity. This involves managing beam profiles, beam arrival times, and the positions of both positron and electron beams at the interaction point.

In the beam position monitoring system, 16 BPMs are used on both sides of the interaction point to gather local and fast acquisition (FA) data. In the fast orbit feedback control logic, the FA data from the local BPMs near the interaction point is combined with the FA data from other BPMs located globally to formulate the control logic design.

### 4.3.7.3 ***Beam Current Measurement***

In the Collider, two types of beam current measurement systems are utilized: average beam current measurement and bunch current measurement. The average current monitor employs Bergoz type DCCT (Direct-current current transformer) sensors. On the other hand, home-made bunch by bunch electronics are used for measuring the bunch current. The details of each subsystem are listed below.

#### *4.3.7.3.1 DC Beam Current Measurement*

Two DCCT sensors are installed in the CEPC collider to measure the average beam current, each serving as a spare one for the other for redundancy.

The In-air NPCT sensor from Bergoz Instrumentation is selected for its low linearity error and high resolution. Figure 4.3.7.17 shows the sensor and front-end electronics [8]. As the beam current at the Z mods is 803.5 mA, the sensor's dynamic range is up to 1.5 A. A vacuum chamber with a special structure needs to be designed and manufactured due to the NPCT sensor's principle.

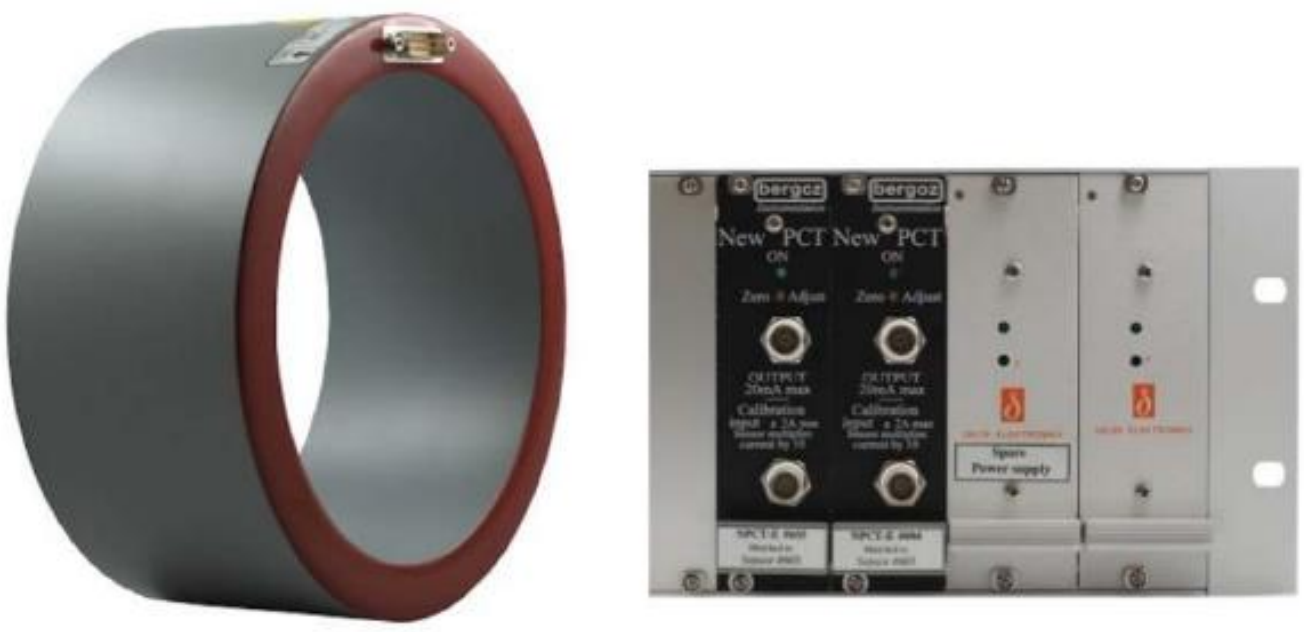

**Figure 4.3.7.17:** Bergoz NPCT senor and front-end electronics

The ceramic gap functions as a capacitor to cut off wall current. The capacitor determines the cut-off frequency, restricting high-frequency signals through the sensor. Loss factor and longitudinal impedance are simulated by CST Particle Studio for different gap sizes.

The vacuum structure is shown in Figure 4.3.7.18 (left), and impedance results for different gap sizes are presented in Figure 4.3.7.18 (right). The simulation reveals that a 2 mm gap results in low impedance at around 0.3 GHz, while a 10 mm gap produces high impedance at about 1 GHz. In this case, a 2 mm gap is chosen for the structure.

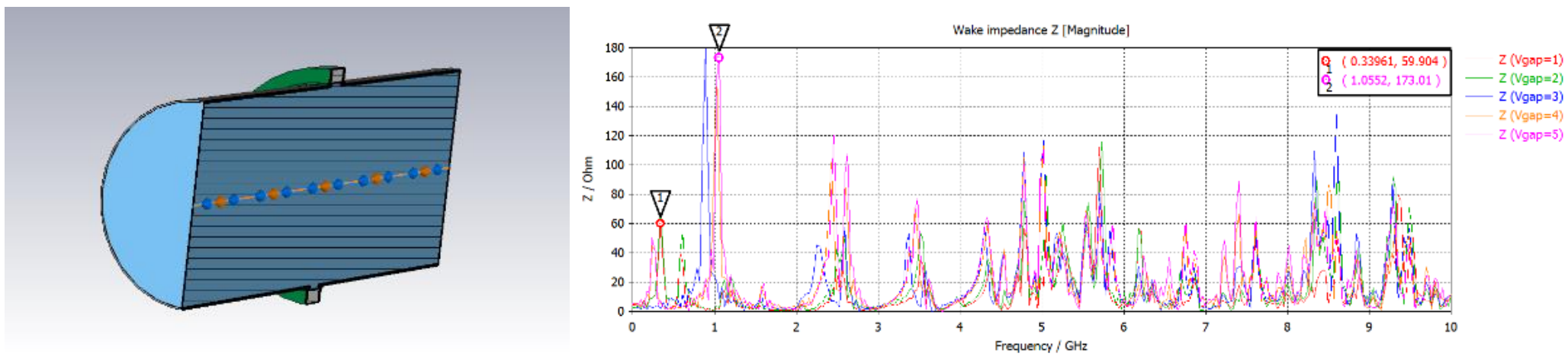

**Figure 4.3.7.18:** Left: Vacuum structure; right: longitudinal wake impedance.

An additional cooling structure is necessary for the gap, as it functions similarly to a small cavity, causing charged particles passing through to lose energy and generate heating on the gap [9].

To achieve efficient cooling, it is necessary to assess the amount of heating power.

To prevent disturbances from external electromagnetic fields, as well as to conduct wall current, a shielding shell for the sensor is required. The use of multiple layers of different materials has proven effective in low current mode [10].

The Keithley DMM7500 7½-Digit multimeter has been selected for A/D conversion in order to achieve high precision measurements. It offers a resolution of 10 nV for DC voltage, with noise and one-year stability that can reach 14 ppm.

#### *4.3.7.3.2 Bunch Current Measurement*

The beam current measurement system, which operates on a bunch-by-bunch basis, is capable of measuring the current of each individual bunch. Subsequently, the central control room can monitor the current value of each bunch in its respective bucket. The detailed visualization of each bunch provides an automated tool to ensure equalization of bunch filling patterns within the main rings, thus offering valuable information for daily operations and facilitating the implementation of top-up injection. This system will be developed in-house and integrated with an FPGA module to achieve more advanced features.

Figure 4.3.7.19 illustrates that the BCM system is composed of three components: the analog frontend circuit (AFE), the digital frontend circuit (DFE), and the software.

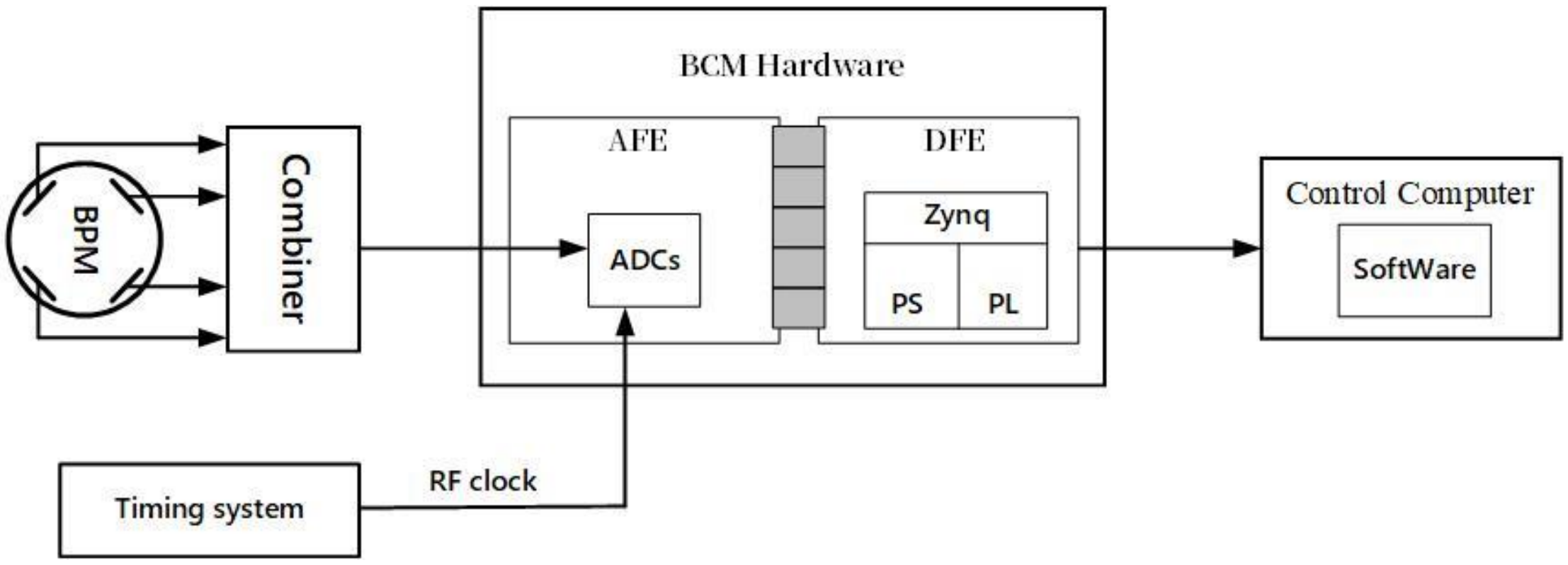

**Figure 4.3.7.19:** Schematic of the bunch current measurement system

The heart of the AFE is its high-speed Analog-to-Digital Converters (ADCs). The front-end circuit receives the sum signal from four button electrodes of a beam position monitor installed in the store ring. The signal is sampled at a frequency of 650MHz, which is the CEPC RF frequency, to extract the peak value of each bunch. Subsequently, the digitized data is instantly transferred to the DFE.

At the heart of the DFE lies a Zynq All-Programmable System-on-Chip (SoC). The defining characteristic of Zynq is that it is comprised of two primary components: a Processing System (PS) built around a dual-core ARM Cortex-A9 processor, and Programmable Logic (PL) that is equivalent to a traditional Field Programmable Gate Array (FPGA).

The digitized data from the AFE is processed by the Zynq PL. To improve the stability of the digitized bunch current data, we digitize multiple turns (specifically, 256 turns) of the bunch current data. The multi-turn data is then subjected to an averaging process, which is written in Verilog Hardware Description Language, to obtain a relatively accurate peak value.

The peak value of each bunch extracted from the sum signal is known as the relative beam current, which is proportional to the real beam current. To obtain the absolute beam current, the relative beam current must be calibrated using interrelated parameters. There are two common methods used for calibrating the relative beam current. The first method, known as the linear mode, is based on the linear relationship between the bunch current and the output signal intensity. The second method, known as normalization based on mathematical principles, calculates the proportion of each bunch current in all the bunches and calibrates it using the DC Current Transformer (DCCT). The individual bunch value is computed as follows:

$$I_i = \frac{A_i}{\sum_{k=0}^{n} A_k} I_{dcct},\ i=0,1,2,3\ldots\ldots n \qquad (4.3.7.11)$$

in which $Ai$ is the peak value of each bunch, and $I_{dcct}$ is the current value of the DCCT obtained from the database of the EPCIS host via LAN, $n$ is the bucket number.

The absolute beam current is then transferred to the control computer for further data processing via the software. Operators in the central control room can remotely access all the bunch current information in each bucket. All data can be stored in the local host computer for offline analysis.

#### 4.3.7.4 *Synchrotron Light Based Measurement*

Two Synchrotron Light Diagnostic Beamlines (SLDB) have been designed for the electron and positron rings, respectively. As shown in Figure 4.3.7.20, SLDB-E is located in the northwest section of the storage ring while SLDB-P is located in the northeast section. The synchrotron light source point is located at the front of the last dipole of the arc, where the downstream is the straight section region. This location allows for easy extraction of light without interference from magnets. Each beamline has two ports: one for x-ray extraction and the other for visible light extraction. The x-ray port is used to measure the transverse beam sizes and calculate the beam emittance of the ring, while the visible light port is used for measuring the bunch length.

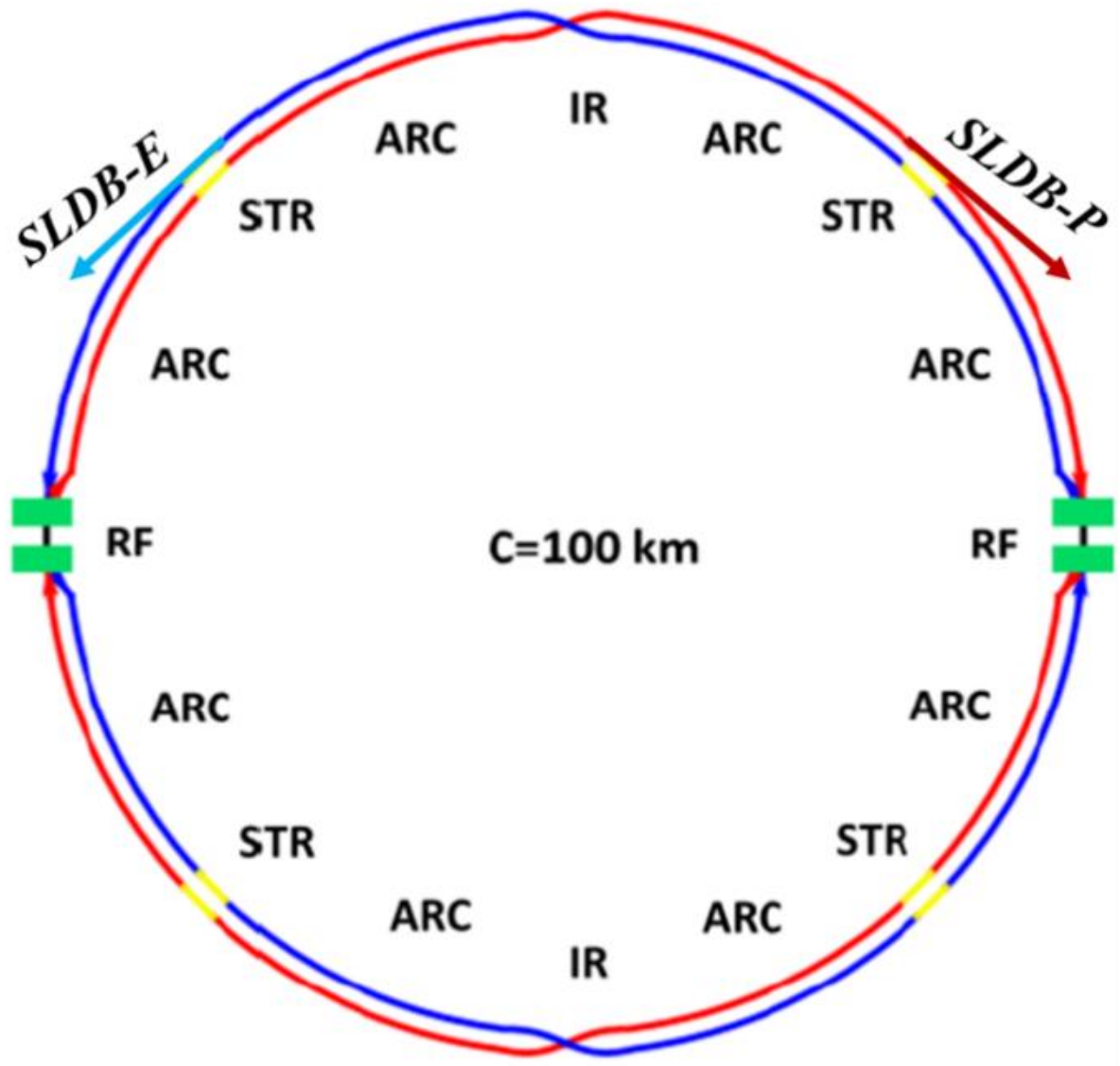

**Figure 4.3.7.20:** Schematic of locations of the synchrotron light diagnostic beamlines

### *4.3.7.4.1 Source Point, Spectrum and Angular Distribution*

The synchrotron light source point is located at the front of the last dipole of the arc. The parameters of the source point are listed in Table 4.3.7.7.

**Table 4.3.7.7:** Parameters of the source point

| Parametr | Unit | Dipole type $B_0$ |
|---|---|---|
| Dipole length | m | 5.737 × 5 |
| Strength of dipole | T | 0.0373 |
| Bending angle | mrad | 2.844 |
| Opening angle | µrad | 4.258 |
| $\beta x$ | m | 100 |
| $\beta y$ | m | 25 |
| Maximum dispersion | m | 0.2 |
| Horizontal beam size | µm | 424 |
| Vertical beam size | µm | 5.7 |

Synchrotron radiation from the dipole produces a wide range spectrum. Figure 4.3.7.21 shows the photon flux from the dipole at 120 GeV beam energy with 17.4 mA current, where the critical energy is 357.2 keV. As shown in Figure 4.3.7.22, the UV and visible light of 200 nm – 700 nm has a vertical opening angle of 0.3 - 0.45 mrad (80% peak flux), while X-rays at 14.2 keV has a vertical opening angle of 21.5 µrad (80% peak flux). Horizontally, the dipole has a bending angle of 2.844 mrad with a total length of 28.685 m, and visible light and X-rays can be extracted separately in the horizontal direction. Therefore, the source point for X-rays is designed at the upstream of the dipole and 2 m to its edge, with a 25 µrad horizontal angle. The visible light port extracts a 0.6 mrad horizontal angle, and the source point corresponds to the dipole position of 4 – 10.54 m to its upstream edge.

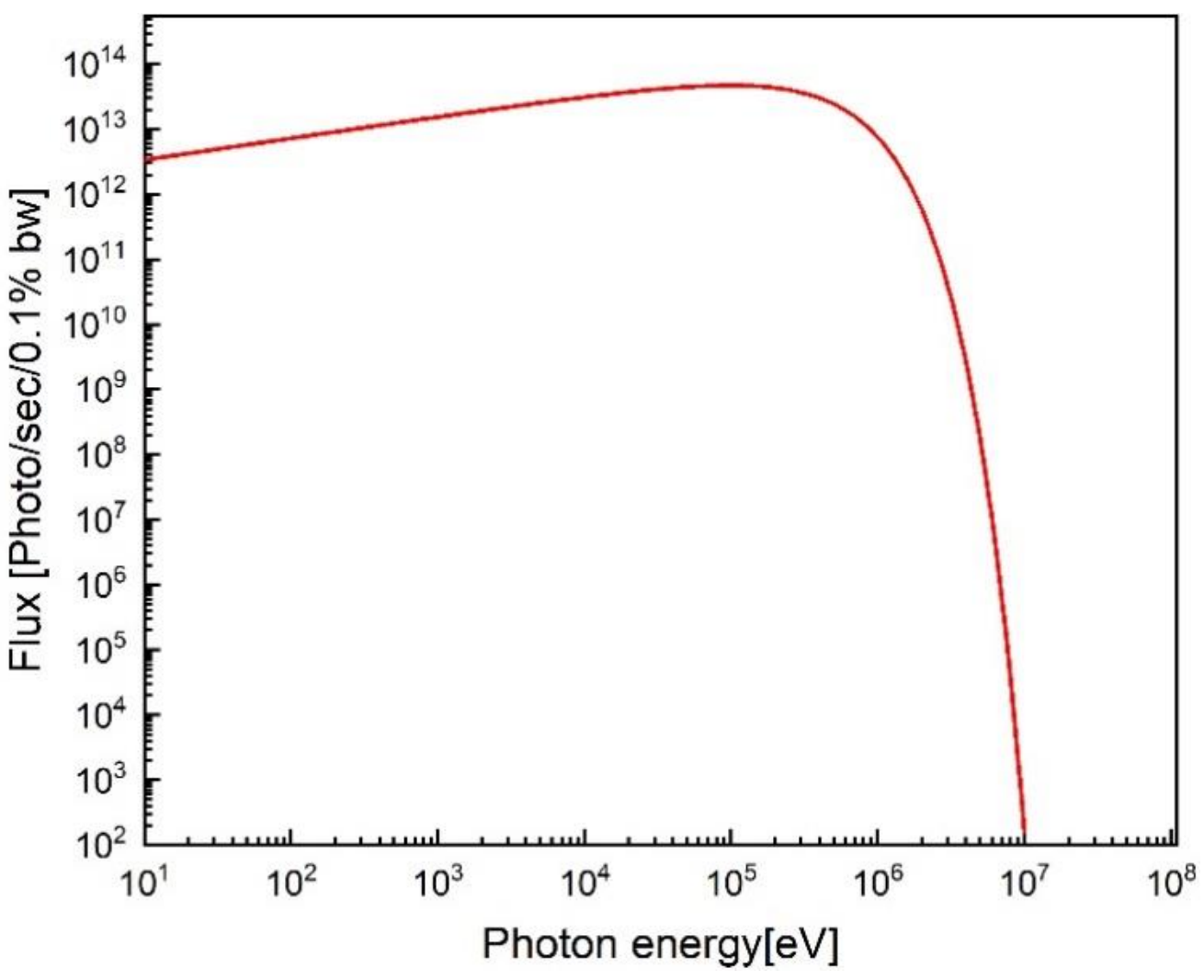

**Figure 4.3.7.21:** Photon flux from dipole at 120GeV beam energy.

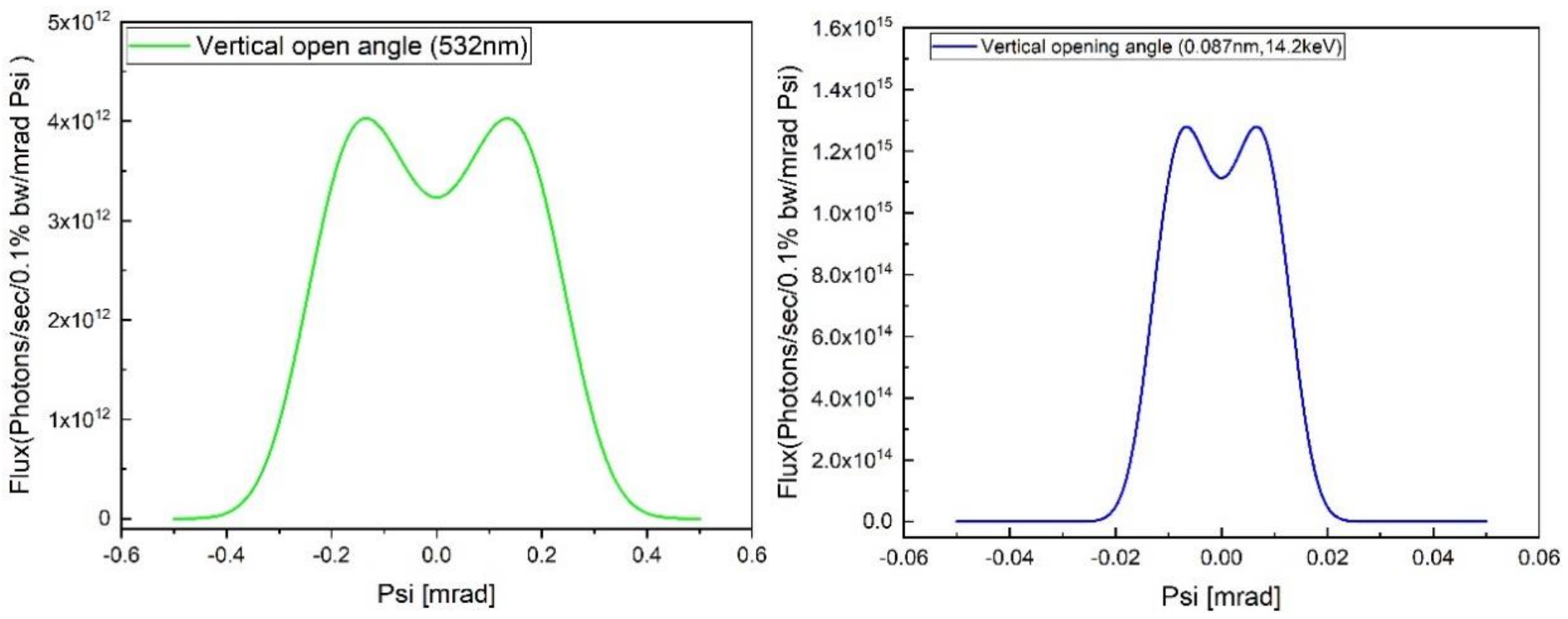

**Figure 4.3.7.22:** Vertical opening angle at 532 nm (left) and 0.087 nm (right)

#### *4.3.7.4.2 X-ray Extraction*

For a storage ring with high beam energy and a long bending radius, it can be challenging to separate SR light from the beam over a long distance. This makes it difficult to place diagnostic elements close enough to the source point. In such cases, traditional beam size measurement methods may not provide sufficient resolution and cannot be applied. To overcome this challenge, the X-ray double slit interferometer [11] is considered the best option for beam size measurement. However, to use the interferometer, monochromatic light is necessary, which requires a crystal monochromator. Unfortunately, the bandwidth of traditional crystal monochromators is limited to $10^{-4}$ – $10^{-5}$, which reduces the light intensity to the detector. Therefore, a Krypton gas filter is chosen as it has a K-edge around 10 keV. The double slit is positioned 100 m away from the source point, and a Rh coating mirror is placed after the double slit to be used as a low-pass filter. The mirror has a grazing incidence angle of 4.5 mrad and reflects only X-rays with energies below 15 keV. This ensures that high energy X-rays cannot reach the X-ray camera, which is located 100 m away from the double slit. Figure 4.3.7.23 shows the spectrum obtained after passing through a 200 µm thick Be window, a Kr filter, and the Rh coated mirror. The peak energy of the spectrum is 14.2 keV.

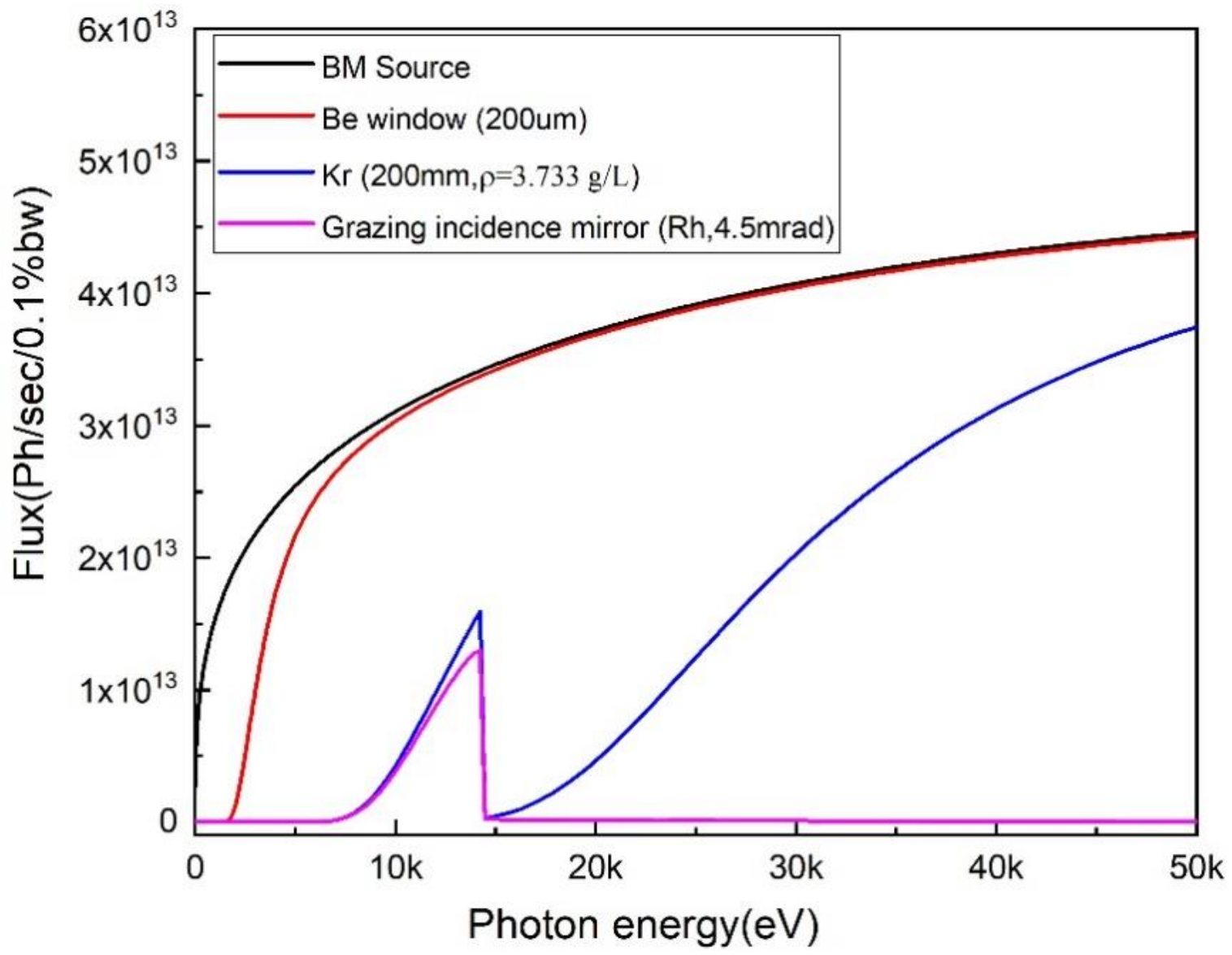

**Figure 4.3.7.23:** X-ray spectrum after a 200 µm Be window, a Kr filter and the mirror.

*4.3.7.4.3 Visible Light Extraction*

The visible light is utilized for measuring the bunch length, whereby a copper extraction mirror is installed at a 45-degree angle to the visible light inside a vacuum chamber. This mirror reflects the light outside the vacuum chamber through a quartz window. The dipole generates an SR power of about 4.7 kW/mrad, with most of it concentrated within the central 20 µrad vertical angle. To safeguard the extraction mirror from the incidence of SR power, only the upper half of the visible light is reflected. The mirror is positioned 98m away from the dipole source and measures 70×35×10 $mm^3$. It extracts 0.6 mrad visible light horizontally and 0.3 mrad vertically.

The CEPC's bunch length ranges from 2.3 to 4.1mm (equivalent to 7.7 – 13.7 ps). To measure this bunch length, a commercial streak camera with a time resolution of 1ps will be utilized.

*4.3.7.4.4 Beam Size Measurement*

Mitsuhashi [12,13] first described the use of a double-slit interferometer to measure small transverse beam sizes. Following initial investigations and calculations, it has been determined that an X-ray double-slit interferometer will be utilized in the X-ray diagnostic beamline at CEPC.

The principle of double-slit interference is rooted in the Van Citterut-Zernike theory. The spatial coherence of a light source with finite size is expressed as the Fourier transform of the light source size as follows:

$$\gamma(\nu) = \int f(x) exp(-i2\pi\nu x) dx \qquad (4.3.7.12)$$

where $\gamma(\nu)$ is the spatial coherence, $f(x)$ is the normalized source distribution, and $\nu$ is the spatial frequency. The intensity of the interference pattern measured in the detector plane can be expressed as:

$$(y_0) = I_0 \left[sinc\left(\frac{2\pi a}{\lambda D} y\right)\right]^2 \left[1 + |\gamma| cos\left(\frac{2\pi d}{\lambda D} y + \varphi\right)\right] \qquad (4.3.7.13)$$

where $a$ is half of the single slit height, $d$ is the distance between the two slits, $D$ is the distance from the double slit to the detector, $\lambda$ is the wavelength of observation, $\phi$ denotes the fringe phase, and $I_0$ is the sum of incoherent intensities from the two slits. The spatial frequency $v$ is expressed as $v = d/\lambda R$, where $R$ is the distance from the source to the double slit.

To calculate the spatial coherence $\gamma$, we fit the intensity distribution of interference fringes using Eq. (4.3.7.13). By taking an inverse Fourier transform of the measured spatial coherence for different double slit spacings $d$ (i.e., different spatial frequencies), we obtain the cross-sectional distribution of the original source. The beam in the storage ring can be approximated as a Gaussian distribution, and to simplify the complex calculation process, we substitute a Gaussian function into the coherence calculation Eq. (4.3.7.12). This allows us to express the relationship between beam size $\sigma$ and spatial coherence $\gamma$ as follows:

$$\sigma = \frac{\lambda R}{\pi d}\sqrt{-\frac{ln\gamma}{2}} \qquad (4.3.7.14)$$

Typically, the beam size is determined by using the spatial coherence measured with a fixed slit spacing, which is calculated using Eq. (4.3.7.14).

Figure 4.3.7.24 illustrates the setup of a double-slit interferometer for measuring beam size at CEPC. The radiation emitted from the dipole first passes through a K-edge filter containing Kr. It then goes through a Be-window, a double slit, and a mirror before forming an interferogram on the imaging screen.

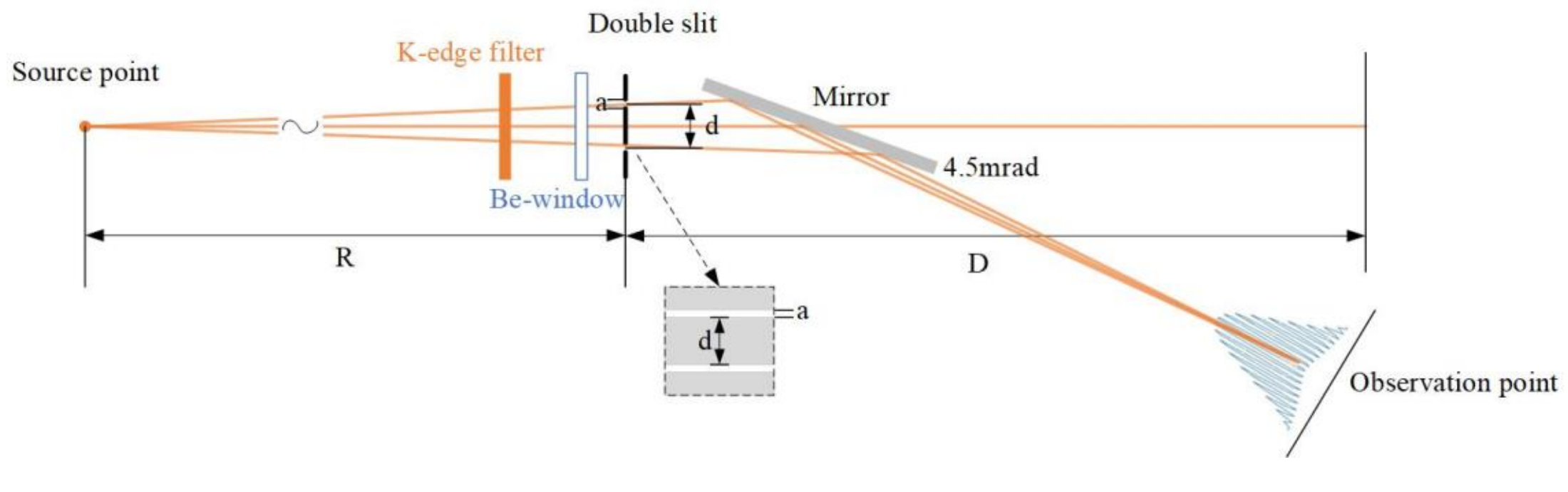

**Figure 4.3.7.24:** A configuration for double slit X-ray interferometer

The expected vertical beam size from the dipole at CEPC is 5.7 µm, corresponding to an apparent angular size of 4.25 µrad. To simulate these beam parameters, we set $R$ to 100 m and use a 14.2 keV X-ray extracted by the K-edge filter containing Kr. Figure 4.3.7.25 shows the spatial coherence as a function of beam size for different double-slit separations. Based on this simulation, a beam size of 5.7 µm will give a spatial coherence of 0.91 with $d$ = 100 µm, 0.72 with $d$ = 200 µm, and 0.47 with $d$ = 300 µm. Although we can measure the light intensity with high precision using an X-ray camera, we choose a double-slit separation of $d$ = 300 µm to account for the divergence angle.

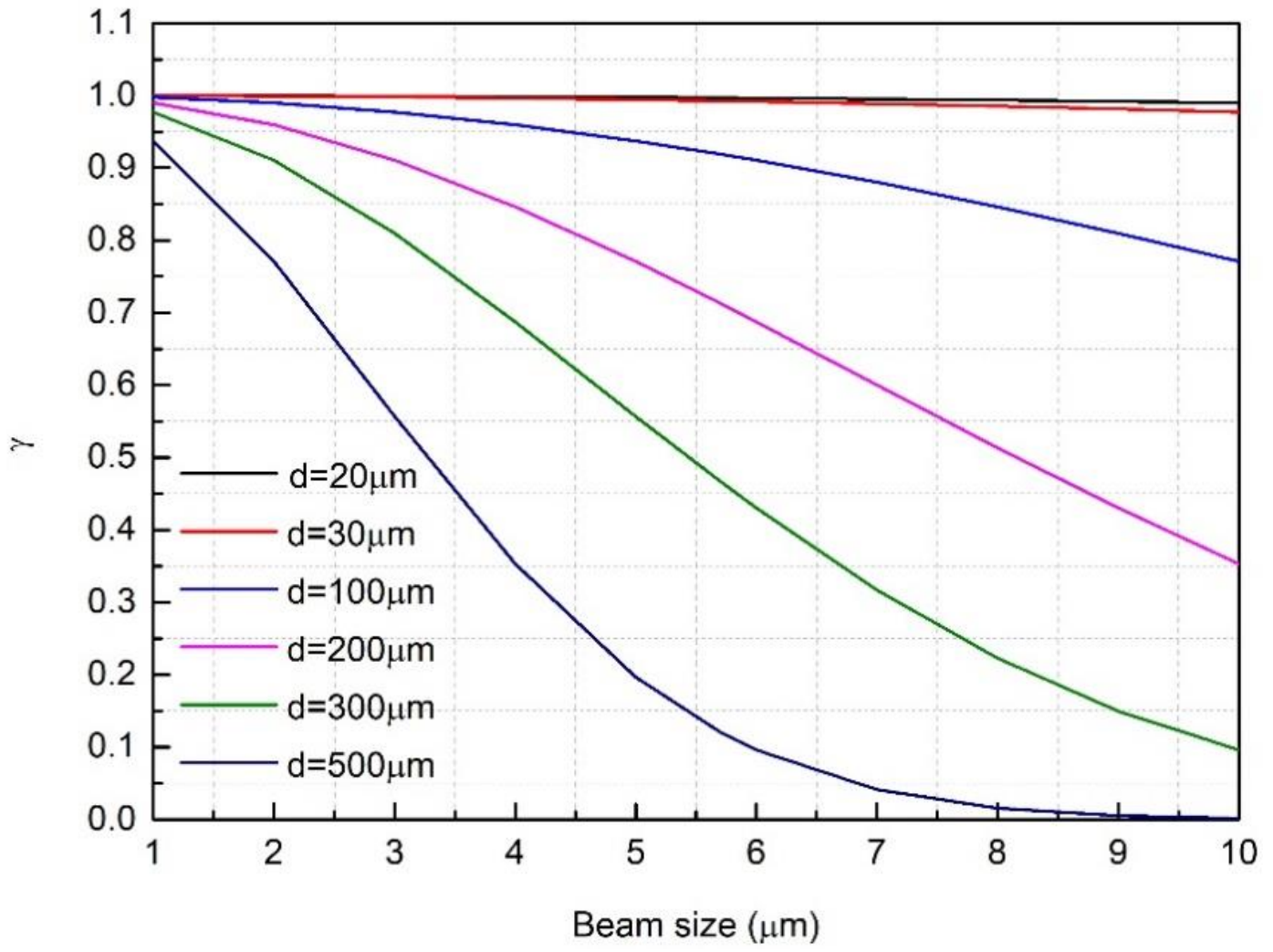

**Figure 4.3.7.25:** The spatial coherence for different double slit separation as a function of beam size.

We simulated an interferogram of synchrotron radiation passing through a double-slit interferometer for the measurement of vertical size. The cross-section of the synchrotron radiation light is approximately Gaussian. We set the horizontal and vertical size of the Gaussian source as 424 µm and 5.7 µm, respectively, with *R* set to 100m, *d* set to 300 µm, and *a* set to 8 µm. To achieve good resolution, we set the distance between the double slit and observation point D to 100m. Figure 4.3.7.26 (left) shows a typical interferogram obtained from the simulation, while Figure 4.3.7.26 (right) shows the intensity distribution curve in the vertical direction. The fringes' contrast is evident, which allows for accurate measurement of the beam size.

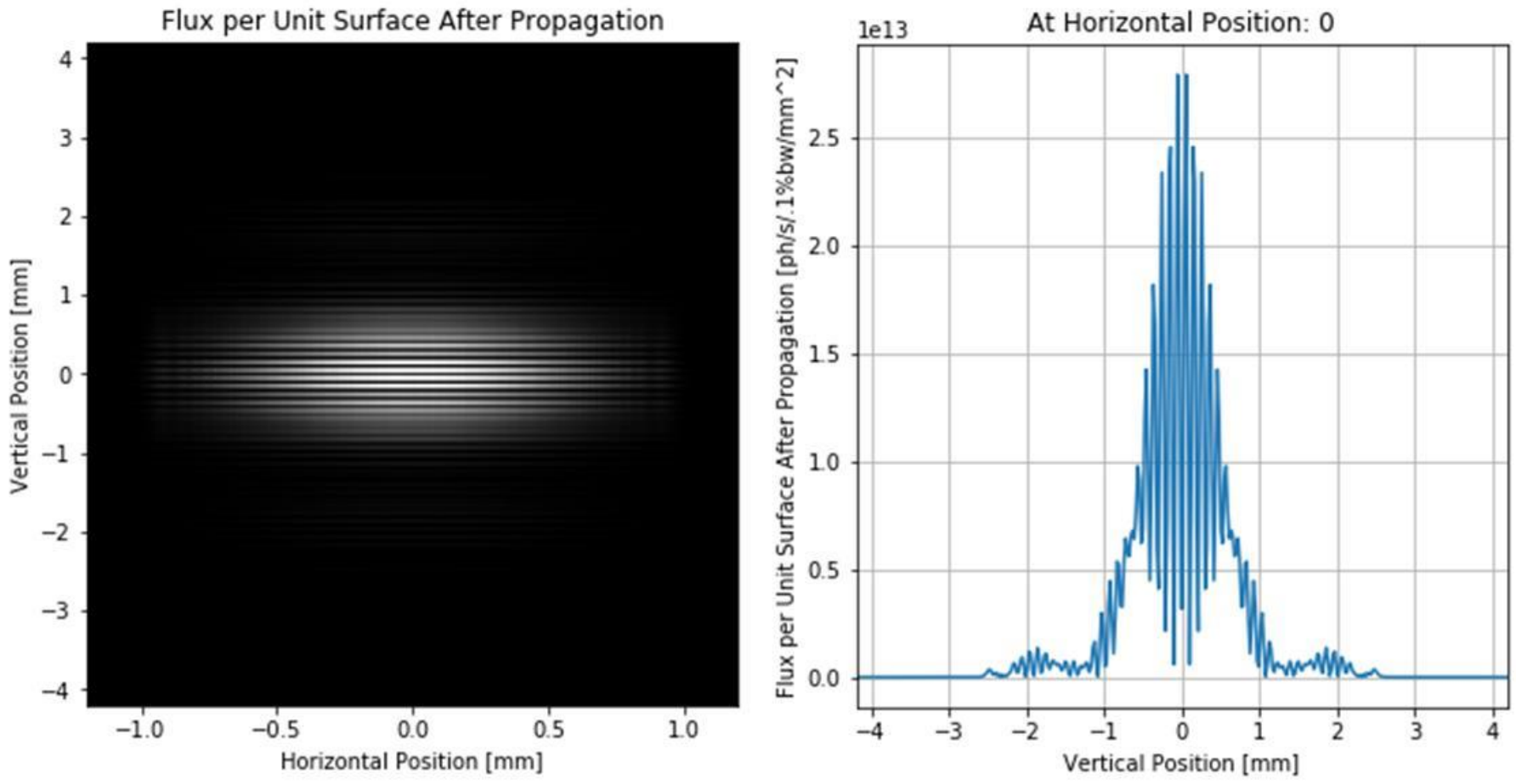

**Figure 4.3.7.26:** A typical interferogram (left) and the intensity distribution curve in vertical direction (right).

### 4.3.7.5 *Beam Loss Monitor System*

A dedicated beam loss monitor (BLM) system is essential for high-energy and high-intensity accelerators to prevent beam loss. As such, BLMs are indispensable for CEPC.

#### *4.3.7.5.1 Selection of the Beam Loss Monitors*

Four options are currently under consideration for the BLM system at CEPC: ionization chambers, Cherenkov counters, PIN-photodiodes, and scintillators + photomultiplier tubes (PMT). To select the appropriate BLM type for a specific application, we focus on several crucial factors, including intrinsic sensitivity, dynamic range, radiation hardness, response time, and sensitivity to synchrotron radiation (SR). Table 4.3.7.8 summarizes the performance parameters of these four BLM options.

**Table 4.3.7.8:** Performance parameters of four BLM types

| Type of BLM | Dynamic range | Response time | Sensitivity (for MIPs) | Radiation resistance | Sensitivity to SR |
|---|---|---|---|---|---|
| Ionization chamber | $10^8$ | 89 μs | 600 ($Elec_{gain}$) (1L) | >100 Mrad | Sensitive |
| PIN-photodiode | $10^8$ | 5 ns [14] | 100 ($Elec_{gain}$) ($1cm^2$) | >100 Mrad | Insensitive |
| Cherenkov counters | $10^5$~$10^6$ | 10 ns | 270 ($PMT_{gain}$)(1L) | 100 Mrad | Insensitive |
| Scintillators+PMT | $10^6$ | 20 ns | ≈$18\cdot10^3$ ($PMT_{Gain}$) | ≈20 Mrad | Sensitive |

Ionization chambers have a considerable dynamic range and are highly radiation hard. However, they have the drawback of being relatively slow and sensitive to synchrotron radiation [15], making them unsuitable for applications in electronic circular accelerators like CEPC.

The scintillators+PMT option is extremely fast, allowing for bunch-to-bunch measurements to be achieved. However, the scintillator can darken when used in high dose level environments, which is a major drawback. Additionally, the gain of PMTs can vary by a factor of 10, necessitating careful intercalibration of their sensitivities. Finally, this detector is costly and also sensitive to synchrotron radiation.

Cherenkov-based fibers are more radiation-hard than scintillators but less sensitive to beam losses. However, with the additional gain of a PMT, their sensitivity exceeds that of an ionization chamber. Cerenkov light is emitted when a charged particle's velocity βc is greater than the light velocity c/n in a medium with an index of refraction n > 1. Cerenkov light is instantaneous, unlike scintillators, and the threshold for light output is several hundred keV, making Cerenkov detectors insensitive to background radiation from synchrotron radiation. For instance, electrons below about 150 keV will not produce any light, while 1 GeV protons or 0.5 MeV electrons produce approximately 169 photons/cm [16]. However, the energy of synchrotron radiation photons strongly depends on the beam energy. In CEPC, the critical energy Ec of synchrotron radiation photons can be calculated as follows: (beam energy E = 120 GeV, bending radius = 6094 m).

$$E_c(keV) = 2.2\frac{E(GeV)^3}{\rho(m)} \approx 628 \text{ keV} \qquad (4.3.7.15)$$

The synchrotron radiation photons produced in CEPC have energies that depend on the beam energy and the bending radius. For example, at a beam energy of 120 GeV and a bending radius of 6094 m, the critical energy $E_c$ of synchrotron radiation photons is calculated to be approximately 628 keV. These photons can undergo photoelectric effect in Cherenkov fibers and generate high energy electrons, which in turn can produce more than 169 photons/cm. Thus, Cherenkov fibers are sensitive to synchrotron radiation and do not offer significant shieldability against it in CEPC.

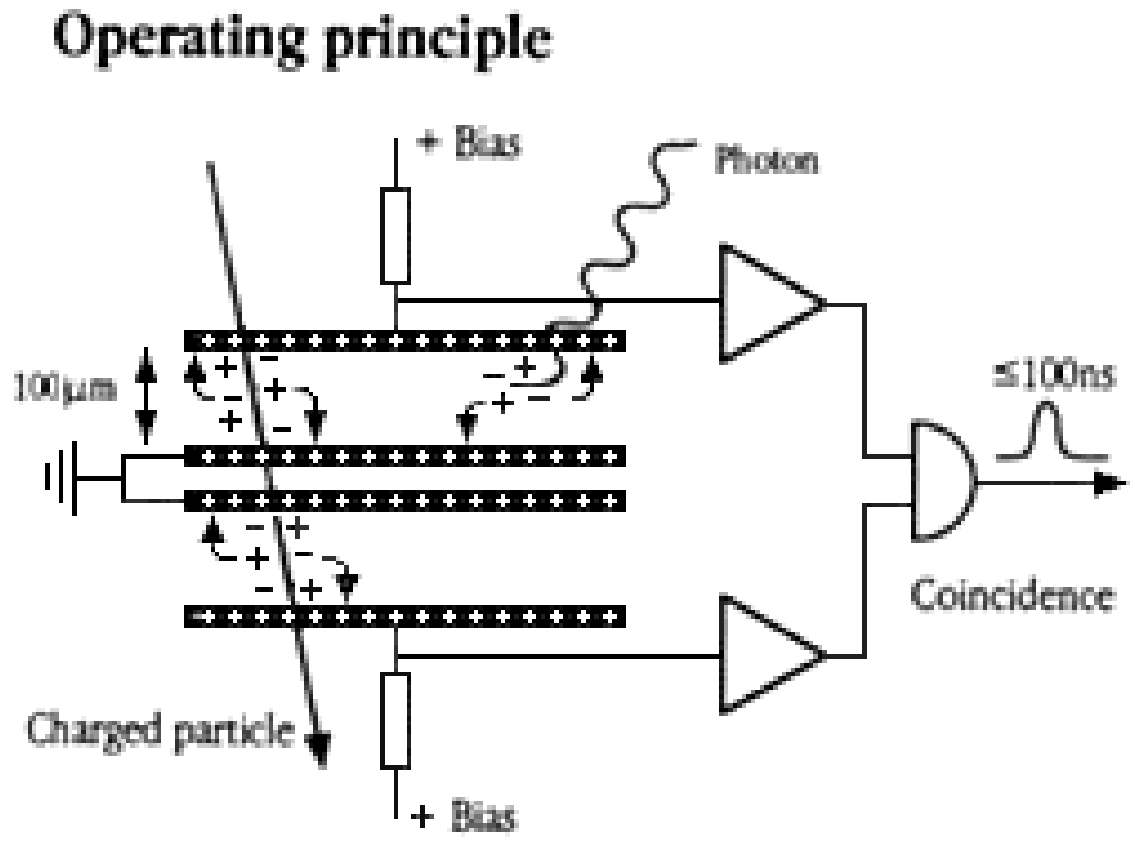

**Figure 4.3.7.27:** principle of PIN-photodiode

The PIN-photodiode is a fast and relatively inexpensive detector with good radiation resistance. It has a large dynamic range and high sensitivity, but it is only available in small sizes. The detector consists of two face-to-face PIN-photodiodes (see Fig 4.3.7.27). High-energy charged particles produce signals in both diodes when passing through, while photons interact in only one diode. Although the PIN-photodiode is relatively insensitive to synchrotron radiation background, synchrotron radiation photons at electron beam energies greater than 45 GeV mainly undergo a photoeffect or a Compton effect, producing emitted electrons that may reach the second diode and result in coincident signals [17]. However, this problem has been elegantly resolved in both HERA and LEP [18,19]. A thin layer of copper or lead between the two diodes (Fig 4.3.7.28) can reduce the probability of the emitted electron reaching the second diode and thus reduce the background counts due to synchrotron radiation. In LEP (electron beam energy E = 80GeV), the copper layer further reduced the background rate by a factor of 10 [18]. The optimal thickness of the layer can be calculated from the range of electrons in matter, where the penetration depth R in which 90-95% of incident electrons are stopped is given by (for Al) [19]:

$$\mathrm{R(Al)} = \mathrm{A}\cdot\mathrm{E}\cdot[1-\frac{B}{(1+C\cdot E)}]mg/cm^{-2} \qquad (16)$$

where A = $0.55\cdot10^{-3}$ g-cm$^{-2}$keV$^{-1}$, B = 0.984, C = $3\cdot10^{-3}$ keV$^{-1}$, E = energy of the electron.

For energies above 100 keV and for materials with higher atomic number (e.g., copper), the range R is approximately 0.6 to 1 times that of aluminum (R(Al)).

In CEPC, synchrotron radiation photons can cause electrons to be emitted through photoeffect, with an energy of 628 keV. The range of these emitted electrons in copper, with a density of ρ(Cu) = 8.96 g/cm3, is approximately R(Cu) ≈ 0.26 mm.

Therefore, a thin layer of approximately 300 μm copper placed between the diodes is sufficient to stop most of the Compton- and photoelectrons caused by synchrotron radiation. This layer will not affect the minimum ionizing particles (MIPs) produced by beam losses, so the detection efficiency of the BLM for beam losses remains unchanged.

A thin layer of a high Z material like lead or copper between the two photodiodes inside the BLM leads to a decrease in background counts due to SR. However, a fraction of the coincidence rate at high dose rates comes from multiple photons that interact simultaneously in the two diodes. To further reduce these coincidence rates, additional lead shielding will be needed around the BLMs [18,19]. The optimal thickness of the lead shield should be determined through testing.

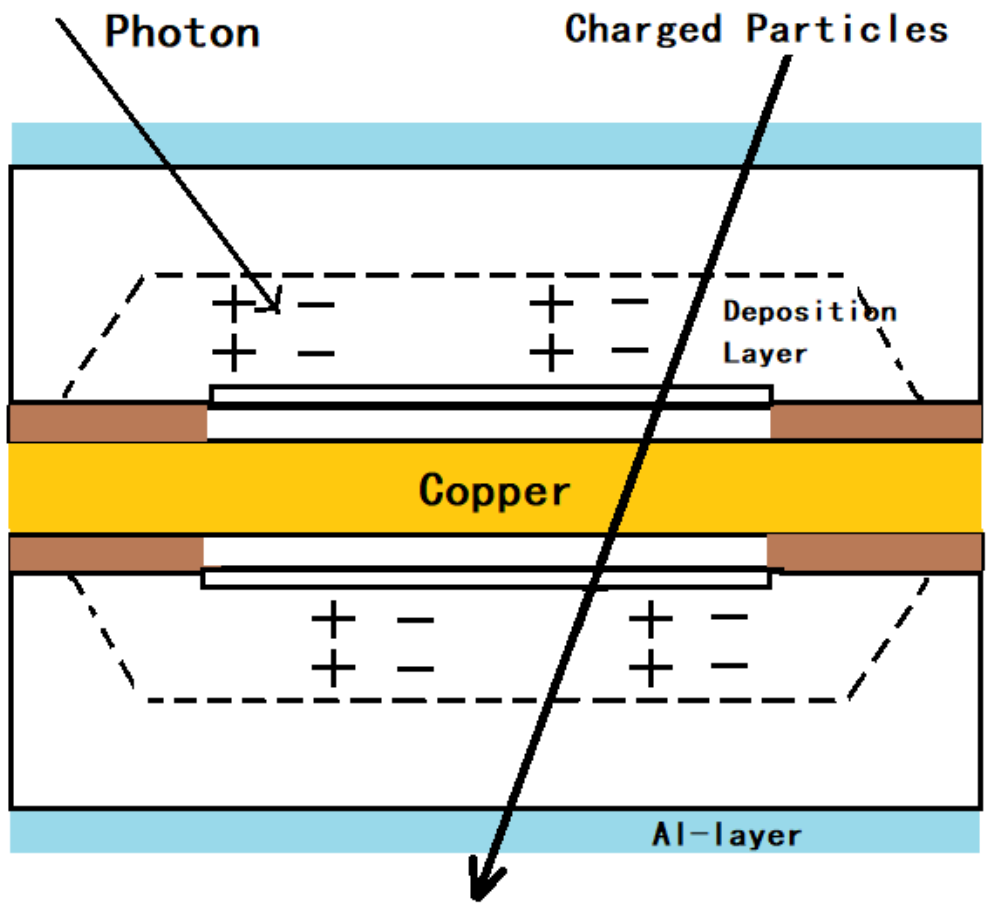

**Figure 4.3.7.28:** Two PIN-photodiodes with a copper layer between them. (Charged particles resulting from beam loss can pass through both diodes, whereas photons generated by synchrotron radiation cannot pass though both diodes due to the presence of inserted copper.)

The comparison between the four BLMs indicates that the improved Pin-Photodiodes detector is the recommended option for installation in the CEPC collider.

### *4.3.7.5.2 Installation Positions of the Beam Loss Monitors*

The detectors will be positioned around the machine at locations where the betatron amplitude functions reach their maximum, specifically in the arcs near each quadrupole. To achieve the highest efficiency of a BLM, it should be placed at the peak of the shower distribution. Given that the shower electrons are concentrated within a narrow range and the detector's effective area is only several $mm^2$, precise placement on the central plane is critical when fixing them onto the ring.

### *4.3.7.5.3 Data Acquizition System of the Beam Loss Monitors*

The front-end amplifies the beam loss signal, which is then digitized and processed by the intelligent chip before being sent to the bus. A PC serves as the console for system control, data acquisition, display, and analysis. To create a distributed data acquisition system, all front-end nodes and the console are linked via an industrial internet bus. This system can provide the required information regarding beam loss intensity, distribution, and timing. See Figure 4.3.7.29 for a schematic of the system's structure.

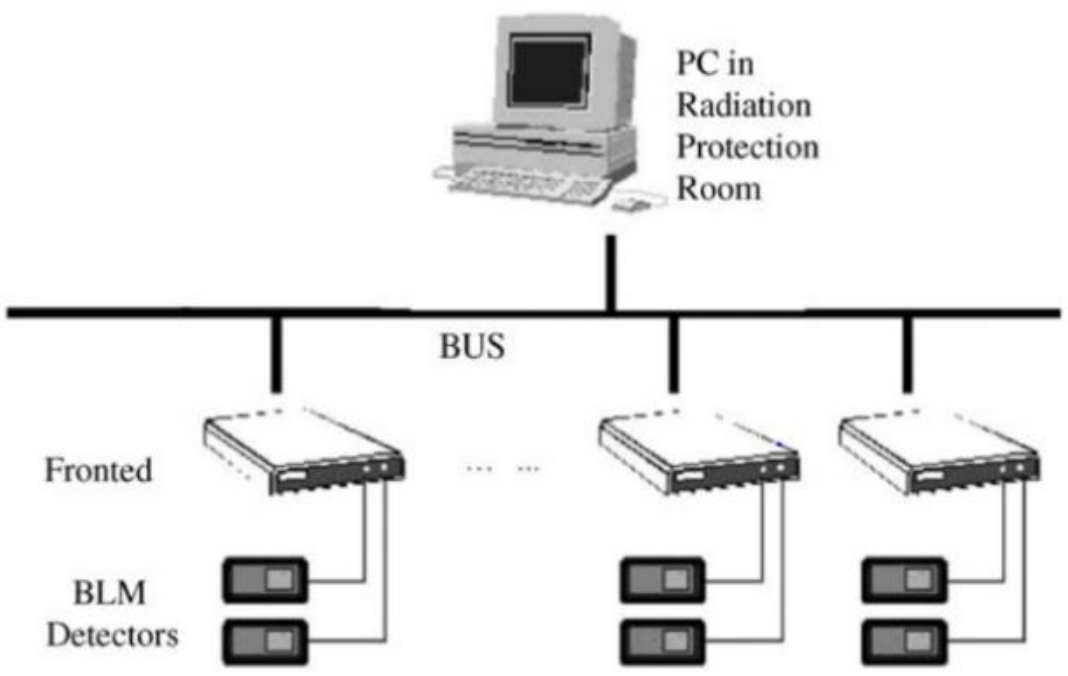

**Figure 4.3.7.29:** Structure of DAQ for BLM at CEPC

### 4.3.7.6 *Tune Measurement*

Measuring and optimizing the betatron tune throughout the operational cycle of a circular accelerator is crucial for ensuring beam quality and longevity, making it one of the most fundamental tasks in the control room. It's worth noting that betatron tunes can vary between positrons and electrons due to the significant orbit separation resulting from the momentum sawtooth effect. Measuring individual bunches can help identify tune shifts related to intensity. Additionally, beam-beam effects can result in various coherent oscillation modes. Therefore, there are several possible solutions for measuring the tune of the CEPC.

#### *4.3.7.6.1 The Kick Method*

For measuring the tune, a pilot bunch can be utilized, and a fast kicker can be employed to excite coherent betatron oscillations [20]. The beam position is monitored turn-by-turn (using broadband processing only) and recorded as a function of time. The Fourier transformation of the displacements provides the fractional part q of the tune, while the width of the Fourier line determines the tune spread with $\Delta q = \Delta Q$ (assuming sufficient resolution).

#### *4.3.7.6.2 The Direct Diode Detection*

A new technology, known as Direct Diode Detection (3D), developed at CERN for LHC tune measurement, will be considered [21]. The fundamental concept behind 3D is to time-stretch the beam pulse from the pickup to increase the betatron frequency content in the baseband. This can be achieved using a simple diode detector, followed by an RC low pass filter [22]. We have tested the 3D method on the BEPCII, and its validity has been demonstrated in a lepton machine [23]. The technique offers many advantages, including simplicity and low cost, robustness against saturation, flattening out the beam dynamic range, and independence from filling patterns. However, it also has a disadvantage, operating in the low-frequency range, which is susceptible to noise.

#### *4.3.7.6.3 The Excitation System*

Measuring the tune via external excitation under feedback operation is challenging due to the damping of the oscillations caused by the feedback system. However, an alternative method has been discovered for measuring the tune with feedback. By examining the signal immediately behind the analog detector of the feedback system, a notch can be observed in the noise spectrum at the tune resonance frequency due to the 180° phase shift of the feedback. These notches can be analyzed, even with operating feedbacks and minimal excitation [24].

### 4.3.7.7 *Beam Feedback System*

CEPC operates in high current and multiple bunch modes, which can lead to instability of coupled bunches. Thus, a feedback system is required to dampen this instability for the main ring. The instability is mainly caused by high-order modes generated by the high-frequency cavity and wall impedance of the beam pipe. Without the feedback system, it can result in severe transverse and longitudinal oscillations of the beam, reducing beam lifetime, and even leading to current threshold.

#### *4.3.7.7.1 Feedback Signal Processing*

Advanced modern accelerators use digital feedback systems on a bunch-by-bunch basis to suppress coupled bunch instability. Typically, a bunch-by-bunch digital feedback system consists of three components: front-end electronics, digital processing electronics, power amplifier, and kicker. The basic structure of the feedback system can be seen in Fig 4.3.7.30.

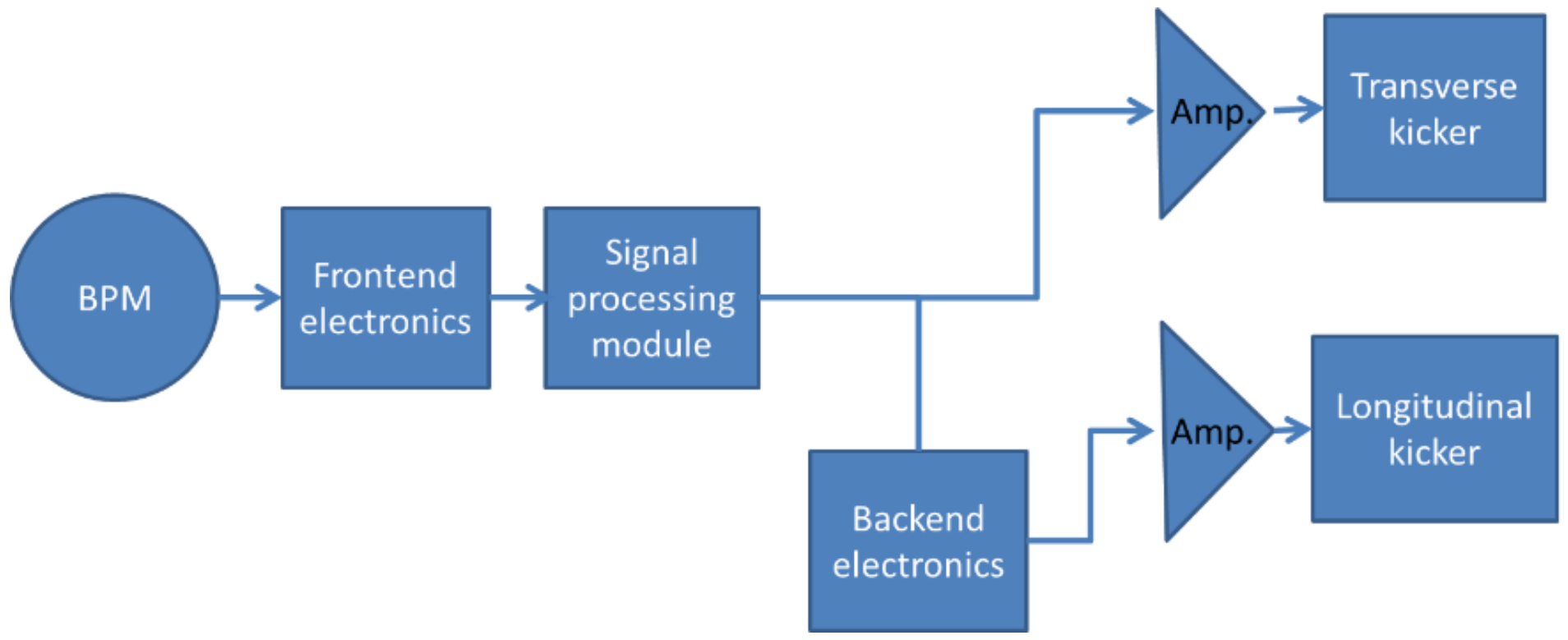

**Figure 4.3.7.30:** Block diagram of feedback system signal

In a transverse feedback system, the front-end electronics converts the BPM oscillation signal and processes it digitally. The digital processing includes a 90-degree phase shift, removal of the closed orbit component, and single turn delay. The resulting analog signal is then sent to the amplifier and kicker to give the beam an angular kick. The power amplifier drives a 50Ω-stripline kicker, which is shorted at one end. The kicker plates are powered differentially using a hybrid power divider driven by the combined amplifier output.

Compared to transverse feedback, longitudinal feedback is more challenging to implement. A longitudinal feedback system requires back-end electronics, and digital signal processing is needed to convert the signal to the carrier frequency. Additionally, a pillbox cavity kicker is necessary for the longitudinal feedback system.

The feedback signal processing electronics perform several important functions, including providing the 90 degrees phase shift required for the feedback signal, providing several laps of time delay, and suppressing the DC component. This helps to filter out the closed orbit component from the beam position oscillation signal.

The transverse kicker changes the angle of the bunch movement, while the pickup electrode detects the position oscillation signal of the beam. The phase difference between the two is 90 degrees, so the signal processing electronics must provide this phase shift. However, since signal processing takes time, it may not be possible to calculate the required kick amount in the current cycle of bunch position signal acquisition. Therefore, additional cycles should be considered when determining the phase shift between the pickup and the kicker. This can be achieved by using the linear combination of two pickup electrode signals with a 90-degree phase difference.

The signal processing electronics also play a crucial role in removing the DC component in the beam signal, also known as the beam closed orbit signal. This is important because the feedback system is designed to suppress the transverse or longitudinal oscillation of the bunch, not the beam closed orbit. If the DC component is not removed from the pickup electrode signal, it would waste feedback power and potentially saturate the power amplifier, resulting in a reduced damping time for the feedback system. To remove the beam closed orbit signal, the signal processing electronics can use a filter to eliminate the cyclotron frequency component in the beam signal.

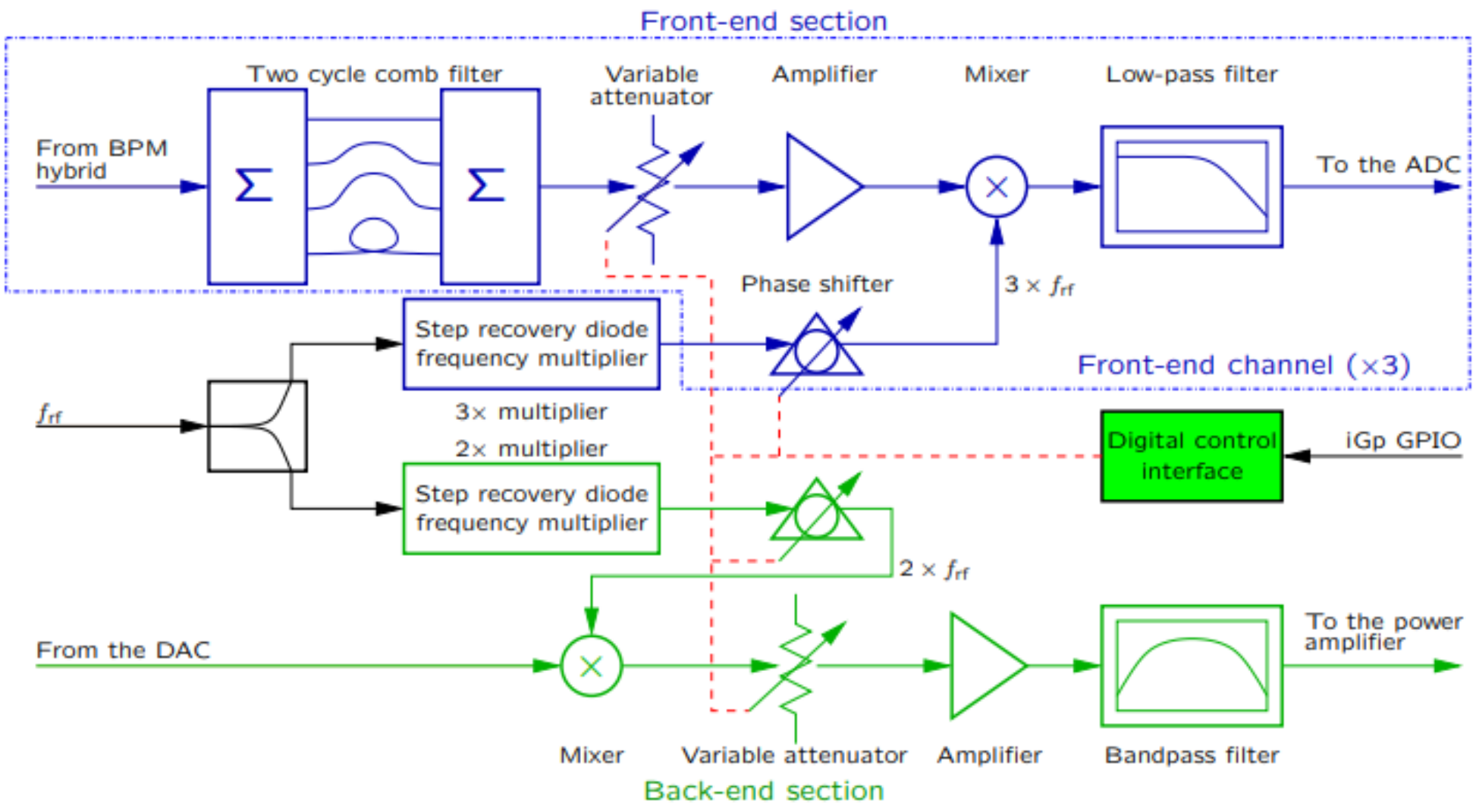

**Figure 4.3.7.31:** Front-back end electronics schematic diagram

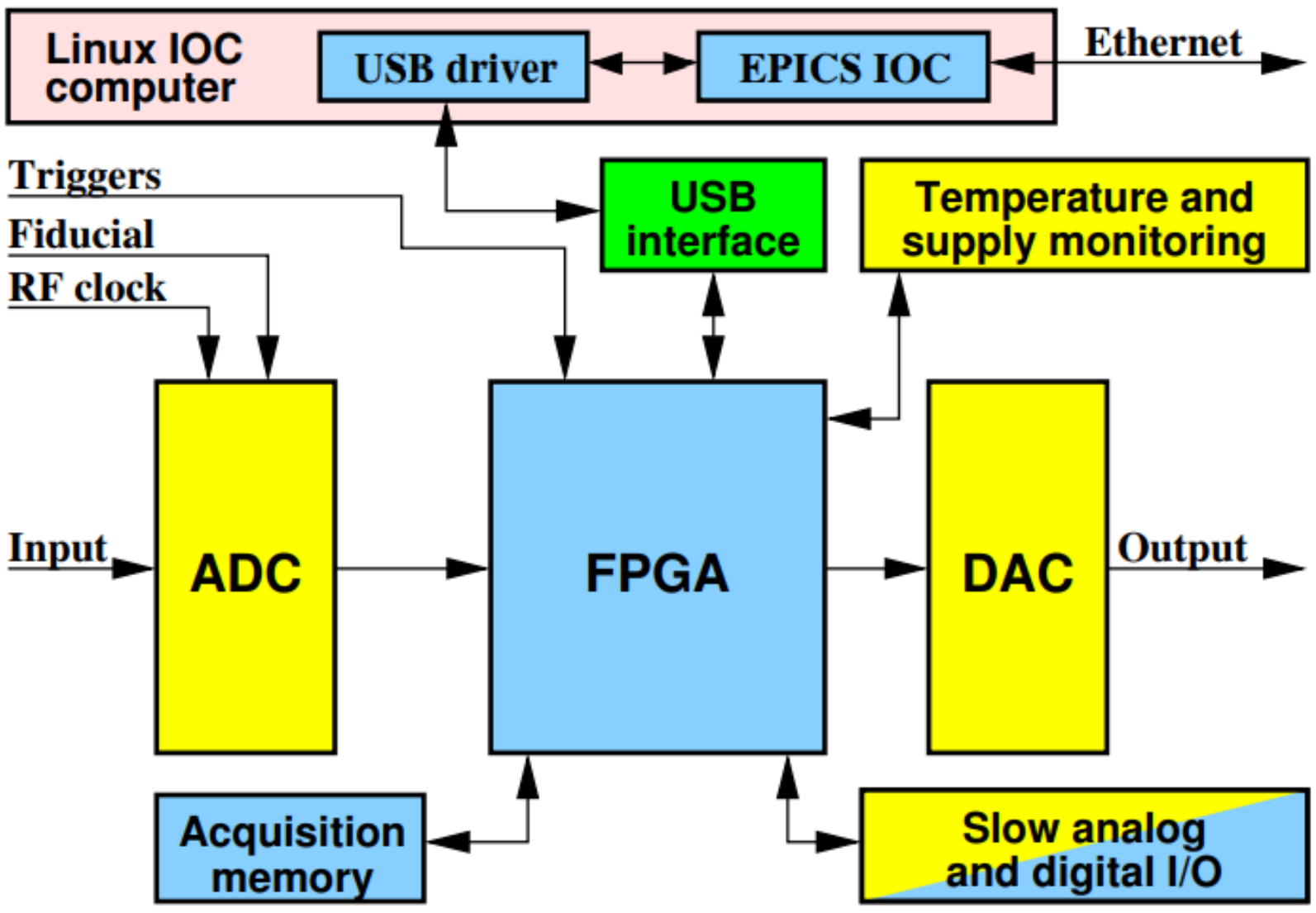

**Figure 4.3.7.32:** Digital signal processing electronics schematic diagram

Currently, many accelerators around the world utilize front-end and back-end electronics, as well as digital signal processing electronics, from Dimtel. The products from Dimtel can be seen in Figure 4.3.7.33 and Figure 4.3.7.34.

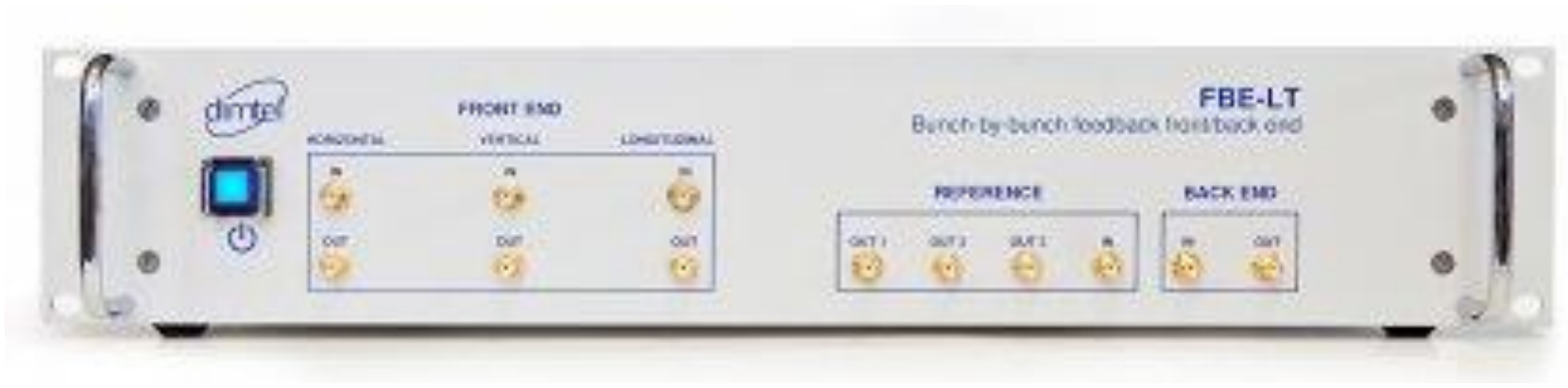

**Figure 4.3.7.33:** Front-back end electronics material object

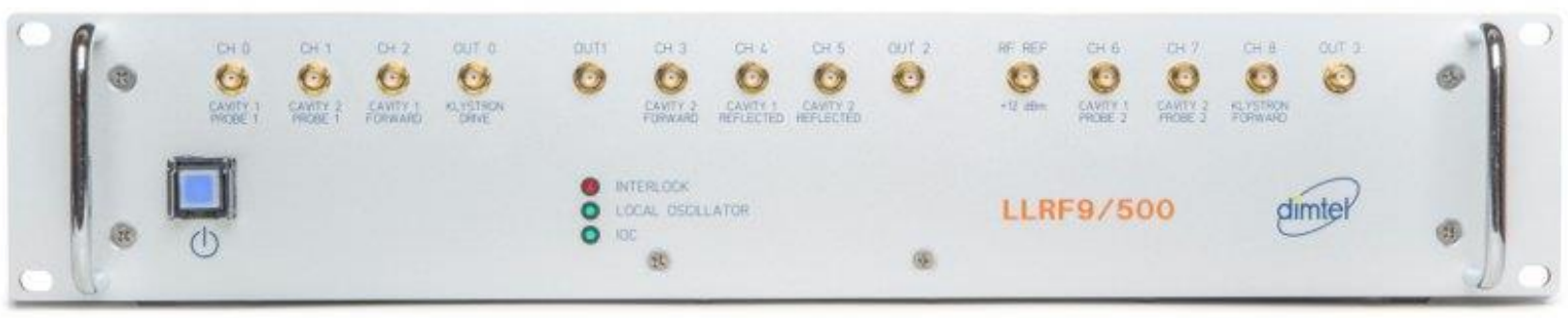

**Figure 4.3.7.34:** Digital signal processing electronics material object

### *4.3.7.7.2 Kicker Design*

The kickers are essential components of the beam feedback system, providing the necessary angular kick to the beam. The stripline type kickers are commonly used for transverse feedback, while the waveguide loaded pillbox cavities are used for longitudinal feedback. In the following sections, we will introduce the design of these two types of kickers.

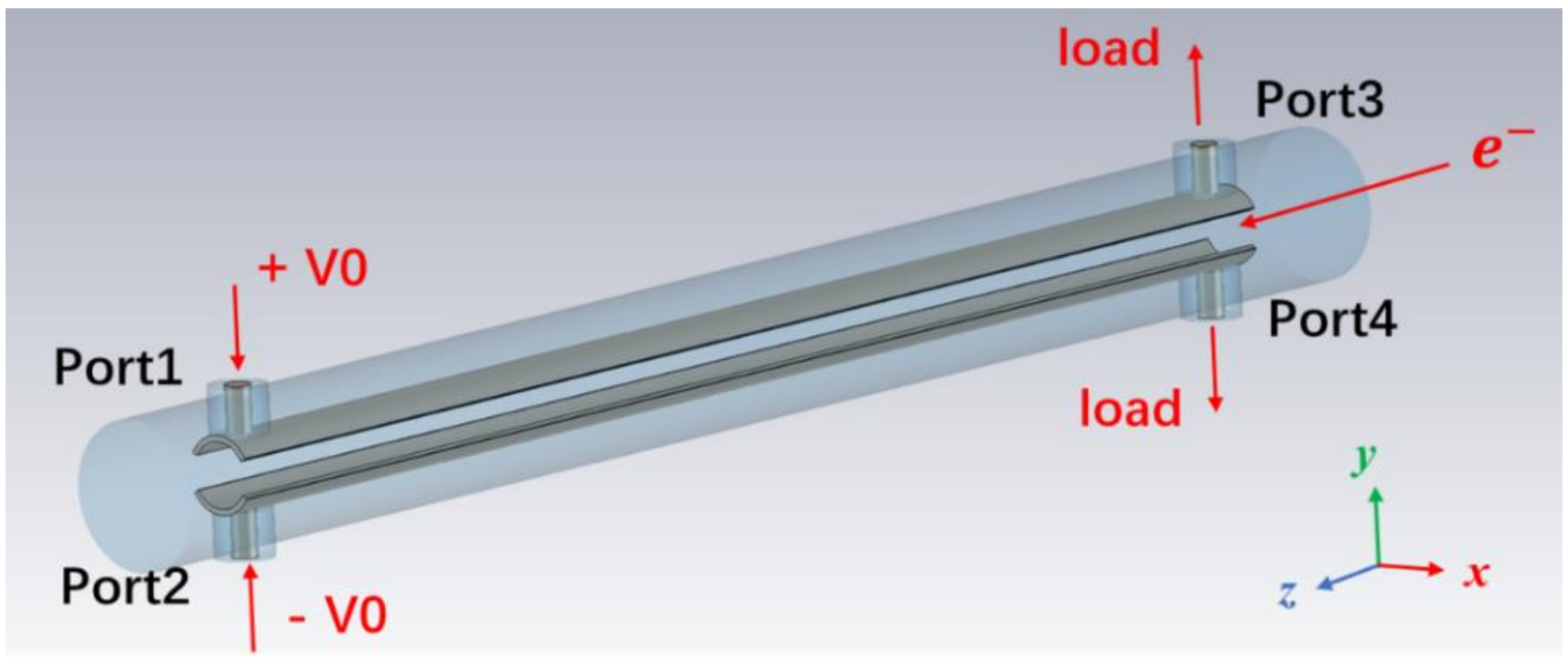

**Figure 4.3.7.35:** The basic structure of transverse feedback kicker

The transverse feedback (TFB) kicker can be seen in Figure 4.3.7.35, which consists of two curved electrodes located inside a round pipe. Each electrode is connected to a coaxial port on both ends. When operating in dipole mode, the kicker is driven by two waves of equal amplitude and opposite phase from the downstream ports. The other two ports are matched to 50 Ohm. The characteristic impedance of the kicker in dipole mode is optimized to 50 Ohm to reduce reflection. The beam pipe has a radius of 28 mm, and two 90-degree electrodes with a thickness of 2mm are located at an inner radius of 17 mm. To suppress coupled bunch instabilities, the required working bandwidth is approximately 25 MHz, for a minimum bunch spacing of 23 ns. The length of the electrode is set to 600 mm, which provides sufficient kick strength and ensures the required bandwidth.

The simulated reflection coefficient S11 of the kicker is below –20 dB within the working bandwidth, as shown in Figure 4.3.7.36.

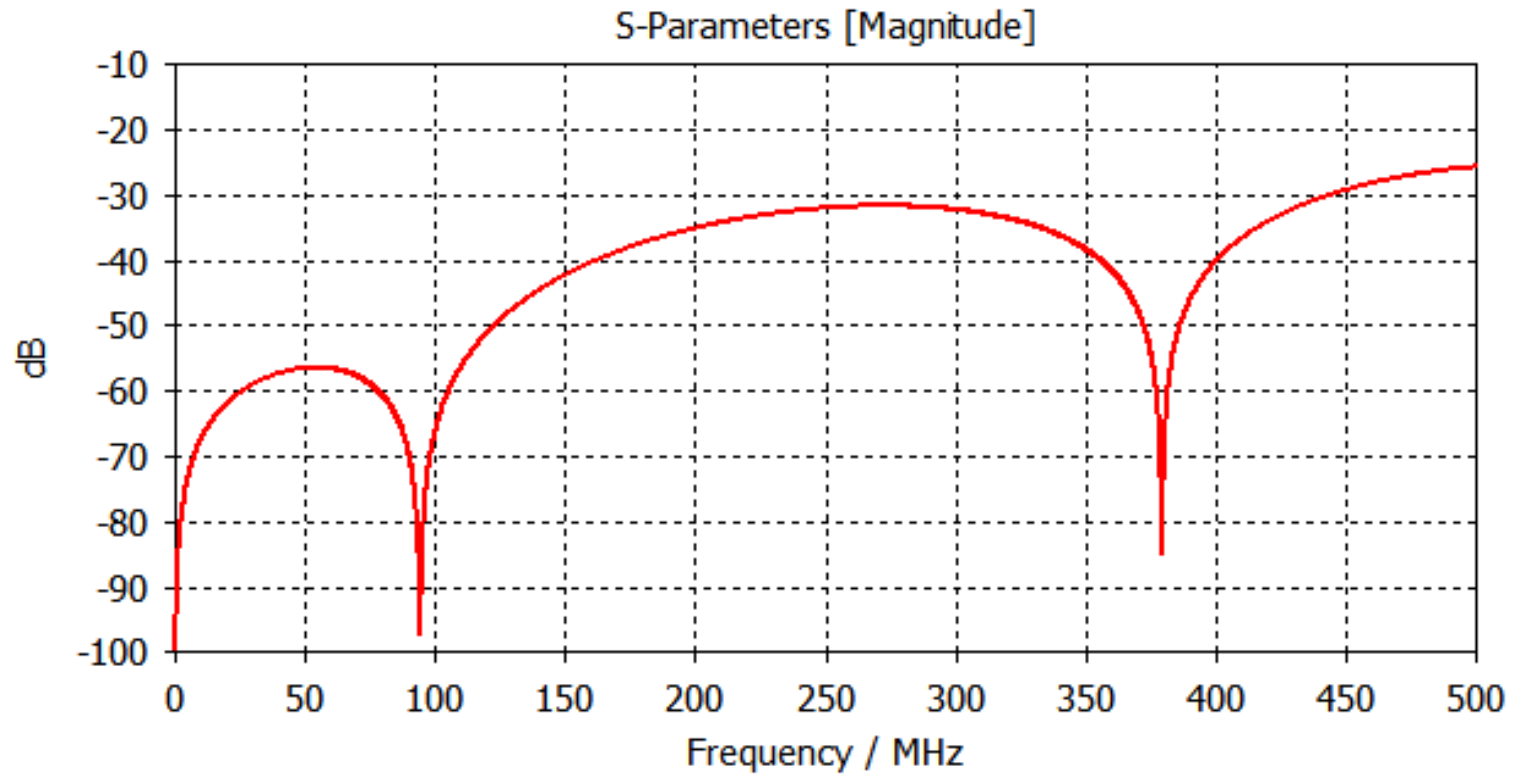

**Figure 4.3.7.36:** Simulated reflection parameter S11 of the transverse feedback kicker

The transverse shunt impedance $R_t$ is another important parameter to measure the efficiency of the kicker. $R_t$ is defined as the ratio between the square of the transverse voltage $V_{\perp}$ and twice the input power $P_{in}$,

$$R_t = \frac{|V_{\perp}|^2}{2P_{in}} \tag{4.3.7.17}$$

where the transverse voltage $V_{\perp}$ can be calculated based on the electromagnetic fields inside the kicker.

$$V_{\perp} = \int (\boldsymbol{E}(z) + \boldsymbol{v} \times \boldsymbol{B}(z)) e^{j(kz+\varphi)} dz \qquad (4.3.7.18)$$

Typically, the $R_t$ of standard stripline type kicker can be evaluated by:

$$R_t = 2Z_{dipole} \left(\frac{g_{\perp} c}{a}\right)^2 \left(\frac{\sin(\omega L/c)}{\omega}\right)^2 \qquad (4.3.7.19)$$

where $a$ is the inner radius of the stripline, $L$ is the stripline length, and $g_{\perp}$is the coverage factor, which can be calculated by the product of the electric field strength $E_{\perp}(0{,}0)$ on the beam axis and the stripline radius $a$. In our case, $g_{\perp}$ is 1.076. The shunt impedance versus frequency of the transverse kicker is shown in Figure 4.3.7.37, with a value of about 140 kΩ within the bandwidth.

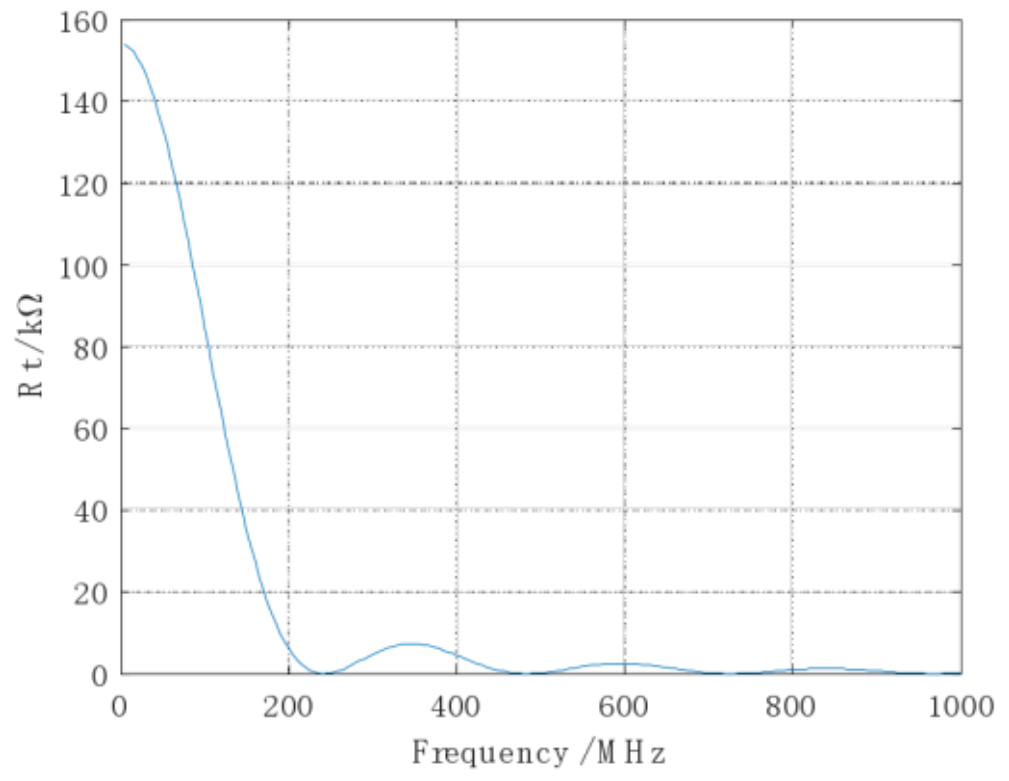

**Figure 4.3.7.37:** Transverse shunt impedance versus frequency.

For the longitudinal feedback (LFB) kicker, the central frequency is typically set to (N/2 + 1/4) × $f_{rf}$ where N is a non-negative integer (i.e., N = 0, 1, 2, ...). In our specific case, the central frequency is set to 1.1375 GHz, and the minimum bandwidth is approximately 25 MHz. Therefore, the working frequency range of the LFB kicker should cover 1.125 – 1.15 GHz.

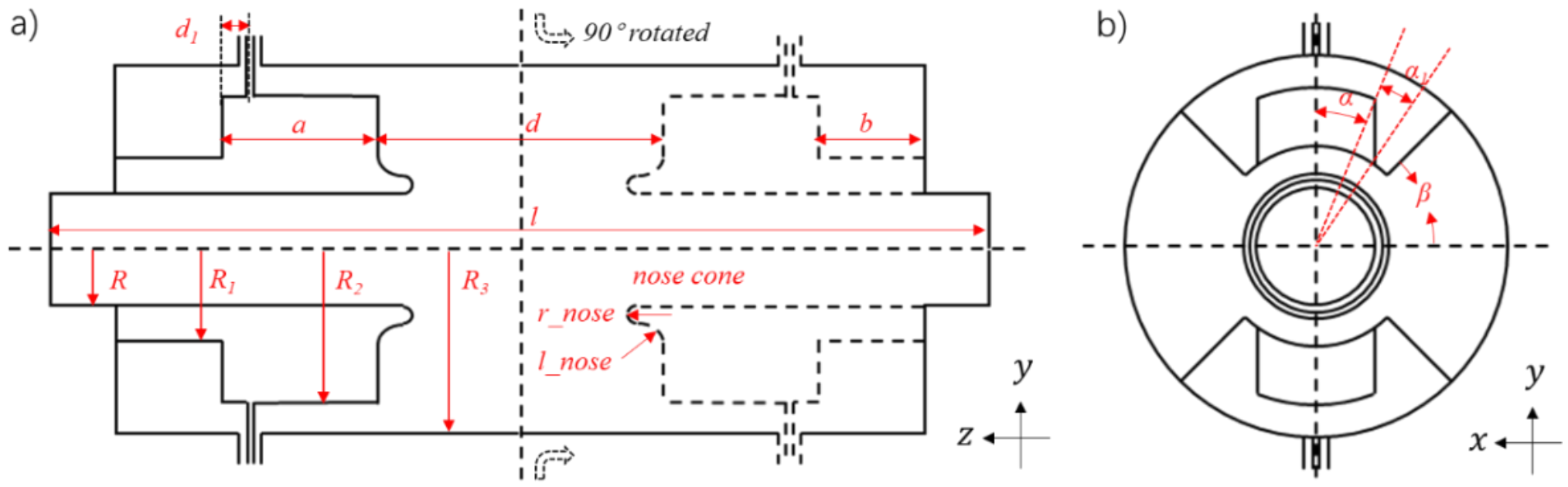

**Figure 4.3.7.38:** The schematic of the longitudinal feedback kicker: (a) the cut view in the yz-plane, the part indicated by the dotted line is rotated by 90 degrees along the beam direction z, and (b) the cut view in the xy-plane.

Figure 4.3.7.38 depicts the schematic model of the longitudinal feedback kicker, which is primarily composed of a pillbox cavity, 4 ridged waveguides, and coaxial ports attached to them. The kicker is driven through 2 coaxial ports located on the +z (or -z) side, with the same phase, while the ports on the opposite side are matched to 50 Ω. The working mode that is excited is the TM010 mode, which provides accumulated voltage along the beam direction to compensate for the energy deviation of bunches. After optimization, the key geometric parameters of the LFB kicker are presented in Table 4.3.7.9.

**Table 4.3.7.9:** Main geometric parameters of the Collider LFB kicker.

| Parameter | Value | Parameter | Value |
|---|---|---|---|
| Beampipe radius *R* | 28 mm | Back cavity length | 27.44 mm |
| Back cavity radius | 56 mm | Cavity length | 303 mm |
| Ridge radius | 82 mm | Port angle | 15.3 deg |
| Pillbox cavity radius | 90 mm | Port base angle | 6.7 deg |
| Cavity gap | 91 mm | Barrier angle | 61 deg |
| Distance of feedthrough | 5 mm | Nose cone radius | 6 mm |
| Ridge length | 53.56 mm | Nose cone length | 7 mm |

The simulated reflection parameter S11 is displayed in Figure 4.3.7.39, which demonstrates a 3 dB bandwidth of 1.115 – 1.168 GHz, covering the necessary frequency range.

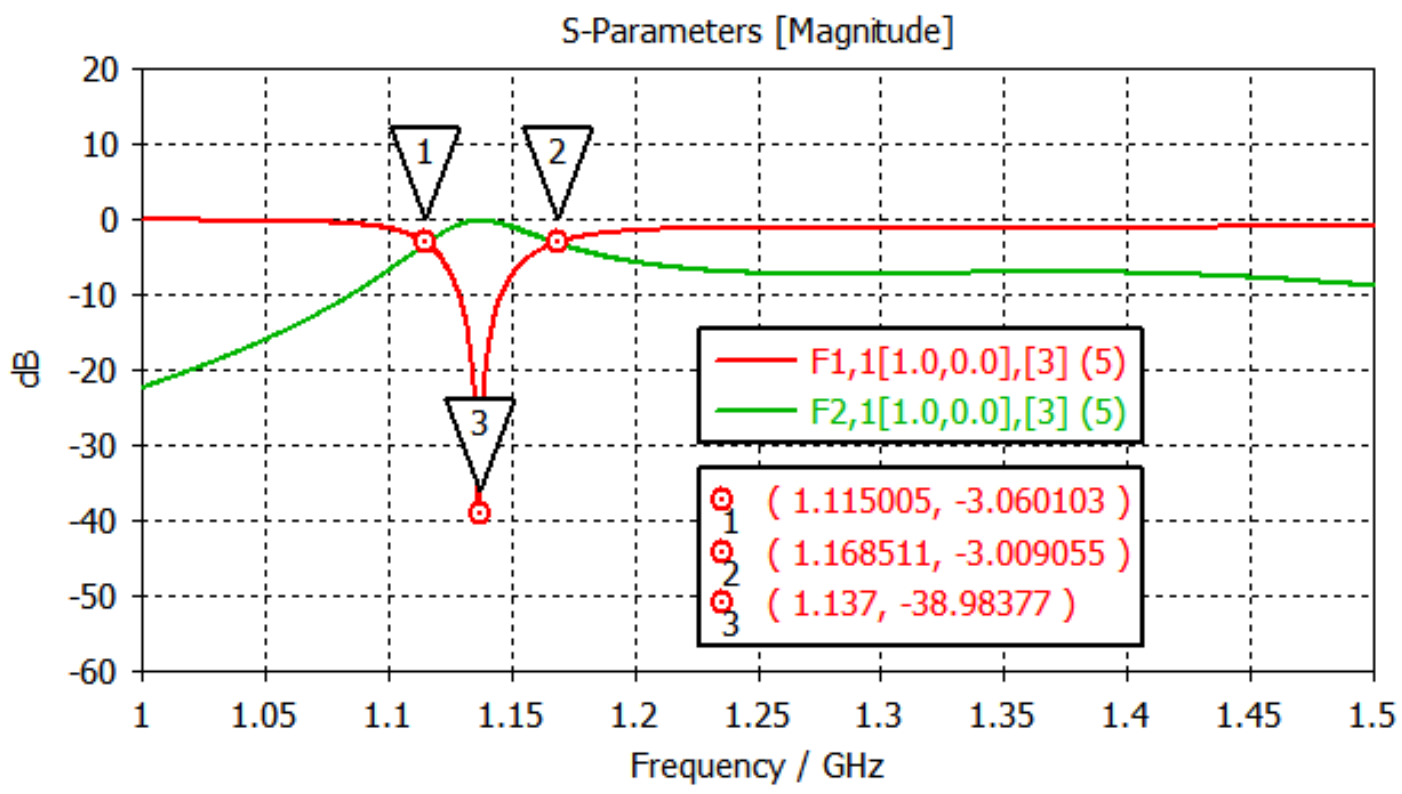

**Figure 4.3.7.39:** The reflection parameter $S_{11}$ of the LFB kicker

The longitudinal shunt impedance is given by $R_s = |V|^2/2P_{in}$, where $V$ represents the longitudinal voltage inside the cavity gap, and $P_{in}$ denotes the input power. Figure 4.3.7.40 illustrates the simulated shunt impedance versus frequency, indicating that the LFB kicker has a minimum shunt impedance of 2600 Ω within the working bandwidth.

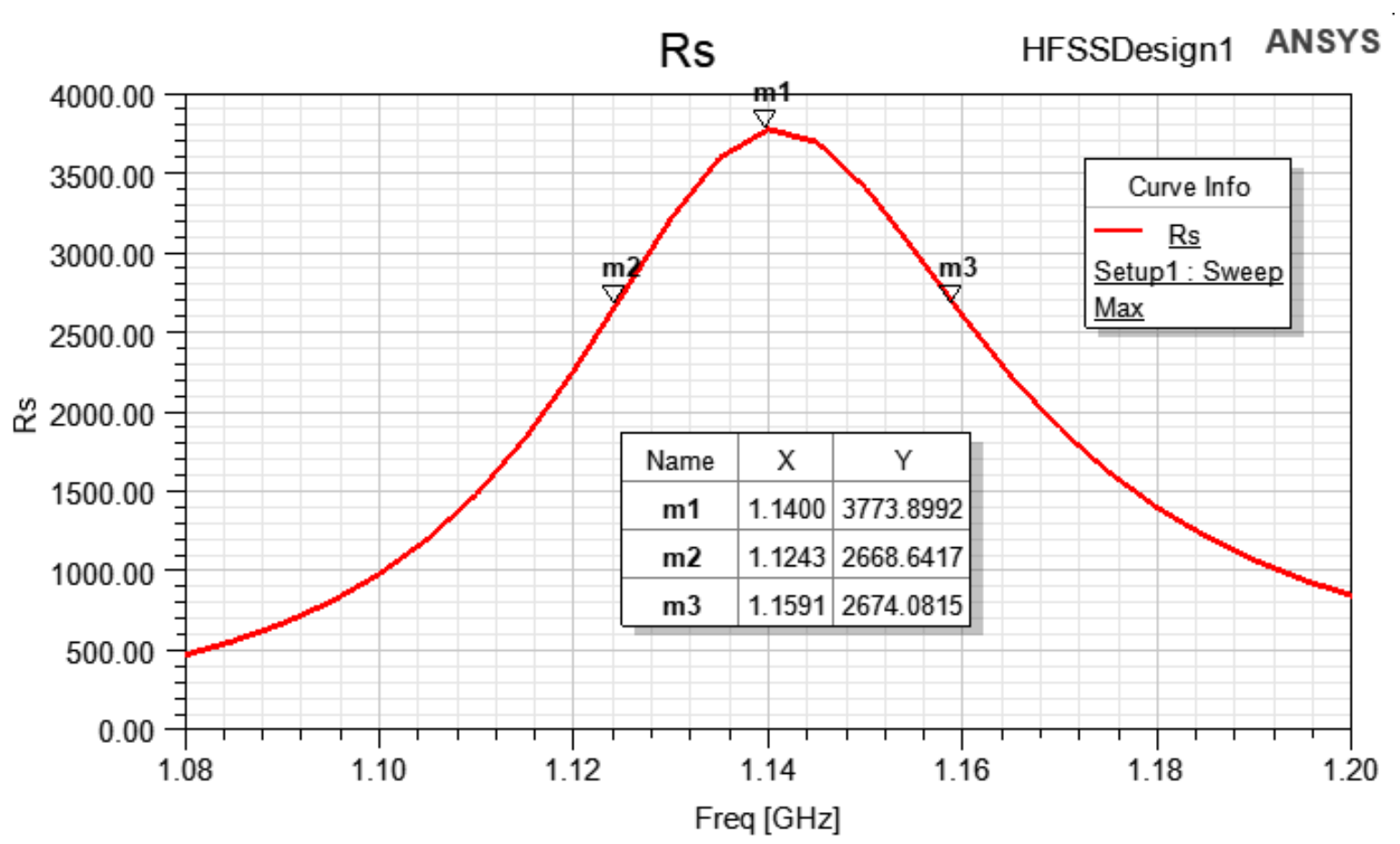

**Figure 4.3.7.40:** The longitudinal shunt impedance of the LFB kicker

### *4.3.7.7.3 Power Calculation and Amplifier Selection*

The ultimate output of the feedback system is the voltage generated by the transverse or longitudinal feedback kicker, which is limited by the power available from the amplifier. The impulse voltage provided by the feedback system to the bunch each turn determines its variation of *x'*, *y'*, or *E*, which ultimately affects the damping rate of the feedback system. During the design of the feedback system, the growth rate of the beam instability oscillation amplitude can be calculated from the coupling impedance on the ring. Based on this, the necessary power can be determined so that the feedback system can provide the required damping rate to counteract the instability.

In the case of a transverse feedback system, the required power for either the horizontal or vertical direction feedback system can be defined as:

$$\frac{1}{\tau_{FB}} = \frac{f_{rf}\sqrt{\beta_p\beta_k}}{2*h*\frac{E}{e}} * G \text{ , P} = \frac{1}{2} * \frac{\Delta V_{FB}^2}{R_k} \qquad (4.3.7.20)$$

where $\tau_{FB}$ is the minimum damping time that the feedback system needs to provide, *G* is gain of the feedback system, i.e. $\frac{\Delta v_{FB}}{\Delta x}$, Δx is the maximum amplitude of the bunch, $\beta_k$ and $\beta_k$ are the beta function at the pickup and kicker location respectively, $R_k$ is the shunt impedance of the kicker, *h* is the harmonic number, and *P* is the total power of the transverse feedback system. These parameters are listed in Table 4.3.7.10 for the Higgs and Z mode operation.

**Table 4.3.7.10:** Transverse feedback system power calculation

| Parameter | Higgs | Z |
|---|---|---|
| Energy（GeV） | 120 | 45.5 |
| Beta functions at pickup (m) | 250 | 250 |
| Beta function at kicker（m） | 250 | 250 |
| Number of kickers | 4 | 4 |
| Revolution time（ms） | 0.33 | 0.33 |
| RF frequency（MHz） | 650 | 650 |
| Kicker shunt impedance | 140 | 140 |
| Amplitude of oscillation | 0.2 | 0.2 |
| Damping time（ms） | - | 1.0 |
| Power(w) | - | 1460 |

Based on the results and considering the Higgs and Z modes together, it seems that the Z mode is the most important factor to consider. This is because if the Z mode is satisfied, then the Higgs mode will also be satisfied.

Given that we require 1460 W of power for both the X and Y directions, we can use a 4-electrode kicker system. Each electrode will need to provide 465 W of power, so we will need eight 500W amplifiers for the transverse feedback system. There are several optional amplifier models available, including the AR 500A250A and Bonn BSA 0125-500.

For the longitudinal feedback system, the power that feedback system needs can be defined as:

$$\frac{1}{\tau_{FB}} = \frac{f_{rf}\alpha}{2*\nu_S*\frac{E}{e}} * G \text{ , P} = \frac{1}{2} * \frac{\Delta V_{FB}^2}{R_k} \tag{4.3.7.21}$$

where $\tau_{FB}$is the minimum damping time that the feedback system can provide, which must be less than the fastest growth time of the longitudinal coupled-bunch (CB) mode, $G$ is the gain of the feedback system, $\frac{\Delta v_{FB}}{\Delta\varphi}$ is the voltage required for correcting unit phase error, $\alpha$ is the momentum compaction factor, $R_k$ is the shunt impedance of the kicker, $P$ is the total power that the longitudinal feedback system. These parameters are listed in Table 4.3.7.11 for the Higgs and Z mode.

**Table 4.3.7.11:** Longitudinal feedback system power calculation

| **Parameter** | **Higgs** | Z |
|---|---|---|
| Energy (GeV) | 120 | 45.5 |
| Momentum compaction ($10^{-5}$) | 0.71 | 1.43 |
| Longitudinal tune | 0.049 | 0.035 |
| Phase acceptance (mrad) | 1.7 | 1.7 |
| RF frequency (MHz) | 650 | 650 |
| Kicker shunt impedance (kΩ) | 2.6 | 2.6 |
| Number of kickers | 4 | 4 |
| Damping time (ms) | - | 65 |
| Power (W) | - | 7720 |

According to the results, when combining the Higgs and Z mode, the Z mode is also a major consideration, similar to the transverse feedback, but requires much more power. Therefore, we need to use two longitudinal kickers to increase the impedance. Each kicker requires 3260 W power. To achieve this, we can use a four-electrode kicker, where each electrode needs to provide 816 W power. For the longitudinal feedback system, we can use eight 1000 W amplifiers. Optional amplifier models include Milmega AS0820-1000 and others.

#### 4.3.7.8 *Vacuum Chamber Displacement Measurement*

Due to heat effects caused by synchrotron radiation and beam loss, the vacuum chamber may experience displacement, leading to a potential decrease in BPM resolution. To calibrate the BPMs, it is necessary to measure this displacement.

The system for measuring the displacement includes a Linear Variable Differential Transformer (LVDT), a signal processing unit, a computer, and a network. This system can be mounted near the BPM, and the sense signal is transmitted via a 600 m cable to the local station.

Research conducted on BPM vacuum box displacement measurement at BEPC II has utilized the DL6230 manufactured by Micro Epsilon in Germany. The DL6230 is an electrical capacitive displacement detector with a dynamic range of 500 μm and an accuracy of 1 nm [25].

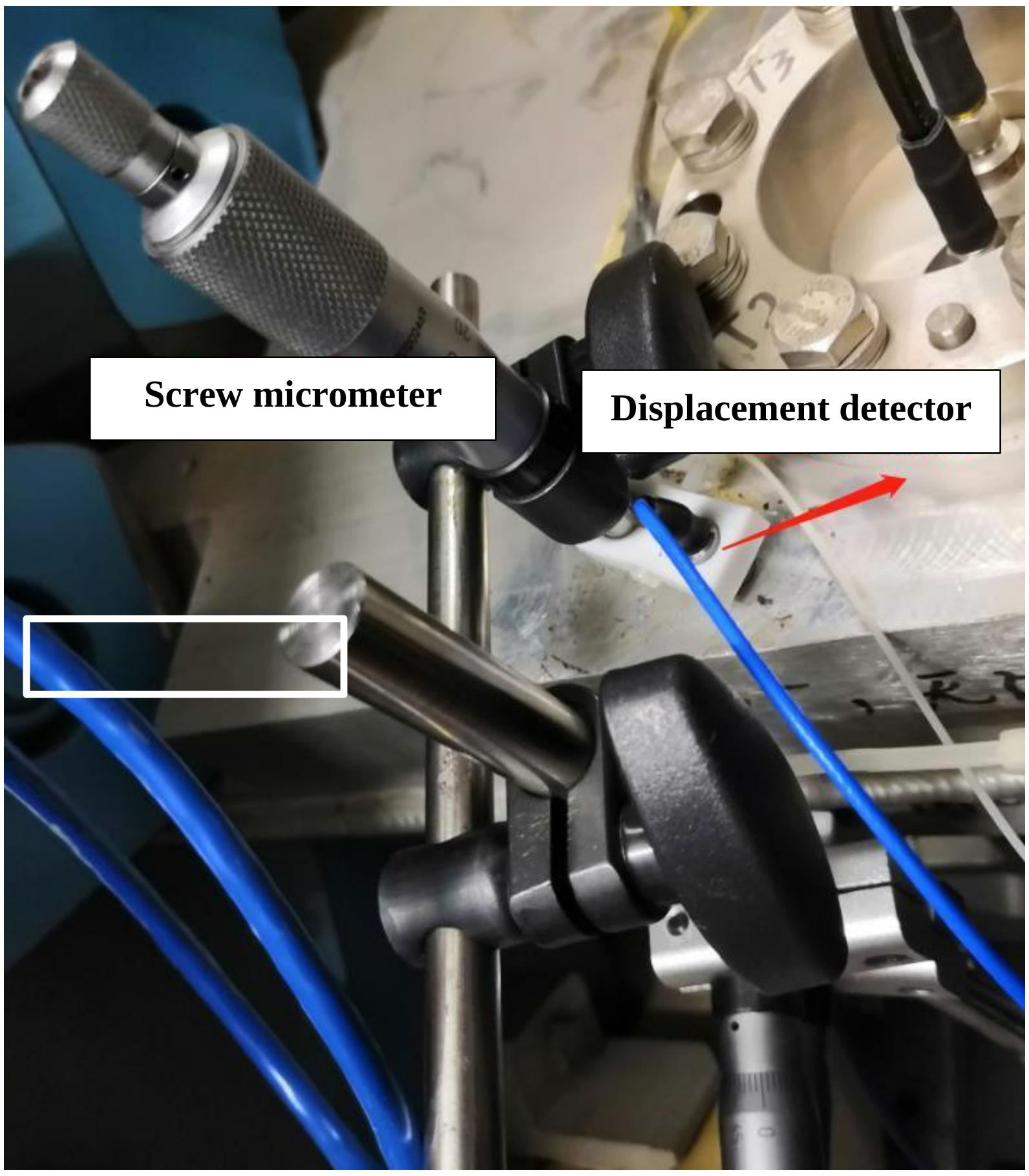

**Figure 4.3.7.41:** The displacement detector installed near one BPM pick-ups in BEPC II.

The screw micrometer, equipped with a displacement detector, is installed in the vacuum box of BPM R1OBPM05 within the BEPC II storage ring. A temperature probe is placed in close proximity to the micrometer. Displacement data from the vacuum box, along with BPM data, beam current intensity, and surrounding temperature changes, are jointly analyzed. Initial results can be seen in Figure 4.3.7.42 (X direction) and Figure 4.3.7.43 (Y direction). In collider mode, the beam injection cycle lasts approximately one hour. Within this cycle, electron injection is completed within one minute, followed by stable beam operation.

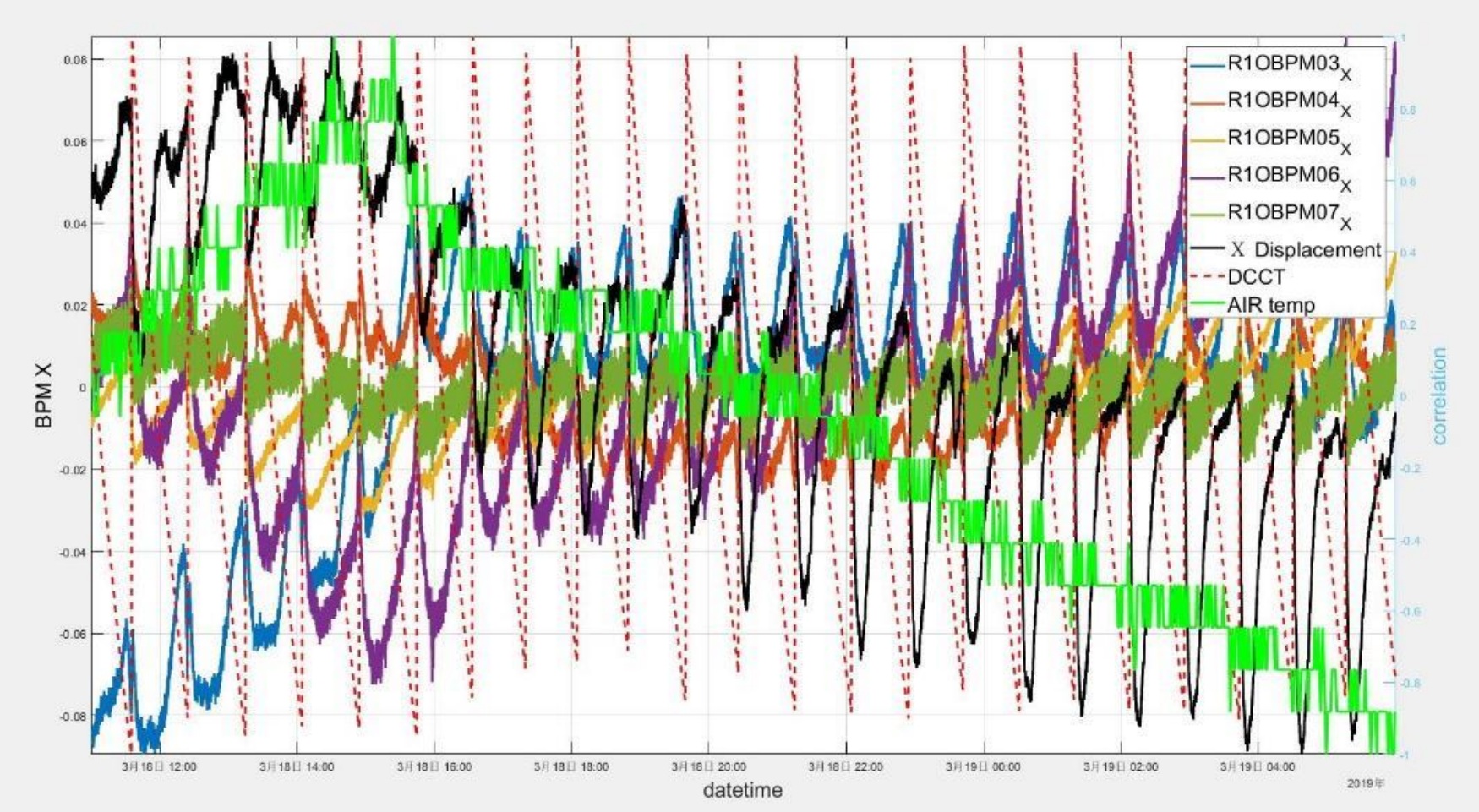

**Figure 4.3.7.42:** BPM data in X direction, vs. vacuum box displacement, beam current, and ambient temperature.

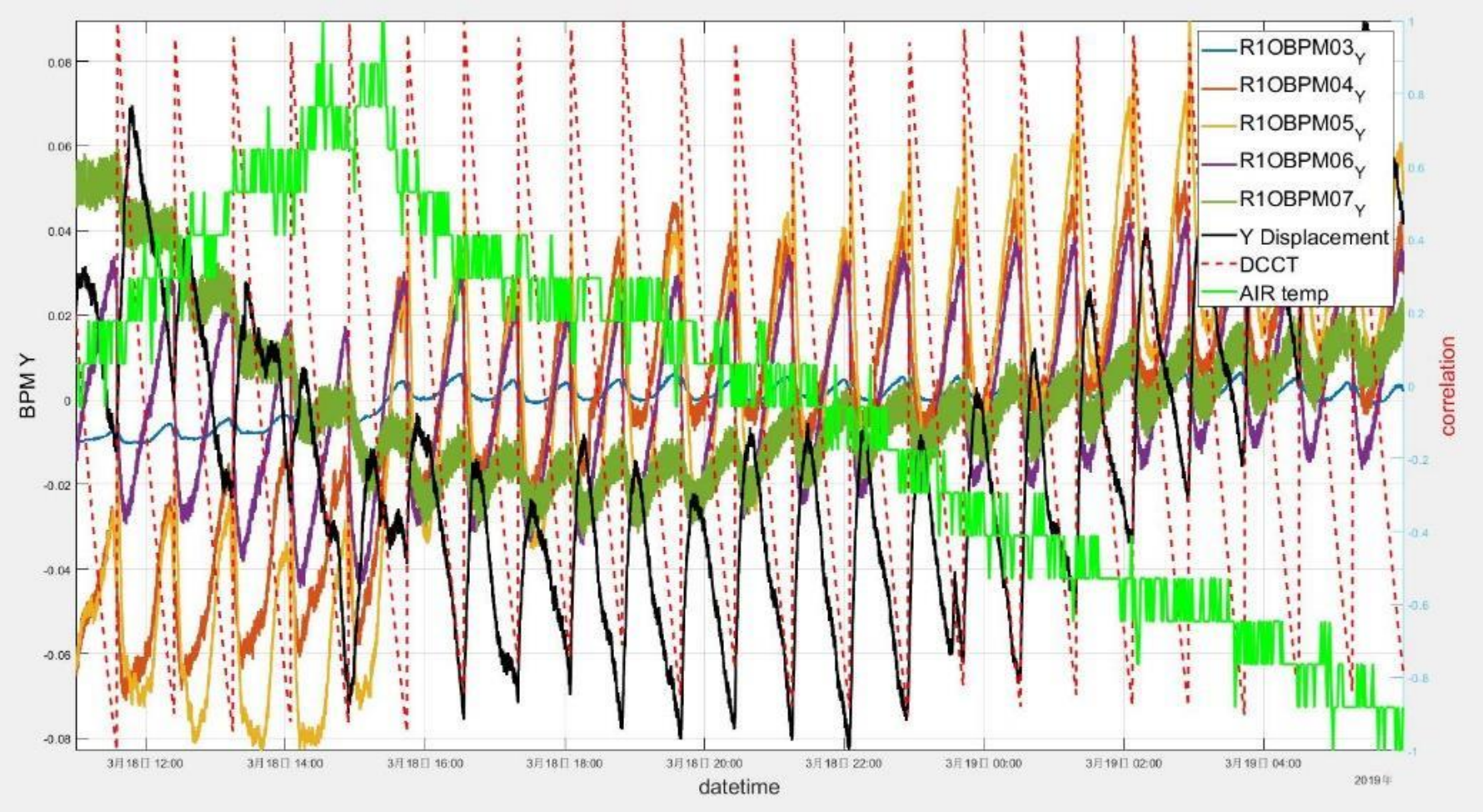

**Figure 4.3.7.43** BPM data in Y direction vs. vacuum box displacement, beam current, and ambient temperature.

## 4.3.7.9 *Other Systems*

Additional instrumentation systems for CEPC include the beam polarization measurement system and the energy spread measurement system.

To measure transverse beam polarization, a Laser Compton Polarimeter can be employed. This system is based on the spin-dependent Compton scattering of circularly polarized photons from polarized electrons and positrons.

### 4.3.8 Injection and Extraction System

The CEPC Collider utilizes a double-ring scheme, which requires specialized injection and extraction systems to accommodate top-up injection at different energy modes. To fulfill this need, an on-axis injection system, an off-axis injection system, and a dump extraction system have been designed for each ring. The overall layout of the CEPC can be seen in Figure 4.3.8.1, while Table 4.3.8.1 and Table 4.3.8.2 provide the injection and extraction physics parameters for each of the Collider's rings.

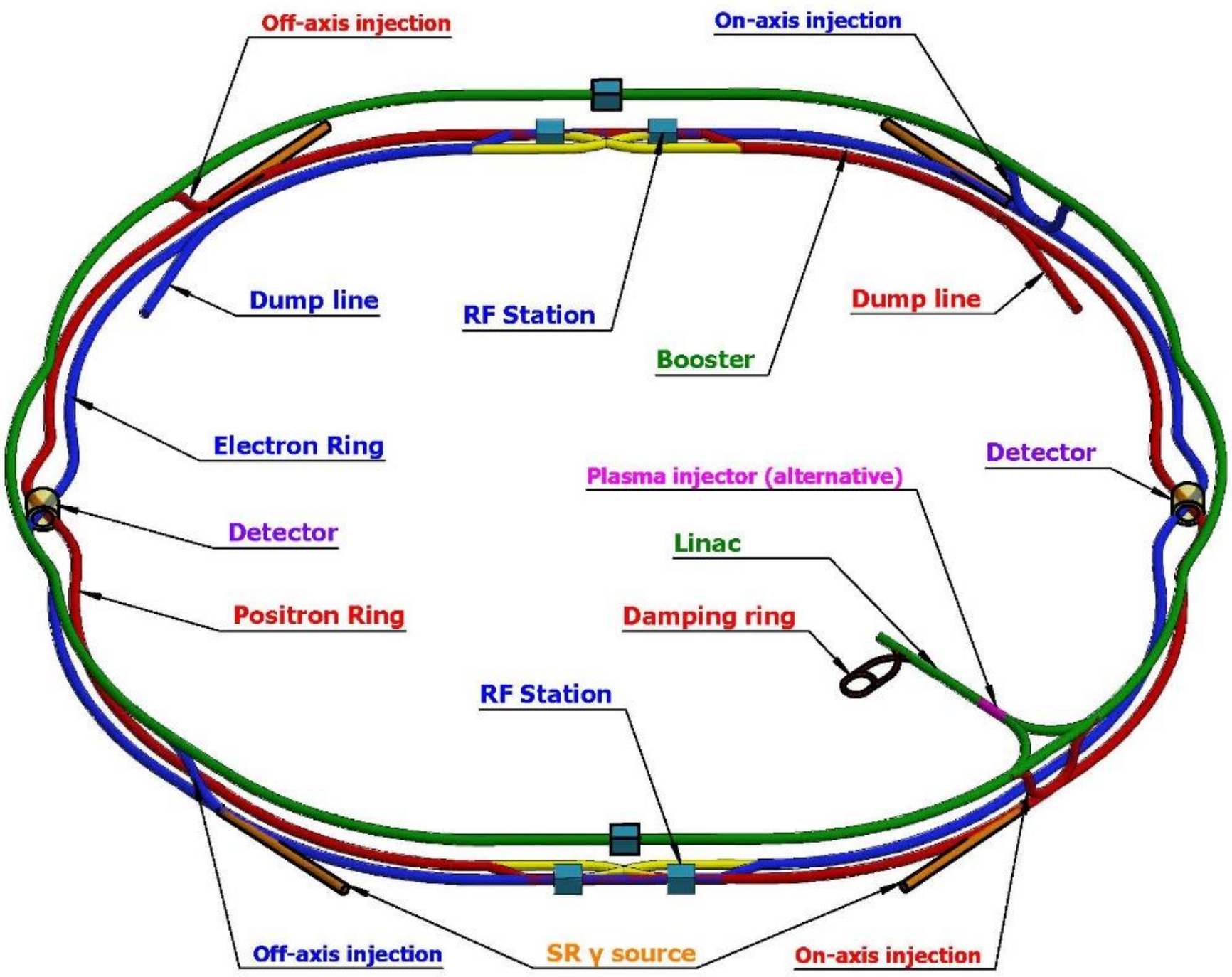

Fig**4.3.8.1**: Overall layout of the CEPC

**Table 4.3.8.1:** Injection and extraction parameter of the CEPC Collider

| **Parameter** | **Higgs** | | | **W** | | **Z** | | $\boldsymbol{t\bar{t}}$ | | |
|---|---|---|---|---|---|---|---|---|---|---|
| Beam energy (GeV) | 120 | | | 80 | | 45.5 | | 180 | | |
| Bunch number | 268 (half ring) | | | 1524 (uniform) | | 12000 | | 28 (half ring) | | |
| Min. bunch spacing (ns) | 680 | | | 220 | | 25 | | 5900 | | |
| Bunch number per train × train number | - | | | - | | 80 × 150 | | - | | |
| Train spacing (ns) | - | | | - | | 245 | | - | | |
| Injection/extraction mode | Bunch by bunch | | | Bunch by bunch | | Train by train | | Bunch by bunch | | |
| Kicker repetition rate (Hz) | 1000 | | | 1000 | | 1000 | | 1000 | | |
| Kicker pulse width (ns) | 1360 | | | 440 | | < 2460 | | 11800 | | |
| Kicker flat top (ns) | - | | | - | | > 1980 | | - | | |
| Kicker rise/fall time (ns) | < 680 | | | < 220 | | < 245 | | < 5900 | | |
| Function | Off-axis Inj. | On-axis Inj. | Ext. Dump | Off-axis Inj | Ext. Dump | Off-axis Inj. | Ext. Dump | Off-axis Inj. | On-axis Inj. | Ext. Dump |
| Kick angle (mrad) | 0.1 | 0.2 | 0.4 | 0.1 | 0.4 | 0.1 | 0.4 | 0.1 | 0.2 | 0.4 |
| Kick integral field strength (T-m) | 0.04 | 0.08 | 0.16 | 0.027 | 0.1 | 0.015 | 0.06 | 0.06 | 0.12 | 0.24 |
| Min. thickness of Septum (mm) | 2 | 2 | 6 | 6 | 6 | 6 | 6 | 2 | 2 | 6 |
| Defletion angle of septa (mrad) | 26 | 35 | 26 | 26 | 26 | 26 | 26 | 26 | 35 | 26 |
| Total integral field strength of septa (T-m) | 10.44 | 14.1 | 10.44 | 6.84 | 6.84 | 3.96 | 3.96 | 15.48 | 21.15 | 15.48 |
| Beam pipe aperture (mm) | 56 | | | | | | | | | |

**Table 4.3.8.2:** Requirements of the CEPC collider injection and extraction systems

| **Parameter** | **Inj.** and ext. **(on-axis)** | **Injection (off-axis)** | **Extraction (dump)** |
|---|---|---|---|
| Kicker repetition rate (Hz) | 1000 | 1000 | 1000 |
| Kicker pulse width (ns) | 1360 | 440~2420 (Adjustable) | 440~2420 (Adjustable) |
| Kicker flat top (ns） | - | 0~1980 (Adjustable) | 0~1980 (Adjustable) |
| Kicker rise/fall time (ns) | <680 | <220 | <220 |
| Injection/extraction period (s) | 0.007 | 1.5 | 1.5 |
| Kick angle (mrad) | 0.2 | 0.1 | 0.4 |
| Kick Integral field strength (T-m) | 0.08/ 0.12 ($t\bar{t}$) | 0.04/ 0.06 ($t\bar{t}$) | 0.16/ 0.24 ($t\bar{t}$) |
| Quantity of kicker group in each subsystem | 1（shared by inj.and ext.） | 4 (4 kicker bump) | 1 |
| Thickness of Septum (mm) | 2 and 6 | 2 and 6 | 6 |
| Defletion angle of septa (mrad) | 35 | 26 | 26 |
| Integral field strength of septa (T-m) | 14.1/21.15 ($t\bar{t}$) | 10.44/15.48 ($t\bar{t}$) | 10.44/15.48 ($t\bar{t}$) |
| Quantity of septa group in each subsystem | 2（inj.and ext.） | 1 | 1 |
| Beam pipe aperture (mm） | 56 | | |

#### 4.3.8.1 *Off-axis Injection System*

The collider rings can utilize off-axis injection in W and Z modes, thanks to their large dynamic aperture (DA), which enables beam accumulation. The hardware design of the collider's off-axis injection system also permits operation in higher energy modes, preserving the potential for off-axis injection in the Higgs or $t\bar{t}$ mode.

Pulsed local bump injection is a classical off-axis injection scheme, as illustrated in Figure 4.3.8.2. Typically, 2 or 4 kickers create a local orbit bump in one turn, which enables the circulating beam's acceptance ellipse to approach the injection septum and capture the injected beam. After several synchrotron radiation damping times, the injected bunch merges with the circulating bunch upon retraction of the local bump, completing the overall accumulation injection process. To avoid distortion of the closed orbit, this injection method requires consistent pulse waveforms for the kicker magnets on both sides of the injection point. Given the maturity of the technology, the CEPC TDR adopts 4-kicker local bump scheme as the baseline.

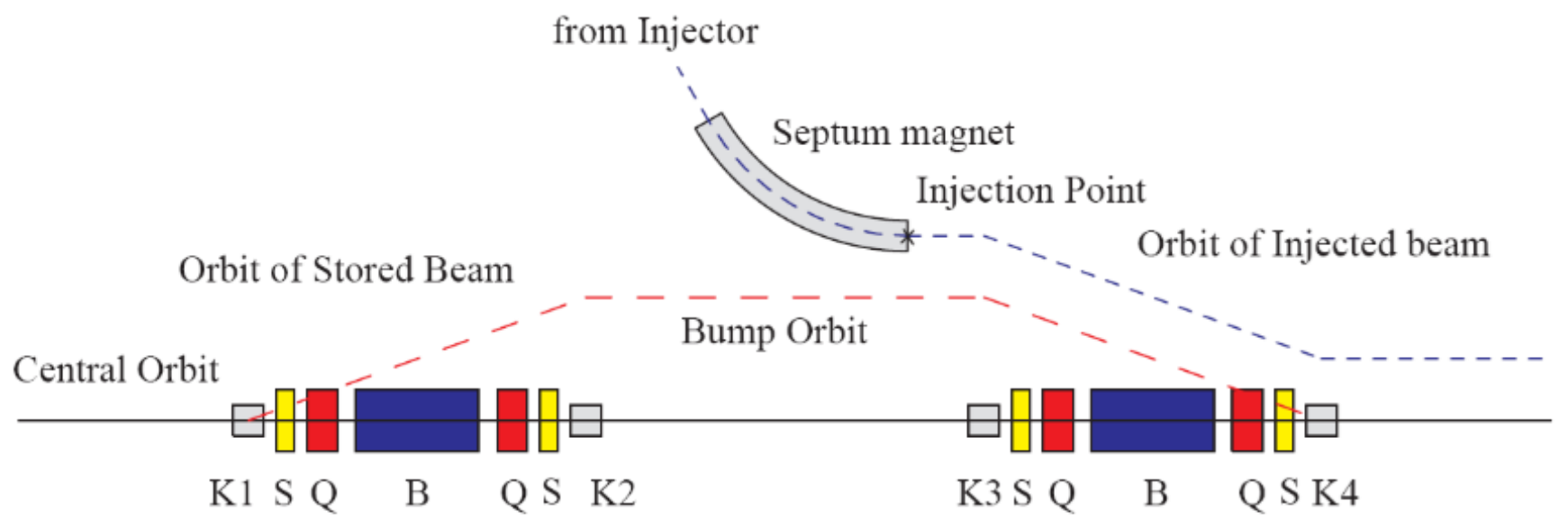

**Figure 4.3.8.2:** Pulsed local bump injection

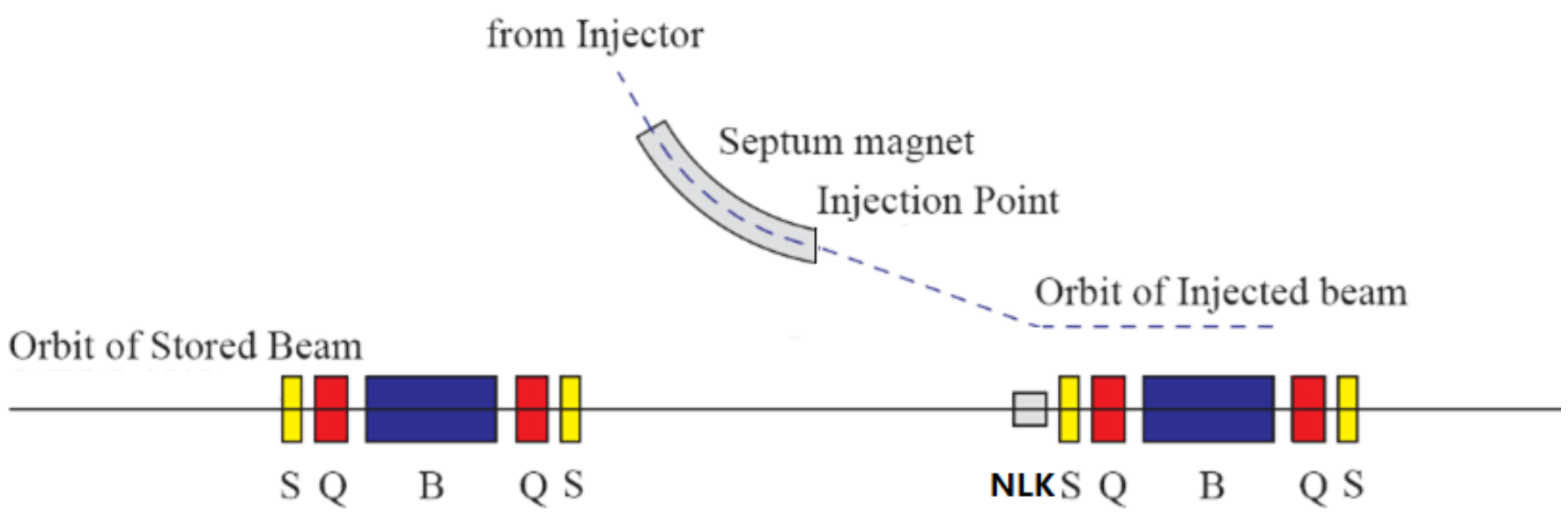

**Figure 4.3.8.3:** Non-linear kicker injection

Another off-axis injection scheme is known as pulsed multipole magnet injection or nonlinear kicker (NLK) injection. [1-5] In this scheme, a pulsed multipole magnet or a nonlinear kicker downstream of the septum magnet is used to deflect the off-axis injected beam, with minimal deflection effect on the on-axis circulating beam. This innovative injection scheme can achieve a more transparent top-up injection process. The layout of the pulsed multipole injection is illustrated in Figure 4.3.8.3. Fewer injection elements are needed, reducing the cost and the contribution to the accelerator's beam impedance. The NLK is equivalent to a septum magnet without a septum plate, making this off-axis injection method less demanding on the machine's dynamic aperture than pulsed local bump injection. It can relax constraints on the accelerator lattice design optimization and ease requirements for processing and installation errors of equipment such as magnets.

The NLK magnet requires research and development but can be considered as an alternative.

In the Z mode, where a large number of bunches are filled in both the collider and the booster rings, off-axis injection from the booster ring to the collider rings is carried out train by train at a repetition rate of 1 kHz to shorten the injection duration. According to the beam filling pattern of the rings in CEPC, the off-axis injection kickers are required to produce a specific trapezoid waveform pulse deflecting magnetic field. Typical specifications for the kicker pulse require the rise time and fall time of the pulse to be ≤ 200 ns, and the flat top width of the pulse to be ≥ 1980 ns. In the W mode, where uniform filling is adopted and the minimum bunch spacing is 220 ns, the kicker pulse bottom width is required to be less than 440 ns to achieve bunch-by-bunch deflection. Thus, to be compatible with both operation energy modes, it is necessary for the kicker pulse width to be adjustable. In fact, from the perspective of off-axis injection physics principles, the requirements for kicker pulse rise time and fall time are not critical in either the pulsed local bump injection scheme or the nonlinear kicker injection scheme, unlike the booster extraction kicker, which must strictly deflect the beam bunch-by-bunch or train-by-train. However, a more perfect trapezoid waveform kick pulse is significant in reducing the disturbance to the circulating beam orbit and suppressing the background of the beam-beam effect.

The distributed parameter type kicker magnet has superior dynamic response characteristics compared to the lumped parameter type. In a fast pulse system, the distributed parameter kicker magnet acts as a delay-line instead of an inductor, which is more conducive to the generation and transmission of trapezoidal waves. Therefore, the delay-line dipole kicker or NLK is the preferred type of kicker for the CEPC off-axis injection system.

The trapezoidal wave pulse is typically generated by a PFN (pulse-forming network) or capacitor bank with a hard switch. Due to the high repetition rate of the kicker system, which will reach 1 kHz, the CEPC kicker systems will adopt solid-state switch technology instead of thyratrons.

Due to the height difference between the Booster ring and the Collider rings in the same tunnel of the CEPC accelerator layout, the Lambertson septa magnet offers unique advantages. The Lambertson magnet is powered by a DC supply without any current flowing through the septum, resulting in a more reliable and maintainable magnet system compared to direct drive septa and eddy current septa systems, which typically rely on pulsed power supplies. However, to achieve traditional off-axis injection within the limitations of the Collider ring's dynamic aperture, the septum thickness must be less than 2 mm. It is possible to combine thin and thick septum plates, with thicknesses of 2 mm and 6 mm. In order to enable Collider off-axis injection, the septa magnets must provide a total deflection angle of 26 mrad across all energy modes.

#### 4.3.8.2 ***On-axis Injection System***

In the Higgs and $t\bar{t}$ modes, the dynamic aperture of the collider ring is smaller, making it difficult to use off-axis injection for beam accumulation. Therefore, a special on-axis swap-out injection scheme is adopted, where the booster ring acts as an accumulating ring due to its larger dynamic aperture [6-7]. The circulating beam is extracted from the collider ring by a fast kicker, which is synchronized with the injected beam from the booster ring. The injected beam then replaces the extracted beam, completing the swap-out injection process. This scheme requires a high-precision and fast kicker system, as well as a synchronization system to ensure the proper timing of the swap-out injection.

In the Higgs energy mode, the CEPC collides 268 uniformly filled bunches in half of the collider ring, with a minimum bunch spacing of 680ns. Due to the smaller beam dynamic aperture in this mode, the on-axis swap-out injection scheme is adopted for beam accumulation. During injection, each bunch in the collision ring is kicked out one by one and re-injected into the booster ring, where it merges with the corresponding bunch with full energy. After four damping times (about 200 ms), the merged bunch is extracted from the booster and injected back into the collider ring. To increase the injection rate of the collision ring and considering the beam current threshold of the booster, seven bunches are continuously extracted at a repetition rate of 1kHz during the first swap-out injection. The number of bunches extracted at one time is gradually increased in subsequent swap-out injections until all 268 bunches are replaced, taking about 4.5 seconds in total.

The on-axis swap-out injection scheme for the collider requires the bottom width of the kicker pulse to be less than 1.36 μs, and the maximum repetition rate is 1 kHz. For this purpose, lumped parameter kicker magnets with ferrite cores and solid-state pulsed power supplies are the preferred solutions for the collider on-axis kicker system.

The Lambertson septa is also suitable for the on-axis injection system of the collider ring, and it is required to provide a total deflection angle of 35 mrad in Higgs mode. The design incorporates a combination of thin and thick septum plates with thicknesses of 2 mm and 6 mm, respectively.

#### 4.3.8.3 *Beam Dump System*

To ensure the safety of the vacuum components in the event of a machine failure, the CEPC Collider ring incorporates a beam dump system capable of swiftly extracting high-energy positron and electron beams from the ring. Similar to the off-axis injection system, the beam dump system utilizes a trapezoid-wave delay-line dipole kicker to deliver the necessary kick for deflecting the beam with an angle of 0.4 mrad in a single turn. Additionally, a set of DC Lambertson septa with a septum thickness of 6 mm is required to steer the beam, resulting in a total deflection angle of 26 mrad across all energy modes.

Within the transport lines leading to the dump, there is a set of dilution kickers designed to decrease the charge density of the extracted bunches. These kickers employ a resonance dilution kicker with a length of 10 m, capable of generating an oscillating magnetic field of 150 Gauss with a 25 μs oscillation period.

#### 4.3.8.4 *Delay-Line Dipole Kicker*

The main parameters of the delay-line dipole kicker magnet for the CEPC off-axis injection system are shown in Table 4.3.8.2. The parameters for the beam dump system are the same as the collider off-axis injection system, except that the total deflection angle is 4 times that of off-axis injection. Therefore, 4 identical kicker magnets are required for each beam dump system.

**Table 4.3.8.2:** Parameter of CEPC collider ring off-axis injection dipole kicker system.

| **Parameter** | **Unit** | **Off-axis Injection** |
|---|---|---|
| Quantity (4 kicker bump, including e-/e+ ring) | - | 8 (2×4) |
| Type | - | In-air delay-line dipole kicker |
| Deflect direction | - | Horizontal |
| Beam Energy | GeV | 120 |
| Deflect angle | mrad | 0.1 |
| Magnetic effective length | m | 1 |
| Magnetic strength | T | 0.04 |
| Integral magnetic strength | T·m | 0.04 |
| Clearance region (H×V) | mm | 56 × 56 |
| Good field region (H×V) | mm | 50 × 50 |
| Field uniformity in good field region | - | ±1.5% |
| Repetition rate | Hz | 1000 |
| Amplitude repeatability | - | ±0.5% |
| Pulse jitter | ns | ≤ 5 |
| Pulse width pulse (5%-5%) | ns | 440~2420 (Adjustable) |
| Pulse flat top | ns | 0~1980 (Adjustable) |
| Pulse Tr/Tf (5%-95%) | ns | <220 |
| Pulse waveform (Trapezoid) | - | 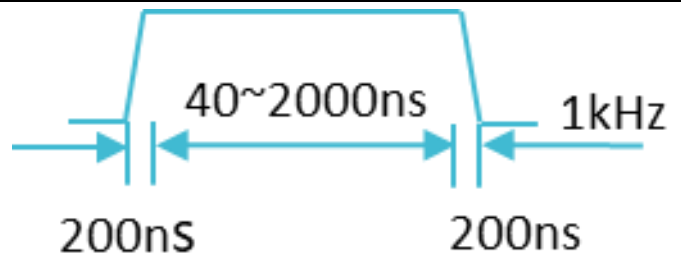 |

When designing the kicker system, it is important to consider the rise time of the delay-line kicker magnet, which is comprised of the rise time of the electrical pulse generated and the filling time (delay time) of the magnet. Two common semiconductor solid-state switches, IGBT and MOSFET, have been investigated and have a typical turn-on time (Tr) and turn-off time (Tf) of not more than 60 ns, as shown in Table 4.3.8.3. Therefore, in the overall design of the kicker system, 80ns is reserved for the rise time of the electrical pulse, while the filling time of the kicker magnet should be controlled within 120 ns to ensure optimal performance.

**Table 4.3.8.3:** Typical performance of the MOSFET and IGBT

| | MOSFET (Cree C2M0045170D) | IGBT ( Infineon FZ600R12KS4) |
|---|---|---|
| Vbr | 1700 V | 1200 V |
| Td (on) | 65 ns | 10 ns |
| Tr | 20 ns | 60 ns |
| Td (off) | 48 ns | 530 ns |
| Tf | 18 ns | 30 ns |

Based on circuit simulation by Pspice, it has been found that a distributed-parameter kicker is better suited for a trapezoid-wave kicker system than a lumped-parameter kicker. Figures 4.3.8.4, 4.3.8.5, and 4.3.8.6 clearly show that the distributed-parameter kicker is better able to generate trapezoid pulse magnetic fields with steep edges, meeting the requirements of the system. Furthermore, the distributed-parameter kicker in the trapezoid-wave discharge system acts like a transmission line with impedance matching, allowing for the output load of the pulsed power supply to be purely resistive. This is essential for generating and transmitting regular trapezoidal wave pulses.

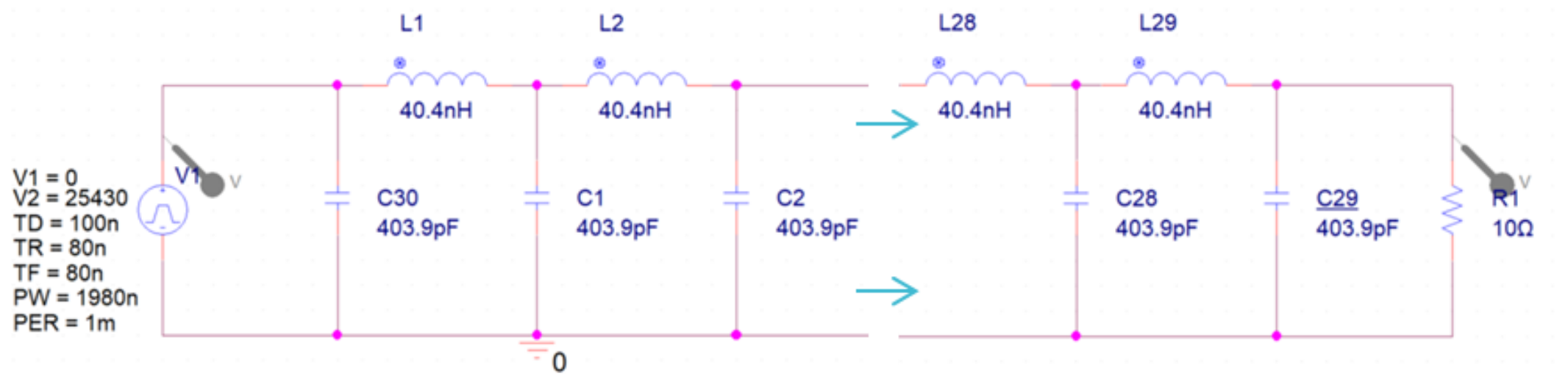

**Figure 4.3.8.4:** The PSpice circuit model of distributed parameter kicker system

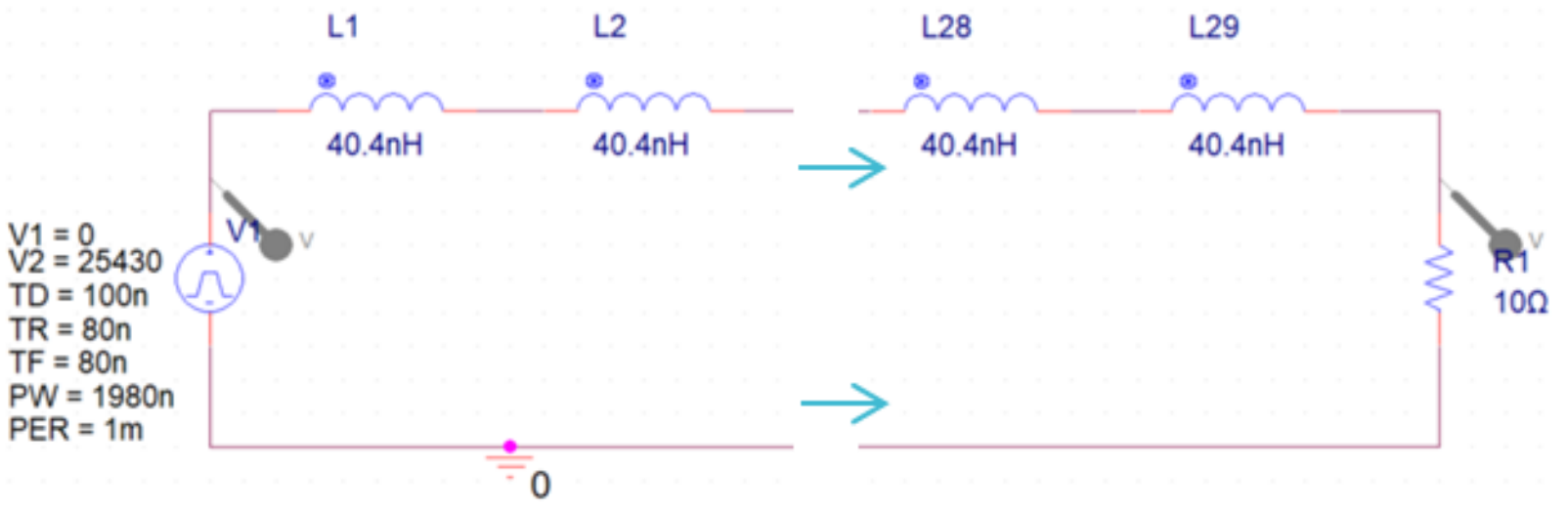

**Figure 4.3.8.5:** The PSpice circuit model of lumped parameter kicker system

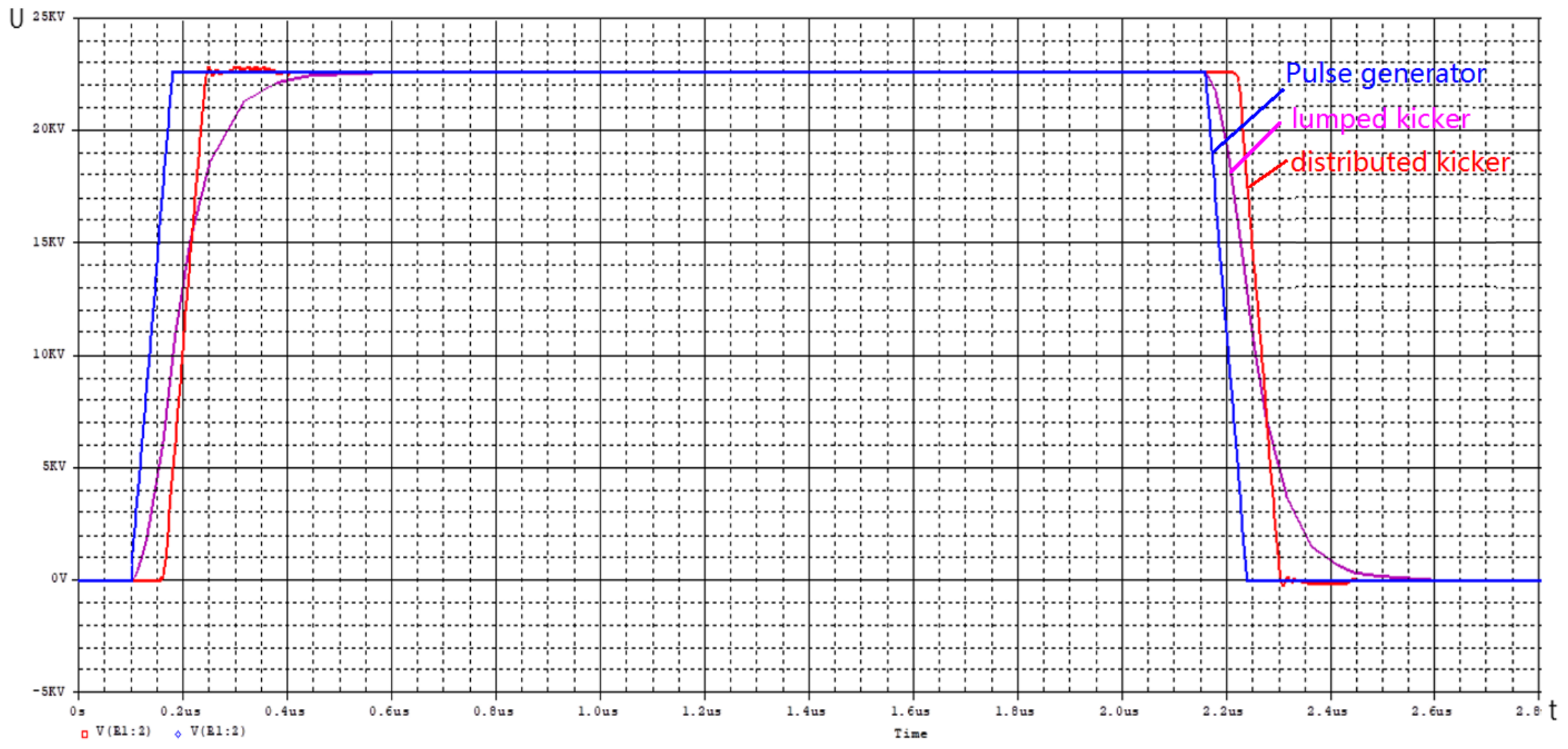

**Figure 4.3.8.6:** Simulation result comparison of distributed kicker and lumped kicker system

To simplify the design of the delay-line kicker magnet, the CEPC project intends to utilize an in-air magnet structure. This approach requires the use of a ceramic vacuum chamber with a metallic coating applied to the inner wall surface. Given that the collider ring adopts a circular vacuum chamber with a diameter of 56 mm, the inner contour of the ceramic vacuum chamber also follows a circular shape with the same diameter. The outer contour, on the other hand, is octagonal, which facilitates positioning during the coating and assembly processes. The profile of the ceramic vacuum chamber is depicted in Figure 4.3.8.7. The material chosen for the ceramic vacuum chamber is 99% alumina oxide ($AL_2O_3$) ceramic.

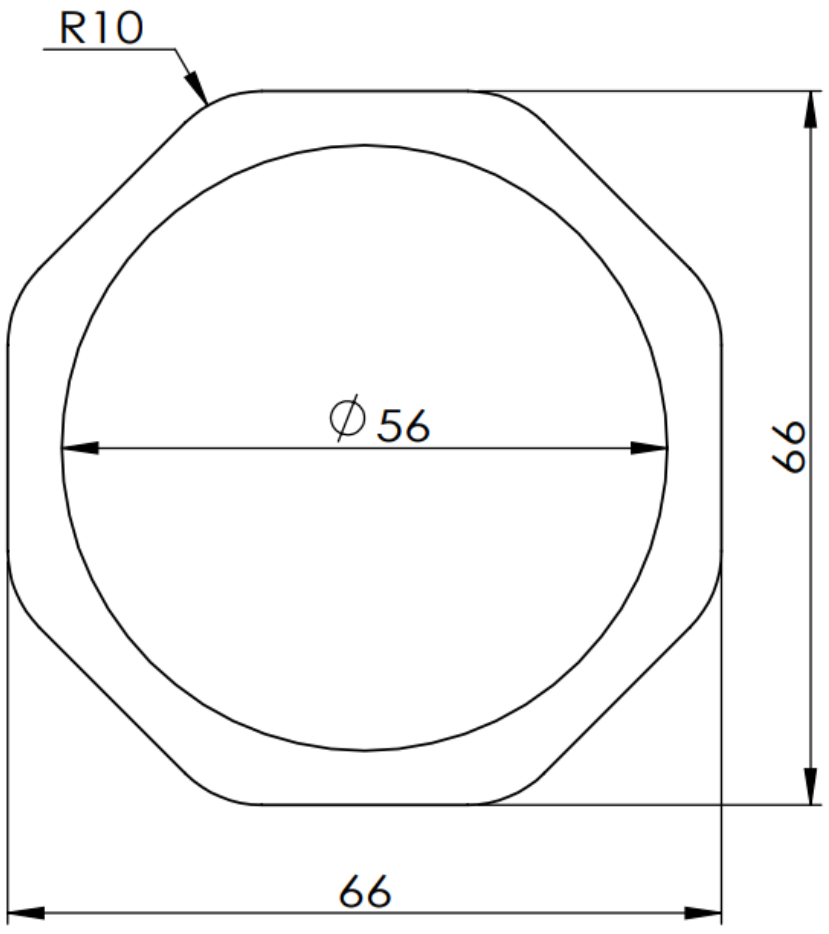

**Figure 4.3.8.7:** The profile of the kicker ceramic vacuum chamber (unit:mm)

The TiN-Ag metallic film will be coated using magnetron sputtering to shield the beam from high frequency wave fields and lower beam impedance. However, research has shown that continuous coating can cause waveform distortion of the pulsed magnetic field due to the eddy current effect. Therefore, a metallic film with a certain pattern can weaken the shielding effect on the pulsed magnetic field, which typically has a spectrum width of less than 100 MHz. Patterns such as comb, ladder, and detour, as shown in Figure 4.3.8.8, were investigated. Simulation results by COMSOL showed that the film with ladder pattern and square resistance of 0.172 Ω is the most effective, as shown in Figure 4.3.8.9. This pattern achieves a balance between the beam's high-frequency wake field and pulsed magnetic field.

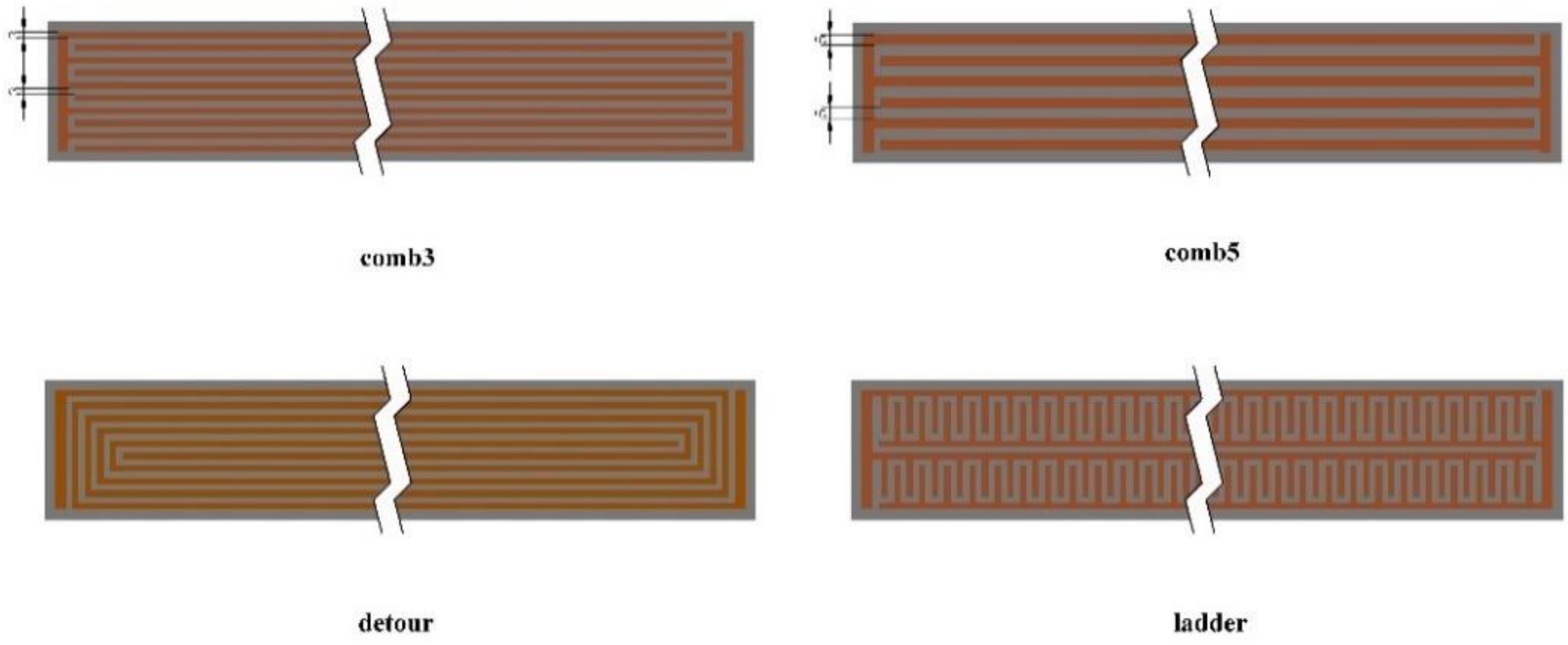

**Figure 4.3.8.8:** Typical film pattern (comb3/5: strip width is 3/5 mm).

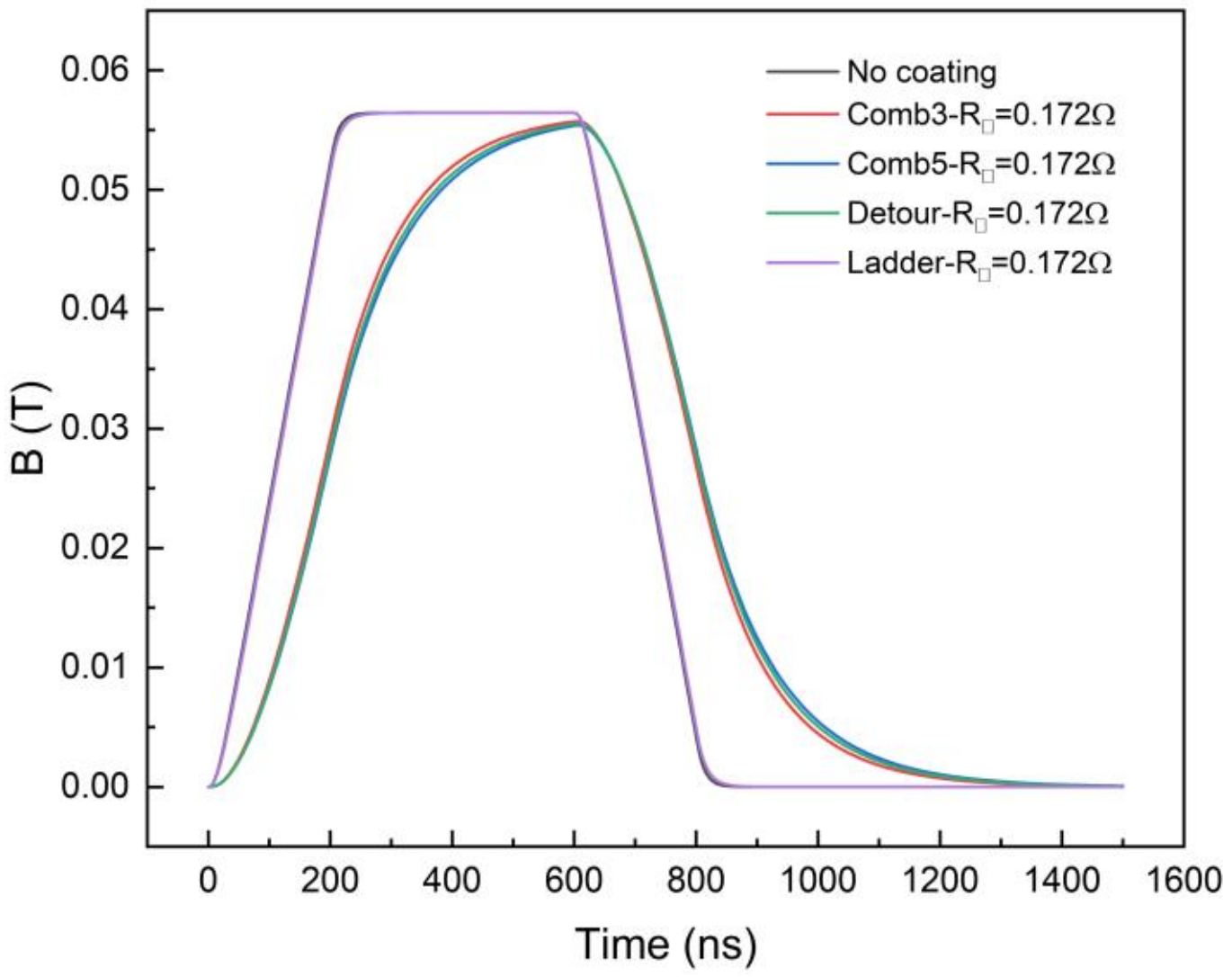

**Figure 4.3.8.9:** Pulsed magnetic field waveform comparison with different films

A dual-C type delay-line dipole kicker magnet structure is adopted for CEPC [8]. The simplified structure of the magnet is shown in Figure 4.3.8.10. The introduced stray capacitance between the +/– high voltage plates and ground potential plates can be used to adjust the characteristic impedance of the magnet. To produce a dipole field, the magnet is excited by differential mode. The equivalent circuit of the dual C-type delay-line kicker is shown in Figure 4.3.8.11.

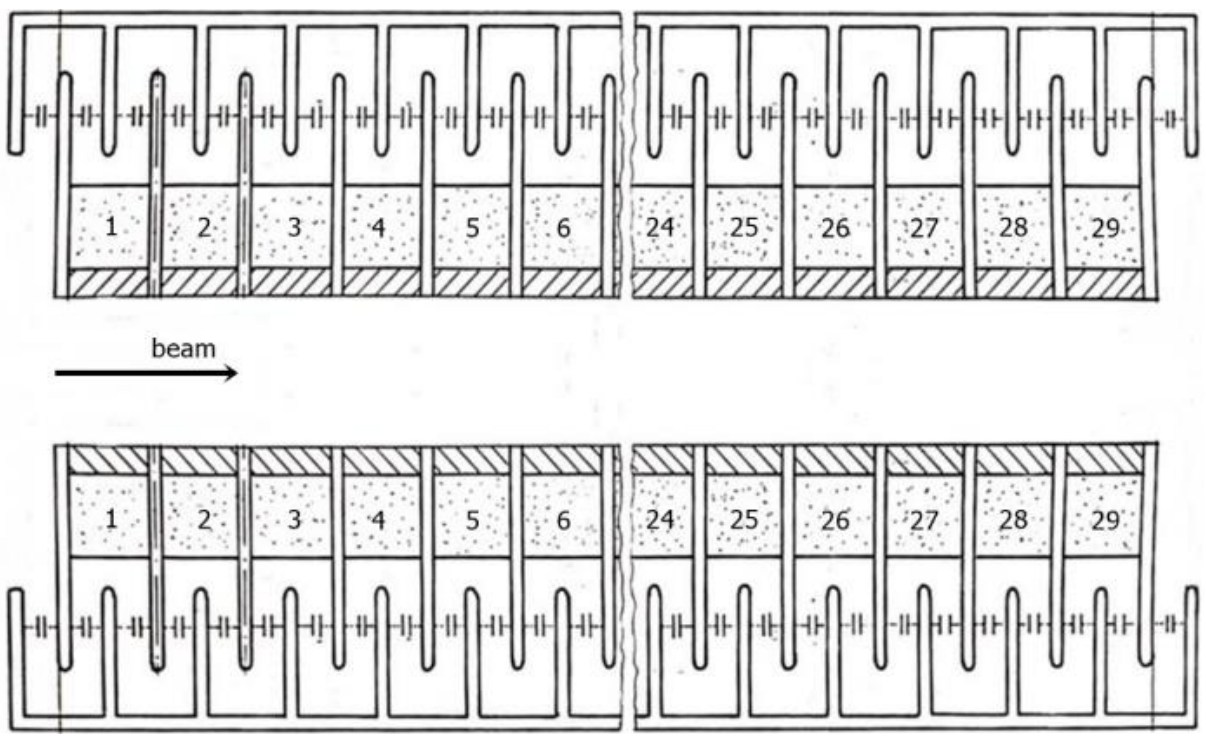

**Figure 4.3.8.10:** Simplified structure of dual C type delay-line kicker magnet

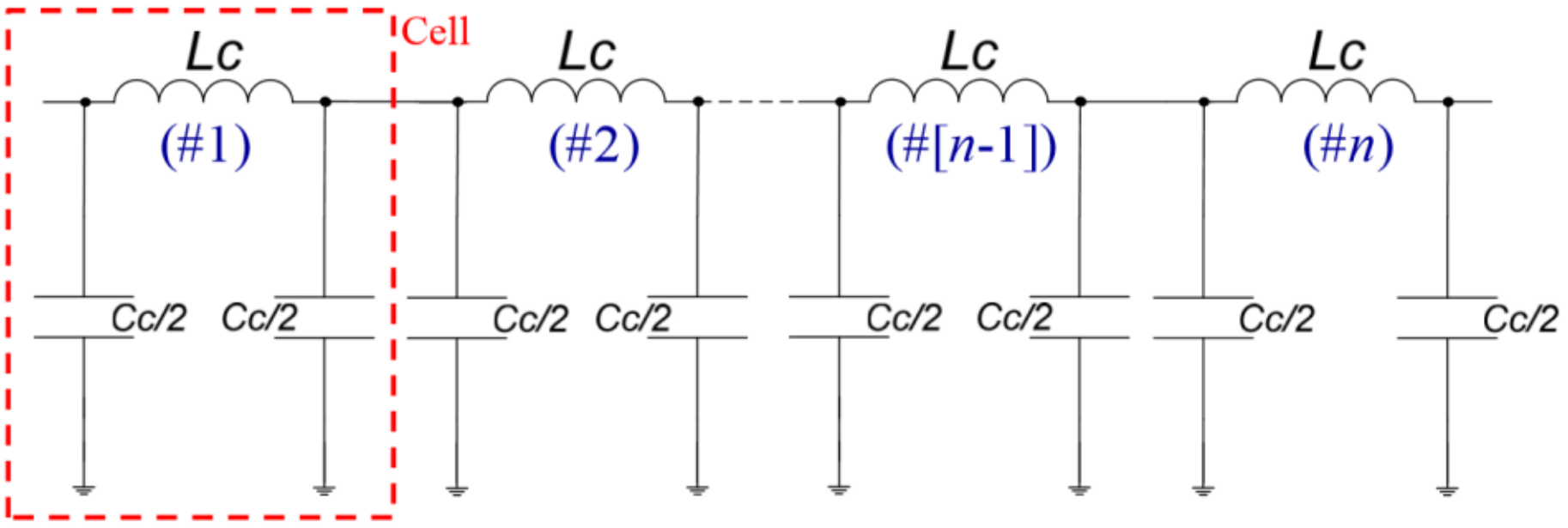

**Figure 4.3.8.11:** Equivalent circuit of dual C-type delay-line dipole kicker

**Table 4.3.8.4:** External parameter input for the CEPC Collier off-axis injection kicker magnet

| Parameters | Value |
|---|---|
| Integral magnetic field strength K (T-m) | 0.04 |
| Rise/fall time (ns) | 120 |
| Inner aperture of ceramic vacuum chamber (mm) | 56 |
| Outer aperture of ceramic vacuum chamber (mm) | 66 × 66 |
| Aperture of magnet (mm) | 100 × 80 |

The input parameters required for the CEPC dual C-type delay-line dipole kicker magnet design are listed in Table 4.3.8.4. Through detailed physical design and calculations, the main parameters of delay-line dipole kicker for CEPC off-axis injection system are listed in Table 4.3.8.5.

**Table 4.3.8.5:** Main parameters of delay-line dipole kicker

| Parameters | Value |
|---|---|
| Aperture of magnet (mm) | 100 × 80 |
| Longitudinal length of magnet cell (mm) | 36 |
| Cell number of magnet | 26 |
| Differential impedance of magnet (Ω) | 12.5 |
| Length of magnet (mm) | 942 |
| Total mechanical Length of magnet (mm) | 1018 |
| Inductance of magnet cell (nH) | 56.6 |
| Total inductance of magnet (nH) | 1471.6 |
| Capacitance of magnet cell (pF) | 362 |
| Total capacitance of magnet (nF) | 9.412 |
| Magnetic strength (Gs) | 425 |
| Exciting current of magnet (A) | 2703 |
| Differential voltage of magnet (V) | 33791 |

Significantly, the capacitance of the magnet cell should be maximized in order to achieve the lowest possible impedance. In this particular case, the characteristic impedance of the magnet structure is 12.5 Ω, necessitating a capacitance of 362 pF for each magnet cell. However, this results in a very large size for the capacitor electrode plate, as shown in Figure 4.3.8.12. This configuration is clearly unsuitable for a double-ring collider where the beam pipes of both rings are in close proximity. To address this issue, the capacitance can be reduced to 1/6 by filling a dielectric material, such as $Al_2O_3$ ceramics with a typical relative permittivity ($\varepsilon_r$) of 10, between the capacitor electrodes.

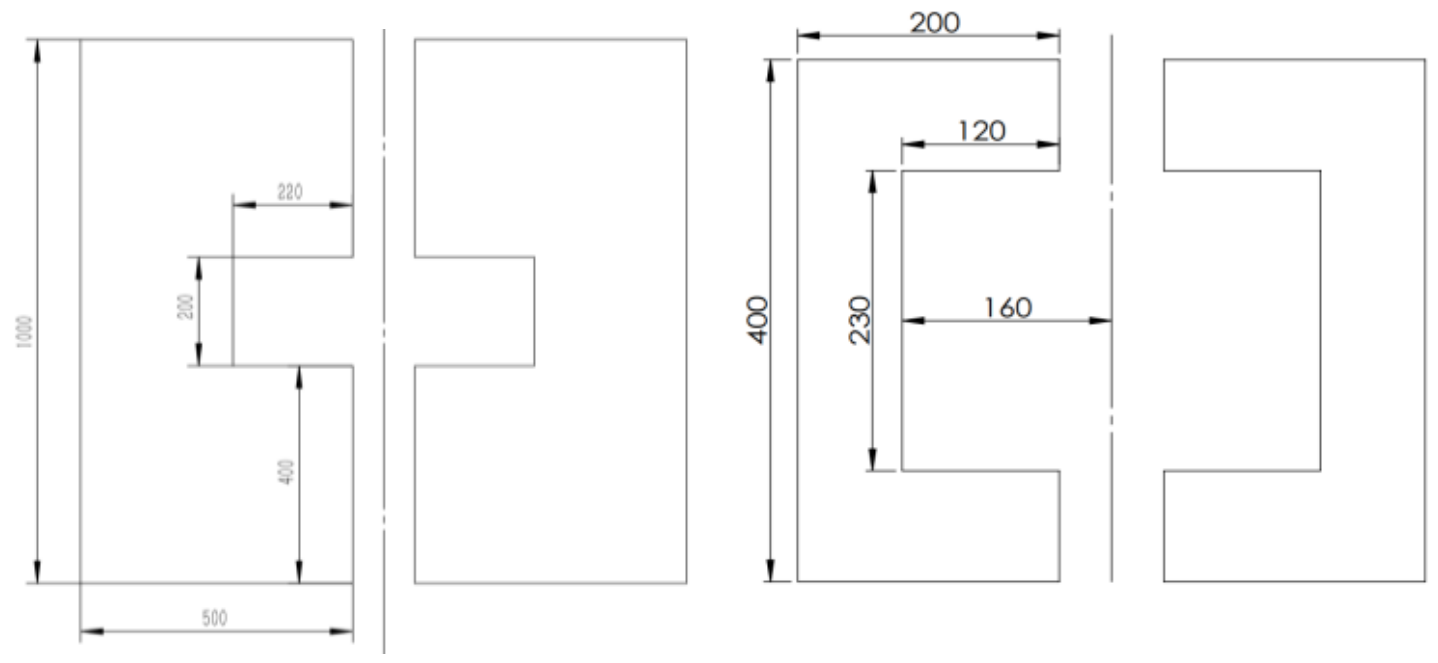

**Fig 4.3.8.12:** Overlapping area of capacitor plates (left: no filling, right: filling ceramic).

The OPERA 2D program is utilized to simulate the electromagnetic field, and the magnetic field line distribution is displayed in Figure 4.3.8.13. Based on the simulation, the kicker magnet's unit length inductance is got as $L_D = 1.51 \times 10^{-9}$ H/mm, slightly lower than the value calculated using empirical formula. Figure 4.3.8.14 displays the magnetic field strength distribution along the X-axis in the range of (–50, 50) mm, while Figure 4.3.8.15 shows the distribution in the range of (–25, 25) mm. The peak field strength is approximately 421 Gs, which is in close agreement with the target field strength of 425 Gs.

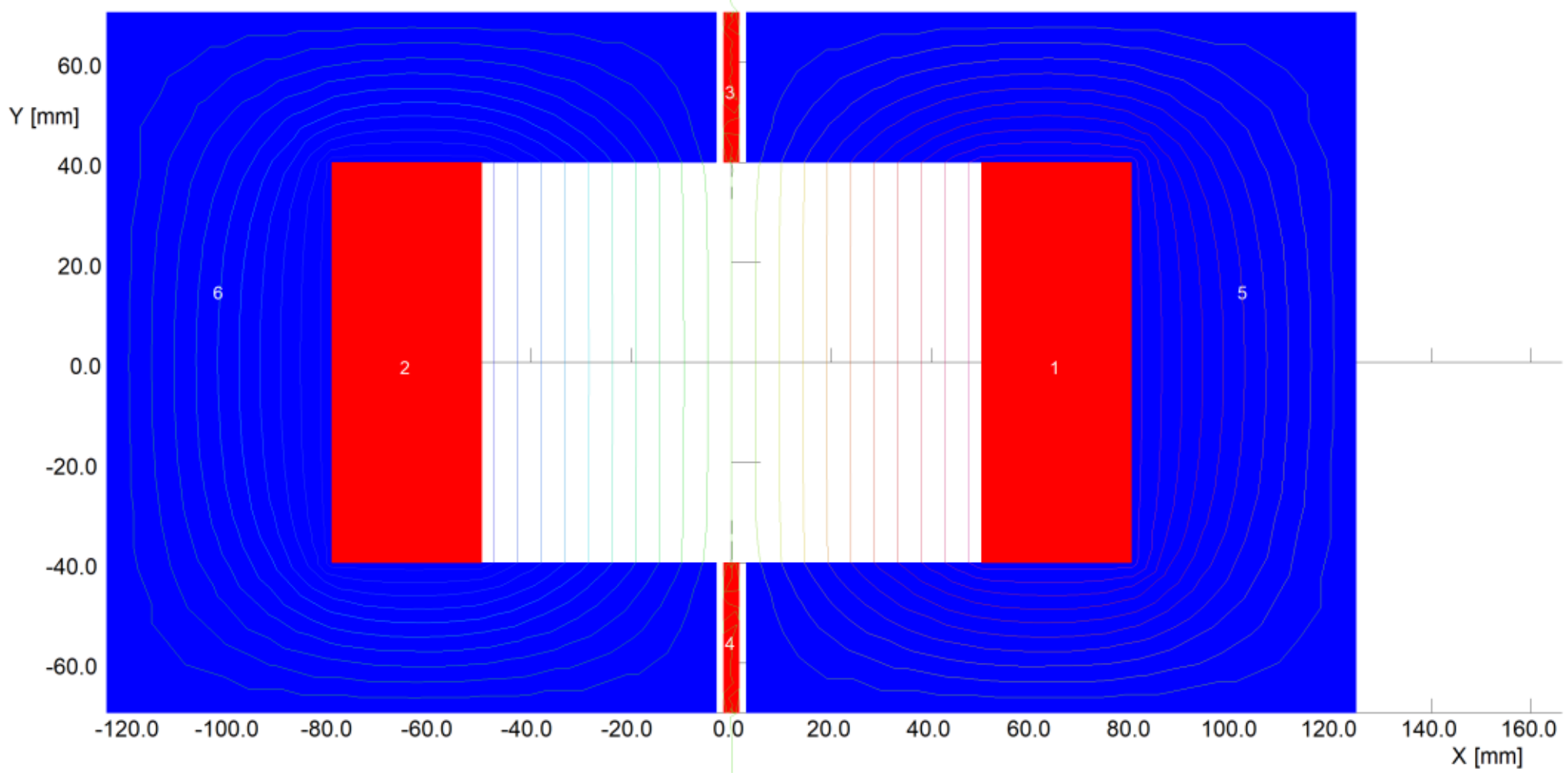

**Figure 4.3.8.13:** Distribution of magnetic field of dipole kicker in Opera2D

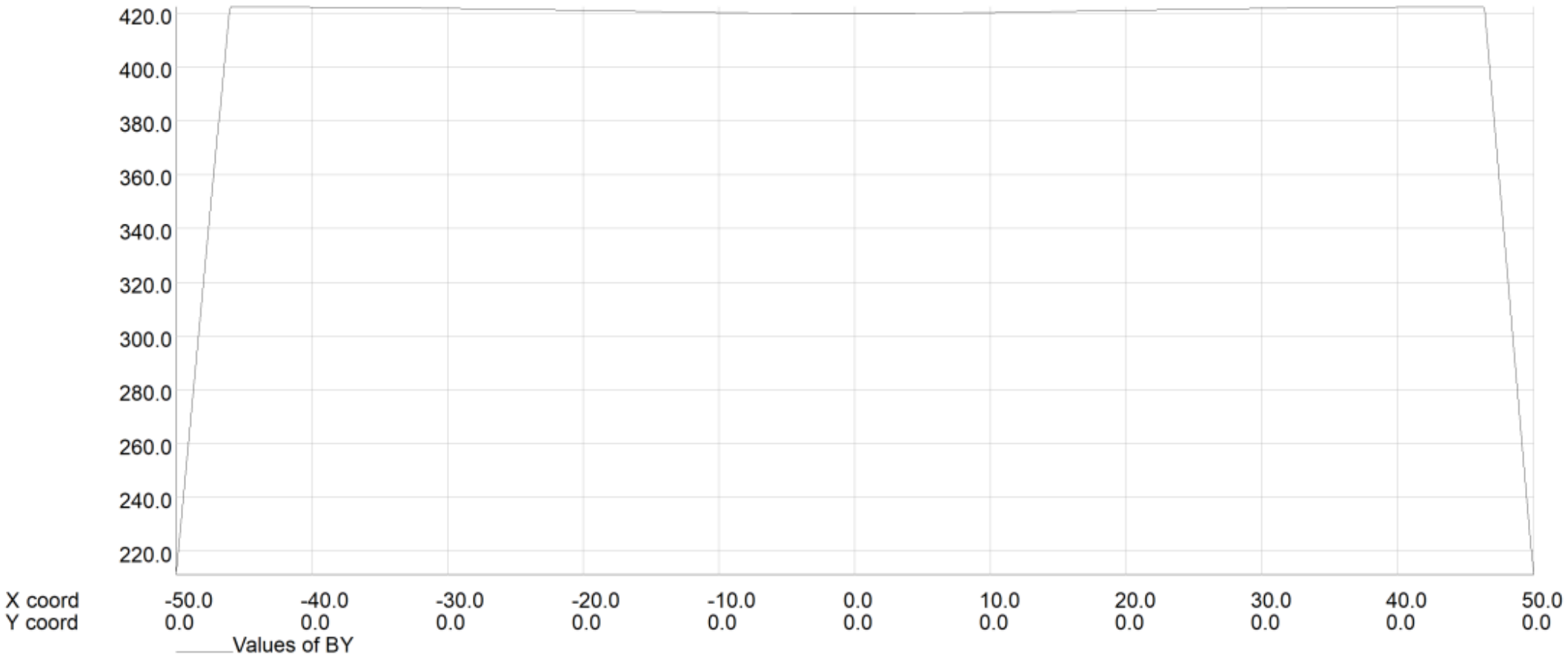

**Figure 4.3.8.14:** Magnetic field strength distribution in x = (–50, 50) mm

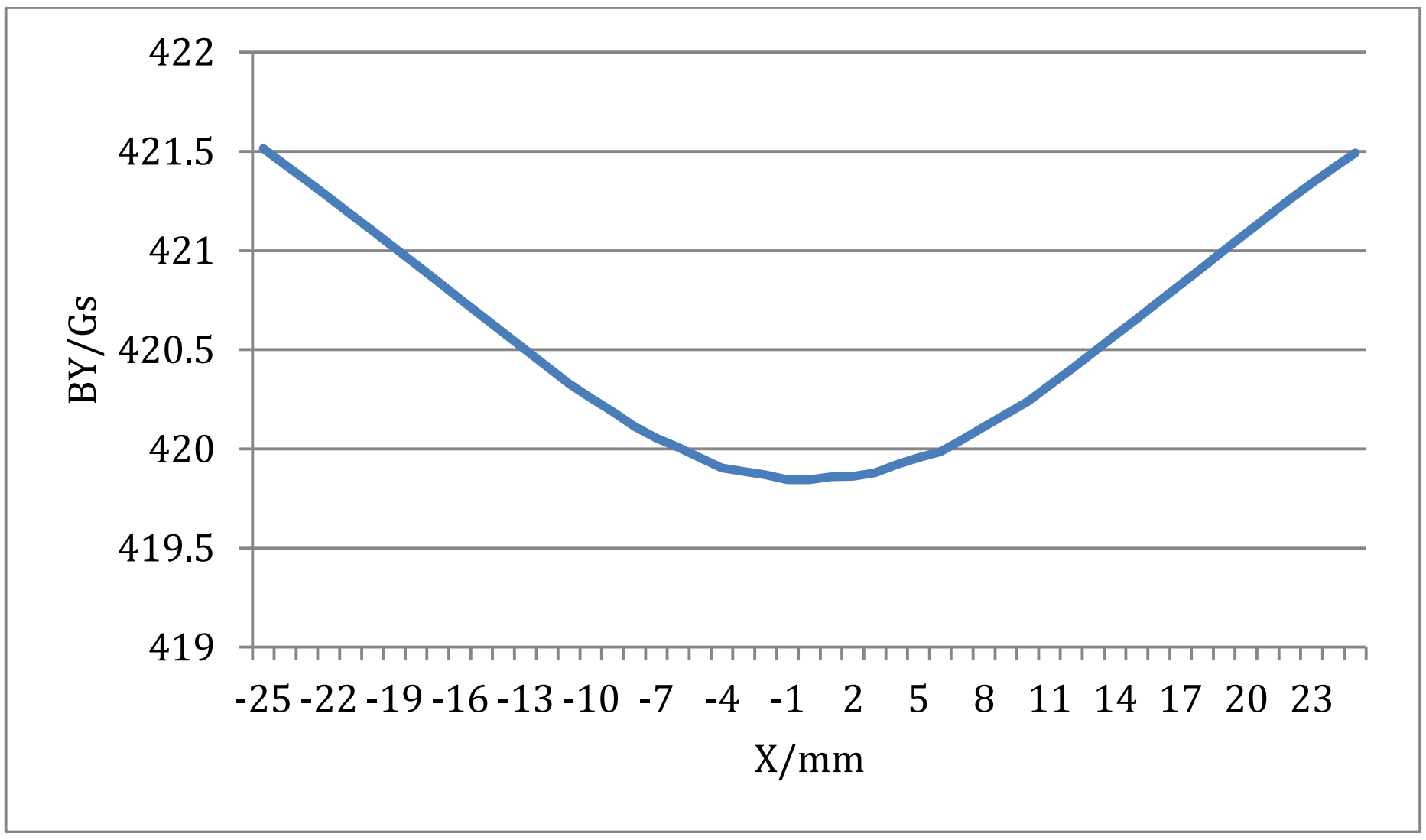

**Figure 4.3.8.15:** Magnetic field strength distribution in x = (–25, 25) mm

The kicker circuit model in PSpice is constructed using the parasitic parameters obtained from the physical design of the magnet, as illustrated in Figure 4.3.8-16. In this setup, the pulse generator is configured with $U_{max}$ = 33791 V, $T_D$ = 100 ns, $T_R$ = 80 ns, $T_F$ = 80 ns, $P_W$ = 1980 ns. The simulation result depicted in Figure 4.3.8.17 indicates that the magnet $U_{MAX}$ = 34412 V, $T_r$ = 203 ns, $T_F$ = 185 ns, which basically satisfies the specifications.

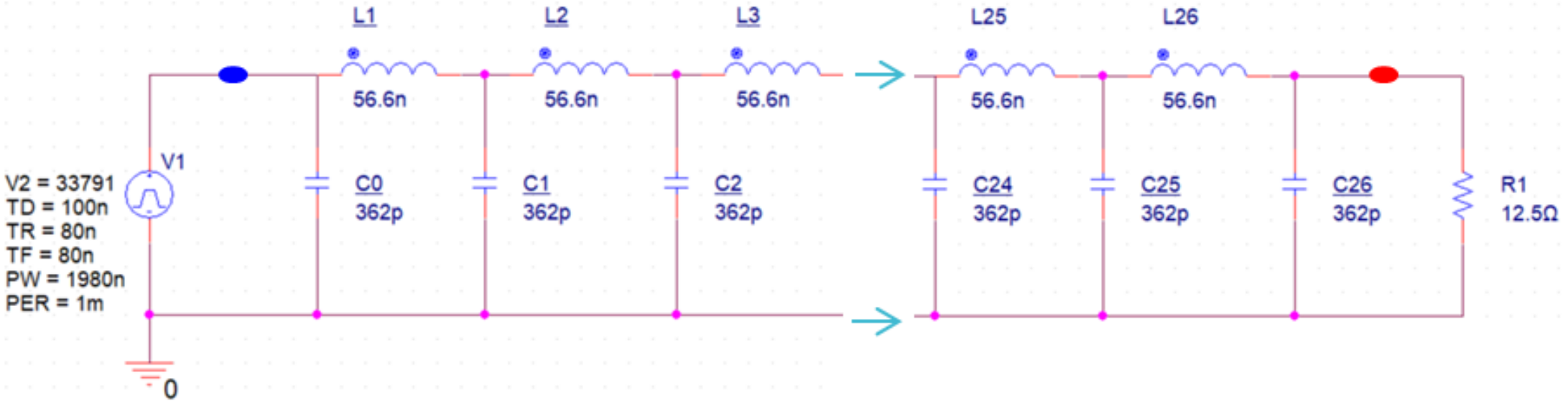

**Figure 4.3.8.16:** Delay-line dipole kicker PSpice circuit model

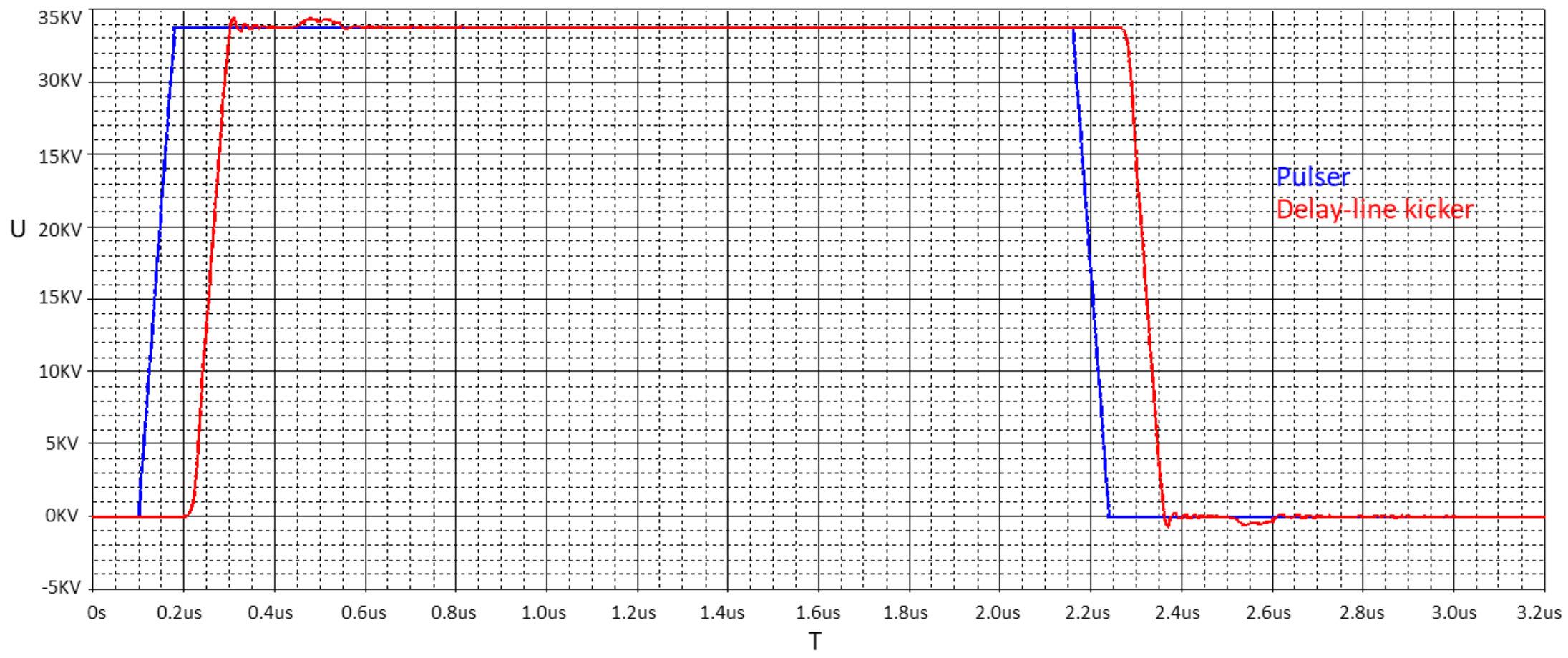

**Figure 4.3.8.17**: Delay-line dipole kicker PSpice simulation result

Figure 4.3.8.18 illustrates the 3D structure model of the CEPC delay-line dipole kicker. Additionally, the magnet profile displayed in Figure 4.3.8.19 provides further insight into the magnet's structural details.

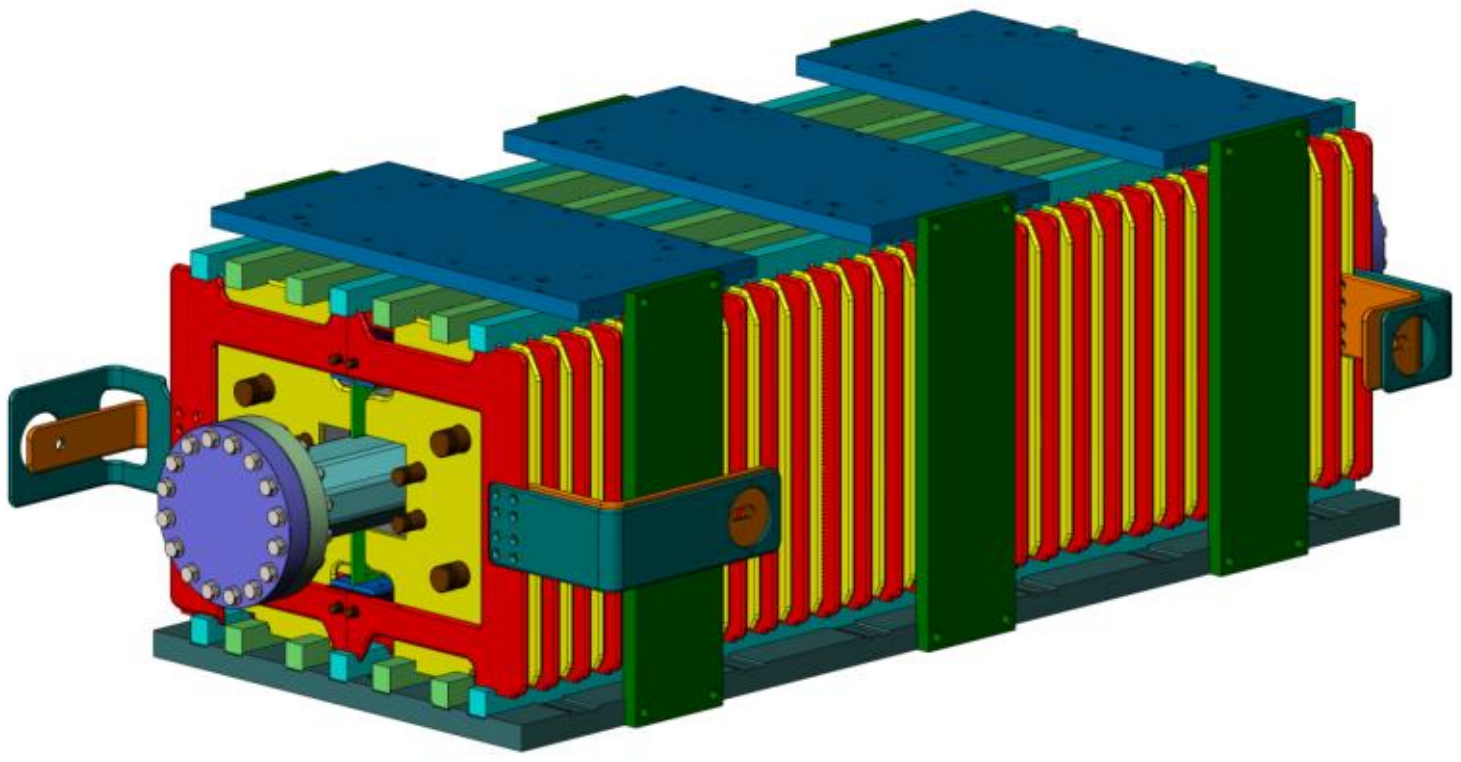

**Figure 4.3.8.18:** 3D structure model of in-air delay-line dipole kicker for CEPC

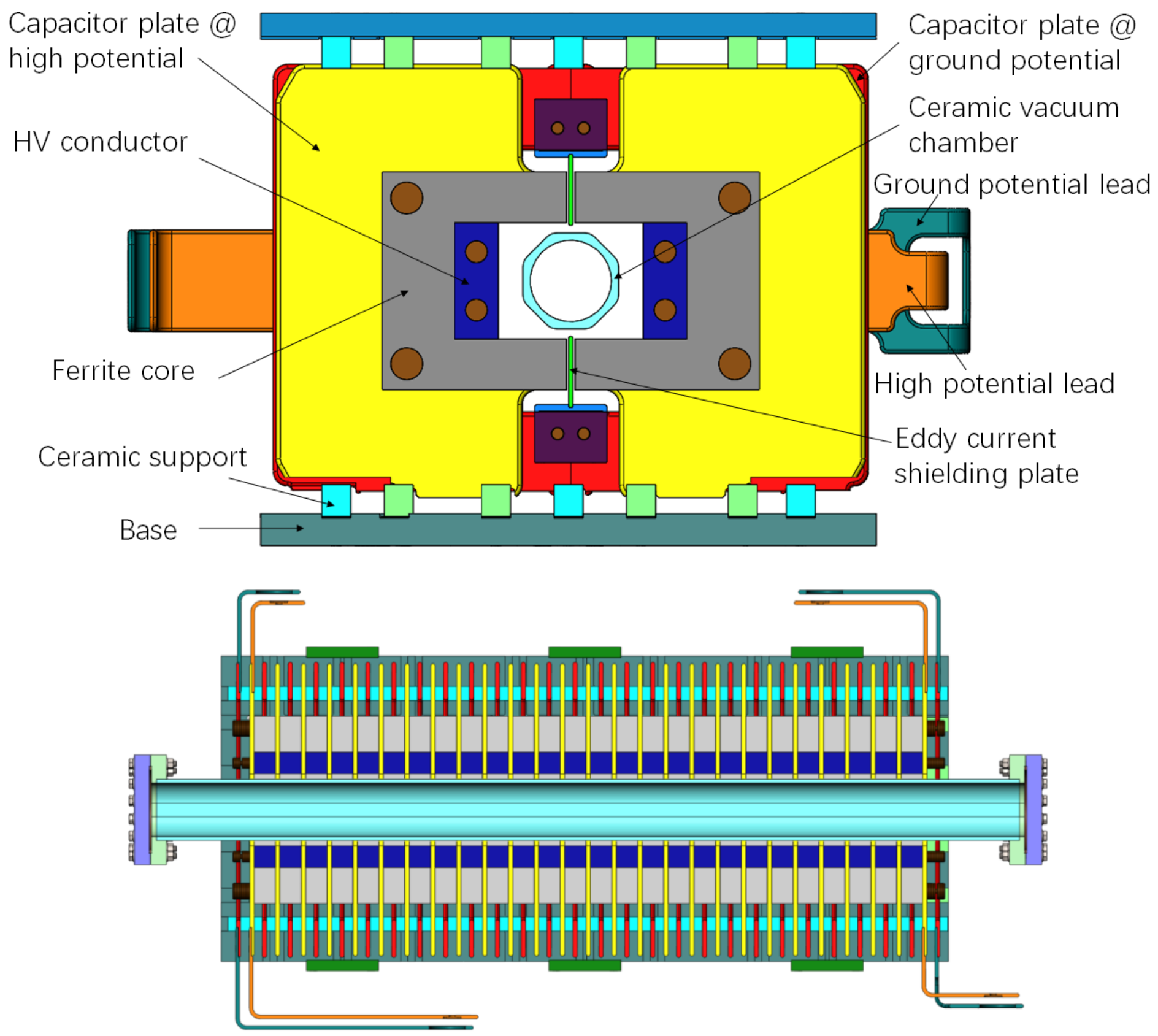

**Figure 4.3.8.19**: Profile of in-air delay-line dipole kicker for CEPC

### 4.3.8.5 *Delay Line Nonlinear Kicker*

Nonlinear kicker (NLK) injection is a recently developed injection scheme derived from pulsed multipole injection. The ideal field distribution necessary for NLK injection is depicted in Figure 4.3.8.20.[5] The stored beam travels through the magnet's center, where the magnetic field is zero, without being disturbed, while the injected beam is located in the strong field region, where it is deflected into the acceptance of the storage ring. Achieving perfect zero field with zero gradient at the magnet's center is difficult as a high field is required to deflect the injected beam. The field distribution of NLK is similar to that of a septum magnet, with no actual septum board separating the strong field and zero field regions. Additionally, the NLK is triggered by a pulser, and the injected beam should be deflected only once during the first turn. The NLK pulse waveform's fall time must be less than one revolution period of the storage ring. Therefore, the NLK is a new injection component with features similar to both a kicker and a septum. However, due to technical immaturity, the NLK injection scheme requires hardware research and development.

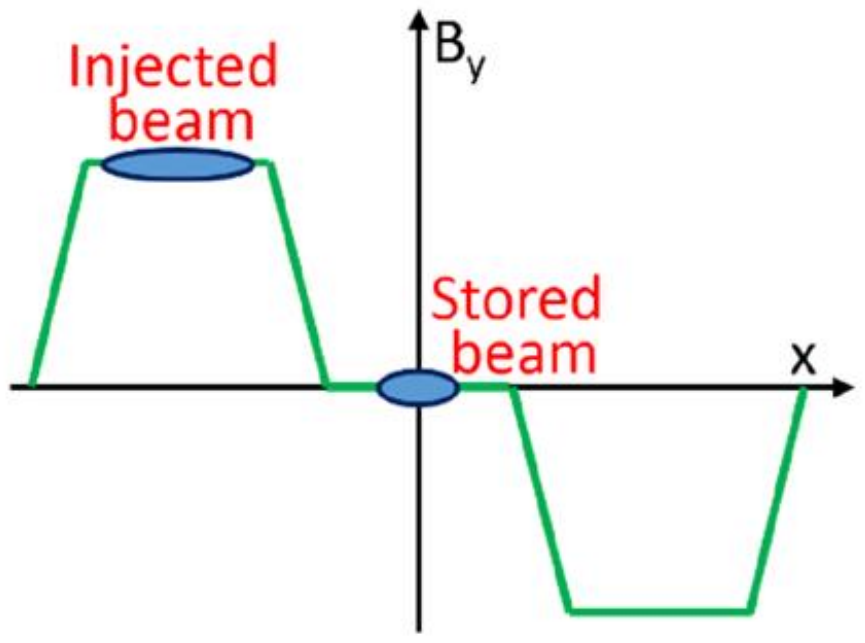

**Figure 4.3.8.20:** Ideal nonlinear kicker field distribution

In principle, the central zero field area of the nonlinear kicker can be achieved in two ways. One method is to cancel the central vector field through the reverse excitation of two groups of centrally symmetric windings, as demonstrated in Figure 4.3.8.21. The other approach is to create a coaxial structure in which four rods carrying positive current are located in the inner area, and another four rods carrying negative current are located in the outer area, as shown in Figure 4.3.8.21(f) and (g). This design produces a coaxial structure in which the field at the center is canceled to zero.

Another approach to achieving a zero-field area in the nonlinear kicker is through the use of eddy current shielding, as illustrated in Figure 4.3.8.22. A traditional ferrite core kicker with two plates can produce a quadrupole field by exciting in common mode instead of differential mode, as shown in Figure 4.3.8.22(b). To further reduce the amplitude and gradient of the magnetic field in the center of the magnet, an improved eddy current shielding plate is introduced into the window of the ferrite core, as shown in Figure 4.3.8.22(c). Compared to the eight-rod NLK without an iron core, the NLK with a ferrite core offers several advantages, including a simpler winding structure, smaller magnet inductance (which improves pulse speed), greater excitation efficiency, and lower requirements for the shape and position errors of the electrodes.

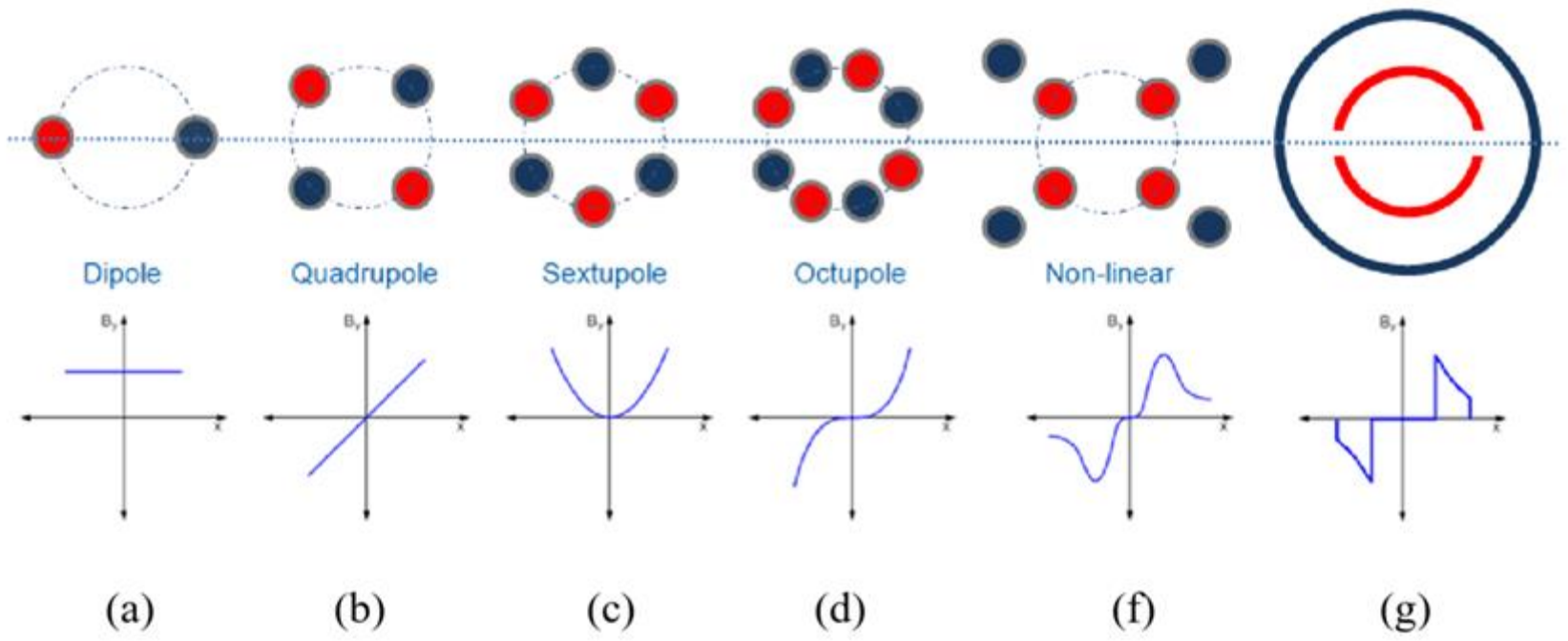

**Figure 4.3.8.21:** Evolution from pulsed Multipole to NLK (red: +direction, blue: –direction).

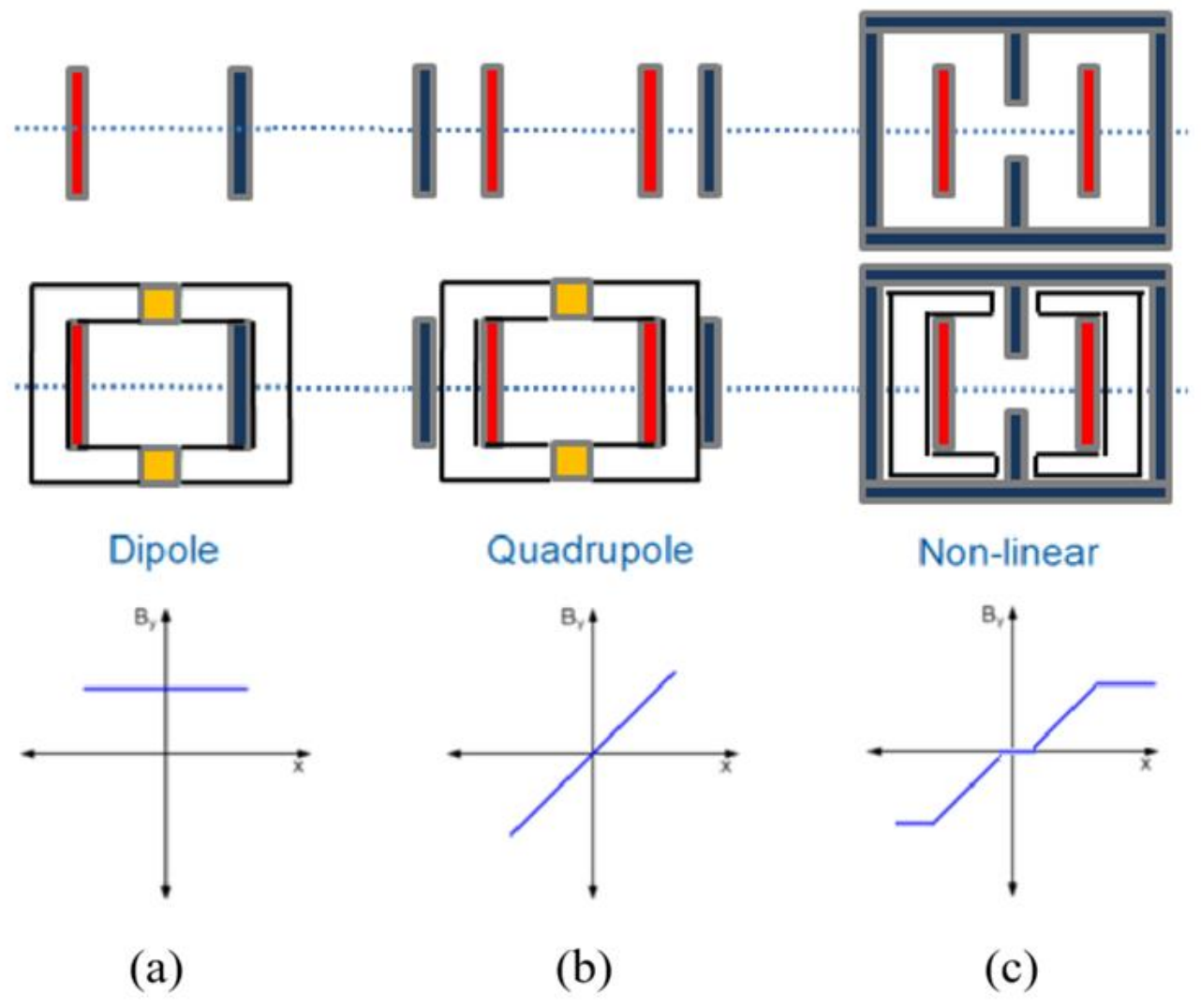

**Figure 4.3.8.22:** Evolution from dipole kicker to NLK

Due to the challenges of achieving a perfect trapezoidal wave pulse in a pulse discharge system when dealing with inductive loads, a novel distributed parameter NLK, known as the delay-line NLK, has been proposed for off-axis injection in the CEPC collider. The design methodology and mechanical structure of the delay-line NLK are similar to those of the dipole delay-line kicker described in the previous section, with the key difference being in the eddy current shield plate design and excitation mode.

**Table 4.3.8.6:** Bunch size and dynamic apertutre (DA) of the Colliders in different energy mode

| Parameter | $t\bar{t}$ | H | W | Z |
|---|---|---|---|---|
| Emittance (nm·rad) | 2.4 | 1.21 | 0.54 | 0.18 |
| Coupling | 0.3% | 0.2% | 0.3% | 2.2% |
| σ(x) (mm) | 2.08 | 1.48 | 0.99 | 0.57 |
| σ(y) (mm) | 0.042 | 0.025 | 0.02 | 0.032 |
| DA(x) (mm) | 20.8 | 13.32 | 17.82 | 13.11 |
| DA(y) (mm) | 1.09 | 0.4 | 0.54 | 0.74 |

The bunch size and dynamic aperture of the CEPC collider rings at different energy modes are critical parameters for the design of the NLK eddy current shielding plate. These parameters are listed in Table 4.3.8.6. The eddy current shielding plate's profile, which is crucial to achieve the required nonlinear field for off-axis injection, is shown in Figure 4.3.8.23. The ideal geometric parameters for the plate, including its shape, radius, width, height, thickness, spacing, material, and other parameters, are determined through careful adjustment and are summarized in Table 4.3.8.7. These parameters are optimized to ensure that the nonlinear field meets the requirements of off-axis injection in all three energy modes. The OPERA-2D model of the NLK is illustrated in Fig. 4.3.8.24, and the horizontal magnetic field distribution on the central plane is displayed in Fig. 4.3.8.25.

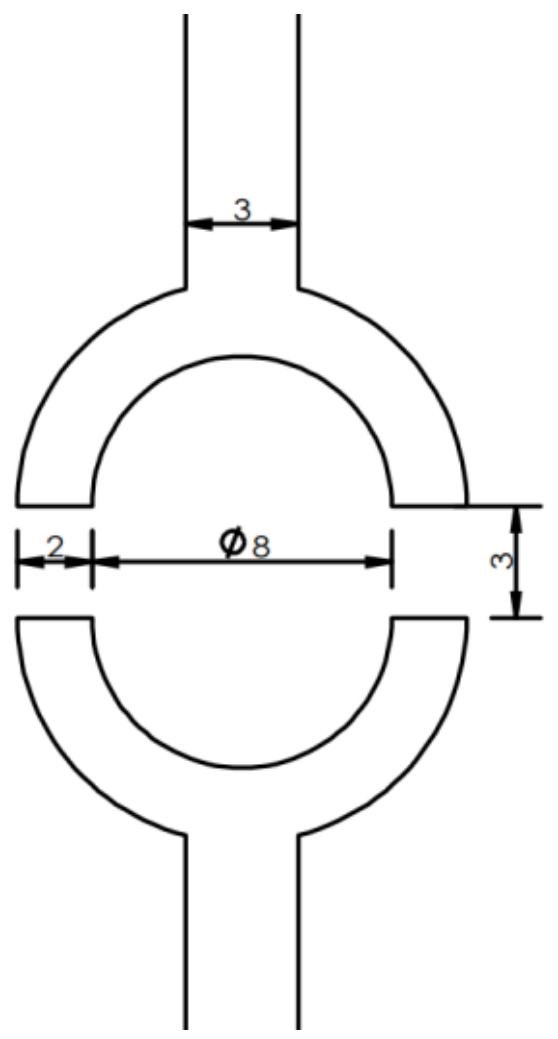

**Figure 4.3.8.23:** The profile of the eddy current shielding plate (unit: mm)

**Table 4.3.8.7:** Geometric parameters of eddy current shielding plate

| Shape | Radius | Width | Height | Thickness | Spacing | Material |
|---|---|---|---|---|---|---|
| Circular arc | 4 mm | 12 mm | 6 mm | 2 mm | 3 mm | Copper |

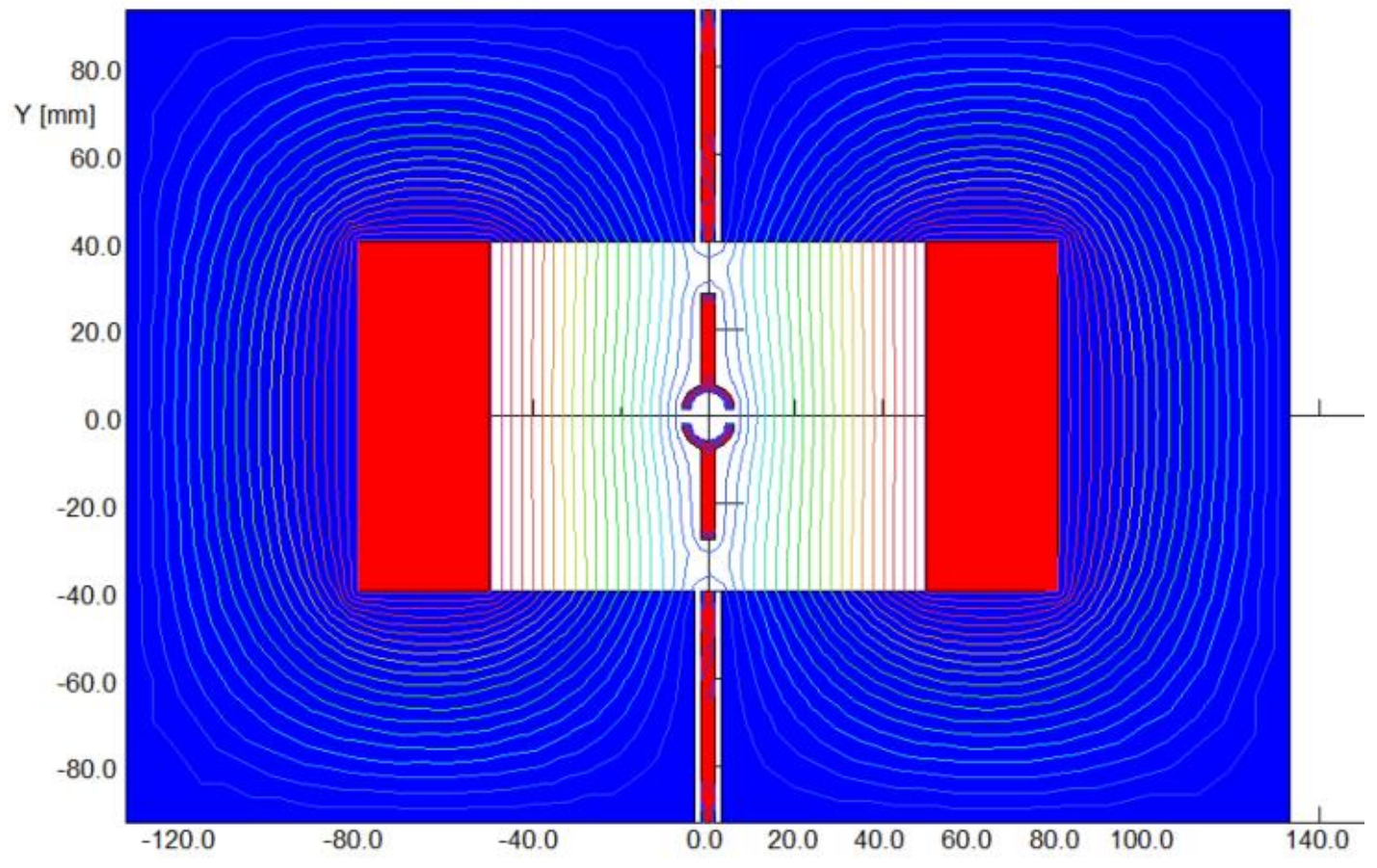

**Figure 4.3.8.24:** OPERA-2D model of the NLK

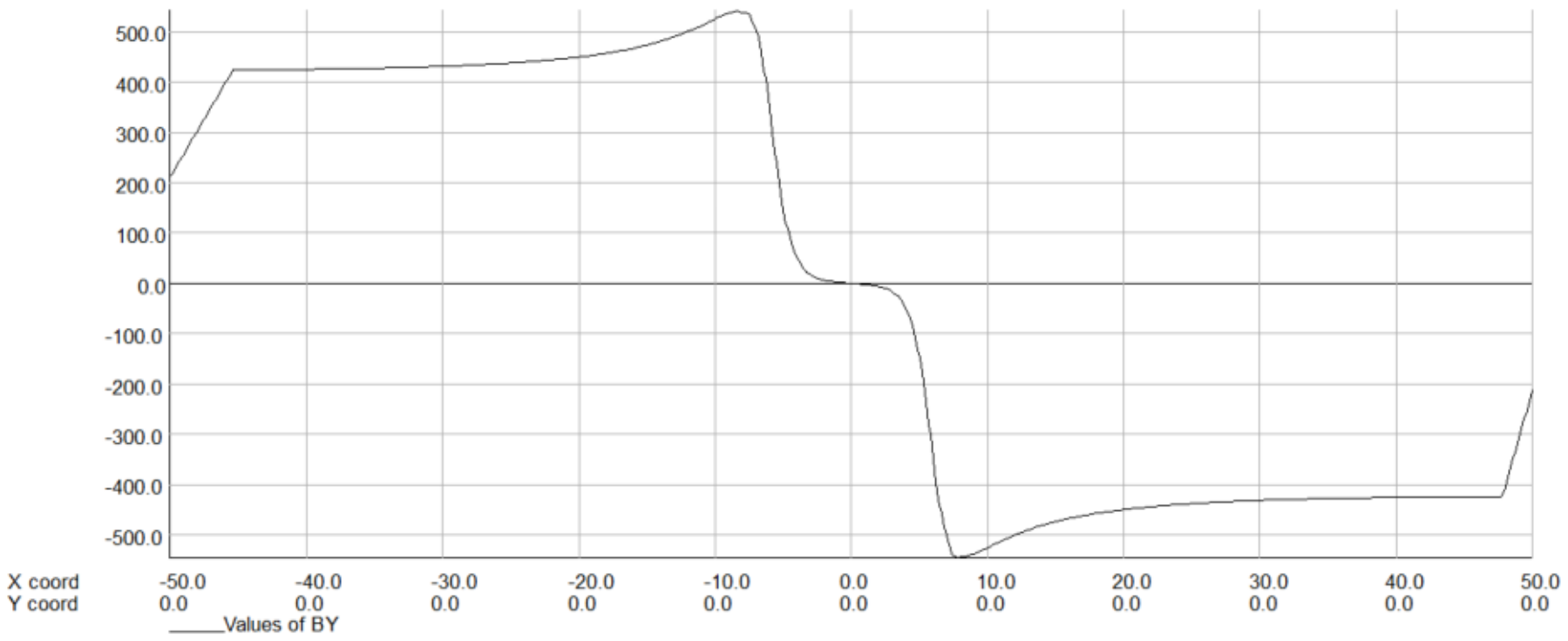

**Figure 4.3.8.25:** Horizontal magnetic field distribution on the central plane

The trapezoidal waveform flat-top width of the NLK is designed to reach 1980ns. However, there is a concern that the eddy current shielding effect may fail due to the long flat-top width. To investigate this, the magnetic field strength of the NLK is simulated with a 200 ns step, and the results are shown in Figure 4.3.8.26. It can be seen that even at 2200 ns, the eddy current shielding effect still exists, and the magnetic field only changes slightly. The maximum magnetic field within the magnetic center at x = ± 2mm is 6.2 Gauss at 200 ns, with a peak magnetic field of 540 Gauss at this time, resulting in a field strength ratio of 1.1%. At 2200 ns, the field strength in the magnetic center is 10.7 Gauss, while the peak magnetic field is 530 Gauss, resulting in a ratio of 2%. Therefore, the eddy current shielding plate can still effectively reduce the magnetic field strength and gradient in the center of the NLK even for long flat-top widths.

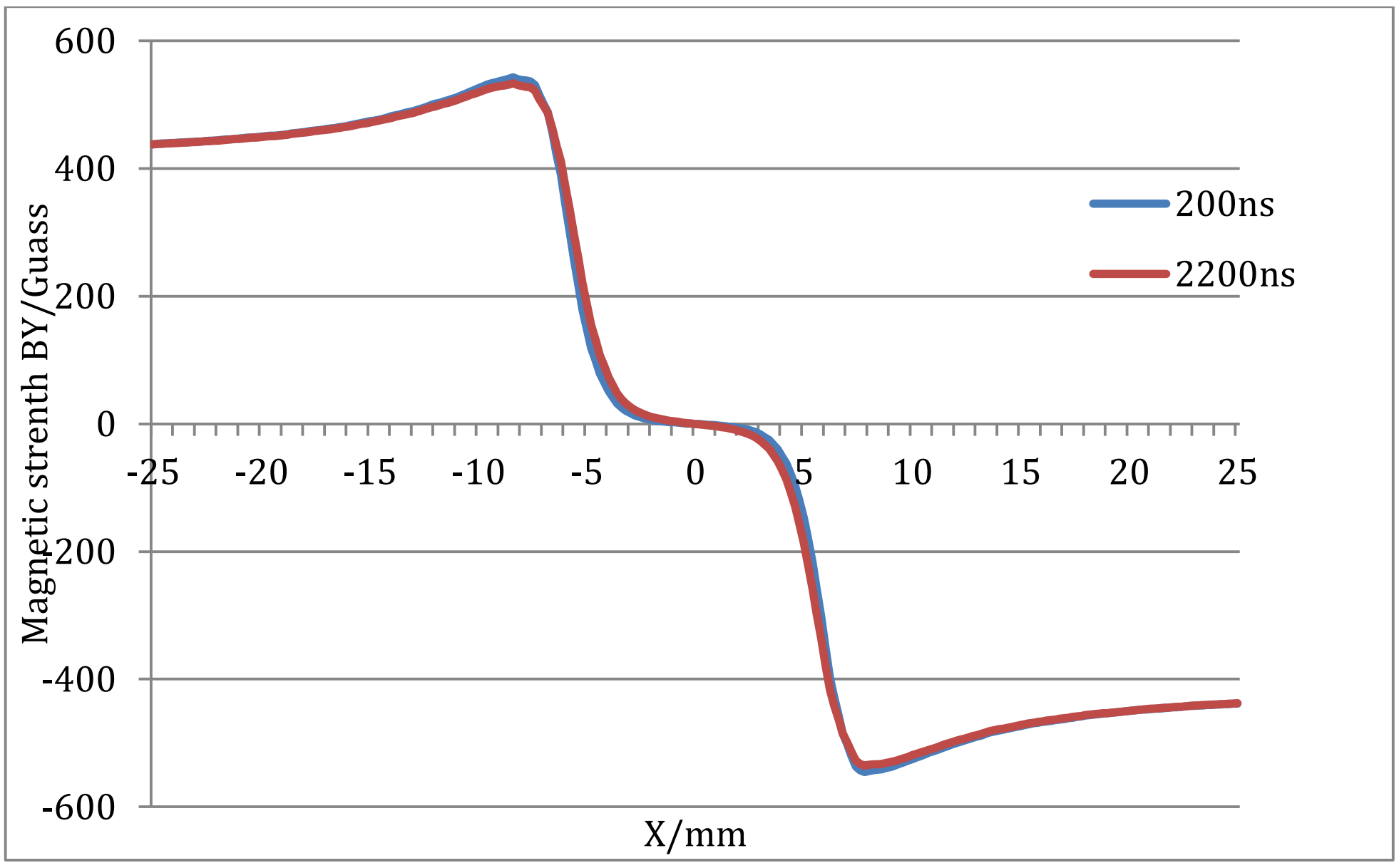

**Figure 4.3.8.26:** Magnetic field distribution of center plane at different time

The structure of the delay-line NLK is derived from the dual C-type delay-line dipole magnet, so the magnet design method is essentially the same. Table 4.3.8.8 and Table 4.3.8.9 provide the initial design parameters and design values, respectively, for the delay-line NLK used in off-axis injection at CEPC.

**Table 4.3.8.8:** Requirements of NLK for CEPC off-axis injection system

| Parameters | Value |
|---|---|
| Integrate magnetic strength (T-m) | 0.04 |
| Rise time of kicker magnet (ns) | 120 |
| Inner aperture of collider ring vacuum chamber (mm) | 56 × 56 |
| Outline dimension of ceramic vacuum chamber (mm) | 85 × 66 |
| Aperture of magnet (mm) | 100 × 80 |

**Table 4.3.8.9:** Main parameters of delay-line NLK

| Parameter | value |
|---|---|
| Aperture of magnet (mm) | 100×80 |
| Longitudinal length of magnet cell (mm) | 34 |
| Cell number of magnet | 26 |
| Even mode impedance of magnet (Ω) | 10 |
| Length of magnet (mm) | 889 |
| Total mechanical Length of magnet (mm) | 963 |
| Inductance of magnet cell (nH) | 24.1 |
| Total inductance of magnet (nH) | 630.1 |
| Capacitance of magnet cell (pF) | 241 |
| Total capacitance of magnet (nF) | 6.3 |
| Magnetic strength (Gs) | 450 |
| Exciting current of magnet (A) | 2255 |
| Voltage between HV plate and ground potential plate (V) | 22550 |

The PSpice circuit model of the delay-line NLK was created for simulation purposes, as depicted in Fig. 4.3.8.27. The pulse generator parameters used were $U_{max}$ = 22550 V, $T_D$ = 100 ns, $T_R$ = 80 ns, $T_F$ = 80 ns, and $P_w$ = 1980 ns. The simulation result, shown in Fig. 4.3.8.28, indicated that the magnet's $U_{max}$ was 22793 V, with $T_{mr}$ and $T_{mf}$ being 146 ns each. These values meet the necessary physics requirements for off-axis injection at CEPC.

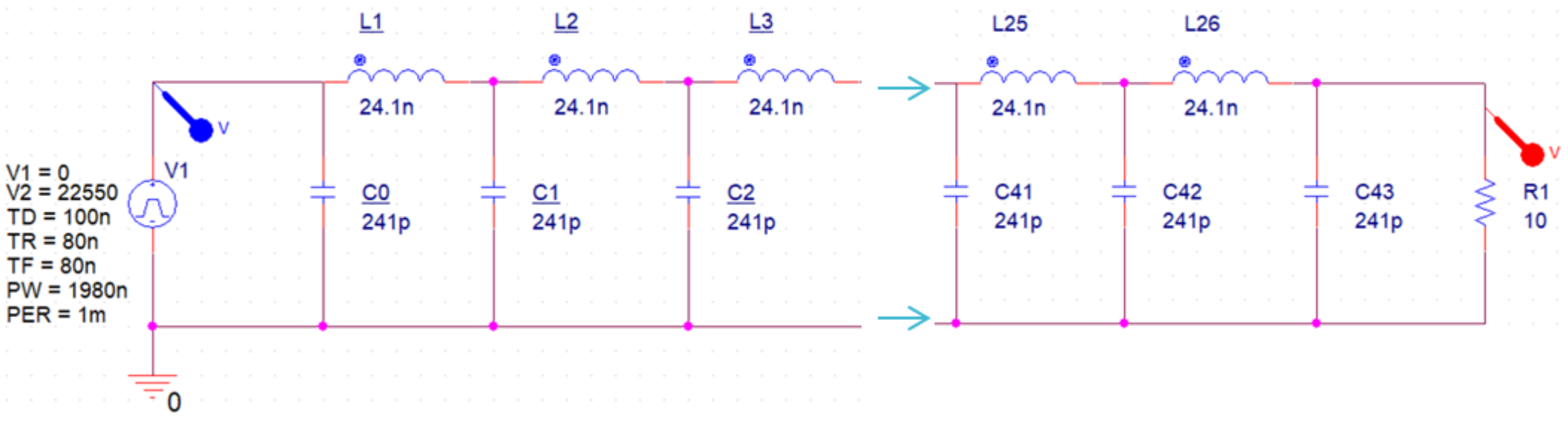

**Figure 4.3.8.27:** Delay-line NLK Pspice model

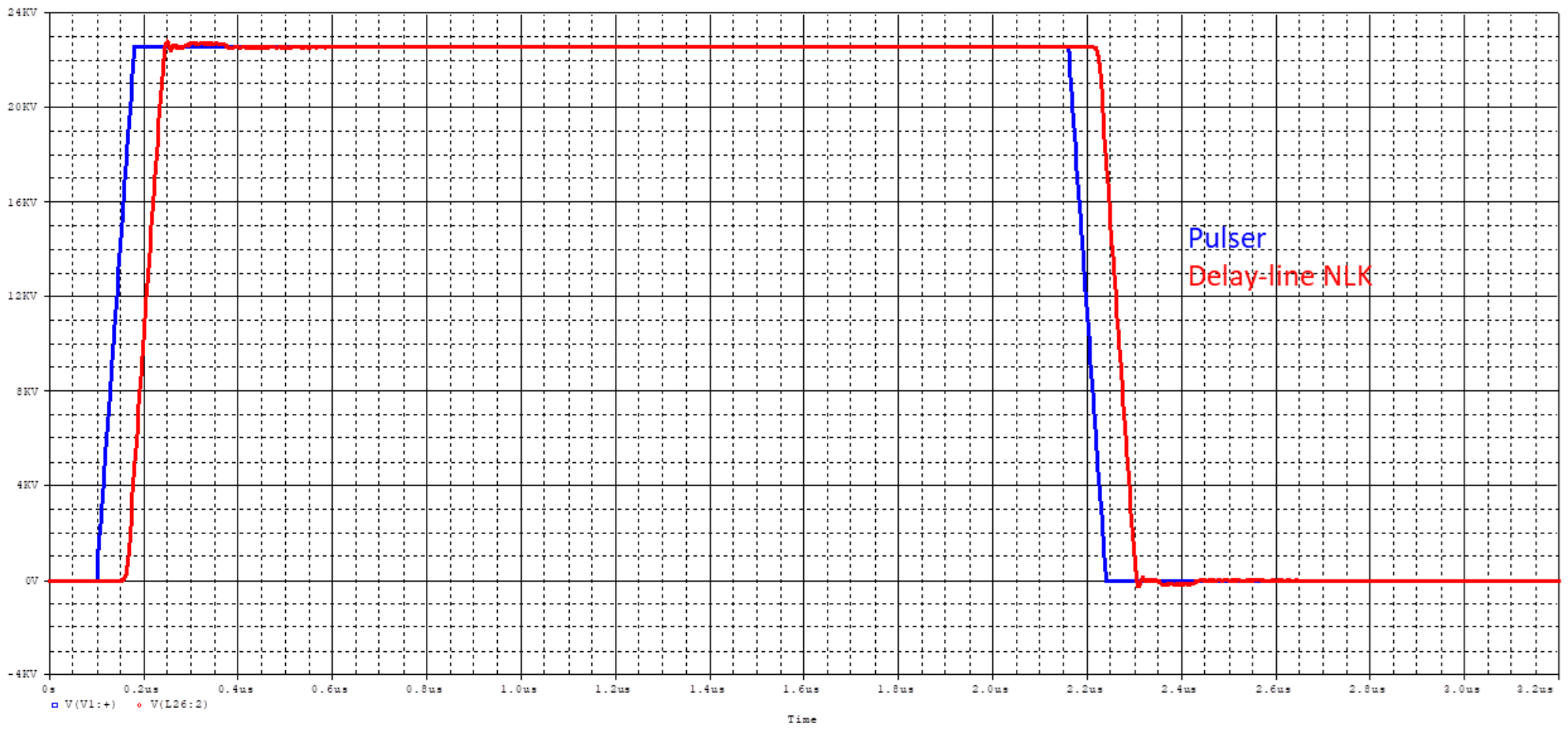

**Figure 4.3.8.28:** Delay-line NLK Pspice simulation result

The 3D mechanical structure model of the delay-line NLK for CEPC is shown in Fig. 4.3.8.29, and profile of the magnet is shown in Fig. 4.3.8.30.

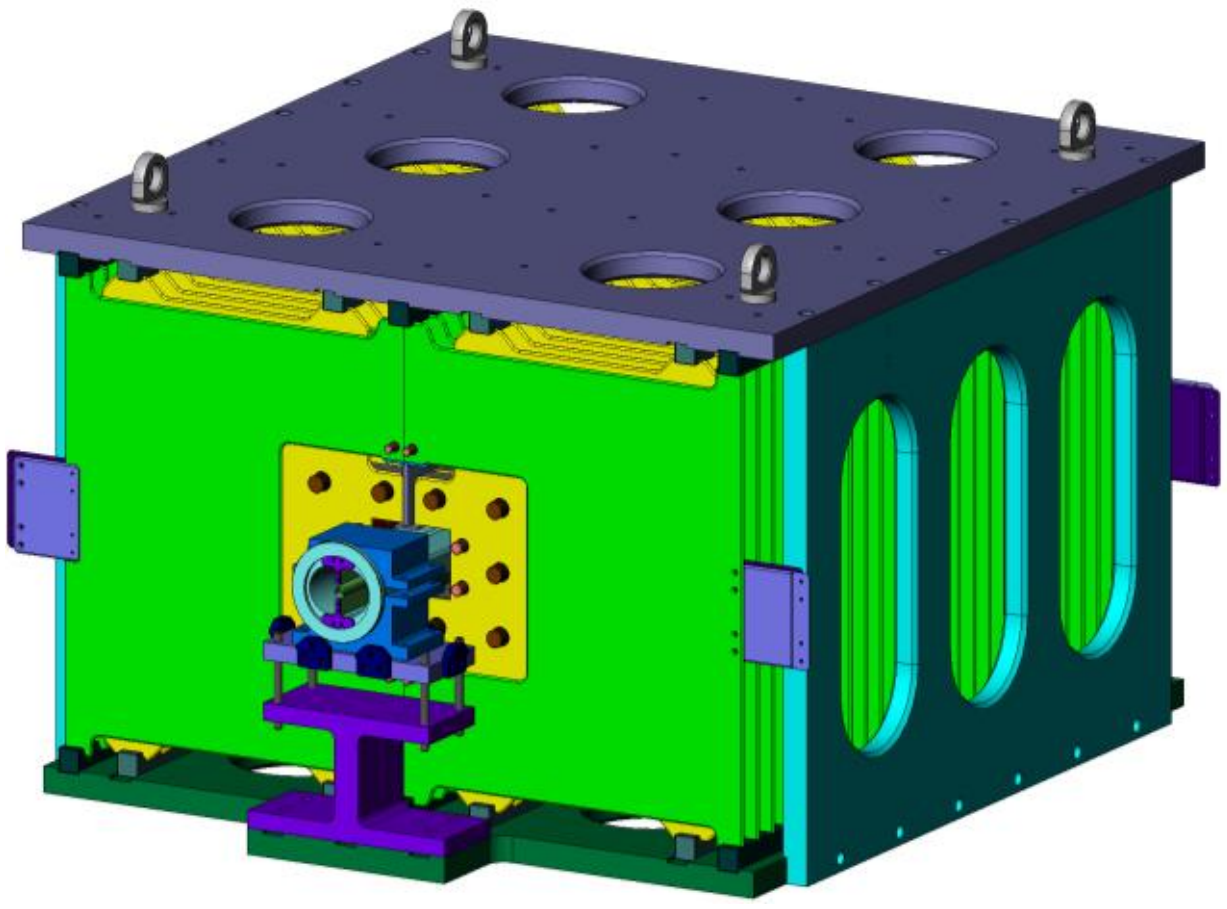

**Figure 4.3.8.29:** 3D mechanical structure model of the delay-line NLK

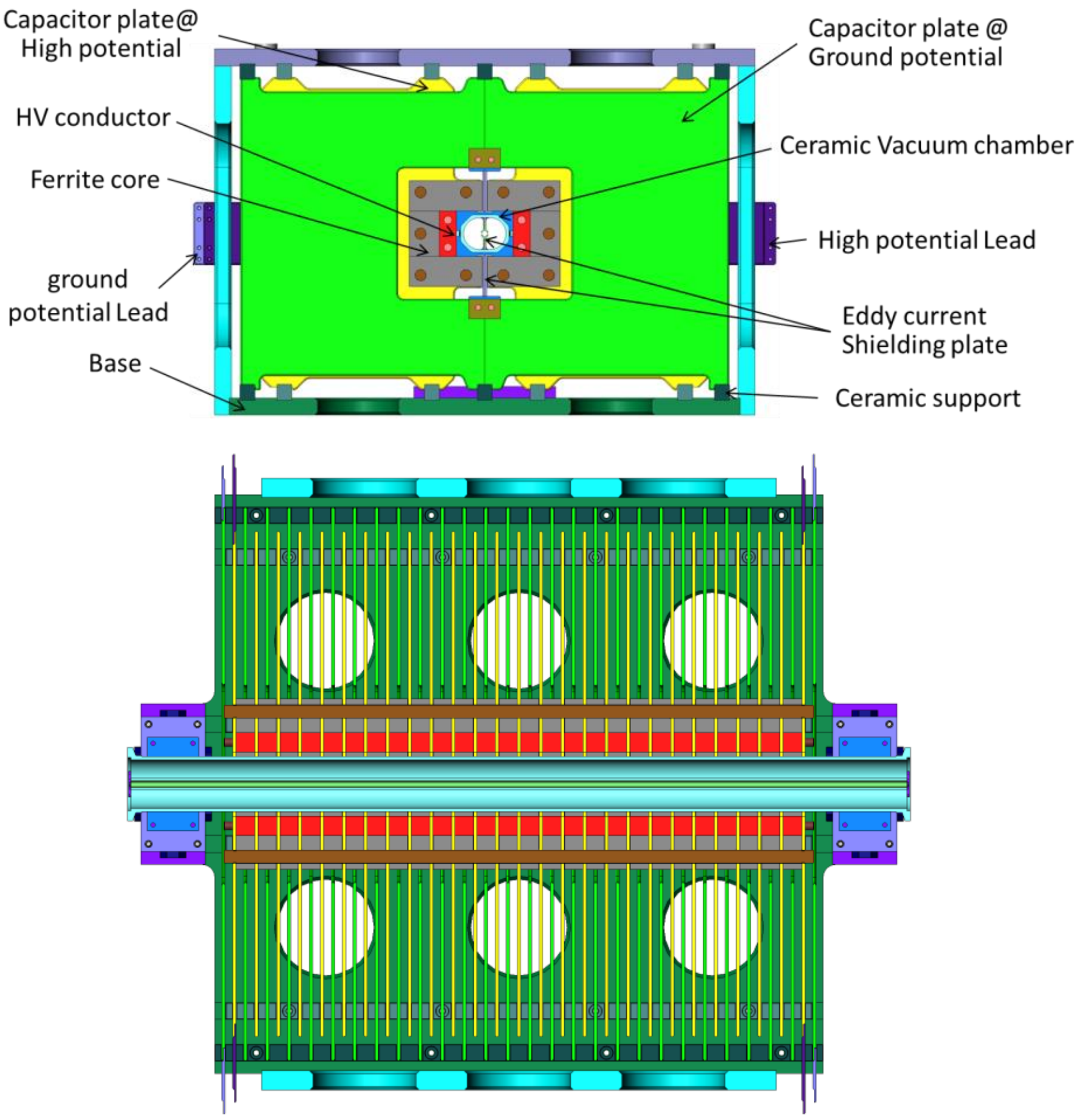

**Figure 4.3.8.30:** Profile structure of NLK

### 4.3.8.6 *R&D on the Delay-line Kicker and Ceramic Vacuum Chamber*

The research and development (R&D) work on the CEPC delay-line kicker and its vacuum chamber began in 2020. The prototype of the ceramic chamber features a racetrack-shaped inner contour with a wall thickness of 5 mm. The inner aperture measures 75 mm (W) × 56 mm (H), aligning with the oval vacuum pipe parameters specified in the CEPC CDR. In comparison to oval and octagon profiles, the racetrack profile of the ceramic vacuum chamber offers better mechanical performance and facilitates achieving a uniform coating. The outer contour of the chamber is octagonal, which simplifies machining and alignment during fabrication. However, it is worth noting that for the CEPC TDR, a round vacuum chamber is used instead, as it is easier to fabricate and coat.

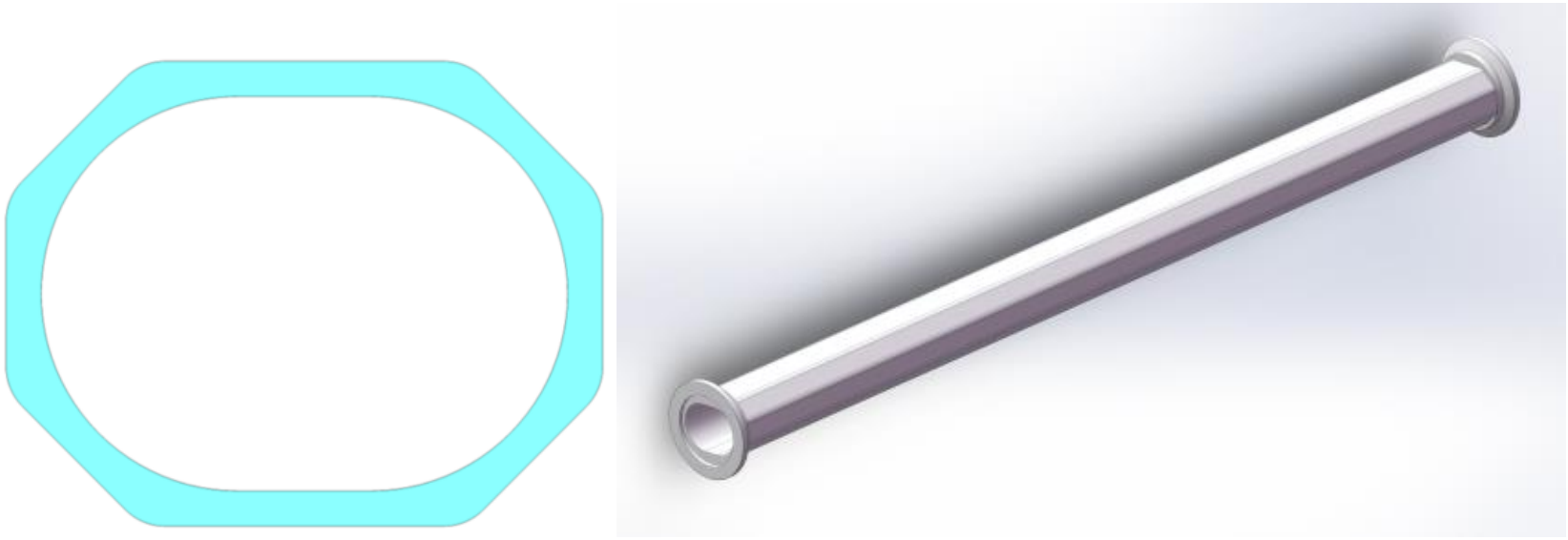

**Figure 4.3.8.31:** Mechanical model of the CDR-type ceramic vacuum chamber prototype.

The developed ceramic vacuum chamber is represented by a mechanical model, as depicted in Figure 4.3.8.31. The chamber has a total length of 1200 mm, which includes the ceramic KF flanges at both ends, and it is formed through integrated molding. To account for the effects of gravity, a racetrack-shaped ceramic vacuum chamber model with a length of 1.2 m and a wall thickness of 5 mm is utilized. The maximum deformation and stress occur at the midpoint of the chamber, with a maximum stress of 4.17 MPa and a maximum displacement deformation of $2.34 \times 10^{-3}$ mm. Both of these values fall within the safe range, as the strength of 99% $AL_2O_3$ ceramic exceeds 100 MPa.

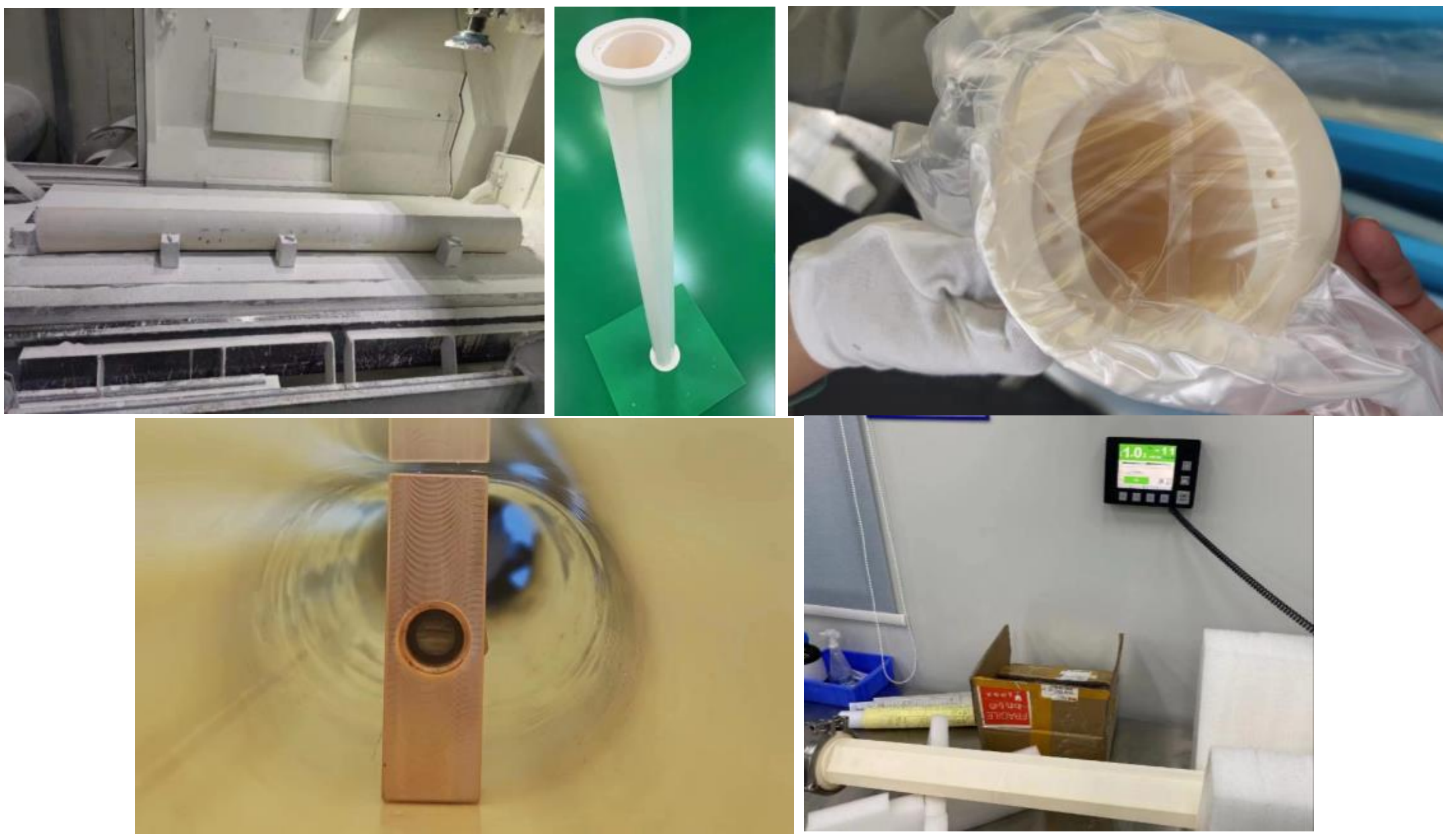

**Figure 4.3.8.32:** Ceramic vacuum chamber prototype fabrication and leakage rate measurement.

The ceramic chamber prototype, as shown in Figure 4.3.8.32, was fabricated from a large alumina billet. Before the sintering process, rough grinding of the billet was necessary. After burning, fine machining and polishing were performed to ensure dimensional tolerances and surface finish, with particular importance placed on minimizing vacuum leakage at the flange locations. The actual measured leakage rate of the final prototype was better than $1.0 \times 10^{-10}\ \mathrm{mbar \cdot L/s}$, demonstrating excellent vacuum sealing capabilities.

Prior to coating the final vacuum chamber, a magnetron sputtering coating test-bench was constructed. This test-bench integrated a cathode wire mover and was used to attempt TiN coating on a round vacuum chamber, as depicted in Figure 4.3.8.33. The conductivity of the TiN film sample, measured using the 4-probe method, was found to be significantly lower than the reference value of $5.8\times10^6$ S/m. To enhance conductivity, a doping process involving metal elements with high conductivity, such as Cu and Ag, was considered. The film morphology, as observed through scanning electron microscopy (SEM), is displayed in Figure 4.3.8.34.

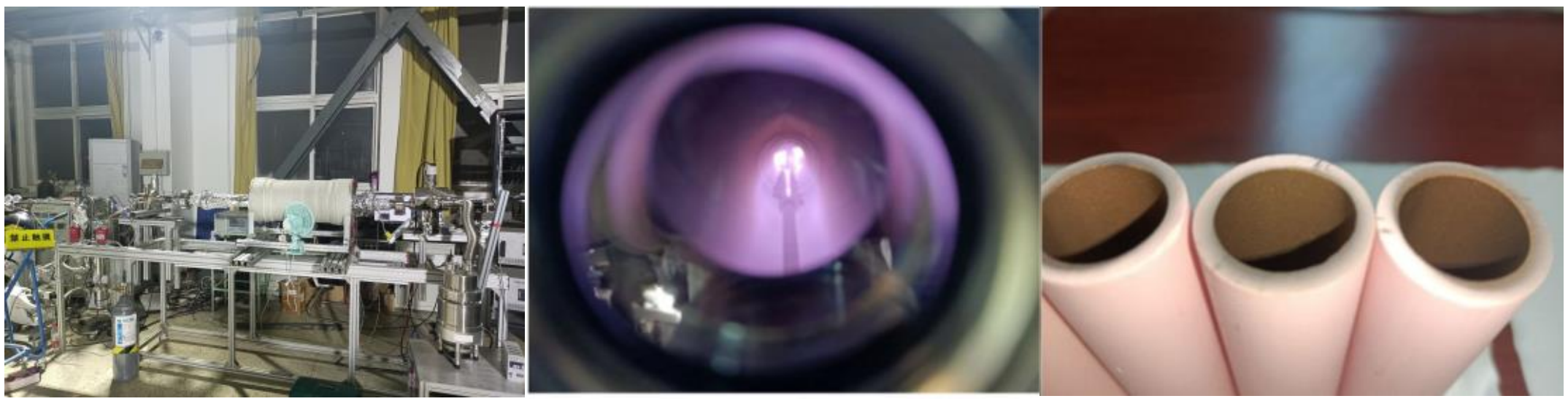

**Figure 4.3.8.33:** Magnetron sputter coating process and Ti-N films.

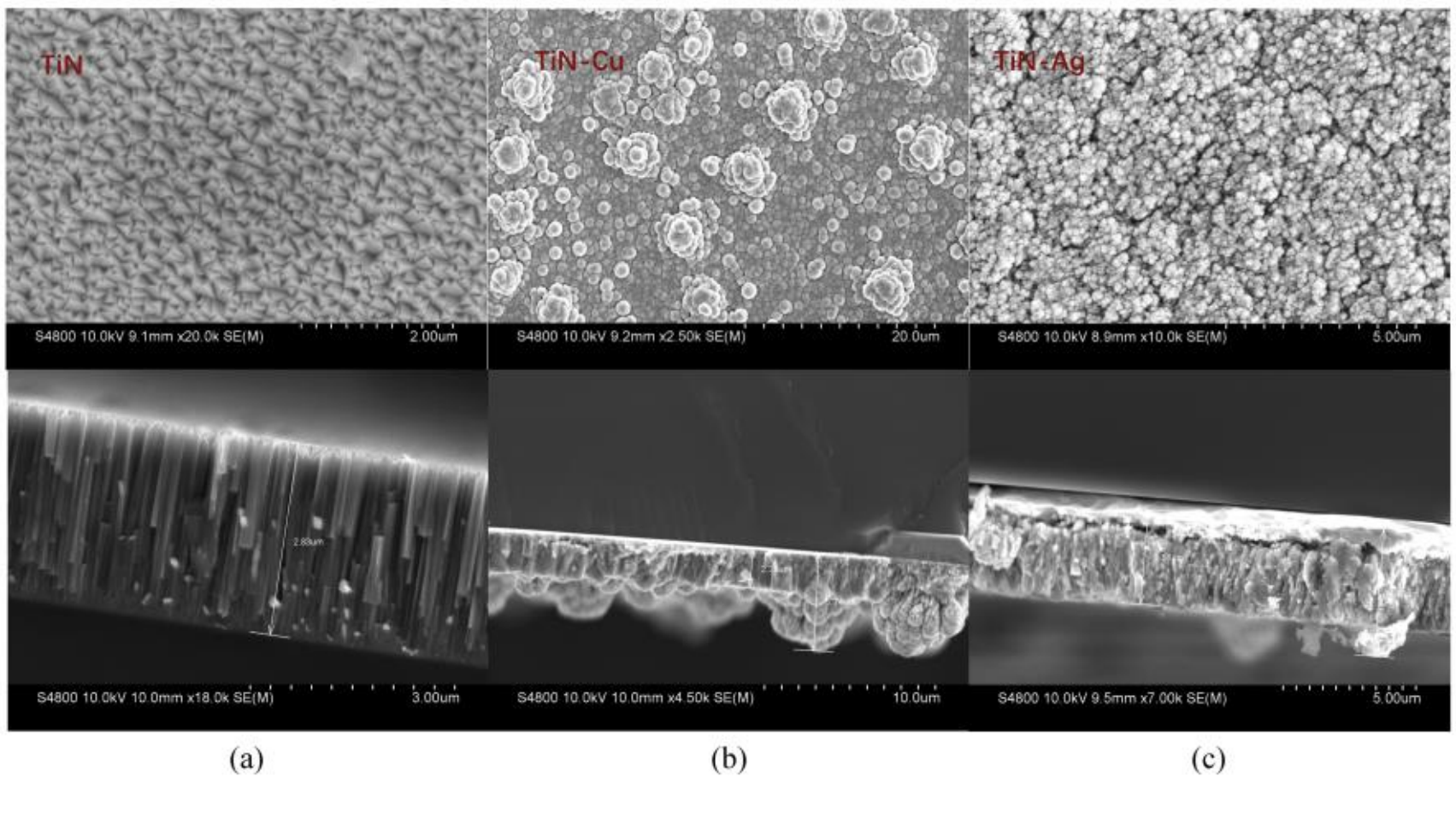

**Figure 4.3.8.34:** Film morphology obtained by scanning electron microscopy (SEM).

As we are aware, sputtering Cu targets in a nitrogen environment can generate CuN. However, it is challenging to form AgN during the sputtering process due to the strict conditions required. Therefore, for subsequent coating experiments, a TiN-Ag combination is adopted. The measured conductivity of the TiN-Ag film, formed under conditions of a coating atmosphere at 10 Pa and a substrate temperature of 180℃, is $8.3 \times 10^6$ S/m.

Research indicates that a ladder pattern coating is the optimal choice for the CEPC in-air trapezoid-wave fast kicker. A designed mask, created through EDM machining using aluminum, is shown in Figure 4.3.8.35. The desired coating pattern was successfully achieved on the inner surface of a splicing racetrack-shaped chamber made from PEEK.

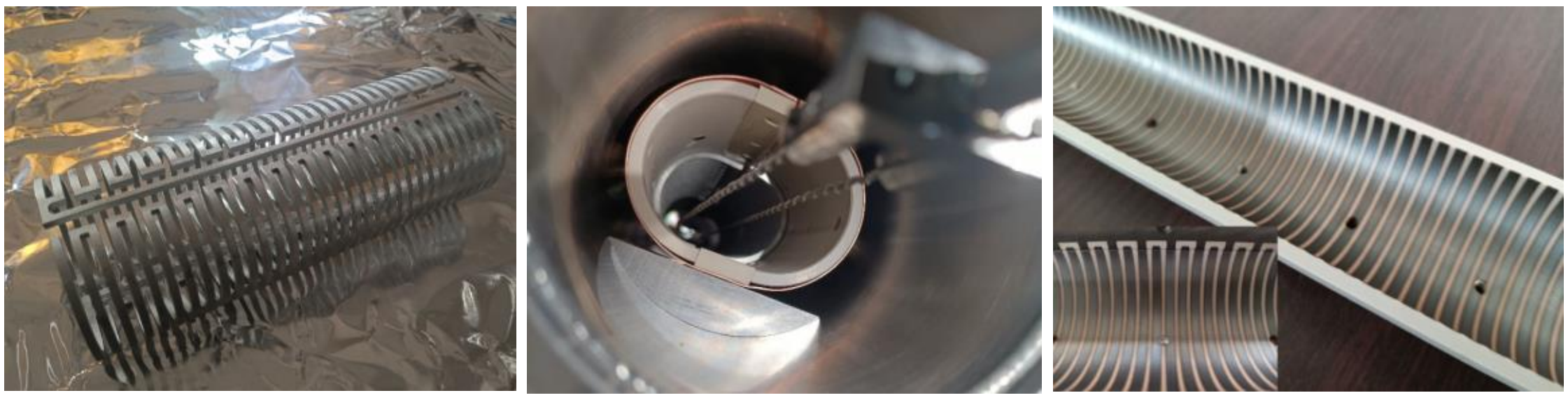

**Figure 4.3.8.35:** Preparation and experiment of coating with ladder pattern.

Based on the preliminary preparation work, the coating process on the formal ceramic vacuum chamber progressed smoothly. Figure 4.3.8.36 displays the final result of the TiN-Ag coating with a ladder pattern applied to the formal racetrack-shaped ceramic vacuum chamber.

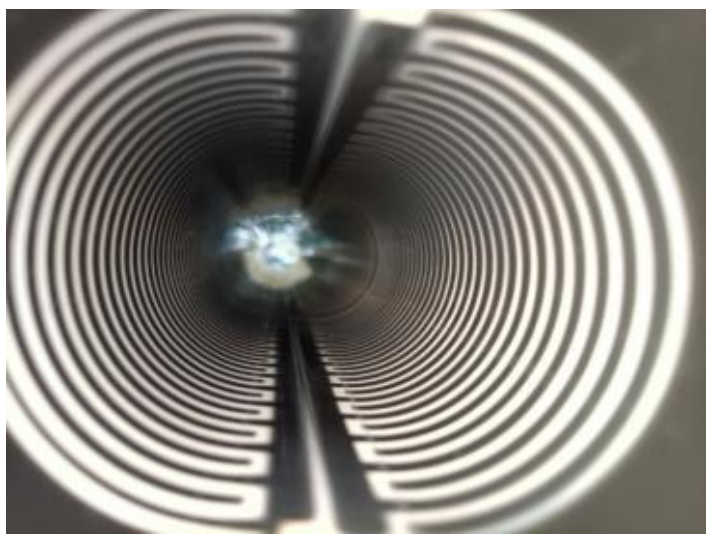

**Figure 4.3.8.36:** Ladder-shaped coating on racetrack-shaped ceramic vacuum chamber.

Beam impedance measurements using the coaxial-line method have been conducted, as depicted in Figure 4.3.8.37. The measured impedance results are found to be in good agreement with the simulations performed in CST, validating the accuracy of the modeling and prediction.

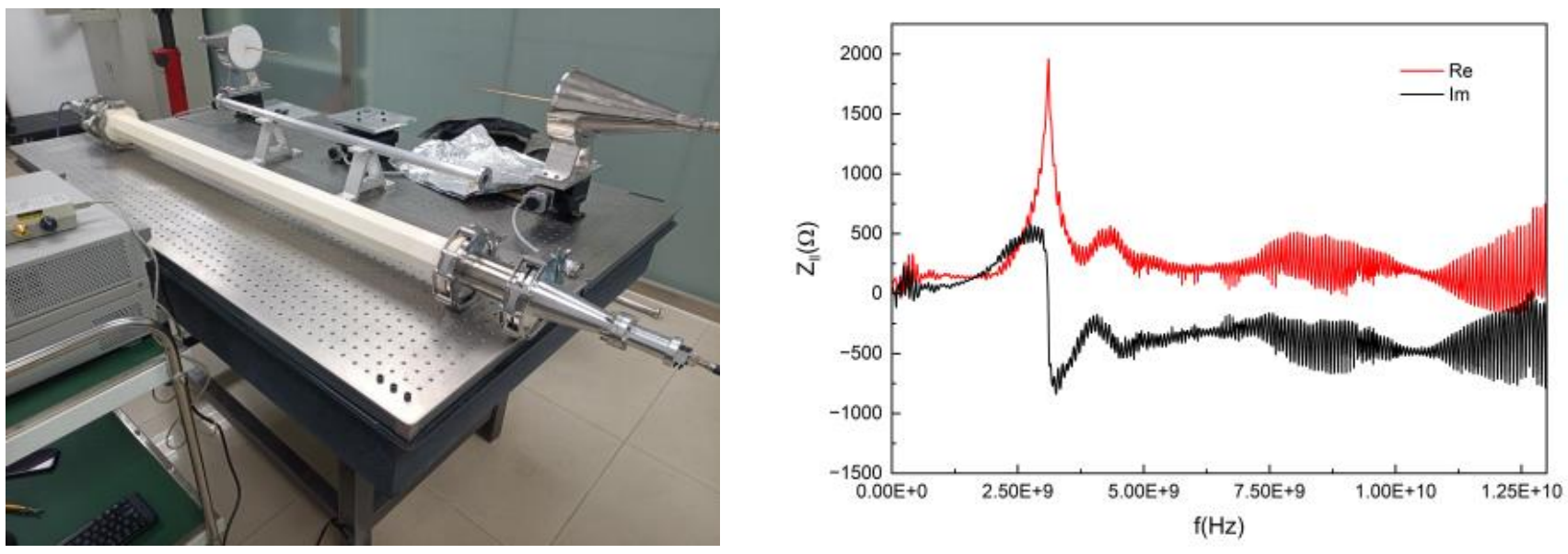

**Figure 4.3.8.37:** Beam impedance measruement by coaxial-line method.

The coated ceramic vacuum chamber prototype was subjected to testing on a pulse magnetic field measurement bench using the inductive coil method. For this purpose, the ferrite core kicker magnet prototype designed for CSNS, featuring a sufficiently large aperture, and the solid-state pulse power supply developed for HEPS booster, were utilized to measure the magnetic field, as shown in Figure 4.3.8.38. By comparing the waveforms captured by an oscilloscope under test conditions with and without the vacuum chamber, it was confirmed that the coated ceramic vacuum chamber allows the kicker trapezoid wave pulse magnetic field, with a rise/fall time of less than 200 ns, to pass through without any distortion in its waveform.

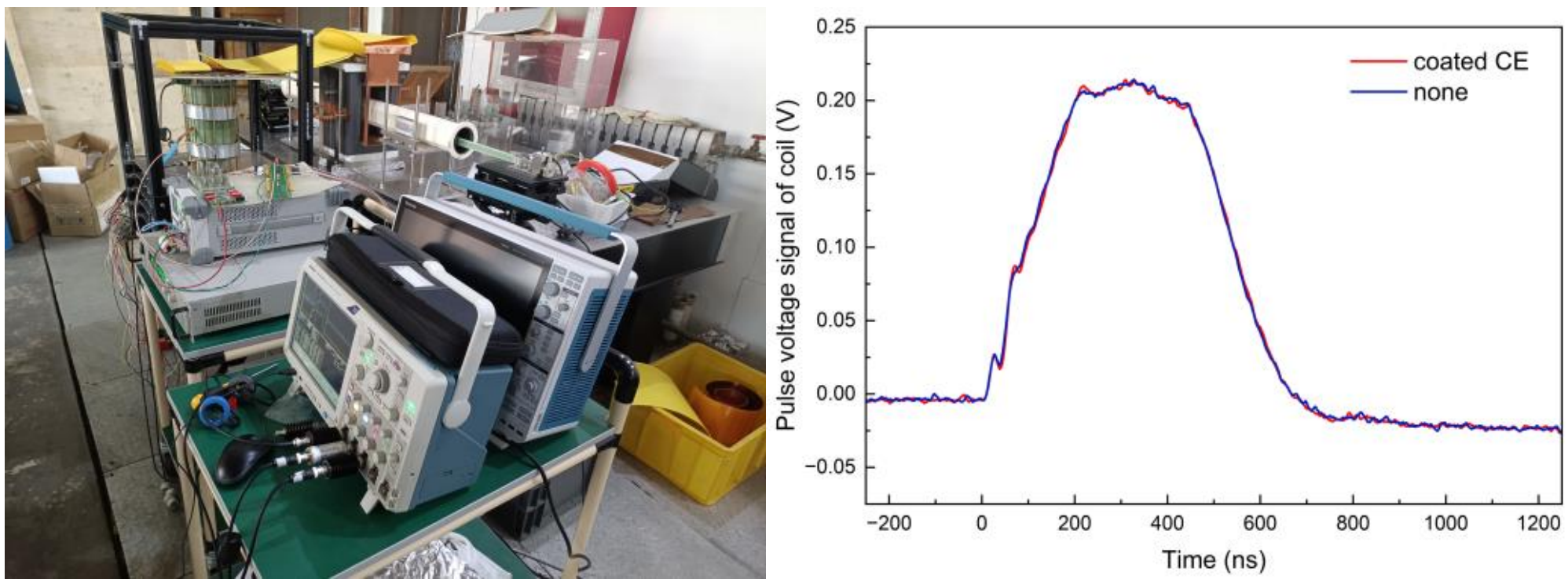

**Figure 4.3.8.38:** Pulse magnetic field measurement and result.

A prototype of the delay-line NLK has undergone R&D, as depicted in Figure 4.3.8.39. Using the Time Domain Reflectometry (TDR) function of a Vector Network Analyzer (VNA), the measured line impedance for the delay-line NLK is 10 Ω, with a common mode impedance of 7 Ω, as illustrated in Figure 4.3.8.40.

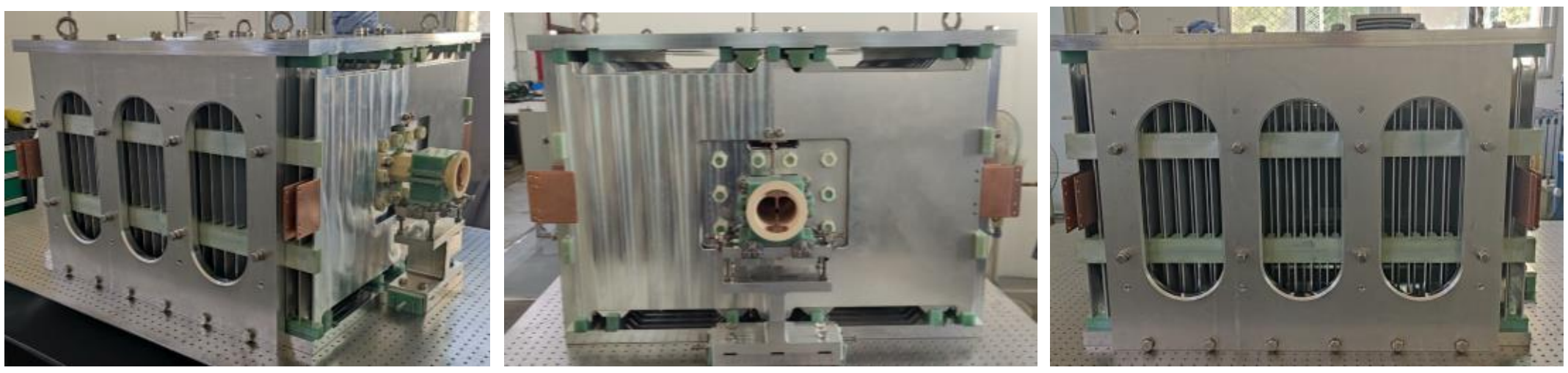

**Figure 4.3.8.39:** Delay-line NLK prototype.

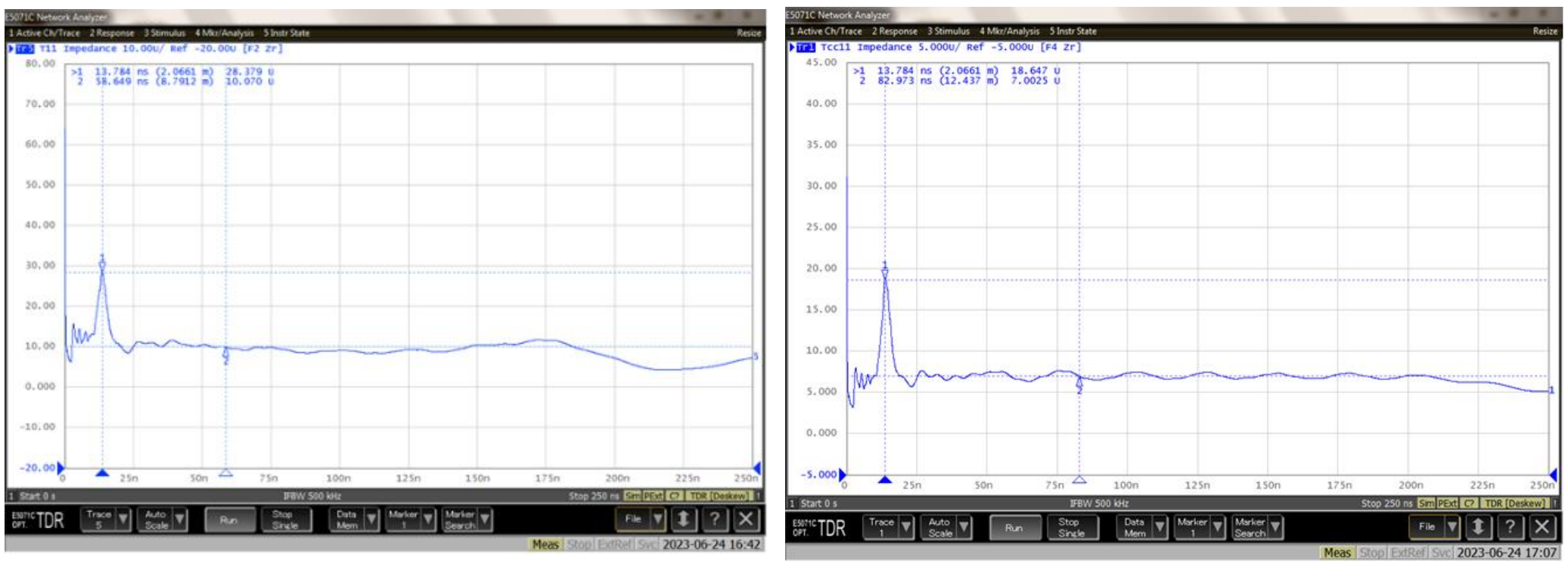

**Figure 4.3.8.40:** Impedance measurement by VNA (left: line impedance, right: common mode impedance)

The magnetic field measurement of the NLK prototype has been successfully conducted using the setup depicted in Figure 4.3.8.41. This setup comprises a 5-stage solid-state inductive adder pulser, a long PCB inductive coil, a movable support, a passive integrator, and an oscilloscope. The NLK was driven in common mode by the pulser with an output voltage of 3 kV. The typical measured waveforms are presented in Figure 4.3.8.42. The measured magnetic field distribution of the prototype closely aligns with the simulation results, as demonstrated in Figure 4.3.8.43.

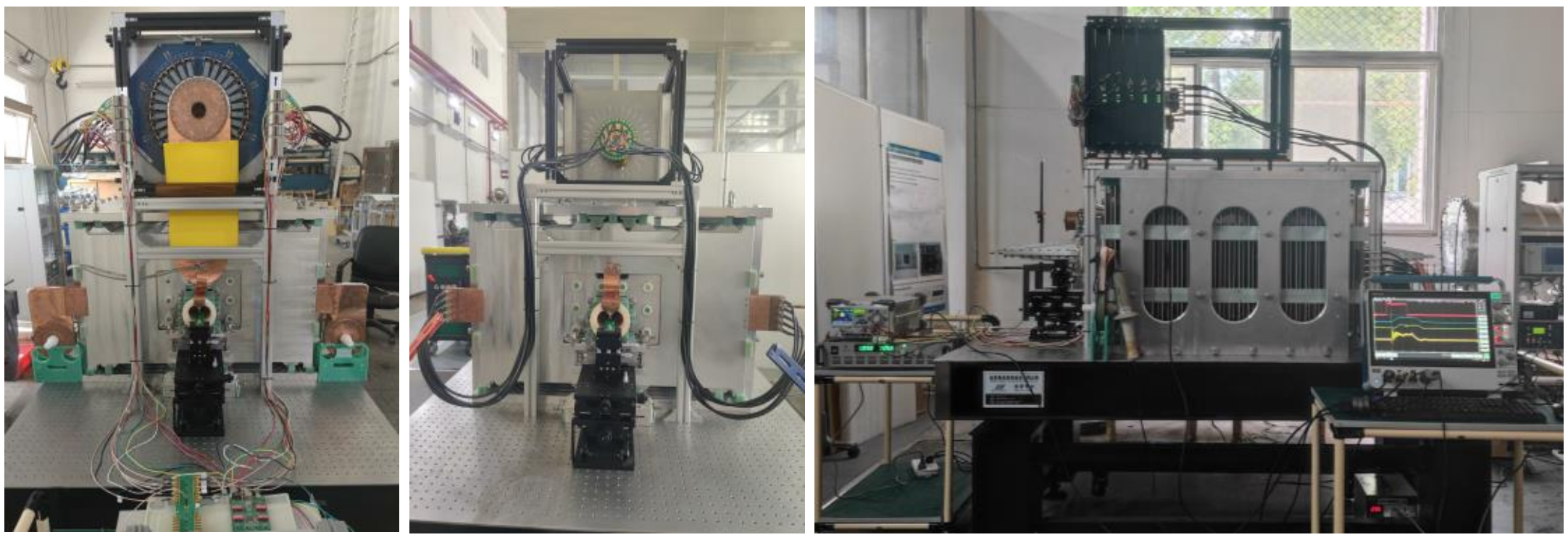

**Figure 4.3.8.41:** Magnetic field mearment bench.

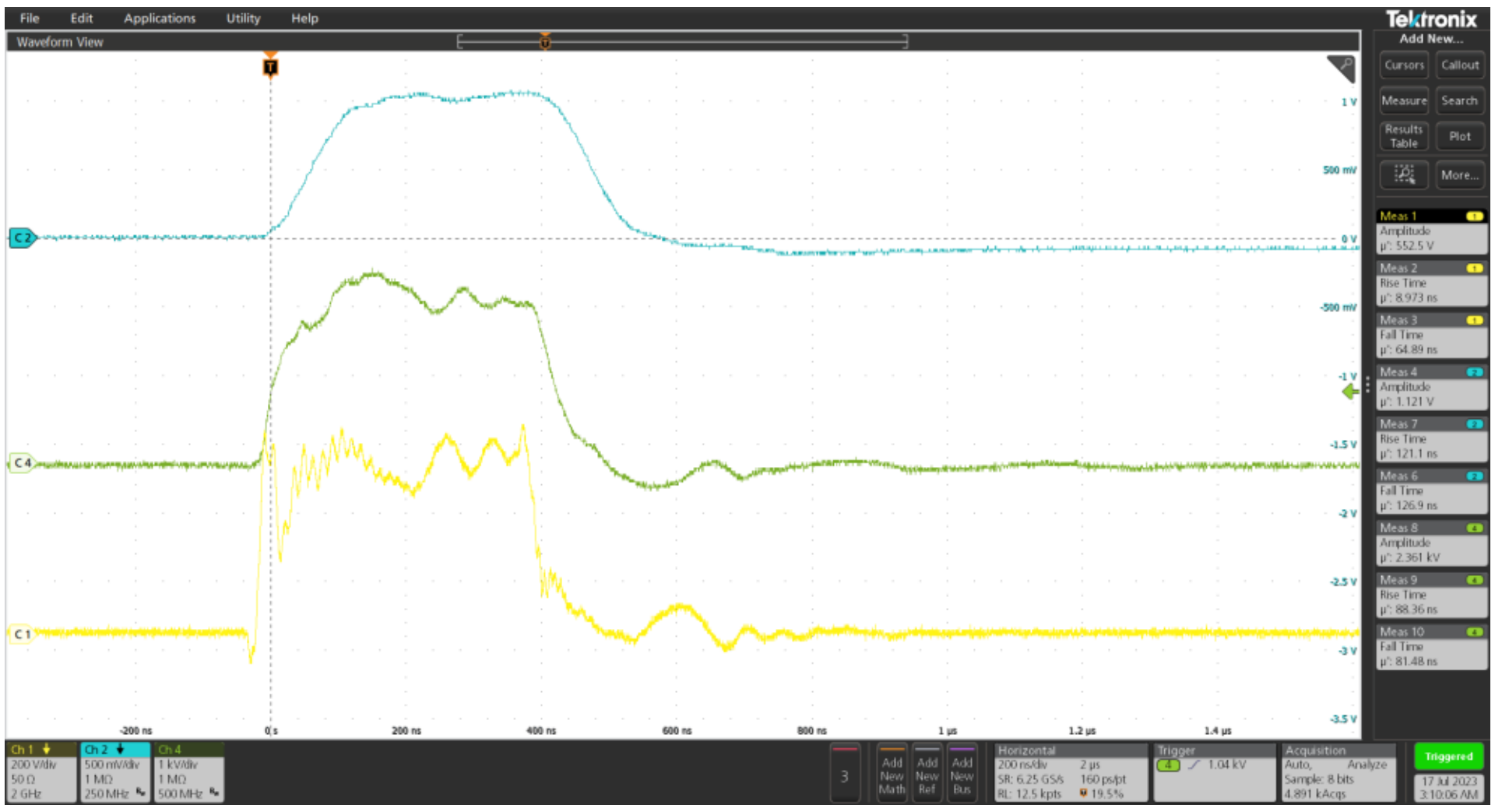

**Figure 4.3.8.42:** Typical waveforms of magnetic field mearment. (top: integral pulsed magnetic field, middle: voltage across terminal resistor, bottom: pulsed current of the pulser)

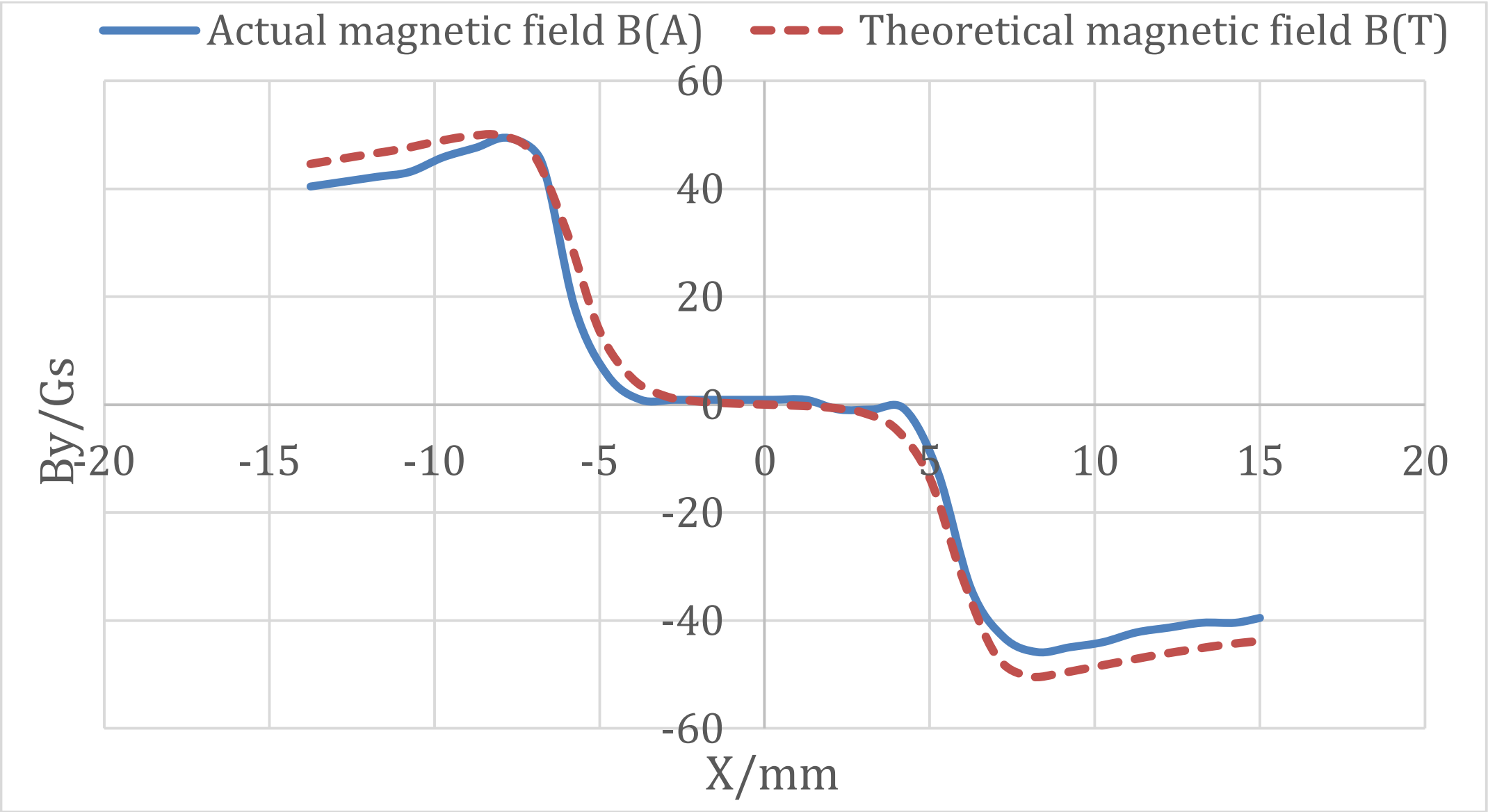

**Figure 4.3.8.43:** Pulsed magnetic field distribution of delay-line NLK prototype.

### 4.3.8.7 *Trapezoidal Wave Pulsed Power Supply System*

To shorten the injection duration, CEPC plans to increase the repetition rate of the kicker system to 1kHz. Therefore, a solid-state pulsed power supply based on IGBT or MOSFET technology will be utilized for the CEPC injection and extraction system. The design parameters of the trapezoidal wave kicker system can be found in Table 4.3.8.10.

**Table 4.3.8.10:** Design parameters of the trapezoidal wave kicker system

| Parameter | Dipole kicker | NLK |
|---|---|---|
| Magnet type | Dual-C delay-line | Dual-C delay-line |
| Aperture of magnet (mm) | 100×80 | 100×80 |
| Characteristic impedance of magnet (Ω) | 6.25（odd mode） | 10（even mode） |
| Cell number of magnet | 26 | 26 |
| Cell inductance of single side (nH) | 28.3 | 24.1 |
| Magnet inductance of single side (nH) | 735.8 | 630.1 |
| Cell capacitance of single side (pF) | 724 | 241 |
| Magnet capacitance of single side (nF) | 18.824 | 6.3 |
| Magnetic field strength (Gs) | 425 | 450 |
| Exciting current (A) | 2703 | 2255 |
| Voltage of magnet (V) | +/-16895.5 | +/+22550 |
| Exciting mode | Differential mode | Common mode |
| Pulse waveform | Trapezoid | Trapezoid |
| Rise/fall time (ns) | 200 | 200 |
| Flat-top width of pulse (ns) | 40~2000 adjustable | 40~2000 adjustable |
| Filling time of magnet (ns) | 117.728 | 63.01 |
| Rise/fall time of pulser (ns) | 80 | 80 |
| Repetition rate (kHz) | 1 | 1 |

The delay-line kicker magnet can be considered as a transmission line in the trapezoidal wave pulse discharge system. If the characteristic impedance of the magnet matches the impedance of the transmission cable and the load resistance $Z_0$, it can achieve perfect matching transmission. Figure 4.3.8.44 shows the simplified circuit diagram.

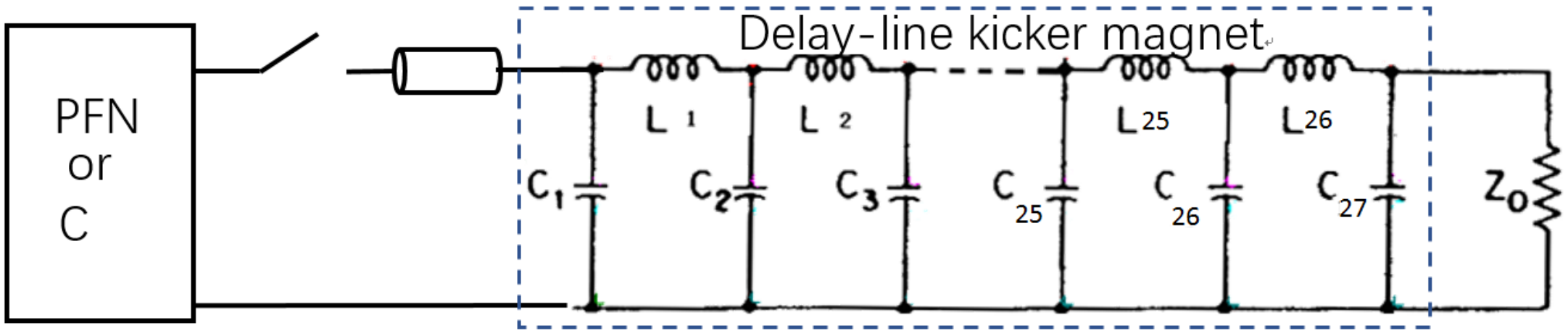

**Figure 4.3.8-44:** The trapezoidal wave pulse discharge system

By using a pulse forming network (PFN) as the energy storage element of the pulsed power supply, an ideal trapezoidal wave pulse with a flat top can theoretically be obtained, as shown in Fig. 4.3.8.45 (left). However, the pulse voltage amplitude is only half of the PFN energy storage voltage. If a capacitor is used for energy storage, the pulse peak voltage can equal the energy storage voltage, but the flat top will experience significant drop for long pulses due to the time constant $\tau = RC$, as shown in Fig. 4.3.8.45 (right). Therefore, the capacitance of the storage capacitor must be large enough to avoid this issue.

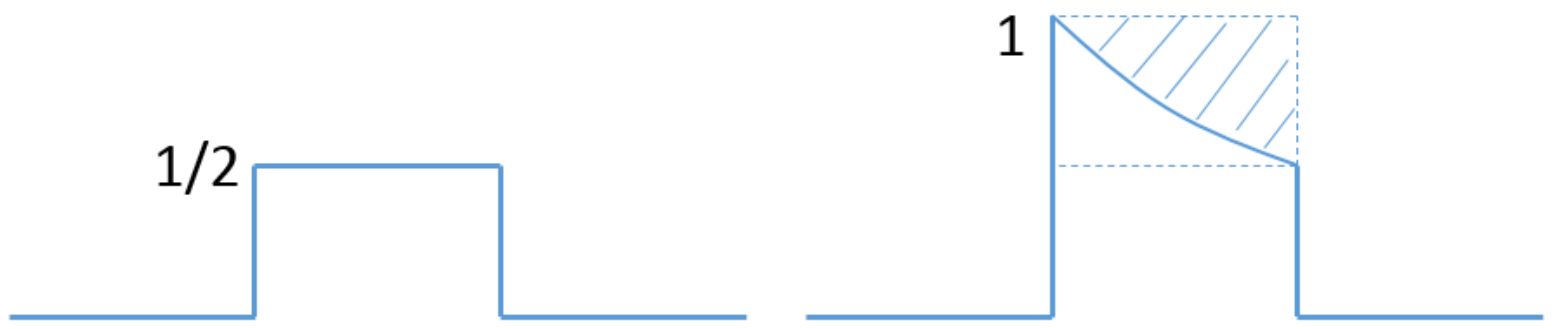

**Figure 4.3.8.45:** Pulse waveform comparison (left: PFN, right: capacitor)

Conventional high-speed high-power semiconductor solid-state switches such as IGBTs and MOSFETs have small power levels for a single chip. As a result, pulse power superposition is often required through some topology circuits. The most common superposition topologies include inductive adder and Marx generator, as shown in Fig. 4.3.8.46 and Fig. 4.3.8.47 [9].

The principle of the inductive adder is essentially the transformer superposition. Its advantages are that the switch circuit on the primary side of the transformer is at ground potential, the switch driver circuit is simple and reliable, and the parasitic inductance can be well controlled due to the coaxial transformer structure design. Its disadvantage is that it is not suitable for long pulses because the transformer core could be saturated.

The principle of the Marx generator is essentially the capacitors discharge in series directly, and it is suitable for applications with different pulse widths. Its disadvantage is that the switch is at a floating potential, and all switch driver signals must be isolated by a transformer or optical fiber. For long pulse applications, optical fiber isolation is more appropriate. Moreover, it is difficult to realize a structure design with low parasitic inductance.

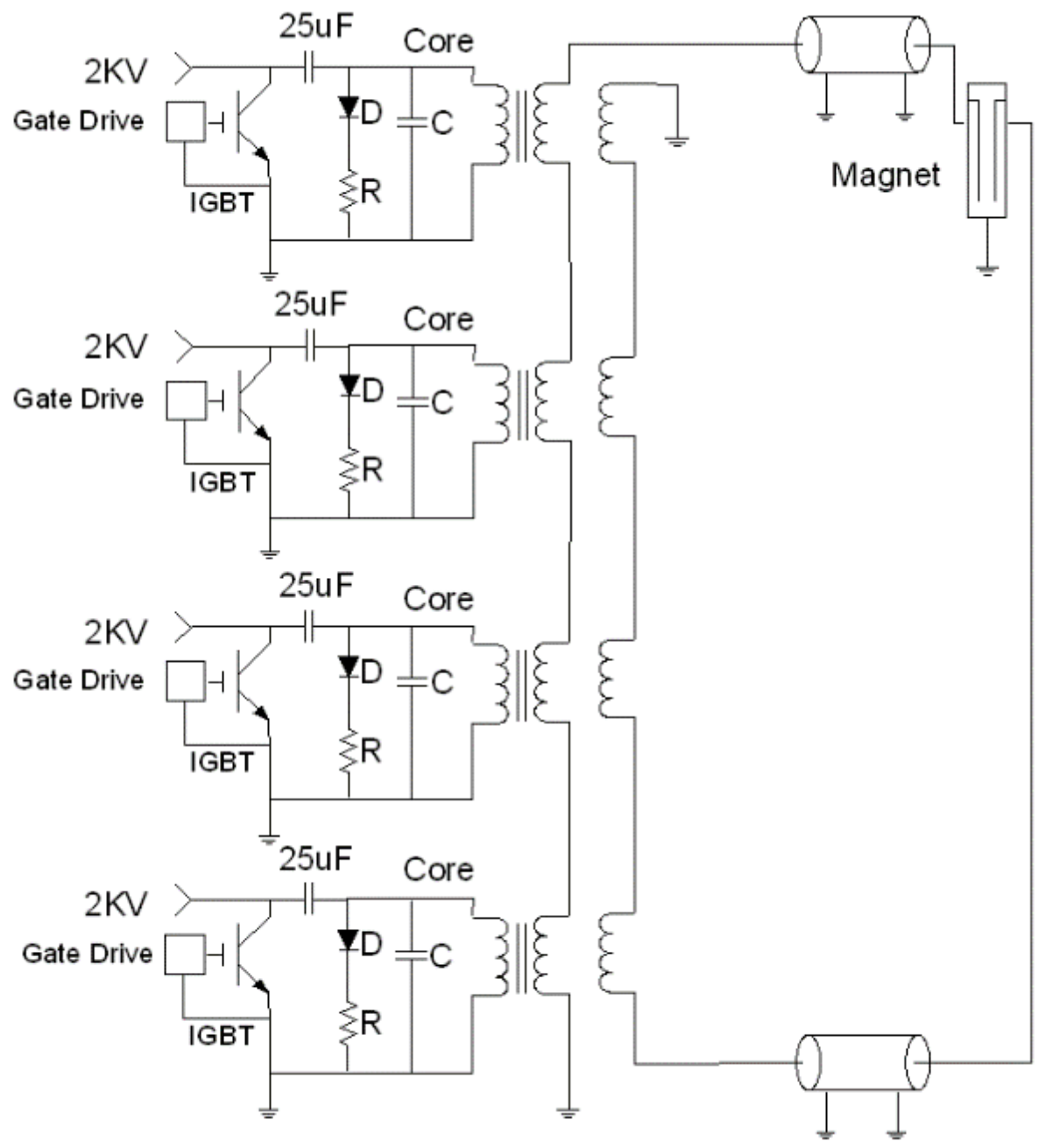

Figure 4.3.8.46: Typical inductive adder topology

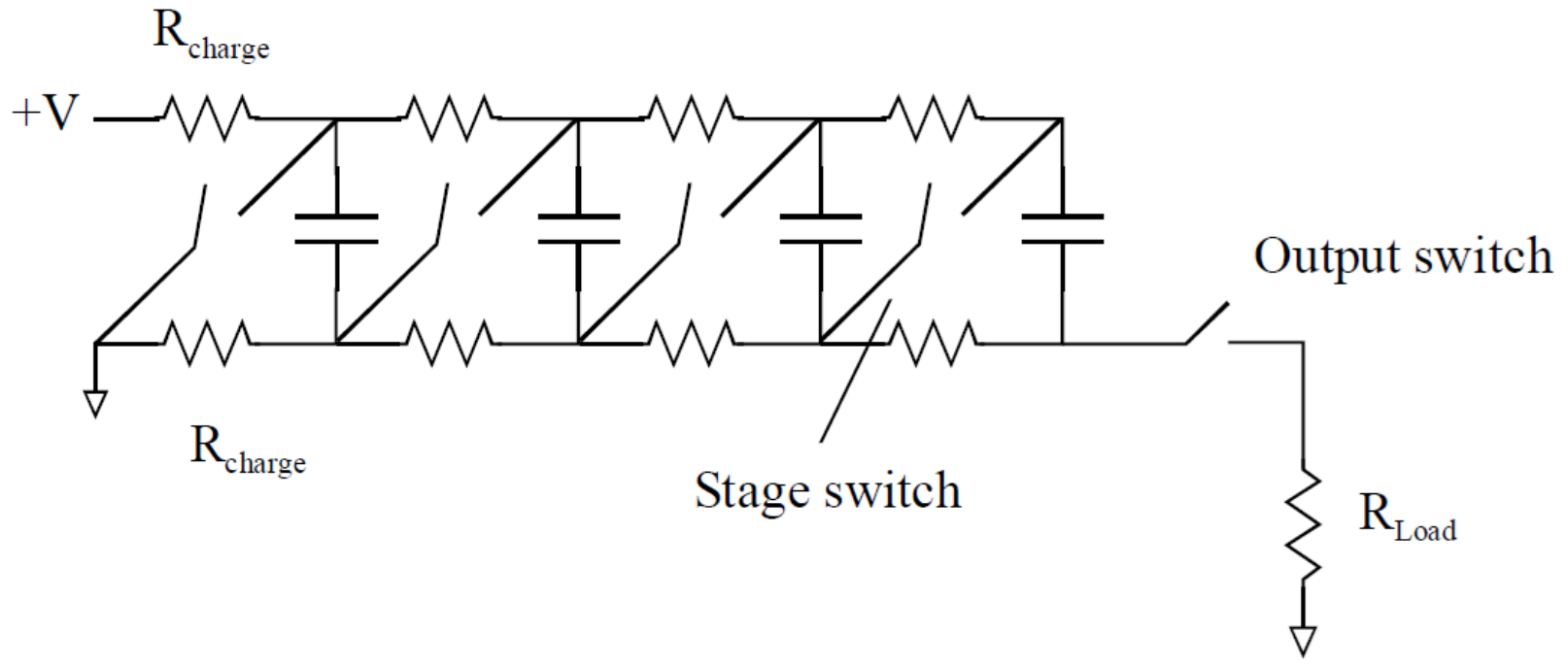

**Figure 4.3.8.47:** Typical Marx generator topology

After conducting a full investigation of the scheme, the proposed topology for the CEPC collider ring injection and extraction trapezoidal wave kicker system is the SiC MOSFET switch-based inductive adder topology, as shown in Figure 4.3.8.48. The polarity of the output high-voltage pulse of the inductive adder can be configured flexibly by changing the connection mode of the secondary end of the coaxial transformer to meet different working modes.

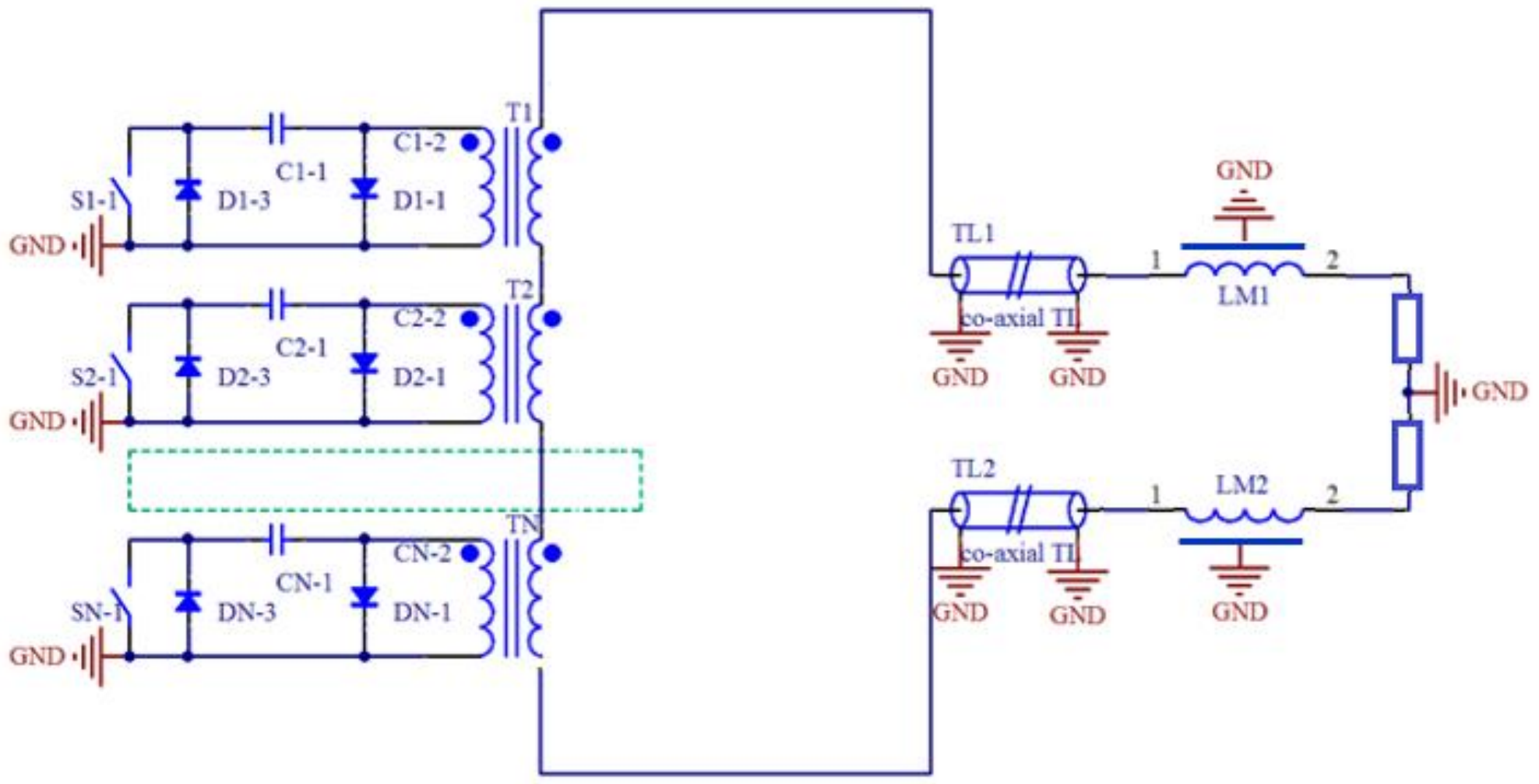

**Figure 4.3.8.48:** Circuit Topology of Trapezoidal Pulse Discharge System in CEPC

The MOSFET switch is the preferred option for the CEPC collider ring off-axis injection kicker system because the IGBT switch has a slower turn-off speed than MOSFET, which cannot meet the minimum pulse width requirement of 440 ns. For this application, the SiC MOSFET Cree C2M0045170D is adopted, whose main performance parameters are shown in Table 4.3.8.3. It has a breakdown voltage of up to 1700 V, a conduction resistance of 45 mΩ, and a pulse surge current larger than 100 A for a single chip.

The inductive adder topology is used for pulse power superposition, with a single-stage output voltage controlled below 1000 V/stage, and the 20-stage stacking output is about 20kV. There are 28~30 MOSFETs in parallel in each stage of the inductive adder, with 3kA pulse current output capacity. Two independent 20-stage inductive adders can be used to excite the two electrodes of each dual C-type delay-line kicker magnet or can be cascaded into a 40-stage adder.

To minimize the voltage drop during the long pulse with a flat top of 2 µs, a discharge circuit with a time constant of $\tau = RC > 20$ µs, 10 times the pulse width, is implemented. To achieve this, the total capacitance C should be greater than 3.2 µF, assuming a load resistance $R$ of 6.25 Ω. For a 20-stage adder, the equivalent capacitance of each stage should be $C_0 > 64$ µF. To achieve this, each capacitor bank of every stage consists of 28~30 2µF/1600 V film capacitors connected in parallel. Compensation technology can also be used to further improve the voltage drop during the long pulse.

In the design of the coaxial transformer, it is recommended to use an iron-based nano-crystalline material for the magnetic core. The sectional area of the magnetic core must be large enough to meet the volt-second product requirement for a rectangular pulse of 2 µs/1kV. The transverse magnetic annealed nano-crystalline magnetic core, of which the typical hysteresis loop is shown in Fig4.3.8.49, has low remanence (Br = 0.2 T) and high saturated magnetic density (Bs = 1.2 T), making it more suitable for unipolar pulse applications. To avoid saturation, the core's sectional area must be larger than S > UT/ΔB = 2 µs × 1 kV / 0.5 T = 40 $cm^2$ if ΔB = 0.5 T. However, a core of this size would be large, heavy, and expensive. Alternatively, a non-magnetic annealed magnetic core can be used, and a DC bias circuit can be added to improve ΔB and reduce the sectional area of the magnetic core. A pulser prototype with pulse width of 300 ns was developed for HEPS.

The magnetic core size of the prototype is Φ160 × 75 × 40 mm, and with a lamination coefficient of 0.8, the effective magnetic core sectional area is 13.6 cm².

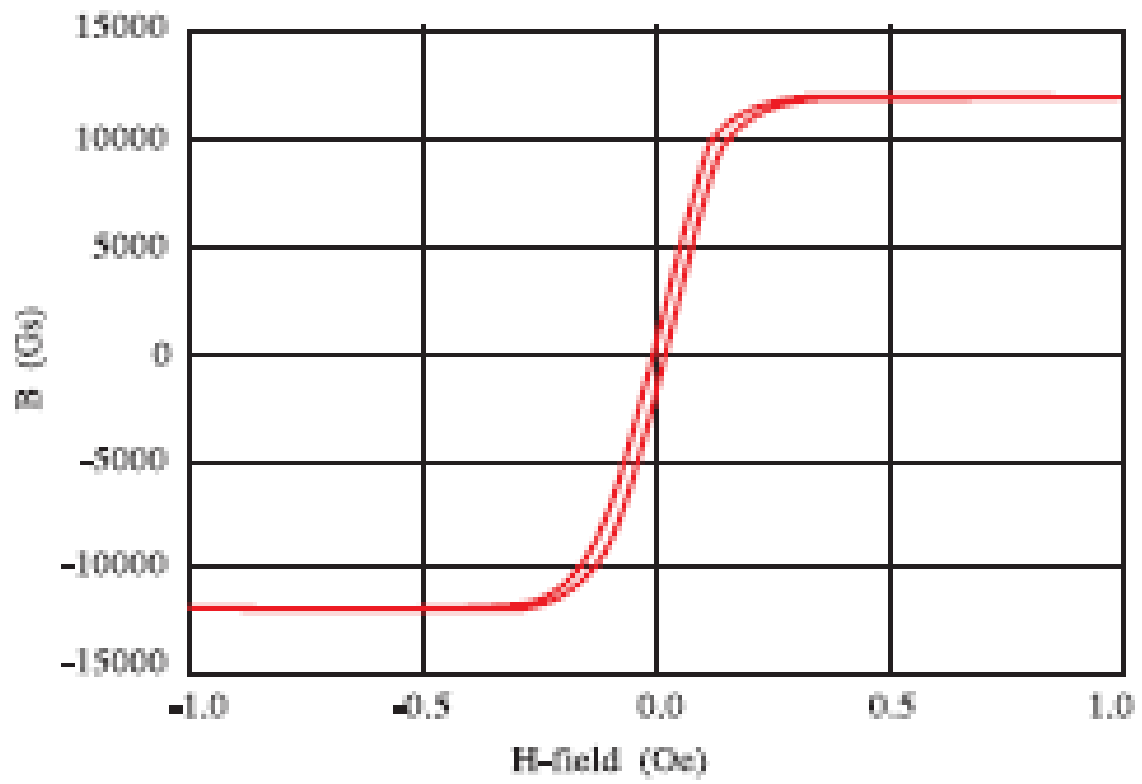

**Figure 4.3.8.49:** Typical hysteresis loop of transverse magnetic annealing nano-crystalline core

In terms of the design of the inductive adder structure, a stacking structure is used for the core and primary winding of the coaxial transformer, while the secondary winding is formed by a copper tube that passes through the center of the core. The switch circuit board is designed as a plug-in unit, which makes maintenance easy. The structure of the multi-stage adder can be mounted either horizontally or vertically, but the horizontal structure is generally preferred for ease of testing, as illustrated in Figure 4.3.8.50.

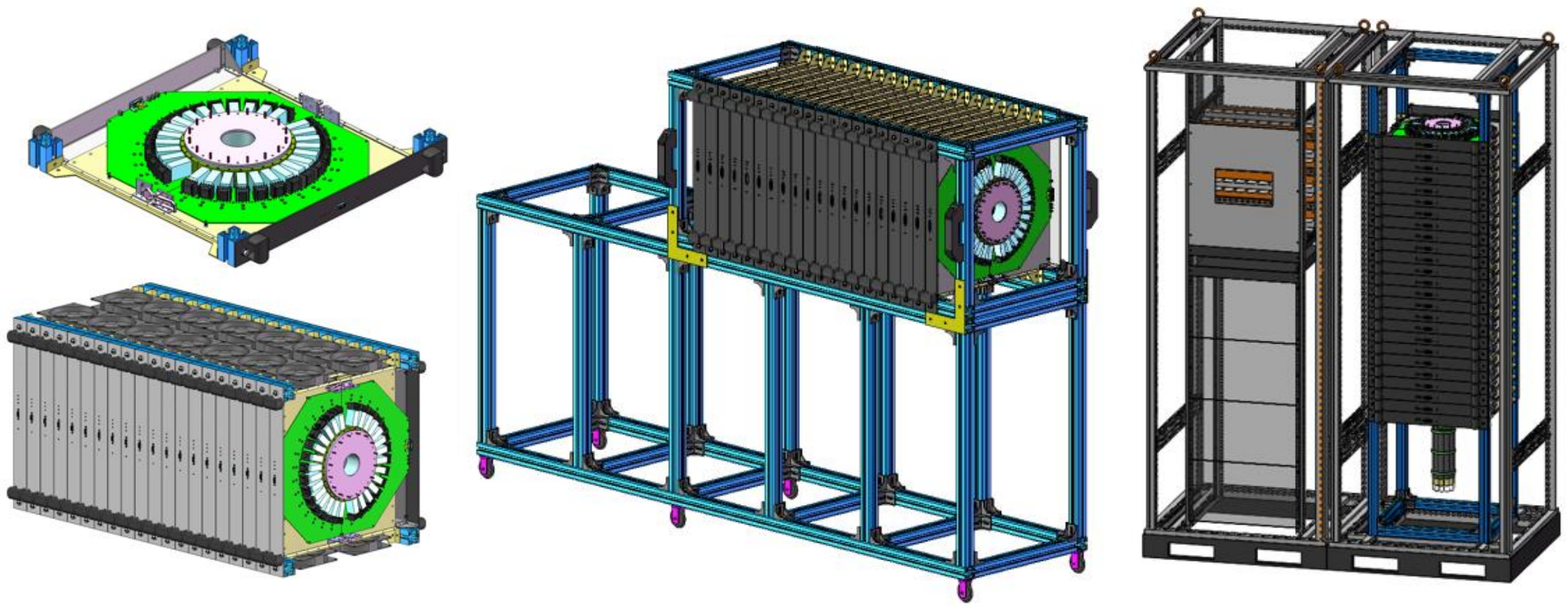

**Figure 4.3.8.50**: Structure design of inductive adder

The R&D single-stage prototype was tested under full power conditions, as shown in Figure 4.3.8.51. At a DC charging voltage of 1400 V, a pulse current of 2800 A was obtained on a 0.5 Ω resistance load. The test waveform is depicted in Figure 4.3.8.52.

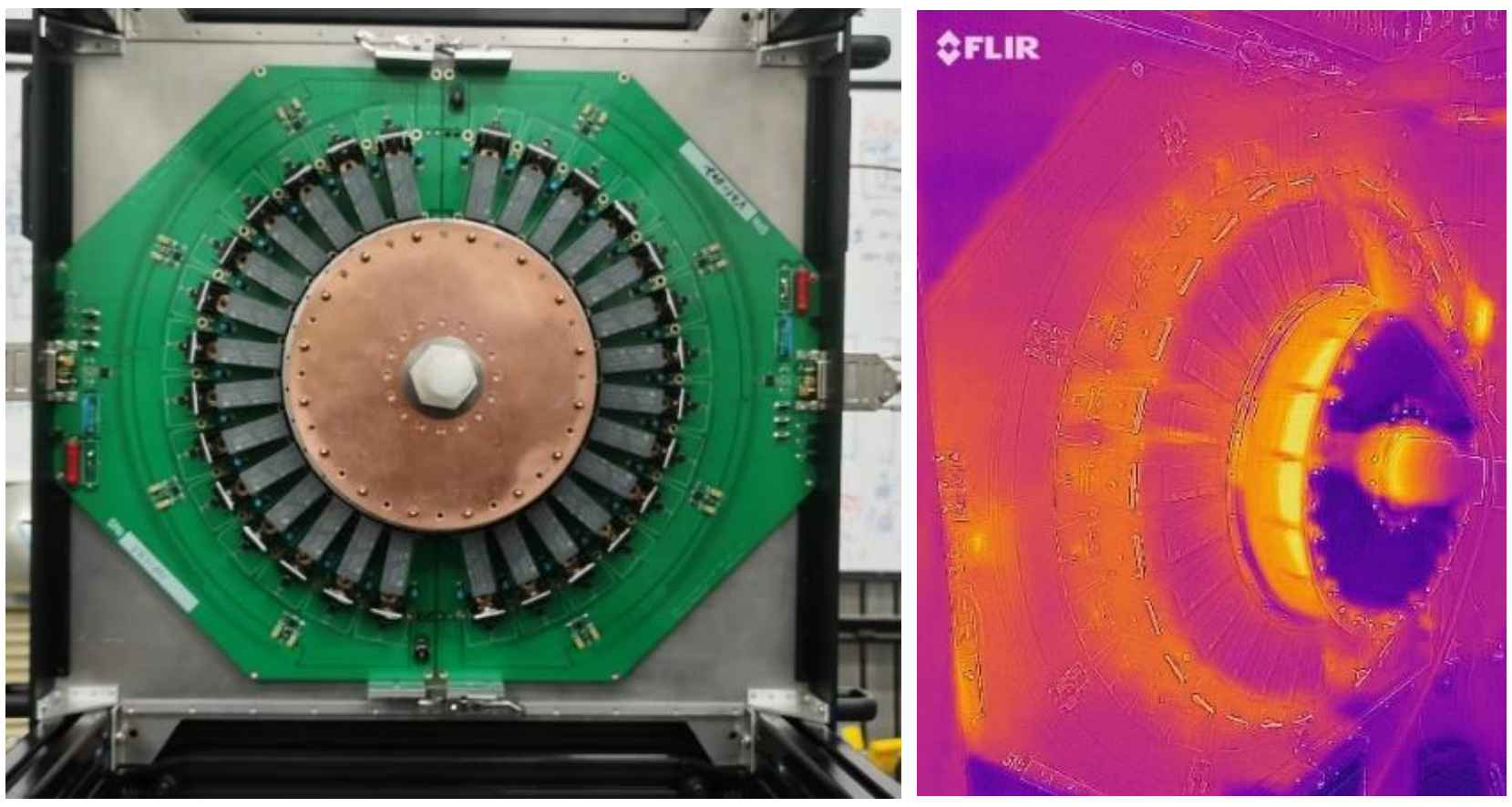

**Figure 4.3.8.51:** Single stage prototype full power test

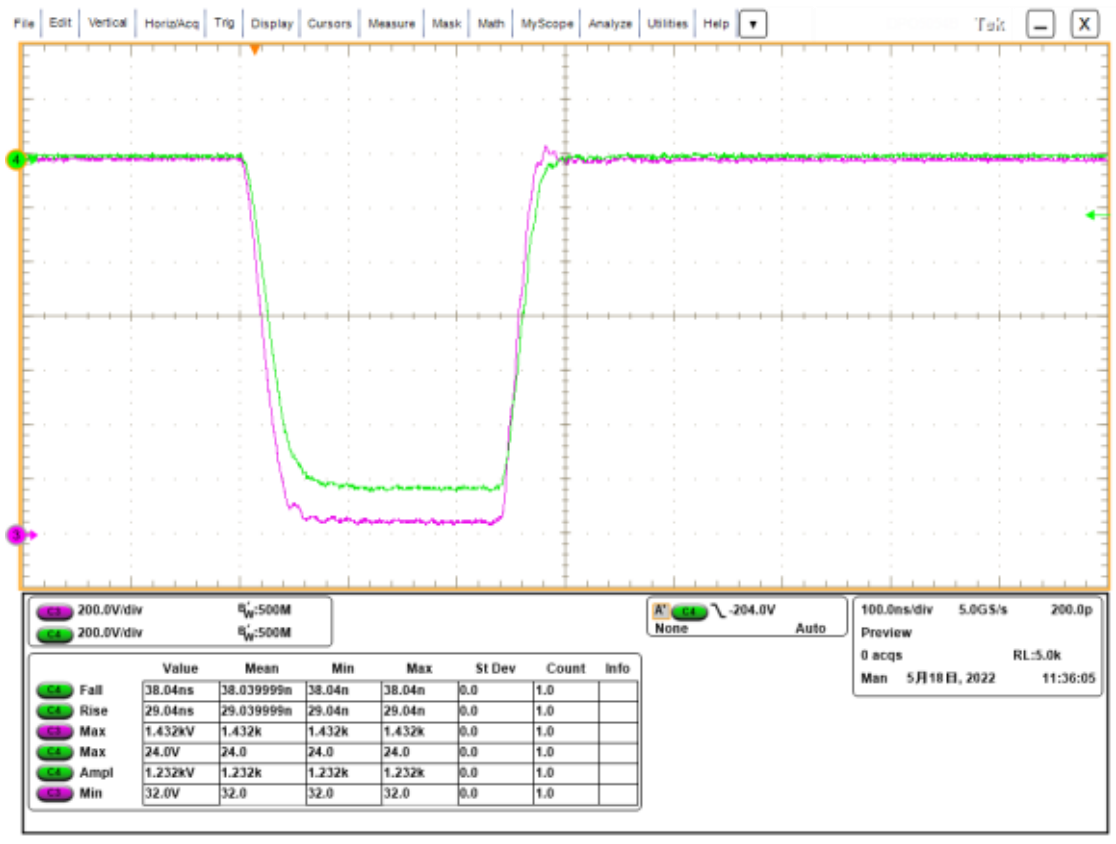

**Figure 4.3.8.52:** Pulse waveform of single stage prototype full power test

The 18-stage inductive adder prototype was constructed and subjected to high voltage testing, as depicted in Figure 4.3.8.53. At a DC charging voltage of 800 V, the adder was able to generate a pulse current of 1.424 kA and a pulse voltage of 14.2 kV when loaded with a 10 Ω resistor. The corresponding pulse waveform is shown in Figure 4.3.8.54. However, under these test conditions, the switching-off of the MOSFETs led to a high reverse voltage peak at the tail of the pulse due to parasitic inductance outside the adder. To prevent overvoltage breakdown of the MOSFET, this peak must be suppressed using a back-peak absorption circuit consisting of high-speed diodes.

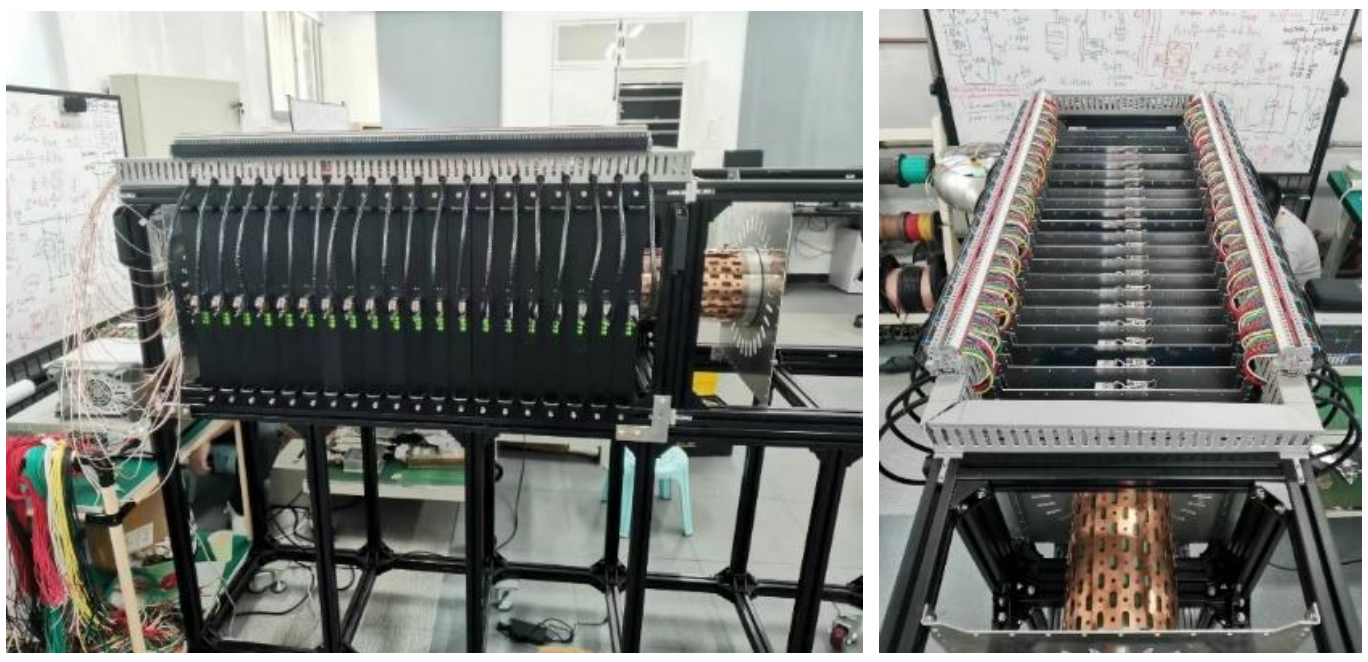

**Figure 4.3.8.53:** An 18-stage inductive adder prototype

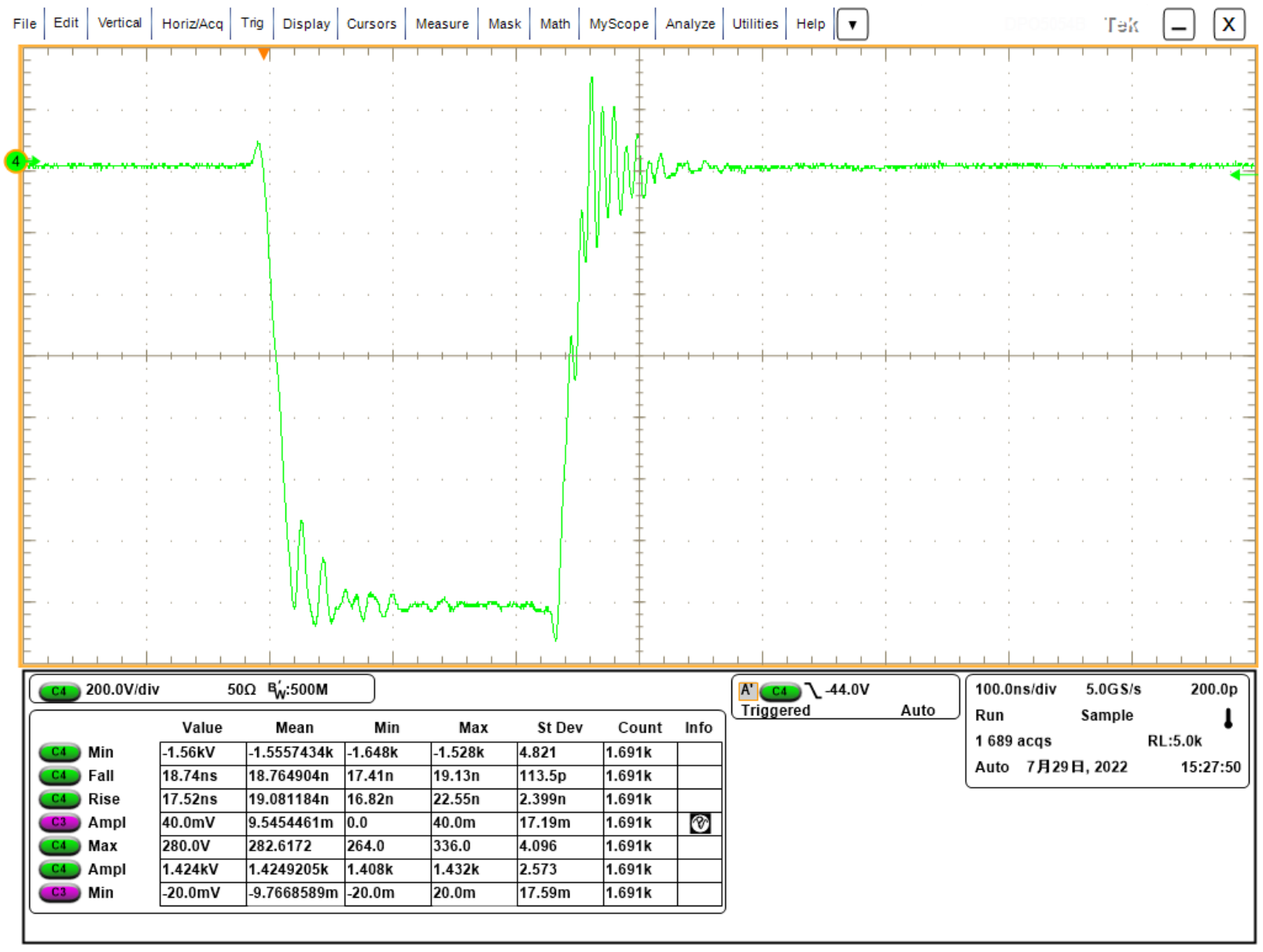

**Figure 4.3.8.54:** The test waveform of 18-stage inductive adder prototype

### 4.3.8.8 *Lumped Parameter Dipole Kicker Magnet*

The on-axis swap-out injection method is exclusively adopted for the Higgs mode in the collider rings, which requires a minimum bunch spacing of 680 ns. In order to deflect the bunches one by one, the injection and extraction kickers must generate a half sine pulse with a pulse bottom width of no more than 1360 ns, and a pulse repetition rate of 1 kHz. The design parameters of the magnet can be found in Table 4.3.8.11. Given the slow speed of the kicker pulse, an in-air dual-C ferrite core kicker magnet is preferred due to its higher excitation efficiency, simpler structure, lower cost, and other advantages. This type of kicker magnet is widely used in synchrotron radiation light source accelerators.

**Table 4.3.8.11:** Design Parameters of kicker magnet for the collider on-axis injection

| Parameter | Unit | On-axis Injection |
|---|---|---|
| Quantity (including e+/e- ring, shared by inj.and ext.system) | - | 4 (2×2) |
| Type | - | In-air lumped parameter dipole kicker |
| Deflect direction | - | Horizontal |
| Beam Energy | GeV | 120 |
| Deflect angle | mrad | 0.2 |
| Integral magnetic strength | T·m | 0.08 |
| Magnetic effective length | m | 1 |
| Magnetic strength | T | 0.04 |
| Clearance region (H×V) | mm | 56 × 56 |
| Good field region (H×V) | mm | 50 × 50 |
| Field uniformity in good field region | - | ±1.5% |
| Repetition rate | Hz | 1k |
| Amplitude repeatability | - | ±0.5% |
| Pulse jitter | ns | ≤5 |
| Bottom width of pulse (5%-5%) | ns | 1360 |
| Tr/Tf (5%-95%) | ns | <680 |
| Pulse waveform | - | 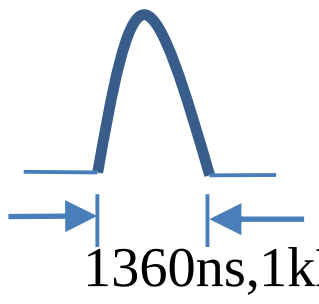 1360ns,1kHz, Half-sine |

The structure of the in-air dual-C ferrite core kicker magnet designed for CEPC is illustrated in Figure 4.3.8.55. The magnet window frame consists of two C-type ferrite cores, which are spliced together. These ferrite cores are separated by copper eddy current shielding plates, which serve to reduce the beam impedance of the magnet. A single-turn exciting coil is formed by the two copper conducting plates in the core window frame. The entire magnet is placed outside the vacuum. To mitigate the shielding effect of eddy currents on the pulse magnetic field, a ceramic vacuum chamber is required. Additionally, to reduce beam impedance and decrease secondary electron emission in the positron ring, a continuous TiN-Ag film coating is necessary on the inner surface of the ceramic vacuum chamber. The kicker magnet can utilize the ferrite core and the ceramic vacuum chamber with the same size as the delay-line kicker magnet for off-axis injection, with a magnet aperture of W × H = 100 × 80 mm and a ceramic vacuum chamber with a 1200mm long, round internal contour size of 56×56 mm and an octagon outer contour size of 85 × 66 mm.

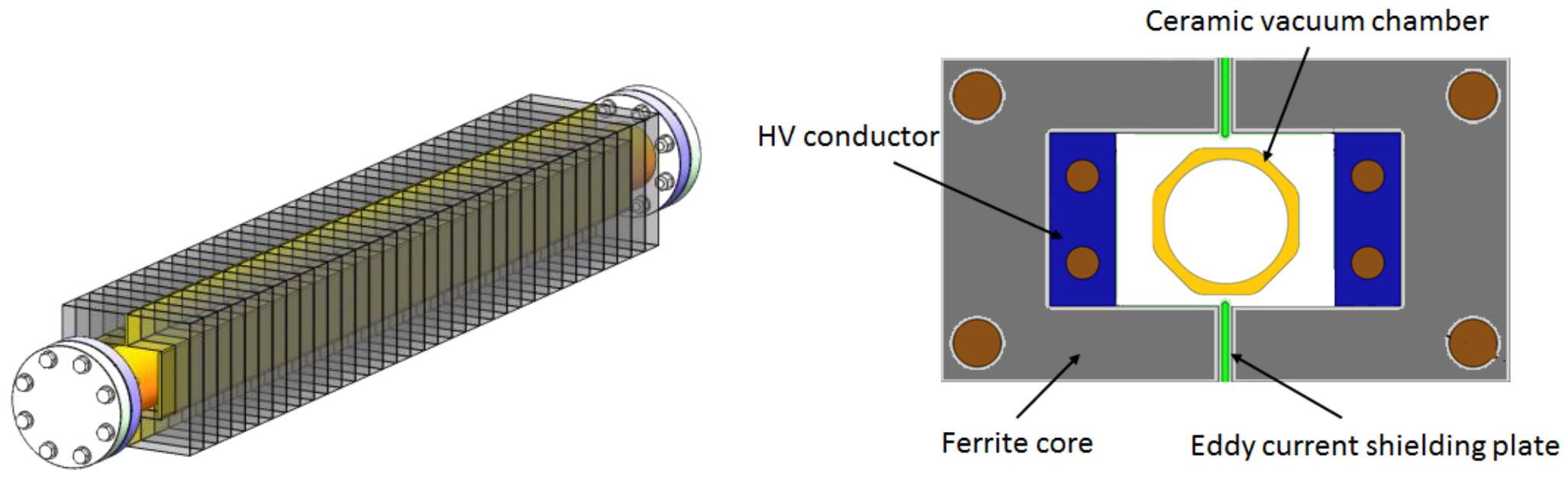

**Figure 4.3.8.55:** In-air dual-C ferrite core lumped parameter kicker magnet

The magnet aperture is set to $W \times H = 100 \times 80$ mm, and the effective length of the magnet is $l = 1$ m, the magnetic field strength is $B = 0.04$ T. Using formula (4.3.8.1) and (4.3.8.2), the total inductance of the kicker magnet is calculated to be $L = 1.6$ μH. The exciting current of the magnet is calculated to be $I = 2.6$ kA.

$$L = \mu_0 \frac{w}{h} l \tag{4.3.8.1}$$

$$I = \frac{B}{\mu_0} h \tag{4.3.8.2}$$

A simple way to generate a half sine pulse is through an LC resonant discharge circuit, as shown in Figure 4.3.8.56. In this circuit, L represents the total inductance of the loop, which includes the kicker magnet inductance and the stray inductance; C represents the capacitance of the energy storage capacitor; and R represents the damping resistance. Under the condition of no damping R=0, the L and C can meet the relations of (4.3.8.3) (4.3.8.4) and (4.3.8.5), in which τ is the pulse bottom width of half-sine pulse:

$$C = \frac{\tau^2}{\pi^2 L} \tag{4.3.8.3}$$

$$Z = \sqrt{\frac{L}{C}} = \frac{\pi L}{\tau} \tag{4.3.8.4}$$

$$U = IZ \tag{4.3.8.5}$$

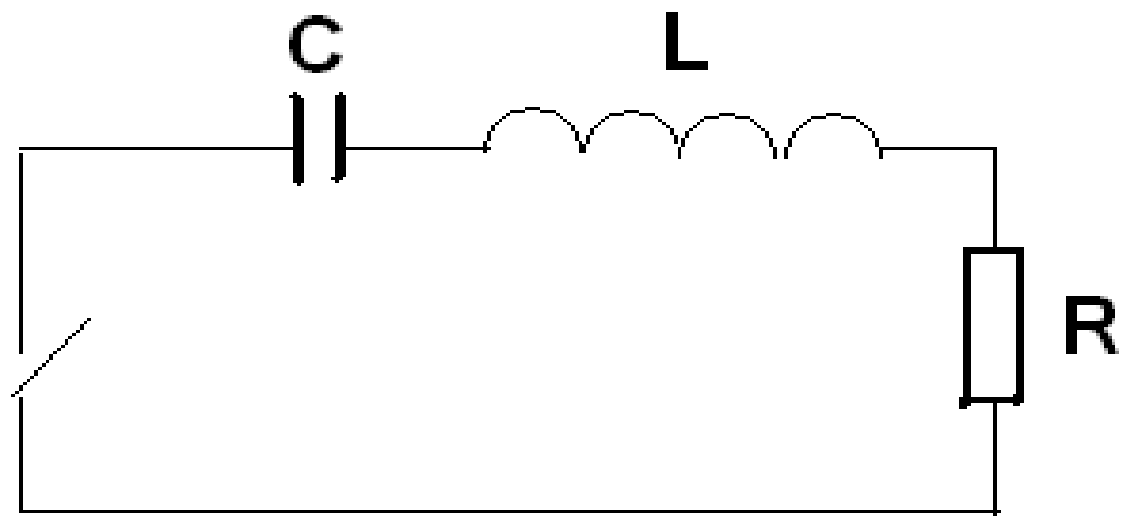

**Figure 4.3.8.56:** Simplified circuit of LC Series Resonant Discharge

Based on the physics requirement of a pulse bottom width τ = 1.36 μs, we can calculate the corresponding values for the components involved. The capacitance can be determined as C = 117 nF. The discharge circuit impedance is found to be Z = 3.7 Ω. Additionally, the working voltage of the LC resonant discharge circuit is U = 9.62 kV.

Table 4.3.8-12 presents the primary design parameters of the lumped parameter ferrite core dipole kicker magnet intended for the on-axis injection system of the collider ring.

**Table 4.3.8.12:** Main parameters of the collider lumped parameter dipole kicker magnet.

| Parameter | Value |
|---|---|
| Magnetic field B (T) | 0.04 |
| Inner aperture of ceramic vacuum chamber (mm×mm) | 56 × 56 |
| Ceramic vacuum chamber outer aperture (mm×mm) | 85 × 66 |
| Magnet aperture w × h (mm×mm) | 100 × 80 |
| Length of magnet *l* (m) | 1 |
| Inductance of magnet L (μH) | 1.6 |
| Magnet exercitation current I (kA) | 2.6 |
| Impedance Z (Ω) | 3.7 |
| Voltage of magnet U (kV) | 9.62 |
| Pules bottom width τ (ns) | 1360 |
| Repetition rate （Hz） | 1000 |

### 4.3.8.9 *Half-sine Wave Solid State Pulsed Power Supply*

The half-sine wave pulsed power supply based on LC series resonance is a mature technology, and a unipolar heavy hydrogen thyratron is usually used as the discharge switch. However, the switching frequency of conventional thyratrons is generally limited to 50 Hz~100 Hz, which cannot meet the 1 kHz repetition frequency requirement of CEPC kicker. Therefore, semiconductor solid-state switch technology must be introduced. A novel topology circuit has been proposed which uses IGBT as the discharge switch, SiC SBD with unidirectional conduction characteristics to obtain unidirectional pulses, and a special inductive adder topology to realize pulse power superposition. The simulation model of the four-stage inductive adder circuit in Pspice is shown in Figure 4.3.8.57, and the simulation results are presented in Figure 4.3.8.58.

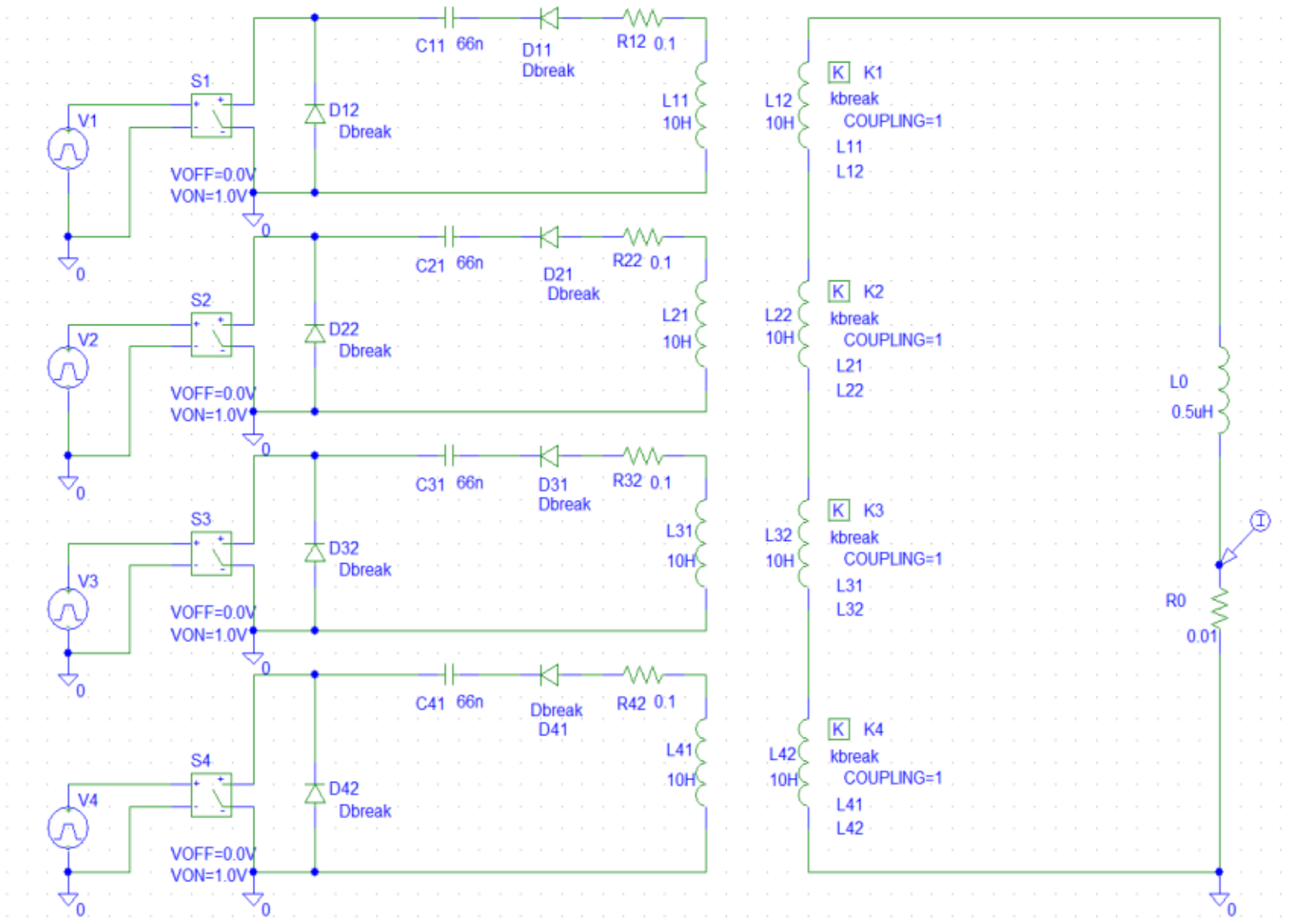

**Figure 4.3.8.57:** Pspice simulation model of the LC resonant discharge inductive adder.

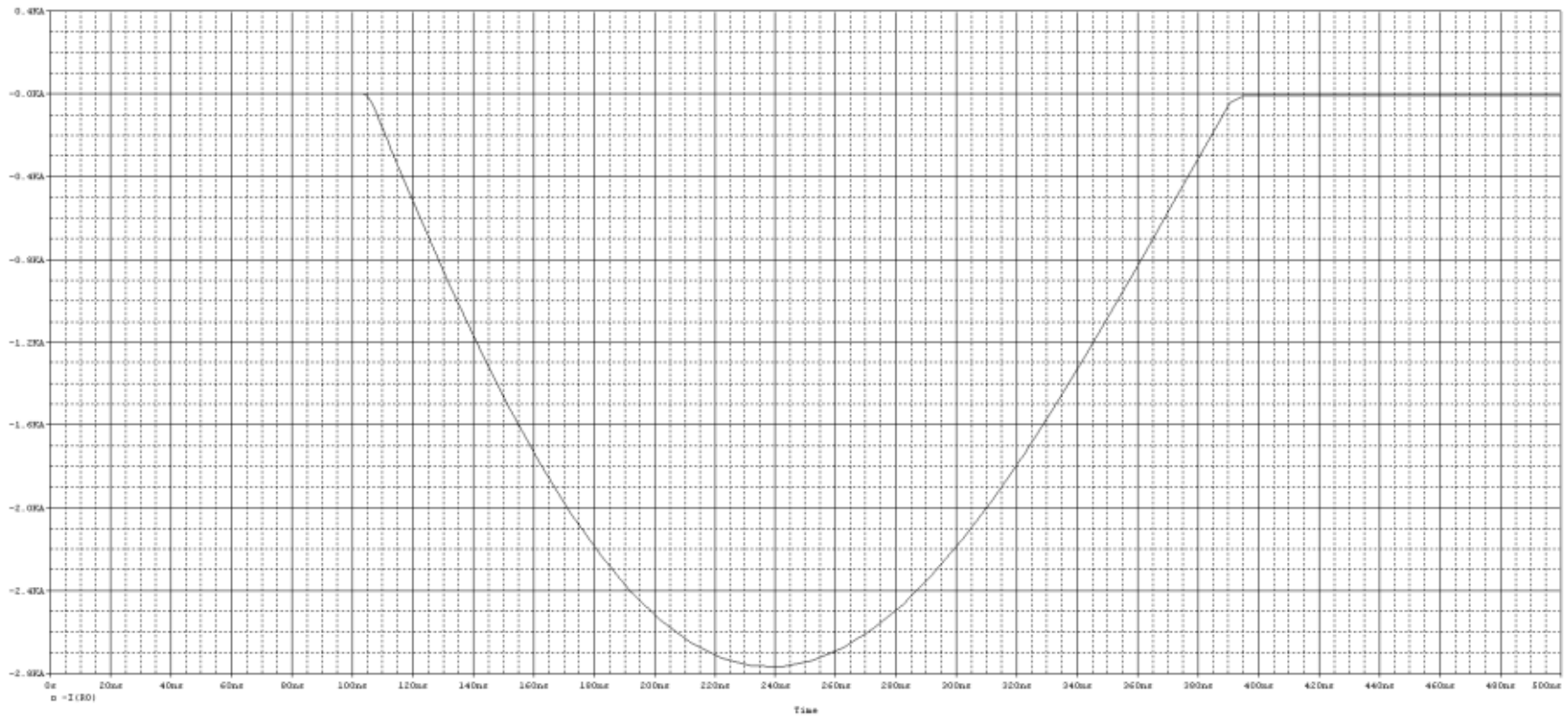

**Figure 4.3.8.58:** Simulation waveform of the LC resonant discharge inductive adder.

The circuit experiment of LC resonant discharge circuit with IGBT and SBD is shown in Figure 4.3.8.59. The experiment results indicate that the proposed circuit topology is suitable for high repetition frequency half-sine pulsed power supply. The stacking structure of the inductive adder is shown in Figure 4.3.8.60.

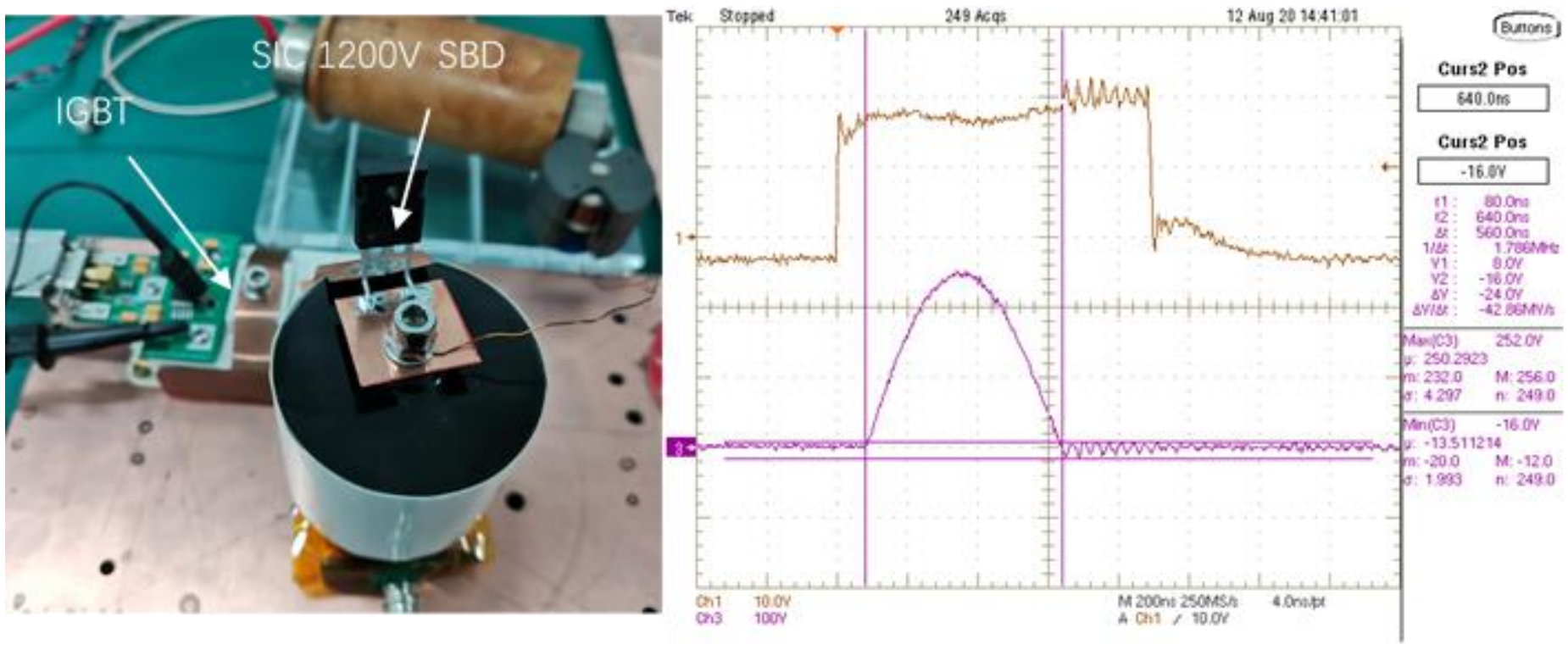

**Figure 4.3.8.59:** The circuit experiment of LC resonant discharge circuit with IGBT and SBD

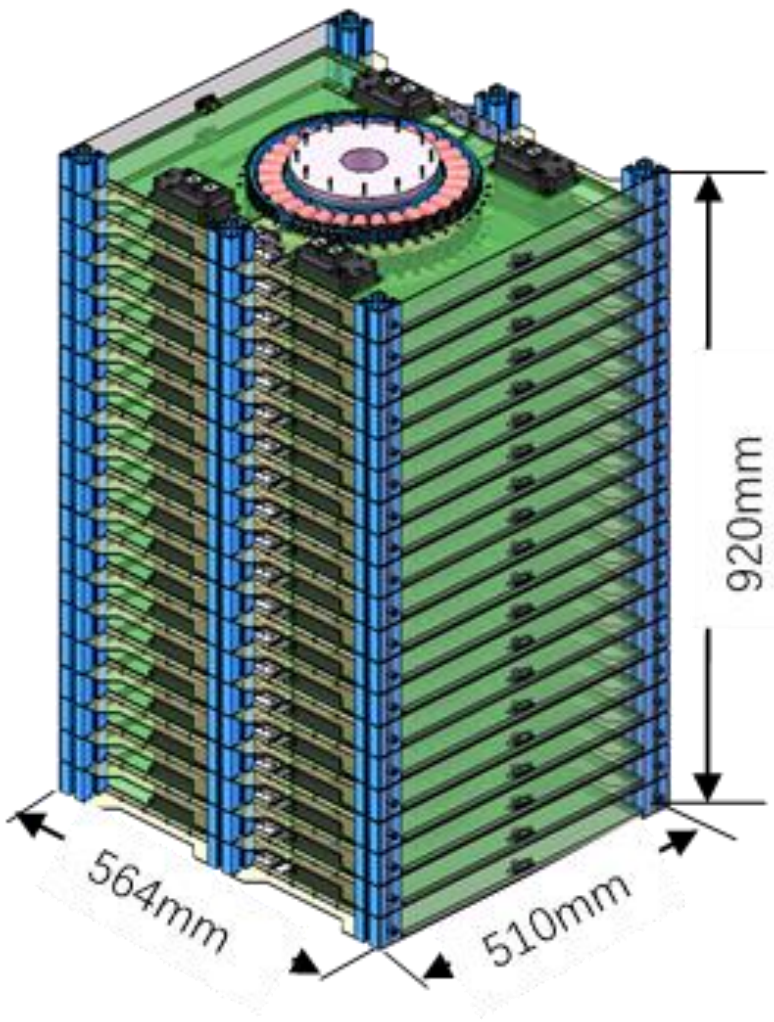

**Figure 4.3.8.60:** Structure schematic of the LC resonance inductive adder

### 4.3.8.10 *Lambertson Magnet*

The dynamic aperture (DA) of the CEPC collider rings is small, which means that the septum magnet must have a thin septum. For the on-axis injection system, the septum thickness must be less than 6mm, while for the off-axis injection system, it must be less than 2 mm.

The most common type of thin septum magnet used is the eddy current plate type septum magnet. This type of magnet uses the eddy current shielding effect of the septum plate to create differential action on the injected beam and the circulating beam, which are separated by the septum plate in space. Therefore, pulse excitation mode must be used for the eddy current septum magnet. The entire magnet, including the laminated iron core and the excitation coil, must be placed in a large vacuum chamber. The thickness of the septum plate is just the thickness of the eddy current shielding plate. For this type of septum, the vector direction of the main deflection magnetic field is parallel to the septum plate, making it more suitable for situations where the injected beam and the circulating beam orbit are on the same horizontal plane.

Lambertson magnet is another type of septum used for beam injection and extraction in particle accelerators. It is actually a special bipolar magnet that operates on DC excitation, which makes it highly reliable and easy to maintain. Unlike the eddy current plate type septum, the vector direction of the main deflection magnetic field of a Lambertson magnet is perpendicular to the septum plate. This makes it more suitable for injection and extraction systems where the injected beam and the circulating beam are not on the same horizontal plane. In the CEPC, the booster ring and collider rings are at different heights, so Lambertson magnet is the preferred option for the injection and extraction process between them. However, designing a Lambertson magnet with a thin septum plate of only 2~6mm is a significant technical challenge because it must include the thickness of the iron septum, the walls of the injection and extraction beam tubes, and the installation margins, as shown in Figure 4.3.8.61.

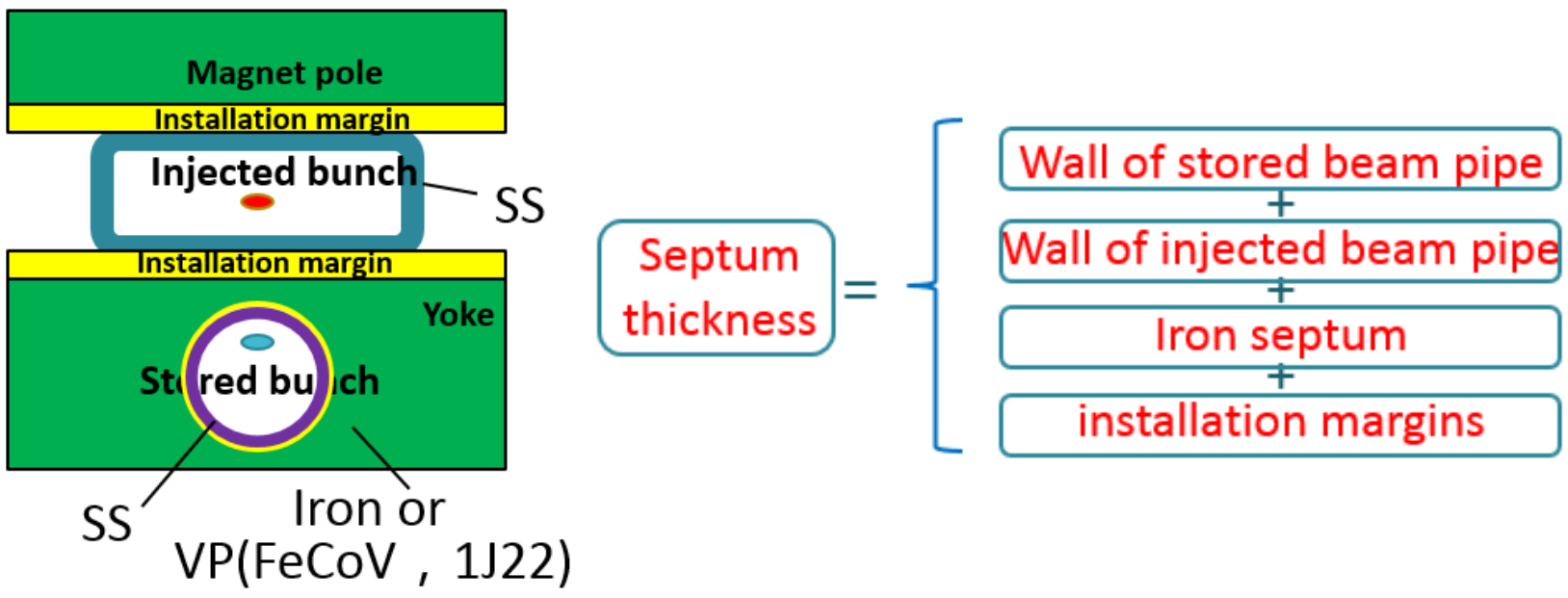

**Figure 4.3.8.61:** Typical structure of conventional Lambertson septa magnet

**Table 4.3.8-13** Design parameters of Lambertson magnets for the Collider off-axis injection.

| Parameters | Unit | CR-LSM-1 | CR-LSM-2 |
|---|---|---|---|
| Quantity | - | 2 × 4 | 2 × 4 |
| Deflection direction | - | Vertical | Vertical |
| Energy | GeV | 120 | 120 |
| Total deflection angle | mrad | 13 | 13 |
| Total Integral field strength of septa | T-m | 5.22 | 5.22 |
| Deflection angle provided by a magnet | mrad | 3.5 | 3.5 |
| Insertion length | m | 1.75 | 1.75 |
| Magnetic field strength for injected/extracted beam | T | 0.8 | 0.8 |
| Min. Septum thickness (including septum board, wall of beam pipes, installation gap) | mm | 6 | 2 |
| Field uniformity | - | < ±0.05% | < ±0.05% |
| Leakage field | - | $\leq 1\times10^{-3}$ | $\leq 1\times10^{-3}$ |
| Clearance of stored beam at lambertson (H×V) (refer to stored beam orbit) | mm | - | - |
| Clearance of inj.&ext. beam at lambertson (H×V) (refer to inj.&ext. beam orbit) | mm | 20 × 20 | 20 × 20 |
| Physical aperture of stored beam vacuum chamber | mm | 56 × 56 | 56 × 56 |

The design parameters of the Lambertson magnets for the CEPC Collider ring's off-axis injection can be referred to in Table 4.3.8.13. However, for the CEPC Collider on-axis injection, a larger quantity of Lambertson magnets is required due to the increased demand for a total deflection angle of 35 mrad.

There is often a trade-off between the thickness of the septum plate and the strength of the leakage field. The key technological challenge for designing thin-septum magnets is to reduce the septum plate thickness as much as possible while still meeting the requirements for the leakage field strength.

Firstly, two proposed magnet structures based on the traditional Lambertson magnet structure are shown in Figure 4.3.8.62. The embedded thin-walled vacuum tube structure minimizes the wall thickness and installation gap of the vacuum tubes for stored beam and injected beam, and the actual septum thickness includes the wall thickness of the tubes and iron septum plate. This kind of in-air magnet structure is relatively simple and is preferred for Lambertson magnets with septum thickness ≥ 3.5 mm. The CEPC collider on-axis injection Lambertson magnets, which require a septum thickness of 6mm, adopted this kind of magnet structure.

The other proposed Lambertson magnet structure is a half in-vacuum scheme, which is a novel design. The upper half of the iron yoke and the coil are located outside the vacuum, and the lower iron yoke is placed in a large stainless steel vacuum chamber as a whole. In this way, the actual thickness of the septum only includes the iron septum, which can be minimized to 2 mm. This magnet structure is an option for the collision ring off-axis injection systems. However, this ultra-thin Lambertson magnet structure is complex, so it can be used in combination with thicker septum Lambertson magnets in the overall layout design of the injection systems.

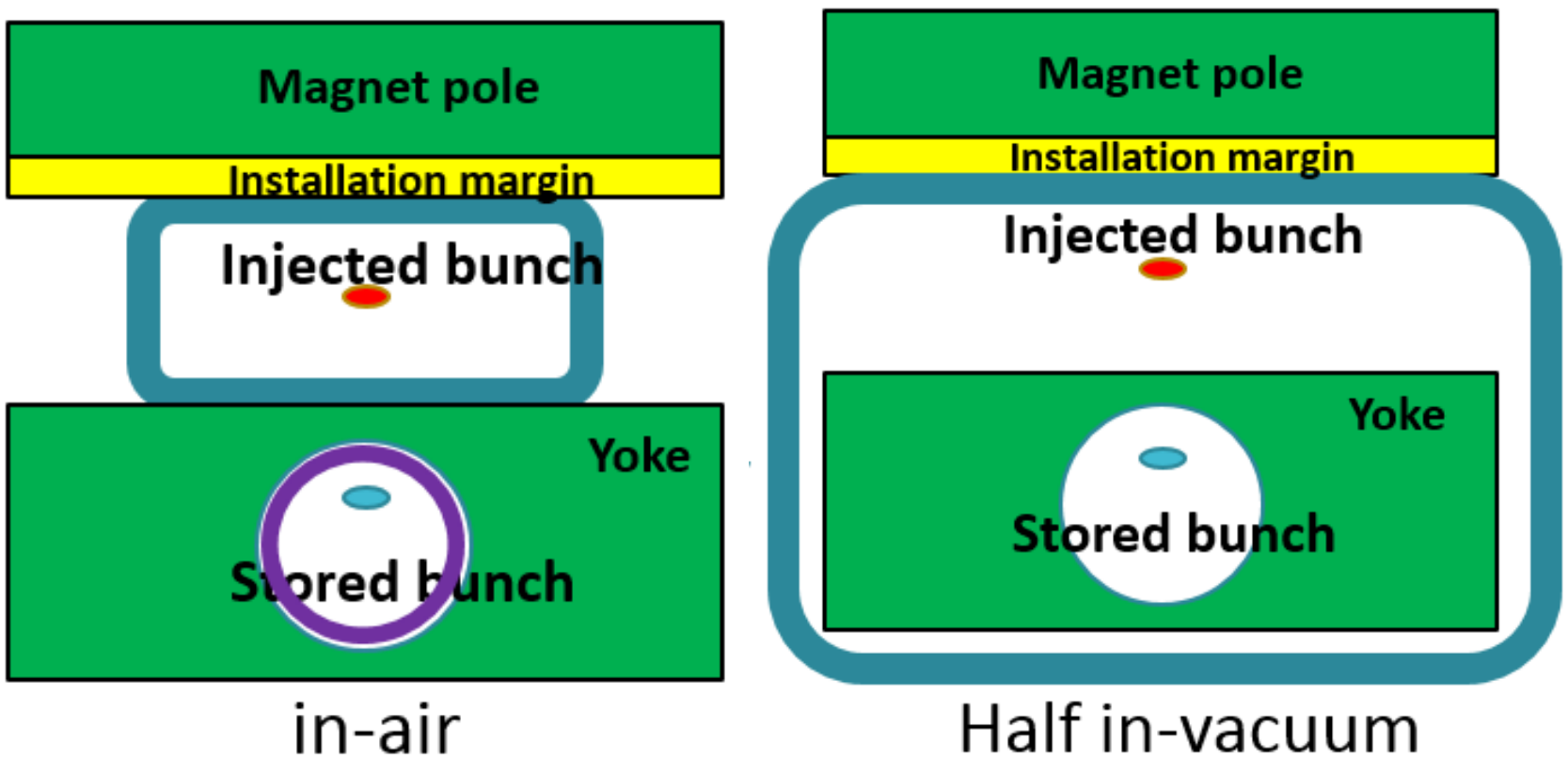

**Figure 4.3.8.62:** Proposed structures of Lambertson magnet

Secondly, according to the principle of the Lambertson magnet, using Fe-Co-V soft magnetic alloy (FeCoV, domestic brand 1J22) for the iron septum can improve magnetic shielding and effectively reduce the absolute leakage field strength in the horizontal and vertical directions. This is due to the higher saturated magnetic flux density (Bs) and permeability ($\mu_r$) of FeCoV compared to pure iron (Fe, domestic brand DT4). Figure 4.3.8.63 shows the comparison curve of the magnetic properties of DT4 and 1J22. The saturation magnetic density of 1J22 can reach Bs = 2.25 T, ensuring that the septum plate does not undergo serious saturation under high field conditions. Figure 4.3.8.64 shows the comparison results of the horizontal and vertical leakage fields under the two different septum plate materials. The absolute leakage field corresponding to the FeCoV alloy septum is more than twice smaller for the same septum thickness. However, FeCoV alloy is expensive and difficult to process, so the magnetic pole, including the septum plate of the Lambertson magnet, can be made by splicing parts made from DT4 and 1J22. It's important to note that the splicing gap on the magnetic pole has a significant impact on both the main field and the leakage field of the magnet.

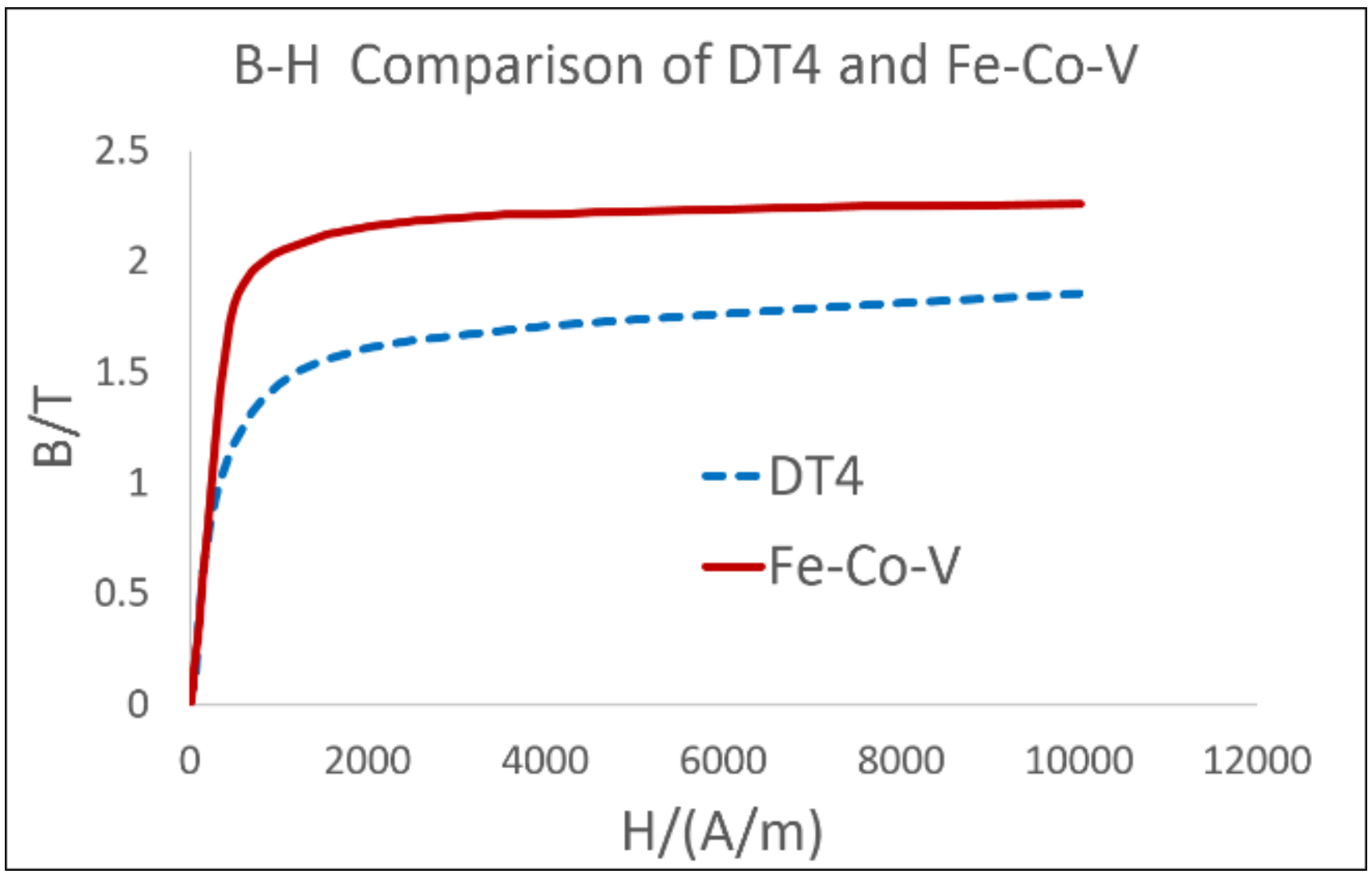

**Figure 4.3.8.63:** Typical B-H curve of Fe (DT4) and FeCoV(1J22)

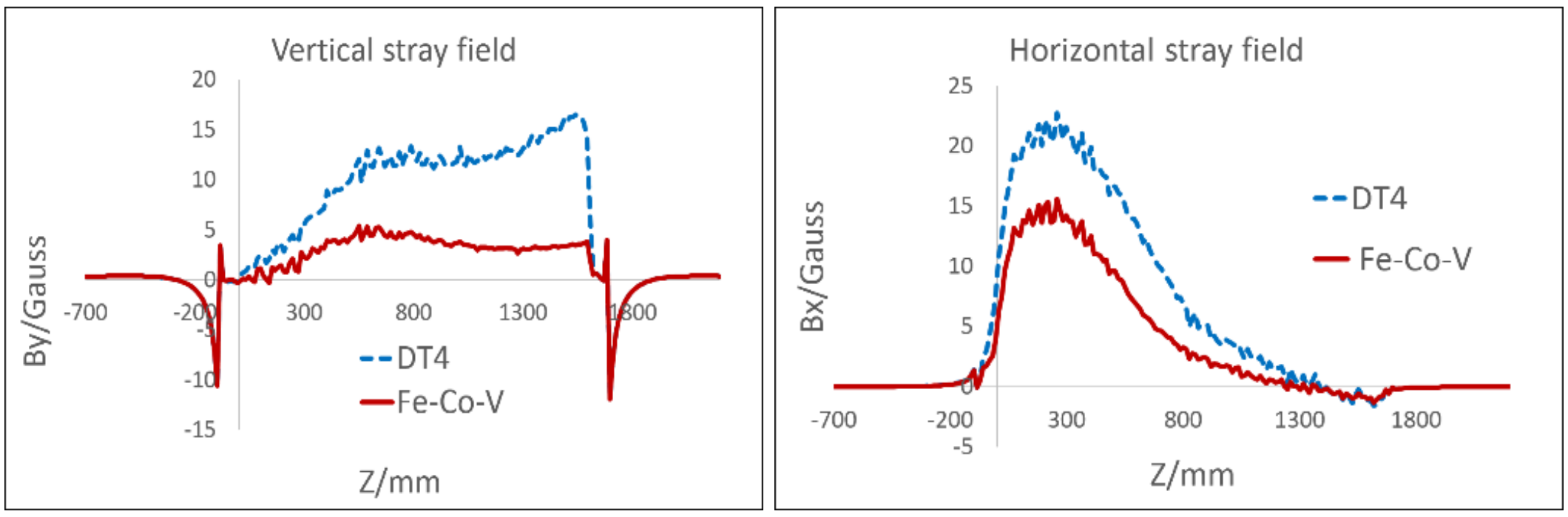

**Figure 4.3.8.64:** Comparison of Vertical and Horizontal Leakage Fields of Two septum Materials

Thirdly, to further reduce the integral leakage field, an approach proposed by APSU involves placing the upstream end of the circulating beam vacuum chamber closer to the side yoke. [10] This results in a leakage field direction opposite to that of the downstream end located just below the injected beam. By cancelling out the vectors with different directions, a lower integral leakage field can be achieved, as shown in Figure 4.3.8.65.

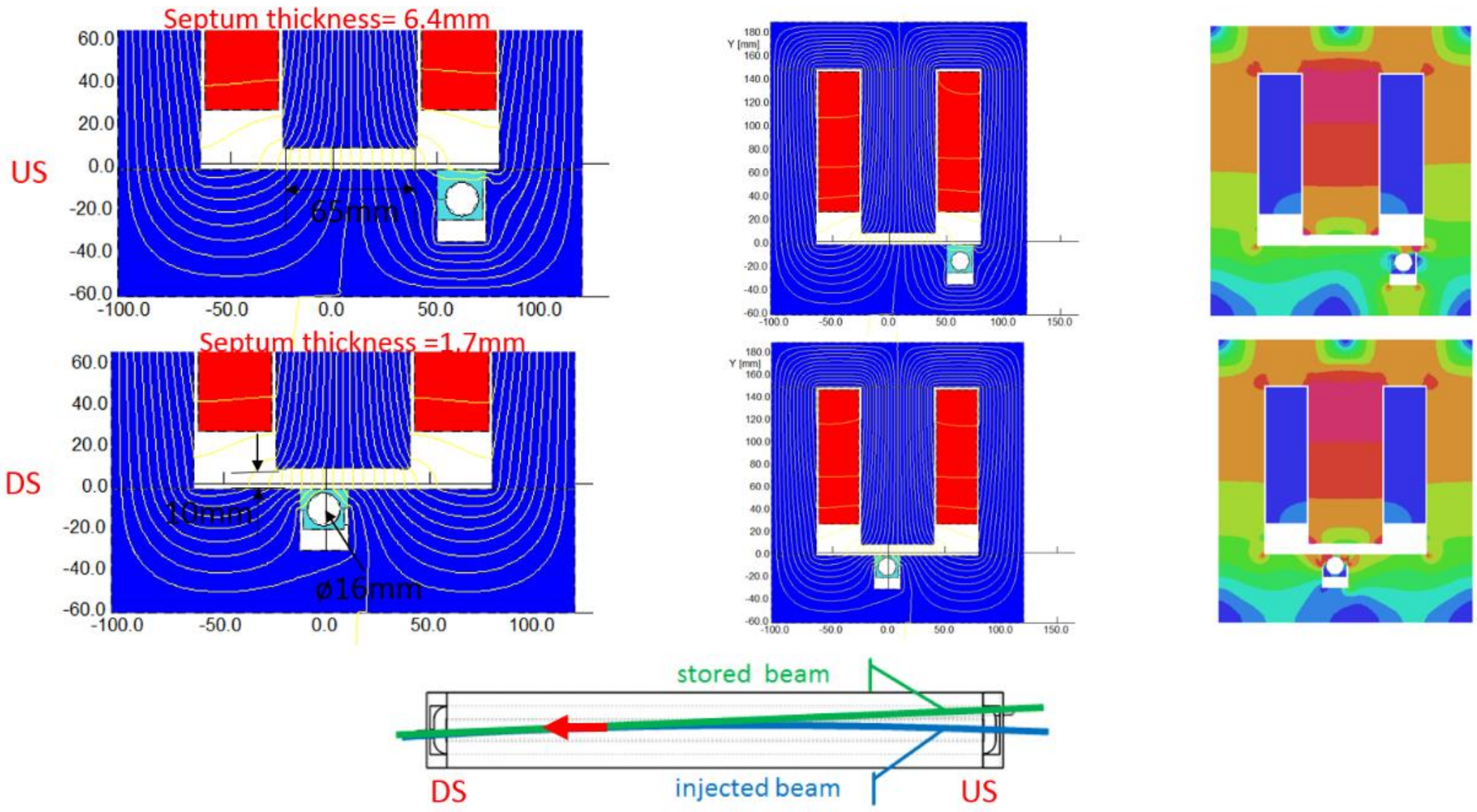

**Figure 4.3.8.65:** Integral Leak Field Cancellation Method

The Lambertson study highlights that the gap surrounding the shielding vacuum chamber of the stored beam can significantly impact the leakage field distribution. As depicted in Figure 4.3.8.66, the different thicknesses of the wings of the vacuum chamber at the magnet entrance affect the direction of the magnetic lines. Thinner wing thickness can lead to easier saturation at this location, causing more magnetic lines to follow Path2 and fewer magnetic lines to follow Path1, thus reducing the leakage field on the stored beam orbit. Reducing wing thickness can decrease both vertical and horizontal leakage fields, particularly the horizontal one, which is the primary means of reducing horizontal leakage, as illustrated in Figure 4.3.8.67.

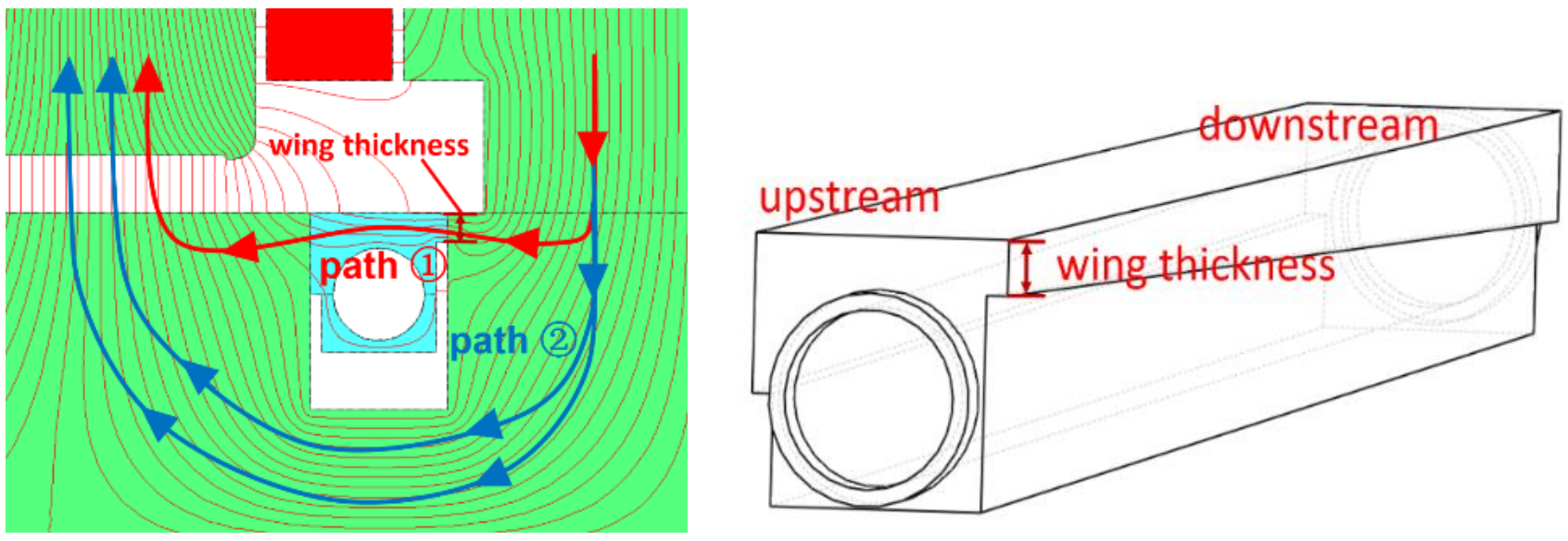

**Figure 4.3.8.66:** Structure model of stored beam shielding vacuum chamber.

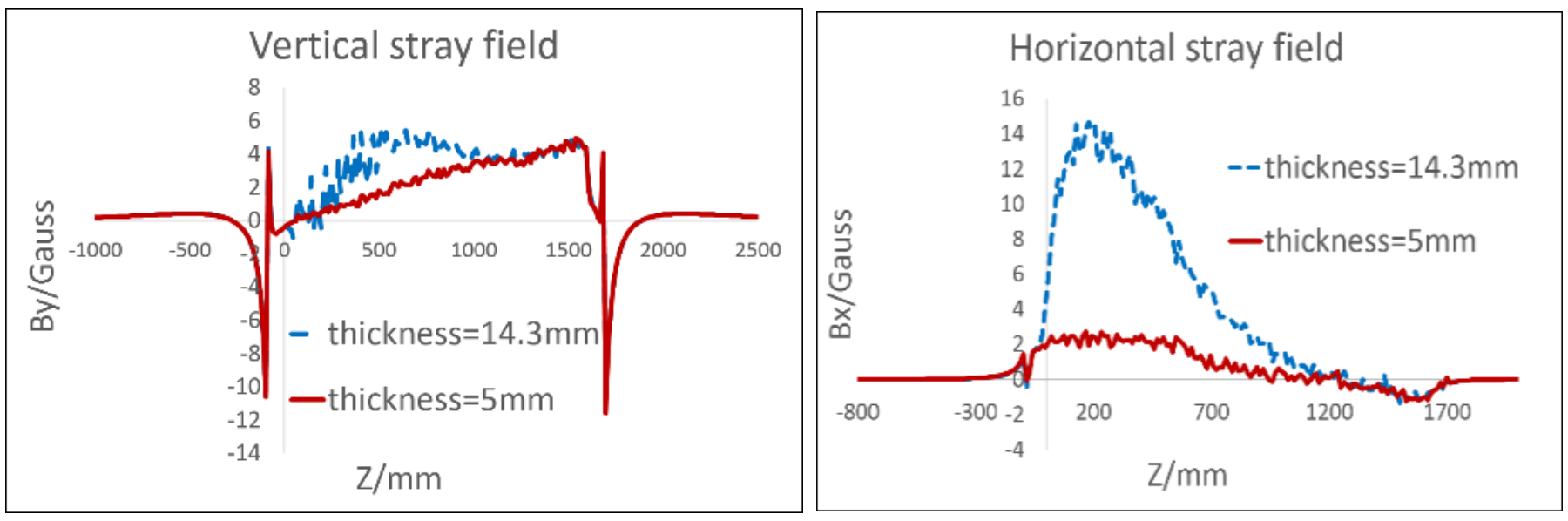

**Figure 4.3.8.67:** Effect on Leakage Field by Air Gap around the shielding Vacuum chamber

Furthermore, the design of the shield plate at the end of the Lambertson magnet is critical in reducing the leakage field. If a shield plate is not present, the leakage field will be generated at both ends of the magnet in the same direction as the main field, resulting in an increase in the integral leakage field. However, with the inclusion of shield plates, the leakage field at both ends of the magnet will be generated in the opposite direction to the main field. This reduces the integral leakage field through positive and negative vector offset. The simulation model and comparison results for the shield plate design are presented in Fig. 4.3.8.68 and Fig. 4.3.8.69.

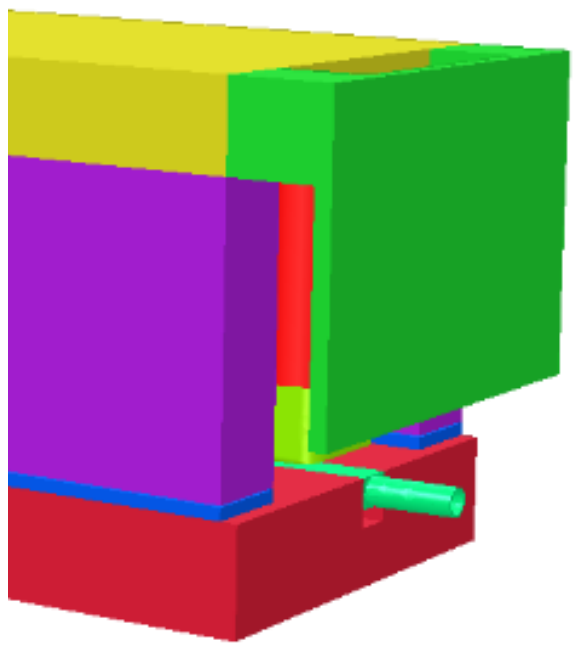

Figure 4.3.8.68: Model of the Shield plate at both end of Lambertson

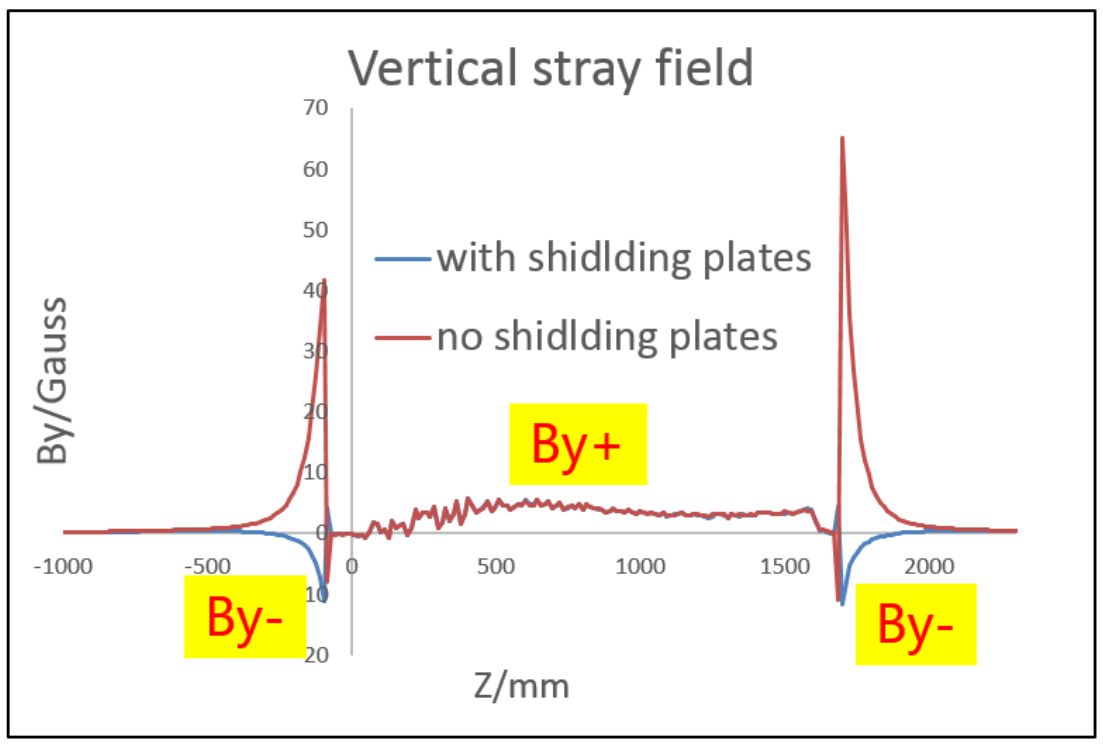

**Figure 4.3.8.69:** Effect on leakage field by shielding plate at both end of Lambertson.

The mechanical design structure of the half-in-vacuum Lambertson magnet prototype developed for the HEPS (High Energy Photon Source) storage ring is illustrated in Figure 4.3.8.70. The magnet is divided into two parts: the upper part, which includes the upper yoke and magnet coil and is located outside the vacuum, and the lower part, which consists of the bottom yoke, the septum plate, and the injected beam passage and is placed in a rectangular stainless steel vacuum chamber. The construction process of the upper part is similar to that of conventional magnets, while the lower part requires vacuum welding technology to ensure its vacuum performance.

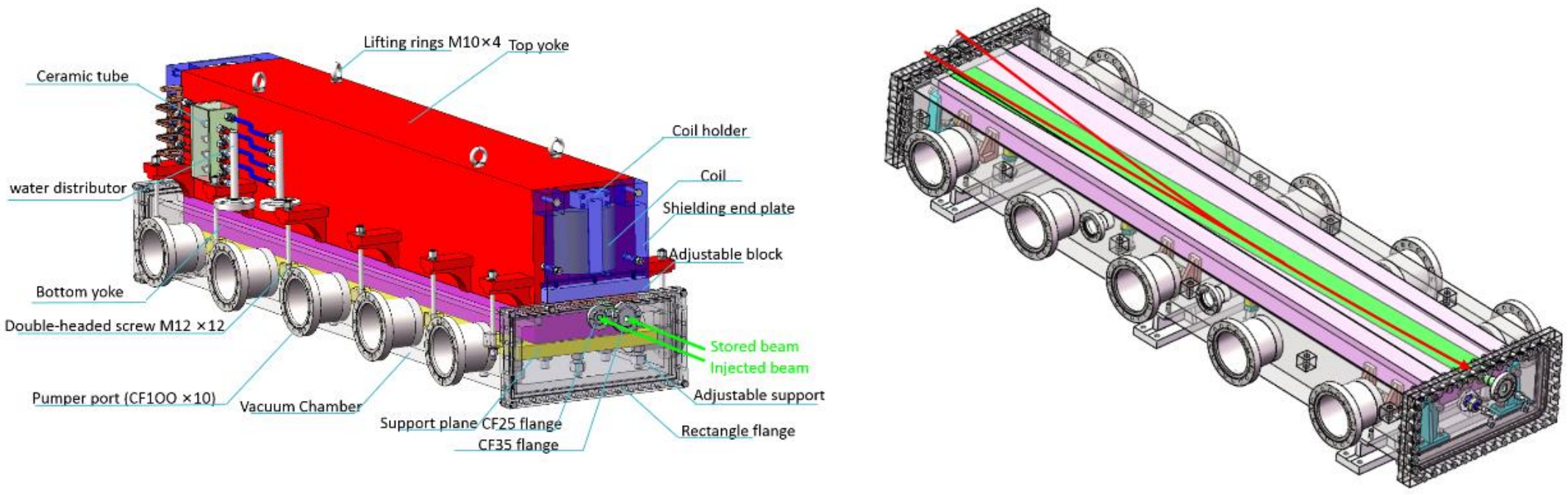

**Figure 4.3.8.70:** Mechanical design of Lambertson magnet for HEPS storage ring

The primary challenge in manufacturing the prototype magnet is the FeCoV alloy (domestic brand: 1J22) shielding block. This material is brittle and hard, making machining extremely risky. The only viable processing method is slow wire cutting (EDM) in small pieces. Additionally, after annealing at 1100 ℃, the material may undergo deformation, leading to flatness deterioration of the pole surface, geometric size changes, and other issues. Unfortunately, annealing deformation is entirely beyond control. Figure 4.3.8.71 depicts the 1J22 shield block machined by EDM. Another hurdle in parts processing is the stainless steel vacuum chamber, which is up to 1.5 meters long, with the thinnest part only 1mm thick. It is crucial to control distortion during machining and welding to ensure proper functionality.

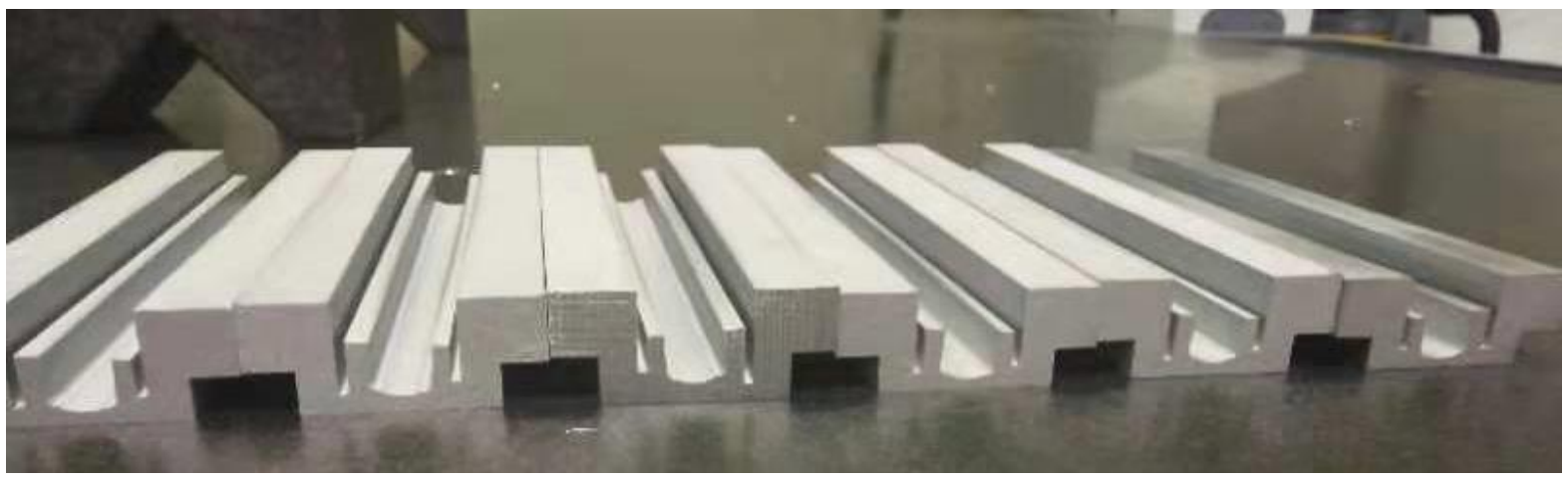

**Figure 4.3.8.71:** 1J22 shielding blocks

Currently, the successful development of the HEPS half-in-vacuum Lambertson prototype has been achieved, and it passed the acceptance process in February 2023. Figure 4.3.8.72 showcases the assembly of the lower half of the magnet structure, while Figure 4.3.8.73 displays the assembly of the complete magnet along with its support system.

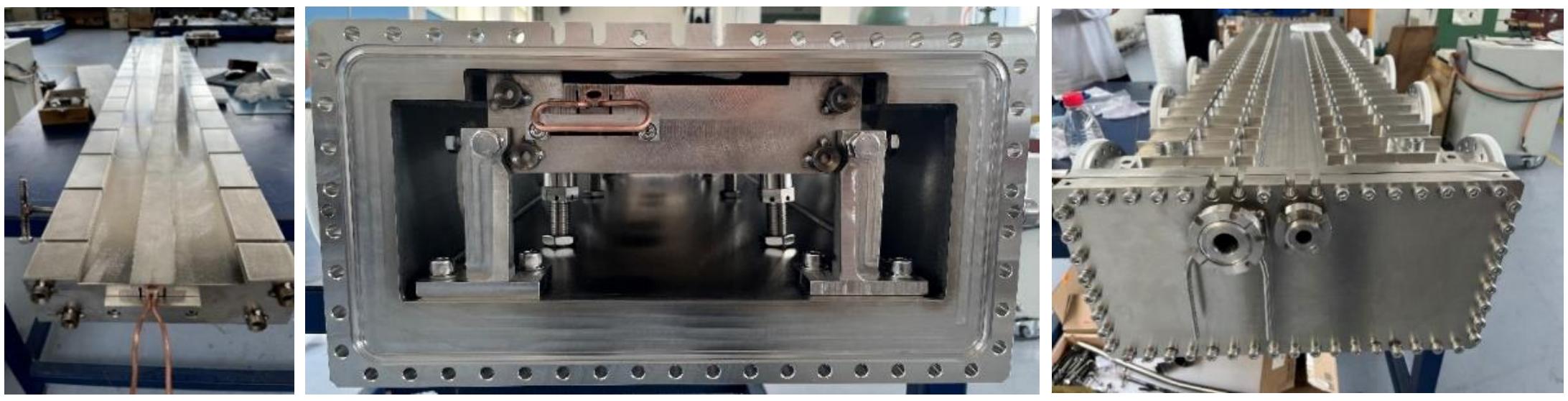

**Figure 4.3.8.72:** Assembly of lower half of the Lambertson magnet prototype for HEPS

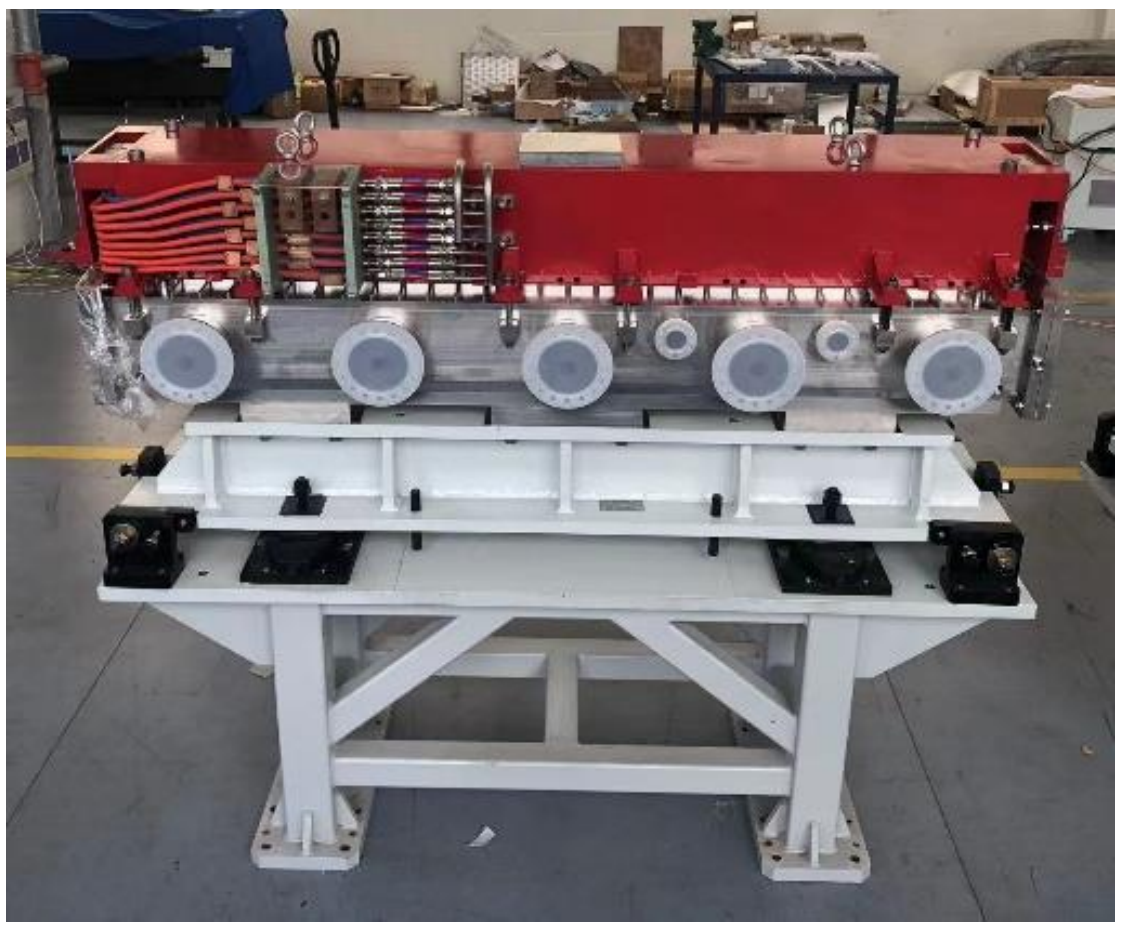

**Figure 4.3.8.73:** Assembly of half-in-vacuum Lambertson magnet prototype for HEPS

The vacuum testing of the lower part of the magnet has been successfully conducted, as depicted in Figure 4.3.8.74. After a period of 7 days, the vacuum pressure inside the chamber reached a value of $5.0 \times 10^{-8}$ Pa. Similarly, the vacuum pressure in the transition section at the end of the magnet reached $2.2 \times 10^{-7}$ Pa, which is in accordance with the simulation results, demonstrating the effectiveness of the vacuum sealing measures.

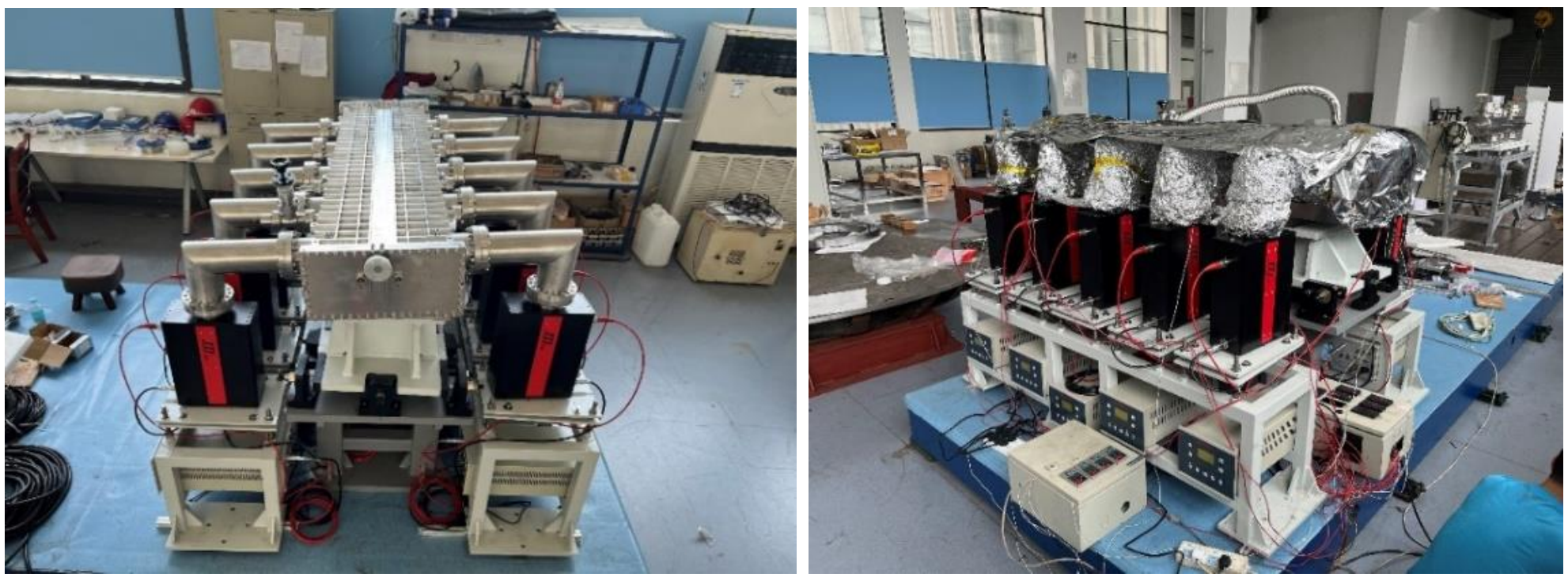

**Figure 4.3.8.74:** The vacuum test of half-in-vacuum Lambertson prototype for HEPS

The magnetic field measurement of the prototype was conducted using a Hall system, as depicted in Figure 4.3.8.75. The measurement results for the main field indicate that at the Y-plane, located at a distance of 1.2 mm from the bottom polar surface, the horizontal field uniformity within the range of x = ±5 mm is 7.22E-04. Additionally, the vertical field uniformity within the range of +1.2 mm to +4 mm (with the bottom polar surface at the Y= 0 plane) is 3.85E-04.

Regarding the measurement results for the leakage field, including the transition part, the integral leakage field in the vertical direction is -6997 Gauss·mm, corresponding to a ratio of 4.3E-04 relative to the integral of the main field. Similarly, the integral leakage field in the horizontal direction is 13464 Gauss·mm, with a ratio of 8.19E-04 to the integral of the main field.

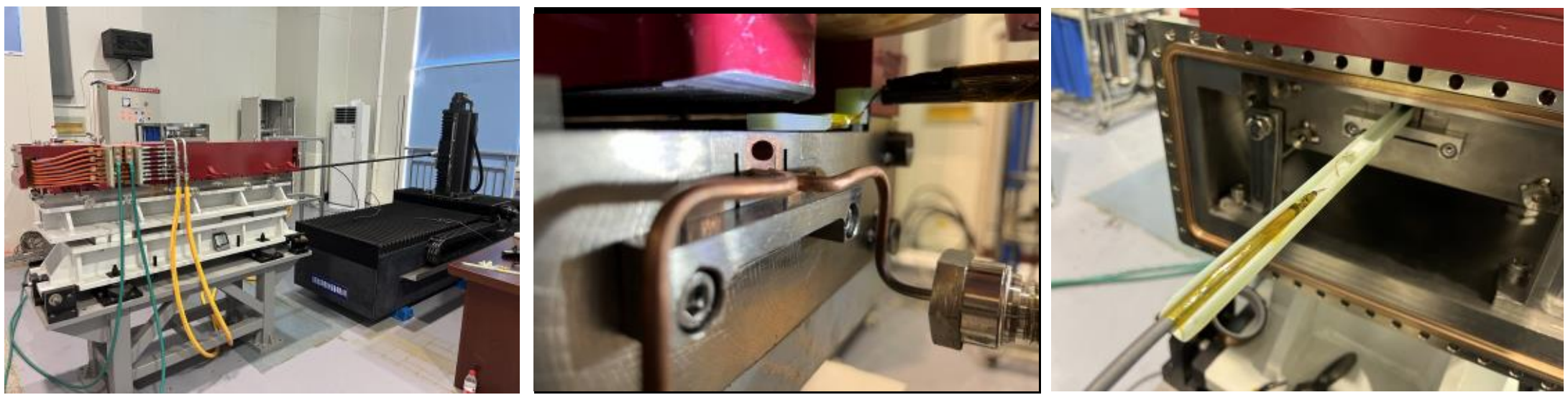

**Figure 4.3.8.75:** The magnetic field measurement of the prototype.

The mechanical design of the in-air Lambertson magnet prototype developed for the HEPS booster injection and extraction systems is depicted in Figure 4.3.8.76. The magnet pole is composed of two materials, pure iorn (DT4) and FeCoV alloy (1J22). The DT4 part is machined using CNC, while the 1J22 part is processed using slow cutting wire EDM in 8 pieces. For the path of the stored beam, a thin-walled stainless steel vacuum tube with a wall thickness of 0.6 mm is embedded in the 1J22 magnet pole parts, while an oval vacuum tube with a wall thickness of 0.6 mm is situated below the pole for injected beam. In addition to the thickness of the iron septum plate, the actual total thickness of the septum plate is ≤ 3.5 mm. Through actual magnetic field measurements and further simulations, it has been discovered that the spliced slits between adjacent 1J22 blocks can amplify the leakage field. Additionally, the spliced slits between the DT4 yoke and 1J22 blocks can significantly impact the quality of the main field signal. To enhance the magnetic field quality, it is necessary to widen the first three 1J22 blocks at the inlet end of the magnet. This adjustment is aimed at preventing the injected beam from crossing the slits between the DT4 yoke and 1J22 blocks, as depicted in Figure 4.3.8.76 (middle and bottom).

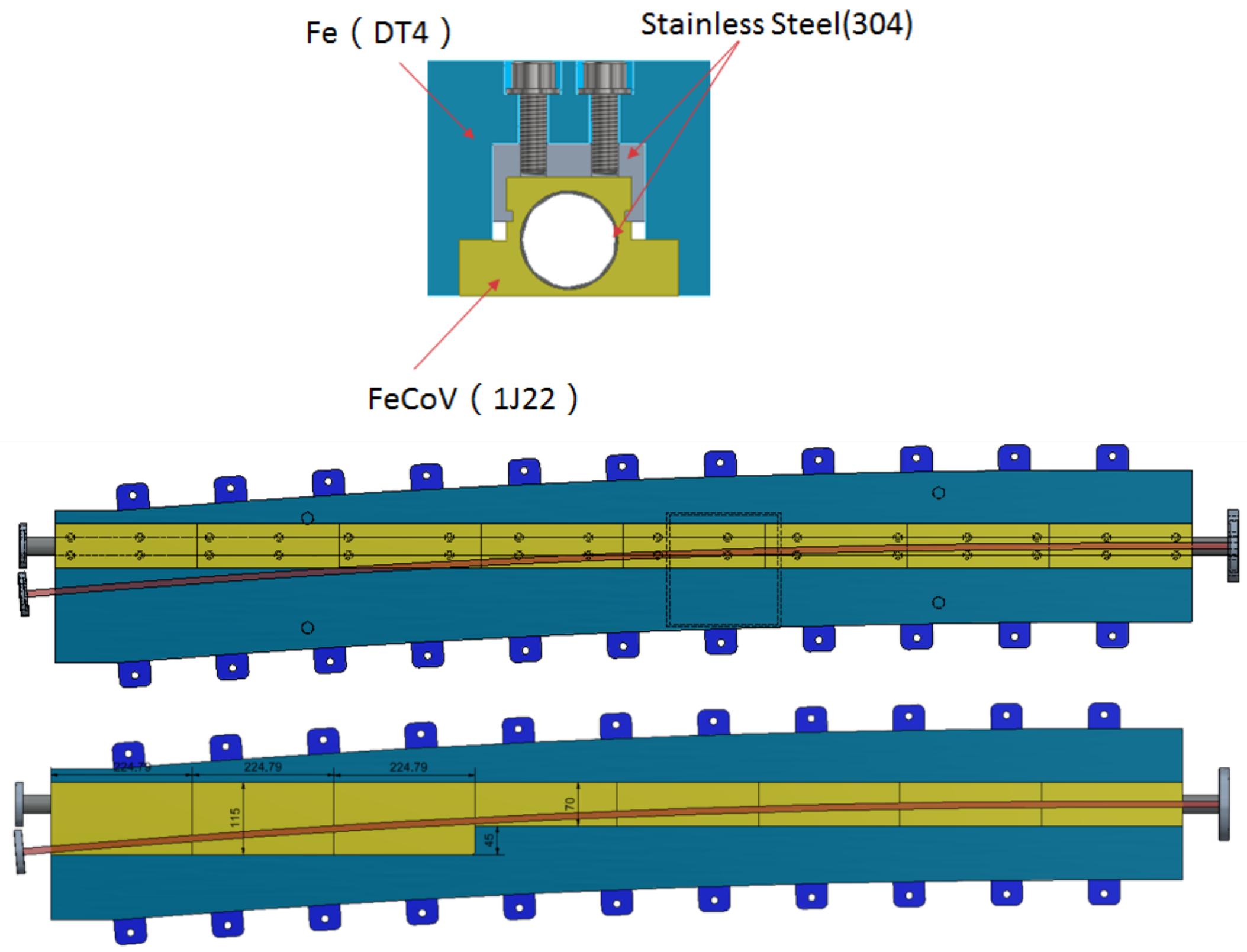

**Figure 4.3.8.76:** The magnet yoke structure of the in-air Lambertson magnet.

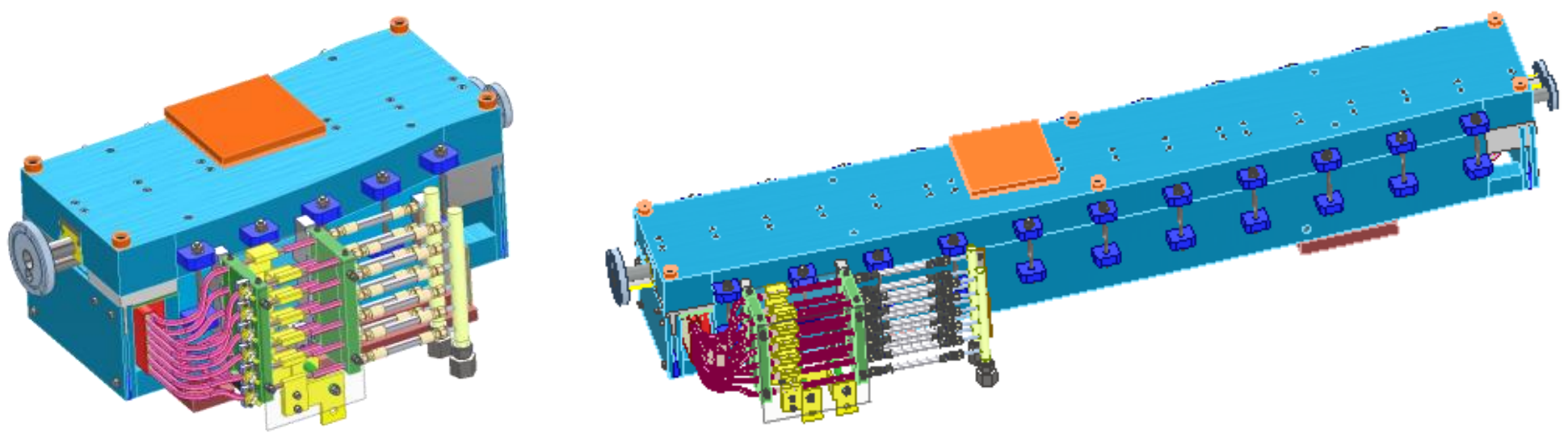

**Figure 4.3.8.77:** Lambertson magnet mechanical design for the HEPS booster (Left: injection; right: extraction)

Figure 4.3.8.77 shows the mechanical design of the Lambertson magnets for HEPS booster injection and extraction systems.

The two types of Lambertson magnets designed for the HEPS booster injection and extraction systems have been successfully developed and passed acceptance in April 2023. To confirm the quality of the magnetic field, Hall point measurements were conducted, as illustrated in Figure 4.3.8.78 and Figure 4.3.8.79. The measurement results align with the simulation model, and both the main field and leakage field meet the design requirements.

The magnet exhibits a center magnetic strength of 1 T, with a main field uniformity of ±0.008% in the horizontal direction and ±0.0016% in the vertical direction. The integral leakage field ratio to the main field integral ($B_y$) is 0.700E-3. These results demonstrate the successful achievement of the desired magnetic field parameters for the Lambertson magnets in the HEPS booster system.

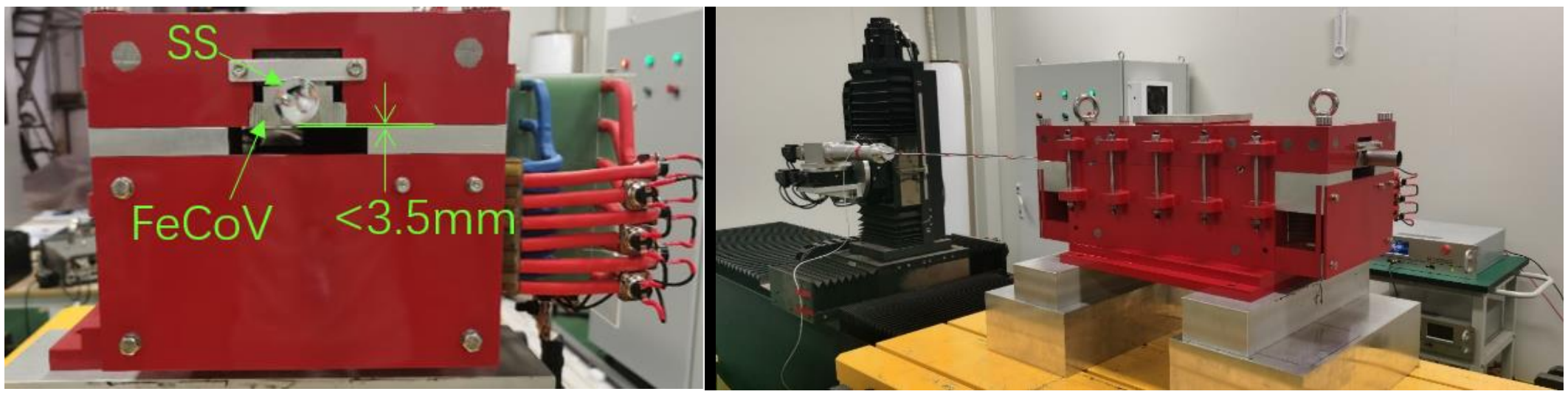

**Figure 4.3.8.78:** Lambertson magnet for the HEPS low energy injection

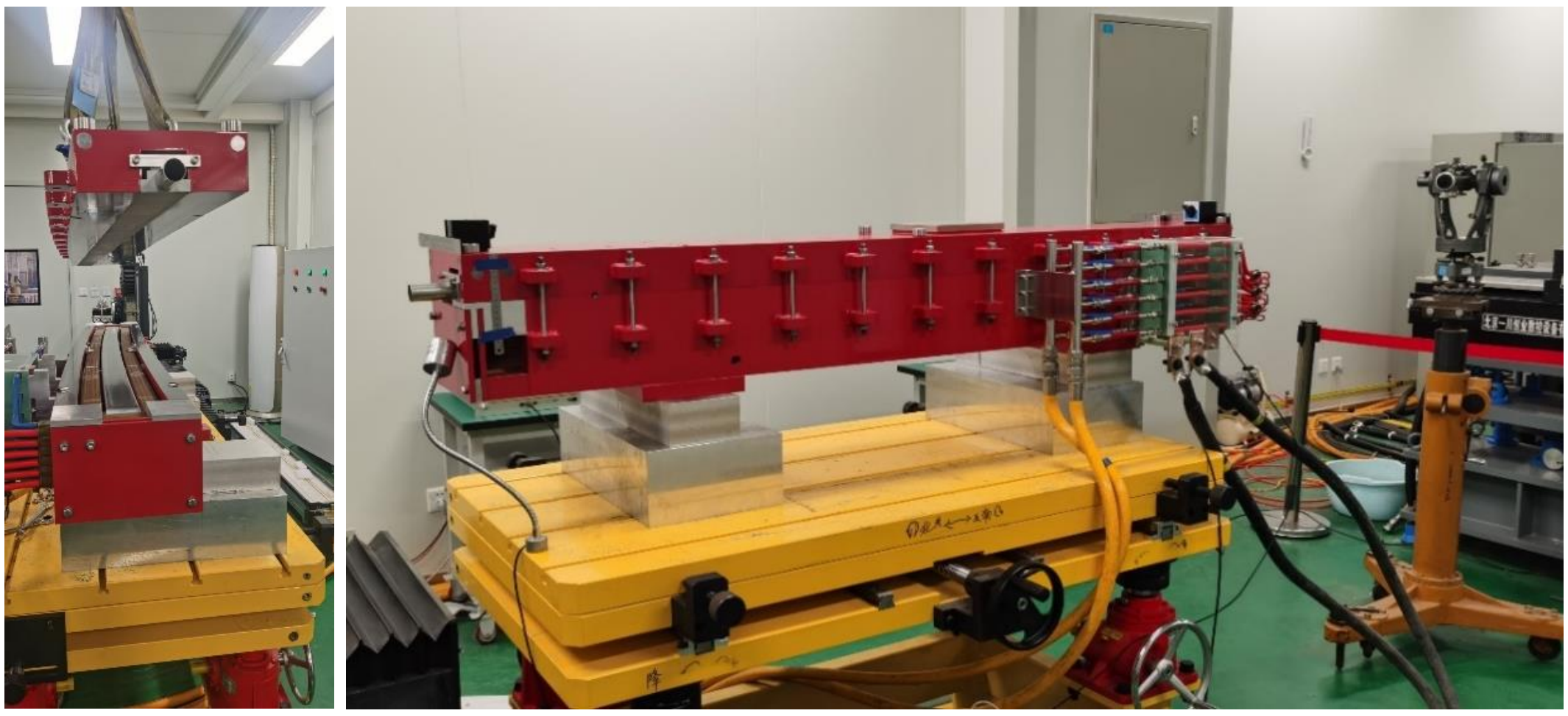

**Figure 4.3.8.79:** Lambertson magnet for the HEPS high energy extraction

### 4.3.9 **Control System**

#### 4.3.9.1 ***Introduction***

The control system for the CEPC must cover a 100 km circumference ring and a 1.8 km Linac injection system. The project consists of three accelerator components: the Linac injection system (including a Damping Ring), the Booster ring for beam accumulation and acceleration, and the Collider ring. The Booster and Collider rings will be housed in the same 100 km tunnel.

The control system for the CEPC should have the capability to control and monitor all equipment with user-friendly operator interfaces (OPIs), reliable and efficient communication channels, advanced beam-tuning tools, and comprehensive application tools to ensure optimal beam source, acceleration, accumulation, and colliding luminosity.

To build such a large-scale control system, it is important to adopt commercial and industrial products and techniques to ensure high quality of the project. However, the distribution of large volumes of control messages and the collection of even larger volumes of monitoring data, as well as the management of system alarms and data archiving across a much larger area than any similar facility, present significant design challenges.

For the entire system to function properly, it is crucial to ensure time synchronization among the widely distributed devices related to the beam source, injection, accumulation, acceleration, extraction, diagnostics, and post-mortem analysis. Specifically, for the booster ring, a synchronization accuracy of several microseconds among the hundreds of related power supplies must be fulfilled.

As electronic techniques continue to evolve, hardware prices rapidly decrease while their performance significantly improves. Therefore, purchases and mass production of hardware should be delayed until later stages of the project. Conversely, technical studies and interface designs between different systems should be conducted as early as possible to facilitate system development, integration, and commissioning. It is advisable to set up a full-scale prototype system first to test for functionality and to aid in the development process.

The CEPC control system is comprised of a global control system, including a Timing system, MPS, and Network, as well as local control systems for various components, such as power supply control, vacuum control, RF control, Injection/Ejection control, and temperature monitoring, shown in Figure 4.3.9.1. This section focuses on both the global control system and the local control systems of the collider ring. Subsequent sections will cover the control systems of the Linac and booster components.

#### 4.3.9.2 ***Control System Architecture***

The entire control system will be divided into three layers: the presentation tier, middle tier, and front-end tier, all based on Ethernet. Ethernet will continue to serve as the backbone of the control system, with 100 Gb/s products already available and 1.2 Tb/s products on the horizon. With the increasing performance of network switches, direct connections to Ethernet with 30Gb/s to 100 Gb/s are a viable option. For systems requiring high-intensity real-time computing, FPGA plus MTCA/XTCA will provide scalability and possible reliability benefits. For most slow-speed applications, a PLC (Programmable Logic Controller)-based system should be used to improve overall reliability.

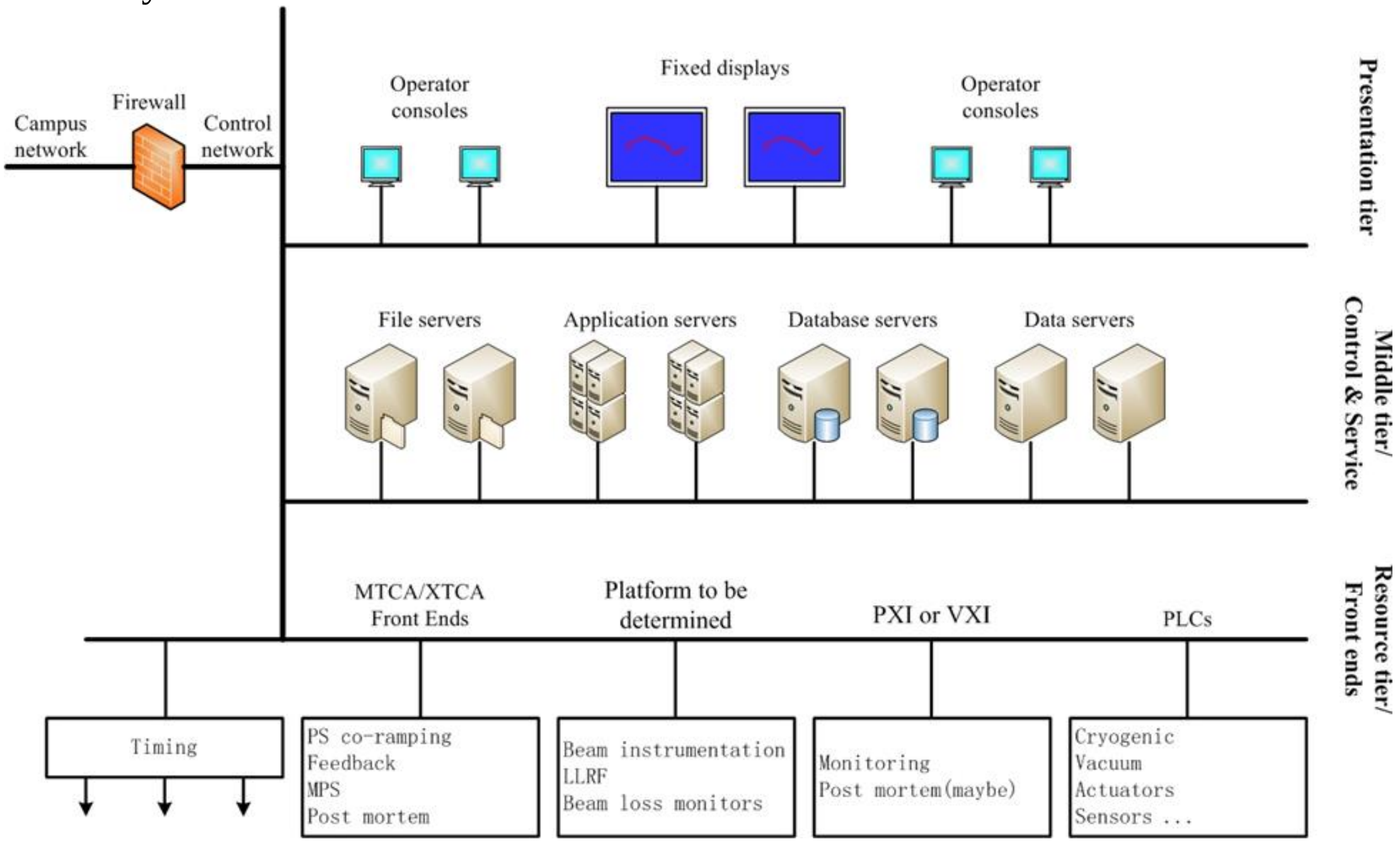

**Figure 4.3.9.1:** Control system architecture

#### 4.3.9.3 *Control Software Platform*

EPICS (Experimental Physics and Industrial Control System) is a widely used control system software platform for large experimental facilities around the world. It was originally developed by Los Alamos National Laboratory and Argonne National Laboratory and has been continuously improved by different laboratories over the years, with many application tools available. EPICS can be installed on most mainstream operating systems and has many device drivers implemented, which makes system integration easier. Due to its successful application in the accelerator control system of the BEPC II (Beijing Electron Positron Collider II) and CSNS (China Spallation Neutron Source), EPICS has been chosen as the control software platform for the CEPC.

EPICS is based on a client-server model and uses Ethernet for communication. The main components of EPICS include the OPI (Operator Interface), IOC (Input Output Controller), and CA (Channel Access). The OPI is the client-side module used by operators, the IOC is the server-side input-output control module, and the CA is the communication module. Figure 4.3.9.2 illustrates the structure of EPICS IOC.

The OPI layer provides database management for the IOC, a graphical user interface, and other control/monitoring software such as system alarms, data archiving, real-time and historical curves. The IOC runs on the server computer to manage the real-time database and communicate with the front-end devices using suitable drivers. The CA module supports TCP/IP protocol to enable clients to access data in the real-time database of the IOCs transparently.

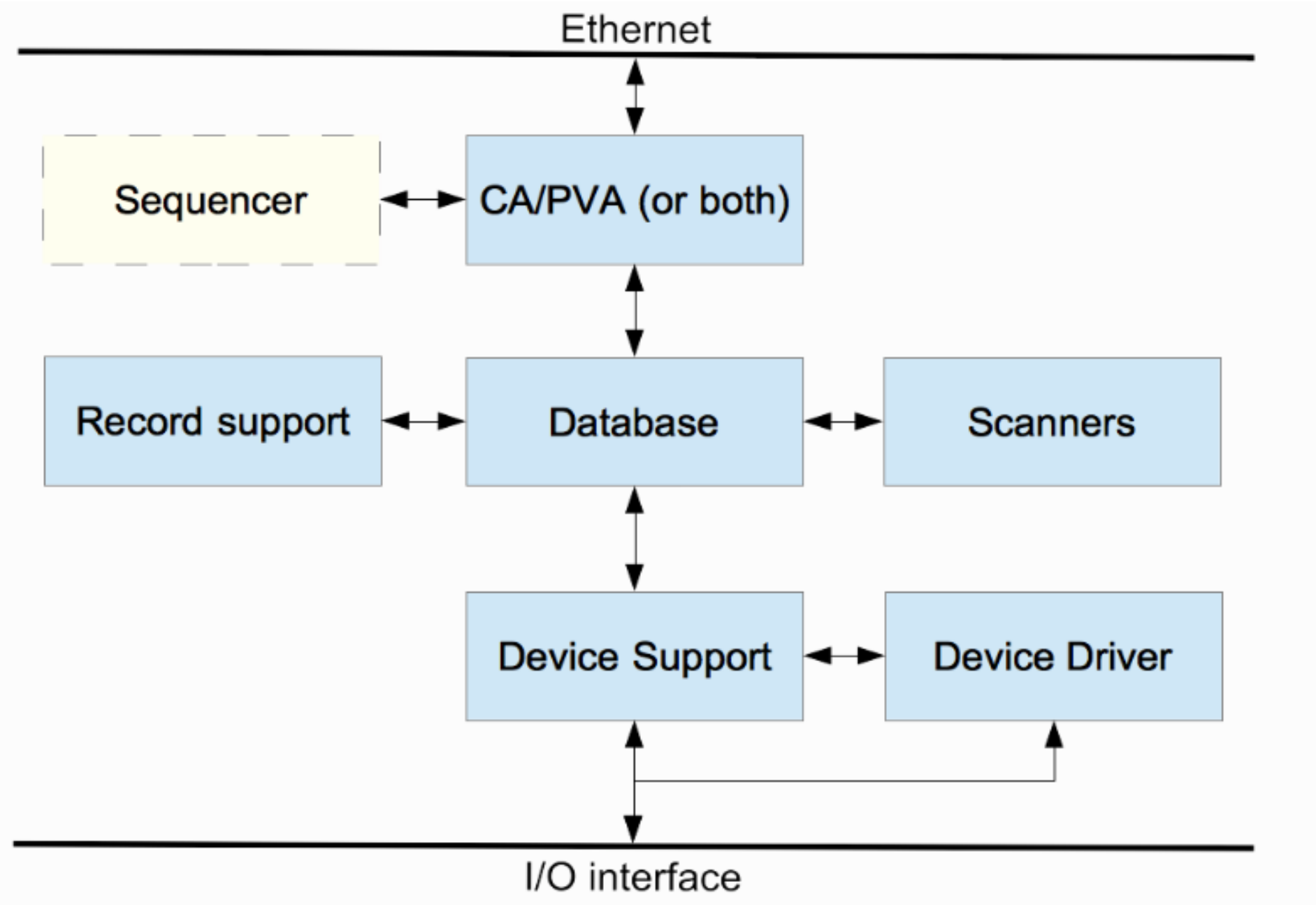

**Figure 4.3.9.2:** Structure of EPICS IOC.

### 4.3.9.4 *Global Control System*

#### *4.3.9.4.1 Computers and Servers*

Cost-effective equipment, such as servers, workstations, and PCs, will be used as operator consoles at the presentation layer, with a user-friendly graphical interface. Operators will be able to monitor and control the accelerator equipment from these consoles, store and recall machine parameters, integrate device data into office products, and define machine-operator sequences. The consoles will also provide tools to display beam status and equipment alarms, access all system data, and plot real-time and historical trends.

The top layer includes a server machine that provides support for accelerator commissioning, database management, data logging and analysis, network management, and general computing resources. The choice of server machine depends on the specific requirements of the accelerator physicists.

#### *4.3.9.4.2 Software Development Environment*

The software development environment should be defined in advance to ensure future compatibility during system integration and maintenance. This includes selecting the operating systems, development tools, software upgrade strategy, and hardware platform. Standardization and version control are essential. A global software and hardware platform should be established for collaborative development.

Upgradability should be a primary consideration during the setup of the development environment and software development. A careful study should be conducted and rules should be established as early as possible to ensure that future upgrades can be carried out smoothly without causing compatibility issues or disrupting system operations.

Some rack servers will be used for data archiving and consultation, accelerator physics software, alarm service, and post-mortem service. Blade servers and RAID disks will be provided for highly reliable soft IOCs and other applications.

#### *4.3.9.4.3 Control Network*

The control-system network will be designed with a redundant core to ensure high availability, with 40 Gb/s or 100 Gb/s interfaces connecting to the aggregation switches. The aggregation switches will then provide communication to the devices through links of 10 Gb/s. Edge switches will be connected to the aggregation switches, providing links of 1 Gb/s and 100 Mb/s to the different devices. This design ensures that the network can handle the high bandwidth and reliability requirements of the control system.

A 40 Gb/s or 100 GB/s backbone is selected due to current technical capabilities and the potential need for large data-transfer rates in both average and sudden cases. With a higher bandwidth, the transmission latency of packets is decreased, thereby increasing the overall system stability.

To balance cost-effectiveness and network performance, a three-layer structure will be implemented with aggregation and edge switches provided at each local control station and the Central Control Room (CCR), shown in Figure4.3.9.3.

As there are around 5000 IPs expected to be used, it is estimated that approximately 130 edge switches will be required.

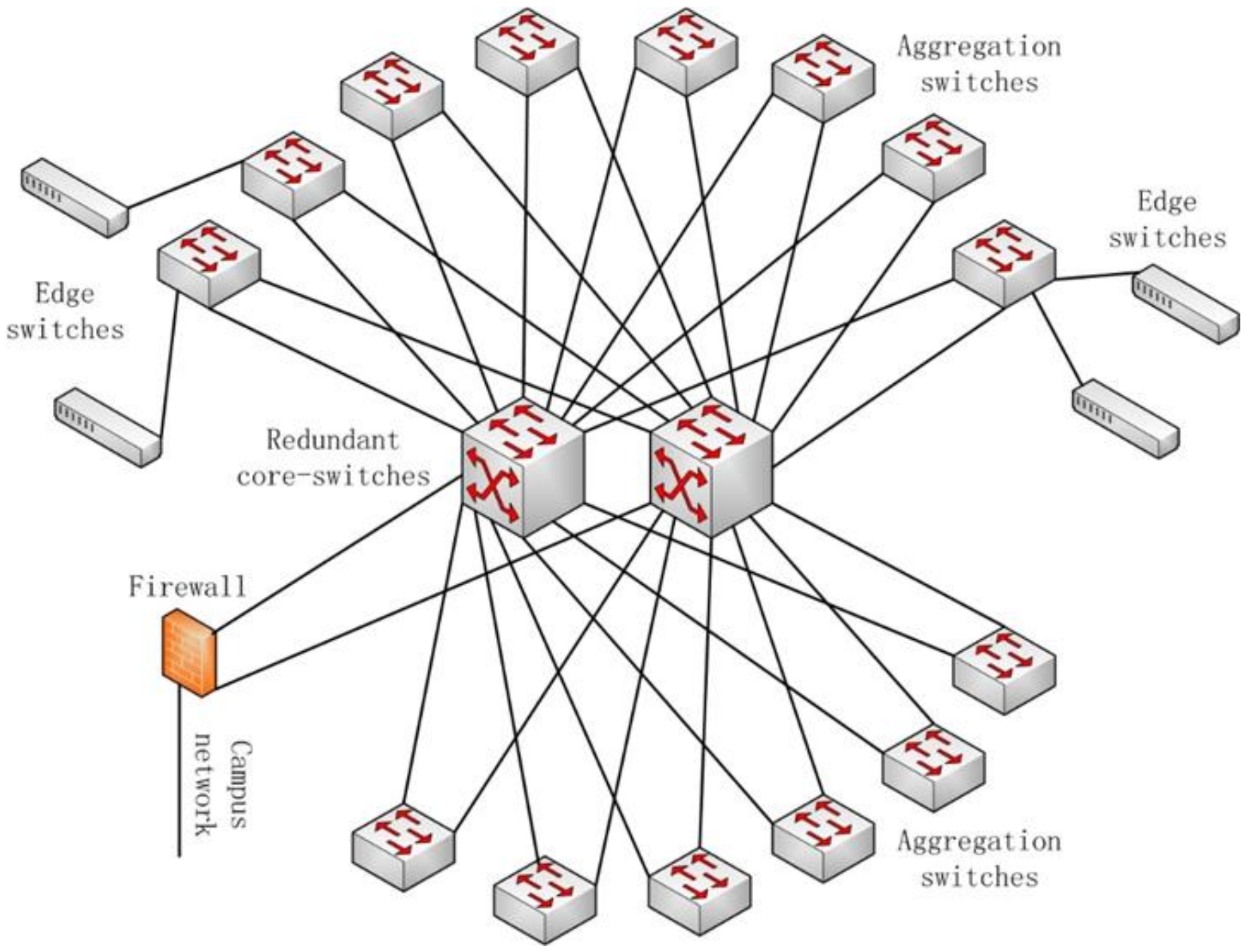

**Figure 4.3.9.3:** The accelerator-control network

#### *4.3.9.4.4 Timing System*

Operation frequencies of the LINAC, the booster and the collider are 2860 MHZ, 1300 MHz and 650 MHz respectively.

The timing system plays a crucial role in synchronizing all the relevant components in the CEPC accelerator complex. It generates a trigger sequence to synchronize the equipment in the LINAC, booster ring, and collider rings, including the electron gun, modulators, pulsed power supplies, injection/ejection kickers, and other relevant components. It matches the delays between the gun triggering and the pulse of the injection kicker so that the bunch can be injected into the corresponding bucket. The timing system also sends timing pulses as reference time triggers to the related beam diagnostic devices for bucket selection and beam-parameter measurements.

*4.3.9.4.5 Machine Protection System*

The CEPC collider ring stores approximately 690 kilojoules of energy for both beams, while the booster injects about 34.5 kilojoules per pulse train. In order to prevent damage to the equipment, the beam in the collider must be directed to the beam dump and injection must be halted when any main component malfunctions or when the dose rates from the lost particles become too high. To provide global protection with an interface to the relevant systems, the MPS (Machine Protection System) will be employed.

The power supplies, RF cavities, beam-loss monitors, pulsed power supplies for kickers, cryogenic system, vacuum system, beam-pipe temperature monitoring system, and timing system are essential components. The need for any additional components should be thoroughly investigated before their involvement.

Redundant controllers will be implemented to improve maintenance and increase system reliability. However, IO connections will not be redundant to minimize the overall cost.

The beam revolution time in the collider ring is approximately 0.17 ms, and the response time of the protection system should be comparable to this. Therefore, it is necessary to design a low-level FPS (Fast Protection System) as part of the MPS.

Fig. 4.3.9.4 shows the architecture of the MPS of one station.

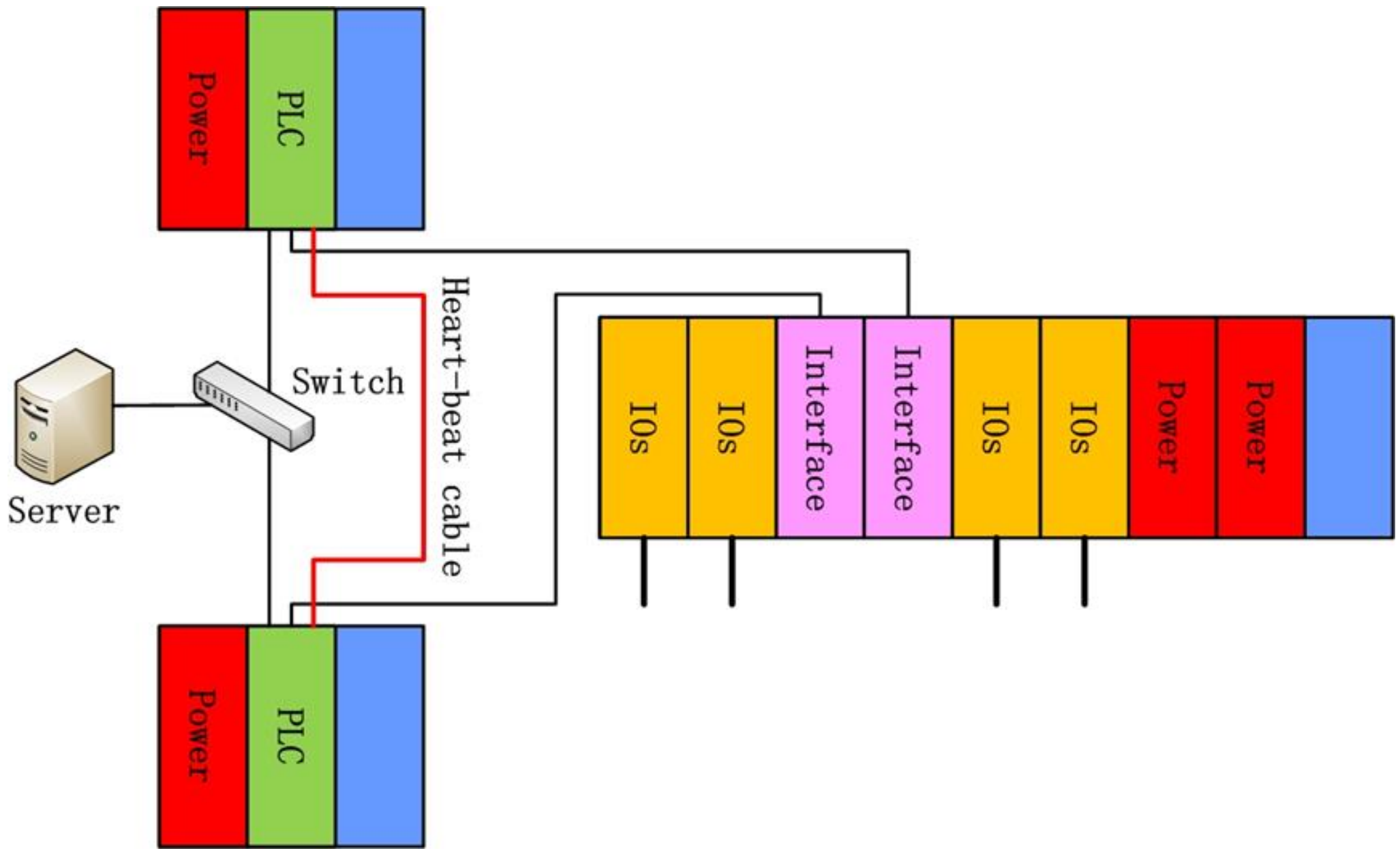

**Figure 4.3.9.4:** Structure of the MPS of one station

*4.3.9.4.6 Data Archiving and Consultation*

The number of signals that require archiving is in the tens of thousands. To improve overall efficiency, several data archivers should be installed with data writing and consultation separated.

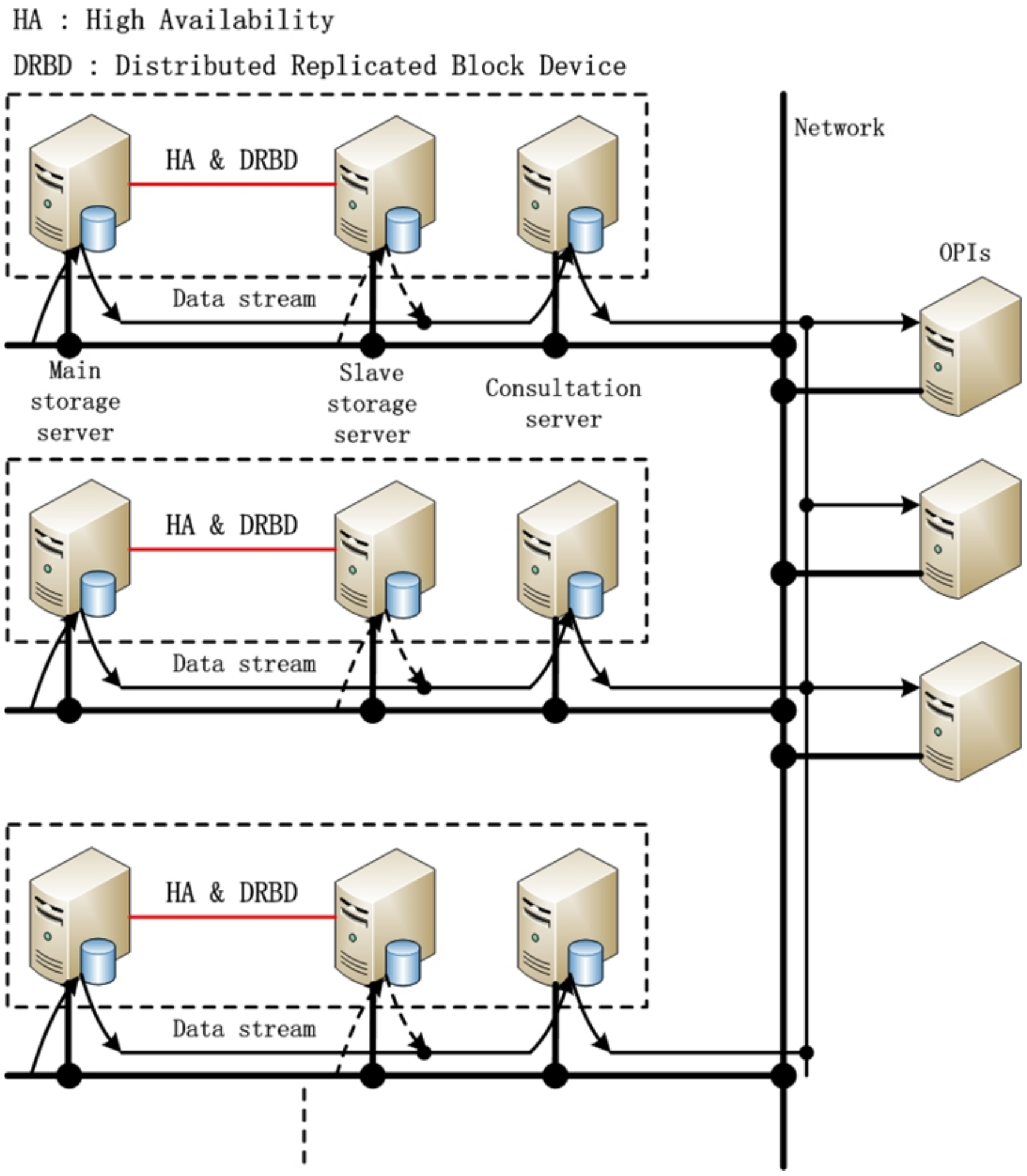

**Figure 4.3.9.5** Hardware configuration and data flow of data archiving

Redundant storage servers will be utilized for data safety and to ensure high data availability. An individual consultation server will also be implemented for fast data retrieval. Both the consultation server and the OPIs (Operator Interface Panels) can be deployed on the campus network to reduce the load on the control network. The system will be developed using JAVA and a relational database.

*4.3.9.4.7 System Alarms*

The alarm system should be capable of collecting, storing, and reporting abnormal situations across the entire system. As there may be thousands of alarm signals generated from different systems, the alarm system should be designed with multiple levels. The first level will be the equipment level where alarm sources will be located, and data with accurate timestamp should be recorded. The overall alarm signals should be transmitted to the network for storage and treatment at the server level. The alarm servers will gather and store the alarm signals from various equipment on the control network, and alarms, along with the corresponding information, should be provided to the operators. The operators should be equipped with OPIs to help them address the problem with the appopriate information. The architecture of the alarm system is depicted in Fig. 4.3.9.6.

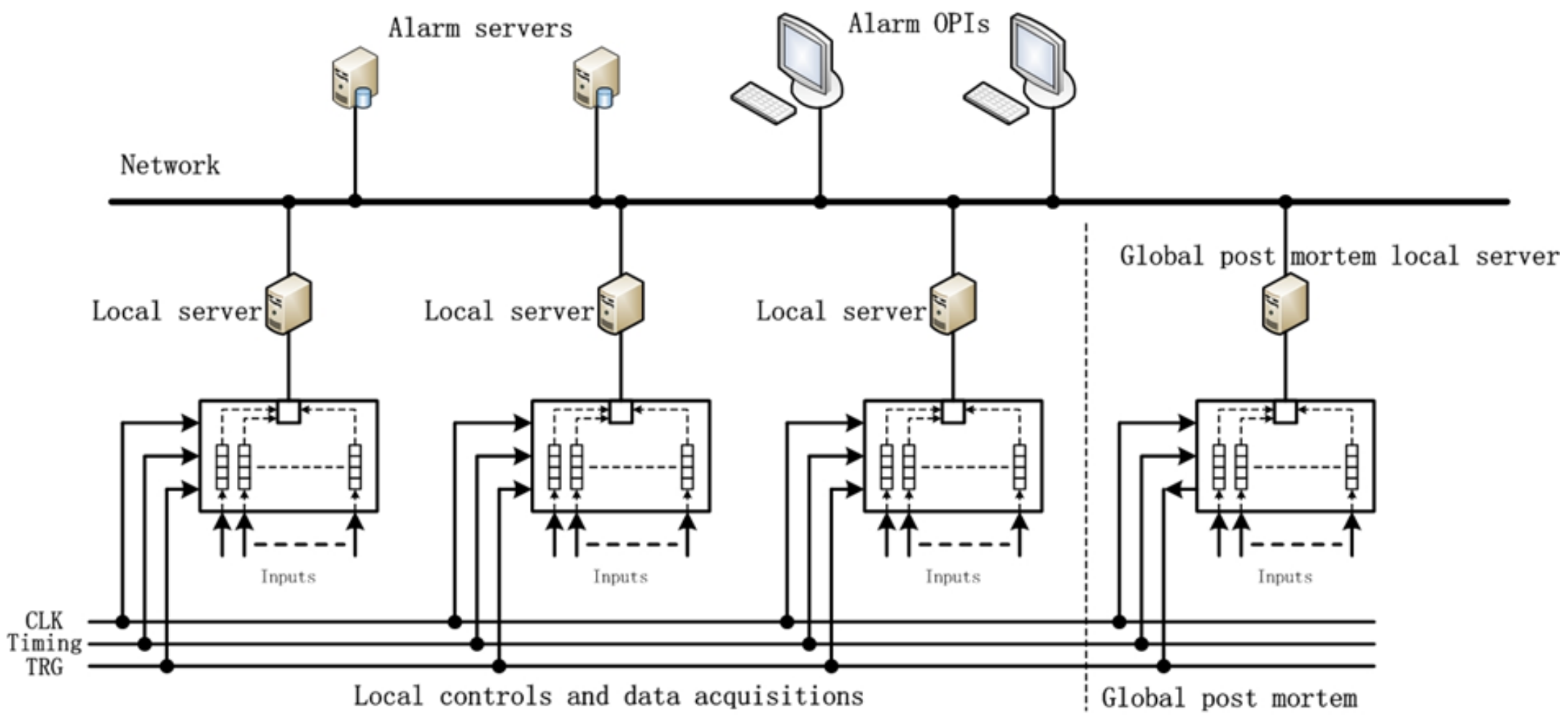

**Figure 4.3.9.6:** Alarm-system architecture

#### *4.3.9.4.8 Post Mortem, Large Data Storage and Analysis*

To ensure proper troubleshooting, each control or data acquisition system of the main components must have the capability and interface to record data for a period with accurate timestamp whenever an issue arises. The time accuracy requirement and time window to record the data may differ for various systems. Some smaller systems may be able to directly provide analog and/or digital signals to the global post mortem system created by the control system. The global post mortem system will then be responsible for recording the data with the necessary time accuracy and window.

The structure of the post-mortem system, which includes local controls and data acquisitions similar to the alarm system, is depicted in Figure 4.3.9.7. Whenever there is a problem, the triggered window data in the local devices will be sent to the post-mortem servers. The post-mortem servers should define and provide a software analysis interface for the data. Post-mortem OPIs should be programmed to analyze the data through this analysis interface.

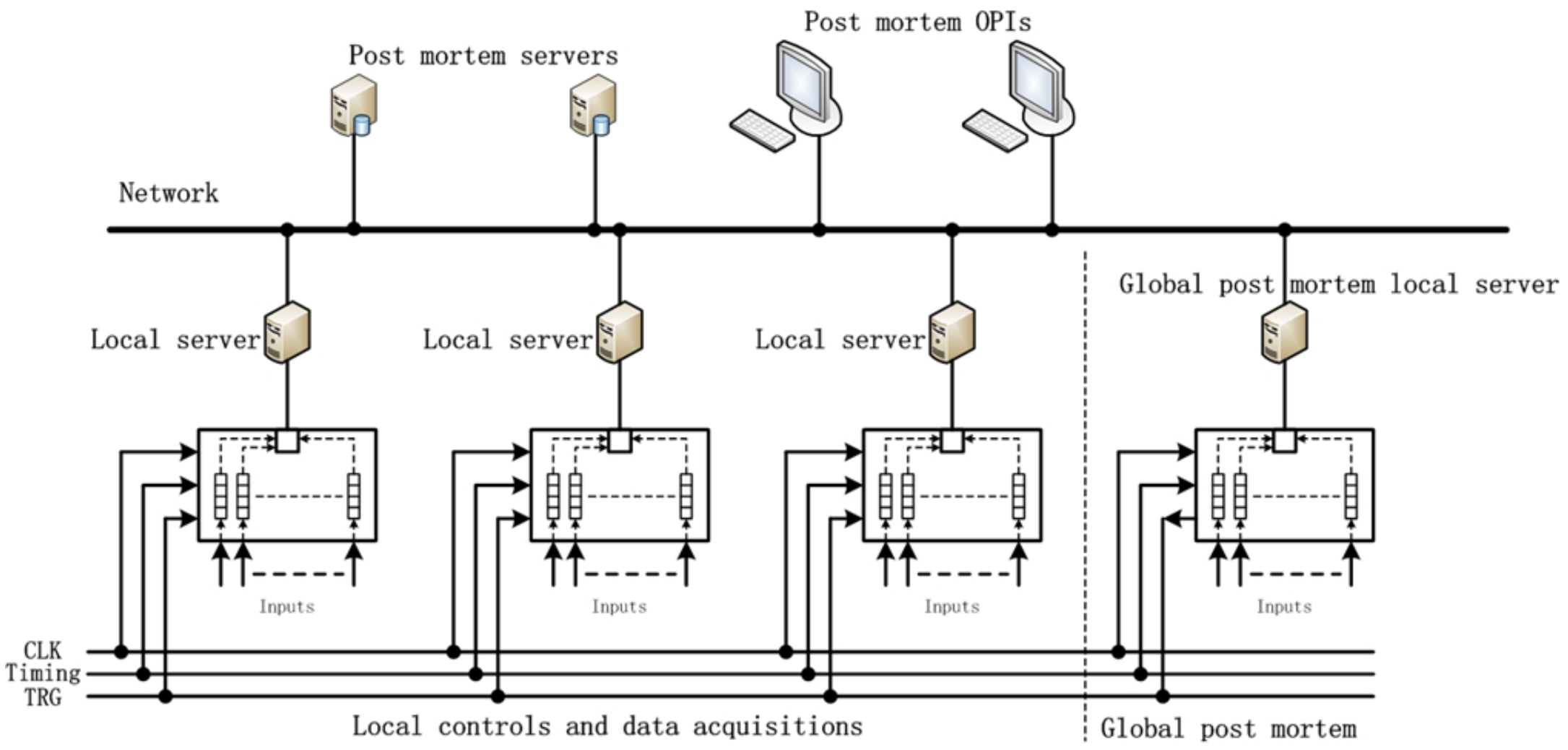

**Figure 4.3.9.7:** Post mortem system architecture

#### *4.3.9.4.9 Event Log*

It is important to develop an automated method to store the status of the main equipment and provide a running report to help operators and experts monitor the overall health of the machine and identify any potential issues before they become serious. A tree structure can be used to differentiate the information from different systems and levels. In addition, comment panels should be provided for each item as well as globally.

### 4.3.9.5 ***Front-end Devices Control***

The front-end control system for the collider ring is comprised of various sub-systems, including power-supply control, vacuum control, temperature monitoring, RF control, cryogenic control, etc.

For the high-bandwidth real-time data exchange between different RF cavities, uTCA crates will likely be used for the LLRF (Low Level Radio Frequency) control.

DPSCM (Digital Power Supply Control Module) modules will be utilized for the approximate 200 power supplies, with co-ramping function integrated into all power supply control modules.

PLCs will be implemented for the control of the cryogenic system, vacuum system, movable collimators, and vacuum chamber temperature monitoring system.

Ethernet integration will be employed to integrate beam instrumentation devices into the overall system.

Post-mortem data acquisition and analysis will be conducted using commercial uTCA crates and modules.

In some cases, industrial computers will serve as the IOC for device-level controls.

#### *4.3.9.5.1 Power-Supply Control*

The collider ring features a total of approximately 4000 power supplies, with over 3000 allocated for correctors. Power-supplies in the booster ring require co-ramping to facilitate beam acceleration, while the remainder only need to be powered to the correct current levels. A power-supply remote control has been designed that can meet both needs, simplifying maintenance with only minor differences in physical connection and software configuration required.

Due to the large size of the CEPC ring, repairing or replacing a broken device within a short timeframe can be challenging. To mitigate these risks, redundant design options will be considered wherever feasible. For example, a self-designed control crate will house two redundant controllers, with each interface card featuring two isolated connectors for connecting to each controller independently. A self-designed passive backplane will facilitate all connections, while two isolated crate power supplies will provide power to the two control routes. To ensure global redundancy, network packets exchange or heart-beat cable connections will be utilized between the two controllers. The redundant connections of the power supply remote control are depicted in Fig. 4.3.9.8, while the preliminary remote control-crate arrangement is illustrated in Fig. 4.3.9.9.

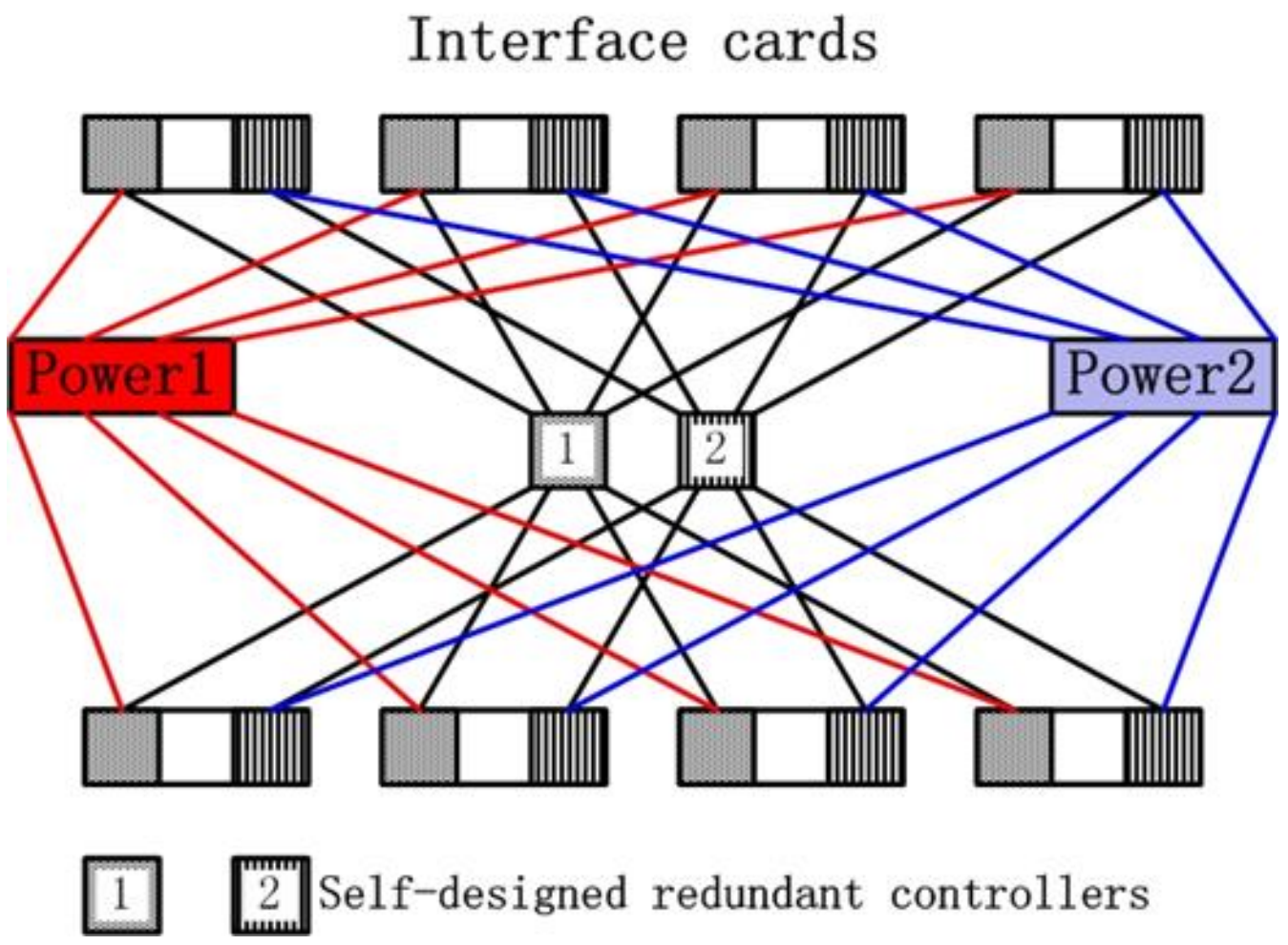

**Figure 4.3.9.8:** Redundant design of the power supply remote control

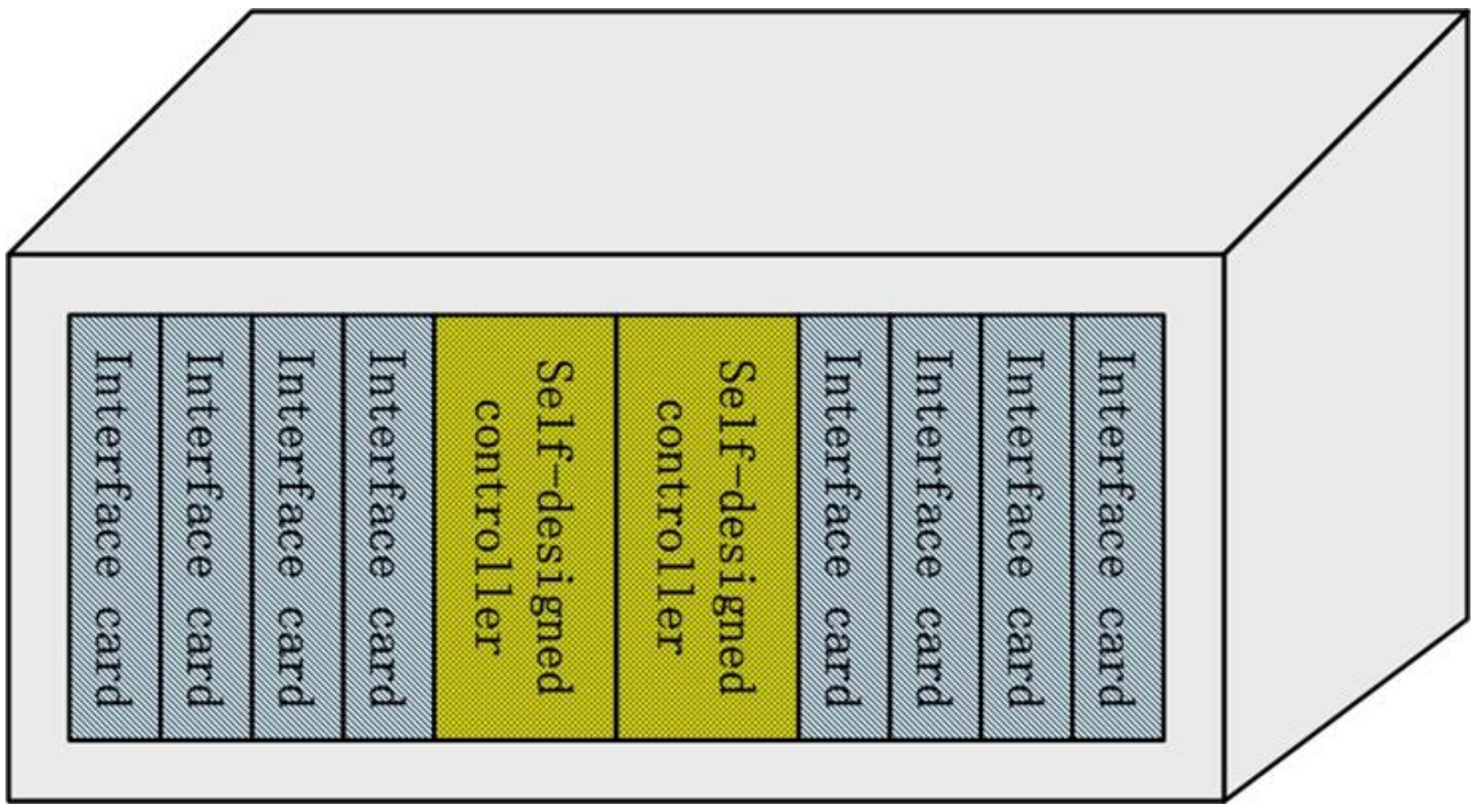

**Figure 4.3.9.9:** Preliminary power supply remote control crate arrangement

*4.3.9.5.2 Vacuum Control*

The interfaces to the vacuum devices typically include relay contacts and RS232/485 ports. Relay contacts are used for on/off control and protection, while RS232/485 are used for monitoring. To ensure reliability, redundant PLCs will be utilized for the relay contacts with a similar structure as shown in Fig. 4.3.8.4, which is also used for the MPS design. The integration of RS232/485 devices will be determined closer to the construction phase, based on the specific products selected.

The CEPC has approximately 1040 vacuum valves and 4320gauges distributed throughout the Collider rings. Due to their critical importance in maintaining the vacuum, complicated interlocks with other systems, such as RFs and kickers, must be designed and prototyped prior to implementation.

*4.3.9.5.3 Vacuum-Chamber Temperature Monitoring*

The mis-steering of the beams can cause an increase in synchrotron-photon heat on the vacuum chambers. Excessive heat may lead to high temperature, poor vacuum, and potential damage to the chambers. It is, therefore, crucial to monitor the vacuum chambers' temperature, especially in the bending sections of the Collider ring. Each dipole joint will require two temperature sensors, resulting in approximately 68000 sensors on the Collider ring.

The relationship between the vacuum-chamber temperature-monitoring system and the MPS will need to be clarified later, given the high number of sensors.

##### *4.3.9.5.4 Integration of Other Subsystems Control*

The Collider ring RF system consists of 240 superconducting accelerating cavities, 120 RF high-power sources, and a low-level control system. To achieve a short bunch length, a high RF voltage and frequency are necessary. The collider ring operates at a frequency of 650 MHz and provides an RF voltage of 2.2 GV.

The interlock system ensures that the cavity tuning is switched off when the RF high-voltage power supplies are at risk or in an unsafe condition. Faults may occur in the cooling-water system, vacuum and temperature of a cavity, or the liquid helium in the cryogenic system. When a fault is detected, the local interlock system sends a warning message and failure signal to the MPS.

## 4.3.10 **Mechanical System**

### 4.3.10.1 ***Introduction***

The magnets span more than 90% of the 100 km ring, as detailed in sub-section 4.3.3. Each magnet requires effective support, focusing on adjustment, performance, and cost-efficiency. The magnet support systems play a crucial role in the mechanical infrastructure of the Collider, Booster, Linac, and Damping Ring, with specific details about the Collider's support system covered in sub-section 4.3.10.2.

The CEPC ring tunnel accommodates both CEPC and the future SPPC. The CEPC Collider ring is positioned near the inner tunnel wall, while the SPPC will be situated near the outer wall. In the RF section, auxiliary devices such as power sources, cryogenic equipment, and cooling water systems are housed in galleries adjacent to the main RF tunnels. Detailed information about the layout of the tunnel cross section and galleries will be provided in sub-section 4.3.10.3.

To reduce beamstrahlung backgrounds from the IR and minimize multi-turn losses, a total of 36 movable collimators are employed for the two IPs in the two rings. The mechanical design of these movable collimators, aimed at decreasing background interference, will be detailed in sub-section 4.3.10.4.

The MDI presents challenges in mechanical design due to space limitations, a compact device layout, and stringent demands for stability and alignment. Sub-section 4.3.10.5 will provide details on the mechanical layout, assembly sequence for accelerator devices within the MDI, and two crucial components: the remote vacuum connector (RVC) and the support system for superconducting magnets.

### 4.3.10.2 ***Magnet Support System***

#### *4.3.10.2.1 Requirements*

Each magnet support consists of pedestal and an adjusting mechanism, with the pedestal being made of either concrete poured during construction or pre-constructed and grouted during construction. The adjusting mechanisms of one magnet have six degrees of freedom (DOFs) when integrated.

The coordinate system for the Collider assumes the +X axis is the horizontal direction perpendicular to the beam and pointing outside of the ring, while the +Y axis is up, and the system follows the right-hand rule. The design goals for the Collider's magnet supports include:

- a range and accuracy of adjustment shown in Table 4.3.10.1,
- stability to avoid creep and fatigue deformation,
- simple and reliable mechanics for safe mounting and easy alignment,
- good vibration performance.

**Table 4.3.10.1:** Adjustment range and accuracy of Collider magnet supports

| Direction | Range of adjustment | Direction | Range of adjustment |
|---|---|---|---|
| X | ≥±20 mm | Δθx | ≥ ±10 mrad |
| Y | ≥±30 mm | Δθy | ≥ ±10 mrad |
| Z | ≥±20 mm | Δθz | ≥ ±10 mrad |

To monitor the deformation of the tunnel floor and ceiling, a hydrostatic leveling system will be implemented. It will be installed on the tunnel floor near collider components and on the magnet supports in the Booster at approximately 500-meter intervals. This sub-section will primarily concentrate on the mechanical design of magnet supports.

The Collider primarily consists of twin-aperture magnets, with the majority being dipole and quadrupole magnets, whereas sextupoles and correctors are single-aperture. Furthermore, two sextupoles at the same location are collectively supported. Table 4.3.10.2 offers details regarding the quantities of these magnets and their corresponding supports.

**Table 4.3.10.2:** Quantities of magnets and their supports in the Collider

| Magnet type | Quantity | Magnet length (mm) | Core number per magnet | No. of supports per core |
|---|---|---|---|---|
| Dipole | 1024 | 19743 (twin-aperture) | 4 | 4 |
| | 1920 | 23142(twin-aperture) | 4 | 4 |
| | 64 | 21700(twin-aperture) | 5 | 4 |
| | 162 | 28450~93378 (single aperture) | 5~17 | 4 |
| Quadrupole | 3008 | 3000(twin-aperture) | 1 | 3 |
| | 8 | 1500 (twin-aperture) | 1 | 2 |
| | 1112 | 500~3500 (single aperture) | 1 | 1~3 |
| Sextupole | 3072 | 1400 | 1 | 3 in common girder |
| | 104 | 300~1000 | 1 | 1 |
| Corrector | 3544 | 875 | 1 | 2 in common girder |

*4.3.10.2.2 Structure of Dipole Magnet Support*

To create a magnet unit, four magnet cores of either 4,738 mm or 5,594 mm in length are interconnected to form a magnet unit with a total length of 19,743 mm and 23,142 mm, respectively. Each core is held in place by four supports, as depicted in Figure 4.3.10.1. The two central supports serve solely for Y-axis adjustments and are referred to as Y supports. One of the end supports can be adjusted in both the Y and X directions and is referred to as an XY support. The remaining support can be adjusted in all three directions (XYZ) and is referred to as an XYZ support. Various adjustment mechanisms are employed, including big screws for vertical adjustments and push-pull bolts for horizontal adjustments.

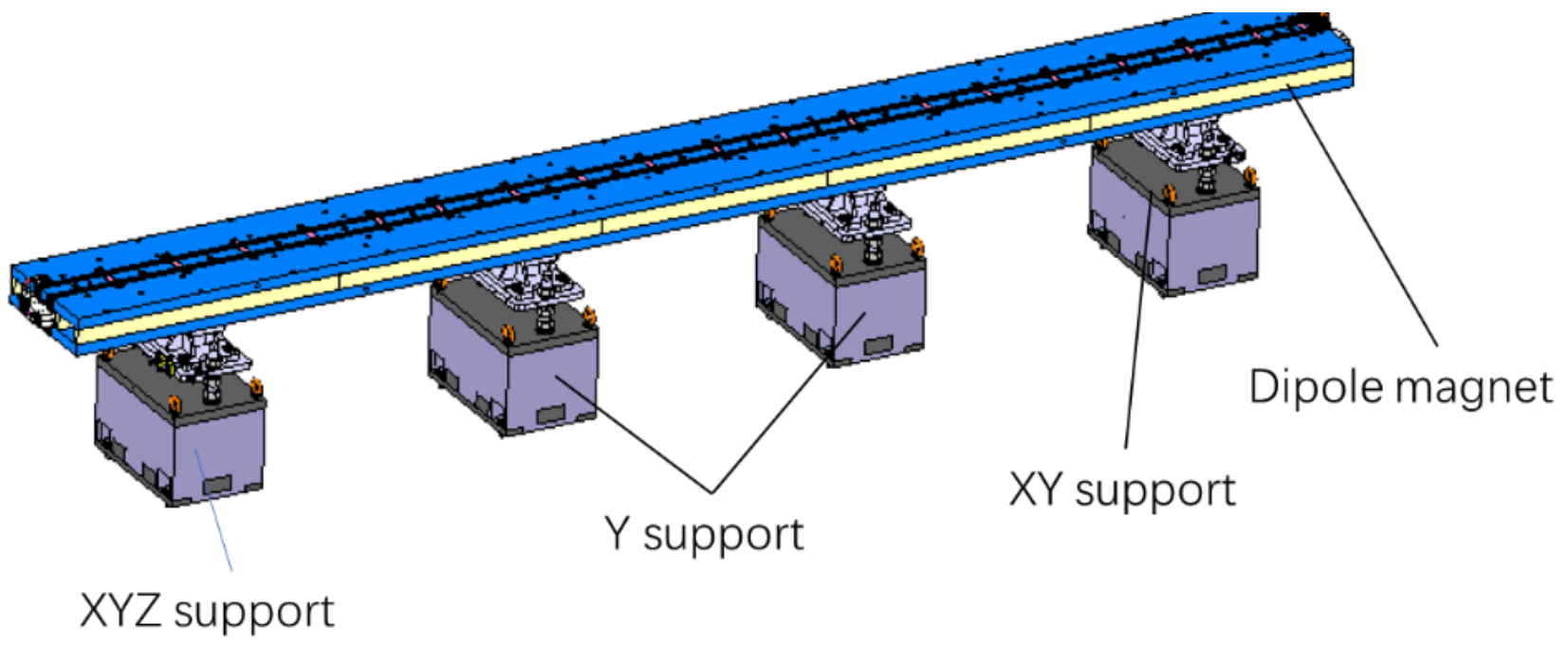

**Figure 4.3.10.1:** Dipole magnet and its supports for each module

Figure 4.3.10.2 depicts the force diagram for the long dipole module. As the magnet is thin and long, we can model it as a slender beam. The deflection of the magnet can be minimized using theoretical calculations or response surface design in ANSYS [1]. To obtain consistent results, the beam length is standardized to 1 m. The maximum deformation of the calculated three points is minimized when l1 equals 0.132 m and l2 equals 0.263 m, as illustrated in Figure 4.3.10.3.

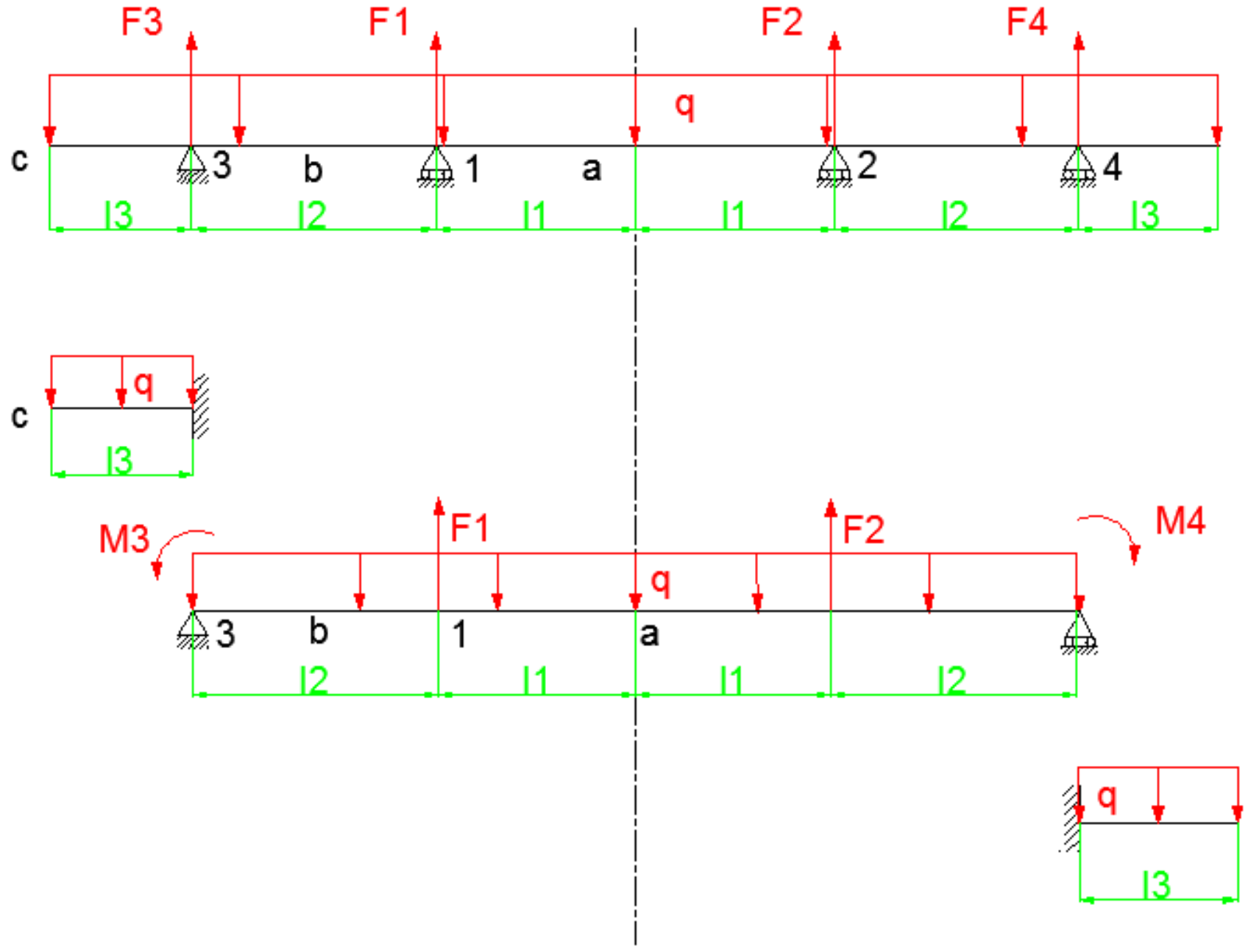

**Figure 4.3.10.2:** Force diagram of 4 support points

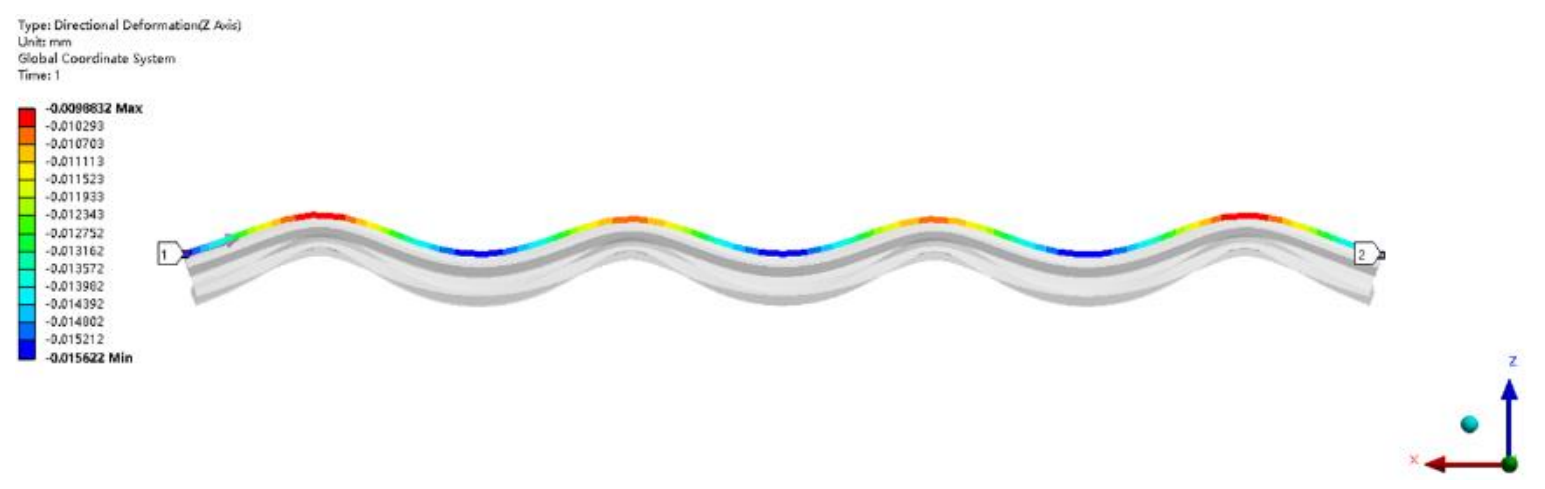

**Figure 4.3.10.3:** Deformation distribution of the optimized slender beam with 4 supports

This method is applicable regardless of the length or cross-section of the magnet core, as long as it can be assumed to be a slender beam. If the length of the magnet changes, the supporting locations can be easily adjusted based on the calculation described above. Using a similar method, the number of supports has also been optimized, as listed in Table 4.3.10.3. Four supports have been chosen.

**Table 4.3.10.3:** Optimization of support quantities for dipole magnet

| Number of supports | 3 | 4 | 5 |
|---|---|---|---|
| Max. deformation (μm) | 19 | 7 | 5 |

After conducting optimization, the 3D dipole model and its support structure were analyzed. It was observed that the maximum uneven deformation of the 5,594 mm long dipole core is 7 μm, which is primarily induced by the influence of gravity and the support system. The natural frequency of the dipole and its support structure exhibits minimal coherence with the stiffness between the pedestal and the ground, as long as the natural frequency of the pedestal itself exceeds 150 Hz. This is because the adjusting screws of the support are much weaker than the grouting joint between the pedestal and the ground, which dominate the serial stiffness.

Assuming the 1st natural frequency of the pedestal is approximately 150 Hz, which is based on previous experience with HEPS, the 1st natural frequency of the magnet and its supports is approximately 50.2 Hz. This mode corresponds to translation in the Z direction, as illustrated in Figure 4.3.10.4. Table 4.3.10.4 provides a list of the first six modes and their respective natural frequencies.

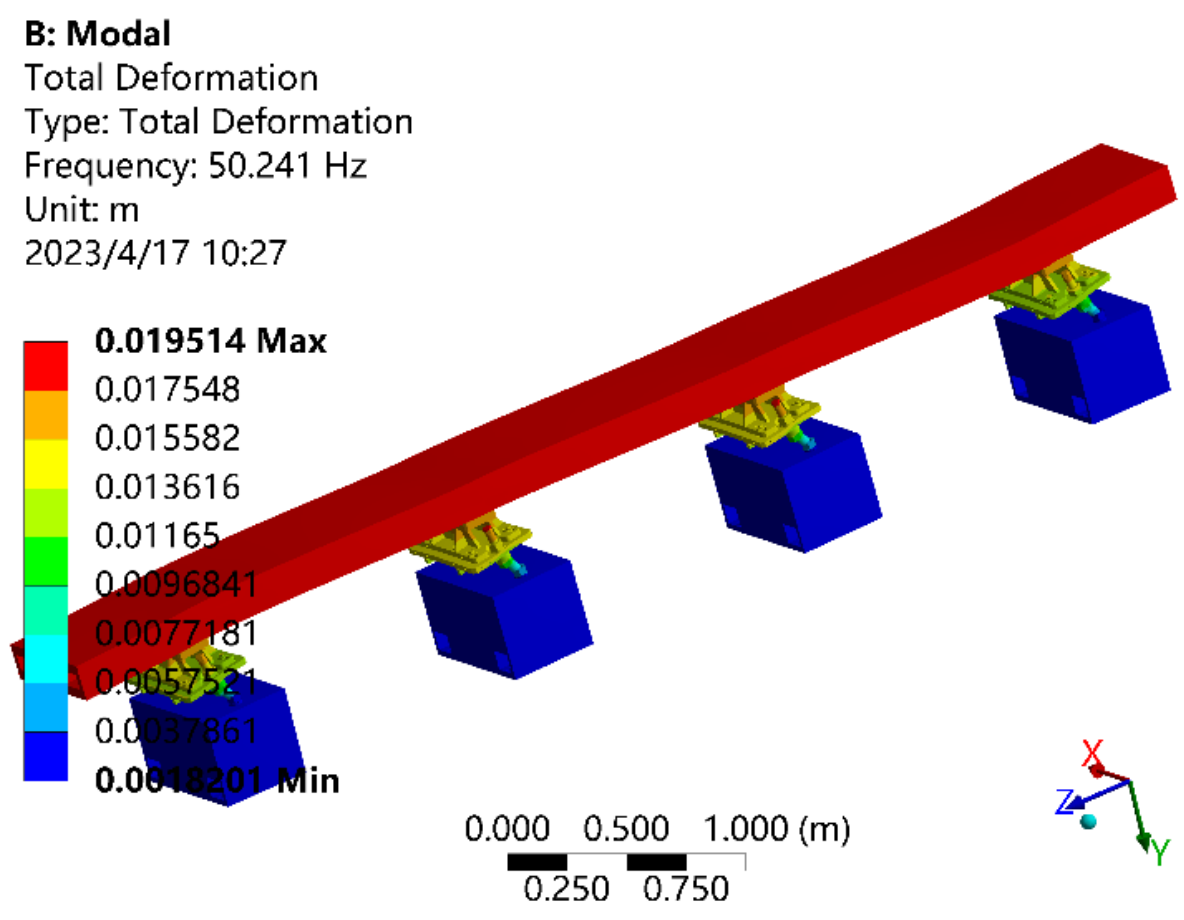

**Figure 4.3.10.4:** The 1st order of mode of the dipole magnet and its support.

**Table 4.3.10.4:** The first six orders of modes for the dipole magnet and its support.

| No. of order | Frequency (Hz) | Mode |
|---|---|---|
| 1 | 50.2 | Z translation |
| 2 | 51.3 | X translation |
| 3 | 59.7 | Yaw rotation |
| 4 | 89.0 | Bend in XZ plane |
| 5 | 117.2 | 1st order bend YZ plane |
| 6 | 123.8 | 2nd order bend YZ plane |

#### *4.3.10.2.3 Quadrupole Support System*

The Collider contains a significant number of quadrupole magnets, with a length of 3,000 mm, which are supported by three supports, as depicted in Figure 4.3.10.5. The support structure for these magnets is similar to that of the dipole magnet. Specifically, magnets with lengths ranging from 1,000 mm to 2,300 mm are equipped with two supports, while magnets measuring 3,000 mm and 3,500 mm in length require three supports. For magnets with lengths of 500 mm and 625 mm, a single pedestal with four supporting screws is designed.

Following optimization, the 3D model of the quadrupole magnet and its support structure was analyzed. The maximum uneven deformation of the 3,000 mm long quadrupole core was found to be less than 2 μm, primarily resulting from the effects of gravity and the support system. Similar to the analysis method used for the dipole magnet and supports, the 1$^{st}$ natural frequency of the quadrupole magnet and its supports is approximately 27.3 Hz. This mode corresponds to translation in the X direction, as depicted in Figure 4.3.10.6. Table 4.3.10.5 presents the first six modes along with their respective natural frequencies.

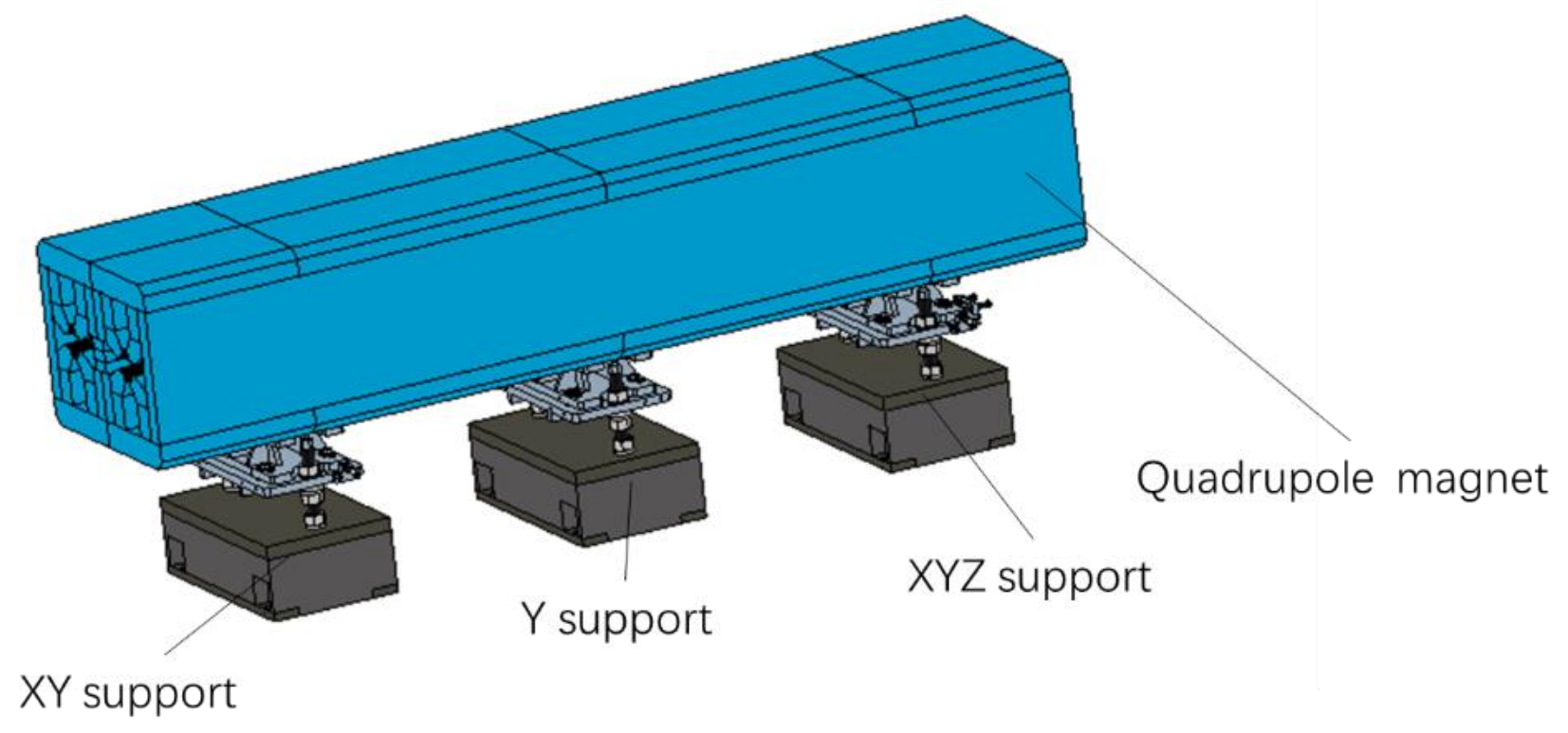

**Figure 4.3.10.5:** Quadrupole supports.

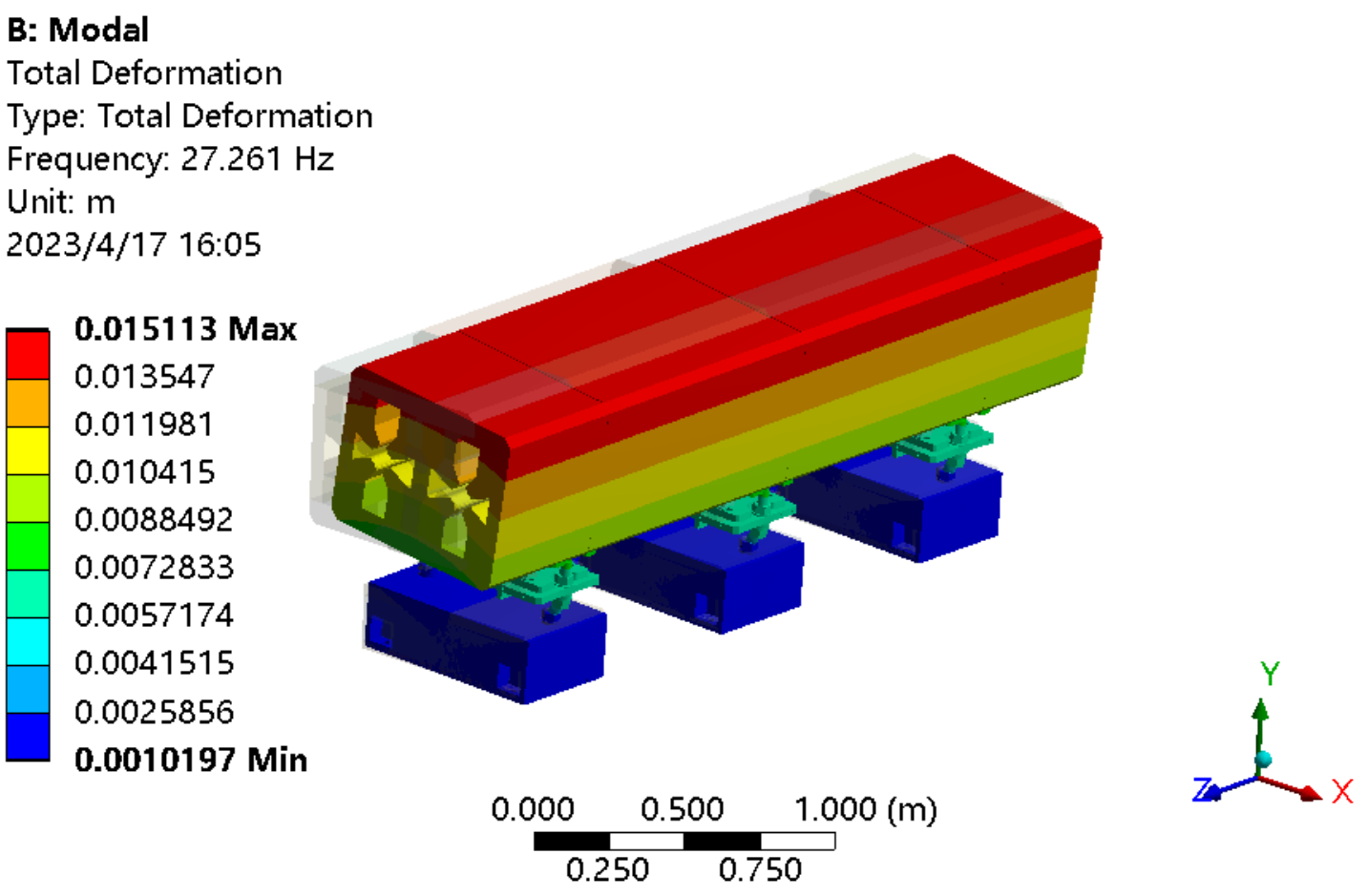

**Figure 4.3.10.6:** The 1st order of mode of the quadrupole magnet and its support.

**Table 4.3.10.5:** The first six modes of the quadrupole magnet and its support.

| No. of order | Frequency (Hz) | Mode |
|---|---|---|
| 1 | 27.3 | X translation |
| 2 | 31.5 | Z translation |
| 3 | 33.2 | Yaw rotation |
| 4 | 65.8 | Pitch rotation |
| 5 | 67.3 | Y translation |
| 6 | 80.4 | Roll rotation |

#### *4.3.10.2.4 Sextupole Support System*

In each cell of the Collider, the two rings feature three corresponding sextupole magnets, each measuring 1,400 mm in length. Two sextupoles are positioned in one ring, while the other ring houses one sextupole. Due to the stringent alignment requirements for sextupoles, the adjacent sextupoles are supported together by a shared girder. The combined length of the magnet and support assembly is approximately 3 meters, as depicted in Figure 4.3.10.7.

For the smaller number of sextupoles located at different positions, they are independently supported using narrower supports. The structure of these supports is similar to those employed for the dipole magnets.

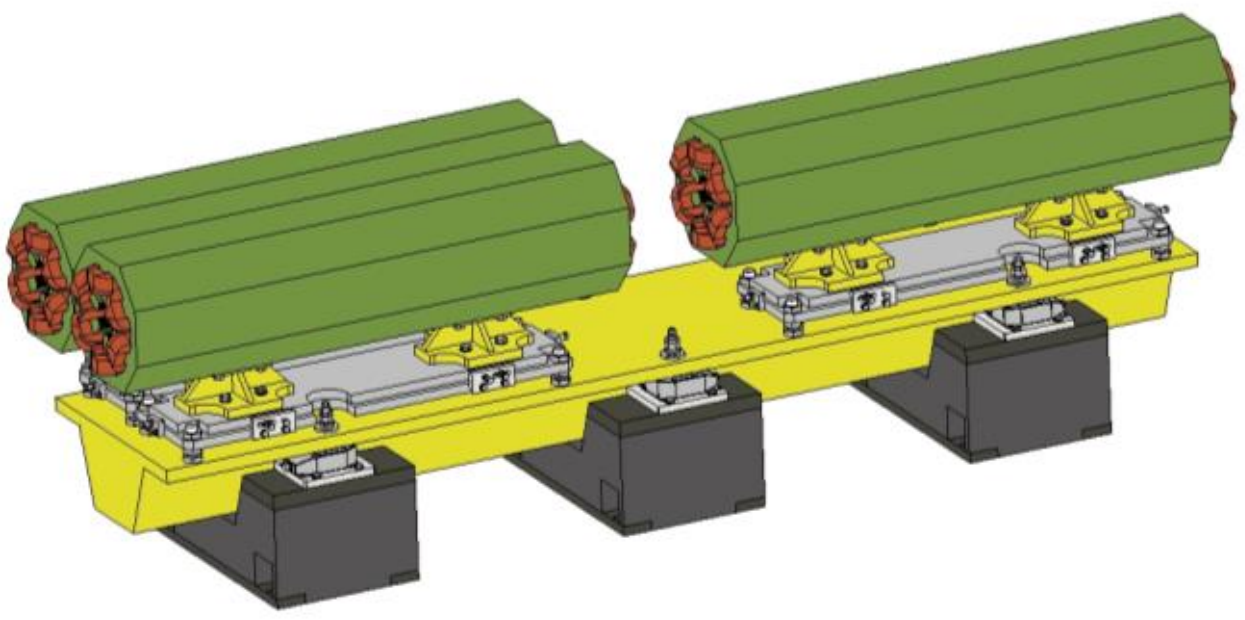

**Figure 4.3.10.7:** Sextupole magnets and their support

*4.3.10.2.5 Corrector Support System*

The Collider has 7,088 correctors, each of which is 875 mm in length. Half of these correctors are horizontal, and the other half are vertical. The supports for the correctors are similar to the main supports used for the dipole magnets, with each magnet being supported by a single support.

*4.3.10.2.6 Magnet Transport Vehicles*

A significant number of magnets are present within the tunnel. To enhance the efficiency of magnet transportation and coarse positioning, a specialized vehicle has been designed, as depicted in Figure 4.3.10.8.

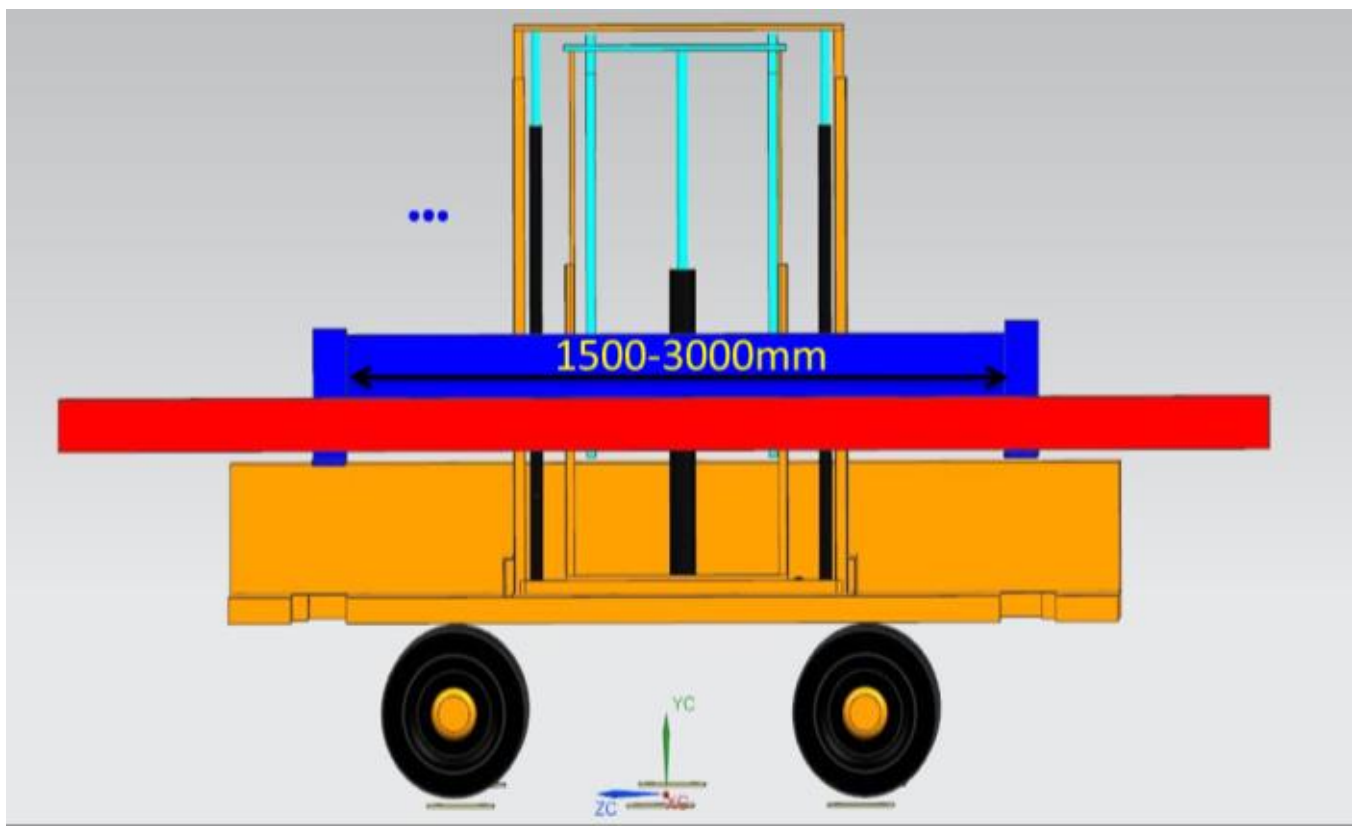

(a)

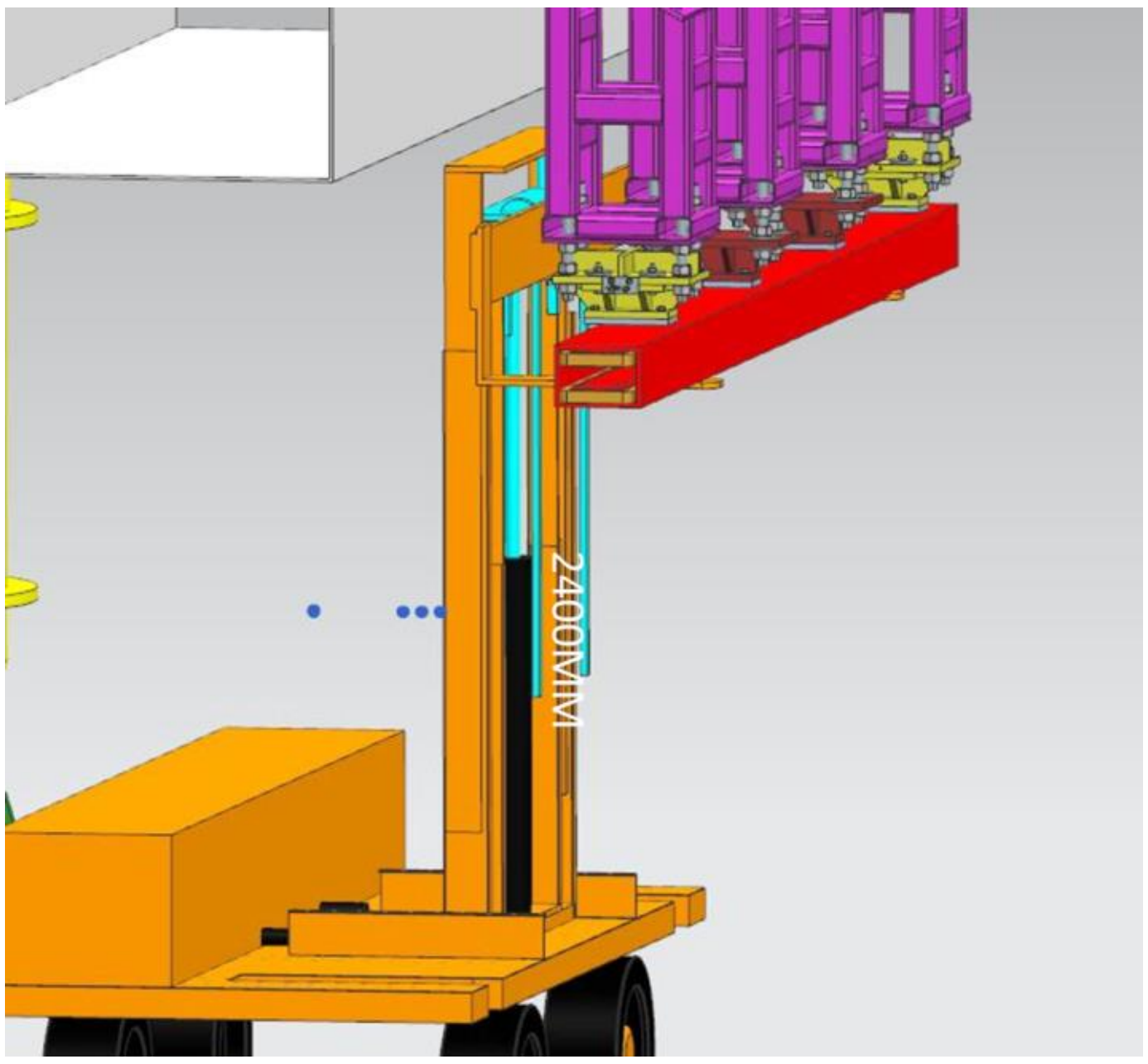

(b)

**Figure 4.3.10.8:** Magnet transport vehicle: (a) transporting dipole in the Collider; (b) transporting dipole in the Booster

The vehicle has a width of 1,600 mm, ensuring smooth movement in the aisle between CEPC and SPPC, even with SPPC devices installed. On the opposite side of the lifting fork, there is a location for a balance weight, adjustable according to the magnet being transported. The lifting fork can extend up to 3,000 mm in the Z direction, accommodating both long magnets like dipoles and quadrupoles and shorter ones. It can also move approximately 900 mm in the X direction and 2,400 mm in the Y direction, facilitating magnet placement on supports in the Collider and Booster. The lifting fork's resolution is designed to be better than 0.5 mm, sufficient for coarse positioning. After installation, magnets can be aligned and adjusted using the adjustment mechanisms on the supports.

#### 4.3.10.3 *Layout of the Tunnel Cross Section and Galleries*

The tunnel of the Collider has a width and height of 6,000 mm and 5,000 mm, respectively, in both the arc and RF sections. The Collider ring is located near the inner wall of the tunnel, while the SPPC is located near the outer wall. The layout of the arc section is shown in Figure 4.3.10.9. There is an operational space of about half a meter near the two walls, which is sufficient for one person. An aisle with a width of 2,400 mm is provided between the CEPC and SPPC for transportation, installation, alignment, and other operations.

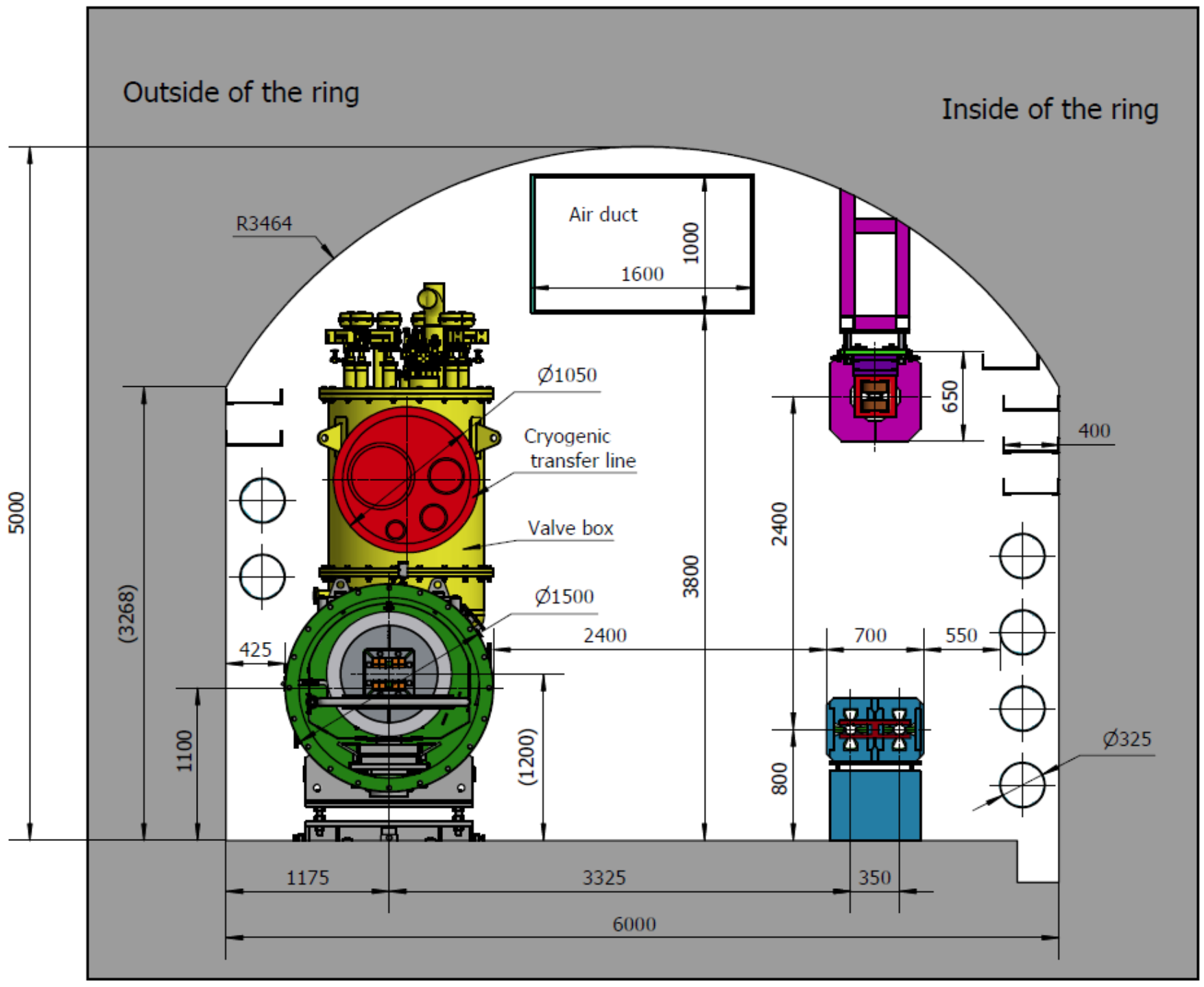

**Figure 4.3.10.9:** Tunnel cross section in the arc-section

CEPC has two RF sections. The RF tunnel is divided into two parts, one with only Collider cryomodules and no Booster cryomodules, and the other is the reverse. The SPPC devices are not present in the RF section since the colliding points of SPPC are close to the RF sections but in a separate bypass tunnel. The cross-sections of the RF tunnel and gallery are illustrated in Figure 4.3.10.10 and Figure 4.3.10.11, respectively.

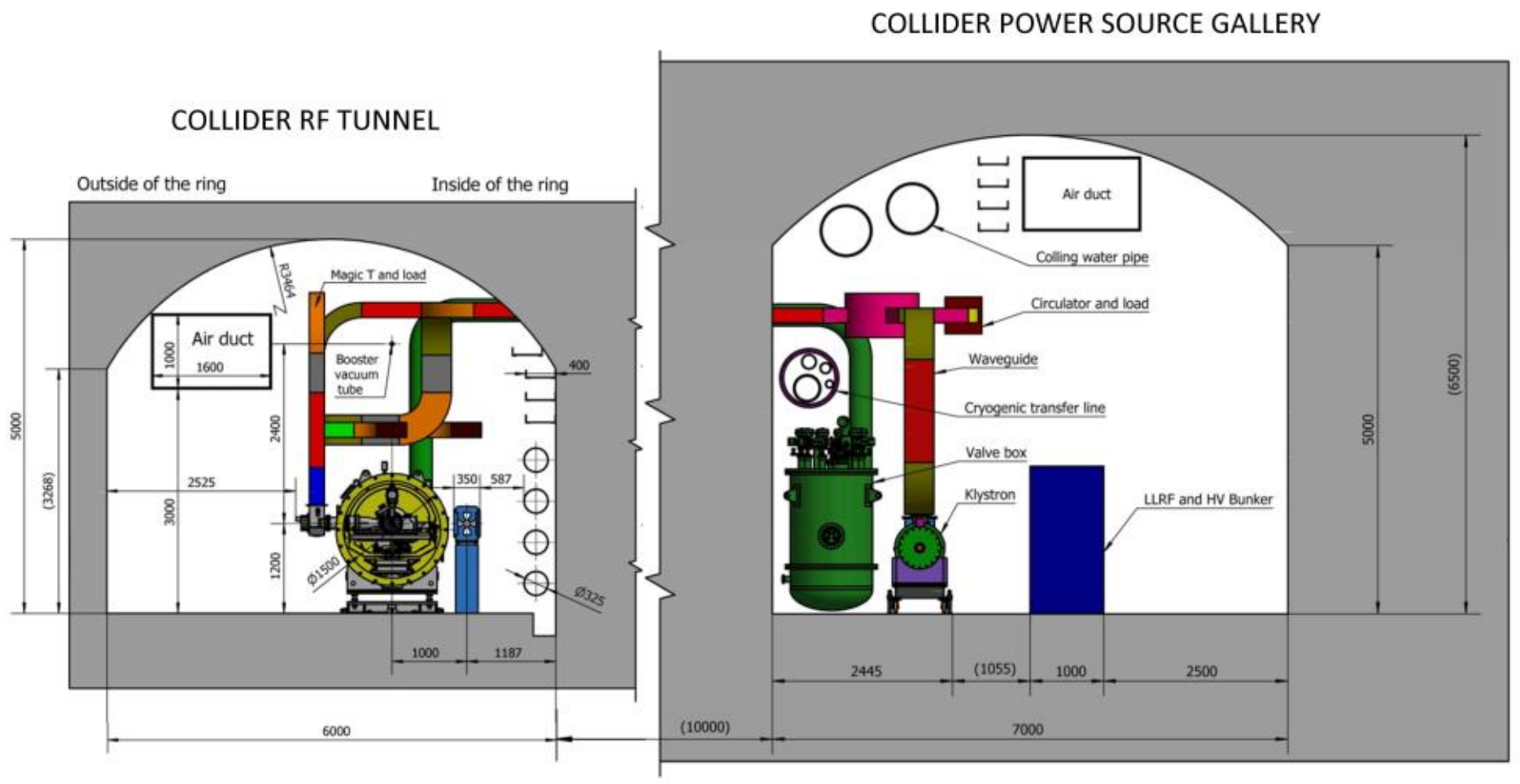

**Figure 4.3.10.10:** Tunnel cross section at the Collider RF-section.

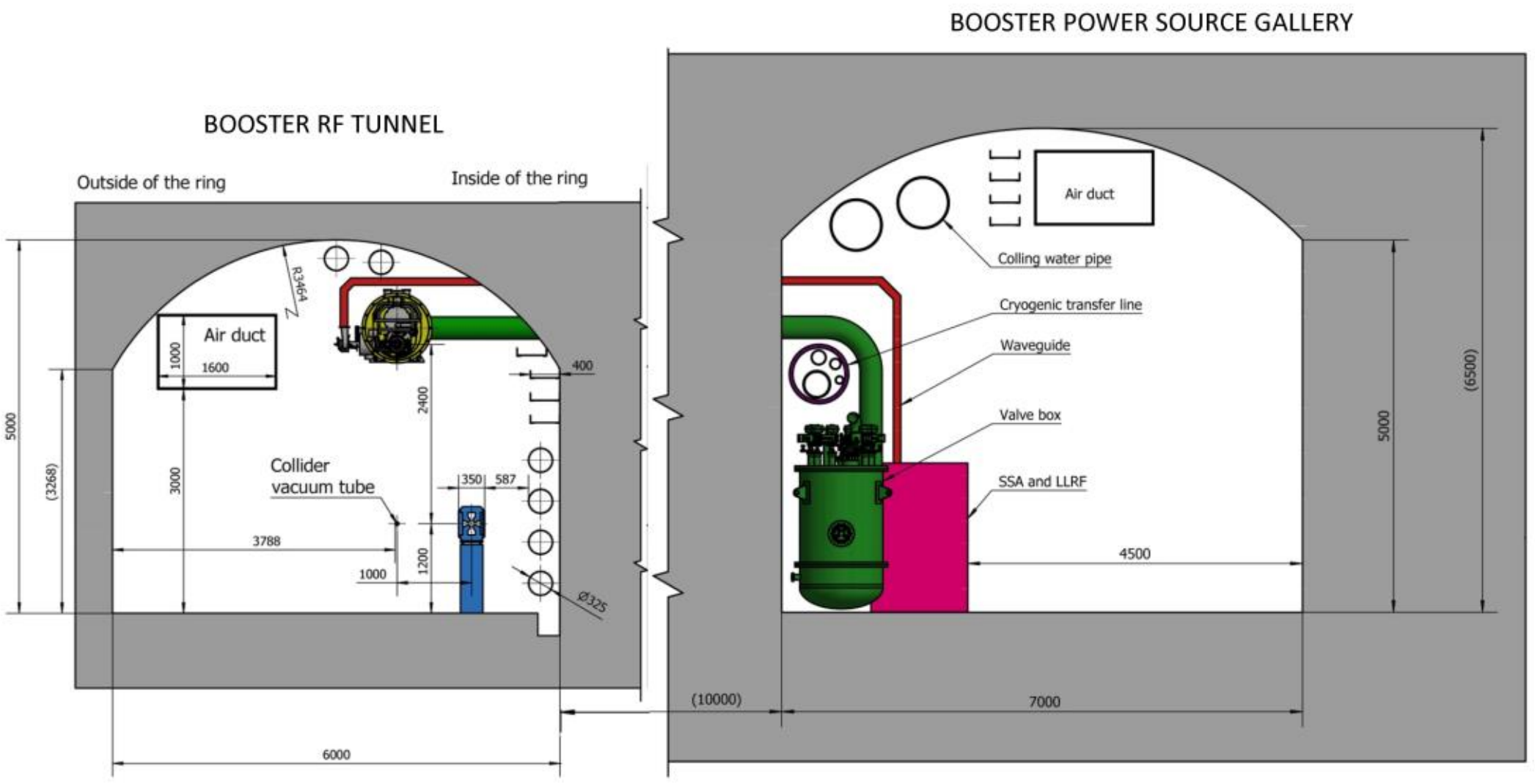

**Figure 4.3.10.11:** Tunnel cross section at the Booster RF-section.

In the RF section, there are several auxiliary devices, including power sources, cryogenic equipment, and cooling water systems. These auxiliary devices are located in the galleries that are situated next to the main RF tunnels. Each ½ RF gallery is 800 meters long, 8 meters wide, and 7 meters high, as shown in Figure 4.3.10.12. The layout of the gallery is symmetrical with respect to the symmetry point of the Collider and Booster. Specifically, there are two galleries for the Collider power sources, each of which is 235 meters in length, two galleries for the Cryogenic systems, each of which is 37 meters in length, two galleries for utilities, each of which is 70 meters in length, and one gallery for the Booster power sources, which is 134.6 meters in length.

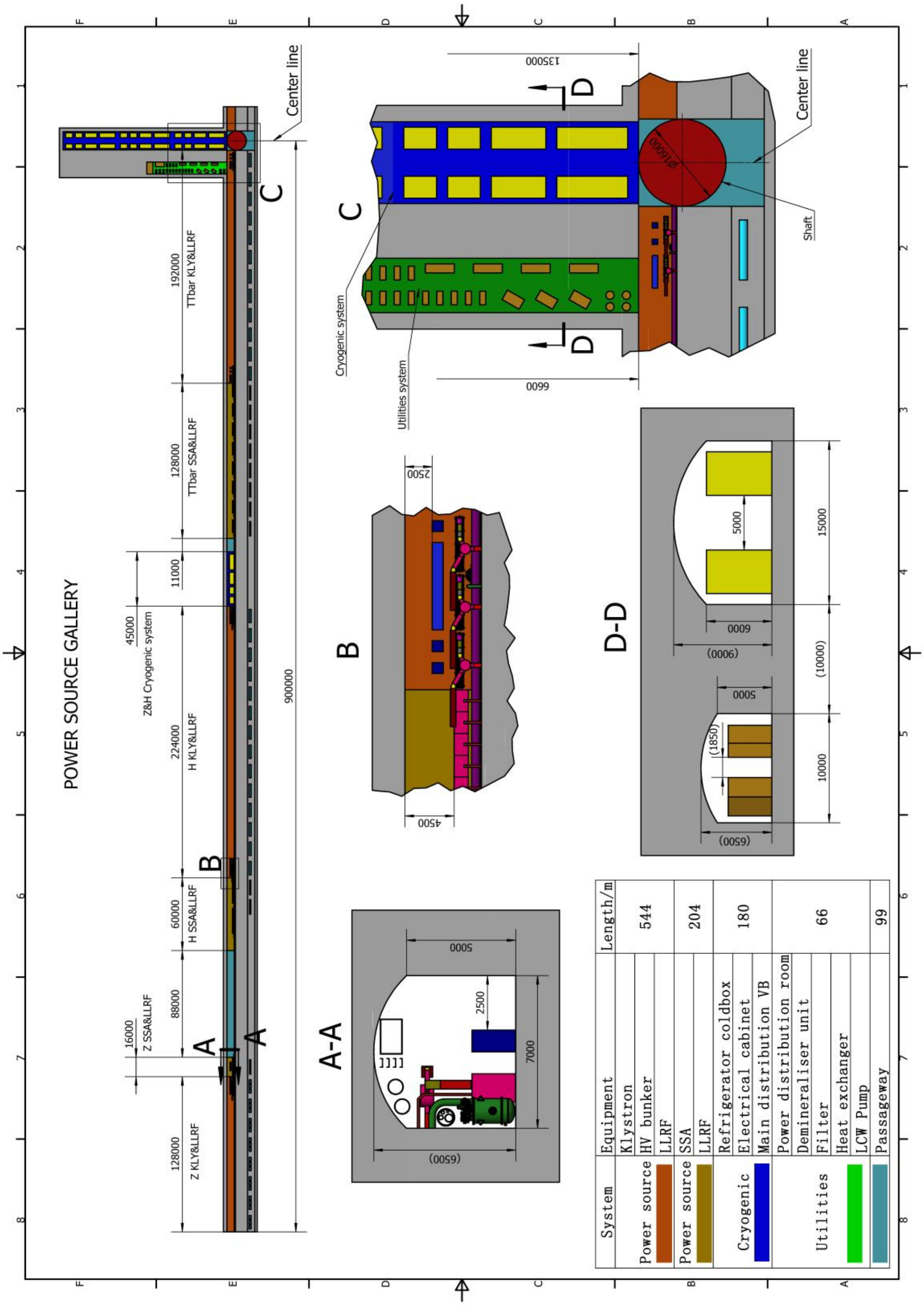

**Figure 4.3.10.12:** Tunnel cross section of the RF gallery.

### 4.3.10.4 *Movable Collimators*

#### *4.3.10.4.1 Introduction*

The collimator is a vacuum component that restricts physical apertures locally in the rings to capture spent electrons/positrons near the beam orbit by bringing jaws in close proximity to the circulating beam. CEPC employs two types of collimators: one for reducing background and another for machine protection. The latter has smaller gaps designed to handle significant beam loss in the event of a beam failure. This sub-section primarily focuses on the former type. The distribution of background-decreasing collimators is detailed in sub-section 4.2.6.3. In this section, we will specifically address the mechanical design of the horizontal movable collimator, noting that the vertical collimator will have a similar design.

The CEPC is characterized by a high beam current, which is approximately 2.5 times that of the highest energy collider LEP, and a high stored energy per beam, which is 3.8 times that of the highest luminosity collider superKEKB. As a result, the SR power incident on the collimator is particularly severe, with a linear load along the x-direction of about 291 W/mm, resulting in a total power of 7.5 kW. Additionally, the impedance needs to be optimized to reduce its impact on the beam stability. To address the new challenges brought by the high energy and luminosity of CEPC, a conceptual design for a horizontal movable collimator was proposed. Four important aspects were studied, including impedance, synchrotron radiation, thermal and mechanical situations, and beam impact.

#### *4.3.10.4.2 Conceptual Design*

In the current lattice layout, the collimator is situated between the quadrupole and the dipole magnet, with only 900 mm of available longitudinal space. The physical aperture must be capable of varying within the range of 4.4–20 mm. The dimensions of the collimator are mainly determined by these two key parameters. A collimator consists of three main subassemblies: the movable jaws that define the physical aperture, the vacuum vessel, and the auxiliary elements, which include the jaw motion system, the water-cooling system, and the support structure, as depicted in Figure 4.3.10.13.

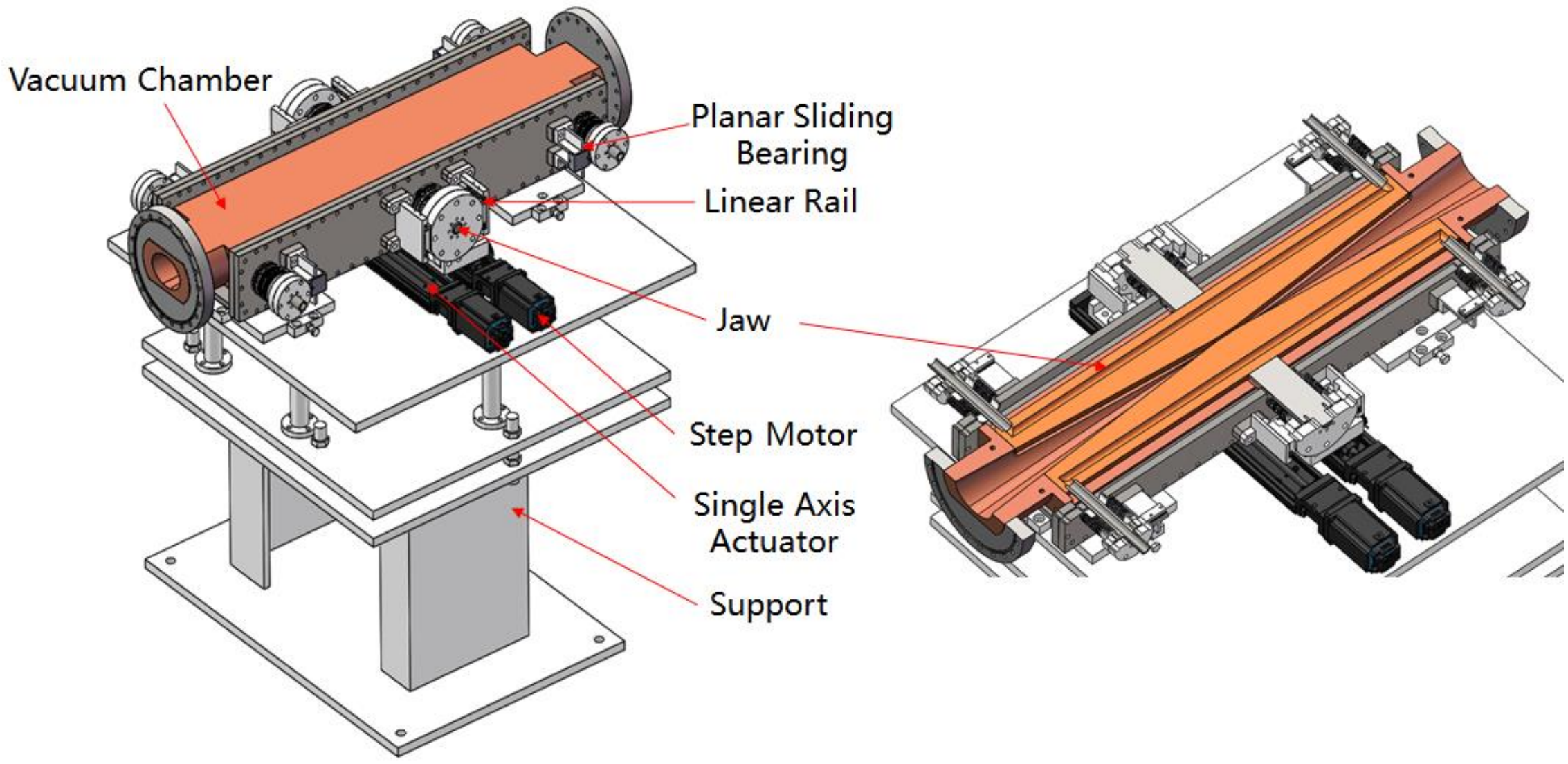

**Figure 4.3.10.13:** Components of the movable collimator

To minimize RF losses, the cross section of the vacuum vessel is gradually altered from a circular section of the beam chamber to a 21 mm wide rectangle that matches the shape of the jaws, as illustrated in Figure 4.3.10.14. The vacuum vessel is primarily constructed from dispersion strengthened copper GlidCopAl-15, with AISI 304L stainless steel flanges. The jaw body is also made from GlidCopAl-15 copper.

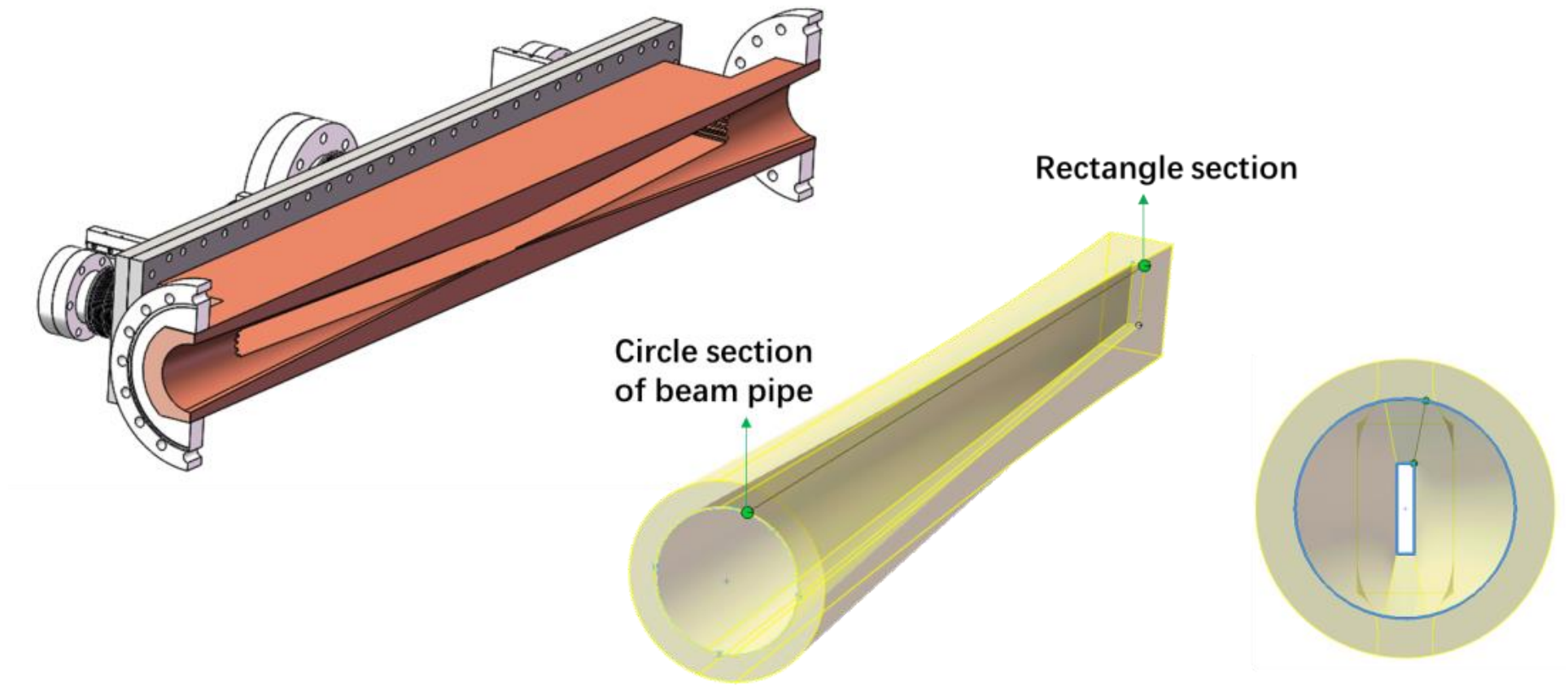

**Figure 4.3.10.14:** The inner contour of the vacuum vessel.

Due to space limitations from the flanges and bolt installation, the available space for the jaw is only 700 mm. The jaw consists of the main body and contact fingers, as depicted in Figure 4.3.10.15. The tip length of the jaw is two radiation lengths, which is a compromise often adopted [2, 3] between two conflicting requirements: to keep the heating of the jaw by showering electrons low enough, and to ensure that electrons passing through the jaw experience a loss of energy sufficiently high to fall out of the machine momentum acceptance. To prevent higher-order modes, contact fingers are attached to the top and bottom of the jaw, while a contactless comb-type RF shield [3, 4] is positioned between the longitudinal ends of the jaw.

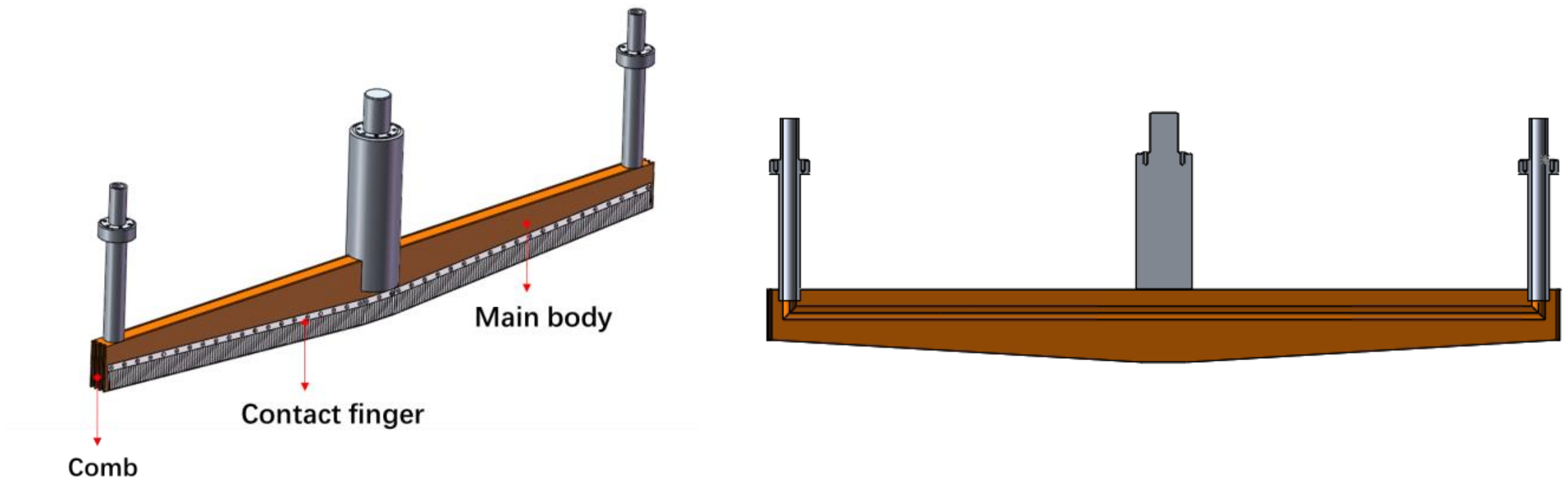

**Figure 4.3.10.15:** Structure of the jaw.

*4.3.10.4.3 Structure Optimization*

The impedance caused by the protruding shape of the jaw has been analyzed. When the physical aperture is 4.4 mm, the loss factor is only 0.016 V/pC, which corresponds to a loss power of only 6.25 W in Higgs mode (i.e., beam current of 17.4 mA with 242 bunches). This low loss power is achieved through the gradual contour of the vacuum vessel and the gentle taper of the jaw.

Due to the excellent impedance characteristics, a series of asymmetric jaws have been analyzed for the possibility of a gentler ramp to allow for a larger SRflux, as shown in Figure 4.3.10.16. The curve of the loss factor with the off-center distance (*d*) of the jaw tip is depicted in Figure 4.3.10.17. The results indicate that the impedance changes only slightly when the tip moves along the beam direction within 100mm, and the impedance increases steeply when moving further away from the center. Therefore, an asymmetric structure with a gentler ramp is feasible.

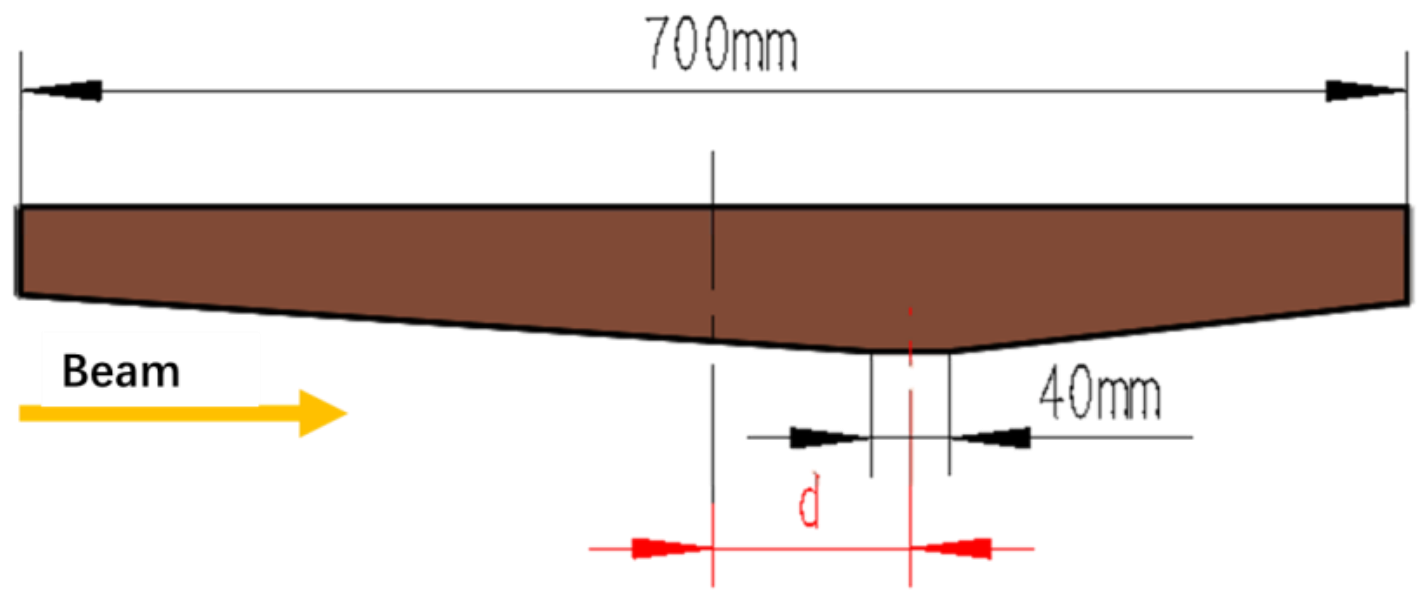

**Figure 4.3.10.16:** The off-center distance (d) of jaw tip.

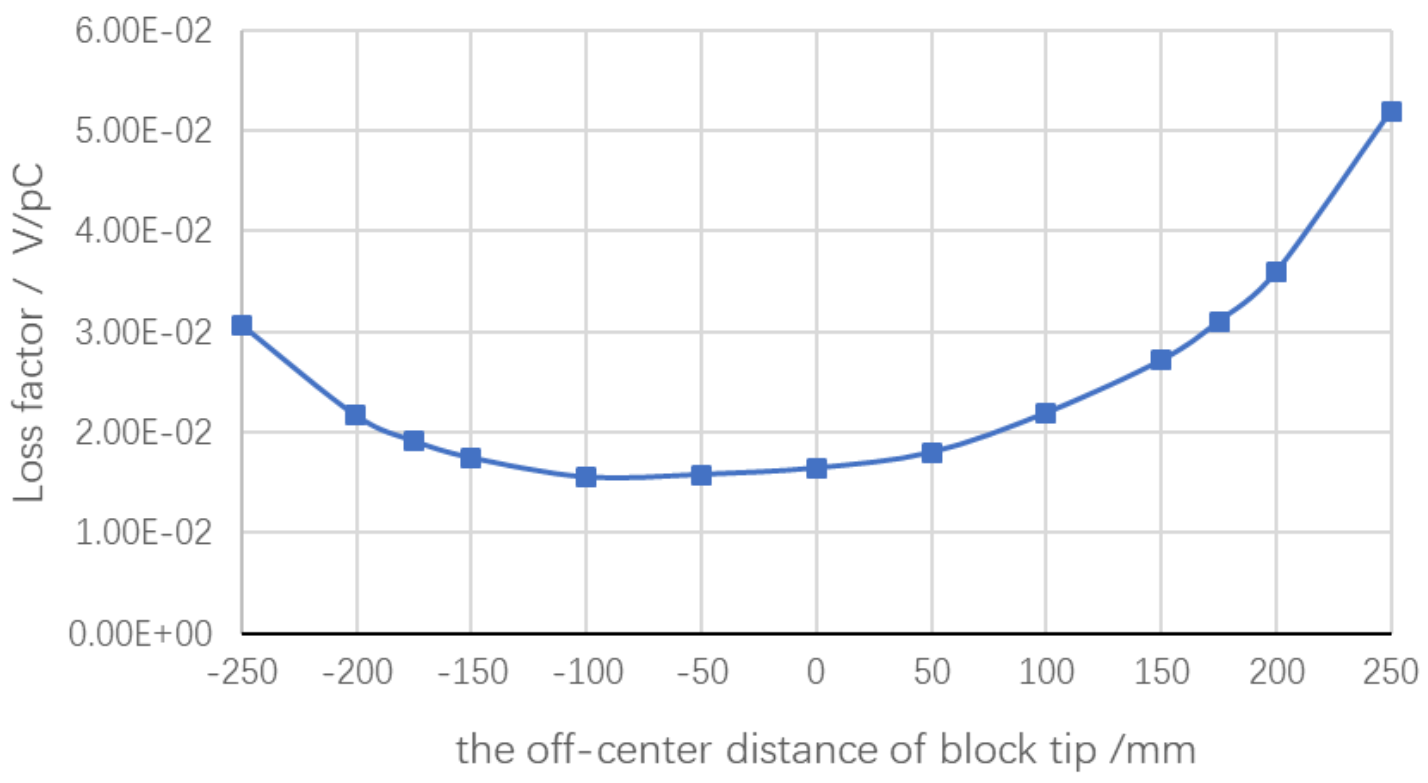

**Figure 4.3.10.17:** Loss factors for collimators as a function of the off-center distance (d) of jaw tip.

The SR power mainly depends on the beam configuration and the deflection radius. According to the formula calculation, the maximum SR power on the collimator is as high as 7.5 kW, and the power density along the x-direction at a vertical wall is about 291 W/mm. The critical energy Ec is equal to 357 keV.

Different interactions occur when the photon hits the jaw, but Compton scattering dominates in the energy range from a few keV to a few hundred keV. A simulation using Geant4 found that only about 70% of the SR energy is deposited in the jaw, as shown in Figure 4.3.10.18. Additionally, the ratio of deposited energy to incident SR energy remains relatively constant regardless of the inclination angle.

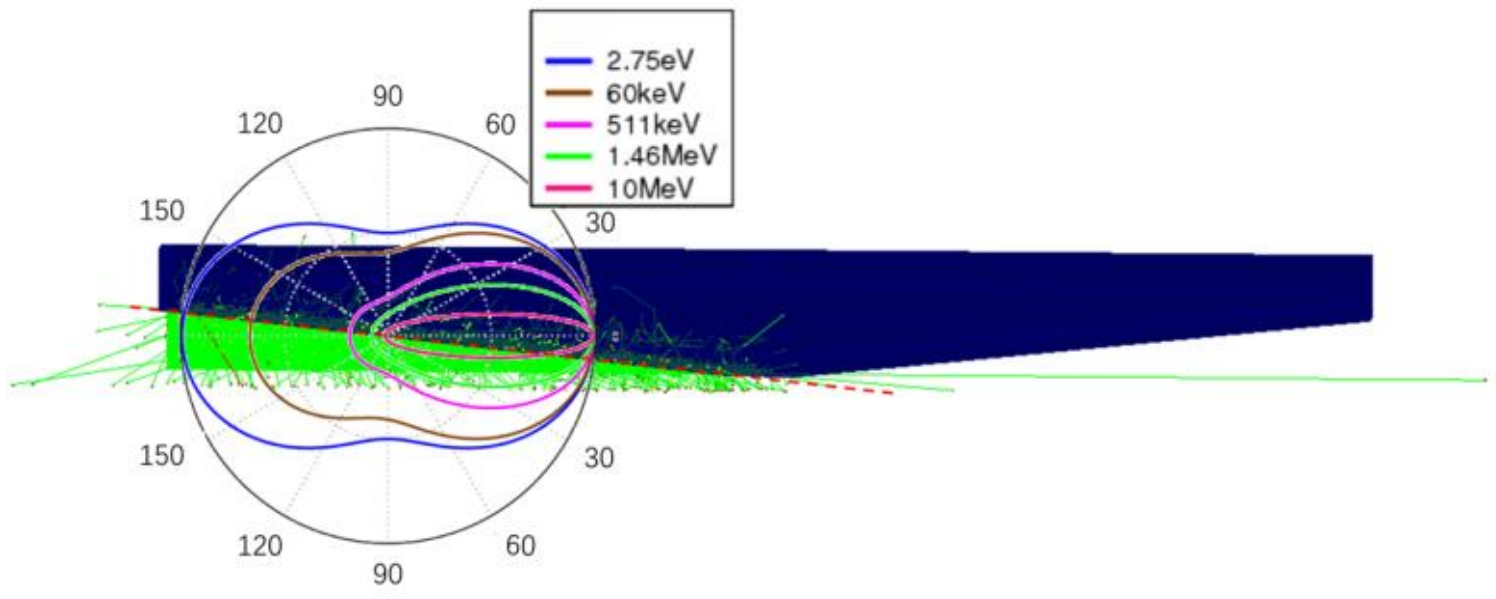

**Figure 4.3.10.18:** Simulation of SR hitting a jaw.

To ensure efficient cooling of the collimator jaw, the distance between the cooling wall and the tip is set to 10 mm, and the convection film coefficient is set to 12,000 W/(m$^2$K). The inclination angle (θ) of the jaw ramp has been optimized to reduce the temperature, as shown in Figures 4.3.10.19 and 4.3.10.20. The results indicate that decreasing the inclination angle is an effective way to reduce the temperature of the jaw.

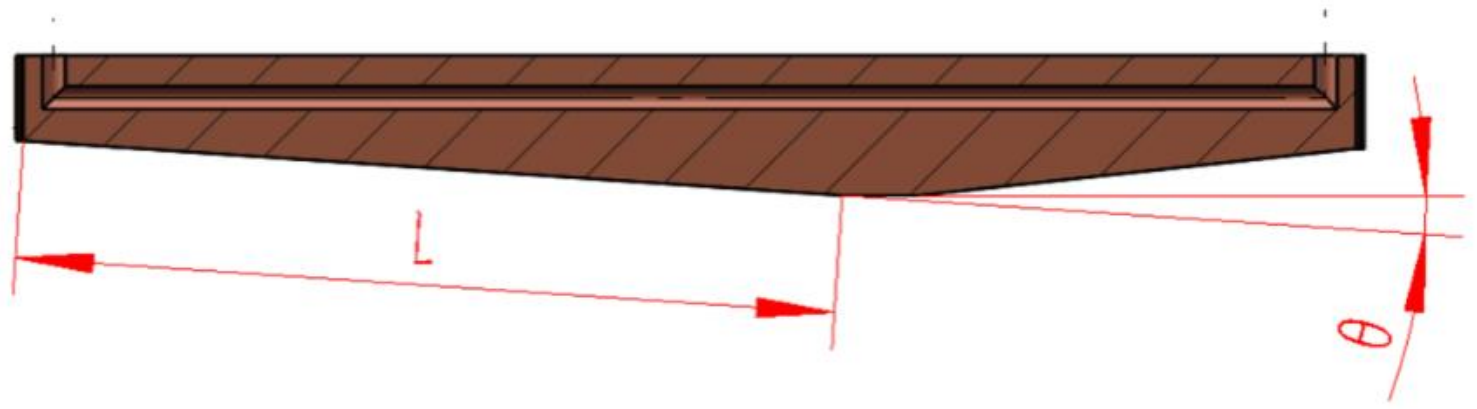

**Figure 4.3.10.19:** The cross section of the jaw

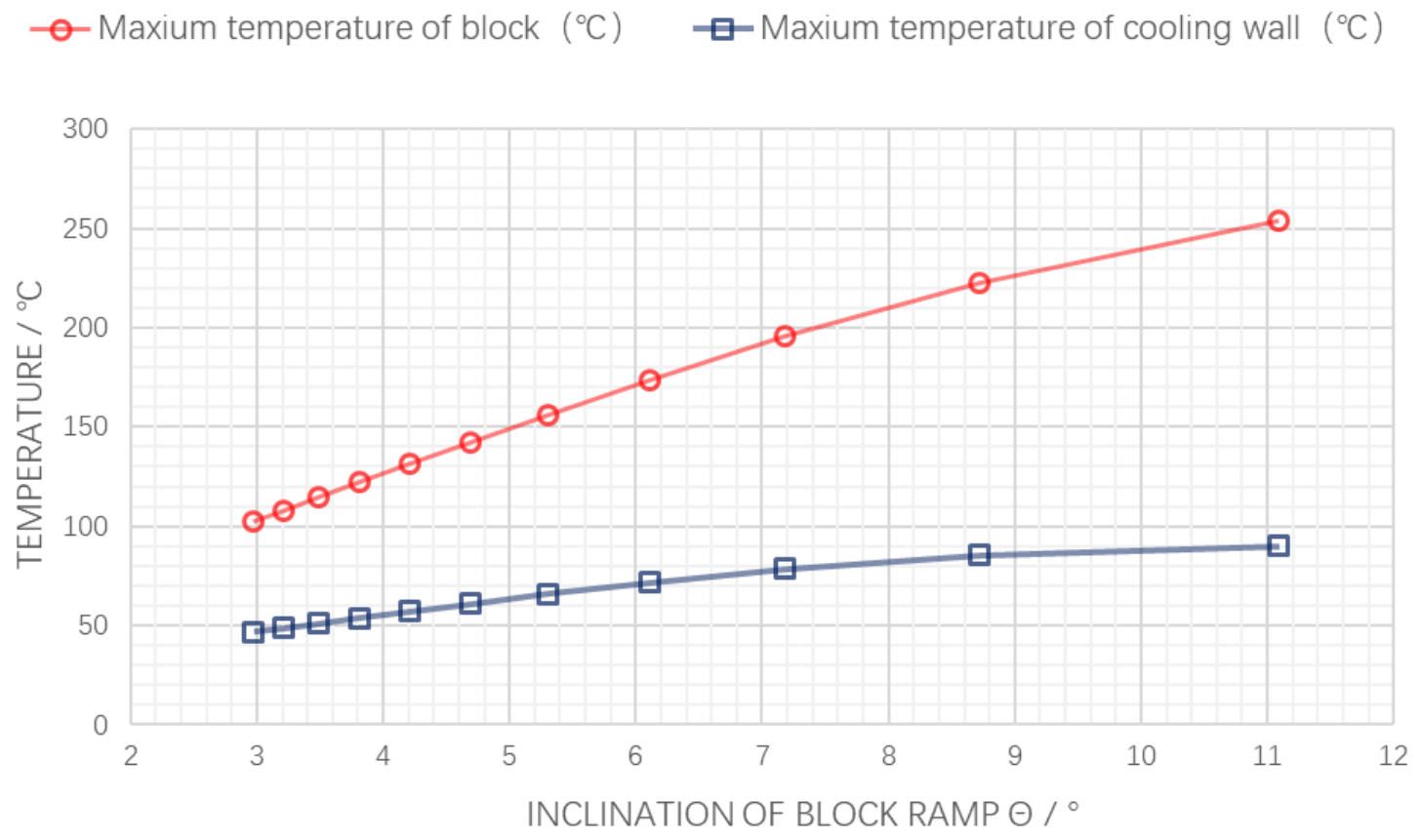

**Figure 4.3.10.20:** Maximum temperature of cooling wall and jaw as a function of the inclination of jaw ramp.

Based on the optimization results mentioned above, it is observed that although the asymmetric jaws perform better in terms of thermal and impedance characteristics, the difference in performance between asymmetric and symmetric jaws is not significant. Therefore, an integrated thermal analysis was conducted for the symmetric jaw, and further changes can be made to the structure if necessary. The maximum temperature of the jaw was found to be 146.7 ℃, and that on the vacuum chamber was 123.9 ℃, as shown in Figure 4.3.10.21. The maximum von Mises stresses were found to be 89.7 MPa and 110.4 MPa on the jaw and the vacuum chamber, respectively, as shown in Figure 4.3.10.22. The maximum deformation of the jaw was found to be 1.2 mm occurring at the jaw end due to thermal deformation mainly in the horizontal direction. The results were considered acceptable for the materials used, which were GlidCopAl-15 copper and AISI 304L stainless steel.

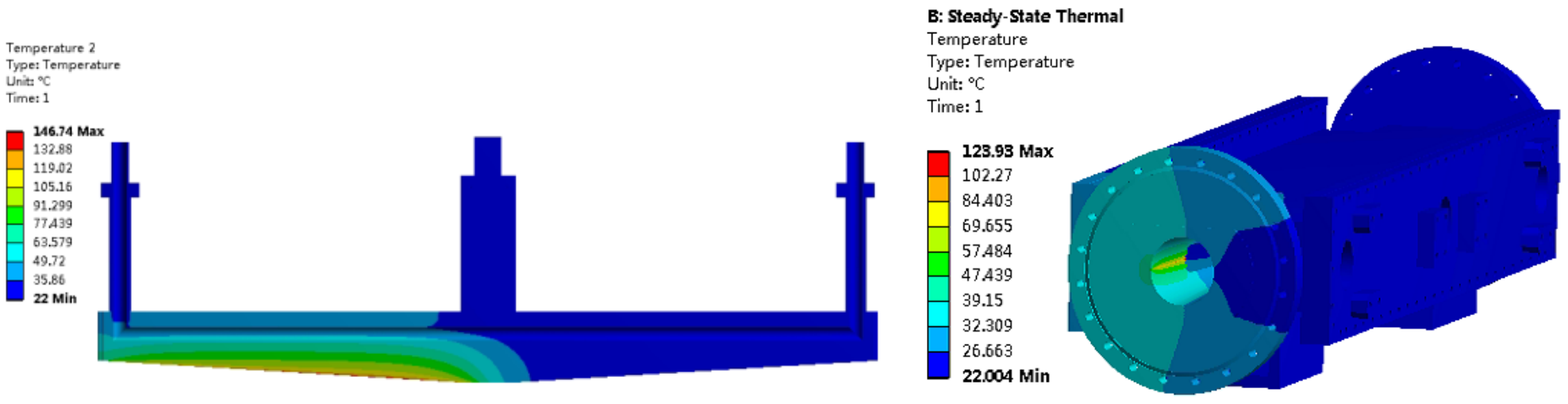

**Figure 4.3.10.21:** The temperature distribution of the collimator.

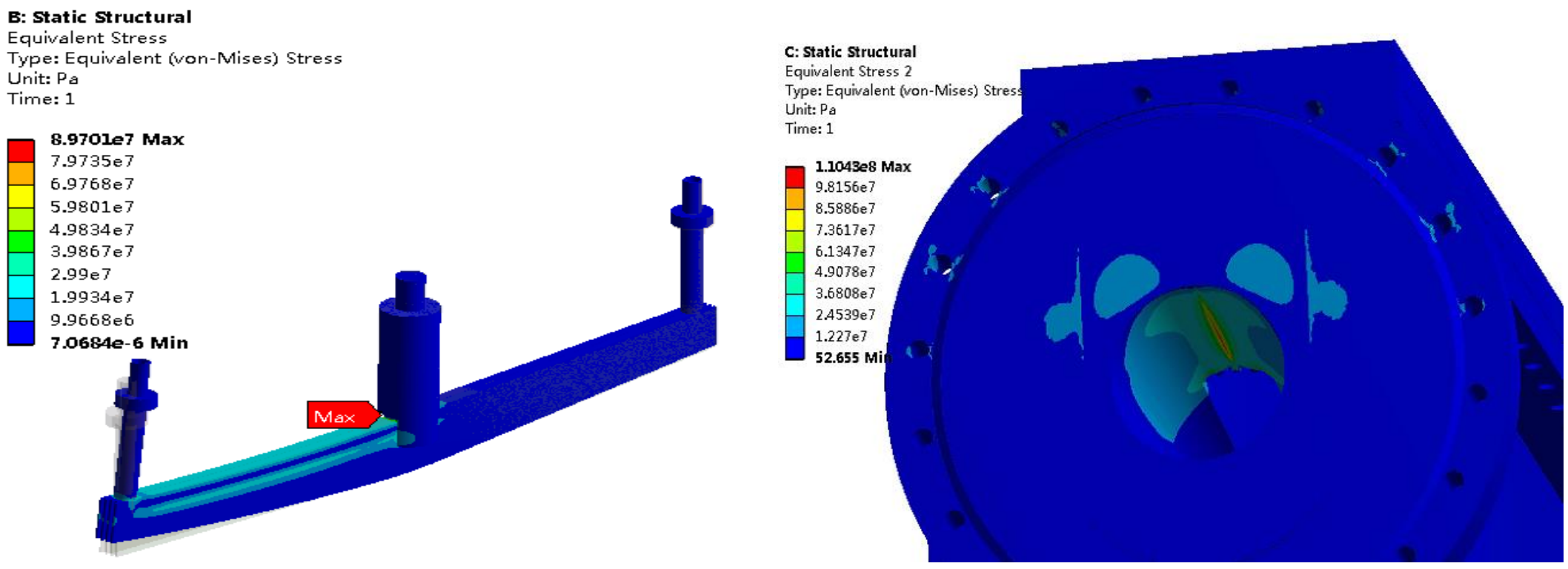

**Figure 4.3.10.22:** The von Mises stress distribution of the collimator

*4.3.10.4.4 Beam Impact*

The temperature load induced by the impact of beam halo under nominal operating conditions and accident scenarios where all 242 beam bunches hit the jaw simultaneously are estimated using GEANT4.

The heating power caused by beam loss from beamstrahlung in the Higgs mode, for multiple turns under nominal operating conditions, is only 21.2 W, which is negligible compared to the SR power. The resulting temperature increase on the jaw is only 1.3 ℃, as shown in Figure 4.3.10.23, and can be ignored.

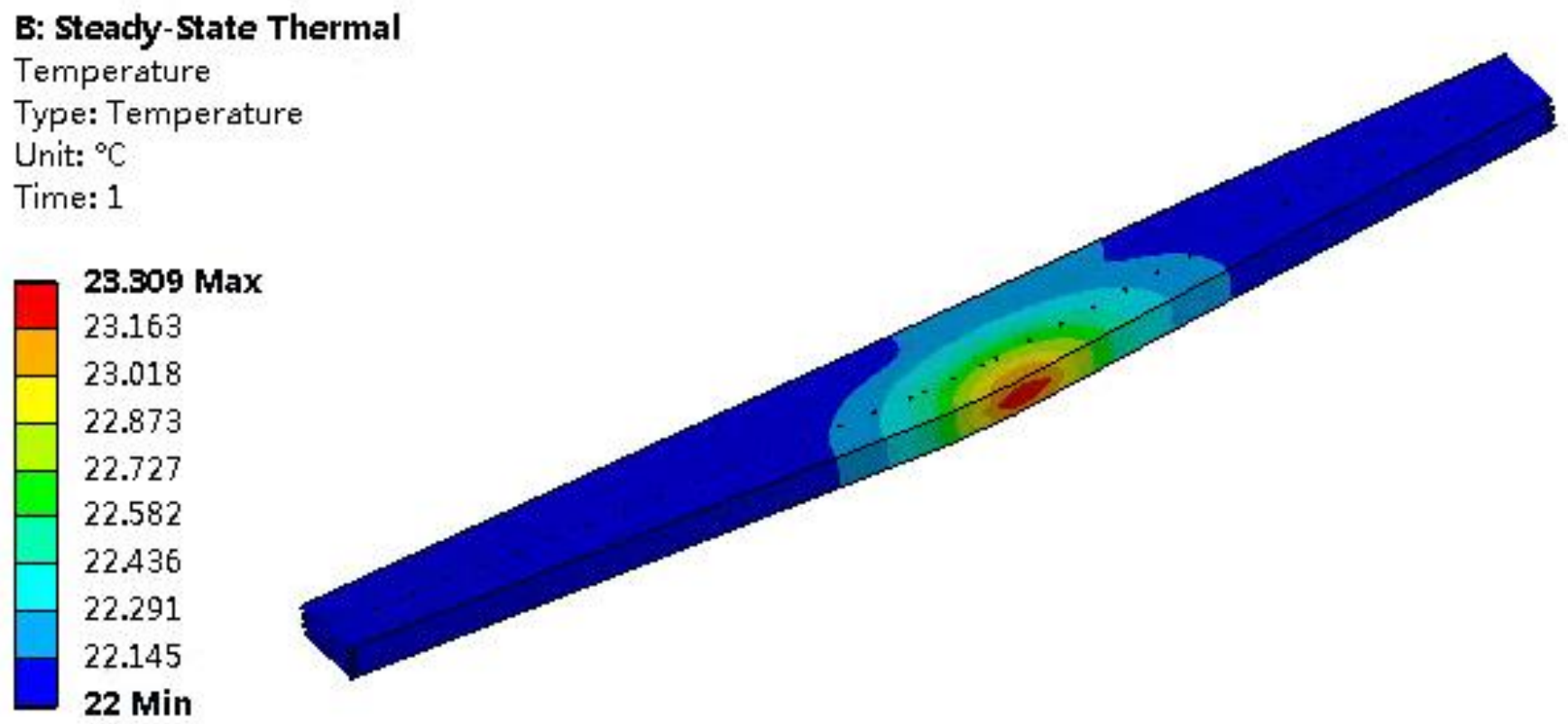

**Figure 4.3.10.23:** Temperature distribution by beam halo.

In an extreme situation where all 242 beam bunches impact the jaw due to beam failure, the stored energy is 698 kJ, and the size of the beam bunches is very small (i.e., $\sigma_x = 165$ µm, $\sigma_y = 35.7$ µm at the location of the collimator). The heat conduction can be ignored as the time for energy deposition is very short. Figure 4.3.10.24 shows the maximum temperature along the projecting direction. As long as the thickness of the copper exceeds about 0.026 mm, the maximum temperature exceeds the melting point. Even carbon cannot withstand such severe beam shocks when the thickness exceeds 1 mm. Hence, not only the collimator but any other device in the facility cannot endure this situation. Hence, it is necessary to explore additional machine protection strategies, where collimators from the machine protection system (MPS) will serve as passive protection devices. This aspect will be investigated in subsequent stages. The thermo-mechanical analysis of the MPS collimator will be undertaken when detailed information about beam loss at collimators is available, and this analysis will encompass aspects such as material selection and structural design.

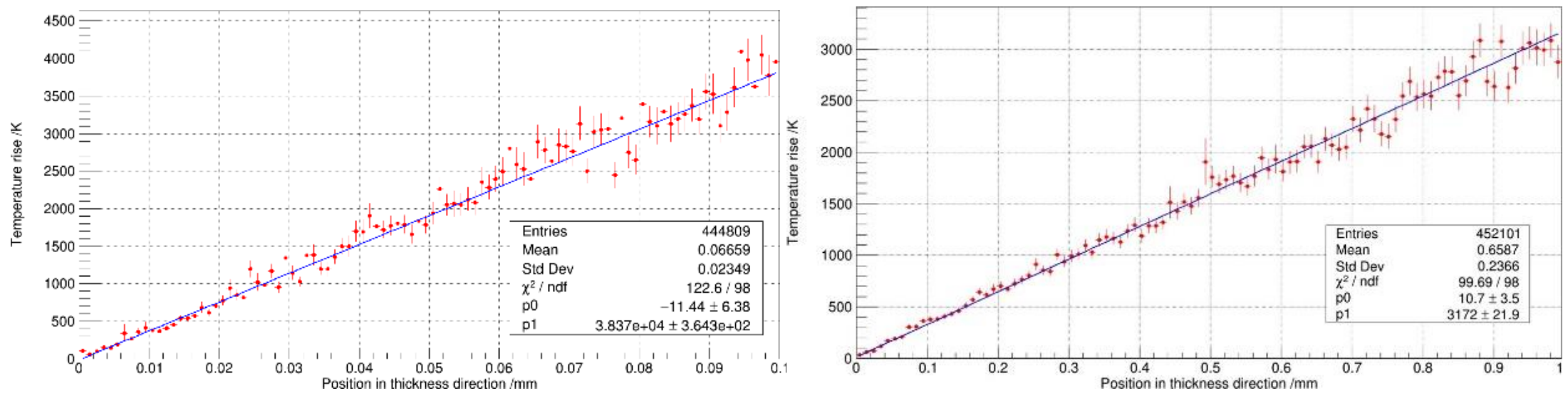

**Figure 4.3.10.24:** Maximum temperature rise along the projecting direction in the extreme beam failure (Left: in copper; Right: in carbon).

### 4.3.10.5 *Mechanical Design in the MDI Area*

The MDI includes various components, such as the detector devices (e.g., VTX, ECAL, solenoid, detector yoke), accelerator devices (e.g., cryostat, superconducting magnets, IP BPM), and vacuum chambers. This sub-section will detail the mechanical layout and assembly sequence for the accelerator devices at the MDI. Additionally, it will cover two critical components: the remote vacuum connector (RVC) and the support system for superconducting magnets.

#### *4.3.10.5.1 MDI Mechanical Layout and Assembly Sequence*

Figure 4.3.10.25 illustrates the cross-section at the MDI. The MDI vacuum chambers have a length of 14,000 mm, while the detector york measures 11,120 mm. Beyond the detector york, the cryostat is supported at the end, resulting in a cantilever length for the cryostat of approximately 4,830 mm.

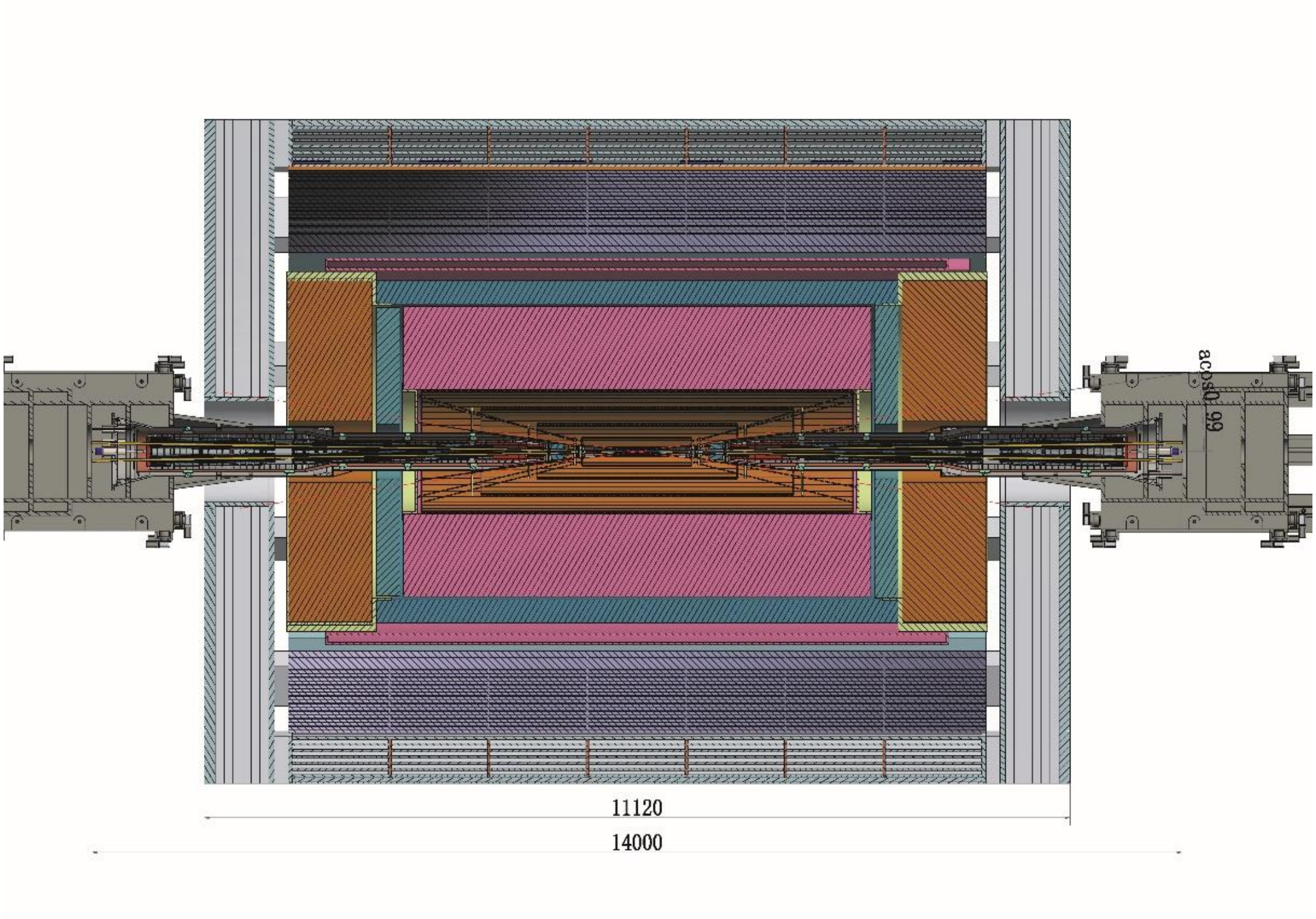

**Figure 4.3.10.25:** Cross section of the MDI region.

Figure 4.3.10.26 provides an enlarged view indicating the distance of each component from the Interaction Point (IP).

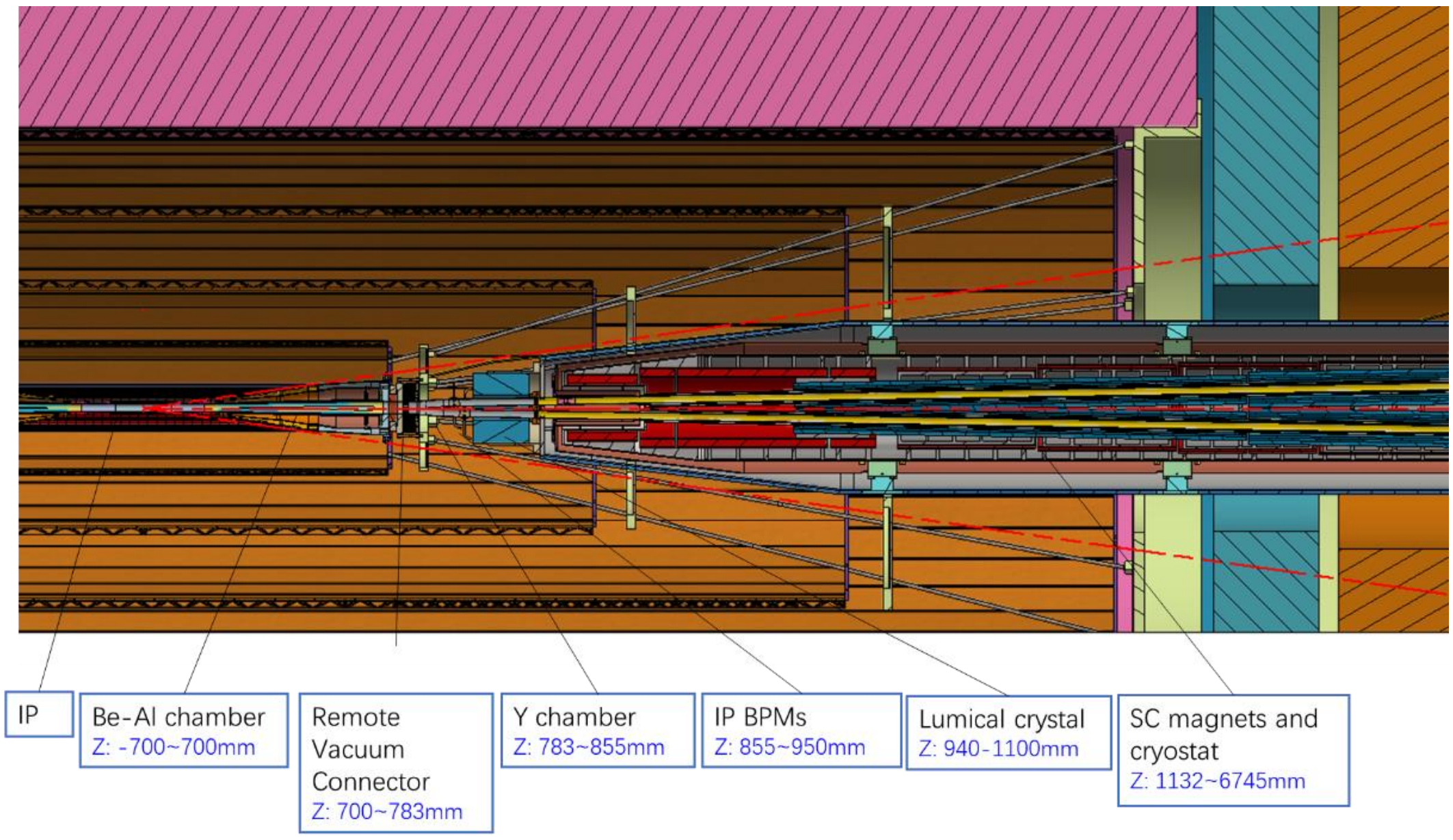

**Figure 4.3.10.26:** An enlarged view of the MDI cross section indicating the distance of each component from the Interaction Point (IP).

The MDI vacuum chambers consist of three parts: the IP chamber, measuring 1,400 mm in length and composed of beryllium near the IP and aluminum at the two ends; and the accelerator chambers, comprising the Y chamber and chambers within the cryostats. The total length is 14,000 mm. The RVC serves as the connector between the IP chamber and the Y chamber, measuring 83 mm in length. Two IP BPMs are located within the single chamber of the Y chamber. The Lumical crystal is positioned at the front end of the cryostat, which is shielded by a 10 mm layer of tungsten. The distances from the IP chamber flange, RVC, and tungsten shielding to the detective angle are 5.3 mm, 14.9 mm, and 16.5 mm, respectively.

Assuming the detectors are assembled, and the IP chamber is installed and aligned, the assembly sequence for the accelerator devices in the MDI region proceeds as follows:

a) Secure the SC magnets with the supports inside the helium vessel.

b) Assemble the helium vessel to the vacuum vessel of the cryostat and perform pre-alignment in the lab.

c) Assemble the vacuum chambers inside the cryostat and align them with the cryostat in the lab.

d) Assemble the Lumical crystal, Y chamber with IP BPM, and perform alignment in the lab.

e) Assemble the cryostat to the cryostat support system on-site, adjusting the cryostat using the adjusting mechanism of the cryostat support system with alignment, and then attach the RVC.

f) Move the cryostat into the detector using the moving mechanism of the cryostat support system to its designated location, connecting the Y chamber with the IP chamber using the RVC.

g) Perform the final alignment.

*4.3.10.5.2Remote Vacuum Connection*

The cryostat's cantilever extends approximately 4,830 mm, providing limited operating space within the detection angle of acos 0.99. Consequently, a remote vacuum connection method is essential.

The RVC needs to meet strict requirements, including a leak rate of less than $2.7 \times 10^{-11}$ Pa·m$^3$/s, high safety and reliability, an all-metal structure for anti-radiation purposes, and it must fit within the limited space available. . Due to the complexity of the design, several candidate structures have been proposed [5-6], including the SuperKEKB-RVC type, the remote chain seal type, the inflatable seal type, and the double knife edges inflatable seal type. Among these, the focus is primarily on the latter one. The SuperKEKB-RVC type was eliminated due to its complicated structure, which cannot fit into the available space, while the remote chain seal type was eliminated due to its weak clamping force over long distances.

Figure 4.3.10.27 is a schematic diagram of the inflatable seals that have been successfully used in high radioactive regions for some accelerators, such as SNS, JSNS, and CSNS. These seals can reach the leak rate of $3.9 \times 10^{-10}$ Pa·m$^3$/s [7], an order of magnitude higher than the requirement for CEPC MDI. The seals are pneumatic controlled and consist of two sets of concentric bellows, gas tubes, vacuum tubes, sealing membranes, and flanges. The leak rate can be calculated using the following equation:

$$Q = C \times \Delta P \tag{4.3.10.1}$$

where $Q$ is the leak rate, $C$ is the conductance and $\Delta P$ is the pressure difference. It is obvious that the leak rate can be improved by decreasing the conductance and the pressure difference.

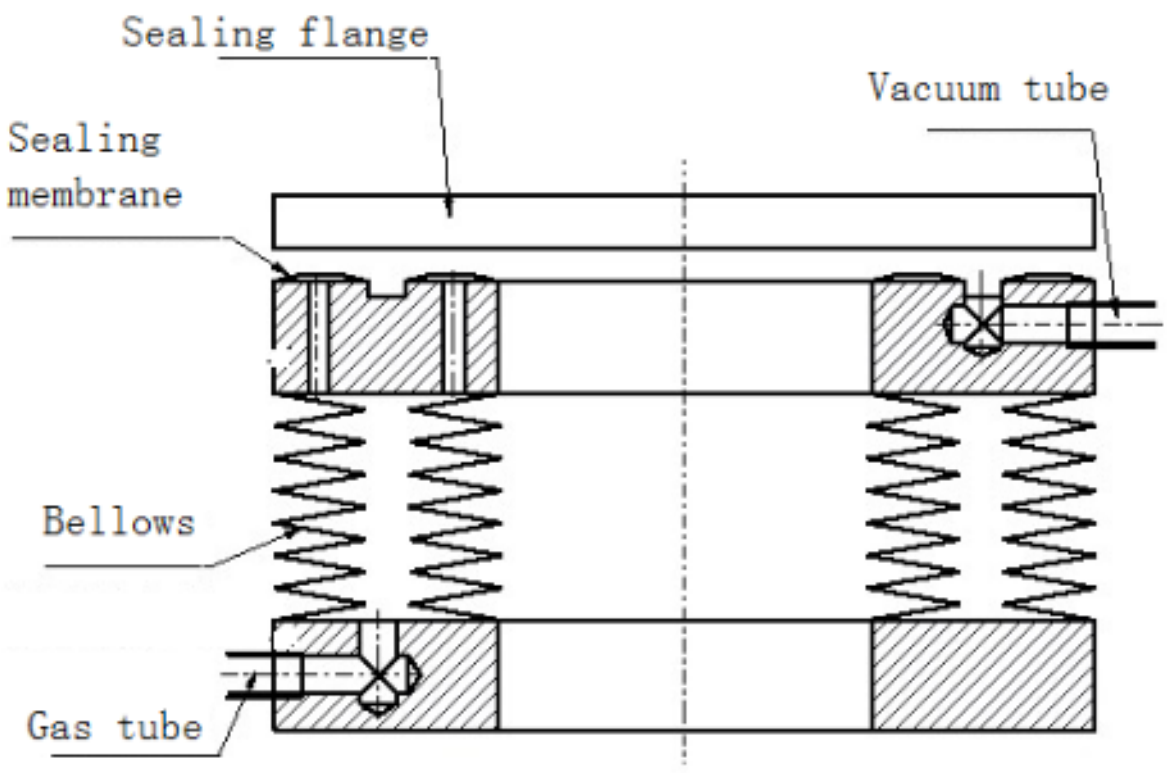

**Figure 4.3.10.27:** Schematic diagram of inflatable seals

The conductance of a single sealing membrane for helium at 25 ℃ and 1 atm pressure difference can be expressed as [8]:

$$C = 34A^2 \ (L/w) \ exp[-3F/(LwR)] \tag{4.3.10.2}$$

where $A$ is the roughness of the contact surface, $L$ is the perimeter of the sealing structure, $w$ is the contact width, $F$ is the force from the bellows, and $R$ is the sealing factor expressing the sealing ability of the material, which is propotional to the material hardness.

In the case of the RVC, if inflatable seals are used, the inner and outer diameters of the seals are almost fixed by the vacuum diameter and the detective angle. The design pressure inside the bellows can be limited by the bellows, so the parameters L and F are nearly determined. To improve the conductance, we can optimize the roughness of the contact surface, the width of the contact surface, and the materials. The knife-edge seal may have better ability because it has a smaller contact width and a smaller sealing factor due to the use of steel-copper sealing instead of steel-steel.

The vacuum groove between the two sealing membranes is an effective way to decrease the leak rate as the pressure difference between the beam vacuum chamber and the groove is much smaller. Thus, the vacuum groove should be included in the RVC design. Experimental results show that the leak rate can be reduced by about 4 levels when the groove is vacuumed. The double knife edges inflatable seal is a candidate design for the RVC with better leak rate ability than the existing inflatable seal, although further development and testing will be done. To address the impedance issue, RF fingers are installed inside the bellows. Figure 4.3.10.28 illustrates the structural design of the RVC with the pressure tube located at the outermost position, about 15 mm from the detective angle.

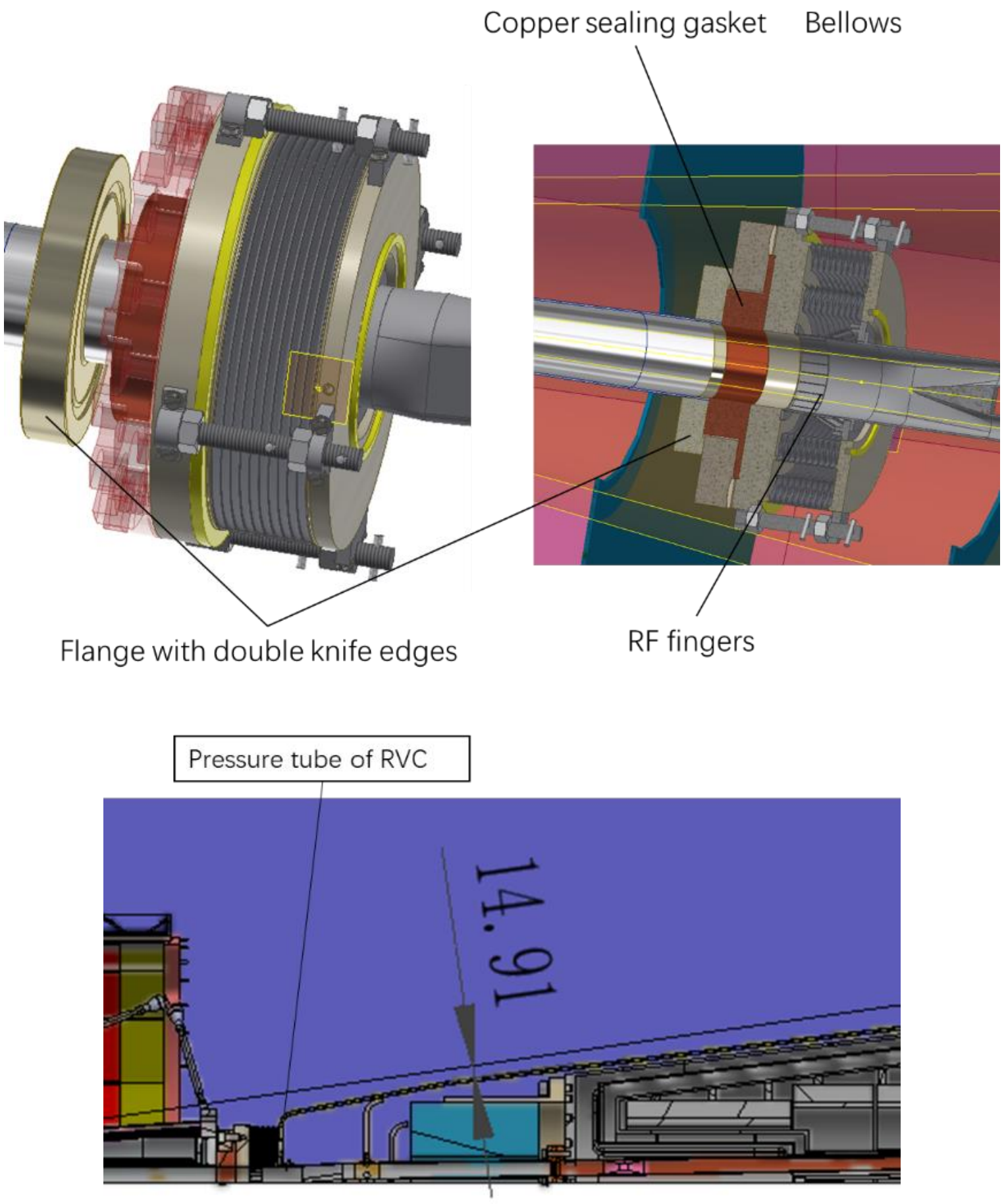

**Figure 4.3.10.28:** Structure design of double knife edges inflatable seal (Up: structure; Down: distance from detective angle)

The inner wall of the RVC is subject to thermal loads from HOM and synchrotron radiation. It is comprised of thin RF fingers and a contact wall, making it difficult to dissipate thermal loads due to the limited ability to add cooling water in close proximity. Based on safety considerations, the structure can endure approximately 1 W/cm$^2$ of HOM power, with a safety margin [9]. Figure 4.3.10.29 shows the temperature distribution during Higgs mode, where the highest temperature recorded was 44.5 ℃, well below the allowable temperature limit.

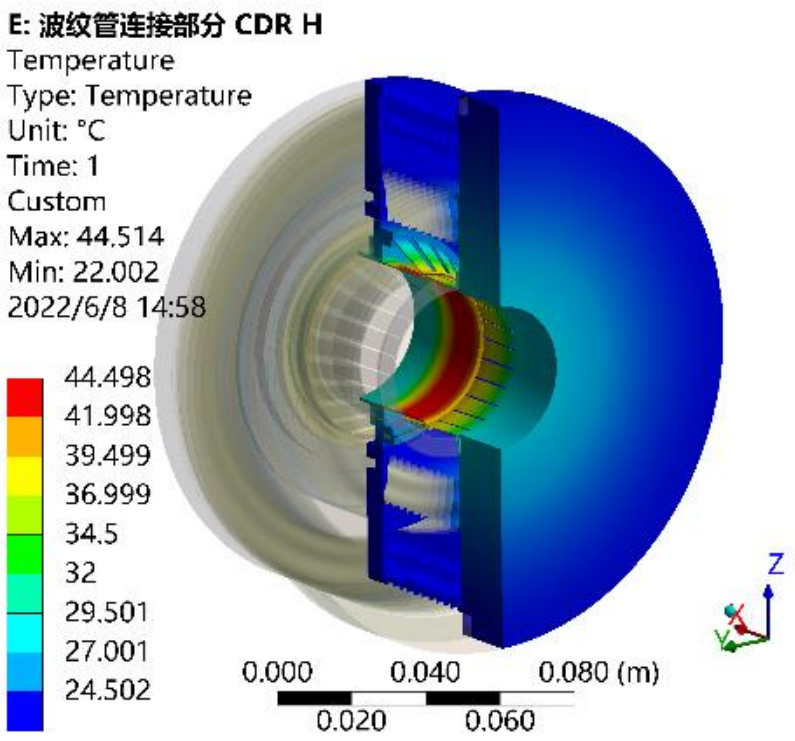

**Figure 4.3.10.29:** Temperature distribution of RVC at Higgs mode

#### *4.3.10.5.3 Superconducting Magnet Support System*

The cryostat for the superconducting (SC) magnet is approximately 5.5 meters in length, and the alignment requirement for the SC magnets is better than 100 μm. The supports for the SC magnets must meet specific requirements, including static and modal stability, accuracy, dimensions, and ease of operation and maitanance.

The support structure of the cryostat is depicted in Figure 4.3.10.30 and includes a movement mechanism, support bed, and adjusting mechanism. The movement mechanism comprises high-precision rails and motor-driven rack and pinion with a motion range of 6 m, slightly longer than the cryostat cantilever. The support bed is a steel structure with internal ribs for added strength, on which the precision rails are mounted. The adjusting mechanism is a steel weldment consisting of two layers. The cryostat is secured to the top layer, while the bottom layer is attached to the moving part of the movement mechanism. Four auto-driven wedge jacks are located between the two layers for vertical adjustment, and three for horizontal adjustment. The movement mechanism guides the cryostat to the working location, after which the alignment and adjustment procedures are carried out using the adjusting mechanism to achieve precise positioning. The resolution requirement for the adjusting mechanism is better than 5 μm. A moniroring alignment system measuring the deformation of the cantilever will also be considered.

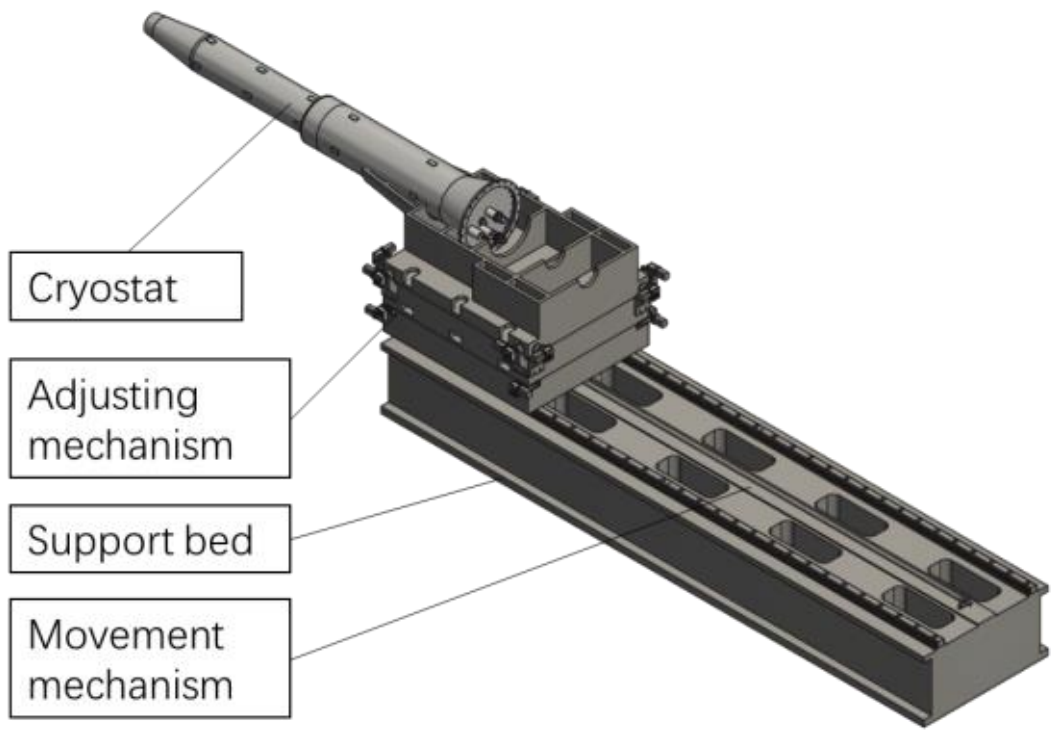

**Figure 4.3.10.30:** Structure of cryostat supports.

Inside the cryostat, there are two types of support: one for securing the SC magnets to the helium vessels, and the other for supporting the helium vessels within the vacuum vessel, as illustrated in Figure 4.3.10.31. Two support schemes are considered for the former, namely the solid support scheme and the skeleton support scheme, as shown in Figure 4.3.10.32. In the latter, support rods with a diameter of 20 mm are used.

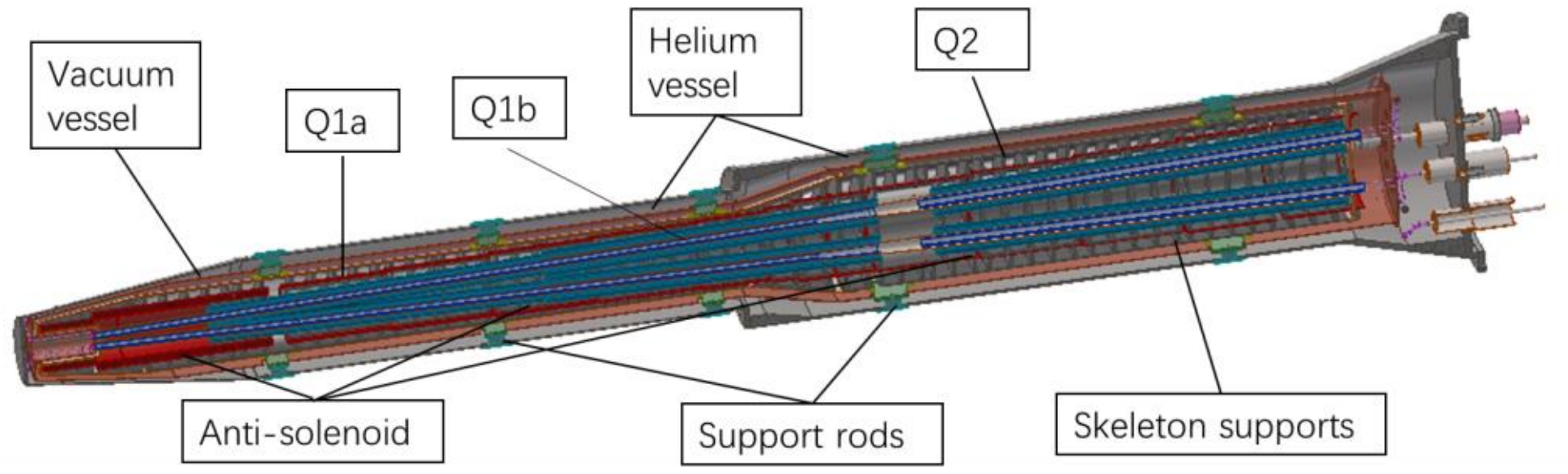

**Figure 4.3.10.31:** Structure of the SC magnets inside cryostat.

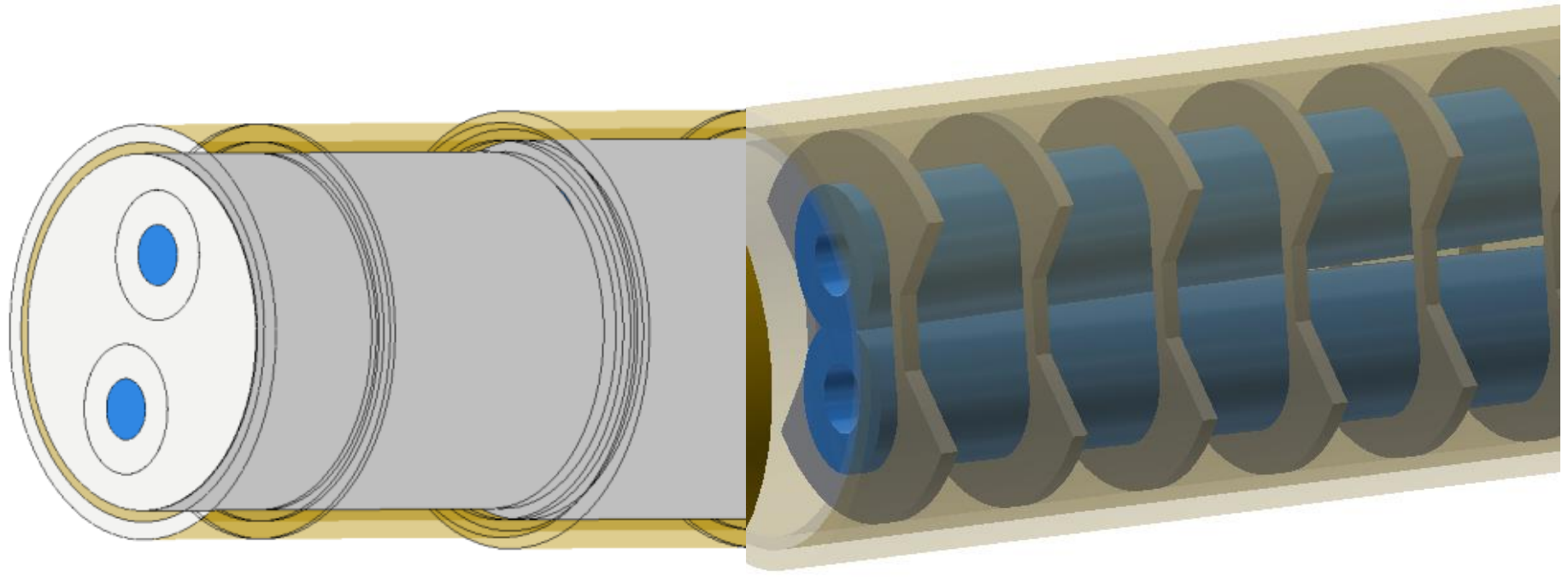

**Figure 4.3.10.32:** Support structure from SC magnet to helium vessel: Left – solid design; Right – skeleton design.

The detector york length is 11,120 mm. To account for conservation considerations, the fixed location of the cryostat is assumed to be at the end of the cryostat. The supports for the cryostat have very little effect on the cryostat deformation, so only the cryostat and structures inside it are used for the analyses.

For a cantilevered cylinder subjected to gravity, the maximum deformation $y_{max}$ and the first natural frequency $\omega_1$ can be expressed as follows:

$$y_{max} = -\frac{ql^4}{8EI} = \frac{2l^4\rho}{\pi ED^2} \tag{4.3.10.3}$$

$$\omega_1 = 3.5\sqrt{\frac{EI}{\rho Al^4}} = \frac{3.5D}{4l^2}\sqrt{\frac{E}{\rho}} \tag{4.3.10.4}$$

where $l$ is the cantilever length, $\rho$ is the density, $D$ is the diameter, $E$ is the elasticity modulus and $I$ is the cross-sectional moment of inertia. The $l$ and $D$ are the shape factors and the specific stiffness $E/\rho$ is the material factor. For a concentric cylindrical cantilever, the maximum deformation $y_{max}$ is also related to the density ratio of the outer and inner structure $\rho'/\rho$, which can be expressed as

$$y_{max} \sim \frac{\rho l^4}{32EI}(D^2 - d^2 + \frac{\rho'}{\rho}d^2) \tag{4.3.10.5}$$

The density and specific stiffness of some common materials are listed in Table 4.3.10.6.

**Table 4.3.10.6:** Density and specific stiffness of some common materials

| Material | Density (kg/m3) | Specific stiffness (Nm/kg $\times 10^3$) |
|---|---|---|
| Stainless steel | 8000 | 25 |
| Ti-6Al-4V | 4430 | 25.7 |
| A6061 (Al) | 2700 | 25.5 |
| Copper (Cu) | 8960 | 12.3 |

The optimization goal is to minimize the uneven deformation of the SC magnets, which should not exceed a certain threshold typically at the sub-millimeter level. The optimization should focus on pre-aligning the magnets to a specific location and then ensuring they remain in their working location under operation forces, such as the magnetic field.

There are two supports in the longitudinal direction of the helium vessel that can be optimized using Response Surface Design in ANSYS. This optimization process is illustrated in Figure 4.3.10.33.

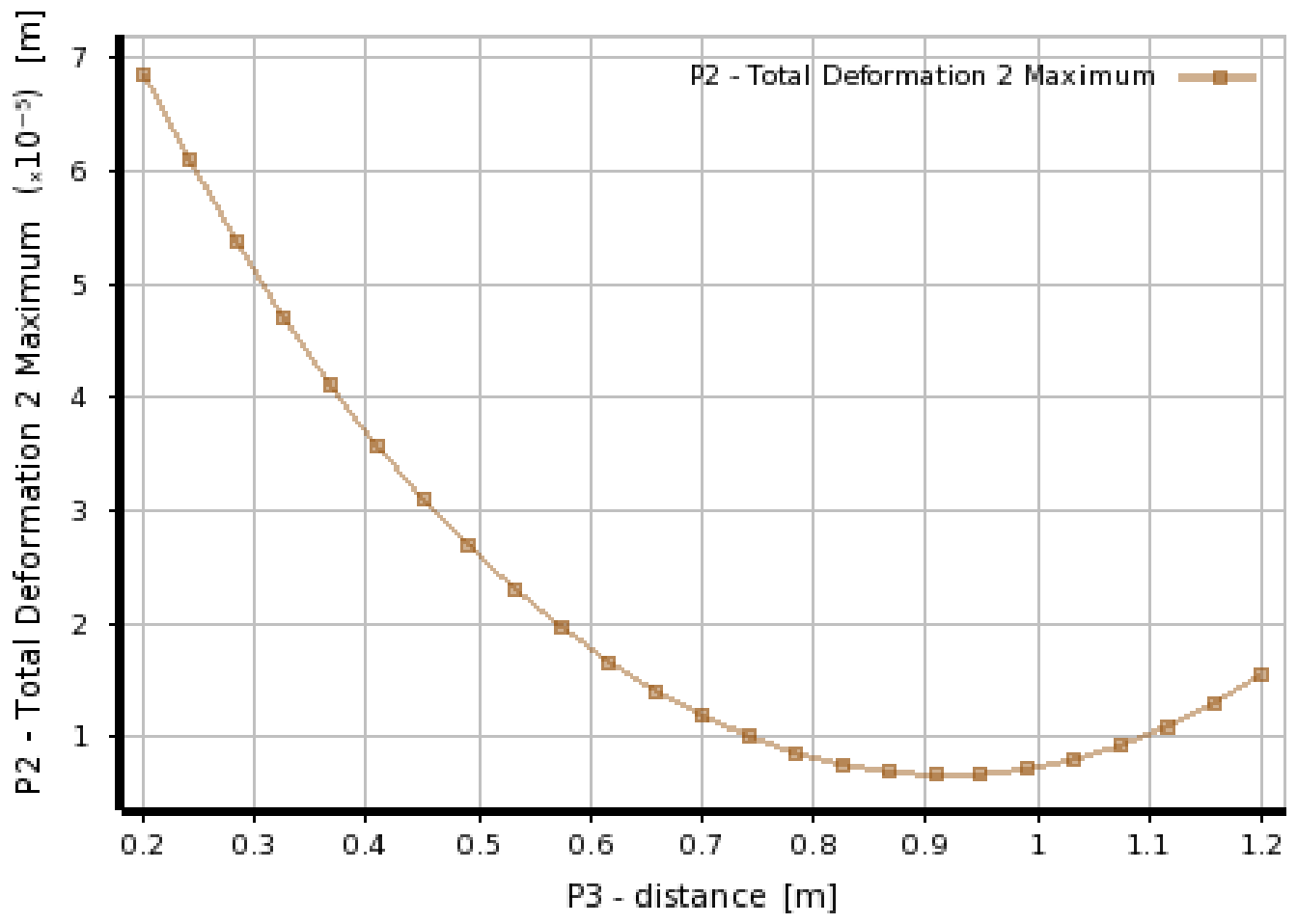

**Figure 4.3.10.33:** Location optimization of support rods

The optimization of materials and supports inside the cryostat has been carried out based on the CDR-designed SC magnets. The selected support material from the SC magnet to the helium vessel is the skeleton stainless steel [10]. After optimizing the supports in the cryostat, the total weight of the cryostat and the devices inside is 2790 kg.

The applied load for integral analyses includes both gravity and the Lorentz force Fz, as shown in Figure 4.3.10.34. The Lorentz force values are $Fz_1$=95 kN, $Fz_2$=81 kN, and $Fz_3$=92 kN, respectively. The results of uneven deformation in the Y direction and displacement in the Z direction are summarized in Table 4.3.10.7, considering both room temperature and a cryogenic temperature of 4.2 K. For the cryogenic case, constant physical properties are assumed for estimation, with plans to recalculate when detailed physical properties become available. For instance, taking Q1b as an example, the maximum vertical deformation is 31 μm, as depicted in Figure 4.3.10.35, and it can persist during temperature fluctuations. The results of modal analysis are shown in Table 4.3.10.8 and Figure 4.3.10.36. The first natural frequency is 14.5 Hz, which represents the integral vertical sway of the cryostat and devices inside. The second-order mode is similar but in the horizontal direction.

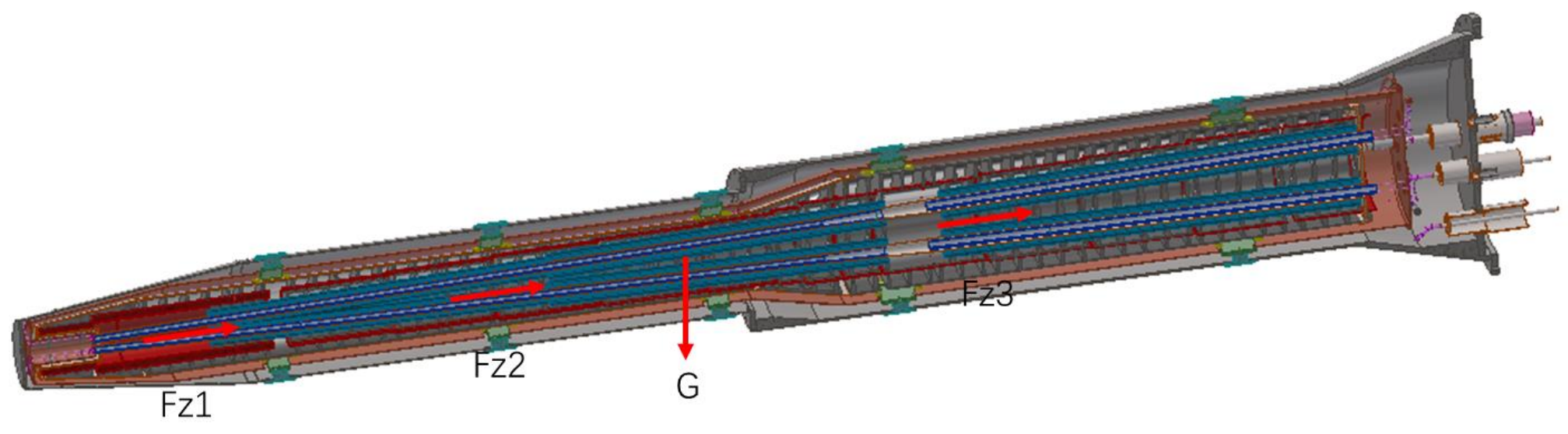

**Figure 4.3.10.34:** The applied forces of the cryostat for integral analyses.

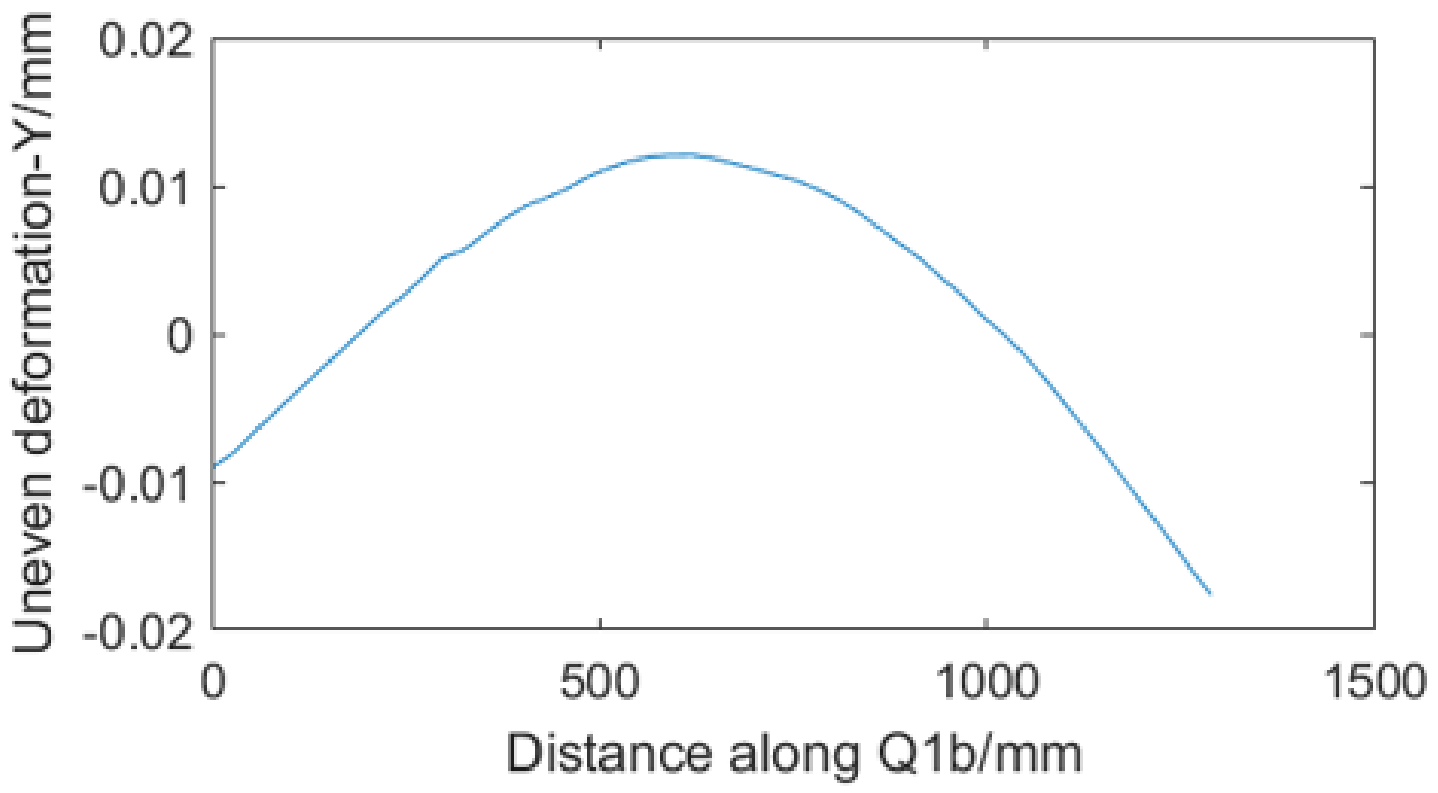

**Figure 4.3.10.35:** Uneven deformation in Y direction of Q1b

**Table 4.3.10.7:** Uneven deformation results.

| Temperature | Uneven deformation in Y direction (μm) | | | Uneven deformation in Z direction (mm) | | |
|---|---|---|---|---|---|---|
| | Q1a | Q1b | Q2 | Q1a | Q1b | Q2 |
| 295 K | 5 | 31 | 14 | 0.3 | 0.26 | 0.24 |
| 4.2 K | 7 | 30 | 12 | Contract with temperature | | |

**Table 4.3.10.8:** The results of modal analyses for the cryostat.

| No. of order | Frequency (Hz) | Mode |
|---|---|---|
| 1 | 14.5 | Integral vertical sway |
| 2 | 14.9 | Integral horizontal sway |
| 3 | 51.7 | Integral vertical wave |
| 4 | 53.7 | Integral horizontal wave |
| 5 | 89.7 | Vertical wave of helium vessel |
| 6 | 91.4 | Horizon wave of helium vessel |

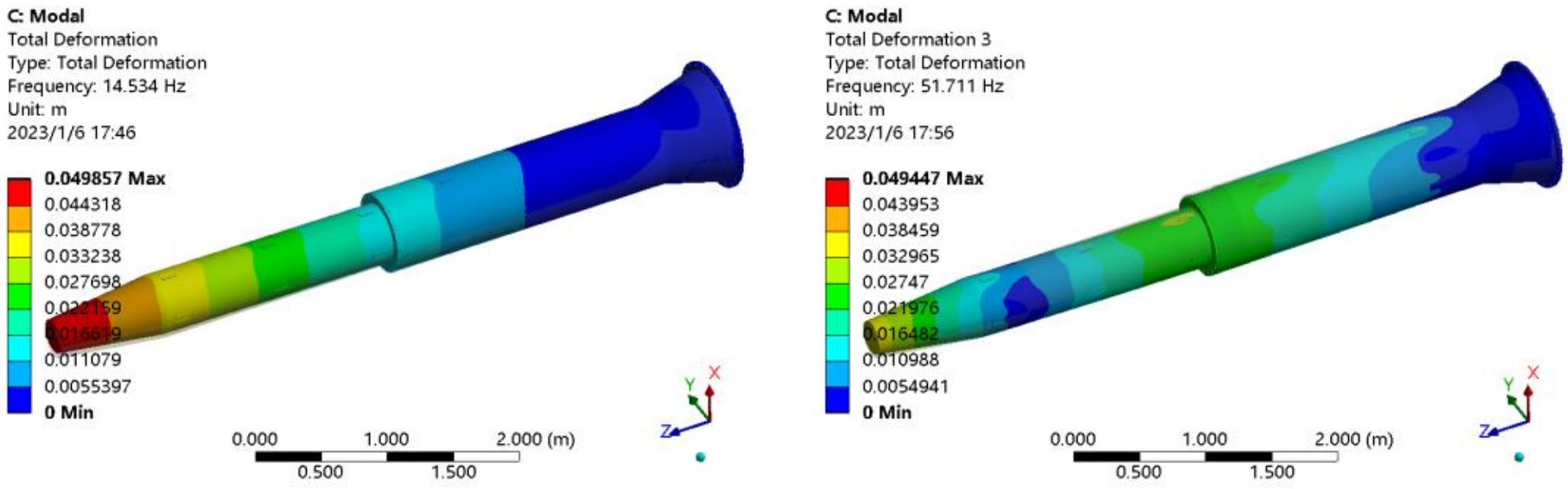

**Figure 4.3.10.36:** The 1$^{st}$ and 3$^{rd}$ modes of the cryostat

An auxiliary support can be introduced for the HCAL to enhance its stability. Figure 4.3.10.37 illustrates a V-groove auxiliary support with translation and rotation degrees of freedom along the longitudinal direction while constraining others. Analyses indicate that incorporating this support can substantially reduce displacement and increase the natural frequency [10]. For instance, when the auxiliary support is positioned 4,000 mm from the IP, the simulated 1$^{st}$ natural frequency of the cryostat can be improved from 14.5 Hz to 29 Hz. Further strategies for stabilization will be explored, including discussions on auxiliary support negotiation and the stiffness of kinematic joints within the support system. Prior to construction, an MDI mock-up will be fabricated to simulate weight distribution of all components and constraints, validating the results of modal analysis. Dynamic analyses will also be conducted once Lorenz force and the time structure of magnet quench become available.

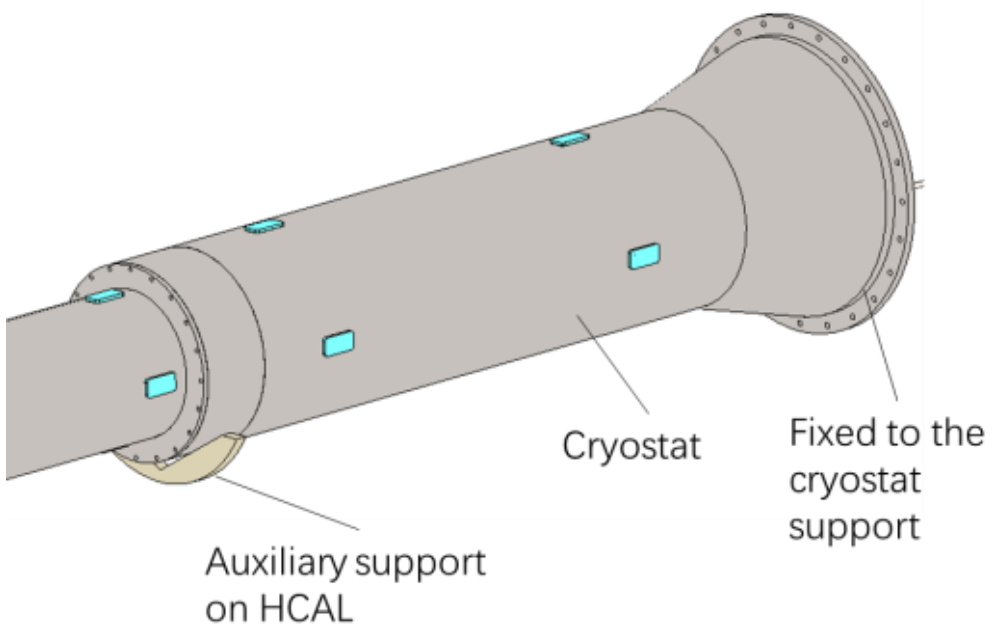

**Figure 4.3.10.37:** An auxiliary support of the cryostat.

### 4.3.11 **Beam Separation System (Electrostatic-Magnetic Separator)**

#### 4.3.11.1 *Introduction*

CEPC is a double-ring collider with two interaction points (IPs), as illustrated in Figure 2.2. The electron and positron beams circulate in opposite directions in the double ring, and the beam energy will reach 120 GeV in the H mode. The electrostatic-magnetic separator is a crucial component required by CEPC to separate electron and positron beams.

### 4.3.11.2 *Electrostatic-Magnetic Separator*

In the RF region, both rings of CEPC share the RF cavities. Each RF station has two sections to bypass half of the cavities when running in W or Z modes. An electrostatic separator, combined with a dipole magnet as shown in Fig. 4.3.11.1, avoids bending the incoming beam.

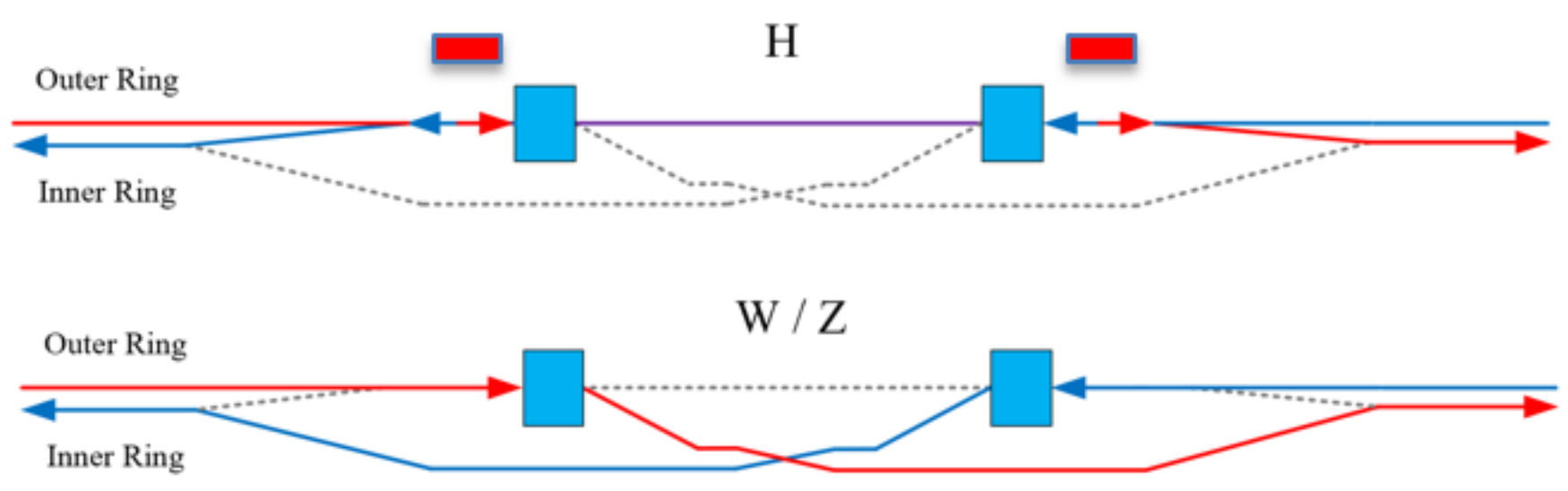

**Figure 4.3.11.1:** Layout of the RF region.

Electrostatic-magnetic separators utilize both electric and magnetic fields to deflect beams in the RF regions. A separator of this type consists of an electrostatic separator and a dipole magnet. In the CEPC, 32 separators are required and are located symmetrically at both ends of the RF regions. Following a set of separators, the deflection distance is approximately 21.2 mm. After this, the beam passes through a 75 m drift to ensure a separating distance of 10 cm between the two beams at the entrance of the quadrupoles. During Higgs mode operation, all RF cavities are shared by the electron and positron beams. For the W and Z modes, the separators in the RF region are turned off so that all bunches can be filled around the entire electron and positron rings. The layout of the RF region, as shown in Fig. 4.3.11.1, depicts the electrostatic-magnetic separators with pink square parts.

There were three schemes for the horizontal deflector:

i. The electrodes and coil are situated within the vacuum chamber, but this arrangement is highly intricate and poses various challenges, such as feedthrough design and high-voltage insulation.
ii. Symmetrically installed at both ends of the electrostatic separator is a pair of magnets. While this structure is straightforward, it introduces additional synchrotron radiation that will strike the RF cavities.
iii. The electrostatic separator is located within the magnet, as depicted in Fig. 4.3.11.2. This configuration is relatively easy to design and construc.

According to the analysis, we decided to choose scheme (iii).

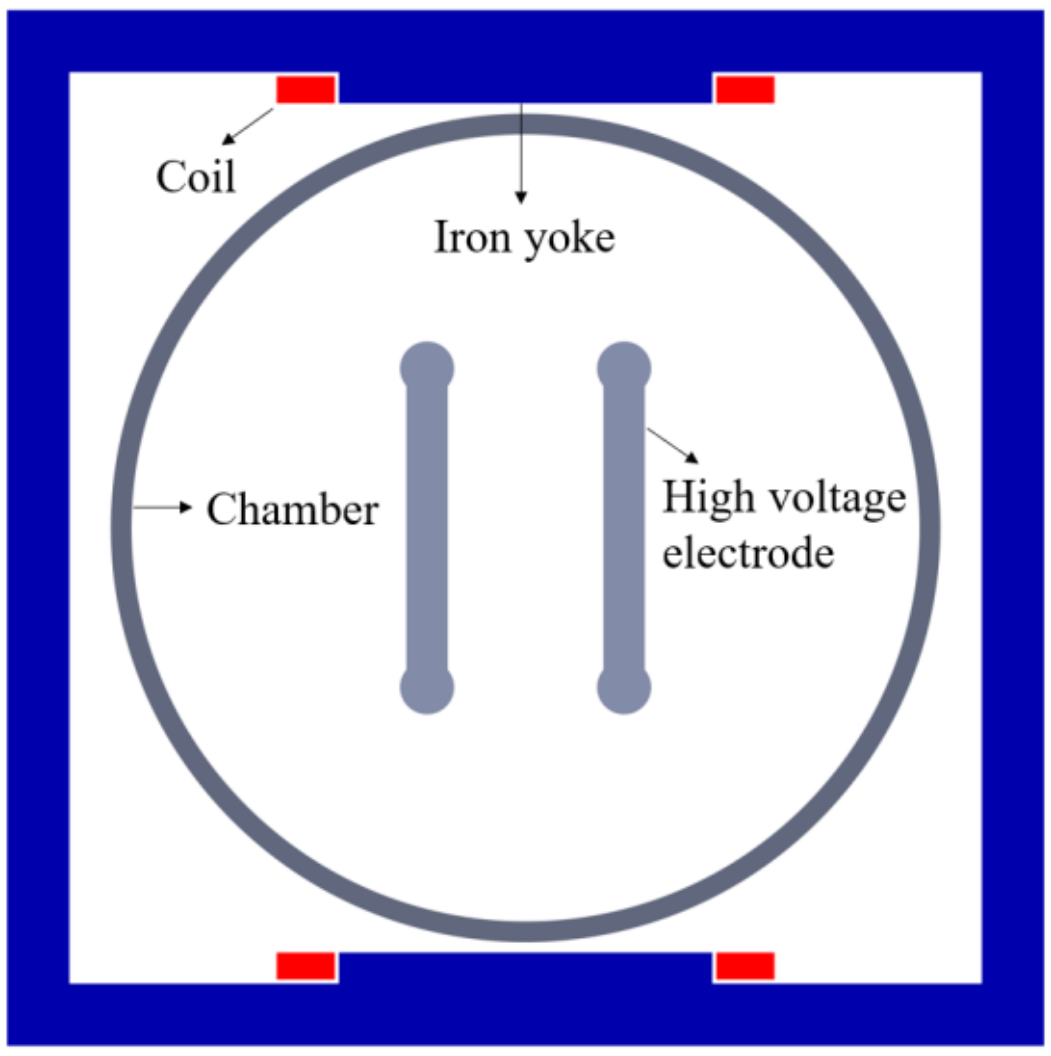

**Figure 4.3.11.2:** Scheme (iii) – The electrostatic separator is placed inside the magnet.

The electric force and magnetic force are equal in magnitude but opposite in direction for the incoming beam. Therefore, the deflecting force Fi, which is the sum of FB and FE, is zero as FB = −FE. However, for the outgoing beam, the electric and magnetic forces are equal in magnitude and direction. Thus, the deflection force Fi is equal to 2FE, where FE is defined as $e{\cdot}E_{\perp}$ and FB is defined as $ec\beta B_{\perp}$, with c being the speed of light, $\beta$ being the relativistic factor ($\beta$ = v/c), $E_{\perp}$ being the electric field, and $B_{\perp}$ being the magnetic field. Figure 4.3.11.3 illustrates the principle of the electrostatic-magnetic separator, while Table 4.3.11.1 presents the parameters of the electrostatic-magnetic separator.

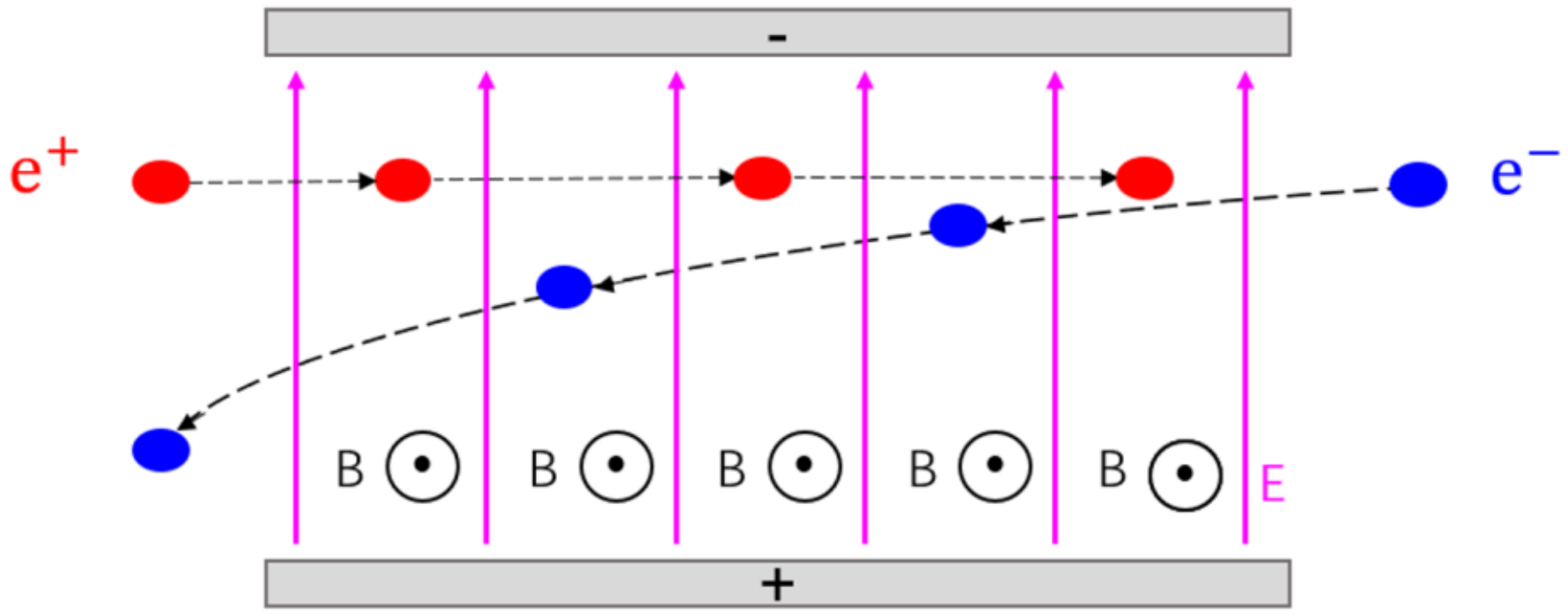

**Figure 4.3.11.3:** Principle of electrostatic-magnetic separaor.

**Table 4.3.11.1:** Parameters of the components of electrostatic-magnetic separator.

| Components | Electrostatic separator | Dipole magnet |
|---|---|---|
| Field strength | 2.0 MV/m | 66.7 Gauss |
| Effective length | 4 m | 4 m |
| Good field region | 46 mm ×11 mm | 46 mm ×11 mm |
| Field homogeneity | ±0.05% | ±0.05% |

### 4.3.11.3 *Electrostatic Separator*

To achieve the desired deflecting angle and deflecting distance, the electric field must reach a minimum of 2 MV/m. The electrostatic separator system consists of two parallel electrodes, a vacuum chamber, insulation support, cooling system, power supply, and ion pump. Each electrode is powered by an equal but opposite high voltage. The effective length of each electrostatic separator is 4 meters.

The vacuum chamber is a stainless-steel tube that is 4.5 meters long with an internal diameter of 380 mm. Titanium is the chosen material for the electrodes due to its good performance in vacuum and mechanical settings, as well as its lower density compared to stainless steel. The support for the electrodes will be manufactured using metal ceramic. All materials are chosen to maintain a good vacuum level of the order of $10^{-10}$ Torr.

Table 4.3.11.2 displays the parameters of the electrostatic separator. The deflecting angle for each electrostatic separator can be calculated using the following formula:

$$\theta \approx \tan\theta = \frac{\boldsymbol{E} \cdot L}{E_0} \tag{4.3.11.1}$$

where $E$, $E_0$, and $L$ represent the electric field, beam energy, and effective length, respectively, with values of 2 MV/m, 120 GeV, and 4 m. The resulting deflecting angle for each electrostatic separator will be approximately 0.067 mrad.

**Table 4.3.11.2:** Parameters of the electrostatic separator.

| | |
|---|---|
| Electrostatic separator length | 4.5 m |
| Inner diameter of the chamber | 380 mm |
| Electrode length | 4.0 m |
| Gap | 75 mm |
| Electric field strength | 2 MV/m |
| Operating voltage | ±75 kV |
| Maximum voltage | ±135 kV |
| Vacuum pressure | $2 \times 10^{-10}$ Torr |

The key challenges include: (i) achieving an electric field homogeneity of ±0.05% in the 46 mm × 11 mm region; (ii) ensuring the electric field matches with the magnetic field to minimize synchrotron radiation; and (iii) achieving a very low beam impedance.

### 4.3.11.4 *Electric Field Homogeneity Optimization*

The electric field homogeneity and maximum field in the aperture are determined by the geometric cross section of the electrostatic separator. After conducting an optimization study of several electrode shapes using Opera, including two parallel electrodes, curved electrodes, and four discontinuous electrodes, the parallel electrodes were found to have the best field homogeneity among the three shapes. While the curved electrodes were difficult to manufacture and showed poor field homogeneity, the four discontinuous electrodes allowed for lower impedance but poor field homogeneity. However, even the ordinary parallel electrodes could not fully meet the required field homogeneity, necessitating modifications to the electrode edge shapes. Fig. 4.3.11.4 shows the four types of electrode cross sections used for the separator.

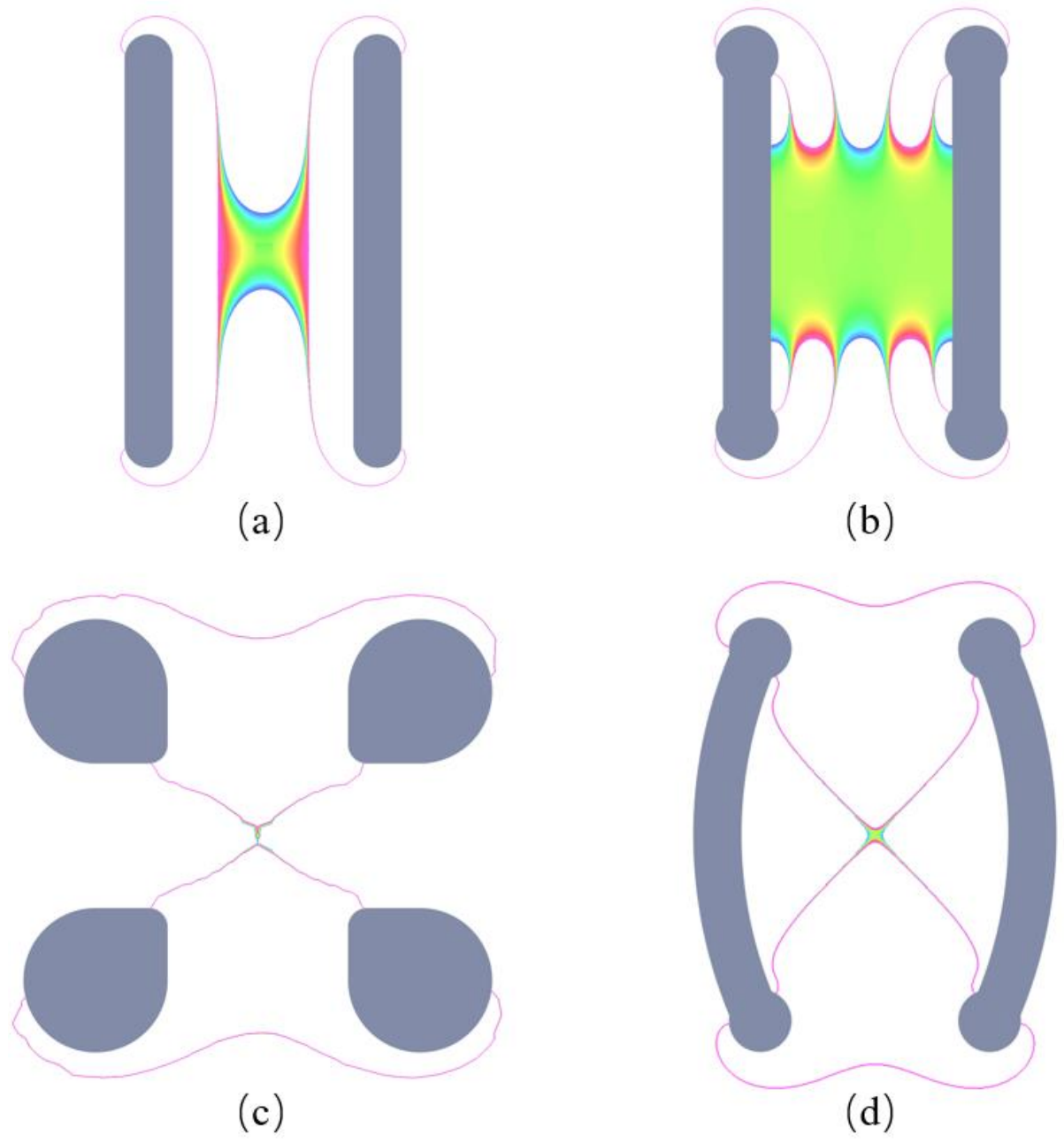

**Figure 4.3.11.4:** Good field region of four kinds of electrode shapes: (a) two parallel electrodes, (b) modified parallel electrodes, (c) four discontinuous electrodes, and (d) curved electrodes

The modified parallel electrodes (Fig. 4.3.11.4 (b)) were ultimately chosen over the other electrode shapes to improve field homogeneity. To decrease beam impedance, ground electrodes were added, but this had a negative effect on electric field homogeneity. Simulations were performed with varying lengths of ground electrodes, and the results are shown in Fig. 4.3.11.5. The good field region decreased as ground electrode length increased, with a maximum length of 100 mm recommended to minimize the effect on field homogeneity. The final geometric cross section of the electrostatic separator is depicted in Fig. 4.3.11.6.

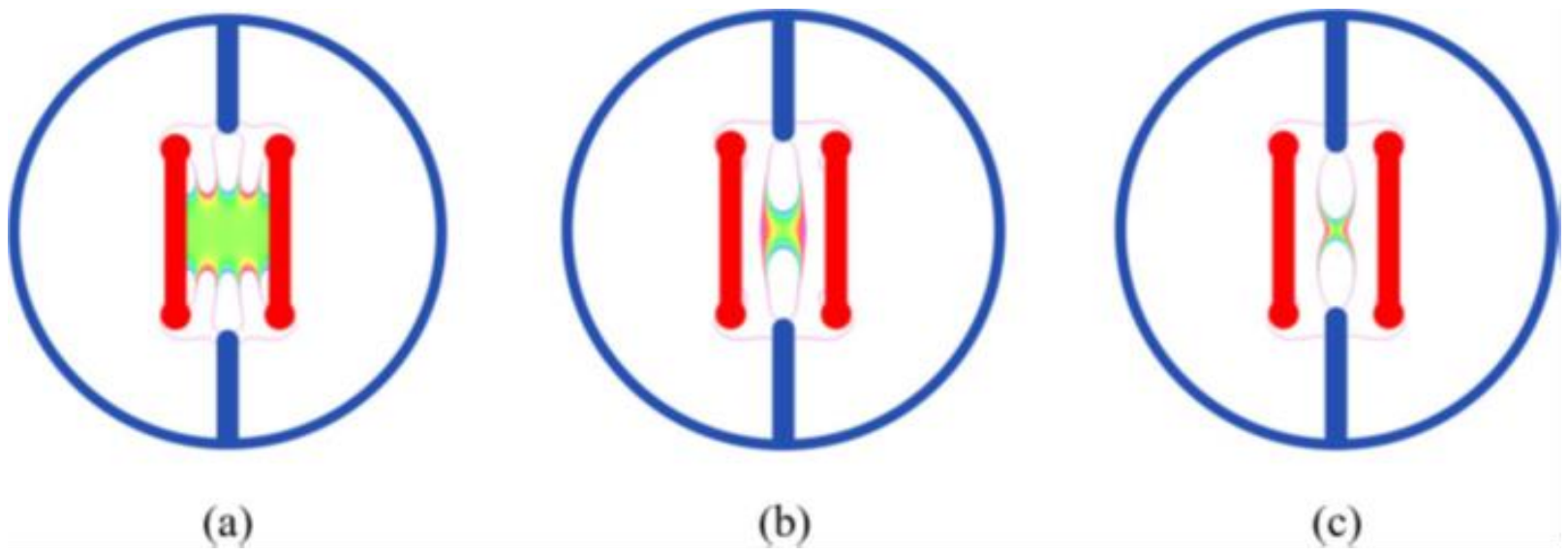

**Figure 4.3.11.5:** Good field region for modified parallel electrodes with three ground electrodes' lengths: (a) 100 mm, (b) 110 mm, (c) 120 mm.

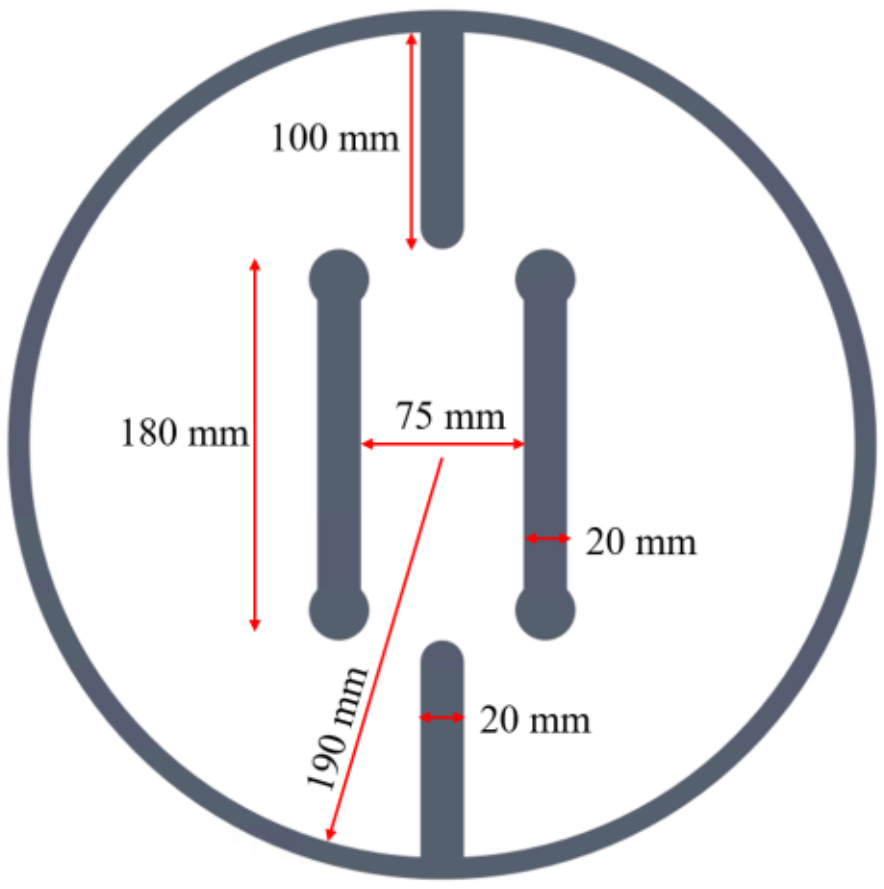

**Figure 4.3.11.6:** Geometric cross section of electrostatic separator.

Additionally, the longitudinally integrated field homogeneity in the region of 46 mm × 40 mm is better than 0.05%, as demonstrated in Fig. 4.3.11.7. Fig. 4.3.11.8 (a) and (b) exhibit the electric field distribution on horizontal and vertical planes, respectively, which indicates that the electric field is uniformly distributed inside the separator. The four "peaks" in Fig. 4.3.11.8 (a) represent the electric field distortion caused by the fringe of electrodes.

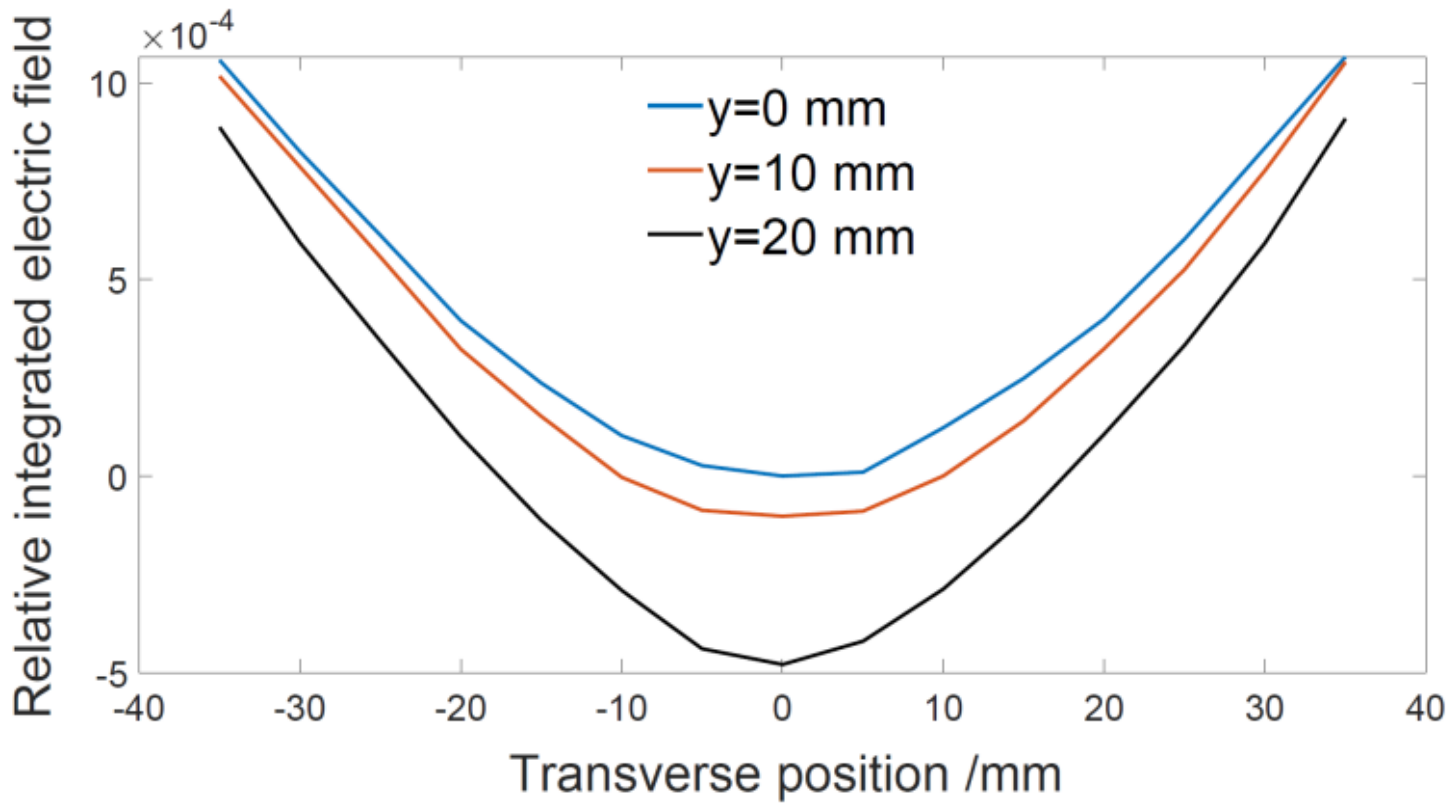

**Figure 4.3.11.7:** Calculated integrated electric field homogeneity with 100-mm ground electrodes.

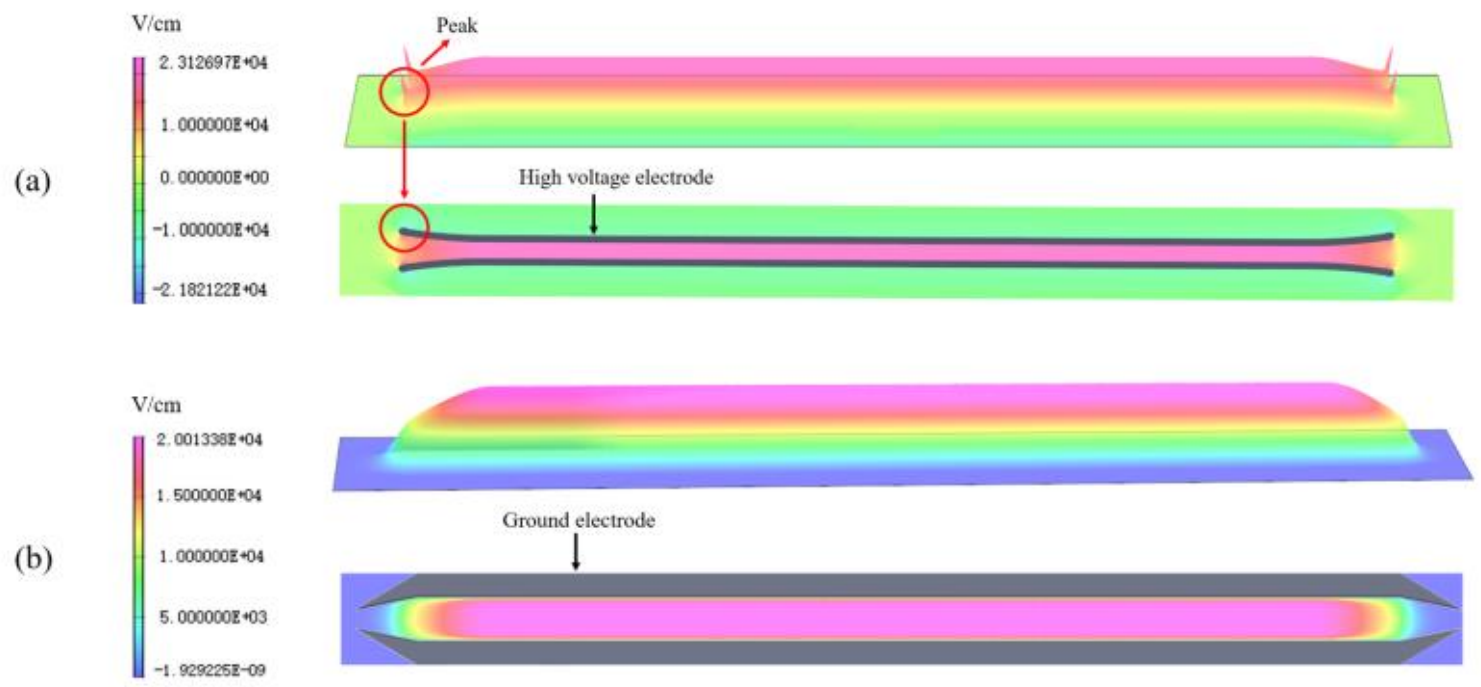

**Figure 4.3.11.8:** Electric field ($E_x$) distribution in (a) horizontal and (b) vertical planes.

From Fig. 4.3.11.9, the vertical ($E_y$) and longitudinal ($E_z$) components of the electric field are shown. The maximum values of $E_y$ and $E_z$ (around 15/30 V/cm) are about three orders of magnitude smaller than that of $E_x$ ($2 \times 10^4$ V/cm, as seen in Fig. 4.3.11.8). Furthermore, Ez has an effect on the beam motion in the z direction, which does not require consideration. The integral of $E_y/E_x$ along the z direction is almost zero ($1.5 \times 10^{-5}$), indicating that $E_y$ has little effect on the beam's motion.

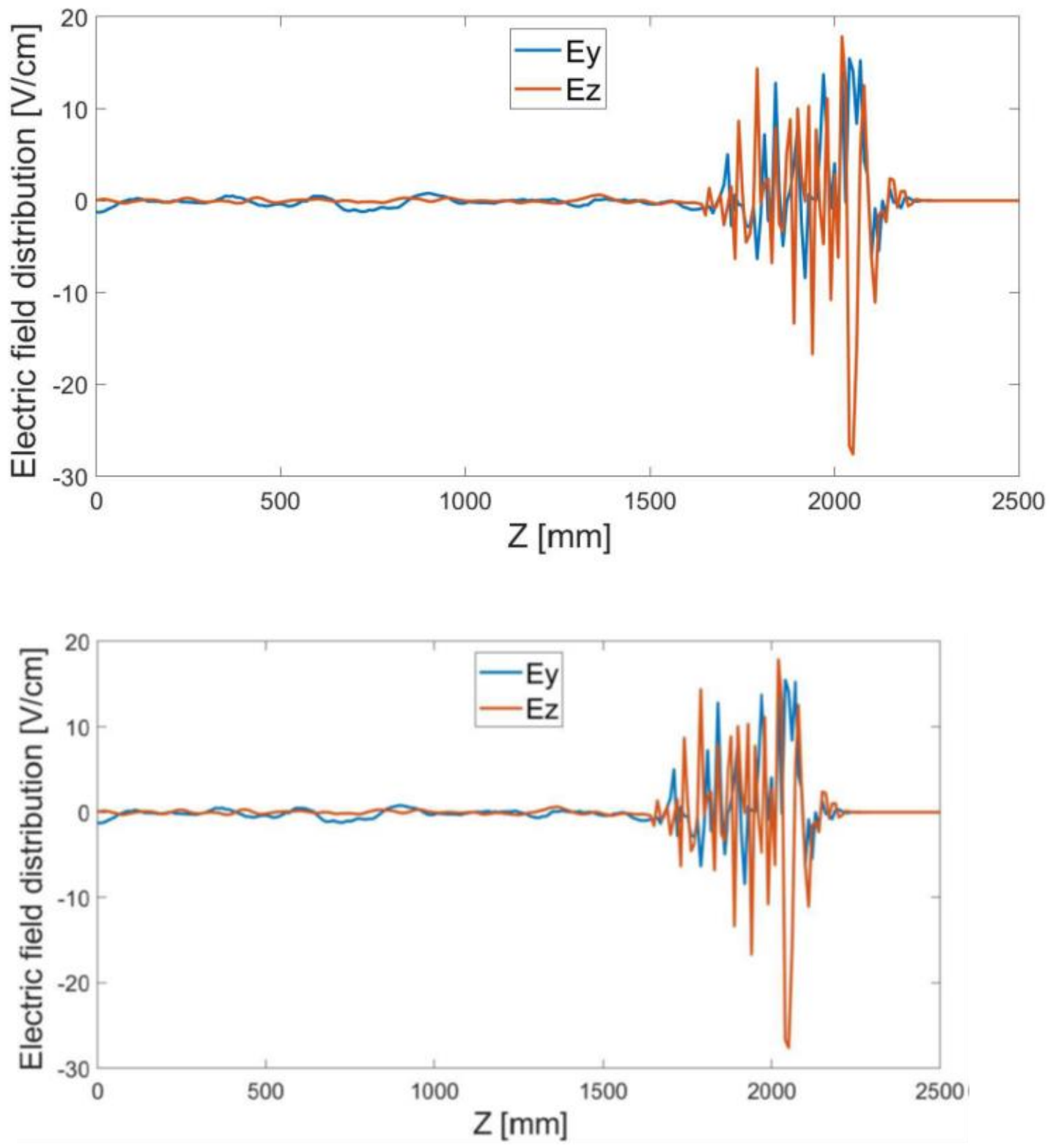

**Figure 4.3.11.9:** Electric field distribution ($E_y$ and $E_z$) in the longitudinal direction of the separator

### 4.3.11.5 *Maximum Electric Field*

Sparking is a significant issue that can result in beam loss and decreased vacuum quality. One common cause of sparking is failing cables or connectors. To minimize the occurrence of sparking, it is important to limit the maximum field strength at the edges and ends of electrodess.

The Kilpatrick limit is used to determine the minimum acceptable distance between the ground electrodes and high-voltage electrodes. It is also necessary to optimize the length of the ground electrode to prevent electric breakdown between the electrodes and vacuum chamber. According to Kilpatrick's criterion, the limit electric field strength $E$ (in V/cm) is calculated using themaximum gap voltage $W$ (in V):

$$WE^2 exp(-1.7 \times 10^5/E) = 1.8 \times 10^{14} \quad (4.3.11.2)$$

For example, when $W$ = 270 kV, the limit electric field is 7.7 MV/m. In simulations, the ground electrode length was varied (100, 110, and 120 mm) with a high voltage of ± 135 kV, and the maximum field and length were limited to less than 7 MV/m and 100 mm, respectively.

### 4.3.11.6 ***Dipole Magnet***

The H-type magnet yoke was chosen for its better field integrals homogeneity and ease of installation with the vacuum system. The center magnetic field needs to reach 66.7 Gauss according to the Lorentz force equation, and the magnet aperture size is 600 mm due to the dimensions of the inner electrostatic separator. The integrated magnetic field homogeneity within a 60 mm × 110 mm area is better than 0.02%, as demonstrated in Fig. 4.3.11.10. The 2D model of the dipole magnet can be seen in Fig. 4.3.11.11.

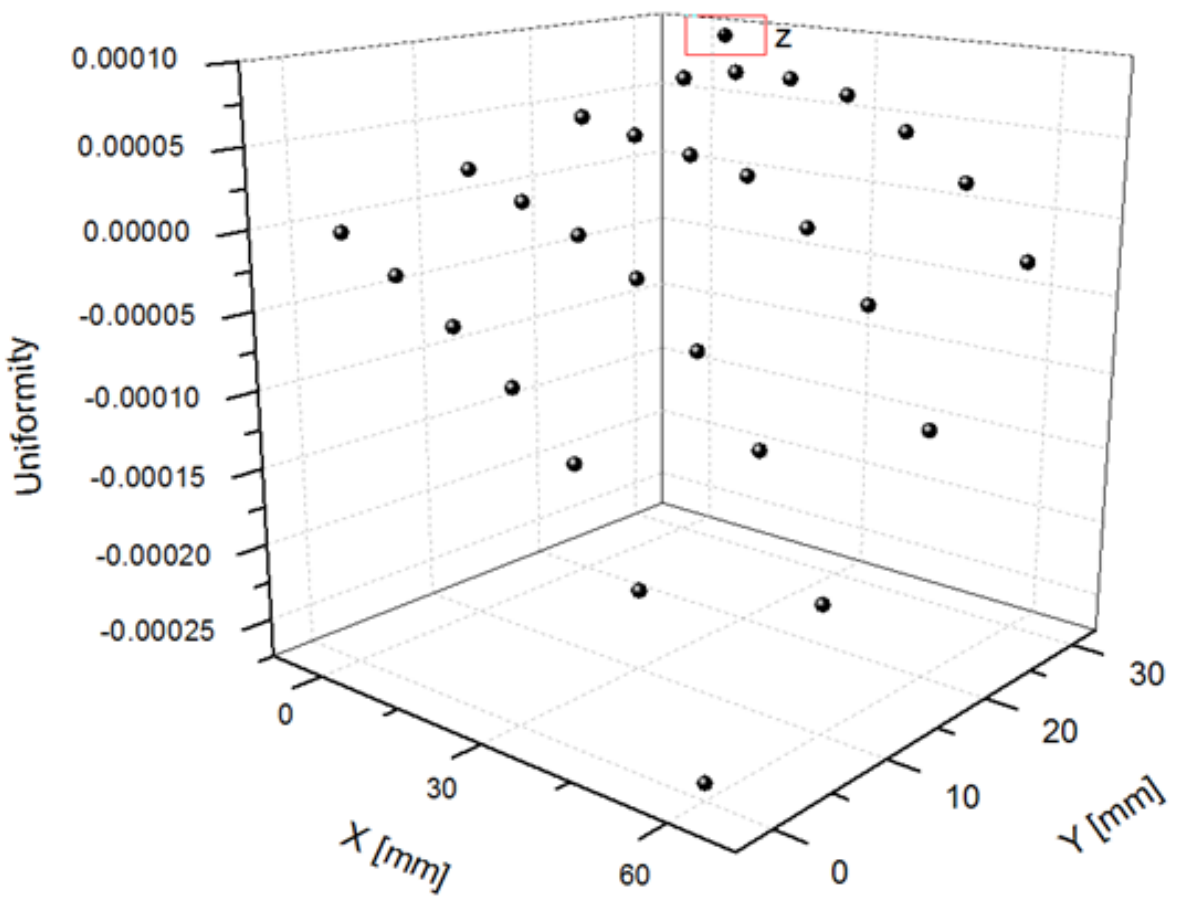

**Figure 4.3.11.10:** Calculated integrated magnetic field homogeneity.

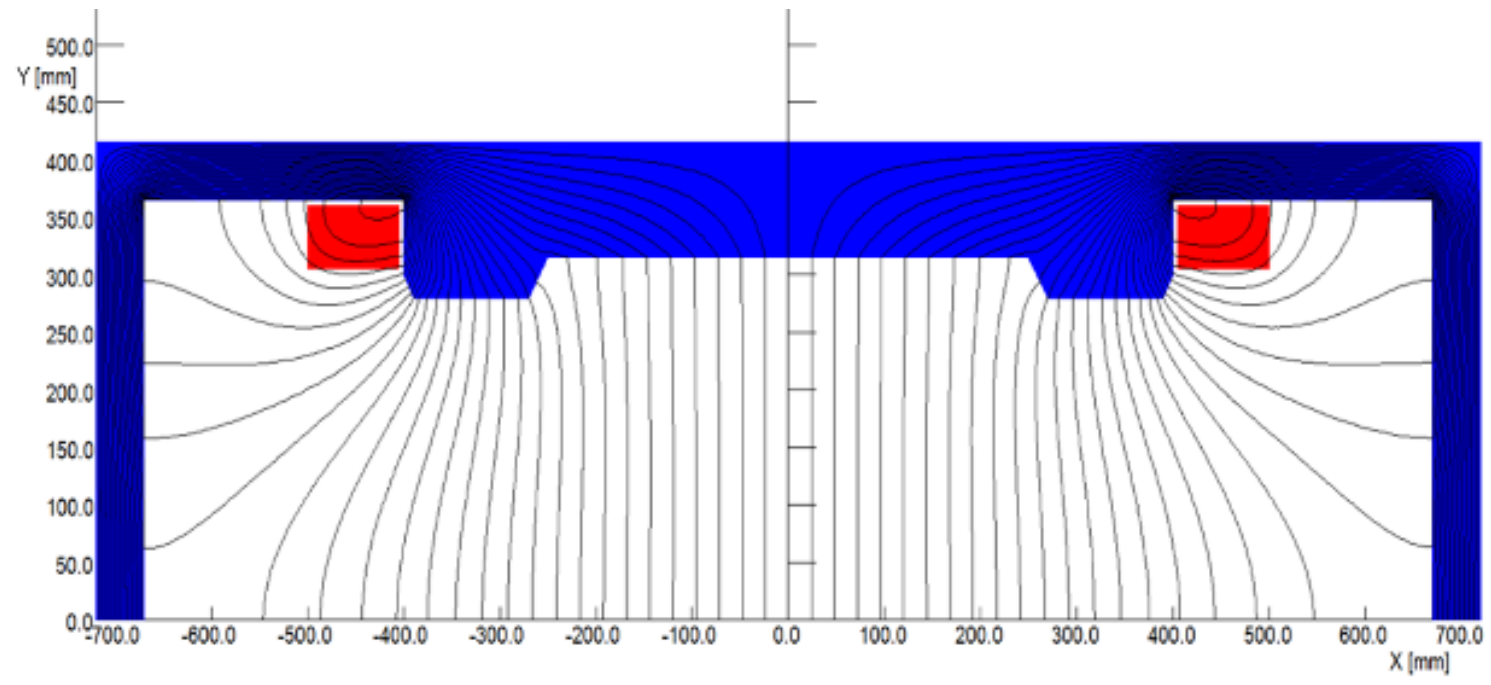

**Figure 4.3.11.11:** 2D model of dipole magnet

The horizontal ($B_x$) and longitudinal ($B_z$) components of the magnetic field can be seen in Fig. 4.3.11.12. As B*z* has the same direction as the beam motion, we do not need to consider its effect. The integral of $B_x$ and B*y* along the z direction is almost zero ($5 \times 10^{-5}$), indicating that $B_x$ has minimal impact on the beam motion.

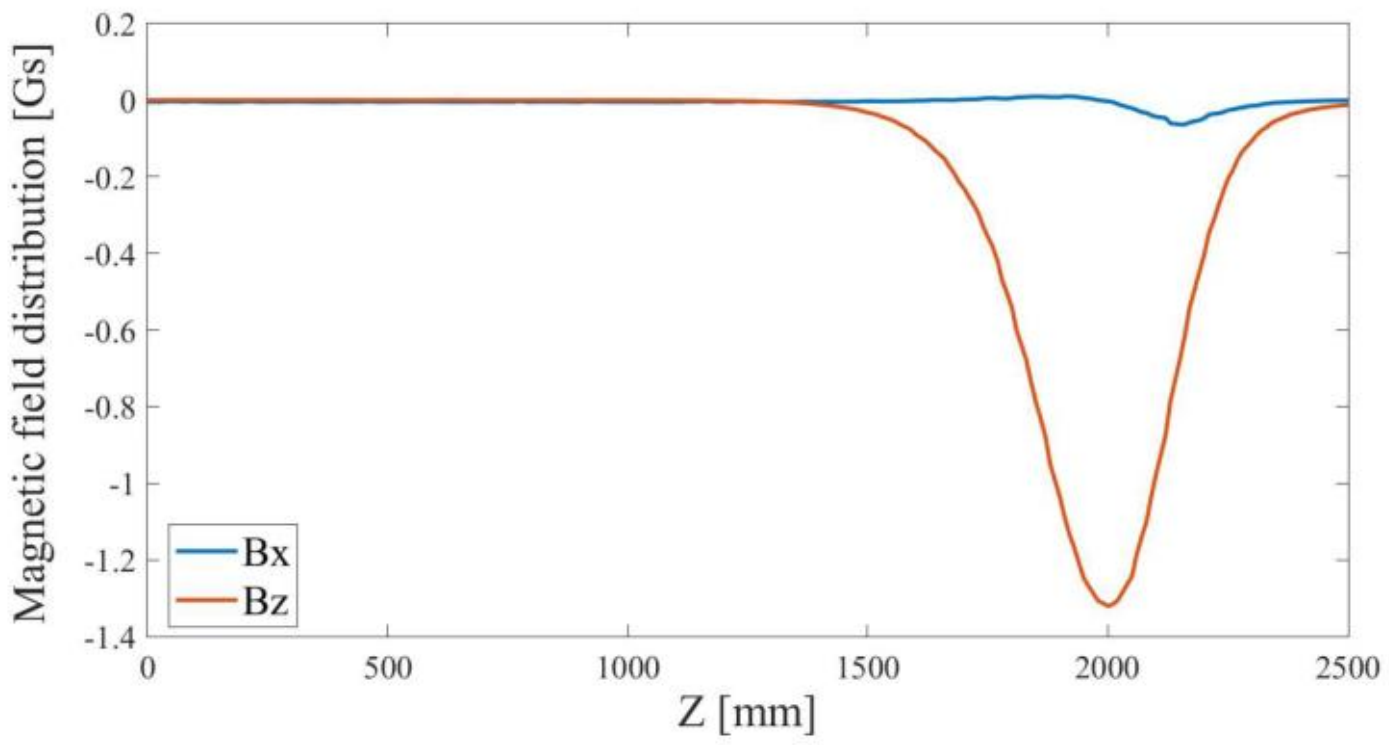

**Figure 4.3.11.12:** Magnetic field distribution ($B_x$ and $B_z$) in the longitudinal direction of the separator.

#### 4.3.11.7 *Field Matching*

The beam experiences both electric and magnetic fields upon entering the separator, and if the electric field does not match the magnetic field, synchrotron radiation from the separation region can damage the RF cavities. To reduce synchrotron radiation and ensure smooth beam passage, the electric and magnetic fields must be perfectly matched, especially at the separator ends. The difference ratio between electric and magnetic forces should not exceed 10%, and their integrals must be equal to prevent the synchrotron from affecting the RF cavities.

The ratio of *E/B*, which represents the electric and magnetic fields, should remain constant throughout the entire length. However, achieving this can be challenging because the magnetic gap is twice as large as the electrode gap. The separator employs two parallel electrodes and a dipole magnet to generate orthogonal electric and magnetic fields. The Lorentz force is defined by the following equation:

$$\boldsymbol{F_L} = q(\boldsymbol{E} + \boldsymbol{V} \times \boldsymbol{B}) \tag{4.3.11.3}$$

in which $q$ refers to the charge of the particle, $V$ represents the velocity vector, and $E$ and $B$ denote the components of the electric and magnetic fields, respectively. In order to achieve $F_L = 0$ for the incoming beam, we aim for $E = -V \times B$. However, the tapering ends of the separator cause the electric field to drop faster than the magnetic field, as observed in Fig. 4.3.11.13. The solution to this issue is to decrease the derivative of the electric field decline while increasing the derivative of the magnetic field decline. To optimize the magnetic field, field clamps and mirror plates are introduced to obtain a more "hard-edge" field distribution. This results in a denser magnetic flux density on the magnetic clamps and mirror plate than that of the core, which plays a crucial role in reducing the magnetic flux divergence at the end and regulating the homogeneity of the integral field. On the other hand, to optimize the electric field, the ends of the electrodes are bent, as shown in Fig. 4.3.11.14, resulting in a perfect match between the electric and magnetic fields. The final optimized result is presented in Fig. 4.3.11.15.

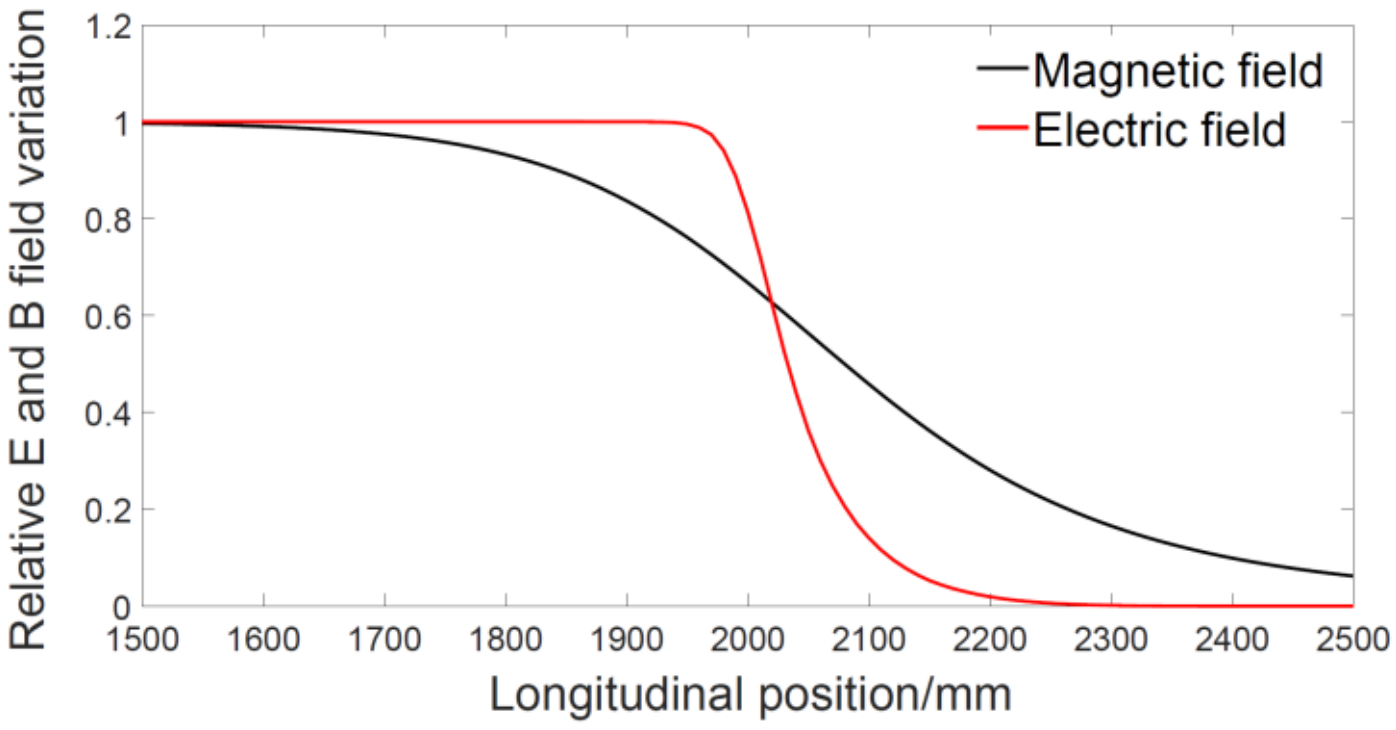

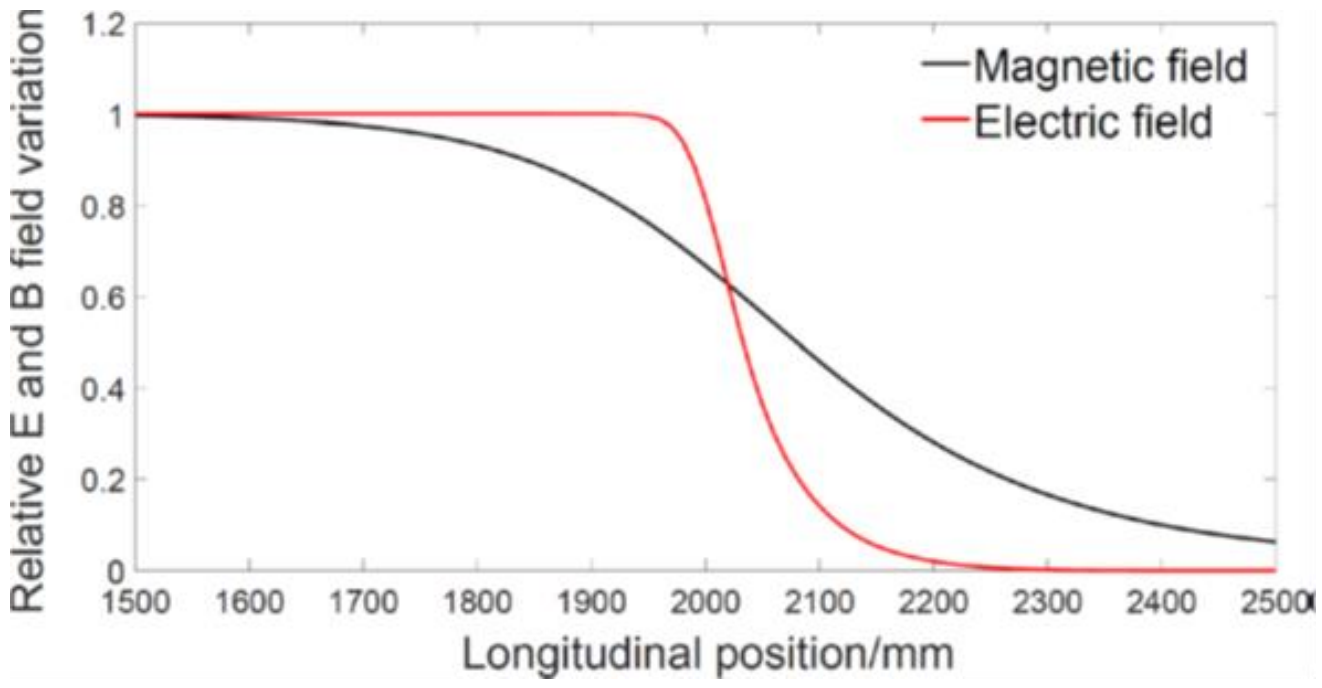

**Figure 4.3.11.13:** The normalized longitudinal E and B field profiles before optimization.

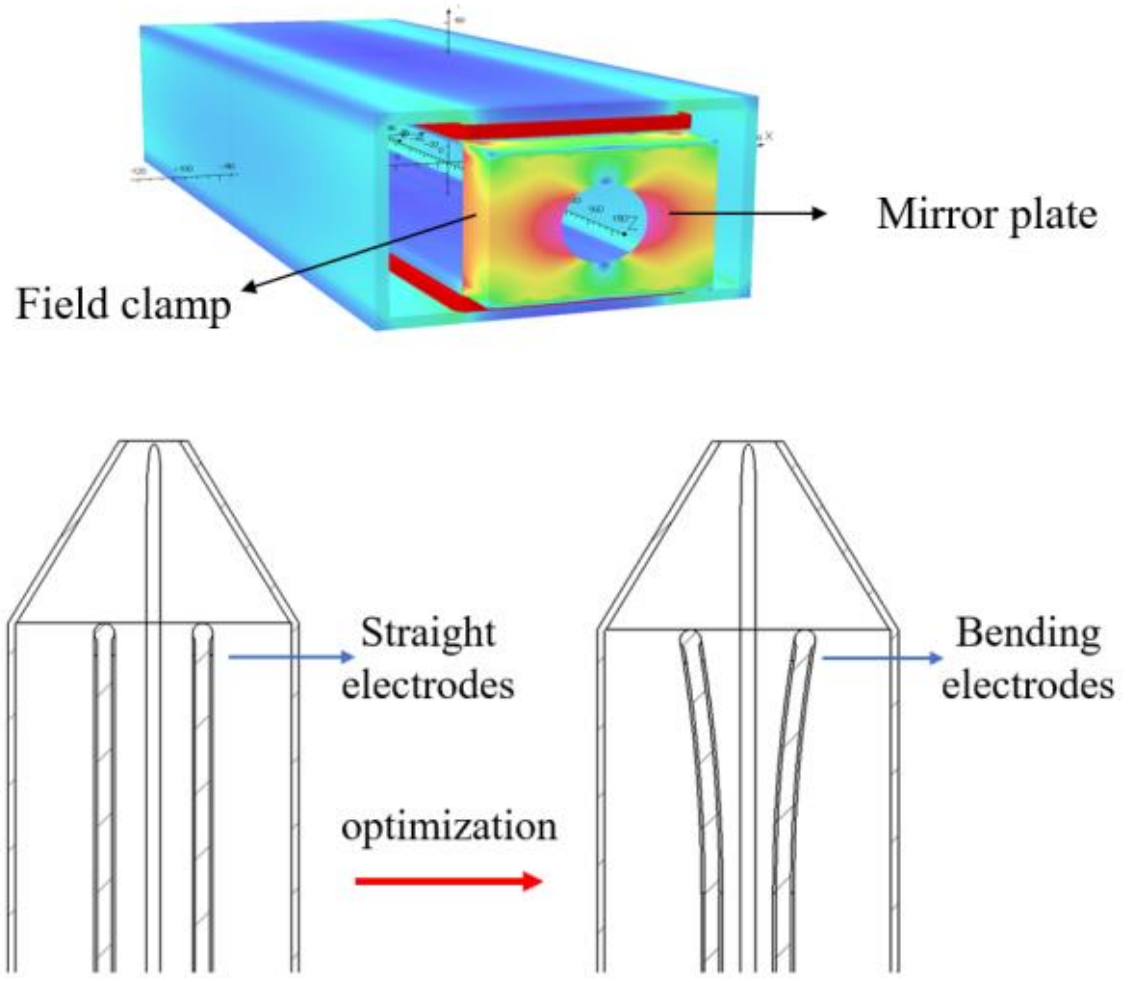

**Figure 4.3.11.14:** Optimization of magnetic field (top) and electric field (bottom)

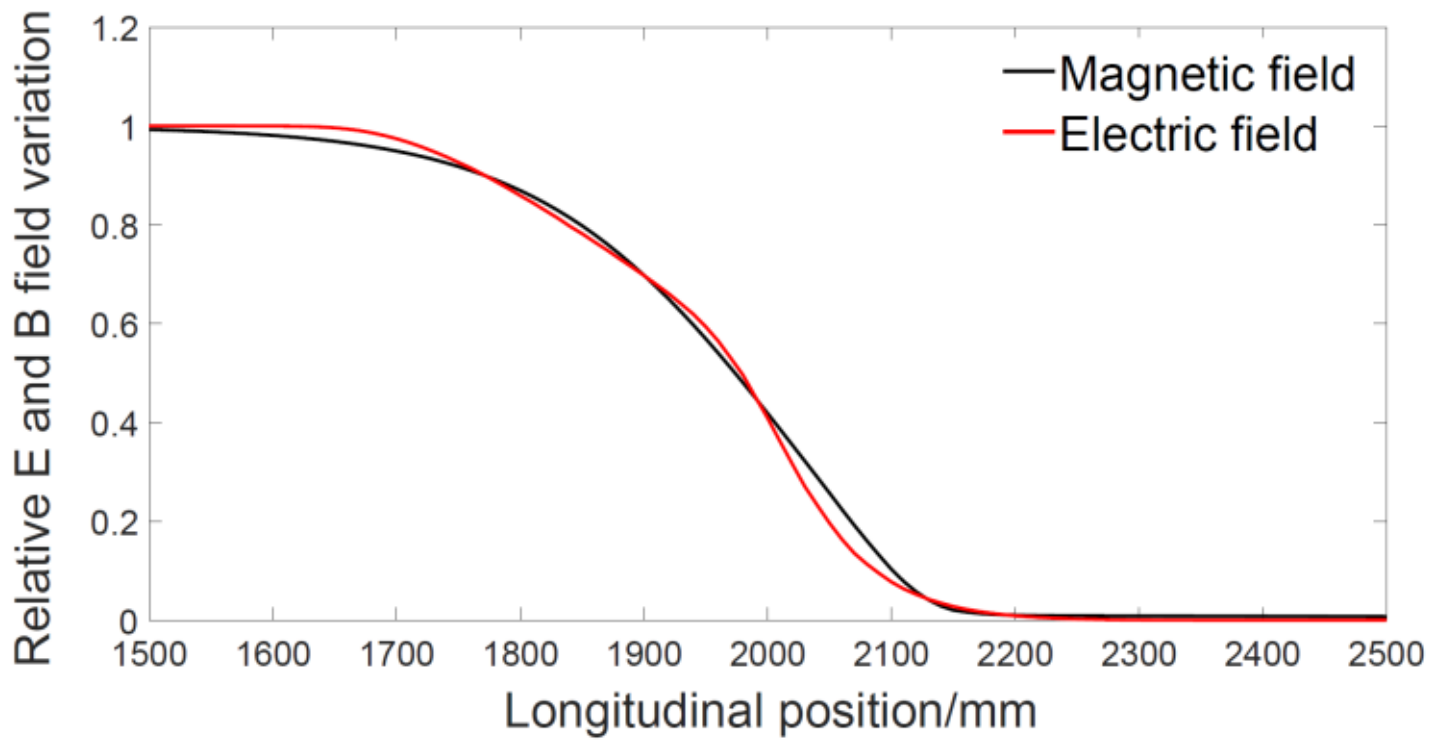

**Figure 4.3.11.15:** The normalized longitudinal E and B field profiles after optimization

### 4.3.11.8 *Beam Impedance*

The RF electromagnetic fields generated by the charged particles traveling through the electrostatic separator create the wake field. The characteristics of the wake field depend on the structure of the separator and can cause the temperature of the vacuum chamber and electrodes to rise due to energy absorption. Therefore, it is essential to maintain a low beam impedance for the separator.

The wake field generated at the discontinuous structures of the separator can impede the smooth passage of beam image current through it. To minimize this impedance, we can draw from the experiences of CESR and BEPC-II, which introduced ground electrodes and tapered ends between the vacuum chamber and the beam pipe. By implementing these measures, we can reduce the loss factor of the separators and ensure a smoother beam current passage.

To reduce energy loss, the ground electrodes with zero potential should have the same length as the separator itself. This ensures that the beam image current travels alongside the ground electrodes, minimizing energy loss. While longer ground electrodes reduce impedance, their proximity to high-voltage electrodes in small apertures can create regions with excessively high electric fields, leading to poor electric field homogeneity. Therefore, a length of 100 mm was chosen as the optimal compromise. Additionally, implementing tapering ends at both sides of the separator can further reduce impedance.

#### 4.3.11.9 *Loss Factor*

The loss factor is a measure of the average electromagnetic energy left behind by a bunch, divided by the square of the bunch charge. To investigate the effect of reducing impedance by incorporating ground electrodes and tapering ends, simulations were conducted with and without these components. The primary results are shown in Table 4.3.11.3.

The beam energy loss can be calculated using the formula:

$$P_{loss} = k n_b f_r {I_b}^2 \tag{4.3.11.4}$$

where $k$ represents the loss factor, which can be computed using CST; $n_b$ is the number of bunches; $fr$ is the revolution frequency of the beam; and $I_b$ is the beam current. Based on the design parameters of the CEPC, a loss of 495 W can be obtained when $k$ = 1.170 in H mode.

The results show that the introduction of ground electrodes and tapering ends can reduce the loss factor $k$ by approximately 20%. This demonstrates that incorporating these components is an effective method for reducing the impedance.

**Table 4.3.11.3:** Loss factor calculated results.

| Description | K [V/pC] |
|---|---|
| Partially tapering + g.e. | 1.394 |
| 3D tapering no g.e. | 1.256 |
| 3D tapering + g.e. | 1.170 |

#### 4.3.11.10 *Mechanical Design*

The mechanical design has been completed, and the entire model can be observed in Fig. 4.3.11.16. The model consists of a dipole magnet, an electrostatic separator, ion pumps, and a shelf. To ensure machining and installation accuracy, the ground electrodes and vacuum chamber of the separator are divided into several sections. The vacuum chamber underwent electrochemical polishing treatment and high-temperature vacuum degassing treatment to achieve ultra-high vacuum. The high-voltage electrode is also divided into three sections, including two bending sections and one straight section, due to its large size. To support the electrodes, insulation supports made of 95% alumina ceramics were used, and they can be fine-tuned to ensure installation accuracy and prevent electrode deformation. Each electrode plate has two feedthrough devices for providing high voltage and cooling liquid. The HV electrodes are cooled using cooling fluid FC-77, and the cooling loop's sketch can be seen in Fig. 4.3.11.16. Bellows were used as soft connections for easy installation and position adjustment. The ceramic tube's interior contains the feedthrough cable and cooling liquid pipe, and it is filled with an atmosphere. The prototype's construction has already been completed.

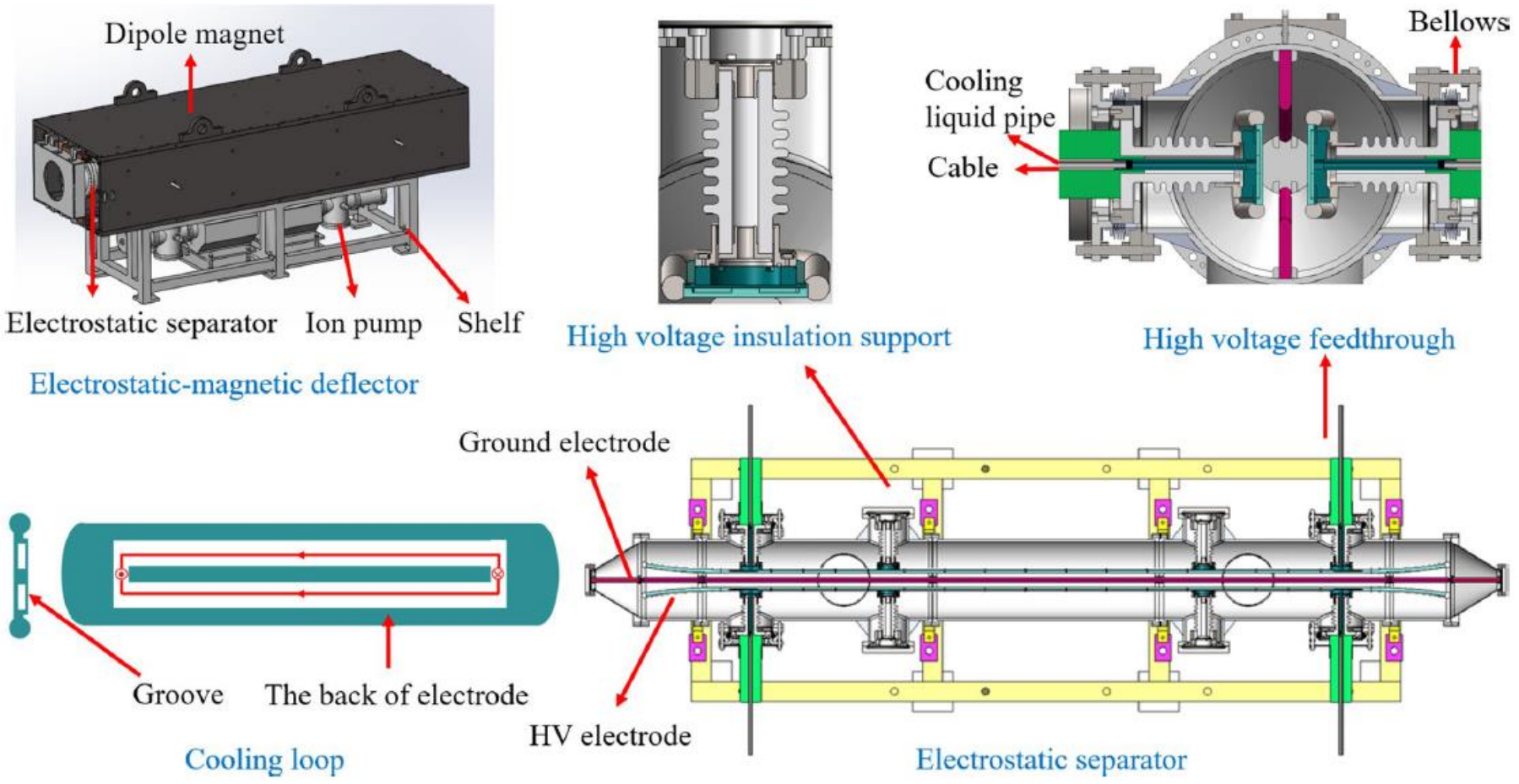

**Figure 4.3.11.16:** Mechanical design of the electrostatic-magnetic separator.

### 4.3.11.11 *Prototype of Electrostatic-Magnetic Separator*

The magnet and the electrostatic separator within the electrostatic magnetic separator were developed separately.

The prototype of the magnet has already been created, as shown in Figure 4.3.11.17. However, due to its substantial size, there is currently no suitable platform for accurate measurements.

We are currently in the process of constructing a new measurement system using a hall probe. Once this system is operational, we will proceed to conduct a precise measurement of the prototype's weak magnetic field.

.

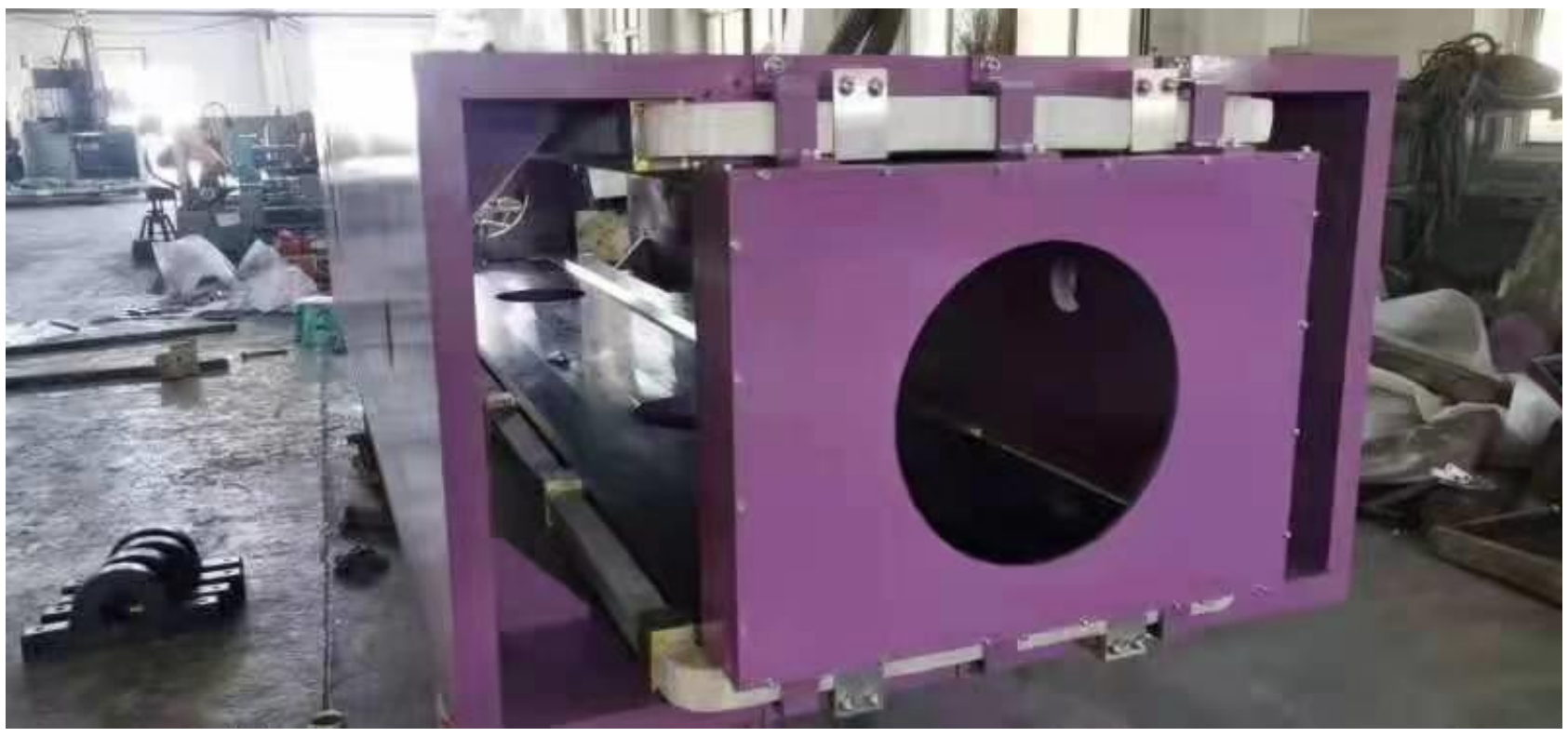

**Figure 4.3.11.17:** Prototype of the magnet.

The prototype of the electrostatic separator has also been developed, as shown in Figure 4.3.11.18. To assess whether the completed statically determined separator system meets our requirements, several tests have been conducted. These tests encompass evaluations of electric field uniformity, vacuum conditions, high voltage performance, and more. Optimization of the electrode plate structure aims to ensure that the simulated electric field uniformity aligns with the design specifications. The vacuum test involves processes such as pumping and baking to enhance vacuum levels and verify compliance with our vacuum requirements. High voltage experiments primarily focus on aging and observing the electrostatic separator's arc behavior under normal working voltage conditions.

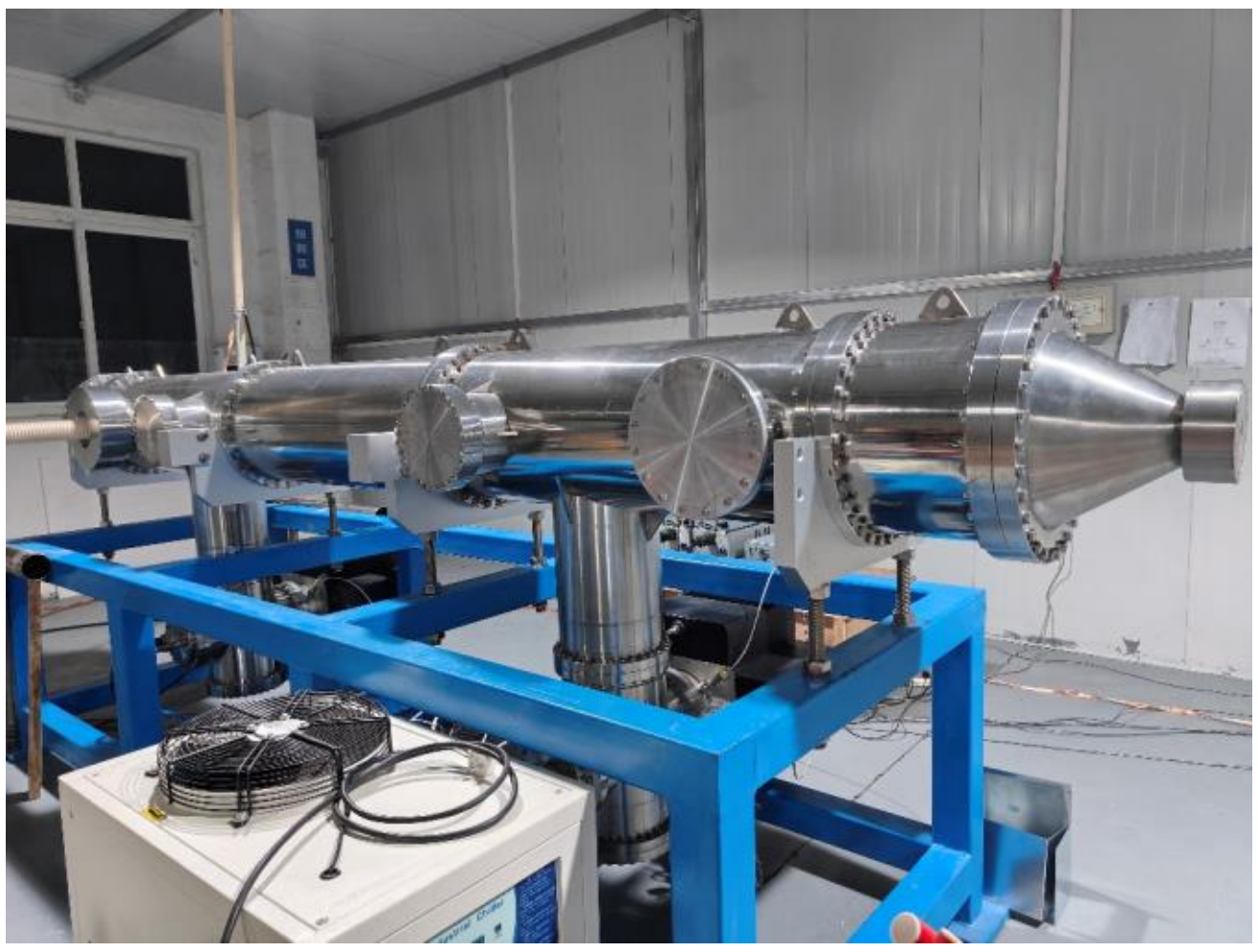

**Figure 4.3.11.18:** Prototype of the electrostatic separator.

The electric field uniformity test was conducted using Opera software simulation. Through adjustments to the electrode plate's shape, width, and edge design, we achieved an electric field uniformity of 46×30 mm² (@ 0.5‰).

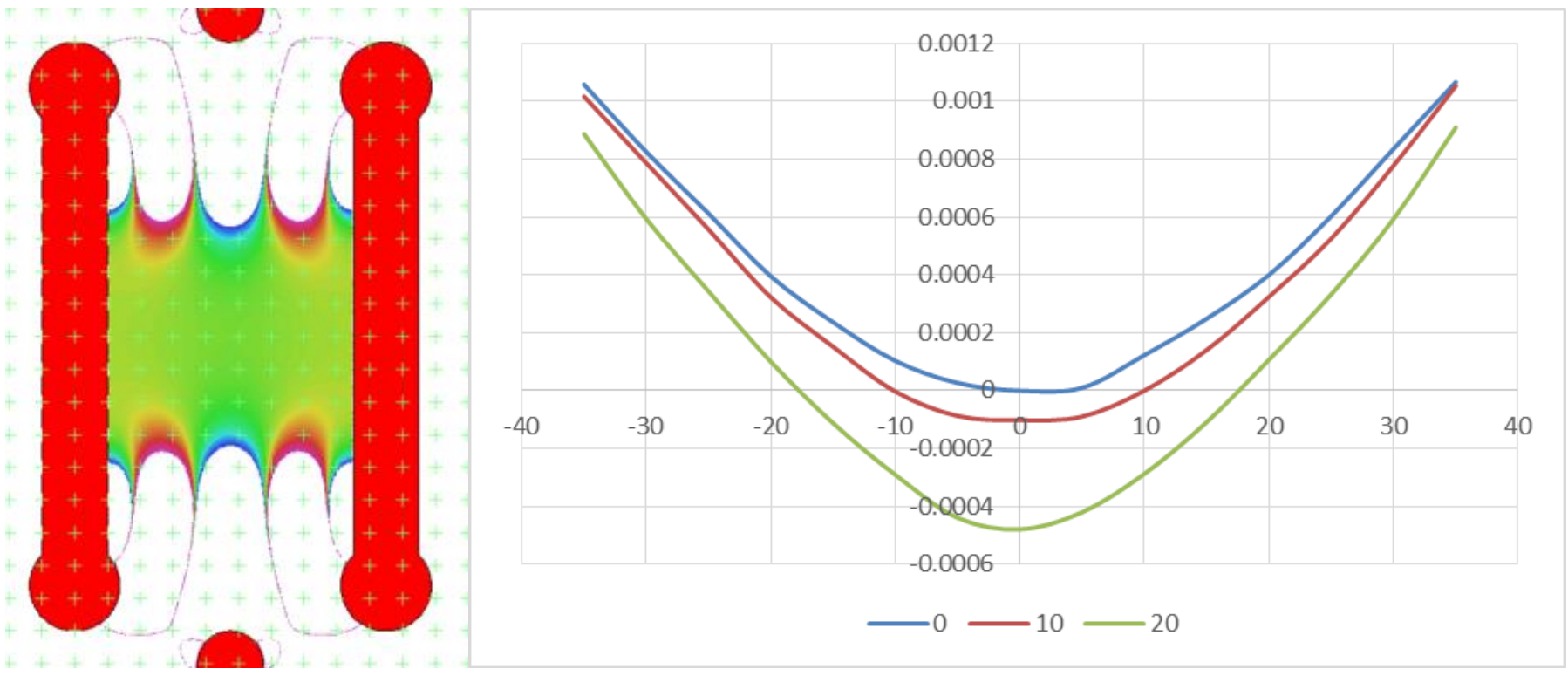

**Figure 4.3.11.19:** Electric field uniformity of the electrostatic separator.

The vacuum test for the electrostatic separator consists of two primary stages. The first stage involves rough pumping and leak detection, followed by high-temperature baking. If no leaks are detected during the baking process, we proceed to initiate ultra-high vacuum pumping. After completing these steps, including rough pumping, baking, and ultra-high vacuum pumping, the vacuum level within the electrostatic separator reaches less than $2.6\times10^{-8}$ Pa, as depicted in Figure 4.3.11.20. This vacuum level fully complies with the specified requirements.

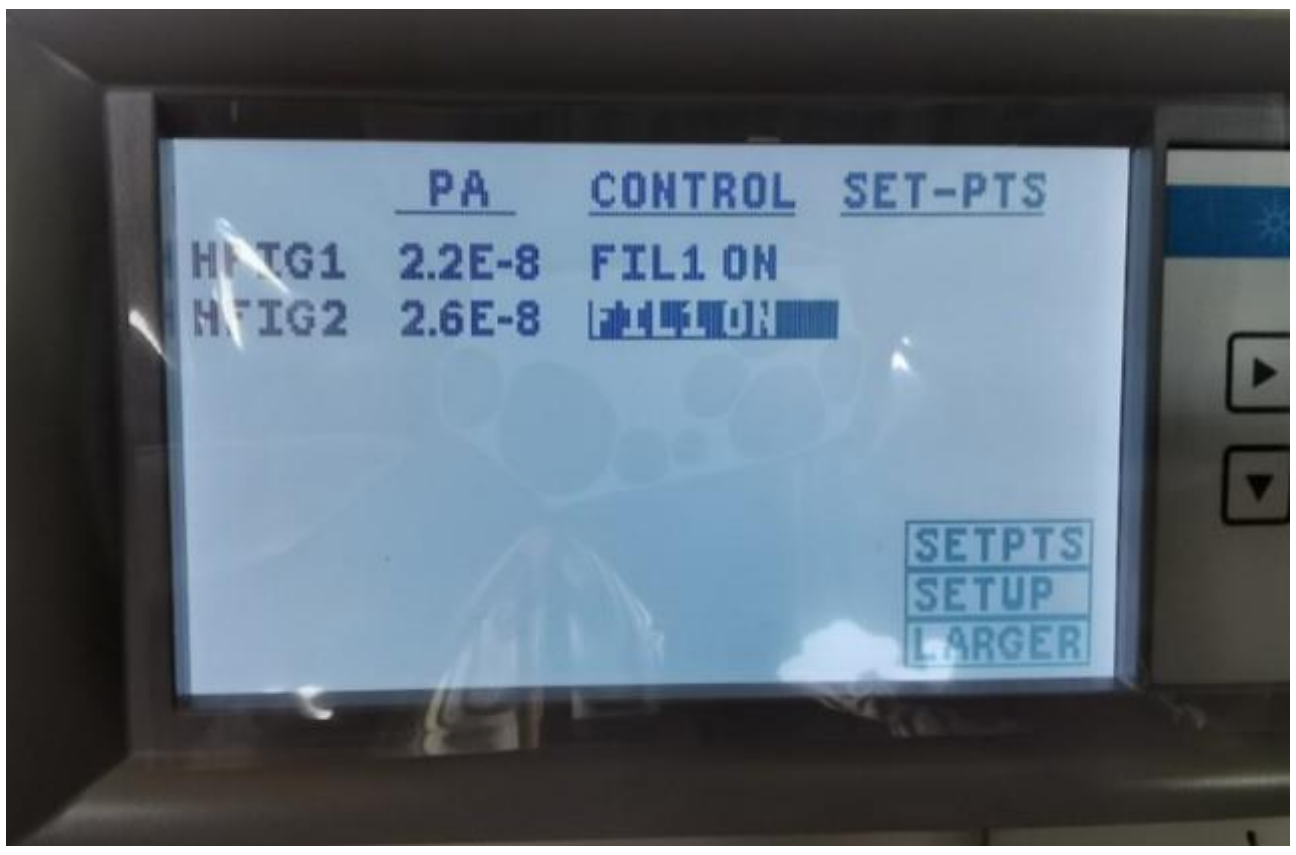

**Figure 4.3.11.20:** Vacuum measurement of the electrostatic separator.

After achieving the required vacuum level, we conducted a high voltage test. During the test, an arc occurred at the air side of the feedthrough. To address this issue, several modifications were made to the HV feedthrough, as illustrated in Figure 4.3.11.21:

1. Increased the length of ceramic parts on the air side.
2. Equipped the outer surface of ceramic parts with umbrella skirts to enhance the creepage distance.
3. Added a smooth cover to the outside of the feedthrough to prevent an increase in field strength at the edge of the mounting screw.

Following these improvements, we were able to attain a high voltage of ±118 kV, and the typical arc rate during testing was approximately 0.1 sparks per hour.

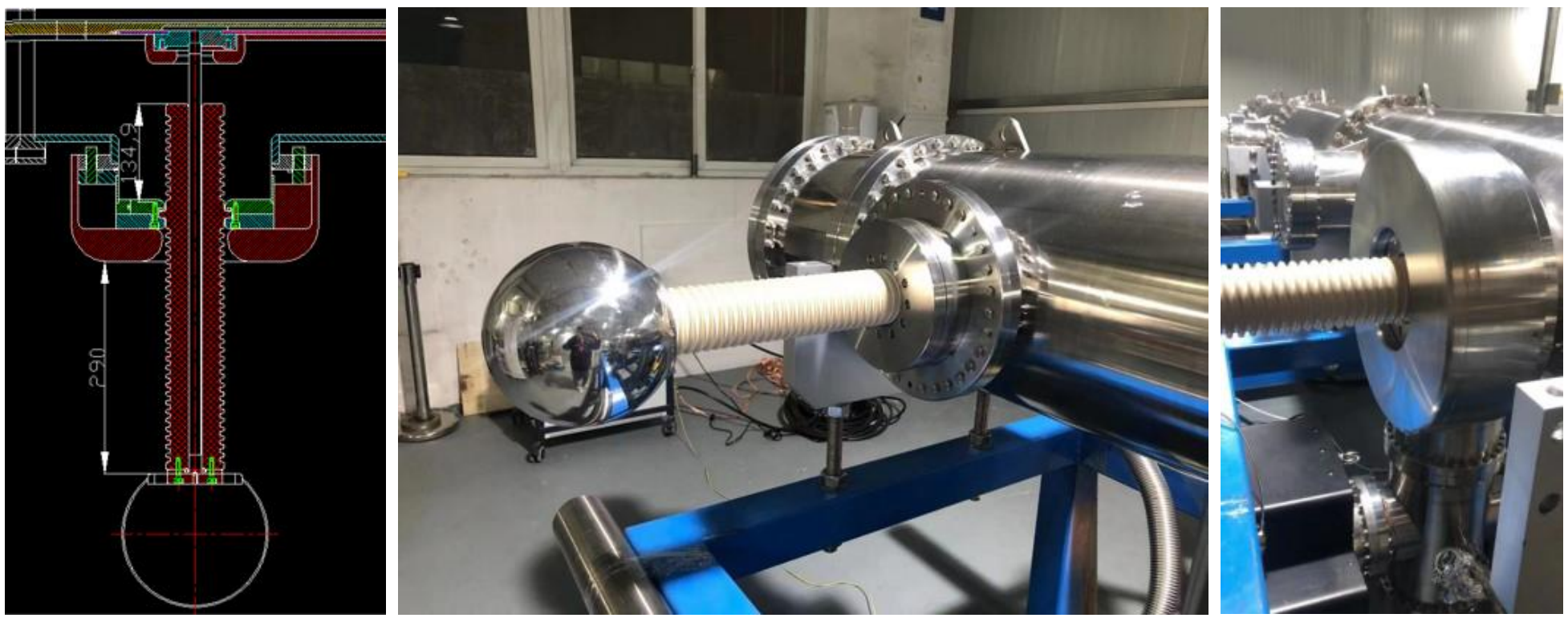

**Figure 4.3.11.21:** Improved feedthrough of the electrostatic separator.

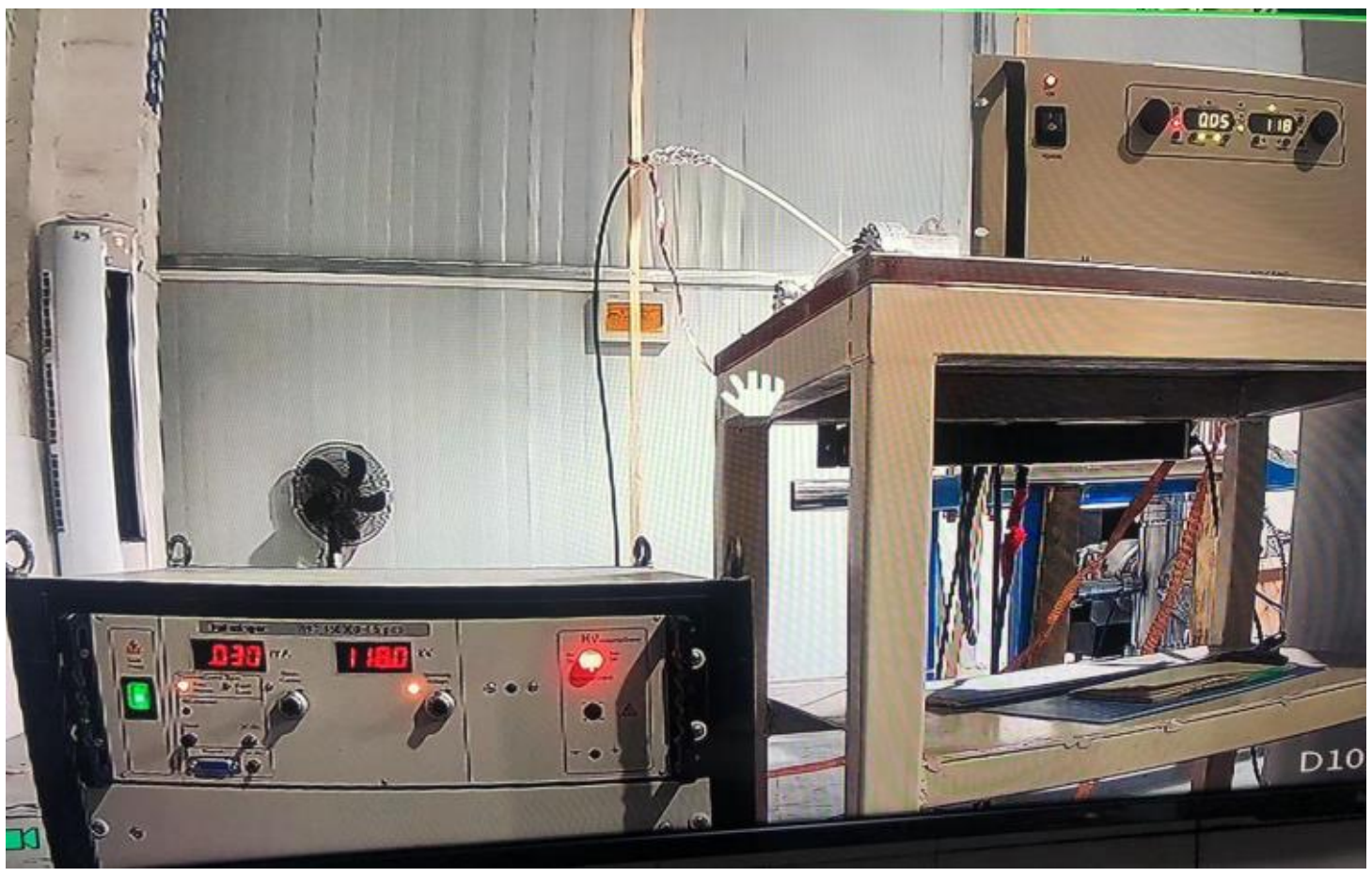

**Figure 4.3.11.21:** High voltage test of the electrostatic separator.

Next, we will proceed to test the Electro-magnetic Separator once the electrostatic separator and magnet are assembled together.

## 4.4 Upgrade Plan

### 4.4.1 Introduction

The physics requirements for the CEPC stipulate a beam energy range from 45.5 GeV to 120 GeV. The luminosity upgrade aims to increase the synchrotron radiation (SR) power per beam from 30 MW to 50 MW. Additionally, there are plans to upgrade the energy to 180 GeV after about 13 years of operation. The TDR has primarily focused on a well-developed and robust design, along with a cost estimate, for a Higgs factory operating at 120 GeV with 30 MW SR power per beam as the baseline configuration. This chapter outlines the strategies for achieving the power and energy upgrades, primarily through the incorporation of additional RF cavities, cryomodules, and the necessary RF power sources and cryogenics.

Table 4.4.1 provides the key parameters for the power and energy upgrade of the CEPC. For the 50 MW power upgrade in the Higgs and W mode, a greater number of 650 MHz 2-cell SRF cavities will be introduced. Furthermore, the 50 MW upgrade in the Z mode will involve the inclusion of additional high-current 1-cell cavities. In order to achieve the energy upgrade, more 650 MHz 5-cell cavities for the Collider and 1.3 GHz 9-cell cavities for the Booster will be incorporated.

Figure 4.4.1 illustrates the layout of the RF section within the Collider, the Booster, and the auxiliary tunnel. It showcases the arrangement of cryomodules dedicated to Higgs/W, Z, and $t\bar{t}$, as well as the associated RF power sources such as klystrons and solid-state amplifiers (SSAs). Additionally, the figure displays the LLRF system, cryogenic refrigerator cold box, auxiliary equipment, and the cooling-water systems essential for the operation of the CEPC.

In addition to the RF and cryogenics systems, the other technical system designs should also be compatible with the power and energy upgrades of the CEPC. The magnet system and magnet power supply system details will be provided in the energy upgrade section. The impact of the upgrades on the beam feedback, vacuum system, radiation protection, and conventional facilities systems are briefly discussed below:

- Beam feedback: The resistive-wall instability is most critical at the Z beam energy, which determines the damping-time requirement for the transverse bunch-by-bunch feedback system. With the Z beam power being upgraded from 30 MW to 50 MW, the transverse feedback system needs more power on kickers for each direction to reach the 0.5 ms requirement. The longitudinal damping time requirement is dominated by the multi-bunch instability induced by the cavity narrow-band impedance. The RF cavities for the 50 MW Z mode will have deep damping of all the Higher Order Modes (HOMs), similar to the 30 MW cavities, eliminating the need for beam feedback in principle. For the energy upgrade, the higher beam energy and lower beam current of $t\bar{t}$ will result in a longer growth time for transverse and longitudinal multi-bunch instability, thereby reducing the requirement on beam feedback time.
- Vacuum system: As the energy and power increase, the gas load due to photo-desorption will also increase. However, the number of sputter ion pumps designed with 11-meter intervals for the Z 30 MW configuration is sufficient to meet the pumping requirement for both the power upgrade (50 MW Higgs/W/Z mode) and the energy upgrade (30/50 MW $t\bar{t}$), thus no improvement is needed for the vacuum system.
- Radiation protection: The synchrotron-radiation shielding design for magnets is based on the Higgs/W/Z 50 MW parameters as the baseline. Less than a two-centimeter lead shielding is required for a 13-year running schedule (10 (H) + 2 (Z) + 1 (W)). Considering $t\bar{t}$ operation with an 18-year running schedule (10 (H) + 2 (Z) + 1 (W) + 5 ($t\bar{t}$)), approximately 2.5-cm lead shielding is needed to shield synchrotron radiation.
- Conventional facilities: The building and tunnel space are designed according to the requirements of the 50 MW Higgs and $t\bar{t}$ modes, while also reserving installation positions for the expansion of auxiliary facilities. The main pipeline and main power cable are designed to meet the highest AC power operation ($t\bar{t}$ mode) and are laid in place at once to accommodate the upgrades.

### 4.4.2 Power Upgrade

In order to upgrade the Collider to operate at 50 MW SR power per beam for the Higgs mode, the input coupler power per cavity will be maintained at the level of 300 kW as in 30 MW mode and 144 more 2-cell cavities will be added, resulting in a total of 336 2-cell cavities for 50 MW Higgs operation. The klystron output power will remain at the 800 kW level, and an additional 72 800 kW klystrons will be added. Therefore, a total of 168 650 MHz 800 kW klystrons will be required for 50 MW Higgs operation. During W mode operation at 50 MW, half of the Higgs cavities per ring will be utilized.

For the high-luminosity Z mode operating at 50 MW SR per beam, an additional 20 KEKB/BEPCII type high-current SRF-cavity cryomodules and corresponding 1.2 MW klystrons for each ring will be added. The high-current beam will bypass the Higgs/$t\bar{t}$ cavities.

The number of Booster cavities will not change for the Collider power upgrade, but the input power per cavity will be increased by augmenting the output RF power of the solid-state amplifiers (SSAs).

Due to the high beam current and HOM power per cavity in the high luminosity Z mode (which can reach up to 30 mA for injection into the empty Collider ring), an ERL-type 1.3 GHz cryomodule with a HOM absorber at 100 K between cavities will be used for the Z mode instead of the low-current TESLA cavities used for the Higgs, W, and $t\bar{t}$ modes. Furthermore, to reduce the impedance seen by the beam, the Z-mode beams will only pass through Z cavities and bypass the Higgs and $t\bar{t}$ cavities.

**Table 4.4.1:** Parameters for CEPC power and energy upgrade. The baseline parameters are included for reference.

| | **Baseline** | | | **Power Upgrade** | | | **Energy Upgrade** | | | |
|---|---|---|---|---|---|---|---|---|---|---|
| | **Higgs** | **W** | **Z** | **Higgs** | **W** | **Z** | $t\bar{t}$ | | | |
| Collider SR power / beam [MW] | 30 | | | 50 | | | 30 | | 50 | |
| Beam energy [GeV] | 120 | 80 | 45.5 | 120 | 80 | 45.5 | 180 | | | |
| Luminosity / IP [$10^{34}$ cm$^{-2}$s$^{-1}$] | 5 | 16 | 115 | 8.3 | 26.7 | 192 | 0.5 | | 0.8 | |
| **Collider 650 MHz cavities** | **2-cell** | | **1-cell** | **2-cell** | | **1-cell** | **Add 5-cell** | **Existing 2-cell** | **Add 5-cell** | **Existing 2-cell** |
| RF voltage [GV] | 2.2 | 0.7 | 0.12 | 2.2 | 0.7 | 0.1 | 10 (6.1 + 3.9) | | 10 (6.1 + 3.9) | |
| Beam current / ring [mA] | 16.7 | 84 | 801 | 27.8 | 140 | 1345 | 3.4 | | 5.6 | |
| Cavity number | 192 | 96×2 | 30×2 | 336 | 168×2 | 50×2 | 192 | 336 | 192 | 336 |
| Cryomodule number | 32 | 32 | 60 | 56 | 56 | 100 | 48 | 56 | 48 | 56 |
| Klystron number | 96 | 96 | 60 | 168 | 168 | 100 | 48 | 168 | 96 | 168 |
| Klystron power [kW] | 800 | 800 | 1200 | 800 | 800 | 1200 | 800 | 800 | 800 | 800 |
| Collider 4.5 K equiv. heat load [kW] | 44.4 | 28.1 | 15.2 | 41.9 | 20 | 20.1 | 128.3 | | 128.3 | |
| **Booster 1.3 GHz cavities** | **9-cell** | | | | | | **Add 9-cell** | **Existing 9-cell** | **Add 9-cell** | **Existing 9-cell** |
| Extraction RF voltage [GV] | 2.17 | 0.87 | 0.46 | 2.17 | 0.87 | 0.46 | 9.7 (7.53 + 2.17) | | 9.7 (7.53 + 2.17) | |
| Beam current [mA] | 1 | 3.1 | 16 | 1.4 | 5.3 | 30 | 0.12 | | 0.19 | |
| Cavity number | 96 | 96 | 32 | 96 | 96 | 32 | 256 | 96 | 256 | 96 |
| Cryomodule number | 12 | 12 | 4 | 12 | 12 | 4 | 32 | 12 | 32 | 12 |
| SSA number | 96 | 96 | 32 | 96 | 96 | 32 | 256 | 96 | 256 | 96 |
| SSA power [kW] | 25 | 25 | 25 | 30 | 30 | 40 | 10 | 10 | 10 | 10 |
| Booster 4.5 K equiv. heat load [kW] | 7.8 | 3.1 | 3.5 | 8.1 | 3.2 | 3.7 | 11.4 | | 11.4 | |
| **Total RF length [m]** | 704 | 704 | 384 | 1088 | 1088 | 608 | 2368 | | 2368 | |
| **Total 4.5 K equiv. heat load [kW]** | 52.2 | 31.2 | 18.7 | 50.0 | 23.2 | 23.8 | 139.7 | | 139.7 | |

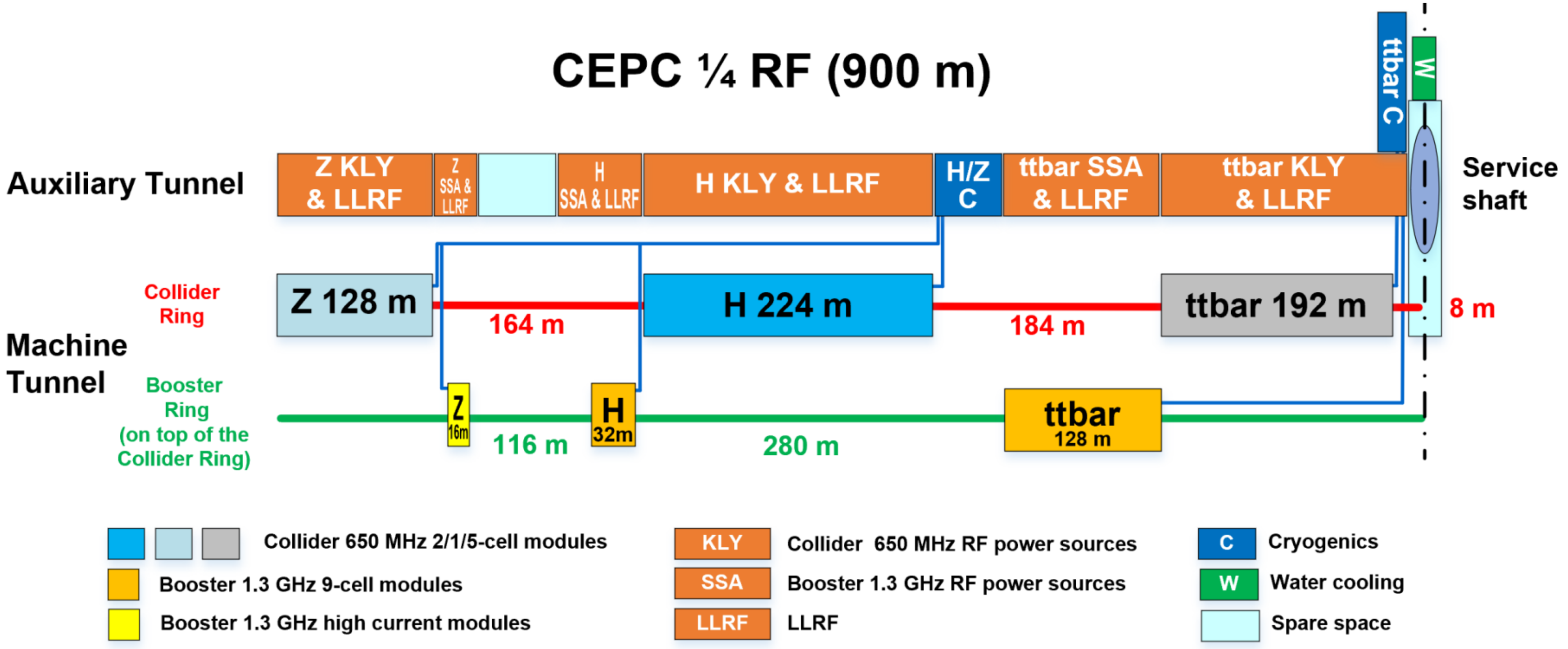

**Figure 4.4.1:** CEPC 1/4 RF section layout.

### 4.4.3 **Energy Upgrade**

For the Collider in the $t\bar{t}$ 30 MW mode, the upgrade entails adding 240 5-cell 650 MHz cavities and 48 sets of 800 kW klystrons in the center of the two RF sections, in addition to the existing 336 Higgs 2-cell cavities. This upgrade increases the RF voltage from 2.2 GV to 10 GV. Each cryomodule will house four 5-cell cavities with HOM couplers on the beam pipe. In the $t\bar{t}$ 50 MW mode, the number of klystrons will be doubled to accommodate the increased input power.

For the Booster, 256 high-gradient 9-cell cavities, each equipped with a 10 kW Solid-State Amplifier (SSA), will be added. This upgrade raises the Booster RF voltage from 2.17 GV to 9.7 GV.

Tables 4.4.2 and 4.4.3 display the estimated heat loads for the $t\bar{t}$ Collider and Booster, respectively. The total required capacity of the cryogenic plant is equivalent to 140 kW at 4.5 K. To meet the demand, eight additional 15 kW cryogenic plants will be installed in addition to the four 15 kW plants for the H, W, and Z modes. All 12 cryogenic plants will be operational for $t\bar{t}$ mode. A large cavern located next to the service shaft of the RF region has been designated for the installation of the new $t\bar{t}$ cryogenic plants.

**Table 4.4.2:** Estimated $t\bar{t}$ cryogenic heat loads for the Collider.

| $t\bar{t}$ 30 / 50 MW | | Unit | Collider | | |
|---|---|---|---|---|---|
| | | | 40-80 K | 5-8 K | 2 K |
| Added $t\bar{t}$ 650 MHz 5-cell cavity modules | Predicted module static heat load | W | 300 | 60 | 12 |
| | Predicted module HOM static heat load | W | 1.6 | 0.8 | 0.1 |
| | Predicted module input coupler static heat load | W | 4 | 2 | 0.2 |
| | Cavity dynamic heat load per module | W | 0 | 0 | 252.1 |
| | Predicted module HOM dynamic heat load | W | 3.2 | 1.6 | 0.3 |
| | Predicted module input coupler dynamic heat load | W | 60.5 | 15.1 | 2.5 |
| | Module dynamic heat load | W | 63.7 | 16.7 | 254.9 |
| | Total heat load per module | W | 369.3 | 79.5 | 267.2 |
| | Cryomodule number | - | 48 | 48 | 48 |
| | EVB heat loss | W | 50 | 10 | 5 |
| | EVB number | - | 48 | 48 | 48 |
| Original Higgs 650 MHz 2-cell cavity modules | Predicted module static heat load | W | 300 | 60 | 12 |
| | Predicted module HOM static heat load | W | 2.4 | 1.2 | 0.1 |
| | Predicted module input coupler static heat load | W | 6 | 3 | 0.3 |
| | Cavity dynamic heat load per module | W | 0 | 0 | 126.8 |
| | Predicted module HOM dynamic heat load | W | 1.9 | 0.9 | 0.2 |
| | Predicted module Input coupler dynamic heat load | W | 34 | 8.5 | 1.4 |
| | Module dynamic heat load | W | 35.9 | 9.4 | 128.4 |
| | Total heat load per module | W | 344.3 | 73.6 | 140.8 |
| | Cryomodule number | - | 56 | 56 | 56 |
| | EVB heat loss | W | 50 | 10 | 5 |
| | EVB number | - | 56 | 56 | 56 |
| | MDVB heat loss | W | 100 | 50 | 20 |
| | MDVB number | - | 8 | 8 | 8 |
| | Total cryogenic transfer line length | m | 2064 | 2064 | 2064 |
| | Cryogenic transfer line heat loss per meter | W/m | 2 | 0.5 | 0.3 |
| | Total cryogenic transfer line heat loss | W | 4128 | 1032 | 619.2 |
| | Total heat load | kW | 47.1 | 10.4 | 22 |
| | Overall net cryogenic capacity multiplier | | 1.54 | 1.54 | 1.54 |
| | 4.5 K equiv. heat load with multiplier | kW | 5.44 | 14.43 | 108.5 |
| | 4.5 K equiv. heat load with multiplier | kW | 128.3 | | |

**Table 4.4.3:** Estimated $t\bar{t}$ cryogenic heat loads for the Booster.

| $t\bar{t}$ 30 / 50 MW | | Unit | Booster | | |
|---|---|---|---|---|---|
| | | | 40-80 K | 5-8 K | 2 K |
| Added new $t\bar{t}$ 1.3 GHz 9-cell cavity modules | Predicted module static heat load | W | 140 | 20 | 3 |
| | Predicted module HOM static heat load | W | 0 | 0 | 0 |
| | Predicted module input coupler static heat load | W | 16 | 8 | 0.8 |
| | Cavity dynamic heat load per module | W | 0 | 0 | 11.25 |
| | Predicted module HOM dynamic heat load | W | 0.3 | 0.2 | 0 |
| | Predicted module input coupler dynamic heat load | W | 11.5 | 1.1 | 0.1 |
| | Module dynamic heat load | W | 11.8 | 1.3 | 11.35 |
| | Total heat load per module | W | 167.8 | 29.3 | 15.15 |
| | Cryomodule number | - | 32 | 32 | 32 |
| | EVB heat loss | W | 50 | 10 | 5 |
| | EVB number | - | 32 | 32 | 32 |
| Original Higgs 1.3 GHz 9-cell cavity modules | Predicted module static heat load | W | 140 | 20 | 3 |
| | Predicted module HOM static heat load | W | 0 | 0 | 0 |
| | Predicted module input coupler static heat load | W | 16 | 8 | 0.8 |
| | Cavity dynamic heat load per module | W | 0 | 0 | 4.5 |
| | Predicted module HOM dynamic heat load | W | 0.1 | 0 | 0 |
| | Predicted module input coupler dynamic heat load | W | 5.8 | 0.6 | 0.1 |
| | Module dynamic heat load | W | 5.9 | 0.6 | 4.6 |
| | Total heat load per module | W | 161.9 | 28.6 | 8.4 |
| | Cryomodule number | - | 12 | 12 | 12 |
| | EVB heat loss | W | 50 | 10 | 5 |
| | EVB number | - | 12 | 12 | 12 |
| | MDVB heat loss | W | 100 | 50 | 20 |
| | MDVB number | - | 8 | 8 | 8 |
| | Total cryogenic transfer line length | m | 1048 | 1048 | 1048 |
| | Cryogenic transfer line heat loss per meter | W/m | 2 | 0.5 | 0.3 |
| | Total cryogenic transfer line heat loss | W | 2096 | 524 | 314.4 |
| | Total heat load | kW | 12.42 | 2.65 | 1.28 |
| | Overall net cryogenic capacity multiplier | | 1.54 | 1.54 | 1.54 |
| | 4.5 K equiv. heat load with multiplier | kW | 1.43 | 3.67 | 6.31 |
| | Total 4.5 K equiv. heat load with multiplier | kW | 11.4 | | |

The design of the Collider magnets is capable of meeting the required working strength and gradient at four different energies: 45.5 GeV, 80 GeV, 120 GeV, and 180 GeV. The maximum field strength and gradient are achieved in the $t\bar{t}$ mode, as indicated in Table 4.4.4. Several new normal-conducting quadrupole magnets, Q3IRU and Q3IRD in Table 4.4.5, which are located outside the superconducting magnets in the MDI area, will be added for the $t\bar{t}$ mode. It is important to note that during the power upgrade, there is no need to change the magnets themselves.

All the magnet power supplies in the Collider are rated for operation at 120 GeV, with an additional 10 % safety margin provided for both current and voltage. When designing the parameters, structural aspects of the power supply, as well as the cable layout, we have taken into consideration the future need for energy upgrading of the accelerator. By upgrading the front-stage DC source and adding modules, the output current and voltage of the power supply can be increased. This allows the operation of the accelerator in the $t\bar{t}$ mode with minimal changes and cost.

**Table 4.4.4:** Field strength and gradient at four energies for the Collider main magnets

| | **Z** | **W** | **Higgs** | $t\bar{t}$ |
|---|---|---|---|---|
| Beam energy [GeV] | 45.5 | 80 | 120 | 180 |
| Field strength of Dipole [T] | 0.015 | 0.0265 | 0.0398 | 0.0597 |
| Field gradient of Quadrupole [T/m]) | 1.9 | 3.3 | 7.1 | 10.6 |
| Field gradient of Sextupole [T/m$^{2]}$ | 209 | 367 | 550 | 825 |

**Table 4.4.5:** Field gradient of Q3IRU and Q3IRD for the Collider

| | **Z** | **W** | **Higgs** | $t\bar{t}$ |
|---|---|---|---|---|
| Beam energy [GeV] | 45.5 | 80 | 120 | 180 |
| Field gradient of Q3IRU [T/m] | 0 | 0 | 0 | -51 |
| Field gradient of Q3IRD [T/m] | 0 | 0 | 0 | 40 |

# 5 Booster

## 5.1 Main Parameters

### 5.1.1 Main Parameters of Booster

The Booster is designed to provide electron and positron beams to the Collider at different energies with the required efficiency. The newest Booster design is consistent with the TDR higher luminosity goals for four energy modes. The main Booster parameters at injection and extraction energies are listed in Table 5.1.1 and Table 5.1.2.

To maintain the required beam current in the Collider, top-up injection is required. The assumptions for the Booster include a 3% current decay in the Collider and a 92% transfer efficiency from the Booster, including 3% beam loss due to the quantum lifetime and 5% beam loss during ramping.

The total beam current in the Booster is limited to less than 0.3 mA for $t\bar{t}$ running, 1 mA for Higgs mode, 4 mA for W mode, and 16 mA for Z. These limits are set by the RF system for different extraction energies.

The beam is injected from the Linac to the Booster using the on-axis scheme and injected from the Booster to the Collider using the off-axis scheme at three different energies for $t\bar{t}$, W, and Z. Additionally, on-axis injection from Booster to Collider has been considered at 120 GeV in case the dynamic aperture of the Collider is not large enough.

The threshold of single bunch current in the Booster due to transverse mode-coupling instability (TMCI) at 120 GeV is 70 mA, which is higher than the maximum single bunch current in the on-axis scheme.

The top-up injection time is 29 seconds for $t\bar{t}$ mode, 31.8 seconds for Higgs on-axis mode, 38.1 seconds for W, and 2.2 minutes for Z. The full injection time from zero current for both beams is 6 minutes for $t\bar{t}$, 10 minutes for Higgs, and 16 minutes for W. The full injection time at the Z pole is 1.8 hours or 50 minutes, depending on different current limitations due to RF system, and the bootstrapping injection starts from 220 mA.

**Table 5.1.1:** Main Booster parameters at injection energy.

| | | *tt* | *H* | *W* | *Z* | |
|---|---|---|---|---|---|---|
| Beam energy | GeV | 30 | | | | |
| Bunch number | | 35 | 268 | 1297 | 3978 | 5967 |
| Threshold of single bunch current | μA | 8.68 | 6.3 | 5.8 | | |
| Threshold of beam current (limited by coupled bunch instability) | mA | 97 | 106 | 100 | 93 | 96 |
| Bunch charge | nC | 1.1 | 0.78 | 0.81 | 0.87 | 0.9 |
| Single bunch current | μA | 3.4 | 2.3 | 2.4 | 2.65 | 2.69 |
| Beam current | mA | 0.12 | 0.62 | 3.1 | 10.5 | 16.0 |
| Growth time (coupled bunch instability) | ms | 2530 | 530 | 100 | 29.1 | 18.7 |
| Energy spread | % | 0.025 | | | | |
| Synchrotron radiation loss/turn | MeV | 6.5 | | | | |
| Momentum compaction factor | $10^{-5}$ | 1.12 | | | | |
| Emittance | nm | 0.076 | | | | |
| Natural chromaticity | H/V | -372/-269 | | | | |
| RF voltage | MV | 761.0 | 346.0 | 300.0 | | |
| Betatron tune $\nu_x/\nu_y$ | | 321.23/117.18 | | | | |
| Longitudinal tune | | 0.14 | 0.0943 | 0.0879 | | |
| RF energy acceptance | % | 5.7 | 3.8 | 3.6 | | |
| Damping time | s | 3.1 | | | | |
| Bunch length of linac beam | mm | 0.4 | | | | |
| Energy spread of linac beam | % | 0.15 | | | | |
| Emittance of linac beam | nm | 6.5 | | | | |

**Table 5.1.2:** Main Booster parameters at extraction energy.

| | | ***tt*** | ***H*** | | ***W*** | ***Z*** | |
|---|---|---|---|---|---|---|---|
| | | Off axis injection | Off axis injection | On axis injection | Off axis injection | Off axis injection | |
| Beam energy | GeV | 180 | 120 | | 80 | 45.5 | |
| Bunch number | | 35 | 268 | 261+7 | 1297 | 3978 | 5967 |
| Maximum bunch charge | nC | 0.99 | 0.7 | 20.3 | 0.73 | 0.8 | 0.81 |
| Maximum single bunch current | μA | 3.0 | 2.1 | 61.2 | 2.2 | 2.4 | 2.42 |
| Threshold of single bunch current | μA | 91.5 | 70 | | 22.16 | 9.57 | |
| Threshold of beam current (limited by RF system) | mA | 0.3 | 1 | | 4 | 16 | |
| Beam current | mA | 0.11 | 0.56 | 0.98 | 2.85 | 9.5 | 14.4 |
| Growth time (coupled bunch instability) | ms | 16611 | 2359 | 1215 | 297.8 | 49.5 | 31.6 |
| Bunches per pulse of Linac | | 1 | 1 | | 1 | 2 | |
| Time for ramping up | s | 7.1 | 4.3 | | 2.4 | 1.0 | |
| Injection duration for top-up (Both beams) | s | 29.2 | 23.1 | 31.8 | 38.1 | 132.4 | |
| Injection interval for top-up | s | 65 | 38 | | 155 | 153.5 | |
| Current decay during injection interval | | 3% | | | | | |
| Energy spread | % | 0.15 | 0.099 | | 0.066 | 0.037 | |
| Synchrotron radiation loss/turn | GeV | 8.45 | 1.69 | | 0.33 | 0.034 | |
| Momentum compaction factor | $10^{-5}$ | 1.12 | | | | | |
| Emittance | nm | 2.83 | 1.26 | | 0.56 | 0.19 | |
| Natural chromaticity | H/V | -372/-269 | | | | | |
| Betatron tune $\nu_x/\nu_y$ | | 321.27/117.19 | | | | | |
| RF voltage | GV | 9.7 | 2.17 | | 0.87 | 0.46 | |
| Longitudinal tune | | 0.14 | 0.0943 | | 0.0879 | 0.0879 | |
| RF energy acceptance | % | 1.78 | 1.59 | | 2.6 | 3.4 | |
| Damping time | ms | 14.2 | 47.6 | | 160.8 | 879 | |
| Natural bunch length | mm | 1.8 | 1.85 | | 1.3 | 0.75 | |
| Full injection from empty ring | h | 0.1 | 0.14 | 0.16 | 0.27 | 1.8 | 0.8 |

Figure 5.1.1 illustrates the Booster ramping scheme. While ramping, the eddy current induces a parasitic sextupole field on the beam pipe inside the dipoles, and the eddy current effect limits the ramping rate [1]. The ramping times are 7.1 s for the $t\bar{t}$ mode, 4.3 s for the Higgs mode, 2.4 s for the W mode, and 1.0 s for the Z mode. Top-up injection requires three cycles for each beam at the Z pole and only one cycle for the other three energy modes [2].

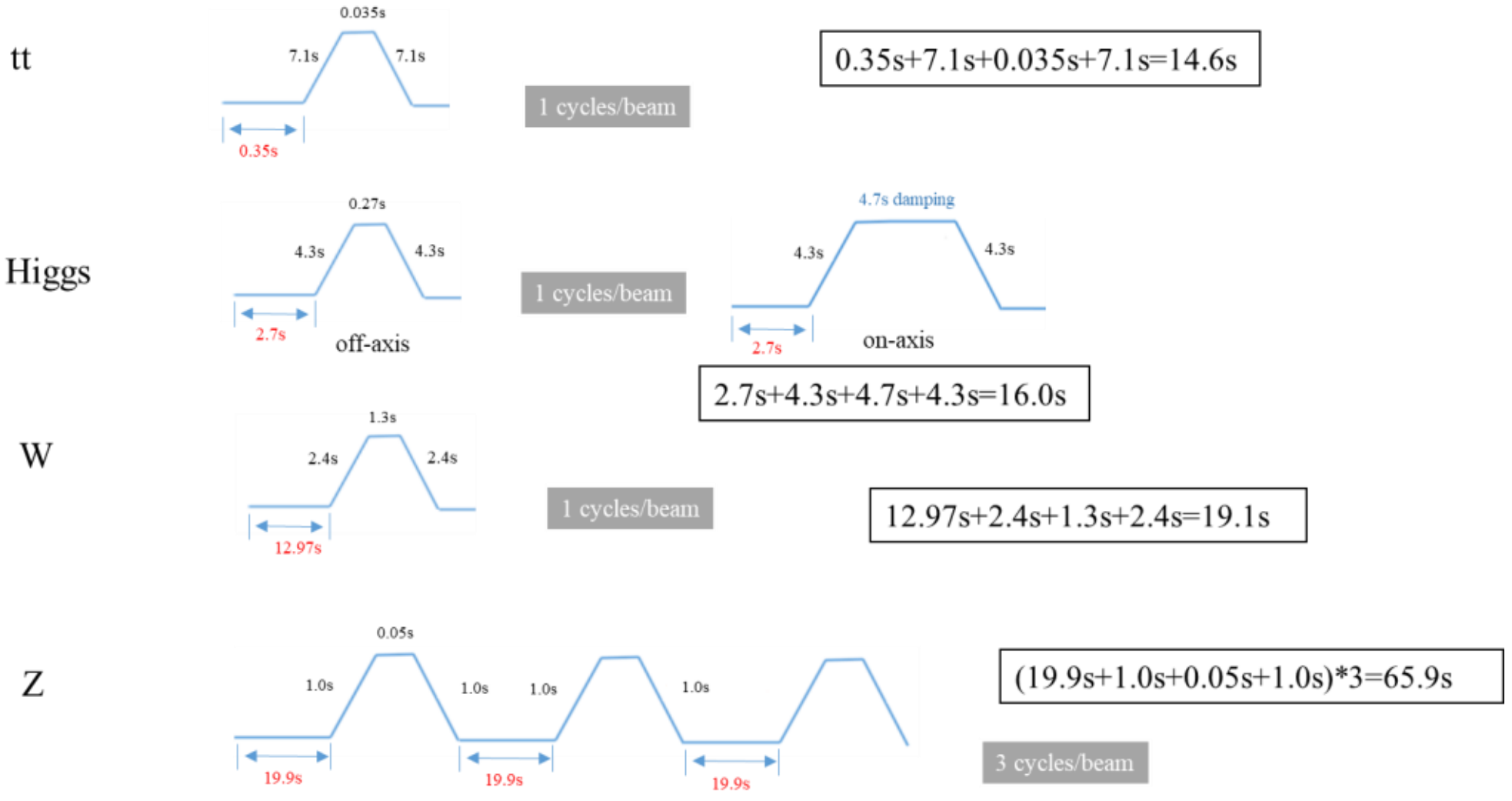

**Figure 5.1.1:** Timing structure for the CEPC Booster.

Different RF ramping curves have been designed for the three energy modes, as shown in Figures 5.1.2 to 5.1.4. The longitudinal tune remains constant at 0.088 during ramping for the Z and W modes. However, for the Higgs and $t\bar{t}$ modes, the longitudinal tune is set to 0.094 and 0.14, respectively, to achieve sufficient energy acceptance and a higher current threshold due to TMCI.

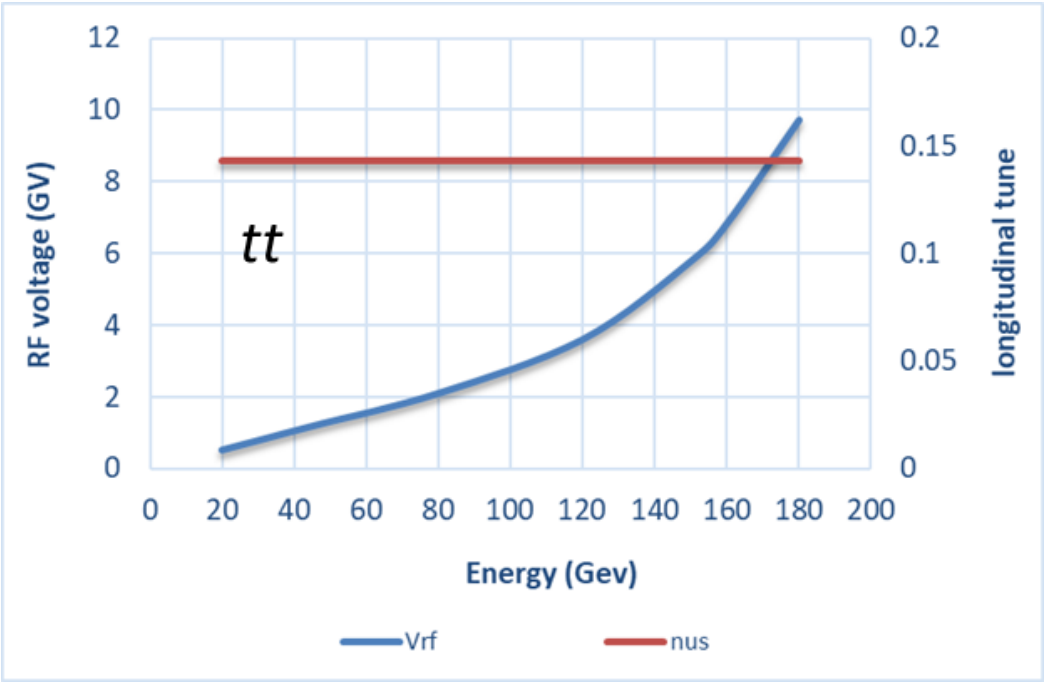

**Figure 5.1.2:** The RF ramping curves for $t\bar{t}$ (180 GeV) mode. The red line is the longitudinal tune and the blue line the RF voltage.

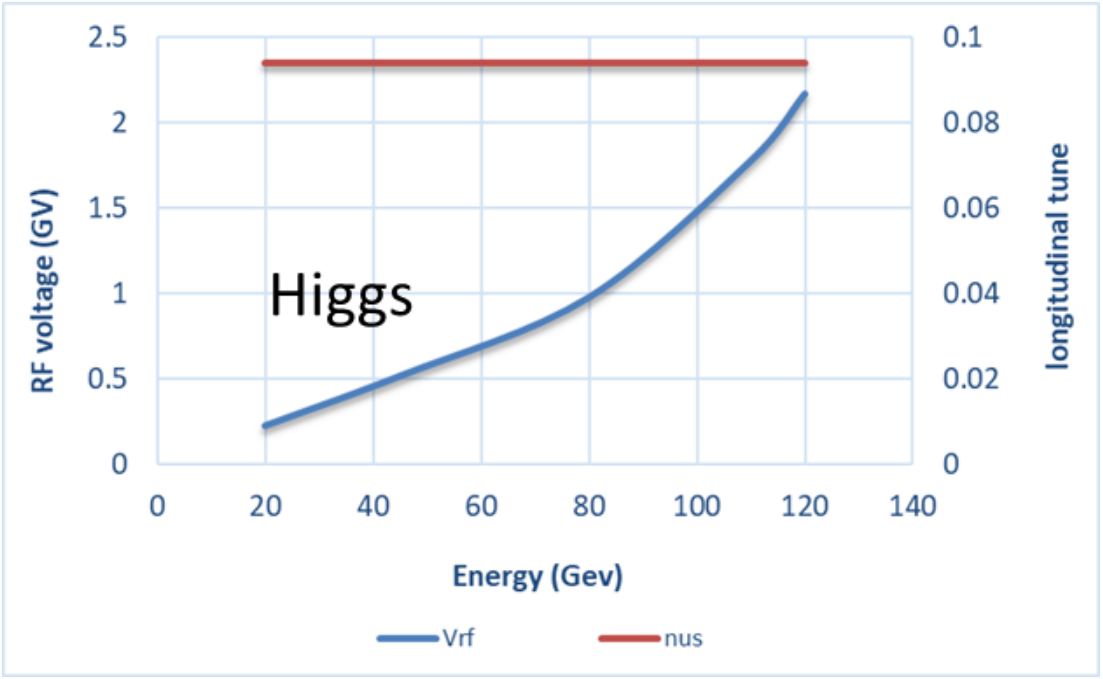

**Figure 5.1.3:** The RF ramping curves for Higgs (120 GeV) mode. The red line is the longitudinal tune and the blue line the RF voltage.

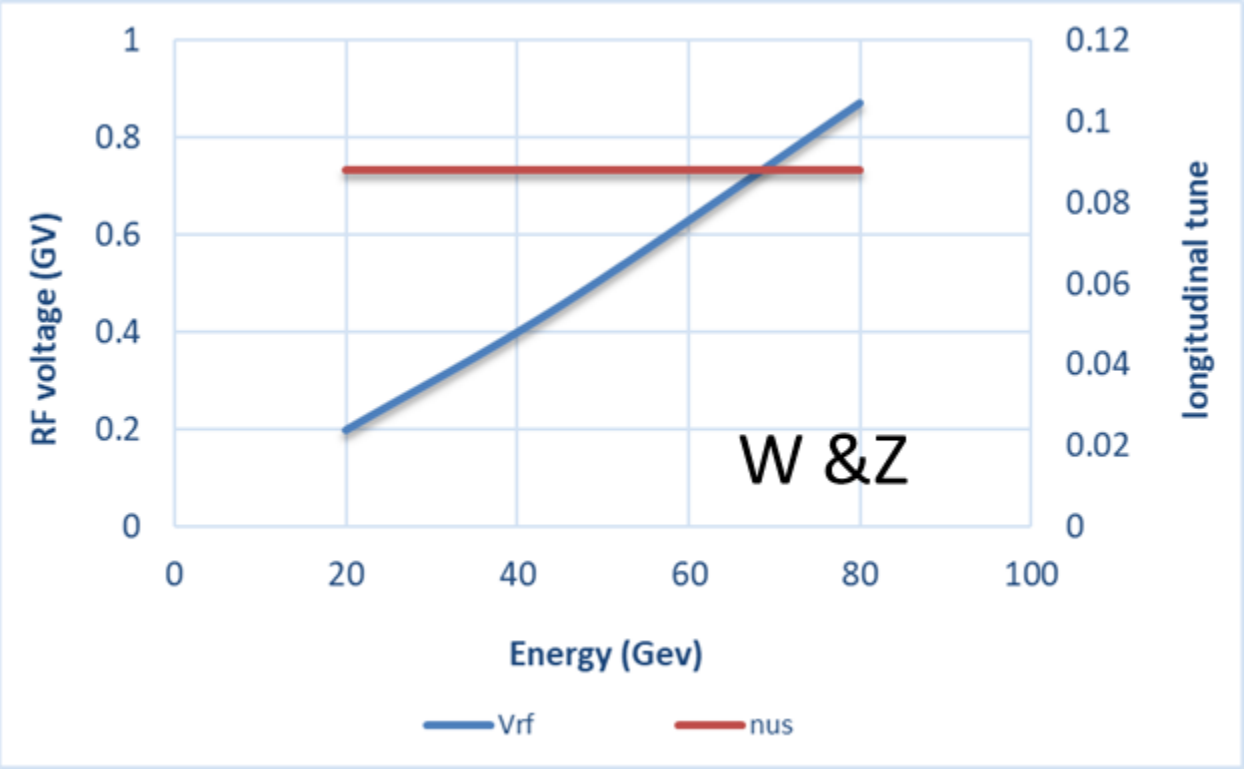

**Figure 5.1.4:** The RF ramping curves for W (80 GeV) and Z (45.5 GeV) mode. The red line is the longitudinal tune and the blue line the RF voltage.

### 5.1.2 RF Parameters

Table 5.1.3 provides an overview of the RF parameters. The selection of superconducting RF cavities was driven by their impressive CW gradient, energy efficiency, and low impedance characteristics. To strike a balance between beam stability and cost considerations, a frequency of 1.3 GHz was selected.

**Table 5.1.3:** Main RF parameters of the Booster

| | $t\bar{t}$ 30/50 MW | | **Higgs** 30/50 MW | **W** 30/50 MW | **Z** 30/50 MW |
|---|---|---|---|---|---|
| | New cavities | Higgs cavities | | | |
| Extraction beam energy [GeV] | 180 | | 120 | 80 | 45.5 |
| Bunch charge [nC] | 1.1 | | 0.78 (20.3)* | 0.73 | 0.81 |
| Beam current [mA] | 0.12 / 0.19 | | 1 / 1.4 | 3.1 / 5.3 | 16 / 30 |
| Injection RF voltage [GV] | 0.761 | | 0.346 | 0.3 | 0.3 |
| Extraction RF voltage [GV] | 9.7 (7.53 + 2.17) | | 2.17 | 0.87 | 0.46 |
| Cavity number (1.3 GHz 9-cell) | 256 | 96 | 96 | 96 | 32 |
| Module number | 32 | 12 | 12 | 12 | 4 |
| Extraction gradient [MV/m] | 28.3 | 21.8 | 21.8 | 8.7 | 13.8 |
| $Q_0$ @ 2 K at operating gradient | 2E10 | 3E10 | 3E10 | 3E10 | 3E10 |
| Average HOM power / cavity [W] | 0.2 / 0.32 | | 10 / 15 | 3.8 / 6.3 | 80 / 150 |
| Input average power / cavity [kW] | 0.3 | 0.2 | 6.5 / 9.2 | 0.3 / 0.5 | 2.5 / 4.5 |
| SSA power [kW] (1 cavity / SSA) | 10 | 10 | 25 / 30 | 25 / 30 | 25 / 40 |

* The small bunch charge number before the parenthesis is for the bunches injected from the linac. The large bunch charge number in the parenthesis is for the bunches injected back from the Collider ring for the swap-out injection at Higgs energy.

## 5.2 Booster Accelerator Physics

### 5.2.1 Optics and Dynamic Aperture

The Booster is situated in the same tunnel as the Collider, placed above the Collider ring, except in the interaction region where there are bypasses to avoid the detectors. After the CDR, the emittance of the Booster at all energy modes is significantly reduced by implementing the modified-TME (theoretical minimum emittance) structure [1]. In order to achieve lower emittance and looser error sensitivity simultaneously, the Booster lattice has been changed from the FODO structure of the CDR to the TME structure in the TDR.

#### 5.2.1.1 *Optics Design*

##### *5.2.1.1.1 Survey Design*

The Booster's horizontal position is carefully designed to align with the center of the Collider's two beams. The horizontal position difference of the Booster relative to the center of the Collider's two rings is rigorously maintained within a tolerance of ±0.17 meters. In the interaction region, the Booster is bypassed on the outer side to prevent interference with the CEPC detectors, with a separation distance of approximately 25 m between the detector center and Booster. To ensure proper injection timing, the Booster has the same circumference as the Collider. Figure 5.2.1.1 illustrates the geometry of the Booster compared to the Collider, as well as the cross section of the tunnel.

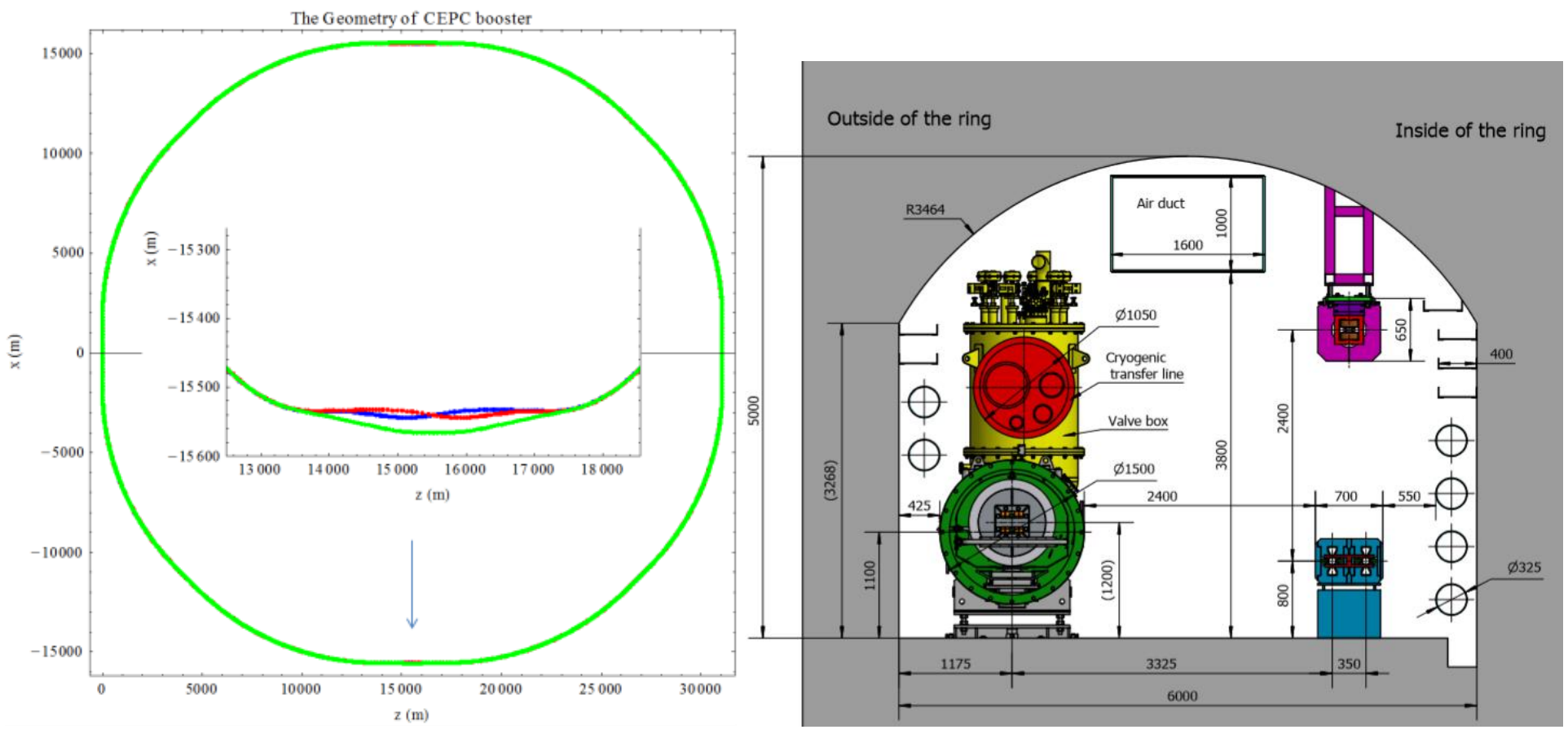

**Figure 5.2.1.1:** Layout of the Booster and Collider and the tunnel cross section. The blue curve in the left plot is the geometry of the collider ring for electron beam, the red curve for positron beam, and the green curve is the geometry of the Booster.

### *5.2.1.1.2 Arc Section*

The arc section of the CEPC is composed of modified-TME cells, with each cell being 78 meters in length. The optics design for the arc cell is shown in Fig. 5.2.1.2. In order to reduce the error sensitivity of the lattice, the quadrupoles are distributed as uniformly as possible. The horizontal phase advance is 100° and the vertical phase advance is 28° for each cell, which has been optimized to balance between emittance and dynamic aperture (DA). The modified-TME structure reduces the emittance of the Booster at 120 GeV from 3.56 nm in CDR to 1.26 nm in TDR. To correct the chromaticity, an interleave sextupole scheme is used, and all the sextupoles are enbeded in the dipoles that are next to the quadrupoles. Additionally, 100 independent sextupoles are added to allow for future chromaticity adjustment and DA optimization. The capability of the chromaticity adjustment by the independent sextupoles is about 10%.

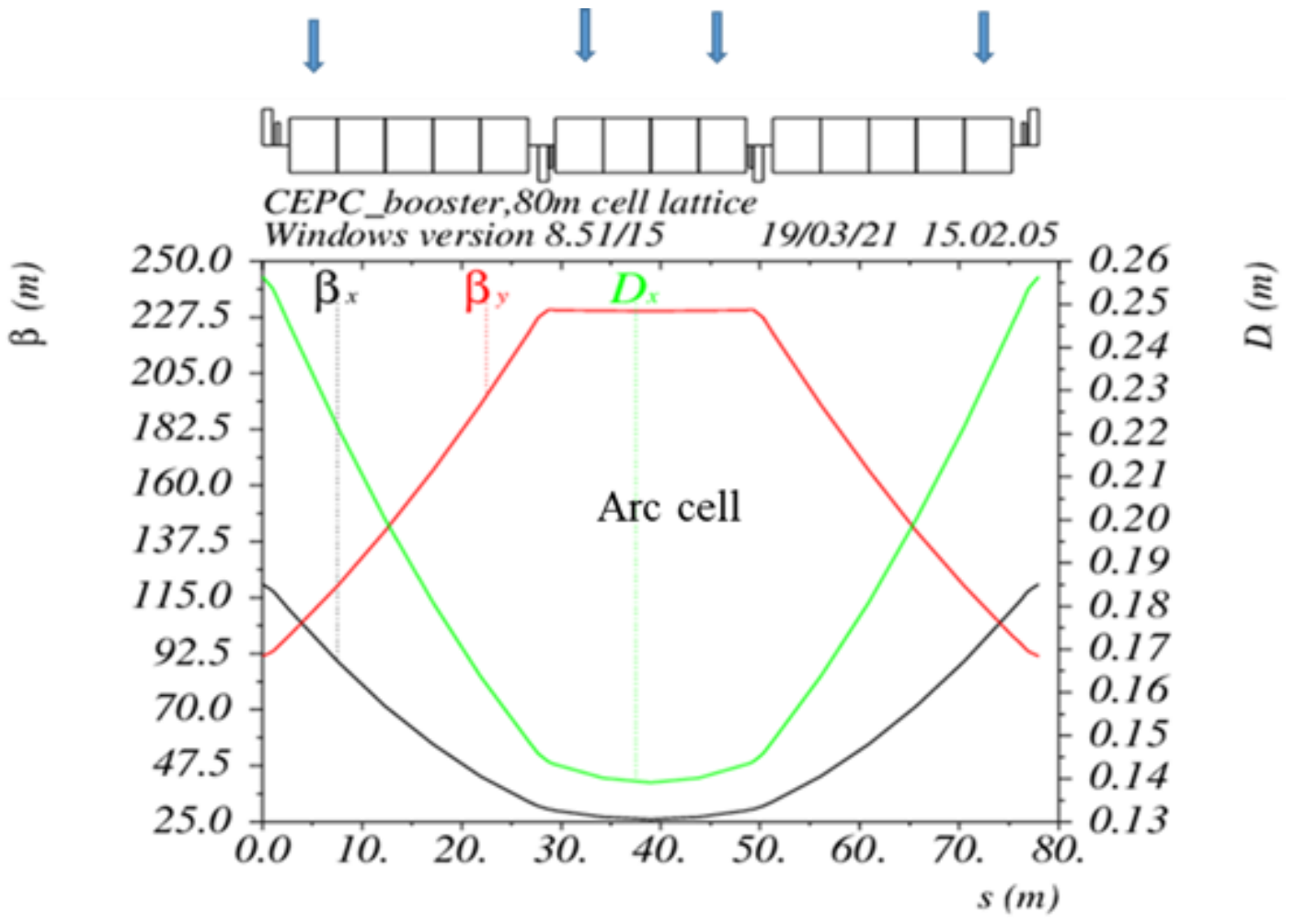

**Figure 5.2.1.2:** The Twiss functions of the TME cell in the arc region.

### *5.2.1.1.3 Dispersion Suppressor*

The function of the dispersion suppressor is to cancel the dispersion induced in the arc section and make a transition between the arc section and the straight section as shown in Fig. 5.2.1.3.

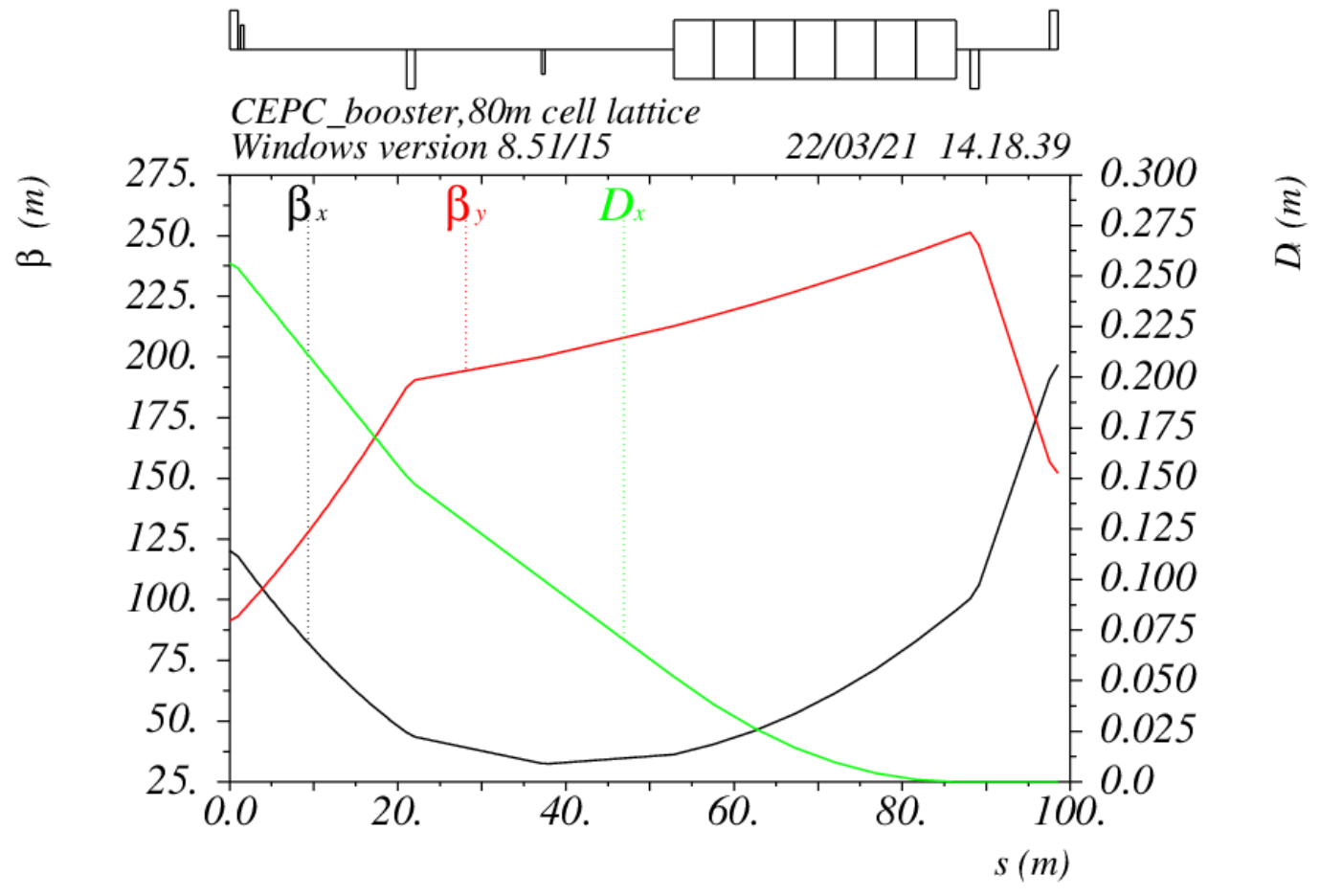

**Figure 5.2.1.3:** The Twiss functions of the dispersion suppressor.

### 5.2.1.1.4 *Straight Sections*

The straight section is made of FODO cell and the length of each cell is 134 m, which is shown in Fig. 5.2.1.4. The optimum phase advances in the arc cell and the straight FODO cell are optimized carefully in order to get the largest dynamic aperture.

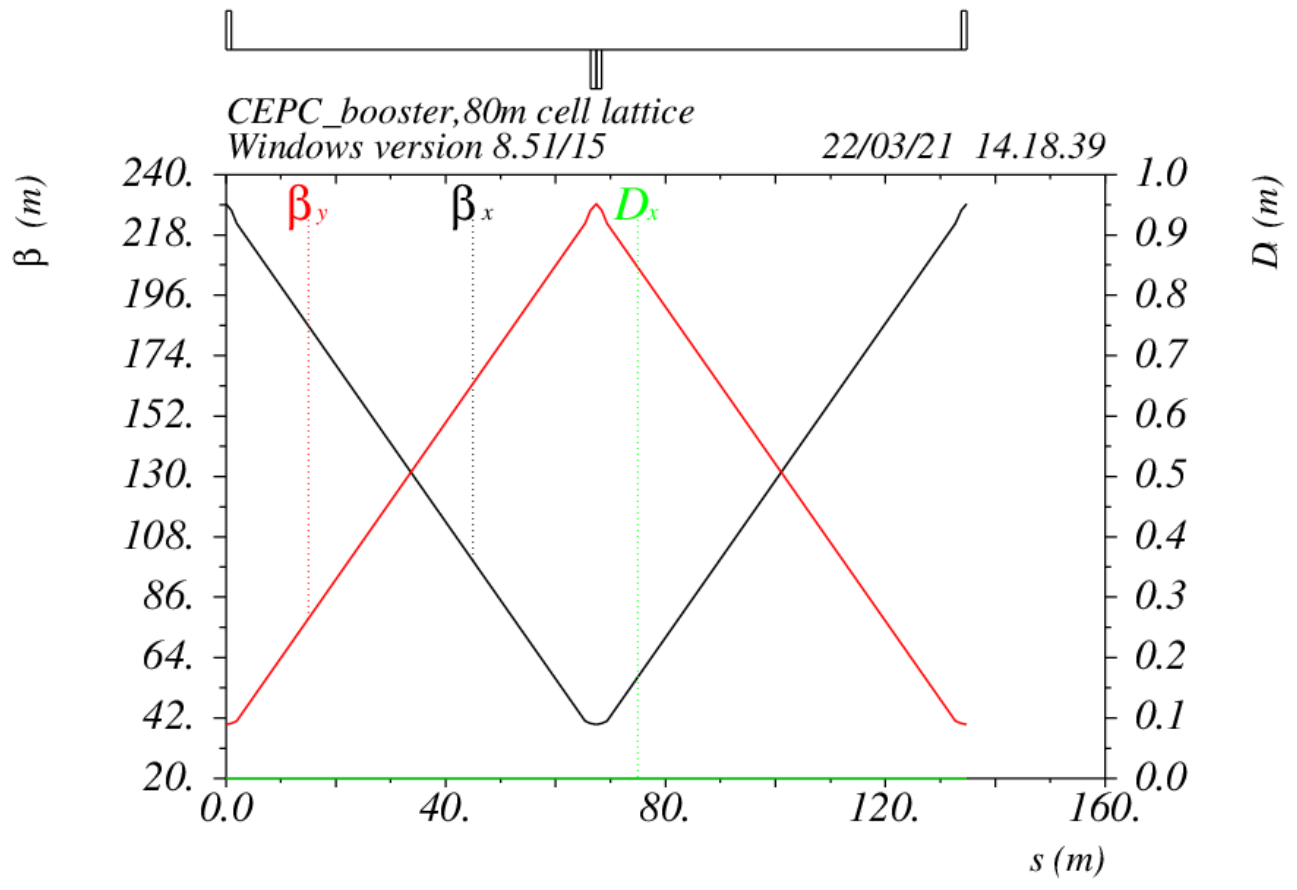

**Figure 5.2.1.4:** The Twiss functions of the FODO cell in the straight section.

### 5.2.1.1.5 *Sawtooth Effect*

The sawtooth effect in the Booster, caused by the fact that there are only two RF stations in the entire ring, results in a maximum orbit distortion of 3 mm at 180 GeV. This off-centre orbit in the sextupoles creates additional quadrupole fields, causing approximately 20% distortion in the optics. The maximum dispersion distortion is around 20 mm and the emittance is increased by 3%. Although the DA reduction due to the sawtooth effect is negligible in the Booster without other errors, the combination of other error effects makes it necessary to taper the quadrupoles individually and taper the dipoles in eight sections while considering the power supply scheme. Figure 5.2.1.5 illustrates the orbit and optics of the booster with sawtooth effect at 180 GeV.

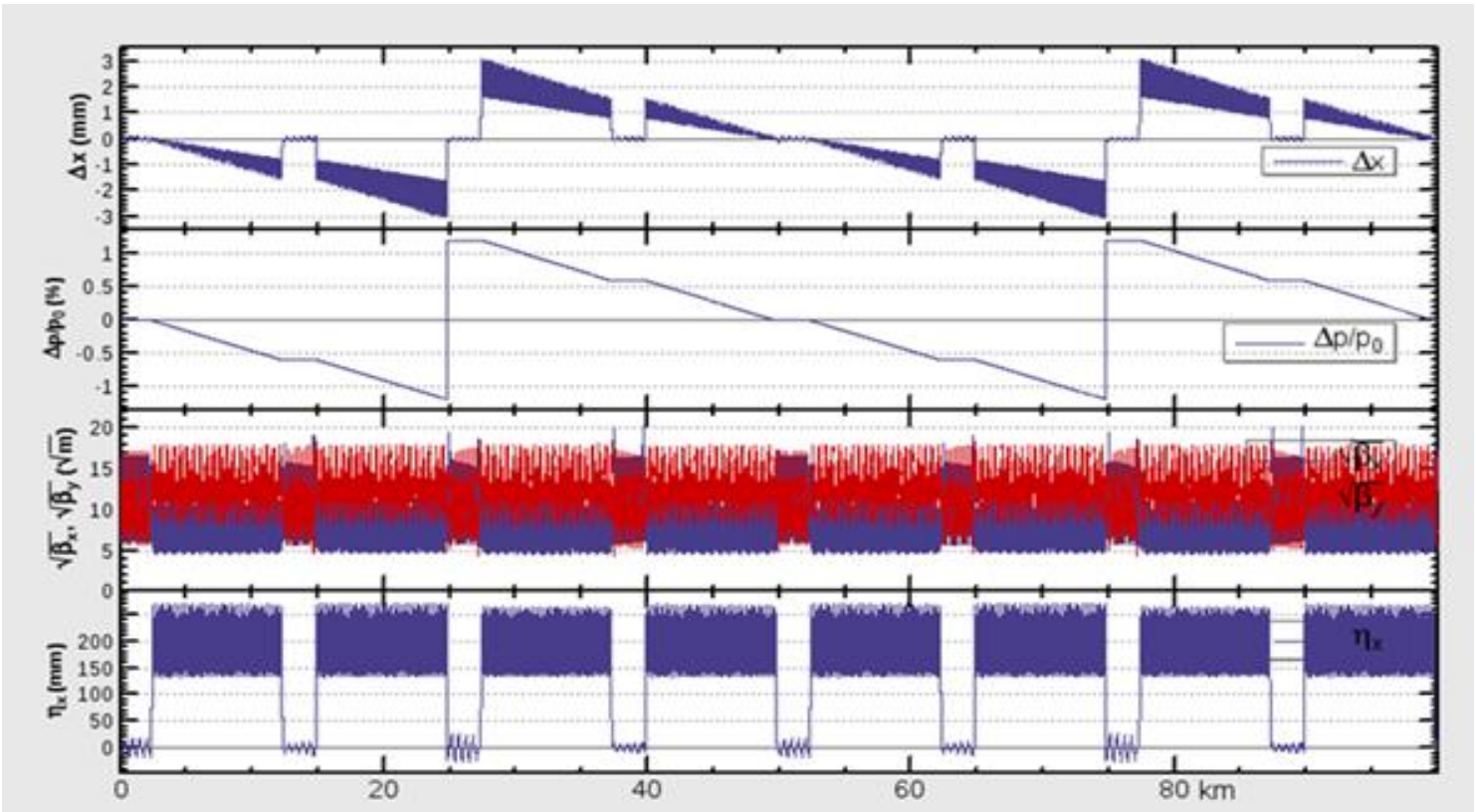

**Figure 5.2.1.5:** Booster orbit and optics with sawtooth effects at 180 GeV.

### 5.2.1.1.6 *Dynamic Aperture*

The dynamic aperture of the bare lattice of the Booster is simulated using SAD with 1500 turns at 30 GeV. Fig. 5.2.1.6 presents the dynamic aperture results of the Booster lattice at 30 GeV without any errors. The left plot in the figure shows the dynamic aperture divided by the injected beam size, while the right plot shows the dynamic aperture divided by the requirement set by the beam stay clear region ($BSC_{x,y} = 4\sigma_{inj_x,y} + O(5\text{mm})$, where $O$ represents the assumed injection orbit error).

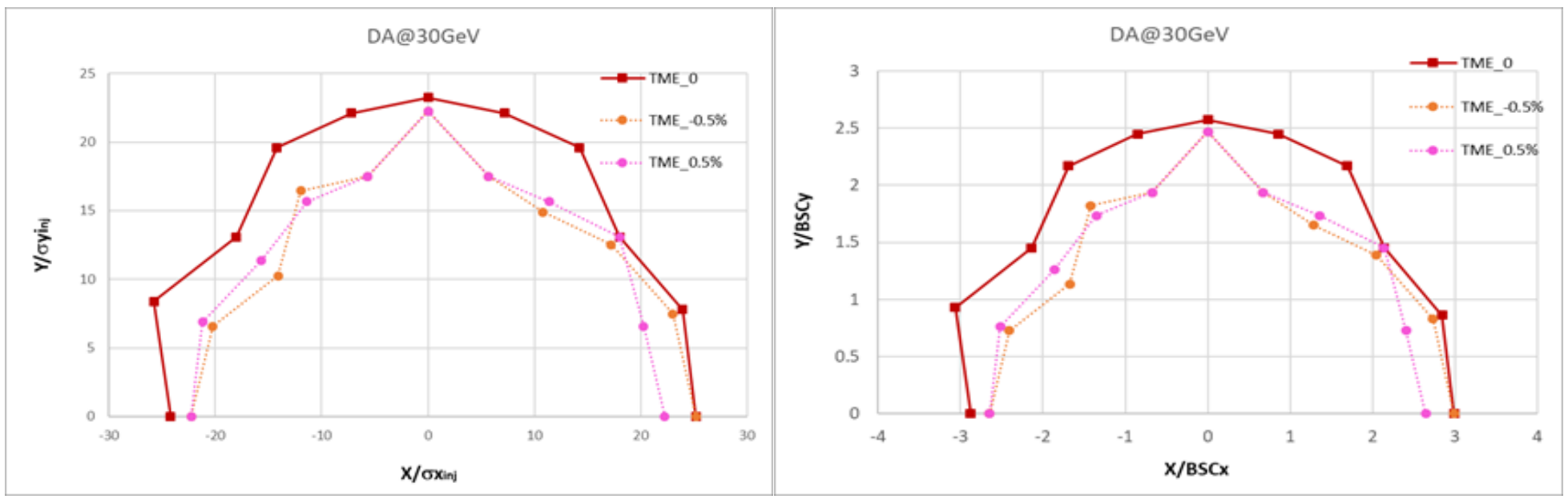

**Figure 5.2.1.6:** The dynamic aperture of the Booster at 30 GeV (left: DA relative to the injected beam size, right: DA relative to the beam stay clear region). The injected emittance from Linac is 6.5 nm for both horizontal and vertical direction. The energy spread of the injected beam is 0.15%.

### 5.2.1.2 *Performance with Errors*

#### 5.2.1.2.1 *Error Analysis*

The details of the error settings are listed in Tables 5.2.1.1, 5.2.1.2 and 5.2.1.3. Gaussian distributions are used for the errors, and they are truncated at 3σ. Due to these errors, the maximum closed orbits are larger than the diameter of the beam pipe, which is 56 mm. Therefore, correction of the first turn trajectory may be necessary. To correct orbit distortions, four horizontal and vertical correctors and eight BPMs are inserted every 2π phase advance. In total, 1218 horizontal correctors and 1220 vertical correctors are used for orbit correction. Figure 5.2.1.7 shows the closed orbit after correction. The maximum orbit is smaller than 0.2 mm after orbit correction.

**Table 5.2.1.1:** Error settings for the magnets.

| Parameters | Dipole | Quadrupole | Sextupole |
|---|---|---|---|
| Transverse shift X/Y (μm) | 100 | 100 | - |
| Longitudinal shift Z (μm) | 100 | 150 | - |
| Tilt about X/Y (mrad) | 0.2 | 0.2 | - |
| Tilt about Z (mrad) | 0.1 | 0.2 | - |
| Nominal field | $1 \times 10^{-3}$ | $2 \times 10^{-4}$ | $3 \times 10^{-4}$ |

**Table 5.2.1.2:** BPM errors.

| Parameters | BPM (10 Hz) |
|---|---|
| Accuracy (m) | $1 \times 10^{-7}$ |
| Tilt (mrad) | 10 |
| Gain | 5% |
| Offset after beam-based alignment (BBA) (mm) | $30 \times 10^{-3}$ |

**Table 5.2.1.3:** Multipole field errors (unit: $1\times10^{-4}$).

| Dipole | Quadrupole |
|---|---|
| $B_1 \leq 2$ | |
| $B_2 \leq 5$ | $B_2 \leq 3$ |
| $B_3 \leq 0.2$ | $B_3 \leq 2$ |
| $B_4 \leq 0.8$ | $B_4 \leq 1$ |
| $B_5 \leq 0.2$ | $B_5 \leq 1$ |
| $B_6 \leq 0.8$ | $B_6 \leq 0.5$ |
| $B_7 \leq 0.2$ | $B_7 \leq 0.5$ |
| $B_8 \leq 0.8$ | $B_8 \leq 0.5$ |
| $B_9 \leq 0.2$ | $B_9 \leq 0.5$ |
| $B_{10} \leq 0.8$ | $B_{10} \leq 0.5$ |

The correction algorithms for the Booster mainly rely on the Matlab-based accelerator toolbox (AT) software. The correction process involves two main steps: orbit correction and optics correction. For orbit correction, the response matrix method is used, with Singular Value Decomposition (SVD) employed to avoid excessive corrector strength and track distortion. Optics correction is achieved through fitting the response matrix with Linear Optics from Closed Orbits algorithm (LOCO), which restores the beta beating and dispersion. The correction process typically converges after three iterations. Table 5.2.1.4 presents statistics of beam parameters after correction iterations using 100 groups of random seeds, while Fig. 5.2.1.7 and Fig. 5.2.1.8 depict the closed orbit distortion and the beta function distortion after corrections.

**Table 5.2.1.4:** Residual errors after correction with 100 random seeds.

| RMS | X | Y |
|---|---|---|
| Orbit (mm) | 0.1300 | 0.0724 |
| Beta beating (%) | 0.57 | 0.19 |
| Dispersion (mm) | 1.82 | 3.5 |

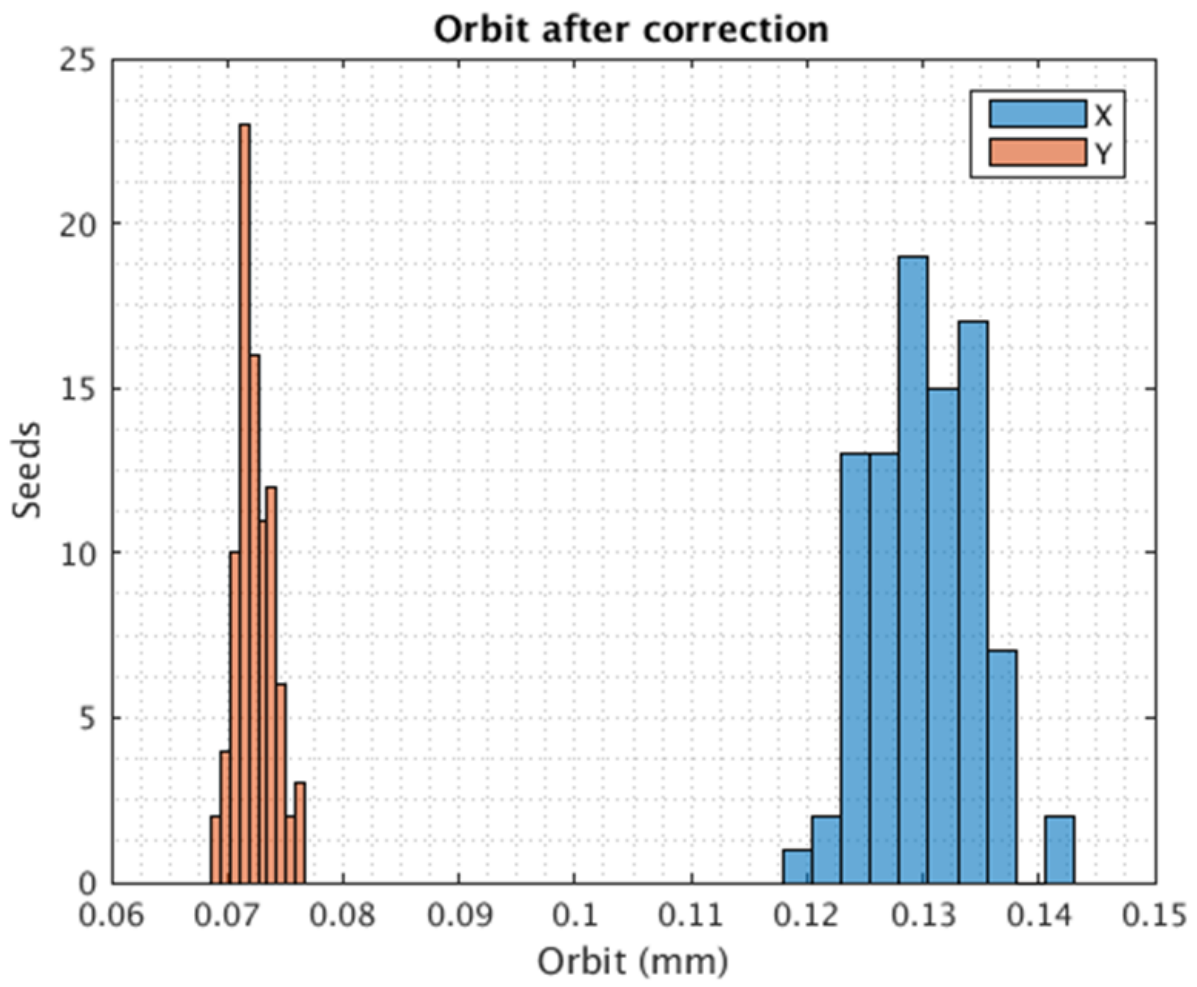

**Figure 5.2.1.7:** Statistical Analysis of RMS closed orbit after COD correction. The RMS values represent orbit distortions at various locations throughout the ring, calculated for each random seed.

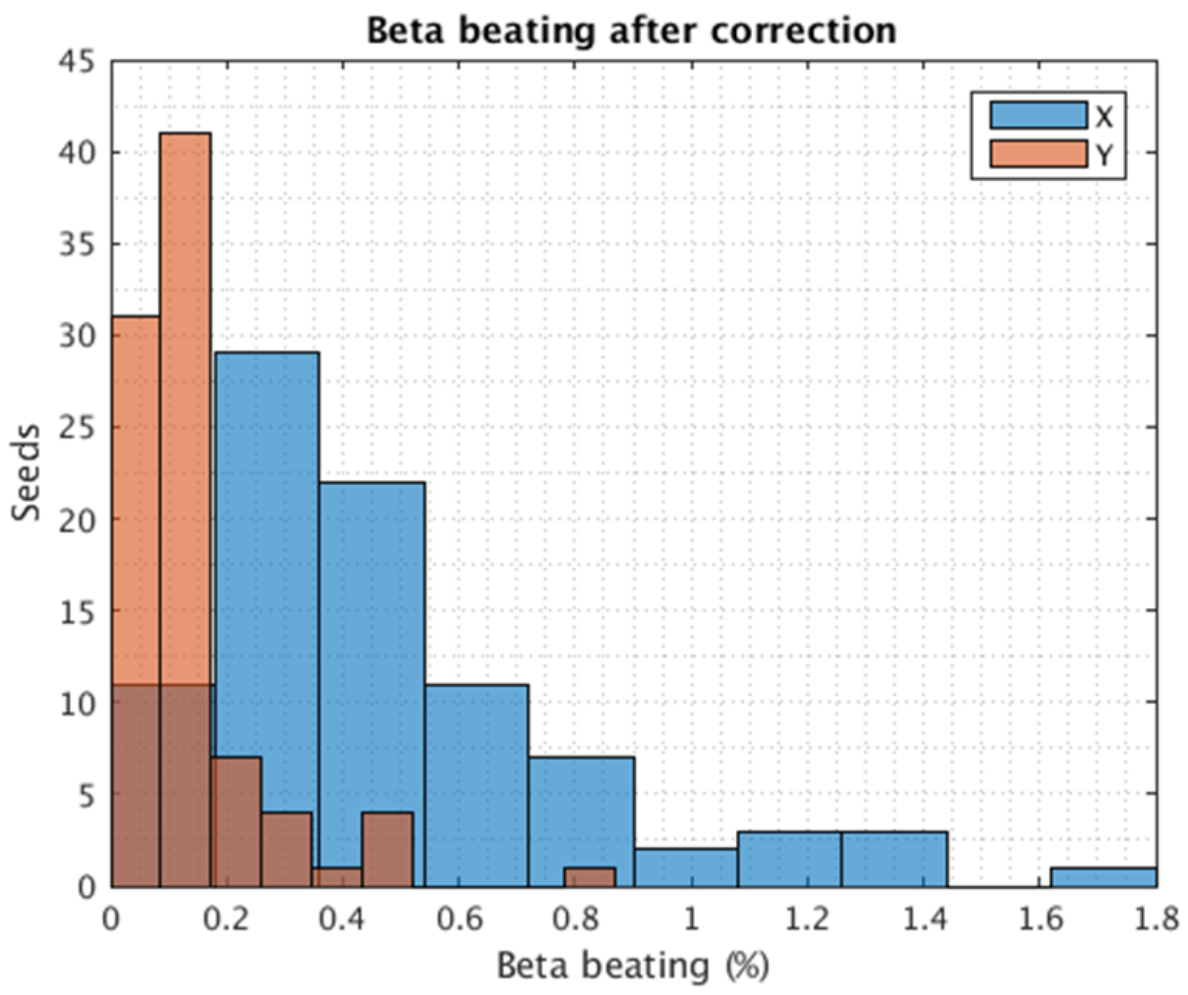

**Figure 5.2.1.8:** Statistical Analysis of RMS beta distortion after optics correction. The RMS values represent beta distortions at various locations throughout the ring, calculated for each random seed.

#### *5.2.1.2.2 Dynamic Aperture with Errors and Corrections*

After transferring the lattice with errors and corrections from AT to SAD, the dynamic aperture is tracked in SAD including the effect of synchrotron radiation. At 30 GeV, the on-axis injection scheme is selected for injecting the beam from the Linac to the Booster. The injected emittance at 30 GeV from the Linac is assumed to be 6.5 nm, and the energy spread is 0.15% at the end of the Linac. The beam stay clear region of the booster at 30 GeV is defined by the equation:

$$BSC_{x,y}(30GeV) = 4\sigma_{inj_x,y} + O \tag{5.2.1.1}$$

where $O$ is the assumed injection orbit error of 5 mm at 30 GeV. The dynamic aperture of the Booster is tracked for 1500 turns at 30 GeV ($\varepsilon_{inj}$ = 6.5 nm), and the results are shown in Fig. 5.2.1.9.

At 45.5 GeV, the beam stay clear region of booster is defined by the equation:

$$BSC_{x,y}(45GeV) = 4\sigma_{x,y} + O \tag{5.2.1.2}$$

where O is the assumed orbit error of 5 mm at 45.5 GeV. The dynamic aperture of the booster is tracked for 1200 turns at 45.5 GeV ($\varepsilon_x$ = 0.18 nm, $\varepsilon_y = \varepsilon_x \times 1\%$), and the results are shown in Fig. 5.2.1.10.

At 80 GeV, the beam stay clear region of the Booster is defined as:

$$BSC_{x,y}(80GeV) = 5\sigma_{x,y} + O \tag{5.2.1.3}$$

where O is the assumed orbit error of 3 mm. The dynamic aperture of the Booster is tracked for 500 turns at 80 GeV ($\varepsilon_x$ = 0.56 nm, $\varepsilon_y = \varepsilon_x \times 1\%$), and the results are shown in Fig. 5.2.1.11.

At 120 GeV, the swap-out scheme is used for the Booster to relax the dynamic aperture requirement of the Collider ring. In the swap-out scheme, off-axis injection from the Collider to the Booster is adopted, and the dynamic aperture requirement of the Booster should be large enough to accept the beam from the Collider ring for both on-momentum and off-momentum particles. The beam stay clear region of the booster at 120 GeV is defined as:

$$\begin{cases} BSC_x(120GeV) = 6\sigma_x + O \\ BSC_y(120GeV) = 39\sigma_y + O \end{cases} \quad (5.2.1.4)$$

where O is the assumed orbit error of 3 mm for both horizontal and vertical directions. The dynamic aperture of the booster is tracked for 250 turns at 120 GeV ($\varepsilon_x$ = 1.26 nm, $\varepsilon_y$ = $\varepsilon_x$ ×1%), and the results are shown in Fig. 5.2.1.12.

At 180 GeV, both on-axis and off-axis injection schemes are considered for the collider ring. For the on-axis injection scheme, the beam stay clear region of the Booster is defined as:

$$\begin{cases} BSC_x(180GeV) = 6\sigma_x + O \\ BSC_y(180GeV) = 50\sigma_y + O \end{cases} \quad (5.2.1.5)$$

For the off-axis injection scheme, the beam stay clear region of the Booster is also defined as:

$$BSC_{x,y}(180GeV) = 5\sigma_{x,y} + O \quad (5.2.1.6)$$

The orbit error, assumed to be 3 mm in both horizontal and vertical directions at 180 GeV, is denoted by O. The dynamic aperture of the Booster is tracked for 150 turns at 180 GeV ($\varepsilon_x$ = 2.84 nm, $\varepsilon_y$ = $\varepsilon_x$ ×1%) and the DA results are shown in Figure 5.2.1.13.

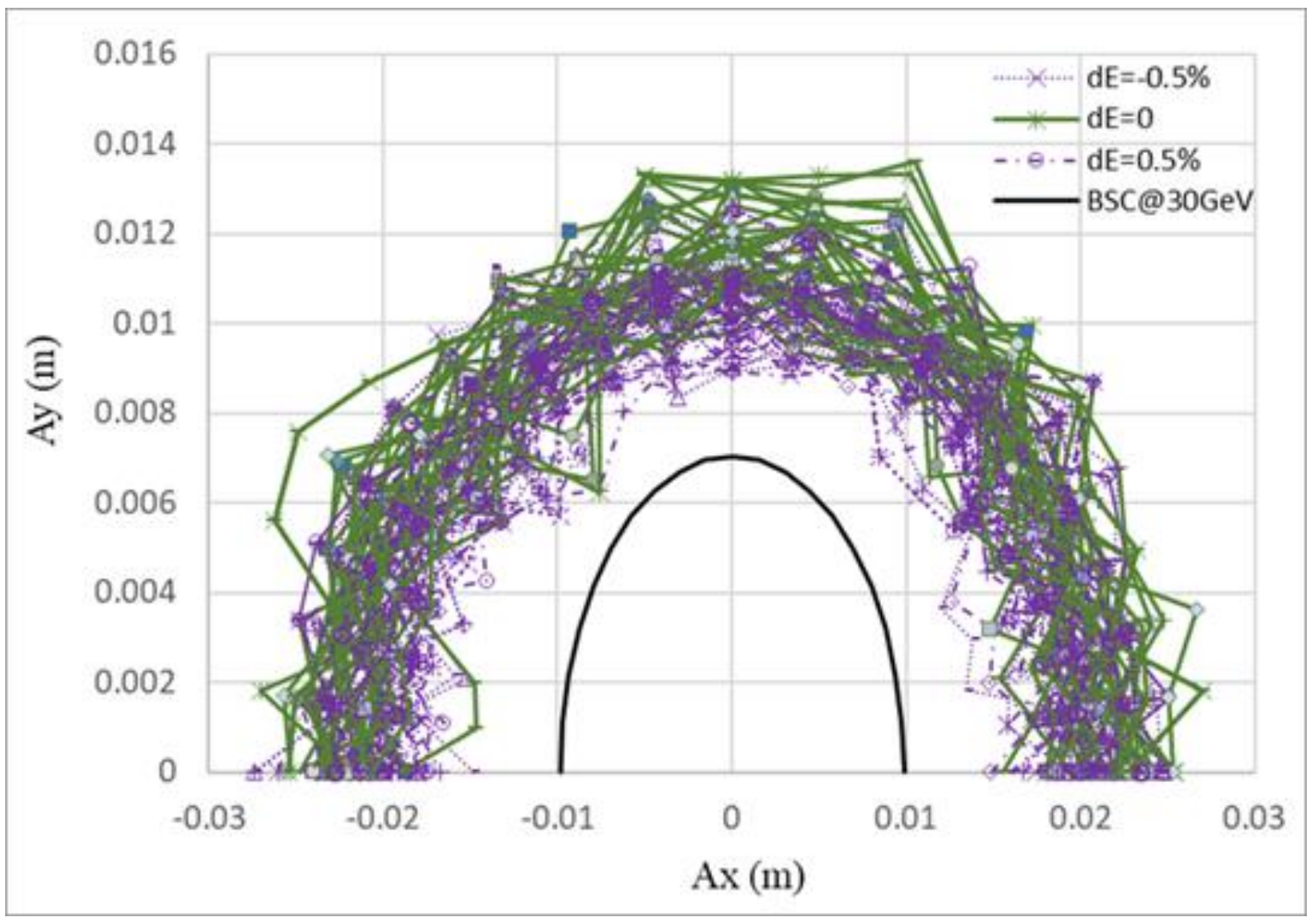

**Figure 5.2.1.9:** Dynamic aperture of the Booster at 30 GeV with errors and synchrotron radiation. The injected emittance from Linac is assumed to be 6.5 nm for both horizontal and vertical directions according to the Linac simulation. An energy deviation of ±0.45% corresponds to 3 times of the injected energy spread from the Linac.

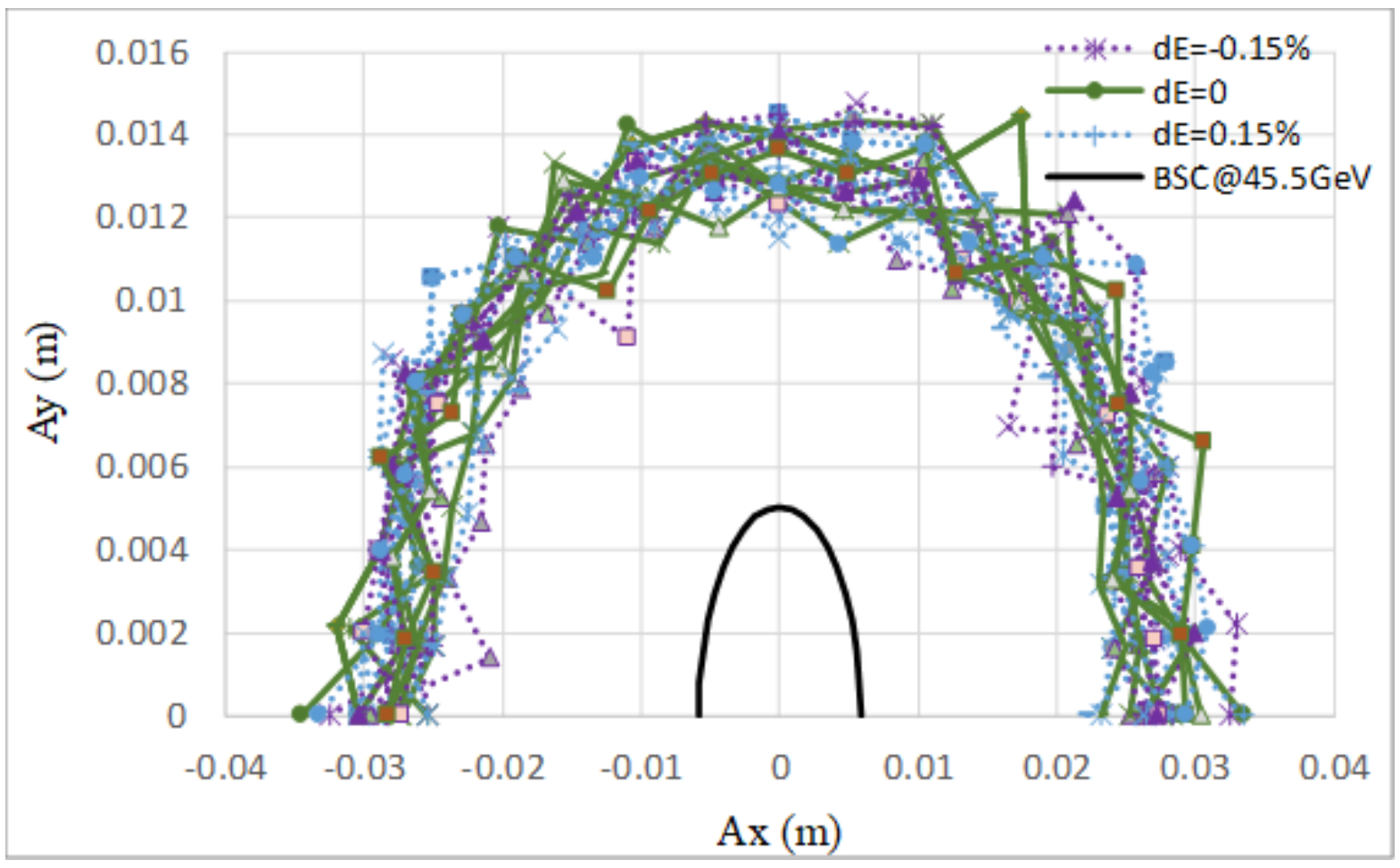

**Figure 5.2.1.10:** Dynamic aperture of the Booster at 45.5 GeV with errors and synchrotron radiation. For the calculation of beam stay clear region in the Booster, the beam emittance is 0.18nm and the coupling is 1%. An energy deviation of ±0.15% corresponds to 4 times of the beam energy spread at this energy.

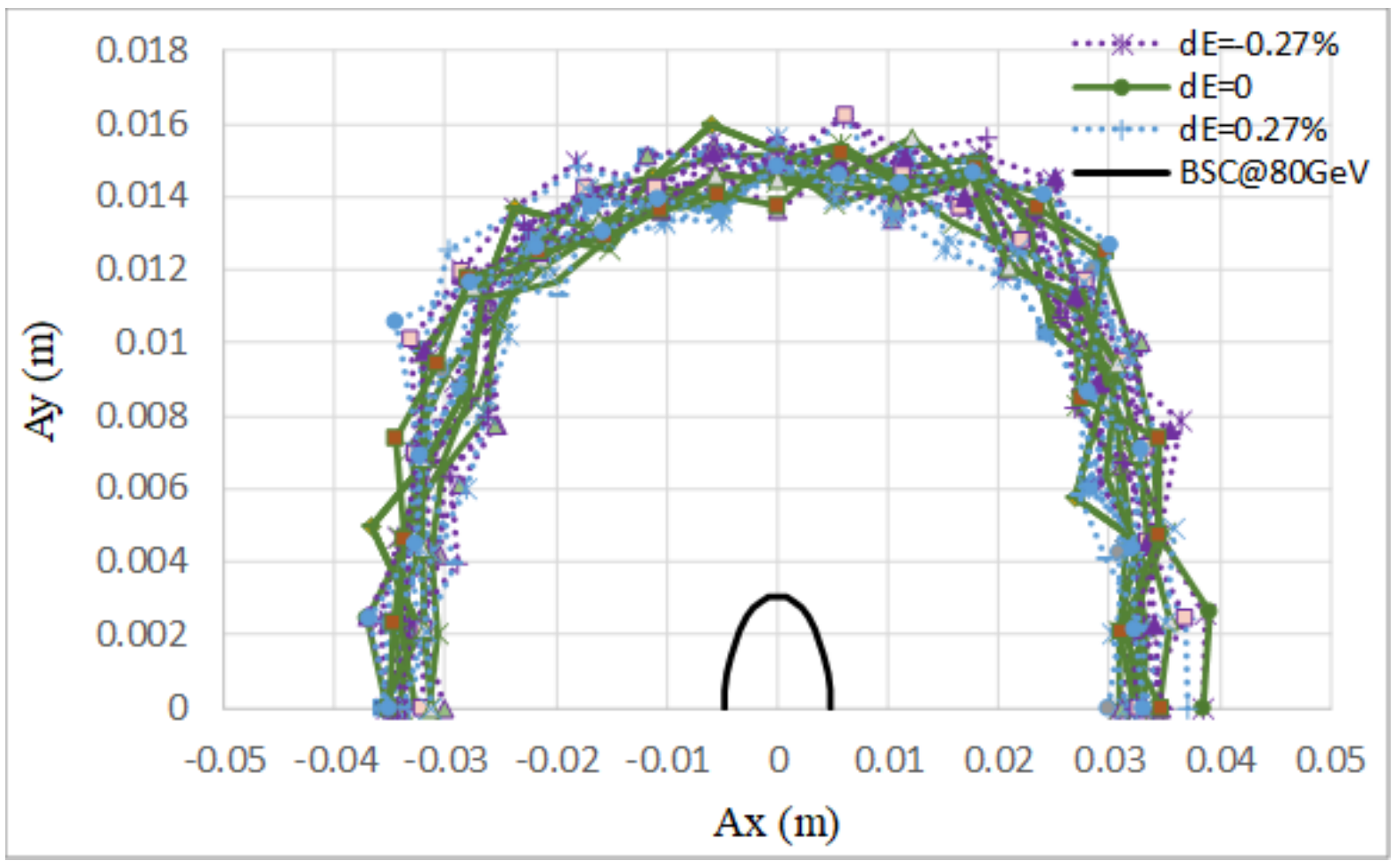

**Figure 5.2.1.11:** Dynamic aperture of the Booster at 80 GeV with errors and synchrotron radiation. For the calculation of beam stay clear region in the Booster, the beam emittance is 0.56 nm and the coupling is 1%. An energy deviation of ±0.27% corresponds to 4 times of the beam energy spread at this energy.

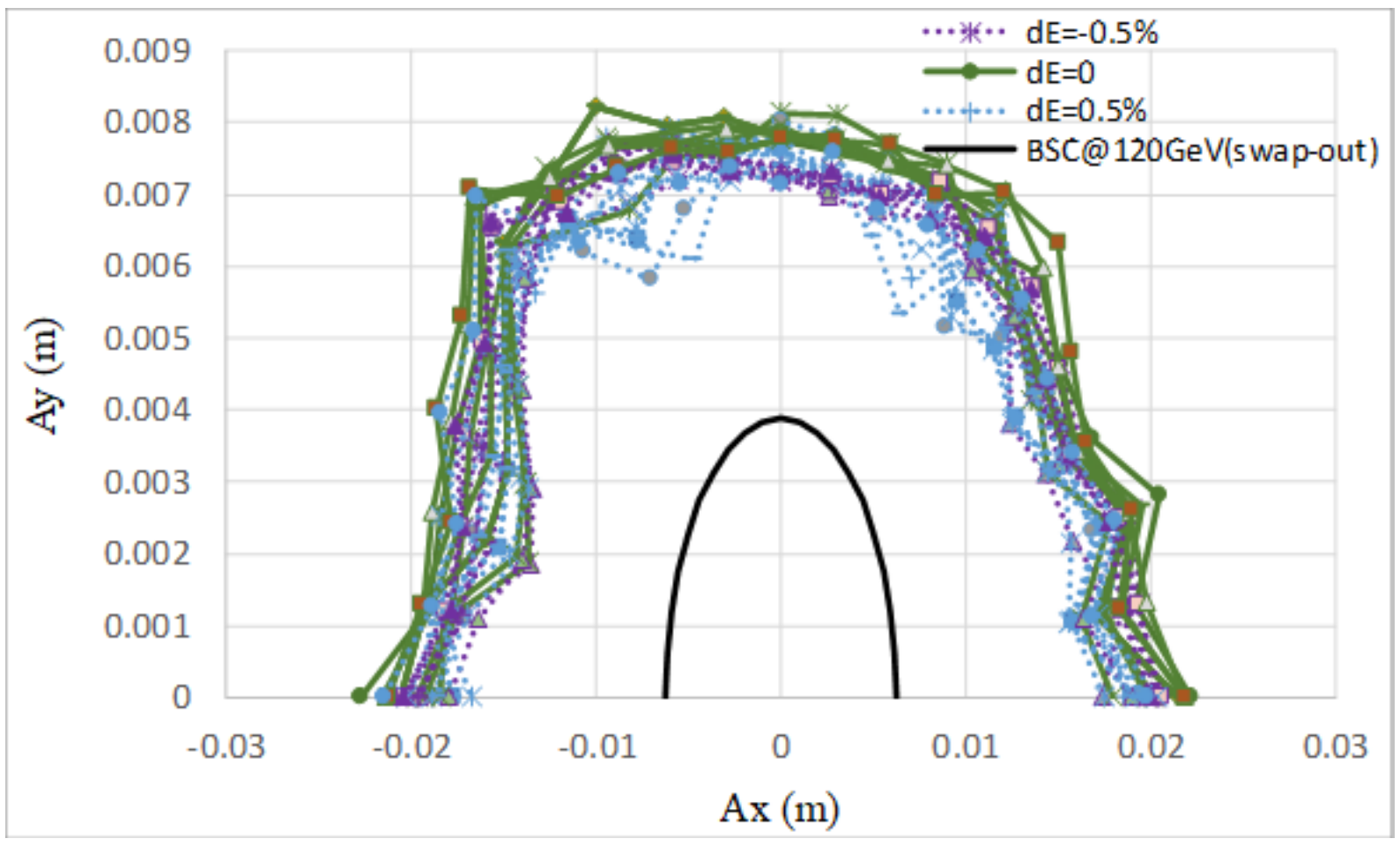

**Figure 5.2.1.12:** Dynamic aperture of the Booster at 120 GeV with errors and synchrotron radiation. At 120 GeV, bunches in the Collider ring need to be injected to the Booster by off-axis injection in the vertical direction. For the calculation of beam stay clear region in the Booster, the beam emittance is 1.26 nm and the coupling is 1%. An energy deviation of ±0.5% corresponds to 5 times of the beam energy spread at this energy.

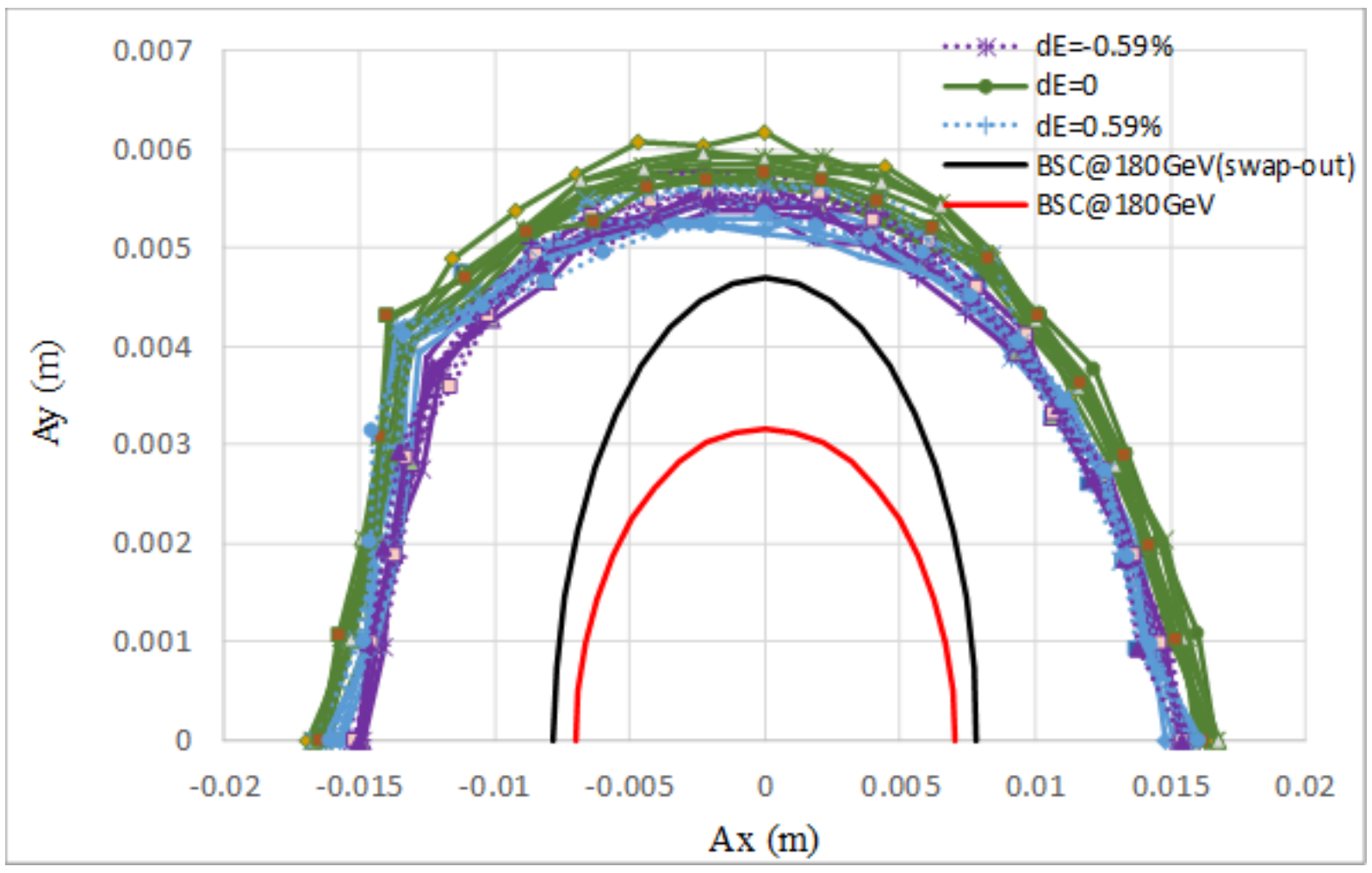

**Figure 5.2.1.13:** Dynamic aperture of the Booster at 180 GeV with errors and synchrotron radiation. At 180 GeV, both swap-out injection and off-axis injection are considered for the Collider ring. For the calculation of beam stay clear regionin in the Booster, the beam emittance is 2.84 nm and the coupling is 1%. An energy deviation of ±0.59% corresponds to 4 times of the beam energy spread at this energy.

#### *5.2.1.2.3 Eddy Current Effect*

An analytical study of the eddy current effect on the Booster has been conducted since the CDR [2], and a ramping curve has been designed to control the maximum sextupole field induced by eddy current, as shown in Eq. (5.2.1.7).

$$B = B_{min} + \frac{B_{max}-B_{min}}{2}(1-\cos[\omega t]) \tag{5.2.1.7}$$

The strength of the normalized sextupole induced on the beam pipe is affected by various factors, such as the beam pipe size, wall thickness, material of the vacuum chamber, and energy ramping speed. In the TDR design, the maximum value of normalized sextupole strength occurs at 45 GeV, reaching 0.000029 $m^{-3}$ (as shown in Fig. 5.2.1.14). Dynamic aperture tracking has been performed in SAD at 45 GeV, including eddy current effects, error effects, and synchrotron radiation effects for 1500 turns. The resulting dynamic aperture at 45 GeV is shown in Fig. 5.2.1.15. To evaluate the effect of dynamic chromaticity during ramping, an extreme case without any corrections was simulated, with only the combined sextupoles open, and the resulting dynamic aperture is shown in Fig. 5.2.1.15.

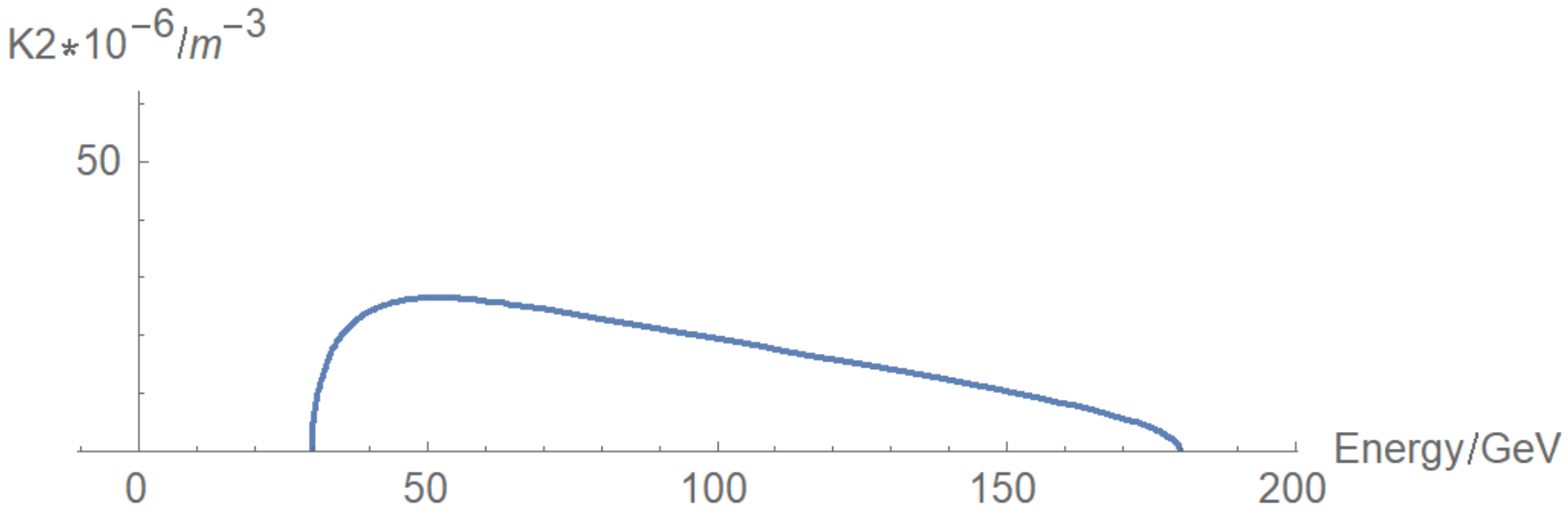

**Figure 5.2.1.14:** The normalized sextupole strength due to eddy current in the Booster vs. beam energy.

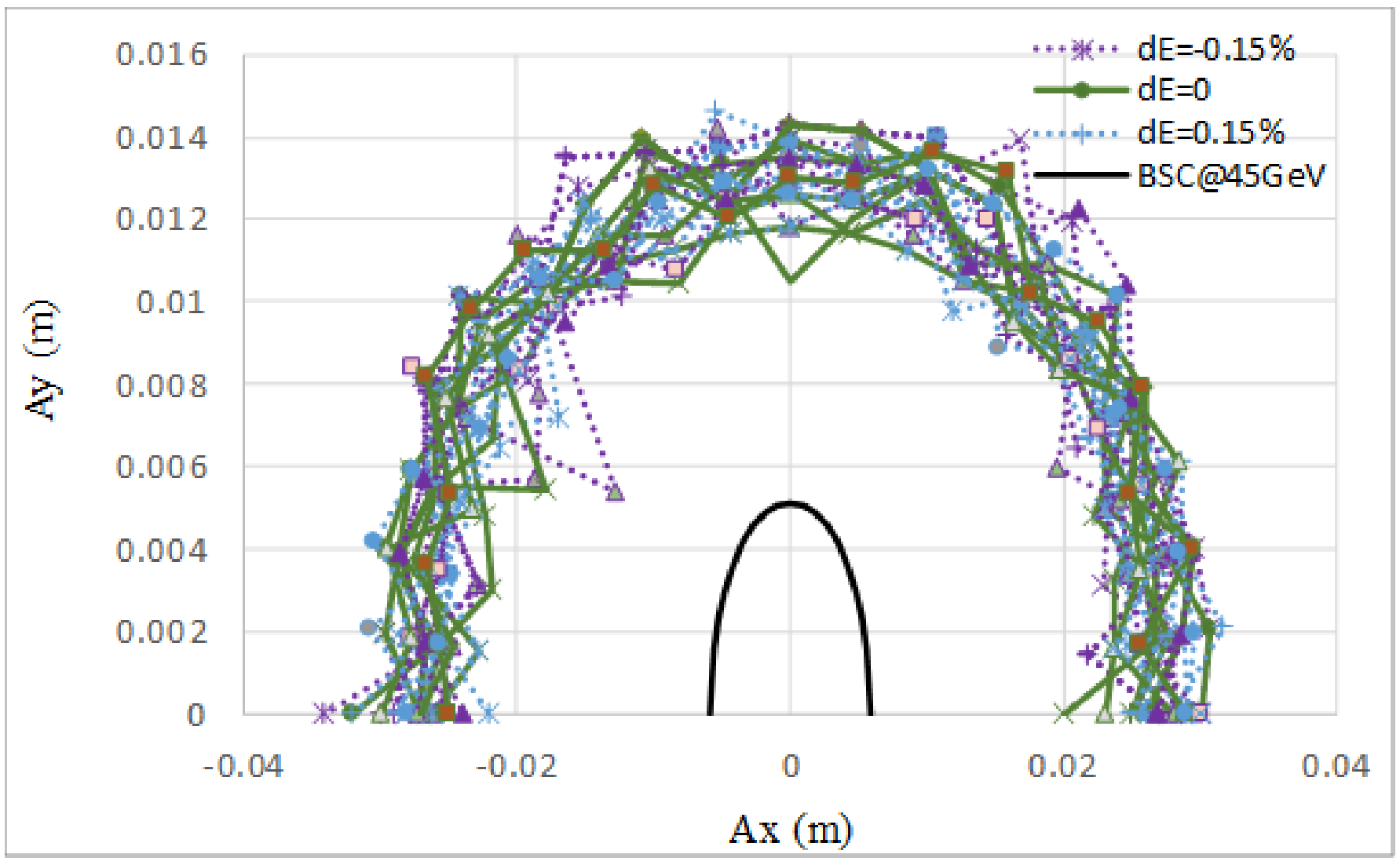

**Figure 5.2.1.15:** Dynamic aperture of the Booster at 45 GeV with eddy current effects, errors, and synchrotron radiation. For the calculation of beam stay clear region ($BSC_{x,y} = 4\sigma_{x,y} + 5$mm), the beam emittance is 0.2 nm and the coupling is 1.5%. An energy deviation of ±0.15% corresponds to 4 times of the beam energy spread at this energy.

### 5.2.1.3 ***Summary***

The new design for a lower emittance Booster is critical for achieving the required luminosity in Higgs mode operation, where the emittance of the Collider ring is halved compared to its value in the CDR. The new optics features a modified-TME lattice, which has reduced the emittance of the booster at 120 GeV from 3.56 nm in the CDR to 1.26 nm. This design has been shown to meet all energy mode requirements, including error effects, synchrotron radiation, and eddy current effects.

### 5.2.2 **Beam Instability**

As an injector for the Collider, the Booster is responsible for providing accumulated electron and positron beams with injection and extraction energies of 30 GeV and 120 GeV, respectively. However, despite the lower beam current in the ring, the coupled impedance from the vacuum components may cause some beam instability, which could lead to deteriorating operation. Therefore, it is crucial to strictly control the impedance.

In this section, we will discuss the impedance budget for the Booster. Based on the impedance studies, we will estimate the beam instabilities at different energies.

#### 5.2.2.1 *Impedance Budget*

Analytical methods are used to estimate the impedance thresholds, which provides a rough criterion for the impedance requirements. The threshold for microwave instability is estimated based on either the Boussard or Keil-Schnell criteria [1]. These criteria impose a limit on the effective broadband impedance, which is estimated to be 88 mΩ.

The transverse mode coupling instability, on the other hand, limits the transverse impedance for a given average bunch current as follows:

$$|Z_{\perp}| \leq \frac{4\nu_s E b}{e I_b R <\beta_{\perp}>} \tag{5.2.2.1}$$

where $\nu_s$ is the synchronous tune, $I_b$ is the bunch current, $R$ and $b$ represent respectively the equivalent ring and beam pipe radius, and $<\beta_{\perp}>$ represents the average transverse beta function. Using the parameters provided in the Table 5.1.1, the threshold $|Z_{\perp}|$ to be approximately 38.76 MΩ/m.

Narrow band impedance limitations can also cause coupled bunch instabilities. A conservative assumption is that the growth rate of the coupled bunch instability for equally spaced bunches should be less than the synchrotron radiation damping. In the longitudinal case, the growth rate is given by [2]:

$$\frac{1}{\tau} = \frac{I_0 \alpha_p}{4\pi(E/e)\upsilon_s} \sum \omega_{pn}\, e^{-(\omega_{pn}\sigma_l)^2}\, Re\, Z_l\,(\omega_{pn}). \tag{5.2.2.2}$$

where $\omega_{pn} = 2\pi f_{rev} \times (pn_b + n + \nu_s)$, $f_{rev}$ is the revolution frequency, and $Z_l$ the longitudinal impedance. This limits the impedance at any resonance frequency of the higher-order-modes (HOMs) in the following way:

$$\frac{f}{\mathrm{GHz}} \frac{Re\, Z}{k\Omega} e^{-(2\pi f \sigma_l / c)^2} < 121.27 \tag{5.2.2.3}$$

Similarly, in the transverse case,

$$\frac{1}{\tau_\perp} = \frac{I_0 f_{rev} \langle\beta\rangle}{(E/e)} \sum \omega_{pn}\, e^{-(\omega_{pn}\sigma_l)^2}\, Re\, Z_\perp\,(\omega_{pn}), \tag{5.2.2.4}$$

where $\omega_{pn} = \omega_{rev} \times (pn_b + n + \nu_{x,y})$. This gives the limit:

$$\frac{Re\, Z}{G\Omega/\mathrm{m}} e^{-(2\pi f \sigma_l/c)^2} < 0.82 \tag{5.2.2.5}$$

The impedance requirements for the Booster are summarized in Table 5.2.2.1. The main sources of impedance are the RF cavities, BPMs, bellows, masks, ports of vacuum pumps, collimators, injection kickers, feedback kickers, valves, and flanges. In the booster ring, the threshold for broadband impedance is more than five times that in colliders. Therefore, the impedance budgets for different components can be scaled accordingly, as shown in Table 5.2.2.2.

**Table 5.2.2.1:** Impedance threshold for the Booster

| Parameter | Symbol | Booster (30 GeV) |
|---|---|---|
| Threshold of broadband $Z_L$ | $\lvert Z_L/n \rvert_{eff}$, | 50.2mΩ |
| Threshold of broadband $Z_T$ | $\lvert Z_\perp \rvert$ MΩ/m | 38.76 MΩ/m |
| Threshold of narrowband $Z_L$ | $\frac{f}{\mathrm{GHz}} \frac{Re\, Z_L}{\mathrm{G\Omega}} e^{-(2\pi f \sigma_l)^2}$ | 121.27 |
| Threshold of narrowband $Z_T$ | $\frac{Re\, Z_\perp}{\mathrm{G\Omega/m}} e^{-(2\pi f \sigma_l)^2}$ | 0.82 |

**Table 5.2.2.2:** Impedance threshold for different components in the Booster

| Components | $Z_{\|}/n$, mΩ | $\lvert Z_\perp \rvert$ MΩ/m |
|---|---|---|
| Resistive wall | 27.64 | 21.68 |
| RF cavities | -4.46 | 0.58 |
| Flanges | 12.48 | 5.37 |
| BPMs | 0.53 | 0.58 |
| Bellows | 9.81 | 5.56 |
| Pumping ports | 0.09 | 1.15 |
| Feedback kickers | 0.09 | 2.49 |
| kickers | 0.45 | 0.38 |
| Taper transitions | 3.57 | 0.96 |
| Total | 50.2 | 38.76 |

### 5.2.2.2 *Instability Threshold*

Due to the lower energy during injection, beam instability caused by impedance can be more severe in the Booster than in the Collider. The resistive wall impedance varies with different vacuum pipe materials. After comparing the impedance of stainless steel, copper, and aluminum, an aluminum chamber with a radius of 28 mm was selected for the Booster vacuum chamber. The dominant factor limiting the bunch current is the transverse mode coupling instability (TMCI) caused by the transverse resistive wall impedance [3]. The formula for calculating the TMCI threshold current is:

$$I_b^{th} \approx 0.7 \frac{4\pi c \nu_s (E/e)}{C} \frac{1}{\sum_i \beta_{t,i} \kappa_{t,i}} \qquad (5.2.2.6)$$

The TMCI threshold current formula gives a bunch current threshold of 6.3 μA, which is higher than the design current of 2.3 μA, for Higgs mode operation at 30 GeV. In the above formula, $\nu_s$ represents the synchrotron tune, $\beta_{t,j}$ is the average beta function in the $j^{th}$ element (dipole, quadrpole, sextupole, etc.), $\kappa_{t,j}$ is the transverse loss factor, and $E$ is the beam injection energy.

Based on the resistive wall impedance for the aluminum chamber with a radius of 28 mm, the bunch current threshold and growth time of multi-bunch coupled instability were calculated for different operation modes at 30 GeV to 180 GeV, and the results are listed in Table 5.2.2.3 and Table 5.2.2.4.

For the Higgs mode, an alternative beam injection scheme for the Collider is the on-axis method. This injection scheme involves injecting some high current bunches from the Collider (7-14 bunches with a current of 61 μA/bunch) into the Booster to combine with the existing bunches, which are then re-injected into the Collider. With this injection scheme, the bunch current threshold is 70 μA, and the multi-bunch instability growth time is about 1215 ms.

**Table 5.2.2.3:** Parameters and instability threshold for the Booster at injection energy

| **Parameters** | **Unit** | ***tt*** | ***H*** | ***W*** | ***Z*** | |
|---|---|---|---|---|---|---|
| Beam energy | GeV | 30 | | | | |
| Bunch number | | 35 | 268 | 1297 | 3978 | 5967 |
| Threshold of single bunch current | μA | 8.68 | 6.3 | 5.8 | | |
| Threshold of beam current (limited by coupled bunch instability) | mA | 97 | 106 | 100 | 93 | 96 |
| Bunch charge | nC | 1.1 | 0.78 | 0.81 | 0.87 | 0.9 |
| Single bunch current | μA | 3.4 | 2.3 | 2.4 | 2.65 | 2.69 |
| Beam current | mA | 0.12 | 0.62 | 3.1 | 10.5 | 16.0 |
| Growth time (coupled bunch instability) | ms | 2530 | 530 | 100 | 29.1 | 18.7 |
| Energy spread | % | 0.025 | | | | |
| Synchrotron radiation loss/turn | MeV | 6.5 | | | | |
| Momentum compaction factor | $10^{-5}$ | 1.12 | | | | |
| Emittance | nm | 0.076 | | | | |
| Natural chromaticity | H/V | -372/-269 | | | | |
| RF voltage | MV | 761.0 | 346.0 | 300.0 | | |
| Betatron tune $\nu_x/\nu_y$ | | 321.23/117.18 | | | | |
| Longitudinal tune | | 0.14 | 0.0943 | 0.0879 | | |
| RF energy acceptance | % | 5.7 | 3.8 | 3.6 | | |
| Damping time | s | 3.1 | | | | |
| Bunch length of linac beam | mm | 0.4 | | | | |
| Energy spread of linac beam | % | 0.15 | | | | |
| Emittance of linac beam | nm | 6.5 | | | | |

**Table 5.2.2.4:** Parameters and instability threshold for the booster at extraction energy

| Parameters | Unit | *tt* | *H* | | *W* | *Z* | |
|---|---|---|---|---|---|---|---|
| | | Off axis injection | Off axis injection | On axis injection | Off axis injection | Off axis injection | |
| Beam energy | GeV | 180 | 120 | | 80 | 45.5 | |
| Bunch number | | 35 | 268 | 261+7 | 1297 | 3978 | 5967 |
| Maximum bunch charge | nC | 0.99 | 0.7 | 20.3 | 0.73 | 0.8 | 0.81 |
| Maximum single bunch current | μA | 3.0 | 2.1 | 61.2 | 2.2 | 2.4 | 2.42 |
| Threshold of single bunch current | μA | 91.5 | 70 | | 22.16 | 9.57 | |
| Threshold of beam current (limited by RF system) | mA | 0.3 | 1 | | 4 | 16 | |
| Beam current | mA | 0.11 | 0.56 | 0.98 | 2.85 | 9.5 | 14.4 |
| Growth time (coupled bunch instability) | ms | 16611 | 2359 | 1215 | 297.8 | 49.5 | 31.6 |
| Bunches per pulse of Linac | | 1 | 1 | | 1 | 2 | |
| Time for ramping up | s | 7.1 | 4.3 | | 2.4 | 1.0 | |
| Injection duration for top-up (Both beams) | s | 29.2 | 23.1 | 31.8 | 38.1 | 134.4 | |
| Injection interval for top-up | s | 65 | 38 | | 155 | 153.5 | |
| Current decay during injection interval | | 3% | | | | | |
| Energy spread | % | 0.15 | 0.099 | | 0.066 | 0.037 | |
| Synchrotron radiation loss/turn | GeV | 8.45 | 1.69 | | 0.33 | 0.034 | |
| Momentum compaction factor | $10^{-5}$ | 1.12 | | | | | |
| Emittance | nm | 2.83 | 1.26 | | 0.56 | 0.19 | |
| Natural chromaticity | H/V | -372/-269 | | | | | |
| Betatron tune $\nu_x/\nu_y$ | | 321.27/117.19 | | | | | |
| RF voltage | GV | 9.7 | 2.17 | | 0.87 | 0.46 | |
| Longitudinal tune | | 0.14 | 0.0943 | | 0.0879 | 0.0879 | |
| RF energy acceptance | % | 1.78 | 1.59 | | 2.6 | 3.4 | |
| Damping time | ms | 14.2 | 47.6 | | 160.8 | 879 | |
| Natural bunch length | mm | 1.8 | 1.85 | | 1.3 | 0.75 | |
| Full injection from empty ring | hr | 0.1 | 0.14 | 0.16 | 0.27 | 1.8 | 0.8 |

### 5.2.3 Injection, Extraction and Transport Lines

#### 5.2.3.1 *Injection into the Boostet*

In the CEPC complex, the Booster is positioned between the Linac and the Collider. The beam produced by the Linac must first be injected into the Booster for acceleration, and then the accelerated beam must be extracted and injected into the Collider. To simplify the injection structure, the design for injection from the Linac to the Booster utilizes a single bunch on-axis injection, as depicted in Figure 5.2.3.1, which illustrates the magnet arrangement in the Booster injection area. The positron and electron beams generated and accelerated by the Linac are transported to the Lambertson magnet in the Booster plane through the Booster injection transport line. Figure 5.2.3.2 illustrates the relationship between the injected bunch and the circulating bunch at the Lambertson magnet. The Booster extraction system is similar to the injection system, consisting of a single kicker and a septum.

The timing pattern of the bunches in the Linac is determined by the frequencies of the linear accelerator and the subharmonic bunchers. To ensure successful injection of all Linac bunches into the Booster buckets, their spacing must be integer multiples of a common time interval. The frequency choice in the Linac is primarily determined by the requirements of the booster's Z-mode, which requires the highest number of injected bunches.

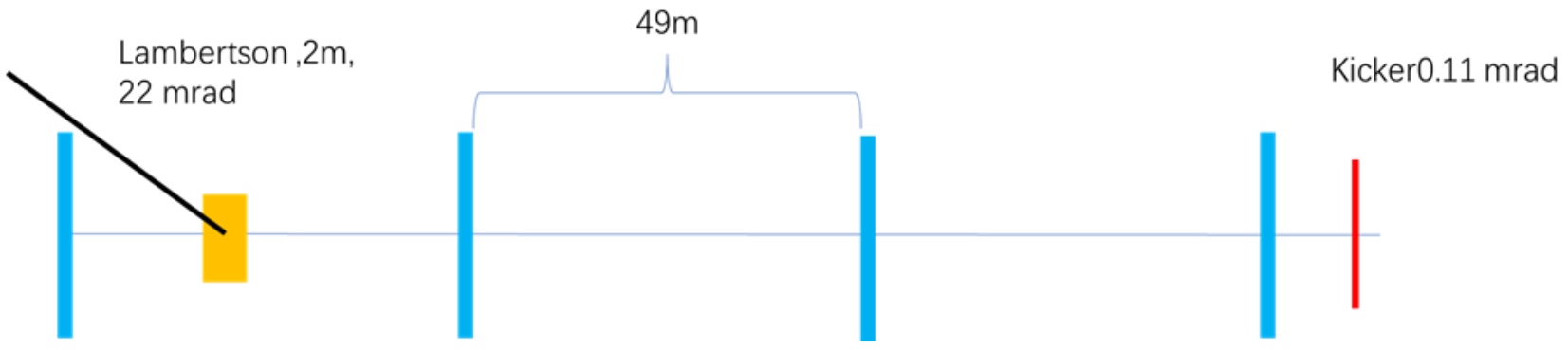

**Figure 5.2.3.1:** Magnet arrangement in the Booster injection area

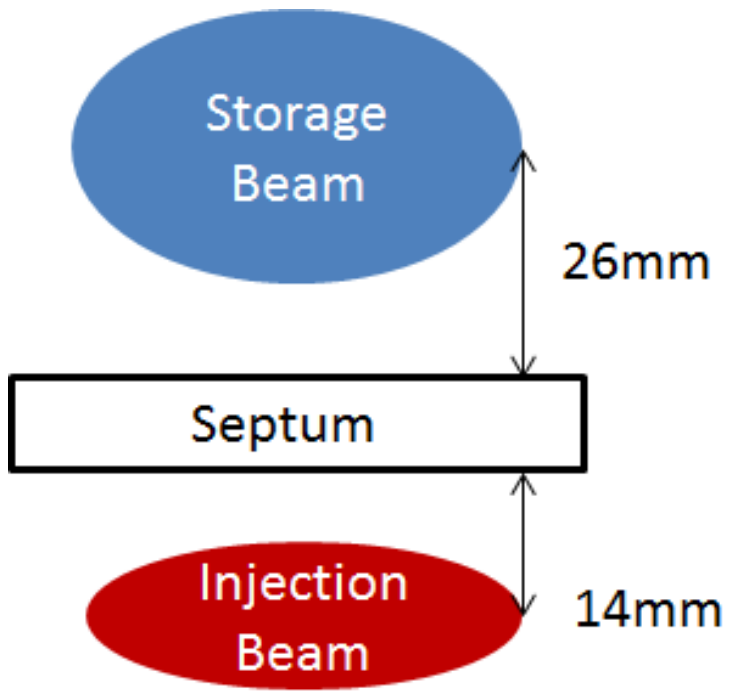

**Figure 5.2.3.2:** Relative bunch positions at the Lambertson magnet.

To ensure smooth injection from the Booster to the Collider, the bunch distribution structure in the Booster is dependent on the bunch pattern in the Collider. In the case of the Higgs, W, and $t\bar{t}$ modes, the Booster utilizes the same bunch distribution as the Collider. For the $t\bar{t}$ and Higgs modes, the number of bunches per beam is split in half in both the Booster and Collider. In contrast, for the W mode, the bunches are evenly distributed throughout the entire ring for both the Booster and Collider.

For the Z mode, a large interval is reserved between adjacent bunch trains in the Collider to account for the rise and fall time of the injection kicker. The Booster also adopts the bunch train mode, with the number of bunches and intervals per bunch train being the same as the Collider. However, due to current limitations in the Booster, the number of bunch trains is only 1/3 of that in the Collider. Table 5.2.3.1 shows the parameters of Booster injection.

**Table 5.2.3.1.** Parameters of Booster injection

| | H | W | Z | $t\bar{t}$ |
|---|---|---|---|---|
| Energy (GeV) | 30 | 30 | 30 | 30 |
| Bunch number | 268 | 1297 | 3978 | 35 |
| Bunch separation (μs) | 0.591 | 0.257 | 0.023 | 4.52 |
| Injection scheme | Bunch by bunch | Bunch by bunch | Bunch train | Bunch by bunch |
| Kicker frequency (Hz) | 100 | 100 | 100 | 100 |
| Kicker rise/fall time (μs) | < 0.59 | < 0.25 | < 5.1 | < 4.5 |
| Injection duration (s) | 2.68 | 12.9 | 0.51 | 0.35 |

The bunches that are sent from the Linac to the Booster always maintain a uniform time structure of either 100 Hz single bunch or 100 Hz double bunch. However, the bunches in the Booster have different spacing in different energy modes. To ensure that the bunches are successfully injected into the bunch bucket in the Booster, a specific relationship needs to be satisfied between the time constants. This relationship is given by $T_{Linac} = N \times T_0 + T_{Booster}$, where $T_0$ is the bunch circulating period in the Booster, $T_{Booster}$ is the bunch interval in the Booster, $T_{Linac}$ is the interval time of the bunches from the Linac, and N is an integer. By slightly adjusting $T_{Linac}$, it is possible to fill the Booster with any pattern that is needed. Table 5.2.3.2 shows the bunch separation for Booster injection.

**Table 5.2.3.2.** Bunch separation for Booster injection

| | H | W | Z | $t\bar{t}$ |
|---|---|---|---|---|
| Energy | 120 | 80 | 45 | 180 |
| Bumch number in Booster | 268 | 1297 | 3978 | 35 |
| Bunch separation (ns) | 591 | 257 | 23 | 4524 |
| Bunch separation from Linac (ms) | 10.0063 | 10.0067 | 10.0069 | 10.0024 |

### 5.2.3.2 *Extraction from the Booster*

The structure design of the magnets in the beam extraction section of the Booster is nearly identical to that in the injection section. To extract the circulating beam out of the Booster, a single kicker deflects it, following which it enters the Lambertson magnet and is then deflected into the outgoing transport line. The extracted bunches are subsequently injected into the Collider. The beam extraction time depends on the bunch structure in the Booster and Collider. For Z energy, this process becomes more complex due to the large number of bunches required in the Collider, with a separation between bunches of only 20~30 ns. To account for the rise time of the kickers, the bunches are arranged train by train in the Collider for Z energy. However, for Higgs, W, and $t\bar{t}$, the bunches are extracted one by one. The arrangement of the bunches in the Collider and Booster for Z energy is shown in Figure 5.2.3.3. The parameters of Booster extraction is shown in Table 5.2.3.3.

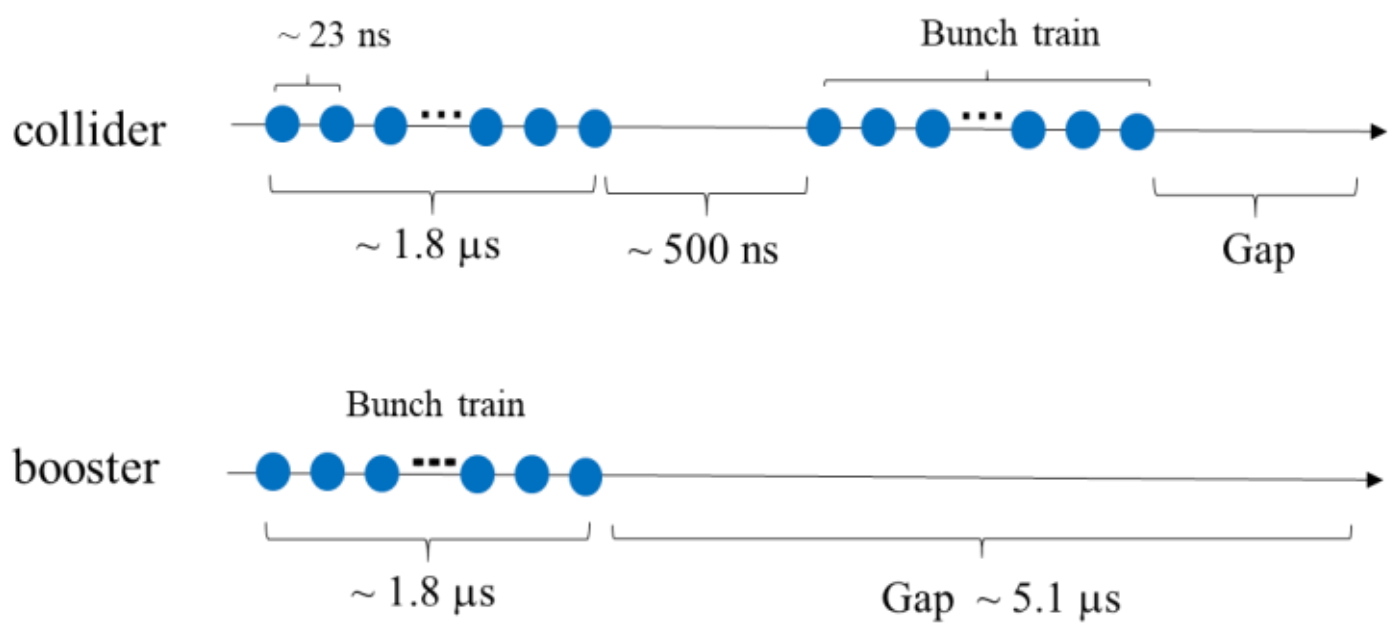

**Figure 5.2.3.3:** Bunch Structure in Booster and Collider.

**Table 5.2.3.3:** Parameters of Booster extraction.

| | H | W | Z | $t\bar{t}$ |
|---|---|---|---|---|
| Energy (GeV) | 120 | 80 | 45 | 180 |
| Bunch number | 268 | 1297 | 3978 | 35 |
| Bunch separation (μs) | 0.591 | 0.257 | 0.023 | 4.52 |
| Extraction scheme | Bunch by bunch | Bunch by bunch | Bunch train | Bunch by bunch |
| Kicker frequency (Hz) | 1000 | 1000 | 1000 | 1000 |
| Kicker rise/fall time (μs) | < 0.59 | < 0.25 | < 5.1 | < 4.5 |
| Extraction duration (s) | 0.268 | 1.297 | 0.051 | 0.035 |

### 5.2.3.3 ***Transport Lines***

The transport line that runs from the straight line to the Booster has the task of separating the positron and electron beams that are transmitted from the Linac, and injecting them into the Booster respectively. To reduce the cost of construction, the Linac is placed on the ground level, while the Booster is positioned 100 meters underground in the Collider tunnel. To accomplish this, the transport line must also deflect the beam vertically to transport it from the Linac to the underground tunnel level. As a result, the transport line to the Booster comprises of three components: beam separation, vertical bending, and horizontal transport. The twiss parameters are shown in Figure 5.2.3.5, some important components included in the transport lines are shown in Table 5.2.3.4.

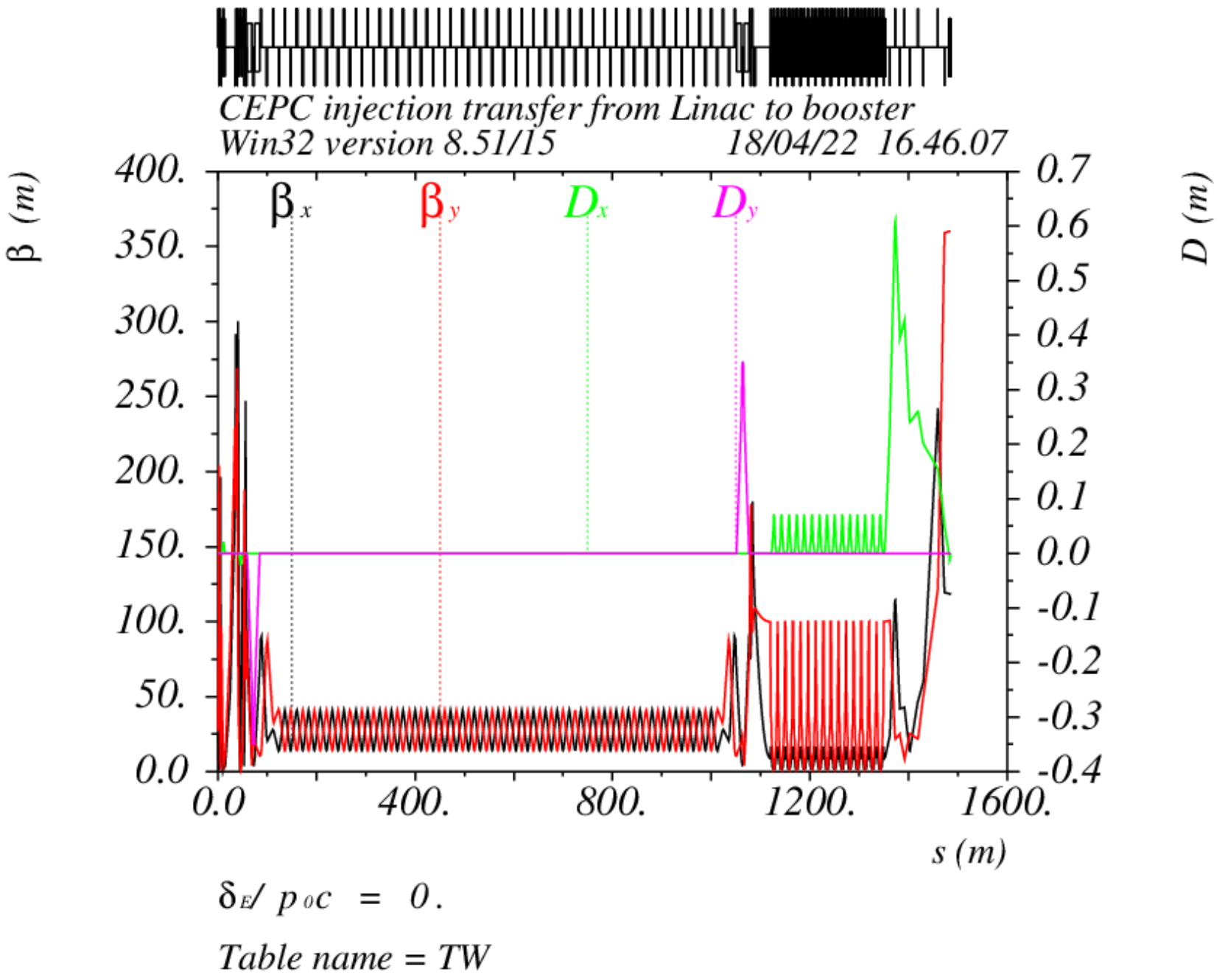

**Figure 5.2.3.5:** Twiss parameters of the transport line.

**Table 5.2.3.4:** Important components in the transport line

| | Numbers | Length (m) | Max Field (T) |
|---|---|---|---|
| Booster injection Kickers | 2 | 1 | 0.011 |
| Booster injection septums | 2 | 2 | 0.46 |
| Booster extraction Kickers | 2 | 1 | 0.12 |
| Booster extraction septums | 2 | 30 | 0.87 |
| Dipoles | 36 | 5 | 1.5 |
| Qudrupoles | 80 | 0.9 | |
| Correctors | 24 | 0.2 | |
| BPMs | 28 | | |
| PR | 2 | | |

## 5.2.4 Synchrotron Radiation and Shield

### 5.2.4.1 *Parameters and Simulation Setup*

The previous sections 4.2.4 and 4.3.6 discussed the 50 MW synchrotron-radiation power in the Collider. This section focuses on the 50 MW case to study the absorbed doses to the Booster. The parameters for this case are listed in Table 5.2.4.1. The beam energy of the Booster varies frequently, and for the Higgs mode, it changes as a function of time, as shown by the blue line in Figure 5.2.4.1. The initial value of the beam energy is 30 GeV when injected from the Linac. According to the parameters in Table 5.2.4.1, the beam energy increases from 30 GeV to 120 GeV in 16.4 seconds before being injected into the Collider. The magnetic field then decreases to prepare for the next injection. The entire ramping process for the Higgs mode takes 38 seconds.

**Table 5.2.4.1:** Booster parameters for 50 MW case.

| | Higgs (on axis) | Z | W | $t\bar{t}$ |
|---|---|---|---|---|
| Beam energy (GeV) | 30~120 | 30~45.5 | 30~80 | 30~180 |
| Current (mA) | 0.98 | 9.5 | 2.85 | 0.11 |
| Injection duration (s) (both beams) | 32.8 | 141.6 | 39.4 | 30 |
| Injection interval (s) | 38 | 153.5 | 155 | 65 |
| Number of SR photons ($s^{-1}m^{-1}$) | $6.6 \times 10^{14}$ | $2.9 \times 10^{13}$ | $3.9 \times 10^{14}$ | $5.8 \times 10^{13}$ |

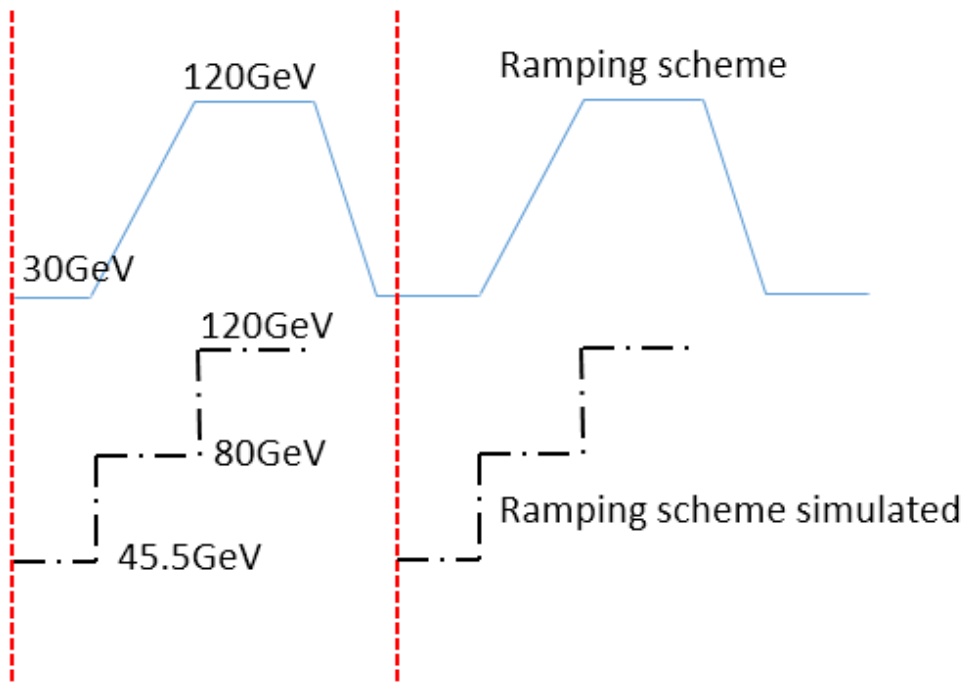

**Figure 5.2.4.1:** Simulation of Booster ramping scheme.

The Equation 4.2.4.6 in Section 4.2.4 can be used to calculate the number of synchrotron radiation (SR) photons emitted from the Booster. However, simulating the ramping process smoothly is difficult. A conservative approximation is to use a step function to simulate the smooth energy ramping, as shown by the black dashed-dotted line in Fig. 5.2.4.1. In FLUKA simulation, the ramping process for the Higgs mode is divided into three intervals, with the beam energy in each interval being constant at 45.5 GeV, 80 GeV, and 120 GeV. This approximation can be generalized to other modes: in the W mode, there are two intervals with beam energy of 45.5 GeV and 80 GeV; in the Z mode, there is one interval with a beam energy of 45.5 GeV; in the $t\bar{t}$ mode, there are four intervals with beam energy of 45.5 GeV, 80 GeV, 120 GeV, and 180 GeV.

Figure 5.2.4.2 shows the top view of the Booster, which includes 4 dipoles, 4 quadrupoles, 4 sextupoles, and 4 vacuum chambers that are not inside the magnets. The geometries of the magnets are illustrated in Figures 5.2.4.3 – 5.2.4.8. The beam-pipes are located at the center of these figures and are made of aluminum.

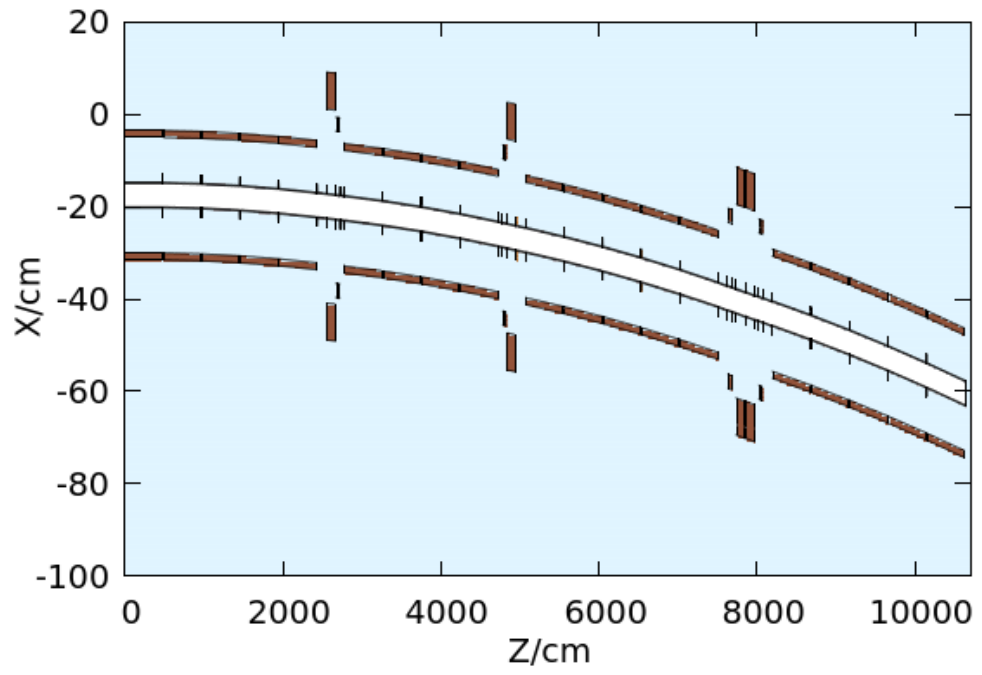

**Figure 5.2.4.2:** Top view of a section in the Booster.

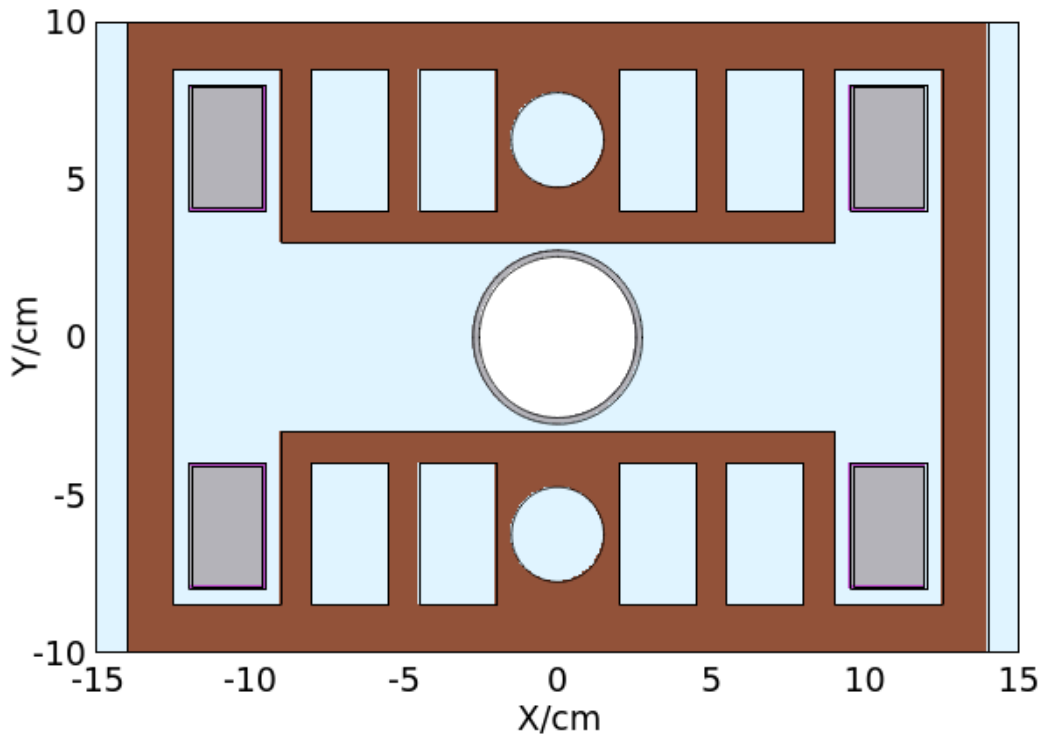

**Figure 5.2.4.3:** The X-Y cross section of the Booster dipole. The magnet cores are represented in dark brown, while the gray parts represent the coils, and the purple parts represent the insulations made of epoxy resin.

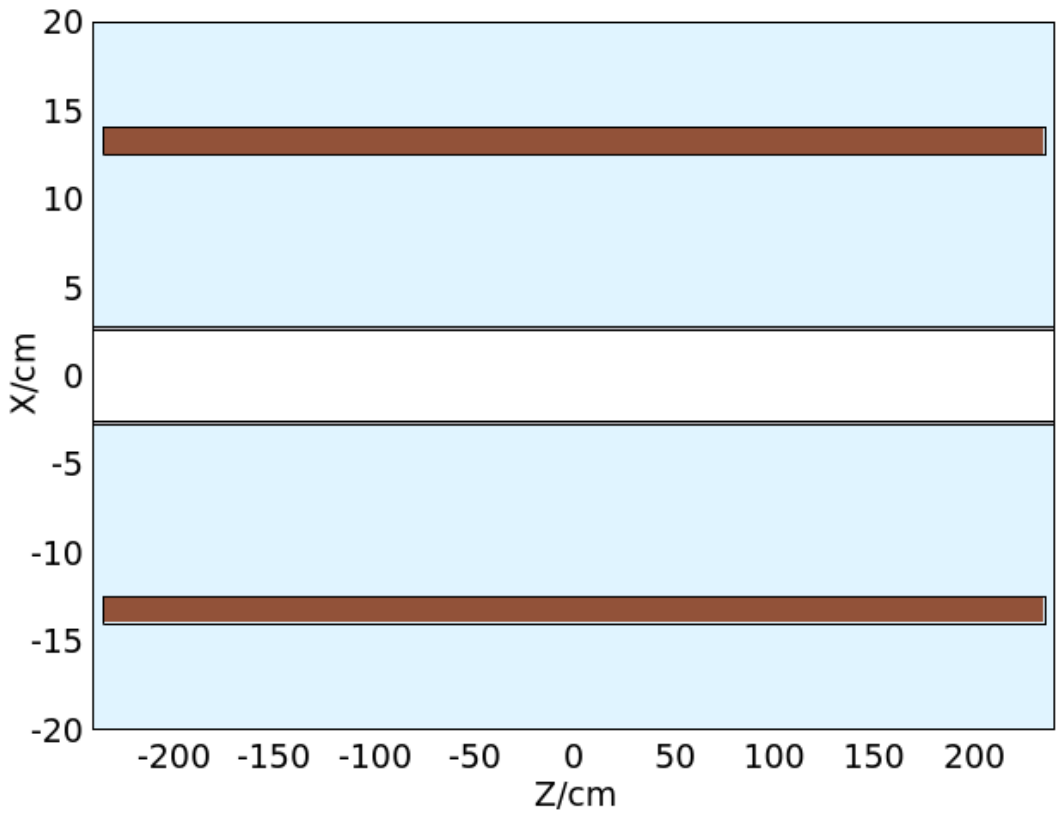

**Figure 5.2.4.4:** The X-Z cross section of the Booster dipole. The colors are explained in Figure 5.2.4.3.

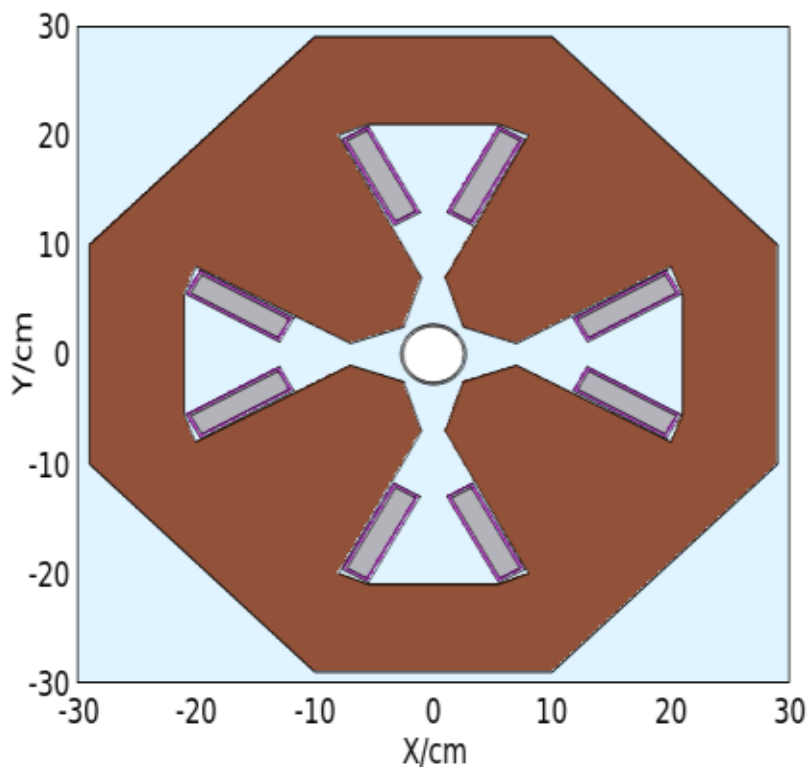

**Figure 5.2.4.5:** The X-Y cross section of the Booster quadrupole. The colors are explained in Figure 5.2.4.3.

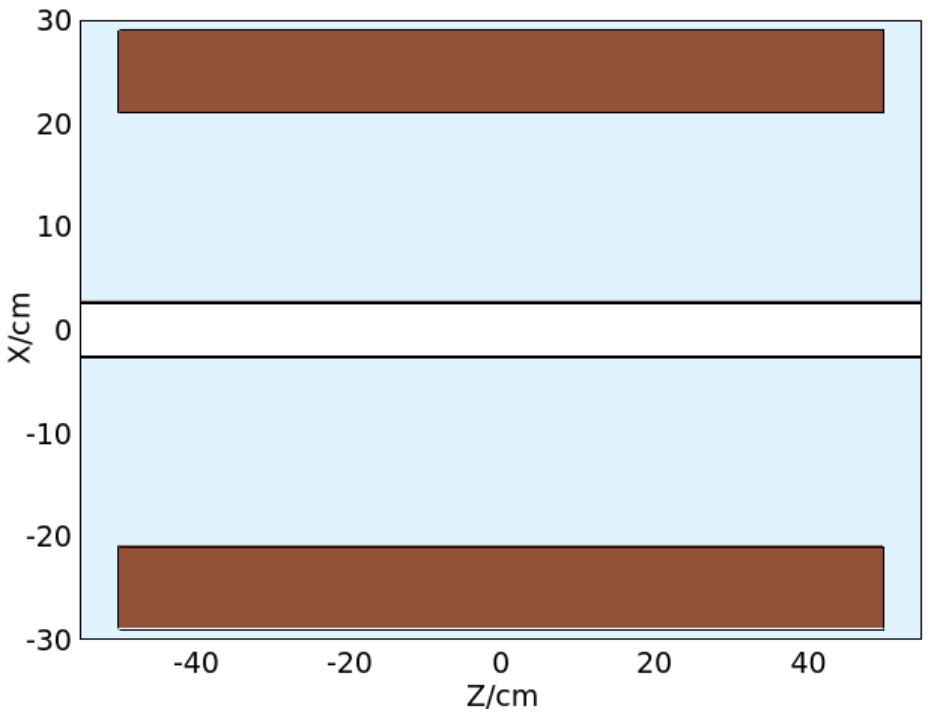

**Figure 5.2.4.6:** The X-Z cross section of the Booster quadrupole. The colors are explained in Figure 5.2.4.3.

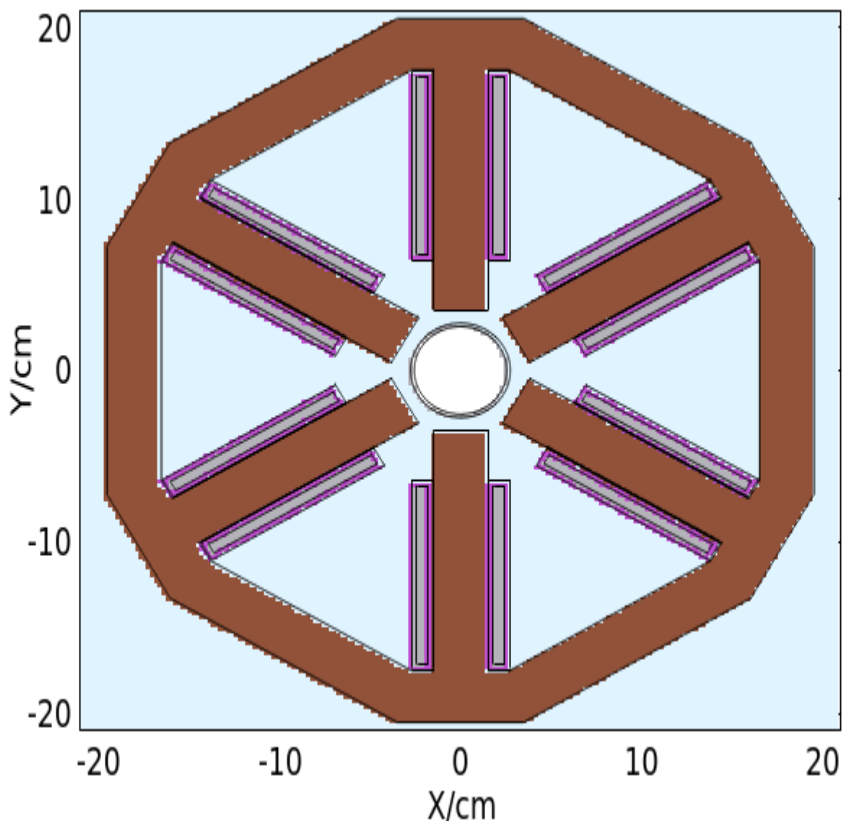

**Figure 5.2.4.7:** The X-Y cross section of the Booster sextupole. The colors are explained in Figure 5.2.4.3.

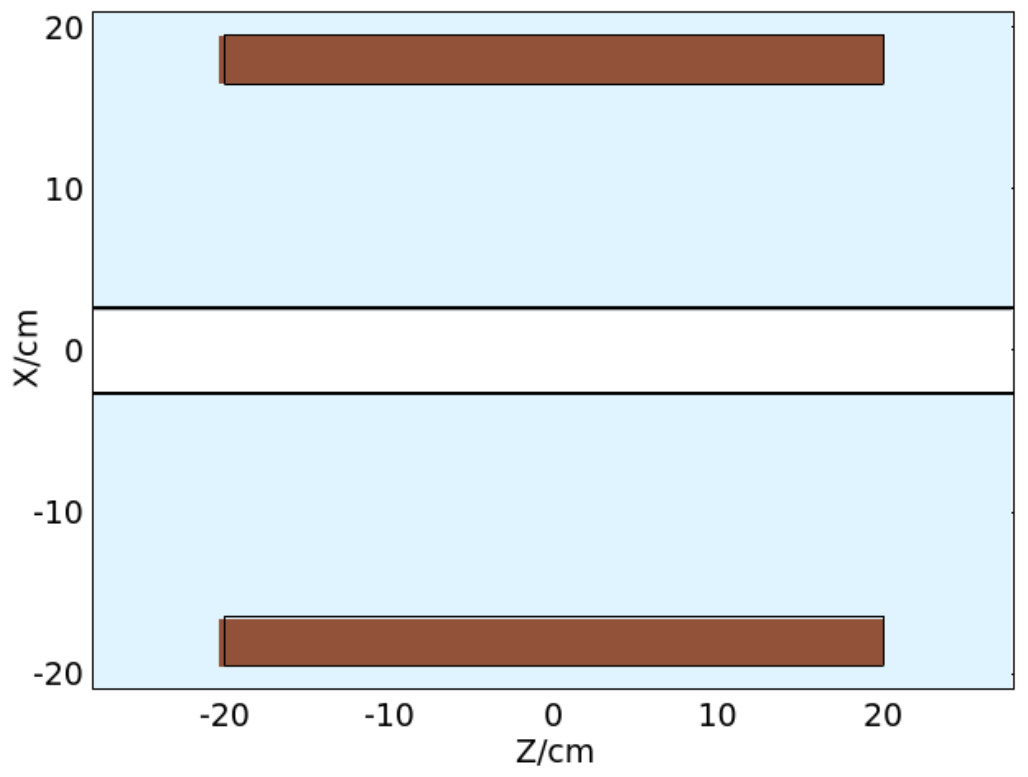

**Figure 5.2.4.8:** The X-Z cross section of the Booster sextupole. The colors are explained in Figure 5.2.4.3.

### 5.2.4.2 *Dose Deposition on the Booster Magnets*

Figure 5.2.4.9 shows the particle fluence in the top view of the Booster, with the red region indicating the area with a higher number of SR photons. The absorbed doses to the magnet insulations are listed in Table 5.2.4.2. Assuming the same operation schedule as in Section 4.2.4.4, with ten years for Higgs mode, two years for Z operation, and one year for WW operation, the accumulted absorbed doses to the magnet insulations are shown in Figure 5.2.4.10. The absorbed doses to the magnet insulations after adding five years of $t\bar{t}$ operation are shown in Figure 5.2.4.11.

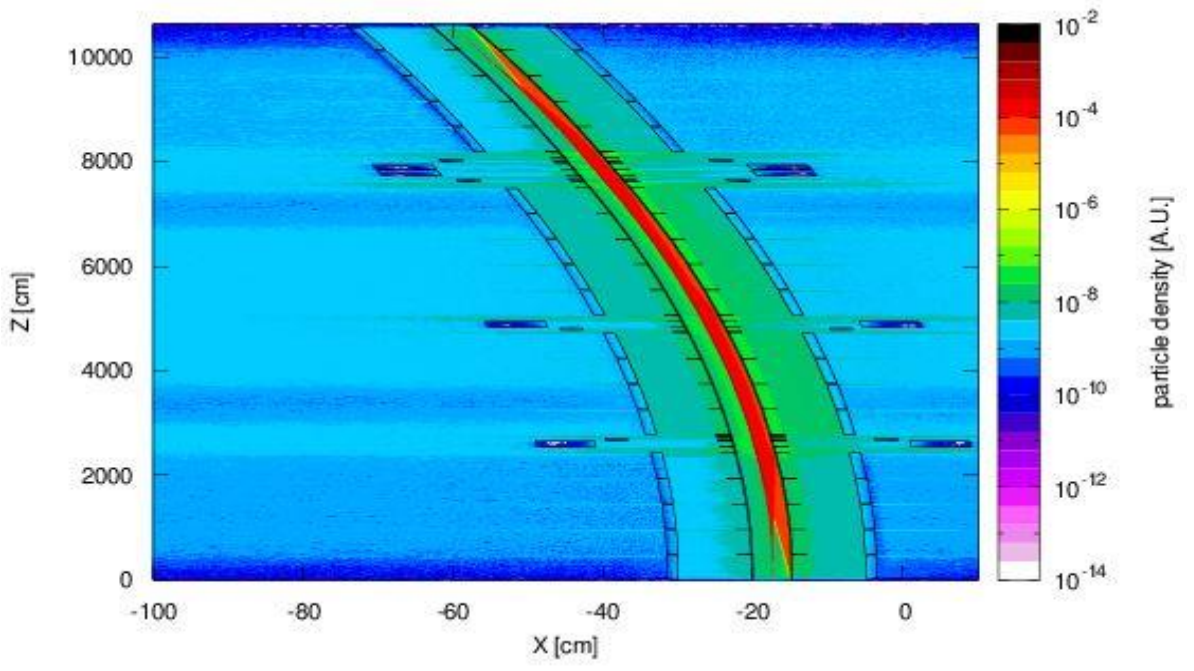

**Figure 5.2.4.9** Particles fluences for the Booster.

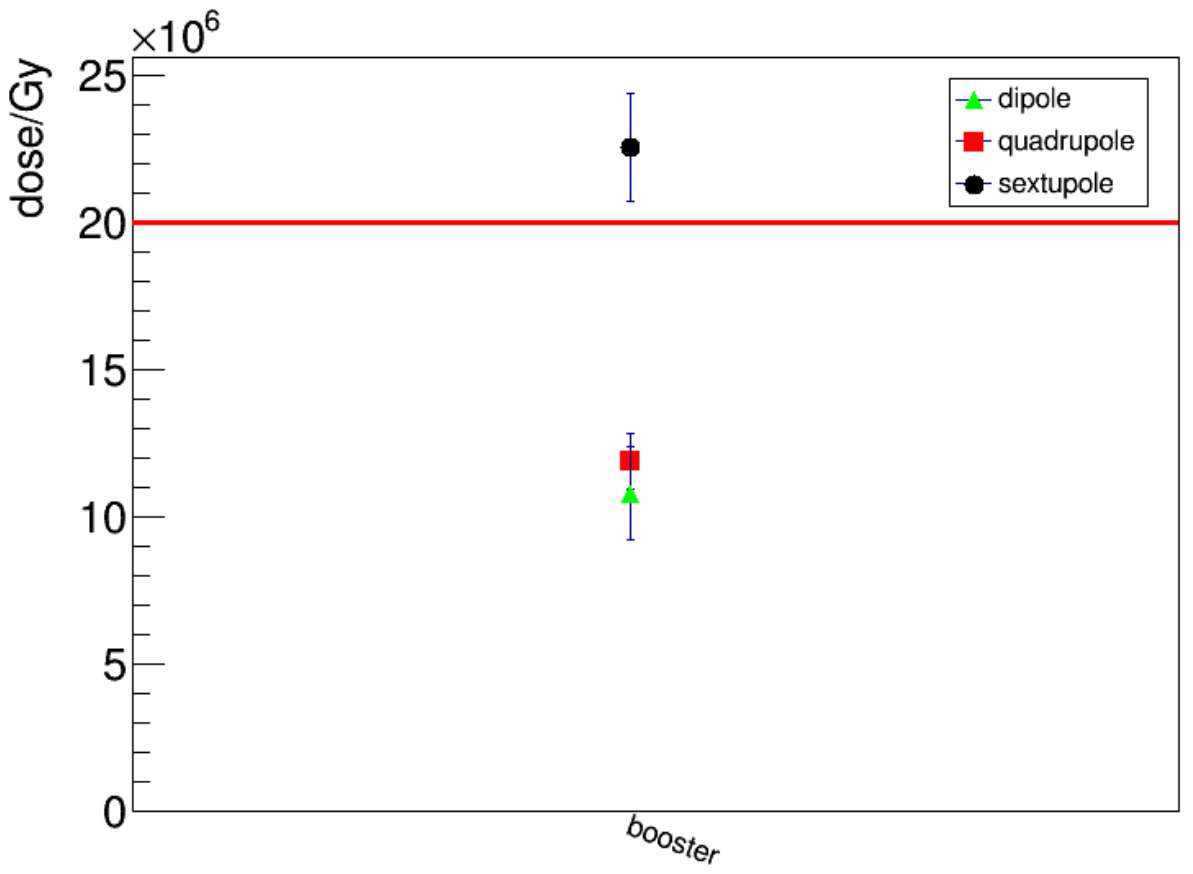

**Figure 5.2.4.10**: Absorbed doses to the Booster magnets after 13-year operation.

After 13 years of running, the doses to the quadrupole and dipole insulation are smaller than the upper limit [1]. However, after 13 years and 5 years of $t\bar{t}$ running, the doses to the Booster dipole and quadrupole insulation will be still lower than the upper limit. This is not the case for sextupoles, as after 13 years of running, the dose to the sextupole insulation will exceed the upper limits. However, the dose rates in the Booster may be over-estimated due to our conservative assumptions in the simulation. In the future, more attention to the absorbed doses of the Booster sextupoles will be required. The percentage of energy deposition is listed in Table 5.2.4.3.

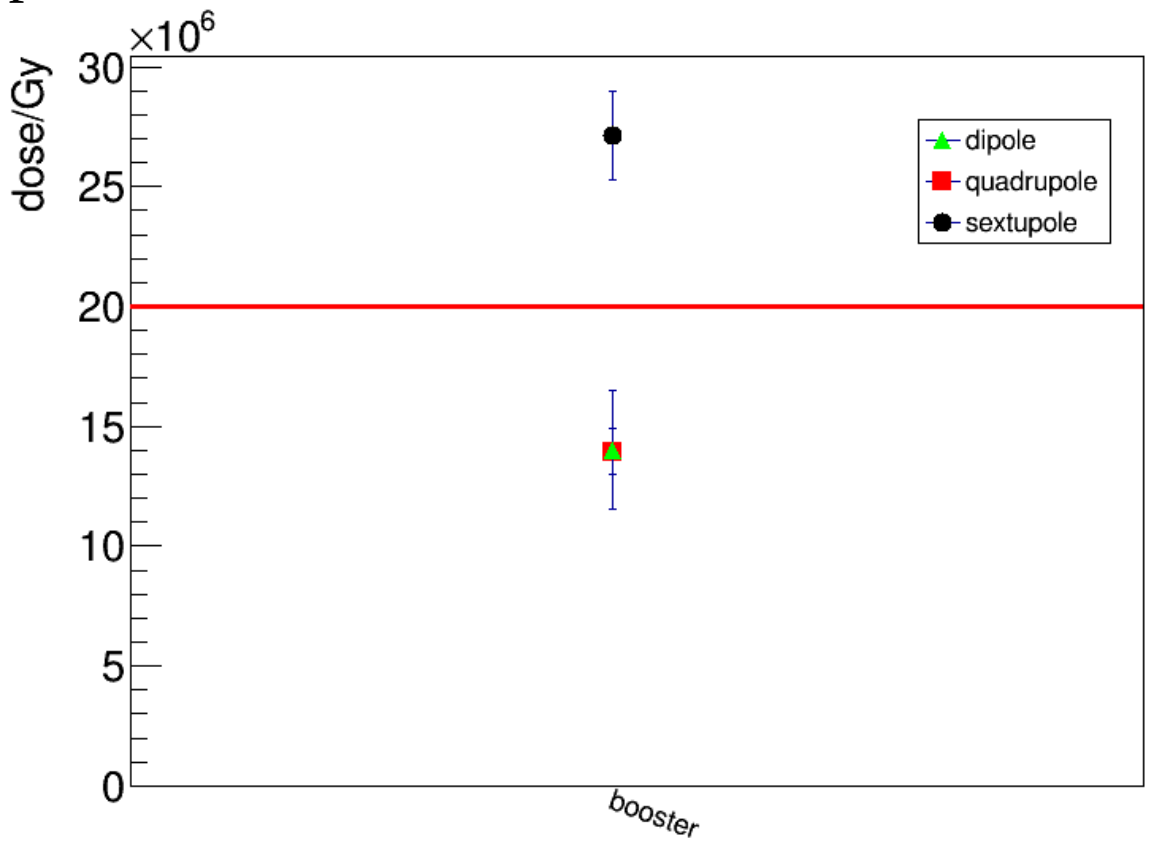

**Figure 5.2.4.11:** Absorbed doses to the Booster magnets after 18-year operation.

**Table 5.2.4.2:** Absorbed doses rate to the Booster magnets (Unit: Gy/Ah)

| | Higgs | Z | WW | $t\bar{t}$ |
|---|---|---|---|---|
| Dipole | $(1.5 \pm 0.3) \times 10^5$ | $(4.6 \pm 2.5) \times 10^3$ | $(1.3 \pm 0.4) \times 10^4$ | $(8.4 \pm 4.9) \times 10^5$ |
| Quadrupole | $(1.6 \pm 0.2) \times 10^5$ | $(6.9 \pm 2.7) \times 10^3$ | $(1.7 \pm 0.4) \times 10^4$ | $(5.3 \pm 0.5) \times 10^5$ |
| sextupole | $(3.2 \pm 0.3) \times 10^5$ | $(10.7 \pm 0.4) \times 10^3$ | $(3.0 \pm 0.6) \times 10^4$ | $(11.9 \pm 0.5) \times 10^5$ |

**Table 5.2.4.3** Ratio of energy deposition power over the total SR power in the Booster.

| | Booster dipole | Booster quadrupole | Booster sextupole |
|---|---|---|---|
| Energy deposition [%] | 0.2 | 0.02 | 0.006 |
| | Beam pipe | | |
| Energy deposition [%] | 0.2 | | |

## 5.3 Booster Technical Systems

### 5.3.1 Superconducting RF System

#### 5.3.1.1 *Booster RF Layout and Parametrs*

The Booster will be equipped with a 1.3 GHz RF system consisting of 96 9-cell cavities that will provide 30 MW of SR power per beam in the Collider's Higgs mode. To upgrade the Collider's Higgs mode to 50 MW of SR power per beam, the number of cavities in the Booster will remain unchanged, and only the peak input power per cavity will be increased from 22 kW to 32 kW. In addition, 256 high-gradient 9-cell cavities will be added to the $t\bar{t}$ mode to raise the RF voltage from 2.17 GV to 9.7 GV. Table 5.3.1.1 is a summary of the Booster's RF system parameters.

Each of the 12-meter-long cryomodules in the Booster contains eight 1.3 GHz 9-cell cavities, with each cavity having two welded HOM couplers on the cavity beam pipe. Additionally, each cryomodule has two beamline HOM absorbers located at room temperature outside the vacuum vessel. Due to the high beam current and high HOM power per cavity in the high-luminosity Z mode (which can be up to 30 mA for injection into the empty Collider ring), an ERL-type 1.3 GHz cryomodule with a HOM absorber at 100 K between cavities will be used instead of the low current TESLA cavities for Higgs, W and $t\bar{t}$ mode. Furthermore, to reduce the impedance seen by the beam, the Z-mode beams will only go through Z-cavities and bypass the Higgs and $t\bar{t}$ cavities.

Booster cavities operate in fast ramp mode, and the beam energy increases almost linearly with time. The total RF voltage must be increased during the energy ramp to keep the synchrotron tune constant. The LLRF system compensates for the large Lorentz force detuning and microphonics compared with the narrow cavity bandwidth (130 Hz) during the voltage ramp. Multipacting may occur in the input coupler and cavity during the voltage ramp, while a counter-phasing ramp of the total effective RF voltage can be used to prevent cavity trips.

To match the injection timing with the Collider ring, the Booster buckets will be half-filled for Higgs and $t\bar{t}$ modes. The transient beam loading of the half-filled ring during both injection and extraction is tolerable. The Higgs parameters presented in Table 5.3.1.1 are for on-axis injection, in which the average Higgs beam current, HOM power, RF duty factor, etc. are all higher than that for off-axis injection. Despite large bunches in the Booster, the transient beam loading of the Higgs on-axis injection is also tolerable.

Due to the low energy injection to the Booster at 30 GeV, the growth time of the coupled bunch instability for all the higher order modes is much shorter than the radiation-damping time during injection and while the Booster is at low energy. With the assumed bunch-by-bunch transverse feedback time of 10 ms and longitudinal feedback time of 200 ms, almost all modes are safe, with enough margin for Higgs, W, and $t\bar{t}$ injection. The HOM absorbers between cavities in the Z-mode high-current 1.3 GHz cryomodule will have deeper HOM damping to suppress CBI. Including the HOM frequency spreads among the cavities will provide more margin (usually more than one to two orders of magnitude).

**Table 5.3.1.1:** CEPC Booster RF Parameters

| | $t\bar{t}$ 30/50 MW | | Higgs 30/50 MW | W 30/50 MW | Z 30/50 MW |
|---|---|---|---|---|---|
| | New cavities | Higgs cavities | | | |
| Extraction beam energy [GeV] | 180 | | 120 | 80 | 45.5 |
| Extraction average SR power [MW] | 0.05 | | 0.5 / 0.67 | 0.02 / 0.04 | 0.05 / 0.1 |
| Bunch charge [nC] | 1.1 | | 0.78 (20.3)* | 0.73 | 0.81 |
| Beam current [mA] | 0.12 / 0.19 | | 1 / 1.4 | 3.1 / 5.3 | 16 / 30 |
| Injection RF voltage [GV] | 0.761 | | 0.346 | 0.3 | 0.3 |
| Extraction RF voltage [GV] | 9.7 (7.53 + 2.17) | | 2.17 | 0.87 | 0.46 |
| Extraction bunch length [mm] | 1.8 | | 1.86 | 1.3 | 0.75 |
| Injection longitudinal damping time [s] | 1.5 | | 1.5 | 1.5 | 1.5 |
| Extraction longitudinal damping time [ms] | 7 | | 24 | 80.8 | 446 |
| Cavity number (1.3 GHz 9-cell) | 256 | 96 | 96 | 96 | 32 |
| Module number (8 cavities / module) | 32 | 12 | 12 | 12 | 4 |
| Extraction gradient [MV/m] | 28.3 | 21.8 | 21.8 | 8.7 | 13.8 |
| Extraction cavity voltage [MV] | 29.4 | 22.6 | 22.6 | 9.0 | 14.3 |
| $Q_0$ @ 2 K at operating gradient | 2E10 | 3E10 | 3E10 | 3E10 | 3E10 |
| $Q_L$ | 4E7 | 4E7 | 1.2E7 | 7.3E6 / 4.4E6 | 1.2E7 / 6.3E6 |
| Cavity bandwidth [Hz] | 33 | 33 | 110 | 178 / 296 | 111 / 208 |
| Peak HOM power per cavity [W] | 0.5 / 0.8 | | ~ 75 / ~ 100 | 11.8 / 19.6 | 146 / 272 |
| Average HOM power per cavity [W] | 0.2 / 0.32 | | ~ 10 / ~ 15 | 3.8 / 6.3 | 80 / 150 |
| Input peak power per cavity [kW] | 8.3 / 9.2 | 5.1 / 5.9 | 22 / 32 | 10.9 / 18.1 | 17 / 32 |
| Input average power per cavity [kW] | 0.3 | 0.2 | 6.5 / 9.2 | 0.3 / 0.5 | 2.5 / 4.5 |
| SSA power [kW] (1 cavity / SSA) | 10 | 10 | 25 / 30 | 25 / 30 | 25 / 40 |
| Total cavity wall loss @ 2 K [kW] | 0.36 | 0.05 | 0.5 | 0.02 | 0.08 |
| Cryogenic dynamic heat-load duty factor | 3.4 % | 3.4 % | 31 % | 6.5 % | 36.6 % |
| RF power duty factor | 3.9 % | 3.9 % | 30 % | 3 % | 14.6 % |
| HOM power duty factor | 42 % | 42 % | 42 % / 49 % | 32 % | 55 % |

* The small bunch-charge number before the parenthesis is for the bunches injected from the linac. The large bunch charge number in the parenthesis is for the bunches injected back from the Collider ring during the swap-out injection for Higgs running.

### 5.3.1.2 *Booster 1.3 GHz SRF Technology*

For the Booster Higgs, W and $t\bar{t}$ modes, TESLA 1.3 GHz 9-cell cavities made of bulk niobium will be used. These cavities will operate at 2 K with $Q_0 = 3 \times 10^{10}$ at 21.8 MV/m for the long term and should reach $Q_0 = 3 \times 10^{10}$ at 24 MV/m using high-Q recipes such as mid-T furnace baking [1] for the vertical acceptance test.

A total of twelve 1.3 GHz 9-cell cavities treated by mid-T furnace baking have demonstrated high $Q_0$ and high gradient, exceeding the CEPC specifications (Figure 5.3.1.1).

An assembled cryomodule at PAPS (Figure 5.3.1.2) consists of eight mid-T high-Q 9-cell cavities, power couplers, tuners, etc. Horizontal testing of the cryomodule revealed a total RF voltage of 190 MV (with an average of 23 MV/m) and an average $Q_0$ of $3.6 \times 10^{10}$ at 21 MV/m. Mid-T technology is considered a superior choice for future high-Q accelerator applications due to its higher performance and a more straightforward and reliable cavity processing recipe compared to nitrogen-doping.

The construction of large projects with hundreds of high-Q 1.3 GHz 9-cell cavities and tens of cryomodules, such as SHINE (Shanghai high repetition rate XFEL and extreme light facility) and $S^3FEL$ (Shenzhen superconducting soft X-ray free electron laser), will offer valuable experience and industrialization insights for the CEPC booster SRF system.

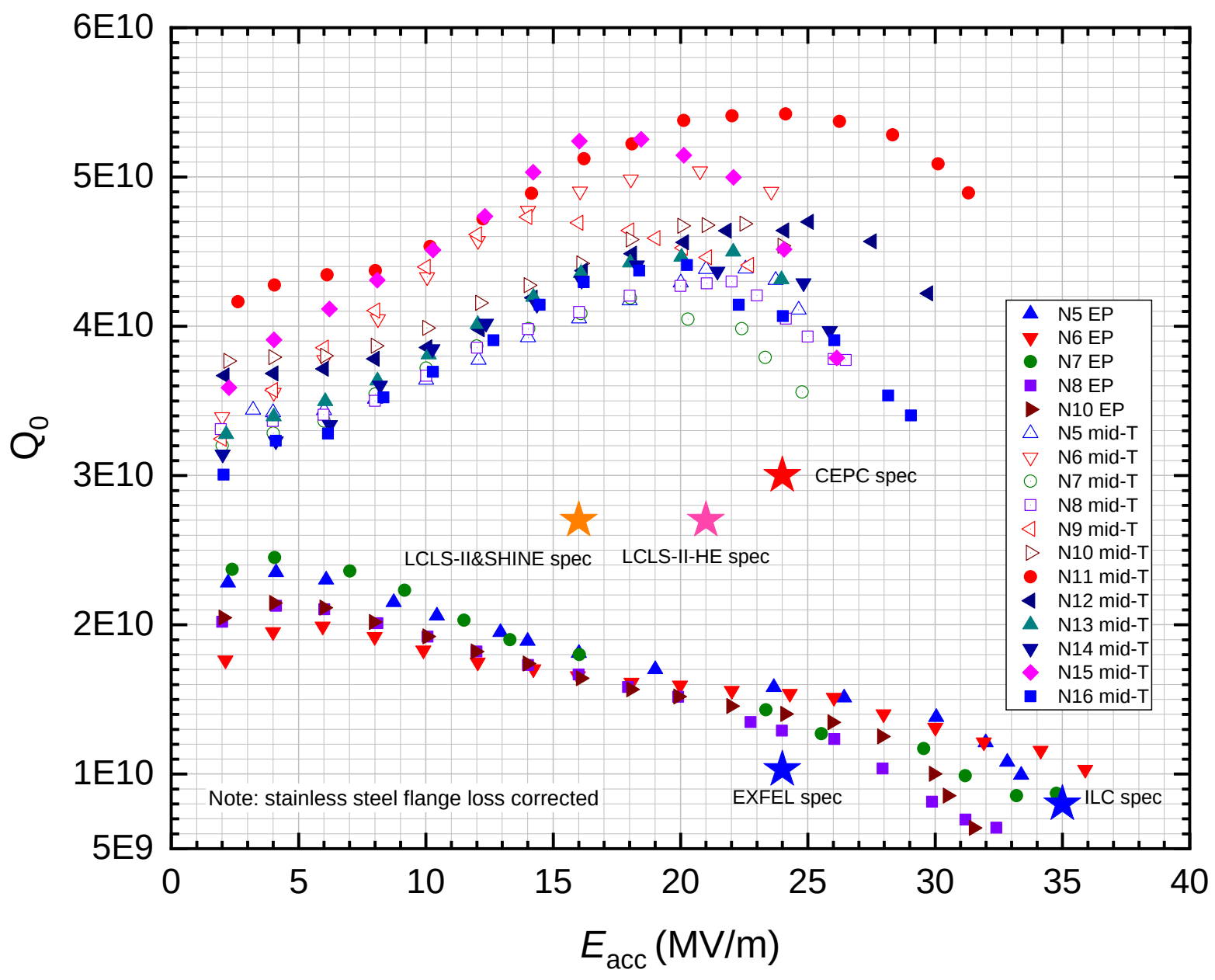

**Figure 5.3.1.1:** Vertical test results of the 1.3 GHz 9-cell cavities

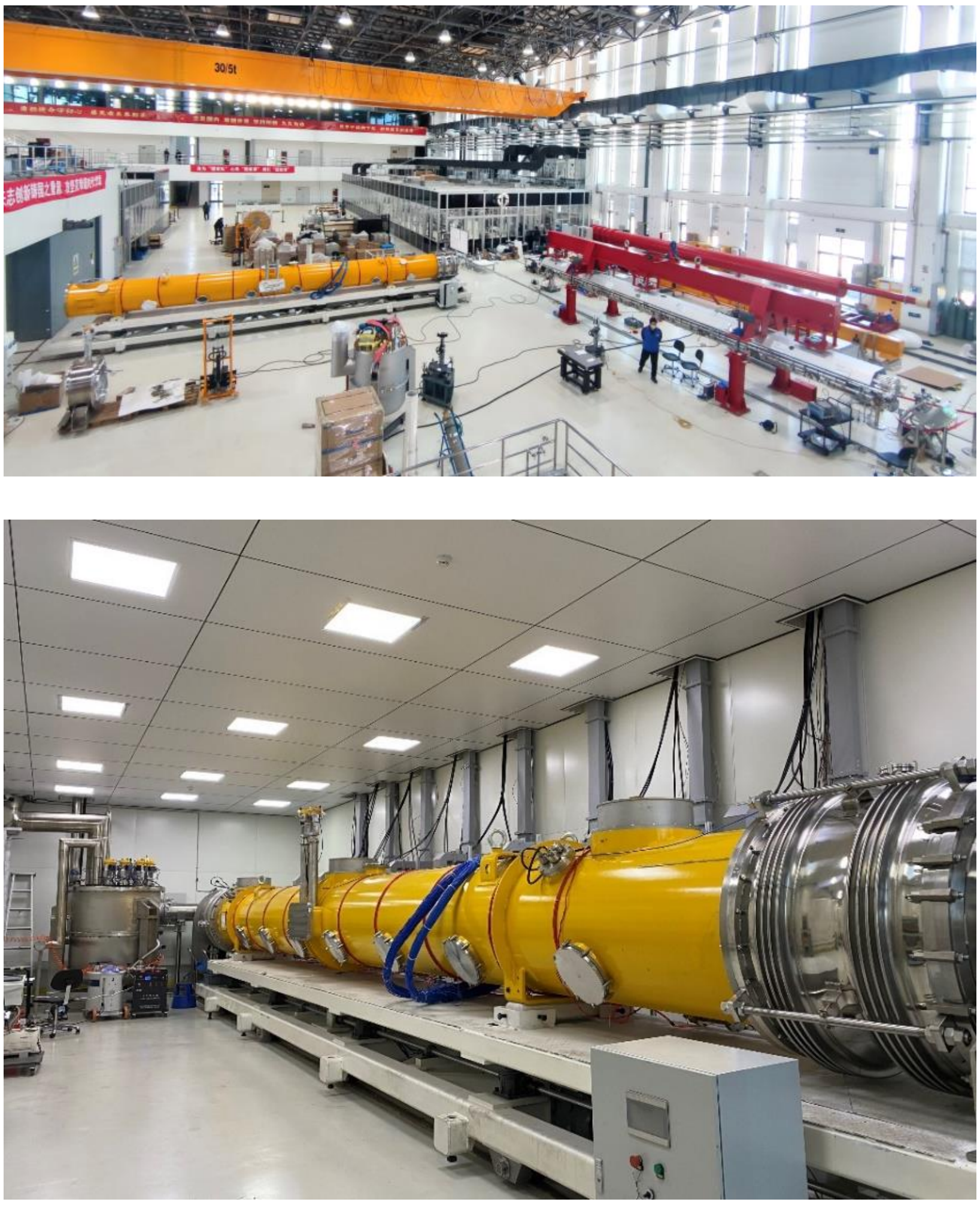

**Figure 5.3.1.2:** Assembly and horizontal test of the 1.3 GHz cryomodule

## 5.3.2 RF Power Source

### 5.3.2.1 *Introduction*

The Booster RF system consists of 96 1300 MHz superconducting RF cavities. For Higgs operation, there are 12 cryo-modules, each containing eight 9-cell superconducting cavities. To provide the necessary power, solid-state amplifiers (SSAs) are used. The parameters for the Booster SRF system are listed in Table 5.3.2.1.

**Table 5.3.2.1:** Booster RF power system parameters

| **Operation frequency (MHz)** | **1300+/-0.5** |
|---|---|
| Cavity number | 96 |
| RF source number | 25kW (SSA) |

Various options were considered for the Booster power sources, taking into account modularity and technology. These options included vacuum tubes such as Klystron, IOT (Inductive Output Tube), and Diacrode, as well as solid-state alternatives [1]. With advances in transistor technology, particularly the sixth generation LDMOSFET, the output power and efficiency of individual transistors has significantly improved. High power can be achieved by combining multiple transistors. As a result, SSAs are now being considered for an increasing number of accelerators, both circular and linear. Their capabilities range from a few kW to several hundred kW and from less than 100 MHz to above 1 GHz. These devices provide reasonable efficiency (~50%), high gain, and a modular design that enhances their reliability.

The SSA offers several advantages, including high reliability which makes it ideal for redundancy design, high flexibility for module design, high stability, and low maintenance requirements. In addition, SSAs operate at low voltages and do not require a warm-up time, making them highly efficient. Considering these advantages, the Booster RF power system has been equipped with SSAs.

#### 5.3.2.2 *Solid-State Amplifier*

Solid state amplifiers (SSAs) have proven to be successful in many accelerators in China, including the accelerator driven sub-critical system (ADS) and the high-energy photon source (HEPS). - Each 9-cell superconducting RF cavity is powered independently with an SSA to simplify the LLRF control and RF transmission system. A total of 96 1.3 GHz SSAs with an RF output power of 25 kW are required. Each SSA consists of two cabinets, as shown in Figure 5.3.2.1. Each cabinet is combined with 24 1.4 kW power modules, which must include circulators and absorbing loads to ensure isolation between modules and withstand the full reflection power. The hot-plug design effectively reduces maintenance time. Water cooling is used to remove the heat loss of transistors and loads of the power module. The status data of power modules, such as voltage, current, and temperature, is monitored. If one power module fails, the SSA can still operate. The 1 dB bandwidth of the SSA needs to be more than 1 MHz to meet the LLRF control requirement. The nominal maximum power is achieved with less than 1 dB compression, and the harmonic power output is less than –30 dBc. The AC to RF efficiency goal for the SSAs at the rated power is at least 40%. The phase and amplitude stability (peak to peak) is less than 1° and 1% during open loop when the cooling-water temperature changes by ±1°C. The mean time between failures (MTBF) should be larger than 30,000 hours with the help of redundancy design, and less than 2% of the power modules should fail per year. An interlock is necessary for external faults, such as water leakage. The output port is a standard WR650 waveguide. The 1.3 GHz/25 kW SSA specifications are listed in Table 5.3.2.2.

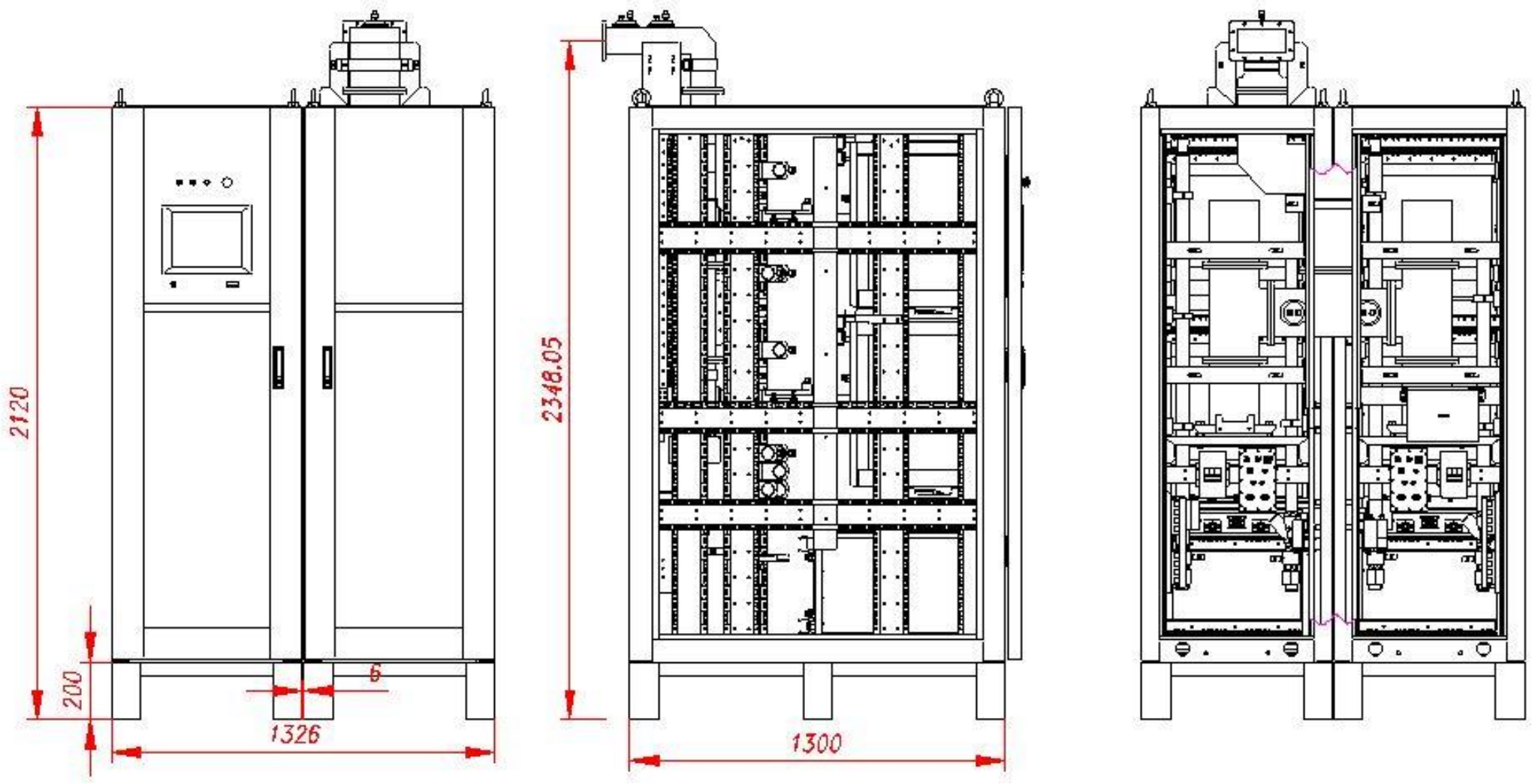

**Figure 5.3.2.1:** Mechanical layout of 1.3 GHz/25 kW solid state amplifier

**Table 5.3.2.2:** 1.3 GHz/25 kW SSA Specifications

| Parameters | Values |
|---|---|
| Frequency | 1.3 GHz |
| Power (< 1 dB compression) | 25 kW |
| Gain | ≥ 65 dB |
| Bandwidth (1 dB) | ≥ 1 MHz |
| Amplitude stability (open loop) | ≤ 1% |
| Phase stability (open loop) | ≤ 1° |
| Phase Variation (1 kW - 25 kW) | ≤ 10° |
| Harmonic | < –30 dBc |
| Spurious | < –60 dBc |
| Efficiency at 25 kW | ≥ 40% |
| MTBF | ≥ 30,000 hrs |
| Redundancy | SSAs can still run with 1 power module |
| Cooling | Deionized water 4 kg/cm$^2$ at 25 $^{O}$C |

### 5.3.2.3 *RF Power Transmission System*

The Booster's superconducting RF system consists of 96 9-Cell cavities, with each cavity being powered by a 25-kW solid-state amplifier operating at 1.3 GHz. This design simplifies the RF distribution system (RFDS), which is illustrated in Figure 5.3.2.2. Standard WR650 waveguide components made of aluminum are used to deliver the power from the solid-state amplifiers to the 9-cell superconducting cavities, with an insertion loss of less than 0.01 dB/m for straight waveguide segments. To monitor forward and reflected power, directional couplers with a directivity of over 30 dB and a coupling coefficient of 60 dB are employed. The waveguide-coaxial transition connects the waveguide system with the coaxial coupler for the 9-cell superconducting cavity; a few short flexible waveguides are also necessary. The placement of the solid-state amplifiers and low-level RF system (SSA+LLRF) in the Booster RFTS is depicted in Figure 5.3.2.2 and 5.3.2.3, respectively.

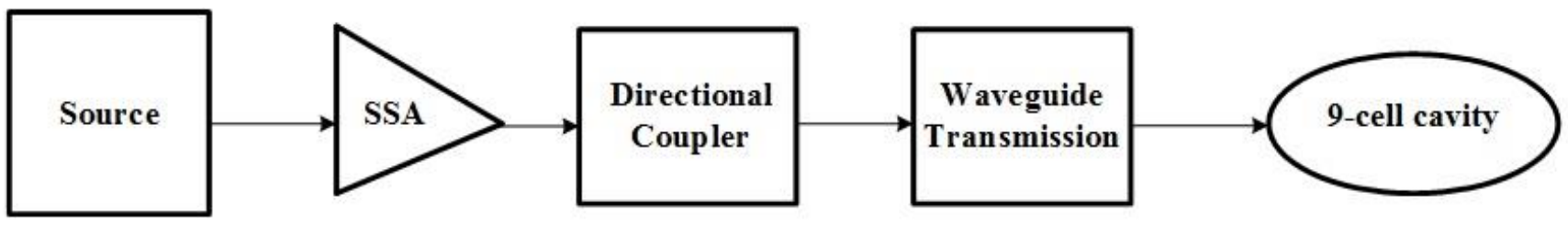

**Figure 5.3.2.2 :** Schematic of the Booster RFTS

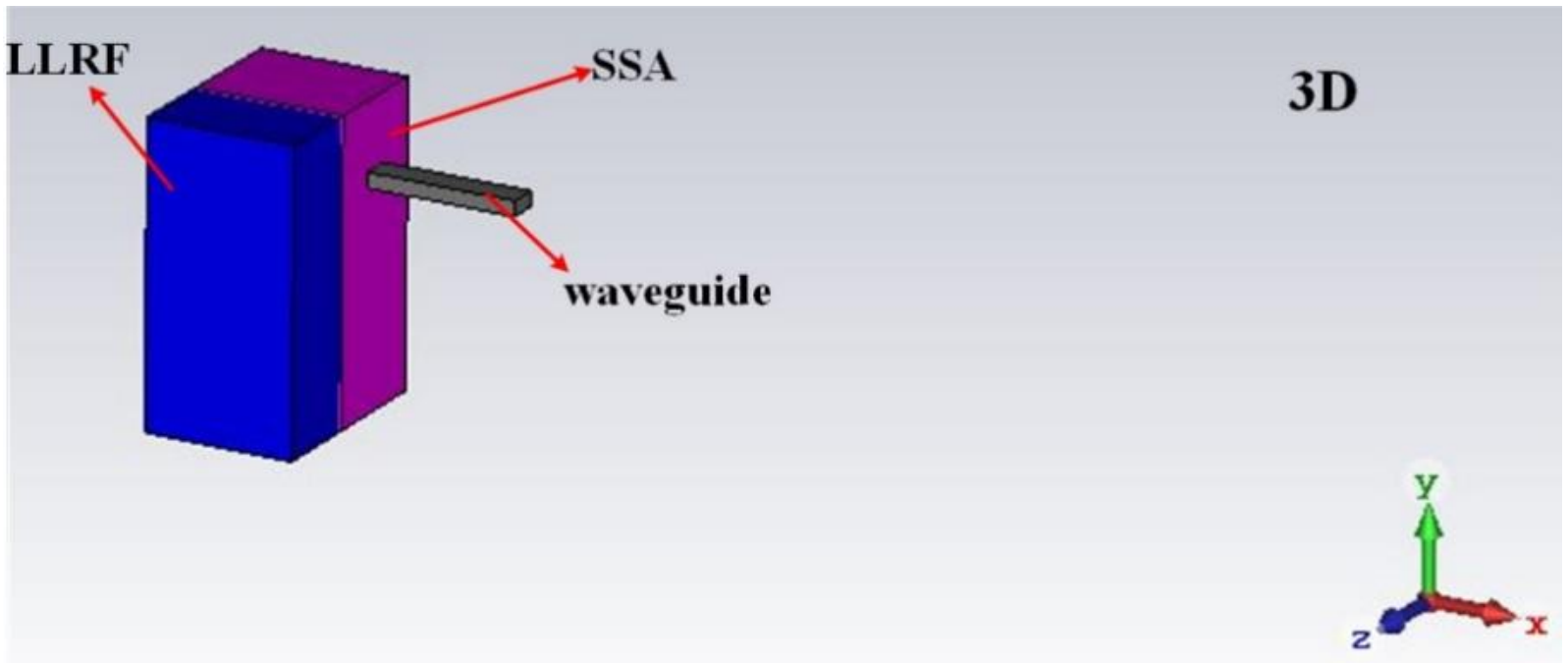

**Figure 5.3.2.3** Placement of the SSA and LLRF

One source per cavity simplifies the RF power delivery. There will be no power splitters or phase shifters in the Booster RFTS system, and the length differences as well as the thermal phase drifts will be easily tracked and compensated by the individual LLRF controls. Besides the waveguides, each power transmission line will have a directional coupler, which will be used to monitor forward and reflected power, and a waveguide-coaxial transition [2]. The waveguide-coaxial transition will connect the waveguide system to the coaxial coupler of the 9-cell superconducting cavity.

### 5.3.2.4 *LLRF System*

The RF power for each of the 96 1300MHz 9-cell cavities in the Booster is delivered by a single solid-state amplifier (SSA); each SSA is accompanied by a set of LLRF hardware. Ten channels of signals, including the cavity-probe pickup signal, need to be down-converted to intermediate frequency (IF) and digitally sampled synchronously.

The LLRF hardware used in the Booster is mostly similar to that used in the Collider, with the main difference being the sampling principle. In the Booster, IF sampling is used instead of direct sampling, as directly sampling the 1.3 GHz signal using a low-frequency clock would result in significant noise due to aperture jitter of the clock. The scheme of the Booster LLRF is shown in Figure 5.3.2.4.

The phase and amplitude error of the LLRF system should be within 0.1 degrees (rms) and 0.1% (rms), respectively.

*Status of LLRF development:*

The development of the universal continuous-wave (CW) LLRF system has been successful and demonstrated on several projects, including PAPS, C-ADS Injector I, and the HEPS Linac. The phase noise of the 1.3 GHz signal is less than 30 fs (10 Hz-10 MHz), and the LLRF for the 1.3 GHz buncher has operated without failure during the conditioning and operation of the normal-conducting cavity. Additionally, the down-converter front-end module made by IHEP has been installed on the HEPS Linac, which covers a frequency range of 100 MHz to 6 GHz.

*Beam commissioning and automation of LLRF for the machine:*

The SSA should work in the linear range as the Booster power ramp up and down with a determined mode, so the LLRF of the Booster control the driver signal of the SSA to generate the desired power waveform.

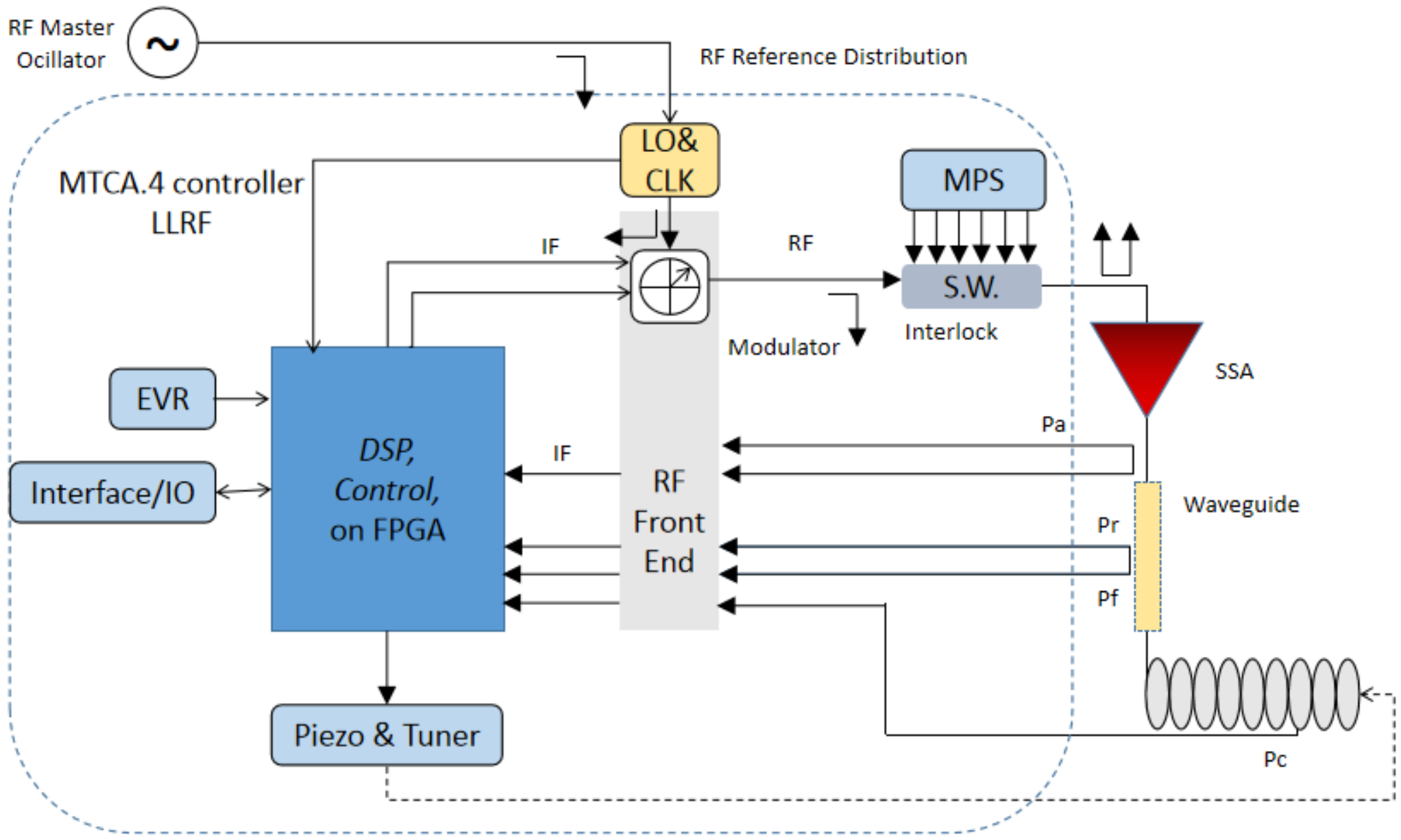

**Figure 5.3.2.4:** LLRF schematic

## 5.3.3 Magnets

### 5.3.3.1 *Introduction*

The circumference of the Booster and the Main Ring is the same, around 100 km. The Booster is equipped with 14,866 dipoles and 3,458 quadrupoles, which means that more than 74% of the Booster's circumference will be covered with magnets. The dipole magnets in the Booster have a length of 4.7 m, and the quadrupole magnets have a length of 2 m. As a result, the cost of magnets is a significant concern in the design, particularly for the dipole magnets.

During the Conceptual Design Report (CDR) stage, the dipole magnets in the Booster were required to have a minimum working field of only 28 Gs @10 GeV, and the field precision better than $\pm 1\times 10^{-3}$. Since this type of dipole magnet has never been designed and produced before, two scaled-down prototype dipole magnets were developed in recent years, one with an iron core and one without. The Cosine-Theta (CT) coil dipole magnet without an iron core is shown in Fig. 5.3.3.1, along with its cross-section, while the iron-core dipole magnet is shown in Fig. 5.3.3.2 with its corresponding cross-section and picture.

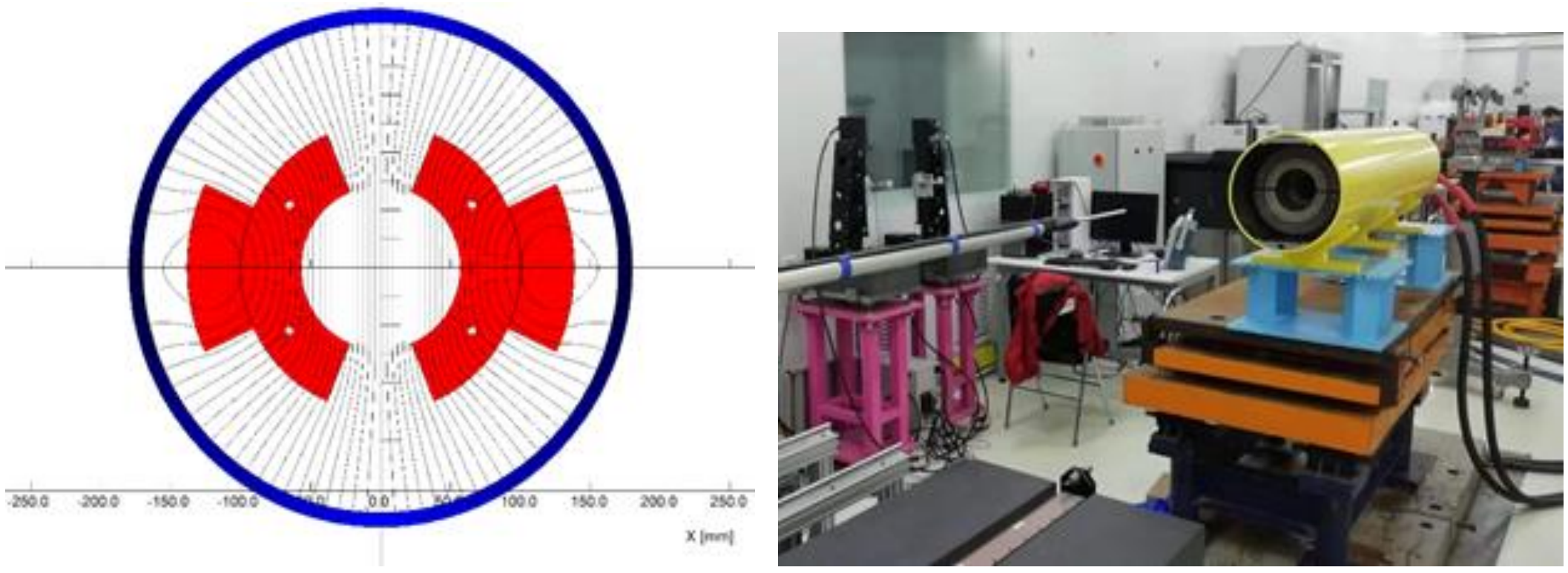

**Figure 5.3.3.1:** The cross section and picture of the CT coil dipole magnet

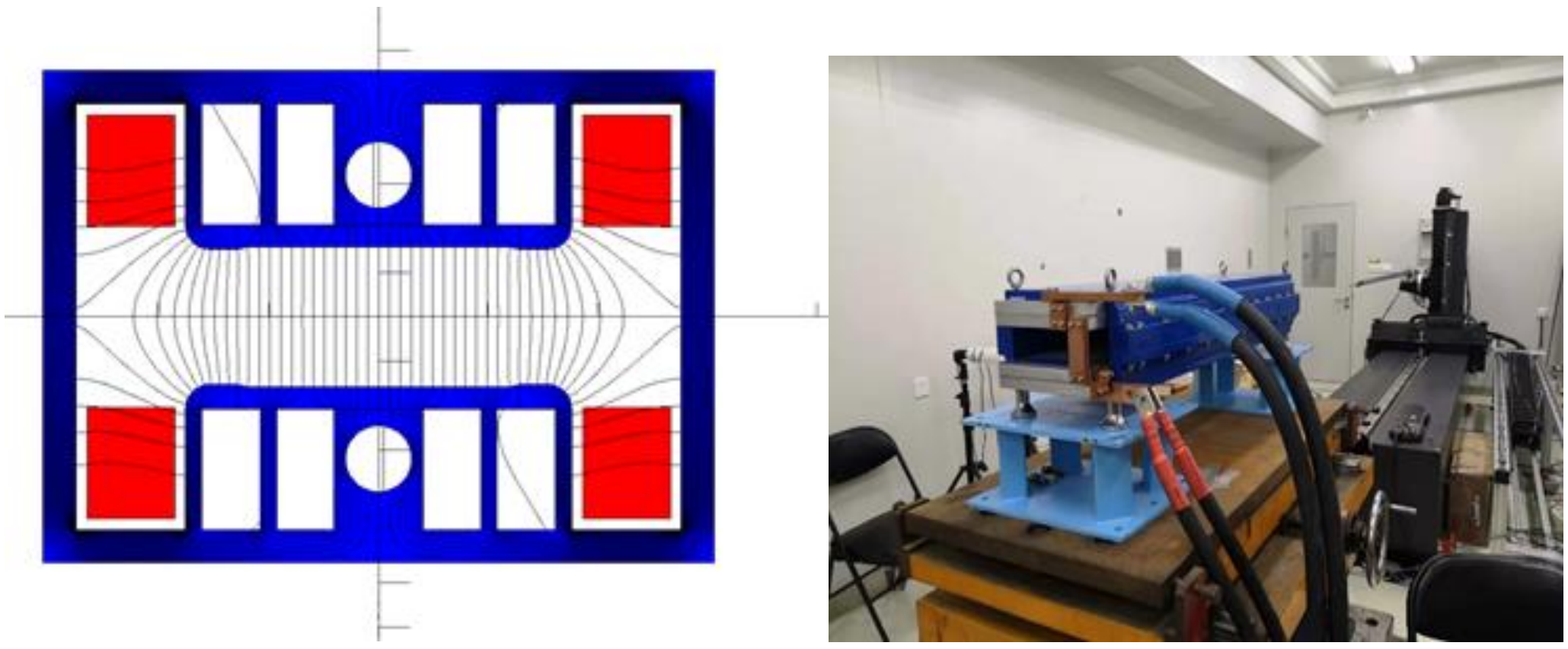

**Figure 5.3.3.2:** The cross section and picture of the iron-core dipole magnet

The CT coil dipole magnet consists of two half coils and a shielding tube. Each half coil has two layers and three turns, which are formed by pure aluminum conductors. The conductors for the coils are directly fabricated from two aluminum tubes with the appropriate diameters. The shielding tube is made from a long iron tube with an inner diameter of 290 mm. Additionally, two long iron plates for field adjustment are placed at the top and bottom of the cylinder. All surfaces of the aluminum conductors are anodized with 50 microns thickness to provide insulation from turn to turn. The inner and outer conductors of the coils are connected by circular conductors at the ends of the magnet, which serve as a bypass. Before connecting the conductors, the anodized film on the connecting surfaces was removed.

The subscale prototype dipole magnet with an iron-core consists of cores and coils. To achieve better shielding effectiveness from the earth's field, a closed H-type core was chosen for the dipole magnet. To increase the field and decrease the weight of the core, as well as fabrication costs, the technology of core dilution in longitudinal direction that was firstly used in LEP's dipole magnets was adopted.

In LEP's case, the minimum working field was 205 Gs, the cores of the magnets were composed of a stack of 1.5 mm thick steel laminations separated by 4 mm gaps which were filled with cement mortart [1].

To simplify the fabrication of the CEPC magnet, both silicon-steel and aluminum laminations were used to stack the magnet cores. Initially, a 1:2 ratio of these laminations was tested in a subscale prototype dipole magnet. However, the test results revealed poor field uniformity at low field levels and low field excitation efficiency compared to a dipole magnet made solely from steel laminations. Consequently, a 1:1 ratio was employed for another subscale prototype dipole magnet, which did not compromise field performance in comparison to the pure steel laminations. This 1:1 ratio was selected for the final design of the iron-core dipole magnet.

To further enhance field strength and reduce core weight, the return yokes' thickness was minimized, and holes were punched in the pole areas of the laminations. To mitigate the impact of the remnant field on low-field precision, grain-oriented silicon-steel laminations with lower coercive force were utilized for core production, which is one-tenth that of non-oriented steel laminations. The magnet coils consist of two turns formed by long aluminum bars without water cooling.

Figures 5.3.3.3 and 5.3.3.4 respectively display the test results of the CT coil dipole magnet. The measured precision of the CT coil dipole magnet is better than the required value of 1E–3 at all field levels, particularly at the low field of 28 Gs. After the magnet is excited three times, from 28 Gs to 338 Gs and back to 28 Gs, the field reproducibility at all field levels is better than the required value of 5E–4.

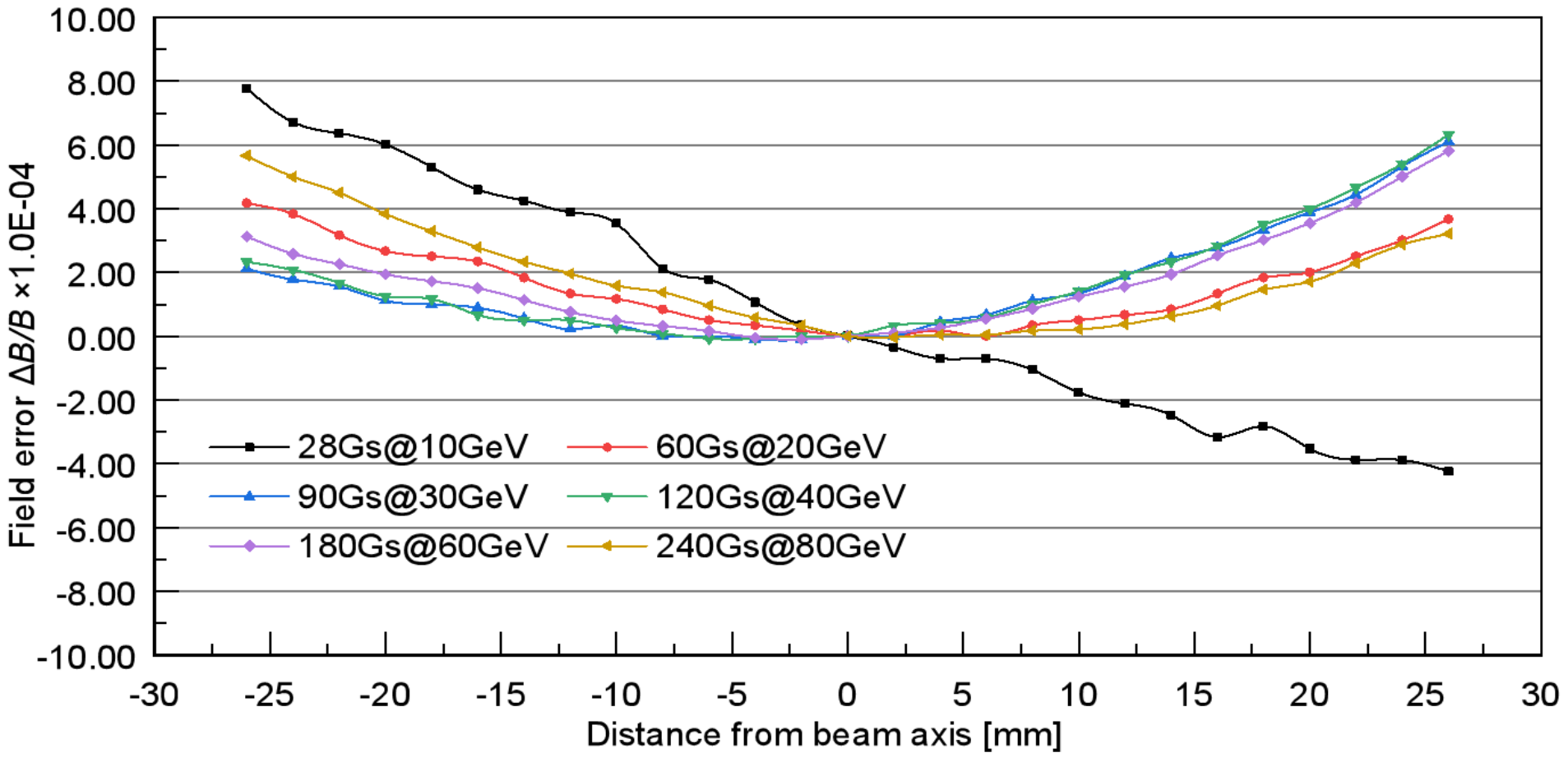

**Figure 5.3.3.3:** The field precision of the CT coil dipole magnet

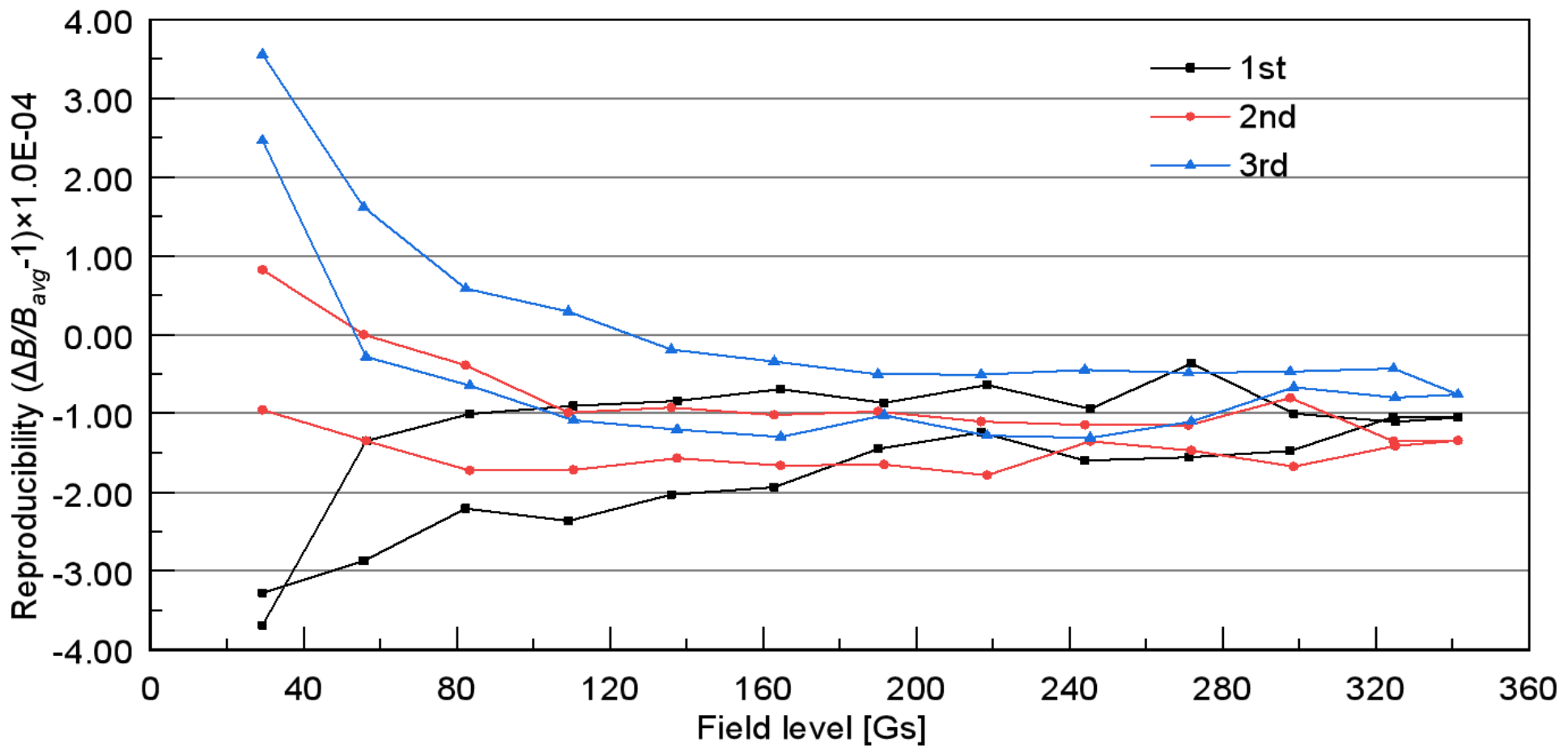

**Figure 5.3.3.4:** The field reproducibility of the CT coil dipole magnet

Figures 5.3.3.5 and 5.3.3.6 show the test results of the iron-core dipole magnet. The measured field precision at 28 Gs is approximately 3E–3, which does not meet the requirements. However, the precision of the magnet is better than 1E–3 after the low field increases to 56 Gs, which corresponds to the injection energy of 20 GeV. The field reproducibility at all field levels is better than the required value of 5E–4 after the magnet is excited three times, from 28 Gs to 338 Gs and back to 28 Gs.

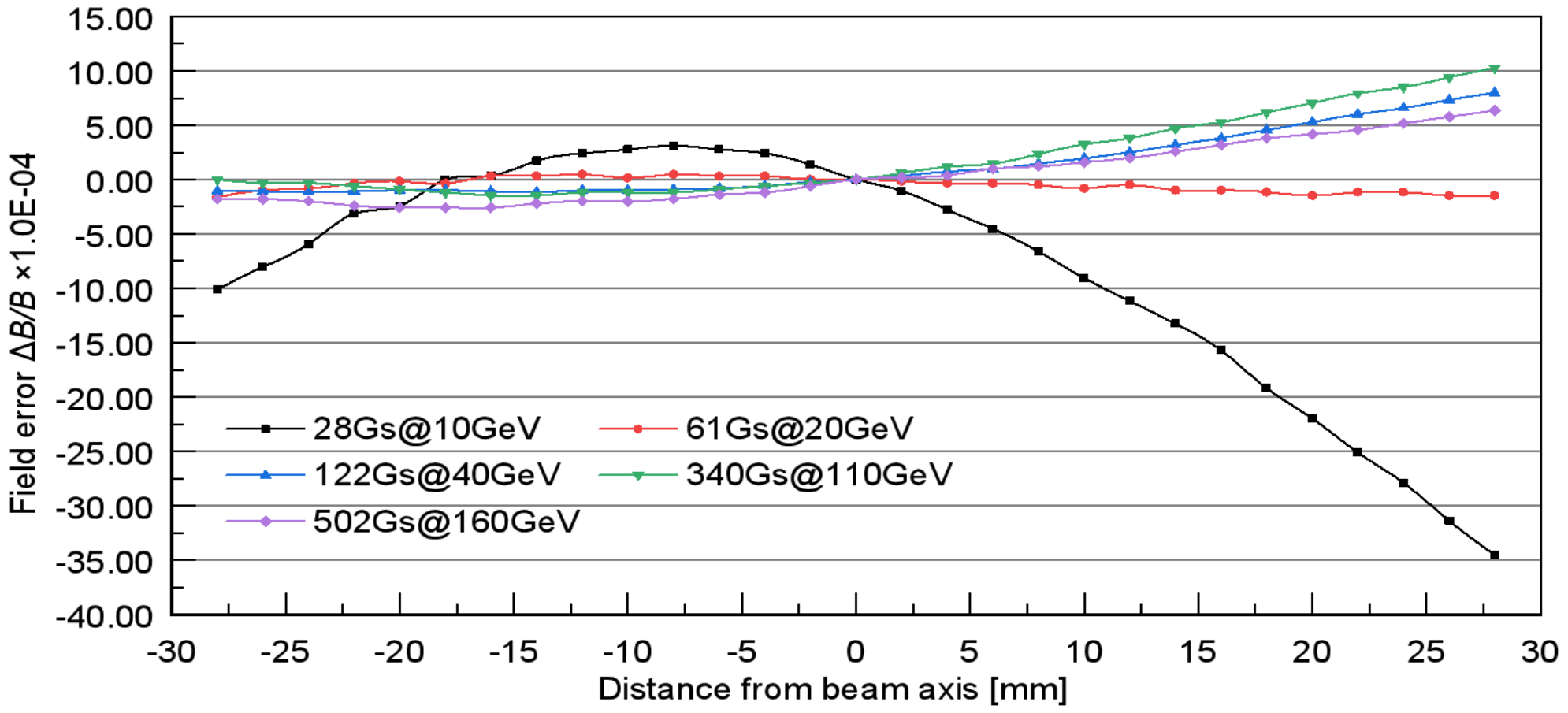

**Figure 5.3.3.5:** The field precision of the iron-core dipole magnet

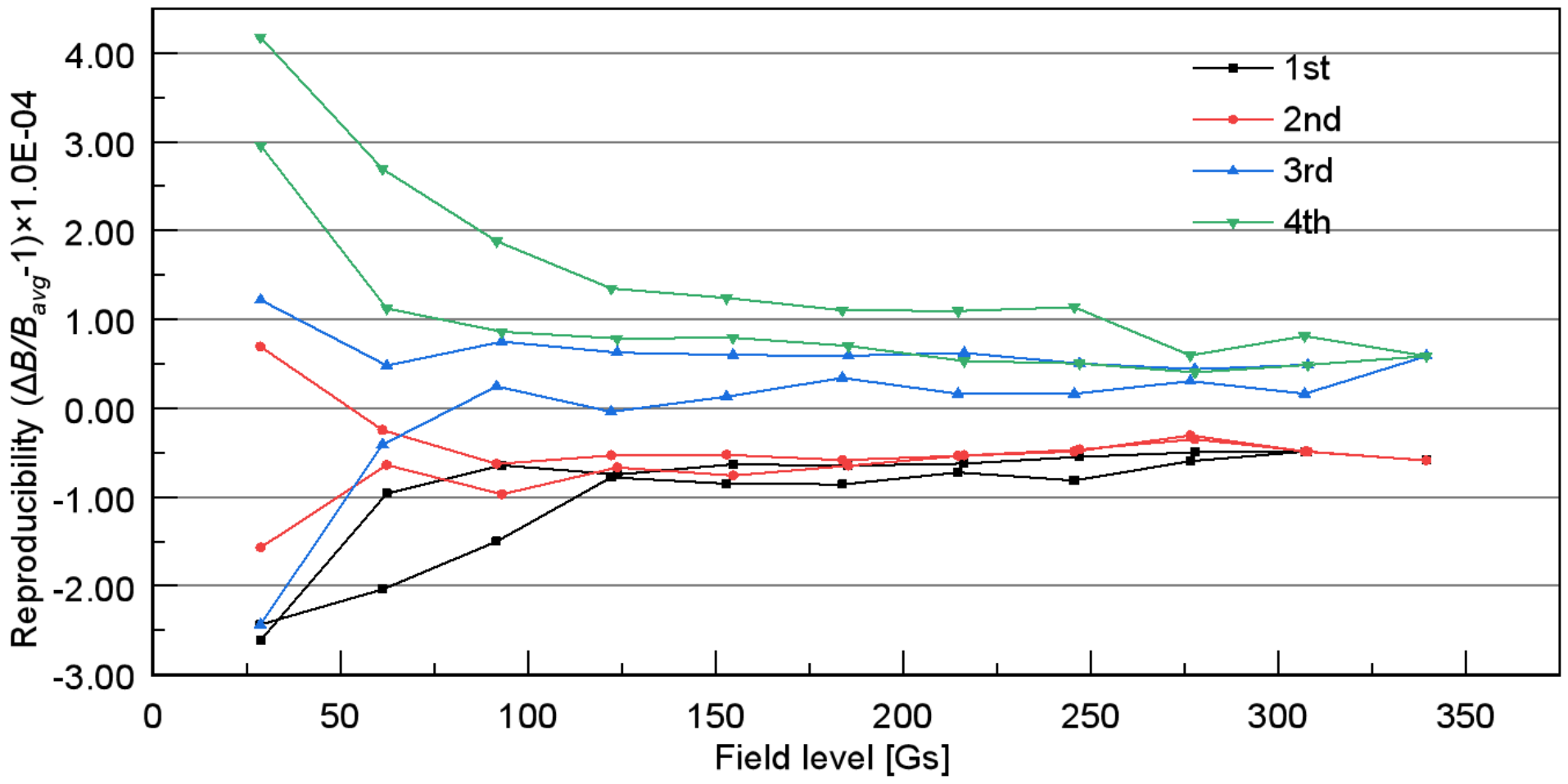

**Figure 5.3.3.6:** The field reproducibility of the iron-core dipole magnet

The cost and power loss of the CT coil dipole magnet are very high, and the key advantages of the iron-core dipole magnet are low cost and low power loss. In order to control the cost and power loss of the magnets in the Booster, a decision was made in 2021 to increase the injection energy of the Booster from 10 GeV to 20 GeV.

However, the cost of the iron-core dipole magnets with cores made of grain-oriented silicon-steel laminations is still higher than those made with non-orientedlaminations. Several years ago, the first scaled-down prototype dipole magnet was made using non-oriented silicon-steel laminations; the test results showed that the field precision could meet the requirements only if the minimum working field of the dipole magnets was increased to 95 Gs. This means that the injection energy of the Booster had to be increased to 30 GeV, as shown in Fig. 5.3.3.7.

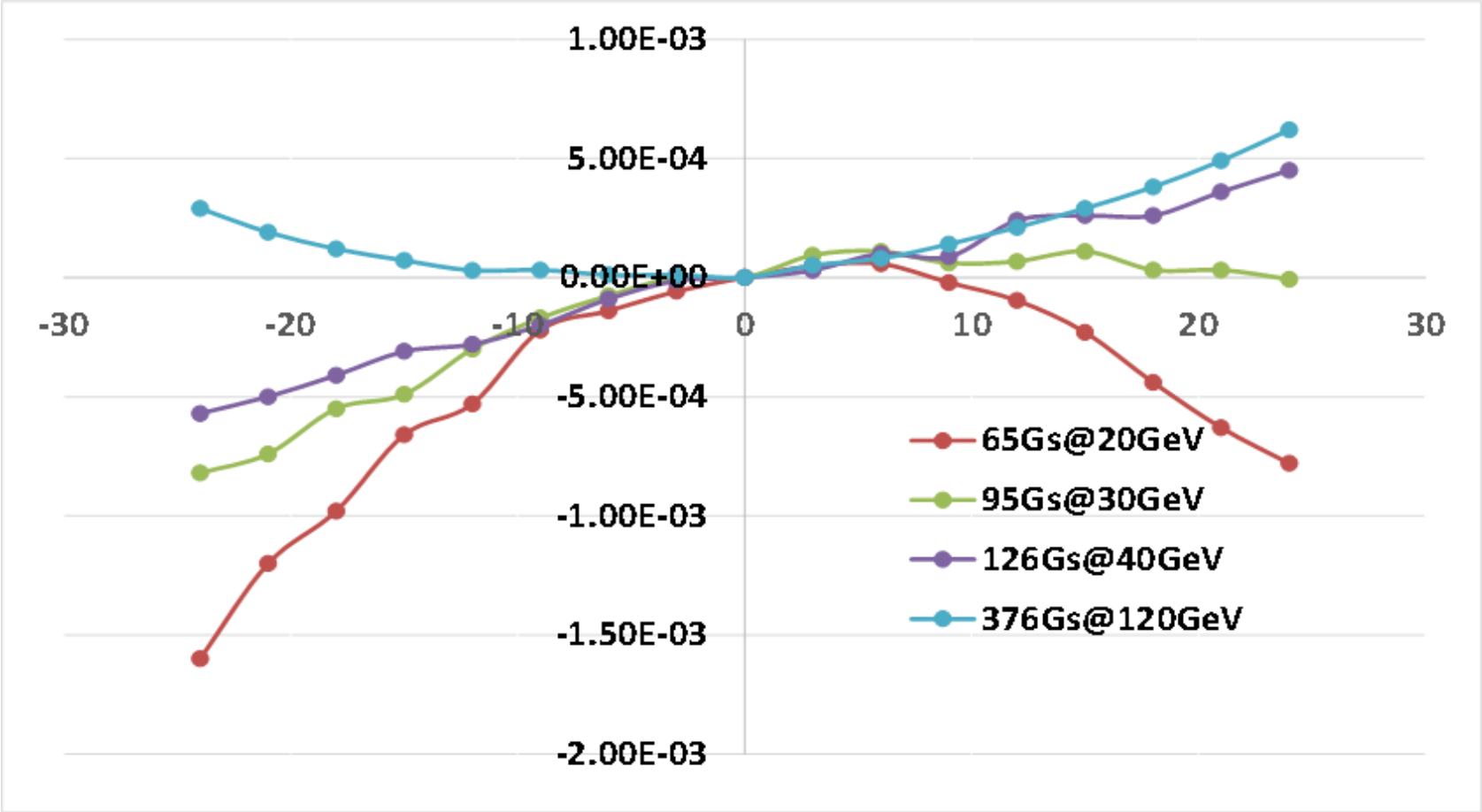

**Figure 5.3.3.7:** The field precision of the first scaled-down prototype dipole magnet

In order to reduce the overall cost of the Booster, based on the test results of the prototype dipole magnets, the injection energy of the Booster has been increased to 30 GeV, and the minimum working field of the dipole magnetsraised to 95 Gs. The dipole magnet with cores made by the non-oriented silicon steel laminations is now the baseline design for the CEPC Booster magnets.

The magnetic field of the dipole and quadrupole magnets of the Booster will change with the beam energy as the particle beams are accelerated from 30 GeV to 45.5 GeV (Z), 120 GeV (Higgs) or 180 GeV ($t\bar{t}$). The duty factors for Z, Higgs, and $t\bar{t}$ are 0.46, 0.47, and 0.28, respectively, which are essential to evaluate the average power loss of the magnets. Figure 5.3.3.8 shows a typical magnetic-field cycle of the CEPC Booster.

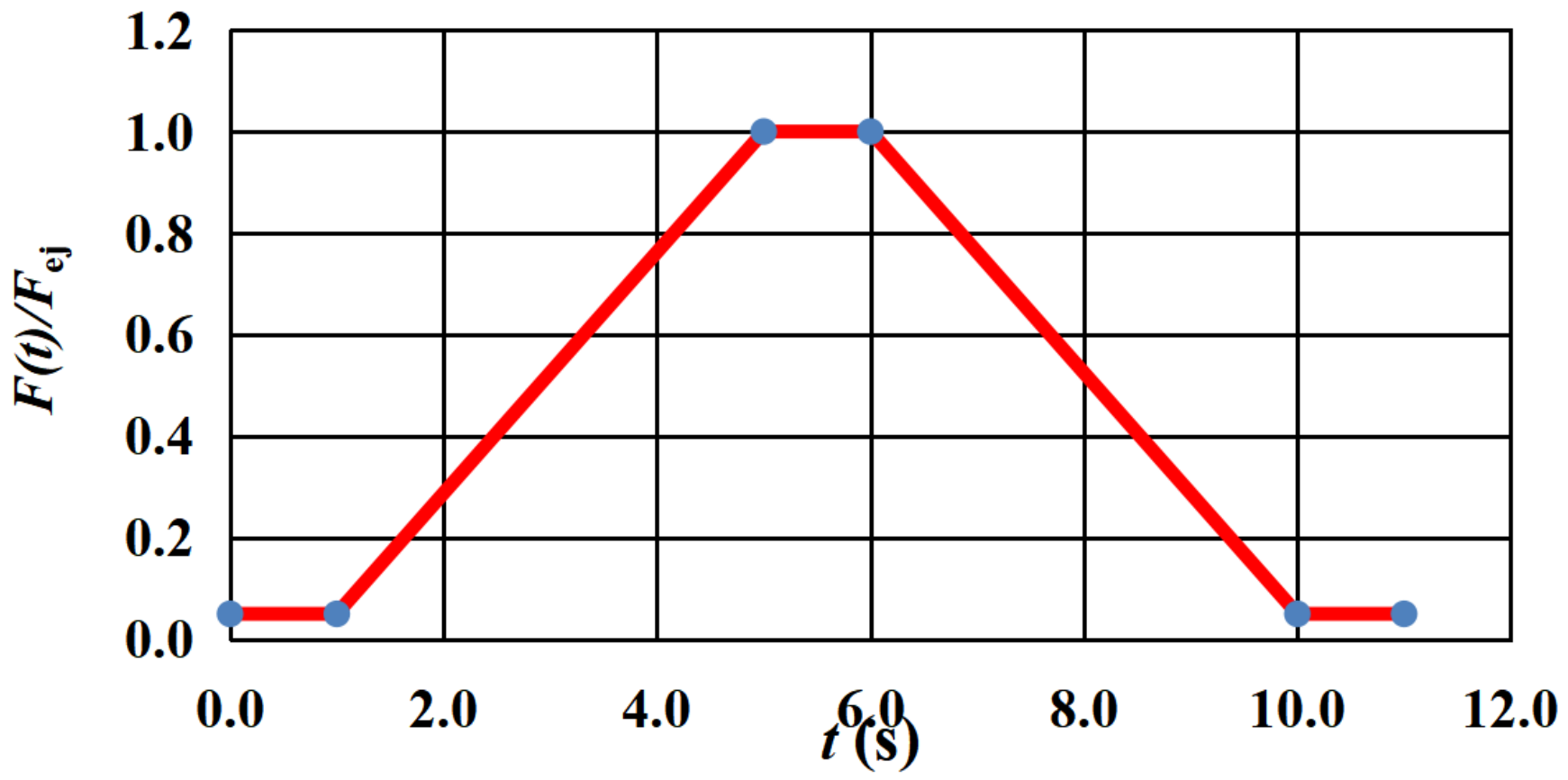

**Figure 5.3.3.8:** The magnetic field cycle of the CEPC booster

### 5.3.3.2 *Dipole Magnets*

The CEPC Booster is designed to have a total of 14,866 dipole magnets, grouped into three families. One family is pure dipoles, while the other two families are dipole and sextupole combined magnets with the sextupole field of 10.69 T/m$^2$ and –12.76 T/m$^2$ respectively at 120 GeV. Most of the magnets in the arc regions are 4.7 meters long, some in the interaction regions are 2.35 meters long. During acceleration from an injection energy of 30 GeV to extraction energy, the magnetic field of the magnets will change from 95 Gs to 376 Gs at 120 GeV or to 564 Gs at 180 GeV. The field precision is required to be better than $1\times10^{-3}$ at all field levels during the acceleration. After several iterations of optimization using OPERA-2D, the cross sections for three families of dipole magnets have been designed and optimized. The resulting magnetic flux lines and field distributions for these magnets are shown in Figures 5.3.3.9 through 5.3.3.11.

During the optimization process, the field quality of both the dipole and sextupole magnets were improved to meet the required specifications. All field optimizations were carried out at a field level of 376 Gs @ 120 GeV, as at this level the influence of the remnant field of the cores can be neglected and the field simulations are relatively accurate.

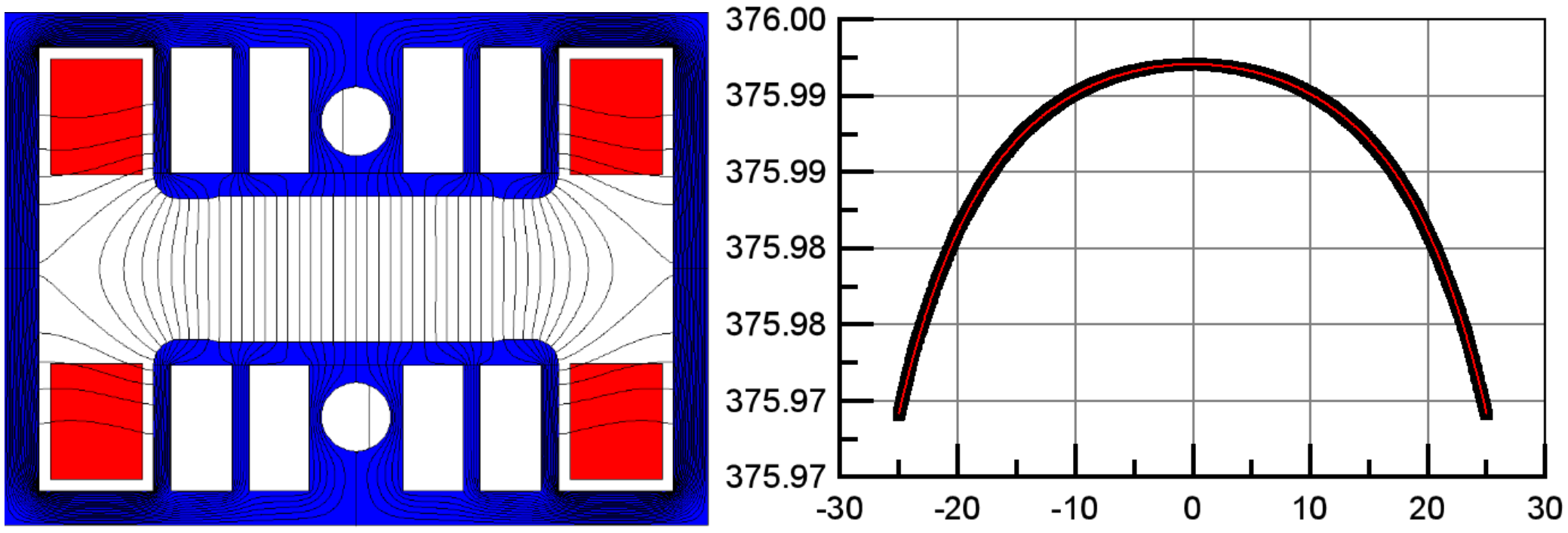

**Figure 5.3.3.9:** The flux lines and field distributions of the pure dipole magnet

The pure dipole magnet has a field distribution curve that is fitted by a polynomial:

$$B_y = -2.7\text{E-}11x^6 - 6.2\text{E-}12x^5 - 1.3\text{E-}08x^4 + 4.3\text{E-}09x^3 - 1.8\text{E-}05x^2 - 2.6\text{E-}07x + 375.99,$$

where $x$ is the distance from the magnet's center in mm, and $B_y$ is the magnetic field strength in Gauss.

The dipole field is 376 Gs, and the total field errors in the good field region is about 0.017 Gs, which is a small fraction of the dipole field (4.4E–5).

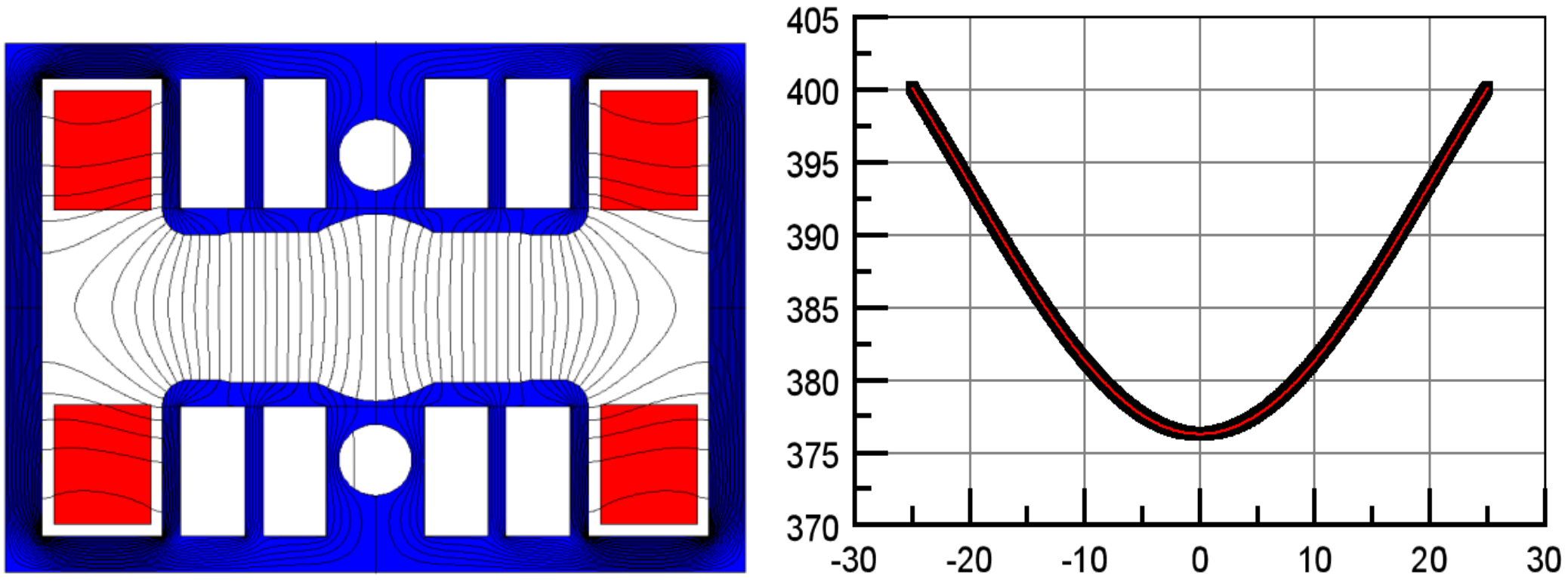

**Figure 5.3.3.10:** The flux lines and field distributions of the dipole-SF combined magnet

For the dipole-SF combined magnet, the magnetic-field distribution can be represented by the following polynomial equation:

$$B_y = 6.1\text{E-}09x^6 + 1.4\text{E-}10x^5 - 2.8\text{E-}05x^4 - 9.7\text{E-}08x^3 + 0.0534x^2 + 1.4\text{E-}05x + 376.28,$$

This equation shows that the dipole field has a strength of 376 Gs and the gradient of the sextupole field is 0.1068 Gs/mm$^2$ (10.68 T/m$^2$).

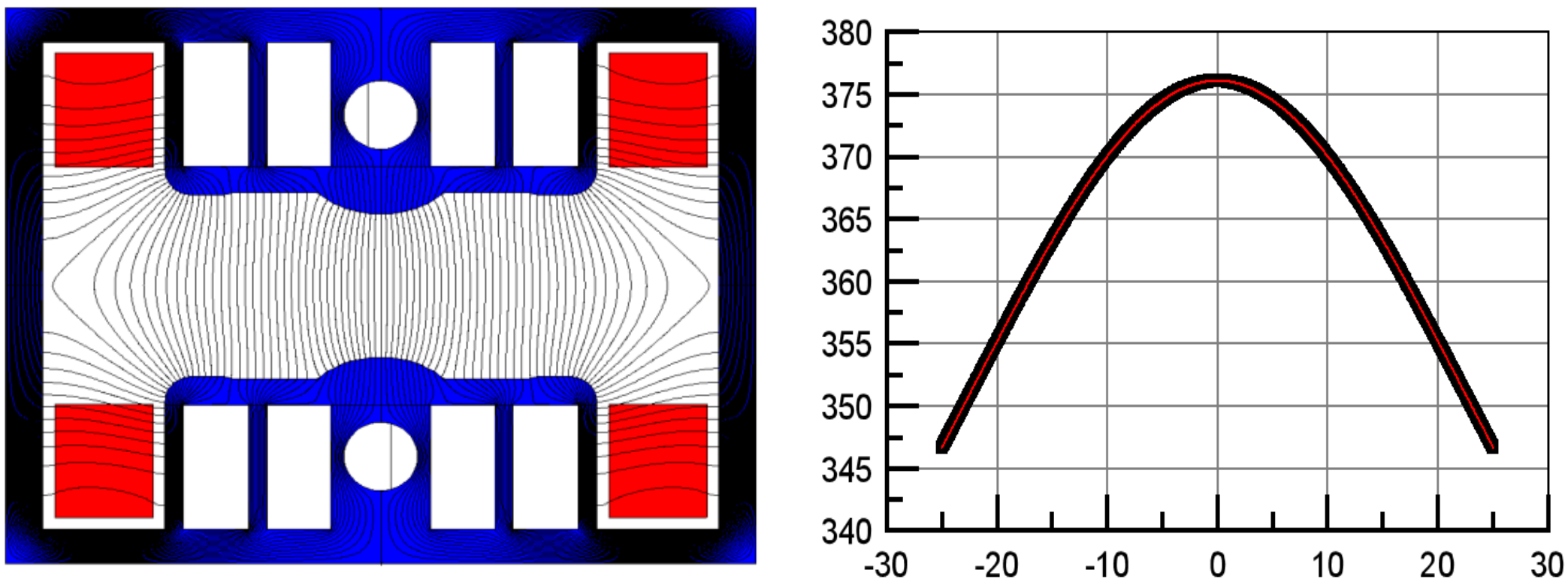

**Figure 5.3.3.11:** The flux lines and field distributions of the dipole-SD combined magnet

For the dipole-SD combined magnet, the field distribution curve can be described by a polynomial equation given below:

$$B_y = -9.2\text{E-}09x^6 + 2.8\text{E-}11x^5 + 3.3\text{E-}05x^4 - 2.7\text{E-}08x^3 - 0.0639x^2 + 7.9\text{E-}06x + 376.09,$$

The dipole field of this magnet is 376 Gs and the gradient of the sextupole field is –0.1278 Gs/mm$^2$ (–12.78 T/m$^2$).

To compensate for the low magnetic field, the cores of the dipole magnets can be constructed using stacks of 1 mm-thick low-carbon steel laminations separated by 1 mm-thick aluminum laminations. Since the magnetic force on the poles is minimal, the return yoke of the core can be made as thin as possible. Additionally, in the pole areas of the laminations, some holes can be stamped to further decrease the weight of the cores and increase the field in the laminations. Utilizing steel-aluminum cores, a thin return yoke and holes in pole areas significantly improve the performance of the iron core, while simultaneously reducing the weight and cost of the magnets.

To enhance the shielding of the earth's magnetic field, the dipole magnets are equipped with closed H-type cores. Since the minimum working field of the magnet is increased to 95 Gs, the impact of remnant field of non-oriented silicon-steel laminations on the low field precision is weak enough to be neglected, this kind of laminations can be used to stack the magnet cores.

The coils installed on the upper and lower poles of the magnets consist of only two turns. To reduce production costs, each turn of the coils is formed using pure aluminum bars with a cross section of 25 × 40 mm$^2$. The coils are cooled by air rather than cooling water due to the low Joule loss in the bars. Instead of conventional vacuum epoxy casting, G10 plates are used for insulation between each turn or between the coil and iron cores. A special structure is utilized to secure the coils onto the iron cores.

Figure 5.3.3.12 displays a 3D rendering and a photograph of the full-scale prototype dipole magnet. Steel bars have been strategically positioned on top and along the sides of the magnet cores to apply a securing force, ensuring the stability of the laminations. To reduce longitudinal deformation, the magnet was carefully positioned on a precisely machined surface of a long I-steel beam. One can refer to Table 5.3.3.1 for the essential design parameters of the dipole magnets.

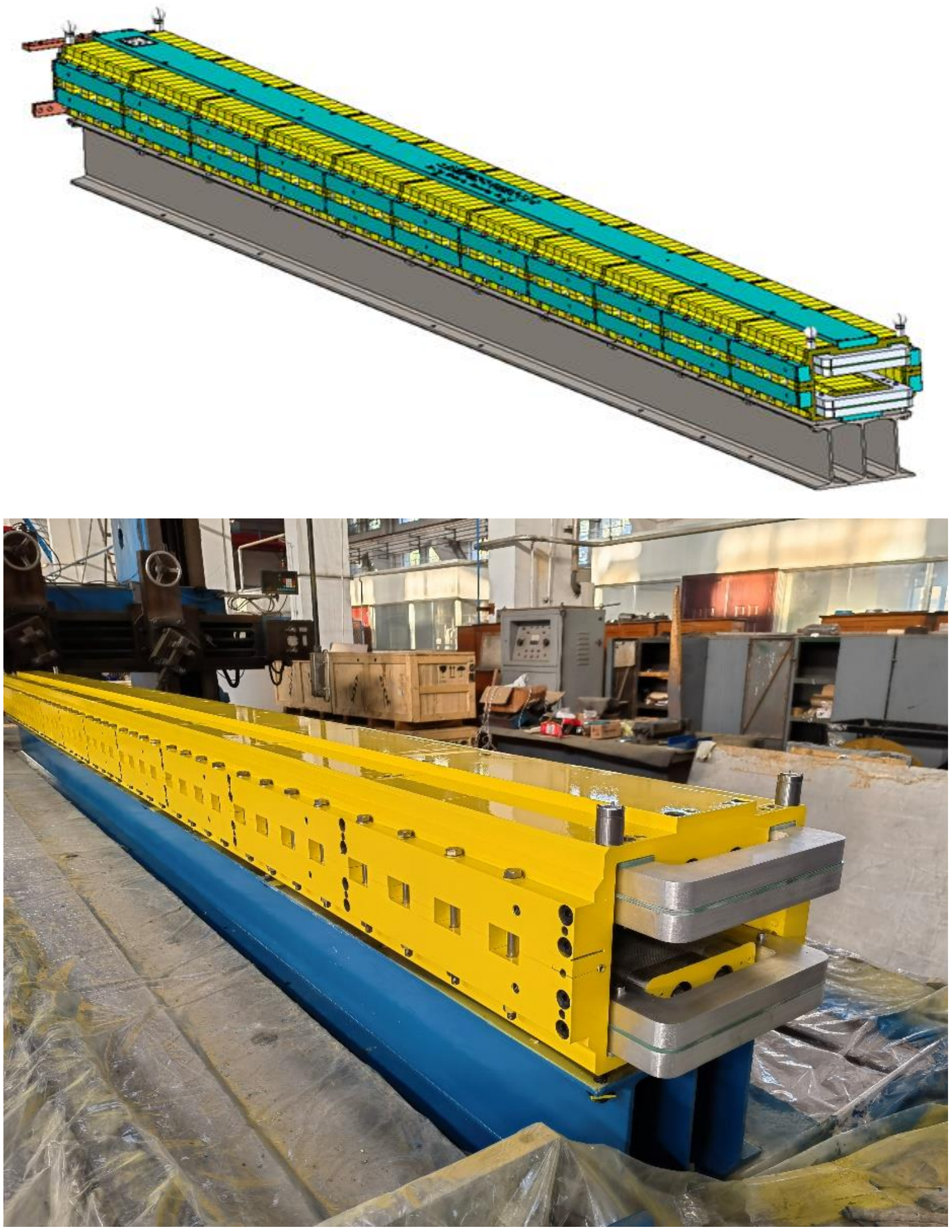

**Figure 5.3.3.12:** The 3D drawing and picture of the full-scale prototype dipole magnet

**Table 5.3.3.1** Main parameters of the Booster dipole magnets

| Magnet name | BST-63B-Arc | BST-63B-Arc-SF | BST-63B-Arc-SD | BST-63B-IR |
|---|---|---|---|---|
| Quantity | 10192 | 2017 | 2017 | 640 |
| Aperture [mm] | 63 | 63 | 63 | 63 |
| Dipole Field [Gs] @180 GeV | 564 | 564 | 564 | 549 |
| Dipole Field [Gs] @120 GeV | 376 | 376 | 376 | 366 |
| Dipole Field [Gs] @30 GeV | 95 | 95 | 95 | 93 |
| Sextupole Field [T/m$^2$] @180 GeV | 0 | 16.0388 | 19.1423 | 0 |
| Sextupole Field [T/m$^2$] @120 GeV | 0 | 10.6925 | 12.7615 | 0 |
| Sextupole Field [T/m$^2$] @30 GeV | 0 | 2.67315 | 3.19035 | 0 |
| Magnetic length [mm] | 4700 | 4700 | 4700 | 2350 |
| GFR [mm] | ±22.5 | ±22.5 | ±22.5 | ±22.5 |
| Field errors | $\pm1\times10^{-3}$ | $\pm1\times10^{-3}$ | $\pm1\times10^{-3}$ | $\pm1\times10^{-3}$ |
| Ampere turns per pole [At] @180 GeV | 1428 | 1583 | 1527 | 1390 |
| Ampere turns per pole [At] @120 GeV | 952 | 1055 | 1018 | 927 |
| Ampere turns per pole [At] @30 GeV | 240 | 266 | 257 | 235 |
| Turns per pole | 2 | 2 | 2 | 2 |
| Current [A] @180 GeV | 714 | 791 | 764 | 695 |
| Current [A] @120 GeV | 476 | 528 | 509 | 463 |
| Current [A] @30 GeV | 120 | 133 | 129 | 118 |
| Size of conductor [mm×mm] | 25×40-Al | 25×40-Al | 25×40-Al | 25×40-Al |
| Max. current density [A/mm$^2$] | 0.71 | 0.79 | 0.76 | 0.69 |
| Resistance of the coil (mΩ) | 1.56 | 1.56 | 1.56 | 0.817 |
| Power loss (W) @180 GeV | 796.1 | 978.4 | 910.7 | 394.7 |
| Power loss (W) @120 GeV | 353.8 | 434.9 | 404.9 | 175.4 |
| Avg. power loss [W] @30-180 GeV | 222.9 | 274.0 | 255.0 | 110.5 |
| Avg. power loss [W] @30-120 GeV | 166.3 | 204.4 | 190.3 | 82.4 |
| Max. DC voltage [V] @180 GeV | 1.12 | 1.24 | 1.19 | 0.568 |
| Max. DC voltage [V] @120 GeV | 0.744 | 0.824 | 0.795 | 0.379 |
| Core height [mm] | 280 | 280 | 300 | 280 |
| Core width [mm] | 400 | 400 | 400 | 400 |
| Core Length [mm] | 4653 | 4653 | 4653 | 2303 |
| Magnet weight [ton] | 3.98 | 3.98 | 4.24 | 1.99 |

A hall probe field measurement system is utilized for assessing the magnetic field of the dipole magnet. This system comprises a 6 m-long slide rail, a trolley, a gauss meter, and a computer. The slide rail is mounted on the magnet's pole surface, with the gauss meter positioned on the trolley, allowing it to traverse the slide rail. To accommodate the 6 m slide rail on the lower pole surface of the magnet, the upper section of the magnet must be displaced. Fig. 5.3.3.13 illustrates the magnet containing the hall probe measurement system.

Fig. 5.3.3.14 displays the integral field distributions, revealing that field uniformity at all field levels is within $\pm1\times10^{-3}$. For assessing harmonic errors in the magnet, a rotating coil field measurement system is planned for future implementation. Fig. 5.3.3.15 depicts the measured field reproducibility, where the magnet was excited four times, and the field reproducibility remained within $\pm5\times10^{-4}$.

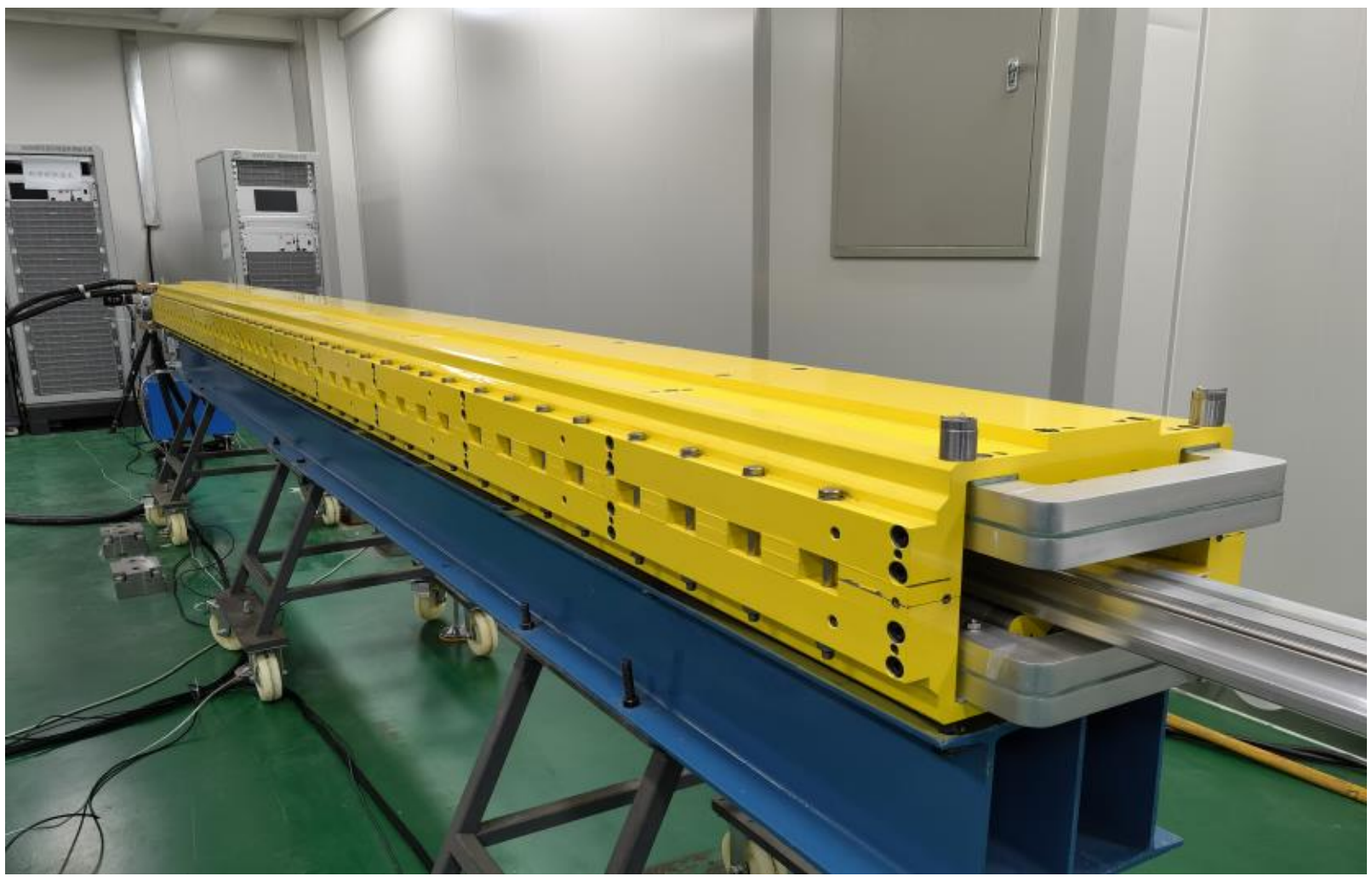

**Figure 5.3.3.13:** The dipole magnet with the hall probe measurement system in it.

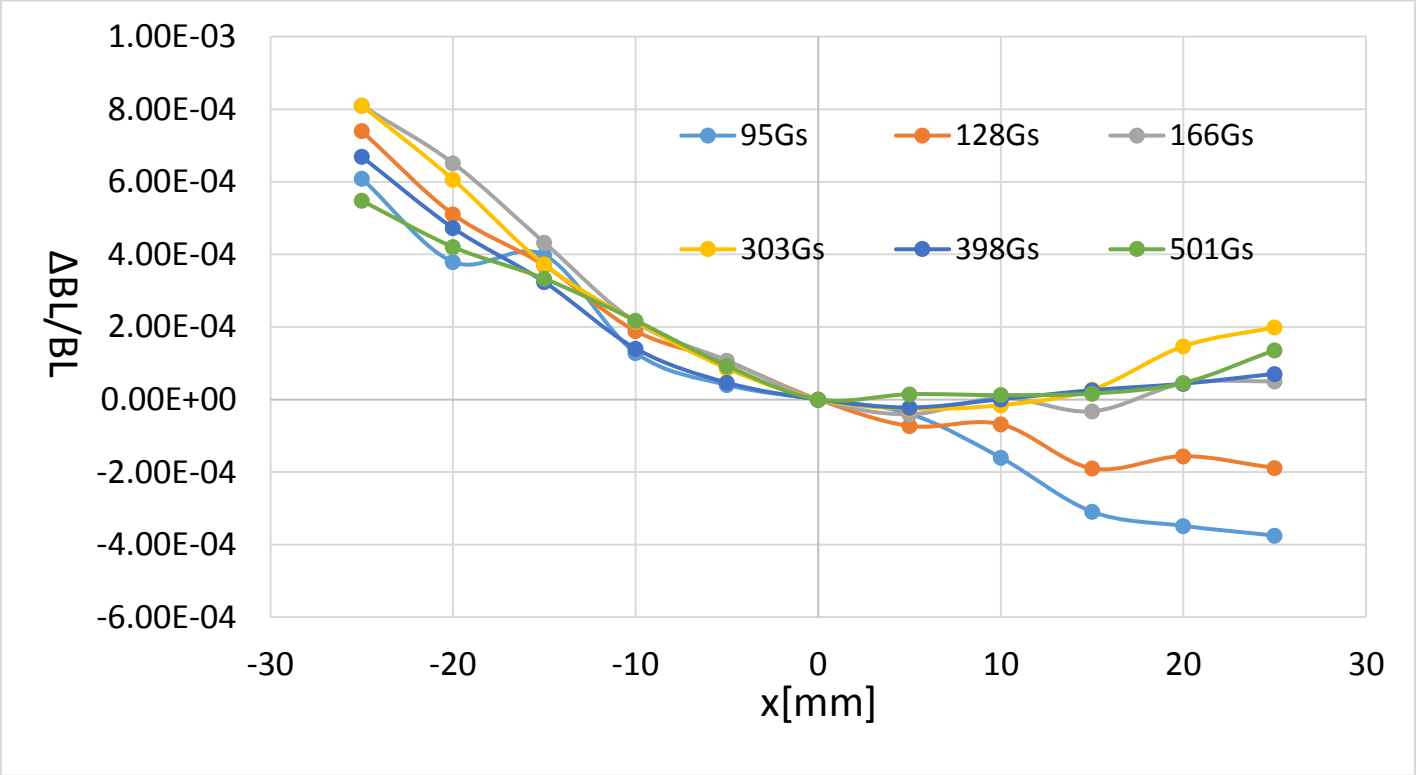

**Figure 5.3.3.14:** The integral field distributions along the x direction.

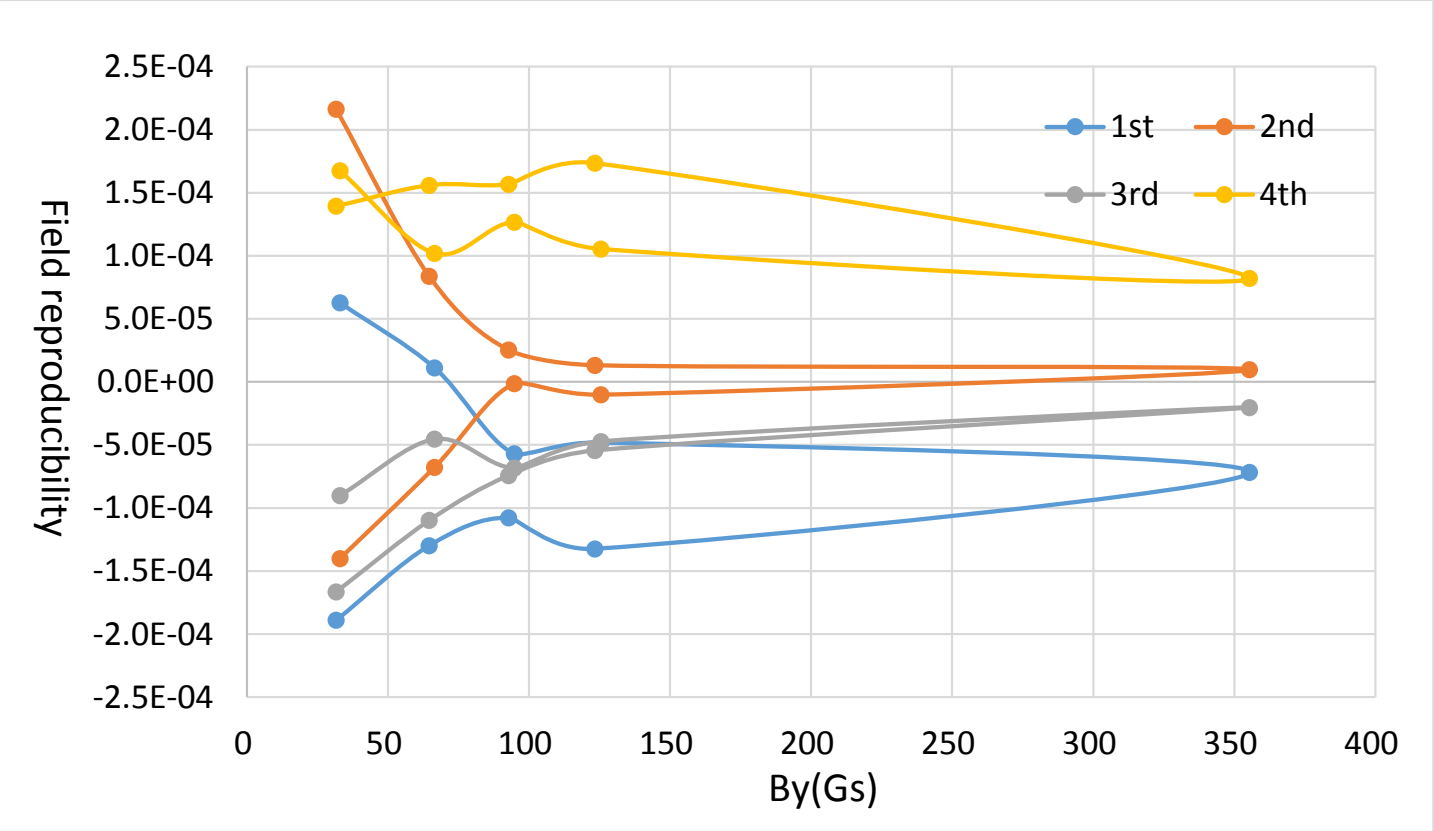

**Figure 5.3.3.15:** The field reproducibility of four excited cycles.

Additionally, to verify the field quality of the combined dipole and quadrupole magnet, a full-scale prototype magnet will be designed, fabricated, and subjected to field quality assessment using a harmonic coil field measurement system during the CEPC EDR stage.

### 5.3.3.3 *Quadrupole Magnets*

As a large number of quadrupole magnets is required for the Booster, reducing the cost of the magnets is of great concern. Additionally, the effective length of the quadrupole magnets can be up to 2,000 mm according to Table 5.3.3.4, so optimized physical and mechanical designs, as well as precise machining and assembly processes, are crucial to ensure the field quality and to reduce the production cost. The following design considerations were taken into account for the quadrupole magnets,

1) A hollow aluminum conductor was selected due to its lower cost and weight, instead of the conventional copper conductor.
2) The iron cores are made of low-carbon silicon-steel sheets with a thickness of 0.5 mm. The whole magnet will be assembled from four identical quadrants for coil installation and can also be split into two halves for vacuum chamber installation.
3) Due to the relatively low gradient of the quadrupole magnets for the CEPC Booster, the pole root of the magnet is rectangular rather than tapered. This allows for a simpler cross-sectional shape for the magnet core, and a simpler racetrack structure for the magnet coil.

Table 5.3.3.2 lists the main design parameters.

**Table 5.3.3.2:** The main parameters of the CEPC Booster quadrupole magnets

| Magnet name | BS-63Q-2000L | BS-63Q-1000L | BS-63Q-700L |
|---|---|---|---|
| Quantity | 1144 | 296 | 2018 |
| Aperture [mm] | 63 | 63 | 63 |
| Quadrupole Field [T/m] @180 GeV | 12.8811 | 15.8575 | 12.0313 |
| Quadrupole Field [T/m] @120 GeV | 8.5874 | 10.5717 | 8.0209 |
| Quadrupole Field [T/m] @30 GeV | 2.147 | 2.643 | 2.005 |
| Effective Length [mm] | 2000 | 1000 | 700 |
| GFR (radius) [mm] | 22.5 | 22.5 | 22.5 |
| Harmonic errors | $\pm 1\times 10^{-3}$ | $\pm 1\times 10^{-3}$ | $\pm 1\times 10^{-3}$ |
| Ampere turns per pole [At] @180 GeV | 5136.4 | 6323.2 | 4797.5 |
| Ampere turns per pole [At] @120 GeV | 3424.2 | 4215.5 | 3198.3 |
| Ampere turns per pole [At] @30 GeV | 856.1 | 1053.9 | 799.6 |
| Turns per pole | 20 | 20 | 20 |
| Current [A] @180 GeV | 257 | 316 | 240 |
| Current [A] @120 GeV | 171 | 211 | 160 |
| Current [A] @30 GeV | 42.8 | 52.7 | 40.0 |
| Size of conductor [mm×mm] | 10×10 D6R1 | 10×10 D6R1 | 10×10 D6R1 |
| Area of conductor [$mm^2$] | 71.74 | 71.74 | 71.74 |
| Max. current density [$A/mm^2$] | 3.58 | 4.41 | 3.34 |
| Resistance of the coil (mΩ) | 135.9 | 69.6 | 49.7 |
| Power loss (kW) @180 GeV | 8.96 | 6.96 | 2.86 |
| Power loss (kW) @120 GeV | 3.98 | 3.09 | 1.27 |
| Avg. power loss [kW] @30-180 GeV | 2.5 | 1.9 | 0.8 |
| Avg. power loss [kW] @30-120 GeV | 1.9 | 1.5 | 0.6 |
| Max. DC voltage [V] @180 GeV | 34.9 | 22.0 | 11.9 |
| Max. DC voltage [V] @120 GeV | 23.3 | 14.7 | 8.0 |
| Height of core [mm] | 520 | 520 | 520 |
| Width of core [mm] | 520 | 520 | 520 |
| Core Length [mm] | 1984 | 984 | 684 |
| Total weight of core [ton] | 6.33 | 3.16 | 2.21 |
| Total weight of magnet [ton] | 6.43 | 3.21 | 2.25 |
| Cooling circuits | 4 | 4 | 4 |
| Water pressure [$kg/cm^2$] | 6 | 6 | 6 |
| Water flow velocity [m/s] | 1.54 | 2.25 | 2.73 |
| Tatol water flow [l/s] | 0.174 | 0.255 | 0.309 |
| Temperature rise of coolant [deg] | 12.3 | 6.5 | 2.2 |

The cross sections for the quadrupole magnets have been designed and optimized using the OPERA-2D program. Since the effective length of the magnet is quite long, the end chamfering has little effect on improving the field quality. After optimization, a special pole profile has been designed, which is composed of a section of a hyperbolic curve and a pole shimming. The pole profile not only meets the field quality requirements better but also reduces the width of the yoke.

The magnetic-flux lines and cloud picture of the $B_{mod}$ are shown in Fig. 5.3.3.16. It can be observed that the largest magnetic-field hotspots are located at the roots of the poles, and the field in other areas is relatively low. Compared with the magnet design of the CDR, the cross-section of the quadrupole magnet of the TDR is smaller. The field quality is specified in a 22.5 mm radius good field region. It can be seen in Table 5.3.3.3 that the 2D field harmonic of the quadrupole magnet can be suppressed below 1E-4.

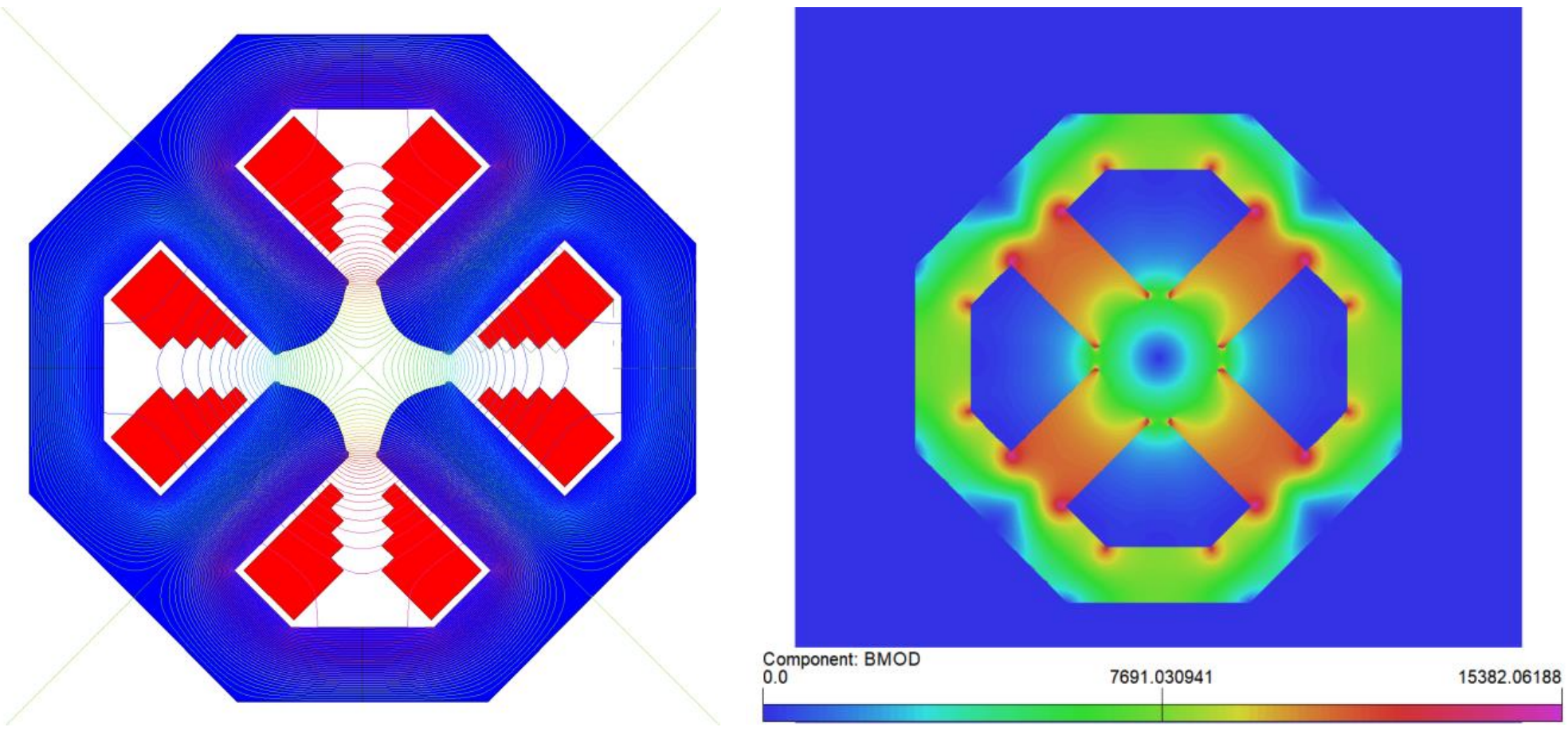

**Figure 5.3.16:** The magnetic-flux lines and cloud picture of the $B_{mod}$

**Table 5.3.3.3:** 2D simulated field harmonics in quadrupole good field region (unit: 1E-4)

| Item | Value |
|---|---|
| $B_6$ | 0.716 |
| $B_{10}$ | 0.317 |
| $B_{14}$ | 0.000835 |

The 3D model of the magnet is shown in Figure 5.3.3.17. The calculated field gradient and effective length meet the requirements at the working current. As the 2D field harmonics have been optimized to be quite small, no additional pole-end chamfer is needed.

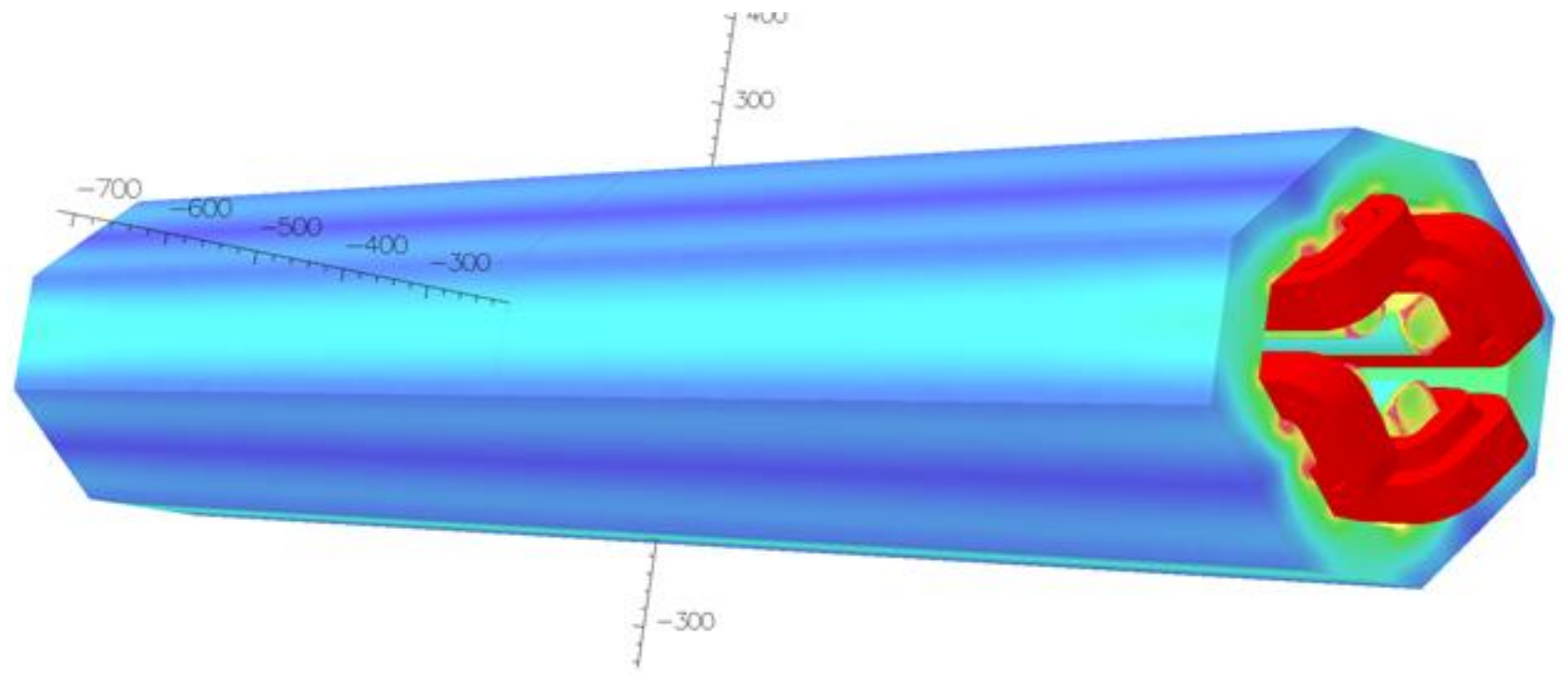

**Figure 5.3.3.17:** 3D module of the quadrupole magnet

The 3D mechanical drawing of the quadrupole magnet and the similar magnet produced in IHEP are shown in Figure 5.3.3.18. The iron core is a four-in-one structure; a single quarter iron core is composed of laminated sheets, an end plate, a puller plate, and a pole-head puller. The end plate, puller plate, and stacked punching sheets are welded together; the three puller plates construct a comprehensive envelope in the periphery of the stacked punching sheets, making the stiffness of the quarter cores stronger. In the pole head position, away from the welding, double-headed bolts are used for fastening. The excitation coil for each pole is made of a simple racetrack coil of 20 turns, each of which is cooled by a dedicated water circuit.

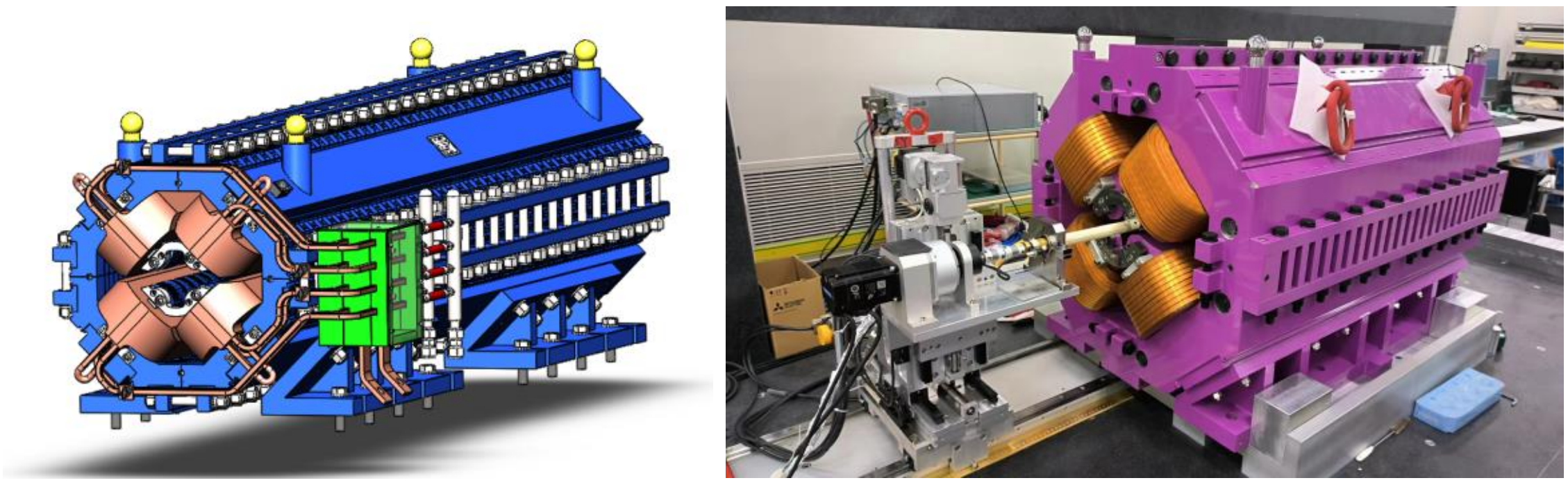

**Figure 5.3.3.18:** 3D drawing of the quadrupole magnet and the similar magnet produced in IHEP.

## 5.3.3.4 *Sextupole Magnets*

There 100 sextupole magnets. The cores of the magnets have a two-in-one structure, made of low-carbon silicon-steel sheets and end plates. By using end chamfering, the field errors can be reduced to meet the strict field requirements. The coils of the magnets have a simple racetrack-shaped structure. The coils are wound from hollow copper conductors.

The cross sections of the sextupole magnets have been meticulously designed and optimized utilizing OPERA-2D. Capitalizing on symmetry, only one-quarter of the magnet is represented in the model. Figure 5.3.3.19 depicts the simulated module of the sextupole magnet, alongside a corresponding magnet fabricated at IHEP.

The main design parameters are listed in Table 5.3.3.4.

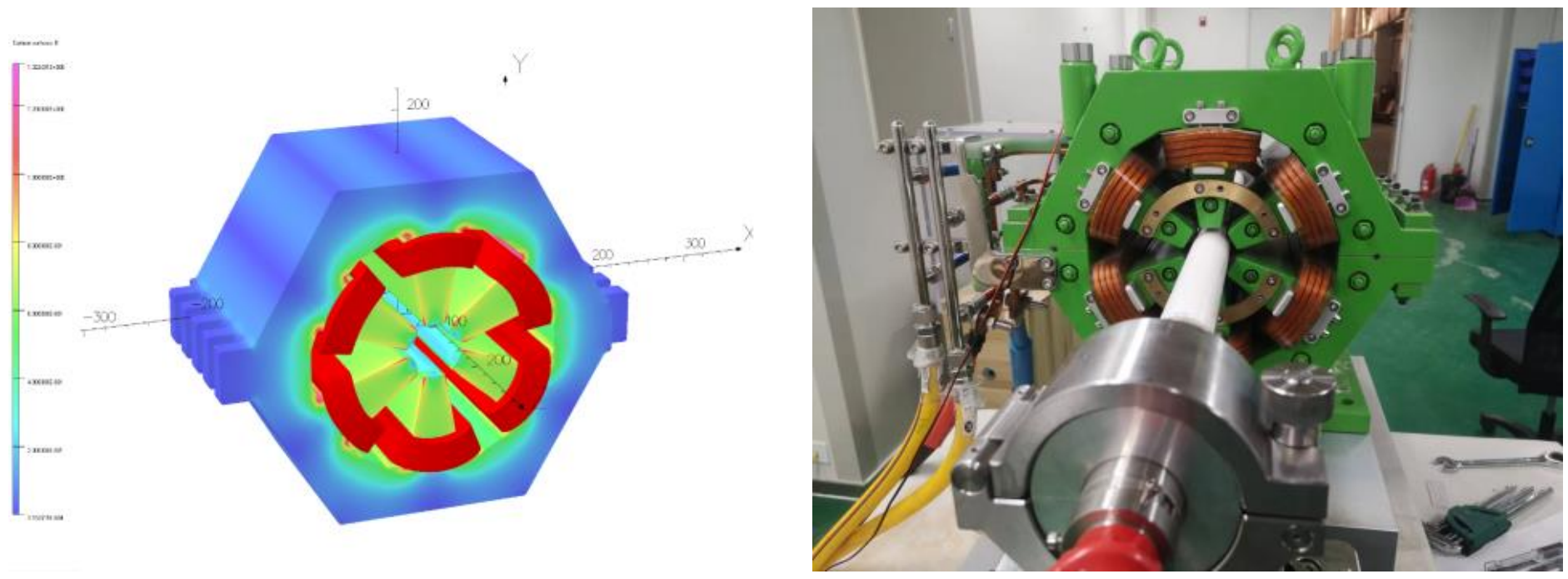

**Figure 5.3.3-19:** Magnetic flux lines of the sextupole magnet and the similar magnet produced at the IHEP.

**Table 5.3.3.4:** Main parameters of the sextupole magnets

| Magnet name | BS-63S |
|---|---|
| Quantity | 100 |
| Aperture diameter (mm) | 63 |
| Magnetic length（mm） | 400 |
| Max. sextupole field (T/m$^2$) | 216.9 |
| Min. sextupole field (T/m$^2$) | 18.1 |
| GFR radius (mm) | 28 |
| Harmonic errors | 1.0E-3 |
| Coil turns per pole | 8 |
| Max. current（A） | 120.4 |
| Cu conductor size（mm） | 6×6D3 |
| Max. current density（A/mm$^2$） | 4.19 |
| Resistance（mΩ） | 27.2 |
| Max Power loss（kW） | 0.40 |
| Avg. Power loss (kW) | 0.16 |
| Inductance（mH） | 1.76 |
| Core length（mm） | 360 |
| Core width & height（mm） | 300 |
| Magnet weight（kg） | 120 |
| Number of water circuits | 2 |
| Water pressure drop（kg/cm$^2$） | 6 |
| Flow velocity (m/s) | 2.12 |
| Water flux (l/s) | 0.03 |
| Temperature rise (℃) | 6.68 |

### 5.3.3.5 *Correction Magnets*

The CEPC Booster uses two types of correctors for correcting the closed orbit in the vertical and horizontal directions. These magnets have the same gap, maximum field, and effective length, so their designs are similar.

To ensure good field quality, the correctors utilize H-type structure cores, which allow for pole-surface shimming to optimize the field. The cores are made by stacking 0.5 mm-thick laminations. The racetrack-shaped coils are wound using solid copper conductor with a size of 5.5× 5 mm$^2$. Each coil has 24 turns, formed into four layers and is air cooled.

OPERA software was employed to simulate the corrector field. the magnetic-flux distributions are presented in figure 5.3.3.20. The 3D drawings of the correctors and the similar magnet produced in IHEP are shown in figure 5.3.3.21.

Table 5.3.3.5 lists the main parameters of the correctors.

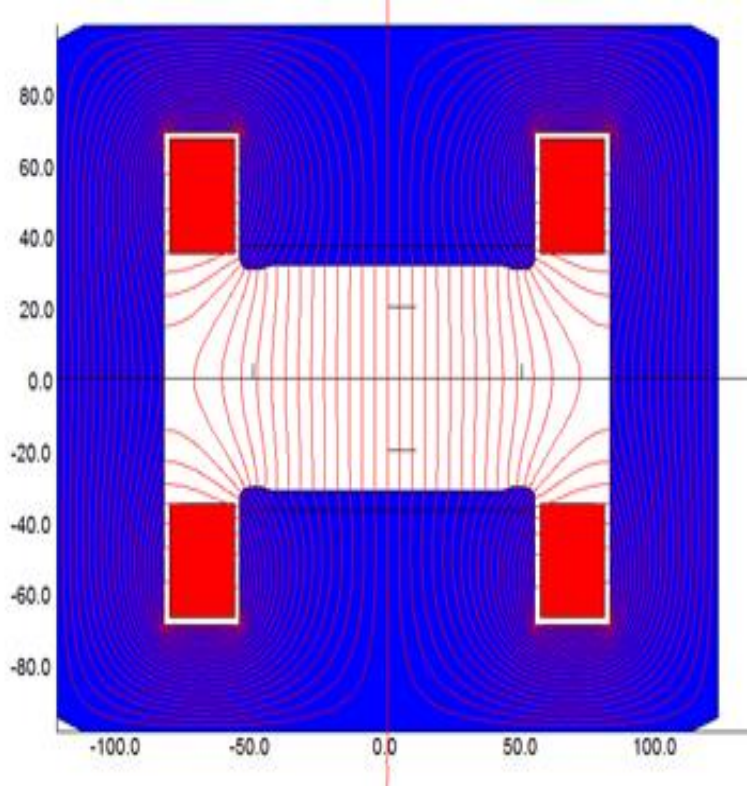

**Figure 5.3.3.20:** Magnetic flux distribution of the Booster correctors

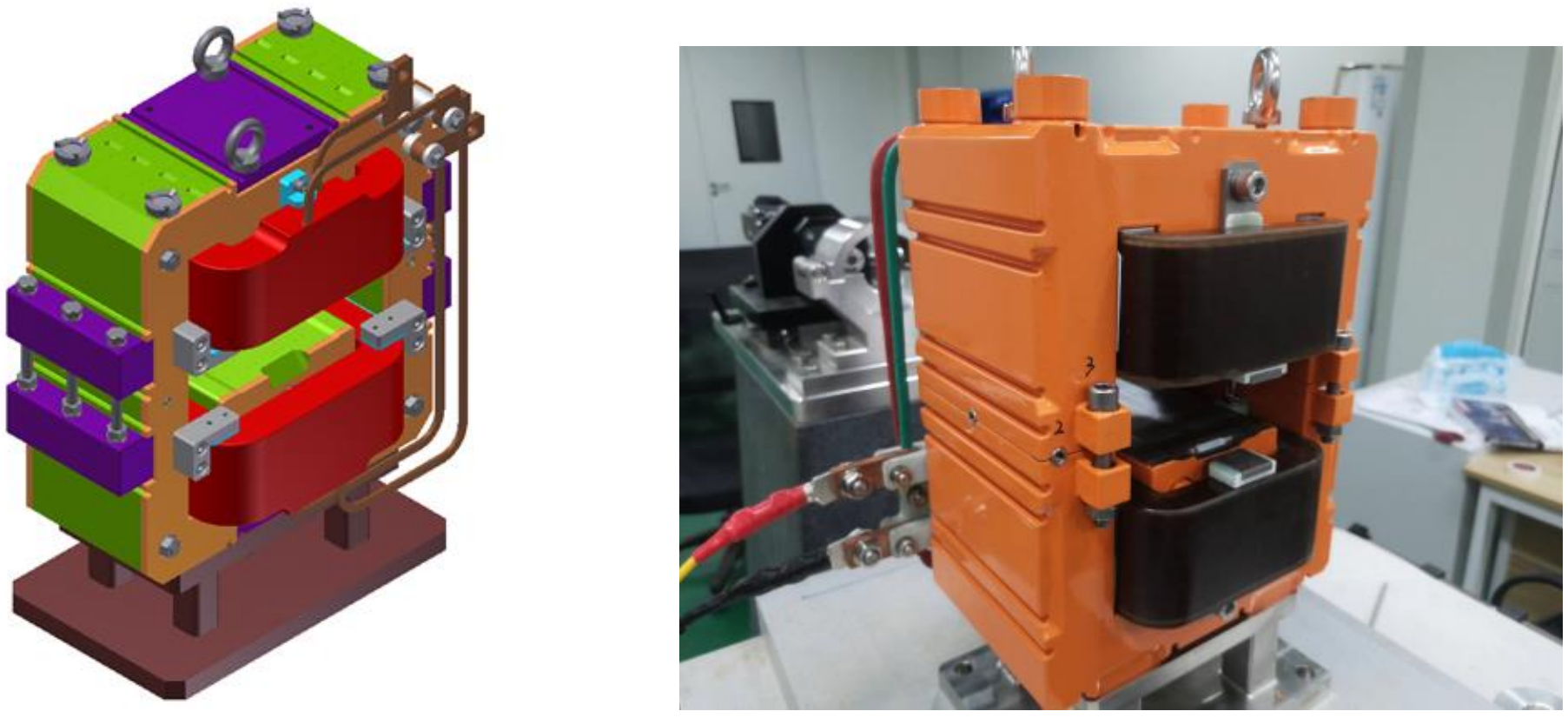

**Figure 5.3.3.21:** 3D drawings of the correctors and the similar magnet produced in IHEP

**Table 5.3.3.5:** The main design parameters of Booster correctors

| Magnet name | BST-63C |
|---|---|
| Quantity | 1200 |
| Magnet gap (mm) | 63 |
| Magnetic field (T) | 0.02 |
| Effective length (mm) | 583 |
| Good field region (mm) | 55 |
| Field errors | 0.1% |
| Ampere turns per pole (AT) | 540 |
| Maximum current (A) | 22.5 |
| Coil turns per pole | 24 |
| Conductor size ($mm^2$) | 5.5×5 |
| Current density ($A/mm^2$) | 0.82 |
| Resistance (Ω) | 0.05 |
| Voltage drop (V) | 1.12 |
| Power loss (W) | 25 |
| Core length (mm) | 550 |
| Core width/height (mm) | 246/198 |
| Magnet weight (kg) | 240 |

### 5.3.3.6 *Dipole Magnets for Transport Lines*

The CEPC Booster has 8 transport lines (TLs), which connect the Linac, the Booster and the Collider: 2 from the Linac to Booster, 2 from the Booster to Collider for on-axis injection, 2 from the Booster to Collider for off-axis injection, and 2 from the Collider to Booster for on-axis injection. These transport lines employ four types of dipole magnets, namely BT0 & BT1, Btv, BT2, and BT3, with lengths of 5 m, 28 m, 35 m, and 15 m, respectively. (Note: BT0 and BT1 are the same type of magnet but with different names.) The highest magnetic field among them is 0.5 T. Before designing the magnets, several considerations were taken into account.

Firstly, all the magnets are operated with DC excitation, which means solid iron can be used to produce the yoke, rather than laminations (without punching die), in order to save costs.

Secondly, it is challenging to fabricate very long yokes, so the dipoles with lengths of 15 m and 35 m need to be divided into three and seven parts, respectively, with each part being 5m long. Thus, they are referred to as 3-in-1 dipoles and 7-in-1 dipoles, respectively. The 28 m long dipoles can also be divided into seven parts, each 4 m long, making them 7-in-1 dipoles.

Thirdly, the design of the coils needs to balance material costs, machining costs, cooling-water flow rate, and temperature rise. Based on these considerations, TU2 copper was chosen as the coil material for all four types of dipoles. The cross-section of the copper conductors for these dipoles are identical, with a size of 12 ×12 $mm^2$ and Φ8.

The magnetic-field optimizations for three types of magnets were performed using OPERA/TOSCA software. The cross-section shapes of these magnets are shown in Figures 5.3.3.22 to 5.3.3.24. The 3D core model of one dipole magnet with end chamfering and integral field distribution curves before and after end chamfering are displayed in Figure 5.3.3.25. The results demonstrate that the uniformity of the integral field is better than 5E-4 after end chamfering. Table 5.3.3.6, 5.3.3.7 and 5.3.3.8 summarize the parameters of the dipoles for the transport lines.

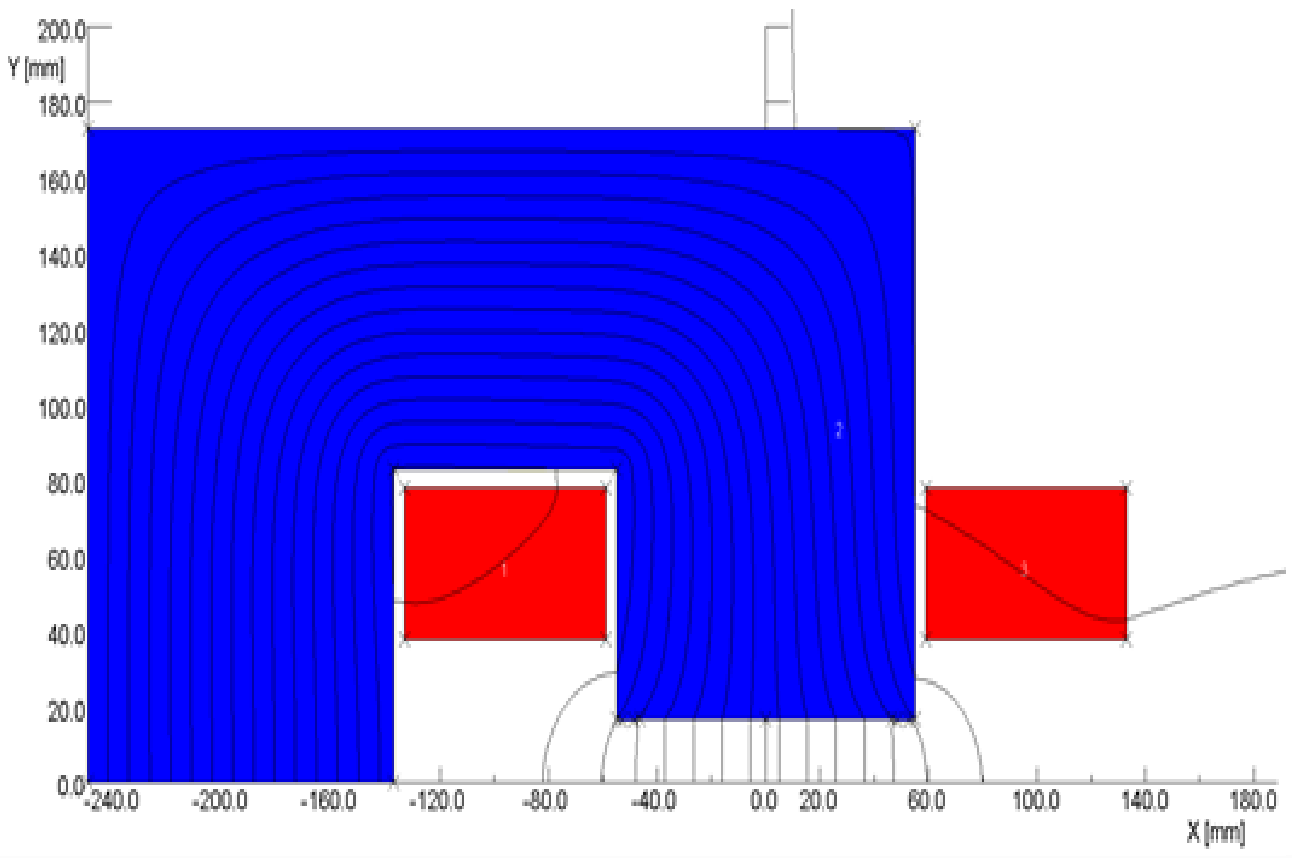

**Figure 5.3.3.22:** The cross-section shape of the BT1 dipole

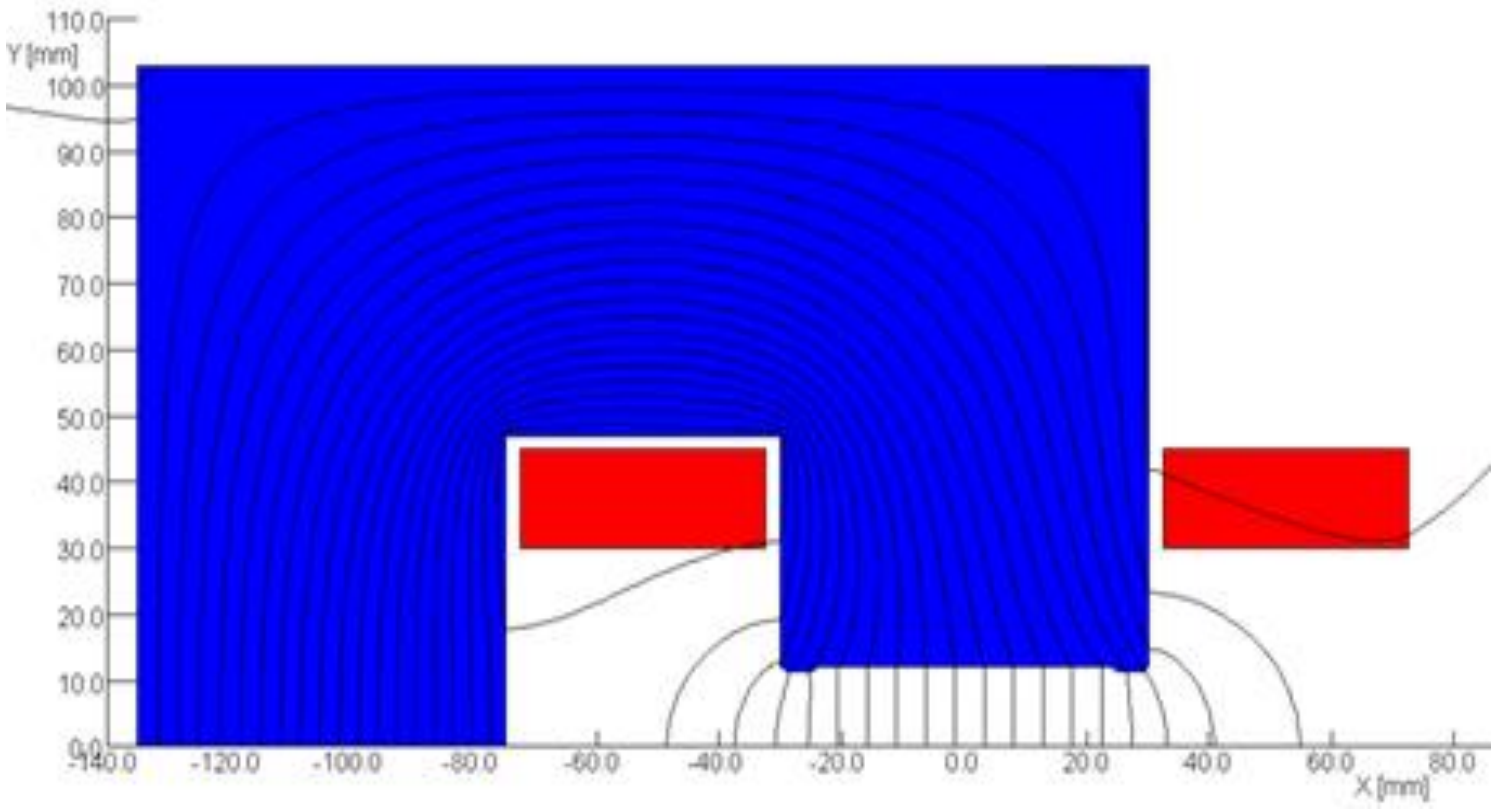

**Figure 5.3.3.23:** The cross-section shape of the BT2 dipole

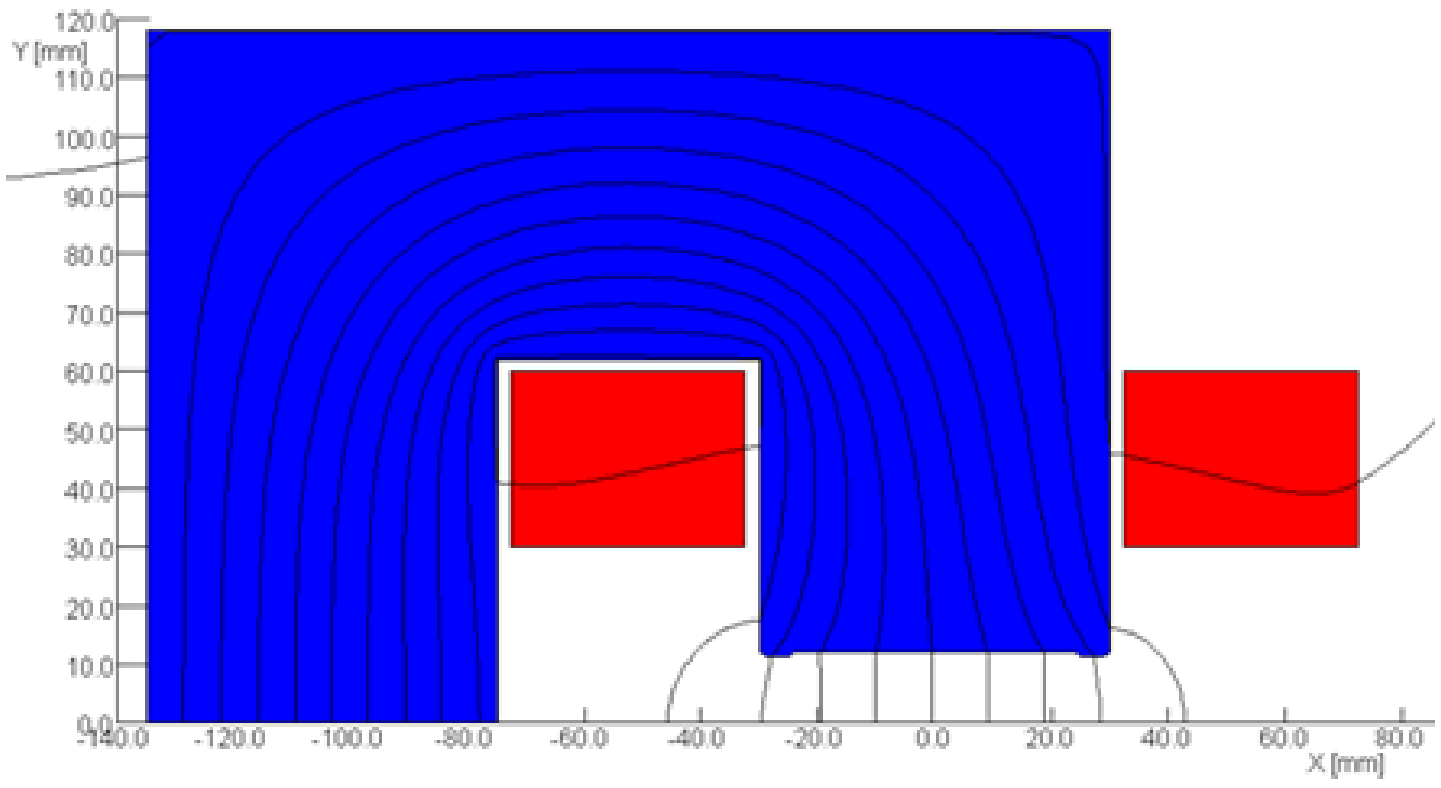

**Figure 5.3.3.24:** The cross-section shape of the BT3 dipole

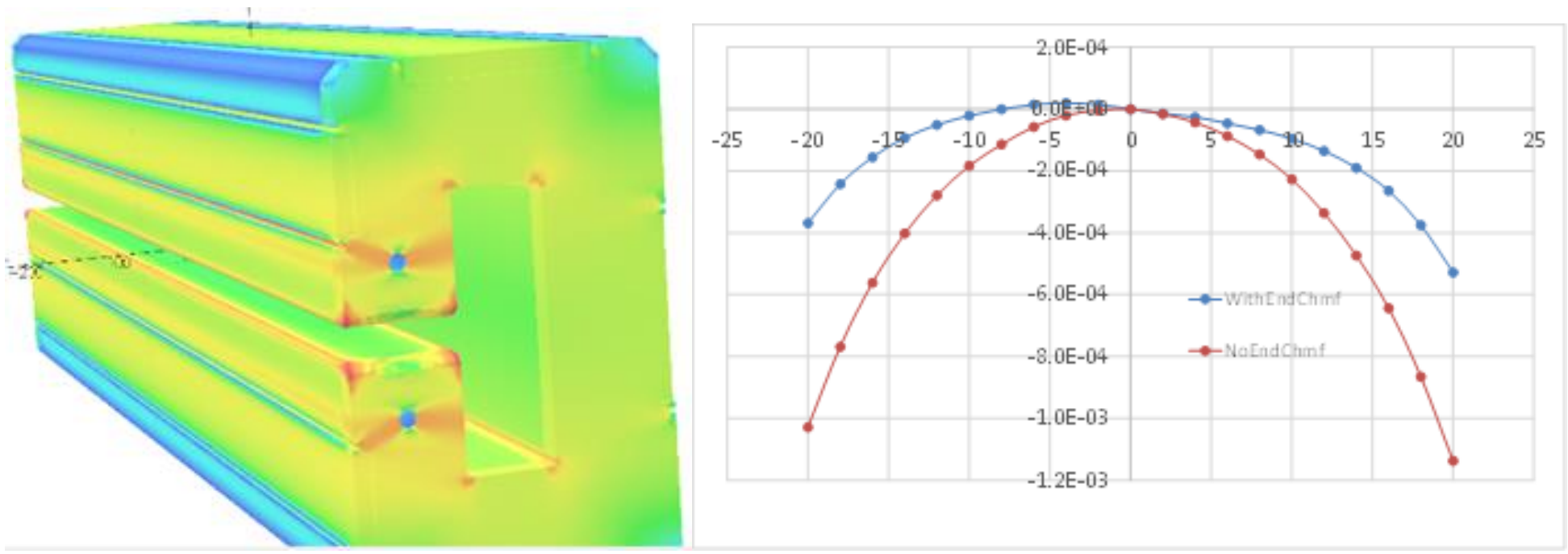

**Figure 5.3.25:** The 3D core model and field distribution of one dipole magnet part

**Table 5.3.3.6** Parameters of the dipoles for the TLs from the Linac to Booster

| Magnet name | BT0 & BT1 | BTv |
|---|---|---|
| Magnet type | whole | 6 in 1 |
| Quantity | 36 | 4 |
| Magnetic Length (m) | 5 | 5×5 + 3 = 28 |
| Strength of field (T) | 1.5 | 1.5 |
| Gap (mm) | 37 | 37 |
| Good field region (mm) | 33 | 33 |
| Field uniformity | 0.1% | 0.1% |
| Excitation amp-turns（At） | 7421 | 7421 |
| Size of conductor（mm） | 12×12 D8 | 12×12 D8 |
| Coils turns on each pole | 24 | 24 |
| Current（A） | 310 | 310 |
| Current density（A/mm$^2$) | 3.33 | 3.33 |
| Number of magnet unit | 1 | 6 |
| Length of magnet unit (m) | 5 | 5/3 |
| Resistance (mΩ）/unit | 218 | 218 |
| Voltage drop (V)/unit | 102 | 102 |
| Power loss (kW)/unit | 47.9 | 47.9 |
| Total power loss of magnet (kW) | 47.9 | 287.6 |
| Width/Height of core (mm) | 320/500 | 320/500 |
| Weight of magnet unit (ton) | 6.4 | 6.4 |
| Total weight of magnet (ton) | 6.4 | 38.4 |
| Number of cooling circuits | 8 | 8 |
| Water pressure drop (kg/cm$^2$) | 6 | 6 |
| Water flow（L/min）/unit | 39.1 | 39.1 |
| Total water flow of magnet | 39.1 | 234.8 |
| Temperature rise (deg) | 17.5 | 17.5 |

**Table 5.3.3.7** Parameters of the dipoles for the TLs from the Booster to Collider

| Magnet name | BT0 & BT1 (on- and off-axis) | BT2 (on-axis) | BT3 (off-axis) |
|---|---|---|---|
| Magnet type | whole | 7 in 1 | 3 in 1 |
| Quantity | 12 | 8 | 4 |
| Magnetic Length (m) | 5 | 5×7 = 35 | 5×3 = 15 |
| Strength of field (T) | 1.5 | 0.426 | 0.81 |
| Gap (mm) | 37 | 24 | 24 |
| Good field region (mm) | 33 | 20 | 20 |
| Field uniformity | 0.1% | 0.1% | 0.1% |
| Excitation amp-turns（At） | 7421 | 1418 | 2840 |
| Size of conductor（mm） | 12×12 D8 | 12×12 D8 | 12×12 D8 |
| Coils turns on each pole | 24 | 4 | 8 |
| Current（A） | 310 | 355 | 355 |
| Current density（$A/mm^2$) | 3.33 | 3.82 | 3.82 |
| Number of magnet unit | 1 | 7 | 3 |
| Length of magnet unit (m) | 5 | 5 | 5 |
| Resistance (mΩ）/unit | 218 | 156 | 156 |
| Voltage drop (V) /unit | 102 | 31 | 54 |
| Power loss (kW)/unit | 47.9 | 5.8 | 17.8 |
| Total power loss of magnet (kW) | 47.9 | 40.5 | 53.4 |
| Width/Height of core (mm) | 320/500 | 220/323 | 220/323 |
| Weight of magnet unit (ton) | 6.4 | 2.8 | 2.8 |
| Total weight of magnet (ton) | 6.4 | 19.2 | 8.3 |
| Number of cooling circuits | 8 | 3 | 6 |
| Water pressure drop ($kg/cm^2$) | 6 | 6 | 6 |
| Water flow（L/min）/unit | 39.1 | 3.5 | 10.4 |
| Total water flow of magnet (L/min) | 39.1 | 24.5 | 31.2 |
| Temperature rise (deg) | 17.5 | 23.7 | 24.5 |

**Table 5.3.3.8:** Parameters of the dipoles for the TLs from the Collider to Booster

| Magnet name | BT0 & BT1 | BT2 |
|---|---|---|
| Magnet type | whole | 7 in 1 |
| Quantity | 36 | 8 |
| Magnetic Length (m) | 5 | 5×7 = 35 |
| Strength of field (T) | 1.5 | 0.426 |
| Gap (mm) | 37 | 24 |
| Good field region (mm) | 33 | 20 |
| Field uniformity | 0.1% | 0.1% |
| Excitation amp-turns（At） | 7421 | 1418 |
| Size of conductor（mm） | 12×12 D8 | 12×12 D8 |
| Coils turns on each pole | 24 | 4 |
| Current（A） | 310 | 355 |
| Current density（$A/mm^2$) | 3.33 | 3.82 |
| Number of magnet unit | 1 | 7 |
| Length of magnet unit (m) | 5 | 5 |
| Resistance (mΩ）/unit | 218 | 156 |
| Voltage drop (V)/unit | 102 | 31 |
| Power loss (kW)/unit | 47.9 | 5.8 |
| Total power loss of magnet (kW) | 47.9 | 40.5 |
| Width/Height of core (mm) | 320/500 | 220/323 |
| Weight of magnet unit (ton) | 6.4 | 2.8 |
| Total weight of magnet (ton) | 6.4 | 19.2 |
| Number of cooling circuits | 8 | 3 |
| Water pressure drop ($kg/cm^2$) | 6 | 6 |
| Water flow（L/min）/unit | 39.1 | 3.5 |
| Total water flow of magnet (L/min) | 39.1 | 24.5 |
| Temperature rise (deg) | 17.5 | 23.7 |

### 5.3.3.7 *Quadrupole Magnet for Transport Lines*

The CEPC transport lines require two types of quadrupole magnets with different aperture diameters. One type has a large aperture of 33 mm and the other has a small aperture of 24 mm. Figure 5.3.3.26 depicts the 2D magnetic-flux distribution for the large-aperture magnet. By optimizing the pole-face design, all high harmonic errors at the reference radii of 13 mm and 10 mm can meet the required field specifications. The design parameters for the two types of quadrupole magnets are summarized in Table 5.3.3.9.

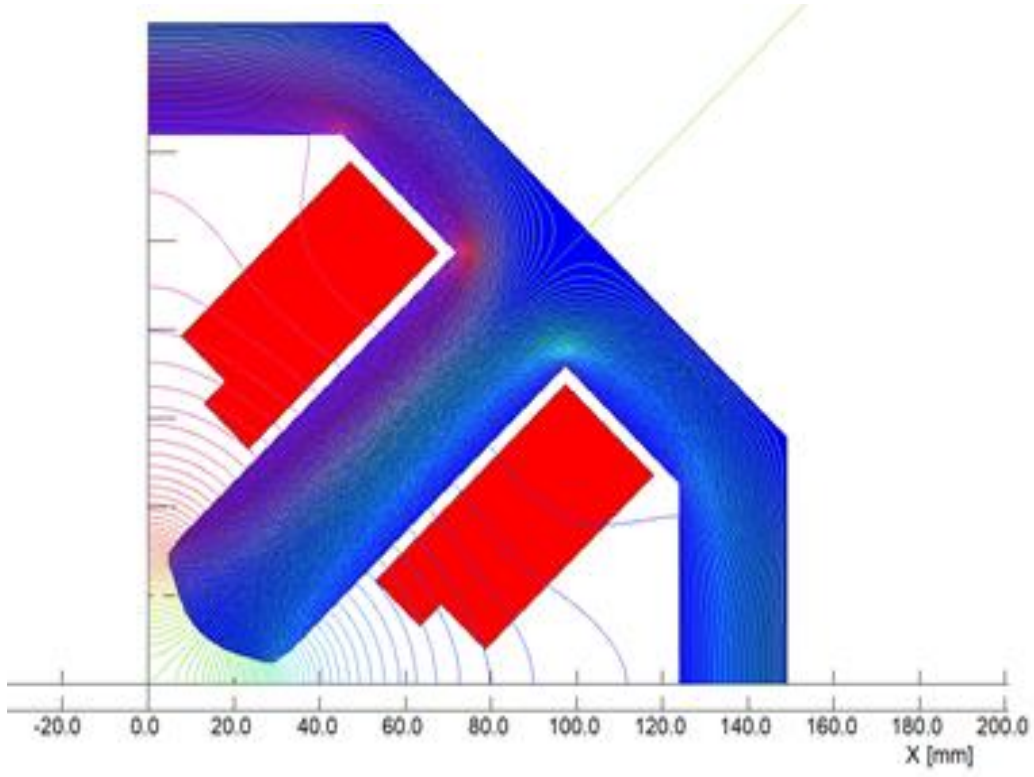

**Figure 5.3.3.26:** The magnetic flux distribution of the quadrupole magnets

**Table 5.3.3.9:** Main parameters of the quadrupole magnets for the transport lines

| Magnet name | 33Q (L-to-B) | 24Q (B-to C and C-to-B) |
|---|---|---|
| Quantity | 80 | 40 |
| Bore diameter（mm） | 33 | 24 |
| Field gradient（T/m） | 14.7 | 27 |
| Magnetic length（m） | 0.9 | 2.0 |
| Ampere-turns per pole（AT） | 1640 | 1593 |
| Coil turns per pole | 72 | 72 |
| Excitation current（A） | 23 | 22 |
| Conductor size（$mm^2$） | 8×3 | 8×3 |
| Current density（$A/mm^2$） | 0.95 | 0.92 |
| Resistance（Ω） | 0.44 | 0.93 |
| Inductance（mH） | 25 | 81 |
| Voltage drop（V） | 10.1 | 20.7 |
| Power loss（W） | 229 | 457 |
| Core width/height（mm） | 420/420 | 400/400 |
| Core length（mm） | 888 | 1991 |
| Magnet weight (ton) | 1.36 | 2.79 |

### 5.3.3.8 *Correction Magnets for Transport Lines*

The CEPC transport lines feature two types of correctors with different gaps: 37 mm and 24 mm. Each magnet is used independently for vertical and horizontal correction of the closed orbit. To optimize the magnetic field, the correctors adopt H-type structure cores with pole surfaces that can be shimmed. The cores consist of stacked 0.5 mm thick laminations. Solid copper conductors are used to wind the racetrack-shaped coils, with a size of 5.5×4 $mm^2$. The current density is lower than 1 $A/mm^2$, so water cooling is unnecessary for the coils.

Magnetic-field simulations of the correctors are performed using the OPERA software. Figure 5.3.3.27 shows the magnetic-flux distribution of one of the correctors. The key parameters of both types of correctors are summarized in Table 5.3.3.10.

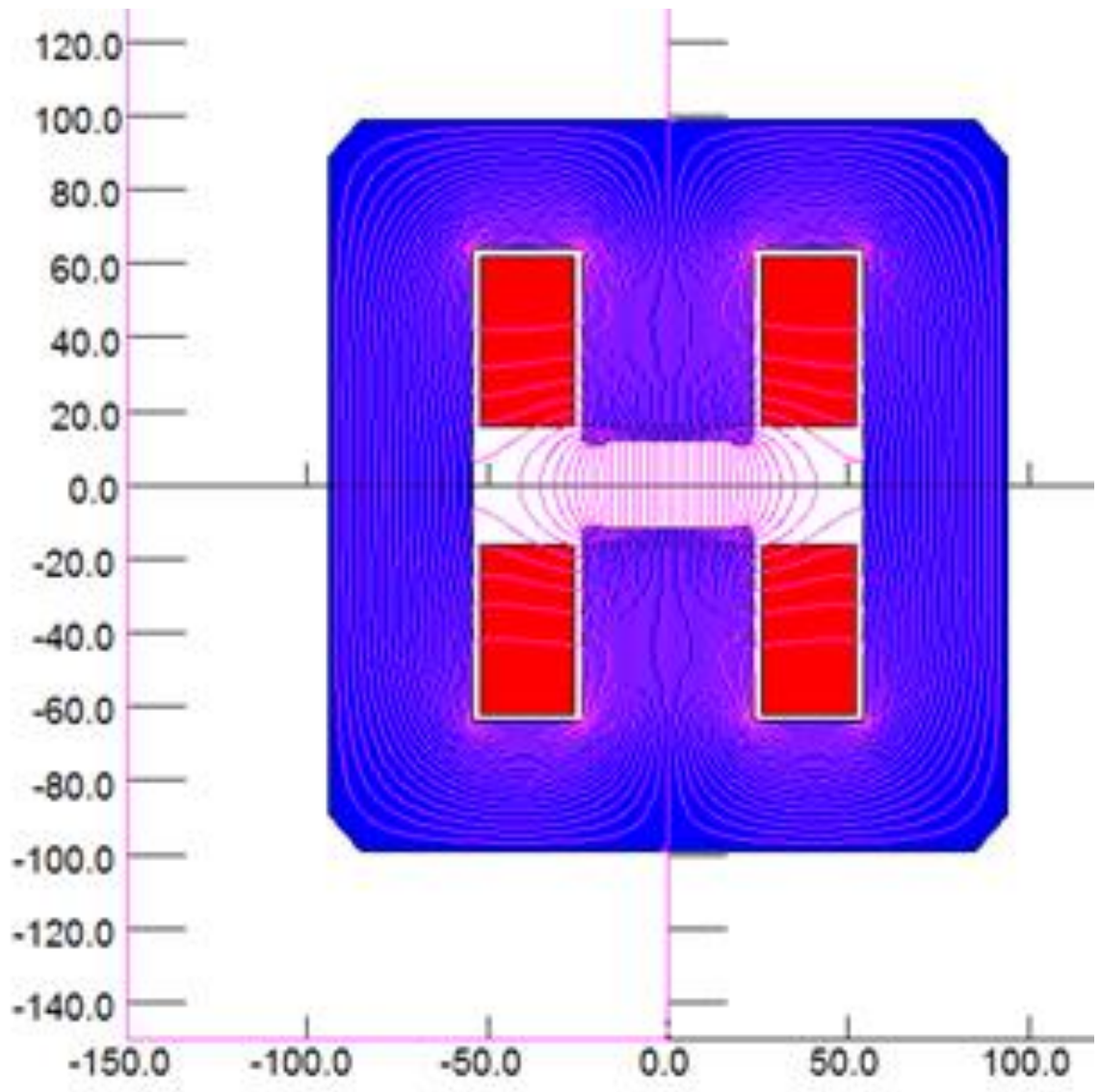

**Figure 5.3.3.27:** Magnetic flux distribution of the correctors for transport lines.

**Table 5.3.3.10:** The Main parameters of correctors for the transport lines.

| Magnet name | 37C（L-to-B） | 24C（B-to-C and C-to-B） |
|---|---|---|
| Quantity | 24 | 30 |
| Gap [mm] | 37 | 24 |
| Max. Field [Gs] | 1680 | 3000 |
| Magnetic Length [mm] | 200 | 300 |
| Good Field Region [mm] | 33 | 20 |
| Field Uniformity | 0.1% | 0.1% |
| Ampere turns per pole [At] | 2473 | 2864 |
| Turns per pole | 96 | 112 |
| Max. current [A] | 26 | 26 |
| Size of conductor [$mm^2$] | 8×3 | 8×3 |
| Current density [$A/mm^2$] | 1.07 | 1.07 |
| Resistance (mΩ) | 96 | 153 |
| Power loss (W) | 64 | 100 |
| Voltage [V] | 2.47 | 3.91 |
| Height of core [mm] | 340 | 320 |
| Width of core [mm] | 280 | 300 |
| Core Length [mm] | 180 | 280 |
| Total weight of magnet [kg] | 144 | 226 |

## 5.3.4 **Magnet Power Supplies**

### 5.3.4.1 *Introduction*

The Booster power supply system is responsible for providing excitation current to the Booster magnets during the particle acceleration process. The working cycle of the Booster involves several stages, including injection, energy raising, energy extraction, and energy lowering. The magnet power supply system is a dynamic power supply that provides current to various magnets during these stages.

The output current of the magnet power supply is proportional to the beam energy, and therefore, all magnet power supplies must change their output synchronously. During the energy raising mode of the Booster, the exciting current rises for 4.3 seconds, followed by a top width of 7.2 seconds and a downtime of 4.3 seconds. The total width of the waveform is 22.2 seconds. Figure 5.3.4.1 shows the current waveform for the Booster magnet.

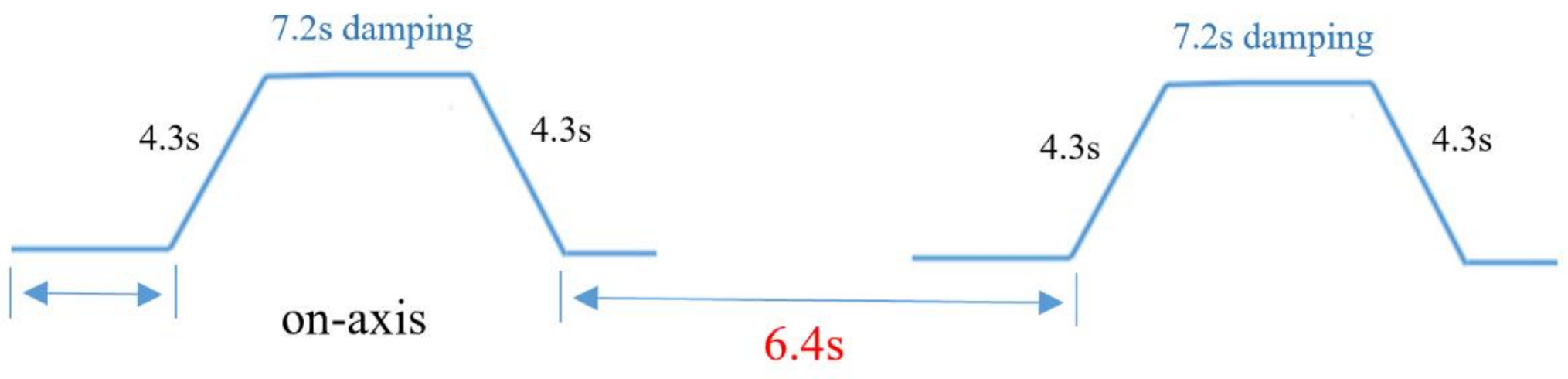

**Figure 5.3.4.1:** The excitation current waveform for the Booster magnet

### 5.3.4.2 *Booster Power Supplies*

The design principles for the Booster power supplies are similar to those for the Collider power supplies. However, there is a difference in that the Collider supplies are DC, while the Booster supplies ramp at a repetition rate of 0.045 Hz.

Table 5.3.4.1 provides a list of the types and specifications of the Booster main magnet and correction magnet power supply.

**Table 5.3.4.1:** Booster power supply requirements

| Magnet | Quantity | Stability/tracking | Output Rating |
|---|---|---|---|
| BST-63B-Arc | 16 | 200 ppm / 0.1% | 560A/ 780V |
| BST-63B-Arc-SF | 7 | 200 ppm / 0.1% | 560A/ 960V |
| BST-63B-Arc-SD | 1 | 200 ppm / 0.1% | 560A/ 960V |
| BST-74B-IR | 4 | 200 ppm / 0.1% | 520A/ 140V |
| QDM-63 | 64 | 100 ppm / 0.1% | 200A/ 50V |
| QFM-63 | 48 | 100 ppm / 0.1% | 250A/ 50V |
| QFM-RF-63 | 8 | 100 ppm / 0.1% | 200A/ 270V |
| QDM-RF-63 | 8 | 100 ppm / 0.1% | 200A/ 270V |
| QFM-ARC-63 | 80 | 100 ppm / 0.1% | 180A/ 360V |
| QDM-ARC-63 | 80 | 100 ppm / 0.1% | 180A/ 280V |
| QF-DIS-63 | 16 | 1100 ppm / 0.1% | 270A/ 50V |
| QD-DIS-63 | 32 | 300 ppm / 0.1% | 250A/ 50V |
| BST-63S | 100 | 500 ppm | 140A/ 35V |
| BST-63C | 1200 | 300 ppm / 0.1% | ±25A/ ±25V |
| Q-Trim | 3200 | 300 ppm / 0.1% | ±6A/ ±12V |
| **Total** | 4864 | | |

##### *5.3.4.2.1 Dipole Magnet Power Supplies*

The Booster is equipped with a total of 14,866 dipole magnets, which can be classified into four families. The BST-63B-Arc family consists of 10,192 magnets, while BST-63B-Arc-SF, BST-63B-Arc-SD, and BST-74B-IR families consist of 2,017, 2,017, and 640 magnets respectively.

The Booster comprises 8 arc units, matching the number in the Collider. For the BST-63B-Arc family, the dipoles in each half arc are connected in series and powered by a single power supply. This arrangement necessitates 16 dipole supplies, each carrying a load of 637 dipoles. The power supplies deliver 560A/780V and provide a power output of 0.44 MW (accounting for cable losses). These supplies will be installed in the surface buildings. The manufacturer's ratings include a safety margin of 10~15% for both current and voltage.

Similar to BST-63B-Arc, the BST-63B-Arc-SF, BST-63B-Arc-SD, and BST-74B-IR families are also powered in series. The BST-63B-Arc-SF family requires 4 power supplies rated at 560A/960V, while the BST-63B-Arc-SD family necessitates 4 power supplies with the same rating. The BST-74B-IR family requires 4 power supplies rated at 520A/140V.

##### *5.3.4.2.2 Quadrupole Magnet Power Supplies*

The Booster consists of 1,016 focusing quadrupoles (QFM-ARC-63) and 2,018 defocusing quadrupoles (QDM-ARC-63) in the arc, divided into 80 focusing families and 80 defocusing families respectively.

Each focusing family has 12 or 13 series-connected magnets powered by a single supply rated at 180A/360V. Similarly, each defocusing family consists of 25 or 26 series-connected magnets powered by a single supply rated at 180A/280V.

In the RF zone, there are 136 QFM-RF-63 and 128 QDM-RF-63 quadrupoles, which are powered in series. Additionally, there are 160 quadrupole magnets powered independently.

All the power supplies for the quadrupoles will be installed in the auxiliary stub tunnel located around the main tunnel.

##### *5.3.4.2.3 Sextupole Magnet Power Supplies*

According to the design of accelerator physics, there are 100 sextupole magnets that are powered independently. The power supplies for these magnets are rated at 140A/35V.

##### *5.3.4.2.4 Corrector Power Supplies*

The Booster contains a total of 1,200 correction magnets, with each correction magnet powered by a separate supply. The power supply for the correction magnets is a conventional high precision DC power supply, and the rating for all correction power supplies is 25A/25V.

Similar to the power supplies for the quadrupoles, all the power supplies for the correction magnets will also be installed in the auxiliary stub tunnel around the main tunnel.

In addition, each of the quadrupoles in the arcs has a trim coil. The trim coils are powered independently.

In summary, the Booster power supply system consists of 28 dipole power supplies, 336 quadrupole power supplies, 100 sextupole power supplies, 1200 corrector power supplies and 3,200 trim power supplies. The total average power consumption for the entire system is 8.46 MW.

#### 5.3.4.3 ***Design of the Power Supply System***

The design of the Booster power supply system is based on the working repetition frequency of the accelerator. Power supplies with a repetition frequency above 10 Hz utilize a White-type resonant power system. On the other hand, power supplies with a repetition frequency below 10 Hz adopt a forced oscillation power supply. Additionally, the power supplies must have energy storage capacity to minimize the impact on the grid. In the case of CEPC, the Booster power supply utilizes the forced oscillation power supply.

Given that the dipole magnet power supply can reach a maximum output voltage of 960 V, a more mature module series mode is implemented for this voltage level. Specifically, the power supply's front stage consists of a DC voltage source combined with a booster circuit, which helps reduce the impact on the grid. The output stage utilizes a two-quadrant chopper connected in series, operating in quadrant I and IV. During the current rise phase, the power supply provides energy, while the magnet absorbs and stores energy. Conversely, during the current fall phase, the magnet releases and consumes energy.

The power supply consists of three power units in series, with the output voltage of each power unit being 350 V. All power units adopt the multi-channel IGBT H-bridge series and phase-shift frequency-doubling drive technology. The topology structure of the power unit is shown in Figure 5.3.4.2.

Furthermore, to reduce the system's voltage to ground, the power supply output is insulated from ground and grounded only at the midpoint.

The power supply is composed of two main parts: the power cabinet and the regulating cabinet. The main power part in the power cabinet is completely insulated from the ground to reduce the system's voltage to ground. To separate the control signals between the power cabinet and the regulating cabinet, optical fibers are used.

Additionally, the control power supply of the power cabinet and the control power supply of the regulating cabinet are separated by transformers. This ensures that the control signals are properly isolated and regulated for safe and accurate operation of the power supply system.

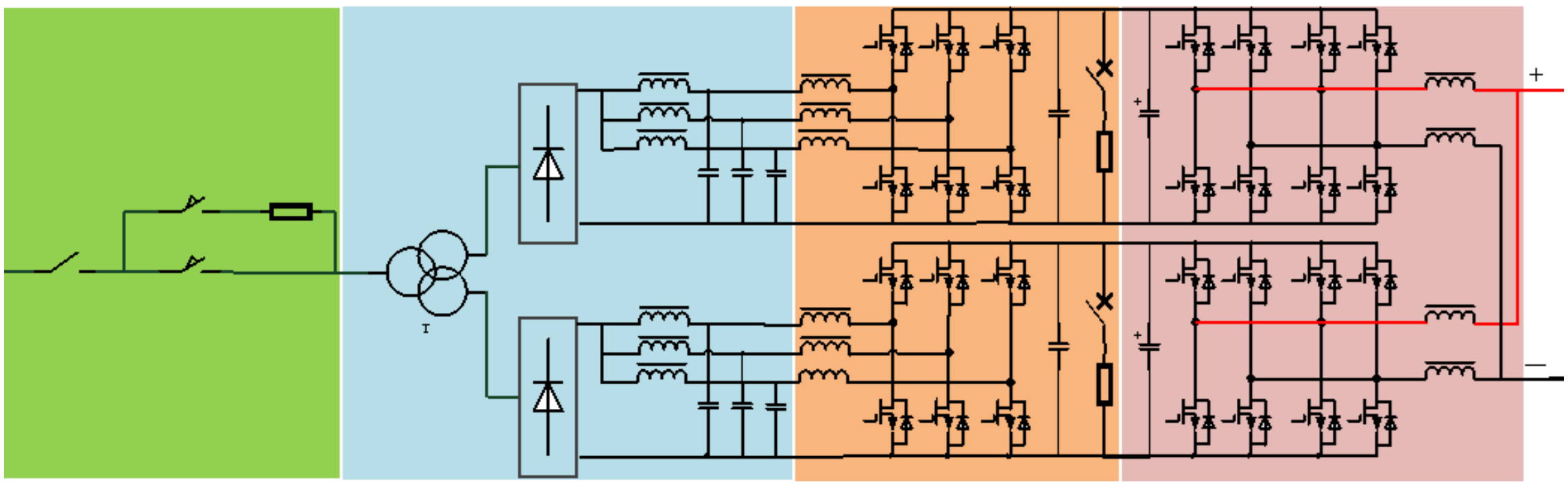

**Figure 5.3.4.2:** Topology structure of a power unit of the dipole magnet power supply.

The quadrupole magnet power supply of the Booster can use the same single power module as the main circuit topology of the bipolar dipole magnet power supply. However, the input voltage limit of the power supply needs to be adjusted to meet the actual output needs of the quadrupole magnets. This adjustment can help improve the consistency of the power module for efficient and reliable operation.

Furthermore, due to the reactive power throughput, it is necessary to consider measures to suppress power grid fluctuations during power supply operation. The power fluctuation suppression circuit should be designed to suppress the 0.045 Hz power grid fluctuation to within ±15% of the average input power, and the front stage energy storage component should suppress the voltage fluctuation of the front stage to within ±15% of the output voltage. This ensures stable and consistent operation of the power supply and minimizes the impact on the power grid.

To ensure the dynamic tracking characteristics of all power supplies in the Booster, the control system uses a full digital technology. The control strategy employs a combination of repetitive control and PID to solve the contradiction between limited loop gain and high-precision dynamic current tracking. This combination ensures the frequency consistency between power sources and the accurate tracking of output current. The control system also works synchronously with other accelerator systems to achieve overall system stability. The circuit topology uses multi-bridge series and phase-out drive to increase the output voltage and synthesis frequency, which improves the dynamic response ability of the system. Figure 5.3.4.3 illustrates the circuit topology.

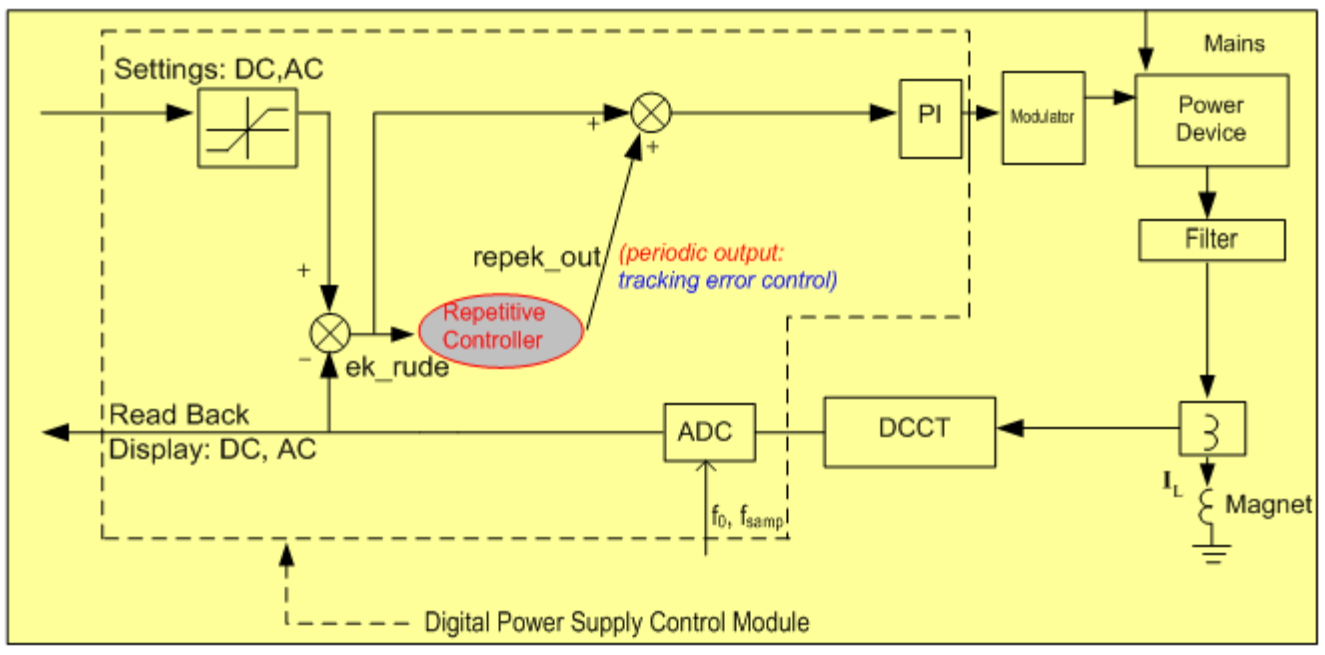

**Figure 5.3.4.3:** Digital controller for the Booster power supplies

The Booster power system will be equipped with DPSCM-II, a fully digital control system that operates based on the same principles as the colliderpower system. To reduce costs, the control software, remote control interface protocol, electronic devices, and physical structure used in the digital control system will be identical, making them interchangeable in different power sources. For more detailed design information, please refer to Sec. 4.3.5 of this report, Collider Magnet Power Supplies.

This design has been implemented for the booster dipole power supply of HEPS. The power supplies are rated at 900A/900V, as shown in Figure 5.3.4.4. Additionally, Figure 5.3.4.5 illustrates the output current and tracking error of the power supply. The performance of the power supply meets the design requirements.

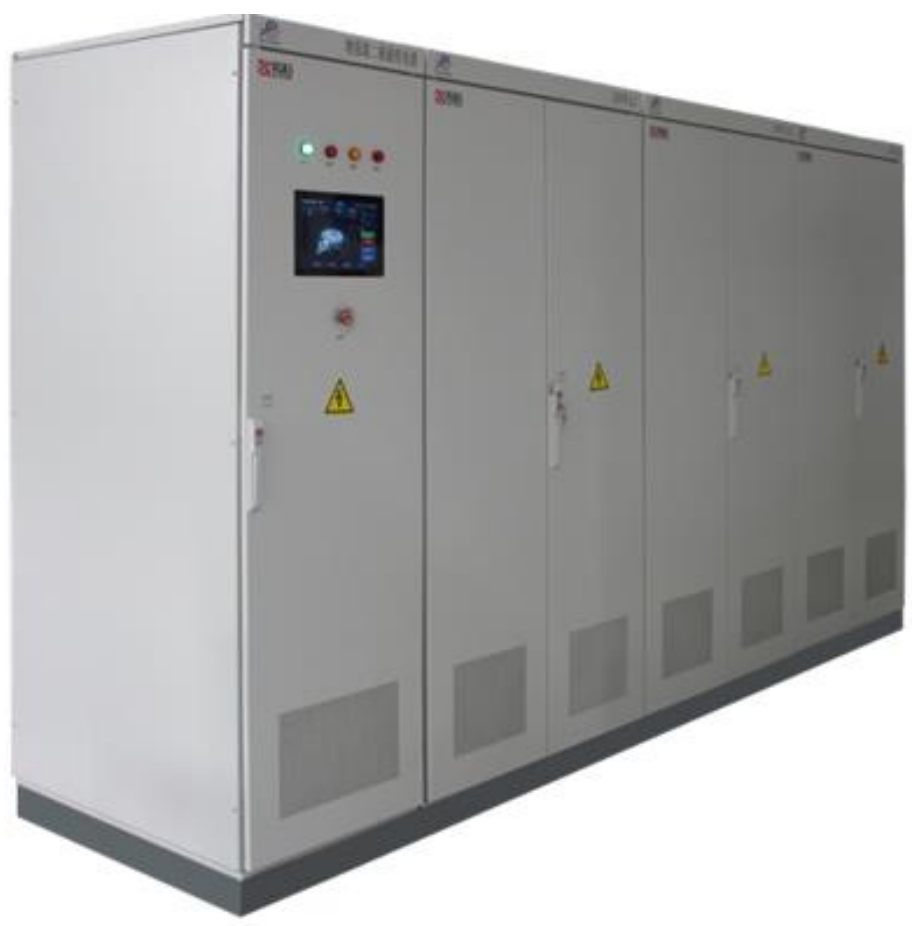

**Figure 5.3.4.4:** The magnet power supply of the HEPS Booster.

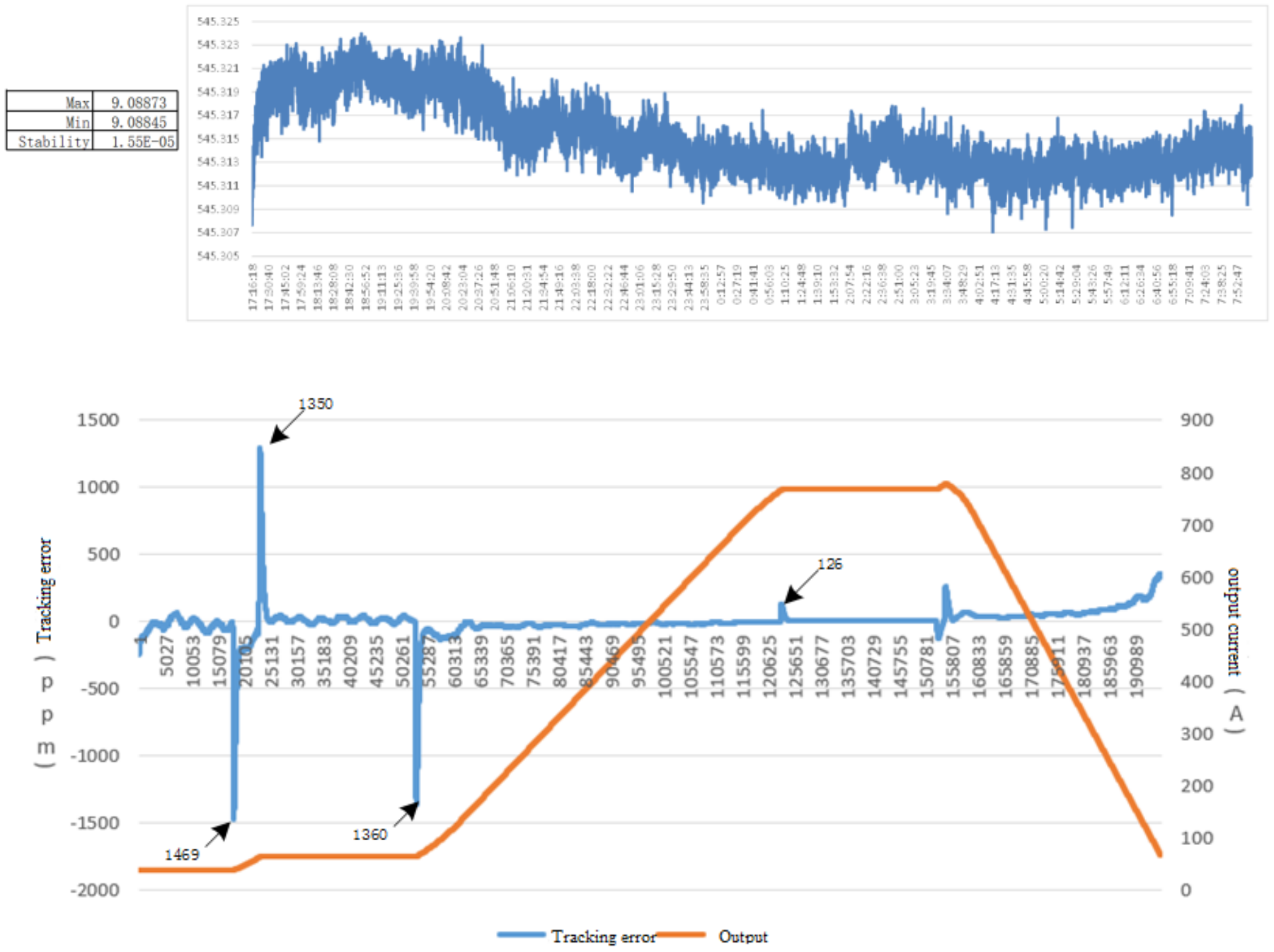

**Figure 5.3.4.5:** The stability and tracking error of the HEPS Booster dipole power supply.

## 5.3.5 **Vacuum System**

The Booster vacuum system comprises of chambers, bellows, pumps, gauges, valves, and others. To minimize beam loss and bremsstrahlung from residual gas, an average pressure of less than $3\times10^{-8}$ Torr is required. However, stainless steel has low electrical conductivity, which can lead to beam instability, and hence aluminum, with its higher conductivity, has been selected for the Booster chambers.

To achieve high vacuum, conventional high vacuum technology will be employed, using small ion pumps distributed around the circumference of the Bosster. This will help achieve the desired pressure and ensure the stability of the beam. As a result of the low beam current in the Booster ring, the shielding of bellows and metal gate valves is unnecessary. Additionally, the secondary electron yield (SEY) of aluminum fulfills the requirements.

### 5.3.5.1 *Heat Load and Gas Load*

Similar to the Collider, the Booster operates in four modes: Higgs, W, Z, and $t\bar{t}$, with two power levels: 30 MW in the baseline and 50 MW in an upgrade plan. In order to calculate the vacuum parameters, the $t\bar{t}$ mode at 50 MW is considered, as it possesses the highest energy and gas load.

Equations (4.3.6.1) and (4.3.6.2) are used to estimate the heat load from synchrotron radiation. For the Booster, $E$ = 180 GeV, $I$ = 0.00020 A, and $\rho$ = 11380.8 m, resulting in a total synchrotron radiation power of $P_{SR}$ = 1.63 MW and a linear power density of $P_L$ = 22.8 W/m, which is negligible.

Moreover, since the magnetic field changes very slowly, the heat load induced by the eddy current is almost zero. Therefore, the overall heat load generated in the Booster is minimal and does not pose any significant challenge.

The gas load in the Booster vacuum system comprises two components: thermal outgassing and synchrotron-radiation-induced photo-desorption. Thermal outgassing mainly contributes to the base pressure in the absence of a circulating beam. Equations (4.3.6.4) and (4.3.6.5) are used to estimate the desorption rate induced by synchrotron radiation. The effective gas load due to photo-desorption is found to be $1.2\times10^{-3}$ Torr·L/s, while the linear gas load due to synchrotron radiation is $1.65\times10^{-8}$ Torr·L/s/m. Assuming the thermal outgassing rate of the vacuum chambers to be $1\times10^{-11}$ Torr·L/s·cm$^2$ for a circular cross-section of 5.6 cm in diameter, the linear thermal gas load ($Q_{LT}$) is estimated to be $1.76\times10^{-8}$ Torr·L/s/m.

To achieve the desired vacuum value of $3.0\times10^{-8}$ Torr, an effective pumping speed of 20 L/s and a 12-meter sputtering ion-pump distribution are required. With these specifications, we can achieve the desired vacuum pressure and ensure stable beam operation.

### 5.3.5.2 *Vacuum Chamber*

Aluminum alloys are a popular choice for electron/positron storage rings, primarily due to their high electric and thermal conductivity, easy extrusion and welding capabilities, and lower cost compared to materials such as stainless steel or copper.

Furthermore, unlike other metals, most aluminum alloys do not become magnetized from machining and welding, and they do not form long lifetime radioisotopes, making them a safer option for use in particle accelerators.

However, the relatively low strength and hardness of aluminum alloys limit their use for all-metal sealing flanges. In such cases, other materials such as stainless steel or copper are preferred.

The dipole vacuum chamber in the Booster has a circular cross-section with a diameter of 56 mm, a length of approximately 6 m, and a wall thickness of 2 mm. These chambers are extruded from Al 6061, and stainless steel conflat flanges are welded onto the ends using transition material.

The finite-element analysis conducted on the vacuum chamber indicates that the highest temperature reached is 28.9°C when the ambient temperature is 25°C and a convective heat-transfer coefficient of $2\times10^{-5}$ W/(mm$^2$·°C) is assumed. The maximum stress and deformation experienced by the chamber are 1.78 MPa and 0.0045 mm, respectively, which are both within the safe limits.

Figure 5.3.5.1 shows the results of the finite-element analysis. The use of Al 6061 and the given dimensions of the chamber ensure that the vacuum chamber is both thermally stable and structurally safe for use in the Booster.

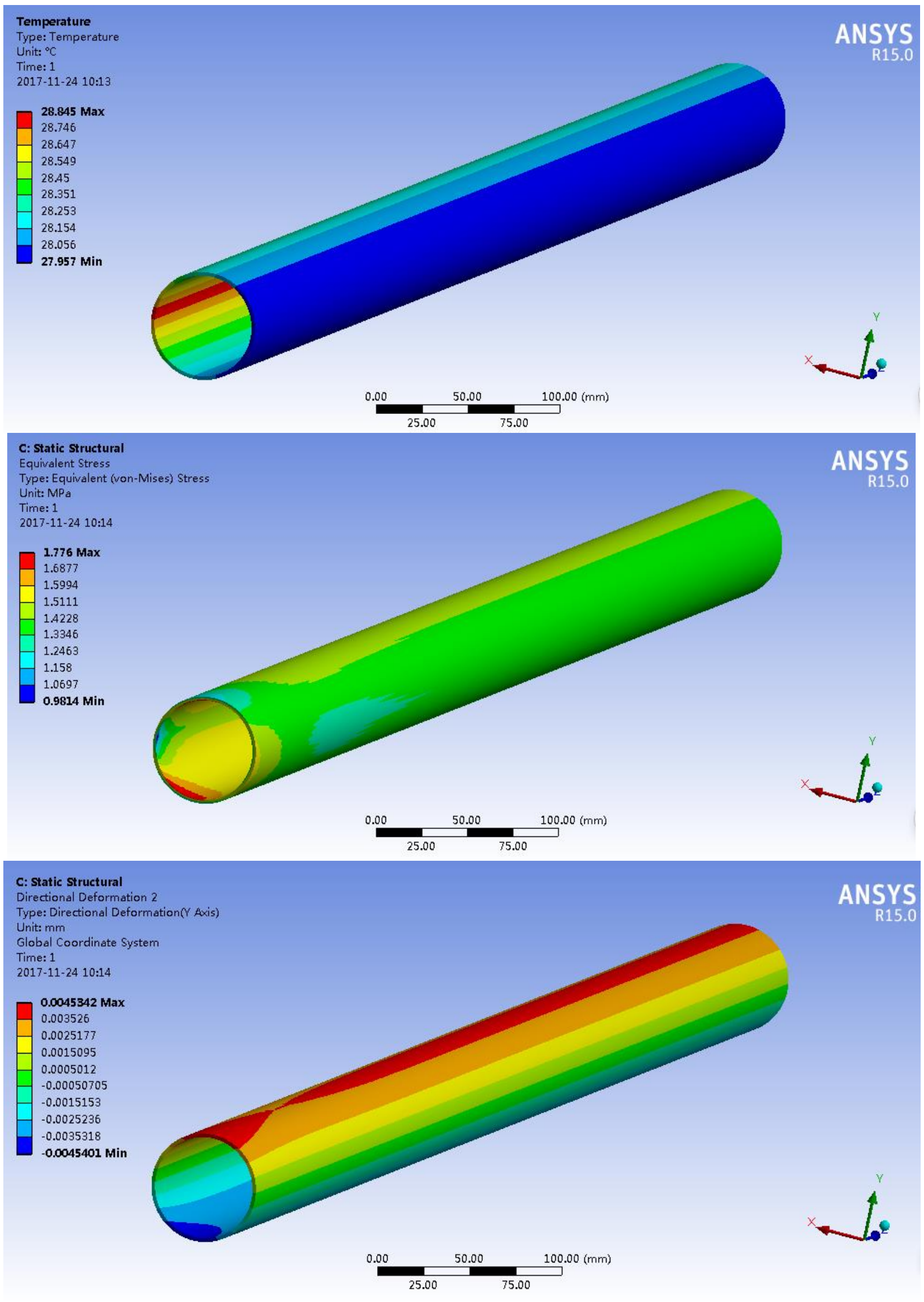

**Figure 5.3.5.1:** Results of the finite-element analysis on the aluminum vacuum chamber

Al-6061 has been selected as the material for the Booster's Al vacuum tubes, with flanges fabricated from explosion-bonded S.S.-Al transition plates. The flanges and tubes are joined using manual AC TIG welding, while the fittings for the water-cooling channels are also made from Al-6061 and welded using TIG. The connection ports of the cooling water channels are machined using a numerically controlled machine tool.

The S.S.-Al transition plates are produced by a domestic company using explosion bonding, and ultrasonic flaw detection is performed prior to machining the flanges. Leak detection is carried out after the flanges have been machined to ensure their quality.

Figure 5.3.5.2 depicts the prototype of the vacuum chamber and S.S.-Al transition flanges. The use of explosion-bonded transition plates and careful welding techniques ensure that the vacuum system is both structurally sound and leak-free.

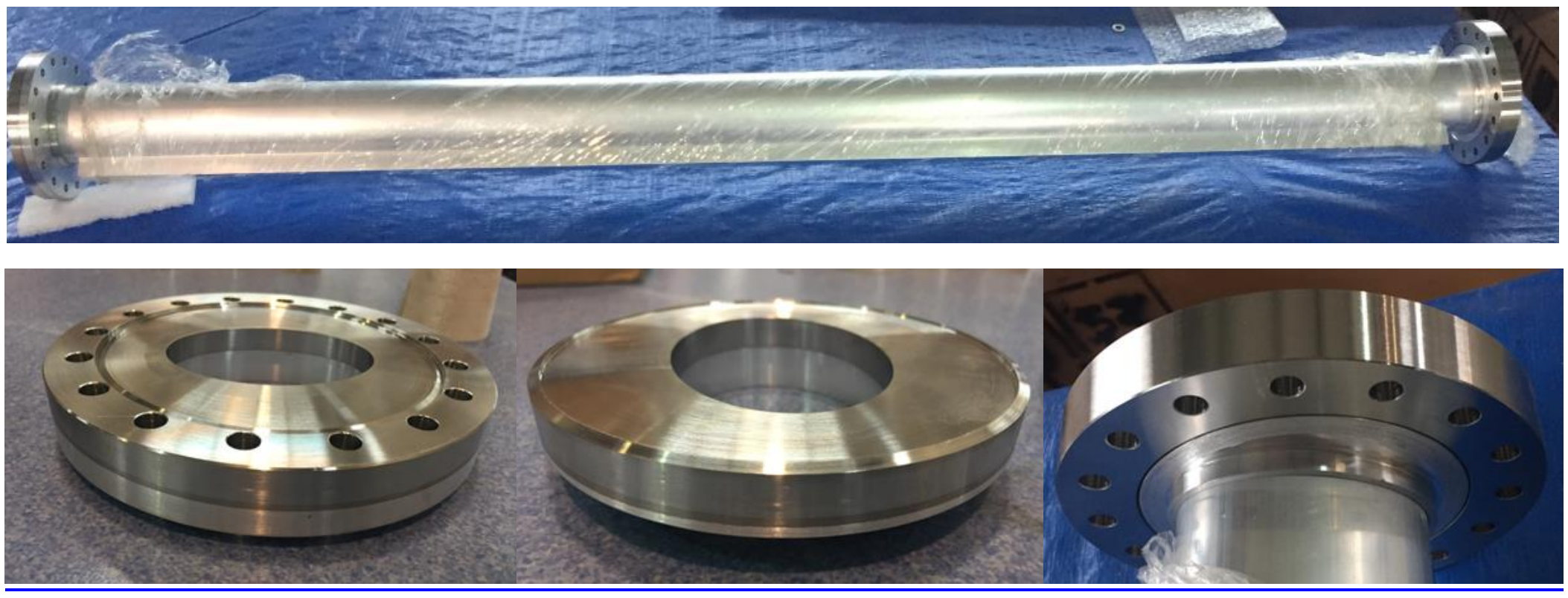

**Figure 5.3.5.2:** Prototype of vacuum chamber and S.S.-Al transition flanges

### 5.3.5.3 *Vacuum Pumping and Measurement*

The circumference of the Booster will be divided into 520 sectors, each with an all-metal gate valve, allowing for manageable length and volume during pumping down, leak detection, bake-out, and vacuum-interlock protection.

Bellows made of stainless steel will be utilized to absorb the expansion and misalignment of vacuum chambers and other vacuum devices during the installation process.

The roughing down process will be carried out using oil-free turbo-molecular pumps, which will bring the pressure down to approximately $10^{-7}$ Torr. The main pumping process will then be followed by ion pumps distributed around the Booster circumference at intervals of about 12 m.

Due to the size of the Booster, it is not feasible to install vacuum gauges at short intervals. However, special sections such as the injection regions, RF cavities, and extraction regions will be equipped with cold cathode gauges and residual gas analyzers. The majority of pressure monitoring will be carried out through the current of the sputter ion pumps, which will be continuously monitored and provide sufficient pressure measurements down to $10^{-9}$ Torr. In instances where mobile diagnostic equipment is required, it can be brought to specific locations during pump down, leak detection, and bake-out when the machine is accessible.

## 5.3.6 Instrumentation

### 5.3.6.1 *Introduction*

The Booster instrumentation system has several requirements, including the quick and accurate monitoring of beam status, efficient measurement and control of bunch current, and the ability to cure beam instabilities. Beam orbit measurement is particularly crucial. Many of the instrumentation components are the same as those used in the Collider, as described in Chapter 4, Section 4.3.7. However, there are some differences between the two systems that we will highlight in this chapter. Table 5.3.6.1 summarizes the requirements of the Booster instrumentation system.

**Table 5.3.6.1:** Main technical parameters of the Booster beam instrumentation systems

| Sub-system | | Method | Parameter | Quantity |
|---|---|---|---|---|
| Beam position monitor | Turn by turn | Button electrode BPM | Measurement area (x × y): ±20 mm × ±10 mm<br>Resolution：< 0.02 mm | 2408 |
| | Bunch by bunch | Button electrode BPM | Measurement area (x× y): ±40 mm × ±20 mm<br>Resolution：0.1 mm | |
| Bunch current | | BCM | Measurement range：<br>10 mA / per bunch<br>Relative precision：1/4095 | 2 |
| Average current | | DCCT | Dynamic range：0.0~1.5A<br>Linearity：0.1 %<br>Zero drift: < 0.05 mA | 2 |
| Beam size | | Double slit interferometer<br>X-ray pin hole | Resolution: 0.2 μm | 2 |
| Bunch length | | Streak camera<br>Two photon intensity interferometer | Resolution: 1ps @10 ps | 2 |
| Tune measurement | | Frequency sweeping method | Resolution: 0.001 | 2 |
| | | DDD | Resolution: 0.001 | |
| Beam loss monitor | | Optical fiber | Space resolution: 0.6 m | 670 |
| Feedback system | | TFB | Damping time: ≤ 1.73 ms | 2 |
| | | LFB | Damping time: ≤ 65 ms | 4 |

### 5.3.6.2 *Beam Position Measurement*

The Booster uses BPMs that are very similar to those in the Collider, with only minor differences in the vacuum pipe. A total of 1,808 BPMs are installed, including several additional ones located in specific locations for special purposes. The front-end and digital electronics are located in the tunnel and placed underneath the magnet girders due to limited space. Radiation shielding for these electronics requires careful design.

Figure 5.3.6.1 depicts the button BPM design, where the electrode radius is 3 mm and the gap between the button and the pipe is 0.25 mm. All parameters of the BPMs are the same with that of the Collider, including the transfer impedance and sensitivity.

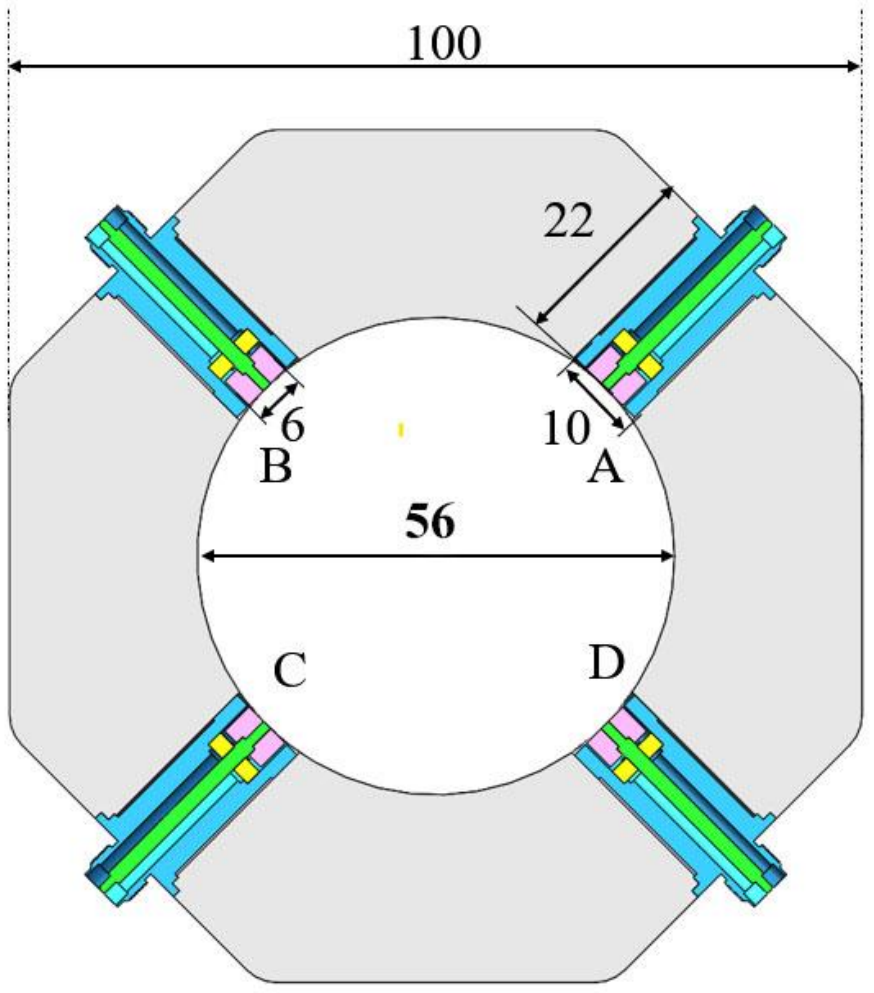

**Figure 5.3.6.1:** Model of the Booster BPM.

The Booster's RF BPM electronics architecture closely resembles that of the Collider. Consequently, we can refer to the BPM electronics design of the Collider for the Booster. The electron and positron beams will ramp up to the required energy for Collider injection from the Booster. The beam current will be lower than that of the Collider, and the beam energy will change turn by turn during the ramping-up process. These are the primary differences between the Collider and the Booster.

In the Collider, the BPM electronics are capable of providing turn-by-turn beam position information, and the resolution will be better than 20 μm in this situation. The BPM electronics consist mainly of two components: the Analog Frontend Electronics (AFE), responsible for amplitude adjustment, frequency filtering, clock reception, ADC sampling of analog signals, and the Digital Front-end Electronics (DFE), which handles AFE data reception, digital BPM algorithm, and data processing and transmission functions. The algorithm logic is implemented in the FPGA of the DFE circuit.

### 5.3.6.3 *Beam Current Measurement*

There are two DCCT sensors in the Booster for measuring beam current. The beam current in the Booster varies depending on the operational mode. For $t\bar{t}$ running, the total beam current is less than 0.3 mA. In Higgs mode, the beam current is 1 mA. For W mode, the beam current is 4 mA, and for Z mode, the beam current is 16 mA. Given that the average current ranges from approximately 0.3 mA to 16 mA, precise measurement and a high signal-to-noise ratio are essential, while minimizing any heating issues. The In-Flange sensor from Bergoz Instrumentation would be the preferred option, and additional shielding is under consideration.

Beam current measurements include both average current and bunch current monitor (BCM). The requirements and types of instrumentation for these two systems are the same as discussed in detail in Chapter 4, Section 4.3.7.3.

### 5.3.6.4 *Synchrotron Light Based Measurement*

The synchrotron light monitor (SLM) in the Booster is simpler than that in the Collider, due to the larger emittance and beam sizes ranging from several hundreds of microns up to millimeters in both *x* and *y* directions. A visible light beam line will be constructed, and the extraction mirror design will be the same as in the Collider. The beam profile will be obtained through visible light imaging, using a telescope to image the source point in the bending magnet onto a CCD camera. A neutral density filter will be placed in front of the camera to attenuate visible light, especially during energy ramping. The emittance will be calculated using the observed beam sizes and lattice parameters of the source point, and the beam profile can be observed at various electron energies.

### 5.3.6.5 *Beam Loss Monitor System*

Beam loss monitors (BLMs) play a crucial role in the machine protection systems of particle accelerators, minimizing losses to protect equipment. Additionally, they are a sensitive tool for localizing losses and determining when they occur in the acceleration cycle. BLMs are mounted outside the accelerator vacuum chamber, and the signal they produce is proportional to the number of lost particles.

Given that the Booster ramps from 30 GeV to 120 GeV, the Booster BLM requires a wide dynamic range. The chosen BLM for the Booster utilizes Cherenkov light generated in a long quartz fiber. Figure 5.3.6.2 shows that two PMTs set at both ends of the fiber, read out with flash ADCs, can be used to determine the beam loss position, as the time for the Cherenkov light to reach the two PMTs is different. The BLM detector in the Booster differs from that of the Collider.

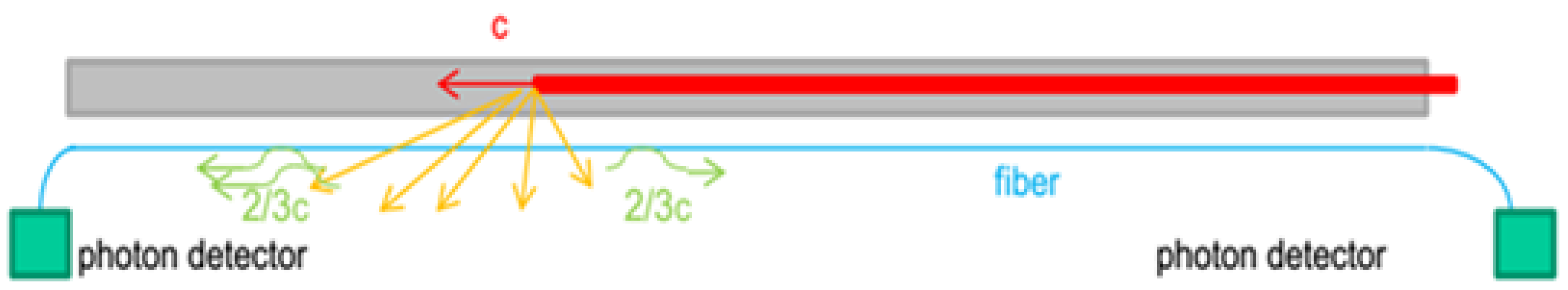

**Figure 5.3.6.2:** Principles of loss detection using optical fibers

After a careful study of different fibers for BLM systems, the SC400 fibers have been chosen for the Booster system, with a fiber length of 150 m. The photomultiplier selected for this system is a Hamamatsu H6780-02 equipped with an FC adapter, which has a typical insertion loss below 0.3 dB. Given that the Booster has a circumference of 100 km, approximately 670 fibers will be installed along the beam pipe.

The signals from the PMTs will be transmitted by coaxial cables to a flash ADC (CAEN V1729A, 4-channel, 14 bits, 2 GS/s) located outside the shield wall. A trigger signal from the accelerator master oscillator will be used as a time reference to determine the position of the beam loss. The flash ADC starts data acquisition when the external trigger arrives. If there is a beam loss in the ($N$+1)th bunch after the arrival of the trigger, the beam loss position can be calculated using the method shown in Fig 5.3.6.3. The upstream and downstream PMTs detect the beam loss pulse signal at times $t_a$ and $t_b$, respectively. $L$ is the length of the fiber that is parallel to the beam pipe, $l$ is the distance from the upstream PMT to the beam loss position, and $t_1 = l/c$ is the time that the bunch travels from the upstream of the fiber to the beam loss location. $t_2$ and $t_3$ are the times that the Cherenkov light signal travels from the beam loss position to the upstream and downstream PMTs, respectively, and $\triangle t$ is the interval time between two bunches. By using Formulas 5.3.6.1 and 5.3.6.2, the beam loss position $l$ and the bunch number $N$ can be calculated as shown in Formulas 5.3.6.3 and 5.3.6.4.

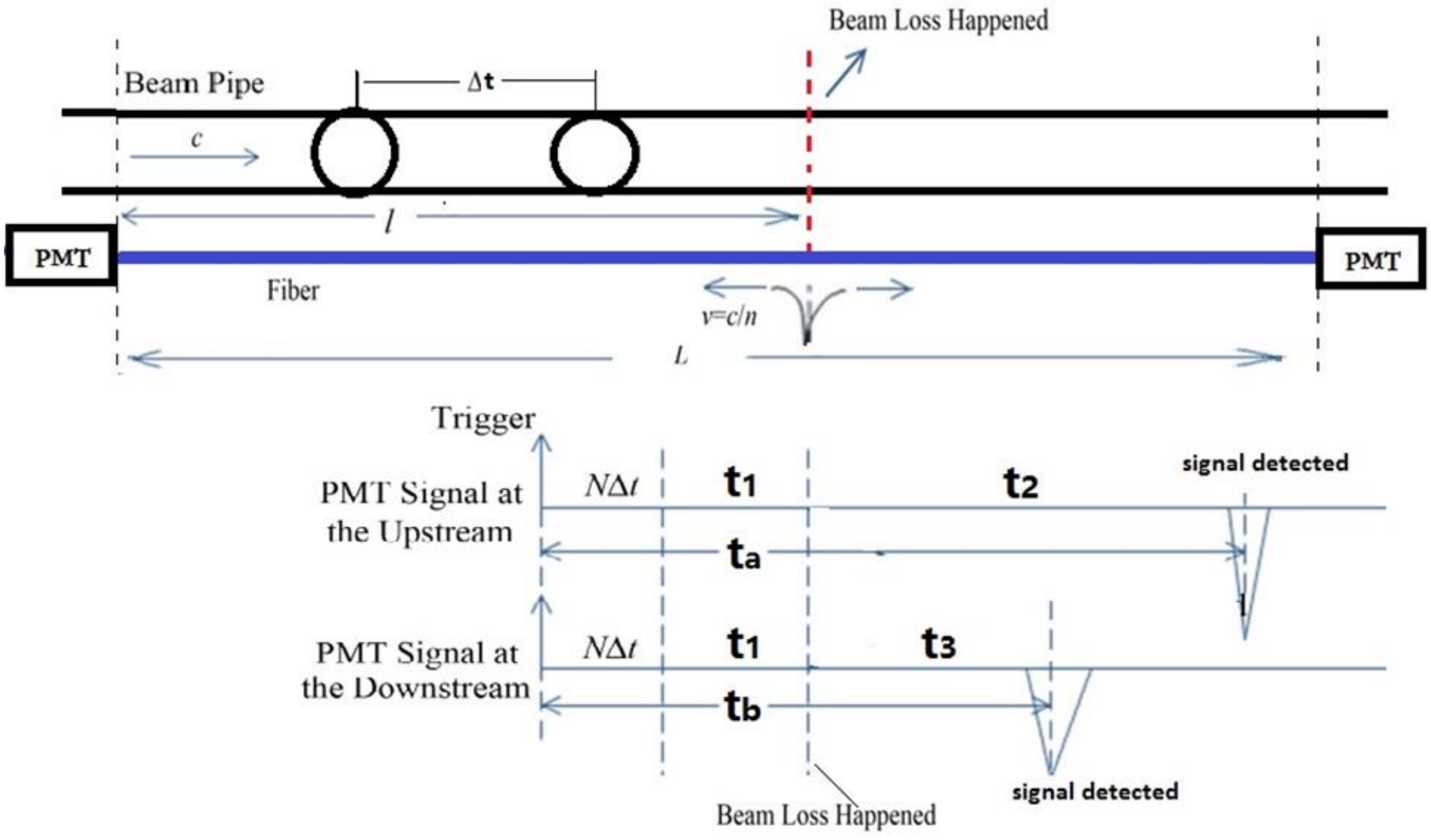

**Figure 5.3.6.3:** Detection principle for beam loss position and bunch number

$$t_a = t_1 + t_2 + N\Delta t = \frac{l}{c} + \frac{l}{v} + N\Delta t \quad (5.3.6.1)$$

$$t_b = t_1 + t_3 + N\Delta t = \frac{l}{c} + \frac{L-l}{v} + N\Delta t \quad (5.3.6.2)$$

$$l = \frac{L-v\ (t_b-t_a)}{2} \quad (5.3.6.3)$$

$$N = \frac{2t_a cv-\ (L+t_a v-t_b v)\ \ (v+c)}{2cv\Delta t} \quad (5.3.6.4)$$

### 5.3.6.6 *Tune Measurement System*

The tune measurement system for the Booster ring can be designed almost the same as the Collider. The details of the system are discussed in Chapter 4, Section 4.3.7.6.

### 5.3.6.7 *Beam Feedback System*

The Booster accelerates from 30 GeV to 120 GeV quickly, causing tune variations during the ramp. The feedback (FB) system must account for these changes. The FB design in the Booster is similar to that of the Collider.

#### *5.3.6.7.1 Feedback Signal Processing*

The feedback signal processing is the same as that in the Collider and the details can be found in Section 4.3.7.7.1.

#### *5.3.6.7.2 Feedback Kicker*

The design of the TFB (Transverse Feedback) kicker in the Booster is identical to that of the Collider, as they share the same beam pipe size. The TFB kicker for the Booster requires a working bandwidth of 20 MHz, with a minimum shunt impedance of approximately 150 kΩ within this bandwidth. For more detailed information, please refer to Section 4.3.7.7.2.

Similarly, the LFB (Longitudinal Feedback) kicker in the Booster has a similar structure to the LFB kicker in the Collider. With an RF frequency of 1.3 GHz in the Booster, the central frequency for the LFB kicker is 1.625 GHz, and the required bandwidth is 1.615 to 1.635 GHz. After optimization, the key geometric parameters of the LFB kicker in the Booster are listed in Table 5.3.6.2. The longitudinal shunt impedance within the specified bandwidth is approximately 1,800 Ω.

**Table 5.3.6.2**. Main geometric parameters of the Booster LFB kicker.

| Parameter | Value | Parameter | Value |
|---|---|---|---|
| Beampipe radius $R$ | 28 mm | Back cavity length $b$ | 36.38 mm |
| Back cavity radius $R_1$ | 38 mm | Cavity length $l$ | 303 mm |
| Ridge radius $R_2$ | 60 mm | Port angle $\alpha$ | 14.3 deg |
| Pillbox cavity radius $R_3$ | 64.2 mm | Port base angle $\alpha_1$ | 6.7 deg |
| Cavity gap $d$ | 64 mm | Barrier angle $\beta$ | 61 deg |
| Distance of feedthrough $d_1$ | 5 mm | Nose cone radius $r_nose$ | 3 mm |
| Ridge length $a$ | 41.62 mm | Nose cone length $l_nose$ | 8 mm |

#### *5.3.6.7.3 Power Calculation and Amplifier Selection*

The principle of the power calculation can be found in Section 4.3.7.7.3.

The results of power requirement in the transverse direction are shown in Table 5.3.6.3. Based on the analysis, considering all modes including injection and extraction, the Z mode is the primary focus due to its higher current and shorter growth time. Z mode extraction, which requires higher energy, satisfies the highest damping time and total power requirements. The calculated power requirement is 188 W for both the X and Y directions. For the transverse feedback system, one two-electrode kicker is utilized for each direction, with each electrode delivering 94 W power. As a result, four 100 W amplifiers can be employed. Optional amplifier models include AR 125A250A, Bonn BSA 0125-100, and others.

**Table 5.3.6.3:** Transverse feedback system power calculation

| Parameter | Z mode - Injection | Z mode - Extraction |
|---|---|---|
| Energy (GeV) | 30 | 45.5 |
| Beta functions at pickup (m) | 120 | 120 |
| Beta function at kicker (m) | 120 | 120 |
| Number of kickers | 2 | 2 |
| Revolution time (ms) | 0.33 | 0.33 |
| RF frequency (MHz) | 650 | 650 |
| Kicker shunt impedance | 150 | 150 |
| Amplitude of oscillation | 0.3 | 0.3 |
| Damping time (ms) | 10 | 10 |
| Power (W) | 81.7 | 188 |

The results of power requirement in the longitudinal direction are shown in Table 5.3.6.4. Based on the evaluation considering all modes, including injection and extraction, Z mode extraction remains the primary focus for the longitudinal feedback system in the Booster. However, the power requirement for the longitudinal feedback system is significantly higher. To address this, two longitudinal kickers are employed to increase the impedance. As a result, each kicker requires 3,030 W power.

To meet this power requirement, four electrode kickers are utilized, with each electrode needing to provide 189 W power. Therefore, a total of sixteen 200 W amplifiers can be used for the longitudinal feedback system. Optional amplifier models that can be considered include Milmega AS0820-250, among others.

**Table 5.3.6.4:** Longitudinal feedback system power calculation

| Parameter | Z mode - Injection | Z mode - Extraction |
|---|---|---|
| Energy (GeV) | 30 | 45.5 |
| Momentum compaction | 1.12 | 1.12 |
| Longitudinal tune | 0.0879 | 0.0879 |
| Phase acceptance (mrad) | 1.7 | 1.7 |
| RF frequency (MHz) | 1300 | 1300 |
| Kicker shunt impedance | 1.8 | 1.8 |
| Damping time (ms) | 200 | 200 |
| Number of kickers | 4 | 4 |
| Power (W) | 1320 | 3030 |

#### 5.3.6.8 *Vacuum Chamber Displacement Measurement System*

The vacuum chamber displacement measurement system in the Booster shares similarities with the measurement systems employed in the Collider. A detailed description of the system can be found in Chapter 4 of the documentation. For more specific information, please refer to Section 4.3.7.8.

### 5.3.7 **Injection and Extraction System**

The CEPC Booster and Collider rings are situated within the same 100 km perimeter tunnel, with the Booster ring located on the tunnel ceiling. The energy of the injected beam from the Linac into the Booster is 30 GeV, and the maximum energy of the extracted beam into the Collider is 120 GeV, with an upgrade potential to 180 GeV. The Booster also acts as a beam accumulation ring for on-axis swap-out injection of the Collider at full energy. Consequently, the injection and extraction systems of the Booster are quite complex, including one low-energy injection system, three high-energy extraction systems (respectively for on-axis injection and off-axis injection to the Collider, and extraction of the Booster beam to the dump, which is shared with the Collider dump system), and one high-energy accumulating injection system, as illustrated in Figure 4.3.8.1 in Chapter 4. Table 5.3.7.1 and Table 5.3.7.2 present the physical parameters related to the injection and extraction of the Booster, as well as the requirements for each injection and extraction subsystem.

**Table 5.3.7.1:** Parameters related to the booster inj. and ext. at different operation mode (on = on-axis, off = off-axis)

| | **Higgs** | | | **W** | | **Z** | | $t\bar{t}$ | | |
|---|---|---|---|---|---|---|---|---|---|---|
| Beam energy (GeV) | 30 | 120 | | 30 | 80 | 30 | 45.5 | 30 | 180 | |
| Function | Inj (on/off) | Extr (off) | Inj/Extr (on) | Inj (off) | Extr (off) | Inj (off) | Extr (off) | Inj (on/off) | Extr (off) | Inj/Extr (on) |
| Bunch number | 242 (half-ring) | | 7 (half-ring) | 1524 (uniform) | | 6000 | | 28 (half-ring) | 1 (half-ring) | |
| Min. bunch spacing (ns) | 680 | | 23800 | 220 | | 25 | | 5900 | 166666.7 | |
| Bunch number per train× train number | - | | - | - | - | 80 × 75 | | - | - | |
| Train spacing (ns) | - | | - | - | - | 2460 | | - | - | |
| Inj/Ext. mode | Bunch by bunch | | | Bunch by bunch | | Bunch by bunch | Train by train | Bunch by bunch | | |
| Kicker repetition rate (Hz) | 100 | 1000 | | 100 | 1000 | 100 | 1000 | 100 | 1000 | |
| Kicker pulse width (ns) | 1360 | | | 440 | | 50 | <6900 | 11800 | | |
| Kicker flat top (ns) | - | | | - | | - | >1980 | - | | |
| Kicker rise/fall time (ns) | <680 | | | <220 | | <25 | <2460 | <5900 | | |
| Timing delay (ns) | <680 | | <23800 | <220 | | <2460 | <4400 | <5900 | | |
| Inj/Extr period (s) | 2.42 | 0.242 | 0.007 | 15.24 | 1.5 | 60 | 0.075 | 0.28 | 0.028 | 0.001 |
| Kick angle (mrad) | 0.11 | 0.2 | 0.1 | 0.11 | 0.2 | 0.11 | 0.2 | 0.11 | 0.2 | 0.1 |
| Kick Integral field strength (T-m) | 0.011 | 0.08 | 0.04 | 0.011 | 0.054 | 0.011 | 0.03 | 0.011 | 0.12 | 0.06 |
| Min. thickness of Septum (mm) | 5.5 | 6 | 6 | 5.5 | 6 | 5.5 | 6 | 5.5 | 6 | 6 |
| Defletion angle of septa (mrad) | 45 | 43 | 43 | 45 | 43 | 45 | 43 | 45 | 43 | 43 |
| Total integral field strength of septa (T-m) | 0.92 | 17.4 | 17.4 | 0.92 | 11.4 | 0.92 | 6.6 | 0.92 | 26.1 | 26.1 |
| Beam pipe aperture (mm) | 56 | | | | | | | | | |

**Table 5.3.7.2**: Requirements of the Booster injection and extraction systems

| Parameter | LE-Injection（on- & off-axis） | HE-Extraction（off-axis） | HE-Inj./Ext.（on-axis） | HE-Extraction (dump) |
|---|---|---|---|---|
| Kicker repetition rate (Hz) | 100 | 1000 | 1000 | 1000 |
| Kicker pulse width (ns) | 50 | 440~2420 (adjustable) | 1360 | 440~2420 (adjustable) |
| Kicker flat top (ns) | - | 0~1980 (adjustable) | - | 0~1980 (adjustable) |
| Kicker rise/fall time (ns) | < 25 | < 220 | < 680 | < 220 |
| Inj./Extr. period (s) | 60 | 1.5 | 0.007 | 1.5 |
| Kick angle (mrad) | 0.11 | 0.2 | 0.1 | 0.2 |
| Kick Integral field strength (Tm) | 0.011 | 0.08/ 0.12 ($t\bar{t}$) | 0.04/ 0.06 ($t\bar{t}$) | 0.08/ 0.12 ($t\bar{t}$) |
| Quantity of kicker group in each subsystem | 1 | 1 | 2(inj. and ext.) | 1 |
| Thickness of Septum (mm) | 5.5 | 6 | 6 | 6 |
| Defletion angle of septa (mrad) | 45 | 43 | 43 | 43 |
| Integral field strength of septa (T-m) | 0.92 | 17.4/26.1($t\bar{t}$) | 17.4/26.1($t\bar{t}$) | 17.4/26.1($t\bar{t}$) |
| Quantity of septa group in each subsystem | 1 | 1 | 2(inj. and ext.) | 1 |
| Beam pipe aperture (mm) | 56 | 56 | 56 | 56 |

### 5.3.7.1 *Injection from Linac*

To optimize the overall cost of the CEPC accelerators, the beam energy injected into the Booster from the Linac is set at 30 GeV. As low-energy injection of the Booster does not require beam accumulation, a single turn injection with an on-axis injection scheme can be adopted.

The circumference of the Booster is 100 km, and the most intensive bunch filling is under the Z mode. The ring can accommodate a maximum of 6000 evenly distributed bunches, divided into 75 trains, with each train containing 80 bunches. The minimum spacing between bunches within a train is 25 ns, and the spacing between trains is 2460 ns. As the bunch time structure of the Linac differs from that of the Booster, low-energy injection into the Booster must be performed bunch by bunch. The injection kicker's pulse bottom width must be less than twice the minimum bunch spacing of 50 ns. The use of stripline kickers driven by super-fast pulsed power supplies can generate nanosecond narrow pulse electromagnetic fields to deflect the injected beam bunch by bunch. The Linac is located above ground, while the Booster is 100 meters underground, connected by two inclined transport lines. Due to the elevation difference, the preferred low-energy injection system for the Booster is the Lambertson septa magnet, with the thinnest part of the septum plate less than 5.5 mm. The layout of the Booster's low-energy injection system is depicted in Figure 5.3.7.1, where the Lambertson magnet provides horizontal deflection, and the kicker provides vertical deflection.

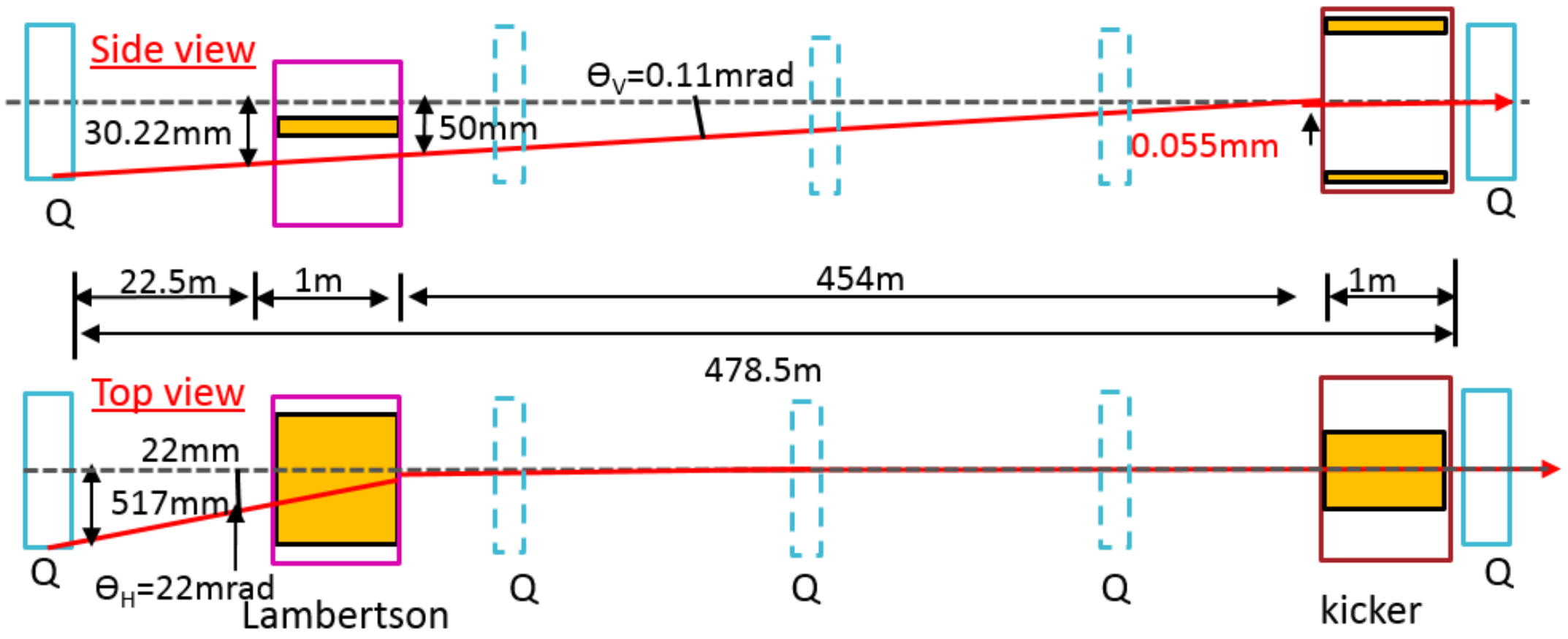

**Figure 5.3.7.1:** Layout of the CEPC Booster low energy injection system

### 5.3.7.2 *Extraction to the Collider in Off-axis Injection Mode*

After being accelerated to collision energy, the beams are extracted and injected into the Collider. In the W and Z modes, off-axis injection scheme is utilized in the Collider. Figure 5.3.7.2 illustrates the layout of the connection area between the Booster and the Collider. The Booster extraction system consists of a horizontally deflecting kicker system and a group of vertically deflecting Lambertson magnets. Under the W and Z modes, the beam is stored in the Booster in the form of a bunch train. The kicker magnet of the Booster extraction system is identical to the off-axis injection system of the Collider. The rise and fall time of the trapezoidal wave pulse of the kicker system should be less than 200 ns, with the flat-top width adjustable from 40 to 2000 ns. The Lambertson magnet's septum thickness should be less than 6 mm.

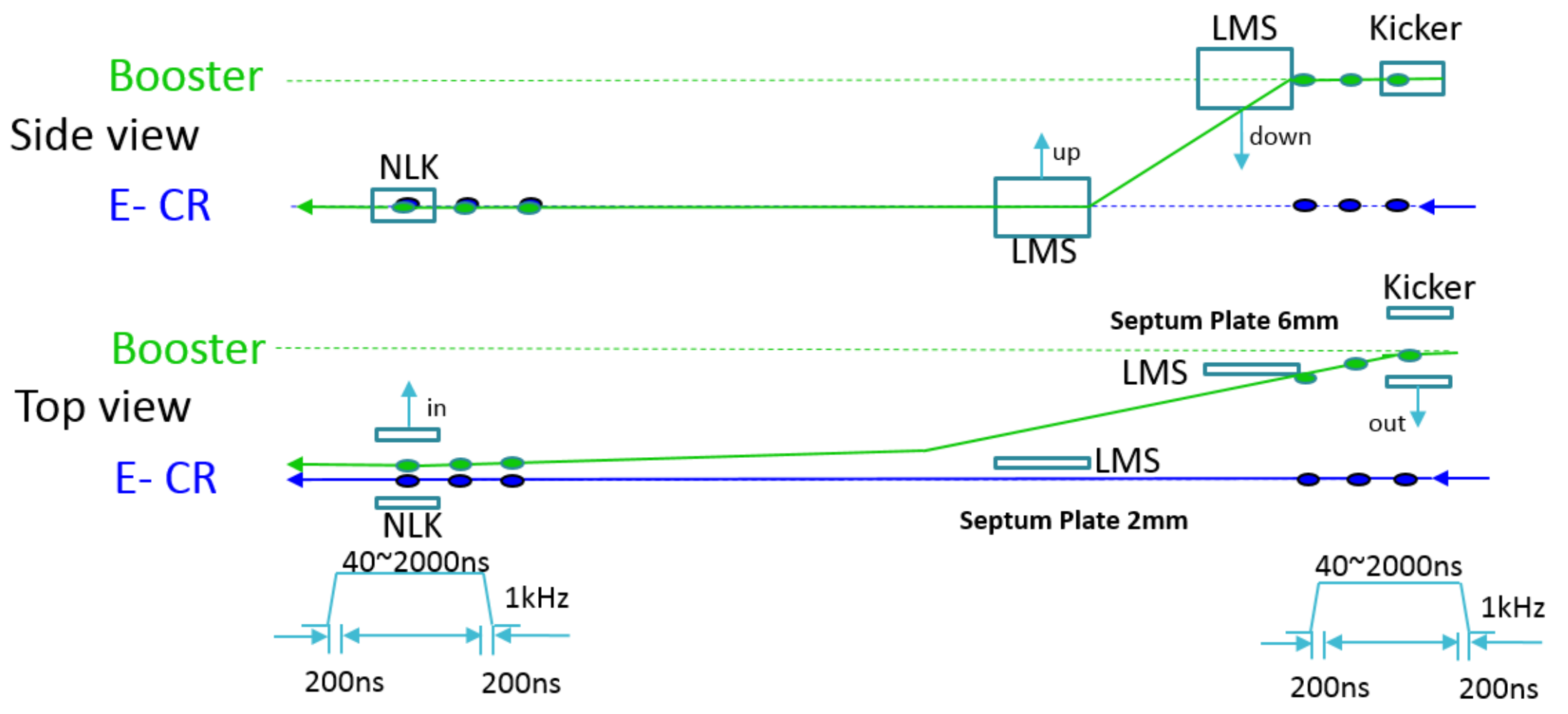

**Figure 5.3.7.2:** Layout of the Booster extraction system for the Collider off-axis injection

### 5.3.7.3 *Extraction and Re-injection to or from the Collider in On-axis Injetion Mode*

In the H mode, the Collider adopt the on-axis injection scheme, and the connection area between the Booster and the Collider is laid out as shown in Figure 5.3.7.3. The extraction system of the Booster also consists of a horizontal deflection kicker system and a group of vertical deflection Lambertson magnets. In the H mode, the minimum bunch spacing is 680 ns, and the bottom width of the kicker pulse is required to be less than 1360 ns. Additionally, the septum thickness of the Lambertson magnet must be less than 6 mm.

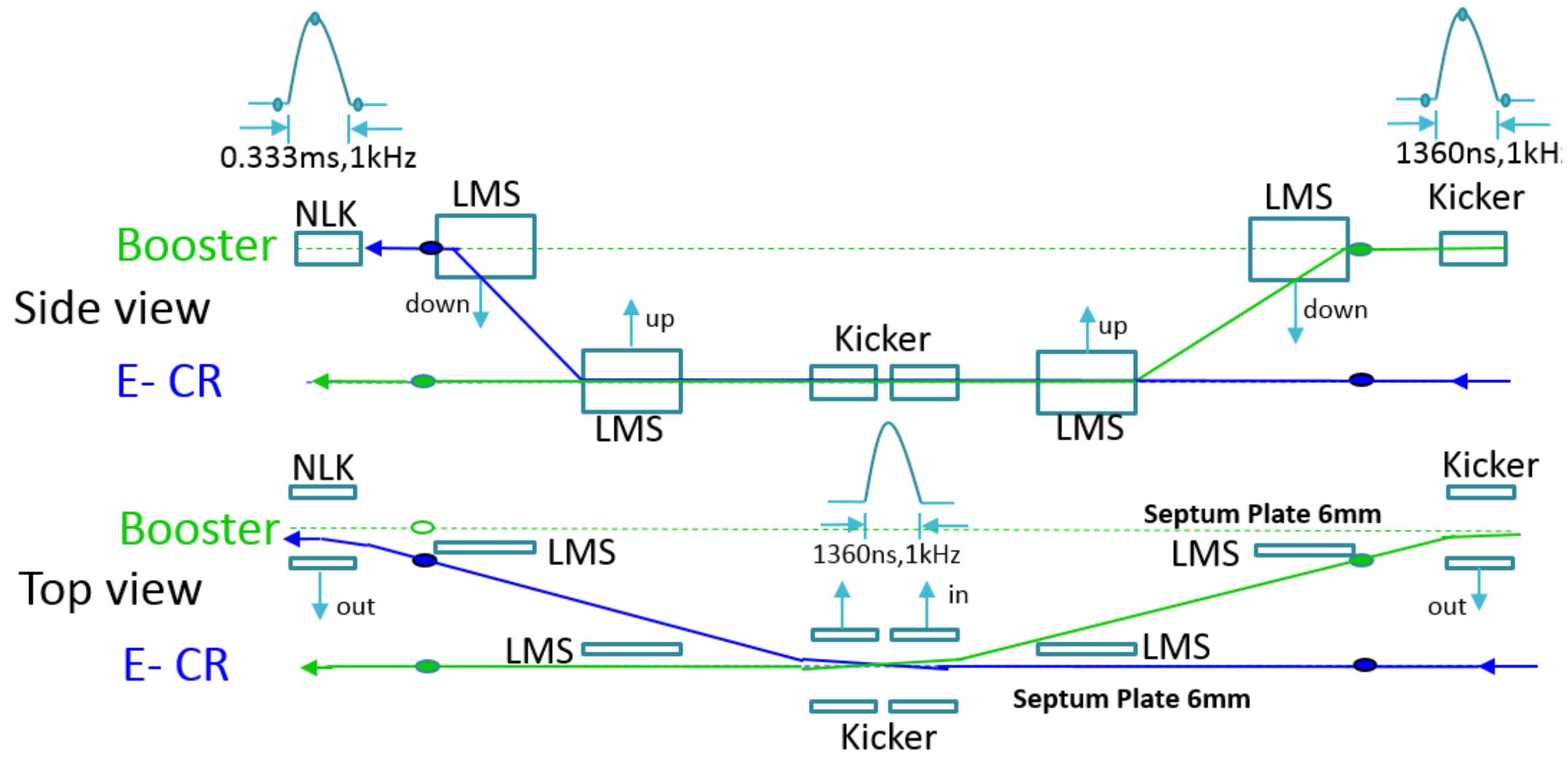

**Figure 5.3.7.3:** Layout of the Booster inj. and ext. system for the Collider on-axis injection

### 5.3.7.4 *Stripline Kicker for Booster Injection from the Linac*

The low-energy injection from the Linac to the Booster employs stripline kickers driven by ultrafast pulsers to deflect the injected beam bunch by bunch. The stripline kicker is essentially an encounter traveling-wave kicker that uses the TEM-mode electromagnetic wave transmitted on the striplines to interact with the charged particle beam flying in the reverse direction. The electromagnetic field distribution of a typical stripline kicker is shown in Figure 5.3.7.4.[1-5]

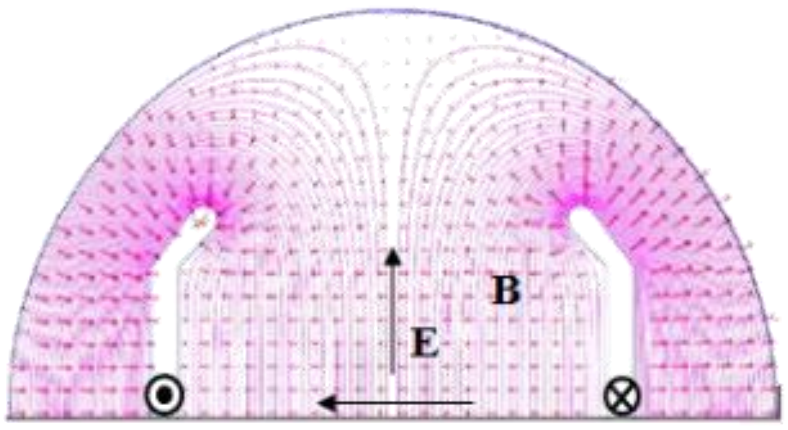

**Figure 5.3.7.4:** Electromagnetic field distribution of a typical stripline kicker

When both the charged particle beam and the TEM-mode electromagnetic wave are moving at the speed of light (*c*), the electric and magnetic forces that act on the particles are equal in magnitude and direction. The interaction force and resulting deflection angle can be expressed as:

$$F_{E+B} = 2\frac{Uq}{d} \tag{5.3.7.1}$$

$$\theta = \theta_E + \theta_B = 2\theta_E = 2g\frac{eU}{E}\frac{l}{d} \tag{5.3.7.2}$$

$$g = \tanh(\frac{\pi w}{2d}) \tag{5.3.7.3}$$

where, $\theta$ is the kick angle (rad), *E* is the beam energy (eV), *U* is the applied pulse voltage for both electrodes (V), *l* is the stripline length, *d* is the distance between the electrodes, *g* is the geometry factor determined by the shape of the electrode, and *w* is the stripline width.

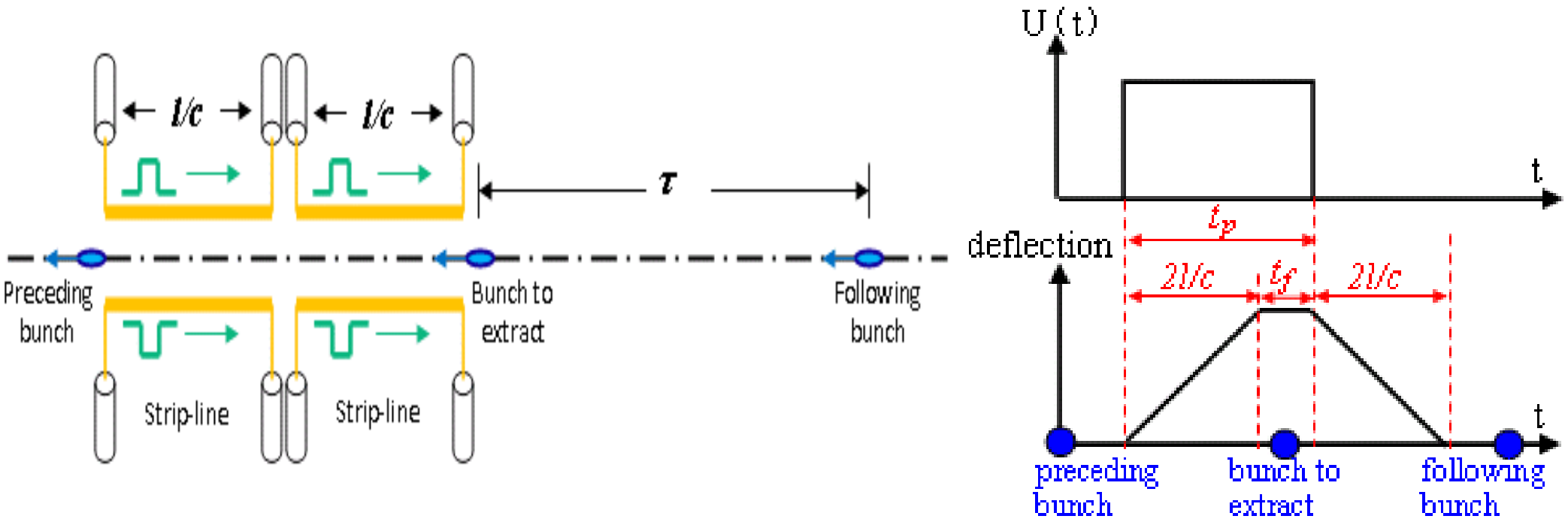

**Figure 5.3.7.5:** The relationship between the electrode length, electric pulse width and pulse field width of a stripline kicker.

As shown in Fig 5.3.7.5, the electrode length of each stripline kicker and electric pulse width must meet the following condition:

$$\begin{cases} l < c\tau/2 \\ 2l/c < t_p < 2\tau - 2l/c \end{cases} \tag{5.3.7.4}$$

where $c$ is the speed of light, $\tau$ is minimum bunch spacing, and $t_p$ is the bottom width of electrial pules.

After optimization, the Booster injection energy is $E$ = 30 GeV, and the total kick angle of the injected beam is $\theta$ = 0.11 mrad. The distance between the electrodes of the stripline kicker is set to $d$ = 44 mm to ensure the required beam clearance region. Using formula (5.3.7.2), when the geometric factor $g$ = 1 and $U \times L$ > 72.6 kV·m, the total length of the stripline electrode is $L$ = 3 m, and the working voltage $U$ > 24.2 kV (or ±12.1 kV).

The minimum bunch spacing of the Booster is $\tau$ = 25 ns. According to formula (5.3.7.4), the electrode length $l$ must be less than 3.75 m. However, due to the difficulty of machining slender electrodes, the length of each electrode is reduced to 1 m, and each injection system includes three 1 m long stripline kickers. Therefore, the bottom width of the kicker electric pulse must be 6.7 ns < $t_p$ < 43.3 ns. In other words, the flat-top width of the electric pulse must be more than 6.7 ns, and the rise/fall time $t_r/t_f$ must be less than 18.3 ns.

**Table 5.3.7.3:** Parameters of the stripline kicker system for Booster low energy injection

| Parameter | Unit | BSTLEIK |
|---|---|---|
| Quantity | - | 2×3 |
| Type | - | Stripline kicker |
| Deflect direction | - | Vertical |
| Beam Energy | GeV | 30 |
| Deflect angle | mrad | 0.11 |
| Length of kicker electrode | mm | 1000 |
| Gap between two electrodes | mm | 44 |
| Odd mode impedance | Ω | 50±1 |
| Even mode impedance | Ω | <65 |
| Clearance region (H×V) | mm | 22×44 |
| Good field region (H×V) | mm | 12×22.6 |
| Integral field uniformity | - | ±2.5% |
| Amplitude of electrical pulse（into 50 Ω） | kV | ±12.25 |
| Repetition rate | Hz | 100 |
| Amplitude repeatability | - | <2% (RMS) |
| Pulse jitter | ns | ≤1 |
| Bottom width of electrical pulse（3%-3%） | ns | < 50 |

The design parameters of the stripline kicker system for the Booster low energy injection are presented in Table 5.3.7.3. Using Formula (5.3.7.3), it is determined that an electrode width of 70 mm is required when the electrode distance is set to 44 mm to obtain the maximum geometric factor of 0.9866. Figure 5.3.7.6 illustrates the 2D model of the kicker electrodes and outer body. To improve the rigidity of the 1-meter-long electrodes, their thickness is set to 10 mm. The inner contour of the outer body consists of four connected elliptical arcs. The ridge at the equator of the outer body is used to isolate the electromagnetic coupling between two striplines, optimizing the matching design of odd-mode and even-mode impedance. The 2D and 3D models of the optimized stripline kicker designed using CST are depicted in Figures 5.3.7.7 and 5.3.7.8, respectively. The odd-mode impedance of the stripline kicker is optimized to $Z_{odd}$ = 50 Ω, and the even-mode impedance is $Z_{even}$ = 60.5 Ω when the distance between ridges is 60 mm. The feedthroughs have an impedance of 50 Ω to match the external RF cables.

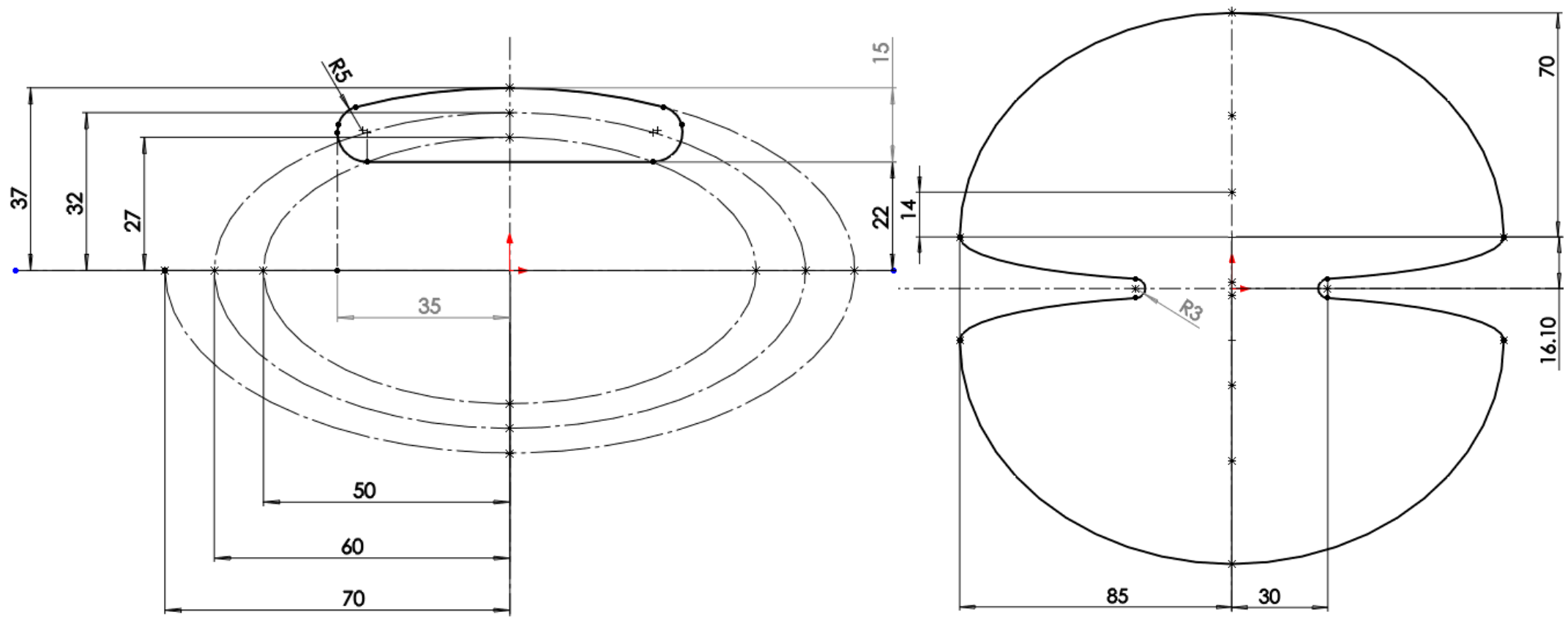

**Figure 5.3.7.6:** Profile of the stripline kicker

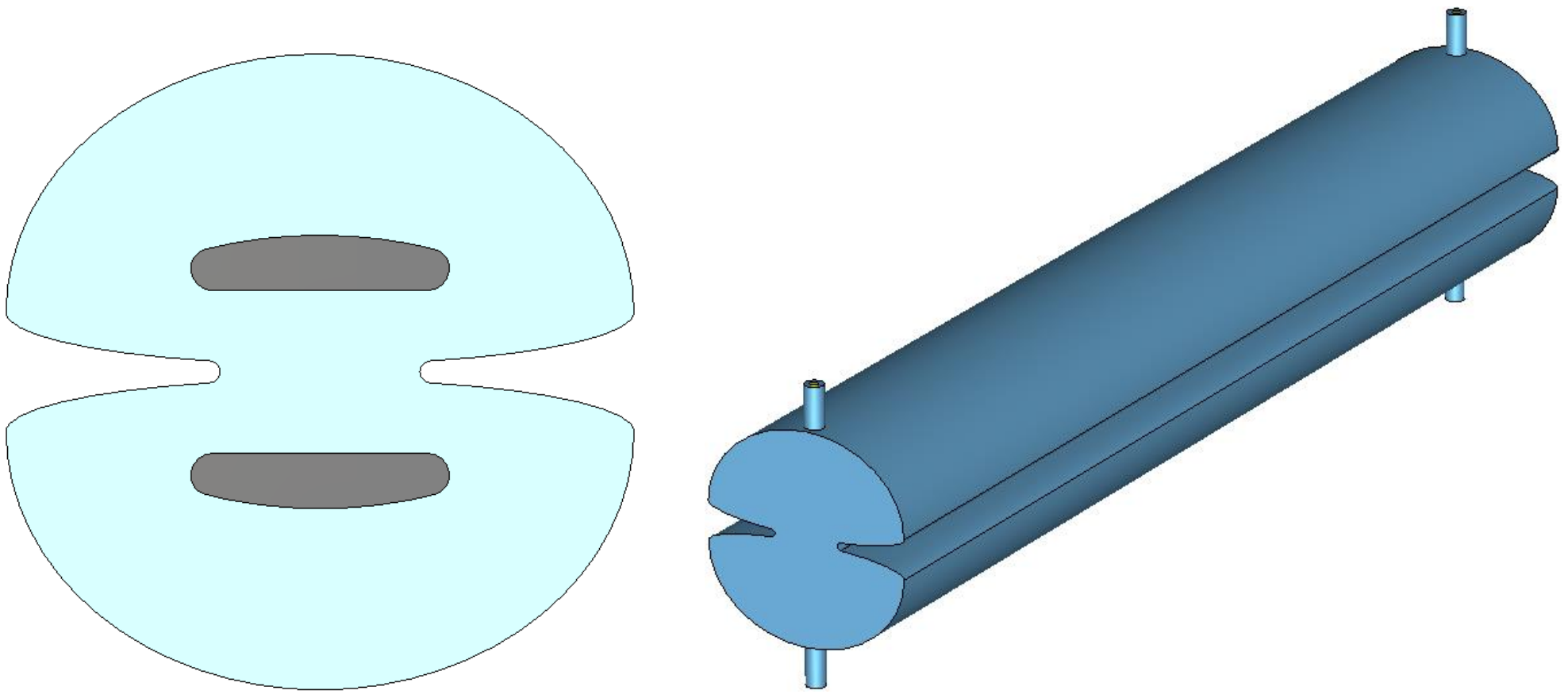

**Figure 5.3.7.7:** Outer body model in CST

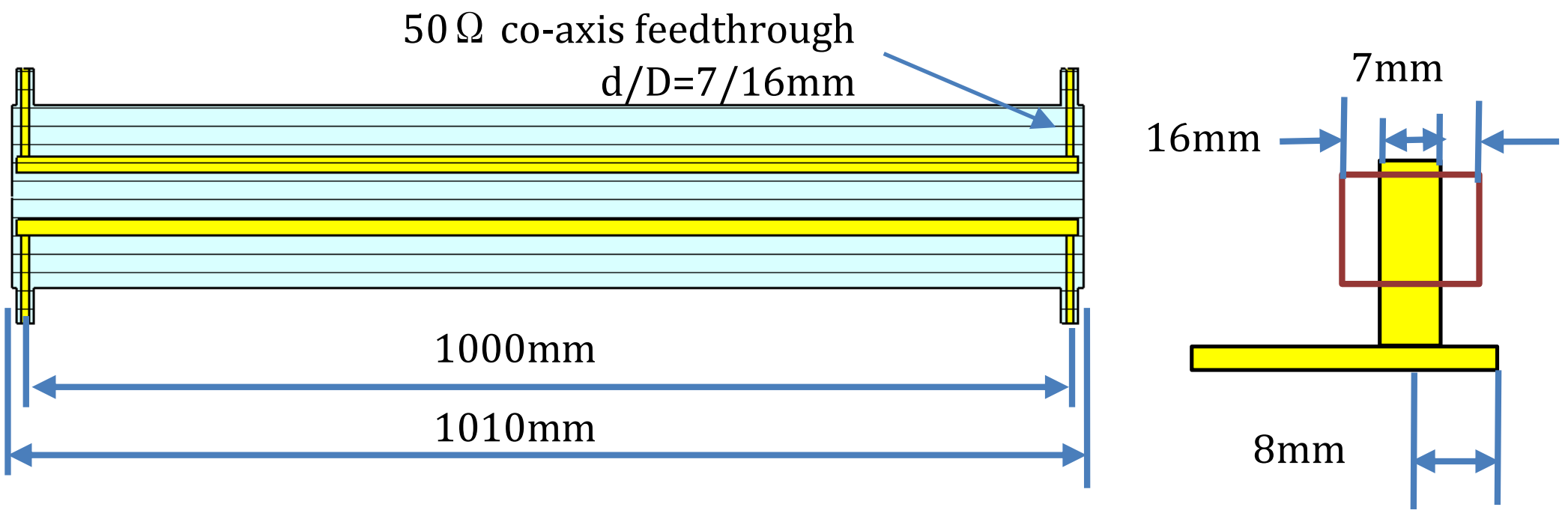

**Figure 5.3.7.8:** The 3D geometric dimensions of the stripline kicker

After optimization, Figure 5.3.7.9 shows the simulation results of the S-parameters (S11, S21) of the stripline kicker. It can be observed that the striplines exhibit excellent microwave transmission characteristics when the frequency is below 500 MHz, while the bandwidth of the electric pulse of the kicker is less than 300 MHz. The TDR (Time

Domain Reflectometry) simulation of the stripline kicker is shown in Figure 5.3.7.10. The odd mode impedance of the stripline has been optimized to 50 Ω. However, due to electromagnetic coupling between electrodes, the even mode impedance is greater than the odd mode impedance, measuring at 60.5 Ω.

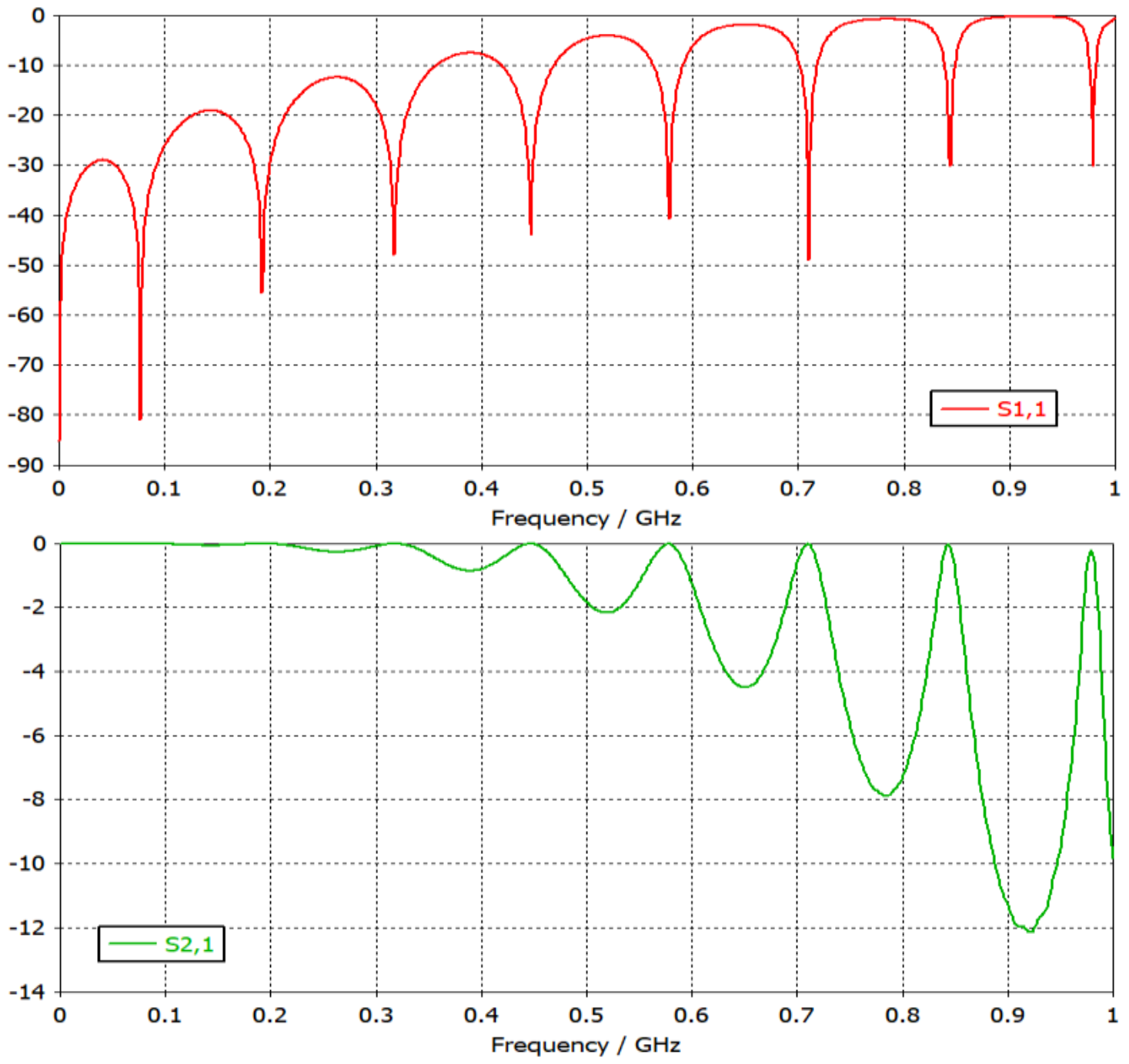

**Figure 5.3.7.9:** S-parameters of the stripline kicker

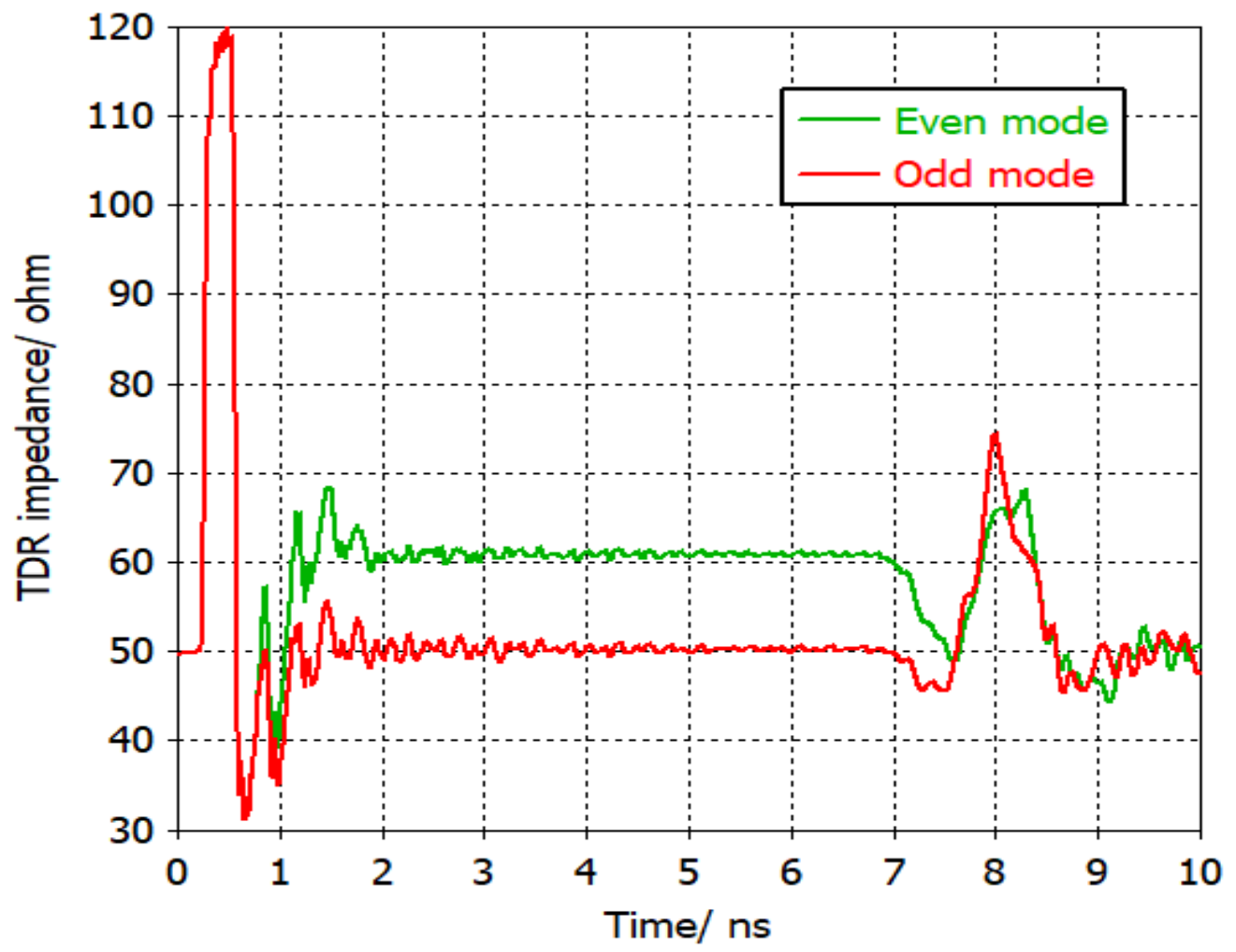

**Figure 5.3.7.10:** TDR (Time Domain Reflectometry) analysis of the stripline kicker.

Figure 5.3.7.11 illustrates the field distribution of the stripline kicker. The field uniformity is approximately 0.52% in the *x* range of (–6, 6) mm and 1.35% in the *y* range of (–11.3, 11.3) mm. Figure 5.3.7.12 shows the local electric field intensity distribution of the kicker, with the maximum local electric field appearing at the end of the electrodes. When the pulse peak is ±15 kV, the maximum local electric field Emax is 9 MV/m. It is noteworthy that this value is less than the breakdown electric field strength of 13 MV/m under ultra-high vacuum conditions.

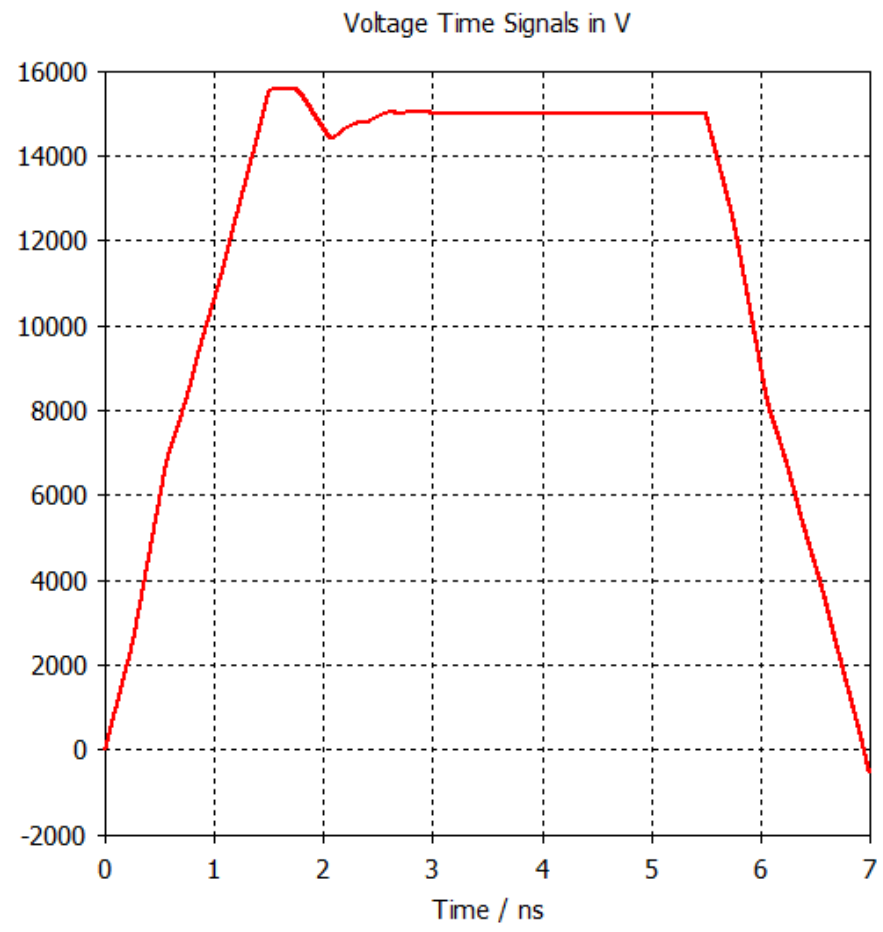

**Figure 5.3.7.11:** Field uniformity of the stripline kicker

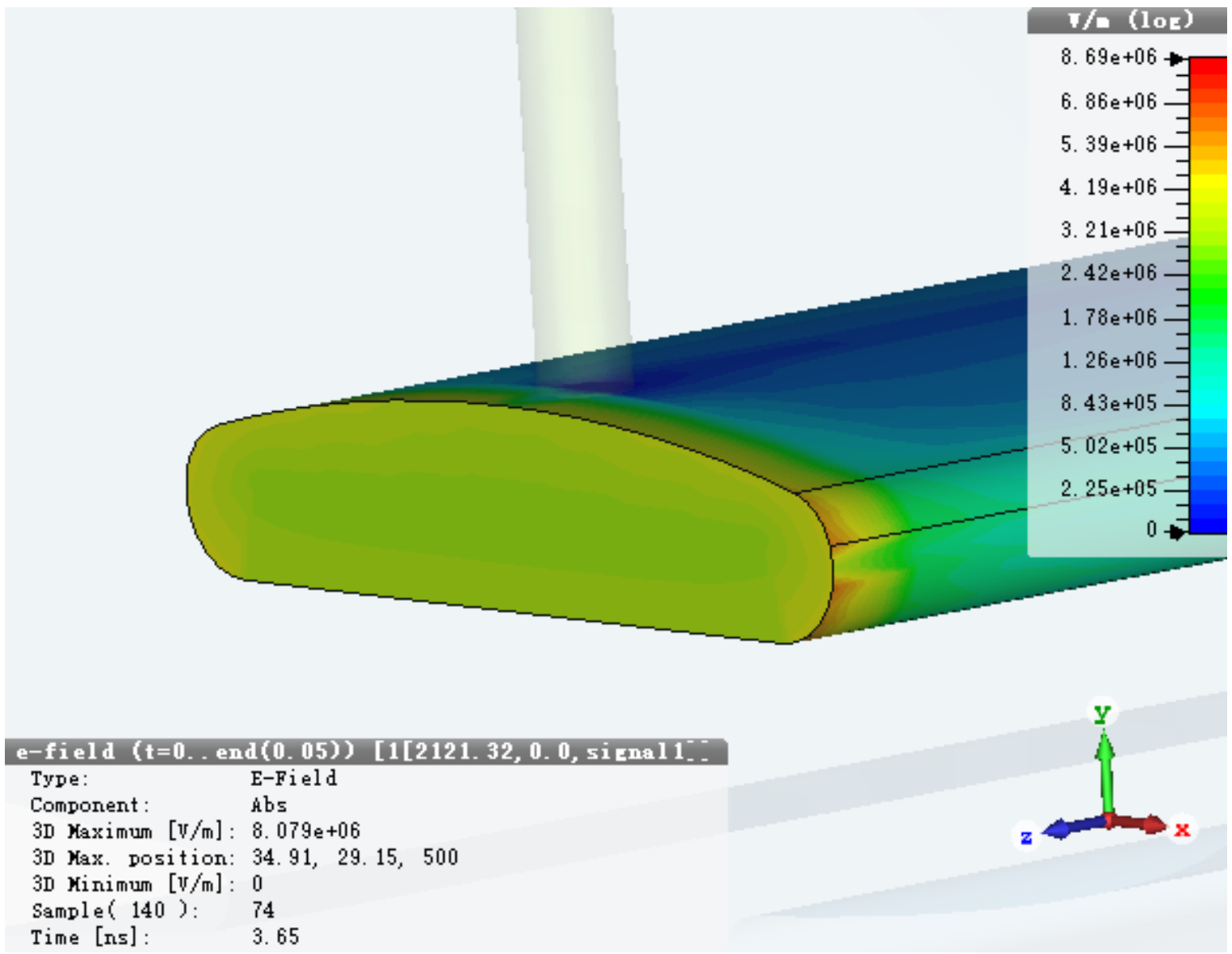

**Figure 5.3.7.12:** Local electric field distribution of the stripline kicker

Figure 5.3.7.13 displays the preliminary mechanical structure design of the stripline kicker, featuring a 1-meter electrode made of stainless steel through CNC processing or cold extrusion. Copper coating on the surface of the eletrodes is necessary to relieve heating effect from the wake-field. Since the electrode is lengthy and subject to large deflection, an insulating support have to be considered to adde in the middle to prevent structural damage. However, the influence of this support on the microwave transmission characteristics of the stripline must be carefully evaluated. The connection between the electrodes and the feedthroughs is a crucial consideration in structural design, and the deformation of the electrodes during baking must be considered.

For the outer body, 316L stainless steel will be used, and it will be produced using EDM in 2~3 sections and then welded together. After rough machining, a solution annealing process must be applied to ensure that the relative magnetic permeability is less than 1.01.

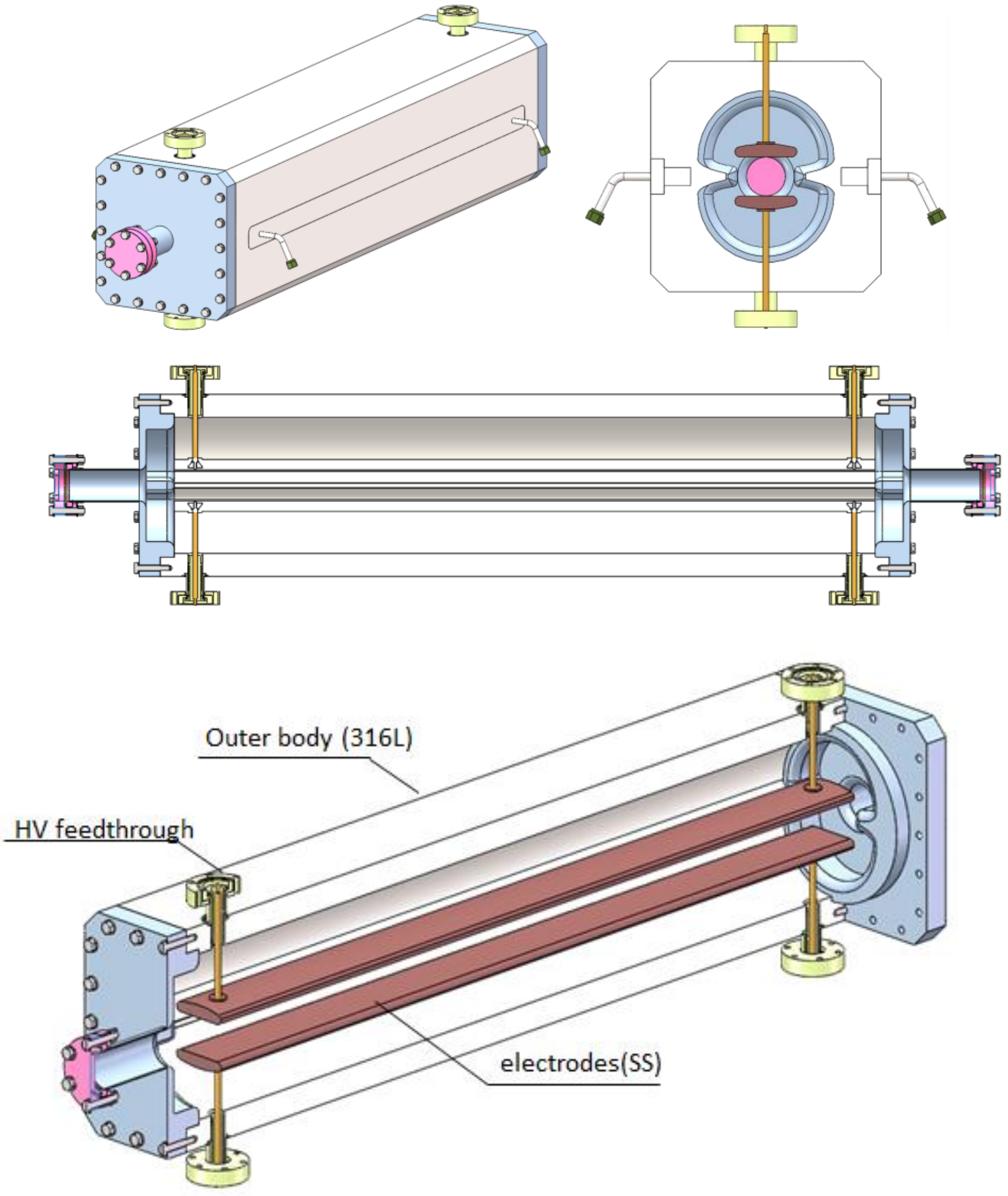

**Figure 5.3.7.13:** Mechanical structure of the stripline kicker.

As part of the HEPS-TF project, two types of strip-line kickers with different electrode lengths were successfully developed in 2019 [6-8]. One type has an electrode length of 750 mm, and the other type has an electrode length of 300 mm.

Figure 5.3.7.14 illustrates the prototype of the strip-line kicker with a 750 mm electrode length. The microwave test and high voltage (HV) test results are presented in Figure 5.3.7.15 and Figure 5.3.7.16, respectively. The tests demonstrate that the prototype satisfies the necessary physical requirements.

During the HV test, a high voltage short pulse with a peak voltage of ±20 kV and a pulse bottom width of 4 ns was transmitted through the strip-line kicker without significant waveform deformation. These findings indicate that the prototype is capable of handling the specified high voltage pulses effectively.

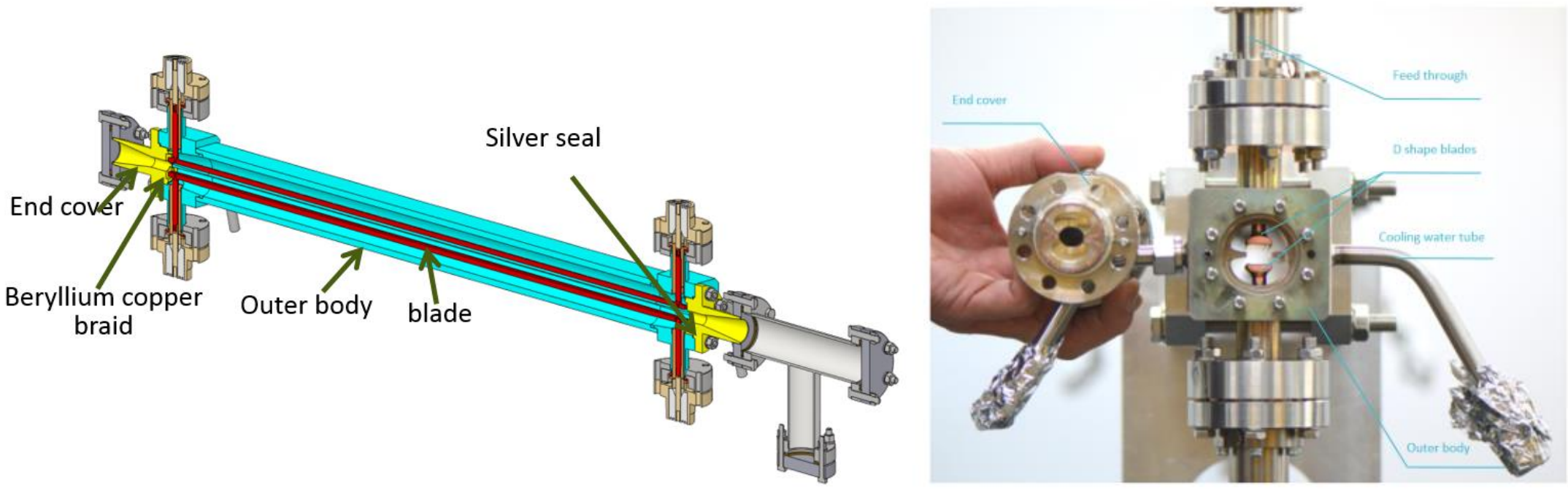

**Figure 5.3.7.14:** A prototype of strip-line kicker with electrode length of 750 mm.

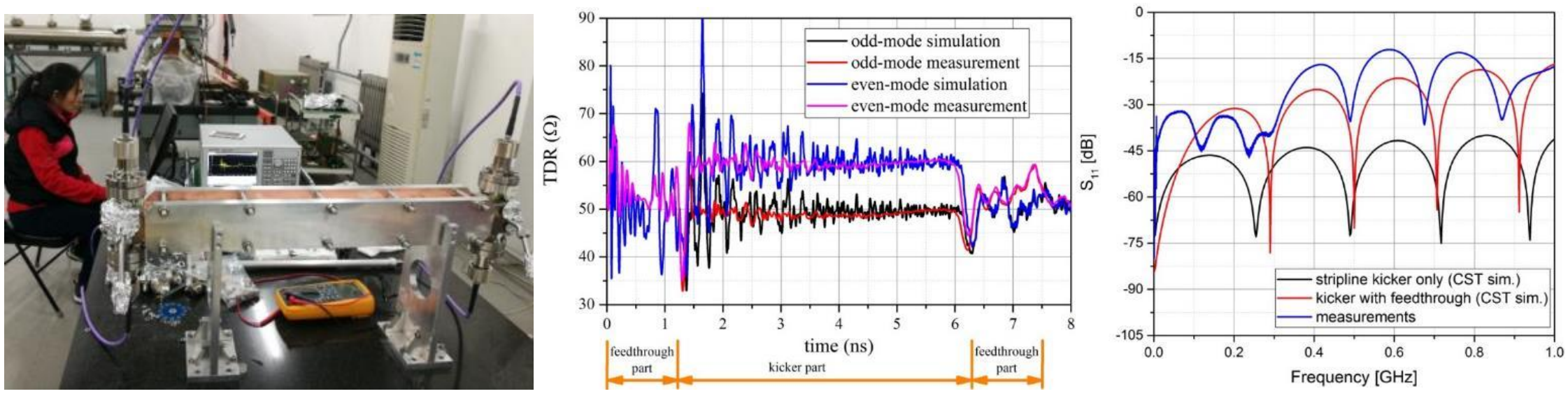

**Figure 5.3.7.15:** Microwave test of the 750 mm-long strip-line kicker prototype.

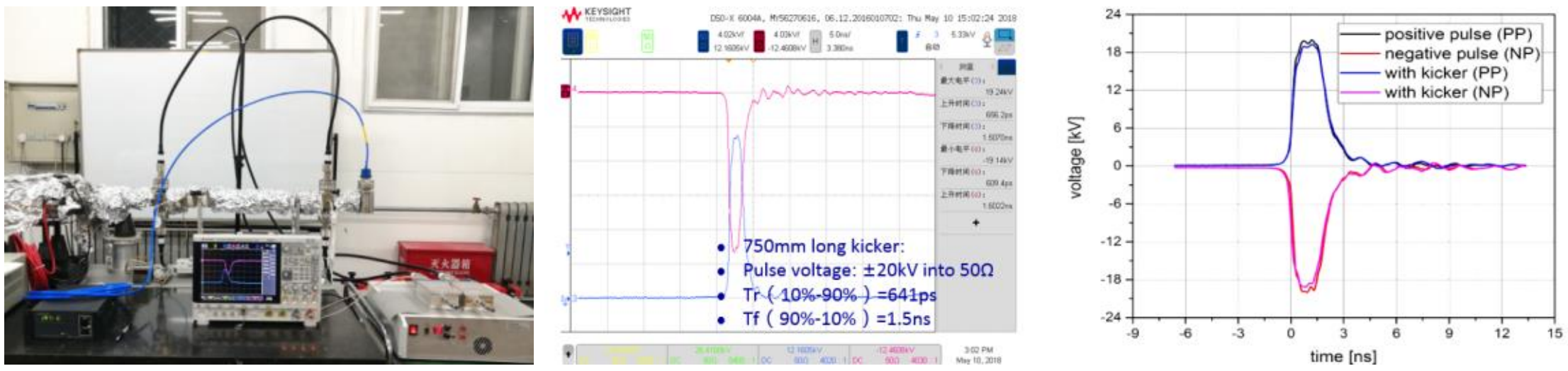

**Figure 5.3.7.16:** HV test of the 750 mm-long strip-line kicker prototype.

Another prototype of strip-line kickers, featuring a 300 mm electrode length, was developed to validate a proposed new structure. In this design, five strip-line kickers were integrated within a large vacuum chamber to minimize the longitudinal distance between adjacent kickers, as depicted in Figure 5.3.7.17. The microwave test and HV test results are displayed in Figure 5.3.7.18 and Figure 5.3.7.19, respectively. The obtained results indicate that the prototype meets the requirements for the HEPS on-axis injection system. Specifically, the kicker pulse bottom width should be less than 10 ns for the swap-out mode and 4 ns for the longitudinal injection mode.

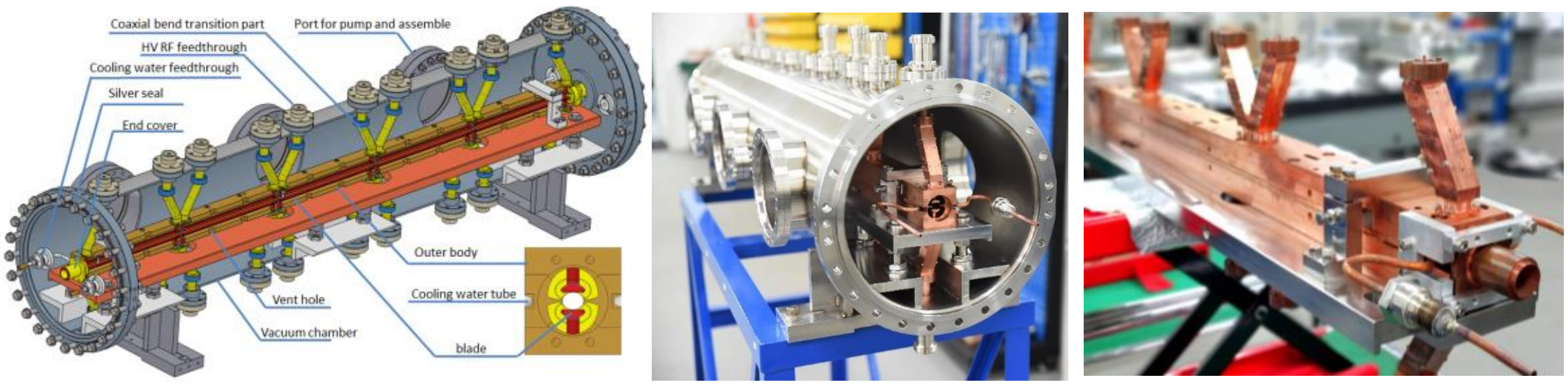

**Figure 5.3.7.17:** A prototype of strip-line kicker with electrode length of 300 mm.

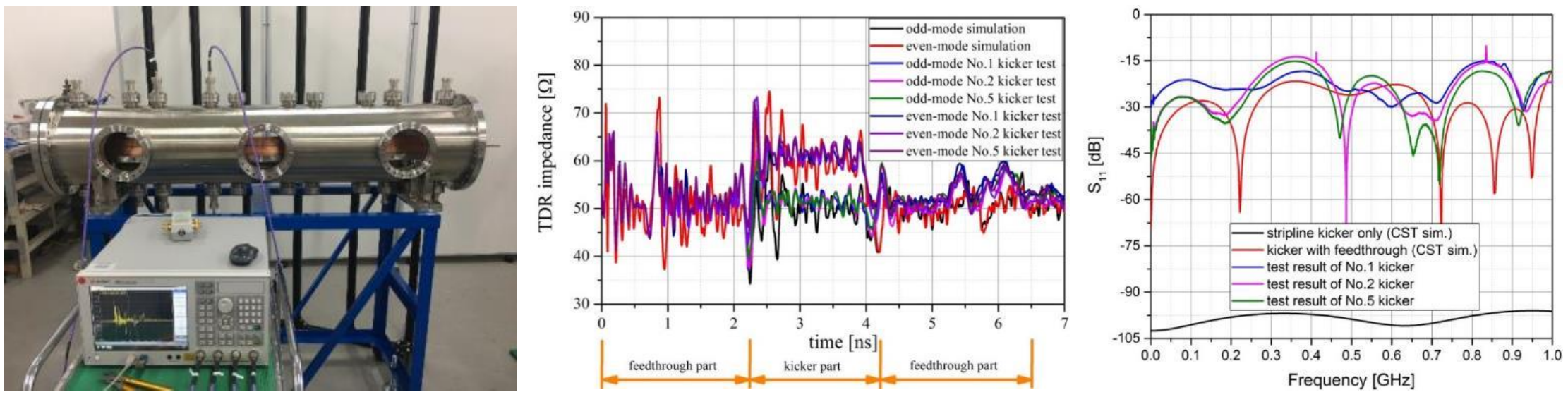

**Figure 5.3.7.18:** Microwave test of the 300 mm-long strip-line kicker prototype.

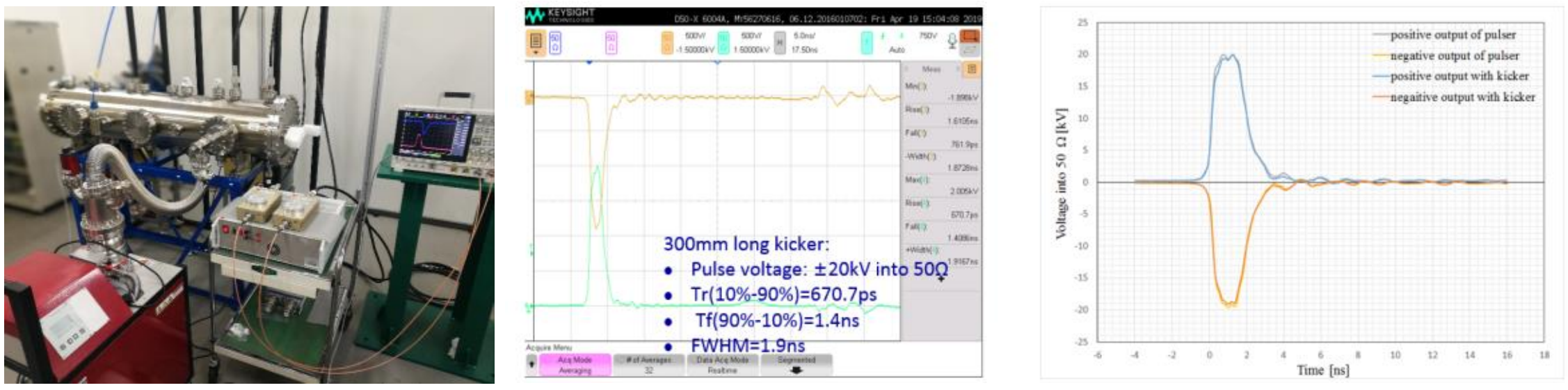

**Figure 5.3.7.19:** HV test of the 300 mm-long strip-line kicker prototype.

### 5.3.7.5 *Fast Pulser for Stripline Kicker in Booster Low Energy Injection*

To meet the requirements of bunch-by-bunch injection under Z energy mode, the stripline kickers of the Booster low energy injection system require pulsed power supplies with specific performance criteria. These criteria include a fast rise time and fall time ($t_r/t_f$ < 18.3 ns), a flat top width of the electric pulse ($t_{top}$ > 6.7 ns), time jitter < 1 ns, bipolar output with a pulse voltage amplitude on a 50 Ω resistive load of greater than ±12.25 kV, and a pulse repetition frequency of 100 Hz. The main design parameters of the pulse power supply are shown in Table 5.3.7.4.

**Table 5.3.7.4:** Parameters of fast pulser for the stripline kicker system

| Parameter | Unit | BSTLEIK pulser |
|---|---|---|
| Quantity | - | 2×3 |
| Stripline kicker odd mode impedance | Ω | 50±1 |
| Stripline kicker even mode impedance | Ω | < 65 |
| Amplitude of electrical pulse（into 50Ω） | kV | ±12.25 |
| Repetition rate | Hz | 100 |
| Amplitude repeatability | - | < 2% (RMS) |
| Pulse jitter | ns | ≤ 1 |
| Bottom width of electrical pulse（3%-3%） | ns | < 50 |
| Rise/fall time of electrical pulse (3%-90%) | ns | < 18.3ns |
| Flat-top of electrical pulse (90%-90%) | ns | > 6.7ns |

The MOSFET based inductive adder is a potential scheme for the fast pulsed power supply; however, it requires a MOSFET with a fast enough switching speed. [9-10] The DE series of RF power MOSFET from IXYS Corporation, such as the DE475-102N21A, has a typical ability to switch on in 5 ns. Unfortunately, this device has been discontinued. The speed of conventional SiC MOSFET on the shelf is slower, such as the C2M0045170D from Cree Corporation, with a switching speed of about 20 ns. This speed is too slow for the Booster low energy injection kicker pulser and therefore, alternative MOSFET options need to be explored.

Another solution for the fast pulsed power supply is the PFL (Pulse Form Line) modulator based on DSRD (Drift Step Recovery Diode), which has been developed in the HEPS-TF project [11-12]. DSRD is a special diode that, under proper pumping conditions, can achieve reverse turn off in nanoseconds. As shown in Figure 5.3.7.20, to generate a very thin and dense electron-hole plasma layer near the PN junction area, the DSRD forward pumping pulse should be less than 150 ns. During the reverse pumping phase, the injected electron-hole plasma should be drawn out as fast as possible to lead to step recovery of the PN junction. DSRDs can be used as a fast-opening switch to release the magnetic field energy stored on the PFL into a resistive load of 50 Ω instantly, with the output pulse width equal to twice the electric length of the PFL.

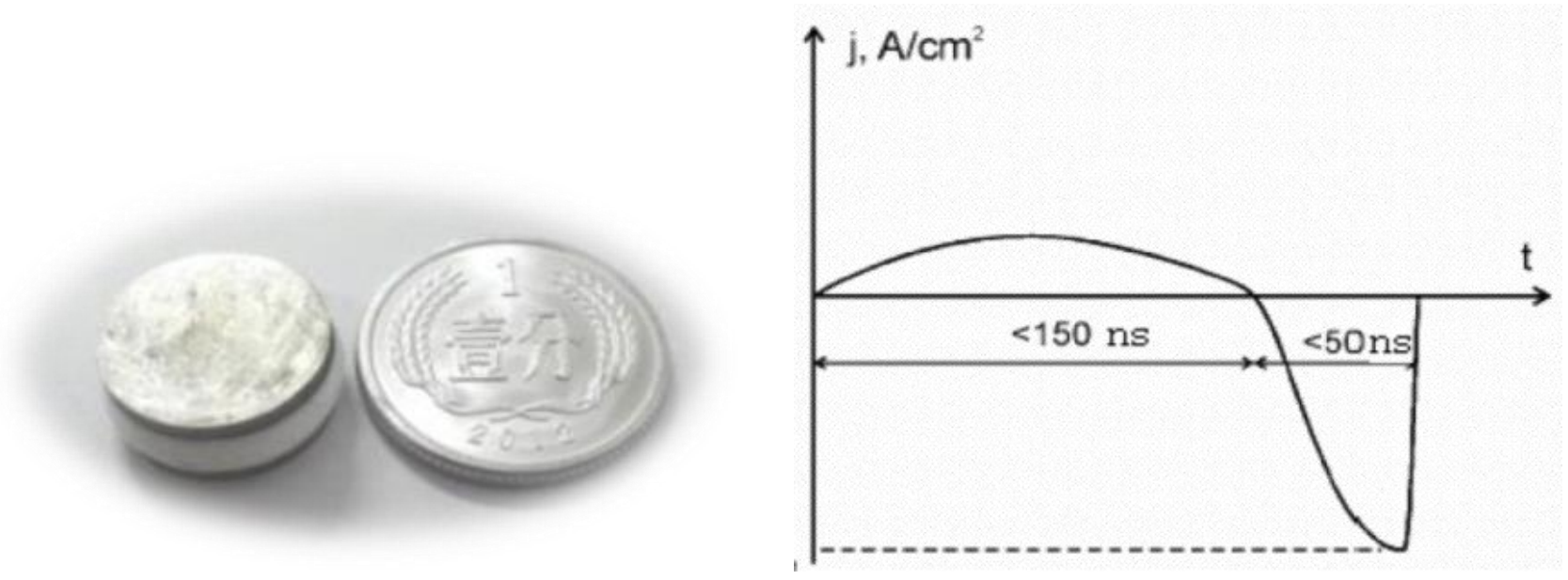

**Figure 5.3.7.20:** DSRD and typical waveform of pumping current

The topology structure of the DSRD pulsed power supply is illustrated in Figure 5.3.7.21.[6] To meet the normal working conditions of DSRD, the pumping circuit is designed as a special 6-stage inductive adder to achieve pulse power superposition. The resonant transformer introduced at the primary of the inductive adder transformer enables two groups of switches to be grounded at the same time, simplifying the circuit design.

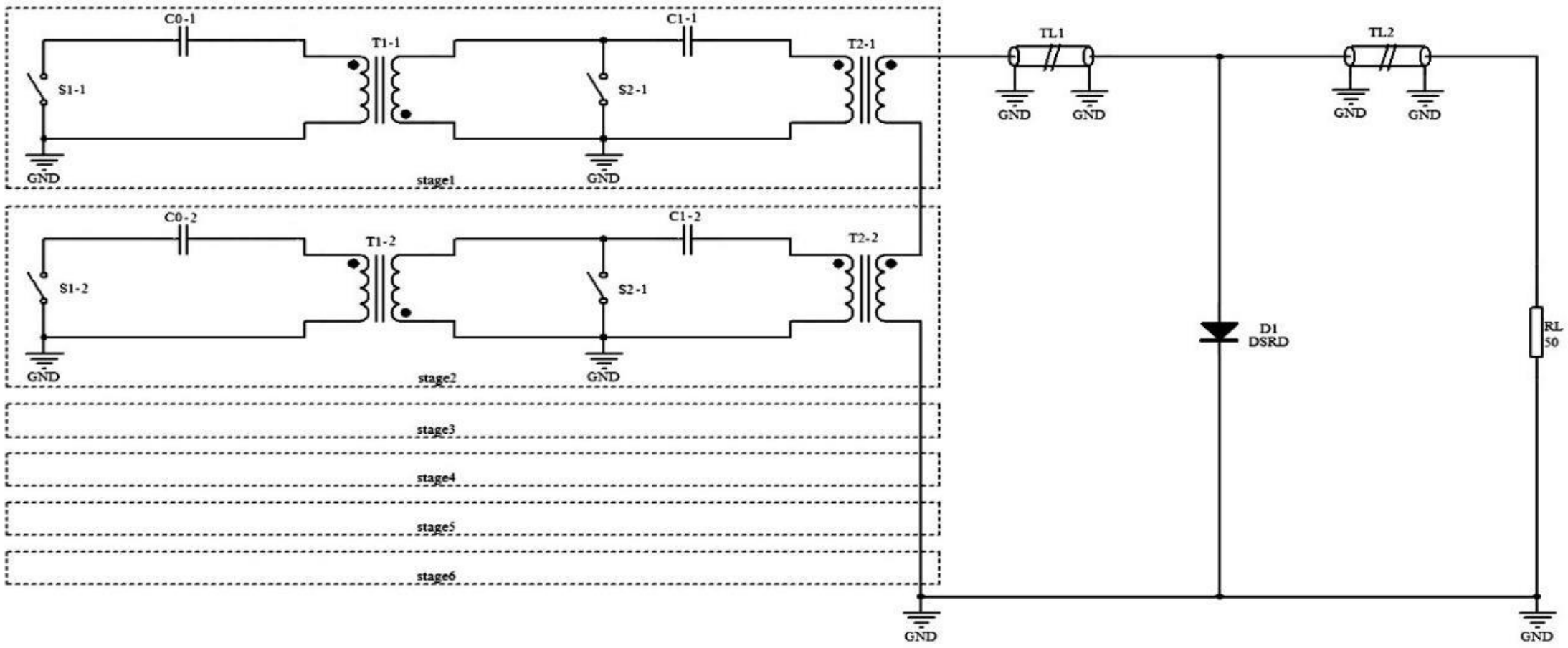

**Figure 5.3.7.21:** Topology of DSRD pulsed power supply

The structure of the DSRD pulsed power supply is depicted in Figure 5.3.7.22, which includes the DSRD pump circuit, pulse forming line (PFL), DSRD components, transmission cable, and terminal resistor. These components are designed in a perfect coaxial structure with an impedance of 50 Ω to generate a clean short electric pulse. Figure 5.3.7.23 shows the DSRD pulsed power supply prototype developed for HEPS. When the PFL length is 300 mm, it can generate a quasi-half-sine pulse waveform with bottom width (3% - 3%) ≤ 10 ns, FWHM ≈ 5 ns, $t_r$ (10-90%) < 2.6 ns, $t_f$ (90-10%) < 3.2 ns, and pulse peak V = 15kV, as shown in Figure 5.3.7.24. The HEPS prototype's performance meets the requirements of the kicker pulsed power supply of the Booster low-energy injection system.

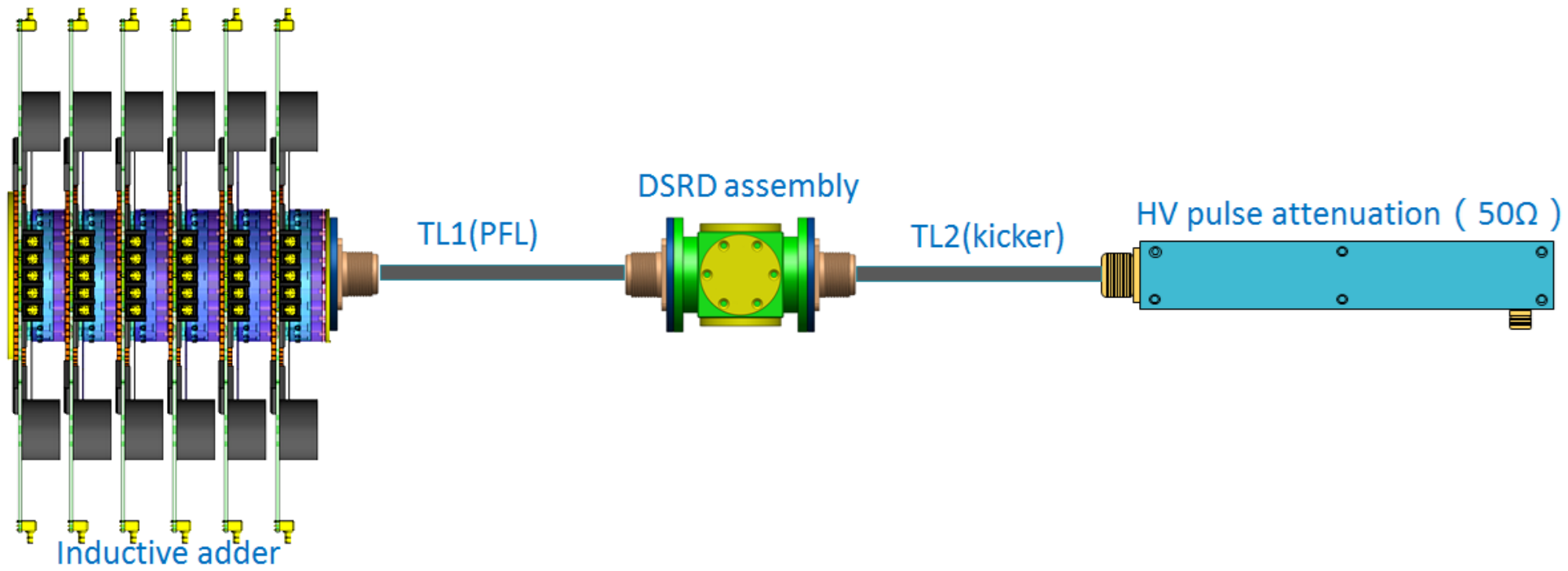

**Figure 5.3.7.22:** Structure of the DSRD pulser

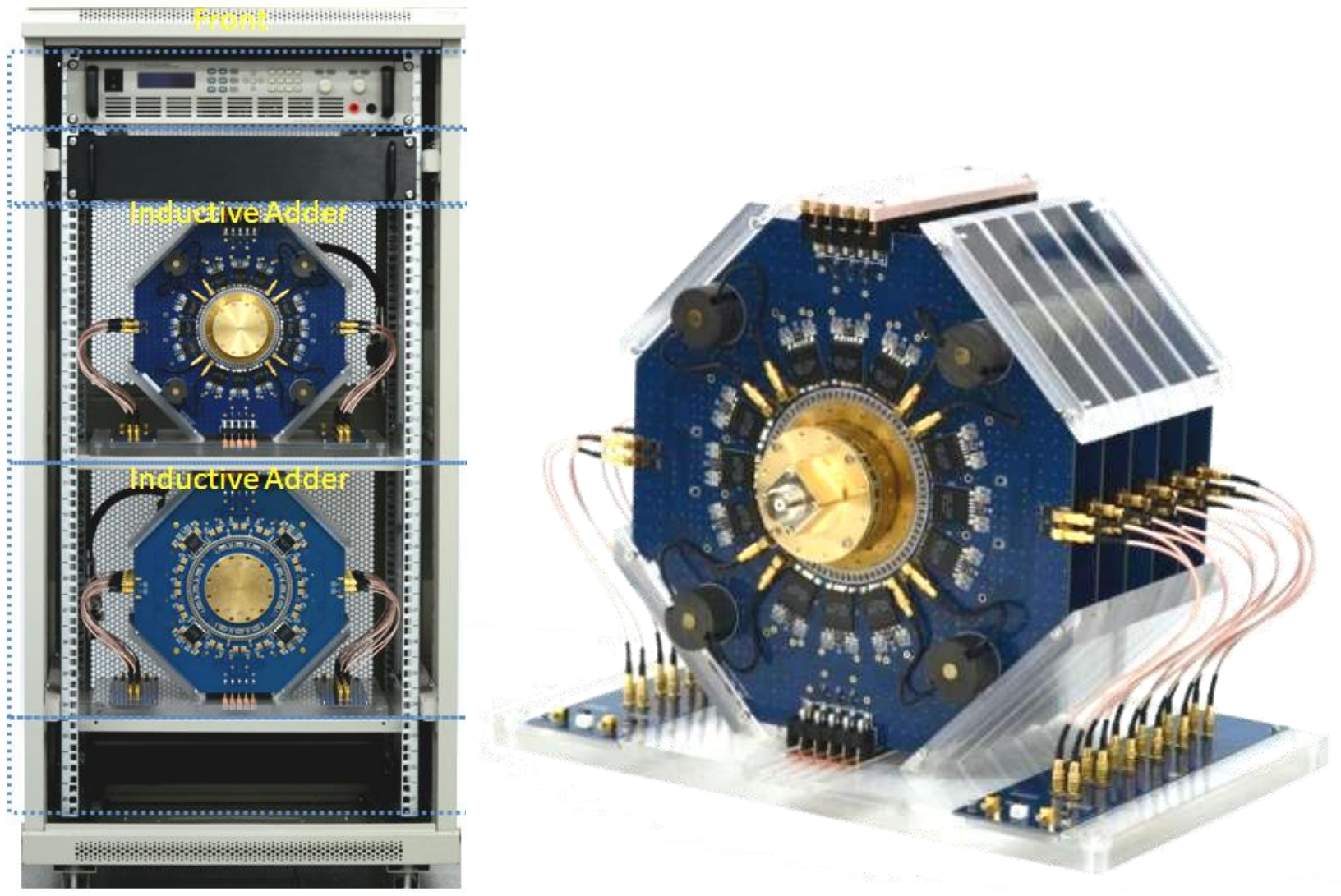

**Figure 5.3.7.23:** DSRD pulser prototype for the HEPS injection kicker system

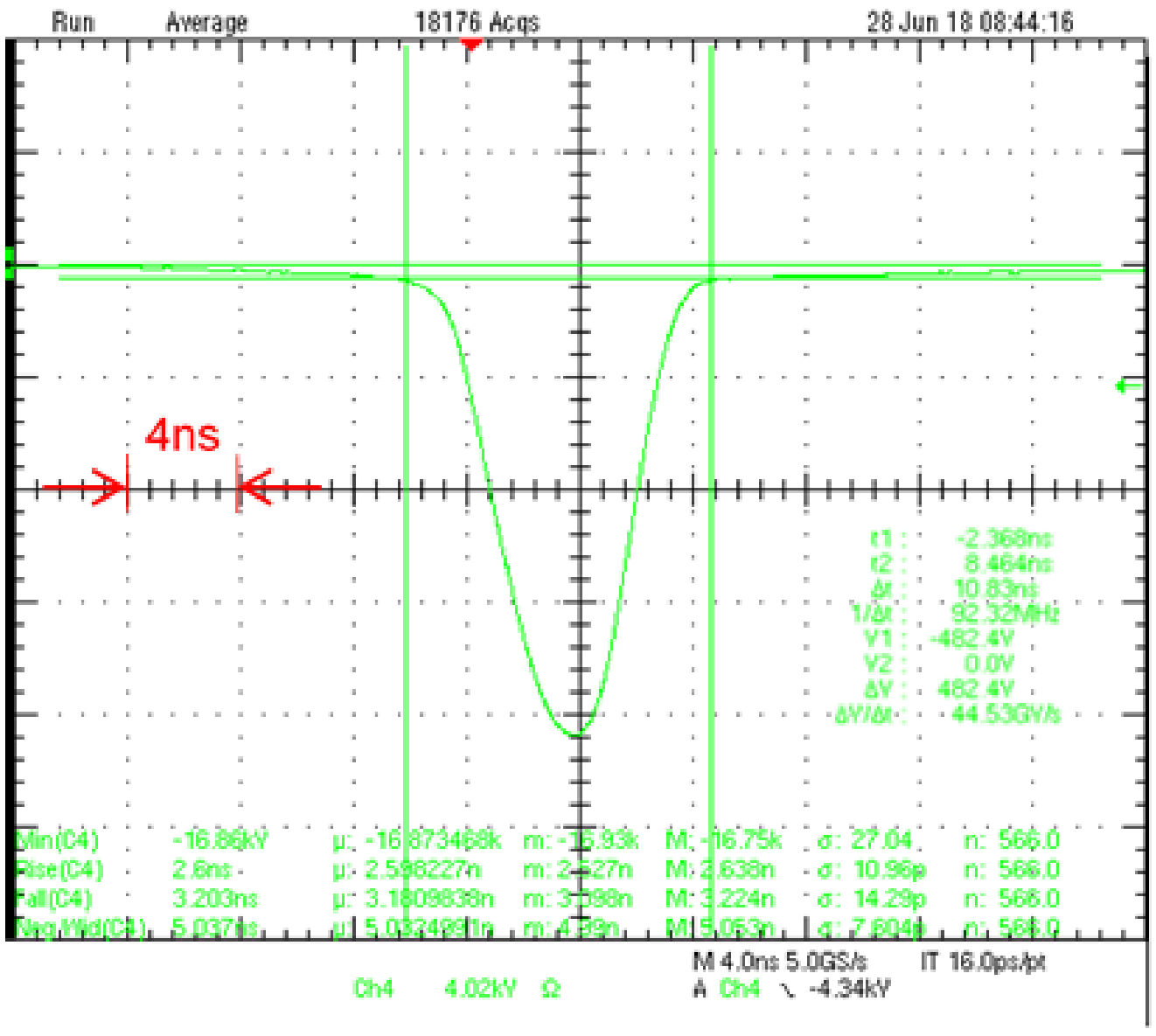

**Figure 5.3.7.24:** Waveform of the DSRD pulser prototype for the HEPS

### 5.3.7.6 *Kicker Magnets for the Booster Extraction System*

The Booster has two high-energy beam extraction systems, one for off-axis injection and the other for on-axis injection of the Collider. The kicker magnet used in the Booster extraction systems is the same as that of the Collider injection kicker, including the same aperture size in the ceramic vacuum chamber and magnet. The main design parameters of the kicker magnet for the Booster extraction systems are shown in Table 5.3.5. Refer to Chapter 4, Sections 4.3.8.4 and 4.3.8.7 for the corresponding magnet design details.

**Table 5.3.7.5:** Parameters of the kicker magnets for the Booster HE ext. and inj. system

| Parameter | Unit | BST-EXT-kicker1 （off-axis inj.） | BST-EXT-kicker2 （on-axis inj.） | BST-INJ-kicker2 （on-axis inj.） | BST- EXT -kicker2 （dump） |
|---|---|---|---|---|---|
| Quantity (e+/e-) | - | 4 + 4 ($t\bar{t}$)) | 2 + 2 ($t\bar{t}$) | 2 + 2 ($t\bar{t}$) | 4 + 4 ($t\bar{t}$)) |
| Type | - | In-air delay-line dipole kicker | In-air lumped parameter dipole kicker | In-air NLK | In-air delay-line dipole kicker |
| Deflect direction | - | Horizontal | Horizontal | Horizontal | Horizontal |
| Beam Energy | GeV | 120/180 GeV ($t\bar{t}$) | 120/180 GeV ($t\bar{t}$) | 120/180 GeV ($t\bar{t}$) | 120/180 GeV ($t\bar{t}$) |
| Deflect angle | mrad | 0.2 | 0.1 | 0.1 | 0.2 |
| Total integral magnetic strength | T·m | 0.08/ 0.12 ($t\bar{t}$) | 0.04/0.06 ($t\bar{t}$) | 0.04/0.06 ($t\bar{t}$) | 0.08/ 0.12 ($t\bar{t}$) |
| Magnetic effective length | m | 1 | 1. | 1 | 1 |
| Magnetic strength | T | 0.04 | 0.04 | 0.04 | 0.04 |
| Clearance region (H×V) | mm | 56 × 56 | 56 × 56 | 56 × 56 | 56 × 56 |
| Good field region (H×V) | mm | 50 × 50 | 50 × 50 | 50 × 50 | 50 × 50 |
| Field uniformity in good field region | - | ±1.5% | ±1.5% | ±1.5% | ±1.5% |
| Repetition rate | Hz | 1k | 1k | 1k | 1k |
| Amplitude repeatability | - | ±0.5% | ±0.5% | ±0.5% | ±0.5% |
| Pulse jitter | ns | ≤ 5 | ≤ 5 | ≤ 5 | ≤ 5 |
| Bottom width of pulse (5%-5%) | ns | Trapezoid: 440~2420 | Half-sine: 1360 | Half-sine: 1360 | Trapezoid: 440~2420 |
| Tr/Tf (5%-95%) | ns | < 200 | < 680 | < 680 | < 200 |

### 5.3.7.7 *Lambertson Magnets for the Booster*

The design parameters for the Lambertson magnet used in the Booster for low energy injection and high energy extraction are provided in Table 5.3.7.6. The key difference between the two types of Lambertson magnets is that the low energy injection Lambertson magnet needs to provide horizontal deflection, and the septum plate thickness should be 5.5 mm, whereas the high energy extraction Lambertson magnet must provide vertical deflection, and the septum plate thickness should be 6 mm. The embedded thin wall vacuum chamber design can be used for both types of magnets, as described in Section 4.3.8.9.

**Table 5.3.7.6:** Parameter of the Lambertson magnet for the CEPC Booster

| Parameters | Unit | BST-LE-LSM | BST-HE-LSM |
|---|---|---|---|
| Quantity | - | 2 × 1 | 2 × 52 |
| Deflection direction | - | Horizontal | Vertical |
| Energy | GeV | 30 | 120 |
| Total deflection angle | mrad | 45 | 43 |
| Total Integral field strength of septa | T-m | 0.92 | 17.4 |
| Deflection angle provided by a magnet | mrad | 45 | 3.5 |
| Insertion length | m | 1.2 | 1.75 |
| Magnetic field strength for injected/extracted beam | T | 0.8 | 0.8 |
| Min. septum thickness (incl. septum board, inj./ext. beam pipe wall, installation gap) | mm | 5.5 | 6 |
| Field uniformity | - | < ±0.02% | < ±0.05% |
| Leakage field | - | $\leq 1\times10^{-3}$ | $\leq 1\times10^{-3}$ |
| Clearance of stored beam at lambertson (H×V) (w.r.t. stored beam orbit) | mm | 30 × 50 | 30 × 50 |
| Clearance of inj.&ext. beam at lambertson (H×V) (w.r.t. inj. & ext. beam orbit) | mm | 18 × 29 | 30 × 30 |
| Physical aperture of stored beam vacuum chamber | mm | 56 × 56 | 56 × 56 |

## 5.3.8 Control System

### 5.3.8.1 *Introduction*

The Booster control system is responsible for controlling and monitoring all functions within the Booster accelerator. The control systems of the Collider, Booster, and Linac accelerators share many common elements, which are discussed in more detail in Chapter 4.

Global control systems, including timing systems, machine protection systems (MPS), network services, and data service systems, provide the same functionality as the control system of the Collider ring. The main stations or service centers of the global control system are typically located in the Collider ring, while the substations or service entrances are located in the Booster ring, Linac, and DR (Damping Ring) respectively.

For example, in the case of the MPS system of the HEPS, when abnormalities in the vacuum system, power supply, beam, or other components are detected, the MPS initiates the cutoff of the RF and disables beam injection from the Linac. This process is depicted in Figure 5.3.8.1 and Figure 5.3.8.2. It is worth noting that the MPS system structure of the CEPC is similar to that of HEPS.

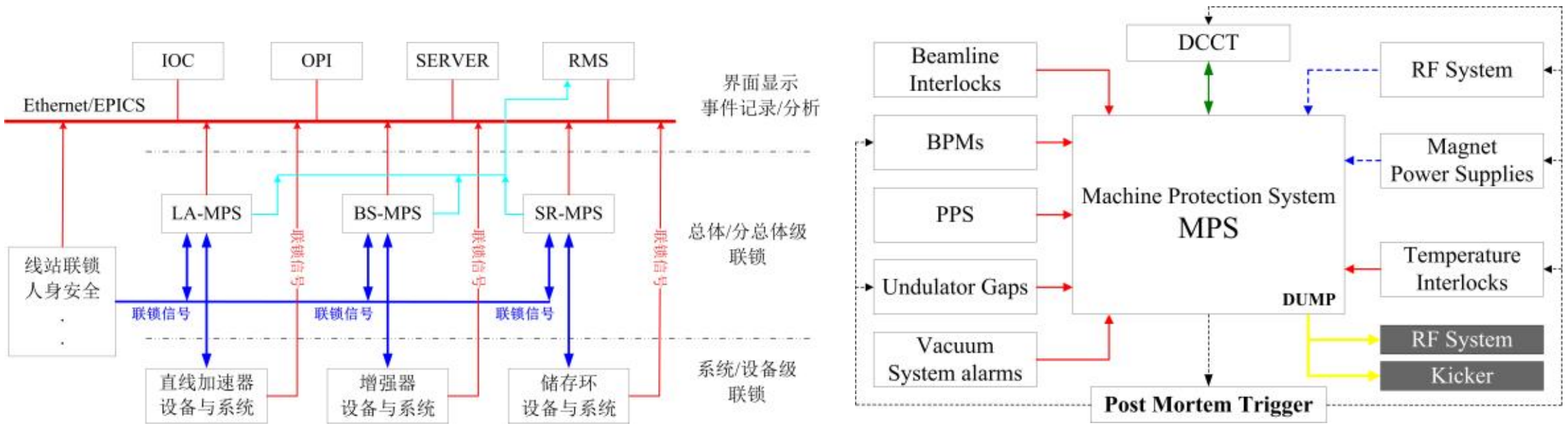

**Figure 5.3.8.1:** MPS of the HEPS.

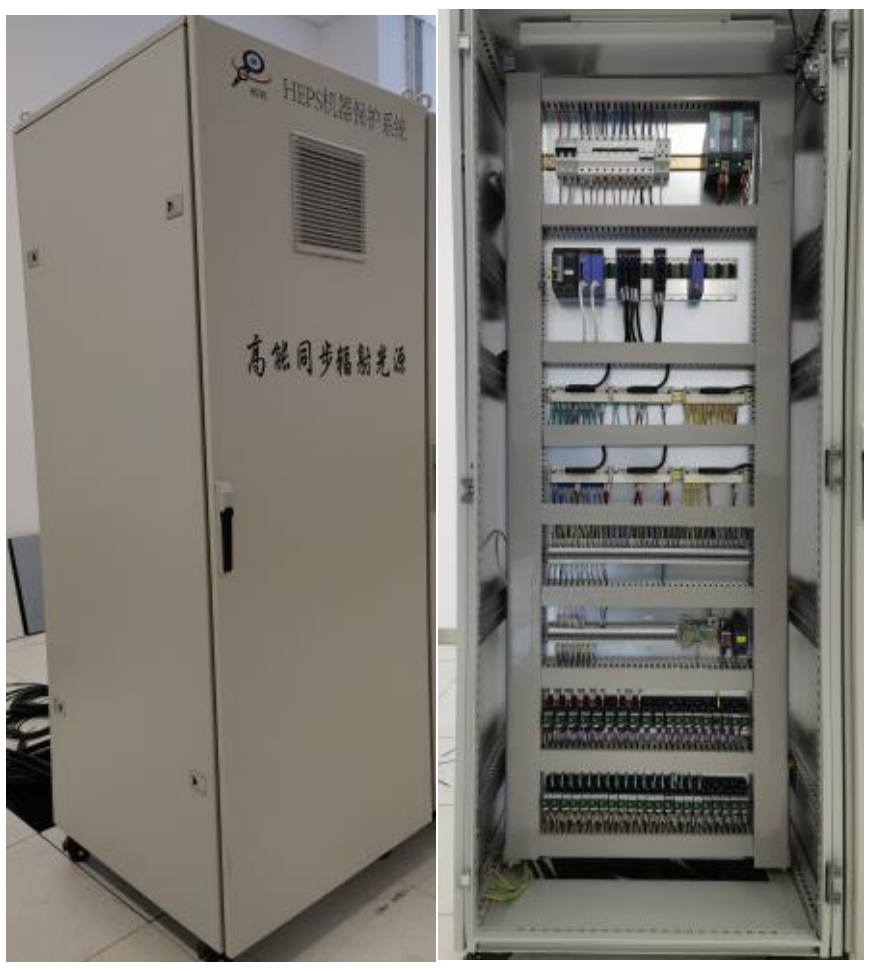

**Figure 5.3.8.2:** MPS control cabinet of the HEPS.

The Booster operation is streamlined through the control system, which offers different graphical user interfaces (GUIs) tailored to various personnel roles. These GUIs include the High-Level Application (HLA) for physicists, Operator Interface (OPI) for routine operators, and Diagnostic system for commissioning and maintenance engineers. Figure 5.3.8.3 provides an illustration of the synoptic GUI of HEPS, showcasing the user interface's visual representation and control elements.

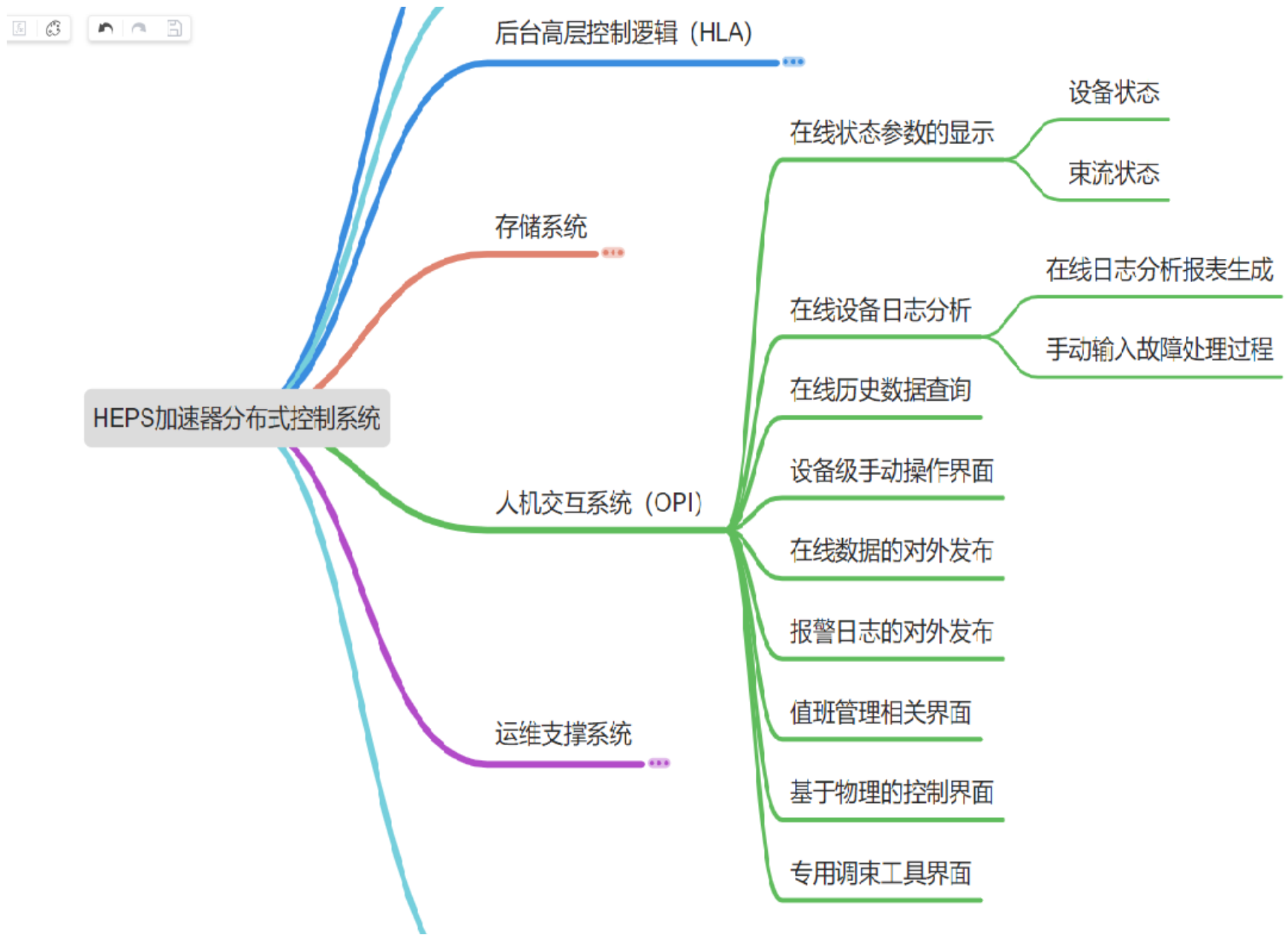

**Figure5.3.8.3:** Graphic User Interface (GUI) of the HEPS.

### 5.3.8.2 *Power Supply Control*

Magnet power supplies for the Booster Ring encompass bending magnets, quadrupole magnets, sextupole magnets, and correctors, as outlined in Table 5.3.8.1. These power supplies are distributed across multiple buildings within the facility to accommodate the diverse magnetic elements throughout the accelerator.

**Table 5.3.8.1:** Type and interface of the Booster magnet power supplies

| Type of PS | Quantitity | Location | Interface to Control System | Timing Signal |
|---|---|---|---|---|
| Dipole | 16 | Evenly distributed in 8 PS halls on the ground | Ethernet | Yes |
| Quadrupole | 32 | Evenly distributed in 8 PS halls on the ground | Ethernet | Yes |
| Sextupole | 32 | Evenly distributed in 8 PS halls on the ground | Ethernet | Yes |
| Corrector | 350 | Evenly distributed underground in 96 auxiliary tunnels | Ethernet | Yes |

The power supply control systems for the Booster and Collider share similarities. In both cases, two redundant controllers are housed within a control crate, and two isolated crate power supplies are employed to deliver power to the two control routes.

However, a notable distinction between the Booster and Collider supplies lies in the requirement for co-ramping of the Booster supplies during beam acceleration with an accuracy of tens of microseconds. To achieve this, the power supply ramping waveform can be pre-downloaded into the front-end controller. Co-ramping is then initiated with a synchronized signal from the Timing system, as depicted in Figure 5.3.8.4. This ensures precise synchronization of the power supply ramping process in the Booster accelerator.

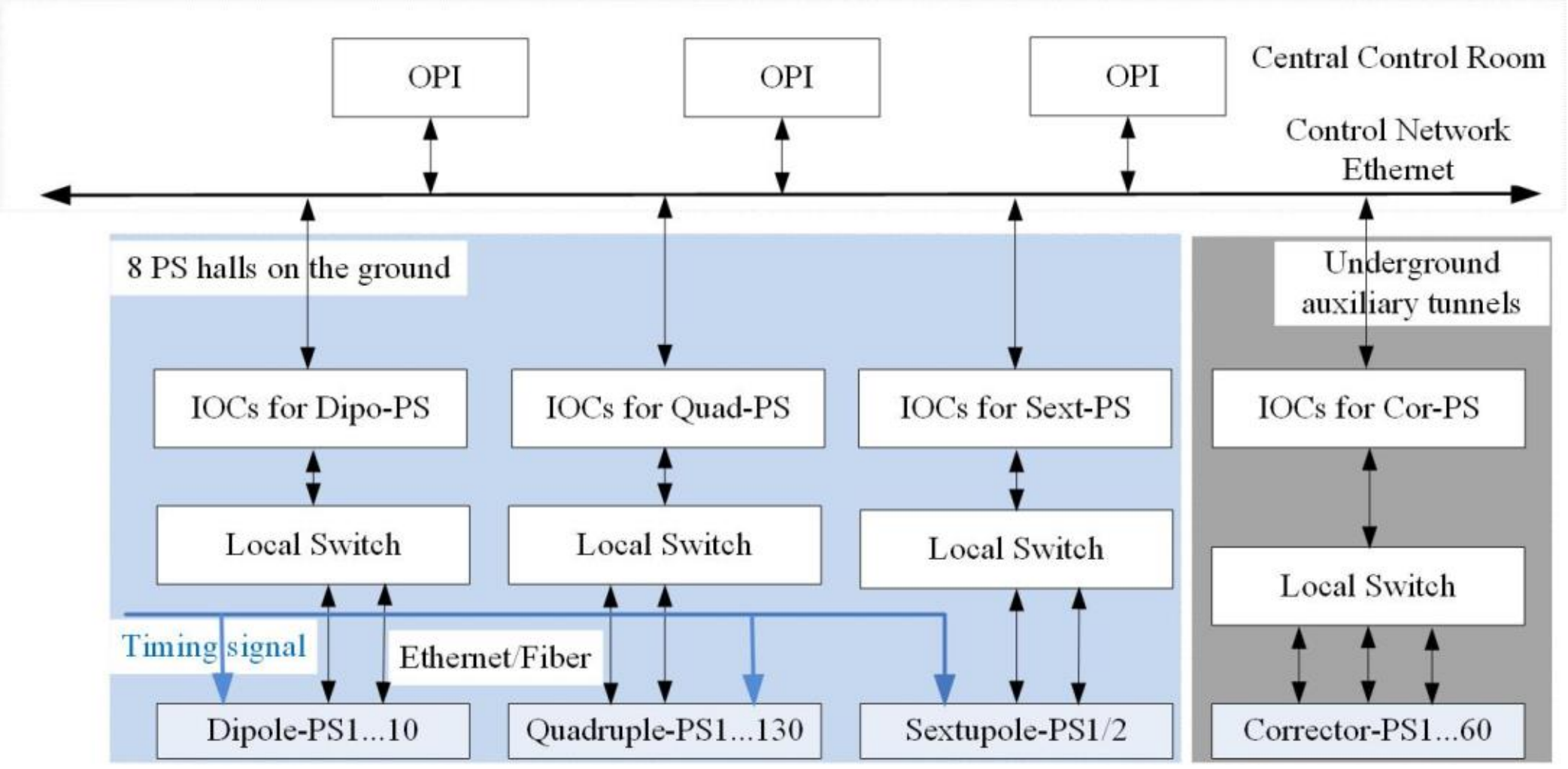

**Figure 5.3.8.4:** Control system of the Booster power supplies.

According to the different types of magnet power supplies, four types of Input/Output Controllers (IOCs) are developed: IOCs of Dipole Power Supplies, IOCs of Quadrupole Power Supplies, IOCs of Sextupole Power Supplies, and IOCs of Corrector Power Supplies. These IOCs are typically located in close proximity to the corresponding power supplies, either in the ground equipment hall or the underground auxiliary tunnel.

To ensure synchronized timing for power supply ramping, the substations of the Timing system play a crucial role by providing synchronous timing trigger signals.

The Power Supply (PS) control system incorporates a user-friendly graphical user interface (GUI) designed for operators. This GUI allows operators to conveniently monitor the current, setpoint, and status of all power supplies within the Booster Ring. Operators can independently control each power supply and also have the option to control all power supplies simultaneously using a single button. This facilitates tasks such as ramping and standardized operations as per the operational requirements. Figure 5.3.8.5 provides a glimpse of the GUI used for power supply control.

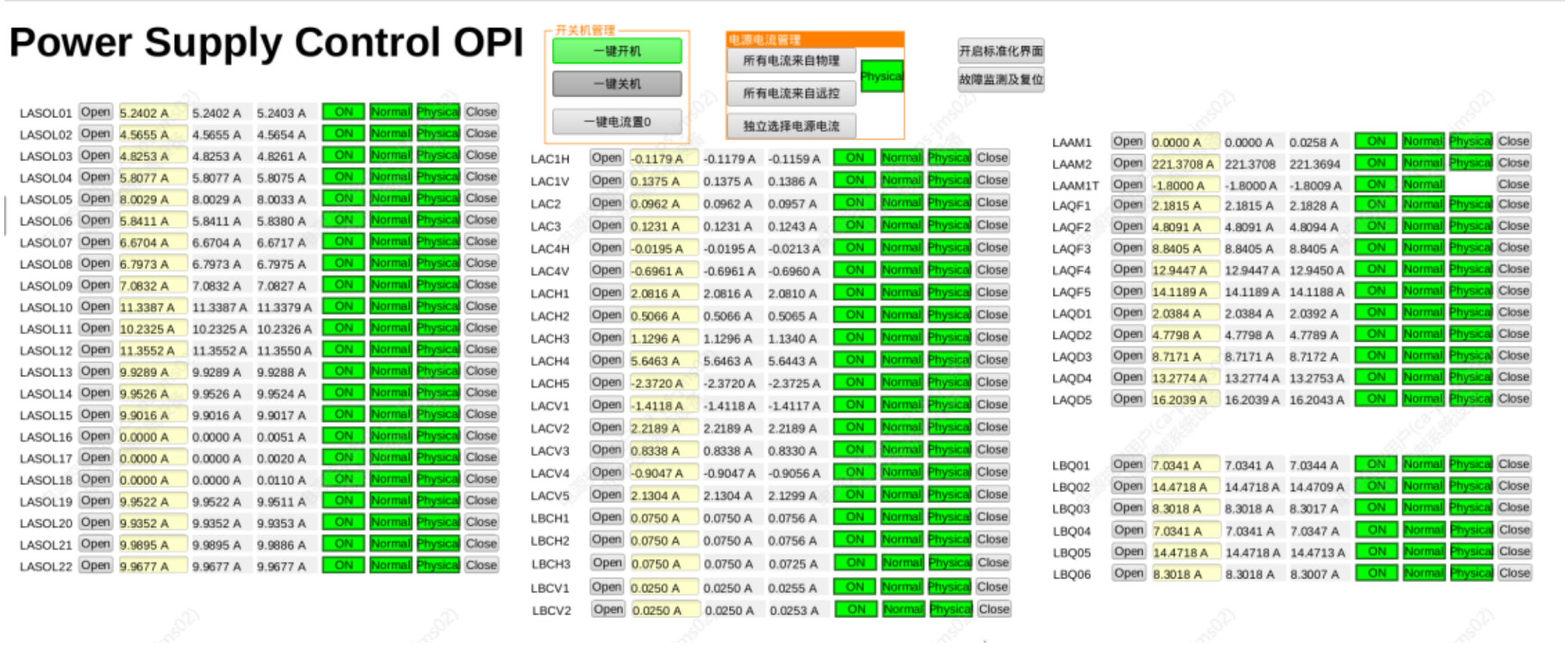

**Figure 5.3.8.5:** GUI of the magnet power supply control.

### 5.3.8.3 *Vacuum System Control*

The Booster is equipped with a total of approximately 520 vacuum valves, 8,400 pump controllers, and 2,160 gauges distributed throughout the ring.

To monitor the gauge set-point outputs and control the sector gate valves, programmable logic controllers (PLCs) will be used as the core of the vacuum protection interlock system. The PLCs will also output interlock signals to MPS, the RF system and other subsystems and receive interlock signals from other subsystems. The hardware structure diagram of the vacuum control system is shown in Figure 5.3.8.6.

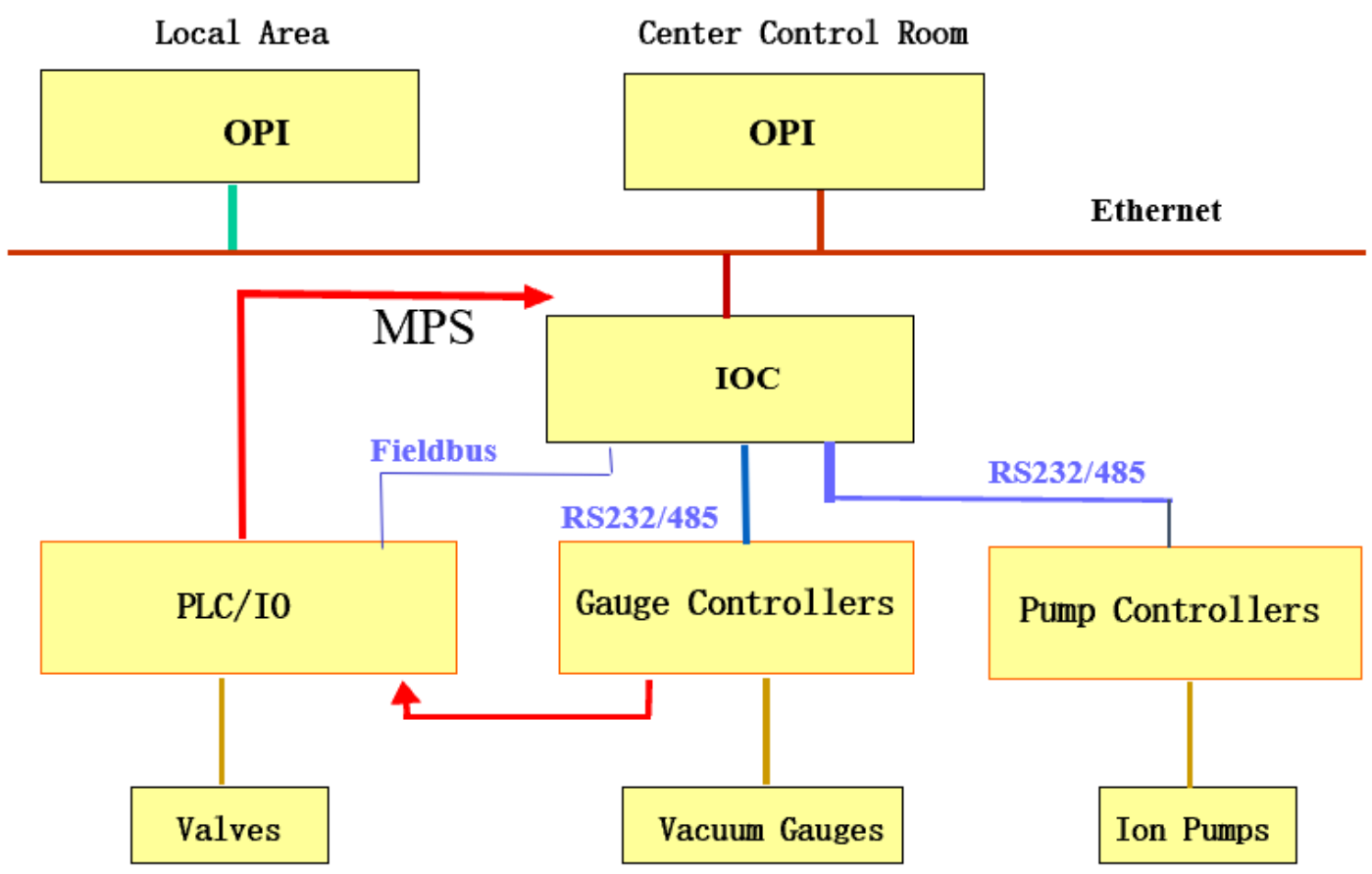

**Figure 5.3.8.6:** Hardware structure diagram of the vacuum control system.

A ladder logic program will be installed and run on the PLC processor module to control the gate valves on a fail-safe basis. A sector valve can only be opened if the adjacent vacuum conditions are satisfied. To close a sector valve, a vote-to-close algorithm will be adopted when the vacuum pressure is above the gauge set-point on both sides of the valve. Additionally, a sector valve will be closed by the PLC under certain conditions, such as power loss, controller failure, or operator input. Figure 5.3.8.7 shows the vacuum control OPI of the HEPS Booster.

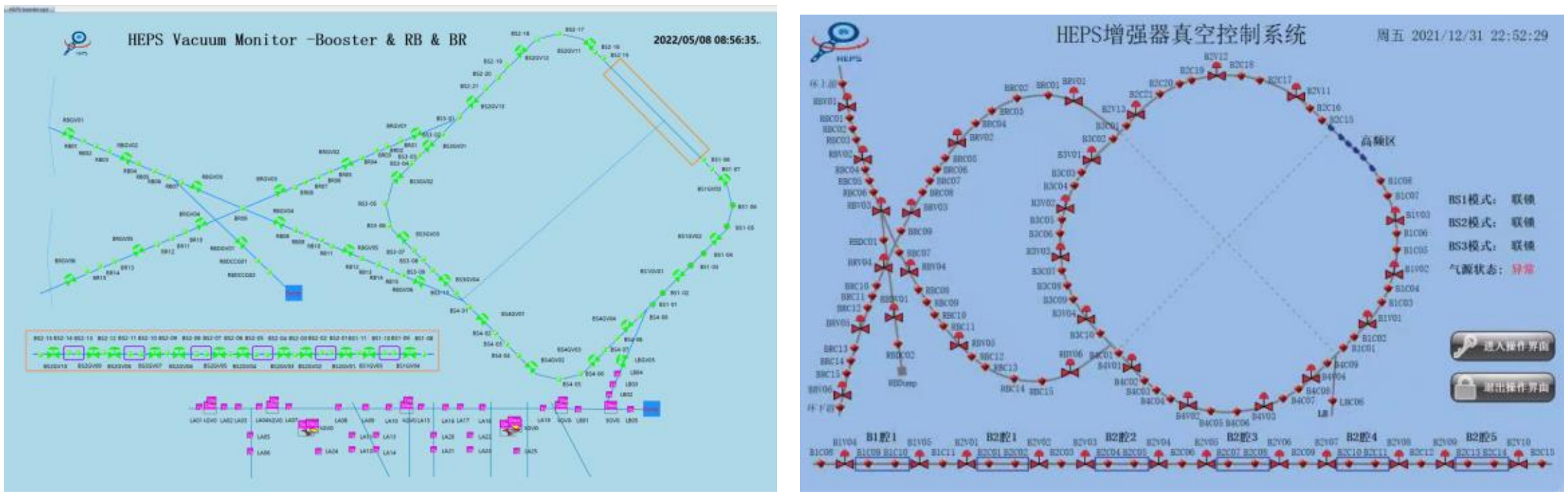

(a) Vacuum monitor OPI (b) Gate valve control OPI

**Figure 5.3.8.7:** Vacuum control OPI of the HEPS Booster.

### 5.3.8.4 *Integration of Other Systems Control*

The Booster features 96 superconducting accelerating cavities, 96 RF high-power source, and a low-level control system. The RF frequency of the Booster is 1300 MHz, and the RF system provides 2.17GV voltage for the Booster. The superconducting cavities will be installed in straight sections of the Booster.

The RF control system is comprised of two main components: a low-level control system utilizing Field-Programmable Gate Arrays (FPGAs), and a device protection system implemented using Programmable Logic Controllers (PLCs). In the event of a fault or unsafe condition, the interlock system switches off the cavity tuning for the RF high voltage power supplies. Faults may occur in the cooling water system, vacuum and temperature of a cavity, or the liquid helium in the cryogenic system. The local interlock system sends a warning message and failure signal to the MPS when a fault has been detected.

The cryogenic system provides the basic cryogenic environment for superconducting RF cavities. Additionally, the cryogenic control system will be integrated into the overall control system.

The RF control system is designed to support both local and remote-control operations. It provides operator interfaces (OPIs) that enable operators to access various functionalities, including displaying RF parameters, controlling RF equipment, accessing interlock information, and receiving alarms. Figure 5.3.8.8 illustrates the data flow within the RF control system, depicting the communication and information exchange between different components.

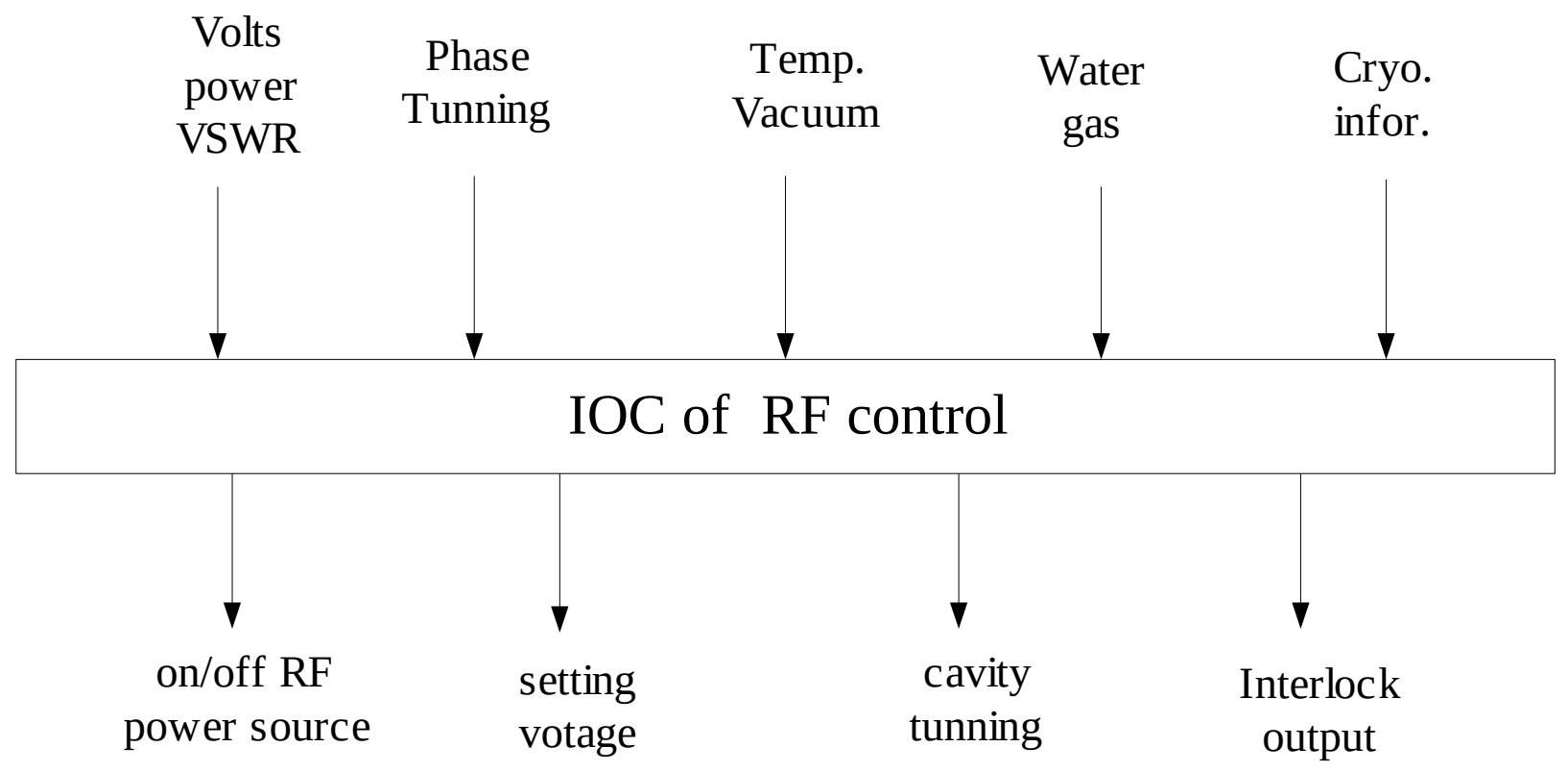

**Figure5.3.8.8:** Data flow of the RF control system.

## 5.3.9 **Mechanical Systems**

### 5.3.9.1 ***Introduction***

The Booster is located above the Collider and has a support structure similar to that of the Collider, with the exception that its pedestal is made of steel frames attached to the tunnel ceiling instead of concrete blocks on the ground. Table 5.3.9.1 lists the quantities of magnets and their supports for the Booster.

**Table 5.3.9.1:** Quantities of magnets and their supports in the Booster

| Magnet type | Quantity | Core length (mm) | No. of supports per core |
|---|---|---|---|
| Dipole | 14226 | 4700 | 4 |
| | 640 | 2350 | 2 |
| Quadrupole | 2314 | 1000/700 | 2 (common frame) |
| | 1144 | 2000 | 2 |
| Corrector | 1200 | 583 | 1 |

The Booster consists of a single ring, and each magnet is individually suspended and supported. The design requirements for the supports in the Booster are similar to those of the Collider. The support design for the Booster must meet the following requirements:

- Range and accuracy of adjustment must be the same as that of the supports in the Collider, which are shown in Table 4.3.10.2.
- Stability to avoid creep and fatigue deformation.
- Simple and reliable mechanics for safe mounting and easy alignment.
- Good vibration performance.
- Good waterproof performance of the tunnel, particularly at the mounting locations.

### 5.3.9.2 ***Topological Optimization of Magnet Support in Booster***

Topology optimization is utilized to design the supporting points in the plane perpendicular to the beam. If there are two supporting points on the tunnel wall and the origin point is the tunnel center, the angle of each point can vary from 0° to 90°. The topology optimization process employs the volume as the constraint and the structure compliance as the objective function, with lower structure compliance indicating higher structure stiffness.

This method is applicable to all types of magnets, and if the magnet or tunnel cross-section changes, the frame structure can be easily redesigned. For instance, the dipole magnets can be optimized using this method. Figure 5.3.9.1 shows the contour of the structure compliance, while Figure 5.3.9.2 illustrates the mesh density distribution with the minimum structure compliance. The blue elements represent low mesh density, signifying their insignificance and potential for elimination in the design, whereas the red elements indicate high mesh density. The two main frames are nearly vertical and are designed with horizontal ribs to achieve a simple structure and a better appearance.

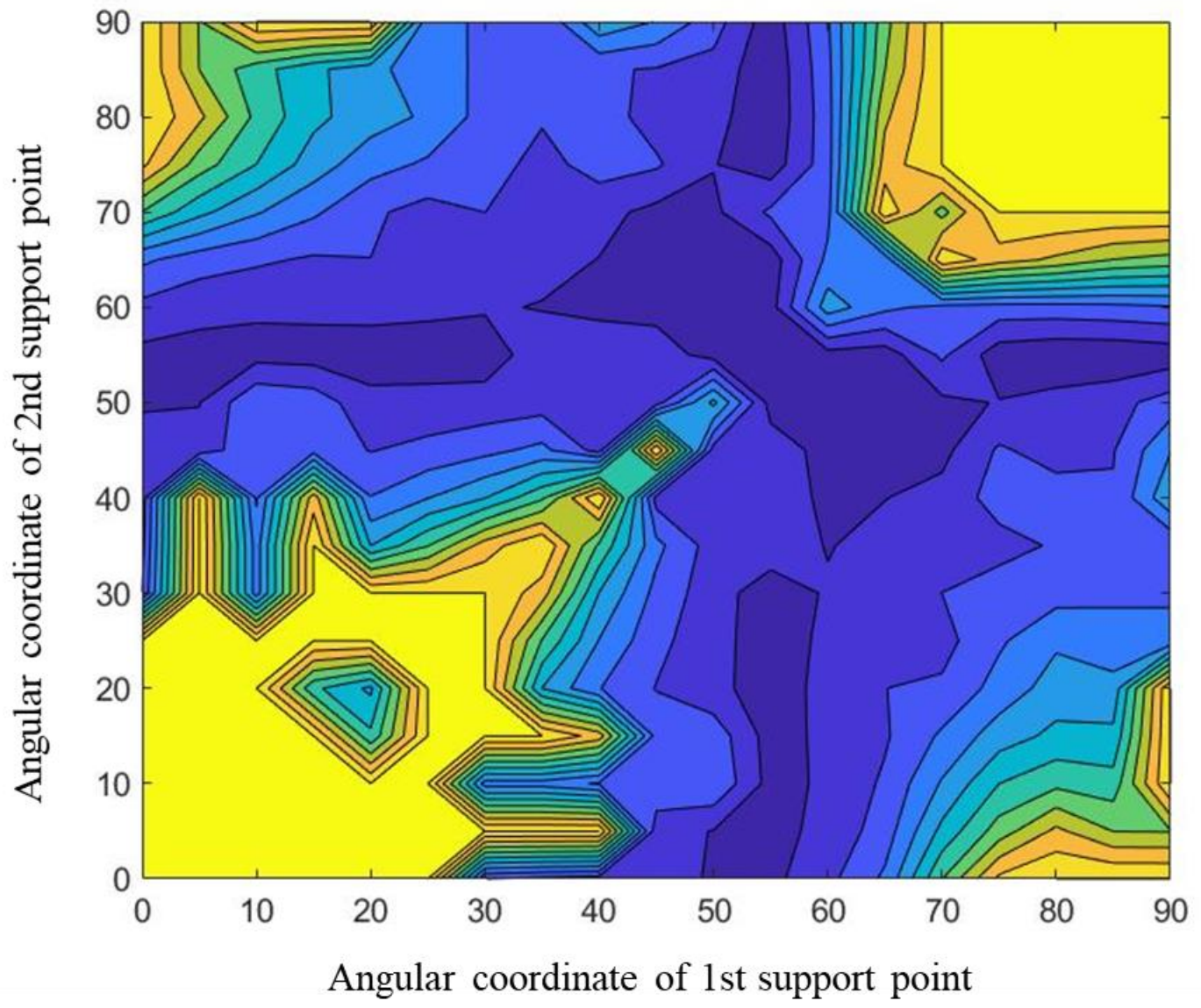

**Figure 5.3.9.1:** Contour of Structure supple degree.

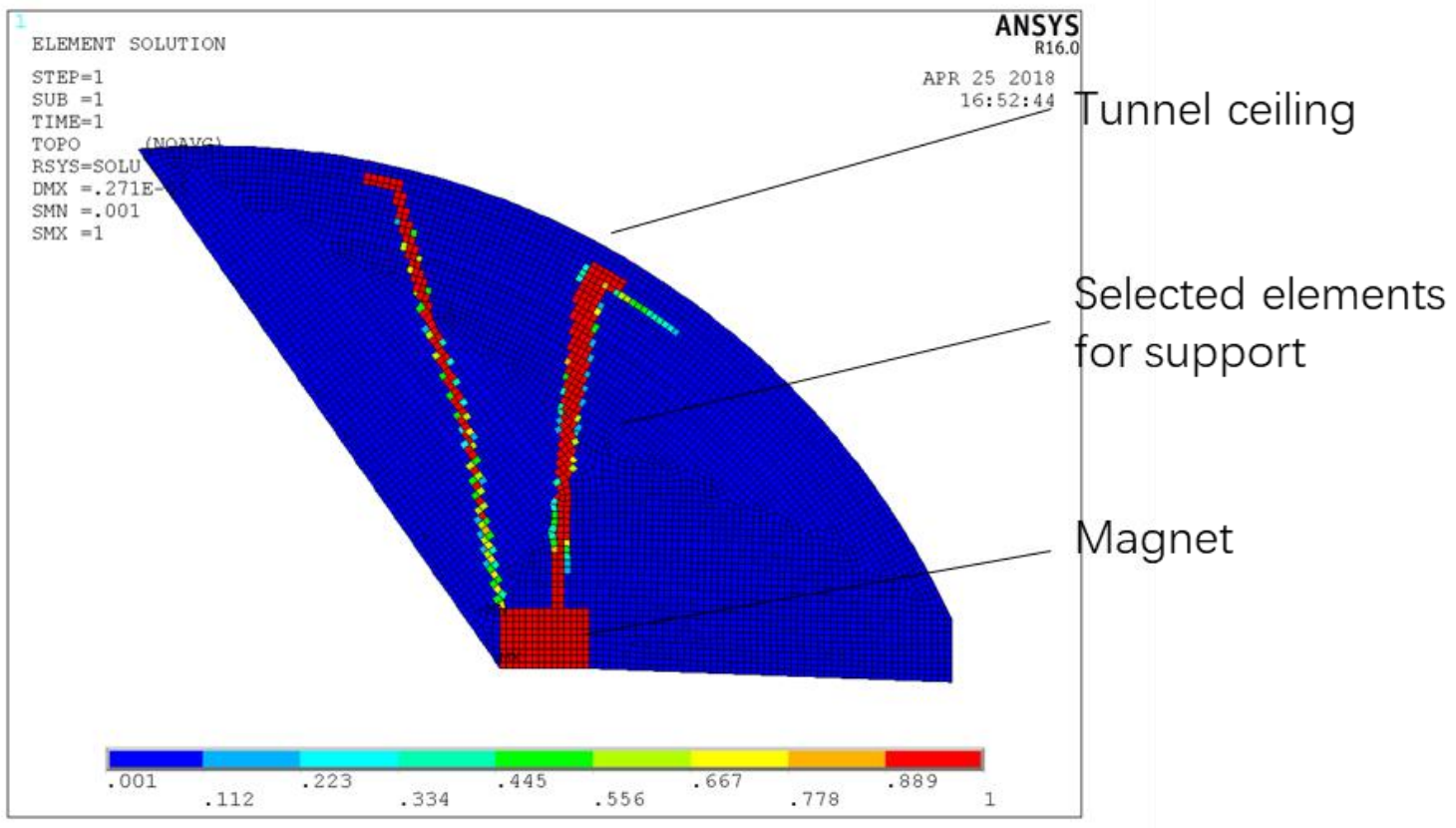

**Figure 5.3.9.2:** Mesh density in the plane perpendicular to the beam.

### 5.3.9.3 *Structure Design of Magnet Supports*

The support location design in the Booster follows a similar approach to that of the Collider. Long dipoles in the Booster are supported by four supports, as shown in Figure 5.3.9.3. The two middle supports are specifically used for Y-axis adjustment and are referred to as Y supports. One of the end supports allows for adjustment in both the Y and X directions, known as the XY support. The remaining support can be adjusted in all three directions (X, Y, and Z) and is referred to as the XYZ support. Various adjustment mechanisms are utilized, including big screws for vertical adjustment and push-pull bolts for horizontal adjustment.

The steel frame pedestals are installed on pre-embedded plates on the tunnel ceiling, providing a sturdy mounting point for the support system.

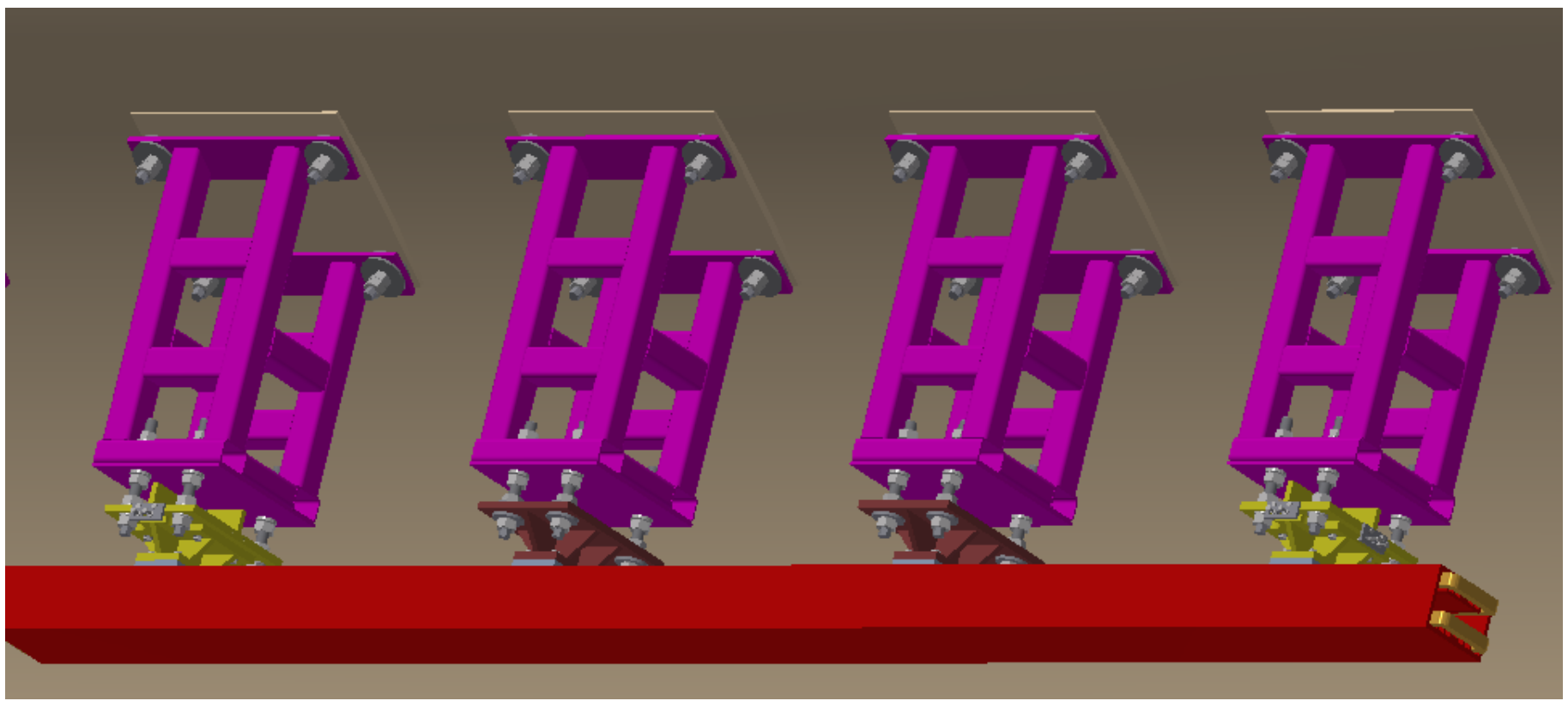

**Figure 5.3.9.3:** Supports of 4,700 long dipole magnets in the Booster.

After optimizing the support locations, the maximum uneven gravitational deformation of the 4,700 long dipole magnets in the Booster is reduced to 4 μm. The first natural frequency of the dipole magnet and its support system is approximately 24.2 Hz when the mounting holes of the pedestal are fixed. This mode corresponds to translation in the X direction, as depicted in Figure 5.3.9.4.

Table 5.3.9.2 provides information on the first six modes and their respective natural frequencies. It is important to note that the weak joints are present at the support screws between the pedestal and the adjusting mechanism, similar to the design in the Collider. Additionally, the relatively tall steel frame pedestals contribute to the weak joints. One approach to improving the natural frequency (dynamic stability) is to shorten the steel frame, which can enhance the overall stability of the system.

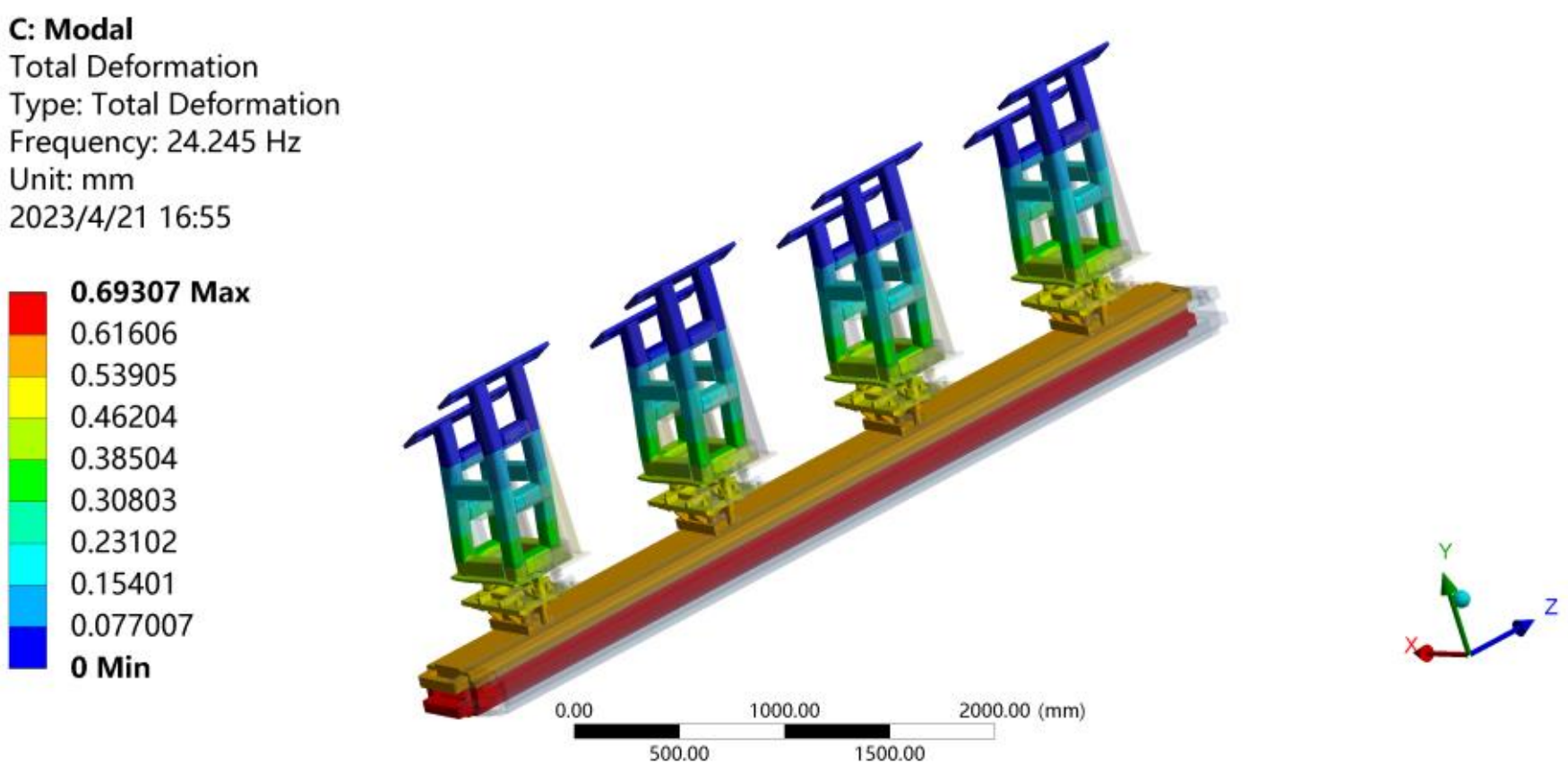

**Figure 5.3.9.4:** The 1st oscillation mode of the Booster dipole magnet and its support.

**Table 5.3.9.2:** The first six oscillation modes for the Booster dipole magnet and its support.

| No. of order | Frequency (Hz) | Mode |
|---|---|---|
| 1 | 24.2 | X translation |
| 2 | 26.6 | Yaw rotation |
| 3 | 28.9 | Z translation |
| 4 | 79.0 | Bend in XZ plane |
| 5 | 117.4 | Roll rotation |
| 6 | 122.2 | Pitch rotation |

Other magnets in the Booster, such as the 1000 mm quadrupole magnets, typically have one or two supports that allow for both horizontal and vertical adjustments. Figure 5.3.9.5 illustrates the support structure for a 1000 mm quadrupole magnet.

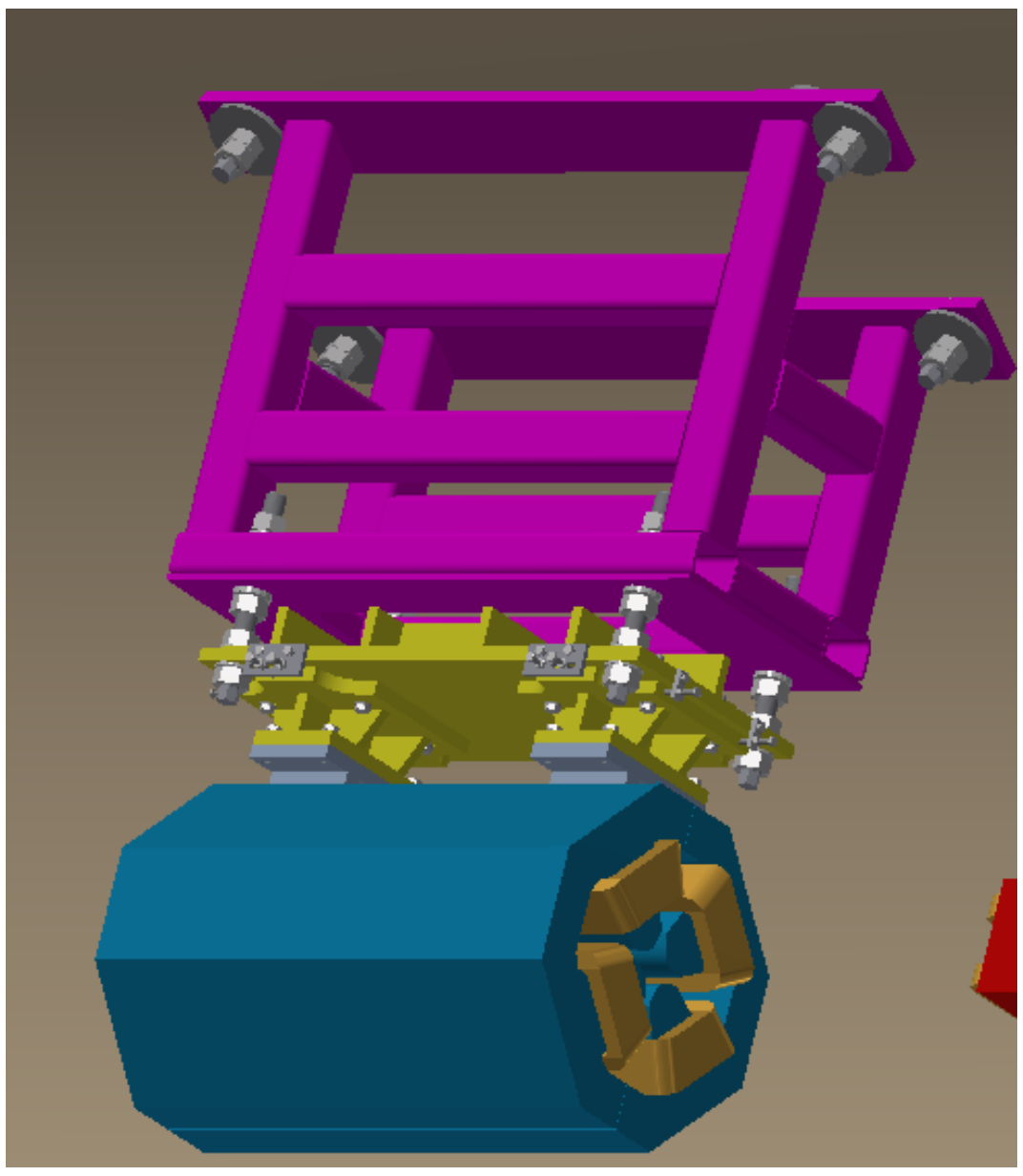

**Figure 5.3.9.5:** Supports of the 1000 mm long quadrupole magnet in the Booster.

Further investigation will be conducted to assess the sensitivity of the jointing stiffness between the pedestal and ceiling. The specifications for the rock ceiling's strength will be determined once the site characteristics are ascertained.

# 6 Linac, Damping Ring and Sources

## 6.1 Main Parameters

The design of the CEPC Linac must take into account three primary considerations: providing electron and positron beams that meet the requirements, high availability for high integral luminosity, and potential for upgrades to meet future requirements.

The design of the CEPC Linac has undergone multiple iterations from the CDR to the TDR stage, resulting in a decrease in beam emittance and an increase in beam energy [1-2]. Due to its large circumference of 100 km and the low magnetic field in injection energy, designing dipole magnets and power supplies for the Booster has been challenging. Increasing the Linac energy was identified as the easiest and most effective solution, resulting in an increase from 10 GeV to 20 GeV, where the Booster dipole magnetic field in injection energy is approximately 60 Gs.

To further reduce the effect of residual magnetism in low injection energy, oriented silicon steel sheet was used for the Booster dipole magnet, despite its high cost. However, if the Linac energy is increased to 30 GeV, non-oriented silicon steel sheet can be used instead. Given the large number of Booster magnets, it was determined in June 2022 that the Linac energy should be 30 GeV to save costs.

To meet the high luminosity scheme requirements, the Linac's emittance was reduced to 10 nm and then further to 6.5 nm. To achieve this, a 1.1 GeV damping ring for positron beam was proposed, and a repetition frequency of 100 Hz was chosen to meet injection speed requirements. According to the injection scheme design, the switching time between the electron accelerator and the positron accelerator must be under 3.5 seconds. To provide a margin of safety, the designated switching time for the linear accelerator is set at 3.0 seconds. The parameters to be switched include the bias voltage and pulser voltage of the electron gun, the accelerating phase of the accelerating unit, the settings of the corrector power supply, and the settings of the dipole magnet power supply. Figure 6.1.1 illustrates the schematic diagram for the dipole magnet power supply switch. Approximately 10 RF pulses are administered before extracting the beam to ensure a smooth switch of the LLRF phase.

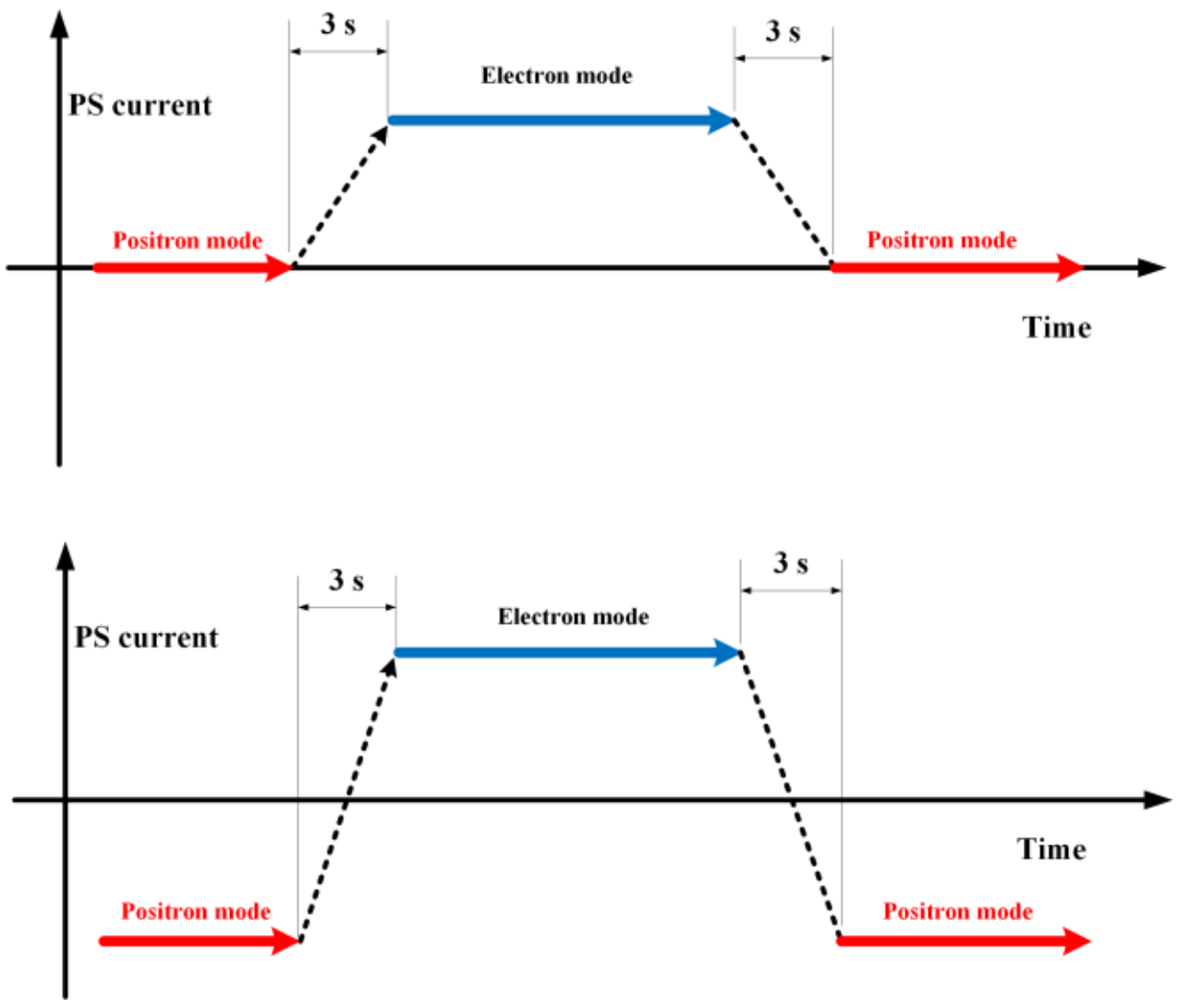

**Figure 6.1.1:** Schematic Diagram of the Positron-Electron Acceleration Mode Switch. The upper figure depicts the operating mode of the dipole magnet power supply in the electron bypass transport line, while the lower figure illustrates the operating mode of the dipole magnet power supply in the bunch length compression section.

The Linac was also designed to accelerate both electron and positron beams in a double-bunch acceleration mode. The specified bunch charge is 1.5 nC, which serves as the baseline value. To ensure readiness for future enhancements and to meet demanding requirements, the maximum allowable bunch charge was set at 3.0 nC. Throughout this document, our primary focus is on the physics design pertaining to a bunch charge of 1.5 nC, while also presenting dynamic results for 3.0 nC. Furthermore, to enhance availability, the design incorporates approximately 15% redundancy in the acceleration units. Refer to Table 6.1.1 for the Linac's parameter details.

**Table 6.1.1:** Main parameters of the Linac.

| Parameter | Symbol | Unit | Design value |
|---|---|---|---|
| Energy | $E$ | GeV | 30 |
| Repetition rate | $f_{rep}$ | Hz | 100 |
| Number of bunches per pulse | | | 1 or 2 |
| Bunch charge | | nC | 1.5 |
| Energy spread | $\sigma_E$ | | $1.5\times10^{-3}$ |
| Emittance | $\varepsilon_r$ | nm | 6.5 |
| Electron energy at target | | GeV | 4 |
| Electron bunch charge at target | | nC | 3.5 |
| Tunnel length | $L$ | m | 1800 |

The CEPC Linac utilizes both S-band and C-band accelerating structures, as indicated in Table 6.1.2. The short-range wakefield of these structures was determined using the wakefield model of periodic structure developed by R. Gluckstern [3], as well as K. Yokoya and K.L.F. Bane [4-5]. Both transverse and longitudinal wakefield effects were considered in all simulations, and the resulting wakefield data is presented in Fig. 6.1.2.

**Table 6.1.2:** Main parameters of the accelerating structures

| Parameter | Unit | S-band | | C-band |
|---|---|---|---|---|
| Frequency | MHz | 2860 | | 5720 |
| Length | m | 3.1 | 2.0 | 1.8 |
| Cavity mode | | 2π/3 | | 3π/4 |
| Aperture | mm | 19~26 | 25 | 12~16 |
| Gradient | MV/m | 22/27 | 22 | 40 |
| Cells (include coupler cells) | | 86 | 55 | 89 |
| Number of Acc. Stru. | | 93 | 16 | 470 |
| Operating temperature | °C | 30 | | 30 |
| Number of Klystron | | 34 | | 236 |
| Klystron Power | MW | 80 | | 50 |

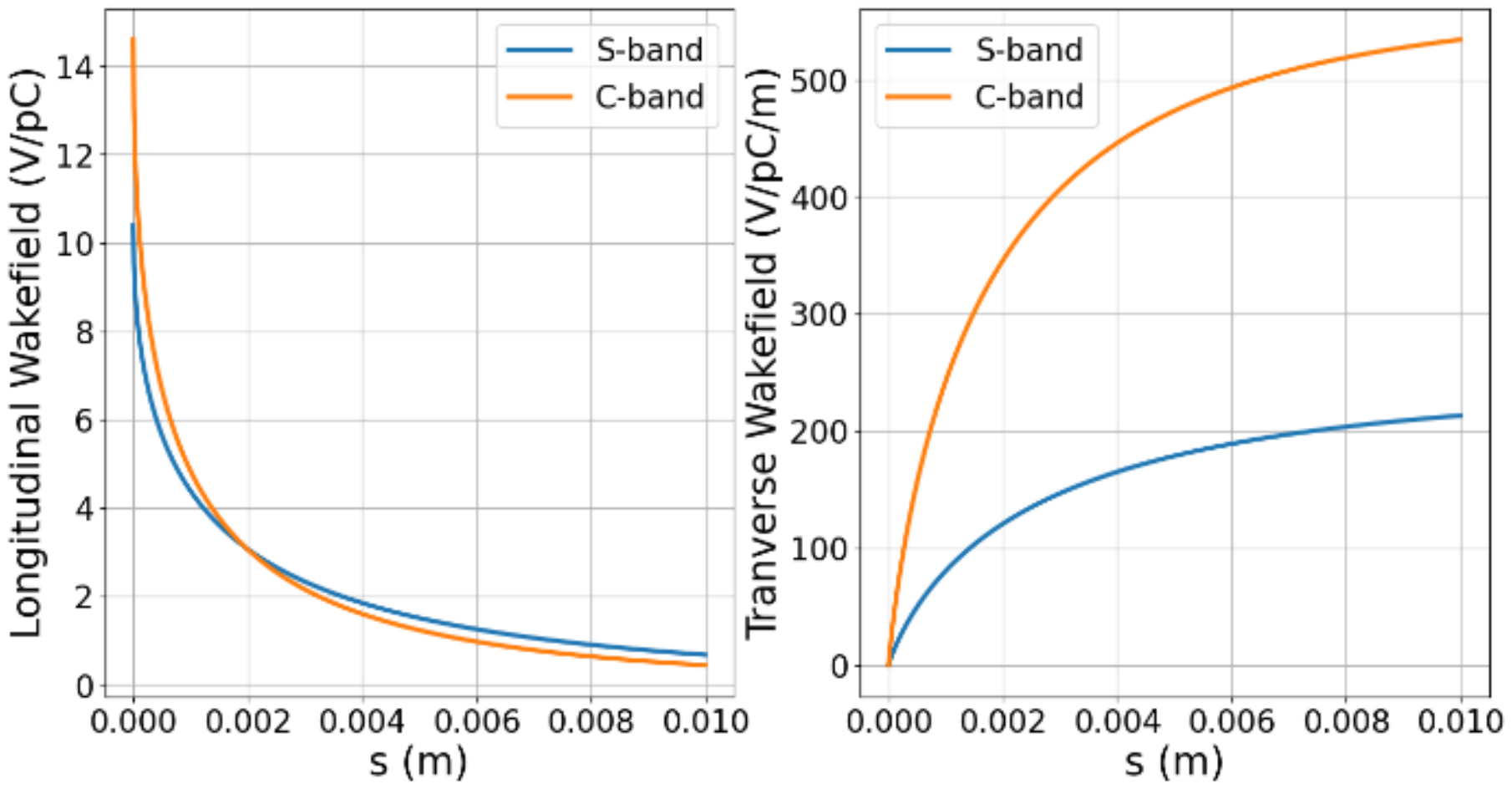

**Figure 6.1.2:** Short-range wakefields of the accelerating structures.

# 6.2 Linac and Damping Ring Accelerator Physics

## 6.2.1 Linac Design – Optics and Beam Dynamics

### 6.2.1.1 *Linac Layout*

The CEPC Linac is a type of linear accelerator that uses normal conducting RF technology and operates at two different frequencies, S-band (2860 MHz) and C-band (5720 MHz). This Linac can produce both electron and positron beams, with a maximum energy of 30 GeV.

The Linac consists of several sections, including the Electron Source and Bunching System (ESBS), which produces and prepares the electron beam for acceleration, and the First Accelerating Section (FAS), where the electron beam is accelerated to 4 GeV, which will be used for positron production.

Additionally, there is a Positron Source and Pre-Accelerating Section (PSPAS) where positron beams are generated and accelerated to 200 MeV, followed by a Second Accelerating Section (SAS) where the positron beam is accelerated to 1.1 GeV.

The Third Accelerating Section (TAS) is where both electron and positron beams are accelerated from 1.1 GeV to their maximum energy of 30 GeV. The Linac also includes an Electron Bypass Transport Line (EBTL) where the electron beam is bypassed in electron mode, and a Damping Ring (DR) where the positron beam is damped to reduce its emittance.

A diagram of the Linac layout can be found in Figure 6.2.1.1.

To prevent interference with other components such as the waveguide, positron source, and transport lines between the Linac and Damping Ring, the deflection direction of the EBTL is set to be vertical with a separation distance of 1.2 meters.

For the FAS, a bunch charge of about 3.5 nC is used for positron production, and an S-band accelerating structure is employed to reduce the wakefield effect. To enable a potential upgrade, the FAS's physical design allows for an electron beam with a bunch charge of up to 7 nC, facilitating the attainment of a positron bunch charge of 3.0 nC.

In the SAS, the emittance of the positron beam is very large, so an S-band accelerating structure with large and nominal apertures is used.

In the TAS, a C-band accelerating structure with a high accelerating gradient is utilized to reduce the Linac's size and cost. The total length of the Linac is 1.6 km with an additional 200 meters as a reserved space, resulting in a total Linac tunnel length of 1.8 km.

To measure the beam energy and energy spread, there are seven energy analyzing stations located throughout the Linac. Each station includes a profile monitor and beam dump to measure the beam properties.

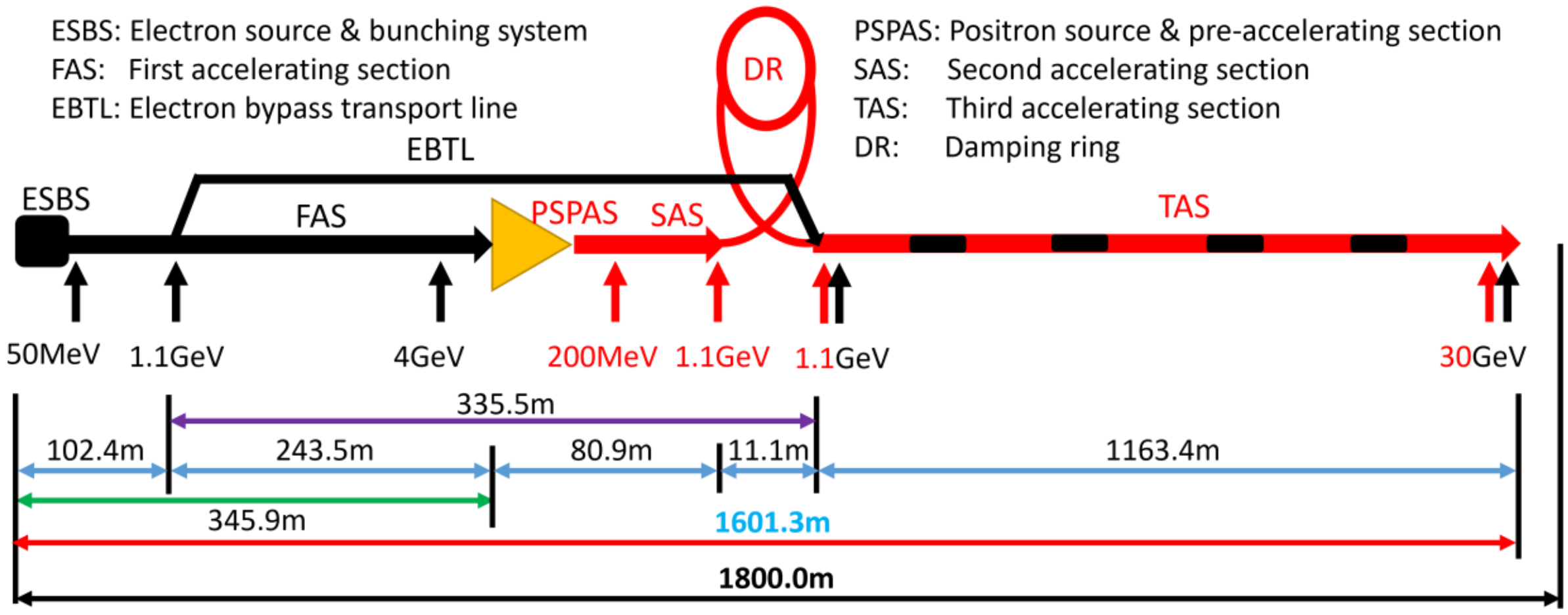

**Figure 6.2.1.1:** The Linac layout.

### 6.2.1.2 *Bunching System*

The Linac is designed to produce electron and positron beams with a bunch charge of 1.5 nC at the exit as the baseline design and 3.0 nC as an upgrade potential design. During positron production mode, the electron beam's bunch charge of the electron gun is increased to 4.5 nC or 10.0 nC. A thermal cathode electron gun is used in the baseline scheme.

The Linac's bunching system consists of two sub-harmonic bunchers (SHBs), an S-band buncher, and a normal S-band accelerating structure, as shown in Fig. 6.2.1.2. The RF frequency of the sub-harmonic bunchers is chosen as 158.89 MHz and 476.67 MHz to meet the requirement of double-bunch acceleration mode and the bunch pattern of the Collider ring. The simulation results with a bunch charge of 10.0 nC, as shown in Fig. 6.2.1.3, indicate that the normalized rms emittance at the bunching system exit is 80 mm-mrad, and the transmission efficiency of the bunching system is greater than 90%. However, based on the Linac beam commissioning results for the High Energy Photon Source (HEPS), the transmission efficiency was approximately 80%, which satisfies the positron production requirements.

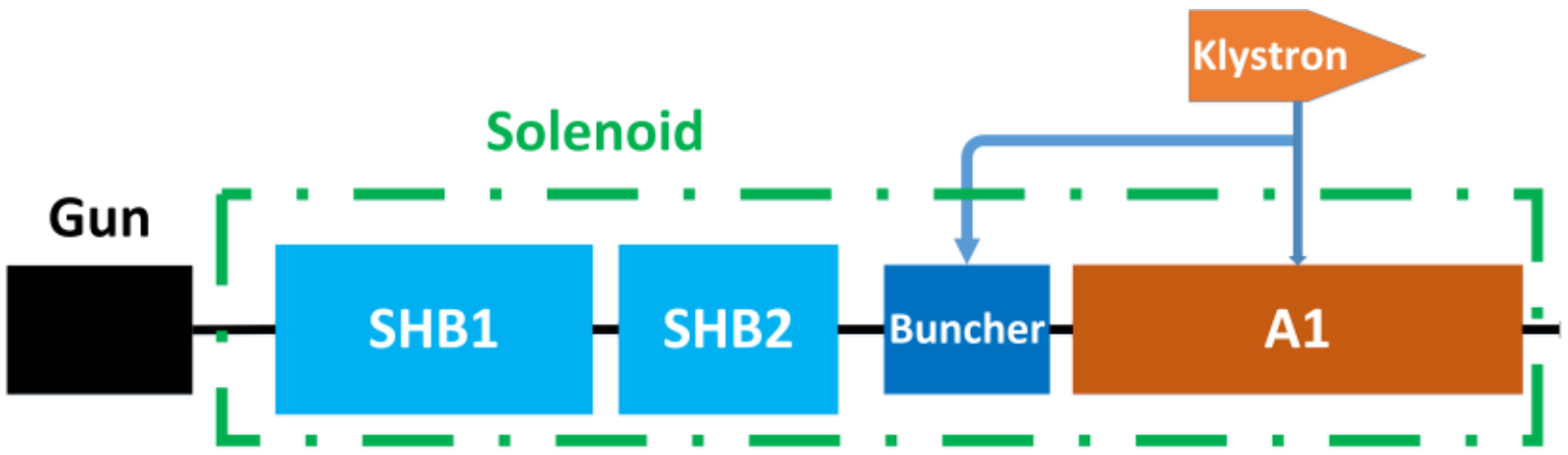

**Figure 6.2.1.2:** Layout of the bunching system.

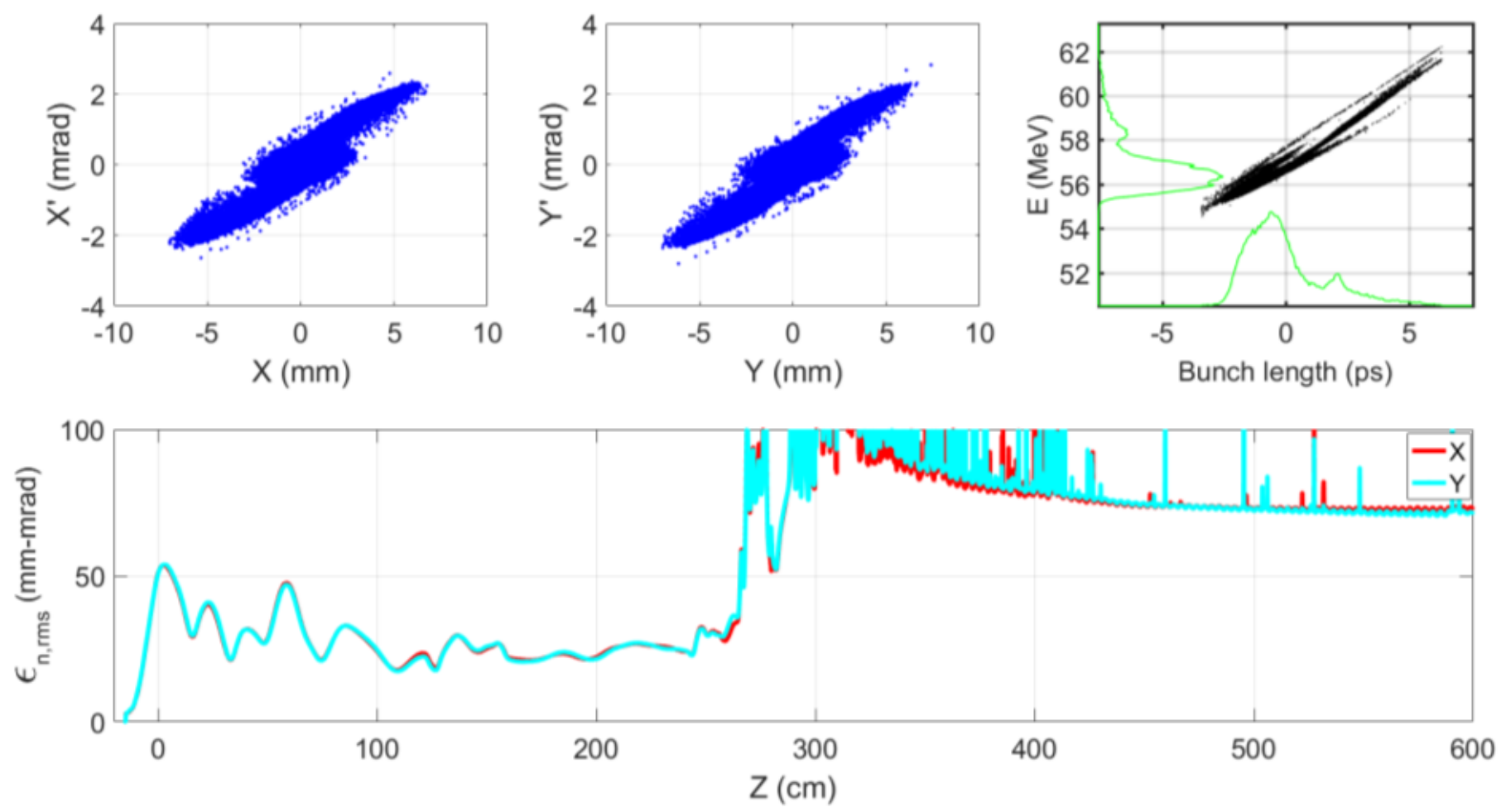

**Figure 6.2.1.3:** Beam distribution at the exit of the bunching system (top) and normalized rms emittance (bottom) along the beam direction.

### 6.2.1.3 *Electron Linac*

The electron Linac is comprised of several key components, including the ESBS, a portion of the FAS with an energy of 1.1 GeV, the EBTL, and the TAS. Within the first part of the FAS, there are six S-band klystrons, each with a power output of 80 MW, and one klystron is reserved as a backup. Each klystron drives four accelerating structures, and the accelerating gradient is set at 22 MV/m. In the TAS, there are 236 C-band klystrons, each with a power output of 50 MW. Of these, 21 klystrons serve as backups, which is about 9% of the capacity, and one klystron is reserved for the deflecting cavity, which is used to measure the bunch length. Each klystron drives two accelerating structures, and the accelerating gradient is set at 40 MV/m, which corresponds to a low RF breakdown rate of $5\times10^{-7}$ /pulse/m.

The EBTL within the Linac has a local achromatic design and must be carefully matched to optimize performance. To simplify the system design, no multipole magnets are employed for higher-order parameter matching. The higher-order dispersion T366 and U3666 are large, so careful control of the energy spread is necessary to limit emittance growth to within 5%. To this end, the energy spread must be kept smaller than 0.4%, as shown in Fig. 6.2.1.4. In our simulation, the energy spread of the beam entering the EBTL section is 0.2%. To ensure that the switching speed of electrons and positrons is met, it is necessary for the rise and fall of the current values of the dipole magnet power supply in EBTL to be completed within 3.0 seconds.

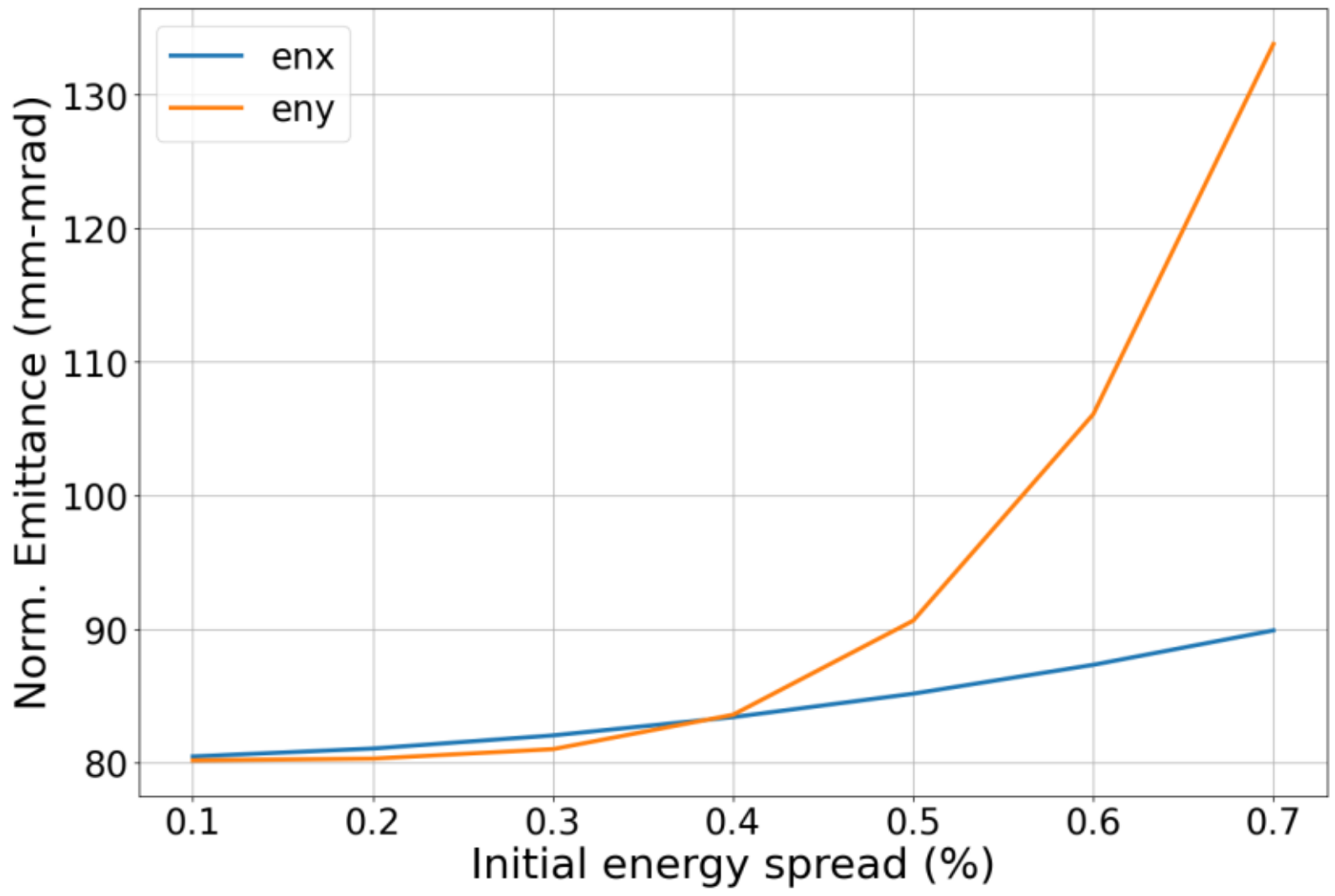

**Figure 6.2.1.4:** Emittance with different initial energy spread of the EBTL.

For a given RF frequency and accelerating gradient, the energy spread in the accelerator is mainly influenced by the Wakefield effect, bunch length, and accelerating phase. To determine the appropriate bunch length, we conducted a parameter scan and analyzed the energy spreads with varying bunch charge and length, as illustrated in Fig. 6.2.1.5. Based on the phase scan simulation results, we selected a bunch length of 0.4 mm. To achieve this desired value, we incorporated a chicane-type bunch compressor at the beginning of the TAS, with the optics function shown in Fig. 6.2.1.6. This includes four dipoles with a bending angle of 10 degrees.

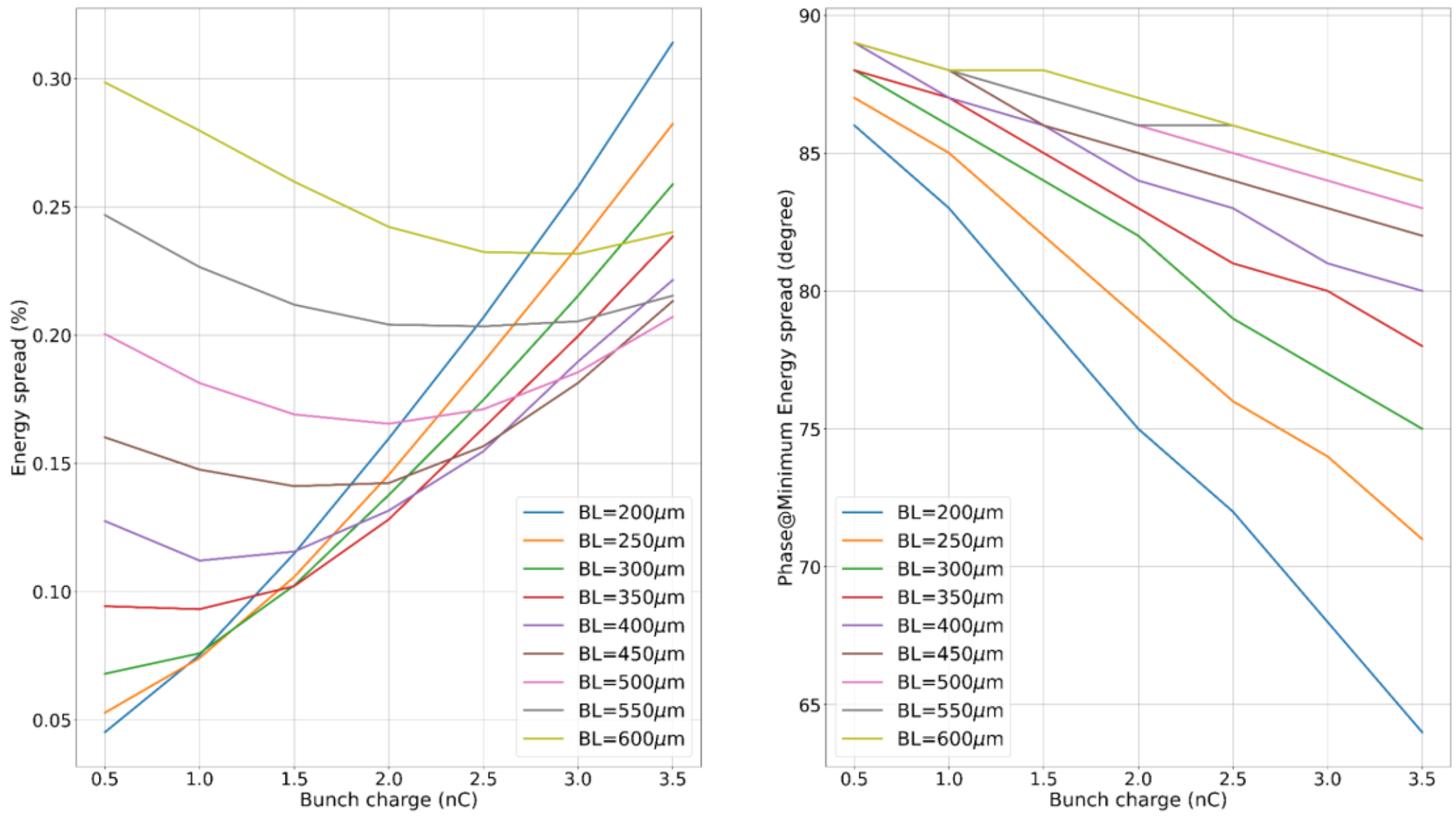

**Figure 6.2.1.5:** Energy spread with phase scan.

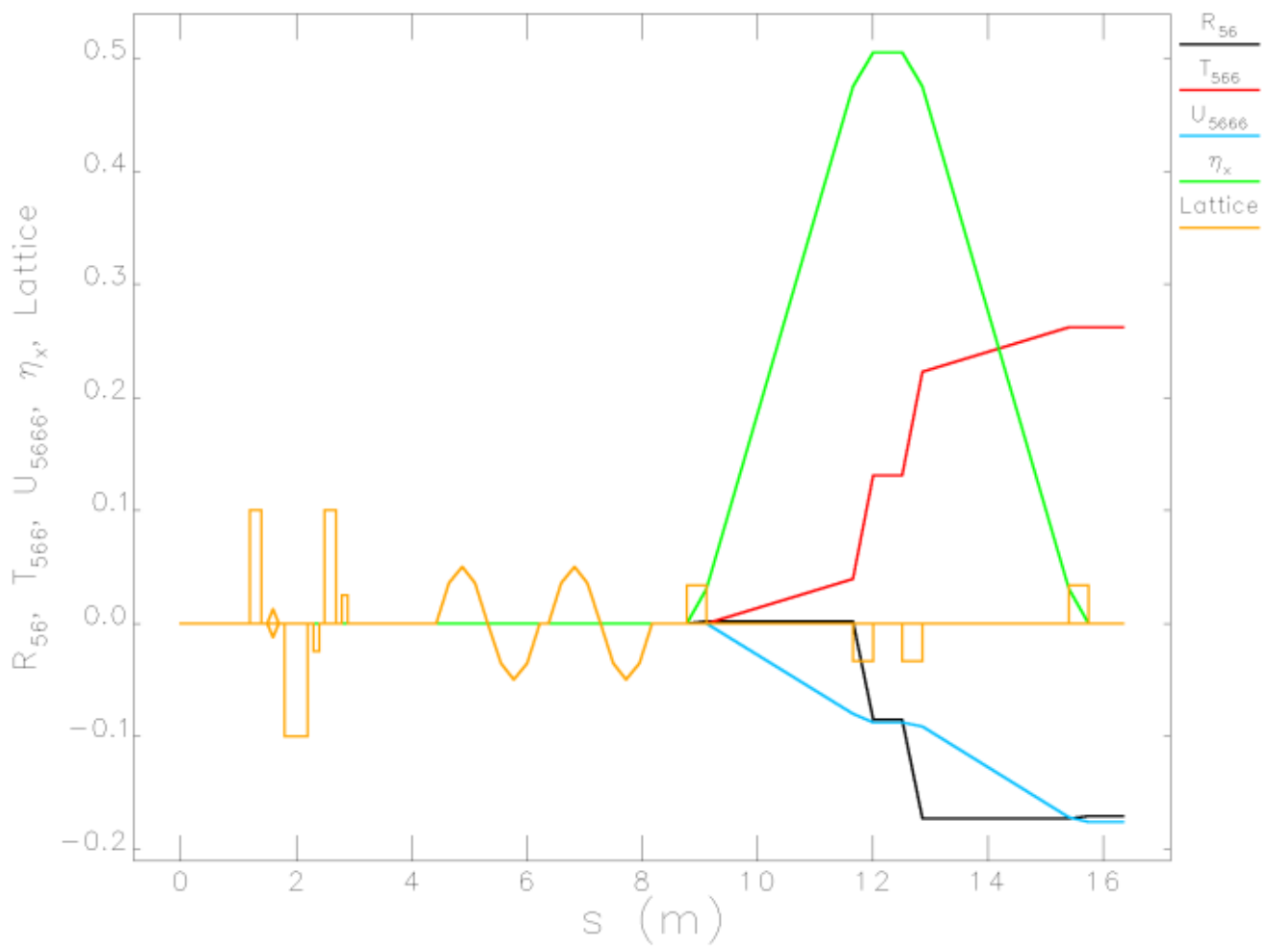

**Figure 6.2.1.6:** Optics function of the bunch compressor. R56 is the momentum compaction factor, T566 is the second-order momentum compaction factor, U5666 is the third-order momentum compaction factor, $\eta_x$ is the horizontal dispersion.

As per the CLIC study [1], the FODO lattice outperforms both the doublet lattice and triplet lattice regarding jitter amplification and emittance growth in the presence of static imperfections. Within the triplet structure, improved symmetry and consistency are achieved for the beam in both horizontal and vertical directions. In the low-energy section, designing the periodic structure is simpler. In the high-energy section (TAS), there is no requirement to adjust the quadrupole magnet settings when switching between positron and electron beams.

For the lattice design, we have chosen a triplet as the primary focusing structure. As the beam energy increases, the period length also increases. In order to reduce the strength requirements for the quadrupole magnet in the high-energy part of the TAS, the designed period phase advance is gradually decreased. The results of the beam dynamics analysis for the electron Linac, including the short-range wakefield and CSR effects, are shown in Fig. 6.2.1.7 and summarized in Table 6.2.1.1. These results meet the required specifications.

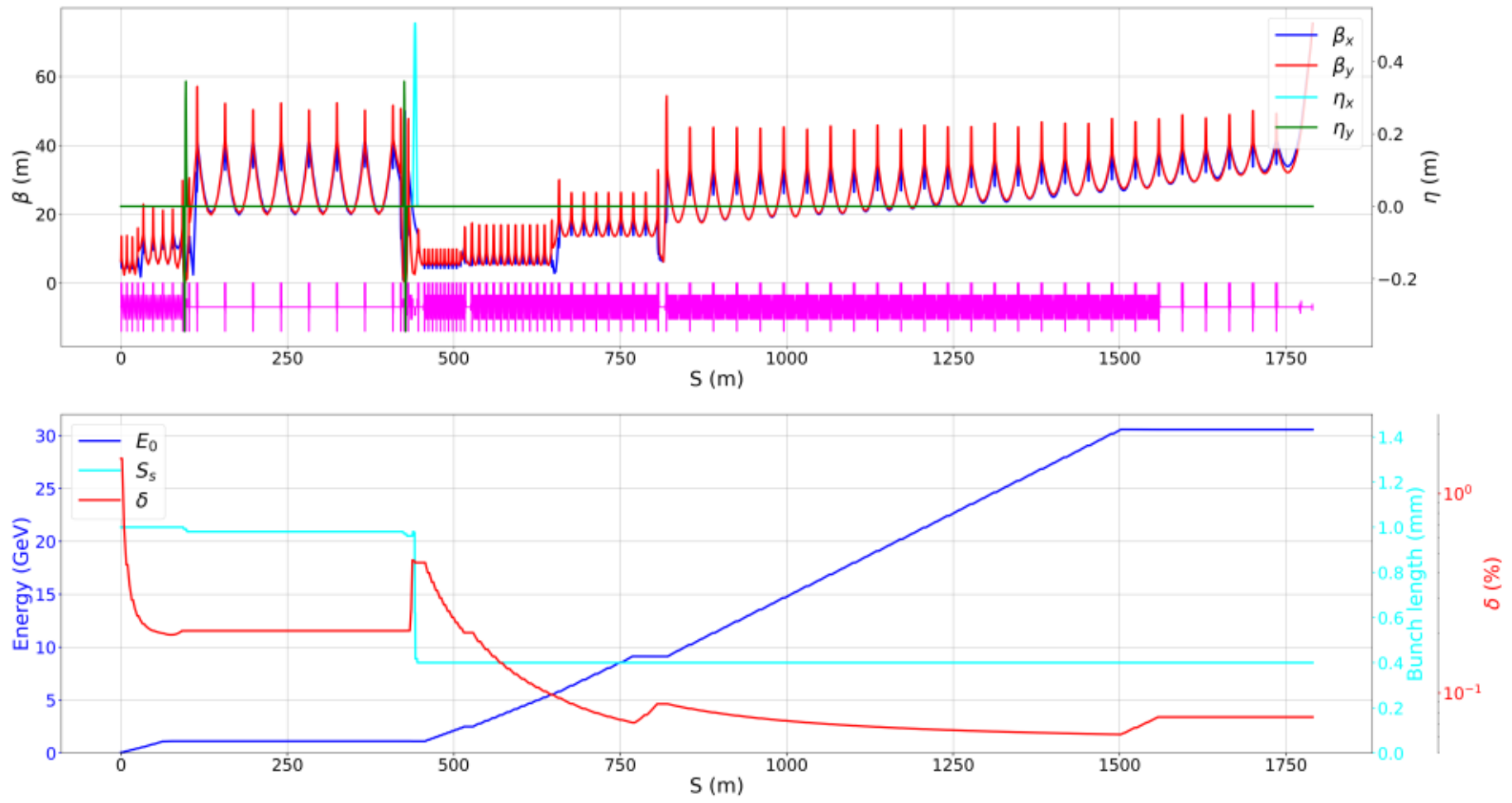

**Figure 6.2.1.7:** Dynamic simulation results of the electron Linac with bunch charge of 1.5 nC and accelerating gradient of 40 MV/m for the C-band accelerating structure.

**Table 6.2.1.1:** Simulation results of the electron Linac with accelerating gradient of 40 MV/m for the C-band accelerating structure.

| **Parameter** | **Unit** | **Design value** | **Simulation result** | |
|---|---|---|---|---|
| | | | **Baseline** | **Upgrade potential** |
| Beam energy | GeV | 30 | 30.56 | 30.06 |
| Bunch charge | nC | 1.5 | 1.5 | 3.0 |
| Energy spread | $10^{-3}$ | 1.5 | 0.76 | 1.34 |
| Emittance (H/V) | nm | 6.5 | 1.38/1.36 | 1.46/1.75 |
| Bunch length | mm | 0.3 ~ 1 | 0.4 | 0.4 |

### 6.2.1.4 ***Positron Linac***

#### *6.2.1.4.1 High Bunch Charge Electron Linac*

The high bunch charge electron Linac is the initial section of the positron beam production, where the electron beam is accelerated from 50 MeV to 4 GeV with a maximum bunch charge of 10 nC. In the FAS, there are 21 S-band klystrons, each with a power output of 80 MW, of which 3 are reserved as backups. To mitigate the short-range wakefield effect, the accelerating phase is set to 75°. Simulation results, as shown in Fig. 6.2.1.8 and Fig. 6.2.1.9, indicate that the energy spread is 0.63% and the normalized rms emittance is 81 mm-mrad for a bunch charge of 10 nC, satisfying the requirements for positron production.

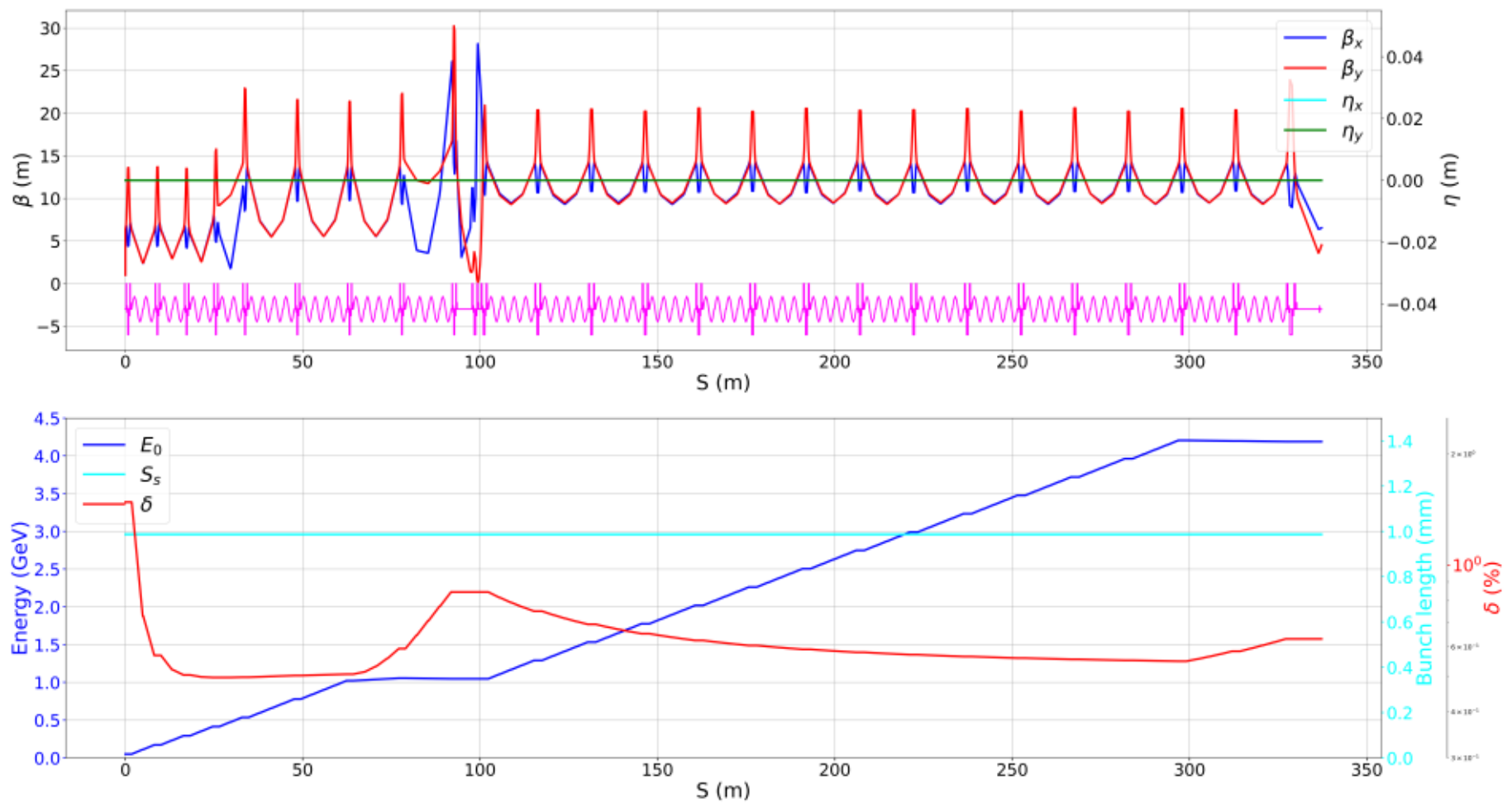

**Figure 6.2.1.8:** Dynamic simulation results of the FAS with high bunch charge.

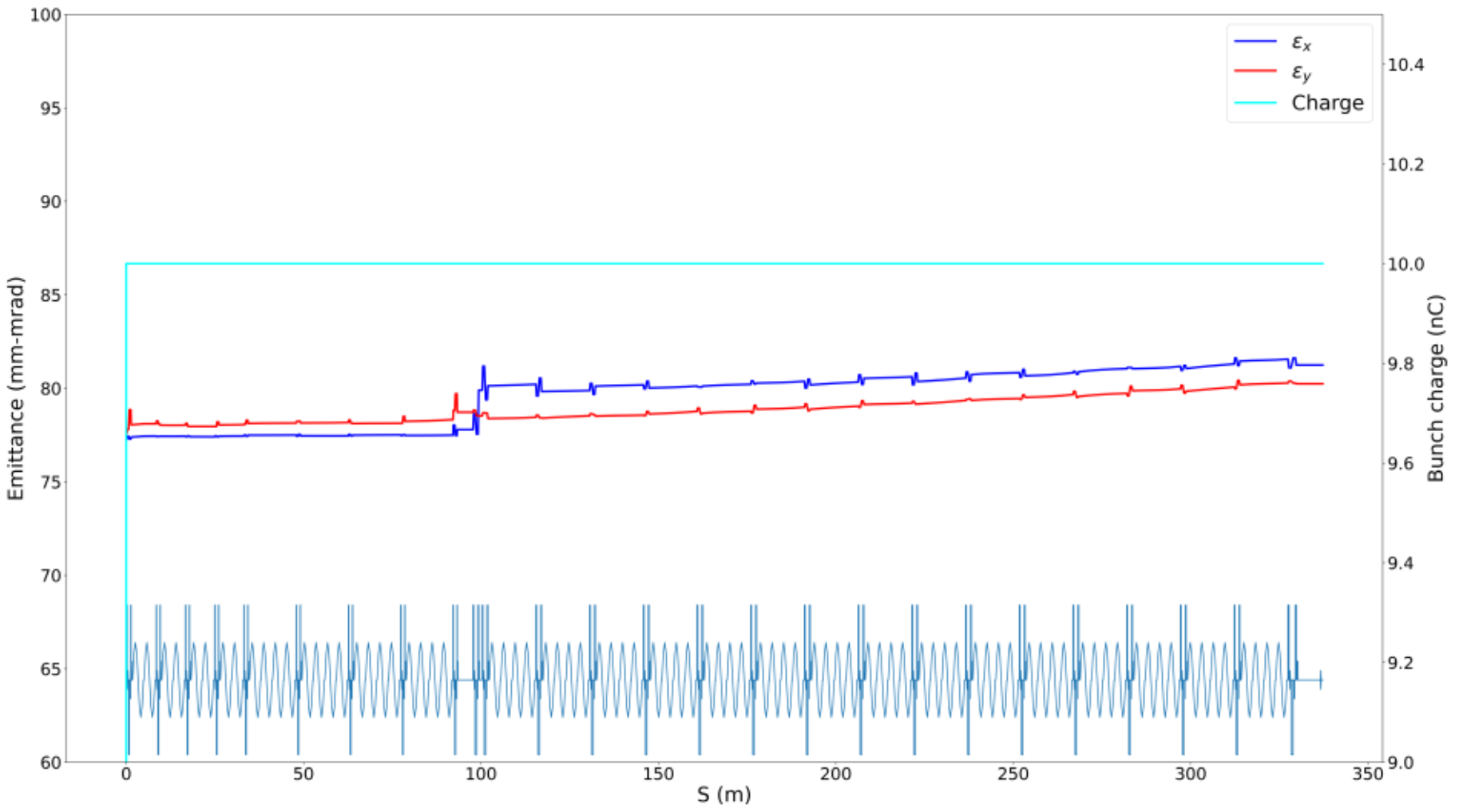

**Figure 6.2.1.9:** The emittance distribution along the FAS in high bunch charge mode.

### *6.2.1.4.2 Positron Source and Pre-accelerating Section*

The CEPC positron source [2] uses a conventional tungsten target to produce positrons through bremsstrahlung and pair conversion. The positron source and pre-accelerating section comprise a target, a flux concentrator (FC), which acts as an adiabatic matching device (AMD), six large-aperture S-band constant-impedance accelerating structures, and a beam separation system. A schematic layout of the PSPAS is shown in Fig. 6.2.1.10. To achieve a positron beam with a bunch charge greater than 3 nC, a 4 GeV primary electron beam with a maximum bunch charge of 10 nC is utilized.

According to the findings at SLAC [3], the threshold value for peak energy deposition density (PEDD) is approximately 35 J/g. In the design, the beam size should be larger than 0.5 mm to maintain PEDD below 24 J/g. An online beam size monitor is in place for interlock purposes to prevent extremely small spot sizes from damaging the target. As a protective measure, a spoiler system, as observed in the SuperKEKB positron source, may be introduced.

The target length of 15 mm is chosen, taking into account the positron yield and energy deposition. The simulated energy deposition is 0.784 GeV/electron, corresponding to a power deposition of about 784 W, necessitating water cooling. After considering the positron capture efficiency, emittance control, and accelerating structure design, a constant impedance acceleration structure with an aperture of 25 mm is chosen. The phase of the accelerating structures is carefully designed, and a chicane is introduced to separate and dump the wasted electrons. The emittance growth is shown in Fig. 6.2.1.11 and simulation results are presented in Table 6.2.1.2.

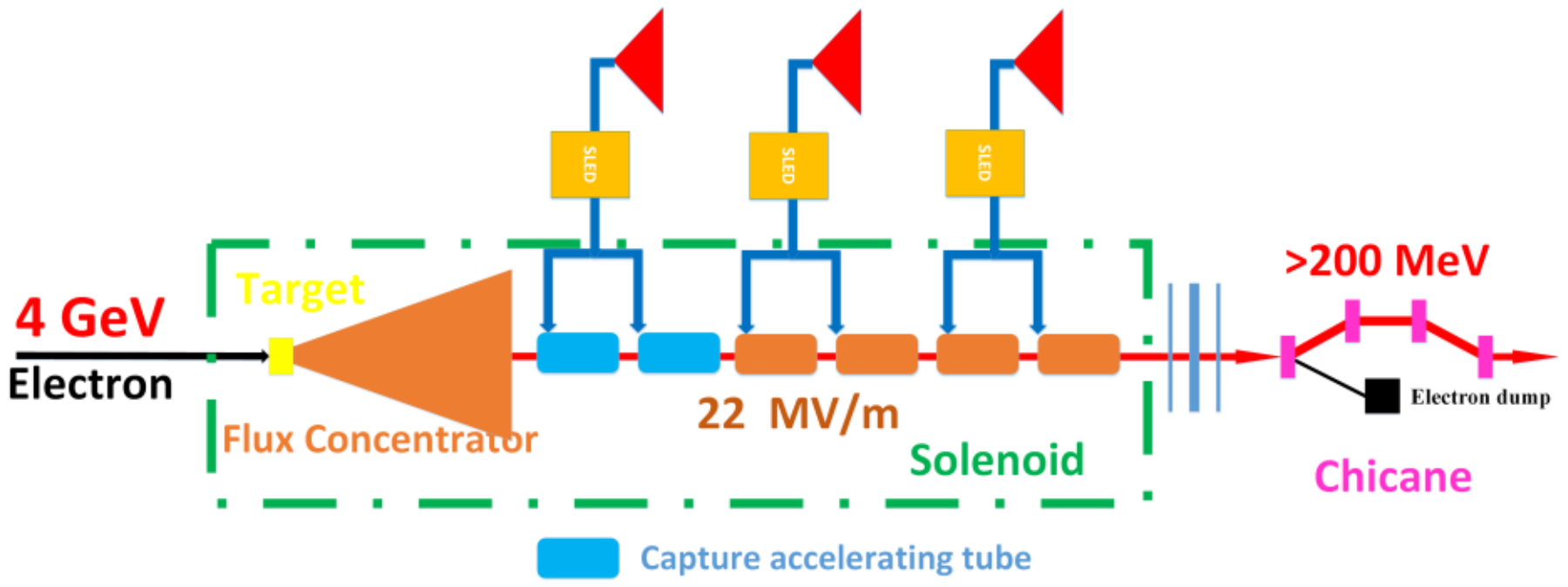

**Figure 6.2.1.10:** Dynamic simulation results of the FAS.

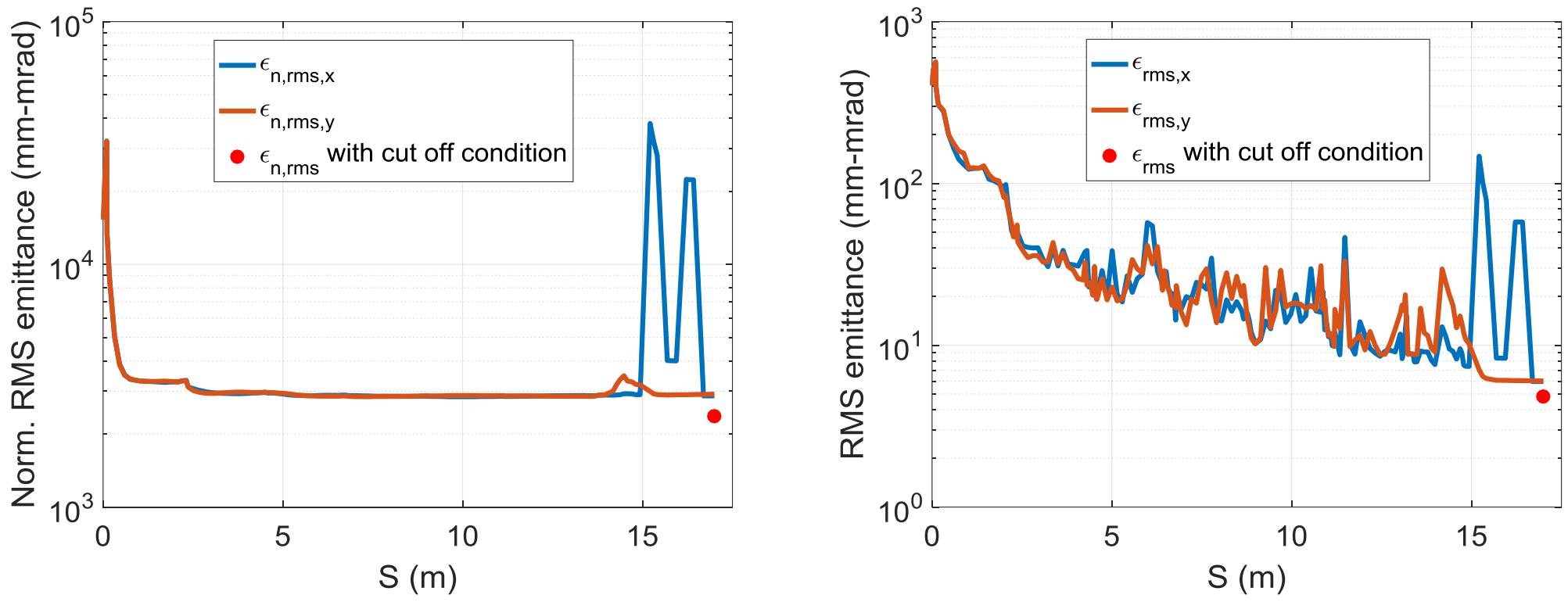

**Figure 6.2.1.11:** The emittance distribution along the PSPAS.

**Table 6.2.1.2:** Simulation results of PSPAS.

| Positron source | Unit | Design value | Simulation result |
|---|---|---|---|
| $e^-$ Beam energy on target | GeV | 4 | |
| $e^-$ Bunch charge on target | nC | 10 | |
| $e^+$ Bunch charge | nC | 3 | 5.5 (with cutoff condition) |
| $e^+$ Energy | MeV | 200 | 250 |
| $e^+$ Norm. RMS emittance | mm-mrad | 2400 | 2370 |

#### *6.2.1.4.3 Second Accelerating Section*

To achieve high transmission efficiency in the SAS, the design incorporates 10 large-aperture accelerating structures with a gradient of 22 MV/m and aperture of 25 mm, as well as 8 normal S-band accelerating structures with a gradient of 27 MV/m. Each accelerating structure is accompanied by an outside magnet triplet structure for transverse focusing.

After simulating the distribution at the exit of the PSPAS, we completed the simulation of the SAS. The results show that the beam energy at the exit of the SAS is greater than 1.1 GeV, with an energy spread of 0.4%, a bunch charge of 4.5 nC, and a normalized rms emittance of approximately 2550 mm-mrad. The bunch charge meets the design requirement, as confirmed by the simulation results.

Figures 6.2.1.12 and 6.2.1.13 show the emittance and bunch charge distribution along the SAS, as well as the beam dynamics results. Additionally, Fig. 6.2.1.14 depicts the beam distribution at the exit of the SAS.

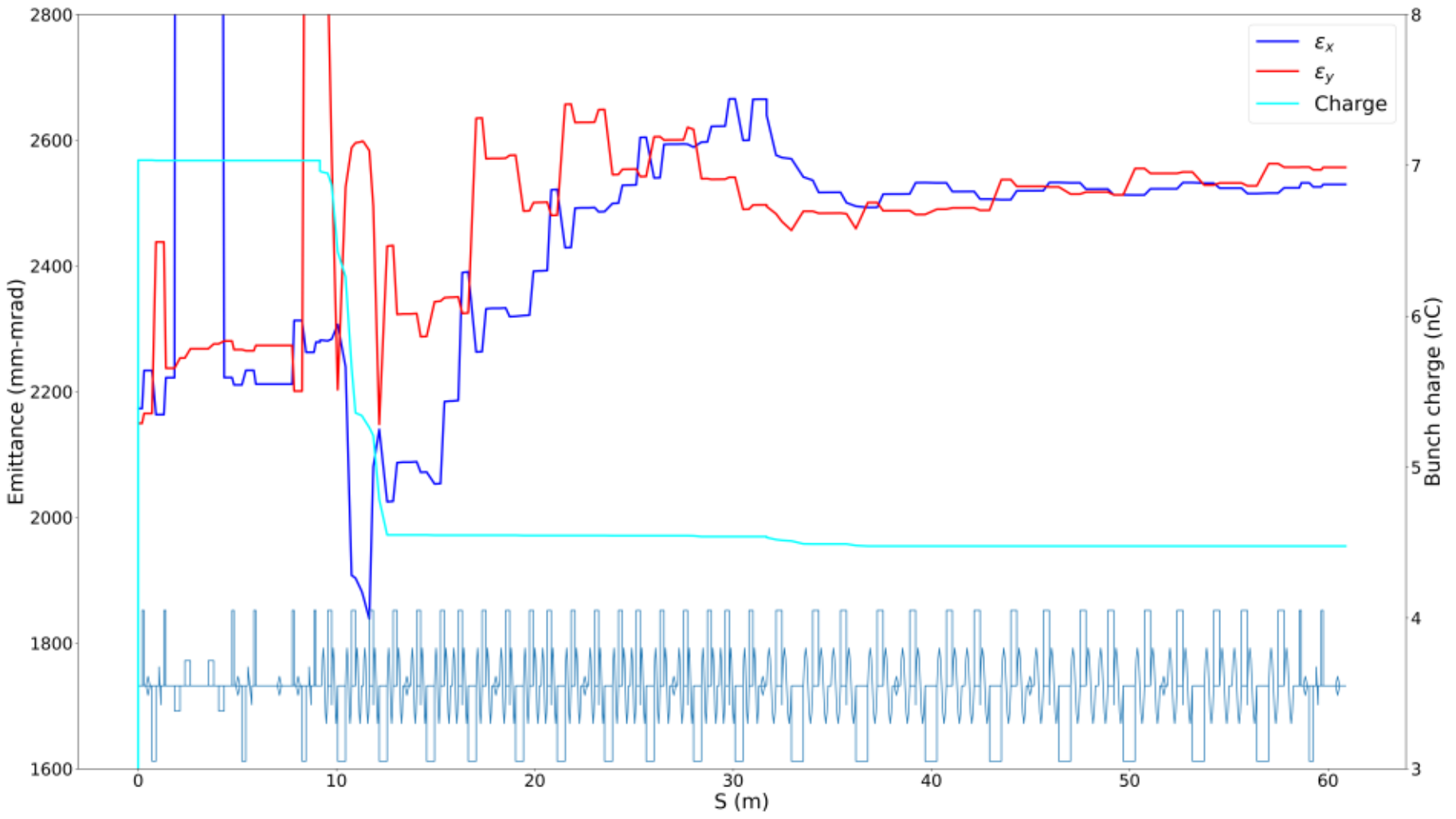

**Figure 6.2.1.12:** The emittance and bunch charge distribution along the SAS.

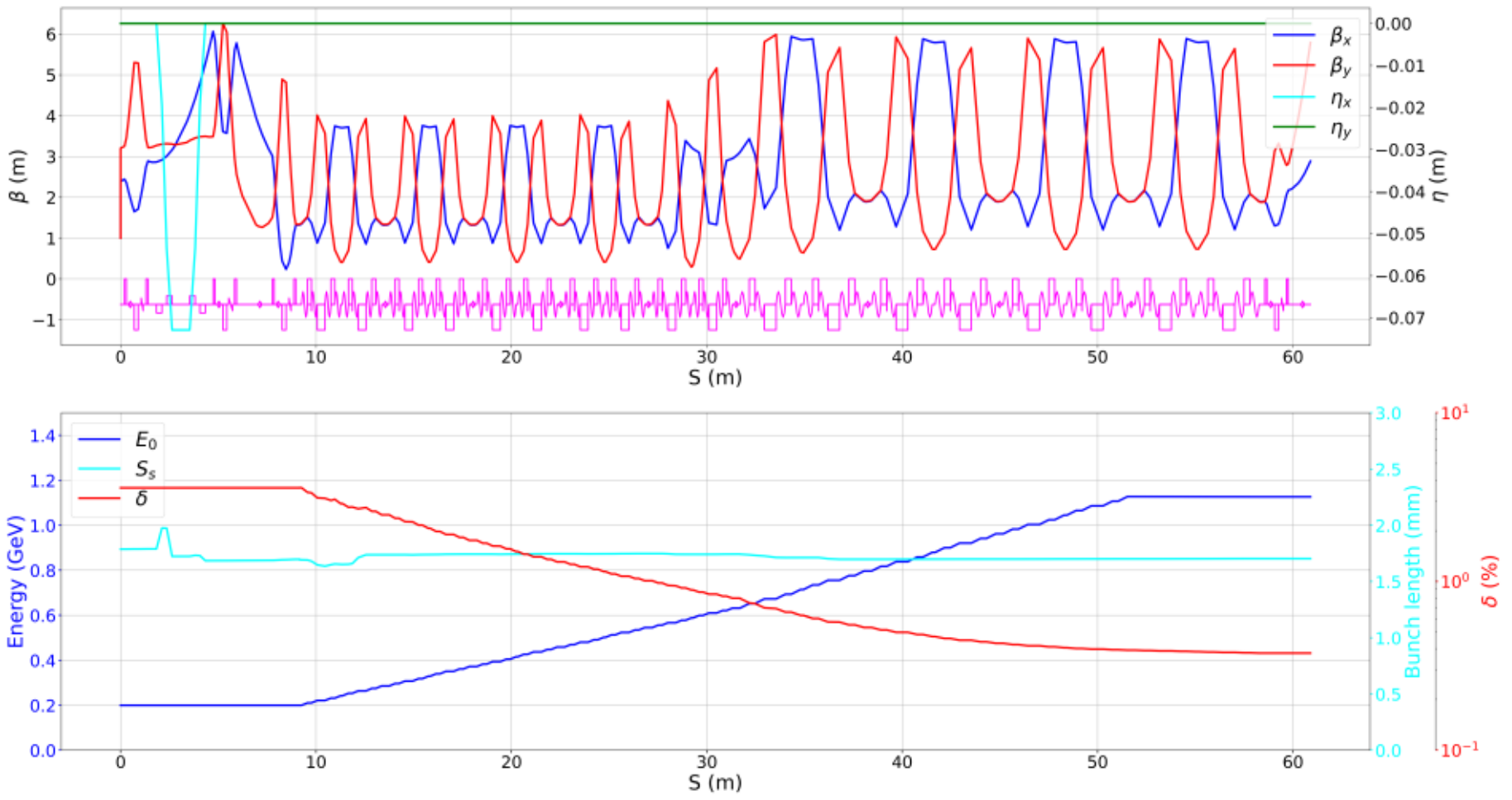

**Figure 6.2.1.13:** Dynamic simulation results of the SAS.

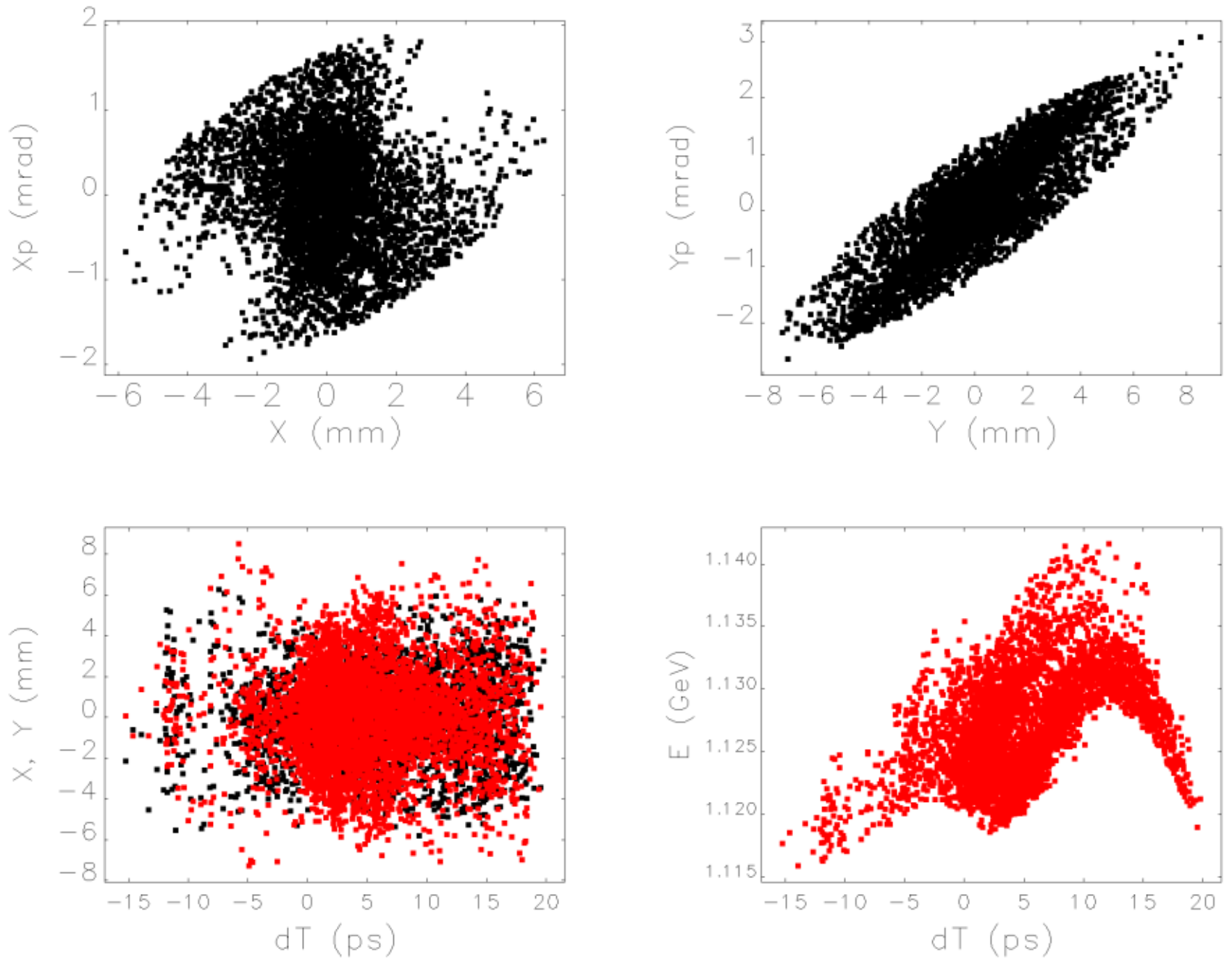

**Figure 6.2.1.14:** Beam distribution at the exit of the SAS.

### *6.2.1.4.4 Third Accelerating Section*

The positron beam is injected into the TAS from the Damping Ring with an energy of 1.1 GeV. The horizontal and vertical normalized rms emittance of the positron beam are 200 mm-mrad and 100 mm-mrad, respectively. The bunch length is compressed to 0.4 mm and the beam is then accelerated to 30 GeV. The simulation results of the beam dynamics in the TAS are shown in Fig. 6.2.1.15 and the relevant parameters are listed in Table 6.2.1.3, which confirm that the design requirements are satisfied.

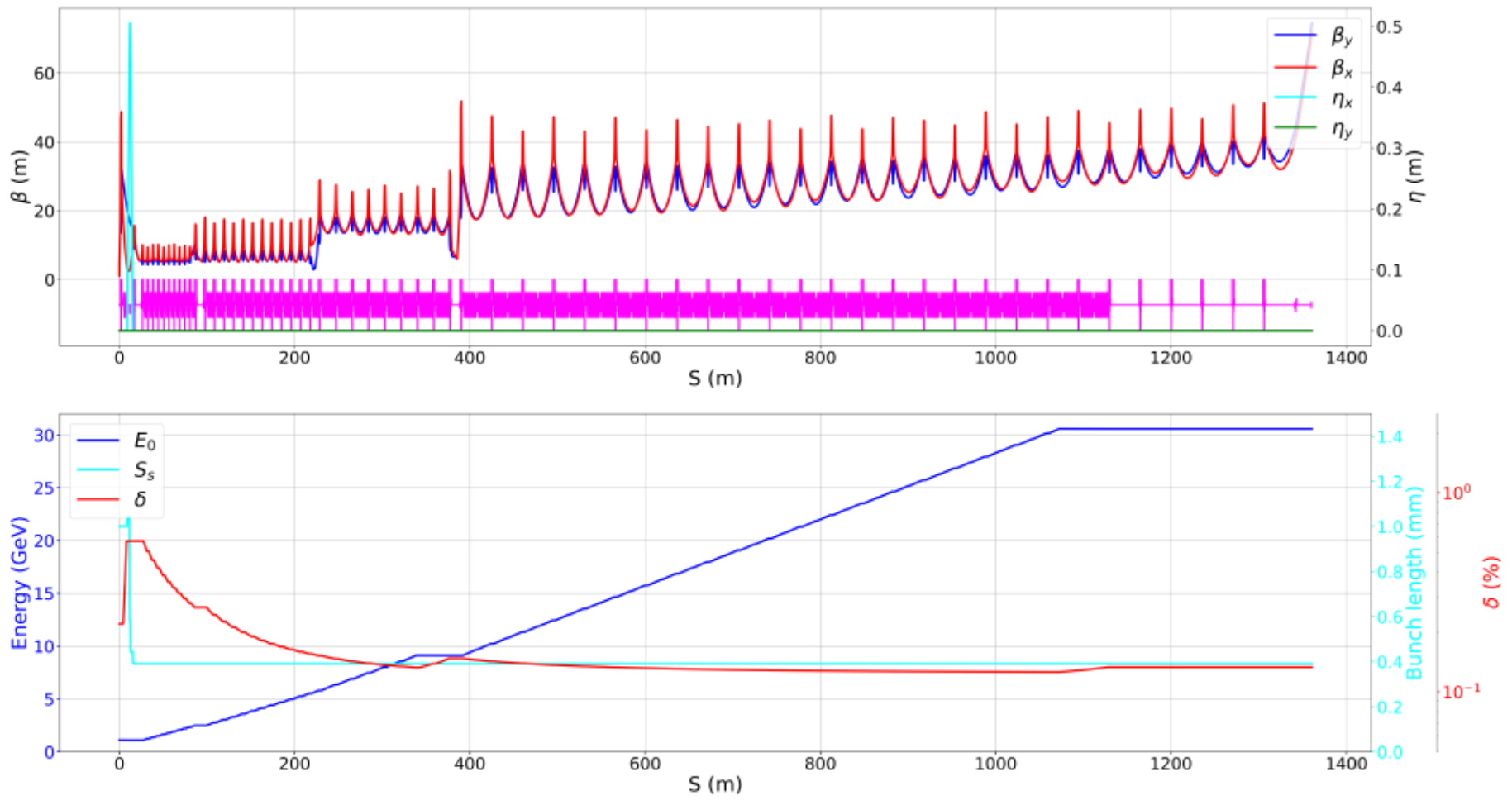

**Figure 6.2.1.15:** Dynamic simulation results of the TAS of the positron Linac with bunch charge of 1.5 nC and accelerating gradient of C-band accelerating structure of 40 MV/m.

**Table 6.2.1.3:** Simulation results of the positron Linac.

| Parameter | Unit | Design value | Simulation result | |
|---|---|---|---|---|
| | | | Baseline | Upgrade potential |
| Beam energy | GeV | 30 | 30.50 | 30.01 |
| Bunch charge | nC | 1.5 | 1.5 | 3.0 |
| Energy spread | $10^{-3}$ | 1.5 | 1.33 | 2.2 |
| Emittance (H/V) | nm | 6.5 | 3.37/1.68 | 4.01/1.71 |
| Bunch length | mm | 0.3 ~ 1 | 0.4 | 0.4 |

### 6.2.1.5 *Error Study*

The error study was performed to evaluate the impact of errors on the beam orbits and energy spread. The errors were modeled as a 3σ truncated Gaussian distribution with settings shown in Table 6.2.1.4, and the BPM resolution was set at 10 μm. The simulation results with correction of the beam orbits are shown in Fig. 6.2.1.16, which demonstrate that the beam orbits can be controlled within 0.5 mm. The phase error and accelerating gradient error of the accelerating structure can lead to energy jitter and energy spread increase. The Booster injection requires an energy jitter of ±0.1% and an energy spread of 0.15%. Fig. 6.2.1.17 shows the simulation results of energy jitter and energy spread with different phase and accelerating gradient errors. The simulation results indicate that a phase error of 0.3 degree and an accelerating gradient error of 0.3% can meet the requirements. The simulation results with errors and corrections are summarized in Table 6.2.1.5, which meet the design requirements.

**Table 6.2.1.4:** Error settings for the error study

| Elements | Number | Transverse misalignment | Longitudinal misalignment | Rotation error |
|---|---|---|---|---|
| | | mm | mm | mrad |
| Electron Gun | 1 | 0.1 | 0.2 | 0.2 |
| Positron source | 1 | 0.1 | 0.2 | 0.2 |
| Large-aperture S-band Acc. Stru. | 16 | 0.1 | 0.1 | 0.2 |
| S-band Acc. Stru. | 93 | 0.1 | 0.1 | 0.2 |
| SHB1 | 1 | 0.1 | 0.15 | 0.2 |
| SHB2 | 1 | 0.1 | 0.15 | 0.2 |
| BUN | 1 | 0.15 | 0.15 | 0.2 |
| C-band Acc. Stru. | 470 | 0.1 | 0.1 | 0.2 |
| C-band defecting cavity | 1 | 0.1 | 0.1 | 0.2 |
| Solenoid | 37 | 0.15 | 0.2 | 0.2 |
| Quadrupole | 372 | 0.1 | 0.2 | 0.2 |
| Dipole | 19 | 0.15 | 0.2 | 0.2 |
| Corrector | 275 | 0.15 | 0.2 | 0.2 |
| BPM | 150 | 0.1 | 0.2 | 0.2 |
| PR | 30 | 0.15 | 0.2 | 0.2 |

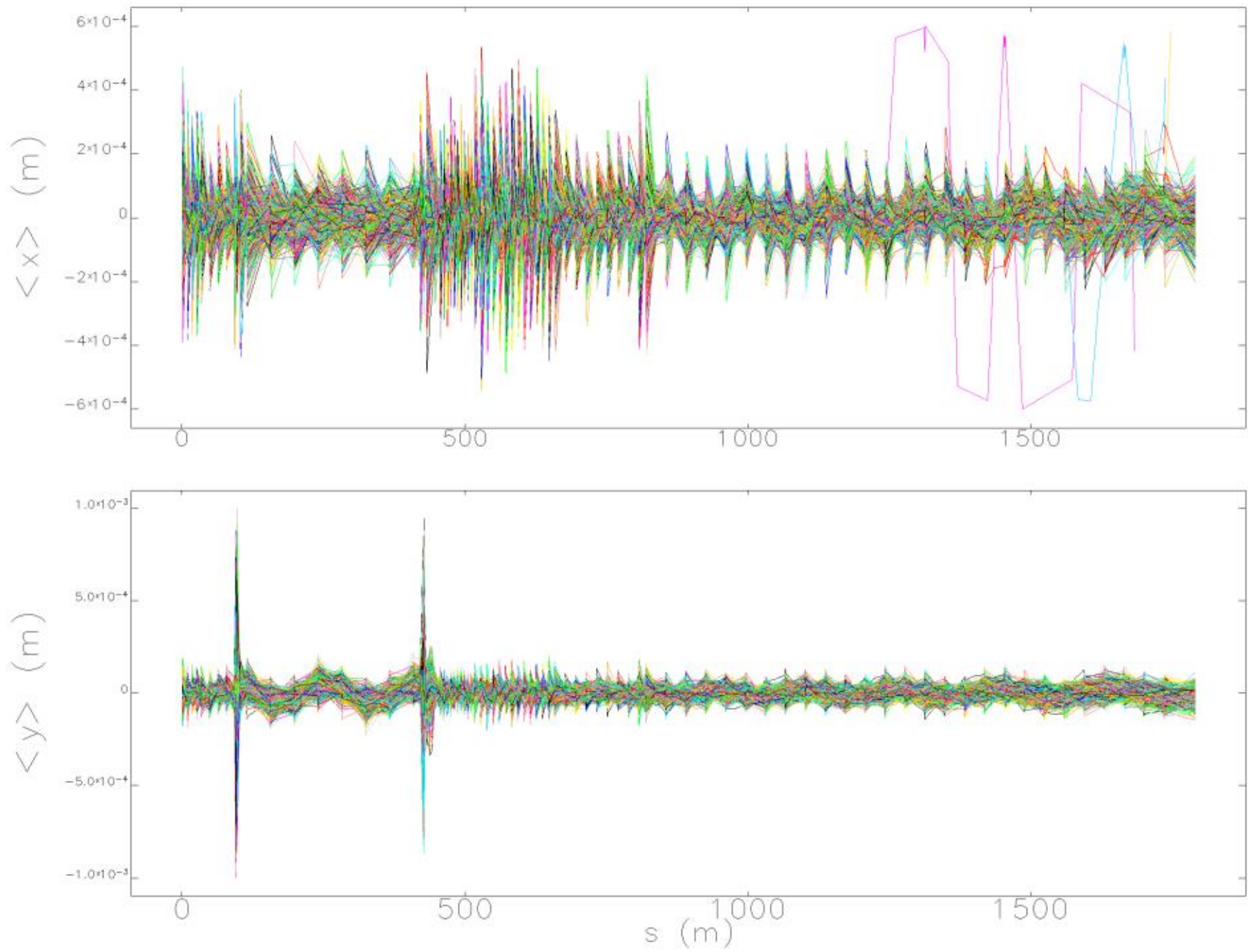

**Figure 6.2.1.16:** Beam orbits of the electron linac with different error seeds and after orbit correction, different colors represent beam orbit of different seeds.

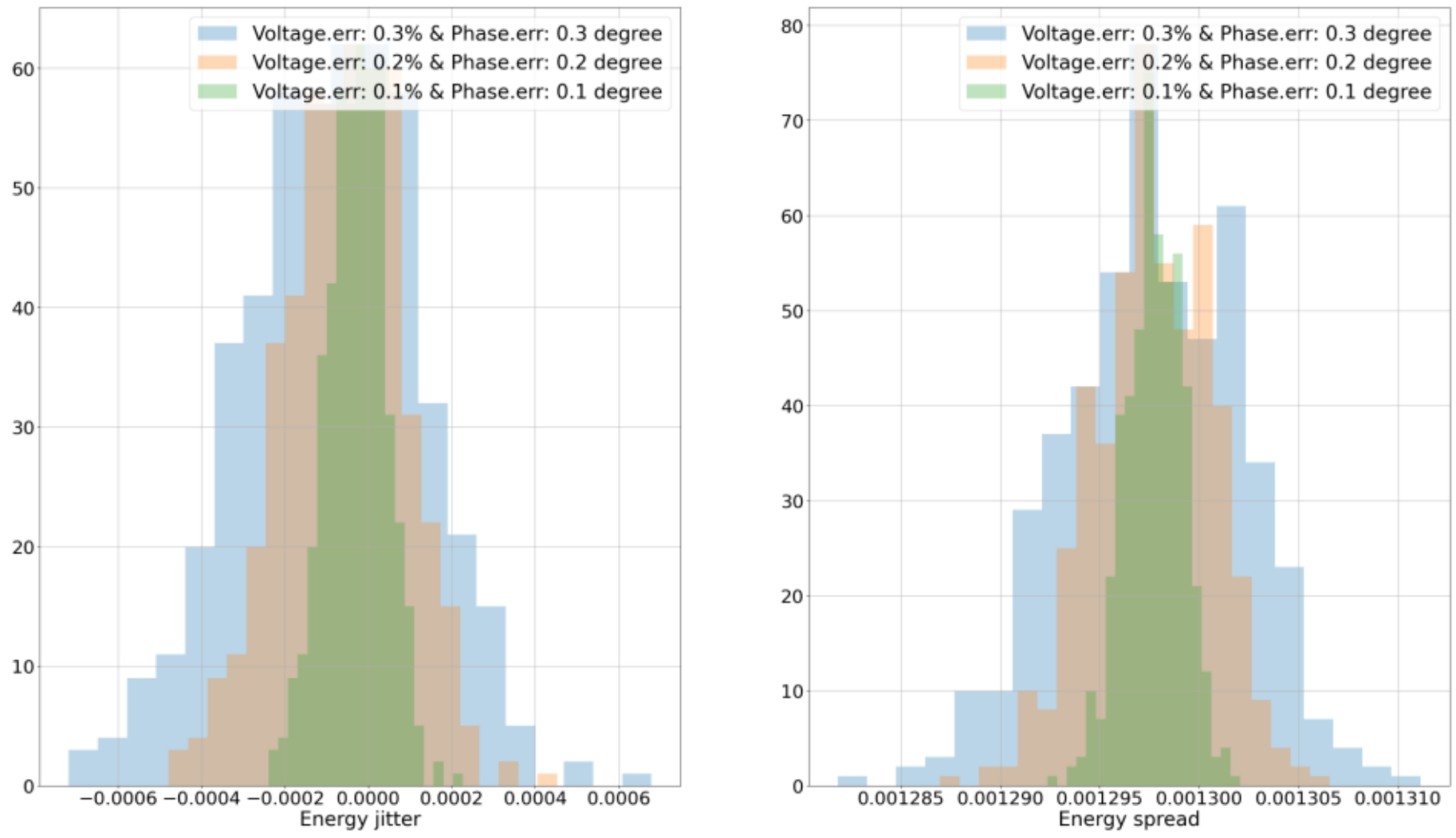

**Figure 6.2.1.17:** Energy jitters of positron Linacs with errors.

**Table 6.2.1.5:** Simulation results with errors

| Parameter | | Unit | Design value | Simulation result | | | |
|---|---|---|---|---|---|---|---|
| | | | | Electron | | Positron | |
| | | | | **Baseline** | **Upgrade potential** | **Baseline** | **Upgrade potential** |
| Beam energy | | GeV | 30 | 30.5 | 30.0 | 30.5 | 30.0 |
| Bunch charge | | nC | 1.5 | 1.5 | 3.0 | 1.5 | 3.0 |
| Energy spread | w/o error | $\times10^{-3}$ | 1.5 | 0.76 | 1.34 | 1.33 | 2.2 |
| | w/ error | | | 0.75±0.14 | 1.45±0.13 | 1.33±0.01 | 2.2±0.01 |
| Energy jitter | | $\times10^{-3}$ | 1.0 | 0.22 | 0.24 | 0.21 | 0.22 |
| Emittance (H/V) | w/o error | nm | 6.5 | 1.38<br>1.36 | 1.46<br>1.75 | 3.37<br>1.68 | 4.01<br>1.71 |
| | w/ error | | | 1.41±0.07<br>1.40±0.06 | 2.11±0.30<br>2.41±0.62 | 3.39±0.08<br>1.69±0.03 | 5.33±1.63<br>2.36±0.56 |

### 6.2.1.6 *Double-bunch Acceleration*

The repetition rate of the Linac is 100 Hz. To meet the injection speed requirement for Z mode, the Linac needs to double the bunch repetition rate. However, increasing the repetition rate to 200 Hz would greatly increase the cost, so we chose the double-bunch acceleration mode instead. Within the same RF pulse, two beam bunches are generated by the electron gun, with a separation of 69.23 ns. The most important issues in this mode are the frequency relationship, which we addressed by designing timing-related parameters, the harmonic number, and beam pattern information of the Collider. These parameters are shown in Table 6.2.1.6. The bunch spacing in the Collider is 23.08 ns, which is acceptable for the detector. Although the bunch spacing in the Linac is about 70 ns, it is still possible to use a pulse compressor [2] even for a C-band accelerating structure, as shown in Fig. 6.2.1.18 and Fig. 6.2.1.19. Another concern arises from the long-range wakefield, where the presence of the first bunch affects the second bunch. In the worst-case scenario, the Q value and shunt impedance for the dipole mode are 14,000 and $2.2 \times 10^6$ Ω/m, respectively. We plan to conduct further investigations while also implementing fine-tuning of the accelerating structure to mitigate this issue.

**Table 6.2.1.6:** Parameters of the timing system and bunch pattern information

| Parameter | Unit | Design value |
|---|---|---|
| Repetition frequency | Hz | 100 |
| Common frequency | MHz | 130 |
| Linac common frequency | MHz | 14.44 |
| Bunch frequency | MHz | 14.44 |
| SHB1 RF frequency | MHz | 158.89 |
| SHB2 RF frequency | MHz | 476.67 |
| LINAC RF frequency | MHz | 2860 |
| | MHz | 5720 |
| Damping ring RF frequency | MHz | 650 |
| Booster RF frequency | MHz | 1300 |
| Ring RF frequency | MHz | 650 |
| Bunch spacing @Collider | ns | 23.08 |
| Bunch spacing @Linac | ns | 69.23 |
| Injection scheme | | bunch-by-bunch |
| Harmonic number | | 45×(2k) + [10, 20, 40], k is integer |
| | | 45×(2k+1) + [5, 25], k is integer |
| Bunch number per train | | 6n, n is integer |

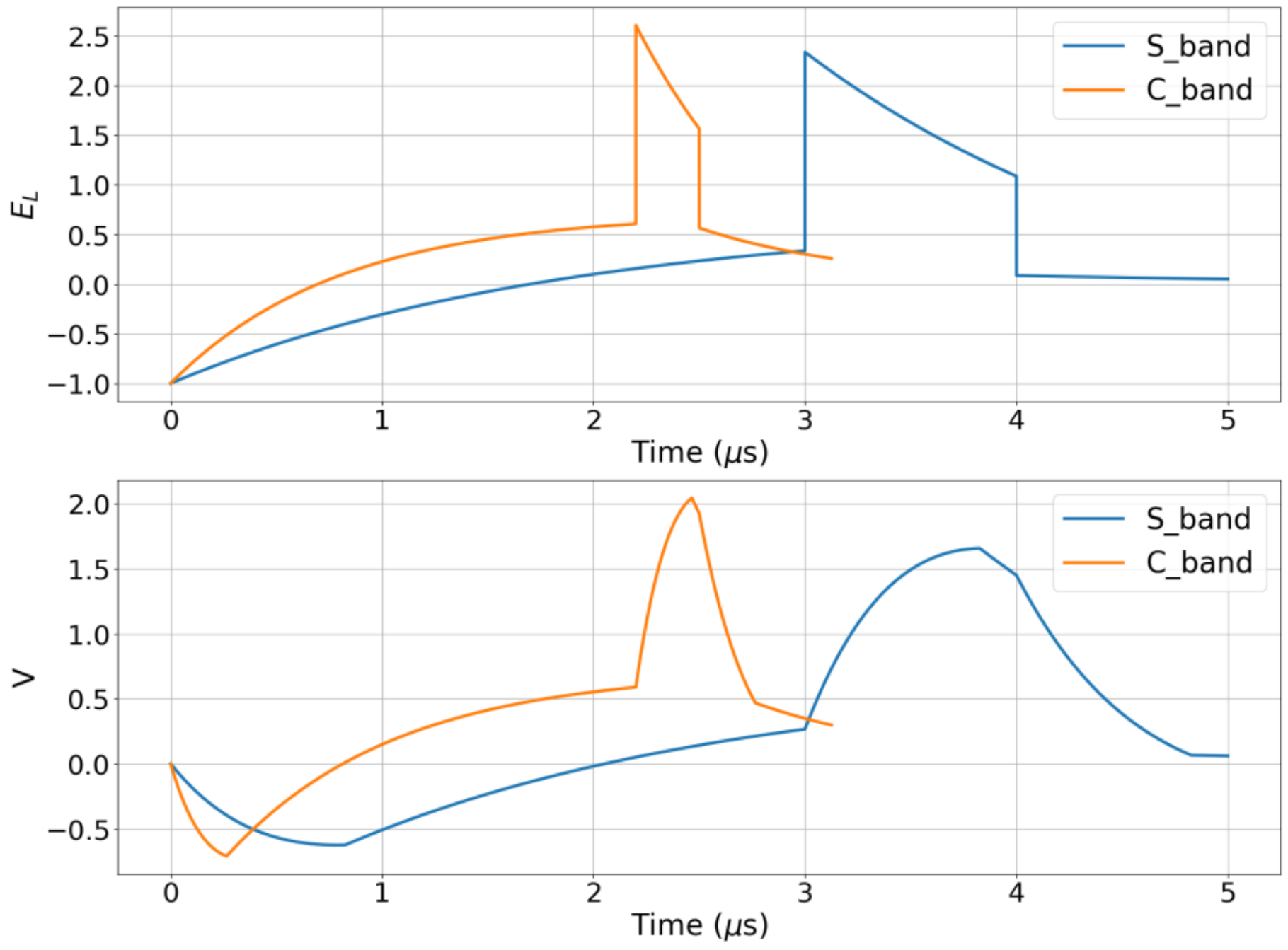

**Figure 6.2.1.18:** Comparison of the relative accelerating gradient and voltage between an accelerating structure with and without a pulse compressor.

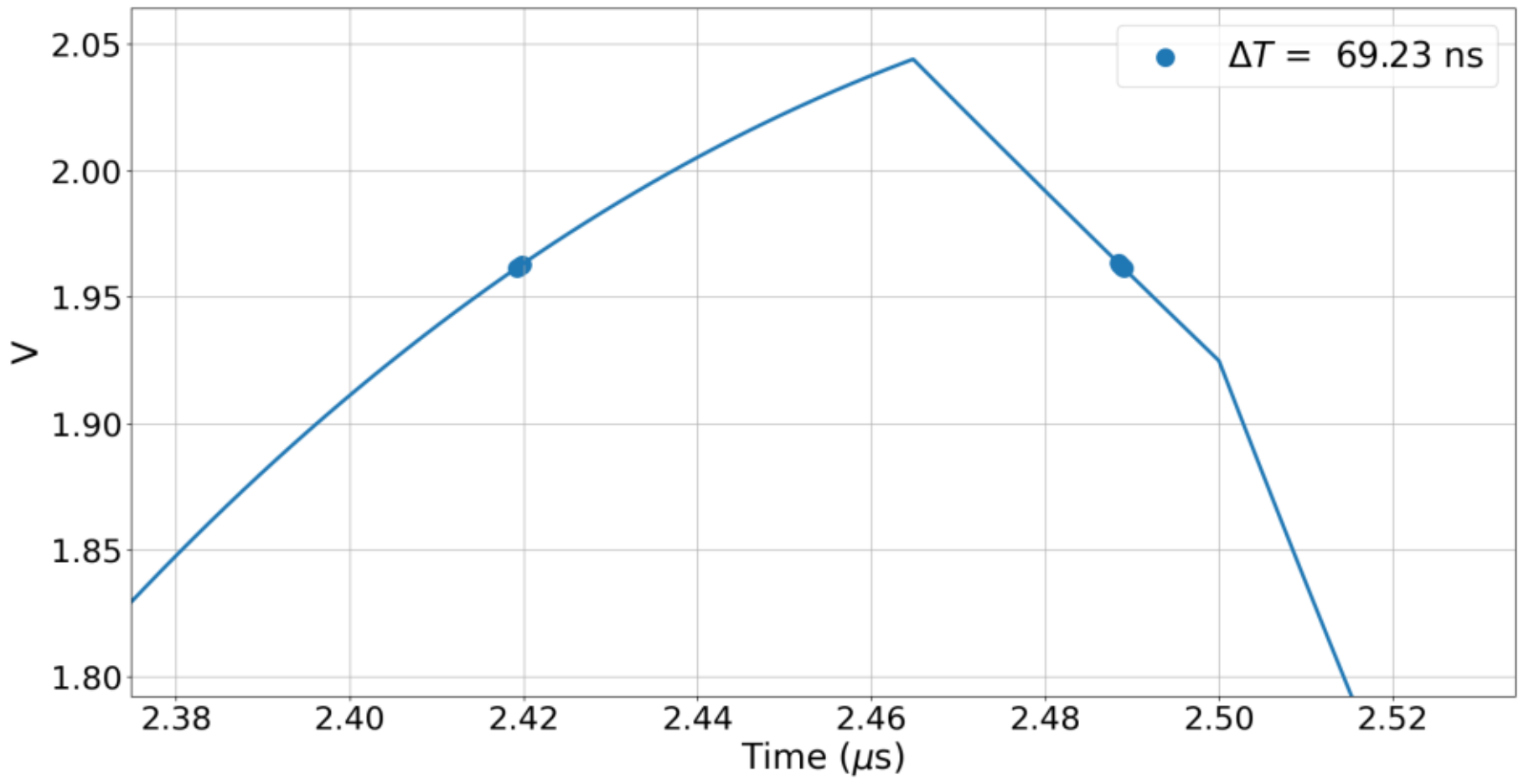

**Figure 6.2.1.19:** Relative voltage of C-band accelerating structure for double-bunch acceleration mode, where "tow point" represents the two bunches.

## 6.2.2 **Damping Ring Design**

The Linac system includes a Damping Ring (DR) with a small 1.1 GeV ring and two transport lines. The main purpose of the DR is to reduce the transverse emittance of the positron beam at the end of the Linac, which will lead to reduced beam loss in the Booster.

The Linac system operates at a repetition rate of 100 Hz, and a one-bunch-per-pulse scheme is used. However, for the high luminosity mode at the Z pole, a double-bunch scheme is also considered.

To generate the positron beam, a 4 GeV electron beam is directed at a tungsten target. The resulting positron beam is then captured by an AMD flux concentrator. Each positron bunch is injected into the damping ring every 10 ms, and two bunches are stored in the ring, resulting in a storage time of 20 ms for each bunch. An upgrade to increase the number of bunches in the ring to four would increase the storage time for each bunch to 40 ms.

The DR system is anticipated to decrease the normalized emittance of the positron beam from 2500 mm-mrad to 166 mm-mrad during a 20 ms storage period based on theoretical estimations. However, if the storage time is extended to 40 ms with four bunches in the ring, the extracted normalized emittance is expected to be 97 mm-mrad.

### 6.2.2.1 *Parameters*

The Damping Ring (DR) operates at an energy of 1.1 GeV and has a circumference of 147 m. The size of the DR was doubled after the CDR to achieve a lower emittance in the Linac and support the higher luminosity goal in TDR. The DR has a racetrack shape, and the arcs were designed with a 60-degree FODO cell and reversed bending magnets.

The injected emittance (normalized) for DR is 2500 mm-mrad, and the injected energy spread is smaller than 0.2%. The positron beam is stored in the DR for either 20 ms or 40 ms, depending on the 100 Hz repetition rate and the multi-bunch storage scheme. The extracted emittance is 166 mm-mrad for the horizontal plane and 75 mm-mrad for the vertical plane, which can be further reduced by storing more bunches.

To ensure injection efficiency, the transverse acceptance of the DR should be larger than five times the injection beam size. Table 6.2.2.1 summarizes the main parameters of the DR.

The main components of the DR are summarized in Table 6.2.2.2.

**Table 6.2.2.1:** Main Parameters of the Damping Ring

| | DR V3.0 |
|---|---|
| Energy (Gev) | 1.1 |
| Circumference (m) | 147 |
| Number of trains | 2 (4)* |
| Number of bunches/trian | 1 (2)# |
| Total current (mA) | 12.4 (24.8)* |
| Bending radius (m) | 2.87 |
| Dipole strength $B_0$ (T) | 1.28 |
| $U_0$ (keV/turn) | 94.6 |
| Damping time $\tau_x/\tau_y/\tau_z$ (ms) | 11.4/ 11.4/ 5.7 |
| Phase/cell (degree) | 60/60 |
| Momentum compaction | 0.013 |
| Storage time (ms) | 20 (40)* |
| Natural energy spread (%) | 0.056 |
| Norm. natural emittance (mm-mrad) | 94.4 |
| Inject bunch length (mm) | 4.4 |
| Extract bunch length (mm) | 4.4 |
| Norm. inject emittance (mm-mrad) | 2500 |
| Norm. extract emittancce *x*/*y* (mm-mrad) | 166 (97)* / 75 (3)* |
| Energy spread inj/ext (%) | 0.18 / 0.056 |
| Energy acceptance by RF (%) | 1.8 |
| $f_{RF}$ (MHz) | 650 |
| $V_{RF}$ (MV) | 2.5 |
| Longitudinal tune | 0.0387 |

* The numbers in parenthises refer to the case of 40 ms storage.

# The number in the parentheses is for the Linac double-bunch operation at Z pole.

**Table 6.2.2.2:** Main components of the Damping Ring

| Name | Type | Number | Length (m) | Value |
|---|---|---|---|---|
| B0 | dipole | 40 | 0.7 | Gap 38 mm, Field 1.3 T |
| Br | dipole | 40 | 0.248 | Gap 38 mm, Field 1.3 T |
| Qarc | quadrupole | 96 | 0.2 | Gap 44 mm, Field 17 T/m |
| QRF | quadrupole | 8 | 0.2 | Gap 38 mm, Field 30 T/m |
| SF | sextupole | 36 | 0.1 | Gap 38 mm, Field 94 T/m$^2$ |
| SD | sextupole | 36 | 0.1 | Gap 38 mm, Field 140 T/m$^2$ |
| CORR | corrector | 60 | 0.07 | Gap 40 mm, Field 20 Gs |
| DRLAM | Lambertson | 2 | 0.5 | Field 0.88 T |
| DRkicker | kicker | 2 | 1.4 | Field 0.028 T |
| MBPM | BPM | 40 | 0.1 | Button BPM, Resolution 20µm @ 5mA |
| RFC | cavity | 2 | 1.4 | 650MHz, 1.25MV/cavity |

### 6.2.2.2 *Optics*

For the DR arc, we have chosen a 60°/60° FODO cell and a reversed bending magnet scheme. The natural emittance of the ring is 44 nm. The DR twiss parameters of the arc cell can be seen in Fig. 6.2.2.1. The length of the FODO cell is 3 m, and the strength of the dipole magnets is 1.28 T.

We have adopted the reversed bending magnet scheme to increase synchrotron radiation damping and reduce the emittance. Each cell contains two dipoles with the same strength but different bending angles. The bending angle of the shorter dipole is less than the longer one, and the angle ratio of these two dipoles is 0.355.

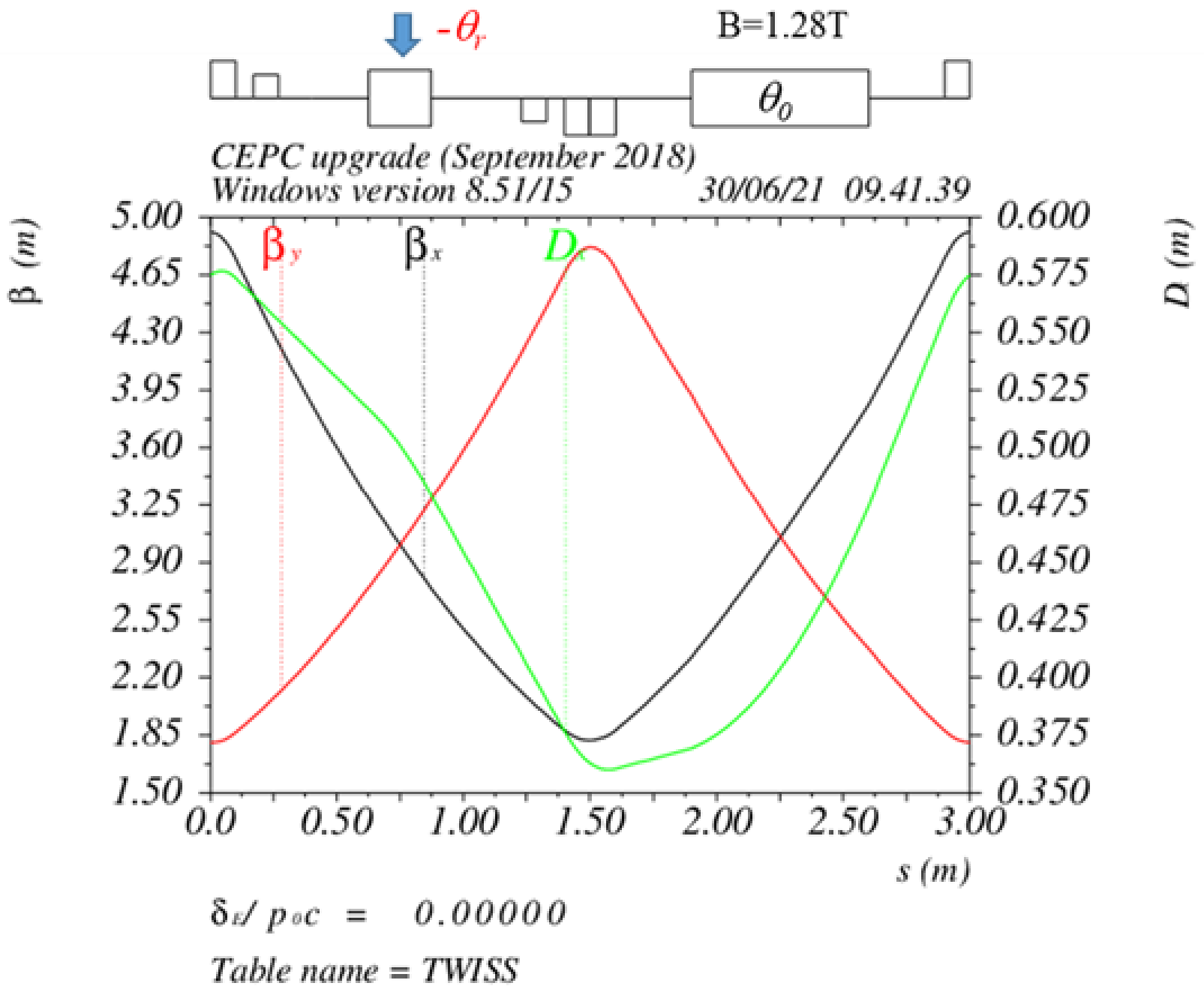

**Figure 6.2.2.1:** The DR twiss parameters for the arc FODO cell.

The dispersion suppressor has the function of canceling out the dispersion induced in the arc section and facilitating a smooth transition between the arc section and the straight section. The straight section is composed of the RF section and the injection/extraction section, where the septum and kicker will be inserted. The twiss parameters for the RF and injection/extraction section can be found in Fig. 6.2.2.2. To address the issue of injection efficiency, the horizontal beta function at the injection point has been designed to be larger than 9 m.

The twiss parameters for the entire ring can be seen in Fig. 6.2.2.3.

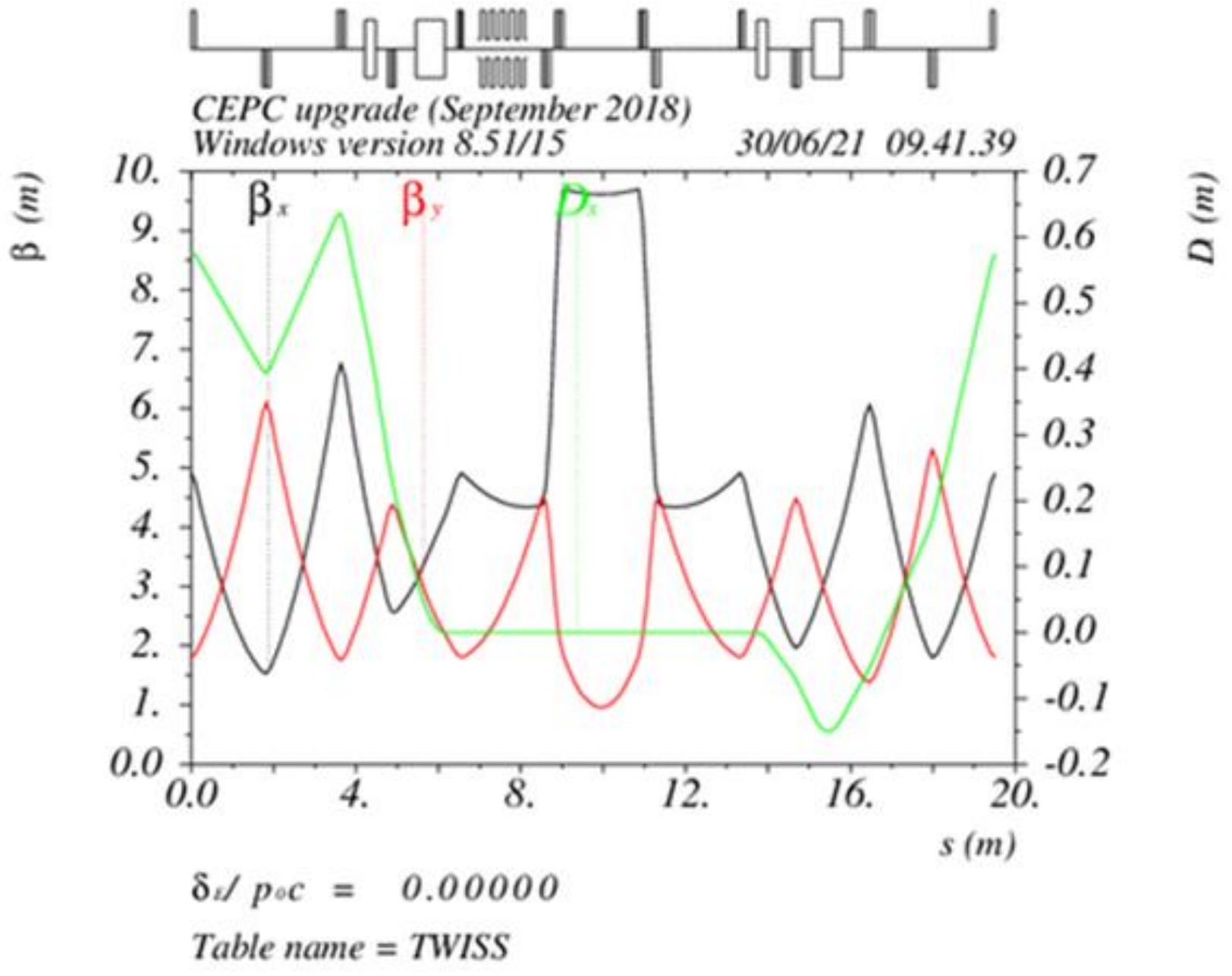

**Figure 6.2.2.2:** The DR twiss parameters for the RF and injection/extraction section.

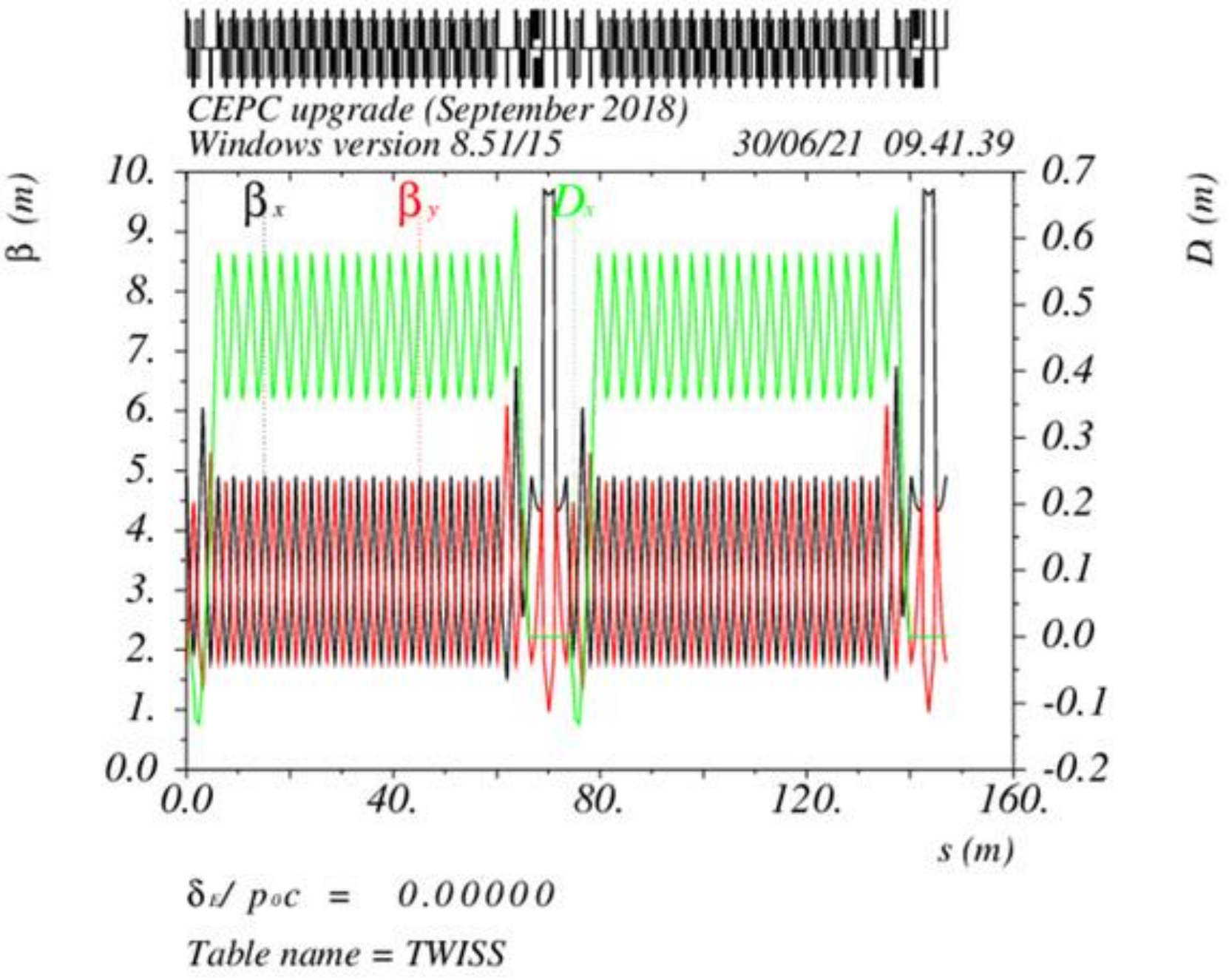

**Figure 6.2.2.3:** The DR twiss parameters for the entire ring.

In addition to the damping ring itself, there are two transport lines connecting the Linac and the DR. The layout of the Damping Ring system can be seen in Fig. 6.2.2.4.

On the transport line entering the DR, the energy spread of the positron bunch is reduced to match the RF acceptance of the Damping Ring. After passing through the DR, the longitudinal bunch size is reduced to minimize the energy spread during the following acceleration in the Linac.

The design of the transport lines can be found in Sec. 6.2.3.

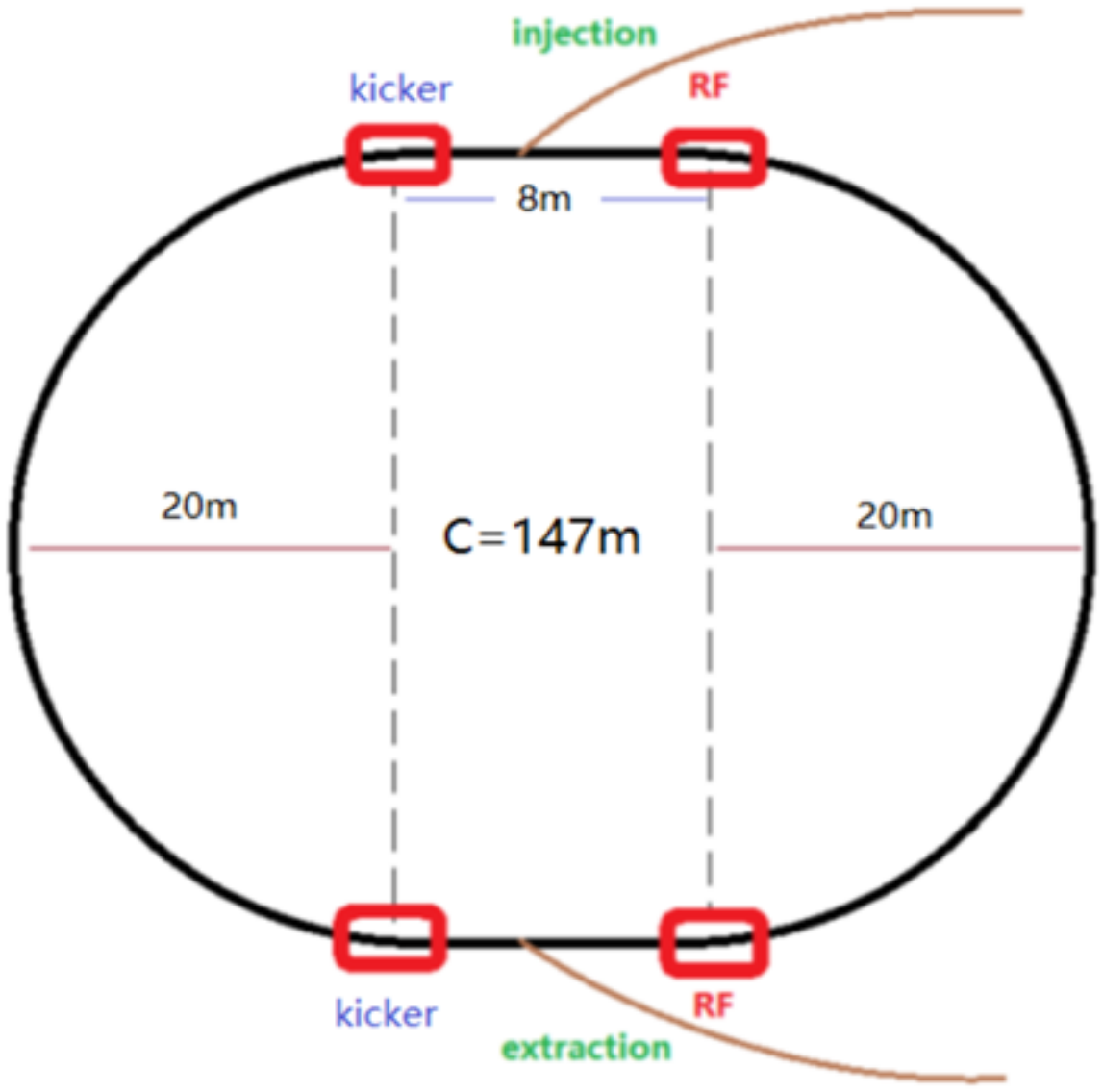

**Figure 6.2.2.4:** Layout of the Damping Ring system.

### 6.2.2.3 *Dynamic Aperture and Error Study*

For the linear chromaticity correction, we have adopted an interleave scheme and two sextupole families. The dynamic aperture of the Damping Ring has been tracked with 100,000 turns by SAD, and the results for the bare lattice can be seen in Fig. 6.2.2.5.

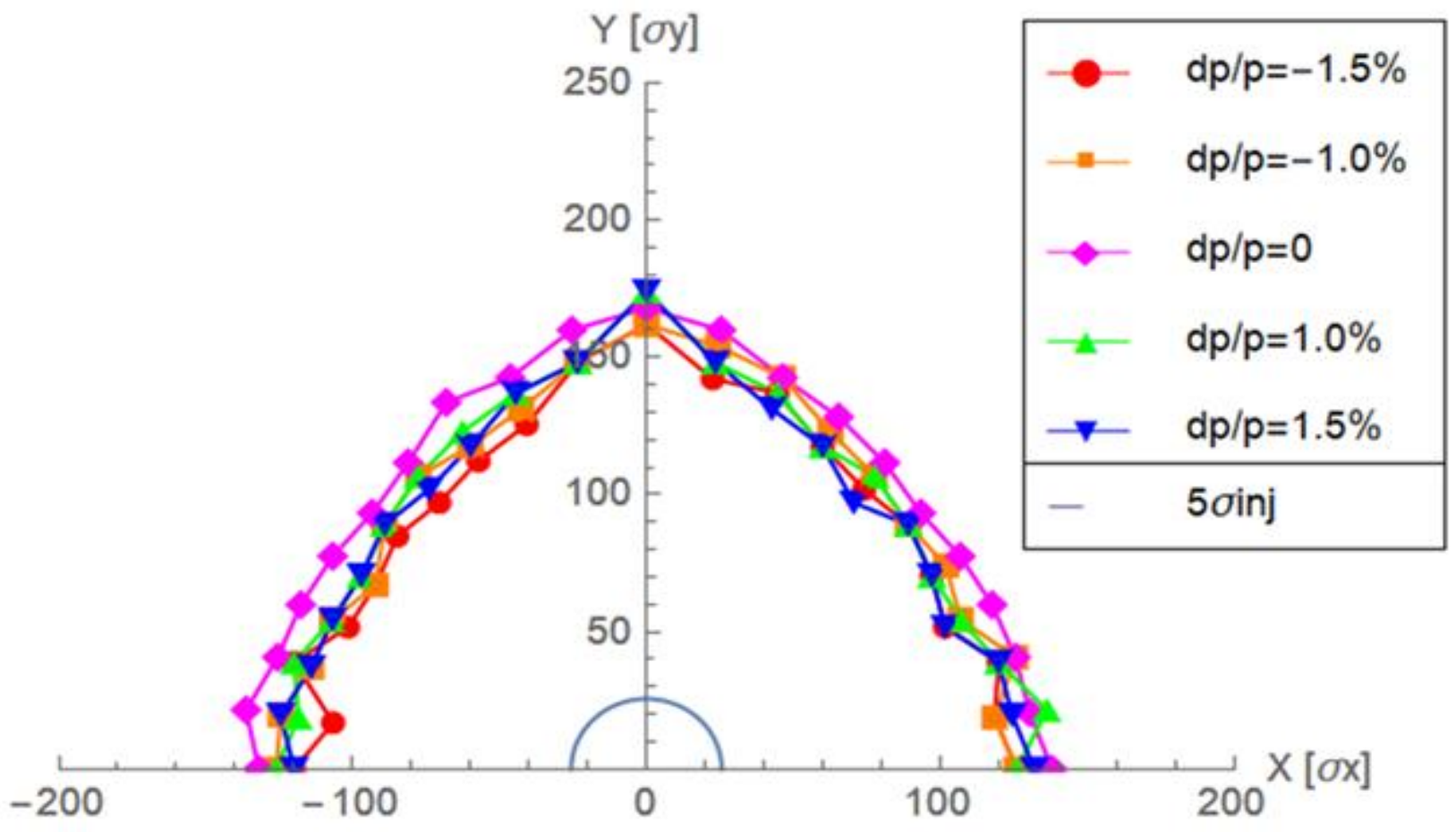

**Figure 6.2.2.5:** Dynamic aperture of the Damping Ring without errors. ±1.5% energy deviation corresponds to ±8.3 times of injected energy spread.

For the error study, we considered misalignment errors, magnet field errors, and BPM errors. The specific error settings for the Damping Ring are provided in Table 6.2.2.3 for misalignment errors, Table 6.2.2.4 for BPM errors, and Table 6.2.2.5 for magnet field errors. By employing only closed orbit distortion (COD) correction techniques, the residual errors for the closed orbit and optics distortion remain within acceptable limits, as depicted in Fig. 6.2.2.6 and Fig. 6.2.2.7, respectively. The dynamic aperture of the Damping Ring, considering errors and orbit correction, is illustrated in Fig. 6.2.2.8.

**Table 6.2.2.3:** Error settings for the magnets.

| Parameters | Dipole | Quadrupole | Sextupole |
|---|---|---|---|
| Transverse shift X/Y (μm) | 100 | 100 | 100 |
| Longitudinal shift Z (μm) | 100 | 150 | 100 |
| Tilt about X/Y (mrad) | 0.2 | 0.2 | 0.2 |
| Tilt about Z (mrad) | 0.1 | 0.2 | 0.1 |
| Nominal field | $1\times10^{-3}$ | $2\times10^{-4}$ | $3\times10^{-4}$ |

**Table 6.2.2.4:** BPM errors.

| Parameters | BPM (10 Hz) |
|---|---|
| Accuracy (m) | $1\times10^{-7}$ |
| Tilt (mrad) | 10 |
| Gain | 5% |
| Offset after beam based alignment (BBA) (mm) | $30\times10^{-3}$ |

**Table 6.2.2.5:** Multipole field errors of the magnet (unit: $1\times10^{-4}$).

| Dipole | Quadrupole | Sextupoles |
|---|---|---|
| $B_1 \leq 2$ | | |
| $B_2 \leq 3$ | $B_2 \leq 3$ | |
| $B_3 \leq 0.2$ | $B_3 \leq 2$ | $B_3 \leq 10$ |
| $B_4 \leq 0.8$ | $B_4 \leq 1$ | $B_4 \leq 3$ |
| $B_5 \leq 0.2$ | $B_5 \leq 1$ | $B_5 \leq 10$ |
| $B_6 \leq 0.8$ | $B_6 \leq 0.5$ | $B_6 \leq 3$ |
| $B_7 \leq 0.2$ | $B_7 \leq 0.5$ | $B_7 \leq 10$ |
| $B_8 \leq 0.8$ | $B_8 \leq 0.5$ | $B_8 \leq 3$ |
| $B_9 \leq 0.2$ | $B_9 \leq 0.5$ | $B_9 \leq 10$ |
| $B_{10} \leq 0.8$ | $B_{10} \leq 0.5$ | $B_{10} \leq 3$ |

As we can see from Fig. 6.2.2.5 and Fig. 6.2.2.8, the dynamic aperture of the damping ring is sufficient to meet the injection requirements for the Damping Ring.

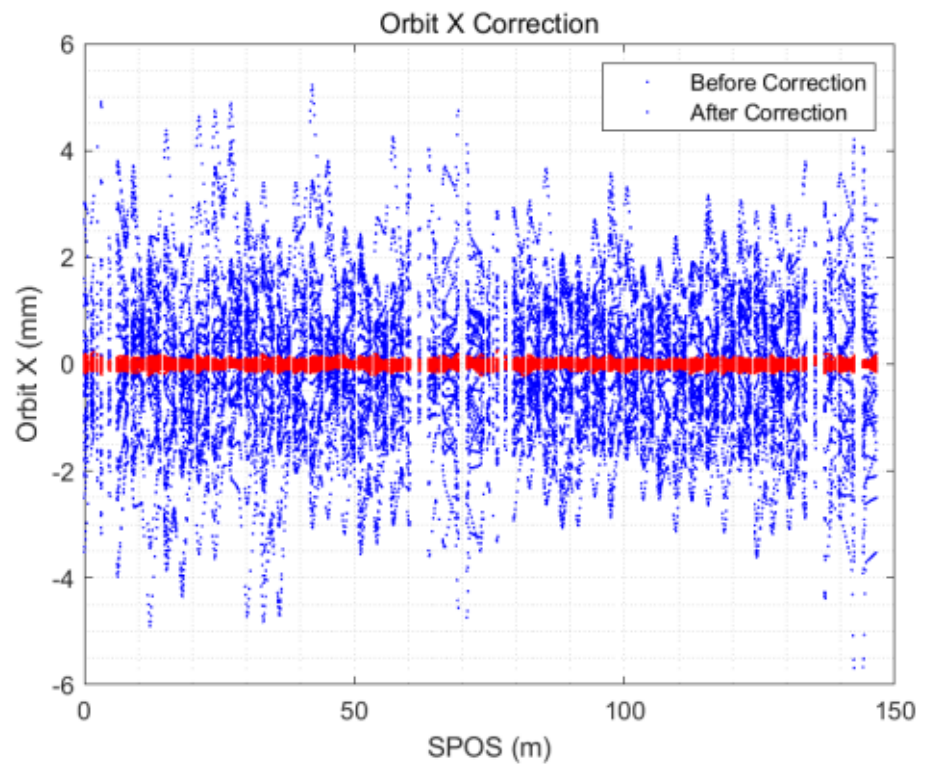

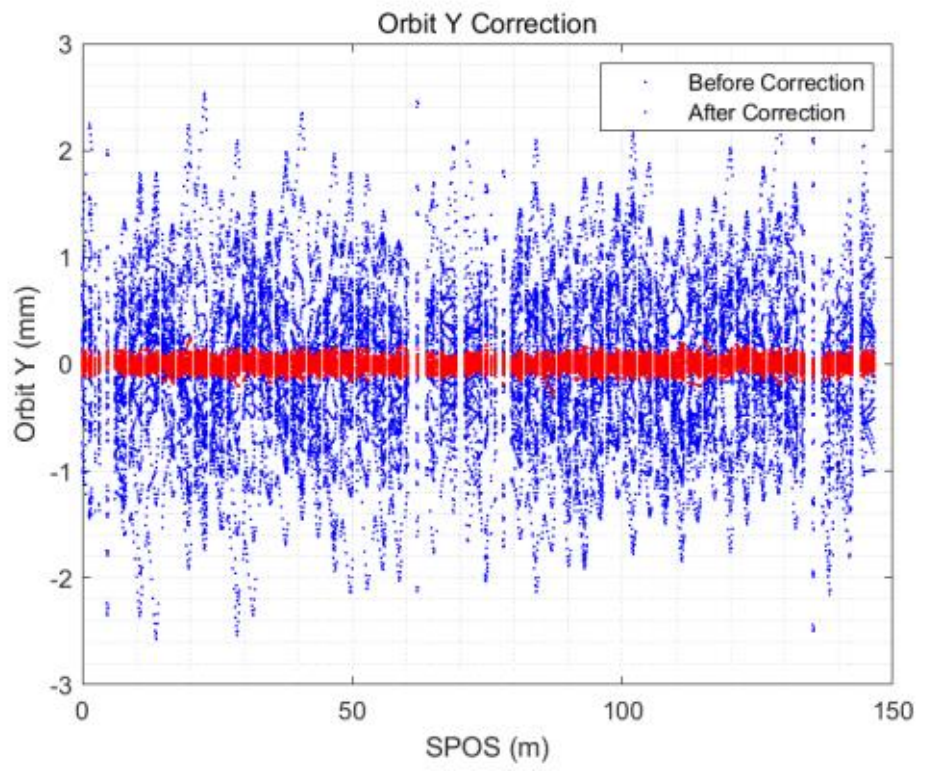

**Figure 6.2.2.6:** Damping Ring closed orbit before and after COD correction.

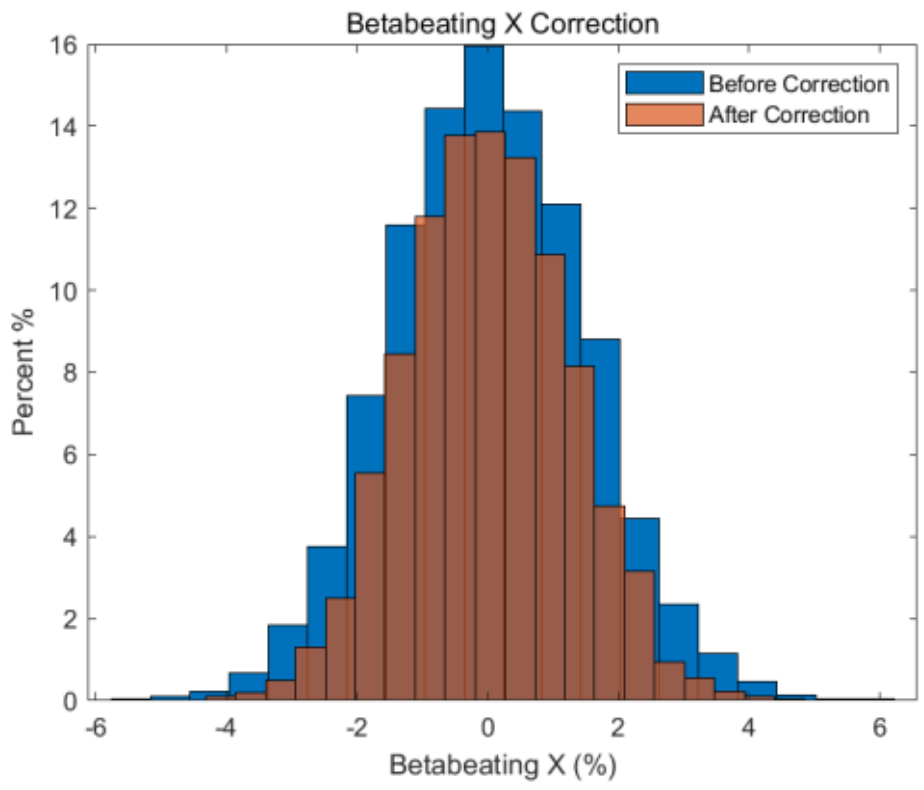

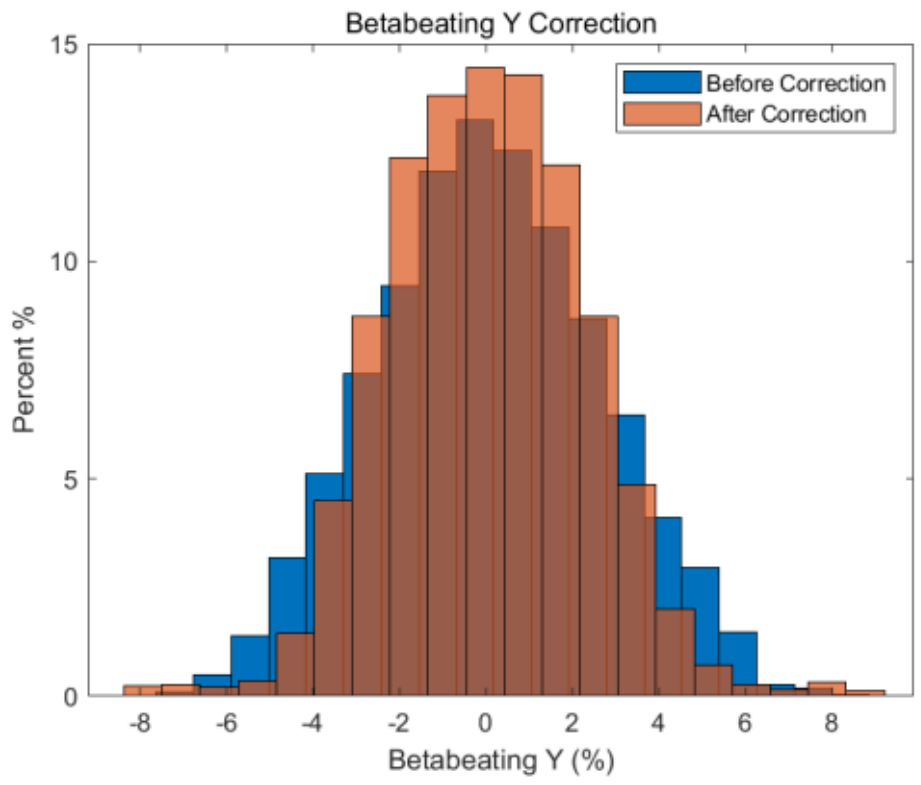

**Figure 6.2.2.7:** Damping Ring beta beating before and after COD correction.

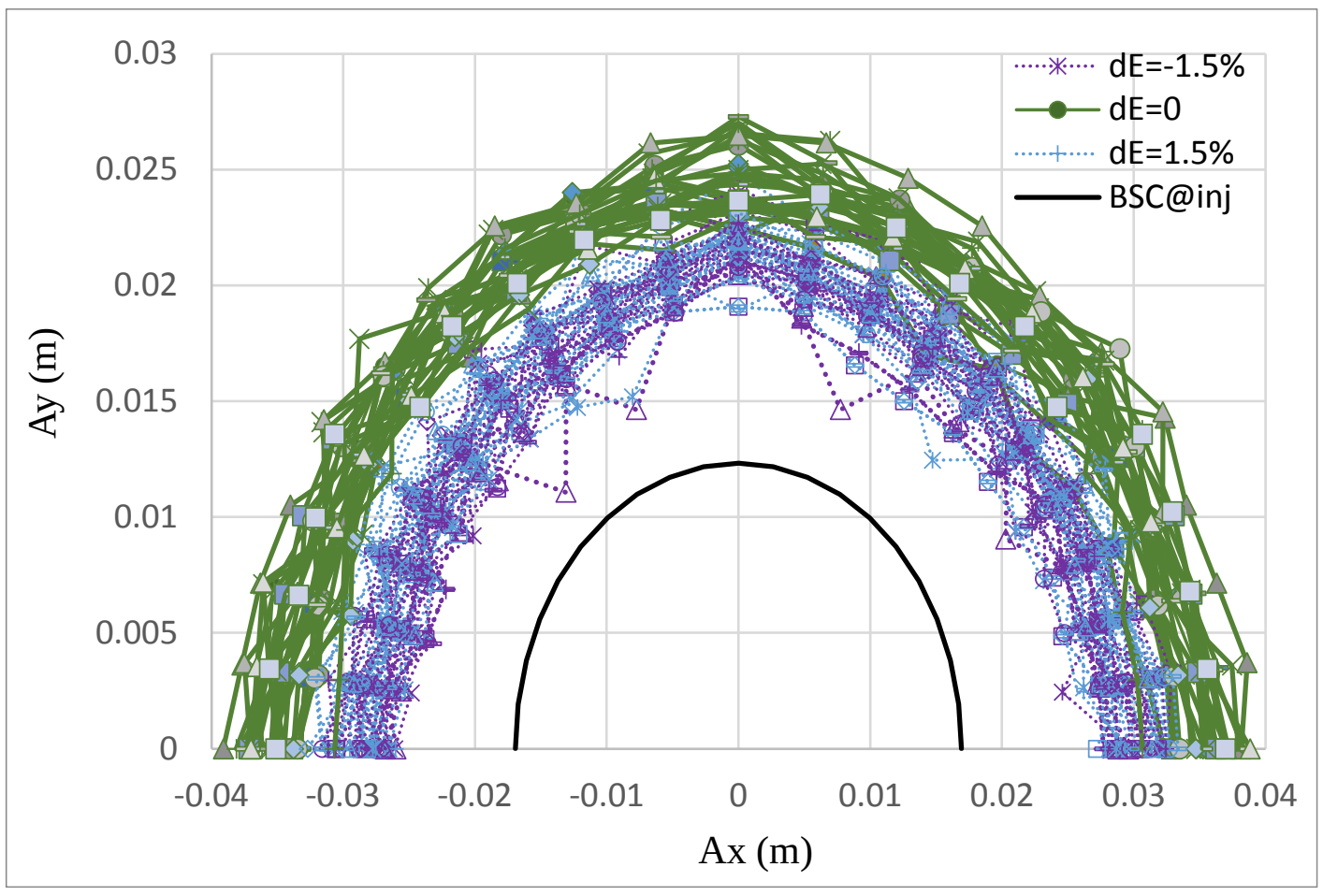

**Figure 6.2.2.8:** Dynamic aperture of the Damping Ring with errors and orbit correction ($BSC_{x,y}$=5 $\sigma_{inj_x,y}$+5mm). ±1.5% energy deviation corresponds to ±8.3 times of injected energy spread.

### 6.2.2.4 *Impedance and Beam Instability*

To reduce the emittance of the positron beam from the S-band Linac, a damping ring (DR) with an energy of 1.1 GeV and a circumference of 147 m is installed between the S-band and C-band Linac. With lower stored beam currents and only 2-4 bunches being filled in the DR, a higher impedance threshold can be achieved. The impedance thresholds are estimated analytically using the same criterion as in Sec. 5.2.2.1, which provides a rough impedance requirement [1-3].

The threshold for microwave instability is estimated using either the Boussard or Keil-Schnell criteria, which implies that the limit on the effective broadband impedance is 55.75 mΩ. The transverse mode coupling instability (TMCI) sets a limit on the transverse impedance threshold of approximately 33.85 MΩ/m. Limitations on narrow band impedance can lead to coupled bunch instabilities. A conservative assumption is that the growth rate of the coupled bunch instability for equally spaced bunches should be less than the damping time, which can be expressed as:

$$\frac{1}{\tau} = \frac{I_0 \alpha_p}{4\pi(E/e)\upsilon_s} \sum \omega_{pn}\, e^{-(\omega_{pn}\sigma_l)^2}\, Re\, Z_l\,(\omega_{pn}). \tag{6.2.2.1}$$

This sets the limit of the impedance at any resonance frequency of the higher-order-modes (HOMs) as follows:

$$\frac{f}{\mathrm{GHz}} \frac{Re\, Z}{k\Omega} e^{-(2\pi f \sigma_l/c)^2} < 239.62 \tag{6.2.2.2}$$

Similarly, in the transverse case, this gives the following limit:

$$\frac{Re\, Z}{G\Omega/\mathrm{m}} e^{-(2\pi f \sigma_l/c)^2} < 2.36 \tag{6.2.2.3}$$

The impedance requirements for the Damping Ring (DR) are consolidated in Table 6.2.2.6. The primary sources of impedance originate from various components such as vacuum chambers, RF cavities, BPMs, bellows, masks, ports of vacuum pumps, injection kickers, valves, and flanges. The number of components in the DR can be estimated by scaling the circumference of the Collider and Booster, enabling us to establish the impedance budget for different components, as detailed in Table 6.2.2.7.

**Table 6.2.2.6:** Impedance threshold in the Damping Ring

| Parameter | Value |
|---|---|
| Broadband $Z_L$ | 55.75 mΩ |
| Broadband $Z_T$ | 33.85 MΩ/m |
| Narrowband $Z_L$ | 239.62 kΩ |
| Narrowband $Z_T$ | 2.36 GΩ/m |

**Table 6.2.2.7:** Impedance budget of different components in the Damping Ring

| **Components** | $Z_{\|}/n$ (mΩ) | $\|Z_{\perp}\|$ (MΩ/m) |
|---|---|---|
| Resistive wall | 30.70 | 18.94 |
| RF cavities | -4.95 | 0.50 |
| Flanges | 13.86 | 4.69 |
| BPMs | 0.59 | 0.50 |
| Bellows | 10.89 | 4.86 |
| Pumping ports | 0.10 | 1.01 |
| kickers | 0.60 | 2.52 |
| Taper transitions | 3.96 | 0.84 |
| Total | 55.75 | 33.85 |

The resistive wall impedance varies with different vacuum pipe radius. After comparing the impedance with stainless steel, copper, and aluminum, an aluminum chamber with an aperture of 30 mm has been chosen for the Damping Ring vacuum chamber. The dominant factor limiting the bunch current is the transverse mode coupling instability (TMCI) caused by the transverse resistive wall impedance [4], as shown in Fig. 6.2.2.9. However, for the Damping Ring, the resistive wall impedance is well below the instability threshold.

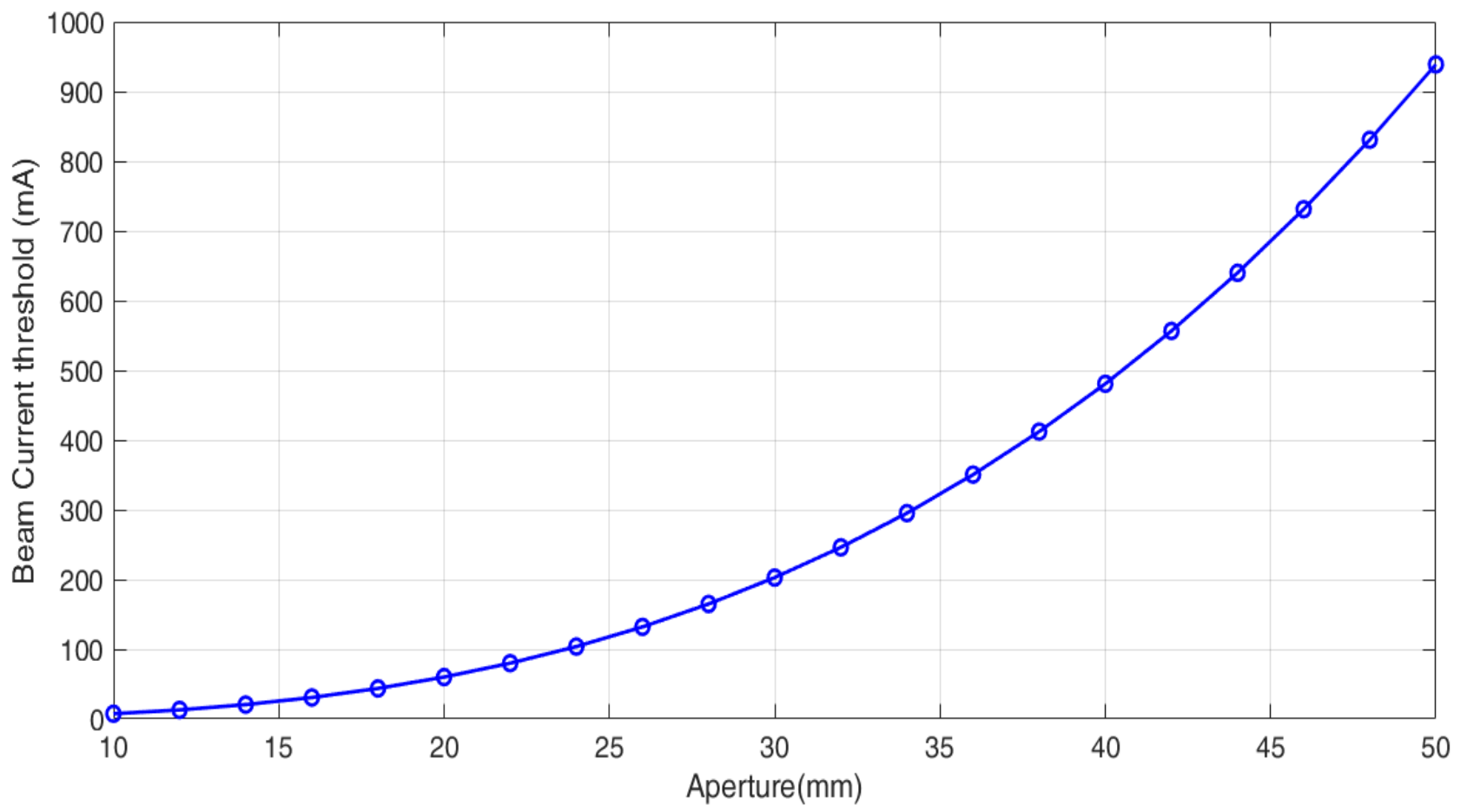

**Figure 6.2.2.9:** Beam current instability threshold *vs.* the beam pipe aperture of the DR.

### 6.2.3 Injection, Extraction and Transport Lines

#### 6.2.3.1 *Injection into Damping Ring*

The CEPC accelerator system is comprised of three main components: the Linac, the Booster, and the Collider. To ensure that the emittance of the positron beam produced by the Linac meets the Booster's requirements, a 1.1GeV Damping Ring (DR) has been incorporated into the positron linac. Figure 6.2.3.1 shows the location of the Damping Ring and Linac. The injection process from the Linac to the Damping Ring is a simple on-axis injection, and the injection section of the Damping Ring is situated in the straight line between the two arc areas. The magnet arrangement in the injection area can be seen in Figure 6.2.3.2. Extraction from the damping ring to the linac is similar to the injection process.

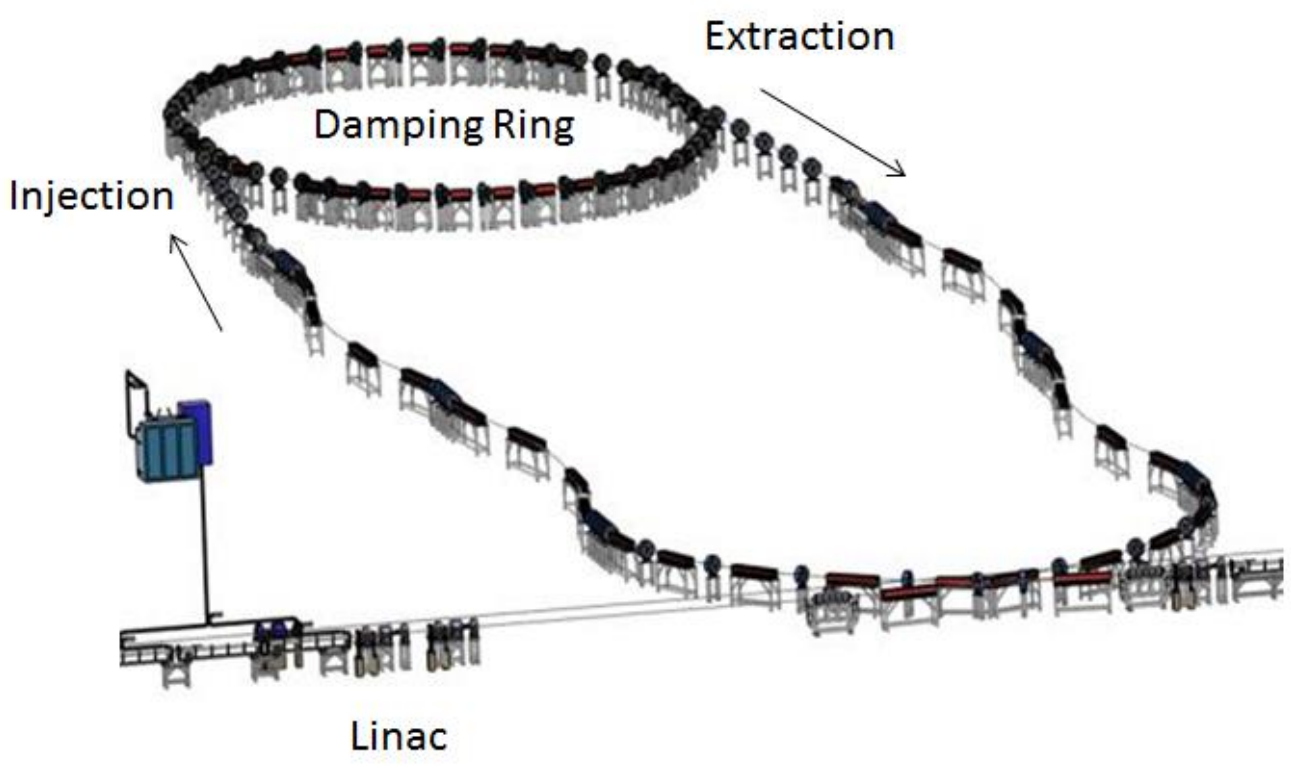

**Figure 6.2.3.1:** Layout of the Damping Ring.

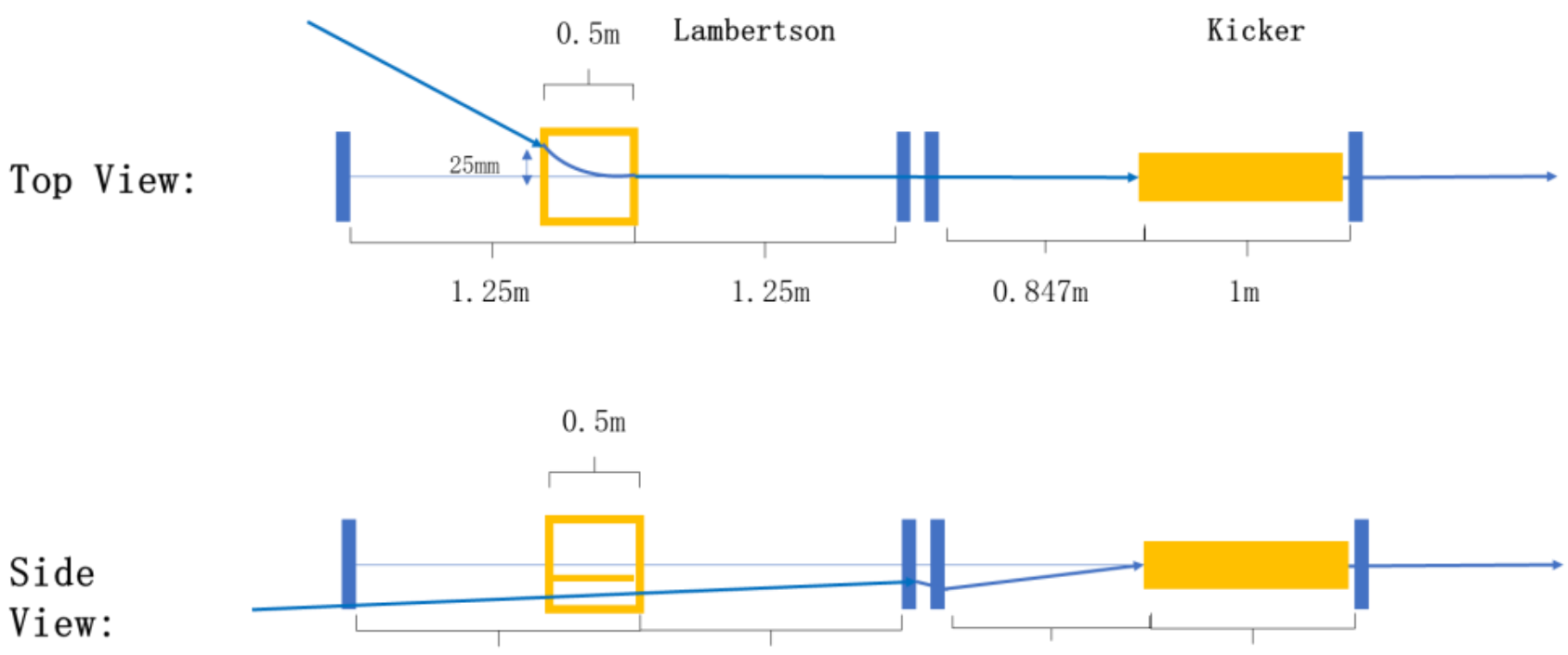

**Figure 6.2.3.2:** Magnet arrangement in the injection section of the Damping Ring

Bunches are injected into the Damping Ring at the same frequency as those in the Linac. Each bunch is damped for 20 ms before being extracted and re-injected into the linac. By holding more bunches in the Damping Ring, the damping time can be increased with minor adjustments. In the Z-energy mode, if the Linac uses a 100 Hz double bunch mode or a 200Hz mode, the injection system to the Damping Ring can operate normally by storing twice as many bunches in the Damping Ring. Figure 6.2.3.3 illustrates the injection process into the Damping Ring.

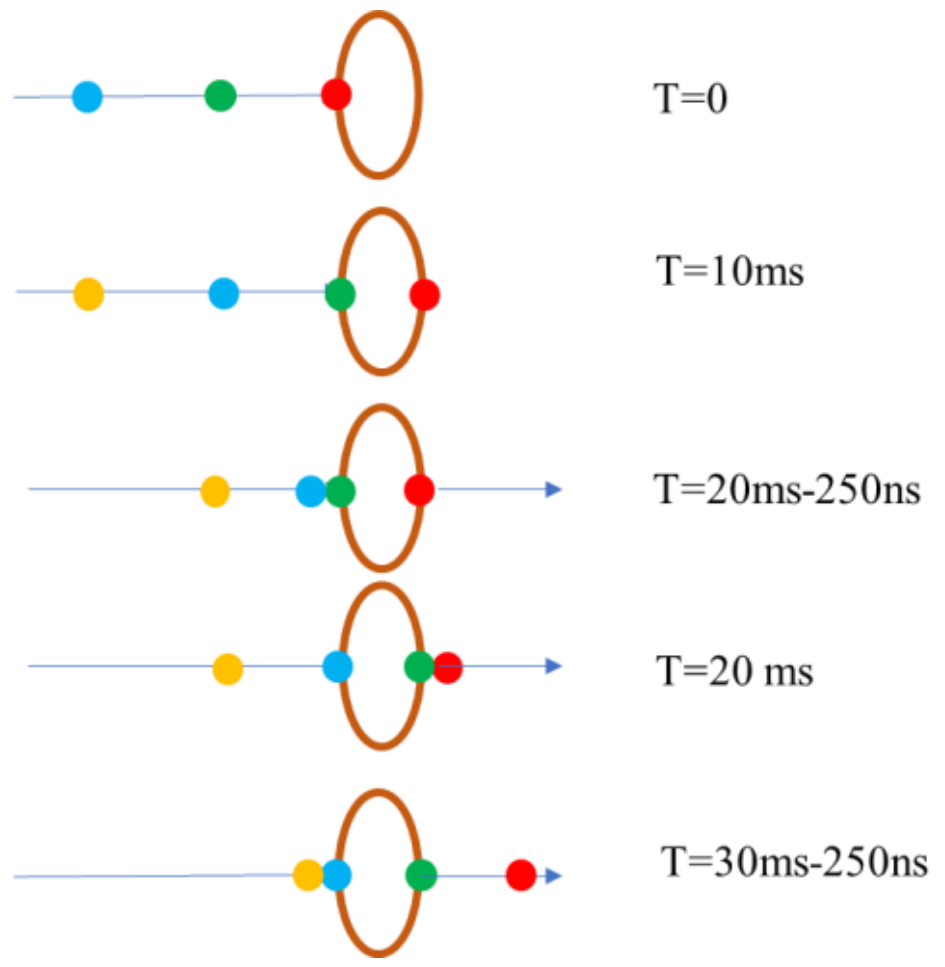

**Figure 6.2.3.3:** Process of injection into the Damping Ring

### 6.2.3.2 *Extraction from the Damping Ring*

The extraction design of the Damping Ring is almost identical to the injection design, and the extraction can be seen as the reverse process of injection.

### 6.2.3.3 *Transport Lines*

Special designs are required for the transport lines from Linac to Damping Ring and from Damping Ring to Linac to meet the injection requirements of the Damping Ring. Prior to entering the Damping Ring, the energy spread of the beam must be reduced to match the RF acceptance of the DR. After exiting the Damping Ring, the bunch length of the beam must be reduced to control the energy spread during acceleration.

The transport line from the Linac to the DR consists of three sections: two chicanes for energy spread compression, an arc section for horizontal deflection, and a matching section for matching the twiss parameters to the DR. Figure 6.2.3.4 displays the twiss parameters of the transport line. The transport line structure from the Damping Ring to the Linac is identical, with only the deflection angle of the chicanes being modified to control the bunch length of the beams being re-injected into the Linac.

The injection and extraction transport lines of the DR have an approximate length of 120 meters each. The primary components incorporated in these transport lines are presented in Table 6.2.3.1. To simulate the beam's trajectory from the DR to the Linac, we utilized the tracking code Elegant. The resulting emittance changes after passing through the transport line are illustrated in Figure 6.2.3.5.

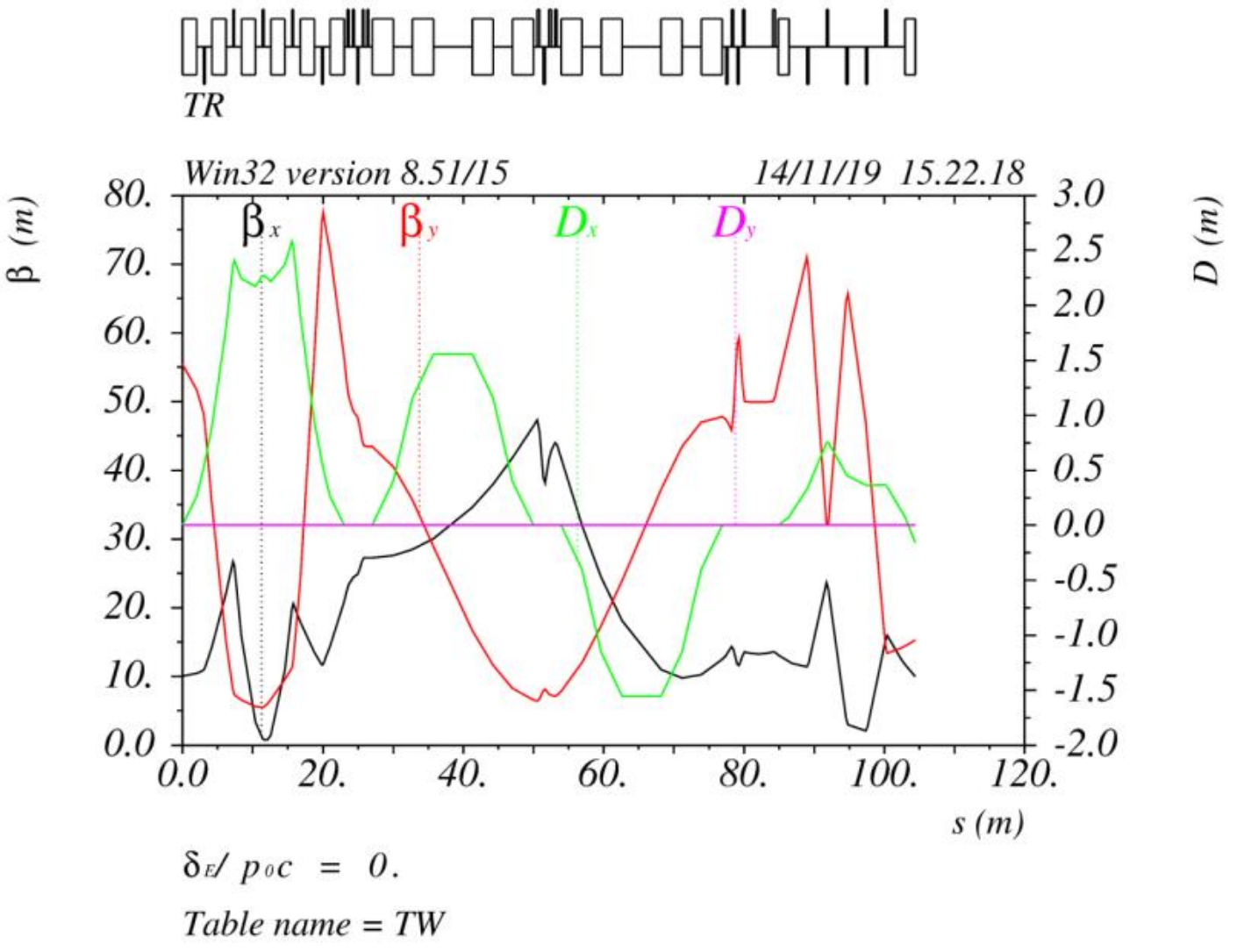

**Figure 6.2.3.4:** Twiss parameters of the transport line from the Linac to the Damping Ring.

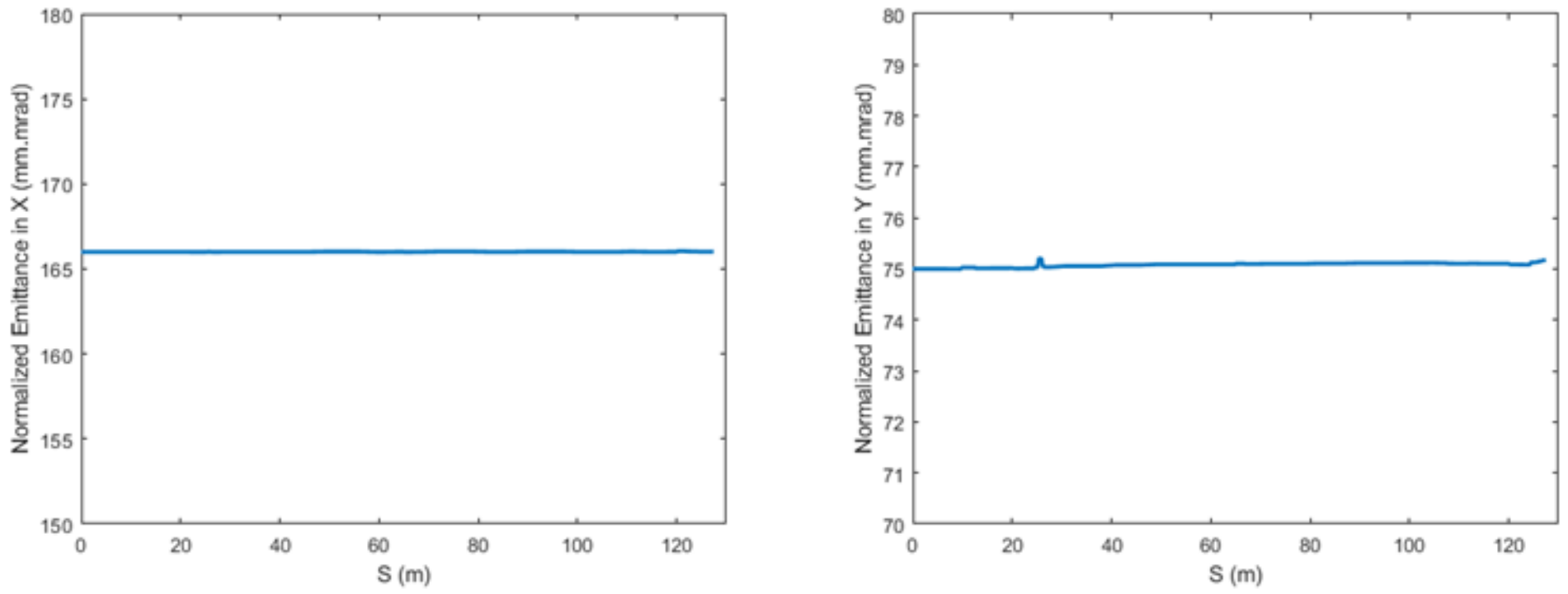

**Figure 6.2.3.5:** Emittance changes along the transport line. The left image shows the emittance in the x direction, and the right image shows the changes in the y direction.

**Table 6.2.3.1.** Main components in the transport line

| | Number | Length (m) |
|---|---|---|
| Dipoles | 14 | 2 |
| Quadrupoles | 24 | 0.3 |
| Correctors | 4 | 0.3 |
| BPMs | 3 | |
| Profile | 1 | |

# 6.3 Linac Technical Systems

## 6.3.1 Electron Source

The electron source is responsible for generating an electron beam with a specific longitudinal distribution that can be utilized for acceleration, injection, and positron production. To cater to the needs of the CEPC Linac, the electron source must operate in two modes: the first mode generates an electron beam with a 1.5 nC bunch charge for electron injection, and the second mode produces a primary electron beam with a 10 nC bunch charge for positron production.

To satisfy the design requirements, a traditional thermionic electron gun has been chosen for electron beam generation. Similar to the BEPC II and KEKB Linacs [1-2], the electron gun in CEPC Linac comprises a cathode-grid assembly, a beam focusing electrode, an anode, a solid-state modulator, a filament power supply, a pulser, a bias power supply, an isolation transformer, and a control system. Figure 6.3.1.1 illustrates the schematic of the electron gun for CEPC Linac, and the main parameters are presented in Table 6.3.1.1.

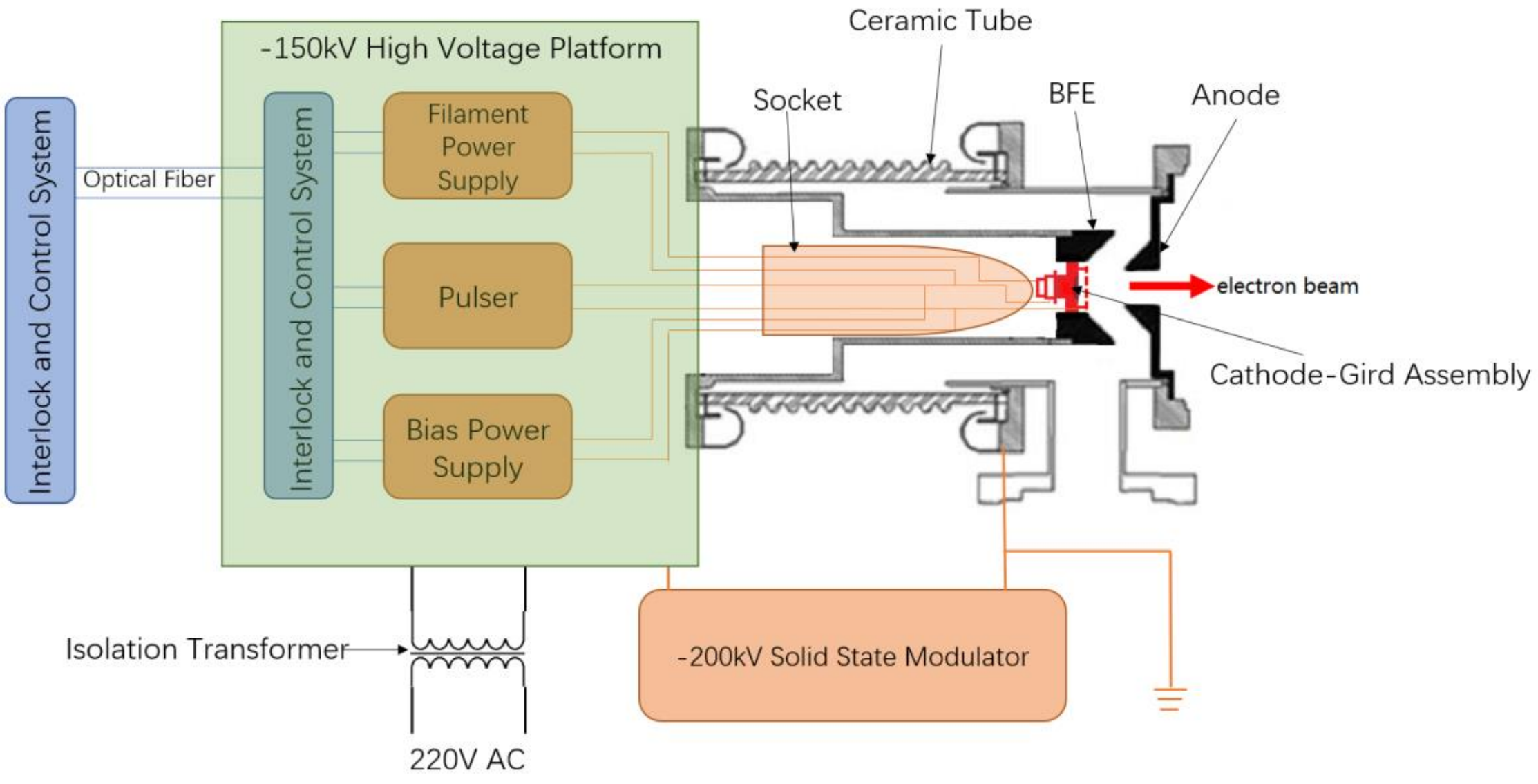

**Figure 6.3.1.1:** Schematic of the thermionic electron gun.

**Table 6.3.1.1:** Main parameters of the electron gun

| Parameter | Unit | Value |
|---|---|---|
| Type | - | Thermionic Triode Gun |
| Cathode | - | Dispenser cathode |
| Beam current | A | > 10 |
| High voltage of anode | kV | 150 |
| Bunch charge 1 | nC | 1.5 ($e^-$ injection) |
| Bunch charge 2 | nC | 10 ($e^+$ production) |
| Repetition | Hz | 100 |

### 6.3.1.1 *Thermal Emission Cathode*

The EGUN code, which is a program for electron beam optics and gun design [3], is utilized to optimize the shape and dimensions of the focusing electrode and anode. The electron trajectories resulting from the optimization process are presented in Figure 6.3.1.2. The beam trajectory is designed to be nearly parallel to achieve low emittance. At a distance of 100 mm from the cathode surface, the x and y emittance values are determined to be 17.84π mm-mrad for a 150 kV high voltage and 10 A beam current. Moreover, the EGUN predicts a beam perveance of 0.169 μP.

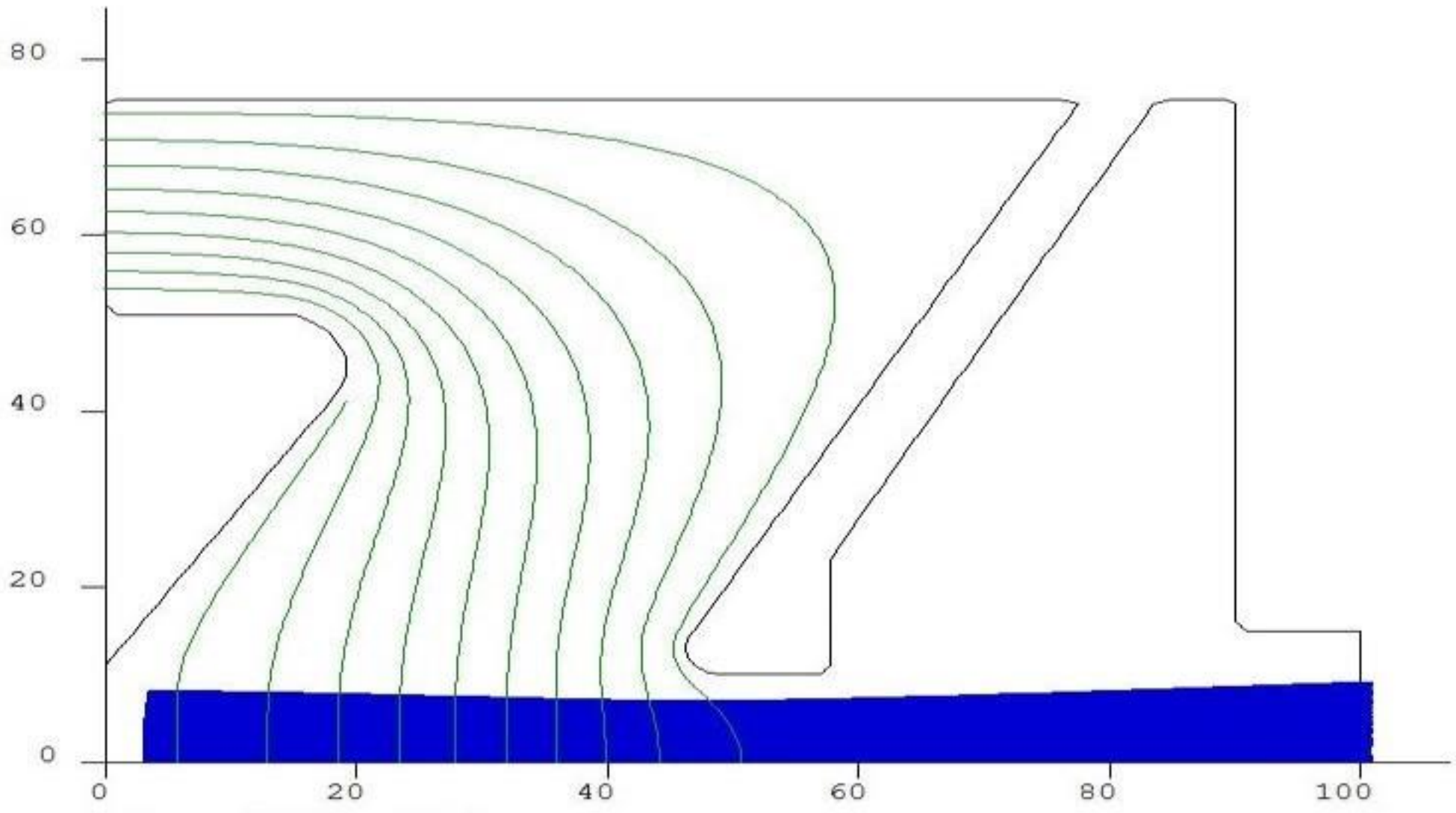

**Figure 6.3.1.2:** Beam trajectory in the electron gun.

The cathode-grid assembly is a crucial component of the electron gun, and its selection is based on its emission current capacity and the anticipated lifetime of the cathode. In recent years, a prototype cathode-grid assembly has been developed at IHEP, which is depicted in Figure 6.3.1.3. This assembly features a cathode area measuring 2 $cm^2$, with a diameter of 16 mm. It is capable of supplying a high current of up to 12 A while maintaining an extended operational lifetime.

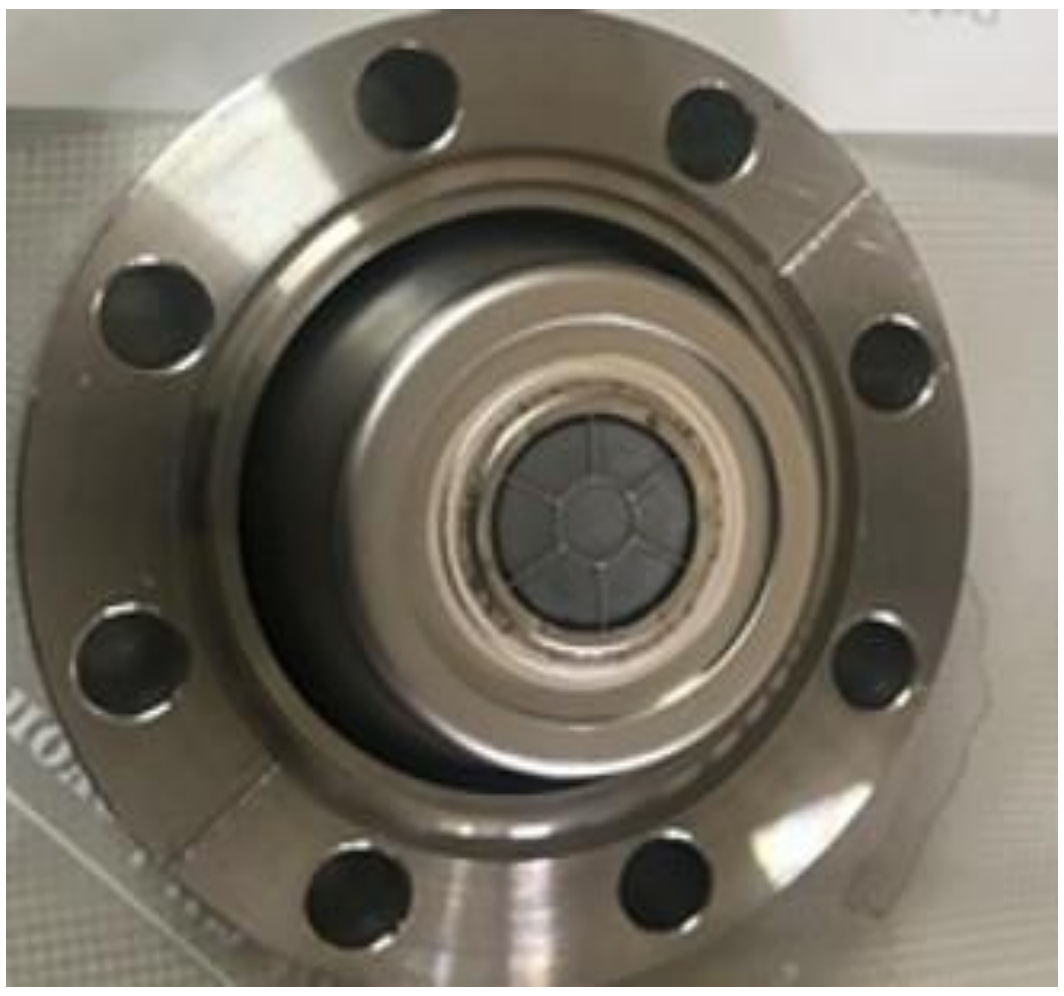

**Figure 6.3.1.3:** A Cathode-grid assembly prototype developed in China.

To generate an electron beam with a large bunch charge (≥10 nC) for positron production, the cathode-grid assembly must have adequate emission capacity and a long lifetime for continuous operation. Experimental tests have been conducted to validate the emission capacity and lifetime of the cathode-grid assembly prototype. As presented in Figures 6.3.1.4 and 6.3.1.5, the emission current of the domestically produced cathode-grid assembly prototype is comparable to that of imported cathodes, and the cathode emission capacity does not deteriorate within 7000 hours of continuous operation. These results indicate that the cathode-grid assembly prototype can fulfill the requirements of the CEPC electron source.

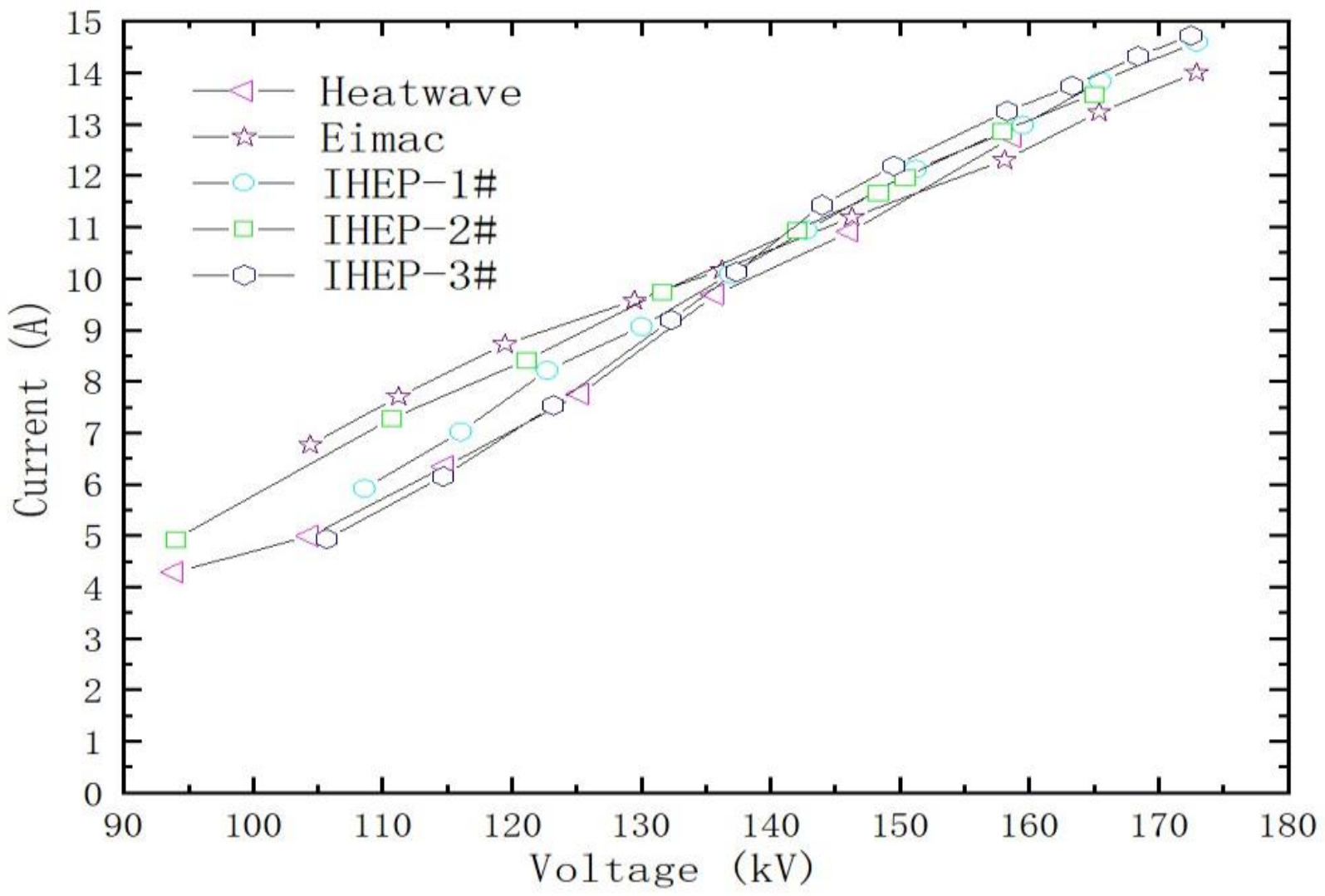

**Figure 6.3.1.4:** Test results of the cathode emission capacity.

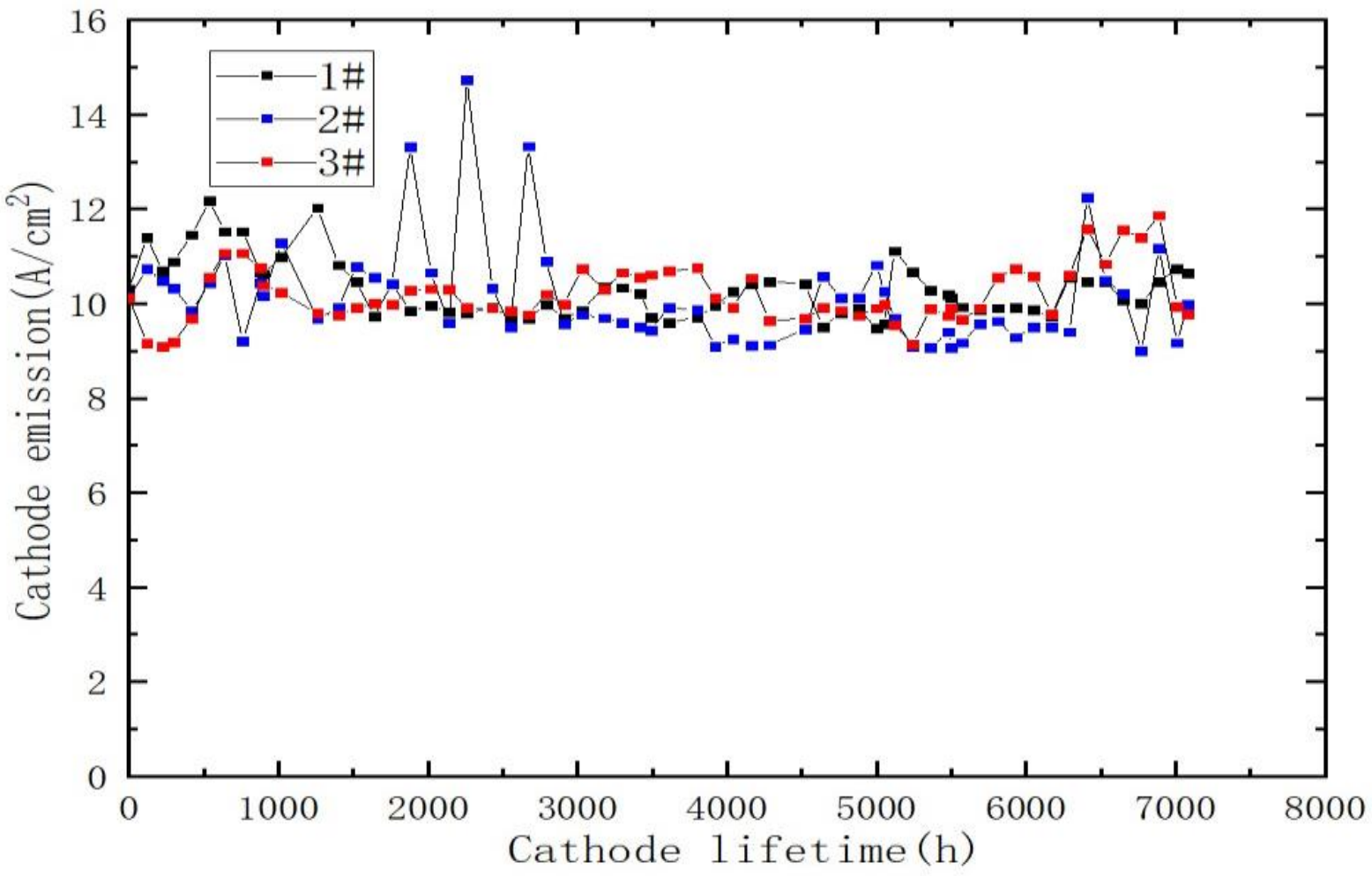

**Figure 6.3.1.5:** Test results of the cathode lifetime.

### 6.3.1.2 *Pulser System*

The electron gun utilized in the CEPC Linac is a thermionic triode gun that consists of a cathode-grid structure with a very short inter-electrode spacing. By utilizing a high-voltage (HV) nanosecond pulser in conjunction with the short inter-electrode spacing, it is possible to generate single or multiple nanosecond electron beam pulses from the gun. The pulser system is a critical element, as it determines the longitudinal bunch length of the emitted electron beam. Table 6.3.1.2 outlines the specifications of the pulser system, which can supply a high voltage of up to 1000 V from an internal power supply with a rise and fall time of <0.8 nanoseconds. The coaxial cable functions as a pulse forming line and energy storage device that is connected to the end of the switch. The pulse width is determined by the length of the coaxial cable, which is matched to a pulse width range of 1 ns to 10 ns.

**Table 6.3.1.2:** Specifications for the pulser system.

| Parameter | Unit | Value |
|---|---|---|
| Pulse voltage | V | 1000 |
| Pulse width | ns | 1-10 |
| Rise time | ns | 0.8 |
| Polarity | | Negative |
| Jitter | ps (RMS) | 20 |
| AC power supply | V | 220 (–10% ~ +10%) |

The pulser system for the CEPC Linac comprises a DC power supply, a control box, and the pulser itself, which are all installed in a high voltage station. The system's configuration is presented in Figure 6.3.1.6. The DC power supply has a voltage range of 0 to 1 kV. The pulser stores and discharges energy with a switch, while the control box regulates the amplitude of the trigger pulse. It samples the trigger pulse, monitors the temperature of the connected structure, processes the trigger signal, and provides an interface between the optical and electrical signals.

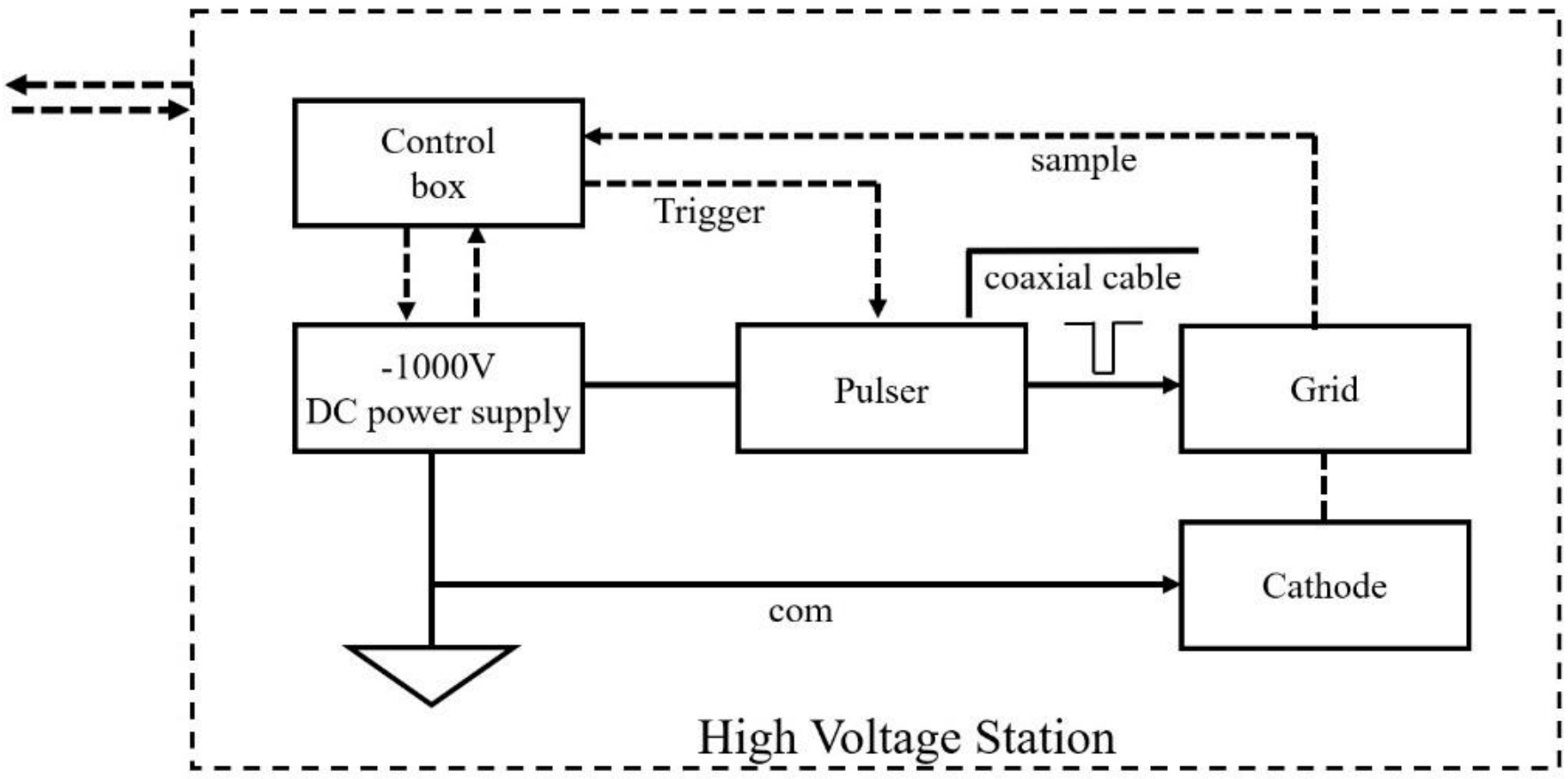

**Figure 6.3.1.6:** Pulser system schematics.

### 6.3.1.3 *High Voltage System*

The electron beam emitted from the thermal cathode is accelerated by a static electric field that is applied between the cathode and anode electrodes.  In the design of the CEPC Linac electron gun, a solid-state modulator capable of delivering a maximum output of 200 kV is being considered. The objective is to provide a stable high-voltage pulse of 150 kV with a pulse width of 2 μs and a repetition rate of 100 Hz between the cathode and anode electrodes. To ensure that the electric field distribution within the gun remains below the breakdown threshold, a simulation was conducted using CST software [4].

For the operating voltage of 150 kV, the surface electric field limit within the electron gun is approximately 10 MV/m for DC operation. In the case of pulsed operation, the limit can be extended up to around 20 MV/m [5]. The simulation results, shown in Figure 6.3.1.7, indicate that when the electron gun is operated at 150 kV, the maximum surface electric field is approximately 12.5 MV/m. This value is significantly lower than the allowable limit of 20 MV/m.

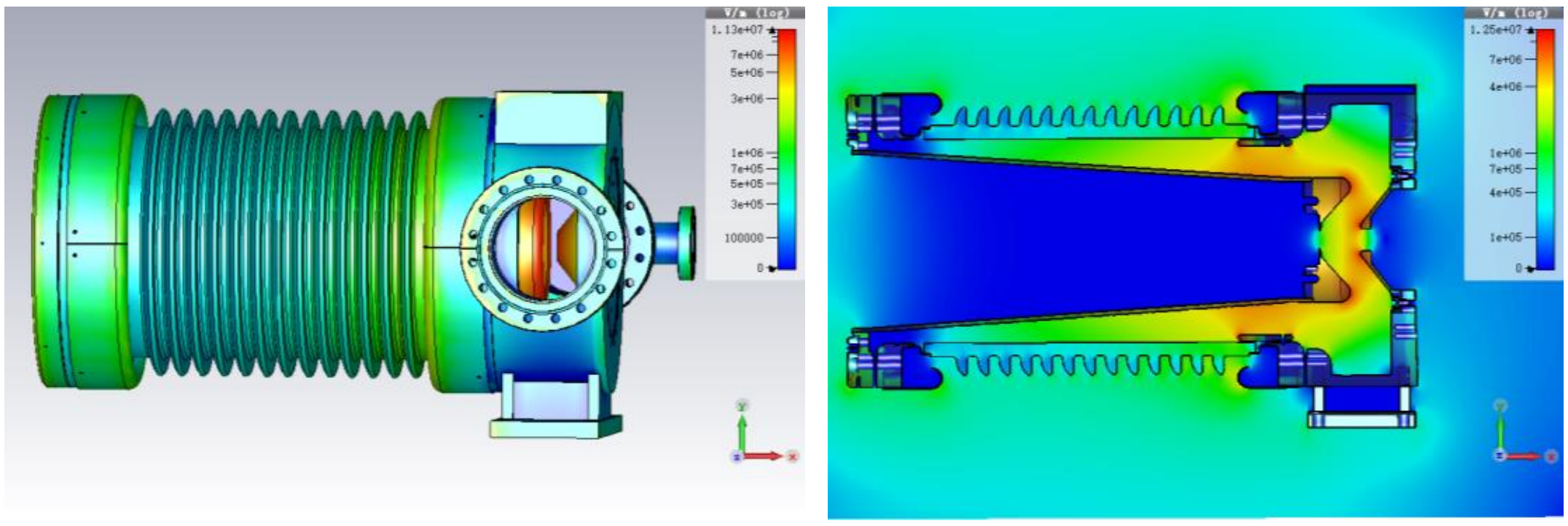

**Figure 6.3.1.7:** Simulation of the E-field distribution in electron gun.

The electron gun's high voltage system comprises a high voltage power supply, a high voltage station, a pulser, a filament power supply, a bias power supply, and a control unit. The primary parameters of the high voltage system are displayed in Table 6.3.1.3. The high voltage system's schematic diagram is presented in Fig.6.3.1.8.

**Table 6.3.1.3:** Main parameters of the high voltage system.

| Parameters | Value |
|---|---|
| Anode High Voltage (kV) | 150 |
| Filament voltage (V) | 6~8 |
| Filament current (A) | 5~7.5 |
| Grid bias voltage (V) | 0~200 |
| Pulse width (ns) | 1 |
| Pulse repetition (Hz) | 100 |

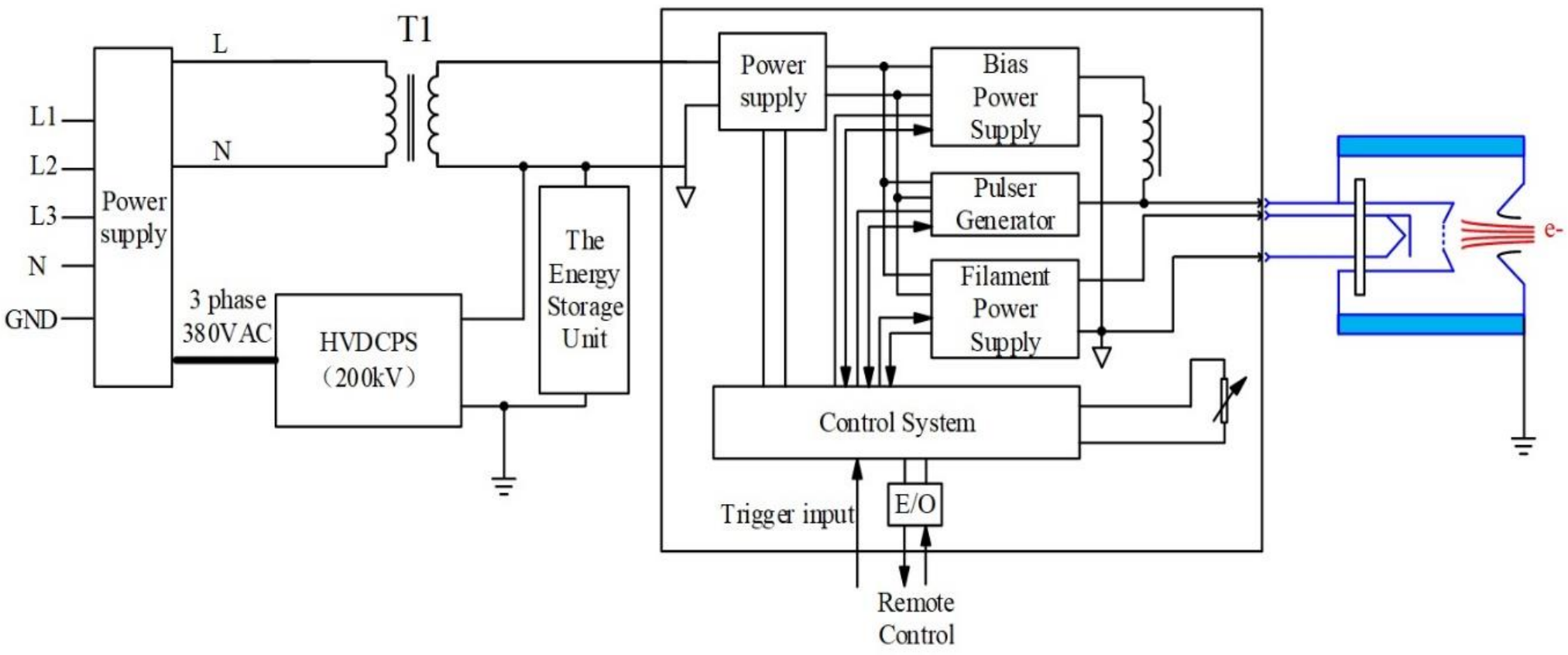

**Figure 6.3.1.8:** Schematic diagram of the high voltage system.

### 6.3.2 Positron Source

The CEPC Linac positron source is responsible for generating the positron beam necessary for experiments. The process of generating positrons uses a conventional scheme where a high energy electron beam collides with a high-Z, high-density converter target of a certain thickness to produce positrons in electromagnetic showers [1-2]. The number of positrons produced per incident electron is roughly proportional to the electron energy, which means that the positron current is proportional to the power of the incident electron beam. To achieve a 1.5 nC bunch charge for the positron beam, a primary electron beam of 4 GeV with an intensity of 10 nC/bunch is required. This amounts to an average beam power of 4 kW at a repetition rate of 100 Hz.

#### 6.3.2.1 *Target*

The design of the positron source target has been studied using G4beamline and FLUKA [3-4]. The optimization of positron yield at the target exit was done by scanning the tungsten target thickness at different electron beam energies, as shown in Fig. 6.3.2.1. The simulation results indicate that the target length between 15 mm and 18 mm has a high positron yield with an electron beam energy of 4 GeV [5].

To consider the peak energy deposition density (PEDD) on the target, the root-mean-square (RMS) beam size of the electron beam is set at 0.5 mm. The target length is adopted as 15 mm, taking into account the energy deposition and cooling requirements. The simulation result of energy deposition is shown in Fig. 6.3.2.2, with a total energy deposition of 0.784 GeV/$e^-$, which corresponds to a power deposition of approximately 784 W. Water cooling is necessary for the target, which is a cylindrical tungsten target embedded in a copper block with a cooling system that can support and cool the target. The intervals of successive shock waves in the CEPC positron source target are 10 ms, and the shock wave diffuse time is about 3.3 μs. Since the interval time is far larger than the shock wave diffuse time, the shock wave effect and thermal dynamics can be damped.

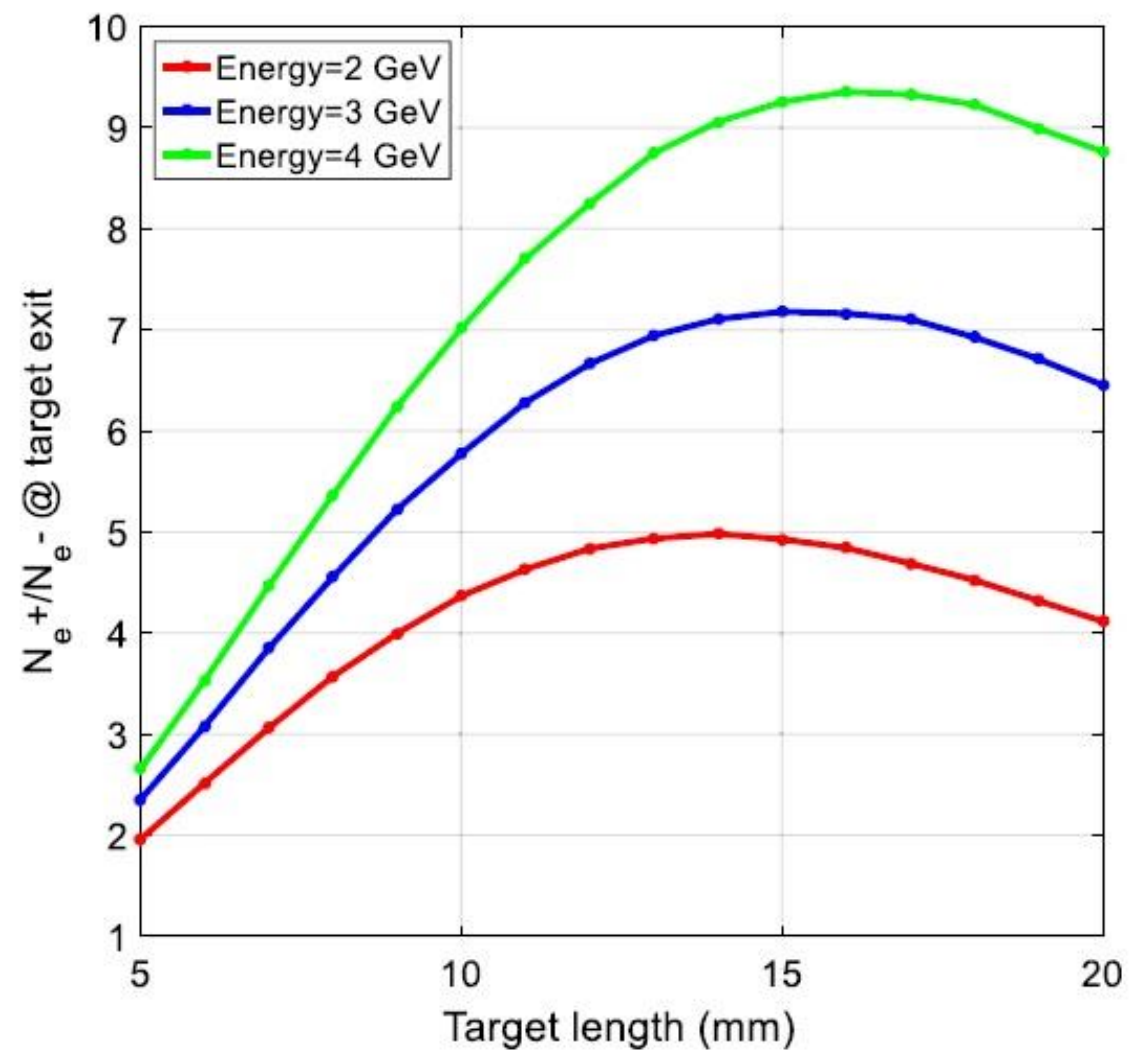

**Figure 6.3.2.1:** Positron yield with different target length and electron energy.

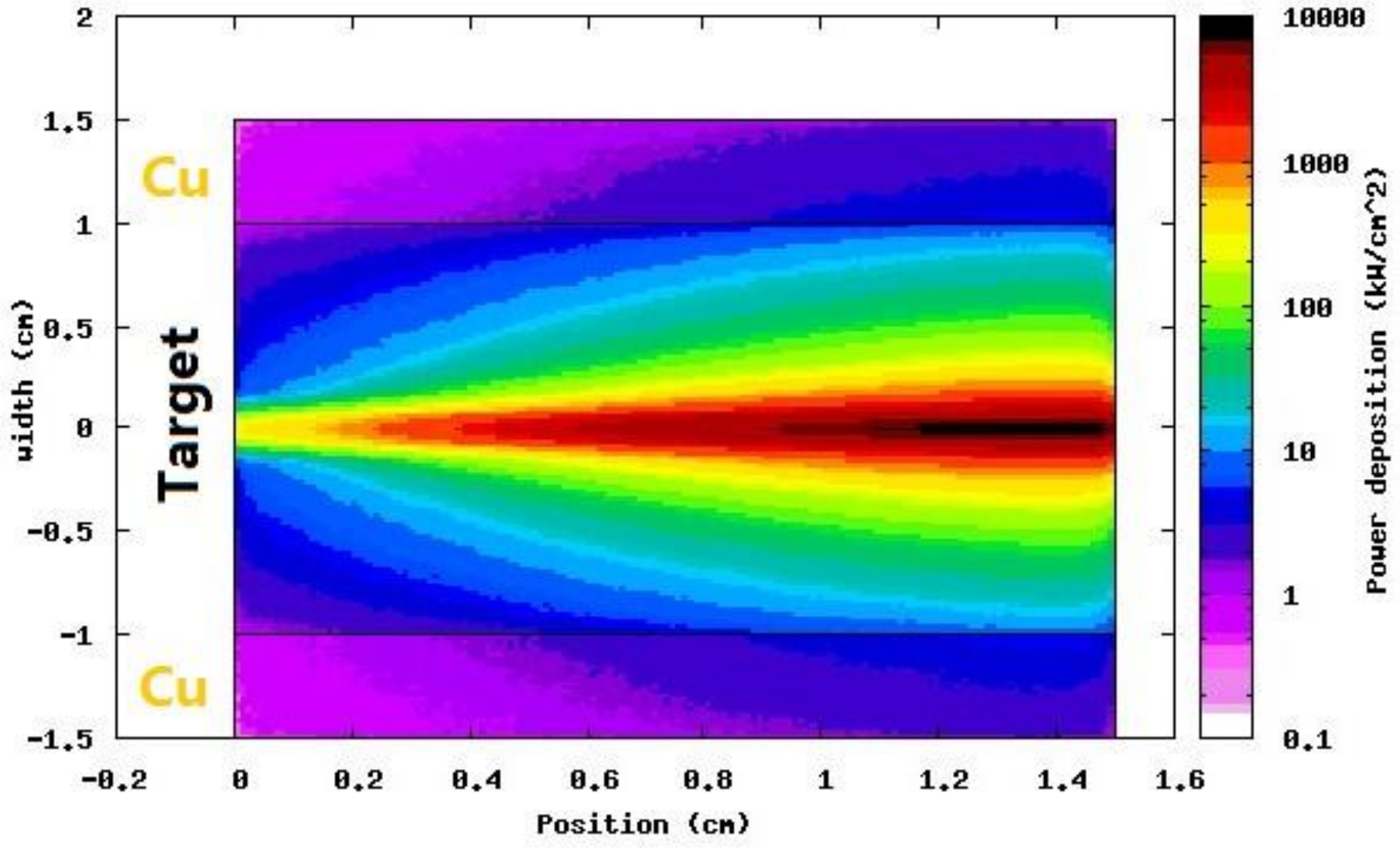

**Figure 6.3.2.2:** Energy deposition on the target.

### 6.3.2.2 *Flux Concentrator*

The target used in the CEPC Linac positron source emits electrons, positrons, and photons across a wide energy spectrum. At the target exit, the positron beam has a small transverse beam size and a high transverse divergence, which necessitates sufficient collimation to obtain a low transverse divergence that matches with the subsequent section.

To improve the beam performance at the downstream capture section, an adiabatic matching device (AMD) is used to transform the transverse phase space by approximately 90 degrees. This transformation results in a larger beam size and a smaller divergence, which is beneficial for downstream capture. The flux concentrator, as shown in Figure 6.3.2.3, is used to achieve this transformation.

Figure 6.3.2.4 shows the solenoid field distribution downstream of the target. The magnetic field changes from a peak value of 6 T to a constant value of 0.5 T, which is produced by the superposition of the flux concentrator and a DC solenoid. Figure 6.3.2.5 shows the beam distribution at the entrance and exit of the AMD.

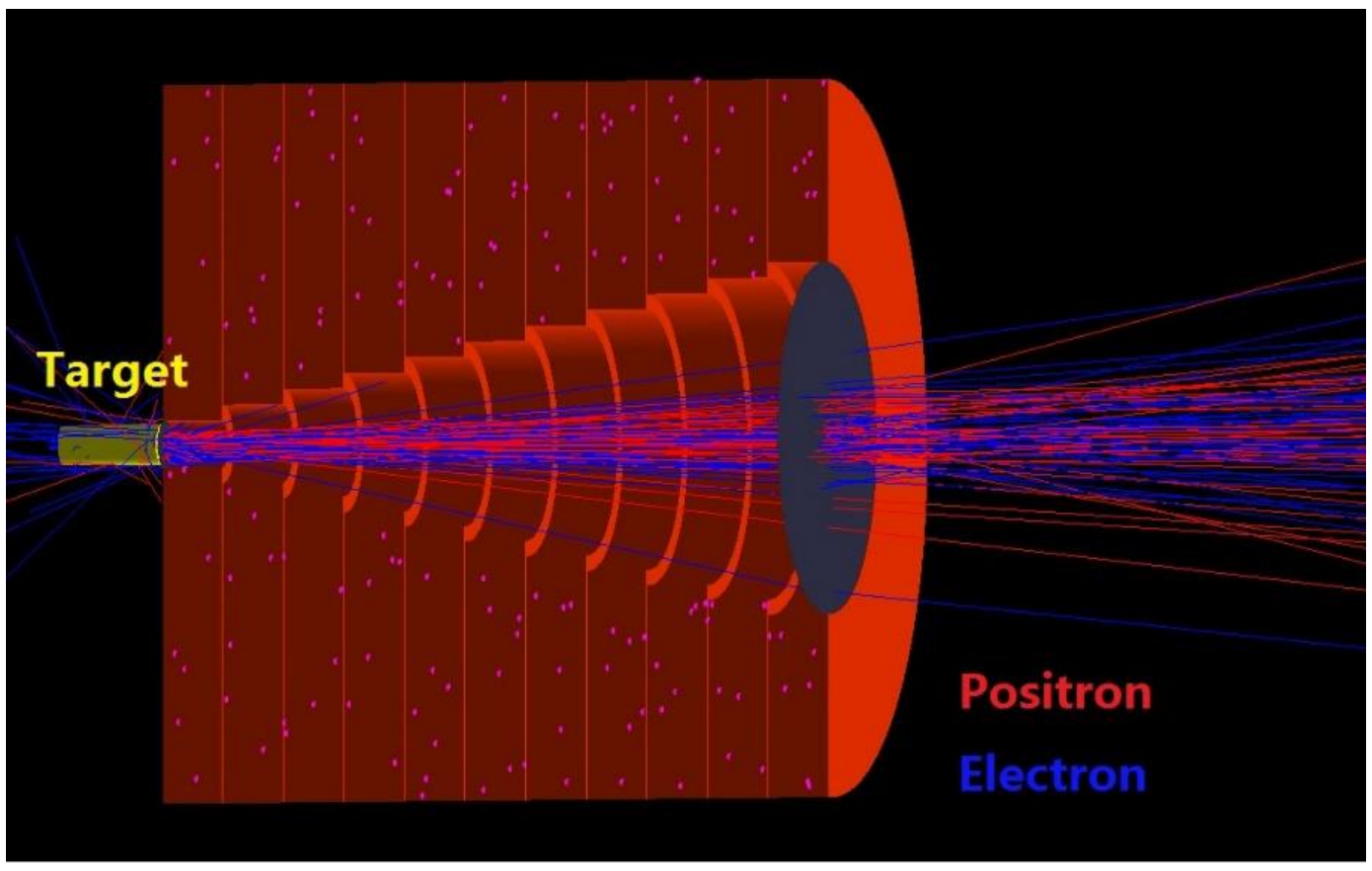

**Figure 6.3.2.3:** Layout of the target and the AMD.

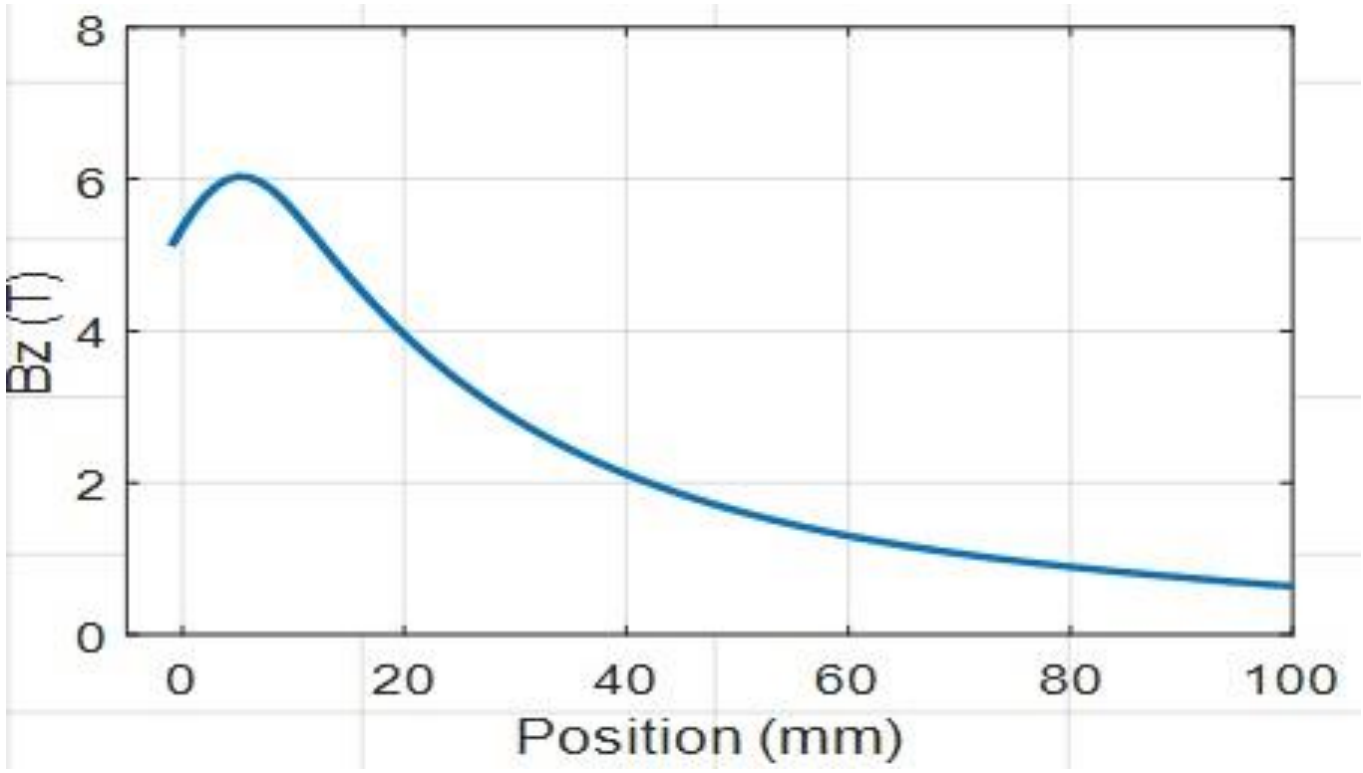

**Figure 6.3.2.4:** Magnetic field of the flux concentrator.

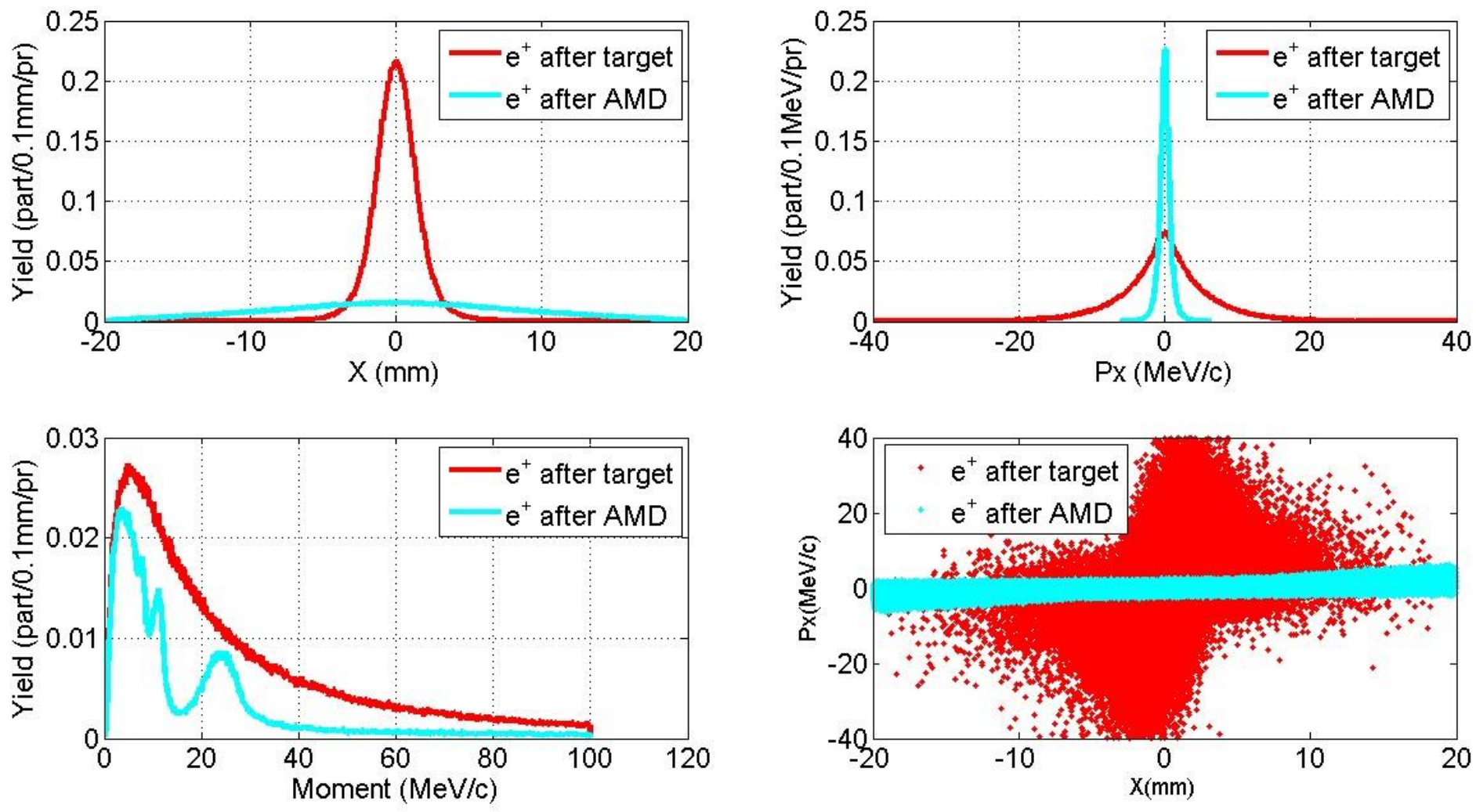

**Figure 6.3.2.5:** Beam transformation by the AMD section.

The flux concentrator is an adiabatic matching device that is placed between the target and the accelerating structure. Its purpose is to produce a magnetic field that has a sharp rise over a distance of less than 5 mm, and then falls off adiabatically over a length of 10 cm. To design the magnetic field of the flux concentrator, simulations were carried out using the Opera software with transient TR modules. The design was based on the positron sources at SLAC and BEPC [6-7], and it comprised a trumpet-shaped copper coil with 12 turns. The coils were separated by a distance of 0.2 mm, and the inner diameters were gradually varied from 7 to 52 mm, while the outer diameters were fixed at 106 mm. The total length of the concentrator was 100 mm. Since the concentrator exhibited rotational symmetry, a two-dimensional model was used in the simulation. The magnetic field was excited by half-sine current pulses with a peak current of 15 kA, a bottom width of 5 μs, and an output time of 2.5 μs. The simulation results showed that the maximum magnetic field on the central axis was 6.3 T, which fulfills the physical design requirement. The simulation results are presented in Fig. 6.3.2.6.

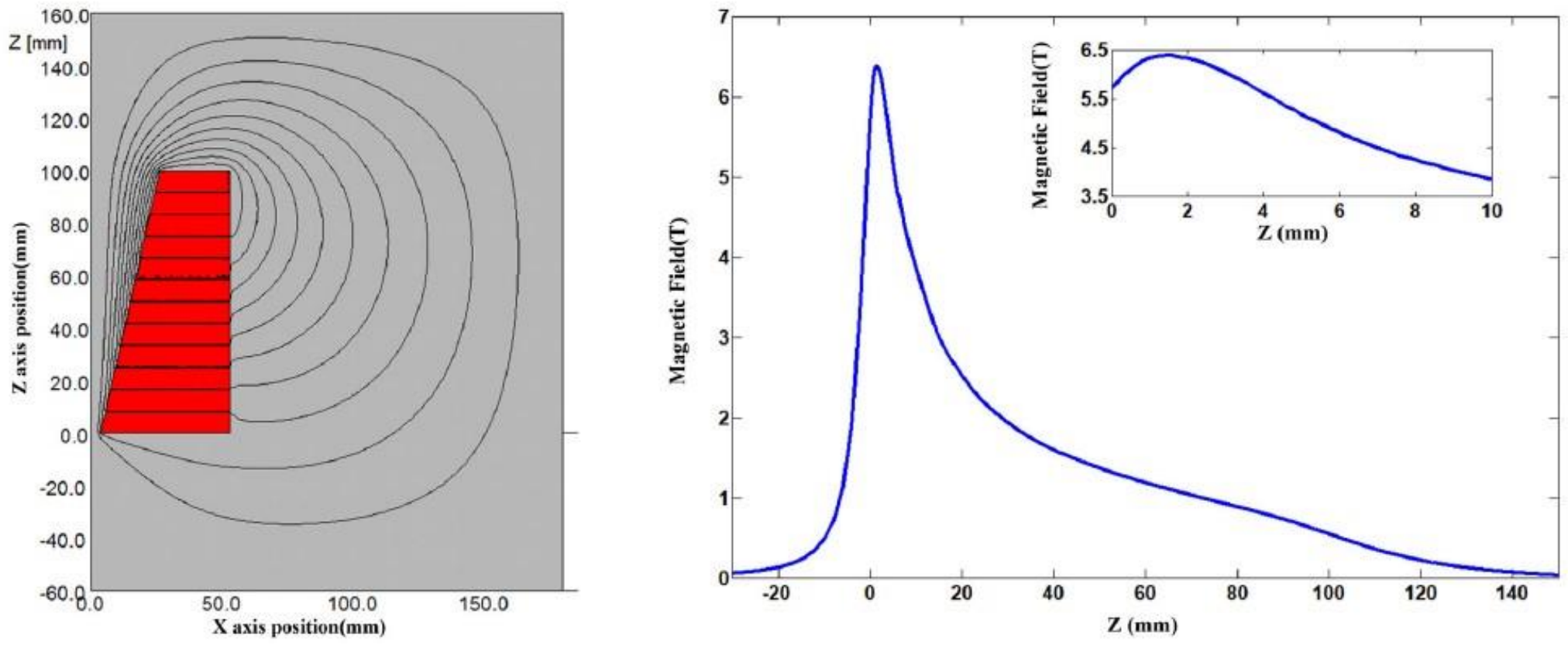

**Figure 6.3.2.6:** Simulation of the magnetic field generation from a flux concentrator.

A flux concentrator prototype has been successfully developed and tested using a 15 kA solid-state modulator, as illustrated in Figure 6.3.2.7. The measured and simulated magnetic fields as a function of the axial position of the flux concentrator were shown in Figure 6.3.2.8, with a peak current of 15 kA. The obtained peak magnetic field of 6.2 T satisfies the CEPC positron source's requirements (>6 T). The measured curve aligned with the designed parameters and was consistent with the simulated results [6].

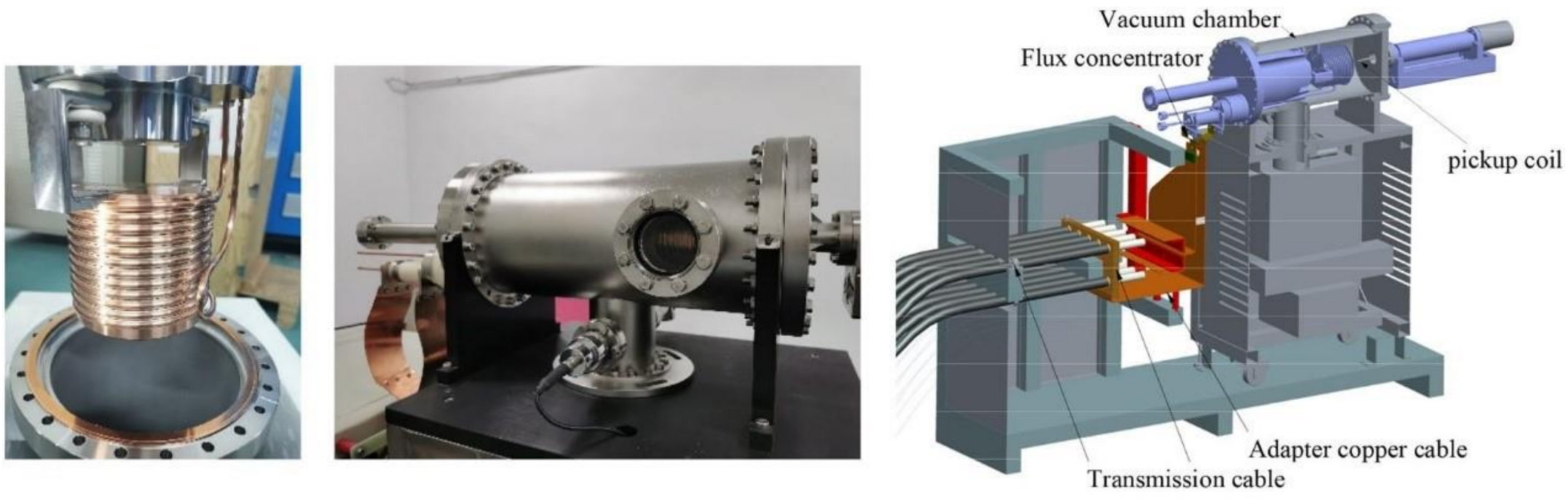

**Figure 6.3.2.7:** Flux concentrator prototype and measurement device.

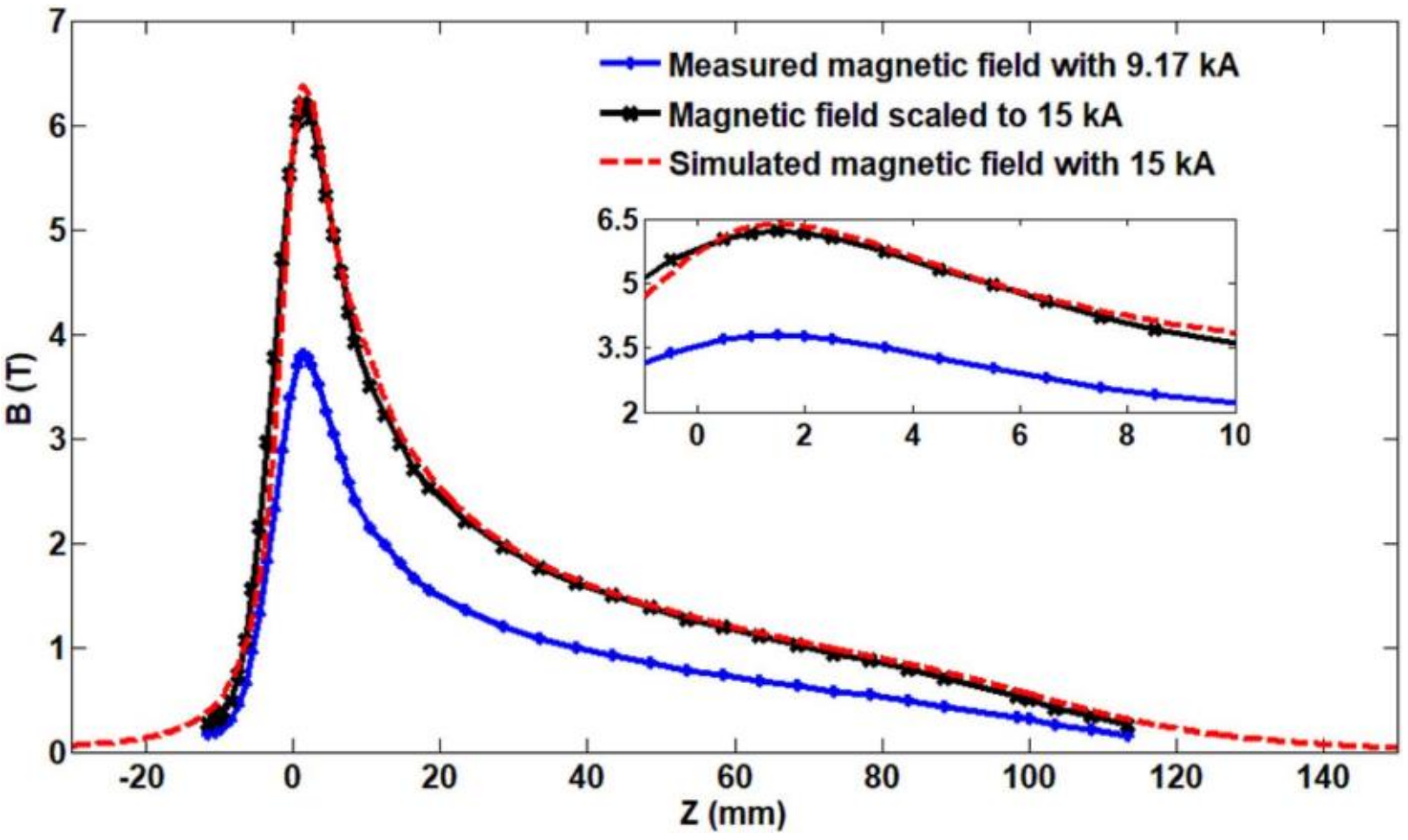

**Figure 6.3.2.8:** Measured and simulated magnetic fields as a function of the axial position.

### 6.3.2.3 *15 kA Solid State Modulator*

The CEPC positron source design utilizes a pulse modulator to generate a high peak current for the flux concentrator, resulting in a strong pulsed magnetic field with a peak value of 6 T to capture positrons. Table 6.3.2.1 displays the specifications of the pulse modulator required for this application. To meet these specifications, a pulse modulator based on all-solid-state switching components with high current, fast rise rate, narrow pulse width, and long-distance and repetitive pulse transmission has been designed, as illustrated in Figure 6.3.2.9.

**Table 6.3.2.1:** Specifications of the pulse modulator

| Parameters | Value |
|---|---|
| Input voltage | 380 V ±10% |
| Output pulse current | 15 kA |
| Pulse width | 5 μs |
| Output waveform | Half sine |
| Capacity peak voltage | 1 ns |
| Current stability | < 0.1% |

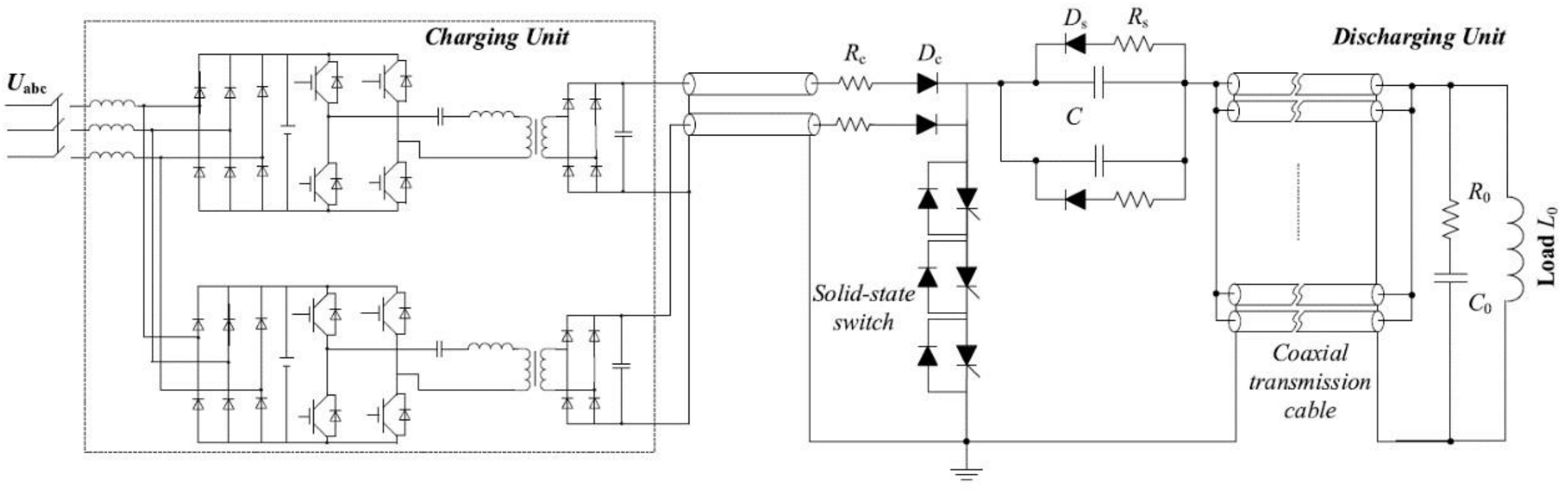

**Figure 6.3.2.9:** Schematics of the high-current pulse modulator.

A prototype of the modulator has been developed using thyristors as solid-state switches [7]. The switch assembly consists of six anti-conduction devices connected in series, as depicted in Figure 6.3.2.10.

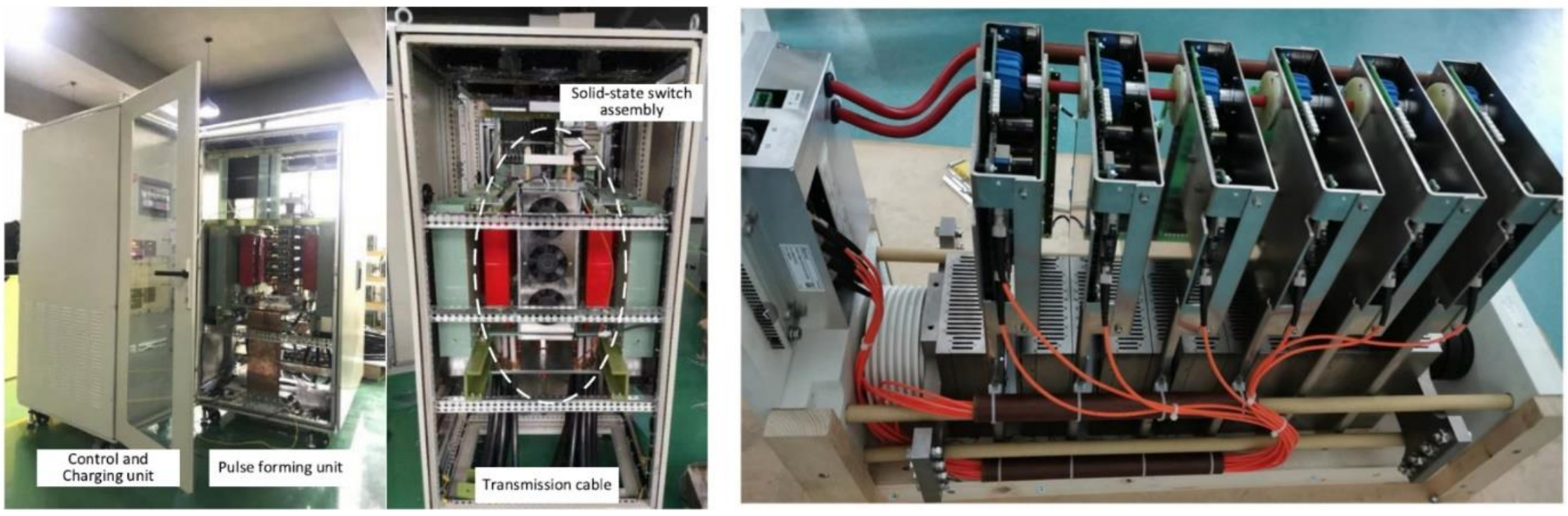

**Figure 6.3.2.10:** Prototype of a pulse modulator and its solid-state switch assembly.

The modulator generated a waveform with a peak current of 15.1 kA and a peak voltage of 15.6 kV after undergoing full power conditioning. For long-term tests, the system was operated at a repetition rate of 50 Hz and exhibited excellent stability. Figure 6.3.2.11 shows the measured current pulse, which exhibited high-frequency ripples typically observed in heavy hydrogen thyratron-based modulators. However, compared to the modulators based on heavy hydrogen thyratrons in the BEPC II project, the solid-state modulator's optimized design yielded an almost ideal half-sine pulse output, resulting in a higher peak current with no high-frequency ripples. This prototype was designed and tested at a repetition of 50 Hz (based on CEPC pre-CDR parameters), and it can be upgraded to 100 Hz.

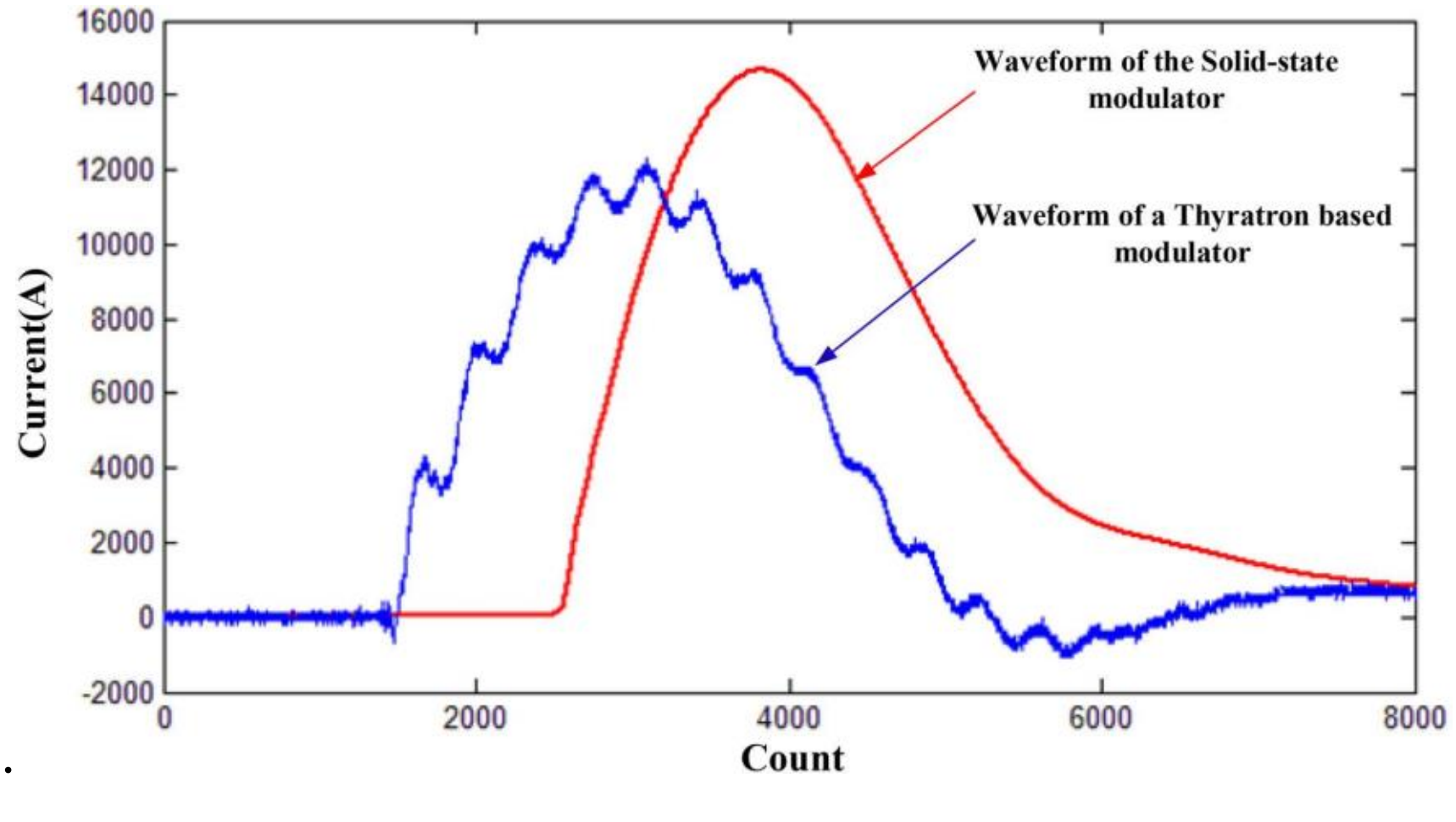

**Figure 6.3.2.11:** Current output waveforms of the pulse modulator prototype.

### 6.3.3 **RF System**

#### 6.3.3.1 ***Introduction***

The Linac provides 30 GeV electrons and positrons at the exit, with a repetition frequency of 100 Hz. For the Higs/W/$t\bar{t}$ operation mode, one bunch per pulse is sufficient. However, for the Z operation mode, a double bunch per pulse is required.

The RF system includes a bunching system and main accelerating structures. The bunching system comprises of two sub-harmonic cavities with frequencies of 158.89 MHz and 476.67 MHz, and one 2860 MHz S-band accelerating structure. For the total 30 GeV linac, there are 5.1 GeV S-band and 28.9 GeV C-band accelerating structures, with frequency of 5,720 MHz. The mature industrial klystron produces 80 MW for S-band and 50 MW for C-band. Pulse compressors are used to increase peak power and save cost. The S-band compressor compresses the 4 μs pulse to 1 μs, and the C-band from 2.5 μs to 0.35 μs.

In the majority of the S-band sections, one klystron is employed to power four accelerating structures. These structures have a gradient of 22 MV/m and a length of approximately 3 meters. However, in certain specific locations, such as after the positron source, there are sixteen larger aperture structures. These structures also operate at a gradient of 22 MV/m, but with a length of 2 meters and a beam aperture of 25mm. After the big-hole structures, there are eight accelerating structures working at 27 MV/m. Here, one klystron powers two accelerating structures to achieve the high gradient.

For the C-band sections, one klystron powers two accelerating structures. These structures operate at an average gradient 40 MV/m, with a length of about 1.8 meters.

The layout of the RF system, including the S-band and C-band components, is depicted in Figure 6.3.3.1. The S-band transmission employs WR284 waveguides, while the C-band transmission utilizes WR187 waveguides. The RF transmission includes components such as hybrid dividers, directional couplers, pumped waveguides, etc. Any remaining RF power at the output of the accelerating structure is absorbed by the dummy load.

In the S-band section, there is a 15% margin allocated for backup, ensuring reliability and redundancy. The total number of S-band accelerating structures is 107.

For the C-band section, the total number of accelerating structures is 470. When the gradient is 40 MV/m, a 9% margin is included.

Additionally, there is one C-band deflecting structure in the linac specifically designed for bunch length measurement purpose.

The LLRF system is built based on the MicroTCA (Micro Telecommunications Computing Architecture) standard. There are two sets of LLRF systems dedicated to the SHBs, 33 sets of LLRF systems for the S-band section and 236 sets for the C-band section, which includes the LLRF system for the C-band deflecting system. The total number of LLRF systems in the Linac is 271 sets. The phase reference line for the Linac is about 1.6 km long. Optical fiber will be used.

In the transport line between the linac and the positron damping ring, there are two sets of S-band RF system (2,860 MHz) used to compress beam energy and bunch length with cavity voltages of 16.5 MV and 20.6 MV, respectively. The length of the accelerating structures is 1 meter long. There is some distance between the two structures. Therefore, each structure is powered by a 30 MW klystron. The RF transmissions consist of the waveguides and pulse compressors.

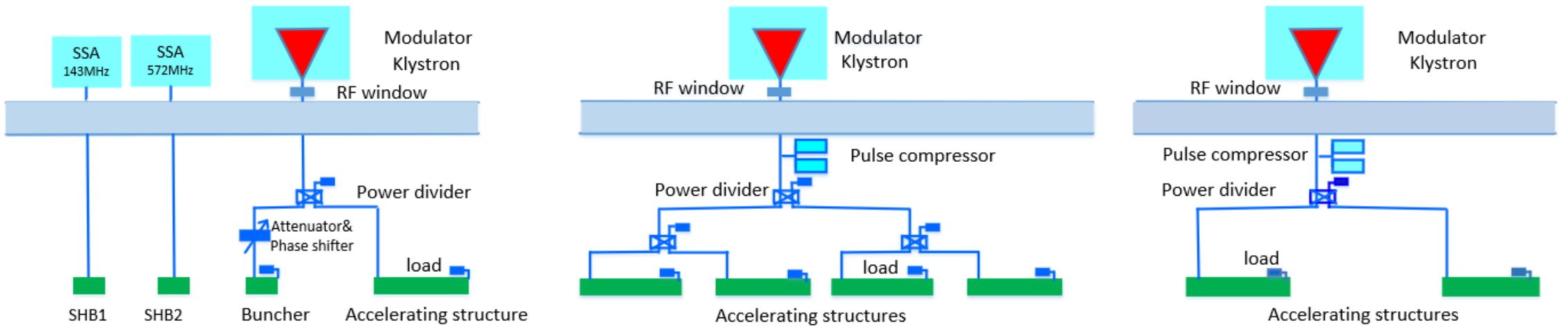

**Figure 6.3.3.1:** Layout of the RF system.

### 6.3.3.2 *Bunching System*

The bunching system is similar to the one employed in BEPC II and HEPS. It comprises two sub-harmonic bunchers (SHB) and a 6-cell constant impedance traveling-wave S-band buncher. The role of this system is to compress the initial 1.0 ns electron bunch at the exit of the electron gun down to a width of 10 ps by means of the traveling-wave buncher. Figure 6.3.3.2 displays a picture of the HEPS SHBs and buncher, which have been successfully employed in HEPS.

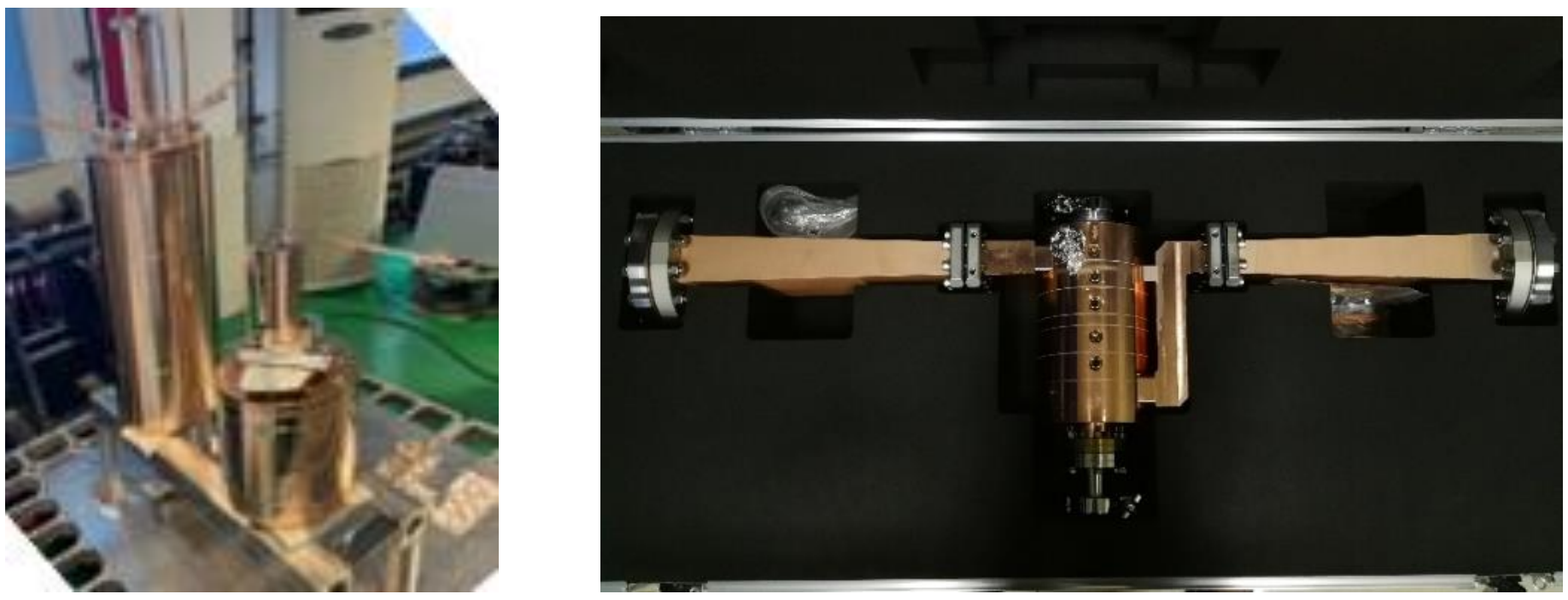

**Figure 6.3.3.2:** SHBs and buncher in the HEPS.

#### *6.3.3.2.1 Sub-harmonic Bunchers*

For both sub-harmonic bunchers, a re-entrant coaxial resonant cavity is utilized. Oxygen free copper (OFC) is employed to reduce the required RF input power. The resonant frequency of the first sub-harmonic cavity (SHB1) is 158.89 MHz and that of the second sub-harmonic cavity (SHB2) is 476.67 MHz. The shunt impedance is 1.46 MΩ and 2.53 MΩ, respectively. The maximum bunching voltage in both cavities is 100 kV. To ensure high reliability and convenient maintenance, each SHB is independently powered with a solid-state amplifier. Taking into account the insertion loss of the coaxial transmission system and providing sufficient power margin, the required peak pulse power of the solid-state amplifiers is 10 kW and 7 kW, respectively. Figure 6.3.3.3 displays the mechanical drawing of the SHBs, which have a cavity diameter of 180 mm, a gap distance of 30 mm, and a total length of 750 mm and 450 mm, respectively. The input coupler, the pickup, the tuner, and the vacuum port are mounted on the end plate of the long drift tube. The frequency tuning range of SHB1 is ±200 kHz, while that of SHB2 is ±500 kHz. The key parameters of the SHBs are presented in Table 6.3.3.1.

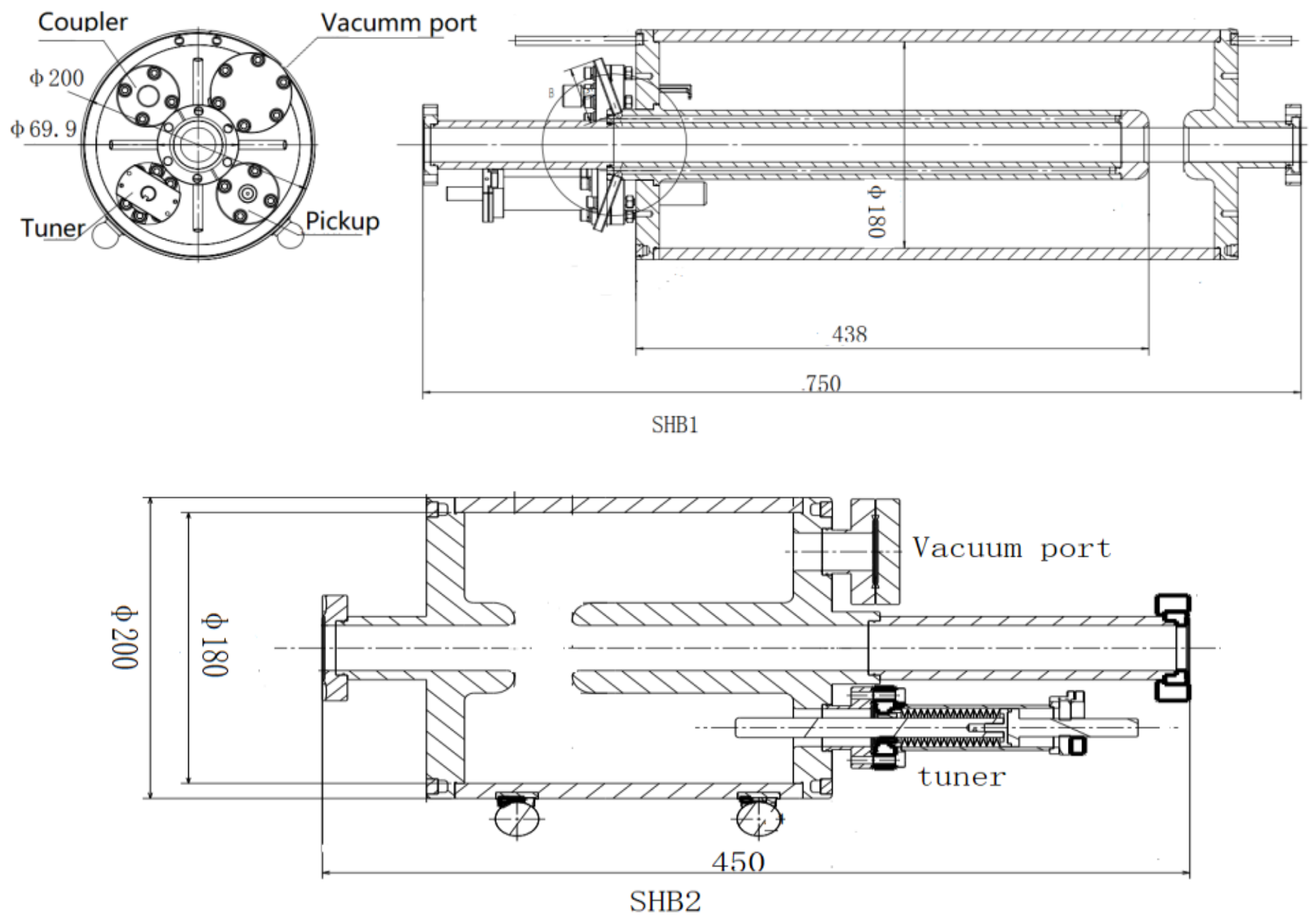

**Figure 6.3.3.3:** Mechanical drawing of the SHBs.

**Table 6.3.3.1:** Main parameters of the SHBs.

| Parameters | Unit | Value | |
|---|---|---|---|
| | | SHB1 | SHB2 |
| Frequency | MHz | 158.89 | 476.67 |
| Shunt Impedance | MΩ | 1.46 | 2.53 |
| Unloaded Q | - | 8475 | 12431 |
| VSWR | - | < 1.05 | < 1.05 |
| $E_{surface,\ max}$ @100 kV | MV/m | 6.4 | 6.1 |
| Required power @100 kV | kW | 10 | 7 |
| RF Structure type | - | Re-entrant SW | Re-entrant SW |

#### *6.3.3.2.2 S-band Traveling-wave Buncher*

The buncher utilized is a 6-cell constant impedance traveling-wave structure operating in the 2π/3 mode at 2860 MHz. The phase velocity is 0.75 times the speed of light, and the shunt impedance is 33.2 MV/m. With an input power of 6 MW, the maximum bunching voltage is 1.2 MV. The primary parameters of the S-band buncher are outlined in Table 6.3.3.2. The mechanical structure is depicted in Figure 6.3.3.4, which has an iris aperture of 24 mm and a total length of 235 mm.

**Table 6.3.3.2:** Main parameters of the S-band buncher.

| Parameters | Unit | Value |
|---|---|---|
| Frequency | MHz | 2860 |
| Phase advance | - | 2π/3 |
| Cell number | - | 6 |
| Phase velocity | - | 0.75 |
| Group velocity | - | 0.0193 |
| Attenuation constant | Np/m | 0.147 |
| Shunt impedance | MΩ/m | 33.2 |
| Unloaded Q | - | 11083 |
| Bunching voltage (Max) | MV | 1.2 |
| VSWR | - | <1.2 |
| Input power | MW | 6 |
| RF Structure type | - | TW/CI |

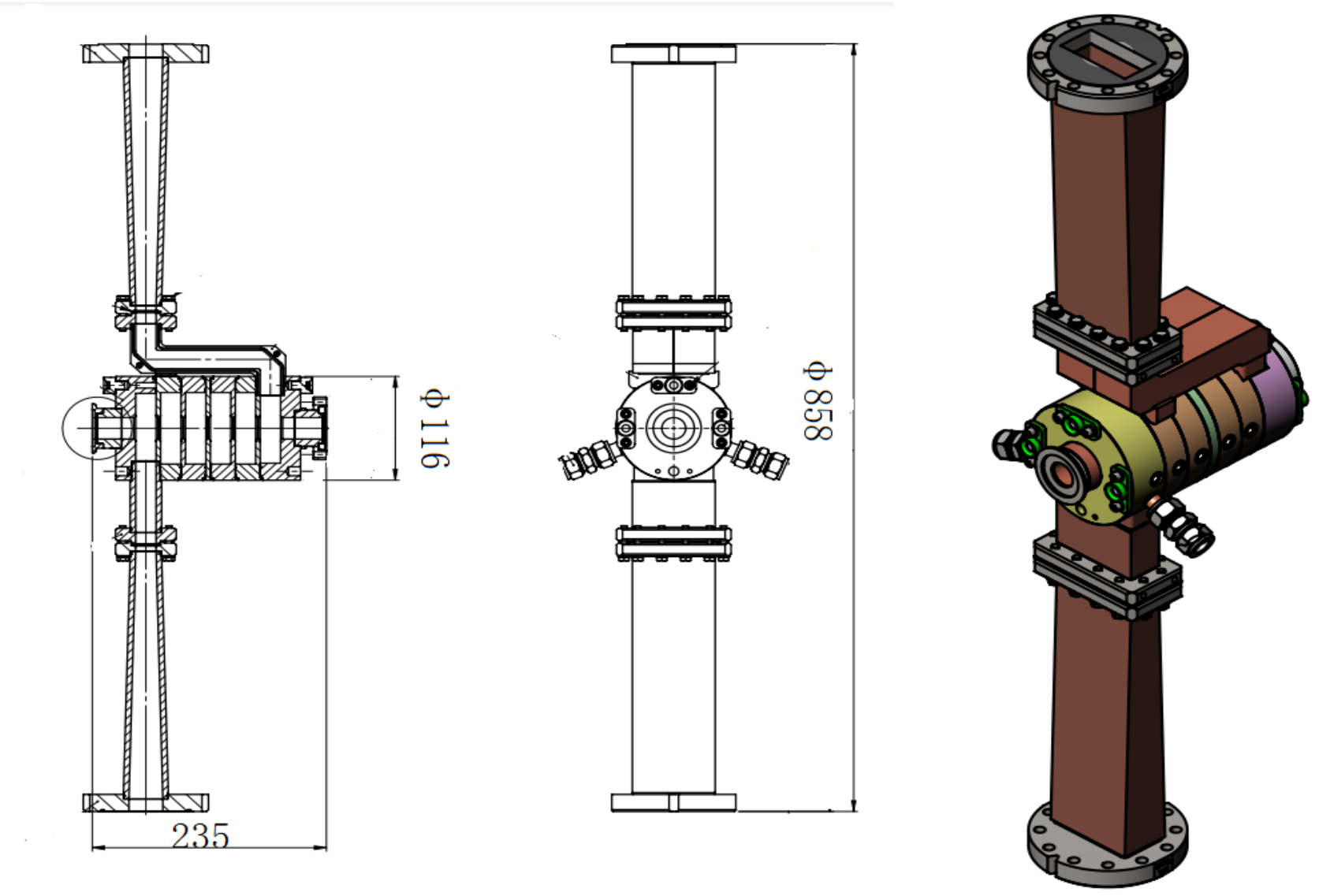

**Figure 6.3.3.4:** Mechanical drawing of the S band buncher.

### 6.3.3.3 ***Accelerating Structure***

#### *6.3.3.3.1 S-band Accelerating Structure*

To accelerate the bunched electron and positron beams to their design energy, S-band constant-gradient copper accelerating structures operating in the 2π/3 mode at 2,860 MHz will be utilized. The parameters of the accelerating structure are shown in Table 6.3.3.3. To minimize emittance growth caused by the asymmetry coupler, a dual-feed racetrack symmetry coupler design will be employed. The operating temperature of the accelerating structure is 30 ℃, and it will be maintained within 0.1˚C to ensure that the phase shift along the entire length of an accelerator section remains within 2˚.

**Table 6.3.3.3**: S-band accelerating structure parameters.

| Parameters | Unit | Value |
|---|---|---|
| Operation frequency | MHz | 2860 |
| Operation temperature | °C | 30.0 ± 0.1 |
| Number of cells | - | 84 +2 coupler cells |
| Section length | m | 3.1 |
| Phase advance per cell | - | 2π/3 - mode |
| Cell length | mm | 34.965 |
| Disk thickness (t) | mm | 5.5 |
| Iris diameter (2a) | mm | 26.23～19.24 |
| Cell diameter (2b) | mm | 83.460～81.78 |
| Shunt impedance ($r_0$) | MΩ/m | 60.3～67.8 |
| Q factor | - | 15465～15373 |
| Group velocity ($v_g/c$) | - | 0.02～0.087 |
| Filling time | ns | 780 |
| Attenuation factor | Neper | 0.46 |

Superfish was used to optimize the single cell, and rounding the cell led to a >12% improvement in the quality factor and reduced wall power consumption. Additionally, the shunt impedance increased by approximately 10.9%. By using irises with an elliptical shape ($r_2/r_1$=1.8), the peak surface field can be reduced by 13%. The tube has a total of 84 cells, and the average Q value is around 15,400. Figure 6.3.3.5 displays the cavity shape optimization and the electromagnetic distribution within the cavity.

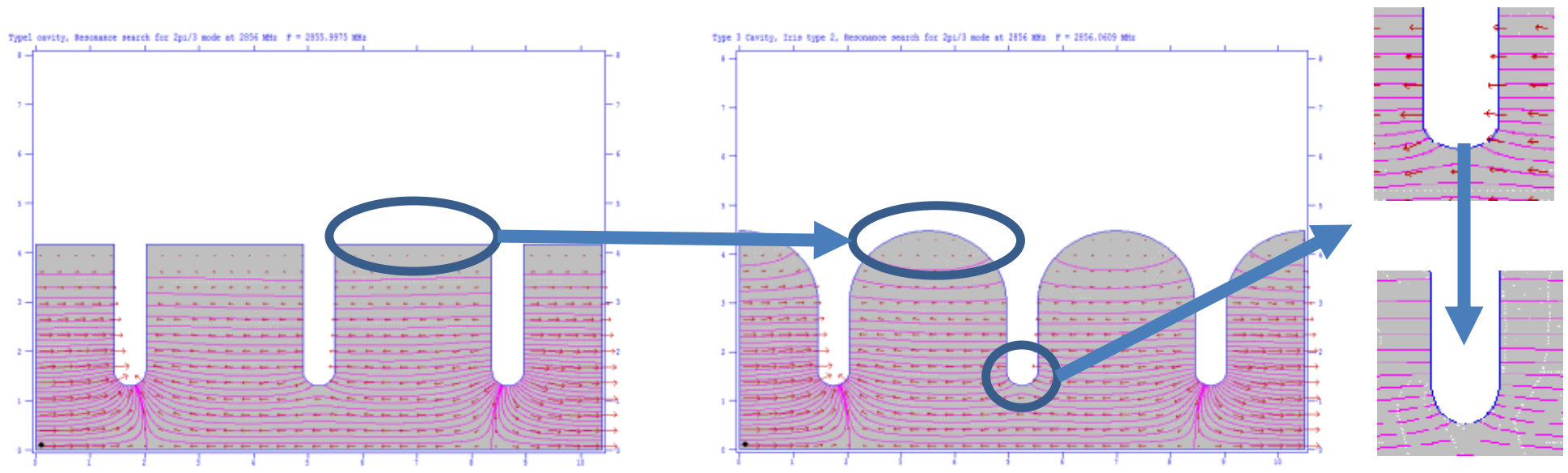

**Figure 6.3.3.5:** Cavity shape optimization and electromagnetic distribution in the cavity.

The coupler used is of the symmetry type, and the simulation model is shown in Figure 6.3.3.6. The input waveguide is a totally symmetrical 3 dB splitter.

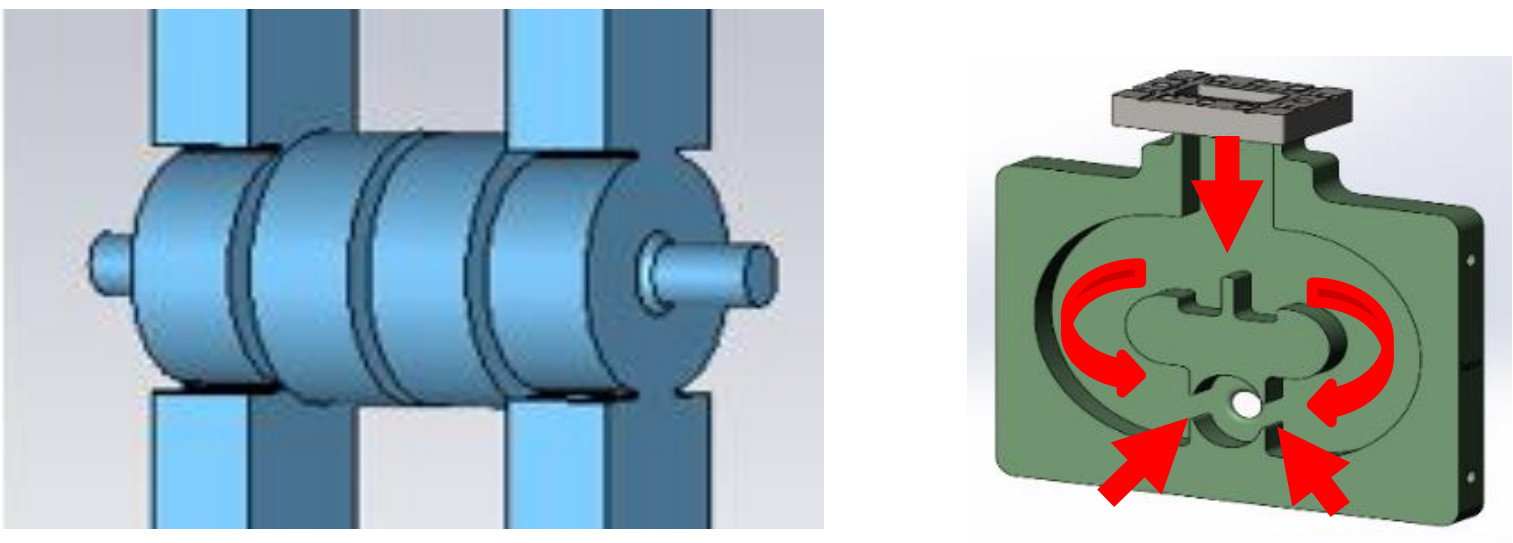

**Figure 6.3.3.6:** Model for the calculation of coupler.

An S-band accelerating structure has been manufactured and high-power tested, with an inner water-cooling system that includes 8 pipes around the cavity. The coupler arrangements are compact, and the splitter is milled together with the coupling cavity, while two tuners are located outside the cavity. Figure 6.3.3.7 shows the structure at the high-power test stand and the power input to the structure. During testing with a pulse compressor, the gradient reached 33 MV/m.

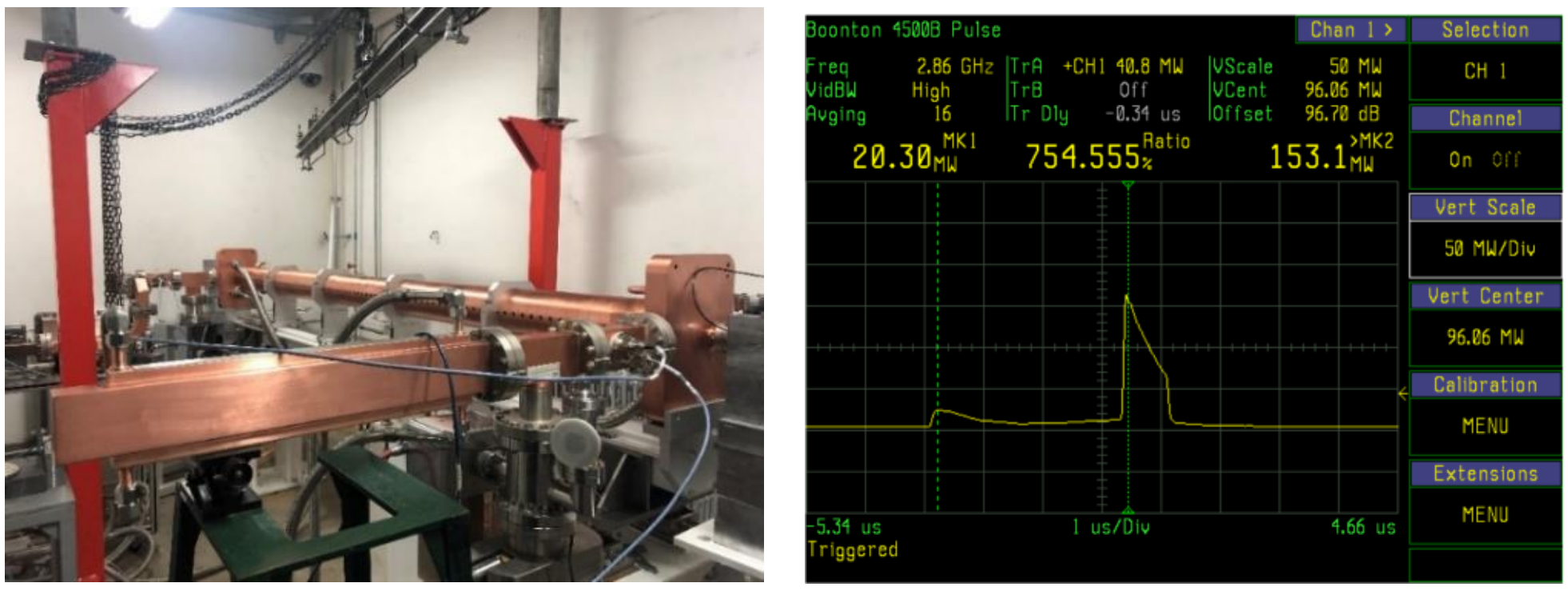

**Figure 6.3.3.7:** The accelerating structure at the high-power test bunch.

The HEPS linac is a 500 MeV S-band normal conducting linac with 9 accelerating structures. Figure 6.3.3.8 depicts the HEPS linac tunnel and equipment corridor. The design of the accelerating structure cavity shape is similar to the CEPC design, with the exception of a frequency of 2,998.8 MHz. The coupler for HEPS utilizes a λg/4 short plan and the coupling cavity employs a racetrack shape to ensure electromagnetic symmetry. This design results in a smaller coupler size compared to a fully symmetric one. Currently, the HEPS linac is operational with a gradient of up to 26 MV/m with the beam. The same coupler design will also be employed for the CEPC project.

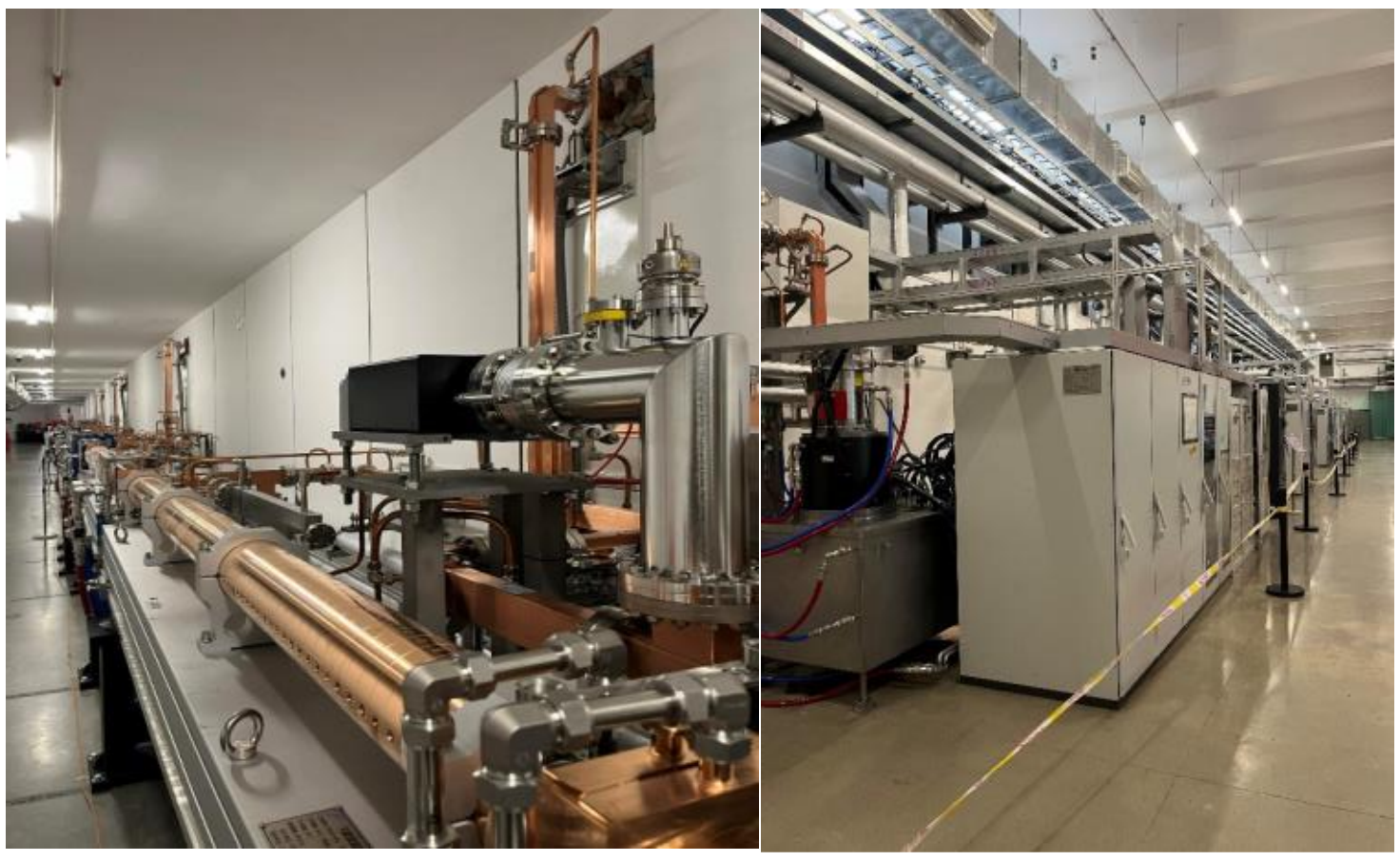

**Figure 6.3.3.8:** The HEPS linac tunnel and equipment corridor.

#### *6.3.3.3.2 C-band Accelerating Structure*

The technology of C-band accelerating structures has been extensively developed and proven successful at KEK and RIKEN/SPring-8. Notably, high field gradients have been achieved in these C-band structures. At KEK, the field gradients in C-band accelerating structures have reached up to 45 MV/m [1], and at RIKEN/SPring-8, C-band accelerating structures have achieved field gradients of 50.1 MV/m at the high-power test bench and 41.4 MV/m with beam at a repetition rate of 120 Hz [2,3]. Furthermore, the C-band RF structure and compressor at SARI (Shanghai Advanced Research Institute) for SXFEL (Soft X-ray Free-Electron Laser) are presently operating stably, confirming the reliability and effectiveness of C-band RF technology in practical applications.

The design of these structures was done with reference to SXFEL, which has 6 C-band RF units. The operation gradient was measured using their system's beam, and they found that the maximum gradient is 41.7 MV/m [4]. The design of the structures is $4\pi/5$, with a length of approximately 1.8 m [5,6,7]. As mentioned earlier, we gave enough margin in our C-band RF system design.

In addition to SARI, IHEP has also designed and manufactured C-band accelerating structures, which operate in $3\pi/4$ mode [8]. The main parameters of this structure are shown in Table 6.3.3.4. Despite CEPC's frequency being 5,720 MHz, 5,712 MHz was chosen for this prototype to match the existing industrial mature power source.

**Table 6.3.3.4:** Main parameters of the C-band accelerating structure.

| Parameter | Unit | Value |
|---|---|---|
| Frequency: f | MHz | 5720 |
| Operation temperature | °C | 30 ± 0.1 |
| No. of Cells | - | 87 + 2 coupler cells |
| Phase advance | - | 3π/4 |
| Total length | m | ~1.8 |
| Length of cell: d | mm | 19.654 |
| Disk thickness: t | mm | 4.5 |
| Average aperture: 2a | mm | 14.04 |
| Average diameter: 2b | mm | 45.6 |
| Shunt impedance: Rs | MΩ/m | 58.4 ~ 73.7 |
| Quality factor: Q | - | 11358~11186 |
| Group velocity: Vg/c | % | 2.8 ~ 0.96 |
| Filling time: tf | ns | 350 |
| Attenuation factor: τ | Np | 0.56 |

Since the C-band accelerating structures are axisymmetric, the RF design can be performed using the 2D electromagnetic code Superfish. It calculates travelling-wave fundamental parameters from standing-wave simulations, making it an ideal tool for this application. The 3π/4 accelerating mode was selected, and a rounded-wall cavity shape was chosen because the quality factor was 10% higher than the disk-loaded design. The cell irises have an elliptical profile to minimize peak surface electric fields. The shunt impedance and $E_{max}/E_0$ were calculated as a function of the disk thickness $t$ for the elliptical average beam radius of 7.02 mm. Figure 6.3.3.9(a) shows the calculation results, with a disk thickness of 4.5 mm selected. Figure 6.3.3.9(b) shows the peak surface electric field normalized to the average accelerating field as a function of the iris ellipticity for the iris radius of 7.02 mm and disk thickness of 4.5 mm, with the optimal value found to be 1.8. The C-band structure at IHEP is shown in Figure 6.3.3.10. Due to the absence of a C-band high-power test bench at IHEP, high-power tests for C-band components will be conducted once a C-band power source becomes available.

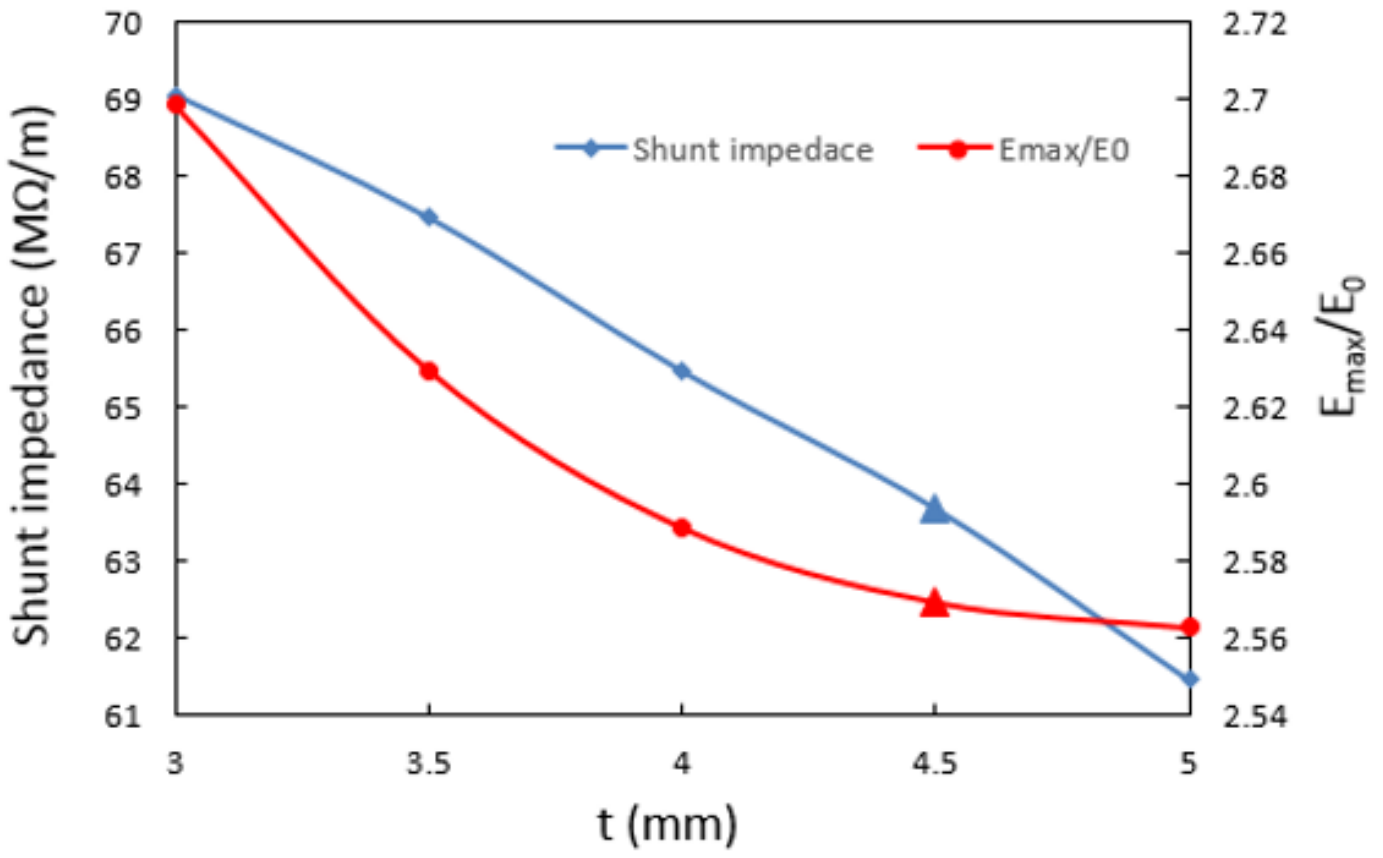

(a)

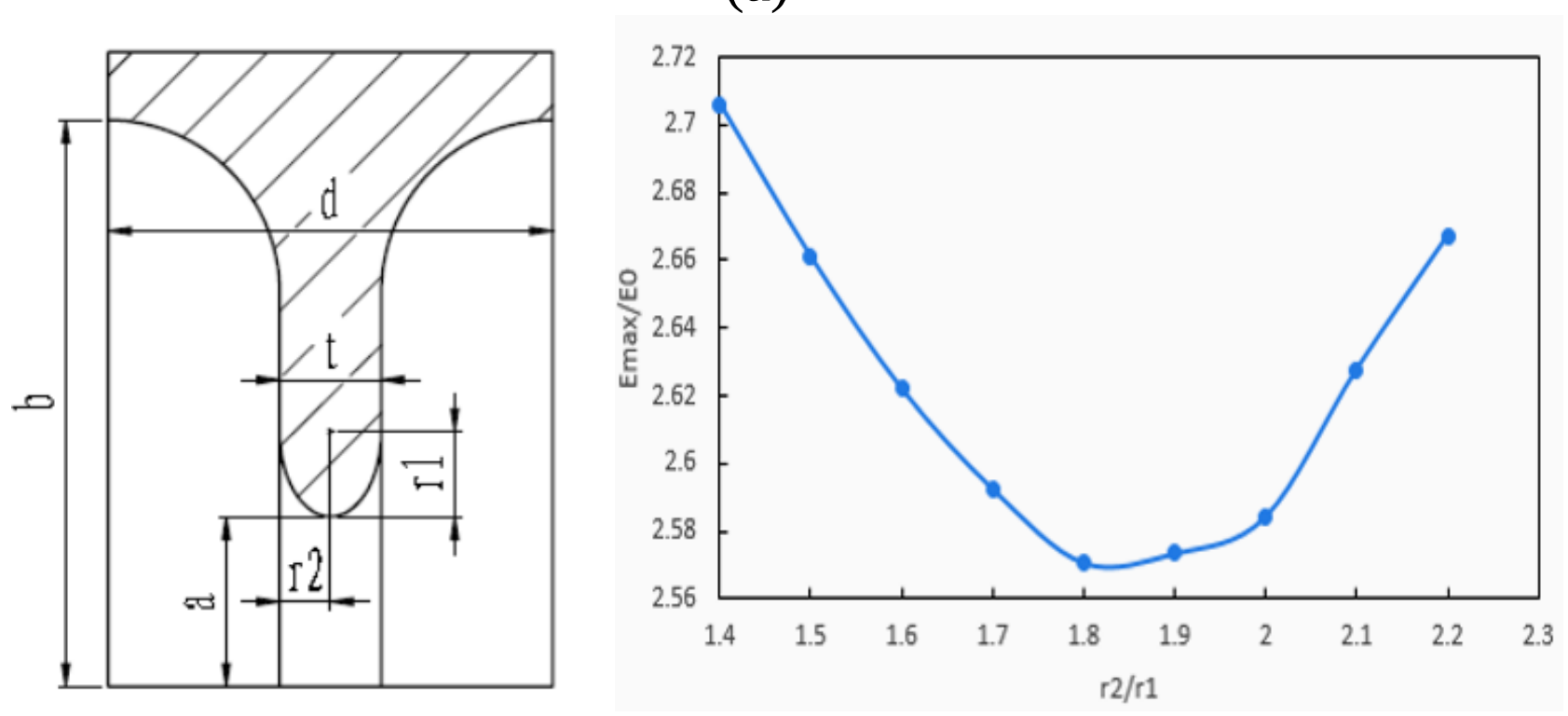

(b)

**Figure 6.3.3.9:** (a) The shunt impedance and $E_{max}/E_0$ vs. iris thickness, (b) $E_{max}/E_0$ as a function of the iris ellipticity.

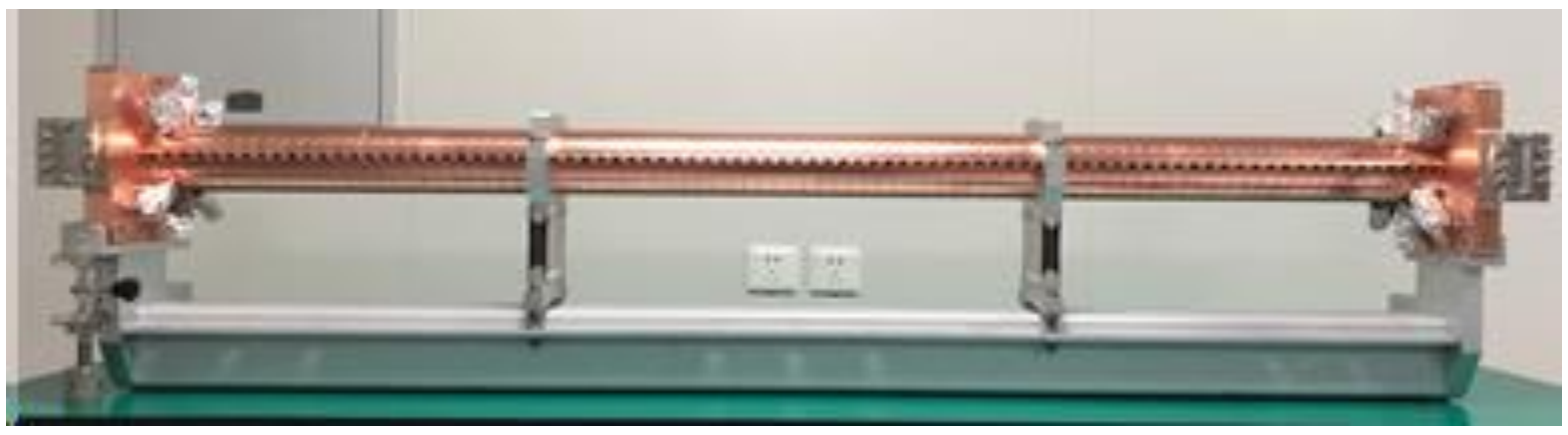

**Figure 6.3.3.10:** C-band accelerating structure at IHEP.

## 6.3.3.4 ***Pulse Compressor***

### *6.3.3.4.1 S-band Pulse Compressor*

To achieve the desired beam acceleration to 30 GeV, we have adopted the experience gained from the operation of the Stanford Linear Accelerator Energy Doubler (SLED) at the maximum klystron output peak power of 80 MW, with a pulse length of 4 μs. The S-band SLED, which is a radiofrequency (RF) pulse compression system that uses high-Q resonant cavities, is one of the critical RF components in the S-band high-power RF station.

The S-band SLED comprises a 3-dB power hybrid and two identical over-coupled cylindrical cavities that resonate at 2,860 MHz. To achieve the pulse compression, a fast-acting triggered π-phase-shifter is inserted into the klystron drive line. During a significant portion of each pulse, the cavities store klystron output power. Then, the phase of the klystron output is reversed, and the cavities release the stored power rapidly into the accelerating section, thereby adding to the klystron output power during the remaining pulse length. This process enhances the peak power at the expense of the pulse length without increasing the average input power consumption.

The specifications of the S-band SLED are presented in Table 6.3.3.5. The cavity is equipped with two coupling slots located between the waveguide and the cavity, which effectively reduce the peak surface field and increase the operating stability in high power conditions. The input pulse has a length of 4 microseconds and undergoes a 180° phase reversal at a time of 3.17 microseconds. Based on operational experience with the BEPC-II Linac, the energy multiplication factor can exceed 1.6.

**Table 6.3.3.5:** Main parameters of the pulse compressor.

| Parameters | Unit | Value |
|---|---|---|
| Operation frequency | MHz | 2860 |
| Resonant mode | - | $TE_{0,1,5}$ |
| Coupling coefficient | - | 5 |
| Peak power gain | - | > 5 |
| Unload Q factor | - | ~100,000 |
| Energy multiplication factor | - | ~1.6 |
| Max. input peak power | MW | 80 |
| Input pulse length | us | 4 |
| Output pulse length | us | 0.83 |
| Repetition rate | Hz | 100 |

The electric and magnetic field distribution of the SLED system can be seen in Figure 6.3.3.11. The detuning rod is positioned at the surface electric field node (as shown in Figure 6.3.3.12) to lower local heating effects. A photograph of the SLED system designed for HEPS is presented in Figure 6.3.3.13.

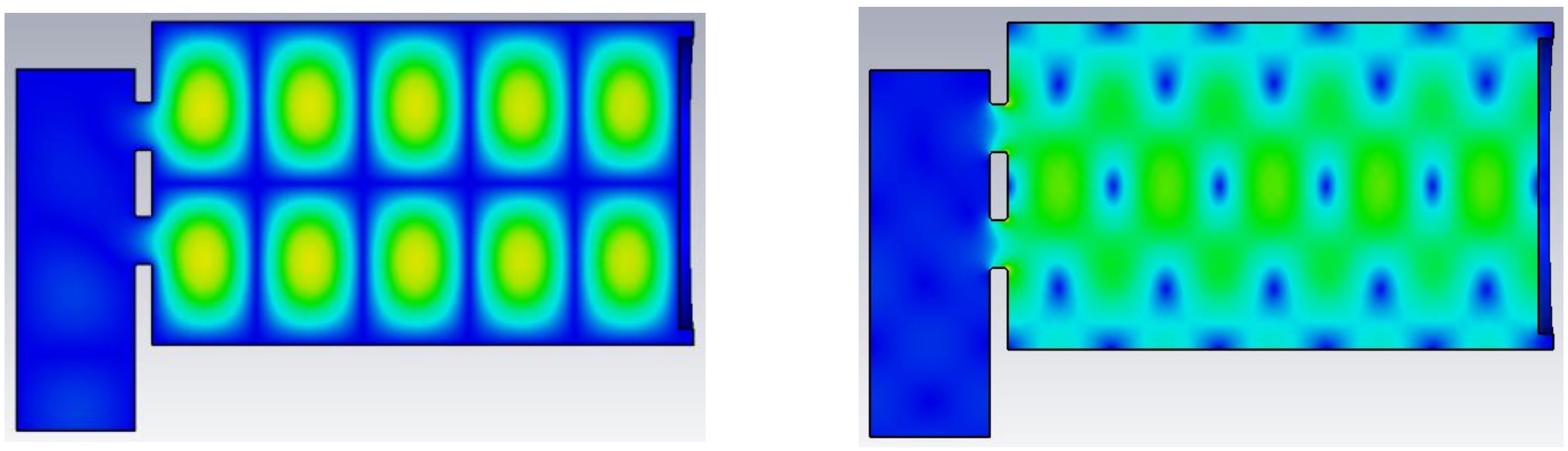

(a) Electric field (b) Magnetic field

**Figure 6.3.3.11:** Electr-magnetic field distribution of the S-band SLED.

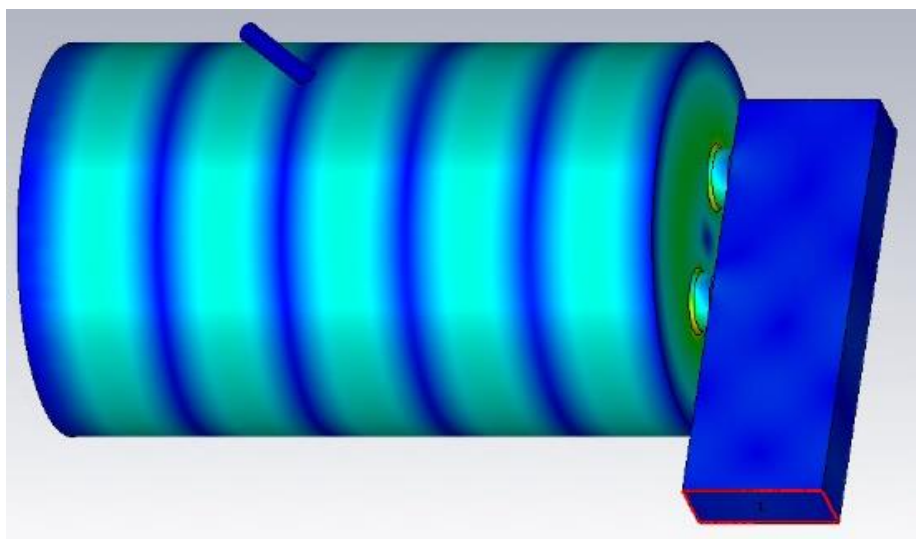

**Figure 6.3.3.12:** The detuning rod in the cavity.

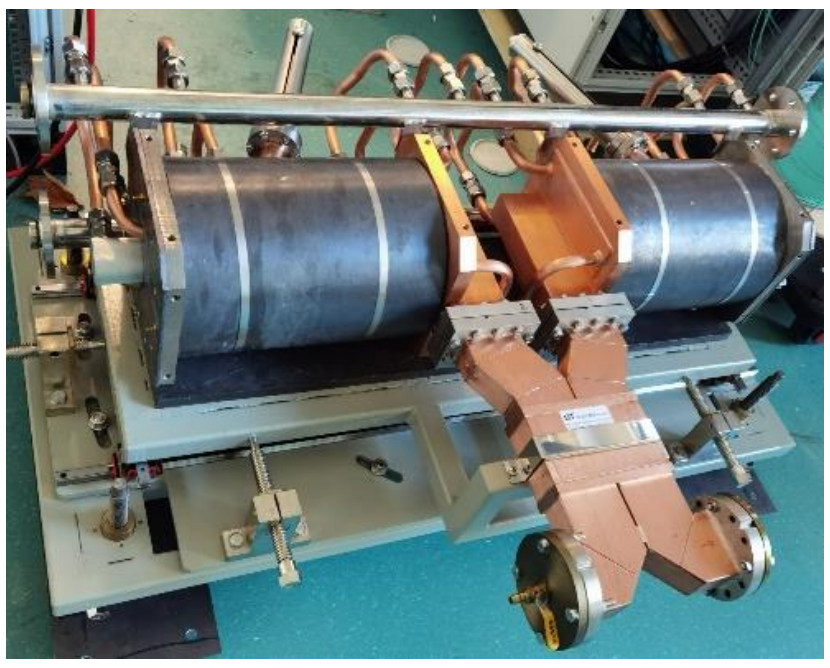

**Figure 6.3.3.13:** SLED for the HEPS.

#### *6.3.3.4.2 C-band Pulse Compressor*

The working principle of C-band pulse compressors is similar to that of S-band pulse compressors. At IHEP, a C-band pulse compressor has been successfully designed [9]. The compressor operates in the $TE_{038}$ mode and has a measured $Q_0$ value of about 150,000. The pulse power gain is 6.2. Figure 6.3.3.14 shows the overall view of the SLED, and Table 6.3.3.6 shows the results of the cold measurements of the cavity.

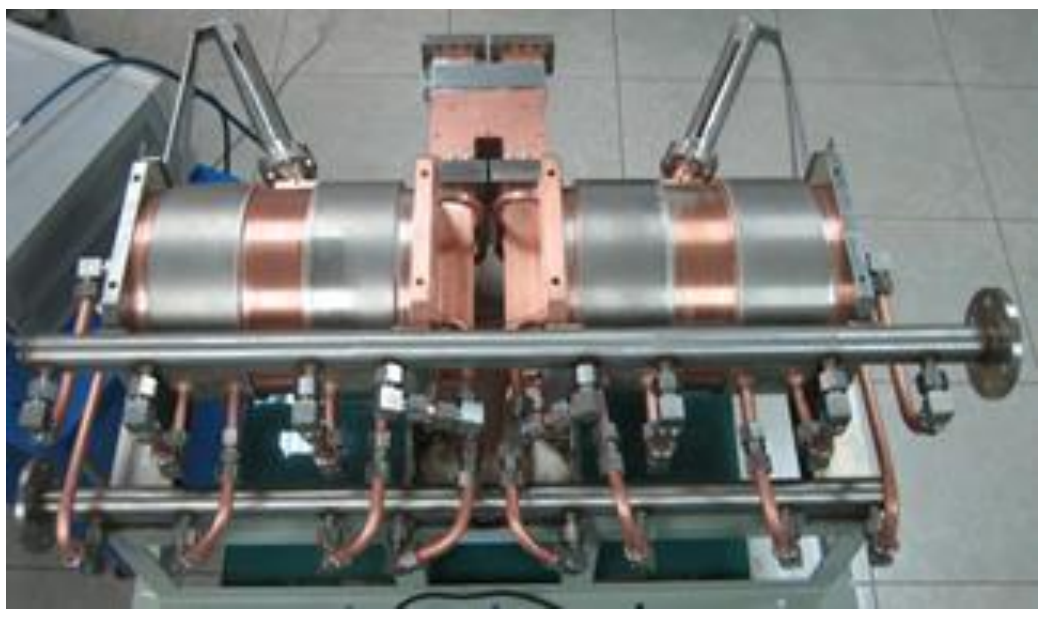

**Figure 6.3.3.14:** C-band SLED at IHEP.

**Table 6.3.3.6:** Main parameters of the pulse compressor.

| Parameter | Unit | Designed | Measured |
|---|---|---|---|
| Frequency | MHz | 5712 | 5712 |
| Mode of operation | - | $TE_{038}$ | $TE_{038}$ |
| Q0 | - | ≥130,000 | about 150,000 |
| β | - | 6.9±0.1 | about 7 |
| Peak power gain | dB | ≥6.8 | 7.1 |
| Operating temperature | °C | 30±0.1 | 30 |

### 6.3.3.5 *RF Power Transfer Line*

The RF Transmission System (RFTS) plays a crucial role in delivering the microwave pulsed power output from the klystron to the accelerating structure and building the electromagnetic field inside it. It comprises various components such as a high-power phase shifter and attenuator, power divider, RF window (ceramic window), straight waveguide, bending waveguide, pumping waveguide, directional coupler, and dummy load.

The rectangular waveguide WR-284 (BJ-32) will be used for the RFTS, which can operate at a pulsed state of 65 MW/4 μs/100 Hz. The cross-sectional inner size of the WR-284 waveguide measures 72.14×34.04 mm.

The high-power phase shifter and attenuator is responsible for adjusting the phase and magnitude of the microwave power. The prototype is shown in Figure 6.3.3.15. The attenuation can be adjusted within a range of 0-40 dB, while the phase can be adjusted for more than 360°. Both the phase and attenuation can be adjusted independently. The vacuum bellows are driven by step motors to facilitate the adjustment of the shorting pistons, and the limit switches are used to control both motor movement directions. Table 6.3.3.7 lists the key parameters of the phase shifter and attenuator.

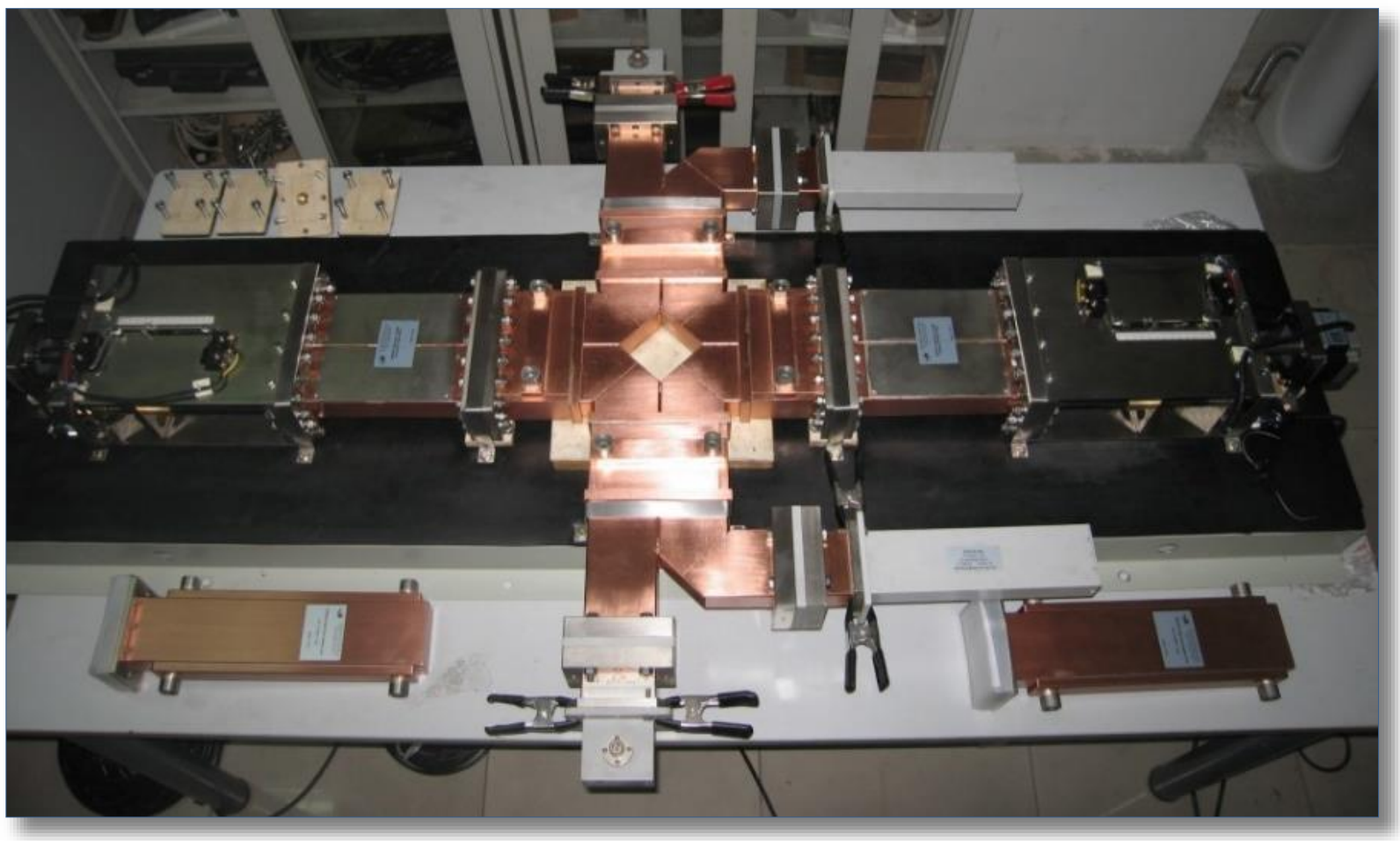

**Figure 6.3.3.15:** Prototype of the high-power phase shifter and attenuator.

**Table 6.3.3.7:** Parameters of the high-power phase shifter and attenuator.

| Parameters | Unit | Value |
|---|---|---|
| Frequency | MHz | 2860 |
| VSWR at central frequency | - | <1.15 |
| VSWR in bandwidth | - | <1.25 |
| Insertion loss | dB | <0.2 |
| Attenuation range | dB | 0-40 |
| Phase shifting range | degree | >360 |

Figure 6.3.3.16 shows the prototypes of the RF window, directional coupler, and bending waveguide. The RF window serves two purposes: it isolates the vacuum to protect the klystron during maintenance, and it can transmit microwave power with very low loss. The pumping waveguide connects to the ion pump, which is responsible for pumping and maintaining the vacuum level within the system. The directional coupler is a signal-picking device that picks up both transmitted and reflected power in a ratio of approximately one in a million. These signals are used for monitoring, interlock protection, closed-loop control, and phase control.

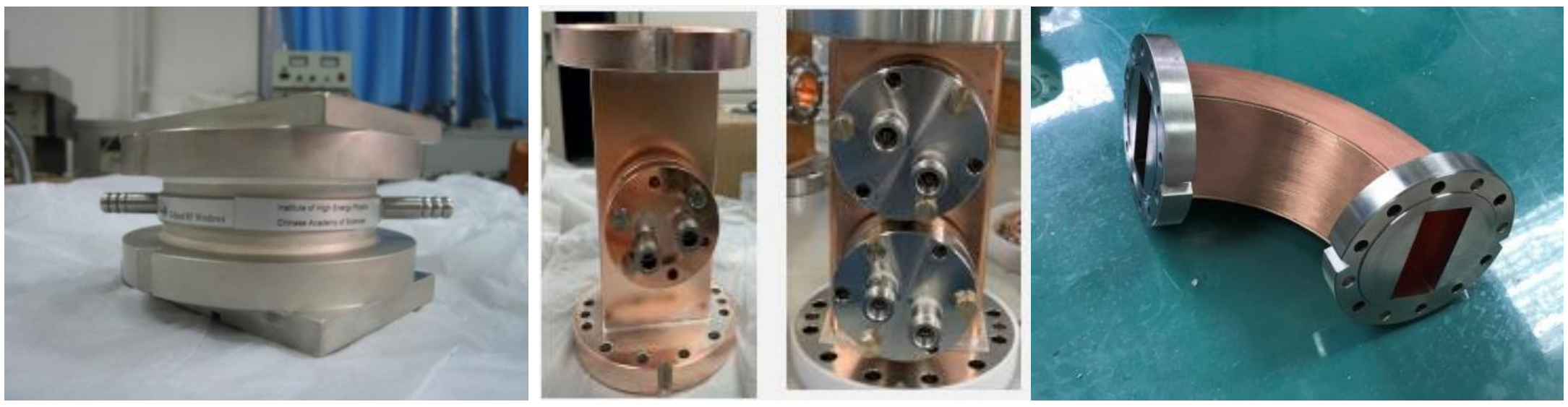

**Figure 6.3.3.16:** Prototype of the RF window, the directional coupler, and the bending waveguide.

The high-power dummy load is designed to absorb microwave power at the necessary ports of the phase shifter and attenuator, power divider, and accelerating structure. It incorporates indirect water cooling and uses brazing rod β-phase SiC ceramics as a peak microwave energy absorber. The SiC dummy load can withstand peak power up to 60 MW in high-power testing and its performance is at the internationally advanced level for similar products. Figure 6.3.3.17 shows the prototype of the dummy load, and Table 6.3.3.8 displays its parameters.

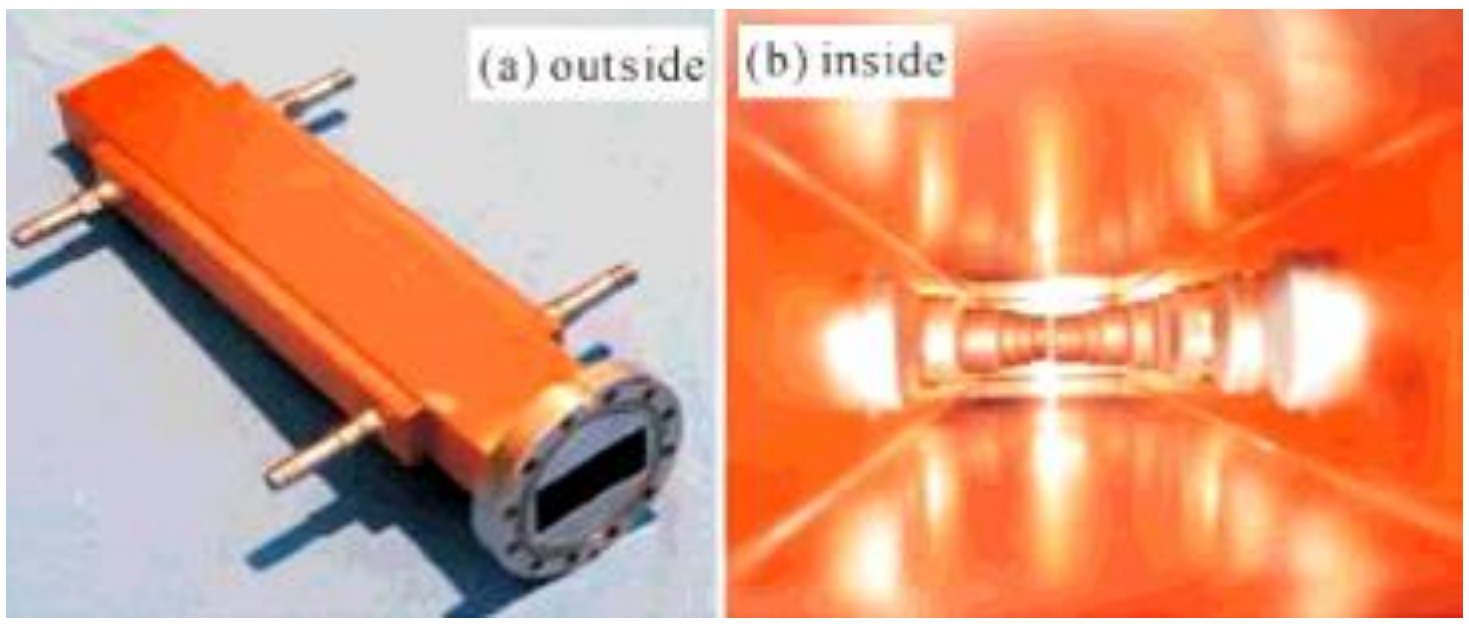

**Figure 6.3.3.17:** Prototype of the dummy load.

**Table 6.3.3.8:** High power SiC dummy load parameters.

| Parameters | Unit | Value |
|---|---|---|
| Frequency | MHz | 2,860 |
| VSWR | | < 1.1 |
| Maximum peak power | MW | 30 (with SLED)<br>10 (without SLED) |
| Repetition frequency | Hz | 100 |
| Pulse width | μs | 4 |

The modular microwave monitoring unit is undergoing a redesign that will consist of several independent parts, including the attenuation unit, filter unit, detection unit, and virtual oscilloscope unit. This modular multi-unit design simplifies maintenance and minimizes the space needed to avoid electromagnetic interference. The microwave monitoring unit performs three functions simultaneously, providing signals for power measurement, observing waveforms on an oscilloscope, and virtual oscilloscope signals for input into the local computer.

For C-band waveguides, the designated waveguide type is WR187. The inner dimensions of the waveguide measure 47.55 mm × 21.55 mm. The operating principles and functionalities of the C-band waveguide components are essentially the same as those used in the S-band.

#### 6.3.3.6 *Low Level RF*

The LLRF control system employs feedback control, which involves measuring the RF signal and comparing it to the desired set-point. The klystron output forward/reflected signal, coupled signal in the RF transmission line, and accelerating structure signal are down-converted to intermediate frequency signals through the RF front-end module. These signals are then digitally sampled, filtered, demodulated, and processed in the LLRF controller before being up-converted to klystron actuation. Real-time analysis of the sampled data allows the low-level system to identify system exceptions and generate interlock signals for machine protection systems. Additionally, the LLRF system provides a communication interface to exchange information with other systems such as general support, beam diagnosis, and upper-level control [10,11].

The LLRF system performs the following functions:

1) Stabilization: ensures the lowest possible rms amplitude, phase errors, and long-term field stability. Feedback is supplemented by feed-forward, which compensates for average repetitive errors, and inter-pulse feedback, which compensates for pulse-to-pulse errors.
2) System monitoring and diagnosis: monitors crucial signals in real-time, such as solid-state amplifier output, klystron output, and acceleration tube forward/reflection. Data is saved in a database for real-time diagnosis and historical record query.
3) Klystron linearization: to achieve maximum efficiency of the Linac, klystrons must operate close to saturation. Due to klystron non-linearity, klystron output power has a strong dependency on feedback gain. The LLRF system provides klystron linearization using digital methods such as look-up tables or polynomials that work over a large range of high voltage settings.
4) Exception handling: in the event of interlock trips or abnormal operating conditions, the LLRF system ensures safe procedures to protect hardware.
5) User interface: provides a unified and flexible control interface for operators.

The LLRF system employs a range of key elements such as modern analogue-to-digital converters (ADCs), digital-to-analogue converters (DACs), powerful Field Programmable Gate-Array (FPGAs), and digital signal processing (DSPs) to process signals. Depending on the processor, clock frequency, and algorithm complexity, the ADC's clock to DAC output can achieve low latencies ranging from a few hundred nanoseconds to several microseconds. High-speed communication links facilitate data exchange between the digital processor and the numerous I/O channels, as well as communication between various signal processing units.

Figure 6.3.3.18 depicts the low-level system scheme, which includes RF signal distribution units, local oscillator signal generators, signal processing platforms, and solid-state amplifiers. The function and design of each module are as follows:

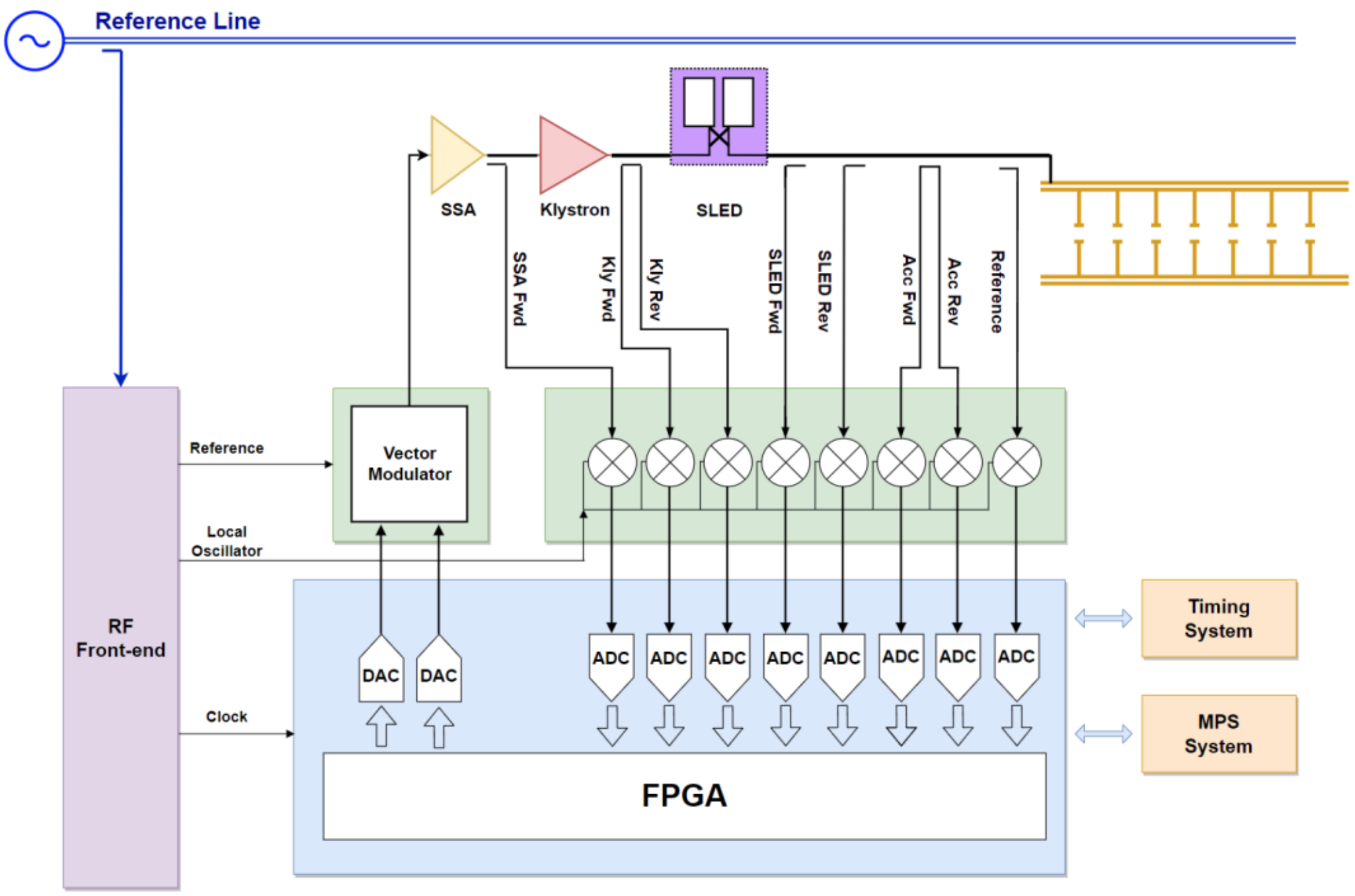

**Figure 6.3.3.18:** Low-level RF system scheme.

1) The RF signal distribution unit performs essential functions such as attenuation, isolation, and signal division for various RF signals, including excitation, acceleration tube forward/reflection, and others, to ensure that the signal amplitude falls within the normal operating range of the hardware system. To enable easy monitoring or processing, the signals are divided into multiple paths.
2) The signal processing platform: To achieve high system availability of up to 99.999%, a robust algorithm, redundancy, and extremely reliable hardware are required for a high-availability LLRF system. Therefore, the MicroTCA architecture will be employed, including chassis, power supply, MCH (MicroTCA Carrier Hub), embedded x86 PC, digital signal processing module, RF front-end, local oscillator generator, and other functional modules.
3) The RF front-end facilitates the multi-channel down-conversion of the RF signal to the intermediate frequency and the up-conversion of the IF to the RF signal. Additionally, it includes multi-channel digital I/O, sensor monitoring, interlock, among others.
4) The local oscillator (LO) signal generator provides a LO for down-conversion of the RF monitor signals, a clock for the digital board, and a reference signal for vector modulation, with a clock frequency of 95.33 MHz and an LO frequency of 2,836.16 MHz for S-band LLRF systems and a clock frequency of 95.33 MHz and an LO frequency of 5,696.16 MHz for C-band LLRF systems. The module will be temperature compensated to account for temperature and humidity changes.
5) The pre-amplifier generates RF excitation for the klystron from the LLRF system.

The LLRF software comprises both firmware and high-level applications. The FPGA firmware can be divided structurally into the framework and control algorithm. The framework implements the basic functions of the board, including PCIe communication, ADC/DAC input and output, DDR memory, clock distribution, hardware configuration, and more. The structure of the framework is depicted in Figure 6.3.3.19.

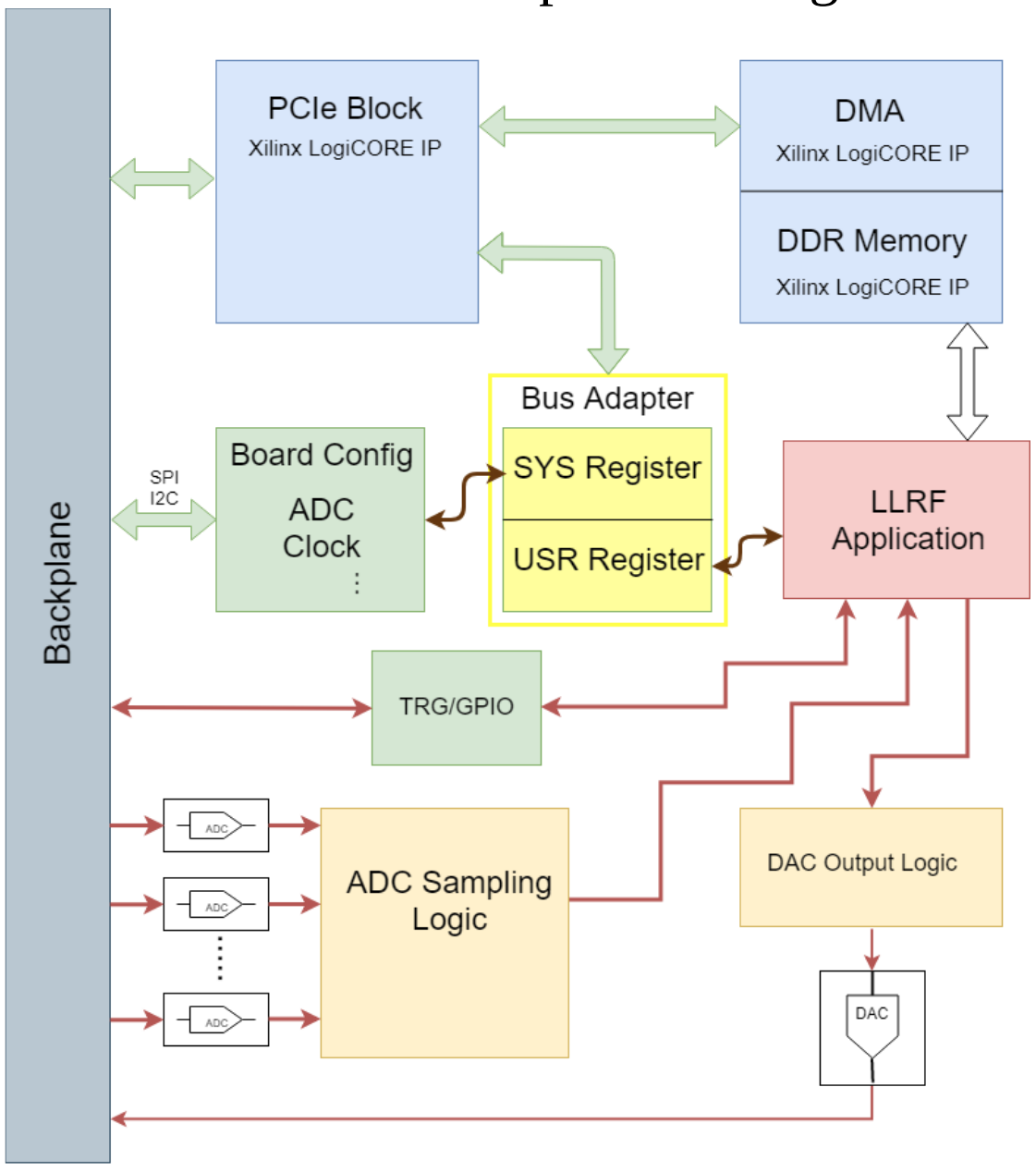

**Figure 6.3.3.19:** LLRF firmware scheme.

The control algorithm is responsible for achieving pulse feedback and feedforward control. The diagram in Figure 6.3.3.120 illustrates this process. The algorithm mainly comprises I/Q feedback control logic, signal amplitude and phase monitoring, trigger logic, system interlock, and other essential functions.

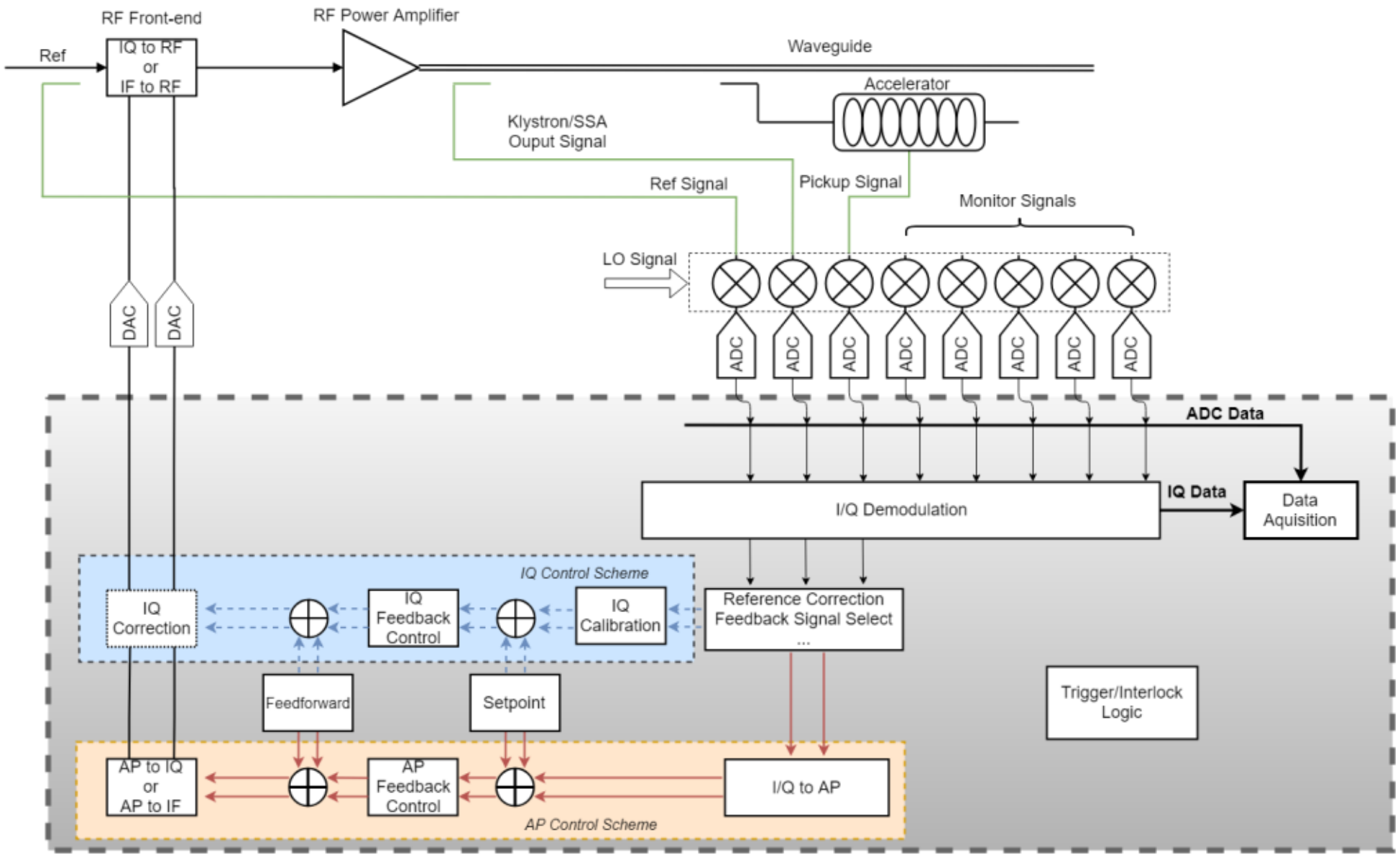

**Figure 6.3.3.20:** LLRF algorithm scheme.

The high-level application of the LLRF system will be built on EPICS and consist of the following components:

1) EPICS Channel Access: This component connects the device control IOC, including the LLRF IOC, with the upper-layer physical application software and user interface.
2) EPICS services: This component provides essential services such as historical data archive, message logging, alarm handling, parameter save/restore, and user interface.
3) Remote hardware management: This component allows for remote monitoring and control of LLRF hardware, such as the MicroTCA chassis and solid-state amplifier.
4) Real-time data acquisition: This component enables high-speed data acquisition and transmission based on Ethernet or high-speed serial communication.

The entire linac spans approximately 1.6 km in length. Using coaxial cables for signal transfer is not feasible due to issues like attenuation and electromagnetic interference (EMI). Therefore, optical fiber will be employed. The Linac operates at an RF frequency of 2,860 MHz, derived from a master oscillator that provides the frequency standard for the Collider (650 MHz), the Booster (1.3 GHz), and the injector (2,860 MHz). RF signals will be transmitted via continuous wave (CW) laser, with detection accomplished by comparing the forward and backward optical phases and subsequently compensated using phase shifters. The phase stability for the 1.6 km transfer will be maintained at less than 0.1 degree. Each LLRF system will require one channel. Figure 6.3.3.20 illustrates the layout of the Linac phase reference line.

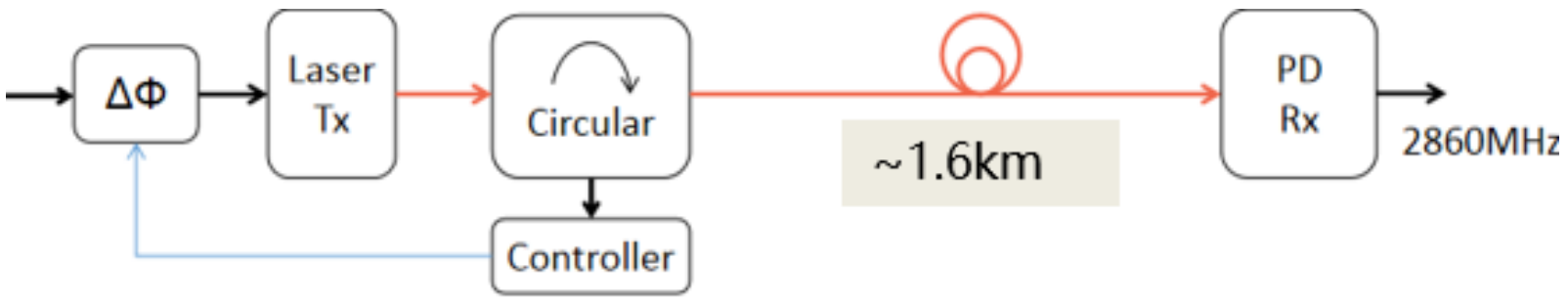

**Figure 6.3.3.20:** CEPC Linac phase reference line

This system has been successfully employed at HEPS. A 700 m phase-stabilized optical fiber was utilized to transfer the 499.8 MHz master oscillator (MO) signal to the Linac and Booster RF systems. In a span of 3 days, out-of-loop residual phase drift of ±0.02° (±110 fs) was achieved by a phase detector between the Transmit-RF and Receive-RF components. A new S-band Transmit-Receive (Tx-Rx) pair for the CEPC is currently under development, requiring only minimal modifications. The 499.8 MHz phase shifter chip was replaced to accommodate the 2,860 MHz frequency, while the laser, photon detector, fiber, and controller bandwidth continue to cover the S-band without alteration. Figure 6.3.3.21 illustrates the Tx-Rx pair at HEPS.

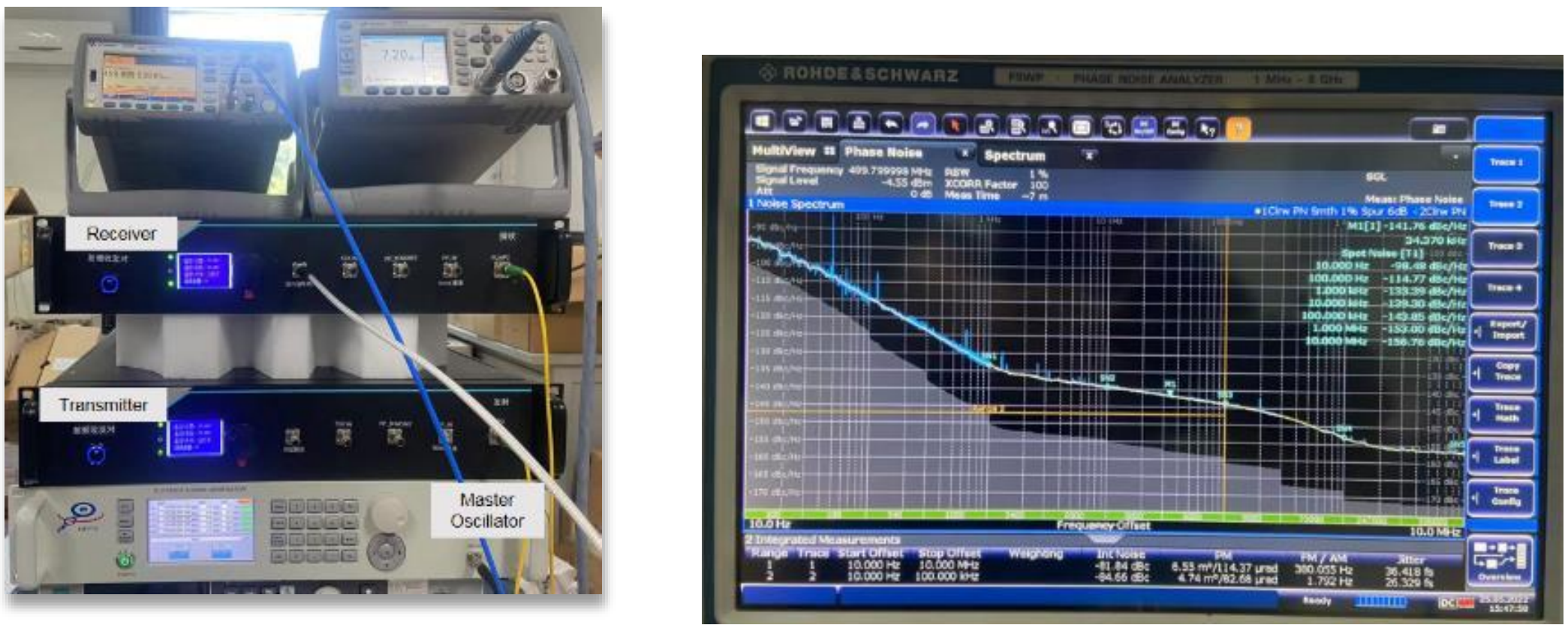

**Figure 6.3.3.21:** The Tx-Rx pair at HEPS.

## 6.3.4 RF Power Source

### 6.3.4.1 *Introduction*

In order to achieve a reasonable length for a linac, it is necessary to operate at high accelerating gradients. However, for copper accelerator structures that are not superconducting, this can result in a high-peak power per unit length and per RF source, particularly when the number of discrete sources is limited. To address this issue and enhance the peak power produced by an RF source, the SLED RF pulse-compression scheme is commonly utilized.

The Linac's primary high-power RF components consist of 33 units of 80 MW S-band (2860 MHz) klystrons and 236 units of C-band (5720 MHz) klystrons, along with conventional solid-state modulators. A waveguide system is employed to transmit power from the klystrons to the accelerating structures. Specifically, the 33 S-band klystrons provide power for 107 accelerating structures with varying accelerating apertures, while the 236 C-band klystrons provide power for 470 accelerating structures and 1 deflection cavity. An RF window is used to ensure vacuum isolation between the klystron and the waveguide transmission system.

#### 6.3.4.2 *Klystron*

The system for RF power source comprises 33 sets of 80 MW pulsed klystrons operating at a frequency of 2860 MHz (S-band). To meet the CEPC power requirement, the BAC method will be utilized to improve the klystron efficiency from 40% to 55%, based on the existing S-band 65 MW klystron used in the BEPC II linac. The BAC (Bunching Alignment and Collection) method comprises three stages: traditional bunching, alignment of the velocity spread of electrons, and collection of external electrons [1]. In comparison to other bunching mechanisms, the BAC method offers the advantage of reducing the length of the klystron while enhancing overall efficiency. The layout and gun parameters remain the same, with cavities being inserted between the $4^{th}$ and $5^{th}$ cavities [2-4]. To reduce both R&D effort and fabrication cost, the other components, such as the gun, coil, and collector, will be reused. The klystron specification is presented in Table 6.3.4.1.

**Table 6.3.4.1**: Specifications of the 2856 MHz / 80 MW klystron

| Parameters | Values |
|---|---|
| Frequency | 2860 MHz |
| Output power | 80 MW |
| Pulse width | 4 μs |
| Voltage | 350 kV |
| Current | 416 A |
| Perveance | 2 μP |
| Gain | >50 dB |
| Efficiency | 55% |

Figure 6.3.4.1 illustrates the 1D AJDISK simulation results for both the original klystron and the proposed BAC-based klystron. The introduction of four new cavities accelerates the processes of electron core oscillation and particle collection from the surroundings. As a result, more external particles are collected by the interaction section of the BAC klystron at the output cavity when compared to the original klystron. The 1D efficiency of the BAC klystron has increased from the original 49% to 64%.

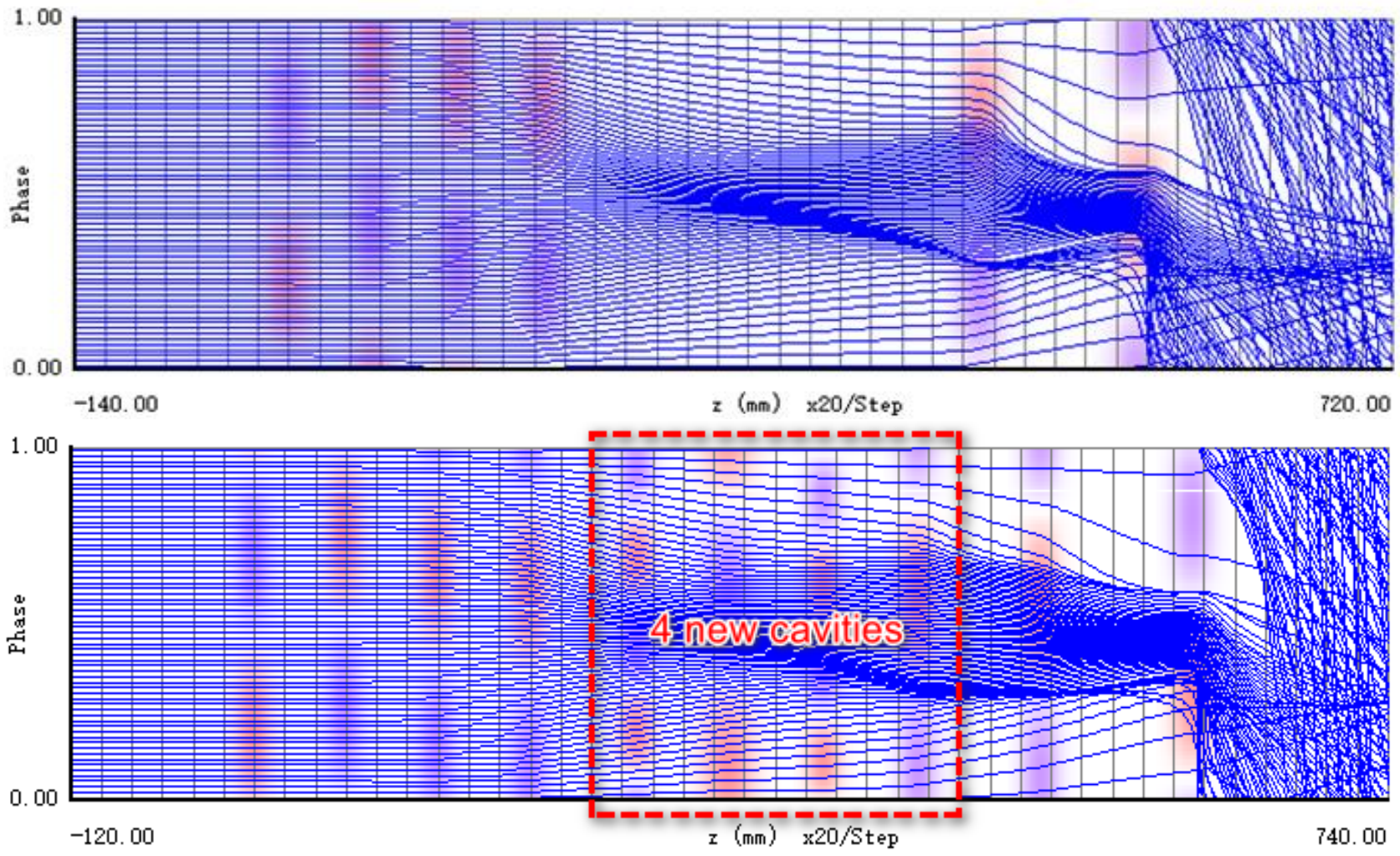

**Figure 6.3.4.1:** 1D simulation result from AJDISK.

To achieve compactness in the Linac, the baseline design also uses klystrons operating at the C-band frequency (5720 MHz). A 50 MW pulsed-power RF source is required to power four accelerating structures, each 1.8 meters in length, at an acceleration gradient of 45 MV/m. To achieve this level of RF power, a high-power electron gun assembly and a solenoid have been designed and simulated. The beam-optics design is conducted using the 2D DGUN code, which results in reduced electric field at the electrodes, reduced cathode loading, and enhanced cathode life. In addition, the 2D POISSON code was utilized to design a high-voltage ceramic seal with reduced electric field to prevent any leakage. Finally, the 3D CST code was used to model the gun assembly and the solenoid, ensuring the results obtained from the 2D codes were verified.

A 2D simulation was conducted in DGUN and POISSON to study beam optics, envelope, and solenoid at 350 kV acceleration potential. The gun design was optimized to obtain a space-charge current of 317.4 A, while reducing the electric field at the electrodes and current density at the cathode. Figures 6.3.4.2 and 6.3.4.3 illustrate the results of the simulation.

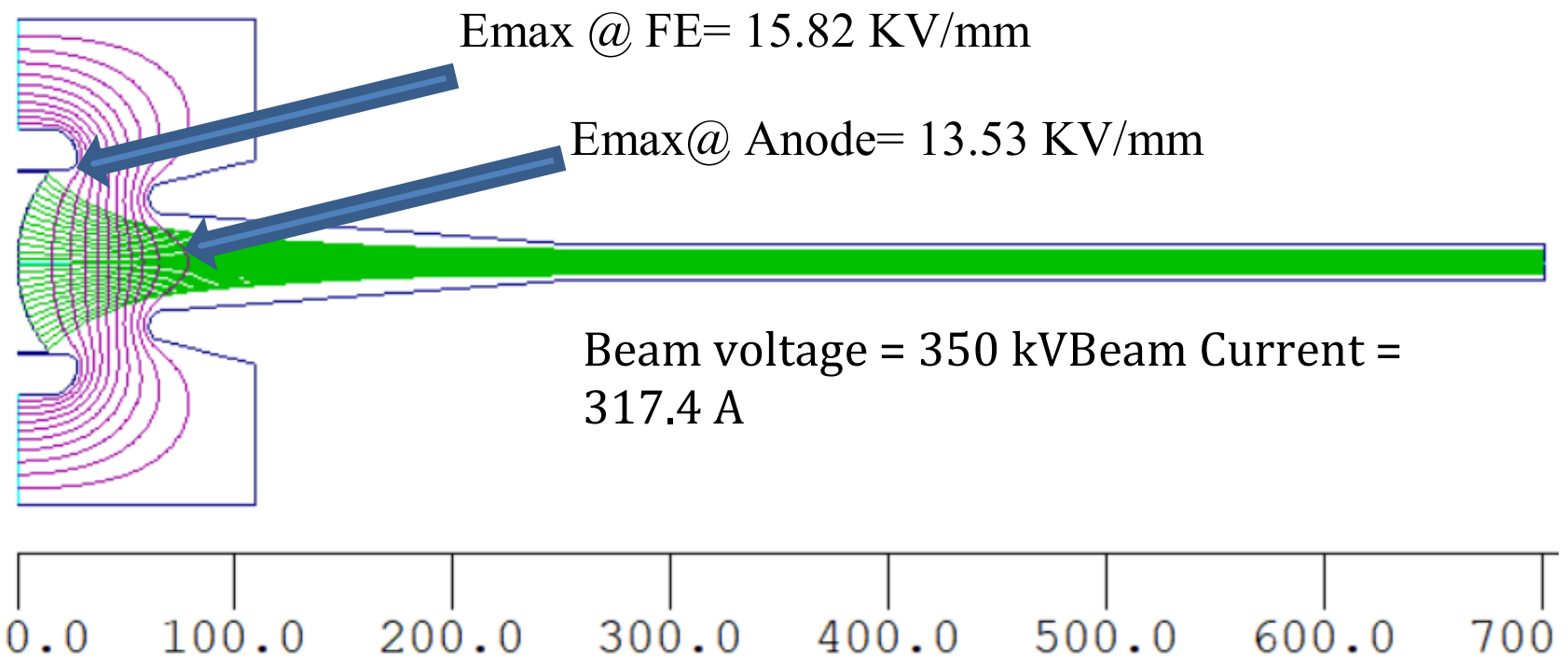

**Figure 6.3.4.2:** Beam optics design and simulation of 50 MW using DGUN.

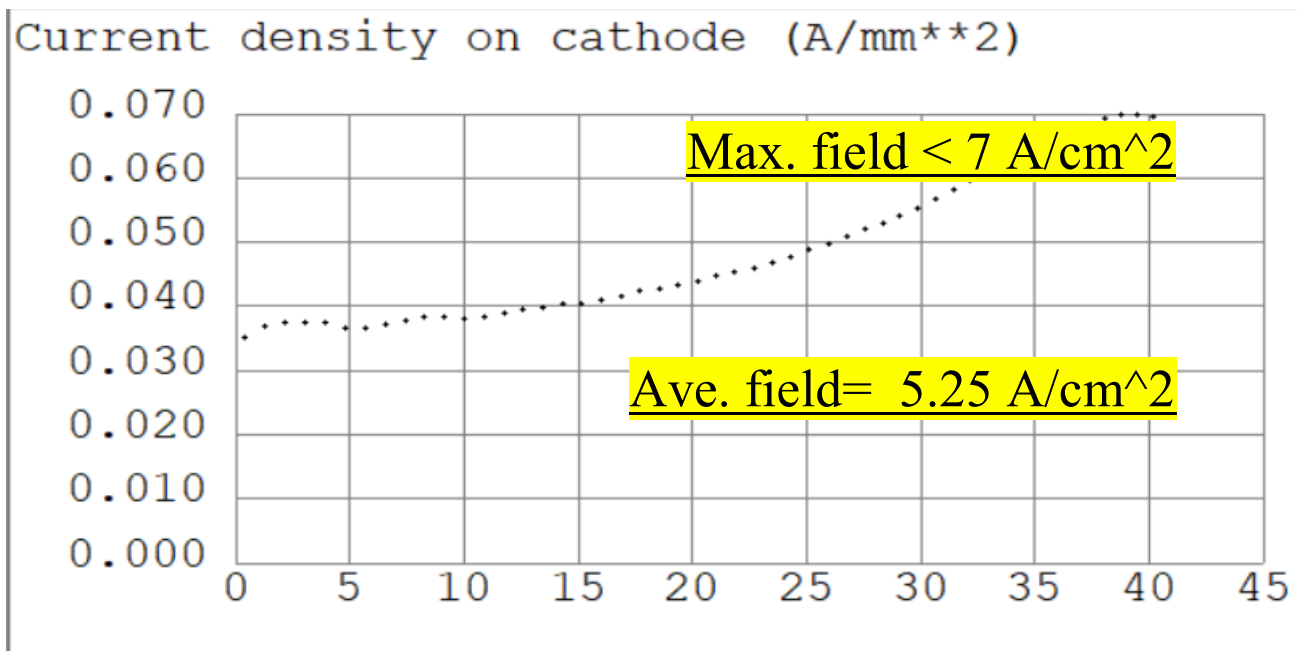

**Figure 6.3.4.3:** Current density on the cathode using DGUN.

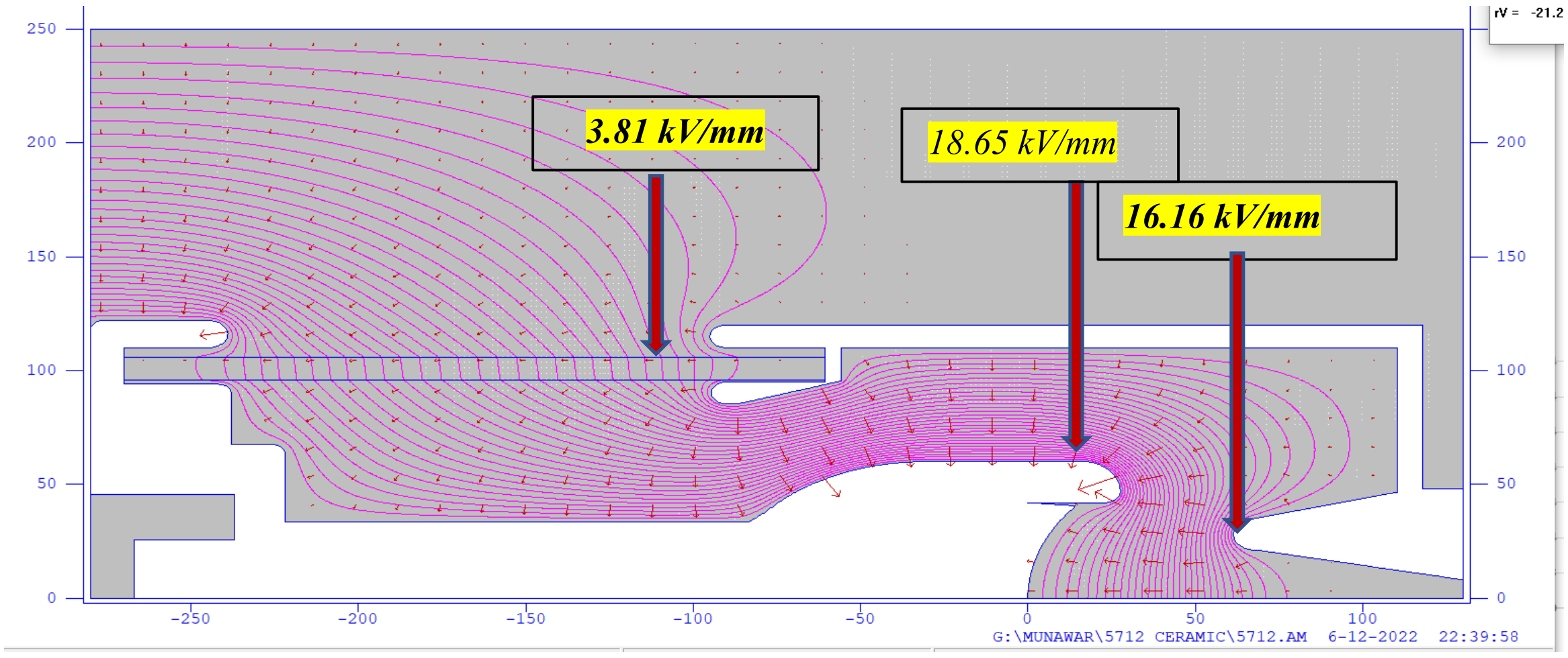

**Figure 6.3.4.4:** Beam optics & ceramic design & simulation of 50 MW gun using POISSON.

The gun envelope was designed and simulated in POISSON. The electric fields were calculated for the entire gun assembly, including the ceramic and optics components, as shown in Figure 6.3.4.4. Table 6.3.4.2 provides the parameters for the CEPC's 50 MW, 5720 MHz klystron. IHEP, in collaboration with a partnering company, is currently developing a prototype of a C-band klystron. This prototype is intended for use in the SXFEL and XFEL projects on the PAPS (Plateform of Advance Proton Source) in the near future.

**Table 6.3.4.2:** Parameters of the 50 MW klystron gun & solenoid

| **Electron Gun Parameters** | **Units** | **50 MW Klystron** |
|---|---|---|
| Cathode diameter | mm | 81.0 |
| Beam tube diameter | mm | 16.0 |
| Beam voltage | kV | 350 |
| Beam current | A | 317.4 |
| Beam perveance | μP | 1.53 |
| Cathode peak loading | A/cm$^2$ | < 7.0 |
| Cathode average loading | A/cm$^2$ | ≤ 5.25 |
| Max. electric field on focusing electrode | kV/mm | 18.65 (POISSON) |
| Max. electric field on Anode | kV/mm | 16.16 (POISSON) |
| Beam radius | mm | 5.42 (average) |
| Ripple rate | % | < 4.80 |
| Beam fill factor | ---- | 0.677 |
| Cathode field | Gs | 33.571 |
| Magnetic field $B_z$ @ tube body, | Gs | 2530 |

### 6.3.4.3 *Solid-State Modulator*

The CEPC Linac RF sources are powered by solid-state modulators (SSMs) with isolated gate bipolar transistor (IGBT) switches. Compared to pulse-forming-network (PFN) modulators with gas-switch thyratrons, the Linac SSM can provide about 8 times more stable pulses. The SSM is composed of high-precision direct-current (DC) power supplies, energy-storage capacitors, IGBTs, low-inductance cables, and a matrix pulse transformer. For the two klystron loads (S-band 2860 MHz /80 MW /100 pps /4 μs and C-band 5720 MHz /50 MW /100 pps /3 μs), the designed parameters for the Linac SSM are listed in Table 6.3.4.3.

**Table 6.3.4.3:** Parameters of SSM

| **Parameters** | **Unit** | **Value** |
|---|---|---|
| Pulse voltage | kV | 400 |
| Pulse current | A | 480 |
| Repeat pulse per second | pps | 100 |
| Pulse top width | μs | 4 |
| Pulse flatness | % (within 4 μs) | 1 |
| Pulse stability | % (RMS, 3000 pulses) | 0.02 |

The Linac SSM modulator schematic is shown in Figure 6.3.4.5.

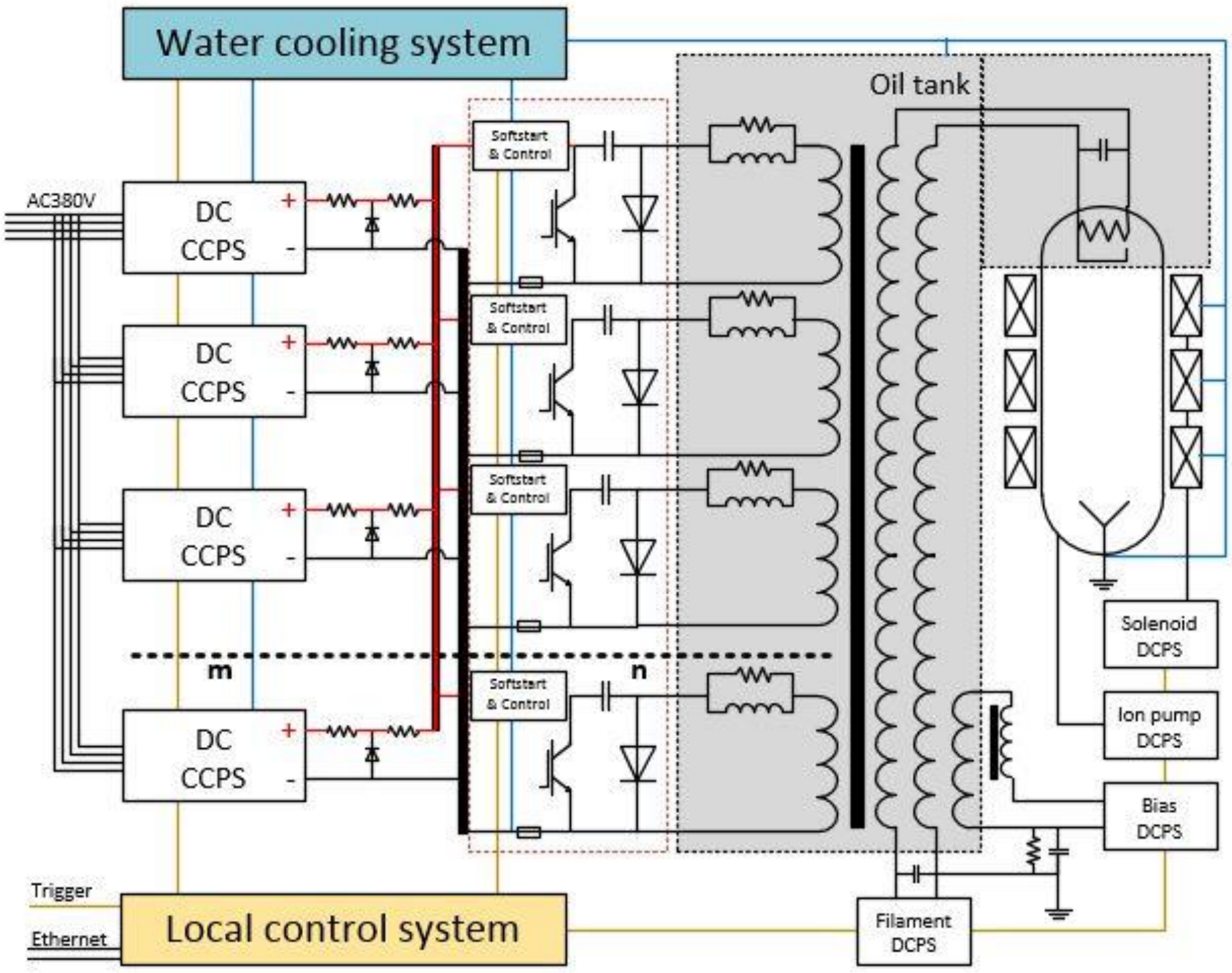

**Figure 6.3.4.5:** Schematic diagram of the SSM modulator.

The pulse top width of the Linac SSM is 4 μs, with a rise time and fall time of approximately 1.8 μs, resulting in a pulse FWHM (full width at half maximum) of 5.8 μs [5]. The average power of the DC power unit is 96 kW, which is calculated using the following formula:

$$400\ \text{kV} \times 480\ \text{A} \times 100 \times 5.8\ \mu\text{s} / 1000000 = 96\ \text{kW}.$$

To ensure redundancy, commercial 1,700V IGBTs are used as discharging unit switches, operating at DC 1,000 V/2,000 A. The total current of all discharging units is 192 kA, and the discharging unit quantity is 96. To maintain a pulse flatness of no more than 1% within 4 μs, a parallel LR compensation circuit is added to the discharging unit.

The Linac SSM pulse transformers are double-cone high-voltage windings that form the DC current loop of the klystron cathode filament. The leakage inductance $L_\delta$ and distributing capacitance $C_d$ give a high frequency $(L_\delta/C_d)^{1/2}$ of the pulse transformer, which must match the impedance of the klystron [6]. Figure 6.3.4.6 shows the SSM pulse transformer for klystron E37302.

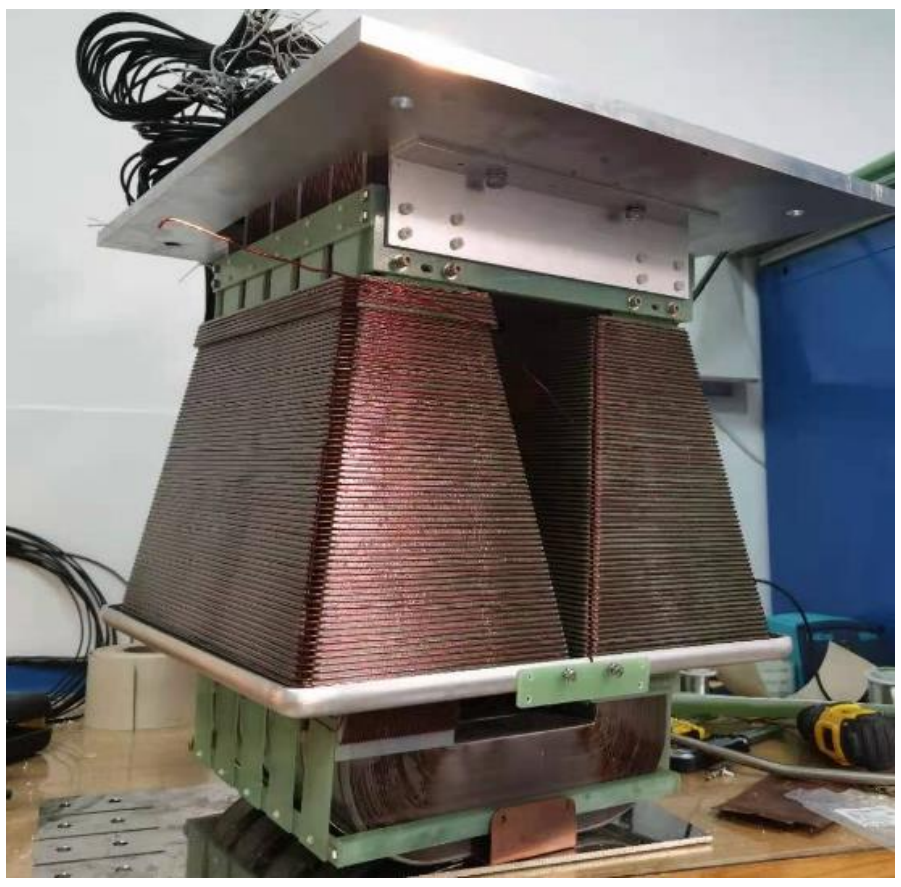

**Figure 6.3.4.6**: SSM pulse transformer (320 kV /380 A /4 μs /50 pps).

### 6.3.5 **Magnets**

#### 6.3.5.1 *Solenoids*

The solenoids of the Linac are classified into four families based on their apertures. SOL-I comprises four solenoids with an aperture of 90 mm, SOL-II has 17 solenoids with an aperture of 210 mm, SOL-III has one solenoid with an aperture of 90 mm, and SOL-IV has 15 solenoids with an aperture of 400 mm. The maximum magnetic fields for SOL-I to SOL-III are 800 Gs, while for SOL-IV, it is 5000 Gs.

Since SOL-I to SOL-III solenoids operate at a relatively low field, they will be wound using enamelled wires with water-cooling plates at the end. On the other hand, SOL-IV solenoids will be wound using hollow copper conductors with water cooling in the conductors, given their high magnetic fields. All the solenoids will be excited by direct current. Figure 6.3.5.1 shows the 3D drawing for one of the solenoids and a similar solenoid made at IHEP.

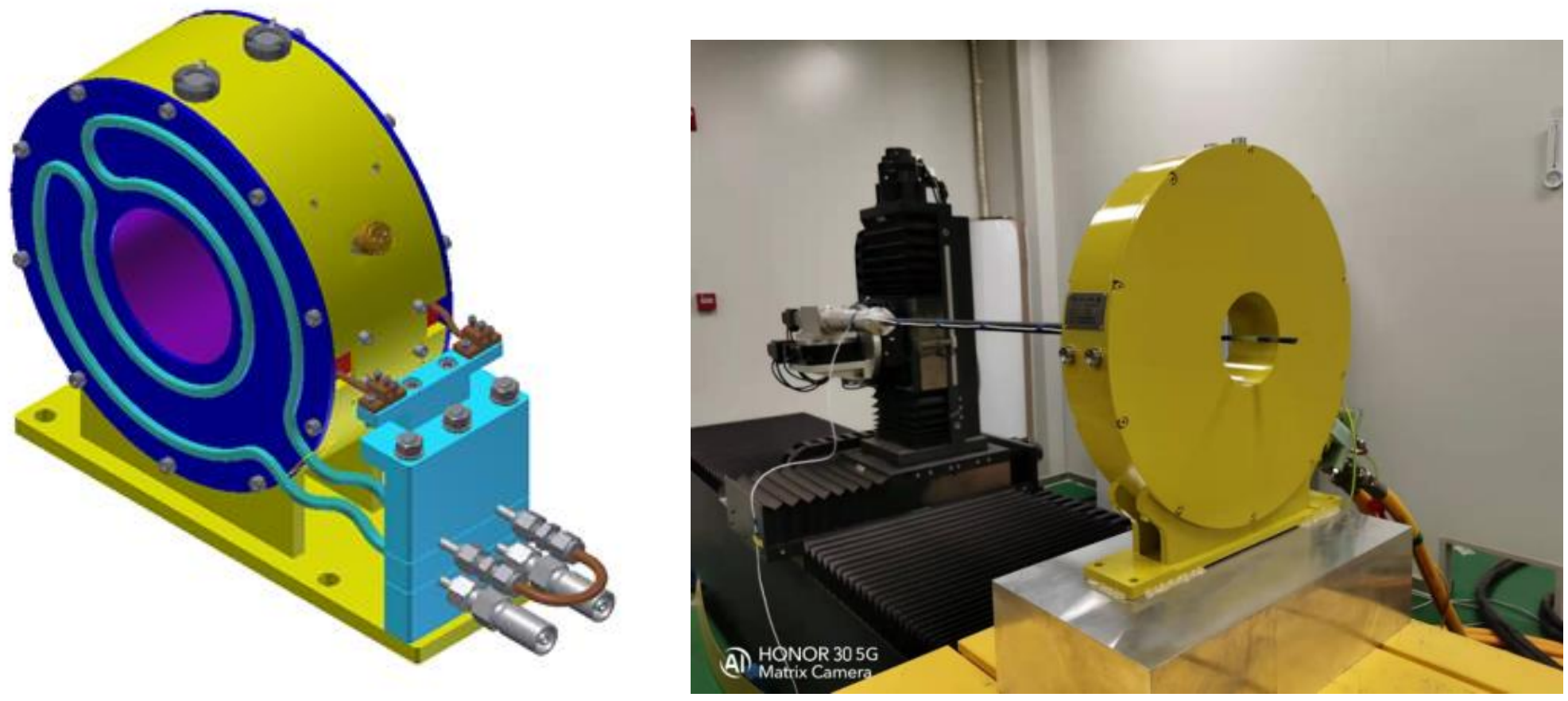

**Figure 6.3.5.1:** One solenoid module for CEPC linac and a similar solenoid made at IHEP.

The main design parameters of the solenoids are listed in Table 6.3.5.1.

**Table 6.3.5.1:** Main design parameters of the Linac solenoids

| Parameters | SOL-I | SOL-II | SOL-III | SOL-IV |
|---|---|---|---|---|
| Quantity | 4 | 17 | 1 | 15 |
| Aperture [mm] | 90 | 210 | 118 | 400 |
| Max. field [Gs] | 400 | 800 | 600 | 5000 |
| Magnetic length [mm] | 80 | 110 | 80 | 1000 |
| Ampere-turns [AT] | 5250 | 25973 | 11250 | 440000 |
| Turns | 500 | 2420 | 1125 | 1400 |
| Current [A] | 10.5 | 10.7 | 10.0 | 314.3 |
| Current density [A/mm$^2$] | 1.44 | 1.47 | 1.37 | 3.91 |
| Conductor size [mm] | 1.95×3.75 | 1.95×3.75 | 1.95×3.75 | 10×10 Φ5 R1 |
| Resistance [Ω] | 0.615 | 9.108 | 2.636 | 0.523 |
| Voltage [V] | 6.5 | 97.8 | 26.4 | 164.2 |
| Power loss [kW] | 0.07 | 1.05 | 0.26 | 51.61 |
| Coil weight [kg] | 15.2 | 225.8 | 65.3 | 1465.0 |
| Max diameter [mm] | 196.9 | 671.0 | 431.0 | 714.0 |
| Water pressure [kg/cm$^2$] | Water-cooling plates at the end | | | 6 |
| Cooling circuits | | | | 10 |
| Length of circuit [m] | | | | 203 |
| Water flow velocity [m/s] | | | | 1.43 |
| Total water flow [l/s] | | | | 1.12 |
| Temperature rise [°C] | | | | 10.9 |

### 6.3.5.2 *Dipole Magnets*

The Linac contains 19 dipole magnets that are classified into five families. To control fabrication risk and cost, the maximum length of the magnet is set to be less than 6 m. All the dipole magnets are excited by DC current. The iron cores can be made of low carbon silicon steel laminations or solid steel, while the coils are wound using hollow copper conductors with water cooling.

The cores of the magnets have a curved structure, allowing for the use of racetrack-shaped coils. All the dipoles use the same H-type structure. To install the vacuum chamber, the whole magnet is split into two halves.

The cross sections for the dipole magnets have been designed and optimized using OPERA-2D. In the simulation, only half of the magnet is modeled. Fig. 6.3.5.2 shows the magnetic flux lines, Fig. 6.3.5.3 shows 3D drawing for one of the dipole magnets and the similar magnet manufactured at IHEP.

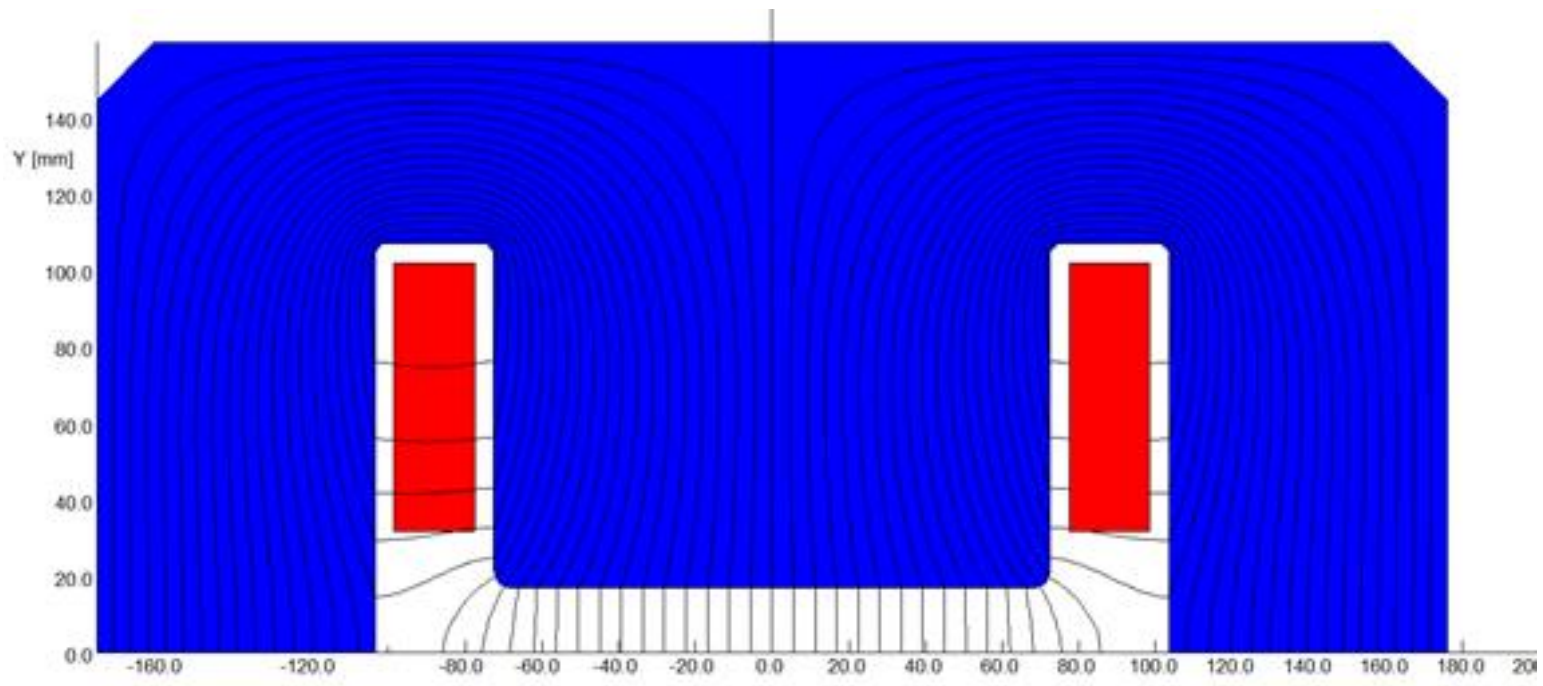

**Figure 6.3.5.2:** 2D flux lines of AM1 dipole magnet (half cross section).

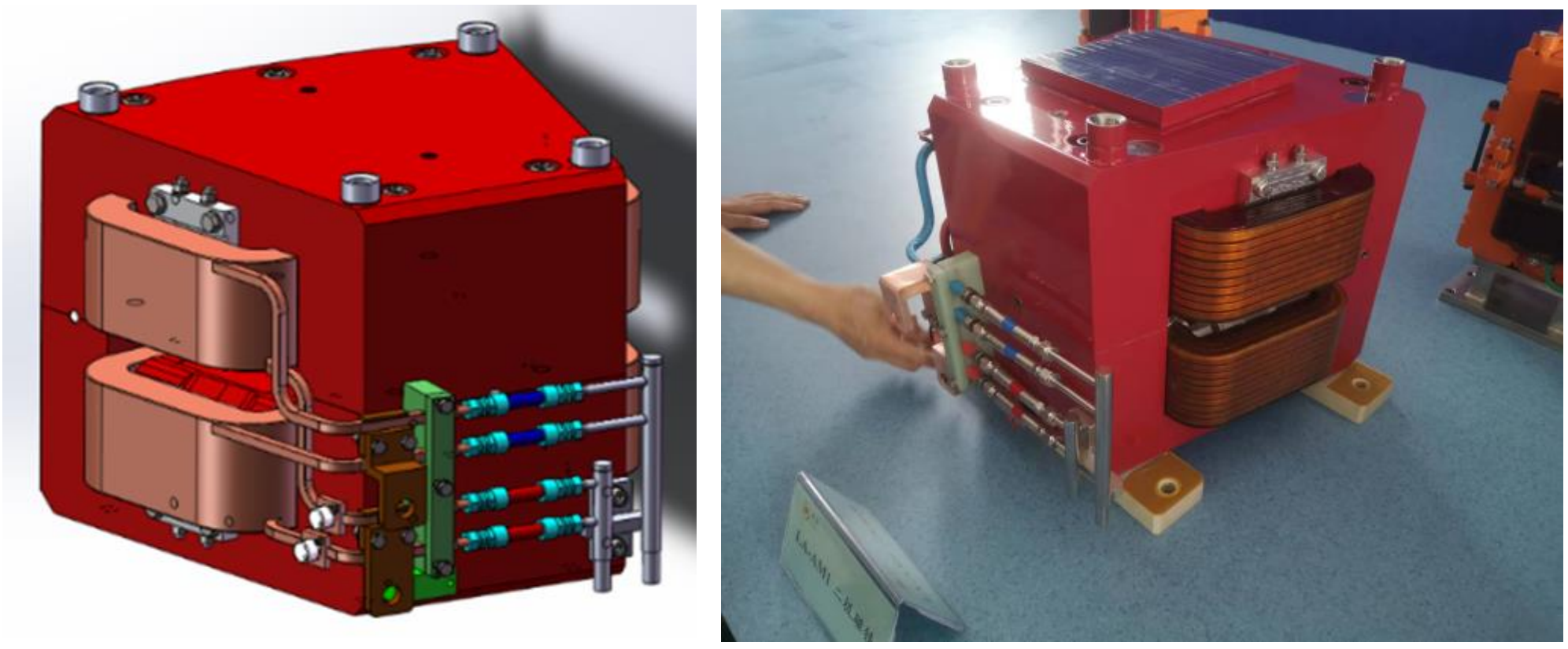

**Figure 6.3.5.3:** 3D drawing of the AM1 magnet and a similar magnet made at IHEP.

The main design parameters of the dipoles are listed in Table 6.3.5.2.

**Table 6.3.5.2:** Main design parameters of the Linac dipole magnets

| Parameters | B | AM1 | AM2/AM5/ AM6/AM7 | AM3/ CB1 | AM4/ CB2 |
|---|---|---|---|---|---|
| Quantity | 4 | 1 | 4 | 5 | 5 |
| Aperture [mm] | 35 | 35 | 24 | 60 | 35 |
| Max. field [Gs] | 10000 | 3300 | 10000 | 4200 | 10000 |
| Magnetic length [mm] | 1047.2 | 260 | 5847 | 698 | 698 |
| Good field region [mm] | 30 | 30 | 20 | 54 | 30 |
| Field errors | 0.1% | 0.1% | 0.1% | 0.1% | 0.1% |
| Ampere turns per pole [At] | 14203 | 4687 | 9739 | 10226 | 14203 |
| Turns per pole | 24 | 24 | 24 | 24 | 24 |
| Max. current [A] | 592 | 195 | 406 | 426 | 592 |
| Size of conductor [mm×mm] | 13×13 D8 | 7×7 D4 | 14×14 D10 | 13×13 D8 | 13×13 D8 |
| Current density [A/mm$^2$] | 4.98 | 5.36 | 3.45 | 3.59 | 4.98 |
| Resistance of the coil (mΩ) | 23.7 | 23.6 | 113.0 | 18.3 | 17.3 |
| Inductance (mH) | 15.8 | 2.8 | 119.2 | 8.6 | 10.5 |
| Voltage [V] | 14 | 5 | 46 | 8 | 10 |
| Power loss (kW) | 8.3 | 0.9 | 18.6 | 3.3 | 6.1 |
| Height of core [mm] | 300 | 250 | 270 | 350 | 300 |
| Width of core [mm] | 400 | 270 | 380 | 500 | 400 |
| Core length [mm] | 1021 | 234 | 5829 | 653 | 672 |
| Weight of magnet [ton] | 1.1 | 0.15 | 5.0 | 1.0 | 0.8 |
| Cooling circuits | 2 | 2 | 4 | 2 | 2 |
| Water pressure [kg/cm$^2$] | 6 | 6 | 6 | 6 | 6 |
| Diameter of cooling hole [mm] | 8 | 4 | 10 | 8 | 8 |
| Water flow velocity [m/s] | 0.115 | 0.034 | 0.129 | 0.133 | 0.137 |
| Total water flow [l/s] | 0.229 | 0.069 | 0.514 | 0.265 | 0.274 |
| Temperature rise [°C] | 8.6 | 3.1 | 8.6 | 3.0 | 5.3 |

### 6.3.5.3 *Quadrupole Magnets*

The Linac's quadrupole magnets are divided into two types: large aperture of 150 mm, and small aperture in three different sizes: 24 mm, 34 mm and 60 mm.

All the magnets are DC-excited and designed and produced in accordance with general accelerator magnet specifications. The core can be made of silicon steel lamination sheets or DT4 solid iron, while the coils are made using hollow copper conductors. The maximum field on the pole tips of most magnets is less than 5000 Gs, allowing the poles to have a rectangle root and the coils to have a simple racetrack structure.

Figure 6.3.5.4 shows the magnetic flux in one of the quadrupole magnets and a similar magnet made at IHEP, while Table 6.3.5.3 and 6.3.5.4 list the main parameters of all quadrupoles.

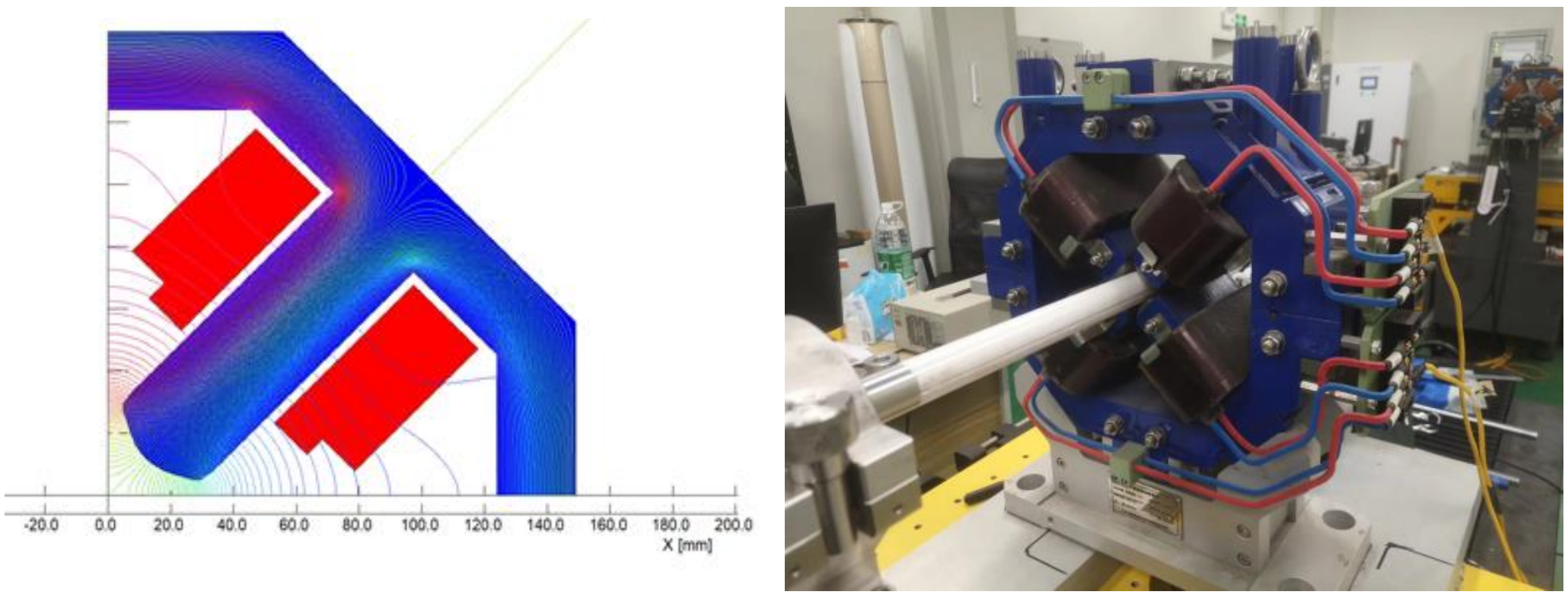

**Figure 6.3.5.4:** The magnetic flux of the magnet and a similar magnet made at IHEP.

**Table 6.3.5.3:** Main parameters of the Linac quadrupole magnets

| Parameters | LA-24Q-300L | LA-24Q-600L | LA-34Q-100L | LA-34Q-200L | LA-34Q-400L |
|---|---|---|---|---|---|
| Number | 104 | 52 | 22 | 87 | 44 |
| Aperture [mm] | 24 | 24 | 34 | 34 | 34 |
| Field gradient [T/m] | 50 | 50 | 22 | 24 | 26 |
| Magnetic length [mm] | 300 | 600 | 100 | 200 | 400 |
| GFR (radius) [mm] | 9.6 | 9.6 | 14 | 14 | 14 |
| Harmonic errors | 1.0E-03 | 1.0E-03 | 1.0E-03 | 1.0E-03 | 1.0E-03 |
| Ampere turns per pole | 2951 | 2951 | 2606 | 2843 | 3079 |
| Turns per pole | 16 | 16 | 20 | 20 | 20 |
| Current [A] | 184 | 184 | 130 | 142 | 154 |
| Size of conductor [$mm^2$] | 7×7D4 | 7×7D4 | 7×7D4 | 7×7D4 | 7×7D4 |
| Current density [$A/mm^2$] | 5.06 | 5.06 | 3.58 | 3.90 | 4.23 |
| Resistance [mΩ] | 22 | 42 | 11 | 20 | 36 |
| Voltage [V] | 4.00 | 7.69 | 1.48 | 2.80 | 5.59 |
| Power loss [kW] | 0.74 | 1.42 | 0.19 | 0.40 | 0.86 |
| Core height/weight [mm] | 288 | 288 | 420 | 420 | 420 |
| Core length [mm] | 291 | 591 | 87 | 187 | 387 |
| Weight of magnet [kg] | 167 | 337 | 104 | 220 | 453 |
| Cooling circuits | 4 | 4 | 4 | 4 | 4 |
| Water pressure [$kg/cm^2$] | 3 | 3 | 3 | 3 | 3 |
| Water flow velocity [m/s] | 2.59 | 1.78 | 3.73 | 2.73 | 1.93 |
| Total water flow [l/s] | 0.130 | 0.090 | 0.188 | 0.137 | 0.097 |
| Temperature rise [°C] | 1.35 | 3.77 | 0.24 | 0.69 | 2.12 |

**Table 6.3.5.4:** Main parameters of the Linac quadrupole magnets (cont'd)

| Parameters | LA-60Q-100L | LA-60Q-200L | LA-150Q-300L | LA-150Q-600L |
|---|---|---|---|---|
| Number | 3 | 6 | 36 | 18 |
| Aperture[mm] | 60 | 60 | 150 | 150 |
| Field gradient [T/m] | 10 | 10 | 7.9 | 7.9 |
| Magnetic length [mm] | 200 | 400 | 300 | 600 |
| GFR (radius) [mm] | 24 | 24 | 60 | 60 |
| Harmonic errors | 1.0E-03 | 1.0E-03 | 1.0E-03 | 1.0E-03 |
| Ampere turns per pole | 3688 | 3688 | 18477 | 18477 |
| Turns per pole | 24 | 24 | 80 | 80 |
| Current [A] | 154 | 154 | 231 | 231 |
| Size of conductor [mm$^2$] | 7×7D4 | 7×7D4 | 9×9D6 | 9×9D6 |
| Current density [A/mm$^2$] | 4.22 | 4.22 | 4.38 | 4.38 |
| Resistance [mΩ] | 26 | 46 | 100 | 169 |
| Voltage [V] | 4.06 | 7.13 | 23.06 | 39.00 |
| Power loss [kW] | 0.62 | 1.10 | 5.33 | 9.01 |
| Core height/weight [mm] | 720 | 720 | 1050 | 1050 |
| Core length [mm] | 178 | 378 | 244 | 544 |
| Weight of magnet [kg] | 593 | 1255 | 1834 | 4008 |
| Cooling circuits | 4 | 4 | 8 | 8 |
| Water pressure [kg/cm$^2$] | 3 | 3 | 3 | 3 |
| Water flow velocity [m/s] | 2.31 | 1.68 | 1.74 | 1.29 |
| Total water flow [l/s] | 0.116 | 0.084 | 0.394 | 0.291 |
| Temperature rise [°C] | 1.28 | 3.09 | 3.22 | 7.36 |

### 6.3.5.4 *Correcting Magnets*

In the CEPC Linac, there are seven types of correctors. Three of these, C1, C2, and C3, have very low fields and are designed as correctors without iron cores. The remaining correctors have relatively high fields and are designed as correctors with iron cores.

The correctors without iron cores have varying structures depending on the coil supports, while all types of correctors with iron cores have a simple C-type frame core made of solid iron bars. The coils for all correctors are wound with solid copper conductors, which are installed on the yokes of the cores. As the correctors operate in DC mode and have a current density lower than 1 A/mm$^2$, the coils do not require water cooling.

Figure 6.3.5.5 shows the coils and 3D drawing of one of the correctors without iron cores, while the magnetic flux distribution and 3D drawing of the correctors with iron cores are depicted in Fig. 6.3.5.6. The main parameters of the seven types of correctors can be found in Tables 6.3.5.5 to 6.3.5.8.

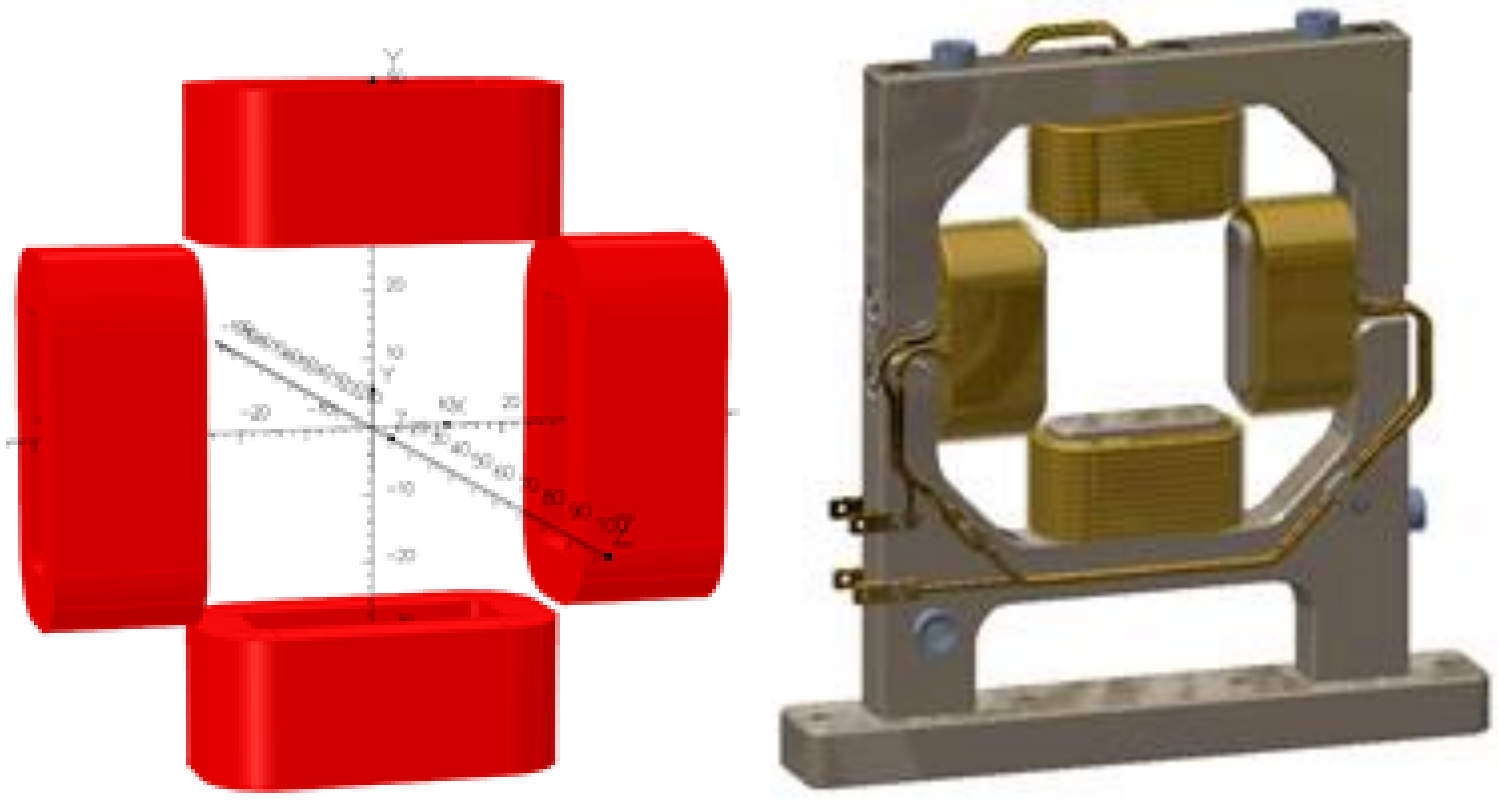

**Fig.6.3.5.5:** The coils and 3D drawing of one corrector without iron cores.

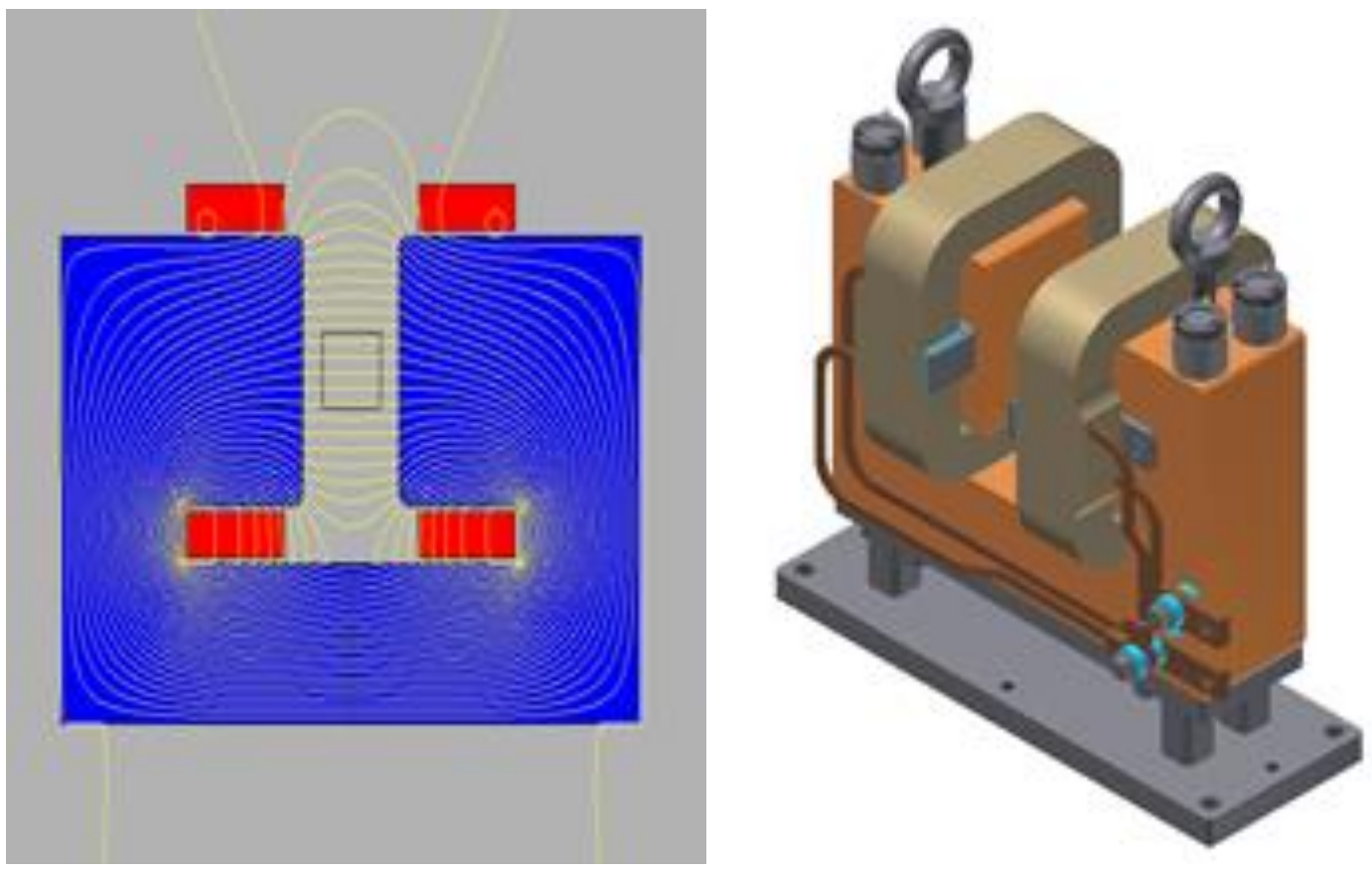

**Fig.6.3.5.6:** The magnetic flux lines and 3D drawing of one corrector with iron cores.

**Table 6.3.5.5:** Main parameters of the C1 corrector

| **Parameters** | **C1** |
|---|---|
| Number | 1 |
| Aperture[mm] | 50 |
| Field [Gs] | 10 |
| Magnetic length [mm] | 40 |
| Mechanical length [mm] | 30 |
| Conductor size [mm] | 1×2 |
| Turns per pole | 80 |
| Current [A] | 2.64 |
| Current density [A/mm$^2$] | 1.5 |
| Voltage [V] | 0.62 |
| Resistance [Ω] | 0.24 |
| Power loss [W] | 1.64 |

**Table 6.3.5.6:** Main parameters of the C2 correctors

| **Parameters** | **C2** |
|---|---|
| Number | 2 |
| Aperture [mm] | 50 |
| Central field [Gs] | 11.2 |
| Effective length [mm] | 53 |
| Mechanical length [mm] | 24 |
| Conductor size [mm] | 1×2 |
| Turns per pole | 4×6=24 |
| Current [A] | 1.5 |
| Current density [A/mm$^2$] | 0.85 |
| Voltage [V] | 0.11 |
| Resistance [Ω] | 0.07 |
| Power loss [W] | 0.17 |

**Table 6.3.5.7:** Main parameters of the C3 correctors

| **Parameters** | **C3** | |
|---|---|---|
| Type of corrector | Horizontal | Vertical |
| Number | 1 | 1 |
| Aperture [mm] | 160 | 186 |
| Central field [Gs] | 50 | 50 |
| Length of coils | 400 | 400 |
| Conductor size [mm] | 1.5×4 | 1.5×4 |
| Turns per pole | 100 (2×50) | 112 (2×56) |
| Current [A] | 6.24 | 6.42 |
| Current density [A/mm$^2$] | 1.04 | 1.07 |
| Voltage [V] | 3.1 | 3.7 |
| Resistance [Ω] | 0.49 | 0.58 |
| Power loss [W] | 19 | 24 |

**Table 6.3.5.8:** Main parameters of the Linac correctors with iron cores.

| Parameters | L100-150C | L100-34C | L200-24C |
|---|---|---|---|
| Quantity | 25 | 100 | 146 |
| Gap[mm] | 150 | 34 | 24 |
| Max. field [Gs] | 50 | 450 | 2500 |
| Magnetic length [mm] | 100 | 100 | 200 |
| Good Field Region [mm] | 30 | 30 | 26 |
| Field uniformity | 1% | 1% | 1% |
| Turns per pole | 30 | 60 | 80 |
| Max. current [A] | 10.1 | 10.3 | 30.3 |
| Size of conductor [$mm^2$] | 2×5 | 2×5 | 3.2×8.5 |
| Current density [$A/mm^2$] | 1.01 | 1.03 | 1.11 |
| Resistance [mΩ] | 82 | 109 | 85 |
| Power loss (W) | 8.4 | 11.6 | 77.7 |
| Voltage [V] | 0.8 | 1.1 | 2.6 |
| Height of core [mm] | 360 | 200 | 200 |
| Width of core [mm] | 300 | 170 | 180 |
| Core Length [mm] | 80 | 90 | 190 |
| Weight of magnet [kg] | 50 | 25 | 82 |

### 6.3.6 Magnet Power Supplies

The total length of the Linac is approximately 1800 meters. The Linac consists of 37 solenoids, 19 dipoles, 372 quadrupoles, and 275 correctors. All of the magnets in the Linac are powered by direct current (DC). For a visual representation of the Linac system's layout, please refer to Figure 6.3.6.1.

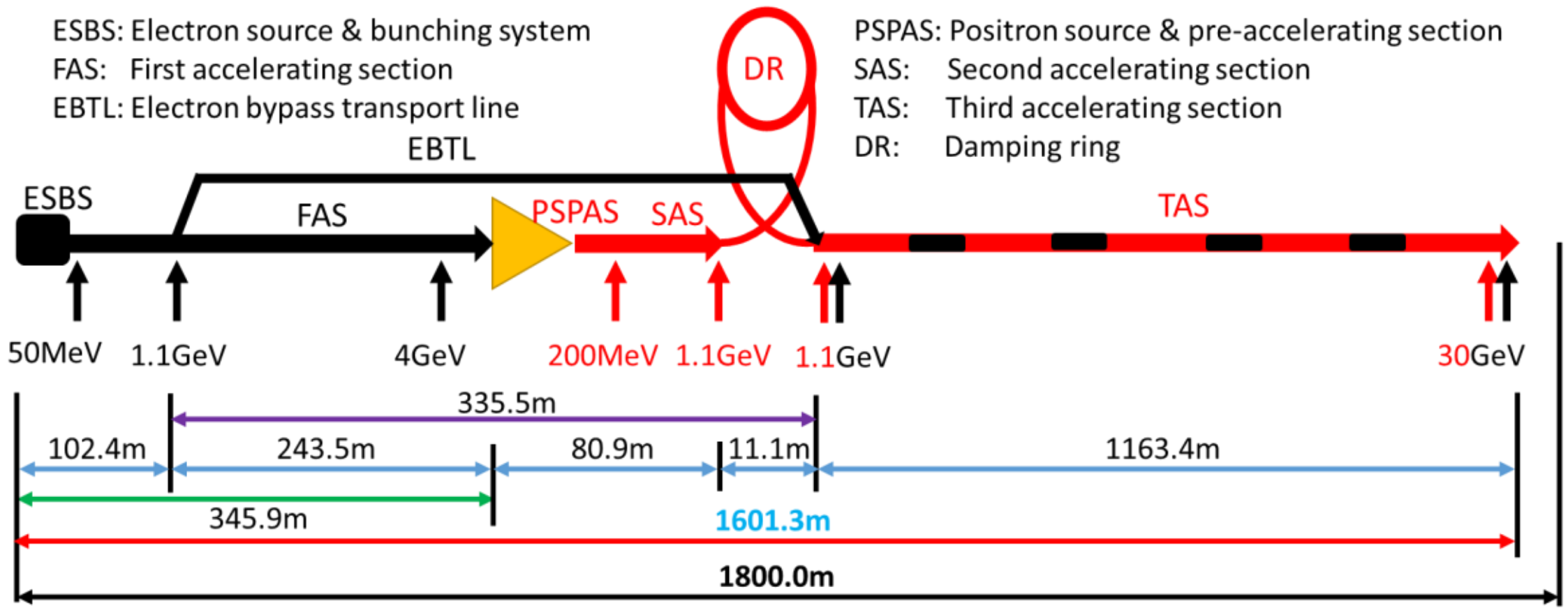

**Figure 6.3.6.1:** Layout of the Linac.

#### 6.3.6.1 Types of Power Supplies

The solenoids in the Linac are categorized into four families based on their apertures. SOL-I consists of four solenoids with a 90 mm aperture, SOL-II comprises 17 solenoids with a 210 mm aperture, SOL-III has one solenoid with a 118 mm aperture, and SOL-IV includes 15 solenoids with a 400 mm aperture. All of the solenoids are powered independently.

In the Linac, there are 19 dipole magnets that are classified into five families. AM1 consists of one magnet, AM2/AM5/ AM6/AM7 includes four magnets, AM3/CB1 has five magnets, AM4/CB2 has five magnets, and B comprises four magnets. All magnets are powered independently.

The Linac is equipped with 372 quadrupole magnets that are divided into nine families. The LA-24Q-300L family consists of 104 magnets, LA-24Q-600L has 52 magnets, LA-34Q-100L includes 22 magnets, LA-34Q-200L has 87 magnets, LA-34Q-400L comprises 44 magnets, LA-60Q-100L consists of 3 magnets, LA-60Q-200L has 6 magnets, LA-150Q-300L includes 36 magnets, and LA-150Q-600L has 18 magnets. Among these families, LA-24Q-300L, LA-34Q-100L, LA-34Q-200L, LA-60Q-100L, and LA-150Q-300L require two magnets to be powered in series, while the other magnets are powered independently.

In the CEPC Linac, there are six types of correctors. C1 and C3 are combination magnet consisting of two sets of horizontal and vertical coils, which requires two power supplies. The remaining corrector magnets are powered independently.

In summary, the Linac system requires a total of 37 solenoid power supplies, 19 dipole power supplies, 250 quadrupole power supplies, and 277 corrector power supplies. All solenoid, dipole, and quadrupole power supplies are unipolar, while the corrector power supplies are bipolar. To minimize cable losses, the power supplies are installed along the klystron gallery near the magnet load. The overall power capacity of the Linac power supply system is 1.57 MW.

Table 6.3.6.1 provides the specifications of the main magnets and correction magnets power supplies for the Linac. However, since the magnet parameters are expected to change in the future, there has been no serious effort made to optimize these parameters and reduce the number of power supply types required.

**Table 6.3.6.1:** Magnet power supply requirements for the Linac

| Magnet | Quantity | Stability /8hours | Output Rating |
|---|---|---|---|
| SOL-I | 4 | 100 ppm | 12A/12V |
| SOL-II | 17 | 100 ppm | 12A/120V |
| SOL-III | 1 | 100 ppm | 11A/35V |
| SOL-IV | 15 | 100 ppm | 350A/190V |
| AM1 | 1 | 100 ppm | 220A/10V |
| AM2/AM5/ AM6/AM7 | 4 | 100 ppm | 470A/55V |
| AM3/CB1 | 5 | 100 ppm | 470A/18V |
| AM4/CB2 | 5 | 100 ppm | 650A/20V |
| AB | 4 | 100 ppm | 650A/20V |
| LA-24Q-300L | 52 | 100 ppm | 200A/13V |
| LA-24Q-600L | 52 | 100 ppm | 200A/13V |
| LA-34Q-100L | 11 | 100 ppm | 150A/7V |
| LA-34Q-200L | 49 | 100 ppm | 160A/10V |
| LA-34Q-400L | 44 | 100 ppm | 170A/13V |
| LA-60Q-100L | 3 | 100 ppm | 170A/13V |
| LA-60Q-200L | 3 | 100 ppm | 170A/13V |
| LA-150Q-300L | 18 | 100 ppm | 250A/55V |
| LA-150Q-600L | 18 | 100 ppm | 250A/55V |
| C1-H/C1-V | 2 | 300 ppm | ±3A/±2V |
| C2 | 2 | 300 ppm | ±3A/±2V |
| C3-H | 1 | 300 ppm | ±7A/±7V |
| C4-V | 1 | 300 ppm | ±7A/±7V |
| L100-150C | 25 | 300 ppm | ±12A/±6V |
| L100-35C | 100 | 300 ppm | ±12A/±6V |
| L200-30C | 146 | 300 ppm | ±33A/±7V |
| **Total** | **583** | | |

### 6.3.6.2 *Design of the Power Supply System*

The design criteria for the Linac power supplies align with those of the Collider supplies. The majority of the Linac power supplies are DC supplies that employ switched mode as the primary topology. Each power supply is specifically designed according to the magnet's parameters and ratings, with a safety margin of 10-15% in both current and voltage.

To ensure convenient maintenance and repair, all Linac power supplies are designed as modular units and incorporate digital control features. Two structural frameworks will be implemented for the Linac power supply as shown in Figure 6.3.6.2. These frameworks will provide the necessary structure and organization for efficient power supply deployment and operation.

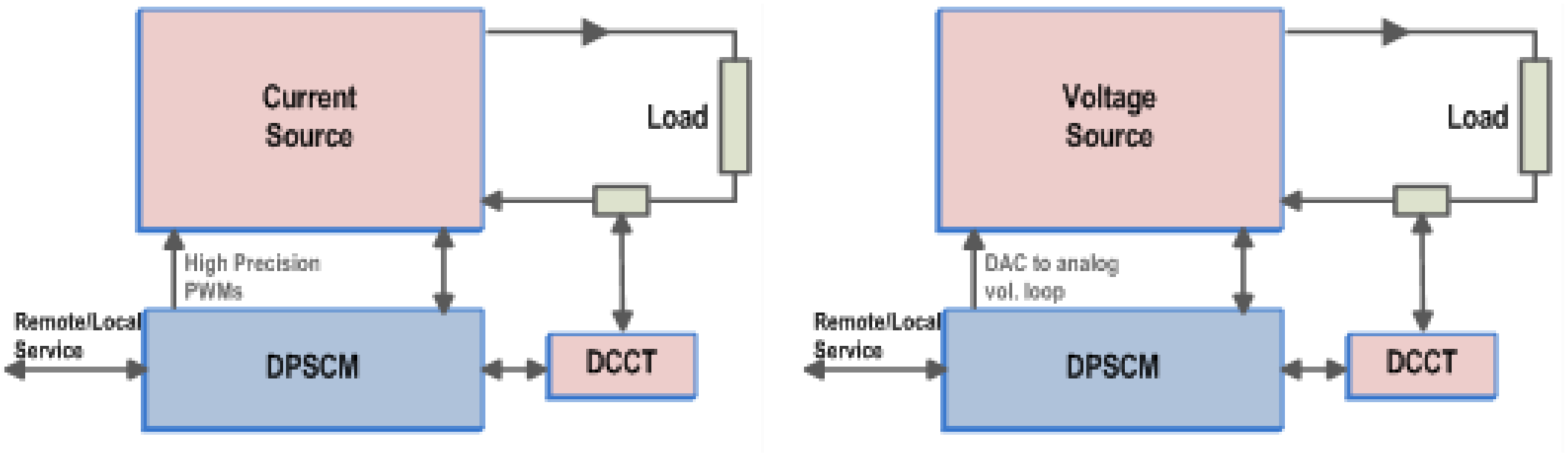

**Figure 6.3.6.2:** Two structural frameworks of the Linac power supply.

In left of Figure 6.3.6.2, the power supply adopts an all-digital + all-switch structure. The digital controller of the power supply enables digital adjustment and control of all control loops. It generates an ultra-high precision PWM signal through the hardware, with a minimum variation of 150 ps. This setup ensures the realization of an ultra-high precision switching power supply.

In the right of Figure 6.3.6.2 (b) illustrates a partial digital + arbitrary topology structure. In this configuration, the power supply's digital controller achieves high-precision digital adjustment and control of the current closed-loop. The output of the current closed-loop control is converted to an analog signal through a digital-to-analog conversion (DAC) circuit, which serves as the reference for the voltage loop. The power part of the supply provides the voltage source, and it can utilize any topology. This control mode leverages the advantages of digital control while overcoming the limitations associated with the dependence of digital control on topological structure and the precision of digital PWM control.

### 6.3.6.3 *Digital Power Supply Control Module Design*

The digital control module of the power supply serves as the executive element for achieving digital control and is crucial for achieving high-precision control.

For the CEPC's power supply system, the second generation of the digital power supply control module (DPSCM-II) will be utilized. This module has been developed specifically for the HEPS project. The DPSCM-II maintains the overall architecture designed by the first generation DPSCM, utilizing a field-programmable gate array (FPGA) as the core for data processing. The DPSCM-II employs a System-on-a-Programmable Chip (SOPC) to realize the digital control of the power supply. Figure 6.3.6-3 provides a block diagram illustrating the power supply structure embedded within the DPSCM-II.

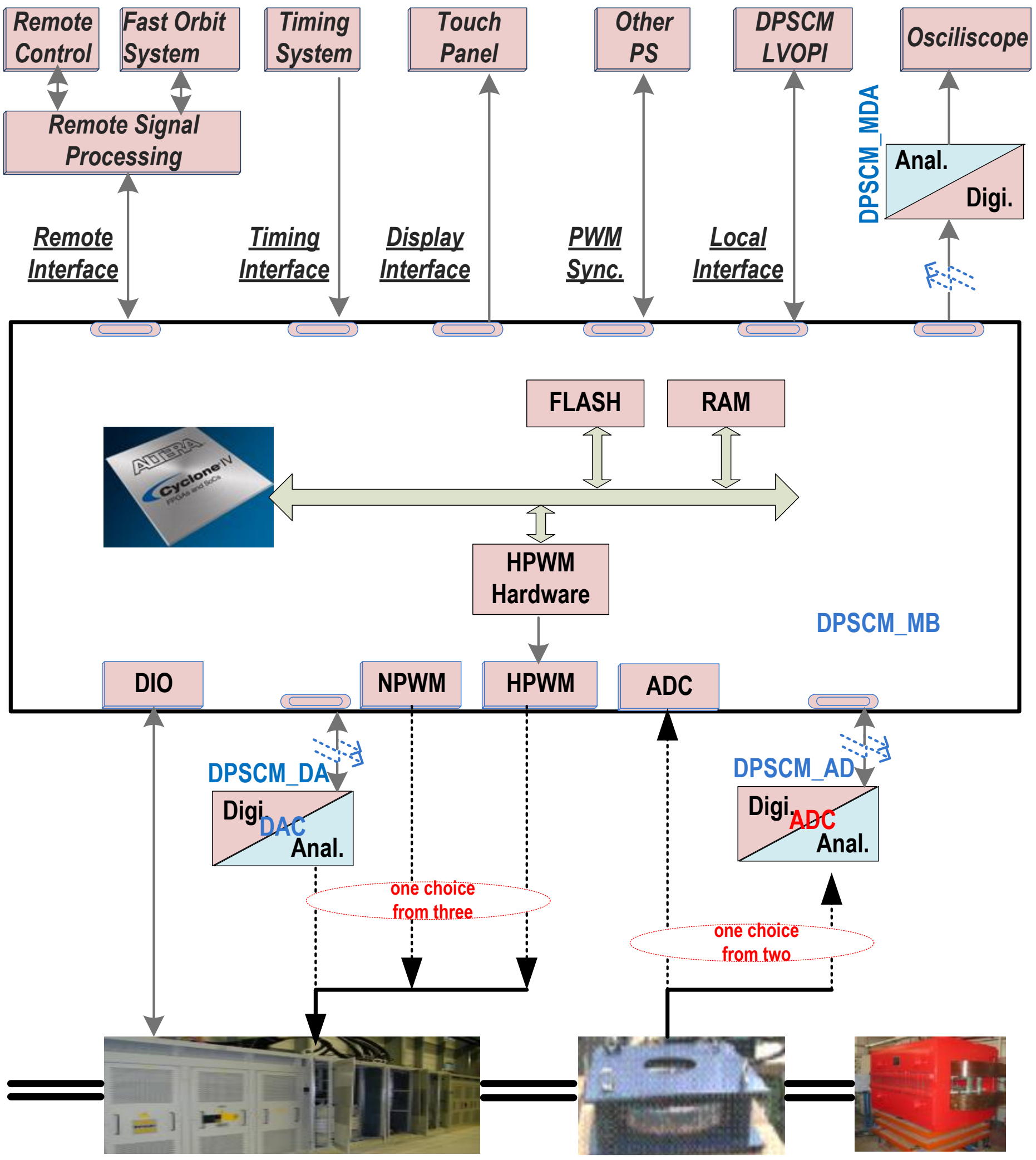

**Figure 6.3.6.3:** Digital power block diagram embedded in DPSCM-II

The DPSCM-II consists of several key hardware components, including the main board (DPSCM_MB), the high-precision ADC control board (DPSCM_AD), the digital-to-analog conversion circuit (DPSCM_DA) for interfacing the digital current closed-loop with other analog control loops of the power supply, and the power supply monitoring interface circuit (DSPCM_MDA) comprised of a multi-channel DAC.

The main board (DPSCM_MB) incorporates various interface circuits that facilitate communication between the power supply and other systems. These interfaces include the optical fiber interface for remote control, the optical fiber interface with the timing system, the control interface for displaying power supply parameters, the PWM synchronization signal interface with other power supplies, and the local debugging man-machine interface.

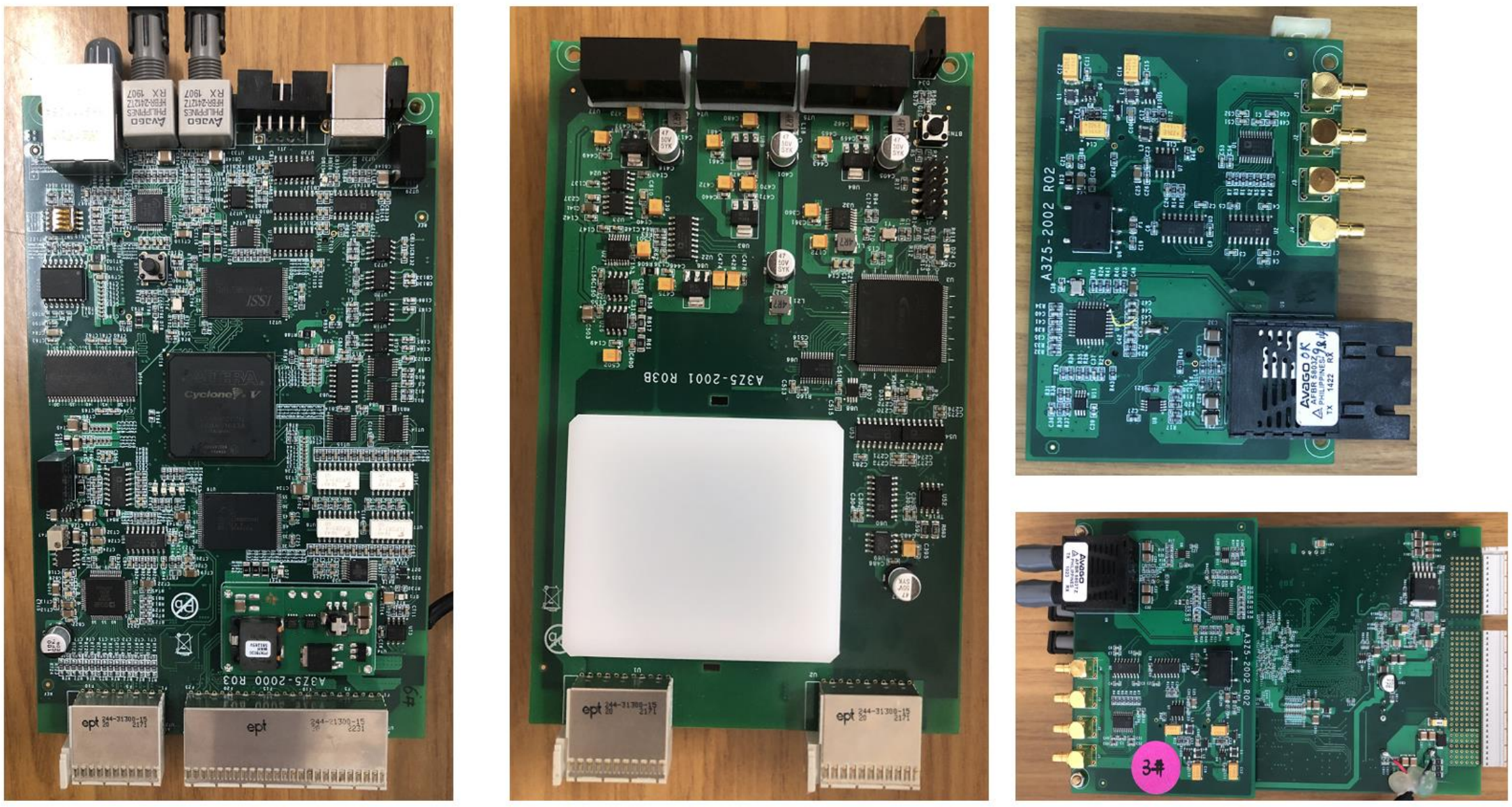

**Figure 6.3.6.4:** Main hardware of DPSCM-II

### 6.3.6.4 *Topology of Power Supplies*

The power supply design for the Linac primarily adopts a DC source combined with switched-mode technology. The DC source utilizes a multiplicative equivalent 12-pulse rectifier to reduce harmonic currents and provide a stable DC voltage at the front-end. To control input power fluctuations, a buck or booster circuit is implemented. The switched-mode converter is responsible for output current control.

The unipolar power supply employs a modular design approach. Figure 6.3.6.4 illustrates the block diagram of two modules in parallel, which have a 300A/50V output and are specifically developed for the HEPS project. This topology increases the effective switching frequency by processing the PWM signal in a series-parallel arrangement of multiple modules. It improves response speed, simplifies output filter design, and reduces switching losses. The modular structure enhances production technology and maintainability.

The modular design utilizes PWM control to achieve constant frequency regulation, optimizing filter design. Based on zero conversion converter technology, a "soft switching" PWM DC/DC full-bridge converter controlled by phase shift is employed. It utilizes the leakage inductance of the high-frequency transformer or primary-side inductance in series, along with the parasitic capacitance of the switching tube, to enable zero-voltage switching. This converter is particularly suitable for medium-power DC power supplies.

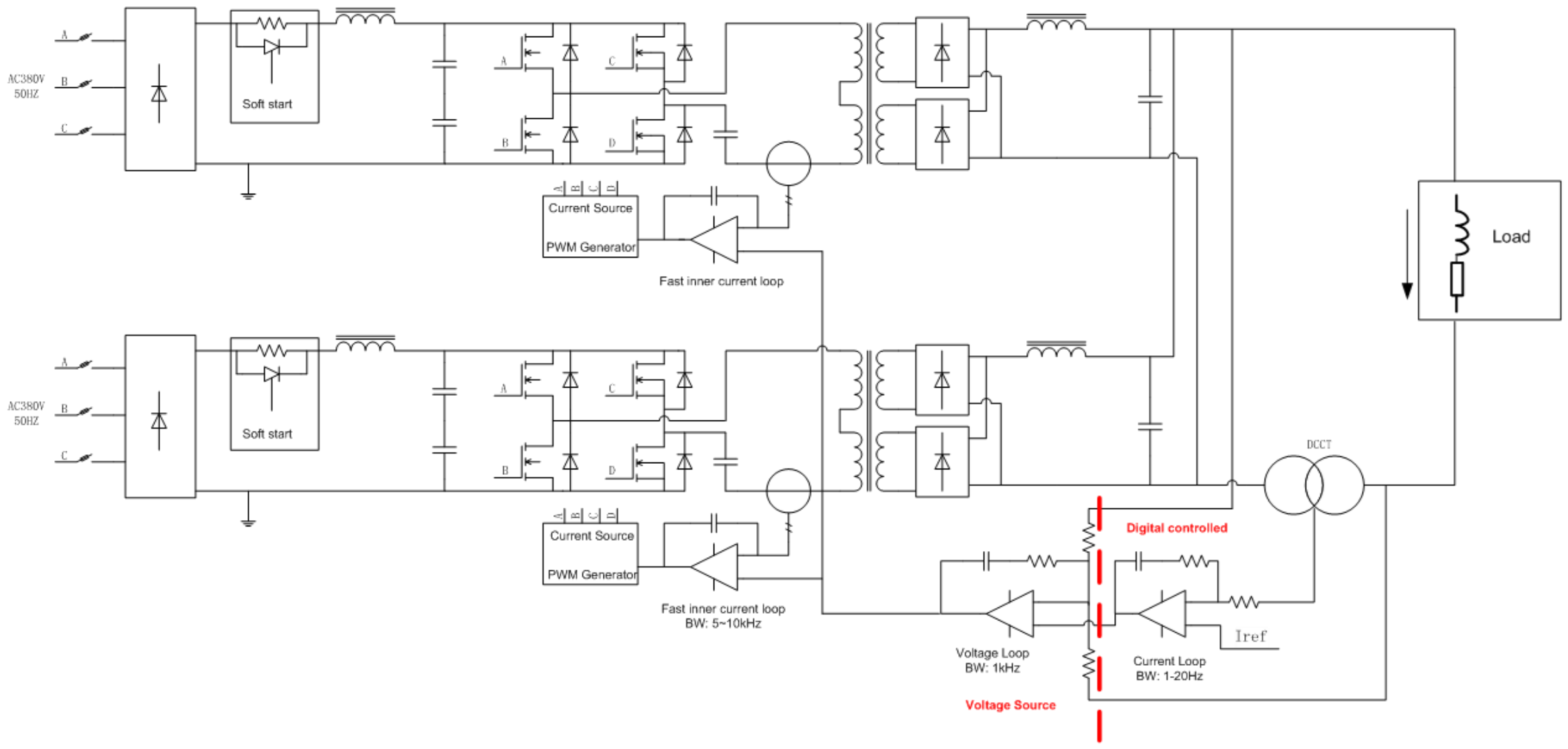

**Figure 6.3.6.4:** Structure diagram of double-module, phase-shifted, zero-voltage switching full-bridge converter.

The digital controller plays a crucial role in ensuring stability and accuracy of the output current by incorporating analog sampling, AD conversion, algorithms, and control parameters.

The power supply for the CEPC corrector magnet utilizes a two-stage control topology. The first stage employs a DC voltage regulator to maintain a stable output DC voltage. The second stage features a bidirectional high-frequency H-bridge structure that serves two functions: a) enabling the output current to be positive or negative, and b) utilizing current feedback in the regulation loop to ensure output current stability.

The circuit topology consists of four high-frequency power switching tubes that form a full-bridge chopper circuit. The two diagonal bridge arm switching tubes work in a complementary manner to switch on and off. This topology guarantees accurate zero-point output and facilitates smooth commutation of positive and negative currents.

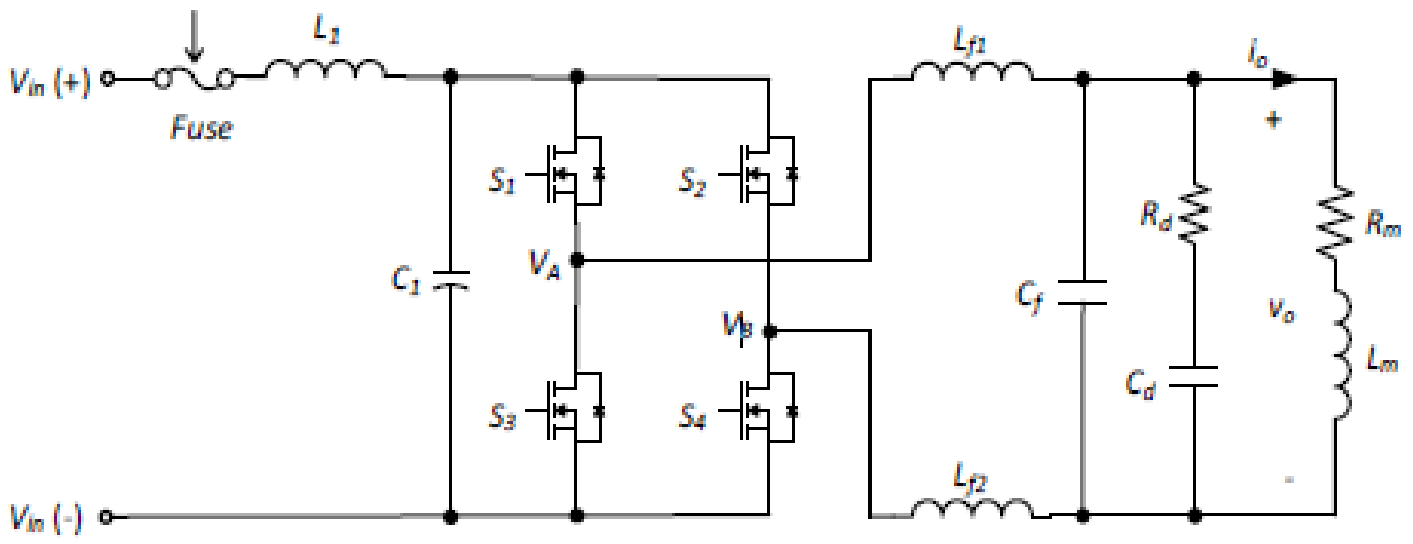

**Figure 6.3.6.5:** Structure diagram of the bipolar power supply.

## 6.3.7 **Vacuum System**

The Linac vacuum system (including Damping Ring) spans approximately 2,100 m in length. It encompasses several components, including the electron gun, positron gun, bunching system, accelerating structures, damping ring, dumps, and the bypass section of the Damping Ring.

### 6.3.7.1 *Vacuum Requirements*

The Linac vacuum system is critical for maintaining a stable and acceptable pressure to ensure optimal beam-transfer efficiency while protecting the waveguides, accelerating tubes, and electron gun from damage caused by high-voltage arcing. To prevent contamination of the e-gun cathodes, a dynamic pressure of less than $2\times10^{-7}$ Torr is required in the Linac and less than $2\times10^{-8}$ Torr is necessary in the electron gun. Table 6.3.7.1 outlines the design specifications for the Linac vacuum system.

**Table 6.3.7.1:** Specifications of the Linac vacuum system

| Devices | Static pressure（Torr） | Dynamic pressure（Torr） |
|---|---|---|
| E-gun | $< 1 \times 10^{-9}$ | $< 2 \times 10^{-8}$ |
| ESBS | $< 5 \times 10^{-8}$ | $< 2 \times 10^{-7}$ |
| Accelerator section | $< 5 \times 10^{-8}$ | $< 2 \times 10^{-7}$ |
| Waveguide section | $< 5 \times 10^{-8}$ | $< 2 \times 10^{-7}$ |

The gas load within the Linac system primarily arises from thermal outgassing, with electron-induced desorption being negligible in practical terms. The vacuum components used in the Linac, such as oxygen-free copper and stainless steel, contribute to the low thermal outgassing rate, which is determined to be $1\times10^{-11}$ Torr·l/s·cm$^2$.

To achieve effective evacuation of the accelerator structure, two ion sputtering pumps with an effective pumping speed of 50 L/s are employed. These pumps facilitate the reduction of the static pressure to approximately $3.2\times10^{-8}$ Torr. Figure 6.3.7.1 presents the calculated distribution of static pressure for the S-band and C-band accelerator structures using Molfow simulations.

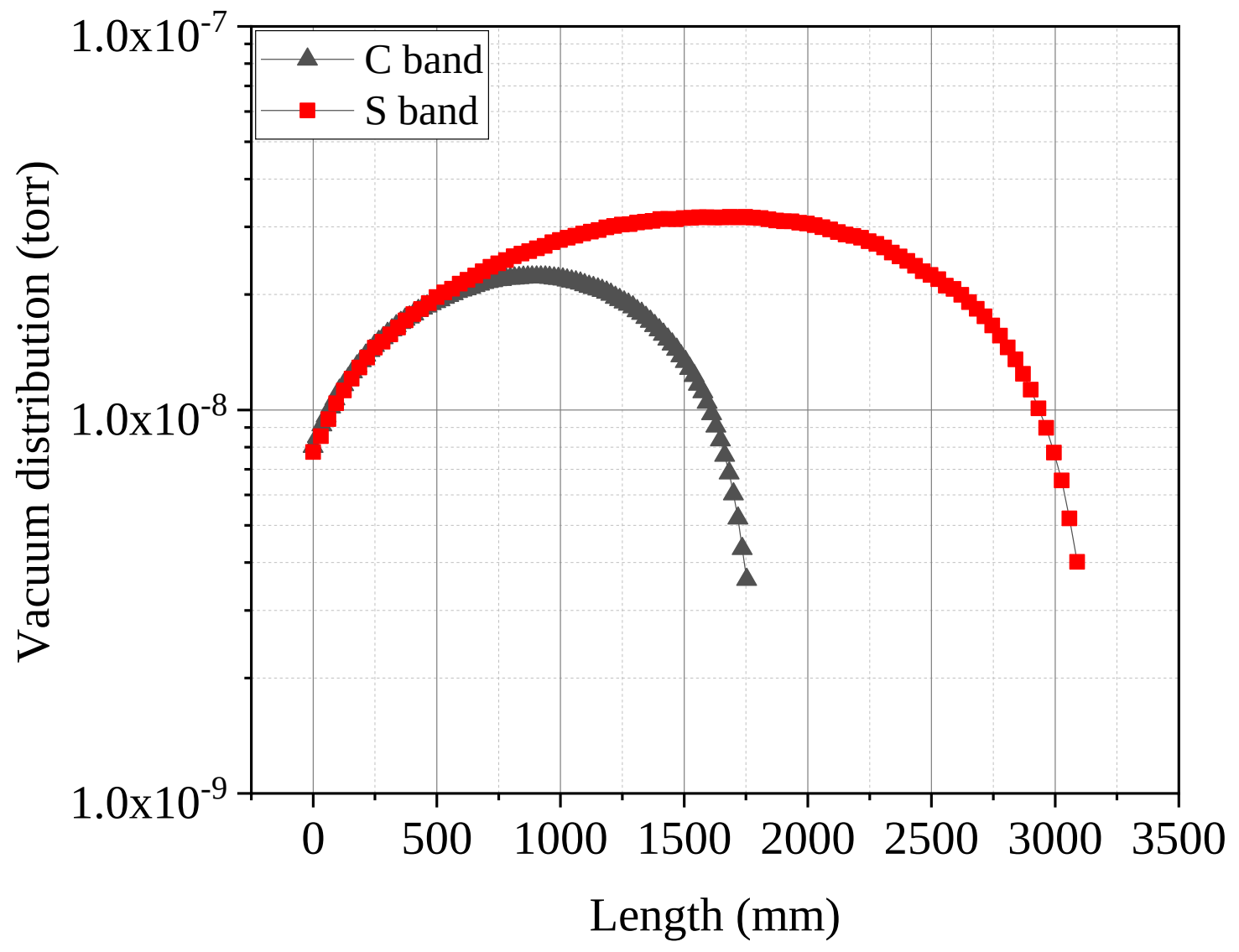

**Figure 6.3.7.1:** The static pressure distrubition of an accelerator structure.

The vacuum chambers for the Linac will be fabricated from stainless steel with a magnetic permeability below 1.05 for the entire Linac and 1.02 for those inside the magnets. These chambers will be equipped with conflat flanges, allowing for reliable and leak-tight connections. The aperture of the vacuum chambers will vary, ranging from 20 to 30 mm.

### 6.3.7.2 *Vacuum Equipment*

The Linac system comprises 271 klystrons and 581 accelerator structures. The total length of the vacuum chambers in the Linac is 931 meters, with a total of 663 bellows installed throughout the accelerator.

Based on the experience gained from BEPC II, the length of the waveguide in the Linac is assumed to be 15 meters.

The number of ion pumps and vacuum gauges required for the Linac system can be calculated according to the specifications provided in Table 6.3.7.2.

**Table 6.3.7.2:** Devices in the Linac vacuum system.

| Devices | Gauge | SIP | Bellow | All metal gate valve |
|---|---|---|---|---|
| Accelerator structure | one | two | / | 1/12 |
| Klystron | one | one | / | / |
| Waveguide | 1/10 | 1/2.5 | one | / |
| Vacuum pipe | 1/10 | 1/2.5 | 2/3 | 1/85 |

In order to achieve high vacuum levels, the Linac system utilizes a combination of double sputtering ion pumps (SIPs) and single vacuum gauges. The SIPs are employed to maintain the vacuum within the klystrons, while the gauges are used for monitoring purposes.

To ensure effective vacuum throughout the Linac, a SIP is distributed along the waveguide at intervals of 2.5 meters, while a vacuum gauge is installed every 10 meters.

Figure 6.3.7.2 provides a vacuum diagram showcasing the arrangement of SIPs and gauges in relation to the klystrons and accelerator structures within the Linac system.

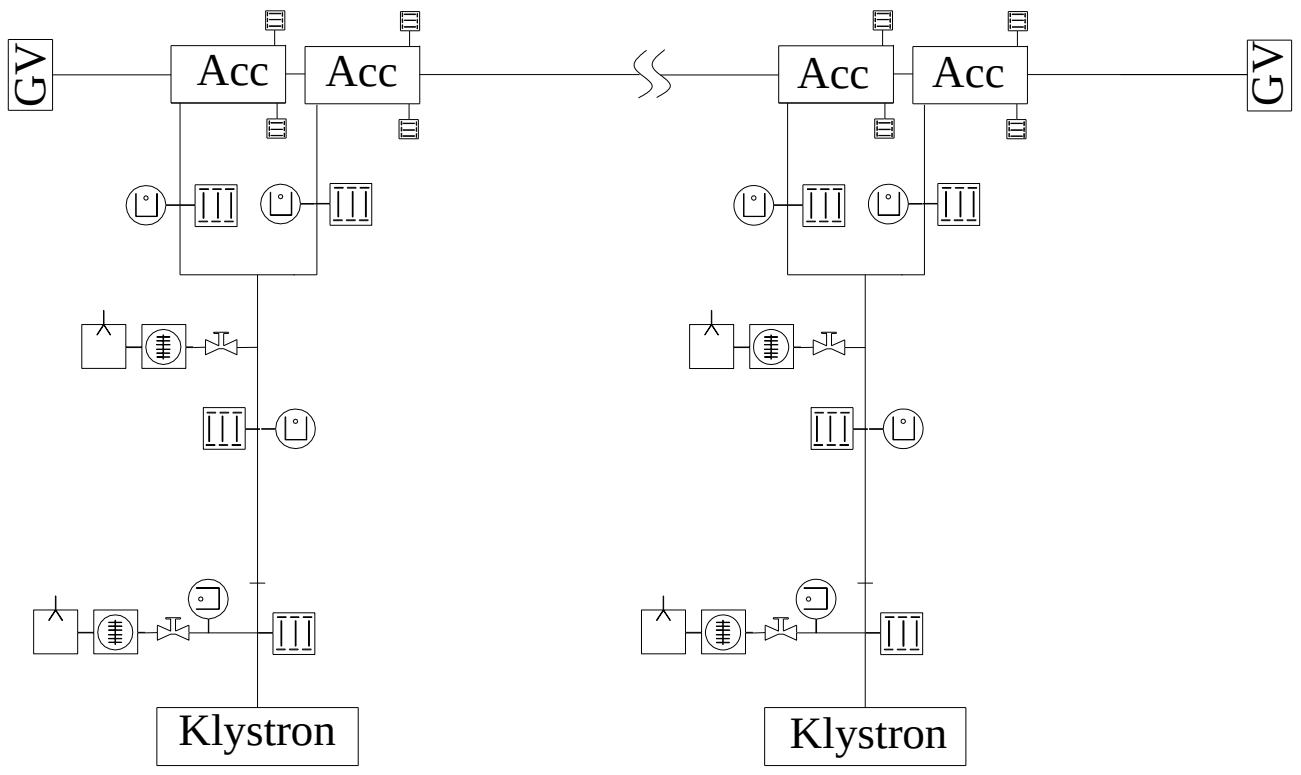

**Figure 6.3.7.2:** Vacuum diagram in relation to the klystron and accelerator structures

The system will utilize 3431 ion pumps for pumping and 1352 cold cathode gauges to measure pressure. The Linac vacuum system is divided into 59 sectors, each with metal gate valves. The vacuum sectors will be roughed down from atmospheric pressure using portable turbo-molecular pumps (TMP) backed by dry scroll pumps. Once a pressure of less than $1\times10^{-6}$ Torr is achieved, the sputter ion pumps will be activated. While the vacuum sector is at high vacuum, the TMP will be manually isolated with all-metal valves to prevent the sector from being exposed to atmosphere in case of pump or power failure.

Due to the high radiation levels in the Linac tunnel, the power supplies and controllers for the Linac vacuum system will be located in the service area. The vacuum devices, such as gauge controllers, pump controllers, gate valves, and residual-gas analyzers, will be interfaced with the machine control system for remote monitoring, operation, and control, with both local and remote capabilities.

#### 6.3.7.3 *Dump Chambers and Membrane Windows*

To ensure the separation between the accelerator vacuum and the dump chamber, a titanium membrane window is utilized. Several prototypes of the dump chamber with a titanium membrane window have been successfully developed and fabricated within the HPES linac vacuum system, as depicted in Figure 6.3.7.3.

The titanium membrane window is in an elliptical shape, measuring 170×10 mm, with a thickness of 0.1 mm. It is welded onto a stainless-steel plate that has a diameter of 183 mm and a thickness of 5 mm.

The deformation test conducted under vacuum conditions revealed that the elliptical window had a deformation of 0.4 mm, while the circular window showed a deformation of 2.3 mm.

The ultimate vacuum achieved within the system is less than $5 \times 10^{-8}$ Pa, ensuring a high-quality vacuum environment for optimal operation.

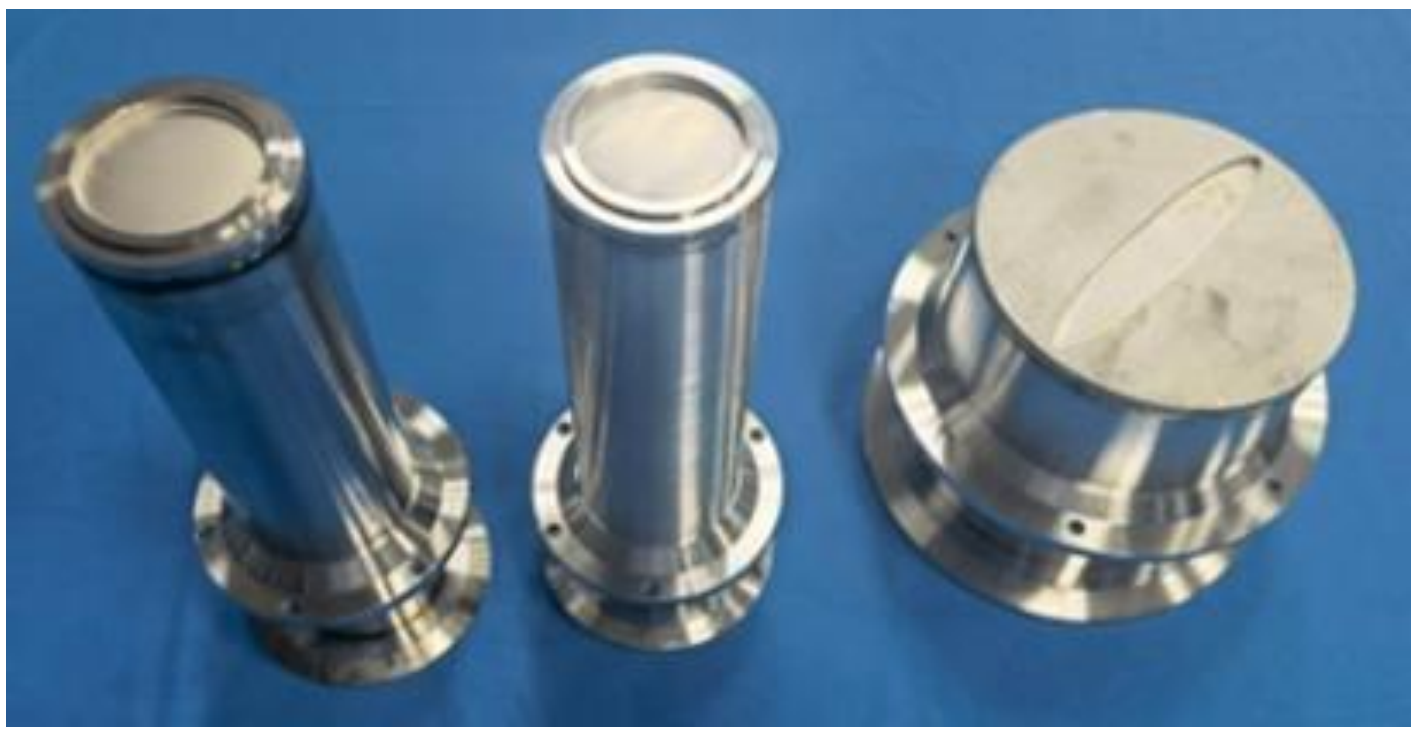

**Figure 6.3.7.3:** Prototypes of dump chamber and membrane windows

### 6.3.8 Diagnostics and Instrumentation

#### 6.3.8.1 *Introduction*

The required instrumentation for the Linac includes beam position monitors, beam profile monitors, and beam current monitors. It is important to note that there are significant differences in the diagnostic signals between electrons and positrons, so the instruments that work well for positrons may be saturated with electrons.

To measure beam positions and angles and determine the beam trajectory, a stripline BPM will be utilized. Additionally, measuring the beam profile can provide information on beam emittance, energy, and energy spread. The integrated charge detectors (ICT) will be used to measure the bunch charge, and the beam diagnostics system will have sufficient dynamic range to accommodate the minimum to maximum single-shot bunch charge.

Table 6.3.8.1 outlines the type, quantity, and function of linear beam detectors required for this project.

**Table 6.3.8.1:** Instrumentation for the Linac.

| Sub-system | Method | Parameter | Amounts |
|---|---|---|---|
| Beam position monitor | Stripline BPM | Resolution: 10 μm | 150 |
| Beam current monitor | ICT | 2.5% @ 1 nC–10 nC | 63 |
| Beam profile measurement | YAG/OTR | Resolution: 30 μm | 30 |

### 6.3.8.2 *Beam Position Measurement*

There are a total of 150 stripline BPMs in the Linac and its transport lines, which are divided into two groups based on the size of the beam-stay-clear area [1]. The first group of BPMs has a beam-stay-clear area with a diameter of 30 mm, while the second group has a diameter of 20 mm.

Figure 6.3.8.1 displays the primary mechanical parameters of the stripline BPM, including the coverage angle $\alpha$, distance $h$ between the strip and pipe, strip thickness $t$, inner radius $r_{in}$, and longitudinal length $L_{strip}$. The inner radius of the stripline has been designed to be equal to the radius of the beam-stay-clear area.

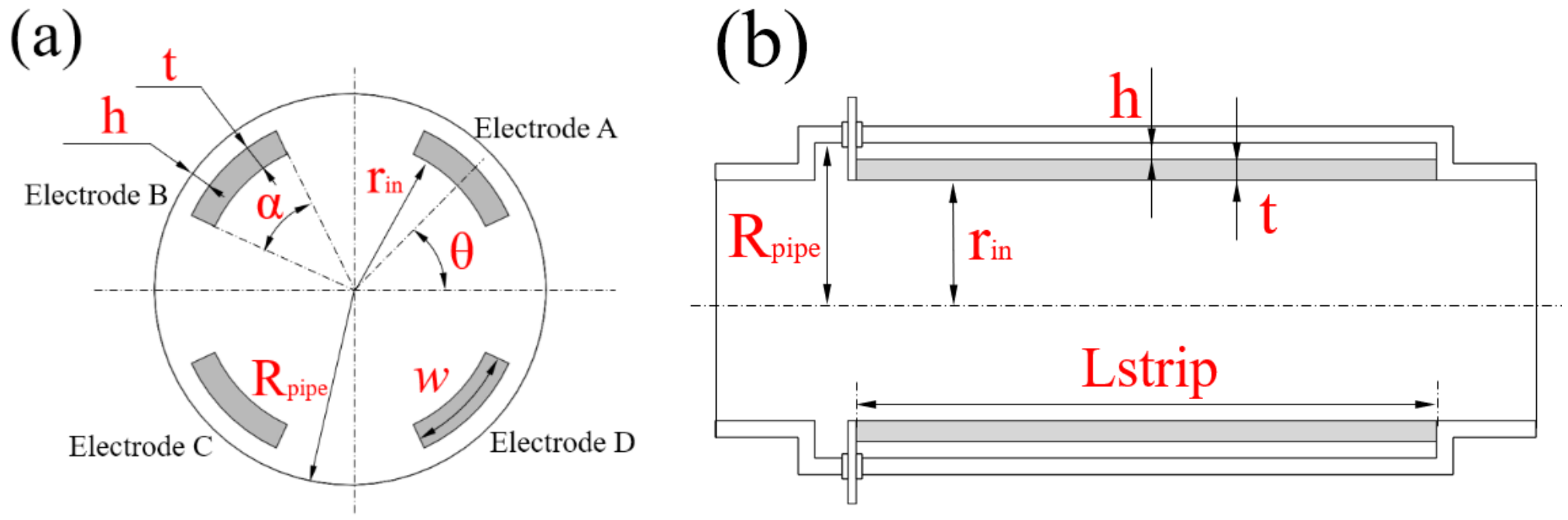

**Figure 6.3.8.1:** Schematic of a stripline BPM: (a) Front view; (b) Side view.

The feedthroughs and coaxial cable used in this design have a characteristic impedance of 50 Ω. To minimize signal reflection, the impedance of the strip electrode was also designed to be 50 Ω. Electromagnetic simulation software such as CST is often utilized to compute the impedance of the stripline [2]. In addition to using simulation software, relevant formulas can also be applied in stripline design.

The characteristic impedance of a single stripline can be calculated using the following formula [3]:

$$Z_{\text{single}} = 60\frac{2\pi}{\varphi}\ln\frac{r_{in}+h+t}{r_{in}+t},\ \varphi=\frac{w}{r_{in}}+\frac{2\pi h}{2\times(r_{in}+h+t)-h},\ w=\alpha\times(r_{in}+\frac{1}{2}t) \qquad (6.3.8.1)$$

Table 6.3.8.1 presents the preliminary design parameters of the Linac. However, further optimization is currently underway due to the fact that the distance h in the Linac-2 is only 0.6 mm to meet the requirement for an impedance equal to 50 Ω.

**Table 6.3.8.1:** Mechanical parameters of the Linac stripline BPM.

| Location | $r_{in}$ | $t$ | $h$ | $\alpha$ | $\theta$* | $Z_{single}$ | $L_{strip}$ | $Z_{CST}$ |
|---|---|---|---|---|---|---|---|---|
| | mm | mm | mm | degree | degree | Ω | mm | Ω |
| ESBS/EBTL/FAS/ PSPAS/SAS | 15 | 1.5 | 2.2 | 30 | 0 | 50.8 | 150 | 50.3±0.5 |
| TAS | 10 | 1.5 | 0.6 | 30 | 0 | 49.9 | 150 | 50.4±0.5 |

* The angle between adjacent electrodes is 90°, the four strips are located at 45°, 135°, 225° and 315°, respectively.

The same beam position monitor electronics utilized in the Collider will be used for the Linac, and further details can be found in Chapter 4, Section 4.3.7.2. The algorithm developed for the Collider will also be adopted for the linac beam.

### 6.3.8.3 *Beam Current Measurement*

Due to the variability of the pulsed beam length (FWHM) ranging from picoseconds to nanoseconds along the Linac in accordance with the physics design, traditional Beam Current Transformers are not suitable due to their bandwidth limitations and bunch length dependency [4]. Instead, an Integrated Charge Transformer (ICT) is selected to measure beam current, as it is non-intercepting and insensitive to bunch length. Bergoz Instrumentation [5] offers an In-Flange ICT sensor, which has a compact structure and shielding and can be directly mounted in the beam line.

In total, there are 42 ICTs installed in the Linac to monitor beam intensity and calculate transmission efficiency. An in-house signal processing electronics system is designed to adjust the signal from the ICT, performing the functions of filtering and amplification. This regulated signal will reduce the sample rate and dynamic range requirements of the data acquisition system. Through post-level electronic system processing, the bunch charge can be accurately calculated. Figure 6.3.8.2 shows the In-Flange ICT sensor from Bergoz.

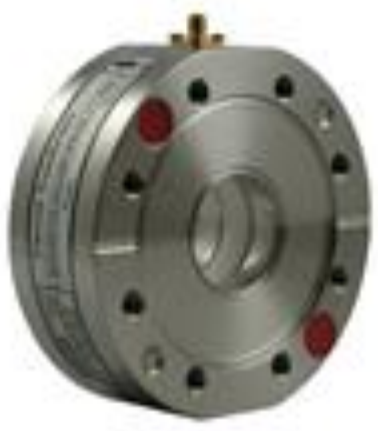

**Figure 6.3.8.2**: In-Flange ICT sensor from Bergoz.

The sensitivity of the ICT sensor must take into account the attenuation caused by the transmission cable to ensure accurate measurement results. For this reason, an on-line functional examination and re-calibration can be carried out using the Cali-winding within the sensor.

### 6.3.8.4 *Beam Profile Measurement*

The beam profile monitor will be a crucial diagnostic tool during the commissioning of CEPC experiments. The most commonly used method for determining the transverse beam profile is based on the interaction of beam particles with a fluorescent screen. When the beam particles interact with the screen, a portion of the deposited energy results in excited electronic states, which partially de-excite via light emission. Therefore, the beam profile can be observed using a CCD camera [6]. In the CEPC Linac, YAG/OTR screens will be used as the beam profile monitors in the straight sections.

#### *6.3.8.4.1 Mechanical Detector*

Our beam profile monitors typically consist of a screen, an insertion mechanism, an illumination system, and a video camera for detection, as shown in Fig. 6.3.8.3. The screen is installed at a 45° angle to the beam, while the camera is positioned at a 90° angle. This setup is advantageous because it minimizes the longitudinal space required for the monitor, and a round beam will appear round in the image.

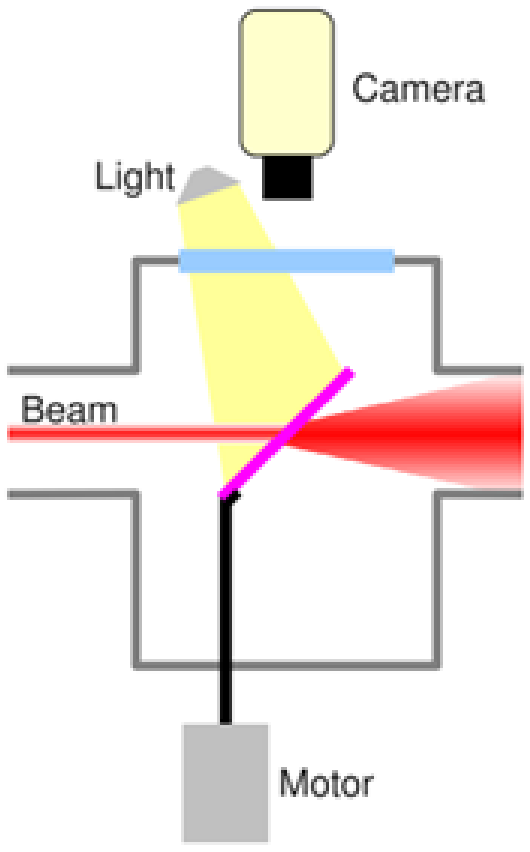

**Figure 6.3.8.3**: Schematic of a scintillating screen monitor.

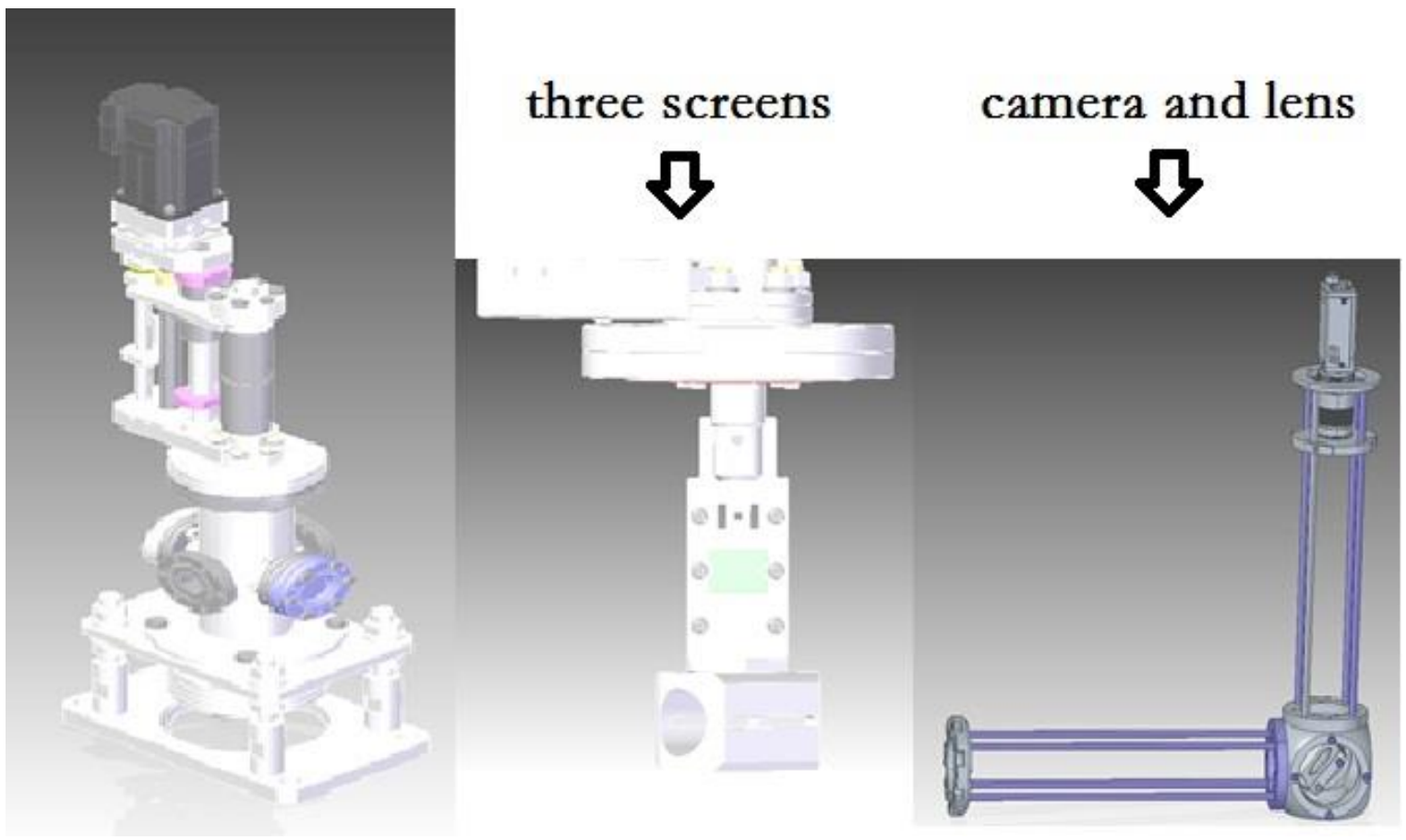

**Figure 6.3.8.4**: Design of the mechanical detector.

The beam profile monitor is designed with three position screens for testing, as shown in Figure 6.3.8.4. During initial commissioning and low beam charge situations, a YAG:Ce crystal plate is used. For high beam charge situations, another screen made of 100 nm aluminum deposited on a polished silicon wafer is used to generate optical transition radiation (OTR) when electrons hit it [7]. This is because OTR represents a linear radiation source, while YAG screens have saturation problems. The last screen is the calibration screen, used to calibrate the optical relay and CCD image acquisition system. The beam image can be obtained from the backside viewport, and the camera is located 1 meter below the beam pipe through a planar-mirror optical relay to reduce radiation damage.

*6.3.8.4.2 Mover Controlling System*

We have selected a new type of motor for the HEPS beam profile monitor design, specifically the SSM24Q-3RG produced in China. An overview of the integrated step servo motor is shown in Figure 6.3.8.5.

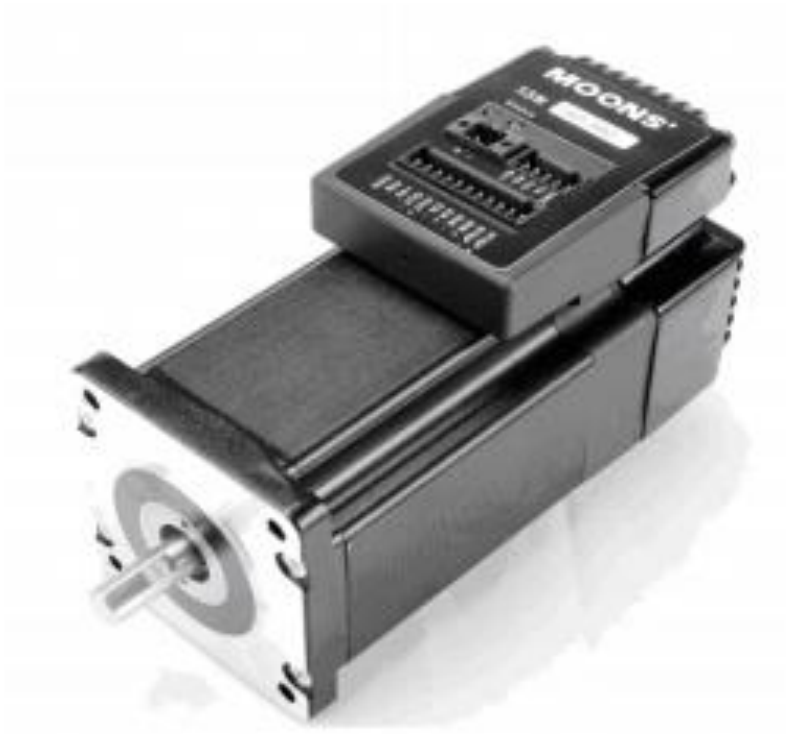

**Figure 6.3.8.5**: Integrated step-servo motor.

Calibration marks on the screen are essential in identifying errors and developing correction algorithms. They can improve the accuracy of the beam profile measurements and help maintain the performance of the beam profile monitor over time.

The high stiffness of the stepper motor, combined with the responsive servo control and 5000-line high-resolution encoder, results in smooth and quiet operation, especially at low speeds. The SSM24Q motor integrates the motor, driver, controller, encoder, and IO, making the mover controlling system very concise. The motor can start working with just DC 24V and DC 48V and RS485 bus. The output torque of 2.4 N-m allows the motor to quickly pull the screen out of the beam line without the need for a deceleration mechanism. The mover works smoothly and stably with low noise to prevent the screen from stopping the beam line [8].

*6.3.8.4.3 Optical Set-up*

The optical system used to capture the YAG/OTR light onto the CCD camera consists of a mirror that deflects the OTR light downwards, a focusing lens, and a CCD camera. These components will be mounted on two rails on a stainless-steel plate, which will be fixed to the machine's support structure. The optical system is protected from stray light, and the CCD camera is shielded with lead to reduce radiation damage. The camera chosen for this purpose is a digital CCD camera (HIKVISION CE-50GM, as shown in Fig. 6.3.8.6) with an Ethernet interface. The camera sensor size is 1280 × 960 pixels, with a pixel dimension of 4.65 μm × 4.65 μm.

The GigE Vision camera was chosen for the image acquisition subsystem due to two main advantages. Firstly, digital signals can be transmitted more reliably than analog signals over long distances. Secondly, the gain parameter of the GigE CCD camera can be remotely adjusted according to the beam intensity during beam commissioning. The HIKVISION CE-50GM CCD cameras were selected as the primary devices for image acquisition. With a 1:1 lens, the system can achieve a spatial resolution of 10 microns. External trigger mode is crucial for the system since capturing the beam images must synchronize with the timing signal of the entire system.

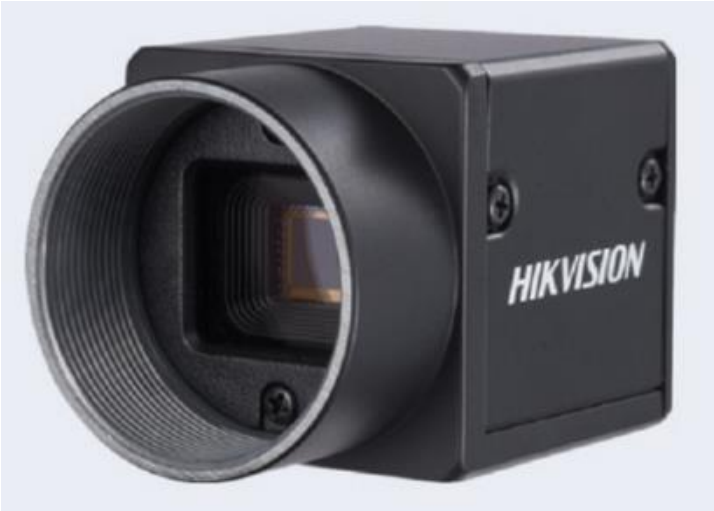

**Figure 6.3.8.6**: HIKVISION CE-50GM CCD camera.

*6.3.8.4.4 Control System*

The control system of the beam profile monitors is designed to be simple and flexible, using two separate field buses. The first is an industrial Ethernet bus used for communication between the control room and cameras, while the second is a 485 serial bus used for communication between the control room and the integrated step-servo motor (as shown in Figure 6.3.8.7). The control system software for the beam profile monitors is based on EPICS, allowing each control system to be designed as a standard software IOC. This design provides greater flexibility for system integration and control, and ensures compatibility with other EPICS-based systems.

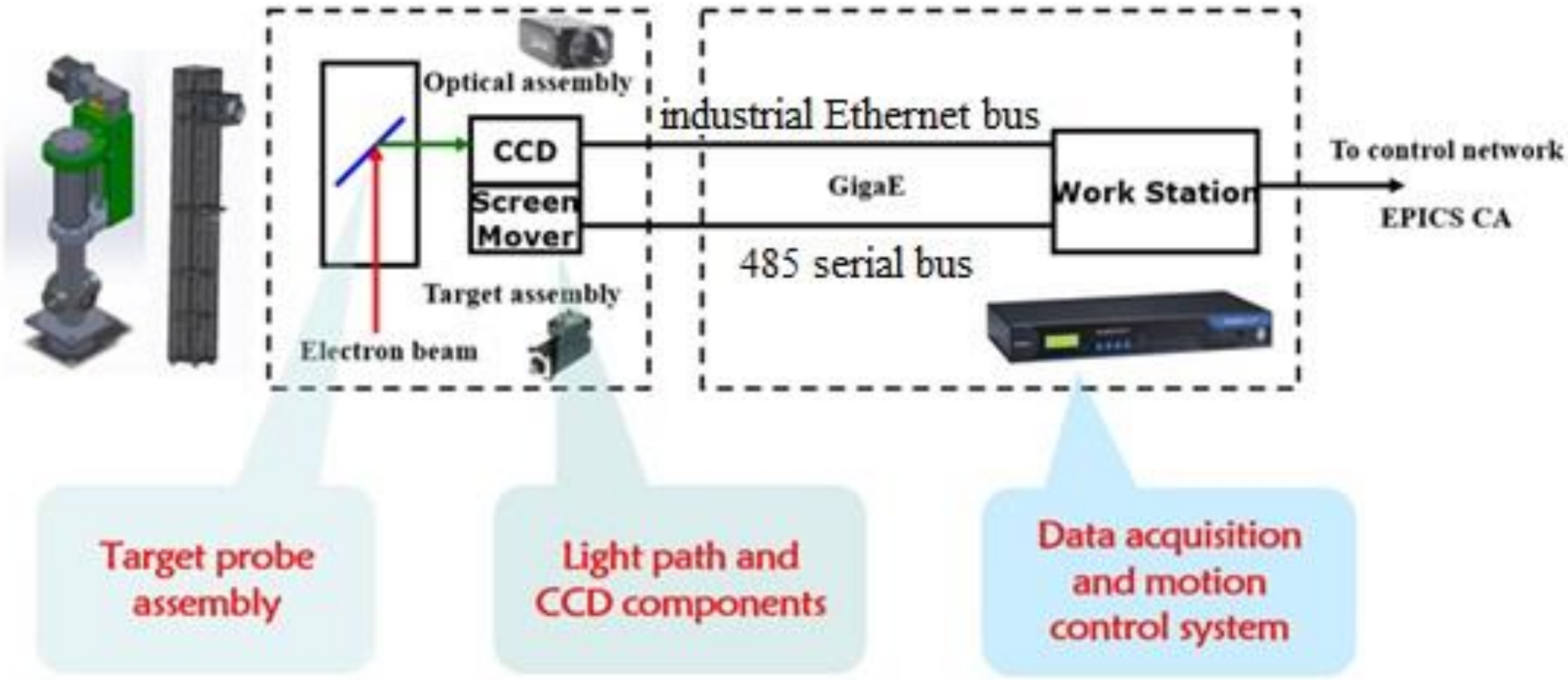

**Figure 6.3.8.7**: The beam profile monitor control system.

### 6.3.8.5 *Beam Energy and Energy Spread*

Beam energy and energy spread can be measured by deflecting the beam with a dipole and using the spot position and size on a fluorescent screen. The measurement are based on the beam profile measurement.

#### 6.3.8.6 *Beam Emittance*

Beam emittance is measured using an upstream quadrupole and observing the changes in the beam profile as a function of quadrupole strength (Q-scan).

### 6.3.9 Control System

#### 6.3.9.1 *Introduction*

The injector of CEPC consists of an electron/positron linear accelerator, with equipment distributed along a 1.8 km gallery. The Linac accelerator is responsible for generating and transferring the electron and positron beams into the Booster ring. The control system enables operators to monitor and control the equipment from both local and central control rooms. To ensure the safety of the devices, a machine protection system is implemented.

All relevant parameters and data are stored in a database for future reference and analysis.

During the construction phase of CEPC, beam commissioning will take place simultaneously with the building process. Therefore, a temporary central control room will be established to facilitate the commissioning activities. Additionally, global control systems such as timing systems, machine protection systems (MPS), and data service systems will also undergo commissioning and trial operation.

Operators will have the capability to adjust the current and select the operating mode of the electron/positron gun. Parameters of the klystrons and modulators will be continuously monitored and displayed, including high voltage, output power, RF phase, and amplitude of the output envelope. Interlock loops are implemented for the klystrons and modulators, which will automatically shut off the high voltage of the corresponding modulator if the pressure outside a vacuum klystron window exceeds a specified limit, ensuring the safety of the system.

#### 6.3.9.2 *Magnet Power Supply Control*

In the Linac of CEPC, there are a total of 663 different types of magnet power supplies. These power supplies are responsible for powering the dipole magnets, quadrupole magnets, solenoids, and correctors, among others. It's important to note that the C1 corrector magnet is a combination magnet with both horizontal and vertical coils, requiring two separate power supplies to operate effectively.

These magnet power supplies are distributed strategically along the linac gallery to ensure efficient and reliable power delivery to the magnets. The control systems used for the power supplies in the Linac are similar to those used in the Collider, as described in Chapter 4.

The power supply control system for the Linac in CEPC is designed to fulfill various user requirements. It incorporates the following functions:

1. Power Supply On/Off Control: The system allows for the local and remote control of turning on/off all power supplies. This feature provides flexibility and convenience in managing the power supply operation.
2. Monitoring of Current and Power Supply Status: The control system continuously monitors the current levels and status of each power supply. It provides information on the on/off status, local/remote mode, and normal/alarm conditions, allowing operators to keep track of the power supply performance.
3. Setting Values: The system supports three modes for making settings:
   a) Direct Setting Mode: This mode enables the operator to provide a single setpoint, and the system ramps the current linearly to reach that setpoint. It is commonly used for initializing, testing, and maintaining power supplies.
   b) Standardization Mode: In this mode, the power supplies are ramped to reset the hysteresis of the magnetic field to the standard position. The current is set along the standard hysteretic loop, ensuring consistent performance.
   c) Knobs: The control system allows for individual adjustment of power supplies using knobs. This feature enables fine-tuning and precise control over the magnet parameters.
4. Interlock System for Protection: The power supply control system incorporates an interlock system to safeguard the magnets and power supplies. It performs regular inspections of cooling waters, magnet temperatures, and power supply conditions. If any abnormalities are detected, the system takes appropriate local actions to address the issues and simultaneously sends out alarm messages to the local and central control stations, ensuring prompt response and protection of the system.

**Table 6.3.9.1:** Control requirements of magnet power supply

| **Magnet PS** | **Quantity** | **Location** | **Interface** | **Communication protocol** |
|---|---|---|---|---|
| Solenoid-PS | 37 | Linac gallery | Ethernet/Fiber | MODBUS-RTU |
| Dipole-PS | 7 | Linac gallery | Ethernet/Fiber | MODBUS-RTU |
| Quadruple-PS | 342 | Linac gallery | Ethernet/Fiber | MODBUS-RTU |
| Corrector-PS | 277 | Linac gallery | Ethernet/Fiber | MODBUS-RTU |

The Linac Power Supplies in CEPC are primarily operated in DC mode. The second generation of the digital power supply control module (DPSCM-II) serves as the core module for the 663 varieties of magnet power supplies. These modules are equipped with optical fiber interfaces that connect to the Input/Output Controllers (IOCs) of the CEPC control system.

To ensure efficient control and management, four types of IOCs have been developed based on different power supply models, as depicted in Figure 6.3.9.1. Each IOC is equipped with two Network Interface Cards (NICs). One NIC is connected to the local switch, enabling control of the power supply through the Modus-RTU communication protocol. The other NIC is connected to the central control system using the EPICS/PVA protocol, facilitating communication and integration with the overall CEPC control infrastructure.

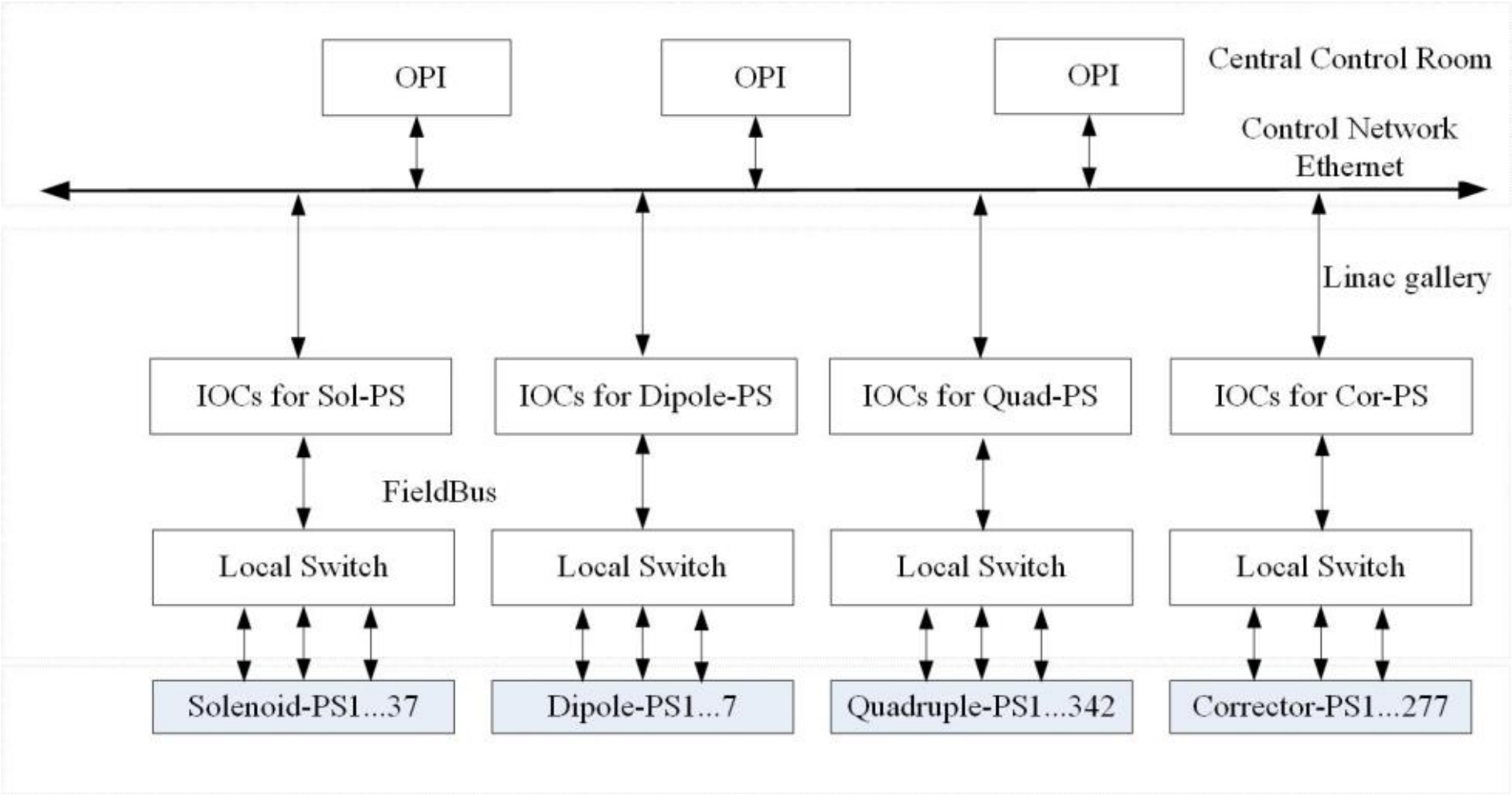

**Figure 6.3.9.1:** Hardware structure of the Linac magnet power supply control

The control system for the Linac power supplies in CEPC includes graphical control panels that display current, setpoint, and status information. Operators can operate the power supplies, make direct settings, perform standardization, and control on/off functions. The system also provides screens for diagnostic information, such as alarm reports and interlock trips. The knob screen allows adjustment of current for individual or multiple power supplies, with the ability to assign knobs to process variables (PV) for incremental or decremental changes.

### 6.3.9.3 *Vacuum Control*

There are total about 30 vacuum valves, 1310 pump and 611 gauges distributed in the Linac tunnel.

The vacuum control system in CEPC's Linac is responsible for monitoring the vacuum pressure inside the vacuum chambers and outside the windows of klystrons. It measures the pressure levels and takes necessary actions to protect the machine in case of a vacuum leak, such as closing valves in the affected Linac sections.

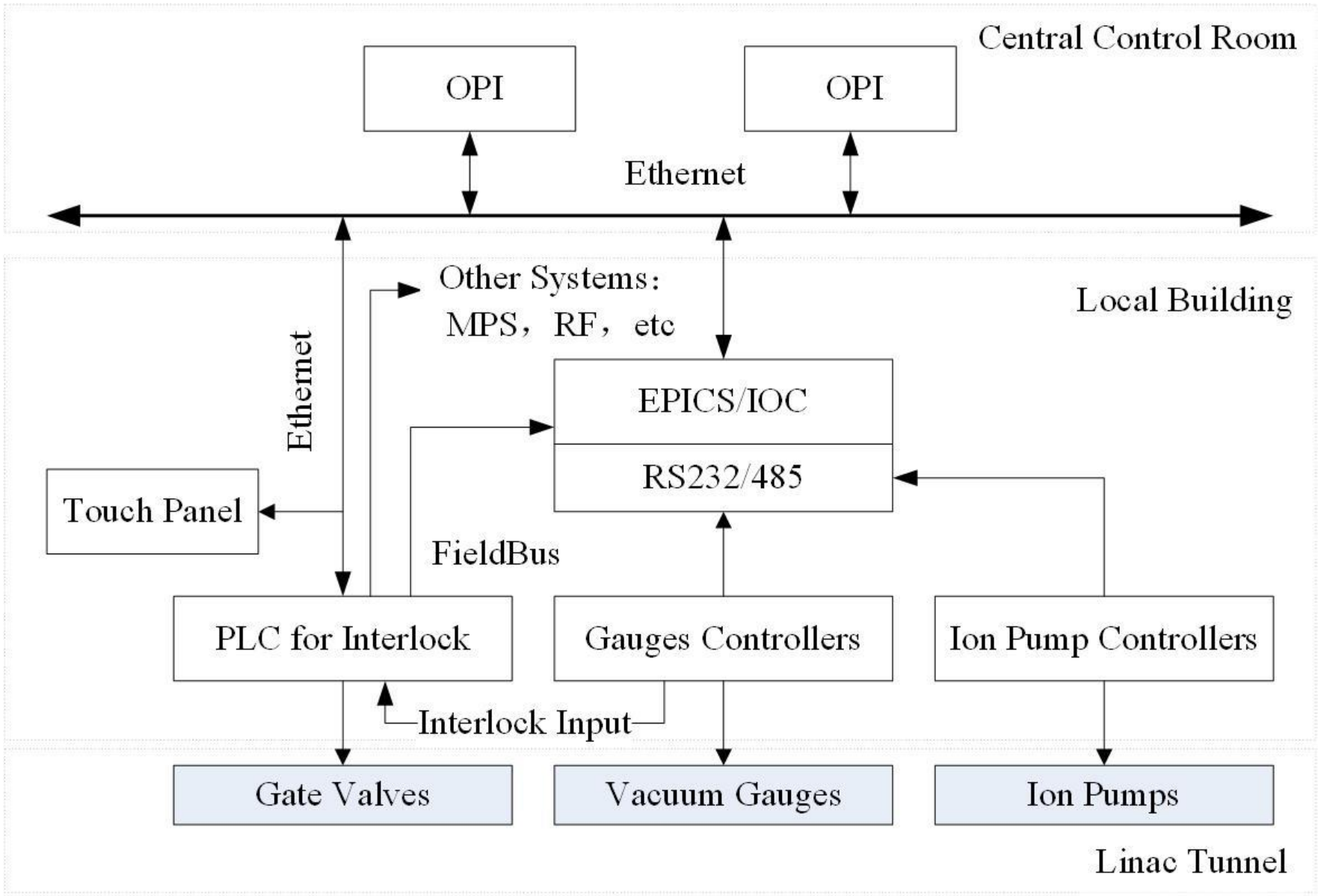

**Figure 6.3.9.2:** Hardware structure of Linac vacuum control

The Linac vacuum control system in CEPC is designed with a master-slave structure to efficiently control the distributed vacuum equipment along the 1.8-kilometer Linac tunnel and gallery. The system consists of nine sub-stations, with each sub-station responsible for controlling vacuum equipment within a 200-meter section.

The vacuum control system is built on the EPICS platform, which provides three levels of functionality: operator interfaces (OPI), input-output controllers (IOC), and device controllers.

The IOCs interface with the device controllers using RS-485 and RS-232 serial buses, as well as with programmable logic controllers (PLCs) via Ethernet. The ion pump controllers, which control the on/off state of the ion pumps and monitor pump current and voltage, communicate with the IOCs through RS-232/485 serial communication.

The gauge controllers, responsible for measuring vacuum pressure and providing interlock signals, also communicate with the IOCs via RS-232/485 interface. The gauge controllers directly output setpoint signals as interlock signals to the vacuum interlock system.

The PLCs serve as the central component of the vacuum protection interlock system. They monitor the gauge setpoint outputs and IOC interlock outputs and control the sector gate valves. The PLCs also exchange interlock signals with the RF system and other subsystems, ensuring coordinated operation and safety.

#### 6.3.9.4 *Integration of Other Sub-systems Control*

The Linac control system in CEPC encompasses various components such as klystrons, modulators, the electron gun system, positron target, and the microwave system. It is responsible for ensuring the proper functioning and interlocking of these components. The main functions of the Linac control system are as follows:

1. Klystron/Modulator Monitor and Interlock:
   - Interlock loops are implemented for klystrons and modulators to ensure safe operation.

- If the vacuum pressure outside or inside the window of a klystron exceeds a specified limit, the high voltage (HV) of the corresponding modulator is turned off.
- Parameters of klystrons and modulators, such as high voltage, output power, RF phase, and amplitude of the output envelope, are monitored and displayed on the console.

2. Phase-Shifter Control:
   - The control system allows for adjustment and monitoring of the stroke of electromotors used in phase-shifters and attenuators. This ensures precise control over the phase of the microwave signal.
3. Electron Gun System:
   - Operators have the capability to adjust the current and select the operation mode of the electron gun.
   - The control system measures and displays parameters such as the filament current and vacuum pressure of the cavity associated with the electron gun.
4. Positron Target Control:
   - The control system is responsible for monitoring and adjusting the position of the positron target, ensuring accurate positioning for optimal performance.
   - The control system includes the capability to display important beam parameters in both the local and central control rooms, such as the beam position, beam loss, beam current, and emittance.

In order to ensure seamless integration and effective maintenance management of the control system, each subsystem is required to adhere to the following standards and rules during the development of the local control system:

1. Compatibility with the specified version of EPICS.
2. Interface integration with PVA (PV Access) of EPICS.
3. Provision of a reserved hardware interface for future expansions or modifications.
4. Compliance with other control system standards, such as EPICS Naming Conventions.

Additionally, each subsystem must support both local control and remote-control operation modes. When operating in central control mode, it should have the capability to not only control various front-end devices but also provide real-time monitoring of their health status, as illustrated in Figure 6.3.9.3.

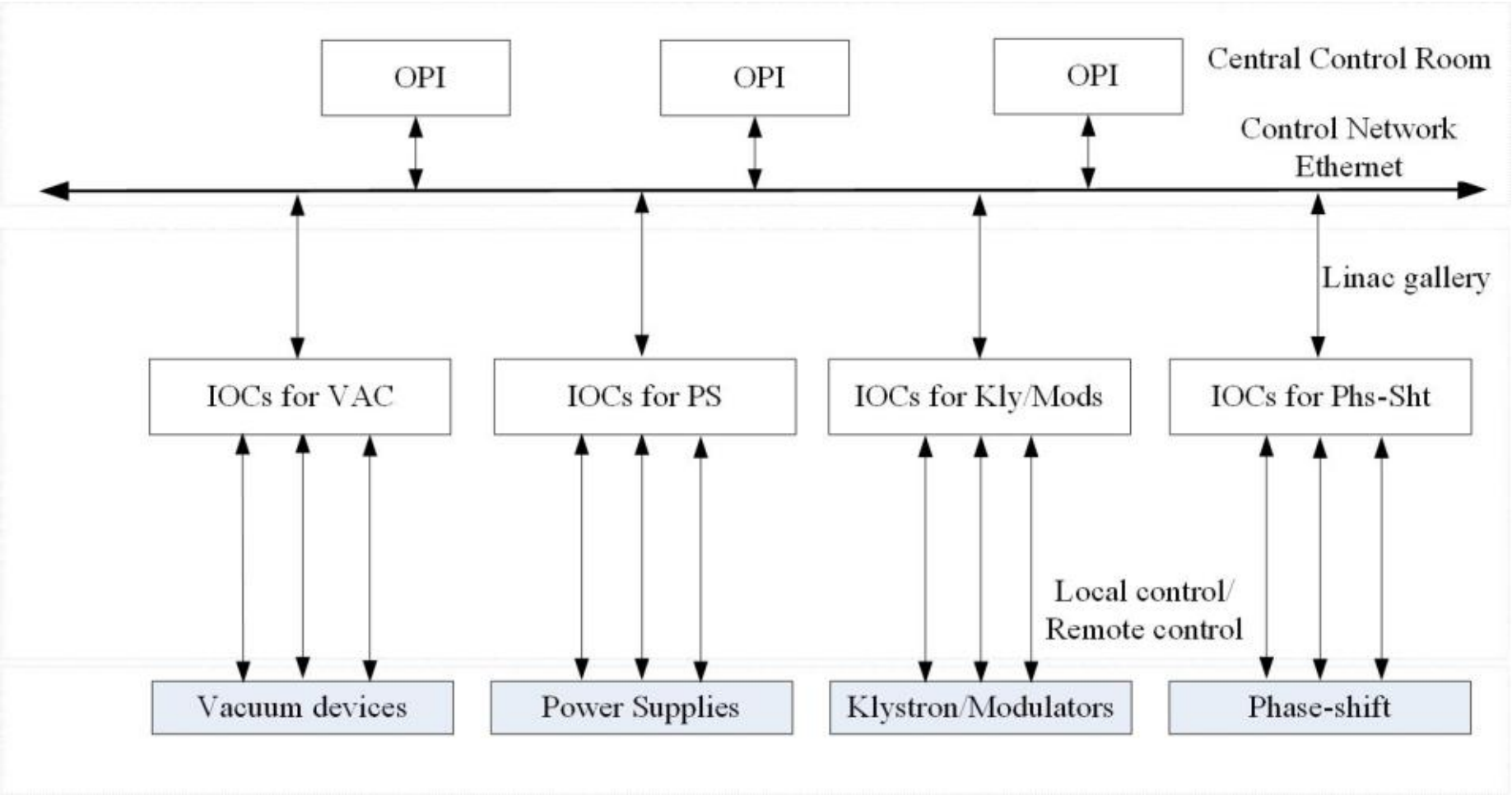

**Figure 6.3.9.3:** Hareware structure of the Linac control system.

### 6.3.10 **Mechanical Systems**

The mechanical system supports the Linac's various devices, including the bunching system, accelerator structures, magnets, vacuum system components, and BPMs. A list of this information is shown in Table 6.3.10.1.

**Table 6.3.10.1:** Mechanical supports in the Linac

| **Supports** | **Quantity (set)** | **Remarks** |
|---|---|---|
| Bunching system support | 1 | Supports of subharmonic buncher, buncher, accelerating structure, solenoid and instruments |
| EBTL support | 2 | |
| Positron accelerating structure support | 1 | Supports of positron source, solenoid, accelerating structure and positron acceleration section |
| Triplet support unit | 104 | Support of the quadruples, correctors and instruments in the triple unit |
| Dipole support | 19 | Supports of 6 kinds of dipoles |
| Accelerating structure support | 579 | Supports of S band and C band accelerating structures |
| Vacuum device support | 7125 | Supports of vacuum chambers, valves and ion pumps |
| BPM support | 260 | |

The support structure for the Linac devices in CEPC is designed with fixed supports that are anchored to the ground. Since there are various types of Linac devices with small batch sizes, flexible steel frames are utilized at the bottom instead of concrete-steel pedestals commonly found in the Collider. Above these steel frames, adjusting mechanisms are incorporated for position adjustment and alignment, similar to those used in the Collider. The requirements for the support structure are detailed in Section 4.3.10.

Figure 6.3.10.1 depicts the support arrangement for the bunching system in the Linac. This system consists of two subharmonic bunchers, one buncher, one accelerating structure, twenty-two solenoids, and related instruments, with a total length of approximately 5.82 meters. Similar to the bunching system and its support in HEPS, the buncher system of CEPC is divided into three groups. Each group features a common steel frame pedestal that is securely mounted to the ground, along with a shared girder for integral adjustment.

The support structure includes two guide rails on the top of the girder, enabling longitudinal movement during assembly. Additionally, each element within the system can be adjusted individually. Vertical adjustments are made using screws, while push-pull bolts are employed for horizontal adjustments, both for the individual elements and the girders.

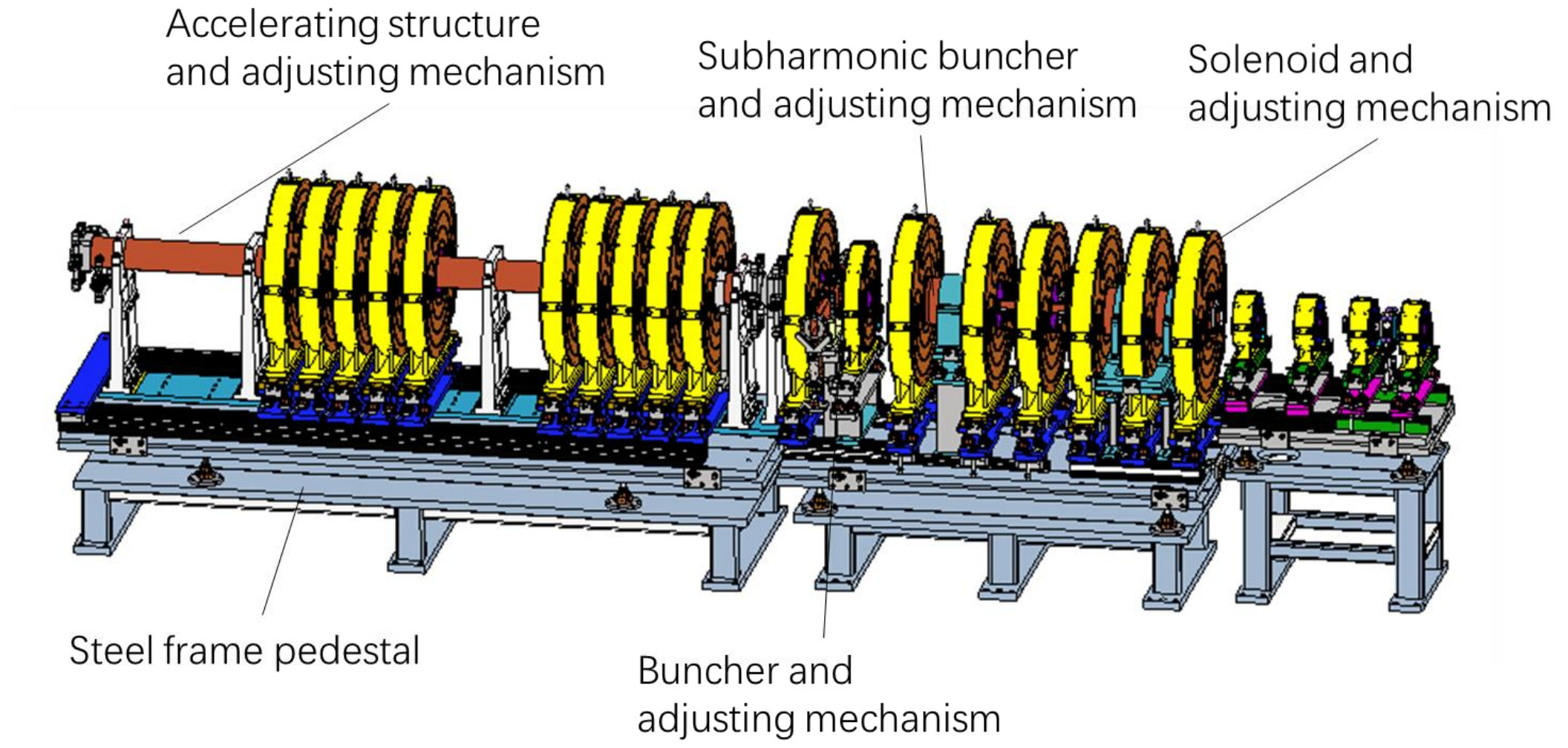

**Figure 6.3.10.1:** Support of the bunching system.

Figure 6.3.10.2 depicts the support structure for an S-band accelerating structure. It consists of vertical supports within the accelerating tube for increased stiffness. The assembly is further supported by a steel frame pedestal mounted to the ground, featuring two adjusting mechanisms.

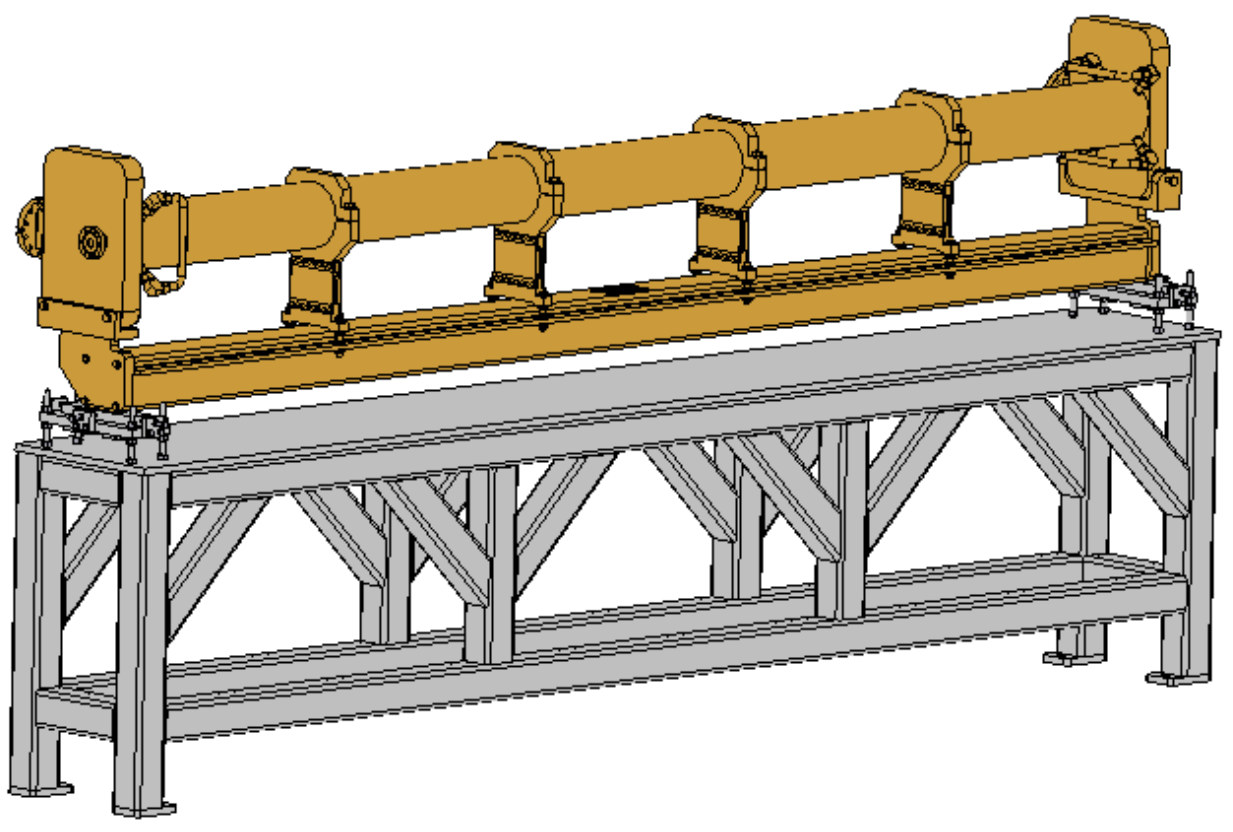

**Figure 6.3.10.2:** Support of the S-band accelerating structure.

Each triplet unit in the Linac consists of three quadrupoles, two correctors, and related instruments. There are three types of triplet units with different apertures: 35 mm, 30 mm, and 24 mm. The unit lengths vary as well, with dimensions of 1.6 meters, 2 meters, and 3.1 meters respectively. The design of each triplet unit follows a similar approach as in HEPS, where a common steel frame pedestal is mounted to the ground, equipped with several adjusting mechanisms. Each element within the triplet unit can be adjusted individually, using screws for vertical adjustments and push-pull bolts for horizontal adjustments. Figure 6.3.10.3 illustrates the support structure for a 1.6-meter long triplet unit.

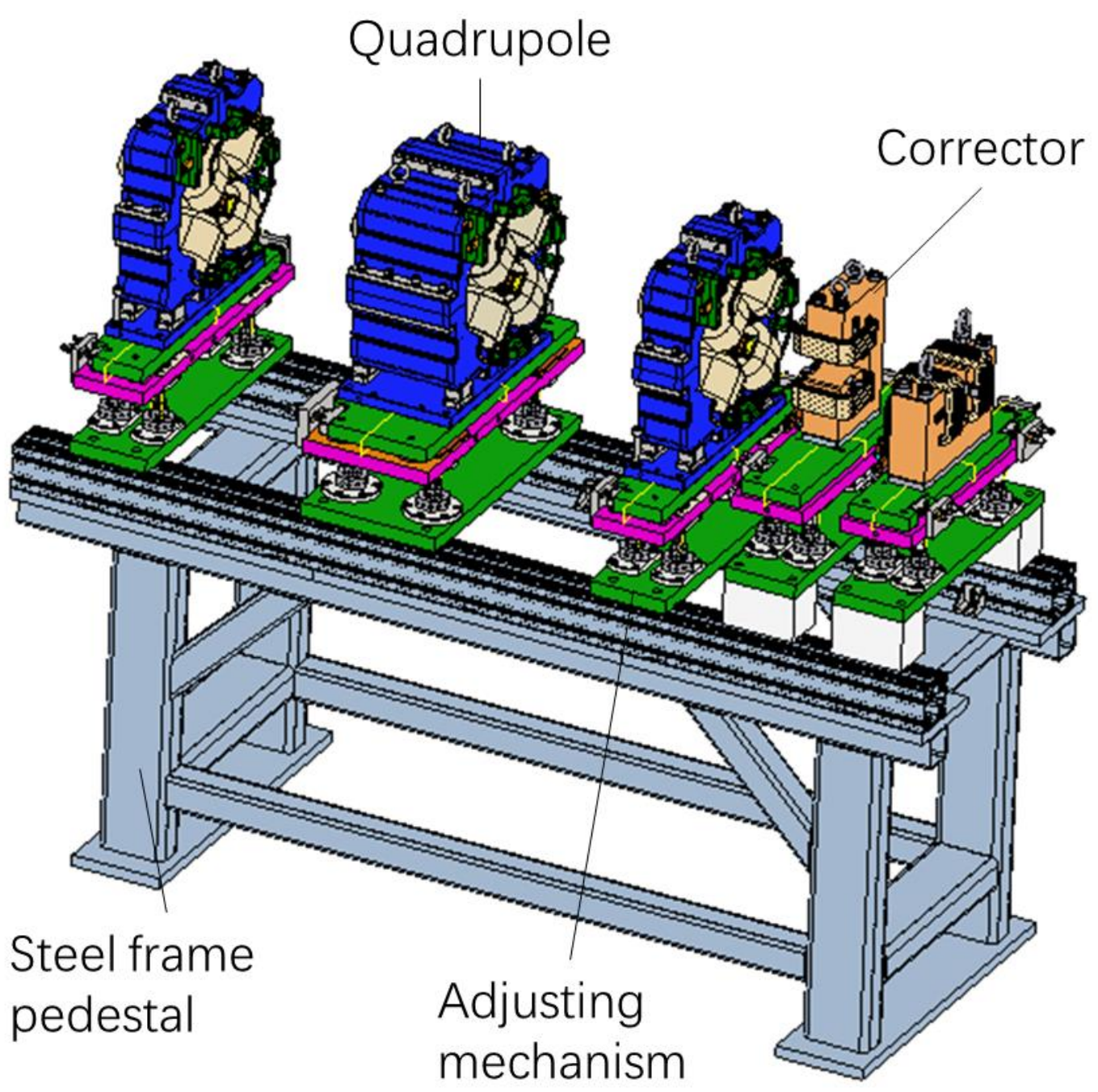

**Figure 6.3.10.3:** Support of 1.6 m-long triplet.

Unlike the underground tunnels where the Collider and Booster are located, the Linac tunnel in CEPC is situated close to the ground surface. The beam within the Linac tunnel is approximately 1.2 meters above ground level. The Linac gallery is positioned above the Linac tunnel, as shown in Figure 6.3.10.4. The tunnel itself has a width of 3.5 meters, but at the bypass location, the width is increased to 5.5 meters.

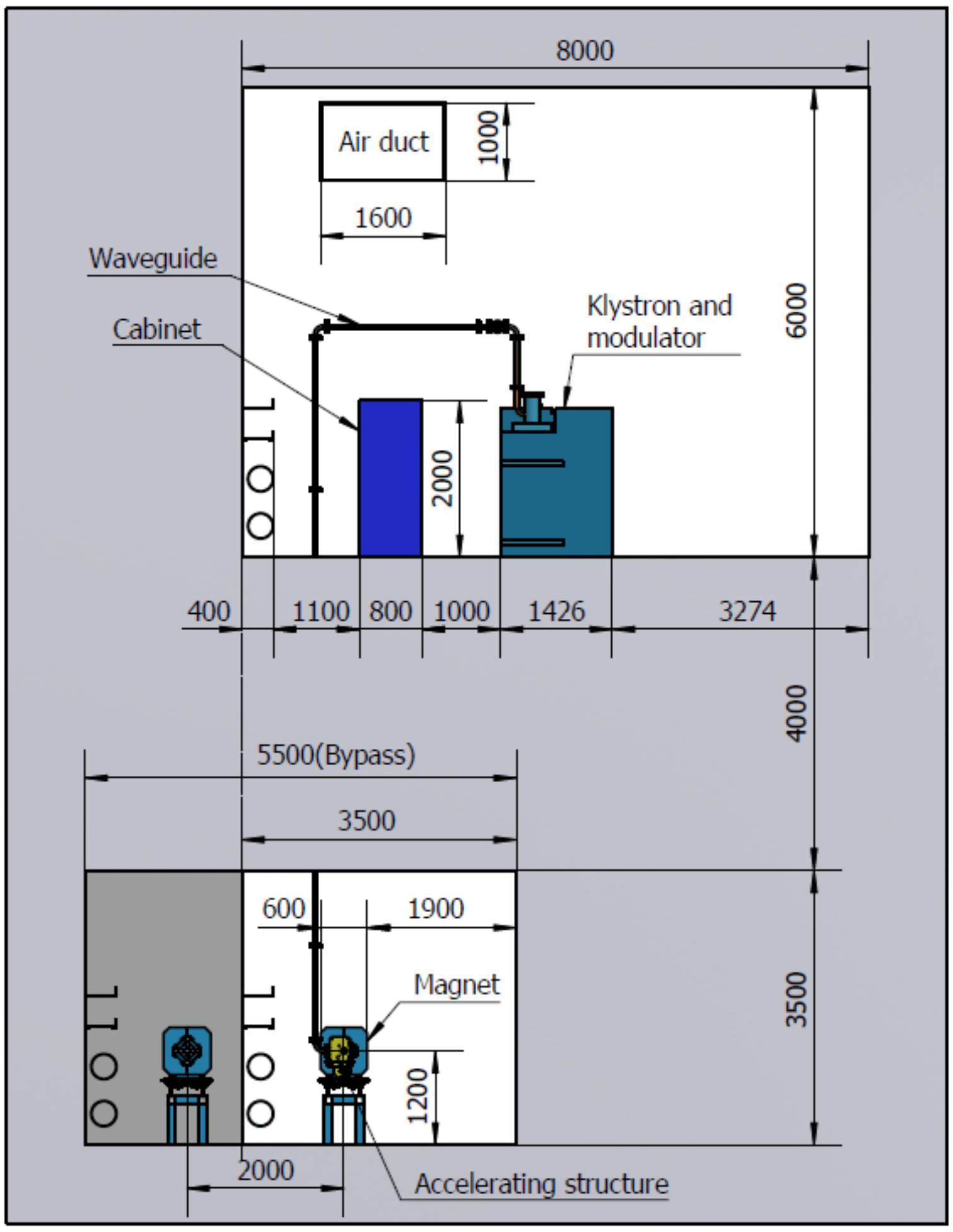

**Figure 6.3.10.4:** Cross section of the Linac and its gallery.

# 6.4 Damping Ring Technical Systems

## 6.4.1 RF System

### 6.4.1.1 *Introduction*

The Damping Ring (DR) is a 1.1 GeV positron damping ring in the Linac complex. It requires two 650 MHz RF cavities working in CW to provide a total of 2.5 MV cavity voltage, with each cavity providing 1.25 MV. A 5-cell cavity at room temperature is designed as the beam instability analysis indicates that, when the aperture of the RF cavity is greater than 90 mm, the Higher-Order-Modes (HOMs) has no effect on the beam. The coupler and door-knob are also designed, and the vacuum of the cavity is calculated to meet the vacuum requirements by reasonably arranging the vacuum pump. The cavity power of 59 kW is required with a 20% margin, and a 90 kW solid-state power source is needed to provide power, considering the attenuation and margin of transmission waveguide. LLRF of DR is also introduced in this section.

### 6.4.1.2 *Higher Order Mode (HOM) Analysis*

In the Damping Ring, the presence of High Order Mode (HOM) from the 5-cell RF cavity can lead to beam instability, both transverse and longitudinal, due to the trapped HOM in the cavity. To ensure stable beam operation, the instability growth rate caused by HOM impedance should be lower than the SR damping time, assuming a uniform filling bunch pattern with damping from the synchrotron radiation. The HOM impedance threshold can be expressed as follows [1]:

$$Z_L = \frac{2E_0 \nu_s}{N_c f_L \alpha_p I_0 \tau_z} \tag{6.4.1.1}$$

where $E_0$ is beam energy, $\nu_s$ is the synchtron tune, $N_c$ is the number of RF cavities, $I_0$ is the beam current, $\alpha_p$ is the momentum compaction factor, and $f_L$ is the frequency of the longitudinal HOM. One critical factor affecting HOM is the pipe aperture, as larger apertures help avoid HOM traps in the cavity. Figure 6.4.1.1 shows the 5-cell RF cavity with two different pipe apertures (44 mm and 90 mm), and Figure 6.4.1.2 compares their respective impedance thresholds.

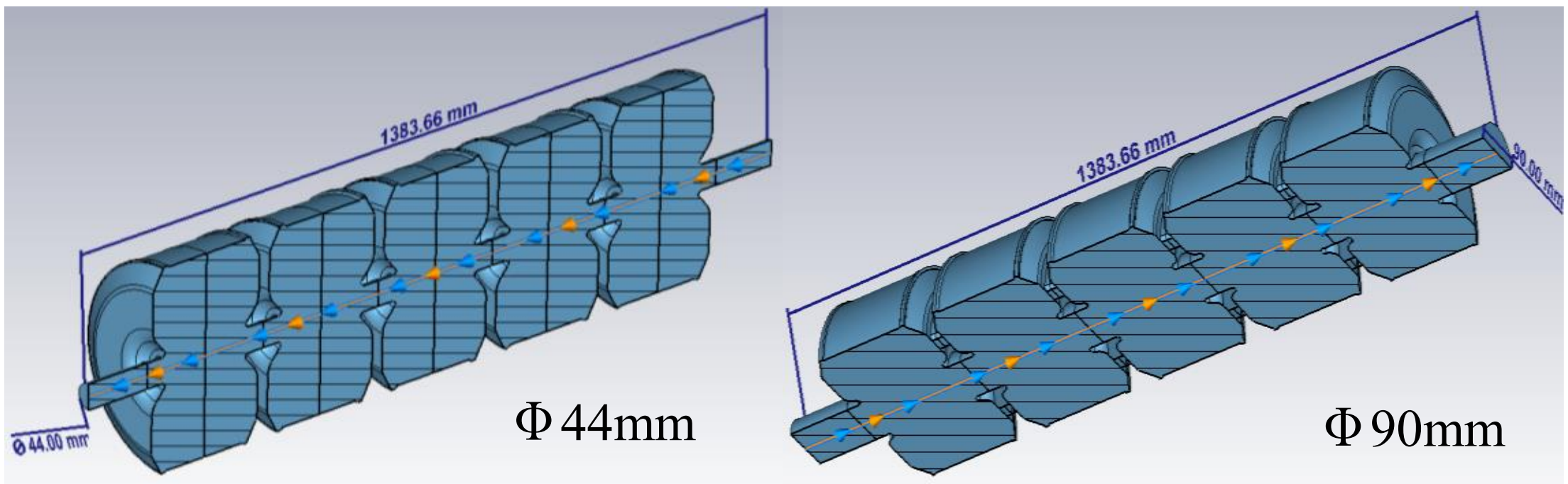

**Figure 6.4.1.1:** RF cavity with two different pipe apertures

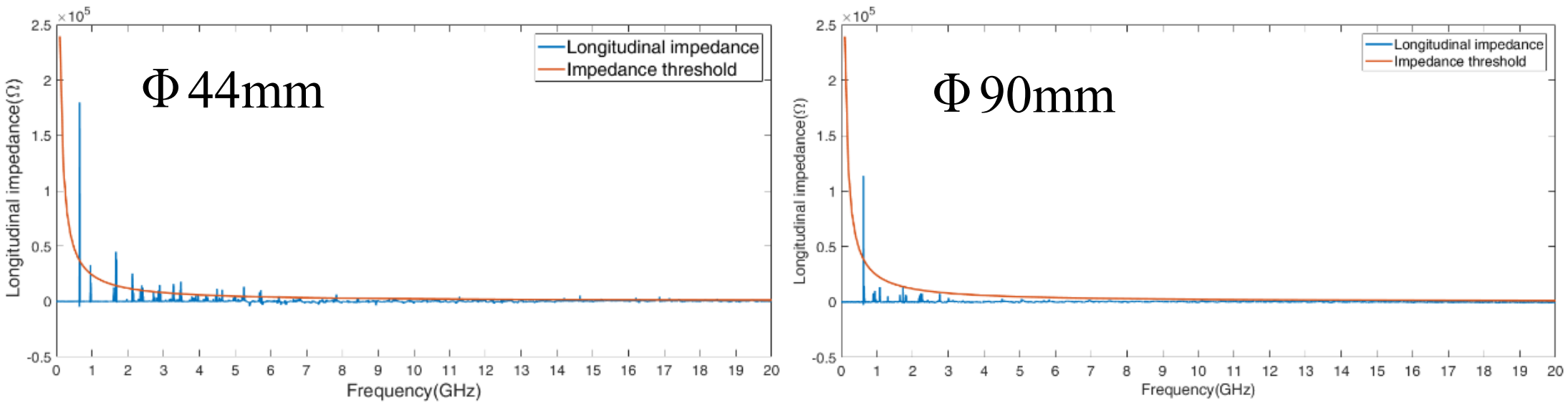

**Figure 6.4.1.2**: Longitudinal impedance and instability threshold for 5-cell cavity with different apertures.

Simulation results show that for the 5-cell cavity with a larger aperture, most of the HOM impedance can be lower than the threshold, effectively eliminating the longitudinal coupled bunch instability. However, the trapped HOM power in the 5-cell cavity remains a serious issue to be studied. The HOM power can be expressed as [2]:

$$P_{HOM} = k_l (eN_b)^2 n_b f_{rev} = k_l I_b^{\ 2} n_b / f_{rev} \tag{6.4.1.2}$$

The calculated loss factor $k_l$ for a bunch length $\sigma_l$ of 5 mm is 5.45 V/pC and 4.12 V/pC for the 5-cell cavity with apertures of 44 mm and 90 mm, respectively. Based on the Damping Ring parameters, the HOM power for the two aperture cavities are, respectively, 68.5 W @10 mA, 137 W @20 mA and 51.8 W @10 mA, 104 W @20 mA. Taking into account the simulation results for impedance threshold and HOM power, the 5-cell cavity with a 90 mm aperture is considered the best choice for the RF system in the Damping Ring.

### 6.4.1.3 ***Normal Conducting Cavity***

For the Damping Ring with an energy of 1.1 GeV, two 5-cell normal conducting cavities have been considered as shown in Figure 6.4.1.3, based on previous studies [3-4]. A 2.5 MV RF voltage is required for these cavities, as indicated in Table 6.4.1.1. The resonant frequency of the 5-cell RF cavity is 650 MHz, with one coupler and two tuners similar to the DESY PETRA cavities [6]. The shunt impedance is 16 MΩ, the Q factor is approximately 32000, and R/Q per meter is approximately 430 Ω/m, with a total cell length of 1153.05 mm. The dissipated power per cavity is estimated to be around 59 kW, considering a 20% margin. The maximum electric field in the cavity is approximately 8.5 MV/m, which is significantly lower than the Kilpatrick field of 23.8 MV/m @ 650 MHz.

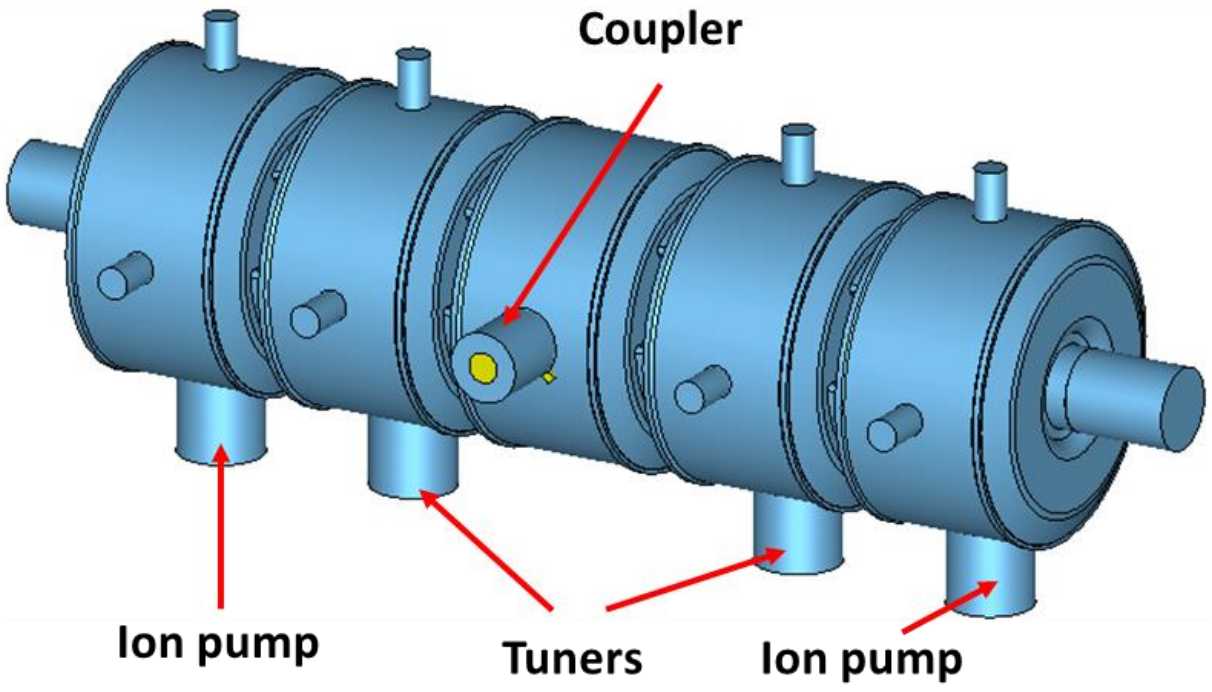

**Figure 6.4.1.3:** A 5-cell RF cavity.

**Table 6.4.1.1:** Main parameters of the 5-cell RF cavity

| Parameters | Unit | Value |
|---|---|---|
| Operation frequency | MHz | 650 |
| Voltage per cavity | MV | 1.25 |
| Cell length | mm | 5×230.61 |
| Shunt impedance | M | 16 |
| R/(Q×l) | Ω/m | 430 |
| Q factor | | 32000 |
| Dissipated power per cavity (20% margin) | kW | 59 |
| Emax | MV/m | 8.5 |

Figure 6.4.1.4 shows the electric and magnetic fields of the 5-cell RF cavity, as simulated using CST MWS. The fundamental mode of the cavity is the π-mode. The optimized field flatness of the longitudinal electric field is shown in Figure 6.4.1.5. The tuning range of the two tuners is 649.64 ~ 650.77 MHz, which covers a range of 30 mm.

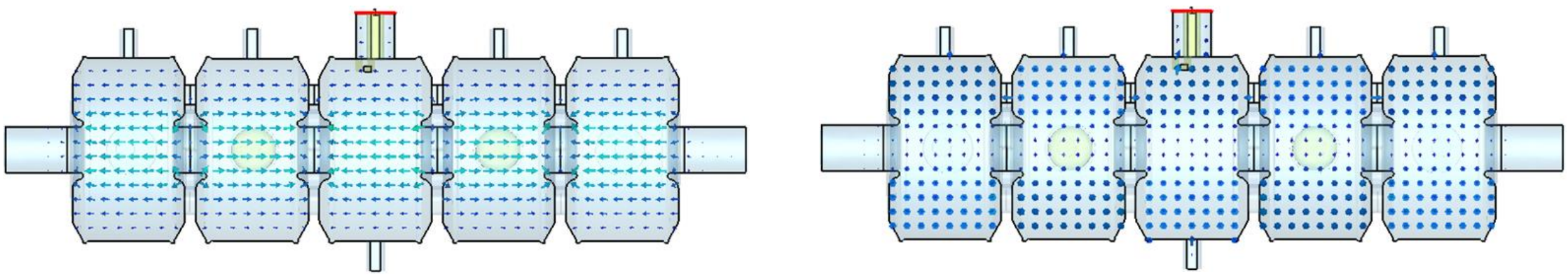

**Figure 6.4.1.4:** Electric field (left) and magnetic field (right) of a 5-cell RF cavity.

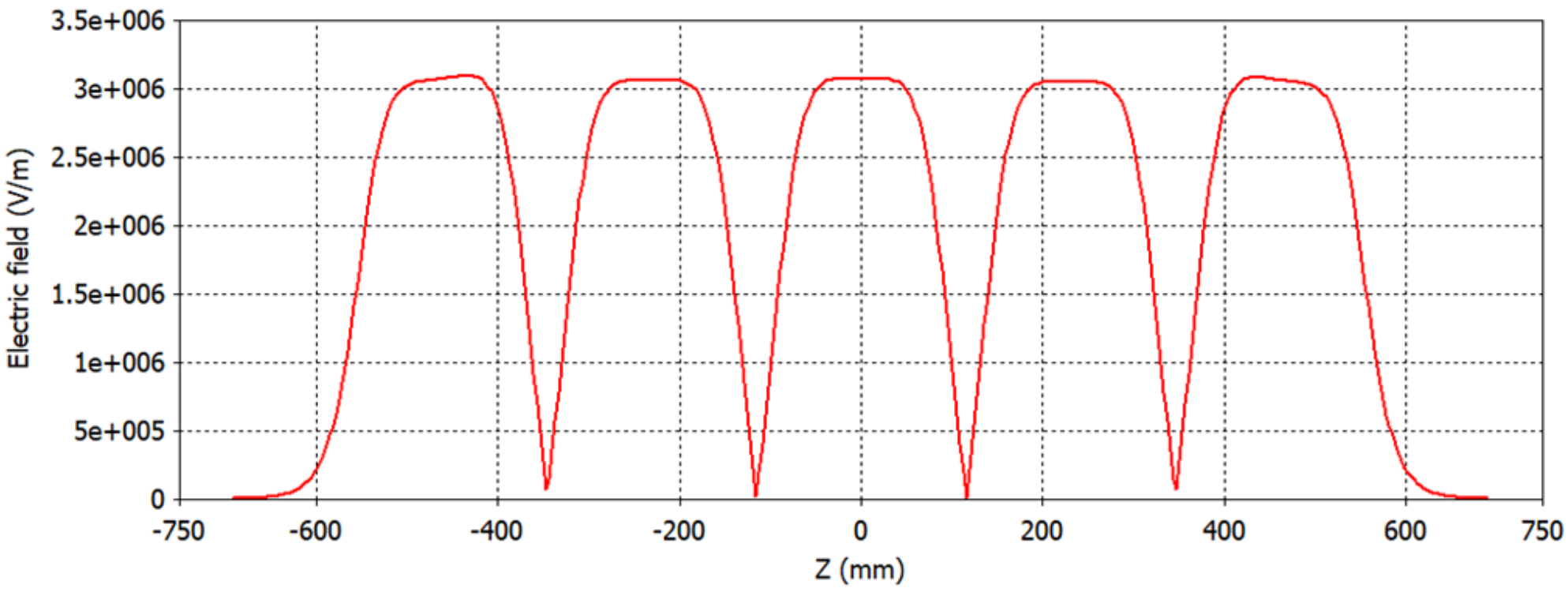

**Figure 6.4.1.5:** Field flatness of the longitudinal electric field of a 5-cell RF cavity.

To feed RF power to the 5-cell RF cavity, a coaxial coupler is utilized, which is composed of a door-knob transition, a disk-type coaxial ceramic window, and a conical coaxial transition. The choke structure is employed to match the impedance of the ceramic window. The calculated reflection coefficient and the electric field distribution with an input power of 59 kW are depicted in Figure 6.4.1.6 and Figure 6.4.1.7, respectively.

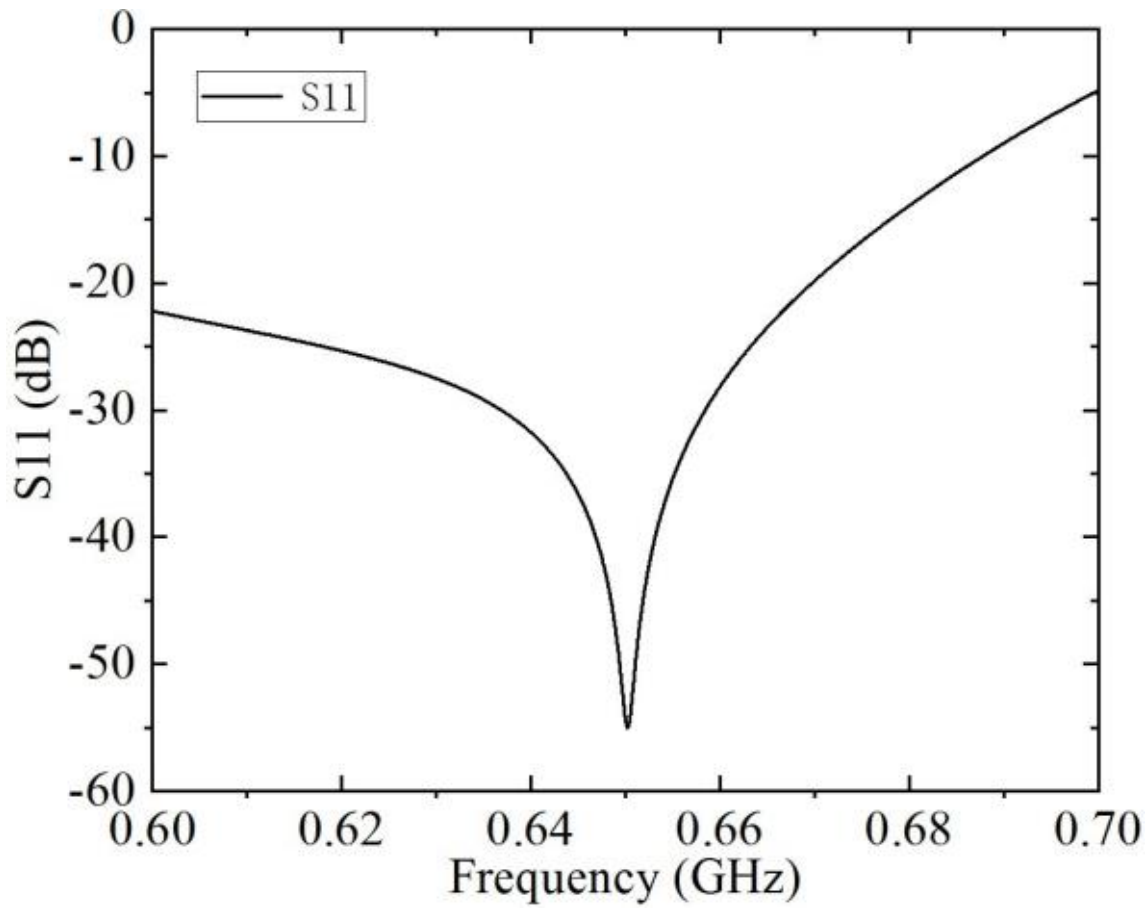

**Figure 6.4.1.6:** The calculated reflection coefficient.

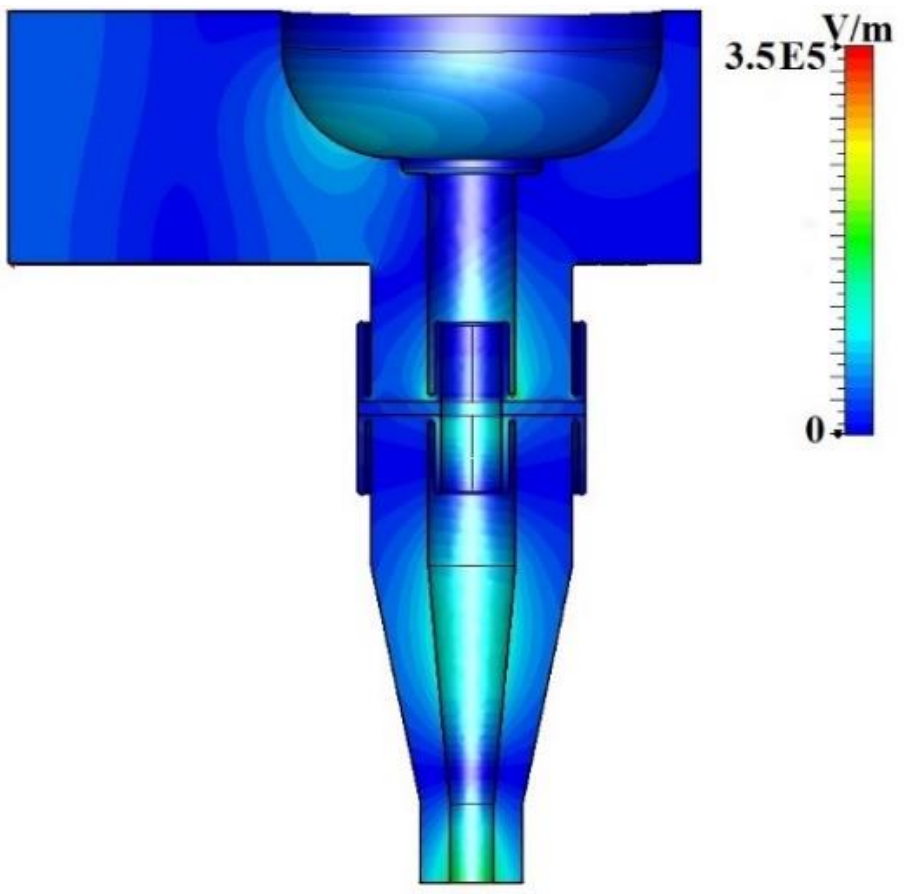

**Figure 6.4.1.7:** The electric field distribution with an input power of 59 kW.

The distribution of vacuum in the 5-cell cavity is evaluated through TPMC simulation using the Synrad+ program developed by CERN. The simulation model is identical to the real 5-cell cavity and is presented in Figure 6.4.1.8. The cavities system utilizes three ion pumps; two pumps, each with a capacity of 100 L/s, are installed on the first and fifth cavity, while the third pump, with a capacity of 15 L/s, is installed on the coupler window. The cavities' interior outgassing rate is set to $2.5\times10^{-11}$ mbar-L/s/cm$^2$, achieved through a treatment process involving oil removal, ultrasonic cleaning, and baking at 60~80°C for degassing, to ensure optimal performance.

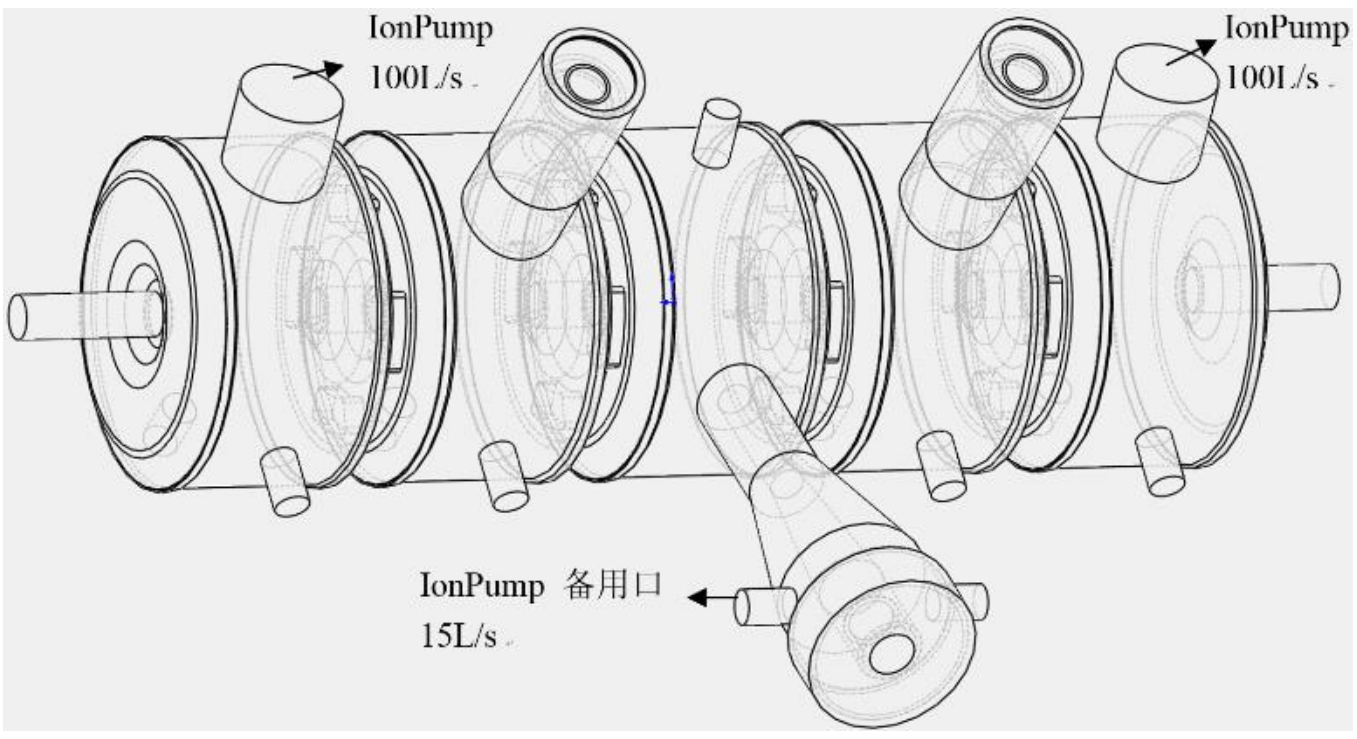

**Figure 6.4.1.8:** Model of a 5-cell cavity for vacuum simulation.

The simulation results for the vacuum distribution in the 5-cell cavity are presented in Figure 6.4.1.9. The vacuum pressure is distributed uniformly in the 5 cavities and their neck tubes, with a range of nearly $3.0\times10^{-7}$ ~ $3.5\times10^{-7}$ Pa. The worst vacuum pressure is $5.2\times10^{-7}$ Pa at the tuner position, which is far from the pumps. However, due to the addition of the 15 L/s ion pump on the coupler window, the pressure at the coupler position is much better, with a value of about $3.4\times10^{-7}$ Pa near the cavity inside. To further improve the vacuum pressure in the cavities under this pump setting, more advanced degassing techniques should be employed to reduce the outgassing rate.

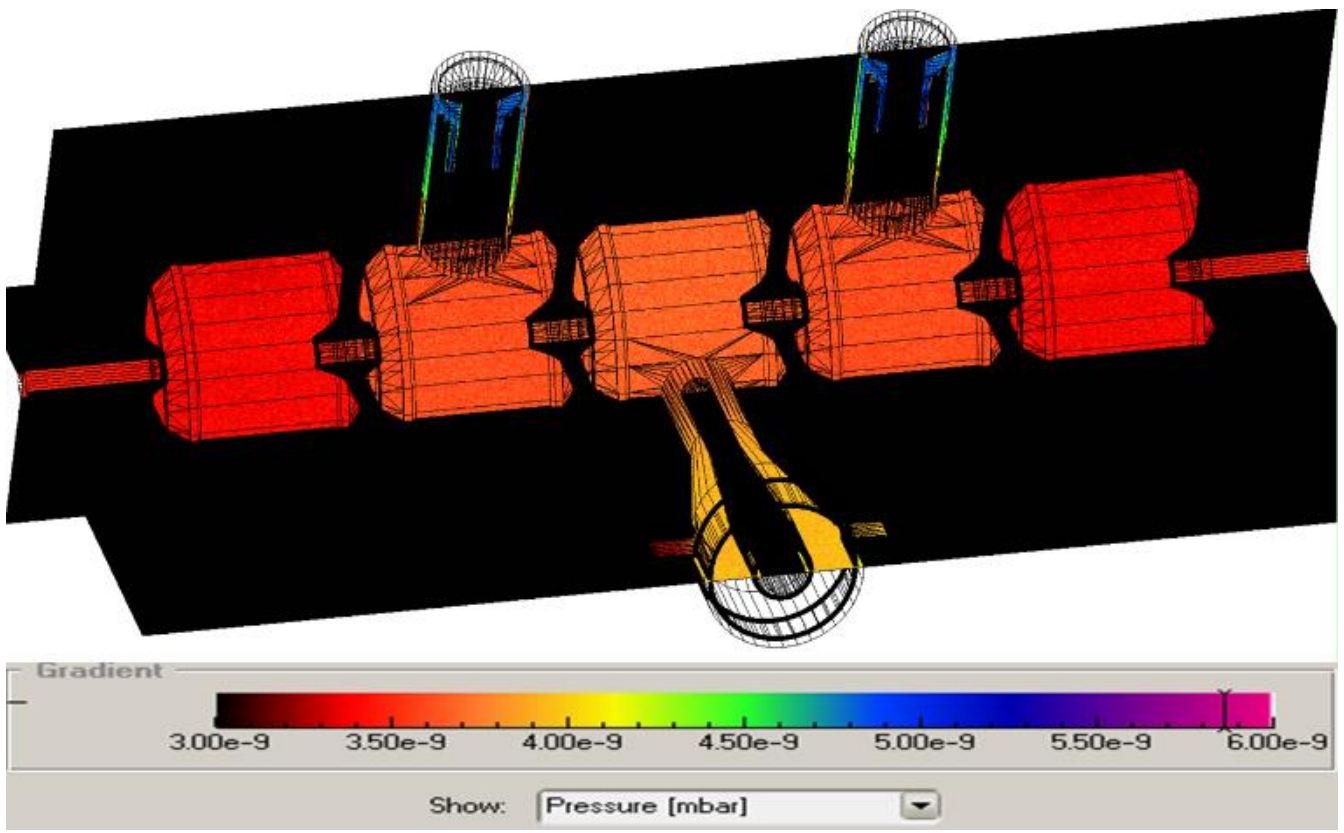

**6.4.1.9:** Vacuum distribution in a 5-cell cavity.

Figure 6.4.1.10 shows the diagram of the 5-cell cavity system, which includes two ion pumps, two tuners, a coaxial coupler, and several ports.

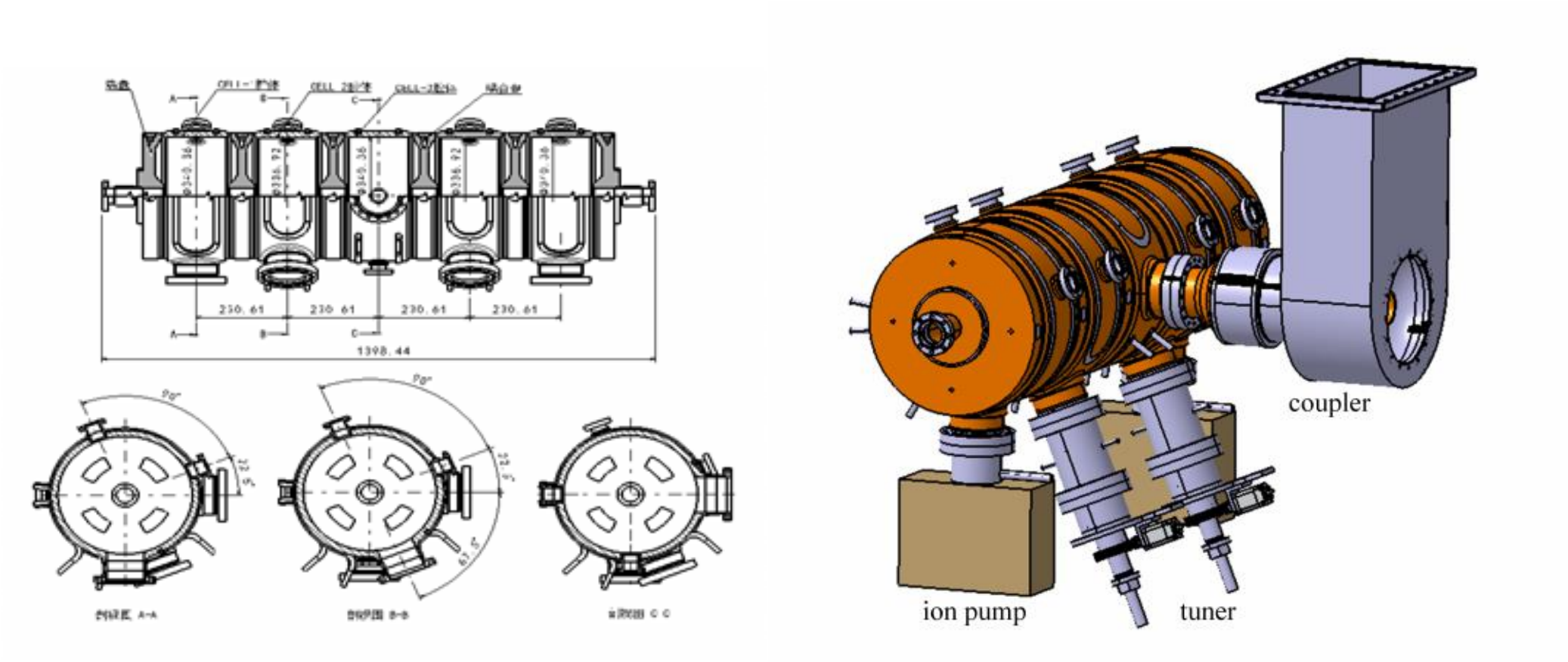

**Figure 6.4.1.10:** Diagram of a 5-cell cavity.

In the HEPS Booster at IHEP, 5-cell cavities are employed to accelerate electrons from 500 MeV to 6 GeV at a frequency of 500 MHz. Figure 6.4.1.11 displays an image of the cavities.

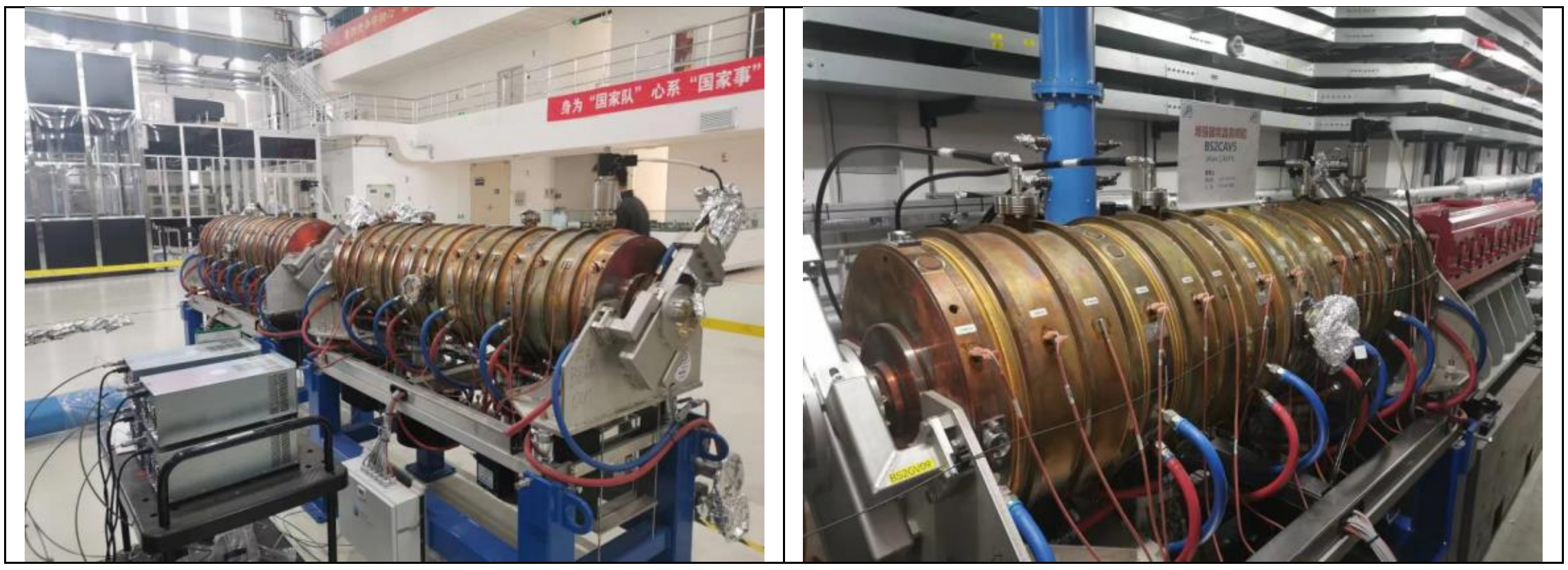

**Figure 6.4.1.11:** 5-cell cavities used in the HEPS Booster at IHEP.

### 6.4.1.4 *RF Power Transfer Line*

The WR1500 aluminum waveguides can be utilized for transferring microwave power from the solid-state amplifier to the RF cavity. The RF transfer line comprises a circulator, a ferrite load, two directional couplers, flexible waveguides, and bending and straight waveguides. As the solid-state amplifier outputs 90 kW power level, and 59 kW of power needs to be delivered to the RF cavity, the transmission efficiency of the RF transfer line is expected to be sufficient.

In order to protect the solid-state amplifier, a CW 90 kW circulator is installed in each feed line to block power that may be reflected or discharged from the cavity. The circulator is a three-port device, where the power from the solid-state amplifier is fed into the first port and transmitted to the second port. However, any power reflected back to the second port will be transmitted to the third port, which is connected to a ferrite load. This way, the solid-state amplifier is protected from any reflected power. After years of hard work, we were able to develop the solid-state amplifier, circulator, and ferrite load in-house.

The directional coupler installed in the RF transfer line will provide both forward and backward (reflected) pick-up signals with directivity better than 30 dB. These signals will be utilized for monitoring, close-loop control, as well as interlock protection during the operation.

In order to eliminate installation errors, reduce vibration transmission, and offset the influence of thermal expansion and contraction in the installation connection, flexible waveguides are adopted in the RF transfer line. These waveguides allow for greater flexibility in the connection, ensuring that any slight movement or deformation does not negatively impact the performance of the system.

### 6.4.1.5 *Low Level RF*

The LLRF system plays a crucial role in ensuring the stability of two normal conducting cavities and power sources by monitoring all signals in the high-level RF system and interacting with other systems. Since the RF frequency is 650 MHz, the basic LLRF can be similar to the Collider LLRF. Directly sampling technique is used to digitize the pickup signals, simplifying the system structure.

As the baseline of the Linac and DR is double bunch accelerating, the waveform for the two bunches needs to be kept stable for a longer period than for a single bunch. Although the bunch charge is relatively small and may not significantly affect beam-loading, feedback with optional feedforward control should be implemented.

To ensure the injection of the bunch to the selected bucket, it is necessary to have event timing from the master oscillator, and the phase relationship between DR and Linac must be adjustable.

The digital LLRF has been successfully demonstrated on the HEPS linac and bunchers during conditioning and commissioning. A directly sampling RF front-end board was developed for sub-harmonic bunchers monitoring and stability control, covering the frequency range from 100 MHz to 700 MHz [7-8]. Figure 6.4.1.12 illustrates the overall design of the LLRF system.

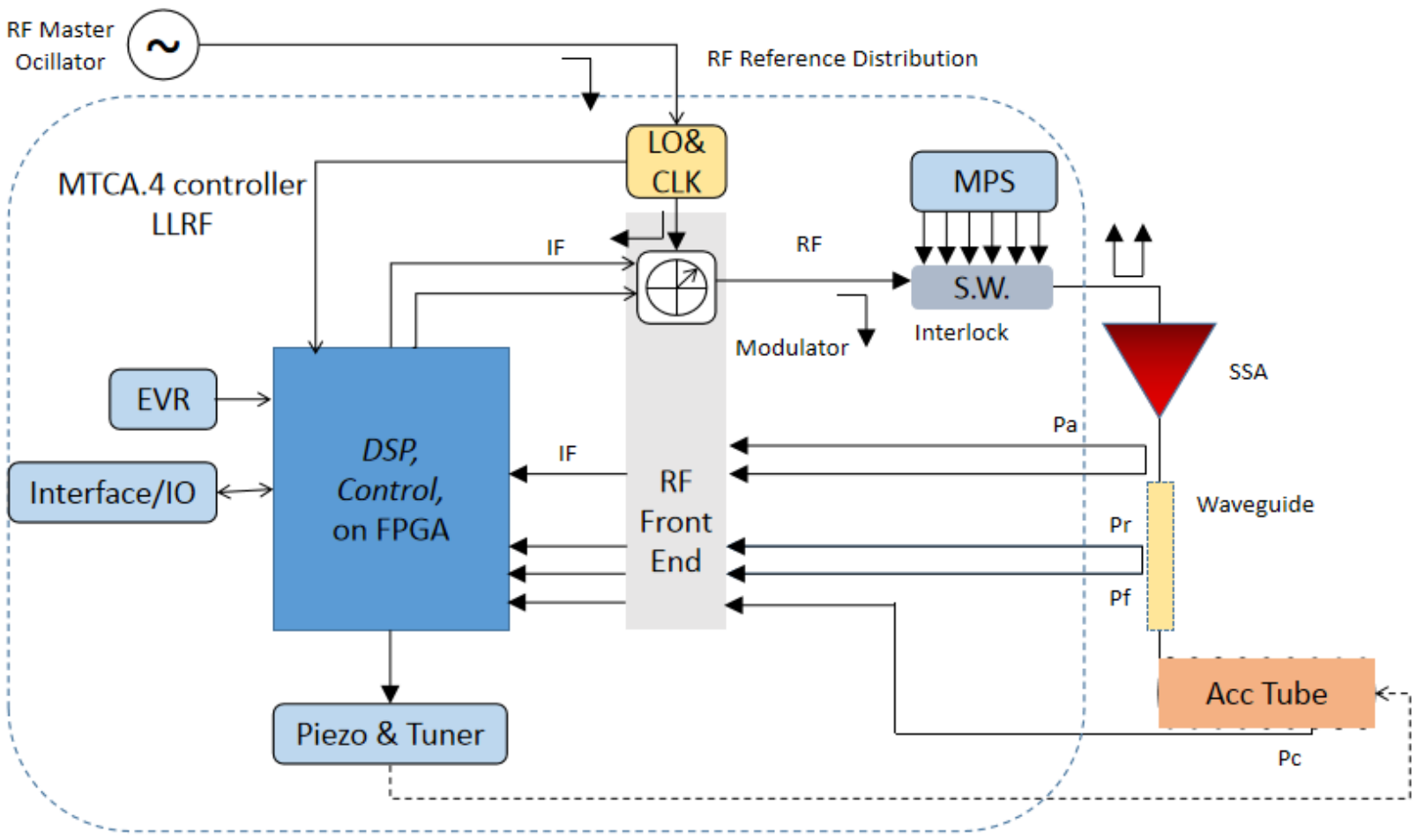

**Figure 6.4.1.12:** Overall design of the Damping Ring LLRF.

## 6.4.2 RF Power Source

To power each 5-cell normal-conducting RF cavity, a solid-state amplifier (SSA) operating at a frequency of 650 MHz is required. Due to RF transmission loss, reflection caused by cavity mismatch, and low-level amplitude-control requirements, each of the two SSAs needs to provide an output power of 75 kW. This output power is combined with 40 2.3 kW power modules, as illustrated in Figure 6.4.2.1. The output port for the SSAs is a standard WR1500 waveguide, and the AC to RF efficiency goal for the SSAs at the rated power is at least 50%.

The 650 MHz/75 kW SSA requires certain design considerations and parameter requirements similar to the 1.3 GHz/25 kW SSA for the Booster. The specifications are listed in Table 6.4.2.1, and a prototype is shown in Fig. 6.4.2.1.

**Table 6.4.2.1:** 650 MHz/75 kW SSA specifications and measurements

| Parameters | Specification | Measurement |
|---|---|---|
| Frequency | 650 MHz | 650 MHz |
| Power (< 1 dB Compression) | 75 kW | 75 kW |
| Gain | ≥ 65 dB | 82 dB |
| Bandwidth (1 dB) | ≥ 1 MHz | 2 MHz |
| Amplitude stability (open loop) | ≤ 1% | 0.8% |
| Phase stability (open loop) | ≤ 1° | 0.5° |
| Phase variation (10 kw-75kW) | ≤ 10° | 8° |
| Harmonic | < –30 dBc | –60 dBc |
| Spurious | < –60 dBc | –70 dBc |
| Efficiency at 75 kW | ≥ 50% | 55% |
| MTBF | ≥ 30000 h | No test |
| Redundancy | SSAs can still run with 1 power module failure for each cabinet | 0.5 |
| Cooling | Deionized water 4 kg/cm$^2$ at 25$^{\circ}$ C | 4 kg/cm$^2$ |

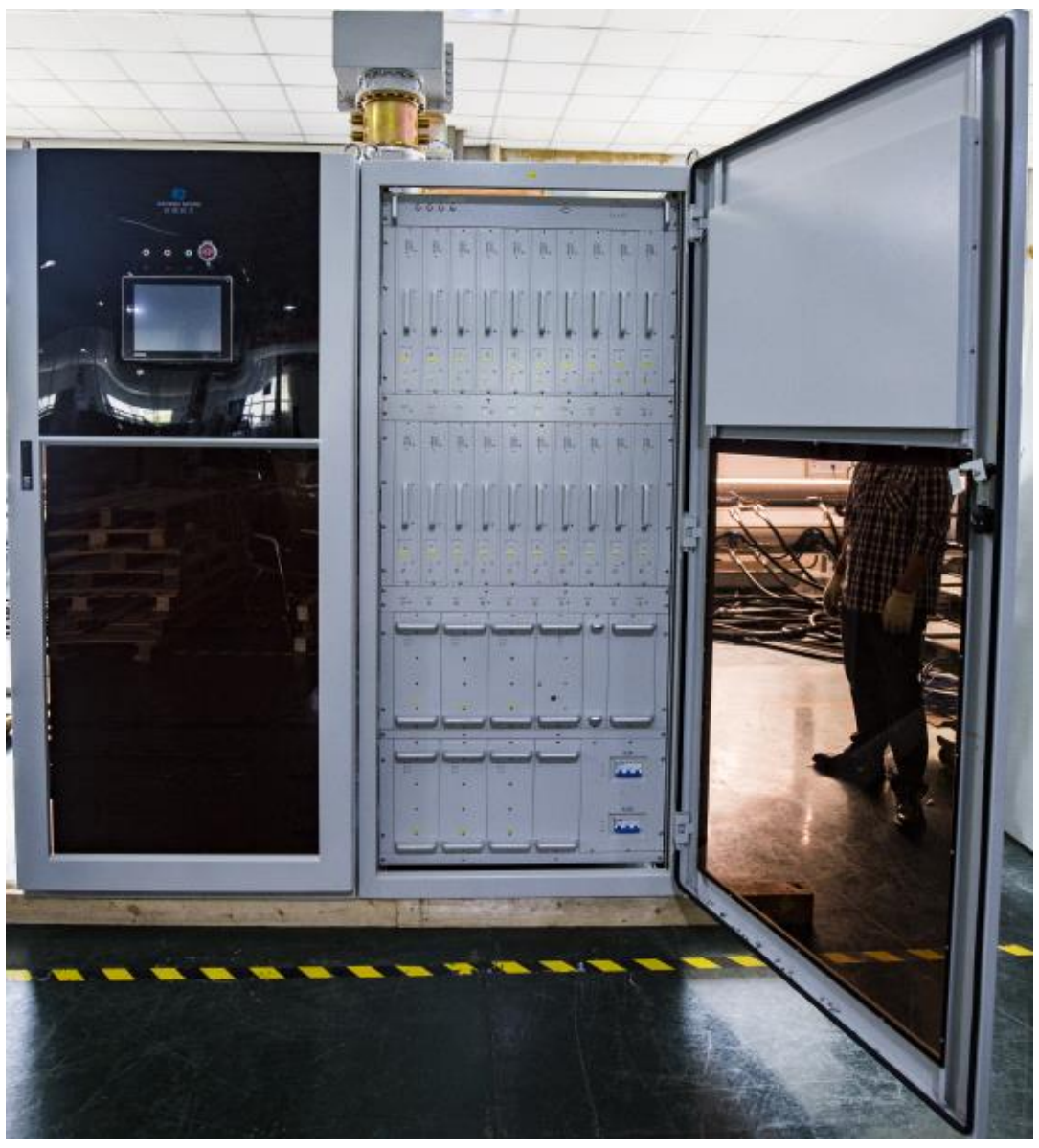

**Figure 6.4.2.1**: 650 MHz /75 kW solid state amplifier.

## 6.4.3 Magnets

### 6.4.3.1 *Dipole Magnets*

There are 80 dipole magnets in the Damping Ring (DR), and 28 dipole magnets for the DR transport line. They are named DR-38B0, DR-38Br, and TL-44B, respectively. These are DC magnets and designed with conventional technologies. The magnets are designed with a C-type structure, and the cross section and magnetic flux line distribution of the DR-38B0 magnet are shown in Figure 6.4.3.1. The main design parameters of the magnets can be found in Table 6.4.3.1.

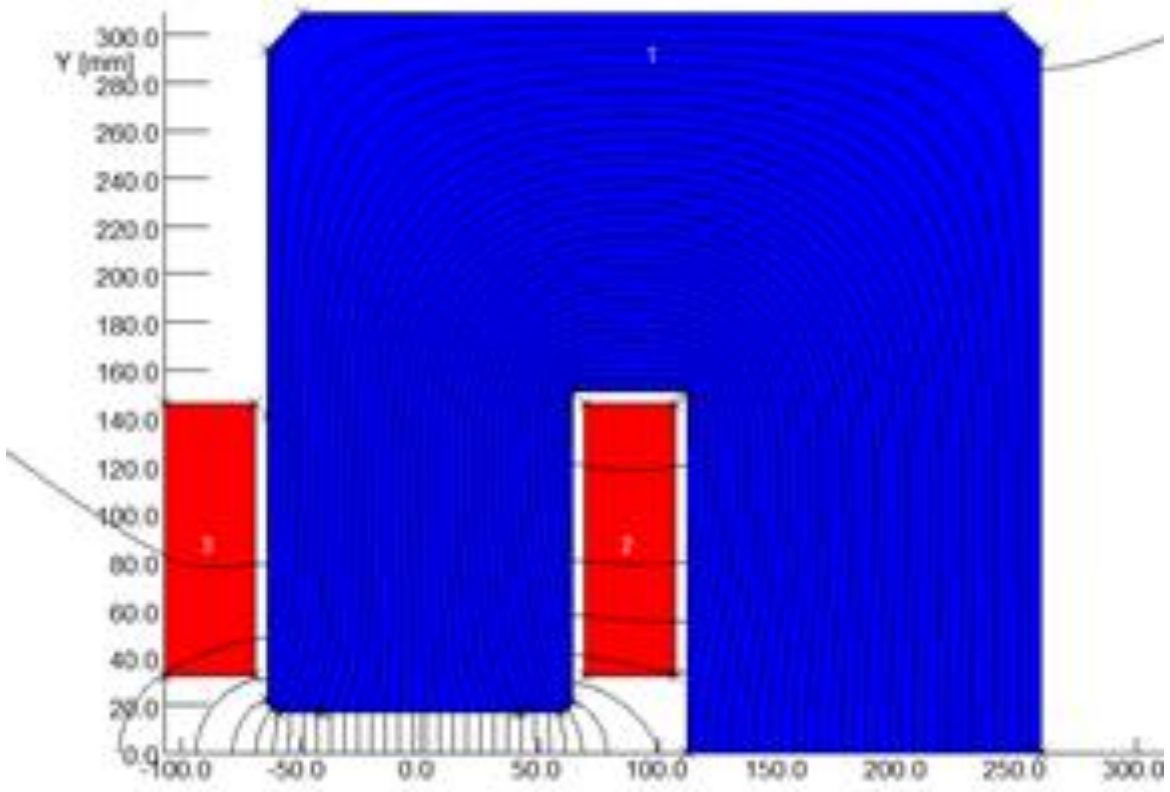

**Figure 6.4.3.1:** Cross section and flux lines of the DR-38B0 dipole magnet.

**Table 6.4.3.1:** Main design parameters of the DR dipole magnets.

| Parameters | DR-38B0 | DR-38Br | TL-44B |
|---|---|---|---|
| Quantity | 40 | 40 | 28 |
| Gap (mm) | 38 | 38 | 44 |
| Max. field (T) | 1.3 | 1.3 | 0.75 |
| Magnetic length (mm) | 700 | 248 | 2000 |
| Good field region (mm) | 60 | 60 | 60 |
| Field errors | 0.1% | 0.1% | 0.1% |
| Ampere-turns per pole (AT) | 20046 | 20046 | 13391 |
| Turns per pole | 48 | 48 | 36 |
| Max. current (A) | 418 | 418 | 372 |
| Conductor size (mm) | 12×12Φ8 | 12×12Φ8 | 12×12Φ8 |
| Current density (A/mm$^2$) | 4.45 | 4.45 | 3.97 |
| Resistance (mΩ) | 42.2 | 23.1 | 72.8 |
| Inductance (mH) | 51.7 | 18.3 | 72.9 |
| Voltage drop (V) | 18.0 | 10.0 | 27 |
| Power loss (kW) | 7.4 | 4.0 | 10.1 |
| Core length (mm) | 672 | 220 | 1970 |
| Core width/height (mm) | 490/423 | 490/423 | 480/440 |
| Magnet weight (ton) | 1.14 | 0.41 | 3.5 |
| Water pressure (kg/cm$^2$) | 6 | 6 | 6 |
| Cooling circuits | 4 | 2 | 4 |
| Water flow velocity (m/s) | 2.79 | 2.65 | 2.04 |
| Total water flow (l/s) | 0.56 | 0.27 | 0.41 |
| Temperature increase (°C) | 3.1 | 3.6 | 5.8 |

### 6.4.3.2 *Quadrupole Magnets*

The quadrupole magnets used in both the DR and its transport beam lines are DC-excited magnets, and their design and manufacturing requirements are the same as those for the CEPC Linac's conventional quadrupoles. To facilitate coil installation, the magnets have a four-in-one structure. The cross section and magnetic flux of the quadrupole magnet can be seen in Figure 6.4.3.2, and the main parameters for all types of magnets are listed in Table 6.4.3.2.

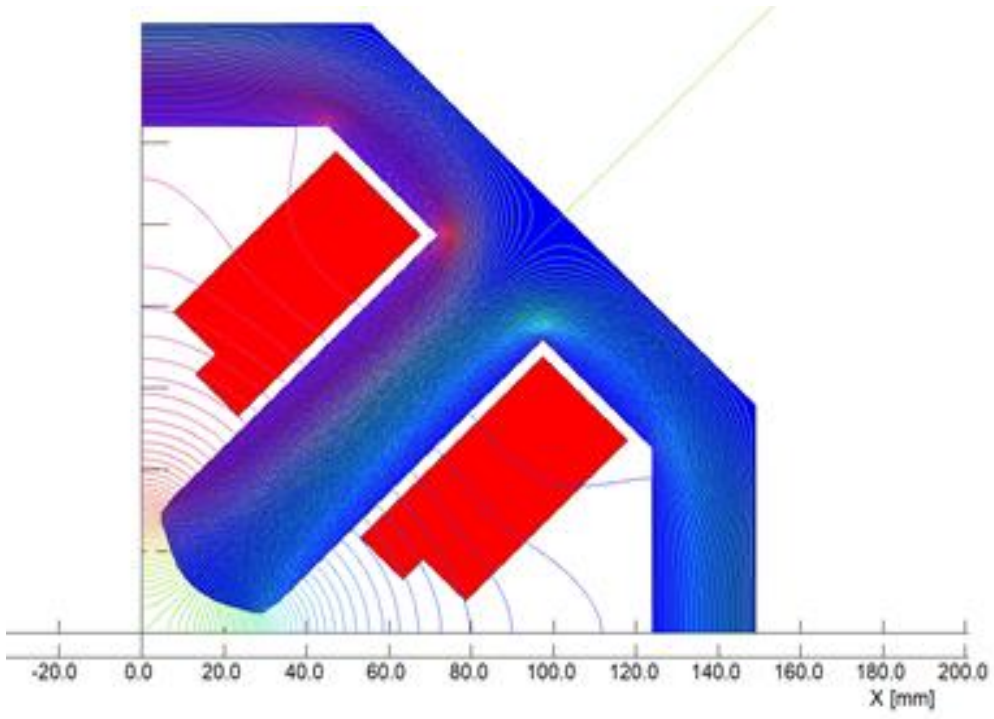

**Figure 6.4.3.2:** Cross section and flux lines of the DR quadrupole magnet.

**Table 6.4.3.2:** Main parameters of the DR quadrupole magnets.

| Parameters | DR-44Q | DR-38Q | TL-54Q |
|---|---|---|---|
| Quantity | 96 | 8 | 28 |
| Aperture (mm) | 44 | 38 | 54 |
| Magnetic length (mm) | 200 | 200 | 300 |
| Field gradient (T/m) | 17 | 30 | 5 |
| GFR-radius (mm) | 17 | 15 | 22 |
| Harmonic errors | 0.1% | 0.1% | 0.1% |
| AT per pole | 3372 | 4438 | 1494 |
| Turns per pole | 20 | 28 | 60 |
| Current (A) | 169 | 159 | 25 |
| Conductor size (mm) | 7×7Φ4 | 7×7Φ4 | 3.2×8.5 |
| Current density (A/mm$^2$) | 4.63 | 4.35 | 0.92 |
| Resistance (mΩ) | 21 | 28 | 120 |
| Voltage (V) | 3.47 | 4.45 | 2.98 |
| Power loss (kW) | 0.58 | 0.70 | 0.1 |
| Core length (mm) | 190 | 190 | 280 |
| Core width/height (mm) | 450 | 410 | 550 |
| Magnet weight (kg) | 250 | 220 | 580 |
| Water pressure (kg/cm$^2$) | 3 | 3 | --- |
| Cooling circuits | 4 | 4 | --- |
| Water flow velocity (m/s) | 2.67 | 2.24 | --- |
| Total water flow (l/s) | 0.13 | 0.11 | --- |
| Temperature rise (°C) | 1.30 | 1.50 | --- |

### 6.4.3.3 *Sextupole Magnets*

There are 72 sextupole magnets used in the DR. Due to the low working magnetic field, the magnet design and parameters are calculated according to the maximum sextupole field requirements. The magnet has a two-in-one structure, and the core material can be made of DT4 solid iron, while the coils are made of solid copper conductors. The magnetic flux of the magnet is illustrated in Figure 6.4.3.3, and the main parameters of the magnets can be found in Table 6.4.3.3.

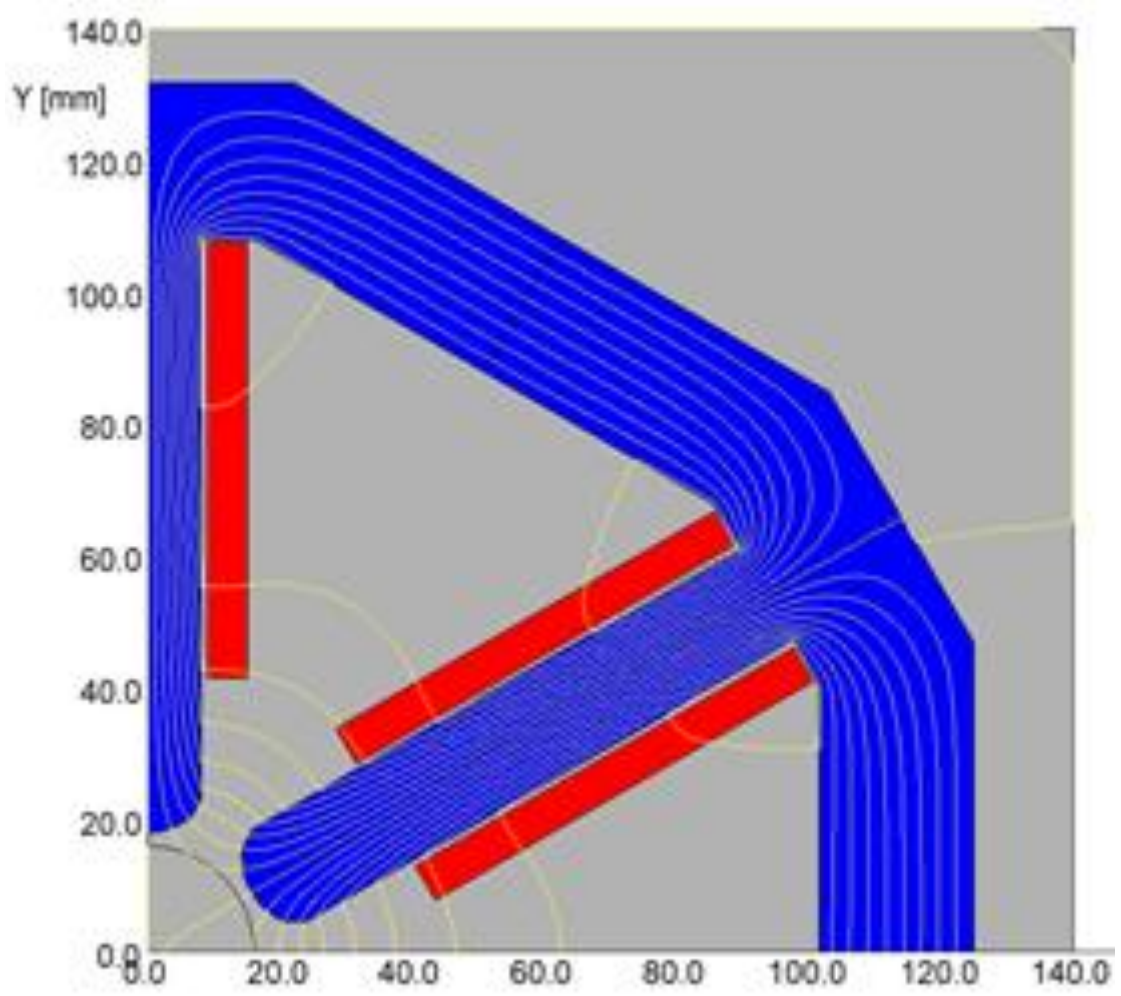

**Figure 6.4.3.3:** Cross section and flux lines of the DR sextupole magnet.

**Table 6.4.3.3:** Main parameters of the DR sextupole magnets.

| Parameters | DR-38S |
|---|---|
| Quantity | 72 |
| Aperture diameter (mm) | 38 |
| Magnetic length (mm) | 100 |
| Max. sextupole field (T/m$^2$) | 140 |
| GFR radius (mm) | 15 |
| Harmonic errors | 0.1% |
| Ampere-turns per pole (AT) | 132 |
| Coil turns per pole | 14 |
| Conductor size (mm) | 2×5 |
| Current (A) | 9.5 |
| Current density (A/mm$^2$) | 0.95 |
| Resistance (mΩ) | 44.5 |
| Voltage (V) | 0.4 |
| Max Power loss (W) | 4.0 |
| Core length (mm) | 90 |
| Core width/height (mm) | 220 |
| Magnet weight (kg) | 30 |

### 6.4.3.4 *Corretor Magnets*

There are a total of 60 correctors used in the Damping Ring. Each corrector has a gap of 40 mm and a magnetic length of 70 mm. To ensure optimal field quality, the corrector magnets are designed without iron cores, as the remnant field of iron cores can adversely affect the field quality. The coils of the corrector magnets are constructed using solid copper conductors in a simple racetrack structure. The coils are supported by H-type structures made of G10 material. The magnetic flux of the corrector magnet is depicted in Figure 6.4.3.4, and further details and key parameters of the magnets are given in Table 6.4.3.4.

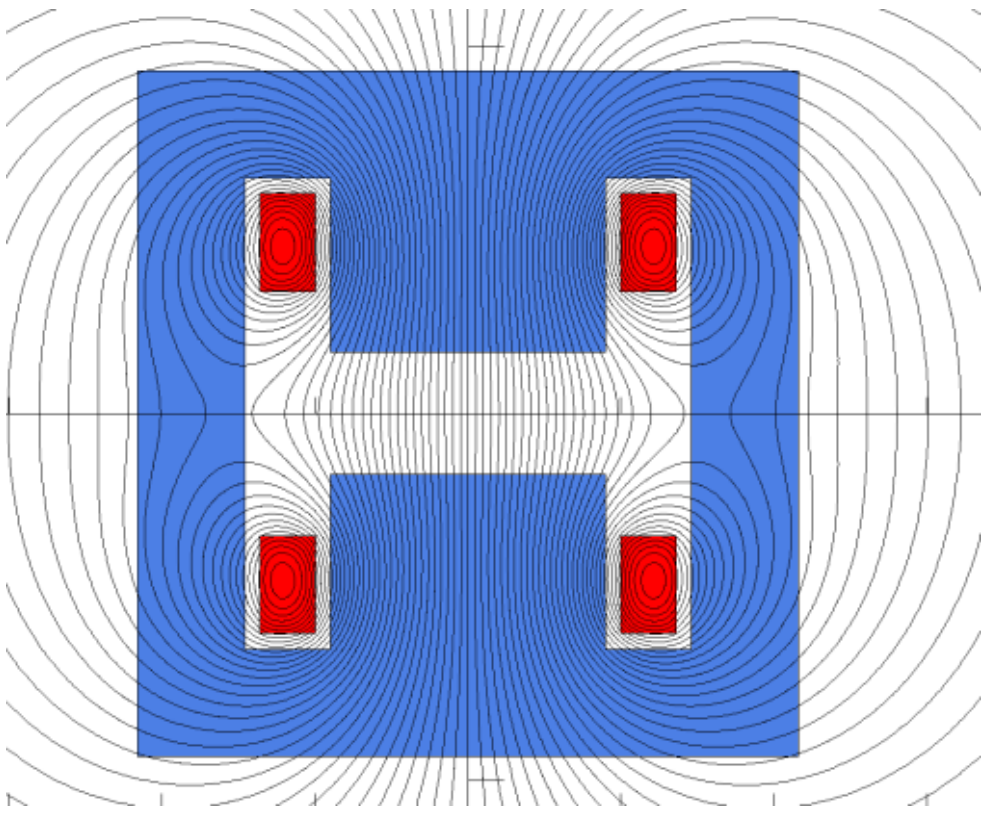

**Figure 6.4.3.4:** Cross section and flux lines of the DR corrector magnet

**Table 6.4.3.4:** Main parameters of the DR corrector magnets.

| Parameters | DR-40C |
|---|---|
| Quantity | 60 |
| Aperture diameter (mm) | 40 |
| Magnetic length (mm) | 70 |
| Max. field (Gs) | 20 |
| GFR (mm) | 30 |
| Field errors | 5% |
| Ampere-turns per pole (AT) | 300 |
| Coil turns per pole | 30 |
| Conductor size (mm) | 3×6.3 |
| Current (A) | 10 |
| Current density (A/mm$^2$) | 0.54 |
| Resistance (mΩ) | 32 |
| Voltage (V) | 0.32 |
| Max Power loss (W) | 3.2 |

### 6.4.4 Magnet Power Supplies

The Damping Ring (DR) in the CEPC operates at an energy of 1.1 GeV and has a circumference of 147 meters. In addition to the DR, there are two transport lines that connect the Linac and the DR. The overall system consists of a total of 108 dipoles, 132 quadrupoles, 72 sextupoles, and 60 correctors. All of these magnets are excited by DC current.

Figure 6.4.4.1 showcases the layout of the Damping Ring system, illustrating the positions and connections of various elements within the DR.

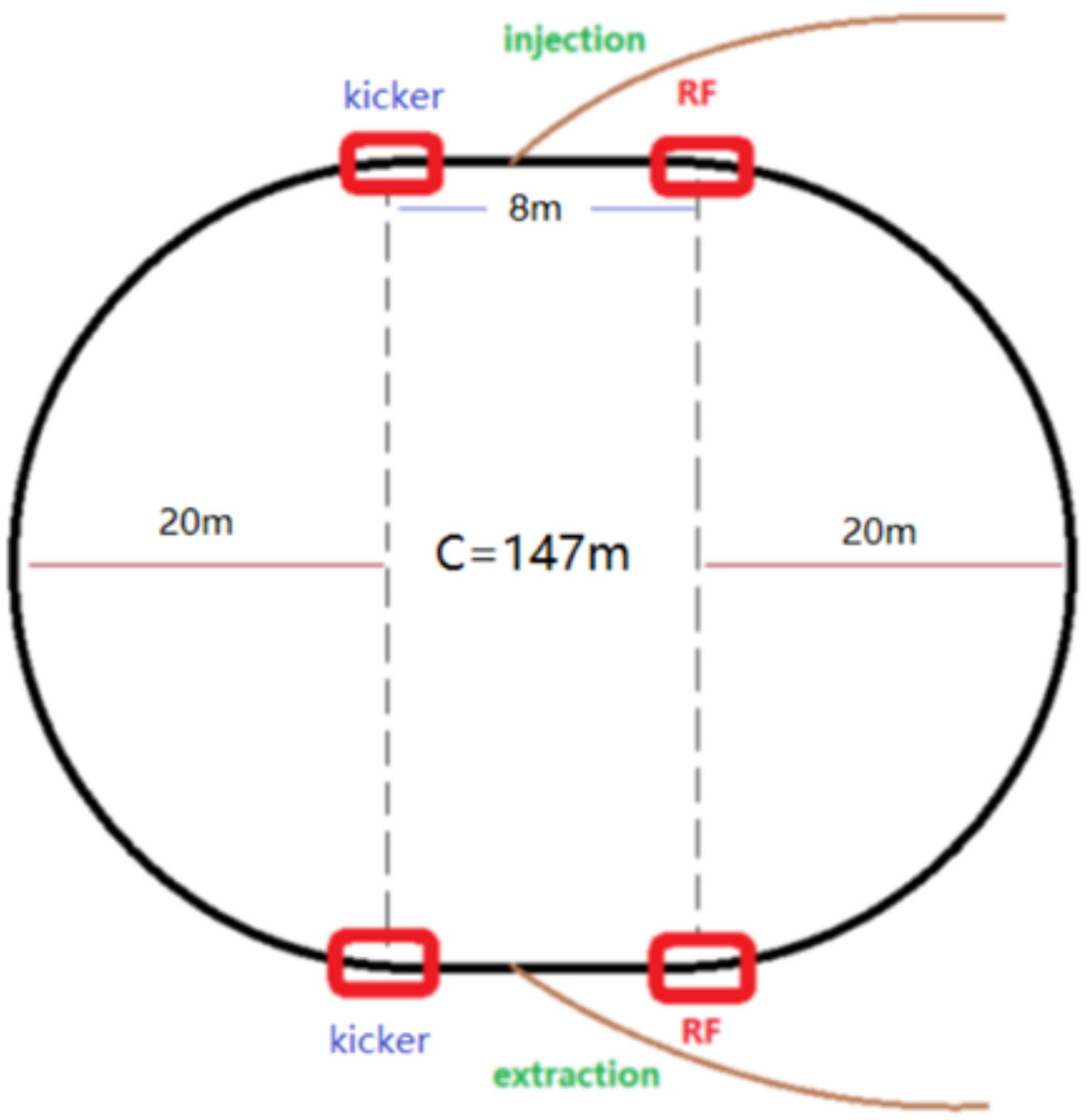

**Figure 6.4.4.1:** Layout of the Damping Ring system.

### 6.4.4.1 *Types of Power Supplies*

In the Damping Ring (DR) of the CEPC, there are a total of 80 dipole magnets. Among them, 40 dipole magnets belong to the DR-38B0 family, and the remaining 40 dipole magnets belong to the DR-38Br family. Additionally, the DR transport line (TL-44B) consists of 28 dipole magnets.

The dipole magnets DR-38B0 and DR-38Br are organized into 8 families, with each family comprising 10 series-connected magnets. Each magnet within these families is powered by an individual power supply. Similarly, the LTD-44B dipole magnets are organized into 4 families, with each family comprising 7 series-connected magnets, each powered by its own power supply.

The damping ring includes 104 quadrupole magnets, which are divided into two families. The DR-44Q family has 96 quadrupole magnets, while the DR-38Q family has 8 quadrupole magnets. Additionally, the DR transport line (LTD-54Q) consists of 28 quadrupole dipole magnets. Each quadrupole magnet, including those in the transport line, is powered independently.

There are 72 sextupole magnets in the Damping Ring, specifically belonging to the DR-38S family. These sextupole magnets are arranged into 2 families, with each family comprising 36 series-connected magnets. Each magnet is powered by its own power supply.

The Damping Ring also incorporates 60 correctors (DR-40C), which are powered independently.

In total, there are 12 dipole power supplies, 132 quadrupole power supplies, 2 sextupole power supplies, and 60 corrector power supplies. The dipole, quadrupole, and sextupole power supplies are unipolar, while the corrector power supplies are bipolar. The overall power consumption for the DR power supply system amounts to 0.90 MW.

For detailed parameters of the main magnet and correction magnet power supplies for the damping ring, please refer to Table 6.4.4.1.

**Table 6.4.4.1:** Magnet power supply requirements for the Damping Ring

| Magnet | Quantity | Stability /8hours | Output Rating |
|---|---|---|---|
| DR-38B0 | 4 | 100 ppm | 460A/200V |
| DR-38Br | 4 | 100 ppm | 460A/110V |
| TL-44B | 2 | 100 ppm | 410A/220V |
| DR-44Q | 96 | 100 ppm | 180A/7V |
| DR-38Q | 8 | 100 ppm | 180A/7V |
| TL-54Q | 26 | 100 ppm | 30A/7V |
| DR-38S | 2 | 100 ppm | 11A/30V |
| DR-40C | 60 | 300 ppm | ±11A/±3V |
| **Total** | **206** | | |

### 6.4.4.2 *Design of the Power Supply System*

The power supplies for the Damping Ring in the CEPC operate in DC mode. To ensure safety and reliable operation, all power supplies are designed with a 10-15% safety margin in both current and voltage. The power supply systems will utilize switching-mode topology, which offers advantages such as high efficiency, reduced size and weight, easy interface with digital controllers, and reduced influence of higher order voltage ripple components on magnet current.

To facilitate maintenance and repair, all power supplies are module-based and feature digital control. This modular design allows for easy replacement and troubleshooting of individual components, improving overall system reliability.

The power supply system for the Damping Ring will adopt two different structural frameworks based on Figure 6.4.4.2. The specific frameworks, as shown in (a) and (b) of the figure, will be implemented to accommodate the requirements and characteristics of the power supply units.

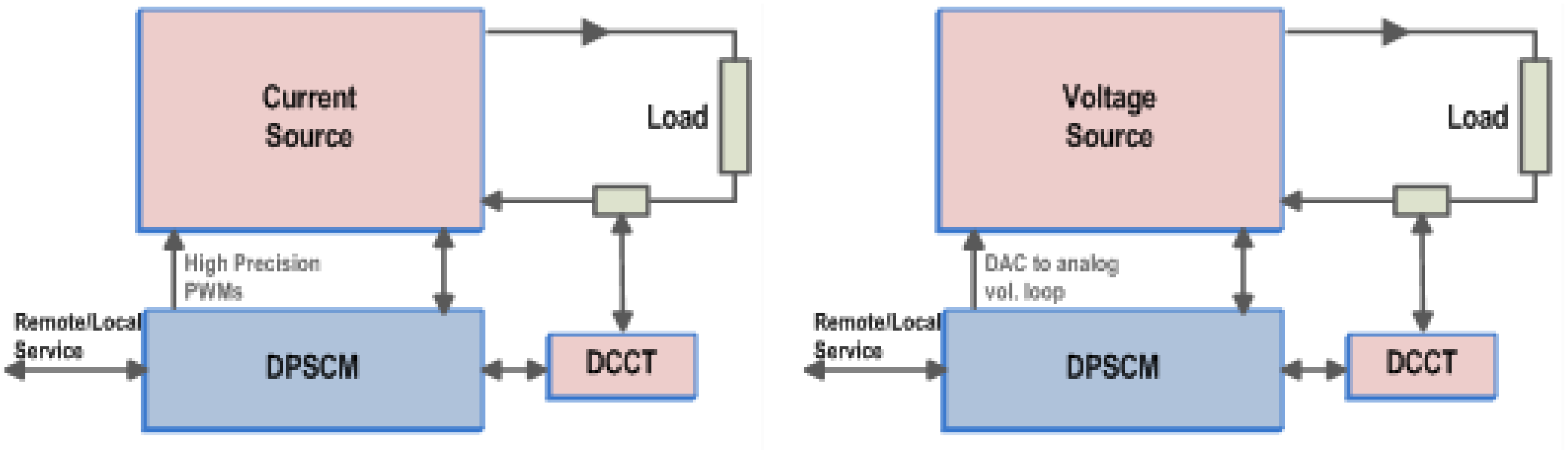

(a) (b)

**Figure 6.4.4.2:** Structural frameworks of the Damping Ring power supply.

Fig. 6.4.4.2. (a) depicts the adoption of an all-digital + all-switch structure for the power supply system. In this configuration, the digital controller of the power supply enables precise digital adjustment and control of all control loops. It generates an ultra-high precision pulse-width modulation (PWM) signal through the hardware, with a minimum variation of 150ps. This enables the realization of an ultra-high precision switching power supply.

Fig. 6.4.4.2. (b) showcases the implementation of a partial digital + arbitrary topology structure. In this case, the power supply's digital controller achieves high-precision digital

adjustment and control of the current closed-loop. The output of the current closed-loop control is converted to an analog signal through a digital-to-analog conversion (DAC) circuit. This analog signal serves as the reference for the voltage loop. The power section of the system provides the voltage source, which can be based on any topology. This control mode leverages the advantages of digital control while overcoming the limitations imposed by the dependence of digital control on topological structure and the precision of digital PWM control.

### 6.4.4.3 *Topology of Power Supplies*

The power supply design for the Damping Ring mainly utilizes a combination of DC source and switched-mode technology. The DC source employs a multiplicative equivalent 12-pulse rectifier to minimize harmonic current and provide a stable DC voltage at the front-end. A buck or booster configuration is implemented to regulate input power fluctuations. The switched-mode converter is responsible for controlling the output current.

The power supply system follows a modular design, as depicted in Figure 6.4.4.3. The module is divided into three stages. The first stage consists of a phase-shift parallel-connected power factor correction (PFC) circuit, which ensures a stable DC voltage output. The second stage employs pulse-width modulation (PWM) control to achieve constant frequency regulation, optimizing the design of the filter. A "soft switching" PWM DC/DC full-bridge converter controlled by phase shift is utilized based on zero conversion converter technology. This converter leverages the leakage inductance of the high-frequency transformer or primary-side inductance in series with the parasitic capacitance of the switching tube to achieve zero-voltage switching. By combining the advantages of resonant and PWM power supplies, this converter is particularly suitable for medium-power DC power supply applications. The output stage includes a 2-Q chopper and LC filter to achieve stable current output.

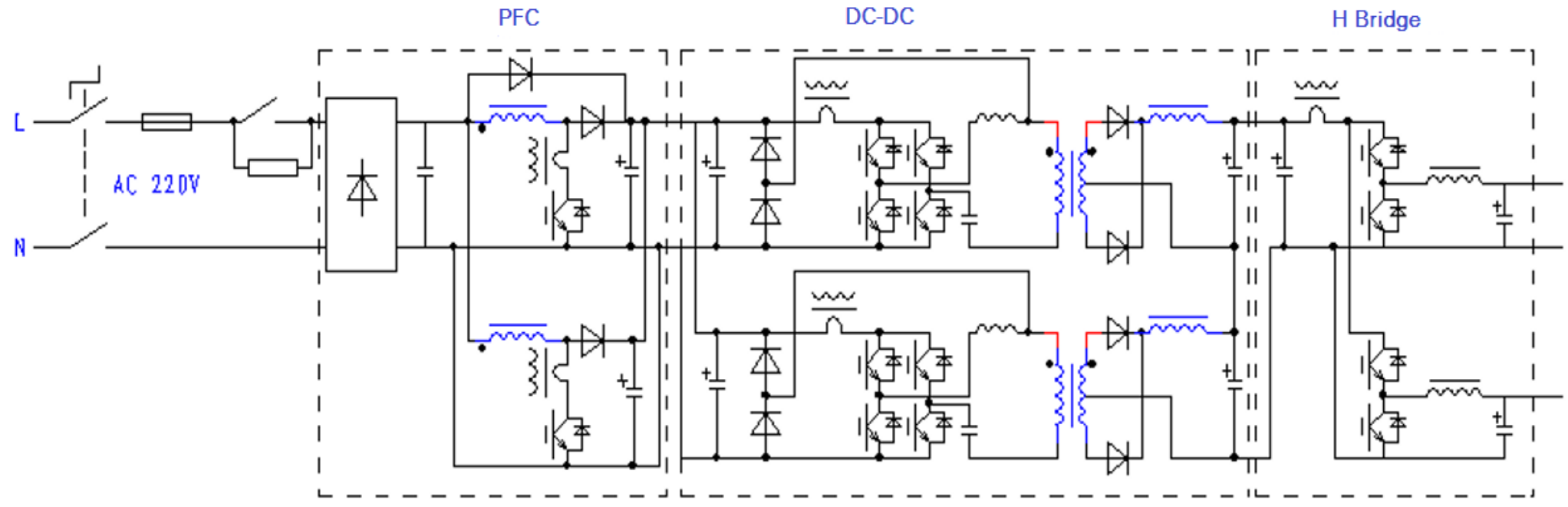

**Figure 6.4.4.3:** Structure diagram of unipolar power supply.

The corrector magnet power supply for CEPC utilizes a two-stage control topology. The first stage incorporates a DC voltage regulator to maintain a stable output DC voltage. The second stage adopts a bidirectional high-frequency H-bridge structure, serving two functions:

a) The output current can be either positive or negative, allowing for bi-directional control.
b) The control loop employs current feedback mode to ensure output current stability.

The circuit topology consists of four high-frequency power switching tubes, forming a full-bridge chopper circuit. Two diagonal bridge arm switching tubes complement each other to enable switching on and off. This topology ensures accurate zero-point output and facilitates smooth positive and negative current commutation.

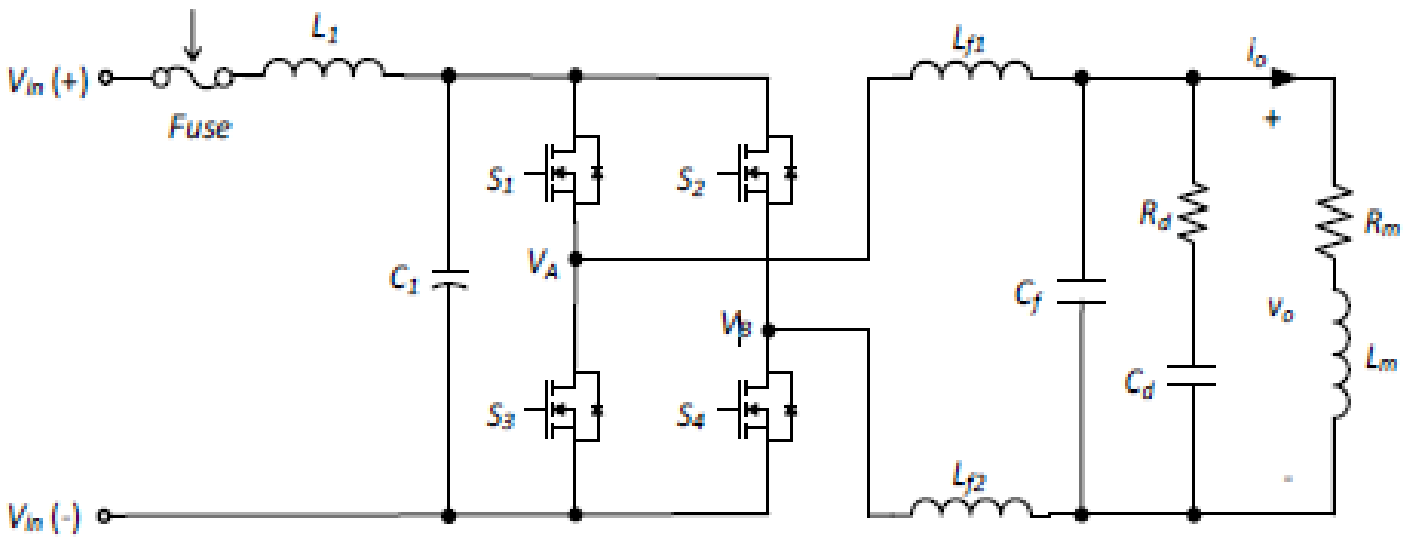

**Figure 6.4.4.4:** Structure diagram of the bipolar power supply.

## 6.4.5 **Vacuum System**

The Damping Ring (DR) is a 147-meter-long storage ring that consists of vacuum chambers, pumps, gauges, and valves. To minimize beam loss and bremsstrahlung radiation resulting from beam residual gas scattering, the DR requires an average pressure of less than $2\times10^{-8}$ Torr.

The vacuum chambers are constructed from aluminum alloys and are designed as round tubes with an inner diameter of 30 mm. To achieve high vacuum, conventional high vacuum technologies will be utilized, and small ion pumps will be distributed throughout the damping ring.

### 6.4.5.1 ***Heat Load and Gas Load***

The heat load within the DR can be assessed using Equations (4.3.6.1) and (4.3.6.2), which describe the synchrotron radiation power emitted by an electron beam engaged in uniform circular motion.

. For the DR, with values of $E$ = 1.1 GeV, $I$ = 0.012~0.024 A, and $\rho$ = 2.87 m, these equations give a total synchrotron radiation power of $P_{SR}$ = 1.08 kW and a linear power density of $P_L$ = 60.1 W/m. These values are considered negligible for the heating load, meaning that the heat generated by the synchrotron radiation is not significant enough to affect the operation of the damping ring.

The gas load in the DR comprises two components: thermal outgassing and synchrotron-radiation-induced photo-desorption. Thermal outgassing primarily contributes to the base pressure in the absence of a circulating beam, whereas photo-desorption is induced by synchrotron radiation. From the experience gained from BEPC II [1], it has been determined that the photodesorption coefficient ($\eta$) for the Damping Ring is equal to $2\times10^{-6}$. This coefficient is determined based on the similar configurations between BEPC II and the DR.

Equations (4.3.6.4) and (4.3.6.5) in Chapter 4 estimate the photo-desorption rate. The effective gas load due to photo-desorption is found to be $1.3\times10^{-6}$ Torr·L/s, and the linear synchrotron radiation gas load is $7.1\times10^{-8}$ Torr·L/s/m.

For aluminum, assuming a good bake-out and careful handling, the thermal outgassing rate is taken as $5\times10^{-12}$ Torr·L/s·cm$^2$. For a vacuum chamber with a circular cross-section and diameter of 30 mm, the linear thermal gas load can be estimated as $Q_{LT}$ = $4.7\times10^{-9}$ Torr·L/s/m. However, compared to the photo-desorption gas load, the thermal outgassing load is negligible.

### 6.4.5.2 *Vacuum Chamber and RF Shielding Bellows*

Like the Booster ring, the Damping Ring vacuum chamber is constructed using aluminum alloy 6061. This material is chosen for its high electric and thermal conductivity, ease of extrusion and welding, and lower cost compared to stainless steel or copper. The manufacturing process for the damping ring vacuum chamber is similar to that of the booster vacuum chambers, with flanges fabricated using explosion-bonded stainless steel-aluminum transition plates. Flanges and tubes are joined together through manual AC TIG welding, while fittings for water-cooling channels are also made from aluminum alloy 6061 and welded using TIG. The connection ports of the cooling water channels are machined using numerically controlled machine tools. A prototype of the aluminum alloy vacuum chamber can be seen in Figure 5.3.5.2.

Due to the low beam current in the damping ring, the secondary electron yield (SEY) of the aluminum vacuum chamber is sufficient to meet the physics requirements, eliminating the need for NEG coating or TiN coating.

To accommodate the extension and misalignment of vacuum chambers and other vacuum devices during installation, RF shielding bellows with springs and contact fingers made of beryllium copper are employed. These bellows are designed to absorb the movement and maintain RF shielding. Additionally, to reduce beam impedance during operation, these modules are equipped with RF bridges to carry the image current.

A mask is designed upstream of the RF bellows to absorb synchrotron radiation (SR) and protect the contact fingers from the heat load caused by SR, as depicted in Figure 6.4.5.1.

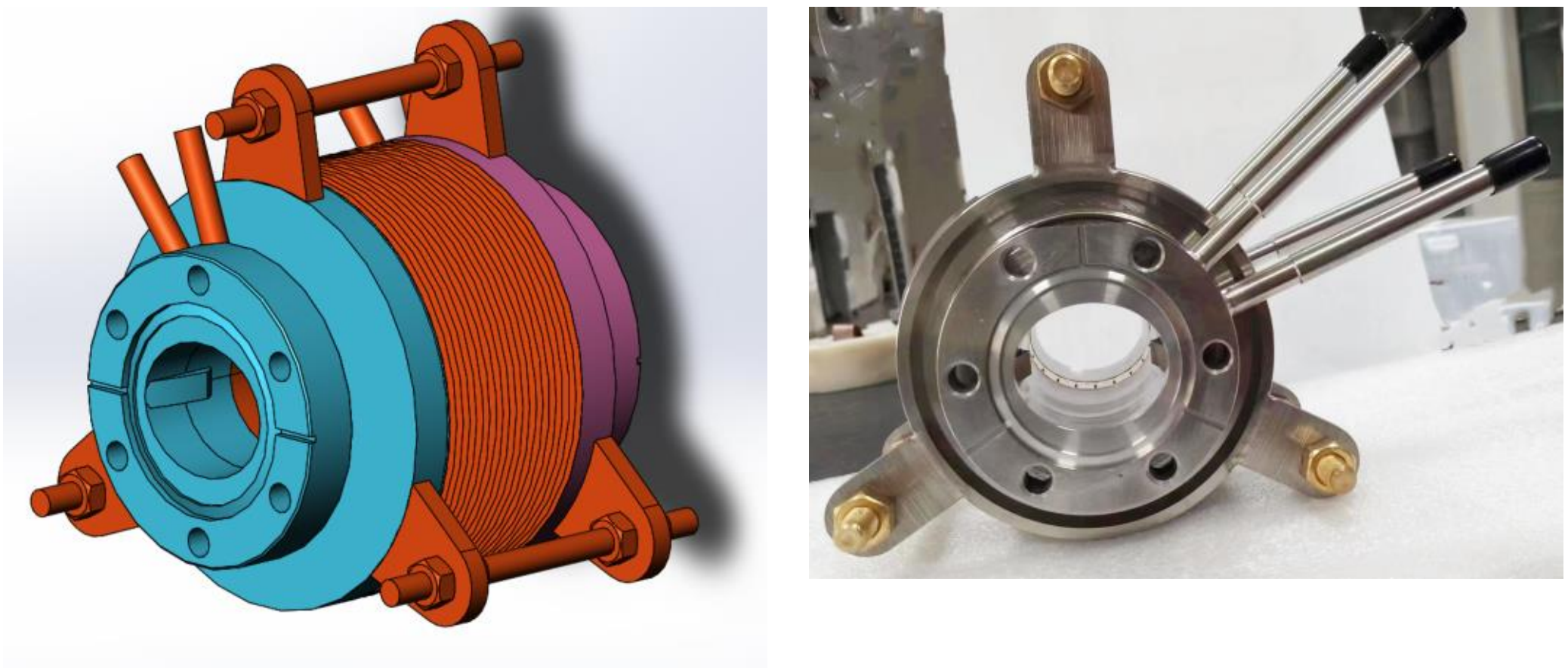

**Figure 6.4.5.1:** A 3D model of RF shielding bellows with mask (left); a prototype of RF shielding bellows (right)

### 6.4.5.3 *Vacuum Pumping and Measurement*

The circumference of the Damping Ring (DR) is divided into 6 sectors, each containing an all-metal RF shielding gate valve. These gate valves serve multiple purposes such as pumping down the sectors, conducting leak detection, and providing vacuum interlock protection. The sectors are specifically designed to isolate different regions of the DR, including the kicker for injection or extraction, arc region, and RF system.

Around the entire circumference of the DR, a total of 100 ion pumps are installed. Each ion pump has a pumping speed of 30 L/s. These pumps are connected to the vacuum chamber through Φ63 pipes that are 50 mm long, limiting the effective pumping speed to approximately 20 L/s. Additionally, RF screens are used to reduce impedance on the pump ports, further reducing the effective pumping speed of each ion pump to around 15 L/s.

The main vacuum devices of the Damping Ring are outlined in Table 6.4.5.1.

**Table 6.4.5.1:** Main devices of the Damping Ring vacuum system

| Device | Specification | Quantity |
|---|---|---|
| Arc vacuum | D30/Al 6061 | 119 m |
| Sputtering ion pump | 30-100 L/s | 100 |
| All metal gate valve | DN40/spring fingers type | 8 |
| RF shielding bellow | DN40 | 75 |
| Vacuum Gauge | CCG | 25 |
| RGA | Mass 100 | 8 |

With an effective pumping speed of 15 L/s and a distribution of 2 meters of sputtering ion pumps, the vacuum value of $1.8\times10^{-8}$ Torr will be reached.

To estimate the thermal outgassing rate of the cavity, a conservative value of $5\times10^{-11}$ Torr·L/s·cm$^2$ is adopted, due to the difficulty of performing an in-situ bake-out of the cavity. In anticipation of a large gas load, two ion pumps with a pumping speed of 400 L/s each are deployed in the RF sectors. However, due to orifice conductance limitations, the effective pumping speed is limited to around 360 L/s.

Given the stronger photo-desorption in the DR, we assume that the residual gas composition is closer to that of the Booster, with a mixture of 75% hydrogen ($H_2$) and 25% carbon monoxide (CO). With this assumption, a conductance is 3.4 times higher than that of air.

For the vacuum design of the transport lines, conventional vacuum technologies will be used. The vacuum chambers will be connected using bellows with conflat-type flanges.

### 6.4.6 Instrumentation

#### 6.4.6.1 *Overview*

The Damping Ring (DR) has a circumference of 147m and operates at 1.1 GeV. A beam instrumentation system is required to diagnose beam properties. Three important beam parameters, namely beam position, beam current, and beam tune factor, need to be measured. This is achieved using specific devices: beam position monitors (BPMs) for position measurement, DC current transformers for beam current monitoring, and the Direct Diode Detector method for betatron tune measurement.

**Table 6.4.6.1:** Parameters of the Damping Ring beam instrumentation systems

| Item | Method | Parameter | Amounts |
|---|---|---|---|
| Average current | DCCT | Resolution: 50μA @ 0.1mA-30mA | 1 |
| Beam position | Button BPM | Resolution: 20μm @ 5mA TBT | 40 |
| Tune | Direct Diode Detection method | Resolution: 0.001 | 1 |

The Damping Ring beam instrumentation system consists of 40 beam position monitors (BPMs), 1 DC current transformer for beam current measurement, and 1 tune measurement device. The design of the beam current measurement and tune monitor is similar to that of the Collider, as described in Section 4.3.7.

### 6.4.6.2 *Beam Position Monitor*

The Beam Position Monitor (BPM) is a crucial system in modern accelerators. It serves the purpose of measuring beam position, monitoring beam orbit, and calculating other important physical parameters. The BPM system consists of pick-ups, cables, and signal processing electronics. The pick-ups are of the button type and have the same diameter as those used in the Collider. However, due to the different beam pipe size, there are slight differences in the characteristics of the BPM pick-ups compared to those in the Collider. The transfer impedance and signal power for a current of 100 mA can be found in Table 6.4.6.2. The electronics of the BPM are designed to be identical to those used in the Collider.

**Table 6.4.6.2:** Parameters of the Damping Ring BPM

| | Transfer impedance | Signal power (I=100 mA) | Sensitivity (in the pipe center) |
|---|---|---|---|
| Collider (b = 28 mm) | 0.093 Ω | –30.6 dBm | 19.82 mm |
| Booster (b = 28 mm) | 0.093 Ω | –30.6 dBm | 19.82 mm |
| Damping ring (b = 15 mm) | 0.169 Ω | –25.3 dBm | 10.68 mm |

The digital BPM electronics consist of two main components: Analog Frontend Electronics (AFE) and Digital Frontend Electronics (DFE) as shown in Fig. 6.4.6.1. The AFE is responsible for tasks such as amplitude adjustment, frequency filtering, clock reception and processing, and analog-to-digital conversion (ADC) sampling of the analog signals. On the other hand, the DFE handles the data processing and data transmission functions, including the reception of data from the AFE, execution of the digital BPM algorithm, and implementation of the algorithm logic in the FPGA of the DFE circuit.

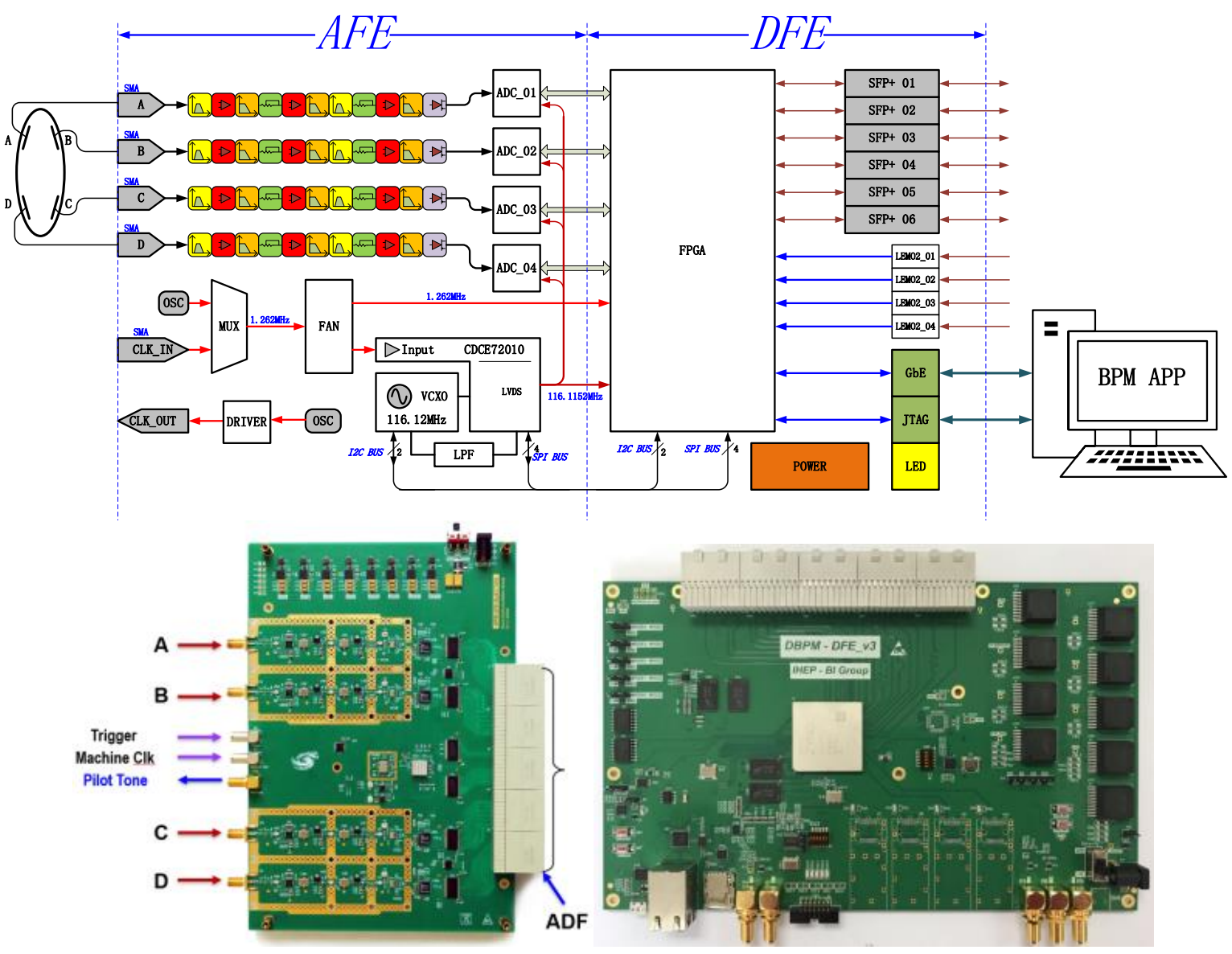

**Fig. 6.4.6.1:** Top: Basic hardware structure and function block diagram of digital BPM electronics. Bottom (left): AFE physical diagram; bottom (right): DFE physical diagram.

The Accelerator Center of the IHEP established a dedicated team to develop digital BPM electronics. Through the collaborative efforts of numerous scientific researchers and postgraduates over the course of eight years, significant progress has been made. The self-developed digital BPM electronics underwent extensive testing in laboratory environments, achieving a remarkable resolution of 10 nanometers. Furthermore, it was successfully deployed in practical beam flow scenarios, demonstrating performance indicators that either met or exceeded those of comparable foreign commercial products.

In 2019, the self-developed digital BPM electronics made its debut in the BEPC II linac accelerator, marking a significant milestone. Subsequently, in 2020, it was successfully implemented in the BEPC II storage ring. Recognizing its exceptional performance, it was decided that the self-developed digital BPM electronics would be utilized in the ongoing construction of the HEPS project.

The R&D team dedicated efforts to address various challenges and enhance the performance of digital BPM electronics. One significant achievement was mitigating the impact of ambient temperature changes on signal processing by implementing thermostat cabinets and crosslink switches. This approach ensured consistent signal processing across all four signaling channels, reducing measurement errors caused by temperature variations.

To address the issue of flow intensity dependence in BPM measurement, the team employed two techniques: "full channel calibration" and "channel automatic baseline removal." These methods helped overcome the intensity-dependent factors, enabling more accurate and reliable BPM measurements.

Clock jitter in ADC sampling was another aspect that required improvement. By utilizing dual-lock-loop clock de-jitter technology, the team achieved enhanced performance in clock jitter, leading to improved signal-to-noise ratio in the ADC sampling data. This advancement contributed to high-precision BPM measurements.

To compensate for measurement errors caused by various factors such as temperature, isolation, and device variations, the team researched and developed "real-time channel calibration" technology. This innovation effectively addressed the issue of slow drift in digital BPM measurement and significantly improved the long-term stability of BPM measurements.

### 6.4.6.3 *Beam Current Monitor*

In the Damping Ring, Bergoz type DC current sensors will be utilized to measure the average beam current. Specifically, the In-flange NPCT sensor from Bergoz Instrumentation has been selected for this purpose. The sensor is chosen for its excellent characteristics, including low linearity error and high resolution, ensuring accurate and reliable beam current measurements.

Figure 6.4.6.2 illustrates the sensor and its accompanying front-end electronics, providing a visual representation of the setup. Considering the beam current in the Damping Ring is relatively low at 24.8 mA, the full-scale range of 200 mA NPCT sensors will be chosen for optimal measurement performance. These sensors exhibit a linearity error smaller than 0.1%, ensuring high precision in capturing the beam current information.

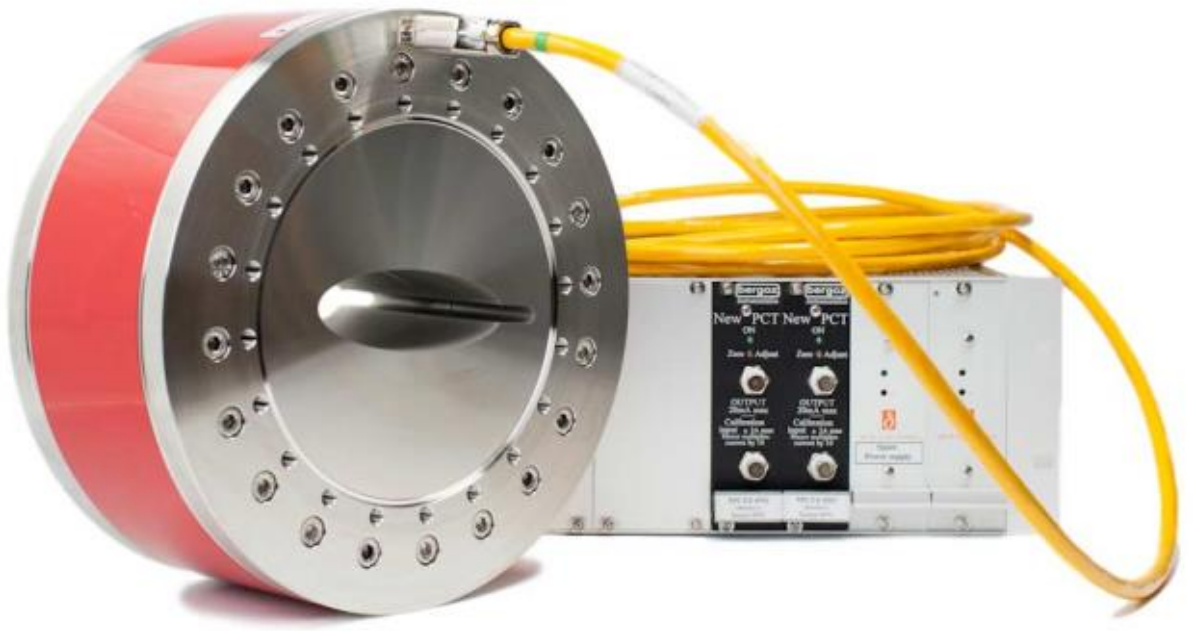

**Figure 6.4.6.2:** The Bergoz in-flange NPCT senor and front-end electronics

For A/D conversion and high-precision measurements in the Damping Ring, the Keithley DMM7500 7½-Digit multimeter has been chosen. This multimeter is renowned for its exceptional performance and accuracy. It provides a resolution of 10 nV for DC voltage measurements. Its noise level is carefully controlled, allowing for accurate measurements even in challenging environments. The one-year stability of the multimeter is as low as 14 ppm (parts per million), indicating its ability to maintain consistent and reliable measurements over extended periods of time.

### 6.4.6.4 *Tune Measurement*

Measuring and optimizing the betatron tune in a circular accelerator is essential for maintaining beam quality and ensuring operational efficiency. The control room relies on accurate tune measurements to make informed decisions. Two common methods for tune measurement are the kick or excitation system-based method and the direct diode detection method.

In the case of the Damping Ring layout, where there is no kicker or excitation system available, the direct diode detection method is the most suitable choice for tune measurement. A technology called Direct Diode Detection (3D), developed at CERN for LHC tune measurement, will be considered for implementation.

The 3D method employs the concept of time-stretching the beam pulse from the pickup, which enhances the betatron frequency content in the baseband. This technique utilizes a simple diode detector followed by an RC low-pass filter to achieve the desired effect. The 3D method has been successfully tested on the BEPC II, a lepton machine, and its effectiveness has been demonstrated. Test results can be seen in Figure 6.4.6.3.

The direct diode detection method offers several advantages, including its simplicity and low cost, robustness against saturation, ability to flatten out the beam dynamic range, and independence from filling patterns. However, it is important to note that the method operates in the low-frequency range, which makes it susceptible to noise interference.

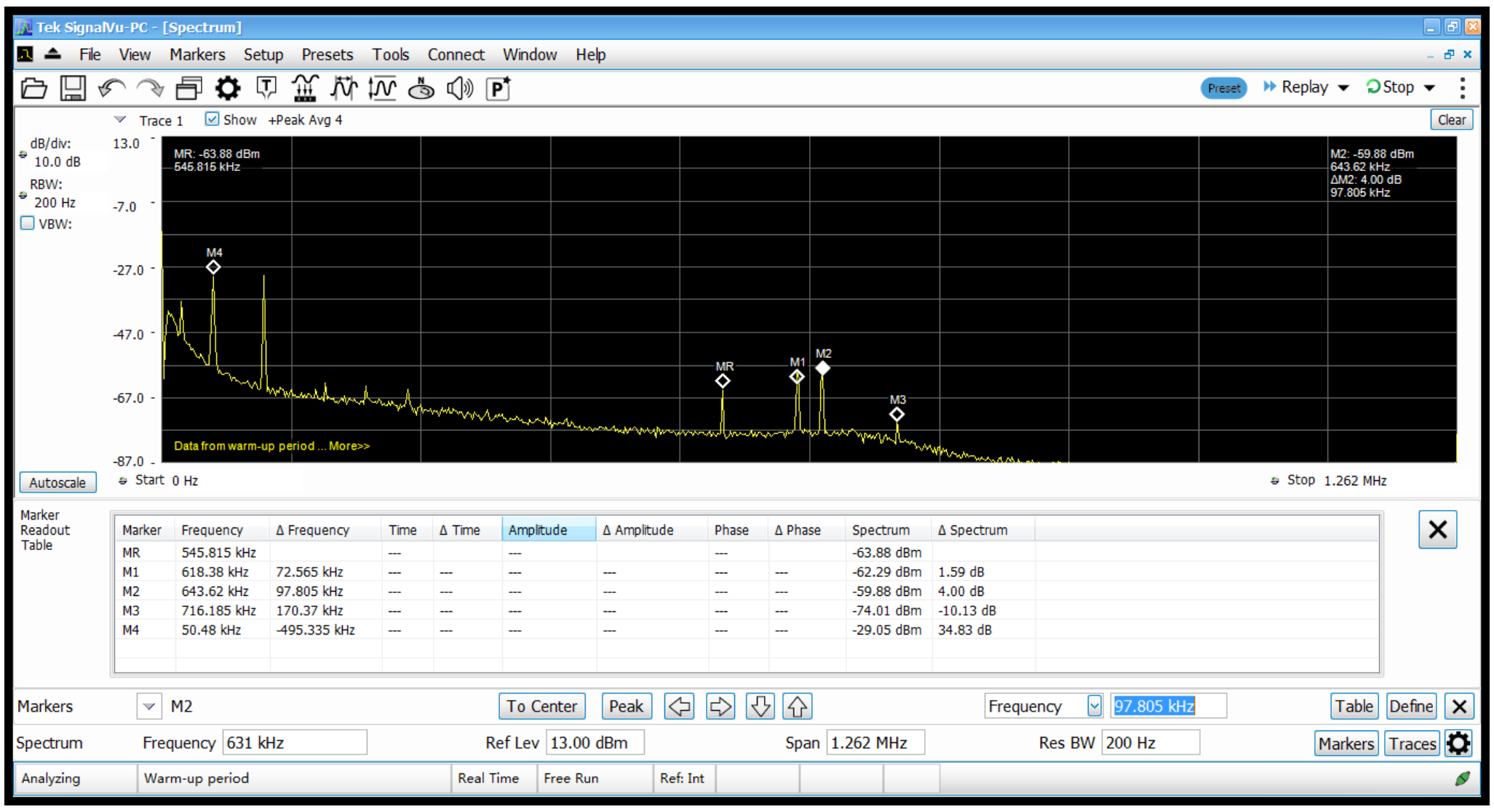

**Figure 6.4.6.3:** Beam test result of the BEPCII 3D tune measurement system

## 6.4.7 Injection and Extraction System

The Damping Ring (DR) is a circular accelerator with a circumference of 147 m and an energy of 1.1 GeV. Its primary function is to decrease the emittance of the positron beam in the Linac. The DR is designed to store 2 or 4 bunches simultaneously by equal interval, with a minimum bunch spacing of 122.5 ns. To achieve this, on-axis bunch-by-bunch injection and extraction techniques are utilized.

The DR's injection and extraction systems consist of a vertically deflecting slotted-pipe kicker magnet and a horizontally deflecting Lambertson magnet, respectively. The layout of these systems is shown in Figure 6.4.7.1, and the relevant physical parameters are detailed in Table 6.4.7.1.

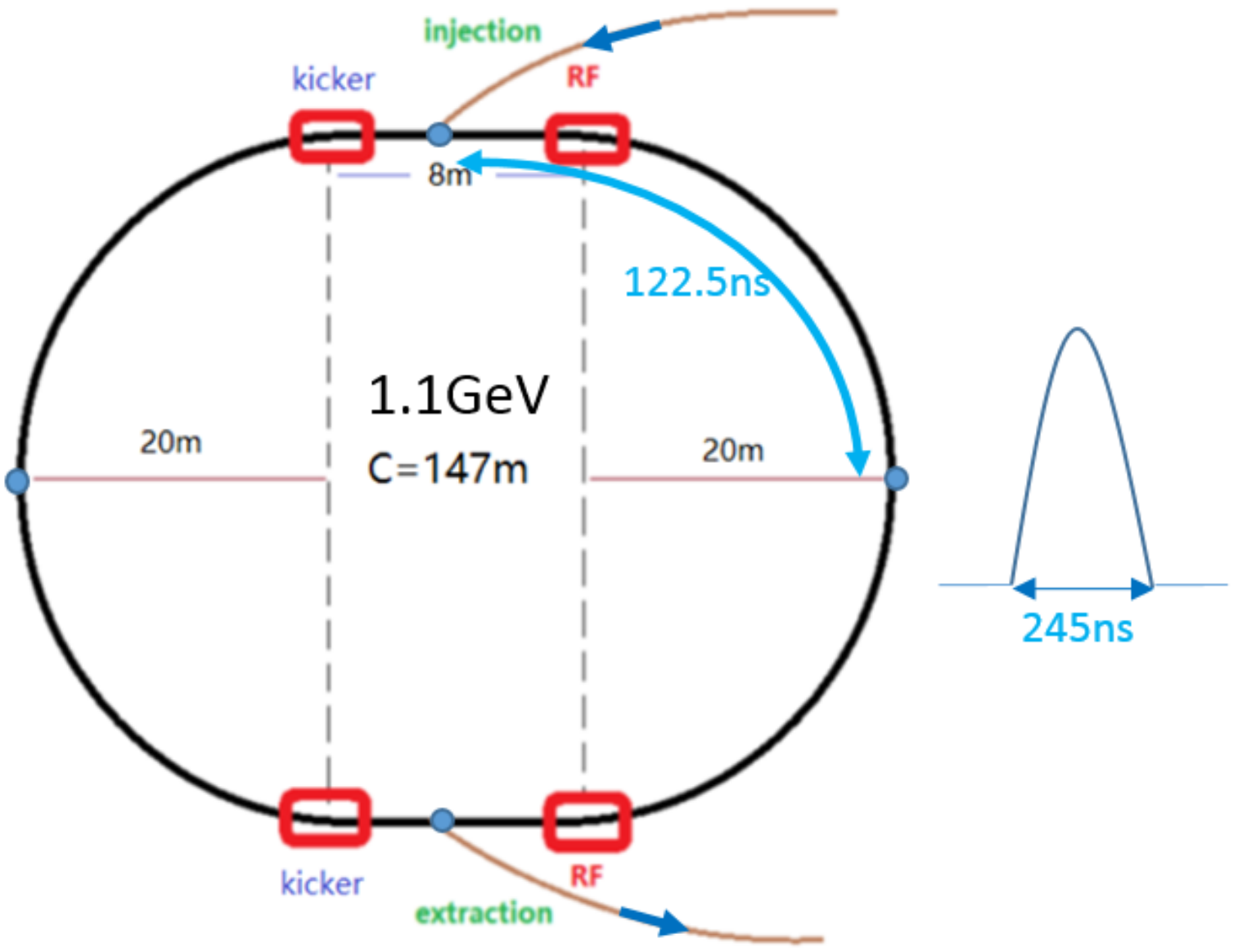

**Figure 6.4.7.1:** Layout of the Damping Ring injection and extraction systems.

**Table 6.4.7.1:** Parameters of the Damping Ring injection and extraction systems.

| **Parameters** | **Injection** | **Extraction** |
|---|---|---|
| Energy (GeV) | 1.1 | 1.1 |
| Bunch number | 2/4 | 2/4 |
| Min. bunch spacing (ns) | 122.5 | 122.5 |
| Injection /extraction mode | Bunch by bunch | Bunch by bunch |
| Kicker repetition rate (Hz) | 100 | 100 |
| Kicker pulse width (ns) | < 245 | < 245 |
| Kicker rise/fall time (ns) | < 122.5 | < 122.5 |
| Timing delay (ns) | < 122.5 | < 122.5 |
| Kicker deflection direction | Vertical | Vertical |
| Kicker deflection angle (mrad) | 10.7 | 10.7 |
| Kick integral field strength (T-m) | 0.0392 | 0.0392 |
| Septa deflection direction | Horizontal | Horizontal |
| Septa deflection angle (mrad) | 120 | 120 |
| integral field strength of septa (T-m) | 0.44 | 0.44 |
| Septa board thickness (mm) | 3.5 | 3.5 |

### 6.4.7.1 *Slotted-pipe Kicker System for the Damping Ring*

The slotted-pipe kicker magnet was chosen for the DR due to its relatively simple structure. Figure 6.4.7.2 depicts the 2D simulation model of the kicker magnet established by OPEAR 2D. The simulation results indicate that the field uniformity in the good field region of 19.8 mm × 16 mm ranges from –0.9% to 1.5%, with the field distribution shown in Figure 6.4.7.3.

The physical design parameters of the magnets are detailed in Table 6.4.7.2.

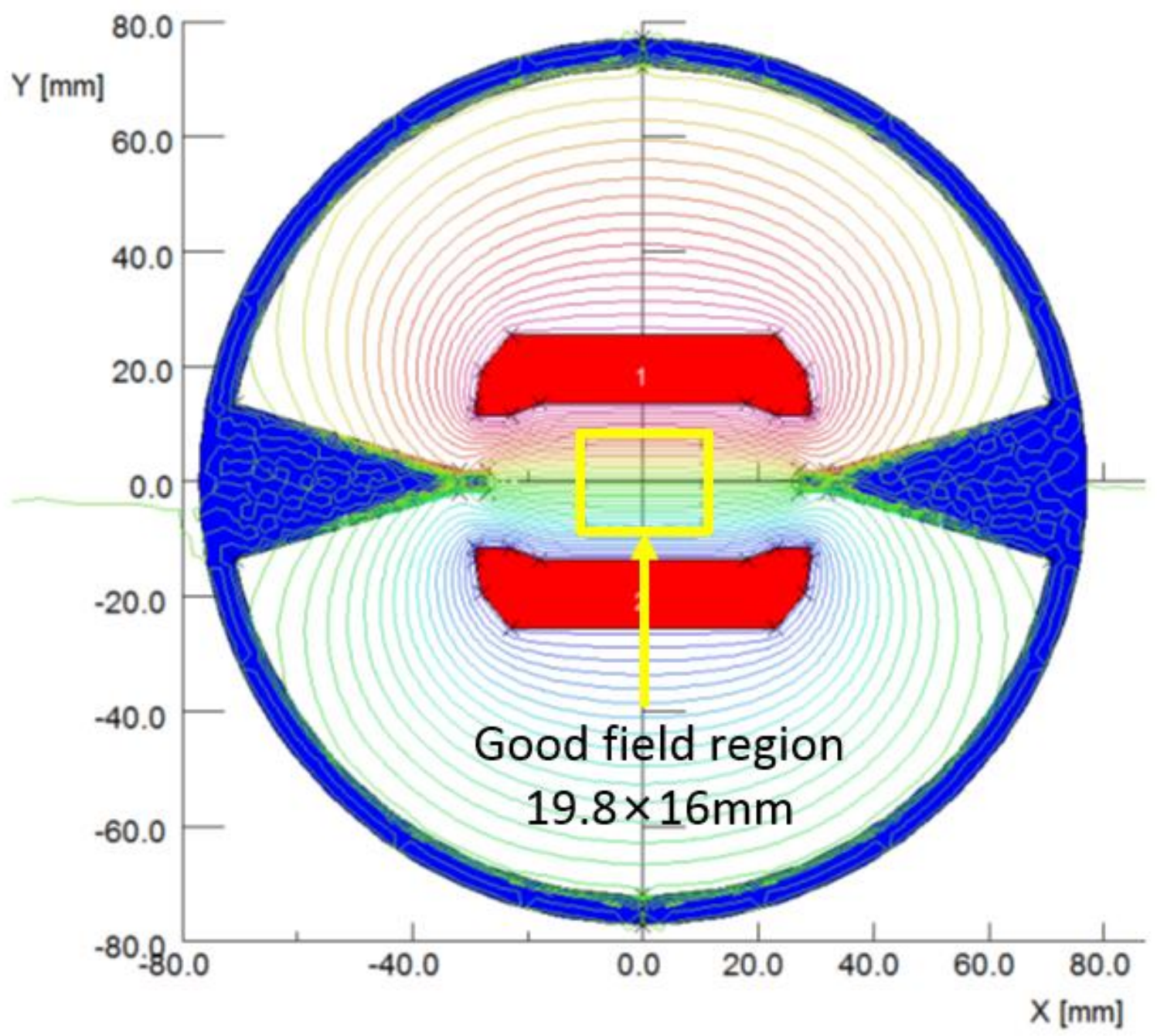

**Figure 6.4.7.2:** 2D simulation model of the slotted-pipe kicker.

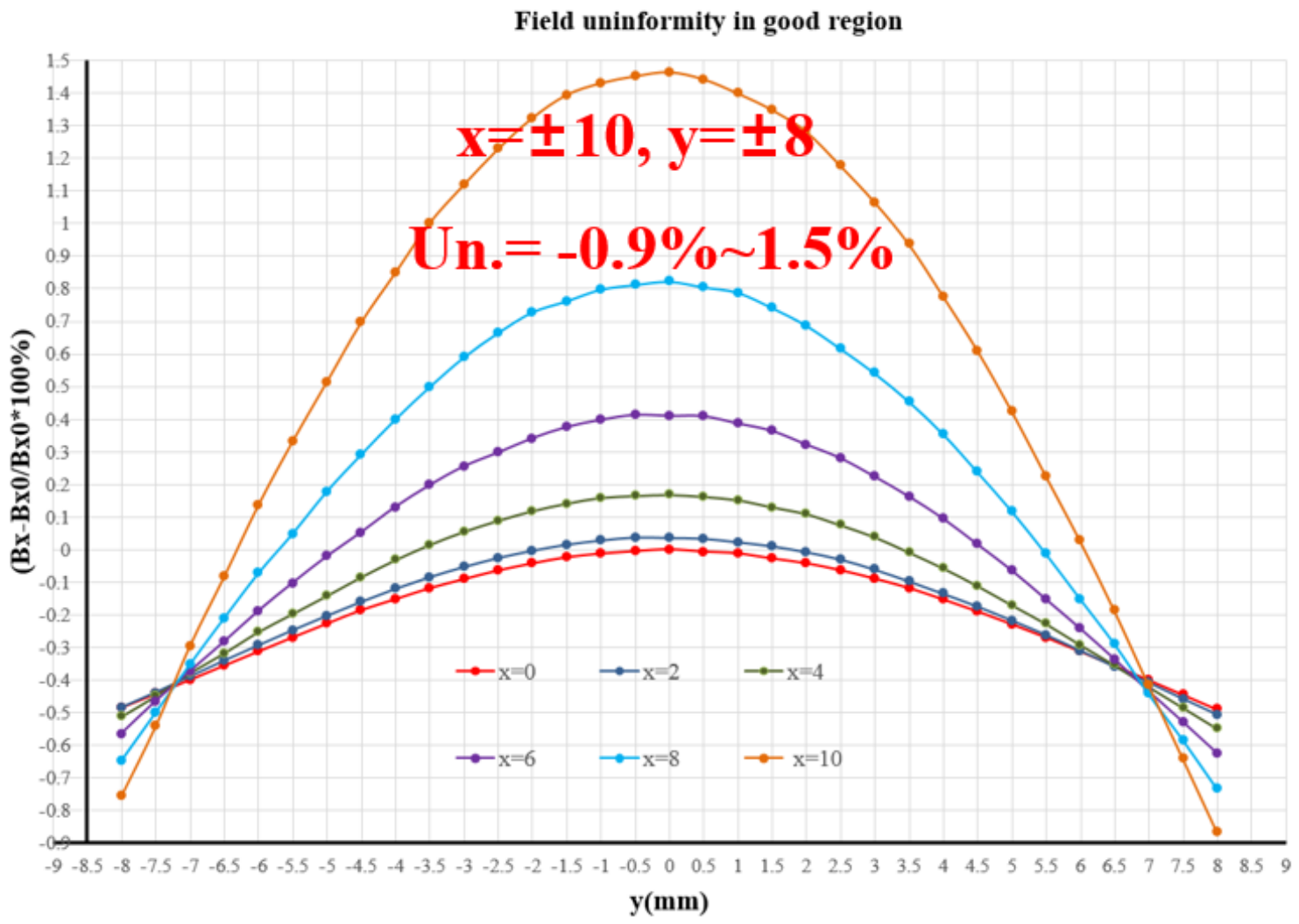

**Figure 6.4.7.3:** Field uniformity of the slotted-pipe kicker.

**Table 6.4.7.2:** Parameters of the DR kicker magnet.

| **Parameters** | **Unit** | **Kicker** |
|---|---|---|
| Quantity | - | 2 |
| Type | - | Slotted-pipe kicker |
| Deflect direction | - | Vertical |
| Beam Energy | GeV | 1.1 |
| Deflect angle | mrad | 10.7 |
| Magnetic effective length | m | 1.4 |
| Magnetic strength | T | 0.0281 |
| Integral magnetic strength | T·m | 0.03934 |
| Clearance region (H×V) | mm | 32.8 × 26.6 |
| Good field region (H×V) | mm | 19.8 × 16 |
| Field uniformity in good field region | - | ±1.5% |
| Magnet coil inductance | nH | 387 |
| Max. voltage of coil | kV | 10.622 |
| Max. exciting current | kA | 2.4 |
| Repetition rate | Hz | 100 |
| Amplitude repeatability | - | ±0.5% |
| Pulse jitter | ns | ≤ 5 |
| Bottom width of pulse (5%-5%) | ns | < 245 |

Figure 6.4.7.4 displays the mechanical design of the kicker magnet for the DR. The outer vacuum chamber is fabricated from 316L stainless steel, which can be segmented using EDM cutting and TIG welding. The electrodes are composed of oxygen-free copper plates and machined with CNC. On the assembly structure, one end of the two electrodes is directly connected to the external vacuum chamber, while the other end is connected to the high-voltage feedthrough, with $AL_2O_3$ ceramic insulation support in the middle. The midpoint of the coil, where the electrodes are connected to the external vacuum chamber, is naturally grounded. Therefore, the magnet must be excited using a bipolar power supply.

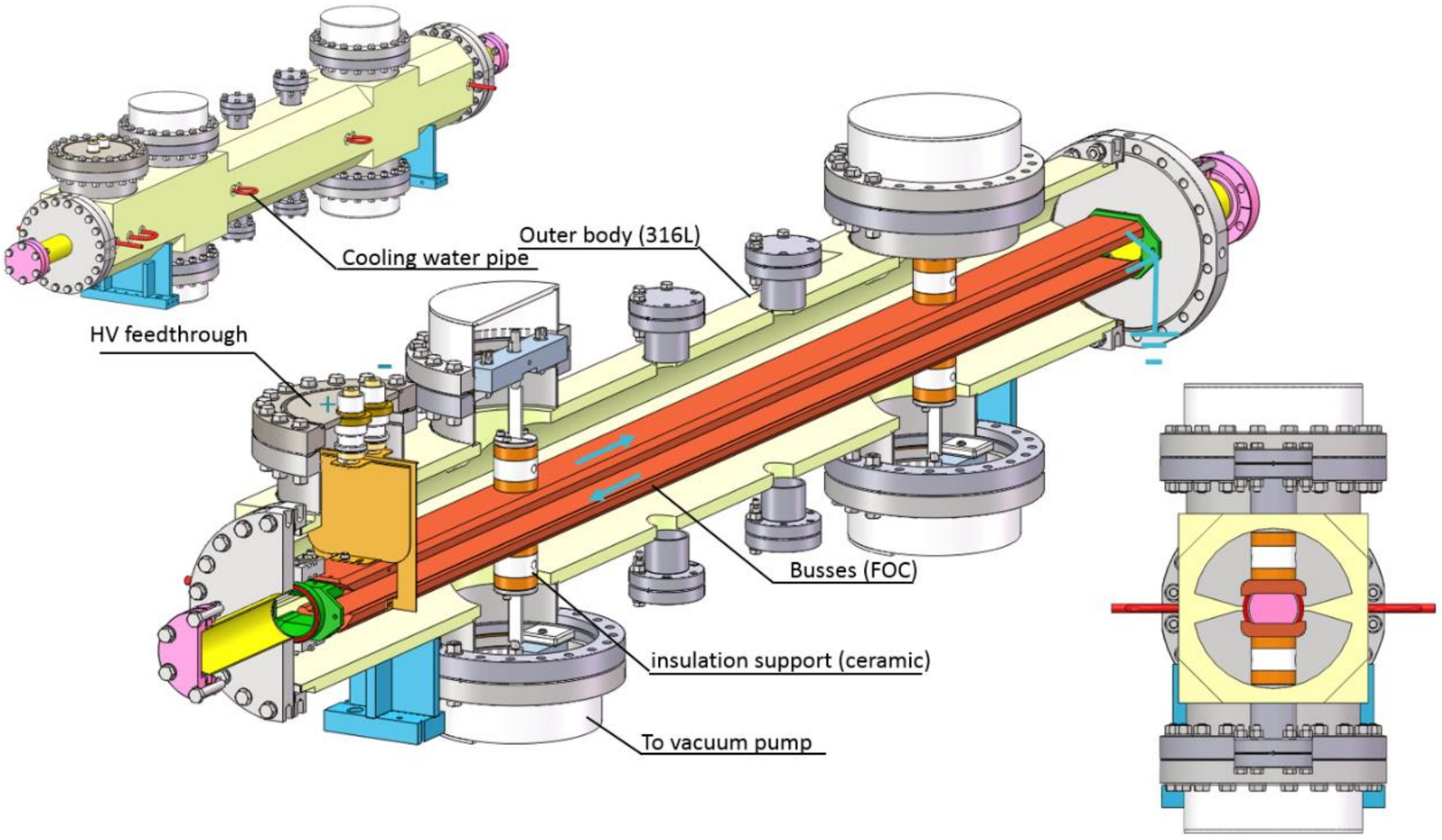

**Figure 6.4.7.4:** Engineer design of the DR slotted-pipe kicker magnet.

The kicker design for the Damping Ring is similar to that of the HEPS Booster. As depicted in Figure 6.4.7.5, the kicker prototype for the HEPS Booster has been successfully developed, and all design requirements, including vacuum, have been met. Figure 6.4.7.6 shows the performance measurement result of the field distribution, which is in good agreement with the Opera simulation results.

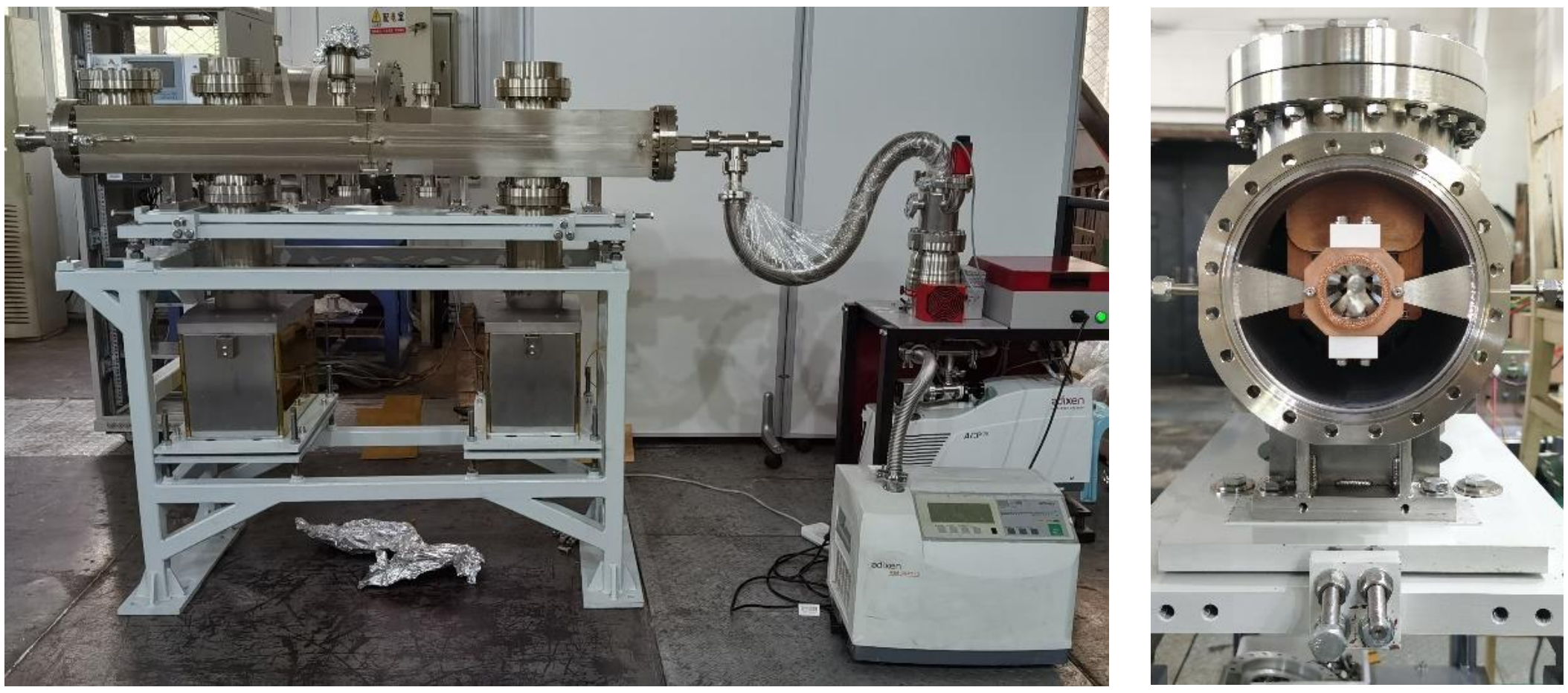

**Figure 6.4.7.5:** Prototype of the HEPS Booster injection kicker.

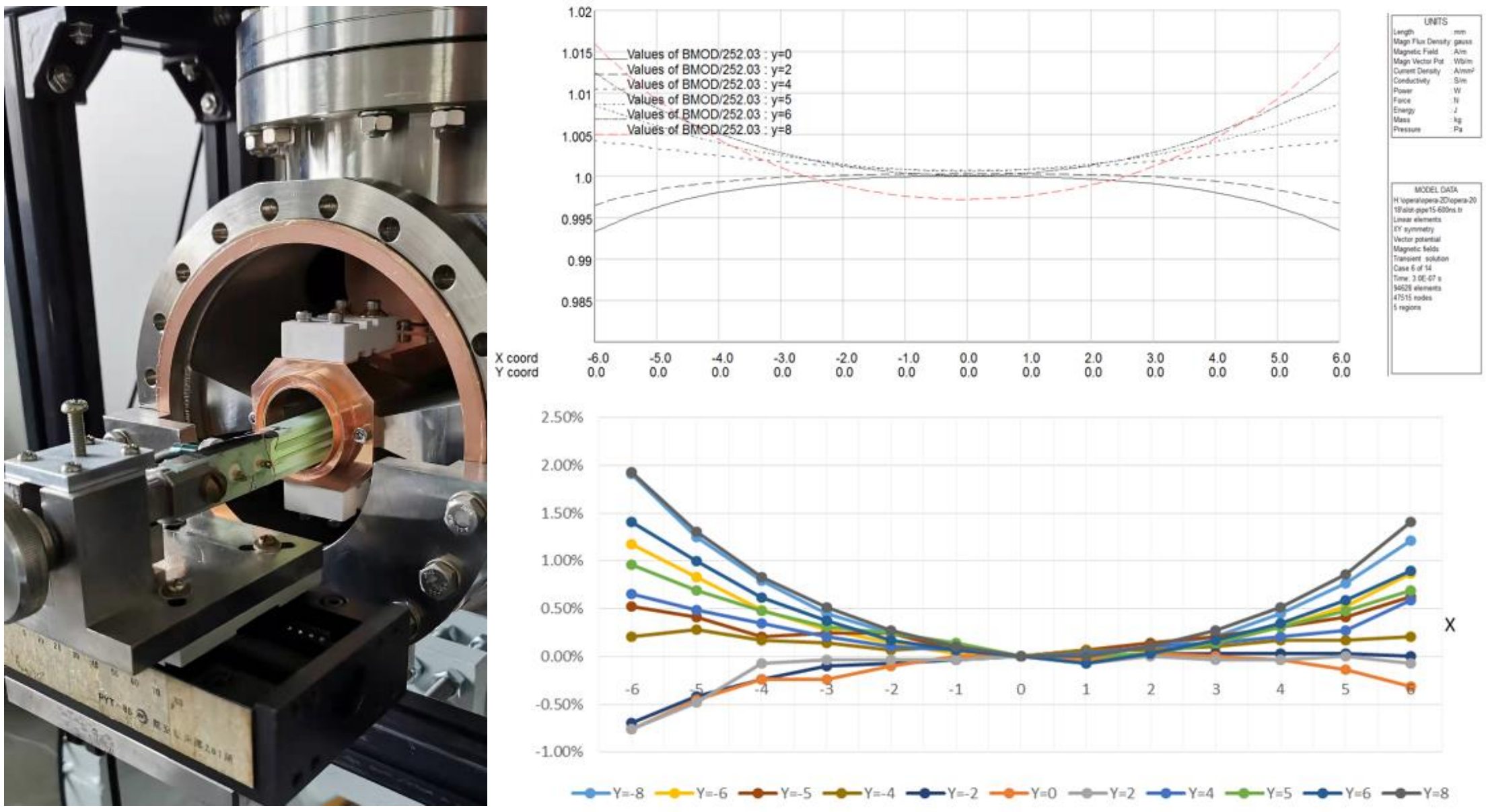

**Figure 6.4.7.6:** Field distribution measurement of the HEPS Booster kicker prototype.

### 6.4.7.2 *Pulsed Power Supply for the Kicker Magnet*

The SiC MOSFET-based inductive adder is a potential solution for the pulsed power supply of the slotted-pipe kicker. The topology of the inductive adder with 20 stages is shown in Figure 6.4.7.7. The co-axial transformer is configured as bipolar output. The pulser is located outside the tunnel, and 10 cables with a length of 30 m are utilized to connect to the kicker. Since the slotted-pipe kicker has a short end, the reflection energy must be returned to the pulser, necessitating the design of a proper absorption circuit at the pulser end. For the design and development of the inductive adder, refer to Chapter 4, Section 4.3.8.6.

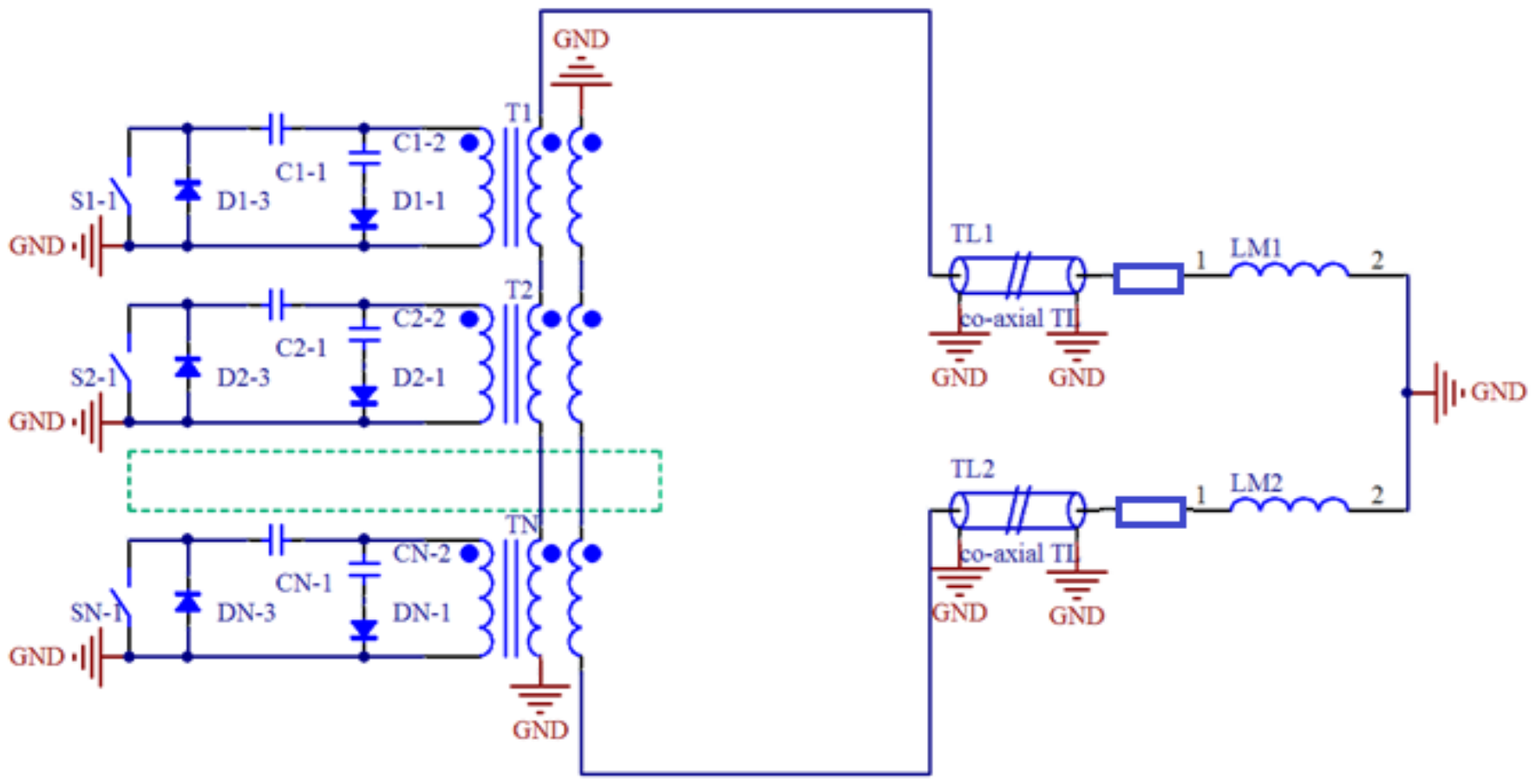

**Figure 6.4.7.7:** Inductive adder based on SiC-MOSFET.

### 6.4.7.3 *Lambertson Magnet for the Damping Ring*

The Lambertson magnet of the DR is a horizontally deflecting magnet with a septum plate thickness of 3.5 mm. The main design parameters can be found in Table 6.4.7.3. The injected beam location coordinates at the inlet and outlet of the magnet are shown in Figure 6.4.7.8.

**Table 6.4.7.3:** Parameters of the DR Lambertson magnet.

| Parameters | Unit | DR-LSM |
|---|---|---|
| Quantity | - | 2 |
| Energy | GeV | 1.1 |
| Deflection direction | - | Horizontal |
| Deflection angle | mrad | 120 |
| Insertion length | m | 0.5 |
| Magnetic field strength for injected/extracted beam | T | 0.883 |
| Min. Septum thickness （including septum board, inj./ext. beam pipe wall, installation gap) | mm | 3.5 |
| Field uniformity | - | < ±0.05% |
| Leakage field | - | $\leq 1\times10^{-3}$ |
| Clearance of stored beam at lambertson (H×V) (w.r.t. the stored beam orbit) | mm | 30× 22 |
| Clearance of inj.&ext. beam at lambertson (H×V) (w.r.t. the inj./ ext. beam orbit) | mm | 22×11 |
| Physical aperture of stored beam vacuum chamber | mm | 30×30 |
| Physical aperture of inj./ext. beam vacuum chanmber | mm | 41×19 |

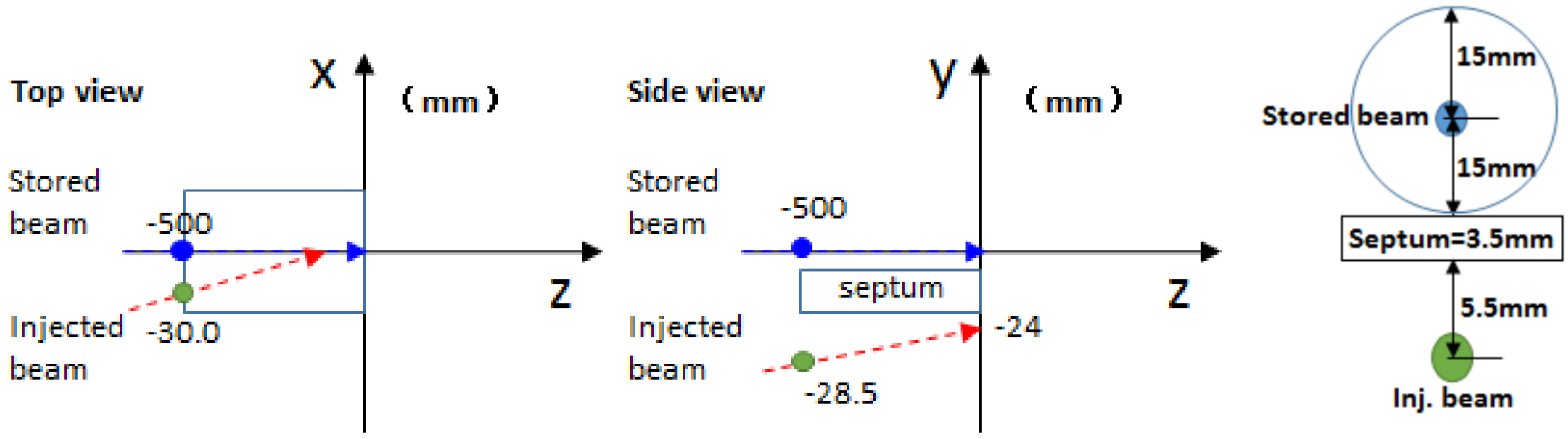

**Figure 6.4.7.8:** Injected beam location coordinates at the inlet and outlet of the magnet.

The physical design and mechanical structure of the Lambertson magnet for the Damping Ring (DR) closely resemble those of the Lambertson magnet used in the Collider, with the main difference being the thickness of the septum plate. In the DR, a 3.5 mm septum plate is employed, while the Collider uses a 6 mm septum plate. The embedded thin-wall vacuum chamber scheme, described in Chapter 4, Section 4.3.8.9, is utilized for the Lambertson magnet in the DR.

Figure 6.4.7.9 displays the 2D profile of the Lambertson magnet. It consists of a circular stainless steel (316L) stored beam pipe with an inner diameter of 30 mm and a wall thickness of 0.6 mm, which is embedded within the magnet's yoke. The yoke is made from a combination of FeCoV alloy (1J22) and pure iron (DT4C). An oval bowed stainless steel beam pipe with an inner aperture of 41×19 mm and a wall thickness of 0.6 mm is included for the injected beam. The total septum thickness, including the 2 mm FeCoV septum wall, is less than 3.5 mm. The magnet's gap is designed to be 22 mm, with a winding size of 70×105 mm and a turn number of 96.

Figure 6.4.7.10 showcases the 3D model of the Lambertson magnet. Its dimensions are 700×344×268 mm, and it possesses an inductance of 0.0228 H. The magnet operates at an exciting current of 166 A.

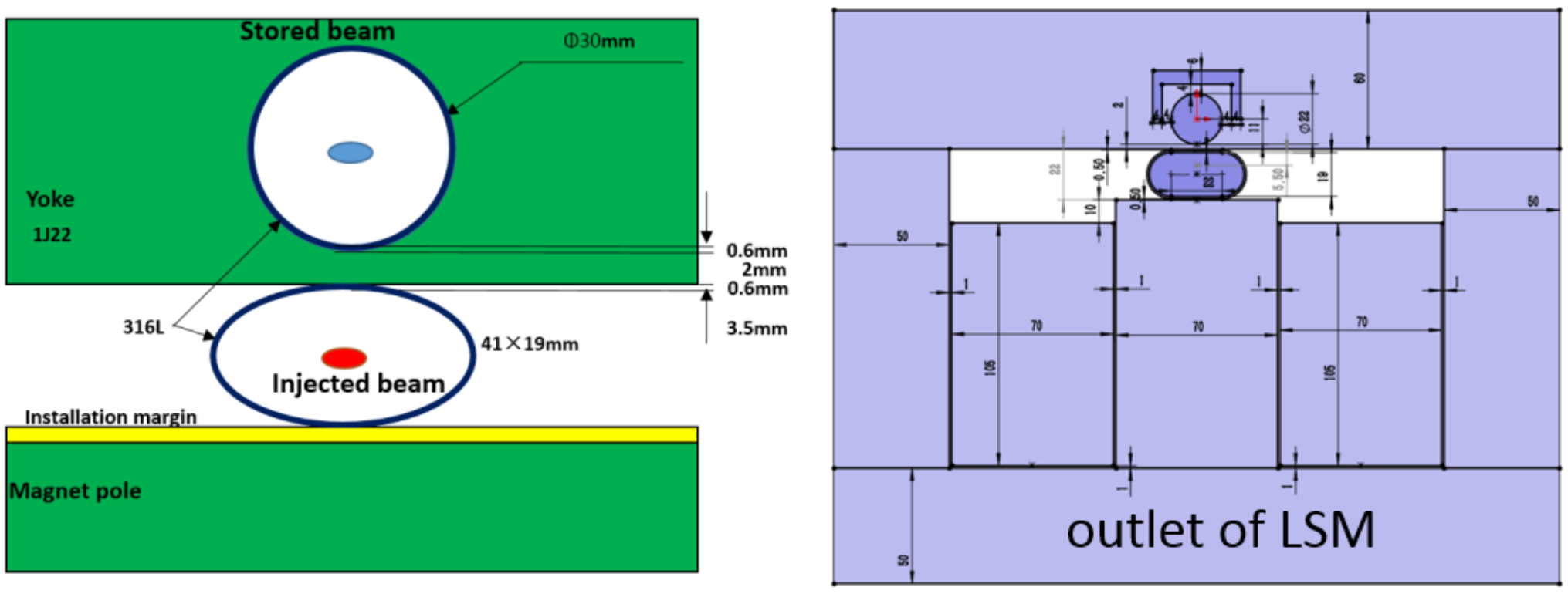

**Figure 6.4.7.9:** 2D profile of the Lambertson magnet for CEPC DR.

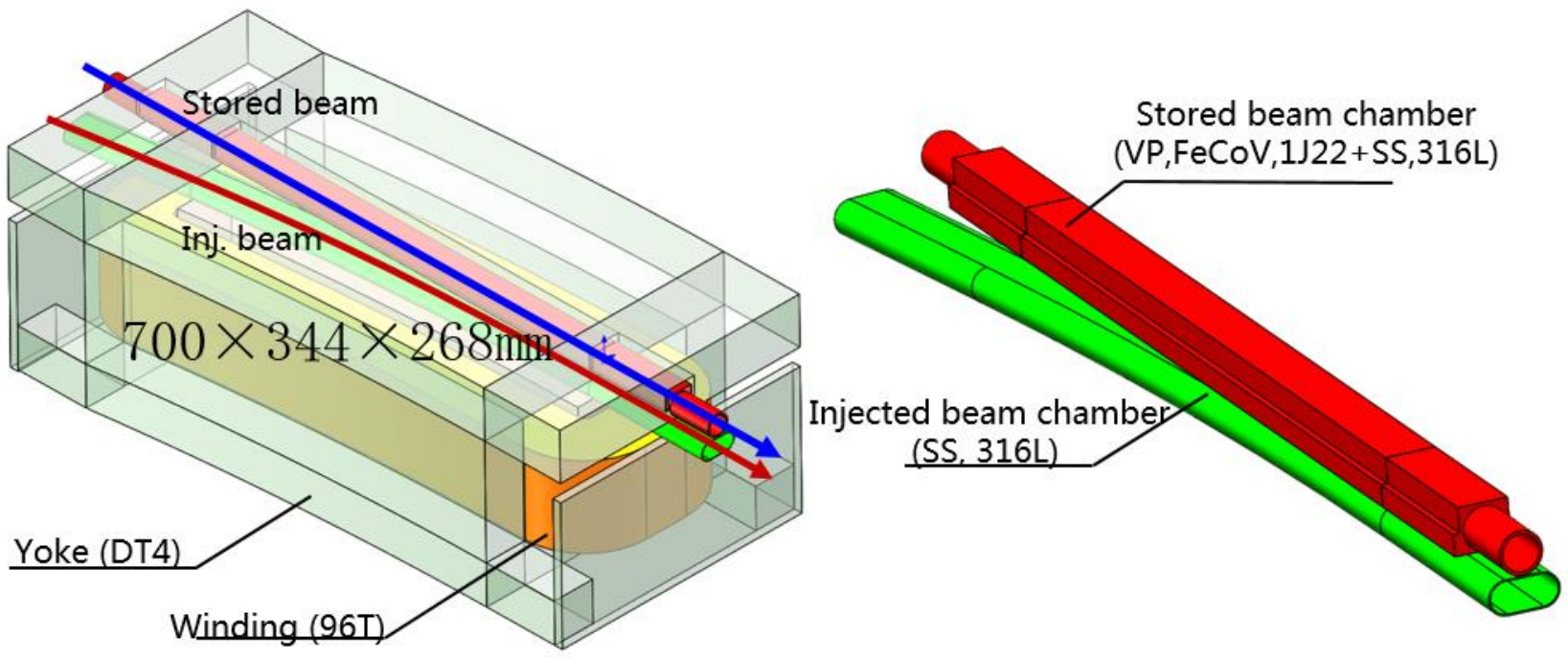

**Figure 6.4.7.10:** 3D model of the Lambertson magnet for CEPC DR.

Figure 6.4.7.11 displays the simulation result of the main field using OPERA-3D. The center magnetic field intensity of the Lambertson magnet is measured to be 8836 Gauss, and the effective length of the magnet is 542 mm.

The simulation result of the leakage field is presented in Figure 6.4.7.12. The vertical leakage is determined to be better than 2.81E-04, while the horizontal leakage is better than 1.79E-06.

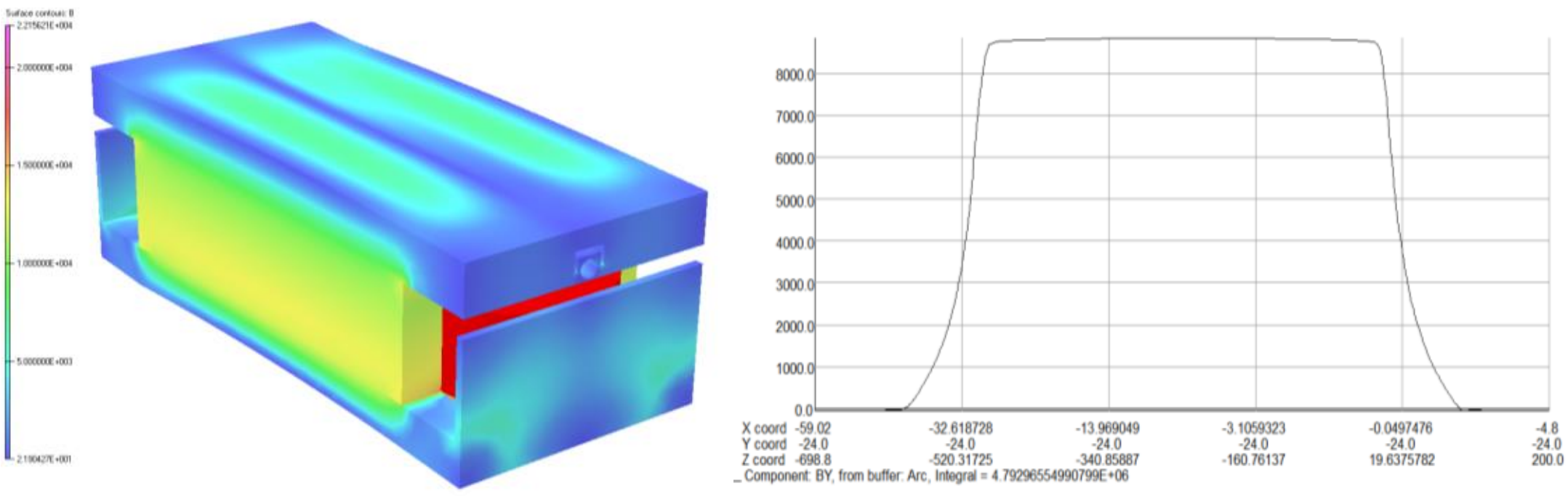

**Figure 6.4.7.11:** Simulation result of the main field.

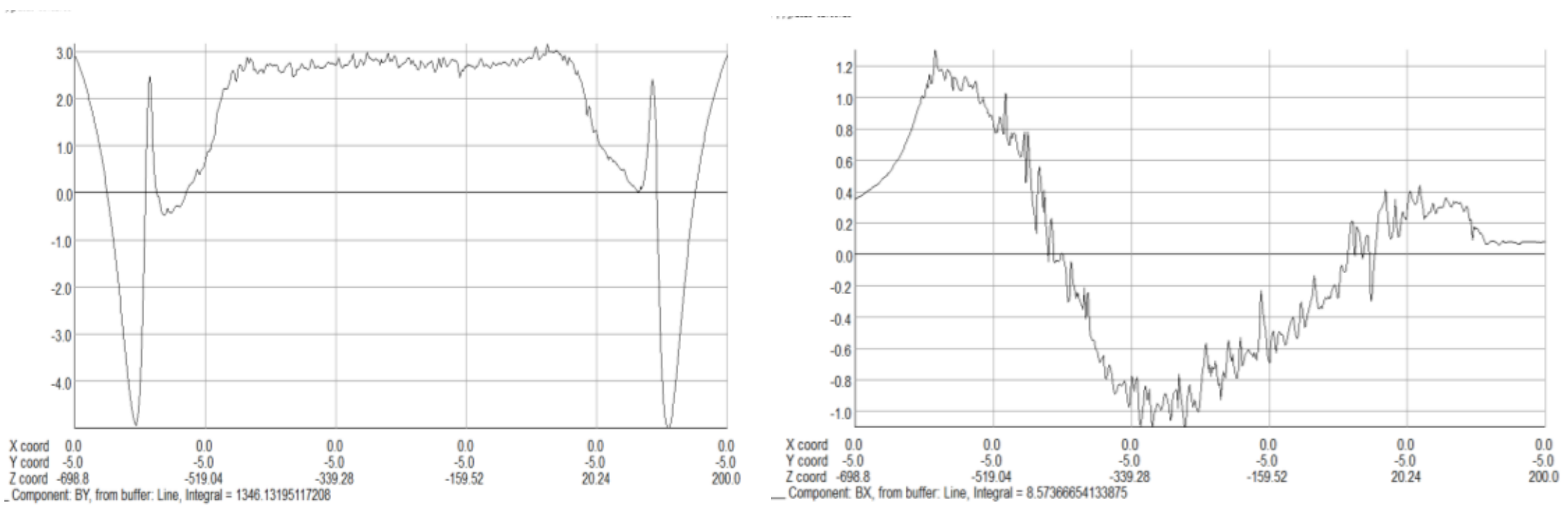

**Figure 6.4.7.12:** Simulation result of the leakage field (left: vert., right: hori.)

A similar magnet design is utilized for the Lambertson magnet used in the HEPS booster injection from the Linac of 0.5 GeV. This magnet was successfully developed and installed in the tunnel in January 2023, as depicted in Figure 6.4.7.13. The septum thickness of this Lambertson magnet is also 3.5 mm, and the inner aperture of the circulating beam vacuum pipe measures 28 mm.

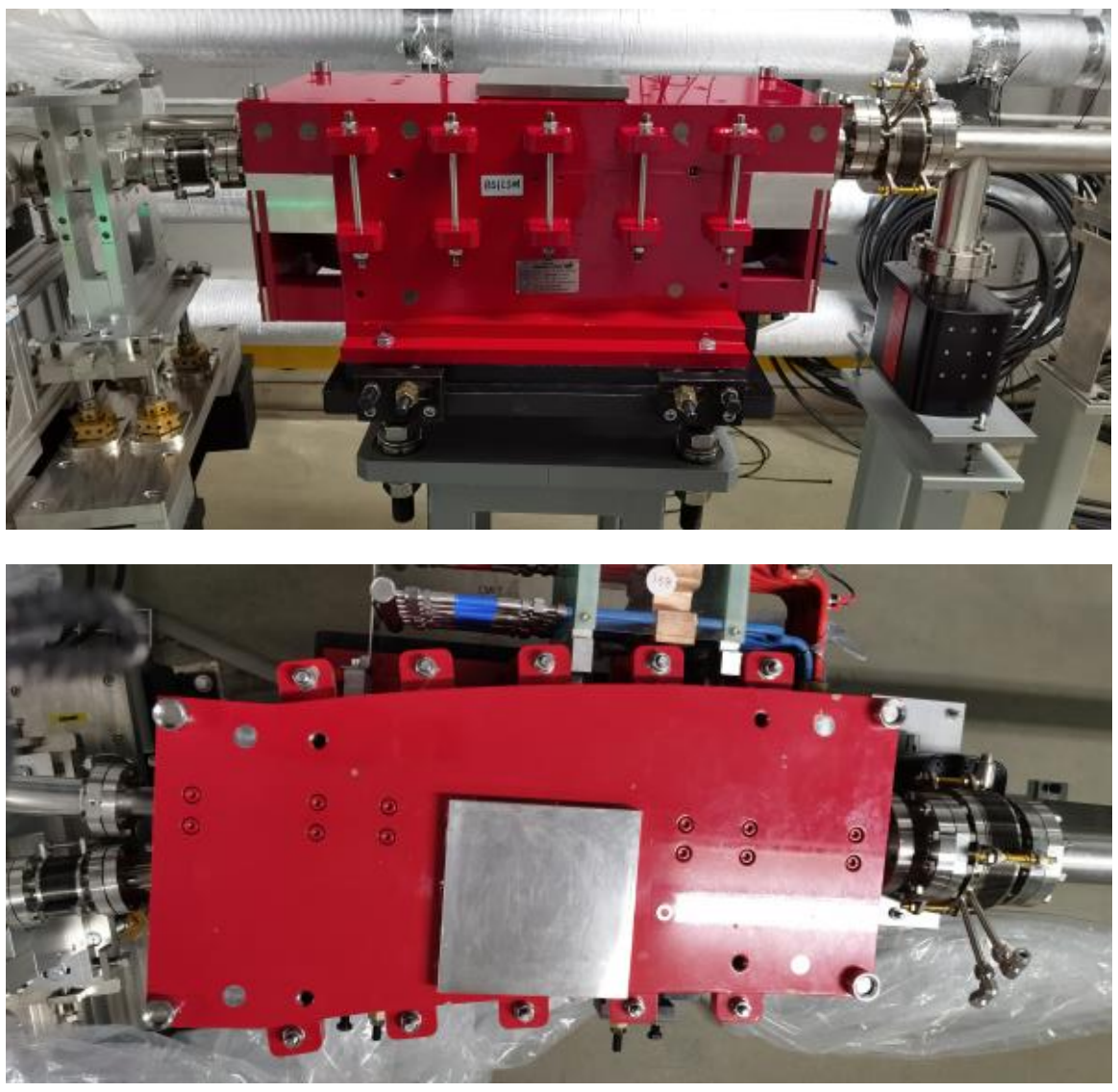

**Figure 6.4.7.13:** The injection Lambertson magnet installed in the HEPS booster tunnel.

## 6.4.8 **Control System**

### 6.4.8.1 ***Introduction***

To reduce the transverse emittance of the positron beam, the Damping Ring has been designed and will be installed as part of the Linac. The control of the Damping Ring encompasses both the global control system and its own specific control requirements.

Similar to the control requirements of the Linac accelerator, the control of the Damping Ring enables operators to monitor and control the equipment from both the Linac temporary control room and the central control rooms. The global control system provides essential functions such as timing synchronization, machine protection, and data server capabilities to support the operation of the Damping Ring.

### 6.4.8.2 *Magenet Poewer Supply Control*

There are 202 different types of magnet power supplies used in the Damping Ring, which include power supplies for dipole, quadrupole, sextupole, and corrector magnets, among others. These power supplies are distributed along the auxiliary tunnel of the Damping Ring. The control requirements and interface specifications for the DR Power Supply are outlined in Table 6.4.8.1.

**Table 6.4.8.1:** Control requirements of magnet power supply

| Magnet PS | Quantity | Location | Interface | Communication protocol |
|---|---|---|---|---|
| Dipole-PS | 10 | Linac gallery | Ethernet/Fiber | MODBUS-RTU |
| Quadruple-PS | 130 | Linac gallery | Ethernet/Fiber | MODBUS-RTU |
| Sextupole-PS | 2 | Linac gallery | Ethernet/Fiber | MODBUS-RTU |
| Corrector-PS | 60 | Linac gallery | Ethernet/Fiber | MODBUS-RTU |

The power supplies in the Damping Ring operate on DC and utilize the DPSCM-II as their core control module, which is equipped with Ethernet/Fiber connectivity. The Input/Output Controllers (IOCs) have been specifically designed for different power supply types, as illustrated in Figure 6.4.8.1.

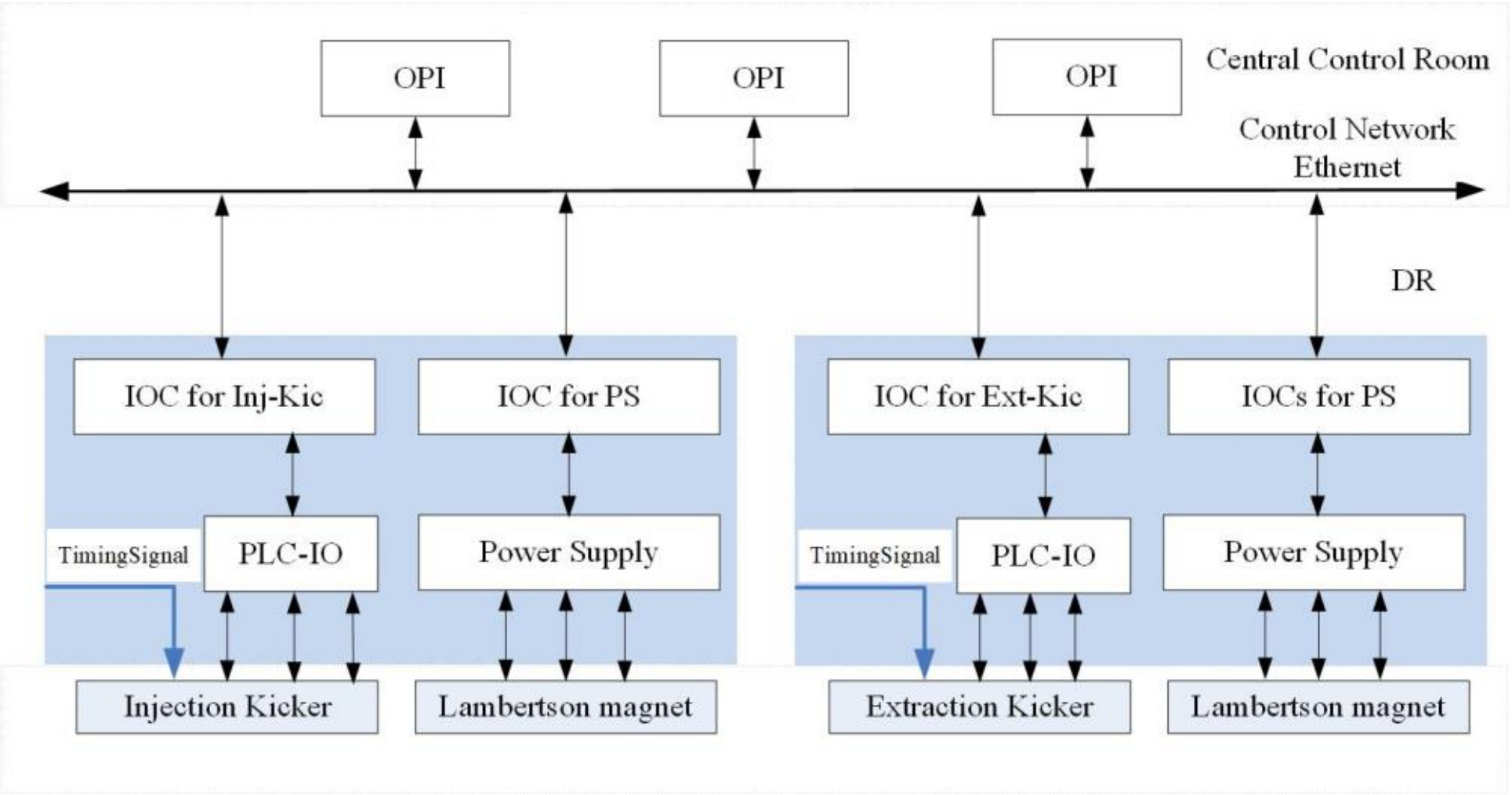

**Figure 6.4.8.1:** Structure of the DR PS control system

The power supply control systems for the Damping Ring and the Linac are designed to support both local and remote-control operations. These control systems offer various functionalities, including the ability to turn on/off power supplies, monitor the current and status of power supplies, and standardize the operation of power supplies.

### 6.4.8.3 *Injection and Extraction Kicker Control*

The control system for the injection and extraction systems in the Damping Ring should have the following functions:

- Monitoring the current, voltage, and running status (on/off, normal/alarm, local/remote) of the kicker power supplies.

- Providing the capability to switch on/off the kicker power supplies and set voltage values both locally and remotely.
- Sending two trigger signals through the timing system to synchronize the bunch injection with the positron beam.
- Performing regular checks on all kicker power supplies and sending alarm messages to the center console in case of faults. Local troubleshooting procedures can be initiated to resolve the issues.
- Monitoring the status and temperature of the kicker magnet cooling system to ensure proper cooling and prevent overheating.
- Storing the status of the devices and alarm information for maintenance and shift staff to review and analyze for maintenance purposes.

The injection and extraction kicker power supplies in the Damping Ring have three operation modes:

1. Check mode: In this mode, the power supplies can be switched on/off locally, the current can be set, and heating can be monitored. This mode is used for routine checks and testing of the kicker power supplies.
2. Positron mode: In this mode, the high voltage is continuously applied to the kicker power supplies to ensure they are preheated and ready for operation. The low voltage remains on throughout the operation. This mode is used when injecting positrons into the Damping Ring.
3. Electron mode: In this mode, after the injection process is completed, the high voltage is reduced to half heating level to save power. The low voltage remains on. This mode is used when injecting electrons into the Damping Ring.

Due to the configuration of the Damping Ring's injection and extraction systems, which include a vertically deflecting slotted-pipe kicker magnet and a horizontally deflecting Lambertson magnet, two Input/Output Controllers (IOCs) were designed for control purposes. One IOC is responsible for controlling the injection kicker and its power supply, as well as the Lambertson magnet associated with it. The other IOC is dedicated to controlling the extraction kicker and its power supply, along with the corresponding Lambertson magnet. The setup and interconnection of these IOCs can be visualized in Figure 6.4.8.2.

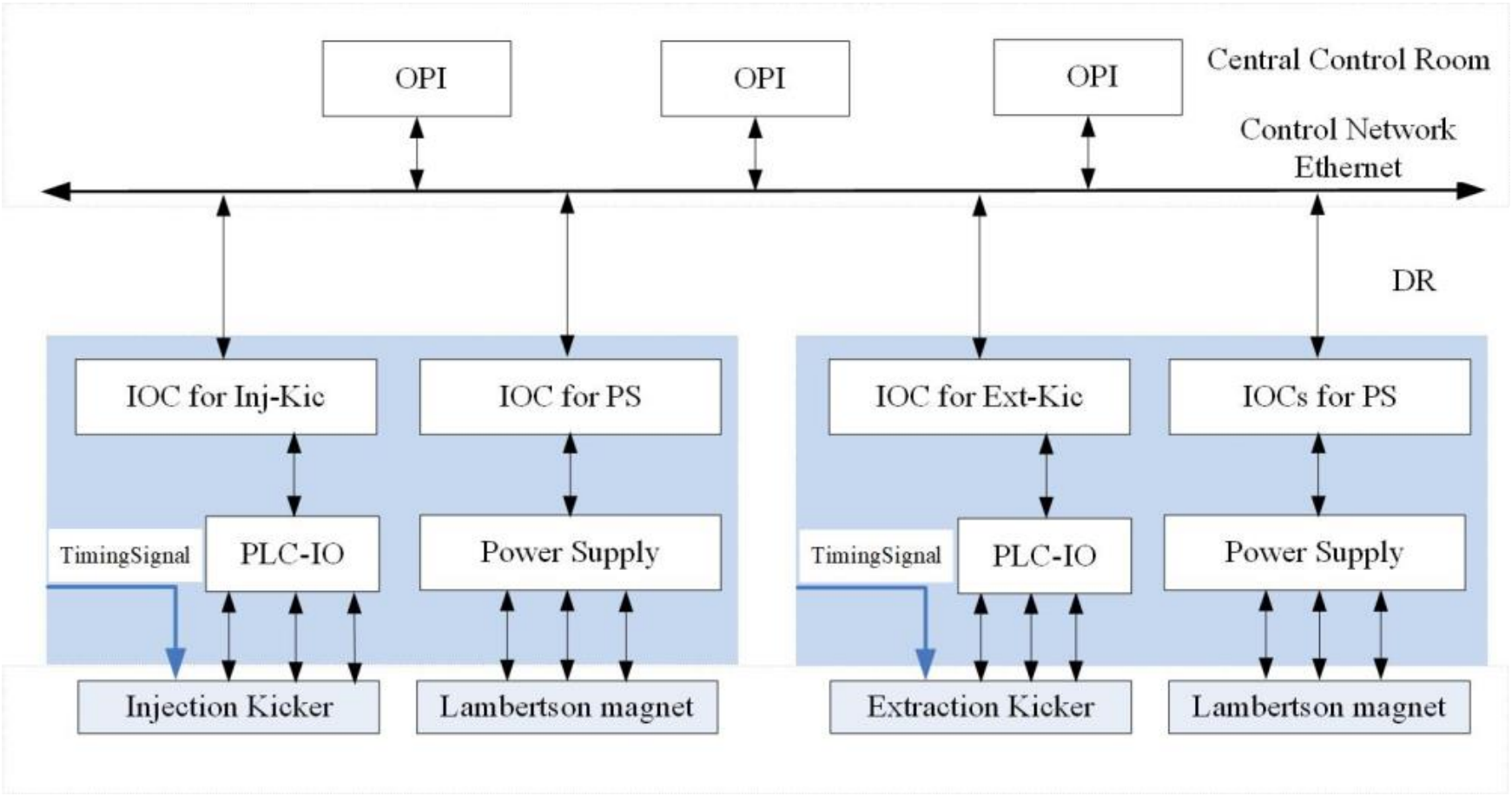

**Figure 6.4.8.2:** Control system of the DR Injection and extraction kicker

### 6.4.8.4 *Vaccum Control*

The Damping Ring is equipped with a total of approximately 4 vacuum valves, 80 pumps, and 6 gauges for maintaining the vacuum conditions. The vacuum control system plays a crucial role in measuring the vacuum pressure within the vacuum chamber and ensuring the protection of the machine by closing valves in the event of a vacuum leak.

Similar to the vacuum control technology employed in the linac, the Damping Ring's vacuum control system utilizes Programmable Logic Controllers (PLCs) to control the opening and closing of vacuum valves. It also communicates with serial port servers to retrieve data from vacuum gauges and ion pumps. Typically, the vacuum control cabinet is located adjacent to the vacuum equipment cabinet.

In Figure 6.4.8.3, the two cabinets on the left represent the vacuum control cabinets, while the remaining cabinets are dedicated to housing the vacuum equipment.

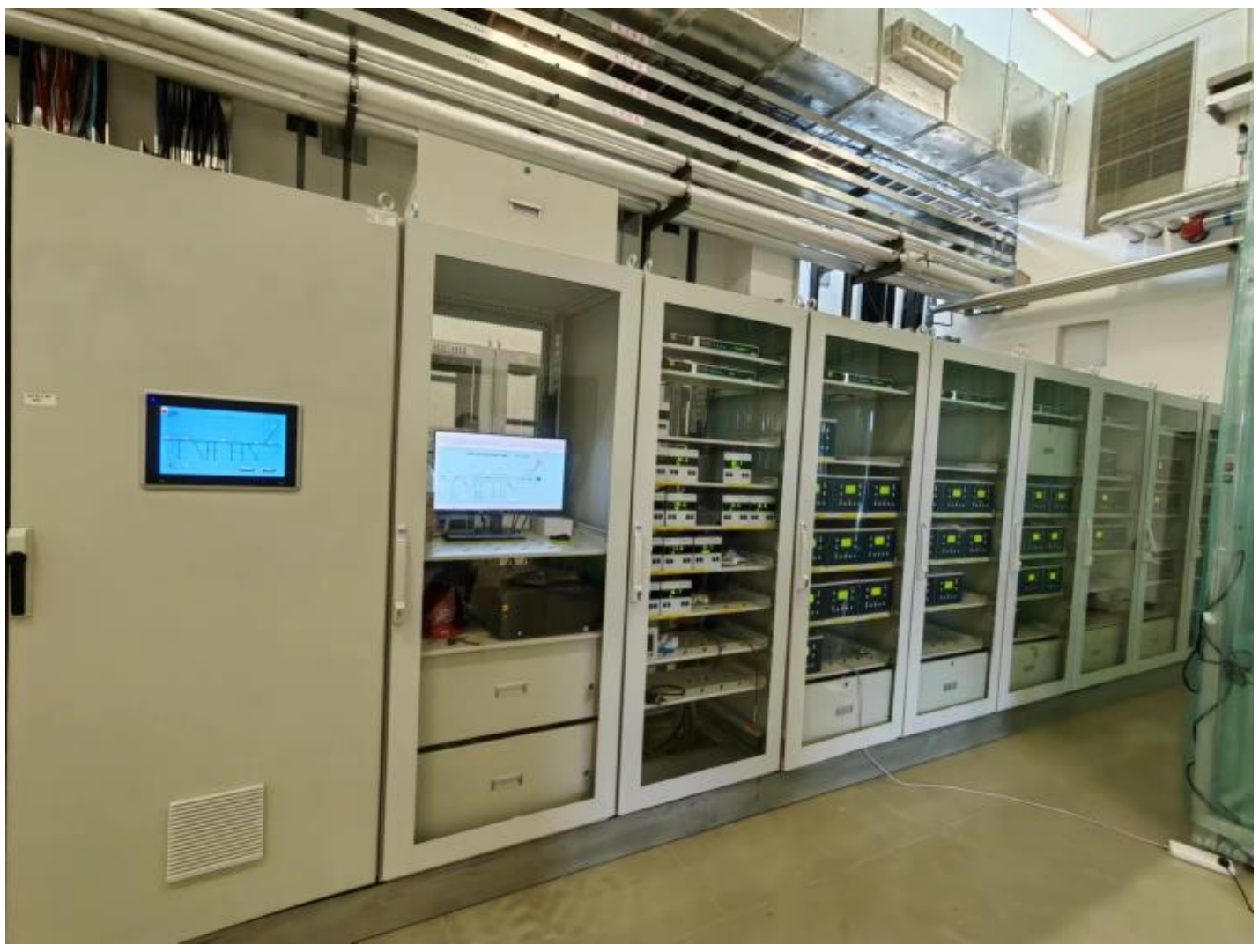

**Figure 6.4.8.3:** HEPS control and vacuum device cabinets.

### 6.4.8.5 *RF Control*

The Damping Ring RF system comprises two accelerating cavities, two RF high-power sources, and a low-level control system. The RF frequency used in the Damping Ring is 650 MHz, and the RF system provides a voltage of 1.25 MV. The RF cavities are installed in the straight sections of the Damping Ring.

The RF control system is divided into two parts: the low-level control system with FPGA and the device protection system with PLC. The RF control system is responsible for the following functions:

1. Monitoring Parameters and Status:
    - RF Gap voltage
    - RF phase
    - Forward/Reflected voltage
    - Forward/Reflected power
    - Tuner position and status of the cavity tuning system
    - Parameters of Higher Order Modes (HOM)
    - Temperature and vacuum levels of the cavities
    - Cooling water and gas system status of the cavities

- Status of klystron, waveguide, RF windows, etc.

2. Controlling RF Equipment:
   - Switching on/off the high-power sources
   - Setting up and adjusting RF voltage and ramp for acceleration, with a control accuracy of 1%
   - Operating the cavity tuning system, including step-by-step adjustment of the tuner position (minimum step of 1 kHz)
   - Continuous adjustment of the RF phase within the range of 0-360 degrees
3. Interlock System: The interlock system ensures the safety of the RF system by disabling the cavity tuning in the event of a fault or unsafe condition in the RF high-voltage power supplies, cooling water system, cavity vacuum, or temperature. The local interlock system sends warning messages and failure signals to the central control room upon detecting a fault.

The RF control system plays a critical role in monitoring and controlling the RF equipment in the Damping Ring, as well as maintaining the safety and operational integrity of the system.

### 6.4.9 **Mechanical Systems**

The mechanical system in the Damping Ring (DR) plays a crucial role in providing support for various devices. The supports are securely fixed to the ground and ensure the stability and proper alignment of the equipment. Steel frame pedestals with adjusting mechanisms are used, similar to those employed in the Linac. The beam height in the DR is consistent with that of the Linac, which is 1.2 meters from the tunnel ground.

Table 6.4.9.1 lists the devices in the DR that require support from the mechanical system. These devices are carefully mounted and positioned to meet the specific requirements of the DR operation. The support structure follows the guidelines outlined in Section 4.3.10, ensuring the structural integrity, precision, and reliability of the supported equipment.

**Table 6.4.9.1:** Supports in the Damping Ring

| Supports | Quantity (set) | Remarks |
|---|---|---|
| Quadrupole & sextuple support | 72 | Common girder for one quadrupole and one sextuple |
| Quadrupole support | 32 | Supports of 2 kinds of quadrupoles |
| Dipole support | 80 | Supports of 2 kinds of dipoles |
| Lambertson support | 2 | |
| Kicker support | 2 | |

The arc section of the Damping Ring (DR) comprises 36 FODO cells, each consisting of specific magnetic elements. The support configuration for these elements varies depending on their type and arrangement.

In the case of the dipoles, each one is individually supported, ensuring their stability and proper alignment within the DR. Adjacent to the dipoles, the quadrupoles and sextupoles are supported together, providing a combined support structure for enhanced stability. This arrangement is depicted in Figure 6.4.9.1.

The RF section of the DR contains a total of 32 quadrupoles and 8 dipoles, which are spaced relatively far apart. Therefore, each quadrupole and dipole in the RF section is supported separately, considering the larger distances between them.

Additionally, the two Lambertson magnets and two kickers in the DR are each supported individually to maintain their stability and precise positioning.

The 60 correctors in the DR are closely associated with the quadrupoles and do not require separate supports, as they are integrated within the overall support structure of the quadrupoles.

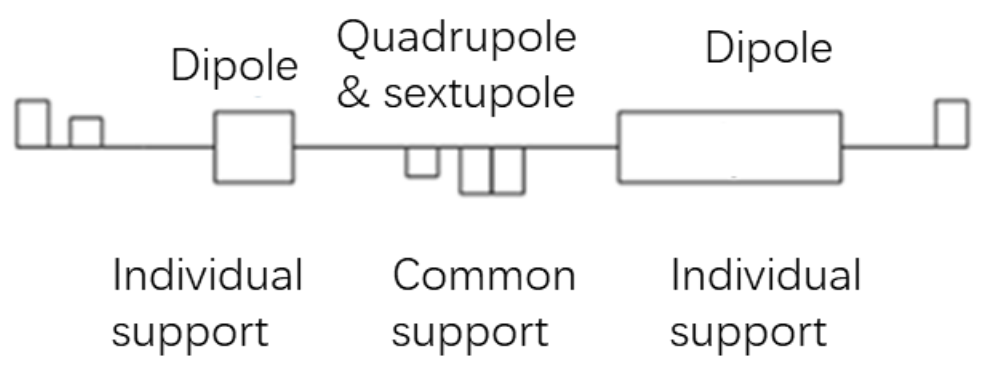

**Figure 6.4.9.1:** Support method of FODO cell

In Figure 6.4.9.2, the support configuration for the adjacent quadrupole and sextupole magnets is shown. A common steel frame pedestal is mounted to the ground, providing a stable base. The individual magnets are adjusted vertically using screws and horizontally using push-pull bolts. This adjustable support system allows for precise alignment and positioning of the magnets.

Figure 6.4.9.3 depicts the support structure for the quadrupole magnets in the RF section. It follows a similar design with a steel frame pedestal mounted to the ground. However, since there is only one magnet in this case, the support structure is simplified accordingly.

Figure 6.4.9.4 illustrates the support of the dipole magnet, which has a length of 700 mm. The support structure for the dipole magnet is also similar to the previous configurations, with a steel frame pedestal providing stability and adjustment mechanisms for precise positioning.

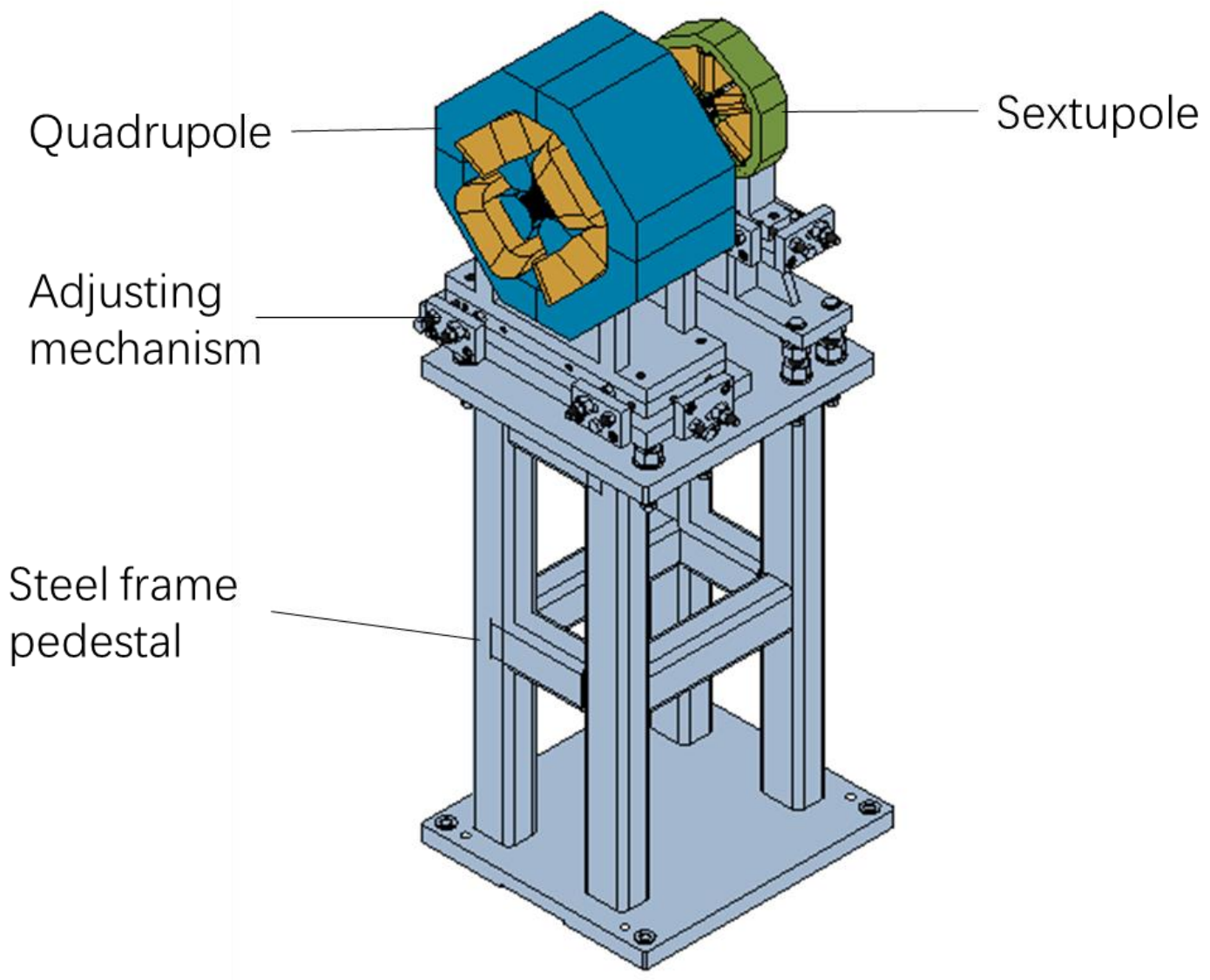

**Figure 6.4.9.2:** Support of adjcent quadrupole and sextupole

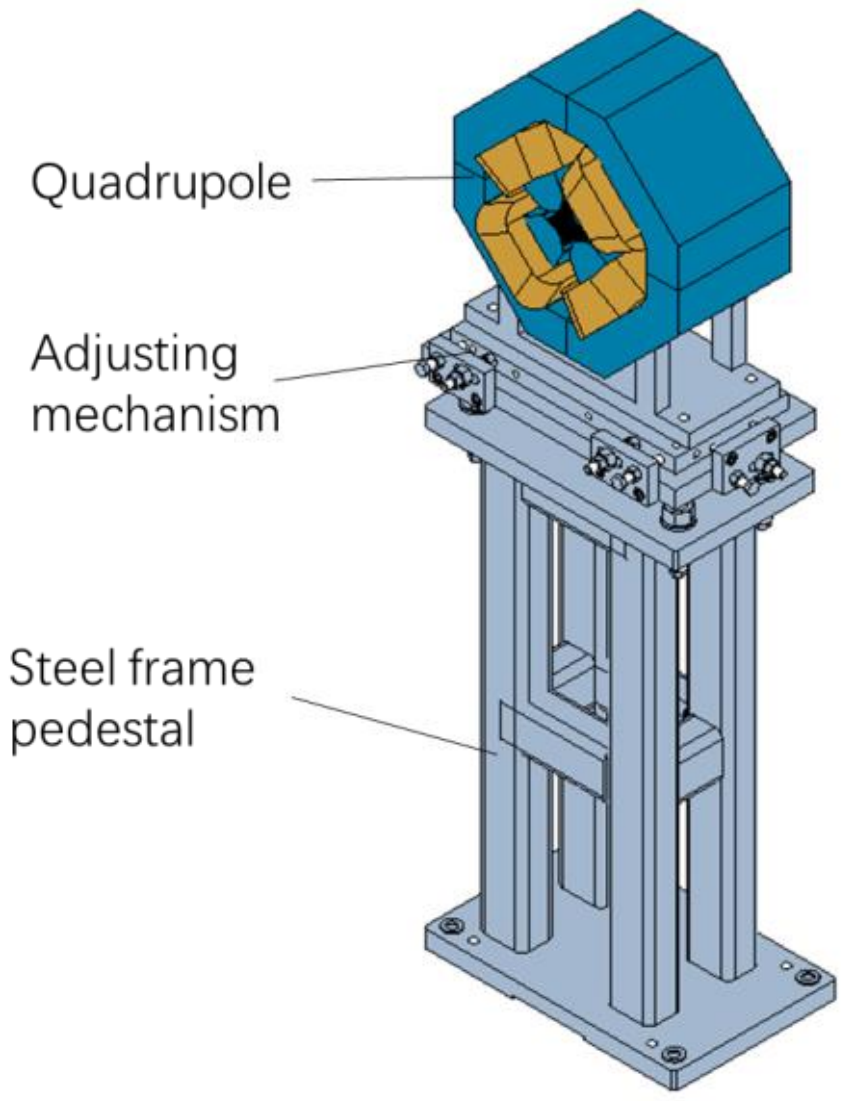

**Figure 6.4.9.3:** Individual support of quadrupole magnet

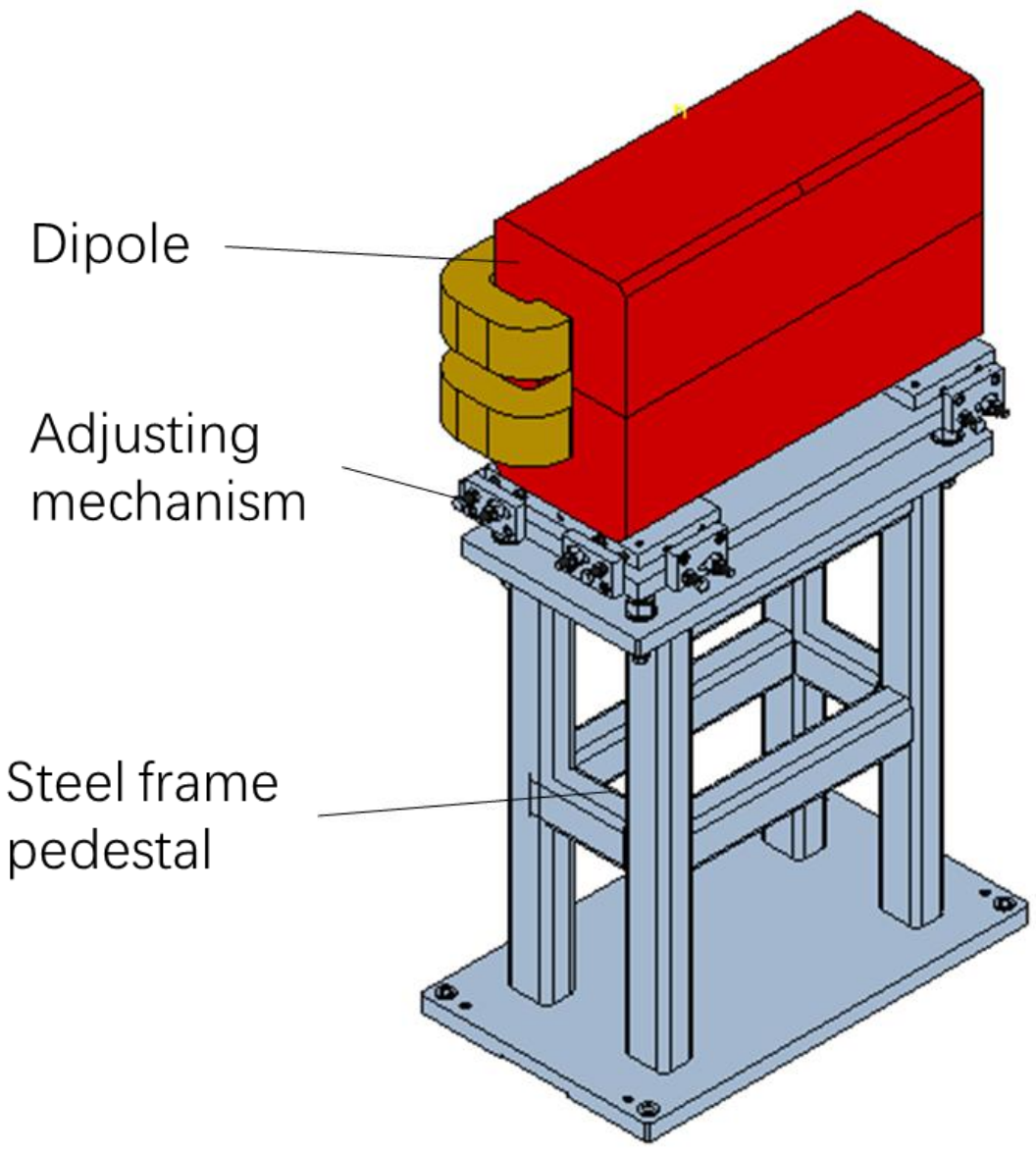

**Figure 6.4.9.4:** Individual support of dipole magnet

# 7 Communal Facilities for CEPC Accelerators

## 7.1 Cryogenic System

### 7.1.1 Overview

The CEPC would use positron and electron beams that are accelerated to 30 GeV after passing through a linear acceleration section, and then further accelerated to 120 GeV using a combination of 96 1.3 GHz 9-cell superconducting RF (SRF) cavities in the Booster and 336 650 MHz 2-cell SRF cavities in the Collider for the Higgs 50 MW mode.

To achieve this, the CEPC will use a cryogenic system that is designed to provide sufficient cooling capacity at design temperature levels, which is necessary to enable the operation of the superconducting RF cavity cryomodules or superconducting magnets cryostats. Specifically, there will be 12 cryomodules for 1.3 GHz 9-cell cavities in the Booster and 56 cryomodules for 650 MHz 2-cell cavities in the Collider for the Higgs 50 MW mode. For the Higgs 30 MW mode, there are 32 cryomodules for 650 MHz 2-cell cavities in the Collider, and the number of cryomodules in the Booster is not changed. [1-2]. The cryogenic system is an essential component of the CEPC as it ensures that the superconducting RF cavities and magnets are kept at the required low temperatures for proper operation.

The CEPC cryogenic system for the Higgs 50 MW mode includes four cryo-stations, each with a capacity of 15 kW @4.5K for the SRF cryogenic system. The SRF cavity cryomodule operates at a temperature of 2K. There are four IR superconducting (SC) magnets working at 4K. The detectors use the thermosiphon cooling method. Two sets of 400 W @4.5K helium refrigerator will be employed for the IR SC magnets. An overview of the CEPC cryogenic system can be seen in Figure 7.1.1.

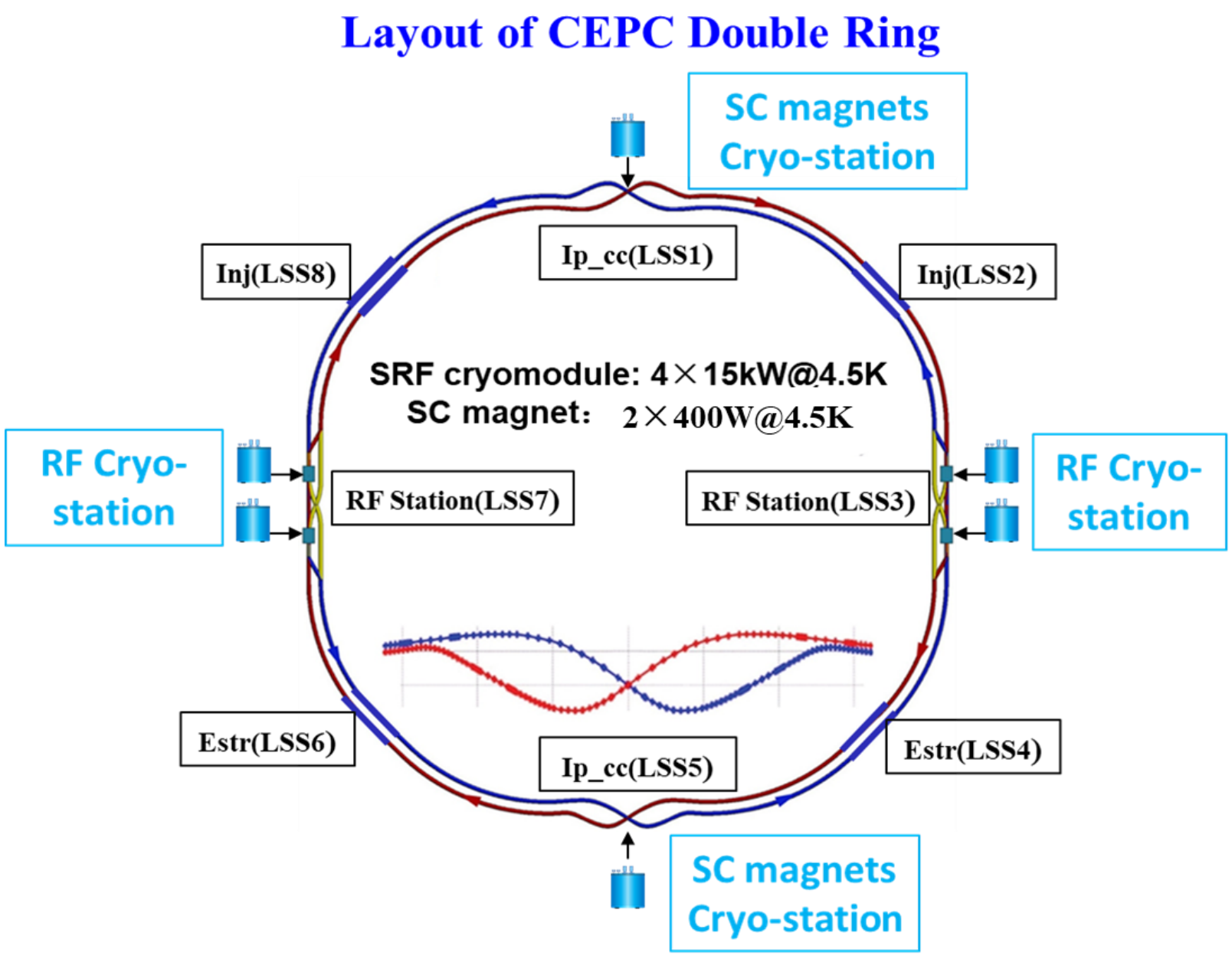

**Figure 7.1.1:** Overview of the CEPC cryogenic system.

The main advances of the CEPC cryogenic system from CDR to TDR are summarized below:

1) The cooling scheme for the SRF cryomodules in both the Booster and Collider has been modified from parallel to series, which improves the performance of the cryogenic system while reducing costs. The design of both cryomodules has been reviewed and improved based on the growing experience of the IHEP cryogenics group. Prototypes for both the Booster and Collider have been built in collaboration with domestic qualified companies and tested in the Platform of Advanced Photon Source (PAPS) infrastructure to further improve the design.
2) The new design of the SC magnet cryogenic system follows a multiple technology strategy under development for the interaction region (IR) magnets, with normal-conducting sextuples. The use of large refrigerators for operational stability and cost was chosen.
3) Several key technologies, including the JT heat exchanger platform, cryo-circulating pump, multi-channel transfer line test platform, and virtual system establishment and automatic control strategy research, have been well studied.

Overall, the CEPC cryogenic system has achieved significant progress from CDR to TDR, laying a strong foundation for further development in the EDR stage.

### 7.1.2 **Cryo-Unit and Cryo-Strings**

Helium cryogenic systems are typically composed of various components, including 4.5K refrigeration systems, 2K refrigeration systems, cryomodules/cryostats for superconducting devices such as SRF cavities or IR SC magnets, tanks zone, helium recovery and purification systems, liquid nitrogen system, control systems, and more. In the case of the CEPC cryogenic system, its design is aimed at providing adequate cooling capacity at the intended temperature levels for the operation of the SRF cavity cryomodules and SC magnets cryostats.

The CEPC cryogenic system's process flow diagram in the SRF system is illustrated in Figure 7.1.2. One Cryo-station will supply the cooling for 14 Collider cryomodules and 3 Booster cryomodules. The cold box produces 3 bara @5K supercritical helium, which is lowered to 4.5K after passing through the subcooler and then further reduced to 2.2K after passing through a 2K heat exchanger. A 2.2K @3 bar subcooled helium will be transported to the JT valve in front of each cryomodule through the multi-channel transfer line. The temperature is then lowered to 2K, and the cavities are immersed in the HeII bath. The helium gas is transferred back to the 2K heat exchanger. After superheating in the counter flow heat exchanger, the gas is compressed in the multiple-stage cold compressors to a pressure in the range of 0.5 to 0.9 bar. It is then separately warmed up to ambient in exchangers and goes back to the warm compressors.

The cryomodules in the CEPC cryogenic system consist of two thermal shields: a 40K~80K shield and a 5K~8K shield. Further details of Strings 1 (Booster) and 2 (Collider) can be found in Figure 7.1.3 and Figure 7.1.4, respectively.

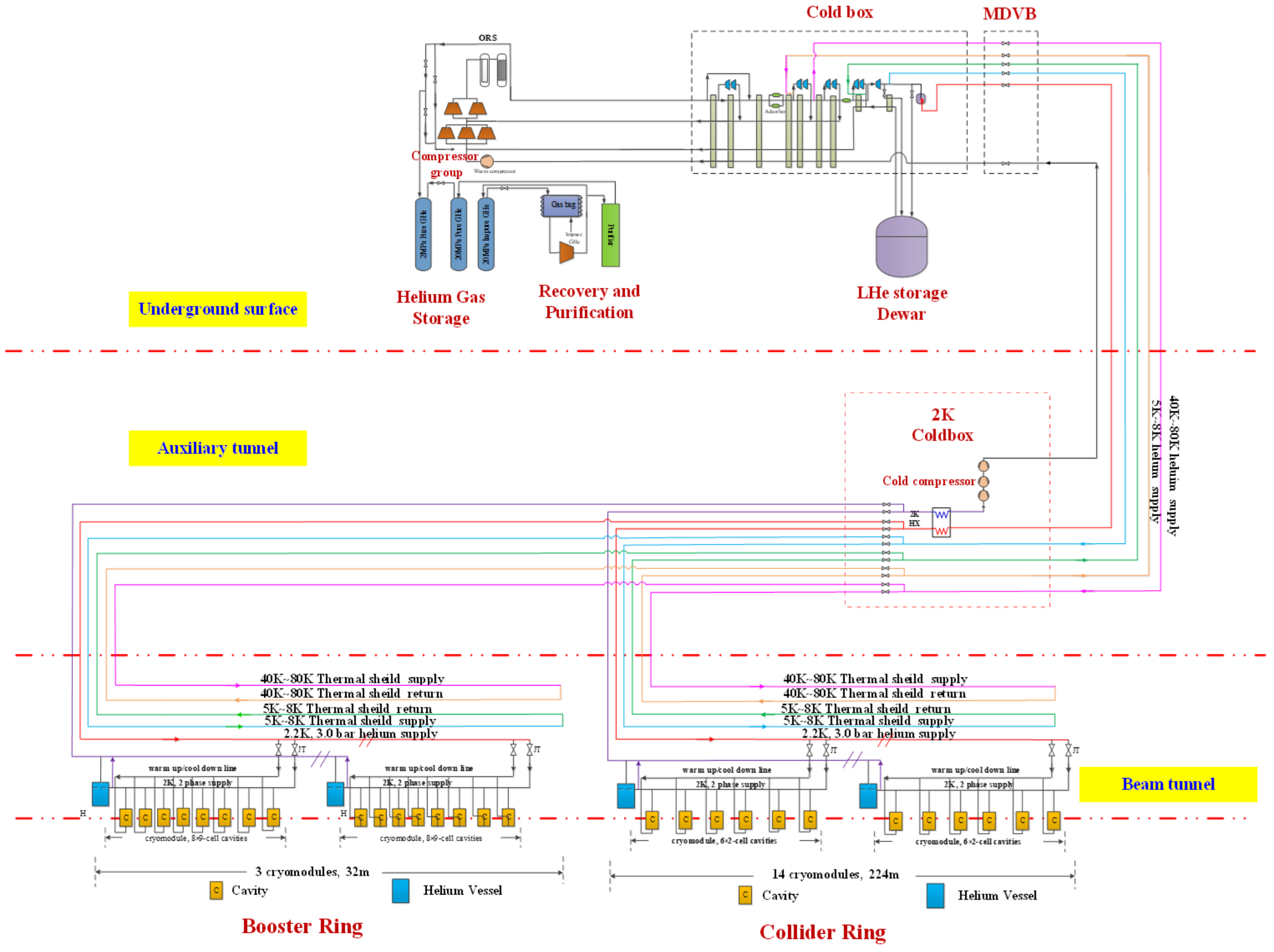

**Figure 7.1.2:** CEPC cryogenic system process flow diagram in the SRF system.

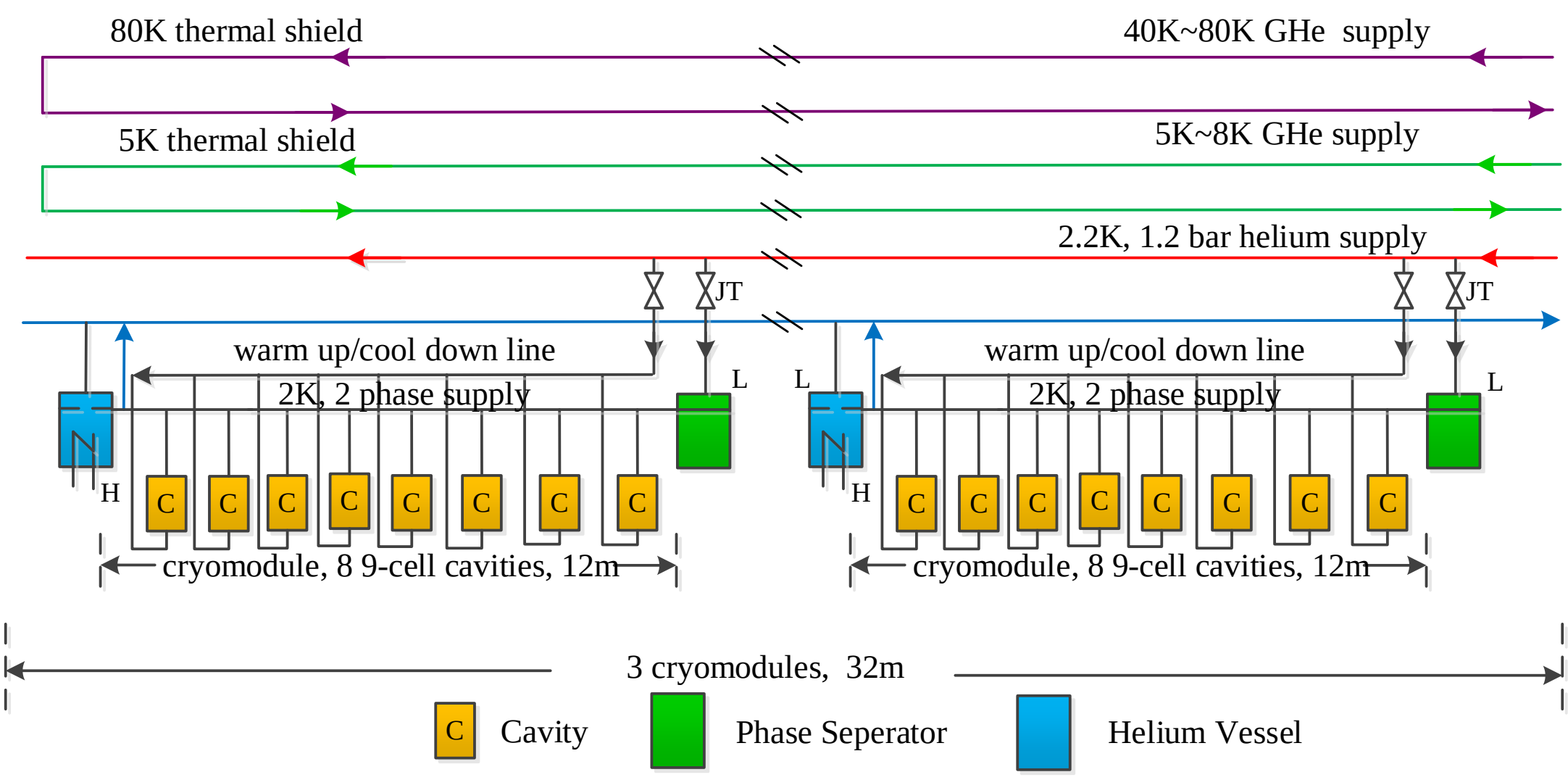

**Figure 7.1.3:** Booster cryomodules for String 1 (Booster).

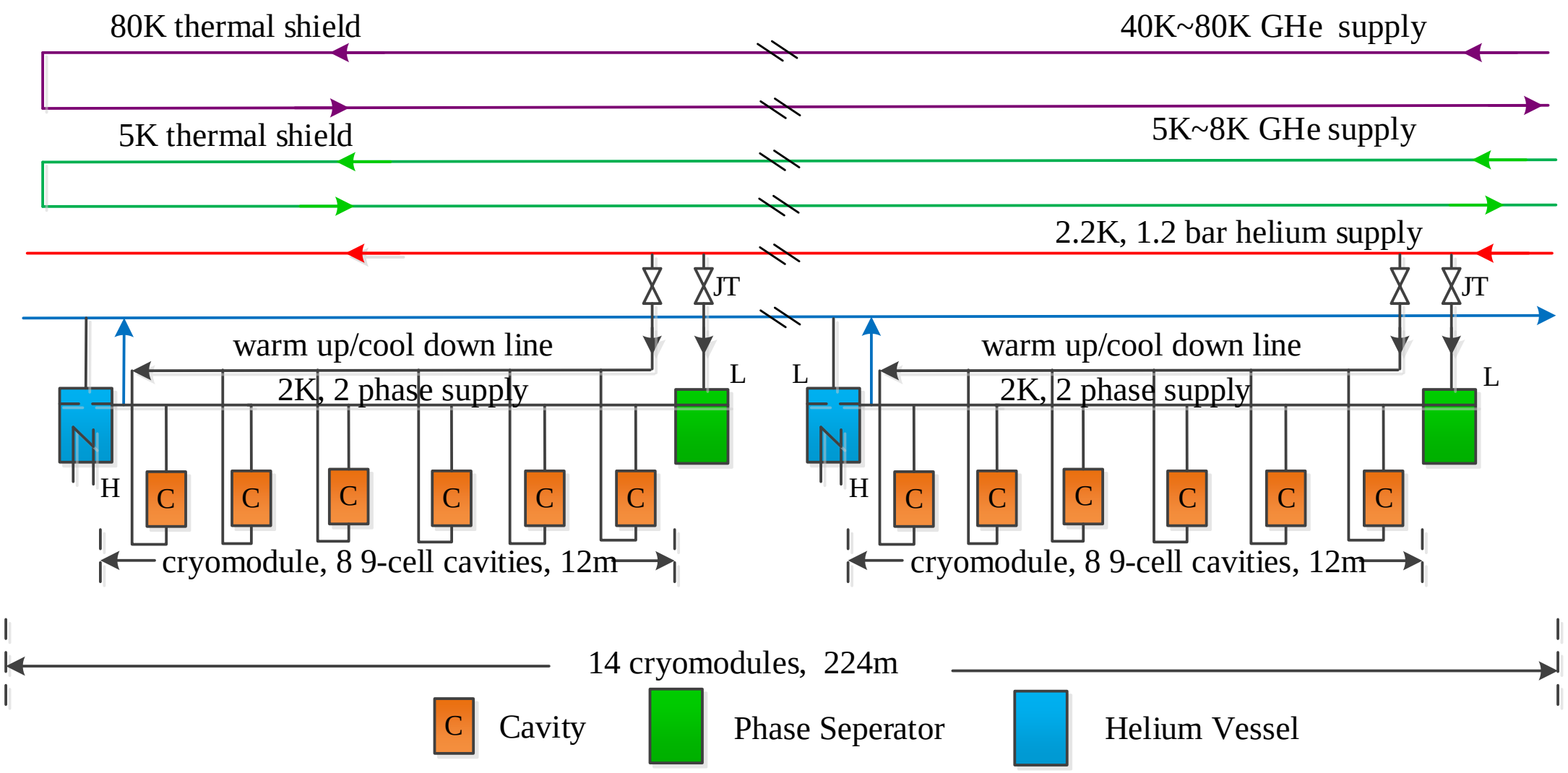

**Figure 7.1.4:** Collider cryomodules for String 2 (Collider).

Each cryomodule in the CEPC cryogenic system has two valves, with the JT-valve being used to expand helium to a liquid helium separator. A two-phase line connects each helium vessel, providing liquid helium supply and helium gas return, and connects to the major gas return header once per-module. Additionally, a small diameter warm-up/cool-down (CD) line connects the bottom of the helium vessels at both ends through a CD valve. The cavities are immersed in baths of 2K saturated superfluid helium.

The cryomodule cooldown process in the CEPC cryogenic system goes through three stages. The first stage is the slow cooldown (from 300K to 150K), where the cryomodule is cooled at a rate of less than 10K per hour (around 7 K per hour) until the cavities reach 150K. The second stage (from 150K to 45K) is the transit zone cooldown, where the rate is increased to approximately 20K per hour once the cavities cross the transient temperature zone to reduce the risk of Q-disease. However, the cavities are hydrogen degassed during processing, which is expected to mitigate this degradation. At around 45K, the cryomodule temperature is maintained through a soaking period of 24 hours to allow various components, including the magnetic shield, to continue their cooldown. Once the temperatures have settled at 45K, the third stage of fast cool down (from 45K to 4.5K) is initiated. During this stage, a minimum mass flow of 30g/s is supplied, and the cooldown time needs less than 10 minutes. The whole cooldown process is about 51 hours.

For the IR SC magnets, the change of the IR sextupole from superconducting (CDR) to normal conducting (TDR) has significantly reduced the heat loads for the cryogenic system, resulting in a modification of the estimated heat loads from the former 8 kW in CDR to just 800 Watts (two sets of 400W@4.5K refrigerator) in TDR. Since the long cryogenic transfer line can be eliminated, the cooling scheme has become simpler. The CEPC cryogenic system process flow diagram in the IR SC magnets side is shown in Figure 7.1.5. The system uses supercritical helium forced flow cooling to cool the IR SC magnets, which has the advantages of stable heat transfer coefficient, cooling individually, and simultaneous parallel cooling between the windings. In addition, two detector magnets adopt the thermosiphon cooling method, which has the benefits of not requiring an external power supply and greatly reducing the operation and maintenance costs of the cryogenic equipment.

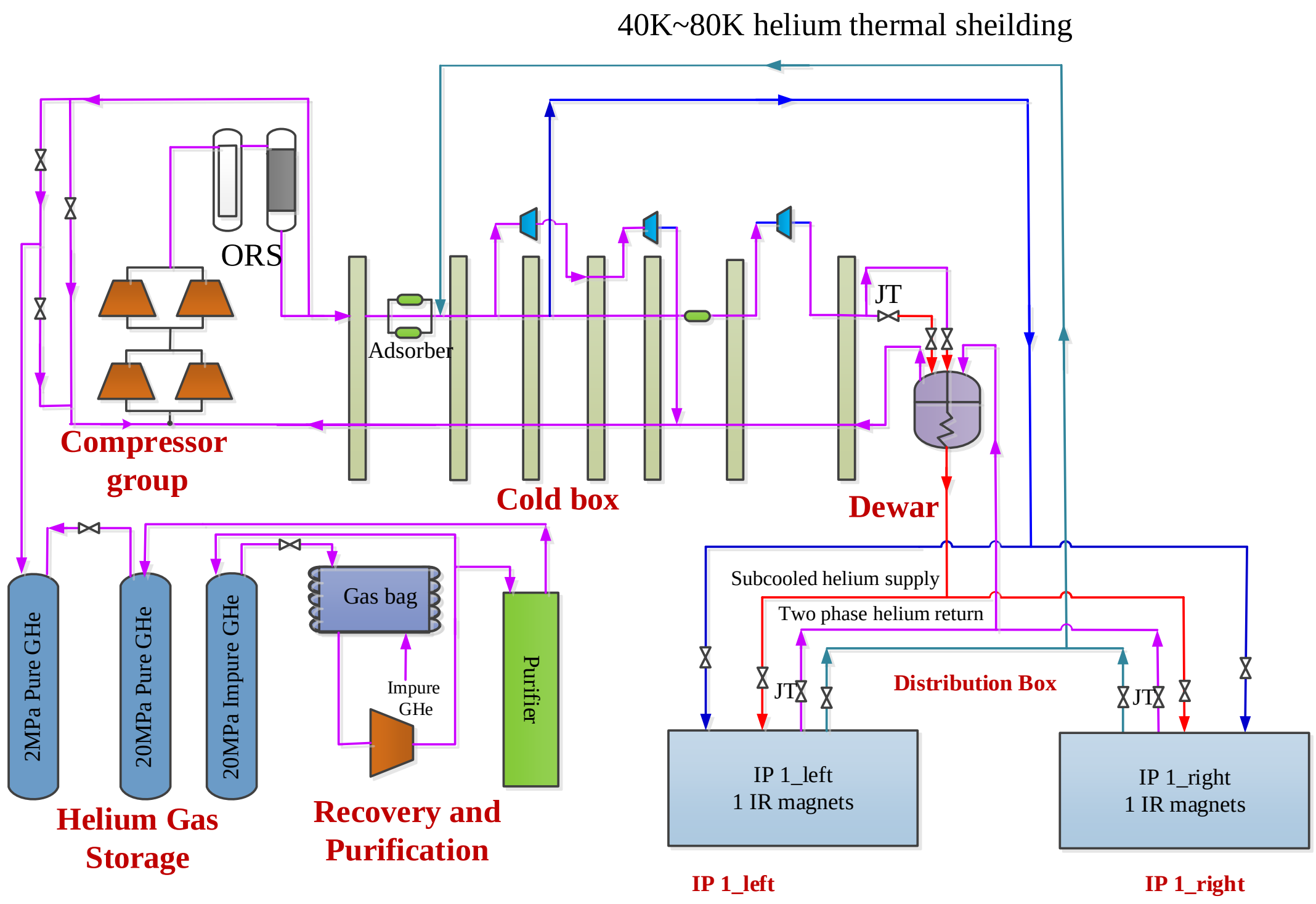

**Figure 7.1.5:** CEPC cryogenic system process flow diagram in the IR SC magnets area.

### 7.1.3 **Layout of Cryo-Unit**

The 2K cryogenic system is comprised of oil-lubricated screw compressors, a liquefied-helium storage vessel, a 2K refrigerator cold box, cryomodules, a helium-gas pumping system, and high-performance transfer lines. The cryogenic stations are located near the RF stations. To provide the necessary cooling power at each RF station, a refrigerator with a capacity of 15 kW @4.5K is installed at each of the four cryogenic stations, which then distributes the cooling to the adjacent superconducting RF cavities.

For the sake of simplicity, reliability, and ease of maintenance, we aim to minimize the number of active cryogenic components distributed around the ring. To achieve this goal, we have chosen equipment locations based on two guiding principles:

1) Whenever possible, we install equipment above ground to avoid the need for excavation. Normal temperature equipment will be installed at ground level.
2) To minimize heat loss, we install low-temperature equipment in close proximity to the cryomodules [3].

The equipment at ground level includes the electric substation, warm compressor station, helium storage tanks, helium purification system, and the 4.5K cold-boxes. In contrast, the cold compressor, 2K heat exchanger, 2K cryomodules, and cryogenic multiple transfer lines are situated underground. One can refer to Figure 7.1.6 for an overview of the cryogenic system's general architecture, Figure 7.1.7 for the overall schematic, and Figure 7.1.8 for a layout.

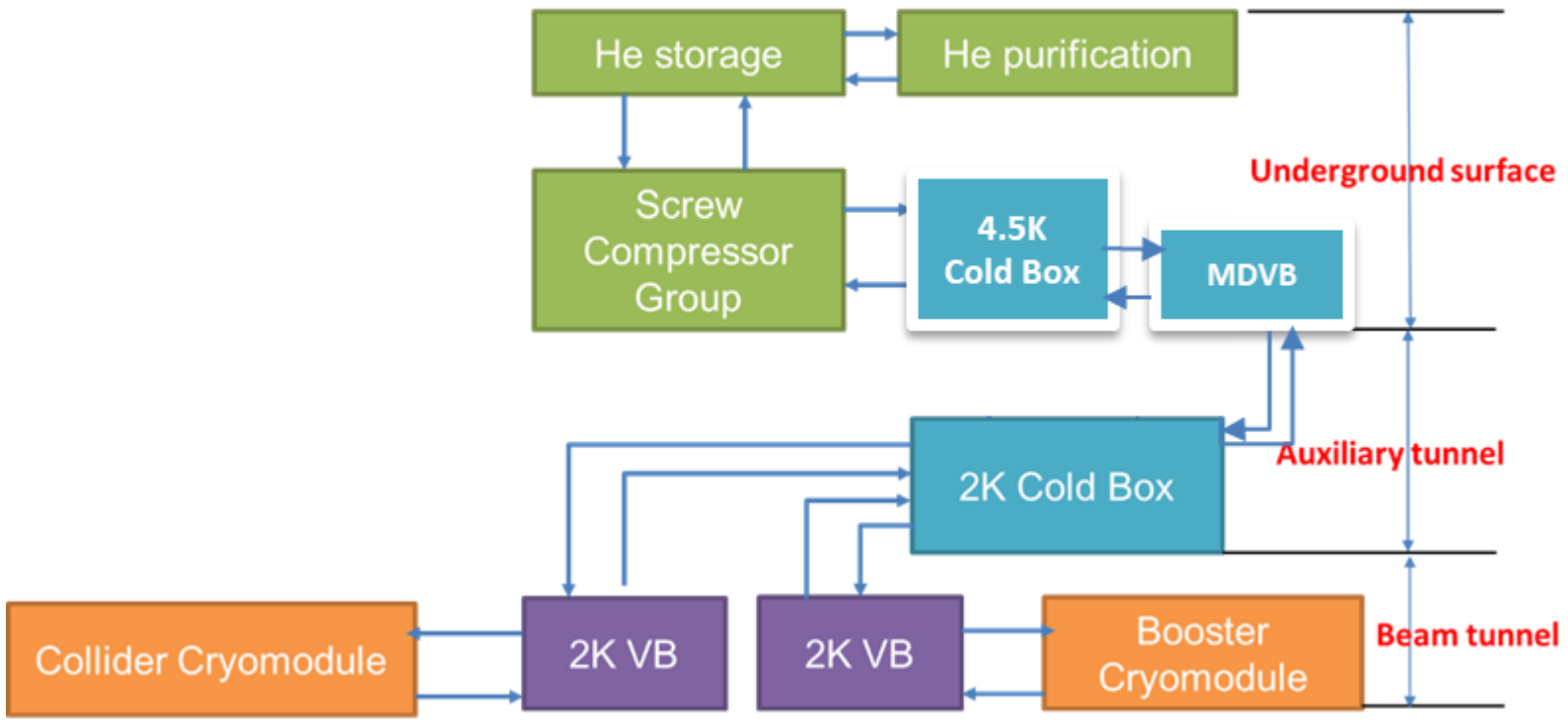

**Figure 7.1.6:** General architecture of the cryogenic system.

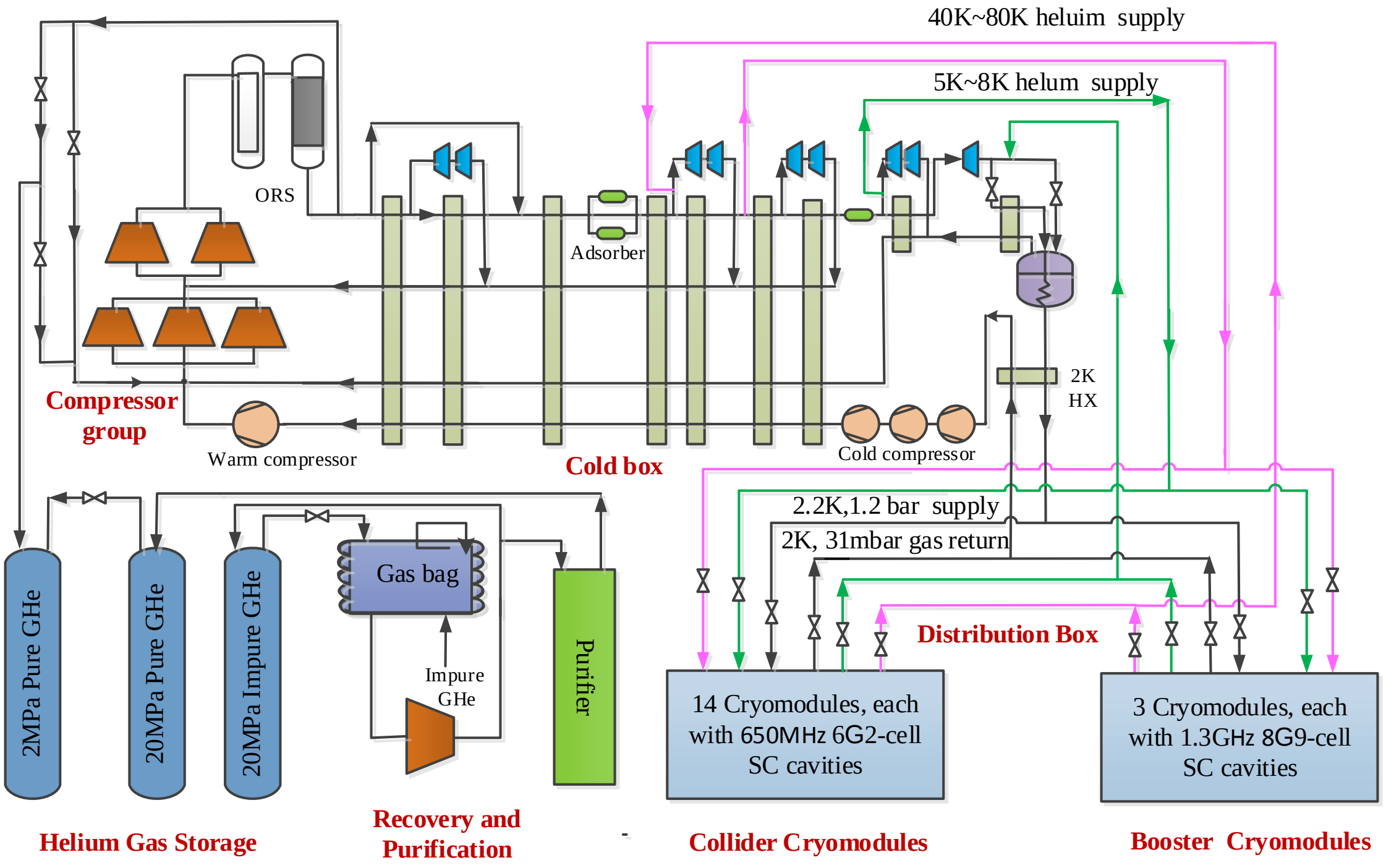

**Figure 7.1.7:** Schematic of the cryogenic system.

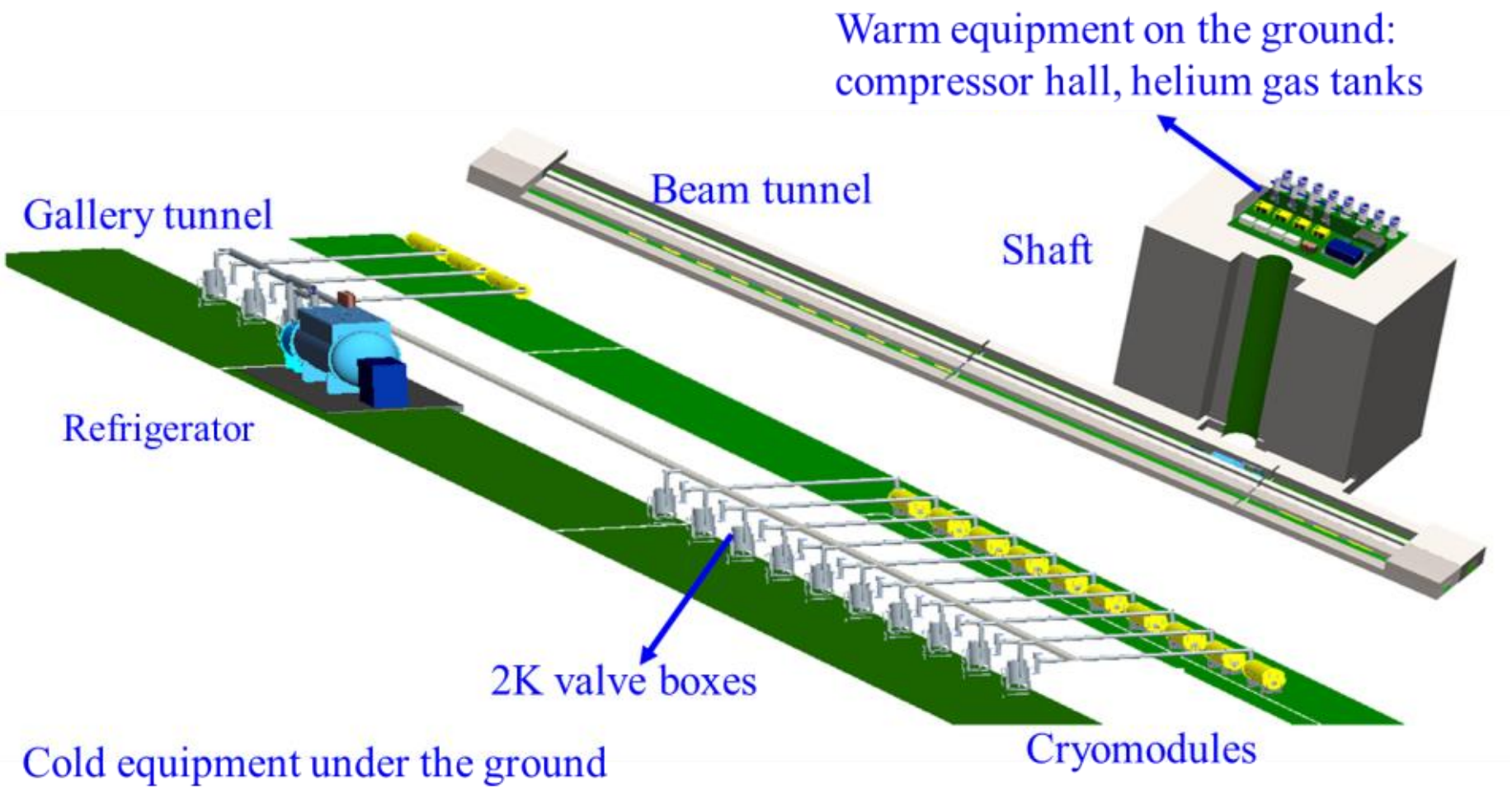

**Figure 7.1.8:** Layout of the CEPC cryogenic system.

Figure 7.1.9 illustrates the layout of the CEPC cryogenic hall and tanks zone. The rotating equipment consists of Piston compressors and Recycle screw compressors. Meanwhile, the tanks zone comprises liquid nitrogen tanks, medium pressure helium tanks, and high-pressure helium cylinders.

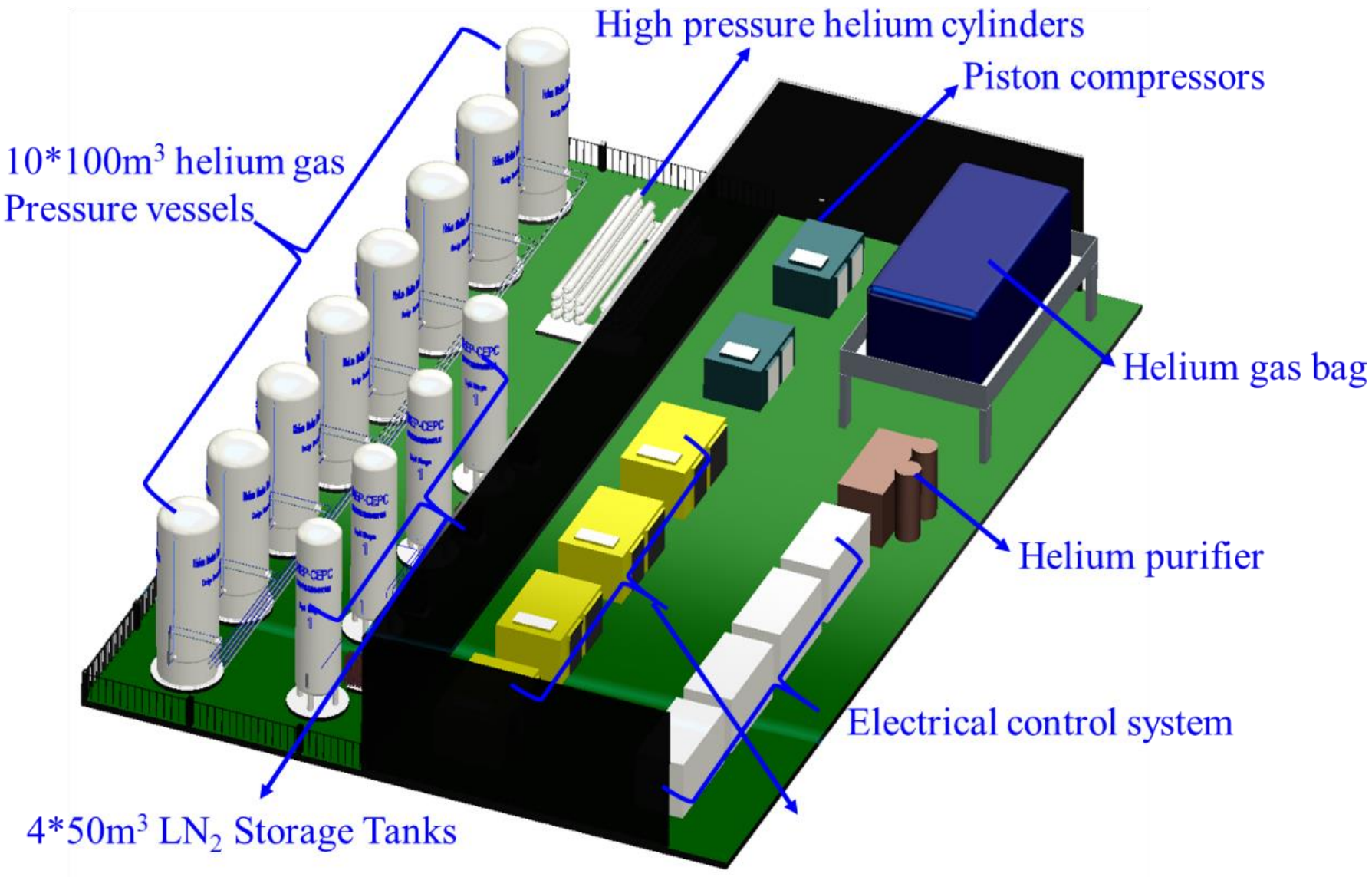

**Figure 7.1.9:** Layout of the CEPC cryogenic hall and tanks zone.

The layout of the tunnel is shown in Figure 7.1.10, and the detailed tunnel cross-section diagram can be found in Section 4.3.10.3 of this report. The distance between the gallery and beam tunnel is 10 meters. The cooling system for the SRF cryomodules in both the Booster and Collider has been modified from parallel in the CDR to series in the TDR to improve cavity performance while reducing costs. As a result, the space required for the gallery tunnel can be significantly reduced, since there are no long transfer lines or many valve boxes.

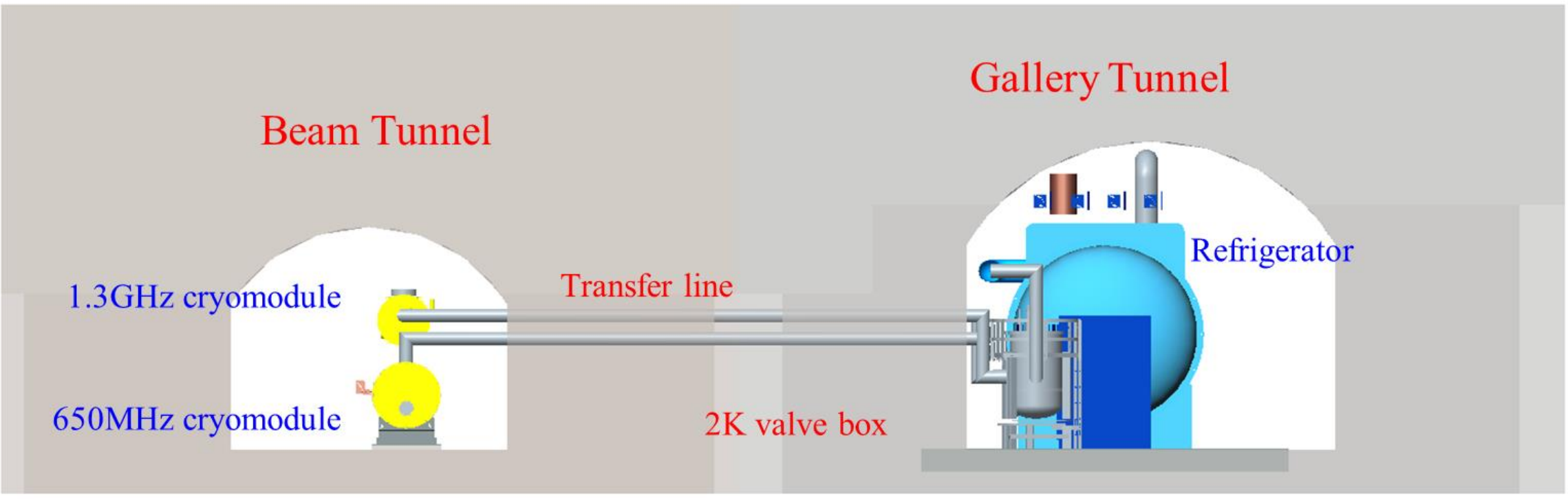

**Figure 7.1.10:** Tunnel layout.

### 7.1.4 Heat Loads

The primary source of heat load comes from the SRF cavities. The Booster employs 1.3 GHz TESLA-type 9-cell cavities with a quality factor of 3.0 x $10^{10}$ @ 21.8 MV/m, while the Collider uses 650 MHz 2-cell cavities with a quality factor of 3.0 x $10^{10}$ @ 14.2 MV/m in the Higgs 50 MW operation mode. The list of RF parameters for the Collider and Booster cavities can be found in Table 4.1.3 in Chapter 4 and Table 5.3.1.1 in Chapter 5.

The cryogenic system has been specifically designed to accommodate the Higgs mode, including the Higgs 50 MW mode. In this mode, the estimated heat loads at 4K are approximately 15 kW at 4.5K for each cryogenic station. To handle these heat loads, four separate refrigeration systems will be used, with each system serving one of the 4 cryogenic stations. The heat loads of the cryogenic system for the SRF cavities in Higgs 50 MW mode are detailed in Table 7.1.1.

**Table 7.1.1:** Heat loads of the cryogenic system for the SRF cavities.

| Higgs Mode 50MW | Unit | Collider | | | Booster | | |
|---|---|---|---|---|---|---|---|
| | | 40-80K | 5-8K | 2K | 40-80K | 5-8K | 2K |
| Predicted module static heat load | (W/module) | 300 | 60 | 12 | 140 | 20 | 3 |
| Predicted module HOM static heat load | (W/module) | 2.4 | 1.2 | 0.1 | 0 | 0 | 0 |
| Predicted module Input coupler static heat load | (W/module) | 6 | 3 | 0.3 | 16 | 8 | 0.8 |
| Predicted module dynamic heat load (RF) | (W/module) | 0 | 0 | 40.1 | 0 | 0 | 40.9 |
| Predicted module HOM dynamic heat load | (W/module) | 4.8 | 2.4 | 0.4 | 9.6 | 4.8 | 0.8 |
| Predicted module Input coupler dynamic heat load | (W/module) | 85.8 | 21.5 | 3.6 | 256 | 25.6 | 3.2 |
| Each Module total heat load | W | 399 | 88.1 | 56.5 | 421.6 | 58.4 | 48.7 |
| Cryomodule number | - | 56 | 56 | 56 | 12 | 12 | 12 |
| MDVB heat loss | W | 50 | 30 | 10 | 50 | 30 | 10 |
| MDVB number | - | 4 | 4 | 4 | 4 | 4 | 4 |
| Total cryogenic transfer line length | m | 936 | 936 | 936 | 168 | 168 | 168 |
| Cryogenic transfer line heat loss per meter | W/m | 2 | 0.5 | 0.3 | 2 | 0.5 | 0.3 |
| Total heat load | kW | 27.22 | 6.08 | 3.76 | 6.19 | 1.02 | 0.73 |
| 4.5K equiv. heat load with 1.54 multiplier | kW | 4.85 | 8.43 | 28.58 | 1.10 | 1.42 | 5.58 |
| 4.5K equiv. heat load with 1.54 multiplier | kW | 41.86 | | | 8.10 | | |
| Total 4.5K equiv. heat load with 1.54 multiplier | kW | 49.96 | | | | | |

In the IR region, the SC sextupole magnets in CDR have been replaced with normal conducting ones, and the cryogenic system for sextupole magnets is eliminated. Table 7.1.2 lists the heat loads on the IR SC quadrupoles and solenoids. In TDR, two sets of 400 W @4.5K helium refrigerators will be used for these IR SC magnets.

**Table 7.1.2:** Heat loads of the cryogenic system for the IR SC magnets.

| | | 4.5K | | | Thermal Shielding 40~80K | | |
|---|---|---|---|---|---|---|---|
| Name | Unit | No. | Heat load for each | Heat load | No. | Heat load for each | Heat load |
| IR SC magnet | W | 4 | 30 | 120 | 4 | 30 | 120 |
| Valve Box of IR SC magnet | W | 4 | 30 | 120 | 4 | 30 | 120 |
| Current lead of IR SC magnet | g/s | 4 | 0.8 | 3.2 | | | |
| Main distribution valve box | W | 4 | 50 | 100 | 2 | 70 | 140 |
| Cryogenic transfer-line | W | 100 | 0.3 | 30 | 100 | 1.5 | 150 |
| Sum of heat load | W | | | 370 | | | 530 |
| Total heat load @4.5K | W | | | 370 | | | 61 |
| Total heat load @4.5K including current lead | W | | | 370W+3.2g/s | | | |
| Total heat load @4.5K | W | | | 440 | | | 57 |
| Coefficient | | | | 1 | | | |
| Total heat load @4.5K | W | | | 440 | | | 57 |
| Total equiv. heat load @4.5K | W | 497 | | | | | |
| Total equiv. heat load @4.5K with multiplier 1.5 | W | 745 | | | | | |

## 7.1.5 Cryogenic System for SRF

### 7.1.5.1 *Cooling Scheme Optimization and Process Calculation*

Conventional superconducting devices such as SRF cavities and IR SC magnets typically operate in the liquid helium temperature region. The simplest and most common way to cool these devices is to immerse them in liquid helium. This requires a cryostat device with good adiabatic properties to minimize heat leakage and preserve the liquid helium as long as possible. The cryostat must also be able to support the weight of the superconducting devices and withstand a certain pressure in the event of a superconducting loss.

With the advancement of cryogenic and superconducting technology, modern accelerators are increasingly built using these technologies to meet the demand for higher beam energy and brightness. However, in many cases, the traditional 4K liquid helium (He I) cryogenic system has been unable to meet the necessary requirements. Therefore,

SRF cavities in cryomodules are being developed to be cooled by the 2K superfluid helium (He II), which is a technically safe and economically reasonable choice. This is due to its large effective thermal conductivity and heat capacity, as well as its low viscosity, which has been successfully applied in the ADS Injector-1 2K superfluid cryogenic system [4-5] and PAPS cryogenic system.

The cooling schemes of superconducting devices (SRF cavity and IR SC magnets) are shown in Figure 7.1.11.

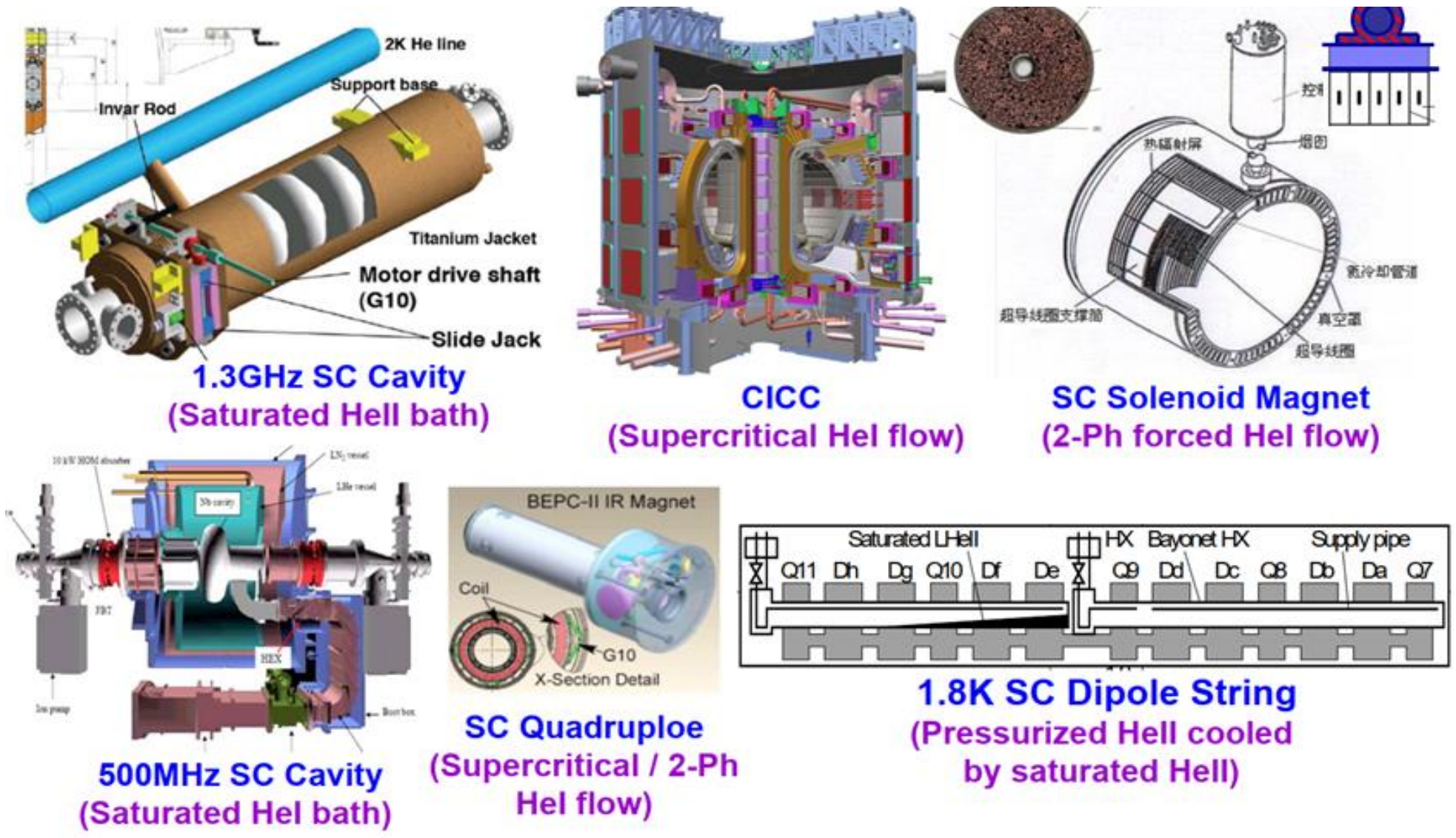

**Figure 7.1.11:** Cooling schemes of the SRF cavity and IR SC magnets.

The RF cavities are cooled using saturated He II at a temperature of 2K. However, cooling at this temperature is associated with a high thermodynamic cost. Therefore, the design of the cryogenic components focuses on intercepting heat loads at higher temperatures. To achieve this, helium-gas-cooled shields intercept both radiation and conduction at two different temperature ranges: 40~80 K and 5~8 K. The 40~80 K thermal shield serves as the first major heat intercept, shielding the cold mass from most of the heat leaks that come from ambient temperature. The 5~8 K shield provides lower temperature heat interception. During operation, a one-phase helium of 2.2K and 1.2 bar is supplied by the refrigerator to all cryomodules.

The cooling scheme for the SRF cryomodules in both the Booster and Collider has been modified from parallel in the CDR to series in the TDR. This modification aims to improve cavity performance while reducing costs. The flow diagrams for the modified cooling scheme are shown in Figure 7.1.12 (CDR) and Figure 7.1.13 (TDR), respectively.

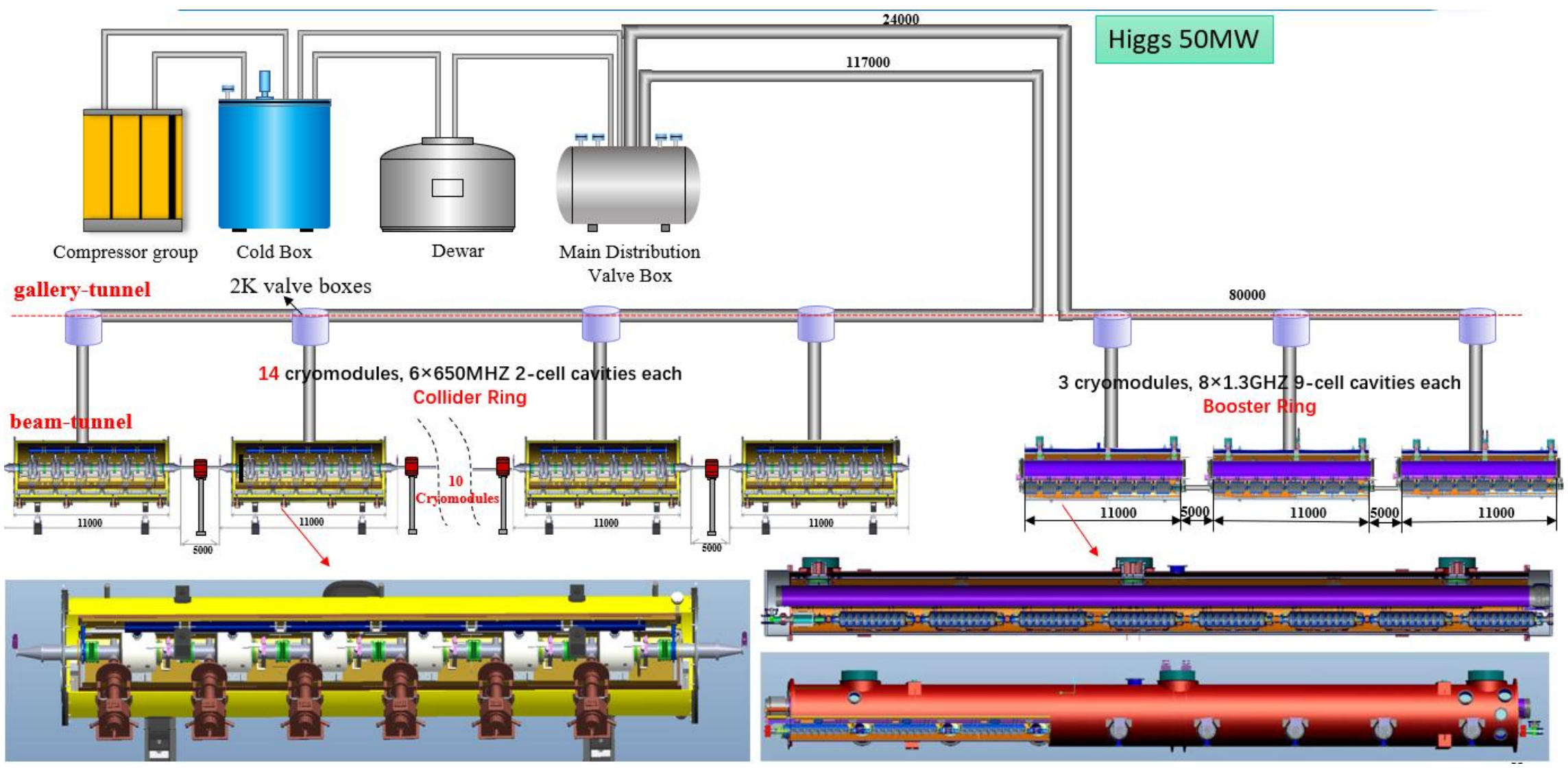

**Figure 7.1.12:** Flow chart of the SRF cryogenic system in CDR.

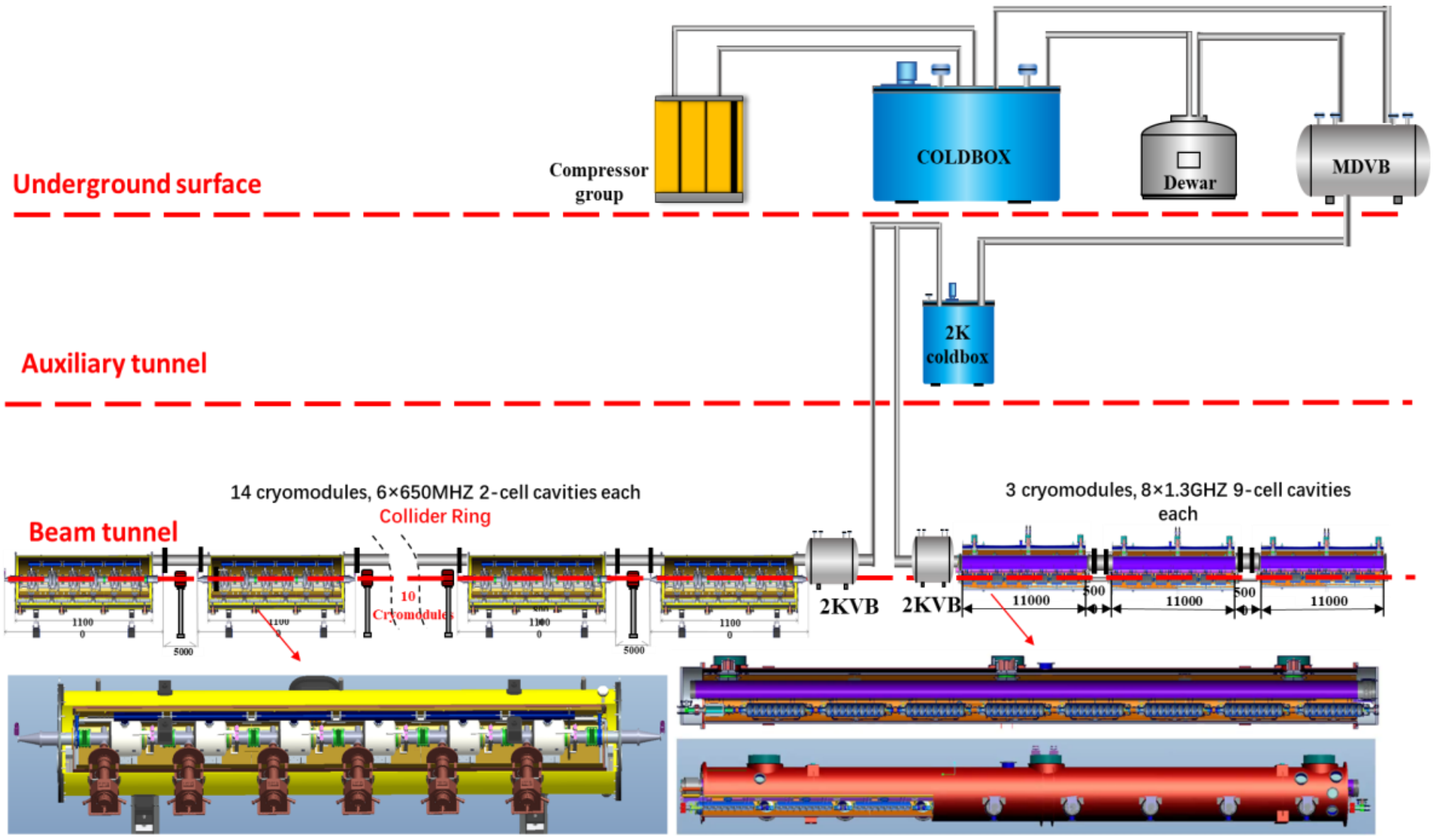

**Figure 7.1.13:** Flow chart of SRF cryogenic system in TDR.

Advantages and characteristics of the series SRF scheme, when compared to the parallel scheme, are outlined below:

(1) Simplified Process and Cost Efficiency: One notable advantage of the series SRF scheme lies in the potential elimination of numerous small valve boxes. Valve boxes play a critical role in cryogenic systems, controlling the flow of cryogenic gases and maintaining specific temperatures. The removal of these valve boxes translates to cost savings, as installation, maintenance, and operation costs can be substantial. Moreover, it streamlines system design and mitigates potential valve-related issues. Furthermore, the series scheme's simpler process can be attributed to the reduced complexity resulting from the removal of these valve boxes.

(2) Maintenance: Within the series SRF scheme, a vacuum partition is strategically placed every four cryomodules to ensure optimal performance and facilitate maintenance. These partitions act as barriers, isolating each section of the system. This isolation is pivotal, ensuring that issues or maintenance activities in one section do not disrupt the entire series of cryomodules. It enhances system stability and reliability by containing potential disruptions. Additionally, in the series scheme, if a specific cryomodule experiences suboptimal cavity performance, it can be temporarily warmed up to 50K while the other cryomodules continue operating at 4.5K. After necessary adjustments are completed, the cryomodule can be swiftly cooled down to 4.5K and then further lowered to 2K.

In summary, the series SRF scheme offers several distinct advantages over the parallel scheme, including reduced process complexity, cost efficiency due to valve box elimination, simplified system design, ease of implementation, and enhanced maintenance capabilities. These characteristics make the series SRF scheme an appealing choice for cryogenic systems utilized in scientific research facilities and particle accelerators, where precision and reliability are paramount.

The EcosimPro software is used to simulate the process calculation of the CEPC SRF cryogenic system, with a heat leakage of 0.15 W/m @ 2K set for the transfer line. The process flow calculation diagrams for both the CDR and the TDR are presented in Figure 7.1.14. Additionally, using the same method, the heat leakage for the 5~8K and 40~80K thermal shields is set to 0.5 W/m and 2 W/m, respectively. Apart from steady-state calculations, non-steady-state processes such as warming-up (WU) and cooldown (CD) are also considered in the design of cryogenic systems. The pressure drop of the HRGP is less than 0.004 bar. By implementing the TDR cooling process scheme, the length of cryogenic transfer lines can be reduced by 15%, and an estimated heat loss of 90 W @2K for each cryo-station can be saved.

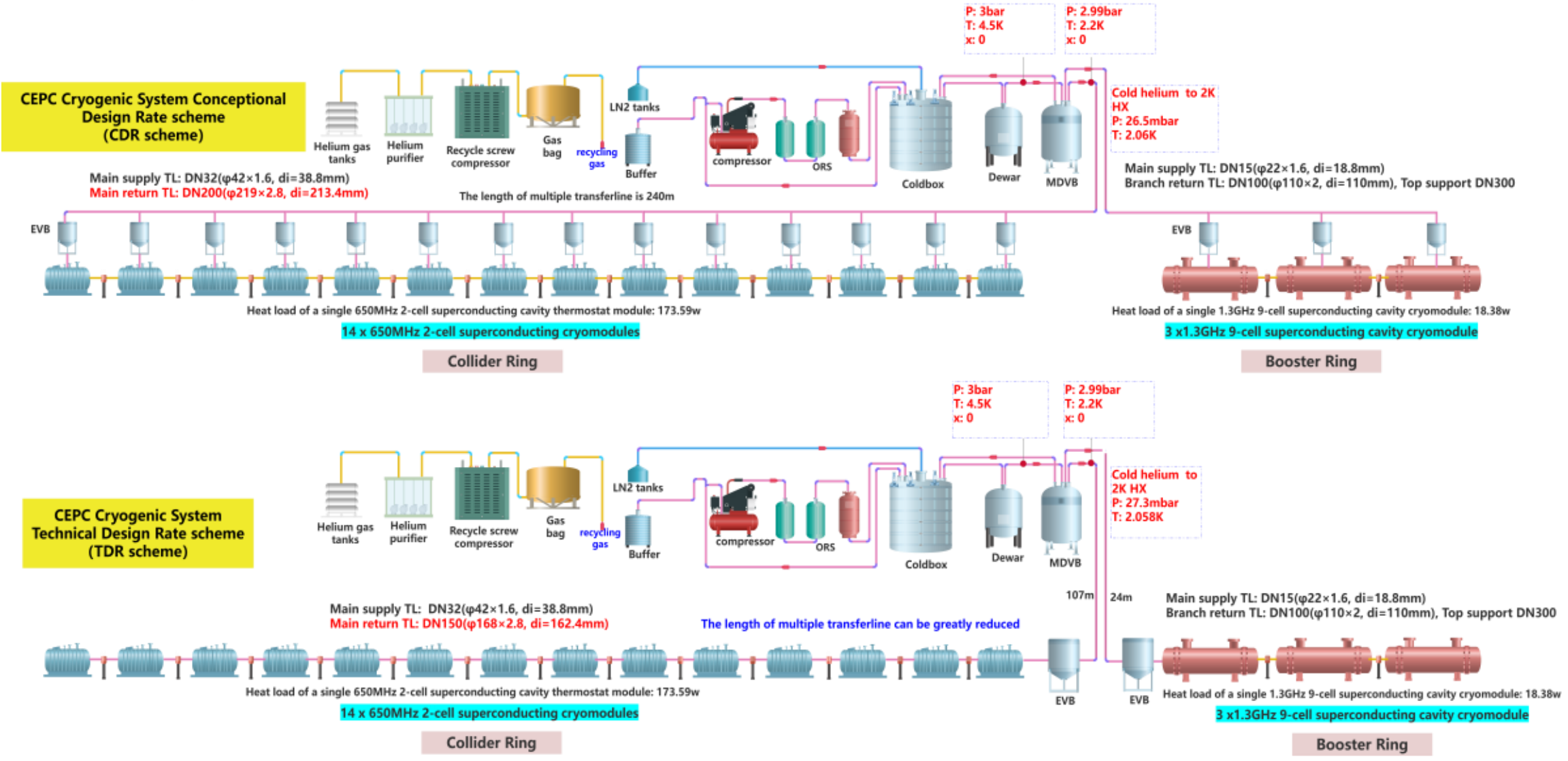

**Figure 7.1.14:** Diagram of process flow calculation in both CDR and TDR.

### 7.1.5.2 *Cryo-strings and Structure Design for SRF Cryomodules*

The RF cavities are cooled by saturated He II at 2K, and due to the high thermodynamic cost of refrigeration at this temperature, the design of the cryogenic components aims to intercept heat loads at higher temperatures. During operation, a one-phase helium of 2.2K and 1.2 bar is provided by the refrigerator to all cryomodules, each of which has two valves. The JT-valve is used to expand helium to a liquid helium separator. A two-phase line, consisting of a liquid-helium supply and concurrent vapor return, connects each helium vessel and is connected to the major gas return header once per module. Additionally, a small diameter warm-up/cool-down (CD) line connects the bottom of the helium vessels at both ends through a CD valve. The cavities are immersed in baths of saturated superfluid helium, which is gravity filled from a 2K two-phase header. The saturated superfluid helium flows along the two-phase header, which is connected to the pumping return line and then to the refrigerator. A cryomodule in a series-connected string diagram is shown in Figure 7.1.15.

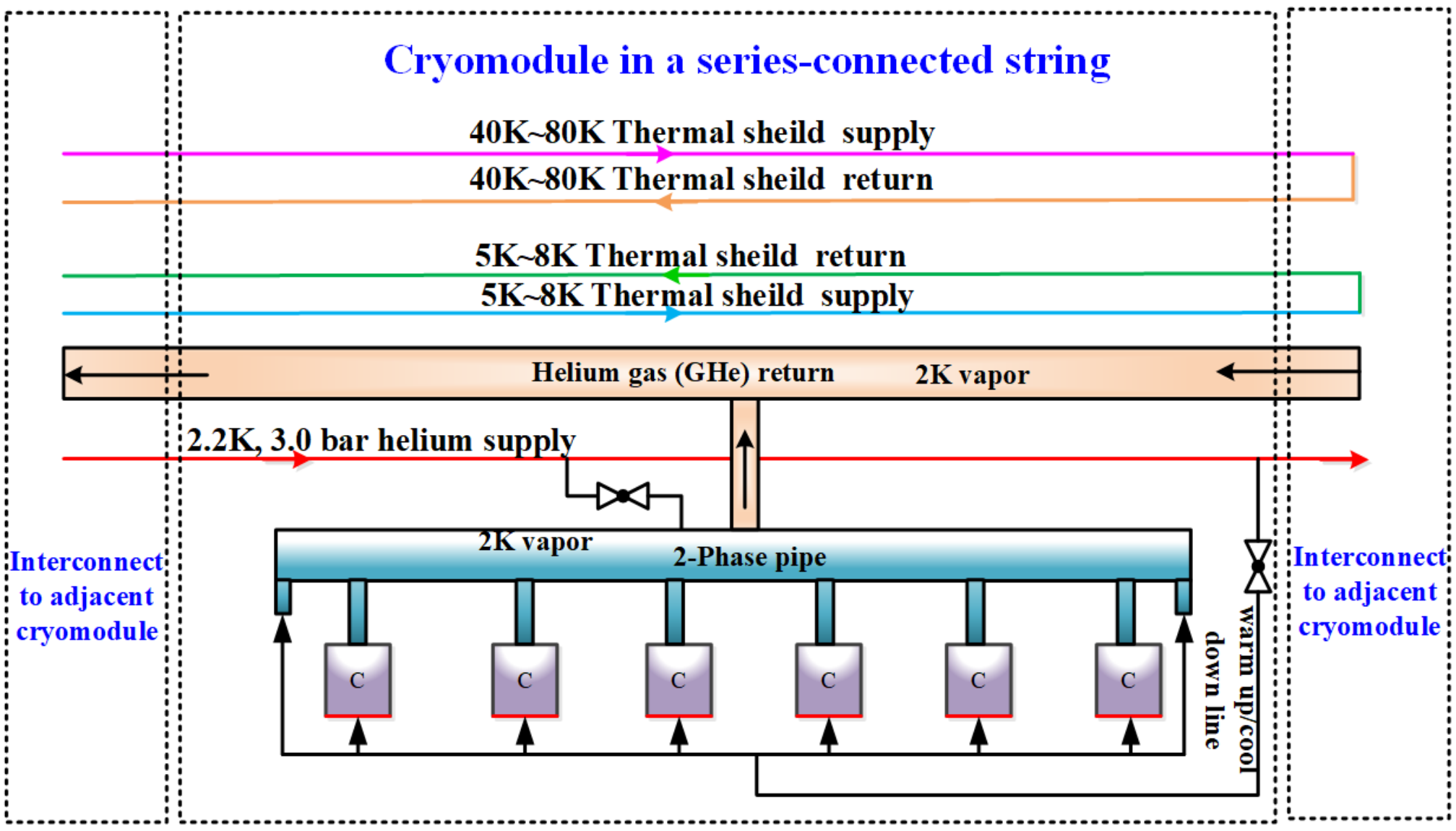

**Figure 7.1.15:** Cryomodule in a series-connected string.

Each set of 1.3 GHz cryomodules in the Booster consists of eight 1.3 GHz 9-cell superconducting cavities, eight primary couplers, eight tuners, a superconducting magnet assembly, a beam position detector (BPM), and a high sub-mode absorber (HOM absorber) located on the beamline between the two modules. The superconducting high frequency cavity operates at a temperature of 2K, and the average unloaded quality factor $Q_0$ required for 16 MV/m accelerated gradient in CW mode should be greater than or equal to $2 \times 10^{10}$. The dynamic thermal load per superconducting cavity is 14 W @ 2K. To ensure proper functioning of the module, the total 2K thermal load per module set, which includes the static thermal load and the main coupler and high secondary mode dynamic thermal load, should be less than 130 W.

For the EXFEL project, IHEP has manufactured 58 1.3 GHz 9-cell cryomodules in cooperation with local companies. This successful collaboration has laid a solid foundation for the design. The manufactured 1.3 GHz cryomodules are depicted in Figure 7.1.16.

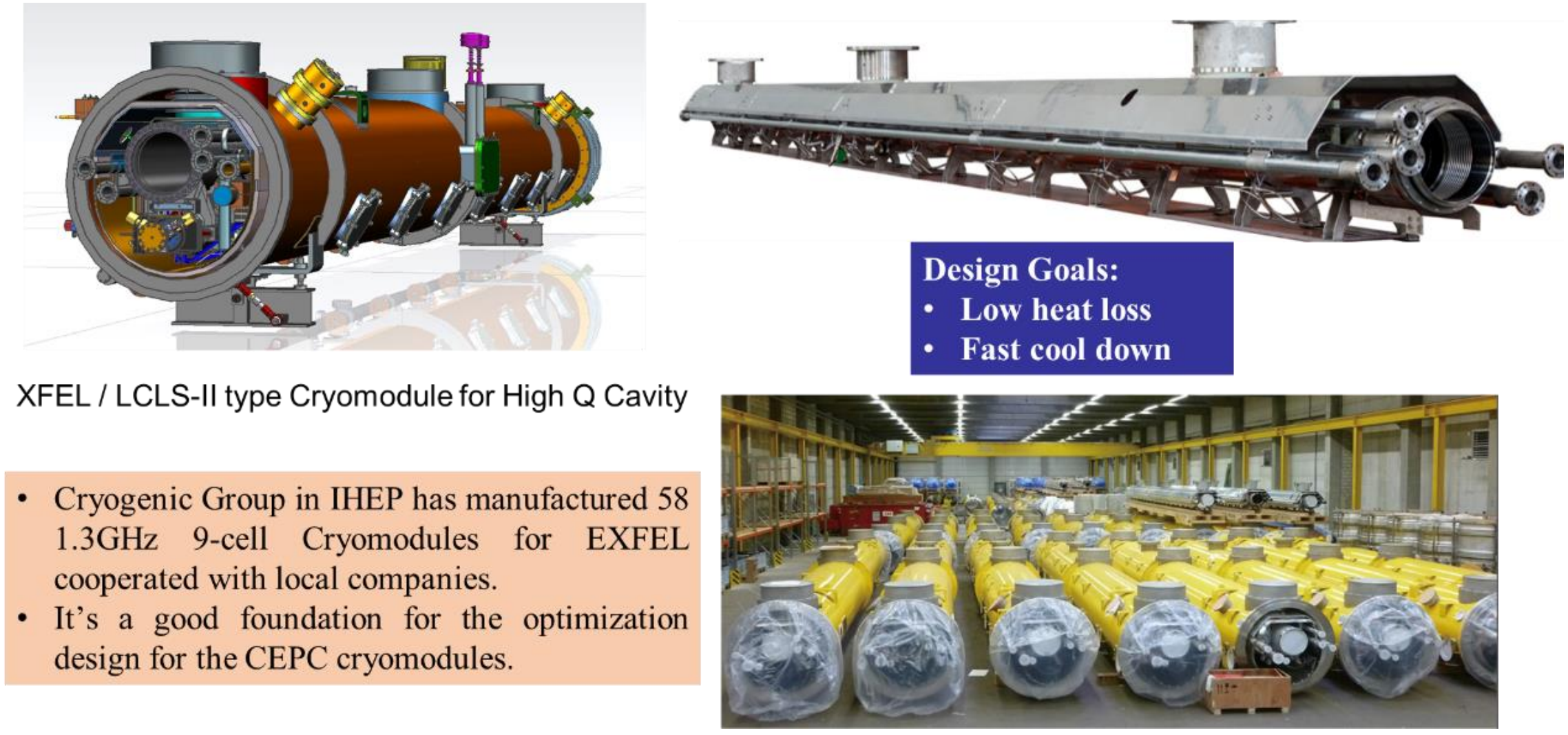

**Figure 7.1.16:** The 58 1.3 GHz 9-cell cryomodules manufactured by IHEP in collaboration with companies in China for the EXFEL.

The cooling requirements for the 1.3GHz 9-cell SRF cavity cryomodule are depicted in Figure 7.1.17 and can be explained as follows:

• First cooling stage: 300K to 150K, which is referred to as "precooling." During this stage, the cooling process must be slow until most of the heat shrinkage is completed. In addition, the following parameters must be met:

- ➢ HRGP: Radial temperature gradient < 15K
- ➢ Axial temperature gradient < 50K
- ➢ Single point cooling rate < 10K/h.

• Second cooling stage: 150K to 45K.

• Third cooling stage: 45K to 4.5K

- ➢ FCD is defined as a cooling rate of (2~3) K/min. Assuming a 3K/min cooldown rate from 45K, the temperature of the SRF cavity can quickly cross the critical temperature point of 9.2K within 10 minutes.

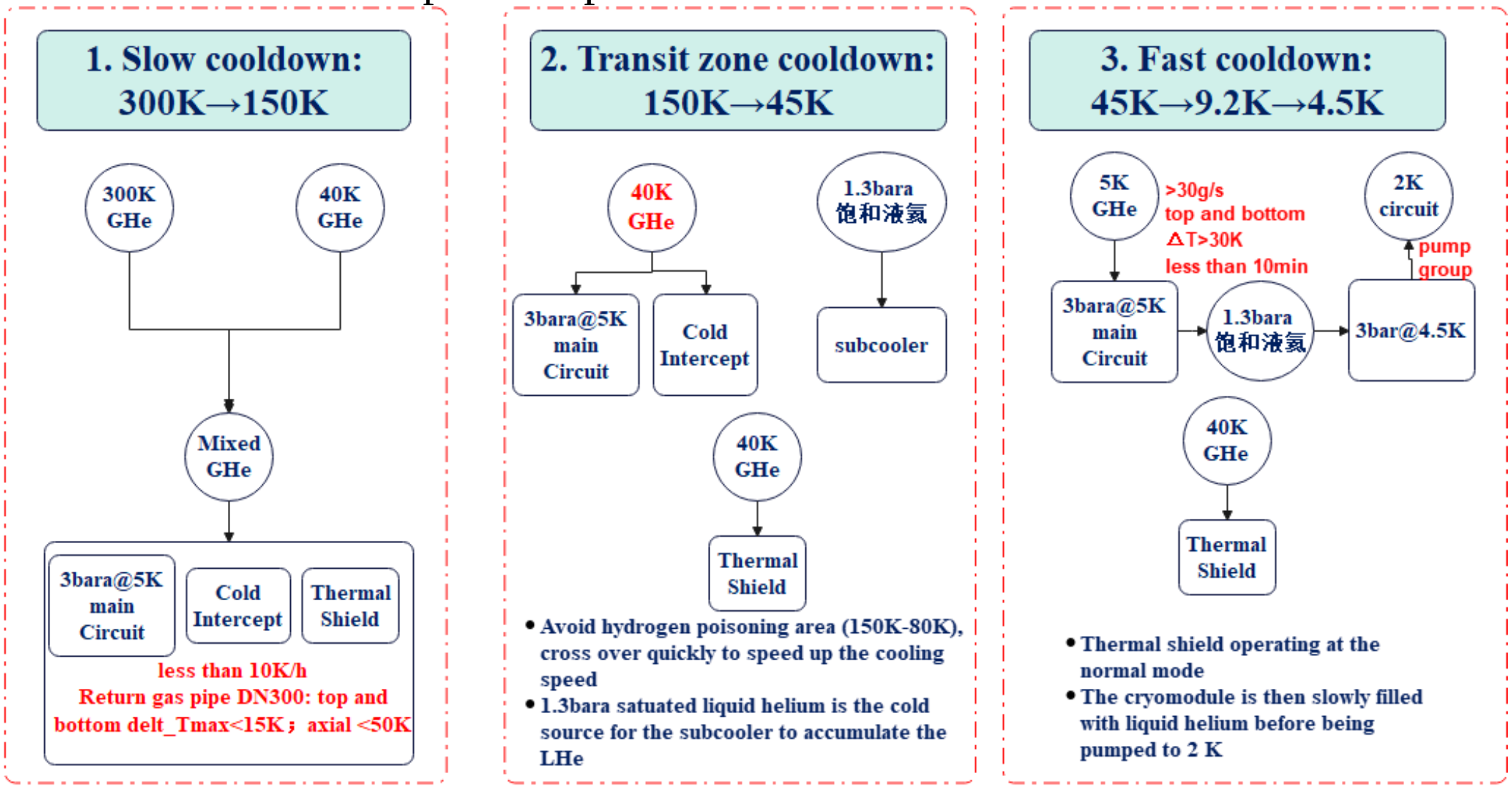

**Figure 7.1.17:** Cooling scheme for the 1.3 GHz 9-cell SRF cavity module.

The unsteady-state simulation of the CM cooldown process for the 1.3GHz 9-cell SRF cavities has been studied and is presented in Figure 7.1.18.

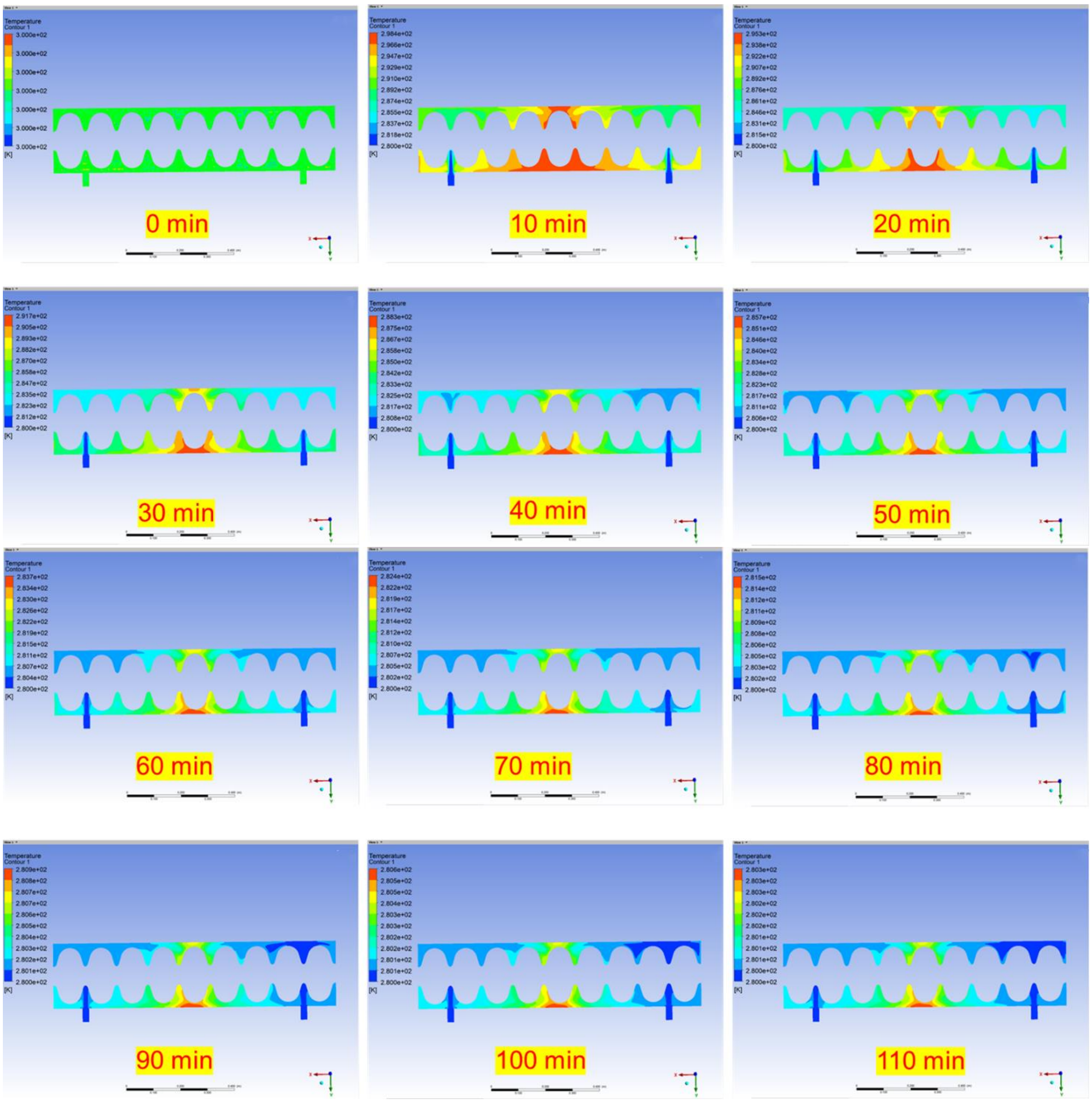

**Figure 7.1.18:** Unsteady-state simulation for the 1.3GHz 9-cell cavities CM cooldown process.

The 650 MHz 2-cell SRF cavities cryomodules in the Collider consist of six 2-cell 650 MHz SRF cavities, six high-power couplers, six mechanical tuners, and two HOM absorbers, with a total length of approximately 9.5 meters. Table 7.1.3 provides the basic mechanical parameters of the 650 MHz 2-cell cavities cryomodule, and Table 7.1.4 presents additional parameters of the same cryomodule. Additionally, Figure 7.1.19 illustrates the mechanical 3D diagram of the 650 MHz 2-cell cavities cryomodule.

**Table 7.1.3:** Basic mechanical parameters of the 650 MHz 2-cell cavities cryomodule.

| **Six 2-cell Cavities Cryomodule** | |
|---|---|
| Overall length (flange to flange, m) | 8.0 |
| Diameter of Vacuum vessel, m | 1.3 |
| Beamline height from floor, m | 1.5 |
| Cryo-system working temperature, K | 2 |
| Number of 200-POST | 6 |

**Table 7.1.4:** Parameters of the 650 MHz 2-cell cavities cryomodule

| **Cryomodule performance** | | **Specification** |
|---|---|---|
| Number of leakages | He →insulation | 0 |
| | He →beam transfer line | 0 |
| | Insulation →coupler | 0 |
| | Insulation →beam transfer line | 0 |
| | Coupler →beam transfer line | 0 |
| Alignment x/y inside (Cavities) | | ±0.5mm |
| Alignment z inside | | within 2 mm |
| Coupler antenna design z | | within 2 mm |

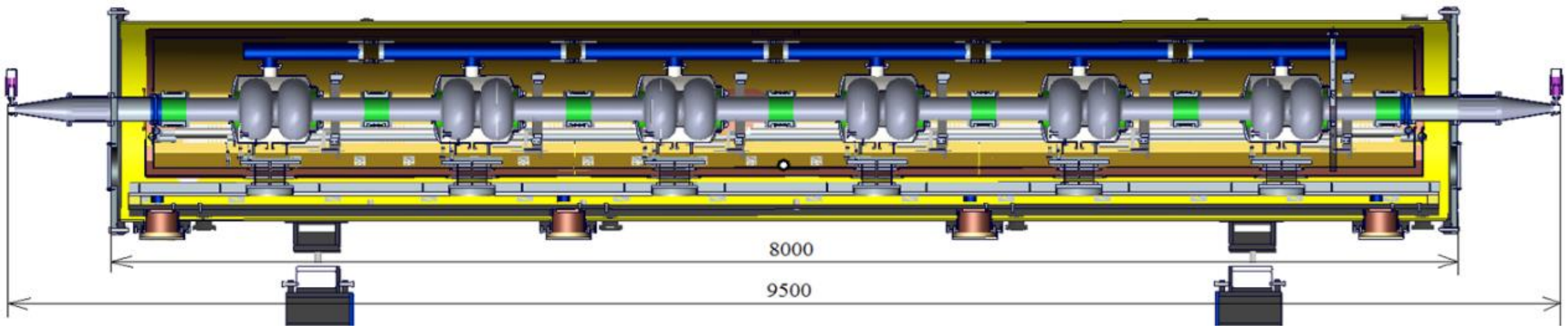

**Figure 7.1.19:** Mechanical 3D diagram of the 650 MHz 2-cell cavities cryomodule.

Unsteady-state simulations have been conducted to study the cooldown process of the 650 MHz 2-cell cavities. It is crucial to strictly control the cooling rate of the superconducting cavity. Hence, a three-dimensional, non-stationary numerical simulation of the cooling process of the 650 MHz 2-cell SRF cavity was performed to establish a simulation basis for achieving automatic cooldown in subsequent experiments. Figure 7.1.20 presents the temperature distribution and velocity domain obtained from the simulation.

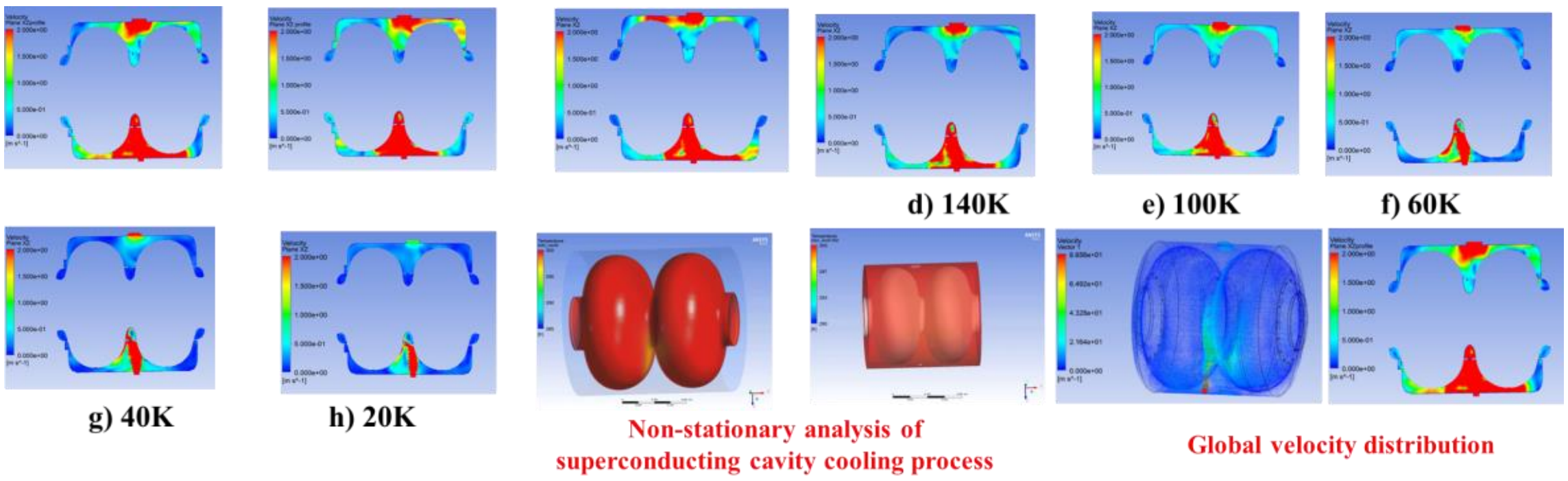

**Figure 7.1.20:** Temperature distribution and velocity domain in the 650 MHz 2-cell SRF cavities cooldown process.

## 7.1.6 Advanced SRF Test Stands and Infrastructure

A superfluid helium cryogenic system named the Platform of Advanced Photon Source Technology R&D (PAPS) was successfully developed to facilitate the performance testing of SRF cavities and cryomodules, as well as research on superfluid helium performance. The PAPS system underwent performance testing on June 18th, 2021, and was designed to provide a strong foundation and ideal conditions for R&D, engineering testing, and verification of the High Energy Photon Source (HEPS) project's superconducting cavities. The PAPS cryogenic system has three vertical test dewars, two horizontal test cryostats, and two beam test cryostats for superconducting cavities, with a heat load of approximately 300 W at 2K. It is a 2.5 KW @4.5K or 300 W@2K superfluid helium cryogenic system designed to support the performance testing of various types of superconducting cavities, such as those operating at 166 MHz, 325 MHz, 500 MHz, 650 MHz, and 1.3 GHz. Figure 7.1.21 shows the process flow diagram (PFD) of the PAPS cryogenic system. The PAPS system successfully passed its commissioning test in June 2021.

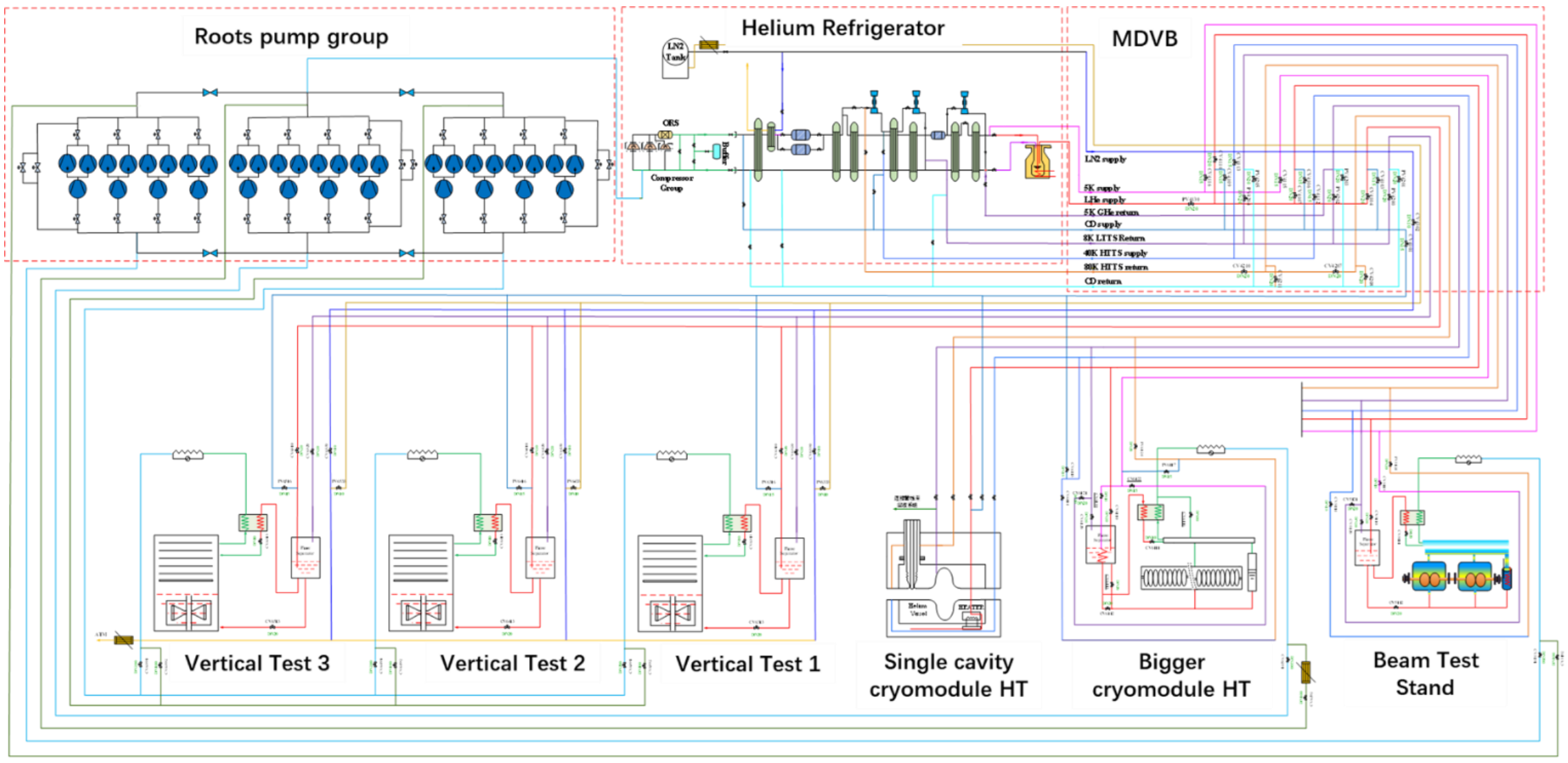

**Figure 7.1.21:** Process flow diagram of the PAPS cryogenic system.

Figure 7.1.22 shows the three-dimensional diagram of the PAPS cryogenic system, which mainly consists of three zones: the cryogenic hall (3#), the radio frequency (RF) hall (2#), and the tanks zone. The cryogenic hall houses the compressor unit, helium recovery and purification system, and other related equipment. The RF hall is equipped with a liquid helium coldbox, vertical/horizontal/beam test cryostats, distribution valve boxes, and other cryogenic equipment. The tanks zone contains two liquid nitrogen tanks (40 m$^3$), five middle pressure helium tanks (100 m$^3$, 1.6 MPa), and four high-pressure helium gas cylinders (20 MPa).

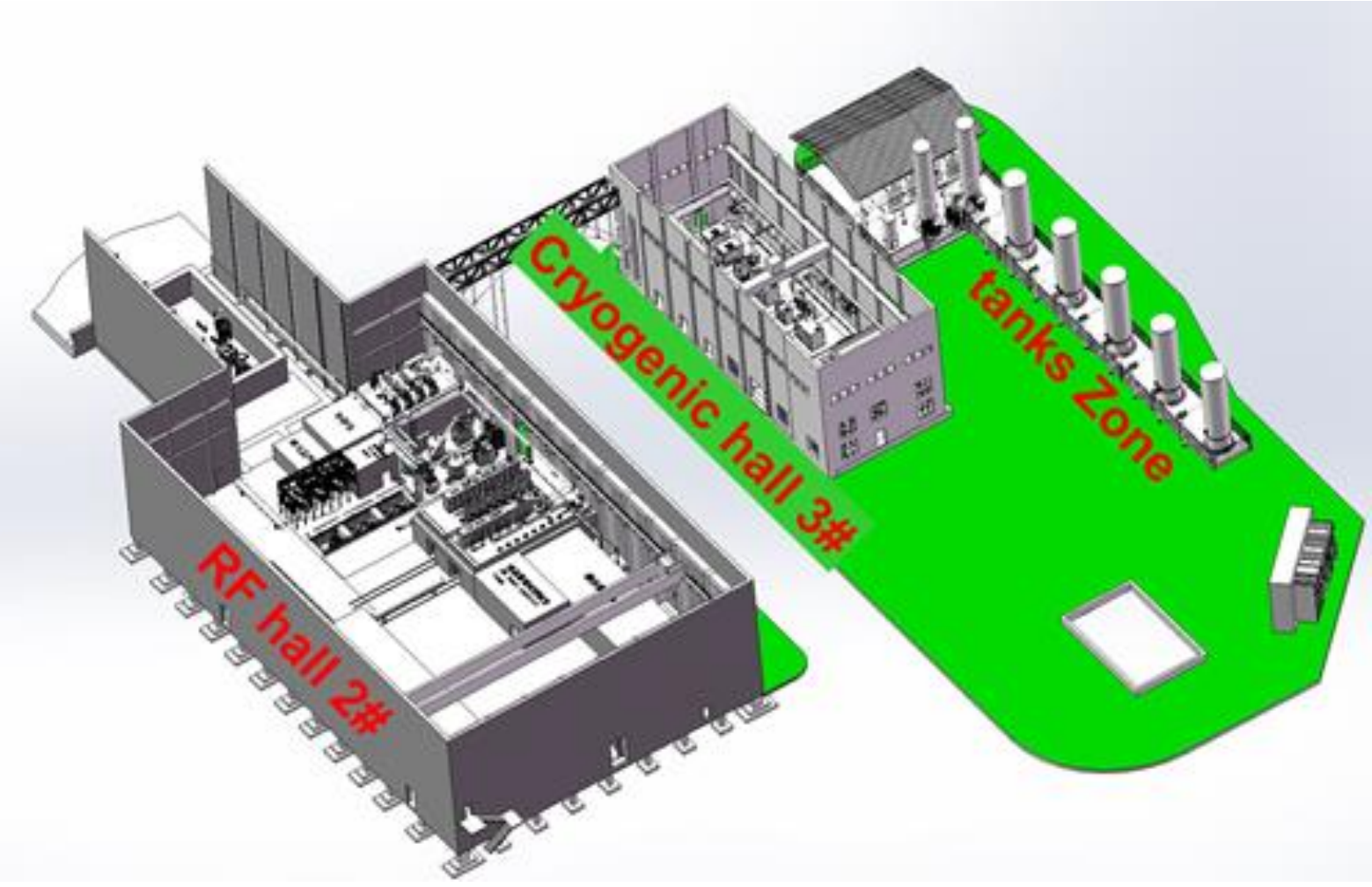

a)

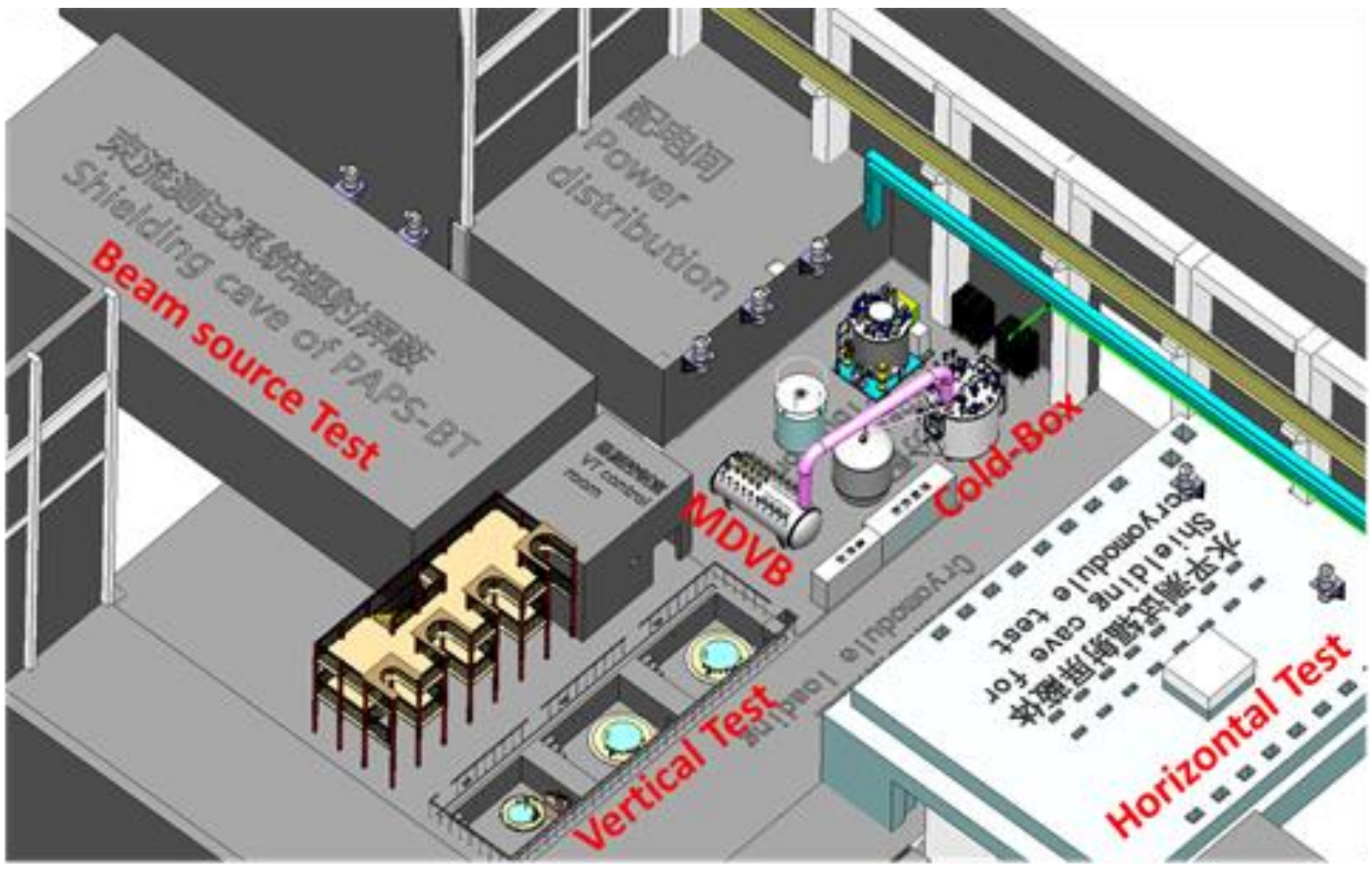

b)

**Figure 7.1.22:** Three-dimensional diagram of the PAPS cryogenic system: a) overall layout, b) RF hall layout.

Figure 7.1.23 shows photos of the cryogenic hall, RF hall, and tanks zone of the PAPS cryogenic system. The system includes three vertical test stands, two horizontal test stands, and a beam test stand.

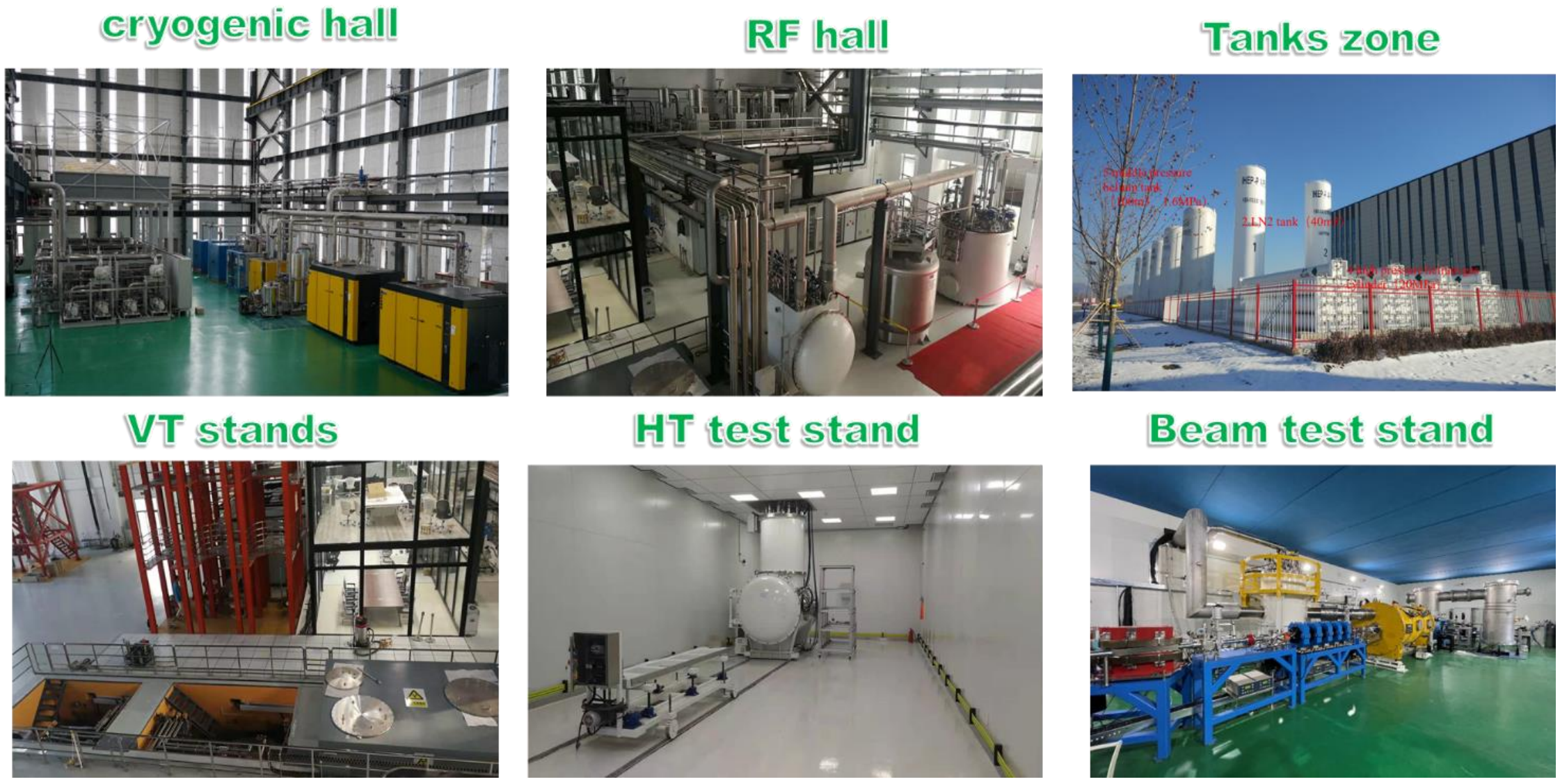

**Figure 7.1.23:** PAPS cryogenic system for RF cavity R&D.

The PAPS facility has successfully tested various types of SRF cavities, including those operating at 166 MHz, 325 MHz, 500 MHz, 650 MHz, and 1.3 GHz. Additionally, prototype cryomodules for the CEPC Collider and Booster ring have been or are planned to be tested in the PAPS system.

The 2K horizontal test for the 1.3 GHz 9-cell cryomodule is scheduled to be conducted in the first half of 2023 at PAPS. The fast cooldown process, which involves a large mass flow rate to force magnetic flux exclusion, can reduce the cooldown time from 40K to 9K to less than 10 minutes. To achieve this, a detailed process scheme has been designed, and the PID for the 1.3 GHz 9-cell cryomodule in PAPS and fast cooldown loop are shown in Figure 7.1.24. The mechanical 3D diagram of the cryomodule, which includes the 2K valve box, feed cap, cryomodule, and endcap, is shown in Figure 7.1.25. To facilitate loading and unloading of the cavities, a flat trolley will be used. As of now, all of the key components of the 1.3 GHz Cryomodule have been completed, and assembly began in October 2022, as depicted in Figure 7.1.26.

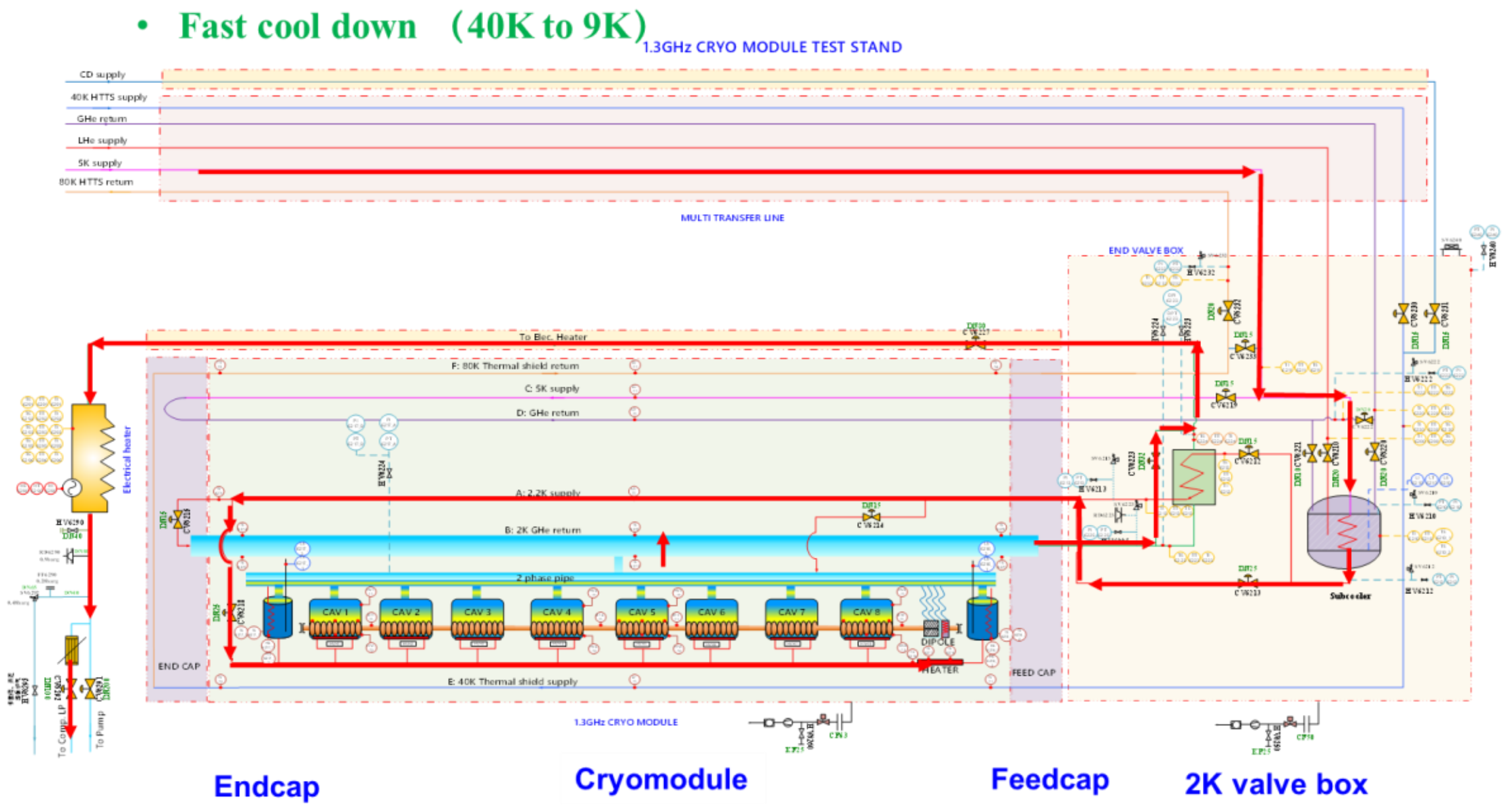

**Figure 7.1.24:** PID diagram for the 1.3 GHz 9-cell cryomodule in PAPS and fast cooldown loops.

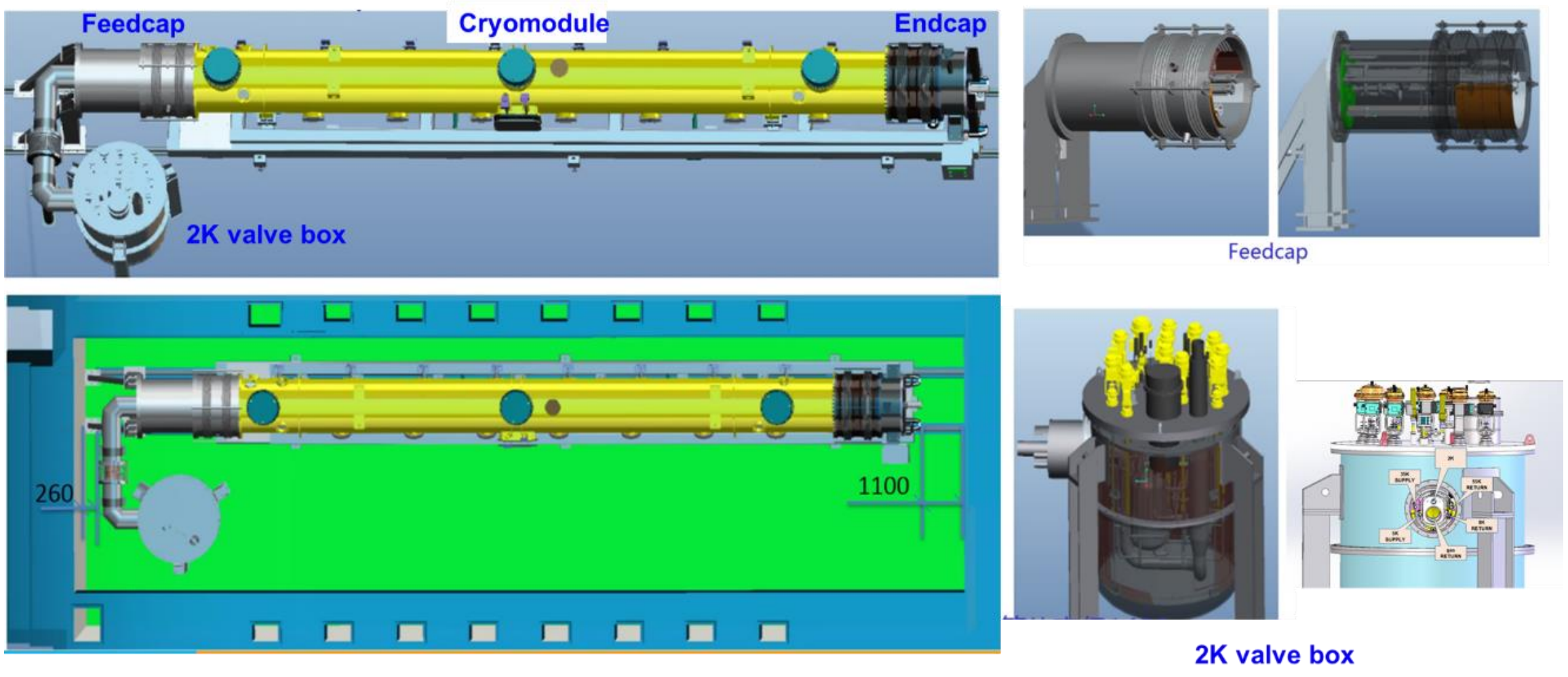

**Figure 7.1.25:** 3D drawings of the 2K valve box, cryomodule, feed cap and endcap.

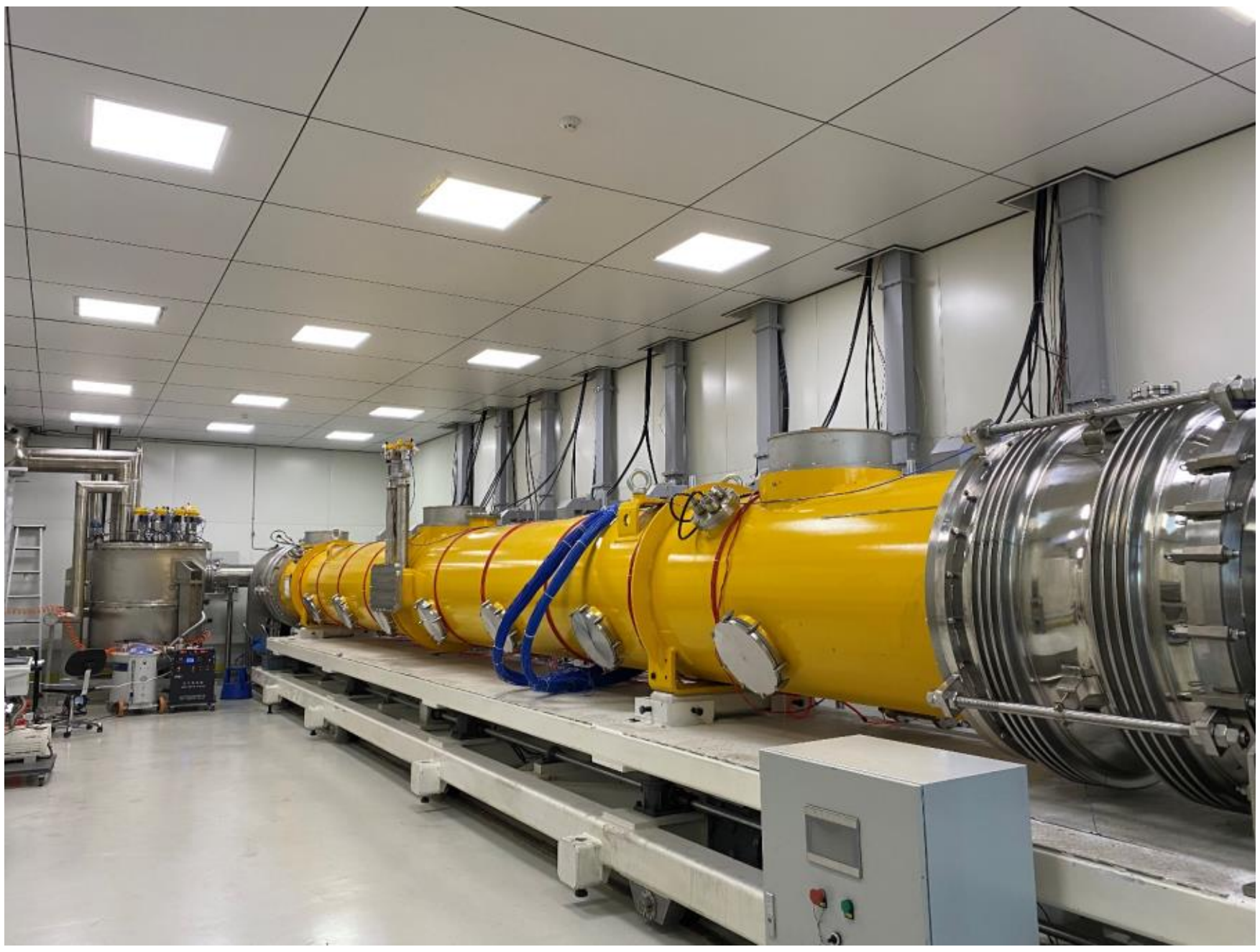

**Figure 7.1.26:** Assembly photo of the 1.3 GHz cryomodule.

A prototype cryomodule with two 2-cell 650 MHz superconducting cavities was tested in the PAPS system beam test stand last year, as shown in Figure 7.1.27. The testing results, including the cooldown curve and 2K stable operation, are shown in Figure 7.1.28.

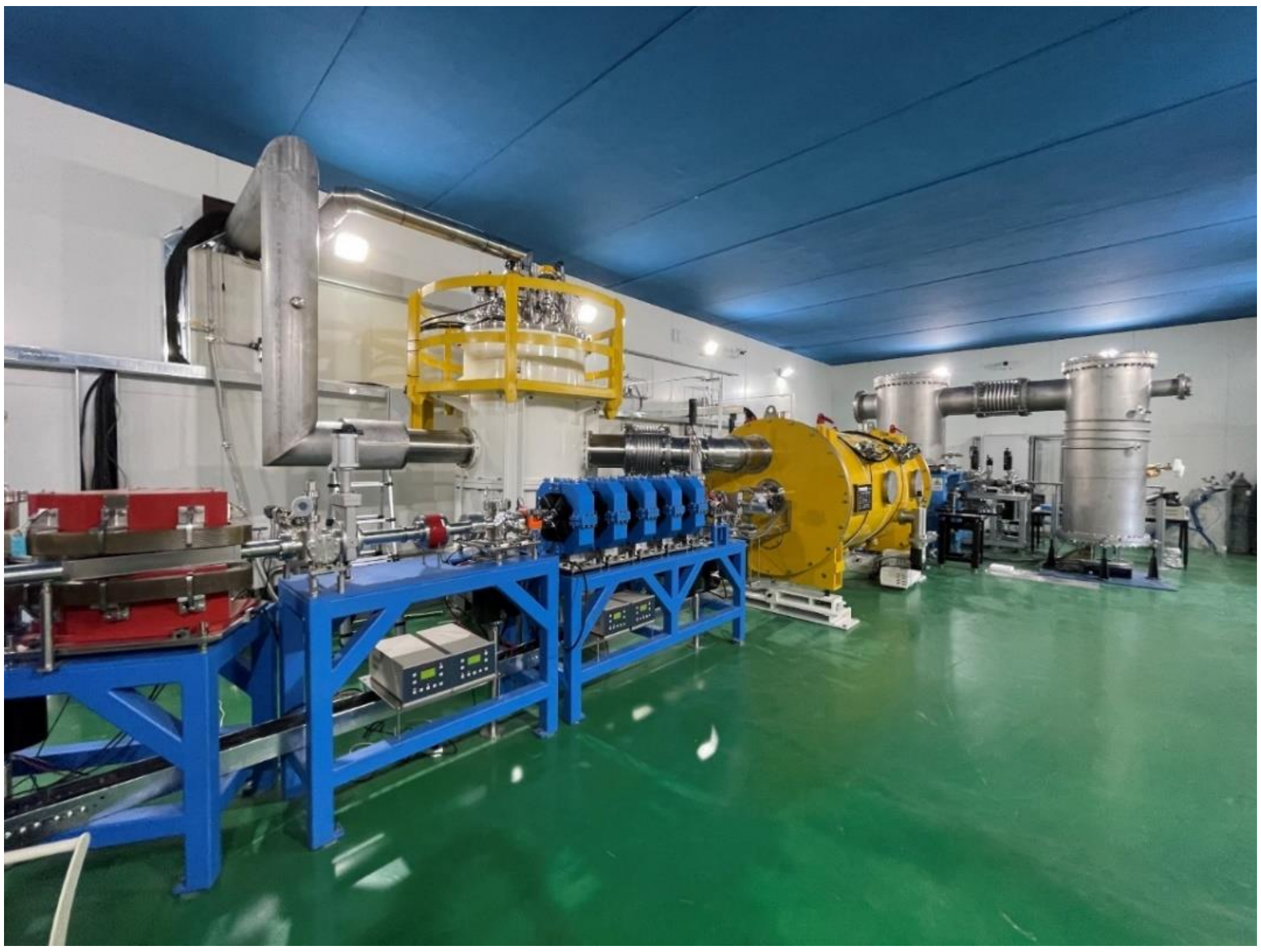

**Figure 7.1.27:** Prototype cryomodule with two 2-cell 650 MHz superconducting cavities on the PAPS beam test stand.

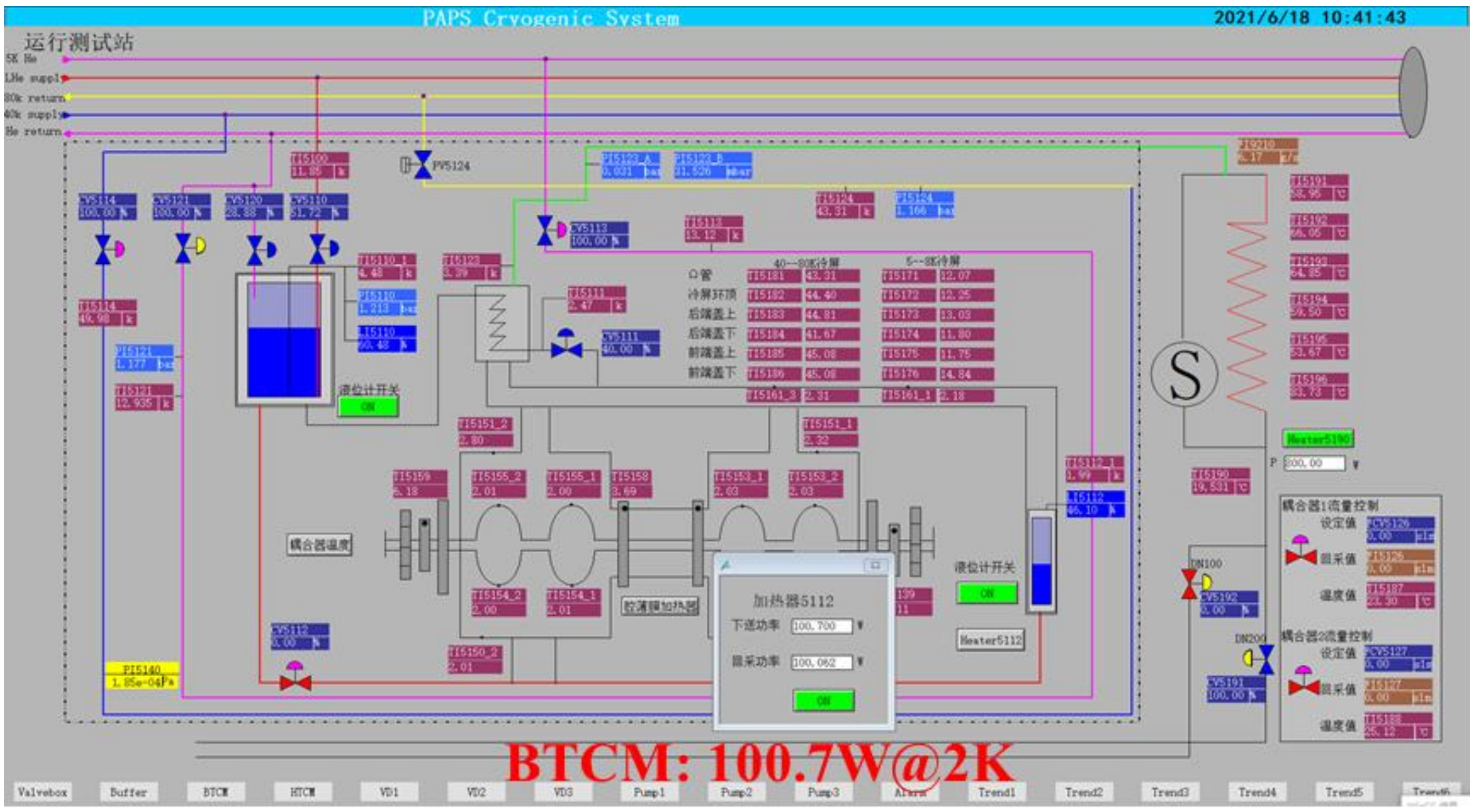

a) BTCM WinCC interface

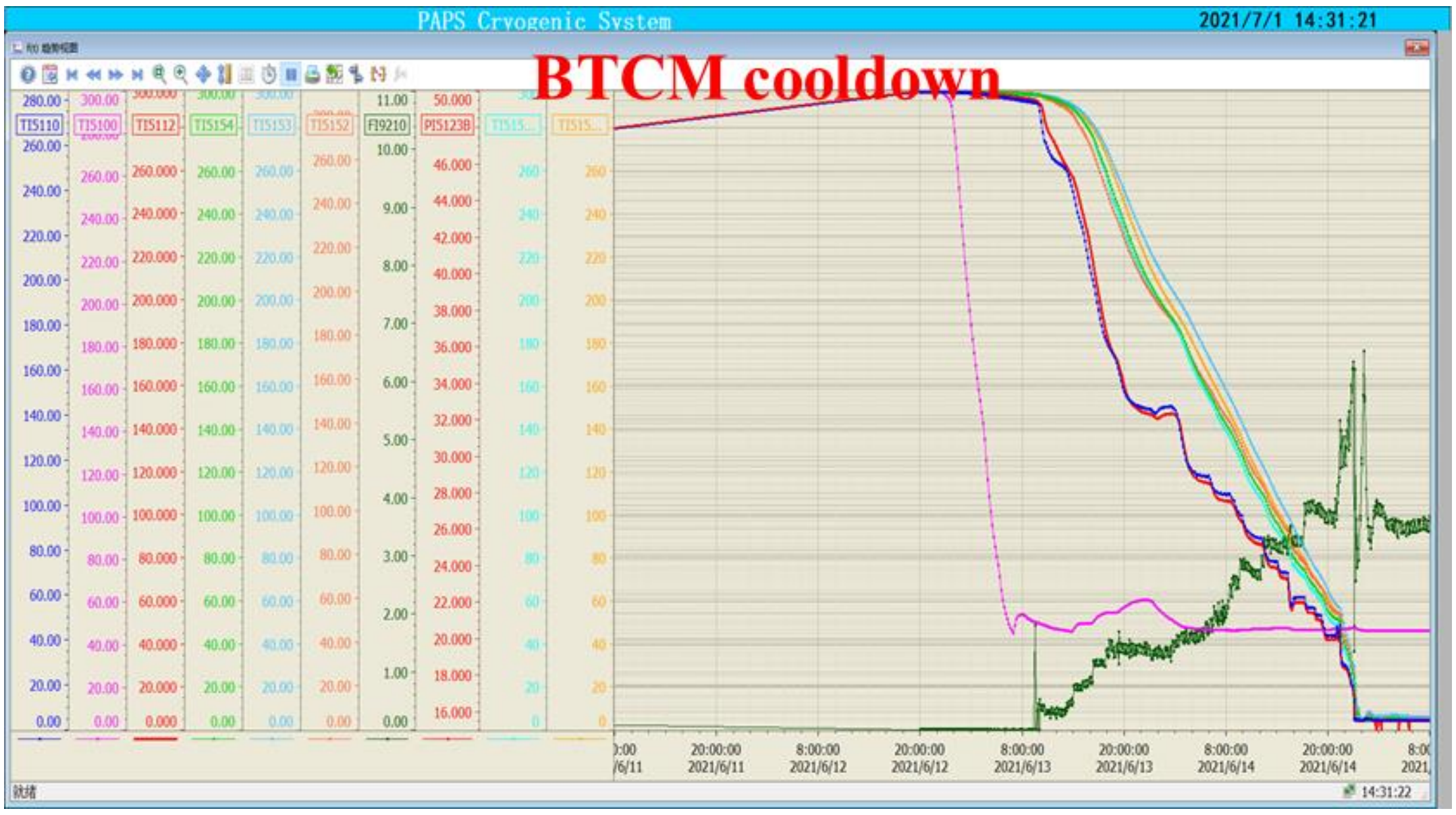

b) BTCM cooldown curve from 300K to 4.5K and 2K

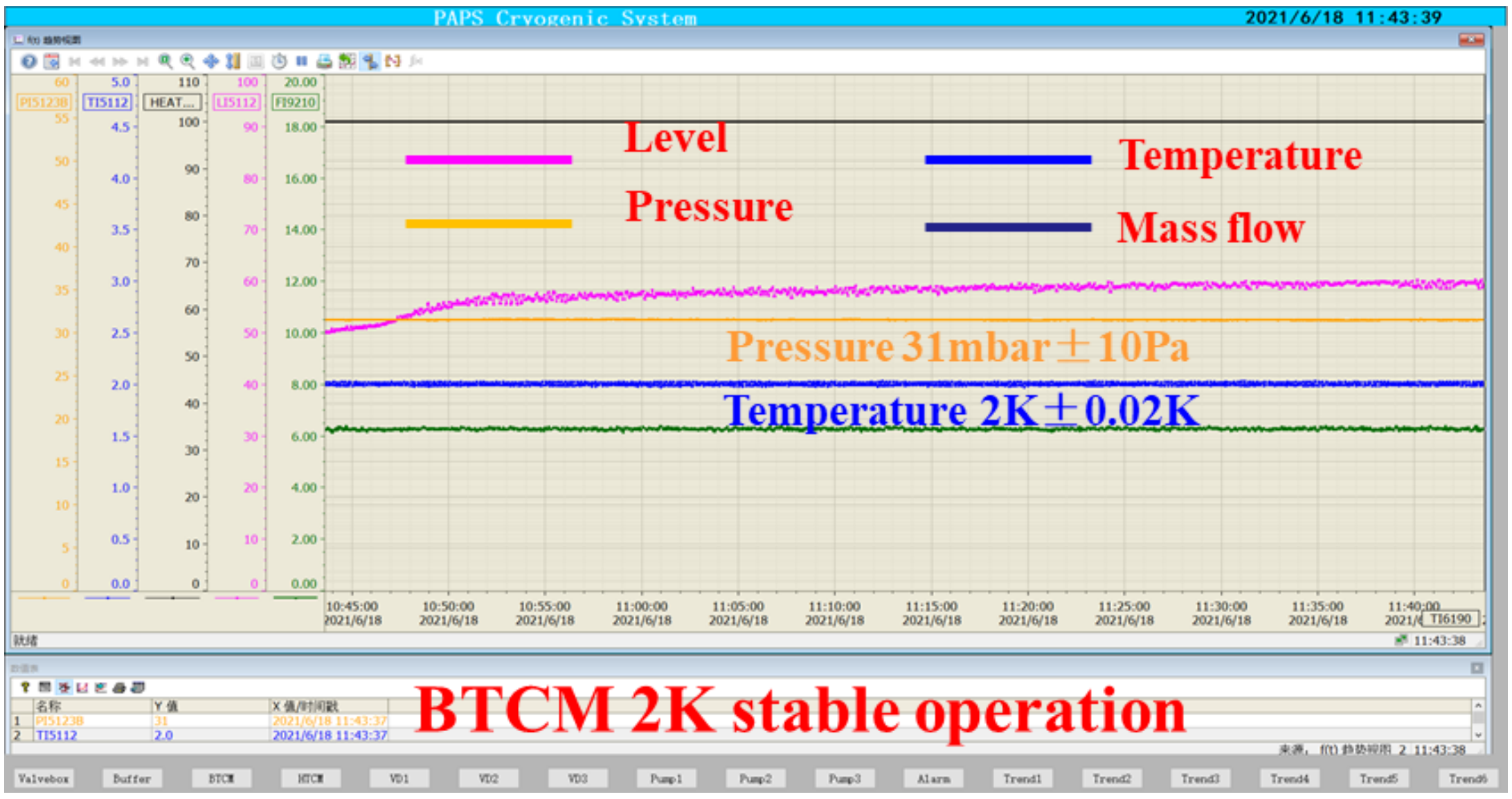

c) BTCM stable operation

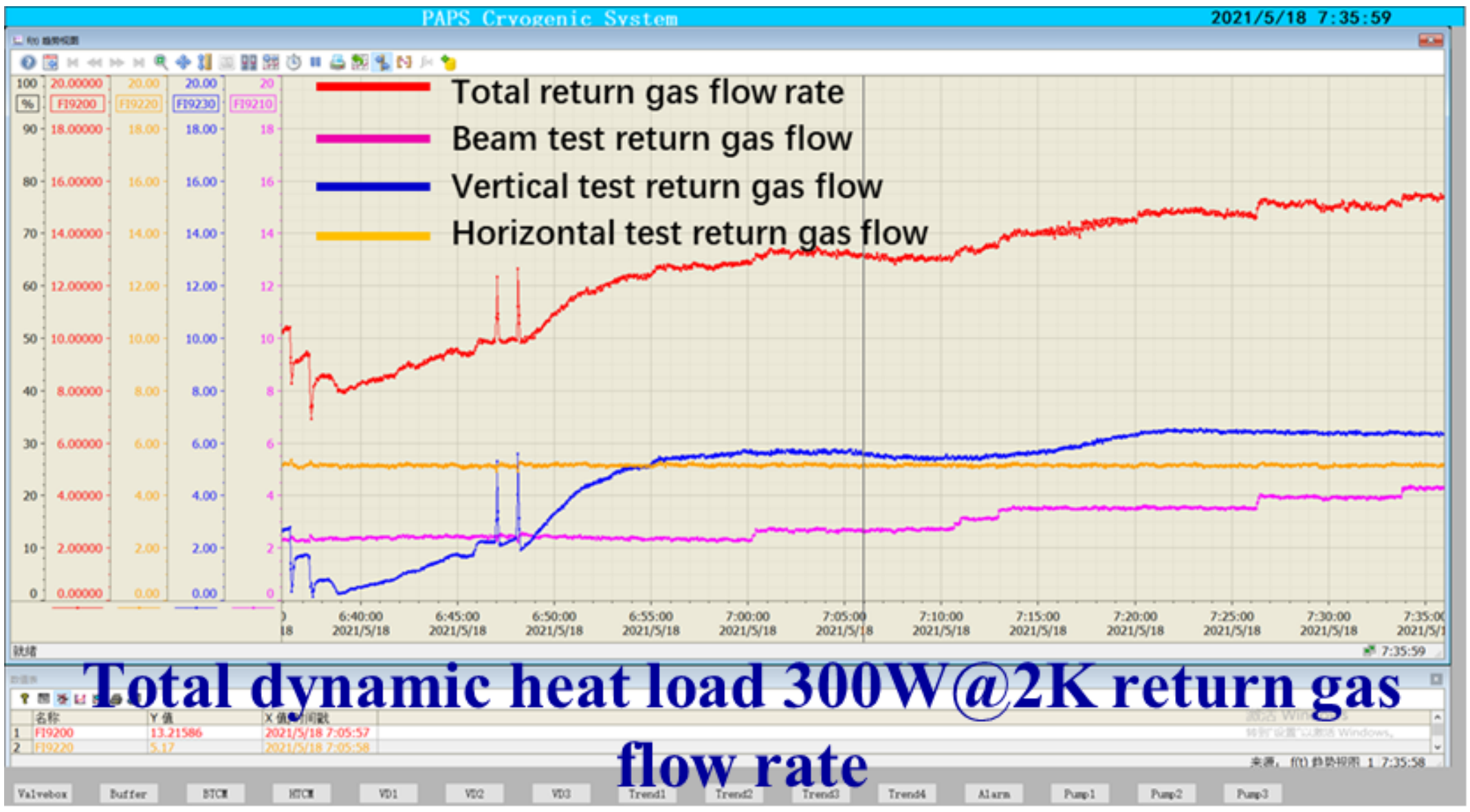

d) Total dynamic heat load 300W@2K return gas flow rate

**Figure 7.1.28:** Testing results of the two 2-cell 650 MHz superconducting cavities cryomodule in PAPS.

### 7.1.7 Large Refrigerator Cryoplant

According to Table 7.1.1, the heat loads necessitate a helium refrigerator cryoplant with a total capacity exceeding 49.96 kW at 4.5 K. To meet these requirements, four separate refrigerators will be used, each with a cryogenic station capacity of 15 kW at 4.5 K, including an operational margin.

Refrigerator design must take into account several aspects, such as cost, reliability, efficiency, maintenance, appearance, flexibility, and convenience of use. Our design focus has been on reducing the initial capital cost of the cryogenic system and the high energy costs associated with its operation over the life of the facility. This is essential as these costs represent a significant portion of the total project budget. Additionally, reliability is a major concern as any unscheduled downtime can significantly impact the experimental schedule.

The main components of the refrigerator include screw helium compressors with associated cooling, oil-removal systems (ORS), a 4.5K coldbox that is vacuum insulated and houses the aluminum plate-fin heat exchangers, and several stages of turbo-expanders. The 2K atmospheric coldbox includes cold compressors and 2K heat exchanger.

The basic cooling process involves expanding compressed helium gas to perform work against low-temperature expansion engines, and then recycling the lower pressure exhaust gas through a series of heat exchangers and subsequent compression. This process is a variant of the Carnot process that has been in use for several decades.

To reach superfluid temperatures, a reduction in helium vapor pressure is necessary, and this is accomplished through the use of multiple-stage cold compressors.

The cryoplant comprises five pressure levels: 20 bara, 4 bara, 1.05 bara, 0.4 bara, and 3 kPa. These are achieved throughthe use of high-pressure screw compressor group, middle pressure screw compressor group, warm compressors, and cold compressors. Additionally, there are six temperature levels in the system, with five-class turbine expansions and 11 heat exchangers.

At the 40K and 5K temperature levels, helium flows are directed to the thermal shields of the cryomodules, and the corresponding return flows are fed back to the refrigerator at appropriate temperature levels. Inside the refrigerator cold-box, switchable absorbers at the 80K and 20K temperature levels are used to purify the helium of residual air, neon, and hydrogen.

Refer to Figure 7.1.29 for the refrigerator process flow diagram.

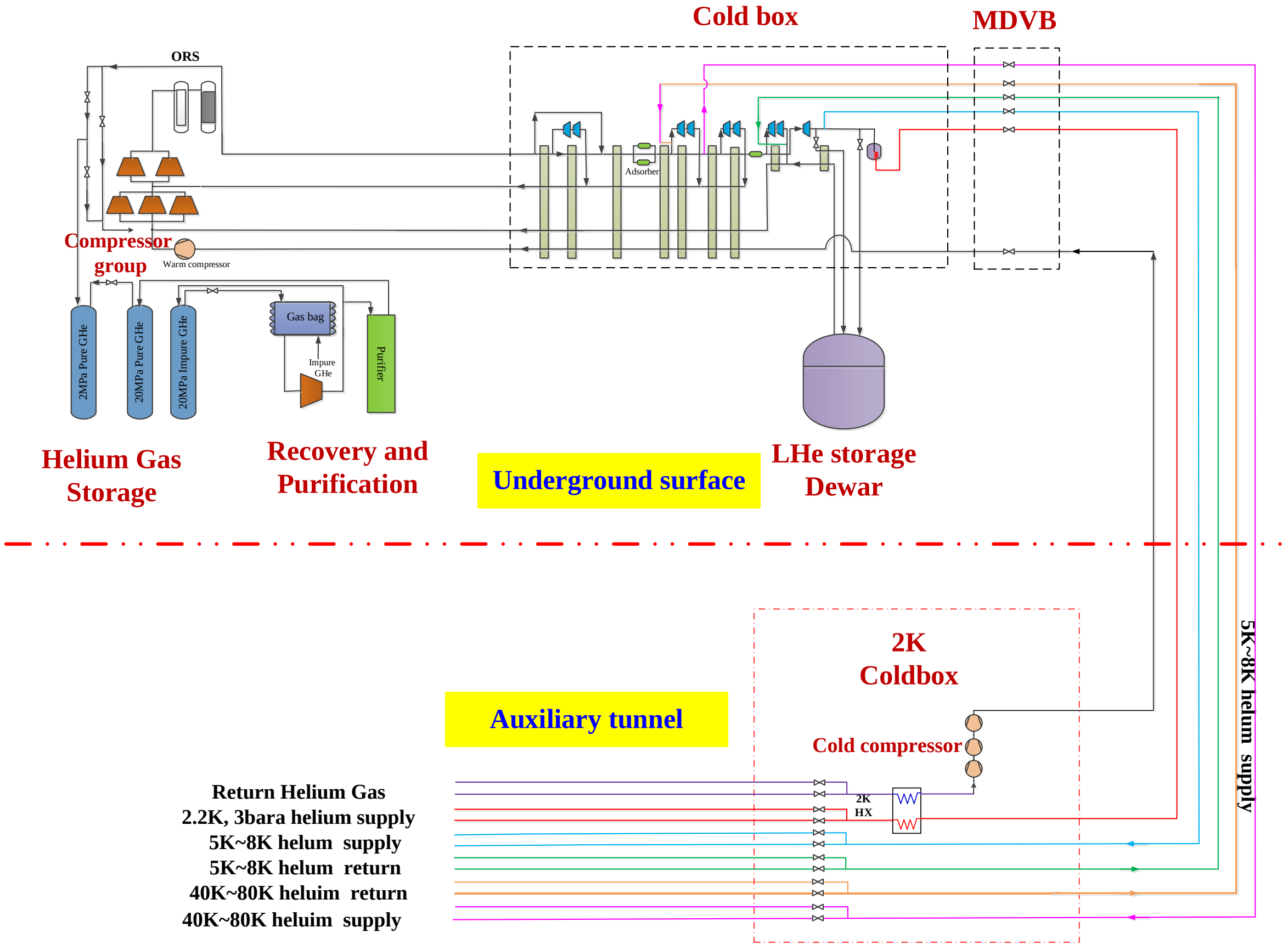

**Figure 7.1.29** Refrigerator process flow diagram.

The cryoplant will provide 4.5K and 2.2K helium to the cryomodules. The helium passes through a phase separator and a 2K counterflow heat exchanger to recover the cooling power. It is then expanded to 31 mbar via a JT-valve in the cryomodule, which results in the production of liquid He II at 2K.

The low-pressure helium vapor from the 2K saturated baths that surround the cavities will return to the refrigerator 2K cold box through the gas return pipe. This vapor is pumped away and then returned to the cryoplant.

Figure 7.1.30 shows three ways to produce 2K superfluid helium: (i) using ambient pumps and an electrical heater solution, (ii) using cryogenic pumps (cold compressors), and (iii) using a combination of ambient and cryogenic pumps. After superheating in the counterflow heat exchanger, the gas is compressed in multiple-stage cold compressors to a pressure in the range of 0.5 to 0.9 bar. This stream is then warmed up to ambient temperature in exchangers before returning to the warm compressors. CERN uses this approach in the LHC [6], which allows for easier restart of the 2K system after a stoppage, and easier adjustment for heat load variations using a warm vacuum compressor. Solutions (ii) or (iii) are typically the best choice for large-scale superfluid helium cryogenic systems, balancing efficiency, reliability, initial investment, and operating costs. Figure 7.1.31 summarizes the application range of cold compressors.

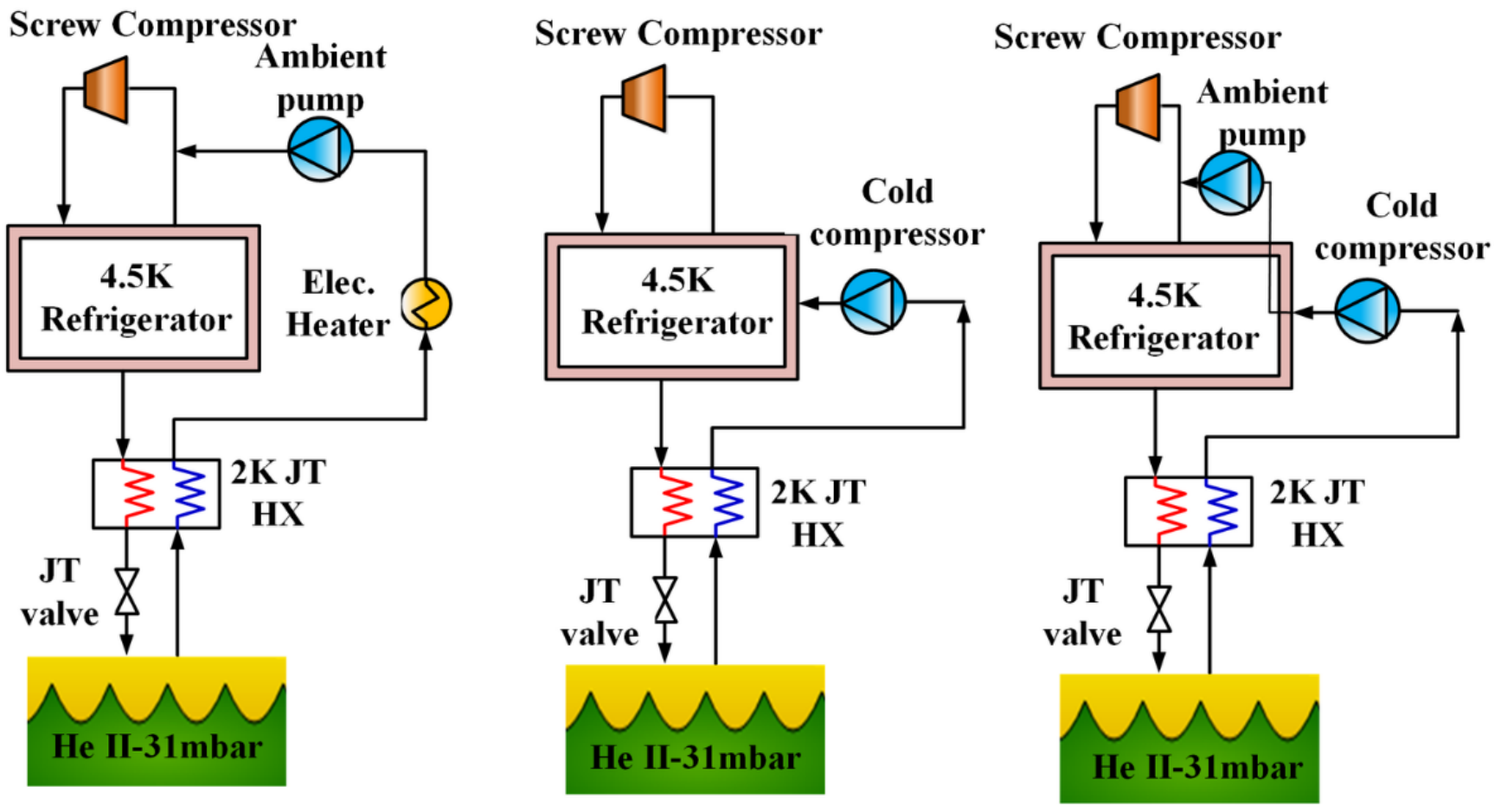

**Figure 7.1.30:** 2K superfluid helium acquisition scheme.

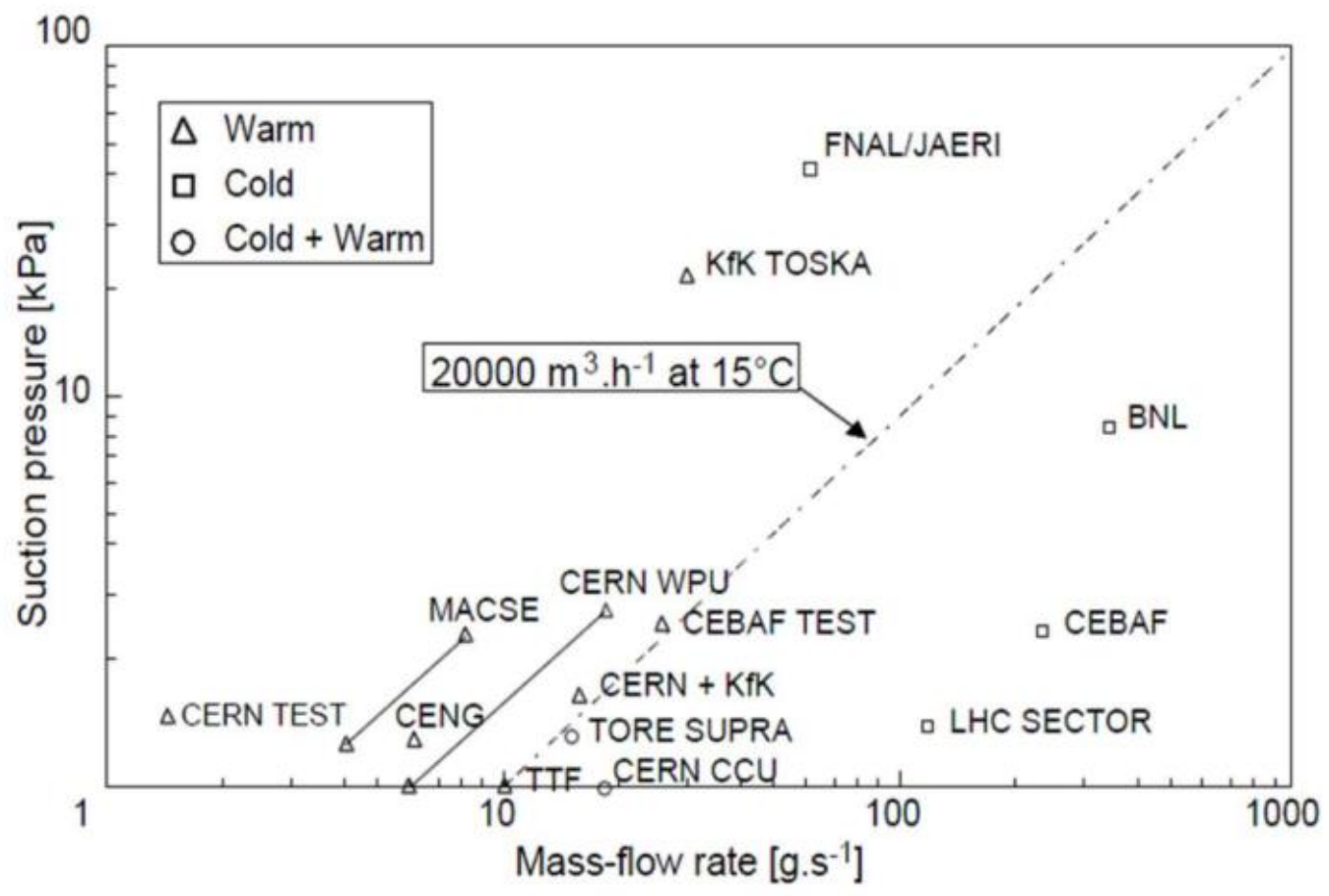

**Figure 7.1.31:** Application range of cold compressors in international cryogenic plants.

CEPC has established an industrial alliance system, led by FULL CRYO company. FULL CRYO company is a state-owned enterprise owned by the IPC, CAS, which focuses on industrial research and development for the 15 kW @4.5K large refrigerator for CEPC. FULL CRYO has made significant advancements in the development of helium refrigerators/liquefiers. They offer a range of mature products, including 2 kW @20K and 10 kW @20K refrigerators/liquefiers. Notably, this year, they achieved acceptance for the CiADS project by the Institute of Modern Physics of the CAS for their 2.5 kW @4.5K or 500W @2K refrigerator. This accomplishment has garnered widespread attention and has been extensively covered by various newspapers. The pilot project for the 10 kW @4.5K or 15 kW @4.5K (required for CEPC) has been approved, and it is expected to be developed and fully prepared for CEPC in the next few years. The coldbox is split into two temperature range sections: a 300K~50K cold box located above the ground, and a 50K~4K coldbox located underground in the tunnel. The 50K~4K coldbox measures approximately 10.63 meters in length, and about 4 meters in width, chosen to minimize the width of the RF auxiliary tunnel and shaft sizes. The 2K coldbox has a lenght of about 10 meters. Table 7.1.4 and Table 7.1.5 provide the parameters of the FULL CRYO refrigerator. The 2K superfluid helium scheme adopted uses a cold compressor.

**Table 7.1.4:** Parameters of the refrigerator from the FULL CRYO

| Parameters | Precooled by Turbin expanders | Precooled by liquid nitrogen |
|---|---|---|
| Consumption of Liquid Nitrogen | 0g/s (0L/h) | 428.7 g/s (2043 L/h) |
| Gas flow for precooling | 100 g/s | 0 g/s |
| Gas flow (g/s) | 1668 g/s | 155 1g/s |
| Power Consumption (KW) | 3935.13 | 3721.3 |

**Table 7.1.5:** Parameters of the cold compressor from the FULL CRYO.

| Parameters | The high pressure compressor | The medium pressure compressor | The low pressure compressor |
|---|---|---|---|
| Gas flow (g/s)* | 1668.00 | 862.80 | 230.7 |
| Pressure of the inlet (bara) | 4.05 | 1.05 | 0.40 |
| Pressure of the outlet (bara) | 18.54 | 4.05 | 1.05 |
| Adiabatic efficiency (%) | 85 | 85 | 85 |
| Power consumption (kW) | 2568 | 1155 | 196.3 |

The structural diagrams of the 4.5K and 2K cold boxes can be found in Figure 7.1.32, with (a) representing the 4.5K cold box and (b) representing the 2K cold box. Additionally, Figure 7.1.32 (c) illustrates the current development progress for the 15 kW helium refrigerator, which includes completion of the following jobs:

- P&ID design and HAZOP analysis,
- structural design of the cooling box,
- development of a 15/8 kW @4.5K prototype helium turbine expander,
- a 4 kW @2K prototype for the large flow cold compressor.

The research plan for the development of the 15 kW @4.5K helium cryoplant is presented in Figure 7.1.33.

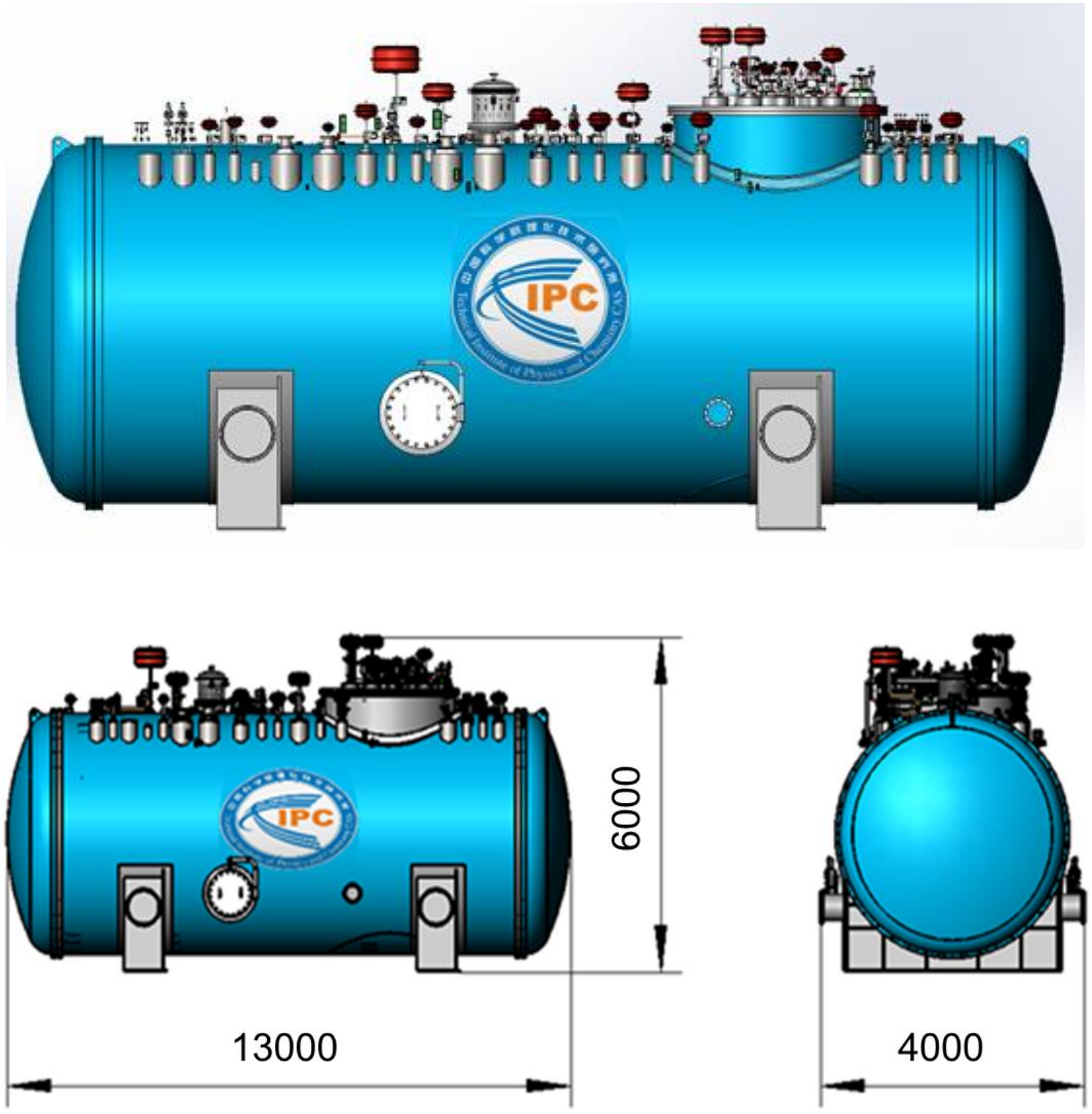

a) 4.5K coldbox

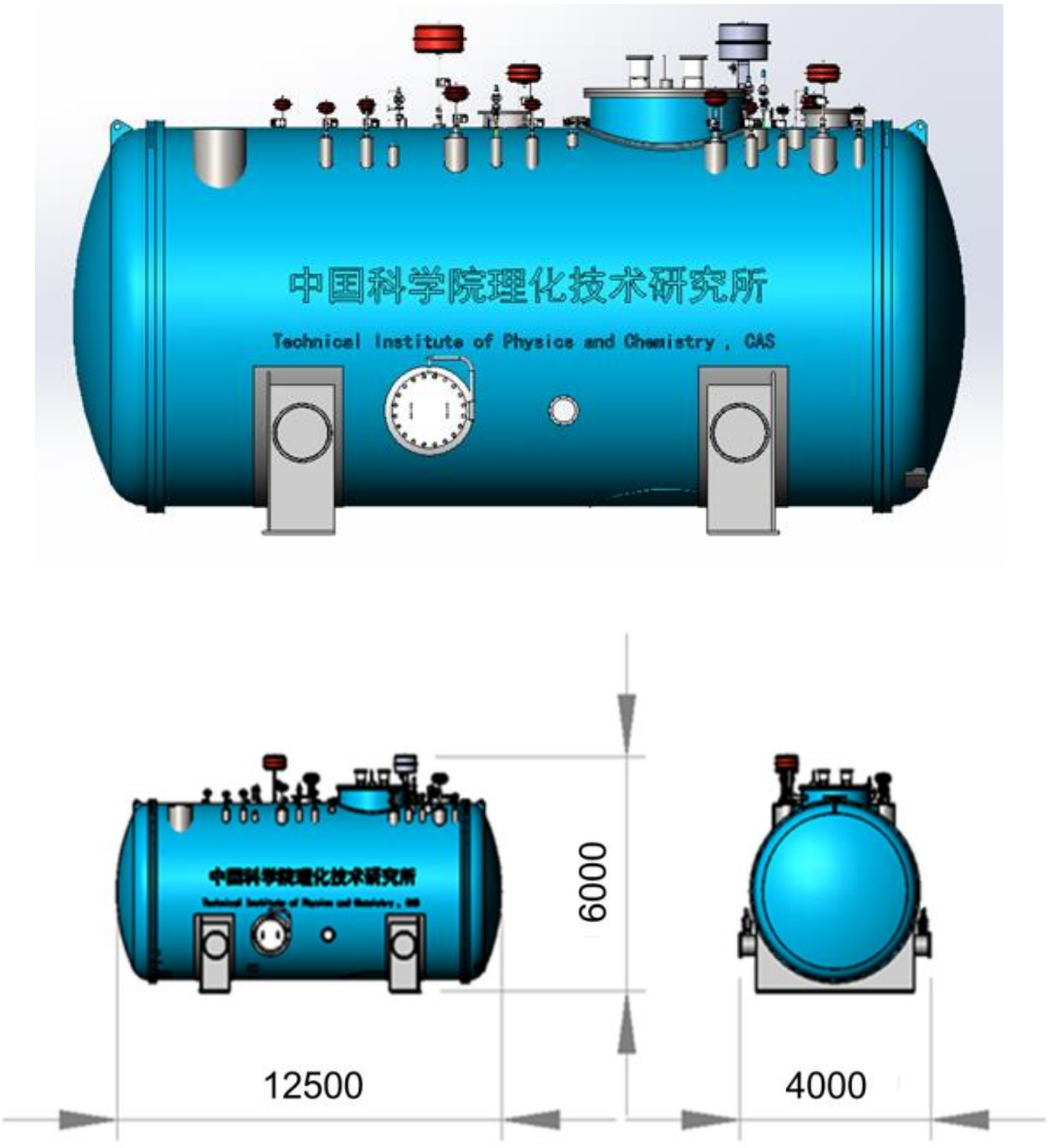

b) 2K coldbox

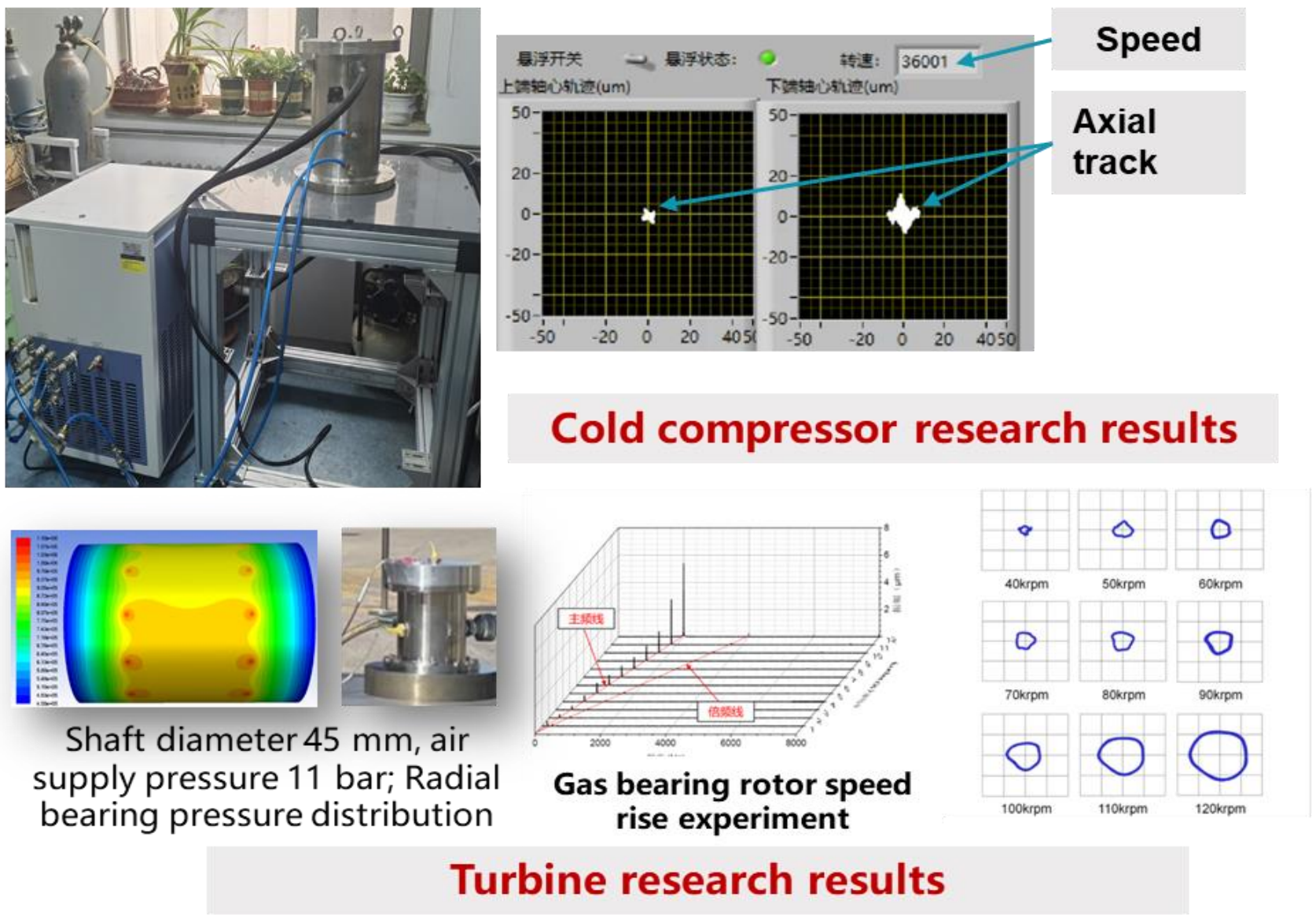

c) current development proress for the 15 kW helium refrigerator

**Figure 7.1.32:** Development progress of a 15 kW@4.5K helium refrigerator.

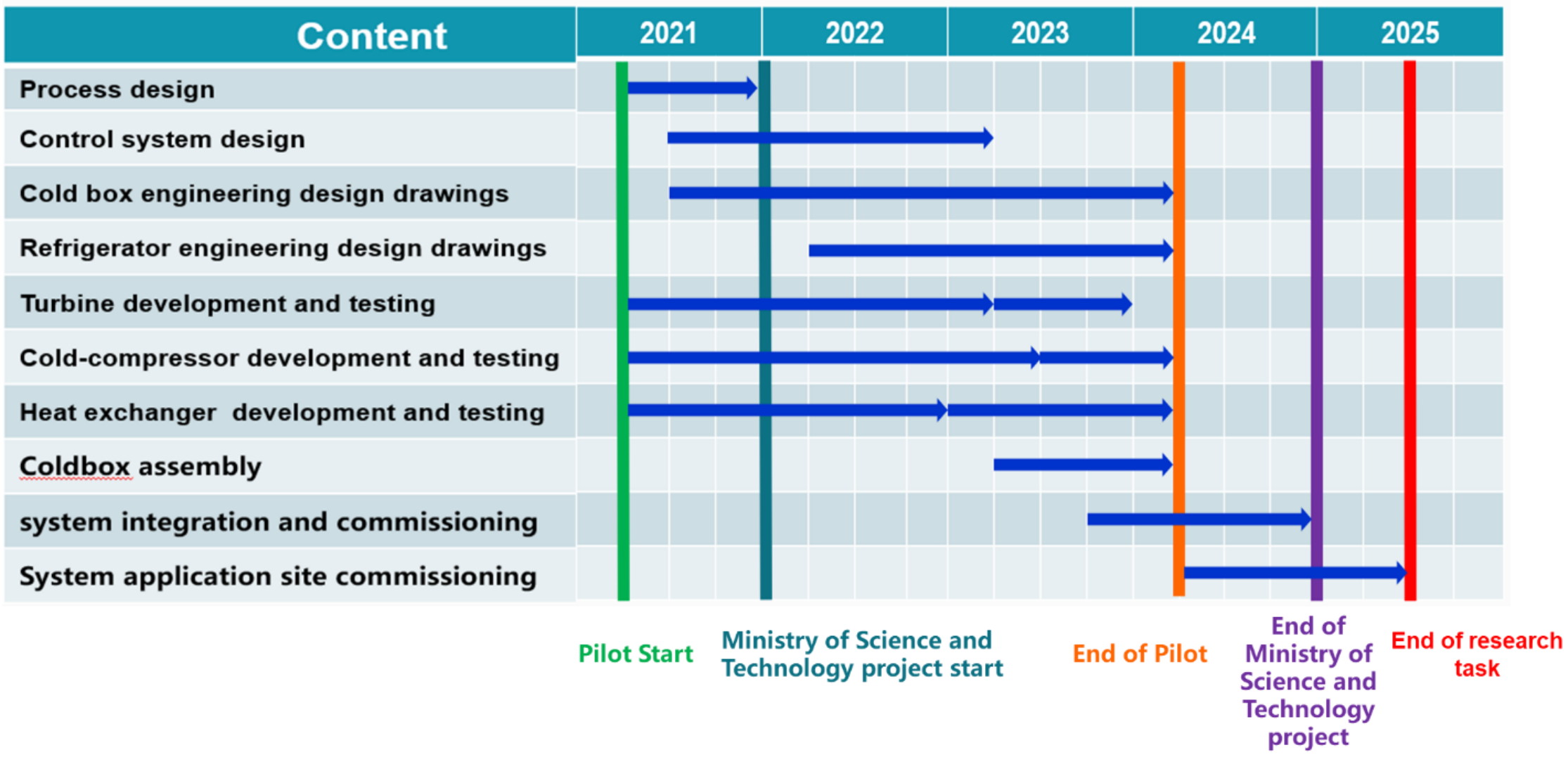

**Figure 7.1.33:** Research plan for the 15 kW@4.5K helium cryoplant development.

## 7.1.8 Cryogenic System for SC Magnets

### 7.1.8.1 *Cooling Scheme Optimization and Process Calculation*

Supercritical helium forced flow cooling is typically used to cool IR SC magnets due to several advantages. Firstly, it allows for direct contact between the superconducting cable and liquid helium, ensuring efficient heat transfer. Secondly, the heat transfer coefficient is stable, ensuring uniform cooling across the magnet. Thirdly, individual coils of the superconducting winding can be cooled simultaneously in parallel between the windings. Fourthly, the mass flow rate of the cryogenic fluid can be adjusted for different cooling branches as needed.

In contrast, the use of saturated or superfluid helium immersion cooling can lead to the existence of a saturated state liquid-vapor helium, as the latent heat of vaporization of liquid helium is very small. This can cause the evaporation of large amounts of liquid helium, leading to thermal disturbance and pressure rise that threatens the stability of the magnetic field. Additionally, supercritical helium is easier to obtain than superfluid helium, making it a more practical choice for forced cooling of IR SC magnets. Cryopumps are typically utilized to circulate the force cold helium in forced cooling systems, as shown in Figure 7.1.34.

Another cooling scheme is the indirect cooling method, where liquid helium does not directly contact the superconductor. Instead, it flows through a coil that is wrapped around the surface of the superconducting magnet, which cools the magnet or removes heat through heat transfer between the coil and the superconductor. This cooling method is mainly used for large, thin-walled superconducting magnets, such as those used in high-energy detectors, which are several meters in diameter and require a magnetic field of only 1 T ~ 2 T, with only one or two layers of coils. This method is more appropriate for cooling the superconducting wire surface using coils to cool the magnet.

In recent decades, the development of GM small chiller/refrigerator and its cooling technology has rapidly progressed. Its cooling power at 4.2K can reach 1 W ~ 3 W. In addition, due to the development of high-temperature superconducting materials, using GM chillers to create current leads eliminates the loss of Joule heat. The thermal conductivity of B2223 at 77K is only 1 W/(m-K), which is about 2 orders of magnitude smaller than the thermal conductivity of copper. This makes it possible to directly cool superconducting magnets with a GM chiller. In 1983, Japan developed a superconductor device cooled by a GM chiller. Direct cooling of superconducting magnets with a chiller has many advantages compared to conventional cooling schemes, including the fact that liquid helium is not required to cool down the magnets. This means that the magnet only needs to be installed in a vacuum vessel, which greatly simplifies the structure of the cryostat and makes it more compact and lightweight. Therefore, this technology of cooling superconducting magnets by direct conduction without helium has attracted significant interest. However, due to the limitation of the cooling power of the chiller, it is currently only used for small superconducting magnet systems.

In the interaction region of CEPC, there are four IR SC magnets cryomodules working at 4K and 32 sextuple magnets working at room temperature, installed at each of the two IPs where detectors are placed. Recently, the requirement for IR sextuple magnets has been updated, with a reduction in the magnet bore and pole field. The conventional sextuple magnet technology with a FeCoV pole can be used, achieving a maximum strength of 3886 T/m$^2$ with a bore aperture of 42 mm, and the maximum current is less than 200 A. As a result, the cryogenic system designed for the 32 sextupole magnets is no longer needed.

However, the forced flow cryogenic cooling of IR SC magnets poses a challenge due to the complex geometries of fluid flow channels and the impingement of heat flux in various modes such as convection, solid and gas conduction, thermal radiation, electrical energy dissipation within the conductor, and ionizing radiation. Cryogenic fluid flow is characterized by fluid flow conditions and flow passage geometry such as simple conduit, annular space, porous medium, bubble plates, external or internal flow in plates.

In TDR, for the IR SC magnets, two sets of 800 W @4.5K helium refrigerators will be employed, and the sub-cooled supercritical helium forced flow cooling scheme will be used due to the limitations of the narrow channels in superconducting magnets, as shown in Figure 7.1.34.

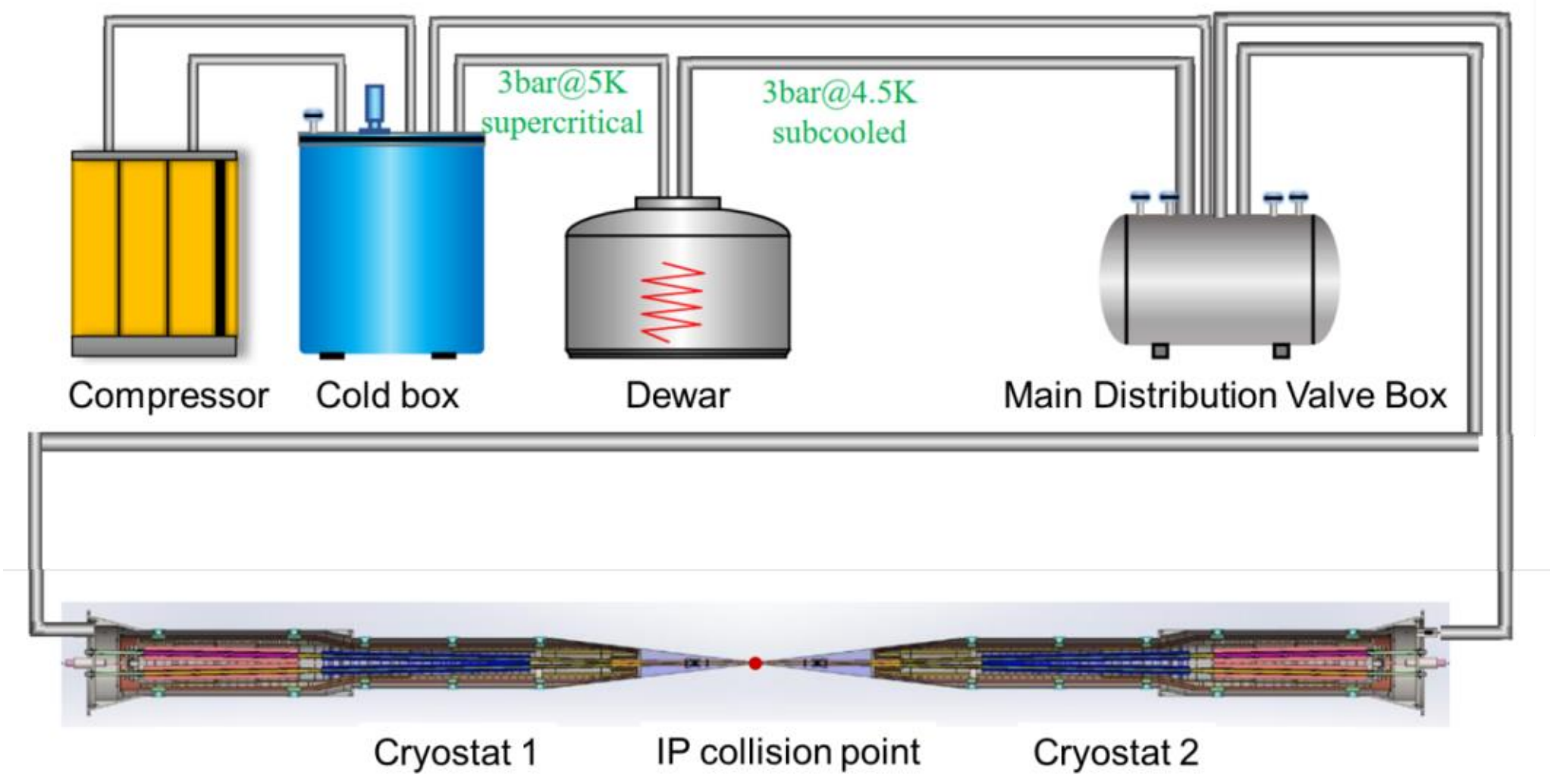

**Figure 7.1.34:** Detailed flow chart for the IR Cryostat (one side).

Similarly, the EcosimPro software is used to simulate the process flow of the IR SC magnets, adopting subcooled supercritical helium at 3 bar and 4.45 K. Assuming a heat leakage of 0.15 W/m for the supercritical helium flow transfer line and 2 W/m for the 40~80K thermal shield line. The process flow calculation of the IR SC magnets cryogenic system is shown in Figure 7.1.35.

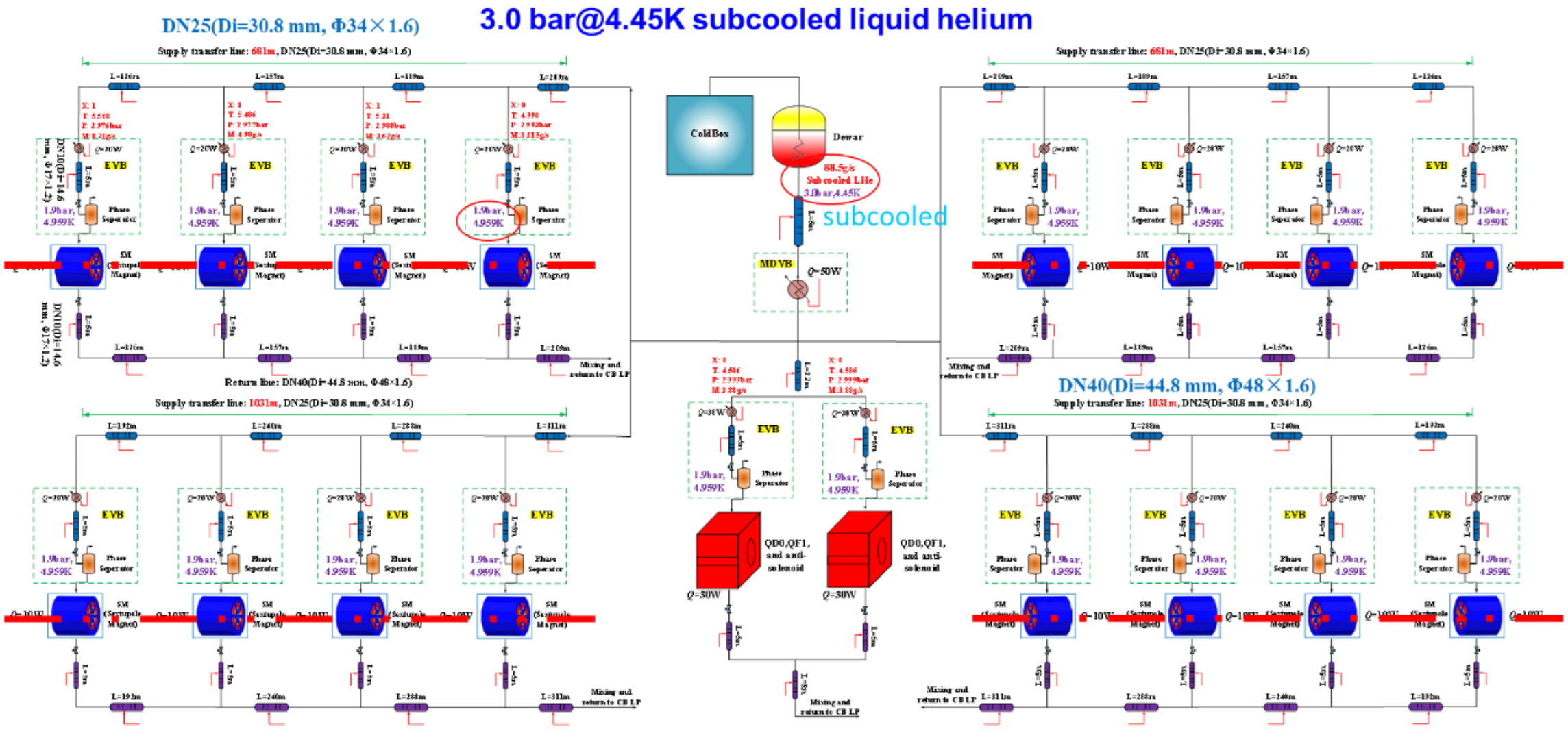

**Figure 7.1.35:** Process flow calculation of IR SC magnets cryogenic system.

### 7.1.8.2 *SC Magnet Cryomodule Design*

In order to meet the requirements for small size and low field leakage, the design of the IR SC magnets cryomodules must prioritize compactness. Typically, a superconducting magnet cryostat is comprised of several components from the inside out:

the cryostat inner barrel room temperature housing, the inner liquid helium annulus, the coil support barrel, the coils themselves, brackets and current leads, the outer liquid helium annulus, the 40~80 K cold screen (thermal shield) with cooling coils, and the cryostat outer barrel room temperature housing. With the exception of the cold screen, which is made of aluminum, the entire cryostat is constructed from stainless steel. Due to the limited space in the radial direction, the magnet is cooled using supercooled helium, which flows through both the outer and inner channels and absorbs heat due to both heat conduction and thermal radiation.

In the CEPC Interaction Region (IR), there are a total of 12 magnets, consisting of 4 QD0 magnets, 4 QF1 magnets, and 4 anti-solenoids. All of these superconducting magnets operate at a temperature of 4.5K.

The cryostat for the IR SC magnets is arranged in groups consisting of 1 QD0 magnet, 1 QF1 magnet, and 1 anti-solenoid. There are a total of 4 IR SC magnets cryostats in the system.

The requirements for the final focusing quadrupoles (QD0 and QF1) are based on a length of 2.2 m and a beam crossing angle of 33 mrad in the interaction region, as indicated in Section 4.3.4 of this report.

Each magnet in the IR is supported radially and axially by stainless-steel supports, which are welded to the outer vacuum shell layer to minimize heat transfer. These support contact points are made of G-10 material, and they also serve to support the helium thermal shielding. Therefore, both heat transfer optimization and structural integrity must be considered when designing these support points.

Based on the mechanical design, each magnet in the CEPC has a total weight of approximately 1 ton. In the event of a magnet quench, approximately 400 kilojoules of energy will be released into the cryogenic system over a period of 2 seconds.

The mechanical design of the CEPC IR cryostat is shown in Figure 7.1.36, with a total weight of approximately 2.5 tons and a length of 6.7 meters. Stress analysis and thermal simulations have been completed as part of the design process. The next step is to secure funding for the manufacturing phase of the project.

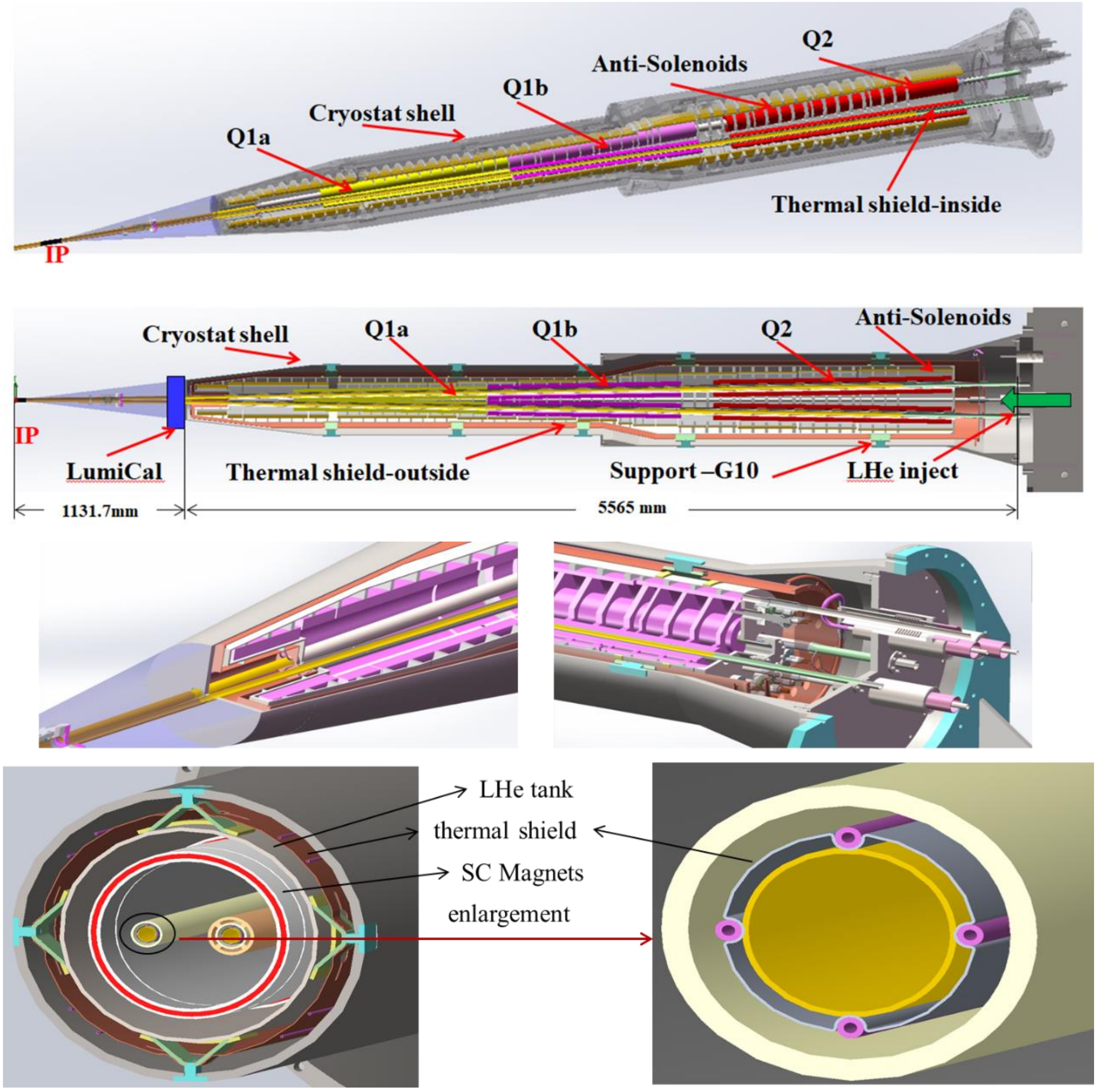

**Figure 7.1.36:** Mechanical design of the MDI cryostat.

The temperature distribution within the vacuum chamber of the CEPC IR cryostat can be seen in Figure 7.1.37, with detailed calculation results presented in Table 7.1.6. The temperature conditions for the calculations were set as follows: the outermost layer was treated as a convective boundary condition, the surface that the support structure contacts with the helium pool was set to 5K, and the surface that contacts the cold screen was set to 60K.

In addition to temperature analysis, stress analysis has also been conducted for the CEPC IR cryostat vacuum chamber, as shown in Figure 7.1.38.

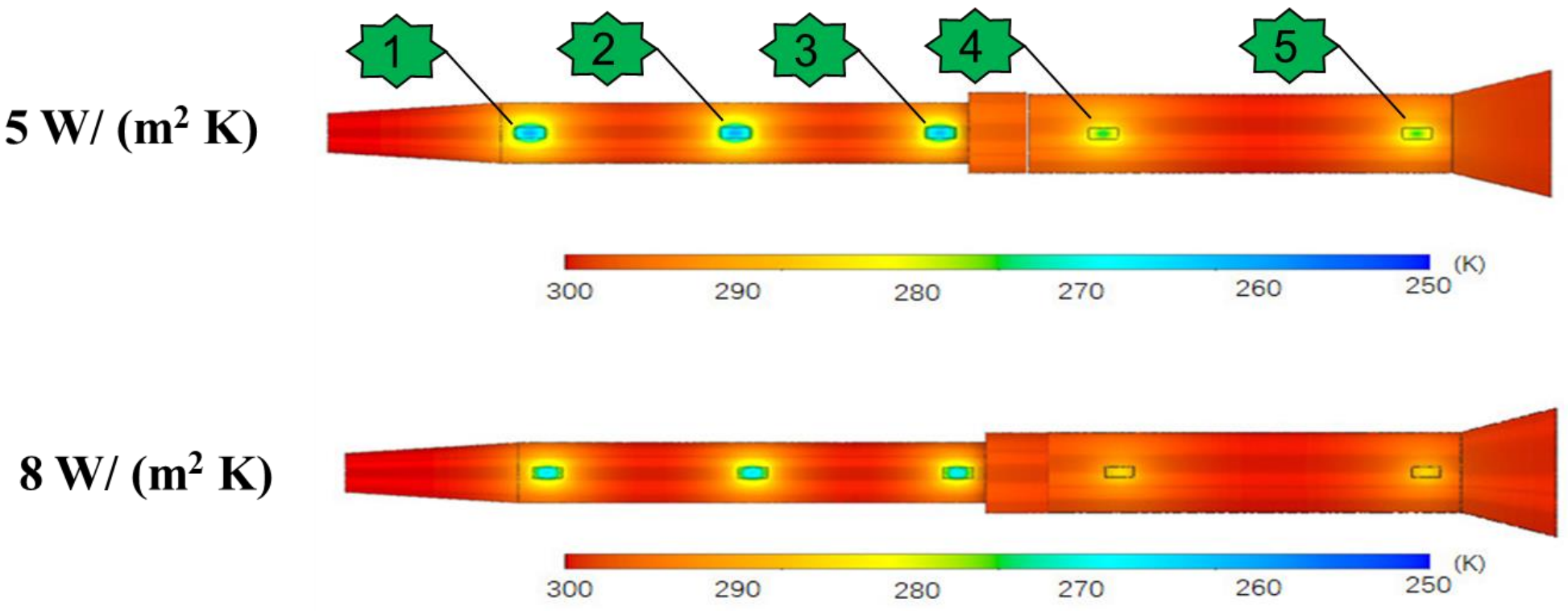

**Figure 7.1.37:** Temperature field of the MDI cryostat vacuum chamber.

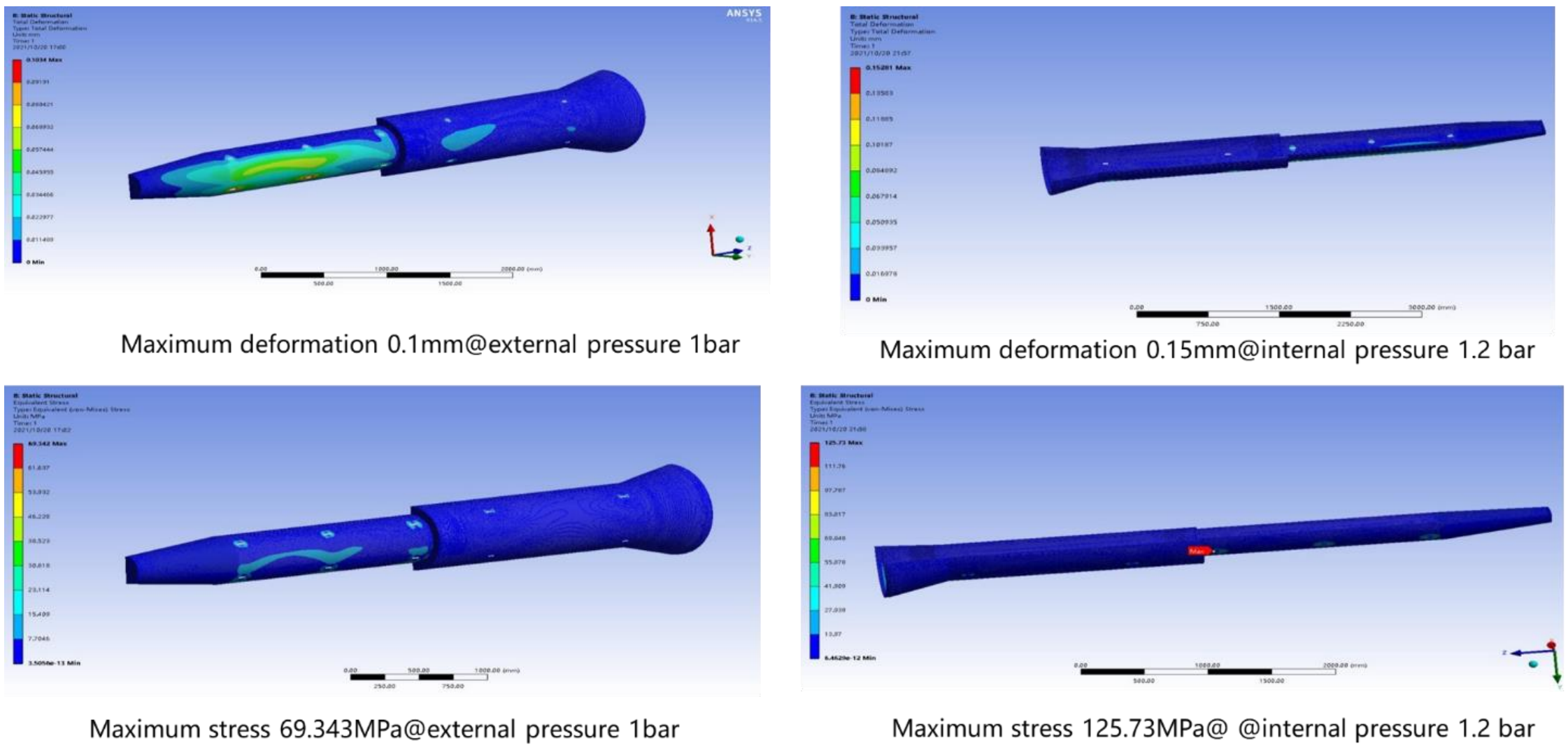

**Figure 7.1.38:** Stress analysis of the MDI cryostat vacuum chamber.

**Table 7.1.6:** Temperature distribution results of the MDI cryostat vacuum chamber.

| Support numbers | | | | | |
|---|---|---|---|---|---|
| Convective heat transfer coefficient | 1 | 2 | 3 | 4 | 5 |
| 5 W/ ($m^2$ K) | 268.97 K | 268.98 K | 270.52 K | 282.76 K | 283.32 K |
| 8 W/ ($m^2$ K) | 273.14 K | 273.14 K | 274.09 K | 285.34 K | 286.65 K |

## 7.1.9 Helium Recovery /Purification System and Liquid Nitrogen System

### 7.1.9.1 *Helium Inventory*

In the Higgs 30 MW/ 50 MW operating mode, the majority of the helium inventory is in liquid form and is used to cool the RF cavities. This liquid helium makes up approximately 70% of the entire system.

The volume of one 1.3 GHz module is about 400 liters, while one 650 MHz module is about 350 liters. The total volume of liquid helium in the system is approximately 20,200 liters. Taking into account the liquid in the Dewar and transfer lines, and adding a safety margin of 1.2, the total liquid volume in the system is estimated to be approximately 52,080 liters. Table 7.1.7 summarizes the helium inventory for different operating modes.

**Table 7.1.7:** Helium inventory in different operation modes.

| | **Higgs 30 MW mode** | | | | | |
|---|---|---|---|---|---|---|
| SRF Cryomodule | Collider | | | Booster | | |
| | 650 MHz 2-cell cryomodule | 400 | Volume/L | 1.3 GHz cryomodule | 350 | Volume/L |
| | Number | 40 | | | 12 | |
| | Total | 16000 | Volume/L | total | 4200 | Volume/L |
| | SRF CM total | 20200 | | | | Volume/L |
| | DW | 4 | 3000 | total | 12000 | Volume/L |
| | Transfer Lines | 8000 | | | | Volume/L |
| | Tota SRF LHe volume | 40200 | | | | Volume/L |
| IR SC Magnets | IR magnets | 150 | Volume/L | | | |
| | Number | 4 | | | | |
| | Total | 600 | | | | |
| | DW | 4 | 400 | total | 1600 | Volume/L |
| | Transfer Lines | 1000 | | | | Volume/L |
| | Tota SRF LHe volume | 3200 | | | | Volume/L |
| Total | 43400 | | | | | Volume/L |
| 1.2 times margin | 52080 | | | | | Volume/L |

| | **Higgs 50 MW mode** | | | | | |
|---|---|---|---|---|---|---|
| SRF Cryomodule | Collider | | | Booster | | |
| | 650 MHz 2-cell cryomodule | 400 | Volume/L | 1.3 GHz cryomodule | 350 | Volume/L |
| | Number | 56 | | | 12 | |
| | Total | 22400 | Volume/L | total | 4200 | Volume/L |
| | SRF CM total | 26600 | | | | Volume/L |
| | DW | 4 | 3000 | total | 12000 | Volume/L |
| | Transfer Lines | 8000 | | | | Volume/L |
| | Tota SRF LHe volume | 46600 | | | | Volume/L |
| IR SC Magnets | IR magnets | 150 | Volume/L | | | |
| | Number | 4 | | | | |
| | Total | 600 | | | | |
| | DW | 4 | 400 | total | 1600 | Volume/L |

| | | | |
|---|---|---|---|
| | Transfer Lines | 1000 | Volume/L |
| | Tota SRF LHe volume | 3200 | Volume/L |
| Total | 49800 | | Volume/L |
| 1.2 times margin | 59760 | | Volume/L |

| **$t\bar{t}$ 50 MW mode** | | | | | | |
|---|---|---|---|---|---|---|
| | Collider | | | Booster | | |
| SRF Cryomodule | 650 MHz 2-cell cryomodule | 400 | Volume/L | 1.3 GHz cryomodule | 350 | Volume/L |
| | Number | 128 | | | 44 | |
| | Total | 51200 | Volume/L | total | 15400 | Volume/L |
| | SRF total | 66600 | | | | Volume/L |
| | SRF total considering 1.5 times margin | 99900 | | | | Volume/L |
| | DW | 4 | 30000 | total | 120000 | Volume/L |
| | Transfer Lines | 40000 | | | | Volume/L |
| | Tota SRF LHe volume | 259900 | | | | Volume/L |
| IR SC Magnets | IR magnets | 150 | Volume/L | | | |
| | Number | 4 | | | | |
| | Total | 600 | | | | |
| | DW | 4 | 3000 | total | 12000 | Volume/L |
| | Transfer Lines | 8000 | | | | Volume/L |
| | Tota SRF LHe volume | 20600 | | | | Volume/L |
| Total | 280500 | | | | | Volume/L |

#### 7.1.9.2 *Helium Recovery and Purification System*

Helium gas is a highly valuable resource with a purity level of over 99.9999%. Therefore, it is essential to have a set of helium recovery and purification systems in place. These systems allow operators to clean the system before start-up, decontaminate it during shutdown periods, and decontaminate bulk helium gas supplies before entering the active system. Additionally, the system enables the recovery of helium gas during unplanned power outages and short maintenance periods, where a full system warm-up is not feasible or when the total gaseous helium inventory of the system exceeds the available helium gas storage capacity.

A gas management panel for the recovery/purifier compressor system is used to regulate the recovery compressor suction and discharge independently of the primary gas management system of the coldbox compressor system. It also allows for the purification of individual gas helium storage tanks and bulk helium gas supplies. The recovery compressor system is interconnected with the coldbox primary high-pressure supply and low-pressure return. When combined with the primary gas management system, the entire system can be scrubbed and purified.

The flowsheet of the Helium Recovery and Purification system, and the 80K helium purifier schematic diagram are shown in Figure 7.1.39 (a) and (b), respectively.

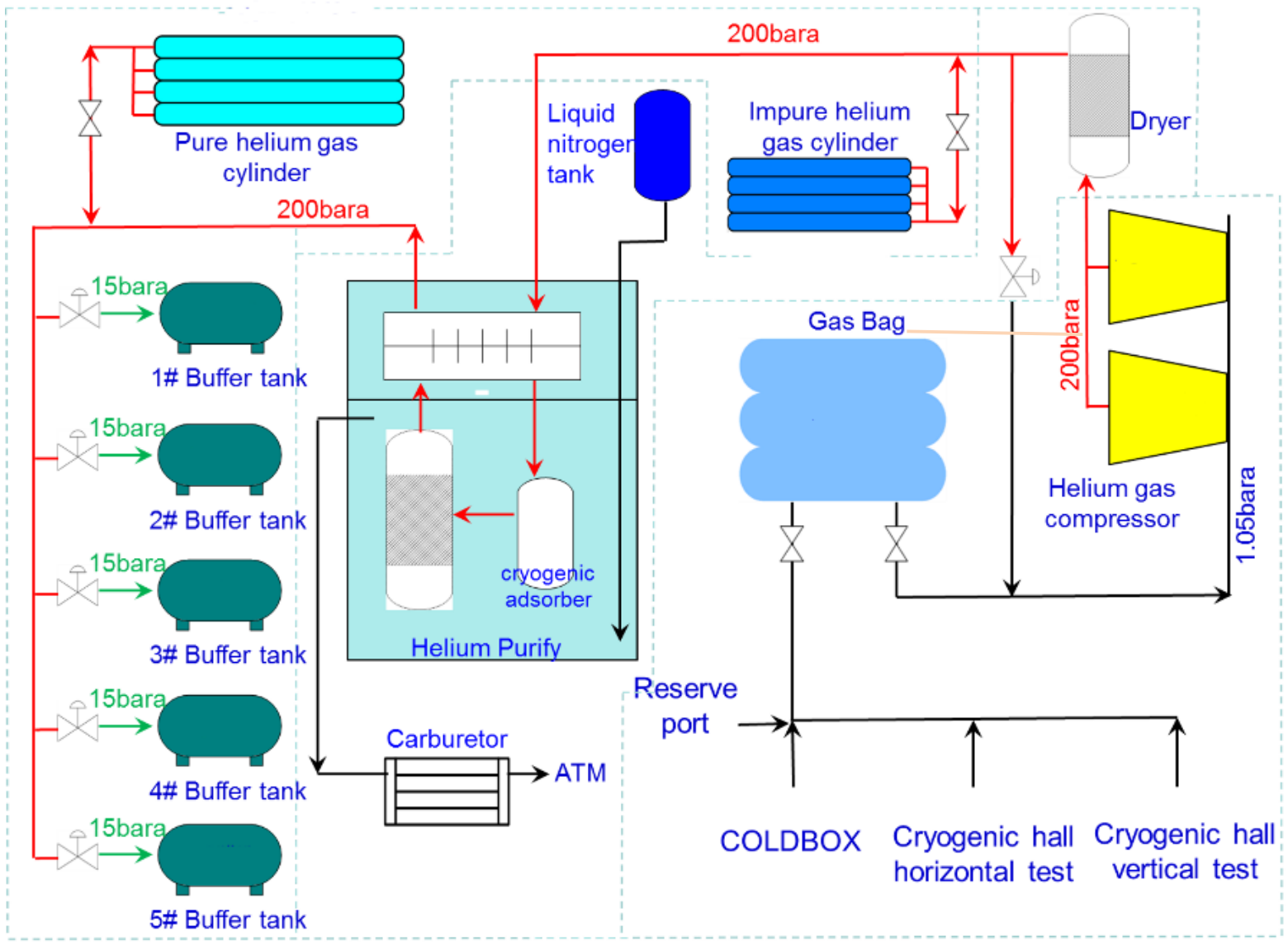

a)

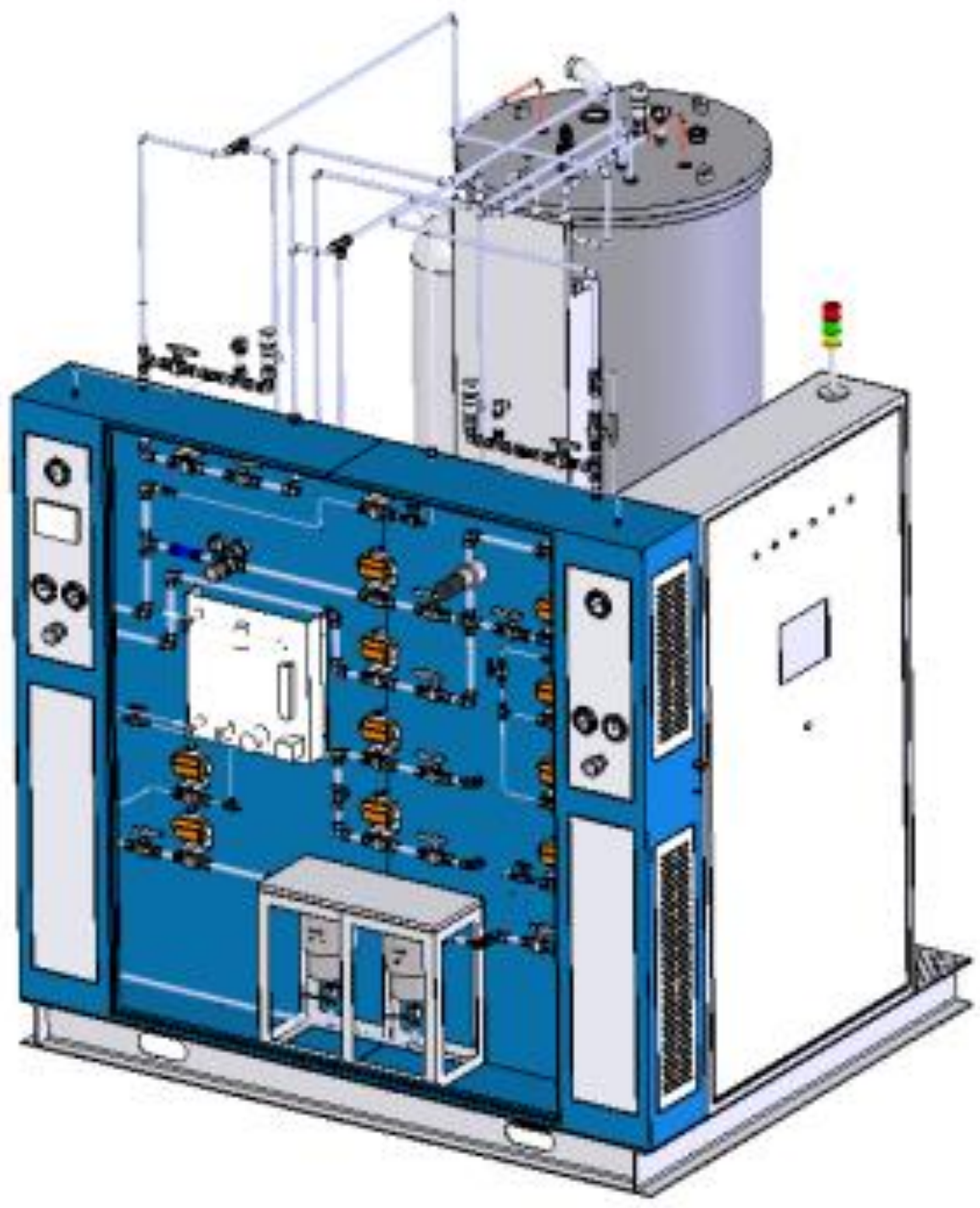

b)

**Figure 7.1.39:** a) Flowsheet of Helium Recovery and purification system; b) Schematic diagram of the 80K helium purifier.

### 7.1.9.3 *Liquid Nitrogen System*

The Helium Recovery and Purification System of CEPC's cryogenic systems require liquid nitrogen for normal operation. This liquid nitrogen is supplied by two double-walled, vacuum-insulated dewars and is fed to the facility through a single circuit, vacuum-jacketed transfer line with multilayer super-insulation to minimize heat leak. IHEP has extensive experience in liquid nitrogen cryogenic system design in recent years.

In 2019, the superconducting magnet side nitrogen cryogenic system of BEPCII underwent an upgrade that involved recycling and reusing the nitrogen cooling capacity of the superconducting magnet cold thermal shielding, as well as the ambient temperature nitrogen from the first stage heat exchanger of the cryogenic helium refrigerator. A set of nitrogen cooling capacity recovery refrigerators was successfully developed, which facilitated the recycling of nitrogen at the superconducting magnet side and reduced the frequency of external liquid nitrogen supply.

### 7.1.10 **Key Technology Research**

#### 7.1.10.1 ***2K JT Heat Exchanger***

The 2K cryogenic negative pressure heat exchanger, also known as the 2K heat exchanger, is a crucial component in the 2K cryogenic system. In the top plot of Figure 7.1.41, it is demonstrated that an efficient HX (efficiency of heat exchanger from 4.2Kto 2K) can increase the yield from 62% to 89%. However, as the scale of the cryogenic system continues to expand, the traditional 2K heat exchanger is no longer adequate to meet the needs of larger 2K cryogenic systems, such as the planned 2.4 kW @2K cryogenic system in CEPC. Thus, a corresponding large flow 2K heat exchanger is required.

Among the various key technologies in the CEPC cryogenic system, we have decided to prioritize investment in the research and development of a 2K negative pressure plate-fin cryogenic heat exchanger at the kW level. This decision was based on factors such as technical importance, technical difficulty, R&D cost, R&D period, and application significance.

The following is a brief description of our R&D and testing work in this direction.

<u>Technical Importance:</u>

The CEPC 2K cryogenic system relies on two core new technologies: the cold compressor and the 2K cryogenic heat exchanger. These components play a crucial role in the system's overall efficiency and performance. The heat transfer efficiency and resistance losses of the 2K heat exchanger are particularly important because they directly impact the system's operating efficiency.

Mastering the technology of the 2K heat exchanger would enable the purchase of separate components instead of purchasing a full set of foreign manufacturers' 2K cryogenic system. For instance, the Shanghai SHINE cryogenic system's entire solution, from the 4K system to the cold press and 2K heat exchanger, is provided by Air Liquide, which can be expensive and creates a high level of technical dependence.

Furthermore, mastering the technology of the high flow rate and high-performance 2K heat exchanger could allow the CEPC to split the entire solution into separate components. This approach would increase supply chain security and reduce technical dependence. The 4K cryogenic system, 2K cryogenic heat exchanger, and decompression machinery could be purchased separately, providing greater flexibility and cost savings.

<u>Technical Difficulty:</u>

Developing a 2K heat exchanger is relatively less challenging than developing a cold compressor. The 2K heat exchanger is a static device that mainly requires design calculations for flow heat transfer and calibration of structural mechanics. Overcoming this technology is possible by investing in two or three experienced technicians.

In contrast, developing equipment like the cold compressor involves more subject areas and requires a larger team with a long-term commitment to overcome the technical difficulties. For example, the Institute of Physical and Chemical Sciences' cold compressor team is a professional team consisting of a dozen people. The complexity of the cold compressor design and the need to overcome technical challenges make it a more challenging component to develop than the 2K heat exchanger.

R&D Cost:

The main R&D cost of the 2K heat exchanger includes the high-performance server required for design calculations, the manufacturing cost of the heat exchanger body, and the testing cost required for conducting tests. With funding of several million RMB, the expected results include high-precision design calculation methods, physical objects that meet performance requirements, improvements in processing technology, high-precision test platforms, and even related high-level papers, patents, and software.

In comparison, the research and development of the cold compressor and other related technologies may require funding in the order of tens of millions.

R&D Period:

Due to the good accumulation in the field of 2K heat exchanger R&D, it is expected that the desired results can be achieved in a relatively short period. It is expected that systematic results can be obtained by the end of 2023, which will lead the work in China and be considered first-class in the international arena. This is an effect that cannot be easily achieved in other research fields.

Application Significance:

The demand for large flow 2K heat exchangers is increasing with the development of more and more kilowatt 2K systems, such as the SHINE and CEPC projects. In the past, only foreign companies like Linde and Air Liquide have mastered the technology for large flow high efficiency 2K heat exchangers. The domestic research in this area has been limited due to a lack of demand, resulting in inadequate design calculation levels and insufficient manufacturing and processing capabilities.

With the support of the CEPC project, we have significantly enhanced our design and calculation capabilities in the field of plate and fin heat exchangers. Furthermore, we have trained several domestic manufacturers to meet the manufacturing process requirements for producing high-performance 2K plate and fin heat exchangers, such as extremely low leakage rates and the ability to withstand large temperature differences. This technology can now be utilized to meet the needs of the CEPC project itself, as well as to elevate the technical levels of domestic manufacturers while supporting the needs of our sister institutes in related fields.

Traditionally, tube-fin and coil-tube heat exchangers have been used in large flow heat exchange systems. However, these structures have limitations that restrict the heat transfer coefficient to 50-150 W/m$^2$. Moreover, if a tube-wound structure is used to design large flow heat exchangers, the resulting volume would be enormous. As a result, the plate-fin heat exchanger has become increasingly popular in such applications due to its high heat transfer coefficient and compact structure. Despite these benefits, designing and calculating plate-fin heat exchangers can be challenging, and they may be prone to leaking under helium conditions. For this reason, the use of plate-fin heat exchangers in the 2K cryogenic field has been limited in the past.

For the CEPC cryogenic system, a 2K heat exchanger with a mass flow rate of 120 g/s (2.4 kW @2K) will be needed. However, due to the limitation of the 2K cooling capacity of the cryogenic system, we will first design and optimize a larger flow (50 g/s, 1 kW @2K) plate-fin heat exchanger for the larger 2K cryogenic system. To ensure its heat transfer efficiency, we will consider a variety of criteria and utilize Computational Fluid Dynamics (CFD) simulation methods to verify its performance. Furthermore, a new heat exchanger test set has been designed based on the 2K cryogenic system of the PAPS to meet the corresponding testing requirements. The heat exchanger setup consists of three types: a) In the vertical test stands, a single-layer coiled finned-tube heat exchanger with a flow rate of 5 g/s is used. You can refer to Figure 7.1.41 (a) for a visual representation of this setup. b) The horizontal test stands employ a double-layer coiled finned-tube heat exchanger with a flow rate of 10 g/s. Figure 7.1.41 (b) illustrates this configuration. c) In the newly designed heat exchanger test platform, a plate fin heat exchanger is utilized. Figure 7.1.41 (c) provides a depiction of this type of heat exchanger.

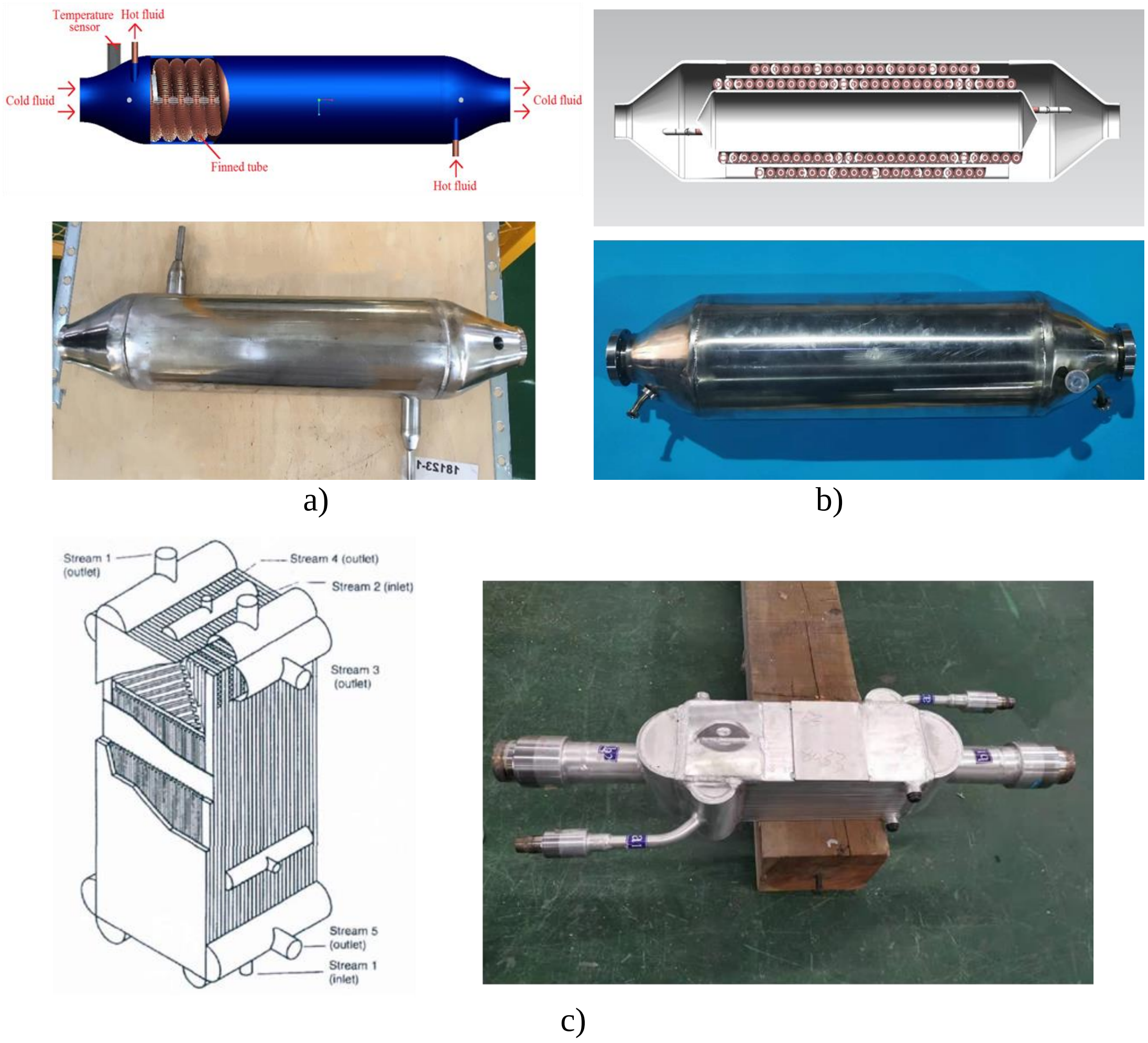

**Figure 7.1.40:** Different types of HX used in PAPS; a) single layer coiled finned-tube HX; b) double layer coiled finned-tube HX; c) plate fin HX.

The structural design and optimization calculations for several large flow heat exchangers have been completed, and manufacturing has begun. Additionally, the design of the corresponding test set has been completed, and the specific structural design phase has started. The process flow for the 2K heat exchanger and the completed test set are depicted in Figure 7.1.41. Figure 7.1.42 illustrates the different fin structures, while Figure 7.1.43 showcases the completed design of the heat exchangers in 3D. Finally, the production process of the 2K heat exchangers is depicted in Figure 7.1.44.

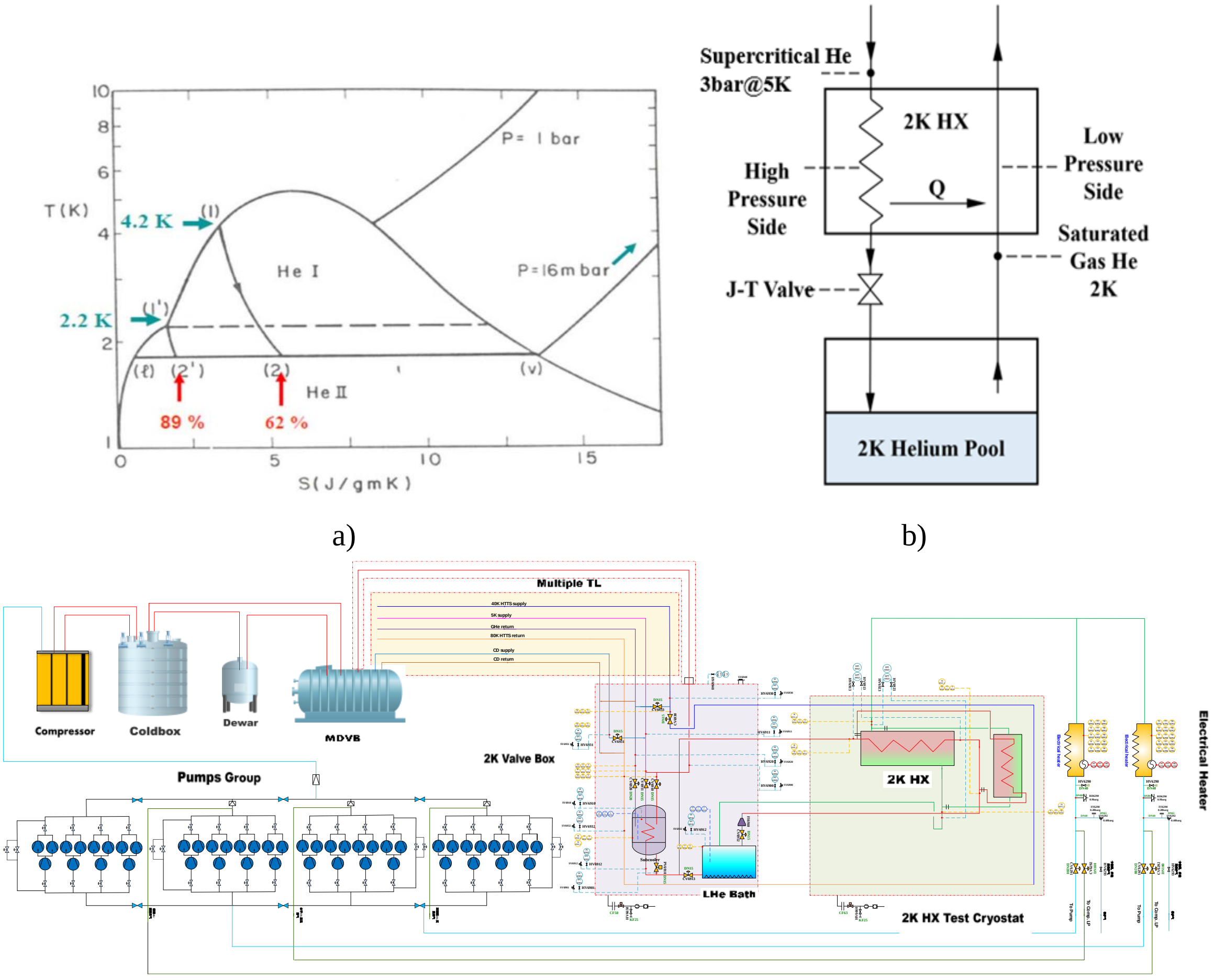

**Figure 7.1.41:** a) Liquefaction yield improved by 2K HX; b) Process flow of the 2K heat exchanger; c) Process flow diagram of the 2K heat exchanger test set in PAPS.

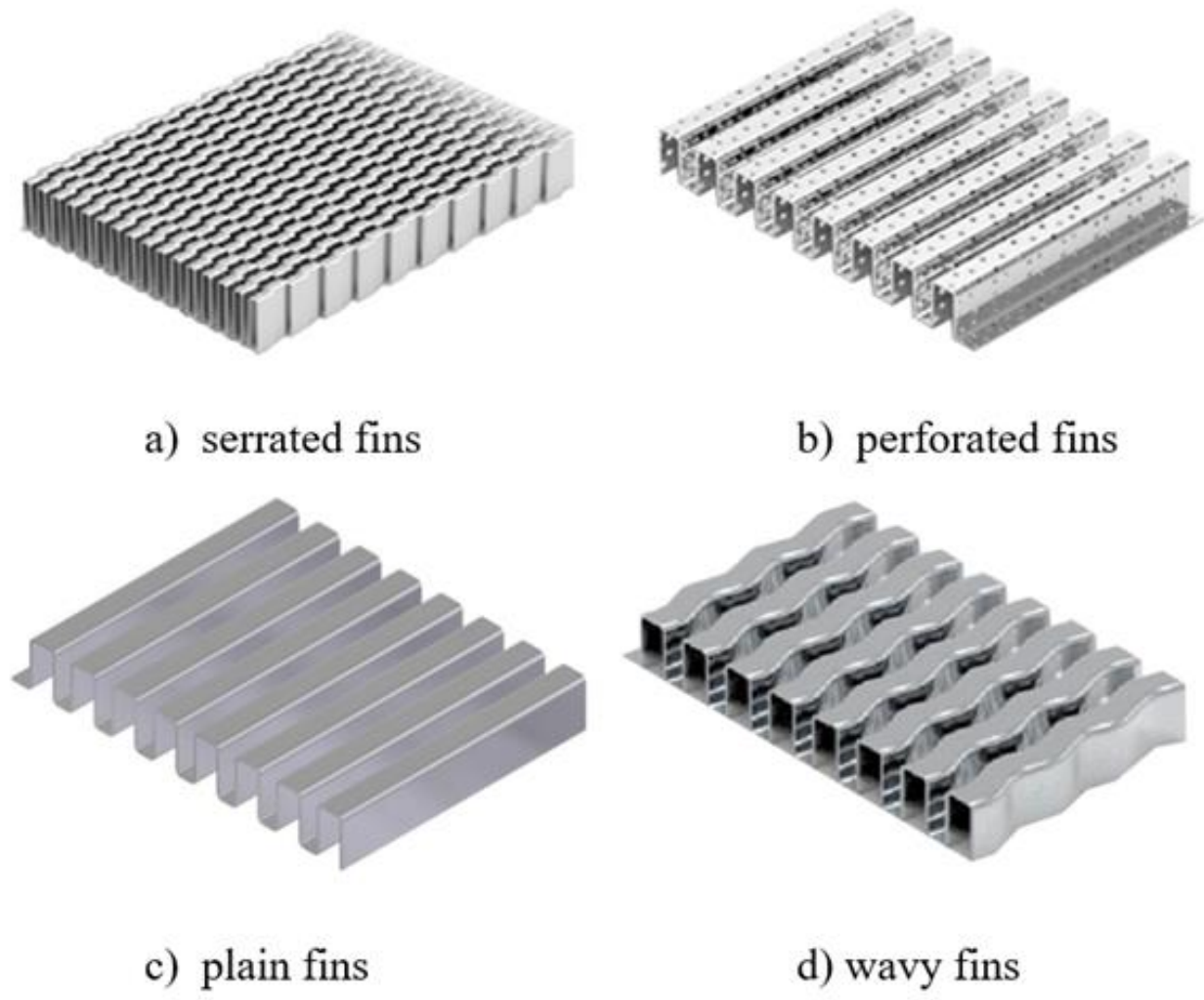

**Figure 7.1.42:** Different fins structure

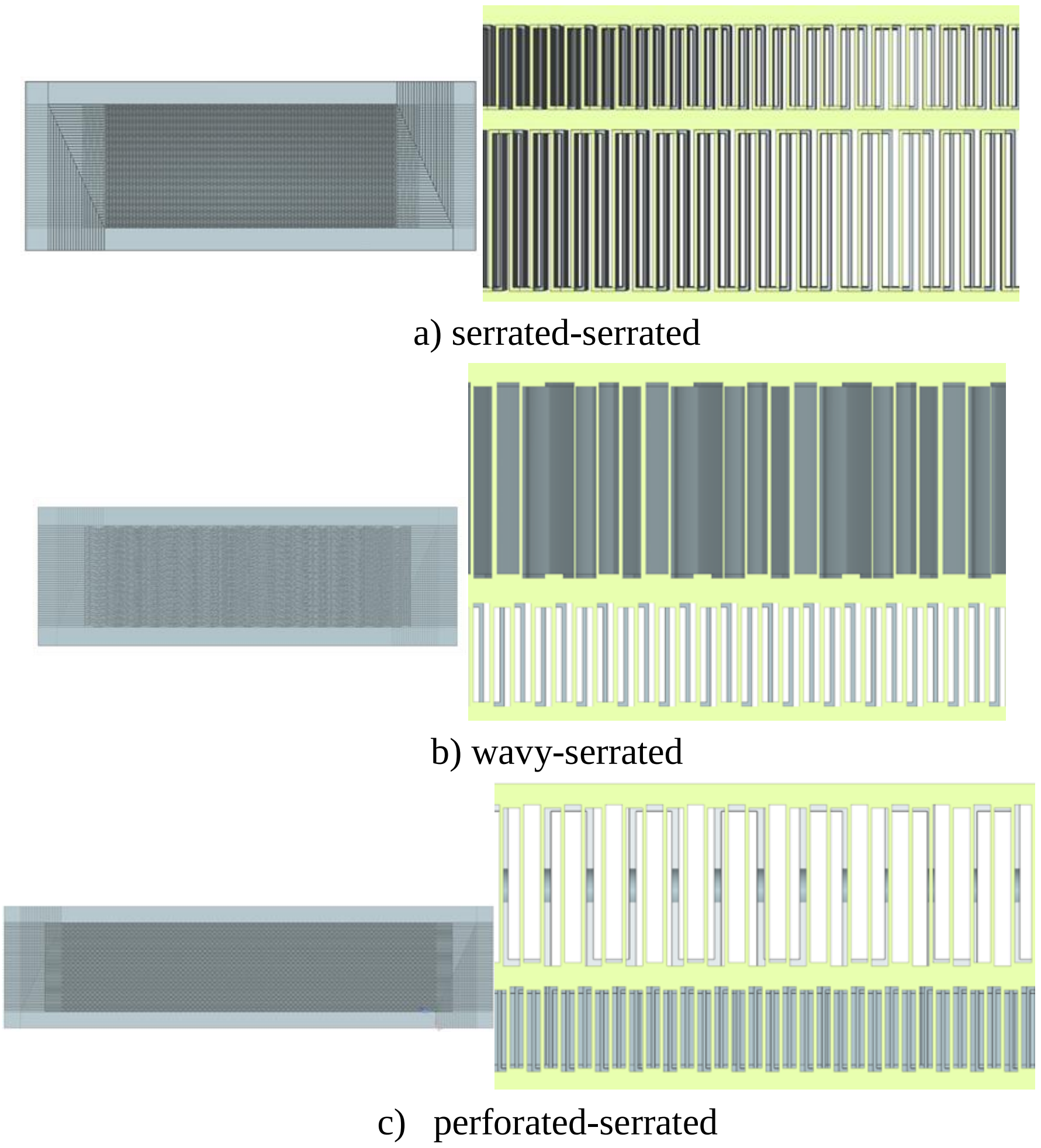

**Figure 7.1.43:** Three-dimensional structures of the complete design for heat exchangers.

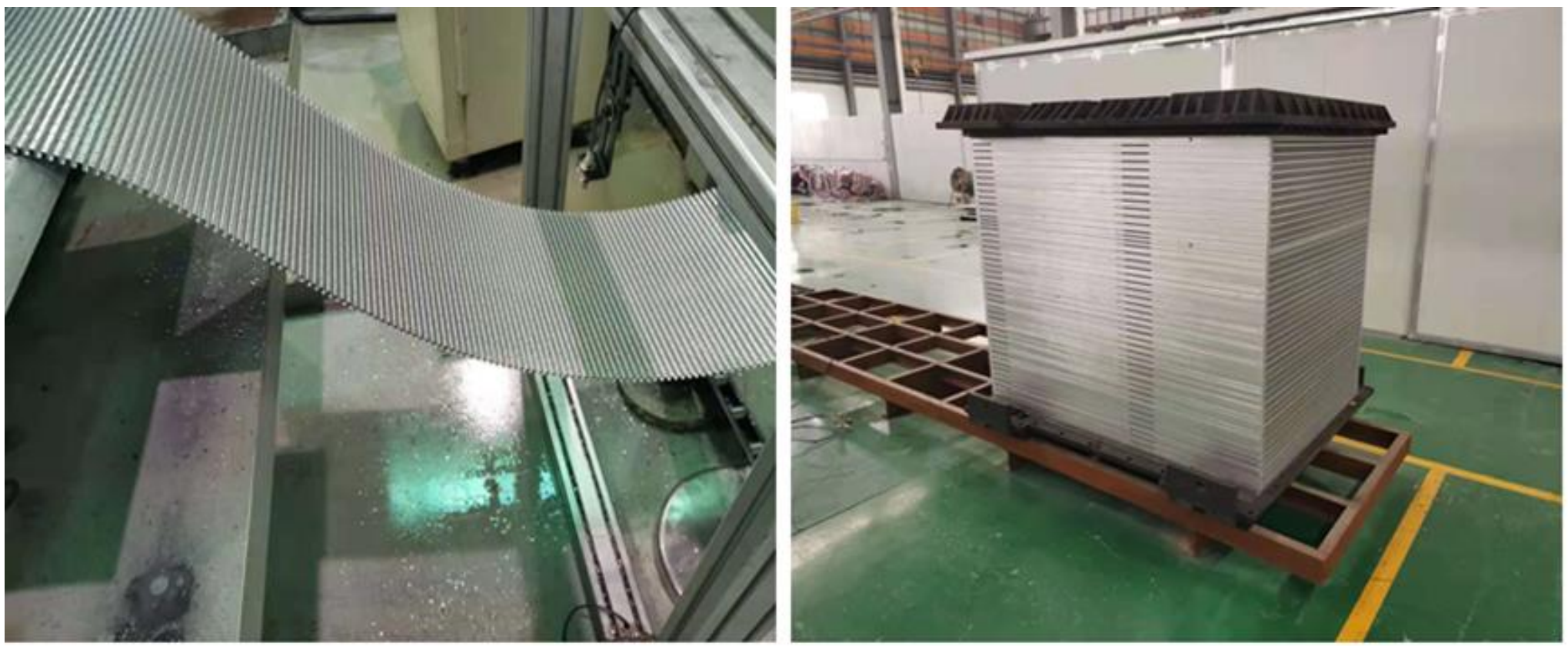

**Figure 7.1.44:** Production process of the 2K heat exchangers.

### 7.1.10.2 *Multiple Cryogenic Transfer Lines Test Platform*

The cryogenic transfer distribution system can be divided into two main parts: the cryogenic distribution valve box and the cryogenic transfer line. The cryogenic distribution valve box is an essential component of the helium cryogenic system, responsible for managing the distribution, flow control, and measurement of refrigerant or cryogenic mass parameters such as temperature, pressure, flow rate, liquid level, and more.

The valve box contains an 80K cold screen and operates within a high vacuum environment of 4.4K. To minimize heat load on the cold screen, a high vacuum multi-layer adiabatic structure is typically used.

Cryogenic transfer lines play a crucial role in transporting cryogenic masses such as liquid or superfluid helium and high-pressure cold hydrogen to different devices and experimental objects. These transfer lines are essential in cooling superconducting magnets in large scientific installations and in hydrogen filling stations. Due to the expensive and easily evaporating nature of liquid helium, cryogenic transfer lines need to have very low heat loss to ensure efficient transportation. A high-performance cryogenic transfer line can significantly reduce experimental operating costs and improve system efficiency, and its design is directly related to the operation of the entire cryogenic system.

To optimize the heat leakage transfer performance, several design elements need to be considered, such as the multi-channel radiation cold screen, the layout structure of multi-channel pipelines, the design of multi-channel support structures, and the number of layers and layer density of multi-layer adiabatic material wrapping for different temperature zones. Multiple cryogenic transfer lines are a key component of the cryogenic transfer and distribution system, and their heat leakage index directly affects the taste of the helium cryogenic cold volume.

In order to design and manufacture a high-performance cryogenic transfer line, a set of test platforms has been built. These platforms can test different channels and different types of multi-channel cryogenic transfer lines, including straight pipes and bent pipes. Figure 7.1.45 illustrates the test platform. The manufacturing works for the cryogenic transfer line have been finished, and experimental testing will be conducted in the next year.

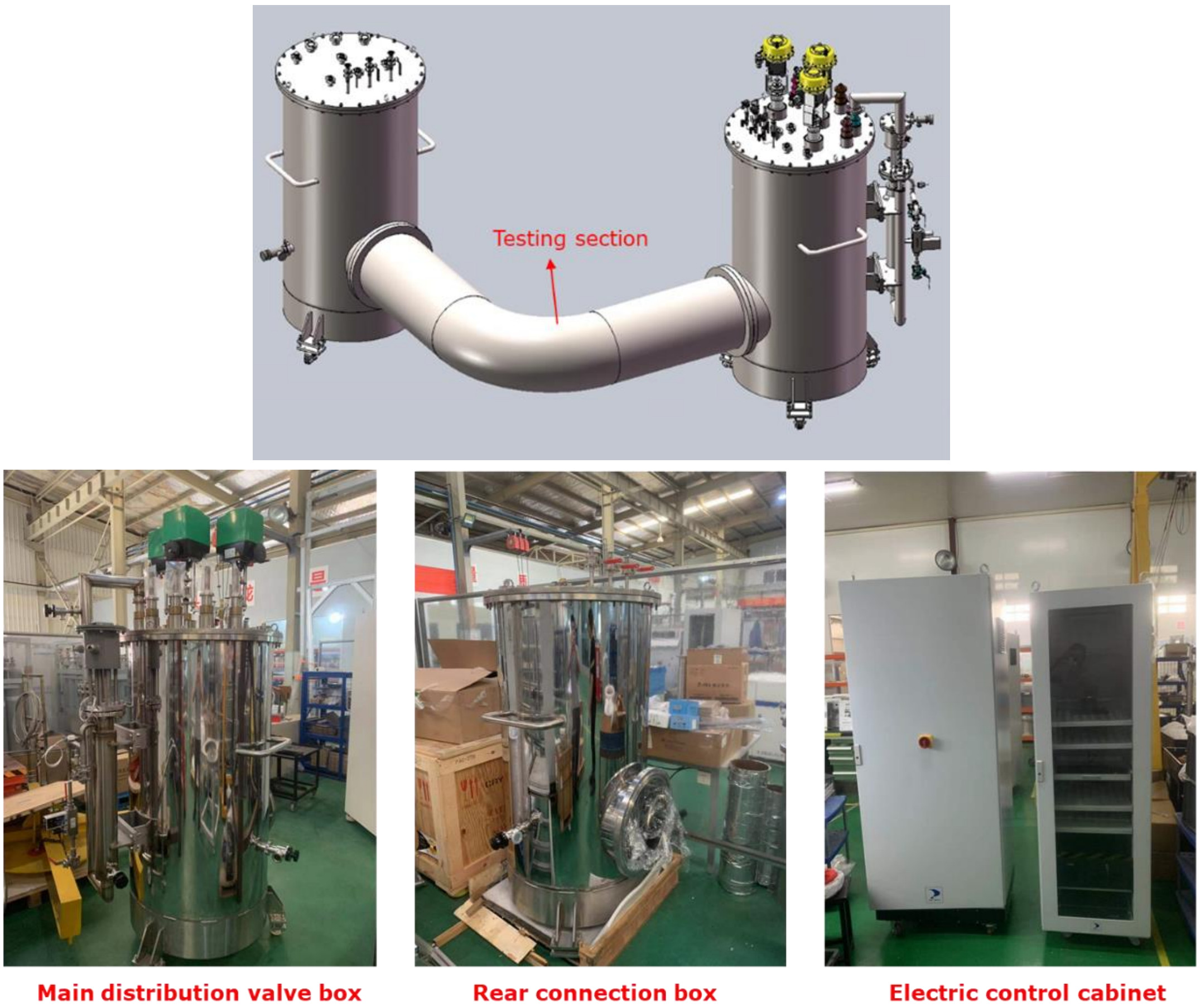

**Figure 7.1.45:** Test platform for a multi-channel cryogenic transfer line.

### 7.1.10.3 *Research on Advanced Control Strategy*

As superconducting accelerator technology advances, longer cryogenic transfer lines will inevitably lead to significant thermal inertia, resulting in non-linear and hysteresis control issues. The classic model free PID method may not be sufficient to properly address the multivariable coupling, nonlinearity, and hysteresis control issues, and additional control strategies may be necessary. CERN's Large Hadron Collider (LHC) project has addressed these issues with a number of studies on dynamic modeling and control techniques, resulting in the construction of the largest helium cryogenic system in the world. CERN has also developed a "virtual" cryogenic system using dynamic simulation models, which has been instrumental in improving control logic, predicting defects, training operators, and enabling online operation. The CERN superconducting accelerator's properties have also been used to construct a number of sophisticated control schemes, which have produced positive results [7-9]. In contrast, there is still a lack of pertinent research, and no similar system has been established in China. Therefore, it is important to perform more extensive research in the areas of dynamic simulation modeling and sophisticated control strategy for the cryogenic systems of superconducting accelerators in China to meet the actual requirements.

The IHEP cryogenic group is currently conducting advanced research on control strategies. We have successfully designed and implemented an automatic cool-down/warming-up procedure for an SRF cavity cryomodule using Model Predict Control (MPC) and Artificial Neural Network (ANN) methods. The main research environment test platform is a beam test cryomodule with two 2-cell 650 MHz superconducting cavities, which were used to verify the automatic cooldown process of the superconducting cavity. Figure 7.1.46 shows a comparison of the environment cooldown test curves between the former manual adjust method and the improved cooldown method. Based on the experimental results, the improved method achieved a quick and smooth cooldown process of the superconducting cavity while meeting the required temperature difference on the cavity. Moreover, the improved automatic cooldown method was more adaptable and saved 40% more time than the original manual control method. This lays the foundation for a more intelligent automated control of future large cryogenic systems or other systems with large hysteresis and non-linear properties. We maintain good discipline by testing all new ideas on prototypes before implementing them in the TDR. We have published two articles on this research [10-11].

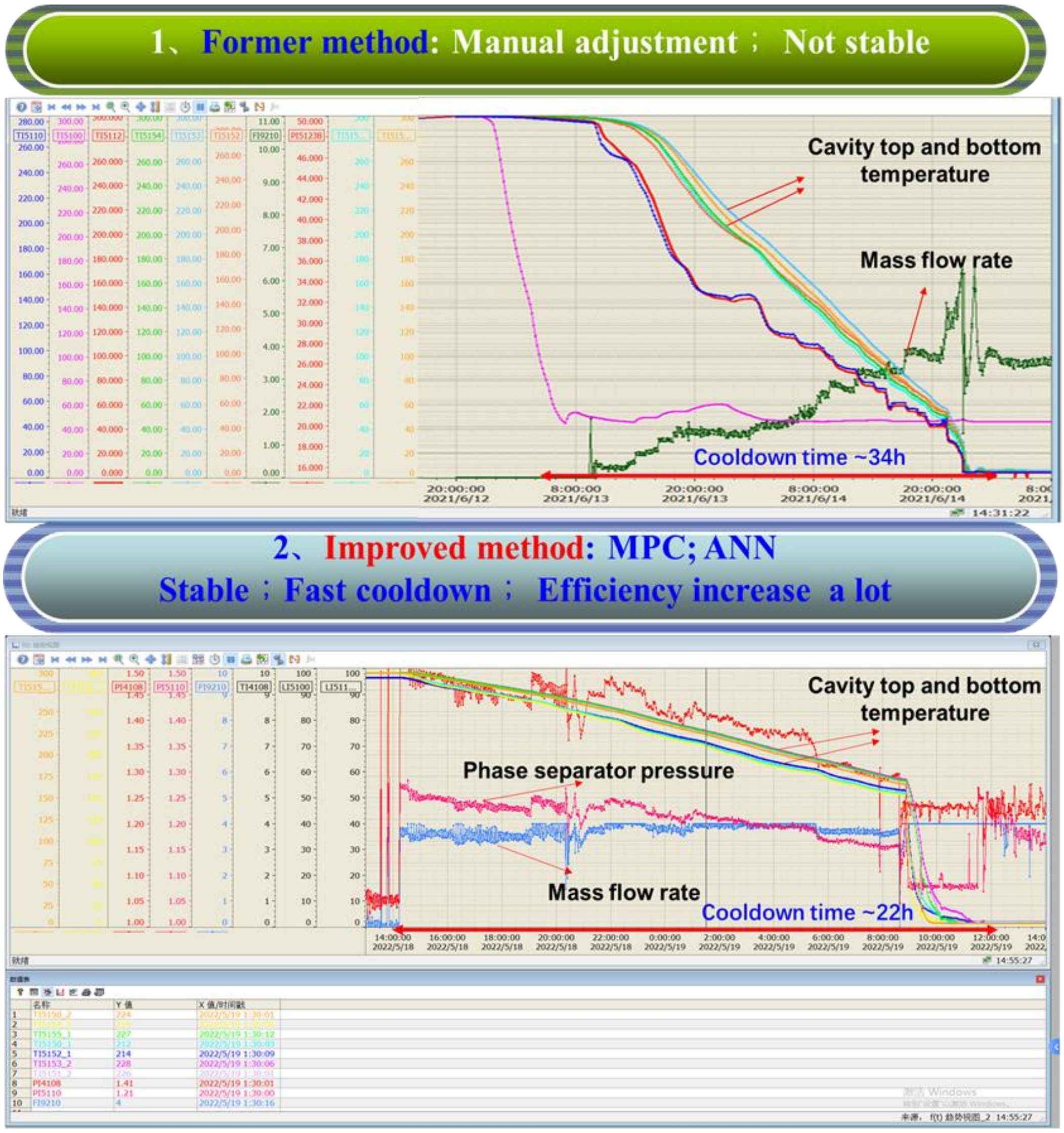

**Figure 7.1.46:** Environment cooldown test curves comparison between the former manual adjust method and improved cooldown method.

## 7.2 Survey and Alignment

### 7.2.1 Overview

The goal of CEPC alignment is to adjust all components to their designed positions within specified tolerances, ensuring a smooth beam orbit under absolute position control [1]. Survey and alignment work can be divided into two periods, namely construction and operation. During the construction period, the focus is on adjusting all components to their designed positions and providing an initial orbit for beam commissioning. This involves establishing the global datum, creating a control network, fiducializing and pre-aligning components, aligning tunnel installations, and ensuring a smooth beam orbit alignment. In the operation period, the beam serves as the datum for alignment, and the aim is to adjust all components according to the beam position, ensuring a smooth orbit for beam running.

The CEPC accelerator physics group has provided the most rigorous alignment accuracy requirements (RMS, 1σ) for certain primary components, as detailed in Table 7.2.1. These requirements pertain to the relative position accuracy between adjacent components.

**Table 7.2.1:** Requirement for the Collider magnet alignment accuracy.

| | **Transverse** (mm) | **Elevation** (mm) | **Roll** (mrad) |
|---|---|---|---|
| Arc Dipole | 0.1 | 0.1 | 0.1 |
| Arc Quadrupole | 0.1 | 0.1 | 0.1 |
| Sextupole | 0.1 | 0.1 | 0.1 |
| IR Quadrupole | 0.05 | 0.05 | 0.1 |

Alignment accuracy requirements for other components are less stringent than those mentioned above and are not included in this list.

Table 7.2.1 provides the total error limitation for primary components in the CEPC alignment process. In order to achieve this level of accuracy, each component undergoes several alignment steps, with the allowable error allocated among them to ensure the final positioning error meets the required accuracy. The alignment steps typically include control network error ($E_C$), fiducialization/pre-alignment error ($E_F$), field measurement (refer only to the measurements conducted during the component installation alignment) error ($E_M$), and installation adjustment error ($E_A$). These errors are independent of each other, so the final positioning error of a component can be calculated as follows:

$$Total\ error = \sqrt{E_C{}^2 + E_F{}^2 + E_M{}^2 + E_A{}^2} \tag{7.2.1}$$

To accurately position a component in the CEPC, it requires defining the component's position in a specific coordinate system and performing the component installation and alignment accordingly. This process involves establishing a control network, component fiducialization, and component installation and alignment. Given the CEPC's substantial size, approximately 32 km in diameter, alignment considerations must account for the irregular undulation of the geoid [2], and the establishment of global datums for measurement and data processing.

Currently, laser tracker is the primary measuring instrument for accelerator alignment. Due to the measurement range of a laser tracker being limited to only dozens of meters, the multi-station overlap method must be used in the tunnel for measurement, relying on common points to transfer spatial position relation. In this situation, controlling error accumulation is crucial for long-distance measurement [3]. With tens of thousands of components and a beam line length of more than 100 km, measurement work for the CEPC is heavy, requiring high efficiency and high accuracy measurement methods.

The mechanical structure and applied forces in the Machine Detctor Interface (MDI) are highly intricate. Achieving precise particle collisions within it requires even greater accuracy than in the arc region. Consequently, specialized alignment strategies are essential.

Uneven settlement of the foundation for such a large machine as the CEPC will be apparent. After the initial installation and operation over a period of time, large deformation is inevitable. To address this, it is essential to conduct regular surveys and alignment checks, as well as install position monitoring equipment in critical areas.

### 7.2.2 Global Datum

All measurement and installation activities for the CEPC are conducted on the Earth's surface, necessitating consideration of the Earth's gravity field. Gravity is used as a reference for instrument leveling during measurements, and observation reductions must also be referenced to the geoid. However, the irregular ellipsoid shape of the Earth and its non-uniform mass distribution result in irregular geoid shapes and non-uniform vertical gravity directions at various surface points. Processing observations based on such surface and directions is challenging. Therefore, the establishment of a quasi-geoid model and a vertical deflection model as a global datum is required to facilitate observation reduction.

Because the geoid's shape closely approximates that of a sphere, we can use a sphere as an approximation for the actual geoid when analyzing the impact of geoid curvature on data reduction, as illustrated in Figure 7.2.1.

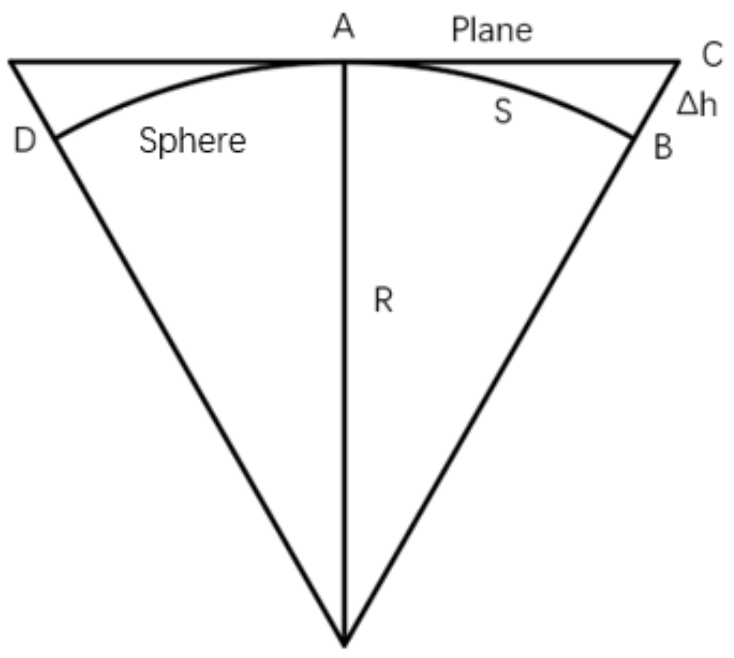

**Figure 7.2.1:** Effect of the geoid curvature on data reduction.

In this figure, AB is an arc of the sphere with a length of S. The Earth mean radius is R = 6371 km. AC is a tangent plane of the curve DB, A is the point of tangency. In the horizontal direction, using plane AC to substitute the geoid AB will generate an error ΔS = AB – AC. According to the following formula [4]:

$$\Delta S = \frac{1}{3}\frac{S^3}{R^2} \tag{7.2.2}$$

It can be calculated that, when S = 5 km, we get ΔS = 1 mm. Therefore, the curvature of the geoid has little influence on the horizontal data reduction. However, in the vertical direction, the error caused by using plane AC substitute the arc AB is BC = Δh. According to the formula below [4]:

$$\Delta h = \frac{S^2}{2R} \tag{7.2.3}$$

the calculated Δh value is 0.78 mm when S = 100 m. Therefore, the geoid curvature has a significant impact on elevation data reduction. When the measurement range exceeds a certain distance, the irregular shape of the geoid must be taken into account. For small accelerators covering a range of tens of meters, local geoid curvature changes can be ignored and treated as a plane. For middle-sized accelerators covering a range within several kilometers, the geoid can be simplified to a sphere or an ellipsoid. However, for CEPC, which is a large-scale machine covering a circumference of 100 km, the irregular curvature variation of the geoid cannot be ignored. Therefore, it is essential to establish accurate models as global datums for measurement and data reduction.

To establish global datums, a network of monuments needs to be built across the entire CEPC area. These monuments are distributed in a grid pattern and are spaced approximately about 5 km apart from each other. Astro-geodetic observations, global navigation satellite system (GNSS) observations, level observations, and gravity measurements are carried out on these monuments.

A gravity field model can be obtained as follows [4]:

$$\mathbf{W}(x,y,z)=\mathbf{U}(x,y,z)+\mathbf{T}(x,y,z) \tag{7.2.4}$$

where W is the gravity potential in point $(x, y, z)$, U is the regular gravity term, which can be derived from gravity field theory, and T is the irregular term, known as the disturbance field. Geoid undulation is the distance between the geoid and the reference ellipsoid along the normal of the earth ellipsoid, denoted by N. The geoid undulation $\Delta N_{AB}^{H_{tnl}}$ of point A and B generated by the disturbance field T can be expressed as Equation (7.2.5) with the vertical deflection $\varepsilon$ along the route from A to B and the positive altitude correction $E_{AB}^{H_{tnl}}$ [4]:

$$\Delta N_{AB}^{H_{tnl}}=-\int_A^B \varepsilon\cdot ds-E_{AB}^{H_{tnl}} \tag{7.2.5}$$

The positive altitude correction $E_{AB}^{H_{tnl}}$ can be obtained by measuring the gravity value, and $\varepsilon$ can be obtained by measuring the vertical deflection value. Therefore, the difference of the geoid undulation between any point and a level benchmark can be obtained by solving the above equation. Combining with the gravity geoid model, the grid-based quasi-geoid model of CEPC can be established.

To establish the global datums for CEPC, we need to construct a new generation of high-precision quasi-geoid refinement model and vertical deflection model using various data sources and modeling techniques. The construction process involves the following data processing strategies:

1) Satellite and ground gravity data will be combined to calculate the global gravity field model and determine the geoid long-wave component with an accuracy better than 1 cm.
2) To enhance the accuracy of the medium-wave component of the geoid, we will employ a more rigorous topographic equilibrium gravity reduction model and algorithm. Our aim is to improve the contribution of the topography equalization correction to the geoid with an accuracy level better than 1 cm. We will study interpolation and estimation methods that are suitable for the discrete gravity data grid and capable of elevating the grid average gravity anomaly to a new level.
3) To achieve high resolution local geoid with an accuracy tolerance of 1cm, we will study the application of the Stokes-Helmert boundary value research achievement to the short-wave component. We will precisely calculate the indirect effects of various topographic positions and corresponding gravity changes using the second Helmert agglutination method. We will also account for the gravitational effect caused by the topography and the mass of the agglutination layer to the Helmert gravity anomaly. In addition, we will fully consider all other corrections that may affect the 1 cm accuracy level.

4) The Laplace equation in the potential theory will be applied to do the best-fit of the gravitation quasi-geoid and GNSS level. Spherical Cap Harmonic Analysis will be used to express the local gravity field, and the difference between two kinds of geoids will be expressed as a non-integer (real) order integer spherical harmonic series expansion in a spherical cap domain, using new concepts and methods of Spherical Cap Harmonic Analysis.

Once the site location is determined, geodetic studies will commence, including the investigation of methods to transfer the quasi-geoid from the surface to the underground tunnel. The goal is to establish a quasi-geoid model with an accuracy better than 5 mm and a vertical deflection model with north-south and east-west component accuracies better than 1.0 arcsecond. The quasi-geoid model and the vertical deflection model will be utilized to reduce height values and determine the elevation directions of the measurement stations.

### 7.2.3 Control Network

The CEPC alignment control network is crucial for ensuring precise installation and alignment of components in the CEPC coordinate system. It serves as a reference for global and local position control, ensuring all components are installed consistently within the unified coordinate frame. To control error accumulation, a three-level control network is designed based on the principle of control from global to local. This includes the primary control network, backbone control network, and tunnel control network.

#### 7.2.3.1 *Primary Control Network*

The function of the CEPC primary control network is to control the shapes and mutual locations of the Linac, Damping Ring, beam transport lines, Booster, and Collider, and to provide high-accuracy constraint data for the lower-level control network. The layout of the primary control network is depicted in Figure 7.2.2.

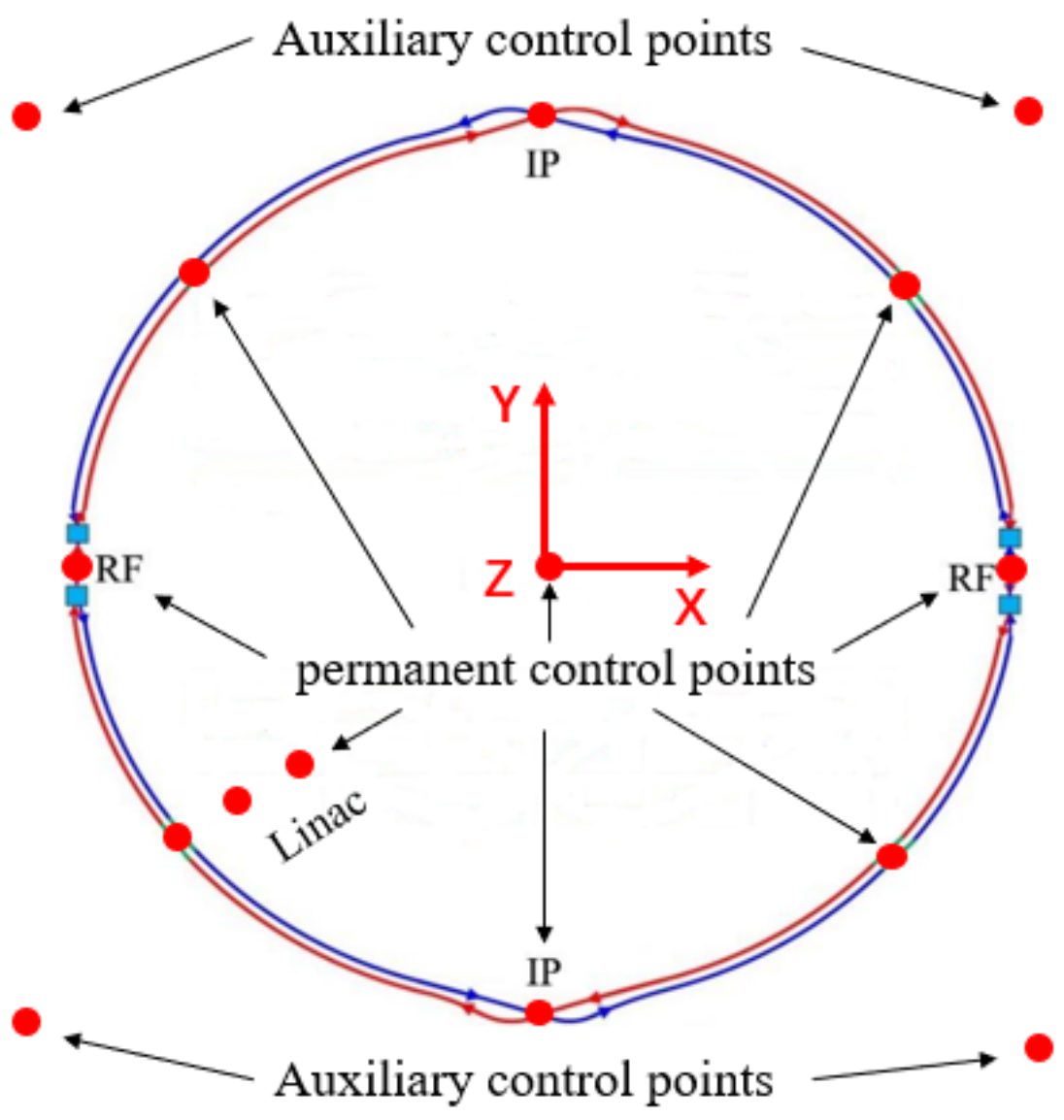

**Figure 7.2.2:** Layout of the CEPC primary control network.

The primary control network consists of 11 permanent control points along with some auxiliary control points. Among these, one permanent point is located at the center of the Collider ring on the ground, and the remaining 10 permanent points are situated in the tunnel. Out of these 10 permanent points in the tunnel, two are positioned at the beginning and end of the Linac, while the other eight are uniformly distributed in the Collider ring. The layout is depicted in Fig. 7.2.2. Above each permanent point in the tunnel, there is a through-hole that connects to either the ground or the roof, allowing the location of these points to be measured accurately from the ground by GNSS [5], as shown in Fig. 7.2.3.

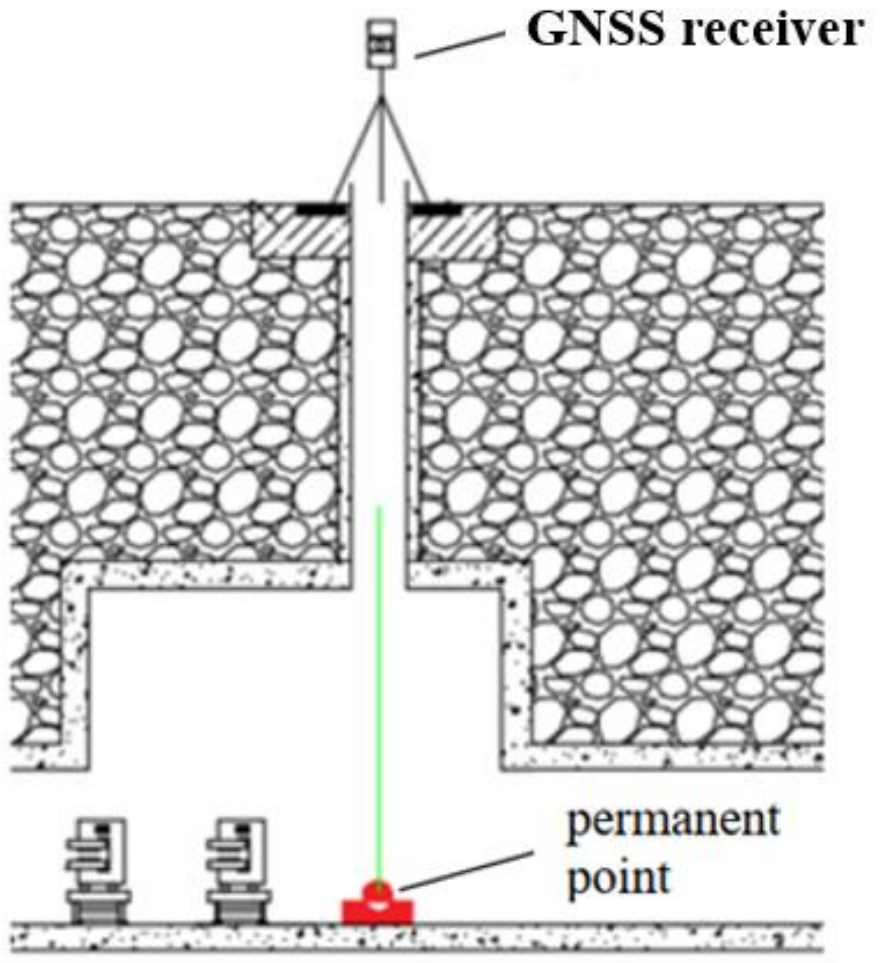

**Figure 7.2.3:** Measurement permanent point from ground.

After studying several layout schemes in the Collider ring, including those with 6, 8, and 10 points in the ring, with or without a center point, it was found that adding more control points can increase the accuracy, but there is a diminishing return on this effect. The simulation results from COSA software show that adding a center point can effectively improve the control network accuracy. After considering the tradeoff between cost and accuracy, the scheme with 8 points plus one center point is selected.

To supplement the primary control network, auxiliary control points will be constructed on the ground. These points must meet specific requirements, including being located at least 200 meters away from any high-voltage cables, avoiding strong reflection of satellite signals in the vicinity, avoiding large water areas to minimize the multi-path effects, and ensuring that there are no obstructions within 10 degrees above the point [6]. An example of an auxiliary control point is shown in Fig 7.2.4, which includes a markstone and a mark. The markstone is constructed using steel bars and poured concrete, and should be built on stable ground with minimal disturbance from road traffic.

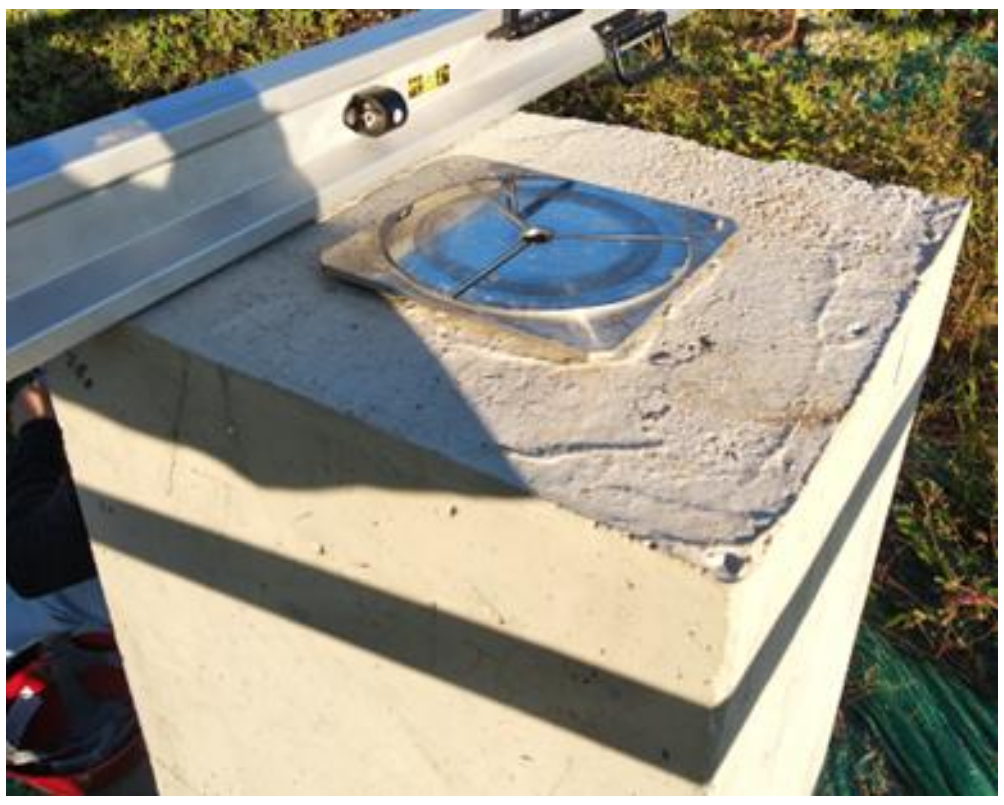

**Figure 7.2.4:** Auxiliary control point

The primary control network for CEPC will be established through GNSS and level measurements. GNSS measurements will involve aligning the GNSS receiver's antenna phase center with a permanent control point using a transit square. The distance from the ground to the permanent control point spans approximately 100-200 meters, and the center accuracy is expected to be better than 2 mm. This will be executed with 16 LEICA GS10 receivers and 16 AR20 choke ring antennas, with one AR20 antenna installed at each control point. GNSS observations will follow the multi-period static synchronous observation method, with each observation period lasting over 24 hours. The static relative positioning accuracy within a 30 km range is anticipated to achieve better than ±6 mm [7-8]. The final horizontal accuracy of the primary control network is expected to be better than 7 mm (1σ).

The level measurement will be conducted using a LEICA DNA03 electronic level measurement system, in accordance with the second-class leveling specification rules. The leveling route will be divided into four sections, including the Linac leveling route, the Linac to ring leveling route, the ring leveling route, and other leveling routes that connect the permanent and auxiliary points. The ring leveling route will be divided into eight sub-leveling routes, each of which will be between adjacent permanent points. The accuracy specifications of the leveling measurement should meet or exceed the tolerances listed in Table 7.2.2.

**Table 7.2.2:** Leveling measurement tolerance (L: length of the levelling route in km)

| Back and forth discrepancy (mm) | Accident standard error (mm/km) | Ring closed error (mm) | Total standard error (mm/km) |
|---|---|---|---|
| $1.8\sqrt{L}$ | 0.45 | $2\sqrt{L}$ | 1.0 |

The data processing of CEPC's primary control network includes two parts: GNSS observations data processing and leveling observations data processing.

The GNSS observations will be processed using the Trimble Business Center (TBC) software. The processing will involve two main calculations: base line calculation and network adjustment. The network adjustment will use the Gauss projection method to calculate the horizontal coordinates of the primary control network, with the CGCS2000 as the reference ellipsoid.

To reduce projection deformation, two methods can be used: the ellipsoid expansion method or the arbitrary projection zone Gauss conformal projection method [9].

The CEPC coordinate system can initially be established within the civil construction coordinate system. In this context, we can define the origin and points along the X and Y axes. By conducting measurements that incorporate both civil control points and the primary control network, we can determine the coordinates of the primary control network points within the civil construction coordinate system. Subsequently, using the designated origin and the points along the X and Y axes, we can establish the CEPC coordinate system and determine the coordinates of the primary control network points within the CEPC coordinate system.

The leveling observations are referenced to the geoid and need to be converted to elevation coordinates in the CEPC coordinate system. This conversion is achieved through the use of the quasi-geoid model, the vertical deflection model, and the approximate coordinates of the measuring points. These components are used to calculate the elevation coordinates of the primary control network.

The processing of leveling observation data consists of two main parts: level adjustment and elevation coordinate calculation. Level adjustment can be performed using various software, and once completed, the leveling elevations of the measuring points can be obtained.

Fig. 7.2.5 displays a vertical vector in the CEPC coordinate system. According to the vertical deflection model and the approximate coordinates of an arbitrary point P, we can get the vertical unit vector $S\left(X_{CS} \quad Y_{CS} \quad Z_{CS}\right)$ of P in CEPC coordinate system. Suppose the angle between CEPC $Z$ axis and the vertical unit vector $S$ is $\alpha = \arccos\left(Z_{CS}\right)$, the angle between $X$ axis and the projection of $S$ in the $X$-$Y$ plane is:

$$\beta = \arccos\left(\frac{X_{CS}}{\sqrt{X_{CS}^2 + Y_{CS}^2}}\right), \tag{7.2.6}$$

The approximate coordinate of P is $\left(x_1 \quad y_1 \quad z_1\right)$ , and the leveling elevation of P is $h$.

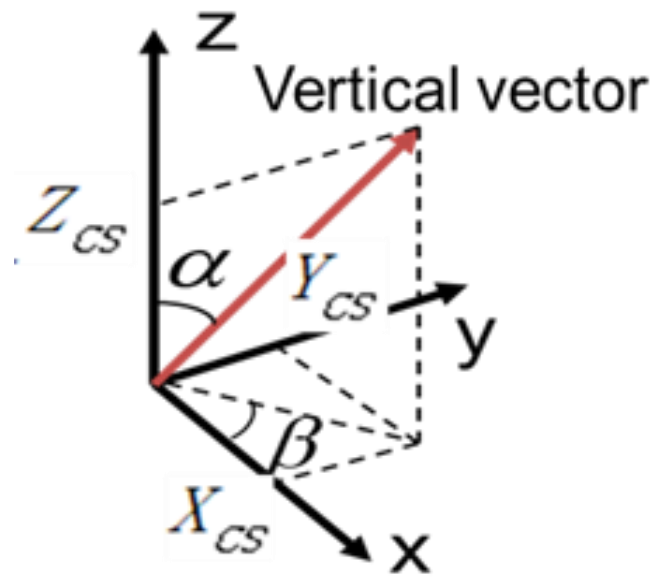

**Figure 7.2.5:** Vertical vector calculation

The projection point of P in the quasi-geoid model is $\left(x_1' \quad y_1' \quad z_1'\right)$, we can get:

$$\begin{cases} x_1' = x_1 + h \cdot \sin\alpha \cdot \cos\beta \\ y_1' = y_1 + h \cdot \sin\alpha \cdot \sin\beta \end{cases} \tag{7.2.7}$$

Plugging ($x_1$', $y_1$') into the quasi-geoid model we can get the elevation coordinate $z_1$' of this projection point. The projection length of $h$ on the CEPC $Z$ axis is $h\cos\alpha$, then the elevation coordinate of P in the CEPC coordinate system is $z_1 = h\cos\alpha + z_1'$.

#### 7.2.3.2 ***Backbone Control Network***

The CEPC backbone control network will be constructed inside the tunnel, connecting the primary control network and the tunnel control network. Its purpose is to reinforce the tunnel control network and reduce error accumulation. Due to the narrow width of the CEPC tunnel, which is only 6 meters, a kind of triangular network structure has been selected to increase control ability and decrease the workload of the measurement. This backbone control network consists of 334 control points, as depicted in Figure 7.2.6. The network comprises short line segments, each roughly spanning 300 meters, and long line segments, each covering approximately 600 meters.

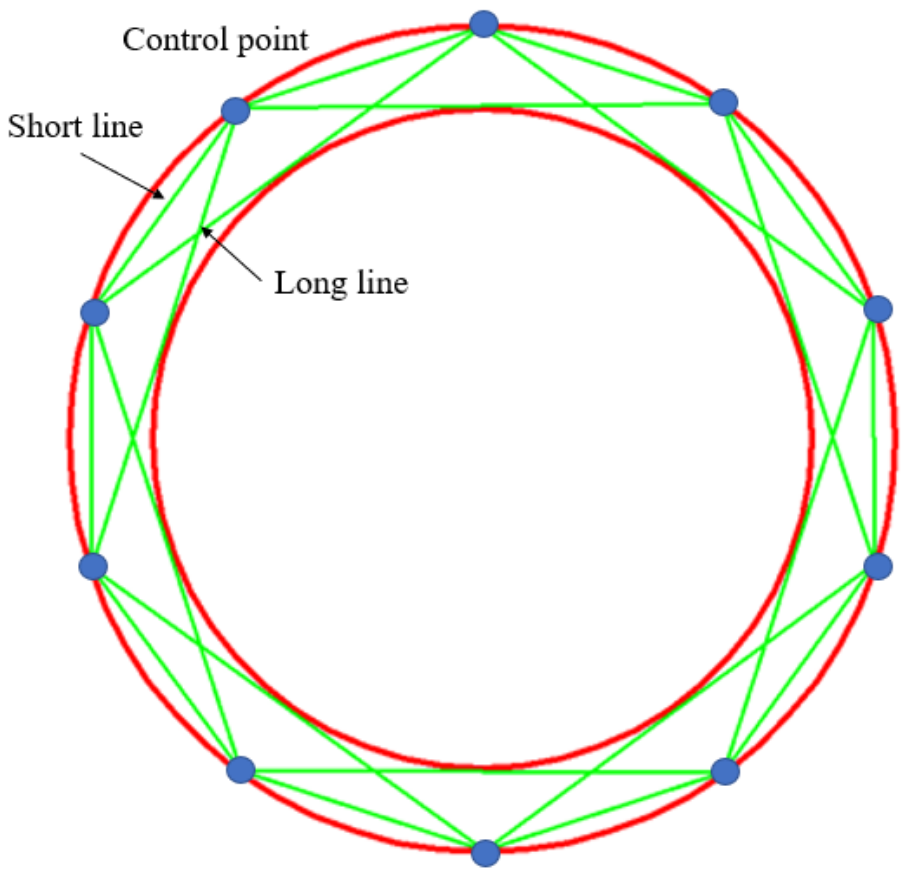

**Figure 7.2.6:** CEPC backbone control network

The CEPC backbone control network will be measured using total stations (a kind of automated measuring instrument that integrates distance and angle measurement functions), levels, and a gyro-theodolite. The total station will use the forced centering measurement method by setting up stations on every control point and measuring the distances and angles of the adjacent points. Within the ring tunnel, the permanent control points, which are integral to the backbone control network, will have gyro-theodolites positioned on each of them to measure 16 azimuths.

The observations obtained from the total stations and the azimuths will be used to calculate the horizontal coordinates of the backbone control network. The transformation parameters between the CEPC coordinate system and the measuring station coordinate system will be calculated using a vertical deflection model, and a 3D adjustment will be applied to calculate the observations obtained from the total stations and the azimuths. The coordinates of the permanent control points will be used as known data to constrain the 3D adjustment. After the 3D adjustment, the 3D coordinates of the backbone control network will be obtained, and the X-Y coordinates will be used as the formal horizontal coordinates. The leveling observations will be used to determine the leveling heights of each point after adjustment. By using the vertical deflection model, quasi-geoid model, and the approximate coordinates of each point, the leveling heights can be corrected to obtain the elevation coordinates of the backbone control network in the CEPC coordinate system.

As an example, a simulation was conducted on a 12.5 km section of the backbone control network between two adjacent permanent control points. The simulation assumed an angle measurement accuracy of 0.5 arcsecond, a distance measurement accuracy of 1 mm + 1 ppm, and an azimuth measurement accuracy of 2.6 arcsecond. The results of the simulation showed that the average point error was better than 4 mm, and the relative point error within 300 m was better than 0.8 mm.

#### 7.2.3.3 *Tunnel Control Network*

The CEPC tunnel control network serves as the position reference for component installation, alignment, and stability monitoring. To achieve high-accuracy relative position control in addition to absolute position control, the tunnel control network must be designed to reference the primary control network for high-accuracy positioning in the CEPC coordinate system, and distribute points in a rational manner with precise measurements to achieve local high-accuracy relative positioning. The CEPC tunnel control network will be evenly distributed along the tunnel by sections with an interval of 6 m, and each section will have 4 points, with 2 located on the floor and 2 on the inner and outer walls, respectively, as shown in Figure 7.2.7. The total length of CEPC tunnel is about 113 km, and it will consist of over 18,800 sections throughout the CEPC complex.

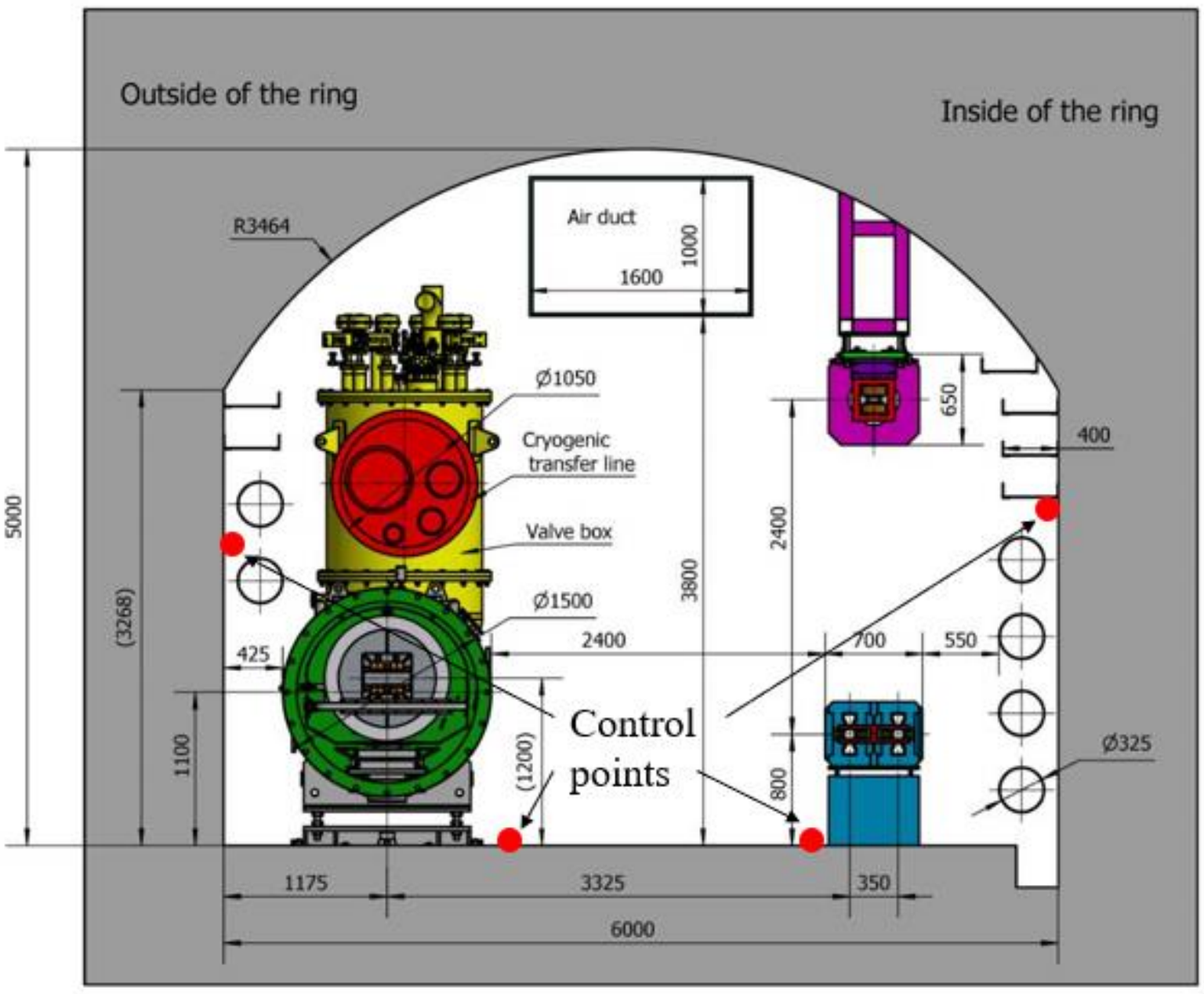

**Figure 7.2.7:** Cross-section of the CEPC tunnel control network.

The CEPC tunnel control network will be measured using laser trackers and levels. To measure the entire tunnel, the multi-station overlap method will be employed for laser tracker measurements. A laser tracker will be positioned in the middle of every adjacent control network section, and a level coordinate system will be established at each station. To improve measurement accuracy, sufficient redundant measurements will be taken between adjacent measuring stations. In each station, 6 network sections will be measured, with a measurement range of approximately 30 meters. With more than 18,800 measuring stations required for the tunnel control network, multiple survey groups can work in different measurement regions simultaneously. Level measurement will only be conducted on the control points located on the floor.

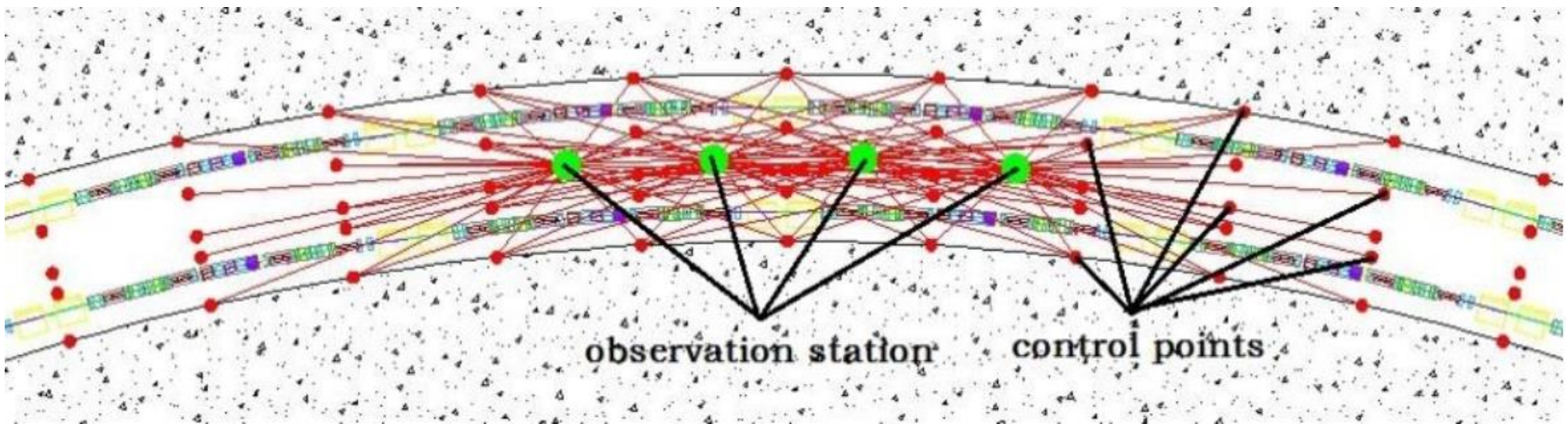

**Figure 7.2.8:** Tunnel control network survey by laser tracker

The observations from the laser tracker and level measurements will be processed using a 3D adjustment. Firstly, the level observations will undergo leveling adjustment and geoid curvature correction to obtain the elevation coordinates of the control points in the CEPC coordinate system. Subsequently, the laser tracker observations will be calculated by a 3D adjustment with elevation constraints, using the elevation coordinates of the control points. The coordinate transformation parameters of the 3D adjustment can be calculated using the vertical deflection model and the approximate coordinates of each station. In the tunnel control network adjustment, the known points will be the backbone network control points, spaced at 300 m intervals. As an illustrative example, a simulation was conducted on a 300 m segment of the tunnel control network between two adjacent backbone network control points. This simulation, following the research of Yang Zhen [10] and Jing Liang [11], utilized the following conditions: angle measurement accuracy of 2 arcseconds, distance measurement accuracy of 0.015 mm + 2 ppm. Using these measurement accuracies, observations for each measuring station were generated. The simulation results indicated an average point error within 300 m of 0.6 mm and a relative point error within 6 m of 0.074 mm.

Table 7.2.3 presents time estimates for the measurement of the backbone control network and tunnel control network, taking into account factors such as workload, measuring speed, and the number of measurement groups.

**Table 7.2.3:** Time estimation for the backbone control network and tunnel control network measurement.

| Instrument | Work load | Speed (1 group per day) | No. of groups (person) | Months |
|---|---|---|---|---|
| Laser tracker | 18,800 stations | 8 stations | 16 (2) | 6 |
| Total station | 334 stations | 4 stations | 8 (4) | 0.7 |
| Level | 113 km | 4 km | 8 (4) | |

An automatic observation scheme has been designed to improve the efficiency of the tunnel control network measurement, which utilizes a visual instrument. The visual instrument is a newly developed instrument that integrates photogrammetry, distance, and angle measurement functions, with high precision and efficiency characteristics. The visual instrument will be installed on a mobile platform, which will move along the strong reflective stripes on the tunnel floor to conduct the measurement, as illustrated in Fig. 7.2.9.

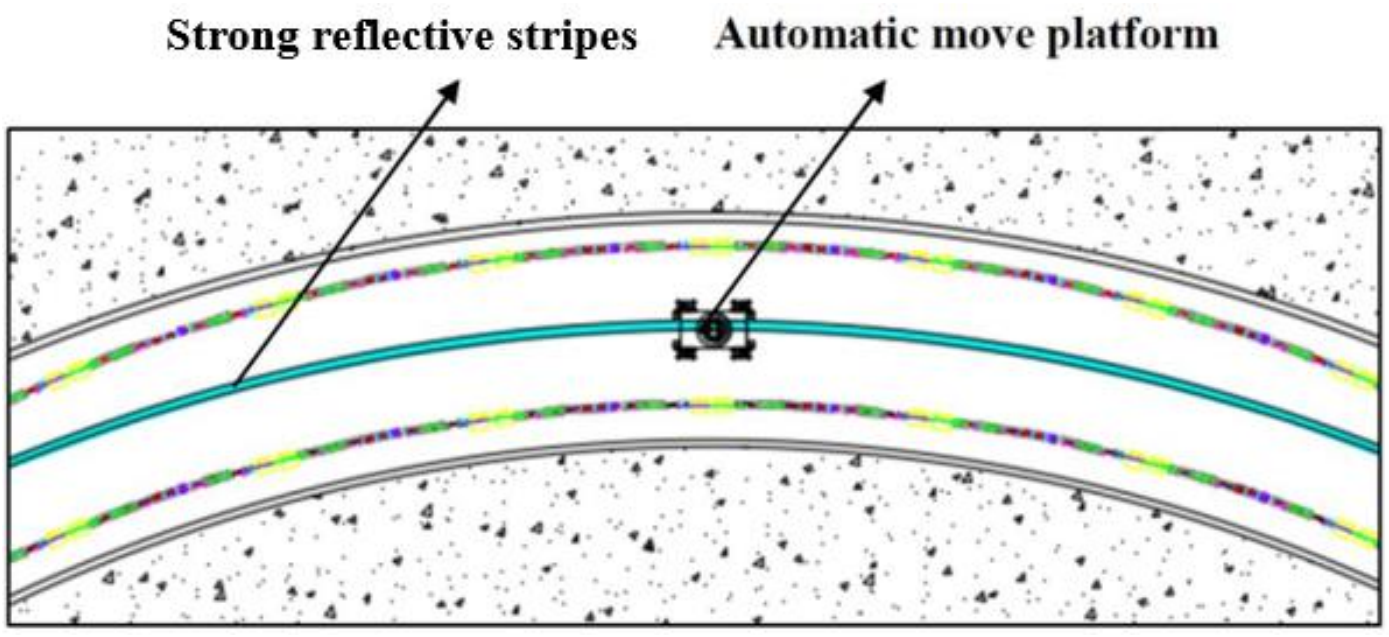

**Figure 7.2.9:** Automatic measurement system

The measurement method is similar to that of a laser tracker. The measurement station will be set up in the middle of every two adjacent sections. When the platform reaches a predetermined station position, the supporting legs will automatically hold up the platform and adjust its level, after which, it will begin to measure the tunnel control network. The observations will be divided into horizontal and vertical observations, as illustrated in Fig. 7.2.10 and Fig. 7.2.11. There will be sufficient redundancy observations between adjacent stations to increase the overlap strength and the point measurement accuracy will be comparable to that of a laser tracker.

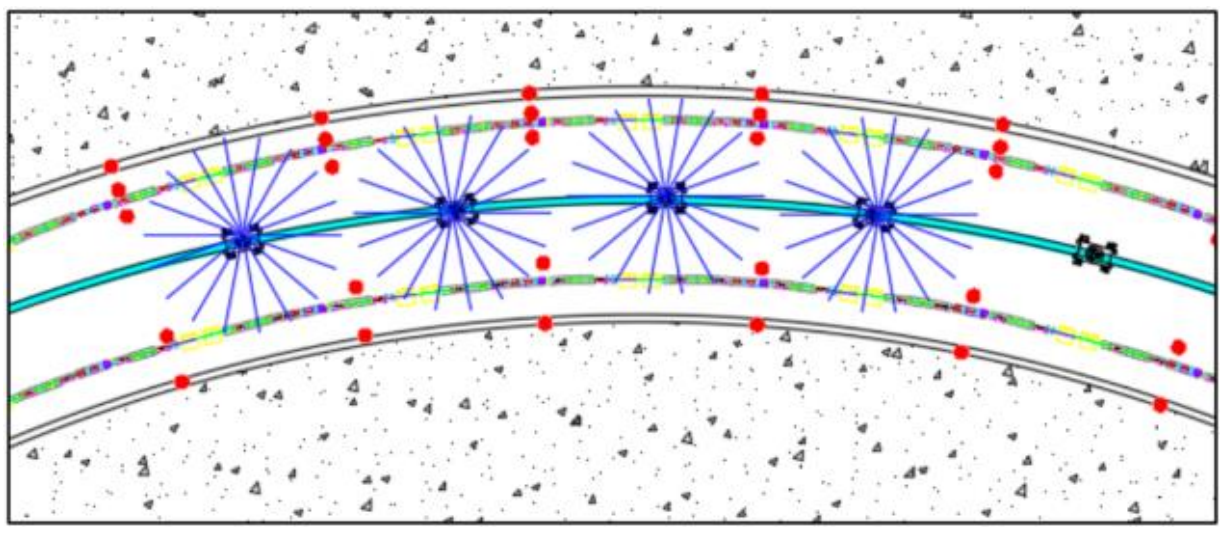

**Figure 7.2.10:** Horizontal observation

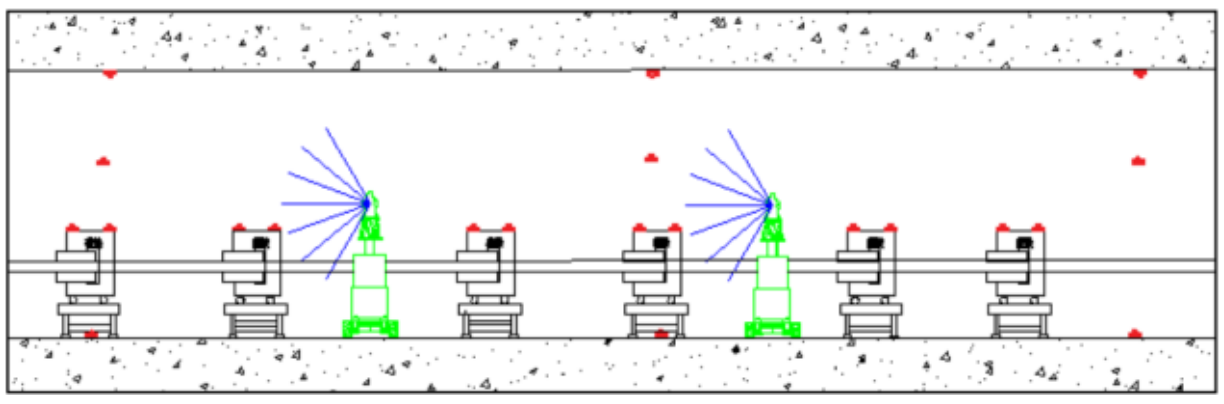

**Figure 7.2.11:** Vertical observation

## 7.2.4 Fiducialization and Pre-alignment

In order to determine the position of a component in CEPC coordinate system, component fiducialization is necessary. Fiducials are located on the top of each component and are used to accurately determine the position of the component. There are two methods for component fiducialization: based on mechanical center or based on magnetic center. The mechanical center or magnetic center of the component is related to its fiducials using these methods.

### 7.2.4.1 *Fiducialization based on Mechanical Center*

The mechanical center of a component is usually its designed beam center, such as the aperture center of a magnet or the centerline of an accelerating structure. The physical surfaces of a component can be measured using a laser tracker or an articulated arm, and best-fit geometrical elements can be derived from the measuring points to obtain the mechanical center of the component. To relate the beam center with the fiducials, a component coordinate system is established, with its origin at the center of the component and using the mechanical center as an axis. The coordinates of the beam entrance point and exit point relative to the physical surfaces can be obtained based on the design. By measuring the fiducials, the spatial position relation between the beam center and the fiducials in the component coordinate system can be established. The fiducialization accuracy is generally as specified in Table 7.2.4.

**Table 7.2.4:** Fiducialization accuracy based on mechanical center.

| Transverse (mm) | Elevation (mm) | Longitudinal (mm) |
|---|---|---|
| 0.05 | 0.05 | 0.08-0.15 |

### 7.2.4.2 *Fiducialization based on Magnetic Center*

The magnetic center of a magnet is its nominal beam center. The magnetic center is not perfectly coincided with its mechanical center. If using the mechanical center as a datum to do component fiducialization, it will introduce errors. To improve the magnet fiducialization accuracy, it needs to use the magnetic center as an axis to establish the component coordinate system.

The magnetic center of a multipole magnet can be measured using a rotating coil. By using a rotating coil, it is possible to measure the magnetic center position relative to the rotation axis of the coil. Once the magnetic center position is known, a laser tracker or a coordinate measuring machine (CMM) can be used to measure the rotation axis position. This information can be used to establish the spatial position relation between the beam center and the fiducials, using the same method as the fiducialization based on the mechanical center. The difference is that the magnetic center is used instead of the mechanical center to build the component coordinate system.

A rotating wire magnet fiducialization system, depicted in Fig. 7.2.12, has been developed and successfully employed for HEPS quadrupole and sextupole fiducialization. This system incorporates a Coordinate Measuring Machine (CMM), a rotating wire magnetic center measuring system, an image probe, two 6-Degree-of-Freedom (6-DOF) stages, and various magnet support and adjustment mechanisms. It facilitates automated magnet fiducialization, where the position of the rotating wire is precisely measured by the image probe. With the assistance of the CMM and the 6-DOF stages, the wire can be automatically aligned with respect to the magnet. This rotating wire magnet fiducialization system will also be used for the fiducialization of CEPC quadrupoles and sextupoles.

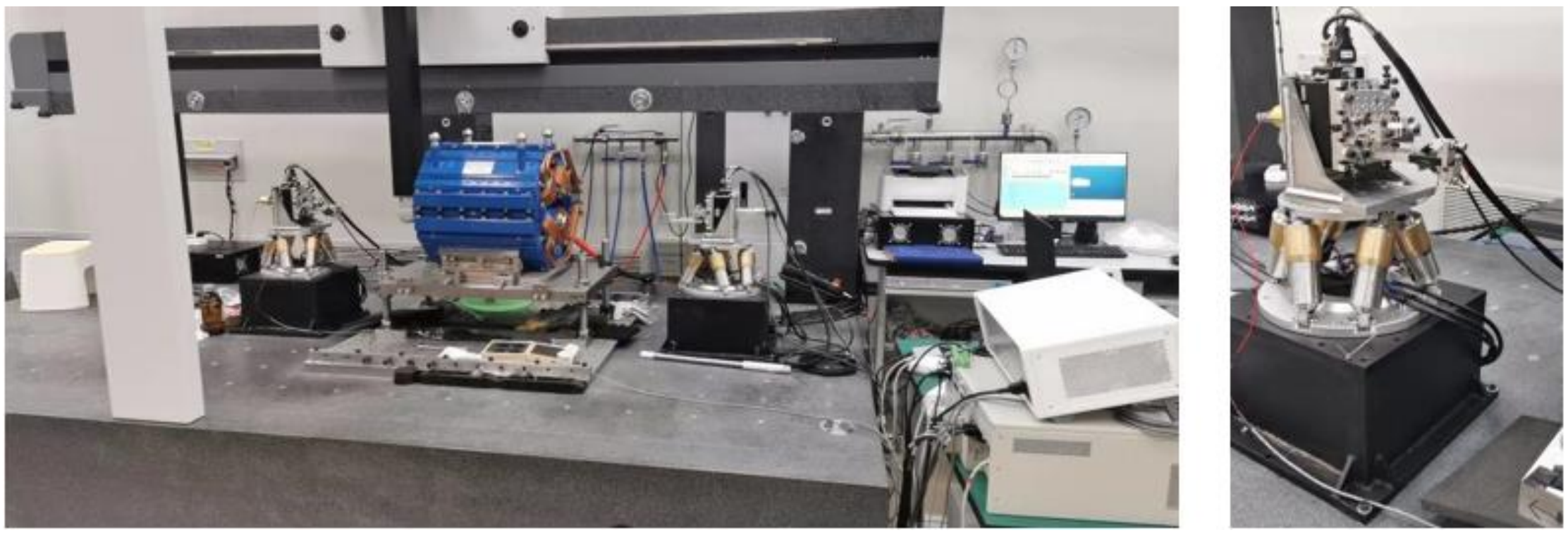

**Figure 7.2.12:** A rotating wire magnet fiducialization system for HEPS.

When employing a rotating wire magnet fiducialization system for magnet fiducialization, the measurement errors consist of the following components:

1. Magnetic center measurement error, which is better than 0.008 mm.
2. Wire rotation axis measurement error, which is better than 0.01 mm.
3. Relating the rotation axis to the fiducials error, which is better than 0.005 mm.

In total, the fiducialization accuracy typically aligns with the values shown in Table 7.2.5.

**Table 7.2.5:** Fiducialization accuracy based on magnetic center.

| Transverse (mm) | Elevation (mm) | Longitudinal (mm) |
|---|---|---|
| 0.014 | 0.014 | 0.05 |

### 7.2.4.3 *Pre-alignment*

To improve the speed of component installation, a pre-alignment scheme can be applied. This involves installing and aligning components on the same girder in advance, so that the cell can be installed and aligned as a single unit in the tunnel. In accordance with the design, the installation of sextupoles in the Collider can utilize a pre-alignment scheme. Figure 7.2.13 illustrates the presence of 1,024 pre-alignment cells within the Collider for this purpose.

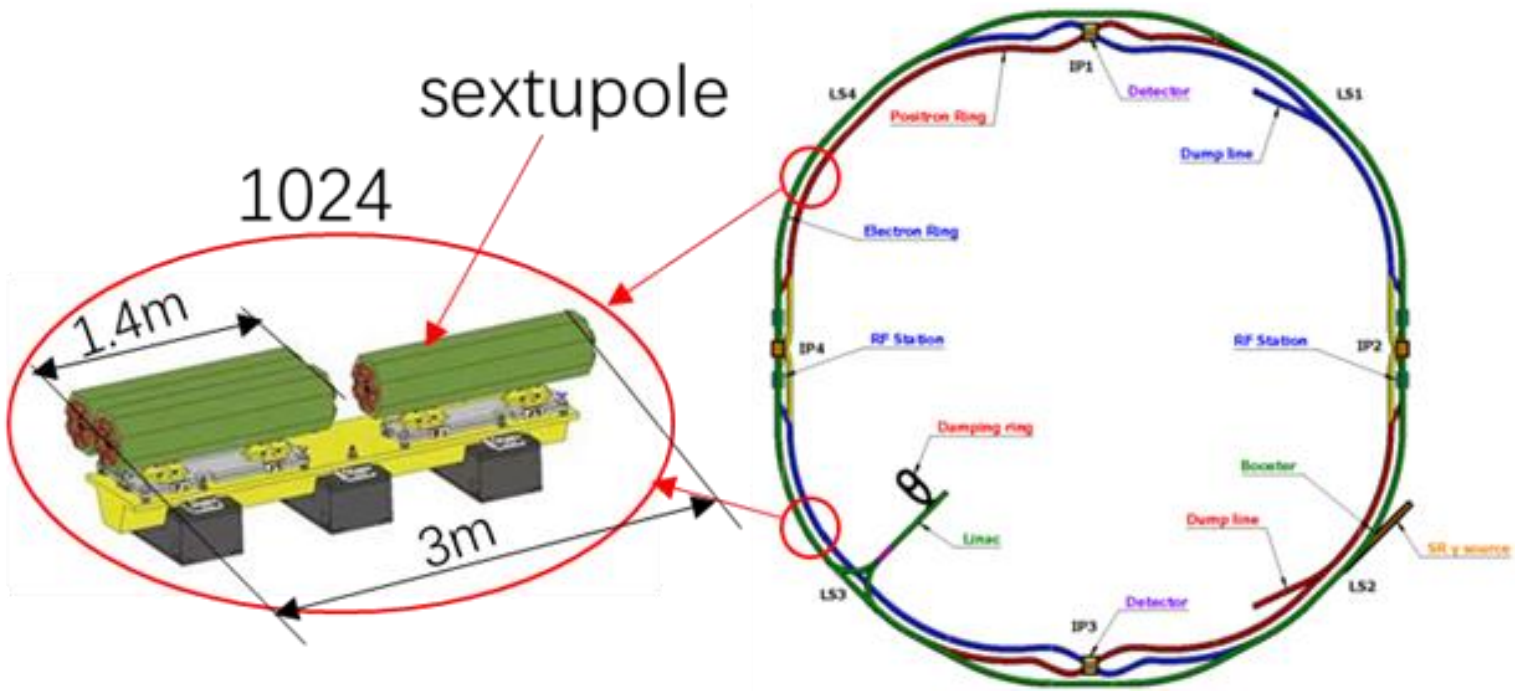

**Figure 7.2.13**: Pre-alignment of sextupoles.

Pre-alignment can be carried out in the test and assembly factory using a laser tracker as the measurement instrument. Each pre-aligned cell must have a cell coordinate system based on the physical surface of the girder. Fiducials are placed on each girder to aid in the recovery of the cell coordinate system. Using coordinate transformation, the nominal coordinates of the components in the cell coordinate system can be calculated based on the component fiducialization results.

By measuring the fiducials of the girder, the cell coordinate system can be recovered. The offsets of the fiducials relative to their nominal coordinates can then be obtained by measurement in the cell coordinate system. Components can be adjusted based on the offsets until their position offset is smaller than the tolerances and then fixed on the girder.

The pre-alignment error comprises the following components:

1. Sextupole fiducialization error: 0.014 mm
2. Laser tracker measurement error: 0.03 mm
3. Adjustment and fixation error: 0.03 mm
4. Error allowance for transportation: 0.02 mm

In total, the accuracy of pre-alignment can be achieved at 0.05 mm in both transverse and elevation dimensions.

Table 7.2.6 provides time estimates for fiducialization and pre-alignment, taking into consideration the component quantity, working speed, and group number. Because the fiducialization and pre-alignment processes operate at a slower pace compared to component installation, these activities will commence 3 to 4 years ahead of the installation phase, ensuring they align with the CEPC construction schedule requirements.

**Table 7.2.6:** Time estimation for the fiducialization and per-alignment

| Component | Instrument | Quantity | Speed (1 group per day) | No. of groups | Months |
|---|---|---|---|---|---|
| Dipole | Laser tracker/Optical instrument | 31259 | 2 | 16 | 39.1 |
| Quadrupole | Rotating wire | 8320 | 3 | 8 | 14 |
| Sextupole | Rotating wire | 3348 | 3 | 8 | 5.58 |
| Corrector | Measuring arm | 9799 | 4 | 16 | 6.12 |
| BPM、PR、DCCT、kicker | Measuring arm | 6132 | 4 | 16 | 3.83 |
| Septum Magnet | Laser tracker | 102 | 1 | 16 | 0.26 |
| Kicker | Laser tracker | 18 | 1 | 16 | 0.05 |
| Electrostatic separator | Laser tracker | 32 | 2 | 16 | 0.04 |
| Collimator /dump | Laser tracker /Optical instrument | 44 | 2 | 16 | 0.06 |
| Solenoid | Measuring arm | 37 | 4 | 16 | 0.02 |
| Accelerating structure | Laser tracker | 577 | 2 | 16 | 0.72 |
| Cavity | Measuring arm | 4 | 2 | 16 | 0.01 |
| Pre-alignment cell | Laser tracker | 1024 | 0.5 | 16 | 5.2 |
| Total | | | | 16 | 75.2 (6.3 years) |

### 7.2.5 Component Installation Alignment

During the tunnel installation phase, the installation and alignment of components will be carried out using the tunnel control network as a reference and laser trackers as the measurement instrument. The component alignment process involves three phases, namely, initial installation alignment, overall survey, and final smooth alignment.

#### 7.2.5.1 *Initial Installation Alignment*

During the initial installation alignment, the first step is to set out the stands and install the components. For the component alignment, a laser tracker station will be set up near the component, as shown in Fig. 7.2.14. The laser tracker is used to measure the control network, and by using a best-fit method, the laser tracker can be located in the tunnel coordinate system. Since the adjustment direction of a component is often not parallel to the axis of the global coordinate system, it is better to establish a local coordinate system for the component adjustment, with an axis parallel to the beam. The laser tracker is used to measure the fiducials on the component, and we can obtain the offset between the actual and nominal positions. This offset is then adjusted to meet the required tolerance, which is typically ±0.05 mm.

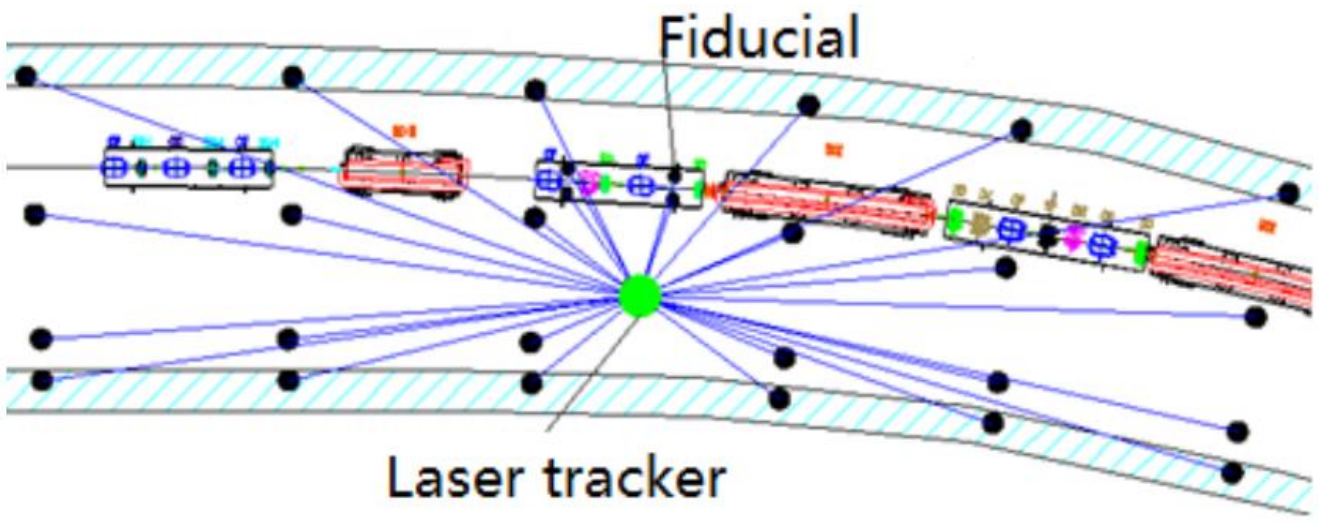

**Figure 7.2.14:** Component alignment by using a laser tracker and tunnel control network.

Our plan involves organizing 16 alignment groups for the Collider, 16 groups for the Booster, and 1 group dedicated to the Linac, Damping Ring (DR), and transport lines (TL) alignment. Tables 7.2.7, 7.2.8, and 7.2.9 provide time estimates for alignment activities in the Collider, Booster, and Linac/DR/TL alignment. These estimates take into account component quantity, working speed, and the number of alignment groups.

**Table 7.2.7:** Time estimation for the Collider alignment

<table>
<tr><th>Component</th><th>Quantity</th><th>Speed<br>(1 group per day)</th><th>No. of groups</th><th>Months</th></tr>
<tr><td>Dipole</td><td>16258</td><td>3</td><td>16</td><td>13.5</td></tr>
<tr><td>Quadrupole</td><td>4148</td><td>4</td><td>16</td><td>2.6</td></tr>
<tr><td>Sextupole</td><td>3176</td><td>4</td><td>16</td><td>2</td></tr>
<tr><td>Corrector</td><td>7088</td><td>6</td><td>16</td><td>3</td></tr>
<tr><td>BPM\PR\DCCT</td><td>3544</td><td>6</td><td>16</td><td>1.5</td></tr>
<tr><td>Septum Magnet</td><td>68</td><td>3</td><td rowspan="6">16</td><td rowspan="6">1.5</td></tr>
<tr><td>Kicker</td><td>8</td><td>4</td></tr>
<tr><td>Cryomodule</td><td>32</td><td></td></tr>
<tr><td>Electrostatic separator</td><td>32</td><td>2</td></tr>
<tr><td>Superconducting Magnets</td><td>4</td><td></td></tr>
<tr><td>Collimator\ dump</td><td>36</td><td>4</td></tr>
<tr><td>Total</td><td></td><td></td><td>16</td><td>24</td></tr>
</table>

**Table 7.2.8:** Time estimation for the Booster alignment

<table>
<tr><th>Component</th><th>Quantity</th><th>Speed<br>(1 group per day)</th><th>No. of groups</th><th>Months</th></tr>
<tr><td>Dipole</td><td>14844</td><td>2</td><td>16</td><td>18.5</td></tr>
<tr><td>Quadrupole</td><td>3458</td><td>4</td><td>16</td><td>2</td></tr>
<tr><td>Sextupole</td><td>100</td><td>4</td><td>16</td><td>0.1</td></tr>
<tr><td>Corrector</td><td>2436</td><td>6</td><td>16</td><td>1</td></tr>
<tr><td>BPM\PR\DCCT</td><td>2408</td><td>6</td><td>16</td><td>1</td></tr>
<tr><td>Cryomodule</td><td>12</td><td></td><td rowspan="3">16</td><td rowspan="3">1.4</td></tr>
<tr><td>Septum Magnet</td><td>32</td><td>2</td></tr>
<tr><td>Kicker</td><td>8</td><td>4</td></tr>
<tr><td>Total</td><td></td><td></td><td>16</td><td>24</td></tr>
</table>

**Table 7.2.9:** Time estimation for the Linac DR, TL alignment

| Component | Quantity | Speed (1 group per day) | No. of groups | Months |
|---|---|---|---|---|
| Dipole | 135 | 3 | 1 | 1.8 |
| Quadrupole | 714 | 4 | 1 | 7 |
| Sextupole | 72 | 4 | 1 | 1 |
| Corrector | 275 | 6 | 1 | 1.8 |
| BPM\PR\DCCT | 180 | 6 | 1 | 1.2 |
| Accelerating structure | 577 | 4 | 1 | 5.8 |
| Solenoid | 37 | 3 | 1 | 0.5 |
| SHB BUN RF Cavity | 4 | 2 | 1 | 0.4 |
| Collimator /dump | 8 | 3 | | |
| Electron Source | 1 | | | |
| Positron Source | 1 | | | |
| Septum Magnet | 2 | | | |
| Kicker | 2 | | | |
| Total | | | 1 | 19.5 |

### 7.2.5.2 *Overall Survey*

Following the initial alignment, it is imperative to conduct an overall survey to assess whether all components fall within the specified tolerance and ensure the smoothness of the beam orbit. This survey includes three parts: the primary control network survey, the backbone control network survey, and the tunnel overall survey. The primary control network survey and the backbone control network survey are the same as above. The tunnel's overall survey encompasses both the survey of the tunnel control network and the assessment of all its components.

To carry out the tunnel overall survey, laser trackers and levels are used. Each station must measure the control network points and component fiducials using laser trackers. To ensure there is enough overlap between adjacent stations, three upstream and three downstream sections of the tunnel control network, along with the components within this range, will be measured at each station. Each station should establish a level coordinate system to incorporate vertical information into the adjustment calculation. For the level measurement, only the floor control points need to be measured, which will be used as constraints in the adjustment calculation.

### 7.2.5.3 *Smooth Alignment*

Although the initial installation aligns components within the specified tolerance relative to the tunnel control network, factors like ground motion and temperature gradients can cause deformation over time. Therefore, after the initial installation and alignment, an overall survey is conducted to assess deformations, followed by realignment to reduce them. Due to the substantial workload of aligning all components precisely, a smooth alignment strategy is favored. This strategy prioritizes relative position alignment between adjacent components, allowing some relaxation in absolute position accuracy. Ground motion and temperature gradients predominantly affect absolute positions globally but minimally impact relative positions locally, making the smooth strategy effective for achieving a smooth beam orbit.

Once the overall survey is completed, the actual position of a component can be compared with its nominal position to examine any offsets. If there are no major errors, an orbit smoothing calculation can be performed. This involves fitting the nominal coordinates of the components to their actual coordinates segment by segment, and calculating the relative position errors between them. The relative position errors of adjacent components will be examined, and if an error larger than the tolerance of 0.08 mm is detected, it means that the beam orbit is not smooth, and the component needs to be adjusted to bring the error within tolerance. This calculation is an iterative process, and after the smoothing calculation is completed, the components that need to be adjusted and the directions and values of the adjustments can be determined. The components can then be adjusted according to the calculation results.

Following the smooth adjustment, a tunnel-wide survey is conducted to verify whether all components meet the specified tolerance criteria. The total relative position error between adjacent components, encompassing fiducialization error (0.014-0.05 mm), laser tracker measurement error (0.05 mm), and smooth alignment error (0.08 mm), totals approximately 0.1 mm. Our plan involves organizing 64 groups to execute this smooth alignment, which will be iterated 2-3 times.

#### 7.2.5.4 *Interaction Region Alignment*

To achieve precise particle collisions, the alignment accuracy requirement within the Interaction Region (IR) is considerably more stringent compared to the arc regions. Additionally, it necessitates that the straightness of the beam orbits in the cryostats on both sides of the Collider be better than 0.1 mm. Meeting these alignment requirements calls for the implementation of specialized alignment strategies. As illustrated in Fig. 7.2.1.5, the cryostats are installed on a movement mechanism that can move along the guide rail. The cryostats will be inserted into the detector, which demands accurate adjustment of the cryostats relative to the detector, and the two cryostats should be aligned coaxially with the required accuracy.

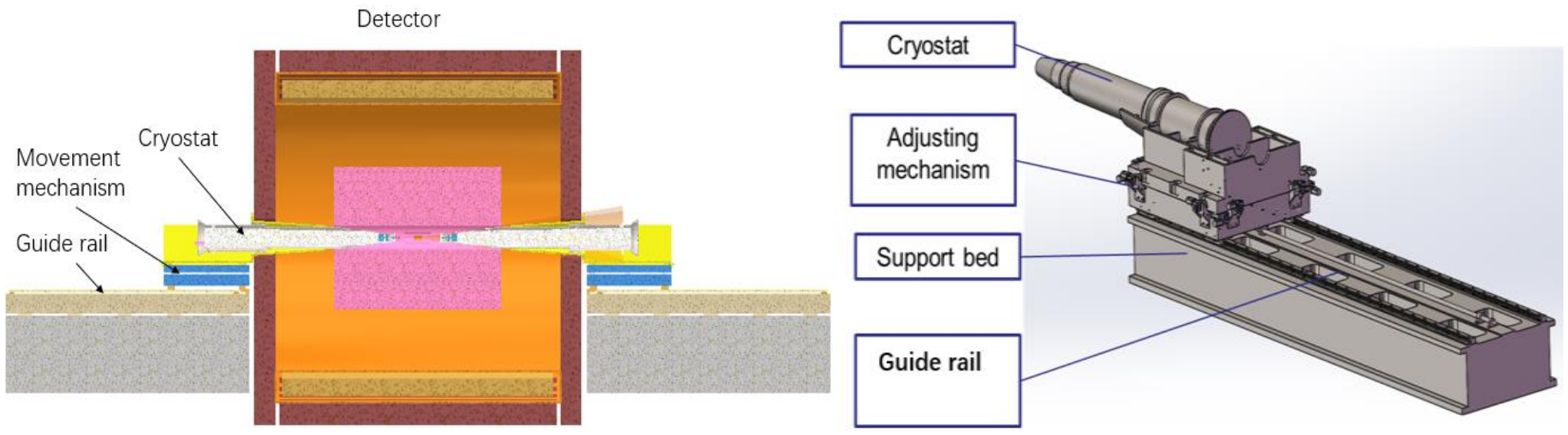

**Figure 7.2.15:** Schematic of the Interaction Region (IR).

Given that the machine detector interface (MDI) scheme is still in the design phase, we can only outline a preliminary alignment strategy. Aligning the MDI effectively requires a multi-step approach to attain the required accuracy. The first step involves cryostat pre-alignment employing a vibrating wire system, which delivers alignment accuracy of around 50 μm in both transverse and elevation dimensions for the magnets within the cryostat. A similar vibrating wire system has been developed and successfully utilized for HEPS.

The second step entails the placement of installation control points on the walls within the MDI area. These control points serve to connect the control networks on either side of the detector. Adjacent to the detector, a laser-based alignment system is installed to establish a straight-line reference for aligning the cryostats on both sides. This laser-based alignment system spans approximately 30 meters, and its accuracy should exceed 0.02 mm. The tunnel control network serves as a reference, and a laser tracker is employed as the measurement instrument to align the laser beam, ensuring its parallelism with the designed beam.

In the third step, the laser-based alignment system is used as a reference to adjust the guide rail and the cryostats to the designed positions in the vertical and transversal directions.

Finally, the cryostats are inserted into the detector along a guide rail. Given that the cryostat is designed as a cantilever structure, it may experience position changes when subjected to electromagnetic forces. Therefore, the development of a monitoring system to address this issue is currently under consideration.

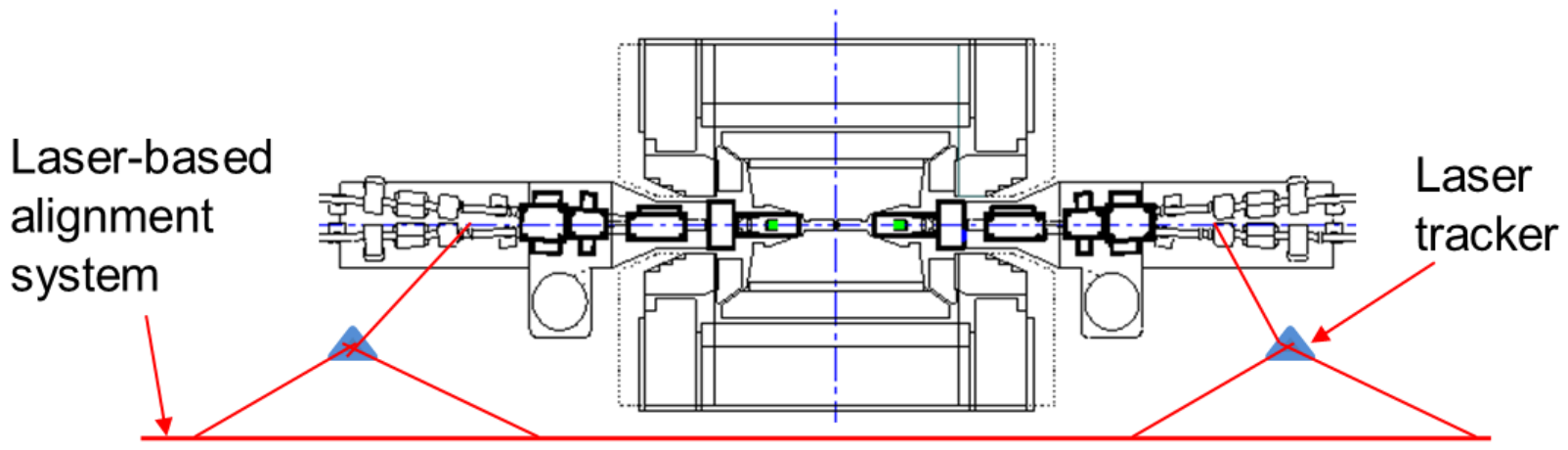

**Figure 7.2.16:** Laser-based alignment system in the IR.

### 7.2.6 Monitoring of Tunnel Deformation and Alignment Strategy

Following the initial installation and alignment, it is imperative to conduct deformation monitoring and deformation alignment to ensure the proper operation of CEPC. For the CEPC, as a newly-built and large-sized machine, deformation is inevitable. The first step in addressing deformation is to determine where it has occurred. Three methods will be applied for identifying the location of deformation.

The first method is to conduct an overall measurement of the control network and components, and compare the results with previous measurements to determine where the deformation has occurred. This method necessitates a shutdown period and a substantial amount of time to complete. During the annual shutdown, an all-encompassing measurement will be carried out.

The second method involves installing monitoring equipment, such as a Hydrostatic Levelling System (HLS) or a Wire Positioning System (WPS), to track settlement and changes in component position, respectively. The third method involves using Beam Position Monitors (BPMs) to measure the beam. Both of these methods can identify deformation immediately, whether during operation or shutdown periods.

A capacitive hydrostatic level sensor, known for its accuracy of better than ±0.01 mm, has been developed and effectively employed for HEPS. To align with CEPC's 0.1 mm requirement, the accuracy of the hydrostatic level sensor can be adjusted to reduce costs. However, due to cost limitations, it is impractical to evenly distribute monitoring sensors across the entire CEPC ring. The plan is to install these sensors in the IRs, which have more stringent alignment requirements. Since the MDI design is not yet finalized, the monitoring strategy is currently under consideration.

Regarding when and how to address deformation, CEPC's component stand does not have an auto-adjust mechanism due to cost considerations. Therefore, any deformation found during operation cannot be adjusted, and adjustments can only be made during shutdown periods. During a shutdown, a measurement will be carried out in the area where deformation has occurred, and the results will be compared with the nominal values to determine the differences. The offset values will then be sent to the physicists to determine which components need adjustment, and what direction and value the adjustment should be made. The alignment will then be adjusted based on the feedback from the physicists.

### 7.2.7 Visual Instrument R&D

Due to the large number of components requiring alignment in CEPC, the measurement task is incredibly time-consuming and arduous. Therefore, it is crucial to conduct research and development in order to create a new measuring instrument that can improve both the measurement precision and efficiency. The visual instrument is a promising new technology that integrates a photogrammetric system, angle measurement system, and distance measurement system to achieve high precision and efficiency in measurements.

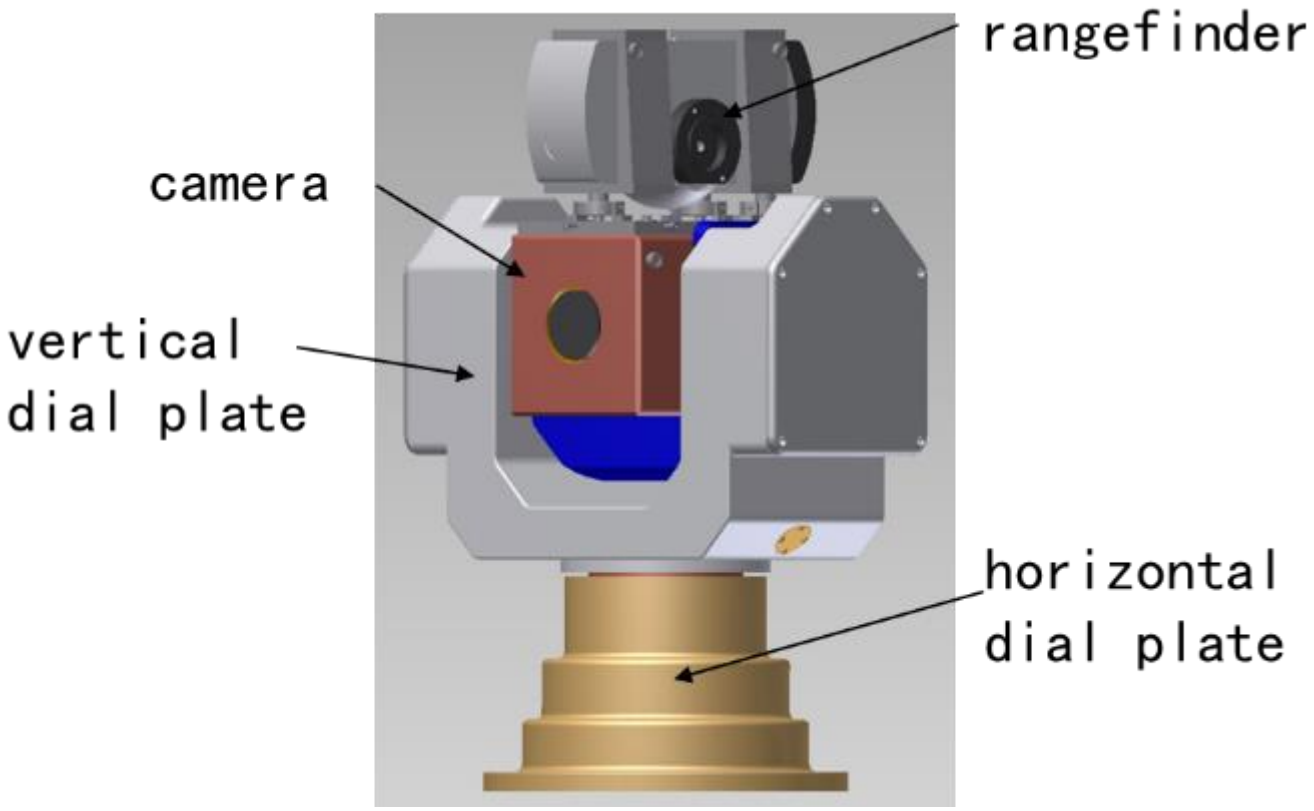

**Figure 7.2.17:** Model of a visual instrument.

#### 7.2.7.1 *Prototype Instrument R&D*

The photogrammetric function is accomplished using a measuring camera, which is made of aluminum alloy to provide adequate strength and stiffness while remaining lightweight. The camera itself is composed of three main parts: the camera unit, the control system, and the light source.

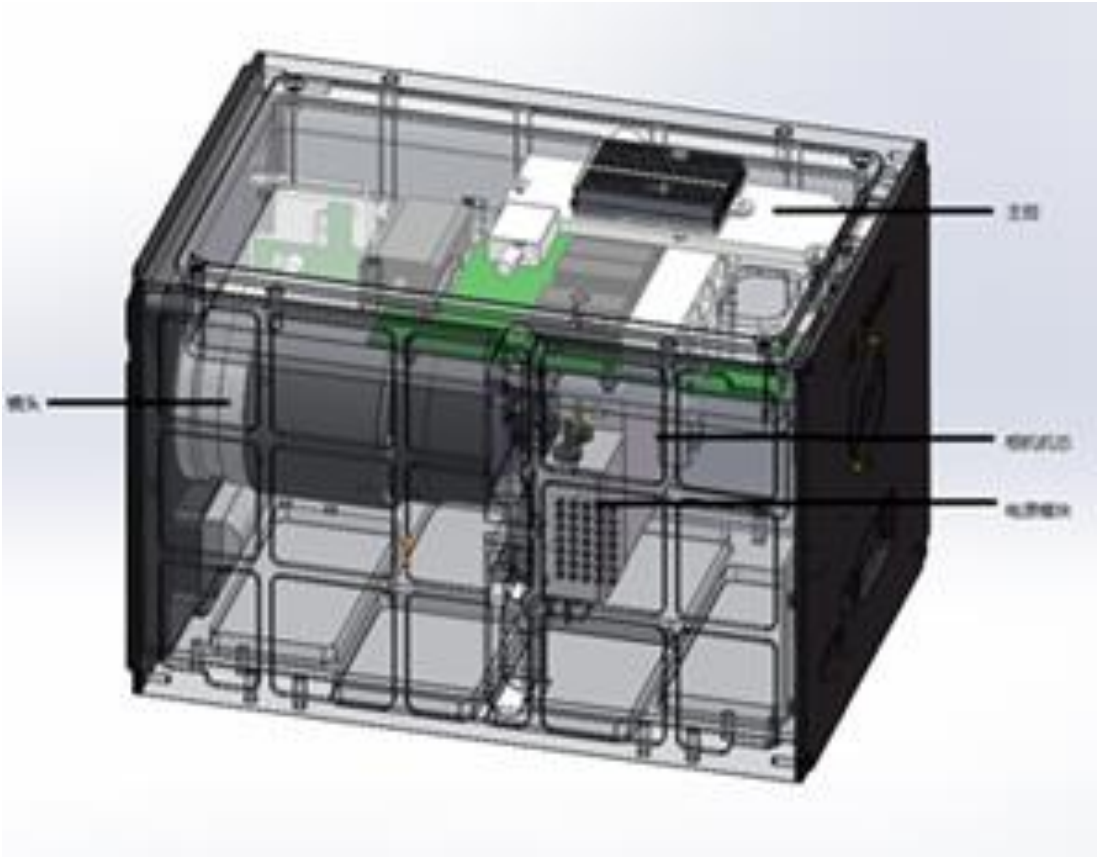

**Figure 7.2.18:** The camera.

The main component of the camera is composed of a lens system with a focal length of 25 mm and a field angle of 49.48°×49.48°. It also includes a CCD unit with a resolution of 5120×5120, a total of 26 million pixels, and a pixel dimension of 4.5μm×4.5μm.

The camera control system consists of a processor module, memory module, camera interface circuit, flash control circuit, shutter control circuit, and host computer interface circuit. During measurement, the camera control system receives control signals from the upper computer system and utilizes them to control the camera and light source for capturing images. The system can preprocess the images in real-time and transmit the processed images to the upper computer system for data calculation. Additionally, the camera controller includes a power circuit that provides power to the camera and light source.

The turntable is a crucial component of the visual instrument. It is where the camera and rangefinder are mounted, and it provides the necessary horizontal and vertical angle observations. The turntable is composed of several elements, including the shafting, piezoelectric ceramic motors, incremental grating, absolute photoelectric angle sensor, and motion controller.

The alt-azimuth shafting structure is employed in the turntable design. The measuring camera and rangefinder are mounted on the pitching axis, which allows for easy addition of counterweights and reduces driving torque. The azimuth axis can rotate 360° in an unlimited manner, while the pitching axis has a rotation range of –60° to +90°. To ensure high precision, large load capacity, and light weight, the mechanical structure of the turntable was designed using 3D modeling and finite element analysis. We used a combination of 3D printing and machining center fabrication processes to create the turntable body, which meets all of the necessary requirements.

The turntable utilizes piezoelectric ceramic motors as the driving equipment, which have an energy density 5-10 times that of ordinary electromagnetic motors. This results in the turntable having low speed and large torque, high resolution and positioning accuracy, sensitive dynamic response, compact structure, and no electromagnetic interference.

To calculate the driving torque, load analysis is necessary. The shafting uses ceramic rings as the rotating part, which has a self-lubrication function, making friction torque negligible. Since the mass on the shafting is not evenly distributed due to the rotating parts on the pitching axis, the largest torque needs to be calculated to select the appropriate motor.

The azimuth shafting is comprised of a rotating shaft, shell, photoelectric encoder, ceramic motor, ceramic ring, flange shaft, connecting shaft, bearing, and other components, as shown in Figure 7.2.19.

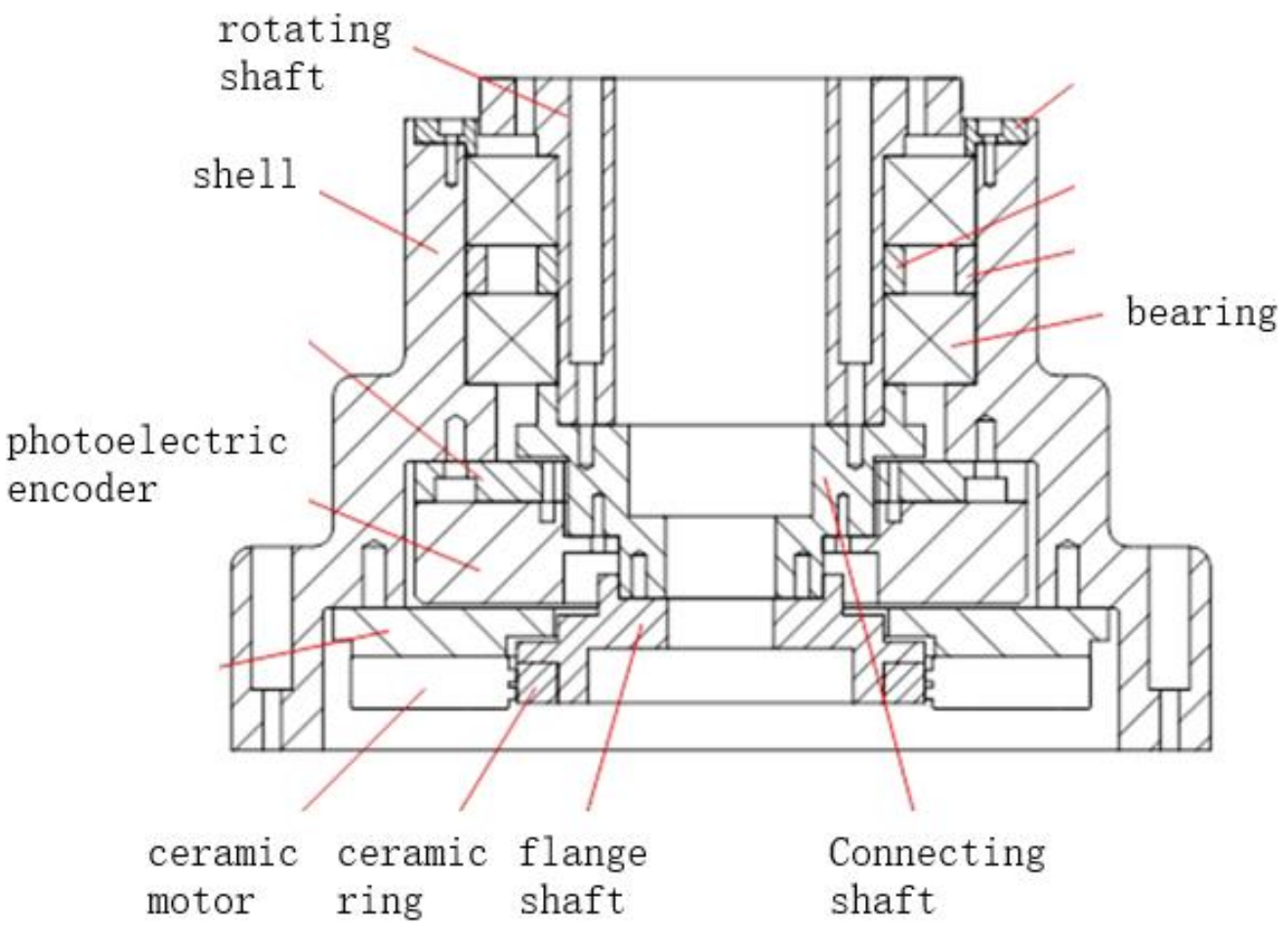

**Figure 7.2.19:** Azimuth shafting

The pitching shafting comprises of components such as ceramic motor, ceramic rings, flange shafts, U-frame, counterweight, left and right shafting, photoelectric encoder, and flange shaft, as shown in Figure 7.2.20.

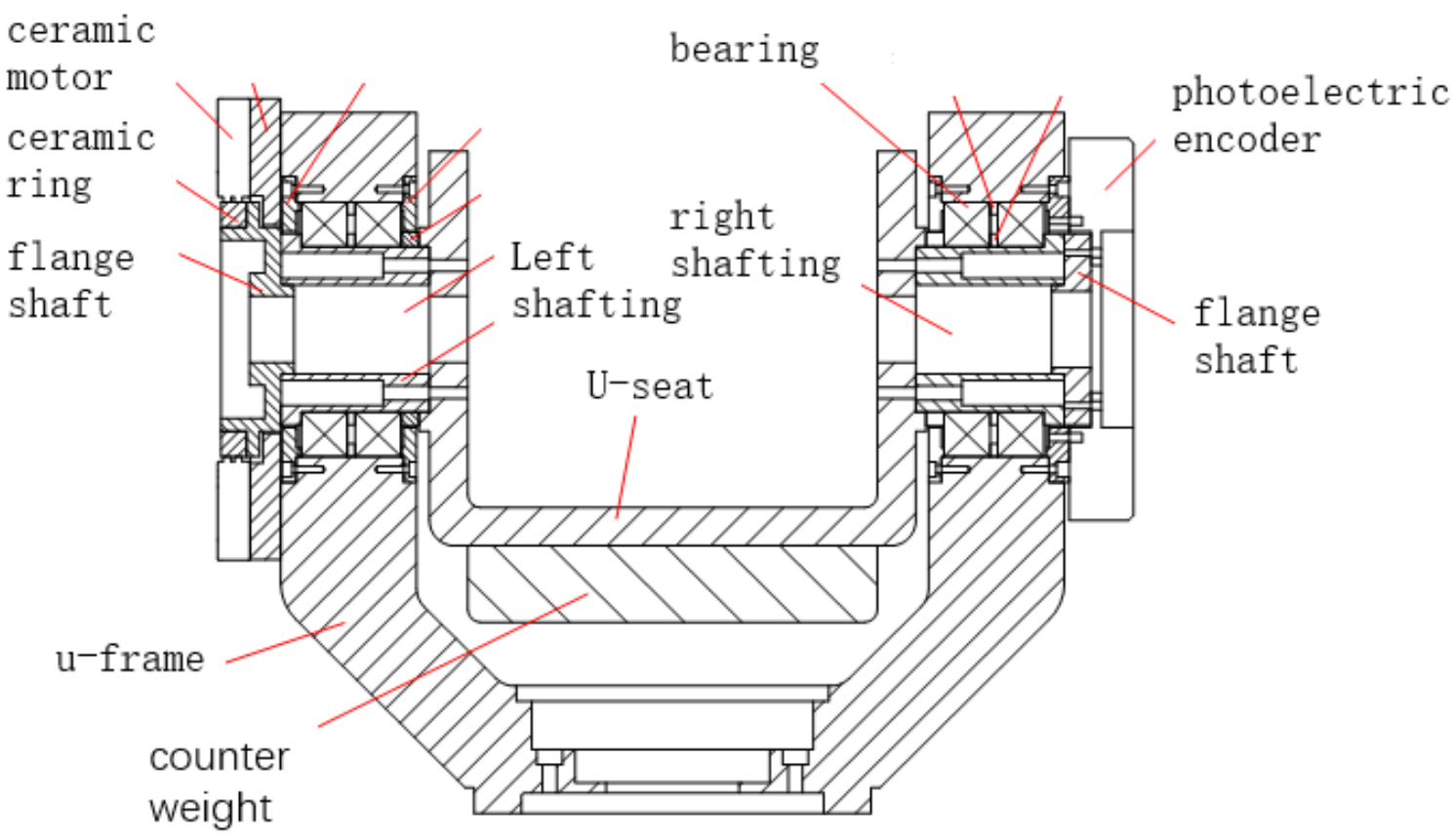

**Figure 7.2.20:** Pitch shafting

The turntable is equipped with a control system, which includes control software, a motion controller, and an actuator. The survey team can operate the turntable from a PC by inputting the rotation angles. The instructions are sent to the motion controller, which controls the motors accordingly. By receiving feedback from the incremental grating, the turntable can accurately control the rotation angles of the shafting. The absolute photoelectric angle sensor can provide real-time measurement of the shafting angles.

The range measurement system of the visual instrument employs the Leica μBase laser absolute rangefinder, which has a distance measurement accuracy of 10 microns. Fig. 7.2.21 shows a prototype of the visual instrument.

.

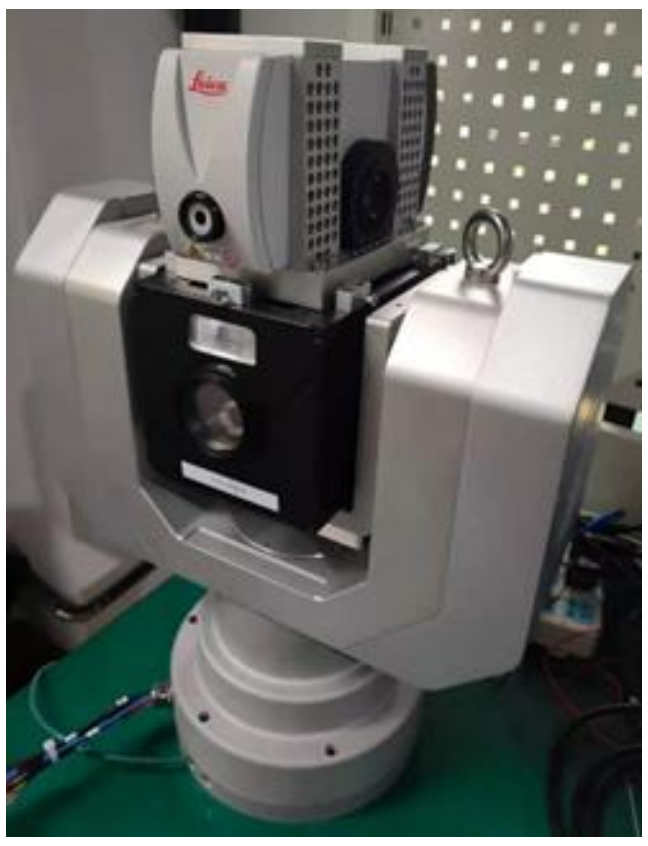

**Figure 7.2.21:** Prototype of a visual instrument.

A five-sided target has been developed for use with the visual instrument measurement, as shown in Fig. 7.2.22. The target's body is a sphere with five perpendicular surfaces cut out of it. Each surface has a circular reflective area coated with glass beads, which give it a retro-reflective function.

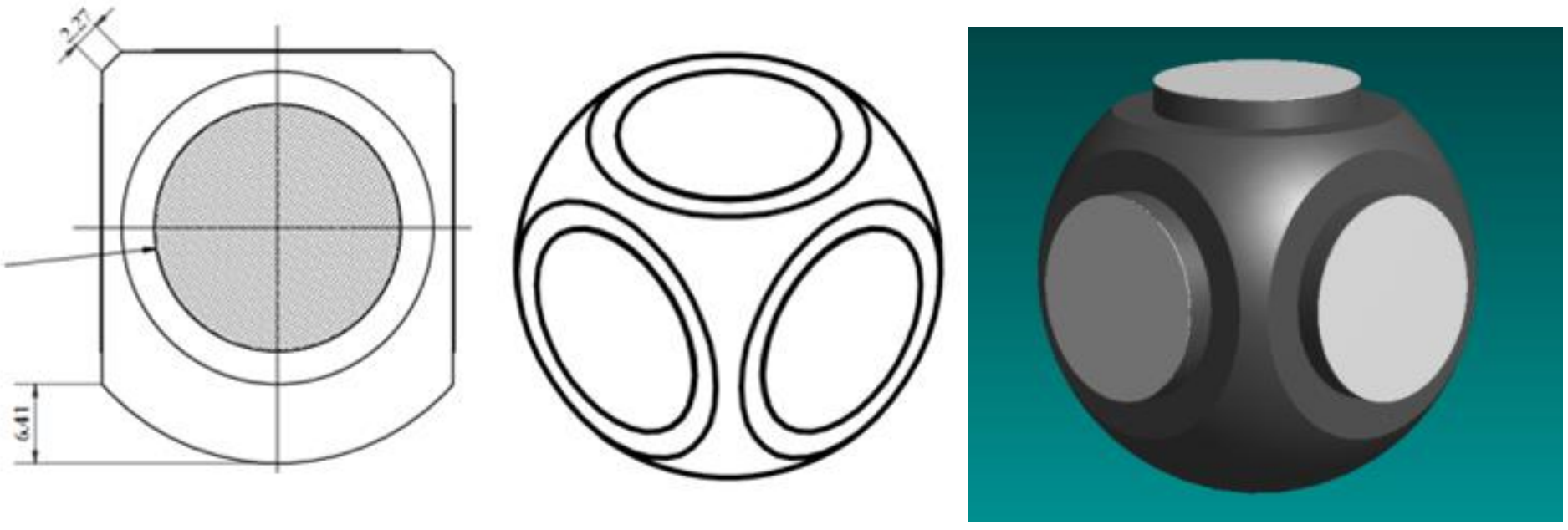

**Figure 7.2.22:** A five-face target

The distance from the center of each circle to the center of the target is a known value. When taking pictures of this target, we can capture images of its circular reflective surfaces. By using image identification, we can calculate the center coordinates of these circles. With these center coordinates and the known distance value, we can calculate the center coordinate of this five-face target. Traditional photogrammetric targets are typically flat and cannot meet the requirements of taking pictures from different angles. This five-face target is a three-dimensional target that can be observed from various angles and is suitable for measuring in narrow tunnel environments.

A coded target is essential for the identification of a large number of points using photogrammetry. Each coded target possesses a unique code and a set of templates designed to facilitate recognition and decoding. Figure 7.2.23 illustrates these templates.

The coded target comprises ten circular marks of identical size. Among these marks, five serve as template points, while the remaining five function as coding points. The template points establish the coordinate system of the coding target. The coding points are strategically positioned at twenty designated locations, with each distribution scheme assigned a specific number.

The decoding process involves reconstructing the positional information of the coding points to determine the code associated with the coded target. This system enables accurate point identification through the application of photogrammetry techniques.

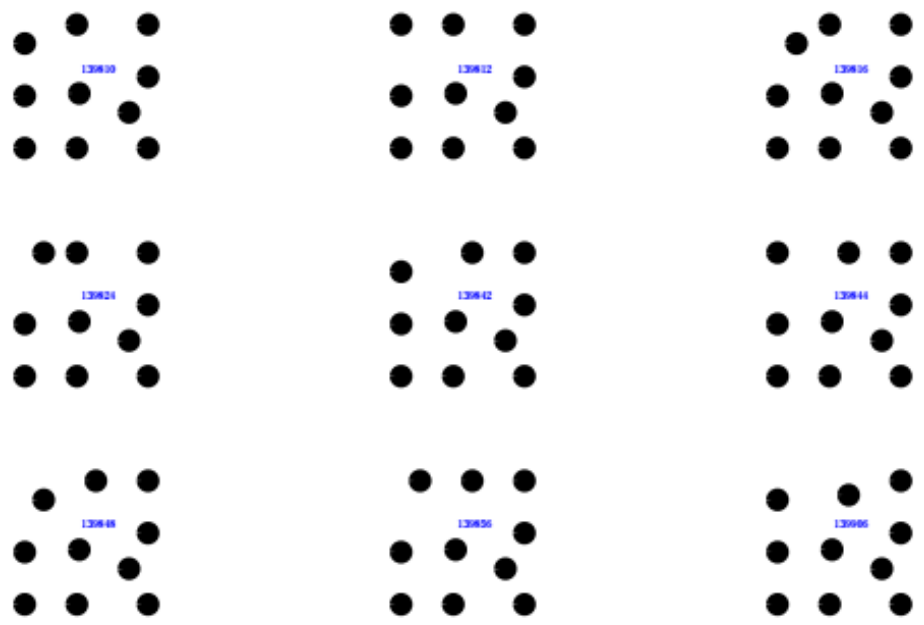

**Figure 7.2.23:** A million capacity coded target.

A million-capacity coded target has been successfully developed along with recognition software. It was tested in a photogrammetry experiment conducted in the CSNS tunnel, where it achieved a recognition rate of over 97%, which can meet the measurement requirement.

### 7.2.7.2 *Measurement Experiment*

Photogrammetry is a crucial function of the visual instrument. Although it has been applied in numerous practical measurements, its suitability for accelerator tunnel measurement and the methodology to carry it out still require further research.

Two photogrammetric experiments were conducted in the CSNS RCS tunnel, using a measuring camera manufactured by Prodetec. In the first experiment, a range of 70 m was covered using hemisphere targets and 8-point coded targets. Various measurement techniques were investigated, targeting the fiducials of 12 components and 32 control points of the tunnel control network. After image processing and adjustment calculations, the coordinates of these points were obtained. The measurement results were compared with those obtained using a laser tracker, and the average point error was found to be 0.4 mm.

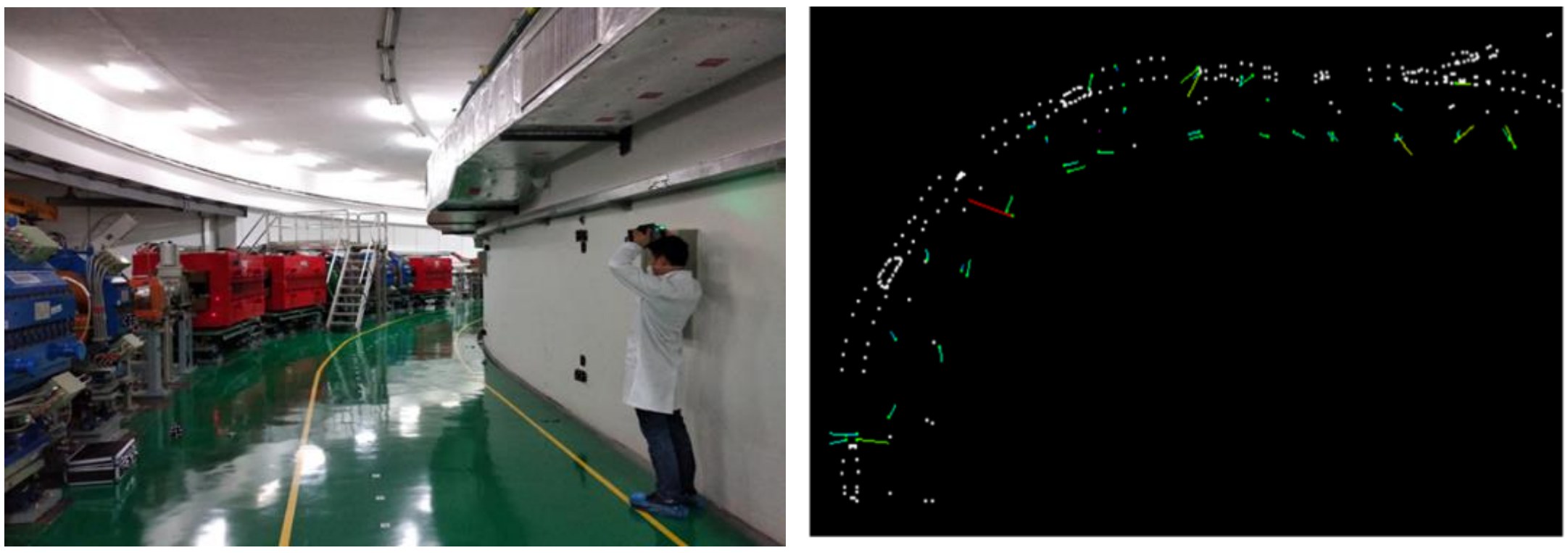

**Figure 7.2.24:** The first photogrammetric experiment in the CSNS RCS tunnel.

The second photogrammetric experiment covered the entire RCS tunnel, which is 240 m long. In this experiment, the five-face target and the million capacity coded target were used. However, the processing technique research for the five-face target had not been completed, resulting in an accuracy of only 0.5 mm. A total of 120 control points and fiducials of 30 components were measured. The error between the photogrammetry and laser tracker was 2.31 mm. The results indicated that photogrammetry is suitable for accelerator tunnel measurement, but in narrow tunnel environments and for long distance measurements, the photogrammetric results are not reliable. In order to improve the measurement accuracy, it is necessary to add angle and distance constraints.

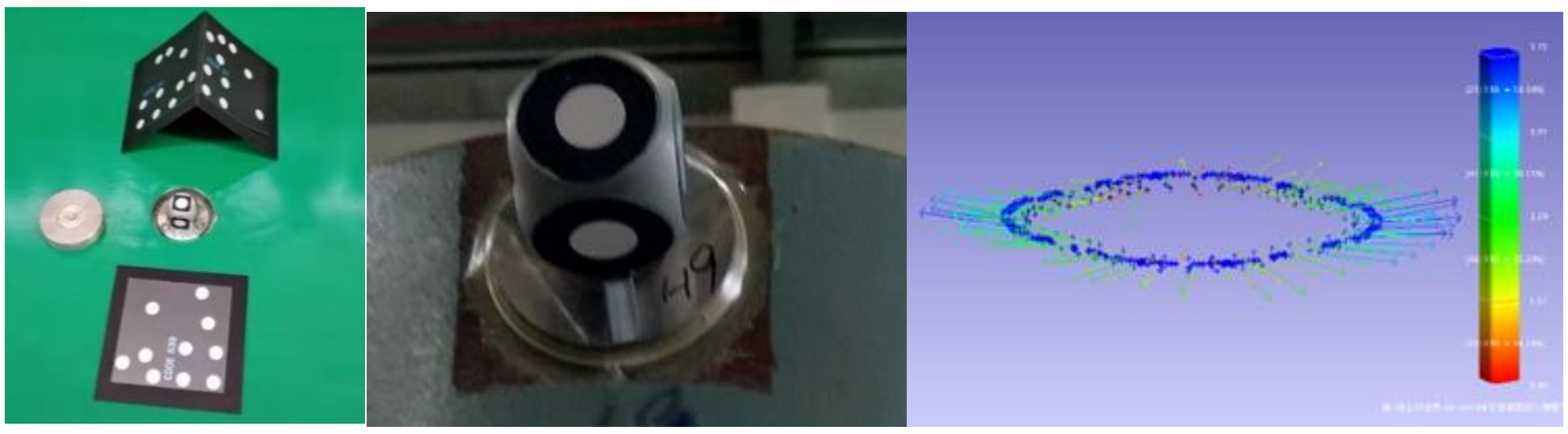

**Figure 7.2.25:** The second photogrammetric experiment in the CSNS RCS tunnel.

Currently, research on the calibration of the visual instrument is underway. Once the calibration process is complete, new measurement experiments will be conducted to verify and validate the accuracy of the measurements.

## 7.3 Radiation Protection and Interlock

This section outlines the expected radiological situation and the measures put in place to minimize the impact on workers and the general public. These measures include adequate shielding, a state-of-the-art radiation monitoring and alarm system, and a strict access-control system.

### 7.3.1 Introduction

#### 7.3.1.1 *Workplace Classification*

Radiation areas are classified into two categories [1]:

a) Radiation-monitored area: Registered radiation workers have unrestricted access to this area at any time. This includes facilities, halls, and areas and surfaces outside concrete shielding that are regularly monitored for radiation levels; The RF auxiliary tunnels and service caverns are also radiation-monitored areas.

b) Radiation-controlled area: Access to this area is restricted, and specific permission and access procedures are required. An example of a radiation controlled area is the auxiliary tunnel.

Clear guidelines for occupancy factors will be established for each structure and area to ensure the safety of workers and the general public.

#### 7.3.1.2 *Design Criteria*

The following standards and rules will be adhered to, as listed below. Table 7.3.1 provides the dose limits from the national standards [2].

1. GB 18871-2002, "Basic Standards for Protection against Ionizing Radiation and for the Safety of Radiation Sources," is the national standard of the People's Republic of China (Table 7.3.1).
2. GB 5172-1985, "The Rule for Radiation Protection of Particle Accelerators," is the national standard of the People's Republic of China.
3. ICRP Publication 103, "The 2007 Recommendations of the International Commission on Radiological Protection," will also be followed.

**Table 7.3.1:** Dose limits in the national standards GB18871-2002.

| | | **Worker** | **Public** |
|---|---|---|---|
| Effective Dose | Average in 5 years | 20 mSv/year | 1 mSv/year |
| | Maximum in a single year of a 5-year period | 50 mSv/year | 5 mSv/year |
| Equivalent Dose | Lens of the eye | 150 mSv/year | 15 mSv/year |
| | Skin | 500 mSv/year | 50 mSv/year |

As per the aforementioned standards, the maximum permissible annual occupational exposure limit is 50 mSv. However, in line with the ALARA ("as low as reasonably achievable") philosophy, efforts will be made to maintain exposures well below this limit. For occupational exposure, the annual dose limit should be kept below 5 mSv, while for the public, the annual dose limit should be below 0.1 mSv. For comparison, CERN's designed area has a dose limit of 6 mSv/yr and its off-site area has a dose limit of 0.3 mSv/yr, with optimization goals of 100 μSv/yr and 10 μSv/yr, respectively. The dose rate limits for shielding design, deduced accordingly, are listed in Table 7.3.2.

**Table 7.3.2:** Prompt dose rate limits for different areas.

| Area | Design Value | Example |
|---|---|---|
| Radiation monitored area | < 2.5 μSv/h | Outside the tunnels, where a worker can stay longer |
| Radiation controlled area | < 25 μSv/h | Outside the tunnels, where a worker can stay occasionally, also including some areas where work is forbidden, such as where the dose rate higher than 1 mSv/h |
| Site boundary | 0.1 mSv/year | |

The dose limits for soil, ground water, and air activation are determined based on GB18871-2002, with detailed descriptions of the radionuclide and its exempted activity or specific activity provided in Appendix A of the standard. If there is more than one type of radionuclide, it is exemptible only if the ratio of activity (or specific activity) to the exempt value of each radionuclide is less than 1, as expressed in Equation (7.3.1):

$$\sum_{i=1}^{n} \frac{S_{i_saturation}}{S_{i_exempt}} < 1 \tag{7.3.1}$$

This approach is convenient for evaluating soil and ground water activation. To ensure that the prompt dose rate is below the exempt value given above [3], the thickness of the soil should be at least 1 meter, with a prompt dose rate of approximately 5.5 mSv/h.

## 7.3.2 Radiation Sources and Shielding Design

### 7.3.2.1 *Interaction of High-Energy Electrons with Matter*

When a high-energy electron beam interacts with matter, it undergoes various processes such as beam-gas, beam-collimator, beam-target, or beam-dump interactions. These interactions result in the production of ionizing radiation fields (prompt, mixed radiation fields) and radioactive nuclei within the target material (induced activity). Electromagnetic cascades and nuclear reactions are the dominant processes that contribute to this production.

### 7.3.2.2 *Radiation Sources*

In a high-energy accelerator, the prompt and residual radiation fields are the two primary sources of radiation, as shown in Fig. 7.3.1.

The prompt radiation fields, also known as mixed radiation fields, mainly consist of neutrons and photons. There may also be some charged hadrons (protons, pions, kaons) and leptons (such as muons) present in the fields. The composition of the fields at a particular point inside or outside the tunnel is heavily influenced by the position relative to beam loss and the type of shielding in place.

In the context of CEPC, there are two primary sources of radiation: random beam losses and synchrotron radiation. Random beam losses can be further classified into linear beam losses and point beam losses.

Linear beam losses occur along the beam lines, encompassing the Linac, transporting lines, Booster, Collider, and extraction lines. On the other hand, point beam losses are concentrated in specific hotspots, including collimators, injection/extraction points, beam dumps, and interaction regions.

In the Linac, measures have been taken to shield the linear and point beam losses by designing beam dumps. The dimensions and thickness of the dumps and walls have been optimized as described in Section 7.3.2.4. Additionally, studies on synchrotron radiation in the Collider and Booster are outlined in Section 4.2.4 and Section 5.2.4. It is worth noting that the prompt doses from linear beam losses in the Collider and Booster are two to three orders of magnitude lower than synchrotron radiation and can be disregarded.

Regarding the point beam losses, the design of Collider dumps is addressed in Section 4.2.5. However, local shielding around collimators, injection/extraction points, and the damping ring are currently pending and will be addressed in the future.

Furthermore, there are several other structures within CEPC, such as shafts, auxiliary tunnels, mazes, shielding doors, and four experiment/RF halls. The shielding design for these structures is not covered in this report but is planned for completion in the future.

The shielding design for the machine-detector interface (MDI) is considered one of the most complex aspects, as it significantly affects the performance of both the accelerator and the detector. Section 4.2.6 in the report provides the main details and discussion related to this topic.

Extensive collaboration between the accelerator and detector teams is underway to study the impacts of MDI design on both the accelerator and detector performance comprehensively. This collaborative effort ensures that the shielding design meets the requirements and considerations of both aspects.

Radioactive isotopes are produced within the accelerator components and tunnel due to nuclear reactions triggered by high-energy primary or secondary particles. These isotopes decay primarily through beta and gamma emission until they reach the "Valley of Stability." Because the half-lives of radioactive isotopes range from fractions of seconds to years or more, radiation fields will persist within the machine once it becomes operational. In Section 7.3.3 of the report, the radioactive isotope productions in the air of the collider tunnel and the linac tunnel have been thoroughly investigated. Additionally, the production of toxic gases has also been studied. It has been determined that the levels of toxic-gas productions remain below the Chinese mandatory standard, indicating compliance with safety regulations.

Furthermore, the specific activities of long-life isotopes in the air, cooling water, and rock are all found to be lower than the values specified in GB18871, which sets standards for environmental radiation protection. While these results provide reassurance, it is important to consider the scale of the entire project. Due to the project's significant magnitude, the overall quantity of isotopes generated may still raise concerns that warrant attention and careful management.

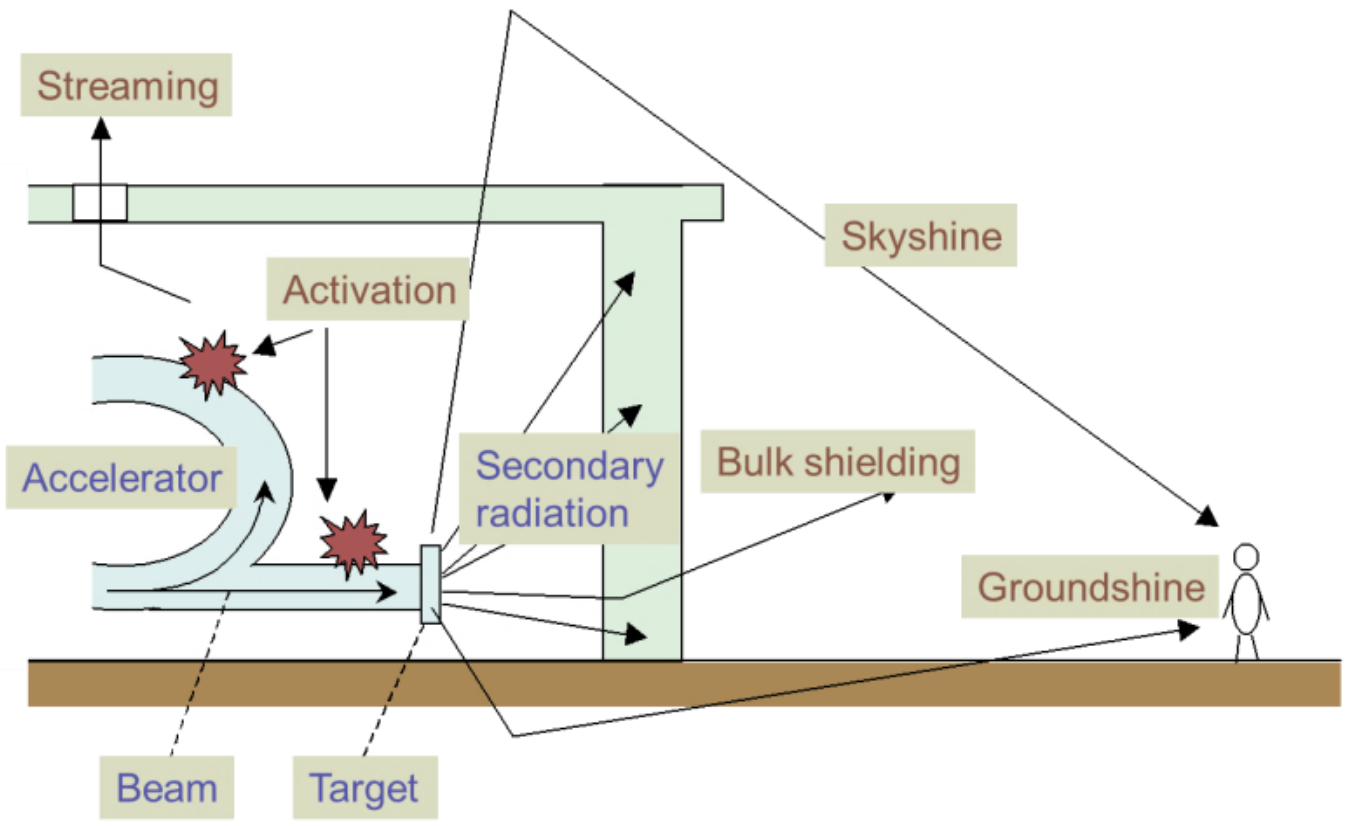

**Figure 7.3.1:** Illustration of the radiation sources around an accelerator.

### 7.3.2.3 *Shielding Calculation Methods*

Monte Carlo (MC) simulations using the FLUKA and MCNP codes are employed to design radiation shielding, and the results are verified using empirical formulas.

### 7.3.2.4 *Radiation-Shielding Design for the Project*

The radiation-shielding design philosophy follows the principle that the thickness of the shielding in the main tunnel should be based on the radiation level resulting from the average beam loss along the tunnel. However, areas such as collision points, injection points, collimation systems, and beam dumps require additional local shielding to reduce the radiation level to match that of the main tunnel.

Dose rate simulations were conducted using 120 GeV electrons with a loss rate of 5~9 mW/m and a 3–mm-thick Cu beam pipe as shielding. The results are shown in Fig. 7.3.2. Conversion coefficients can be used to determine radiation levels for other beam loss parameters.

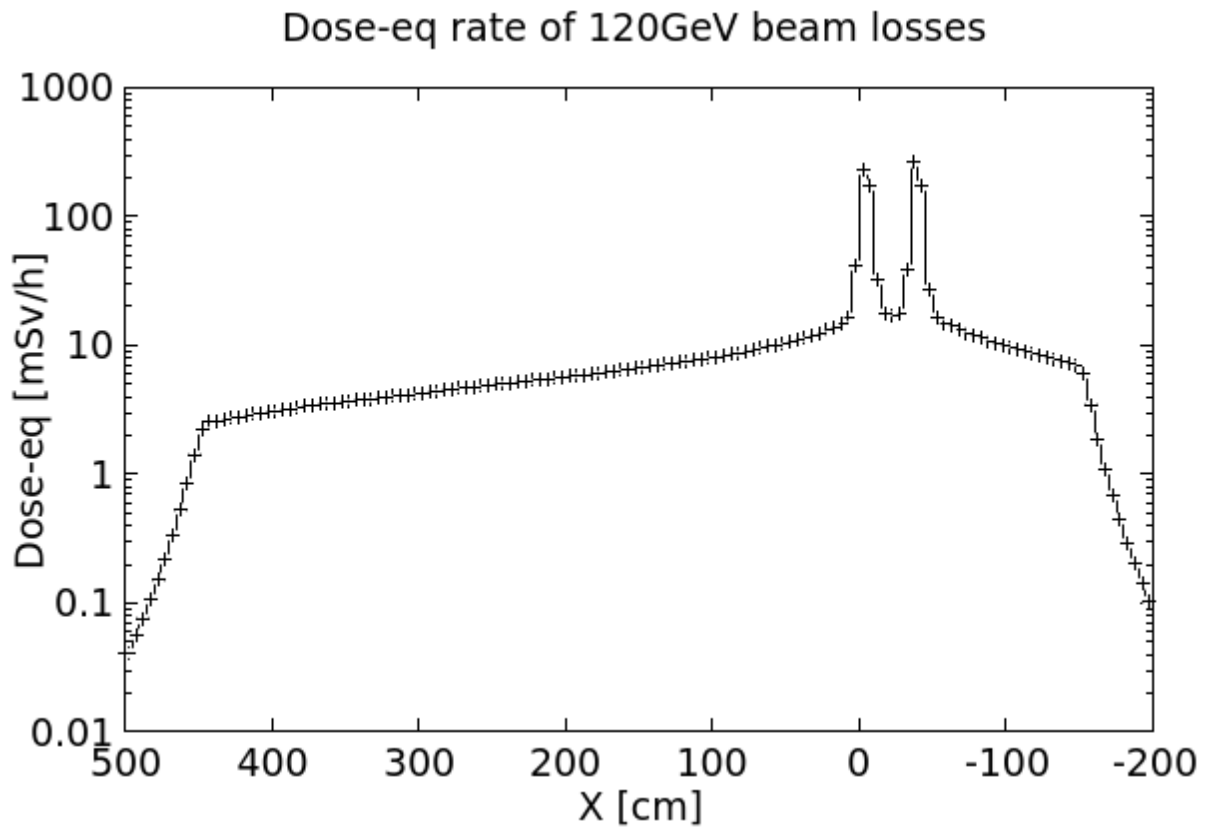

**Figure 7.3.2:** The distribution of prompt radiation dose rate (from –150 cm to 400 cm is the space in the tunnel).

The absorbers are the critical components of the dump, and their material selection is crucial. To ensure the best thermal conductivity and melting temperature, several materials such as aluminum, iron, copper, and graphite have been investigated. For the Linac beam absorber, iron that has been chemically, thermally, and mechanically processed and forged is selected. The high-energy absorber in the Collider will be made of graphite.

Monte Carlo simulations have been employed to find the optimal dimensions for the absorbers. Using FLUKA, the energy deposition and ambient dose equivalent distribution caused by the primary beam were calculated. The FLUKA simulations were conducted with two examples of beam parameters, as shown in Table 7.3.3. The design of the collider dump can be found in Chapter 4, Section 4.2.5.3. The Collider dump is designed to absorb both the Collider beam and the Booster beam. Currently, the design for the transporting line from the Booster to the Collider dump is in progress. Fig. 7.3.3 illustrates the positions of the Linac dumps within the linac tunnel. A total of seven dumps are strategically placed along the linac tunnel to safely terminate beams with varying energy levels. The preliminary dimensions of the absorber were set to 200 cm in radius and 1,000 cm in length. It's essential to emphasize that the maximum ambient dose-equivalent outside the dump should not exceed 5.5 mSv/h to ensure a consistently safe dose rate under all conditions.

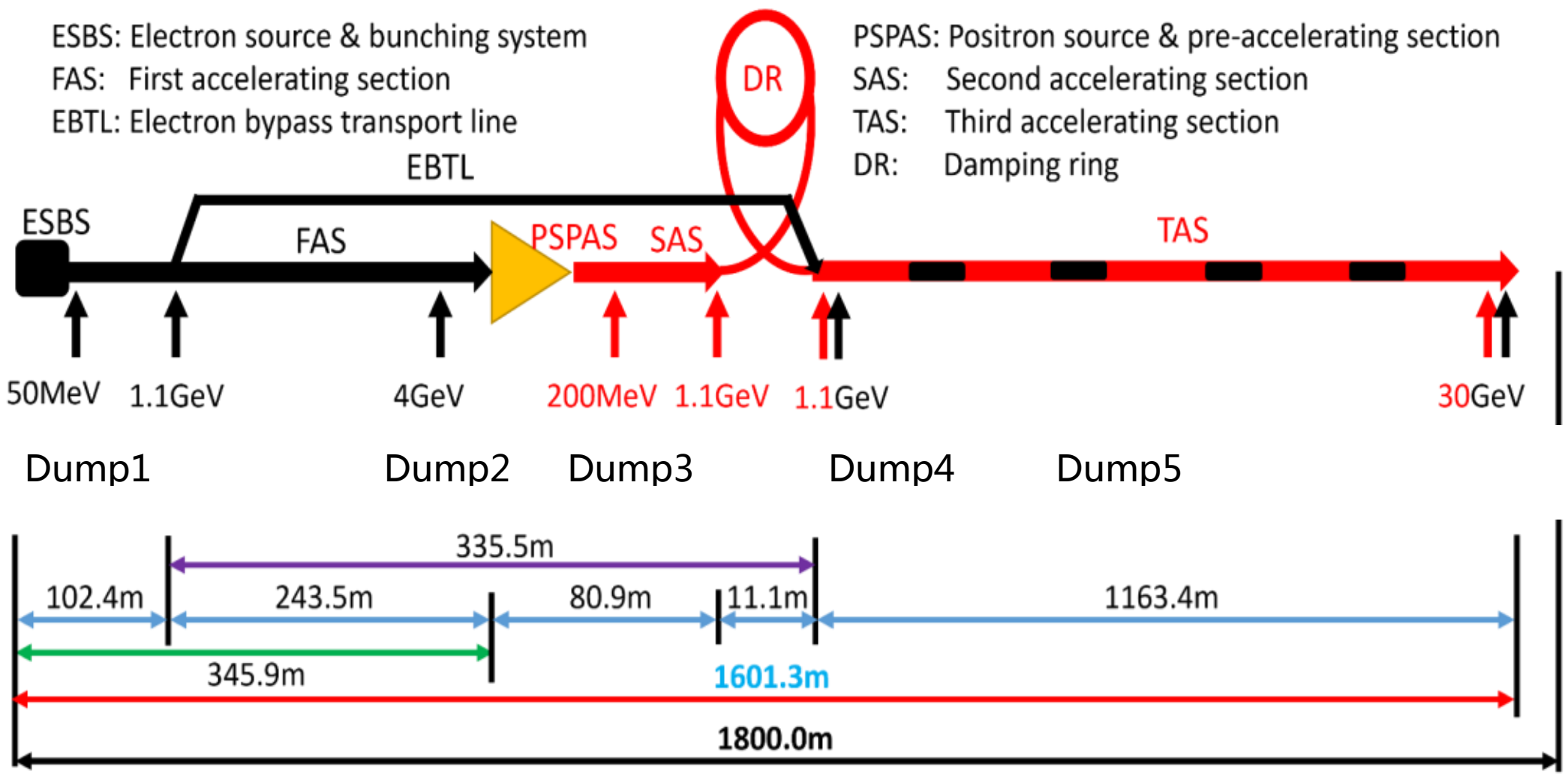

**Figure 7.3.3**: The sketch of Linac.

**Table 7.3.3:** Parameters used in the FLUKA simulations for the dumps.

| | **Linac** | **Collider** |
|---|---|---|
| Energy | 0.06~30 GeV | 45.5 GeV |
| No. of Electrons per second | $(3.7\text{~}12.5) \times 10^{10}$ | $2.8 \times 10^{15}$ |
| Absorbed energy | 4~432 KJ/h | 20 MJ/h |

The preliminary dimensions for the 30 GeV Linac dump core were selected to be 5 cm in radius and 260 cm in length, with carbon as the material. Due to the dominance of photons in the center of the dump core, iron is an appropriate choice as the shielding material. For neutron shielding outside the iron shell, polyethylene is used. Figures 7.3.4 and 7.3.5 depict the longitudinal and transverse ambient dose equivalent distribution curves with both the iron and polyethylene shielding. The iron shell has a radius of 120 cm and a length of 110 cm, while the polyethylene shell has a radius of 10 cm and a length of 10 cm.

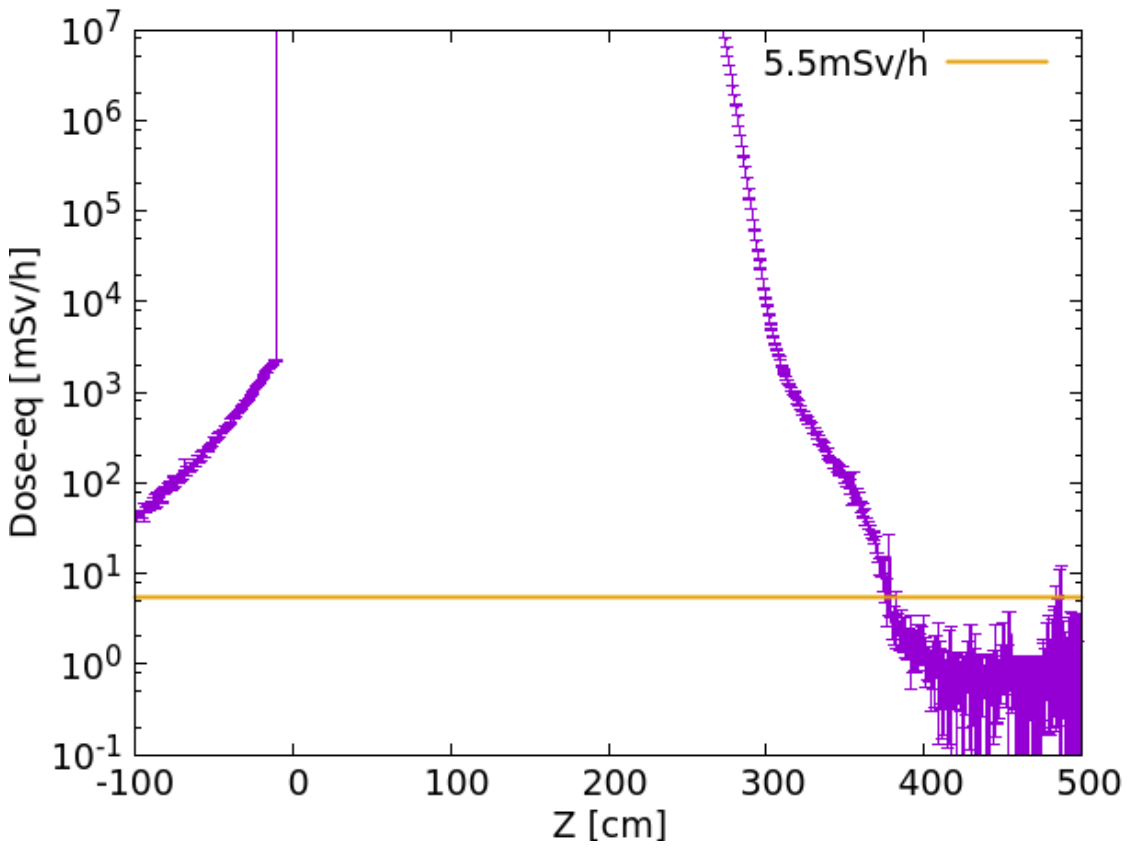

**Figure 7.3.4:** Longitudinal ambient dose equivalent distribution of the Linac absorber.

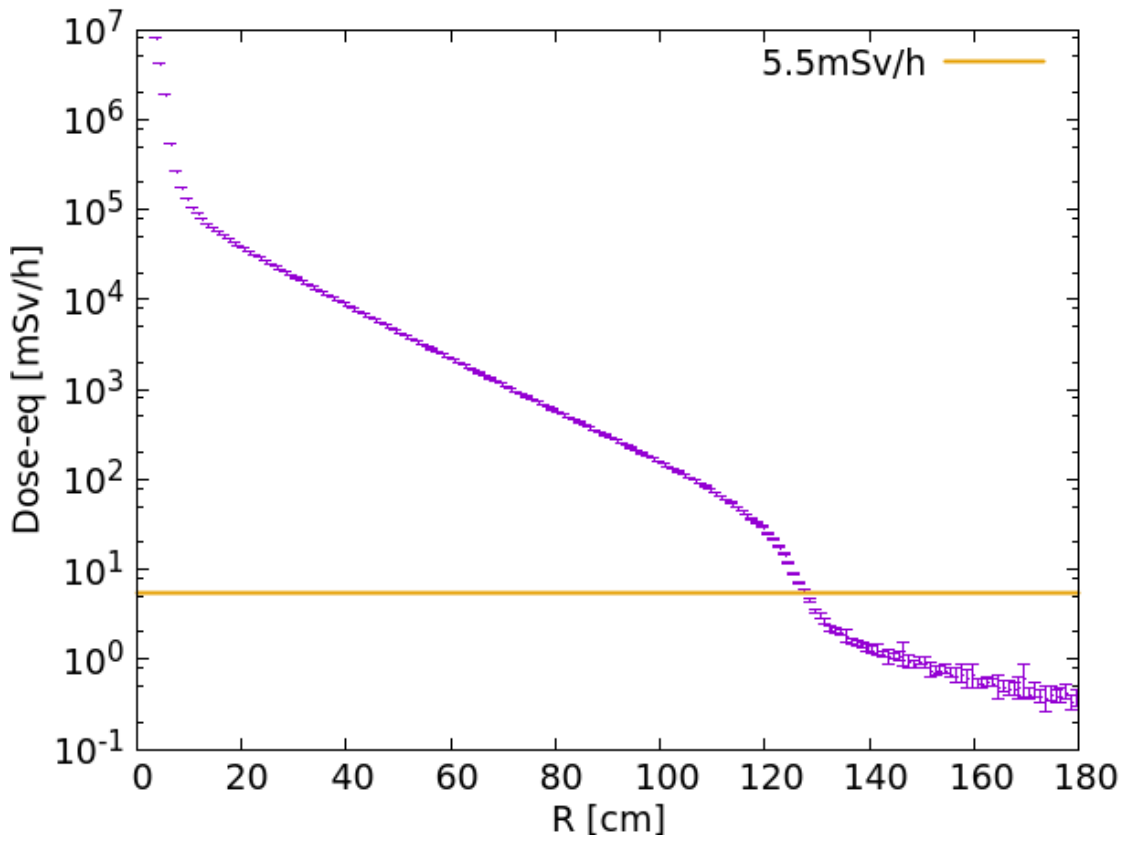

**Figure 7.3.5:** Transverse ambient dose equivalent distribution of Linac absorber.

All other Linac dumps have a three-layer structure: carbon+iron+polyethylene. The dimensions of the Linac dumps are summarized in Table 7.3.4.

**Table 7.3.4:** The radii and lengths of the Linac dumps.

<table>
<tr><th rowspan="2"></th><th rowspan="2">Beam energy</th><th colspan="2">C</th><th colspan="2">C+Fe</th><th colspan="2">C+Fe+Polyethylene</th></tr>
<tr><th>R/m</th><th>Length/m</th><th>R/m</th><th>Length/m</th><th>R/m</th><th>Length/m</th></tr>
<tr><td>Dump 1</td><td>60 MeV</td><td rowspan="6">0.05</td><td>0.15</td><td>0.7</td><td>1.0</td><td>---</td><td>---</td></tr>
<tr><td>Dump 2</td><td>4 GeV</td><td>1.25</td><td>1.1</td><td>2.4</td><td>1.2</td><td>2.6</td></tr>
<tr><td>Dump 3</td><td>250 MeV</td><td>0.4</td><td>0.5</td><td>0.9</td><td>0.55</td><td>1.0</td></tr>
<tr><td>Dump 4</td><td>1.1 GeV</td><td>0.75</td><td>0.75</td><td>1.6</td><td>0.85</td><td>1.7</td></tr>
<tr><td>Dump 5</td><td>6 GeV</td><td>1.45</td><td>0.9</td><td>2.4</td><td>1</td><td>2.5</td></tr>
<tr><td>Dump 6/7</td><td>30 GeV</td><td>2.6</td><td>1.2</td><td>3.7</td><td>1.3</td><td>3.8</td></tr>
</table>

The beam loss assumptions for the Linac are presented in Table 7.3.5. By utilizing this information, it is possible to simulate the dose equivalent. An example of such a simulation in the second accelerating section can be found in Figure 7.3.6. To ensure radiation levels remain below the upper limit of 5.5 mSv/h, the required wall thickness can be determined. The corresponding thickness values for the wall are listed in Table 7.3.6.

**Table 7.3.5**: Beam-loss assumptions for the Linac.

| Position | Length | Beam energy | Number of bunches $[s^{-1}]$ | Beam losses/bunch [nC] | Number of particles $[10^{10}/s]$ |
|---|---|---|---|---|---|
| FAS | 200m | 300MeV | 200 | 0.5 | 62.5 |
| Position target | 15mm | 4GeV | | 10 | 1250 |
| PSPAS | 15m | 5~200MeV | | 10 | 1250 |
| SAS | 3m | 300MeV | | 2 | 250 |
| | 30m | 600MeV | | 0.2 | 25 |
| TAS | 1163 | 1.1~30GeV | | 0.1 | 12.5 |

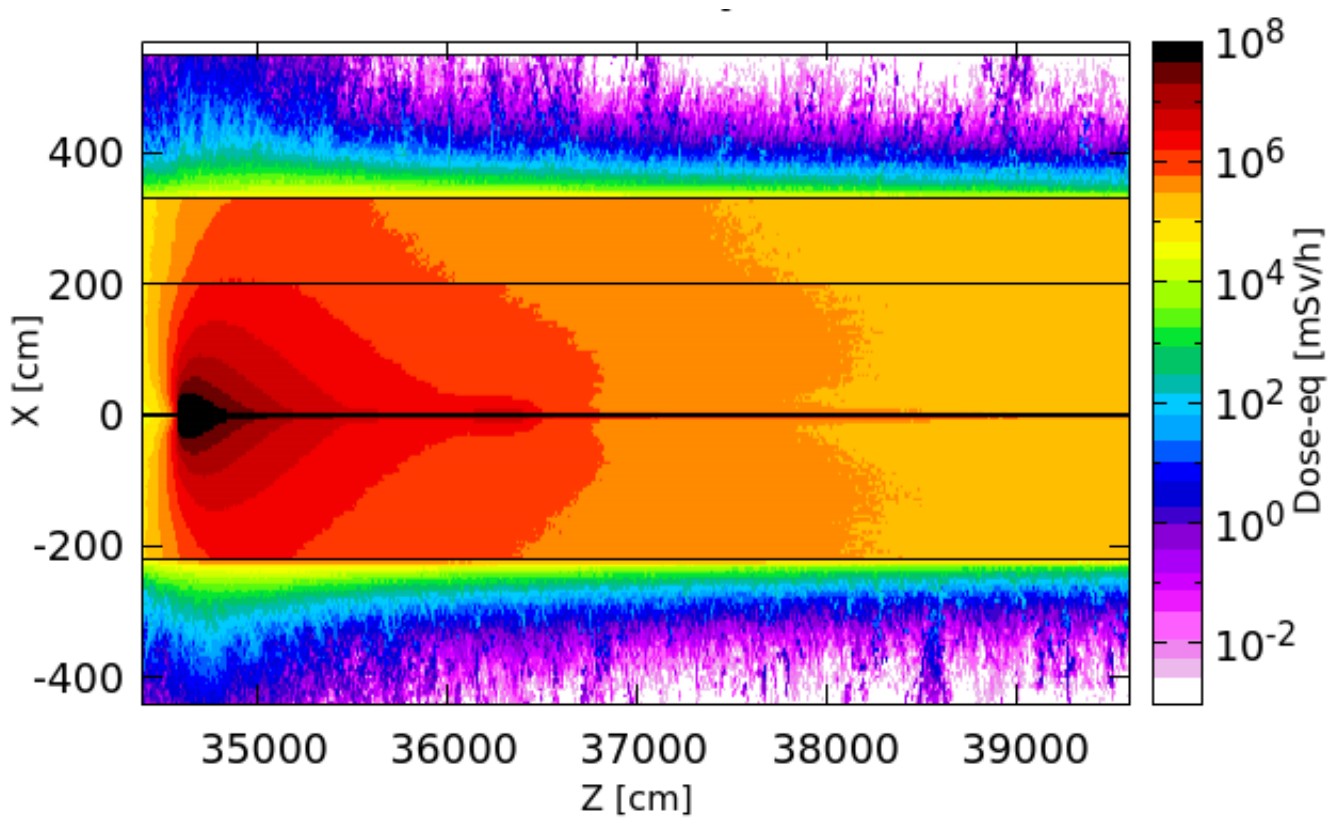

**Figure 7.3.6:** Dose equivalent distributions in the second accelerating section.

**Table 7.3.6:** The thickness of the linac wall

| Wall thickness | FAS | SAS | TAS |
|---|---|---|---|
| Left | 0.3 m | 1.9 m | 0.3 m |
| Right | 0.2 m | 1.9 m | 0.3 m |
| Bottom | 0.3 m | 2.1 m | 0.3 m |
| Top | 1.3 m | 4.1 m | 2.0 m |

Numerous electronics are located in the auxiliary tunnel, some in close proximity to the beam pipes. These electronics may be impacted by the electromagnetic shower and neutron radiation within the tunnel. The 1 MeV neutron equivalent fluence during Higgs operation is illustrated in Figure 7.3.7.

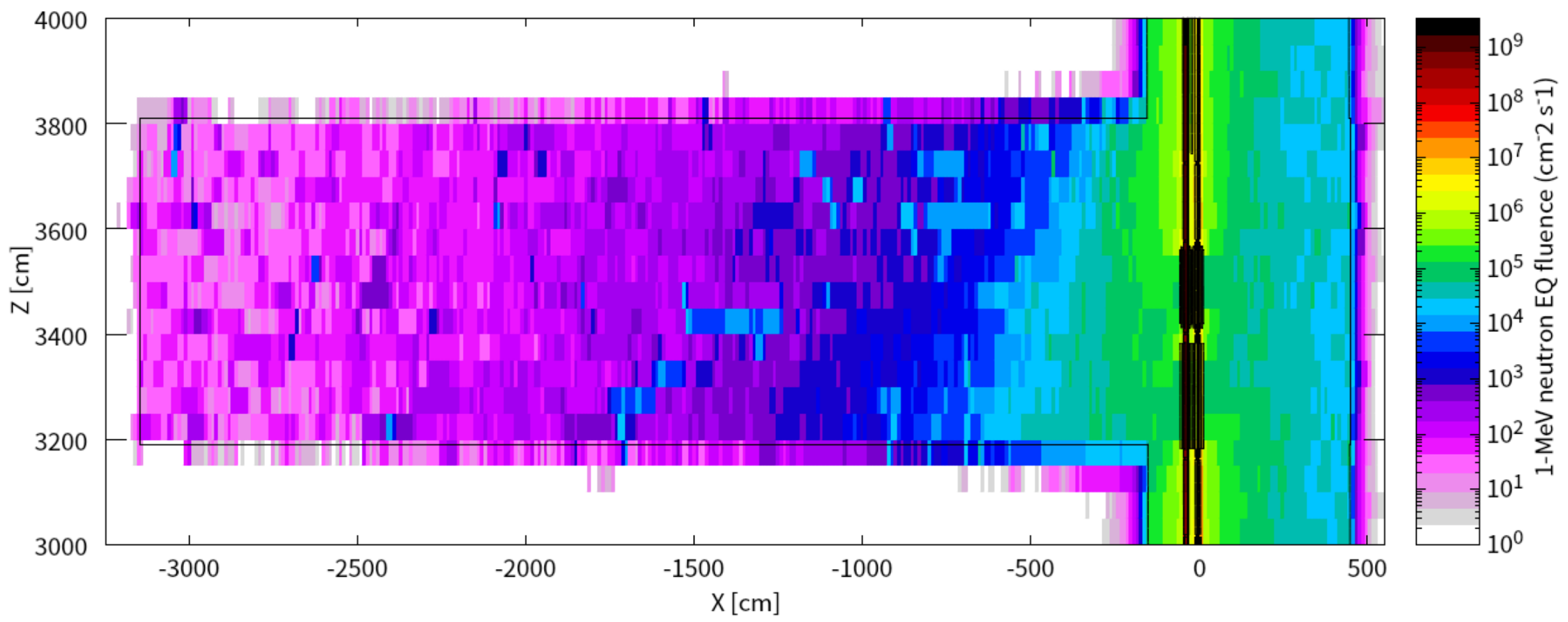

**Figure 7.3.7:** The 1MeV neutron equivalent fluence in the tunnel and the nearby auxiliary tunnel.

The electronics installed in the tunnel (including auxiliary tunnels) will be shielded as follows:

1. Cabinets and electronics for magnet power sources and vacuum systems will be placed in the auxiliary tunnel. Labyrinth and shielding doors will be installed at the entrances of the auxiliary tunnels to enhance protection.
2. Kicker power sources and power sources for trim coils will be situated near the beam pipe, approximately one meter away. These power sources will be shielded using lead and organic materials like polyethylene or paraffin wax surrounding the electronic cabinets.
3. BPM (Beam Position Monitor) electronics, known for their radiation resistance, will be located in alcoves on the tunnel floor. These alcoves will be covered with metal and organic materials to provide shielding.

Given the potential coexistence of SPPC and CEPC within the same tunnel, it is essential to account for their mutual influence. The simultaneous operation of SPPC and CEPC necessitates consideration of cumulative radiation effects within the tunnel. CEPC's radiation may increase the thermal load and the risk of quenching for SPPC's superconducting magnets, while SPPC's radiation may also pose a risk to CEPC equipment. As SPPC is still in the preliminary design stage, these issues will be thoughtfully addressed during the detailed refinement of the SPPC design process.

## 7.3.3 **Induced Radioactivity**

### 7.3.3.1 ***Specific Activity and Calculation Methods***

Lost beam can interact with surrounding components and induce the production of radionuclides. In addition, the secondary γ-rays are absorbed by the air, leading to the generation of harmful gases such as $O_3$ and $NO_X$. The main isotopes that are activated in this process include:

- Concrete shielding: $^{24}$Na
- Air activation: $^{11}$C, $^{13}$N, $^{15}$O

- Cooling water activation: $^{11}C$, $^{13}N$, $^{15}O$, $^{7}Be$, $^{3}H$
- Soil activation: $^{7}Be$,$^{3}H$

FLUKA and traditional methods of folding particle fluence with energy-dependent cross sections were used to calculate the specific activity induced by electron interactions in the beam line, shielding components, and the environment.

The tunnel geometry used is the same as in Sec. 4.2.4. To investigate the effect of water content in rocks, radionuclide production was simulated twice: one with a rock wall and the materials listed as medians in Table 7.3.7, and the other with a water wall. The major element compositions of granitic samples near the CEPC candidate sites were also collected and listed in Table 7.3.7.

**Table 7.3.7:** Major element composition of different rocks (wt %).

| | | | | | |
|---|---|---|---|---|---|
| C | --- | N | --- | O | 30~70 |
| Na | 0.1~2.9 | Mg | 0.4~3.7 | Al | 3.5~9.7 |
| Si | 26~39 | P | 0.02~0.16 | K | 1.8~3.7 |
| Ca | 0.2~4.5 | Ti | 0.09~0.8 | Mn | 0.02~0.12 |
| Fe | 0.8~6.3 | | | | |

The random beam losses and synchrotron radiation (SR) can produce radionuclides, with the SR power of 50 MW during ttbar operation and random beam losses with SR power of 50 MW being the most critical cases. We conducted separate simulations for these cases. The number of particles lost was $5.9 \times 10^{6}$ per second, and the number of SR photons was $1.2 \times 10^{16}$ per second. We calculated the ratios of radionuclide production to limits in GB18871 and listed them in Tables 7.3.8 (cooling water), 7.3.9 (air), and 7.3.10 (water in the wall). For the rock, only the leachable nuclides are listed in Table 7.3.11.

**Table 7.3.8:** Ratios of radionuclides production to the limits in GB18871 for the cooling water.

| | | Specific activity / GB18871 |
|---|---|---|
| Beam losses | O15 | 2.44 |
| | C14 | $3.5 \times 10^{-7}$ |
| | Be7 | $1.3 \times 10^{-2}$ |
| | H3 | $2.3 \times 10^{-6}$ |
| SR | None | |

**Table 7.3.9:** Ratios of radionuclides production to the limits in GB18871 for the air.

| | | Specific activity/GB18871 |
|---|---|---|
| Beam losses | Ar41 | $1.4 \times 10^{-3}$ |
| | Ar37 | $6.1 \times 10^{-9}$ |
| | P33 | $1.9 \times 10^{-8}$ |
| | O15 | $2.7 \times 10^{-4}$ |
| | C14 | $7.7 \times 10^{-7}$ |
| | Be7 | $1.1 \times 10^{-5}$ |
| | H3 | $3.5 \times 10^{-9}$ |
| SR | Ar41 | $6.5 \times 10^{-6}$ |
| | C14 | $1.2 \times 10^{-2}$ |

**Table 7.3.10:** Ratios of radionuclides production to the limits in GB18871 for the water in the wall.

| | | Specific activity/GB18871 |
|---|---|---|
| Beam losses | O15 | $2 \times 10^{-3}$ |
| | C14 | $5\times 10^{-10}$ |
| | Be7 | $3\times 10^{-5}$ |
| | H3 | $6\times 10^{-9}$ |
| | F18 | $5\times 10^{-6}$ |
| SR | C14 | $2\times 10^{-12}$ |
| | H3 | $1\times 10^{-10}$ |

**Table 7.3.11:** Ratios of radionuclides production to the limits in GB18871 for the rock in the wall.

| | | Specific activity/GB18871 |
|---|---|---|
| Beam losses | Mn54 | $6.9 \times 10^{-4}$ |
| | Ca45 | $5.5 \times 10^{-6}$ |
| | Na22 | $7.2 \times 10^{-4}$ |
| | H3 | $5.9 \times 10^{-9}$ |
| SR | H3 | $1.0 \times 10^{-10}$ |

A similar method can be used to obtain radionuclide production in the air of the Linac area. The results are listed in Table 7.3.12.

**Table 7.3.12:** Ratios of radionuclides production to the limits in GB18871 in the air of the Linac area.

| | Specific activity/GB18871 |
|---|---|
| Ar41 | 0.13 |
| Ar37 | $3.4 \times 10^{-7}$ |
| Cl38 | 0.41 |
| Cl36 | $3.1 \times 10^{-10}$ |
| S35 | $3.4 \times 10^{-5}$ |
| P33 | $4.6 \times 10^{-6}$ |
| P32 | $4.6 \times 10^{-5}$ |
| Si31 | $4.3 \times 10^{-4}$ |
| F18 | $2.7 \times 10^{-4}$ |
| O15 | 2.58 |
| C14 | $7.8 \times 10^{-5}$ |
| Be7 | $3.8 \times 10^{-2}$ |
| H3 | $1.3 \times 10^{-5}$ |

The specific activity of O15 in the cooling water is slightly higher than the limit specified in GB18871. However, the half-life of O15 is 122 seconds, which means that several minutes after the machine shuts down, the specific activity will decrease to less than half of its initial value. Hence, it is crucial to wait for an adequate duration to ensure that the specific activity of O15 is sufficiently low before discharging water and air. Thus, the slightly elevated ratio of O15 will not be a problem. Additionally, all other ratios of specific activities of radionuclide production are less than one, and the sum of these ratios is also less than one, which meets the requirements of GB18871.

#### 7.3.3.2 *Estimation of the Amount of Nitrogen Oxides*

Gamma rays have the ability to decompose atmospheric oxygen into free radicals which then react with $O_2$ molecules to form $O_3$ (ozone). Ozone can further react with NO in the air to form $NO_X$, while $NO_2$ combines with $H_2O$ to form $HNO_3$. The production rate of $O_3$, $NO_X$, and $HNO_3$ is 10, 4.8, and 1.5 molecules per 100 eV energy of the γ ray absorbed, respectively. To simplify calculations, we focus only on the production of $O_3$, and the amount of $NO_X$ can be derived from the aforementioned proportions.

In an irradiation space with a volume of V, the chemical decomposition of ozone and ventilation for its removal are taken into account. The number of ozone molecules, *N*, can be described by the following equation:

$$\frac{dN}{dt} = PG - \left(\alpha' + \frac{KF}{V}\right)N. \tag{7.3.2}$$

Solving this differential equation gives:

$$N = \frac{PG}{\alpha' + \frac{KF}{V}}\left[1 - e^{-(\alpha' + \frac{KF}{V})t}\right], \tag{7.3.3}$$

in which:

*N* – the number of ozone molecules produced;

*P* – the power absorbed by the air in eV/s;

*G* – the production of $O_3$, 0.06 molecules per eV is used in the calculation;

*F* – the ventilation rate of the irradiated area in $cm^3/s$;

*V* – the volume of the irradiated area in $cm^3$;

*K* – the mixing uniformity coefficient, K = 1/3;

*α'* – the chemical decay constant of $O_3$, $2.3 \times 10^{-4} s^{-1}$, corresponding to a 50 minutes chemical half-life of $O_3$;

*t* – the irradiation time in seconds.

One ppm of $O_3$ in the air is equivalent to $2.463 \times 10^{13}$ $O_3$ molecules in one $cm^3$ of air:

$$C_p = \frac{N}{2.463 \times 10^{13} V} \tag{7.3.4}$$

According to the basic concept of energy transmission when γ rays pass through the air, it is easy to deduce the power absorbed by the air using the following formula:

$$P = 6.25 \times 10^{18} \sum_i \left[E_{\gamma i} \Psi_{\gamma i} (K/\Phi)_i\right] \rho_{air} V_{air} \tag{7.3.5}$$

In this formula:

*P* – power absorbed by the air, in J;

$E_{\gamma i}$ – energy interval of the γ ray, in eV;

$\Psi_{\gamma i}$ – average flux in the energy interval of the γ ray, in $cm^{-2}s^{-1}eV^{-1}$;

$(K/\Phi)_I$ – the conversion coefficients from mono-energetic photon flux to air kerma (the quotient of dE by dm is defined as the sum of the initial kinetic energies of all the charged ionizing particles liberated by uncharged ionizing particles within a volume element of mass dm), in Gy-cm$^2$;

$\rho_{air}$ – air density in standard conditions, in kg/m$^3$;

$V_{air}$ – volume of air, in m$^3$;

$6.25\times10^{18}$ – conversion coefficient, in eV/J.

In which $E_{\gamma i}\Psi_{\gamma i}$ is the fluence rate of photons, $\rho_{air}V_{air}$ is the mass of air. $E_{\gamma i}$ and $\Psi_{\gamma i}$ could be obtained using FLUKA simulations; $(K/\Phi)_i$ can be interpolated from the ICRP74 report [4].

Hence, the $O_3$ concentration can be expressed by:

$$C_p = \frac{PG}{2.463\times10^{13}V(\alpha' + \frac{KF}{V})}\left[1 - e^{-(\alpha' + \frac{KF}{V})t}\right] \tag{7.3.6}$$

The density of $O_3$ is $1.964\times10^{-3}$g/cm$^3$, so the concentration of $O_3$ in the air can be expressed in g/cm$^3$:

$$C_g(t) = 7.97\times10^{-17}\frac{PG}{V(\alpha' + \frac{KF}{V})}\left[1 - e^{-(\alpha' + \frac{KF}{V})t}\right] \tag{7.3.7}$$

Because the chemical half-life of $O_3$ is only 50 minutes, the concentration of $O_3$ in the tunnel could become easily saturated. The saturated concentration is (in g/cm$^3$):

$$C_g(t) = 7.97\times10^{-17}\frac{PG}{V(\alpha' + \frac{KF}{V})}. \tag{7.3.8}$$

Combining the parameters in Chapter 4, the concentrations of $O_3$ in the tunnel for the four operation modes are listed in Table 7.3.13. The $O_3$ concentrations in Table 7.3.13 are the same for different operation modes because the concentrations are saturated. The concentration of $O_3$ in the Linac is$1.0\times10^{-6}$ μg/m$^3$. The concentrations are lower than the limit of mandatory standards. Similarly, the concentrations of $NO_X$ are also below the mandatory limit.

**Table 7.3.13:** Ozone concentration for four different operation modes.

| | Number of SR photons | Deposited energy in the air per SR photon [GeV/cm$^3$] | $O_3$ mass [μg/m$^3$] |
|---|---|---|---|
| Higgs | $4.7\times10^{18}$ | $2.8\times10^{-8}$ | $8.3\times10^{-6}$ |
| Z | $8.4\times10^{19}$ | $1.8\times10^{-9}$ | $8.3\times10^{-6}$ |
| WW | $1.6\times10^{19}$ | $6.0\times10^{-9}$ | $8.3\times10^{-6}$ |
| $t\bar{t}$ | $1.4\times10^{18}$ | $7.6\times10^{-8}$ | $8.3\times10^{-6}$ |

### 7.3.4 Personal Safety Interlock System (PSIS)

#### 7.3.4.1 *System Design Criteria*

The PSIS was designed with several criteria in mind to ensure safe operation of the CEPC:

1. All critical device interlock signals are generated by hardware to ensure reliability;
2. The PSIS has top priority to shut off the beam in the CEPC Central Control System at the highest interlock level;
3. Fail-safe mechanisms ensure that the beam will be shut off when a critical device experiences a breakdown;
4. Redundancy measures are in place to increase reliability, reduce fault time, and preserve upgrade possibilities;
5. The PSIS incorporates multilayer protection measures, such as an interlock key, emergency shut-off button, emergency door-open button, acousto-optic alarm, patrol search and secure before beam start-up, and surveillance cameras to ensure personal safety;
6. The PSIS is designed to be people-oriented, with the primary goal being personal safety. It is also designed to be convenient to operate and maintain, with a user-friendly human-computer interface.

#### 7.3.4.2 *PSIS Design*

The PSIS is composed of two main systems: the Programmable Logic Controller (PLC) and the Access Control System (ACS). The PLC is responsible for monitoring the interlocked equipment, while the ACS administers interlock information. Access to interlocked areas is strictly regulated, requiring the identification of all persons who enter to be recorded. The layout of the PSIS is shown in Figure 7.3.8.

The PLC system comprises several components, including the access controller, interlock key, emergency/patrol button, emergency door-open button, acousto-optic alarm, and interlock equipment. These devices provide multilayer personnel protection by programming interlock signals. The system also features host double backup and double lines for signal transmission to ensure its reliability.

The ACS includes a camera, LED display, and data server. It allows the PSIS to monitor interlock areas, display interlock information, and store data. This adds an additional layer of security to the system.

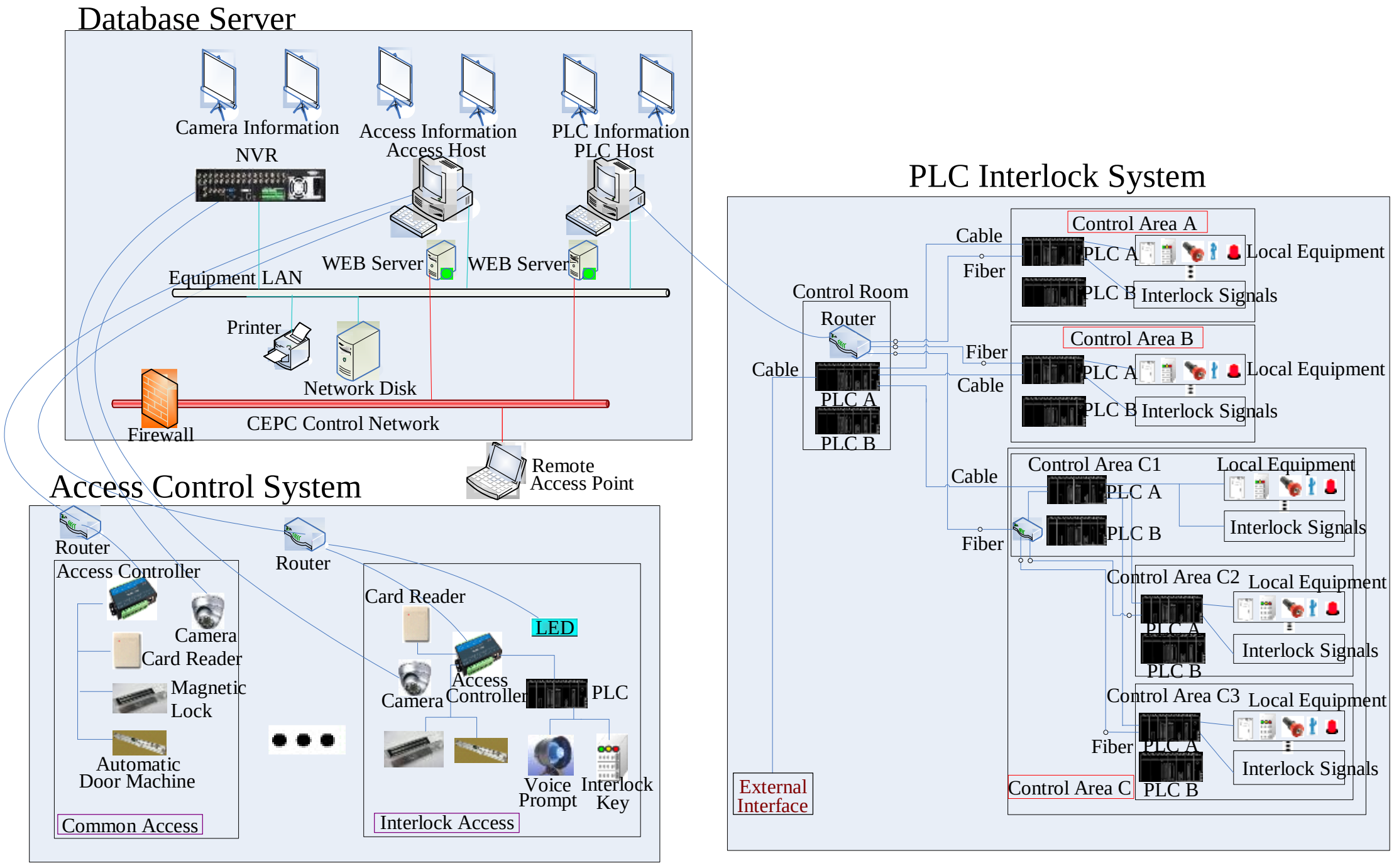

**Figure 7.3.8:** Layout of the PSIS.

Before start-up, a patrol search and security check must be conducted in every interlock area to ensure that no one is left behind. An acousto-optic alarm will alert individuals to evacuate the area when the "ready" signal from the Central Control System (CCS) is received. During this time, the access control system will still be in a "shutdown" state. In the event of an accident, the emergency button will immediately turn off the beam. Figure 7.3.9 illustrates the operation flowchart of the PSIS.

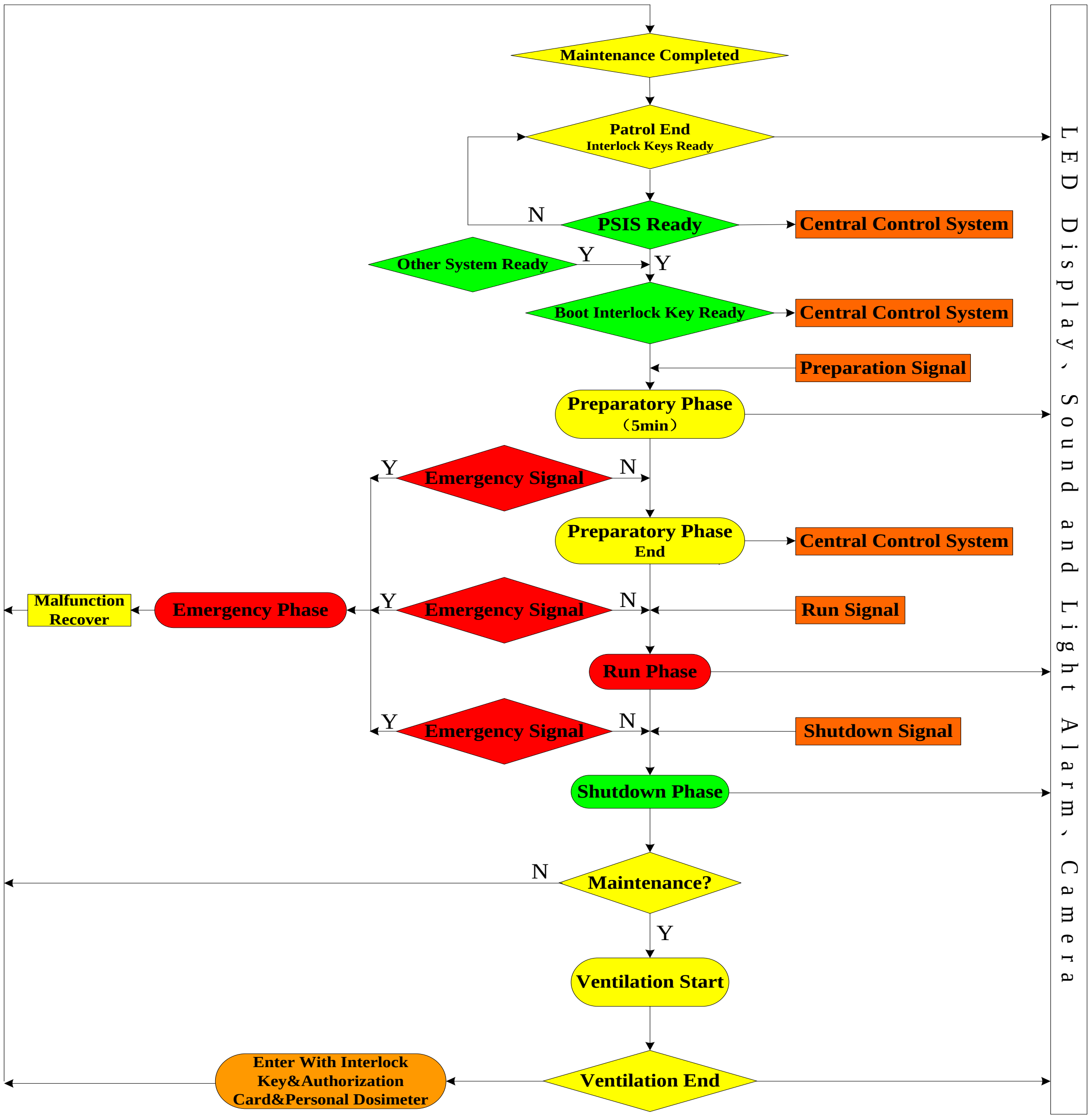

**Figure 7.3.9:** PSIS operation flow chart.

## 7.3.5 Radiation-Dose Monitoring Program

### 7.3.5.1 *Radiation-Monitoring System*

The radiation-monitoring system is designed to meet the latest legal requirements and international standards. It will be based on the results of the preliminary hazard analysis, the latest technical developments, and specific requirements such as the time structure and composition of the radiation fields. This new state-of-the-art system will provide reliable and accurate continuous measurements of the ambient dose equivalent and the ambient dose-rate equivalent in the underground areas, as well as the surface areas inside and outside the project perimeter.

If preset radiation levels are exceeded within radiation controlled areas, an alarm will be triggered and transmitted, and remote alarms will sound in the control rooms. The radiation-monitoring system will also permanently monitor the level of radioactivity in water and air released from the project, and include hand-foot monitors, site-gate monitors, tools, and material monitors. It will provide remote supervision, long-term database storage, and off-line data analysis to ensure the safety of personnel and the environment.

A typical frame diagram of the radiation monitoring system is shown in Figure 7.3.10. The data will be accessible via the internet, allowing for convenient and timely monitoring of the radiation levels in the project area.

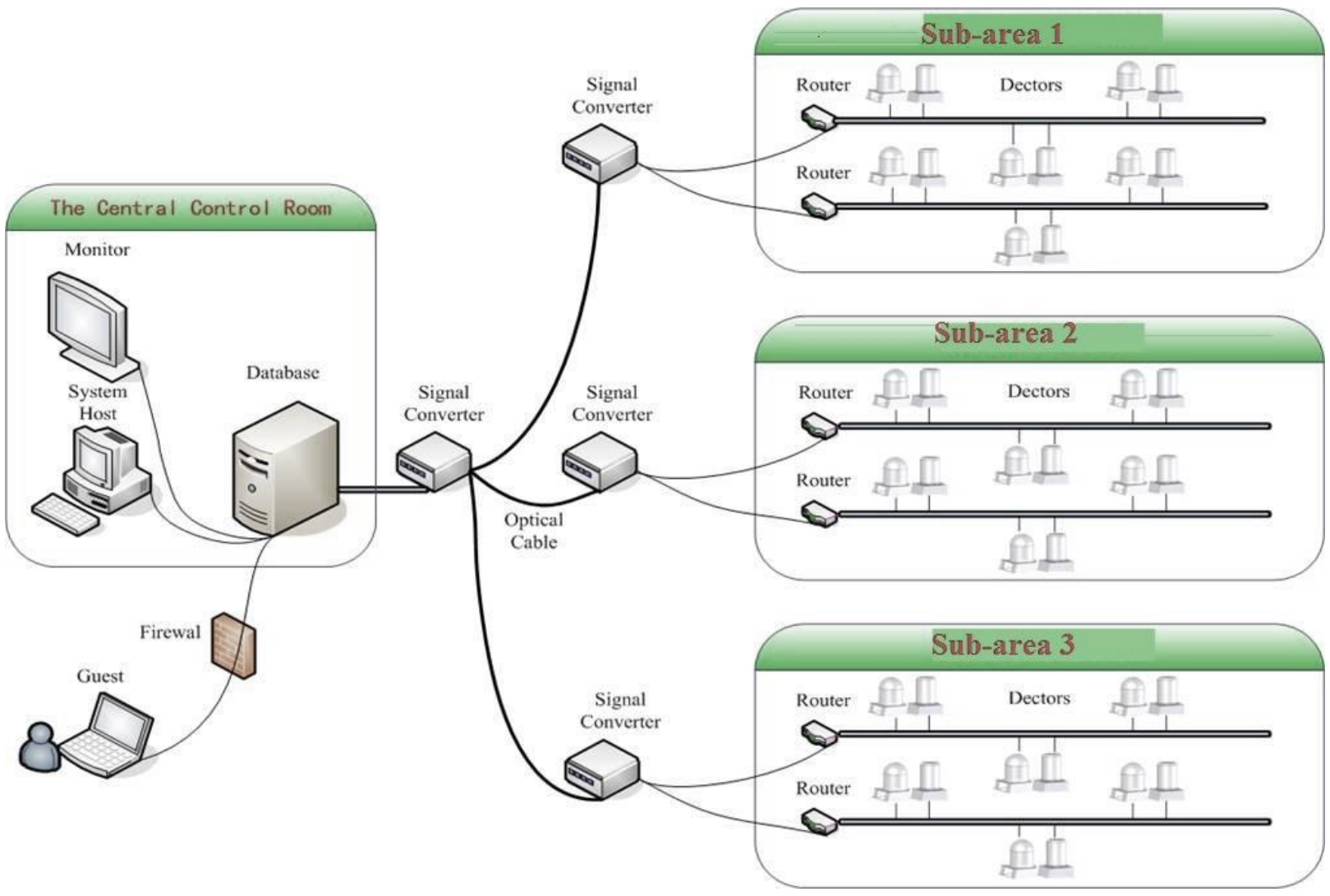

**Figure 7.3.10:** Frame diagram of the radiation monitoring system.

To ensure reliable control and data communication under different conditions, the communication system is designed with four communication paths: Ethernet, wireless, GPRS (General Packet Radio Service), and offline record.

### 7.3.5.2 *Workplace Monitoring Program*

The workplace monitoring program is designed to ensure that radiation levels in the workplace comply with relevant regulations. Monitoring sites are established at all main entrances to high-radiation-level areas and in workplaces close to the accelerator or radioactive sources. If a radiation level exceeds the set value, monitors will sound an alarm to inform people to evacuate. Each site is equipped with one gamma detector and one neutron detector to detect the presence of ionizing radiation.

### 7.3.5.3 *Environmental Monitoring Program*

The objective of the environmental monitoring program is to ensure compliance with regulatory limits and provide early warning of any imminent violation. To achieve this, dose rates will be monitored in various environmental mediums such as air, water, plants, and soil. A baseline will be established with a complete background survey conducted 2-3 years before equipment installation, and monitoring will continue for 3-5 years after operation begins to evaluate the facility's impact on the environment.

Monitoring stations will be located at critical or representative places to measure the dose rate and integrated dose. Each station will consist of a pressurized ionization chamber for gamma monitoring and a rem-counter for neutrons, which will generate alarms when dose-rate thresholds are exceeded. Fluid or gas extraction ducts that may contain radioactivity produced by the facility will be equipped with on-line real-time monitors for short-lived radioactive substances, along with a sampler. The monitors' readings will be stored in a database, and the filters will be periodically replaced and analyzed in an off-line laboratory for longer-lived beta and gamma activity. These measurements will be carried out, particularly after facility upgrades.

#### 7.3.5.4 *Personal Dose Monitoring Program*

Every staff member is required to participate in the personal dose monitoring program. For gamma dose monitoring, OSL (Optically Stimulated Luminescence) will be utilized, while the CR-39 solid track dosimeter will be utilized for neutron-dose monitoring. Furthermore, individuals entering the tunnel for maintenance must use an electronic personal dose alarm.

### 7.3.6 **Management of Radioactive Components**

Radioactive contamination can occur in accelerator components and maintenance tools and fixtures due to their exposure to radiation during operation. The degree of contamination depends on various factors, including the material composition, location, and irradiation history. In addition, the specific activity of these items changes over time due to radioactive decay. To ensure proper disposal and management of radioactive items, all contaminated materials will be transferred to temporary storage and then disposed of or sent to long-term repositories in accordance with relevant legislation and regulations.

# 8 SPPC

## 8.1 Introduction

### 8.1.1 Science Reach of the SPPC

The SPPC (Super Proton-Proton Collider) is an ambitious project aimed at surpassing the capabilities of the LHC (Large Hadron Collider). It is envisioned to be an extremely powerful machine with a center of mass energy of 125 TeV, far exceeding the energy reach of its predecessor. Additionally, it is expected to achieve a peak luminosity of 1.0 $\times 10^{35}$ $cm^{-2}s^{-1}$ per interaction point (IP), and an integrated luminosity of approximately 30 $ab^{-1}$, assuming two interaction points and a runtime of 20-30 years [1-4].

The SPPC, in conjunction with the CEPC, will provide unprecedented precision in studying Higgs physics [2]. However, what the scientific community anticipates even more is the SPPC's potential to explore a significantly larger region of the new physics landscape, propelling our understanding of the physical world to new heights.

Energy-frontier physics encompasses various critical questions that the SPPC aims to address. These include elucidating the mechanism of Electroweak Symmetry Breaking (EWSB) and the nature of the electroweak phase transition, tackling the naturalness problem, and advancing our comprehension of dark matter. While these three inquiries can be interconnected, they also present diverse paths of exploration, leading to the exploration of more fundamental physics principles. By venturing into uncharted territory, the SPPC holds tremendous potential for groundbreaking discoveries that could shed light on all of these questions and revolutionize our understanding of the universe.

The CEPC serves as a "Higgs factory," enabling high-precision measurements of various properties of the Higgs boson. For instance, using the recoil mass method, the CEPC can accurately determine the absolute Higgs couplings to Z bosons (g(HZZ)) and the invisible decay branching fraction, both at the sub-percent level. It can also measure the couplings to gluons (g(Hgg)), W bosons (g(HWW)), and heavy fermions [g(Hbb), g(Hcc), and g(H)] with a precision within a few percentage points. Furthermore, the CEPC can measure the rare decay couplings [g(H and g(H))] at the 10% level. However, due to its center of mass energy limitations, the CEPC cannot directly measure the couplings g(Htt) and g(HHH), which are crucial for understanding Electroweak Symmetry Breaking (EWSB) and naturalness [5].

Expanding on the CEPC's Higgs factory program, the SPPC will produce billions of Higgs bosons, offering significant physics opportunities, particularly in the realm of rare but relatively clean channels. The SPPC can enhance the measurement of the Higgs-photon coupling, observe the coupling g(H), and investigate other rare decays such as $t \rightarrow Hc$ and $H \rightarrow \mu\tau$. With a higher energy threshold than the CEPC, the SPPC could measure g(HHH) with a precision of 10% [6], and directly determine the coupling g(Htt) at a sub-percentage level [7]. The Higgs self-coupling, g(HHH), is considered the holy grail of experimental particle physics, not only due to the experimental challenges involved but also because it provides insights into the form of the Higgs potential. By measuring g(HHH), the SPPC can contribute to answering questions related to the nature of the electroweak phase transition, whether it is of the first or second order, which is crucially linked to the concept of electroweak baryogenesis.

As an energy frontier machine, the SPPC has the potential to discover new particles in the O(10 TeV) range and uncover new fundamental physics principles. The naturalness problem, arising from the disparity between the electroweak and fundamental scales, could be addressed through the SPPC's exploration. Discovering new particles near the electroweak scale would be a significant advancement for our understanding of the naturalness principle. The SPPC's capabilities could surpass the current concept of naturalness, probing fine-tuning levels down to $10^{-4}$, beyond what the LHC has achieved at $10^{-2}$.

Dark matter remains one of the most intriguing enigmas in particle physics and cosmology. Among the various candidates, Weakly Interacting Massive Particles (WIMPs) remain the most compelling candidates for dark matter. If dark matter particles have interactions with Standard Model particles comparable to the strength of weak interactions, it is plausible that their mass could fall within the TeV range, a range that could be explored at the energy of the SPPC. By combining the constraints obtained from both direct (underground) and indirect (astroparticle) dark matter searches, the SPPC could significantly expand the parameter space for WIMP models.

At the energy regime of the SPPC, the Standard Model particles effectively become "massless," and both electroweak symmetry and flavor symmetry are restored. In this environment, the top quark and electroweak gauge bosons are expected to exhibit parton-like behavior in the initial state and resemble narrow jets in the final state. Understanding the behavior of Standard Model processes in such a unique and unprecedented environment presents new challenges and exceptional opportunities for refining our techniques in the quest for new physics at higher energy scales.

### 8.1.2 The SPPC Complex and Design Goals

The SPPC is a comprehensive accelerator facility designed to support research across various fields of physics, similar to the multi-use accelerator complex at CERN. In addition to the energy frontier physics program conducted in the collider, each of the four accelerators in the injector chain can also facilitate independent physics programs. These stages include the proton linac (p-Linac), rapid cycling synchrotron (p-RCS), medium-stage synchrotron (MSS), and the final stage synchrotron (SS), as depicted in Figure 8.1.1 and described in more detail in Section 8.4. These stages can engage in research when beam availability is not required for the subsequent accelerator stage.

The versatility of the SPPC extends beyond its p-p collider capabilities, with the option to explore heavy ion collisions, which opens doors to deeper investigations into nuclear matter studies. Furthermore, there is potential for electron-proton and electron-ion interactions, broadening the scope of research possibilities. In essence, the SPPC will assume a pivotal role in experimental particle physics in a post-Higgs discovery era. It serves as the natural progression of the circular collider physics program following the CEPC. The combination of these two world-class machines represents a significant milestone in our ongoing quest to comprehend the fundamental laws of nature.

With the 100 km circumference tunnel, jointly defined by the CEPC and SPPC projects, our goal is to achieve proton-proton collisions in the middle of the century. The anticipated accelerator technology and cost considerations are taken into account in this endeavor. The magnetic field required to bend the protons around the ring is projected to be 20 T, utilizing magnets built with a new type of high-temperature superconductor (HTS), the iron-based superconductor (IBS).

Considering the expected advancements in detector technology, it is anticipated that a peak luminosity of approximately $1 \times 10^{35}$ cm$^{-2}$s$^{-1}$ will be achievable. However, the potential for even higher luminosities is also being considered for the long-term operation. The project plans to have at least two interaction points (IPs) available for experiments. Table 8.1.1 highlights some key parameters relevant to the project.

**Table 8.1.1:** Key parameters of the SPPC baseline design

| Parameter | Value | Unit |
|---|---|---|
| Center of mass energy | 125 | TeV |
| Peak luminosity | $1.0 \times 10^{35}$ | cm$^{-2}$s$^{-1}$ |
| Number of IPs | 2 | |
| Circumference | 100 | km |
| Injection energy | 3.2 | TeV |
| Overall cycle time | 5-12 | Hours |
| Dipole field | 20 | T |

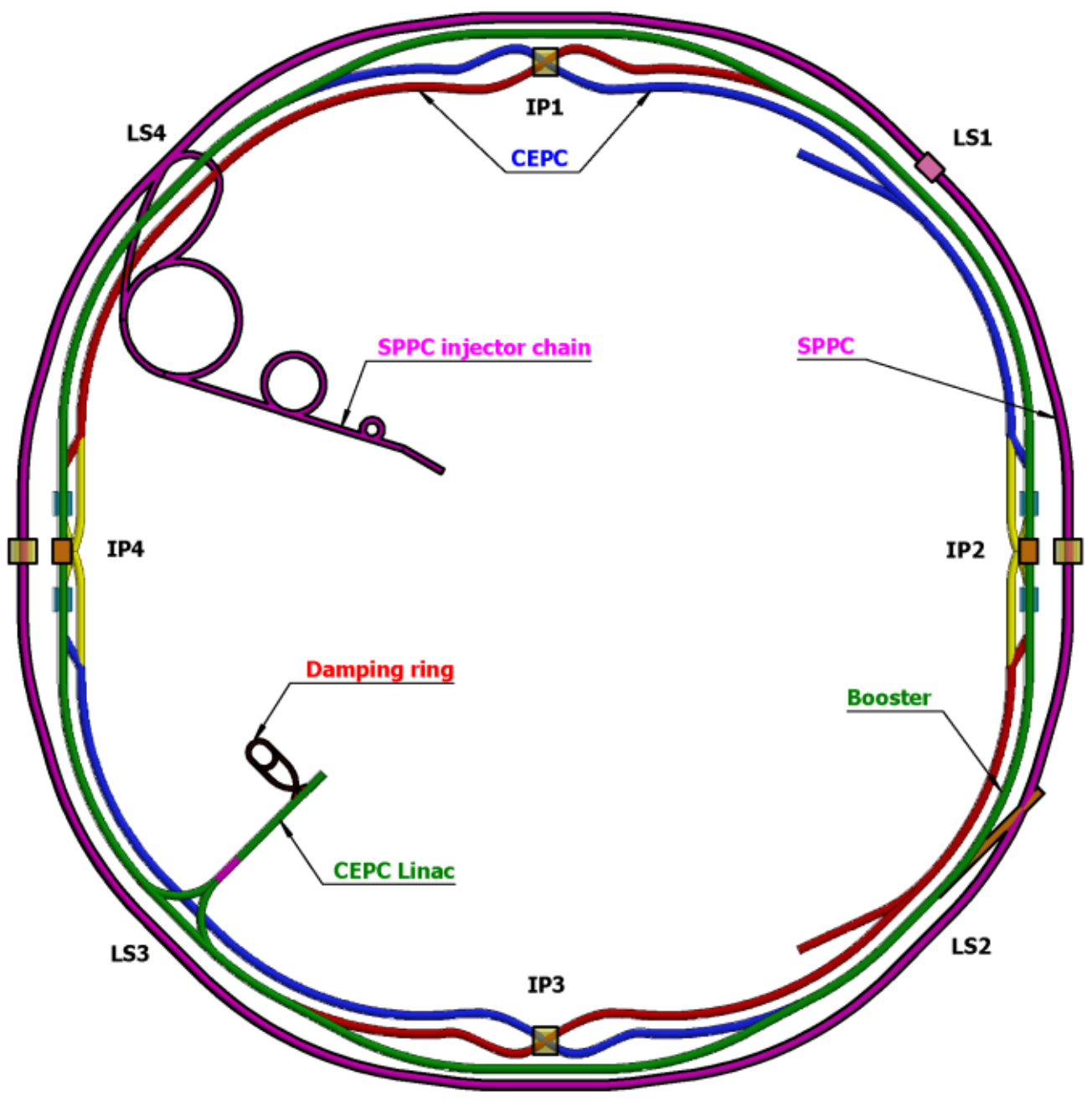

**Fig. 8.1.1:** SPPC and CEPC accelerator complex.

### 8.1.3 **Overview of the SPPC Design**

The SPPC will be situated in the same 100 km circumference tunnel that is to be constructed for the CEPC. The tunnel's shape and symmetry have been designed as a compromise between the requirements of the two colliders. However, the SPPC necessitates relatively longer straight sections, which will be discussed in more detail later. The tunnel consists of eight arcs and eight straight sections that accommodate two large detectors, injection and extraction systems, RF (radio frequency) stations, and a complex collimator system.

Based on the expected advancements in High-Temperature Superconducting (HTS) technology, particularly the IBS technology, and high-field magnet technology in the coming 20-30 years, it is anticipated that the main dipole magnets will be able to achieve a magnetic field strength of 20 T. This can be achieved at a reasonable cost or even at a lower cost compared to using $Nb_3Sn$ superconductors. Twin-aperture magnets will be utilized for the two-ring collider design. A filling factor of 78% in the arcs, similar to the LHC, is assumed. The SPPC is expected to provide beams with a center of mass energy of 125 TeV.

By utilizing a circulating beam current of approximately 0.19 A and implementing small beta functions (β*) of 0.5 m at the collision points, the initial luminosity of the SPPC can reach $4.3 \times 10^{34}$ $cm^{-2}s^{-1}$ per IP. With a luminosity leveling scheme, this can be further increased to a peak luminosity of $1.0 \times 10^{35}$ $cm^{-2}s^{-1}$ at its maximum operation.

The high beam energy and beam current in the SPPC will generate intense synchrotron radiation. This places critical demands on the vacuum system, which is based on cryogenic pumping. To overcome this technical challenge, efficient beam screens are being developed to extract the substantial heat load generated by the synchrotron radiation. Additionally, reducing the electron cloud density within the beam apertures is crucial. It is expected that these technical challenges will be resolved within the next two decades.

If the need arises to reduce the power of synchrotron radiation, alternatives can be explored. This may involve decreasing the bunch population or the number of bunches, alongside efforts to achieve a smaller beta function (β*) in order to maintain the luminosity.

Similar to other proton colliders that employ superconducting magnets, the injection energy of the SPPC is primarily determined by the field quality of the magnets at the lower end of their operational range. The presence of persistent currents in the coils can cause distortions in the field distribution at the injection energy.

Several factors support the utilization of a relatively higher injection energy. One such factor is the coupling impedance, which plays a significant role in addressing collective beam instabilities. Additionally, a smaller geometrical emittance is desirable to reduce the apertures of the beam screen and magnet. Furthermore, there are requirements regarding the good-field region of the magnets.

Taking inspiration from the LHC, where the ratio of top to bottom fields is 15, an injection energy of 4.17 TeV can be considered. However, a larger ratio of 20 could also be contemplated, leading to an injection energy of 3.13 TeV. Opting for the latter ratio would result in a more cost-effective injector chain. In this report, a compromise has been adopted, with an injection energy of 3.2 TeV being selected.

The injector chain plays a crucial role in pre-accelerating the beam to the desired injection energy while maintaining the required bunch current, bunch structure, and emittance. It also determines the duration of the beam fill period, which is significant for the overall operation of the SPPC.

To achieve the injection energy of 3.2 TeV, a four-stage injector chain is proposed. The first stage is the p-Linac, which accelerates the beam to 1.2 GeV. Subsequently, the beam is further accelerated in the p-RCS (rapid cycling synchrotron) to reach 10 GeV. The third stage, known as the MSS (medium-stage synchrotron), continues the acceleration process up to 180 GeV. Finally, the SS (final stage synchrotron) completes the pre-acceleration, bringing the beam energy to the desired injection energy of 3.2 TeV.

Having high repetition rates for the lower energy stages is advantageous as it helps to reduce the cycling period of the SS. This is particularly important because the SS relies on superconducting magnets for its operation. Furthermore, the beams with high repetition rates from the lower energy stages can be utilized for other research applications when the accelerators are not actively preparing the beam for injection into the SPPC.

In order to maintain the beam emittances and control excessive beam-beam tune shifts, it is necessary to address the effects of synchrotron radiation cooling. If left uncontrolled, synchrotron radiation would rapidly reduce the beam emittances and lead to undesirable tune shifts caused by the beam-beam interaction.

To mitigate these effects, noise in transverse deflecting cavities is introduced. This noise helps limit the minimum transverse emittances, preventing excessive emittance reduction (emittance heating) and minimizing the resulting tune shifts.

Without implementing measures to control these effects, the luminosity of the collider would decay exponentially from its initial peak. The decay time, approximately 8.1 hours in this case, determines the lifetime of the luminosity.

To maximize the integrated luminosity over the course of operation, it is desirable to have a short turnaround time. The turnaround time refers to the period in the machine cycle that excludes the collision period. Ideally, the turnaround time should be shorter than the beam decay time. The initially assumed average turnaround time of 2.3 hours is acceptable, resulting in an optimized complete cycle time of approximately 7-11 hours. However, a shorter turnaround time, such as 0.8 hours, would be more preferable to maximize the integrated luminosity.

The peak and average luminosities can be increased by utilizing synchrotron damping to lower the transverse emittance and accepting higher but manageable tune shifts. However, without leveling, excessive peak luminosities and the number of interactions per beam crossing may occur. Limiting the peak luminosity through leveling helps control this number while allowing for an increase in average luminosity. Using more and closely spaced bunches can reduce the number of interactions per bunch crossing without compromising peak luminosities. If the beam current remains constant, reducing the number of protons per bunch proportionally becomes necessary. To maintain the desired luminosity, allowing further synchrotron damping to lower the emittances without increasing the tune shifts is important.

Designing and constructing the collider, including the injector chain, presents numerous technical challenges. However, the two most formidable challenges are the development and production of 20-T magnets utilizing IBS technology and the implementation of an effective beam screen to mitigate the effects of intense synchrotron radiation. Resolving these challenges will require substantial research and development efforts over the next decades.

### 8.1.4 **Compatibility with CEPC and Other Physics Prospects**

The current proposal includes the retention of the CEPC collider rings and its full energy booster in the tunnel alongside the SPPC. This configuration allows for a potential future electron-proton (e-p) collision program utilizing one beam from each of the CEPC and SPPC facilities. However, there is a more ambitious plan being considered, which involves keeping the entire CEPC complex, including the detectors, during the construction of the SPPC. This would require significant modifications to the tunnel. At IP1 and IP3, the SPPC would bypass the CEPC detectors in a separate tunnel that is also used for the CEPC booster bypass. At IP2 and IP4, the SPPC bypass tunnel is connected to the SPPC experimental halls as depicted in Fig. 8.1.1.

Furthermore, the SPPC is also designed to accommodate heavy ion collisions. Therefore, a dedicated interaction point (IP) and ion beam acceleration from the injector chain will be included in the plans for the SPPC.

## 8.2 Key Accelerator Issues and Design

### 8.2.1 **Main Parameters**

#### 8.2.1.1 ***Collision Energy and Layout***

In order to achieve the desired center of mass energy of 125 TeV with a 100 km circumference, an extremely high magnetic field of 20 T is required. This exceeds the capabilities of current magnet technology using $Nb_3Sn$ superconductors. However, there is optimism regarding the development of IBS-HTS (iron-based high-temperature superconductors) technology, which is expected to become available and more cost-effective within the next 10-15 years. IBS-HTS magnets are expected to generate the required 20 T field over a longer period of time, making them a suitable choice for the SPPC.

Given the long circumference, it is important to design the arc sections to be as compact as possible, in order to provide sufficient space for the necessary long straight sections. While the lattice study is still ongoing, it is assumed that a traditional FODO scheme will be used throughout the accelerator, except at the interaction points (IPs) where triplets are employed to achieve very small beta functions ($\beta^*$). The arcs make up the majority of the circumference, with a filling factor of 0.78, similar to the LHC.

A key consideration is determining the number and lengths of the long straight sections. These sections are necessary to accommodate the placement of large physics detectors, as well as the beam injection and extraction systems, collimation systems, and RF stations. Approximately 18.2 km of total length is allocated for the long straight sections, consisting of eight sections in total. Two sections (IP1 and IP3) are about 3.3 km long each; another two sections (IP2 and IP4) are about 3.8 km long each, while the remaining four sections are about 1 km long each.

With this configuration, the top beam energy is 62.5 TeV, which results in a collision energy of 125 TeV. The main parameters of the SPPC are summarized in Table 8.2.1.

**Table 8.2.1:** Main parameters of the SPPC

| Parameter | Value | Unit |
|---|---|---|
| General design parameters | | |
| Circumference | 100 | km |
| Beam energy | 62.5 | TeV |
| Lorentz gamma | 66631 | |
| Dipole field | 20.3 | T |
| Dipole curvature radius | 10258.3 | m |
| Arc filling factor | 0.79 | |
| Total dipole magnet length | 64.455 | km |
| Arc length | 81.8 | km |
| Number of long straight sections | 8 | |
| Total straight section length | 18.2 | km |
| Energy gain factor in collider rings | 19.53 | |
| Injection energy | 3.2 | TeV |
| Number of IPs | 2 | |
| Revolution frequency | 3.00 | kHz |
| Physics performance and beam parameters | | |
| Initial luminosity per IP | $4.3\times10^{34}$ | $cm^{-2}s^{-1}$ |
| Beta function at collision | 0.50 | m |
| Circulating beam current | 0.19 | A |
| Nominal beam-beam tune shift limit per IP | 0.015 | |
| Bunch separation | 25 | ns |
| Number of bunches | 10082 | |
| Bunch population | $4.0\times10^{10}$ | |
| Accumulated particles per beam | $4.0\times10^{14}$ | |
| Normalized rms transverse emittance | 1.2 | μm |
| Beam lifetime due to burn-off | 8.1 | hours |
| Total inelastic cross section | 161 | mb |
| Reduction factor in luminosity | 0.81 | |
| Full crossing angle | 73 | μrad |
| rms bunch length | 60 | mm |
| rms IP spot size | 3.0 | μm |
| Beta at the first parasitic encounter | 28.6 | m |
| rms spot size at the first parasitic encounter | 22.7 | μm |
| Stored energy per beam | 4.0 | GJ |
| SR power per beam | 2.2 | MW |
| SR heat load at arc per aperture | 27.4 | W/m |
| Energy loss per turn | 11.6 | MeV |

### 8.2.1.2 *Luminosity*

The initial luminosity of the SPPC is expected to reach 4.3 × $10^{34}$ $cm^{-2}s^{-1}$, which would increase to a range from 1.0 × $10^{35}$ $cm^{-2}s^{-1}$ to 1.3 × $10^{35}$ $cm^{-2}s^{-1}$ during th course of collision. This is significantly higher than the luminosities achieved in previously built machines such as the LHC [1], Tevatron [2], and designs such as SSC [3], HE-LHC [4], and comparable to FCC-hh [5].

With the same bunch spacing, the SPPC will have a higher number of interactions per bunch crossing compared to current detectors' capabilities. However, it is anticipated that ongoing research and development efforts in detector technology, as well as general technical advancements, will address this challenge. The belief is that advancements in detectors and overall technical evolution will enable the detectors to effectively handle the increased number of interactions at the SPPC's high luminosity levels.

The average, or integrated, luminosity is an important parameter to consider in collider experiments. Several factors affect the integrated luminosity, including the loss of stored protons from collisions, the cycle turnaround time, and the shrinkage of the transverse emittance due to synchrotron radiation.

As synchrotron radiation causes the transverse emittance to shrink over time, it can help maintain or even increase the instantaneous luminosity after the collision starts. However, if left uncontrolled, it can also lead to an unacceptable increase in the beam-beam tune shift. To counteract the emittance shrinkage and limit the tune shift to an acceptable level, an emittance blow-up system is employed.

Another method to increase the luminosity is by adjusting the beta function ($\beta^*$) during the collisions. By taking advantage of emittance shrinking while keeping the beam-beam tune shift constant, the $\beta^*$ can be optimized to enhance the luminosity.

### 8.2.1.3 *Bunch Structure and Population*

In order to achieve high luminosity operation, it is desirable to have many bunches with a relatively small bunch spacing. However, there are limitations to the bunch spacing that need to be considered. One limitation is the presence of parasitic collisions in the proximity of the interaction points (IPs), and another limitation is the electron cloud instability. The ability of the detector trigger systems to cope with short bunch spacing is also an important consideration. The trigger systems must be able to effectively identify and record the collision events within the short time interval between bunch crossings.

For the SPPC, a nominal bunch spacing of 25 ns has been adopted, similar to the LHC. This value is determined by the RF system in the medium-stage synchrotron (MSS) of the injector chain and is maintained throughout the accelerator. However, it is possible to achieve shorter bunch spacing through a bunch splitting mechanism, such as a five-fold bunch splitting scheme, implemented in the MSS [6].

Time gaps between bunch trains are necessary for beam injection and extraction in both the SPPC and the injector chain. The lengths of these gaps depend on the practical design of the injection and extraction systems, as well as the rise time of the kickers used for beam extraction from the SPPC. Currently, a few microseconds are assumed for the rise time of the kickers.

The bunch filling factor, which represents the number of bunches relative to the ring circumference, is taken to be about 76% at maximum in the SPPC [7]. This value is smaller than in the LHC due to the need for more injection gaps in the SPPC design.

Bunch population is determined in the p-RCS of the injector chain, where the beam from the p-Linac fills the RF buckets using both transverse and longitudinal paintings. With the nominal bunch number and bunch population, the circulation current will be about 0.19 A in the collider rings.

#### 8.2.1.4 *Beam Size at the IP*

The beam sizes are determined by the β* of the insertion lattice and the beam emittance. The initial normalized emittance is predefined in the p-RCS of the injector chain and maintained with a slight increase as the SPPC reaches its top energy, accounting for factors such as nonlinear resonance crossings. However, at the top energy of 62.5 TeV and during the later stages of acceleration, synchrotron radiation will come into play, causing damping times of approximately 0.51 hours for the transverse emittance and 0.25 hours for the longitudinal emittance. This will result in significantly smaller emittances after the collision starts compared to their initial values when the beams reach the top energy. However, the emittances cannot be allowed to decrease without limit due to considerations of beam-beam tune shift and instant luminosity. Therefore, a stochastic emittance heating system is necessary to control the synchrotron radiation cooling and maintain the desired emittance level during the collision process.

#### 8.2.1.5 *Crossing Angle at the IP*

To minimize background from parasitic collisions near the IPs and ensure the separation of the two beams, a crossing angle is employed. The crossing angle is chosen to prevent beam overlap at the first parasitic encounters, which occur at a distance of 3.75 m from the IPs when the bunch spacing is 25 ns. The separation at these locations is maintained to be at least 12 times the rms beam size. At the collision energy of the SPPC, the crossing angle is approximately 73 μrad. This non-zero crossing angle results in a reduction of luminosity by approximately 19% compared to head-on collisions. However, this reduction can be compensated for by using crab cavities. It's important to note that with smaller bunch separations, the crossing angle needs to be larger, resulting in a larger reduction in luminosity. Additionally, the crossing angle at injection may differ due to variations in lattice settings and larger emittance.

#### 8.2.1.6 *Turnaround Time*

Turnaround time in the context of the SPPC refers to the total duration of a machine cycle when the beams are not in collision. It encompasses various stages, including the programmed countdown checking time prior to injection, the final check with a pilot shot, the time required for beam filling using SS beam pulses, the ramping up or acceleration period, and the ramping down time. For filling one SPPC ring, approximately 2.5 SS beam pulses are needed, resulting in a minimum filling time of approximately 9.3 minutes, including pilot pulses. The ramping up and down times are 19.5 minutes and 10 minutes, respectively. Thus, the minimum turnaround time is calculated to be 47.8 minutes, or approximately 0.78 hours.

However, based on the experience gained from operating the LHC and other proton colliders, it has been observed that only around one-third of the operations, from injection to reaching the top energy, are successful. As a result, the average turnaround time is closer to 2.3 hours. This value is considered acceptable, and when combined with a luminosity run time of 4-8 hours during which the beams are in collision, it leads to a total cycle time of approximately 7-11 hours.

### 8.2.2 Key Accelerator Physics Issues

#### 8.2.2.1 *Synchrotron Radiation*

Synchrotron radiation (SR) power plays a crucial role at multi-TeV energies, especially when using high-field superconducting dipoles. The power of synchrotron radiation is directly proportional to the fourth power of the Lorentz factor and inversely proportional to the radius of curvature in the dipoles. In the case of SPPC, with a beam current of 0.19 A and a magnetic field of 20 T, the synchrotron radiation power reaches approximately 26.3 W/m per aperture in the arc sections. This value is about 120 times higher than that observed at the LHC. The critical photon energy is approximately 8.4 keV.

It is worth noting that synchrotron radiation effects are not limited to the dipoles; they also occur in the high-gradient superconducting quadrupole magnets used in the SPPC.

At the SPPC, synchrotron radiation poses significant technical challenges, particularly in relation to the vacuum system and the maximum achievable circulating current. If the synchrotron radiation is absorbed at the temperature of liquid helium in the magnet bores, it would result in an excessive heat load. Therefore, an alternative approach is needed, where the synchrotron radiation is absorbed at a higher temperature.

To achieve this, a beam screen or a similar capture system must be positioned between the beam and the vacuum chamber. However, the presence of the beam screen restricts the aperture of the beam tube, increases the beam impedance and necessitates a larger superconducting magnet bore radius. The operating temperature of the beam screen becomes a critical parameter in the design process.

In addition to managing the heat load, the beam screen also plays a crucial role in controlling the coupling impedance and minimizing the impact of the electron cloud effect.

The synchrotron radiation affects the beam dynamics as the energy approaches the top level. The longitudinal and transverse emittances naturally decrease over time, with lifetimes of about 0.51 and 0.25 hours, respectively. The short damping times can assist in eliminating collective beam instabilities. Exploiting this characteristic can potentially improve machine performance by allowing the transverse emittances to decrease, leading to an increase in luminosity. However, to prevent excessive beam-beam tune shift, stochastic transverse field noise systems are necessary to regulate the reduction in emittance.

#### 8.2.2.2 *Intra-beam Scattering*

Intra-beam scattering (IBS) refers to the phenomenon where particles within a bunch scatter off each other, causing longitudinal momentum to couple into transverse motion. This scattering process can lead to an increase in the transverse emittances and can slow down the cooling effect of synchrotron radiation on the emittance. At the collision energy of SPPC with the given parameters, IBS has a negligible effect, and the emittance growth lifetime is more than a hundred hours. However, as the emittance shrinks due to synchrotron cooling, the impact of IBS becomes significant and eventually limits further reduction of the emittance.

#### 8.2.2.3 *Beam-beam Effects*

Beam-beam effects play a crucial role in determining the luminosity of a collider and can have various consequences such as emittance growth, reduced lifetime, and instabilities. These effects arise from both head-on interactions and long-range or parasitic interactions between the colliding beams. In the case of a proton-proton collider, there are different types of beam-beam effects that impact its performance.

The incoherent beam-beam effects influence the beam lifetime and dynamic aperture. The PACMAN effects introduce variations in the interaction between individual bunches, leading to fluctuations in the luminosity. Additionally, coherent beam-beam effects can result in beam oscillations and instabilities.

It is common practice to select a nominal beam-beam parameter to guide the design and operation of the collider. A value of 0.015 for one interaction point or 0.030 for two interaction points nowadays is considered reasonable. This choice is supported by the successful operation of the LHC, which reported stable performance with a total beam-beam parameter value ($DQ_{tot}$) of around 0.03 when having three interaction points [9].

#### 8.2.2.4 *Electron Cloud Effect*

The electron cloud (EC) phenomenon is known to cause beam instability in particle colliders. It occurs when accumulated photo-electrons and secondary electrons interact with the circulating beam, leading to several detrimental effects on the beam dynamics and overall luminosity of the collider. This phenomenon has been a significant challenge in previous colliders such as PEP II, KEKB, LHC, and BEPC.

The EC effect can induce coupled bunch instability, where the motion of subsequent bunches becomes linked, further exacerbating the instability. It can also result in emittance blow-up, which causes an increase in the beam size and reduces the luminosity of the collider [10].

In the case of next-generation super proton colliders like SPPC, with a higher bunch population exceeding $10^{11}$ and a smaller bunch spacing of 25 ns or less, the impact of the EC effect becomes even more critical to achieve the desired luminosity level of $1 \times 10^{35}$ $cm^{-2}s^{-1}$.

Furthermore, the use of low-temperature beam pipes for the superconducting magnets at SPPC presents an additional challenge. The deposition of power on the beam screen from secondary electron multipacting can become a significant issue. In the dipole magnets of the LHC, it has been observed that the deposited power increases exponentially to approximately 10 W/m when the Secondary Electron Yield (SEY) exceeds 1.4. To mitigate this problem, it is necessary to control the SEY at SPPC and ensure it remains below 1.4, or even lower, through the application of coatings such as TiN or NEG (Non-Evaporable Getter) on the internal walls of the beam screen and other devices within the vacuum system.

#### 8.2.2.5 *Beam Loss and Collimation*

Beam losses play a crucial role in ensuring the safe operation of a machine like the SPPC, where the stored beam energy can reach a maximum of 9.1 GJ per beam in the intermediate stage (refer to Section 8.2.6). These losses can be categorized into two classes: irregular and regular.

Irregular beam losses are avoidable and often occur due to beam misalignment or faults in accelerator elements. For instance, a trip in the RF system can lead to a loss of synchronization during acceleration and collisions. Vacuum-related issues also fall under this category. While a well-designed collimator system can collect most of the lost particles, even a fraction of these losses can cause problems elsewhere in the machine as they are distributed.

On the other hand, regular losses are non-avoidable and localized primarily in the collimator system or other aperture limits. They occur continuously during operation and reflect the beam's lifetime and transport efficiency in the accelerator. Various effects, such as the Touschek effect, beam-beam interactions, collisions, transverse and longitudinal diffusion, residual gas scattering, halo scraping, and instabilities, determine the lowest achievable losses.

In high-power or high-stored-energy proton accelerators, there is a risk of halo particles reaching the vacuum chambers and becoming lost. The consequences of such losses can be severe, including triggering quenching of superconducting magnets, generating unacceptable background in detectors, damaging radiation-sensitive devices, and causing residual radioactivity that hinders hands-on maintenance. To mitigate these issues, collimation systems are employed to confine particle losses to specific locations with better shielding and heat-load transfer capabilities.

For proton-proton colliders with high beam energies and very high stored energy, such as the SPPC, the situation becomes even more complex due to the need for extremely high collimation efficiency. In the case of SPPC, where the stored energy in the beam can reach up to 9.1 GJ per beam, the cleaning inefficiency should be lower than $3.0 \times 10^{-6}$ for the same beam loss power. Achieving this level of efficiency necessitates a more advanced and efficient collimation method compared to the one employed in the LHC.

### 8.2.3 **Preliminary Lattice Design**

#### 8.2.3.1 ***General Layout and Lattice Considerations***

Various lattice designs are examined [1,2] for proton-proton colliders, primarily focusing on arc configurations. There are typically three types of arc designs for such colliders, each employing different methods for dispersion suppression. These include full bend suppressors (with the same number of dipoles per cell as in the regular arc cells), half bend suppressors (with reduced dipole strength per cell), and LHC-like suppressors [3] (featuring decreased dipole strength and shorter cell length). The third design is preferred for the baseline configuration due to its compatibility with an electron-positron collider in the same tunnel.

#### 8.2.3.2 ***Arcs***

The arcs consist of conventional FODO cells, with each FODO cell to have a basic phase advance of 90 degrees in both transverse planes. It's important to note that larger beta functions and dispersions result in larger magnet apertures, which can significantly impact magnet costs. Additionally, the dispersion function plays a crucial role in the design of momentum collimation.

Fig. 8.2.1 illustrates the magnet arrangement and beam optics within a standard FODO cell in the current design. In this configuration, the horizontal beta function and dispersion function reach their maximum values at the midpoint quadrupole. Within each cell, there are 12 bending magnets, each measuring 14.45 meters in length and possessing a strength of 20 Tesla. The quadrupole length is 6 meters. The total length of a standard FODO cell spans 213.4 meters, and each arc comprises a total of 44 such cells.

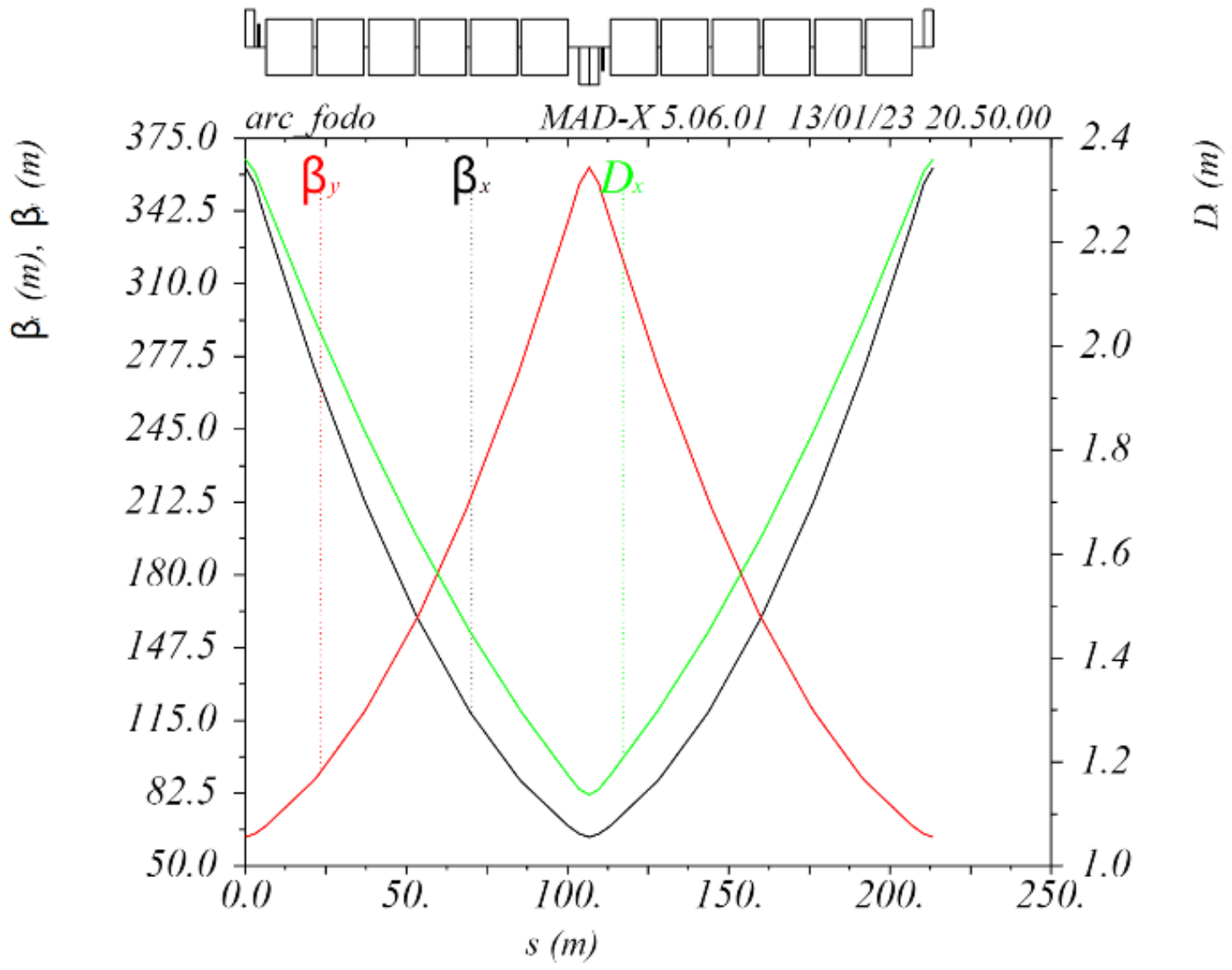

**Figure 8.2.1:** Lattice functions of a standard arc FODO cell

#### 8.2.3.3 *Dispersion Suppressor*

The dispersion suppressor (DS) aligns the dispersion functions and beta function between the arc section and the contiguous long straight section. In addition to achieving optical alignment, the SPPC dispersion suppressor is engineered with the capability to make slight adjustments to the SPPC layout to fulfill compatibility requirements with CEPC. Fig. 8.2.2 illustrates the left portion of the dispersion suppressor, with the right portion exhibiting similar characteristics. It consists of two FODO cells, each shorter than a standard arc cell, containing only four dipoles or left empty in each half-cell. The first half of the DS cell features a flexible drift space, ranging from 30 meters to 80 meters. The last half-cell remains devoid of dipoles to simplify betatron function matching.

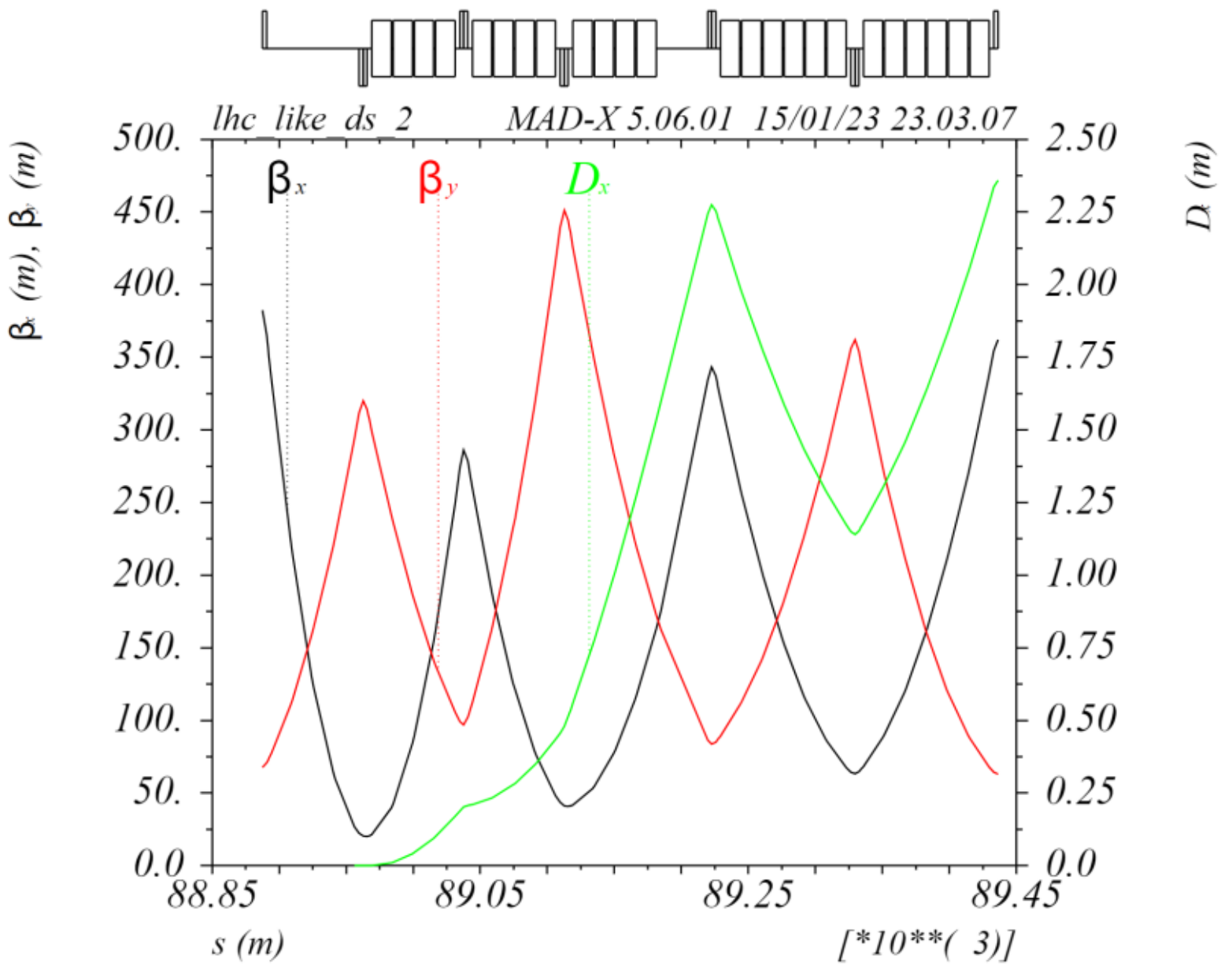

**Figure 8.2.2:** Left portion of the dispersion suppressor

### 8.2.3.4 *High Luminosity Insertions*

The long straight sections IP2 and IP4 are dedicated to proton-proton (p-p) collisions. The lattice configurations for IP2 and IP4 are identical and exhibit anti-symmetry, facilitating the generation of crossing angles at the interaction points (IPs). This configuration enables the beams to transition from the outer ring to the inner ring, or vice versa. Preliminary lattice designs for both IP2 and IP4 are depicted in Fig. 8.2.3.

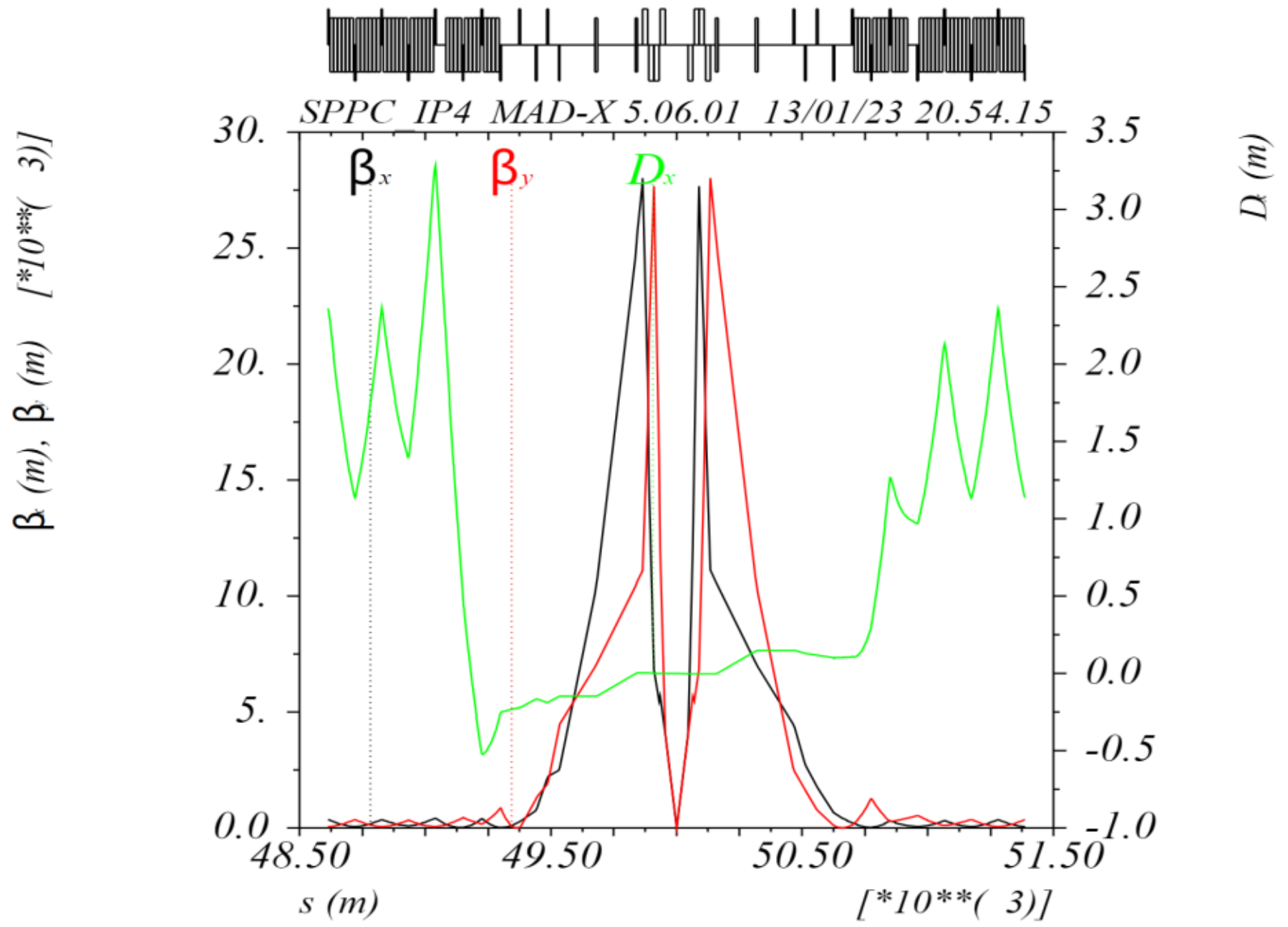

**Figure 8.2.3:** Lattice function in IP2 and IP4

### 8.2.3.5 *Other Stright Sections*

The collimation sections in the bypass of IP1 and IP3 contain both the betatron and momentum collimation systems. The lattice design is shown in Fig. 8.2.4. More details will be given in Section 8.2.5.

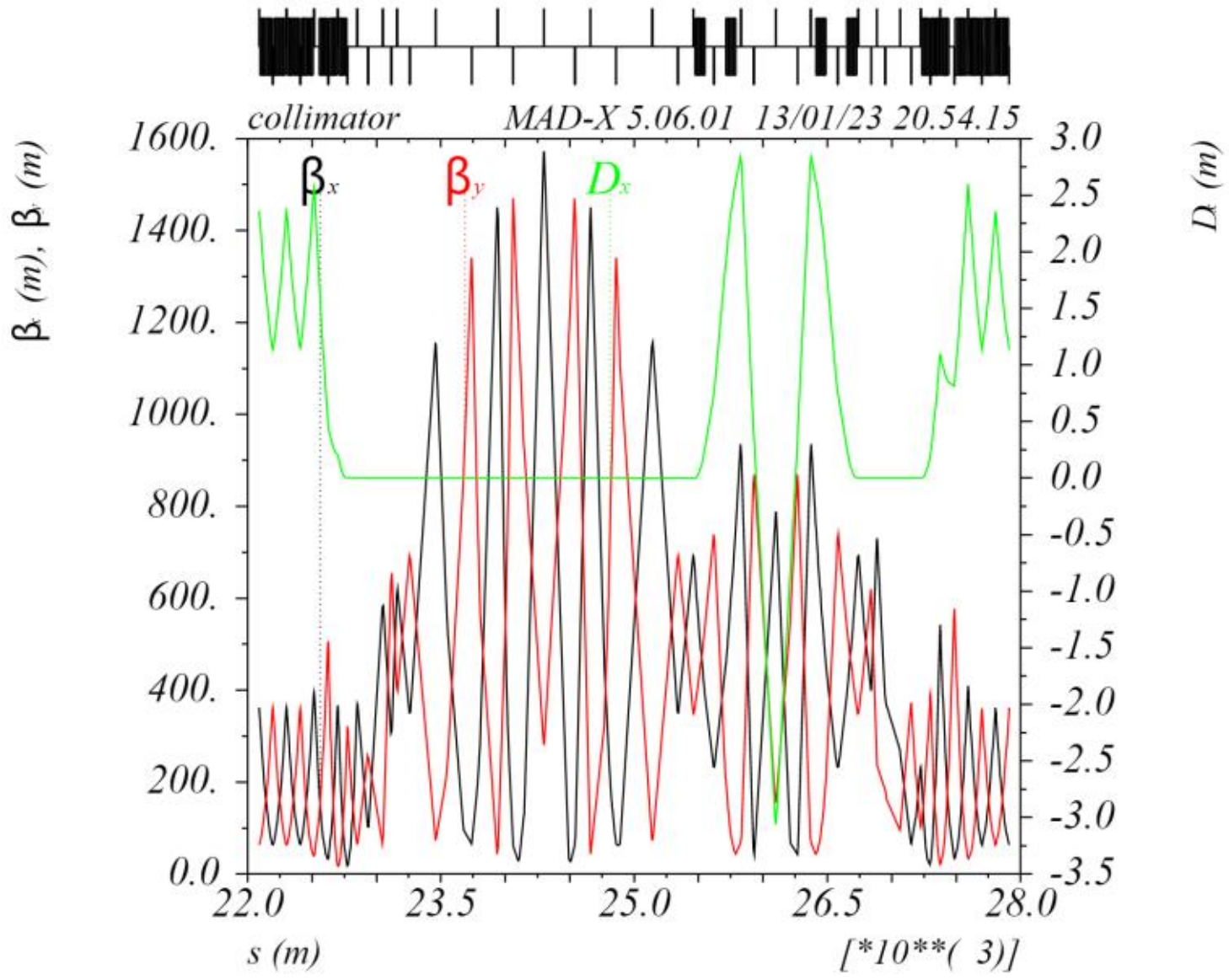

**Figure 8.2.4:** Lattice function at the IP1 bypass

Apart from the arcs, high-luminosity insertions, and the collimation sections, lattice designs have not been completed for the remaining straight sections, which encompass injection (LS4), extraction (LS2), RF stations (LS1), and a reserved straight section (LS3). To close the ring temporarily, these sections are substituted with a standard simple FODO lattice, as illustrated in Fig. 8.2.5.

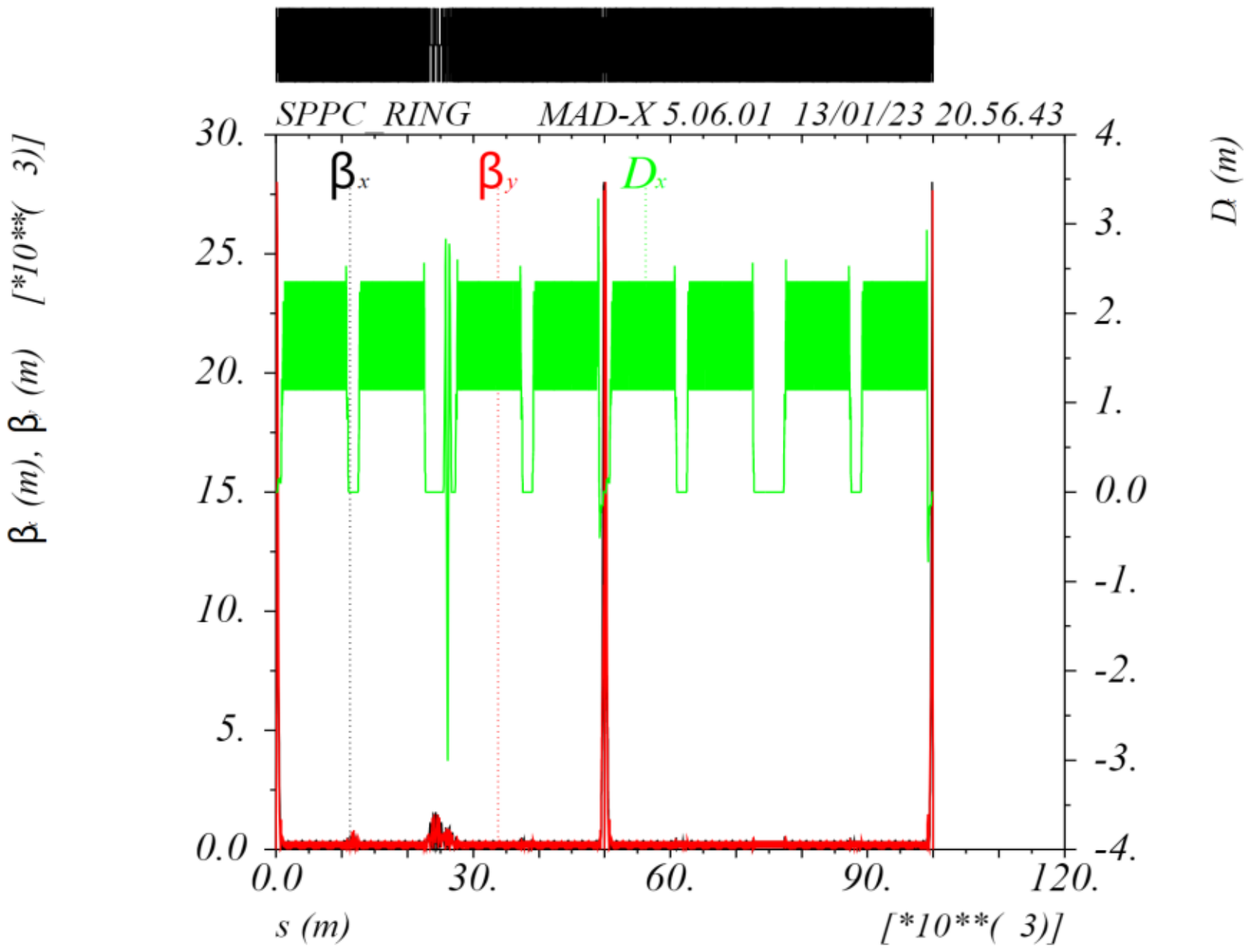

**Figure 8.2.5:** Lattice function of other straight sections.

### 8.2.3.6 *Nonlinearity Optimization and Dynamic Aperture*

The chromaticity in the SPPC collider ring primarily arises from the powerful final focusing quadrupoles and a significant number of arc quadrupoles. While arc sextupoles are employed to address chromaticity issues across the entire ring, they also introduce tune shifts, acting as resonance driving terms.

An interleaved sextupole scheme, where each arc quadrupole is paired with a sextupole of the same polarity, is under investigation. A periodic arrangement of 4 FODO cells, each with a phase advance of 90°/90°, plays a pivotal role in canceling out key nonlinearities [4]. Notably, no sextupoles are present in the quasi-FODO cells within the dispersion suppressors. To counteract the momentum-dependent tune shifts caused by the sextupoles, two pairs of octupoles per arc are employed [5]. Adjustments to the phase advances between different sections are made to reduce the chromatic amplitude function (W-functions) across the entire ring, with particular focus on the IPs. The chosen working point, (0.12, 0.13), is carefully selected to mitigate beam-beam effects.

Dynamic aperture calculations are conducted using SIXTRACK simulations [6]. In Fig. 8.2.6, the dynamic apertures are depicted, accounting for nonlinear magnets but excluding field errors in the main magnets. Off-momentum dynamic apertures are simulated with a relative momentum deviation of 0.05%, exceeding the momentum acceptance limit imposed by synchrotron motion.

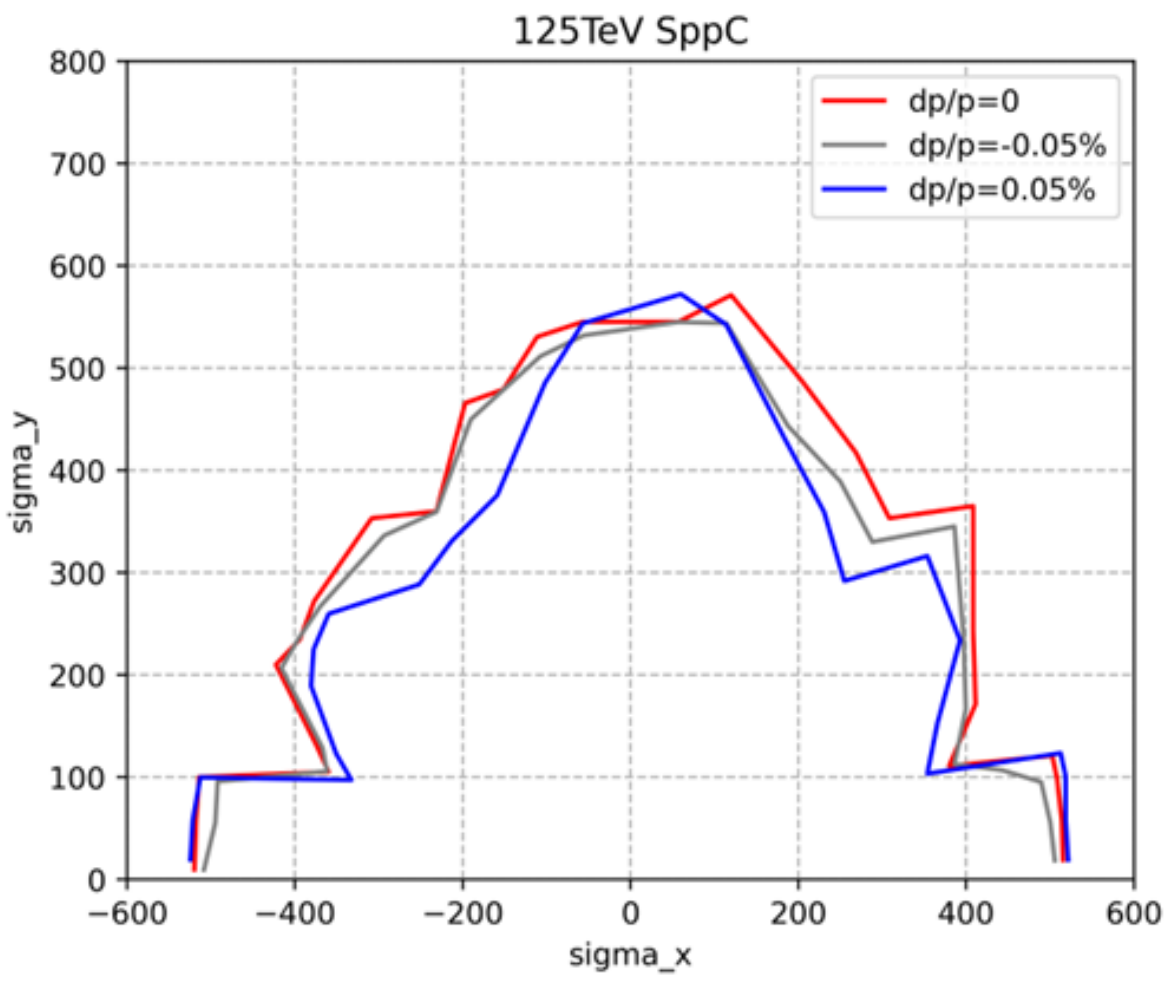

**Figure 8.2.6:** Dynamic aperture for the on- and off-momentum particles.

### 8.2.3.7 *Magnet Gradients and Strengths*

The magnet gradients and strengths derived from the lattice design are detailed in Table 8.2.2. Magnet aperture diameters are categorized into two groups: (1) for insertions or matching sections, with D = 60 mm; (2) for the arcs, with D = 45 mm. It's noteworthy that the pole-tip fields of both dipole and quadrupole magnets remain within or close to the 20 T limit.

**Table 8.2.2:** Gradients and strengths of magnets in the SPPC lattice

| Magnet type | Length | Gradient | Strength of $B_{pole}$ |
|---|---|---|---|
| Main dipole ($\phi$ = 45mm) | 14.45192 m | | 20 T |
| Main quadrupole ($\phi$ = 45mm) | 6 m | 477 $Tm^{-1}$ | 10.74 T |
| Strongest quadrupole ($\phi$ = 60mm) | 6 m | 676 $Tm^{-1}$ | 20.28 T |
| Sextupole SF/SD ($\phi$ = 45mm) | 0.5 m | 6711 $Tm^{-2}$/13588 $Tm^{-2}$ | 1.70 T / 3.44 T |
| Octupole OC1/OC2 ($\phi$ = 60mm) | thin-lens | 254371 $Tm^{-3}$/279808 $Tm^{-3}$ | 1.14 T / 1.26 T |

### 8.2.3.8 *Error Correction*

Magnet misalignments and field errors can introduce distortions into the closed orbit and optics of the system. Table 8.2.3 provides a comprehensive overview of the misalignment errors and field errors considered for the simulations. The correction process encompasses several essential steps: commencing with closed orbit correction, with sextupoles deactivated to eliminate the feed-down effect from strong sextupoles, followed by beta function and dispersion correction with sextupoles activated, and culminating in betatron coupling correction. These optics corrections are executed using the Accelerator Toolbox [7], employing the response matrix fit methodology.

Following the correction procedures, 31 out of the initial 34 seeds remain viable. Fig. 8.2.7 presents a visual representation of the corrected closed orbit, dispersion, beta beating, and coupling. Ongoing work includes a dynamic aperture study under the influence of these errors.

**Table 8.2.3:** Error tolerances of SPPC elements (reference radius of the field error is 17 mm)

| Element | Error | Error desc. | Units | Main dipole | Separation dipole |
|---|---|---|---|---|---|
| Dipole | $\sigma(\psi)$ | roll angle | mrad | 0.5 | 1 |
| | $\sigma(\delta B/B)$ | random b1 | % | 0.1 | 0.05 |
| | $\sigma(\delta B/B)$ | random b2 | $10^{-4}$ units | 0.92 | 0.1/1.1 |
| | $\sigma(\delta B/B)$ | random a2 | $10^{-4}$ units | 1.04 | 0.1/0.2 |
| | $\sigma(\delta B/B)$ | uncert. a2 | $10^{-4}$ units | 1.04 | TBD |
| | | | | | |
| | | | | Main quadrupole | IR triplet / other |
| Quad. | $\sigma(x), \sigma(y)$ | | mm | 0.5 | 0.2/0.5 |
| | $\sigma(\psi)$ | roll angle | mrad | 1 | TBD/0.5 |
| | $\sigma(\delta B/B)$ | random b2 | % | 0.1 | TBD/0.05 |
| | | | | | |
| BPM | $\sigma(x), \sigma(y)$ | | mm | 0.3 | 0.3 |
| | $\sigma(\text{read})$ | | mm | 0.2 | 0.05/0.2 |

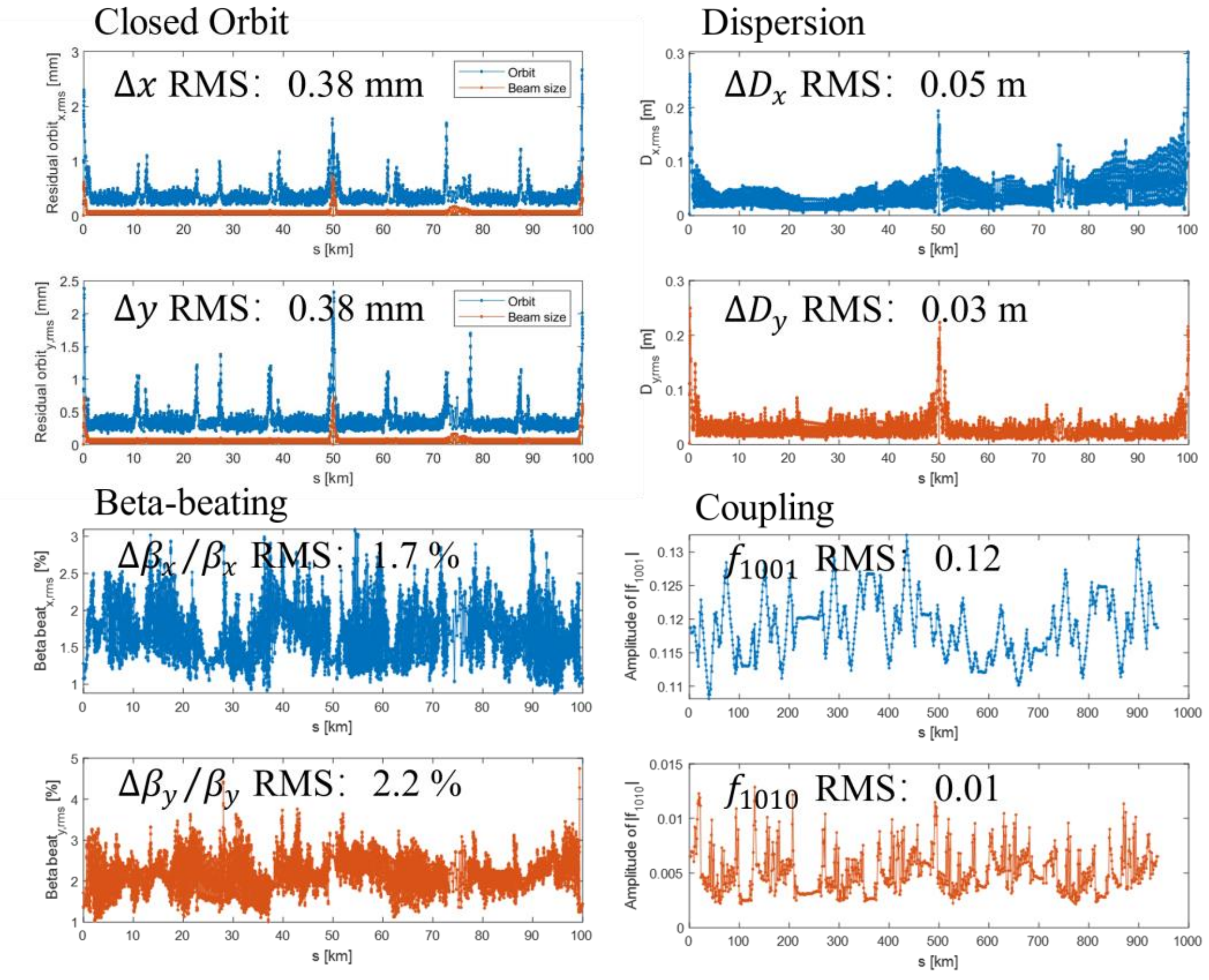

**Figure 8.2.7**: The closed orbit, dispersion, beta beating and coupling after error correction for the SPPC collider ring.

## 8.2.4 Luminosity and Leveling

The initial luminosity of $4.3 \times 10^{34}$ cm$^{-2}$s$^{-1}$ for the next-generation proton-proton collider may seem modest compared to other colliders. It is lower than the HL-LHC [13] while comparable to the FCC-hh [5, 11-12]. However, this design allows for the instantaneous luminosity to reach up to $1.3 \times 10^{35}$ cm$^{-2}$s$^{-1}$ during the collision process, with the potential for even higher luminosity in future upgrades.

In addition to the synchrotron radiation power limiting the circulation current and luminosity, the number of interactions per bunch crossing also poses a constraint on the achievable luminosity. Ongoing research and development efforts on detectors, along with general technological advancements, are expected to address the challenge of data pile-up. However, it remains important to increase the average and integrated luminosity while maintaining the maximum instantaneous luminosity [14,15]. To achieve this, various luminosity leveling schemes can be employed, as illustrated in Figure 8.2.8, taking into account factors such as the loss of stored protons from collisions, cycle turnaround time, shrinkage of transverse emittance due to synchrotron radiation, and beam-beam shift.

To control the shrinkage of emittance, an emittance blow-up system is necessary. Another method to enhance luminosity is by adjusting β* during collisions, utilizing the benefits of emittance shrinkage while keeping the beam-beam tune shift constant.

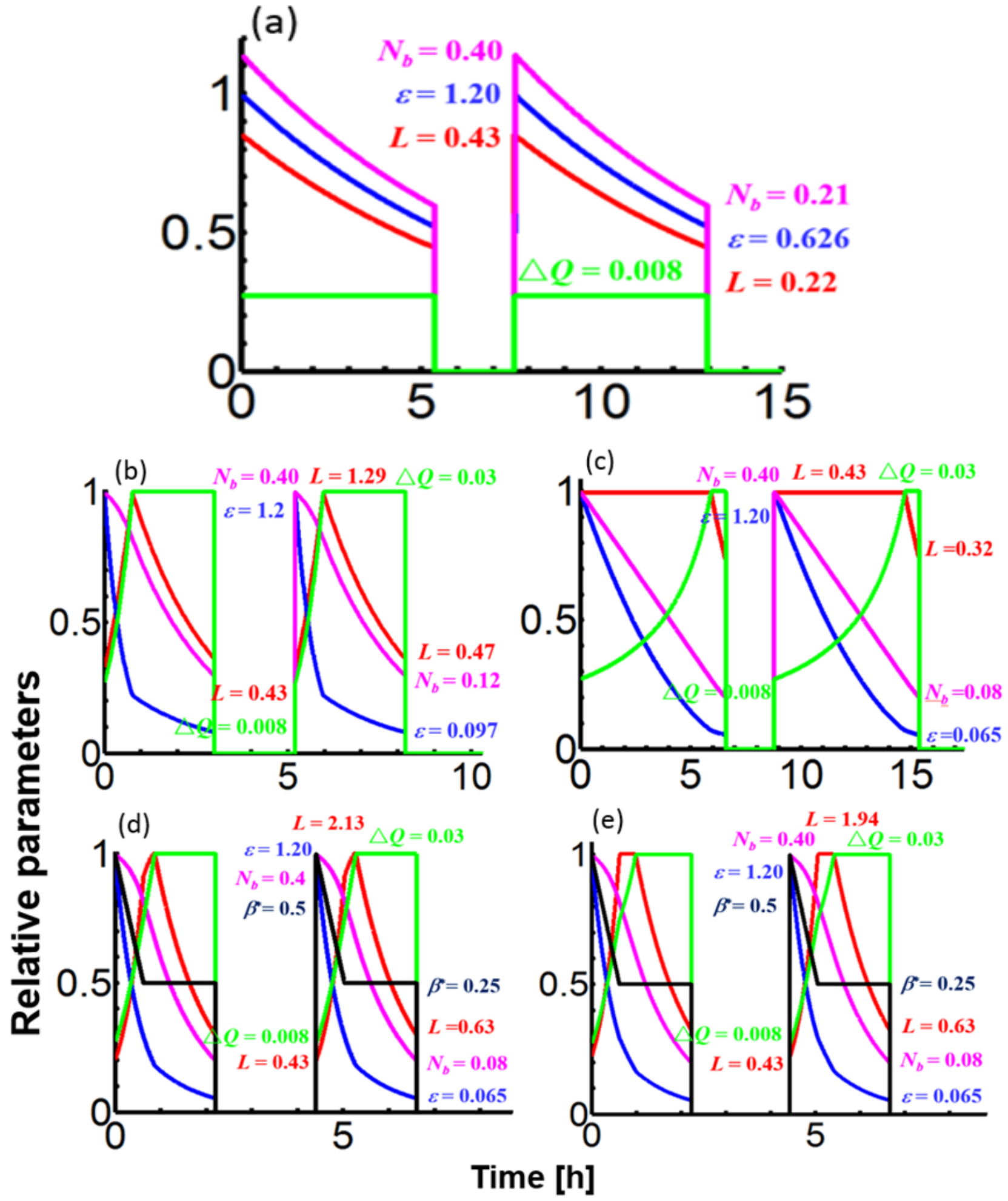

**Figure 8.2.8:** Evolution of parameters vs time with a turnaround time of 2.3 hours and bunch spacing of 25 ns. Red: luminosity, magenta: number of protons per bunch, blue: transverse emittance, green: beam-beam tune shift, black: beta* at the IP. (a) keeping the tune shift constant; (b) allowing the tune shift to rise to 0.03; (c) keeping the instantaneous luminosity constant; (d) dynamically changing the beta* down to 0.25 m; (e) limiting the peak luminosity below 1.94. In plots a), b) and c), beta* is kept constant at the nominal 0.5 m.

### 8.2.5 Collimation Design

The proposed collimation method for the SPPC involves combining the betatron and momentum collimation systems within the same insertion. This arrangement aims to address the issue of particles experiencing significant energy loss in the transverse collimators due to the Single Diffractive effect (SDE) [16]. By employing the momentum collimation system, particles that would otherwise be lost in the downstream cold region can be effectively cleaned, resulting in a substantial improvement in cleaning efficiency. Our studies indicate that this method is highly efficient in resolving this issue.

Unlike the momentum collimation section at the LHC [17-18], where intentional non-zero dispersion is designed between adjacent DS (dispersion suppressor) sections, the joint between the momentum collimation and transverse collimation sections at SPPC requires an achromatic end. This design is crucial to achieve the desired cleaning inefficiency of only $3.0 \times 10^{-6}$. To implement this approach, a five-stage collimation system is necessary for both betatron and momentum collimations. In order to provide the required phase advances within the collimation section, both dipoles and quadrupoles utilized in the system are superconducting magnets, deviating from the conventional use of warm magnets. Additionally, ensuring adequate protection from beam loss and radiation for the cold dipole and quadrupole magnets is indispensable [19]. The layout of the entire cleaning insertion is depicted in Figure 8.2.9, while Figure 8.2.10 illustrates the betatron and dispersive functions, along with the loss map within the collimation section. Relevant parameters are listed in Table 8.2.4. It is important to note that in order to align with the CEPC layout, the existing design, which necessitates a lengthy straight section of 4.3 km, should be adjusted to a length of 3.3 km. This modification is deemed achievable by implementing higher magnetic fields for the superconducting quadrupoles employed in that section.

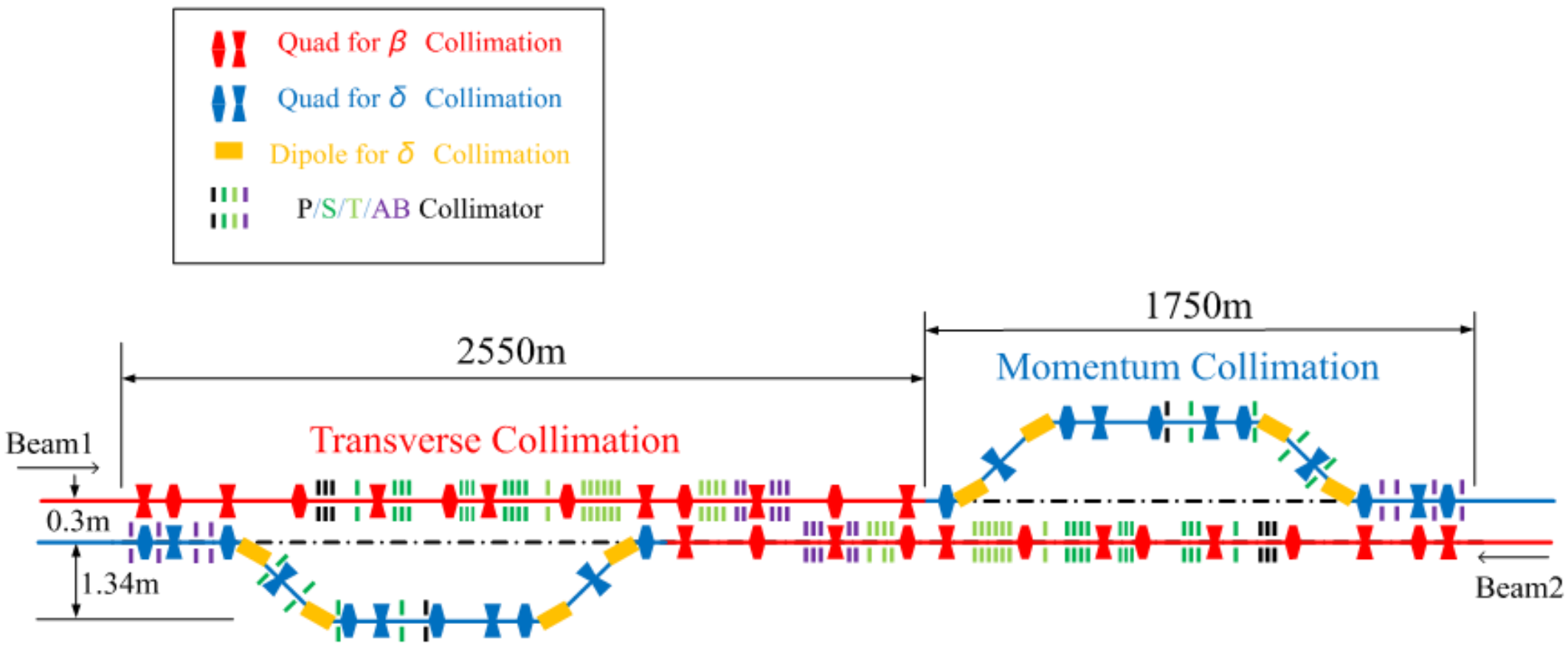

**Figure 8.2.9:** Layout of the collimation insertion. P/S/T/AB denote primary collimator, secondary collimator, tertiary collimator and absorber (to be modified for a shorter straight section of 3.3 km).

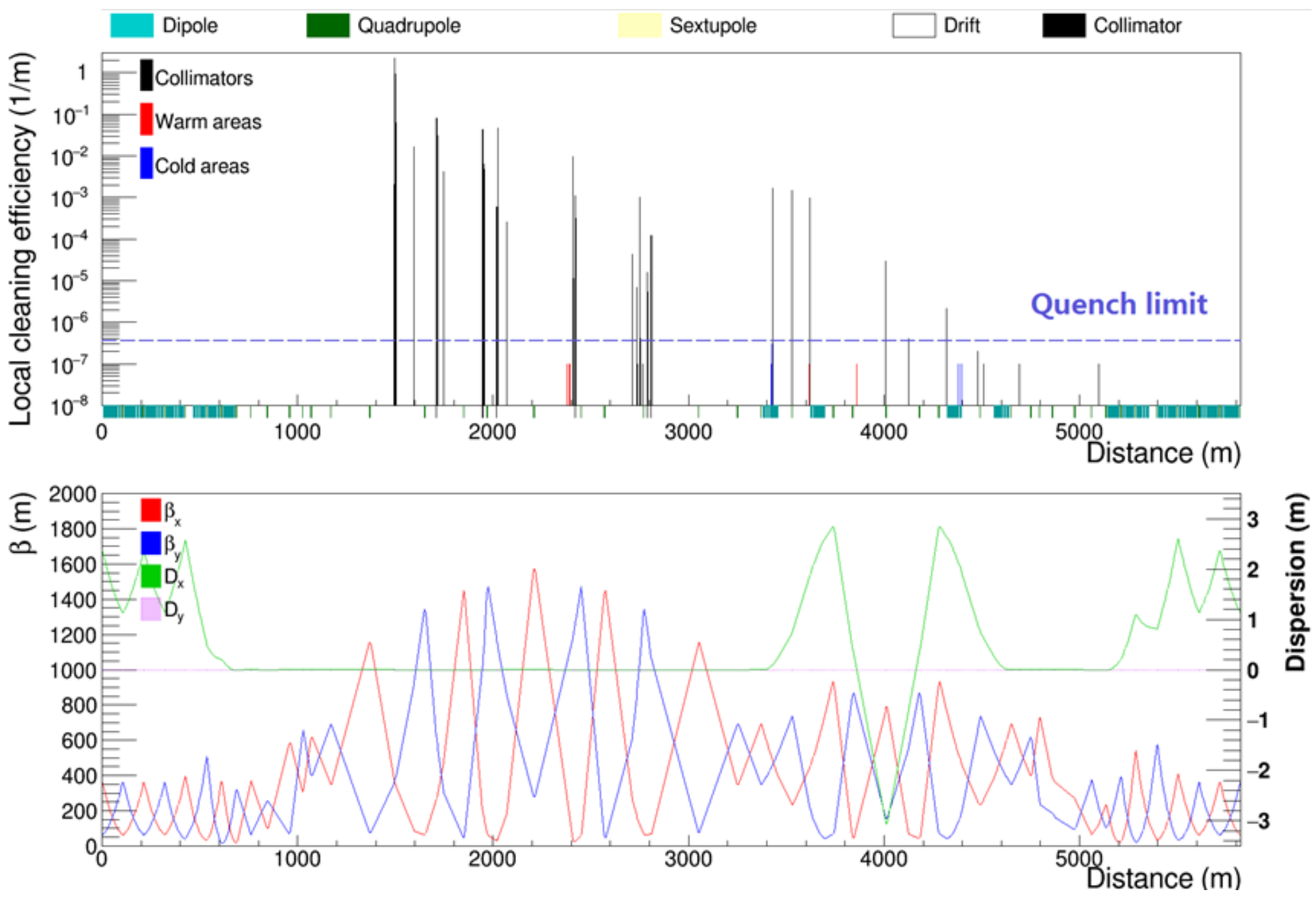

**Figure 8.2.10:** Proton loss map in the collimation insertion with an initial vertical halo distribution (upper) and the lattice (lower) (to be modified for a shorter straight section of 3.3 km).

**Table 8.2.4:** Basic parameters of the collimation insertion at the SPPC

| Parameter | Value | Unit |
|---|---|---|
| Total cleaning insertion length | 4.3(*) | km |
| Length of betatron / momentum collimation section | 2.55 / 1.75 | km |
| Horizontal phase advance of betatron / momentum | 3.54π /2.27π | rad |
| Warm/cold quadrupole length | 6.0 | m |
| Maximum quadrupole strength | 0.00022 | $m^{-2}$ |
| Dipole length in the momentum cleaning section | 14.45 | m |
| Number of dipoles per group | 4 | |
| Number of dipole groups | 4 | |
| Dipole field | 20 | T |
| Maximum beta function ($\beta_x$ / $\beta_y$) | $1.59/1.47\times10^3$ | m |

(*) To be modified for a shorter straight section of 3.3 km.

Multi-particle simulations have been performed using the Merlin code, utilizing the lattice parameters and collimator settings. As an initial approach for the betatron collimator settings, similar physical gaps and phase advances as those used in the LHC were employed to verify the effectiveness of the novel collimation method. In order to enhance simulation efficiency with a large number of particles ($10^8$), a ring-type distribution was chosen for the horizontal plane, while a Gaussian distribution was utilized for the vertical plane. Based on the positions of the lost particles, three protective collimators made of Tungsten, with apertures matching that of the primary momentum collimator, were strategically placed to intercept these particles. The maximum energy spread of particles that can pass through the primary collimator was found to be approximately 0.1%.

The results depicted in Figure 8.2.10 illustrate that the proposed collimation method exhibits an extremely low cleaning inefficiency in the downstream DS regions, achieving a level better than $5 \times 10^{-7}$, thereby meeting the requirements of SPPC.

### 8.2.6 Cryogenic Vacuum and Beam Screen

#### 8.2.6.1 *Vacuum*

SPPC features three distinct vacuum systems to facilitate its operations: an insulation vacuum for the cryogenic system, a beam vacuum for the low-temperature sections, and another beam vacuum for the chambers in the room-temperature sections.

#### *8.2.6.1.1 Insulation Vacuum*

The primary objective of this vacuum system is to prevent convective heat transfer rather than achieving high vacuum levels. Therefore, there is no requirement for extremely low pressures. Prior to the cool-down process, the room-temperature pressure within the cryostats only needs to be better than 10 Pa. Once cooled, the pressure will naturally stabilize around $10^{-4}$ Pa, assuming there are no significant leaks.

Given the substantial volume of insulation vacuum required for SPPC, cost-effective design considerations are crucial.

#### *8.2.6.1.2 Vacuum in Cold Sections*

Principle of HTS magnets allows for higher cold bore temperatures, exceeding 4 K, compared to LTS magnets, thereby reducing the cost of the extensive cryogenic system. However, achieving a very high vacuum level is crucial to limit beam loss and maintain beam quality, as the pumping speed of hydrogen gas is strongly influenced by temperature. Currently, we are evaluating two temperature options: either sticking to the conventional temperature of 1.9 K used at LHC or exploring a more aggressive approach at 3.8 K. In the latter case, the use of auxiliary pumping systems, such as cryosorbers employed at LHC, becomes necessary [1-2]. Further research and analysis are currently underway to make a decision.

In regions where superconducting quadrupoles are utilized, such as interaction regions or areas surrounding experiments, it is essential to maintain a very high-quality vacuum (less than $10^{13}$ $H_2$ per $m^3$) to prevent background interference in the detectors. However, in these areas, the beams travel in straight lines, resulting in relatively low synchrotron radiation levels.

In the arcs of SPPC, the requirement for vacuum is primarily driven by the desired beam lifetime, which is influenced by nuclear scattering of protons on the residual gas. To achieve a beam lifetime of approximately 100 hours, it is necessary to maintain an equivalent hydrogen gas density below $10^{15}$ $H_2$ per $m^3$. However, a significant challenge arises due to the substantial synchrotron radiation power generated in these regions.

Directly exposing the magnet bore, operating at a temperature of 3.8 K (or 1.9 K), to the synchrotron radiation would result in an excessively high power load that needs to be dissipated. To address this issue, a beam screen is employed, operating at a higher temperature range, typically around 40-60 K. The beam screen is positioned between the beam and the cold bore. Its purpose is to intercept the synchrotron radiation, preventing it from directly impacting the magnet bore. However, at such temperatures, the beam screen tends to desorb hydrogen gas, particularly when subjected to the intense synchrotron radiation. On the other hand, the space outside the beam screen is cryopumped due to the low temperature of the bore. Slots are introduced in the shield to facilitate pumping of the beam space, ensuring effective gas removal.

#### *8.2.6.1.3 Vacuum in Warm Sections*

The warm regions of SPPC serve as the housing for the beam collimation, injection, and extraction systems. To mitigate the risk of superconductor quenching caused by inevitable beam losses in these areas, warm magnets are employed instead of superconducting magnets. However, the vacuum pumping requirements in these regions are challenging due to desorption resulting from beam losses.

In order to maintain the required vacuum levels, the implementation of Non-Evaporable Getter (NEG) technology is likely necessary. NEG materials can effectively absorb and retain gas molecules, thereby assisting in achieving the desired vacuum conditions. Fortunately, these warm sections are typically shorter in overall length or more limited in extent compared to the cold sections of the accelerator. As a result, any issues arising from the vacuum pumping requirements can be more easily managed and controlled within these specific regions.

### 8.2.6.2 *Beam Screen*

The primary role of a beam screen is to provide shielding for the cold bore of the superconducting magnets against the intense synchrotron radiation (SR) [3]. Due to the exceptionally high beam energy and magnetic field strength in the arc dipoles of SPPC, the synchrotron radiation generated is significantly stronger compared to that of the LHC. The estimated power of synchrotron radiation in the arc dipoles at SPPC is approximately 26.3 W/m per aperture, considerably higher than the 0.22 W/m at LHC [4]. This heightened SR power level poses significant challenges for the design of the beam screen.

The design of the beam screen at SPPC necessitates finding a compromise that effectively addresses multiple considerations. This includes managing the heat load extraction, minimizing obstruction to the bore aperture, providing vacuum pumping capability, reducing coupling impedance, mitigating the electron cloud effect, and more. An ideal beam screen design might involve separating the functions of the screen itself, as explored in studies conducted for FCC-hh and SPPC [5]. The beam screen could encompass the beam while having a slot on the outer side, operating at a relatively lower temperature to control impedance. Simultaneously, the absorption structures through which synchrotron radiation passes via the slot would operate at a higher temperature to minimize the wall power required for extracting synchrotron radiation power.

The operating temperature of the beam screen needs to strike a balance. It must be sufficiently high to prevent excessive wall power requirements for heat removal, while avoiding excessively high resistivity in the high-temperature superconducting material or copper coating on its internal surfaces. This ensures that impedance remains within acceptable limits and minimizes the risk of excessive power radiation onto the 3.8 K bore.

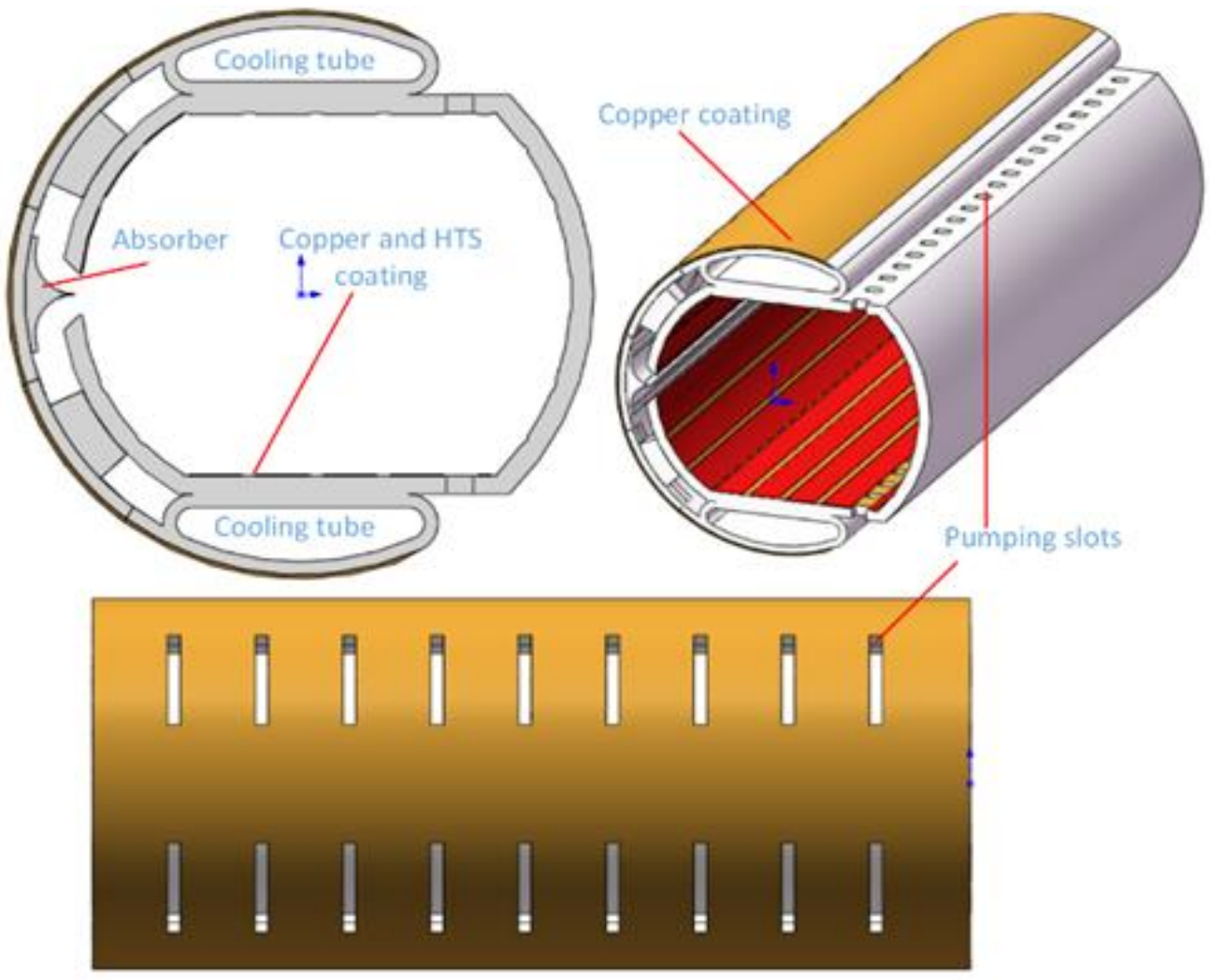

**Figure 8.2.11:** Schematic for the beam screens with inner HTS coating [5].

The design of the beam screen at SPPC must successfully address various requirements, including vacuum stability, mechanical support, impact on beam dynamics, and refrigeration power. To meet these demands, a conceptual schematic for a basic beam screen, akin to the idea developed for FCC-hh, is under consideration, as depicted in Fig. 8.2.11.

The beam screen faces several key challenges at SPPC, which include:

1. Synchrotron Radiation:
   The deposition of synchrotron radiation power in the main dipoles of SPPC is approximately sixty times higher than that at LHC. Absorbing this power directly at 3.8 K would impose an excessive and costly cryogenic load. Therefore, a beam screen positioned between the beam and the cold bore is crucial. The operating temperature of the beam screen should be chosen to strike a balance between wall power economy, technical feasibility, and impedance control. Coating the inner surface of the screen with a layer of high-temperature superconducting material (e.g., IBS-HTS, YBCO) or copper helps reduce resistive impedance. However, higher temperatures can increase this impedance due to higher electrical resistivity, leading to a potential increase in collective beam instability. The operating temperature also needs to be carefully selected to minimize heat radiation and conduction to the cold bore, while considering desorption. Different refrigerants, such as liquid neon or liquid oxygen, can be explored.
2. Electron Cloud:
   A well-designed beam screen structure can mitigate the generation of photo-electrons that contribute to electron cloud formation. The proposed design from FCC-hh incorporates a slit in the outer mid-plane of the screen and curved interception surfaces that redirect synchrotron radiation into confined absorption structures, where photo-desorption is not problematic. However, this design requires more aperture space and adds complexity to the mechanical structure. Even without direct exposure to synchrotron radiation, the inner surface of the screen should be coated with a thin film of low secondary electron yield to minimize electron production.
3. Vacuum:
   Achieving and maintaining the appropriate vacuum within the beam screen depends on factors such as beam structure, beam energy, beam population, critical photon energy, and synchrotron radiation power. The beam structure significantly affects the buildup of electron and ion clouds, which can lead to vacuum instability. The pumping speed plays a crucial role in vacuum stability. The beam screen must be designed to allow effective pumping by ensuring sufficient transparency to the cold vacuum duct. However, increasing transparency by adding more slots can also increase resistive impedance, potentially causing beam instabilities.
4. Magnet Quenches:
   The beam screen should possess sufficient strength to withstand the pulsed electromagnetic forces generated during a superconducting magnet quench event [6]. The use of stainless steel as the base structure material can help reduce such forces. However, applying a thick copper film (around 75 µm) to decrease wall impedance can create a strong source of electromagnetic force. Thinner films can reduce the force but result in higher resistive impedance. Coating the beam screen with an HTS film appears to be a promising solution, albeit with increased technical complexity.

5. Impedance:
   Optimizing the shape and size of the beam screen structure, in addition to the inner surface coating, is necessary to minimize transverse wall impedance.

### 8.2.7 Other Technical Challenges

In addition to the two key technologies mentioned earlier, high-field magnets and vacuum/beam screens, several other important technologies must be developed in the next decades to successfully build SPPC. These include:

1. Machine Protection System:
   The machine protection system is crucial for ensuring the safe operation of SPPC. It requires the development of highly efficient collimation techniques to minimize beam losses and protect the accelerator components. Additionally, a reliable beam abort system is essential for safely dumping the enormous energy stored in the circulating beams in the event of a magnet quench or other abnormal operating conditions.
2. Feedback System:
   A complex feedback system is necessary to maintain beam stability. This system plays a critical role in managing emittance blow-up in the main ring, which is important for controlling beam-beam induced instabilities and maintaining a consistent integrated luminosity.
3. Beam Loss Control and Collimation:
   Controlling beam loss and implementing effective collimation techniques pose additional challenges, especially in high-power accelerators within the injector chain. The introduction of a 10 GeV proton RCS (Rapid Cycling Synchrotron) with power levels reaching a few MW is a new and demanding task that requires special attention.
4. Cryogenic System:
   The design and implementation of a massive cryogenic system for magnets, beam screens, and RF cavities is a critical aspect of SPPC. The cryogenic system needs to be carefully considered to ensure efficient and reliable operation.

### 8.2.8 Optional High Luminosity Operation Scenario

The baseline design for the SPPC specifies an energy of 125 TeV and a magnetic field of 20 T. However, it is foreseeable that certain physics studies may require the collider to operate at lower energies but with higher luminosity. Therefore, the CDR design for operating at 75 TeV is kept as an optional operation scenario, which can provide significantly higher luminosity by employing a higher bunch population and circulating beam current. To preserve this operational flexibility, the SPPC is designed to store a higher energy in the beams, which has implications for the collimation design. Additionally, the bunch population from the injector chain is also higher for the optional operation scenario. Table 8.2.5 presents the key parameters for both the baseline design and the optional operation scenario.

**Table 8.2.5:** Main parameters of the baseline dsign and an optional operation scenario

| Parameters | @75 TeV | @125 TeV | Unit |
|---|---|---|---|
| Dipole field | 12 | 20 | T |
| Initial luminosity per IP | $1.0\times10^{35}$ | $4.3\times10^{34}$ | $cm^{-2}s^{-1}$ |
| Annual integrated luminosity | 1.24 | 0.65 | $ab^{-1}$ |
| Circulating beam current | 0.73 | 0.19 | A |
| Bunch population | $1.5\times10^{11}$ | $4\times10^{10}$ | |
| Normalized emittance | 2.4 | 1.2 | mm-mrad |
| Stored energy per beam | 9.1 | 4.0 | GJ |
| SR power per beam | 1.1 | 2.2 | MW |
| SR heat load at arc per aperture | 12.8 | 26.3 | W/m |

## 8.3 High-field Superconducting Magnet

To explore new physics beyond the standard model, it is crucial to push the energy level of high-energy particle colliders to the highest possible levels. Currently, the highest collision energy achieved is 14 TeV by the LHC [1-2]. However, in order to further increase the energy, it is necessary to develop accelerator magnets that can provide a higher magnetic field.

Recognizing this need, the Enhanced European Collaboration Project (EuCARD-2) was initiated by the European Commission with the specific goal of conducting research and development for particle accelerators, focusing on the development of high-field magnets [3]. Similarly, the U.S. Department of Energy established the Magnet Research and Development Program (MDP) in 2017 [4]. This program is dedicated to developing high-field magnets required for next-generation particle accelerators.

As a result of these collaborative efforts, 16 T superconducting dipole magnets have been developed for the pre-research phase of the FCC [5]. Additionally, in China, the IHEP and its collaborators are actively engaged in the research and development of next-generation 12-24 T dipole magnets as part of the pre-study for the SPPC [6-8].

For magnetic fields above 15 T, HTS materials are preferred. The non-uniformly distributed superconducting current, including the varying-field-induced persistent current, is responsible for complex magnetization in HTS high-field coils and magnets. Experimental and numerical studies confirm that the temporal drift and spatial distortion caused by the persistent current can degrade the overall field quality. Considering the persistent current, non-uniformly distributed currents in HTS tapes can significantly affect stress and strain distribution, posing risks to the stable operation of the magnet.

In the case of high-field dipole magnets, it is highly advantageous to wind the coils with a compact large-current cable consisting of tens or even hundreds of superconductors. However, the use of HTS conductors like ReBCO poses challenges due to their large aspect ratio, which makes it difficult to apply the well-established Rutherford cable design to ReBCO. In recent years, several new concepts for ReBCO cables have been proposed [9]. Among them, the Roebel cable is the only compact and fully transposed option that can achieve a current density comparable to that of a single tape [10-11]. However, the high performance of the Roebel cable comes at the cost of using half of the raw ReBCO tapes.

To address this issue, we are currently developing a new large-current cable specifically designed for ReBCO-coated conductors. This innovative cable design involves directly bending ReBCO stacks, enabling the achievement of a very high current density while keeping costs within a reasonable range [12].

Additionally, iron-based superconductors (IBS) have emerged as a rapidly developing alternative. These materials exhibit ultra-high critical fields and have a large capacity for carrying current, while also possessing low anisotropy [13]. In 2016, a significant milestone was achieved when the Institute of Electrical Engineering, Chinese Academy of Sciences (IEE-CAS) successfully manufactured the first 100-meter long 7-filamentary Sr1−xKxFe2As2 (Sr122) IBS tape using the powder-in-tube method. This breakthrough opens up great potential for large-scale applications of IBS [14].

To explore the application of IBS in high-field magnets, the collaboration between the IHEP and the IEE has focused on the fabrication and optimization of IBS tapes and coils.

The project aims to tackle important scientific and technological challenges related to advanced superconducting accelerator magnets. Specifically, the objectives are to develop high-field accelerator magnets surpassing 16 T and explore methods for controlling persistent current in HTS coils. Over the past few years, notable progress has been made, as outlined below:

1. The project successfully developed the first high-field IBS racetrack and solenoid coils. These coils were tested and demonstrated their performance at magnetic fields of up to 10 T and 32 T, respectively. This experimental achievement provides evidence for the feasibility and advantages of utilizing IBS materials in high-field applications.
2. The team also accomplished the development of the first dual-aperture model dipole magnet in China, capable of reaching the 12-T class. Furthermore, work is currently underway to fabricate a 16-T hybrid model dipole magnet.
3. Additionally, the project successfully developed 2.2-meter CCT magnets, for the HL-LHC project.

## 8.3.1 Development of High Field Model Dipole Magnets

### 8.3.1.1 *Development of LPF1 Series Dipoles*

A high-field twin-aperture model dipole magnet called LPF1 (Let the Proton Fly) was successfully designed, fabricated, and tested. The magnet incorporated 4 NbTi racetrack coils on the outer side and 2 $Nb_3Sn$ racetrack coils on the inner side. Its purpose was to generate a 12-T main magnetic field in two apertures at a temperature of 4.2 K. In the initial test, LPF1 achieved a magnetic field of 10.2 T at 4.2 K, which corresponds to 72% of the critical current of the short sample (refer to Fig. 8.3.1).

During this test, most instances of quenching occurred in one of the outermost NbTi coils. The limited performance was attributed to inadequate pre-stress in the magnet. Recognizing this issue, the magnet was reassembled with higher pre-stress and given the new name LPF1-S. In subsequent testing, LPF1-S reached a maximum field of 10.7 T, equivalent to 77% of the critical current of the short sample. However, in contrast to LPF1, the majority of quenching events in LPF1-S originated from one of the $Nb_3Sn$ coils in the peak field region.

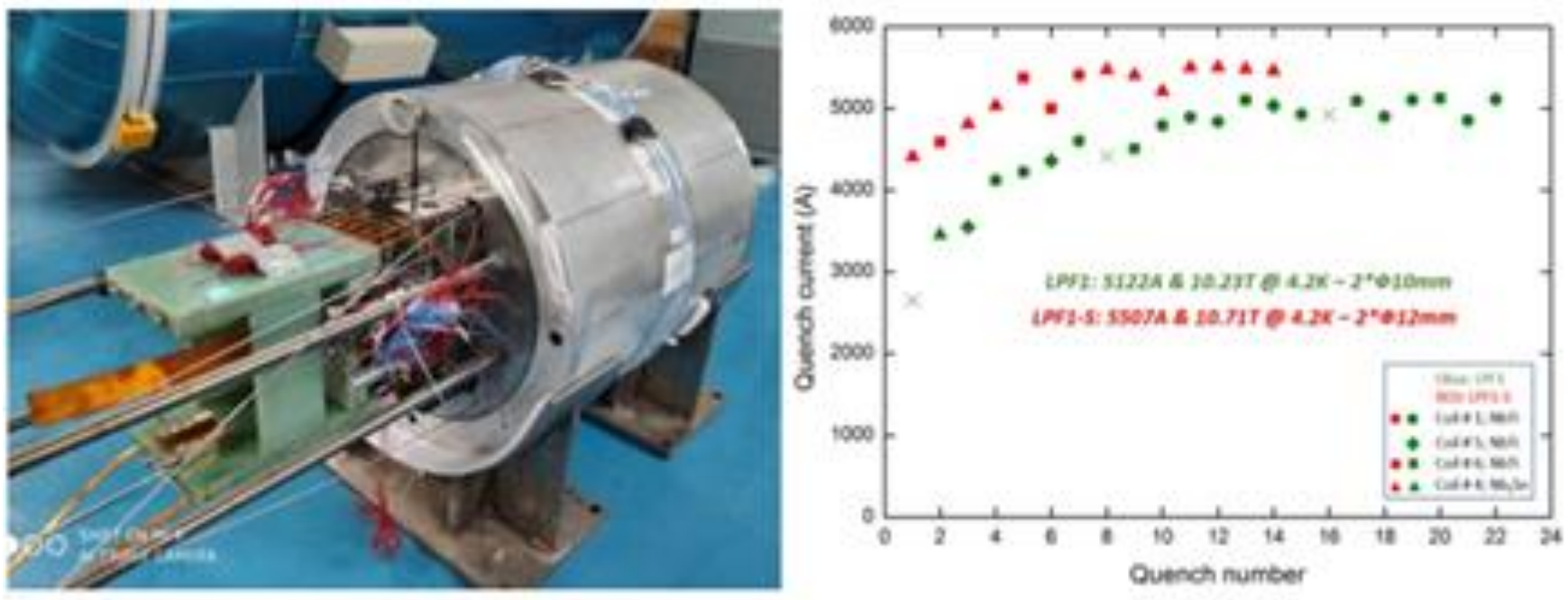

**Figure 8.3.1:** Left: LPF1 model dipole after assembly; Right: Performance of LPF1 / LPF1-S at 4.2 K.

The primary reason for these quenches in LPF1-S was identified as imperfect epoxy impregnation of the $Nb_3Sn$ coil. This flaw in the epoxy impregnation process likely led to localized areas of reduced stability and increased susceptibility to quenching.

A new dipole magnet named LPF1-U has been developed based on the knowledge gained from the previous magnets, LPF1 and LPF1-S. In 2021, LPF1-U successfully achieved a main magnetic field of over 12 T within two 14-mm apertures [15]. Unlike its predecessors, LPF1-U incorporated two newly fabricated $Nb_3Sn$ coils that utilized recently developed Rutherford cables. This collaboration between IHEP and Toly Electric Works Company Ltd. in Wuxi, China facilitated the production of high-quality superconducting Rutherford cables. The newly fabricated cables exhibited no signs of intersection or over-pressing on their surfaces, and the extracted strand tests demonstrated good current-carrying capacity.

Each pancake of the $Nb_3Sn$ coils in LPF1-U consisted of 28 turns of 20-strand conductor wound around an aluminum-bronze island. The total length of the cable used in each coil was approximately 27 m. The cable had a rectangular cross-section measuring 8.6 mm × 1.45 mm, and it featured a 316 L steel sheet (5 mm × 20 um) inserted between the two strand layers to minimize electrical coupling losses during powering.

The outermost two NbTi coils in LPF1-U were also replaced with newly fabricated coils wound with wider NbTi cables. The coil design was re-optimized with longer straight sections and lower load line ratios. Each layer of the NbTi coil consisted of 32 turns of 31-strand cables with a width of 14.3 mm. The middle two NbTi coils were reused from LPF1 and LPF1-S.

The magnetic design of LPF1-U was based on the utilization of a $Nb_3Sn$ strand capable of delivering a critical current density of approximately 2400 A/mm² at 12 T and 4.2 K. For the NbTi strands, the critical current density was 2613 A/mm² and 2816 A/mm² at 5 T and 4.2 K, respectively. All the $Nb_3Sn$ and NbTi strands were supplied by Western Superconducting Technologies Company Ltd. in Xi'an, China. Table 8.3.1 provides detailed parameters of the cables and strands.

The magnetic design of LPF1-U incorporates a graded coil configuration to optimize the performance of both the $Nb_3Sn$ and NbTi coils in different field regions. The total width of the coils is increased to 83.1 mm, which is 8.6 mm wider compared to LPF1. The coil height remains unchanged at 57.6 mm. To allow for the insertion and testing of ReBCO and IBS coils under high field and stress conditions, the gap between the left and right coils is expanded to 14 mm. This provides sufficient space for the additional coils while maintaining the desired magnetic field configuration.

LPF1-U is specifically designed to achieve a bore field of 12 T within the two apertures, with an operating margin of approximately 13% at 4.2 K. This corresponds to an operating current of 6575 A. The magnet exhibits a stored energy of 545 kJ/m and an inductance of 24.61 mH/m at the nominal current. The Lorentz force experienced in each quadrant is 4.96 MN/m in the horizontal direction and 0.57 MN/m in the vertical direction. These forces tend to exert a separating effect, pushing the left and right coils apart and away from the winding islands. Table 8.3.2 provides detailed specifications of the LPF1-U magnet.

The LPF1-U magnet adopted most of the mechanical features from the LPF1-S magnet and retained the bladder and key concept. The shell and yoke were reused, along with the vertical and horizontal pads, albeit thinned to accommodate the broader coil pack of LPF1-U. To facilitate the insertion of HTS (ReBCO and IBS) coils with balloon ends, a 14-mm-thick SSL plate was manufactured and positioned between the inner two $Nb_3Sn$ coils. This SSL plate featured two rectangular channels (12 mm × 25 mm) on the upper and lower sides to accommodate the HTS coils. Additionally, the SSL plate contained two through bores along the axial direction with a diameter of 10 mm for magnetic field measurement purposes.

To assess the mechanical behavior of the magnet, a 3D finite element analysis (FEA) model was developed using the ANSYS code. The FEA results indicated that during assembly at room temperature, the pressurization of the bladders with 33 MPa could create a 0.2 mm gap between the yoke and the nominal keys. This gap would be subsequently filled by interference keys. As the magnet cooled down, the shrinkage of the shell would increase the lateral preload on the coil package. The increased pre-stress would be sufficient to establish contact between the coils and the SSL plate, thereby preventing any motion of the coils. The peak stress experienced during the assembly, cooling down, and excitation to the nominal current would remain below 110 MPa, ensuring the structural integrity of the magnet.

A total of four powering test runs were conducted on the LPF1-U magnet using a 13 KA power supply at IHEP. During the first run, the magnet was energized to 8.5 T, corresponding to a current of 4400 A. Subsequently, the inserted ReBCO coil underwent preliminary testing under the 8.5 T magnetic field. However, the experimental data revealed a significant increase in resistance in the ReBCO coil when the operating current reached only about 30% of Iss.

The reason behind this drastic degradation was attributed to an undesirable support structure between the HTS (ReBCO) and LTS ($Nb_3Sn$) coils. This support structure led to stress concentration at certain parts of the HTS and central $Nb_3Sn$ coils. Upon disassembling the magnet, clear mechanical damage was observed in the ReBCO coil, confirming the initial suspicion. Additionally, a dent was discovered on the surface of one of the $Nb_3Sn$ coils.

In response, the support structure was modified, the damaged $Nb_3Sn$ coil was replaced with a new one, and the magnet was re-assembled for further testing and evaluation.

Figure 8.3.2 displays the complete quench histories of three test runs. In the second powering test run, a total of 13 training quenches were performed, as indicated by the green marks in the figure for the first 13 data points. The quench protection criteria for this run were set at 100 mV for the differential voltages, with a validation time window of 30 ms. As for the direct voltage signals, the threshold was 300 mV, also with a validation time of 30 ms.

During the second test run, the very first training quench occurred at 5111 A, corresponding to a main field strength of 9.64 T, which was 78% of the nominal current (In). Subsequently, the quench current increased rapidly in the following 12 training tests, with the 13th test reaching a quench current of 5860 A, corresponding to a main field strength of 10.85 T.

Most of the quench events (10 out of 13) were localized in the transition parts (two-layer jumps) of coils 1# and 6#. These coils were the newly fabricated NbTi coils. Following these occurrences, the test had to be temporarily halted due to occasional breakdowns of the helium liquefier. As a result, the temperature of the magnet had to be restored to room temperature (RT) during this interruption.

Following the recovery of the helium liquefier, the third run of LPF1-U powering tests was conducted one month later. This time, the magnet underwent an extensive training process, resulting in the recording of 103 quenches. This allowed for the identification of two distinct training phases and two elastic plateaus.

In the first training phase, the criteria for quench protection remained the same as in the second run. Although a de-training phenomenon was observed initially, the first training phase spanned from the 14th to the 25th ramping tests, during which the quench current steadily increased from 5538 A (10.33 T) to 6002 A (11.08 T). Subsequently, a plateau phase followed, consisting of 29 training tests where the magnet experienced quenches at fluctuating currents around 6050 A (11.15 T). Upon analyzing the voltage signals, spikes were detected prior to the quenches, indicating that these voltage spikes were occurring before the quenches were identified by the protection system based on the pre-defined criteria. It was suspected that flux jumps in the $Nb_3Sn$ coils were the most likely cause of these voltage spikes, as the majority of these judged "quenches" originated from the $Nb_3Sn$ coils.

To prevent the protection system from being triggered by these flux-jump-induced voltage spikes, adjustments were made to the triggering criteria. The threshold for differential voltages was increased from 100 mV to 300 mV, with a validation time of 30 ms, and the threshold for direct voltage signals was increased from 300 mV to 500 mV, also with a validation time of 30 ms. Simulation results indicated that even when quenched at the design current with more than 70% of the energy extracted to the dump resistor and varistor, the peak voltage in LPF1-U would still be lower than 800 V with the modified triggering criteria.

The modification had a significant effect, leading to the occurrence of the second training phase. As shown in the figure, the quench currents increased again from 6152 A (11.32 T) to 6449 A (11.8 T) during the 55th to 74th ramping tests. Subsequently, the magnet exhibited instability, with the quench current fluctuating irregularly, albeit showing a general upward trend. Finally, a maximum quench field of 12.15 T at a current of 6664 A was achieved during this run.

**Table 8.3.1:** Main parameters of cables and strands [15].

| Parameter | $Nb_3Sn$ coil | NbTi coil | Outer NbTi coil |
|---|---|---|---|
| Cable reference | IHEPNS1 | IHEPWN1 | IHEPNT1 |
| Number of turns | 28 | 32 | 32 |
| Cable width (mm) | 8.6 | 16 | 14.3 |
| Cable height (mm) | 1.5 | 1.5 | 1.5 |
| Number of strands | 20 | 38 | 31 |
| Insulation material | S-glass | Kapton | Kapton |
| Insulation thickness (mm) | 0.28 | 0.15 | 0.15 |
| Strands diameter (mm) | 0.818 | 0.82 | 0.825 |
| N-Sc/Sc | 1 | 1 | 1.3 |
| *RRR* | 100 | 130 | 130 |
| Reference field ($R_f$) (T) | 12 | 5 | 5 |
| $I_c$ @ $R_f$ /4.2K (A) | 630 | 690 | 655 |
| $J_c$ @ $R_f$ /4.2K (A/mm$^2$) | 2398 | 2613 | 2816 |

**Table 8.3.2:** Main design parameters of LPF1-U magnet [15].

| Parameters | Unit | Value |
|---|---|---|
| Number of apertures | - | 2 |
| Aperture diameter | mm | 14 |
| Inter-aperture spacing | mm | 179 |
| Designed operating current (LTS) | A | 6350 |
| Designed main field | T | 12 |
| Peak field in coil | T | 12.2 |
| Operating temperature | K | 4.2 |
| Margin along the load line | % | ~13 |
| Width of coil (one quadrant) | mm | 83.1 |
| Height of coil (one quadrant) | mm | 57.6 |
| Stored energy | kJ/m/bore | 545 |
| Inductance | mH/m/bore | 24.61 |
| Inner Nb3Sn coil – No. | - | 2 |
| -Straight section length | mm | 200 |
| Middle NbTi coil – No. | - | 2 |
| -Straight section length | mm | 200 |
| Outer NbTi coil – No. | - | 2 |
| -Straight section length | mm | 340 |
| Outer diameter of the iron yoke | mm | 500 |
| Outer diameter of the magnet | mm | 620 |
| Total length of the magnet | mm | 630 |
| Minimum bending radius of coils | mm | 60 |
| Lorentz force Fx/Fy (one quadrant) | MN/m | 4.96 /0.57 |
| Fringe field (@ r = 500 mm) | T | 0.052 |

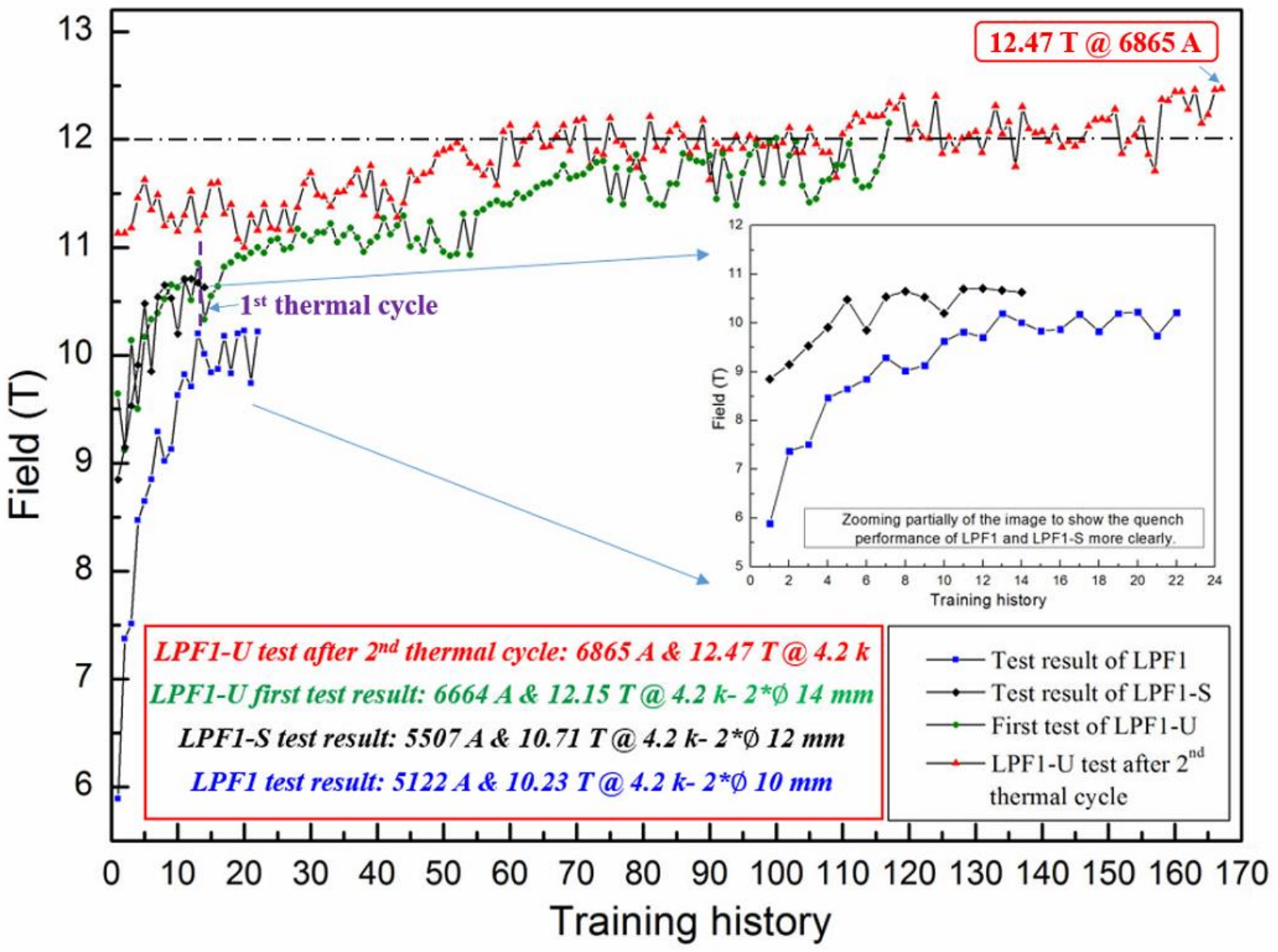

**Figure 8.3.2:** Performance test results of LPF1, LPF1-S and LPF1-U [15].

During the entire training process, the bridge and terminal voltage variations were carefully examined and utilized to assess the occurrence of quench events. Additionally, to enhance precision, each coil was divided into nine segments using eight Voltage Transformers (VTs), allowing the recording of voltage variations in these individual segments. By analyzing the voltage signals within these segments, it was possible to achieve a more precise determination of the origin of quench events.

Upon analyzing all the relevant data, it was observed that the majority of quenches occurred in coil 3#, which is an $Nb_3Sn$ coil situated in the central area of the straight section adjacent to the region with the highest magnetic field. As previously explained, this particular coil had experienced stress concentration issues during the first test run but had not been replaced. Furthermore, a slight dent was discovered on the coil's surface, coinciding with the location of origin for most of the quenches. This indicated the presence of permanent damage in part of coil 3#. The over-compaction of the coil can result in irreversible degradation of the Ic in $Nb_3Sn$ cables. It can also lead to significant coupling between filaments, thereby increasing the persistent current effect and diminishing the cable's stability in the face of growing flux jumps.

These factors, namely the stress concentration and the permanent damage in coil 3#, were identified as the primary causes of the instability observed in LPF1-U and the occurrence of a large number of training quenches.

Further analysis of the instability of LPF1-U at high field revealed an intriguing phenomenon. By examining the data obtained from FBGs embedded in coils to track strain state changes within the magnet during the training process, a strong correlation was observed between the changes in wavelength curves of FBGs and variations in quench currents. Specifically, during the training quenches from the 95th to the 104th (conducted in a single day under identical test conditions and with typical irregular quench currents), the adjacent waveforms (quenches) remained the same or changed regularly as the quench current increased gradually. However, they changed irregularly when the quench current unexpectedly decreased.

This irregular change in waveform indicated an irregular change in the strain state of the magnet during excitation, which could be attributed to an unstable mechanical structure or inappropriate stress distribution, as discussed previously. The deformation of the support structure prevented the transmission of compressive stress uniformly from the shell to the coils to counteract the Lorentz force. Upon disassembling the magnet after warming up, pressure-sensitive papers (such as "Fuji" paper traces) placed between coils during assembly were examined. These traces revealed inhomogeneity with lower stress loads in the coil ends but higher stress loads in the straight section. This observation suggested that the mechanical instability of the magnet was possibly another reason for the irregular variations in the quench current, particularly at high field (current) stages.

Additionally, monitoring with FBGs and RSGs attached to the rods and shell showed a slight increase in strain during excitation, indicating that the applied pre-stress was still insufficient for operating the magnet at fields beyond 12 T. A ratcheting effect, resulting from cumulative residual deformation of the structure, was observed during the run test and subsequent tests after the thermal cycle. After quenches and the disappearance of Lorentz forces, a positive residual strain ranging from a few $\mu\varepsilon$ to a few tens $\mu\varepsilon$ was measured in the rods. This effect has also been observed in tests of other dipole magnets, underscoring the importance of applying higher but safer pre-stress to enable the magnet to withstand large Lorentz forces at high fields.

To assess the long-term operational stability of LPF1-U following the training study, the magnet was subjected to consecutive energizations at 6500 A (11.88 T) and 6600 A (12.05 T). These stable currents were maintained for a duration of over 30 minutes in each test, and no quench events occurred during the experiments.

Throughout the entire training test process, the bore field of LPF1-U was primarily measured using a Hall probe. The Hall probe was fixed on a removable rod and positioned at the longitudinal center of one aperture. As the iron and the coil were in close proximity, the saturation effect became noticeable at approximately B~2.5 T.

The magnet's transfer function (TF) decreased from 2.896 T/kA at low fields to 1.611 T/kA at a bore field of approximately 12 T. This represents a reduction of 44.4% in the TF value. Additionally, the B-I curve of LPF1-U was plotted in the figure, with red markers indicating the measured field values and a blue line representing the simulated results. The measured field strength obtained using the Hall probe exhibited excellent repeatability at the same operating current during the performance tests. Furthermore, there was a high level of consistency observed between the simulated and tested results, demonstrating the accuracy of the simulations in predicting the field strength.

Despite the challenges faced during the test of LPF1-U, including damage to a portion of the $Nb_3Sn$ coil 3# due to the initial imperfect support structure and mechanical disturbances causing prolonged training and quench current plateaus, the primary objectives of the test were successfully achieved. The magnet was able to attain a main field exceeding 12 T in the last two test runs, and it demonstrated improved stability after undergoing the thermal cycle.

In particular, a main field strength of 12.47 T (with a peak field in the coils reaching 12.7 T) was achieved at an operating current of 6865 A, with a load line ratio of 90%. This accomplishment represents a significant milestone for LPF1-U. Additionally, the other coils, especially the newly fabricated NbTi coils, exhibited excellent performance with only a few training quenches and demonstrated upward potential.

#### 8.3.1.2 *Development of LPF3 Dipole*

The LPF series dipole magnets have played a vital role in enabling the research team to acquire essential fabrication skills for high-field superconducting accelerator magnets. Building upon the knowledge gained from previous experiences, the team is currently in the process of fabricating a 16-T hybrid high-field dipole magnet called LPF3, as illustrated in Figure 8.3.3.

LPF3 adopts a hybrid design, featuring six racetrack $Nb_3Sn$ coils on the outside to generate a main field strength of 12-13 T. Additionally, it incorporates several HTS insert coils inside the magnet to further enhance the field strength to 16-19 T, while maintaining two apertures. The HTS insert coils are designed in a block-type configuration, aligning the tapes with the magnetic flux to fully utilize their enhanced current-carrying capability when the field is parallel to the wider surface.

The mechanical components for LPF3 have already been fabricated, marking a significant milestone in the project. The coil fabrication and pre-assembly are currently underway, with plans to conduct performance tests in 2023.

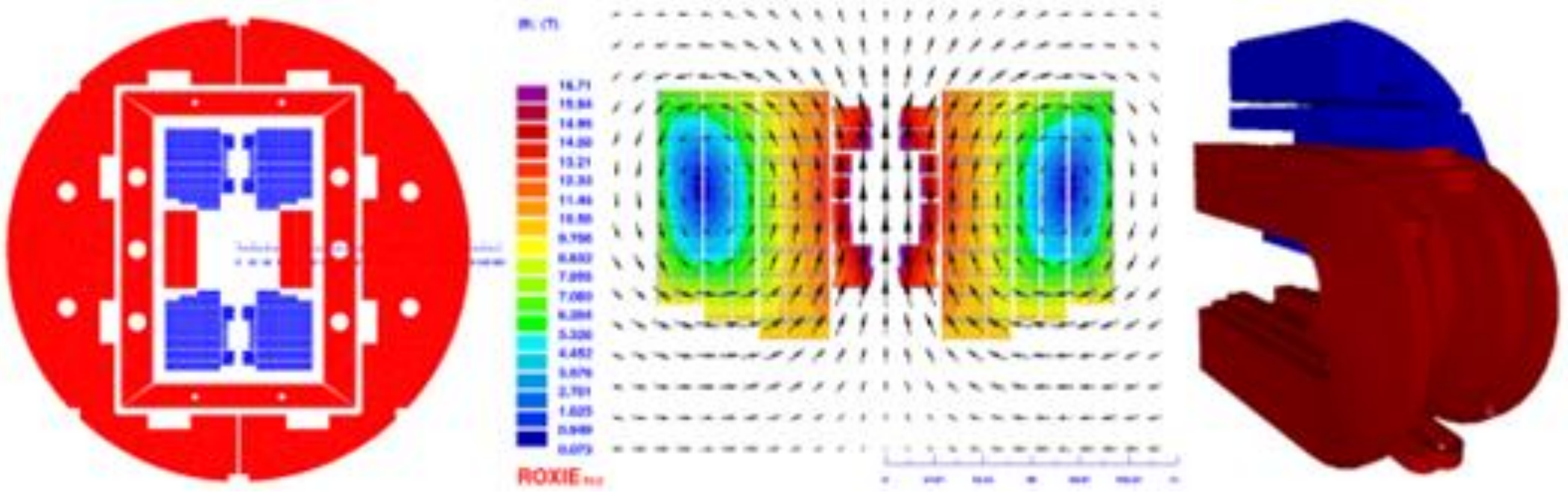

**Figure 8.3.3:** Left: Cross-section of LPF3; Middle: Flux distribution; Right: Coil ends with yoke.

As part of the fabrication process for the HTS insert coils of LPF3, a novel HTS transposed cable is currently being developed [16]. The cable manufacturing involves two main steps. Firstly, several HTS tapes are stacked and wrapped with thin copper tapes, forming what is referred to as a "strand." Then, multiple strands are transposed and cabled together, as depicted in Figure 8.3.4. This entire process is carried out on an industrial production line, allowing for the efficient production of cables that can span hundreds of meters.

During the cable manufacturing process, the HTS tapes undergo in-plane bending multiple times. Consequently, we conducted tests to determine the critical current of the HTS tapes at different in-plane bending radii. The critical bending radius is defined as the radius at which the critical current (Ic) of the sample degrades by 5%. The test results are presented in Figure 8.3.5.

For 4-mm-width Samri and SST tapes, the critical bending radii were found to be 400 mm and 500 mm, respectively. It's worth noting that the Ic values of all samples remained stable during the thermal cycle tests. Figure 8.3.5(b) demonstrates the repeatability of the in-plane bending performance of Samri tapes. When the in-plane bending radius was reduced to 500 mm or 400 mm, a significant reduction in the critical current was observed, consistent with the results shown in Figure 8.3.5(a). The difference in the critical in-plane bending radii among the three samples may be attributed to fatigue effects.

Combining the results from both parts of the study, it is recommended that the in-plane bending radius should not be smaller than 500 mm to ensure that the REBCO tape's performance does not degrade by more than 5% after cabling.

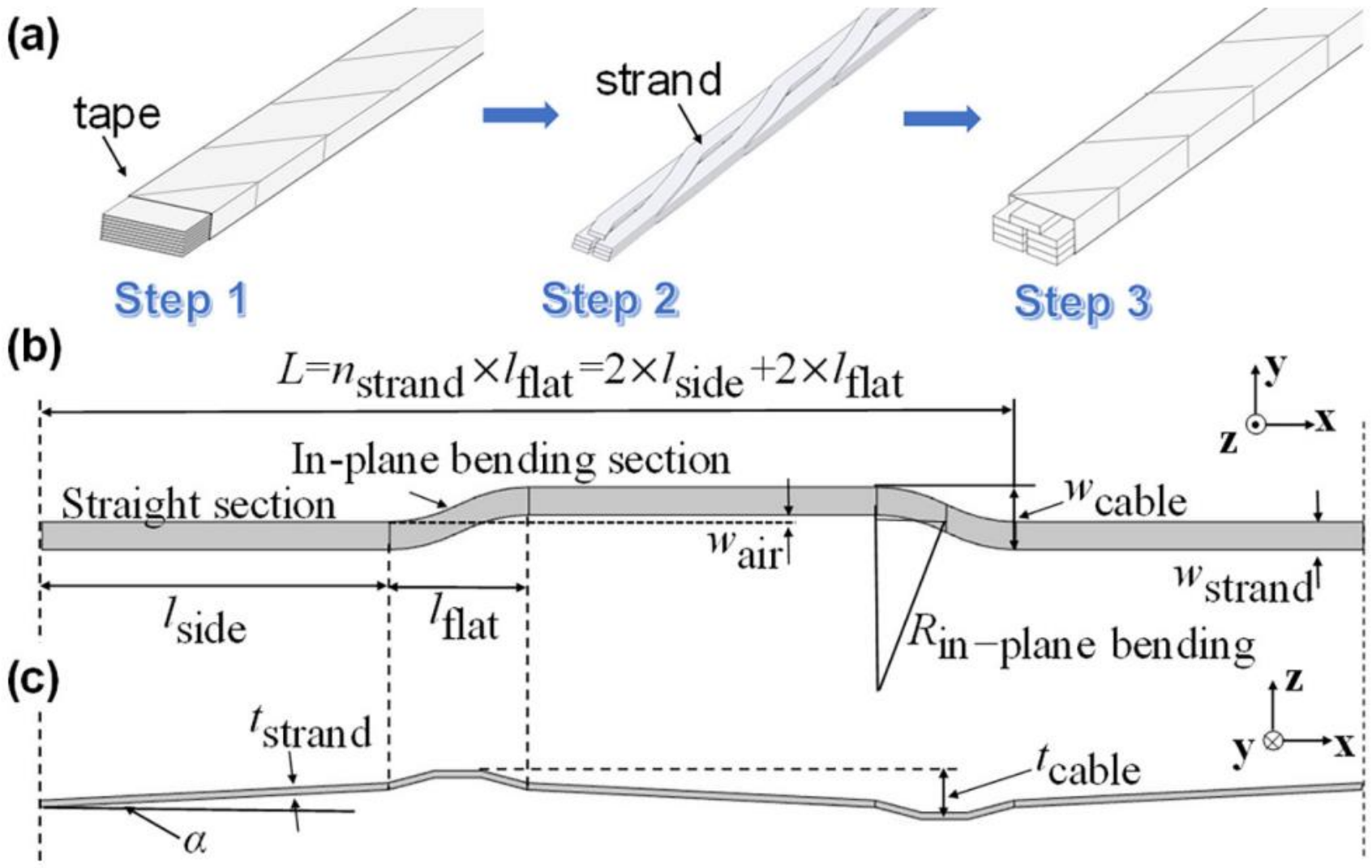

**Figure 8.3.4:** (a) Schematic illustration and fabrication steps of the X-cable. (b) Top view and (c) Side view of one strand in the X-cable.

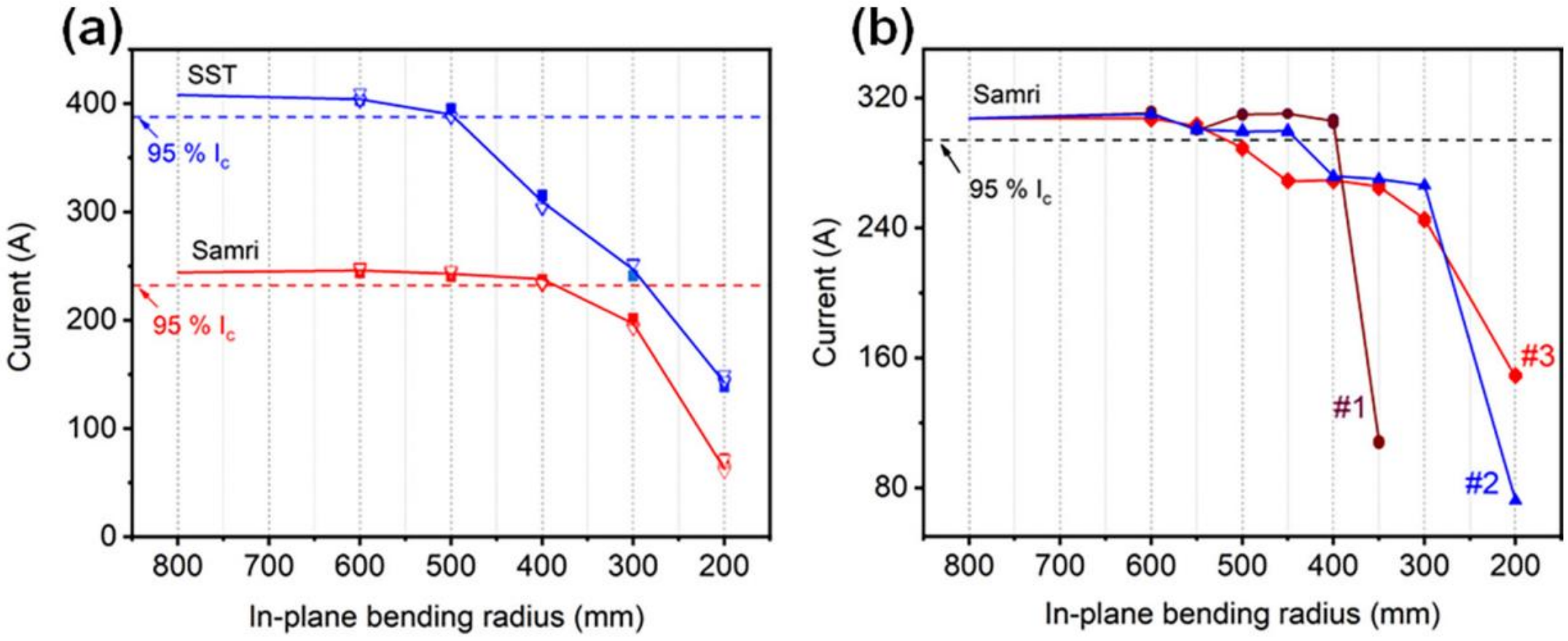

**Figure 8.3.5:** Critical currents of 4-mm-width REBCO tapes at different in-plane bending radii. (a) Each sample is only bent to a specific radius. (b) Each sample is tested at different in-plane bending radii from large to small.

Figure 8.3.6 depicts the magnetic field distributions measured using MCorder 2D, along with the calculated current density distributions, for the SST tapes after undergoing an in-plane bending performance test. The X-axis represents the width direction of the tested tape, while the Y-axis represents the length direction.

In an undamaged REBCO tape, the loop current, centered around the width, symmetrically distributes on both sides of the tape. Additionally, the loop current flows in opposite directions. However, when a defect is present in the superconducting film, the loop current is unable to pass through the defect and instead generates a circulating current in the undamaged region.

Comparing the test results, when the in-plane bending radius is 200 mm, numerous defects appear in the superconducting film. As the bending radius increases, the number of defects decreases. At a radius of 500 mm, no defects are detected on the superconducting film, which aligns with the results of the in-plane bending performance test.

The successful manufacturing of the prototype X-cable was achieved based on the in-plane bending properties of REBCO tapes. For the initial design, a full-size cable was envisioned to comprise nine strands, with each strand containing eight tapes, resulting in a total of seventy-two tapes in the cable. However, in the prototype cable, only twenty REBCO tapes were utilized, distributed among four strands in different patterns. The remaining spaces were filled with Hastelloy tapes. The cross-section of the prototype cable can be observed in Figure 8.3.7 (a). Each REBCO tape within the strand is numbered consecutively from the top tape (T1) to the bottom tape (T8).

Specifically, Strand S1 consists of two REBCO tapes, positioned at the top and bottom (S1T1 and S1T8). Strand S2 and S3 contain four and six REBCO tapes, respectively, with the REBCO tapes situated in the middle of the two strands (from S2T3 to S2T6, and S3T2 to S3T7). Strand S4 encompasses eight REBCO tapes (from S4T1 to S4T8). This arrangement allows the prototype cable to effectively represent an all-superconducting cable while maintaining a reasonable cost.

Table 8.3.3 and Table 8.3.4 provide the parameters of the prototype cable and the REBCO tapes utilized. The Hastelloy tapes have a width and thickness of 4 mm × 90 μm, while the copper foil used in the cable measures 5 mm × 50 μm.

The prototype X-cable has a total length of approximately 10 meters and is manufactured in a continuous manner. The schematic diagram of the cable is illustrated in Figure 8.3.7 (b). In order to transport the cable from the company to the test lab, it was wound into a disc shape with a diameter of 840 mm. Prior to the winding process, two 1.2-meter samples (C-Sample 1 and C-Sample 2) were taken to assess any potential degradation caused by the bending with a diameter of 840 mm. These samples were chosen such that all strands included at least one in-plane bending section, which is the area most susceptible to degradation.

Upon arrival at the laboratory, two additional samples (C-Sample 3 and C-Sample 4) were cut from the wound cable. The lengths of these two samples are 1.77 meters and 1.8 meters, respectively. Unfortunately, during the cabling process, a failure occurred at the 4.5-meter position due to a local defect in one of the strands.

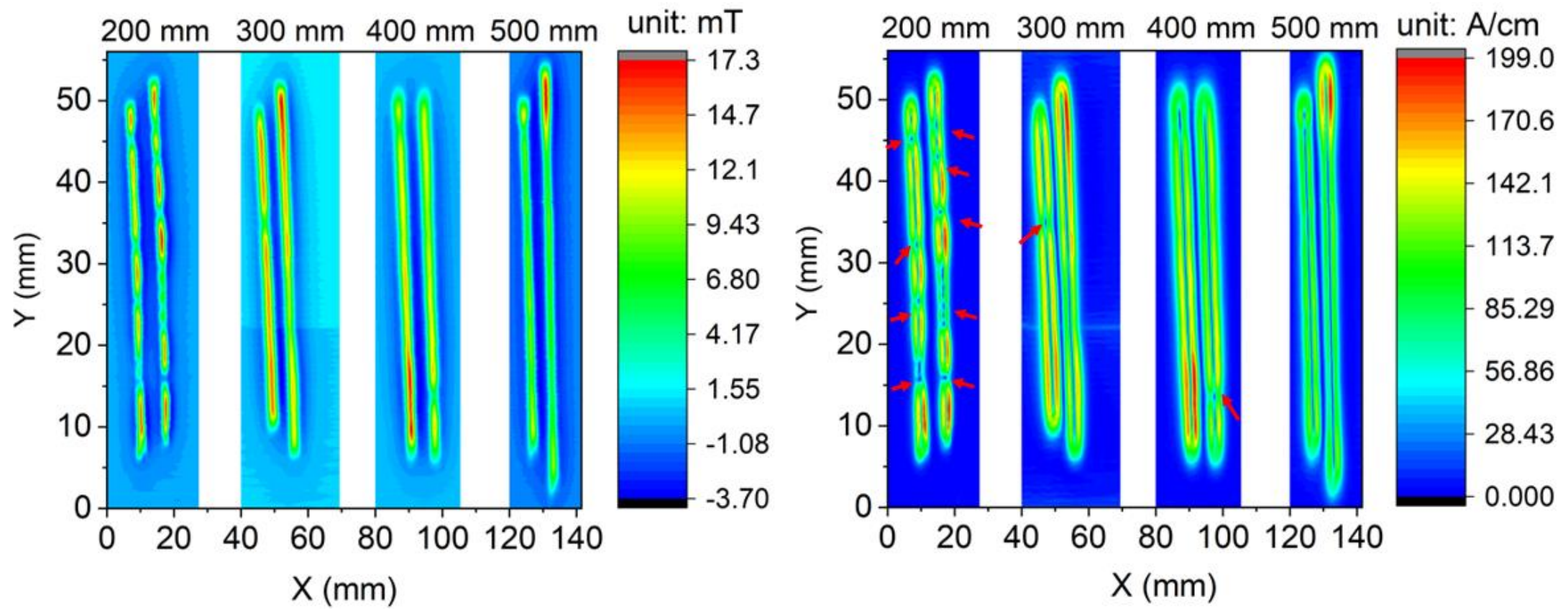

**Figure 8.3.6:** a) Measured magnetic field distributions and (b) calculated current density distributions for the SST tapes after the in-plane bending performance test scanned by MCorder 2D. From left to right are samples with in-plane bending radii of 200, 300, 400, and 500 mm. The defects are marked with red arrows.

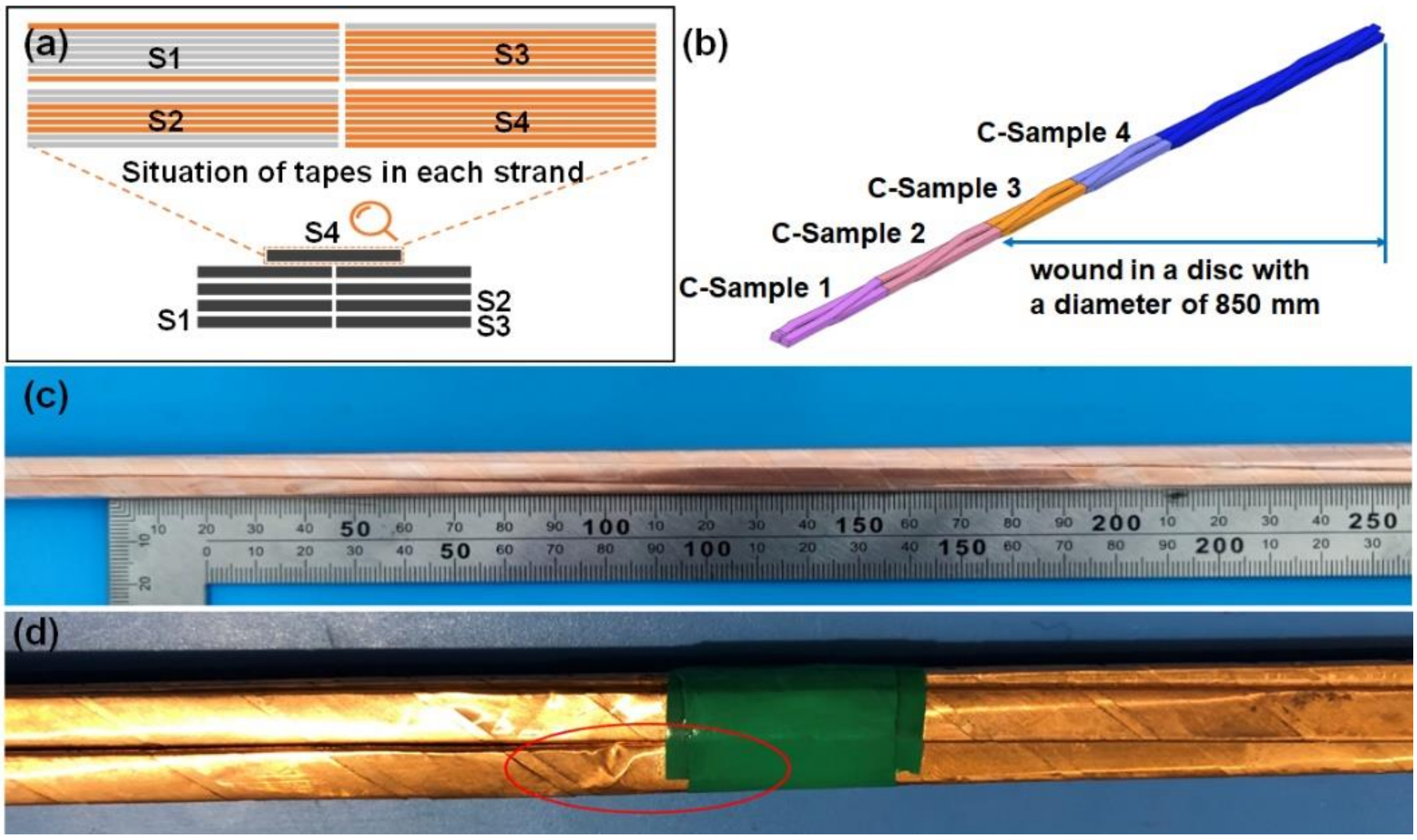

**Figure 8.3.7:** (a) Cross-section of the prototype cable: orange and grey rectangles represent REBCO tapes and Hastelloy tapes, respectively. (b) Schematic illustration of the prototype cable. (c) Picture of the prototype cable showing the in-plane bending section. (d) Picture of the failure occurred in the prototype cable, and the failure is marked with a red circle

**Table 8.3.3:** Design parameters of the prototype cable

| Parameter | Value |
|---|---|
| Cable length | 10 m |
| Cable width | 9.76 mm |
| Cable thickness | 7.4 mm |
| transposition period length $L$ | 1.6 m |
| In-plane bending section length $l_{\text{flat}}$ | 180 mm |
| Number of strands $n_{\text{strand}}$ | 9 |

**Table 8.3.4:** Characteristics of REBCO tape contained in the prototype cable.

| Parameter | REBCO tape |
|---|---|
| Manufacture | Samri |
| Width and thickness | 4 mm × 76 μm |
| Critical current at self-field (1μV·cm$^{-1}$) | ≥140 A @ 77 K |
| The thickness of the silver cap layer | ~2 μm |
| The thickness of the superconducting layer | ~1 μm |
| The thickness of the Hastelloy substrate layer | ~60 μm |
| The thickness of the copper stabilization layer | ~5 μm |

Figure 8.3.8 (a) displays the Ic and n-values of REBCO tapes contained in C-Sample1-4. These values are obtained by fitting the measured E-I curves using a power-law model with a critical electric field of 1μV·cm$^{-1}$. It is evident that the Ic values of all REBCO tapes in C-Sample 1, C-Sample 2, C-Sample 4, and most of the tapes in C-Sample 3 align with the test report provided by Samri, with a minimum Ic value of 140 A along the length. However, the tapes S4T1, S4T2, S4T3, and S4T4 in C-Sample 3 exhibit lower Ic values due to the aforementioned failure. These results demonstrate that the tapes perform well if no failures occur during the cabling process. Additionally, the performance of the cable is not affected by the 840 mm bending diameter.

To investigate the cause of the failure, we disassembled C-Sample 3 and discovered that the copper foil wrapped around S4 at the position of the failure is thicker than normal. This was due to the breakage and subsequent manual repair of the copper foil during strand assembly. The variation in thickness along the length resulted in the failure of that particular strand during the transposition process. To prevent such incidents in future production, it is recommended to use higher strength copper foil.

Figure 8.3.8 (b) illustrates the measured and simulated E-I curves of C-Sample 4. The measured Ic and n-value are 1939.8 A and 20.76, respectively, obtained by fitting the measured E-I curve with a critical electric field of 1 $\mu V \cdot cm^{-1}$. It is important to note that an approximately linear voltage was observed due to the soldering of the voltage taps into the copper terminals along with the cable. This linear voltage is caused by a constant resistance of 2.76 mΩ, which was determined by fitting the experimental E-I curve.

The measured Ic value represents 93.6% of the simulated value. This discrepancy can be attributed to non-uniform current distribution among the twenty REBCO tapes, caused by different terminal resistances (Rt) for each tape. Figures 8.3.8 (c, d) display the simulated critical current and transport current variations for each tape, revealing an evident unbalanced current distribution resulting from the field distribution. It should be noted that the simulation assumes the same Rt for all tapes, while in reality, the Rt of each tape differs. This further exacerbates the imbalance in current distribution.

Consequently, some tapes reach their Ic earlier than others as the Icable increases, leading to a lower current-carrying capacity than the cable is actually capable of carrying. This observation aligns with our previous tests where the termination soldering length was varied. When the soldering length was 3 cm and 10 cm, the measured Ic values were 1200 A and 1830 A, respectively. As the soldering length increases, the absolute values and fluctuations in Rt for each tape decrease, resulting in a more evenly distributed current throughout the cable and ultimately an increase in the measured Ic.

These findings emphasize the significance of employing appropriate termination technology to ensure optimal cable performance when applied in coils.

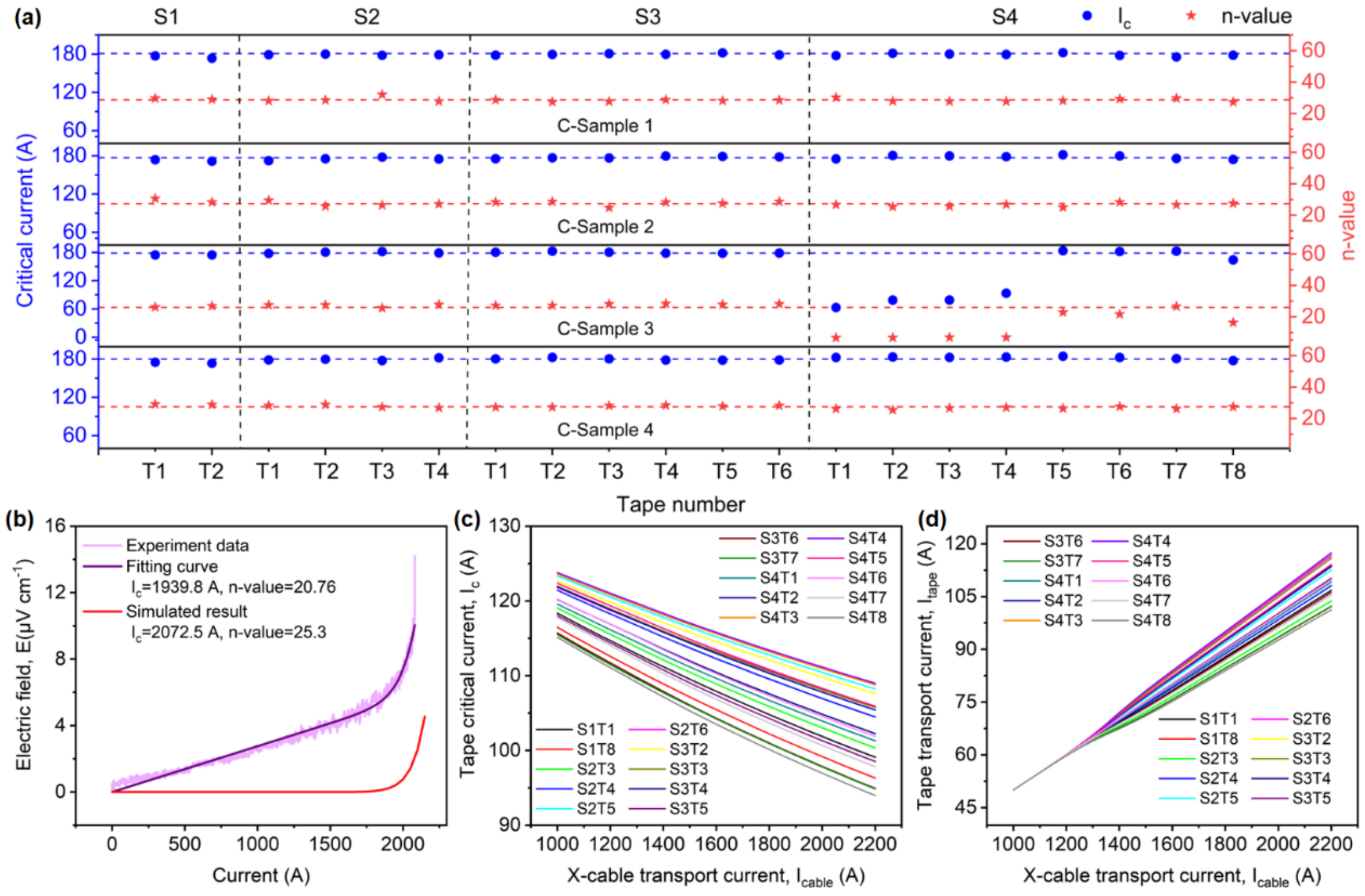

**Figure 9.3.8:** (a) Ic and n-values of REBCO tapes. The blue and red dash lines are the average values of Ic values and n-values, respectively. (b) Measured and simulated E-I curves of C-Sample 4 at 77 K and self-field. (c) and (d) individually REBCO tape critical current and transport current by simulation.

A preliminary investigation was conducted on the persistent current effect in the HTS insert for LPF3. By employing previously established simulation methods, the distribution of superconducting currents and magnetic fields could be estimated, as depicted in Figure 8.3.9. The results indicate that the magnetic field induced by persistent currents would decrease the center field strength. Analysis of the Lorentz forces further reveals non-uniform force loads concentrated primarily around the sides of the windings.

In order to minimize the magnetic field component perpendicular to the tape surface and enhance the current margin, a genetic algorithm was employed to optimize the alignment of the magnets. This optimization process aims to improve the field quality and manage stress more effectively within the LPF3 magnet. Taking into account the aforementioned effects will be the next step in the optimization process for field quality and stress management in the LPF3 magnet.

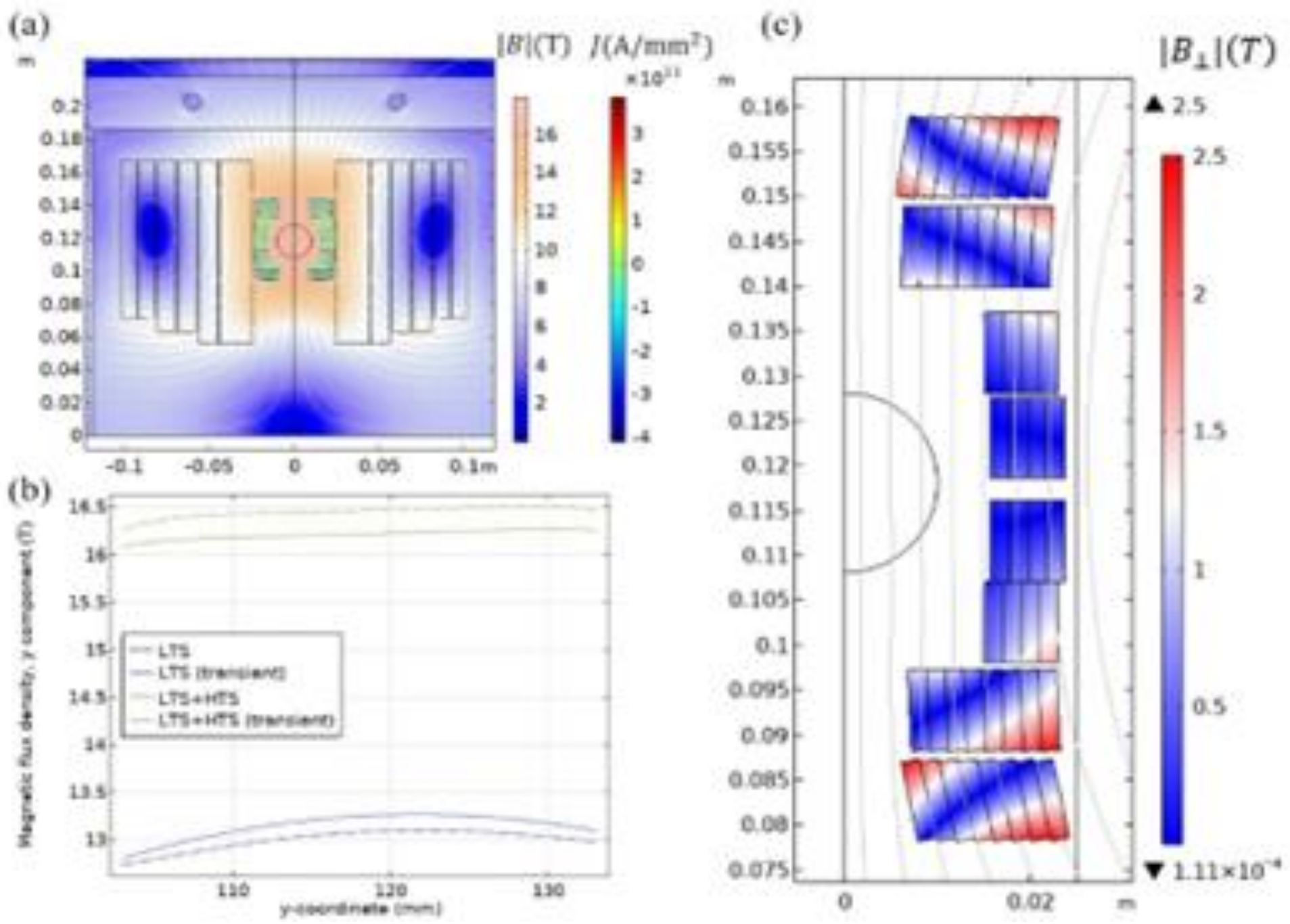

**Figure 8.3.9:** (a) Simulated field and current distribution (a) of the insert prototype and (b) along the center line of bore R=15 mm. (c) Perpendicular field with adjusted magnet alignments.

## 8.3.2 **Progress of High Field Magnet R&D with IBS**

### 8.3.2.1 ***Development and Performance Test of the IBS Racetrack Coils***

Two IBS racetrack coils were successfully manufactured using the 100-m 7-filamentary $Ba_{1-x}K_xFe_2As_2$ (Ba122/Ag) tapes, employing the wind-and-react method. To withstand the high stress experienced under a high magnetic field, the IBS tape was co-wound with stainless steel tape, as illustrated in Fig. 8.3.10. These two coils were subsequently inserted into the LPF series dipole magnets mentioned earlier. Experimental tests were conducted on the IBS racetrack coils under background fields of up to 10 T.

Fig. 8.3.11 illustrates the quench currents of the two IBS racetrack coils, along with the $I_c$ of the IBS short sample. The first IBS coil achieved a quench current of 45.9 A at 4.2 K and 7.5 T, which corresponds to approximately 64% of the $I_c$ of the short sample. The limited performance was likely due to the heat generated at the inner joint. On the other hand, the second IBS coil achieved a quench current of 65 A at 4.2 K and 10 T, representing around 86.7% of the $I_c$ of the short sample at 10 T, and 81.25% of the quench current at self-field.

These tests demonstrate that the high-field performance of the IBS coils is less influenced by the background field compared to other materials. This suggests that IBS holds great promise as a material for high-field magnet applications.

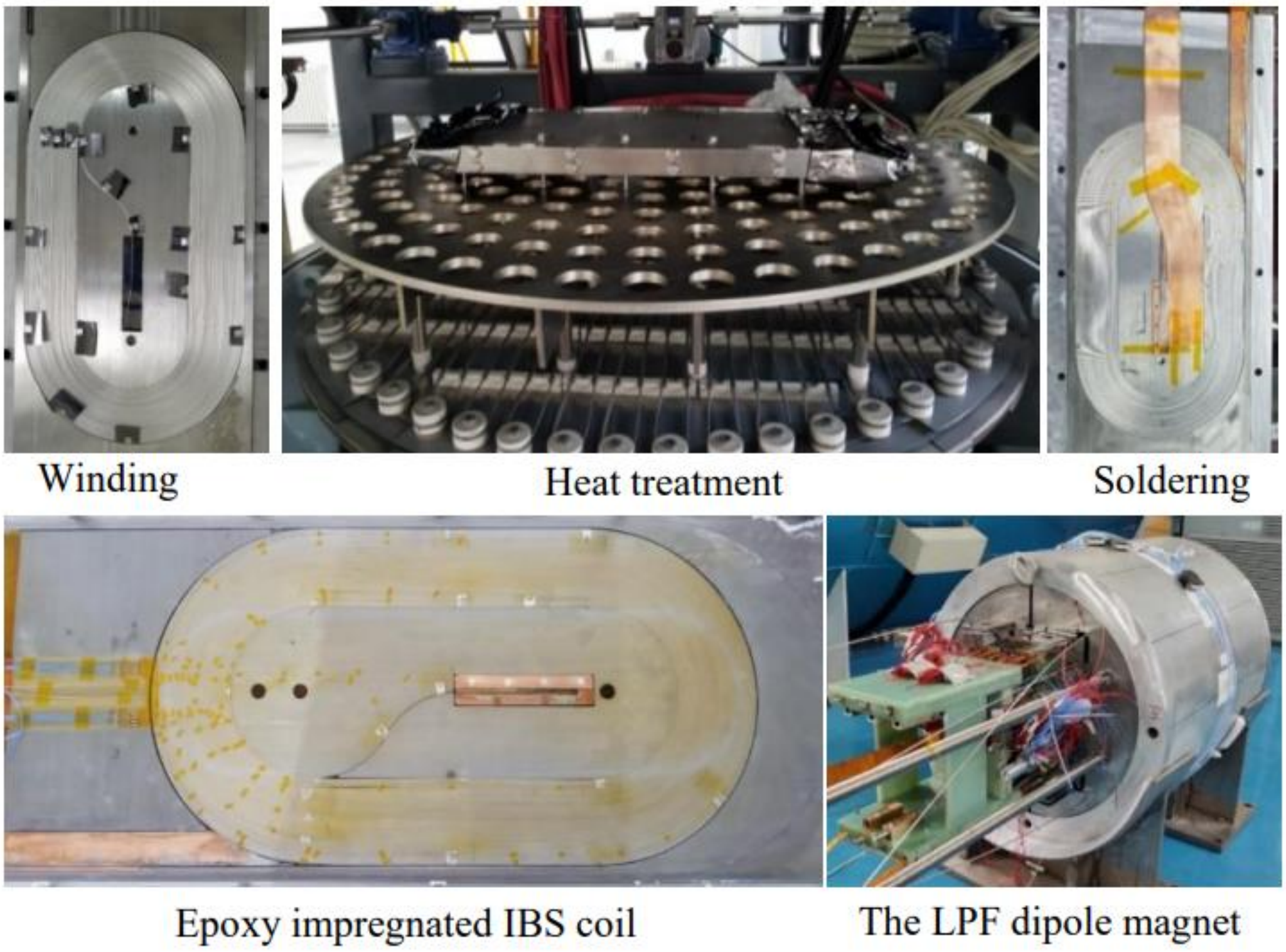

**Figure 8.3.10:** Fabrication of IBS racetrack coil and the dipole magnet inserted with IBS coil.

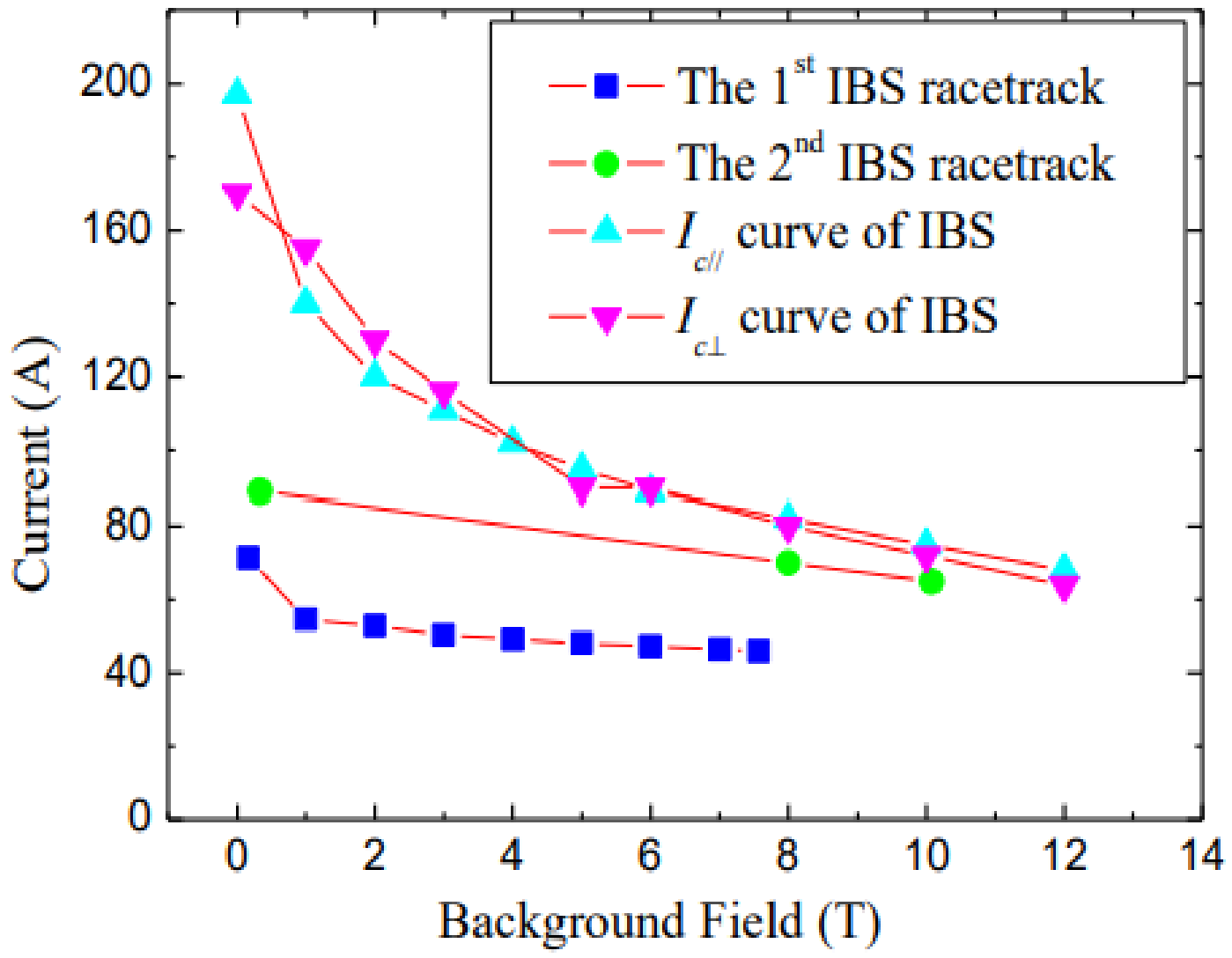

**Figure 8.3.11:** Quench current vs. background field of the 1st and 2st IBS racetrack coils

### 8.3.2.2 *Development and Performance Test of the IBS Solenoid Coils*

In addition to the racetrack coils, we have successfully fabricated IBS solenoid coils and tested their performance under high background fields. The first IBS single pancake coil, constructed using seven-filamentary $Ba_{1-x}K_xFe_2As_2$ (Ba122/Ag/AgMn) tape, achieved a current of 26 A when subjected to a 24 T background field. This value represents approximately 40% of the current at zero external field, as depicted in Fig. 8.3.12.

Over the past two years, there have been significant improvements in the critical current density of Ba122-IBS tape. To capitalize on these advancements, a new Ba122/Ag/AgSn tape was utilized to fabricate the latest IBS solenoid coils. However, it should be noted that the Ba122 superconducting core within the IBS tape is brittle, and the Ag/AgSn sheath lacks sufficient strength during the winding process. Consequently, when the bending diameter falls below 30 mm, the transport $I_c$ is unavoidably reduced due to cracks observed in the Ba122 superconducting cores.

To mitigate this issue, the Ba122/Ag/AgSn tape was employed to manufacture new double pancake coils via the wind-and-react method. Fig. 8.3.13 (a) displays the external view of a coil with an inner diameter of 30 mm. The quench current reached 122 A at 4.2 K and 10 T, which corresponds to 96% of the $I_c$ of the 30 mm-diameter bent sample and 89% of the $I_c$ of the straight short sample, as shown in Fig. 8.3.13 (b). These results indicate that the hard-way bending of the innermost tape has minimal impact on reducing the coil performance.

Moving forward, six IBS coils will be fabricated in series to create an insert solenoid coil operating at 35 T. This development aims to verify the application of IBS at super-high fields.

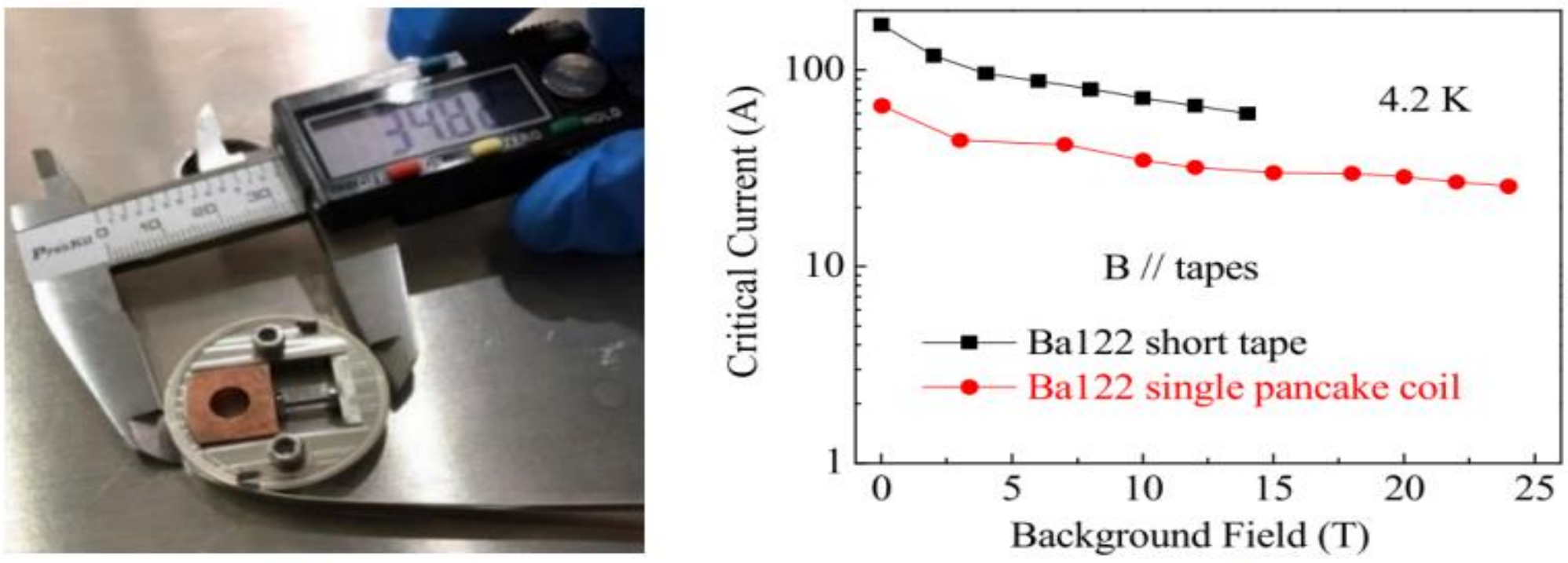

(a) (b)

**Figure 8.3.12:** The outer view of the IBS single pancake coil with an inner diameter of 30 mm (a), Magnetic field dependence of transport critical current at 4.2 K for the IBS short tape and SPC.

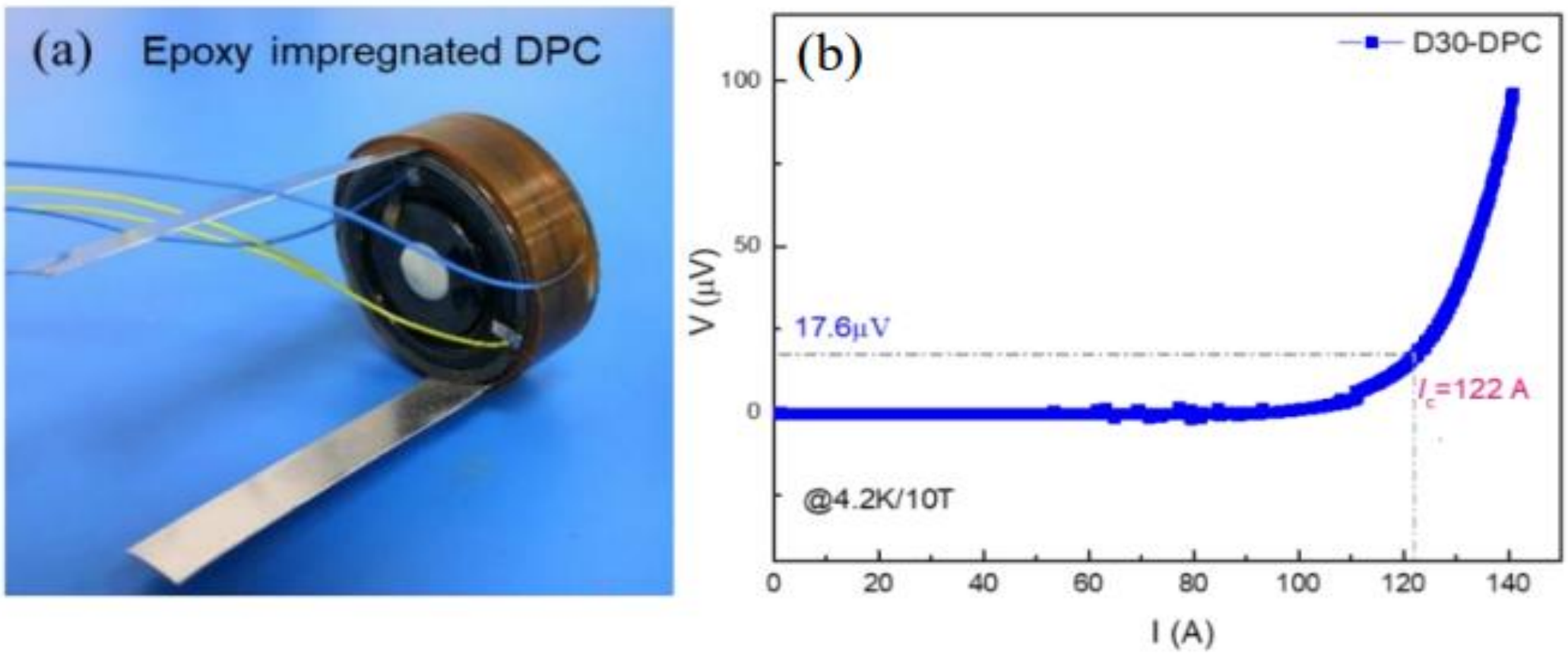

**Figure 8.3.13:** (a) The IBS double pancake coil with an inner diameter of 30 mm, (b) I-V curve of the coil at 4.2 K and 10 T.

### 8.3.3 Development of HL-LHC CCT Magnets

HL-LHC, the high-luminosity upgrade project for the LHC, aims to significantly increase the peak luminosity of the collider by a factor of 5. To achieve this ambitious goal, several key innovative technologies are required, particularly in the development of higher performance superconducting magnets. The HL-LHC project involves the replacement of the superconducting magnets at the interaction region.

The CCT superconducting magnets developed specifically for HL-LHC will play a crucial role in providing a 5 T-m integral dipole field in two perpendicular directions within the two 100-mm diameter apertures. These magnets have a length of 2.2 m. In September 2020, a full-scale prototype of the CCT magnet, developed collaboratively by Chinese institutions including IHEP, IMP, WST, and others, successfully passed performance tests at 4 K. The prototype was subsequently shipped to CERN, and in November 2020, its qualified performance was confirmed through tests conducted at 1.9 K, as shown in Fig. 8.3.14.

Currently, the fabrication of 12 series CCT magnets is underway, and they are scheduled to be delivered to CERN by the end of 2023. This represents a significant milestone, as it will mark the first application of CCT superconducting magnets in an operational accelerator.

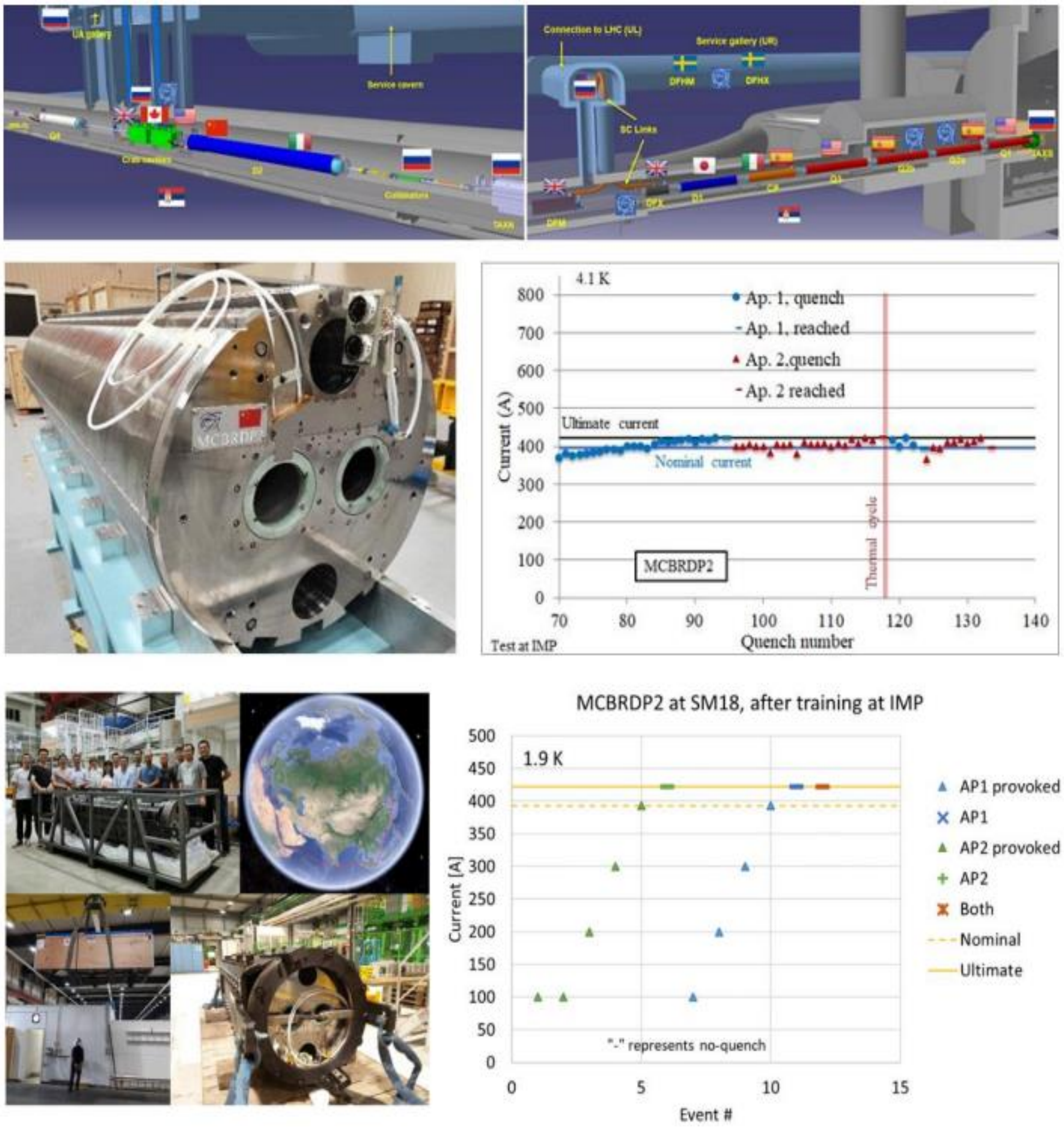

**Figure 8.3.14:** Top: International teamwork for developing superconducting magnets of HL-LHC. Center left: CCT magnet developed by the Chinese team. Center right: Performance test of the 1st prototype magnet at 4 K in China. Bottom left: Prototype shipped to CERN. Bottom right: CERN retest at 1.9 K reached the design target without quench.

## 8.4 Injector Chain

### 8.4.1 General Design Considerations

The injector chain is a complex accelerator system in its own right, and it plays a crucial role in reaching the required beam energy of 3.2 TeV for injection into the SPPC. This four-stage acceleration system is designed to provide energy gain per stage between 8 and 18. Its primary function is not only to accelerate the beam to the necessary energy for injection into the SPPC but also to prepare the beam with specific properties such as bunch current, bunch structure, emittance, and beam fill period.

The four stages of the injector chain are illustrated in Fig. 8.4.1, and additional parameters are provided in Table 8.4.1. It's worth noting that the lower the stage, the higher its repetition rate. The p-Linac, which is a superconducting linear accelerator, operates at a repetition rate of 50 Hz. The p-RCS operates at a repetition rate of 25 Hz. The MSS has a relatively lower repetition rate of 0.5 Hz. The SS, which utilizes superconducting magnets with a maximum dipole field of approximately 8 Tesla, operates at an even slower repetition rate. The higher repetition rates for the earlier stages help minimize the SS cycling period, thus reducing the overall SPPC beam fill time.

To facilitate maintenance, cost efficiency, and accommodate the physics programs, the first three stages of the injector chain will be constructed at a relatively shallow underground level, around –15 m. On the other hand, the SS, with its larger circumference, will be built either at the same level as the SPPC or at a deeper underground level ranging from –50 m to –100 m.

As outlined in Table 8.4.1, the different stages of the SPPC injector chain are required to operate for only fractions of the time when serving the SPPC collisions. However, there is potential for these stages to operate with longer duty cycles or continuously to generate high-power beams for other research applications during periods when they are not involved in the SPPC operations. Currently, the bunch population at the SPPC is primarily limited by the synchrotron radiation (SR) power. This implies that the accelerators in the injector chain have the capacity to accommodate a higher number of accumulated particles per pulse or deliver increased beam power for their own distinct applications when not specifically serving the SPPC. This capability is not only valuable for their standalone research purposes but also holds significance for future upgrades of the SPPC.

Given the complexity of such an injector system, the construction and commissioning process is expected to span approximately 10 years, advancing stage-by-stage. Therefore, it is desirable to initiate the construction of the injector accelerators several years prior to the SPPC construction. This allows for an overlap between the operation of the CEPC and the construction of the injector chain, ensuring efficient use of time and resources.

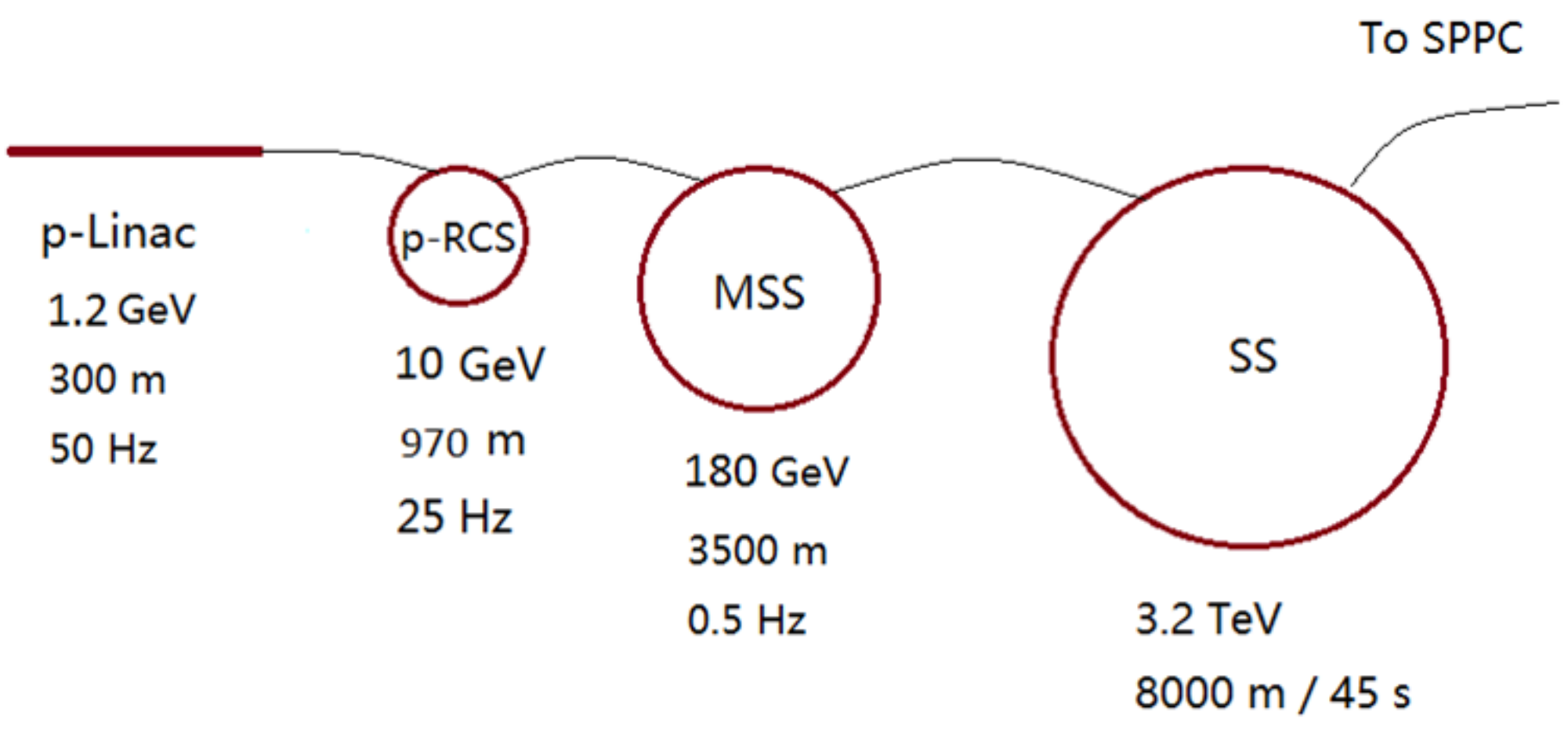

**Figure 8.4.1:** Injector chain for the SPPC.

## 8.4.2 **Preliminary Design Concepts**

### 8.4.2.1 ***Linac (p-Linac)***

Superconducting linacs have experienced remarkable advancements over the past two decades [1], and it is anticipated that they will continue to progress significantly in the coming decades. With this in mind, we have chosen a design for the p-Linac with an energy of 1.2 GeV and a repetition rate of 50 Hz. The continuous beam power is estimated to be 1.63 MW, with the potential for at least half of this power to be allocated for various other applications beyond the primary purpose.

#### 8.4.2.2 *Rapid Cycling Synchrotron (p-RCS)*

The p-RCS is designed to deliver a continuous beam power of 3.4 MW, which is a significant achievement. In fact, only one other proton driver study, specifically for a future Neutrino Factory, has demonstrated comparable performance [2]. The high repetition rate of 25 Hz is a crucial factor in reducing the beam filling time in the MSS. Since only a fraction of this power is required to fill the MSS, the majority of the beam pulses generated by the p-RCS can be effectively utilized for other physics programs. The p-RCS will employ well-established accelerator technology but on a larger scale than existing rapid-cycling proton synchrotrons. To facilitate its operation, high-Q ferrite-loaded RF cavities are planned to provide a very high RF voltage of approximately 3 MV. The RF frequency swing of 36-40 MHz has been selected to achieve the desired bunch spacing of 25 ns, which is essential for the SPPC's operational requirements.

#### 8.4.2.3 *Medium Stage Synchrotron (MSS)*

The MSS is designed to deliver beam power similar to that of the p-RCS, but with significantly higher beam energy and a much lower repetition rate. The design of the MSS draws inspiration from successful accelerators such as the SPS at CERN and the Main Injector at Fermilab. However, due to the higher beam power, stringent control measures must be implemented to minimize beam loss. To ensure efficient operation, the MSS will employ the same RF system as the p-RCS. Additionally, a 200-MHz RF system has been reserved for future bunch splitting, which will enable a 5-ns bunch spacing. This advanced RF system will enhance the versatility of the MSS, allowing it to serve additional physics programs beyond its role as an injector for the SS.

#### 8.4.2.4 *Super Synchrotron (SS)*

The SS will utilize superconducting magnets similar to those employed at the LHC, albeit with a higher ramping rate. Since the energy of the SS is significantly lower than that of the LHC, the impact of synchrotron radiation is negligible and does not need to be taken into consideration. Currently, there are no evident critical technical risks associated with the construction of the SS. However, it remains uncertain whether the beam from the SS will have dedicated physics programs beyond its role as an injector for the SPPC. Further exploration is required to determine additional potential applications for the SS.

**Table 8.4.1:** Main parameters of the injector chain for the SPPC

| | Energy | Average current | Length/ Circum. | Repetition Rate | Max. beam power or energy | Dipole field | Duty factor for next stage |
|---|---|---|---|---|---|---|---|
| | GeV | mA | Km | Hz | MW/MJ | T | % |
| p-Linac | 1.2 | 1.4 | ~0.3 | 50 | 1.6/ | - | 50 |
| p-RCS | 10 | 0.34 | 0.97 | 25 | 3.4/ | 1.0 | 6 |
| MSS | 180 | 0.02 | 3.5 | 0.5 | 3.7/ | 1.7 | 13.3 |
| SS | 3200 | - | 8.0 | 1/50 | /51.6 | 11.3 | 1.3 |

To accommodate heavy-ion beams at the injection energy of the MSS, the implementation of a dedicated heavy-ion linac (i-Linac) and a new heavy-ion synchrotron (i-RCS) is essential. This infrastructure will run in parallel to the existing proton linac/RCS. The i-Linac will be responsible for accelerating heavy-ion beams to the desired injection energy, while the i-RCS will facilitate their subsequent acceleration to achieve a beam rigidity of approximately 36 T-m. This beam rigidity is equivalent to that of a 10 GeV proton beam. By incorporating the i-Linac and i-RCS into the accelerator complex, the SPPC will be capable of accommodating and studying heavy-ion beams alongside proton beams.

# 9 Conventional Facilities

## 9.1 Introduction

In the pre-CDR and CDR phases of the CEPC project, preliminary studies of geological conditions for potential site locations were carried out in several provinces in China. These provinces included Hebei, Guangdong, Shanxi, Zhejiang, Jiangsu, Hunan, and Jilin as shown in Fig. 9.1.1. Geological surveys were conducted in six specific sites: Qinhuangdao in Hebei, Shen-Shan in Guangdong, Huangling in Shanxi, Changchun in Jilin, Huzhou in Zhejiang, and Changsha in Hunan.

In the TDR, the focus of site selection has narrowed down to three specific sites. These sites are Qinhuangdao in Hebei, Huzhou in Zhejiang, and Changsha in Hunan. This chapter discusses in detail each of the three sites. They all meet the general site requirements of the CEPC project.

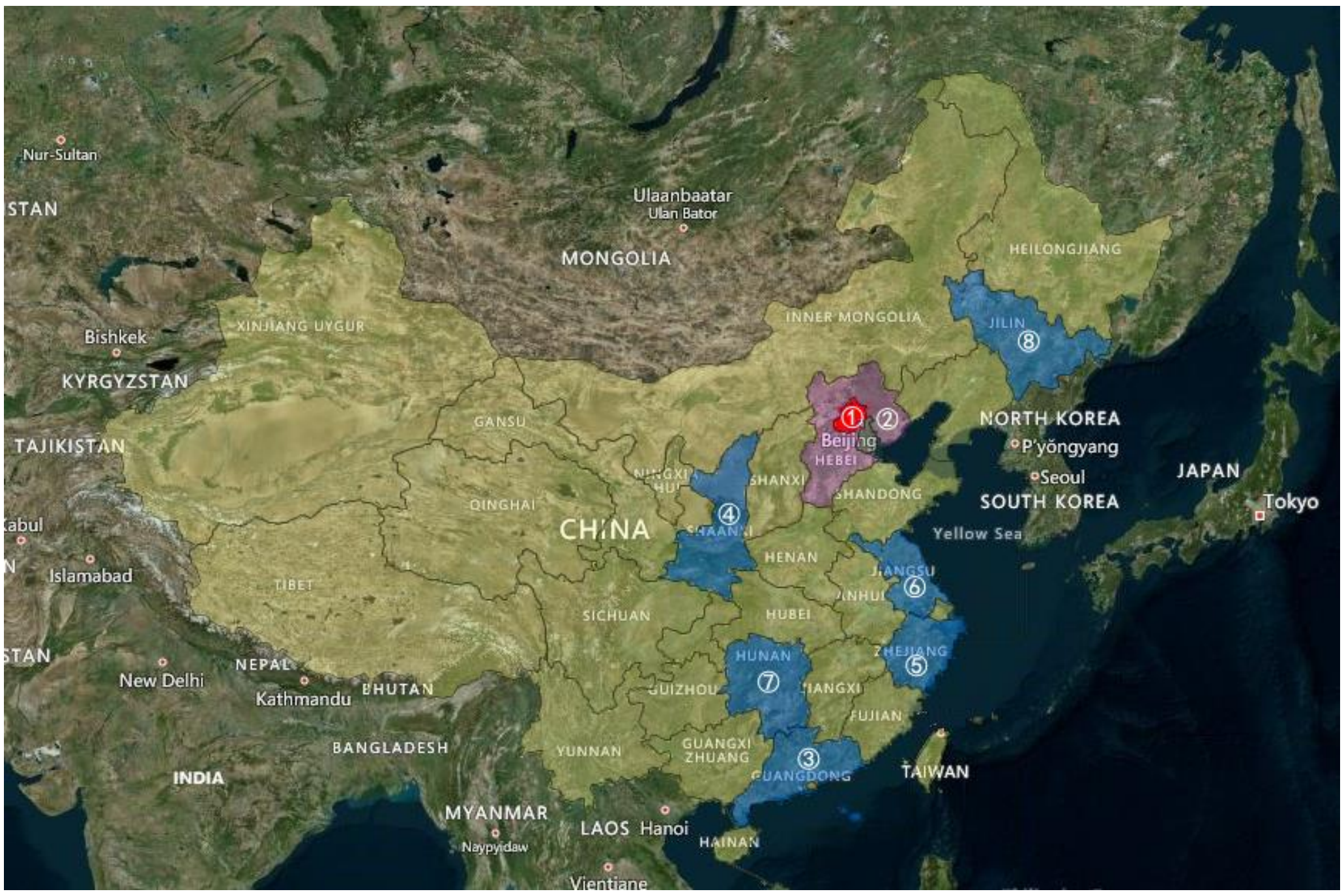

**Figure 9.1.1:** Geographical representation of CPEC project potential sites – the key regions related to the CEPC project, encompassing the Capital Beijing and seven provinces. Our report concentrates on three specific regions: Qinhuangdao (located in Hebei, near Beijing), Huzhou (in Zhejiang), and Changsha (in Hunan).

① Beijing ② Hebei ③ Guangdong ④ Shaanxi ⑤ Zhejiang ⑥ Jiangsu ⑦ Hunan ⑧ Jilin

## 9.2 Overall Layout and Common Design Criteria

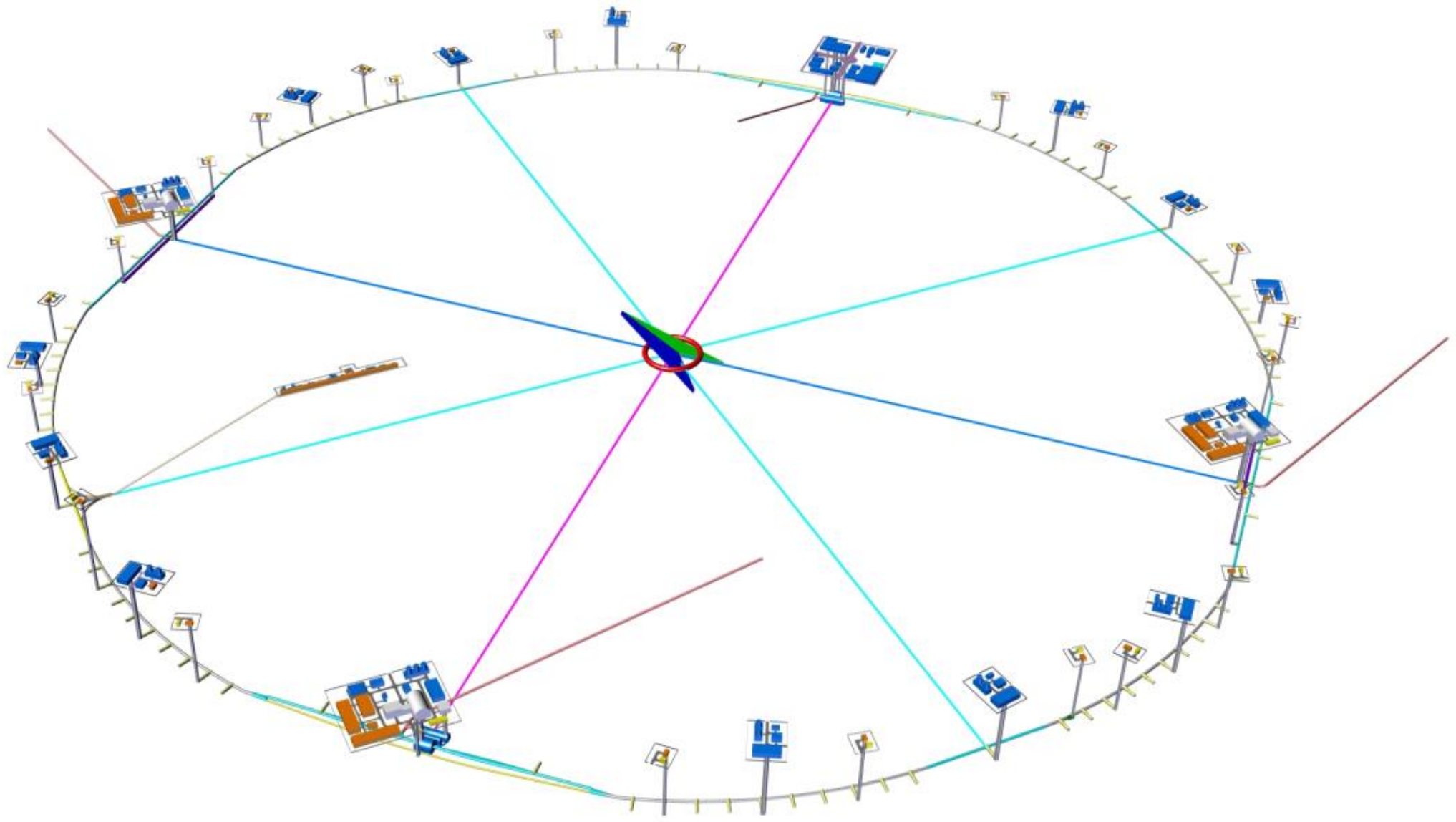

**Figure 9.2.1:** Layout of surface and underground CEPC structures.

CEPC is composed of a Collider, an injection system into the Collider (consisting of a Linac, a Booster, a Damping Ring, and transport lines), and two large physics detectors, all housed in civil construction as illustrated in Fig. 9.2.1. The design and construction of each part of the CEPC are determined by factors such as their geometric relationships, environmental conditions, and safety considerations. Practicality, adaptability, and operating efficiency are key criteria that must be carefully considered in the design of the civil construction.

The scope of work for the CEPC project includes several requirements that must be met. These requirements are:

1. Construction of a main tunnel with a circumference of 100 km and about 100 m below ground to house the Collider and Booster, auxiliary tunnels for the Booster bypass and RF equipment, the Linac tunnel (on surface) and equipment gallery, and transport line tunnels.
2. Creation of experiment halls that span 30-40 m and are located about 100 m below ground, along with additional chambers for power sources, cryogenics, water cooling systems, etc.
3. Provision of accesses to the experiment halls, including access tunnels, transport shafts, and emergency exits.
4. Construction of ancillary structures at ground level with a total area of 140,450 $m^2$, such as structures near the shaft openings, substations, electric distribution, cryogenics rooms, and ventilation fan rooms.
5. Space for staging construction equipment and materials, as well as dumping sites.
6. Provision of related lifting equipment, conveyance systems, electric supply systems, drainage systems, ventilation and air conditioning systems, communication systems, control and monitoring systems, safety escape systems, and firefighting systems, including fire alarms, hydrants, gas fire-extinguishing systems, and a smoke exhaust system. Maintenance and potential future upgrades of these systems are fully considered in their design.

## 9.3 General Site Requirements

When selecting the site for the CEPC, several factors beyond engineering technology conditions, such as topography and geology, must be taken into account. These factors include the location of the site, local government support, the social and cultural environment, regional development, and the potential environmental impact of the construction. These external construction conditions may ultimately prove to be the decisive factor in site selection.

The site selection for the CEPC should take into consideration the following factors:

1. Geography:
   The site should be large enough and appropriately located to accommodate the future development of the Institute of High Energy Physics of the Chinese Academy of Sciences. It should also promote the CEPC project and the construction of an international science city.
2. Natural conditions:
   a) The site should have good structural stability conditions and avoid deep faults and motions and deformations that are recent in geologic time. Seismic peak acceleration should generally be less than 0.10 g.
   b) Good rock conditions are necessary. Large areas of hard rock with stable lithology are suitable for construction of underground caverns.
   c) The site should not have large height differences, and mostly low mountains and hilly areas are preferred.
   d) The quaternary overburden should not be thick.
   e) The permeability of rock should be relatively low.
   f) External dynamic geological phenomena should be relatively small.
3. Access conditions:
   To minimize capital costs and accelerate the progress in the initial stages of the construction, the site should be located where access is convenient.
4. Environmental factors:
   The site should have few environmental impact problems and should not involve environment sensitive zones, such as nature preserves, parks, military areas, or other environmental constraints such as wetlands.
5. Economic factors:
   The site should have good construction conditions and related economic factors.

## 9.4 Qinhuangdao Site

### 9.4.1 Siting Studies

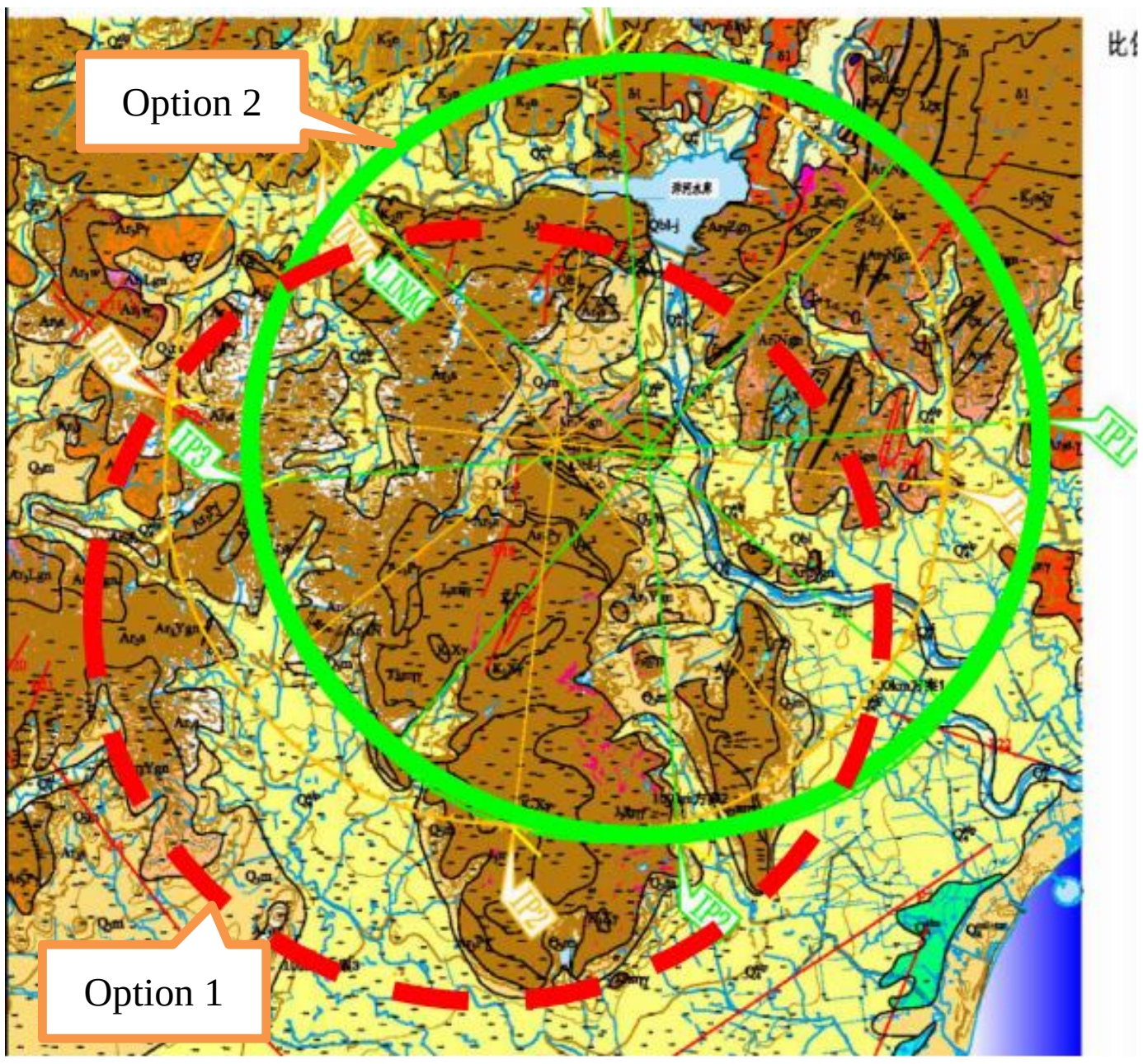

**Figure 9.4.1:** Two options of the Qinghuangdao site.

There are two options for the CEPC project at the Qinghuangdao site, as shown in Fig. 9.4.1. During the CDR stage, Option 1 was chosen as the representative scheme through comparison. However, in recent years, the northern part of Changli County has undergone rapid development. In the south of Option 1, a new main road has been constructed, and many residential areas have been built on both sides of the road, resulting in a high population density. Therefore, after comparing and selecting route lines in the TDR stage, Option 2 was selected as the representative site scheme.

The Qinhuangdao site is situated in the Funing District of Qinhuangdao City, in Hebei Province, including the Beidaihe District, Changli and Lulong Counties. The region is hilly, with elevations ranging from 0 to 600 m, and consists of Archaean gneiss, Mesozoic magmatic rocks, volcanic rocks from the Yanshan period, and some Mesozoic sand shale. The rock is predominantly hard with no thick overburden and has a basic seismic intensity of degree VII, making it suitable for constructing large underground caverns and tunnels. The depths of the underground caverns are relatively consistent.

### 9.4.2 Site Construction Condition

#### 9.4.2.1 *Geographical Position*

The Qinhuangdao site is situated in the northeast of Hebei province and the northwest of Qinhuangdao city. It is located 478 km north of Shijiazhuang, the provincial capital, 240 km east of Beijing, and 23 km west of Qinhuangdao.

### 9.4.2.2 *Transporttion Condition*

Funing District benefits from various convenient transportation options. It is well-connected by the Tianjin-Qinhuangdao passenger dedicated line, the main national railway, and local railway trunk lines such as the Beijing-Qinhuangdao high-speed railway, as well as the Beijing-Harbin, Tianjin-Shanhaiguan, Datong-Qinhuangdao, and Qinhuangdao-Shanhaiguan railways. Additionally, the district is easily accessible by road via the Beijing-Shenyang, Coastal, and Qinhuangdao-Chengde expressways, as well as two national roads (102 and 205) and five provincial roads. Qinhuangdao port is just 35 km away, while Shanhaiguan airport and Qinhuangdao's new airport are 45 km and 25 km away, respectively.

### 9.4.2.3 *Hydrology and Meteorology*

The climate of Qinhuangdao is classified as semi-humid continental monsoon and falls within the warm temperate zone. It is also influenced by the marine climate, resulting in four distinct seasons, ample sunshine, and abundant rainfall. The mean air temperature is 10.2℃, and there is a frost-free period of 177 days. The average annual rainfall is 730.7 mm, and there are 2,745 hours of sunshine per year.

Qinhuangdao is home to several perennial rivers, including the Luanhe, Yinma, Yanghe, Daihe, and Tanghe. All of these rivers flow from north to south and empty into the Bohai Sea.

The Yanghe Reservoir, located in Funing District, Qinhuangdao City, is approximately 2.8 km away from CEPC. It serves as a large-scale water conservancy hub for flood control, irrigation, power generation, and fish farming. The reservoir is situated on the mainstream of Yanghe River, with a basin area of 755 $km^2$ and a total storage capacity of $3.86 \times 10^8$ $m^3$.

Qinhuangdao has numerous small reservoirs in its vicinity.

### 9.4.2.4 *Economics*

As of the end of 2020, Qinhuangdao had a registered population of 3.0018 million people.

Qinhuangdao is a robust economic city in Hebei Province, and it was one of the first 14 coastal cities to open up to foreign investment in China. It has earned a place among the top 40 cities in the country for investment attractiveness, and it is home to several economic and technological development zones, including the national Qinhuangdao Economic and Technological Development Zone, Qinhuangdao Export Processing Zone, and Yanshan University Science Park.

In 2021, Qinhuangdao's GDP reached 184.376 billion yuan, reflecting a 6.8% increase over the previous year. The city's per capita GDP rose to 58,774 yuan, representing a 6.6% increase over the previous year.

### 9.4.2.5 *Engineering Geology*

The chosen site is situated in a hilly area, with higher elevation in the west and lower elevation in the east. The Yanghe River and Yinma River systems are the main river networks in this region. The dominant geological formations consist of metamorphic rocks from the Archaeozoic era, including gneiss and schist, Mesozoic magmatic rocks such as granite, and volcanic rocks such as tuff, along with a smaller amount of sandstone and mudstone intercalated with sandstone from the Mesozoic era. The rocks are primarily hard, and the surface overburden is relatively thin, while the thickness of river alluvium ranges from 15 to 20 m. The selected site area does not have deep regional fractures, with a ground motion peak acceleration of 0.10~0.15 g and a basic seismic intensity of degree VII, indicating a relatively stable area. Two types of groundwater are present, including pore water in loose rocks and fissure water in the bedrock weathered zone, with the latter being less abundant. The site area does not have significant exogenic geological processes and the weathered zone thickness is 20 to 30 m, making it suitable for large-scale underground construction projects without major engineering geological restrictions.

The engineering geological considerations for the project include:

1. Water ingress into the tunnel:
   The possibility of water bursting should be considered in areas passing through the Yanghe River downstream of the Yanghe Reservoir, in the southeast of the Yanghe Reservoir, and corresponding alluvial plains. Local fault fractured zones may also be susceptible to water ingress, especially in areas with a thick partially weathered zone.
2. Stability of surrounding rocks:
   Most of the tunnel sections consist of slightly weathered to fresh rock mass, indicating relatively stable surrounding rocks. However, in some tunnel sections, moderately weathered rock mass may cause stability issues. In areas where the tunnel passes through fault fractured zones, surrounding rock stability is poor and remedial measures are necessary. The inlet section of the vertical shaft consists of moderately weathered rock mass with poor stability, requiring corrective measures. The experiment halls have a large span and high side walls, and whether underground excavation or open excavation methods are used, block stability issues of the side walls should be taken into account. If open excavation is used, the surrounding rocks above the upper moderately weathered zone are of poor stability.

### 9.4.3 Project Layout and Main Structure

#### 9.4.3.1 *General Layout of the Tunnel and Surface Structures*

##### *9.4.3.1.1 General Layout Principles and Requirements*

The general layout principles and requirements are as follows:

1. The tunnel's layout, length, and buried depth are designed to meet the specific needs of the accelerator and detectors.
2. The operation of the tunnel must prioritize security, easy management, and convenient traffic flow.
3. The geologic structure surrounding the tunnel is simple, and the hydrogeological conditions are suitable for safe construction.
4. Access to hydroelectricity is readily available for the project.
5. Shafts and adits will serve as entry points to the tunnel.
6. Shafts will avoid densely populated areas, and auxiliary facilities, such as cooling towers and substations, will be located near access shafts.
7. The tunnel layout must meet the requirements for transportation and installation of experimental equipment.

8. The number and length of construction adits will be determined by terrain, geologic conditions, construction methods, and external traffic, optimizing person-hours and time requirements for each tunnel section.
9. Efforts will be made to minimize the impact on the local ecology, and surface facilities will avoid existing buildings.
10. The project must adhere to government regulations and norms.

#### *9.4.3.1.2 General Layout*

Figures 9.4.2 and 9.4.3 depict the plan and profile of the 100-km circumference tunnel. The longitudinal slope of the tunnel is set at 0.3% to meet topological and drainage needs during both construction and operation. The surrounding rocks of the tunnel primarily comprise granite, gneiss, schist, and tuff, and are mainly classified as Class II to III.

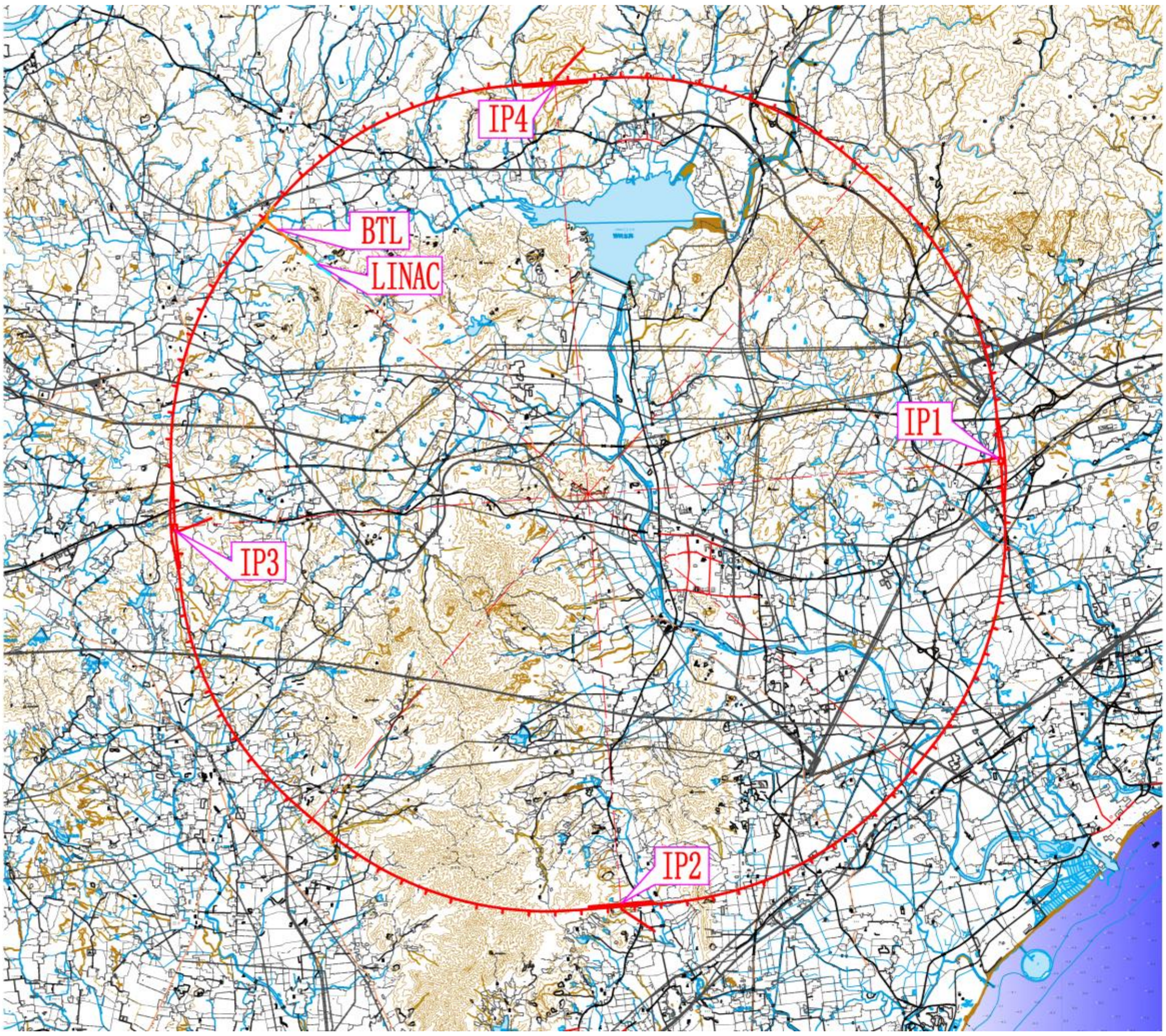

**Figure 9.4.2:** Layout plan of the CEPC tunnel.

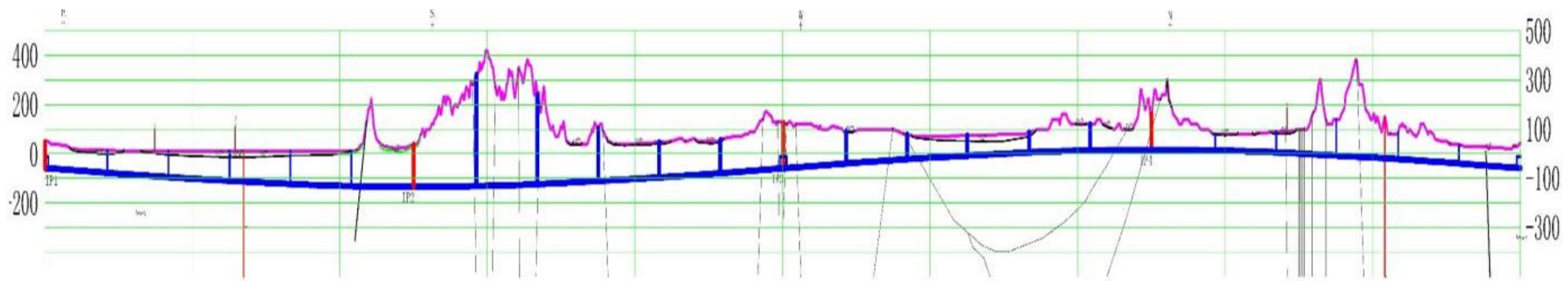

**Figure 9.4.3:** Longitudinal profile of the CEPC tunnel.

Underground structures consist of the following as shown in Fig. 9.4.4:

- Collider and Booster ring tunnel.
- Experiment halls (includes main and service caverns): IP1 and IP3 are experiment halls for CEPC, and IP2 and IP4 are RF sections for CEPC, which will also be the locations for future experiment halls for SPPC.
- Linac and BTL tunnels: Linac tunnel, hall for the Damping Ring, BTL tunnel and its branch tunnels
- Auxiliary tunnels: RF auxiliary tunnels near IP2 and IP4, Booster bypass tunnels near IP1 and IP3, and many short auxiliary tunnels.
- Vertical shafts in experiment halls and RF zones and along the ring tunnel for personnel and delivery of equipment to tunnels and halls, and for providing channels for ventilation, refrigeration, cooling and control and monitoring lines.

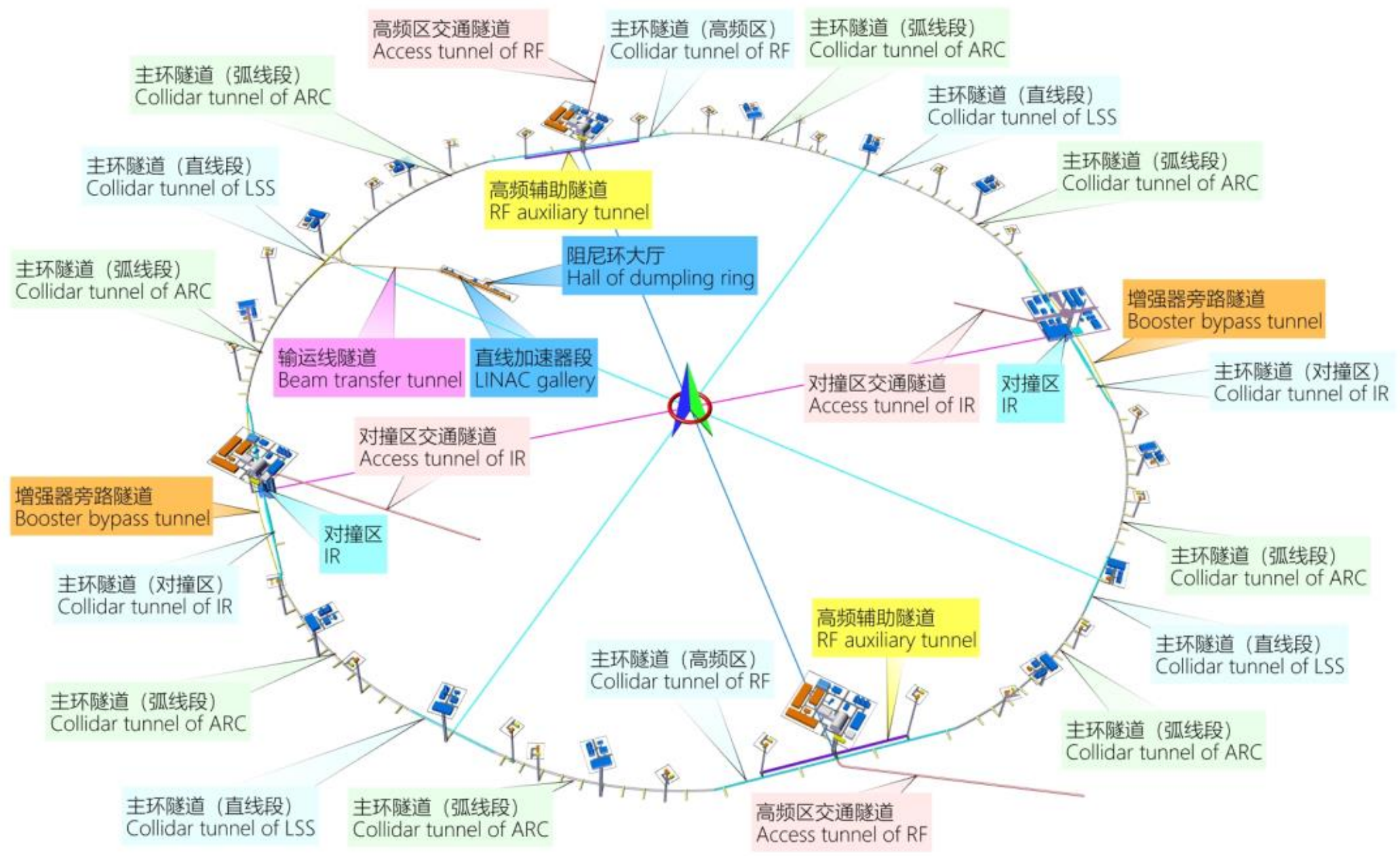

**Figure 9.4.4:** Layout of the underground structure.

Surface structures such as auxiliary equipment structures, cooling towers, substations, and ventilation systems located within the main ring area are strategically placed near the vertical shafts. This ensures easy access to the underground network for personnel and equipment, as well as facilitating channels for ventilation, refrigeration, cooling, and control and monitoring lines.

## 9.4.3.2 *Underground Structures*

### *9.4.3.2.1 Collider Tunnel*

The Collider tunnel consists of the following sections:

- 4 curved arc sections, each with a length of 10,270.44 m
- 4 curved arc sections, each with a length of 10,185.70 m
- 2 interaction regions (IRs), IP1/IP3, each with a length of 3,337.13 m
- 2 linear sections for RF, IP2/IP4, each with a length of 3,776.90 m
- 4 linear sections (LSs), each with a length of 986.84 m

The total tunnel length is 100 km. The tunnel structures are shown in Fig. 9.4.5 and itemized in Table 9.4.1.

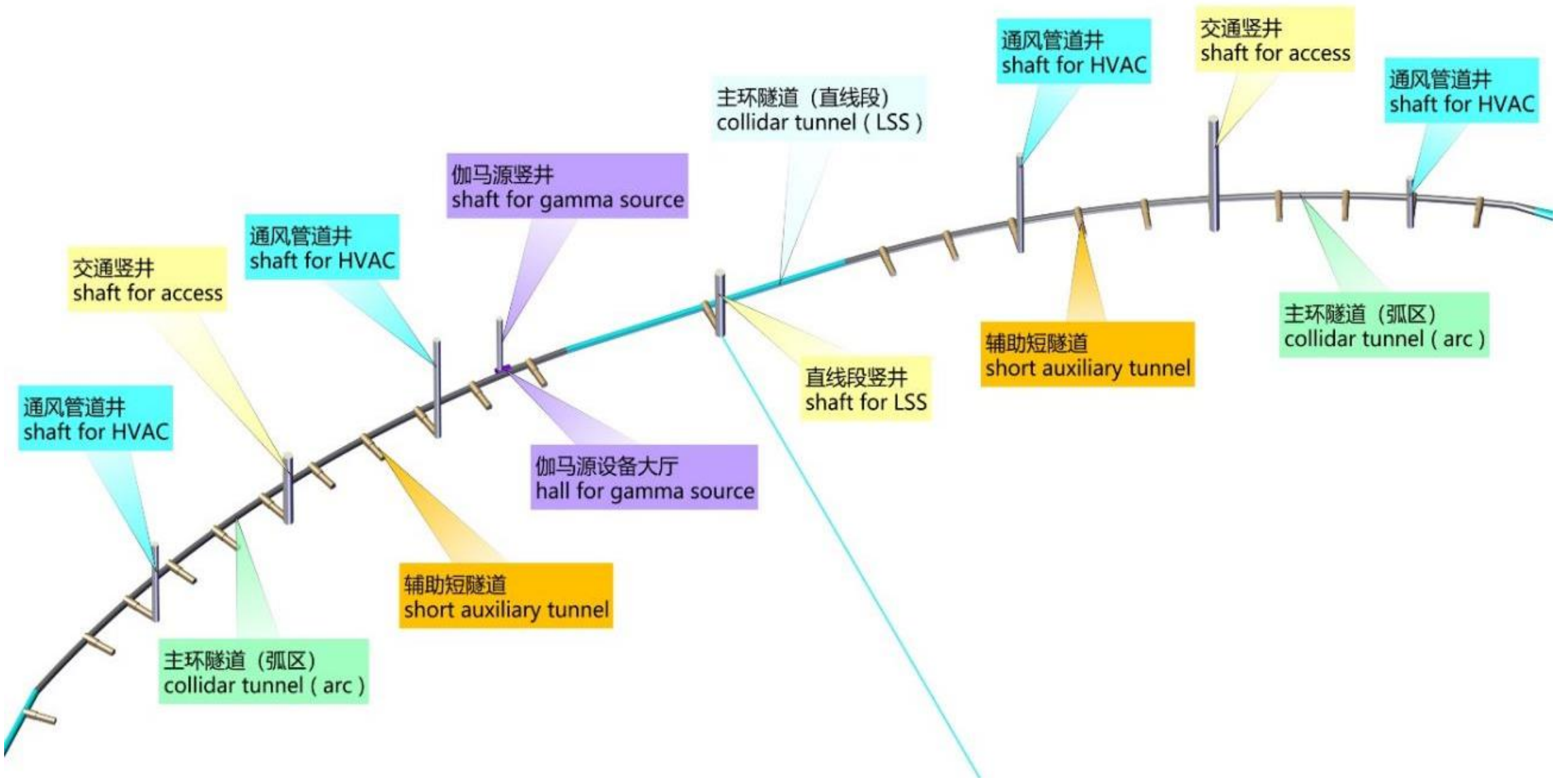

**Figure 9.4.5:** Layout of the Collider ring tunnel.

**Table 9.4.1:** Characteristics of the Collider tunnel structures.

| Item | | Unit | Qty | | | Remarks |
|---|---|---|---|---|---|---|
| Linear section | Qty | Section | 2 | 2 | 4 | IR = 3337.13 m<br>RF = 3776.90 m<br>LS = 986.84 m<br>Width varies from 6.00 to 11.40 m. H is 4.5 m |
| | Length | m | 3337.13 | 3776.90 | 986.84 | |
| | Dimension (∩-shaped) | m | 6.00 | × | 5.00 | |
| Arc section | Qty. | Section | 4 | | 4 | IR-LSS = 10270.44 m<br>LSS-RF = 10185.70 m |
| | Length | m | 10270.44 | | 10185.70 | |
| | Dimension (∩-shaped) | m | 6.00 | × | 5.00 | |
| Collider ring diameter | | m | 31831 | | | |
| Total length of tunnel | | km | 100.00 | | | |
| Longitudinal gradient of tunnel | | | 0.30% | | | |

The tunnel cross section is divided into three parts as shown in Fig. 9.4.6:

- Outer side, reserved for SPPC equipment.
- Inner side, where CEPC equipment and components will be installed.
- Central part of the tunnel for equipment operation and transport.

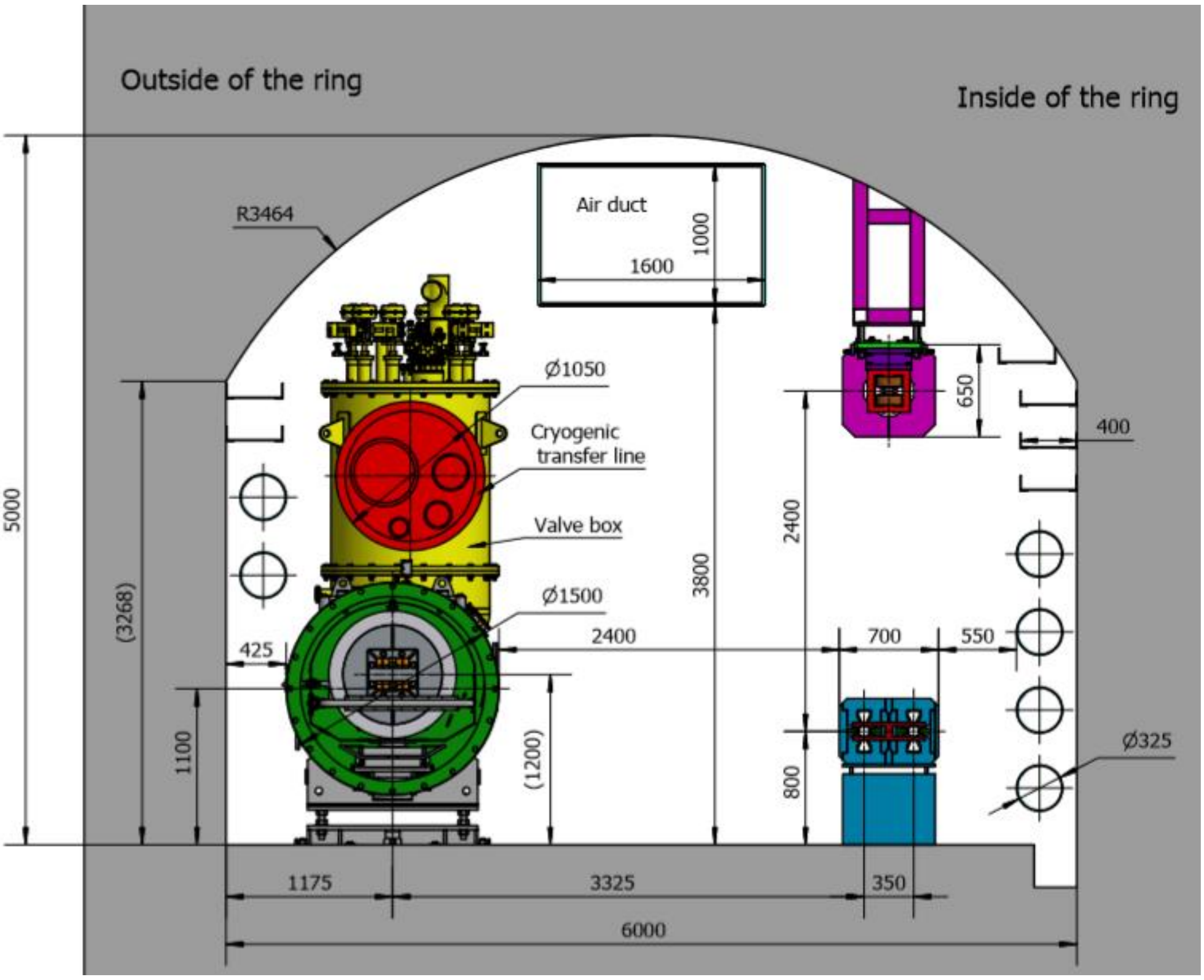

**Figure 9.4.6:** Tunnel cross section in the arc sections.

#### *9.4.3.2.2 IR Sections IP1 and IP3*

IP1 and IP3 are where the large detectors will be located. Therefore, in these straight sections, the Booster beam must bypass around the detectors. Zones within 1,509.30 m of both ends are divided into two tunnels. One tunnel is for the e+/e- colliding beams, 6.00 m to 11.40 m wide, 4.50 m high, and 1509.30 m long in each direction. The tunnel for the Booster bypass is 3.50 m wide, 3.50 m high, and 3,018.60 m long. Additional details can be found in Table 9.4.2, and Figs. 9.4.7 and 9.4.8 illustrate the arrangement. The dimensions are determined based on the requirements of local control, electrical systems, beam instrumentation, vacuum, and cooling equipment, as well as space for the passage of personnel and maintenance equipment.

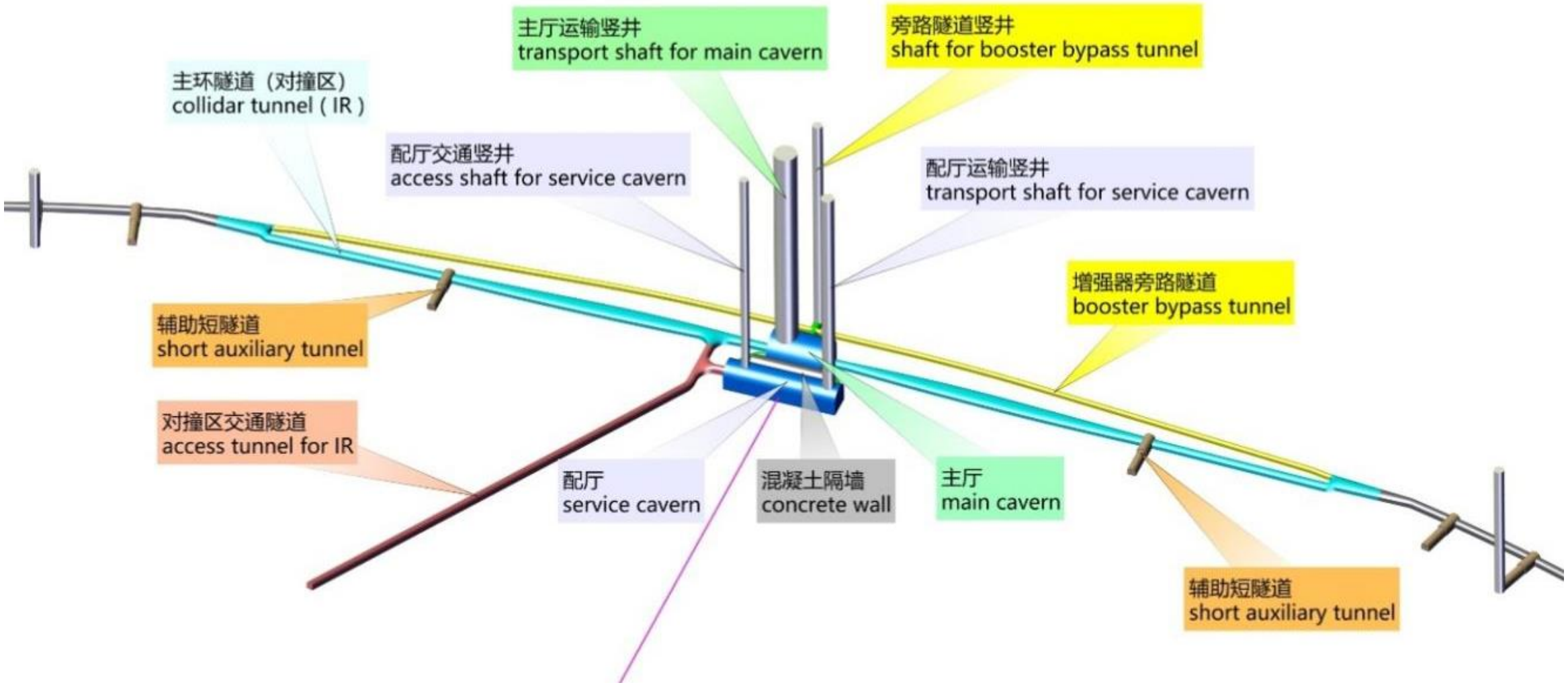

**Figure 9.4.7:** Tunnel layout in the IR area.

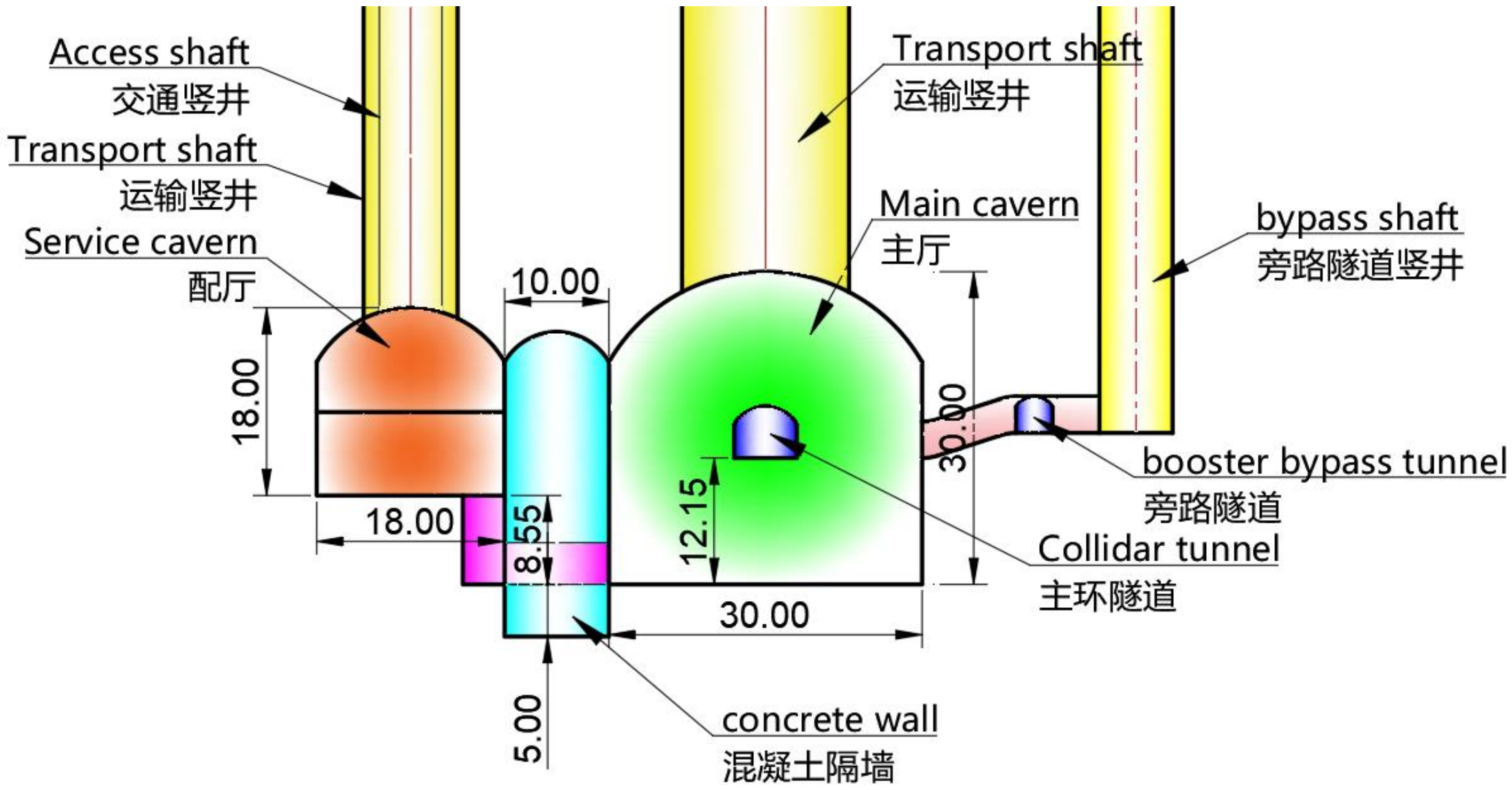

**Figure 9.4.8:** Tunnel cross section in the IR area.

**Table 10.4.2:** Structures of the Booster bypass tunnel in the IR area.

| Item | | Unit | Qty |
|---|---|---|---|
| Bypass tunnels in the IR area | Qty | Nos. | 2 |
| | Length of individual tunnel | m | 3018.60 |
| | Dimension (∩-shaped) | m | 3.50 × 3.50 |

*9.4.3.2.3 RF Sections IP2 and IP4*

The RF cavities are located in IP2 and IP4, which are also the locations for future SPPC detectors. The straight sections in these areas will serve as the RF tunnels. The tunnel is 6.00 m wide, 5.00 m high, and 3,776.90 m long. An auxiliary tunnel of length 1,800.00 m is parallel to the RF tunnel and symmetric around the RF system midpoint, which is 8.00 m wide and 7.00 m high. The RF tunnel, illustrated in Figs. 9.4.9 and 9.4.10 with additional details in Table 9.4.3, will be used to house RF power sources, water cooling equipment, vacuum, and cooling equipment, and will have sufficient room for personnel and maintenance equipment.

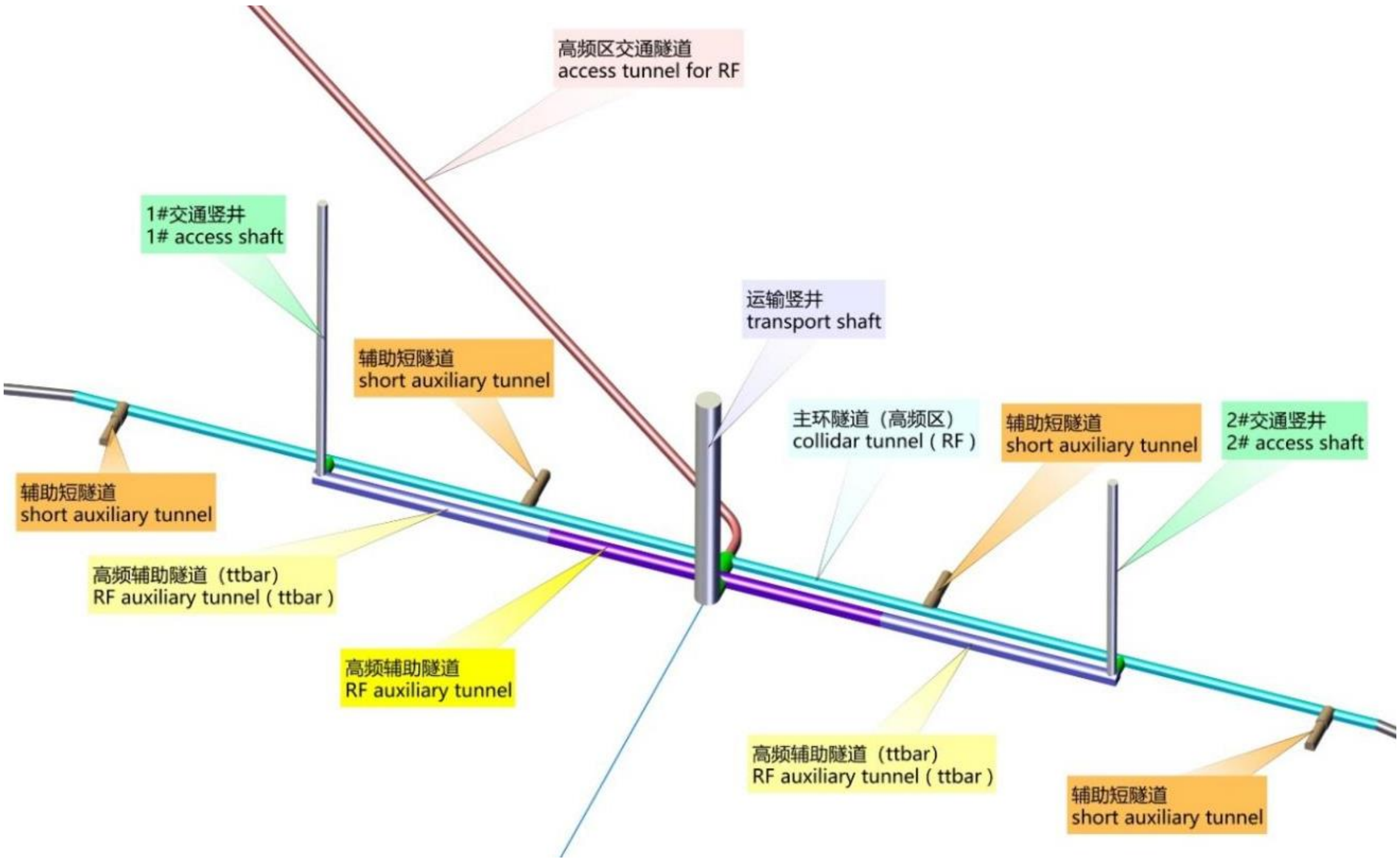

**Figure 9.4.9:** Tunnel layout in the RF Zone.

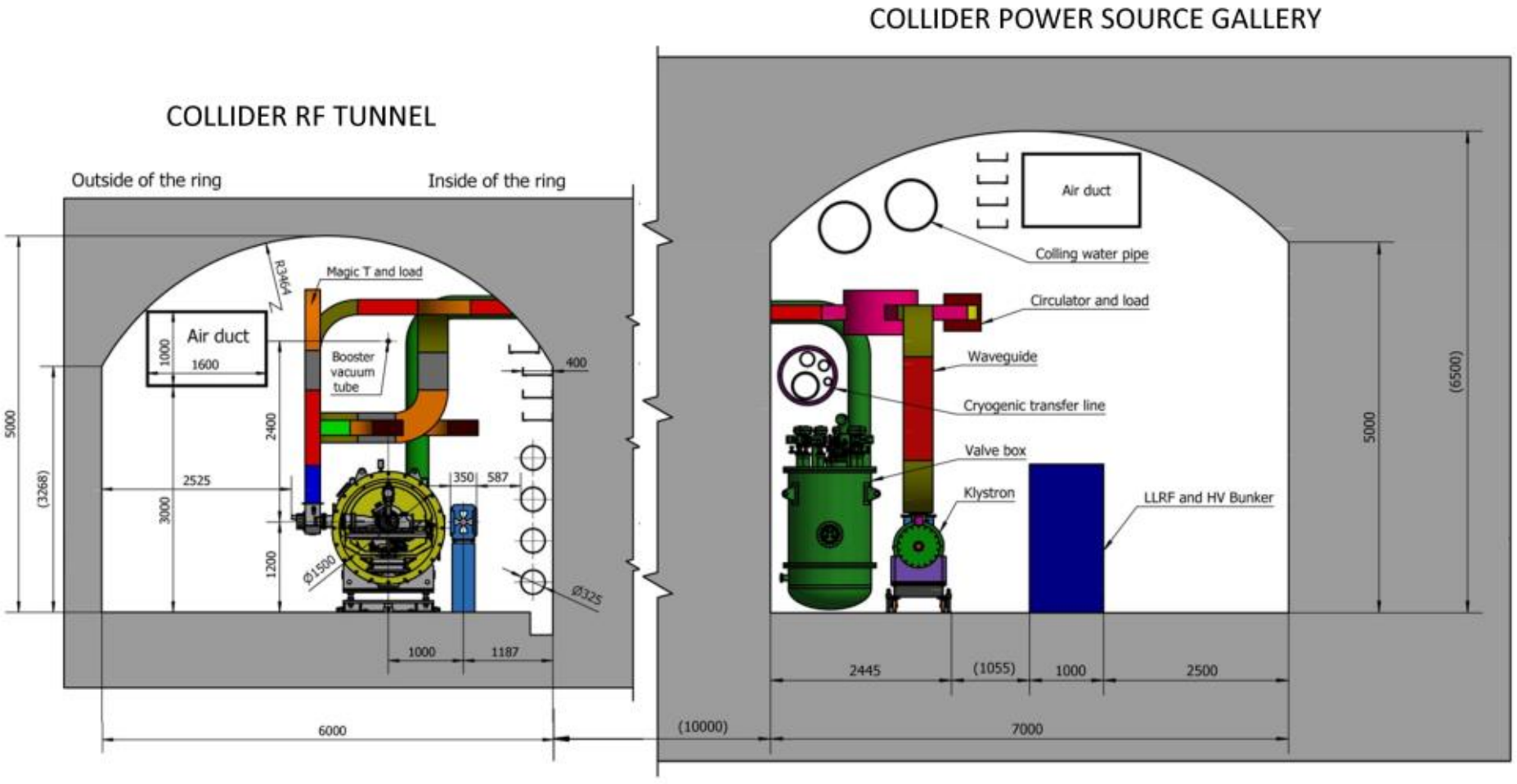

**Figure 9.4.10:** Tunnel cross Section in the RF Zone.

**Table 9.4.3:** Structure of the service tunnel in the RF zone.

| Item | | Unit | Qty |
|---|---|---|---|
| Service tunnels in the RF zone | Qty | Nos. | 2 |
| | length of individual tunnel | m | 1800 |
| | Dimension | m | 8.00 × 7.00 |

### *9.4.3.2.4 Linear Section of the Collider Tunnel*

There are four linear sections (LSS) symmetrically located around the ring. Each linear section has a length of 986.84 m and the same cross section as in the arc sections, 6.00 m wide and 5.00 m high.

### *9.4.3.2.5 Auxiliary Tunnels*

There are 96 auxiliary tunnels uniformly distributed around the ring for the underground power transmission system, electronic equipment, and other auxiliary components, as shown in Table 10.2.4. Each auxiliary tunnel is 30.00 m long, and has a cross section of 7.00 m × 7.00 m for the first 10.00 m and 6.00 m × 6.00 m for the remaining 20.00 m.

**Table 10.4.4:** Auxiliary tunnels

| Item | | Unit | Qty |
|---|---|---|---|
| Auxiliary tunnels | Qty | Nos. | 96 |
| | length of individual tunnel | m | 30 |
| | Dimension (∩-shaped) | m | 7.00 × 7.00 (Front 20 m) |
| | | m | 6.00 × 6.00 (Rear 10 m) |

### *9.4.3.2.6 Collider Experimental Hall*

The experimental halls have a large size of 50 m × 30 m × 30 m (L×W×H), while the service caverns for the detectors are 80 m × 18 m × 18 m in size. The main cavern floor level is located 12.15 m below the collider ring tunnel and 8.55 m below the service cavern base, as shown in Fig. 10.4.8 and listed in Table 10.4.5.

**Table 10.4.5:** Experimental halls

| Item | | Unit | Qty |
|---|---|---|---|
| Cavern Qty | | Nos. | 2 |
| Main cavern | Dimension (L×W×H) | m | 50 × 30 × 30 |
| Service cavern | Dimension (L×W×H) | m | 80 × 18 × 18 |

### *9.4.3.2.7 Linac to Booster Transfer Tunnel*

The Linac to Booster transfer (BTL) tunnel connects to the inner side of the linear section between IP3 and IP4. This BTL tunnel consists of three parts: two $e^+/e^-$ separation tunnels of 60 m long each, followed by an inclined downward tunnel of 1,020 m in length, which is connected to two branching tunnels, each of 460 m long. The cross section of the BTL tunnel has a width of 4.5 m and a height of 3.50 m.

The Linac tunnel is 3.50 m wide and 3.50 m high, with local width expanded to 5.50 m. The klystron gallery is 6.00 m high and 8.00 m wide.

At the middle of the Linac, there is a Damping Ring tunnel with dimensions of 3.50 m wide, 3.50 m high, and a circumference of 147 m. There are also two transport lines between the Linac and Damping Ring, each with a length of 120 m.

As with the other tunnels, the transfer tunnel has room for cooling, electrical and local control equipment, with clear space for pedestrians and room for moving and using maintenance equipment. Further details can be found in Fig. 9.4.11 and Table 9.4.6.

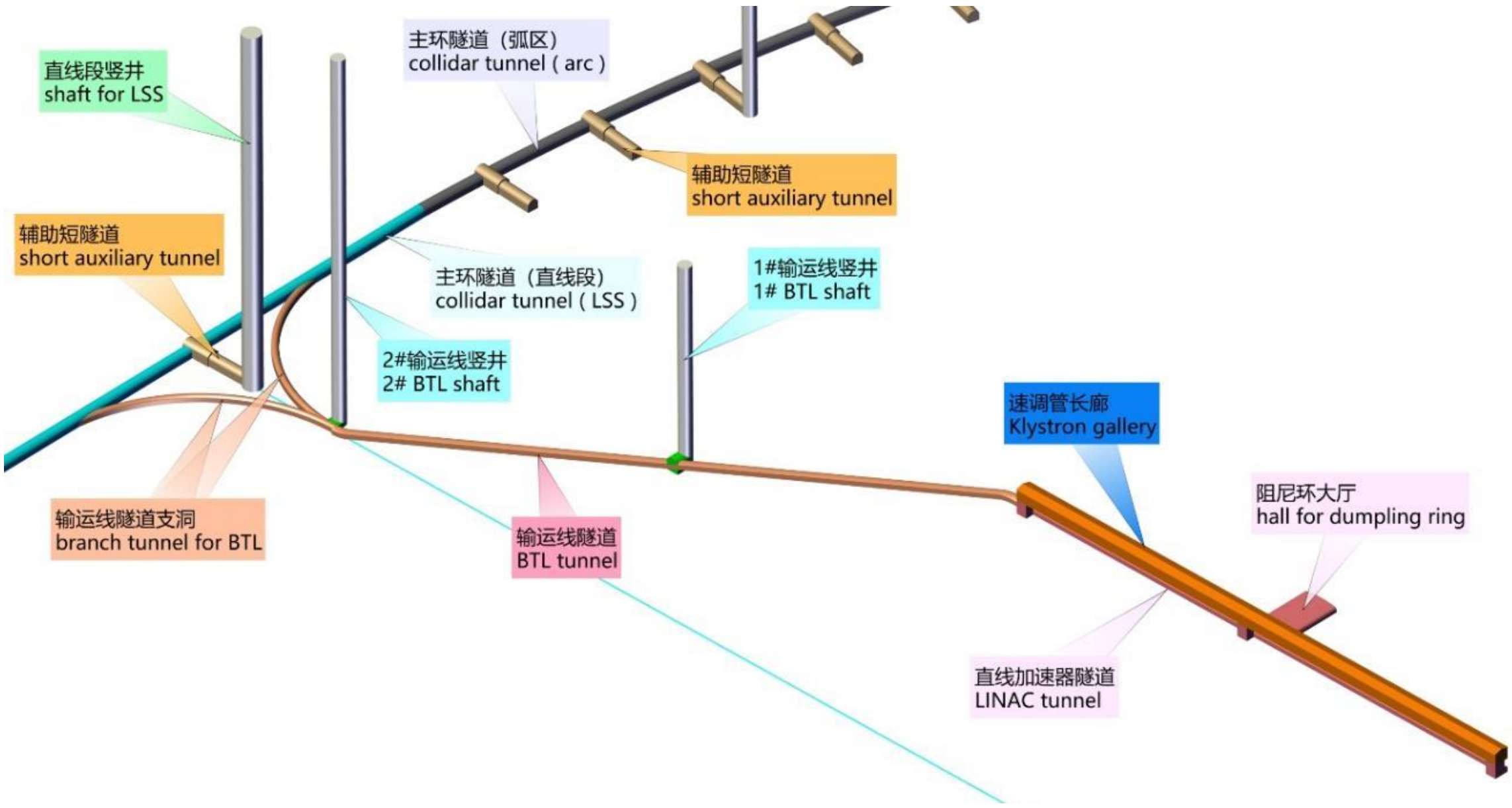

**Figure 9.4.11:** Layout Plan of the Linac and BTL tunnels.

**Table 9.4.6:** Linac and BTL tunnels.

| Item | | Unit | Qty | Remarks |
|---|---|---|---|---|
| Linac tunnel | Qty | Nos. | 1 | Local width 5.5 m |
| | length of individual tunnel | m | 1800 | |
| | Dimension (∩-shaped) | m | 3.50 × 3.50 | |
| Klystron gallery | Qty | Nos. | 2 | |
| | length of individual tunnel | m | 1800 | |
| | Dimension | m | 6.00 × 8.00 | |
| Damper Ring tunnel | Qty | Nos. | 1 | |
| | length of individual tunnel | m | 147 | |
| | Dimension (∩-shaped) | m | 3.50 × 3.50 | |
| DR transport line tunnel | Qty | Nos. | 2 | |
| | length of individual tunnel | m | 120 | |
| | Dimension (∩-shaped) | m | 3.50 × 3.50 | |

| Item | | Unit | Qty | Remarks |
|---|---|---|---|---|
| Linac beam separation tunnel | Qty | Nos. | 2 | |
| | length of individual tunnel | m | 60 | |
| | Dimension (∩-shaped) | m | 3.50 × 3.50 | |
| BTL tunnel before branch tunnel | Qty | Nos. | 1 | |
| | length of individual tunnel | m | 1020 | |
| | Dimension (∩-shaped) | m | 4.50 × 3.50 | |
| Branch tunnel of BTL | Qty | Nos. | 2 | 2 branch lines |
| | length of individual tunnel | m | 460 | |
| | Dimension (∩-shaped) | m | 3.00 × 3.00 | |
| Transport shaft for LINAC | Qty | Nos. | 3 | Transport shaft for LINAC shows a square shape for connecting the LINAC tunnel to klystron gallery |
| | Dimension | m | 6.00 × 6.00 | |
| Shaft for BTL | Qty | Nos. | 2 | |
| | Diameter | m | 7.00 | |

*9.4.3.2.8 Gamma-ray Source Tunnel and Experimental Hall*

The γ-ray source tunnel is situated at the anti-clockwise end of LSS2 and LSS4, serving as a link between the collider ring tunnel and the gamma source equipment hall. Detailed dimensions are provided in Table 9.4.7.

**Table 9.4.7:** Gamma source civil construction structures

| Item | | Unit | Qty | Remark |
|---|---|---|---|---|
| Gamma source tunnel | Qty | Nos. | 2 | Positioned at LSS2 and LSS4 |
| | length of individual tunnel | m | 300 | |
| | Dimension (∩-shaped) | m | 4.00 × 4.00 | |
| Equipment hall | Qty | Nos. | 2 | |
| | length | m | 15 | |
| | Dimension (∩-shaped) | m | 10.00 × 6.00 | |
| Shaft | Qty | Nos. | 2 | |
| | Diameter | m | 6.00 | |

*9.4.3.2.9 Access Tunnels*

Access tunnels are located on the outer side of IP1 and IP3, and the inner side of IP2 and IP4, mainly for personnel transportation and small experimental equipment and consumables. The cross-section is 5.00 m wide and 5.50 m high. Please refer to Table 9.4.8 for further details.

**Table 9.4.8:** Access tunnels

| Item | Unit | Qty |
|---|---|---|
| Qty | Nos. | 4 |
| Dimension (∩-shaped) | m | 5.00×5.50 |
| Longitudinal gradient | | 8% |

*9.4.3.2.10 Vehicle Shafts*

Each interaction region (IR) is equipped with a main cavern transport shaft that has a diameter of 16.00 m, a service cavern transport shaft with a diameter of 9.00 m, a service cavern access shaft with a diameter of 6.00 m, and a bypass tunnel shaft with a diameter of 7.00 m.

Each RF section has one vertical shaft with a diameter of 15.00 m for transport and two shafts with a diameter of 6.00 m for access.

An access and pipe shaft with a diameter of 10.00 m is provided for each linear section of the collider ring.

At each gamma source equipment hall, a shaft with a diameter of 6.00 m is located for transportation and pipeline purposes.

Each curved section has one access and pipe shaft with a diameter of 10.00 m and two ventilation shafts with a diameter of 7.00 m.

Two access and pipe shafts with a diameter of 7.00 m are located at the end and midpoint of the slope section of the transfer tunnel.

Table 9.4.9 summarizes these 42 shafts.

**Table 9.4.9:** List of shafts

| Region | Item | Qty | Diameter(m) |
|---|---|---|---|
| IR | Transport shaft | 2 | 16.00 |
| | Bypass tunnel access shaft | 2 | 7.00 |
| | Auxiliary shaft | 2 | 9.00 |
| | Auxiliary access shaft | 2 | 6.00 |
| RF | Transport shaft | 2 | 15.00 |
| | Transport shaft | 4 | 6.00 |
| Linear tunnel | Access & pipe shaft | 4 | 10.00 |
| Curve sections | Access & pipe shaft | 8 | 10.00 |
| | Ventilation shaft | 16 | 7.00 |

*9.4.3.2.11 Design of the Underground Structure*

1) Tunnel shape:

   Three tunnel shapes have been considered: circular, gate, and horseshoe. If the TBM method is used for tunneling, the circular shape will be selected. However, if the drill-blast tunneling method is used, the shape and dimensions will be determined based on construction and transportation requirements, as well as equipment layout and accessibility during installation and operation. The selection of the tunnel shape and construction method will be made through a comprehensive analysis of technical and economic factors. At this stage, the gate-shape, as shown in Fig. 9.4.13, has been selected.

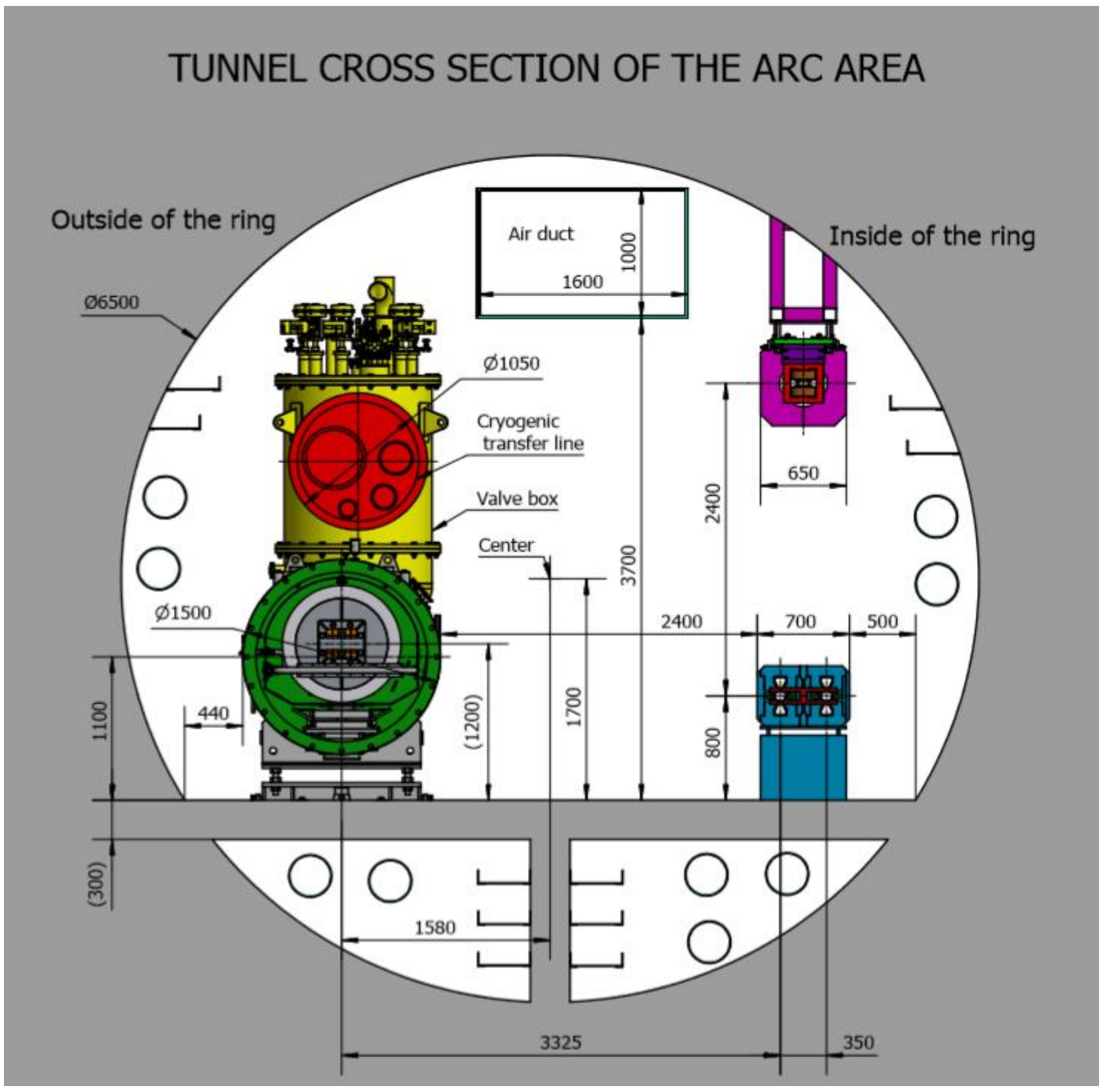

**Figure 9.4.12:** Circular option in the Collider arc section.

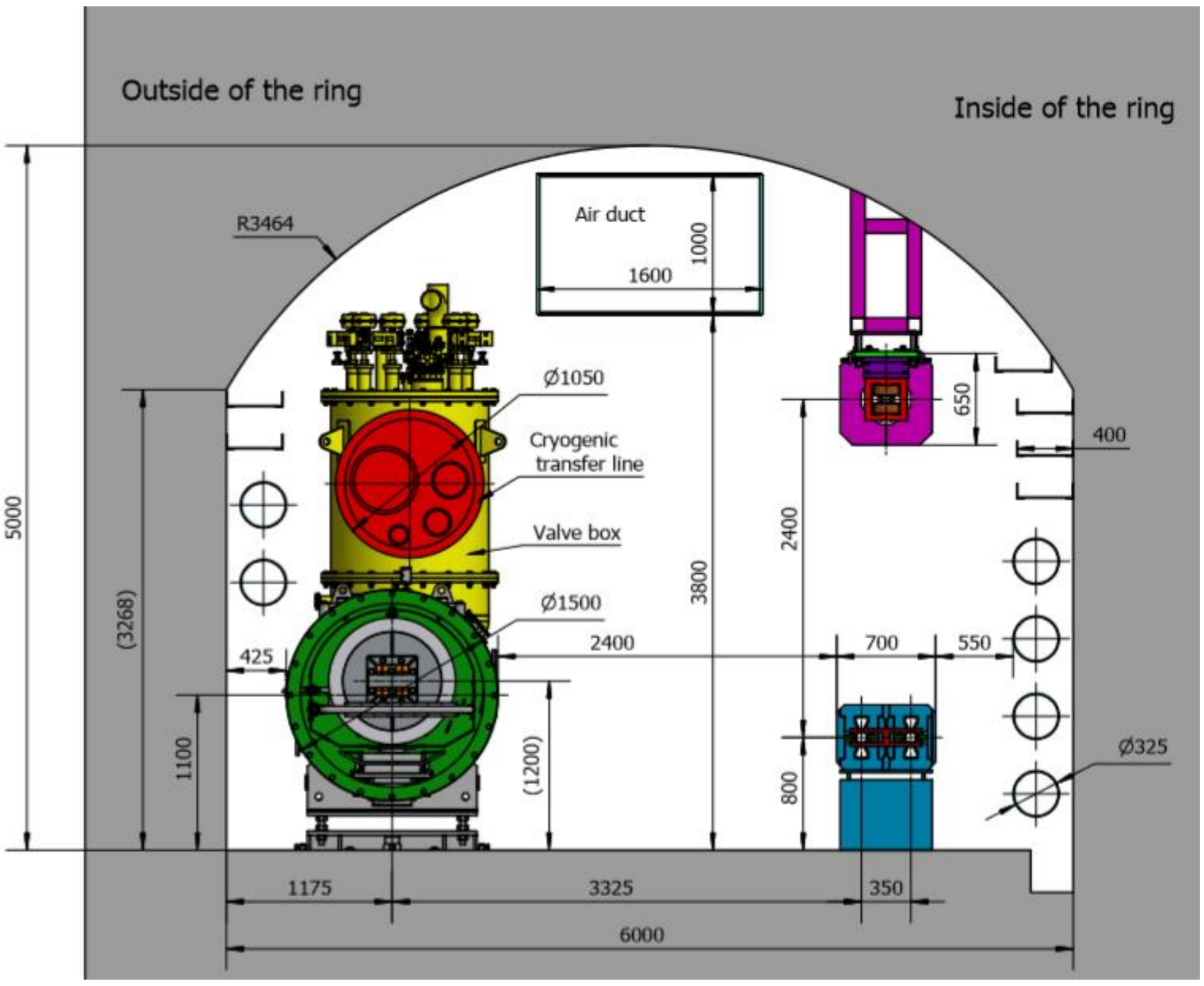

**Figure 9.4.13:** Gate-shape option in the Collider arc section

2) Tunnel lining and waterproofing:

The waterproofing of the underground caverns is classified as Grade I. Support and lining structures must meet both structural and waterproof requirements. Various types of linings are available, such as bolt-shotcrete, reinforced concrete, steel fiber concrete, and steel structure. Waterproof materials include waterproof membranes, coatings, rigid waterproof materials, and concrete admixtures. Since the selection of lining structure and waterproof material has a significant impact on the project's economy, it will be determined through a comprehensive technical and economic comparison, considering structural and waterproof requirements. Currently, the following options are being considered: drain holes and profiled steel sheets for the crown in Class II surrounding rocks, drain holes, profiled steel sheets for the crown, and damp-proof decorative partitions in Class III surrounding rocks, and waterproof membranes/boards/coatings along with 25 to 50 cm thick waterproof concrete lining in Class IV to V surrounding rocks.

3) Shaft structure:

The tunnels are accompanied by a number of shafts, each sized according to their specific functions. For instance, the transport shaft's diameter is determined based on the size of the equipment to be transported, pipeline layout, evacuation passage requirements, and necessary thickness for support. Similarly, the diameter and depth of the shaft for construction and ventilation purposes is determined by construction ventilation needs.

The support for these shafts consists of sprayed anchor and reinforced concrete lining. The thickness of shotcrete and lining concrete is dependent on the shaft's diameter and depth, surrounding rock type, and groundwater, among other factors.

4) Experiment halls:

Large spans require the selection of class I and II surrounding rocks as much as possible for cavern locations. Regions with large geological tectonic belts, fault fracture zones, joint fissure development zones, high in-situ stress zones, goaf zones, and abundant groundwater must be avoided. The depth of caverns should be determined through comprehensive analysis of lithology, rock mass completeness, weathering unloading degree, in-situ stress magnitude, groundwater situation, construction conditions, and experimental requirements. Generally, the overburden thickness should be at least twice the excavation width of the cavern.

Due to the small depth of the experiment halls and strict waterproof requirements, a combination of flexible support and reinforced concrete lining is used. Flexible support consists of one or several combinations of shotcrete, rock bolts, and anchor cables.

#### 9.4.3.3 *Surface Structure*

All surface structures should be located as close to the access shafts as possible. These buildings house water cooling facilities, low-temperature facilities, ventilation systems, air compression systems, power transformer substations, and electrical transmission and distribution and DC power supplies. The floor area of each surface structure is listed in Table 9.4.10.

**Table 9.4.10:** Area of surface structures

| Surface Structure | Area (m2) | | | | | | | | | | | | Total(m2) |
|---|---|---|---|---|---|---|---|---|---|---|---|---|---|
| | P1 | P2 | P3 | P4 | P5 | P6 | P7 | P8 | C1-8 | V1-16 | LINAC | BT | |
| Control and duty rooms | 1200 | 300 | 300 | 300 | 1200 | 300 | 300 | 300 | | | 400 | | 4600 |
| Magnet power source | 2000 | 1500 | 1500 | 1500 | 2000 | 1500 | 1500 | 1500 | | | 500 | 400 | 13900 |
| High-frequency power source | | | 4000 | | | | 4000 | | | | 8400 | | 16400 |
| 110kV substation | 1500 | 1500 | 2500 | 1500 | 1500 | 1500 | 2500 | 1500 | | | | | 14000 |
| 10kV substation | 600 | 500 | 1000 | 500 | 600 | 500 | 1000 | 500 | 2400 | 3200 | 400 | | 11200 |
| HVAC system | 800 | 600 | 800 | 600 | 800 | 600 | 800 | 600 | 4800 | 3200 | 600 | 300 | 14500 |
| Cryogenic system (helium compression system) | 1000 | | 4000 | | 1000 | | 4000 | | | | | | 10000 |
| Cooling water system | 800 | 600 | 1500 | 600 | 800 | 600 | 1500 | 800 | 3200 | | 500 | 200 | 11100 |
| Experimental assembly hall | 1500 | | | | 1500 | | | | | | | | 3000 |
| Magnet assembly hall | 3000 | | | | | | | | | | | | 3000 |
| Transfer system | 200 | 200 | 200 | 200 | 200 | 200 | 200 | 200 | 1600 | | 200 | 150 | 3550 |
| Air compression system | 150 | 150 | 150 | 150 | 150 | 150 | 150 | 150 | | | 150 | | 1350 |
| Water cooling system | 1500 | 1200 | 1200 | 1200 | 1500 | 1200 | 1200 | 1200 | 8000 | | 500 | | 18700 |
| Electronic room | 1000 | 600 | 600 | 600 | 1000 | 600 | 600 | 600 | | | 450 | 100 | 6150 |
| Miscellaneous | 500 | 500 | 500 | 500 | 500 | 500 | 500 | 500 | 2400 | 2400 | 200 | | 9000 |
| Total | 15750 | 7650 | 18250 | 7650 | 12750 | 7650 | 18250 | 7850 | 22400 | 8800 | 12300 | 1150 | 140450 |

Note: C1-8 refer to surface structures corresponding to the ventilation and access shaft in the middle of each cruved section, and V1-16 refer to surface structures corresponding to the ventilation shaft at each curved section.

## 9.4.4 **Construction Planning**

### 9.4.4.1 *Main Construction Conditions*

The project is situated 30.5 km away from Qinhuangdao City and can be conveniently accessed as discussed in section 10.4.2. Additionally, national policies for poverty alleviation have led to the construction of village-to-village roads and improved household access, further enhancing local transportation conditions.

The project area is characterized by low mountains and hills, providing ample space for a distributed arrangement of construction sites. It is a suburban area with rapid socio-economic growth, and the placement of quarries/borrow areas and spoil areas for the project will need to align with the overall local planning, which may present some challenges.

The project area is located near downtown Qinhuangdao and is serviced by 110 kV and 220 kV substations in the Funing District, as well as 35 kV substations in the surrounding townships. Each construction area will be powered through connection to these 35 kV substations, with plans to establish a more permanent power supply arrangement at a later stage in the construction process.

The project area intersects the Yanghe River twice. Fortunately, the area enjoys a reliable water supply due to the presence of multiple rivers and reservoirs in the region. However, as part of the project area passes through residential areas with a large population and numerous buildings, there will be significant disruptions to local residents and businesses during the construction period.

### 9.4.4.2 *Construction Scheme of the Main Structures*

#### *9.4.4.2.1 Collider Tunnel Construction*

After considering factors such as geological conditions, topography, transportation, electric power availability, construction duration, and layout, the drill and blast tunnelling method was chosen for the construction of the experiment halls, auxiliary tunnels, and access shafts. The 46 vertical shafts, ranging in diameter from 16 m to 7 m, are intended for equipment transportation, ventilation, and emergency egress.

When it comes to large and long tunnels, the two main methods used are the mining method and the tunnel boring machine (TBM) method. Both methods have their own advantages and disadvantages, and a comprehensive analysis will be carried out to determine the most suitable method for each of the various underground structures. Factors such as tunnel dimensions (including depth), geology (including rock compressive strength, structure, and fragmentation degree), groundwater and environmental conditions, traffic capacity, water and power supply, and adit layout, along with the construction timeline that best matches the production and installation of the CEPC technical components, will all be considered in making this decision.

For medium and short tunnels, the conventional method used is drill and blast, which offers the advantage of flexible construction and the ability to adapt to varying geology. When used for the Collider ring, this method utilizes various vertical shafts arranged around the circumference, with a total of 32 construction zones and 64 working faces. Each working face has a control length of approximately 1.6 km. The first step in construction involves the excavation of the vertical shafts.

The Collider ring is then excavated using sectionalized full-face smooth blasting via drilling with self-made platform hand drills. The spoil is loaded into 4.0 $m^3$ mine carts using crawler loaders and transported to spoil bunkers at the bottom of the vertical shafts via electro mobile traction mine carts. The spoil is then loaded into 4.0 $m^3$ buckets using crawler loaders and lifted to the openings of the vertical shafts with winches. From there, the muck is directly loaded onto 10-ton dump trucks using chutes and transported to the spoil area.

The tunnel lining is done with formwork jumbo, and concrete is transported horizontally using a concrete mixer truck and delivered using a concrete pump.

While the drill and blast method has been selected for the Collider ring tunnel construction, some may wonder why TBMs were not chosen instead. China has made significant progress with this technology, using it in the construction of large high-speed rail systems and new subways. A TBM excavates using mechanical energy to break rocks and avoids the problems encountered with traditional drill and blast methods such as difficult construction adit placement, unreliable construction times due to excavation surface constraints, and challenges with debris disposal and ventilation.

TBMs have advantages in favorable geological conditions, including high excavation efficiency, minimal disturbance to surrounding rocks, minimal over excavation, a smooth excavation surface, good quality, a relatively safe working environment, small impact on the surrounding environment, and minimal disturbance to residents. If the Collider ring tunnel were to be constructed using TBMs, 8 open-type TBMs would be installed around the circumference with maximum control length of a single working face at 17.2 km. The excavated spoil would be removed using a 900 mm belt conveyor and then transported by trucks. Locomotive-traction railcars would be used for transportation within the tunnel.

##### *9.4.4.2.2 Shaft Construction*

The vertical shafts described earlier have an average depth of around 129 m, with significant variation ranging from a minimum of 50 m to a maximum of 245 m. They are all supported by a combination of bolt-shotcrete support and reinforced concrete lining.

The equipment transportation shafts for the experiment halls, with diameters of 16 m and 15 m, are constructed in three phases. Phase I involves constructing a 5 m pilot shaft using the top-down shaft-sinking method. Once the pilot shaft is bottomed, layer enlargement is carried out in phase II from the top down to the top elevation of the experiment hall, followed by placing of the shaft wall concrete using the slip form from the bottom up in phase III.

The construction of other smaller diameter shafts is carried out using the full-face shaft-sinking method through drilling and masonry from the top down.

The pilot shafts are excavated using full-face sectionalized drilling and blasting, using the smooth blasting technique, from the top down. Blast holes are drilled with an umbrella drilling rig and rock drill. The spoil is loaded onto 4 $m^3$ hook-type buckets using central rotary grab loaders. These buckets are then lifted to the openings of vertical shafts using winches, with the spoil stirred automatically using hooks. The rock ballast is loaded directly onto 10-ton dump trucks through chutes and transported to the spoil area.

The enlargement of the experiment hall shafts is carried out through bench blasting and excavation in layers, with protective layers reserved for pre-splitting. The bench height ranges from 3 to 5 m. Presplit holes are drilled with a down-the-hole drill, and main blast holes are drilled with a hydraulic crawler drill. The rock ballast is stirred with a 1 $m^3$ hydraulic backhoe excavator, and then transported from the shaft bottom through the pilot shaft. Finally, the ballast is loaded onto 10-ton dump trucks through 2 $m^3$ side-dump loaders. Equipment and materials are lifted using a movable gantry crane at the shaft opening.

For the vertical shafts described above, the construction is supported by bolt-shotcrete support + reinforced concrete lining. The concrete lining is done after excavation is complete, with the concrete placed using slip form from the bottom up.

The construction of concrete lining, excavation, and support of other shafts are carried out through drilling and masonry. The concrete is placed from the bottom up in a sectionalized manner using YJM type integrated steel forms moving downwards.

Concrete is produced by a surface concrete batching plant, transported horizontally by a concrete mixer truck, vertically transported by a vacuum chute, and then manually vibrated and compacted by an immersion vibrator.

##### *9.4.4.2.3 Experimental Hall Construction*

As previously mentioned, the two experiment halls have specific dimensions and locations. They are buried at a depth of approximately 160 m at maximum and 80 m at minimum above the foundations. The shafts and access tunnels are constructed for the purpose of construction only. The access tunnel has an inverted U-shape with dimensions of 5 × 5.5 m and an average slope gradient of 8%.

The construction of the experiment halls is divided into two phases: phase I involves the excavation of the chute shaft, while phase II involves the excavation of the experiment hall itself. The chute shaft has an excavation diameter of 5.0 m and follows the same construction method as the transport shafts. The excavation is carried out in layers from the top down, with layer I having an excavation height of 8 m and the other layers having heights of 5 to 7 m.

Layer I is divided into three construction zones, and each zone is excavated using sectionalized full-face smooth blasting. The excavation process starts with the pilot drift tunnel, followed by adjacent zones. Each working zone in the layer is further divided into four parts for excavation and support. A 2.0 m thick protective layer is reserved close to the structural plane to ensure safety during the excavation process.

To prepare for the blasting, presplit holes are drilled using a 100B down-the-hole drill, and main blast holes are drilled using a ROCD7 hydraulic crawler drill. After the blasting, equipment and materials are lifted using a movable gantry crane at the shaft opening or transported using trucks through the access tunnels.

The spoil, or excavated material, is removed from the construction site using the same method as in the shaft construction process described earlier.

#### 9.4.4.3 *Construction Transportation and General Construction Layout*

##### *9.4.4.3.1 Construction Transportation*

The majority of the sections along the 30.5 km route from Qinhuangdao City are currently connected by simple roads, which can be reconstructed and expanded to serve as on-site access roads. However, to meet the transportation requirements for construction and experiment equipment, the site access roads in the areas of the experiment halls are constructed to meet national standard Class III, with a concrete or asphalt concrete pavement and a subgrade width of 8.5 m. Site access roads in high frequency zones are constructed to meet national standard Class IV, with a concrete or asphalt concrete pavement and a subgrade width of 6.5 m. In contrast, site access roads in the areas of the other access shafts and ventilation shafts are Class III single-lane roads for mines, with a clay-bound macadam pavement and a subgrade width of 4.5 m.

For construction transportation, the aim is to use local trunk roads as much as possible, while for shaft construction, rural roads will be used wherever possible. Transportation for the construction of experiment halls will take place on permanent roads, and the possibility of constructing new roads remains open.

##### *9.4.4.3.2 General Construction Layout*

Construction zones are distributed near the inlets of shafts and construction adits for all experiment halls, while temporary construction facilities such as workshops, warehouses, and construction camps are centrally located within each construction zone. The preliminary general construction layout plan specifies 32 centralized construction zones, distributed at an average interval of 3.2 km.

The project will utilize existing local resources, including roads, bridges, aggregate processing plants, concrete batching plants, production and living facilities, drainage facilities, and power transmission and communication lines.

The selection of spoil areas will be done in collaboration with local cities and towns to ensure the most appropriate location is chosen.

##### *9.4.4.3.3 Land Occupation*

The permanent land occupation will mainly include surface structures and permanent roads, totaling approximately 1.28 million m$^2$. The temporary land occupation will include the material yard, spoil area, temporary roads, and construction sites, totaling approximately 2.70 million m$^2$, with 1.60 million m$^2$ occupied temporarily by the spoil area.

#### 9.4.4.4 ***General Construction Schedule***

##### *9.4.4.4.1 Comprehensive Indices*

The construction will be completed in phases. The first phase will involve preparation, which includes land acquisition and resettlement, establishing supplies of water, power, and compressed air, road connection and communications, and site leveling. This phase is expected to last for 6-8 months.

The construction of vertical shafts will involve excavation, support, and installation of lining. For a vertical shaft with a diameter of less than 10 m, the monthly advance will typically be between 20 to 40 m in depth through the upper overburden and then 60 to 80 m for the rock section. For larger vertical shafts, such as those with a 15 m diameter, the shaft-sinking method will be used, with an advance of 30 to 50 m each month.

The progress of tunnel excavation by drill and blast is dependent on the length of the working face and the classification of the rock. Most of the rock in the collider ring tunnel is classified as Class II and Class III. With a working face length of 1.7 km, the average advance is expected to be between 80 to 100 m/month. Taking into account the excavation of auxiliary caverns, the entire excavation period is estimated to be about 24 months.

Alternatively, if open-type TBMs are used, the average advance is expected to be between 500 to 800 m/month, with a working face length of 17.2 km. In this case, the tunnel excavation period is estimated to be about 26 months. The excavation of auxiliary caverns will be carried out after the completion of the main cavern excavation with multiple working faces, and is expected to take about 3 months.

The lining of the ring tunnel will involve the completion of one concreting berth (12 m long) every 2-3 days, with a rate of approximately 150 m/month. In the tunnel lining phase, the length of the working face will be 3.4 km, with waterproofing done concurrently. If there is a single working face, the period for tunnel lining and waterproofing is expected to be about 10 months. If progress is too slow, the quantity of equipment or the number of working faces can be increased.

The construction of the experiment halls will involve the pilot shaft and cavern excavation, which will take approximately one month. The excavation and support of the top layer of the cavern will take 4-5 months, while the excavation and support of each remaining layer will take about 2 months. In total, the excavation and support will take an estimated 10-13 months.

*9.4.4.4.2 Proposed Total Period of Construction with Drill-Blast Tunneling Method*

The total construction period for the project is 54 months, which includes 8 months for construction preparation, 43 months for the construction of main structures, and 3 months for completion.

The controlling project construction item is the collider ring tunnel from IP3 to LSS3. The critical path for this construction item is as follows: construction preparation (8 months) → construction of vertical shafts (5 months) → tunnel excavation (24 months) → tunnel lining and waterproofing (10 months) → installation of ventilation equipment and access equipment of the shaft (4 months) → completion (3 months).

It is important to note that the construction of surface structures will be carried out concurrently with the underground work, so as not to lengthen the construction timeline.

*9.4.4.4.3 Proposed Total Period for Construction with TBM Methos (Using Open-Type TBMs)*

If tunnel boring machines (TBMs) are used instead of drill and blast, the length of the construction phases will be somewhat different. The construction of the launching and arriving TBM shafts will be combined with the construction of the permanent vertical shafts. The designing and manufacturing of the TBMs will be done during the construction preparation phase. The TBMs will take 10 months for design and manufacturing, 2-3 months for transportation to the site, and 2 months for installation and commissioning.

The total period of construction with the TBM method will be 51 months, which includes 8 months for construction preparation, 40 months for the construction of main structures, and 3 months for completion. The critical path for this construction method is as follows: Design, manufacture, and transportation of TBM equipment (13 months) → TBM assembly (2 months) → tunnel excavation with TBM (22 months) → collider ring tunnel short auxiliary tunnel as well as second-time expansion excavation (3 months) → tunnel lining and waterproofing (8 months) → completion (3 months).

It is important to note that the construction of surface structures will be carried out concurrently with the underground work.

The following measures can be taken to optimize the overall construction schedule for collider ring tunnel construction with the TBM method:

1) Streamline the design and manufacturing process of TBMs to reduce the time required for their production.

2) Adopt advanced and reasonable TBM construction equipment to improve the speed of TBM boring. This includes selecting the most appropriate TBM for the specific geological conditions and using efficient cutting tools.
3) Employ double-shield TBMs that allow tunnel excavation and lining to be carried out simultaneously, thereby reducing the time required for lining the tunnel.
4) Use a combined approach of TBM boring and drill-blast tunneling methods. The drill-blast tunneling method can be used to deal with adverse geological segments, reducing the time required for the TBM to cope with these defects and improving the overall boring efficiency of the TBM.
5) Develop high-speed continuous vertical transport equipment for vertical shaft and vertical transport equipment for duct pieces. This equipment can be used to transport materials vertically, reducing the amount of work required for adits and shortening the TBM boring length, thereby reducing the total construction time with the TBM method.

By combining the TBM and drill and blast tunneling methods, the advantages of each method can be utilized to improve the overall construction process. The fast boring speed, low environmental impact, safe construction, and flat and smooth excavation face associated with the TBM can be combined with the flexibility, adaptability to geological conditions, improved safety, and lower capital cost associated with drill and blast.

This hybrid approach allows the TBM to be used for most of the tunneling work, with the drill and blast method being used selectively to deal with adverse geological conditions such as hard rock, fault zones, or areas with high water inflows. This can help reduce the time required for tunneling in challenging conditions and improve overall efficiency.

Additionally, the drill and blast method can be used to construct larger caverns or sections of the tunnel that are not amenable to the TBM method. This flexibility in construction methods allows for a more efficient and adaptable construction process that can be customized to the specific geological conditions of the project.

Overall, by combining the advantages of the TBM and drill and blast tunneling methods, the construction process can be optimized for both cost and time, leading to timely project completion and improved overall project outcomes.

## 9.5 Changsha Site

### 9.5.1 Characteristics of Changsha Site

Hunan, often referred to as "Xiang" for short, is situated in the south-central region of China and the southern part of the middle reaches of the Yangtze River. The provincial capital of Hunan is Changsha. The province lies on the transition zone between the eastern coastal regions and the central and western regions, serving as the junction of the Yangtze River open economic belt and the coastal open economic belt. Therefore, it plays a vital role as a significant component of the Yangtze River economic belt. The city cluster of Changsha-Zhuzhou-Xiangtan is a crucial part of the urban agglomeration in the middle reaches of the Yangtze River. Furthermore, Hunan has exceptional construction and operation conditions for the CEPC project.

1) Remarkable location advantages and a fast and convenient transportation system:

Hunan Province is situated in the middle reaches of the Yangtze River and south of Dongting Lake, covering a land area of 211,800 $km^2$. As of the end of 2021, the province had a permanent resident population of 66.22 million and a GDP of RMB 4,606.3 billion, ranking 10th in China.

Hunan Province lies on the transition zone between the eastern coastal regions and the central and western regions, serving as the junction of the Yangtze River open economic belt and the coastal open economic belt. It has a highly developed infrastructure system comprising high-speed rail, aviation, and expressways. The province also boasts a comprehensive transportation system of water, land, and air, interconnected and crisscrossed in three dimensions. This makes it an essential hub province, connecting the east with the west, the south with the north, reaching rivers and seas, and communicating with the world.

The Changsha-Zhuzhou-Xiangtan region, where the proposed site for the CEPC project is located, is the core growth pole of Hunan's economic development, enjoying remarkable location advantages.

2) Stable regional structure and favorable natural conditions:

Hunan Province is not located on any of the 23 main seismic belts in China, and weak historical and recent seismic activities have been observed in the province. The Changsha site, where the proposed project is located, has a basic seismic intensity of degree VI, a dynamic peak acceleration of 0.05 g, and good regional structural stability. Furthermore, there are no active faults in the area since the Late Pleistocene.

The Changsha-Zhuzhou-Xiangtan region has a large continuous and complete granite rock mass, with the slightly weathered ~ fresh rock mass being shallowly buried. This rock mass has hard rock and weak water permeability, making it suitable for constructing large underground caverns.

Hunan Province is situated in the continental subtropical monsoon humid climate zone, characterized by a mild climate and little impact from sandstorms, typhoons, and extreme weather conditions.

3) World-leading supercomputing capability:

The National Supercomputing Center in Changsha is the sole national supercomputing center in central and western China, authorized by the Ministry of Science and Technology of the People's Republic of China. It began operations at Hunan University in November 2014. Tianhe-2, a $10^{16}$-flop supercomputer, won the global supercomputing championship for six consecutive years. This supercomputer can provide valuable services for CEPC research and is strongly complementary to and integrated with CEPC. It can also fuel scientific and technological innovation and inject new momentum into the project.

4) Solid and safe power grid with sufficient and reliable power supply:

The province's main grid framework consists of "two north-south, four east-west, and two ring networks". Additionally, two 500,000-volt substations will be built in Wangcheng District to meet the 220,000-volt power demand and 500 MW-load required for the CEPC project. The Heimifeng Pumped Storage Power Station, located nearby, can assist the power grid in balancing the impact, enhancing the load capacity and major accident response capability, and ensuring the power safety of CEPC.

5) Harmonious and peaceful society with a pleasant living environment:

Hunan is located in the strategic hinterland of China, far from the surrounding sensitive areas, and is known for ethnic unity and social stability. The province is also famous for its beautiful mountains and rivers, profound cultural connotations, and a pleasant living environment. The regional innovation system led by the Changsha-Zhuzhou-Xiangtan National Independent Innovation Demonstration Zone is becoming increasingly mature. The integrated growth of Changsha-Zhuzhou-Xiangtan and the development of the pan-Changsha-Zhuzhou-Xiangtan city cluster offer various options for project site selection and provide excellent working and living conditions for high-end talent.

6) Vibrant innovation ecosystem and exceptional research capabilities:

Hunan Province has made significant investments in scientific and technological innovation in recent years, establishing itself as a leading hub for innovation and economic growth. The province has created an environment conducive to innovation and open economic development, resulting in numerous groundbreaking scientific and technological advancements, as well as independent innovative products. Hunan's innovation capabilities rank 11th in China, while its overall patent strength ranks 7th in the country, facilitating the transition of key industries towards the middle and high ends of the value chain. These achievements, coupled with an exceptional research workforce, make Hunan an ideal location for CEPC.

7) Supportive competitive industries and strong construction foundation:

Hunan Province is well-equipped to support the construction of CEPC. With the construction of a long underground tunnel being one of the main works of the project, Hunan's construction machinery industry is at the forefront of the field in China. The province is the most representative cluster of the six major industry bases of construction machinery in China, with the highest degree of industrial agglomeration, the most complete product categories, and the largest industrial scale in the country. The construction machinery industry in Hunan has formed an industrial pattern with Changsha as the flagship, Changsha and Xiangtan as the centers for whole machine manufacturing, and Zhuzhou and Hengyang as the auxiliary supporting for spare parts. Hunan produces over 400 models and specifications of products across 12 categories and over 100 subcategories, accounting for 70% of the total types of construction machinery in China. This industry covers a range of sectors, including concrete machinery, excavation machinery, lifting machinery, rock drilling machinery, and TBM. Changsha is the only city in the world that has four of the world's top 50 construction machinery enterprises. Since 2010, the scale of Hunan's construction machinery industry has continued to rank first in the country, with an output value accounting for around 23% of the country's total. With such a solid foundation and supportive industries, the construction of CEPC in Hunan can be carried out smoothly and efficiently.

### 9.5.2 **Engineering Geological Conditions**

#### 9.5.2.1 ***Survey and Test***

In December 2018, the CPC Hunan Provincial Committee, the People's Government of Hunan Province, and the Department of Science and Technology of Hunan Province led the organization of relevant departments and scientific research institutions in the province to conduct site selection for the CEPC project in Hunan. After analyzing various factors such as geographical location, geomorphic features, engineering geology, traffic conditions, planning, and livable environment of the Science City, three sites in Changsha, Xiangtan, and Longhui were preliminarily selected.

Further analysis revealed that the Changsha site outperformed the Xiangtan and Longhui sites in all the above-mentioned aspects. Hence, the Changsha site was selected for further survey and design.

Between June and October 2019, PowerChina Zhongnan conducted a 1:50,000 engineering geological plane survey and mapping of 100 $km^2$. A total of 538.81 meters were drilled for exploration in 5 boreholes, and a total of 435.61 meters were tested for acoustic waves in 5 boreholes at the Changsha site. The survey report was completed through comprehensive analysis and compilation of the collected data.

**Table 9.5.1:** Main survey and test workload of the Changsha site.

| Field | Item | Unit | Quantity | Remarks |
|---|---|---|---|---|
| Data collection | 1/10,000 topographic map | Nr. | 100 | Total area: about 2800$km^2$ |
| | 1:50,000 regional geological map | Nr. | 6 | Qingshanpu map sheet, Zhangshugang map sheet, Tongguan map sheet, Tongpensi map sheet, Jinjing map sheet and Xinshi Town map sheet |
| | Project information for the surrounding area of the site | Set | 2 | Heimifeng and Pingjiang Pumped Storage Power Stations |
| Surveying and mapping | 1:1000 topographic survey | $km^2$ | 1.26 | 4 underground experimental halls |
| Geology | 1:50,000 engineering geological plane surveying and mapping | $km^2$ | 100 | Collider tunnel and shaft |
| | 1:1000 engineering geological plane surveying and mapping | $km^2$ | 1 | 4 underground experimental halls |
| | 1:10,000 geological route survey | km | 100 | Collider tunnel and shaft |
| | 1:1000 engineering geological section surveying and mapping | km/Nr. | 2.0/4 | Underground experimental hall |
| | Borehole log and record | m/hole | 538.81/5 | - |
| | Pit (trench) exploration | $m^3$ | 142 | - |
| Drilling | Core drilling | m/hole | 538.81/5 | - |
| Geophysical prospecting | Acoustic wave test of borehole | m/hole | 435.61/5 | - |
| | Borehole video recording | m/hole | 78.3/1 | Lengjiaxi Group sandy slate stratum |
| | EH-4 section exploration | km/Nr. | 2.281/4 | 2 stratigraphic contact zones and 2 faults in the east of Tongpensi |
| In-situ test | Water pressure test in borehole | Segment-times | 53 | - |
| | Borehole water level observation | Time(s) | 8 | Drilling of underground experimental hall, 2 times each |
| Test | Identification of optical properties of rocks | Group | 12 | 12 groups of granite |
| | Physical and mechanical tests of rocks | Group | 14 | 12 groups of granite and 2 groups of sandy slate |
| | Simple analysis of water quality | Nr. | 5 | 4 Nos. of groundwater and 1 nos. of surface water |

### 9.5.2.2 ***Regional Geology and Seismicity***

#### *9.5.2.2.1 Regional Geology*

The Changsha site is located in the transition zone between the Changsha-Pingjiang Basin and Dongting Lake Alluvial Plain. An outcrop of black mica (two-mica) monzonitic granite of Wangxiang rock mass, which is nearly elliptical and relatively hard, is exposed to the surface. On the east side of the rock mass, metamorphic sandstone and slate of the Lengjiaxi Group of the Qingbaikou System are mainly distributed, and the rock is relatively hard. The west side of the rock mass is mainly covered by Quaternary strata. The site is situated under the anticline of the Jiangnan Anteclise of the Yangtze Platform, located in the uplift area of the Mufushan Fault Block, which is a secondary unit of the uplift area of East Hunan-East Hubei Fault Block. Additionally, it is located on the Jinjing-Tongpensi Neo-Cathaysian structural belt, as shown in Figure 9.5.1.

The geological structure of the site is relatively simple, and it is dominated by NNE-trending and NE-trending fault structures, as depicted in Figure 9.5.2. Since the Late Pleistocene, there have been no active faults in and around the site area. The main regional faults include Miluo-Ningxiang-Xinning Fault, Tongpensi East Fault, and Jinpenling Fault.

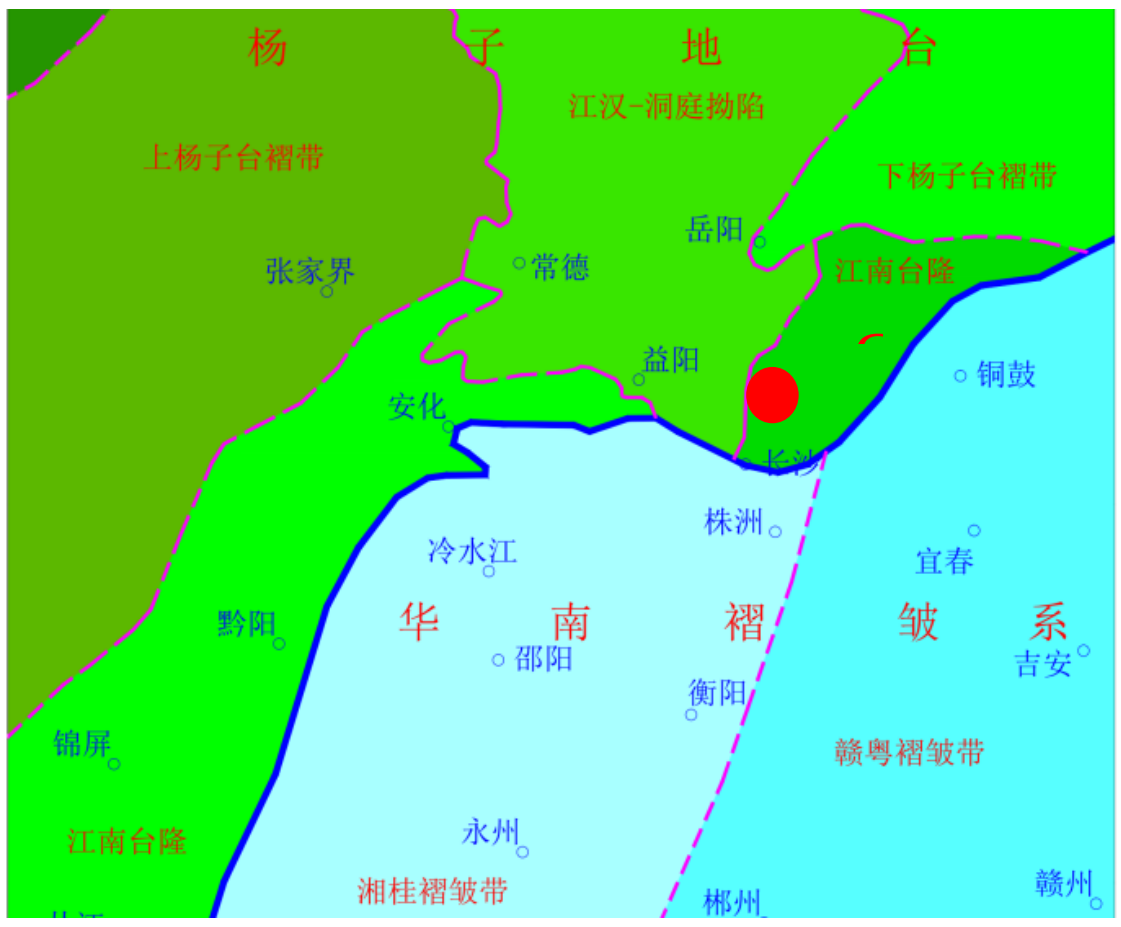

Figure 9.5.1: Geological structure of the Changsha site.

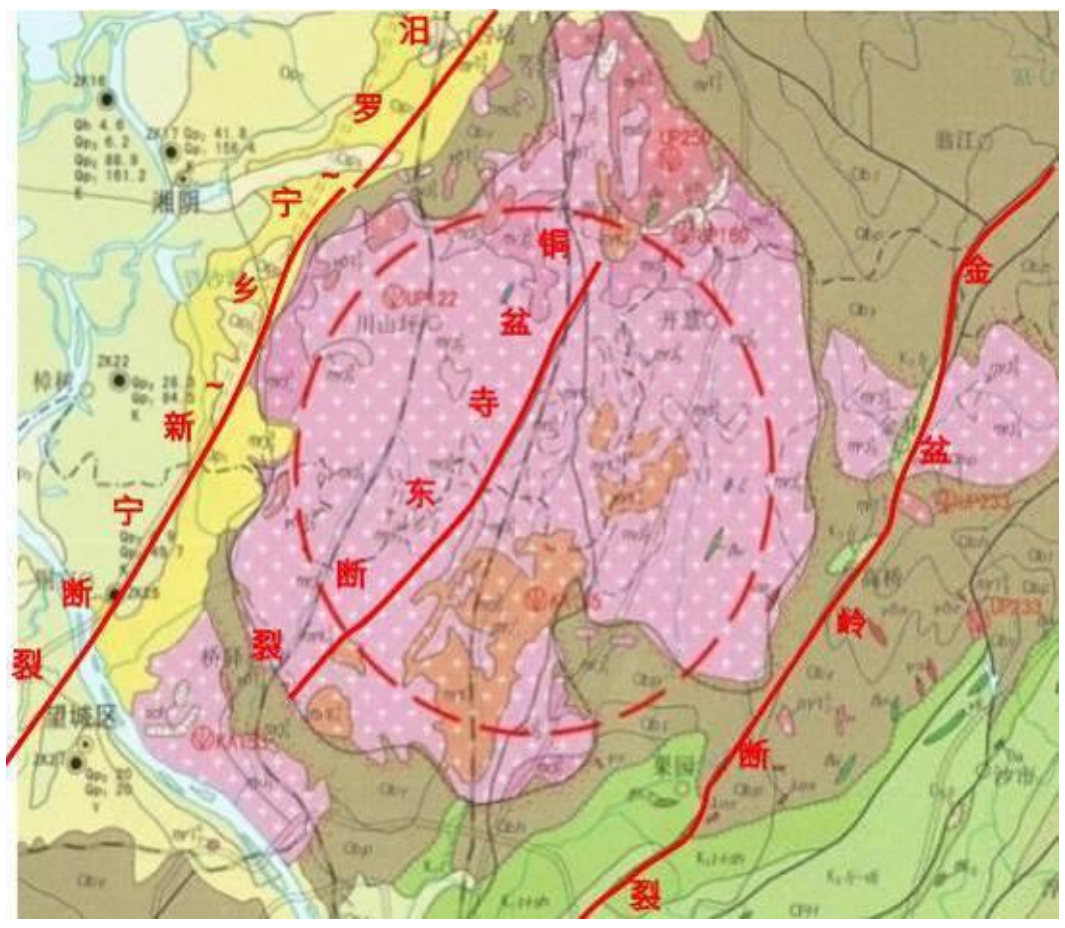

Figure 9.5.2: Regional fault distribution of the Changsha site.

*9.5.2.2.2 Seismicity*

There have been no earthquakes above 4.75 magnitudes recorded at or around the Changsha site. Since 1970, only a few earthquakes have been recorded by instruments, and the largest modern earthquake recorded has a magnitude of ML 2.1, as depicted in Figure 9.5.3 and Figure 9.5.4.

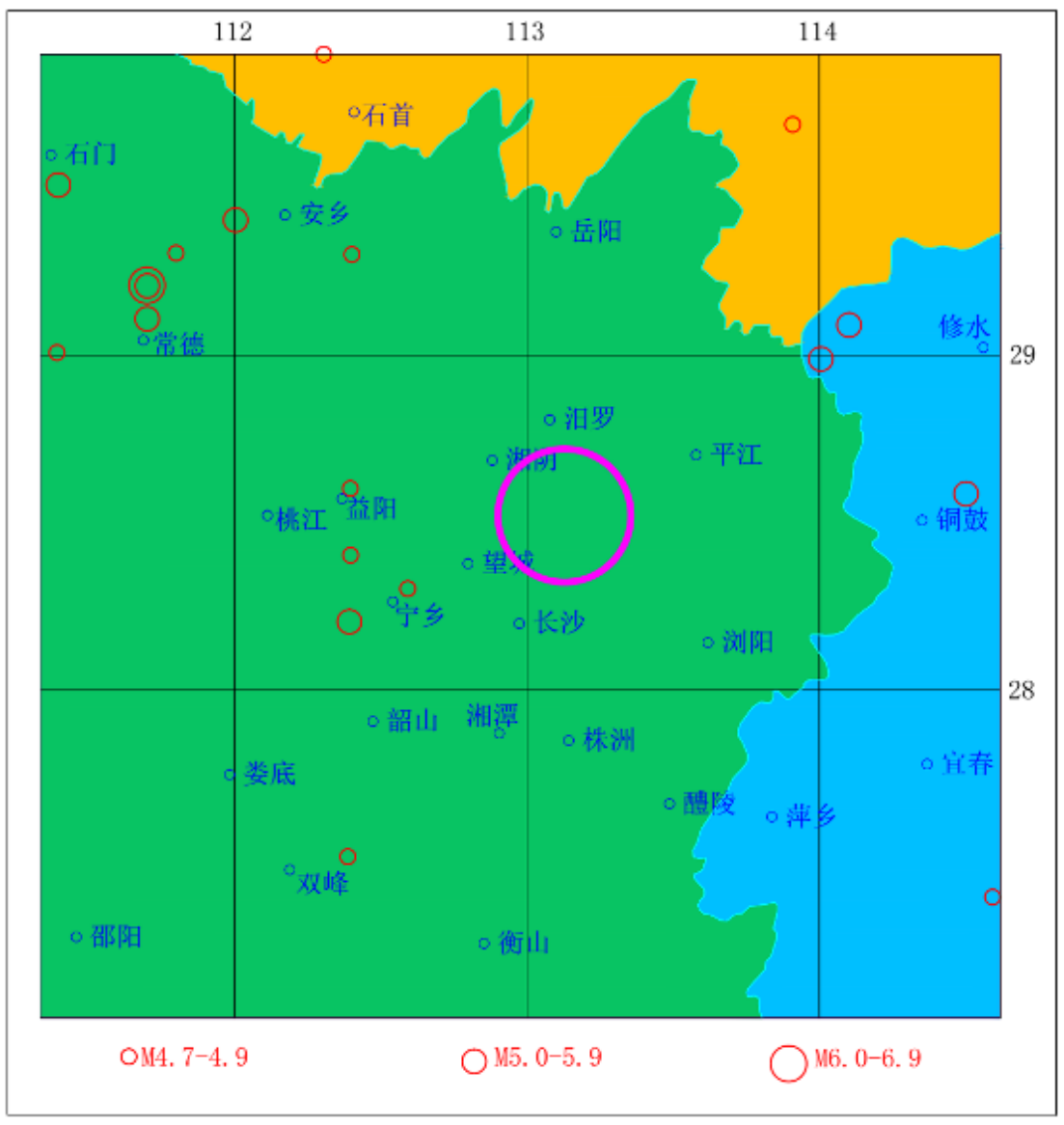

Figure 9.5.3: Epicenter distribution of regional moderately strong earthquakes.

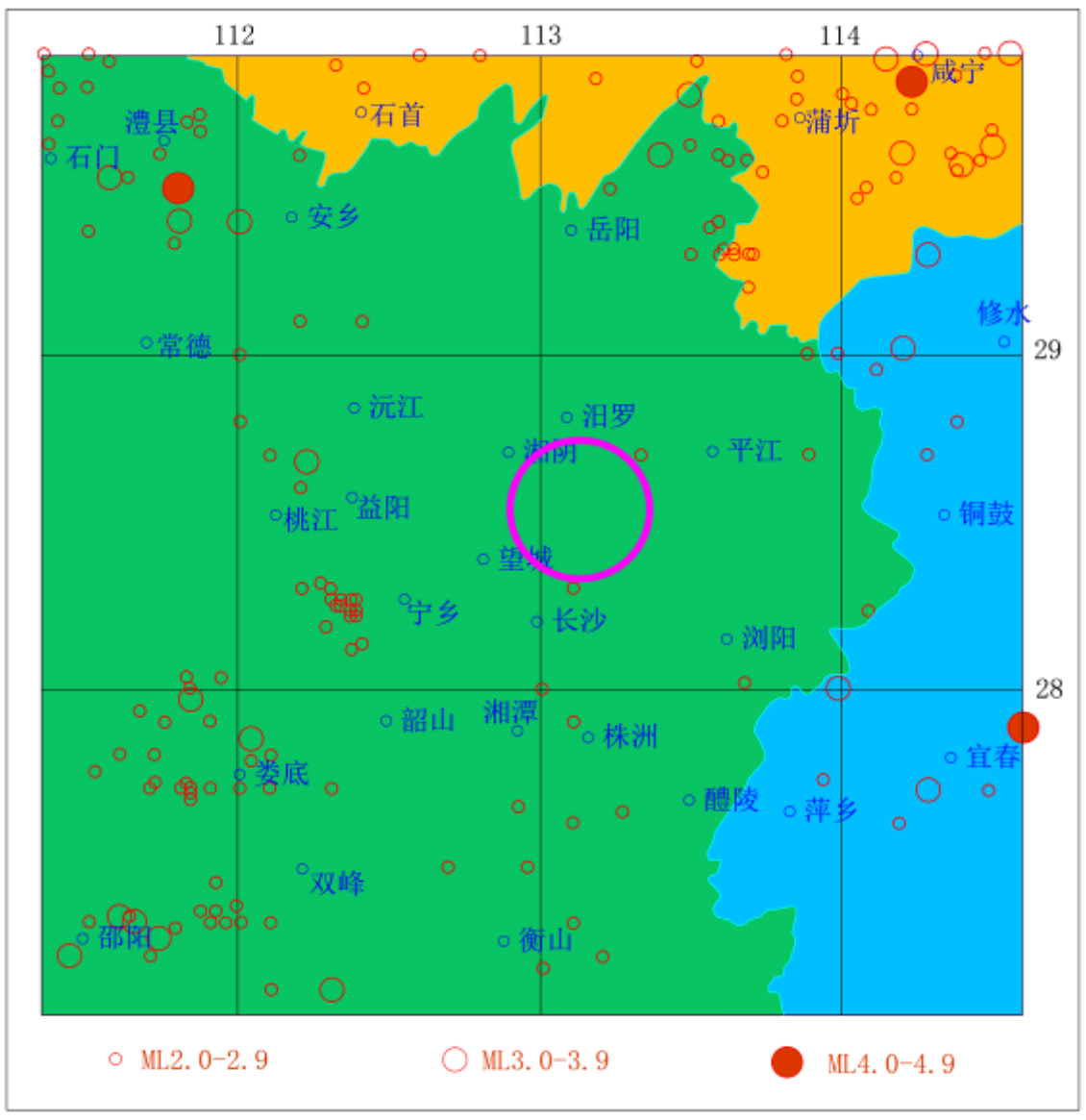

Figure 9.5.4: Epicenter distribution of regional earthquakes.

Based on the neotectonic movement of the site, it appears that there is mainly intermittent crustal uplift movement. Although there are faults distributed at the site and its surroundings, the last activity is at least before the Late Pleistocene, indicating that there is no active fault in and around the site area since the Late Pleistocene. Additionally, no destructive historical earthquakes have been recorded in and around the site, and the largest modern earthquake recorded by the instrument only had a magnitude of ML 2.1.

According to the "ZZAP [2003] No. 35" issued by the China Earthquake Administration, the horizontal peak acceleration value of the bedrock of Heimifeng Pumped Storage Power Station in the site ring is 0.0471 g when the lower reservoir has a 10% probability of exceedance in 50 years. A supplementary assessment in 2009 found that the horizontal peak acceleration value of bedrock is 0.0547 g when the lower reservoir of the hydropower plant has a 10% probability of exceedance in 50 years. Fig. 9.5.5 shows a seismic ground motion parameter zonation map of the Changsha site.

Based on the Seismic Ground Motion Parameters Zonation Map of China (GB 18306-2015), the seismic peak ground motion acceleration is 0.05 g when the site area has a 10% probability of exceedance in 50 years. The characteristic period of the seismic ground motion response spectrum is 0.35 s, and the corresponding basic seismic intensity of the site is degree VI.

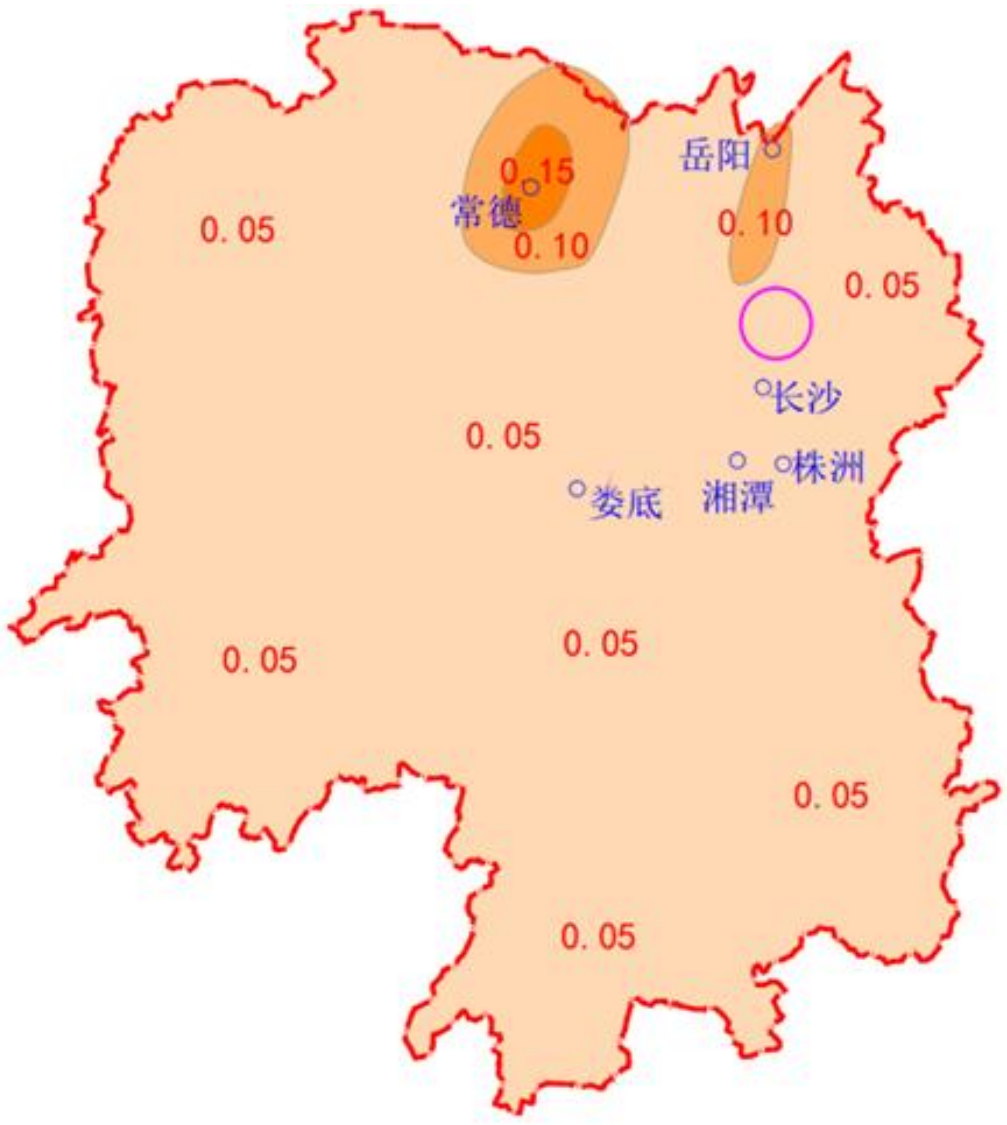

**Figure 9.5.5:** Seismic ground motion parameter zonation map of the Changsha Site (unit: g).

To summarize, the seismic peak ground motion acceleration at the Changsha site is 0.05 g, indicating a moderate level of seismic hazard. The corresponding basic seismic intensity of the site is degree VI, which is considered to be a moderate level of seismic intensity. Overall, the regional tectonic stability is good, as there have been no active faults or destructive historical earthquakes recorded in the area since the Late Pleistocene.

### 9.5.2.3 *Basic Geological and Topographic Conditions of the Project Area*

The Changsha site is situated in the Yuchi Mountain area, which is an extension of Mufu Mountain. The terrain in this area is characterized by a higher elevation in the middle and lower elevation around it. The highest peak in the area is Damo Mountain, which is 777.5 m above sea level. Heimifeng, located in the southeast, has an elevation of 587.2 m above sea level, while the surrounding ground elevation is typically about 70 m.

Exing Mountain, Jiufeng Mountain, Yuchi Mountain, Yingzhu Mountain, and Hanjia Mountain are situated in the middle of the site area, running from west to east. These ridges are primarily distributed in a northeast direction, forming a low mountain and hilly terrain. The west and north of the site area are near the Dongting Lake Plain, where four terraces are developed, resulting in a valley terrace and floodplain landform. The east and south of the site area belong to the Changping (Changsha-Pingjiang) basin, characterized by hilly terrain.

The site area is situated in the Xiangjiang River basin, where streams and ditches are abundant. In the south of the site area, the Shahe River, Baisha River, Xunlong River, and Malin River are developed from west to east, and the overall flow direction is in a southwest direction. In contrast, Baishui River and Chedui River are developed in the north of the site area from west to east, with a flow direction that is mostly towards the north.

The collider tunnel is situated along the foot of low mountains such as Heimi Peak, Jiufeng Mountain, Exing Mountain, and Piaofeng Mountain. The tunnel route is characterized by hilly terrain, with a surface elevation ranging from 45 m to 75 m and a surface elevation difference of 10 m to 30 m, featuring slight terrain undulation. Topographic and geomorphic features of the site area can be referred to from Figure 9.5.6 to Figure 9.5.7.

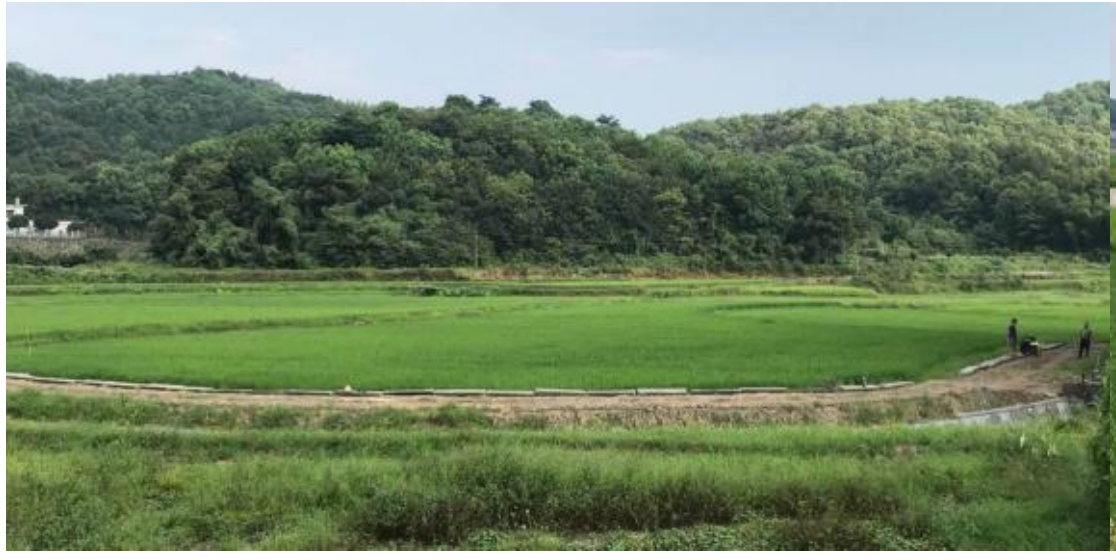

**Figure 9.5.6:** West of the site area.

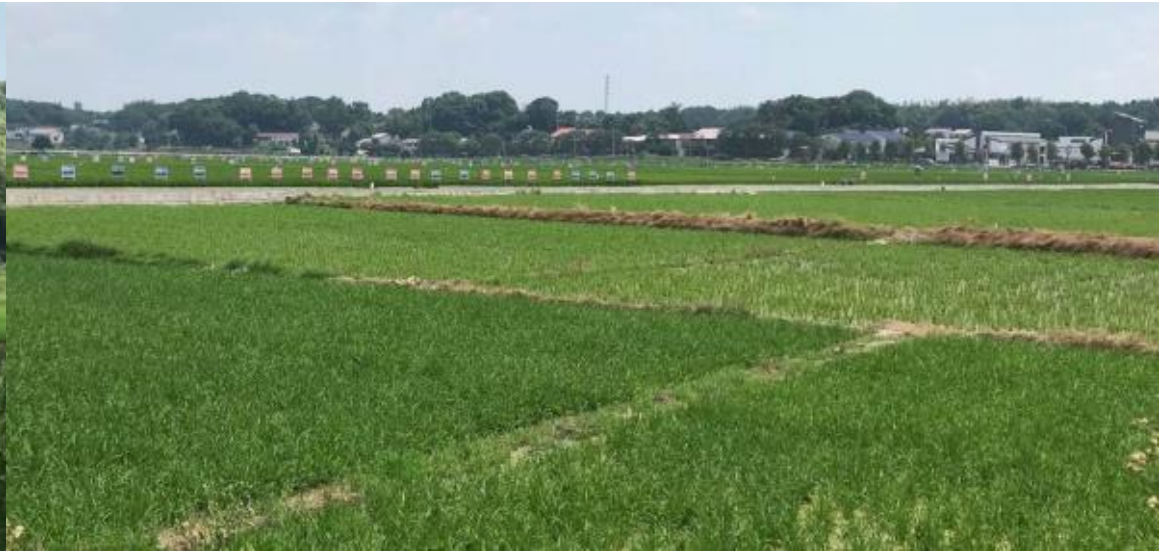

**Figure 9.5.7:** East of the site area.

### 9.5.2.4 *Assessment of Engineering Geological Conditions*

#### *9.5.2.4.1 Underground Experimental Hall*

There will be four underground experimental halls, two for the CEPC experiments and two for the CEPC RF systems, which in the future will accommodate experiments of the SPPC. They are numbered from 1# to 4#. These underground experimental halls are situated in the rock mass of Wangxiang granite as shown in Figs. 9.5.8 – 9.5.11. The boreholes for acoustic wave tesing are depicted in Figs. 9.5.12 – 9.5.15.

The surrounding rocks of the cavern are slightly weathered to fresh Late Jurassic granite, which is relatively hard to hard, with an average saturated compressive strength of 51.4 MPa to 64.2 MPa. The layout area of each hall is not affected by large faults or stratum contact zones. The comprehensive Rock Quality Designation (RQD) index of slightly weathered to fresh rock mass revealed by exploration borehole is 86.3% to 90.6%, with relatively complete to complete rock mass accounting for about 99.0%, indicating good rock mass integrity. The groundwater in the layout area of each hall is shallowly buried, with the 1# to 3# underground experimental halls located between 75 m to 85 m below the groundwater level and the 4# underground experimental hall about 60 m below the groundwater level. The surrounding rock of the cavern is a weakly to slightly permeable rock mass.

Based on the crustal stress test results of the Heimifeng Pumped Storage Power Station, it is speculated that the crustal stress fields of the 1# to 4# underground experimental halls are all low crustal stress fields controlled by gravity stress. The caverns have good overall stability, and the engineering geological conditions are suitable for building a large underground cavern group. The types of surrounding rock in the caverns of the 1# to 4# underground experimental halls are shown in Table 9.5.2. The comprehensive proportion of Class Ⅱ to Class Ⅲ surrounding rock is about 99.9%, and Class Ⅳ surrounding rock is only distributed in a local zone with dense joints, accounting for about 0.1%.

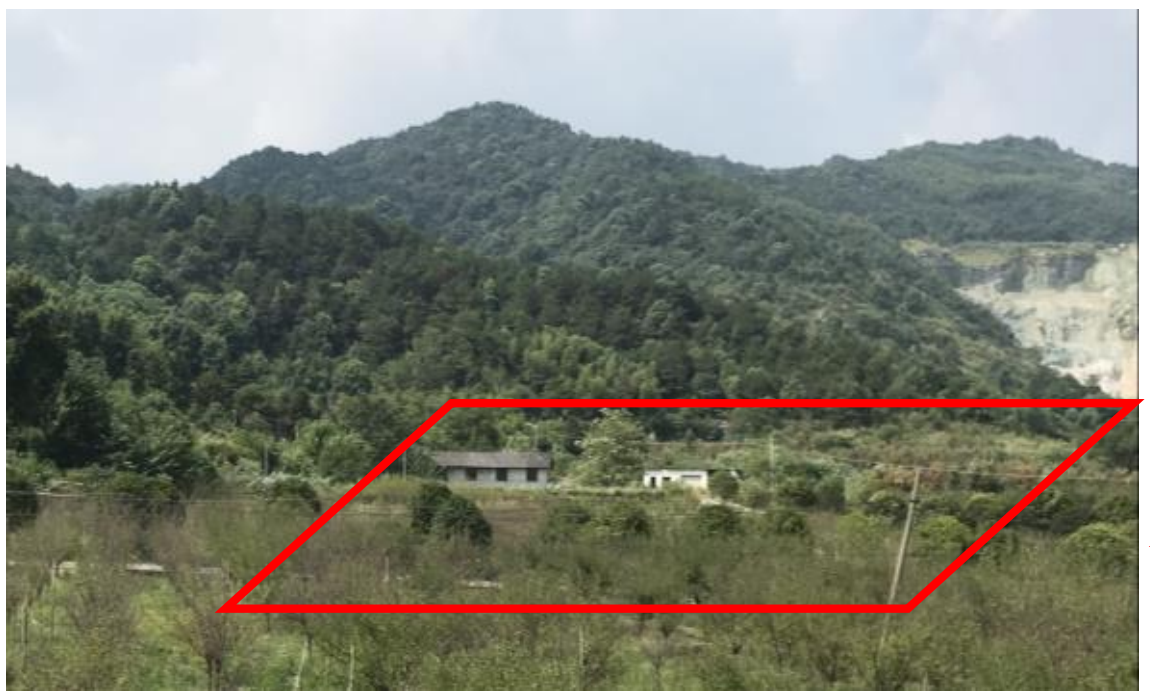

**Figure 9.5.8:** Appearance of 1# underground experimental Hall

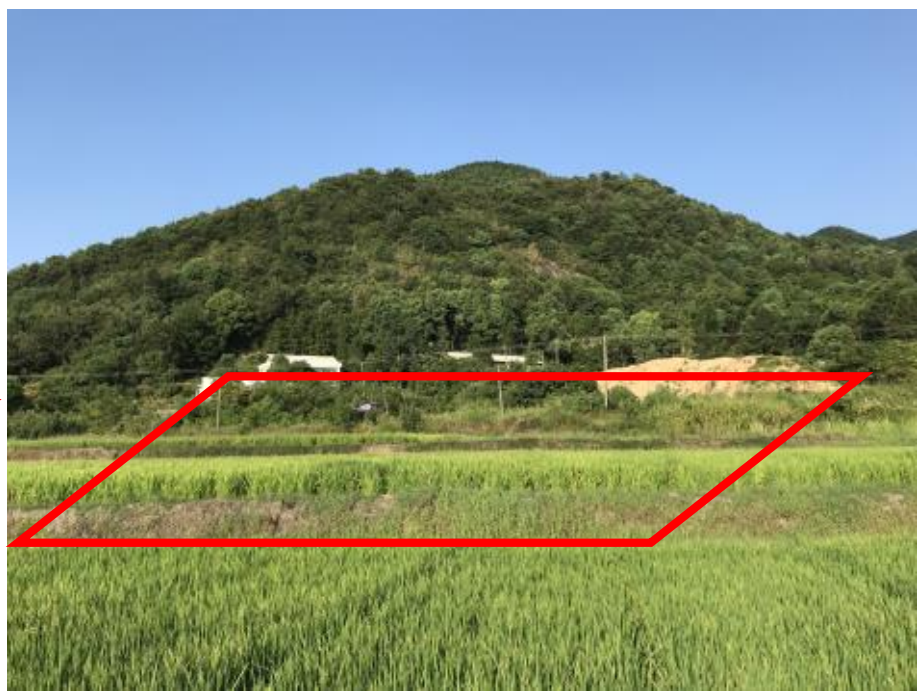

**Figure 9.5.9:** Appearance of 2# underground experimental Hall

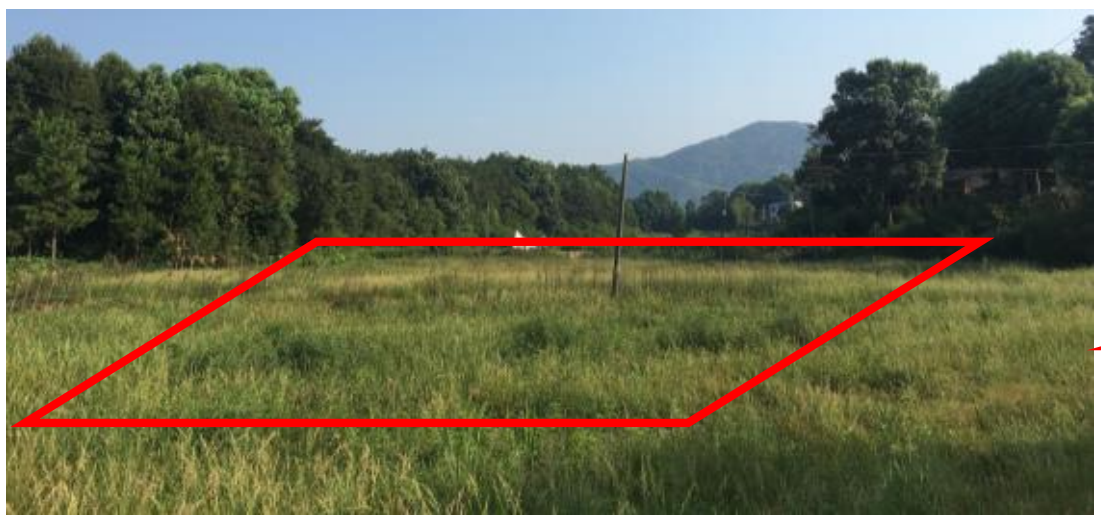

**Figure 9.5.10:** Appearance of 3# underground experimental hall

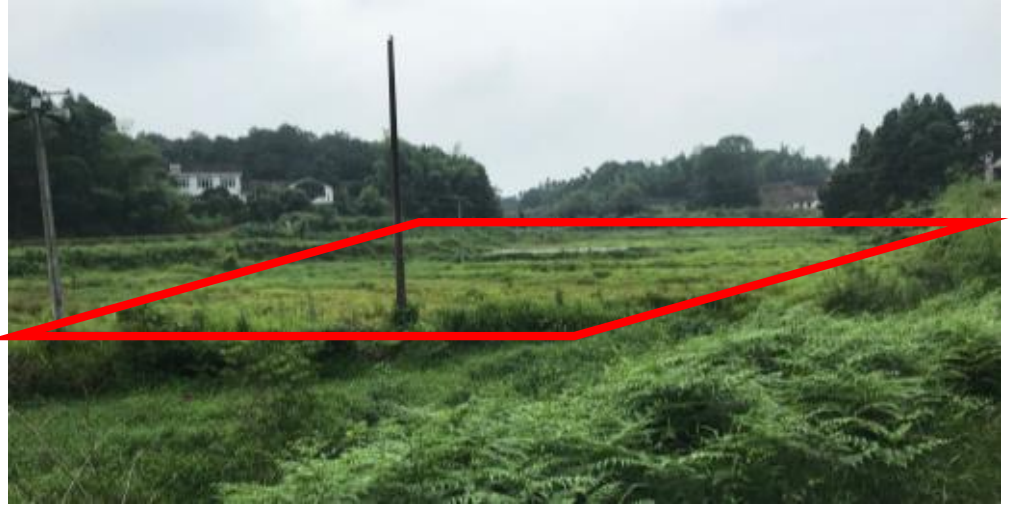

**Figure 9.5.11:** Appearance of 4# underground experimental hall

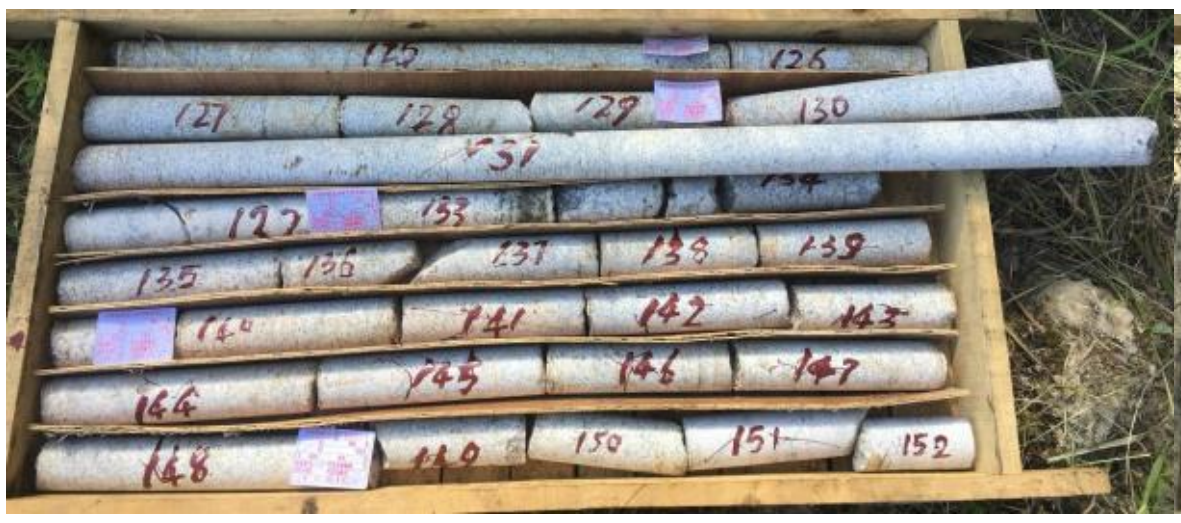

**Figure 9.5.12:** ZK1 borehole of 1# underground experimental hall

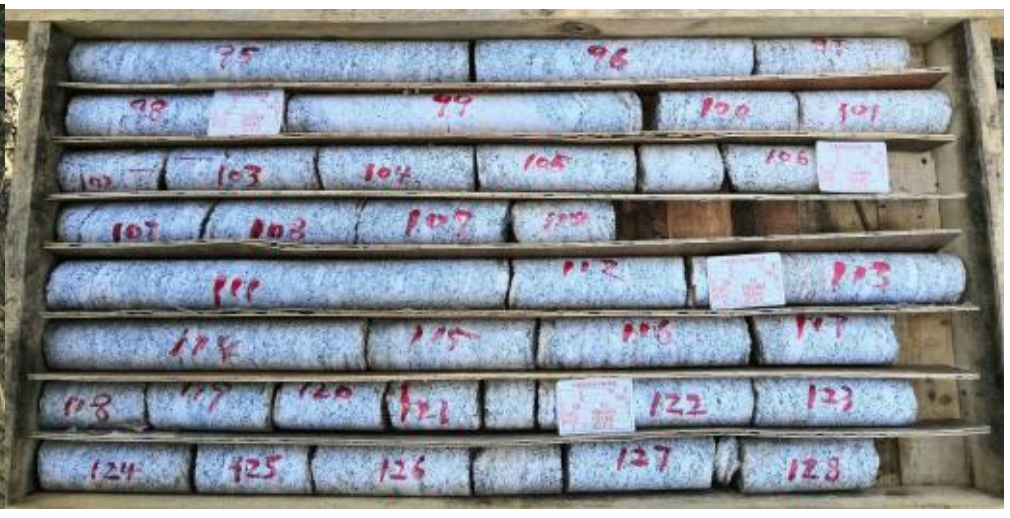

**Figure 9.5.13:** ZK2 borehole of 2# underground experimental hall

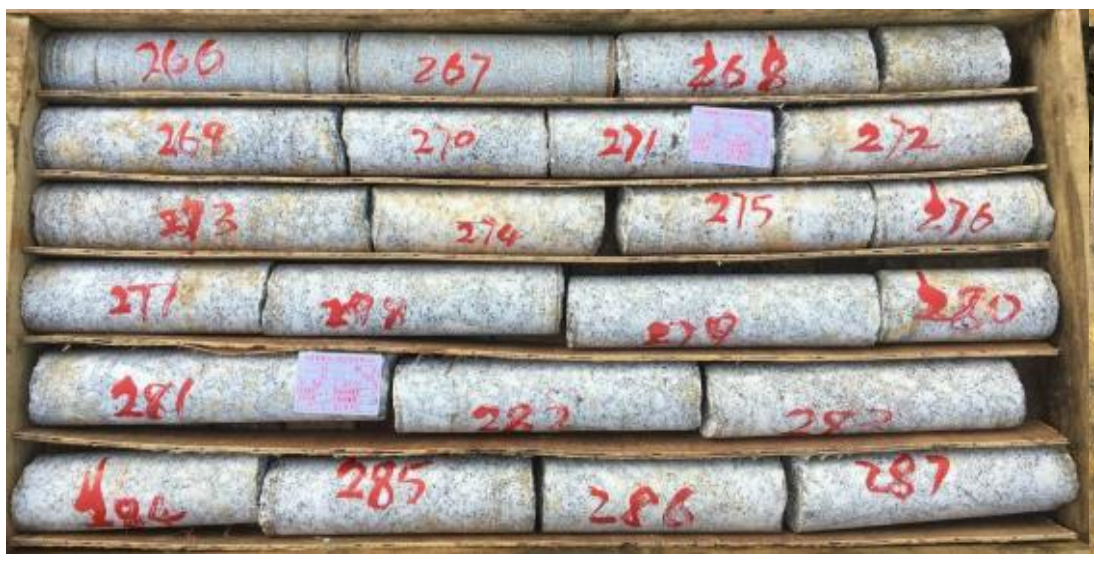

**Figure 9.5.14:** ZK3 borehole of 3# underground experimental hall

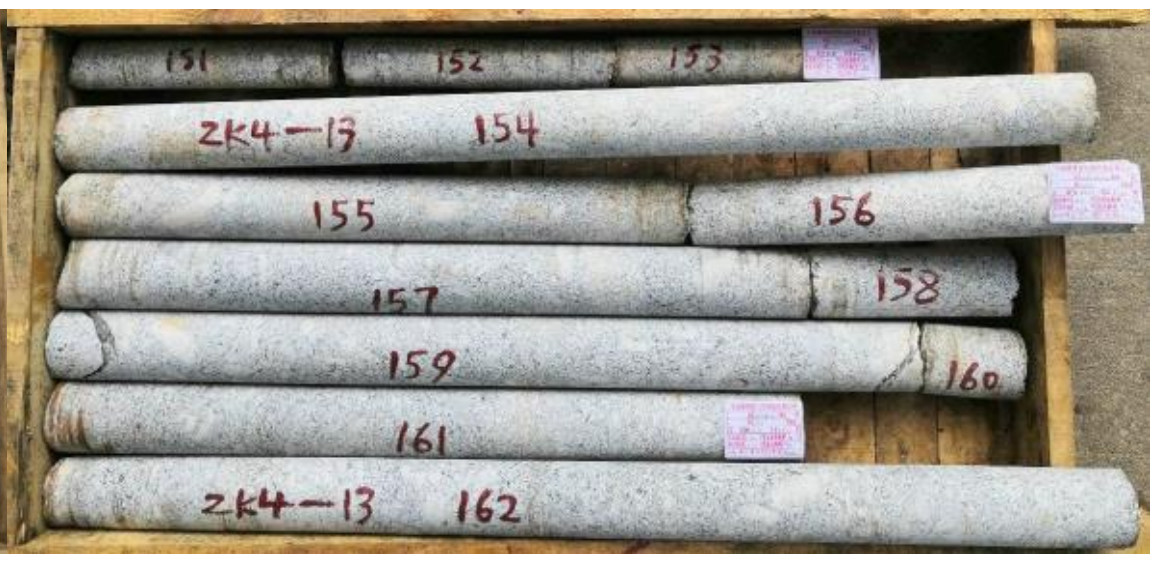

**Figure 9.5.15:** ZK4 **b**orehole of 4# underground experimental hall

**Table 9.5.2:** Classification of surrounding rocks in the caverns of the underground experimental halls.

| Location | Proportion of surrounding rock/% | | |
|---|---|---|---|
| | Class Ⅱ surrounding rock | Class Ⅲ surrounding rock | Class Ⅳ surrounding rock |
| | 0.70<Kv | 0.35<Kv≤0.70 | 0.15<Kv≤0.30 |
| 1# underground experimental hall | 83.0 | 17.0 | 0.0 |
| 2# underground experimental hall | 84.3 | 15.3 | 0.4 |
| 3# underground experimental hall | 70.0 | 30.0 | 0.0 |
| 4# underground experimental hall | 82.0 | 18.0 | 0.0 |
| Average | 79.8 | 20.1 | 0.1 |

Note: The integrity coefficient of rock mass is determined based on the results of the acoustic wave test conducted on the boreholes.

##### *9.5.2.4.2 Collider Tunnel*

The collider tunnel runs along the base of low mountains such as Heimifeng, Exing Mountain, and Piaofeng Mountain. The terrain along the tunnel route is hilly with slight undulations. The surface elevation ranges from 45 m to 95 m, with a vertical burial depth of 55 m to 105 m, and an average burial depth of around 80 m. The Baisha River area in the south section of the tunnel has the minimum vertical burial depth at approximately 45 m, while the Heimifeng area in the southwest section has the maximum burial depth at around 325 m.

The Wangxiang granite rock mass makes up approximately 93% of the tunnel sections, while the Lengjiaxi Group stratum comprises around 7%. The surrounding rocks of the cavern consist mainly of slightly weathered to fresh granite and sandy slate. The average saturated compressive strength of the higher half values of granite is 66.8 MPa, while the average value is 58.6 MPa, and the average of the lower half values is 51.0 MPa.

Most of the rocks are relatively hard to hard. The Lengjiaxi Group stratum's saturated compressive strength is generally 30 MPa to 35 MPa, and the rocks are relatively hard.

There are a total of 23 large faults developed along the tunnel, mainly comprising steep dip and medium dip faults. The width of the fault fracture zone is generally 1 m to 15 m and comprises cataclastic granite, granitic cataclastic rock, crushed rock, and fault breccia, with alteration signs of silicification, chloritization, and other factors. The joints and fissures along the tunnel are weakly developed, dominated by steep dip angles, and the rock mass is relatively complete.

The rock mass along the collider tunnel is generally not deeply weathered, with intensely weathered rock mass occurring at a lower burial depth of 15 m to 30 m, and the overlying moderately weathered to fresh rock mass with a thickness of generally 40 m to 80 m.

The tunnel is located more than 45 m below the groundwater level, and the surrounding rock of the cavern is mainly weakly to slightly permeable rock mass. The fault and dense joint zones are the primary water-bearing bodies and permeable channels.

Based on the crustal stress test results of the Heimifeng Pumped Storage Power Station, it is speculated that the crustal stress field exists along the tunnel, controlled by gravity stress. As the vertical burial depth along the tunnel is not too large, it is generally a low crustal stress field.

According to preliminary classification and statistics, which is listed in Table 9.5.3, the complete granite in the surrounding rock of the collider tunnel cavern is Class II surrounding rock, accounting for approximately 74.5% comprehensively. The granite with poor to relatively complete integrity and the sandy slate with relatively complete integrity are Class III surrounding rock, accounting for about 24.7% comprehensively. The relatively broken to broken granite, poorly intact to broken sandy slate, and developed part of Class II fault are Class IV surrounding rock, accounting for about 0.6% comprehensively. The developed part of Class I fault is Class V surrounding rock, accounting for about 0.2%. Through comprehensive analysis, Class II to III surrounding rocks account for about 99.2%, and Class IV to V surrounding rocks account for about 0.8%.

**Table 9.5.3:** Statistics of classification of surrounding rocks of the collider tunnel cavern

| Class of Surrounding Rocks | Description | Length/km | Proportion/% | Comprehensive proportion/% | Remarks |
|---|---|---|---|---|---|
| II | Slightly weathered ~ fresh granite | 74.45 | 74.5 | 74.5 | The tunnel section dominated by slightly weathered ~ fresh granite is 92.48km long in total (excluding faults), 80.2% Class II surrounding rock. |
| III | Slightly weathered ~ | 18.19 | 18.2 | 24.7 | The tunnel section dominated by |
| | fresh granite | | | | slightly weathered ~ fresh granite is 92.48km long in total (excluding faults), 19.6% Class III surrounding rock. |
| | Slightly weathered ~ fresh sandy slate | 6.53 | 6.5 | | The tunnel section dominated by sandy slate is 6.85km long in total, 95.3% Class III surrounding rock. |
| Ⅳ | Slightly weathered ~ fresh granite | 0.19 | 0.2 | 0.6 | The total length of granite tunnel section is 92.48 km (excluding faults), 0.2% Class IV surrounding rock. |
| | Slightly weathered ~ fresh sandy slate | 0.32 | 0.3 | | The tunnel section dominated by sandy slate is 6.85km long in total, 4.7% Class IV surrounding rock. |
| | Developed part of Class II fault | 0.11 | 0.1 | | For faults whose extension length is less than 10km, the width of the fracture zones is considered as 10m per zone, and there are 11 zones in total. |
| Ⅴ | Developed part of Class I fault | 0.24 | 0.2 | 0.2 | For faults whose extension length is more than 10km, the width of the fracture zones is considered as 20m per zone, and there are 12 zones in total. |

#### 9.5.2.4.3 Shafts

A total of 40 shafts are positioned along the collider tunnel, ranging from 6 to 15 meters in diameter with a depth of around 50 to 120 meters. The total depth of all the shafts combined is about 3,904 meters, with an average depth of 98 meters. The shafts are located in hilly areas, with the mouth of the shaft being relatively flat to slightly steep in a few cases. More than 95% of the shafts are situated in relatively hard to hard granite rock mass, with only two shafts (34# and 35#) situated in relatively hard sandy slate of the Lengjiaxi Group. The geological profile of the shafts is depicted in Fig. 9.5.16.

The geological structure in the area of each shaft is relatively simple, and the rock mass is slightly weathered. The intensely weathered rock mass is buried at a depth of 5 to 30 meters, with an average burial depth of approximately 20 meters. The groundwater in the shaft area is generally shallow, and the upper overburden and completely to intensely weathered rock mass are weak to moderately permeable. The lower moderately weathered to fresh rock mass is generally slightly permeable.

The surrounding rock of the shaft cavern consists of mostly overburden and completely/intensely weathered granite or sandy slate from top to bottom, with moderately weathered/fresh granite or sandy slate also present. The overburden and completely/intensely weathered rock mass have relatively poor physical and mechanical properties, strong water-bearing capabilities, and weak to moderate water permeability, resulting in a generally Class IV to V surrounding rock classification. Moderately weathered/fresh granite is mostly hard to relatively hard, with a relatively integral rock mass, poor water-bearing capability, weak water permeability, and generally classified as Class II to III surrounding rock. Only the local relatively crushed rock mass is distributed in Class IV to V surrounding rock. The moderately weathered/fresh sandy slate is generally a hard rock with poor integrity, poor water-bearing capability, and weak water permeability, mainly classified as Class II to III surrounding rock. However, some densely developed joints or foliations may fall under Class IV to V surrounding rocks.

In summary, the upper overburden, completely/intensely weathered rock mass, and relatively broken granite and sandy slate comprise about 24% of the overall surrounding rock classification, classified as Class IV to V. The lower part of the shaft is located in the moderately weathered/fresh rock mass, where the surrounding rock of the cavern is mainly Class II to III, comprising about 76% of the overall surrounding rock classification.

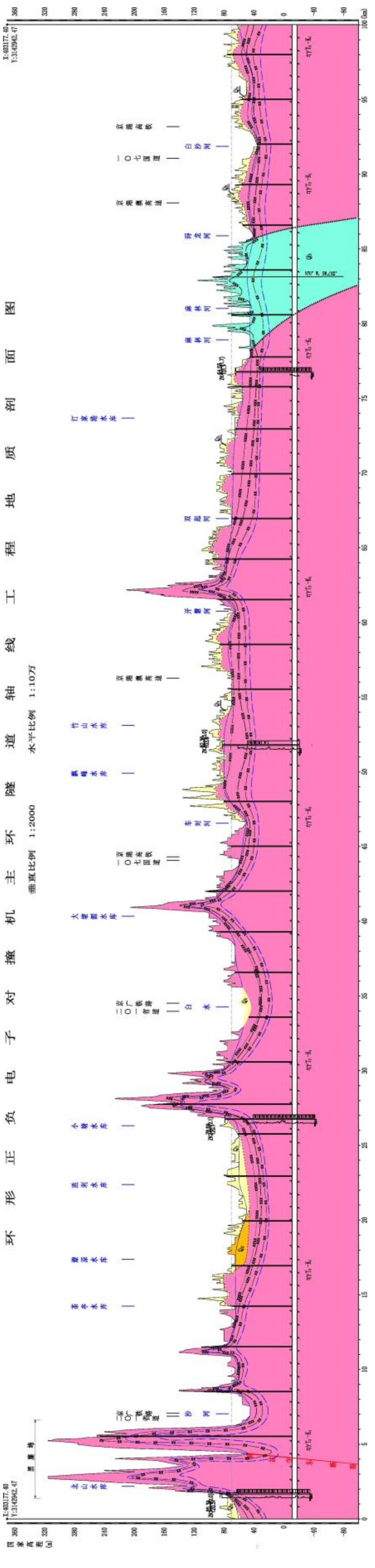

**Figure 9.5.16:** Engineering geological profile of the collider tunnel axis at the Changsha site.

#### *9.5.2.4.4 Ancillary Works*

a) Ground laboratory

The 1#-4# ground laboratory sites do not exhibit any unfavorable physical or geological phenomena, such as active faults, collapses, landslides, or mudslides, indicating a stable geological environment. The site is located in a hilly area with generally gentle terrain and ground slopes ranging from 2° to 15°. The upper part of the site is mainly composed of eluvium and diluvium overburden and completely/intensely weathered granite, with a thickness ranging from 2.6 m to 33.7 m, exhibiting good physical and mechanical properties and bearing capacity of more than 150 kPa on average. Additionally, there are no soft rock strata present that could affect the site's stability. The site and foundation have good stability.

Although the groundwater in the site area is shallowly buried, the drainage conditions on the site are adequate, ensuring that groundwater has no significant adverse effects on project construction. Based on the Specification for Reservoir Area Engineering Geological Investigation of Hydropower Projects (NB/T 10131-2019), the 4 ground laboratory sites are deemed suitable for project construction.

b) Highway works

The ground investigation has revealed the presence of county highways or village roads along the collider tunnel and within 200 m of the shafts, making transportation very convenient. The routes planned for construction are generally short and located near the collider tunnel. Both the collider tunnel and the shafts are located in hilly areas, with mostly flat and open terrain nearby. No large-scale active faults, unfavorable geological masses, or special rock and soil masses that could restrict highway construction were found during the investigation of the collider tunnel route.

The eluvium, diluvium, alluvium, proluvium overburden, and completely/intensely weathered granite and sandy slate near the collider tunnel exhibit good physical and mechanical properties, making them suitable natural foundations for meeting the bearing requirements of highway subgrade and retaining walls. Consequently, the construction of highways near the collider tunnel and shafts is deemed feasible.

c) Power transmission and transformation works

The investigation has identified existing 500 kV substations near the collider tunnel, which include the Shaping, Luocheng, and Dinggong substations. Shaping and Dinggong substations are located on the south side of the collider tunnel, with a horizontal distance of 3.8 km and 5.0 km from the tunnel respectively. Luocheng substation is situated on the north side of the collider tunnel, with a horizontal distance of approximately 5.8 km from the tunnel.

The area between the substation site and the collider tunnel, as well as along the collider tunnel, is a hilly terrain that is generally flat and open. The upper overburden and completely/intensely weathered rock mass exhibit good physical and mechanical properties, and there is no distribution of unfavorable geological masses that could restrict the transmission line. Thus, it is deemed suitable for the construction of substations and transmission lines.

d) Science City

The site of Science City is situated on the eastern side of Xiangjiang River. The area has not witnessed any active fault movements since late Pleistocene, and there are no unfavorable physical and geological phenomena like landslides, collapses, and debris flows in the region. The geological environment and regional structure are relatively stable. The site area comprises a hilly terrain with a relatively flat topography and small relative surface elevation differences. The surface of the site area is covered by Quaternary overburden. The underlying bedrock is of late Jurassic granite and is relatively hard. The overburden and weathered rock mass have good physical and mechanical properties, and the site and foundation are stable. Though the site area has shallow groundwater, the drainage conditions are favorable, and the groundwater has negligible impact on the project construction. Overall, the site of the Science City is suitable for the project construction.

#### 9.5.2.5 ***Water Supply Source***

##### *9.5.2.5.1 River*

According to the investigation of the collider tunnel route, the tunnel crosses a total of 7 large rivers, namely the Shahe River, Baishui River, Chedui River, Malin River (north), Malin River (south), Xunlong River, and Baisha River. The riverbed elevation ranges from 35 m to 65 m. Among these rivers, the Baisha River, Malin River (south), and Chedui River are the larger ones. The water in these rivers is abundant and of good quality, making it suitable as a cooling water source during the operation of the collider.

##### *9.5.2.5.2 Reservoir*

Based on the investigation, there are 7 small Type I reservoirs and 1 medium-sized reservoir located along the collider tunnel. The design water level of the reservoirs ranges from 63.0 m to 89.0 m, with a reservoir capacity of $120\times10^4$ $m^3$ to $1024.8\times10^4$ $m^3$. The total reservoir capacity is $2624.12\times10^4$ $m^3$. The Liaoyuan Reservoir, located on the west side of the collider tunnel, is the largest reservoir near the tunnel, with a total storage capacity of $1024.8\times10^4$ $m^3$. The water quality of all eight reservoirs along the collider tunnel is generally good, and the water quantity is sufficient to meet the demand for cooling water during the operation of the collider. Therefore, these reservoirs can be used as the water source for the collider.

##### *9.5.2.5.3 Municipal Water Supply*

Based on the investigation, there are 12 towns located near the collider tunnel, with a straight-line distance ranging from 0.3 km to 4.5 km from the tunnel. Currently, the municipal water supply pipe network has been completed in these towns, and they can be considered as a source of water replenishment during the operation period of the collider. Water can be supplied from the nearby towns through the municipal water supply system.

##### *9.5.2.5.4 Water Supply Line*

Based on the investigation, there are 7 large rivers, 8 small (Type I) reservoirs, and 12 market towns located along and near the collider tunnel. These sources can be considered as potential water sources for water supply during the operation period. All water source points are located above the collider tunnel or in close proximity to it. The distance of the water supply line is short, and the topographic and geological conditions are similar to those of the collider tunnel and its nearby highways, power transmission and transformation lines, etc., which are mostly hilly landforms. The physical and mechanical properties of the upper overburden and completely to intensely weathered granite are good. Generally, there are no unfavorable geological masses and special rock and soil masses that restrict the construction of water supply facilities, such as water supply pump stations and pipelines. Therefore, the area is suitable for the construction of water supply projects.

#### 9.5.2.6 *Natural Construction Material*

##### *9.5.2.6.1 Stone*

Based on the test results, it appears that the granite in the project area is suitable for use as concrete aggregate. The test results indicate that the granite has no alkali reaction capacity and is somewhat inactive as an aggregate. Additionally, the exposed granite in the project area is dense and hard with high saturated compressive strength.

It is also noted that the excavated materials from underground caverns of the Project are mostly slightly weathered to fresh granite, with a total excavation volume of about $310\times10^4$ $m^3$ and abundant material sources. This suggests that there is a sufficient supply of suitable aggregate for the construction of the Project.

##### *9.5.2.6.2 Natural Gravel*

The natural gravel materials near the project area are located in the Xiangjiang River and Dongting Lake area and are abundant in supply. The horizontal distance from the collider tunnel to these areas is approximately 12 km and 20 km, respectively, making transportation convenient. However, environmental protection regulations and guidelines must be followed to ensure water and soil conservation.

### 9.5.3 **Project Layout and Buildings**

The Changsha Site is situated in the northern part of the Changsha-Zhuzhou-Xiangtan Integrated Economic Belt, which falls under the jurisdiction of Changsha City. It is adjacent to Yueyang City in the north and comprises areas in Wangcheng District, Changsha County, Xiangyin County, Yueyang City, and Miluo City. The site is in close proximity to the central urban area of Changsha, approximately 20 km from Huanghua International Airport, 25 km from Changsha South Railway Station, and 2.5 km from the existing Heimifeng Pumped Storage Power Station. The region is well connected with several transportation routes, including the Beijing-Hong Kong high-speed railway, Beijing-Guangzhou Railway, Beijing-Hong Kong-Macao Expressway, and National Highway 107. Moreover, Changsha urban roads like Furong North Road and Xiangjiang North Road pass through the western side of the site, ensuring convenient access.

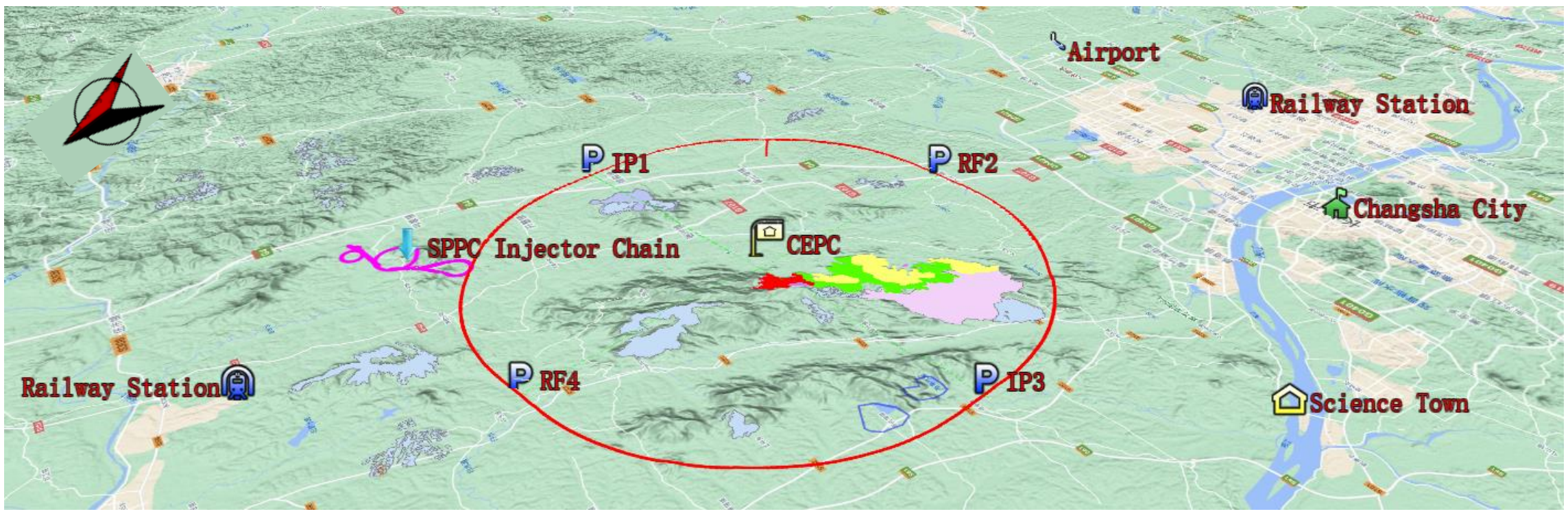

**Figure 9.5.17:** Geographical Location Map of Changsha Site

### 9.5.3.1 *Project Rank and Design Sytandard*

Based on the Standard for Flood Control (GB 50201-2014), the main works of the Project adhere to a flood control standard where the design basis flood level at the portal or shaft mouth of the external passage of the underground tunnel must be equal to or greater than the highest flood level with hydrological records or history. As for ground buildings, the protection level and flood control standard are established considering the 110 kV substation, with protection level III and a flood return period of 50 years. Additionally, the Project's main building structures are regarded as important structures and are subject to safety level I and a design service life of 100 years, according to the GB 50153-2008 Unified Standard for Reliability Design of Engineering Structures.

Moreover, the waterproofing standard of the Project is designed as Grade I, which is outlined in the GB50108-2008 Technical Specification for Waterproofing of Underground Works. This specification specifies that the structure should have no water seepage or wet stains on its surface.

In terms of seismic considerations, the Changsha site has a seismic peak ground acceleration of 0.05g and a basic seismic intensity of VI. The seismic fortification classification of beam line tunnel buildings is Class A, while the supporting power supply buildings, cryogenics systems, cooling water systems, water cooling systems, electronic rooms, etc., have the same seismic fortification classification as that of the main buildings. Additionally, the seismic fortification classification of other secondary ground buildings is Class C.

### 9.5.3.2 *General Layout of Underground Cavern*

The underground structures of the Project encompass various components such as the collider ring passage, experimental hall, linac and BTL tunnel, auxiliary tunnel, shaft, and access tunnel along the tunnel. The specific parameters of the cavern are outlined in Table 9.5.4, while the general layout is displayed in Figure 9.5.18.

**Table 9.5.4:** Characteristics of Underground Cavern

| S/N | Part | Cavern name | Clearance (W × H) | Length of cavern (m) | Remarks |
|---|---|---|---|---|---|
| 1 | LSS1 IR | Collider tunnel | ∩-shaped 6.5m(12.0m)×6.0m | 3337.13 | Section varies |
| 2 | | Bypass tunnel | ∩-shaped 3.5m×3.5m | 3337.13 | |
| 3 | 1#ARC | Collider tunnel | ∩-shaped 6.0m×5.0m | 10270.44 | |
| 4 | LSS2 | Collider tunnel | ∩-shaped 6.0m×5.0m | 986.84 | |
| 5 | 2#ARC | Collider tunnel | ∩-shaped 6.0m×5.0m | 10185.70 | |
| 6 | LSS3 RF | Collider tunnel | ∩-shaped 6.0m×5.0m | 1692 | |
| 7 | | Bypass tunnel | ∩-shaped 8.0m×7.0m | 1500 | |
| 8 | 3#ARC | Collider tunnel | ∩-shaped 6.0m×5.0m | 10270.44 | |
| 9 | LSS4 | Collider tunnel | ∩-shaped 6.0m×5.0m | 986.84 | |
| 10 | 4#ARC | Collider tunnel | ∩-shaped 6.0m×5.0m | 10185.70 | |
| 11 | LSS5 IR | Collider tunnel | ∩-shaped 6.5m(12.0m)×6.0m | 3337.13 | Section varies |
| 12 | | Bypass tunnel | ∩-shaped 3.5m×3.5m | 3337.13 | |
| 13 | 5#ARC | Collider tunnel | ∩-shaped 6.0m×5.0m | 10270.44 | |
| 14 | LSS6 | Collider tunnel | ∩-shaped 6.0m×5.0m | 986.84 | |
| 15 | 6#ARC | Collider tunnel | ∩-shaped 6.0m×5.0m | 10185.70 | |
| 16 | LSS7 RF | Collider tunnel | ∩-shaped 6.0m×5.0m | 1692 | |
| 17 | | Bypass tunnel | ∩-shaped 8.0m×7.0m | 1500 | |
| 18 | 7#ARC | Collider tunnel | ∩-shaped 6.0m×5.0m | 10270.44 | |
| 19 | LSS8 | Collider tunnel | ∩-shaped 6.0m×5.0m | 986.84 | |
| 20 | 8#ARC | Collider tunnel | ∩-shaped 6.0m×5.0m | 10185.70 | |

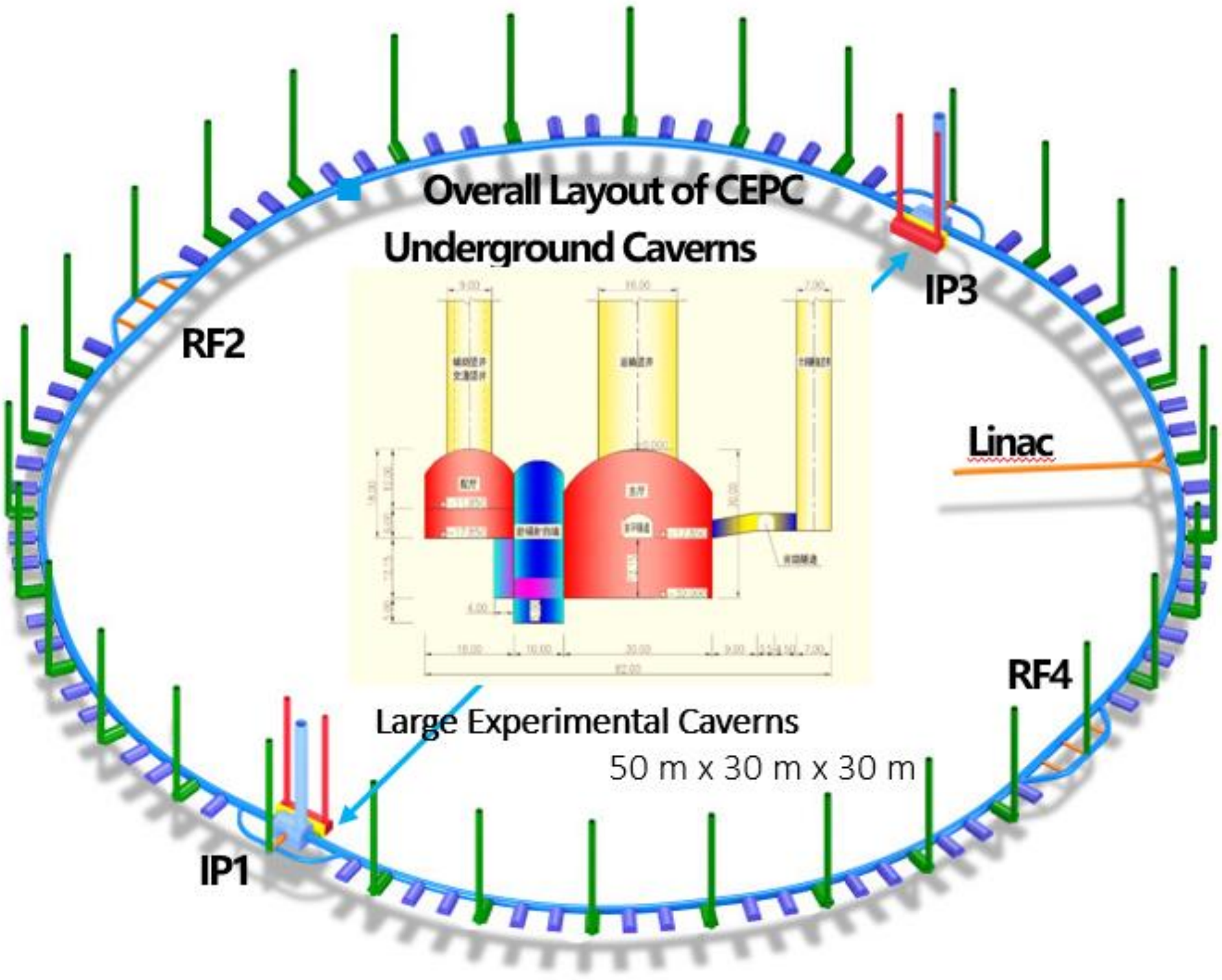

**Figure 9.5.18:** General layout of underground buildings.

### 9.5.3.3 *Surface Structures*

#### *9.5.3.3.1 Comparison of Section Types of Collider Tunnel*

The collider tunnel serves multiple purposes, including accommodating the electron-positron collider, booster, and future SPPC (Super Proton-Proton Collider). It also provides space for air ducts, water pipes, wire racks, and other equipment. The size and passage width of the CEPC and SPPC equipment supported on the tunnel ground determine the clearance width of the cavern, while the height is determined by the necessary safe distance between CEPC and the booster.

To optimize space utilization, the ∩-shaped section is the most suitable due to its proximity to the construction clearance with least excavated materials. However, excavating the ∩-shaped cavern can only be achieved through drilling and blasting methods. In cases where the tunnel length is extensive, traditional drilling and blasting techniques can result in a significant construction period. In such situations, the use of a tunnel boring machine (TBM) becomes more appropriate for efficient construction of long tunnels.

For the 100 km-long collider tunnel, the construction period and investment are crucial factors to consider. Therefore, two cross-section forms provided by IHEP, namely the ∩-shaped section and circular section, are compared from a technological and economic perspective. This comparison will help determine the representative option of the Changsha site. Figure 9.5.19 shows the size and layout of the ∩-shaped section, while Figure 9.5.20 displays the circular section.

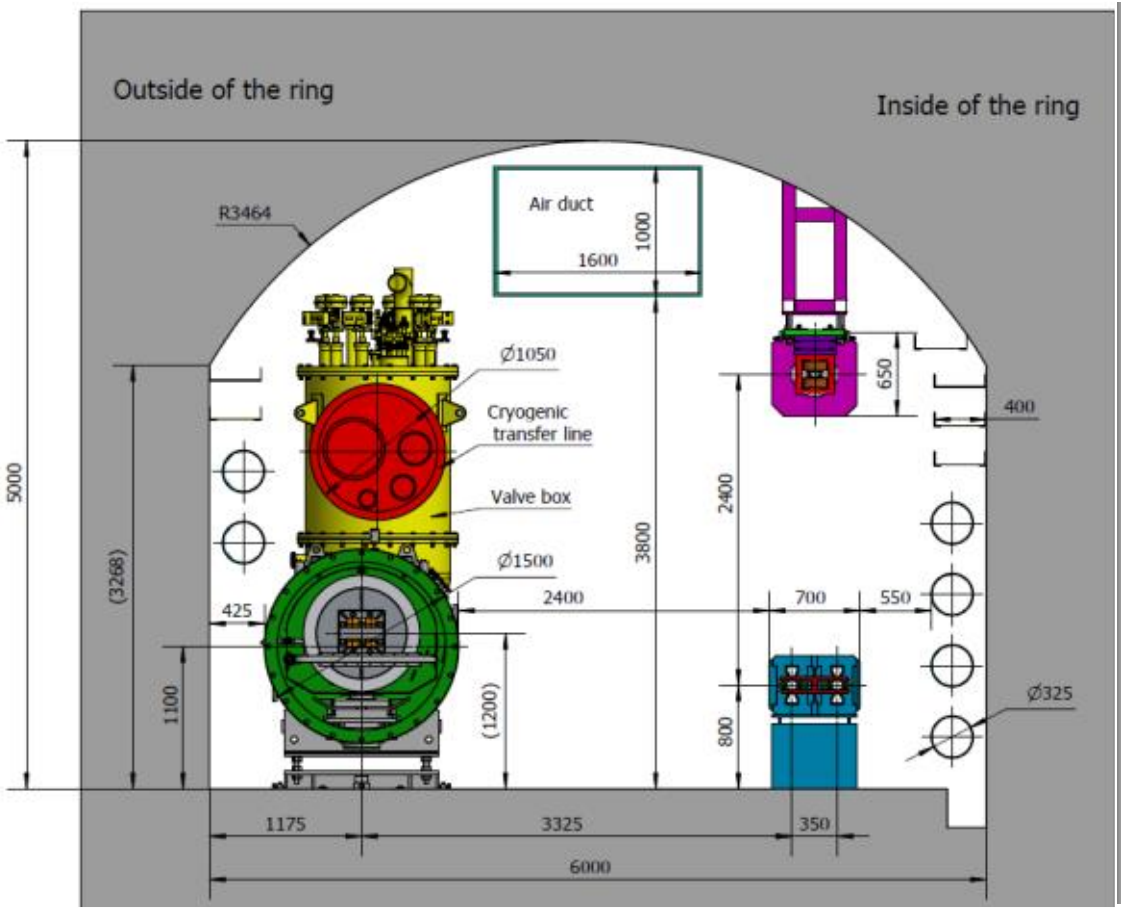

**Figure 9.5.19:** Typical layout of ∩-shaped section

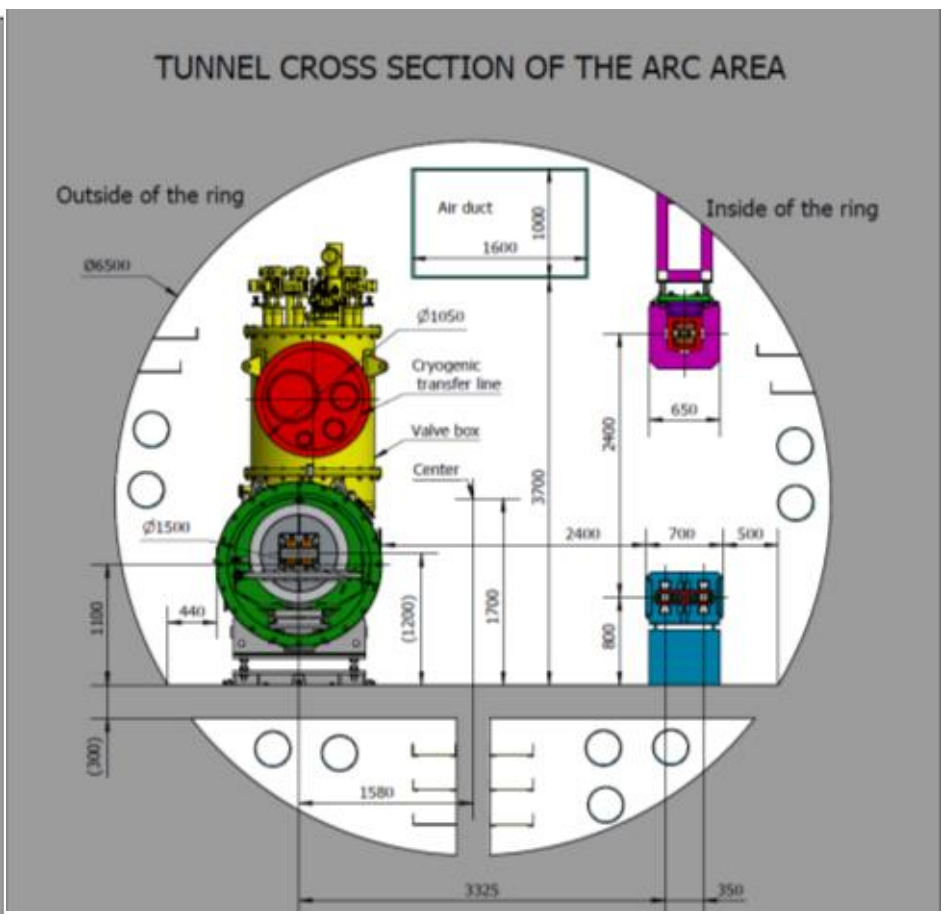

**Figure 9.5.20:** Typical layout of circular section

1) Quantities and investment

The ∩-shaped and circular sections of the collider tunnel have spans of 6.0 m and 6.5 m, respectively, with consistent support parameters and cavern length during the comparison of quantities and investment. The larger excavation area of the circular section results in larger quantities of rock tunnel excavation, shotcrete, and bottom slab concrete than the ∩-shaped section.

The main difference in unit price is the cost of rock tunnel excavation. Table 9.5.5 shows a comparison of quantities of different sections of the collider tunnel.

**Table 9.5.5:** Comparison of Quantities of Different Sections of Collider Tunnel

| S/N | Name of Works | Unit | Quantities of Circular Section | Quantities of Inverted U-shaped Section | Difference (Circular - inverted U-shape) |
|---|---|---|---|---|---|
| 1 | Tunnel excavation of stonework | 10,000 $m^3$ | 389.02 | 320.51 | 68.51 |
| 2 | C25 shotcrete | 10,000 $m^3$ | 21.22 | 14.40 | 6.82 |
| 3 | Mesh rebar | t | 3981.74 | 2699.15 | 1282.59 |
| 4 | Underground anchor stock Φ22, L=3m | 10,000 pcs. | 18.10 | 16.86 | 1.24 |
| 5 | Drain hole D56 | 10,000 m | 17.62 | 16.28 | 1.34 |
| 6 | Drain pipe | 10,000 m | 36.25 | 24.61 | 11.65 |
| 7 | Waterproof slab | 10,000 $m^2$ | 4.32 | 2.98 | 1.34 |
| 8 | Lining concrete C30P12 | 10,000 $m^3$ | 1.30 | 1.19 | 0.10 |
| 9 | Lining rebar | t | 778.20 | 953.85 | -175.65 |
| 10 | Waterproof canopy | 10,000 $m^2$ | 125.92 | 125.74 | 0.18 |
| 11 | Bottom slab concrete C30P12 | 10,000 $m^3$ | 58.39 | 13.95 | 44.44 |
| 12 | Bottom slab rebar | t | 4953.71 | 4953.71 | 0.00 |
| 13 | Rubber waterstop | m | 3602.78 | 2483.98 | 1118.80 |
| 14 | Backfill grouting | 10,000 $m^2$ | 1.08 | 1.13 | -0.05 |
| 15 | Consolidation grouting | 10,000 m | 0.84 | 0.84 | 0.00 |

2) Analysis on construction period

The analysis of the construction period reveals that both the TBM and conventional drill-blast tunnelling methods have their own advantages and disadvantages. The TBM method is faster, with a tunneling rate 3~10 times faster than the conventional drill-blast tunnelling method. It also has high quality, safety, environmental protection, and labor-saving benefits. However, the TBM is not as adaptable to geological conditions, has a large main unit weight, and requires high technical and management levels of construction personnel. Moreover, the TBM's advantages cannot be fully utilized for short tunnels.

On the other hand, the drill-blast tunnelling method is more flexible in adapting to geological conditions, and it can fully utilize the many shafts distributed along the line for construction access. Although the construction period is longer, it is still economical in general. Therefore, for short tunnels, such as the access hole, bypass tunnel, short auxiliary tunnel, linac tunnel, BTL tunnel, access tunnel, gamma source tunnel, and other caverns in the Project, the drill-blast tunnelling method should be adopted.

The construction of the collider tunnel with the drill-blast tunnelling method will make full use of 32 shafts evenly distributed along the construction access line. This method utilizes a total of 64 working faces, each approximately 1,600 m in length. The entire construction process with the drill-blast tunnelling method spans 50 months. This duration includes 6 months for construction preparation, 41 months dedicated to the construction of the main works, and a final 3 months for completion.

To carry out the construction using the TBM method, eight open-type tunnel boring machines are utilized. The TBM launch shaft and reception shaft align with the permanent building shaft. The overall construction period using the TBM method lasts for 52 months. This duration includes 6 months for construction preparation, 43 months for the construction of the main works, and a final 3 months for completion.

##### *9.5.3.3.2 Design Scheme for Waterproofing and Drainage of Underground Cavern*

The waterproof grade of the underground caverns is Grade I, with a specific standard that water seepage is unacceptable, and the structural surface must be free of wet stains. A comprehensive technical and economic comparison is required to determine the most appropriate waterproofing and drainage scheme due to the significant economic impact of the lining structure and waterproofing materials, and according to the structural and waterproofing requirements.

The type of support used for the tunnel is closely related to the waterproofing and drainage measures that need to be implemented. Bolting and shotcreting support shall be adopted for the Class I and Class II surrounding rocks of the Project to ensure the stability of the cavern, while reinforced concrete lining shall be adopted for the Class IV and Class V surrounding rocks.

As the underground tunnel works are located below the groundwater level where the hydrogeological conditions are complex, waterproofing and drainage measures shall be formulated according to different support types.

1. Waterproofing scheme for a tunnel lined with reinforced concrete:
   The tunnel section lined with reinforced concrete shall implement a self-waterproofing system for the main body of the reinforced concrete structure. Additionally, an additional waterproof layer will be applied to achieve a combination of rigidity and flexibility, along with multi-channel waterproofing. This comprehensive approach ensures the tunnel meets the necessary safety standards for its intended use. For the construction joints of the inside lining, a mid-buried waterstop will be employed to provide waterproofing. Similarly, for the deformation joints of the inside lining, both a mid-buried waterstop and an externally bonded waterstop will be utilized to ensure effective waterproof treatment.

2. Waterproofing measures for a tunnel with bolt-shotcrete support:
   Shotcrete and anchor stock support are highly effective lining forms for underground engineering in rock strata. In the case of ensuring bonding strength, an 80 mm thick shotcrete layer can withstand seepage pressures exceeding 1.0 MPa. The Project's surrounding rocks primarily consist of hard and intact granite, with a permeability rate generally below 1 Lu. For tunnel sections with Class II and III surrounding rocks, it is feasible to directly apply shotcrete onto the cleaned excavation face. This method achieves a superior anti-seepage effect, enhancing the overall integrity and stability of the tunnel structure.

3. Waterproofing for the tunnel ceiling:
   Given the complex hydrological and geological conditions of underground engineering, water leakage can be expected in certain tunnel sections even after implementing bolt-shotcrete support. Our institute has successfully addressed such issues using a specific method involving a waterproof ceiling and damp-proof partition wall within the tunnel. For instance, in the case of the access hole of a domestic pumped storage power station, bolt-shotcrete support is utilized as the permanent support type. Following excavation, the local crown experiences significant water yield with a wide distribution range, resulting in a wet pavement with exposed water. To enhance the operational environment within the tunnel and ensure the safety of vehicles entering and exiting, a waterproof treatment is implemented in the water seepage tunnel section. This treatment involves affixing an 8 mm thick sunlight board canopy onto a galvanized square tube framework. After this waterproof treatment, the tunnel not only appears aesthetically pleasing but also remains dry, resulting in a highly effective outcome in terms of improved appearance and moisture control.

#### *9.5.3.3.3 Excavation and Support of Large-span Underground Cavern*

The Project comprises of two experimental halls located in IP1 and IP3 sections, respectively, comprising the main cavern, radiation-proof retaining wall, and service cavern. The cavern's maximum excavation span is 60.0 m. Please refer to Table 9.5.6 for details on the cavern characteristics of the experimental halls and Figure 9.3.22 for a typical cross-section of the CEPC experimental hall.

**Table 9.5.6:** Characteristics of Caverns in Experimental Hall

| | Name | Quantity | Dimensions (L×W×H) |
|---|---|---|---|
| CEPC experimental hall | Main cavern | 2 | 50.0m × 30.0m × 30.0m |
| | Radiation-proof retaining wall | 2 | 60.0m × 10.0m × 35.0m |
| | Service cavern | 2 | 80.0m × 18.0m × 18.0m |

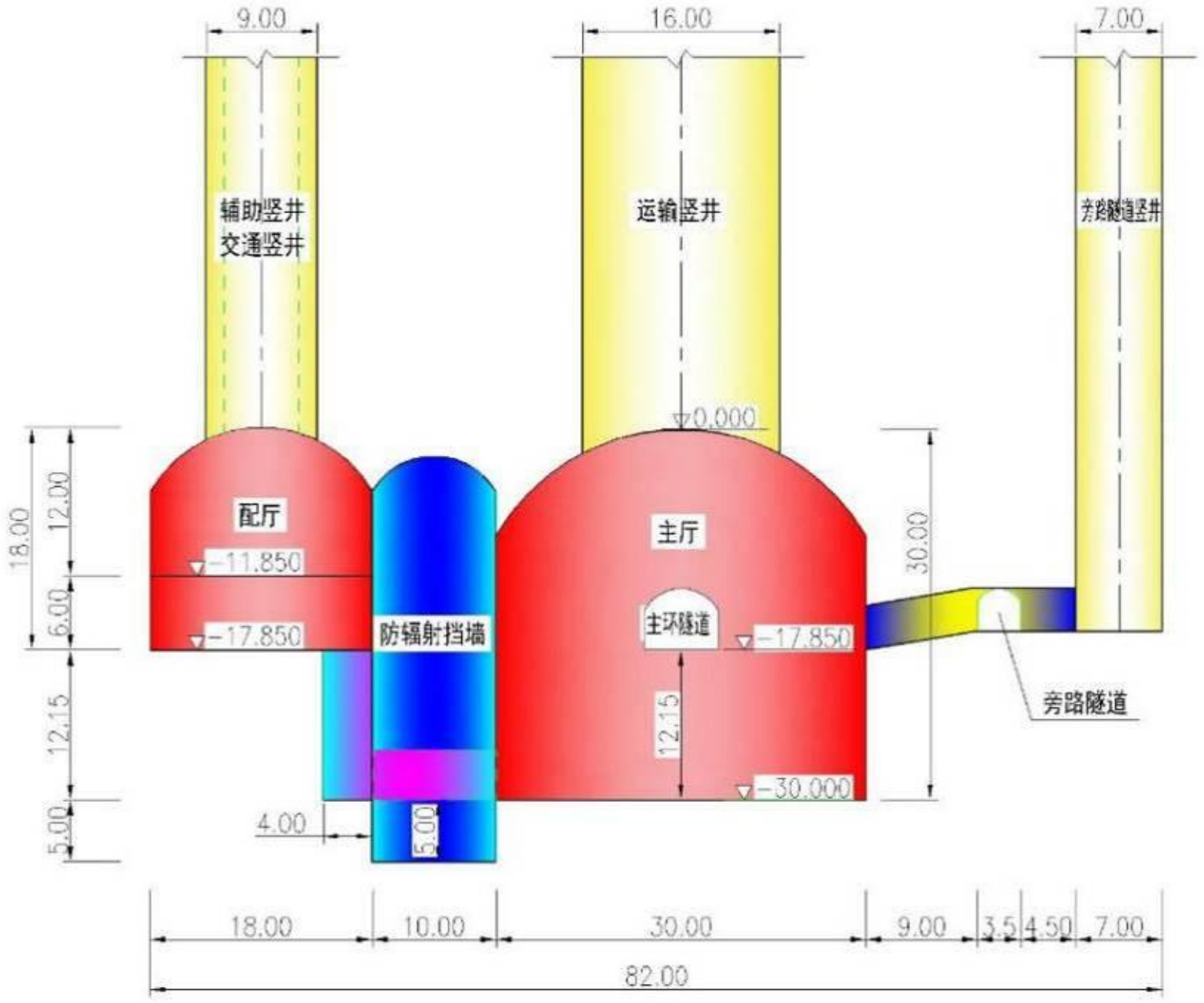

**Figure 9.5.21:** Typical Profile of CEPC Experimental Hall

**Table 9.5.7:** Experimental Hall Dimensions and List of Main Equipment

| Device Name | Parameter | Number | Installation position | Notes |
|---|---|---|---|---|
| Experimental Hall | 50.0m×30.0m×30.0m | 1 | / | |
| Experimental auxiliary hall | 80.0m×18.0m×18.0m | 1 | / | The ground of the auxiliary hall is flush with the ground of the accelerator main ring tunnel(Two layers) |
| Detector body | 14.0m×14.0m | 1 | Experimental Hall | |
| No.1 Crown block | Across the hall, lifting weight of 300t, lifting height of 20 meters | 1 | Experimental Hall | Remove the width of the platform on both sides of the span |
| No.2 Crown block | Across the hall, lifting weight of 20t, lifting height of 20 meters | 1 | Experimental Hall | Remove the width of the platform on both sides of the span |
| No.3 Crown block | Cross the auxiliary hall, lifting weight 10t | 1 | Experimental auxiliary hall | / |
| Gantry Crane | Lifting weight 1000t, span 20m | 1 | Assembly Hall | Shaft 1 is located in the experimental hall, with a diameter of 16m |
| Freight Elevator | 10t | 1 | Inside shaft 2 | Shaft 2 is located in the experimental auxiliary hall, with a diameter of 9 meters |
| Elevator | 15 Peoples | 1 | Inside the elevator shaft | The elevator shaft is located in the experimental auxiliary hall, with a diameter of 6 meters |
| Lift Platform | 1.5m× 2.0m | 2 | Experimental Hall | |
| Traffic Tunnel | / | / | / | Connect to the main ring tunnel and leave a branch to connect to the auxiliary hall |

## 9.5.4 Construction Organization Planning

### 9.5.4.1 *Construction Transportation*

The Project benefits from good site access conditions, with a range of national, provincial, county and village-to-village roads leading to the experimental hall and shaft construction area. The Beijing-Hong Kong-Macao Expressway, National Highway 107 and Provincial Highway 201 all pass through the site area, while the Yueyang-Lincwu Expressway, Provincial Highway 207, Furong North Avenue and Changsha Ring Expressway are all adjacent to the site area. To ensure effective transportation of construction and experimental equipment, the access roads for the experimental hall and RF zones will be located on permanent roads, with a new road built to connect to local trunk roads nearby. This new road will meet national standard Class III, with asphalt concrete pavement, a subgrade width of 8.5 m and a pavement width of 7.0 m.

For the transportation of other equipment, such as ventilation and transportation shafts, rural roads will be used wherever possible. A new local road will be built to meet national standard Class IV, with concrete pavement, a subgrade width of 4.5 m and a pavement width of 3.5 m.

During the construction period, materials will primarily be transported to the site by road. However, a small proportion may be transported first by railway from either Yangqiao Railway Station or Changsha Railway Station on the Beijing-Guangzhou Railway, before being transported by road to the site. Major electromechanical equipment will be transported by water to Changsha Xingang Port (Xianing Port), before being transported to the construction site by road.

### 9.5.4.2 *General Construction Layout*

To best utilize the layout of engineering buildings, topographic and geological conditions, and construction characteristics, the general construction layout will be a combination of centralized and decentralized arrangements.

Local resources will be fully utilized for temporary construction facilities. Utilities such as the aggregate processing system, concrete mixing station, and laboratory will be centrally located. Two aggregate processing systems will be set up in positions close to the two experimental halls, and four mixing stations and four laboratories will be centrally located near the experimental hall and RF zones.

Other temporary construction facilities will be scattered throughout each construction zone. Based on the preliminary general construction layout planning and construction characteristics, 32 construction zones will be specified, distributed at an average interval of 3.2 km. Each construction zone will have its own production and living facilities, such as concrete mixing stations, workshops, warehouses, and construction camps, or will share them with adjacent construction zones.

The spoil area will be selected in combination with local town planning. Priority will be given to existing planned spoil areas in the local area, and a new spoil area will only be considered when there is no existing one available.

### 9.5.4.3 *Land Occupation*

The land used for construction is divided into two categories: permanent and temporary occupation. The Project primarily features underground structures, with the permanent land occupation limited to areas where there are surface structures and permanent roads, amounting to a total area of approximately 1.85 million m$^2$. The temporary land occupation includes the spoil area, temporary roads, and construction sites, encompassing a total area of about 1.69 million m$^2$, of which 590,000 m$^2$ is designated for the spoil area.

#### 9.5.4.4 *Construction Duration*

Based on the construction organization design, the total construction period is estimated to be 50 months. This period includes 6 months for construction preparation, 41 months for the main works construction, and 3 months for completion.

### 9.5.5 Land Acquisition and Resettlement

Based on the current CEPC layout plan, the Project does not involve large-scale residential areas, large-scale enterprises, or Class I protected forest lands. However, it is considered a significant construction project and falls within the scope of prequalification for major projects that permanently use basic farmland, as required by policy. Some sections of the Project will require adjustments to avoid planned mining areas, spotty-pattern cultural relics, and historic sites located on the land. This can be achieved by modifying the layout of ground buildings.

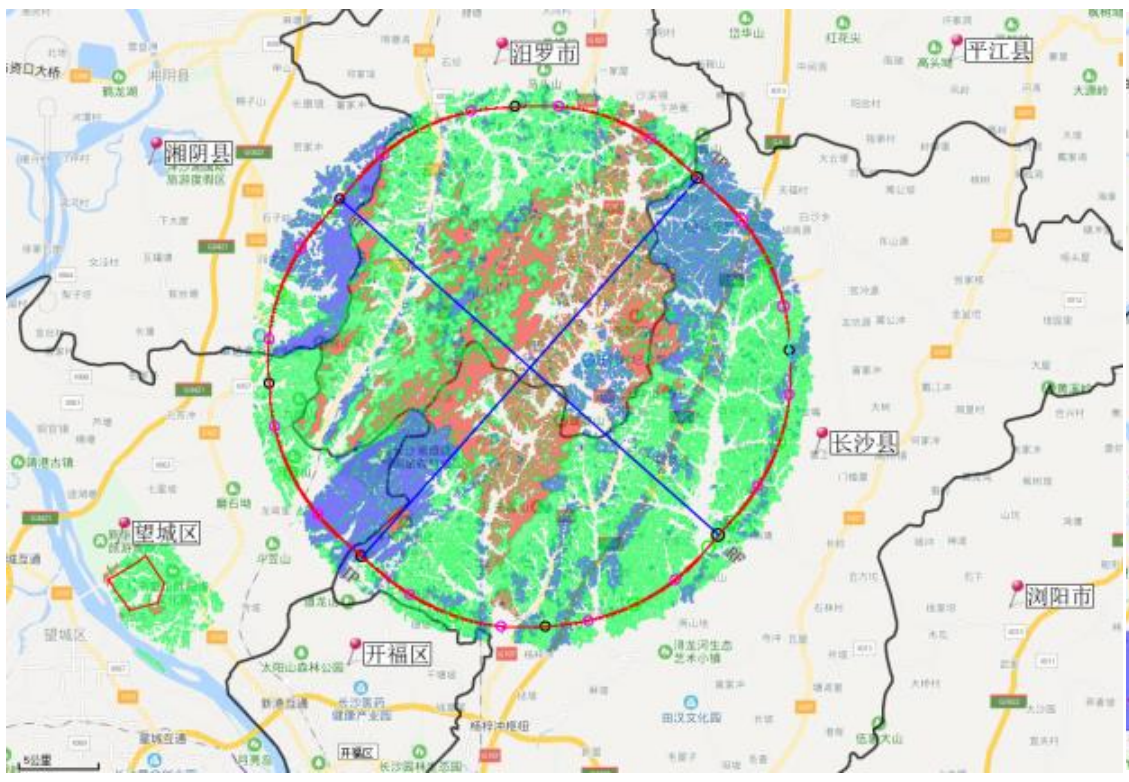

**Figure 9.5.22:** Forest land protection classification

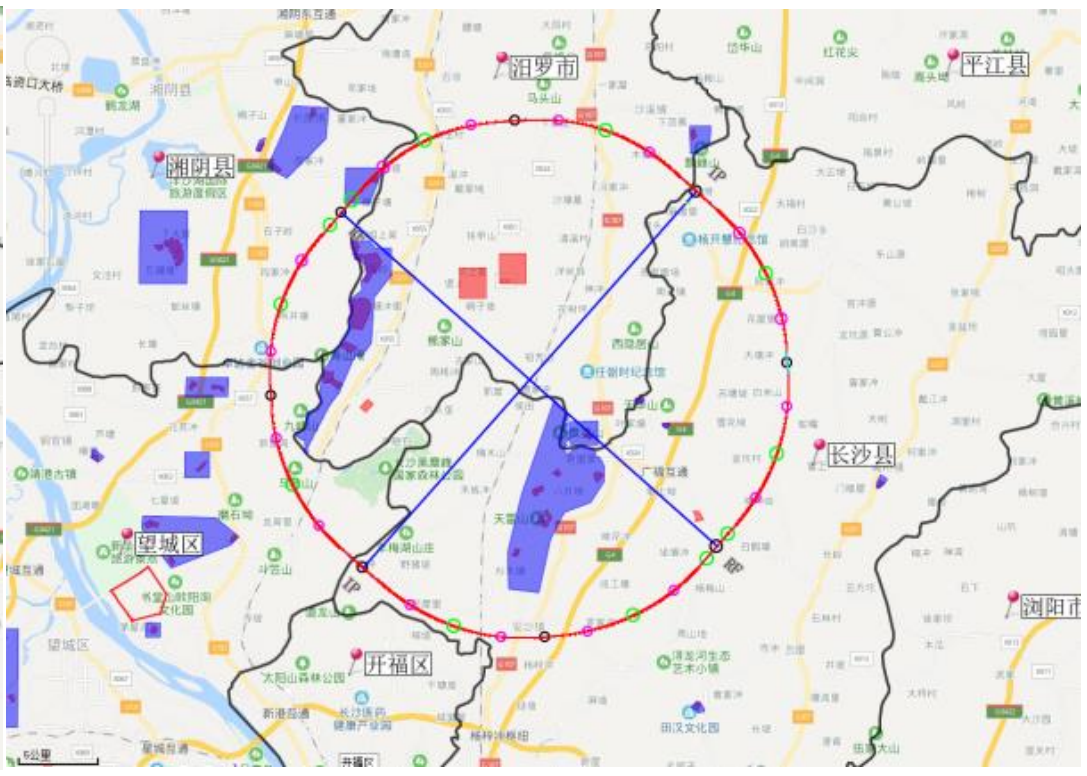

**Figure 9.5.23:** Mineral resources exploitation plan

### 9.5.6 Environmental Impact Assessment

The CEPC Project's main ring line in the southwest passes through the underground of the ecological protection red line, as well as Heimi Peak Forest Park, Beishan Mountain Provincial Forest Park, and Exing Mountain Provincial Forest Park. However, the ground buildings and shaft exits do not encroach upon the ecological protection red line. The Kwun Yam Temple Scenic Area that is affected is not a core scenic area. Additionally, the construction of the project is not subject to regulatory constraints by the ecological protection red line and forest parks.

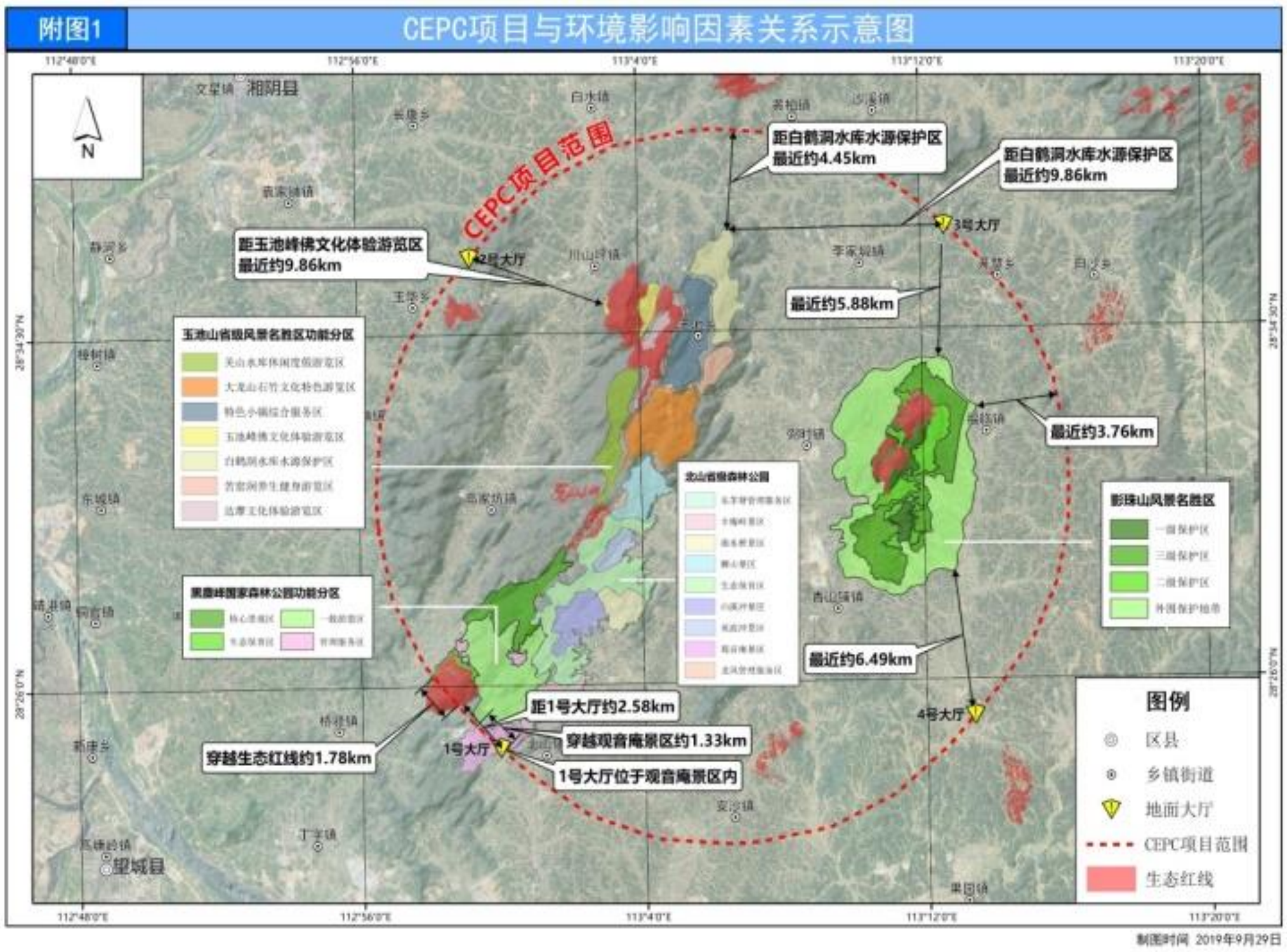

**Figure 9.5.24:** Schematic diagram of the relationship between the CEPC project and environmental impact factors

### 9.5.7 Science City Planning

According to the current plans, the Changsha Xiangjiang Science City Base will be located in Wangcheng District, Changsha City, on the east bank of the Xiangjiang River, at the southeast corner of the intersection of Xiangjiang North Road and Huangqiao Avenue. The area is primarily farmland but will be transformed into a vibrant, modern city with excellent transportation connections, including a 10-minute drive to the underground project. The city will also benefit from a complete set of municipal supporting infrastructure, providing easy access to downtown and the airport.

The Science City will cover an area of 10,000 mu and will be built in phases. It is expected to attract 6,000 world-class scientists to settle in the city and host more than 10,000 partner scientists and visiting scholars from around the world every year. The city aims to become an international hub for scientific research, business, entertainment, and residence.

The planning concept for the Science City is "surrounded by mountains and rivers, shining in the star city," which emphasizes the integration of wisdom in a livable ecology and the creation of an atmosphere suitable for business. The city will foster a smart science environment with international influence, promoting scientific research and innovation, open interaction, and humanities and science.

Overall, the Changsha Xiangjiang Science City Base will be a rare and livable landscape resource in the city, with a beautiful environment, surrounded by Matan Mountain and located beside the Xiangjiang River. The city's unique features, combined with its advanced infrastructure and planning, make it an exciting destination for world-class scientists and a promising location for innovative scientific research and business development.

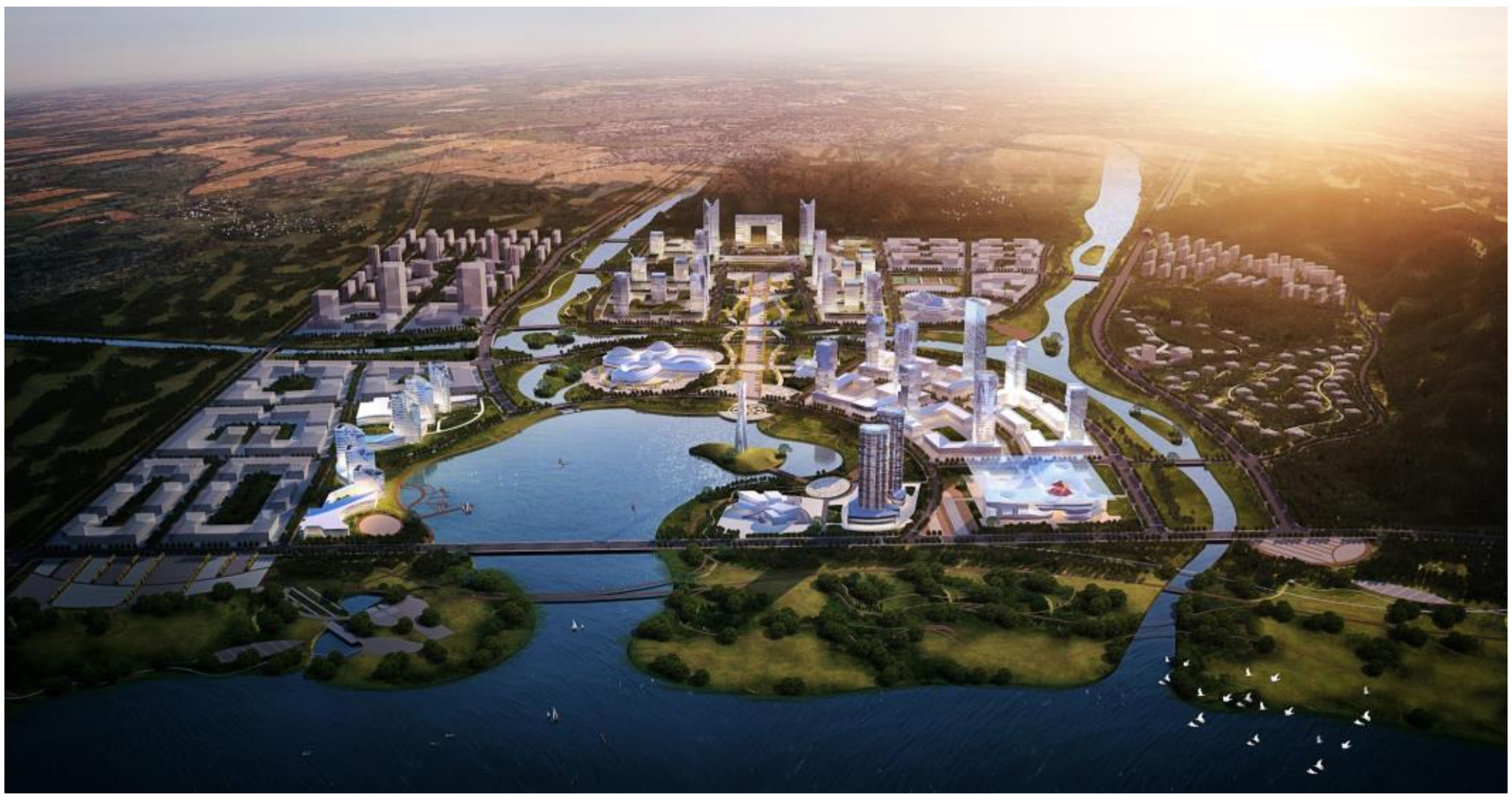

**Figure 9.5.25:** Aerial view for the planning of Xiangjiang Science City

### 9.5.8 Comprehensive Assessment

1. The site selection for the CEPC construction requires a location with convenient transportation, a beautiful environment, rich tourism resources, good cultural and geological conditions, a high internationalization level, local government support, and potential for future development, in order to establish an international science city. The Changsha-Zhuzhou-Xiangtan region, located in Hunan Province, offers superior construction and operation conditions for the project for several reasons.

   First, the region has relatively stable geological structures and meteorological conditions. Hunan Province is not located on any of the 23 main seismic belts in China and has not experienced a destructive earthquake with an intensity over 5 for more than 2,000 years, as evidenced by the archaeology at Mawangdui in Changsha. This stability provides a favorable foundation for construction. Additionally, the region experiences fewer sudden natural disasters such as landslides, collapses, and debris flow.

   Second, the region has a relatively solid security guarantee. Hunan is far from surrounding sensitive areas and is not vulnerable to military, violent, or terrorist attacks, which provides a relatively reliable safety guarantee for large-scale infrastructure facilities.

   Third, the region offers a world-leading industrial foundation for project construction. The Project involves excavating underground tunnels, which require a large number of high-quality construction machinery. Hunan Province is at the forefront of the construction machinery field globally, and relevant construction contractors in the province have accumulated rich experience in large-scale tunnel construction.

   Fourth, the region boasts a fast and convenient three-dimensional transportation network. Located on the transition zone between the eastern coastal areas and the central and western regions and the junction of the Yangtze River open economic belt and the coastal open economic belt, Hunan Province sees developed infrastructure systems of high-speed rail, aviation, and highways, making it an essential hub province to connect the east with the west and the south with the north.

Finally, the region has livable conditions that integrate the advantages of modern cities and natural landscapes. Hunan has beautiful mountains and rivers and profound cultural connotations. The regional innovation system led by the Changsha-Zhuzhou-Xiangtan National Independent Innovation Demonstration Zone is increasingly mature. The integrated growth of Changsha-Zhuzhou-Xiangtan and the development of the pan-Changsha-Zhuzhou-Xiangtan city cluster broaden the options for project site selection and provide superior working and living conditions for high-end talent.

2. In the Changsha site, a comprehensive collection of regional geological data along the collider tunnel has been conducted. This data has been used to investigate the distribution and characteristics of the Wangxiang granite rock mass. The investigation was carried out through plane geological surveying and mapping, as well as geological mapping, drilling, geophysical prospecting, and testing, which allowed for the preliminary identification of the geological environmental conditions around the underground experimental hall and the surrounding rock conditions of the underground cavern.

   The Changsha site is situated in South China, where neotectonic movement is relatively inactive, ensuring good regional tectonic stability. The prevailing hilly landform along the collider tunnel provides suitable conditions for constructing large underground caverns. A continuous and expansive Wangxiang granite rock mass, which is relatively intact, is exposed in the site area.

   Based on the preliminary investigation, it has been determined that the 1# to 4# underground experimental halls are all located in slightly weathered to fresh Wangxiang granite rock mass. This rock mass is characterized as relatively hard to hard, with approximately 99% of the rock mass being relatively intact. The majority of the rock mass exhibits slight permeability, and the surrounding rocks are predominantly classified as Class II to III, accounting for approximately 99.9% of the total. These conditions offer favorable tunneling conditions, with a total seepage amount of 293.1 $m^3$/d.

   The vertical burial depth of the collider tunnel typically ranges from 55 m to 90 m. About 93% of the tunnel section is situated within the Wangxiang granite rock mass, while only a small portion of the southeast section is located in the Presinian sandy slate. The lower limit burial depth of the intensely weathered rock mass along the line ranges from 20 m to 35 m. The caverns are all located in slightly weathered to fresh hard rock, characterized by relatively intact rock mass and low water permeability. Class II to III surrounding rocks account for approximately 99.2%.

   The shafts along the collider tunnel are positioned within hilly areas, with a relatively flat topography at the shaft mouth. More than 95% of the shafts are located within the granite rock mass. The geological structure of the shaft arrangement area is relatively simple, and the rock mass is mildly weathered. The overall proportion of Class II to III surrounding rock is approximately 76%, with a total seepage amount of 3,458 $m^3$/d for the shafts.

3. In this phase, preliminary proposals for the layout scheme and main building types of the collider tunnel, experimental hall, and ground buildings are made for the Changsha site. Additionally, a planning and design scheme for the Science City is put forward. Preliminary proposals for the main parameters and layout of lifting equipment, electrical system, water cooling system, ventilation and air conditioning system, fire protection, and water supply and drainage are also made. A site access scheme is selected, and a construction method and general construction layout of the main works are proposed, taking into account the controlled construction period. Any important physical indexes affecting the project construction in the area are identified, and policy documents for land acquisition compensation for the areas involved in land acquisition are collected. An evaluation of the impact of the project construction on the environment is conducted, and the feasibility of the project construction is preliminarily demonstrated from an environmental perspective. The main construction and installation quantities, as well as equipment quantities, are prepared, and the project investment is estimated.

## 9.6 Huzhou Site

### 9.6.1 Project Overview

CEPC is a proposed circular electron-positron collider with a circumference of 100 km. It is planned to be built in a tunnel that is about 100 m deep underground. Table 9.6.1 provides information on the civil construction aspect of the project, which includes the construction of the underground tunnel, surface structures, and various mechanical and electrical facilities required for the infrastructure.

**Table 9.6.1:** CEPC project characteristics.

| S/N | Description | Dimensions L×W×H (m) / Area ($m^2$) | Quantity | Remarks |
|---|---|---|---|---|
| | Representative site | Huzhou city, Zhejiang province | | |
| **I** | **Underground tunnel** | | | |
| 1 | Linac | | | |
| | Linac (ground) | 1800×8×6 (Above the ground） | 1 | |
| | | 1800×3.5×3.5 (Below the ground） | 1 | |
| | Damping ring | 147×20×3.5 | 1 | |
| | Beam transfer tunnel | 1600×3.5×4.375 | 1 | |
| 2 | Interaction region (IR) | | | |
| | Experimental main cavern | 52×33.5×32 | 2 | |
| | Transport shaft | 70×Φ16.0 | 2 | |
| | Collider ring tunnel (eIRD & pIRU) | 3337.13× (6-11.4) × 5 | 2 | |
| | Booster bypass tunnel | 3018.6×3.5×3.5 | 2 | |
| | Service cavern | 82×20×19 | 2 | |
| | Access tunnel for IR | 45×3.8×3.7 | 4 | |
| | Access shaft | 70×Φ6.0 | 2 | |
| | Auxiliary shaft | 70×Φ9.0 | 2 | |
| 3 | Radiofrequency zone (RF) | | | |
| | RF auxiliary tunnel | 3747.3×8×7 | 2 | |
| | Transport shaft | 70×Φ15 | 2 | |
| | Access shaft | 70×10×10 | 2 | |
| | Transport tunnel for RF | 10×7×5.55 | 4 | |
| | Collider ring tunnel | 3776.90×6×5 | 2 | |
| | Access tunnel for RF | 1200×7.5×7.5 | 2 | |
| 4 | Collider ring tunnel | | | |
| | Collider ring tunnel | 10270.44×Φ6.5 | 4 | 4 arcs |
| | | 10185.70×Φ6.5 | 4 | 4 arcs |
| | | 986.84×Φ6.5 | 4 | 4 linear sections |
| 5 | Auxiliary cavern | | | |
| | Auxiliary tunnel | 74.3×6.8×5.4 | 85 | |
| | Access & pipe shaft | 4.5×4.5×70 | 28 | |
| | Access tunnel | 1200×8×7.5 | 3 | |
| **II** | **Surface structures** | | | |
| 1 | IR (PA1) surface structure complex | 17450 | 1 | |
| 2 | RF (PA2, PA4) surface structure complex | 14150 | 2 | |
| 3 | IR (PA3) surface structure complex | 30050 | 1 | |
| 4 | Surface structures at linear accelerator section (PA5) | 12950 | 1 | |
| 5 | Substation, cooling and ventilation station (PA9, PA16, PA23, PA30) surface structure complex | 7980 | 4 | |
| 6 | Substation, cooling, ventilation station and other surface structure complex | 1980 | 24 | |
| **III** | **Construction** | | | |
| 1 | Quantity of main works | | | |
| | Soil excavation | 10,000 $m^3$ | 137.04 | |
| | Rock open excavation | 10,000 $m^3$ | 385.60 | |
| | Rock tunnel excavation | 10,000 $m^3$ | 671.86 | |
| | Concrete (including shotcrete and segment) | 10,000 $m^3$ | 174.81 | |
| | Reinforcement | 10,000 t | 9.11 | |
| | Steels | 10,000 t | 2.78 | |
| | Anchor bolt/bar | 10,000 Nr. | 154.09 | |
| | Waterproof mortar | 10,000 $m^2$ | 292.75 | |
| | Drain hole | 10,000 m | 63.04 | |
| | Drainage floral tube | 10,000 m | 31.07 | |
| | Composite waterproof plate | 10,000 $m^2$ | 12.99 | |
| 2 | Main construction materials | | | |
| | Cement | 10,000 t | 54.20 | |
| | Timber | 10,000 t | 1.75 | |
| | Explosive | 10,000 t | 1.82 | |
| | Reinforcement and steels | 10,000 t | 11.89 | |
| 3 | Labor force | | | |
| | Total man-days | 10,000 days | 240 | |
| | Average number of workers | Nr. | 1600 | |
| | Number of workers at peak | Nr. | 2200 | |
| 4 | Temporary land occupation | | | |
| | Combined tunneling | 10,000 $m^2$ | 49 | |
| 5 | Construction period | | | |
| | Drilling-blasting method | Month | 52 | Excluding preparation period |
| | TBM | Month | 40 | Excluding preparation period |
| | Combined tunneling method | Month | 52 | Excluding preparation period |

## 9.6.2 Engineering Construction Conditions

### 9.6.2.1 *Geographical Location and Transportation*

The CEPC site is located in the Yangtze River Delta Region, which is undergoing significant development under various national strategies, such as the Yangtze River Economic Belt (YREB) and the Yangtze River Delta city cluster plans. The region is focused on the Belt and Road Initiative (BRI) and has established a network of high-speed transportation systems, including high-speed railways, Yangtze River shipping, and international airport hubs, to promote the development of an international city cluster.

The Huzhou Site of the CEPC project is conveniently located with excellent transportation infrastructure. The site benefits from a dense network of interlaced and staggered railways, as well as expressways extending in all directions. In addition, there are seven airports located around the alternative site, including Shanghai Hongqiao International Airport, Shanghai Pudong International Airport, Hangzhou Xiaoshan International Airport, Nanjing Lukou International Airport, Sunan Shuofang International Airport, Changzhou Benniu International Airport, and Ningbo Lishe International Airport. Among these airports, Shanghai Pudong International Airport is one of the largest airports in China.

### 9.6.2.2 *Hydrology and Meteorology*

The CEPC site area enjoys a subtropical humid monsoon climate with four distinctive seasons. In general, spring experiences frequent cold-air outbreaks, with temperatures rebounding considerably. Summers are hot, with a prevailing southeast monsoon and frequent thundershowers. Autumn droughts are common, and winters are cold and dry, with a prevailing northwest wind.

From 2005 to 2016, the average annual temperature in the Huzhou City area was 15.5-16 °C, and the average monthly temperature was 24.6 °C. The frost-free period lasts for 224-246 days. The region has abundant rainfall, with an average annual number of rainy days of 142-155 days and an average annual precipitation of 1273.7 mm. The western hilly and mountainous areas receive an annual rainfall of 1335.6 mm, while the eastern plain receives an average of 1246.4 mm. Rainfall is concentrated between May and September, which accounts for 58.7% of the annual precipitation, while from October to the next April, it accounts for 41.3%.

Additionally, the interannual variation of precipitation is significant, with the ratio of the maximum annual precipitation to the minimum one being 2.44. This ratio is 2.52 in the western mountainous area and 2.40 in the eastern plain. However, the typhoon-triggered flood period has little influence on the selected site.

### 9.6.2.3 *Water Supply and Energy*

As the CEPC Project has a high electricity demand, the Huzhou Site located in Zhejiang will rely on the Zhejiang Power Grid. By the end of 2016, the total installed capacity in Zhejiang Province had reached 83.31 million kW. Moreover, the power supply was diversified, with renewable energy accounting for 19% and non-fossil energy for 27%. In 2016, the energy output was 319 billion kWh, including 10% from renewable energy and 25.8% from non-fossil energy sources, as shown in Fig. 9.6.1.

The total capacity required for the CEPC Project is 340.7 MW, which is less than 0.5% of the total installed capacity in Zhejiang Province. It is preliminarily estimated that the power consumption (excluding Science City) is 3.07 billion kWh, less than 1% of the total energy output in Zhejiang Province. Furthermore, Zhejiang Province is continuing to promote power construction projects and build a clean and diversified power supply to guarantee the power demand of the CEPC project.

The Zhejiang Power Grid is closely interconnected with the power grids of Anhui, Jiangsu, and Shanghai in East China, with a “two AC lines and one DC line” UHV grid framework provided. Furthermore, the grid is continuously improved to be intelligent, efficient, and reliable, featuring sufficient power supply, solid grid structure, coordinated transmission and distribution networks, overall planning of urban and rural power grids, and flexible dispatching and operation. The “six-transverse and two-longitudinal” 500kV grid serves as the backbone, with 220kV and 110kV power grids providing support and a base respectively. The Zhejiang Power Grid is equipped with large-scale pumped storage power stations and has rich operation experience in large-scale power supply, which can effectively adapt to the power consumption changes of the CEPC project.

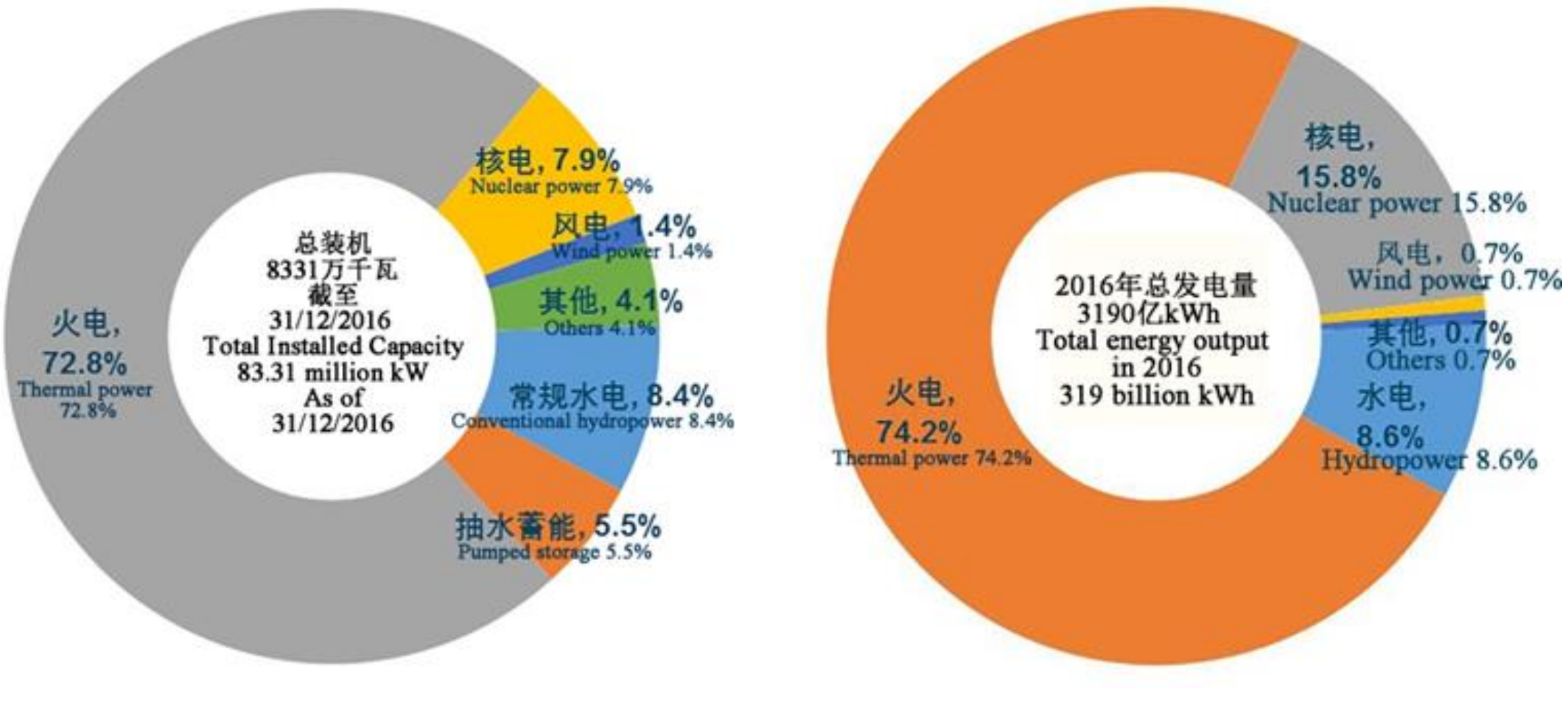

**Figure 9.6.1:** Installed capacity and energy output in Zhejiang Province in 2016.

The Huzhou site benefits from excellent power receiving conditions, relying on the 500kV Miaoxi Substation and 1000kV UHV Substation in Northern Zhejiang. The CEPC project's power consumption reliability and stability are ensured by the proximity of these regions to the intersection of the "two AC lines and one DC line" UHV grid framework of Zhejiang Power Grid, as well as the 1800 MW installed capacity of Tianhuangping Pumped Storage Power Station to the west.

In addition, since the construction and operation of Science City and experimental equipment require a substantial amount of water for cooling, sufficient water supply near the site area is necessary. Many reservoirs and lakes surround the alternative site, such as Taihu Lake, Laohutan Reservoir, Duihekou Reservoir, Dongtiaoxi River, and Xitiaoxi River. Thanks to abundant rainfall and water flow, the alternative site can guarantee an adequate water supply.

#### 9.6.2.4 *Regional Economy, Educational and Human Environment*

The Huzhou site has undergone a preliminary analysis in terms of its capability for scientific research and innovation, economic conditions, as well as ecological and cultural factors.

(1) Scientific research innovation capability:

The Yangtze River Delta Region is an area with one of the most comprehensive science and technology strengths in the country. It leads in scientific and technological development level and innovation capability, thanks to its 1/5 R&D institutes in universities, 1/4 of national key laboratories, 1/5 of academicians from the Chinese Academy of Sciences and Chinese Academy of Engineering, and various top Chinese universities and scientific research institutes such as Zhejiang University, Fudan University, Nanjing University, Shanghai Jiao Tong University, Tongji University, University of Science and Technology of China, and Shanghai Branch of the Chinese Academy of Sciences.

Moreover, the alternative site in Huzhou is adjacent to Hangzhou and located close to universities and scientific research institutes in Shanghai, Nanjing, and Hefei. This proximity promotes scientific and technological cooperation and facilitates talent cultivation.

(2) Economic conditions:

The Yangtze River Delta Region is a highly developed region with a strong advanced manufacturing industry in China, contributing to over 1/5 of China’s GDP despite only occupying 2.1% of the national land area. It is at the forefront of the development of the information economy, intelligent equipment, and other industries. The region is a key gateway to the Asia-Pacific region and has extensive international exposure, accumulating significant experience in foreign cooperation and exchange. In fact, foreign experts in Shanghai account for approximately 1/6 of all foreign experts in China.

The site in Huzhou is strategically located at the junction of Huzhou, Changxing, Deqing, and Anji, at the heart of the Yangtze River Delta region. The city of Huzhou has a solid industrial foundation and excellent financial reform and innovation conditions, making it a pilot demonstration city for "Made in China 2025" and a financial reform and innovation pilot zone. Its proximity to major economic hubs such as Shanghai, Hangzhou, and Nanjing promotes scientific and technological cooperation and talent cultivation, which is highly beneficial for the CEPC project.

(3) Ecological and cultural conditions:

Huzhou City, where the CEPC site is located, is widely known as the "Capital of Silk, Land of Fish and Rice, and Land of Culture". It is a unique prefecture-level city in China that is fully covered by ecological counties. The city is renowned for its beautiful natural scenery, with picturesque lakes, mountains, and forests. It is also the cradle of the concept of "Lucid waters and lush mountains are invaluable assets" and is listed as an Eco-Civilization Demonstration Area.

### 9.6.2.5 *Engineering Geologial Conditions*

#### *9.6.2.5.1 Overview of Regional Geology*

The CEPC site is located in the Qiantang fold belt within the Yangtze platform, with a crustal thickness of around 30 km. The neotectonic movement in the area is mainly characterized by differential up-and-down movements of fault blocks, with NE and NNE-trending faults being the most developed, followed by NW and near EW-trending faults, most of which were active faults prior to the Late Pleistocene. While regional earthquakes primarily occur in the north, the earthquakes that have significantly impacted the engineering site in history were mainly in the East China region. The far-field earthquakes that had the most significant impact on the engineering site had an intensity of degree VI. According to the Seismic Ground Motion Parameter Zonation Map of China (GB18306-2015), the proposed site has a peak ground acceleration of 0.05g, and the basic seismic intensity is degree VI, indicating good regional tectonic stability.

#### *9.6.2.5.2 Geological Condition*

(1) Topography

The project area is situated in the middle mountainous and hilly region in western Zhejiang, characterized by higher elevations in the southwest and lower elevations in the northeast. The highest peak in the area is Moganshan, which rises to an elevation of 719 m and has a NE strike. The Dongtiaoxi River, located in the east, flows towards the northeast, and most secondary gullies in the area drain into it. Meanwhile, the Xitiaoxi River in the west also flows towards the northeast, merging with the Dongtiaoxi River near Huzhou before flowing into Taihu Lake in the northern plain region.

The Moganshan mountains belong to the Zhongshan hilly landform in western Zhejiang, with elevations ranging from 50-400m. The eastern parts of the mountains, such as Huzhou-Deqing-Pingyao, are part of the northern Zhejiang plain, characterized by a dense network of rivers and lakes, with elevations ranging from 1.5-4.5m.

(2) Formation lithology

There is completely exposed strata at the proposed site (as shown in Fig.9.6.2) but it is controlled by NE-trending faults (as shown in Fig.9.6.3).

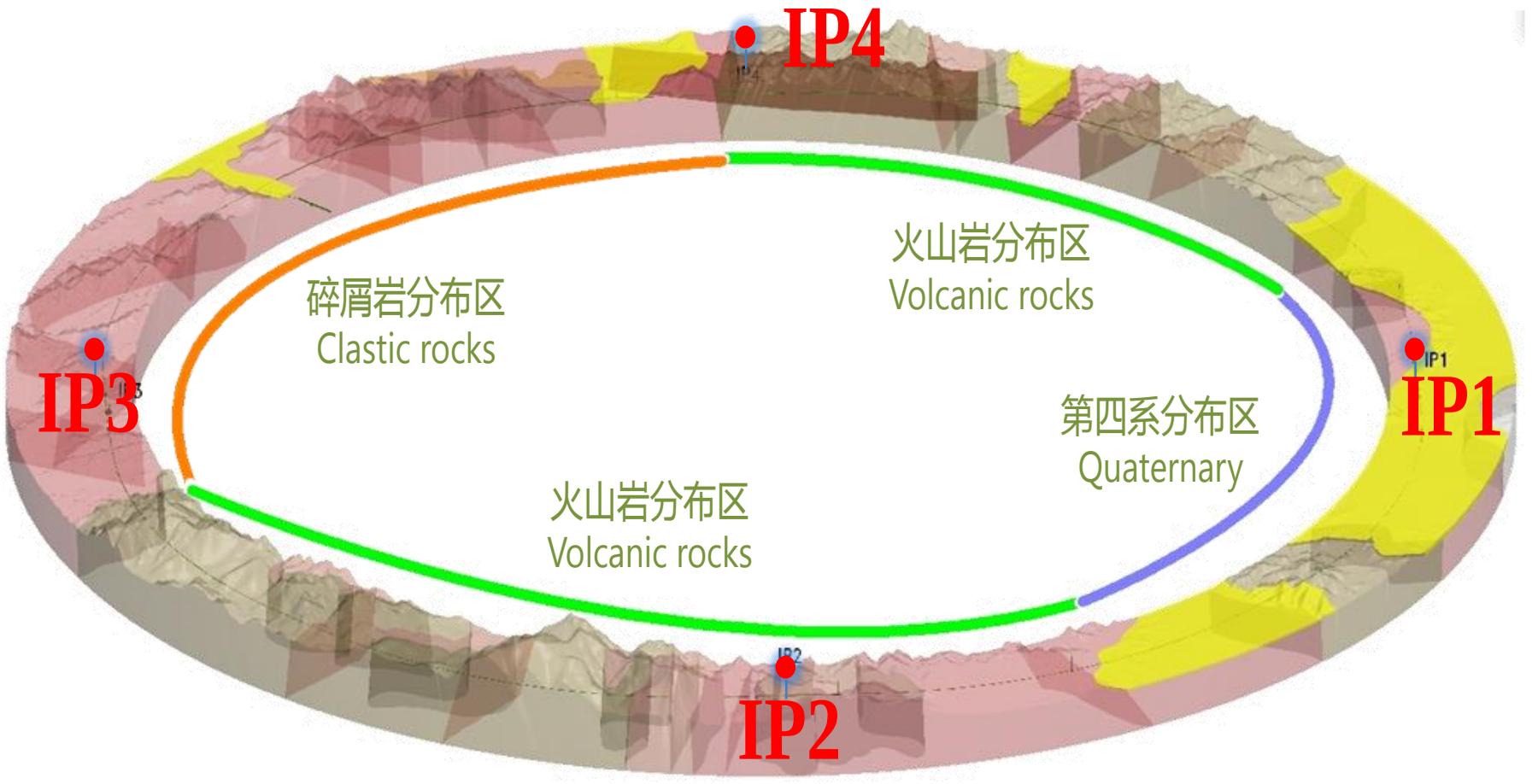

**Figure 9.6.2:** 3D presentation of the engineering geology.

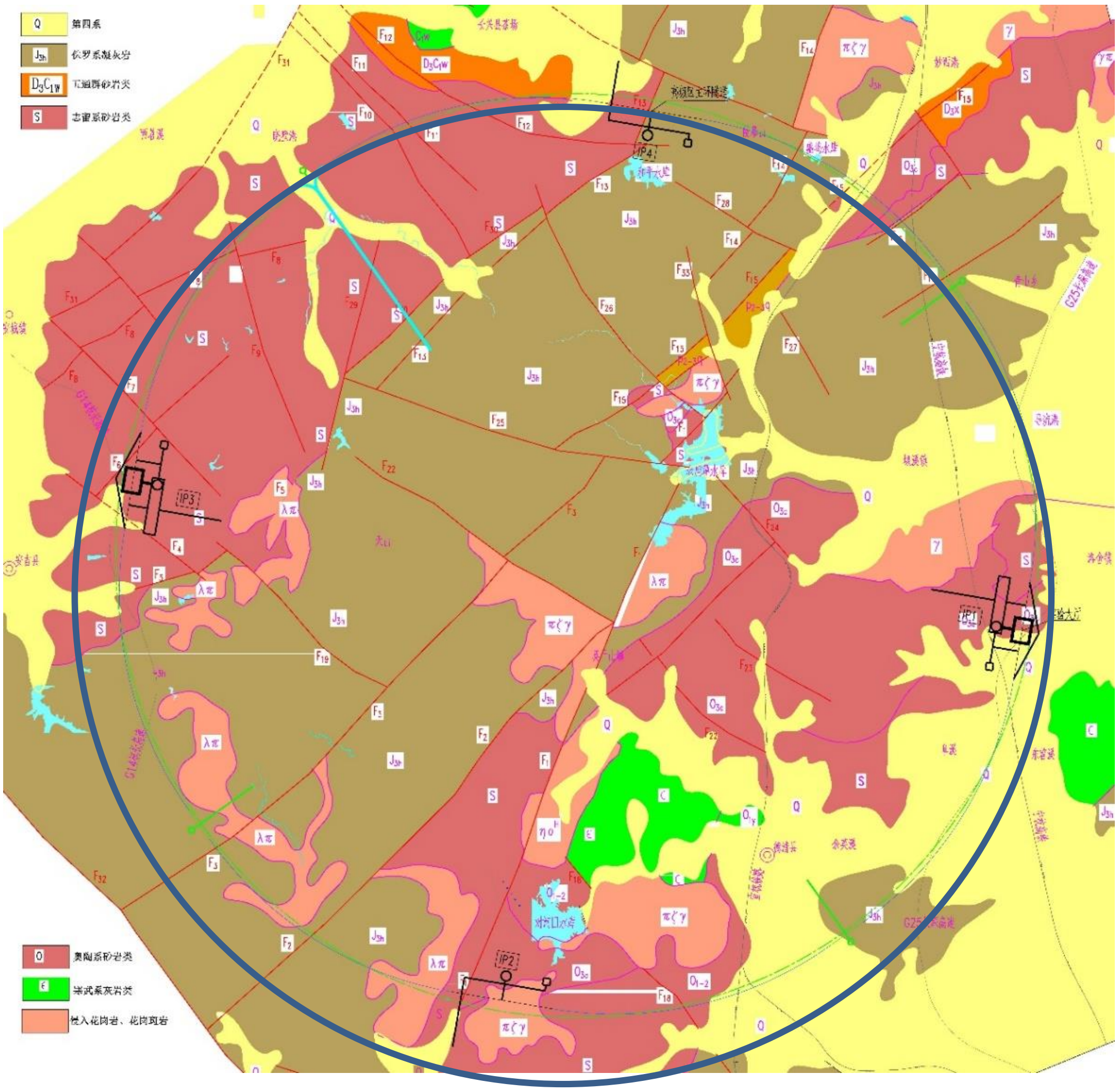

**Figure 9.6.3:** Engineering geological map of the proposed site (1: 1 million).

(3) Geologic structure:

The geological structure of the site is dominated by faults, with small folds also present. The faults can be categorized into two groups based on their attitude: (i) significant NE faults with high dip angles that extend far but are mostly dislocated by NW faults, and (ii) moderately dipped NW-NWW faults with short extensions.

Joints at the site are mostly similar to these two groups of faults. Bedding joints in sedimentary rocks in the hilly area have a NE strike and low dip angle, and are cut by NW-NWW faults. The strata are generally medium to thick, with local areas of medium to thin strata. The rock masses at the site are relatively broken.

(4) Hydrogeologic condition:

The groundwater in the project area consists mainly of bedrock fracture water and pore diving water, which are recharged from atmospheric precipitation and surface runoff. These waters seep along the fissures and discharge into the low valley. The sandstone interbedded with mudstone mainly contains layered fissure water, which is influenced by the degree of fracture development of the rock mass and the fault fracture zone. The thickness of the fully weathered layer is generally 10-15m, and water storage and permeability are good. Water gushing phenomena easily occur when encountering fault fracture zones. In addition to the development of structural joints, bedding joints are also present, but they are relatively thin. The intrusive and volcanic rock has a relatively complete massive structure, with a thickness of 5-15m in the fully weathered layer, and reticulated fissures are developed. The limestone is thick-layered, and local karst water may appear, causing gushing water or mud phenomena.

Shallow groundwater is submerged in the overburden and is significantly affected by atmospheric precipitation and surface water. Aquifers in the mountain foothills and valleys are distributed in strips, mainly consisting of sand and gravel, and are loose. Shallow groundwater in the eastern plain area is directly connected to surface water, and the water level is 1-3m in the dry season. The deep fine sand or round gravel layer has the characteristics of confined water. Near Qianyuan Town in the east, the burial depth becomes larger, generally 25.0-30.0m, and the thickness of the aquifer remains relatively constant.

(5) Physico-mechanical property of rock mass:

The empirical recommended values of physical and mechanical parameters of rock mass with different weathering degrees and lithology are shown in Table 9.6.2.

**Table 9.6.2:** Recommended parameters for rock physics and mechanics

| Rock name | Weathering degree | Gravity (kN/m$^3$) | | Proportion | Compressive strength (MPa) | | Softening index |
|---|---|---|---|---|---|---|---|
| | | Dry | Wet | | Dry | Wet | |
| Lithic quartz sandstone | Weak weathered | 25.2 | 25.5 | 2.73 | 101 | 39 | 0.40 |
| Lithic sandstone | Slightly weathered | 25.0 | 25.4 | 2.79 | 89 | 38 | 0.44 |
| | Weak weathered | 26.3 | 26.4 | 2.75 | 126 | 76 | 0.59 |
| Muddy siltstone | Slightly weathered | 25.1 | 25.2 | 2.73 | 44 | 24 | 0.54 |
| Silty mudstone | Slightly weathered | 25.0 | 25.5 | 2.76 | 77 | 17 | 0.30 |
| | Weak weathered | 25.9 | 26.4 | 2.80 | 86 | 22 | 0.35 |
| Granite porphyry | Slightly weathered | 25.9 | 25.7 | 2.70 | 150 | 100 | 0.65 |
| Rhyolitic crystal debris fused tuff | Slightly weathered | 25.7 | 25.8 | 2.68 | 156 | 123 | 0.80 |
| | Weak weathered | 26.0 | 26.1 | 2.71 | 160 | 131 | 0.79 |
| Rhyolite | Slightly weathered | 25.4 | 25.7 | 2.68 | 126 | 100 | 0.78 |
| | Weak weathered | 25.7 | 25.9 | 2.70 | 281 | 141 | 0.55 |
| Huanglong formation limestone | Weak weathered | 27.4 | 27.7 | 2.74 | 84 | 61 | 0.73 |

#### *9.6.2.5.3 Geological Conditions and Evaluation of Site Engineering*

The project area is characterized by a low hilly landform, and the valley is situated on the east and west sides of Mogan Mountain, having a minor impact on the project. The lithology of the project area mainly consists of clastic rocks such as Ordovician and Silurian sandstone with mudstone, along with volcanic rocks such as Jurassic fused tuff, tuff, rhyolite, and granite. Cambrian and Carboniferous limestone formations are also sporadically exposed, and the recommended scheme avoids them to prevent complications during the construction phase.

Faults dominate the geological structure of the engineering area, and folds are not developed. The NNE-NE-trending faults control the distribution of strata, which are large in scale and mainly extensional. The sandstone strata are dominated by NW-trending faults, which are also extensional in nature.

Based on regional geological analysis, the engineering area exhibits low in-situ stress levels, good regional structural stability, simple formation lithology, and a medium geological structure. No geological condition can restrict the project's construction, and the engineering geological and hydrogeological conditions are suitable for constructing large underground caverns. It is, therefore, recommended to conduct subsequent engineering investigations promptly to determine the layout of the cavern group.

## 9.6.3 **Project Construction Scheme**

### 9.6.3.1 ***Engineering Site Selection***

Based on the above analysis, the selection of plans depends on factors such as topography, geology, construction layout, and traffic conditions. Three site schemes are depicted in Fig. 9.6.4. The comparison in Table 9.6.3 shows that Scheme 3 has similar topographic conditions to Scheme 2 but better geological conditions. It is located further away from the Taihu Plain water network and Dongtiao River Basin, resulting in less groundwater impact. Additionally, Scheme 3 avoids core sensitive areas such as Deqing County Town and the planned Science & Technology City, making it less restrictive for construction. Therefore, after comprehensive consideration, Scheme 3 is deemed the optimal option and is recommended as the site selection for Huzhou. Further investigations are needed to determine the final construction layout.

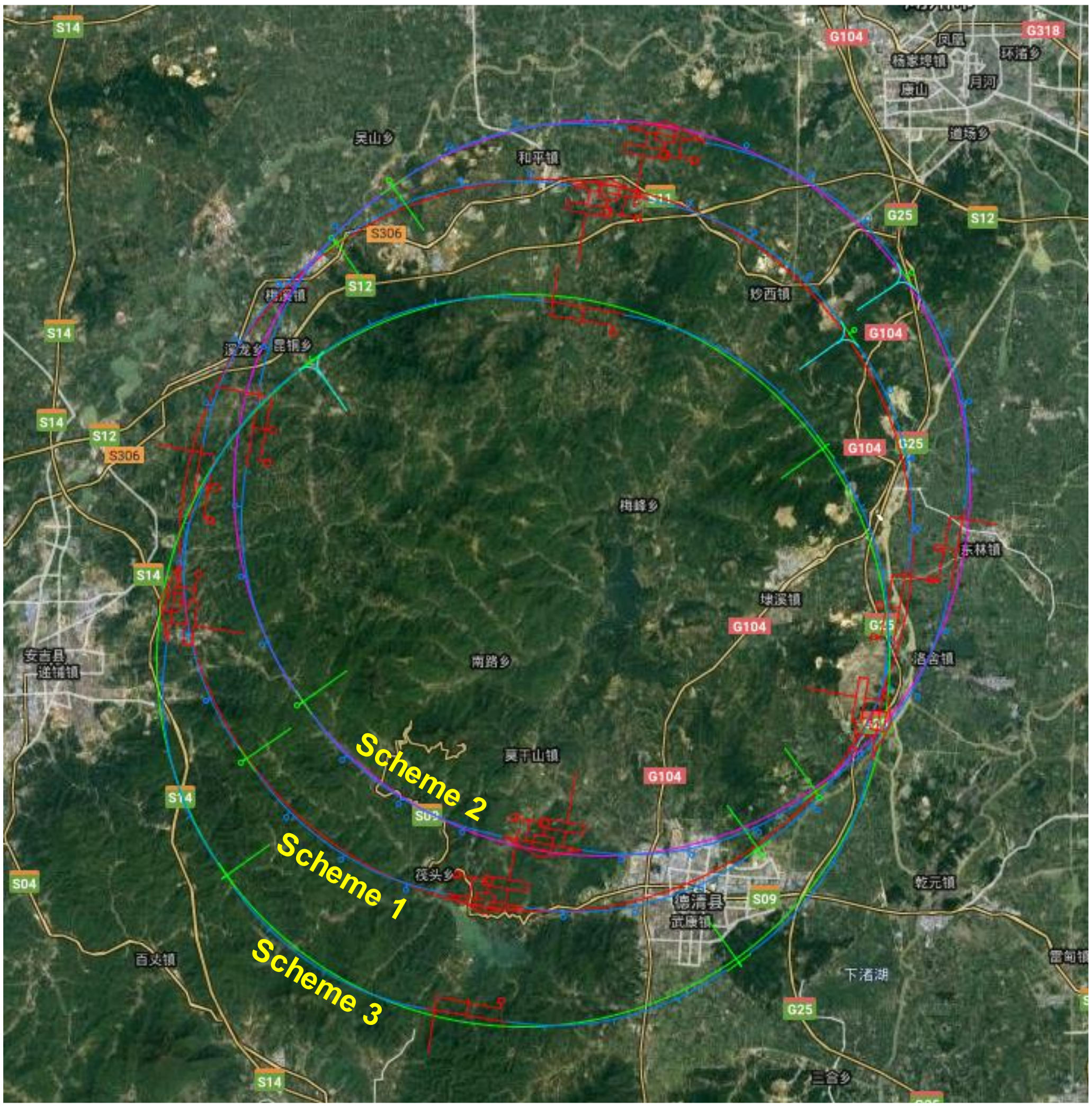

**Figure 9.6.4:** Project site selection schemes.

**Table 9.6.3:** Table of CEPC project site selection factors.

| Site-selection scheme | | Scheme 1 | Scheme 2 | Scheme 3 | Scheme comparison |
|---|---|---|---|---|---|
| Basic information | Typical geographical features | The main ring is located on the edge of the mountain between Dongtiaoxi and Xitiaoxi. The main ring passes through Deqing County, is located north of Duihekou Reservoir, and is adjacent to the planned Huzhou Science and Technology City on the north side. | It is located on the northeast side of Scheme 1, avoids Deqing County on the south side, intersects with the Science City on the north side, and overlaps with the Taihu Plain on the east side. | Located on the south side of Scheme 1, it encloses Deqing County and Duihekou Reservoir in the main ring. The main ring is situated in the mountain's rock foundation, except for a few parts near Deqing County on the east side with low terrain (deep overburden). | |
| | Center coordinates of main ring | X=3395887.82<br>Y=490538.99 | X=493013.86<br>Y=3398362.70 | X=489563.46<br>Y=3390905.92 | |
| Engineering construction conditions | Topographic condition | It is in low mountainous areas, accounting for about 75%, and the ground elevation is 20-500m. The highest ground elevation along the line is Siping Mountain on the southwest side, with an elevation of about 525m. The southeastern section along the line near Deqing County is a plain area, with a ground elevation of 0-30m, and the overburden along the line is developed, about 20-50m. The most extensive water system designed along the line is Dongtiao River, and near the South Hall is Duihekou Reservoir. | The southwest section comprises low mountainous areas, and the ground elevation is 20-270m. The highest ground elevation along the line is Ditang Mountain on the west side, with an elevation of about 370m. The northeastern section is in the plain area, accounting for about 50% of the total. The ground elevation is 0-30m, and the overburden along the line is primarily developed, about 10-50m. The largest water systems along the line are Dongtiaoxi, Miaoxigang, and Xitiaoxi. | It is in low mountainous areas, accounting for about 80 %. The ground elevation is typically 20-430 m. The highest ground elevation along the line is about 500 m in the faucet on the northwest side. The area near Deqing County is plain; the ground elevation is 0-10m, and the overburden is deep, about 20-50m. The most extensive river system along the line is Dongtiaoxi River. | Scheme 3 and Scheme 1 are similar, with low mountainous areas as the central part, and Scheme 2 with low mountains and plain terrain, each accounting for half. |
| | Geological condition | The rock layers cut along the tunnel mainly include Upper Jurassic ($J_3h$) rhyolite, dacite crystal vitreous tuff, and intrusive rocks, accounting for about 30%; Silurian, Ordovician, Devonian sandstone lithology accounted for about 36 %; Cambrian and Ordovician limestone lithology account for about 10 %, and 24% of the area passes through the deep overburden.<br>The structure of the penetrating faults along the line is composed of NE and NW-trending mid-steep dip angle faults, mostly normal faults, extensional, and a small number of reverse faults, compression-torsion. It intersects with the ring tunnel at a large angle. The groundwater level along the line is shallow, composed of bedrock fissure water and local pore water in the overburden. The water content of sandstone and limestone is high, and the water inflow in the cave may be significant. | The main rock strata cut along the tunnel include Upper Jurassic ($J_3h$) rhyolite, dacite crystal glass tuff, and intrusive rocks, accounting for about 16 %; Silurian, Ordovician, Devonian sandstone lithology accounted for about 42 %; Cambrian and Ordovician limestone lithology account for about 7 %, and 35% of the area passes through the deep overburden.<br>The structure and groundwater along the line are the same as in Scheme 1. | The main rock strata cut along the tunnel include Upper Jurassic ($J_3h$) rhyolite, dacitic crystal glass tuff and intrusive, accounting for about 40 %; Silurian, Ordovician, and Devonian sandstone lithology accounted for about 41 %; Cambrian and Ordovician limestone lithology account for about 3 %, and 16% of the area passes through the deep overburden.<br>The structure and groundwater along the line are the same as in Scheme 1. | In terms of lithological conditions, Scheme 3 is the best, Scheme 1 is the second, and Scheme 2 is poor. The three schemes have no obvious advantages or disadvantages in structure. There are no apparent advantages and disadvantages in hydrogeological conditions, and there is a possibility of sudden water inflow and large water inflow.<br>In a comprehensive comparison, Scheme 3 is better. Scheme 2 is the worst. |
| | Execution conditions | Scheme 1 passes through Deqing County, and the construction channel layout and shaft construction impact the surrounding built urban areas. In addition, the east hall and the adjacent main ring are close to the East Tiaoxi River, and the groundwater is abundant, which is challenging to construct. | The high buried depth section of the main ring in Scheme 2 is short, and the layout conditions of the main ring and the experimental hall are good. However, the east side of the main ring is in the plain area, the covering layer is thick, and the river network and water system are developed, so there are specific difficulties in construction. | The main ring of Scheme 3 is close to the existing transportation infrastructure, and the construction conditions are reasonable. However, it is the longest through the mountainous region on the southwest side, and the layout of the construction passage in the hall on the south side is challenging. The temporary road layout conditions for the construction of the working face are poor, and the structure of the shaft is tough. | Scheme 2 has the best construction conditions, while Scheme 1 and Scheme 3 have their own advantages and disadvantages. |
| Operational conditions | Traffic condition | G25 Changshen expressway, S14 Hangchang expressway, S12 Shenjiahu expressway, and S306 provincial highway can be reached nearby. | It is far from the main expressways and provincial roads, and the traffic conditions are slightly worse. | G25 Changshen expressway, S14 Hangchang expressway, and S12 Shenjiahu expressway can be reached nearby. | Scheme 1 and Scheme 3 are better. Scheme 2 is worst. |
| | Power supply conditions | It is close to Deqing County and about 5km from the center of Anji County. The overall terrain | It is close to Deqing County, about 8km from the center of Anji County, and about 8km | It is close to load concentration places such as Deqing County and Anji County, and the | Small difference among the three schemes. |
| | | along the main ring is flat, and the power supply conditions are good. The Southwest side, about 1/6 to 1/8 of the main ring section, is in the mountains, and power supply conditions are poor. | from Huzhou. The terrain along the main ring is flat, and the power supply conditions are good. About 1/5-1/4 of the main ring section on the southwest side is located in mountainous areas, and the power supply conditions are relatively poor. | power supply conditions for the main parts are good. About 1/5-1/4 of the main ring section on the southwest side is located in a mountainous area, and the power supply conditions are relatively poor. | |
| | Water supply condition | Municipal water supply to the project area, including the internal and drainage design. | Municipal water supply to the project area, including the internal and drainage design. | Municipal water supply to the project area, including the internal and drainage design. | Small difference among the three schemes. |
| Relationship with the surrounding environment | Relationship with surface drainage | The west and north sides of the main ring are adjacent to Xitiao River, the east side partially intersects with Dongtiao River, and the south side is located on the north bank slope of Duihekou Reservoir. | The north side of the main ring is adjacent to Xitiao River, and the east side intersects with Dongtiao River in an extensive range. | The east side of the main ring is close to Dongtiao River, and the others are far away from existing water systems and reservoirs. | In terms of seepage prevention and drainage construction difficulty and leakage risk, Scheme 3 is the smallest, Scheme 1 is larger, and Scheme 2 is the largest. |
| | Relationship with surrounding settlements | (1) The north hall is in the hilly area near Heping Town, with a small distribution of settlements.<br>(2) The south hall is located on the north bank slope of Duihekou Reservoir, with several settlements.<br>(3) The west hall is located on the hilly west side of Anji County, with fewer settlements.<br>(4) The east hall is in the plain area between Daixi Town and Luoshe Town, with several settlements.<br>(5) The linac segment is located in the hilly mountainous area on the southeast side of Miaoxi Town, with fewer residential areas;<br>(6) The southeast side of the main ring segment passes through Deqing County, the east side passes through Daixi Town/Luoshe Town and the plain area along the East Tiaoxi River, and the northwest side passes through Meixi Town, Xilong Township, Heping Town and the plain area along the West Tiaoxi River. The above-mentioned area is densely distributed with residential houses and settlements. | (1) The north hall is in the farmland area on the east side of Heping Town, with several settlements.<br>(2) The south hall is located northwest of Deqing County, with several settlements.<br>(3) The west hall is in the hilly area south of Xilong Township, with a small distribution of settlements.<br>(4) The east hall is in the aquatic product breeding plain area on the west side of Donglin Town, with several settlements.<br>(5) The linac segment is in the hilly and mountainous area south of Miaoxi Town, with several settlements.<br>(6) The south and east side of the main ring segment passes through Deqing County in a small amount, and the east side passes through Luoshe Town, Donglin Town, and the plain farming area along the Dongtiao River. The northeast side intersects with the planned Huzhou Science and Technology City, and the northwest side passes through Meixi Town, Xilong Township, Heping Town, and the plain area along the Xitiao River. The above-mentioned area is densely distributed with residential houses and settlements. | (1) The north hall is located on the northwest side of the Peace Reservoir, with a small distribution of settlements;<br>(2) The south hall is in the hilly area to the southeast of Duhekou Reservoir, with a small distribution of settlements.<br>(3) The west hall is in the hilly eastern area of Anji County, with a small distribution of settlements.<br>(4) The east hall is located on the hillside where the Longgang Temple is located on the west side of Luoshe Town, with a small distribution of settlements.<br>(5) The linac segment is in the hilly and mountainous area to the south of Miaoxi Town, with a small distribution of settlements.<br>(6) The south and east sides of the main ring section pass through Deqing County and the plain area along the Dongtiao River, and the west side passes through the rural areas around Anji County. The above-mentioned area is densely distributed with residential houses and settlements. | In terms of the impact on the surface settlements, Scheme 3 is the least, and both Scheme 1 and Scheme 2 are more. |
| | Relationship with surrounding transport infrastructure | The north hall and west hall are near the S306 provincial road, the south hall is near the S09 provincial road, and the east hall is near the G25 Changshen Expressway. | The four halls are far away from expressways and provincial roads. | The north hall is adjacent to S12 Shenjiahu Expressway, the west hall is adjacent to S14 Hangzhou-Changzhou Expressway, the east hall is adjacent to G25 Changshen Expressway, and the south hall is approximately 5km away from G104 National Highway. | Scheme 3 can make the closest use of the built transportation infrastructure, and the later operation stage is the most convenient, followed by Scheme 1 and Scheme 2. |

## 9.6.3.2 *Layout and Structural Design of Underground Building*

### *9.6.3.2.1 Layout of Underground Building*

The CEPC's underground infrastructure comprises a 100 km long main ring, with even distribution of two IR test areas and two RF high-frequency areas along its length. The remaining length is divided into eight arc segments and four straight segments, with the linac segment connecting to one of the linear segments. There are a total of eight permanent traffic tunnels connecting to the ground, arranged in the four test areas. In addition, cable and ventilation shafts are placed every 3 km along the main ring in the four straight segments, and a short auxiliary tunnel is arranged every 1 km.

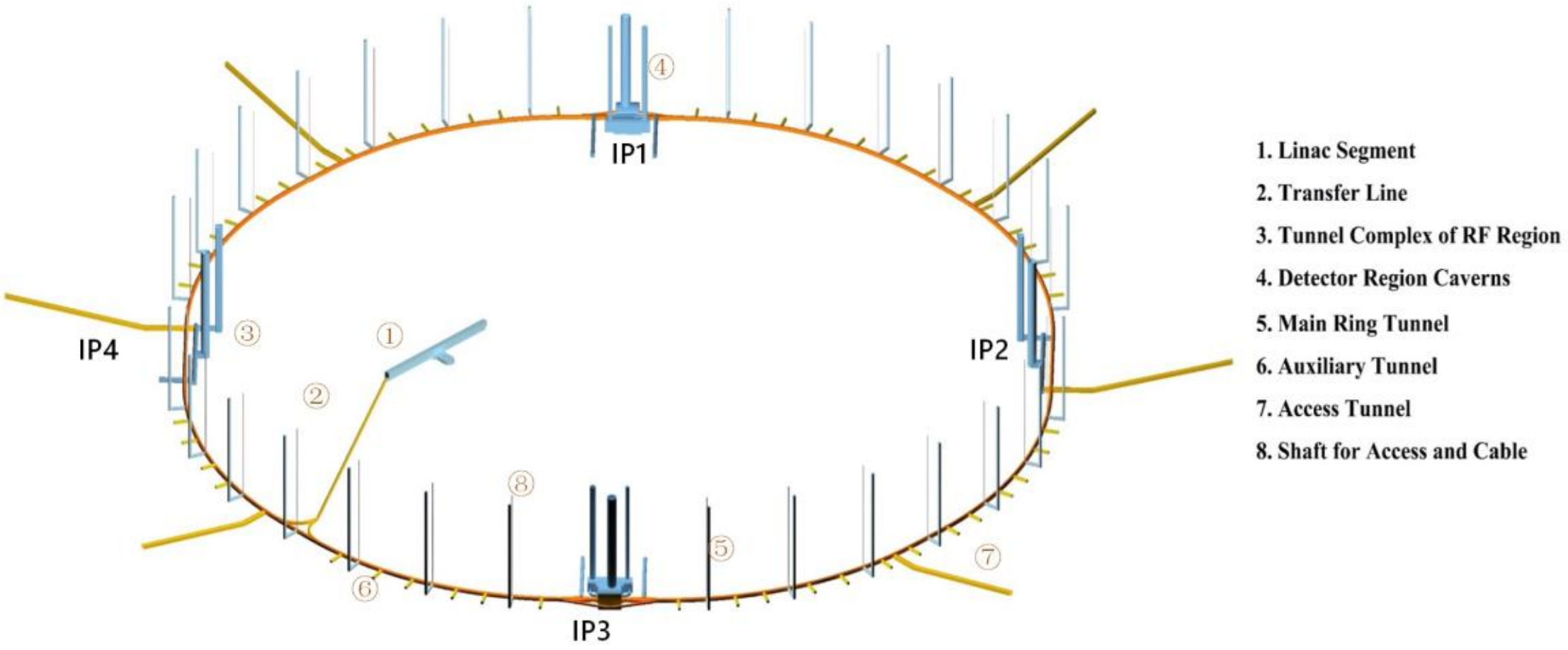

**Figure 9.6.5:** Layout of underground buildings.

#### *9.6.3.2.2 Design of Underground Structures*

(1) Cavern support design:

The design of support for the underground caverns is based on the actual geological conditions of the project site. The support system consists of bolt-shotcrete support supplemented with a rigid support, and system support with local treatment. Reasonable support parameters are determined based on the size of the caverns and surrounding rock class.

In this stage of the design, the support parameters are initially determined through surrounding rock classification and engineering analogy, as shown in Table 9.6.4. Considering the geological conditions of the main caverns in the project, and based on the engineering analogy, the support parameters for the underground caverns are preliminarily determined, as shown in Fig. 9.6.6.

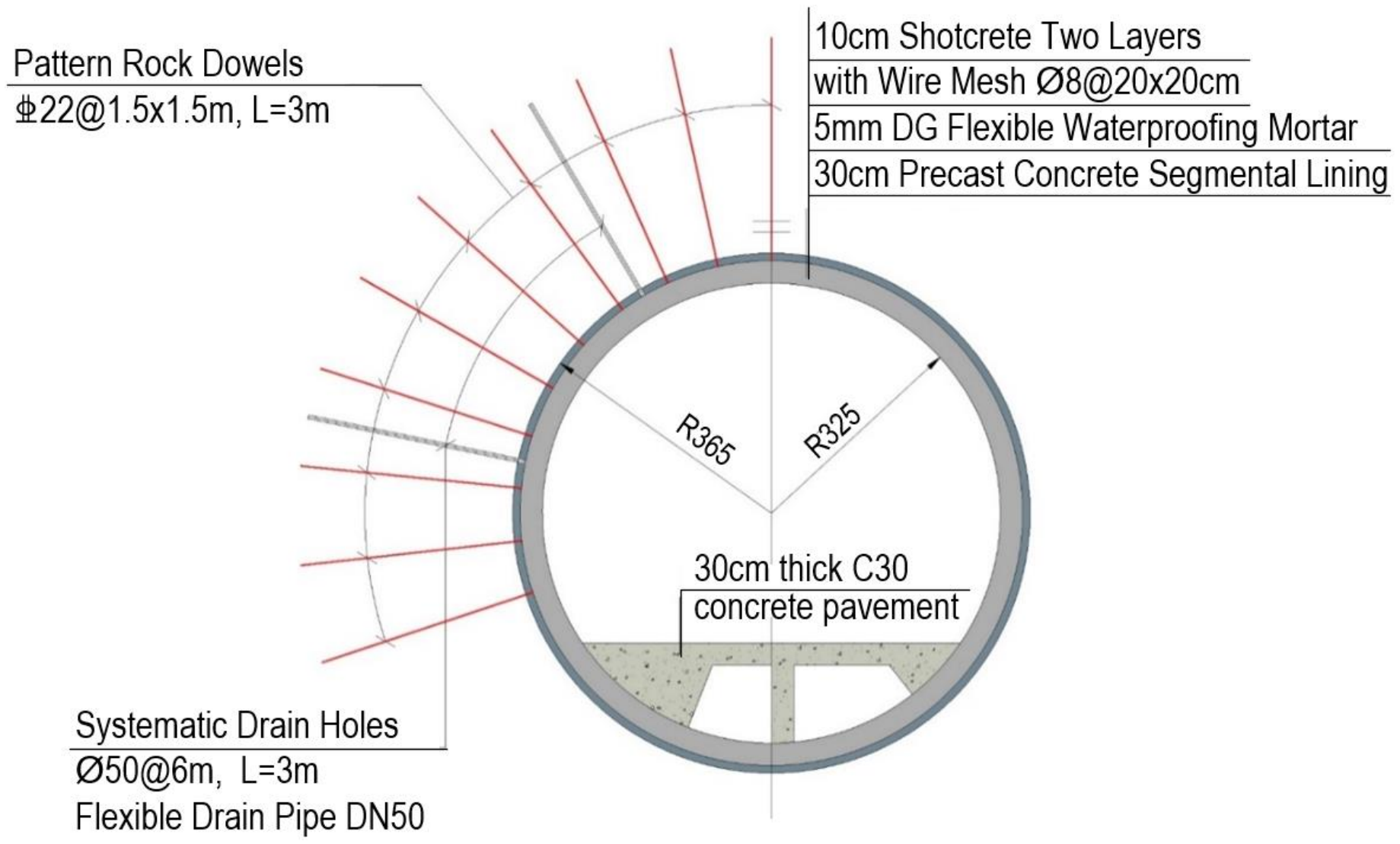

**Figure 9.6.6:** Schematic diagram of typical excavation and support of main ring tunnel (TBM method).

**Table 9.6.4:** Preliminarily determined support parameters for experimental main cavern and service cavern.

| Support parameter / Tunnel description and support part | | Type of anchor bolt (mortar bolt) | | | |
|---|---|---|---|---|---|
| | | Diameter (mm) | Spacing (m) | Length (m) | Remarks |
| Experimental main cavern and service cavern | Arch crown | C25 | 3.0×1.5 | 6 | Arranged at intervals |
| | | C28 | 3.0×1.5 | 8 | |
| | | C28 | 1.5 | 8 | Arch support: two rows |
| | Sidewall | C28 | 3.0×1.5 | 8 | Arranged at intervals |
| | | C25 | 3.0×1.5 | 6 | |
| Spraying layer and mesh | Arch crown, side wall, and end wall: shotcrete thickness of 15 cm; wire mesh with spacing 20×20cm; steel-fiber shotcrete, with the thickness of 15cm, applied to local fracture zones. | | | | |
| Additional measures | Local unstable area of the arch crown: spot mortar bolt (C28, L=8m) or anchor bundle (3C28, L=10m); Local unstable area of side wall: spot prestressed anchor (C36, L=12m); Spot prestressed tendon (L=12-15 m, 100t) | | | | |

Due to the location of experimental equipment along the collider ring tunnel, the waterproofing requirements for the project are stringent, with Grade I waterproofing being necessary. The plain area with a dense surface water network and a deep covering layer presents a critical challenge to the waterproof design of the project. The design elevation of the main ring is -70m, and there is no consideration for groundwater reduction. The external water pressure could reach around 0.7 MPa based on the 0 elevations of surface water. Following project investigation, it is recommended to use DG flexible polymer waterproof mortar for waterproofing and to employ grouting and seepage cut-off measures for areas with significant local seepage.

### 9.6.3.3 *Layout of the Ground Building*

The ground buildings include cooling facilities, low-temperature facilities, ventilation facilities, air compression facilities, power transmission and transformation facilities, maintenance facilities, fire protection facilities, management facilities, test facilities, and others. Their locations are shown in Figs. 9.6.7 and 9.6.8. There are five primary surface structure complexes: the IR surface structure complexes (PA1 and PA3) above the underground IR; the RF surface structure complexes (PA2 and PA4) above the underground RF; and the linac surface structure complex (PA5), which is located close to the linac.

Additionally, there are 28 secondary surface structure complexes, which are the access shaft surface structure complexes (PA6-PA33) located above the access and pipe shafts on the side of the collider ring tunnel. The spacing between these complexes is approximately 3 km.

The total ground building area is about 168,190 $m^2$, and the land area is approximately 620.39 mu (as shown in Table 9.6.5).

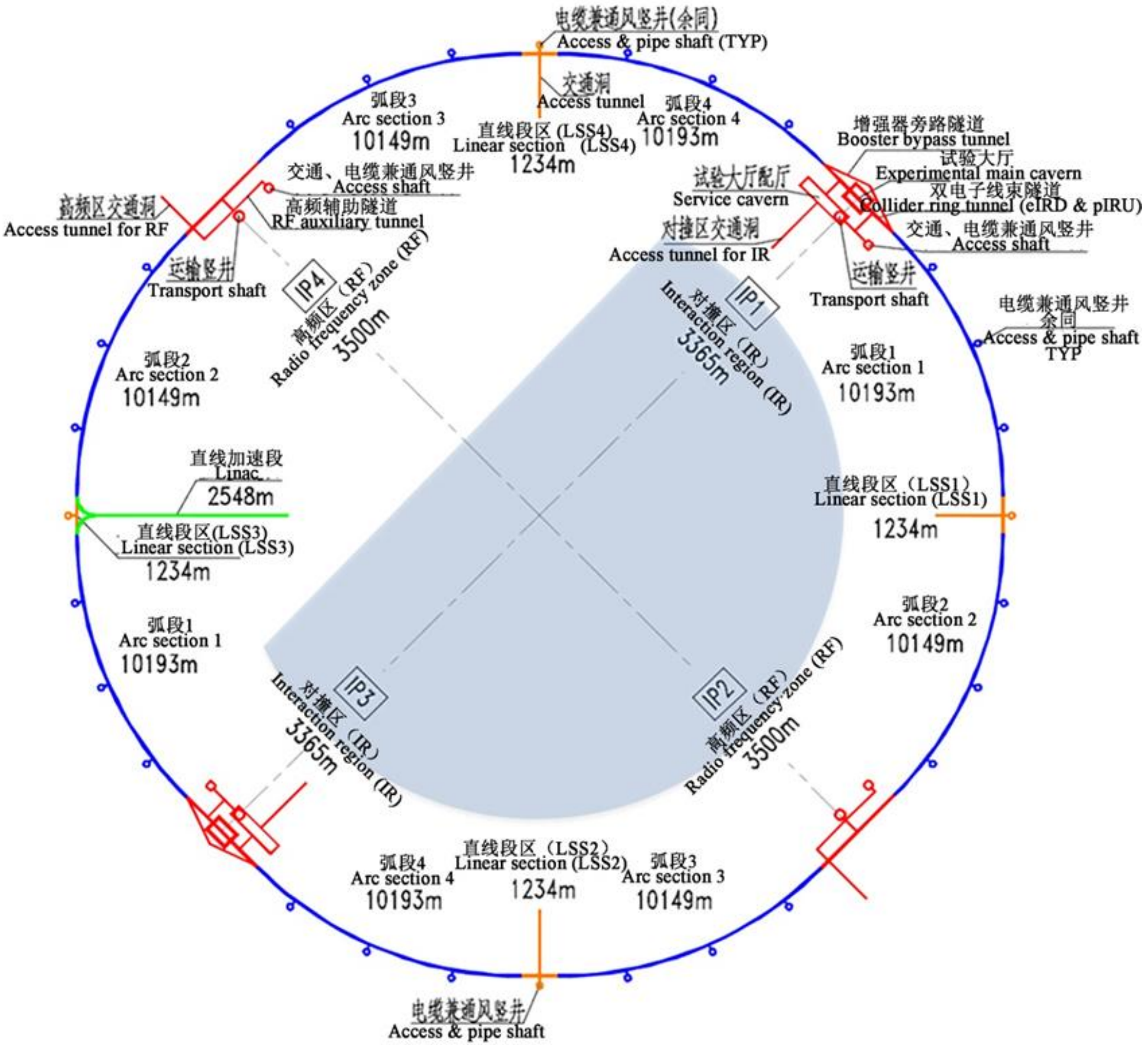

**Figure 9.6.7:** Layout of cavern complex for CEPC.

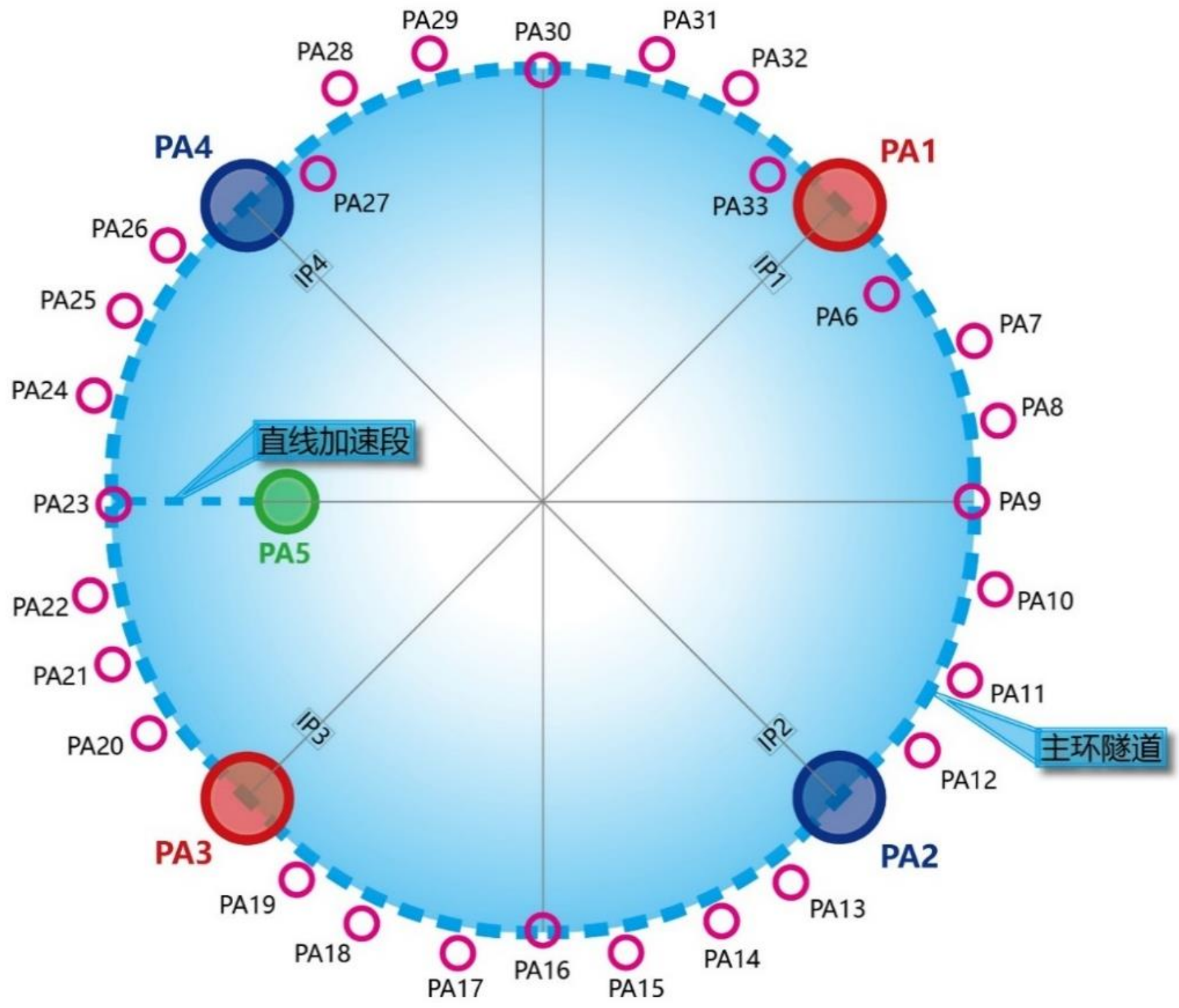

**Figure 9.6.8:** 2D position diagram of CEPC ground building complex.

**Table 9.6.5:** Area of surface structures and land use.

| Surface structures | Unit | Location | | | | | | | Total |
|---|---|---|---|---|---|---|---|---|---|
| | | PA1 | PA2 | PA3 | PA4 | PA5 | PA9, PA16, PA23, PA30 | Other locations | |
| Ground building area | $m^2$ | 17450 | 14150 | 30050 | 14150 | 12950 | 31920 | 47520 | 168190 |
| Land area | $m^2$ | 43173 | 34080 | 64872 | 34080 | 19320 | 59664 | 158400 | 413589 |

## 9.6.4 Construction Organization Design

### 9.6.4.1 *Construction Conditions*

The project benefits from convenient external transportation as it is located near roads that connect to Hangzhou, Huzhou, and other places. Additionally, railway transportation is easily accessible with the nearby Shilai Railway Station and Hangzhou North Railway Station, both of which have existing roads connecting to the project site. The nearest ports to the site are the Shanghai Port and Beilun Port of Ningbo. Table 9.6.6 provides a breakdown of the main civil works quantities.

**Table 9.6.6:** Main quantities of civil works.

| No. | Item | Unit | Quantity | Remarks |
|---|---|---|---|---|
| 1 | Soil excavation | 10,000$m^3$ | 137.04 | |
| 2 | Rock open excavation | 10,000$m^3$ | 385.60 | |
| 3 | Earthworks backfill | 10,000$m^3$ | 8.52 | |
| 4 | Rock tunnel excavation | 10,000$m^3$ | 671.86 | Including temporary construction access |
| 5 | Concrete | 10,000$m^3$ | 174.81 | |
| 6 | Reinforcement | 10,000t | 9.38 | |
| 7 | Steels | 10,000t | 2.86 | |
| 8 | Anchor bolt | 10,000 Nr. | 152.01 | |
| 9 | Waterproof mortar | 10,000$m^2$ | 279.75 | DG flexible polymer |
| 10 | Drain hole | 10,000m | 66.25 | |
| 11 | Drainage floral tube | 10,000m | 31.07 | |
| 12 | EVA waterproof plate | 10,000$m^2$ | 12.99 | |
| 13 | Copper waterstop | 10,000m | 12.26 | |
| 14 | Sealing rod | 10,000m | 38.23 | Water swelling |

#### 9.6.4.2 *Construction of Major Works*

The project involves large-scale underground caverns as the main construction content and essential construction items. The length of the main ring tunnel is about 100 km, passing through various strata blocks with a buried depth ranging from 0 to 500 m and significant variation. Due to such variations, a single excavation method cannot meet the construction needs of this project, and a combination of drilling and blasting method and TBM method is recommended for construction. The tunnel boring machine method is proposed for tunnel sections with a significant buried depth, complex branch tunnel layout, and vicinity of urban areas. The drilling and blasting method is recommended for tunnel sections with shallow buried depth, rich groundwater, and complex geological conditions. Additionally, considering the construction period, geological conditions, and tunnel section structure, the double shield TBM is recommended for the construction of this project.

The project utilizes eight permanent access tunnels, which are evenly distributed along the main ring tunnel, as the construction passages. The primary ring tunnel has a gate-shaped cross-section, and the distance between two adjacent access tunnels is approximately 12.5 km. If the drill-blasting method is adopted, a branch tunnel needs to be constructed between two adjoining access tunnels, and the cross-section of the branch tunnel should be consistent with that of the permanent access tunnel.

#### 9.6.4.3 *General Construction Layout Planning*

The project requires both permanent and temporary land for construction. The total land area needed for the project is 856,600 $m^2$ (1434.9 mu), out of which 168,190 $m^2$ (249.9 mu) is permanent land. The temporary land requirement includes various facilities such as construction factories, warehouses, office and living areas, temporary construction roads, etc. The temporary construction facilities cover an area of approximately 490,000 $m^2$ (735.0 mu), and the temporary construction road covers an area of around 300,000 $m^2$ (450.0 mu).

#### 9.6.4.4 *General Construction Progress*

Due to the project's scale and complexity, it is divided into two construction periods: the civil engineering construction period and the physical equipment installation construction period. The civil engineering phase utilizes subsection, TBM, and drill-and-blast methods for construction. The construction of the main ring tunnel and pre-control network for the physical equipment project are division nodes between the two phases. Concurrently with physical engineering construction, arrangements are made for the construction of civil engineering supporting infrastructure, such as power distribution, communication, HVAC, and fire protection engineering.

The construction of the CEPC is yet to be approved. Assuming the project would start in 2027, the project's preparatory period will be scheduled from January 2027 to February 2028, with the main construction period taking place from March 2028 to March 2031. The completion time for the TBM method civil construction of the main ring tunnel is set for the end of March 2031. The construction and measurement of the control network for the first-phase main ring tunnel and the supporting civil works will take place from January to December 2031. From early January 2032 to the end of August 2033, physical equipment, including supports and magnets for the main ring, enhancer, and linear damping ring, will be aligned, installed, and calibrated in the main ring tunnel.

The drilling and blasting method civil construction completion time is scheduled for the end of March 2032. The construction and measurement of the control network for the second-stage main ring tunnel will take place from January to December 2032. The alignment, installation, and calibration of physical equipment, including the main ring, booster, linear damping ring support, and magnet, will take place in the main ring tunnel of the second phase from January 2033 to August 2034. Equipment alignment calibration and debugging will be conducted from September to December 2034. The CEPC project is scheduled to be completed by the end of 2034, including both the civil and physical engineering phases.

#### 9.6.4.5 *Main Technical Supply*

The project's peak civil works will involve 2,200 individuals, with an average of 1,600 individuals working on the project. The entire construction phase will require approximately 2.08 million man-days. During the peak period of rock excavation, the average intensity will be 197,000 cubic meters per month. Meanwhile, during the peak period of concrete pouring, the average intensity will be 125,000 cubic meters per month. The primary mechanical equipment utilized for the project will include five double-shield TBMs with a diameter of 7.0 meters.

### 9.6.5 **Science City**

#### 9.6.5.1 *Overview of Science City*

The CEPC Science City is situated in the urban area of Huzhou, and it has been built using the CEPC large-scale discipline device. The city performs various functions such as scientific research, learning and education, conference communication, related equipment manufacturing, and scientific research achievements transformation. Its location is in the southwest of Huzhou's central city and the south of Huzhou science & technology city. It is approximately 9 kilometers away from the center of Huzhou and 5 kilometers away from Huzhou high-speed railway station. The city's planning scope covers a land area of about 3.92 square kilometers and extends to Nankang Avenue in the north, Lujiayang Road in the south, Zhanqian Road in the west, and Huanyang Avenue in the east.

The base where CEPC Science City is built has a flat terrain, and its proximity to the river allows for efficient drainage. The geological environment conditions both within and around the area are stable, providing excellent conditions for development and construction.

The base boasts of abundant water and vegetation, with Miaoxi Port flowing through the southeast. The area has an excellent natural ecological resource endowment, making it ideal for shaping the urban blue-green space.

#### 9.6.5.2 ***Science City Planning Scheme***

(1) Functional Structure:

The urban development and environmental landscape corridors will be constructed along main roads and water systems, with the ecological lake surface serving as the landscape core. Three groups will be formed to meet the functional requirements: the scientific research core area, the public exchange area, and the supporting living area (as shown in Fig. 9.6.9). The planning structure will have one center and multiple corridors, and the group will be integrated.

The research area will consist of the core research park of the Institute of High Energy Physics, the storage, assembly, and maintenance functions of CEPC-related components, and the main functions of research institutions of top international institutions. The public exchange area will take the ecological lake as the landscape core and plan for public service functions such as scientific exhibitions, urban parks, conference exchange centers, central business districts, commercial pedestrian streets, and talent apartments. The supporting living area will be based on high-quality communities and provide various living supporting functions such as medical care, education, culture, sports, commerce, etc., to improve the urban role of Science City.

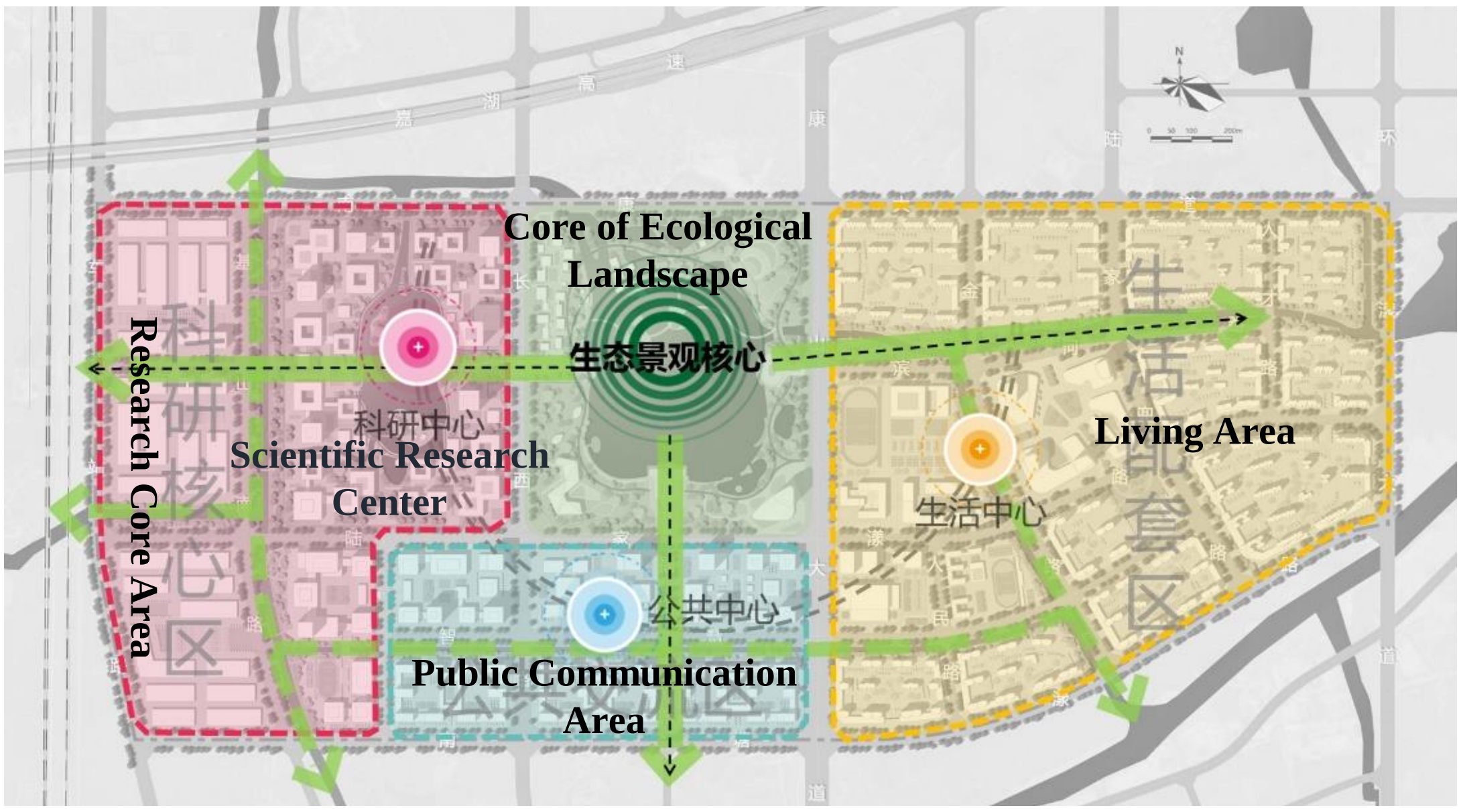

**Figure 9.6.9:** Functional planning structure of Science City.

(2) General layout

The general layout of the Science City is shown in Fig. 9.6.10.

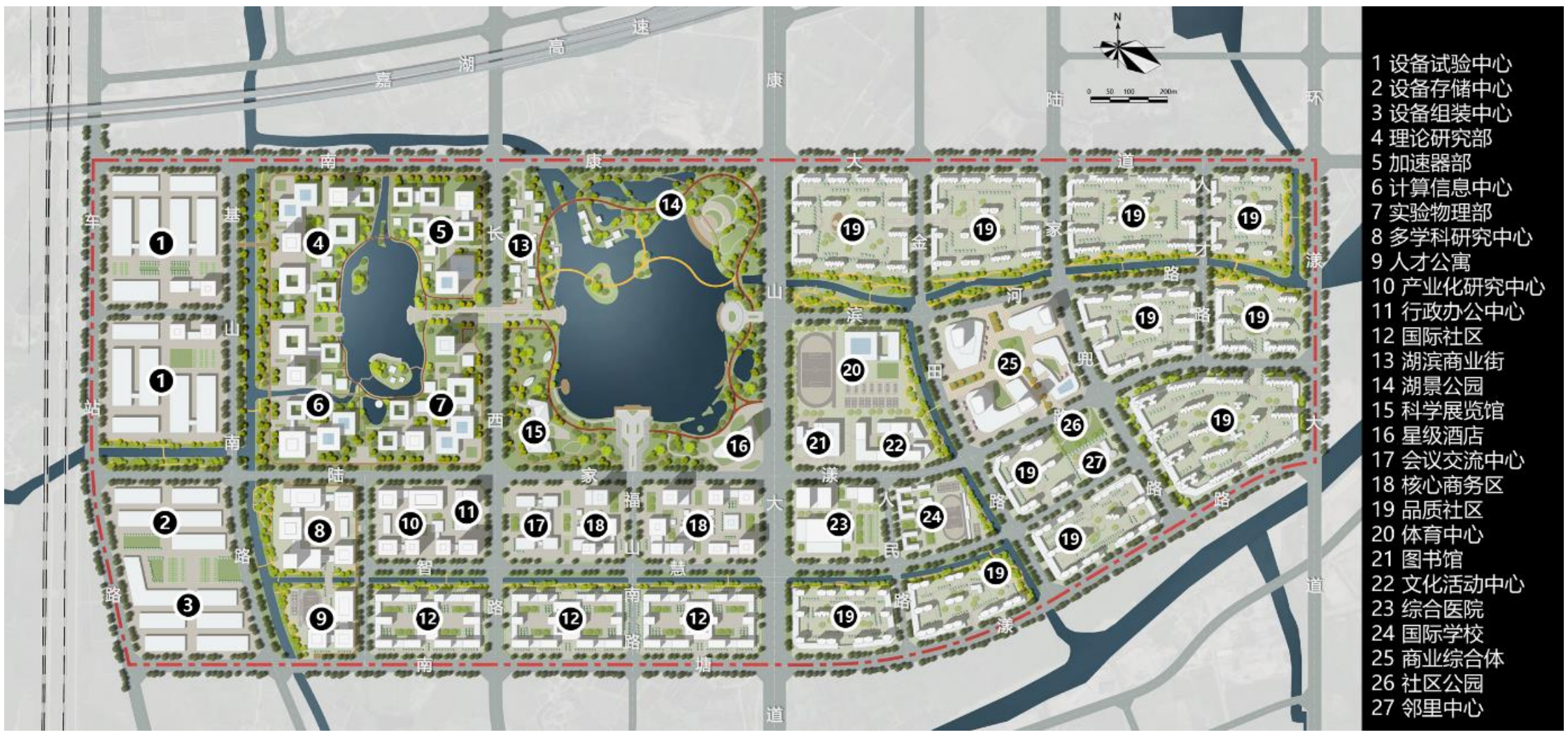

**Figure 9.6.10:** General layout plan of Science City.

(3) Land Use Planning:

The International Science City has a planning area of 3.92 square kilometers, which includes 3.49 $km^2$ of urban construction land and 0.43 $km^2$ of water area, resulting in a water surface rate of 11.0%. The residential land covers 98.01 hectares, including second-class residential land and mixed commercial and residential land, accounting for 28.07% of the urban construction land. The land for public management and public service facilities covers 72.47 hectares, accounting for 20.76% of the urban construction land. The commercial service industry facilities cover 26.13 hectares, accounting for 7.48% of the urban construction land. The industrial and scientific research mix covers an area of 38.86 hectares, accounting for 10.99% of the urban construction land. The urban road covers an area of 62.32 hectares, accounting for 17.85% of the urban construction land. The square covers an area of 51.84 hectares, accounting for 14.85% of the urban construction land, with green park space occupying 46.30 hectares, accounting for 13.26%.

(4) Architectural Rendering:

Figure 9.6.11 depicts an architectural rendering of the Science City.

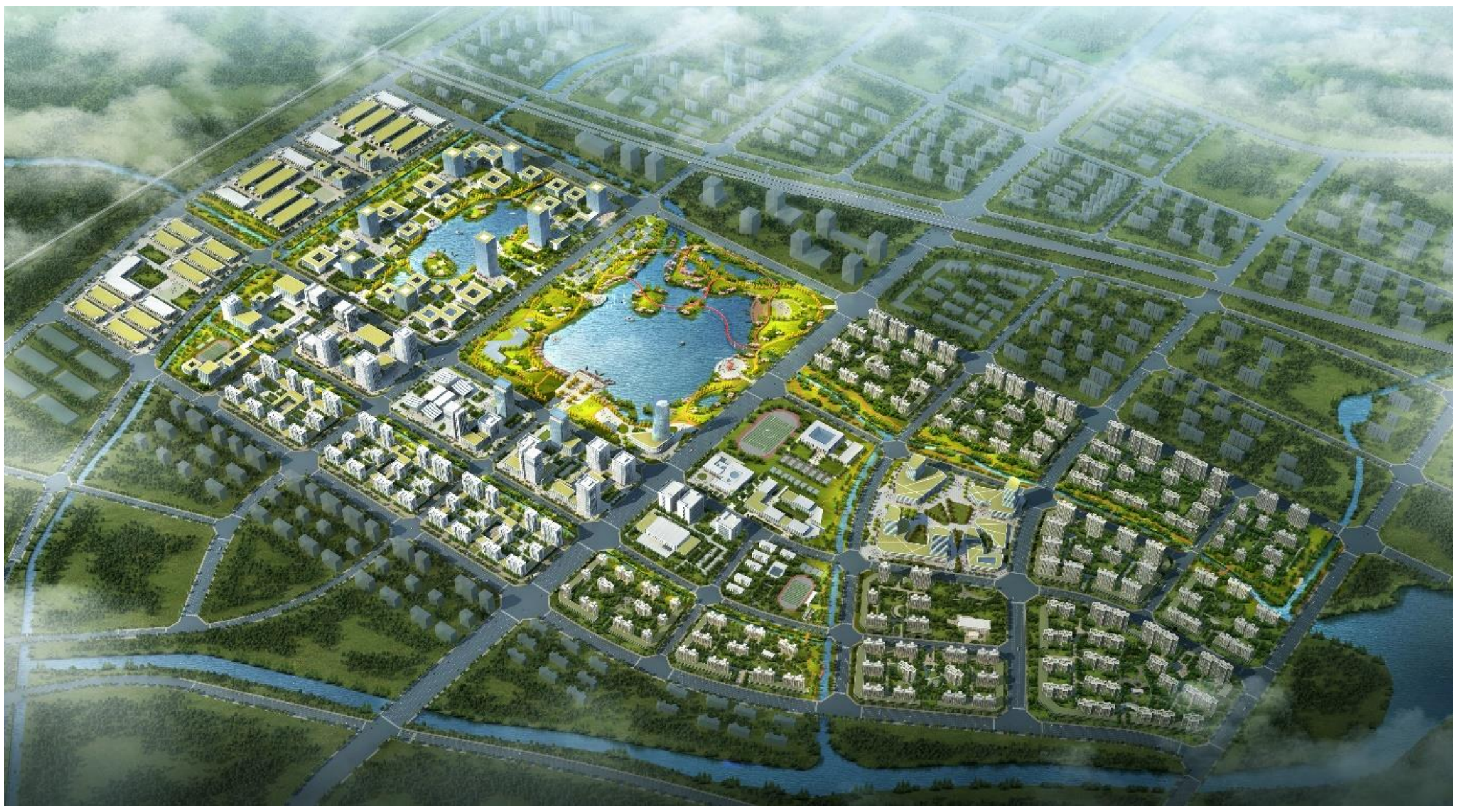

**Figure 9.6.11:** Architectural rendering of the Science City.

## 9.7 Electrical Engineering

### 9.7.1 **Electrical System Design**

#### 9.7.1.1 ***Power Supply Range and Main Loads***

The electrical power system at the Collider complex serves to power both the high-energy physics experiments and the general facilities, including ventilation, air conditioning, lighting, elevators, and other common needs. During operation for Higgs physics, the total electrical load for physical experiments and general facilities is approximately 262 MW. Table 9.7.1 provides a summary of the estimated power loads.

**Table 9.7.1:** Main power loads for the CEPC (at Higgs /30 MW)

| SN | System for Higgs (50 MW /beam) | Location and power Requirement (MW) | | | | | | Total (MW) |
|---|---|---|---|---|---|---|---|---|
| | | Collider | Booster | Linac | BTL | IR | Surface building | |
| 1 | RF Power Source | 96.90 | 0.15 | 12.26 | | | | 109.31 |
| 2 | Cryogenic System | 9.72 | 1.71 | | | 0.16 | | 11.59 |
| 3 | Vacuum System | 5.40 | 4.20 | 0.60 | | | | 10.20 |
| 4 | Magnet Power Supplies | 42.16 | 8.46 | 2.15 | 4.89 | 0.30 | | 57.96 |
| 5 | Instrumentation | 1.30 | 0.70 | 0.20 | | | | 2.20 |
| 6 | Radiation Protection | 0.30 | | 0.10 | | | | 0.40 |
| 7 | Control System | 1.00 | 0.60 | 0.20 | | | | 1.80 |
| 8 | Experimental Devices | | | | | 4.00 | | 4.00 |
| 9 | Utilities | 37.80 | 3.20 | 1.80 | 0.60 | 1.20 | | 44.60 |
| 10 | General Services | 7.20 | | 0.30 | 0.20 | 0.20 | 12.00 | 19.90 |
| | **Total** | 201.78 | 19.02 | 17.61 | 5.69 | 5.86 | 12.00 | 261.96 |

#### 9.7.1.2 *Power Supplies*

The project proposes using a 220 kV electrical system, with two central substations (220 kV/ 110 kV/ 10 kV) located in the project area. Each substation will have two 240 MVA transformers.

Four key substations (110 kV/ 10 kV) will be placed near the shaft ground exits of the IR and RF areas (IP1-IP4), with a total of 4 step-down substations. IP1 and IP3 will have three 80 MVA transformers and twenty-four (24) 10 kV outgoing feeders each, while IP2 and IP4 will have two 50 MVA transformers and twenty-six (26) 10 kV outgoing feeders each. These substations will mainly power the shaft exits at ground level and the underground tunnels. The electrical distribution system network is shown in Figure 9.7.1.

The ancillary buildings at the exit of each ground shaft will receive power from 10 kV/ 0.4 kV transformers located at the ground exit of each shaft. There will be a total of 36 such systems, and the 10 kV power will come from the 110 kV/ 10 kV step-down substation.

To create a 10 kV looped network in the tunnel, underground 10 kV cables will run from the 110 kV/ 10 kV step-down substation to the underground tunnel via each shaft. A 10 kV/ 0.4 kV transformer and distribution system will be installed near each load point, and there will likely be 96 such systems.

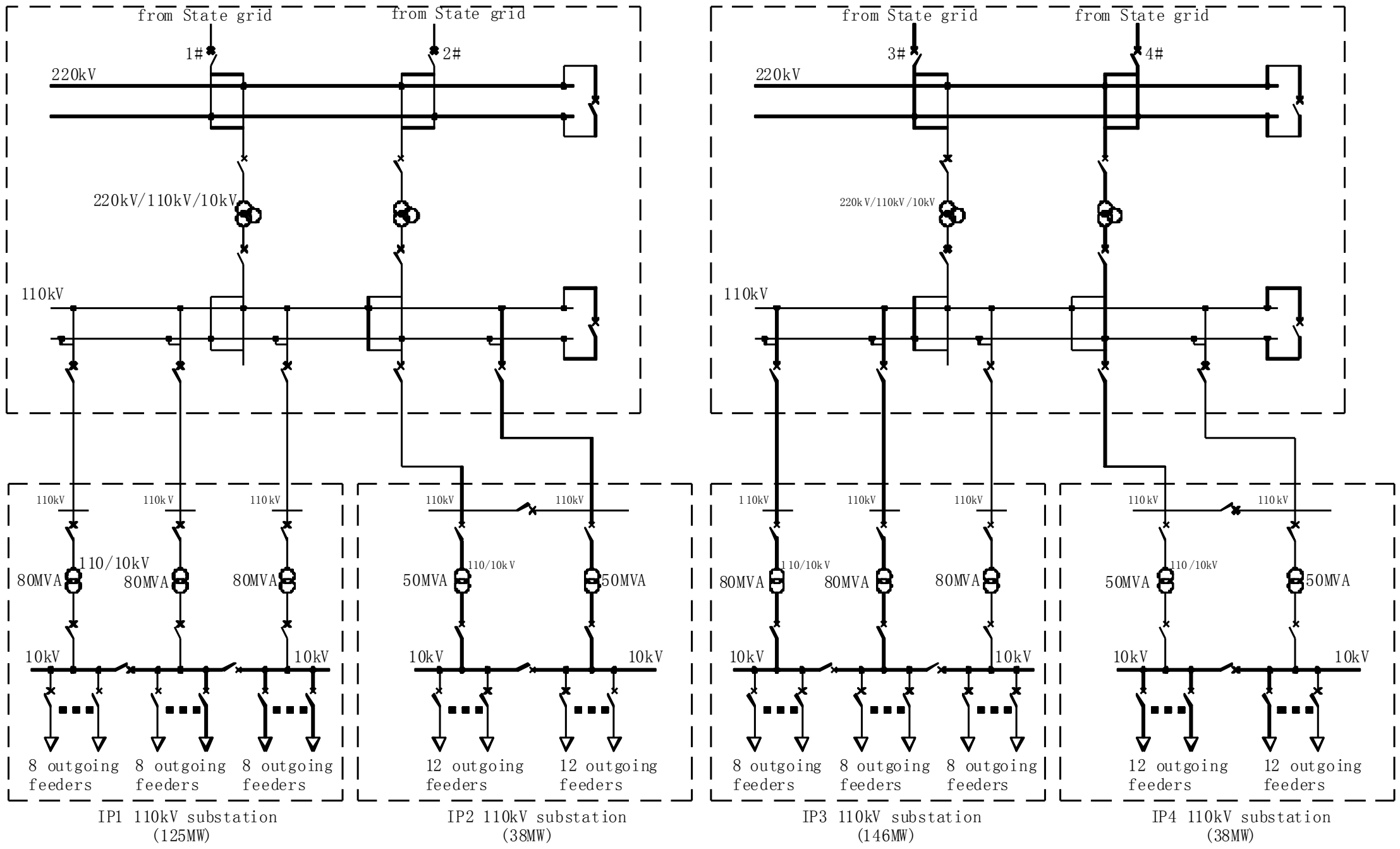

**Figure 9.7.1:** Schematic diagram for the power supply network.

### 9.7.1.3 *Additional Details on the 220 kV, 110 kV and 10 kV Systems*

#### (1) 220 kV Power Supply System

Each of the two 220 kV central substations will have two 220 kV incoming lines and four 110 kV outgoing lines, along with twelve (12) 10 kV outgoing lines for a total of 24 outgoing lines. The 220 kV system will use double-bus wiring, while the 110 kV system will use double-sectionized double-bus wiring, and the 10 kV system will use sectionalized single-bus wiring.

The transformers will be three-phase, three-volume, natural oil-circulating, air-cooled, and on-load regulating. For both the 220 kV and 110 kV systems, outdoor GIS equipment will be used. For the 10 kV system, there will be withdrawable metal-enclosed high-pressure vacuum switch cabinets.

The proposed design for the substation includes laying out the outdoor 220 kV GIS equipment in a single-row configuration, while the outdoor 110 kV system and indoor 10 kV systems will be arranged in a double-row layout. These substations are planned to be unattended and will instead rely on a supervisory computer control system that utilizes micro-processor-based protection relay and automatic safety devices.

To ensure protection of the 220 kV systems, a dual-configuration protection system will be employed. The DC systems will feature a dual-changer, dual-storage configuration and will be sectionalized into a single-bus configuration with 110 V or 220 V. Communication with the electric power grid distribution station will be facilitated via optical fiber.

#### (2) 110 kV Power Supply System

There will be a total of four (4) 110kV/ 10kV step-down substations to be installed, with one in each of the IP1~IP4 areas. Each substation will include two or three transformers, with a capacity of 50 MVA or 80 MVA, along with 4 to 6 110 kV incoming lines and 24 to 26 10 kV outgoing lines.

The 110 kV system will feature a double-bus wiring configuration, while the 10 kV system will include both double-bus and sectionalized single-bus wiring. The transformers will be three-phase with double-volume natural oil circulation and air-cooled on-load regulating transformers.

For the outdoor 110kV system, GIS equipment will be used, and there will be withdrawable metal-enclosed high-pressure vacuum switch cabinets in place.

The main 110 kV transformer will be located indoors and arranged in a single-row layout. It will have overhead incoming and outgoing lines. The 10 kV indoor equipment, on the other hand, will be arranged in a double-row layout and will be equipped with cable outlets.

The substations will be unmanned and will instead rely on a supervisory computer control system utilizing micro-processor based protection relay and automatic safety devices. For the 110 kV system, a dual-configuration protection system will be employed. The DC systems will have a dual-changer, dual-storage configuration and will be sectionalized into a single-bus configuration with 110 V or 220 V.

Communication with the electric power grid distribution station will be facilitated through optical fiber.

**(3) 10 kV Power and Distribution System**

A total of thirty-six (36) 10kV/ 0.4kV transformers and distribution systems will be installed at ground level. Additionally, there will be approximately one hundred 10kV/ 0.4kV transformers and distribution systems located near each load point in the underground tunnel. All 10 kV power supplies will be fed from the 110 kV/ 10 kV step-down substations.

Sectionalized single-bus wiring will be used for both the 10 kV and 0.4 kV systems. However, filtering devices will be required to suppress the large number of higher harmonics produced by the numerous large-capacitance rectifying components. A centralized reactive compensation mode will be employed to accomplish this.

Dry-type transformers have been selected for use, with withdrawable metal-enclosed high voltage switch cabinets for the 10 kV systems and low voltage extraction type switch cabinets for the 0.4 kV systems. Active power filters will also be included.

The transformers will be placed indoors in a double row and provided with cable outlets.

To protect against critical loads, where a power failure could cause damage, diesel generators, EPS power supplies, or UPS will be installed.

### 9.7.1.4 *Lighting System*

The project includes normal lighting and emergency lighting systems for both the ground facilities and the underground facilities. The emergency lighting system is essential to provide good, visible lighting in critical areas for personnel during excavation. In the event of a power failure, the emergency lighting system can be powered by a diesel generator or EPS.

Energy-saving lighting fixtures will be used, and fluorescent lamps will be utilized throughout the project. Mining lamps will be employed in the experiment and assembly halls, while moisture-proof lamps will be used in the tunnels to withstand the damp environment.

### 9.7.2 Automatic Monitoring System

The project requires monitoring of various systems such as power supply systems in shafts and underground tunnels, ventilation and air-conditioning systems, and other common facilities. The 110 kV substations will be monitored through the substation integrated automation system included in the power supply system.

To facilitate monitoring, a ground monitoring center will be established. It will consist of a data server, operator workstation, printer, network equipment, and UPS. Each controlled system will have a local control unit to collect information. Redundant communication will be provided using industrial Ethernet and optical fiber to ensure system reliability.

### 9.7.3 Communication System

#### 9.7.3.1 *Service Objects*

Internal communication encompasses voice, network communication, and other forms of communication within each experiment hall and equipment room in the shafts and tunnels. However, it should be noted that the system described in this section does not include the communication systems for the buildings located on the ground level.

#### 9.7.3.2 *Communication Mode*

Optical cables will be installed in cable trays within the shafts and tunnels. Multi-Service Transport Platform (MSTP) optical communication equipment will also be installed in each machine room, forming a self-correcting and multi-service ring-type optical fiber communication network with a bandwidth of 10 Gb/s.

For each user, one soft switch system in each communication center will provide service. Telephone communication between users and external parties will be facilitated through the soft switch system.

Additionally, a leakage coaxial cable communication system will be implemented around the tunnel, providing wireless communication for personnel.

#### 9.7.3.3 *Computer Network*

The computer network is structured with a core layer, convergence layer, and access layer. The core layer is located in the ground-level control center, while the convergence layer is situated in each experiment hall. In the access layer, an Ethernet switch is provided for each user.

Redundancy is incorporated into the Ethernet switch and other equipment located in the core and convergence layers. The layers are interconnected with fiber optics, which enables a transmission rate of gigabit-class.

Furthermore, the router and internet security equipment facilitate the connection between the core layer equipment and external internet.

#### 9.7.3.4 *Communication Power Supply and UPS*

Within the machine room, there is one 48 V high frequency switching communication power supply that is equipped with two groups of 48 V storage batteries. This power supply is responsible for providing power to the optical communication equipment and the soft switch system. Its communication power supply time must not be less than three hours.

Furthermore, a UPS (uninterruptible power supply) is also present in the machine room, supplying power to the network equipment. This UPS is designed to provide power for a minimum of three hours.

### 9.7.4 Video Monitoring System

The video monitoring system is comprised of network high-definition cameras, transmission networks, and monitoring center equipment. These cameras will be installed in the shafts, tunnels, and experiment halls.

The monitoring center will be equipped with hard disk video recorders, storage devices, video servers, and large-screen video monitoring devices to facilitate the management and storage of the video footage.

### 9.7.5 Power Quality and Energy-saving

Power quality is a crucial consideration for CEPC, with the main issues being voltage stability, harmonic filtering, and reactive power compensation. Additionally, transient voltage dips, which are often caused by lightning strikes on the transmission network overhead lines, must be mitigated to prevent equipment downtime.

Efficient energy usage, storage, and recovery must be a primary focus for the design of CEPC. As a technology driver, this project presents an opportunity to push for more efficient and sustainable ways to use electrical and thermal energy. Therefore, a strict emphasis on energy efficiency and sustainability must be incorporated into the design of the powering system.

## 9.8 Cooling Water System

### 9.8.1 Overview

Cooling water plays a critical role in CEPC as most of the electrical power consumed is absorbed by it. Along with its cooling function, it is essential for maintaining a constant operating temperature in some subsystems.

The cooling water system comprises a low-conductivity water (LCW) closed-loop circuit, a cooling tower water (CTW) circuit, and a deionized water make-up system. The LCW system absorbs heat from various devices, which is then transferred through heat exchangers to the cooling tower water circuits (CTW). Finally, the CTW rejects the heat into the atmosphere through cooling towers. A flow diagram of a typical cooling water system is shown in Figure 9.8.1.

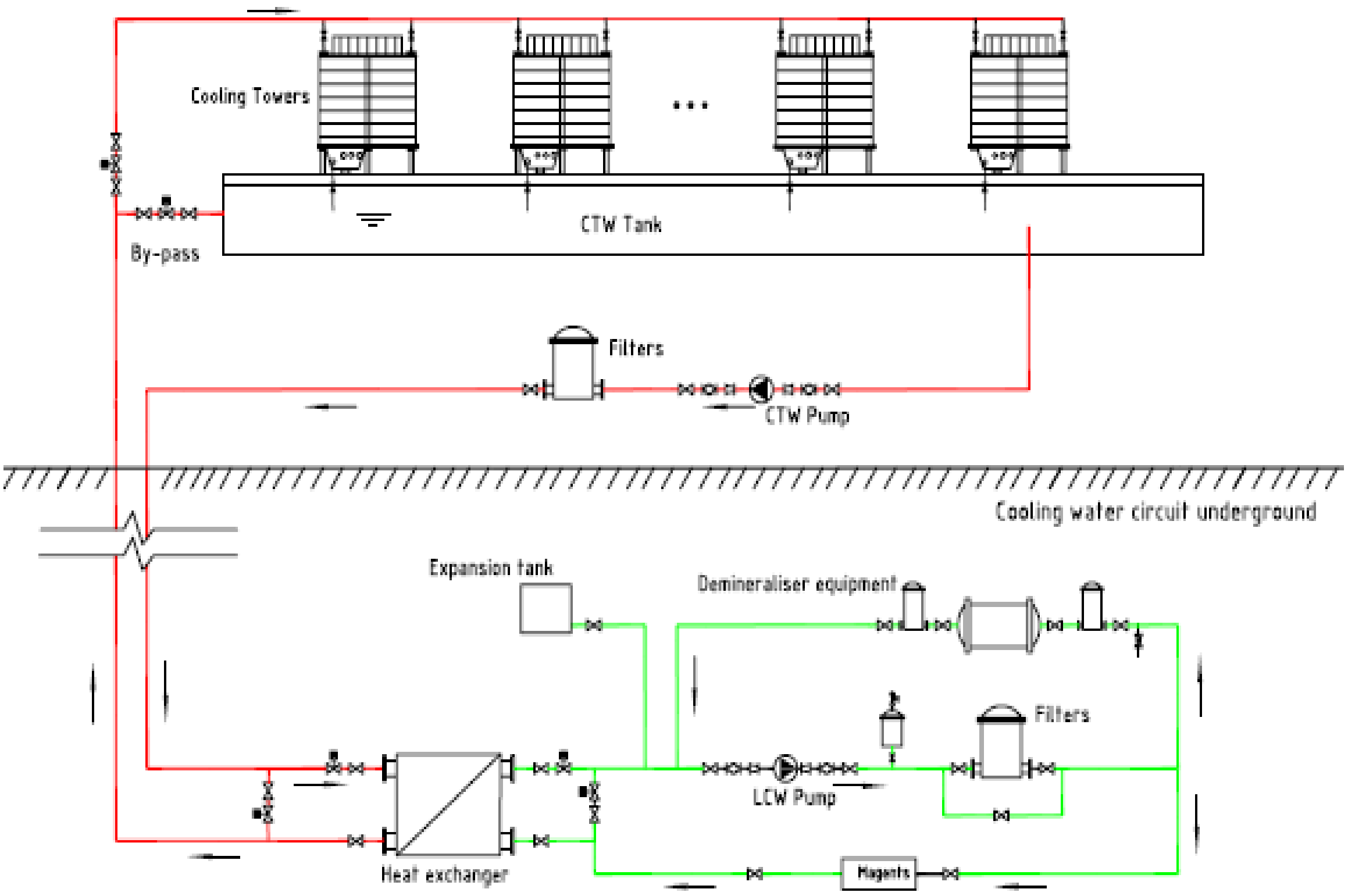

**Figure 9.8.1:** Flow diagram of a typical cooling water system.

The major heat sources in CEPC are RF power sources, magnets, vacuum chambers, cryogenic compressors, and power converters (magnet power supplies). During operation for Higgs physics, the total heat load dissipated is approximately 221 MW. The estimated heat loads for each component in the Higgs operation mode are summarized in Tables 9.8.1.

**Table 9.8.1:** Estimated cooling water heat loads (at Higgs /30MW)

| System | Location and heat loads (MW) | | | | | |
|---|---|---|---|---|---|---|
| | Collider | Booster | Linac | BTL | IR | Total |
| Accelerating tube / Waveguide | | | 1.36 | | | 1.36 |
| Power source | 36.90 | 0.15 | 9.18 | | | 46.23 |
| Cryogenics | 9.50 | 1.60 | | | 0.16 | 11.26 |
| Experimental devices | | | | | 3.60 | 3.60 |
| Magnets | 31.22 | 5.18 | 1.76 | 4.31 | | 42.47 |
| Vacuum chamber of ring | 64.00 | 6.20 | 0.50 | | | 70.70 |
| Power convert for magnets | 4.05 | 0.81 | 0.18 | 0.47 | 0.03 | 5.54 |
| Condenser in stub tunnel | 13.20 | | | | | 13.20 |
| Pump | 21.65 | 1.46 | 1.65 | 1.32 | 1.05 | 27.13 |
| Total | 180.52 | 15.40 | 14.63 | 6.10 | 4.84 | 221.49 |

There will be 16 cooling tower water (CTW) pump stations, with one at each of the 16 points around the CEPC ring, along with an additional CTW system for the Linac. To reduce pipe pressure, the CTW equipment will be installed near the access shafts, while the low-conductivity water (LCW) circuits will be located underground at machine level. Each point of the ring and the Linac will have a deionized water plant to supply LCW makeup water for the circuits in the area. Figure 9.8.2 illustrates the layout of the cooling water circuits around the ring.

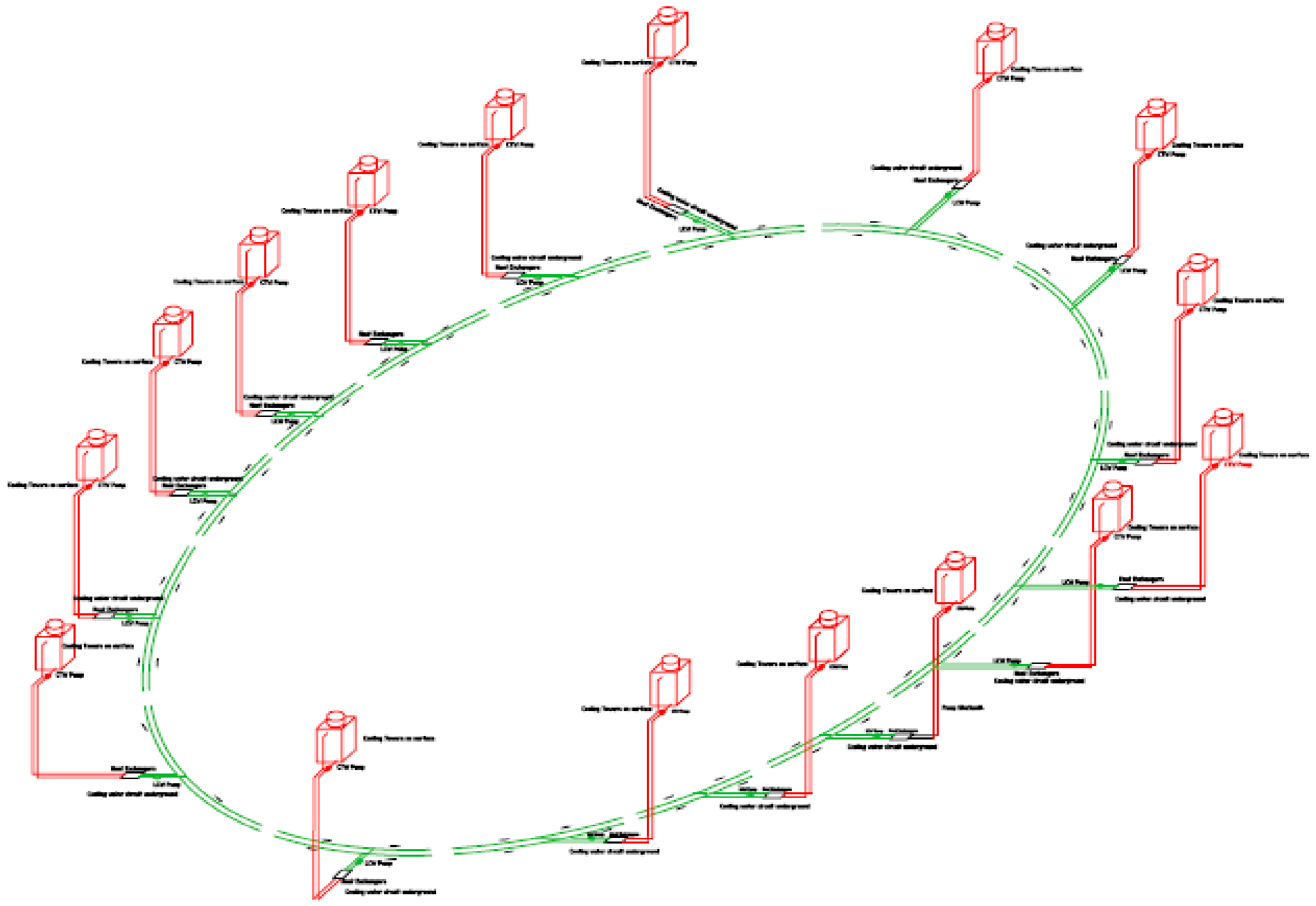

**Figure 9.8.2:** Layout of cooling water circuits around the ring.

### 9.8.2 **Cooling Tower Water Circuits**

The cooling tower water system is an essential component of the LCW circuits, providing coolant for the heat exchangers at each point of the ring and Linac area where service buildings will house pumps, filters, and other necessary components. The supply water temperature is set at 29°C, based on a wet-bulb air temperature of 26°C ambient.

Table 9.8.2 outlines the main parameters of these circuits, including flow rates and pressures, to ensure that the cooling tower water system operates efficiently and effectively.

To introduce make-up water, an automatic valve is connected to the raw water pipeline. The flow diagram of the cooling tower water system is depicted in Figure 9.8.3, illustrating the various components and their interconnections within the system.

**Table 9.8.2:** Parameters of the cooling tower water system (at Higgs /30 MW)

| Parameters | Collider | Booster | Linac | BTL | IR | Total |
|---|---|---|---|---|---|---|
| Heat load (MW) | 180.52 | 15.40 | 14.63 | 6.10 | 4.84 | 221.49 |
| Supply water temperature | 29°C | | | | | |
| Temperature rise | 5°C | | | | | |
| Flow rate ($m^3$/h) | 31044 | 2648 | 2516 | 1049 | 832 | 38089 |

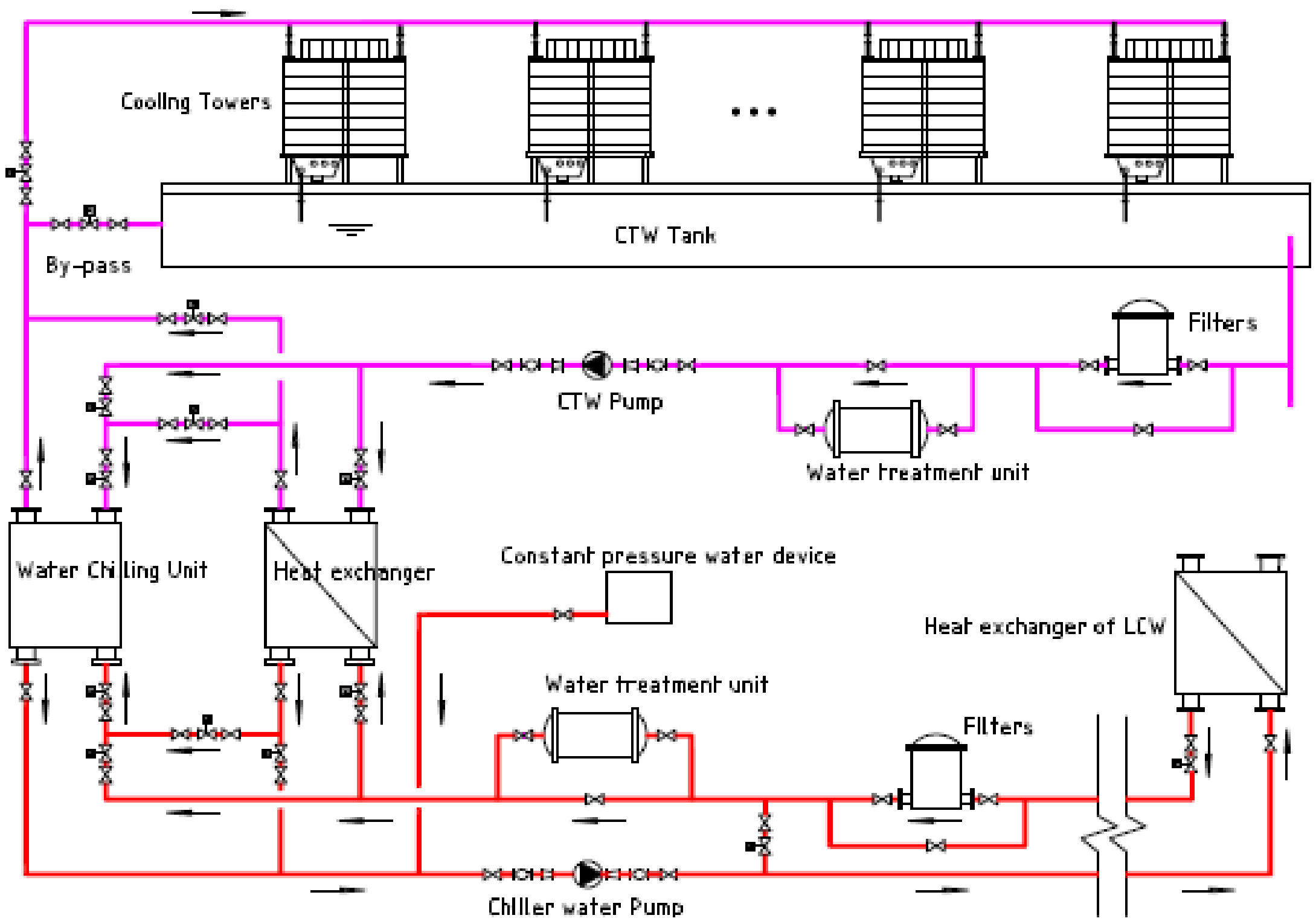

**Figure 9.8.3:** Flow diagram of CTW.

### 9.8.3 Low-conductivity Circuits

The LINAC, BTL, main ring, and experimental areas contain several closed-loop LCW subsystems that are defined by equipment characteristics, operational requirements, and location. These subsystems include:

- The LINAC area subsystem, which consists of the accelerating tube, waveguide circuit, klystron circuit, and the beam transfer line circuit.
- The main ring area subsystem, which comprises magnet circuits, vacuum circuits, RF circuits, and power resource circuits at each point. Each magnet and vacuum circuit serves two half-adjacent sixteenth sections of the ring.
- The experimental area subsystems, which are responsible for removing the heat generated by the experimental equipment. These subsystems are bypassed with about 1% flow rate through a demineralizer to maintain the low conductivity water at over 1 MΩ-cm specific resistance.

The main design parameters of the LCW subsystems in the LINAC, BTL, main ring, and experimental areas are provided in Tables 9.8.3. The total designed circulation flow of LCW is approximately 26,751 $m^3/h$, and the total designed cold load is 197 MW.

The supply water temperature for the ring is set to 31°C, based on the outlet temperature of the cooling tower water. It is worth noting that this temperature may be exceeded in the summer months. The relatively high temperature is a deliberate choice to avoid the use of expensive refrigeration equipment.

**Table 9.8.3:** Parameters of low-conductivity circuits (at Higgs /30 MW)

| System | location | Heat loads | Flow rate |
|---|---|---|---|
| | | MW | $m^3/h$ |
| Accelerating tube / Waveguide circuit | Linac | 1.428 | 246 |
| Power source circuit（Linac） | | 9.639 | 1658 |
| Power convert for magnet circuit（Linac） | | 0.189 | 33 |
| Magnet circuit（Linac） | | 2.373 | 408 |
| Power convert for magnet circuit（BTL） | BTL | 0.494 | 85 |
| Magnet circuit（BTL） | | 4.526 | 778 |
| Magnet and condenser circuit | Ring | 3.255×16 | 560×16 |
| Vacuum chamber circuit | | 4.067×16 | 396×16 |
| Power source circuit（Ring） | | 19.451×2 | 3345×2 |
| Power coupler circuit（Ring） | | 0.189×2 | 16×2 |
| Power convert for magnet circuit（Ring） | | 0.638×8 | 110×8 |
| Circuit for experiment area | IP | 1.89×2 | 325×2 |
| **Total** | | 196.602 | 26751 |

Numerous measures have been implemented to maintain the water quality of the primary water system. These include adjusting the flow quantity to control the resistivity and pH of the water, equipping the main pipeline with exhaust filters to ensure the smooth exhaust of the circulating circuit, adopting deoxygenation measurements, and implementing nitrogen protection. A post-1 µm filter has also been installed to the side flow exchange column to prevent any leakage of resin particles. Additionally, the material of the water-contacted pipeline is strictly controlled to prevent any contamination.

A flow diagram of the cooling tower water system can be found in Fig. 9.8.4.

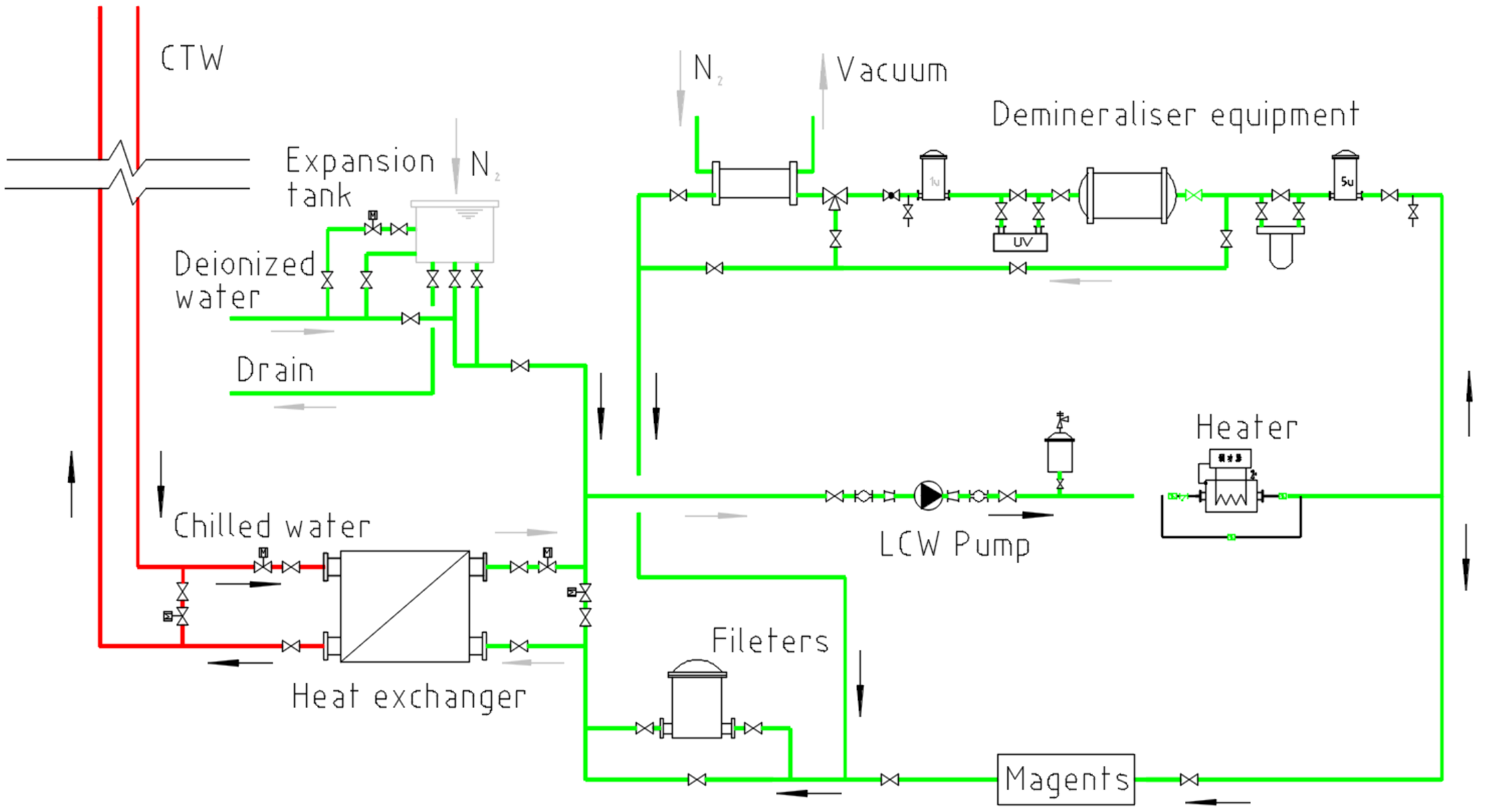

**Figure 9.8.4:** Flow diagram of a typical LCW.

### 9.8.4 Deionized Water System

There will be a deionized water system installed at sixteen different locations around the ring, as well as at the Linac. The purpose of this system is to provide low-conductivity makeup water for each of the LCW (Low-Conductivity Water) circuits in the area. The specific resistivity of the supply water used by the system is more than 10 MΩ-cm, indicating that it is a high-purity water source.

To achieve the desired level of purity, a membrane process will be used to demineralize the water. Each deionized water system will be capable of producing 10 t/m$^3$ of deionized water. This water will be piped to each point of use and circulated in the main pipe to maintain its quality.

For water points that are located far away from the main pipe, small filters and ion exchange columns can be added to improve the quality of the water. This will ensure that water quality is maintained throughout the entire system, and that all LCW circuits are supplied with water of the required purity.

### 9.8.5 Wastewater Collection and Discharge System

During the operation of CEPC, processed wastewater may be generated due to accidents, cooling water replacement, equipment disassembly, and air conditioning condensation. Process wastewater is divided into two categories: ordinary wastewater and low-level radioactive wastewater.

Ordinary wastewater can be discharged directly or recycled after treatment. Low-level radioactive wastewater will either be collected, stored, and recycled, or discharged after it is naturally attenuated to meet the relevant national standards.

Due to the distribution of low-level radioactive wastewater discharge points and the volume of discharge, it is necessary to establish multiple facilities to collect and store the wastewater.

To accomplish this, drainage ditches and catchment wells will be installed along the tunnel, with one catchment well (V = 2 m$^3$) for every 500 meters in length. The drainage ditch will be slope-oriented towards the catchment well with a slope of 1‰.

In addition, there will be 18 storage areas for low-level radioactive wastewater located underground. Each storage area will have a volume equivalent to the maximum volume of the cooling water subsystems in the corresponding area. The total storage capacity for all 18 storage areas is 5,400 m$^3$. The storage capacity of each area is provided in Table 9.8.4.

**Table 9.8.4:** Parameters of low-level radioactive wastewater collection and discharge system

| Parameters | Collider | Booster | Linac | BTL | IR |
|---|---|---|---|---|---|
| Storage volume (m$^3$) | 16×300 | — | 100 | 100 | 2×50 |
| Total (m$^3$) | 5100 | | | | |

### 9.8.6 Cooling Water Control System

The cooling water control system for CEPC has been designed to meet multiple requirements beyond just thermostatic control and process needs. Long-term stable operation, high reliability, and easy maintenance are key considerations, as is the ability to achieve high-precision measurement control in the presence of strong electromagnetic interference and high radiation.

The system is also designed to facilitate network communication with overall and hierarchical monitoring, while remaining state-of-the-art and allowing for future expansion. To achieve these goals, all process parameters will be collected, well displayed, and stored, including on-line trend graphs, historical trend graphs, and tabular data. The process program and run status of the equipment will be dynamically displayed to operators, and an alarm system will be implemented to ensure timely response to any issues that arise.

The system will also include functions for data communication with the central control room, remote monitoring, and data management. These features will enable the cooling water control system to be integrated with the broader control and monitoring systems of CEPC, ensuring that the entire facility operates efficiently and effectively.

## 9.9 Ventilation and Air-Conditioning System

### 9.9.1 Indoor and Outdoor Air Design Parameters

The HVAC system for CEPC maintains temperature and humidity for buildings and tunnels, provides fresh air for tunnels before personnel enter, and ensures safety in different accidental scenarios. The system is divided into several subsystems: air-conditioning, ventilation, heating and cooling source, and control.

#### 9.9.1.1 *Outdoor Air Parameters*

The proposed Funing site has the following outdoor parameters:

Winter:

- Dry-bulb temperature for winter heating: –9.6°C
- Temperature for winter ventilation: –4.8°C
- Temperature for winter air conditioning: –12°C
- Relative humidity for winter air conditioning: 51%
- Wind speed during winter: 2.5 m/s

Summer:

- Dry-bulb temperature for summer air conditioning: 30.6°C
- Wet-bulb temperature for summer air conditioning: 25.9°C
- Temperature for summer ventilation: 27.5°C
- Relative humidity for summer ventilation: 55%
- Daily mean temperature for summer air conditioning: 27.7°C
- Wind speed during summer: 2.3 m/s

#### 9.9.1.2 *Indoor Design Parameters*

The temperature of the tunnel at CEPC is maintained within the range of 30~34°C and is kept below 35°C. The relative humidity is controlled within 50%~60% and is kept below 65%.

The temperature of the 4 experiment halls is controlled at around 26°C and the relative humidity is maintained within the range of 50%~60%, and is kept below 65%.

### 9.9.2 **Layout of the HVAC System**

The equipment for the HVAC system at CEPC is located in over 34 surface sites and underground structures, such as experimental halls and additional chambers.

For surface sites, HVAC and compressed air rooms are utilized to house the heating and cooling source, the compressed air source, the central-station air handling units for the main tunnel, and large-scale ventilation fans.

In the underground areas, the HVAC equipment room is installed adjacent to the air conditioning field.

### 9.9.3 **Tunnel Air-Conditioning System**

The air conditioning cold load for the underground collider ring tunnel at CEPC is approximately 14 MW.

During operation, the heat generated by the electrical equipment is primarily removed by the low-conductivity cooling water system. Any remaining heat can be removed by the tunnel ventilation and air-conditioning system.

The ventilation and air-conditioning system is also crucial for several other purposes, such as providing fresh air for personnel, controlling equipment temperatures, dehumidifying to prevent condensation, discharging gas generated in accidents, removing air in the tunnel before personnel enter, and filtering exhaust gas.

The tunnel is divided into 32 sections by the shafts located around the circumference of the machine, with an approximate spacing of 3 kilometers between each section. Each section is treated independently by the ventilation and air-conditioning system, with each shaft serving as an air supply and exhaust.

There are 32 ventilation and air-conditioning equipment rooms at ground level for the 32 vent shafts. Each room contains 3 combined air conditioning units, with 2 units in use and 1 on standby. The processed air is directed to the chambers connected to the underground tunnel via air vents and then to the air pipe in the tunnel via a relay blower. The air is subsequently exhausted through the relay air exhaust blowers located at the bottom of the shafts and is then either directed to the inlet of a combined air conditioning unit or vented outside.

When the outdoor air temperature is lower than the air supply temperature of the air conditioner, the cooling water unit is shut down, and the outdoor air is sent directly to the air conditioning area without cooling. All exhaust air is discharged out of the loop tunnel.

The main tunnel in CEPC is the largest single space, spanning over 100 km. The air conditioning system used for the main tunnel is complex, and all air handling units (AHUs) are located on the surface for easy access and maintenance. There are three primary challenges faced when designing air conditioning systems for large tunnels: dehumidification, space air diffusion, and integration with the ventilation system. Efficient dehumidification is critical due to the heavy humidity load, and factors such as cooling water temperature, cooling coil heat transfer area, and total airflow must be calculated to ensure effective dehumidification. The ring tunnel is divided into 32 sections, each around 3,125 m in length. Computational fluid dynamics (CFD) is used to ensure a uniform temperature field when the system is running. The duct system is used for both air conditioning and ventilation, providing fresh air and exhausting harmful air in emergencies. The air conditioning layout for the main tunnel is shown in Fig. 9.9.1.

Besides the main tunnel, there are other underground structures, such as the experimental hall and its additional chambers, auxiliary tunnels, and equipment gallery, which all have their own locally arranged air conditioning equipment. Depending on the specific process requirements of each area, there are three types of air conditioning systems used: the all-air system is used for halls and large spaces, the air-water system is used for compact rooms where the relative humidity needs to be kept below 65%, and the all-water system is used for rooms that do not require any specific control of relative humidity.

In surface sites, industrial air conditioning systems, comfort air conditioning systems, and heating systems are used depending on the different operating conditions of halls and rooms. Unlike in the underground, where the indoor conditions are almost solely related to the accelerator's operating mode, outdoor temperature and relative humidity, especially during transitioning seasons, are additional factors that can affect the indoor conditions.

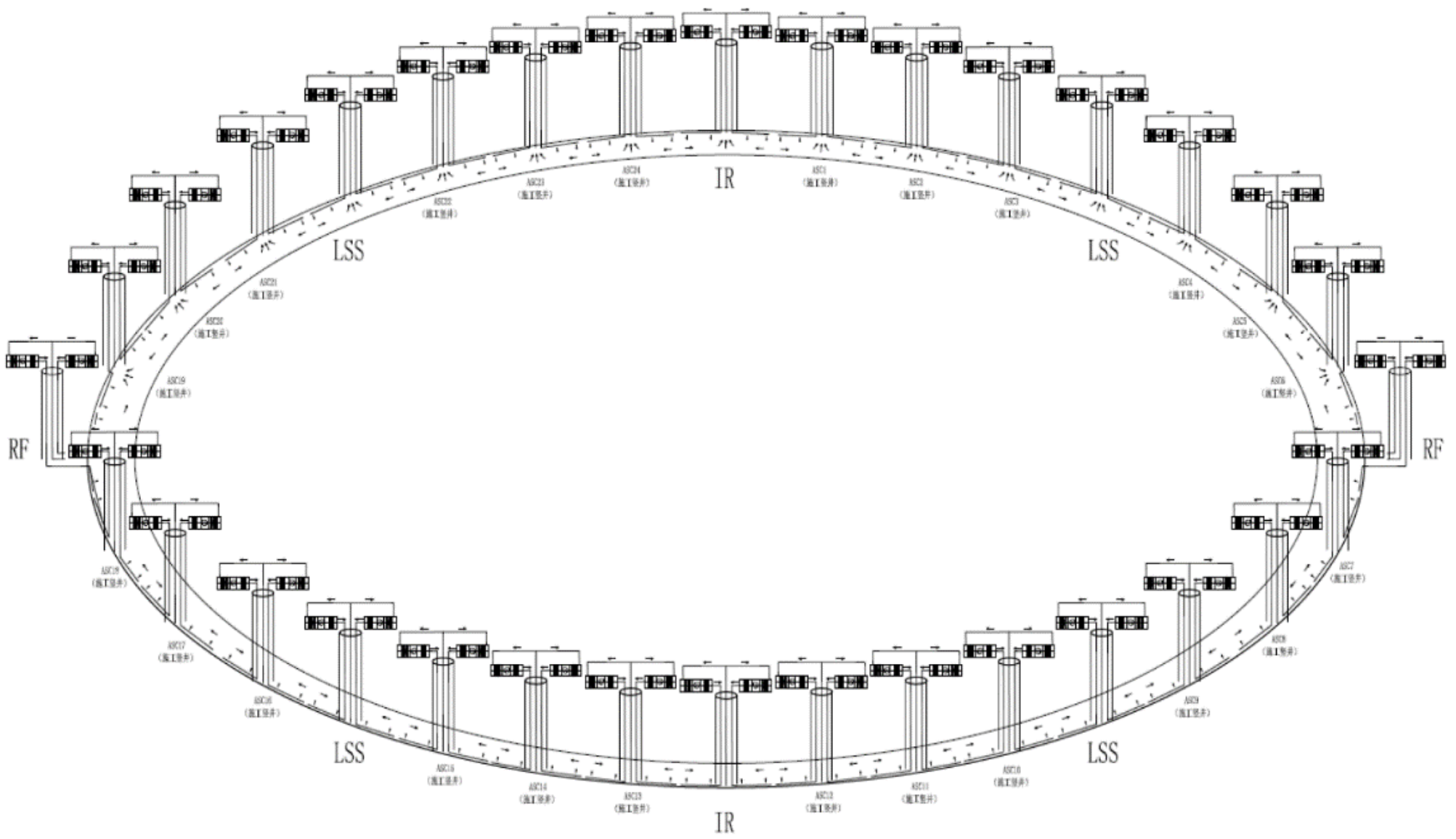

**Figure 9.9.1:** Layout of the air conditioning system in the main tunnel.

### 9.9.4 Air-Conditioning System in Experimental Halls

Independent ventilation and air conditioning systems will be installed in the IP1 and IP3 experiment halls, as well as in adjacent assembly areas. Additionally, specialized smoke, argon, and gas emission systems will be installed in these areas. In enclosures where gas emission is required, the atmospheric pressure is always kept higher than in adjacent cavities, due to the continuous operation of the gas emission system. The main air discharge is achieved through the upper air exhaust and collection chamber, while the air conditioning unit is located in ground level buildings and connected to the air supply and discharge point via the vent pipe in the vent shaft.

In the event of a fire, the ventilation system will stop automatically, and the smoke extraction system may be activated manually after personnel have been evacuated and the fire is under control. Therefore, manual instructions for the fire brigade will be provided at both ground and underground locations to ensure proper procedures are followed in case of emergency.

### 9.9.5 Ventilation and Smoke Exhaustion System

The ventilation system in CEPC serves two essential functions: providing intake air for personnel and emergency ventilation in case of accidental scenarios. During operation, the air intake system is shut down, and the corresponding exhaust airflow controls the negative pressure, with the ventilation airflow minimized in this mode. When CEPC is shut down, air exchange is required before personnel can enter the underground, with negative pressure controlled by intake airflow and ventilation air controlled by exhaust airflow. If necessary, all exhausted air should be filtered based on environmental evaluation.

In the event of an accidental scenario such as a fire, oxygen deficiency hazard, or cryogenic gas leak, reliable and rapid measures must be taken immediately. The dedicated intake fan and exhaust fan will run as soon as possible once the emergency is detected. To ensure personnel safety during an emergency, the diffusion of space air must be calculated, with assistance from CFD in the verification process.

The emergency smoke control and exhaust system is an essential safety feature for personnel working underground. It will be integrated with the mechanical air exhaust system to ensure that any smoke or harmful gases are quickly removed from the tunnel and experiment halls in the event of an emergency.

In addition, depending on the final site selection, it may be possible to utilize the significant amount of waste heat generated by the facility to support the implementation of national rural revitalization efforts in surrounding villages. Alternatively, the waste heat could be repurposed for biological applications. This would not only promote sustainable energy use, but also provide potential economic benefits to the local community.

### 9.9.6 Heating and Cooling Source

CEPC relies on its heating and cooling facilities to provide hot water and chilled water for AHUs. These essential facilities are distributed across 34 sites above ground, and together they provide a total heating load of 14 MW and a total cooling load of 24.4 MW.

The heating source primarily utilizes the city heating network, with local boilers serving as a secondary source. In addition to these conventional heating sources, a heat recovery system is employed to achieve significant energy savings. When operating chillers can supply hot water in certain conditions, the heat recovery system is activated. All AHUs for industrial air conditioning systems are equipped with electrical reheating for relative humidity adjustment.

The water-cooled centrifugal chiller is used to produce chilled water for air conditioning as the cooling source. Some chillers are fitted with frequency converters that can operate at lower loads and achieve a better coefficient of performance (COP).

CEPC uses a four-pipe system for its hot water and chilled water circuits. The hot water system operates at temperatures ranging from 30 to 50℃, while the chilled water system operates at temperatures ranging from 5 to 11℃ or 6 to 12℃. To ensure optimal electricity consumption, various measures such as the use of frequency converters for pumps and hydraulic computation trade-offs are implemented to regulate the transfer of cooling and heat quantity ratios. Accurate pressure calculations are also essential for each water system, given that the elevation difference can range from 50 m to 110 m, thus determining the nominal pressure of pipes and equipment.

Given the higher dehumidification load of the tunnel, any condensed water from underground AHUs is collected and stored in local tanks as a buffer and delay. The collected water is then tested and, if deemed safe, it can be pumped to the surface and discharged.

### 9.9.7 **Control System**

The HVAC system in CEPC uses a distributed control system (DCS) as its control structure, with a local programmable logic controller (PLC) managing sensors, actuators, and equipment in the field. Each site is equipped with supervisory and management computers, which collect data from the system's processor nodes and provide operators with control screens. The management computer also logs operating data and triggers alarms during emergencies.

To ensure the safety of personnel in CEPC, multiple safety measures are in place, including software protection and hardwire interlock. In normal conditions, the software protection provides all necessary protection functions. However, in the event of a control system malfunction, the hardwire interlock provides critical protection for personnel and prevents potential harm.

Overall, the controllers and computers in the system connect and communicate using standard Industrial Ethernet (IE), allowing for efficient and reliable communication and control of the HVAC system in CEPC.

## 9.10 Compressed Air System

Vacuum valves, cryogenic valves, and other pneumatic equipment used in CEPC extensively rely on compressed air to function. Compressed air is considered a safe and reliable source of power for pneumatic equipment, such as vacuum valves. The following requirements should be met for compressed air usage in CEPC:

- Air pressure: 0.6 MPa
- Pressure dew point: –40℃
- Particle removal efficiency: 99.98% @MMPS
- Oil carry-over：0.01 ppm

In CEPC, a compressed air system will be installed at sixteen points of the ring and at the Linac to supply clean gas for the machine in the area. Each compressor system will consist of twin screw compressors, tanks, refrigerant dryers, adsorption dryers, filters, a pressure control unit, cooling systems, pipe systems, and control systems. Standby equipment should be connected to the system to ensure uninterrupted operations during repair and maintenance. For better pressure stability and air quality, storage tanks and local filters will be installed underground.

The process of compressed air production is illustrated in Figure 9.10.1.

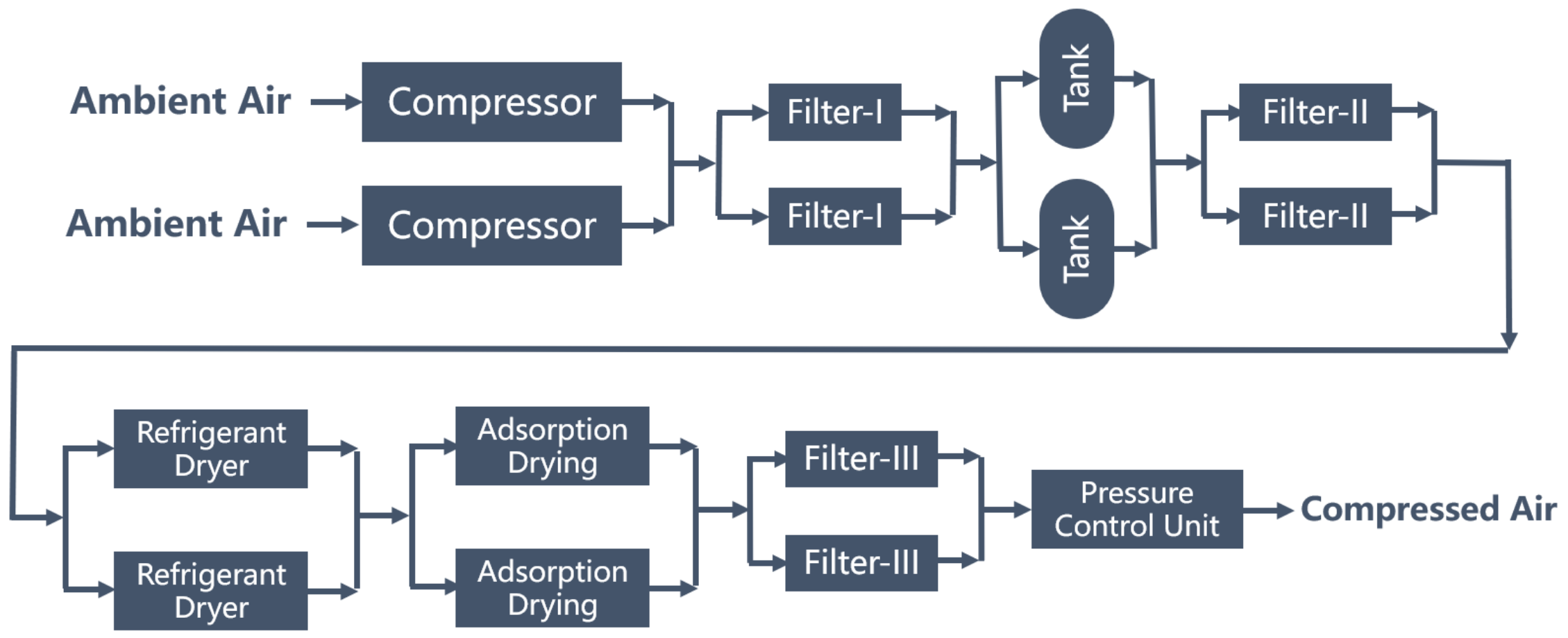

**Figure 9.10.1:** The process of compressed air production.

## 9.11 Fire Protection and Drainage Design

### 9.11.1 Fire Protection Design

#### 9.11.1.1 *Basis for Fire Protection Design*

The fire protection plan for this project is designed to prioritize fire prevention, with fire fighting measures as a supplement. The plan includes four steps: prevention, cut-off, extinguishing, and exhaust. To minimize fire hazards, fire prevention and exhaust systems, hydrant and fire extinguisher systems, and fire detection and fire alarm systems are combined with building fire prevention and evacuation.

The design of the fire protection plan is based on relevant codes, including the Design on Building Fire Protection and Prevention, Design Code for Scientific Experiment Buildings, Design Code for Office Building, and Design of Extinguisher Distribution in Buildings.

Given the presence of computers and electromagnetic equipment in the tunnel and experiment hall, and the fact that the area is unmanned during operation with only a few familiar personnel entering for maintenance, there is a small possibility of spontaneous combustion of internal equipment. The plan considers at most one ignition point at the same time and does not consider the possibility of an explosion in case of an internal fire. However, all personnel who enter the area are expected to be familiar with evacuation procedures.

The experiment hall is classified as an underground equipment room, while the ring tunnel is of a unique nature with no industry standard available in China. The fire protection design for this project refers to the Standard on Fire Protection for City Underground Tunnels, the Code of Design on Building Fire Protection and Prevention, and various underground engineering specifications, including the Technical Code for Urban Utility Tunnel Engineering, the Code for Design of Prevention of Mine Fires in Coal Mines, and the Coal Mine Safety Code. Special measures can also be considered based on specific conditions, such as the characteristics of the internal equipment and personnel job descriptions.

### 9.11.1.2 ***Fire-fightig for Buildings***

#### *9.11.1.2.1 Fire Resistance of Building Structures*

The ground level buildings have a fire protection rating of either Grade I or Grade II, and non-combustible or refractory materials are used in their construction to ensure their fire safety.

To reduce the fixed fire load in the tunnel and experiment halls, all building and interior decoration materials used for tunnel lining, experiment hall, and other parts of the building should be non-combustible, except for joint materials. The air ducts of the ventilation system should also be made of non-combustible materials, with flexible joints made of refractory materials. Cables in the tunnels should be flame-retardant cables or mineral-insulated cables, while cable trays and conduits should be fire-resistant.

For vertical and inclined shafts, as well as independent refuge rooms connected to the main tunnel and used for safety evacuation, emergency refuge, and other purposes, their load-bearing structure's fire resistance limit should not be lower than the fire limit requirement of the main structure of the tunnel. This ensures that all areas of the project have sufficient fire resistance to minimize the risk of fire spreading.

#### *9.11.1.2.2 Safety Escape*

Each experiment hall in the project must meet the requirements outlined in Clause 12.1.10 of the Code for Fire Protection Design of Buildings. The maximum allowable building area for each fire zone in underground equipment rooms is 1500 $m^2$, which the area of all experiment halls in this project satisfies. Each experiment hall is equipped with two or more vertical access shafts that directly connect to the surface outdoors to meet the requirement for two safe exits in every fire zone. Additionally, fire cuts are arranged around each hall with an area based on the maximum number of people simultaneously in the fire zone, meeting the requirement of 5 persons per $m^2$. This ensures that in the event of a fire, all personnel have a clear path to evacuate safely.

For the ring tunnel, the evacuation vertical shaft is combined with the air shaft to maximize efficiency and space. There is one vertical shaft directly connected to the outside at approximately every 3,000 m, and the evacuation staircase leading to the outside is inside the air shaft. A smoke-free cavern and fire cut are located at the entrance of the evacuation staircase to ensure the safety of the vertical shaft and create space and time for individuals unable to escape on their own to wait for rescue. These measures provide a clear and safe evacuation route for personnel in the event of an emergency.

#### *9.11.1.2.3 Arrangement of Fire Partitions*

The distance from the farthest point to the evacuation exit is 1,500 meters for staff. In Europe, a series of simulation tests have determined that the maximum distance for people inside a tunnel to escape without issue from smoke density is 250 meters. However, the existing vertical access shafts may not be enough to meet this requirement, so additional facilities must be added. Fire-proof boards are used as fire partitions at intervals, and fire-proof materials are used to seal locations where pipes cross the fire-proof boards. Class A fire doors are placed between every two fire zones, creating relatively independent fire zones. Every fire zone can use the fire zones on both sides as emergency escape channels. Fire partitions can be installed at 500-meter intervals to ensure personnel safety, protect nearby equipment, and minimize fire loss, considering the maximum escape distance from the farthest point.

If it is difficult to arrange fire partitions, fire cuts can be added inside, as per Clause 12.1.8 of the Code for Fire Protection Design of Buildings. This clause specifies that a single tunnel must have refuge facilities, such as personnel evacuation exits leading directly to the outside or independent shelters. Short auxiliary tunnels independent of the collider ring tunnel are present at approximately 1,000-meter intervals. Adding independent fire cuts can help meet personnel refuge requirements within the maximum evacuation distance.

#### 9.11.1.3 *Water-based Fire-fighting*

1. The necessary items for fire protection include indoor fire hydrants and extinguishers underground, municipal water supply for outdoor fire protection, gas fire-extinguishing systems for the control room, a seepage drain pump based on experience from similar projects, and an estimate of leakage.
2. Fire hydrants are designed following the regulations of the Code for Fire Protection Design of Buildings (GB50016-2014). The spacing between hydrants should not exceed 50 meters. As a result, there will be 2,100 sets of type SNJ65 pressure-stabilizing and reducing hydrants. The hydrant mouth is perpendicular to the wall surface and 1.10 meters above the floor. A combined type fire hydrant box (04S202-P21) with an extinguisher is selected, complete with one DN65 hydrant, one Ø19 water gun, one 25-meter rubber-lined hose, and one fire-fighting coiled hose. Building fire extinguishers are also provided.
3. All fire pipes leading to the hydrants are connected to the fire cistern, which is located on the surface and combined with the air shaft. Water automatically flows into the tunnel through two main pipes. After decompression by a reducing valve, the water is connected to the hydrants. There are 32 fire cisterns, each with a volume of 200 cubic meters, and the water is supplied by the municipal water supply. Each fire cistern serves about 3.2 kilometers of the collider ring. The main fire pipe is arranged in a ring, forming an independent system. The water supply volume of hydrants inside the tunnel is 20 L/s, and for a fire duration of 2 hours, the water consumption is 144 cubic meters. The fire cistern can supply this amount.
4. Each hydrant system comprises one fire cistern, 65 hydrants, about 6,400 meters of plastic-coated steel pipe, and 15 valves. The entire ring tunnel and the experiment halls consist of 32 such systems.
5. The distribution of mobile extinguishers shall follow the relevant requirements of the Code for Design of Extinguisher Distribution in Buildings (GB50140-2005).
6. The tunnel and halls house the accelerator and experiments, which include a significant amount of electromagnetic equipment. According to the Code for Design of Extinguisher Distribution in Buildings (GB50140-2005), if a fire occurs, it may be a Class D or Class E electric fire. The entire tunnel is considered one fire zone and has a total of 6,500 distribution points, each with 2 sets of Grade 3A MF/ABC6 ammonium phosphate dry powder extinguishers, for a total of 13,000 extinguishers. The two 1,200 square meter experiment halls each have 4 MFT/ABC20 wheeled fire extinguishers.

### 9.11.1.4 *Smoke Control*

To ensure safe personnel evacuation in case of a fire, a combination of mechanical smoke extraction and mechanical exhaust systems will be used for the smoke control system.

In the event of a fire, the mechanical smoke extraction system will promptly extract smoke and provide fresh air in the tunnel, experiment halls, and auxiliary caverns to ensure safe evacuation. Smoke will be exhausted during the accident in these areas, while in the accessory rooms of the auxiliary cavern, smoke will be extracted after the accident.

To divide the collider ring tunnel into 32 sections, the vertical shaft of the experiment hall, the air shaft, and the vertical shaft at the RF section will be utilized. Additionally, an exhaust duct will be installed in the upper part of the collider ring tunnel, and a smoke exhaust fan will be installed at the bottom of each vertical shaft.

To ensure the safety of personnel and equipment during a fire, the collider ring tunnel is divided into several smoke control zones. Smoke screens are installed between these zones, and a smoke exhaust port is located in the middle of each zone. If a fire occurs in the tunnel, the smoke exhaust port in the affected zone will be immediately opened to exhaust the smoke. Simultaneously, the smoke exhaust fan at the bottom of the vertical shaft will activate to remove the smoke. A fire damper is installed before the smoke exhaust fan, and if the temperature of the smoke reaches 280°C, the temperature fuse will activate, the fire damper will close, and the smoke exhaust fan will stop running.

In the event of a fire in the experiment hall, the smoke exhaust fan will immediately start to remove the smoke, and the smoke will be exhausted outside the facility through the smoke exhaust system. This combination of a mechanical smoke extraction system and mechanical exhaust system provides prompt removal of smoke, allowing for safe personnel evacuation and protection of equipment.

### 9.11.1.5 *Electricity for Fire-fighting*

1. The fire-fighting equipment is powered by a second-class load and emergency power supplies from the 0.4 kV distribution system. In case of an accident, a backup emergency power supply such as a diesel generator, EPS power supply, or DC system power supply will be utilized to ensure reliable electric power for fire-fighting.
2. Separate and independent circuits will be used for power distribution in fire-fighting. Fire-resistant cables will be used to ensure safety.
3. Fire-fighting equipment that requires power includes pumps, smoke-control and exhaust equipment, automatic fire alarms and extinguishers, emergency lighting, evacuation signs, and fire-resistant rolling shutter doors.
4. Battery backups will be used to power emergency lighting and evacuation signs in case of power failure, and they should be able to hold a charge for at least 30 minutes.
5. Fire accident lighting and safe evacuation signs will be installed to mark evacuation passages, staircases, and safe exits.
6. The minimum illumination intensity for evacuation accident lighting should not be less than 0.5 Lx, and lights should be installed on the wall or the ceiling.
7. Evacuation signs should be placed above each safe exit. In each passage and around corners, evacuation signs should be installed on the wall 0.8 m above the ground and spaced no more than 20 m apart. Inflammable covers should be installed for accident lighting and evacuation signs to ensure safety.

#### 9.11.1.6 *Fire Auto-alarms and Fire-ffighting Coordinated Control System*

1. The fire alarm and fire-fighting control system can automatically or manually send alarm signals and fire-fighting coordinated control directions, record, display, and print fire and coordinated control information based on different signals given by the fire detection equipment.
2. The fire alarm system covers a wide range of areas, including the ring tunnel, experiment halls, auxiliary ring tunnels, transportation passages, auxiliary chambers, auxiliary buildings at the surface of shaft openings, buildings housing substations and electric power distribution, the control room, electronics rooms, power supply hall, assembly hall, refrigerating machine room, and ventilation fan room.
3. Control equipment such as audible and visual alarms, fire hydrants, heating, ventilation and air conditioning, smoke exhaust fans, smoke control valves, and fire-resistant rolling shutter doors are included in the fire alarm system.
4. The fire auto-alarm system is powered mainly by AC 220 V in the 0.4 kV fire-fighting power supply. It is equipped with a UPS and backup batteries for power outage situations.
5. The signal and power transmission lines of the fire auto-alarm system are made of fire-retardant copper core shielded cables. All cables are laid out to go through metal or flexible metal conduits or metal sealed ducts to ensure safety.
6. The fire auto-alarm system is grounded through the common grounding network with a maximum grounding resistance of 4 Ω to ensure proper grounding.

#### 9.11.1.7 *Drainage Design*

Based on previous similar projects and initial water leakage estimates, the water seepage amount into the sumps has been estimated. Each sump has a provisional volume of 1,000 $m^3$ and there will be 24 water-collecting wells in total.

Each sump will be equipped with 4 horizontal multistage clean water pumps with synchronous discharge and suction. The chosen sump pump is a TPD200-280-43X7 pump, with a flow rate of 288 $m^3$/h and can raise water to a maximum height of 286 m. It requires 260 kW of power to operate. There will be three pumps for regular use and one for standby at each sump, resulting in a total of 96 pumps required for the project. In addition, valves and plumbing will also be needed to complete the system.

The drain header chosen for the project is DN350, which refers to a 350 mm pipe. This size was chosen for economical flow velocity.

## 9.12 Permanent Transportation and Lifting Equipment

The installation of accelerator components and massive detectors in the experimental halls will closely follow the completion of the civil construction. Specialized transportation and handling equipment such as cranes will be required to move both components and personnel. Some of the equipment will be transported to the underground experiment hall for assembly by an overhead traveling crane after being assembled into larger units in the assembly hall.

During operations, personnel and materials or equipment necessary for experiment maintenance, overhaul, and replacement will be transported at a lower frequency.

There will be a total of 46 access shafts from the surface to the underground experiment halls, with depths ranging from 50.6 m to 244.38 m and an average depth of approximately 129.17 m.

Eight access shafts will serve the halls at IP1 and IP3. At each international region (IR), there are various shafts with different diameters for different purposes. These include a 16-meter diameter shaft and a 9-meter diameter shaft specifically for material transportation, a 6-meter diameter shaft for personnel transportation, and a 7-meter diameter shaft that leads down to the Booster bypass tunnel.

At IP2 and IP4, where the RF cavities are located, personnel transportation is done through two 6-meter access shafts, while one access shaft with a diameter of 15 meters is used for materials.

For each linear section (LS) in the Collider ring, a 10-meter access shaft is mainly used for personnel transportation.

There are eight 10-meter shafts in the arc sections. They are mainly used for material transportation during the construction period and for personnel transportation during operations. The sixteen 7-meter ventilation shafts in the arc sections are mainly used for material transportation during the construction period and for ventilation during operation.

Two 7-meter diameter access shafts will be constructed for the BTL tunnel.

There are two magnet assembly halls at IP1 and IP3, which will use gantry cranes for transportation and assembly. Overhead cranes will be used in the main and service caverns underground, and elevators will transport personnel from the surface to the underground areas.

There will be 20 electric station wagons in the tunnel, stationed at each access shaft area, and an additional truck at the IP1 and IP3 experimental halls.

The permanent lifting and transportation equipment includes 5 gantry cranes, 2 overhead cranes, 24 elevators, and 20 electric trucks, which are itemized in Table 9.12.1.

**Table 9.12.1:** Quantities of Transportation Lifting

| Project section | Equipment and specs | QTY | Unit weight (t) | Total weight (t) |
|---|---|---|---|---|
| Transport shafts and ventilation shafts | Access shaft elevator | 18 sets | 6 | 108 |
| | Rail for Access shaft elevator | 18 sets | | 20+38+22+28+36+36+42+40+40+40+43+43+26+25+25+18+16+33 |
| IP2, IP4 | Equipment transportation gantry crane for shaft with diameter of 15m 30t crane | 2 sets | 60 | 120 |
| | Rail for shaft gantry crane | 2 sets | | 72+50 |
| IP1, IP3, Bypass tunnel access shafts | Elevator for shaft with diameter of 7m | 2 sets | 6 | 12 |
| | Rail for shaft elevator | 2 sets | | 20+40 |
| IP1, IP3 auxilliary access shafts | Elevator for shaft with diameter of 6m | 2 sets | 6 | 12 |
| | Rail for shaft elevator | 2 sets | 46 | 25+31 |
| IP1, IP3 ground assembly hall | Ground 1500 ㎡ experiment assembly hall 1500t gantry crane | 1 set | 2000 | 2000 |
| | Ground 1500 ㎡ experiment assembly hall 1000t gantry crane | 1 set | 1200 | 1200 |
| | Ground 1500 ㎡ experiment assembly hall 80t gantry crane | 2 sets | 120 | 240 |
| | Gantry crane rail QU70 | 2 sets | 12 | 24 |
| IP1, IP3 auxilliary shafts | Elevator for equipment transportation gantry crane of shaft with diameter of 9m | 2 sets | 6 | 12 |
| | Equipment Transportation shaft elevator rail | 2 sets | | 20+40 |
| IP1, IP3 underground main cavern, service cavern | IP1 and IP3 underground main cavern overhead crane 20t overhead crane L=28m | 1 set | 50 | 50 |
| | Main cavern overhead crane rail QU80 | 1 set | 5 | 5 |
| | IP1 and IP3 underground service cavern overhead crane 10t overhead crane L=18.5m | 1 set | 30 | 30 |
| | Service cavern overhead crane rail QU70 | 1set | 3 | 3 |
| Underground vehicle | Trucks for both passenger and goods (Power driven) | 20 sets | 5 | 100 |
| Total | | | | 4785 |

## 9.13 Green Design

### 9.13.1 Energy Consumption

The CEPC is projected to have a total power consumption of approximately 346 MW during operation, which represents a significant amount of energy. Based on a preliminary estimate, the project is expected to consume around 2 billion kWh, equivalent to 720,000 tonnes of standard coal. This level of energy consumption would result in the emission of approximately 1,886,400 tonnes of carbon dioxide. As a result, it is critical that the design of the CEPC takes into account energy conservation, consumption reduction, and sustainability as crucial social responsibilities.

### 9.13.2 Green Design Philosophy

Green Design, also known as Ecological Design, Design for Environment, and Environment Conscious Design, is a design methodology that considers environmental attributes as design objectives, rather than constraints. Its goal is to incorporate these attributes into the design without compromising performance, quality, or functionality, and to make use of natural materials. Green Design aims to achieve ecological balance between humans and nature by taking into consideration environmental benefits in decision-making during the design process to minimize damage to the environment.

For the CEPC, the reduction of energy and material consumption will be a primary goal during the design stage, and environmental factors and pollution prevention measures will be incorporated. The core principles of Green Design are the "3R's": Reduce, Recycle, and Reuse. The aim is not only to reduce the consumption of substances and energy and minimize the emission of harmful substances, but also to fully consider recycling or reuse throughout the entire project.

We strongly advocate for the CEPC to prioritize simplicity, practicability, energy efficiency, material consumption reduction, cyclic utilization, and environmental friendliness. By incorporating these principles, the CEPC can be designed to be more sustainable and contribute to a healthier planet.

### 9.13.3 Green Design Implementation

#### 9.13.3.1 *Reduce – Reduce Environmental Pollution and Energy Consumption*

Optimize the design parameters of physical test facilities, improve the energy efficiency of physical test equipment and reduce the energy consumption of the equipment. In civil engineering and infrastructure, building energy-saving design considers factors such as building orientation, spacing, solar radiation, wind direction and external space environment using climate-sensitive design and energy-saving methods. Buildings are designed with low energy consumption and eco-environmental protection using environmental protection materials and green construction to reduce pollution and damage caused by the project.

#### 9.13.3.2 *ReReuse and Recycle – Recycling, Regeneration and Reuse*

1. The cooling system generates a significant amount of residual heat, typically discharged into the atmosphere due to its low temperature (30°C ~ 40°C). Recycling this residual heat can result in significant energy savings.
2. Heat pump technology can upgrade the residual heat, followed by an advanced residual heat utilization technology that incorporates various heat exchange, thermal power conversion, and residual heat refrigeration technologies. This can effectively utilize CEPC residual heat resources, for example, by providing domestic heating for science and technology parks, refrigeration and air conditioning, and agricultural greenhouses.
3. "Sponge City" concept can be introduced to collect and process rainwater and waste water, establish a recycling system, and build a water-saving "sponge technology park" that promotes waste water utilization.
4. Incorporating renewable energy sources such as wind power and solar photovoltaic power stations into the project can increase renewable energy utilization and reduce carbon emissions. There is ample space for large solar panel arrays.

#### 9.13.3.3 *Advanced Energy Management System*

The CEPC information system will collect and process energy consumption data, including electricity, fuel gas, and water, to analyze building energy consumption. Energy planning, monitoring, statistics, and consumption analysis will be conducted to conserve energy. Key energy consumption equipment and energy metering equipment will be managed to ensure energy efficiency.

# 10 Environment, Health and Safety Considerations

Environmental, safety, and health (ES&H) considerations are an integral part of the design, construction, and operation of the CEPC. Lessons learned from existing accelerators have allowed us to identify key hazards and associated risks. This chapter provides an overview of various aspects related to ES&H, including work planning and training (Section 10.2), environmental impact (Section 10.3), ionizing radiation (Section 10.4), fire safety (Section 10.5), cryogenic and oxygen deficiency hazards (Section 10.6), electrical safety (Section 10.7), non-ionizing radiation (Section 10.8), and general safety issues (Section 10.9).

The preparation of environmental impact and occupational hygiene assessment documents will be conducted once the project is approved by the government. Certain requirements, such as monitoring the background radiation level of the site and conducting public participation investigations, which are essential for the implementation of safety-related documents, can only be finalized at that stage. This chapter provides a general overview of environmental impact factors and proposes countermeasures to mitigate their effects. It also introduces the technical measures and preliminary designs related to radiation safety.

It's important to note that natural hazards such as seismic disturbances or extreme weather events require specific emergency planning procedures. These aspects are not addressed in detail within this TDR. Separate protocols and guidelines will be developed to address these natural hazards and ensure the safety and resilience of the CEPC.

Furthermore, surveys will be conducted to identify sites of historic and prehistoric significance. Measures will be taken to preserve and mitigate the project's impact on these sites in coordination with national historic preservation laws.

## 10.1 General Policies and Responsibilities

The construction and operation of the CEPC will prioritize prevention and safety above all else. We are committed to strengthening environmental protection management throughout the project. A primary objective of the project is to avoid any accidents during both the construction and operation phases. To achieve this, comprehensive training will be provided to all personnel, including staff and contractors, to foster a strong safety consciousness and a dedicated and responsible work attitude.

The fundamental principles and expectations guiding our efforts are as follows:

- No safety incidents
- No casualties
- No environmental pollution

The CEPC's environmental, safety, and health plans fully comply with all applicable government laws and regulations. Every employee shares the responsibility for maintaining a safe working environment. We firmly believe that through the unwavering persistence and collective efforts of all individuals involved, the CEPC can successfully achieve its goals of environmental protection, safety, and health.

## 10.2 Work Planning and Control

### 10.2.1 Planning and Review of Accelerator Facilities and Operation

Once the CEPC project receives government approval, the compilation of the Environmental Impact Assessment Document (EIAD) becomes a crucial step. Prior to the commencement of project construction, the EIAD must undergo a comprehensive assessment and obtain approval from the relevant competent authorities. This document assesses the potential environmental impact of the project and outlines measures to mitigate any adverse effects.

Upon completion of the main building construction and installation of safety-related accelerator components within the tunnel, a comprehensive safety analysis document will be compiled. This document will include a detailed analysis of the actual construction process, highlighting any deviations from the original EIAD and providing justifications for these changes. Prior to initiating a physics run, a thorough review of the safety analysis document will be conducted by competent authorities. Additionally, an on-site inspection will be carried out to verify the implementation of all safety measures, including the examination of shielding components, radiation dose monitoring systems, personal protection systems, and other safety-related aspects.

These measures and procedures are implemented with the primary goal of preventing accidental access to radiation areas. Adequate shielding measures are put in place to ensure that areas outside the machine are safely accessible, thus minimizing the risk of exposure to radiation.

### 10.2.2 Training Program

The CEPC project is committed to adhering to relevant national laws and regulations pertaining to ES&H practices. In order to ensure the safety and well-being of personnel, ES&H training programs will be implemented based on the specific operational environment, hazard factors, and ES&H management requirements. The aim is to continuously enhance the ES&H awareness and skills of all employees.

The training programs will encompass various components, including lectures, practical exercises (where applicable), and examinations that individuals must successfully complete. Certain training sessions will be one-time events, primarily for new employees, while others will require periodic renewal, often on an annual basis.

ES&H training will cover a range of topics, including but not limited to:

- Radiation safety and protection training: All personnel entering radiation areas will receive comprehensive training on radiation safety and protection measures.
- Occupational health training: Management personnel will undergo specific occupational health training to address their unique responsibilities.
- Special job training: Personnel involved in potentially hazardous tasks, such as confined space entry or handling hazardous materials, will receive specialized safety training prior to undertaking these assignments.
- Special equipment operation training: Personnel who are required to operate potentially hazardous equipment, such as forklifts or overhead cranes, will participate in specialized training programs before being authorized to use such equipment.

- Other relevant ES&H training: A variety of courses will be available, some of which may be voluntary while others are mandatory based on job descriptions. Examples include emergency preparedness management, first aid training, and computer security.

### 10.2.3 Access Control, Work Permit and Notification

Chapter 7.3 of this report defines controlled access areas. Entry into these areas is restricted based on personnel's training experience, availability of personal protective equipment (PPE), and their accumulated radiation dose.

It is mandatory to carefully plan and optimize all activities conducted within controlled radiation areas. This includes conducting an assessment to estimate the collective dose and individual effective doses that participants may be exposed to. Workers must be thoroughly informed about the potential hazards associated with their tasks, including any abnormal situations that may arise. Furthermore, they should receive training specifically tailored to the nature of their work.

## 10.3 Environment Impact during Construction Period

Chapter 9 of the TDR provides a detailed overview of the buildings and structures, both surface and underground. One potential site under consideration is located in Funing District, approximately 30 km away from the nearest major city, Qinhuangdao. The surrounding area comprises mainly of villages, without any large-scale buildings or structures. It is important to identify and avoid any existing underground pipeline networks during the excavation process.

During the construction phase, the main environmental impacts to consider are related to water and noise. Measures need to be taken to mitigate the impact of blasting vibrations and noise generated during underground construction. Additionally, the construction activities should not compromise the water quality, and appropriate measures must be implemented to manage domestic sewage and wastewater produced during construction.

The key focus areas for environmental protection during construction are water and noise protection. This includes optimizing the sewage treatment process to minimize water pollution and implementing noise protection measures to safeguard the local ecosystem and the well-being of the nearby population.

### 10.3.1 Impact of Blasting Vibration on the Environment and Countermeasures

The project is primarily situated in suburban areas, with a portion of the facility located in residential areas and another portion in non-residential areas. One of the main concerns is the potential damage to houses due to blast shock waves, as well as the residents' tolerance towards vibration frequency and intensity. Fortunately, the impact on the ground environment is minimal since the majority of the working faces are deep underground.

To mitigate the potential negative effects, it is crucial to exercise control over the blasting process by implementing appropriate blasting parameters during construction. By carefully selecting these parameters, the potential damage to houses can be minimized.

The impact on residents will largely depend on the specific underground construction methods employed, whether it involves drill-blast techniques, Tunnel Boring Machines (TBMs), or a combination of both. Each method has its own set of advantages and considerations regarding their impact on nearby residents.

### 10.3.2 **Impact of Noise on the Environment and Countermeasures**

The impact on the sound environment is primarily caused by excavation blasting, crushing of sand and gravel, mixing of concrete, construction transport, and operation of heavy machinery. To address this, low noise equipment and necessary workforce protection should be implemented during construction.

### 10.3.3 **Analysis of Impact on the Water Environment**

Sewage and wastewater generated in the project construction area consist of construction wastewater and domestic sewage. Discharging sewage directly into nearby watercourses without treatment can have a significant impact on water quality. Therefore, it is essential to conduct treatment of both construction wastewater and domestic sewage. This treatment should ensure that the discharges meet the required standards or, alternatively, implement recycling methods to minimize environmental impact.

### 10.3.4 **Water and Soil Consevation**

Water loss and soil erosion can occur due to inadequate design of living quarters, construction roads, and disposal areas. The thick overburden exacerbates the impact on surrounding surface water. Thus, it is necessary to implement engineering and biological measures to prevent scouring of rainfall runoff to the construction site and disposal area, reducing water loss and soil erosion.

## 10.4 Environment Impact during Operation Period

The main ring of the project will be located at a depth of 50 to 150 meters underground. Apart from access points, the surface area will remain accessible to the general public, allowing customary activities such as agriculture and residential uses to continue. When evaluating the environmental impact, the primary concern lies in the release of radiation and radioactivity to areas outside the project's control.

### 10.4.1 **Groundwater Activation and Cooling Water Release Protection**

Groundwater activation can only occur in areas where water comes into close proximity to the tunnel and where the rock is not waterproof. Fortunately, most of the ring is composed of waterproof and dry rock, minimizing the potential for groundwater activation. However, it is essential to note that radioactivity may still be generated in the groundwater outside the tunnel or leached out from activated rock or soil by the groundwater. Special attention needs to be given to monitoring and assessing the impact of this radioactivity, especially with regards to off-site wells that may be contaminated with H-3 (tritium) due to the movement of underground aquifers.

The radioactivity produced in the cooling water circuits is primarily a result of high-energy radiation and the high-energy tail of synchrotron radiation. To mitigate any potential risks, the cooling circuits will be designed as closed systems, ensuring that no release of radioactivity occurs during normal operation. In the event of maintenance or accidental draining of the circuits, the cooling water will be collected in the general ring-drainage system. Before release, the accumulated water will be sampled and closely monitored for activity levels. If necessary, the release from the ring drains can be delayed for a significant period. Additionally, in cases of flooding, immediate pumping of water will take place, as dilution will reduce the concentration of radioactivity to negligible levels.

### 10.4.2 **Radioactivity and Noxious GasesReleasedinto Air**

Section 7.3.3 of this report provides a comprehensive discussion on the radionuclides generated in the tunnel air, notably N-13 and O-15, as well as the production of noxious gases such as O-3 and NOx.

To mitigate the release of these isotopes and noxious gases, the vacuum chamber will be adequately shielded. This shielding effectively minimizes the amount of synchrotron radiation that escapes into the air, thereby reducing the production of these harmful components. Additionally, the design considerations include determining the height of the air release point and controlling the release velocity within the air and ventilation system (Section 9.5). These measures are implemented to further decrease the concentration of these detrimental substances in the environment.

It is imperative to conduct monitoring activities, both at the vent ports and ground level, to ensure proper assessment and control of these hazardous components throughout the operation of the project. Regular monitoring provides essential data to evaluate the effectiveness of mitigation measures and facilitates prompt action if necessary.

### 10.4.3 **Radioactive Waste Management**

Materials that are anticipated to undergo activation should be carefully designed or selected to minimize their volume and weight. This strategy is aimed at reducing the overall amount of radioactive material generated during the project. National regulations provide explicit guidelines regarding the threshold at which materials are classified as non-radioactive. Any items found to be radioactive will be managed in accordance with the procedures detailed in Section 7.3.6.

## 10.5 Radiation Impact Assessment to the Public

Synchrotron radiation and radiation induced by lost beam are two main sources of ionizing radiation. The shielding design and radiation dose caused by these are calculated and listed in Section 4.2.4, Section 5.2.4, and Section 7.3.

### 10.5.1 **Radiation Level during Normal Operation**

The radiation inside the tunnel consists of synchrotron radiation and radiation induced by random beam loss. Synchrotron radiation has a soft spectrum and is difficult to transport through the soil or shaft to the environment. When considering the radiation caused by random beam loss, the radiation level near the tunnel wall is significantly below 5.5 mSv/h, as per our design for the shaft construction.

If there is a shaft that directly connects to the tunnel horizontally and then turns vertically towards the ground surface, creating a two-legged maze-like geometry, the distance from the tunnel to the ground surface exceeds 100 m. Additionally, we can incorporate blocks or shielding doors between the shaft and the tunnel. Based on our preliminary calculations, these measures can effectively reduce the radiation to a very low level, well below the annual control limits.

### 10.5.2 Radiation Level during Abnormal Condition

The most severe concern arising from accelerator operation is the potential entry of individuals into the tunnel while the accelerator is running. However, this risk can be mitigated through the implementation of a personal protection system. The design ensures that unauthorized personnel cannot enter the tunnel, and strict access control management further minimizes the possibility of public access to certain areas of the tunnel.

Another abnormal condition that may occur during accelerator operation is the sudden loss of all electrons/positrons at a specific point within the accelerator components. Although such incidents are rare, they are of utmost concern and are addressed by the machine protection system (MPS). The MPS is designed to detect such situations, and even in the event of such an occurrence, the radiation level resulting from it remains very low due to the enclosure of the machine within the tunnel, located 100 meters underground.

## 10.6 Fire Safety

Fire protection is primarily focused on fire prevention and complemented by firefighting measures. The approach encompasses four key steps: prevention, cutoff, extinguishing, and exhaust. Further details regarding these measures can be found in the previous chapter.

The fire protection measures for buildings include provisions for firefighting. Water-based firefighting systems are employed, along with strategies for smoke control. Additionally, a ventilation and smoke exhaustion system are implemented to manage the spread of smoke and provide a safe environment.

Power supplies designated for fire protection purposes are categorized as Grade II loads. Automatic switch-over devices with dual power supplies are installed at the location of the last-level power distribution, ensuring uninterrupted electrical power for fire prevention equipment. All electrical equipment utilized for fire prevention is equipped with independent power supply circuits, and these circuits are constructed using fire-resistant cables.

To facilitate safe evacuation, emergency lighting and signs are installed in each evacuation passage, staircase room, and exit. Emergency lights are equipped with battery backup systems, ensuring their functionality during power outages or emergencies.

## 10.7 Cryogenic and Oxygen Deficiency Hazards

### 10.7.1 Hazards

Cryogens encompass substances such as liquid nitrogen and liquid helium. Hazards associated with cryogenic materials include the risk of cold burns (frostbite), explosions, and oxygen deficiency leading to asphyxiation. Oxygen deficiency hazard (ODH) occurs when the oxygen content drops below 19.5% and can result in asphyxiation. It is important to note that certain safety rules and procedures applicable to cryogens can also be relevant to confined spaces, even in the absence of cryogenic substances.

#### 10.7.2 **Safety Measures**

Personnel Controls: It is important to carry out targeted safety education and provide technical training for staff in accordance with cryogenic and oxygen deficiency laws and regulations. Eye, hand, and body protection should be provided to prevent potential hazards. Implementing a two-person rule or a three-person rule in cryogenic and oxygen deficiency work areas is advisable. Personnel entering ODH areas must undergo medical certification and training, with periodic renewal of ODH certification. Adequate ventilation should be ensured in these areas, and personnel can carry oxygen monitors in addition to the permanently installed monitors and warning signs.

#### 10.7.3 **Emergency Controls**

Developing an accident emergency rescue plan for cryogenic and oxygen-deficient working areas and conducting regular emergency rescue drills are of utmost importance.

## 10.8 Electrical Safety

Having a thorough understanding of safety protocols and recommended practices is essential for safeguarding against electrical hazards, such as electrical shock, burns, or other delayed effects. The implementation of an effective electrical safety program is crucial as it not only promotes awareness of potential hazards but also provides valuable information to individuals working with electrical equipment. Additionally, conducting "lock-out tag-out" training ensures that personnel are well-informed and adequately trained to carry out their assigned electrical tasks safely.

## 10.9 Non-ionization Radiation

Since 1974, the International Radiation Protection Association (IRPA) has been actively addressing the issue of protection against non-ionizing radiation (NIR). As a result, many countries have entrusted their radiation protection authorities with the responsibility of ensuring protection against hazards associated with the use of NIRs.

Various forms of non-ionizing radiation, such as electromagnetic fields, microwaves, RF (Radio Frequency), lasers, and strong magnets, have the potential to pose health risks. However, our understanding of some of these hazards remains limited and controversial.

RF and microwave radiation can be absorbed by body tissues, leading to an increase in temperature. In the case of the CEPC, high-power RF systems utilizing klystrons are employed. As these klystrons are connected to the RF cavities through waveguides, the primary concern is limited to potential leakage radiation. During startup, strict monitoring of RF power levels is planned to be implemented. It is worth noting that historical evidence indicates that the primary hazard associated with RF systems is X-ray exposure rather than microwaves.

The "GBZ1-2002, Hygienic Standard for Industrial Enterprise Design" regulation establishes the contact limit for high-frequency radio waves in working areas. This regulation ensures that adequate measures are in place to mitigate potential risks arising from exposure to high-frequency radio waves in occupational settings.

## 10.10 General Safety

### 10.10.1 Personal Protective Equipment

The CEPC will prioritize the provision of necessary, reliable, and appropriate personal protective equipment (PPE) for its staff. Regular checks and maintenance will be conducted to ensure the effectiveness of the PPE. The range of PPE includes, but is not limited to:

- Safety Helmets (hard hats)
- Safety Shoes: Puncture-proof boots, insulated boots, oil-proof boots, acid-resistant boots, and anti-static shoes
- Eye and Face Protection: Goggles and protective covers
- Hearing Protection: Earplugs, earmuffs, anti-noise caps, and ear muffs
- Respiratory Protection: Dust masks, gas masks, and air packs for firefighters
- Hand Protection Equipment: Chemical-resistant gloves, insulating gloves, handling gloves, and fire-resistant gloves
- Workwear: Work clothes (long sleeves), special work clothes (such as anti-static overalls, wading overalls, waterproof overalls), and easily identifiable nighttime reflective vests.

For radiation-related operations:

- Radiation workers will be equipped with personal dosimeters, personal dose alarms, and various types of lead protective equipment, including lead aprons, lead glasses, and lead gloves. Those entering areas with high radioactivity will also wear disposal clothing.

Regarding traffic and vehicle-related PPE:

- Special transport operators will be provided with dust protection facilities and anti-noise equipment.

By providing comprehensive PPE, CEPC aims to ensure the safety and well-being of its staff in various operational scenarios.

### 10.10.2 Contractor Safety

CEPC is a large-scale construction project involving a substantial workforce, including numerous contractors and subcontractors. To ensure smooth operations, safety, and minimal environmental impact, it is imperative that all tasks are completed in adherence to established policies and regulations. In order to achieve this, the following measures should be implemented:

1. Familiarization with Policies: All contractors and subcontractors must be well-versed in the project's environmental protection, safety and health policies, plans, site safety regulations, emergency handling and evacuation procedures, and other relevant safety regulations.

2. Compliance and Qualifications: Contractors and subcontractors must comply with the supervision and management of CEPC's security personnel and possess the necessary qualifications to carry out their designated responsibilities within the project. They should adhere to national environmental protection, safety, and health regulations and standards.
3. Responsibility System: Safety management personnel will establish a hierarchical responsibility system that encompasses all levels of personnel involved in the project. This ensures accountability and promotes a culture of safety throughout the construction site.
4. Equipment Safety: All equipment brought to the site by contractors and subcontractors must undergo thorough inspection to ensure its safety, reliability, and proper working condition. Any equipment found to be faulty or unsafe should not be used and should be promptly repaired or replaced.
5. Subcontractor Responsibility: Subcontractors bear the responsibility for the safety and health of their employees. They should take proactive measures to prevent unsafe work practices and provide adequate training, supervision, and necessary personal protective equipment (PPE) to their workforce.

### 10.10.3 Traffic and Vehicular Safety

All vehicles involved in CEPC operations must be equipped with seatbelts to ensure personnel protection. Additionally, special transport operators will require dust protection facilities and anti-noise equipment to mitigate potential hazards.

The project's traffic tools encompass a range of vehicles, including new energy cars, trams, lifting systems (such as shaft hoist winches), accident rescue vehicles, fire trucks, engineering and material handling vehicles, forklifts, crane trucks, and earth-moving vehicles. Traffic and vehicle safety considerations revolve around four aspects: station and hub safety design, vehicle safety control, traffic facilities, and traffic safety management.

A comprehensive traffic safety control system will be established, including a signal control system that enables coordinated management of vehicles with similar or different road rights. A dispatch and command center will be utilized for the supervision, command, dispatch, and emergency management of all types of vehicles.

CEPC's route and station design prioritize safety functions such as well-maintained road surfaces, clearly marked entrances and exits, designated safety exits, appropriately managed intersections, rights management protocols, and designated emergency evacuation sites.

Transportation facilities within CEPC will guarantee the personal safety of vehicle operators and other traffic participants. This will be achieved through the installation of traffic signs, the implementation of isolation facilities, effective access control management, the incorporation of on-board safety facilities, and the provision of tire chain facilities when necessary.

### 10.10.4 Ergonomics

Certain activities carried out within CEPC, such as working in restricted spaces, adopting awkward or static postures, performing repetitive motions, handling vibrating tools, or engaging in forceful exertions, have the potential to cause injuries and diminish worker effectiveness.

To mitigate ergonomic risks, it is crucial for workers and supervisors to proactively assess activities and evaluate workplace conditions. They are encouraged to actively engage with their ES&H coordinator or contact the program manager for support and assistance. This can range from informal consultations to formal evaluations, ensuring that appropriate measures are taken to address ergonomic concerns and promote worker well-being.

# 11 Project Planning

Leading nations worldwide prioritize the advancement of significant research infrastructure, and international strategic dialogues have been convened to chart the course for the development of particle physics and forthcoming large scientific facilities. The deliberation document of the 2020 European Strategy for Particle Physics Update (ESPPU) explicitly underscored that "an electron-positron Higgs factory stands as the top-priority next collider"[1], encouraging the European particle physics community to explore "the realization of a proton collider at the highest attainable energy." This consensus closely aligns with the CEPC roadmap, underscoring the scientific significance, technological innovation, and forward-thinking strategy of the CEPC initiative.

Meanwhile, in the United States, using Higgs as a tool for discovery is one of the five science drivers identified by the P5 in 2014. Snowmass 2021 summary reaffirmed that those 5 drivers are still appropriate for the next decade and emphasized the need for US participation in the construction of any $e^+e^-$ Higgs factory that has a firm commitment to go forward.

## 11.1 Project Management

### 11.1.1 Project Development and Organization

The CEPC project development plan comprises three distinct stages, each accompanied by a corresponding organizational structure and management system. In this section, we will delve into the primary objectives, organizational framework, and management strategies for each of these stages.

While the CEPC project originates in China and will be hosted there, its vision extends to becoming an international endeavor. The project's organization and management will mirror the active involvement of international stakeholders. The vital role played by international participation was evident in the successful delivery of the CEPC conceptual design report, technical design report, and is expected to gain even greater significance in the project's future phases.

### 11.1.2 Stage I

The primary objectives for this stage include: (1) finalizing the accelerator design while conducting R&D on advanced technologies and components; (2) cultivating essential expertise and assembling a proficient scientific and engineering team to ready CEPC for construction; (3) preparing for site selection and formulating civil enegineering designs; (4) securing government approval for the commencement of construction.

The structure of the CEPC Study Group, as depicted in Figure 11.1.1, has remained unchanged since its inception on September 13, 2013, during the CEPC-SPPC inauguration meeting held in Beijing, China. The Institutional Board (IB) comprises representatives from universities and research institutes. The IB formulates regulations, establishes policies, and appoints the Steering Committee (SC), which is responsible for overseeing the CEPC project's strategic direction and making key executive decisions. The SC, in turn, designates Project Directors who are responsible for the management of CEPC's operations and activities.

Within the CEPC framework, three divisions—Accelerator, Detector, and Theory—are led by groups of conveners. Additionally, the hosting laboratory, IHEP, provides administrative staff and experienced managers who engage with government bodies and funding agencies and report to the Chinese Academy of Sciences. This organizational structure, while straightforward, has proven to be highly effective in driving CEPC forward.

The CEPC International Advisory Committee (IAC) was established in 2015 with the purpose of providing guidance to the CEPC Study Group and steering the overall development of the project. Comprising eminent experts in accelerator, detector, and particle physics, as well as former leaders of major high-energy physics laboratories and international collaborations, the IAC has proven to be an invaluable and highly successful resource. In further support of the Technical Design Report (TDR) for the CEPC accelerator, an International Accelerator Review Committee (IARC) was constituted to offer counsel on all aspects related to accelerator design and R&D, encompassing considerations for the Machine-Detector Interface (MDI) and upgrade capabilities. This committee has played a pivotal and effective role in the project.

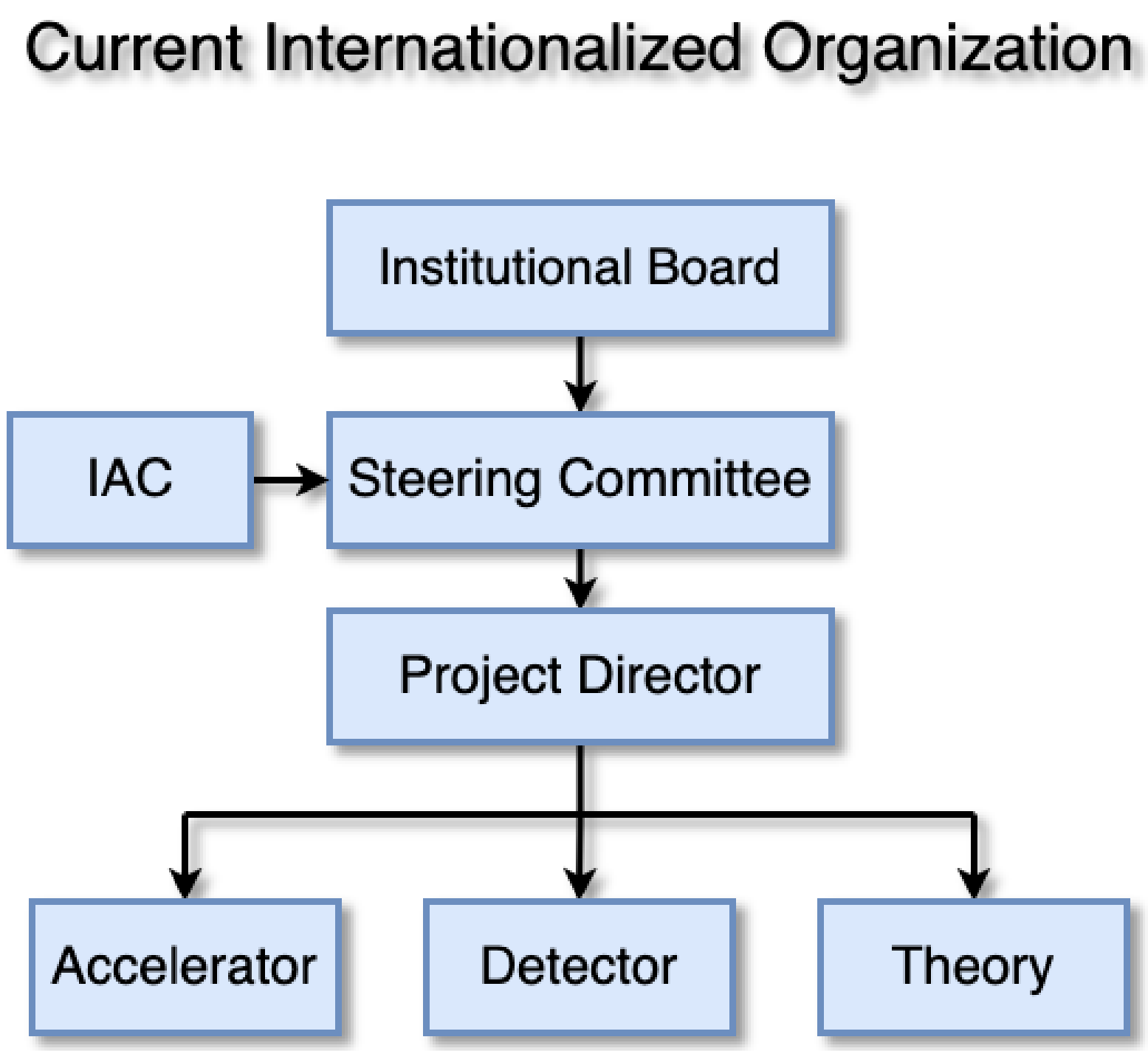

**Figure 11.1.1:** Current CEPC Organizational Chart.

International collaboration stands as a cornerstone of the CEPC project. Robust partnerships have been forged between the domestic CEPC study groups and numerous international research institutions. These collaborative efforts have not only identified critical challenges in CEPC detector design and physics investigations but also initiated dedicated R&D programs to address them effectively.

The CEPC team finalized the Concept Design Report (CDR) for both the $e^+e^-$ collider and the detectors in 2018. Subsequently, they have undertaken numerous refinements and optimizations to the CDR design, which are now being incorporated into the CEPC Technical Design Report (TDR), scheduled for completion in 2023. The team remains on course to conclude the Engineering Design Report (EDR) phase with necessary technological and engineering preparations by 2025. The CEPC Study Group has devised a strategic plan aimed at securing approval from the central government within the 15th Five-Year Plan (2026-2030) and aims to commence construction as early as 2027.

### 11.1.3 **Stage II**

#### 11.1.3.1 ***Organizational Chart in Stage II***

Following the approval of the CEPC project in Stage I, CEPC will initiate the formation of international collaborations pertaining to the CEPC $e^+e^-$ collider and the accompanying experiments. Simultaneously, construction of the CEPC accelerator will commence and be subsequently finalized. In parallel with the accelerator's construction, the experimental collaboration will undertake the development of two advanced, general-purpose detectors. The overall CEPC complex will then undergo the commissioning phase.

In Stage II, a revised organizational structure has been proposed to facilitate and embrace future international involvement, as depicted in Figure 11.1.2. This structure has been shaped by the discussions held during the 2018 CEPC workshop and takes into account the insights of the International Advisory Committee. Within this framework, all constituent elements will be designed to incorporate international participation.

The Institution Committee is entrusted with the formulation of bylaws and pivotal decisions concerning organizational matters. National representatives serve as liaisons with the national funding agencies, and the respective institutions they represent hold positions within the Institution Committee. The Coordination Committee holds responsibility for the oversight of CEPC and the enactment of executive decisions. Key management appointments will be proposed by the Coordination Committee, subject to ratification or rejection by the Institution Committee.

Assisted by the Accelerator Review Committee and the Detector R&D Committee, the Project Director assumes the role of coordinating activities and plans across each group.

Accelerator construction can commence promptly following approval. Simultaneously, plans are in place for issuing a call for detector Letters of Intent. Subsequently, two detectors will be chosen, and International Collaborations will be established in alignment with these selections. These collaborations are expected to furnish their Detector Technical Design Reports within two years from the commencement of accelerator construction.

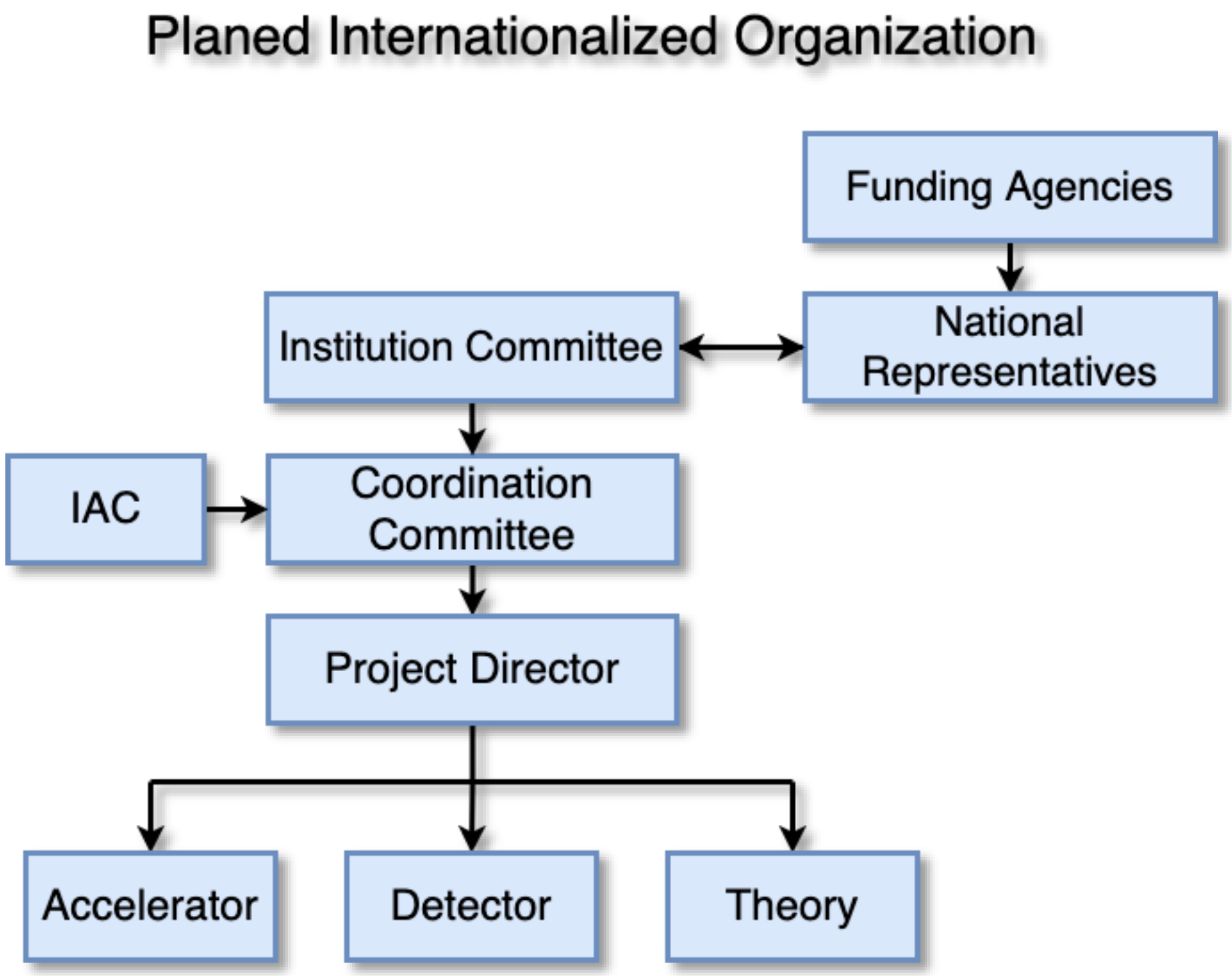

**Figure 11.1.2:** Planned CEPC Organizational Chart in Stage II.

### 11.1.3.2 *Roles and Responsibilities for the Planned Organization in Stage II*

The organizational structure of the CEPC is stratified into multiple tiers for R&D, as well as for construction. Each level delineates specific responsibilities and authorizes individuals or groups to initiate and document actions.

#### *11.1.3.2.1 Representation in the Institution Committee*

The Institution Committee (IC) serves as the highest governing body and supervisory authority for the CEPC project, comprising international representatives from their respective funding agencies. These funding agencies will collaboratively establish an agreement delineating their authority and responsibilities. The CEPC agreement is designed with a fixed term, encompassing an approximate construction period of 8 years, followed by 13 years of operation. This fixed term can be extended by mutual agreement among the Institution Committee members.

Withdrawal from the project should follow the procedural rules after the agreement takes effect. Generally, an application should be submitted at least six months or one year in advance and executed after a vote by all committee members. If necessary, a vote by all committee members may be conducted to enforce the mandatory withdrawal of a member unit. This stringent approach to withdrawal is deemed essential to ensure the stability of membership among international states, enabling prudent long-term planning for the project.

The Institution Committee is entrusted with several key responsibilities, including the formulation of regulations for the CEPC study group, decisions related to accepting or removing cooperative units and members, and the election of Coordination Committee members. Each member unit is entitled to have two official delegates, and in cases where a member state boasts a significant number of participating units, two representatives may be appointed to serve as facilitators. These facilitators are integral members of the Institution Committee and are accountable to their respective funding agencies.

Of the two delegates, one is expected to be a particle physicist, thereby contributing a scientific perspective to the CEPC's work, while the other represents the relevant government administration. The Institution Committee exercises its functions by convening institutional committee meetings, where each member unit or state wields a single vote.

The Institution Committee is expected to have a chairperson who assumes responsibility for the committee's operations. The chairperson will be elected by the committee members through a specified election process designed for the Institution Committee.

For the election to be valid, the representatives attending the Institution Committee meeting should number no less than half of the total members of the Institution Committee. Candidates for the position of chairperson are to be anonymously recommended by the representatives of the Institution Committee. Candidates must also be members of the Institution Committee, and each representative is entitled to nominate one candidate at most. Candidates who receive nominations from over half of the representatives will be elected directly.

In cases where only one candidate is nominated but does not secure more than half of the votes, a second round of voting is required for that candidate to receive over half of the votes before being elected. If this threshold isn't met, the initial round of elections will be deemed invalid, necessitating a re-election. When multiple candidates are nominated, the two candidates with the highest number of nominations will proceed to a vote, with each representative selecting at most one candidate. Candidates who obtain more than half of the votes will be elected directly. If the majority threshold isn't achieved, the election round will be considered invalid, and a repeat election will be arranged.

The proposed term of chairperson can be 4 to 5 years, with the possibility of re-election for subsequent terms.

The committee also includes a vice-chairperson (IC Deputy) who collaborates with and supports the chairperson in their responsibilities. The vice chairperson of the Institution Committee is nominated by the IC chairperson and officially appointed following approval by the Institution Committee meeting.

For new member units seeking to join, they are required to submit an application to the chairperson of the IC. Membership is granted only upon approval by the IC during their meetings. In the case of international representatives, should a member unit express their intention to withdraw from the committee, it is their responsibility to inform the chairperson of the Institution Committee. Subsequently, the chairperson will then notify all members of the Institution Committee regarding the withdrawal request.

##### *11.1.3.2.2 Coordination Committee*

The Coordination Committee (CC) serves as the executive body primarily responsible for the day-to-day management of the CEPC project, as well as critical matters in the decision-making and implementation processes. The Coordination Committee comprises the chairperson of the Coordination Committee, the chairperson and vice chairperson of the Institution Committee, in addition to some number of members of the Coordination Committee.

The chairperson of the Coordination Committee is elected by the Institution Committee using the Coordination Committee's designated election process. The chairperson holds their position for a five-year term, with the possibility of being re-elected for subsequent terms. Meanwhile, the members of the Coordination Committee are nominated by the chairperson of the Coordination Committee and subsequently approved by the Institution Committee during their meetings. For such approvals, it is required that the representatives attending the meeting amount to at least half of the total members of the Institution Committee, with votes received not being less than two-thirds of the members present.

A similar procedure is followed in instances involving the expansion or reduction of members in the Coordination Committee. Any significant issues that concern all unit members of the project must be presented by the Coordination Committee to the Institution Committee for ratification.

During the R&D phase, the International Advisory Committee (IAC) plays a pivotal role in offering expert insights and guidance throughout the management and execution processes of the Coordination Committee. Comprising distinguished scientists and accomplished laboratory and project leaders with extensive experience in project management, planning, and strategy execution, the IAC is an invaluable resource. Members are appointed by the Institution Committee solely based on their scientific eminence, without regard to their nationality. Furthermore, some members are elected from Non-Member States [2].

The IAC is tasked with providing technical advice, fostering engagement with international academic institutions and organizations, informing strategic decisions, facilitating employment placements, and undertaking various other activities related to the project. It serves as a conduit for establishing connections between the project and organizations worldwide, creating opportunities for global collaboration.

##### *11.1.3.2.3 Project Director*

The Project Director assumes full responsibility for orchestrating the scientific research activities within the project. The Project Director is nominated by the chairperson of the Coordination Committee from among its members and subsequently requires approval and appointment by the Institution Committee during their meetings. For such appointments, it is necessary that the representatives participating in the meeting number at least half of the total members of the Institution Committee and that the votes received are not less than two-thirds of the attending members. This same procedure is also followed for the replacement of the Project Director.

To aid in the Project Director's tasks, 1-2 deputy Project Directors may be nominated. These deputies are nominated to assist the Project Director in their duties. The chairperson of the Institution Committee is responsible for notifying all members of the Institution Committee after obtaining approval from the Coordination Committee.

The Project Director holds the responsibility for formulating preliminary research activity plans, nominating conveners for each study group, and executing these plans once they gain approval from the Coordination Committee. Furthermore, the Project Director oversees the coordination of physics design, technical innovation, and prototype development during the R&D phase.

To enhance the efficiency and effectiveness of scientific policymaking, project design, construction, and operation, two International Committees have been established. The first is the Accelerator Review Committee, which offers guidance on all matters related to accelerator design and R&D. The second committee, the Detector R&D Committee, assesses and approves Detector R&D proposals from the international community. This enables international participants to seek funding from their respective funding agencies and make substantial, sustained contributions to the project. The specific focus of each review committee may evolve as the project progresses through its engineering phases. These committees will serve as valuable sources of information and ideas for the Project Director's team, aiding in their management and coordination efforts.

*11.1.3.2.4Accelerator, Detector and Theory groups*

The CEPC R&D is presently overseen and directed by a team of eminent scientists, which includes academicians. These individuals have substantial involvement in experimental collaborations and projects such as BEPC, DayaBay, JUNO, LHC, HL-LHC, LEP experiments, SuperKEKB, ILC, and more. They possess extensive experience spanning project initiation, design, construction, operation, maintenance, management, and the enhancement of large-scale experiments.

The project study group is comprised of three main divisions: the accelerator group, the detector group, and the theory group, each led by respective conveners. These primary divisions are further organized into several secondary subsystems. This project study group holds the responsibility for various aspects, including technical design, installation, and commissioning, as well as defining and managing the technical scope, cost, and project schedule. Its role is to finalize the design and implement it in coordination with configuration management. The specific subsystems within the group are tasked with creating detailed integrated plans essential for efficient task completion while ensuring alignment with the overarching project plan.

### 11.1.4 **Stage III**

#### 11.1.4.1 ***Organizational Chart in Stage III***

In this stage, both the hosting laboratory and a dedicated CEPC Council will have been established. As the CEPC accelerator progresses into the commissioning and regular operation phase, the experiments will install detectors, commence their scientific investigations, and shift their focus towards generating valuable scientific output. Simultaneously, the management team formulates plans and executes strategies for future upgrades and the optimization of the entire complex.

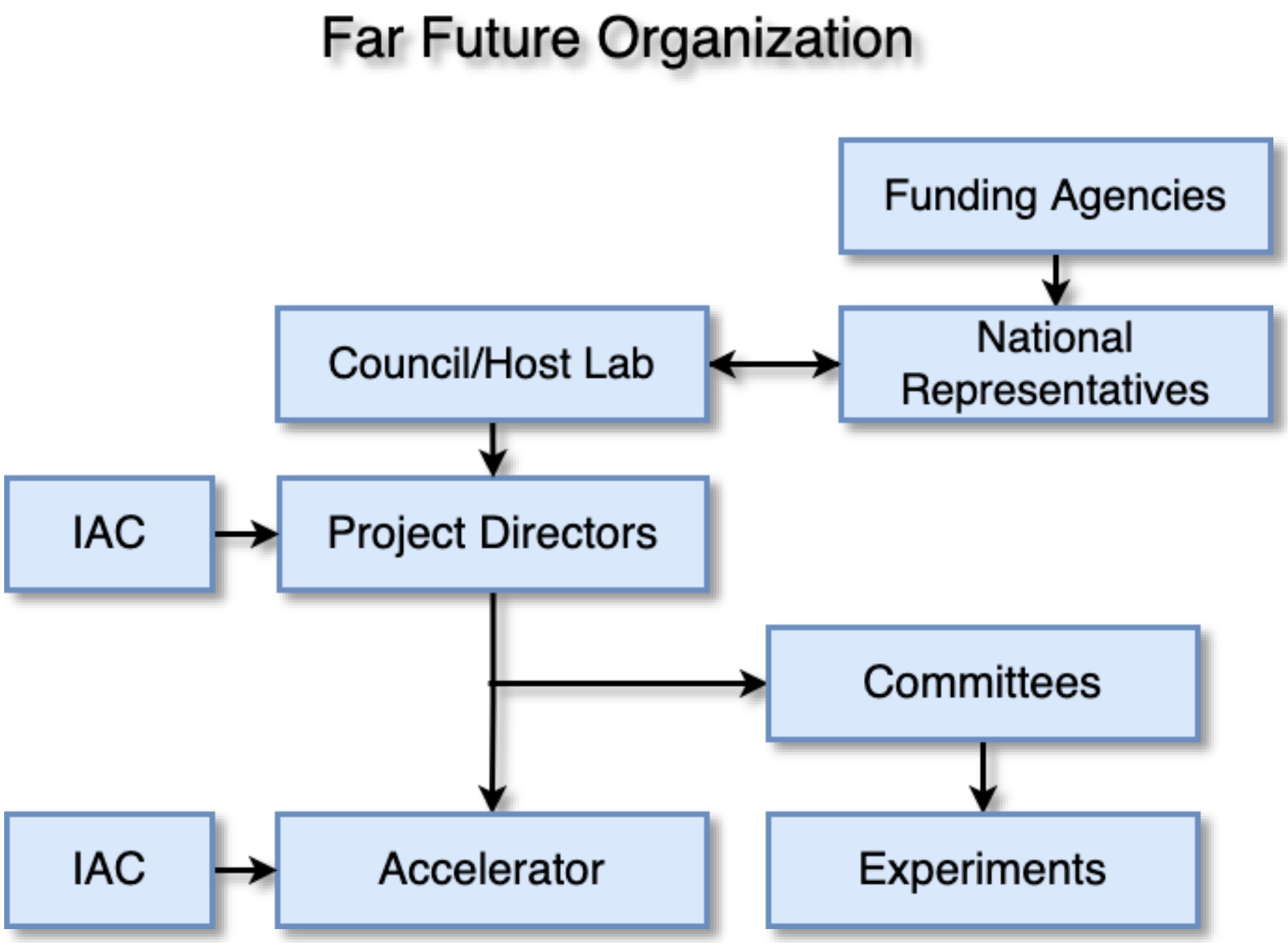

**Figure 11.1.3:** Far Future CEPC Organizational Chart in Stage III.

The organizational structure will naturally evolve over time. As the construction phase concludes, the organizational structure will naturally expand to encompass a council representing the participating countries, along with host laboratory management, which will provide supervision and oversight for the project.

### 11.1.4.2 *Roles and Responsibilities for the Future organization in Stage III*

Organizational units will continually adapt to evolving circumstances. The revised units and their functions are outlined as follows:

#### *11.1.4.2.1 Representation in the Council and the Host Lab*

The CEPC will establish a robust Council as its highest governing body, holding ultimate authority over the project and responsibility for pivotal decisions. The Council is composed of international representatives designated by the funding agencies. In cases where a country is supported by multiple funding institutions, two representatives may be elected to participate in the Council. The Council exercises control over project activities in scientific, technical, and administrative matters, including the approval of strategic programs, budget adoption, expenditure review, determination of necessary staffing levels, and more. Council delegates should possess the requisite stature to make timely decisions.

The Council will elect a president and two vice-presidents from its members, with each serving a predetermined term. The number of participants must not be fewer than half of all Council member states and the election requires approval from at least two-thirds of the participants. Naturally, as the host country of the project, the selection of the president should prioritize factors related to the construction site to facilitate multi-party coordination.

The CEPC's operations will be overseen by the Council member states, with each Council member state appointing two official delegates, who may be accompanied by advisers during Council meetings. One of the delegates should be a particle physicist, while the other should represent the government. Each Member State holds a single vote, and the Council members should convene an international conference annually.

As the project's construction site, the host laboratory assumes full responsibility for organizing and executing the project. This responsibility encompasses reviewing the comprehensive department establishment plan, implementing management regulations, formulating task and budget plans, supervising and assessing overall progress and target attainment, and addressing significant issues that arise during the implementation process, among other tasks. As the designated national laboratory for the future, the chairperson of the host laboratory, in practice, is typically expected to be of Chinese nationality.

The project will adhere to the safety standards and environmental regulations of the host laboratory. It is imperative that operational activities follow established procedures and that all project facets comply with both the host's regulations and the long-term regional development plans.

##### *11.1.4.2.2 Project Directors*

The CEPC will implement a project director responsibility system, which includes a project director, chief engineer, chief craftsman, chief economist, and their corresponding deputy positions. The daily management structure is organized around the core team comprising the project director, deputy project directors, chief managers, and the director of the project office.

The primary duties of this core team encompass organizing the project to meet predefined objectives within specified timelines and budget constraints. This involves task and fund allocation, the formulation of phased implementation plans and goals, coordination of project performance, verification of the status of plan and goal completions, and the formulation of solutions for significant issues that arise during the implementation process, which are then submitted to the host laboratory for approval.

Furthermore, the International Advisory Committee, as previously described, will offer technical insights, advice on collaboration, and guidance on membership to the project directors.

##### *11.1.4.2.3 Accelerator*

In the operational phase, the accelerator primarily focuses on maintenance and operational support. This encompasses tasks such as overseeing technology management, securing funding, ensuring quality standards, and monitoring progress. Any issues arising during the process should be escalated to the project directors for approval.

##### *11.1.4.2.4 Experiments*

Various large-scale experiments conducted through international collaboration, utilizing the project's infrastructure, will attract domestic and international particle scientists for joint scientific research. To facilitate the management of day-to-day collaboration affairs, a steering committee will be formed to support the spokesperson.

Furthermore, collaboration regulations will be established to govern the management methods and decision-making mechanisms of the collaboration. These regulations will be subject to discussion and approval during collaboration meetings. Additionally, several committees will be set up to guide future work within the experimental collaboration. These committees include the technical committee, operational committee, international conference committee, publishing committee, physics committee, and finance committee.

Since CEPC is a fundamental scientific research project, data obtained will be shared among all participants, and physical results will be officially published. Any new technologies independently developed by a research institution during the process will remain the property of the respective research institution.

### 11.1.5 Environment of Investment and Management

While aiming to establish effective supervision, we will create a comprehensive governance and operational framework around CEPC's large research infrastructure. This framework will feature a reasonable and comprehensive institutional arrangement, an efficient governance system, and clear division of responsibilities. There are three key aspects to this:

1. **Facilitation of Funds and Goods:** To support the inflow of funds and goods from funding agencies for large research infrastructure, we will adopt models like CERN in Europe and the Hainan Free Trade Zone in China. Funding agencies will be encouraged to engage and participate. We will establish a streamlined licensing process and make it convenient for funding agencies to get involved. These agencies will pledge to meet specific requirements and submit relevant materials for documentation before joining collaborative research. To ensure a convenient experience for various funding agencies, the construction and operation of the CEPC project will apply for special preferential policies from the host, covering factors such as acquisition, standard formulation, access permits, and other preferential policies.
2. **Liberalization and Facilitation of Funds and Goods:** The construction and operation tasks will be opened up in phases, with an account platform system established for the facilitation of cross-border goods and fund settlement. The smooth flow of funds between the large research infrastructure area of CEPC and overseas locations will be promoted systematically. Cooperation agreements with trade ports will be put in place to simplify customs procedures. Funded items will be shipped from overseas, processed and consolidated through trade ports, and then transported to the CEPC comprehensive area. The duration and location of goods storage at the trading port will be negotiated as necessary.
3. **Property Rights Protection:** We will create a property rights protection system to safeguard the funds, goods, intellectual property rights, and corresponding rights of funding agencies and individuals in accordance with the law. This system will clarify the allocation of rights among multiple parties during collaborative research and the transformation of outputs. It will also enhance penalties for property rights infringement and improve mechanisms such as credit classification supervision for various entities in the property field. Over time, we will work to establish an open and transparent process regulatory environment.

These measures will help ensure effective governance and operations around CEPC's large research infrastructure, encouraging international participation and making the collaboration process more efficient and convenient.

### 11.1.6 Personnel Policy

In line with the development requirements of the comprehensive governance zone surrounding the large research infrastructure of CEPC, we will establish a robust talent attraction and training system. This talent attraction strategy primarily encompasses both direct employment [3] and the secondment of personnel from participating institutions, offering them flexibility and ease of mobility. The CEPC project is committed to implementing an inclusive talent acquisition and residency policy, fostering an environment conducive to attracting and retaining top talent.

The organization will actively bring in scientific researchers and highly skilled professionals, while establishing a comprehensive talent training and support system. This system will encompass the creation of a fair and rational mechanism for talent recruitment, recognition, utilization, and competitive compensation. Additionally, there will be enhancements to the international talent assessment framework.

Looking ahead, CEPC will leverage its existing talent pool to expand both in terms of scope and headcount. Direct employment will be a means of ensuring a consistent supply of human resources throughout the project's duration. Participating organizations will have the opportunity to devise their long-term human resource strategies, including seconded personnel within the CEPC international collaboration as a valuable resource pool. The formation of a robust and stable central team, operating within a straightforward and efficient management structure, is a realizable goal.

We will enhance the talent service management system to facilitate the seamless sharing of work permits, visa information, residence details, and collaborative inspections. Moreover, talent service centers will be set up to offer a comprehensive array of services covering work, employment, education, and daily life support for foreign scientific researchers, professional and technical experts, and their families, thereby safeguarding their lawful rights and interests.

As for participants in the international collaboration, the responsibility for managing employment benefits and social security will rest with the participating organizations. When managed effectively, this approach will ensure the stability and continuity of the talent pool.

The organization will pursue more favorable entry and exit management policies, including the introduction of renewable long-term visas or residence and work permits for visitors and their accompanying family members. Additionally, efforts will be undertaken to streamline the visa application process and to establish a dedicated green channel for the CEPC delegation to expedite their visa applications, ensuring a seamless and efficient customs clearance process for high-level foreign talents.

### 11.1.7 **Management Tools**

In terms of selecting management tools, we will develop a comprehensive CEPC management system tailored to the specific requirements of the CEPC project. In doing so, we will draw inspiration from two key systems: the ARP (Academia Resource Planning) system utilized by the Chinese Academy of Sciences (CAS) and the Material Management System of the East China Academy. They are described as follows:

- The Chinese Academy of Sciences' next-generation ARP system comprises six central applications: the human resources system, a comprehensive financial system, a scientific research project system, a scientific research condition system, an electronic document system, and an international cooperation system. These core applications serve as the foundation for the system, and it further extends to encompass a range of microservices tailored to address the specific business management requirements of the academy.

- The Material Management System at the East China Academy primarily encompasses five core modules: an e-mall for production and office materials, management of engineering project equipment and materials, asset management for equipment, system administration, and a task center.

The CEPC management system will cover various aspects, including document management, human resource management, funding management, engineering management, and material management.

Central to this system will be project planning and execution management. We aim to apply innovative management principles and leverage cutting-edge information technology to enhance the administration of human resources, financial resources, and the allocation of resources such as scientific research infrastructure, while optimizing and expanding management processes. Our ultimate goal is to create an efficient management service information technology platform.

Simultaneously, the implementation of this platform will facilitate further advancements in CEPC management practices, continually enhancing the effectiveness and efficiency of our management efforts. This, in turn, will promote the maximization of benefits in terms of technological innovation and talent development.

## 11.2 Regulations for Procurement and Bidding

Equipment and service procurement play a crucial role in the establishment and ongoing operation of large research infrastructure. In the case of CEPC, the imperative is to secure top-tier products and services timely and cost-effectively, all at equitable and reasonable prices. The procurement process primarily centers around supplier sourcing, contract negotiation, and contract management.

Procurement management is intricately tied to the quality, cost, and progress of large research infrastructure construction and operation, leading to the establishment of a standardized process for equipment procurement management. To facilitate this process, a dedicated procurement management office and a team of administrators will be established to serve as the primary professionals for contract management, modifications, changes, and potential terminations.

The procurement management will place a strong emphasis on selecting suppliers offering the best overall value through competitive processes, promoting the use of competitive methods for contract agreements. The project also encourages procurement to incorporate commercial products and technologies through centralized government procurement. Furthermore, we advocate for continuous improvement methods and the initiation of innovative procedures to ensure procurement is both timely and cost-effective.

Procurement can be categorized primarily into two types: domestic procurement and import purchasing. Domestic procurement offers flexibility and straightforward procedures, with options including centralized government procurement, open tendering, or self-bidding. Import purchasing comes into consideration when foreign materials demonstrate lower prices, superior quality, excellent performance, and an overall cost advantage compared to domestic procurement.

## 11.2.1 **Domestic Procurement**

### 11.2.1.1 ***Scope of Bidding***

The project's construction and operation will establish the bidding scope in accordance with the Tendering and Bidding Law of the People's Republic of China (referred to as the Tendering and Bidding Law), the Government Procurement Law of the People's Republic of China (referred to as the Government Procurement Law), and the Notice of the General Office of the State Council on the Catalogue and Standards of Centralized Government Procurement of Central Budget Units (2020 Edition) (GBF [2019] No. 55) (referred to as the Catalogue and Standards of Centralized Government Procurement) [1]. Reference practices from the construction experiences of projects like BEPC, Daya Bay, CSNS, ADS, HEPS, and JUNO will inform the bidding process.

Tendering and bidding procedures will be implemented in compliance with the provisions of the Tendering and Bidding Law when the project (comprising the project itself, associated goods, and services related to project construction) meets the specified financial thresholds. Typically, this includes single contracts for construction, vital equipment, materials, goods, surveying, design, supervision, and related services. Furthermore, the tendering and bidding process will apply if the total estimated value of combined procurement contracts for the aforementioned projects meets or exceeds the specified standards. (Please note that this regulation became effective on June 1, 2018).

Items falling under the purview of the centralized procurement agency's procurement project directory, as specified in the government's centralized procurement directory and standards, shall be procured through the agency in accordance with the established regulations. Items listed within the departmental centralized procurement projects directory, as outlined in the government's centralized procurement catalogue and standards, will be procured following the relevant departmental guidelines.

Additionally, goods and services exceeding an individual or cumulative amount of more than RMB 1 million, along with projects exceeding RMB 1.2 million, shall adhere to the stipulations set forth in the Government Procurement Law. Open tendering must be utilized when the procurement of goods and services by the government surpasses an individual amount of more than RMB 2 million.

Project items concerning matters of national security, state secrets, emergency rescue and disaster relief, or those that fall under special circumstances, such as the utilization of poverty relief funds for work relief or the necessity of employing migrant workers, and are deemed unsuitable for the bidding process, may not be subjected to bidding in accordance with the provisions outlined by the State [2].

The project's construction will entail engagement with a multitude of procurement and outsourcing processing businesses. These businesses can be broadly categorized into the procurement and outsourcing processing of civil engineering, scientific equipment and instruments, materials, design, and supervision services, all of which will be acquired in accordance with the aforementioned requirements.

#### 11.2.1.2 *Organization Form for Bidding*

Self-bidding vs. Agent bidding:

As per Article 12 of the Tendering and Bidding Law of the People's Republic of China, the party inviting tenders (tenderee) retains the right to select a bidding agency to manage the bidding processes. It is not permissible for other entities or individuals to designate a bidding agency on behalf of the tenderee. A tenderee with the capability to create bidding documents and conduct bid evaluations is authorized to manage the bidding processes independently. No external entity or individual should compel the tenderee to employ a bidding agency. In cases where the law mandates a project to undergo the tendering process, the tenderee, if opting to manage the tender on its own, is required to report this decision to the relevant administrative and supervisory authorities for record-keeping.

IHEP possesses the capability for self-organized bid invitations. Since 2006, IHEP has independently managed its bidding processes, overseen by the Financial Assets Department. This department presently comprises five full-time personnel dedicated to bidding and procurement, two of whom hold professional qualification certificates for tendering, and two are qualified in legal matters. The CEPC project's approval will lead to a substantial increase in the number of staff dedicated to procurement management.

The institute has established a comprehensive set of guidelines to standardize bidding and procurement activities. Over the years, it has assembled a proficient team with expertise spanning technical, budgetary, financial, and engineering management disciplines. This multidisciplinary approach has been especially valuable in the construction of significant research infrastructure projects such as BEPCII, CSNS, JUNO, and others. IHEP has successfully conducted numerous self-organized bidding exercises, amassing substantial experience in tendering and bidding procedures.

#### 11.2.1.3 *Bidding Mode*

CEPC will adhere to the principles of fairness, impartiality, competition, and merit. The project will employ the bidding process for procuring essential instruments, equipment, materials, etc., with some exceptions for projects that do not meet the criteria for bidding, like domestic and international cooperation items or cases with a limited number of capable units.

CEPC is a project characterized by a high level of expertise, a substantial volume of non-standard equipment for accelerators and detectors, the requirement for specialized materials processing, and a limited number of research units both domestically and internationally that can undertake such tasks. The procurement of non-standard equipment, particularly those demanding high levels of expertise, primarily follows an invitation to tender approach, with supplementary domestic and international cooperation.

### 11.2.2 **Import Purchasing**

The CEPC project aims to improve the oversight of government procurement for imported products by implementing a dedicated procurement platform within the CEPC comprehensive governance zone. This functionality will be seamlessly integrated into the overall management tool. Purchasers will be required to file applications based on the specific circumstances of planned imported products. The project will further streamline the process for reviewing imported products, establishing scientifically grounded collection cycles and frequencies for import product applications. It will implement a "package" approval system and create a mechanism that prioritizes efficiency in screening and approval timelines.

Purchasers should adhere to the principle of "whoever purchases is responsible," fully embrace their primary responsibilities, enhance internal control management systems, conduct comprehensive market research and price assessments, and thoughtfully define the requirements for government procurement of imported products. Simultaneously, it is vital to reinforce the performance management of imported product procurement activities.

In the context of import purchasing, it is imperative to place special emphasis on in-depth supply market research and supply market analysis. A comprehensive evaluation should encompass factors such as source stability, interest rates, inflation rates, exchange rates, business cycles, industrial transfer trends, population dynamics, lifestyle changes, cultural norms, living and working environments, scientific and technological considerations, technological advancements, product and technological innovations, legal regulations, environmental sustainability, and social responsibility.

In terms of project construction and operation, the task force should proactively seek superior global supply resources, engage in ongoing management, and conduct qualification assessments of industry suppliers and agents. This should be coordinated with equipment installation and service timelines while implementing planned material and procurement management practices.

## 11.3 Financial Models

### 11.3.1 **Construction Costs**

Internationally, CERN employs a regional approach tied to GDP for funding, given the strong coordinating institution in the European Union. Similarly, ILC plans to rely on regional contributions with intricate multilateral negotiations within a defined contract. The 'share' funding model is commonly used for international research infrastructure projects, allowing countries to specify their expected contributions. These projects fall under the category of basic research, primarily funded through central government budgets. Due to their substantial size and long investment cycles, it's recommended that the costs be separately listed as a line item in the state’s budget. Furthermore, local governments are encouraged to increase their investment in project-related resources, such as land, staffing, infrastructure, and research and development funds, to support the establishment of project headquarters, core equipment, databases, and other essential components.

For CEPC, China will serve as the primary contributor, while other countries can participate as either members of regional consortia or by supporting specific activities or deliverables. Historically, the host country has been responsible for providing all civil engineering and a significant portion of accelerator and detector technologies. However, there is potential for other nations to explore opportunities to fund their own "high-tech" companies to supply a larger share of the required equipment.

International contributions to the overall project cost can take the form of funds or in-kind contributions. In-kind contributions have the potential to be converted into monetary value for cost ratio assessment. Different funding agencies may employ varying investment methods and proportional designs. The commitment and allocation of costs from all contributing agencies may change as the project progresses through construction, operation, upgrade, and retirement phases. If there are deviations from the project's implementation goals in later stages, there may be a heightened reliance on the host lab for fundraising arrangements.

To ensure a comprehensive evaluation, a system will be established that encompasses four dimensions: goods, financial, human, and academic contributions, along with their respective sub-indicators. This system will facilitate a methodical approach to calculate, summarize, and present relevant indicators. Participants will formalize binding agreements based on the physical donation model, clearly outlining the responsibilities, allocation, and management of technical risks, costs, and progress among collaborators. All of these considerations will be addressed during project approval negotiations.

For items that are not suitable for in-kind contributions, such as installation, testing, and support for the project team during hardware commissioning (pre-operations), funding will be provided through a Project Fund directly contributed in cash by the collaboration members [1]. The creation of this Fund serves to facilitate responses to emergencies and the management of issues as the project advances.

The construction of the project differs from typical infrastructure projects, as it carries a high level of unpredictability and risk. To maintain standardization while accommodating unforeseen expenses, it's essential to establish a dedicated budget category for "unforeseeable" costs. This allows for flexibility in process innovation and the supply of engineering technology.

During the project initiation phase, the CEPC organization will conduct a comprehensive budget estimate that spans the entire project lifecycle. This estimate should consider all future expenses, including labor costs, spare parts expenses, upgrade and renovation costs, and decommissioning costs throughout the operational phase.

### 11.3.2 **Operational Costs**

The project integrates aspects of infrastructure, science, technology, and processes, and will actively seek support from international collaborations, encompassing various forms of assistance, such as funding, goods, and more. We will coordinate domestic and foreign research institutions and enterprises to engage in collaborative research related to equipment production and installation during both the construction and operational phases.

In China the operational funds for large research infrastructure are provided by the National Development and Reform Commission (NDRC), and these funds are exclusively dedicated to local facility operations. In accordance with international norms, research personnel involved in the CEPC project will be obligated to contribute an annual maintenance and operation fee. These fees will serve as crucial sources of funding to cover the operational costs of the project.

## 11.4 Siting Issues

The technical criteria for selecting the site of the CEPC are approximately quantified as follows:

1. Earthquake intensity less than seven on the Richter scale.
2. Earthquake acceleration less than 0.1 g.
3. Ground surface vibration amplitude less than 20 nm at frequencies ranging from 1 to 100 Hz.
4. Presence of granite bedrock at a depth of approximately 50–100 meters, among other considerations.

The site selection process commenced in February 2015. Preliminary geological studies for potential CEPC site locations have been conducted in various regions, including Qinhuangdao and Xiongan in Hebei province, Huangling county in Shanxi province, Huzhou in Zhejiang province, Changchun in Jilin province, and Changsha in Hunan province. All of these sites meet the construction requirements for CEPC. An illustrative example of a site location and its geological conditions is presented in Figure 11.4.1.

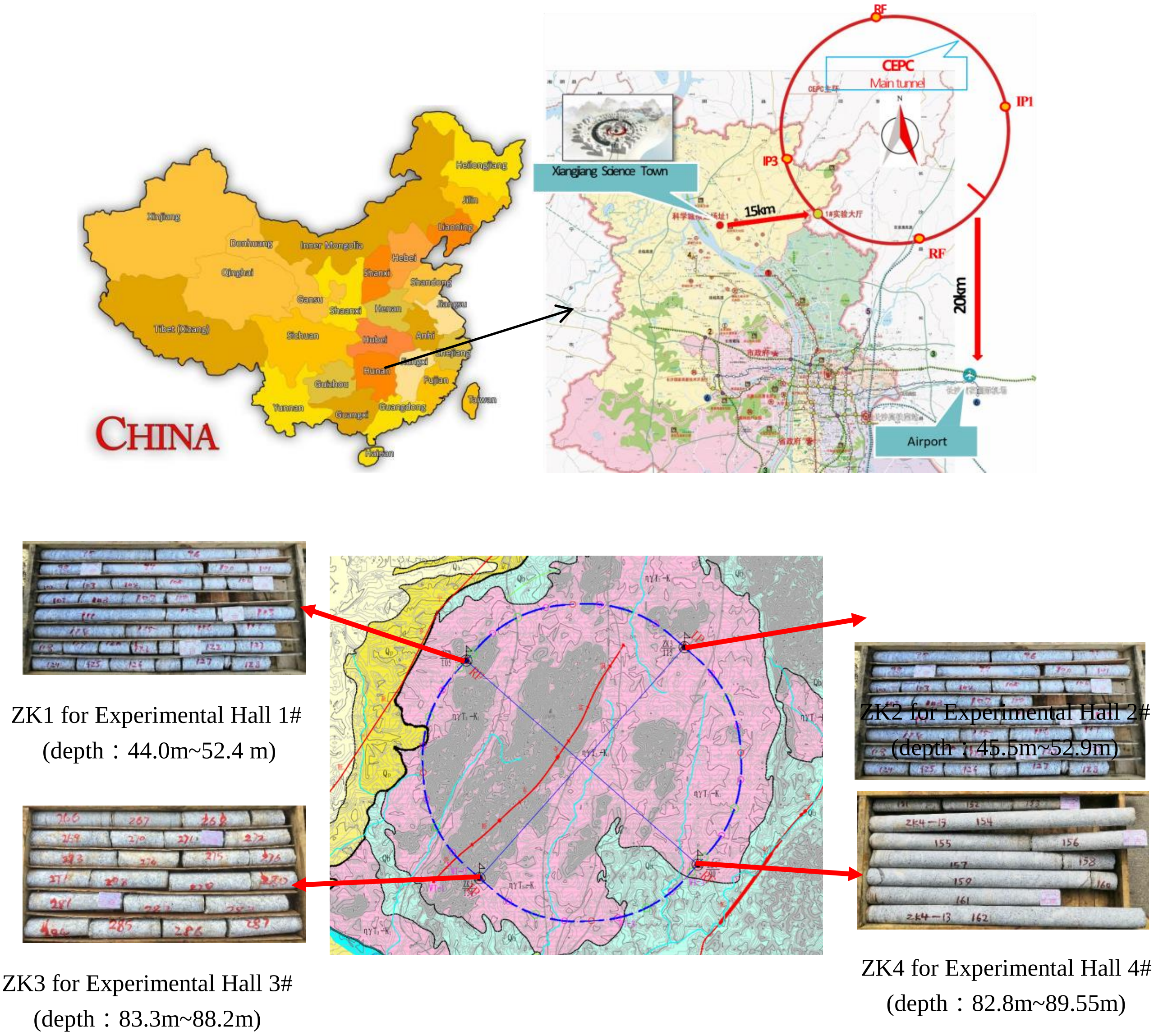

**Figure 11.4.1:** CEPC Changsha site, Hunan province and geological condition investigation (an example of the possible sites).

The purpose of this chapter is not to detail the process for conducting site selection or to outline the specific steps for choosing the final site. Instead, it aims to provide clear and validated criteria, along with a comprehensive evaluation of a site's suitability for the construction and operation of the CEPC.

### 11.4.1 Layout Principles

1. Fulfill the engineering geological conditions, encompassing site-specific geological attributes, geological structure, hydrogeological conditions, and the physical and mechanical properties of rock and soil masses.
2. Ensure compliance with safety standards, streamlined management, and accessible transportation for engineering facilities.

3. Minimize adverse impacts on the local ecological environment.
4. Establish convenient access to water and electricity.
5. Optimize tunnel routes to minimize construction duration and reduce overall engineering volume.
6. Accommodate transportation needs for experimental equipment.
7. Adhere to national regulations and specifications in force.

### 11.4.2 **Layout of Underground Structures**

The components of the CEPC accelerator primarily consist of a ring housed in a circular underground tunnel with an approximate total length of 100 kilometers. This main ring is complemented by two Interaction Region (IR) areas and two Radio-Frequency (RF) system areas, evenly distributed. The remaining sections of the main ring are further segmented into eight arc sections and four straight sections.

The linear accelerator (Linac) sits on the surface and is connected to one of the straight sections underground via transport lines. A total of eight permanent traffic tunnels are strategically positioned, four in the IR and RF areas and four in the straight sections, linking to the surface. For every 3 kilometers along the main ring, a cable and ventilation shaft is incorporated, and an auxiliary short tunnel is added every 1 kilometer.

### 11.4.3 **Site Infrastructure**

The surface buildings primarily encompass various facilities, including cooling, low-temperature, and ventilation facilities, as well as air compression systems, power transmission and transformation facilities, maintenance infrastructure, fire-fighting equipment, management buildings, and testing facilities.

A project as extensive and large-scale as the CEPC will significantly increase the demand on regional utility infrastructure capacity. Specifically, its operational electrical power requirement will range from 260 to 340 MW, contingent on the operation mode. Additionally, it is important to investigate how extreme winter and summer temperatures might affect the water-cooling systems for certain accelerator components.

The construction site must fulfill the prerequisites for water supply, rainwater and sewage drainage, electricity, heating, telecommunications, road infrastructure, and site preparation. This encompasses the provision of water supply, drainage, electricity, communication facilities, roads, gas supply, and heat supply, collectively known as the "seven supplies," along with site leveling, referred to as the "one leveling." These essential conditions enable the CEPC working group to promptly initiate construction. All infrastructure development must strictly adhere to regulations concerning natural ecological services, environmental quality and safety, and the responsible utilization of natural resources, ensuring an environment conducive to project development.

Drawing from the experiences of CERN and ILC, the hosting organization must create an environment conducive to the success of this major international research facility [1]. A science city will be developed around the research infrastructure, providing essential social amenities such as housing, schools, medical facilities, and shopping centers. The Science City will encompass three distinct urban functions: the scientific research core area, a public exchange area, and a living support area. Designed to accommodate over 10,000 individuals for both work and residence, its size will expand gradually. Consequently, the hosting organization must shoulder the crucial responsibility of establishing a policy framework with international appeal and flexibility. This should encompass housing security through purchase and rental options, as well as swift permit processing for spouses seeking employment opportunities, among other considerations.

### 11.4.4 **Environmental Impacts**

The collider will be installed at a depth of 50-100 meters under the ground surface, and the presence of a 10-meter-thick rock layer is adequate to effectively block and reduce radiation to negligible levels. This ensures that the surface area remains unaffected by immediate radiation exposure. Our commitment to safety extends to employing well-established design methodologies, honed over years of hands-on experience in constructing and operating colliders. These methods are tailored to safeguard both operators and equipment.

Environmental impact assessment and acceptance procedures will be conducted at the outset of the project's construction and during its operation. The oversight of environmental protection authorities will ensure comprehensive and reliable monitoring and supervision of both construction and operational phases. The selection of the collider's location will prioritize avoiding underground water sources to prevent any radiation pollution of these water sources. This will be achieved through thorough geological investigation and research.

### 11.4.5 **Risk Factors**

The primary risk during construction pertains to radiation exposure. The CEPC main ring is situated at a depth of 50-100 meters underground, and strict controls will be implemented to limit radiation exposure in areas where people live to less than 0.1 mSv per year.

During accelerator operation, while radiation levels within the tunnel may be relatively high, a comprehensive set of interlocking safety measures will be designed and continuously improved to prevent personnel from approaching the operational area. Once the power is turned off, radiation levels will quickly return to safe levels, enabling personnel to enter for maintenance and other tasks.

The CEPC project will have a total operating power of 260 MW, and full power operation will be achieved at the end of the 8-year construction period. To support the operation of the infrastructure, dedicated power lines will be established.

Local governments will incrementally increase electricity supply each year, well exceeding the CEPC's consumption needs. It's important to note that during the later years of operation, when the Collider operates in Z and W modes, the power requirements will be significantly reduced.

The investment and financing risks during CEPC's operation primarily revolve around operating costs. First, the Ministry of Finance commits to fully supporting the annual operating costs of nation’s large research infrastructure, irrespective of the local fund allocation. Second, substantial revenue will be generated from the consumption of food, lodging, transportation, and tourism activities by scholars residing in or visiting the local area. Additionally, the $\gamma$-ray light source equipment will be subject to international-standard user charges for enterprise users. Finally, the completion of the CEPC project will not only stimulate the growth of the local high-tech industry but also create opportunities for industry and science and education bases, resulting in both direct and indirect revenue streams.

Collectively, these funding sources are expected to comprehensively cover operation and maintenance costs, potentially resulting in annual surplus funds.

## 11.5 Industrialization and Mass Production of the Accelerator Components

### 11.5.1 Introduction

The CEPC project represents one of the most advanced future colliders, featuring a large-scale accelerator complex comprising a 1.8 km linear accelerator and two 100 km circular accelerators, including a booster and a collider. This endeavor necessitates large-volume components for various accelerator subsystems, including magnets and power supplies, vacuum and mechanical systems, Superconducting Radio-Frequency (SRF) components, Radio-Frequency (RF) power systems, cryogenics, instrumentation, control, survey and alignment, radiation protection, and electron and positron sources, among others. For example, the project will involve the production of approximate 37,000 magnets, more than 20,000 magnet power supplies, approximately 200 km of copper with NEG film, 100 km of aluminum, and 6 km of stainless-steel vacuum pipes, over 80,000 mechanical supports, about 192 650-MHz two-cell SRF cavities, and 96 1.3-GHz nine-cell SRF cavities, around 96 high efficient klystrons, to name a few.

In terms of industrialization, two important characteristics should be considered for these components. The first characteristic pertains to magnet-like components, which not only demand high precision in fabrication but also a remarkable quantity, given that they will populate most of the 100 km tunnels. The second characteristic is related to components similar to SRF cavities, which require complex technological support and may not offer final performance guarantees from vendors. For instance, SRF cavities, as high-tech state-of-the-art components, necessitate meticulous preparation and assembly of subcomponents, involving processes like electron-beam welding, controlled electro-chemical polishing, high-pressure rinsing, annealing, assembly of complete cavities into cryomodules, all of which must be executed in clean or semi-clean room environments, adhering to well-defined procedures. Consequently, the project will heavily rely on the industry to provide cost-effective production of these large-volume components while maintaining a high technological standard.

Fortunately, when contemplating the mass production of such high-technology components, there is much to be learned from the experiences of projects like the LHC, the European XFEL, and earlier works such as the ILC Project Implementation Plan (PIP). Notably, the recent High-Luminosity LHC project achieved several key factors that have been successfully used in the industrialization process. Some of these key elements include:

1. Clear governance and formal engagement of all project partners.
2. A robust Product Breakdown Structure (PBS) and Technical Design Report (TDR).

3. Well-defined schedule and cost baselines.
4. The utilization of configuration management, integration, documentation, and records tracking tools.
5. An early, pragmatic assessment of whether components should be manufactured in-house or sourced externally.
6. Early involvement of the industry with robust sourcing campaigns.
7. The ability to insource the production of components with low Technology Readiness Levels.
8. Rigorous review policies at all project stages, among other practices.

These lessons from past projects, including the European XFEL, which produced approximately 80 SRF cryomodules (equivalent to ~640 cavities) and remains the largest deployment of SRF technology to this day, are valuable. The European XFEL was constructed by a European consortium of laboratories and industrial partners. Based on the successful experience of the European XFEL, the ILC PIP summarized key points and mass production models, such as the "hub laboratory" approach, similar to the operation method employed by LCLS II through the collaboration of Fermilab, JLab, ANL, and Cornell University. These ideas can also be beneficial for the CEPC project. Further details will be discussed in subsequent sections.

Regardless, a fundamental objective of any strategy for industrialization and mass production is to minimize unit costs and mitigate risks to the greatest extent possible. It is imperative to comprehend the implications of different challenges and approaches in achieving this goal.

### 11.5.2 The Challenges

Typically, the challenges in the development of a new accelerator stem from the following factors:

1. The technologies needed and their level of technological maturity.
2. The industrialization of short series of components that are not commercially available off the shelf (COTS).
3. The crucial trade-off between cost and time.

Following the Technical Design Report (TDR), most of the technologies for the CEPC meet the specified requirements, with the exception of certain alternative technologies like plasma acceleration. The status and maturity levels of CEPC accelerator technologies are presented in Table 11.5.1. However, it is imperative to establish definitive mass production standards for components in the industry, particularly those anticipated to become future products in their catalog, such as high-efficiency klystrons.

**Table11.5.1:** Status and maturation of CEPC accelerator technologies

| Accelerator System | Specification Met | Prototype Completed |
|---|---|---|
| Magnets | √ | |
| Vacuum | √ | |
| RF power | | √ |
| Mechanics | √ | |
| Magnet power supplies | √ | |
| SRF | √ | |
| Cryogenics | √ | |
| Linac and sources | √ | |
| Instrumentation | √ | |
| Control | | √ |
| Survey and alignment | √ | √ |
| Radiation protection | √ | |
| SC magnets | | √ |
| Damping ring | √ | |

During the period of 2021-2025 (the 14th China Five-Year Plan), it is expected that the CEPC will complete all R&D programs and seek approval for construction from the central government. Upon approval, the project's construction period would be scheduled from 2027 to 2034. By 2036, the construction and commissioning phases are projected to conclude. This means that we need to complete all mass production, installation, and commissioning within a ten-year span, from 2027 to 2036.

As a point of reference, the High Energy Photon Source (HEPS) has a significantly smaller circumference, approximately 1.3 km, compared to the 100 km circumference of the CEPC, which also roughly corresponds to the number of components involved in the project. On the other hand, the construction timeline for the CEPC, approximately 10 years, is not significantly longer than the construction time of around 6.5 years for HEPS. The formidable timing challenges, in some respects, center around the industrialization and mass production aspects of the project.

A more efficient approach involves simultaneous production and installation of equipment. Civil engineering and campus construction surveys were initiated even during the Conceptual Design Report (CDR) phase. Furthermore, additional site investigations were conducted during the Technical Design Report (TDR) phase, in collaboration with various company partners such as Huanghe Company, Huadong Company, and Zhongnan Company.

As of now, three preferred site candidates for the CEPC have been identified, each located in Hebei province, Zhejiang province and Hunan province, overseen by the aforementioned companies. Strategies for accelerator installation were also explored in collaboration with these company partners. The installation timeline will be primarily based on the capabilities of the survey and alignment teams, but numerous approaches will require careful consideration.

Regarding the timeline, we have a 4-year window during the Engineering Design Report (EDR) phase, spanning from 2024 to 2027, to focus on detailed preparations for industrialization and mass production.

### 11.5.3 Strategies to Address the Challenges

Drawing from the experiences of both LHC [1] and ILC PIP, several key points have surfaced to address the challenges mentioned earlier and manage cost-effective manufacturing.

#### 11.5.3.1 *Mass-Production Model*

In the ILC PIP report, several potential mass-production models were proposed, drawing from the experiences of the LHC and the European XFEL project. These models can be valuable for the CEPC in terms of reducing production time and cost. Among these models, a globally distributed approach, based on the 'hub laboratory' concept, is one of the preferred options as identified by the ILC.

This model typically assumes that certain equipment (e.g., cryomodule) is an attractive in-kind contribution, which leads to the division of component production across various hub laboratories distributed globally, spanning the three regions. As the name implies, the hub laboratory serves as the central coordinating body for regional cryomodule production. These hub laboratories establish a robust collaboration with the primary laboratory (CEPC Project) through the adopted governance mechanism.

However, it is important to note that the actual approach to mass production may be influenced by governance-related factors and the project's funding structure, which can be challenging to predict in advance.

Another potential cost-effective approach to industrial contracts involves centralizing the procurement of individual components produced by the industry to the greatest extent possible. These contracts could be overseen either by designated laboratories, if available, or by CEPC groups, with a preference for in-kind contributions. Such an approach logically stems from addressing the matter of laboratory collaboration.

An alternative to the aforementioned model involves the potential for a centralized production facility, or even several large plants. Such a monolithic facility could be managed by industry or a consortium, with the goal of minimizing management overhead and potentially achieving cost savings by consolidating production into a single location. This approach also provides the advantage of facilitating the exchange of best practices among collaborating industries.

In all the potential scenarios under consideration, it's essential to emphasize the pivotal role of laboratories in ensuring overall cost control.

#### 11.5.3.2 *Make or Buy Process*

Drawing from the experience of HL-LHC, when dealing with components at a Technology Readiness Level (TRL) [2] ranging between 2 and 4, it is critical to assess the feasibility of elevating them to TRL 9 within a few years. A systematic evaluation of each component is essential, referred to internally as the 'Make or Buy' process. This early classification of components enables data-driven decisions, such as initiating competitive R&D efforts for fast-track development or transferring production R&D to collaborative efforts. This process also forms the foundation of the sourcing strategy, helping identify market companies capable of responding to component tenders.

In terms of technology transfer methods, it's essential to determine which components necessitate comprehensive process guidance for companies and which ones can be effectively managed through a meticulously specified production process, often referred to as the 'build-to-print' process. The latter approach involves clear documentation and rigorous sign-off at each stage of the production process. For instance, during the TDR phase, extensive R&D was conducted for SRF cavity development in close collaboration with companies. Several employees from these companies received thorough training in various processes, such as cavity RF frequency control and tuning, electro-chemical surface treatment, cleanroom operations, and more. This collaborative effort led to the formulation of detailed documentation and established sign-off procedures for every step of the process.

#### 11.5.3.3 *Infrastructures*

We must have a clear vision of components that may not be suitable for production in industry and instead require testing infrastructure or more efficient internal production. Given the scale of CEPC, particularly during the installation phase, some components may encounter assembly challenges that can be more efficiently addressed through internal production. Testing infrastructure, on the other hand, may require additional support from laboratories or prove to be too expensive in comparison to the component itself. Hence, it's crucial to consider whether testing infrastructure is adequate or if it could potentially become a bottleneck in the mass production process.

#### 11.5.3.4 *Collaboration with Industry and CIPC*

The early involvement of industry, particularly in the case of the most critical materials and components, is of paramount importance. Building upon the experience gained at IHEP, including projects such as BEPC II and HEPS, CEPC has engaged with a wide array of outstanding domestic accelerator-related companies. To facilitate this collaboration, CEPC established the CEPC-SPPC Industrial Promotion Consortium (CIPC), inviting participation from both domestic and international industrial entities.

Through CIPC, numerous strategic studies have been conducted in collaboration with company partners at an early stage. These studies encompassed areas such as site investigations, alignment, and project installation schemes. Additionally, several critical technology research projects were undertaken in conjunction with company partners. For example, in the development of superconducting RF technology, training was provided to personnel in cavity RF frequency tuning, surface treatment, cleanroom operations, and other key areas. Another instance is the R&D focused on high-efficiency klystrons, which can significantly reduce CEPC's power consumption. This comprehensive research effort was carried out in collaboration between IHEP and Kunshan Guoli company.

As of the year 2022, the CIPC membership exceeded 70, with over 20 company partners actively participating in CEPC's R&D efforts during both the CDR and TDR phases. Table 11.5.1 provides a summary of the main CIPC members and the accelerator technical systems they were involved with during the TDR period.

**Table 11.5.1:** CIPC member companies and their involvement in the CEPC R&D

| | Companies/ Acc. systems | Magnet | Power supplier | Vacuum | Mechanics | RF Power | SRF/ RF | Cryogenics | Survey and alignment | Radiation protection | Civil |
|---|---|---|---|---|---|---|---|---|---|---|---|
| 1 | Beijing HE Racing | Participate | | | | | Participate | | | | |
| 2 | Xingxia OCIT | | | | | | Participate | | | | |
| 3 | Kunshan Guoli | | Participate | | | Participate | | | | | |
| 4 | Huadong Guangdian | | | | | | Participate | | | | |
| 5 | Huiyu Vacuum | | | Participate | Participate | | | | | | |
| 6 | Hefei Keye | Participate | | Participate | Participate | | | | | | |
| 7 | Wuxi Creative Tech. | | | | | | Participate | | | | |
| 8 | Beijing Zhongkefuhai | | | | | | | Participate | | | |
| 9 | Jiangsu Cryote | | | | | | | Participate | | | |
| 10 | Anhui Wangrui | | | | | | | Participate | | | |
| 11 | SCSC (Shanghai Kechuang) | Participate | | | | | | | | | |
| 12 | Beijing Gaoneng Tech. | | | | | | | | | Participate | |
| 13 | Jiangsu Chenxin | | | | | | | | | Participate | |
| 14 | Hefei Juneng | Participate | | Participate | | | Participate | | | | |

| | Companies/ Acc. systems | Magnet | Power supplier | Vacuum | Mechanics | RF Power | SRF/ RF | Cryogenics | Survey and alignment | Radiation protection | Civil |
|---|---|---|---|---|---|---|---|---|---|---|---|
| 15 | Huaqiao University | | | | | | | | | | Participate |
| 16 | Beijing Puda Ditai | | | | | | | | Participate | | |
| 17 | Chengdu Puda Ditai | | | | | | | | Participate | | |
| 18 | Wuhan University | | | | | | | | Participate | | |
| 19 | Shanghai Kelin | Participate | | | | | | | | | |
| 20 | Suzhou Superconducting tech. | Participate | | | | | | | | | |
| 21 | Shanghai Superconducting tech. | Participate | | | | | | | | | |
| 22 | Yellow river Corp. | | | | | | | | | | Participate |
| 23 | Huadong Corp. | | | | | | | | | | Participate |
| 24 | Zhongnan Corp. | | | | | | | | | | Participate |

Such measures will prove beneficial in enhancing clarity and reducing costs. Simultaneously, efforts are underway to assess international vendors and gather more comprehensive insights into the industry. To facilitate this, consideration is being given to the creation of a dedicated website. This website would serve as a platform to present future requirements in a highly technical and industry-oriented format, along with tendering documents. The aim is to provide companies with the necessary information about upcoming tenders several weeks or months before the procurement process commences, thereby affording them ample time to prepare and engage their industry stakeholders effectively.

#### 11.5.3.5 ***Sourcing***

Maximizing competition among vendors is crucial to maintaining the lowest reasonable price. Generally, cost reductions are associated with large-volume production, making investments in additional infrastructure and the application of aggressive mass-production techniques cost-effective. Consequently, this approach significantly benefits large-scale projects like CEPC by reducing unit costs and, in turn, the overall project expenses.

However, it's important to acknowledge that many accelerator components are often highly specialized, which may limit the number of available industrial producers. Therefore, careful consideration is required to strike a balance between cost-saving measures and the risks associated with too few suppliers. One potential solution involves breaking down components into more common design sub-components to expand the pool of potential vendors. Identifying subcontractors, categorizing these sub-components, and issuing project-level bids can also increase the number of vendors and help reduce costs.

Organizing industrial gatherings with global presentations and engaging in discussions with project engineers is another effective approach. The goal here is to prevent the over-reliance on too few suppliers and stimulate vendor interest in participating in CEPC procurement. The transfer of technology and knowledge also represents one of the project's most valuable assets.

#### 11.5.3.6 ***Tools to Ensure the Follow-up***

Tools play a crucial role in the procurement and production of components for a large-scale project. In the case of CEPC, the accelerators are expected to operate for several decades or undergo upgrades to 50 MW mode and $t\bar{t}$ mode. As a result, what transpires during production is not only important during the manufacturing phase but also throughout the entire operational lifetime of the accelerator. Nearly all components will, at some point, require maintenance or repairs.

To ensure effective production follow-up and smooth operation in the future, the project should mandate the use of standardized tools for documentation and records management. These tools track components from their design stage through the various manufacturing steps, allowing for the retrieval of any non-conformities or test results in a straightforward and long-term manner. Importantly, these tools are uniform regardless of the origin of the supply, whether it's from industry, in-kind contributions, or internal production. For reference, the first Quality Plan for HL-LHC was established in 2013.

CEPC also developed a project management platform called DeepC during the TDR period. This platform is constructed based on the concept of multi-dimensional factor decomposition and is designed to store and manage engineering data for CEPC projects. It aims to assist CEPC's R&D and design teams in achieving unified and efficient project and data management. Currently, the platform is in the debugging phase and is anticipated to be put into use during the Engineering Design Report (EDR) period. For more details, readers are referred to Appendix 9 of this report.

#### 11.5.3.7 *Summary*

Addressing the challenges of industrialization and mass production is a common concern for a large-scale and specialized project like CEPC. While there are no magical solutions, there are several strategies that can enhance resilience in the face of adverse events. The primary objective of any approach is to reduce the unit cost, thereby minimizing the overall project cost.

Optimizing technology and organizational methods, and fostering healthy competition among vendors, can be highly effective in achieving this goal. Additionally, beyond the commercial benefits, generating vendor interest and facilitating knowledge transfer represent crucial avenues. Our laboratory's wealth of knowledge is one of our most valuable assets.

Furthermore, in the industrialization process, it is possible to envision the development of new equipment, such as high-efficiency klystrons, which may become future products in the industry's catalog. Such initiatives can also contribute to the rapid development of various industries, including magnet technology, vacuum systems, and more.

## 11.6 Impact on Science, Economy, Culture and Society

As a world-class research facility, CEPC will serve as a hub for conducting research and promoting innovation in related fields. Furthermore, it will extend its purpose beyond research to encompass education and public services [1-2]. In this section, we explore the significant advantages that CEPC will bring and provide an overview of the evaluation criteria used to assess its impact.

### 11.6.1 Benefits of the CEPC

#### 11.6.1.1 *Scientific Development and Output*

The scientific objectives of the CEPC encompass the exploration of new physics through the studies of Higgs, W, Z bosons, and the top quark [2]. Notably, the CEPC's discovery potential surpasses that of the Large Hadron Collider (LHC) by an order of magnitude, enabling the investigation of new physics with energy scales of up to 10 TeV. Additionally, the CEPC holds promise in potentially unveiling dark matter by leveraging the Higgs boson as a portal. Moreover, the CEPC provides a unique platform for in-depth exploration of flavor physics, including the examination of its underlying structures, by investigating the coupling of the Higgs boson.

Furthermore, the CEPC offers the prospect of studying the phase transitions of the early universe, thereby addressing fundamental questions in particle physics. The resolution of any of these challenges would undoubtedly constitute a significant breakthrough in basic research [2]. The research carried out at the CEPC is poised to position China at the forefront of collider particle physics research and contribute to the advancement of other fundamental scientific disciplines.

The primary benefit of the CEPC for scientists would be the generation of research publications. Figs. 11.6.1 and 11.6.2 display the number of LHC publications and their citations, with the peak in Fig. 11.6.2 marking the 2012 Higgs discovery. According to a prior study, the average value of the benefits stemming from LHC scientific publications is approximately 280 million euros [3].

Furthermore, CERN has adapted and made available software originally designed for high energy physics to the public domain, proving valuable in various other application domains. Notable examples include ROOT, a toolset for data analysis and visualization, and Geant4, which simulates particle interactions with matter, finding applications in medicine, DNA radiation damage simulation, and other industrial uses [4].

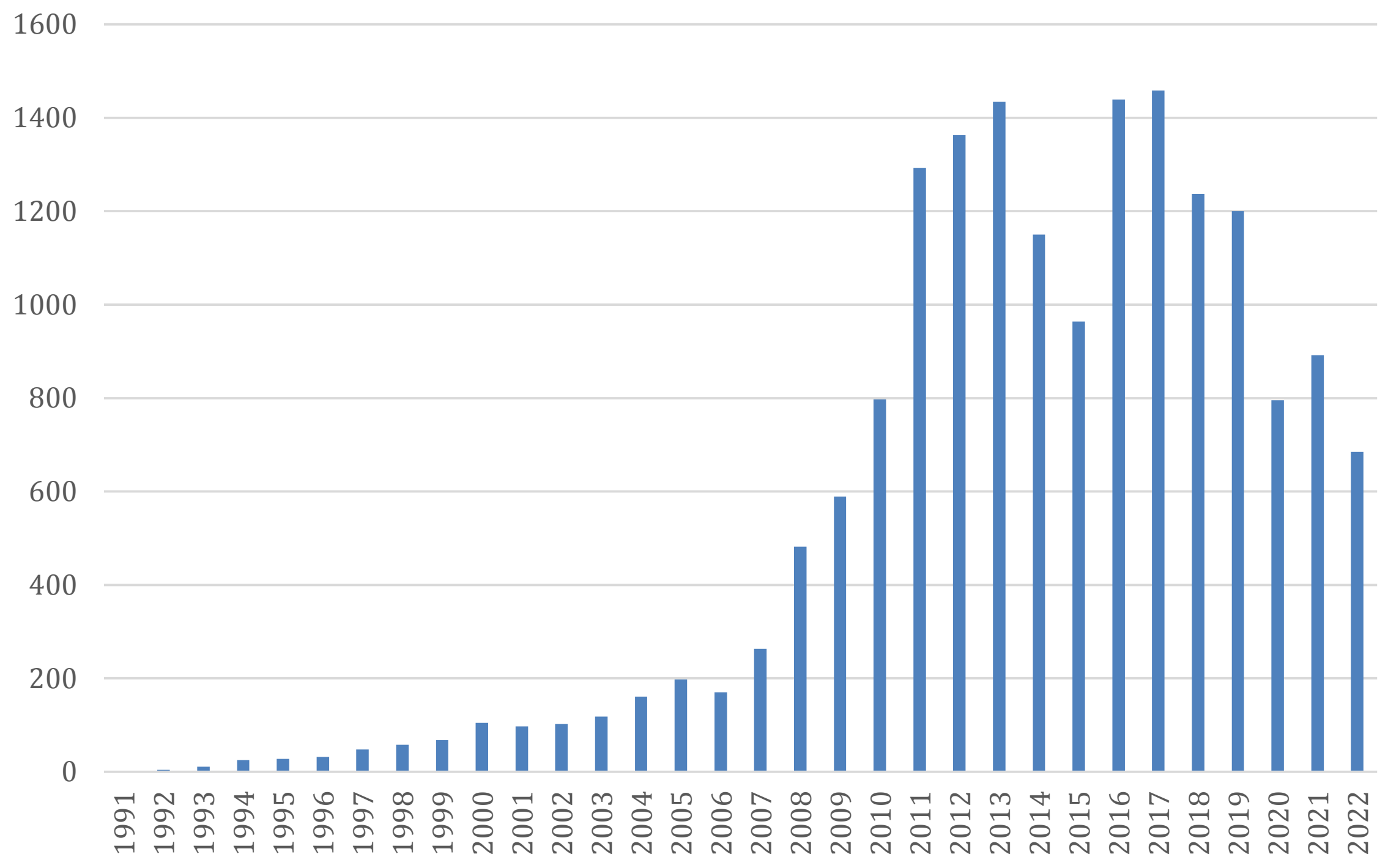

**Figure 11.6.1:** Publications by the LHC collaborations.

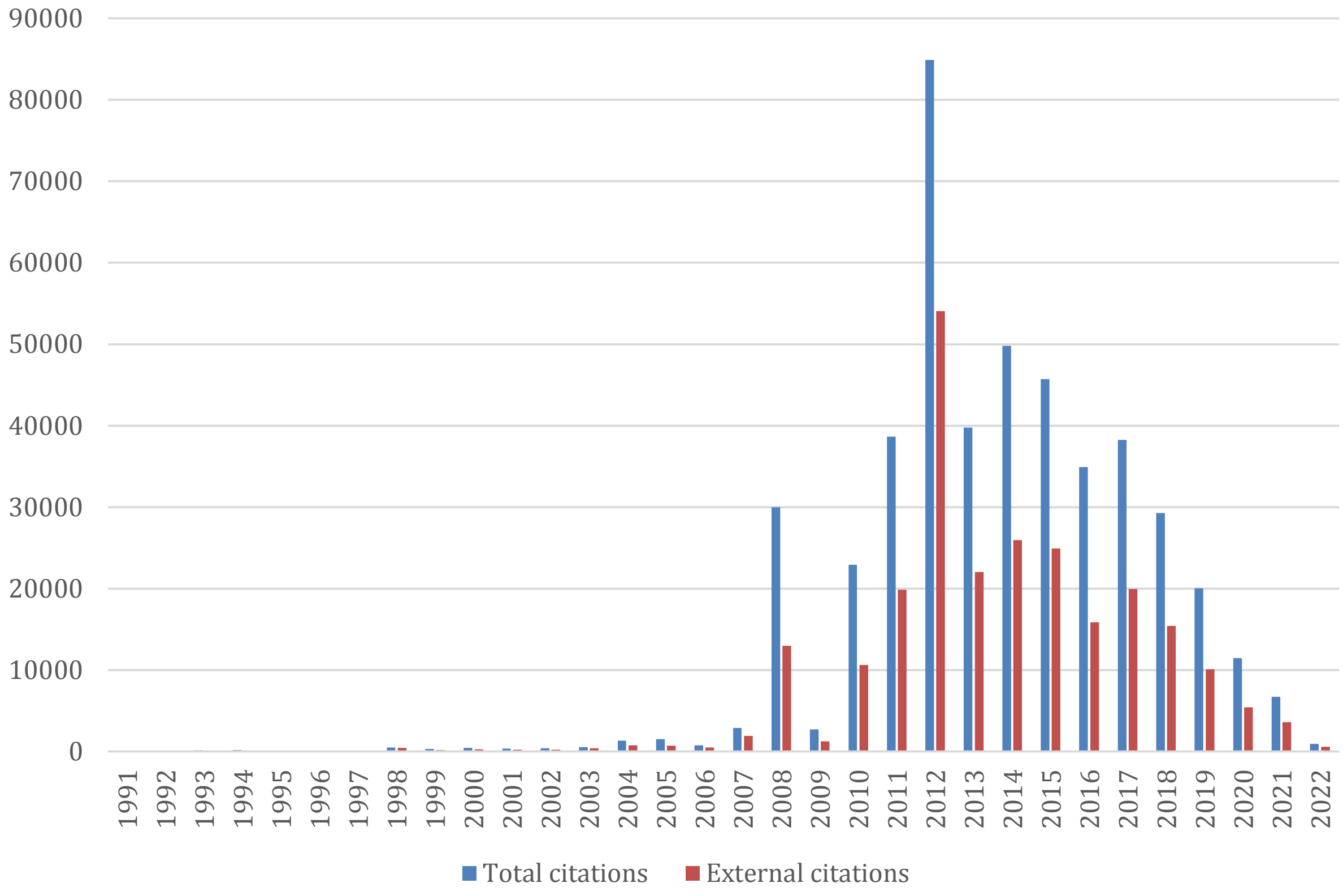

**Figure 11.6.2:** Citations for the LHC publications.

#### 11.6.1.2 *Economic Benefits*

The CEPC has the potential to significantly enhance the regional economic landscape by creating opportunities linked to its construction and operation. This, in turn, will stimulate the growth of diverse industries in the region.

The CEPC has the capacity to generate additional employment opportunities for society and encourage the formation of associated businesses, thus fostering both direct and indirect job growth. The CEPC project will engage in collaborative efforts with its suppliers and other related companies to undertake medium and long-term research and development initiatives.

The continuous advancement of science and technology is poised to lead to a steady expansion of the regional tertiary sector. Consequently, the local employment landscape will experience optimization. Furthermore, nearby commercial enterprises and other businesses stand to gain from the increased customer traffic. The presence of the CEPC is expected to stimulate the growth of local tourism and service industries, such as hotels, conference venues, and catering services.

The CEPC is expected to draw substantial financial investments, enhancing the local investment climate. Simultaneously, it will contribute to the growth of associated intangible income sources, including the commercial utilization of big data and service-related earnings. Collectively, these developments are poised to significantly bolster local economic growth.

#### 11.6.1.3 *Technological Spillovers*

The establishment and operation of a research infrastructure yield a partial return on the social costs incurred, primarily through the dissemination of knowledge to suppliers, other firms, and organizations. Technology transfer from scientific research transpires through both formal and informal channels, enabling the transmission of technology, skills, methodologies, and expertise from research institutions and universities to businesses and government entities. This, in turn, results in the creation of economic value and the advancement of various industries.

The science and technology harnessed in the CEPC are so cutting-edge that they push the boundaries of current engineering and technical solutions. In fact, the creation of the CEPC has given rise to entirely new technologies, with direct applications in the development of improved and novel industries. For example, the CEPC will drive advancements in high-power and high-efficiency klystron technology, while the research and development of 1.3 GHz and 650 MHz superconducting radio-frequency (SRF) systems will propel progress in superconducting RF technology, large-scale liquid helium cryogenic systems, superconducting materials, and related technologies. The development of superconducting high-field magnet technology, apart from benefiting high-energy physics, will have wide-ranging applications in industries such as healthcare and beneficiation.

Furthermore, high-precision silicon pixel detectors, as employed in the CEPC, hold potential for utilization in experiments involving synchrotron radiation light sources, nuclear industry applications, medical imaging, and various other fields.

In addition to the notable technological advancements mentioned, the CEPC is proactively championing innovation in accelerator design, offering the potential for added value. An example is the R&D of novel plasma wakefield acceleration, which has the capacity to reduce construction costs and complexity. Furthermore, the CEPC accelerator has the capability to produce high-energy $\gamma$-rays through synchrotron radiation, thereby fostering the advancement of multidisciplinary science. Collectively, these initiatives are poised to significantly drive technological progress, benefiting not only China but also contributing to global technological growth.

The CEPC is set to foster the cultivation of competitive high-tech enterprises. Already, over 70 high-tech enterprises have become part of the research team, leading to the establishment of the CEPC Industrial Promotion Consortium (CIPC). (For details, please refer to Sec. 11.5.3.4.) The CIPC is actively engaged in the R&D of critical technologies. Upon approval of the CEPC, a substantial increase in the number of participating enterprises and research institutes is expected.

In comparison, the CERN Openlab has initiated more than 50 collaborative projects since 2010. CERN's computer scientists collaborate with leading tech companies in joint R&D efforts. These partnerships allow companies to test their latest products in CERN's state-of-the-art research environment, while CERN gains the opportunity to explore emerging technologies. For instance, ongoing projects with Intel and Micron are investigating the application of machine learning to enhance data processing from particle collisions. Concurrently, projects with Oracle and Siemens are leveraging these technologies to enhance control systems for the LHC. It is reasonable to anticipate that the CEPC will similarly facilitate a spillover effect, fostering advancements in technological development.

### 11.6.1.4 *Training for Students and Talents*

The CEPC project is poised to attract a multitude of scientists, engineers, and supporting personnel from across the globe. It will facilitate the establishment of exceptional teams, nurturing a substantial pool of high-level talents. The CEPC is expected to act as a magnet for graduate and doctoral students, postdoctoral fellows, trainees, and young scientists in general. For these individuals, involvement in the dynamic environment of a leading laboratory offers direct access to firsthand scientific data, opportunities to grapple with intriguing and complex challenges arising from large-scale experiments, and inclusion within a network of brilliant minds.

This hands-on experience will significantly influence the skills acquired during the early stages of their careers, exerting a profound impact on the job market. This, in turn, will have far-reaching effects on their professional trajectories and future earning potential.

A survey [4] undertaken by the CERN Alumni initiative involved current and former members of LHC experiments, primarily consisting of doctoral students (85 percent). These participants represented a diverse array of academic disciplines, with the majority having backgrounds in experimental physics (69 percent), followed by theoretical physics (11 percent), and others (20 percent).

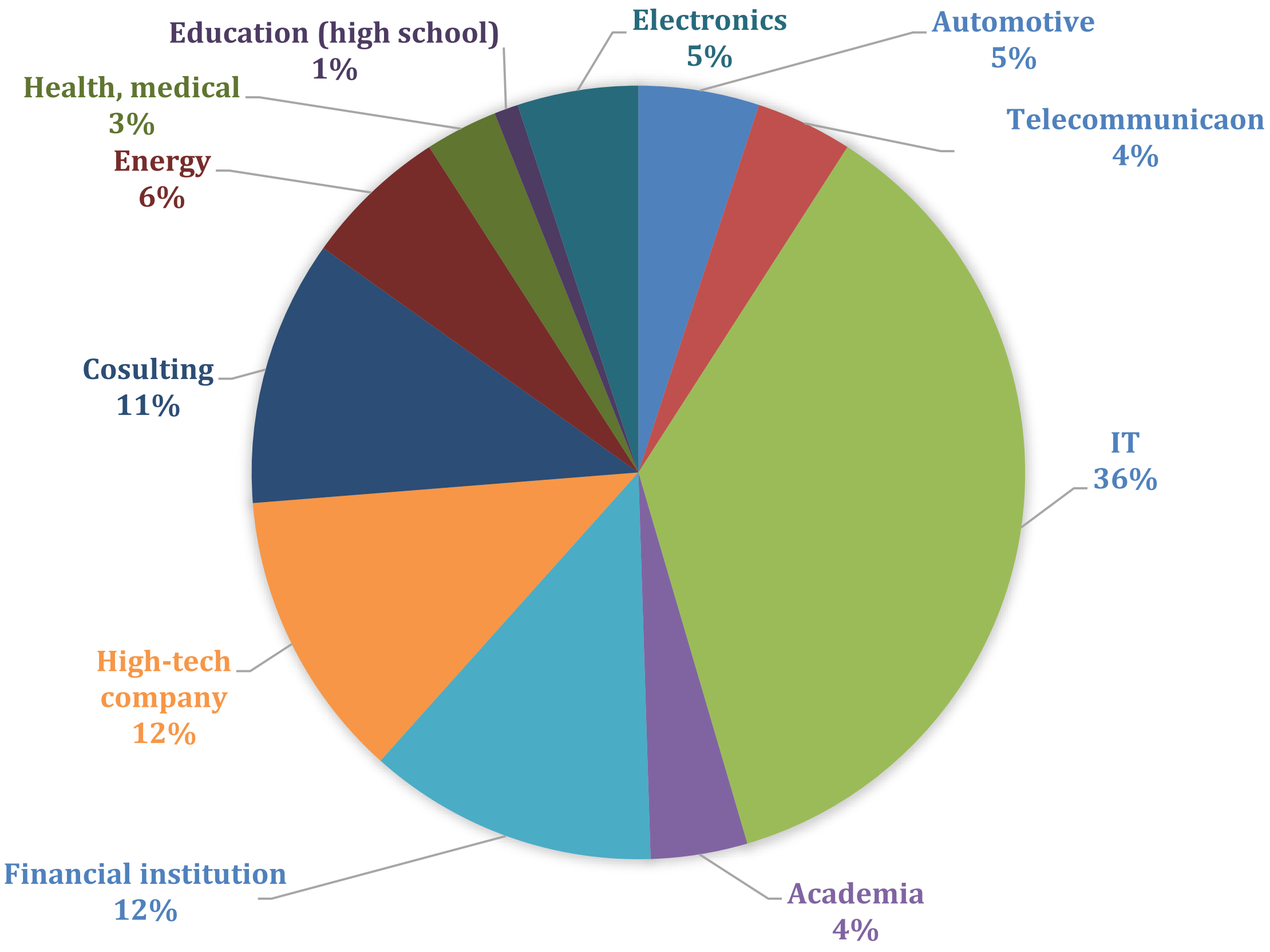

**Figure 11.6.3:** Respondents of the CERN Alumni survey (2018).

Figure 11.6.3 indicates that among the 2,859 respondents, a significant portion is currently employed in the private sector, particularly in roles related to engineering, consulting, information technology (IT), and other fields. Respondents have reported that their experiences at CERN have equipped them with a diverse skill set that they deem crucial in their current positions. These skills encompass programming, collaborating within international teams, data analysis, logical reasoning, and effective communication. Notably, those who spent more time at CERN expressed greater satisfaction with their present roles and acknowledged the acquisition of a broader and more varied skill set.

It has been estimated [3] that the beneficiaries of human capital formation [6,7] at LHC from 1993 to 2025 include 37,000 young researchers: 19,400 students and 17,000 post-docs (excluding participants in schools or short training programs). The LHC benefit has been quantified as the additional salary earned throughout one's entire career.

#### 11.6.1.5 ***Increase of International Scientific Influence***

As a symbol and flagship of contemporary scientific and technological progress, the CEPC embodies qualities of originality and leadership. It is poised to assume a pivotal role in securing the international scientific and technological forefront, establishing a strategic national scientific and technological force, propelling reforms within the scientific and technological system, fostering an open and innovative ecosystem, drawing global scientific and technological talent, actively engaging in global scientific and technological governance, enhancing national soft power, and facilitating non-governmental diplomacy. This aligns perfectly with the objectives and requirements of a major national scientific and technological infrastructure and holds significant strategic importance.

Scientists and the global industrial community associated with electron-positron Higgs factories have collectively forged an ecosystem. The CEPC has garnered robust international support for its scientific research, comprehensive designs, and the R&D of pivotal technologies from this worldwide collaborative network. Looking ahead, the establishment of a science city will be founded upon the CEPC's initiatives. The CEPC team will persist in playing an active role in research across various projects and will effectively propel the advancement of a diverse array of crucial technologies.

The CEPC will bolster global academic communication and collaboration, laying the groundwork for international partnerships in the field of particle physics. In the words of a former ATLAS spokesperson, "*The enthusiasm and motivation to explore particle physics at the high-energy frontier knows no borders between the nations and regions of the planet. It is shared between physicists of widely different cultures and origins. This is evident today when looking around the large but still overcrowded auditoria where the latest results from the LHC are presented, as with the announcements of the Higgs-boson discovery. Such results are, in turn, presented by speakers on behalf of LHC collaborations that span the globe, with physicists from all inhabited continents*."

#### 11.6.1.6 ***Public Goods***

As an international center for high-energy physics research, the CEPC will cater to the public's various needs, including education, the dissemination of scientific knowledge, and other social services. This outreach effort encompasses organized scientific tourism both on-site and online, as well as the communication of scientific advancements through conventional and social media channels. The web presence and social media engagement of the research infrastructure attract a broad audience of external users, creating virtual communities of interested citizens. These communities, in turn, serve as sources of inspiration for science fiction movies and documentary films. The Internet is inundated with news and commentary stemming from the most prominent projects, ultimately generating cultural content as a byproduct of research.

The construction of essential infrastructures, such as roads, communication networks, power grids, and water conservancy systems, in the vicinity of the CEPC, is set to provide substantial benefits to the local population. Furthermore, as previously mentioned, the technologies developed at the CEPC have significant applications in precision mechanics, vacuum technology, microwave systems, superconductivity, cryogenic systems, magnets, and more. The outcomes of this research effort have the potential to not only generate substantial profits for industrial enterprises but also to enhance the quality of life by ensuring safety, comfort, and health for the people. The impact of the CEPC is poised to be multifaceted, influencing various aspects of society.

The impact of the CEPC will be particularly significant in the medical field. One notable example is the use of radio-pharmaceuticals for diagnostic purposes, as seen in the case of PET scans. The detectors originally designed for particle physics experiments have found applications in medical imaging, enabling the identification of cancerous tumors. Beyond aiding in tumor detection, this technology has the potential to contribute to cancer treatment, holding the promise of improving medical outcomes for individuals with cancer.

### 11.6.2 **Evaluation System**

As previously outlined, there are several metrics for assessing the advantages of the CEPC. These metrics are consolidated in Table 11.6.1. Some of these metrics can be estimated by drawing parallels with the LHC, while others require evaluation by experts or the acquisition of data from public surveys or statistics within related industries.

**Table 11.6.1:** Evaluation system (indices) of socioeconomic benefits of the CEPC

(Analysis method: ① To be evaluated by experts; ② Analogy with the LHC; ③ Statistics of corresponding industries; ④ Construction and operation plan; ⑤ Public survey; ⑥ Estimation made by the operating institutions; ⑦ Comprehensive evaluation.)

| **Primary index** | **Secondary index** | **Tertiary index** | **Evaluation method (Value)** |
|---|---|---|---|
| Scientific development and output (A) | Scientific development (A1) | Major scientific breakthroughs and impacts (A11) | ① |
| | | Making China a world leader in high energy physics (A12) | ① |
| | | Promoting the development of other basic frontier fields (A13) | ① |
| | | Major technological breakthroughs and impacts (A14) | ① |
| | Scientific output (A2) | Number of scientific papers published in international scientific index (ISI) journals (A21) | ② (17266) |
| | | Citations for publications (A22) | ② (Total: 466034; External:249002) |
| | | Number of books published (A23) | ② (32) |
| | | Number of international patent authorizations and public patent applications (A24) | ② (112) |
| | | Number of doctoral dissertations completed based on or partially based on the CEPC (A25) | ② (4588) |
| | | Frequency and type of scientific events (A26) | ② (12) |
| | | Number of joint projects to be implemented (A27) | ② (>50) |
| | | Software and hardware released by the CEPC (A28) | ② ($10^7$ lines of code) |
| | | Innovations with Chinese characteristics (A29) | ① |
| Economic benefits (B) | Economic structure (B1) | Increasing the proportion of regional tertiary industry (B11) | ③ |
| | | Development of industries (B12) | ③ |
| | Employment (B2) | Increasing job opportunities directly/indirectly (B21) | ③ |
| | | Optimization of the employment structure (B22) | ③ |
| | | Number of full-time employees (classified by age, sex and nationality) (B23) | ② (3459) |
| | | Costs of employment, operation and maintenance (B24) | ④ |
| | | Number of companies (classified by department, activity area, scale, technical progress level and ownership) (B25) | ② |
| | Economic benefits during the construction period (B3) | Cumulative GDP created during the construction period (B31) | ④ |
| | | Jobs created or retained during the construction period (B32) | ④ |
| | | Procurement amount (B33) | ④ |
| | | Contracts signed with suppliers and other companies (B34) | ④ |
| | | Social contributions (e.g. tax payments) made by the CEPC and its employees (B35) | ④ |
| | | Number of medium and long-term cooperative R&D contracts signed with business partners (B36) | ④ |
| | | Number and type of spin-off companies established due to the operation of the CEPC (B37) | ④ |
| | | Total number of visitors and users (B38) | ②, ③ |
| | | Increasing the number of customers using the services of local tourism companies (hotels, conference venues, catering) (B39) | ②, ③ |
| | Economic benefits during the operation period (B4) | Attracting a large number of people and funds (B41) | ②, ⑤ |
| | | Improvement of local investment environment and regional investment attraction (B42) | ②, ⑤ |
| | | Development of local tourism (B43) | ③, ⑤ |
| | | Intangible asset income (data commercial application, service income, etc.) (B44) | ①, ④ |
| Technological spillovers (C) | Technical progress (C1) | Innovative technical solutions proposed during project construction (C11) | ①, ④ |
| | | Number of contracts signed to develop new equipment to meet specific needs (C12) | ④ |
| | | Number of patent licenses (C13) | ④ |
| | | Number of patents licenses completed in cooperation with companies (C14) | ④ |
| | | Number of international patent licenses (C15) | ④ |
| | | Number of joint R&D projects with enterprises; Number of new products and service prototypes jointly developed with enterprises (C16) | ④ |
| | | Number of cooperative projects directly participated by enterprises (C17) | ④ |
| | | Number of R&D projects commissioned by enterprises (C18) | ④ |
| | | Number of technical prototypes and designs (jointly developed by enterprises and the CEPC) for industrial production (C19) | ④ |
| | Technological breakthroughs (C2) | Breakthroughs in key, core technologies (superconducting accelerator technology, microwave power supply and refrigeration technology, etc.) (C21) | ④ |
| | | Applications of nuclear technology; Development of relevant instruments and equipment (C22) | ④ |
| | | Optimization of critical equipment (C23) | ④ |
| | Spillover effect (C3) | Technology spillover benefits at industrial level (C31) | ① |
| | | Technology spillover benefit at spatial level (C32) | ① |
| Training for students and talents (D) | Student training (D1) | Master students trained by CEPC (D11) | ②, ⑥ |
| | | Doctoral students trained by CEPC (D12) | ②, ⑥ |
| | | International students trained by the CEPC (D13) | ②, ⑥ |
| | | Post-doctoral trained by the CEPC (D14) | ②, ⑥ |
| | Technical training (D2) | Grants awarded to trainees who participate in CEPC training activities (D21) | ②, ⑥ |
| | | Number of trainees (D22) | ②, ⑥ |
| | | Number of master's and doctoral dissertations based on the CEPC (D23) | ②, ⑥ |
| | | Job positions created by attracting foreign researchers and technicians (D24) | ②, ⑥ |
| | High-level talents (D3) | Number of outstanding scientists, engineers and research teams cultivated based on the CEPC (D31) | ②, ⑥ |
| | | Number of laureates of national/international awards (D32) | ②, ⑥ |
| | | Number of international top talents (D33) | ②, ⑥ |
| | Other talents (D4) | Number of construction talents (D41) | ②, ⑥ |
| | | Number of management talents (D42) | ②, ⑥ |
| | | Number of engineers (D43) | ②, ⑥ |
| | | Number of technicians (D44) | ②, ⑥ |
| | | Number of employees (classified by age, sex and nationality) (D45) | ②, ⑥ |
| | Benefits to universities (D5) | Development of disciplines (D51) | ① |
| | | Training for teachers (D52) | ① |
| Increase of international scientific influence (E) | Academic communication and cooperation (E1) | The number of international academic conferences organized (E11) | ②, ⑥ |
| | | International scientific and technological cooperation mechanism established with other countries (E12) | ②, ⑥ |
| | International Center (E2) | Promoting the construction of an international science center (E21) | ① |
| | | Communication and cooperation between the CEPC and foreign research infrastructures (E22) | ②, ⑥ |
| | Improvement of national image (E3) | Social influence on foreign public (E31) | ① |
| | | The influence of international cooperation on national relations (E32) | ① |
| | | The influence on journals (E33) | ① |
| Public good (F) | Infrastructure(F1) | Mileage of new roads (F11) | ③, ④ |
| | | Relocation and resettlement of local residents (F12) | ④ |
| | | Construction of communication network, power grid and water conservancy infrastructure (F13) | ④ |
| | Health and medical services (F2) | Direct contributions to the development of health and medical services (F21) | ④ |
| | | Improvement of medical service conditions(F22) | ⑦ |
| | Public education (F3) | Improvement of the conditions and quality of higher education (F31) | ④, ⑥ |
| | | Scientific education/popularization activities (F32) | ⑤, ⑥ |
| | | Improvement of the public's scientific awareness and scientific confidence (F33) | ⑦ |
| | Public safety (F4) | Improvement of local public security (F41) | ④ |
| | | Ecological protection and construction (F42) | ④ |
| | | Improvement of the capability to predict geological disasters (F43) | ④ |

### 11.6.3 **Summary**

The CEPC will address crucial scientific questions in particle physics with unparalleled precision and discovery potential. It aligns with the national commitment to fundamental research, elevates China's standing in the global particle physics community, and holds strategic significance. Upon realization, the CEPC will emerge as a premier global platform for science and technology, yielding significant advantages in the realms of science, economics, culture, and society.

# 12 Cost and Schedule

## 12.1 Construction Cost Estimate

### 12.1.1 Introduction

The construction cost analysis of the CEPC adopts a bottom-up approach, utilizing a widely recognized Work Breakdown Structure (WBS) commonly used in global accelerator projects. This method commences at the project's lowest tier and progressively consolidates expenses upward to derive the overall project cost. Within the CEPC's WBS, seven hierarchical levels are established, with the current breakdown spanning Level 3 to 6, contingent on specific WBS elements. The top-level WBS, as demonstrated in Table 12.1.1, comprises 13 distinct items.

Item 1 pertains to project management, while Items 2-7 encompass the constituents of the accelerator complex: Collider, Booster, Linac, Damping Ring, electron and positron sources, transport lines, and systems common to accelerators. These costs encompass fabrication, assembly, transportation and handling, and testing. A detailed list of accelerator components can be found in Appendix 2. Item 8 encompasses conventional facilities, inclusive of civil engineering and utilities. Item 9 comprises two $\gamma$-ray beamlines. Item 10 accounts for two detectors and related equipment, such as data acquisition, computing, networking, and software development. Items 11 and 12 cover the installation and commissioning expenses, typically 3% each of the construction cost in China. Item 13 represents a contingency allocation, currently set at 8%, within the required range of 6-9% set by the funding agency, the Development and Reform Commission (DRC).

**Table 12.1.1:** Level 1 of the CEPC Work Breakdown Structure (WBS)

| Level 1 | WBS Element Title |
|---|---|
| | **TOTAL** |
| 1 | Project Management |
| 2 | Accelerator Physics |
| 3 | Collider (Ch 4) |
| 4 | Booster (Ch 5) |
| 5 | Linac, Damping Ring and Sources (Ch 6) |
| 6 | Transport Lines (Ch 4, 5, 6) |
| 7 | Systems Common to Accelerators (Ch 7) |
| 8 | Conventional Facilities (Ch 9) |
| 9 | Gamma Ray Sources (Appendix 5.1) |
| 10 | Experiments |
| 11 | Installation (3% of items 3-7) |
| 12 | Commissioning (3% of items 3-7) |
| 13 | Contingency (8% of items 1-12) |

Notably, the cost estimation excludes manpower expenses (distinct from construction costs in China), project timeline-driven escalation, or operational outlays. Additionally, expenses linked to project engineering and R&D prior to construction start are not factored into the estimate. Furthermore, as per DRC guidelines, the expenses for spare components are to be encompassed within the operational costs.

Given the volatility of material and labor prices, our assessment employs the average value spanning 2018 to 2022.

It's important to emphasize that during the cost review conducted by an international committee, the committee suggested adopting a more cautious approach by setting the contingency at 20% at this stage. Following thorough consideration, we opted to maintain the contingency at 8% to align with the DRC requirement. Our strategy involves refining the cost estimates and minimizing uncertainties in the upcoming engineering design (EDR) phase before embarking on construction.

### 12.1.2 **Project**

The CEPC project's complete construction expense amounts to 36.4 billion yuan (approximately USD 5.2 billion). The distribution is presented in Table 12.1.2 and Figure 12.1.1. Predominantly, the accelerators constitute the largest expenditure segment, accounting for 52% of the overall construction cost. Civil construction corresponds to around 28% of the total, while experiments encompass 11%.

**Table 12.1.2:** CEPC project cost breakdown, (Unit: 100,000,000 yuan)

| **Total** | **364** | **100%** |
|---|---|---|
| Project management | 3 | 0.8% |
| Accelerator | 190 | 52% |
| Conventional facilities | 101 | 28% |
| Gamma-ray beam lines | 3 | 0.8% |
| Experiments | 40 | 11% |
| Contingency (8%) | 27 | 7.4% |

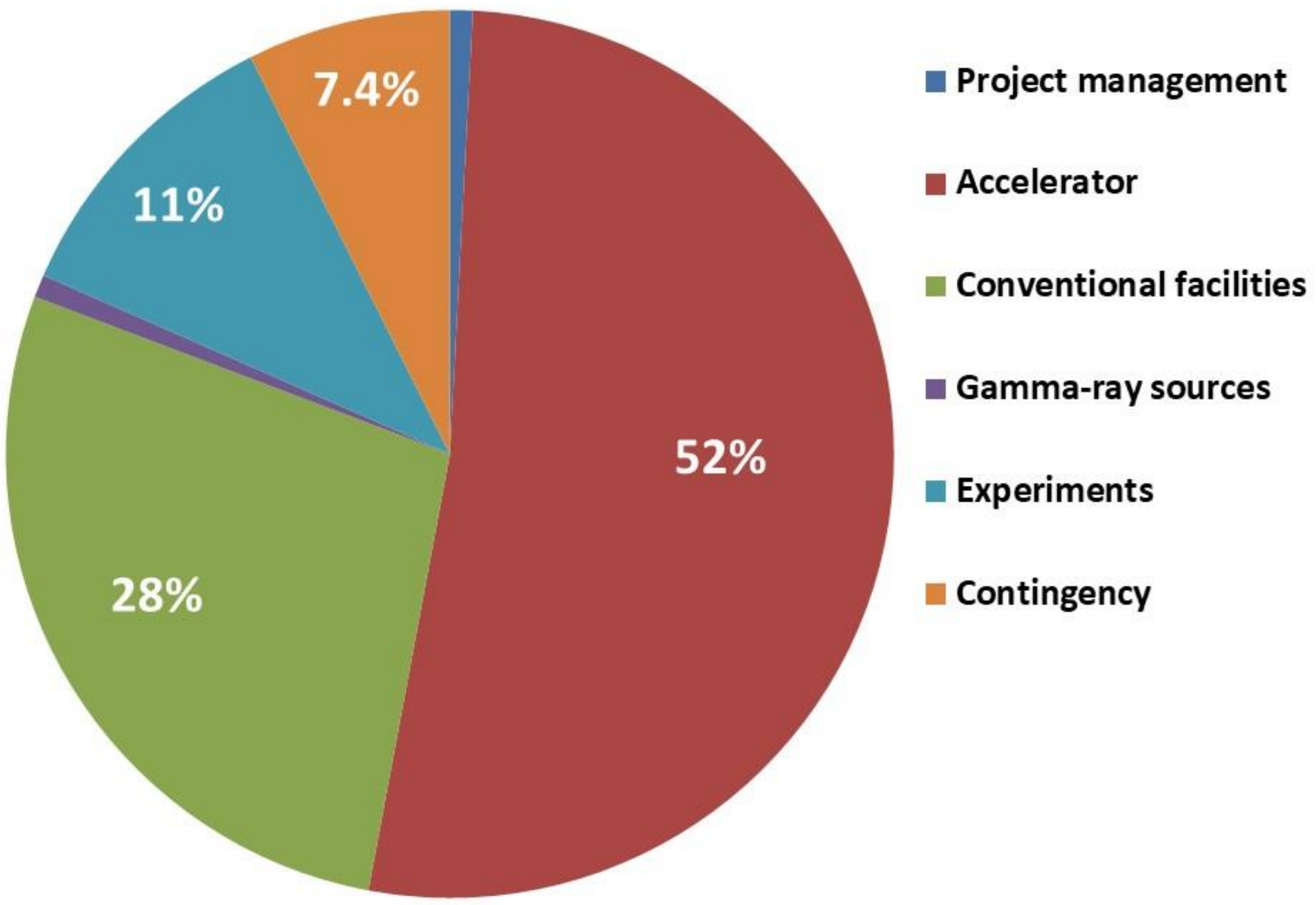

**Figure 12.1.1:** CEPC project cost breakdown.

### 12.1.3 **Accelerator**

The construction expenditure for the accelerators amounts to 19 billion yuan (approximately USD 2.7 billion). The distribution is outlined in Table 12.1.3 and depicted in Figure 12.1.2, presenting a detailed division at the machine level. Further granularity is provided in Table 12.1.4 and illustrated in Figure 12.1.3, offering an in-depth breakdown at the technical system level.

Within the accelerator complex, the Collider section accounts for 53% of the total cost, with the subsequent allocations being the Booster (22%), the Linac and sources (9.7%), and the common systems (8.4%). Installation and commissioning are 3% each.

**Table 12.1.3:** CEPC accelerator cost breakdown to the machine level.
(Unit: 100,000,000 yuan)

| **Accelerator Total** | **190** | **100%** |
|---|---|---|
| Accelerator physics | 0.80 | 0.42% |
| Collider | 100.0 | 53% |
| Booster | 41.4 | 22% |
| Linac and sources | 18.4 | 9.7% |
| Damping ring | 0.65 | 0.34% |
| Transport lines | 1.64 | 0.86% |
| Common systems | 15.9 | 8.4% |
| Installation (3%) | 5.34 | 2.8% |
| Commissioning (3%) | 5.34 | 2.8% |

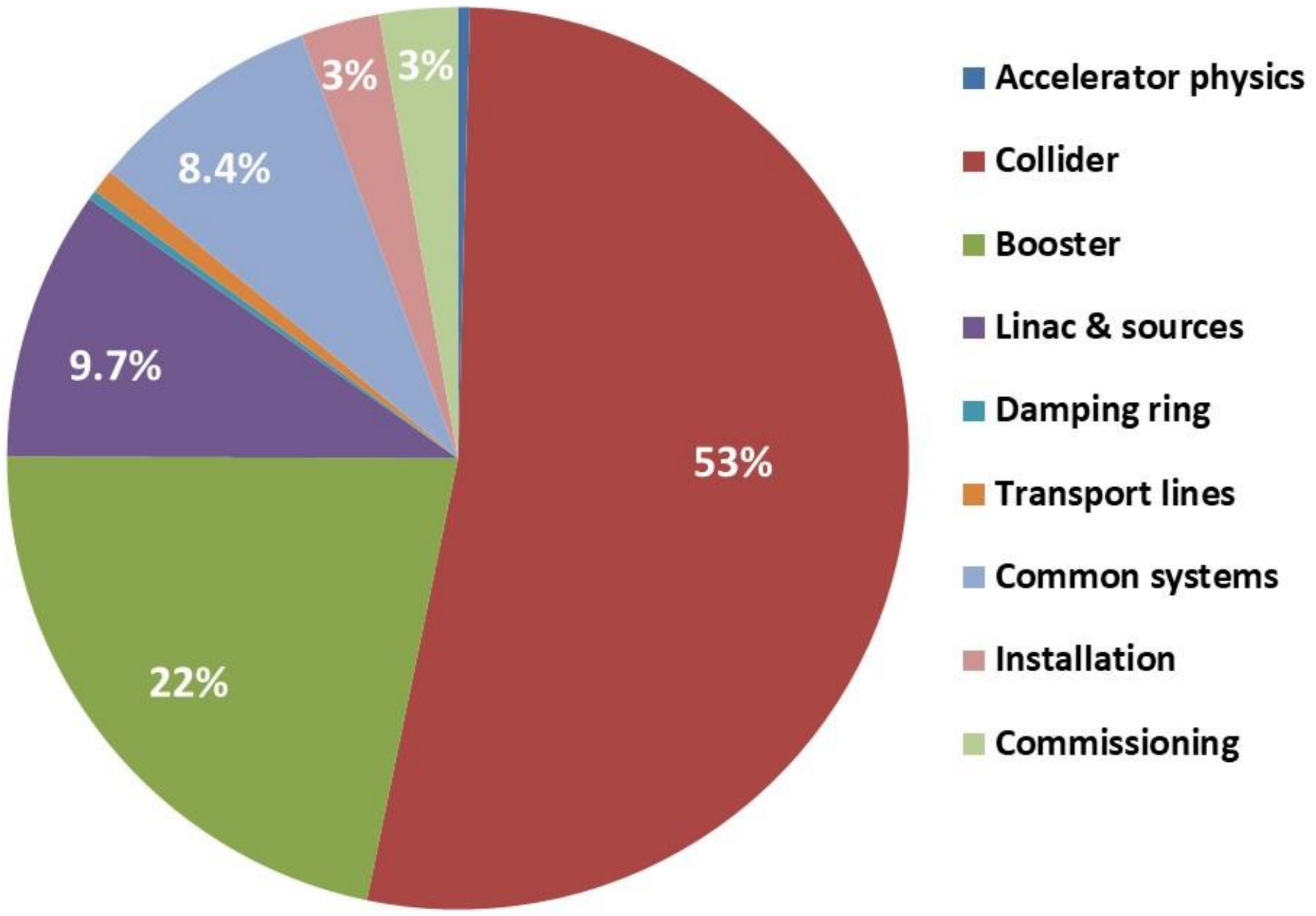

**Figure 12.1.2:** CEPC accelerator cost breakdown to machine level.

Within the array of accelerator systems, the trio of significant components, namely SRF, RF power source, and cryogenics, collectively contribute around 15% of the total expense. Notably, the magnets' outlay (27%) and the vacuum system's cost (17%) hold substantial weight due to their extensive span, extending over hundreds of kilometers. Additionally, the 30 GeV Linac constitutes about 10% of the overall cost.

**Table 12.1.4:** CEPC accelerator breakdown to the technical system level. (Unit: 100,000,000 yuan)

| **Accelerator Total** | **190** | **100%** |
|---|---|---|
| Accelerator physics | 0.80 | 0.42% |
| SRF | 7.3 | 3.9% |
| RF power source | 11 | 5.8% |
| Magnets | 51 | 27% |
| SC magnets | 0.88 | 0.46% |
| Magnet power supplies | 9.0 | 4.8% |
| Vacuum | 32 | 17% |
| Instrumentation | 11 | 5.6% |
| Inj. / Extr. | 1.2 | 0.61% |
| Control | 5.5 | 2.9% |
| Mechanics | 13 | 6.6% |
| Beam separation system | 0.80 | 0.42% |
| Linac and sources | 18 | 9.7% |
| Damping ring | 0.65 | 0.34% |
| Transport lines | 1.6 | 0.86% |
| Cryogenics | 9.9 | 5.2% |
| Survey and alignment | 4.4 | 2.3% |
| Radiation protection | 1.6 | 0.86% |
| Installation (3%) | 5.3 | 2.8% |
| Commissioning (3%) | 5.3 | 2.8% |

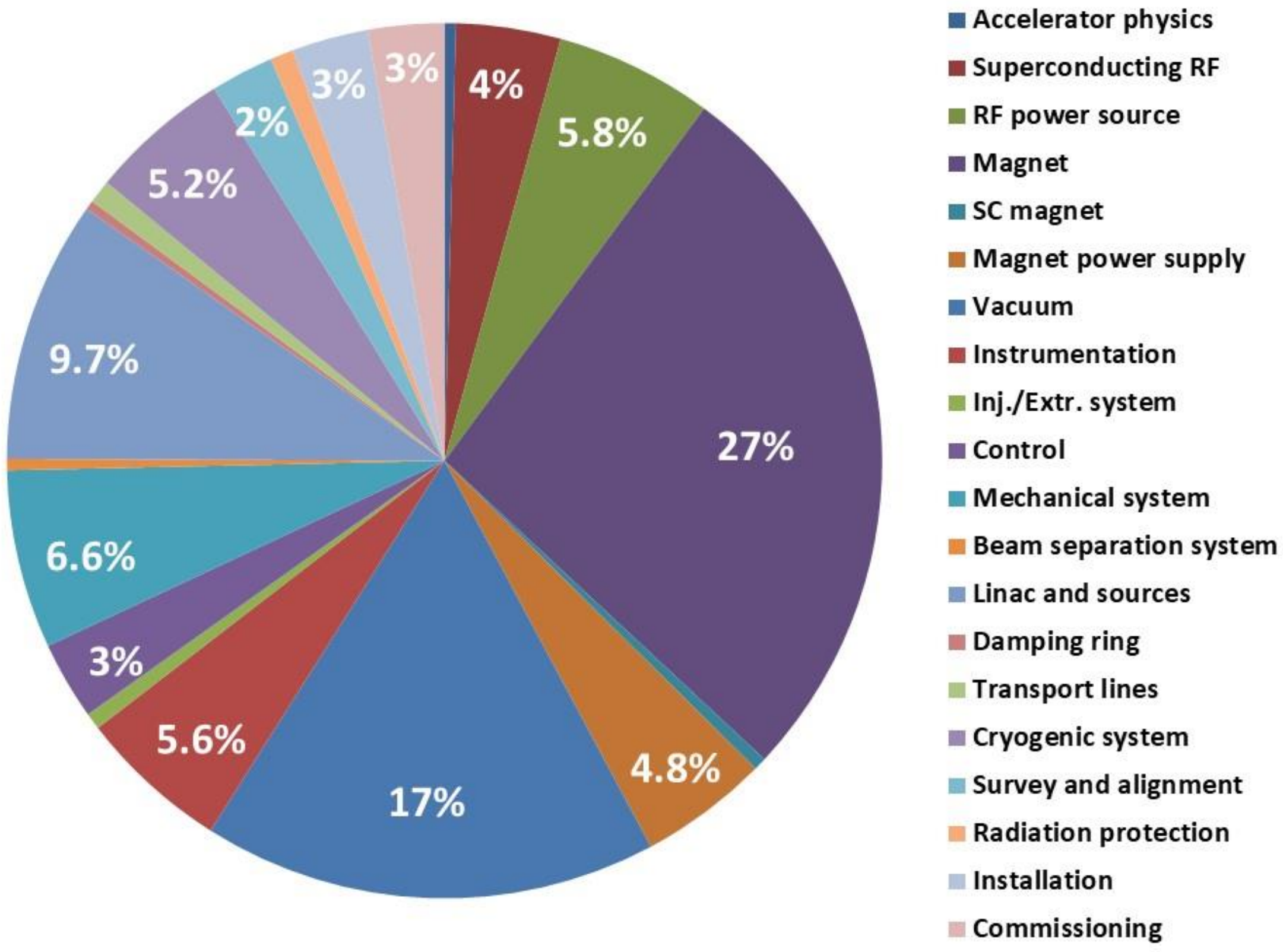

**Figure 12.1.3:** CEPC accelerator cost breakdown to the technical system level.

### 12.1.4 **Utilities**

Utilities encompass electrical and mechanical frameworks that furnish electricity, water, HVAC, cooling water, fire-fighting systems, drainage, compressed air, and more. These utilities play a vital role in sustaining the functionality of the accelerator complex. A detailed expenditure breakdown is presented in Table 12.1.5, visually depicted in Figure 12.1.4. Among the major cost components, two standout categories are the electrical system (48%) and the cooling water system (30%).

**Table 12.1.5:** CEPC utilities cost breakdown. (Unit: 100,000,000 yuan)

| **Utilities Total** | **30.0** | **100%** |
|---|---|---|
| Electrical system | 14 | 48.% |
| HVAC system | 2.6 | 8.5% |
| Fire protection, water supply and drainage | 1.9 | 6.3% |
| Water cooling system | 9.0 | 30% |
| Compressed air system | 0.71 | 2.4% |
| Other equipment | 1.4 | 4.8% |

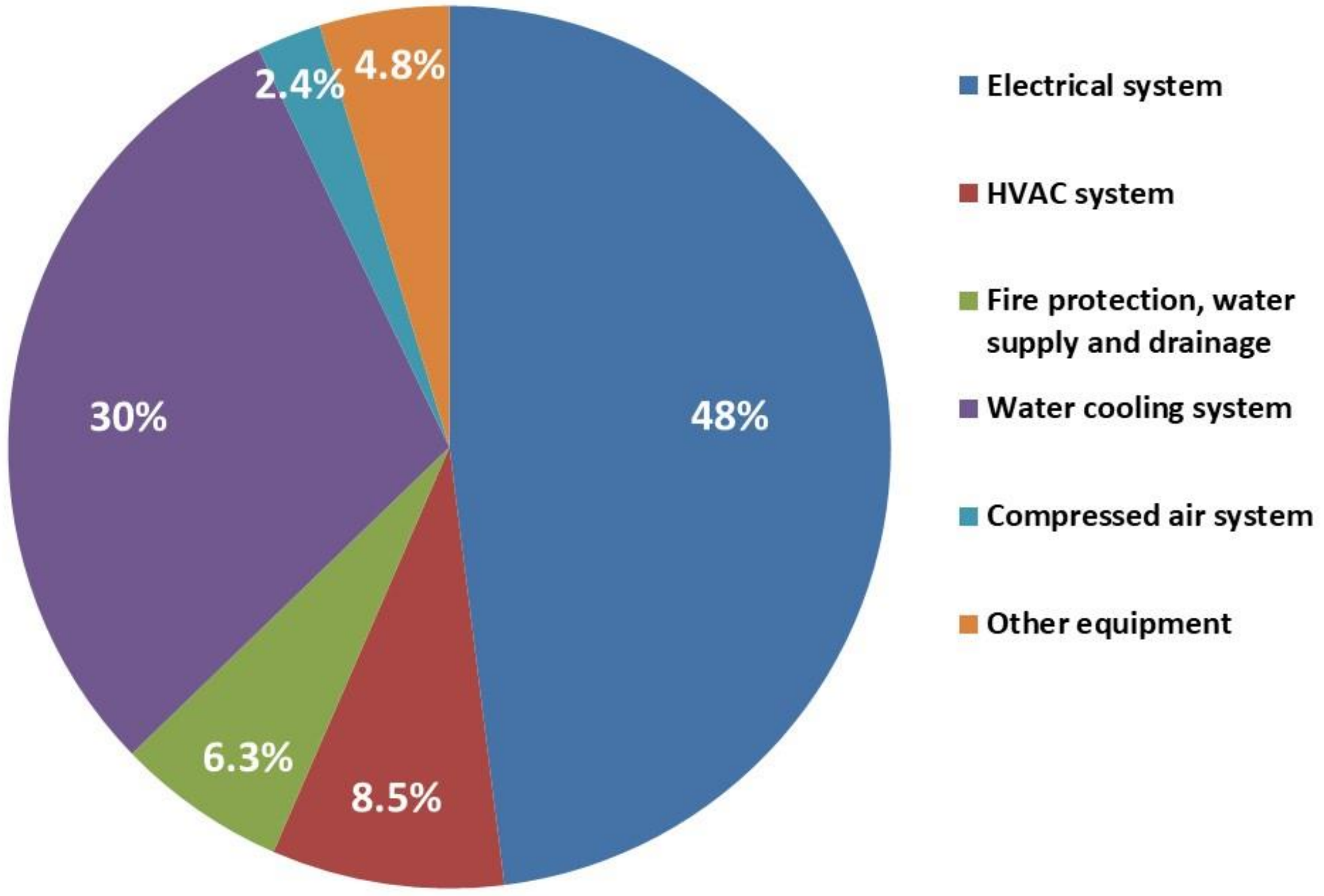

**Figure 12.1.4:** CEPC utilities cost breakdown.

### 12.1.5 Civil Engineering and Related Activities

This category includes the following components:

1. Civil construction: This comprises both subterranean structures (main tunnel, auxiliary tunnels, shafts, adits, experimental halls, service caverns, etc.) and above-ground buildings.
2. Mechanical structures and equipment: Encompasses cranes, elevators, tracks, and associated machinery.
3. Temporary constructions: Involves roads and buildings necessary during the construction period, intended for subsequent demolition.
4. Independent expenses: Cover overheads, profits, documentation fees, permit fees, and similar costs.
5. Other expenses: Encompass fees for environmental evaluation, water and soil preservation.

It's important to note that costs associated with land acquisition and population relocation are not included, as they fall under the coverage of the local government.

To ensure an accurate cost estimate for civil construction, three reputable construction companies were engaged to independently assess three potential sites: Yellow River Consulting Co., Ltd. (YREC) for Qinhuangdao, Huadong Engineering Corporation Limited (HDEC) for Huzhou, and Zhongnan Engineering Corporation Limited (ZNEC) for Changsha. These companies are distinguished engineering firms with a track record of similar projects across China. Each company received identical civil engineering design parameters, including tunnel size and length, shaft number and size, experimental hall specifications, and temporary road length, utilizing 2022 Q4 labor and material prices. The three estimates ranged between 6.6 billion and 7.1 billion yuan, showing a variation of around 8%.

Following the completion of these estimates, an impartial domestic committee was convened to review the assessments. This committee affirmed the appropriateness of the costing methodology and deemed the estimates reasonable.

The findings of the domestic committee were compiled into a concise report and submitted to an international civil cost review sub-panel, and subsequently forwarded to an international TDR cost review committee and the CEPC International Advisory Committee (IAC) for their reference.

## 12.2 Upgrade Cost Estimate

The cost estimate detailed in Section 12.1 pertains to the CEPC baseline design, encompassing a 30 MW synchrotron radiation (SR) power per beam for Higgs and W mode operation, and 10 MW SR power per beam for Z mode operation. Multiple upgrade plans have been outlined:

1. Upgrading power to 30 MW per beam for Z mode operation.
2. Enhancing power to 50 MW per beam for H, Z, and W mode operation.
3. Implementing an energy upgrade for $t\bar{t}$ mode operation at 30 and 50 MW per beam.

Table 12.2.1 showcases the additional cost for each of these upgrades. Combining all three upgrades indicates a total expenditure of approximately 7.5 billion yuan (roughly USD 1.1 billion).

The estimated costs for these upgrades anticipate a significant portion being covered by international contributions.

**Table 12.2.1:** Cost estimate for power and energy upgrade. (Unit: 100,000,000 yuan)

<table>
<tr><th rowspan="2"></th><th rowspan="2">Z<br>30 MW</th><th colspan="2">H/W/Z</th><th colspan="2">$t\bar{t}$</th><th></th></tr>
<tr><th>H/W<br>50 MW</th><th>Z<br>50 MW</th><th>30 MW</th><th>50 MW</th><th>SUM</th></tr>
<tr><td>SRF system</td><td>3.2</td><td>3.9</td><td>1.6</td><td>11.8</td><td>0</td><td>20.5</td></tr>
<tr><td>RF power source</td><td>7.2</td><td>7.6</td><td>5.0</td><td>7.6</td><td>4.9</td><td>32.3</td></tr>
<tr><td>Cryogenics system</td><td>1.0</td><td>0.3</td><td>0.3</td><td>16.2</td><td>0</td><td>17.8</td></tr>
<tr><td>Vacuum system</td><td>1.8</td><td>0</td><td>0</td><td>0</td><td>0</td><td>1.8</td></tr>
<tr><td>Magnet</td><td>0</td><td>0</td><td>0</td><td>0.1</td><td>0</td><td>0.1</td></tr>
<tr><td>Magnet power supply</td><td>0</td><td>0</td><td>0</td><td>2.5</td><td>0</td><td>2.5</td></tr>
<tr><td rowspan="2">SUM</td><td rowspan="2">13.2</td><td>11.8</td><td>6.9</td><td>38.2</td><td>4.9</td><td rowspan="2">75</td></tr>
<tr><td colspan="2">18.7</td><td colspan="2">43.1</td></tr>
</table>

## 12.3 Operational Cost Estimate

The primary expense for operating the CEPC is its electricity consumption. Appendix 3 outlines that for Higgs mode operation using 30 MW of synchrotron radiation (SR) power per beam, the CEPC requires a total of 262 MW of electricity. With the CEPC scheduled to operate 6,000 hours annually, the estimated yearly power consumption amounts to approximately 1.6 TW-hr. Electricity rates vary by region; in the Beijing vicinity, prices range between 0.85-0.89 yuan per kW-hr. Negotiating the electricity price with the local power company during site selection is a significant consideration. Our aim is to secure a rate of 0.50 yuan per kW-hr by directly purchasing electricity from the local provider rather than the grid. This effort targets an annual electricity cost of about 800 million yuan for the CEPC.

The CEPC's maximum operational capacity is 7,500 hours per year, necessitating an estimated annual power usage of about 2 TW-hr, resulting in an electricity cost of roughly one billion yuan.

Drawing from operational data of BEPC II and various international accelerators, it's evident that electricity expenses typically constitute around half of the total operational expenditure. Hence, the CEPC's projected total operational cost is estimated to range between 1.6 to 2 billion yuan per year.

## 12.4 Project Timeline

Chinese civil construction companies engaged in the site selection process estimate that a 100 km tunnel can be excavated in less than five years using drill-and-blast methods, followed by the installation of the accelerator and detectors. The entire civil construction period for the CEPC tunnel spans 54 months, comprising an 8-month preparation phase, 43 months for constructing the main structures, and a final 3-month completion phase. For details regarding the CEPC tunnel layout, refer to Figure 12.4.1.

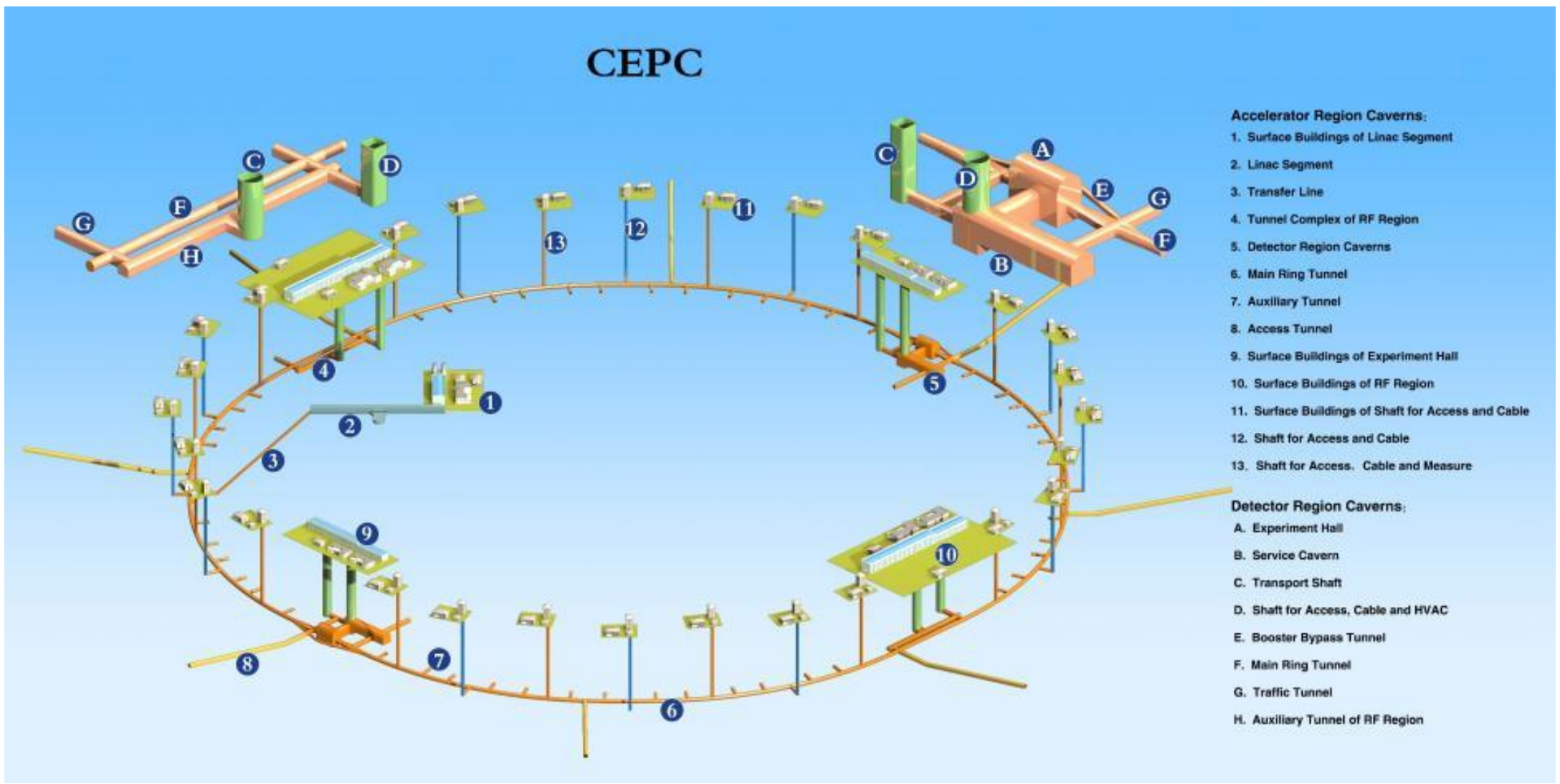

**Figure 12.4.1:** CEPC tunnel layout

Following the completion of the TDR in 2023, a subsequent phase, the engineering design (EDR) period (2024-2026), will ensue. Throughout this phase, ongoing R&D efforts for key technical components will persist, prototypes will be constructed, infrastructure will be established for the industrialized manufacturing of the necessary components, and the final site selection will be concluded.

Anticipated construction for the CEPC is set to commence around 2027, aligning with China's 15th five-year plan, and is projected to span 8 years, reaching completion by approximately 2035. Post-commissioning, the tentative operational plan entails a 10-year duration dedicated to Higgs physics exploration, followed by successive 2-year and 1-year operations in Z mode and W mode, respectively.

The SPPC, an integral facet of the CEPC-SPPC initiative, aims to spearhead energy-frontier discoveries over a prolonged development period exceeding two decades. Given its intricate nature and numerous technical hurdles, ongoing research endeavors are imperative to address critical physics and pivotal technological challenges. Among these challenges, a paramount focus lies on developing iron-based high-field superconducting magnets capable of achieving a minimum of 20 T. These magnets are instrumental in facilitating proton–proton collisions at a center-of-mass energy up to 125 TeV and sustaining a luminosity level of $4.3\times10^{34}$ $cm^{-2}s^{-1}$.

It's premature to establish a definitive timeline for the SPPC project. An initial estimate suggests that the R&D phase for the HTS magnet might span approximately 20 years. Consequently, the earliest projection for construction commencement could be around 2050.

The CEPC project's preliminary timeline is depicted in Table 12.4.1, and the Collider's timeline is illustrated in Table 12.4.2. For a more comprehensive timeline, please refer to Appendix 10 in this report.

**Table 12.4.1:** A preliminary timeline of the CEPC project

| 标识号 | 任务模式 | 任务名称 | 工期 | 开始时间 | 完成时间 |
|---|---|---|---|---|---|
| 1 | | Collider | 104.35 months | 2027/1/1 | 2034/12/31 |
| 2 | | Superconducting RF System | 78.3 months | 2027/1/1 | 2032/12/31 |
| 7 | | RF Power Source | 58.6 months | 2027/1/1 | 2031/6/30 |
| 12 | | Magnets | 65.25 months | 2027/1/1 | 2032/1/1 |
| 17 | | Superconducting Magnets in the IR | 78.3 months | 2027/1/1 | 2032/12/31 |
| 23 | | Magnet Power Supplies | 63.05 months | 2027/1/1 | 2031/10/31 |
| 27 | | Vacuum System | 65.25 months | 2027/1/1 | 2032/1/1 |
| 32 | | Instrumentation | 93.55 months | 2027/11/1 | 2034/12/31 |
| 37 | | Inj. & Extr. | 104.35 months | 2027/1/1 | 2034/12/31 |
| 42 | | Control System | 65.25 months | 2027/1/1 | 2032/1/1 |
| 49 | | Mechanical Systems | 65.25 months | 2027/1/1 | 2032/1/1 |
| 54 | | Booster | 104.35 months | 2027/1/1 | 2034/12/31 |
| 55 | | Superconducting RF System | 78.3 months | 2027/1/1 | 2032/12/31 |
| 60 | | RF Power Source | 45.6 months | 2027/1/1 | 2030/6/30 |
| 65 | | Magnets | 84.75 months | 2027/1/1 | 2033/6/30 |
| 69 | | Magnet Power Supplies | 63.05 months | 2027/1/1 | 2031/10/31 |
| 73 | | Vacuum System | 65.25 months | 2027/1/1 | 2032/1/1 |
| 78 | | Instrumentation | 93.55 months | 2027/11/1 | 2034/12/31 |
| 83 | | Inj. & Extr. | 104.35 months | 2027/1/1 | 2034/12/31 |
| 88 | | Control System | 65.25 months | 2027/1/1 | 2032/1/1 |
| 95 | | Mechanical Systems | 65.25 months | 2027/1/1 | 2032/1/1 |
| 100 | | Linac, Damping Ring and Sources | 104.35 months | 2027/1/1 | 2034/12/31 |
| 101 | | e- source | 52.15 months | 2027/1/1 | 2030/12/31 |
| 105 | | e+ source | 52.15 months | 2027/1/1 | 2030/12/31 |
| 109 | | Linac | 104.35 months | 2027/1/1 | 2034/12/31 |
| 110 | | RF System | 52.15 months | 2027/1/1 | 2030/12/31 |
| 116 | | RF Power Source | 45.6 months | 2027/1/1 | 2030/6/30 |
| 121 | | Magnets | 44.65 months | 2028/8/1 | 2032/1/1 |
| 125 | | Magnet Power Supplies | 37 months | 2029/1/1 | 2031/10/31 |

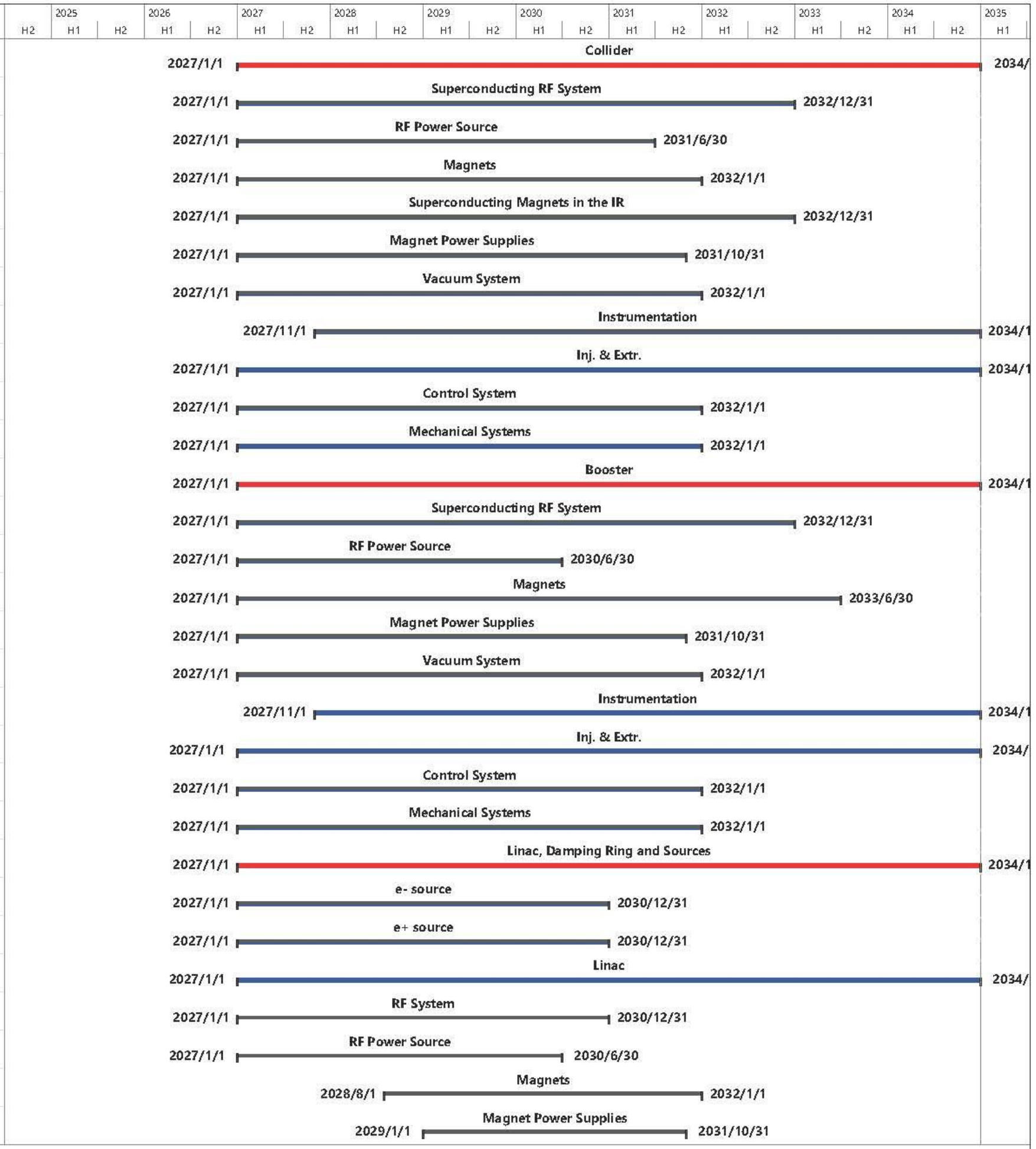

Project: Simple Project Plan
Date: 2023/10/18

| | | | | | |
|---|---|---|---|---|---|
| Task | Summary | Inactive Milestone | Duration-only | Start-only | External Milestone |
| Split | Project Summary | Inactive Summary | Manual Summary Rollup | Finish-only | Deadline |
| Milestone | Inactive Task | Manual Task | Manual Summary | External Tasks | Progress |

Manual Progress

Page 1

| 标识号 | 任务模式 | 任务名称 | 工期 | 开始时间 | 完成时间 |
|---|---|---|---|---|---|
| 129 | | Vacuum System | 65.25 months | 2027/1/1 | 2032/1/1 |
| 134 | | Instrumentation | 93.55 months | 2027/11/1 | 2034/12/31 |
| 139 | | Control System | 45.6 months | 2027/1/1 | 2030/6/30 |
| 146 | | Mechanical Systems | 32.75 months | 2029/7/1 | 2032/1/1 |
| 151 | | Damping Ring | 104.35 months | 2027/1/1 | 2034/12/31 |
| 152 | | RF System | 52.15 months | 2027/1/1 | 2030/12/31 |
| 156 | | RF Power Source | 32.6 months | 2027/1/1 | 2029/6/30 |
| 161 | | Magnets | 44.65 months | 2028/8/1 | 2032/1/1 |
| 165 | | Magnet Power Supplies | 37 months | 2029/1/1 | 2031/10/31 |
| 169 | | Vacuum System | 65.25 months | 2027/1/1 | 2032/1/1 |
| 174 | | Instrumentation | 93.55 months | 2027/11/1 | 2034/12/31 |
| 179 | | Inj. & Extr. | 91.35 months | 2027/1/1 | 2033/12/31 |
| 184 | | Control System | 39.15 months | 2027/1/1 | 2030/1/1 |
| 191 | | Mechanical Systems | 32.75 months | 2029/7/1 | 2032/1/1 |
| 196 | | Transport lines | 104.35 months | 2027/1/1 | 2034/12/31 |
| 197 | | RF system | 52.15 months | 2027/1/1 | 2030/12/31 |
| 203 | | RF Power Source | 32.6 months | 2027/1/1 | 2029/6/30 |
| 208 | | Booster to collider e+ transport line | 104.35 months | 2027/1/1 | 2034/12/31 |
| 209 | | Magnets | 71.7 months | 2027/1/1 | 2032/6/30 |
| 213 | | Magnet Power Supplies | 37 months | 2029/1/1 | 2031/10/31 |
| 217 | | Vacuum System | 65.25 months | 2027/1/1 | 2032/1/1 |
| 222 | | Instrumentation | 93.55 months | 2027/11/1 | 2034/12/31 |
| 227 | | Control System | 65.25 months | 2027/1/1 | 2032/1/1 |
| 234 | | Mechanical Systems | 32.75 months | 2029/7/1 | 2032/1/1 |
| 239 | | Booster to collider e- transport line | 104.35 months | 2027/1/1 | 2034/12/31 |
| 240 | | Magnets | 71.7 months | 2027/1/1 | 2032/6/30 |
| 244 | | Magnet Power Supplies | 37 months | 2029/1/1 | 2031/10/31 |
| 248 | | Vacuum System | 65.25 months | 2027/1/1 | 2032/1/1 |
| 253 | | Instrumentation | 93.55 months | 2027/11/1 | 2034/12/31 |

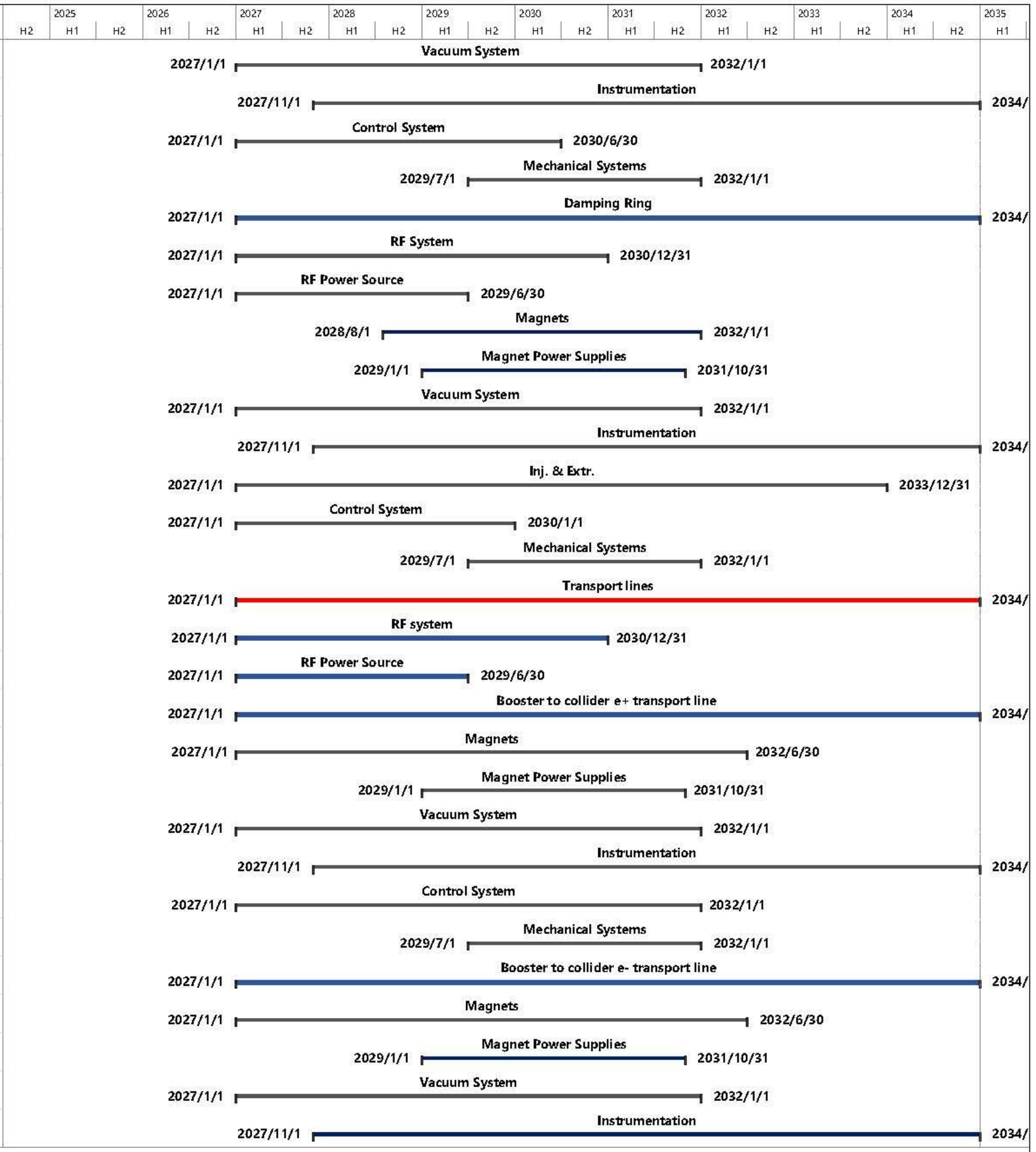

Project: Simple Project Plan
Date: 2023/10/18

Task | Split | Milestone | Summary | Project Summary | Inactive Task | Inactive Milestone | Inactive Summary | Manual Task | Duration-only | Manual Summary Rollup | Manual Summary | Start-only | Finish-only | External Tasks | External Milestone | Deadline | Progress | Manual Progress

Page 2

| 标识号 | 任务模式 | 任务名称 | 工期 | 开始时间 | 完成时间 |
|---|---|---|---|---|---|
| 258 | | Control System | 65.25 months | 2027/1/1 | 2032/1/1 |
| 265 | | Mechanical Systems | 32.75 months | 2029/7/1 | 2032/1/1 |
| 270 | | Linac to booster e+ transport line | 104.35 months | 2027/1/1 | 2034/12/31 |
| 271 | | Magnets | 71.7 months | 2027/1/1 | 2032/6/30 |
| 275 | | Magnet Power Supplies | 37 months | 2029/1/1 | 2031/10/31 |
| 279 | | Vacuum System | 65.25 months | 2027/1/1 | 2032/1/1 |
| 284 | | Instrumentation | 93.55 months | 2027/11/1 | 2034/12/31 |
| 289 | | Control System | 65.25 months | 2027/1/1 | 2032/1/1 |
| 296 | | Mechanical Systems | 32.75 months | 2029/7/1 | 2032/1/1 |
| 301 | | Linac to booster e- transport line | 104.35 months | 2027/1/1 | 2034/12/31 |
| 302 | | Magnets | 71.7 months | 2027/1/1 | 2032/6/30 |
| 306 | | Magnet Power Supplies | 37 months | 2029/1/1 | 2031/10/31 |
| 310 | | Vacuum System | 65.25 months | 2027/1/1 | 2032/1/1 |
| 315 | | Instrumentation | 93.55 months | 2027/11/1 | 2034/12/31 |
| 320 | | Control System | 65.25 months | 2027/1/1 | 2032/1/1 |
| 327 | | Mechanical Systems | 32.75 months | 2029/7/1 | 2032/1/1 |
| 332 | | Common systems | 104.45 months | 2025/1/1 | 2033/1/1 |
| 333 | | Cryo | 65.25 months | 2027/1/1 | 2032/1/1 |
| 339 | | Survey & alignment | 78.35 months | 2027/1/1 | 2033/1/1 |
| 343 | | Radiation protection | 91.35 months | 2025/1/1 | 2032/1/1 |
| 349 | | Conventional facilities | 65.25 months | 2027/1/1 | 2032/1/1 |
| 350 | | Civil | 58.65 months | 2027/1/1 | 2031/7/1 |
| 387 | | Mechanical and Electrical Systems | 65.25 months | 2027/1/1 | 2032/1/1 |
| 388 | | Mechanical Structure and Equipment | 65.25 months | 2027/1/1 | 2032/1/1 |
| 389 | | Temperary Construction | 65.25 months | 2027/1/1 | 2032/1/1 |
| 390 | | Independent Expenses | 65.25 months | 2027/1/1 | 2032/1/1 |
| 391 | | Other Expenses | 65.25 months | 2027/1/1 | 2032/1/1 |

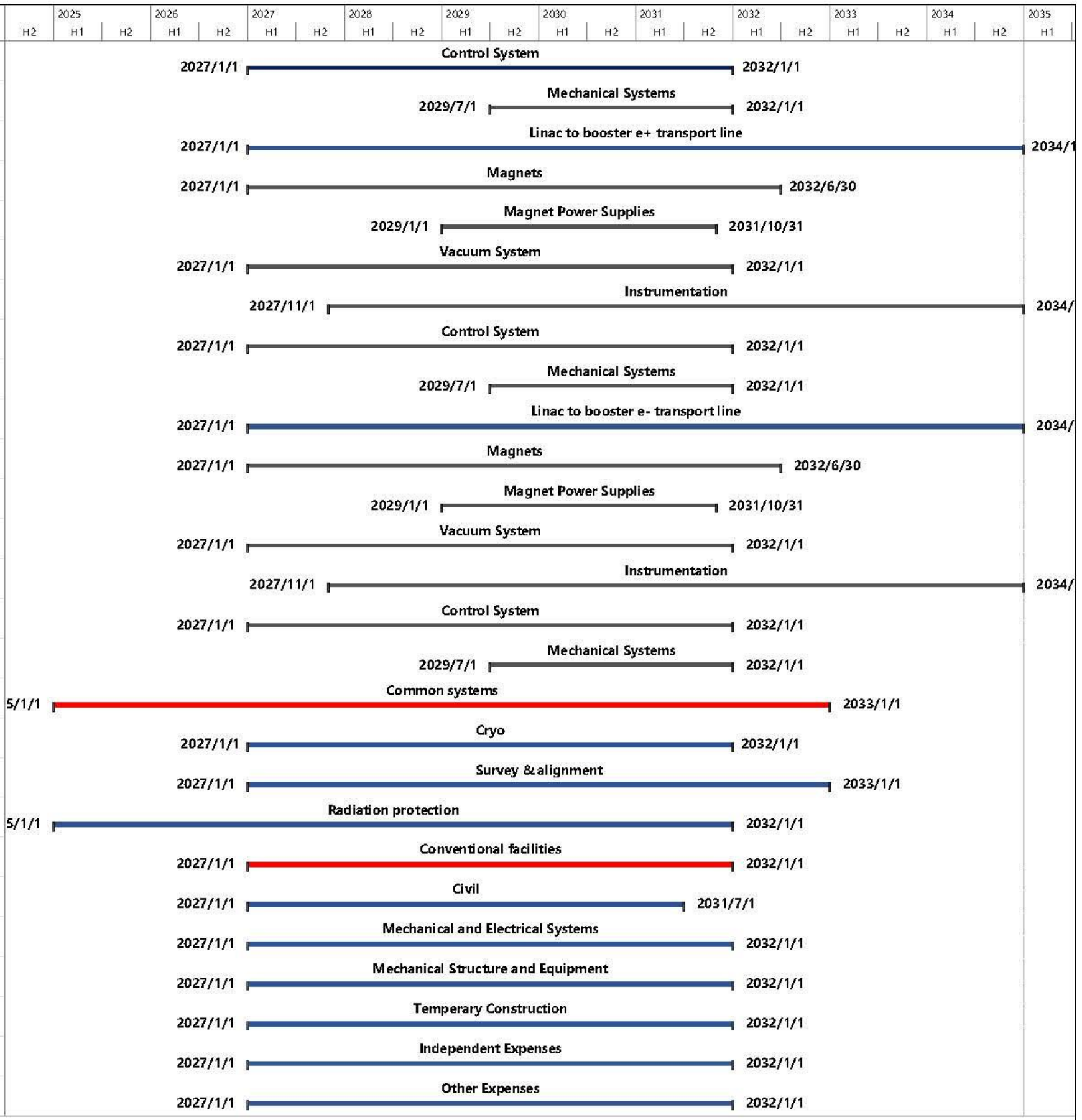

Project: Simple Project Plan
Date: 2023/10/18

Task | Split | Milestone | Summary | Project Summary | Inactive Task | Inactive Milestone | Inactive Summary | Manual Task | Duration-only | Manual Summary Rollup | Manual Summary | Start-only | Finish-only | External Tasks | External Milestone | Deadline | Progress | Manual Progress

**Table 12.4.2:** A preliminary timeline of the CEPC Collider

| 标识号 | 任务模式 | 任务名称 | 工期 | 开始时间 | 完成时间 |
|---|---|---|---|---|---|
| 1 | | **Collider** | **104.35 months** | **2027/1/1** | **2034/12/31** |
| 2 | | **Superconducting RF System** | **78.3 months** | **2027/1/1** | **2032/12/31** |
| 3 | | Bidding procurement, material procurement | 13.05 months | 2027/1/1 | 2027/12/31 |
| 4 | | Processing and manufacturing | 26.1 months | 2028/1/1 | 2029/12/31 |
| 5 | | Integration, testing | 39.15 months | 2029/1/1 | 2031/12/31 |
| 6 | | Transportation, installation, and configuration | 13.1 months | 2032/1/1 | 2032/12/31 |
| 7 | | **RF Power Source** | **58.6 months** | **2027/1/1** | **2031/6/30** |
| 8 | | Bidding procurement | 6.45 months | 2027/1/1 | 2027/6/30 |
| 9 | | Processing and manufacturing | 39.15 months | 2027/7/1 | 2030/6/30 |
| 10 | | Integration, testing | 6.6 months | 2030/7/1 | 2030/12/31 |
| 11 | | Transportation, installation, and configuration | 6.45 months | 2031/1/1 | 2031/6/30 |
| 12 | | **Magnets** | **65.25 months** | **2027/1/1** | **2032/1/1** |
| 13 | | Bidding procurement | 45.6 months | 2027/1/1 | 2030/6/30 |
| 14 | | Processing and manufacturing | 47.95 months | 2027/5/1 | 2031/1/1 |
| 15 | | Integration, testing | 48 months | 2027/6/1 | 2031/2/1 |
| 16 | | Transportation and installation | 58.8 months | 2027/7/1 | 2032/1/1 |
| 17 | | **Superconducting Magnets in the IR** | **78.3 months** | **2027/1/1** | **2032/12/31** |
| 18 | | Procurement and production of superconducting cables, processing, manufacturing, and debugging of magnet | 13.05 months | 2027/1/1 | 2027/12/31 |
| 19 | | Engineering design and bidding of superconducting magnet | 13.05 months | 2027/1/1 | 2027/12/31 |
| 20 | | Development of magnetic field measurement and quench protection system | 26.1 months | 2027/1/1 | 2028/12/31 |
| 21 | | Single superconducting magnet processing and manufacturing, low-temperature vertical test | 39.15 months | 2028/1/1 | 2030/12/31 |
| 22 | | Assembly of superconducting magnet and cryostat, low temperature horizontal test, transportation and installation | 26.15 months | 2031/1/1 | 2032/12/31 |
| 23 | | **Magnet Power Supplies** | **63.05 months** | **2027/1/1** | **2031/10/31** |
| 24 | | Plan review, bidding procurement | 6.45 months | 2027/1/1 | 2027/6/30 |
| 25 | | Processing and manufacturing | 52.15 months | 2027/7/1 | 2031/6/30 |
| 26 | | Integration, testing | 43.55 months | 2028/7/1 | 2031/10/31 |

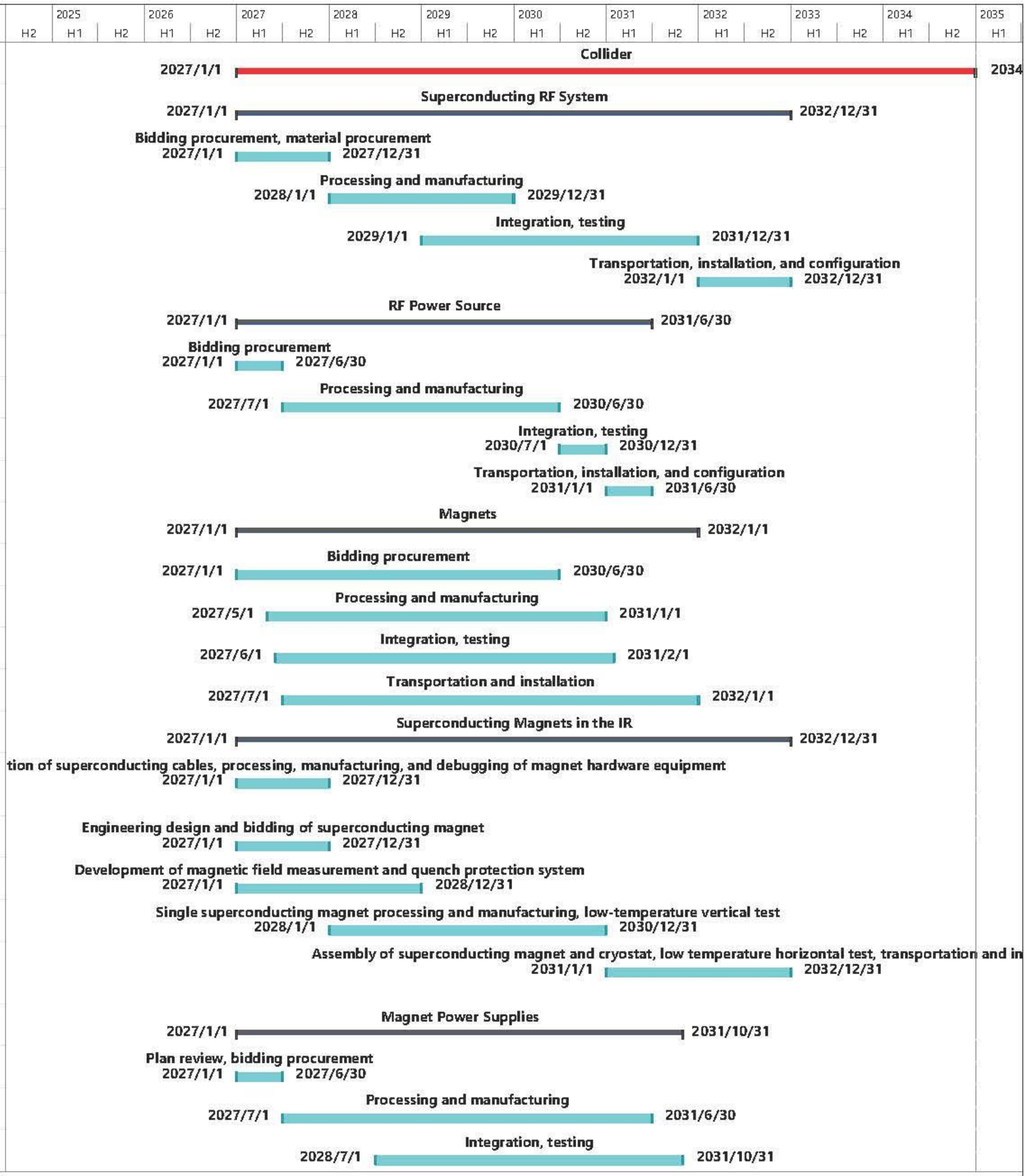

Project: Simple Project Plan
Date: 2023/10/18

Task, Split, Milestone, Summary, Project Summary, Inactive Task, Inactive Milestone, Inactive Summary, Manual Task, Duration-only, Manual Summary Rollup, Manual Summary, Start-only, Finish-only, External Tasks, External Milestone, Deadline, Progress, Manual Progress

Page 1

| 标识号 | | 任务模式 | 任务名称 | 工期 | 开始时间 | 完成时间 |
|---|---|---|---|---|---|---|
| 27 | | | **Vacuum System** | **65.25 months** | **2027/1/1** | **2032/1/1** |
| 28 | | | Design and prototype verification | 13.1 months | 2027/1/1 | 2028/1/1 |
| 29 | | | Bidding procurement | 6.6 months | 2028/1/1 | 2028/7/1 |
| 30 | | | Processing and manufacturing | 45.75 months | 2028/7/1 | 2032/1/1 |
| 31 | | | Integration, testing | 39.2 months | 2029/1/1 | 2032/1/1 |
| 32 | | | **Instrumentation** | **93.55 months** | **2027/11/1** | **2034/12/31** |
| 33 | | | Design and prototype verification | 21.8 months | 2027/11/1 | 2029/6/30 |
| 34 | | | Bidding procurement | 5.35 months | 2028/12/1 | 2029/4/30 |
| 35 | | | Processing and manufacturing | 52.25 months | 2029/5/1 | 2033/4/30 |
| 36 | | | Installation and offline debugging | 52.2 months | 2031/1/1 | 2034/12/31 |
| 37 | | | **Inj. & Extr.** | **104.35 months** | **2027/1/1** | **2034/12/31** |
| 38 | | | Design and prototype verification | 39.1 months | 2027/1/1 | 2029/12/31 |
| 39 | | | Processing and manufacturing | 59.85 months | 2029/6/1 | 2033/12/31 |
| 40 | | | Integration, testing | 39.2 months | 2031/1/1 | 2034/1/1 |
| 41 | | | Installation and offline debugging | 13.1 months | 2034/1/1 | 2034/12/31 |
| 42 | | | **Control System** | **65.25 months** | **2027/1/1** | **2032/1/1** |
| 43 | | | Control scheme design and laboratory prototype testing | 13.1 months | 2027/1/1 | 2028/1/1 |
| 44 | | | Equipment procurement | 19.6 months | 2028/1/1 | 2029/6/30 |
| 45 | | | software development | 13.1 months | 2028/7/1 | 2029/6/30 |
| 46 | | | Laboratory software and hardware testing | 8.65 months | 2028/12/1 | 2029/7/31 |
| 47 | | | On site installation and debugging | 18.5 months | 2029/8/1 | 2030/12/31 |
| 48 | | | Delivery, trial operation, joint debugging, etc | 13.1 months | 2031/1/1 | 2032/1/1 |
| 49 | | | **Mechanical Systems** | **65.25 months** | **2027/1/1** | **2032/1/1** |
| 50 | | | Bidding procurement | 4.3 months | 2027/1/1 | 2027/4/30 |
| 51 | | | First machine processing, assembly, and acceptance | 43.6 months | 2027/5/1 | 2030/8/31 |
| 52 | | | Batch processing and manufacturing | 56.6 months | 2027/7/1 | 2031/10/31 |
| 53 | | | Testing, transportation, acceptance | 52.25 months | 2028/1/1 | 2032/1/1 |

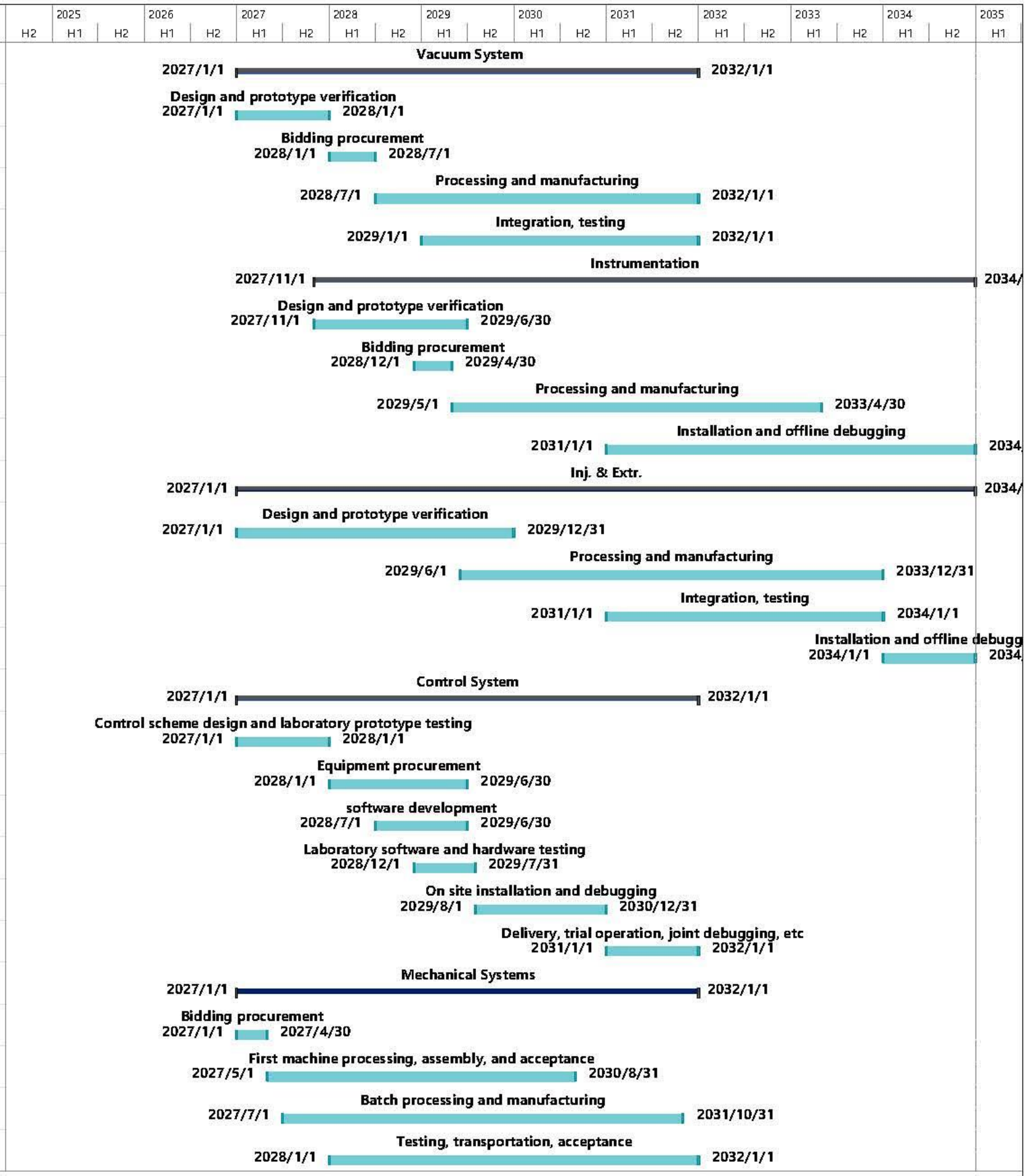

Project: Simple Project Plan
Date: 2023/10/18

Task, Split, Milestone, Summary, Project Summary, Inactive Task, Inactive Milestone, Inactive Summary, Manual Task, Duration-only, Manual Summary Rollup, Manual Summary, Start-only, Finish-only, External Tasks, External Milestone, Deadline, Progress, Manual Progress

# Appendix 1: Parameter List

## A1.1: Collider

| **Fundamental constants** | **Unit** | **H** | **Z** | **W** | $\boldsymbol{t\bar{t}}$ |
|---|---|---|---|---|---|
| electronic charge | C | | 1.60E-19 | | |
| speed of light | m/s | | 3.00E08 | | |
| $C_q$ | | | 3.83E-13 | | |
| fine structure constant | α | | 0.0073 | | |
| classical radius of the electron [$r_e$] | m | | 2.82E-15 | | |
| Euler's constant [$\gamma_E$] | | | 0.577 | | |
| electron Compton wavelength [$\lambda_e$] | m | | 3.86E-13 | | |
| rest mass energy of the electron | MeV | | 5.11E-01 | | |
| | | | | | |
| **Accelerator Parameters** | | | | | |
| Beam energy [E] | GeV | 120 | 45.5 | 80 | 180 |
| Circumference [C] | km | | 100 | | |
| Luminosity [L] | $cm^{-2}s^{-1}$ | 5E34 | 115E34 | 16E35 | 0.5E34 |
| SR power/beam [P] | MW | | 30 | | |
| Bending radius [ρ] | m | | 10700 | | |
| Number of IPs $N_{IP}$ | | | 2 | | |
| Bunch number $n_b$ | | 268 | 11934 | 1297 | 35 |
| Filling factor [κ] | | | 0.67 | | |
| Lorentz factor [γ] | | 234834 | 89041 | 156556 | 352250 |
| Revolution period [$T_0$] | s | | 3.33E-04 | | |
| Revolution frequency [$f_0$] | Hz | | 3003 | | |
| Magnetic rigidity [Bρ] | T·m | 400.27 | 151.77 | 266.85 | 600.42 |
| Momentum compaction factor [$\alpha_p$] | | 0.71E-05 | 1.43E-05 | 1.43E-05 | 0.71E-05 |
| Energy acceptance requirement [η] | % | 1.6 | 1.0 | 1.05 | 2.0 |
| Cross-section for radiative Bhabha scattering [$\sigma_{BB}$] | $cm^2$ | 1.27E-25 | 1.34E-25 | 1.36E-25 | 1.32E-25 |
| Lifetime due to radiative Bhabha scattering [$\tau_L$] | min | 40 | 90 | 60 | 81 |
| build-up time of polarization [$\tau_{BKS}$] | hr | 2.0 | 256 | 15 | 0.26 |
| | | | | | |
| **Beam Parameters** | | | | | |
| Beam current [I] | mA | 16.7 | 803.5 | 84.1 | 3.3 |
| Bunch population [$N_e$] | $10^{10}$ | 13.0 | 14.0 | 13.5 | 20.0 |
| emittance-horizontal [$\varepsilon_x$] | nm·rad | 0.64 | 0.27 | 0.87 | 1.40 |
| emittance-vertical [$\varepsilon_y$] | pm·rad | 1.3 | 1.4 | 1.7 | 4.7 |
| coupling factor [κ] | | 0.002 | 0.005 | 0.002 | 0.0034 |
| Bunch length SR [$\sigma_{s.SR}$] | mm | 2.3 | 2.5 | 2.5 | 2.2 |
| Bunch length total [$\sigma_{s.tot}$] | mm | 4.1 | 8.7 | 4.9 | 2.9 |
| | | | | | |
| **Interaction Region Parameters** | | | | | |
| Betatron function at IP-vertical [$\beta_y$] | mm | 1 | 0.9 | 1 | 2.7 |
| Betatron function at IP-horizontal [$\beta_x$] | m | 0.30 | 0.13 | 0.21 | 1.04 |
| Transverse size [$\sigma_x$] | μm | 14 | 6 | 13 | 39 |
| Transverse size [$\sigma_y$] | nm | 36 | 35 | 42 | 113 |
| Beam-beam parameter [$\xi_x$] | | 0.015 | 0.004 | 0.012 | 0.071 |
| Beam-beam parameter [$\xi_y$] | | 0.11 | 0.127 | 0.113 | 0.1 |
| Hourglass factor | $F_h$ | 0.90 | 0.97 | 0.90 | 0.89 |
| Lifetime due to Beamstrahlung [$\tau_{BS}$] | min | 40 | 2800 | 195 | 23 |
| **RF Parameters** | | | | | |
| RF voltage [$V_{rf}$] | GV | 2.20 | 0.12 | 0.70 | 10 |
| RF frequency [$f_{rf}$] | MHz | 650 | | | |
| Harmonic number [h] | | 216816 | | | |
| Synchrotron oscillation tune [$\nu_s$] | | 0.049 | 0.035 | 0.062 | 0.078 |
| Energy acceptance RF [$\eta$] | % | 2.2 | 1. 7 | 2.5 | 2.6 |
| **Synchrotron Radiation** | | | | | |
| SR loss/turn [$U_0$] | GeV | 1.8 | 0.037 | 0.357 | 9.1 |
| Damping partition number [$J_x$] | | 1 | | | |
| Damping partition number [$J_y$] | | 1 | | | |
| Damping partition number [$J_z$] | | 2 | | | |
| Energy spread SR [$\sigma_{\delta SR}$] | % | 0.10 | 0.04 | 0.07 | 0.15 |
| Energy spread total [$\sigma_{\delta tot}$] | % | 0.17 | 0.13 | 0.14 | 0.20 |
| Transverse damping time [$\tau_x$] | ms | 44.6 | 816 | 150 | 13.2 |
| Longitudinal damping time [$\tau_\varepsilon$] | ms | 22.3 | 408 | 75 | 6.6 |
| **Ring Parameter** | | | | | |
| Circumference [C] | km | 100 | | | |
| Revolution period [$T_0$] | s | 3.34E-04 | | | |
| Revolution frequency [$f_0$] | Hz | 2997.4344 | | | |
| Betatron tune [$Q_{x/y}$] | | 445/445 | 317/317 | 317/317 | 445/445 |
| Damping time [$\tau_{x/y/s}$] | ms | 44.6/44.6 /22.3 | 816/816/ 408 | 150/150/ 75 | 13.2/13.2 /6.6 |
| Number of arc regions | | 8 | | | |
| Number of interaction regions | | 2 | | | |
| Number of straight section regions | | 4 | | | |
| Number of RF regions | | 2 | | | |
| Length of arc regions | m | 10270.44 / 10185.70 | | | |
| Length of interaction regions | m | 3337.13 | | | |
| Length of straight section regions | m | 986.84 | | | |
| Length of RF regions | | | | | |
| Total number of dipoles | | 3776.90 | | | |
| Total number of quadrupoles | | 3170 | | | |
| Total number of sextupoles | | 4140 | | | |
| Total number of correctors | | 3176 | | | |
| | | 7088 | | | |
| **Regular lattice period parameters** | | | | | |
| Lattice type | | FODO | | | |
| Cell numbers in each period | | 23 | | | |
| Phase advance/2$\pi$ (x/y) | | 5.75/5.83 | 3.83/3.83 | 3.83/3.83 | 5.75/5.75 |
| Period length | m | 1256.38 | | | |
| Dipole type in regular lattice | | $B_0$ | | | |
| Maximum $\beta$ value | m | 92.4 | 94.0 | 94.0 | 92.4 |
| Minimum $\beta$ value | m | 16.1 | 31,7 | 31.7 | 16.1 |
| Maximum dispersion | m | 0.156 | 0.288 | 0.288 | 0.156 |
| Number of dipoles [$B_0$] | | | 46 | | |
| Dipole length (average) | m | | 21.4 | | |
| Strength of dipole | T | 0.0398 | 0.0151 | 0.0265 | 0.0597 |
| Quadrupole type in regular lattice | | | QF/D | | |
| Number of quadrupoles [QF/D] | | | 46 | | |
| Quadrupole length | m | | 3 | | |
| Strength of quadrupole | $Tm^{-1}$ | 7.1 | 1.9 | 3.3 | 10.6 |
| Sextupole type in regular lattice | | | SF/D | | |
| Number of sextupoles [SF/D] | | | 8/8 | | |
| Sextupole length | m | | 1.4/2.8 | | |
| Strength of SF | $Tm^{-2}$ | 550 | 209 | 367 | 825 |
| Strength of SD | $Tm^{-2}$ | 565 | 214 | 376 | 847 |
| **Main RF parameter** | | | | | |
| Frequency | GHz | | 0.65 | | |
| Harmonic number | | | 216816 | | |
| Cavity type | | 2-cell | 1-cell | 2-cell | 5-cell |
| Cavity operating voltage | MV | 11.5 | 4.0 | 7.3 | 31.8 |
| Cavity operating gradient | MV/m | 24.9 | 17.4 | 15.9 | 27.6 |
| Number of cavities per cryomodule | | 6 | 1 | 6 | 4 |
| Cavity active length | m | 0.46 | 0.23 | 0.46 | 1.15 |
| Cryomodule length | m | 11 | 3 | 11 | 11 |
| Total number of cryomodules | | 32 | 60 | 32 | 48 |
| Total number of cavities | | 192 | 60 | 192 | 192 |

## A1.2: Booster

| Accelerator Parameters | Unit | Injection | Extraction | | | |
|---|---|---|---|---|---|---|
| | | | $t\bar{t}$ | Higgs | W | Z |
| Beam energy [E] | GeV | 30 | 180 | 120 | 80 | 45.5 |
| Circumference [C] | km | | | 100 | | |
| Revolution frequency [$f_0$] | kHz | 2.99 | 2.99 | 2.99 | 2.99 | 2.99 |
| $N_B$ / beam | | | 35 | 268 | 1297 | 3978 |
| Lorentz factor [$\gamma$] | | 5.87E4 | 3.52E5 | 2.35E5 | 1.57E5 | 8.90E4 |
| Magnetic rigidity [$B\rho$] | T.m | 100 | 600 | 400.28 | 266.85 | 151.77 |
| Beam current [I] | mA | | 0.11 | 0.98 | 2.85 | 9.5 |
| Bunch charge [$N_e$] | nC | | 0.99 | 0.71 | 0.73 | 0.8 |
| emittance-horizontal [$\varepsilon_x$] | nm.ra d | 6.5 | 2.83 | 1.26 | 0.56 | 0.19 |
| RF voltage [$V_{rf}$] | GV | 0.35 (H) | 9.7 | 2.17 | 0.87 | 0.46 |
| RF frequency [$f_{rf}$] | GHz | 1.3 | 1.3 | 1.3 | 1.3 | 1.3 |
| Harmonic number [h] | | | | 433633 | | |
| SR loss / turn [$U_0$] | GeV | 6.5E-3 | 8.45 | 1.69 | 0.33 | 0.034 |
| Transverse damping time [$\tau_x$] | ms | 3.1 E3 | 14.2 | 47.6 | 160.8 | 879 |
| Betatron tune $\nu_x$ | | 321.23 | 321.23 | 321.23 | 321.23 | 321.23 |
| Betatron tune $\nu_y$ | | 117.18 | 117.18 | 117.18 | 117.18 | 117.18 |
| Momentum compaction factor | | 1.12E-5 | 1.12E-5 | 1.12E-5 | 1.12E-5 | 1.12E-5 |
| Synchrotron oscillation tune [$\nu_s$] | | | 0.14 | 0.0943 | 0.0879 | 0.0879 |
| Energy acceptance RF [η] | % | 3.8 (H) | 1.78 | 1.59 | 2.6 | 3.4 |
| Energy spread [$\sigma_\delta$] in equilibrium | % | 0.025 | 0.15 | 0.099 | 0.066 | 0.037 |
| Energy spread [$\sigma_\delta$] from Linac | % | 0.15 | | | | |
| Bunch length [$\sigma_z$] in equilibrium | mm | | 1.8 | 1.85 | 1.3 | 0.75 |
| Bunch length [$\sigma_z$] from Linac | mm | 0.4 | | | | |
| Number of arcs | | 8 | | | | |
| Number of short straight sections | | 4 | | | | |
| Number of long straight sections | | 2 | | | | |
| Number of straight sections with RF | | 2 | | | | |
| Number of bypass at IP | | 2 | | | | |
| Total number of dipoles | | 14866 | | | | |
| Total number of quadrupoles [QF/D] | | 3458 | | | | |
| Total number of sextupoles [SF/D] | | 100 | | | | |
| **Regular lattice period parameters** | | | | | | |
| Lattice type | | TME | | | | |
| Phase advance (horizontal/vertical) | | 100°/28° | | | | |
| Cell length | m | 78 | | | | |
| Number of dipoles in a cell | | 14 | | | | |
| Dipole length | m | 4.7 | | | | |
| Defection angle of dipole | mrad | 0.44 | | | | |
| Magnetic field of the dipole at injection | T | 0.00945 | | | | |
| Magnetic field of the dipole at ejection | T | | 0.0564 | 0.0376 | 0.0252 | 0.0143 |
| Number of quadrupoles | | 3 | | | | |
| Quadrupole length | m | 2.0/0.7 | | | | |
| Strength of QF | $m^{-2}$ | 0.02 | | | | |
| Strength of QD | $m^{-2}$ | -0.02 | | | | |
| Sextupole length (combined) | m | 4.7 | | | | |
| Strength of SF | $m^{-3}$ | 0.0267 | | | | |
| Strength of SD | $m^{-3}$ | -0.0319 | | | | |
| Length of BH/BV | m | 0.3 | | | | |
| Maximum strength of BH | T | 0.02 | | | | |
| Maximum strength of BV | T | 0.02 | | | | |
| Maximum horizontal β value | m | 120 | | | | |
| Minimum horizontal β value | m | 28 | | | | |
| Maximum Vertical β value | m | 230 | | | | |
| Minimum Vertical β value | m | 91 | | | | |
| Maximum dispersion | m | 0.256 | | | | |
| **Dispersion suppressors** | | | | | | |
| Length | m | 99 | | | | |
| Horizontal phase advance/2π | | 0.27 | | | | |
| Vertical phase advance/2π | | 0.078 | | | | |
| Number of dipoles | | 7 | | | | |
| Dipole length | m | 4.7 | | | | |
| Strength of dipole at injection | T | 0.00945 | | | | |
| Strength of dipole at ejection | T | | 0.0564 | 0.0376 | 0.0252 | 0.0143 |
| Number of quadrupoles | | 3 | | | | |
| Quadrupole length | m | 2.0/1.0 | | | | |
| Strength of SF | $m^{-3}$ | 0.0 | | | | |
| Strength of SD | $m^{-3}$ | -0.0 | | | | |
| **Arcs** | **Unit** | **Value** | | | | |
| Length | m | 10219 | | | | |
| Number of cells per ARC | | 131 | | | | |
| Number of dispersion suppressors per arc | | 2 | | | | |
| Horizontal phase advance/2π | | 35.4 | | | | |
| Vertical phase advance/2π | | 10.2 | | | | |

| **Type** | **Length[m]** | **Strength** |
|---|---|---|
| B | 4.7 | 0.00945~0.0564 [T] |
| QF/D | 2.0/1.0 | 0.02[$m^{-2}$] |
| SF | 4.7 | 0.0267 [$m^{-3}$] |
| SD | 4.7 | -0.0319[$m^{-3}$] |
| **Short Straight section** | | |
| Length | m | 1201 |
| Horizontal phase advance/2π | | 2.24 |
| Vertical phase advance/2π | | 2.25 |
| **Type** | **Length[m]** | **Strength** |
| QF/DM1 | 2 | 0.021~0.027 [$m^{-2}$] |
| **Long Straight section with RF** | | |
| Length | m | 3421 |
| Horizontal phase advance/2π | | 16.82 |
| Vertical phase advance/2π | | 15.87 |
| **Type** | **Length[m]** | **Strength** |
| QF/DRF | 1.5 | 0.0608 [$m^{-2}$] |
| QF/DM4 | 1 | 0.02~ 0.036 [$m^{-2}$] |
| **Bypasses to IP** | | |
| Length | m | 3929 |
| Number of cells with dipoles | | 8 |
| Number of dispersion suppressors | | 0 |
| Horizontal phase advance/2π | | 10.02 |
| vertical phase advance/2π | | 10.00 |
| **Type** | **Length[m]** | **Strength** |
| B2 | 2.4 | 0.0084~0.053 [T] |
| QF/DM2 | 1 | 0.0281~0.0296 [$m^{-2}$] |

| **The main RF system parameter** | | | | | |
|---|---|---|---|---|---|
| Frequency | GHz | 1.3 | | | |
| Harmonic number | | 433633 | | | |
| Cavity type | | 9-cell cavity | | | |
| RF voltage (extraction) | GV | 9.7 | 2.17 | 0.87 | 0.46 |
| Cavity operating gradient | MV/m | 28.3/ 21.8* | 21.8 | 8.7 | 13.8 |
| Number of cavities per cryomodule | | 8 | | | |
| Cavity active length (nine-cells) | m | 1.038 | | | |
| Cryomodule length | m | 12 | | | |
| Total number of cryomodules | | 44 | 12 | 12 | 4 |
| Total number of cavities | | 352 | 96 | 96 | 32 |

* Cavity gradient of added $t\bar{t}$ cavities / cavity gradient of existing Higgs cavities

## A1.3: Linac, Damping Ring and Sources

| **Main parameter of Linac** | **Unit** | **Value** |
|---|---|---|
| Energy [$E_{e-}$/ $E_{e+}$] | GeV | 30 |
| Repetition rate [$f_{rep}$] | Hz | 100 |
| Bunch charge ($e^-$ / $e^+$) | nC | 1.5 |
| Energy spread ($e^-$ / $e^+$) [$\delta_E$] | $10^{-3}$ | 1.5 |
| Emittance ($e^-$ / $e^+$) | nm | 6.5 |
| Tunnel length [L] | m | 1800 |
| **Power source** | | |
| S-band power source (number, power, frequency) | 34, 80MW, 2860MHz | |
| C-band power source (number, power, frequency) | 236, 50MW, 5720MHz | |
| Solid-state power source (number, power, frequency) | 1, 20kW, 158.89MHz | |
| | 1, 20kW, 476.67MHz | |
| **Accelerating structure** | | |
| Large-aperture S-band accelerating structure (number, gradient, length) | 16, 22MV/m, 2.0m | |
| Normal S-band accelerating structure (number, gradient, length) | 8, 27MV/m, 3.1m | |
| | 85, 22MV/m, 3.1m | |
| C-band accelerating structure (number, gradient, length) | 470, 40MV/m, 1.8m | |
| Subharmonic buncher (number, voltage, length) | 1, 10 kV, 0.8m | |
| | 1, 120kV, 0.5m | |
| C-band deflecting cavity (number, voltage, length) | 1, 20MV, 1.8m | |
| **Magnet** | | **PS** |
| Solenoid (number, aperture, field, length) | 4, 90mm, 400Gs, 80mm | 4 |
| | 17, 210mm, 800Gs, 110mm | 17 |
| | 1, 118mm, 600Gs, 80mm | 1 |
| Quadrupole (number, aperture, gradient, length) | 15, 400mm, 0.5T, 1m | 15 |
| | 22, 34mm, 22T/m, 100mm | 11 |
| | 87, 34mm, 24T/m, 200mm | 49 |
| | 44, 34mm, 26T/m, 400mm | 44 |
| | 104, 24mm, 50T/m, 300mm | 52 |
| | 52, 24mm, 50T/m, 600mm | 52 |
| | 6, 60mm, 10T/m, 100mm | 3 |
| | 3, 60mm, 10T/m, 200mm | 6 |
| | 36, 150mm, 7.9T/m, 300mm | 18 |
| | 18, 150mm, 7.9T/m, 600mm | 18 |
| Dipole (number, gap, field, length) | 1, 35mm, 0.33T, 262mm | 1 |
| | 4, 24mm, 1.0T, 5.847m, bidirectional power supply | 4 |
| | 5, 60mm, 0.42T, 698mm, bidirectional power supply, switching within 3s | 5 |
| | 5, 35mm, 1.0T, 698mm | 5 |
| | 4, 35mm, 1.0T, 1.047m, power supply switching within 3s, vertical bending | 4 |
| | 1, 50mm, 10Gs, 30mm | 2 |
| | 1, 136mm, 50Gs, 400mm | 2 |
| Corrector (number, gap, field, length) | 25, 150mm, 50Gs, 100mm | 50 |
| | 2, 50mm, 10 Gs, 24mm | 2 |
| | 100, 34mm, 450Gs, 100mm | 100 |
| | 146, 24mm, 0.25T, 200mm | 146 |
| **Beam diagnostics** | | |
| BPM (number, aperture, resolution, length) | 150, 30/20mm, 10μm, 0.2m | |
| ICT (number, aperture, length) | 63, 35mm, 40mm | |
| PR (number, aperture, resolution, length) | 30, 35mm, 20μm, 0.3m | |
| **Beam dump** | | |
| | 1, 60MeV, 2mm*1mm, 10nC, 1Hz | |
| | 1, 4GeV, 0.5mm*0.5mm, 10nC, 1Hz | |
| | 1, 250MeV, 1.5mm*1.5mm, 5nC, 1Hz | |
| | 1, 1.1GeV, 0.5mm*0.5mm, 5nC, 1Hz | |
| | 1, 6GeV, 0.5mm*0.3mm, 3nC, 1Hz | |
| | 2, 30GeV, 0.3mm*0.3mm, 3nC, 1Hz | |
| Number, energy, beam size, bunch charge, repetition rate | 1, 250MeV, 1.5mm*1.5mm, 5nC, 100Hz | |
| **Electron Gun** | | |
| Gun type | Thermionic Triode Gun | |
| Cathode | Dispenser cathode | |
| Beam Current (max.) | A | 15 |
| High Voltage of Anode | kV | 150-200 |
| Bias Voltage of Grid | V | 0 ~ -200 |
| Pulse length（FWHM） | ns | 1 |
| Repetition Rate | Hz | 100 |
| **Positron source** | | |
| Electron beam energy on the target | GeV | 4 |
| Electron bunch charge on the target | nC | 3.5~7.0 |
| Target material | / | W |
| Target thickness | mm | 15 |
| Flux Concentrator magnetic field | T | 5.5 |
| Positron bunch charge after capture | nC | 1.9~3.9 |
| Positron energy after capture section | MeV | 200 |
| **Sub-harmonic buncher (SHB1/SHB2)** | | |
| Frequency | MHz | 158.89/476.67 |
| Shunt Impedance | MΩ | 1.46/2.53 |
| Unloaded Q | / | 8475/12431 |
| VSWR | / | <1.05 / <1.05 |
| $E_{surface,\ max}$ @100 kV | MV/m | 6.4/6.1 |
| Required power@100 kV | kW | 10/7 |
| RF Structure type | / | Re-entrant SW |
| **Buncher** | | |
| Frequency | MHz | 2860 |
| Phase advance | / | 2π/3 |
| Cell number | / | 6 |
| Phase velocity | / | 0.75 |
| Group velocity | / | 0.0193 |
| Attenuation constant | Np/m | 0.147 |
| Shunt impedance | MΩ/m | 33.2 |
| Unloaded Q | / | 11083 |
| Bunching voltage (Max) | MV | 1.2 |
| VSWR | / | <1.2 |
| Input power | MW | 6 |
| RF Structure type | / | TW/CI |
| **Accelerating structure (S-band)** | | |
| Frequency | MHz | 2860 |
| Operation temperature | $^{o}$C | 30.0 ± 0.1 |
| Number of cells | - | 84 +2 coupler cells |
| Section length | m | 3.1 |
| Phase advance per cell | - | 2π/3 - mode |
| Cell length | mm | 34.965 |
| Disk thickness (t) | mm | 5.5 |
| Iris diameter (2a) | mm | 26.23 ~ 19.24 |
| Cell diameter (2b) | mm | 83.460 ~ 81.78 |
| Shunt impedance ($r_0$) | MΩ/m | 60.3 ~ 67.8 |
| Q factor | - | 15465 ~ 15373 |
| Group velocity ($v_g$/c) | - | 0.02 ~ 0.087 |
| Filling time | ns | 780 |
| Attenuation factor | Neper | 0.46 |
| **Accelerating structure (C-band)** | | |
| Operation frequency | MHz | 5720 |
| Operation temperature | $^oC$ | 30.0 ± 0.1 |
| Number of cells | - | 87 +2 coupler cells |
| Section length | m | 1.8 |
| Phase advance per cell | - | $3\pi/4$ - mode |
| Cell length | mm | 19.654 |
| Disk thickness (t) | mm | 4.5 |
| Average iris diameter (2a) | mm | 14.04 |
| Average cell diameter (2b) | mm | 45.6 |
| Shunt impedance ($r_0$) | MΩ/m | 58.4 ~ 73.3 |
| Q factor | - | 11358 ~ 11186 |
| Group velocity ($v_g$/c) | - | 0.028 ~ 0.00096 |
| Filling time | ns | 350 |
| Attenuation factor | Neper | 0.56 |
| **Damping Ring** | | |
| Energy | GeV | 1.1 |
| Circumference | m | 147 |
| Repetition frequency | Hz | 100 |
| Bending radius | m | 2.87 |
| Dipole strength $B_0$ | T | 1.28 |
| $U_0$ | keV | 94.6 |
| Damping time x/y/z | ms | 11.4/11.4/5.7 |
| $\delta_0$ | % | 0.056 |
| Nature $\varepsilon_0$ (normalized) | µm.rad | 94.4 |
| Nature $\sigma_z$ | mm | 4.4 |
| $\varepsilon_{inj}$ (normalized) | µm.rad | 2500 |
| $\varepsilon_{ext\ x/y}$ (normalized) | µm.rad | 166/75 |
| $\delta_{inj}$ / $\delta_{ext}$ | % | 0.18 /0.056 |
| Energy acceptance by RF | % | 1.8 |
| $f_{RF}$ | MHz | 650 |
| $V_{RF}$ | MV | 2.5 |

# Appendix 2: Technical Component List

| | System | Name | Number | Typical parameters | Remarks |
|---|---|---|---|---|---|
| **1** | **Magnet** | | | | |
| | Collider | | | | |
| | | B0A | 1024 | Dual aperture dipole, Gap 66mm, Field 597Gs, Length 19.7m | |
| | | B0B | 1920 | Dual aperture dipole, Gap 66mm, Field 597Gs, Length 23.1m | |
| | | B1 | 64 | Dual aperture dipole, Gap 66mm, Field 298.5Gs, Length 23.1m | |
| | | BMV01IRD | 4 | Dipole, Gap 36.9mm, Field 152Gs, Length 61m | |
| | | BMVIRD | 20 | Dipole, Gap 66mm, Field 729.6Gs, Length 44.2m | |
| | | BMHIRD | 16 | Dipole, Gap 66mm, Field 316.5Gs, Length 28.5m | |
| | | BGM1 | 32 | Dipole, Gap 66mm, Field 300.8Gs, Length 31.8m | |
| | | BGM2 | 32 | Dipole, Gap 66mm, Field 72.4Gs, Length 29m | |
| | | BMV01IRU | 4 | Dipole, Gap 36.9mm, Field 71Gs, Length 93.4m | |
| | | BMVIRU | 20 | Dipole, Gap 66mm, Field 306.2Gs, Length 69m | |
| | | BMHIRU | 16 | Dipole, Gap 66mm, Field 129.6Gs, Length 45m | |
| | | BRF0 | 2 | Dipole, Gap 66mm, Field 180.1Gs, Length 50m | |
| | | BRF | 16 | Dipole, Gap 66mm, Field 419.9Gs, Length 34.1m | |
| | | | | | |
| | | Q | 3008 | Dual aperture quadrupole, Gap 66mm, Field 10.6T/m, Length 3m | |
| | | QH | 8 | Dual aperture quadrupole, Gap 66mm, Field 10.6T/m, Length 1.5m | |
| | | Q1AIR | 4 | Superconducting quadrupole, Gap 40mm, Field 142T/m, Length 1.21m | |
| | | Q1BIR | 4 | Superconducting quadrupole, Gap 52mm, Field 85.5T/m, Length 1.21m | |
| | | Q2IR | 4 | Superconducting quadrupole, Gap 62mm, Field 97.7T/m, Length 1.5m | |
| | | Q3AIR | 8 | Quadrupole, Gap 40mm, Field 40T/m, Length 1.5m | |
| | | Q3BIR | 8 | Quadrupole, Gap 40mm, Field 40T/m, Length 1.5m | |
| | | QDVHIRD | 32 | Quadrupole, Gap 66mm, Field 18.4T/m, Length0.5m | |
| | | QFVIRD | 16 | Quadrupole, Gap 66mm, Field 18.2T/m, Length 1m | |
| | | QFHHIRD | 32 | Quadrupole, Gap 66mm, Field 22.4T/m, Length 0.6m | |
| | | QDHIRD | 16 | Quadrupole, Gap 66mm, Field 22T/m, Length 1.3m | |
| | | QCMIRD | 24 | Quadrupole, Gap 66mm, Field 22.4T/m, Length 3m | |
| | | QDCIRD | 16 | Quadrupole, Gap 66mm, Field 19.8T/m, Length 0.5m | |
| | | QFCIRD | 8 | Quadrupole, Gap 66mm, Field 19.7T/m, Length 1m | |
| | | QMIRD | 32 | Quadrupole, Gap 66mm, Field 23.9T/m, Length 3.5m | |
| | | QMIRD | 16 | Quadrupole, Gap 66mm, Field 21.9T/m, Length 0.6m | |
| | | QMIRD | 8 | Quadrupole, Gap 66mm, Field 21.2T/m, Length 1.3m | |
| | | QDGMH | 64 | Quadrupole, Gap 66mm, Field 22.1T/m, Length 0.6m | |
| | | QFGM | 32 | Quadrupole, Gap 66mm, Field 21.9T/m, Length 1.3m | |
| | | QFVIRU | 16 | Quadrupole, Gap 66mm, Field 11.9T/m, Length 1m | |
| | | QDVHIRU | 32 | Quadrupole, Gap 66mm, Field 12T/m, Length 0.5m | |
| | | QDHIRU | 16 | Quadrupole, Gap 66mm, Field 17.9T/m, Length 1m | |
| | | QFHHIRU | 32 | Quadrupole, Gap 66mm, Field 18.1T/m, Length 0.5m | |
| | | QCMIRU | 24 | Quadrupole, Gap 66mm, Field 21T/m, Length 3m | |
| | | QDCIRU | 16 | Quadrupole, Gap 66mm, Field 13.1T/m, Length 0.5m | |
| | | QFCIRU | 8 | Quadrupole, Gap 66mm, Field 13T/m, Length 1m | |
| | | QMIRU | 16 | Quadrupole, Gap 66mm, Field 17.8T/m, Length 1m | |
| | | QMIRU | 8 | Quadrupole, Gap 66mm, Field 20.7T/m, Length 2m | |
| | | QMIRU | 8 | Quadrupole, Gap 66mm, Field 13.7T/m, Length 2m | |
| | | QMIRU | 16 | Quadrupole, Gap 66mm, Field 13.7T/m, Length 1m | |
| | | QSEP | 20 | Quadrupole, Gap 66mm, Field 24T/m, Length 1.2m | |
| | | QRF | 132 | Quadrupole, Gap 66mm, Field 24.1T/m, Length 2.3m | |
| | | QFSTRH | 8 | Quadrupole, Gap 66mm, Field 9.9T/m, Length 1.2m | |
| | | QFSTR | 12 | Quadrupole, Gap 66mm, Field 9.9T/m, Length 2.3m | |
| | | QSEP | 12 | Quadrupole, Gap 66mm, Field 15.8T/m, Length 1m | |
| | | QSTRH | 4 | Quadrupole, Gap 66mm, Field 11.4T/m, Length 1m | |
| | | QSTR | 16 | Quadrupole, Gap 66mm, Field 11.4T/m, Length 2m | |
| | | QAI | 40 | Quadrupole, Gap 66mm, Field 11.6T/m, Length 2m | |
| | | QHAI | 4 | Quadrupole, Gap 66mm, Field 11.4T/m, Length 1m | |
| | | QHAO | 4 | Quadrupole, Gap 66mm, Field 11.4T/m, Length 1m | |
| | | QAO | 44 | Quadrupole, Gap 66mm, Field 13.6T/m, Length 2m | |
| | | QSTR | 32 | Quadrupole, Gap 66mm, Field 11.4T/m, Length 2m | |
| | | QSTRH | 4 | Quadrupole, Gap 66mm, Field 11.4T/m, Length 1m | |
| | | QSEP | 12 | Quadrupole, Gap 66mm, Field 15.8T/m, Length 1m | |
| | | QDI | 48 | Quadrupole, Gap 66mm, Field 13.3T/m, Length 2m | |
| | | QSTRHI | 8 | Quadrupole, Gap 66mm, Field 12.6T/m, Length 1m | |
| | | QINJI | 32 | Quadrupole, Gap 66mm, Field 5.3T/m, Length 2m | |
| | | QFI | 44 | Quadrupole, Gap 66mm, Field 12.6T/m, Length 2m | |
| | | QDO | 48 | Quadrupole, Gap 66mm, Field 13.3T/m, Length 2m | |
| | | QFSTRHO | 8 | Quadrupole, Gap 66mm, Field 12.6T/m, Length 1m | |
| | | QINJO | 32 | Quadrupole, Gap 66mm, Field 5.3T/m, Length 2m | |
| | | QFO | 44 | Quadrupole, Gap 66mm, Field 12.6T/m, Length 2m | |
| | | | | | |
| | | SFI | 512 | Sextupole, Gap 66mm, Field 825T/m^2, Length 1.4m | |
| | | SDI | 1024 | Sextupole, Gap 66mm, Field 846.8T/m^2, Length 1.4m | |
| | | SFO | 512 | Sextupole, Gap 66mm, Field 825T/m^2, Length 1.4m | |
| | | SDO | 1024 | Sextupole, Gap 66mm, Field 846.8T/m^2, Length 1.4m | |
| | | VSCIRD | 16 | Sextupole, Gap 66mm, Field 38.9T/m^2, Length 0.6m | |
| | | VSIRD | 8 | Sextupole, Gap 66mm, Field 1901.7T/m^2, Length 0.6m | |
| | | HSCIRD | 16 | Sextupole, Gap 66mm, Field 280T/m^2, Length 0.6m | |
| | | HSIRD | 8 | Sextupole, Gap 66mm, Field 1888.6T/m^2, Length 0.6m | |
| | | SC0IRD | 4 | Sextupole, Gap 66mm, Field 315T/m^2, Length 1m | |
| | | VSCIRU | 16 | Sextupole, Gap 66mm, Field 36.9T/m^2, Length 0.6m | |
| | | VSIRU | 8 | Sextupole, Gap 66mm, Field 1804.8T/m^2, Length 0.6m | |
| | | HSCIRU | 16 | Sextupole, Gap 66mm, Field 218.8T/m^2, Length 0.6m | |
| | | HSIRU | 8 | Sextupole, Gap 66mm, Field 1476.3T/m^2, Length 0.6m | |
| | | SC0IRU | 4 | Sextupole, Gap 66mm, Field 315T/m^2, Length 1m | |
| | | | | | |
| | | CH | 3544 | Horizontal corrector, Gap 66mm, Field 225Gs, Length 0.88m. | |
| | | CV | 3544 | Vertical corrector, Gap 66mm, Field 225Gs, Length 0.88m. | |
| | Booster | | | | |
| | | BDISARC | 640 | Dipole, Gap 63mm, Field 338Gs, Length 2.35m. | |
| | | BARC | 14226 | Dipole, Gap 63mm, Field 338Gs, Length 4.71m. | |
| | | QDM | 140 | Quadrupole, Gap 63mm, Field 11.8T/m, Length 1.0m. | |
| | | QFM | 156 | Quadrupole, Gap 63mm, Field 12.2T/m, Length 1.0m. | |
| | | QFMRF | 4 | Quadrupole, Gap 63mm, Field 12.4T/m, Length 2.0m. | |
| | | QDMRF | 4 | Quadrupole, Gap 63mm, Field 15.3T/m, Length 2.0m. | |
| | | QFARC | 1018 | Quadrupole, Gap 63mm, Field 11.1T/m, Length 2.0m. | |
| | | QDARC | 2018 | Quadrupole, Gap 63mm, Field 11.1T/m, Length 0.7m. | |
| | | QFRF | 60 | Quadrupole, Gap 63mm, Field 16.6T/m, Length 2.0m. | |
| | | QDRF | 58 | Quadrupole, Gap 63mm, Field 16.6T/m, Length 2.0m. | |
| | | SF | 40 | Sextupole, Gap 63mm, Field 217T/m^2, Length 0.4m. | |
| | | SD | 60 | Sextupole, Gap 63mm, Field 437T/m^2, Length 0.4m. | |
| | | CH | 68 | Horizontal corrector, Gap 63mm, Field 0.02T, Length 0.58m. | |
| | | CV | 1152 | Vertical corrector, Gap 63mm, Field 0.02T, Length 0.58m. | |
| | | CHb | 1130 | Horizontal corrector (attached to dipoles), Gap 66mm, Field 0.0025T, Length 4.71m. | |
| | Linac | | | | |
| | | B | 4 | Dipole, Gap 34mm, Field 1T, Length 2.356m. | |
| | | CB1 | 4 | Dipole, Gap 54mm, Field 0.5T, Length 0.279m. | |
| | | AM1 | 2 | Dipole, Gap 34mm, Field 0.3T, Length 0.262m. | |
| | | AM2 | 1 | Dipole, Gap 44mm, Field 0.8T, Length 5.236m. | |
| | | AM3 | 1 | Dipole, Gap 44mm, Field 1T, Length 5.847m. | |
| | | 100Q | 48 | Quadrupole, Gap 100mm, Field 10T/m, Length 0.3m. | |
| | | 60SQ | 6 | Quadrupole, Gap 60mm, Field 15T/m, Length 0.1m. | |
| | | 60LQ | 3 | Quadrupole, Gap 60mm, Field 15T/m, Length 0.2m. | |
| | | 40SQ | 88 | Quadrupole, Gap 40mm, Field 28T/m, Length 0.2m. | |
| | | 40LQ | 50 | Quadrupole, Gap 40mm, Field 28T/m, Length 0.4m. | |
| | | 32SQ | 38 | Quadrupole, Gap 32mm, Field 36T/m, Length 0.3m. | |
| | | 32LQ | 19 | Quadrupole, Gap 32mm, Field 36T/m, Length 0.6m. | |
| | | S1 | 1 | Solenoid, Aperture 100 mm, Field 0.1T, Max.Length 80mm | |
| | | S2 | 1 | Solenoid, Aperture 100 mm, Field 0.1T, Max.Length 120mm | |
| | | S3 | 20 | Solenoid, Aperture 100 mm, Field 0.1T, Max.Length 50mm | |
| | | S4 | 15 | Solenoid, Aperture 400 mm, Field 0.5T, Max.Length 1m | |
| | | FS1 | 4 | Solenoid, Aperture 90 mm, Field 0.06T, Max.Length 80mm | |
| | | 100C | 17 | Corrector x and y, Gap 100mm, Field 0.015T, Length 0.25m. | |
| | | L60C | 3 | Corrector x and y, Gap 60mm, Field 0.015T, Length 0.1m. | |
| | | L40C | 46 | Corrector x and y, Gap 40mm, Field 0.08T, Length 0.1m. | |
| | | L32C | 19 | Corrector x and y, Gap 32mm, Field 0.085T, Length 0.2m. | |
| | Damping Ring | | | | |
| | | B0 | 40 | Gap 38mm, Field 1.3T | |
| | | Br | 40 | Gap 38mm, Field 1.3T | |
| | | Qarc | 96 | Gap 44mm, Field 17T/m | |
| | | QRF | 8 | Gap 38mm, Field 30T/m | |
| | | SF | 36 | Gap 38mm, Field 94T/m^2 | |
| | | SD | 36 | Gap 38mm, Field 140T/m^2 | |
| | Transport Lines | | | | |
| | DR | B | 28 | Gap 44mm, Field 0.75T | |
| | | Q | 28 | Gap 54mm, Field 5T/m | |
| | L to B | BT0 & BT1 | 36 | Gap 37mm, Field 1.5T | |
| | | BTv | 4 | Gap 37mm, Field 1.5T | |
| | | 33Q | 80 | Gap 33mm, Field 14.7T/m | |
| | | 37C | 24 | Gap 37mm, Field 0.168T | |
| | B to C | BT0&BT1 | 12 | Gap 37mm, Field 1.5T | |
| | | | | | |
| **2** | **SRF** | | | | |
| | Collider | | | | |
| | | 650 MHz 2-cell Nb cavity | 192 | Q0 > 3E10 at 25 MV/m, max Eacc > 28 MV/m | |
| | | 650 MHz power coupler | 192 | > 350 kW CW | |
| | | 650 MHz HOM coupler | 384 | > 1 kW CW | |
| | | 650 MHz cavity tuner | 192 | slow tuning > 340 kHz, fast tuning > 1.5 kHz | |
| | | 650 MHz cavity HOM absorber | 64 | > 5 kW CW at room temperature | |
| | | 650 MHz cryomodule | 32 | 11 m | |
| | | 650 MHz cavity vaccum system | 32 | < 1E-7 Pa | |
| | Booster | | | | |
| | | 1.3 GHz 9-cell Nb cavity | 96 | Q0 > 3E10 at 22 MV/m, max Eacc > 24 MV/m | |
| | | 1.3 GHz power coupler | 96 | > 35 kW peak, 10 kW average | |
| | | 1.3 GHz cavity tuner | 96 | slow tuning > 420 kHz, fast tuning > 0.5 kHz | |
| | | 1.3 GHz cavity HOM absorber | 24 | > 10 W at 80 K | |
| | | 1.3 GHz cryomodule | 12 | 12 m | |
| | | 1.3 GHz cavity vaccum system | 12 | < 1E-7 Pa | |
| | | | | | |
| **3** | **RF Power Source** | | | | |
| | Collider | | | | |
| | | Klystron | 96 | 650MHz/800kW | |
| | | PSM Power Supply | 96 | 120kV/16A | |
| | | Circulator and Load | 96 | 650MHz/800kW | |
| | | Waveguide | 96 | 650MHz/800kW | |
| | | LLRF | 96 | 0.10% | |
| | | | | | |
| | Booster | | | | |
| | | SSA | 96 | 1.3GHz/25kW | |
| | | Waveguide | 96 | 1.3GHz/25kW | |
| | | LLRF | 96 | 0.10% | |
| | | | | | |
| | Injector | | | | |
| | | S Band Klystron | 34 | 2860MHz/80MW | |
| | | Modulator for S Band Klystron | 34 | 400kV/500A | |
| | | Load and auxillary power supply for S band klystron | 34 | | |
| | | S Band Klystron gain amplifier | 34 | 2860MHz/1000W | |
| | | C Band Klystron | 236 | 5720MHz/50MW | |
| | | Modulator for C Band Klystron | 236 | 350kV/400A | |
| | | Load and auxillary power supply for C band klystron | 236 | | |
| | | C Band Klystron gain amplifier | 236 | 5720MHz/1000W | |
| | Damping Ring | | | | |
| | | SSA | 2 | 650MHz/90kW | |
| | Transport Line | | | | |
| | | S Band Klystron | 2 | 2856MHz/30MW | |
| | | Modulator for S Band Klystron | 2 | 270kV/280A | |
| | | Load and auxillary power supply for S band klystron | 2 | | |
| | | S Band Klystron gain amplifier | 2 | 2856MHz/800W | |
| **4** | **Survey and Alignment** | | | | |
| | | GPS | 16 | 0.5mm/km | |
| | | Zenith plummet | 8 | 1／200000 | |
| | | Gyrotheodolite | 1 | 3" | |
| | | Relative gravimeter | 2 | 1μgal | |
| | | Laser tracker | 69 | 0.015mm/m | |
| | | FARO arm | 8 | 0.025-0.036mm/2.4M | |
| | | Optical level | 32 | 0.1mm/km | |
| | | Digital level | 8 | 0.2mm/km | |
| | | Transit square | 32 | 0.0254mm | |
| | | Total station | 16 | 0.5" | |
| | | Electronic gradienter | 32 | 0.001mrad | |
| | | Visual instrument | 19 | 1μm+1ppm | |
| | | HLS | 48 | 0.03mm | |
| | | Laser collimator | 2 | 0.1mm/100m | |
| | | Alignment telescope | 2 | CZW | |
| | | Tool microscope | 4 | 0.01mm | |
| | | CMM | 4 | 0.005mm | |
| | | Laser interferometer | 6 | 0.005mm | |
| | | Absolute multiline | 3 | 0.005mm | |
| | | Baseline calibration device | 2 | 0.2mm/60m | |
| | | Pick-up truck | 4 | | |
| **5** | **Instrument-ation** | | | | |
| | Collider | | | | |
| | | BPM | 4252 | The inner diameter 56mm，diameter of the button 5mm with the gap 0.3mm; | |
| | | | | Measurement area (x/y)：±20mm×±10mm<br>COD Resolution：<0.6um<br>Measurement time of COD：< 4 s | |
| | | BLM | 5800 | Dynamic range:120 dB<br>Maximum counting rates ≥10 MHz | outside the vacuum chamber |
| | | DCCT | 2 | Dynamic measurement range：0.0~1.5A<br>Linearity：0.1 %<br>Zero drift: <0.05mA | |
| | | BCM | 2 | Measurement range：10mA /bunch<br>Relatively precision：1/4095 | |
| | | Transverse Feedback | 3 | Damping time<=1ms | |
| | | Longitudinal Feedback | 2 | "Dynamic range:120 dB | |
| | | Synchrotron light monitor | 4 | Beam size resolution:0.2 μm<br>Bunch length resolution：1ps@10ps | |
| | | Tune measurement | 2 | Resolution:0.001 | |
| | Booster | | | | |
| | | BPM | 2408 | The diameter of the button 5mm with the gap 0.3mm;<br>Measurement area (x/y)：±20mm×±10mm;<br>TBT Resolution：<0.02mm; | |
| | | BLM | 670 | The optical fiber-based monitor, Space resolution: 0.6m | outside the vacuum chamber |
| | | DCCT | 2 | Dynamic measurement range：~1.5A<br>Resolution: 50uA@0.6-8mA<br>Linearity：0.1 %<br>Zero drift: <0.05mA | |
| | | BCM | 2 | Measurement range：10mA /bunch<br>Relatively precision：1/4095 | |
| | | Transverse Feedback | 2 | Damping time<=3ms | |
| | | Longitudinal Feedback | 2 | Damping time<=65ms | |
| | Linac + 2 transport lines | | | | |
| | | BPM | 140 | Length 200mm, stripline 150mm | |
| | | Profile | 80 | YAG/OTR;<br>Resolution:30um | |
| | | ICT | 42 | Length 44mm | |
| | | Beam energy and spread | 3 | 0.10% | |
| | | beam emittance | 3 | 10% | |
| | Damping Ring | | | | |
| | | DCCT | 1 | Resolution: 50uA@0.1mA-30mA | |
| | | BPM | 40 | Button BPM Resolution: 20um @ 5mA TBT | |
| | | Tune measurement | 1 | Frequency sweeping; Resolution:0.001 | |
| **6** | **Vacuum** | | | | |
| | Collider | | | | |
| | | Vacuum chamber | 20337 | DN56， CuCrZr （C18150） | |
| | | Bellow | 21350 | RF shielding | |
| | | Pump | 13333 | 100L/s | |
| | | Gauge | 4320 | CCG | |
| | | Valve | 1040 | RF-CF63 | |
| | Booster | | | | |
| | | Vacuum chamber | 17000 | DN56， Al6061 | |
| | | Bellow | 12000 | 316L | |
| | | Pump | 8400 | 50L/s | |
| | | Gauge | 2160 | CCG | |
| | | Gate Valve | 520 | DN63 | |
| | Injector | | | | |
| | | vacuum chamber | 300 | 304/316L | |
| | | Bellow | 2517 | 304/316L | |
| | | Pump | 4221 | 200L/s | |
| | | Gauge | 1549 | CCG | |
| | | Gate Valve | 60 | CF100 | |
| | Damping Ring | | | | |
| | | vacuum chamber | 147 | DN30， Al6061 | |
| | | Bellow | 75 | 304/316L | |
| | | Pump | 100 | 100L/s | |
| | | Gauge | 25 | CCG | |
| | | Gate Valve | 8 | CF100 | |
| | Transport line | | | | |
| | | vacuum chamber | 3680 | 304/316L | |
| | | Bellow | 736 | 304/316L | |
| | | Pump | 736 | 200L/s | |
| | | Gauge | 530 | CCG | |
| | | Gate Valve | 70 | CF100 | |
| **7** | **Linac** | | | | |
| | Source | | | | |
| | | **Electron source** | | | |
| | | Gun body | 1 | 150kV, L=0.5m | |
| | | High Voltage Power Supply | 1 | 200kV | |
| | | Cathode grid | 1 | 10A | |
| | | Accessory equipment | 1 | 200kV | |
| | | **Positron source** | | | |
| | | Positron conversion device | 1 | 6 Tesla, L=0.5m | |
| | | Flux concentrator power supply | 1 | 15kA/15kV | |
| | | Positron focusing coil | 10 | 0.5 Tesla | |
| | | DC power supply | 6 | 600A/500V | |
| | RF system | | | | |
| | | Sub-harmonic buncher | 2 | 158.89MHz & 476.67MHz， L1=0.75m, L2=0.45m | |
| | | Buncher | 1 | 2860MHz, L=0.235mm | |
| | | S-band accelerating structure | 93 | Gradient：22MV/m&27MV/m, L=3.1m | |
| | | S-band Big-hole accelerating structure | 16 | Gradient：22MV/m @φ25mm, L=2m | |
| | | S-band RF Pulse compressor | 32 | Energy multiplication factor: 1.6 | |
| | | S-band Waveguide | 33 | VSWR≤1.1 | |
| | | C-band accelerating structure | 470 | $f_0$: 5720MHz, Gradient:40MV/m, L=1.8m | |
| | | C-band deflecting structure | 1 | $f_0$: 5720MHz, L=1.8m | |
| | | C-band RF Pulse compressor | 235 | $f_0$: 5720MHz， Energy multiplication factor: 1.6 | |
| | | C-band Waveguide | 235 | $f_0$: 5720MHz | |
| | | RF measurement system (S+C) | 269 | VSWR≤1.1 | |
| | | Low level RF (S+C) | 269 | Phase & Amplifier stability(rms): 0.3/0.3% | |
| | | Phase reference line | 269 | phase stability: ±0.1° | |
| | DR Cu cavity | | | | |
| | | 650 MHz 5-cell Cu cavity | 2 | 2.5MV/m, L=2m | |
| | | 650 MHz power coupler | 2 | 90 kW CW | |
| | | 650 MHz tuner | 4 | plunger range -20 mm to +40 mm | |
| | | cavity vacuum and cooling system | 2 | | |
| | | Waveguide | 2 | 650MHz/90kW | |
| | | LLRF | 2 | Phase & Amplifier stability(rms): 0.1/0.1% | |
| | Transport line RF | | | | |
| | | Accelerating structure (1m) | 2 | 2860MHz, L=1m | |
| | | pulse compressor | 2 | Energy multiplication factor: 1.6 | |
| | | Waveguide system | 2 | VSWR≤1.1 | |
| | | load | 2 | VSWR≤1.1 | |
| | | Low level RF&SSA | 2 | Phase & Amplifier stability(rms): 0.3/0.3% | |
| **8** | **Cryogenic System** | | | | |
| | Cryogenics for SC cavities | | | | |
| | | Refrigerator (15 kW@ 4.5 K) | 4 | 18kW@4.5K | |
| | | Liquid Helium tank | 4 | heat loss: ≤30W, vacuum insulation leakage rate : ≤1E-8 Pa.m3/s | |
| | | Master valve box | 4 | heat loss: ≤15W, vacuum insulation leakage rate: ≤1E-8 Pa.m3/s | |
| | | Connection valve box | 32 | heat loss: ≤0.5 W/m, vacuum insulation leakage rate: ≤1E-8 Pa.m3/s | |
| | | Low temperature pipeline | 1100 | working pressure: 2 MPa, volume: 100m3, helium purity:99.999% | |
| | | Medium pressure Helium tank | 52 | working pressure: 20 MPa, volume: 25m3 | |
| | | High pressure high purity helium tank | 4 | working pressure: 20 MPa, volume: 25m3 | |
| | | High pressure impure helium tank | 4 | flow rate:100m3/h, working pressure: 20MPa | |
| | | High pressure helium gas recycling compressor | 8 | flow rate:200m3/h, working pressure: 20MPa | |
| | | High pressure helium purifier | 4 | | |
| | | Pure helium gas | 80000 | 3kW@4.5K | |
| | | Control system | 4 | 5000L/tank | |
| | | Vacuum system | 4 | heat loss: ≤30W, vacuum insulation leakage: ≤1E-8 Pa.m3/s | |
| | | Room temperature pipeline system | 4 | heat loss: ≤15W, vacuum insulation leakage: ≤1E-8 Pa.m3/s | |
| | | 2K valve box | 8 | heat loss: ≤0.5 W/m, vacuum insulation leakage rate: ≤1E-8 Pa.m3/s | |
| | Cryogenics for IR SC magnets | | | working pressure: 1.6 MPa, volume: 100m3, helium purity:99.999% | |
| | | Refrigerator (700W, 4.5 K) | 2 | working pressure: 20 MPa, volume: 25m3 | |
| | | Liquid Helium tank | 2 | working pressure: 20 MPa, volume: 25m3 | |
| | | Master valve box | 2 | flow rate:100m3/h, working pressure: 20MPa | |
| | | cryostat for SC Magnets | 4 | flow rate:100m3/h, working pressure: 20MPa | |
| | | current lead | 6 | | |
| | | Connection valve box | 4 | | |
| | | Low temperature pipeline | 40 | | |
| | | Medium pressure Helium tank | 4 | | |
| | | High pressure high purity helium tank | 2 | | |
| | | High pressure impure helium tank | 6 | | |
| | | High pressure helium gas recycling compressor | 4 | | |
| | | High pressure helium purifier | 2 | | |
| | | Pure helium gas | 80000 | | |
| | | Control system | 2 | | |
| | | Vacuum system | 2 | | |
| | | Room temperature pipe line system | 2 | | |
| **9** | **Magnet Supports** | | | | |
| | Collider | | | | |
| | | SPT_DT_C | 12096 | Range and accuracy of adjustment: X≥±20mm， △X≤0.02mm, Y≥±30mm， △X≤0.02mm, Z≥±20mm， △X≤0.02mm, θX≥±10mrad， △θX≤0.05mrad, θY≥±10mrad， △θY≤0.05mrad, θZ≥±10mrad， | |
| | | SPT_DS_C | 1218 | | |
| | | SPT_Q3000T_C | 3008 | | |
| | | SPT_Q1500T_C | 8 | | |
| | | SPT_Q1000S_C | 152 | | |
| | | SPT_Q500S_C | 128 | | |
| | | SPT_Q625S_C | 112 | | |
| | | SPT_Q1250S_C | 56 | | |
| | | SPT_Q3000S_C | 48 | | |
| | | SPT_Q3500S_C | 32 | | |
| | | SPT_Q2000S_C | 396 | | |
| | | SPT_Q1150S_C | 28 | | |
| | | SPT_Q2300S_C | 144 | | |
| | | SPT_Q1500S_C | 16 | | |
| | | SPT_Scommon_C | 1024 | | |
| | | SPT_S600_C | 96 | △θZ≤0.05mrad | |
| | | SPT_S1000_C | 8 | | |
| | | SPT_C875_C | 3544 | | |
| | | SPT_vacuum_C | 54350 | | |
| | | SPT_vacuum_m_C | 30256 | | |
| | | SPT_valve_C | 1040 | | |
| | | SPT_pump_C | 13333 | | |
| | | SPT_BPM_C | 4252 | | |
| | | SPT_DCCT_C | 2 | | |
| | | SPT_BCM_C | 2 | | |
| | | SPT_RFCryo_C | 100 | | |
| | | Movable collimator | 40 | | |
| | | Fixed collimator | 80 | | |
| | | SPT-SC | 4 | resolution<0.005mm | |
| | | RVC | 4 | leak rate<2.7e-11 Pa· $m^3/s$ | |
| | Booster | | | | |
| | | SPT_D4700_B | 10192 | Range and accuracy of adjustment:<br>X≥±20mm，<br>△X≤0.02mm,<br>Y≥±30mm，<br>△X≤0.02mm,<br>Z≥±20mm，<br>△X≤0.02mm,<br>θX≥±10mrad，<br>△θX≤0.05mrad,<br>θY≥±10mrad，<br>△θY≤0.05mrad,<br>θZ≥±10mrad，<br>△θZ≤0.05mrad | |
| | | SPT_DS4700_B | 4034 | | |
| | | SPT_D2350_B | 640 | | |
| | | SPT_Q1000_B | 2306 | | |
| | | SPT_Q2000_B | 1152 | | |
| | | SPT_S400_B | 100 | | |
| | | SPT_C583_B | 1200 | | |
| | | SPT_vacuum_B | 30000 | | |
| | | SPT_vacuum_m_B | 30244 | | |
| | | SPT_valve_B | 520 | | |
| | | SPT_pump_B | 8400 | | |
| | | SPT_BPM_B | 2408 | | |
| | | SPT_DCCT_B | 2 | | |
| | | SPT_BCM_B | 2 | | |
| | | SPT_RFCryo_B | 12 | | |
| | Linac | | | | |
| | | SPT_SHB_B_L | 1 | | |
| | | SPT_BUN_B_L | 1 | | |
| | | SPT-AS_B_L | 1 | | |
| | | SPT_SOL_B_L | 1 | | |
| | | SPT_BPM_B_L | 1 | | |
| | | SPT-Q1600_L | 14 | | |
| | | SPT-Q2000_L | 38 | | |
| | | SPT-Q3100_L | 52 | | |
| | | SPT_Q_EBTL_L | 2 | | |
| | | SPT_C_EBTL_L | 2 | | |
| | | SPT_BPM_EBTL_L | 2 | | |
| | | SPT_PS_L | 1 | | |
| | | SPT_PSSOL+Acc_L | 6 | | |
| | | SPT_Acc2000_L | 10 | | |
| | | SPT_Acc3100_L | 8 | | |
| | | SPT_AM1_L | 1 | | |
| | | SPT_AM2/AM5_L | 2 | | |
| | | SPT_AM3/CB1_L | 5 | | |
| | | SPT_AM4/CB2_L | 9 | | |
| | | SPT_AM6/AM7_L | 2 | | |
| | | SPT_Corrector_L | 21 | | |
| | | SPT_SAcc_L | 109 | | |
| | | SPT_CAcc_L | 470 | | |
| | | SPT_vacuum_L | 2000 | | |
| | | SPT_vacuum_m_L | 98 | | |
| | | SPT_valve_L | 60 | | |
| | | SPT_pump_L | 3431 | | |
| | | SPT_BPM_L | 150 | | |
| | | SPT_ICT_L | 63 | | |
| | Damping Ring | | | | |
| | | SPT_B0_DR | 40 | | |
| | | SPT_Br_DR | 40 | | |
| | | SPT_Q_S_DR | 72 | | |
| | | SPT_Qarc_DR | 24 | | |
| | | SPT_QRF_DR | 8 | | |
| | | SPT_vacuum_DR | 225 | | |
| | | SPT_valve_DR | 10 | | |
| | | SPT_pump_DR | 100 | | |
| | | SPT_BPM_DR | 40 | | |
| | | SPT_DCCT_DR | 1 | | |
| | Transport line | | | | |
| | | SPT_D5000_BTC | 48 | | |
| | | SPT_Q1000_BTC | 48 | | |
| | | SPT_Q2000_BTC | 12 | | |
| | | SPT_BPM_BTC | 32 | | |
| | | SPT_vacuum_BTC | 480 | | |
| | | SPT_vacuum_m_BTC | 108 | | |
| | | SPT_pump_BTC | 192 | | |
| | | SPT_valve_BTC | 8 | | |
| | | SPT_D5000_CTB | 24 | | |
| | | SPT_Q1000_CTB | 24 | | |
| | | SPT_Q2000_CTB | 6 | | |
| | | SPT_BPM_CTB | 16 | | |
| | | SPT_vacuum_CTB | 240 | | |
| | | SPT_vacuum_m_CTB | 54 | | |
| | | SPT_pump_CTB | 96 | | |
| | | SPT_valve_CTB | 4 | | |
| | | SPT_D5000_LTB | 20 | | |
| | | SPT_D3000_LTB | 4 | | |
| | | SPT_D4000_LTB | 64 | | |
| | | SPT_Q1000_LTB | 76 | | |
| | | SPT_Q2000_LTB | 48 | | |
| | | SPT_BPM_LTB | 28 | | |
| | | SPT_vacuum_LTB | 1500 | | |
| | | SPT_vacuum_m_LTB | 220 | | |
| | | SPT_pump_LTB | 600 | | |
| | | SPT_valve_LTB | 30 | | |
| | | SPT_D2000_LTD | 14 | | |
| | | SPT_Q300_LTD | 24 | | |
| | | SPT_BPM_LTD | 3 | | |
| | | SPT_vacuum_LTD | 60 | | |
| | | SPT_vacuum_m_LTD | 14 | | |
| | | SPT_pump_LTD | 24 | | |
| | | SPT_valve_LTD | 2 | | |
| | | SPT_D2000_DTL | 14 | | |
| | | SPT_Q300_DTL | 24 | | |
| | | SPT_BPM_DTL | 3 | | |
| | | SPT_vacuum_DTL | 60 | | |
| | | SPT_vacuum_m_DTL | 14 | | |
| | | SPT_pump_DTL | 24 | | |
| | | SPT_valve_DTL | 2 | | |
| **10** | **Radiation Protection** | | | | |
| | | PLC Interlock sub-system | 1 | | PPS |
| | | Interlock Key | 1 | | PPS |
| | | Access control sub-system | 1 | | PPS |
| | | Integrated information display device | 80 | | PPS |
| | | Video surveillance sub-system | 152 | | PPS |
| | | Interlock doors and Shielding Doors | 38 | | PPS |
| | | PPS test platform | 1 | | PPS |
| | | Area Dose Monitroing System | 267 | | Dose monitoring system |
| | | Environmental Dose Monitoring System | 65 | | Dose monitoring system |
| | | Water Monitorning System | 16 | | Dose monitoring system |
| | | Air Monitorning System | 16 | | Dose monitoring system |
| | | Personal Dose Monitoring System | 1 | | Dose monitoring system |
| | | Radiation Monitorning Test System | 1 | | Dose monitoring system |
| | | Portable Neutron Monitoring Detector | 10 | | Environmental Sample Measurement System |
| | | Portable Gamma Monitoring Detector | 10 | | Environmental Sample Measurement System |
| | | Surface Contamination Detector | 10 | | Environmental Sample Measurement System |
| | | Electron Personal Dose Monitoring Detector | 825 | | Environmental Sample Measurement System |
| | | Gamma Sepctrometer | 4 | | Environmental Sample Measurement System |
| | | Liquid scintillation counter spectrometer. | 4 | | Environmental Sample Measurement System |
| | | Beam Dump | 2 | | Radiation Shielding |
| | | Hot Spots Shielding | 7 | | Radiation Shielding |
| | | Local shielding for holes | 556 | | Radiation Shielding |
| **11** | **Power Supply** | | | | |
| | Collider | | | | |
| | | **Name** | **Number** | **Stability /8hours** | **Output Rating** |
| | | B0A+B | 16 | 10ppm | 1300A/500V |
| | | B1 | 16 | 10ppm | 150A/50V |
| | | BMV01IRD | 4 | 10ppm | 90A/40V |
| | | BMVIRD | 4 | 10ppm | 250A/90V |
| | | BMHIRD | 4 | 10ppm | 160A/50V |
| | | BGM1 | 4 | 10ppm | 150A/60V |
| | | BGM2 | 4 | 10ppm | 80A/60V |
| | | BMV01IRU | 4 | 10ppm | 40A/50V |
| | | BMVIRU | 4 | 10ppm | 160A/70V |
| | | BMHIRU | 4 | 10ppm | 70A/50V |
| | | BRF0 | 2 | 10ppm | 90A/50V |
| | | BRF | 4 | 10ppm | 210A/60V |
| | | Q | 320 | 10ppm | 120A/450V |
| | | QH | 8 | 10ppm | 120A/50V |
| | | Q3IRD | 4 | 10ppm | 160A/50V |
| | | Q4IRD | 4 | 10ppm | 170A/50V |
| | | Q5IRD | 4 | 10ppm | 40A/30V |
| | | QDVHIRD | 4 | 10ppm | 120A/180V |
| | | QFVIRD | 4 | 10ppm | 120A/150V |
| | | QFHHIRD | 4 | 10ppm | 160A/240V |
| | | QDHIRD | 4 | 10ppm | 160A/190V |
| | | QCMIRD | 4 | 10ppm | 140A/510V |
| | | QDCIRD | 4 | 10ppm | 100A/120V |
| | | QFCIRD | 4 | 10ppm | 90A/100V |
| | | QMIRD | 4 | 10ppm | 120A/810V |
| | | QMIRD | 4 | 10ppm | 110A/150V |
| | | QMIRD | 4 | 10ppm | 100A/120V |
| | | QDGMH | 4 | 10ppm | 110A/440V |
| | | QFGM | 4 | 10ppm | 110A/350V |
| | | Q3IRUJ2 | 4 | 10ppm | 90A/40V |
| | | Q4IRUJ2 | 4 | 10ppm | 90A/40V |
| | | QFVIRU | 4 | 10ppm | 90A/110V |
| | | QDVHIRU | 4 | 10ppm | 80A/130V |
| | | QDHIRU | 4 | 10ppm | 120A/140V |
| | | QFHHIRU | 4 | 10ppm | 80A/170V |
| | | QCMIRU | 4 | 10ppm | 100A/450V |
| | | QDCIRU | 4 | 10ppm | 70A/90V |
| | | QFCIRU | 4 | 10ppm | 70A/80V |
| | | QMIRU | 4 | 10ppm | 100A/140V |
| | | QMIRU | 4 | 10ppm | 100A/140V |
| | | QMIRU | 4 | 10ppm | 70A/110V |
| | | QMIRU | 4 | 10ppm | 70A/110V |
| | | QSEP | 20 | 10ppm | 140A/60V |
| | | QRF | 12 | 10ppm | 70A/410V |
| | | QFSTRH | 8 | 10ppm | 50A/30V |
| | | QFSTR | 12 | 10ppm | 70A/60V |
| | | QSEP | 12 | 10ppm | 90A/40V |
| | | QSTRH | 4 | 10ppm | 80A/40V |

| | | | | | | | |
|---|---|---|---|---|---|---|---|
| | | QSTR | | 16 | 10ppm | | 100A/50V |
| | | QAI | | 40 | 10ppm | | 100A/50V |
| | | QHAI | | 4 | 10ppm | | 60A/40V |
| | | QHAO | | 4 | 10ppm | | 60A/40V |
| | | QAO | | 44 | 10ppm | | 80A/60V |
| | | QSTR | | 32 | 10ppm | | 60A/50V |
| | | QSTRH | | 4 | 10ppm | | 60A/40V |
| | | QSEP | | 12 | 10ppm | | 90A/50V |
| | | QDI | | 48 | 10ppm | | 70A/60V |
| | | QSTRHI | | 8 | 10ppm | | 70A/40V |
| | | QINJI | | 32 | 10ppm | | 90A/30V |
| | | QFI | | 44 | 10ppm | | 70A/60V |
| | | QDO | | 48 | 10ppm | | 70A/60V |
| | | QFSTRHO | | 8 | 10ppm | | 70A/40V |
| | | QINJO | | 32 | 10ppm | | 70A/40V |
| | | QFO | | 44 | 10ppm | | 70A/60V |
| | | SFI | | 256 | 20ppm | | 200A/110V |
| | | SDI | | 256 | 20ppm | | 200A/200V |
| | | SFO | | 256 | 20ppm | | 200A/110V |
| | | SDO | | 256 | 20ppm | | 200A/200V |
| | | VSCIRD | | 16 | 20ppm | | 100A/30V |
| | | HSCIRD | | 16 | 20ppm | | 150A/30V |
| | | SC0IRD | | 4 | 20ppm | | 150A/50V |
| | | VSCIRU | | 16 | 20ppm | | 100A/30V |
| | | HSCIRU | | 16 | 20ppm | | 120A/30V |
| | | SC0IRU | | 4 | 20ppm | | 150A/50V |
| | | CH | | 3544 | 100ppm | | ±20A/±20V |
| | | CV | | 3544 | 100ppm | | ±20A/v20V |
| | | B0A-TRIM | | 2048 | 100ppm | | ±6A/±12V |
| | | B0B-TRIM | | 3840 | 100ppm | | ±6A/±12V |
| | | Q-TRIM | | 6016 | 100ppm | | ±15A/±10V |
| | | SCQ1a | | 4 | 10ppm | | 2200A/12V |
| | | SCQ1b | | 4 | 10ppm | | 1800A/12V |
| | | SCQ2 | | 4 | 10ppm | | 2200A/12V |
| | | Anti-solenoid | | 4 | 10ppm | | 1500A/20V |
| | | SC-Corrector | | 72 | 10ppm | | ±90A/±11V |
| | | | | | | | |
| | | **Name** | **Number** | **Electrode length/width** | **Nominal gap** | **Maximum operating field strength** | **Nominal vacuum pressure** |
| | | Electrostatic -Magnetic separator | 32 | 4.0m/180mm | 75mm | 2MV/m | 2.7e-8 Pa |
| | Booster | | | | | | |

| | | Name | Number | Stability /tracking error | Output Rating |
|---|---|---|---|---|---|
| | | BST-63B-Arc | 16 | 200ppm / 0.1% | 560A/780V |
| | | BST-63B-Arc-SF | 7 | 200ppm / 0.1% | 560A/960V |
| | | BST-63B-Arc-SD | 1 | 200ppm / 0.1% | 560A/960V |
| | | BST-74B-IR | 4 | 200ppm / 0.1% | 520A/140V |
| | | QDM-63 | 64 | 100ppm / 0.1% | 200A/50V |
| | | QFM-63 | 48 | 100ppm / 0.1% | 250A/50V |
| | | QFM-RF-63 | 8 | 100ppm / 0.1% | 200A/270V |
| | | QDM-RF-63 | 8 | 100ppm / 0.1% | 200A/270V |
| | | QFM-ARC-63 | 80 | 100ppm / 0.1% | 180A/360V |
| | | QDM-ARC-63 | 80 | 100ppm / 0.1% | 180A/280V |
| | | QF-DIS-63 | 16 | 100ppm / 0.1% | 270A/50V |
| | | QD-DIS-63 | 32 | 100ppm / 0.1% | 250A/50V |
| | | BST-63S | 100 | 300ppm / 0.1% | 140A/35V |
| | | BST-63C | 1200 | 300ppm / 0.1% | ±25A/±25V |
| | | Q-TRIM | 3200 | 300ppm / 0.2% | ±6A/±12V |
| | Transport Line | | | | |
| | | **Name** | **Number** | **Stability /8hours** | **Output Rating** |
| | TL L-B | BT0&BT1 | 6 | 100ppm | 350A/1130V |
| | | Btv | 4 | 100ppm | 350A/1130V |
| | | TL-33Q | 80 | 100ppm | 30A/20V |
| | | T-37C | 24 | 300ppm | ±30A/±20V |
| | TL B-C OFF-axis | BT0&BT1 | 2 | 100ppm | 400A/360V |
| | | BT3 | 2 | 100ppm | 400A/360V |
| | | TL-24Q | 12 | 100ppm | 30A/30V |
| | | T-24C | 10 | 300ppm | ±30A/±20V |
| | TL B-C ON-axis | BT0&BT1 | 4 | 100ppm | 400A/360V |
| | | BT2 | 4 | 100ppm | 400A/360V |
| | | TL-24Q | 28 | 100ppm | 30A/30V |
| | | T-24C | 20 | 300ppm | ±30A/±20V |
| | TL D-L | TL-44B | 4 | 100 ppm | 410A/220V |
| | | TL-54Q | 28 | 100 ppm | 30A/7V |
| | Linac | | | | |
| | | **Name** | **Number** | **Stability /8 hours** | **Output Rating** |
| | | AM1 | 1 | 100ppm | 220A/10V |
| | | AM2/AM5/AM6/AM7 | 4 | 100ppm | 470A/55V |
| | | AM3/CB1 | 5 | 100ppm | 470A/18V |
| | | AM4/CB2 | 5 | 100ppm | 650A/20V |
| | | B | 4 | 100ppm | 650A/20V |
| | | LA-24Q-300L | 52 | 100ppm | 200A/13V |
| | | LA-24Q-600L | 52 | 100ppm | 200A/13V |
| | | LA-34Q-100L | 11 | 100ppm | 150A/7V |
| | | LA-34Q-200L | 49 | 100ppm | 160A/10V |
| | | LA-34Q-400L | 44 | 100ppm | 170A/13V |
| | | LA-60Q-100L | 3 | 100ppm | 170A/13V |
| | | LA-60Q-200L | 3 | 100ppm | 170A/13V |
| | | LA-150Q-300L | 18 | 100ppm | 250A/55V |
| | | LA-150Q-600L | 18 | 100ppm | 250A/55V |
| | | SOL-I | 4 | 100ppm | 12A/12V |
| | | SOL-II | 17 | 100ppm | 12A/120V |
| | | SOL-III | 1 | 100ppm | 11A/35V |
| | | SOL-IV | 15 | 100ppm | 350A/190V |
| | | C1-H, C1-V | 2 | 300ppm | ±3A/±2V |
| | | C2 | 2 | 300ppm | ±3A/±2V |
| | | C3-H | 1 | 300ppm | ±7A/±7V |
| | | C3-V | 1 | 300ppm | ±7A/±7V |
| | | L100-150C | 25 | 300ppm | ±12A/±6V |
| | | L100-35C | 100 | 300ppm | ±12A/±6V |
| | | L200-30C | 146 | 300ppm | ±33A/±7V |
| | Damping Ring | | | | |
| | | **Name** | **Number** | **Stability /tracking error** | **Output Rating** |
| | | DR-38B0 | 4 | 100 ppm | 460A/200V |
| | | DR-38Br | 4 | 100 ppm | 460A/110V |
| | | DR-44Q | 96 | 100 ppm | 180A/7V |
| | | DR-38Q | 8 | 100 ppm | 180A/7V |
| | | DR-38S | 2 | 100 ppm | 11A/30V |
| | | DR-40C | 60 | 300 ppm | ±11A/±3V |
| **12** | **Injection and Extraction** | | | | |
| | Collider | | | | |
| | | CR-OFFA-kicker | 8 | in-air delay-line dipole kicker, horizontal, 0.1mrad, 0.04T·m, L=1m, B=0.04T(H), inc. ion pump | |
| | | CR-OFFA-kicker pulser | 8 | PFN, solid-state, 440~2400ns, inc. cable | |
| | | CR-OFFA-thin-septa | 8 | Lambertson, vertical, DC, L=1.75m, B=0.8T, Septum=2mm, main field clearance =20*20mm, 13mrad, 5.22T·m, inc. ion pump | |
| | | CR-OFFA-thick-septa | 8 | Lambertson, vertical, DC, L=1.75m, B=0.8T, Septum=6mm, main field | |
| | | | | clearance =20*20mm, 13mrad, 5.22T·m, inc. ion pump | |
| | | CR-OFFA-septa power supply | 4 | DC switch power supply, inc. cable | |
| | | CR-ONA-kicker | 4 | in-air ferrite-core dipole kicker, horizontal, 0.2mrad,0.08T· m, L=1m, B=0.04T(H), inc. ion pump | |
| | | CR-ONA-kicker pulser | 4 | LC, solid-state,1360ns, halfsine, inc. cable | |
| | | CR-ONA-thin-septa | 20 | Lambertson, vertical, DC, L=1.75m, B=0.8T, Septum=2mm, mainf ield clearance =20*20mm,17.5mrad, 7.05T·m, inc. ion pump,2 groups | |
| | | CR-ONA-thick-septa | 20 | Lambertson, vertical, DC, L=1.75m, B=0.8T, Septum=6mm, main field clearance =20*20mm,17.5mrad, 7.05T·m,inc. ion pump,2groups | |
| | | CR-ONA-septa power supply | 8 | DC switch power supply, inc. cable | |
| | | CR-DUMP-kicker | 8 | in-air delay-line dipole kicker, horizontal, 0.4mrad, 0.16T·m, L=1m, B=0.04T(H), inc. ion pump | |
| | | CR-DUMP-kicker pulser | 8 | PFN, solid-state, 440~2400ns, inc. cable | |
| | | CR-DUMP-septa | 16 | Lambertson, vertical, DC, L=1.75m, B=0.8T, Septum=6mm, main field clearance =20*20mm,26mrad, 10.44T·m,inc. ion pump | |
| | | CR-DUMP-septa power supply | 4 | DC switch power supply, inc. cable | |
| | | CR-DUMP-dilution kicker | 10 | CW full-sine， L=2m, 150Gauss， T=25us, total length=10m | |
| | | CR-DUMP-dilution kicker pulser | 10 | CW full-sine, T=25us | |
| | | CR-local control system | 6 | PLC, touch screen, oscilloscope | |
| | | Field measurement devices | 1 | Hall point measurement system for Lambertson | |
| | Booster | | | | |
| | | BST-LEI-kicker | 6 | Strip-line kicker, vertical, 46ns, L=1m, inc. ion pump | |
| | | BST-LEI-kicker pulser | 6 | 46ns， DSRD pulser, inc. cable | |
| | | BST-LEI-septa | 2 | Lambertson, horizontal, DC, L=1.2m, B=0.8T, Septum =5.5mm, mainfield clearance=50*50mm, 45mrad, 0.92T-m, inc. ion pump | |
| | | BST-LEI-septa power supply | 2 | DC switch power supply, inc. cable | |
| | | BST-HEE-OFFA-kicker | 4 | in-air delay-line dipole kicker, horizontal, 0.2mrad, 0.08T·m, L=1m, B=0.04T(H), inc. ion pump | |
| | | BST-HEE-OFFA-kicker pulser | 4 | PFN, solid-state, 440~2400ns, inc. cable | |
| | | BST-HEE-OFFA-septa | 26 | Lambertson, vertical, DC, L=1.75m, B=0.8T, Septum =6mm, mainfield clearance=30*30mm, 43mrad, 17.4T-m,inc. ion pump | |
| | | BST-HEE-OFFA-septa power supply | 2 | DC switch power supply, inc. cable | |
| | | BST-HEE-ONA-kicker | 2 | in-air ferrite-core dipole kicker, horizontal, 0.1mrad, 0.04T·m, L=1m, inc. ion pump | |
| | | BST-HEE-ONA-kicker pulser | 2 | LC, solid-state,1360ns, halfsine, inc. cable | |
| | | BST-HEE-ONA-septa | 26 | Lambertson, vertical, DC, L=1.75m, B=0.8T, Septum =6mm, mainfield clearance=30*30mm, 43mrad, 17.4T-m,inc. ion pump | |
| | | BST-HEE-ONA-septa power supply | 2 | DC switch power supply, inc. cable | |
| | | BST-HEI-ONA-kicker | 2 | in-air Nonlinear kicker or pulsed sextupole, horizontal, 0.1mrad, 0.04T·m, L=1m, inc. ion pump | |
| | | BST-HEI-ONA-kicker pulser | 2 | LC, solid-state,0.3ms, halfsine, inc. cable | |
| | | BST-HEI-ONA-septa | 26 | Lambertson, vertical, DC, L=1.75m, B=0.8T, Septum =6mm, mainfield clearance=30*30mm, 43mrad, 17.4T-m,inc. ion pump | |
| | | BST-HEI-ONA-septa power supply | 2 | DC switch power supply, inc. cable | |
| | | BST-DUMP-kicker | 4 | in-air delay-line dipole kicker, horizontal, | |
| | | | | 0.2mrad, 0.08T·m, L=1m, B=0.04T(H), inc. ion pump | |
| | | BST-DUMP-kicker pulser | 4 | PFN, solid-state, 440~2400ns, inc. cable | |
| | | BST-DUMP-septa | 26 | Lambertson, vertical, DC, L=1.75m, B=0.8T, Septum =6mm, mainfield clearance=30*30mm, 43mrad, 17.4T-m,inc. ion pump | |
| | | BST-DUMP-septa power supply | 2 | DC switch power supply, inc. cable | |
| | | BST-local control system | 10 | PLC, touch screen, oscilloscope | |
| | | Field measurement devices | 1 | Hall point measurement system for Lambertson | |
| | Damping Ring | | | | |
| | | DR-kicker | 2 | Slotted-pipe kicker, vertical,250ns, L=0.25m, DR beam pipe φ=30mm, inc. ion pump | |
| | | DR-kicker pulser | 2 | solid-state, inductive adder, 250ns, inc. cable | |
| | | DR-septa | 2 | Lambertson, horizontal, DC, L=0.5m, Septum=6mm, DR beam pipe φ=30mm, inc. ion pump | |
| | | DR-septa power supply | 2 | DC switch power supply, inc. cable | |
| | | DR-local control system | 2 | PLC, touch screen, oscilloscope | |
| | | Field measurement devices | 1 | inductive coil measurement system for kicker | |
| **13** | **Control System** | | | | |
| | | CCR (central control room) | | | |
| | | EPICS | | | |
| | | IOC/OPI | | | |
| | | Alarming system/Data Archiver/Elog | | | |
| | | Core Switch | | | |
| | | Aggregation Swtich | | | |
| | | Edge Swtich | | | |
| | | Server computer | | | |
| | | Workstation computer | | | |
| | | Personal computer | | | |
| | | Industrial Personal computer | | | |
| | | PLC (Programmable Logic Controller) | | | |
| | | Power Supply controller | | | |
| | | MPS（Machine protection system） | | | |
| | | Serial device server | | | |
| | | EVG (Event Generator) | | | |
| | | EVR (Event Receiver) | | | |
| | | ATCA/μTCA (Advanced Telecom Computing Architecture) | | | |
| | | Paperless recorder/Temperature Controller | | | |
| | | UPS | | | |
| | | Cabinet | | | |

# Appendix 3: Electric Power Requirement

The primary power consumption is attributed to the RF system, magnet system, cryogenics system, vacuum system, and heat removal devices.

## A3.1: RF System

The wall-plug power consumption for RF is outlined in Table A3.1 for the Collider and in Table A3.2 for the Booster. Additionally, the power consumption of the Linac klystrons is estimated and documented in Table A3.3.

**Table A3.1:** Collider RF wall plug power efficiency.

| | |
|---|---|
| Wall to PSM power supply/modulator | 95% |
| Modulator to klystron | 96% |
| Klystron to Waveguide | 75% |
| Waveguide to coupler | 95% |
| Coupler to cavity | ~100% |
| Cavity to beam | ~100% |
| Overall efficiency | ~65% |
| LLRF control | 5% more |

Required wall plug power for Collider RF = beam power × 1.05 / 0.65 = 96.9MW. (beam power = 30 MW per beam or 60 MW total)

**Table A3.2:** Booster RF wall plug power efficiency.

| | |
|---|---|
| Wall to SSA power supply | 95% |
| SSA number | 96% |
| Operation pulsed power (H mode) | 18.2 kW |
| Duty factor (H mode) | 3.8% |
| SSA to waveguide | 50% |
| Waveguide to coupler | 98% |
| Coupler to cavity | ~100% |
| Cavity to beam | ~100% |
| LLRF control | 5% more power |

Required wall plug power for Booster RF =1.05*96*18.2*3.8%/50%/95%/98%= 0.15 MW.

**Table A3.3:** Linac RF wall plug power efficiency.

| | |
|---|---|
| Wall to Modulator | 95% |
| Modulator to klystron | 70% |
| Klystron to accelerator structure | 40% |
| Klystron operation pulsed power | (40*147+65*35) MW |
| Duty factor | 0.04% |
| Klystron number | 147 C-band + 35 S-band |

Required wall plug power for Linac RF = (40*147+65*35)*0.04%/40%/70%/95%= 12.26 MW.

## A3.2: Magnet System

The power consumption attributable to the magnet system across the four distinct operating modes is consolidated in Tables A3.4, A3.5, A3.6, and A3.7. It's noteworthy that magnet power consumption constitutes approximately 70% of the total system loss. The magnet power supplies operate via switched mode, demonstrating an estimated efficiency of around 90% under full load and 80% under low load conditions. To calculate the power loss of the power supplies, the formula used is: (cable loss + magnet loss) * (1 / 0.9 / 0.95 - 1).

Cable loss: Magnets within two neighboring half-arcs will interconnect in series, drawing power from one or possibly two power supplies located within ground-level halls. For single magnet loads, power supplies will be positioned in auxiliary stub tunnels adjacent to the main tunnel. These cables are constructed from copper, with a cable current density maintained below $2A/mm^2$, resulting in cable loss accounting for approximately 20% of the overall system loss.

**Table A3.4:** Power consumption of the magnet system in H mode (30 MW /beam)

| Location | Magnet | Magnet cable | Power supply | Total |
|---|---|---|---|---|
| Collider | 31.22 | 6.72 | 4.22 | 42.16 |
| Booster | 5.18 | 2.43 | 0.85 | 8.46 |
| Linac and BTL | 6.08 | 0.25 | 0.70 | 7.04 |
| IR | 0.0 | 0.026 | 0.274 | 0.3 |
| Total | 42.49 | 9.43 | 6.04 | 57.96 |

**Table A3.5:** Power consumption of the magnet system in W mode (30 MW /beam)

| Location | Magnet | Magnet cable | Power supply | Total |
|---|---|---|---|---|
| Collider | 13.88 | 3.17 | 1.89 | 18.94 |
| Booster | 2.34 | 1.08 | 0.38 | 3.80 |
| Linac and BTL | 6.08 | 0.25 | 0.70 | 7.04 |
| IR | 0.0 | 0.02 | 0.1 | 0.12 |
| Total | 22.30 | 4.52 | 3.01 | 29.90 |

**Table A3.6:** Power consumption of the magnet system in Z mode (30 MW /beam)

| Location | Magnet | Magnet cable | Power supply | Total |
|---|---|---|---|---|
| Collider | 4.77 | 1.26 | 0.67 | 6.71 |
| Booster | 0.80 | 0.25 | 0.13 | 1.28 |
| Linac and BTL | 6.08 | 0.25 | 0.70 | 7.04 |
| IR | 0.0 | 0.02 | 0.1 | 0.12 |
| Total | 11.65 | 1.89 | 1.6 | 15.15 |

**Table A3.7:** Power consumption of the magnet system in $t\bar{t}$ mode (30 MW /beam)

| Location | Magnet | Magnet cable | Power supply | Total |
|---|---|---|---|---|
| Collider | 69.08 | 14.64 | 9.30 | 93.03 |
| Booster | 11.58 | 5.47 | 1.89 | 18.94 |
| Linac and BTL | 6.08 | 0.25 | 0.70 | 7.04 |
| IR | 0.0 | 0.026 | 0.274 | 0.3 |
| Total | 86.74 | 20.39 | 12.18 | 119.31 |

## A3.3: Cryogenic System

The cryogenic system responsible for the SRF cavities delivers 2K superfluid helium to the cryomodules in both the Collider and the Booster. In operation at the Higgs 30 MW mode, with a safety margin of 1.54 times, the combined 4.5K equivalent heat load for the Booster and Collider amounts to 49.46 kW. To manage this load, four individual 15 kW refrigerators operating at 4.5K will be utilized. The large refrigerator within the CEPC exhibits a Coefficient of Performance (COP) at 4.5K of approximately 219 W/W. With these parameters, the installed power for the Collider and Booster in the Higgs 30 MW mode is estimated at 9.72 MW and 1.71 MW, respectively.

The cryogenic system dedicated to the SC magnets provides 4.5K helium for the cryomodules utilized by the IR magnets. In total, the 4.5K equivalent heat load for both the Booster and Collider stands at 0.745 kW. With a Coefficient of Performance (COP) at 4.5K approximately at 219 W/W, the necessary installed power is calculated to be 0.16 MW.

During operation modes of H, W, Z, and $t\bar{t}$, the computed 4.5K equivalent heat load for the Booster and Collider is detailed in Tables A3.8 and A3.9.

**Table A3.8:** Electrical consumption during operation for Cryogenics (30 MW /beam)

| Mode | Location and electrical consumption (MW) | | | | Total (MW) |
|---|---|---|---|---|---|
| | Collider | Booster | Linac and BTL | IR | |
| H | 9.72 | 1.71 | | 0.16 | 11.59 |
| W | 4.39 | 0.69 | | 0.16 | 5.24 |
| Z | 4.4 | 0.8 | | 0.16 | 5.36 |
| $t\bar{t}$ | 27.53 | 2.32 | | 0.16 | 30.01 |

**Table A3.9:** The calculated 4.5K equiv. heat load of Booster and Collider (30 MW /beam)

| | Collider | | | Booster | | |
|---|---|---|---|---|---|---|
| | H | W | Z | H | W | Z |
| Equiv. heat load at 4.5 K (kW) | 44.39 | 20.16 | 15.15 | 7.82 | 2.44 | 3.52 |

## A3.4: Vacuum System

The electrical components within the vacuum system primarily consist of ion pumps and vacuum gauges, with their respective electricity consumption detailed in Table A3.10.

**Table A3.10:** Electrical consumption during operation for the vacuum system (30 MW /beam)

| Mode | Location and electrical consumption (MW) | | | | Total (MW) |
|---|---|---|---|---|---|
| | Collider | Booster | Linac and BTL | IR | |
| H | 5.4 | 4.2 | 0.6 | | 10.2 |
| W | 9.78 | 3.8 | 0.65 | | 14.23 |
| Z | 9.6 | 3.8 | 0.65 | | 14.05 |
| $t\bar{t}$ | 9.9 | 4.2 | 0.65 | | 14.75 |

## A3.5: Heat Removal Devices

The power demand for heat removal devices, encompassing water cooling, ventilation, and air conditioning, is projected to account for approximately 15% of the overall electric power. Achieving this target necessitates meticulous design considerations.

## A3.6: Total Facility

We factor in additional power consumption for instrumentation, control, radiation protection, detectors (referred to as "experimental devices"), and general services to compute the total power consumption for CEPC during its operation in the H, W, Z, and $t\bar{t}$ modes. The summarized data is presented below in Tables A3.11, A3.12, A3.13, and A3.14.

**Table A3.11:** Total facility power consumption in Higgs mode (30 MW/beam)

| | System for Higgs (30 MW /beam) | Location and power Requirement (MW) | | | | | | Total (MW) |
|---|---|---|---|---|---|---|---|---|
| | | Collider | Booster | Linac | BTL | IR | Surface building | |
| 1 | RF Power Source | 96.90 | 0.15 | 12.26 | | | | 109.31 |
| 2 | Cryogenic System | 9.72 | 1.71 | | | 0.16 | | 11.59 |
| 3 | Vacuum System | 5.40 | 4.20 | 0.60 | | | | 10.20 |
| 4 | Magnet System | 42.16 | 8.46 | 2.15 | 4.89 | 0.30 | | 57.96 |
| 5 | Instrumentation | 1.30 | 0.70 | 0.20 | | | | 2.20 |
| 6 | Radiation Protection | 0.30 | | 0.10 | | | | 0.40 |
| 7 | Control System | 1.00 | 0.60 | 0.20 | | | | 1.80 |
| 8 | Experimental Devices | | | | | 4.00 | | 4.00 |
| 9 | Utilities | 37.80 | 3.20 | 1.80 | 0.60 | 1.20 | | 44.60 |
| 10 | General Services | 7.20 | | 0.30 | 0.20 | 0.20 | 12.00 | 19.90 |
| | Total | 201.78 | 19.02 | 17.61 | 5.69 | 5.86 | 12.00 | 261.96 |

**Table A3.12:** Total facility power consumption in W mode (30 MW/beam)

| | System for W (30 MW /beam) | Location and power Requirement (MW) | | | | | | Total (MW) |
|---|---|---|---|---|---|---|---|---|
| | | Collider | Booster | Linac | BTL | IR | Surface building | |
| 1 | RF Power Source | 96.90 | 0.15 | 12.26 | | | | 109.31 |
| 2 | Cryogenic System | 6.16 | 0.68 | | | 0.16 | | 7.00 |
| 3 | Vacuum System | 9.78 | 3.80 | 0.65 | | | | 14.23 |
| 4 | Magnet System | 18.94 | 3.80 | 2.15 | 4.89 | 0.12 | | 29.90 |
| 5 | Instrumentation | 1.30 | 0.70 | 0.20 | | | | 2.20 |
| 6 | Radiation Protection | 0.30 | | 0.10 | | | | 0.40 |
| 7 | Control System | 1.00 | 0.60 | 0.20 | | | | 1.80 |
| 8 | Experimental Devices | | | | | 4.00 | | 4.00 |
| 9 | Utilities | 29.60 | 2.80 | 2.20 | 0.60 | 1.20 | | 36.40 |
| 10 | General Services | 7.20 | | 0.30 | 0.20 | 0.20 | 12.00 | 19.90 |
| | Total | 171.18 | 12.53 | 18.06 | 5.69 | 5.68 | 12.00 | 225.14 |

**Table A3.13:** Total facility power consumption in Z mode (30 MW/beam)

| | System for W (30 MW /beam) | Location and power Requirement (MW) | | | | | | Total (MW) |
|---|---|---|---|---|---|---|---|---|
| | | Collider | Booster | Linac | BTL | IR | Surface building | |
| 1 | RF Power Source | 96.90 | 0.15 | 12.26 | | | | 109.31 |
| 2 | Cryogenic System | 3.32 | 0.77 | | | 0.16 | | 4.25 |
| 3 | Vacuum System | 9.60 | 3.80 | 0.65 | | | | 14.05 |
| 4 | Magnet System | 6.71 | 1.28 | 2.15 | 4.89 | 0.05 | | 15.08 |
| 5 | Instrumentation | 1.30 | 0.70 | 0.20 | | | | 2.20 |
| 6 | Radiation Protection | 0.25 | | 0.10 | | | | 0.35 |
| 7 | Control System | 1.00 | 0.60 | 0.20 | 0.005 | 0.005 | | 1.81 |
| 8 | Experimental Devices | | | | | 4.00 | | 4.00 |
| 9 | Utilities | 25.80 | 2.80 | 2.00 | 0.60 | 1.20 | | 32.40 |
| 10 | General Services | 7.20 | | 0.30 | 0.20 | 0.20 | 12.00 | 19.90 |
| | Total | 152.08 | 10.10 | 17.86 | 5.70 | 5.62 | 12.00 | 203.35 |

**Table A3.14:** Total facility power consumption in $t\bar{t}$ mode (30 MW/beam)

| | System for $t\bar{t}$ (30 MW /beam) | Location and power Requirement (MW) | | | | | | Total (MW) |
|---|---|---|---|---|---|---|---|---|
| | | Collider | Booster | Linac | BTL | IR | Surface building | |
| 1 | RF Power Source | 96.90 | 0.15 | 12.26 | | | | 109.31 |
| 2 | Cryogenic System | 27.53 | 2.32 | | | 0.16 | | 30.01 |
| 3 | Vacuum System | 9.90 | 4.20 | 0.65 | | | | 14.75 |
| 4 | Magnet System | 93.03 | 18.94 | 2.15 | 4.89 | 0.30 | | 119.31 |
| 5 | Instrumentation | 1.30 | 0.70 | 0.20 | | | | 2.20 |
| 6 | Radiation Protection | 0.30 | | 0.10 | | | | 0.40 |
| 7 | Control System | 1.00 | 0.60 | 0.20 | | | | 1.80 |
| 8 | Experimental Devices | | | | | 4.00 | | 4.00 |
| 9 | Utilities | 47.20 | 4.80 | 2.50 | 0.60 | 1.20 | | 56.30 |
| 10 | General Services | 7.20 | | 0.30 | 0.20 | 0.20 | 12.00 | 19.90 |
| | Total | 284.36 | 31.71 | 18.36 | 5.69 | 5.86 | 12.00 | 357.98 |

## A3.7: Electric Power Requirements for Beam Power Upgrade

In the preceding sections A3.1 – A3.6, the electric power requirements are detailed for a synchrotron radiation (SR) power of 30 MW per beam. When the SR power is upgraded to 50 MW per beam, the comprehensive power consumption for CEPC during its operation in the H, W, Z, and $t\bar{t}$ modes is provided in Tables A3.15 – A3.18 below.

**Table A3.15:** Total facility power consumption in Higgs mode (50 MW/beam)

| | System for Higgs (50 MW /beam) | Location and power Requirement (MW) | | | | | | Total (MW) |
|---|---|---|---|---|---|---|---|---|
| | | Collider | Booster | Linac | BTL | IR | Surface building | |
| 1 | RF Power Source | 161.60 | 1.73 | 14.10 | | | | 177.43 |
| 2 | Cryogenic System | 9.17 | 1.77 | | | 0.16 | | 11.10 |
| 3 | Vacuum System | 5.40 | 4.20 | 0.60 | | | | 10.20 |
| 4 | Magnet System | 42.16 | 8.46 | 2.15 | 4.89 | 0.30 | | 57.96 |
| 5 | Instrumentation | 1.30 | 0.70 | 0.20 | | | | 2.20 |
| 6 | Radiation Protection | 0.30 | | 0.10 | | | | 0.40 |
| 7 | Control System | 1.00 | 0.60 | 0.20 | | | | 1.80 |
| 8 | Experimental Devices | | | | | 4.00 | | 4.00 |
| 9 | Utilities | 46.40 | 3.80 | 2.50 | 0.60 | 1.20 | | 54.50 |
| 10 | General Services | 7.20 | | 0.30 | 0.20 | 0.20 | 12.00 | 19.90 |
| | Total | 274.53 | 21.26 | 20.15 | 5.69 | 5.86 | 12.00 | 339.49 |

**Table A3.16:** Total facility power consumption in W mode (50 MW/beam)

| | System for W (50 MW /beam) | Location and power Requirement (MW) | | | | | | Total (MW) |
|---|---|---|---|---|---|---|---|---|
| | | Collider | Booster | Linac | BTL | IR | Surface building | |
| 1 | RF Power Source | 161.60 | 0.10 | 14.10 | | | | 175.80 |
| 2 | Cryogenic System | 4.39 | 0.69 | | | 0.16 | | 5.24 |
| 3 | Vacuum System | 9.90 | 4.20 | 0.60 | | | | 14.70 |
| 4 | Magnet System | 18.94 | 3.80 | 2.15 | 4.89 | 0.12 | | 29.90 |
| 5 | Instrumentation | 1.30 | 0.70 | 0.20 | | | | 2.20 |
| 6 | Radiation Protection | 0.30 | | 0.10 | | | | 0.40 |
| 7 | Control System | 1.00 | 0.60 | 0.20 | | | | 1.80 |
| 8 | Experimental Devices | | | | | 4.00 | | 4.00 |
| 9 | Utilities | 37.60 | 2.87 | 2.50 | 0.60 | 1.20 | | 44.77 |
| 10 | General Services | 7.20 | | 0.30 | 0.20 | 0.20 | 12.00 | 19.90 |
| | Total | 242.23 | 12.96 | 20.15 | 5.69 | 5.68 | 12.00 | 298.71 |

**Table A3.17:** Total facility power consumption in Z mode (50 MW/beam)

| | System for Z (50 MW /beam) | Location and power Requirement (MW) | | | | | | Total (MW) |
|---|---|---|---|---|---|---|---|---|
| | | Collider | Booster | Linac | BTL | IR | Surface building | |
| 1 | RF Power Source | 169.30 | 0.34 | 14.10 | | | | 183.74 |
| 2 | Cryogenic System | 4.40 | 0.80 | | | 0.16 | | 5.36 |
| 3 | Vacuum System | 9.90 | 4.20 | 0.60 | | | | 14.70 |
| 4 | Magnet System | 6.71 | 1.28 | 2.15 | 4.89 | 0.12 | | 15.15 |
| 5 | Instrumentation | 1.30 | 0.70 | 0.20 | | | | 2.20 |
| 6 | Radiation Protection | 0.30 | | 0.10 | | | | 0.40 |
| 7 | Control System | 1.00 | 0.60 | 0.20 | | | | 1.80 |
| 8 | Experimental Devices | | | | | 4.00 | | 4.00 |
| 9 | Utilities | 32.64 | 2.43 | 2.50 | 0.60 | 1.20 | | 39.37 |
| 10 | General Services | 7.20 | | 0.30 | 0.20 | 0.20 | 12.00 | 19.90 |
| | Total | 232.75 | 10.35 | 20.15 | 5.69 | 5.68 | 12.00 | 286.62 |

**Table A3.18:** Total facility power consumption in $t\bar{t}$ mode (50 MW/beam)

| | System for $t\bar{t}$ (50 MW /beam) | Location and power Requirement (MW) | | | | | | Total (MW) |
|---|---|---|---|---|---|---|---|---|
| | | Collider | Booster | Linac | BTL | IR | Surface building | |
| 1 | RF Power Source | 161.60 | 0.24 | 14.10 | | | | 175.94 |
| 2 | Cryogenic System | 27.53 | 2.32 | | | 0.16 | | 30.01 |
| 3 | Vacuum System | 9.90 | 4.20 | 0.60 | | | | 14.70 |
| 4 | Magnet System | 93.03 | 18.94 | 2.15 | 4.89 | 0.30 | | 119.31 |
| 5 | Instrumentation | 1.30 | 0.70 | 0.20 | | | | 2.20 |
| 6 | Radiation Protection | 0.30 | | 0.10 | | | | 0.40 |
| 7 | Control System | 1.00 | 0.60 | 0.20 | | | | 1.80 |
| 8 | Experimental Devices | | | | | 4.00 | | 4.00 |
| 9 | Utilities | 55.00 | 4.80 | 2.50 | 0.60 | 1.20 | | 64.10 |
| 10 | General Services | 7.20 | | 0.30 | 0.20 | 0.20 | 12.00 | 19.90 |
| | Total | 356.86 | 31.80 | 20.15 | 5.69 | 5.86 | 12.00 | 432.36 |

# Appendix 4: Risk Analysis and Mitigation Measures

## A4.1: Machine-Detector Interface (MDI)

An assembly sequence for the interaction region (IR) is crucial to demonstrate the feasibility of the design.

The principles for mechanically supporting the various machine and detector components in the shared central regions are particularly important for the cantilevered cryostat housing the final doublet. This should take into account the mechanical interface with the central beam pipe through a remote vacuum connector.

Considering the integrated electromagnetic forces, which can be very large, a mechanical stabilization strategy is necessary, especially in the event of a quench. A monitoring alignment system that measures the deformation of the MDI should be implemented once the MDI is finalized.

Further studies on collimation are required to control beam-induced backgrounds near the detectors and protect critical accelerator components. This includes evaluating beam-loss tolerances, analyzing collimator survival and damage under realistic conditions, and assessing the impact of chosen collimation apertures on beam lifetimes and the impedance budget.

Specification, design, and evaluation of tuning procedures are important to maximize luminosity performance under realistic error conditions. This involves bringing and maintaining beams in collision using feedback mechanisms (such as the beam-beam deflection method), correcting for optical aberrations through local beam size tuning knobs, and mitigating beam blow-up caused by the beam-beam interaction.

Synchrotron radiation generated at the IR by the solenoidal fringe fields will reach the vacuum chamber as far as 213 m from the IP. The impact of synchrotron radiation on the MDI area is significant at all energy levels, especially at the ttbar threshold where the critical energy is highest.

## A4.2: Survey and Alignment

In the case of a newly built 100 km tunnel, ground motion is inevitable and can potentially disrupt existing alignment results, leading to component position errors exceeding tolerances. To address this issue, periodic surveys must be conducted, and sensors should be installed to monitor the position of components.

Temperature differences during component fiducialization, installation, and machine operation can impact alignment results. Therefore, it is crucial to control the temperature during these three periods. Ideally, the temperatures should be maintained at the same level, and the effects of temperature changes on component alignment should be thoroughly studied.

The schedule for survey and alignment is tight, and any accidents can cause delays in completing the work on time. To mitigate this risk, it may be necessary to prepare additional instruments and allocate more human resources.

In the MDI, the cryostat is designed with a cantilevered structure. However, this design poses a significant risk to the mechanical stabilization due to the integrated electromagnetic forces, which can be substantial. Therefore, implementing a monitoring alignment system that measures the deformation of the MDI is of utmost importance in this regard.

## A4.3: Machine Protection and Collimation System

The energy stored in the CEPC is on the order of MJ, as indicated in Table A4.3.1. In the event of an uncontrolled beam striking the accelerator beam-pipe, it can lead to damage to the detectors, accelerator components and surrounding equipment. Therefore, the machine protection and collimation system play a crucial role and can have potential impacts on the project. This includes considerations for robustness against injection failures, asynchronous beam-dump kicker firing, "dust" events, sudden beam-loss phenomena (similar to the SuperKEKB), and collimator survival in case of reduced beam lifetime.

**Table A4.3.1:** Beam current and stored energy for different energy modes.

| | | Higgs | Z | WW | $t\bar{t}$ |
|---|---|---|---|---|---|
| Beam energy (GeV) | | 120 | 45.5 | 80 | 180 |
| 30MW | Current (mA) | 16.7 | 803.5 | 84.1 | 3.3 |
| | Energy (MJ) | 0.66 | 12.0 | 2.2 | 0.20 |
| 50MW | Current (mA) | 27.8 | 1339.2 | 140.2 | 5.5 |
| | Energy (MJ) | 1.1 | 20 | 3.7 | 0.33 |

Several principles for machine protection, including "protect the machine," "protect the beam," and "provide evidence," must be meticulously considered. The top priority is to prevent damage to detectors and accelerator components. In the event of any unusual activity, the system should promptly assess whether the machine needs to be shut down. It is crucial to compile a list of potential failures that could result in beam loss and subsequent machine breaks. At a minimum, the most probable failure modes, especially worst-case scenarios and their probabilities, must be taken into consideration.

Simultaneously, efforts should be made to maximize operational time, striking a balance between protection and operation. If the protection system necessitates halting operations, clear diagnostics should be provided to enable a thorough understanding of the situation. It is essential to maintain both long-term logging of parameters with low frequency and transient records. This dual approach ensures comprehensive monitoring and documentation of system behavior.

There are two types of protection schemes: active protection and passive protection. In active protection, an action is initiated upon detection of a failure signal, such as extracting a beam to dumps. Active protection involves a response time between the occurrence of failures and the initiation of the protection system. Conversely, passive protection mechanisms utilize specialized equipment, such as collimators and shields, to localize beam loss. This localization helps significantly reduce beam loss at critical locations, such as detectors and superconducting magnets.

In designing the collimator, beam loss simulations are crucial and should encompass various scenarios, including:

1. Optimum/acceptable/particular operating conditions:
   - Detector background control, beam halo/tails, top-up injection.
   - Changes in optics, tuning, collimator aperture setting, etc.
2. Fast beam loss:
   - Rapid equipment failure, injection failure, beam abort failure, unexpected tuning beam failure, sudden beam loss (comprehension is essential).
3. Different operation modes, suc as Higgs, Z, W and $t\bar{t}$.

Simulations also play a crucial role in determining the optimal design parameters for collimators by considering beam-matter interactions. This includes determining the location, multi-stage design, gap, length, quantity, taper, material, energy deposition, and other relevant parameters for each collimator. The design of local shielding for the collimators should be optimized based on simulations of beam-matter interactions, similar to the approach used in the design of HEPS (High Energy Photon Source) collimators. The impact on downstream components caused by the collimators will be analyzed through a combination of tracking simulations and beam-matter interaction simulations. Based on these analyses, local shielding will be implemented in all downstream regions that are affected by the collimators to ensure optimal performance and protection of the system. The dose at the location of manual maintenance and driven devices for movable collimators should also be carefully simulated. In addition, it is important to reduce the impedance caused by collimators, as impedance can limit the accelerator performance.

In the preliminary CEPC collimator design, both the betatron collimators and the momentum collimators are installed in the Collider rings. Simulation results reveal that beam losses resulting from failures in bending magnets and sextupole magnets can result in a significant background in IRs. Dedicated collimators are installed to shield the beam loss in the superconducting magnets near IPs. Meanwhile, masks upstream of IP are installed to protect against synchrotron radiation. As a result, approximately 50 collimators per ring are installed in the Collider.

Continuous design studies for the machine protection and collimation system will be conducted as part of the ongoing EDR program. This includes relevant R&D efforts and international collaboration with currently operational facilities like SuperKEKB and LHC.

## A4.4: Ground Motion

With small beam sizes at the Interaction Point (IP) positions (several tens of nanometers in the vertical direction) and the presence of strong final focusing system (FFS) quadrupoles in the CEPC, beam orbit and luminosity are highly sensitive to vibrations. Among all types of vibrations, ground motion is the primary contributor. Controlling ground motion is crucial, and the most critical step is site selection, especially for a large infrastructure to be built 100 meters underground over a circumference of approximately 100 kilometers. For seeking a relative complete rock base (100 m underground as an average, being away from the earthquake zone), inviting a geologist involved in the site selection is necessary. Additionally, the selected site should be far from transportation routes, areas with high cultural noise, rivers or reservoirs, regions prone to frequent rainstorms, and large buildings to avoid wind-induced vibrations. Conducting year-round vibration monitoring with a complete seasonal cycle on candidate sites is essential to determine the dominant vibration sources and finalize the site location.

Mechanical vibrations caused by ground motion, particularly affecting the IR quadrupoles, can lead to lateral beam jitter, resulting in a shift in the beam centroids and incomplete overlap of the colliding beams at the IP. If this induced offset becomes significant compared to the beam size, it can significantly degrade the luminosity. Even for offsets smaller than the beam size, luminosity degradation can still occur due to additional sensitivity through the hourglass effect and beam-beam blow-up effects.

To ensure optimal beam collision conditions and maximize luminosity in the presence of ground motion, beam orbit feedback systems are essential at the IP. Two potential methods exist for the IP orbit feedback system. The first method relies on the beam-beam deflection driven approach, utilizing measurements from beam position monitors (BPM) upstream and downstream of the IP. The second method is based on luminosity measurement.

Preliminary assessments have evaluated the accuracy requirements of BPMs and the performance of the beam-beam deflection feedback. The current plan is to employ the beam-beam deflection method for vertical beam orbit feedback and utilize fast luminosity monitors for horizontal beam orbit feedback.

Minimizing the transfer function of the vibration transmission path is an effective approach to maintaining beam stability. The magnet-support assembly plays a crucial role in achieving this goal. The stability of the magnet-support system is typically assessed based on the assembly's natural frequency and vibration amplification factor. The requirements for magnet-support stability will be determined through beam simulations, and the support system techniques should be developed accordingly.

# Appendix 5: Applications in Photon Science

## A5.1: Operation as a High Intensity γ-ray Source

### A.5.1.1 Parameters and Design of the γ-ray Source

#### *A5.1.1.1: Parameters*

During their motion within the CEPC, high-energy electron beams are capable of generating high-quality synchrotron radiation (SR). This section delves into the design of the γ-ray source and explores its extensive applications in photonuclear physics, nuclear science and technology, nuclear astrophysics, as well as in the life sciences and cultural heritage preservation. We also delve into the detailed examination of beamline design and the γ-ray focusing system, along with a comprehensive discussion of various photon-counting gamma-ray spectrometers. To ensure the success of this project, a series of five workshops were conducted, engaging experts from relevant fields who offered invaluable insights and guidance [1].

The design philosophy underlying the synchrotron radiation lines places the utmost importance on preserving the integrity of particle collisions. Within this framework, CEPC is intended to function as a γ-ray source while minimizing any adverse impact on the concurrent high-energy physics operations. The existing bends in the arc region will be harnessed to generate γ-rays with a critical energy in the range of hundreds of keV. Additionally, a wiggler will be strategically installed to produce γ-rays with energies surpassing the mega-electron volt threshold, and its length will be deliberately kept short to curtail the growth of one-turn synchrotron radiation losses.

The synchrotron radiation lines for both the wiggler and bending magnets will be housed within a shared tunnel. Consequently, the specific locations for this wiggler and bending magnets have been chosen near the entrance of an arc region in the electron ring, as illustrated in Figures A5.1.1 and A5.1.2.

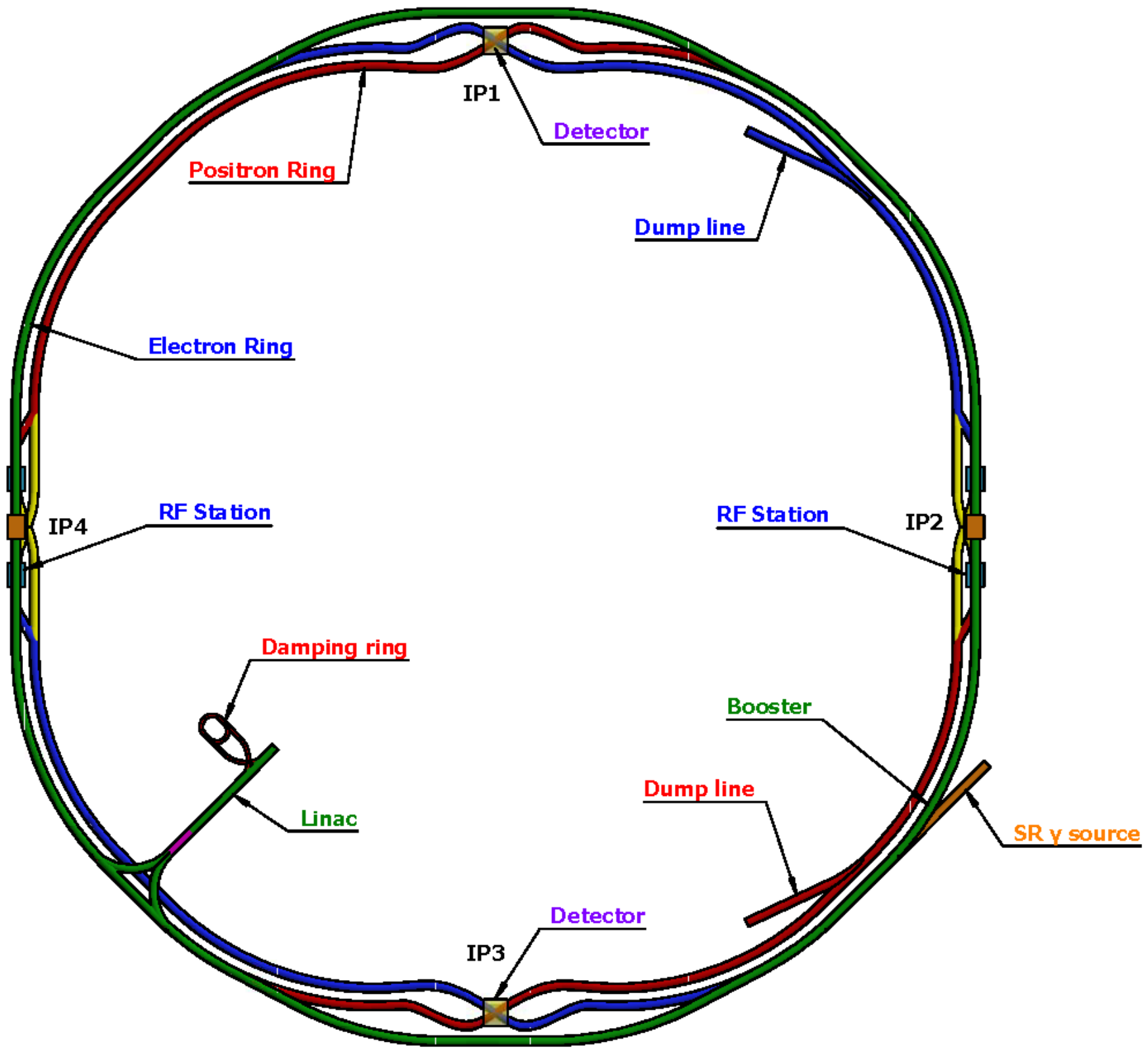

**Figure A5.1.1:** Location of the CEPC SR stations.

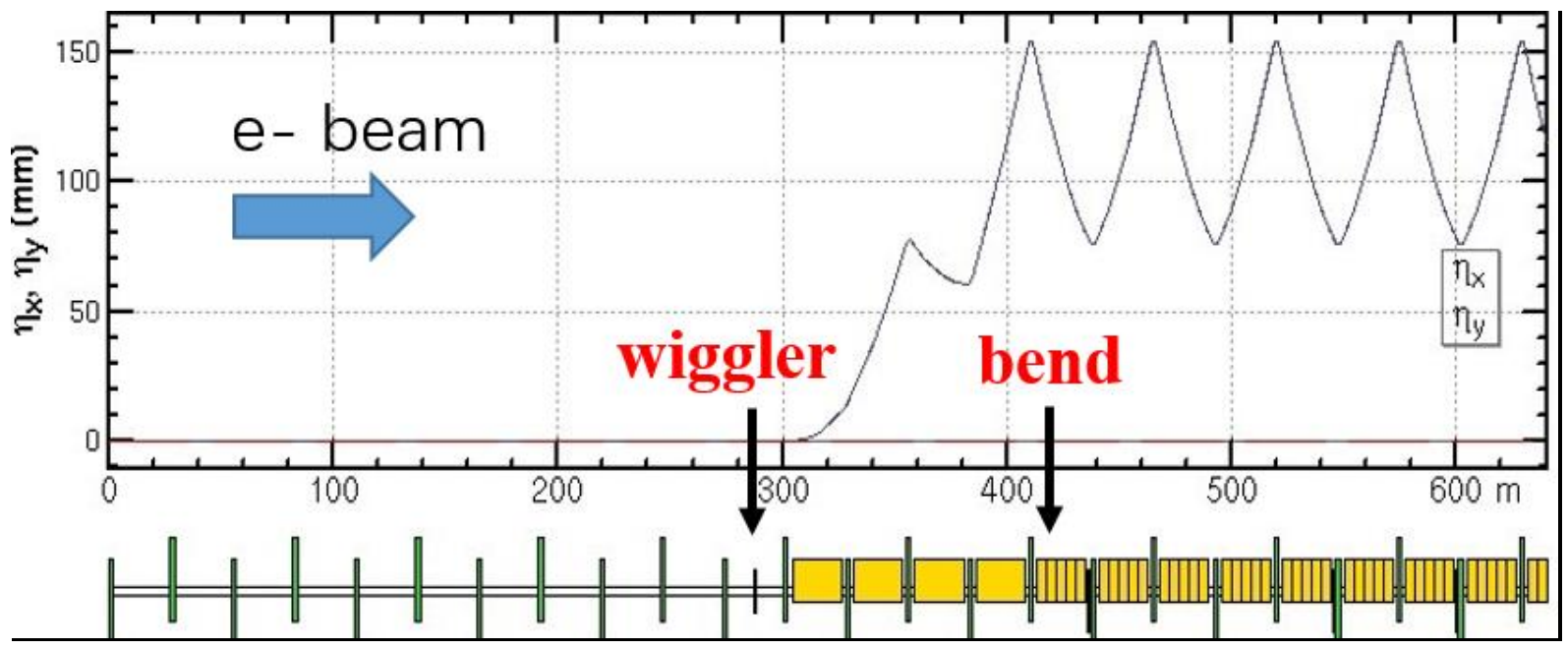

**Figure A5.1.2:** Locations of the wiggler and bending magnets.

The magnetic field strength of the wiggler is precisely 2.0 T, a requirement essential for the synchrotron radiation experiment. To minimize the increase in synchrotron radiation power, the wiggler consists of just one period and measures a mere 0.32 meters in length. This wiggler boasts a distinctive gamma energy value of 19.2 MeV, as indicated in Table A5.1.1.

In contrast, the bending magnet operates with a magnetic field strength of 400 Gs and spans approximately 20 meters in length. For a comprehensive view of the γ-ray source's spectral brightness in the CEPC, please refer to Figure A5.1.3. Detailed parameters for the CEPC can be found in Table A5.1.2.

**Table A5.1.1:** CEPC wiggler and bending magnet parameters.

| **Wiggler parameters** | |
|---|---|
| B (T) | 2 |
| Total length (m) | 0.32 |
| Magnetic period Length (m) | 0.32 |
| Period number | 1 |
| Characteristic gamma energy (MeV) | 19.2 |
| **Bending magnet parameters** | |
| B (T) | 0.04 |
| Length (m) | 20 |
| Critical energy (keV) | 383 |

**Table A5.1.2:** CEPC electron beam parameters

| | |
|---|---|
| Energy (GeV) | 120 |
| Current (mA) | 16.7 |
| Circumference (km) | 100 |
| Bunches | 268 |
| Bunch length (mm) | 4.1 |
| Coupling constant | 0.2% |
| Natural emittance (nm-rad) | 0.64 |
| Energy spread | 0.17% |

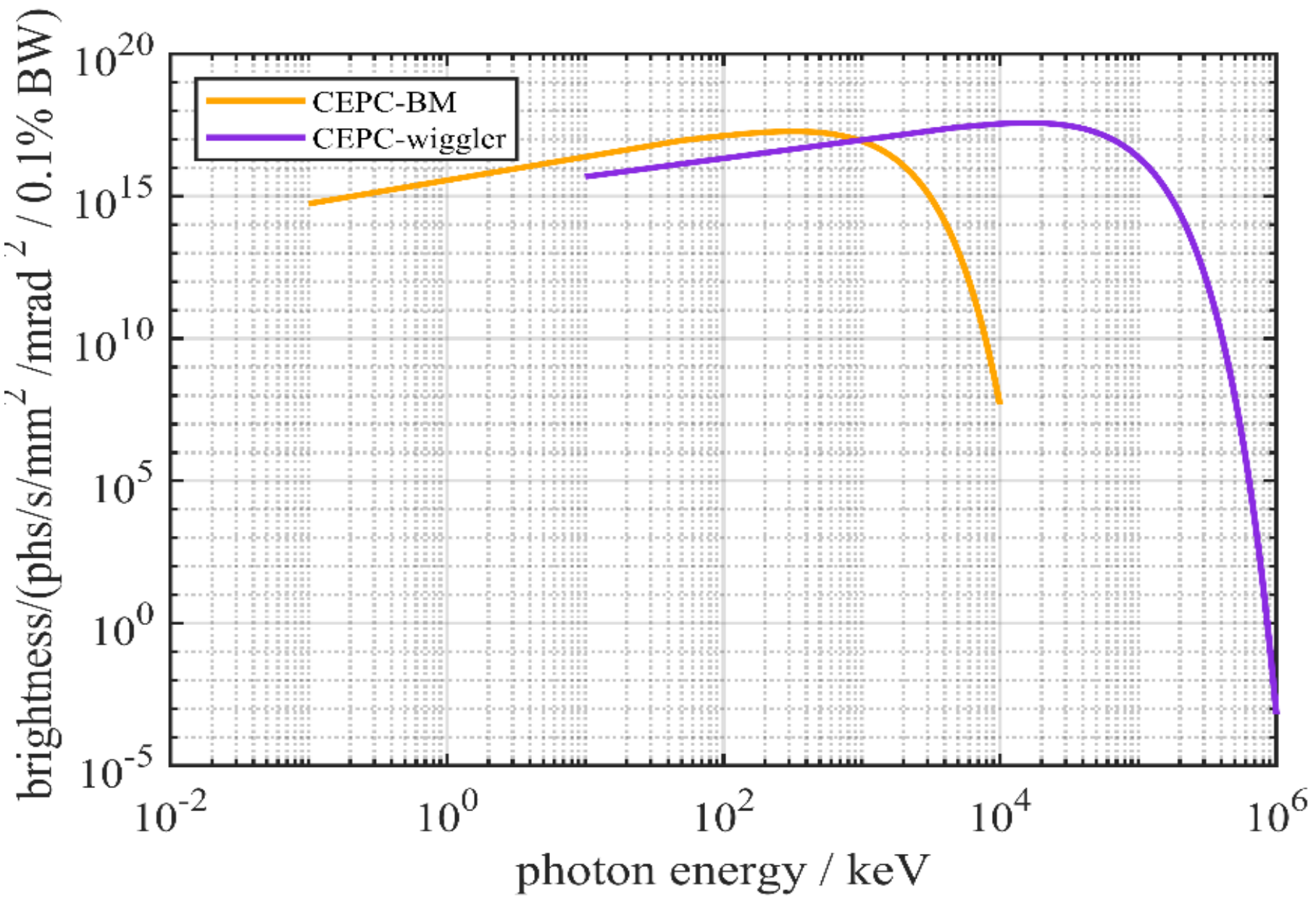

**Figure A5.1.3:** Spectral brightness of the CEPC, evaluated by synchrotron radiation from the bending magnet and wiggler.

Furthermore, in the energy range exceeding 100 keV, CEPC's synchrotron source demonstrates notable advantages in terms of brightness and flux. Figure A5.1.4 provides a visual comparison of the spectral brightness among the CEPC, HEPS [2], and SSRF [3].

The diagram illustrates the on-axis brightness for all beamlines, except for the APPLE-Knot beamline, during the high-brightness mode of the HEPS storage ring. A comparison between the flux and energy distribution of CEPC's gamma rays and those of the forthcoming high-energy Compton scattering gamma light source [4] is graphically represented in Figure A5.1.5.

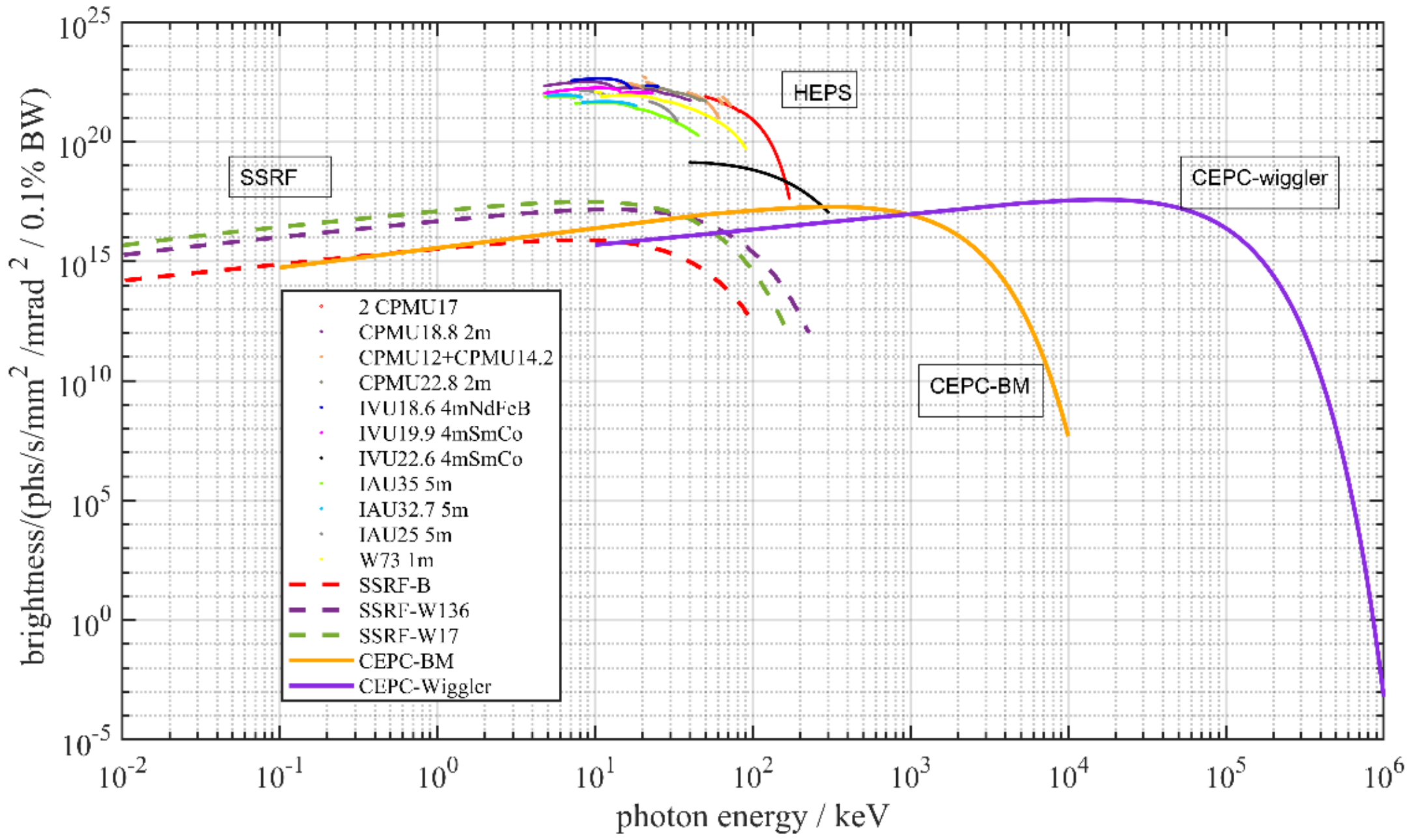

**Figure A5.1.4:** Comparison of spectrum brightness of the CEPC (bending magnet and wiggler), HEPS (13 beamlines at the high-brilliance mode) [2] and SSRF (3 beamlines).

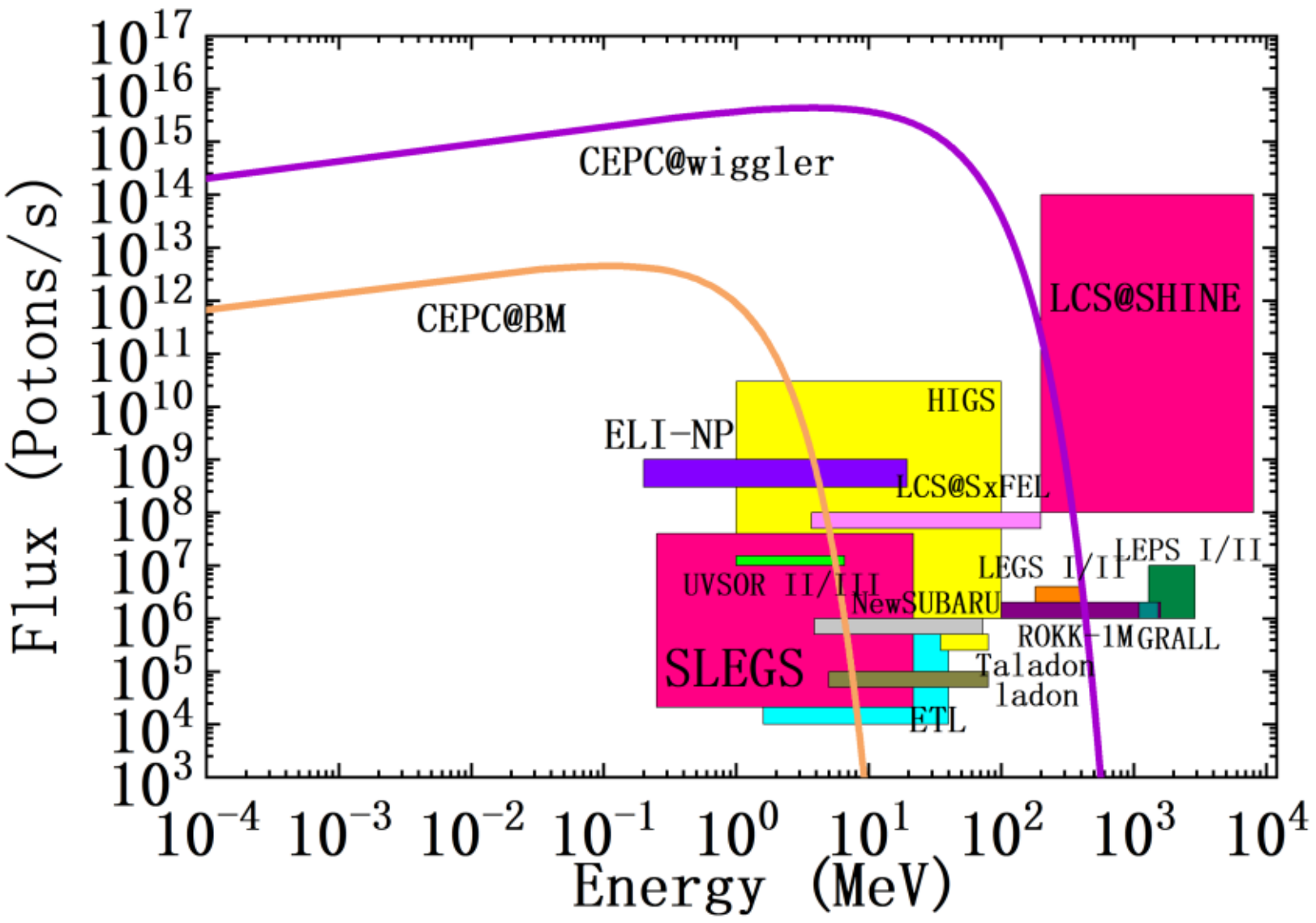

**Figure A5.1.5:** Comparison of the flux and energy region distribution between the CEPC γ-rays and the future high-energy Compton scattering gamma light source [4].

The performance of several important gamma sources is detailed in Table A5.1.3 and compared with the capabilities of the CEPC. Notably, the γ-ray flux generated by the CEPC significantly surpasses that of all other high-energy Compton scattering gamma light sources worldwide.

**Table A5.1.3:** Performance comparison between the CEPC gamma source and leading high-energy Compton scattering gamma light sources worldwide.

| Source | CEPC BM | CEPC Wiggler | SSRF (China) | TUNL HIGS (USA) | TERAS (Japan) | ALBL (Spain) |
|---|---|---|---|---|---|---|
| Gamma energy range (MeV) | 0.1~5 | 0.1~100 | 0.4-20<br>330-550 | 2-100 | 1-40 | 0.5-16<br>16-110<br>250-530 |
| Eenergy resolution (ΔE/E) | continuous | continuous | 5% | 0.8~10% | | |
| Flux | $> 10^{12}$@0.1% | $> 10^{16}$@0.1% | $10^6$ | $10^8$ | $10^4$-$10^5$ | $10^5$-$10^7$ |

In the development of the high-energy synchrotron light source for CEPC, three pivotal challenges must be addressed:

i. When designing the wiggler, factors such as radiation power, baseline layout, cost, and polarization for Z/W modes must be taken into account. The wiggler, featuring a 2 T magnetic field strength necessary for producing high-energy gamma beams, should not introduce excessive radiated power that could disrupt the normal operation of the CEPC collision mode.
ii. The design of an ultra-high energy synchronous beamline is essential.
iii. To ensure the quality of the generated gamma beams at the megaelectron volt scale and the proper functioning of the beamlines, ultra-hard X-ray focusing lenses will be employed.

The introduction of an additional wiggler for the SR γ-ray source leads to a slight increase in energy loss per turn, rising from 1.812 GeV to 1.823 GeV, representing a $\Delta U_0/U_0$ increment of 0.6%. An in-depth examination of the synchrotron radiation effects of the wiggler on beam dynamics has been conducted. These effects can be effectively mitigated by adjusting the magnet strength to accommodate the beam energy at each magnet. Following this tapering process, the closed orbit is confined to a mere 2 µm, as depicted in Figure A5.1.6.

Given the relatively short length of the wiggler, its nonlinear impact on single beam dynamics is negligible, and the dynamic aperture remains almost identical, both with and without the wiggler, as illustrated in Figure A5.1.7. The synchrotron radiation induced by the wiggler results in a modest increase in the natural energy spread, from 0.1% to 0.116%. Furthermore, the total energy spread, inclusive of the beam-beam effect, rises from 0.17% to 0.175%, thereby increasing the total bunch length by approximately 3%.

The reduction in luminosity due to the synchrotron radiation from the wiggler is approximately 3%, a relatively minor decrease, and further investigation to restore luminosity is a viable avenue. These findings are succinctly summarized in Table A5.1.4.

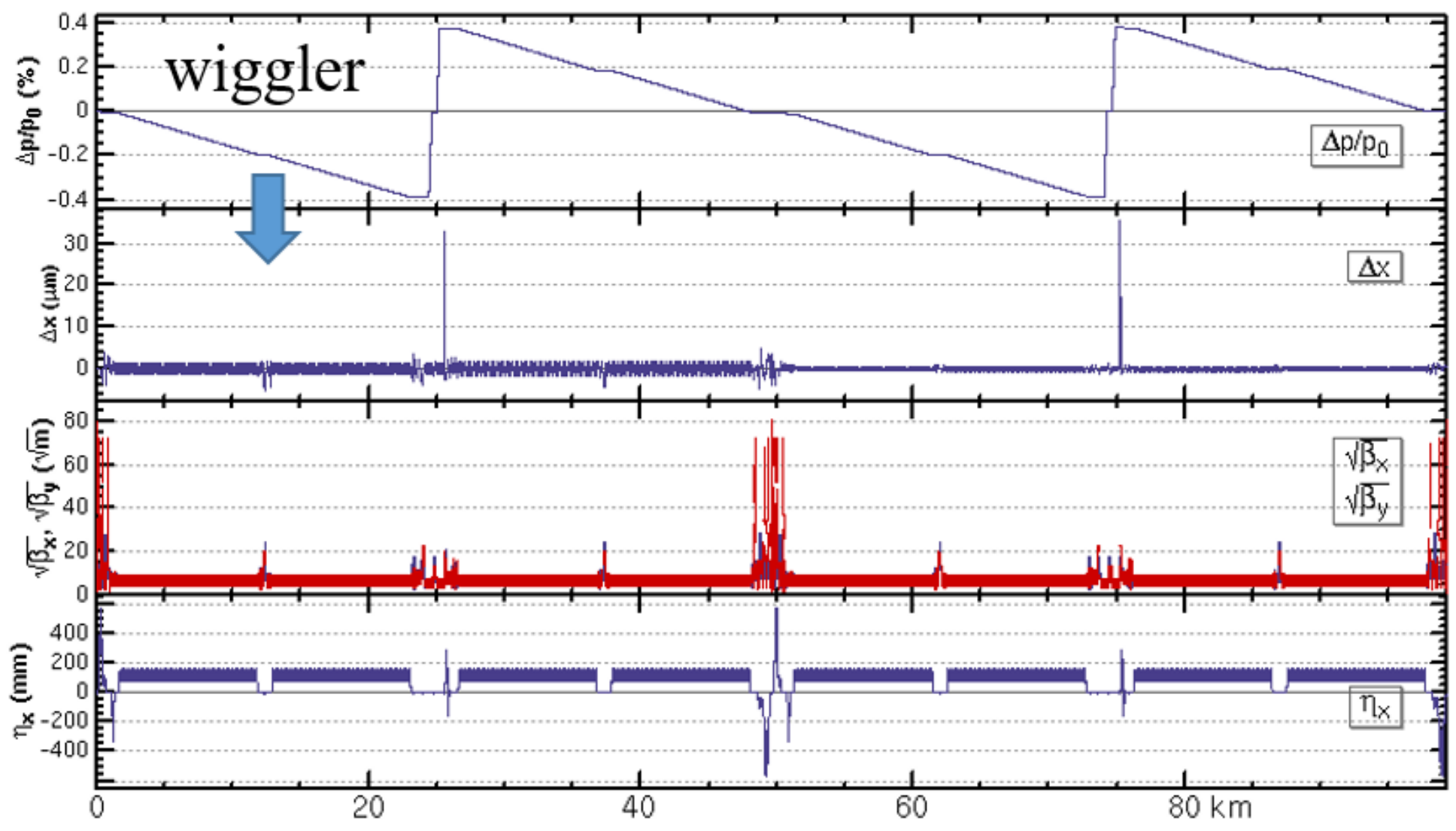

**Figure A5.1.6:** Orbit and optics with wiggler after magnet strength tapering in the CEPC collider ring at Higgs energy.

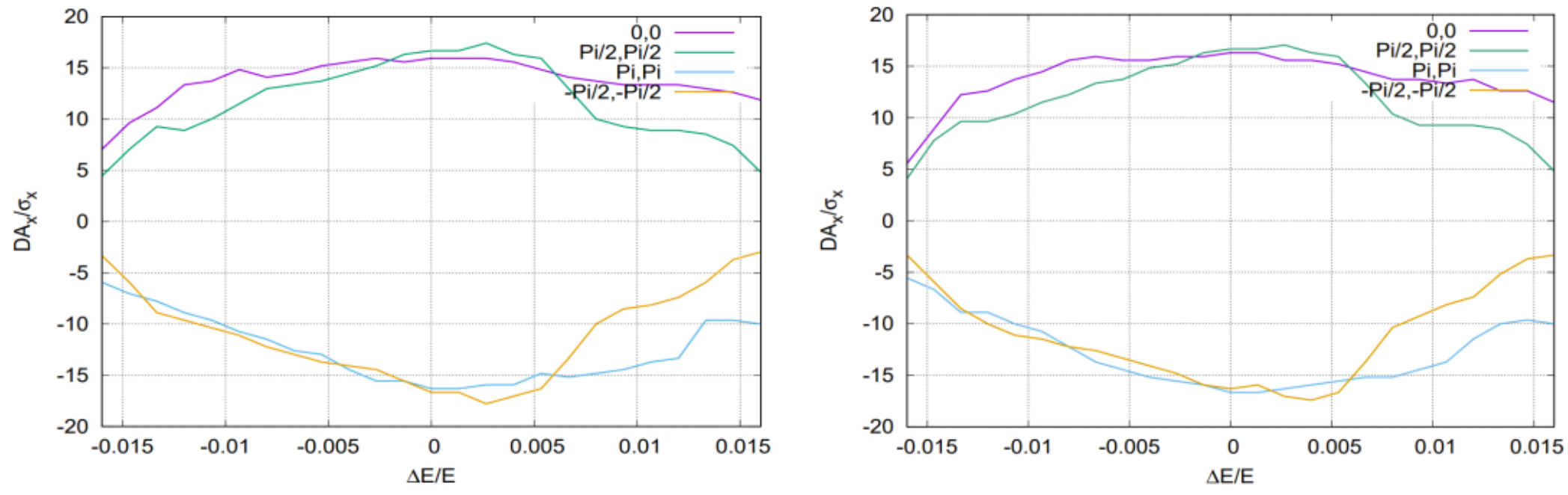

**Figure A5.1.7**: Comparison of the dynamic aperture with (left) and without (right) wiggler at Higgs energy.

**Table 5.1.4:** CEPC key beam parameters without and with wiggler.

| | **w/o wiggler** | **w/ wiggler** |
|---|---|---|
| Beam energy [GeV] | 120 | |
| Energy loss per turn (GeV) | 1.812 | 1.823 |
| Natural energy spread [%] | 0.1 | 0.116 |
| Energy loss per turn (GeV) | 1.812 | 1.823 |
| Total energy spread [%] | 0.17% | 0.175% |

### *A5.1.1.2: Design of the Vacuum Chamber*

The critical aspect of the beam extraction design involves creating a configuration that diverts the synchrotron beam from the primary beam path. The dipole magnet has a total length of about 20 meters with a magnetic field of 400 Gs. The wiggler is positioned at the forefront of the first dipole. The generation of synchrotron radiation commences at the first dipole and extends through to the last dipole. The separation of the photon beam from the electron beam is accomplished at the conclusion of the last dipole, as visualized in Figure A5.1.8. As illustrated in Figure A5.1.9, the vacuum box for the first meter remains unaltered, whereas the vacuum box for the second meter must be extended by an additional 150 mm.

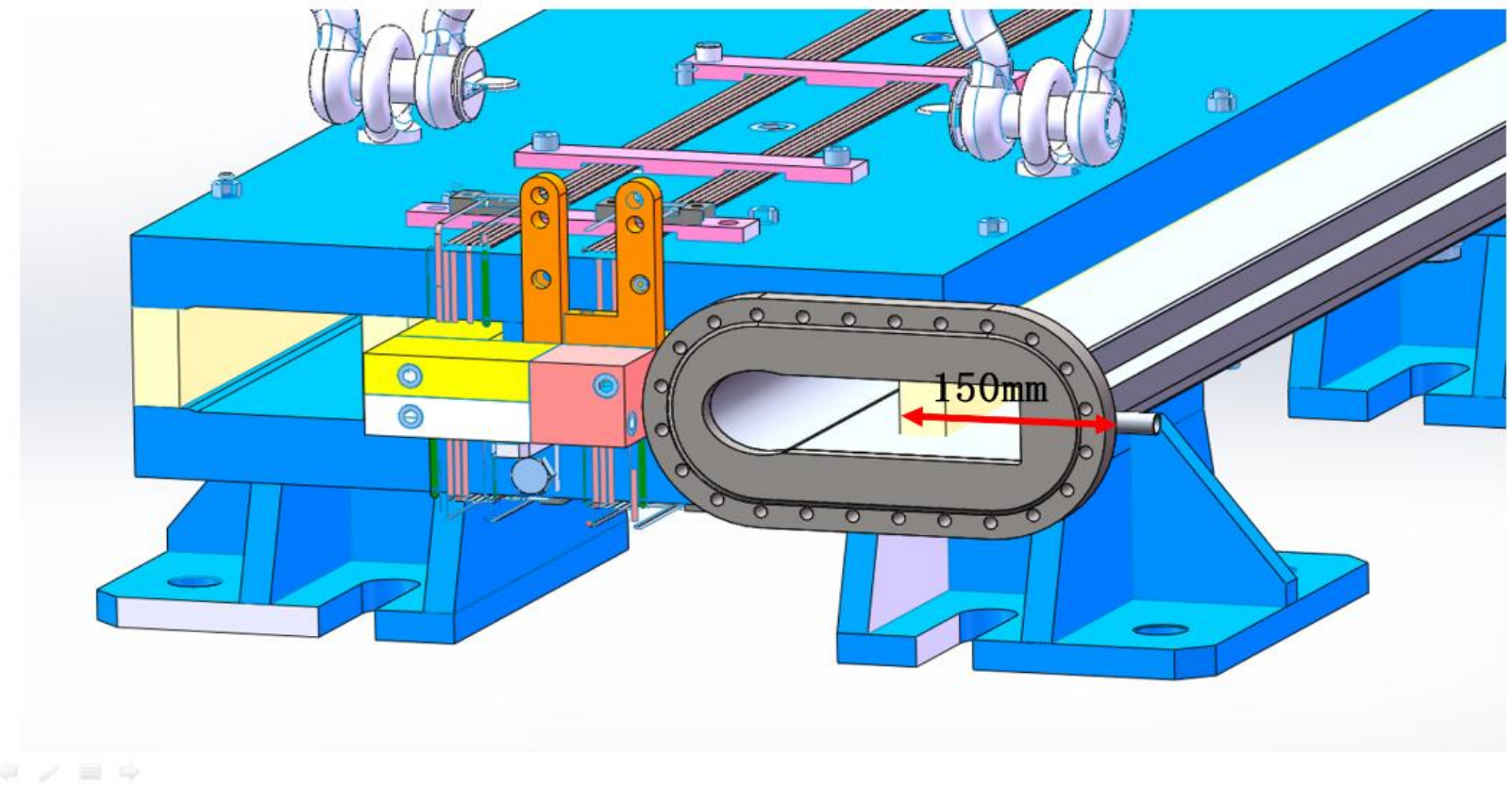

**Figure A5.1.8:** Vacuum pipe in the dipole and requirements for aperture extension at the end of the last dipole.

The separation of the synchrotron light from the electron beam takes place at the terminus of the dipole iron, where the synchrotron light is guided from the right side, as illustrated in Figure A5.1.9. A triangular photon absorber is positioned between the two conduits to capture any synchrotron light that is not directed outward. There is also a corresponding cooling water pipe behind the flat box buttressed with a 100 mm diameter synchrotron light lead-out pipe.

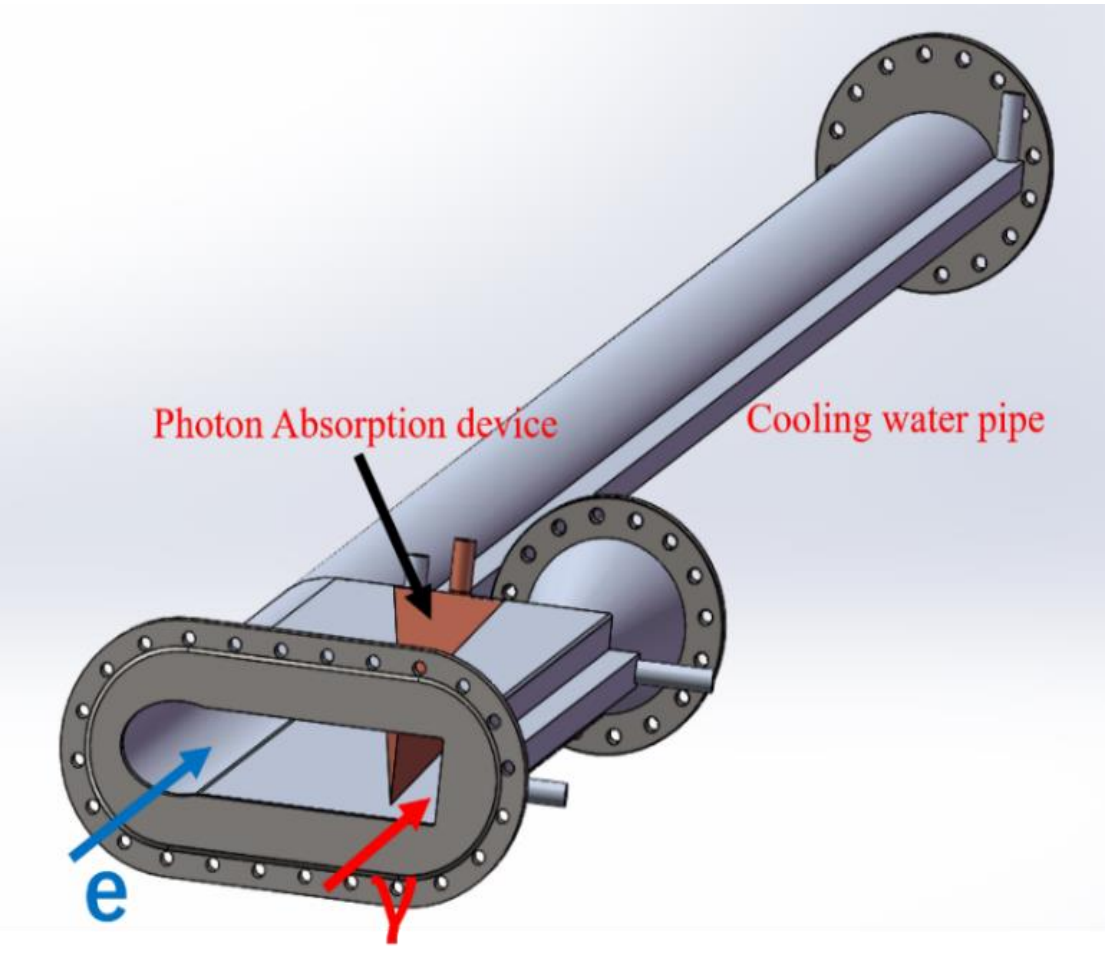

**Figure A5.1.9:** Schematic diagram of separation of the synchrotron light from the electron beam.

Figure A5.1.10 illustrates the structural configuration of the synchrotron radiation extraction pipe. The inability to accommodate a larger vacuum pipe stems from the proximity of the dipole iron's end to the quad magnet, necessitating a narrower section of the pipe to pass through the quad magnet's gap. The lead-in pipe is notably extended, and an ion pump has been incorporated to maintain the requisite vacuum level within the pipe.

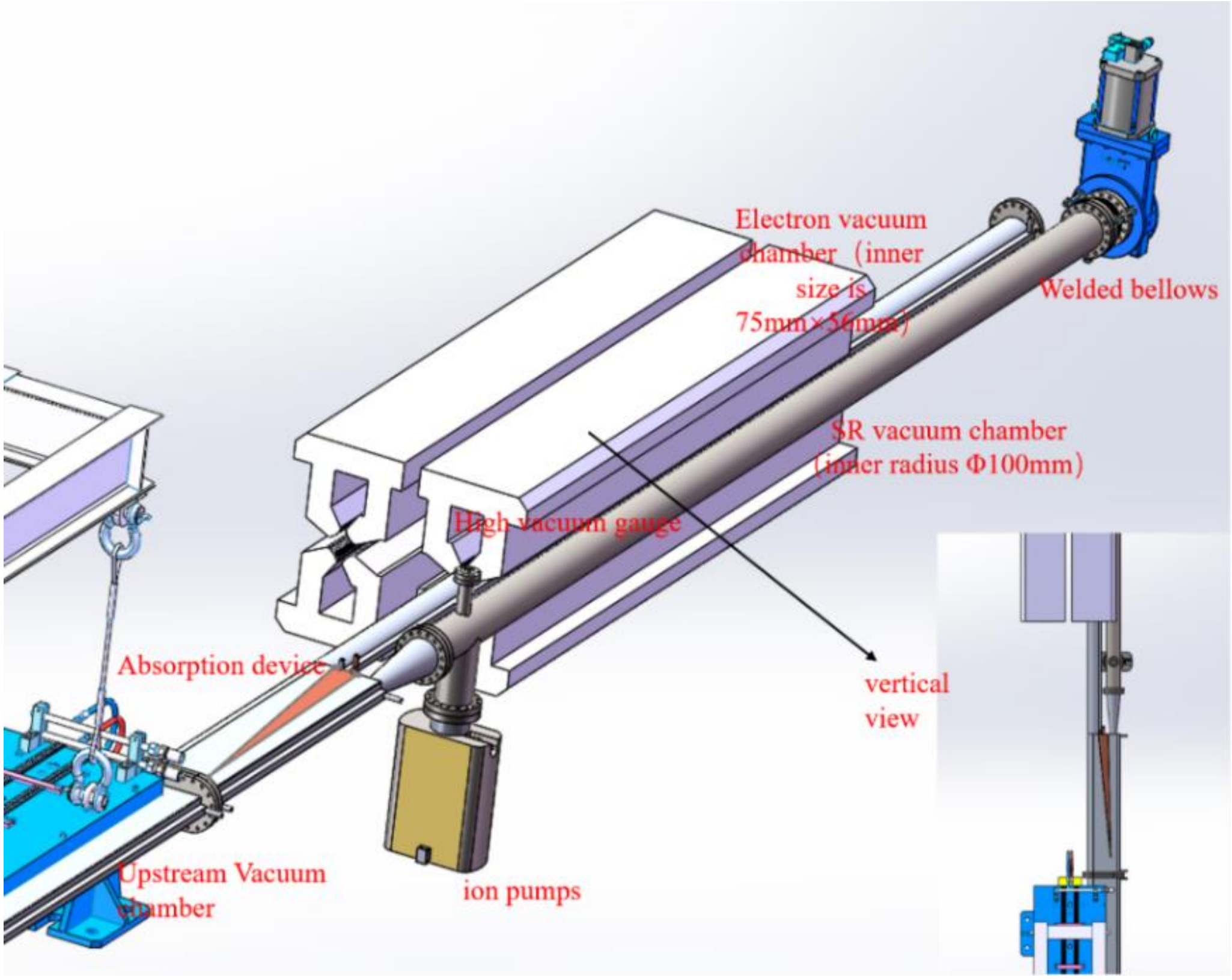

**Figure A5.1.10:** A structural configuration of the synchrotron radiation extraction pipe

The vacuum system design for the photon beamline necessitates attention to the following aspects:

i. Considering the substantial length of the beamline, which extends beyond 700 meters (comprising roughly 300 meters to traverse the tunnel and an additional 400 meters to access the SR experimental stations), the recommended vacuum system includes a combination of getter films (NEG films) and ion pumps.
ii. The beamline can be divided into eight sections, with each section separated by all-metal ultra-high vacuum valves (a total of 9 valves per section). Each section is equipped with distinct rough suction and gas filling ports, and 5-10 ion pumps are evenly distributed within each section. These ion pumps are employed to evacuate gases that are challenging to remove with getter films, such as $CH_4$. The inner walls of all vacuum boxes are coated with NEG films.
iii. The vacuum boxes can be constructed using 304 stainless steel, with an inner diameter of Φ100 mm and a wall thickness of 3 mm.
iv. Vacuum measurements in each section should involve two measurement points, and cold cathode vacuum gauges are employed for this purpose.
v. In the photon collimation area, it is imperative to position temperature measurement points within the adjacent vacuum boxes to monitor temperature fluctuations in the vacuum box's outer wall.

vi. Comprehensive lead block shielding is imperative to safeguard the entire beamline.
vii. The transportation of materials and personnel across this extensive beamline within the tunnel necessitates special consideration and well-organized arrangements.

### *A5.1.1.3: Focalization of γ-ray*

Gamma ray techniques offer the capability to perform non-destructive assessments of both natural and man-made materials. They supply valuable elemental, chemical, and structural insights into the internal composition of these materials. The primary challenge in utilizing ultra-high energy synchrotron radiation light source technology lies in producing a high-quality gamma beam. This necessitates efficient transmission and precise focusing of the synchrotron radiation beam. Consequently, there is a critical need for the development of a MeV ray focusing system [5].

The attainment of a nanometer-scale focused spot of hard X-ray is made possible through the use of Laue lenses. A Laue lens functions as a single reflection concentrator, relying on Bragg diffraction from numerous small crystal slabs that are carefully oriented to direct incoming radiation toward a central focal point [6]. The quality of this focusing system is contingent upon factors such as the crystal's size, the precision of its orientation, and its mosaicism [7].

Crystals form the essential foundation of the Laue lens, as this lens concentrates incident gamma rays by utilizing Bragg diffraction from the crystal structure.

The initial concept of a Laue lens for gamma-ray focusing relied entirely on a design centered around mosaic crystals [8]. Subsequently, building upon the groundwork involving mosaic crystals, innovations such as the introduction of curved crystals and the utilization of the Quasi-mosaicity effect have represented significant advancements. Notably, it has been demonstrated that Quasi-mosaicity crystals offer the potential to achieve both high diffraction efficiency and the successful focalization of a diffracted beam [9].

The selection of an appropriate crystal material is a crucial consideration for the Laue lens, as the choice directly impacts its performance. To maintain optimal diffraction intensity, it is preferred to use pure crystals, consisting of no more than two elements. The materials suitable for Laue lenses must fulfill several prerequisites. They should naturally exist in a crystalline state at standard room temperature and pressure, exhibit minimal reactivity in ambient air, and should not be radioactive or highly toxic. Additionally, they must efficiently diffract rays. Eighteen crystal types meet these criteria, including elements such as Al, Si, V, Cr, Ni, Cu, Ge, Mo, Rh, Pd, Ag, Ba, Ta, W, Ir, Pt, Au, and Pb [10].

Maximum peak reflectivity for a given mosaicity can be expressed as in this formula [11-12]:

$$R_{peak} = \frac{1}{2}[1 - \exp(-2W(0)QT_0)] \exp\left(-\frac{\mu T_0}{\cos\theta_B}\right) \qquad \text{(A5.1.1)}$$

where $W$ represents the Gaussian function, which accounts for the slight misalignment of crystallites concerning each other, as described by an angular distribution. $Q$ denotes the diffraction efficiency, $T_0$ is the thickness traversed by the beam, $\mu$ signifies the linear absorption coefficient of the crystal, and $\theta_B$ represents the Bragg angle.

Figure A5.1.11 displays the peak reflectivity of 18 crystal types at various incident energies. At an energy level of 0.1 MeV, Al, Si, V, Cr, and Ge exhibit high reflectance exceeding 36%, with Si and Ge being particularly favorable choices due to their abundant industrial production. For an energy of 0.5 MeV, Ni, Cu, Mo, Rh, Pd, Ag, Ta, W, Ir, Pt, Au, and Pb display enhanced reflectance exceeding 30%. Among these, Cu, Ni, Ag, Rh, and Pb are more cost-effective selections, as they possess a mosaic structure. At energies of 1 MeV and 1.5 MeV, Ta, W, Ir, Pt, Au, and Pb continue to demonstrate superior reflectance. However, Ta, W, and Ir are hindered by their extremely high melting points of 2996°C, 3410°C, and 2454°C, respectively, rendering them expensive and challenging to obtain in large quantities as pure crystals. Pt is prohibitively costly and not feasible for large-scale production. As depicted in Figure A5.1.13, at approximately 1 MeV, Au crystals exhibit the highest peak reflectivity, establishing them as the optimal choice.

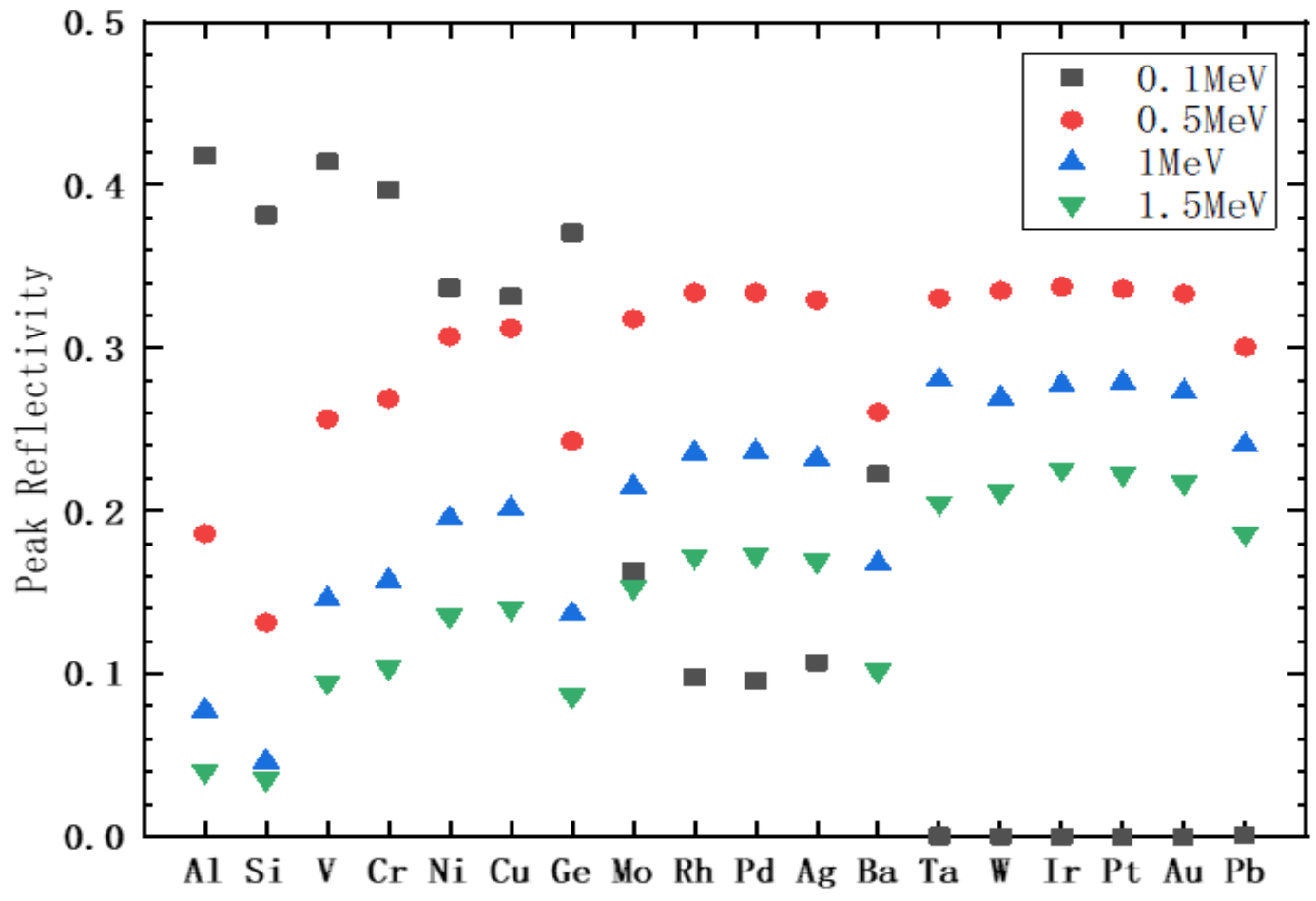

**Figure A5.1.11:** The peak reflectance of the crystal at four different energy levels: 0.1 MeV, 0.5 MeV, 1 MeV, and 1.5 MeV. These calculations assume the crystal to be a mosaic crystal with a mosaic degree of 30 arcseconds, and the crystal thickness is constrained within the range of 1 to 25 millimeters.

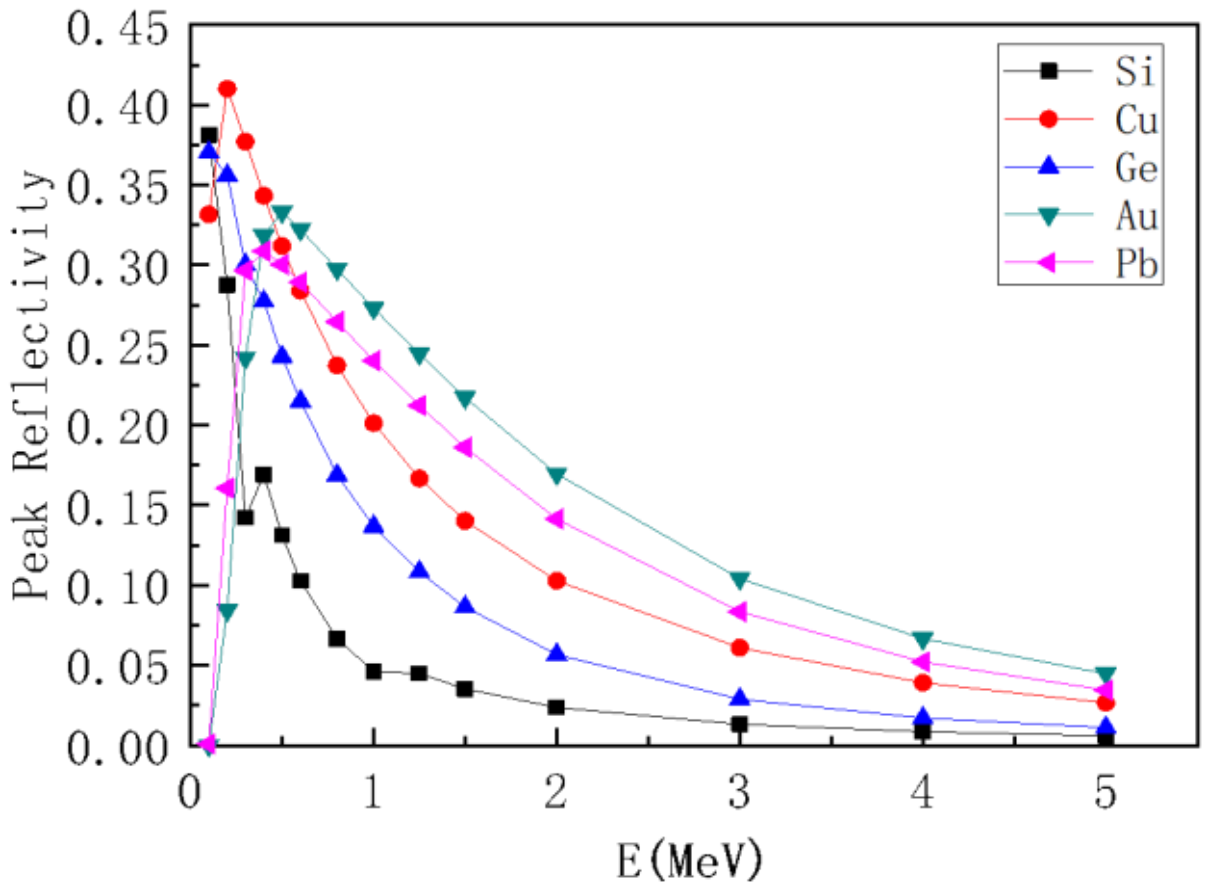

**Figure A5.1.12:** Variation of peak reflectance with energy.

The current initial design of the Laue lens, intended for focusing γ-rays within the energy range of 0.8-1.2 MeV, involves the selection of concentric rings containing Au crystals of varying radii. This choice leverages the fact that rings with different radii can effectively diffract distinct energy ranges, and the energy ranges diffracted by adjacent rings overlap. This design enables the Laue lens to cover a broader energy spectrum. Gold (Au) has been chosen as the crystal material for diffraction within the Laue lens.

The relationship between radius and energy can be obtained from Eq. (A5.1.2) [13]:

$$r = \frac{h_p c f}{d_{hkl} E} \tag{A5.1.2}$$

where $f$ is called the focal length. $d_{hkl}$ is the spacing of the lattice plane (hkl), $h_p$ is the Plank constant, $c$ is the speed of light in vacuum, and $h_p c = 1.24 \times 10^{-6}$ eV-m.

With a focal length of 20 meters, the radius of the Laue lens measures 131.66 mm when the energy is set at 0.8 MeV and 87.78 mm when the energy is at 1.2 MeV.

The Laue lens exhibits smoother energy diffraction when the crystal rings are closely arranged, and there is significant overlap in energy between adjacent rings. Consequently, the Laue lens is designed with five crystal rings, and each ring accommodates as many crystals as possible, maximizing the lens's effective area. The crystal arrangement on each ring is depicted in Figure A5.1.13.

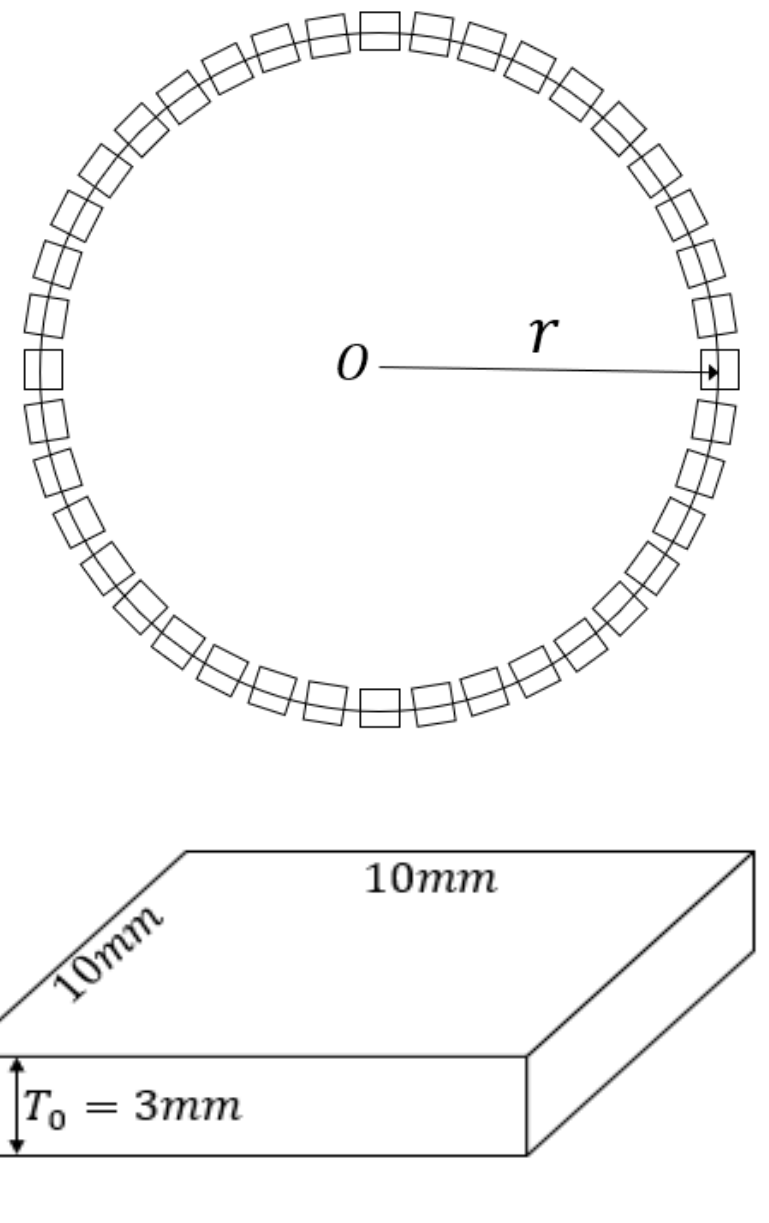

**Figure A5.1.13:** (a) Arrangement of Au crystals on each ring; (b) Au crystal size: 3×10×10 $mm^3$

To optimize diffraction efficiency, the crystal employs the (111) plane as the diffraction plane. All crystals are uniform in size, measuring 10×10 $mm^2$, and a smaller size is selected to better approximate a spherical profile. At a diffraction energy of 1 MeV, with a crystal thickness of 3 mm, the Laue lens can be approximated as spherical, and its radius measures 40 meters, which is double the focal length. The geometric shape of the Laue lens is depicted in Figure A5.1.14.

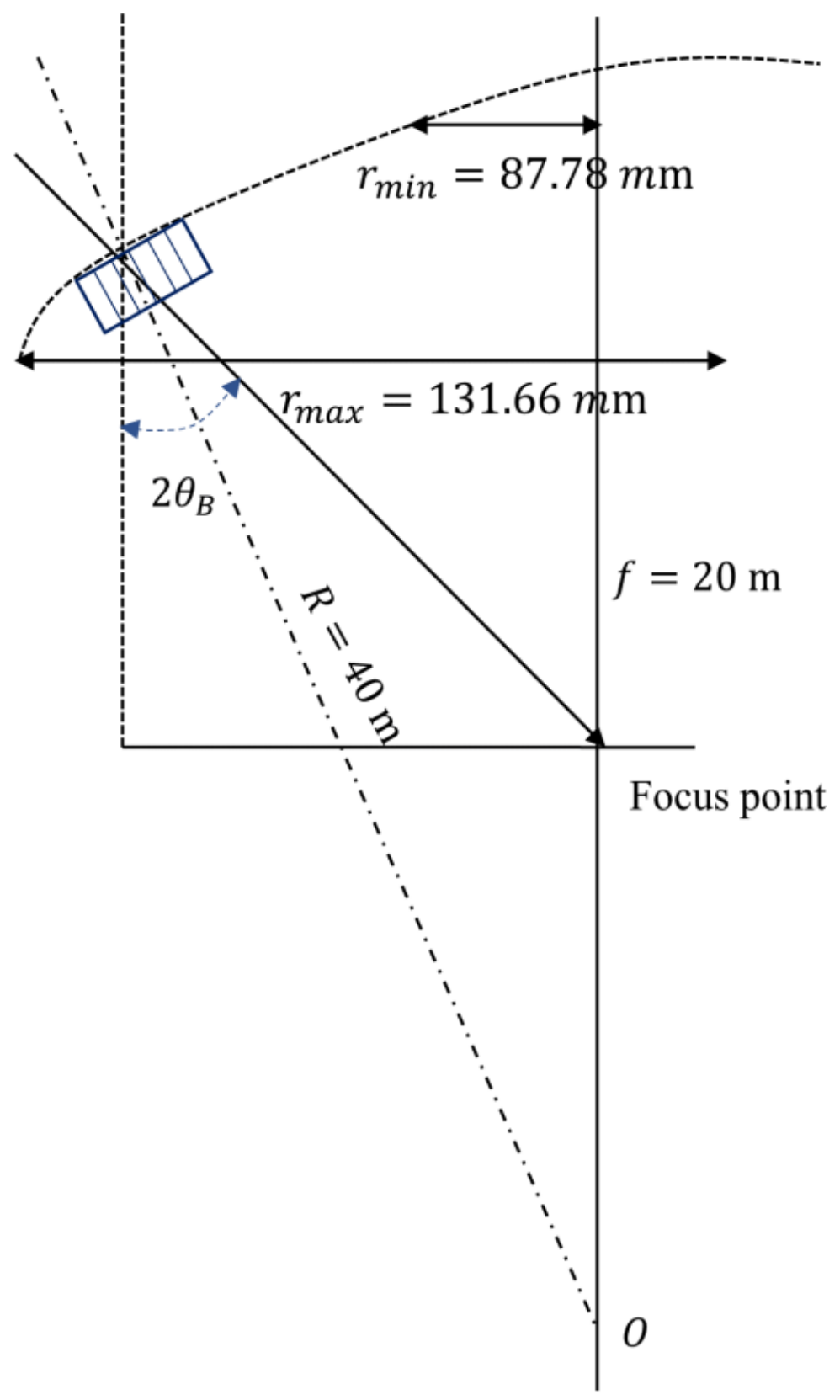

**Figure A5.1.14:** Geometry of 0.8-1.2MeV Laue lens

## A5.1.2: Applications of γ-ray Source

Powerful γ-rays, with energies surpassing the mega-electron volt (MeV) threshold, find versatile applications across a wide spectrum of fields. These applications span photon-nuclear physics, nuclear science and technology, nuclear astrophysics, quantum electron dynamics (QED), and even extend to diverse uses in the life sciences and the preservation of cultural heritage.

Photonuclear reactions are crucial for examining nuclear reaction mechanisms and nuclear structure, with practical applications including (i) radiation shielding and radiation transport analyses, (ii) calculating absorbed doses during radiotherapy, (iii) activation analyses, (iv) safeguards and inspection technologies, (v) nuclear waste transmutation, (vi) fission and fusion reactor technologies, and (vii) astrophysical nucleosynthesis [14].

### *A5.1.2.1: Giant Resonance*

Photonuclear action operates like a valve-like reaction, where the valve energy for (γ, n) and (γ, p) reactions corresponds to the binding energy of the last neutron and proton in the target nucleus. The last neutron's binding energy can be accurately determined by measuring the threshold of the (γ, n) reaction. Nuclear giant resonances, such as those depicted in Fig. A5.1.15, represent collective motion within atomic nuclei and are a focal point of research. By measuring the excitation function of photonuclear reaction giant resonances, valuable insights into nuclear deformation can be gained [15]. This excitation function is directly linked to the extent of nuclear deformation [16]. The availability of high brightness and high energy γ-ray beams facilitates experimental measurements of the photonuclear reaction's excitation function [17].

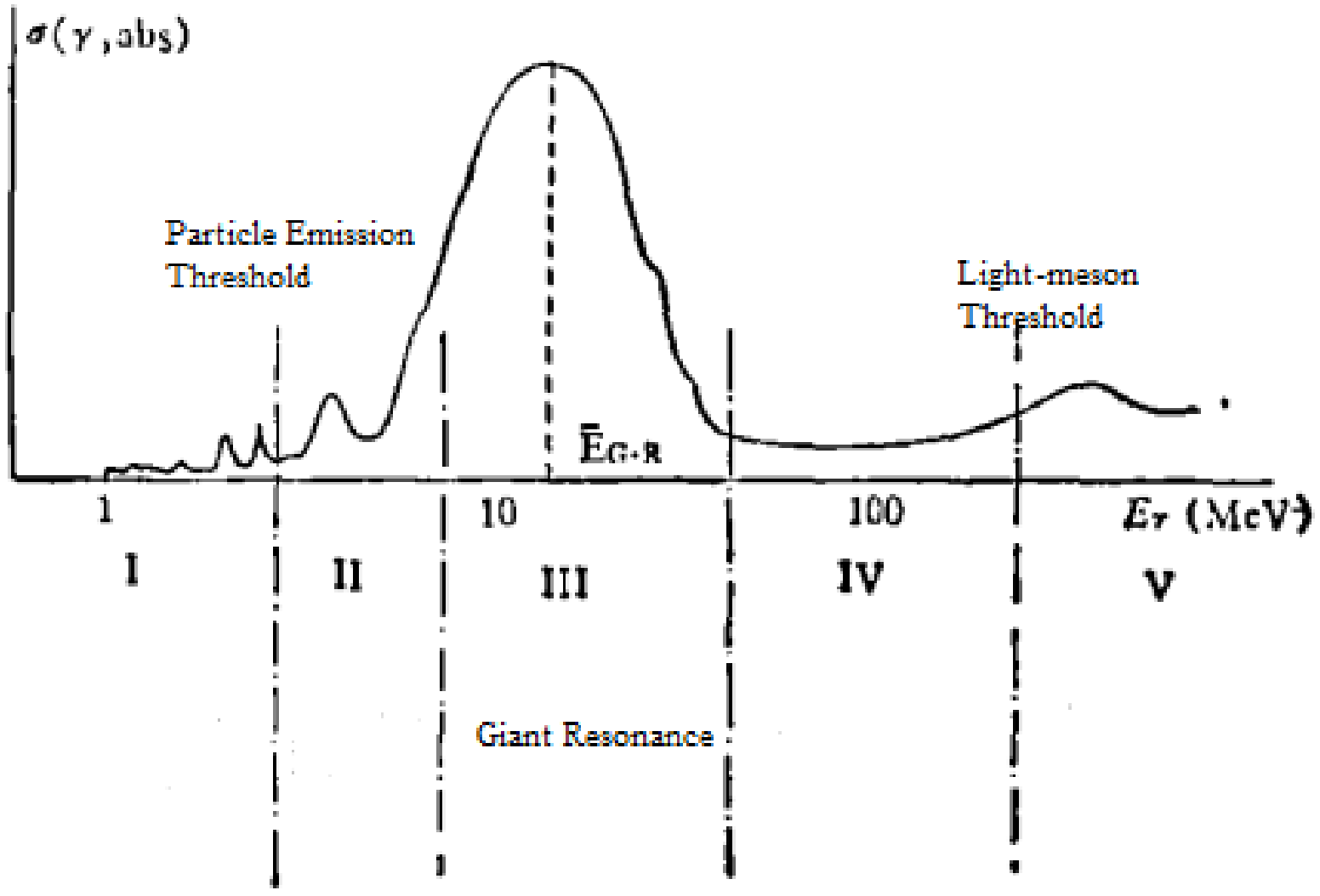

**Figure A5.1.15:** Generalization of the behavior of nucleus after absorption of photons [18].

Giant resonance absorption is not the sole mechanism of photonuclear reactions; the Y(γ, p) / Y(γ, n) reaction yield ratio exceeds predictions based on the resonance absorption mechanism. This suggests the presence of a direct interaction process within photonuclear reactions, indicating that giant resonance absorption is not the exclusive process [19].

### *A5.1.2.2: Medical Isotope Molybdenum-99*

99mTc, as the decay daughter of 99Mo with a half-life of 66.02 hours, stands as the most widely utilized radioisotope in the realm of nuclear medicine. The growing demand for 99Mo/99mTc has presented challenges for nuclear medicine's advancement. The production of 99Mo via the photonuclear reaction, 100Mo(γ, n)99Mo, emerges as a potential solution [20]. This reaction exhibits a threshold at 9 MeV and a peak cross-section of 150 mb at 14.5 MeV, as illustrated in Figure A5.1.16. The wiggler γ-ray energy range of CEPC aligns well with the requirements for this process.

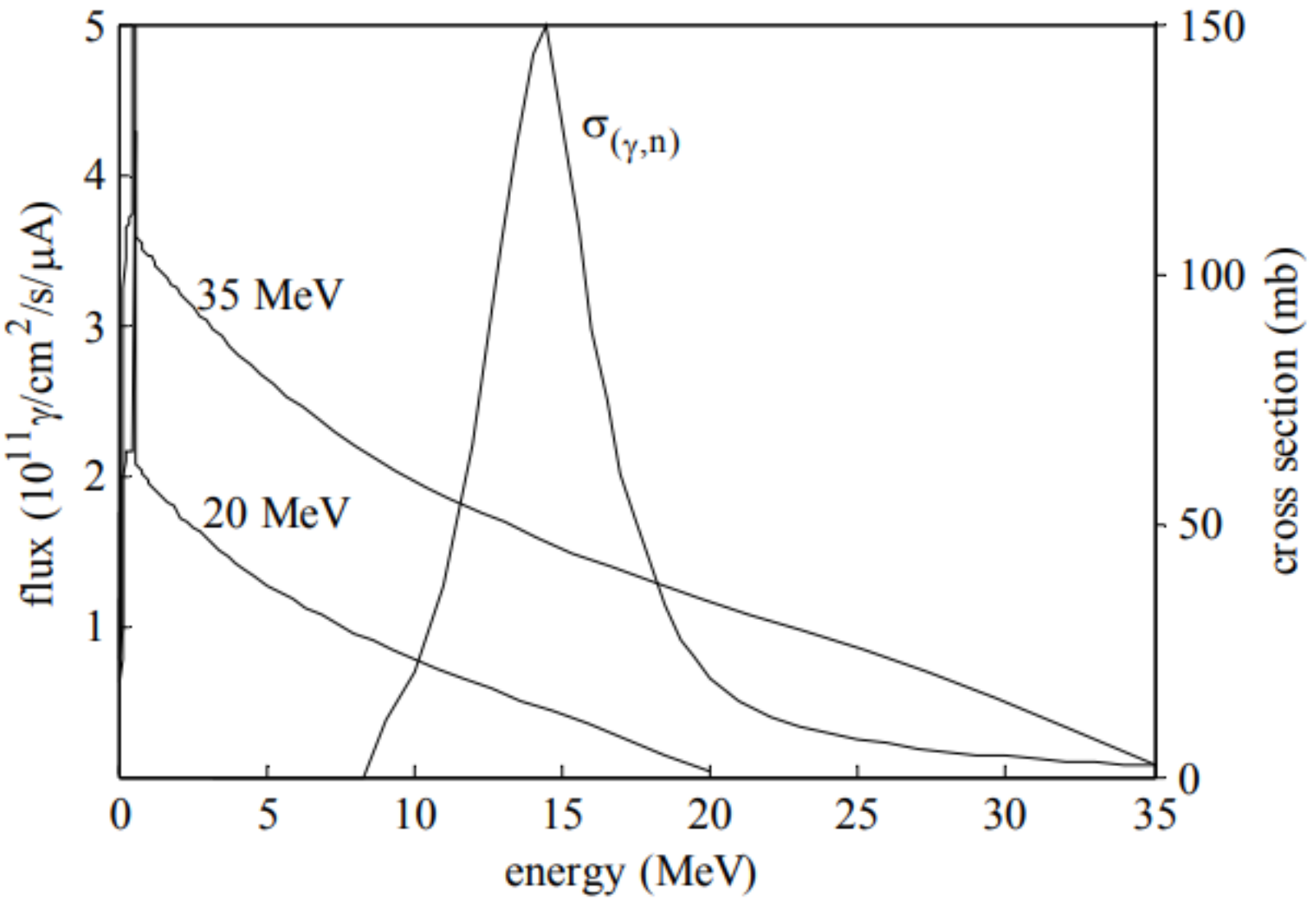

**Figure A5.1.16:** Photonuclear cross section of 100Mo and average bremsstrahlung photon spectra produced with a 20- and 35-MeV electron beam in a Mo target [21].

### *A5.1.2.3: Photonuclear Reaction Evaluation Data*

The anticipated development of advanced gamma-beam facilities is poised to address persistent discrepancies observed in data obtained from quasi-monoenergetic beam measurements.

### *A5.1.2.4: Nuclear Astrophysics*

#### *A5.1.2.4.1: Input Information for Nuclear Physics Study*

Nuclear astrophysics seeks to comprehend the origin and evolution of cosmic elements, the timeframes governing stellar development, the conditions within stars, energy generation from thermonuclear processes, and their impact on stellar evolution, as well as the mechanisms underlying astrophysical phenomena and the attributes of compact stars [22]. These inquiries heavily rely on nuclear physics data, including reaction cross-sections, nuclear masses, and other nuclear structure properties [23], which serve as essential empirical foundations for astrophysical models [24]. Enhanced by the advent of the CEPC synchrotron radiation source, the availability of high-energy gamma beams, approximately 5 to 6 orders of magnitude more powerful than existing sources, promises a substantial expansion of the experimental platform in photonuclear physics.

#### *A5.1.2.4.2: The 12C(α, γ) 16O Reaction*

The 12C(α, γ)16O reaction, in conjunction with the triple-α process, plays a pivotal role in determining the absolute abundance of carbon and oxygen in the universe [25]. The ratio of 12C to 16O abundance sets the initial conditions for the subsequent evolution of massive stars and significantly impacts both medium-mass and heavy-mass nuclides [26]. However, the "Holy-grail" reaction, 12C(α, γ)16O, occurs in stars at $E_{c.m.} = 0.3$ MeV, as predicted by Gamow theory, resulting in a cross-section of about $10^{-17}$ barn, which is too minuscule to be accurately measured. Most estimations fall far short of achieving the sub-10% uncertainty required by stellar models [27].

Employing the inverse reaction 16O(γ, α)12C proves to be an efficient research approach, given that the cross-section of 16O(γ, α)12C is approximately 100 times greater than that of 12C(α, γ)16O at the center of mass energy, as demonstrated in the study [28].

The gamma beam under development at the Laser-Compton Scattering gamma source at IHEP (LCS/IHEP) facilitates the study of this critical reaction using the time-reversal reaction 16O(γ, α)12C, guided by the well-established detailed-balance principle. In this type of photodissociation experiment, the 13C(α, n)16O background is not anticipated. Furthermore, the ability to measure detailed angular distributions allows for precise determination of the ratio, a crucial factor for accurate extrapolation to stellar energies.

The "Holy-grail" reaction, 12C(α, γ)16O at the Gamow peak, stands as the most significant application for a CEPC synchrotron radiation beam generated by a wiggler magnetic field. The designed central energy of 7.16±0.3 MeV specifically caters to this reaction.

*A5.1.2.4.3: p-Process*

Another valuable application in nuclear astrophysics involves measuring photo-disintegration reactions, which serve as a probe for the p-process. The p-process has been postulated to account for the creation of stable, neutron-deficient nuclides heavier than iron found in the solar system. These nuclides cannot be generated through slow or rapid neutron captures (s-process or r-process) but can arise from seed nuclides created in the s-process or r-process through photodisintegration reactions or capture reactions. The synthesis of p-nuclei in stars, known as the p-process, is thought to be primarily driven by radiative proton captures and (γ, n) photon-disintegrations on existing, more neutron-rich species [29-30].

The photoinduced transformation reaction associated with the p-process necessitates the integration of nuclear physics inputs (including reaction cross sections, gamma intensity functions, nuclear mass, and nuclear energy level density) derived from theoretical calculations and experiments to ensure precise results [31]. These nuclear physics parameters must be determined through experimental means. The gamma source system, serving as a tool for investigating photo-induced decay reactions, is expected to furnish updated nuclear physics data essential for grid computing related to the p-process. A CEPC-SR gamma source is capable of delivering high-intensity γ-rays spanning the 1 MeV to 20 MeV range, thus enabling the generation of requisite photodisintegration reactions.

### *A5.1.2.5: Industrial and Material Science Applications: Non-destructive Testing*

Angular distribution measurements in photoemission have proven to be a valuable tool for determining the electronic structure of both bulk solids and surfaces, as well as for research on thin epitaxial layer systems. These thin layer systems hold significance in comprehending semiconductor heterostructures and thin magnetic films. For instance, the application of photoelectron diffraction has yielded successful insights into the epitaxial growth of metal and semiconductor layers, significantly advancing our understanding of the growth mechanisms in these systems. Moreover, these measurements find utility in various applications, including internal structure assessment of precision workpieces in aerospace and other industries, ammunition packing density inspection, quality assessment of critical weapon components in the defense sector, non-destructive testing of vital automotive components, online monitoring and quality control in the steel industry (e.g., engine blades), as well as sample evaluation in geology and archaeology [32-33].

### A5.1.3: Detection Methods of High Intensity γ-ray

In $\gamma$-ray spectroscopic imaging systems, achieving high energy resolution is critical for a variety of industrial, medical, and scientific applications. As a result, a meticulous design of both detectors and readout electronics is imperative. Numerous semiconductor materials have been explored for the development of photon-counting $\gamma$-ray spectrometers.

Si-PIN radiation detectors have been a subject of study for several decades among researchers across a spectrum of fields, including medical, industrial, space, environmental, security, high-energy physics, and γ-ray spectroscopy. Nevertheless, silicon's low atomic number and elevated leakage currents constrain its utility primarily to low-energy γ-ray and charged-particle detection [34]. In the case of Si-PIN detectors, the detection energy range spans from 5 to 30 keV, with the potential to achieve an energy resolution as fine as 3 keV at 59.5 keV (at –5°C).

In applications necessitating $\gamma$-rays with energies exceeding tens of keV, CdTe and CdZnTe emerge as top-tier material choices. This preference stems from their high absorption efficiency resulting from their atomic numbers and their wide bandgap (approximately 1.5 eV), which facilitates low dark current [35-37]. While CdZnTe detectors are considered room temperature devices, cooling to temperatures ranging from –5°C to –40°C is typically employed when striving for high resolution at photon energies below 100 keV [38].

HPGe remains the optimal choice for low-energy discrete γ-ray spectroscopy (with photon energies ranging from 0.1 MeV to 5 MeV). In contrast, the LaBr3:Ce crystal exhibits superior peak efficiency for high-energy γ-ray detection [39], accompanied by significantly improved time resolution when compared to HPGe detectors. The recent advancement of LaBr3:Ce as a scintillator material represents a noteworthy enhancement in the realm of scintillation detectors designed for high-energy γ-ray measurements [40]. LaBr3:Ce demonstrates commendable energy resolution (approximately 3% at 662 keV) and outstanding time resolution (less than 1 ns). Moreover, these scintillators exhibit minimal sensitivity to thermal effects, maintaining consistent light output even during temperature fluctuations [41-42].

4H-SiC stands out as one of the most established wide bandgap semiconductor materials, offering distinct advantages over silicon (Si). These advantages have paved the way for the development of highly efficient 4H-SiC detectors for γ-ray spectroscopy, particularly in demanding environments where instruments need to function at elevated temperatures or in the presence of intense radiation. Notably, 4H-SiC-based radiation detectors have been reported with remarkably low leakage current densities, reaching as low as 0.1 pA $cm^{-2}$ at 25 °C, in stark contrast to typical Si radiation detectors with leakage currents that are four orders of magnitude higher, approximately 1 nA $cm^{-2}$ [43].

Another merit of 4H-SiC is its relatively low electron affinity, measuring at 3.17 eV [44]. This property aids in reducing the thermionic emission current component, thus minimizing the parallel white noise emanating from the 4H-SiC radiation detector. This, in turn, contributes to the overall noise in the photon counting spectroscopic system. Additionally, 4H-SiC exhibits a high resistance to radiation-induced damage [45]. In terms of energy resolution, a spectrometer built using Commercial-off-the-shelf 4H-SiC

UV p–n photodiodes and operating at 20 °C has demonstrated impressive results, ranging from 1.66 keV ± 0.15 keV at 5.9 keV to 1.83 keV ± 0.15 keV at 59.5 keV [46].

For higher-energy γ-rays, the selection is a CsI detector, akin to the one employed in the BES electromagnetic calorimeter (BESEMC), with calibration spanning from 300 MeV to 2 GeV. The crucial range of interest lies below 500 MeV, with an energy resolution of approximately 2.5%/√(GeV). An alternative option involves Cerenkov detection.

### A5.1.4: References

# A5.2: Operation as a High Energy Free Electron Laser (FEL)

## A5.2.1: Introduction

Differing from conventional lasers, the Free-Electron-Laser (FEL) generates light through the interaction of high-energy electron beams traveling within a vacuum through an undulator [1]. Here, the electron beam engages with the magnetic field of the undulator via the Lorenz force, inducing a back-and-forth motion along its propagation path. The transverse acceleration of electrons, driven by the undulator's influence, initiates spontaneous emission. With sufficient undulator length, electrons substantially interact with the spontaneously generated radiation field, experiencing energy gains or losses in response to the phase. This correlated energy manipulation naturally results in density modulation on the wavelength scale of the radiation, facilitated by the $R_{56}$ factor in the undulator transfer matrix. The outcome is the creation of coherent radiation, characterized by significantly narrower bandwidth and higher intensity compared to spontaneous radiation. This FEL mechanism serves to amplify spontaneous emission, and following the terminology of quantum lasers (amplified spontaneous emission, ASE), it is known as Self-Amplified Spontaneous Emission (SASE) FEL [2,3].

The undulators, where FELs originate, are composed of a repetitive magnet structure. The magnetic field within the planar undulator follows a sinusoidal pattern with odd harmonics. Undulator radiation arises from "free" electrons moving within a vacuum. As there are no quantized electron energy states, there exist no inherent restrictions on the photon energy of the emitted radiation. Consequently, there is no intrinsic limitation to photon energy, enabling the generation of coherent hard x-ray radiation. The on-axis radiation wavelength in the undulator is determined by factors such as the undulator period, field strength, and the beam energy, as governed by the resonance condition:

$$\lambda_r = \frac{\lambda_u}{2\gamma^2}\left(1 + \frac{K^2}{2}\right) \tag{A5.2.1}$$

where $\lambda_r$ is the radiation wavelength, $\lambda_u$ the undulator period, $\gamma$ indicates the beam energy and $K$ is so called undulator parameters: $K = \frac{e}{2\pi mc^2}\lambda_u B_0$, $B_0$ is the magnetic peak field.

Advanced undulator technology can furnish a minimum undulator period, typically around 10 mm or slightly longer. As indicated by Eq. (A5.2.1), the radiation wavelength (i.e., photon energy) is predominantly reliant on the beam energy. In the case of a hard x-ray FEL aiming for a photon energy of approximately 10 keV, the typical beam energy would be in the vicinity of 10 GeV. Should a higher photon energy be necessary, a commensurate increase in the beam energy is warranted.

The inaugural hard x-ray FEL was realized at the Linear Coherent Light Source (LCLS) at the SLAC National Accelerator Laboratory in 2009 [4]. This achievement marked a significant milestone for x-ray user facilities, ushering in a transformation from storage-ring-based synchrotron light sources to linac-based light sources. X-ray FELs offer laser-like radiation in the x-ray spectrum, boasting peak power and brightness levels that are 10 billion times greater than those of existing coherent x-ray sources. Moreover, these FELs can compress the beam pulse into femtosecond time scales. The ultra-bright, short x-ray pulses empower scientists to capture images of atoms and molecules with femtosecond time resolution, making them a unique tool for multidisciplinary research.

This tool is instrumental in the study of atomic and electron dynamics within various materials, such as metals, semiconductors, ceramics, polymers, catalysts, plastics, and biological molecules, among others [5]. Following the debut of the first XFEL at LCLS, several more facilities have been established, achieving further advances in coherent x-ray technology. Notable examples include the SACLA-xFEL in Japan [6], PAL-xFEL in Korea [7], the European XFEL in Germany [8], and the Swiss-XFEL. Furthermore, high repetition rate megahertz XFELs powered by Continuous Wave (CW) superconducting linacs are in development, with projects like LCLS II in Stanford and the SHINE project in Shanghai leading the way.

Existing x-ray FELs have exhibited limitations, particularly in achieving a maximum photon energy of approximately 20-25 keV. To broaden the scope of scientific applications for FELs, which include the exploration of matter under extreme conditions, advancements in industry and society, and the facilitation of physics research in x-ray photonics, it is imperative to consider higher electron beam energies. As an example, the CEPC linear accelerator, as described in Chapter 6 of this report, will be capable of producing beam energies reaching up to 30 GeV. This makes it a highly promising facility for the generation of high-energy FELs in the range of 50 to 100 keV.

### A5.2.2: Schematic Plan of an Ultra-high Energy X-ray FEL

The design and construction of x-ray FELs have witnessed rapid progress over the past few decades, resulting in a well-established understanding of how to build an FEL facility. A typical x-ray FEL accelerator complex comprises essential components such as an electron gun, bunch compression system, main linac, bunch distribution and collimation system, and the undulator system. Thanks to extensive international collaboration, the key parameters for each system, such as emittance, peak current, and energy spread, are well-defined, and the related critical technologies have been significantly advanced.

Nevertheless, the existing facilities exhibit limitations in achieving a maximum photon energy, typically capped at around 20-25 keV. To fully unlock the potential offered by 30 GeV beams, it is recommended to embark on the development of an ultra-high x-ray FEL at the CEPC, reaching an energy level of 100 keV. This ambitious endeavor aims to explore new frontiers in scientific research enabled by such cutting-edge capabilities.

In accordance with the resonance condition of the undulator radiation formula, as outlined in Eq. (A5.2.1), achieving a photon energy of 100 keV necessitates a beam energy of approximately 20 GeV. However, it's essential to strike a balance as excessively high energy levels can lead to undesirable energy spread growth due to spontaneous radiation. Therefore, a beam energy of 25 GeV is selected as a prudent compromise.

Figure A5.2.1 illustrates the conceptual layout for harnessing the primary section of the linac complex at CEPC to produce ultra-high energy x-ray FELs. In the baseline design, the current electron source employs a thermal cathode gun, characterized by an emittance exceeding the typical requirements of an x-ray FEL. To address this, an independent electron gun using a photon cathode is introduced. The photon cathode gun can generate a bunch charge of 0.4 nC with an emittance of approximately 0.4 mm-mrad. The beam's energy at the conclusion of the gun and subsequent pre-accelerator unit reaches 5 MeV. Subsequently, the beam transits through a 3rd harmonic cavity, enabling quasi-linear chirping of the energy.

The first bunch compression chicane is positioned at a beam energy of 137 MeV, resulting in a substantial boost in beam current, multiplying it by approximately threefold, from 40 A to 120 A. The subsequent bunch compression chicane is situated at a beam energy of 550 MeV, where the compression factor reaches about 4, leading to a remarkable increase in the bunch current to 500 A. The third and final bunch compression occurs at an energy level of 5 GeV, marked by a compression factor of 10, enabling the beam current to reach its maximum capacity of 5000 A.

The 25 GeV beam is subsequently directed from the main linac into the undulator system through a dog-leg transport line. The undulator field can be fine-tuned by adjusting its gap, facilitating the generation of variable wavelengths. The total length of the undulator exceeds 200 meters, a critical factor in allowing the SASE FELs to attain saturation, particularly at the high photon energy of around 100 keV.

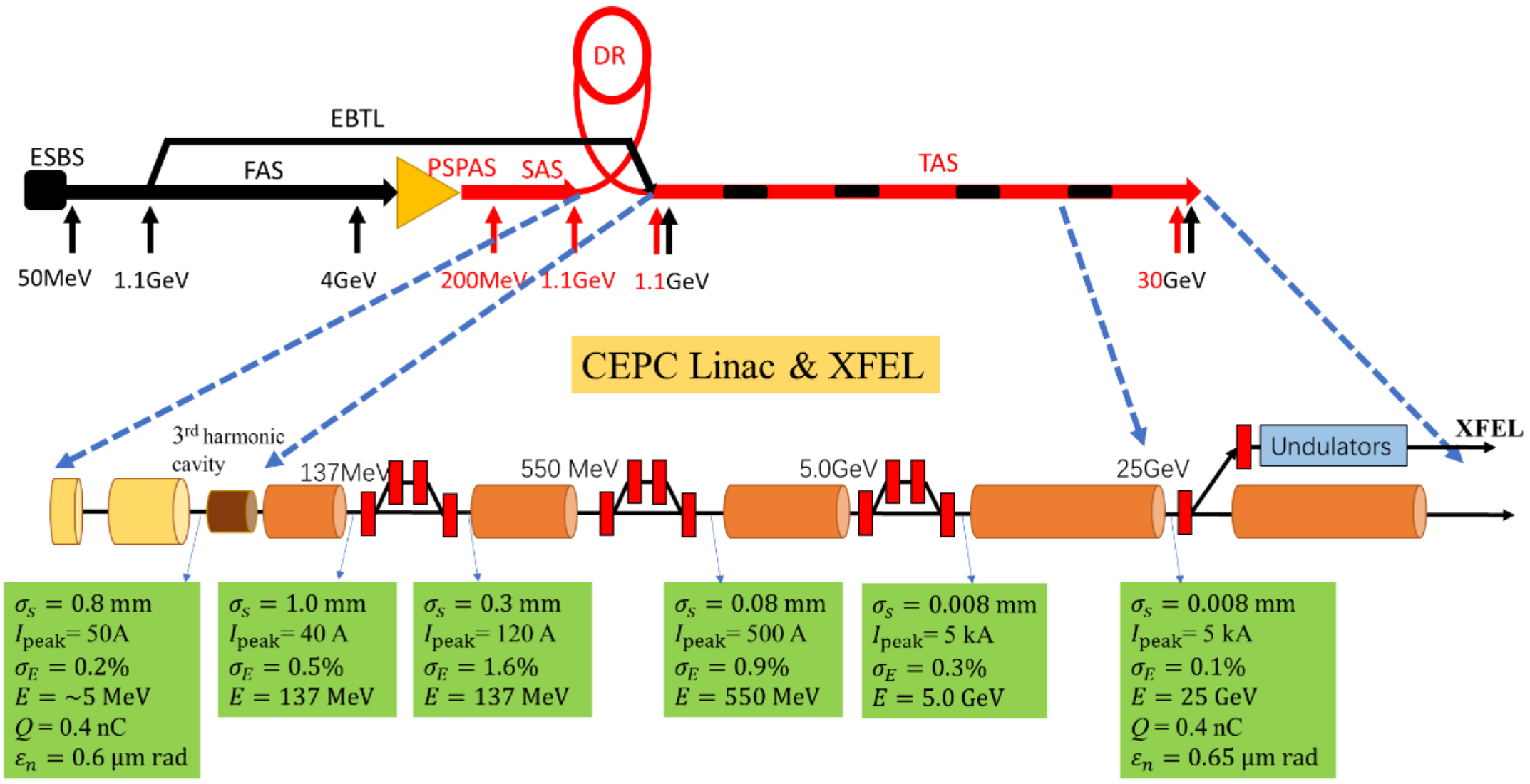

**Figure A5.2.1:** Schematic plan of x-ray FEL at the CEPC

### A5.2.3: Electron Parameters at the End of Linac

A comprehensive start-to-end simulation was conducted to assess the quality of the extracted beam. As depicted in Figure A5.2.2, the distributions of beam energy and current are showcased. Notably, the energy exhibits a spread within the range of 25.265 GeV to 25.290 GeV, with a width of 25 MeV. The slice energy spread, which holds particular significance in the context of FEL generation, is approximately 2-5 MeV. The beam current registers at a level of 5 kA, with a bunch length of around 30 μm, equivalent to 100 fs.

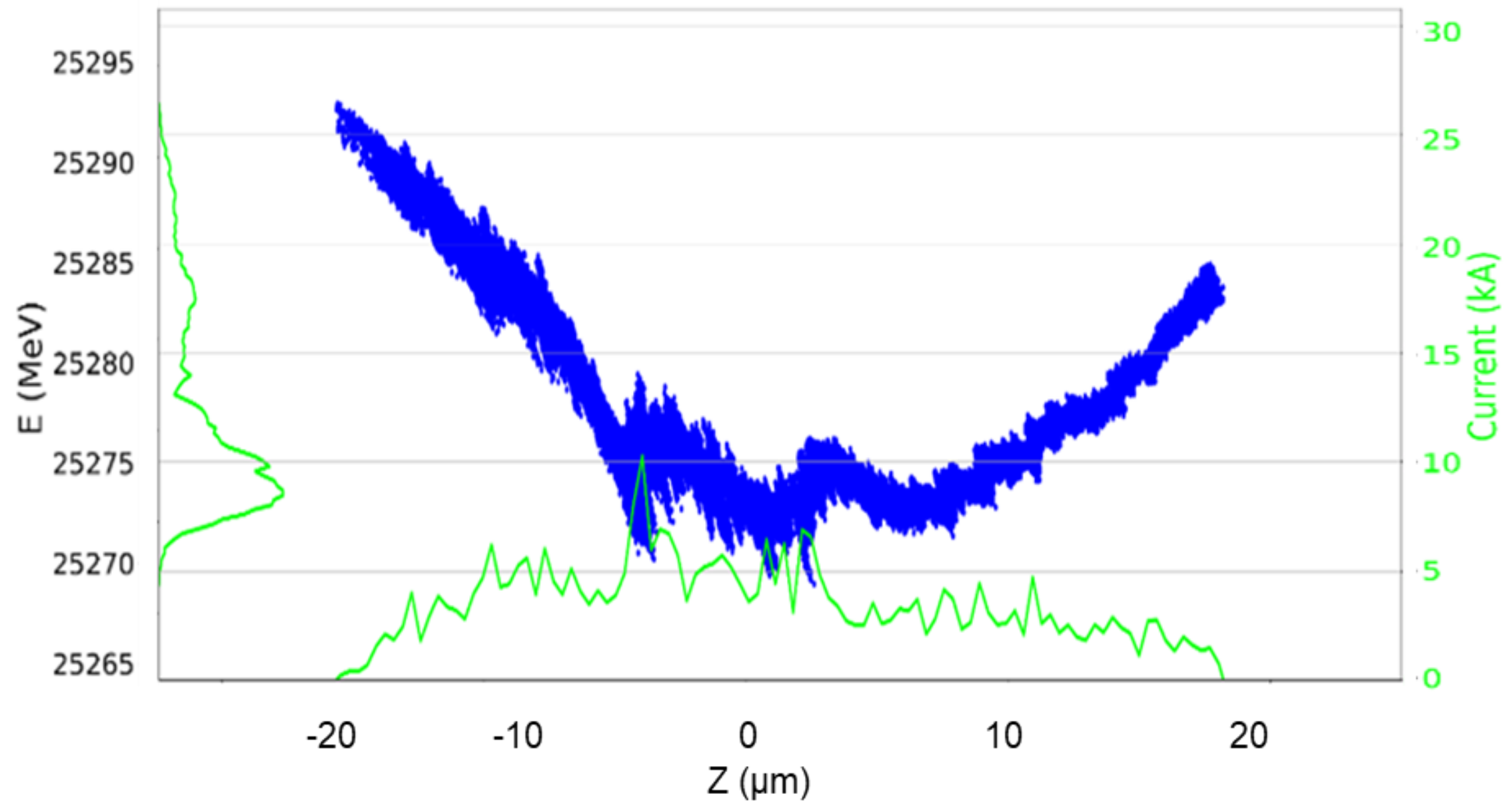

**Figure A5.2.2:** Simulated beam energy and current distribution at the end of the linac. Blue dots represent the phase space distribution of the simulated macro-particle, while the left and bottom curves illustrate the energy and current distributions, respectively.

The designated design parameters for the electron beam are summarized in Table A5.2.1. Both the normalized emittance and energy spread consider sliced values within the collaborative range involving the electron and optical fields.

**Table A5.2.1:** Electron parameters used for the x-ray FEL at the CEPC

| Parameter | Value | Unit |
|---|---|---|
| Electron energy | 25 | GeV |
| Bunch charge | 0.4 | nC |
| Peak current | 5000 | A |
| Bunch length | 8 | μm |
| Normalized emittance(slice) | 0.65 | mm mrad |
| Energy spread (slice) | 2.0 | MeV |
| Max. Repetition rate | 100 | Hz |

### A5.2.4: Design Criteria of the Undulator System

The selection of undulator parameters is based on the global optimization of the parameter space, as defined by the resonance condition. Meticulous optimization is imperative to achieve a reasonable saturation length and acceptable wavelength tunability. Figure A5.2.3 illustrates the magnetic field characteristics of the undulator concerning the undulator period, showcasing both superconducting and permanent magnet undulators. The $Nb_3Sn$ superconducting undulator produces the highest field, exceeding that of the NbTi superconducting undulator by approximately 30%. The permanent magnet in-vacuum undulator generates a relatively lower field, even with a narrow gap of 6 mm, which is 2 mm smaller than that of the SCUs. The NdFeB undulator can yield approximately 15% more field than the SmCo undulator, although the latter exhibits superior radiation resistance. The dashed line denotes the requisite undulator field across various periods to generate 100 keV FELs utilizing 25 GeV beams.

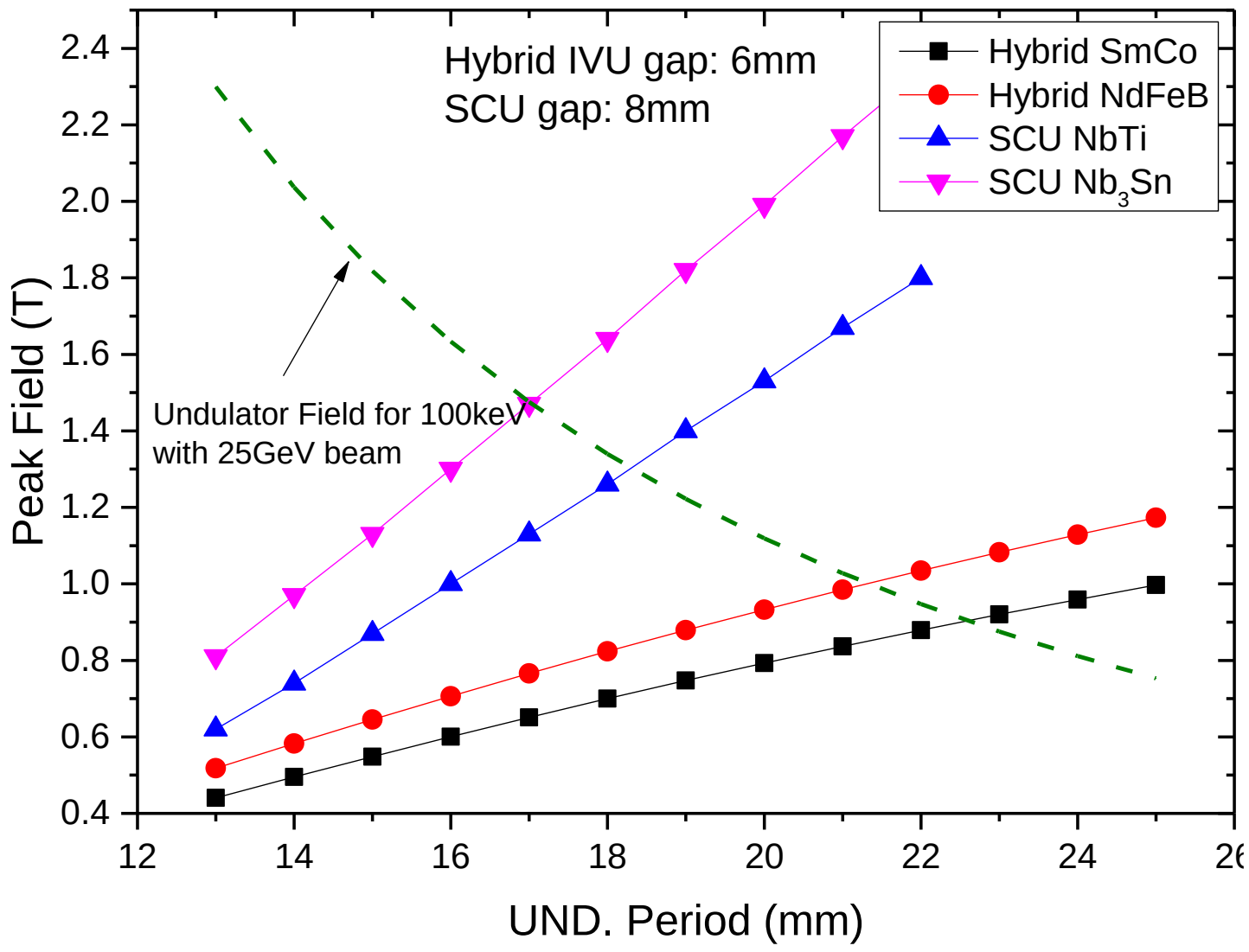

**Figure A5.2.3:** Undulator field as a function of period. The dashed line shows the required field to generate 100 keV FELs by 25 GeV beam energy.

While superconducting undulator technology is advancing rapidly, it's worth noting that the permanent magnet undulator is a more mature and widely adopted option. Consequently, a favorable choice is to opt for the NdFeB undulator. As depicted in Figure A5.2.3, the undulator period should surpass 21 mm, as any shorter period would not permit the generation of lower photon energy FELs. A period of 25 mm is selected to deliver a magnetic field of approximately 1.2 T, surpassing the minimum requirement by 0.5 T. This range offers a degree of photon energy tunability.

In practical applications, the undulator field can be estimated by fitting parameters in accordance with the scaling law. The field of the permanent magnet NdFeB in-vacuum undulator, as illustrated in the plot above, is estimated using the following formula: [16]

$$B = 3.10487\exp[-4.24914\left(\frac{g}{\lambda_u}\right) + 0.80266\left(\frac{g}{\lambda_u}\right)^2] \qquad \text{(A5.2.2)}$$

The NbTi superconducting udnulator field can be estimated by for following equation:

$$B = (0.28052 + 0.05789\lambda_u - 0.0009\lambda_u{}^2 + 5\times10^{-6}\lambda_u{}^3)\exp[-\pi(\frac{g}{\lambda_u} - 0.5)] \qquad \text{(A5.2.3)}$$

The selected undulator parameters are detailed in Table A5.2.2.

**Table A5.2.2:** Undulator parameters for 100 keV FELs

| Parameter | Value | Unit |
|---|---|---|
| Undulator Period | 25 | mm |
| Minimum Gap | 6 | mm |
| Maximum Field | 1.2 | T |
| Undulator parameter, K | 2.8 | |
| Undulator type | IVU, hybrid | - |
| Magnet material | NbFeB | - |
| Polarization | Linear | |

### A5.2.5: FEL Radiation Parameters and SASE FEL Simulation

A zero$^{th}$-order estimation of FEL characteristics is typically conducted using a one-dimensional model of the Pierce parameter, $\rho_{1D}$:

$$(2\rho_{1D})^3 = \frac{\mu_0 e^2 n_0 K^2 [JJ]^2 {\lambda_u}^2}{16\pi^2 m\gamma^3} \quad \text{(A5.2.4)}$$

In which $n_0$ is the beam density, $K = 0.934\lambda_u[cm]B[T]$ is the non-dimensional undulator parameter. The coupling factor $[JJ] = J_0(\frac{K^2}{2(2+K^2)}) - J_1(\frac{K^2}{2(2+K^2)})$, where $J_n$ is the Bessel function of the first kind. $[JJ]$ equals to 1 for a helical undulator. The basic SASE FEL parameters can be estimated by $\rho_{1D}$:

- One dimensional power gain length: $L_{g,1D} = \frac{\lambda_u}{4\sqrt{3}\pi\rho_{1D}}$
- Saturation length: $L_{sat} \approx 20L_g$
- Saturation power: $P_{FEL} \approx \rho_{1D}P_e$
- Spectrum bandwidth: $2\rho_{1D}$
- Co-operating time: $t = \frac{1}{\omega\rho_{1D}}$

These are one-dimensional estimations. However, when considering factors such as finite beam emittance, energy spread, and radiation diffraction, it becomes necessary to account for three-dimensional effects [9-11]. This leads to the calculation of the 3D gain length, as defined by the following equation: [12]

$$L_{g,3D} = L_{g,1D}(1 + \delta) \quad \text{(A5.2.5)}$$

where $\delta$ can be calculated by the Ming Xie's formula:

$$\delta = a_1{\eta_d}^{a_2} + a_3{\eta_\varepsilon}^{a_4} + a_5{\eta_\gamma}^{a_6} + a_7{\eta_\varepsilon}^{a_8}{\eta_\gamma}^{a_9} + a_{10}{\eta_d}^{a_{11}}{\eta_\gamma}^{a_{12}} + a_{13}{\eta_d}^{a_{14}}{\eta_\varepsilon}^{a_{15}} + a_{16}{\eta_d}^{a_{17}}{\eta_\varepsilon}^{a_{18}}{\eta_\gamma}^{a_{19}} \quad \text{(A5.2.6)}$$

The parameters $\eta_d = \frac{L_{g,1D}}{2k_r{\sigma_x}^2}$, $\eta_\varepsilon = 2L_g\beta k_r\varepsilon$, $\eta_\gamma = 2L_g\lambda_u\sigma_\gamma$, where $\sigma_x$ is the beam transverse size, $\beta$ is the beam beta function, $\varepsilon$ is the emittance, $\sigma_\gamma$ is the energy spread, and $a_n$ denotes to the polynomial fitting coefficients:

$$a_1 = 0.45;\ a_2 = 0.57;\ a_3 = 0.55;\ a_4 = 1.6;\ a_5 = 3;\ a_6 = 2;\ a_7 = 0.35;$$
$$a_8 = 2.9;\ a_9 = 2.4;\ a_{10} = 51;\ a_{11} = 0.95;\ a_{12} = 3;\ a_{13} = 5.4;\ a_{14} = 0.7;$$
$$a_{15} = 1.9;\ a_{16} = 1140;\ a_{17} = 2.2;\ a_{18} = 2.9;\ a_{19} = 3.2$$

The equations mentioned above, especially the Peirce parameter and the 1D/3D gain length, serve as crucial and valuable tools for quickly establishing design parameters, reducing the dependence on extensive numerical simulations.

Utilizing the beam parameters outlined in Table A5.2.1 and the undulator characteristics provided in Table A5.2.2, we can estimate the properties of the 100 keV SASE FELs. Assuming a beta function of 80 m and a 1D Pierce parameter of $1.6\times10^{-4}$, we can estimate the gain length and the saturation length. The left plot in Figure A5.2.4 presents the projected saturation length as a function of beam energy. It becomes evident that employing a low beam energy of less than 20 GeV results in an undesirably long undulator system, exceeding 300 meters. However, the saturation length experiences a rapid reduction as beam energy increases. Beyond 25 GeV, the saturation length stabilizes at approximately 270 meters. Considering that the European XFEL has successfully implemented an undulator system within the length range of 200-300 meters, a 270 m long undulator system appears to be feasible. This observation affirms the potential applications of CEPC, including the capacity to drive a 100 keV-level FEL as a byproduct of the 30 GeV linac.

The right plot illustrates the 3-dimensional gain length as a function of the beta function. Enhancing the beta function leads to a shorter 3D gain length, but it also results in beam density dilution and an increase in the Pierce parameter. Consequently, an optimized beta function of 80 m is selected.

Comprehensive calculations were carried out using numerical simulations conducted with the Genesis 1.3 code [13], a three-dimensional tool capable of performing time-dependent simulations. These simulations encompass critical factors, including diffraction, slippage, energy spread, and emittance. In the numerical simulations, the FEL wavelength is established at 0.0134 nm, corresponding to a photon energy of 93 keV.

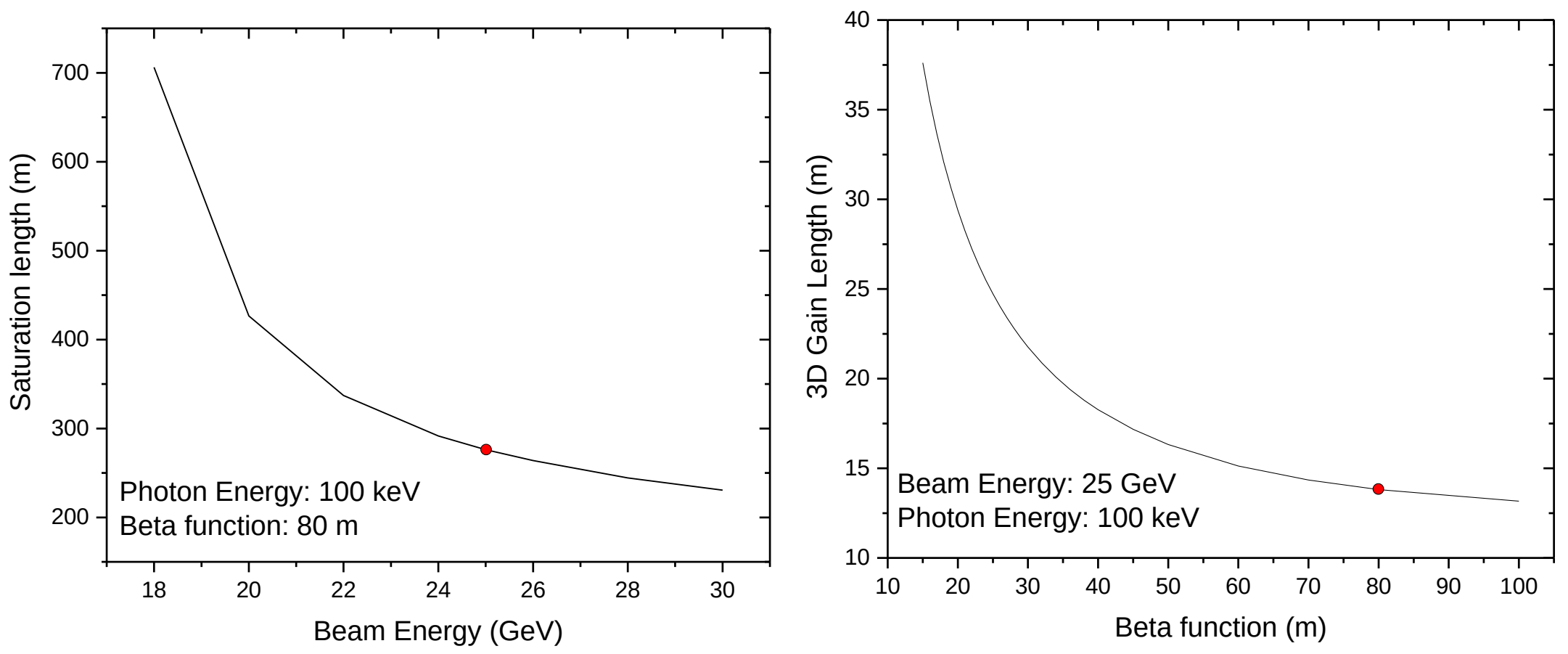

**Figure A5.2.4:** Left – saturation length as a function of beam energy; Right – 3D gain length with respect to the beta function.

Figure A5.2.5 provides insights into the characteristics of the 93 keV SASE FEL at an undulator length of 300 meters, where the FEL reaches saturation. For the sake of generality, the beam current is presumed to exhibit a uniform distribution of 5000 A. The suggested beam length is 14 femtoseconds. The left plot illustrates the temporal power distribution, with a maximum slice power reaching as high as $1\times10^{10}$ W. The power temporal distribution is Fourier-transformed into the frequency domain to calculate the spectrum, as depicted in the right plot. It's evident that the central wavelength is approximately 0.01339 nm, corresponding to a photon energy of 93 keV. Fitting the spectrum using a Gaussian function reveals that the sigma is approximately $2.18\times10^{-6}$ nm, resulting in $\Delta\omega/\omega \approx 3.26\times10^{-4}$, which closely aligns with the Pierce parameter calculated by Eq. (A5.2.4).

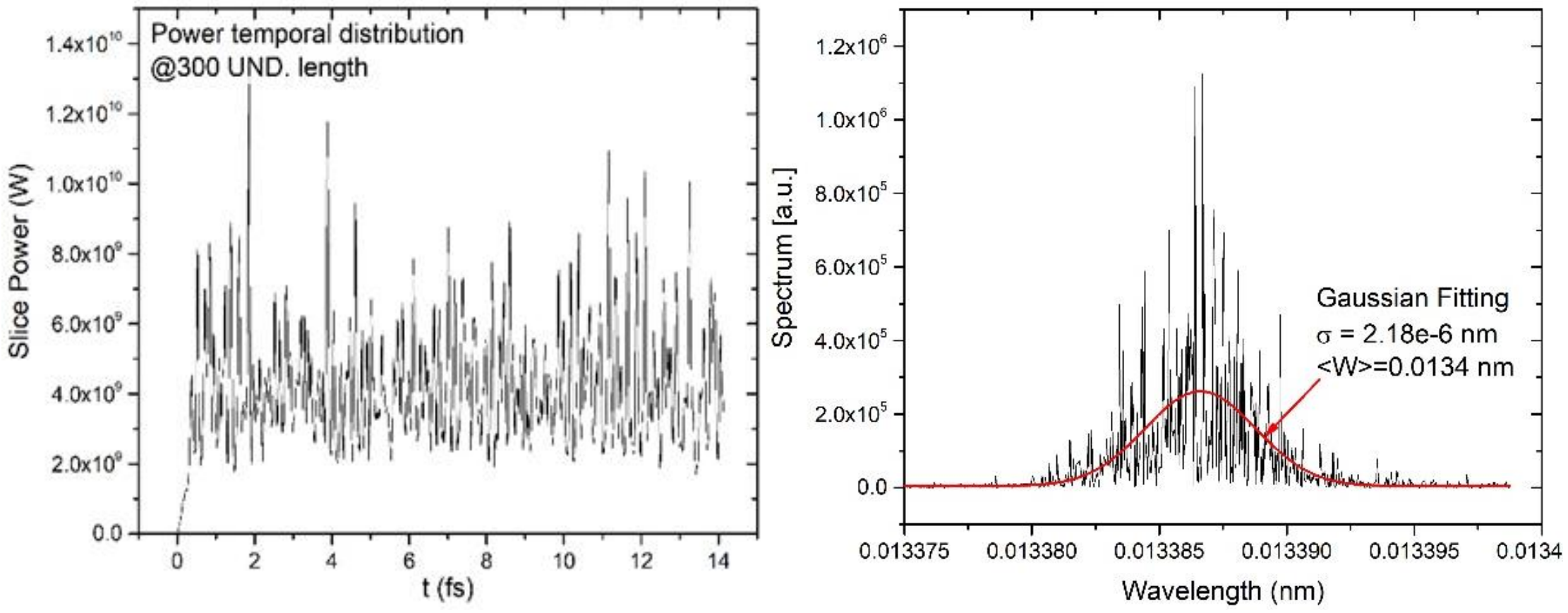

**Figure A5.2.5:** SASE FEL characters at the undulator position 300 m. Left – temporal power distribution; Right – spectrum in arbitrary unit.

Figure A5.2.6 illustrates the SASE FEL exponential gain curve concerning the undulator length. The FEL gain shows a tendency to reach saturation within the range of 250-300 meters. At this point, the maximum peak power reaches approximately $5\times10^9$ W.

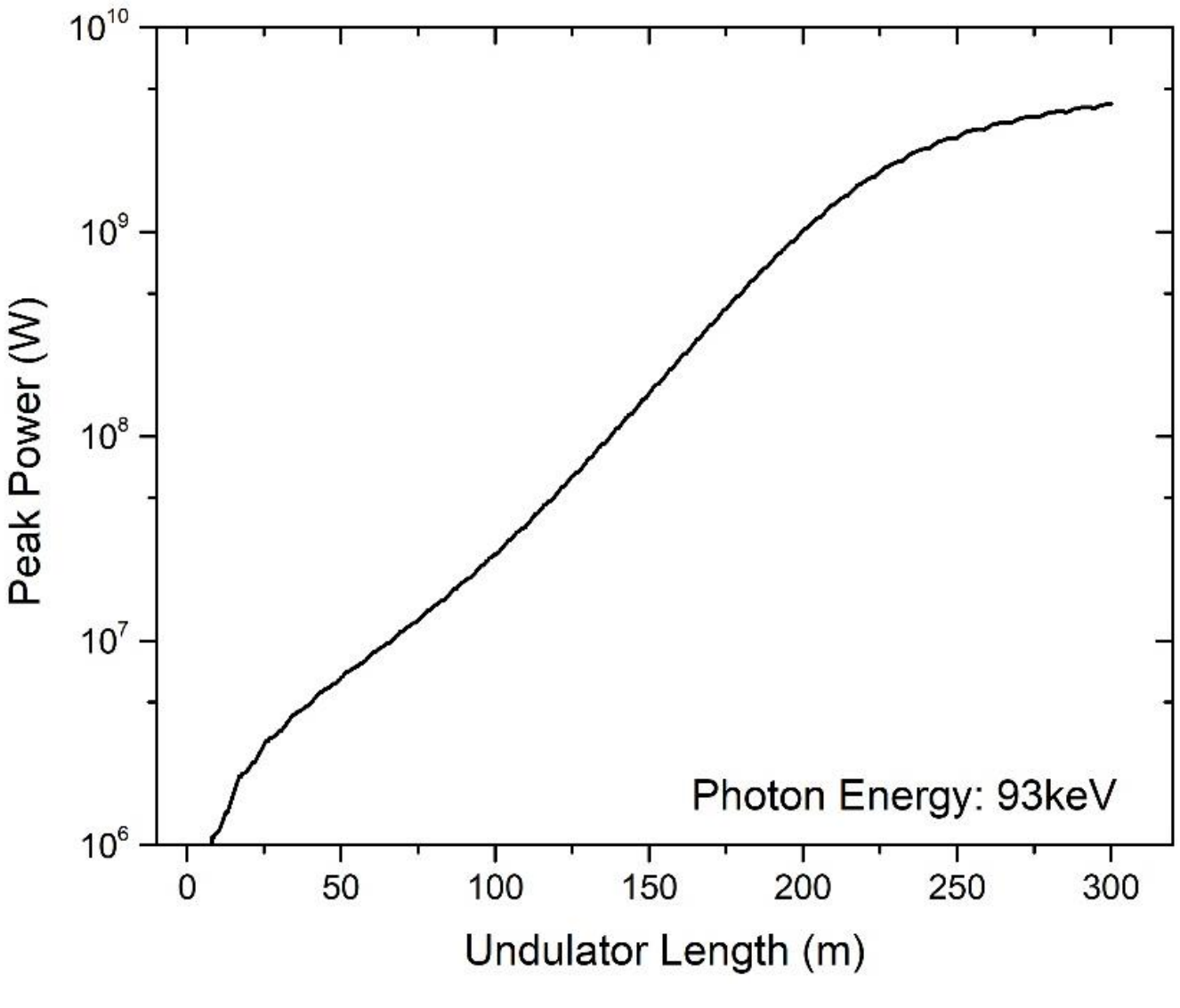

**Figure A5.2.6:** Exponential gain of the peak power versus undulator length.

Table A5.2.3 compiles the parameters and numerical simulation results for the anticipated 93 keV FEL to be generated at the CEPC linac.

**Table A5.2.3:** Expected FEL parameters at CEPC

| Parameter | Value | Unit |
|---|---|---|
| Beta function | 80 | m |
| 1 D Pierce Parameter | $1.6 \times 10^{-4}$ | |
| 3 D gain length | 13.8 | m |
| Saturation length | 250 | m |
| Bandwidth $\Delta\omega/\omega$ | $3.26 \times 10^{-4}$ | |
| Photon energy | 93 | keV |
| Radiation wavelength | 0.013 | nm |

### A5.2.6: Undulator Technology R&D

It's worth noting that undulator technology, while well-developed in facilities related to synchrotron radiation and Free Electron Lasers, is not a standard component of colliders. To demonstrate the readiness of undulator technology, we present an overview of the undulator technology study at IHEP.

IHEP has actively engaged in the development of undulator technologies for the construction of the High Energy Photon Source (HEPS), a 4th generation light source [14]. Currently, several types of undulators are under construction, and they are slated for deployment in HEPS within the next few years [15]. Table A5.2.4 provides details about two types of in-vacuum undulators designed for HEPS. The Cryogenic Permanent Magnet Undulators (CPMU) have a length of two meters and employ PrFeB as the magnet material. The In-Vacuum Undulators (IVU) designed for normal temperatures have a length of four meters and use NdFeB and SmCo as magnet materials for different devices. These undulators have varying periods ranging from 12 mm to 22.8 mm and operation gaps that span from 5.2 mm to 15.2 mm, as detailed in Table A5.2.4.

**Table A5.2.4:** The CPMU and IVU undulator built for HEPS and their indicators.

| Undulator Type | Length [m] | Period [mm] | Performance Gap [mm] | Max. Peak Field [T] |
|---|---|---|---|---|
| CPMU | 2 | 12.0 | 5.2-7.0 | 0.81 |
| CPMU | 2 | 14.2 | 5.2-9.9 | 1.00 |
| CPMU | 2 | 16.7 | 5.2-10.0 | 1.19 |
| CPMU | 2 | 18.8 | 5.2-13.1 | 1.35 |
| CPMU | 2 | 22.8 | 7.2-16.0 | 1.18 |
| IVU (NdFeB) | 4 | 18.8 | 5.2-13.1 | 1.35 |
| IVU (SmCo) | 4 | 19.9 | 5.2-14.0 | 0.97 |
| IVU (SmCo) | 4 | 22.6 | 5.2-15.2 | 1.10 |

Figure A5.2.7 illustrates one representative In-Vacuum Undulators (IVUs) developed at IHEP. The left plot presents the inner magnet structure during a field measurement conducted using a dedicated system. The right plot showcases a fully assembled IVU, prepared for installation.

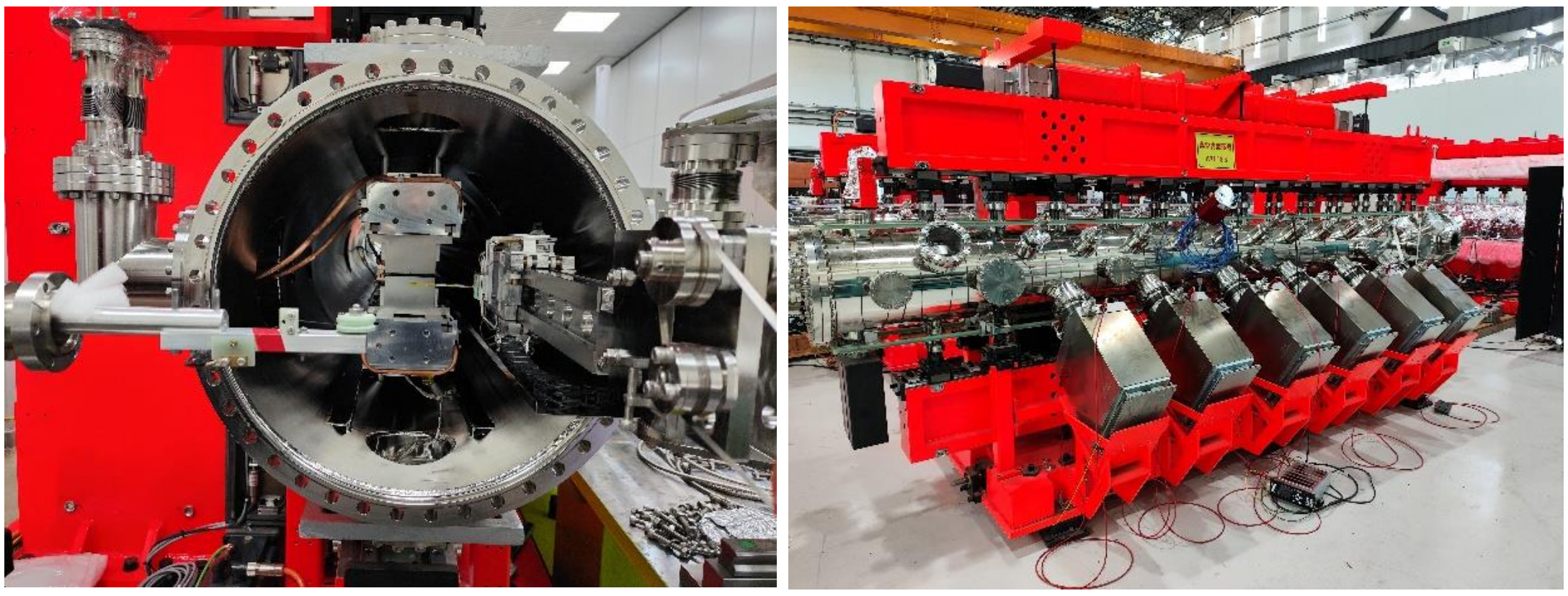

**Figure A5.2.7:** An in-vacuum undulator developed at IHEP.

## A5.2.7: References

# Appendix 6: CEPC Injector based on a Plasma Wakefield Accelerator

In the CEPC Booster outlined in the Conceptual Design Report (CDR), the circumference spans 100 km, and it is tasked with elevating beam energy from 10 GeV to 45.5 GeV or potentially even higher (up to 180 GeV). Consequently, the dipole magnetic field in the Booster must possess a wide adjustable range. Furthermore, if the beam begins with an initial energy of 10 GeV before entering the Booster, the lower limit of the dipole magnetic field in the Booster stands at approximately 28 Gauss. Notably, the requirements for dipole field error (< 0.029 Gauss) and dipole field reproducibility (< 0.015 Gauss) in this scenario are significantly smaller than Earth's magnetic field (around 0.5 Gauss). These factors collectively pose significant challenges in the manufacturing and measurement of the dipole magnet. To mitigate the dipole magnet issue in the Booster, one potential solution is to introduce the CEPC Plasma Injector (CPI) to enhance beam energy prior to injection into the Booster.

## A6.1: Level of Maturity of the CPI Conceptual Design

As of the present, plasma wakefield acceleration (PWFA) techniques have not reached a level of maturity suitable for a high-energy physics project, and it appears that their development may not align with the current CEPC TDR/EDR timeline. However, it's worth noting that the precise construction schedule for the CEPC project has not been finalized. Additionally, implementing the plasma injector is not expected to significantly impact the fundamental infrastructure, allowing for some flexibility in the timeline compared to other hardware systems or the overall physics design. Furthermore, it's essential to emphasize that the plasma injector serves as an alternative approach to address the challenge of the low field dipole, and its readiness will not impede the overall CEPC project timeline, even if it is not prepared in time.

## A6.2: Overall Design of CPI

The CPI concept was initially introduced in 2017, followed by the release of the conceptual design V1.0 in 2018. In this context, we now introduce the most recent iteration of CPI, V3.0, as illustrated in Fig. A6.2.1.

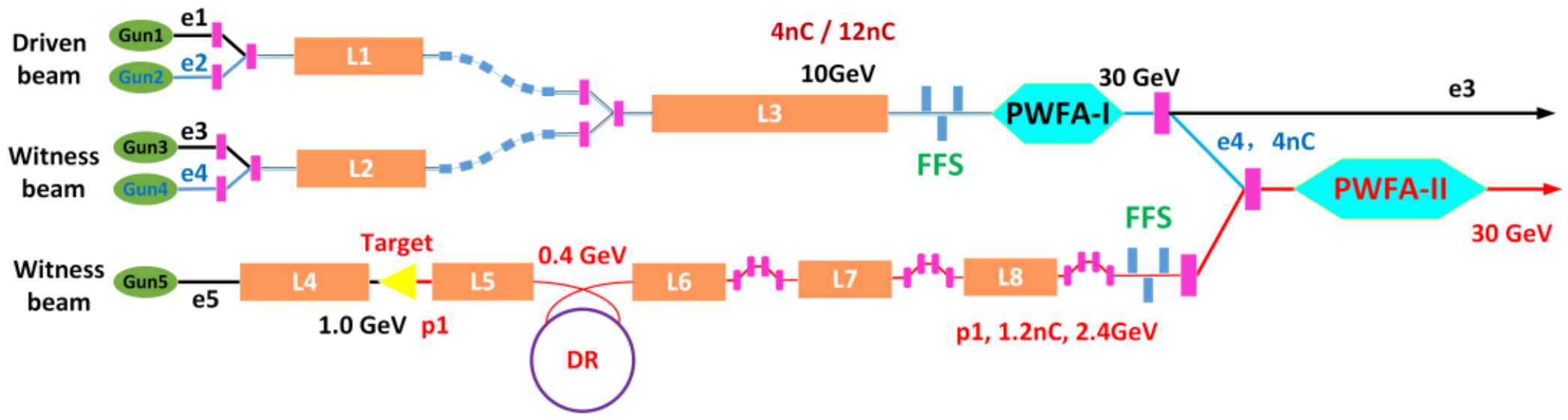

**Figure A6.2.1:** Conceptual Design V3.0 of the CEPC Plasma Injector (CPI). Gun1, Gun3, and Gun4 are S-band photocathode electron guns, producing 4 nC, 1 nC, and 3.6 nC electrons per bunch, respectively. Gun2 is an L-band photocathode gun delivering 12 nC electrons per bunch, while Gun5 is a thermionic cathode gun capable of generating 1 ns / 10 nC electrons per bunch. The linacs labeled L1-L8 utilize S-band technology to accelerate both electron and positron beams. The "Target" is designated for positron production and shares materials and structures with the baseline positron source. The acronyms "DR" and "FFS" stand for damping ring and final focusing system, respectively, with detailed explanations provided in sections A6.5.1.2 and A6.5.2. PWFA-I and PWFA-II represent the plasma acceleration sections for electrons and positrons, respectively. Among these components, "e1" to "e5" denote electron bunches, while "p1" represents a positron bunch.

The electron and positron beams reach 30 GeV in PWFA-I and PWFA-II, respectively. In the case of electron acceleration, a 4 nC electron bunch, with specific shapes, originates from RF Gun1, serving as the driver in PWFA-I. Simultaneously, another 1.2 nC electron bunch is produced from RF Gun 3 to act as a trailer in PWFA-I. These two bunches undergo compression in L1 and L2, respectively, and are then combined and accelerated to 11 GeV in L3. In PWFA-I, the trailer bunch achieves 30 GeV, while the driver bunch expends nearly all of its energy.

Regarding positron acceleration, "e4," with 30 GeV and a charge of 3.4 nC after PWFA-I, serves as the driver beam, while "p1," with 2.4 GeV and a charge of 1.2 nC, functions as the trailer beam within the PWFA-II section. These two beams are merged into the same bunch upon injection into PWFA-II, with an approximate spacing of about 170 μm. The positron bunch gains energy from the electron driver bunch, resulting in an acceleration to 30 GeV. The positron bunch is sourced from the positron beamline, comprising a positron target, a damping ring, and the compression chicane. The CPI design must meet the requirements of the CEPC Booster, as outlined in Table A6.2.1:

**Table A6.2.1:** Requirement of the CEPC Booster parameters.

| Parameter | Symbol | Unit | Requirement |
|---|---|---|---|
| $e^-$ /$e^+$ beam energy | $E_{e-}/E_{e+}$ | GeV | 30 |
| frequency | $f_{rep}$ | Hz | 100 |
| $e^-$/$e^+$ bunch population | $N_{e-}/N_{e+}$ | nC | > 1.0 |
| Energy spread ($e^-$/$e^+$) | $\sigma_e$ | | < 2×10$^{-3}$ |
| Emittance ($e^-$/$e^+$) | $\varepsilon_r$ | nm· rad | < 10 |
| Bunch length ($e^-$/$e^+$) | $\sigma_l$ | mm | 0.2~ 2 |
| Switch time $e^-$/$e^+$ | | s | < 2 |
| Energy stability | | | < 2×10$^{-3}$ |
| Longitudinal stability | | mm | < 2 |
| Orbit stability | | mm | <3 (H) / 3 (V) |
| Failure rate | | % | < 1 |

## A6.3: Electron Acceleration in PWFA-I with High Transformer Ratio

### A6.3.1: Basic Theories of Electromagnetic Field in a Blowout Regime

In the blowout regime of PWFA, the drive beam exhibits higher density than the background plasma, resulting in the formation of a series of bubble-like electron evacuated regions with a positively charged background following the drive beam. The blowout regime offers several advantages over non-blowout scenarios, including a higher acceleration gradient and improved beam quality for the accelerated beam. While the blowout regime is characterized by strong non-linearity and relativistic effects, its electromagnetic (EM) field distribution remains predictable due to its straightforward charge and current distribution.

To initiate the theoretical derivation, we express Maxwell's equations in terms of potentials:

$$\frac{\partial^2}{\partial t^2} - \nabla^2 \begin{bmatrix} \mathbf{A} \\ \varphi \end{bmatrix} = \begin{bmatrix} \mathbf{J} \\ \rho \end{bmatrix}. \tag{A6.3.1}$$

In the above expression, A and $\varphi$ represent the vector and scalar potentials, respectively, while J and $\rho$ denote the current and charge densities, respectively. Typically, in a PWFA setup, the drive beam is highly relativistic, essentially frozen in the frame comoving with the beam. As a result, the wakefield also experiences a nearly frozen state similar to the drive beam. This condition is referred to as the quasi-static approximation (QSA), which can be expressed mathematically as:

$$\frac{\partial}{\partial s} \ll \frac{\partial}{\partial \xi} \tag{A6.3.2}$$

for any field variables such as $\mathbf{A}$, $\mathbf{J}$, $\varphi$ and $\rho$, where

$$\xi = \mathrm{t} - \mathrm{z}, \tag{A6.3.3}$$

$$\mathrm{s} = \mathrm{z}, \tag{A6.3.4}$$

is the comoving frame. With the QSA, the Maxwell's equations can be simplified as

$$-\nabla_{\perp}^{2}\begin{bmatrix}\mathbf{A}\\ \varphi\end{bmatrix}=\begin{bmatrix}\mathbf{J}\\ \rho\end{bmatrix}. \quad \text{(A6.3.5)}$$

where $\nabla_{\perp}^{2}=\partial^2/\partial x^2+\partial^2/\partial y^2$, and x and y are the two transverse coordinates. We use the cylindrical coordinate system $(z, r, \theta)$ and assume the rotational symmetry $\partial/\partial\theta=0$. Within the blowout region the current and charge have the simple form (in the normalized units where density is normalized to the plasma density and charge is normalized to the elementary charge)

$$J_z=-n_b, \quad \text{(A6.3.6)}$$
$$J_r=0, \quad \text{(A6.3.7)}$$
$$\rho=1-n_b, \quad \text{(A6.3.8)}$$

in which $n_b$ is the density of the electron beam that has non-zero value for $r<\sigma_r$, and $\sigma_r$ is the radius of the beam. Thus, we can write the expressions of the potentials within the blowout region:

$$A_z=A_{z0}+\lambda\ln r, \quad \text{(A6.3.9)}$$
$$\phi=\phi_0-\frac{r^2}{4}+\lambda\ln r, \quad \text{(A6.3.10)}$$

where $A_{z0}=A_z|_{r=0}$ and $\phi_0=\phi|_{r=0}$ are the axial longitudinal vector potential and scalar potential, respectively, $\lambda=\int_0^\infty r\,n_b dr$ is the normalized beam current, and we have assumed $\sigma_r\ll r$. Define the pseudo-potential as:

$$\Psi\equiv\phi-A_z, \quad \text{(A6.3.11)}$$

which in the blowout region has the form:

$$\Psi=\Psi_0-\frac{r^2}{4}, \quad \text{(A6.3.12)}$$

where $\Psi_0=\phi_0-A_{z0}$. It is clear that $\partial\Psi/\partial\xi=d\Psi_0/d\xi$ is not a function of $r$ within the blowout region. Furthermore, from the Lorenz gauge condition $\nabla\cdot A+\partial\varphi/\partial t=0$ and the QSA, we have the following relation:

$$A_r=-\frac{r}{2}\frac{\partial\Psi}{\partial\xi}. \quad \text{(A6.3.13)}$$

Use $E=-\nabla\varphi-\partial A/\partial t$, $B=\nabla\times A$ and the rotational symmetry, the EM field within the blowout region can be expressed as

$$E_z=E_{z0}=\frac{d\Psi_0}{d\xi}, \quad \text{(A6.3.14)}$$
$$E_r=\left(1+\frac{dE_{z0}}{d\xi}\right)\frac{r}{2}-\frac{\lambda}{r}, \quad \text{(A6.3.15)}$$
$$B_\theta=\frac{dE_{z0}}{d\xi}\frac{r}{2}-\frac{\lambda}{r}, \quad \text{(A6.3.16)}$$

where $E_{z0}=E_z|_{r=0}$. The transverse EM field provide the focusing force for the electron beam trapped in the wakefield:

$$F_r = -\frac{r}{2} - (1 - v_z)\frac{dE_{z0}}{d\xi}\frac{r}{2} + (1 - v_z)\frac{\lambda}{r}, \tag{A6.3.17}$$

which is generally simplified *to* $F_r = -r/2$ for ultrarelativistic ($v_z \approx 1$) beam. The focusing force is a conservative force that generally results in the conservation of normalized emittance. The longitudinal force, denoted as $F_z$ and equal to $-E_z$, is solely a function of the longitudinal coordinate $\xi$ and remains independent of $r$ as long as the beam remains within the blowout region. This characteristic is advantageous in the blowout regime since the uniformity of the longitudinal force across the transverse plane enhances the monochromaticity of the accelerated beam and improves energy efficiency. However, it's important to note that the uniformity of the longitudinal force in the longitudinal direction is not always guaranteed.

To optimize the energy efficiency of PWFA, a critical requirement for an accelerator operating in production mode, it is imperative to minimize the longitudinal non-uniformity of the longitudinal force. This ensures that all slices of the drive beam lose energy synchronously, while all slices of the trailing beam gain energy synchronously. To accomplish this, we can leverage the beam loading effect, where the presence of the charged beam within the plasma wakefield alters the field structure. According to a simplified model, the axial pseudo-potential in the blowout regime can be approximated as:

$$\Psi_0 \approx \frac{r_b^2}{4}, \tag{A6.3.18}$$

where $r_b$ is the blowout radius. Which can be estimated according to

$$r_b \approx 2\sqrt{\lambda}, \tag{A6.3.19}$$

if the electron beam has an adiabatical current profile. Thus, to achieve a uniform deceleration/acceleration force for the drive/trail beam, we can use a linearly increasing/decreasing beam current profile:

$$\frac{d}{d\xi}\lambda_d = E_+, \tag{A6.3.20}$$

$$\frac{d}{d\xi}\lambda_t = -E_-, \tag{A6.3.21}$$

where $\lambda_d$ is the normalized current of the drive beam, $\lambda_t$ is the normalized current of the trail beam, $E_+$ and $E_-$ are the absolute value of the uniform acceleration and deceleration field. This indicate that the current profile shape should have a triangular or trapezoidal form. Such type of electron/positron ($e^-/e^+$) beams can be produced using a photocathode gun through the manipulation of the longitudinal laser profile. Alternatively, they can be generated by implementing the "emittance exchange" method within the beamline [1].

### A6.3.2: Baseline Design for CPI Electron Acceleration

In PWFA, the transformer ratio (TR) is characterized as the average energy gain (per single particle) of the trailer beam relative to the energy loss of the driver beam [2]. In the case of the CPI, the TR should be maintained at a minimum value of (30-10)/10 = 2. In In Tables A6.3.1 and A6.3.2, you can find the beam and plasma parameters for the electron acceleration stage, along with the beam quality following the PWFA-I stage.

**Table 6.3.1:** Beam and plasma parameters for CPI electron acceleration

| Beam | Driver | Trailer |
|---|---|---|
| Plasma density $n_p(\times 10^{16} cm^{-3})$ | 0.50334 | |
| Driver energy $E$(GeV) | 10 | 10 |
| Normalized emittance $\epsilon_n(mm \cdot mrad)$ | 20 | 10 |
| Length (μm) | 350 | 90 |
| Spot size (μm) | 3.89 | 2.75 |
| Charge (nC) | 4 | 1.2 |
| Energy spread $\delta_E$(%) | 0 | 0 |
| Beam distance (μm) | 180 | |

**Table 6.3.2:** Beam qualities after PWFA-I

| | |
|---|---|
| Accelerating distance (m) | 6.3 |
| Trailer energy $E$(GeV) | 30.0 |
| Normalized emittance $\epsilon_n(mm \cdot mrad)$ | 10 |
| Charge (nC) | 1.2 |
| Energy spread $\delta_E$(%) | 0.32 |
| R | 2.0 |
| Efficiency (%) (driver -> trailer) | 66.0 |

The beam qualities generally fulfill the CEPC Booster's requirements, except for the energy spread, which stands at 0.32%, slightly exceeding the Booster's specified limit of 0.2%. However, it's worth noting that the slice energy spread is considerably smaller than the overall beam energy spread. To address this, we can incorporate an additional plasma dechirper section to reduce the energy spread to a level below 0.2%.

As previously discussed in the last section, it is necessary to modulate the profiles of the driver and trailer beams to attain a high transformer ratio (≥ 2). Figure A6.3.1 illustrates the beam current profile and the wake structure within the plasmas.

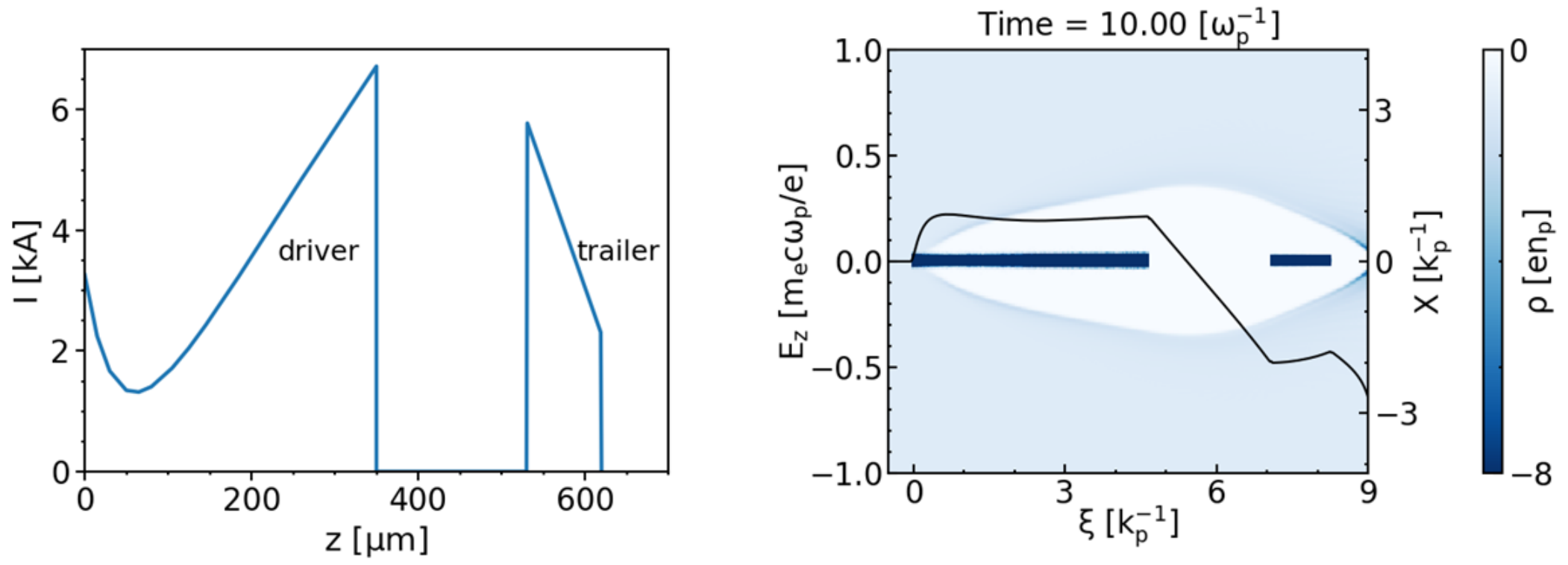

**Figure A6.3.1:** Beam current profile (left) and plasma wake structure (right). In both figures, the driver and the trailer travel from right to left. In the left figure, the blue line represents the beam current as it varies along the longitudinal position. In the right figure, the black line represents the longitudinal wakefield along the axis.

### A6.3.3: Preliminary Error Tolerance Analysis for CPI Electron Acceleration

Within the CPI parameter domain, the acceleration structures in plasmas operate on a 100-μm-scale. Consequently, the error tolerance requirements are significantly more stringent compared to conventional accelerators, posing potential challenges to the linac. A comprehensive examination of error tolerances has been conducted, encompassing factors such as the driver-trailer transverse/longitudinal offset, driver/trailer tilt, driver/trailer energy and energy spread, driver/trailer bunch charge, plasma density, among others. In this context, we will illustrate the driver-trailer transverse offset and the driver's tilt as examples:

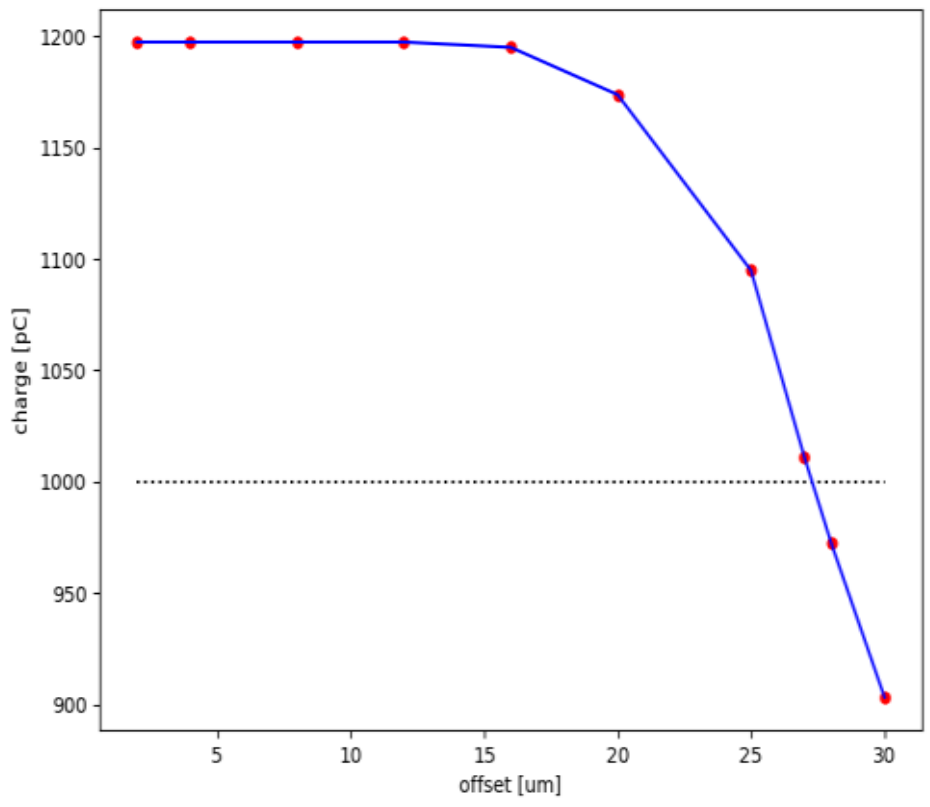

**Figure A6.3.2:** Error tolerance for the driver-trailer transverse offset. The horizontal axis represents the horizontal distance between the driver's beam center and the trailer's, while the vertical axis indicates the particles remaining in the trailer following the PWFA-I section. The black dashed line signifies the required bunch charge of the booster ring, which must not fall below 1 nC.

Figure A6.3.2 presents the results of the error tolerance analysis concerning the driver-trailer transverse offset. As the offset in the horizontal direction increases, the bunch charge of the trailer after the PWFA diminishes. To align with the Booster's stipulation of not less than 1 nC bunch charge following plasma acceleration, our numerical simulations indicate that the transverse offset should not exceed 27 μm.

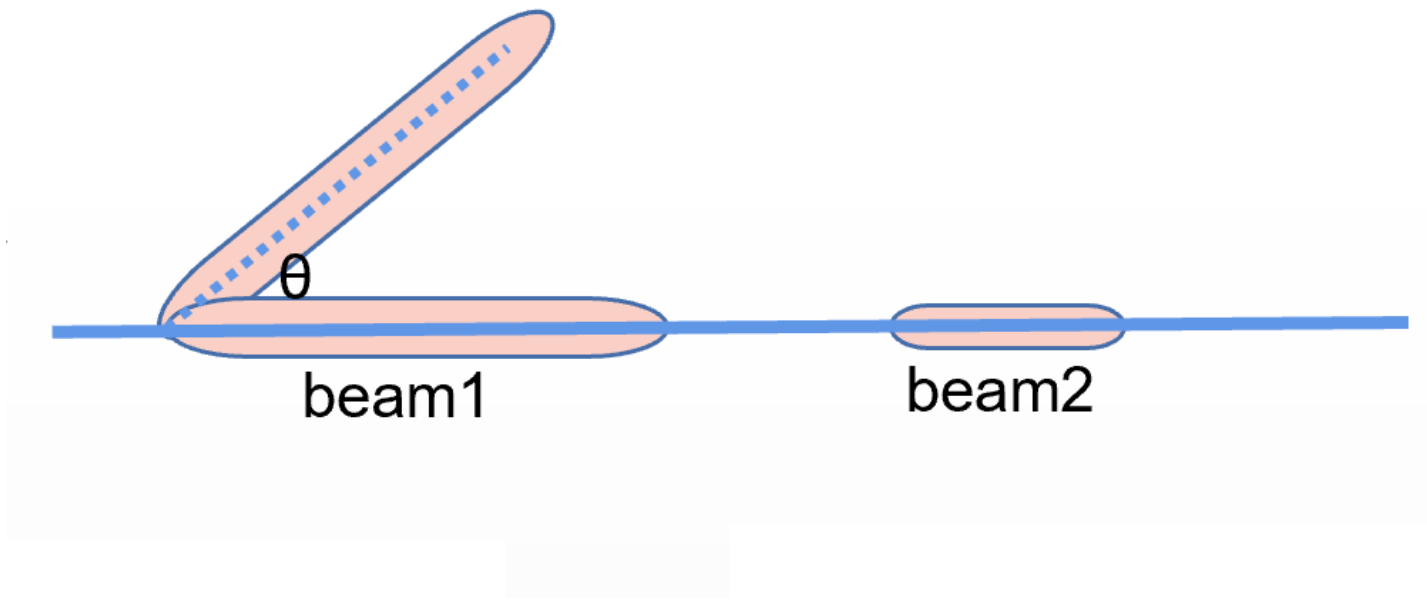

**Figure A6.3.3:** Driver’s tilt angle

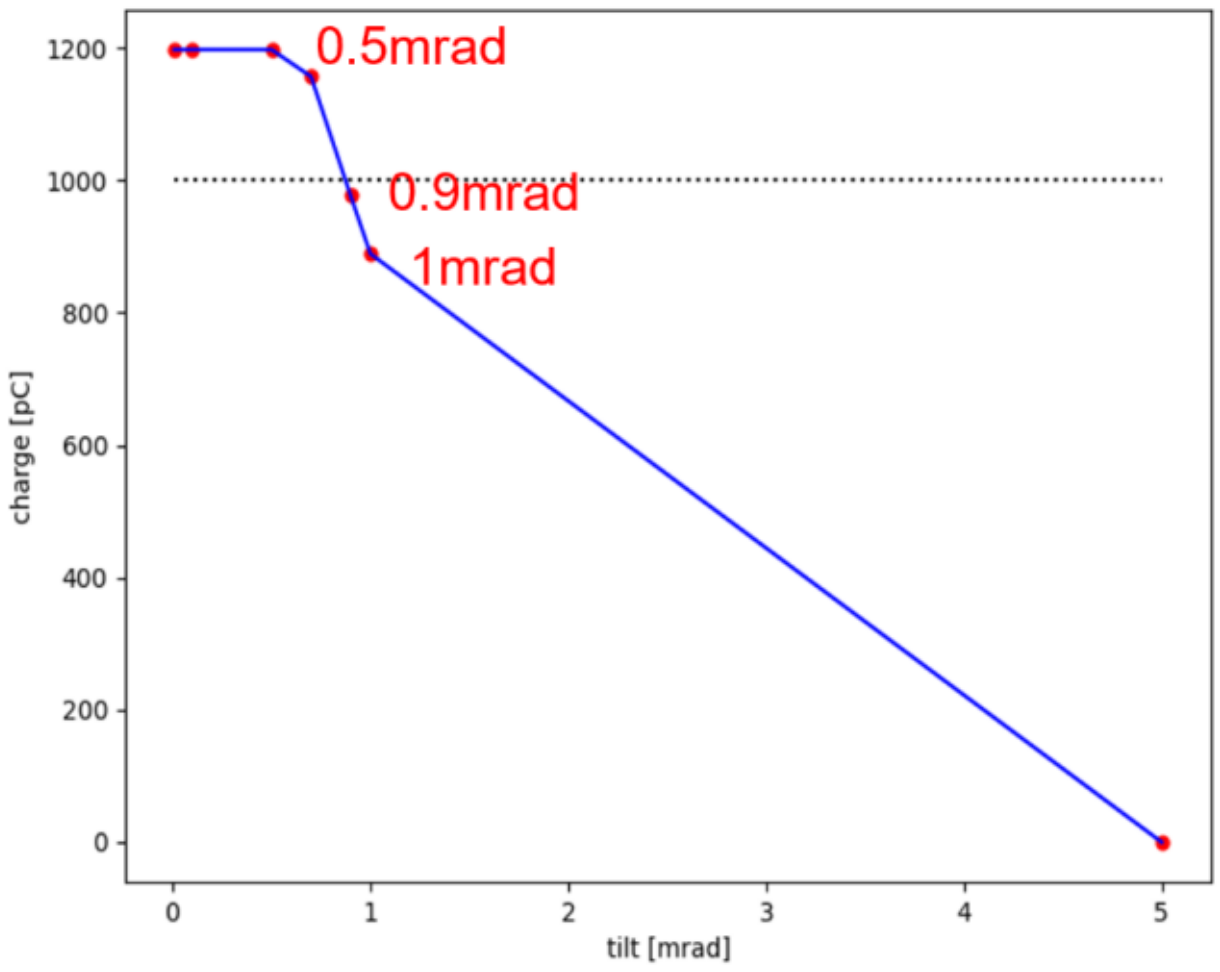

**Figure A6.3.4:** Error analysis for the driver’s tilt angle. The horizontal axis represents the longitudinal tilt angle between the driver and the trailer, as depicted in Figure A6.3.3. The vertical axis and the black dashed line correspond to those in Figure A6.3.2.

We define the driver's tilt angle as illustrated in Fig. A6.3.3 and perform error analysis using PIC simulations. Preliminary results, depicted in Fig. A6.3.4, indicate that based on our simulations, the threshold for the driver's tilt angle should be approximately 0.9 mrad if we aim to retain 1 nC of electrons after the acceleration.

These error analyses demonstrate that CPI imposes more stringent demands on the linac in comparison to conventional injectors or boosters, thereby significantly increasing the challenges associated with linac design.

# A6.4: Positron Acceleration Scheme in CPI

## A6.4.1: A New Positron Acceleration Scheme with High Beam Quality and High Efficiency

In the PWFA community, the most significant breakthroughs have been achieved using high-intensity electron bunches to generate a nonlinear wakefield that accelerates a second electron bunch. However, this method is not effective for positron acceleration due to the extremely limited volume at the rear of such wakes, where the wakefield accelerates and focuses positrons. Positron acceleration in plasmas is a recognized challenging problem, and various approaches have been proposed to overcome this limitation. One such approach involves using hollow plasma channels generated by readily available electron beams, as they offer a uniform accelerating field in transverse planes and zero focusing force within the channels. However, any misalignment of the drive and trailing bunches can trigger a pronounced beam-breakup instability, resulting in beam emittance growth and, ultimately, the loss of positrons. Consequently, this instability has constrained further research into the use of hollow channels for high-energy positron acceleration.

For CPI positron acceleration, we introduce a nonlinear scheme that offers field structures suitable for the stable acceleration and focusing of a positron bunch within a hollow plasma channel [3]. In this scheme, the focusing field exhibits nearly linear variation, and the longitudinal field remains largely independent of the transverse dimensions throughout the channel region where the positron beam is situated. This particular field structure facilitates guided propagation and efficient acceleration of a high-charge positron beam. Additionally, the positron beam effectively loads the wake, acquiring energy while maintaining a narrow energy spread. Although the efficiency of this scheme is not yet on par with electron PWFA, it currently stands as the most promising approach for achieving high-efficiency, high-quality positron acceleration within PWFA.

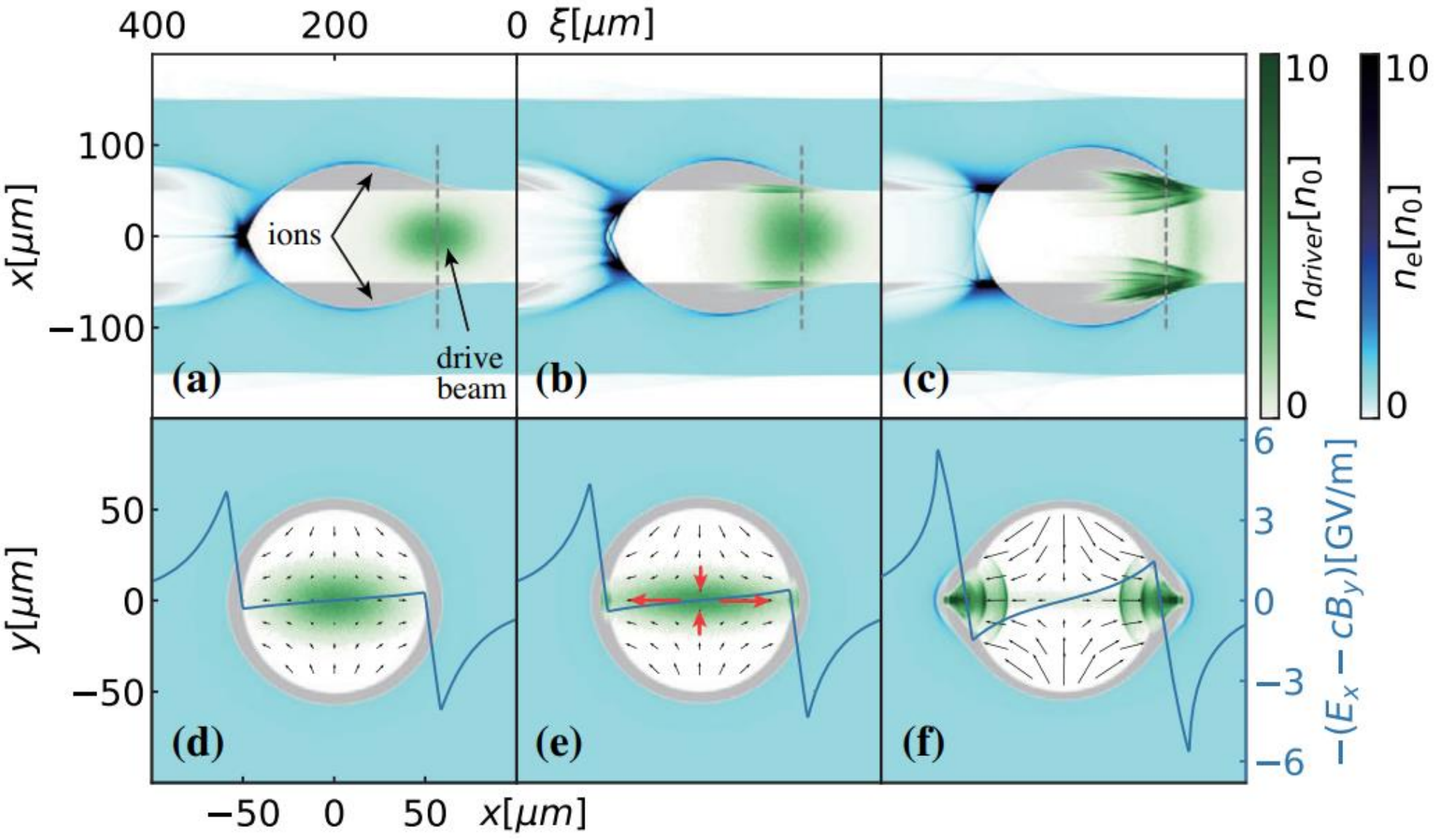

**Figure A6.4.1:** Evolution of an asymmetric electron beam and wakefields in a hollow plasma channel. Plots (a)–(c) depict plasma and beam densities in the x–ξ plane (with the beam moving from left to right), while plots (d)–(f) focus on the cross-section of the beam's center slice, indicated by the gray dashed line, and transverse wakefields. The black arrows represent the vector flow of the transverse wakefield, denoted as $-\overrightarrow{W}_{\perp} = -(E_x - cB_y)\hat{x} - (E_y - cB_x)\hat{y}$. The blue line signifies the lineout of $-\overrightarrow{W}_{\perp}$ at $y = 0$. Additionally, the red arrows indicate the spot size evolution of the drive beam. The propagation distances are 0, 2.5, and 10 cm, moving from left to right.

Here, we provide a summary of how our scheme operates. The key lies in inducing a quadrupole transverse wakefield within the hollow plasma waveguide using an asymmetric electron beam driver (where $\sigma_x \neq \sigma_y$, with $\sigma_{x;y}$ representing the beam's rms spot sizes in the transverse directions). As the beam progresses into the hollow channel, the evolution of its spot size is governed by the self-excited quadrupole wakefields, as depicted in Fig. A6.4.1. In one plane, the narrow part of the beam undergoes focusing, while the wider part experiences defocusing until a majority of electrons reach the inner plasma channel boundary in that particular plane. Conversely, in the other plane, the electrons are tightly focused. A dense drive bunch repels plasma electrons from the wall, leaving the much more massive ions behind. Once this equilibrium state is achieved, the electron beam propagates with minimal further evolution, as the defocusing quadrupole fields are counteracted by focusing forces from the exposed ions. These ions subsequently attract plasma electrons, forming a sheath. This process resembles the creation of a half-bubble on opposite sides of the channel wall. However, in this case, the returning sheath electrons overshoot the initial channel boundary and occupy the entire cross-section of the hollow channel. This results in a focusing force that guides the trailing positron bunch in both planes when it is positioned in this region. The positron bunch can extract a significant amount of energy by flattening or beam-loading the longitudinal accelerating field. Additionally, some plasma electrons are drawn toward the axis by the positrons, forming a denser electron filament that reinforces the focusing force.

## A6.4.2: Baseline Design of CPI Positron Acceleration

### *A6.4.2.1: Design for CPI Positron Acceleration*

To attain high-quality positron acceleration, we employ transversely separated electron drivers to minimize the evolution process, as illustrated in Figure A6.4.2. In this configuration, two electron drive beams share identical beam parameters with a transverse separation of 315 μm, approximating the equilibrium state described in the preceding section. The current profile of the loaded positron beam is selected to optimize its energy spread.

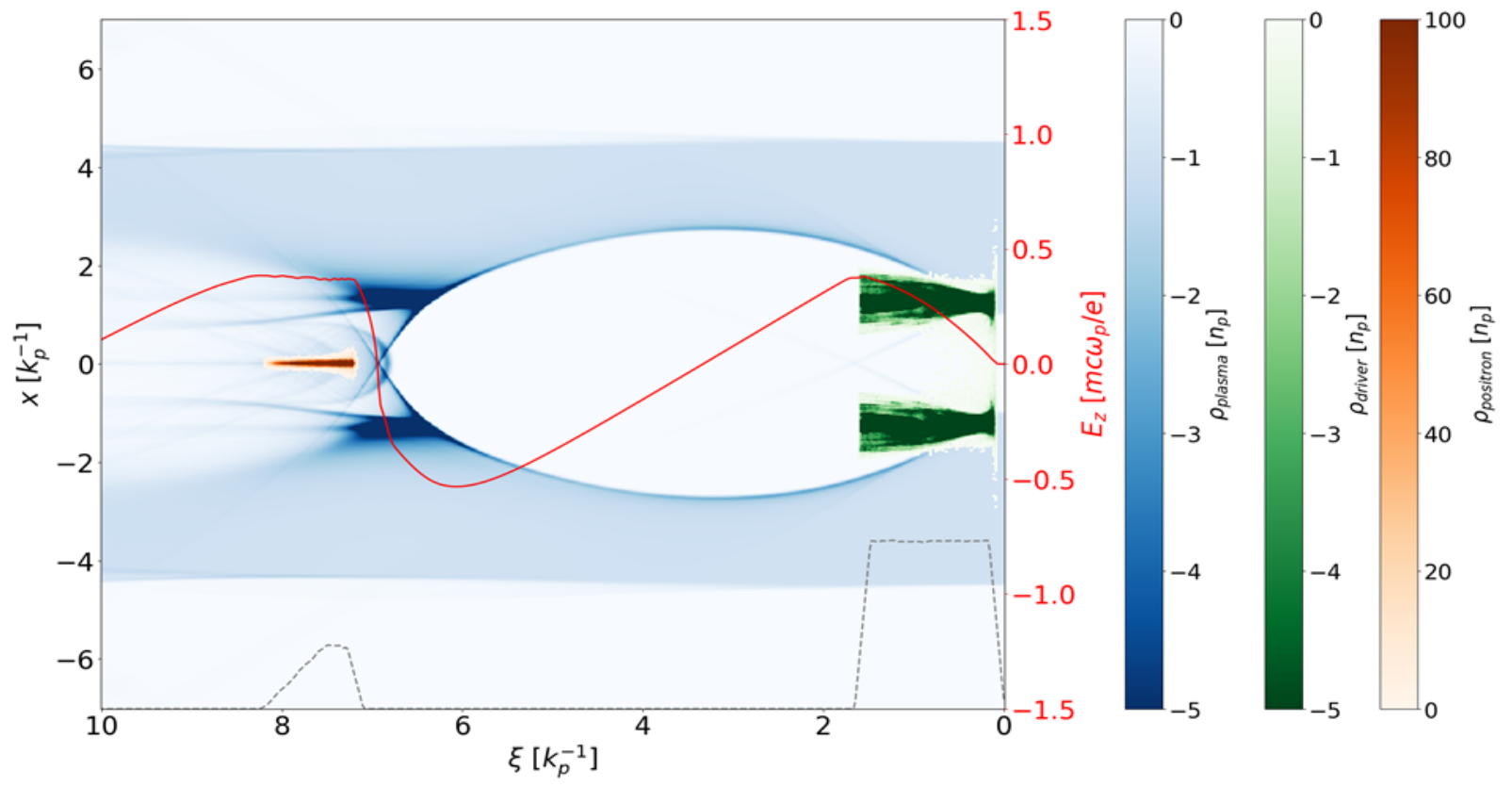

**Figure A6.4.2:** Illustration of positron beam acceleration in hollow plasma channel. Comprehensive parameters for the hollow plasma channel and the beams are provided in Table A6.4.1. Following a 20.8-meter acceleration distance, the witness positron beam reaches an energy of 30.1 GeV, corresponding to an average accelerating gradient of 1.3 GV/m. The beam loading efficiency, transitioning from plasma wake to the positron beam, is measured at 22.6%.

**Table A6.4.1:** Parameters of the hollow plasma channel and the beams

| Plasma parameters | | |
|---|---|---|
| Density ($cm^{-3}$) | 1.133e15 | |
| Inner radius (μm) | 158 | |
| Outer radius (μm) | 711 | |
| Beam parameters | Driver | Trailer |
| Charge (nC) | 6.45 | 1.1 |
| Energy (GeV) | 30 | 3 |
| Transverse size (μm) | 32 | 6 |
| Normalized emittance $\epsilon_n$(mm·mrad) | 32 | 16 |
| Length (μm) | 237 | 153 |
| Energy spread $\delta_E$(%) | 0 | 0 |
| Beam longitudinal distance (μm) | 885 | |

Table A6.4.2 reveals that, with the exception of the final energy spread, all other beam parameters of the witness positron beam meet the necessary criteria for the plasma injector. The longitudinal phase space of the positron beam is depicted in Figure A6.4.3.

**Table A6.4.2:** Results of positron beam acceleration

| Positron beam parameters | |
|---|---|
| Charge (nC) | 1.1 |
| Energy (GeV) | 30.1 |
| Normalized emittance $\epsilon_n$ (mm·mrad) | 41.6 (x) |
| | 18.7 (y) |
| Energy spread $\boldsymbol{\delta_E}$(%) | 0.68 |
| Acceleration properties | |
| Acceleration length (m) | 20.8 |
| Acceleration gradient (GV/m) | 1.3 |
| Beam loading efficiency (%) | 22.6 |

In actuality, the primary factor contributing to the energy spread of the positron beam is the correlated energy spread, often referred to as the energy chirp. However, the slice energy spread within specific regions adheres to the requirements of the injector. The reduction of energy chirp can be accomplished through the precise adjustment of the current profile and loading phase of the positron beam. Furthermore, we introduce a novel method for compressing the slice energy spread in the subsequent section.

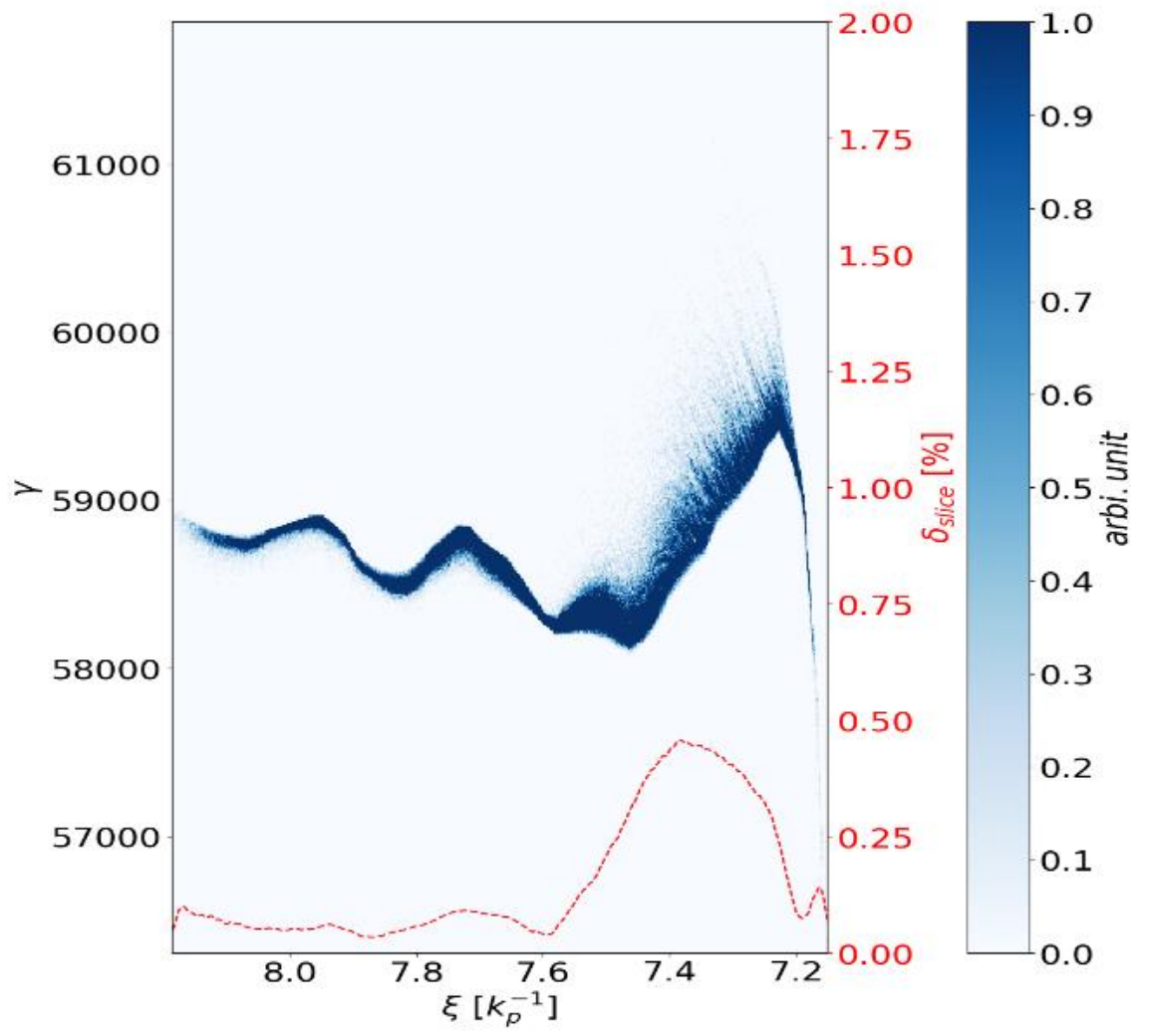

**Figure A6.4.3:** Longitudinal phasespace and distribution of slice energy spread of the positron beam.

### *A6.4.2.2: Design for CPI Positron Energy Compression*

The approach to further minimize the energy spread of the positron beam involves the initial conversion of the slice energy spread into energy chirp using a dispersive section. Subsequently, we employ an appropriate plasma wakefield to eliminate the energy chirp, a process referred to as a plasma dechirper. A schematic representation of this energy compression system is presented in Figure A6.4.4. In this setup, a conventional chicane is utilized to establish a correlation between the longitudinal position and energy of the positron beam. Matching sections have been meticulously designed to modify the transverse dimensions of the positron beam, thereby preserving its emittance within the plasma dechirper.

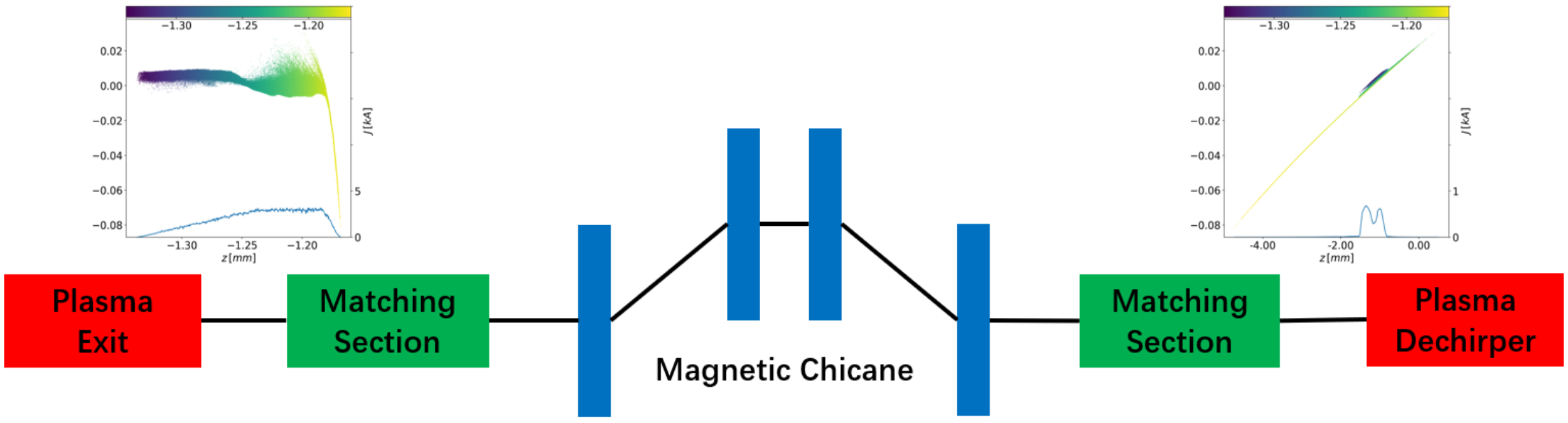

**Figure A6.4.4:** Sketch of the energy decompression system.

Following the matching transport lines and the chicane, the beam parameters for the positron beam are outlined in Table A6.4.3. Notably, the longitudinal dispersion ($R_{56}$) within the chicane measures 5 cm. When compared to the positron beam described in the previous section, it's evident that the bunch length has increased approximately tenfold, and there is a slight elevation in the emittance of the positron beam at the plasma dechirper's entrance.

**Table A6.4.3:** Positron beam parameters at the entrance of plasma dechirper

| Parameter | Symbol | Unit | Value |
|---|---|---|---|
| Beam charge | $Q_{e+}$ | nC | 1.1 |
| Beam energy | $E_{e+}$ | GeV | 30.1 |
| Energy spread | $\sigma_e$ | % | 0.68 |
| Emittance | $\varepsilon_n$ | mm·mrad | 151(x) / 35.1(y) |
| Bunch length | $\sigma_l$ | mm | 0.322(rms) |
| Peak Current | I | kA | 0.647 |

Next, the linear chirp of the positron bunch can be effectively eliminated through the application of an Active Plasma Dechirper (APD), powered by an electron beam. In this context, the plasma features a uniform density profile with a density of 1e13 $cm^{-3}$. The positron beam is loaded just behind the first bubble generated by the intense electron bunch, as illustrated in Figure A6.4.5. Following a propagation distance of 5.05 meters, the final energy spectrum of the positron beam is displayed in the same figure. The primary segment of the elongated positron beam contains a charge of 1.05 nC, with a slight reduction in the mean energy to 30 GeV. Furthermore, the energy spread has been notably reduced to 0.156%, and the emittance of this portion is effectively controlled.

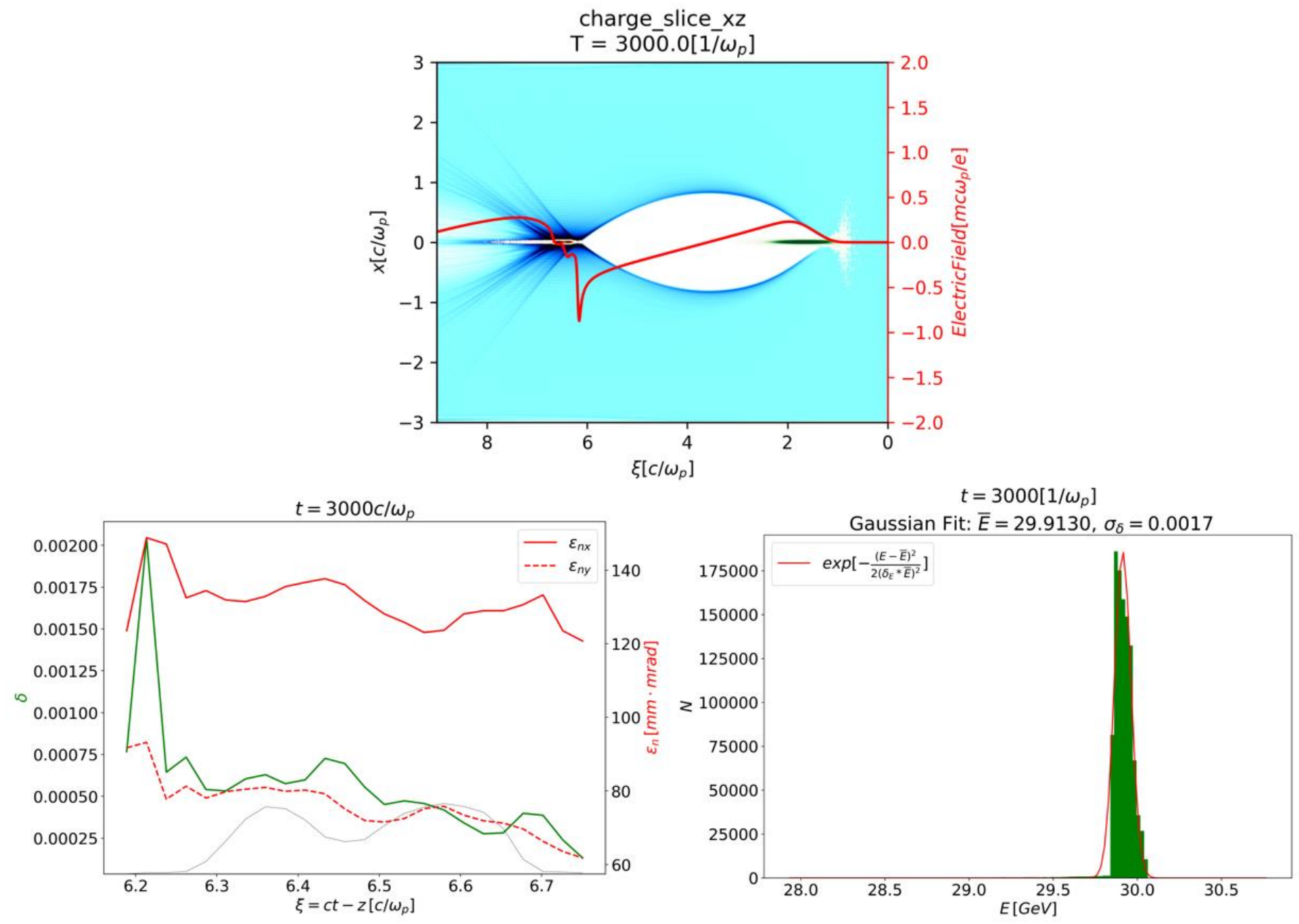

**Figure A6.4.5:** Top: Snapshot of APD. Bottom: left – positron beam parameters; right – energy spectrum after APD.

**Table A6.4.4:** Positron beam parameters after plasma dechirper

| Parameter | Symbol | Unit | Value |
|---|---|---|---|
| Beam Charge | $Q_{e+}$ | nC | 1.05 |
| Beam energy | $E_{e+}$ | GeV | 30.0 |
| Energy spread | $\sigma_e$ | % | 0.156 |
| Emittance | $\varepsilon_n$ | mm·mrad | 131 (x) / 76.2 (y) |

# A6.5: Beamline Design for CPI

## A6.5.1: Electron Beamline for CPI

### *A6.5.1.1: Electron Source and 11 GeV Linac for CPI*

According to plasma acceleration simulations, the necessary beam parameters and longitudinal profile are outlined in Table A6.5.1.1 and illustrated in Figure A6.5.1.1. The Linac comprises two injectors, a bunch pair merge system, a main Linac, and a final focusing system (FFS). The injectors supply the driver and trailer beams with the required longitudinal profile. The bunch pair merge system is designed as an isochronous merge system, capable of merging two bunches with different energy levels using a common dipole. To account for BNS damping, the energy of the trailer bunch is lower, while the driver bunch has higher energy. We have designed beam energies of 100 MeV and 200 MeV to accommodate these requirements.

The main Linac must accommodate both the driver and trailer beams with varying energies. Due to the driver's higher velocity compared to the trailer, the bunch separation increases as it propagates through a drift with negative longitudinal dispersion $R_{56}$≈-$L/\gamma^2$, where $L$ represents the length of the drift and $\gamma$ is the Lorentz factor. Consequently, at the entrance of the main Linac, the bunch distance and length should be smaller than the PWFA's required values.

The input beam distribution at the entrance of the main Linac is depicted in Fig. A6.5.1.2, and a doublet focusing structure is employed within the main Linac. To account for the wakefield effect, an S-band accelerating structure is chosen. The main Linac comprises two sections. In the first section, there are 12 periods, each consisting of a doublet and two accelerating structures. One klystron drives two accelerating structures, providing an accelerating gradient of 27 MV/m. In the second section, there are 20 periods, each featuring a doublet and eight accelerating structures, with three periods serving as acceleration backups. Here, one klystron drives four accelerating structures, and the accelerating gradient is 22 MV/m. It's evident that the two bunches can be matched.

Simulation results of the main Linac are presented in Figure A6.5.1.3 and Table A6.5.1.2. The emittance distribution of the main Linac is displayed in Figure A6.5.1.4, demonstrating that the simulation results align with the PWFA requirements. The distribution at the exit of the main Linac is shown in Figure A6.5.1.5, indicating that the bunch length, distance, and longitudinal profile largely fulfill the necessary criteria. Beam size control will be managed in the FFS. Given the substantial energy spread and the requirement for a small beam size, the FFS has been meticulously designed to address these considerations.

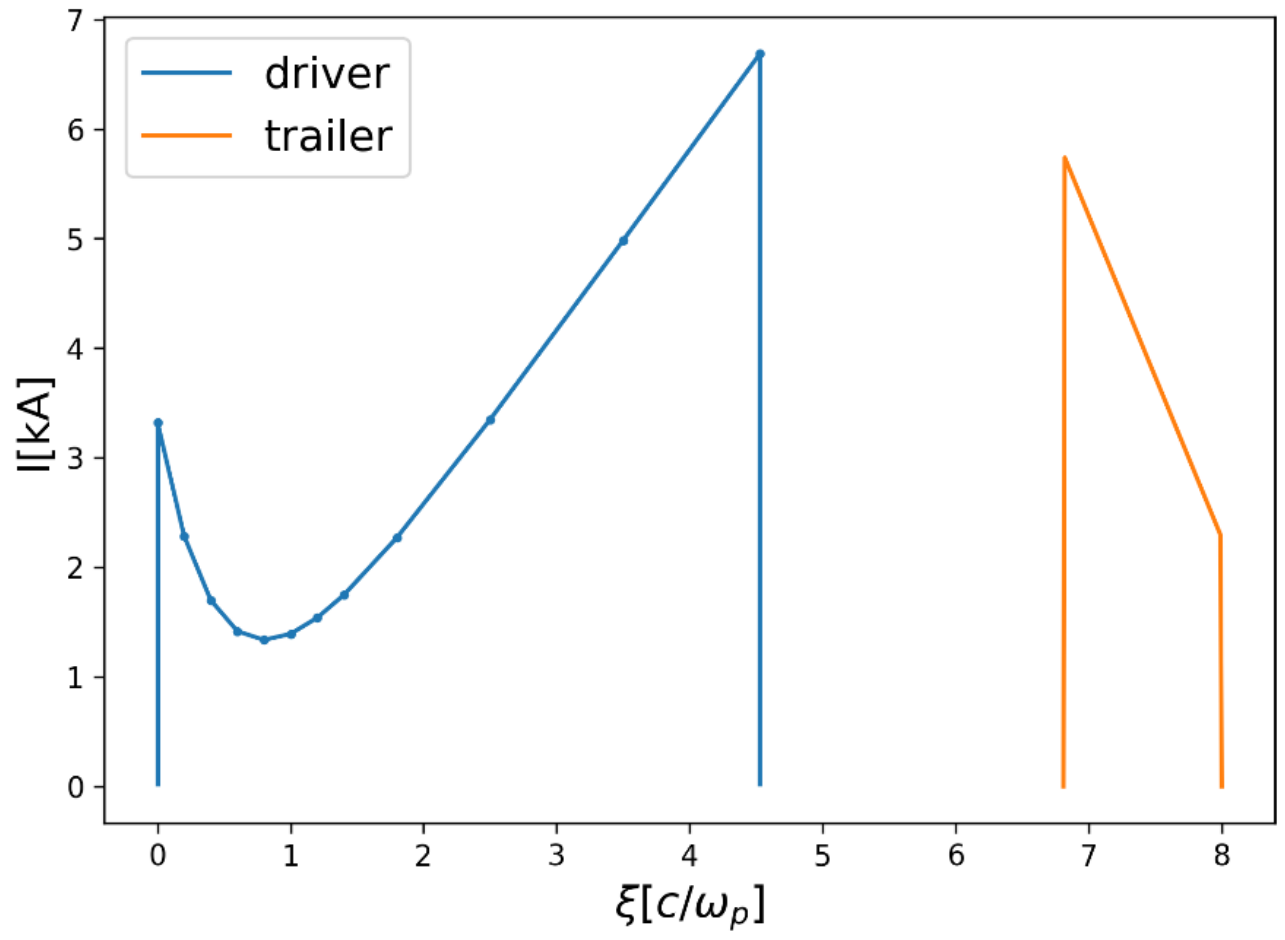

**Figure A6.5.1:** Longitudinal profile of the PWFA beam, $c/\omega_p = 74.903$ μm.

**Table A6.5.1:** e1 and e3 beam parameters at the end of L3 (e1, e3 and L3 are shown in Figure A6.2.1)

| **Parameter** | **Driver** | **Trailer** |
|---|---|---|
| Energy *E* (GeV) | 11 | 11 |
| Normalized emittance $\epsilon_n$ (mm-mrad) | 20 | 10 |
| Bunch length (μm) | 340 | 89.2 |
| Beam size (μm) | 3.89 | 2.75 |
| Charge (nC) | 3.87 | 1.19 |
| Beam distance (μm) | 171 | |

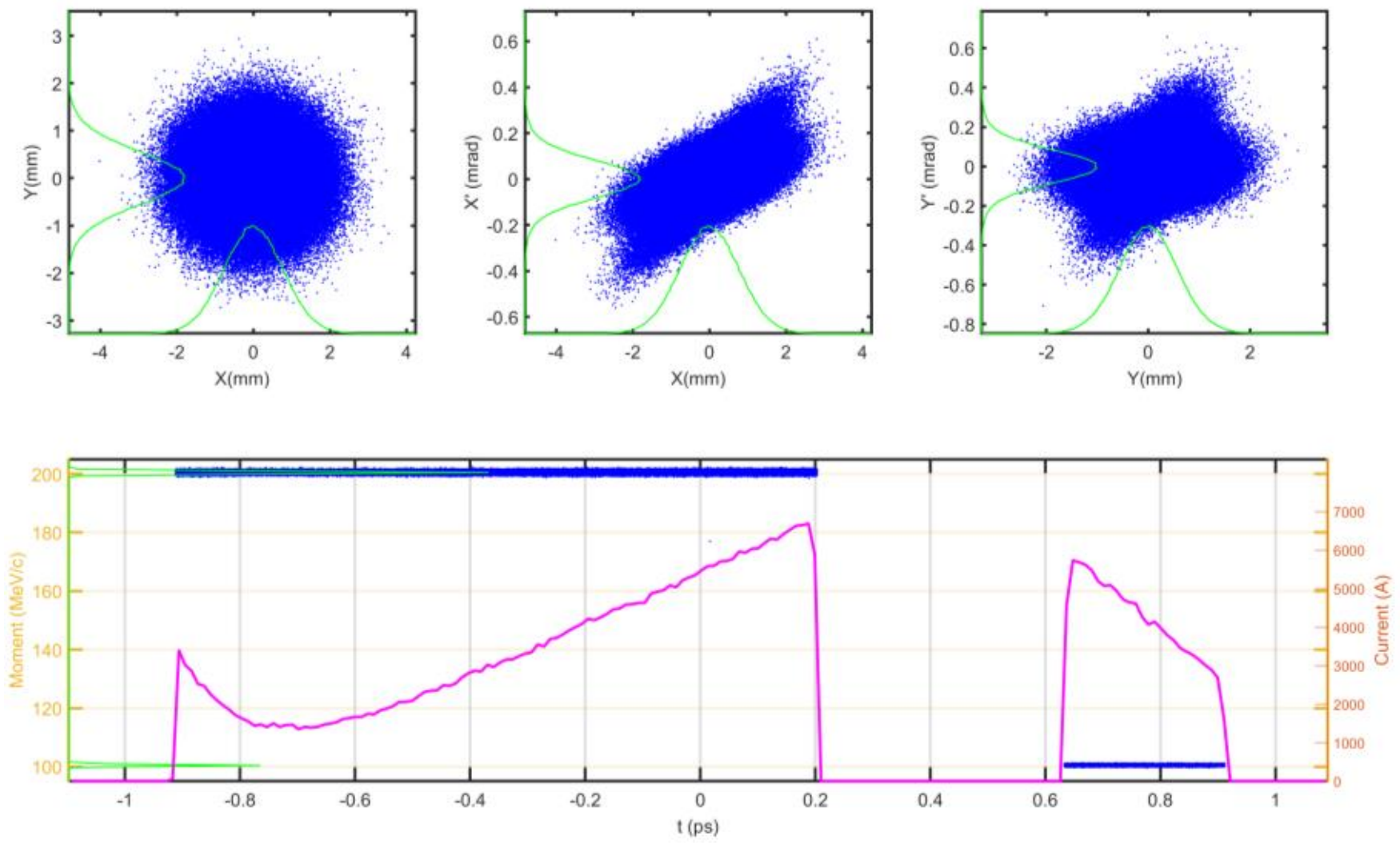

**Figure A6.5.2:** Beam distribution at the entrance of the main Linac: upper plot – transverse, lower plot – longitudinal.

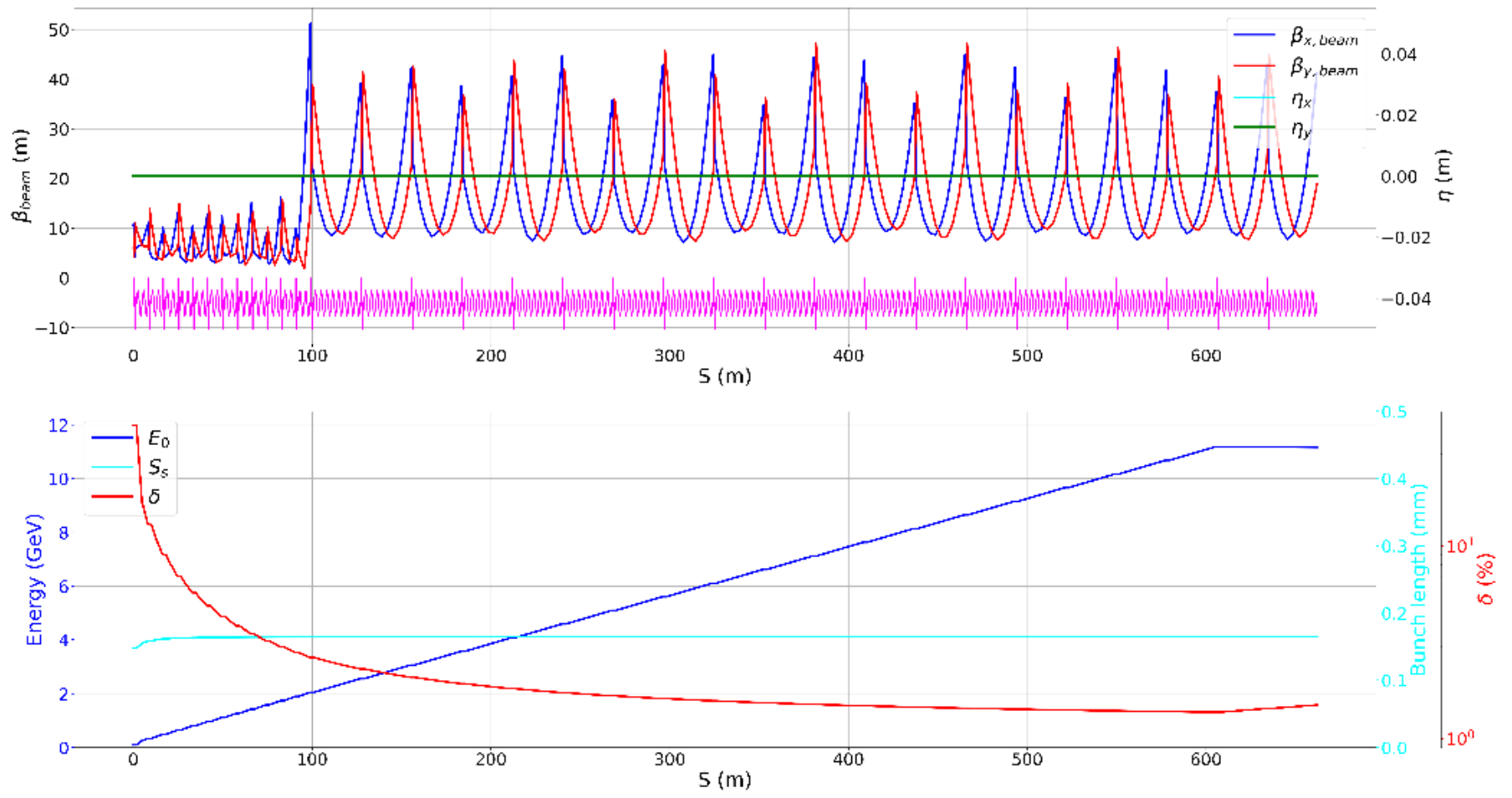

**Figure A6.5.3:** Simulation Results of Beam Dynamics in the Main Linac. The upper plot displays the beam optics function along the primary linac, while the lower plot illustrates the alterations in energy (dark blue), energy spread (red), and bunch length (light blue) along the main linac.

**Table A6.5.2:** Simulation results of the main Linac

| Parameter | Driver | Trailer | Total |
|---|---|---|---|
| Energy $E$ (GeV) | 11.23 | 10.94 | 11.16 |
| Normalized emittance $\epsilon_n$ (mm-mrad) (H/V) | 20.6/20.2 | 10.6/10.2 | 18.8/18.0 |
| Bunch length (μm) | 339.9 | 88.9 | 599.2 |
| Beam size (μm) (H/V) | 192/132 | 178/97 | 189/124 |
| Charge (nC) | 3.87 | 1.19 | 5.06 |
| Energy spread | 1.14% | 0.34% | 1.5% |
| Beam distance (μm) | 170.4 | | / |

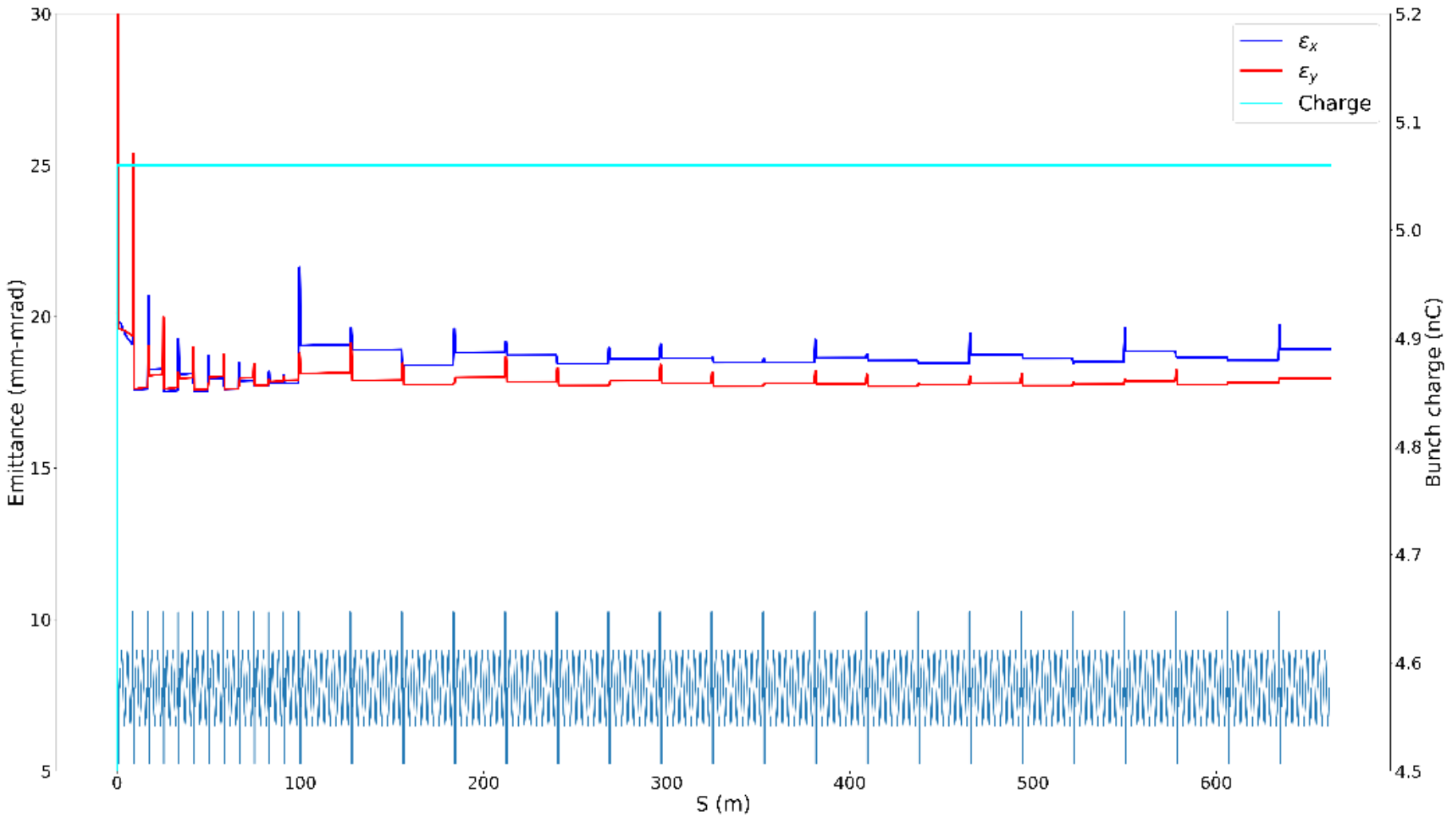

**Figure A6.5.3:** Emittance distribution along the main Linac

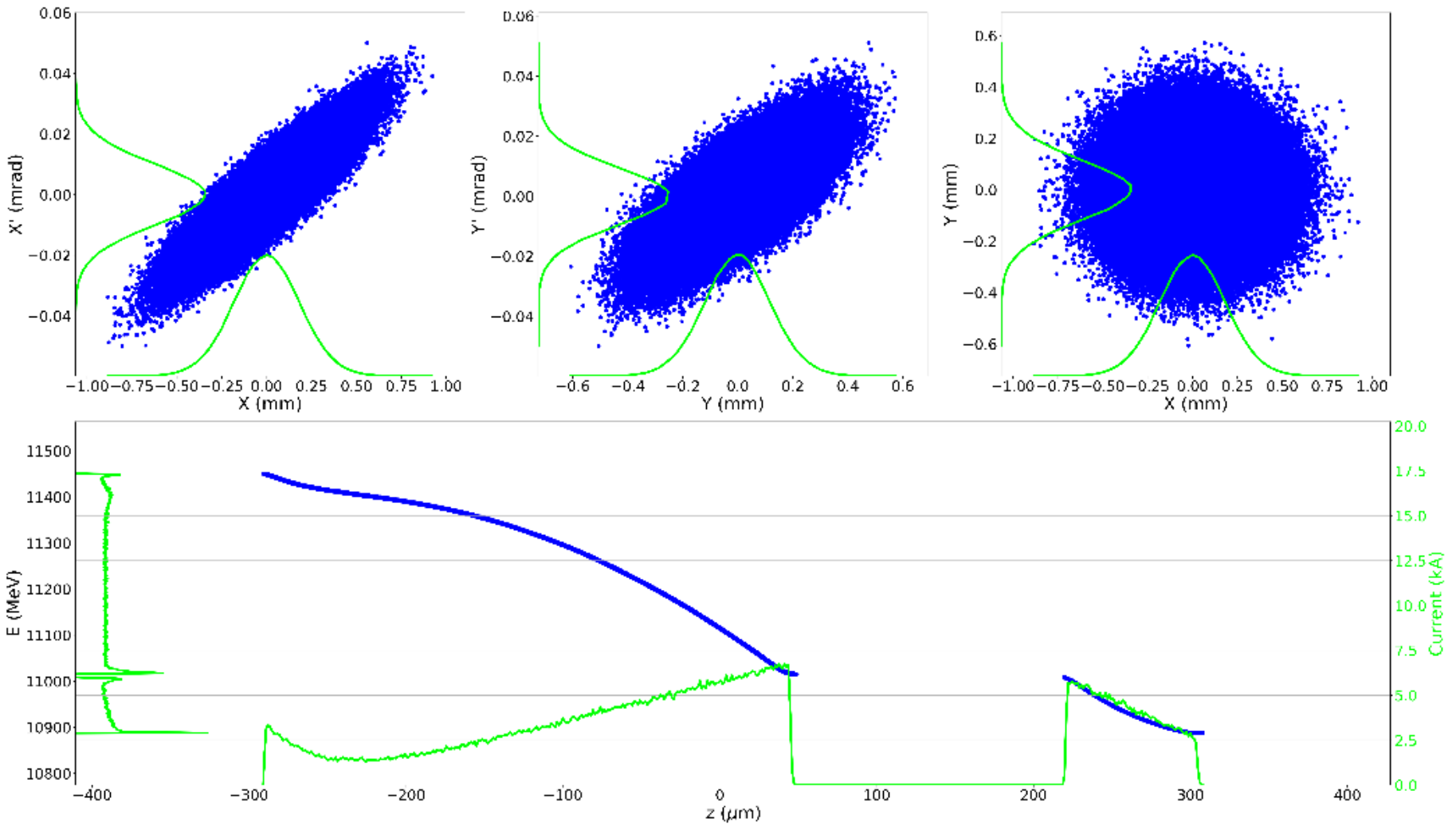

**Figure A6.5.4:** Beam distribution at the exit of the main Linac: upper plot – transverse, lower plot – longitudinal.

### A6.5.1.2: Final Focusing System for Electron Beamline

To achieve a transverse beam size in the micrometer range at the entrance, the PWFA necessitates small beta functions at the focal point for both the horizontal and vertical planes ($\beta^*$ = 1.63 cm). This requirement can be met by implementing a final focusing system prior to the PWFA. The design of the final focusing system for electrons adheres to the following specifications:

- Provide the desired small beam sizes at the focal point.
- Allocate sufficient longitudinal space from the final quadrupole to the focal point to accommodate the insertion of PWFA devices.
- Compensate for the chromaticity generated by the final triplet to achieve a large energy acceptance. Without chromaticity correction, the transverse beam size at the end of the final focusing cannot attain the desired goal [4].
- Ensure good performance for both the driver and trailer beams.
- Maintain a predominantly straight beamline.

As a result of the triangular density z-distribution inherent to PWFA and the acceleration process in the Linac, the beam at the entrance of the final focusing exhibits a significant beam energy spread and a nearly uniform distribution, as illustrated in Fig. A6.5.1.1. This necessitates an energy acceptance range as wide as ±2.5%, and the performance at large energy deviations should be sufficiently accurate due to the quasi-uniform energy distribution.

The final focusing system preceding the PWFA, which bears a striking resemblance to FFTB [5] but with the addition of a final triplet to accommodate a round beam profile, is comprised of modular sections including the final transformer (FT), chromaticity correction for the vertical plane (CCY), chromaticity correction for the horizontal plane (CCX), and matching transformer (MT) [6−8]. To enhance its performance, the chromaticity correction sections are kept separate from the bunch compressor, diverging from the approach of FACET-II [9]. Reverse bending is implemented in the CCX and CCY sections to ensure a primarily straight beamline. The lattice design for the final focusing system for electrons is depicted in Fig. A6.5.1.6. Key parameters for the final focusing system for electrons are outlined in Table A6.5.1.3. Achieving small transverse beam sizes necessitates positioning the final-triplet quadrupoles as close to the focal point as possible to minimize chromaticity and maintain minimal beta functions at the final-triplet quadrupoles. The distance from the last quadrupole to the focal point, denoted as L*, is restricted to 3 meters, primarily to accommodate the insertion of the PWFA devices.

**Table A6.5.3**: Key parameters the final focusing system for electron.

| Parameter | Driver | Trailer |
|---|---|---|
| Energy $E$ [GeV] | 11.23 | 10.94 |
| Normalized emittance $\epsilon_n$ [mm-mrad] (H/V) | 20.6/20.2 | 10.6/10.2 |
| Target beam size [μm] | 3.89 | 2.75 |
| Energy spread [%] | 1.14 | 0.34 |
| Beta functions at the focal point β* [cm] | 1.63 | |
| Distance from last quadrupole to the focal point L* [m] | 3 | |

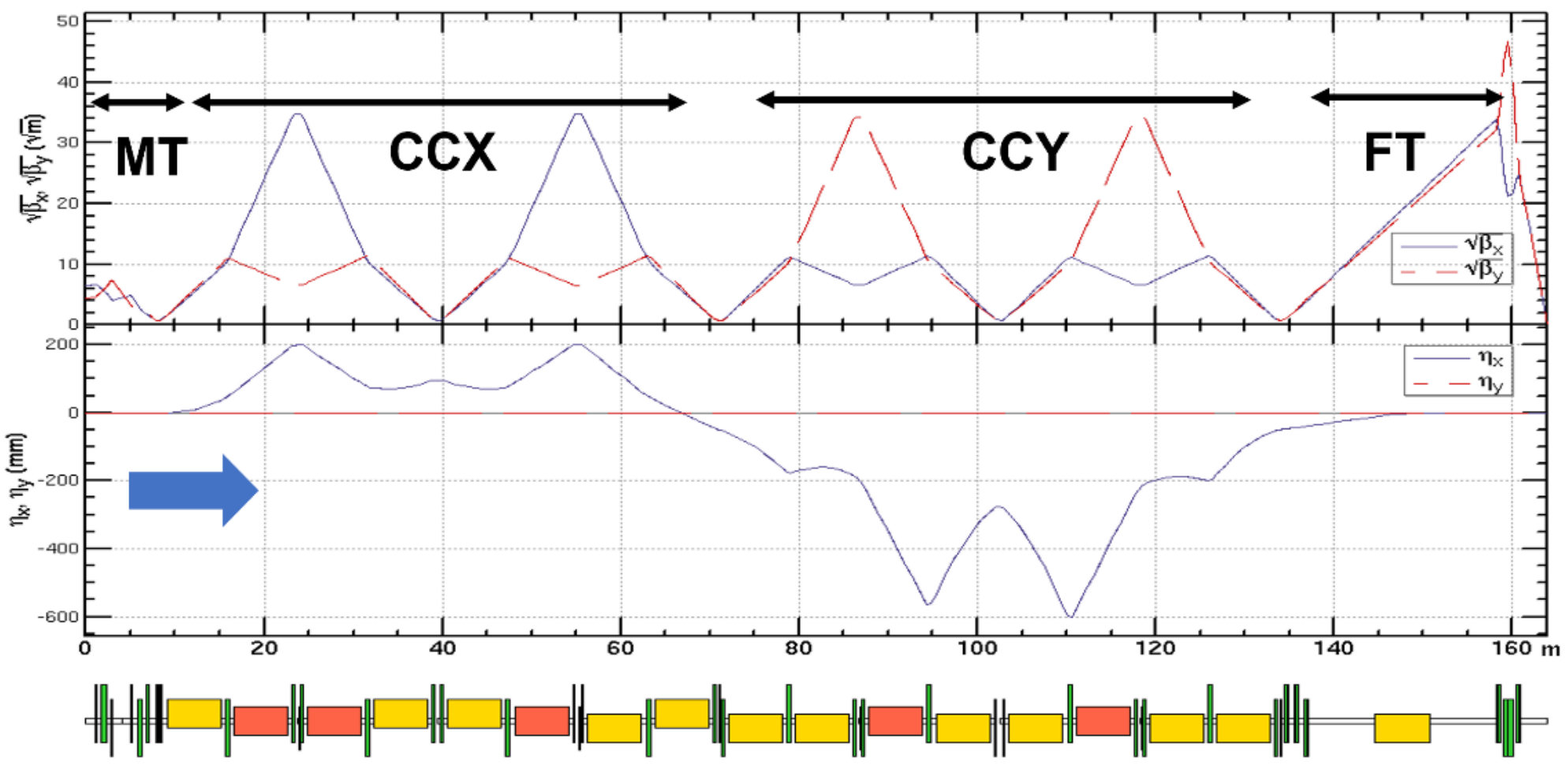

**Figure A6.5.6:** Lattice functions for the final focusing system in the electron beamline. The focal point is positioned at the right end. The final focusing system encompasses a matching transformer (MT), chromaticity correction for the horizontal plane (CCX), chromaticity correction for the vertical plane (CCY), and a final transformer (FT).

Two pairs of sextupoles are strategically positioned at the four beta peaks to counteract the first-order vertical and horizontal chromaticity induced by the final-triplet quadrupoles. The geometric aberrations stemming from the thin sextupoles are effectively neutralized through the application of a paired sextupole setup with a –I transformation. Furthermore, any geometric aberrations arising from the finite length of the sextupoles are rectified through the incorporation of additional weak sextupoles [10]. With the combined use of the primary sextupoles, phase tuning, and supplementary sextupoles, the correction of up to third-order chromaticities is achieved [7, 11, 12].

The chromatic aberrations at the focal point, following the applied corrections, are illustrated in Fig. A6.5.1.7. The relative beta and alpha distortions resulting from energy deviations exhibit favorable performance within ±2.5%. However, it's important to note that the third-order dispersion becomes substantial when the energy deviation exceeds ±2%. This necessitates a more effective correction for high-order dispersion, along with related aberration correction, and a reduction in the beam energy spread originating from the Linac. To minimize the transverse beam size, additional sextupoles are strategically employed to correct high-order dispersion while simultaneously controlling the growth of other aberrations.

To decrease the deviation of the trailer beam's energy relative to the lattice's design energy, adjustments are made by aligning the mean energy of the trailer beam with that of the driver beam. The beam sizes were determined through tracking using ELEGANT, accounting for synchrotron radiation effects. Figs. A6.5.1.8 and A6.5.1.9 depict the transverse beam distributions of the trailer and driver beams, respectively. The target beam sizes for both beams are achieved, except for the horizontal size of the driver beam due to significant dispersion and beam energy spread.

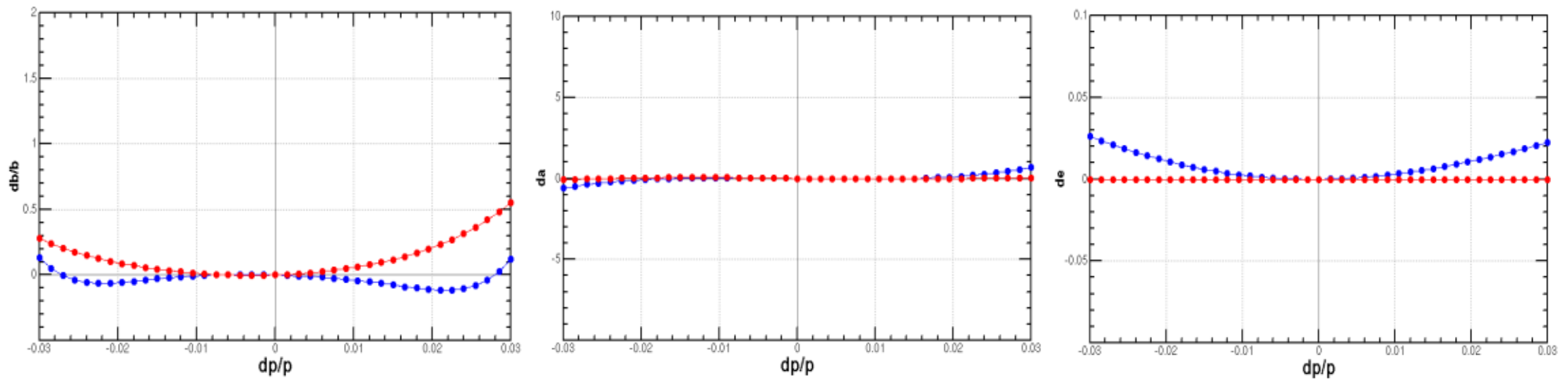

**Figure A6.5.7**: Chromatic aberrations at the focal point: (a) relative beta (Δβ /β), (b) relative alpha (Δα/α), and (c) relative dispersion distortions due to energy deviation.

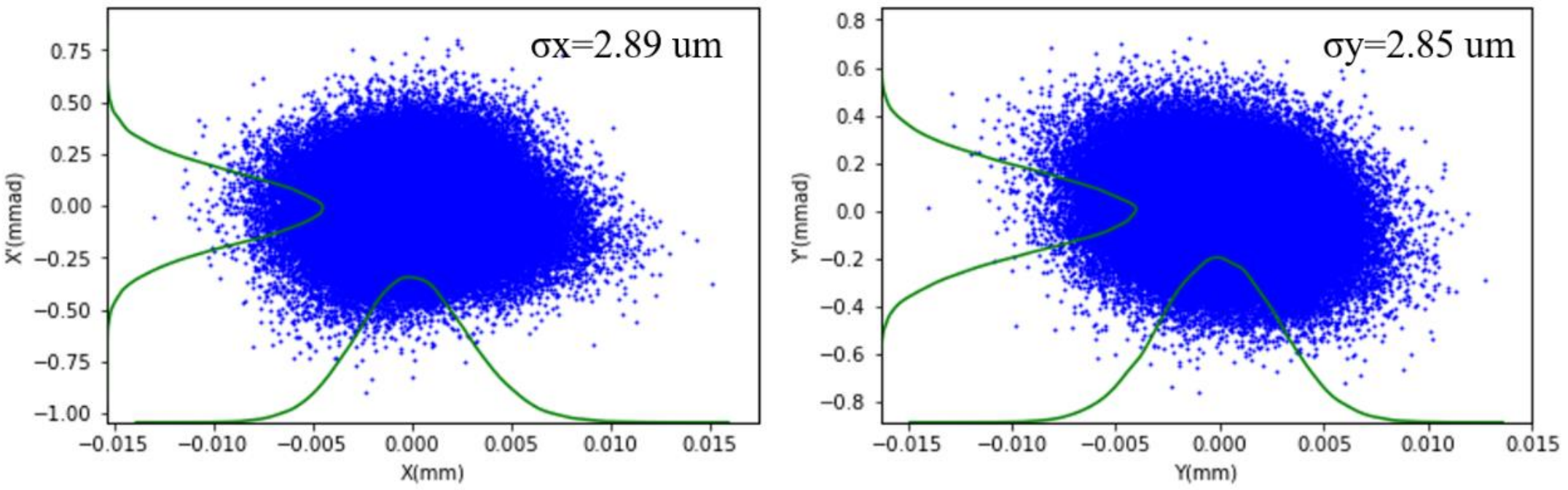

**Figure A6.5.8**: Transverse distribution of the trailer beam at the focal point.

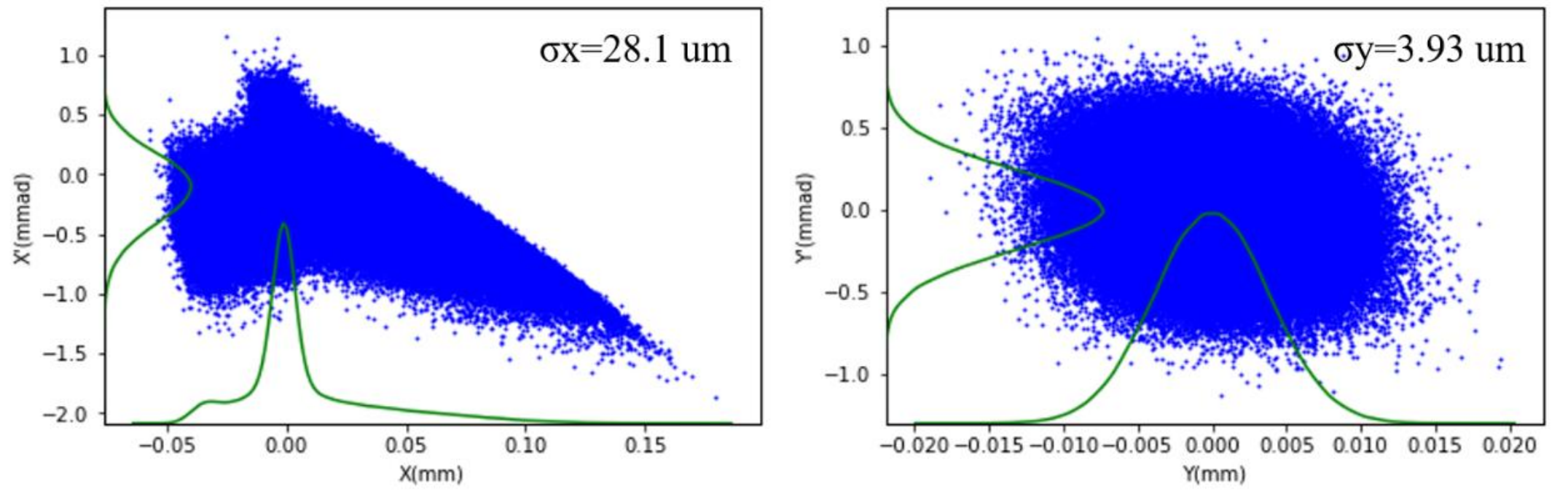

**Figure A6.5.9**: Transverse distribution of the drive beam at the focal point.

## A6.5.2: Positron Beamline for CPI

### *A6.5.2.1: Positron Damping Ring*

#### *A6.5.2.1.1 Damping Ring Parameters*

The positron damping ring (DR), as shown in A6.5.10, operates at an energy of 400 MeV with a circumference of 20.5 meters, featuring a racetrack shape.

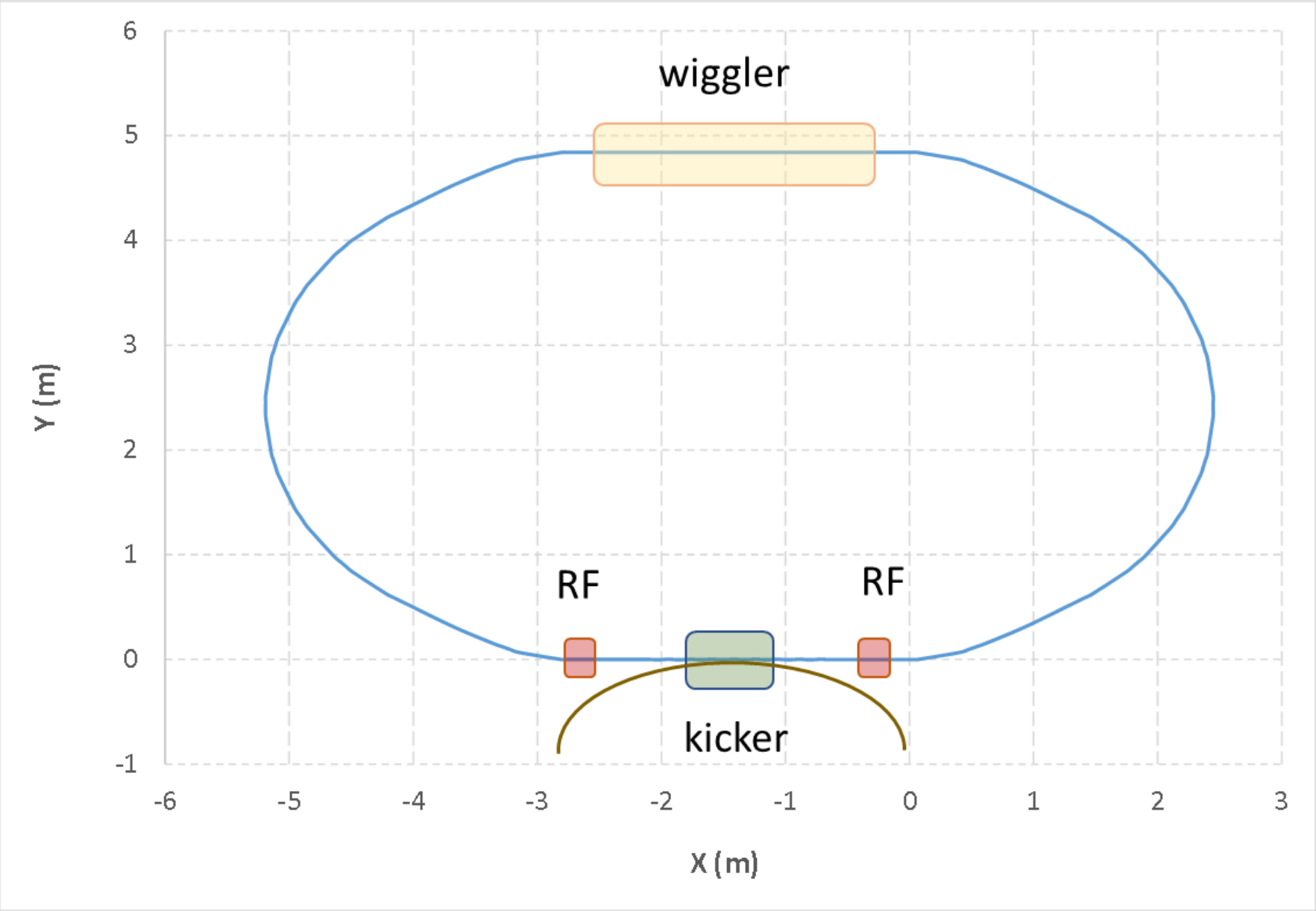

**Figure A6.5.10**: Layout of the damping ring.

The arcs are constructed with a 60-degree FODO cell and combined quadrupole/sextupole magnets. The injected emittance (normalized) for the DR is set at 2500 mm·mrad, and the injected energy spread is maintained below 1.0%. Depending on the 100 Hz repetition rate of the Linac and the multi-bunch storage scheme, the positron beam will be stored in the DR for either 20 ms or 30 ms. The extracted emittance stands at 68 mm·mrad in the horizontal plane and 57 mm·mrad in the vertical plane, with potential for further reduction through longer bunch storage time. Superconducting wigglers are employed to simultaneously decrease the emittance and damping time. To ensure efficient injection, the transverse acceptance of the DR should exceed five times the injection beam size. Key parameters for the damping ring are outlined in Table A6.5.4, while the parameters for the wiggler are summarized in Table A6.5.5:

**Table A6.5.4:** Main Parameters of the positron damping ring for PWFA injector.

| | **DR V2.1** |
|---|---|
| Energy (MeV) | 400 |
| Circumference (m) | 20.5 |
| Bunch number | 2 (3) |
| $B_0$ (T) | 0.97 |
| $U_0$ (keV/turn) | 5.0 |
| Damping time x/y/z (ms) | 10.9/10.9/5.4 |
| $\delta_0$ (%) | 0.054 |
| $\varepsilon_0$ (mm-mrad) | 11 |
| Natural $\sigma_z$ (mm) | 4.3 |
| Extract $\sigma_z$ (mm) | 4.4 |
| $\varepsilon_{inj}$ (mm-mrad) | 2500 |
| $\varepsilon_{ext}$ x/y (mm-mrad) | 68/57 (14/9) |
| $\delta_{inj}$ /$\delta_{ext}$ (%) | 0.6 /0.054 |
| Storage time (ms) | 20 (30) |
| RF frequency (MHz) | 500 |
| RF voltage (MV) | 1.5 |
| Cavity number (single cell) | 2 |
| Energy acceptance by RF (%) | 2.3 |
| Harmonic number | 34 |
| Cavity length (m) | 0.5 |

**Table A6.5.5:** Wiggler Parameters of the positron damping ring.

| **Parameters** | |
|---|---|
| Dipole strength (T) | 4.61 |
| Magnetic period (m) | 0.176 |
| Total length (m) | 1.42 |
| Average $\beta x$ (m) | 1.3 |

*A6.5.2.1.2 Damping Ring Optics*

We selected a 60°/60° FODO cell and implemented the combined magnet scheme (quadrupole + sextupole) denoted as (Q+S) for the arcs to minimize the cell length effectively. The natural emittance of the ring is determined to be 11 mm·mrad. Twiss parameters for the arc cell of the DR are presented in Fig. A6.5.11. The FODO cell measures 1.1 meters in length, with dipole magnets exerting a field strength of 0.97 Tesla. To enhance synchrotron radiation damping and reduce emittance, superconducting wigglers (SC wigglers) are employed.

One of the straight sections comprises the RF section and injection/extraction section, where septum and kicker devices will be incorporated. Twiss parameters for the RF and injection/extraction section are depicted in Fig. A6.5.12. The design ensures a larger horizontal beta function at the injection point, exceeding 4 meters, to optimize injection efficiency. In the other straight section, where the wigglers are installed, a small horizontal beta function is implemented to control the emittance induced by the wigglers. The working point for the entire ring is set at 3.16 for the horizontal direction and 3.21 for the vertical direction. Twiss parameters for the wiggler section can be found in Fig. A6.5.13, while those for the entire ring are illustrated in Fig. A6.5.14.

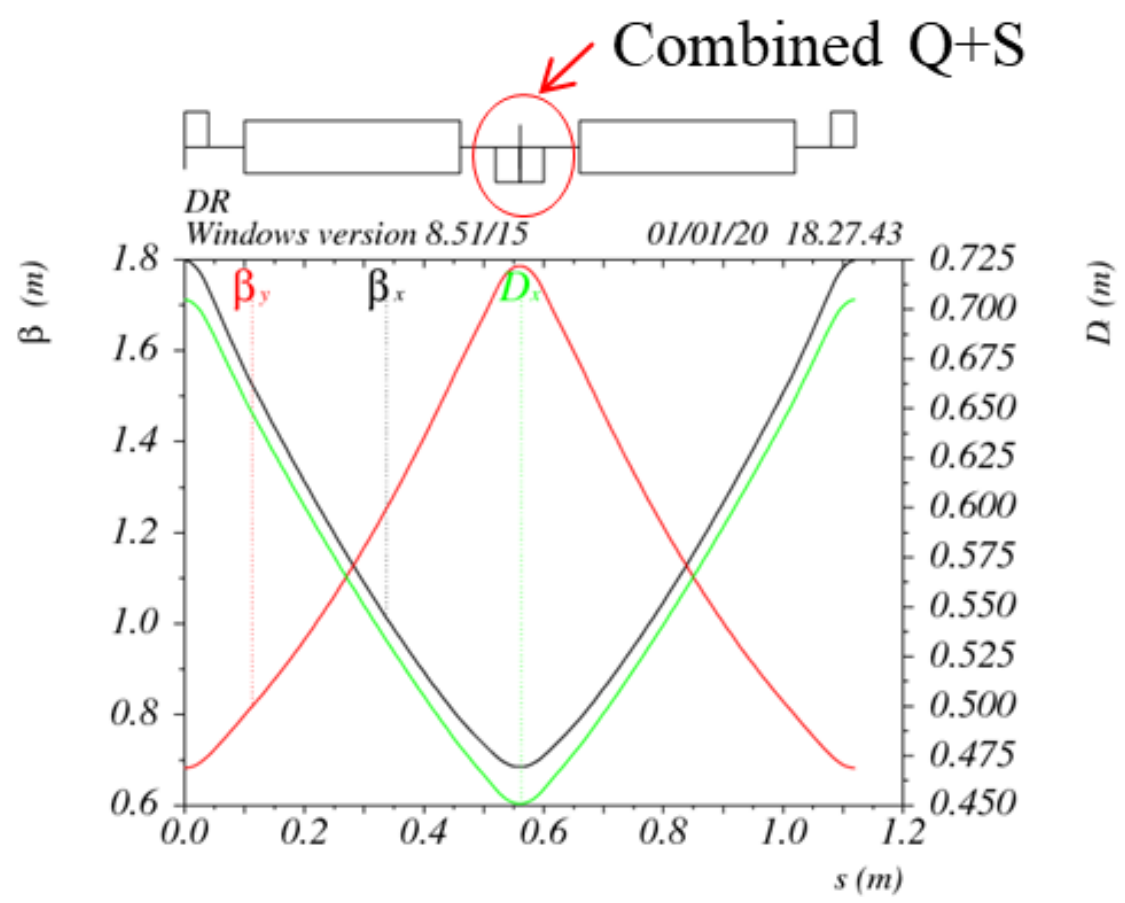

**Figure A6.5.11:** DR twiss parameters for the arc FODO cell.

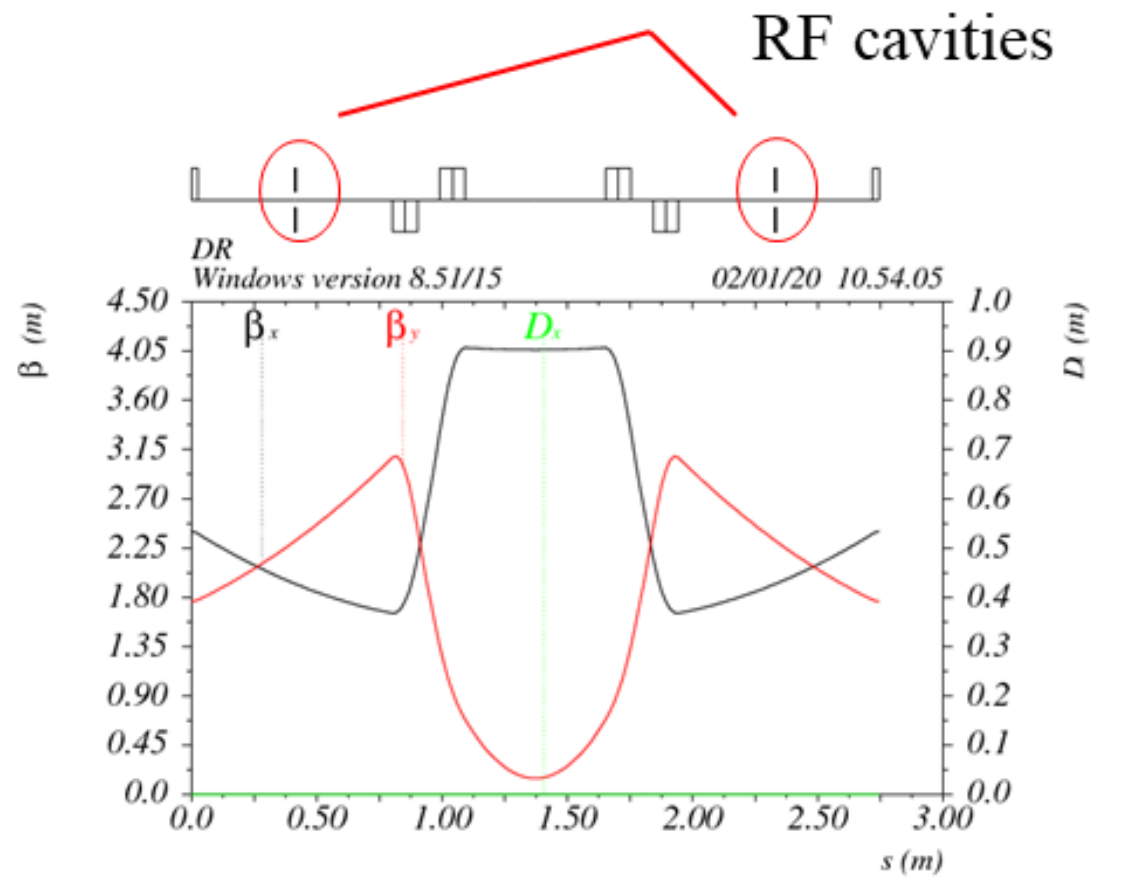

**Figure A6.5.12:** DR twiss parameters for the RF and injection/extraction section.

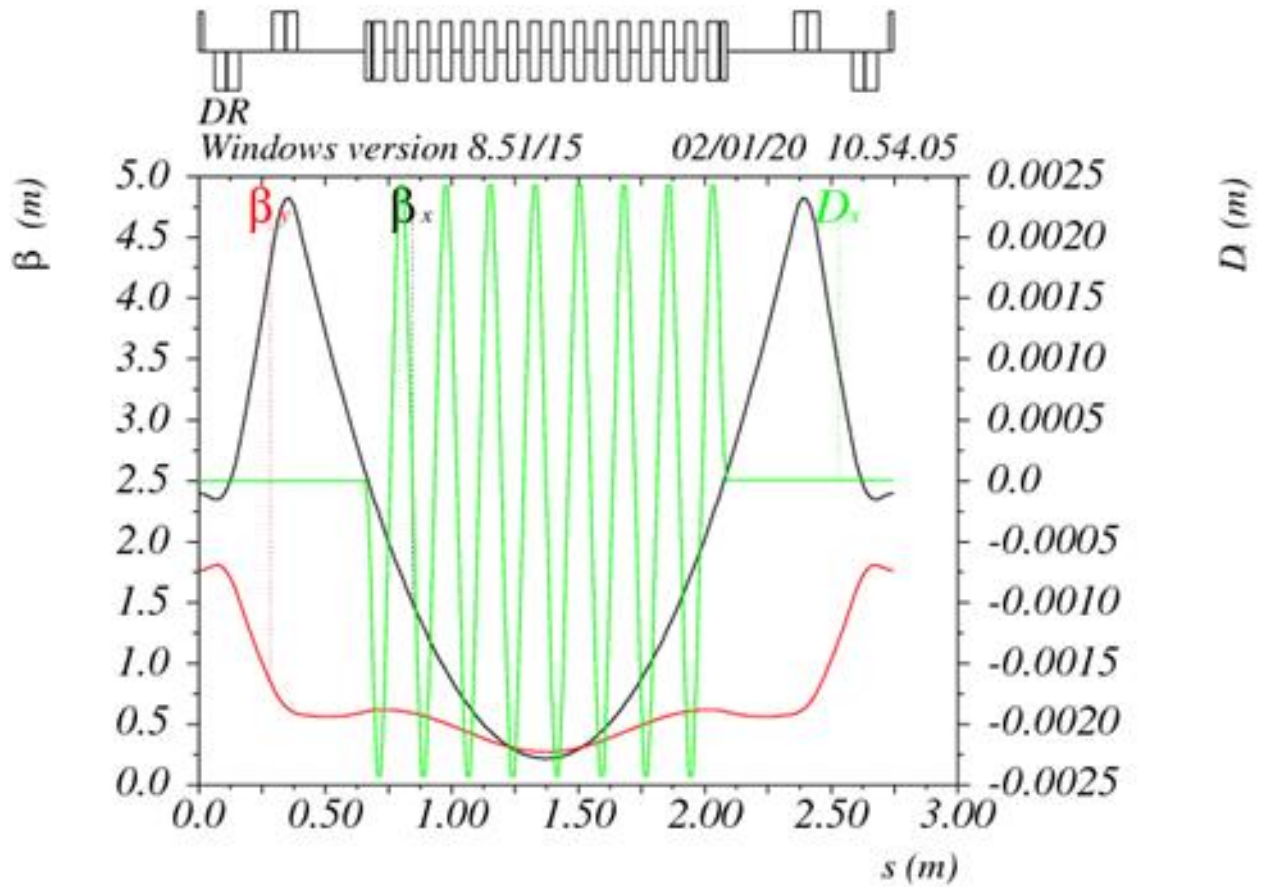

**Figure A6.5.13:** DR twiss parameters for the wiggler section.

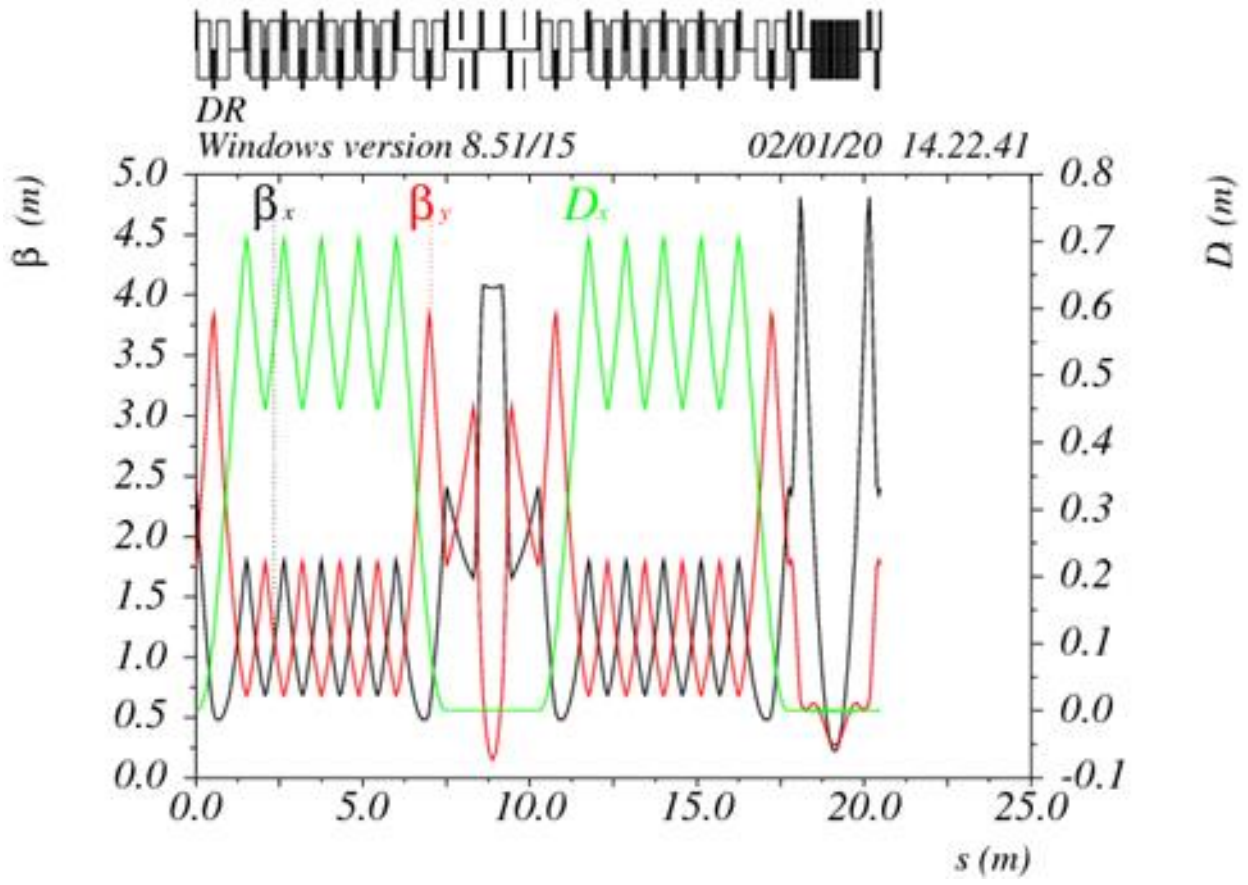

**Figure A6.5.14:** DR twiss parameters for the entire ring.

An interleaved scheme and two sextupole families are implemented for linear chromaticity correction. The dynamic aperture of the damping ring is tracked over 100,000 turns using SAD. The dynamic aperture results for the bare lattice are depicted in Fig. A6.5.15. Based on these results, it is evident that the dynamic aperture of the damping ring satisfies the injection requirements.

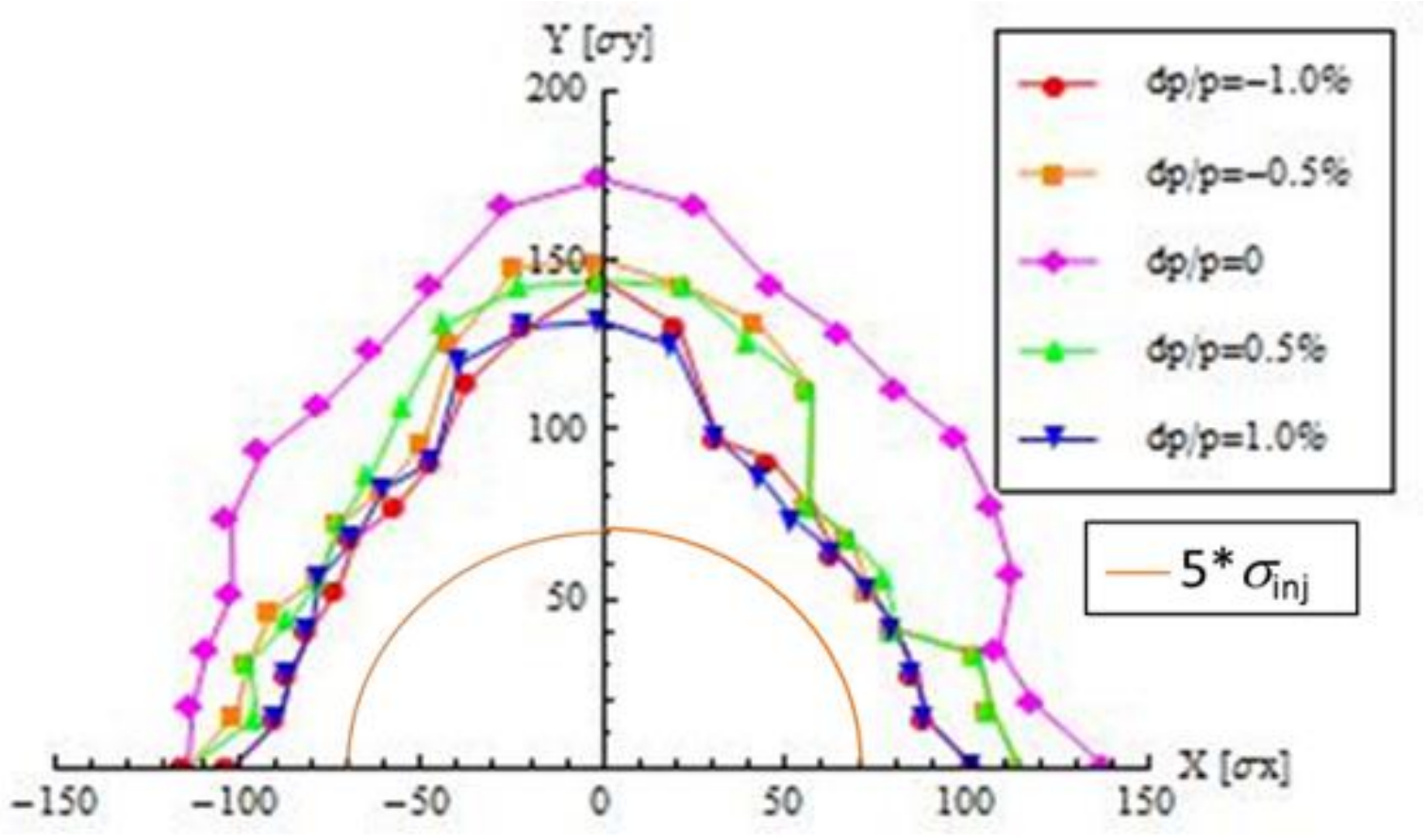

**Figure A6.5.15:** Dynamic aperture of the damping ring for PWFA injector without errors.

### *A6.5.2.2: Bunch Compression System*

#### *A6.5.2.2.1 Parameters Design for EC/BC*

The manipulation of the longitudinal phase space is achieved through two transport lines between the Linac and the damping ring. Prior to entering the damping ring, the energy spread of the positron bunch is compressed to match the RF acceptance of the damping ring. After exiting the damping ring, a reduction in longitudinal bunch length is necessary to facilitate the transition to the plasma accelerator.

The injection energy spread for the energy compressor is 3%, and the injection bunch length is 1.5 mm. Following the energy compressor, the energy spread needs to be reduced to less than 0.7% to be compatible with the damping ring's RF system. The specific parameters for the energy compressor are detailed in Table A6.5.6.

**Table A6.5.6**: Parameters of the Energy Spread Compressor.

| **Parameter** | **EC** |
|---|---|
| Initial energy (MeV) | 395 |
| $\delta_{inj}$ (%) | 3 |
| Initial $\sigma_z$ (mm) | 1.5 |
| $f_{RF}$ (MHz) | 2860 |
| RF voltage (MV) | 28.5 |
| $\phi_{RF}$ (degree) | -69 |
| $R_{56}$ (mm) | 244 |
| Final energy (MeV) | 400 |
| $\delta_{ext}$ (%) | 0.5 |
| final $\sigma_z$ (mm) | 7.3 |

After the damping ring, three bunch compressors have been designed for the PWFA injector to achieve an ultra-short bunch length. The injection bunch length for the bunch compression system is 4.4 mm, and the injection energy spread is 0.054%. Subsequently, after passing through the bunch compressors, the beam energy is increased to 2.4 GeV to control the final energy spread, and the bunch length is reduced to 20 μm, enabling injection into the plasma accelerator. The detailed parameters for the 3-stage bunch compressor system are provided in Table A6.5.7.

**Table A6.3.7:** Parameters of the 3-stage Bunch Compressor

| **Parameter** | **BCI** | **BCII** | **BCIII** |
|---|---|---|---|
| Initial energy (MeV) | 400 | 400.1 | 405 |
| $\delta_{inj}$ (%) | 0.054 | 0.367 | 2.17 |
| Initial $\sigma_z$ (mm) | 4.4 | 600 | 100 |
| $f_{RF}$ (GHz) | 2.860 | 5.712 | 5.712 |
| RF voltage (GV) | 0.0056 | 0.12 | 4.18 |
| RF gradient (MV/m) | 20 | 40 | 40 |
| L (m) | 0.28 | 3 | 104 |
| $\phi_{RF}$ (degree) | 89 | 88 | 61.5 |
| $R_{56}$ (mm) | 1200 | 27.6 | 5.5 |
| Final energy (MeV) | 400.1 | 405 | 2400 |
| $\delta_{ext}$ (%) | 0.367 | 2.17 | 1.83 |
| final $\sigma_z$ (μm) | 600 | 100 | 20 |

*A6.5.2.2.2 Optics Design for the Bunch Compressors*

The optics for the 3-stage bunch compressor has been designed and is illustrated in Fig. A6.5.16. Each compressor comprises an RF section and a dispersive section, with the total length of the 3-stage bunch compressor spanning 270 meters, inclusive of four chicanes.

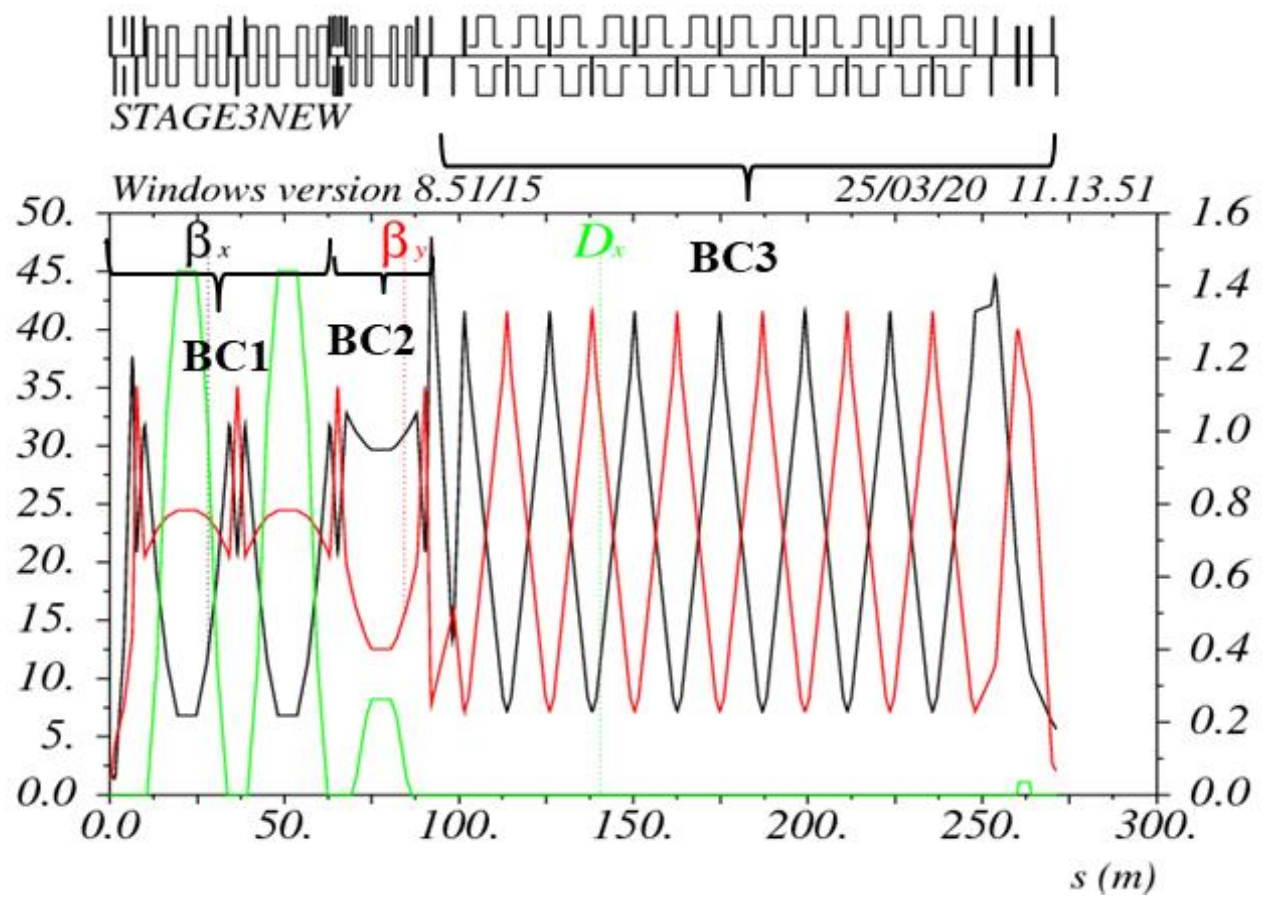

**Figure A6.5.16:** Optics of the 3-stage bunch compressor.

*A6.5.2.2.3 Beam Simulation for the Bunch Compressors*

After configuring the RF voltage and phase accordingly, we conducted particle tracking through the 3-stage bunch compressor using an initial Gaussian distribution. This ensured that the desired bunch length and energy spread could be achieved. The simulation results for the longitudinal beam distribution after each bunch compressor are presented in Fig. A6.5.17 through Fig. A6.5.19.

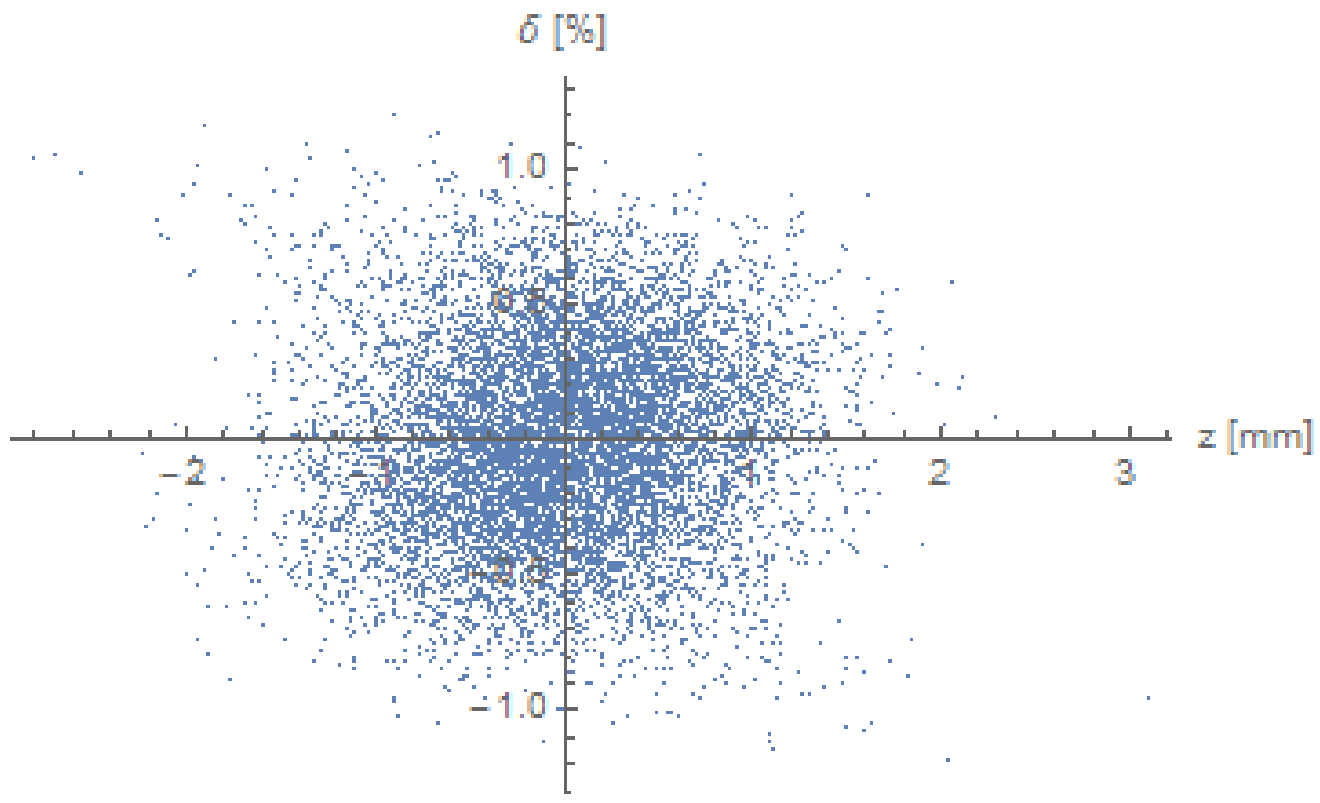

**Figure A6.5.17:** Longitudinal beam distribution after the first bunch compressor.

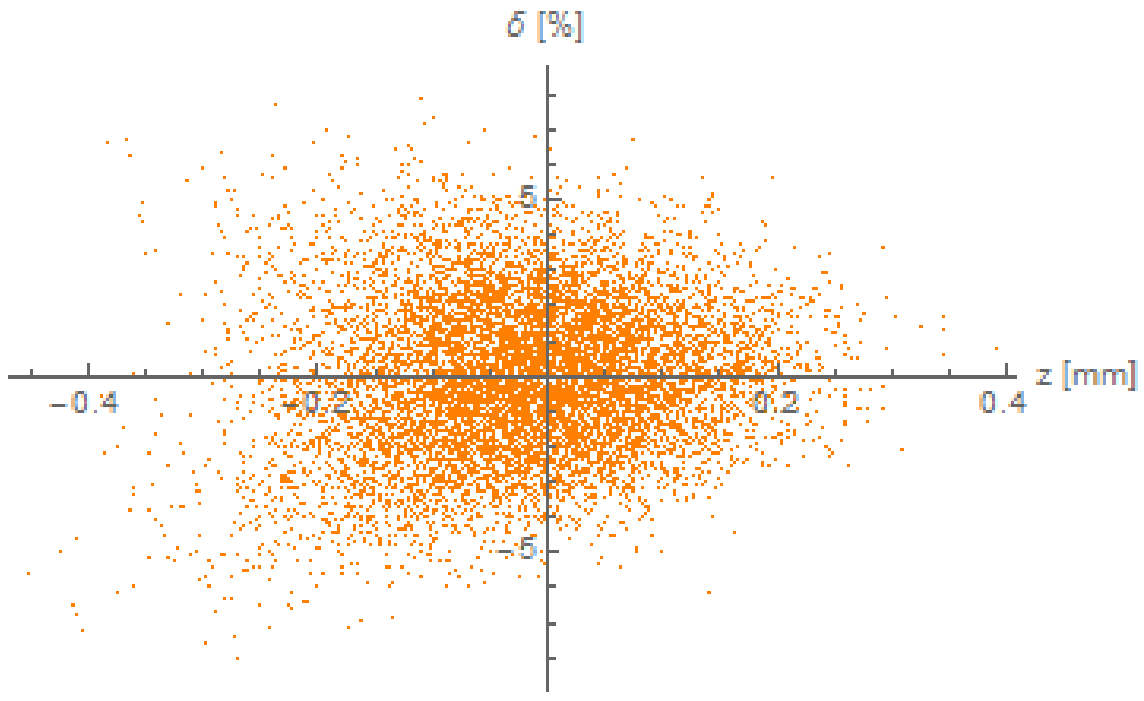

**Figure A6.5.18:** Longitudinal beam distribution after the second bunch compressor.

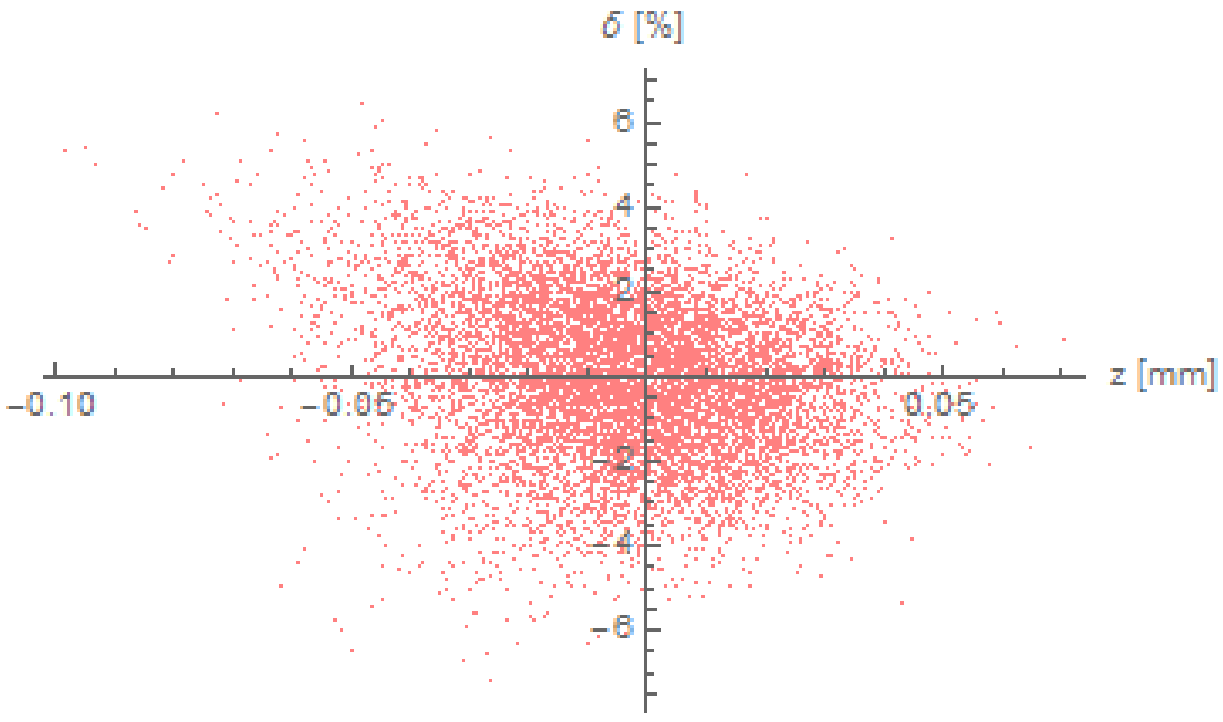

**Figure A6.5.19:** Longitudinal beam distribution after the third bunch compressor.

### *A6.5.2.3: Final Focusing System*

#### *A6.5.2.3.1 Optics Design for the Final Focusing System*

The primary objective of the Final Focusing System (FFS) design is to achieve a transverse positron beam size at the 20 μm level for both horizontal and vertical directions. The optics configuration of the FFS for the positron beam is depicted in Fig. A6.5.20. A triplet structure is employed to focus the transverse beam size, and a local chromaticity correction scheme is implemented to mitigate the sensitivity to energy spread. Without this correction, the transverse beam size at the focal point would exceed 100 μm. Fig. A6.5.21 illustrates the chromaticity function of the final focus system with local correction.

The FFS operates at an energy of 2.4 GeV, and the distance from the final quadrupole to the focal point is set at 3.0 m. Careful control of the $R_{56}$ parameter in the FFS is essential to prevent bunch lengthening while the beam passes through this system. Additionally, the dipole strength in the dispersion section is 88 Gs with a length of 2.0 m, corresponding to a critical energy of synchrotron radiation at 33.6 eV.

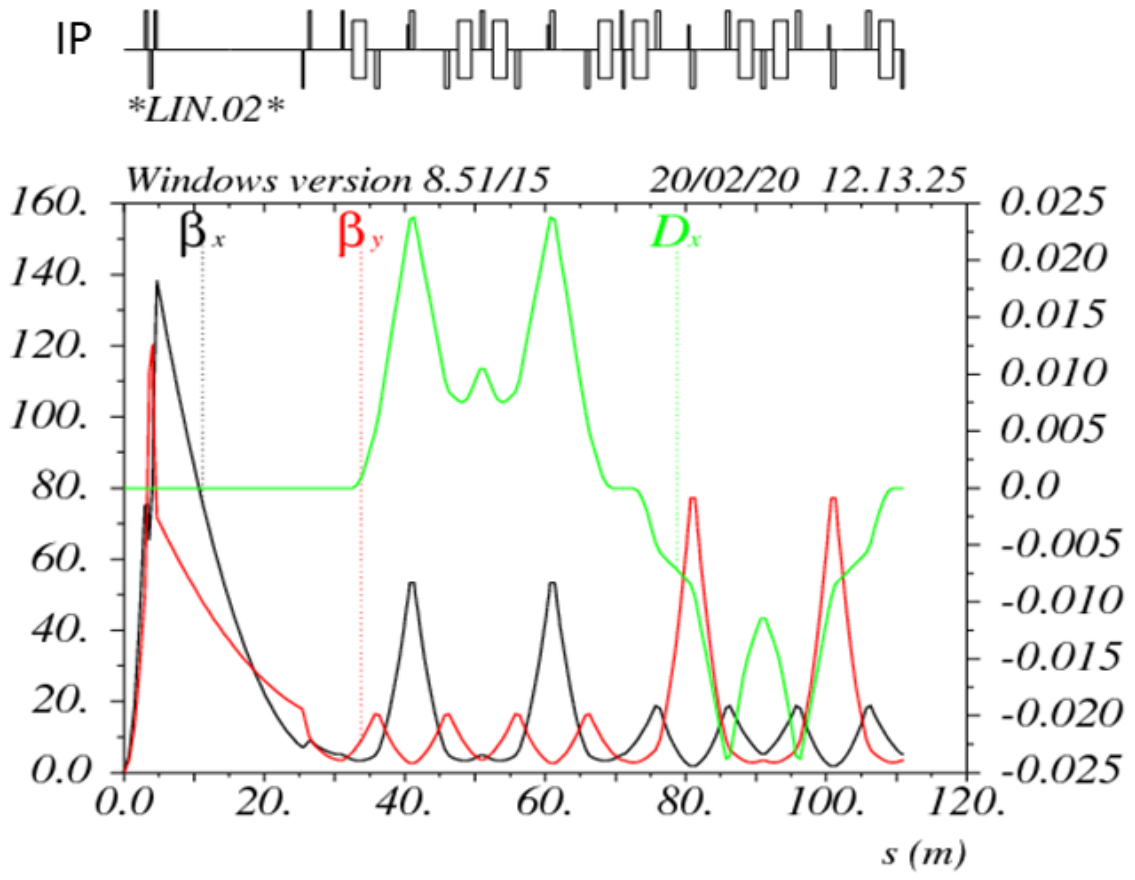

**Figure A6.5.20:** Optics of the final focusing system for the positron beam.

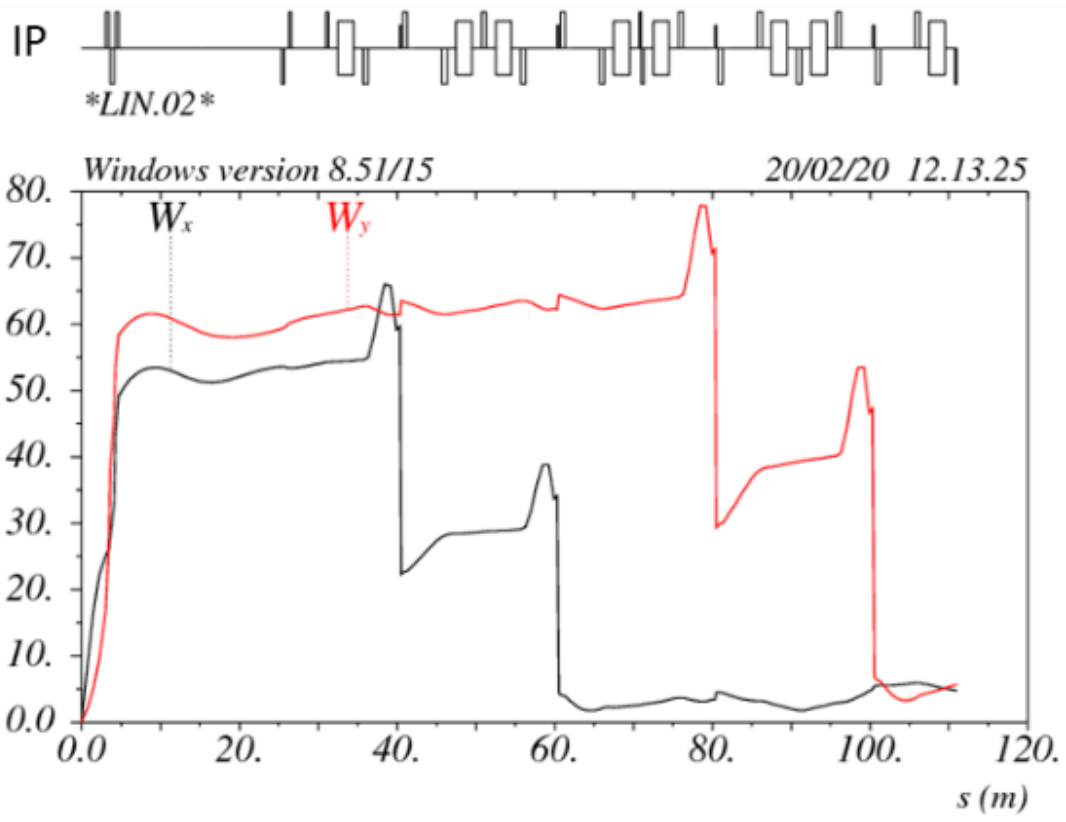

**Figure A6.5.21:** Chromaticity function of the final focusing system with local correction.

*A6.5.2.3.2 Beam Simulations for the Bunch Compressors and FFS*

The start-to-end simulations, spanning from the exit of the damping ring to the entrance of the plasma accelerator, have been conducted using SAD with initial Gaussian distributions. Fig. A6.5.22 provides an overview of the entire lattice, encompassing bunch compressors, the final focusing system, and the matching section in between. In these simulations, it is assumed that the normalized emittance after the damping ring is 15 mm·mrad for the horizontal direction and 10 mm·mrad for the vertical direction.

The results of the beam simulations are presented in Fig. A6.5.23 to Fig. A6.5.25. According to these simulations, the positron beam size can be reduced to 20 μm in all three directions at the entrance of the plasma accelerator.

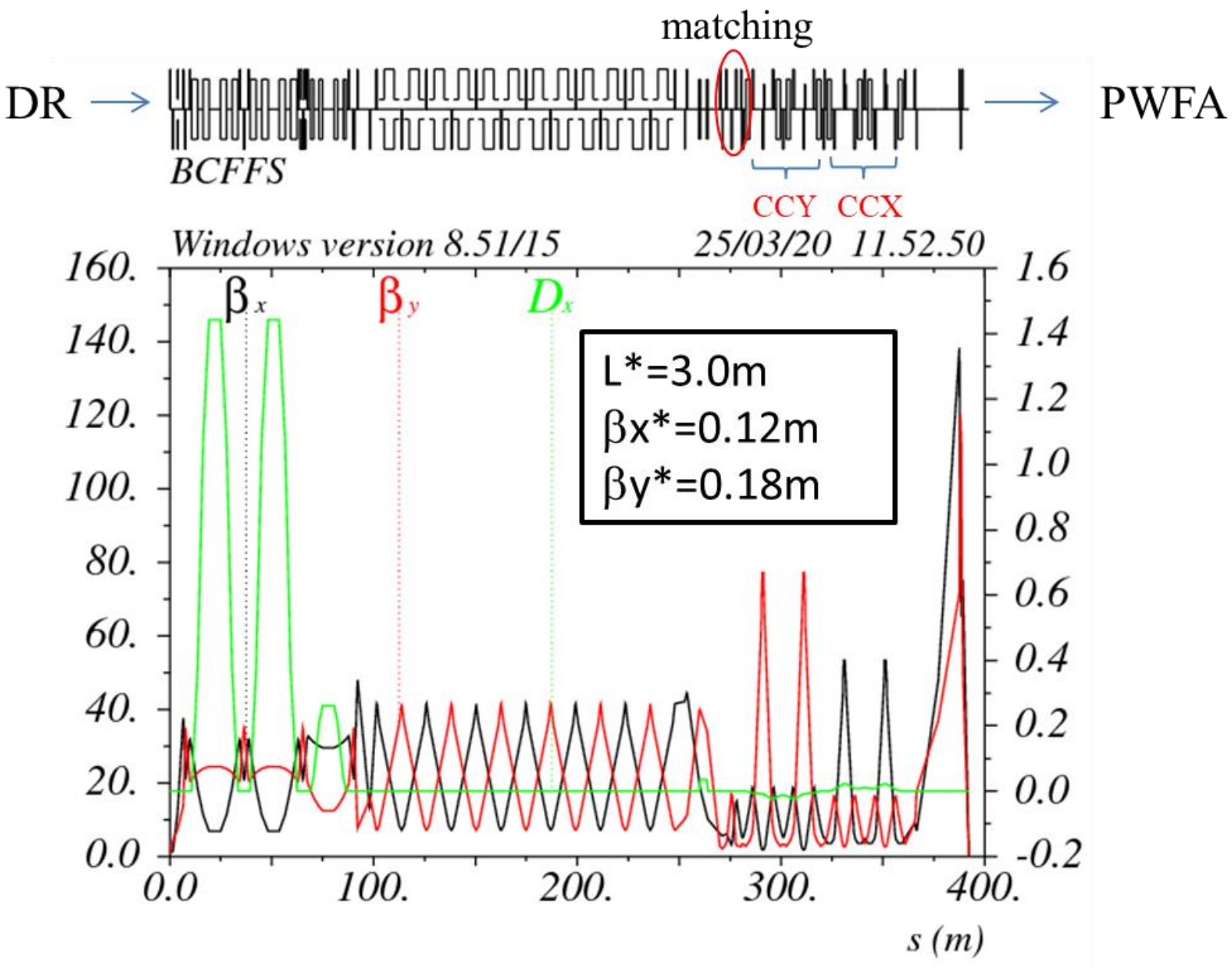

**Figure A6.5.22:** Optics of the transport line including bunch compressors and final focusing system.

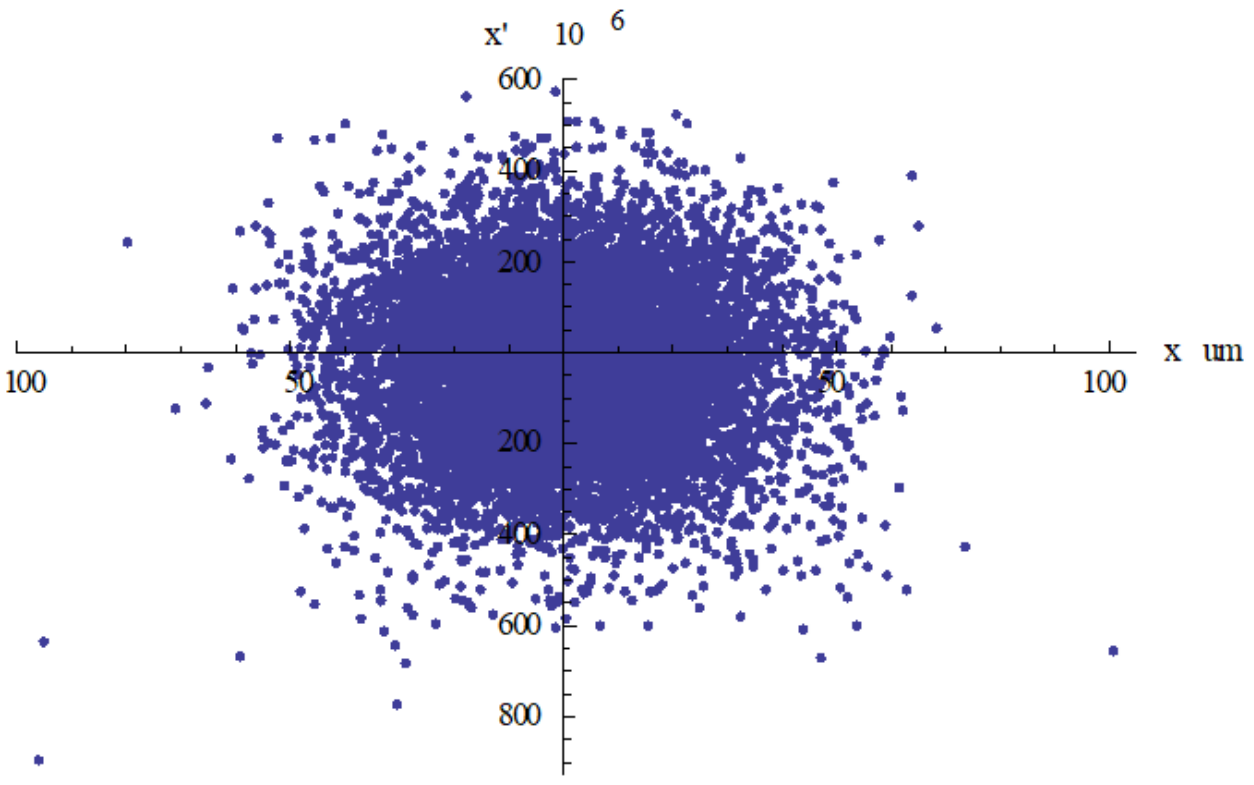

**Figure A6.5.23:** Horizontal beam distribution at the focusing point.

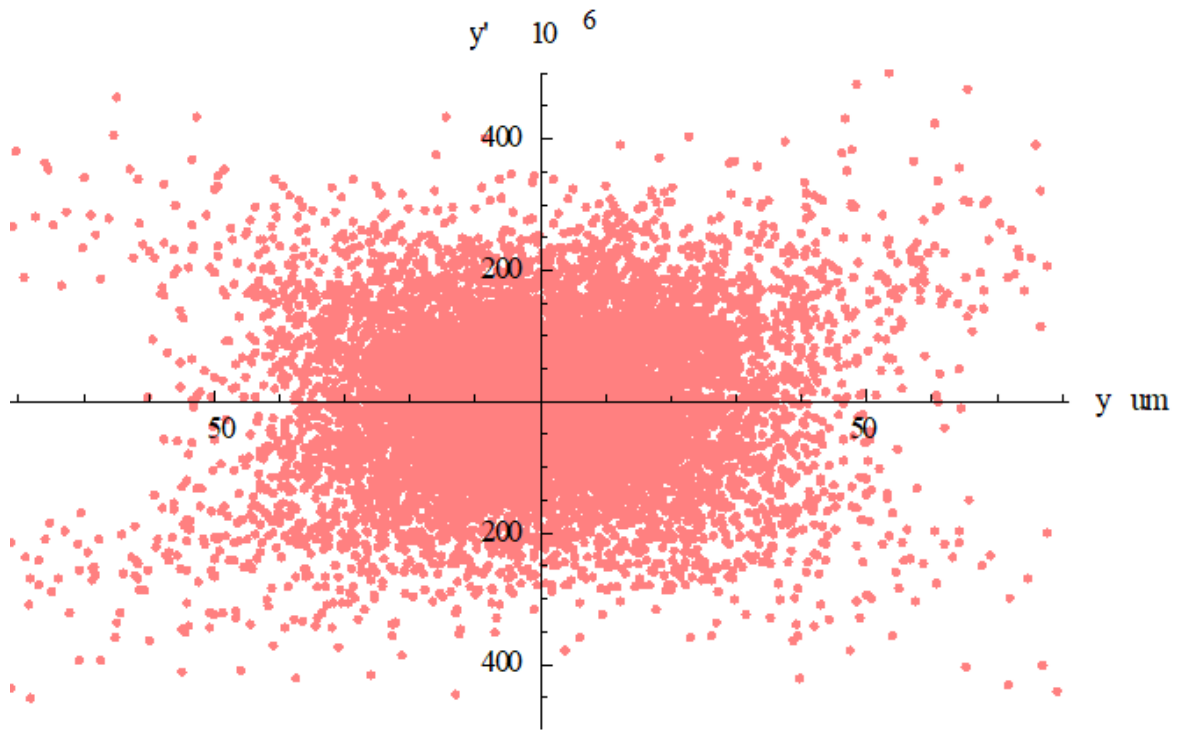

**Figure A6.5.24:** Vertical beam distribution at the focusing point.

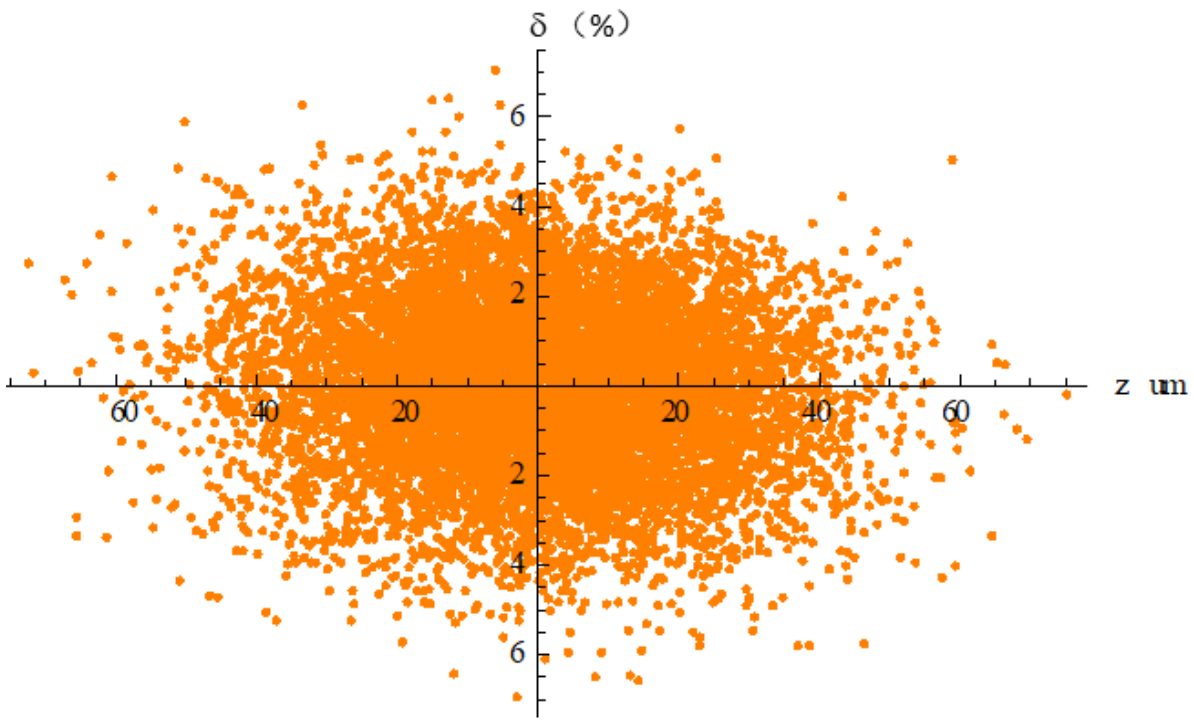

**Figure A6.5.25:** Longitudinal beam distribution at the focusing point.

## A6.6: References

# Appendix 7: Operation for *e-p*, *e-A* and Heavy Ion Collision

## A7.1: Introduction

The SPPC explores the energy frontier with accelerators, while the CEPC provides essential precision for Higgs study, shedding light on mass generation and spontaneous symmetry breaking. However, neither is ideal for examining nucleon and nucleus internal structure, despite their ability to create hadronic matter through collisions. Constructing both in a common complex opens an opportunity for ultra-high energy *e-p* or *e-A* collisions (where *e* stands for either $e^-$ or $e^+$), with a center-of-mass energy of 4.2 TeV (for 37.5 TeV protons) and 5.5 TeV (for 62.5 TeV protons), surpassing the Electron-Ion Collider (EIC, currently under construction in the US) and other proposed lepton-hadron colliders. The table below shows energy ranges for various under-construction and envisioned lepton-hadron colliders.

**Table A7.1:** Center-of-mass energy of future *e-p* or *e-A* colliders

| | EIC | LHeC | FCC-*he* | CEPC-SPPC *e-p* |
|---|---|---|---|---|
| $E_{c.m.}$ (TeV) | 0.02 – 0.14 | 1.3 | 3.5 | 4.2 – 5.5 |

Precise measurements of scattered leptons in such a lepton-hadron facility exploring ultra-deep inelastic scattering (DIS) offer a controlled probe of a proton's inner structure and quantum fluctuations, down to an unprecedented distance of $10^{-4}$ fm (1/10,000th of a proton's size). This study has the potential to detect dynamics related to restoring spontaneously broken symmetries of the standard model and quantum fluctuations caused by physics beyond the standard model. These measurements involve a proton with a momentum transfer exceeding one TeV while keeping the proton intact, resulting in detailed tomographic images of quark and gluon spatial distributions, covering momenta ranging from one-tenth to one-thousandth of the proton's momentum. This information is sensitive to QCD's color confinement.

Replacing proton beams with heavy ion beams in the SPPC will generate the hottest quark-gluon plasma ever, conditions resembling the early microseconds of our universe. Additionally, the possibility of colliding leptons with heavy ions in an *e-A* collider means that heavy ions of varying atomic weights can serve as the world's smallest vertex detectors. This allows for mapping the dynamics of color neutralization and investigating the emergence of hadrons, a crucial phase in the universe's evolution. The combination of the CEPC and the SPPC within a common accelerator complex provides a unique global facility to explore the fundamental structure of matter, its birth, and its evolution.

Aligning a lepton beam from CEPC and a hadron beam from SPPC for collision at one or multiple interaction points (IP) is a relatively straightforward process. Projections estimate a luminosity of several times $10^{33}$ cm$^{-2}$s$^{-1}$ at each detector in *e-p* or *e-A* collisions. However, several challenges need to be addressed. First, when considering *e-p* or *e-A* collisions, it's crucial that CEPC and SPPC have matching circumferences. Currently, SPPC, situated against the outer tunnel wall, has a slightly larger circumference than CEPC, which is positioned along the inner wall. Adjustments to CEPC's configuration will be necessary when converting it into an *e-p* or *e-A* collider. Another challenge involves the design of an interaction region (IR) that optimizes performance while effectively managing facility resources, including site power, when both lepton and hadron facilities are operating concurrently. Furthermore, a third challenge pertains to routing the non-colliding beams away from the detectors when running *e-p* or *e-A* collisions in conjunction with SPPC programs.

In this appendix, we provide an overview of a preliminary design concept for an *e-p* or *e-A* collider utilizing the CEPC and SPPC. It's important to note that the positron beam can be employed for *e-p* or *e-A* collisions with minimal alteration of the design parameters for positron-proton or positron-ion collisions.

## A7.2: *e-p* or *e-A* Accelerator Design Considerations

It is assumed that no major upgrades will be necessary for either the CEPC or SPPC facilities to enable *e-p* or *e-A* collisions. The design for the *e-p* or *e-A* collider is constrained by the operational limits of CEPC and SPPC, including the synchrotron radiation power budget and fundamental beam effects. Within these constraints, some beam and machine parameters can be adjusted to optimize *e-p* or *e-A* collider performance.

Since *e-p* or *e-A* collisions are proposed as an additional capability of the CEPC-SPPC facility, questions arise about the feasibility of simultaneous operation with $e^+e^-$ or *pp* collisions while maintaining acceptable performance for multiple physics programs.

The two primary distinctions between the beams generated by CEPC and SPPC pertain to their bunch structure and beam emittance. In the current baseline design for the Higgs factory program, CEPC employs an electron beam with just 268 bunches. This limitation is due to the synchrotron radiation power budget. In contrast, SPPC's proton or ion beams consist of 10,080 bunches. It's evident that maintaining the original bunch structures of both CEPC and SPPC would be highly inefficient, as the majority of proton or ion bunches wouldn't collide with electron or positron bunches. This effectively rules out the possibility of simultaneous $e^+e^-$ and *e-p*/*A* collision operations within the facility.

It's important to note that the analysis assumes a lepton energy of 120 GeV for *e-p* or *e-A* collisions. However, for physics conducted at lower lepton energies, such as 45.5 GeV, the conclusion may differ, and it might be possible to concurrently operate $e^+e^-$ and *e-p* or *e-A* collisions. Nevertheless, this appendix concentrates solely on high lepton energy scenarios for $e^+e^-$ and *e-p*/*A* collisions, with CEPC operating in H mode.

When not constrained by the simultaneous operation of $e^+e^-$ and *e-p*/*A* collisions, the electron beam is no longer limited to 268 bunches. It can, in fact, be adjusted to match the bunch numbers of a proton or ion beam from SPPC. Moreover, since only one lepton beam is required in the CEPC ring during *e-p* or *e-A* collisions, the electron beam current can be doubled, reaching 33.4 mA while adhering to the same synchrotron radiation power limit of 30 MW per beam. This represents a significant advantage that has the potential to double the *e-p* or *e-A* luminosities.

The choice of beam focusing parameters is also influenced by the requirements of interaction region designs. While the proton beam in SPPC has a fundamentally round shape, it is possible to shape the electron beam into a round form by employing transverse optical coupling. This approach should substantially streamline the process of aligning the beam spot sizes, ultimately leading to a noteworthy enhancement in *e-p* or *e-A* luminosity.

Given the proton beam energy in SPPC, which stands at 37.5 TeV (optional) or 62.5 TeV (design goal), the influence of synchrotron radiation on beam emittance becomes increasingly significant. The damping time for a proton or heavy ion beam is comparable to or even shorter than the time the beam is stored. Consequently, the emittance of the proton or ion beam will tend toward an equilibrium value, reflecting a balance between synchrotron radiation damping, quantum excitation, and intra-beam scatterings throughout the storage period. This equilibrium state will have implications for both peak luminosity and integrated luminosity, a topic that will be explored further in Section A7.4.

The *e-p* or *e-A* collider, which is built upon the CEPC-SPPC framework, exhibits a remarkable level of asymmetry, featuring an energy ratio between the two colliding beams of 312.5 (or 521) for a proton energy of 37.5 TeV (or 62.5 TeV) and an electron beam energy of 120 GeV. This energy contrast surpasses that of any previously constructed, designed, or studied lepton-hadron colliders. As a result, particles resulting from these collisions tend to be densely concentrated around zero scattering angles concerning the direction of the hadron beam. Consequently, it is anticipated that the forward detection of particles with exceedingly small scattering angles will be a critical requirement in the design of the detectors.

The scientific program that relies on deep inelastic scattering as a probing method typically necessitates the collection of experimental data across a range of energy levels. This energy scan imposes the need for a collider design that is optimized to cover a wide center-of-mass energy range. In this appendix, we introduce an initial conceptual design with nominal parameters established at a singular representative energy point, specifically 120 GeV for the electron energy and 37.5 TeV for the proton energy (equivalent to roughly 14 TeV per nucleon for fully stripped lead ions).

## A7.3: *e-p* Collisions

Table A7.2 outlines the nominal parameters for *e-p* collisions involving 120 GeV electrons colliding with 37.5 TeV protons. The electron beam current is set at 33.4 mA, which comfortably remains within the operational limit of 60 MW for synchrotron radiation power. In order to enhance luminosity and optimize the interaction region design, the electron beam emittance will be adjusted to a round shape through the introduction of transverse optical coupling. It's worth noting that *e-p* luminosities at different energies can be estimated using a comparable design approach and adhering to similar parameter limits.

**Table A7.2:** CEPC-SPPC *e-p* and *e-A* design parameters

| Particle | | Proton | Electron | Lead ($^{208}Pb^{82+}$) | Electron |
|---|---|---|---|---|---|
| Beam energy | TeV | 37.5 | 0.12 | 14.8 | 0.12 |
| CM energy | TeV | 4.2 | | 2.7 | |
| Beam current | mA | 730 | 33.4 | 730 | 33.4 |
| Particles per bunch | $10^{10}$ | 15 | 0.7 | 0.18 | 0.7 |
| Number of bunches | | 10080 | | 10080 | |
| Bunch filling factor | | 0.756 | | 0.756 | |
| Bunch spacing | ns | 25 | 25 | 25 | 25 |
| Bunch repetition rate | MHz | 40 | 40 | 40 | 40 |
| Norm. emittance, (x/y) | μm rad | 2.35 | 282 | 0.22 | 282 |
| Bunch length, RMS | Cm | 7 | 0.5 | 7 | 0.5 |
| Beta-star (x/y) | Cm | 75 | 3.7 | 75 | 0.88 |
| Beam spot size at IP (c/y) | Mm | 6.6 | 6.6 | 3.25 | 3.25 |
| Beam-beam per IP(x/y) | | 0.0004 | 0.12 | 0.0016 | 0.12 |
| Crossing angle | mrad | ~0.95 | | ~0.95 | |
| Hour-glass (HG) reduction | | 0.77 | | 0.34 | |
| Luminosity/nuclei per IP, with HG reduction | $10^{33}$ cm$^{-2}$s$^{-1}$ | | | 1.0 | |
| Luminosity/nucleon per IP, with HG reduction | $10^{33}$ cm$^{-2}$s$^{-1}$ | 4.5 | | 23.6 | |

In Table A7.2, the proton beam parameters align with those of the SPPC design, as outlined in Chapter 8. The bunch numbers for both the electron and proton beams are set at 10,080, with a 40 MHz repetition rate (equating to a 25 ns bunch spacing) and a gap (or multiple gaps) of 24.4 km within the beam bunch trains. The final focusing for the proton beam remains consistent with that of *pp* collisions. However, to match the beam spot sizes at the collision point, the electron $\beta^*$ is increased to 3.67 cm. In comparison, the vertical $\beta^*$ for CEPC $e^+e^-$ collisions is just 1.0 mm.

Geometric correction factors affecting the *e-p* collision luminosity include the crab crossing and hour-glass effects. To address the challenges posed by the small bunch spacing, a crossing angle is introduced to enable rapid beam separation near the interaction point, mitigating the parasitic beam-beam effect. A minimum crossing angle of 0.95 mrad is required, providing a horizontal separation of 5 ($\sigma_e$+$\sigma_p$) ≈ 3.6 mm at the first parasitic collision point, which is located 3.75 m from the interaction point. Here, $\sigma_e$ and $\sigma_p$ represent the root-mean-square sizes of the electron and proton beams at that specific location.

Our proposal involves implementing SRF crab cavities on both sides of the collision point to restore head-on collisions, as the luminosity loss due to a crab crossing angle would otherwise be substantial. The estimated transverse kick voltages are approximately 63 MV for the proton beam and around 1 MV for the electron beam, assuming a 650 MHz RF frequency and modest values for the betatron functions (400 m and 200 m) at the location of the crab cavities. We assume there is no luminosity reduction resulting from crab collisions after compensation.

It's important to note that due to a significant increase in the electron $\beta^*$, the luminosity reduction factor attributed to the hour-glass effect reaches 77%.

## A7.4: *e-A* Collisions

The conceptual design for *e-A* collisions at the CEPC-SPPC facility adheres to the same design principles as for *e-p* collisions. However, it's essential to recognize that the synchrotron radiation damping effect on heavy ion beams is considerably stronger compared to proton beams. Consequently, this necessitates some additional considerations and adjustments to the beam parameters.

It can be demonstrated through scaling that the synchrotron radiation damping time for an ion beam is significantly shorter, with a factor of $A^4/Z^5$, compared to a proton beam in a storage ring of equal magnetic rigidity. Here, A represents the atomic number, and Z represents the number of stripped electrons from the ion. Taking a fully stripped lead ion ($^{208}Pb^{82+}$) as an example, this reduction factor results in approximately 0.5. Consequently, the damping of the lead ion beam is approximately twice as fast as that of a proton beam.

The equilibrium emittance of an ion beam demonstrates an even greater reduction factor, $Z^3/A^4$, resulting in a value of 0.0003 for fully stripped lead ions. This implies that the equilibrium emittance of the lead ion beam (maintained through a balance of synchrotron radiation damping and quantum excitations) is approximately four orders of magnitude smaller than that of the proton beam. It's essential to note that this outcome is not physically plausible, as it doesn't account for intra-beam scattering, which can lead to emittance growth.

In reality, after a few damping times, the ion beam's emittance will settle into an equilibrium value, considering the interplay of radiation damping, quantum excitations, and the heating effect induced by intra-beam scatterings. It's estimated that the actual equilibrium emittance of a fully stripped lead ion beam in the SPPC is around 0.22 μm rad, which is roughly 10 times smaller than the proton emittance. This value is utilized to estimate the luminosity of *e-A* collisions, as presented in the fifth and sixth columns of Table A7.2. However, it's important to acknowledge that the precise estimation of ion beam emittance relies on the lattice design of the SPPC ring.

The remaining parameters in Table A7.2, such as bunch length, number, and final focusing, align with the *e-p* collision design presented in the third and fourth columns. This results in a luminosity of $1.0\times10^{33}$ $cm^{-2}s^{-1}$ per nuclei and $2.4\times10^{34}$ $cm^{-2}s^{-1}$ per nucleon at each detector.

## A7.5: Additional Comments

The design of the interaction region holds paramount importance in the context of an *e-p* or *e-A* collider. Forward particle detection is an essential component for the detector system. Moreover, accommodating both lepton and hadron beams necessitates a substantial detector space.

In the case of the proton or ion beams, this requirement aligns with that of the SPPC. Conversely, the CEPC detector space is intentionally minimal to facilitate an exceptionally small vertical $\beta^*$ (~1 mm) for high luminosity. Expanding this space significantly will pose one of the primary challenges in the design process.

Another pivotal concern is ensuring an adequate separation of the colliding beams at the positions of the final focusing magnets. With the separation stemming from the crab crossing angle being only a few centimeters, assuming a detector space of ±25 m, this separation is smaller than the physical dimensions of warm or superconducting magnets.

Consequently, supplementary strategies and schemes must be put in place to prevent any interference between the beam transport and these magnets.

In the current CEPC-SPPC baseline configuration, within the tunnel, the lepton rings are positioned on the inner side, while the hadron rings are situated on the outer side, with a separation of a few meters between them. This setup results in a discrepancy of up to 20 meters in the circumferences of the lepton and hadron collider rings. To ensure proper beam synchronization at multiple interaction points, this difference must be addressed in the ring design.

It is imperative to perform estimations of beam and luminosity lifetimes, and to conduct a comprehensive evaluation of various factors that can limit these lifetimes. This includes the critical investigation of nonlinear and collective beam dynamics, with a specific emphasis on the beam-beam effect.

# Appendix 8: Opportunities for Polarization in CEPC

Beam polarization is a critical aspect in CEPC's design. On one hand, polarized beams can be used for beam energy calibration at Z-pole and WW threshold, which is essential for precision measurements of the properties of Z and W bosons [1]. This application requires at least 5% to 10% beam polarization for both electron and positron beams. Non-colliding pilot bunches are essential for this calibration to avoid energy spread caused by beamstruhlung during collisions.

On the other hand, longitudinally polarized colliding beams are crucial for precision tests of the standard model and exploring new physics through colliding beam experiments. Achieving 50% or more longitudinal polarization at the interaction points (IPs) without sacrificing luminosity is necessary. The SLC's experiments with longitudinally polarized electron beams already offered unique capabilities for specific measurements [2] and adding polarization for both beams can further enhance luminosity, as demonstrated in ILC studies [3]. These applications require substantial design efforts to tackle challenges in generating, maintaining, and manipulating polarized electron (positron) beams.

In this context, we investigate the potential for beam polarization at CEPC using the parameters and lattice designs from the CEPC Conceptual Design Report (CDR) [1], unless stated otherwise. The findings from these studies have not been incorporated into the baseline design in this CEPC Technical Design Report (TDR). Applying the baseline beam parameters and lattice designs directly for evaluating beam polarization performance results in suboptimal outcomes. Nonetheless, these studies support the feasibility of beam polarization applications at CEPC, provided future accelerator chain designs incorporate necessary modifications.

## A8.1: Introduction

In this introductory section, we first provide an overview of spin dynamics in electron (positron) circular accelerators, establishing the groundwork for subsequent discussions. Following that, we outline our general considerations regarding the preparation of polarized beams for utilization in the CEPC Collider.

### A8.1.1: Fundamental Spin Dynamics

The precession of the spin expectation value $\vec{S}$ and of $\hat{S} = \frac{\vec{S}}{|\vec{S}|}$ of a relativistic charged particle in an electromagnetic field follows the Thomas-BMT equation [4,5], which for a circular accelerator can be expressed as:

$$\frac{d\hat{S}}{d\theta} = \left[\vec{\Omega}_0(\theta) + \vec{\omega}(\vec{u};\theta)\right] \times \hat{S} \quad \text{(A8.1.1)}$$

with

$$\vec{\Omega}_0(\theta) = \vec{\Omega}_{00}(\theta) + \Delta\vec{\Omega}(\theta) \quad \text{(A8.1.2)}$$

where $\vec{\Omega}_0(\theta)$ is the spin-precession vector on the closed orbit at the azimuthal angle $\theta$, $\vec{\Omega}_{00}(\theta)$ denotes the contribution of the fields on the design orbit, and $\Delta\vec{\Omega}(\theta)$ represents the contribution of the fields due to the magnet errors and correction coils. $\vec{\omega}(\vec{u};\theta)$ is due to the orbital oscillation $\vec{u}$ relative to the closed orbit. We denote the right-handed orthonormal set of unit-length solution of Eq. (A8.1.1) as $\left(\hat{n}_0(\theta), \hat{m}_0(\theta), \hat{l}_0(\theta)\right)$, where $\hat{n}_0(\theta)$ is the periodic solution satisfying $\hat{n}_0(\theta + 2\pi) = \hat{n}_0(\theta)$. Then a spin vector $\hat{S}$ perpendicular to $\hat{n}_0$ precesses by a rotation angle $2\pi\nu_0$ in one revolution around $\hat{n}_0$, and $\nu_0$ is called the closed-orbit spin tune. We define $\hat{k}_0(\theta) = \hat{m}_0(\theta) + i\hat{l}_0(\theta)$, then it follows that $\hat{k}_0(\theta + 2\pi) = \exp(i2\pi\nu_0)\hat{k}_0(\theta)$. In a planar ring without spin rotators, $\hat{n}_0$ is close to vertical, and $\nu_0 \approx a\gamma$, where $a$= 0.00115965219 for electron (positron), and $\gamma$ is the relativistic factor for the design energy. The concept of $\hat{n}_0$ and $\nu_0$ on the closed orbit can be extended to the more general phase-space coordinates [6–8], with their counterparts being the invariant spin field $\hat{n}(\vec{u};\theta)$ with $\hat{n}(\vec{u};\theta + 2\pi) = \hat{n}(\vec{u};\theta)$, and the amplitude-dependent spin tune $\nu_s(\vec{I})$ describing the spin precession rate around $\hat{n}$, where $\vec{I}$ denotes the actions of the orbital motion. The projection of the spin vector of a particle on the invariant spin field $J_S = \hat{S} \cdot \hat{n}$ is an adiabatic invariant of its spin motion. Machine imperfections and orbital oscillations contribute to $\Delta\vec{\Omega}(\theta)$ and $\vec{\omega}(\vec{u};\theta)$ respectively and may perturb the spin motion in a resonant manner when the following condition of spin-orbit coupling resonances (spin resonances in short) is nearly satisfied,

$$\nu_s = K = k + k_x\nu_x + k_y\nu_y + k_z\nu_z, \qquad k, k_x, k_y, k_z \in \mathbb{Z} \tag{A8.1.3}$$

where $K$ represents the spin resonance location, $\nu_x$ and $\nu_y$ represent the horizontal and vertical betatron tunes, and $\nu_z$ denotes the synchrotron tune. These values are conventionally used when dealing with weak couplings, as opposed to the more comprehensive orbital tunes $\nu_I$ , $\nu_{II}$ and $\nu_{III,}$ which are derived from the orbital eigenanalysis [9]. $\hat{n}(\vec{u};\theta)$ deviates from $\hat{n}_0(\theta)$ near these spin resonances. In a misaligned ring, $\hat{n}_0(\theta)$ can deviate strongly from the design direction near integer spin resonances $\nu_0 = k, k \in \mathbb{Z}$ . Spin resonances with $|k_x| + |k_y| + |k_z| = 1$ and $|k_x| + |k_y| + |k_z| > 1$ are called the first-order spin resonances and higher-order spin resonances, respectively. As we will discuss in more detail later, depolarization can occur during the acceleration process in the Booster due to spin resonance crossings, and it can also happen in the Collider due to radiative depolarization, with spin resonances playing a significant role in both scenarios.

There are two primary methods for producing polarized electron (positron) beams. In a storage ring, $e^+$ and $e^-$ beams can naturally become vertically polarized from scratch through the Sokolov-Ternov effect [10], driven by spin-flip synchrotron radiations. In a perfectly aligned ring without spin rotators, the polarization aligns vertically along $\hat{n}_0$. However, stochastic photon emissions disrupt the orbital motion, leading to radiative depolarization. These opposing effects collectively establish an equilibrium beam polarization $P_{\mathrm{DK}}$, approximated by [11]:

$$P_{\mathrm{DK}} \approx \frac{P_\infty}{1 + \frac{\tau_{\mathrm{BKS}}}{\tau_{\mathrm{dep}}}} \tag{A8.1.4}$$

where $P_\infty$ and $\tau_{\mathrm{BKS}}$ are, respectively, the equilibrium beam polarization and the time constant of the Sokolov-Ternov effect without considering the radiative depolarization effect [12]. $P_\infty$ can reach as high as 92.4% in an ideal planar ring without spin rotators, $\tau_{\mathrm{BKS}}[s] \approx \frac{99}{2\pi}\frac{C[\mathrm{m}]\rho[\mathrm{m}]}{E[\mathrm{GeV}]^5}$ with $C$ and $\rho$ denoting the circumference and average radius of bending magnets of the ring. $\tau_{\text{dep}}$ is the time constant of radiative depolarization, and $\tau_{\text{DK}}$ is the polarization build-up time to reach $P_{\text{DK}}$, satisfying $\tau_{\text{DK}}^{-1} = \tau_{\text{BKS}}^{-1} + \tau_{\text{dep}}^{-1}$.

Additionally, polarized $e^+$ and $e^-$ beams can be produced initially at the source, followed by acceleration and transmission through the injector chain, ultimately leading to their injection into the Collider.

Next, we must differentiate the polarization evolution for two types of bunches in the Collider. For non-colliding pilot bunches, following injection, the polarization evolves as follows:

$$P_{\text{t}}(t) = P_{\text{DK}}\left(1 - e^{-\frac{t}{\tau_{\text{DK}}}}\right) + P_{\text{inj}}e^{-\frac{t}{\tau_{\text{DK}}}} \quad \text{(A8.1.5)}$$

In contrast, for colliding bunches under top-up injection, employed to address the relatively short beam lifetime and sustain a high average luminosity, the polarization undergoes a sawtooth-like evolution. The time-averaged beam polarization, denoted as $P_{\text{avg}}$, represents a compromise between the equilibrium beam polarization, $P_{\text{DK}}$, in the Collider, and the injected beam polarization, $P_{\text{inj}}$:

$$P_{\text{avg}} = \frac{P_{\text{DK}}}{1 + \frac{\tau_{\text{DK}}}{\tau_b}} + \frac{P_{\text{inj}}}{1 + \frac{\tau_b}{\tau_{\text{DK}}}} \quad \text{(A8.1.6)}$$

where $\tau_b$ is the beam lifetime.

### A8.1.2: General Considerations of Polarized Beam Preparation

For the CEPC Collider, owing to the weak bending magnets and the larger circumference designed to minimize synchrotron radiation power, the time constant associated with the Sokolov-Ternov effect is significantly extended at lower energies. Specifically, $\tau_{\text{BKS}}$ stands at 256 hours at 45.6 GeV, 15.2 hours at 80 GeV, and 2 hours at 120 GeV, in sharp contrast to the short beam lifetime, which is on the order of approximately 1 hour for colliding bunches. As we will discuss in more detail later, simulations suggest that with meticulous closed orbit and optics corrections, achieving an equilibrium beam polarization, $P_{\text{DK}}$, exceeding 10% is relatively straightforward at 45 GeV, feasible at 80 GeV, but becomes increasingly challenging at higher energies such as 120 GeV, and may even be impossible. For energies corresponding to Z and W bosons, the utilization of the Sokolov-Ternov effect can yield adequate beam polarization for non-colliding pilot bunches dedicated to beam energy calibration. It has been proposed that at Z energies, the incorporation of asymmetric wigglers [13] into the Collider [14] would significantly reduce the polarization build-up time to approximately 10-20 hours, thus facilitating timely beam energy calibration.

However, relying solely on the Sokolov-Ternov effect, the time-averaged beam polarization ($P_{\text{avg}}$) for the colliding bunches is additionally diminished by a factor $\frac{\tau_{\text{DK}}}{\tau_b}$ relative to $P_{\text{DK}}$. At 45.6 GeV, this factor falls within the range of 10 to 100. To reduce this factor, one would need to shorten $\tau_{\text{DK}}$ by implementing more powerful asymmetric wigglers and extend $\tau_b$, but achieving both conditions simultaneously is unattainable without compromising luminosity.

Instead, by preparing beams with a substantial initial polarization at the injector and subsequently injecting them into the Collider, it becomes feasible to attain a high beam polarization level for both the non-colliding pilot bunches and the colliding bunches [15]. For instance, assuming $P_{\text{inj}}$= 80% and $P_{\infty}$= 92%, we can calculate the minimum required $P_{\mathrm{DK}}$ to achieve $P_{\mathrm{avg}} \geq 50\%$ for the three operational energies, as outlined in Table A8.1.1. Notably, due to the relationship between $\tau_{\mathrm{BKS}}$ and $\tau_b$, higher beam energies necessitate a larger $P_{\mathrm{DK}}$, presenting a more significant challenge in mitigating radiative depolarization in the Collider. Nonetheless, injecting highly polarized beams into the Collider holds the potential to attain a substantial $P_{\mathrm{avg}}$ alongside high luminosity, which is crucial for realizing colliding beams with a high longitudinal polarization. This approach can also enhance beam energy calibration using non-colliding pilot bunches.

**Table A8.1.1:** CEPC beam parameters related to $P_{\mathrm{avg}}$ for colliding bunches.

| Beam energy | 45.6 GeV (Z) | 80 GeV (W) | 120 GeV (Higgs) |
|---|---|---|---|
| $\tau_b$ (hour) | 2.5 | 1.4 | 0.43 |
| $\tau_{\mathrm{BKS}}$ (hour) | 256 | 15.2 | 2.0 |
| $P_{\mathrm{DK,min}}$ | 1.5% | 14.2% | 33.0% |

The CEPC injector chain, as described in the CEPC TDR, comprises an unpolarized electron and positron source, a 30 GeV Linac, a full-energy Booster, and transfer lines. Figure A8.1.1 illustrates the proposed modifications to the CEPC accelerator complex to incorporate polarized beams.

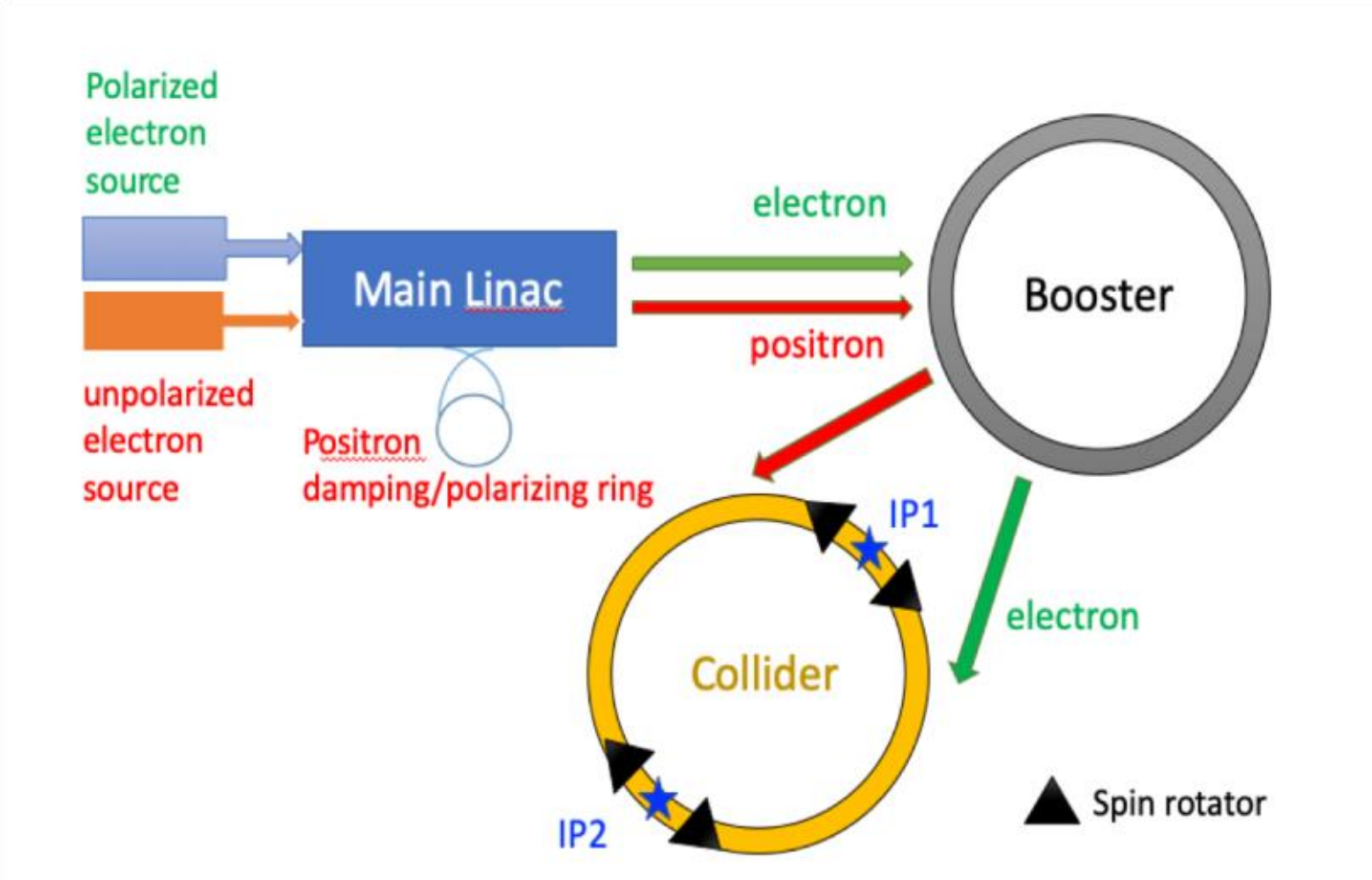

**Figure A8.1.1:** Proposed modifications to the CEPC accelerator complex to incorporate polarized beams.

In Section A8.2, we will delve into the polarized beam sources. The addition of a polarized electron gun allows for the production of electron bunches with 80% or higher polarization [16–18]. However, the technical challenges associated with developing polarized positron sources to meet the requirements of top-up injection for colliding bunches remain substantial. As a provisional approach, we assume that electron beams are polarized while positron beams remain unpolarized in the colliding beam experiments. Nevertheless, it remains feasible to generate 20% or more beam polarization through the Sokolov-Ternov effect for beam energy calibration in the positron Damping Ring, with the assistance of asymmetric wigglers [19].

Subsequently, the polarized beams are conveyed through the injector chain, and it is crucial to maintain the beam polarization throughout this process. Prior investigations conducted for the SLC [20] and ILC [3] have indicated minimal polarization loss in the linac and transfer lines. In Section A8.3, we assess depolarization effects during Booster acceleration. Our earlier investigations indicated that with an alternative Booster lattice [21], polarization loss remains minimal when accelerating to 45.6 GeV and 80 GeV, but the possibility of increased depolarization becomes more evident at higher beam energies. In the case of the TDR Booster lattice, however, a more significant depolarization risk is observed during the acceleration to 45.6 GeV and 80 GeV. We also consider potential future enhancements to address and mitigate this depolarization issue.

Moving forward, we turn our attention to beam polarization considerations in the Collider. In Section A8.4, we delve into the investigations of radiative depolarization effects, while Section A8.5 explores the utilization of spin rotators to attain longitudinal polarization. Section A8.6 offers preliminary insights into beam energy calibration, and Section A8.7 discusses operational scenarios concerning polarized beams. Additionally, some initial studies regarding the Compton polarimeter are presented in A8.8. In conclusion, we summarize the primary findings and outline future endeavors for the post-TDR phase.

## A8.2: Polarized Beam Source

### A8.2.1: Polarized $e^-$ Source

Polarized electron bunches can be produced through photocathode electron guns employing Negative Electron Affinity-activated GaAs-based cathodes, which absorb circularly polarized photons [22]. Superlattice structures demonstrate superior polarization performance compared to bulk materials, owing to the inherent splitting of heavy-hole (HH) and light-hole (LH) subbands under strain, thereby enhancing the polarization ratio. Currently, the highest polarization performance is attained using GaAs/GaAsP Superlattice structures grown on GaAs substrates, making it the prevalent cathode material in polarized electron sources.

The GaAs cathode surface is coated with O(NF3)-Cs, which is highly sensitive to gas contamination, including carbon monoxide, carbon dioxide, oxygen, and other gases. Residual gases in the gun chamber can degrade the quantum efficiency (QE) either through chemical poisoning or ion back bombardment. Consequently, this cathode material typically operates in an extremely high vacuum environment ($10^{-12}$ Torr) to maintain QE and ensure a long lifespan. Compared to RF guns, achieving an extremely high vacuum is more straightforward in a DC gun, making it the preferred choice for a polarized electron source. Table A8.2.1 provides a comparison of beam requirements for the polarized electron source for CEPC relative to other accelerators, while Table A8.2.2 outlines the parameter specifications for a polarized electron gun designed for CEPC.

**Table A8.2.1:** Comparison of electron gun parameter requirements between CEPC and other accelerators [23]

| Parameter | CEPC | SLC | ILC | CLIC | EIC |
|---|---|---|---|---|---|
| Polarization [%] | >80 | 85 | >80 | >80 | 85 |
| Bunch charge [nC] | 3.3 | 9-16 | 3.2 | 1 | 7 |
| Number of microbunches | 1 | 1 | 1312 | 312 | 8 |
| Repetition rate [Hz] | 100 | 120 | 5 | 50 | 100 |
| Average current from gun [μA] | 0.33 | 1.1-1.9 | 21 | 15.6 | 5.6 |

**Table A8.2.2:** Specifications of polarized electron gun for CEPC

| Parameter | Specification |
|---|---|
| Gun type | Photocathode DC gun |
| Cathode material | Superlattice GaAs/GasAsP |
| Voltage | 150-200kV |
| Quantum Efficiency | 0.5% |
| Polarization | >85% |
| Bunch population | $2.1 \times 10^{10}$ |
| Drive laser | 780 nm ($\pm$20 nm), 10μJ@1ns |

The polarized electron gun produces longitudinally polarized electron beams, which are subsequently directed through a Wien filter. The polarization direction is rotated to a vertical orientation to facilitate polarization measurements using a Mott polarimeter. The electron beam, now vertically polarized, continues its journey through the injector chain. The specific design of the beamline and its integration with the Linac will be subject to further investigation and refinement.

IHEP has successfully developed a photocathode DC high-voltage electron gun, illustrated in Fig. A8.2.1 [24]. This DC gun can operate reliably at 400 kV in an extremely high vacuum environment of $2\times10^{-12}$ torr within the gun chamber. The photocathode material employed in this electron gun is bulk GaAs, activated by a 530 nm green laser. Additionally, a photocathode featuring a GaAs/GaAsP Superlattice structure can be introduced into the electron gun through a load-lock system to enable high-polarization research. This experimental platform with the electron gun serves as a robust foundation for future investigations into polarized electron sources.

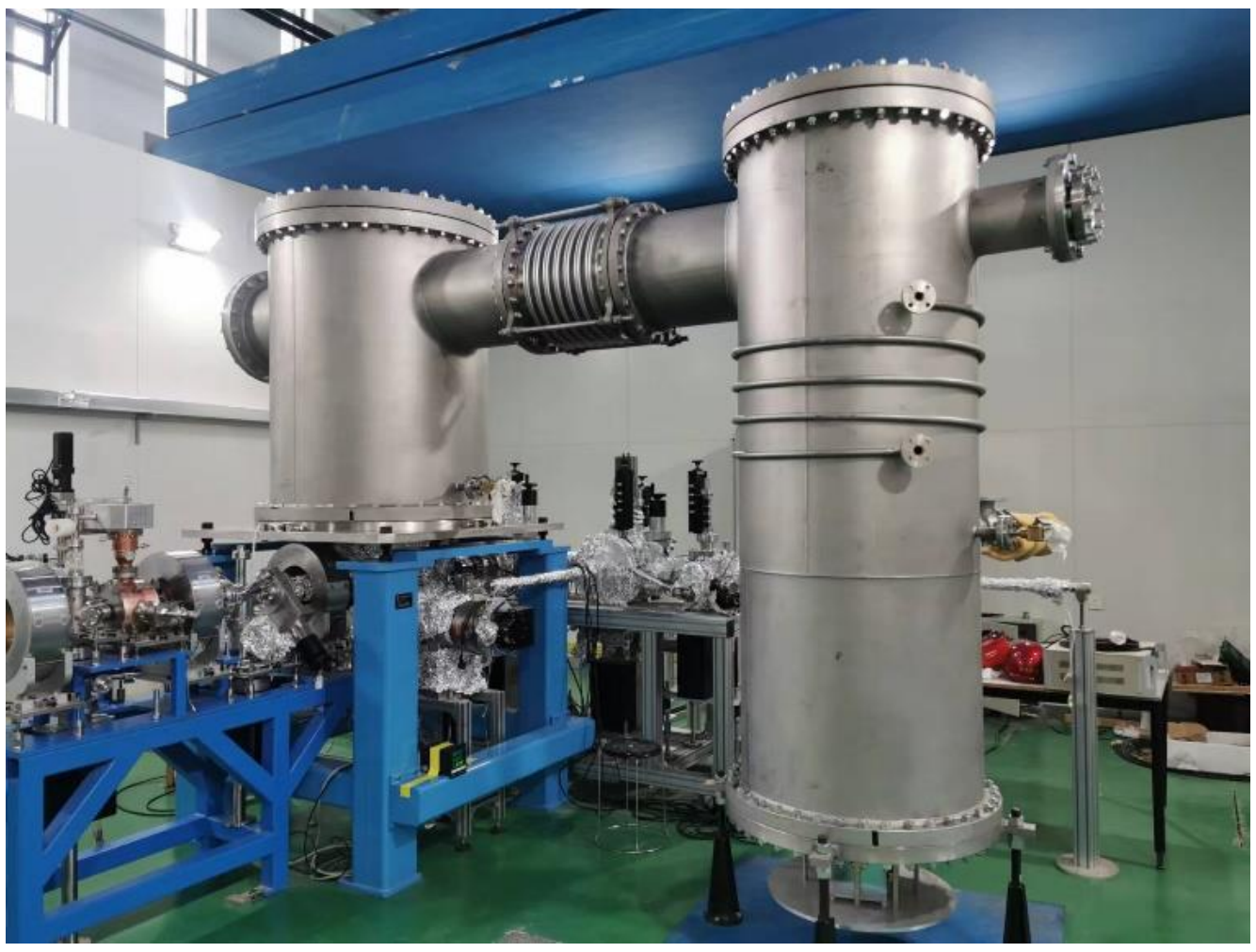

**Figure A8.2.1:** A photocathode DC high-voltage electron gun developed at IHEP.

## A8.2.2: Polarized $e^+$ source

Conversely, the development of polarized positron sources remains technically challenging [25]. Employing the ILC polarized positron source necessitates an electron drive beam with an energy exceeding 100 GeV [26], significantly complicating the design of the CEPC injector. Furthermore, alternative polarized positron source methods suffer from inadequate yield for application in the CEPC.

In the injector design outlined in the CEPC TDR, 1.9~3.9 nC of unpolarized positron bunches are generated through the interaction of a 4 GeV, 3.5~7.0 nC unpolarized primary electron bunch with a target. After pre-acceleration, these positron bunches are injected into a positron Damping Ring to attain the required beam quality for subsequent transportation. By default, 4 positron bunches remain in the positron Damping Ring for 20 ms to meet the requirements for filling the colliding bunches, resulting in extracted unpolarized bunches.

The prospect of polarizing the positron bunches using the Sokolov-Ternov effect within the positron Damping Ring or another dedicated ring of similar size has been previously explored. However, achieving this would demand exceedingly strong asymmetric wigglers to polarize all positron bunches while adhering to the necessary timing constraints for filling all colliding bunches, presenting significant challenges.

Nonetheless, it is more feasible to generate polarized positron bunches to facilitate beam energy calibration. Suppose we store one or two positron bunches in the positron Damping Ring for an extended duration, such as 10 minutes, in addition to the other bunches used for top-up injection. In that case, aiming for the generation of over 20% beam polarization requires achieving a self-polarization build-up time $\tau_{\mathrm{DK}}$ of approximately 30 minutes. It is noted that the self-polarization build-up time for the positron Damping Ring in the CEPC TDR is approximately 197 minutes, which is deemed unsatisfactory.

A different design for the positron Damping Ring was investigated in reference to [19], featuring a higher beam energy of 1.542 GeV and the incorporation of asymmetric wigglers. These modifications were aimed at decreasing the polarization build-up time to 34 minutes. Simulations of the dynamic aperture and equilibrium beam polarization, accounting for machine imperfections, affirmed the viability of this approach. This study implies that further enhancements to the positron Damping Ring are possible, with the potential to generate the required positron beam polarization for facilitating beam energy calibration.

## A8.3: Polarized Beam Acceleration in the Booster

As the electron (positron) beam undergoes acceleration in a booster synchrotron, both the closed-orbit spin tune $\nu_0$ and the amplitude-dependent spin tune $\nu_s$ of beam particles experience changes, given that $\nu_s \approx \nu_0 \approx a\gamma$. Consequently, this results in the crossing of underlying spin resonances, which has the potential to cause beam depolarization. The polarization loss during the passage through a single, isolated spin resonance can be estimated using the Froissart-Stora formula [27]:

$$\frac{P_{\mathrm{f}}}{P_{\mathrm{i}}} = 2e^{-\frac{\pi|\widetilde{\omega}_K|^2}{2\alpha}} - 1 \tag{A8.3.1}$$

where $P_{\mathrm{i}}$ and $P_{\mathrm{f}}$ are the beam vertical polarization before and after crossing the resonance $\nu_0 = K$, $\widetilde{\omega}_K$ is the spin resonance strength which depends on the lattice design and machine imperfections, $\alpha = \frac{da\gamma}{d\theta}$ is the spin resonance crossing rate. We can identify three parameter regimes in the value of $|\widetilde{\omega}_K|/\sqrt{\alpha}$.

- First, if the resonance is very strong, or the acceleration very slow, say $|\widetilde{\omega}_K|/\sqrt{\alpha} > 1.84$, then $\left|\frac{P_{\mathrm{f}}}{P_{\mathrm{i}}}\right| > 99\%$ but with a spin flip, and we call this the “slow crossing” regime.

- Second, if the resonance is very weak, or the acceleration very fast, say $|\widetilde{\omega}_K|/\sqrt{\alpha} < 0.056$, then $\frac{P_\text{f}}{P_\text{i}} > 99\%$, and we call this the “fast crossing” regime.
- Third, if both conditions are not satisfied, $\left|\frac{P_f}{P_\text{i}}\right|$ is reduced, and we call this the “intermediate” regime.

Two families of significant spin resonances are noteworthy in this context:

1. The integer spin resonances, denoted as $\nu_0 = k, k \in \mathbb{Z}$, pertain to the tilt of $\hat{n}_0$. These resonances are primarily induced by horizontal magnetic fields stemming from vertical orbit offsets in quadrupoles and dipole roll errors. They are commonly referred to as "imperfection resonances." The strength of an imperfection resonance is the same among beam particles.
2. The first order "parent" spin resonances, $\nu_0 = k \pm \nu_y$, are driven by horizontal magnetic fields arising from vertical betatron oscillations in quadrupoles. These resonances are conventionally known as "intrinsic resonances." The strength of an intrinsic resonance scales with the square root of the vertical betatron motion of beam particles.

Adjacent imperfection resonances are spaced at intervals of 440 MeV. During the acceleration of the CEPC Booster from its injection energy at 30 GeV ($a\gamma = 68.1$) to extraction energies of 45.6 GeV ($a\gamma = 103.5$) , 80 GeV ($a\gamma = 181.5$) , and 120 GeV ($a\gamma = 272.3$), it is evident that tens or even hundreds of spin resonances from these two families will be traversed.

In the case of a circular accelerator characterized by an ultra-high effective lattice periodicity, the contributions of strengths from intrinsic and imperfection resonances within each periodic structure tend to cancel each other out for small values of $a\gamma$. However, they combine coherently for what are termed "super-strong spin resonances" [28]. These super-strong spin resonances occur near $(mPM \pm \nu_B)/\eta_{arc}, m \in \mathbb{Z}$, where $P$ and $M$ represent the periodicity of the lattice and the number of dipoles per superperiod, $\nu_B$ is the vertical betatron phase advance in all arc dipoles, and $2\pi\eta_{arc}$ denotes the total bending angle in all arc sections. For a large value of $PM$, these super-strong spin resonances are well separated, and the first super-strong spin resonances are located near $\nu_B/\eta_{arc}$.

Table A8.3.1 provides a comprehensive listing of lattice parameters pertaining to the spin resonance structure for the CEPC Booster, as documented in both the CEPC TDR and CDR. Additionally, it includes an alternative design of the CEPC Booster lattice (Alternative). The first occurrences of super-strong spin resonances are found as follows: below 45.6 GeV for the TDR lattice, within the energy range of 80 GeV to 120 GeV for the CDR lattice, and beyond 120 GeV for the Alternative lattice.

**Table A8.3.1:** Parameters relevant for the spin resonance structure

| Parameters | TDR | CDR | Alternative |
|---|---|---|---|
| $\nu_y$ | 116.83 | 261.2 | 353.28 |
| Basic arc cell structure | TME | FODO | FODO |
| Vertical phase advance per arc cell | 28 degree | 90 degree | 90 degree |
| $P$ | 8 | 8 | 8 |
| $M$ | 126 | 97 | 140 |
| $\eta_{arc}$ | 126/127 | 97/99 | 140/142 |
| $\nu_B$ | 78.4 | 194 | 280 |
| $PM$ | 1008 | 776 | 1120 |
| $\nu_B/\eta_{arc}$ | 79.0 | 198 | 284 |

In the case of the Alternative lattice, characterized by a substantial vertical betatron phase advance, spin resonances tend to exhibit a general weakness. These resonances typically manifest at beam energies beyond the operational range, a fact corroborated by a meticulous assessment of the spectral strength of spin resonances. Consequently, the crossing of these resonances comfortably falls within the "fast crossing" regime, resulting in minimal depolarization during the acceleration process toward the Z and W energies. This assertion is substantiated by both analytical calculations of depolarization using Eq. (A8.3.1) and extensive multi-particle tracking conducted with the Bmad code [29] throughout the entire acceleration process. This scenario implies the potential to accelerate polarized electron (positron) beams to exceptionally high beam energies without necessitating the implementation of Siberian snakes [19], further bolstering their applicability for subsequent uses.

In the context of the TDR lattice, it's worth noting that the initial encounter with super-strong spin resonances occurs early in the acceleration process. This observation is visualized in Fig. A8.3.1, which depicts the spectrum of intrinsic resonance strength within the operational beam energy range. This spectrum was calculated using the DEPOL code [30] and based on a vertical normalized amplitude ($\epsilon_{y,norm}$) of $10\pi$ mm·mrad. It's significant to highlight that the very first super-strong intrinsic resonance is situated at $a\gamma = 76.83$, a finding that closely aligns with the theoretical predictions.

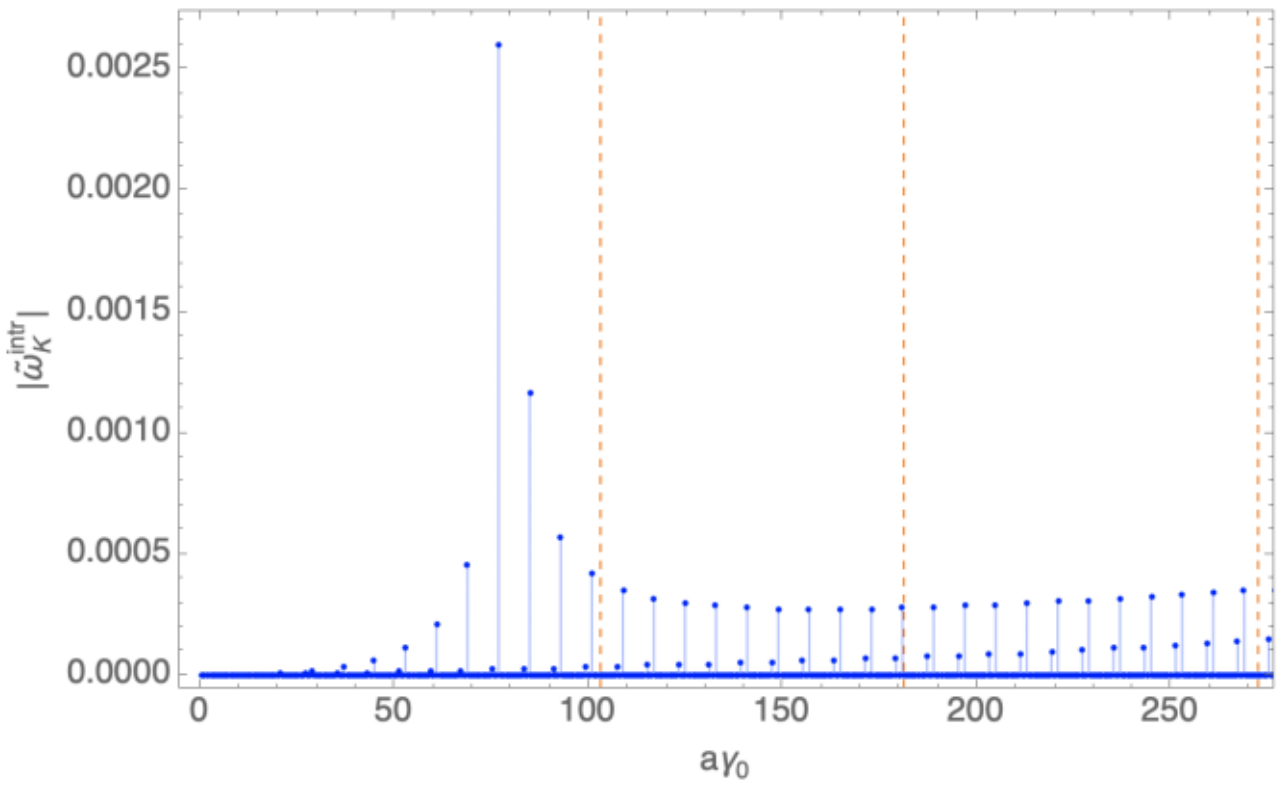

**Figure A8.3.1:** Spectrum of intrinsic resonance strength in the TDR lattice, using a vertical normalized amplitude of $10\pi$ mm · mrad. The three vertical dashed lines (from left to right) represent the Z, W, and Higgs energies, respectively.

For a Gaussian beam with a rms vertical emittance of $\epsilon_{y,rms}$, the depolarization when crossing a single intrinsic resonance at $\nu_0 = K$ is:

$$\frac{P_f}{P_i}\left(K, \epsilon_{y,rms}, \alpha\right) = \frac{1 - \frac{|\widetilde{\omega}_K^{intr,\pm}(\epsilon_{y,rms})|^{2^2}}{\alpha}}{1 + \frac{|\widetilde{\omega}_K^{intr,\pm}(\epsilon_{y,rms})|^{2^2}}{\alpha}} \tag{A8.3.2}$$

Numerous intrinsic resonances are encountered during the acceleration process. For an initial 100% vertically polarized beam, the vertical beam polarization at a given time $t$ during the acceleration, denoted as $P_{trans}^{intr}(t)$, can be reasonably approximated by multiplying the remaining polarization resulting from each intrinsic resonance encountered by the beam before time $t$:

$$P_{trans}^{intr}(t) \approx \prod_{K \le G\gamma(t)} \frac{P_f}{P_i}(K, \epsilon_{y,rms}, \alpha) \tag{A8.3.3}$$

Then, the final polarization after the acceleration will be $P_{trans,f}^{intr} = P_{trans}^{intr}(t_{ramp})$.

We began with an initial vertical rms emittance of 6.5 nm at an injection energy of 30 GeV and achieved an equilibrium vertical rms emittance of 2 pm at 120 GeV following dedicated orbit and optics correction. We conducted an analysis of the evolution of the vertical rms emittance throughout the acceleration process. For each intrinsic resonance within the working beam energy range, we adjusted its strength based on the vertical rms emittance at the time of crossing. Subsequently, we employed Eq. (A8.3.2) and (A8.3.3) to estimate the polarization loss after traversing all intrinsic resonances during the acceleration. The crossing of spin resonances was determined using a cosine-shaped energy ramping curve. As a result, $P_{trans,f}^{intr}$ stands at 51.5%, 47.7%, and 41.5% after the acceleration to 45.6 GeV, 80 GeV, and 120 GeV, respectively.

We also conducted an assessment of the strengths of imperfection resonances within a single, imperfect TDR Booster lattice following comprehensive error modeling and subsequent corrections. This evaluation was executed utilizing the DEPOL algorithm, which incorporates orbit and optics parameters computed through the SAD code [31]. As depicted in Fig. A8.3.2, the spectrum illustrates the strength of imperfection resonances throughout the working beam energy range. Generally, the strengths of these imperfection resonances exhibit an increasing trend with energy, with the notable exception of the first super-strong intrinsic resonance at $a\gamma$ = 76, which aligns well with theoretical predictions. Notably, no imperfection resonances induce more than 1% depolarization in the Z and W modes. Nevertheless, it becomes apparent that there exist several imperfection resonances at ultra-high energies that could potentially lead to significant depolarization.

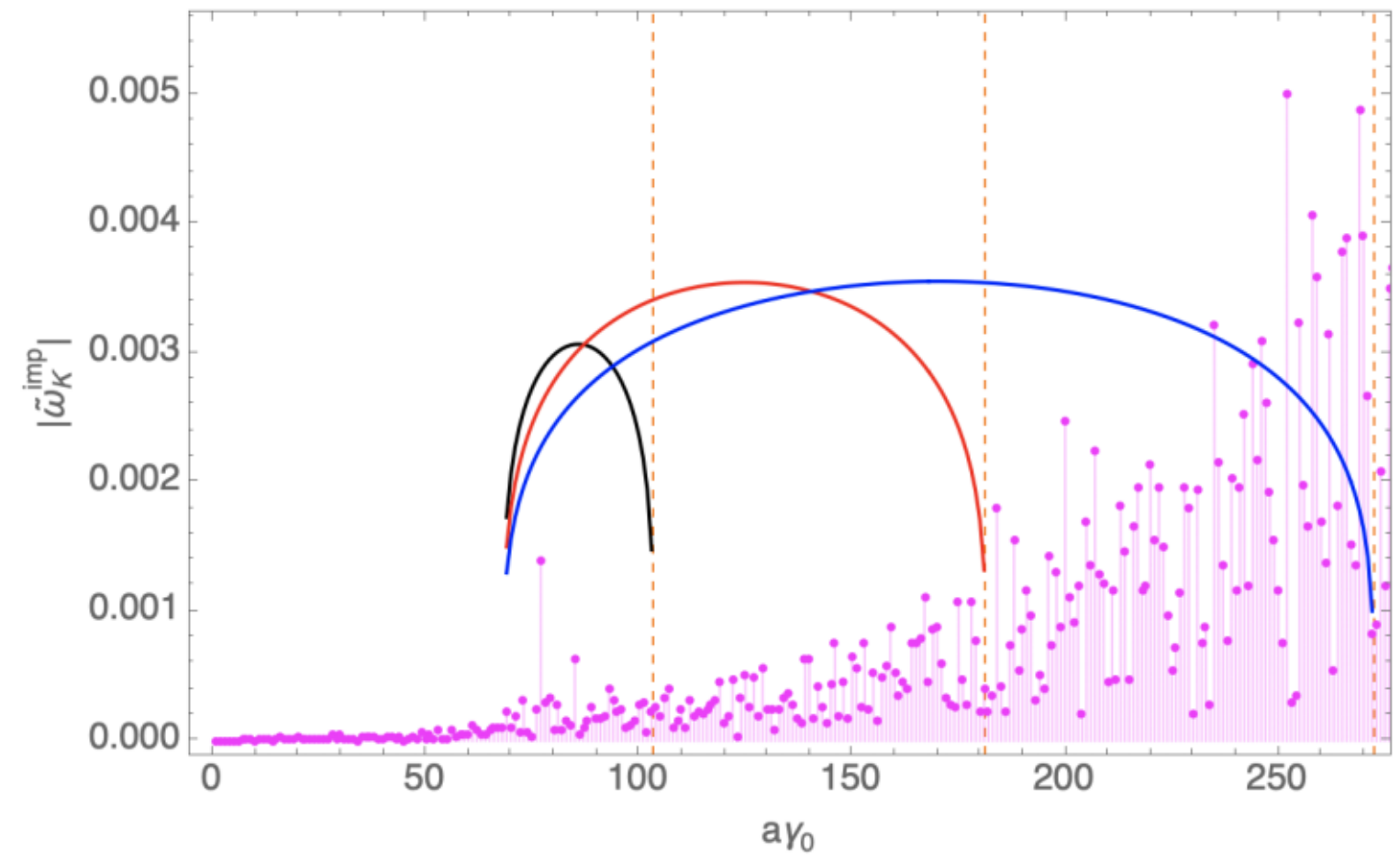

**Figure A8.3.2:** Spectrum of imperfection resonances for an imperfect Booster lattice, spanning the working energy range. The three vertical dashed lines correspond to the three extraction energies. The green, magenta, and brown curves represent the upper strength limit for each imperfection resonance, with the depolarization upon its crossing remaining below 1% for the Z-mode, W-mode, and H-mode, respectively.

We quantified the depolarization resulting from the crossings of imperfection resonances through a method akin to that employed for intrinsic resonances. Beginning with an initial 100% vertical polarization, we can approximate the vertical polarization at a given time $t$ during acceleration, denoted as $P_{trans}^{imp}(t)$, by successively multiplying the polarization retained after each imperfection resonance encounter that occurs prior to time $t$:

$$P_{trans}^{imp}(t) \approx \prod_{K \leq G\gamma(t)} \frac{P_f}{P_i}(K, \alpha) = \prod_{K \leq G\gamma(t)} \left[2e^{-\frac{\pi\left|\tilde{\omega}_K^{imp}\right|^2}{2\alpha}} - 1\right] \tag{A8.3.4}$$

Subsequently, the ultimate vertical polarization following the entire acceleration process can be expressed as $P_{trans,f}^{imp} = P_{trans}^{imp}(t_{ramp})$. As a result, $P_{trans,f}^{intr}$ is 99.5%, 97.6%, and 56.7%, upon reaching the energy levels of 45.6 GeV, 80 GeV, and 120 GeV, respectively.

By combining the influences of both intrinsic and imperfection resonances, we can now assess the comprehensive depolarization effects arising from both types of resonances:

$$P_{trans}(t) \approx P_{trans}^{intr}(t) \times P_{trans}^{imp}(t) \tag{A8.3.5}$$

Consequently, the final beam polarization after the entire acceleration process is denoted as $P_{trans,f} = P_{trans}(t_{ramp})$. Subsequently, $P_{trans,f}$ is found to be 51.2%, 46.6%, and 23.5% after reaching energy levels of 45.6 GeV, 80 GeV, and 120 GeV, respectively.

While a more comprehensive, element-by-element tracking of the entire acceleration process provides a higher precision in determining the polarization transmission efficiency, the above estimation already highlights the primary factors contributing to depolarization. On one hand, enhancing the vertical focusing to shift the first super-strong resonances to a higher beam energy is advantageous. This adjustment is crucial for mitigating depolarization during the acceleration to the Z and W energies. On the other hand, it is essential to investigate the lattice structure's sensitivity to misalignment errors concerning the strengths of imperfection resonances. This study is a vital component of lattice design and optimization.

## A8.4: Radiative Depolarization in the Collider

It is well known that the stochastic nature of synchrotron radiation causes radiative depolarization, namely the spin diffusion effect [32]. However, there are two different theories that describe the underlying mechanism of spin diffusion. On one hand, a photon emission perturbs the orbital motion and leads to a change in $\hat{n}$ and thus a slightly different $\hat{S} \cdot \hat{n}$. The stochastic nature of synchrotron radiation then leads to a diffusion and thus a reduction of beam polarization, namely "non-resonant spin diffusion." This is the basis of the Derbenev-Kondratenko formula [10] (D-K formula) on the equilibrium beam polarization.

$$P_{eq} = \frac{-\frac{8}{5\sqrt{3}} \times \oint d\theta \left\langle \frac{1}{|\rho|^3} \hat{b} \cdot \left( \hat{n} - \frac{\partial \hat{n}}{\partial \delta} \right) \right\rangle}{\oint d\theta \left\langle \frac{1}{|\rho|^3} \left[ 1 - \frac{2}{9} (\hat{n} \cdot \hat{s})^2 + \frac{11}{18} \left( \frac{\partial \hat{n}}{\partial \delta} \right)^2 \right] \right\rangle} \tag{A8.4.1}$$

The time constant of the spin diffusion effect is:

$$\tau_d^{-1} = \frac{5\sqrt{3}}{8} \frac{r_e \gamma_0^5 \hbar}{m_e} \frac{1}{2\pi} \oint d\theta \left\langle \frac{1}{|\rho|^3} \left[ \frac{11}{18} \left( \frac{\partial \hat{n}}{\partial \delta} \right)^2 \right] \right\rangle \tag{A8.4.2}$$

The key to evaluating the DK formula lies in assessing the spin-orbit coupling function $\frac{\partial \hat{n}}{\partial \delta}$. On the other hand, in the proximity of spin resonances where spin motion becomes far from adiabatic, $\hat{S} \cdot \hat{n}$ is no longer an invariant of motion. The repetitive fast but "uncorrelated" crossings of the spin resonances, caused by stochastic photon emissions, become another source of radiative depolarization, referred to as "resonant spin diffusion" [33]. Especially at ultra-high beam energies, the combined influence of synchrotron oscillation and synchrotron radiation can result in "repetitive fast but uncorrelated crossings" of the underlying spin resonances [34]. The influence of resonant spin diffusion extends beyond the immediate proximity of the underlying spin resonances.

Regardless of the underlying mechanism of radiative depolarization, the strength of the depolarization effects can be quantified by the ratio $\frac{\tau_{\mathrm{BKS}}}{\tau_{\mathrm{d}}}$, and the equilibrium polarization can be estimated using Eq. (A8.1.4).

As previously mentioned, stochastic photon emissions induce a random walk-in particle energies, resulting in an rms relative beam energy spread $\sigma_\delta$ and a spread in spin precession frequencies denoted as $\sigma_0 = \nu_0 \sigma_\delta$. This quantity also signifies the amplitude of variation in the spin precession rate of any particle within the beam due to synchrotron radiation and synchrotron oscillation. The modulation index $\sigma = \frac{\sigma_0}{\nu_z}$, where $\nu_{\mathrm{z}}$ represents the synchrotron tune, characterizes the spread of spin precession frequencies in the beam relative to the spacing of adjacent synchrotron sideband spin resonances.

At low energies, typically when $\sigma_0 \ll 1$, especially when the modulation index $\sigma \ll 1$, the design energy can be selected such that $\nu_s$ of beam particles falls between significant spin resonances. This is the regime where the theory of non-resonant spin diffusion applies. In this regime, $\frac{\partial \hat{n}}{\partial \delta}$ can be solved perturbatively, revealing the influence of spin resonances. Subsequently, the depolarization effect can be evaluated as described in [33]:

$$\frac{\tau_{BKS}}{\tau_d} \approx \frac{11}{18} \sum_{k=n-l}^{n+l} \sum_{m=-\infty}^{\infty} \left( \frac{\nu_0^2 |\tilde{\omega}_k|^2 e^{-\sigma^2} I_m(\sigma^2)}{[(\nu_0 - k - m\nu_z)^2 - \nu_z^2]^2} \right) \tag{A8.4.3}$$

where $I_m$ is the modified Bessel function,

$$\tilde{\omega}_k = \frac{1}{2\pi} \int_0^{2\pi} \Delta\Omega_x e^{i\nu_0(\Phi(\theta') - \theta') + ik\theta'} \, d\theta'$$

is the strength of the integer spin resonance $\nu_0 = k, k \in \mathbb{Z}$, which the tilt of $\hat{n}_0$, with $\Delta\Omega_x$ representing the radial component of $\Delta\vec{\Omega}(\theta)$ due to magnet errors and corrector fields, $\Phi(\theta') = R \int_0^{\theta} \frac{1}{\rho_x} d\,\theta'$ for the average bending radius $R$ and local bending radius $\rho_x$. The integer part of $\nu_0$ is denoted by $n$, and only a few Fourier harmonics of $\tilde{\omega}_k$ with $|k - n| \leq l$ have a strong influence, where $l$ is a small positive integer to be determined as a result of check of convergence.

For very small values of modulation index σ, $I_m(\sigma^2)$ vanishes for $m \neq 0$, and Eq. (A.8.4.3) reduces to

$$\frac{\tau_{BKS}}{\tau_d} \approx \frac{11}{18} \sum_{k=n-l}^{n+l} \frac{\nu_0^2 |\tilde{\omega}_k|^2}{[(\nu_0 - k)^2 - \nu_z^2]^2} \tag{A8.4.4}$$

which exhibits the first-order "parent" synchrotron spin resonances $\nu_0 = k \pm \nu_z, k \in \mathbb{Z}$, the width of the dip in equilibrium polarization near these resonances scales with $\nu_0^2$ and the amplitude of $\tilde{\omega}_k$, which also generally increase with energy. Depolarization caused by these first-order "parent" synchrotron spin resonances becomes more pronounced at higher beam energies, often outweighing the impact of first-order "parent" betatron spin resonances $\nu_0 = k \pm \nu_x, k \in \mathbb{Z}$, and $\nu_0 = k \pm \nu_y, k \in \mathbb{Z}$. To mitigate the influence of these spin resonances, specially designed correction coils, forming closed-orbit vertical bumps, can be employed to generate "anti-harmonics" that partially counteract these nearby integer resonances [35]. With a larger value of $\sigma$, the depolarization effect displays resonant behavior at the higher-order synchrotron sideband spin resonances $\nu_0 = k + k_z\nu_z, k, k_z \in \mathbb{Z}$. For a fixed $\nu_z$,, σ scales with $\gamma^2$, and higher-order synchrotron sideband spin resonances become more prominent as beam energies increase. To attain a higher equilibrium polarization level, it is a common practice to select $[a\gamma] \approx 0.5$, where $[x]$ denotes the fractional part of $x$.

At higher energies, $\sigma_0$ typically increases, causing some beam particles to inevitably intersect with important spin resonances during the process of synchrotron radiation and synchrotron oscillation, resulting in resonant spin diffusion. Notably, in Ref. [33], it was argued that there exists a distinct parameter space related to the impact of stochastic photon emissions on the crossings of an underlying spin resonance during precession-rate oscillations driven by synchrotron oscillations. This parameter space is characterized by the "correlation index" $\kappa = \frac{\nu_0^2 R}{\nu_z^3 c\tau_{\text{BKS}}}$. The perturbative depolarization theory of higher-order synchrotron sideband spin resonances is applicable when $\kappa \ll 1$. This regime will be referred to as the "correlated regime" for brevity.

In contrast, when the radiation is highly intense or the synchrotron tune is extremely small, the condition $\kappa \ll 1$ no longer applies. Furthermore, if $\sigma \gg 1$ is satisfied, the successive crossings of the underlying spin resonance in synchrotron oscillations become entirely uncorrelated. In such a scenario, it has been suggested that the depolarization effect can be assessed as follows:

$$\frac{\tau_{BKS}}{\tau_d} \approx \frac{c\tau_{BKS}\sqrt{\pi/2}}{R}\sum_{k=n-l}^{n+l}\frac{|\widetilde{\omega}_k|^2}{\sigma_0}exp[-\frac{(\nu_0-k)^2}{2\sigma_0^2}] \quad \text{(A8.4.5)}$$

This spin diffusion theory will hereafter be denoted as the "uncorrelated regime." The theories of the "correlated regime" and "uncorrelated regime" yield significantly different depolarization effects, and the precise definition of the conditions for their application remains somewhat unclear at this point.

We carry out simulation studies for the CEPC CDR Collider lattice and parameters. To obtain an initial estimate of radiative depolarization in the Collider, we introduced typical misalignment errors and relative field errors of magnets into the Collider lattice. These errors were assumed to follow a Gaussian distribution truncated at ±3σ. Subsequently, we conducted a detailed closed orbit and optics correction process to restore the lattice's performance. Additionally, we adjusted the vertical emittance to the target value of 1.6 pm by deliberately tilting all quadrupoles in the four straight section regions by 0.2 degrees. We also introduced zero-length skew quadrupoles adjacent to the final focusing quadrupoles to account for the solenoid fringe field contribution.

The rms values for the horizontal and vertical closed orbit errors are 37 μm and 28 μm, respectively. The rms values for the horizontal and vertical beatings are 0.36% and 3.4%, respectively. It's important to note that we did not model the relative offset errors between BPMs and adjacent quadrupole magnets.

Subsequently, we conducted Monte-Carlo simulations to assess the depolarization effects using PTC [36,37] with this imperfect lattice. The simulation results were then compared with the theoretical predictions, as illustrated in Fig. A8.4.1 [38].

For the Z and W energies, the Monte-Carlo simulation results exhibit a closer agreement with the theory of the correlated regime, indicating a clear presence of the fine structure of higher-order synchrotron sideband spin resonances. However, at the Higgs energy, despite the very low equilibrium polarization level, the simulation results closely align with the predictions of the theory of the uncorrelated regime. Generally, they are higher than the predictions of the theory of the correlated regime, with no clear indications of higher-order synchrotron sideband spin resonances.

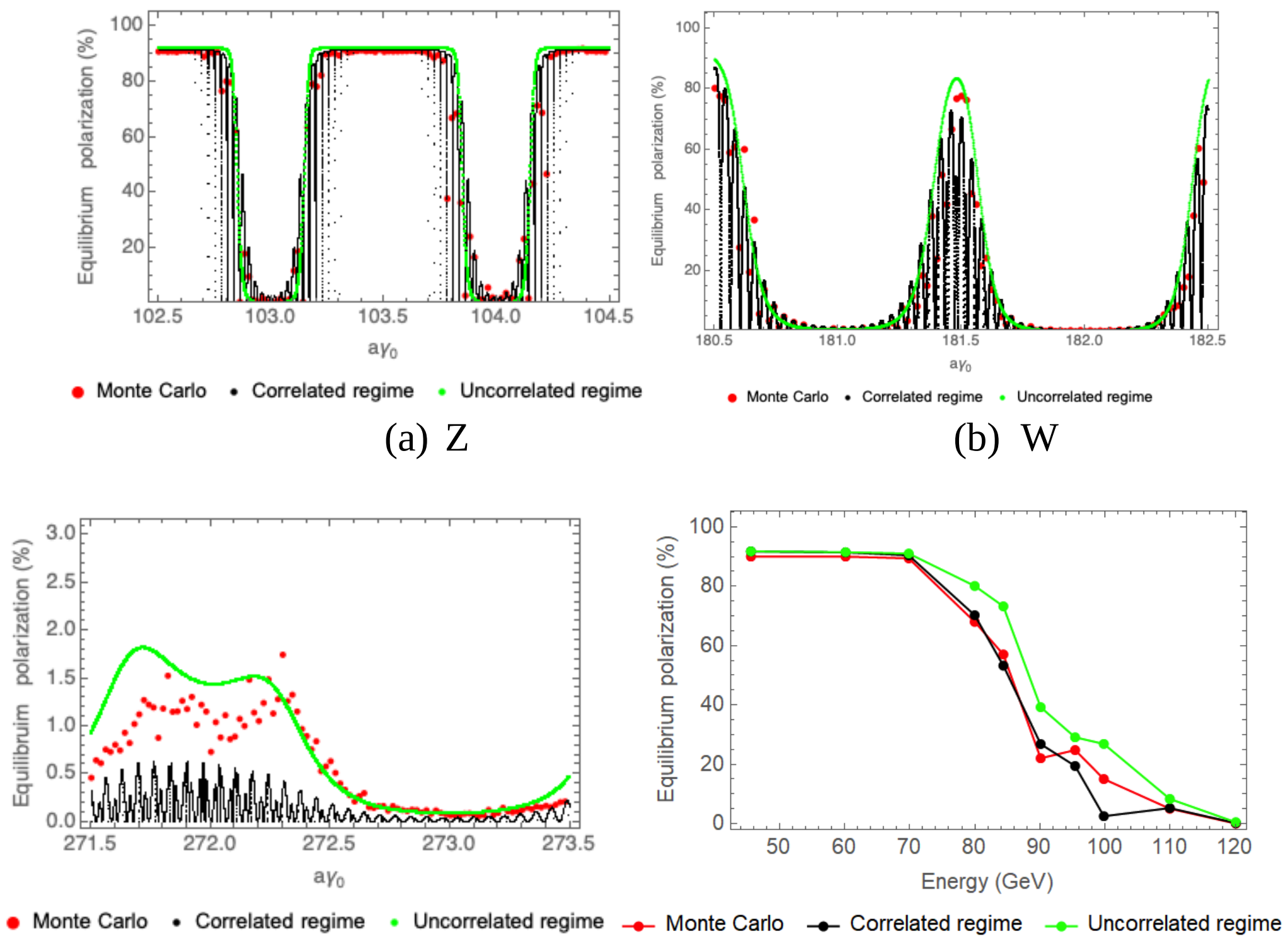

(a) Z (b) W

(c) H (d) a range of beam energies

**Figure A8.4.1:** The equilibrium polarization as a function of $a\gamma_0$ near three working energies of the CEPC in the first three figures, and the equilibrium polarization as a function of beam energy in the last figure.

Furthermore, up to 90 GeV, the simulated equilibrium polarization closely aligns with the theory of the correlated regime and remains smaller than that predicted by the theory of the uncorrelated regime. However, as beam energies increase further, noticeable discrepancies emerge between the simulations and the theoretical predictions. Nevertheless, this outcome underscores the greater challenge in achieving a high level of equilibrium polarization at higher beam energies.

The results presented above pertain to a single imperfect lattice, and further investigations involving multiple error seeds are warranted to provide a more comprehensive understanding. Additionally, future studies should encompass a broader range of depolarizing contributors, such as the impact of detector solenoids, and incorporate a more comprehensive set of machine imperfection sources. These efforts will aid in predicting achievable beam polarization levels and in devising strategies to achieve high beam polarization.

Finally, as indicated by the comparison between simulation outcomes and theoretical predictions, it is conceivable that radiative depolarization may enter the "uncorrelated regime" at the ultra-high beam energies to be explored in the CEPC. Figure A8.4.2 illustrates the equilibrium beam polarization estimated using Eq. (A8.4.5), assuming $[a\gamma_0] = 0.5$, with the assumption that the two adjacent integer spin resonances are of equal strength, independent of beam energy. While the strength of integer spin resonances generally increases with energy, Eq. (A8.4.5) implies that radiative depolarization may weaken at higher beam energies. It is pertinent to investigate the applicability of harmonic spin matching at Higgs and $t\bar{t}$ energies and explore the feasibility of achieving a few percent beam polarization in these regimes.

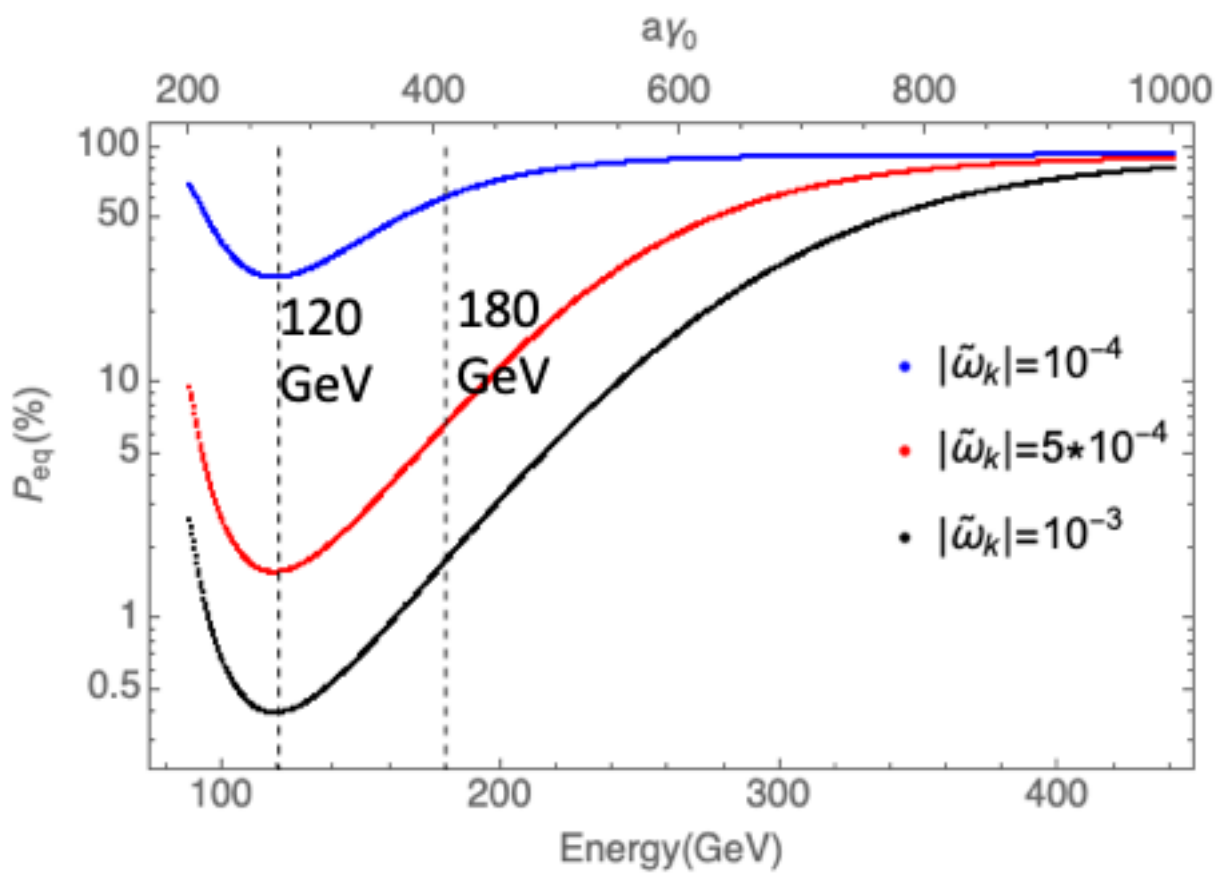

**Figure A8.4.2:** The equilibrium polarization as a function of beam energy for $[a\gamma_0] = 0.5$, for three different sets of adjacent integer spin resonance strengths.

## A8.5: Spin Rotator in the Collider

We conducted an investigation into the feasibility of achieving longitudinal polarization within the Collider at the Z energy, utilizing the CEPC CDR Collider lattice and its associated parameters. It's important to highlight that the spin rotator design we have explored is modular and does not disrupt the interaction region. Furthermore, these spin rotators can be seamlessly incorporated into the TDR lattice or any potential future redesign of the CEPC collider ring lattice.

To achieve longitudinal polarization at the IPs in the Collider, a pair of spin rotators must be inserted around each of the two IPs. We have opted for a solenoid-based spin rotator design [39]. Each spin rotator comprises a solenoid magnet section and a horizontal bending magnet section. In the spin rotator situated upstream of the IP (RotatorU), the solenoid section rotates the spin vector from the vertical direction to the radial direction by π/2 radians. Subsequently, the horizontal bending magnet section further rotates the spin vector by an odd multiple of π/2, effectively aligning the spin vector with the longitudinal direction at the IP. Conversely, the rotation of a spin vector from the longitudinal direction at the IP back to the vertical direction within the arc can be accomplished using a spin rotator downstream of the IP (RotatorD). This RotatorD employs a similar configuration to RotatorU but in a reversed sequence. This pair of spin rotators helps maintain vertical polarization for most of the Collider and mitigates significant depolarization.

Two potential arrangements exist for the placement of a pair of spin rotators around each IP, depending on whether the spin rotators on both sides of the IP share the same polarity (symmetric arrangement) or have opposite polarities (anti-symmetric arrangement), as depicted in Fig. A8.5.1 [39]. The anti-symmetric configuration has the advantage of precisely realigning the spin vector to the vertical direction within the arc, even when operating at energies other than the design energy. Furthermore, this arrangement aligns with the geometry of the Collider in the interaction region (IR), and thus has been selected for our study.

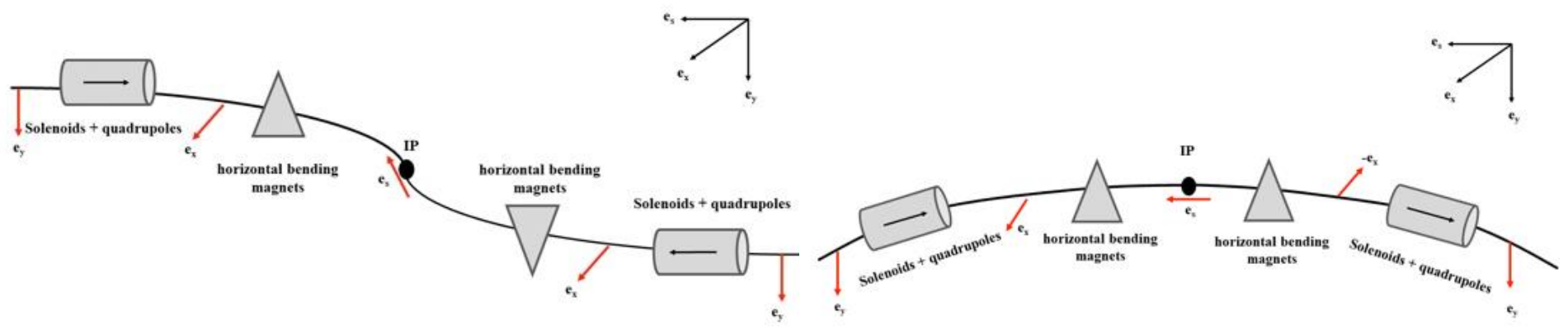

**Figure A8.5.1:** Two possible arrangements of the spin rotator magnets: anti-symmetric (left) and symmetric (right). The red arrow indicates the direction of polarization.

Given the presence of two IPs in the Collider, there exists flexibility in the solenoid helicity, allowing for the Sokolov-Ternov effect in the two arcs to either accumulate or cancel each other out. We acknowledge that the layout depicted in the right plot of Fig A8.5.2 can mitigate the radiative depolarization effects in the arc regions. However, our primary emphasis remains on the layout illustrated in the left plot to maintain the condition $\nu_0 \approx a\gamma$, which is essential for beam energy calibration.

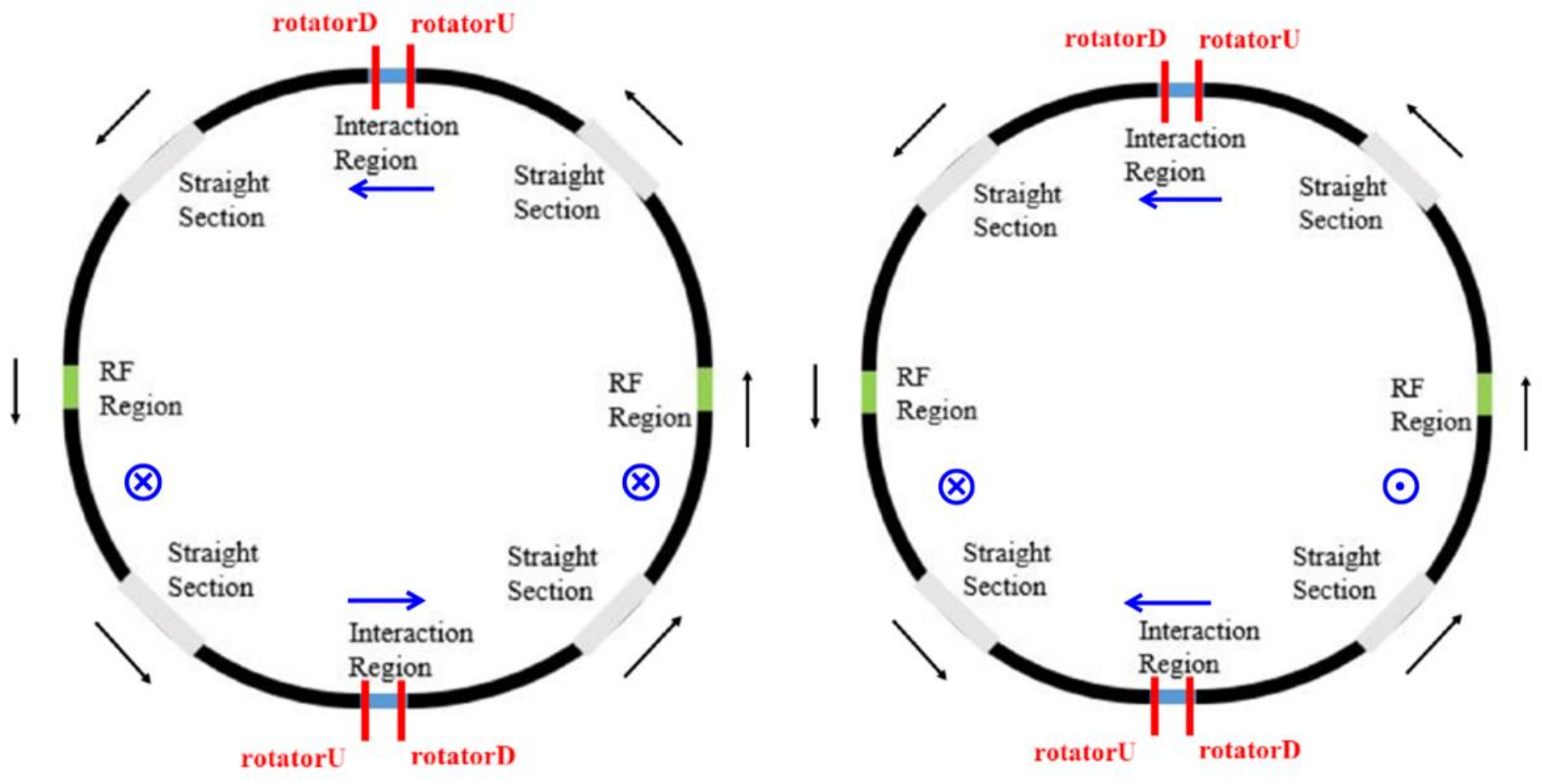

**Figure A8.5.2:** The locations of spin rotators in the Collider, taking the positron ring as an illustration. Black arrows indicate the direction of beam motion, Blue arrows indicate the direction of $\hat{n}_0$. The two figures differ in the helicity of the solenoids, leading to the same direction of $\hat{n}_0$ in both arcs (left plot) and opposite directions of $\hat{n}_0$in both arcs (right plot).

The necessary integral strength of the solenoid is approximately 240 T·m to achieve a π/2 spin vector rotation at a beam energy of 45.6 GeV. This equates to a total length of 30 meters when utilizing superconducting solenoid magnets with a field strength of 8 T. To counteract transverse coupling effects, the solenoid magnets are interspersed with quadrupoles. The lattice configuration and optical parameters for one such solenoid section are depicted in Fig. A8.5.3.

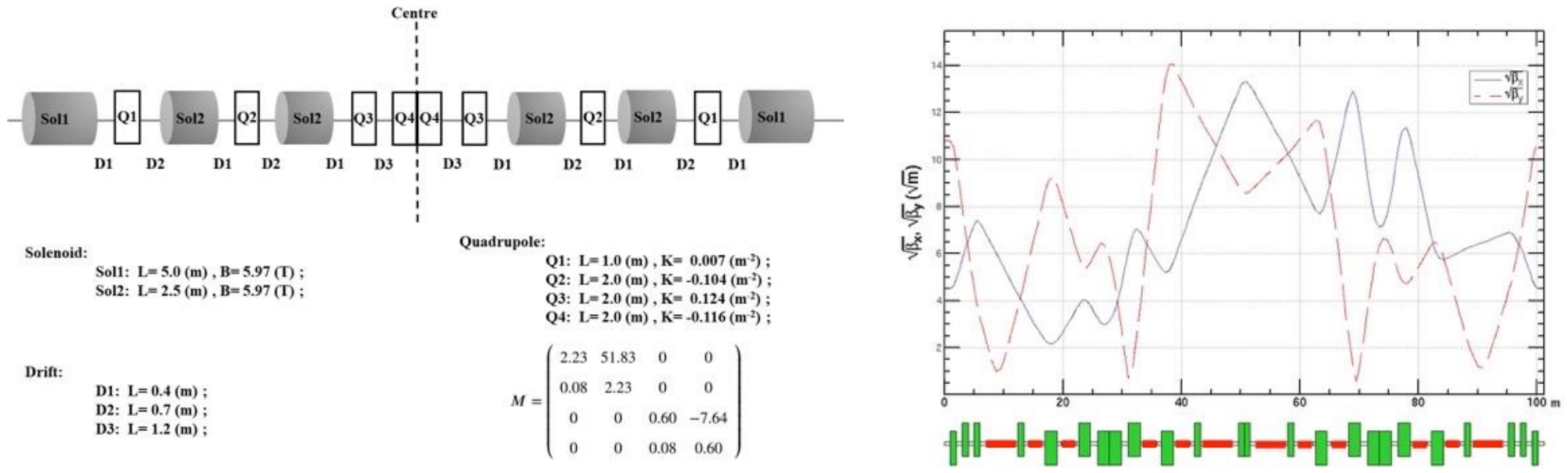

**Figure A8.5.3:** The left plot shows the lattice setup of one solenoid decoupling module, each solenoid section contains two such modules with addition quadrupoles for optics matching, the beam envelope function is shown in the right plot for a solenoid section.

To rotate the spins from the longitudinal to the radial direction within a horizontal bending magnet section, the total spin rotation angle must be an odd multiple $k$ of $\pi/2$, equivalent to an orbital bending angle of at least 15.18 mrad at 45.6 GeV. Given a half crossing angle of 16.5 mrad at the IP, the total bending angle required from the IP to the exit of the interaction region (IR) closely approximates the necessary value. Consequently, our choice is to insert the solenoid sections immediately adjacent to both sides of the IR, as illustrated in Fig. A8.5.4. To achieve the required spin rotation angle, additional bending magnet sections ($\Delta\theta 1$=1.39 mrad and $\Delta\theta 2$=2.65 mrad at 45.6 GeV) are essential in both spin rotators, positioned next to the solenoid sections. Each of these bending magnet sections comprises four identical 90-degree FODO cells, resembling the standard arc FODO cells but with reduced bending angles to form an achromat. In the counterpart region of the positron collider ring, the solenoid sections are replaced by straight sections with quadrupoles to maintain the necessary geometry and optics.

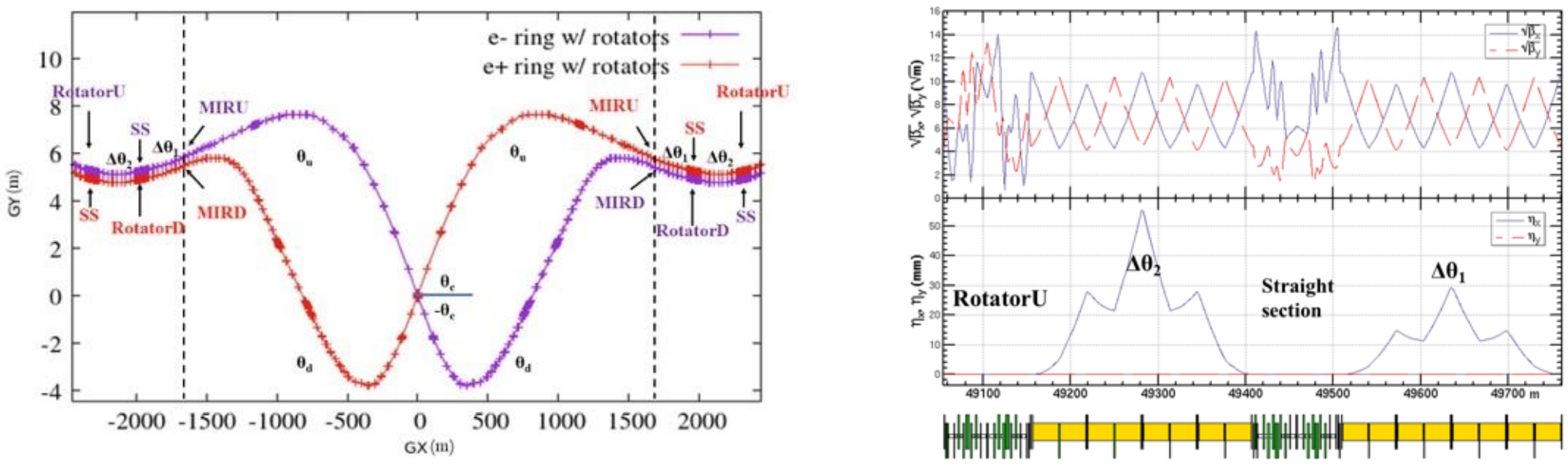

**Figure A8.5.4:** Layout of the spin rotators near the IR region (left plot), and the optical functions in the upstream rotator region (right plot).

Subsequently, we performed retuning of several quadrupoles in the straight sections to restore the fractional part of betatron tunes. Additionally, the deflection angles of the dipoles in the remaining segments of the Collider were reduced accordingly to ensure that the total horizontal deflection angle across the entire ring amounts to $2\pi$. To address the leading-order chromaticity issues, adjustments were made to the sextupoles in the arc sections, utilizing the SAD software.

To assess the impact of the spin rotators on the orbital motion performance, we examine three distinct scenarios. The first scenario involves the Collider lattice without spin rotators, denoted as the CDR lattice. The second scenario entails the Collider lattice with spin rotator insertions to achieve longitudinal beam polarization, referred to as Solenoid On. The third scenario involves the Collider lattice with spin rotator insertions, but with the solenoids deactivated and the quadrupoles retuned to restore the lattice, designated as Solenoid Off. Table A8.5.1 provides a comparison of key orbital parameters among these three cases. Notably, the latter two cases exhibit an increase in the integer parts of the betatron tunes and minor differences in the momentum compact factor and U0. Additionally, the presence of the spin rotators leads to an approximately 2.8 km increase in the collider ring's circumference. Importantly, there are no distinctions in the orbital parameters between the Solenoid Off and Solenoid On scenarios.

**Table A8.5.1:** The comparison of several key lattice parameters between the CDR Collider lattice and the lattices with spin rotators (Solenoid On/Off)

| Parameters | CDR lattice | Solenoid On | Solenoid Off |
|---|---|---|---|
| Tune $\upsilon_x/\upsilon_y/\upsilon_z$ | 363.11/365.22/0.028 | 381.11/383.22/0.028 | 381.11/383.22/0.028 |
| Emittance $\epsilon_x/\epsilon_z$ | 0.18nm/0.886μm | 0.18nm/0.886μm | 0.18nm/0.886μm |
| Momentum compaction factor $\alpha_c$ | $1.11 \times 10^{-5}$ | $1.07 \times 10^{-5}$ | $1.07 \times 10^{-5}$ |
| Circumference (m) | 100016.35 | 102841.95 | 102841.95 |
| SR energy loss per turn $U_0$ (MeV) | 35.47 | 35.91 | 35.91 |
| $\beta$ function at IPs $\beta_x^*/\beta_y^*$ | 0.2/0.001 | 0.2/0.001 | 0.2/0.001 |

Simulation results indicate a moderate reduction in the dynamic aperture, as illustrated in Fig. A8.5.5. This reduction can be further optimized through the inclusion of additional sextupole families, a conventional step in lattice optimization. However, for the colliding beams at the design luminosity, it is crucial to evaluate beam lifetime, considering limitations imposed by the dynamic aperture. This assessment accounts for the combined effects of beam-beam interactions, beamstrahlung, radiation damping, and quantum excitation.

For this purpose, we performed element-by-element tracking in SAD, simulating the behavior of 1,000 particles over 100,000 turns (equivalent to 34 seconds), taking these effects into account. With sextupoles set to an optimized configuration, the particle loss during the tracking is negligible even at the nominal bunch population of $8\times10^{10}$. This observation indicates that the beam lifetime is entirely satisfactory in the presence of the spin rotators.

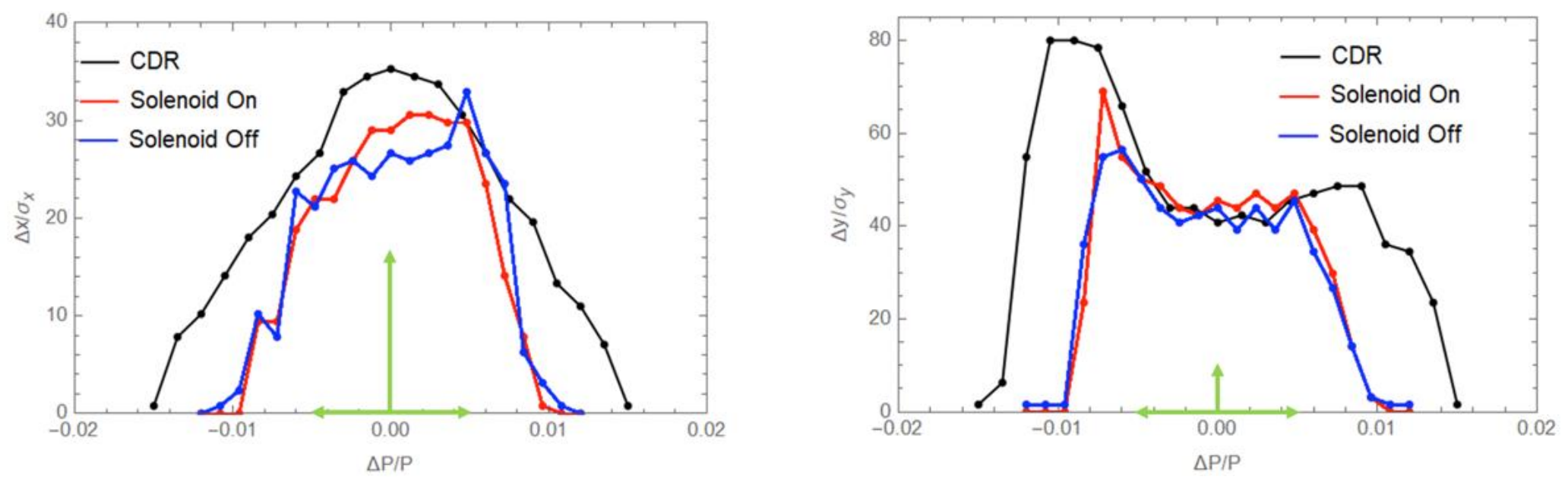

**Figure A8.5.5:** Comparison of the dynamic apertures between the CDR lattice and the CDR lattice with spin rotators (Solenoid On/Off)

We also conducted a numerical assessment of the spin motion performance using the BMAD/PTC code. In a practical storage ring, the solenoid magnetic field may not be perfectly compensated due to inherent magnet errors. These magnet errors induce spin resonances and consequently result in a diminished equilibrium beam polarization. To account for these imperfections, we introduced relative field errors for solenoids and quadrupoles in the solenoid sections, with a root-mean-squared value of 0.05%. Additionally, relative roll errors were introduced for the quadrupoles, with a root-mean-squared value of 0.01%.

We conducted a comparison between the longitudinal projection of beam polarization at the IP and the equilibrium polarization at the Z-pole. As depicted in Fig. A8.5.6, the longitudinal projection of $\hat{n}_0$ at the IP exhibits a symmetric distribution on both sides of the designated energy point, aγ = 103.5. Importantly, the influence of magnet errors is found to be negligible.

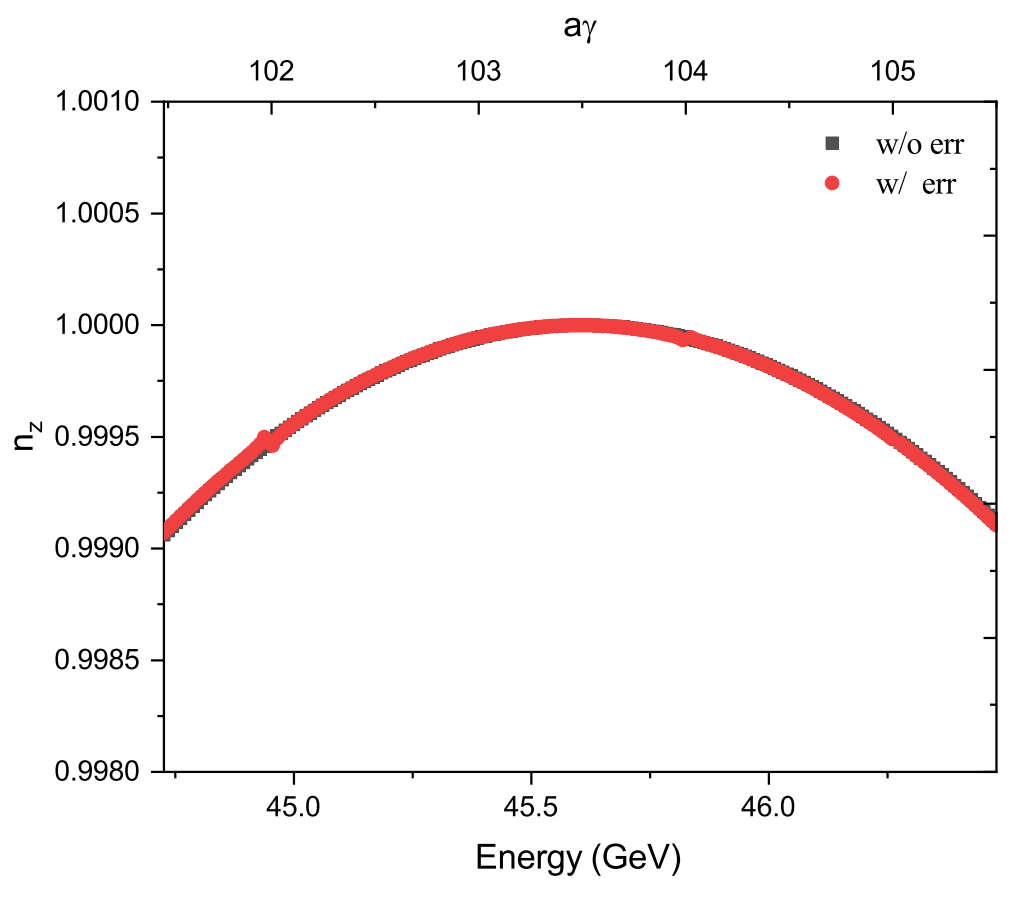

**Figure A8.5.6:** The longitudinal projection of $\hat{n}_0$ at the IP for different beam energies, for the lattices with and without magnet errors. The two curves overlap with each other indicating that the longitudinal projection of $\hat{n}_0$ is not sensitive to the magnet errors.

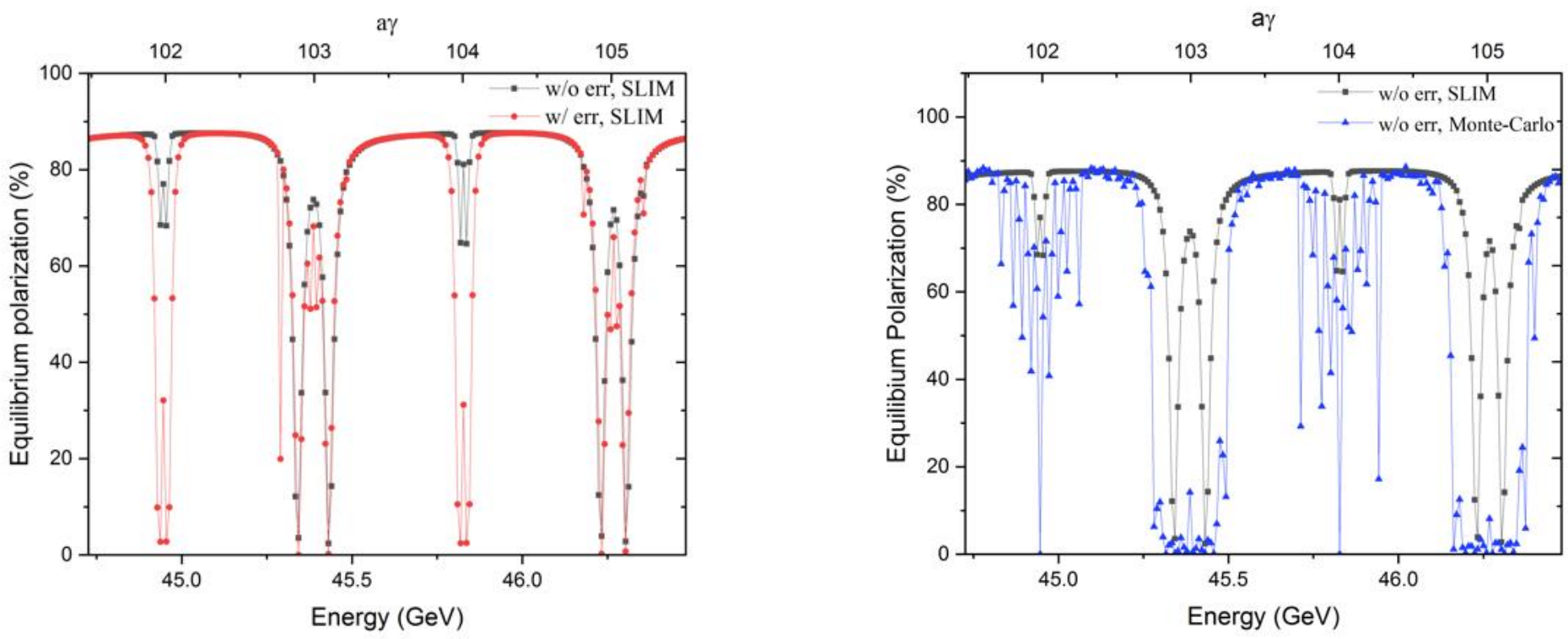

**Figure A8.5.7:** Simulations of the equilibrium beam polarization at different beam energies. The left plot shows the results of SLIM simulation for the lattices with and without magnet errors, the right plot shows the comparison between the SLIM and Monte-Carlo simulations for the lattice without magnet errors.

Fig. A8.5.7 illustrates a comparison between the equilibrium beam polarization simulated using SLIM within BMAD, both in the absence and presence of magnet errors. The equilibrium polarization level can reach very high values when the beam energy is far from spin resonances. Magnet errors tend to enhance depolarization near spin resonances; however, their effect on the polarization degree is negligible when the fractional part of aγ is close to 0.5. This underscores the robustness of the design against imperfections in the solenoid sections.

Additionally, Fig. A8.5.7 provides insights into the influence of higher-order spin resonances on equilibrium beam polarization, relative to simulation results obtained with SLIM. Here, the depolarization effects of higher-order spin resonances are simulated using a Monte Carlo method based on PTC. Notably, the spin resonance regions, where the equilibrium polarization level is low, become broader, and higher-order synchrotron sideband spin resonances become apparent. These simulations affirm that the depolarization effects introduced by the spin rotators can be largely mitigated within such an anti-symmetric structure. In future studies, we will introduce machine imperfections for the entire Collider equipped with spin rotators and evaluate the performance following dedicated orbit and optics corrections.

The presented spin rotator design is optimized for the Z energy but cannot achieve longitudinal polarization at the W and Higgs energies. Exploring the feasibility of a spin rotator design capable of covering a broader energy range, potentially encompassing the W and Higgs energy zones, is under consideration. In this tentative exploration, we assume that the solenoid magnets in the spin rotators will be deactivated when operating in the W and Higgs energy regimes. Subsequently, the quadrupoles within the solenoid sections need to be retuned to establish a standard uncoupled optics configuration. It's important to note that the same machine layout is retained for different beam energies, with fine adjustments required in regions like the IR to achieve varying $\beta^*$ values at IPs. These adjustments have a moderate impact on beam parameters, but they lead to a reduction in momentum acceptance, necessitating further optimization with additional families of sextupoles.

Furthermore, there is still room for optimizing the length of the spin rotators. For instance, the current design assumes longitudinal polarization for both beams. However, if we were to implement spin rotators exclusively for the electron beam, it would be possible to reduce the length of the bending angle compensation sections, resulting in a reduced increase in circumference, down to 1.18 km. Moreover, if the crossing angle at the IP could be altered to 30.35 mrad in a revised optics design, then the circumference increase could be further reduced to approximately 400 m for longitudinal polarization of either the electron beam alone or both beams.

## A8.6: Beam Energy Calibration

The calibration of beam energy using a polarized electron (positron) beam in a storage ring relies on the relationship that the closed orbit spin tune $\nu_0 \approx a\gamma$, where $a$ is very precisely known through quantum electrodynamics calculations and experimental measurements. Once $\nu_0$ can be accurately measured, the central beam energy can be determined with precision, with some systematic errors accounting for the discrepancy between $\nu_0$ and $a\gamma$, as well as the extrapolation from the measured beam energy (an average around the ring in terms of spin precession rate) to the center-of-mass energy at the interaction points.

The resonant depolarization (RD) method [40], which employs a pulsed "depolarizer" with frequency-scan capability to induce a narrow artificial spin resonance for beam depolarization, and measures $\nu_0$ according to the location of depolarization with the precisely known depolarizer frequency, offers the most precise measurements of beam energy thus far. Its implementation at the CEPC would necessitate achieving transverse polarization levels of at least 5% to 10% for both beams. Additionally, a laser-Compton polarimeter capable of measuring vertical polarization, drawing from the experiences at LEP [41], would be required.

Alternatively, the implementation of one or a few laser-Compton polarimeters in the Collider to monitor the turn-by-turn evolution of the longitudinal/radial component of the beam polarization could offer an alternative method. Through Fourier analysis, this approach can provide a spectrum of spin precession frequencies and enable the measurement of $\nu_0$. Known as the "free spin precession" approach, it has the potential to offer a rapid and independent measurement of beam energy alongside resonant depolarization. However, this method requires a relatively substantial longitudinal/radial polarization.

Regarding the RD method, we have conducted initial simulations using a Mathematica implementation of the simplified model by I. Koop [42]. This model includes one-turn spin precession considering the impact of underlying integer spin resonance, an RF depolarizer, synchrotron oscillation, and synchrotron radiation. The simulations involved tracking 120 particles for approximately 100,000 turns, which corresponds to minutes in the real world. These simulations were executed on a 120-core cluster and required only a few minutes to complete.

For the three distinct cases outlined in Table A7.6.1, we conducted simulations of the RD process, as depicted in Fig. A7.6.1. In the case of a non-colliding pilot bunch at Z energy, the figure shows a sharp decline in beam polarization near the spin resonance. However, at W energy, where the modulation index is increased, the polarization dip becomes less distinct due to the combined influence of synchrotron oscillation and synchrotron radiation. This increased modulation index has the potential to limit the precision of beam energy calibration.

In the case of colliding bunches at the Z-pole, the modulation index is even larger, making it exceedingly challenging, if not impossible, to employ resonant depolarization as a precise means of measuring the beam energy.

**Table A7.6.1:** Three cases for simulations of the resonant depolarization process.

| Cases | Z pilot bunch | Z colliding bunch | W pilot bunch |
|---|---|---|---|
| $a\gamma$ | 103.4921 | 103.4921 | 181.5123 |
| $\sigma_\delta$ | 3.847e-4 | 1.3e-3 | 6.79e-4 |
| $\nu_z$ | 0.03448 | 0.03448 | 0.062 |
| Modulation index | 1.15 | 3.9 | 2.0 |
| Resonance strength $\omega_K$ | 1e-4 | 1e-4 | 1.5e-4 |
| Scan rate $\varepsilon'=d\nu/dn$ | 1e-8 | 1e-8 | 2e-8 |
| $\omega_K^2/\varepsilon'$ | 1 | 1. | 1.125 |
| Scan time (s) | 130 | 130 | 65 |

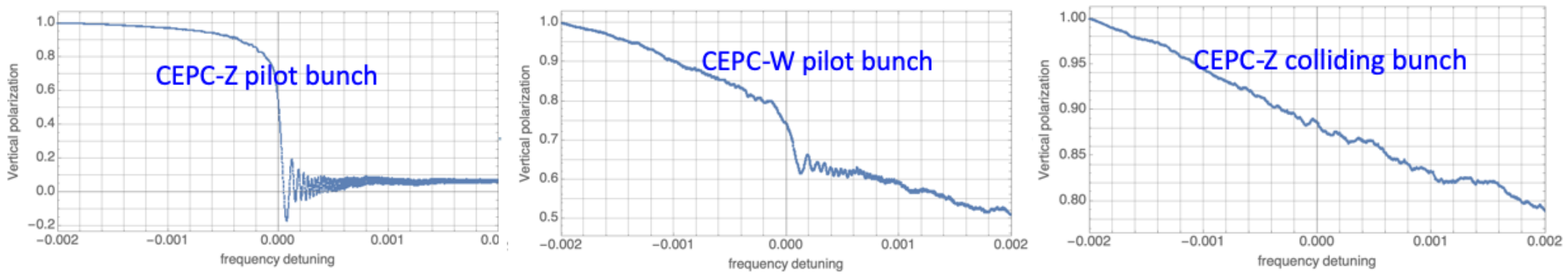

**Figure A7.6.1:** Simulations of the resonant depolarization process for three cases.

Further work is in the planning stages to enhance our comprehension of the precise procedures required for conducting these experiments. This includes carefully selecting parameters for key hardware components such as the RF depolarizer, as well as conducting a thorough analysis of error sources in the beam energy calibration process. Additionally, experiments are currently underway to test the resonant depolarization procedures at the BEPCII, and similar efforts can be pursued in the HEPS booster and HEPS storage ring in the forthcoming years.

## A8.7: Operation Schemes of Polarized Beams

### A8.7.1: Injection of Polarized Beams

In our colliding beam experiments, we operate under the assumption that polarized electron ($e^-$) bunches collide with unpolarized positron ($e^+$) bunches. This choice arises from the unavailability of a reasonable source for polarized positrons that could meet the requirements for top-up injection. As illustrated in Fig. A8.7.1, the commencement of a physics fill involves injecting all the colliding bunches. Subsequently, the colliding beam experiment unfolds with all these bunches in top-up injection mode. At this point, we can prepare polarized $e^+$ and $e^-$ bunches from the source and introduce them into the collider rings to conduct the resonant depolarization (RD) measurements.

For the purpose of beam energy calibration, we opt to utilize only one bunch per species. This chosen bunch can be efficiently depolarized, and once depolarized, it can be readily removed and replaced with a new polarized bunch. This process aligns with the capabilities of the injector chain and ensures a streamlined calibration procedure.

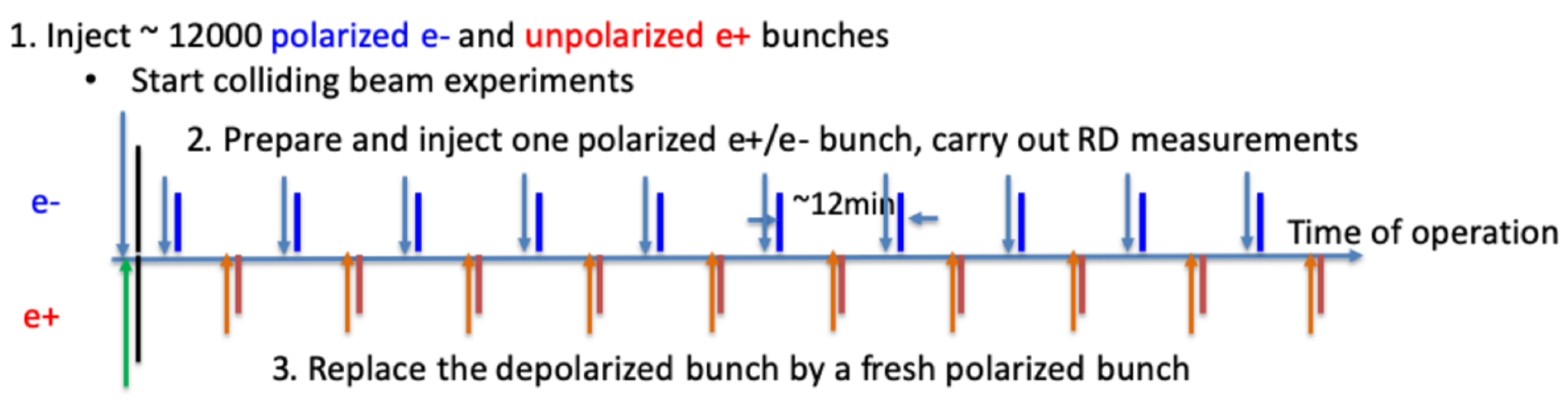

**Figure A8.7.1:** Operation scenario for the resonant depolarization scheme at the Z-energy.

## A8.7.2: Alternative Scheme for Beam Energy Calibration Only

In the context of beam energy calibration, a beam polarization of 5% to 10% is regarded as sufficient. An alternative approach under consideration involves harnessing the self-polarization build-up in the Collider to generate polarized electron and positron beams.

### *A8.7.2.1: Operation Scenario at the Z-pole*

At the Z-energy level, the Collider have $\tau_{BKS}$ durations extending over 250 hours, necessitating the introduction of asymmetric wigglers to enhance polarization build-up. Table A8.7.1 outlines a typical parameter configuration for these asymmetric wigglers and their impact on beam parameters, notably reducing $\tau_{BKS}$ to 20 hours. A total of 10 units of asymmetric wigglers were incorporated into the Collider lattice. While these asymmetric wigglers contribute to an increase in the beam energy spread, SAD simulations involving 1,000 particles, including element-by-element synchrotron radiation effects, indicated no particle loss within 200,000 turns. Consequently, the lifetime constrained by the dynamic aperture is not a concern for this set of wigglers. On the flip side, the presence of these wigglers amplifies higher-order synchrotron sideband spin resonances, and their influence on equilibrium beam polarization was assessed through Monte-Carlo simulations. The given parameter settings suggest that achieving over 10% beam polarization is feasible within a time frame of 2.6 hours.

**Table. A8.7.1:** Parameters of asymmetric wigglers in the Collider for self-polarization at Z-energy

| Parameters | Wigglers off | Wigglers on |
|---|---|---|
| Wiggler parameters $B_+$(T)/$B_-$(T)/$L_+$(m)/$L_-$(m) | --- | 0.6/-0.15/1/2 |
| SR radiation per turn $U_0$(MeV) | 37 | 49 |
| Rms energy spread $\sigma_\delta$ | 4e-10 | 1.2e-3 |
| Sokolov-Ternov time $\tau_{BKS}$ (h) | 253 | 20 |
| Equilibrium polarization w/o considering radiative depolarization $P_\infty$ | 92% | 82% |

The operational scenario is outlined in Fig. A8.7.2. It is assumed that there are 144 non-colliding bunches in each beam designated for dedicated resonant depolarization measurements. At the beginning of each physics run, these unpolarized bunches are introduced into the collider ring. Subsequently, the asymmetric wigglers are activated to expedite the polarization build-up. After approximately 2.6 hours, the beam polarization of these bunches reaches 10%. Following this, the wigglers are deactivated, and approximately 12,000 colliding bunches in each beam are injected into the Collider to commence physics data collection. Resonant depolarization measurements can then be conducted every 12.5 minutes for each beam on a single bunch; after that this unpolarized bunch is refilled. These measurements proceed in a bunch-by-bunch manner, allowing for continuous monitoring of the beam energies.

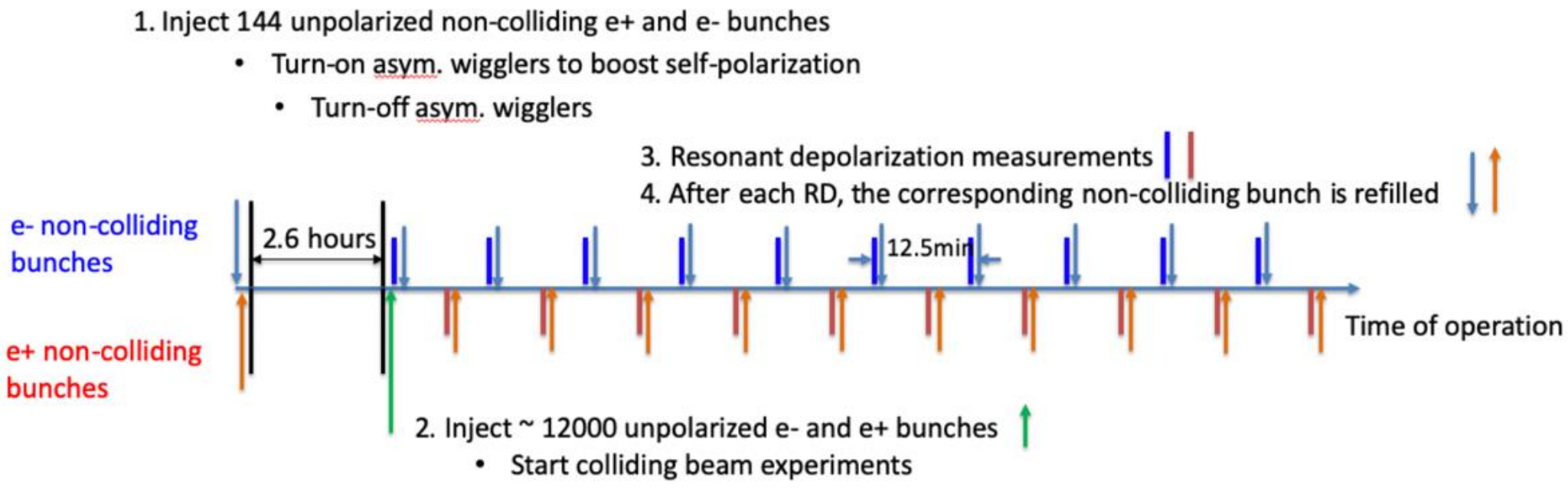

**Figure A8.7.2:** Operation scenario for the resonant depolarization scheme at the Z-energy, based on self-polarization build-up in the collider ring.

Therefore, there is a 30-hour interval between consecutive resonant depolarization measurements on a particular bunch, allowing ample time for a refilled bunch to achieve 10% beam polarization. To facilitate this, we contemplate a non-colliding bunch population of $4 \times 10^{10}$ particles immediately following refilling. The process of refilling a non-colliding bunch requires approximately 16 seconds and occurs every 6 minutes, representing only a fraction of the injection workload compared to refilling the colliding bunches.

### *A8.7.2.2: Operation Scenario at the W-energy*

At the W energy level, $\tau_{BKS}$ is approximately 15 hours, and it takes just 2 hours to attain a 10% beam polarization. Consequently, there is no requirement for the use of asymmetric wigglers. In this scenario, we assume the presence of 12 non-colliding bunches per beam, while there are approximately 1,200 colliding bunches per beam. Following the initial injection of all these bunches, the beam polarization reaches 10% within 2 hours, enabling resonant depolarization measurements to be conducted every 10 minutes for each beam. There is a 2-hour interval between successive resonant depolarization measurements on a specific bunch, providing ample time to restore sufficient beam polarization. It's worth noting that the beam lifetime for non-colliding bunches is in the range of tens of hours, meaning that only a small fraction of the depolarized bunch charge needs to be recovered every 2 hours.

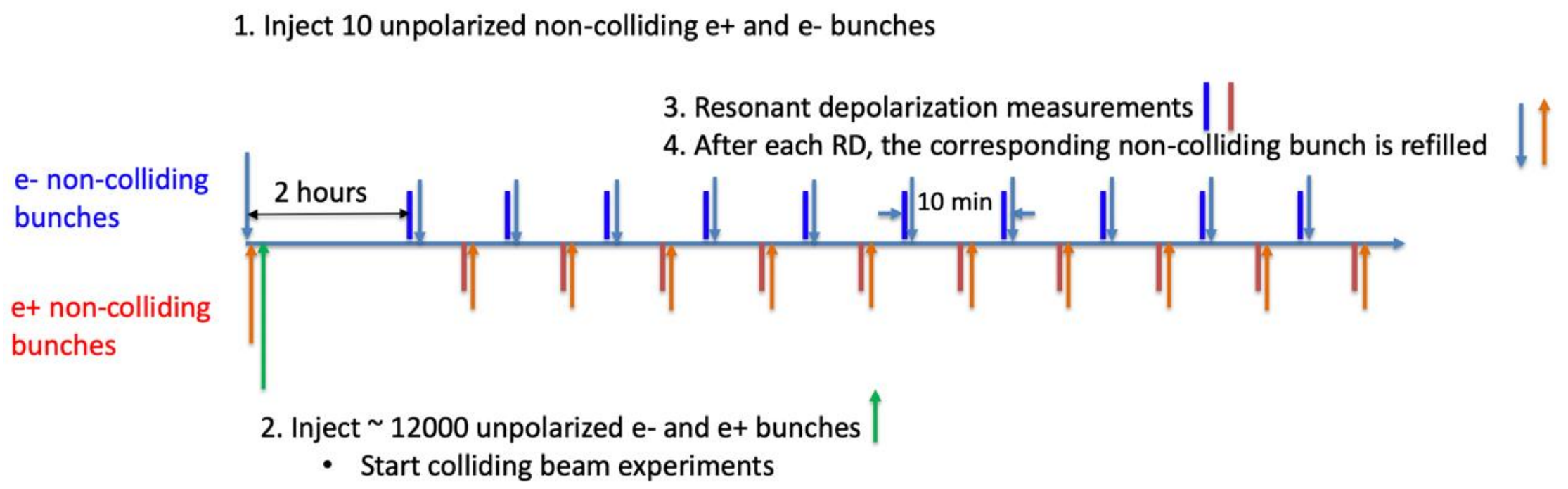

**Figure A8.7.3:** Operation scenario for the resonant depolarization scheme at the W-energy.

### *A8.7.2.3: Comparison between Two Schemes of Polarized Beam Generation*

Table A8.7.2 provides a preliminary comparison between the two schemes for polarized beam generation intended for beam energy calibration. Further refinement and elaboration of both schemes will be undertaken in subsequent stages.

**Table. A8.7.2:** Comparison between the two schemes of polarized beam generation for energy calibration

| Polarized beam generation scheme | | Sokolov-Ternov effect in the collider rings | Injection of polarized beams |
|---|---|---|---|
| Hardware | Polarized e- gun | None | Yes |
| | Asymmetric wigglers | In collider rings (Z-energy) | In the e+ damping ring |
| Polarization level | | 5%~10% | >60% for e-, >20% for e+ |
| Dead time for physics | | Initial ~2 hours in each fill | None |
| Frequency of RD measurements | | Every ~ 10 min per beam | More frequent for e- beam |
| RD on colliding bunches | | None | Possible |
| Horizontal polarization for the free spin precession method | | Strong spin flipper in the collider | Spin rotator in the transfer line, much higher polarization |

## A8.8: Compton Polarimetry

The applications of polarized beams at the CEPC necessitate highly precise measurements of beam polarization. Initially, it is anticipated that resonant depolarization (RD) measurements will involve an extensive frequency sweep of the depolarizer in just a few minutes. Within this timeframe, tens of measurements of vertical beam polarization are required, each with an accuracy better than 1%. Additionally, the realization of longitudinally polarized beams demands the precise determination of longitudinal beam polarization at the interaction points (IPs), preferably with an accuracy much better than 1%. This level of precision is essential to fully leverage the vast statistical potential offered by collision events at the Z-pole. Furthermore, the free spin precession measurement represents another promising method for beam energy calibration. This approach necessitates accurate, turn-by-turn measurements of the radial and/or longitudinal components of beam polarization from several independent polarimeters.

Moreover, for the default scheme involving the generation of polarized beams from the source, there is a requirement for polarimeters to measure polarization immediately after the polarized electron source, as well as in the positron Damping Ring and the Booster synchrotron. These multifaceted demands pose significant challenges in the R&D of polarization measurement techniques at the CEPC.

The Compton polarimeter is a well-established and non-invasive instrument ideally suited for electron beam energies ranging from a few GeVs and beyond. In this setup, the scattered electrons and photons can be independently measured and analyzed to accurately determine the components of the beam polarization vector for both $e^+$ and $e^-$ beams [43].

Our current focus has been on designing a Compton polarimeter capable of measuring the transverse polarization of the electron beam by detecting the spatial distribution of scattered electrons [44]. This specific aspect of polarization measurement will be presented in this section, with investigations into other aspects of polarization measurements planned for the future.

### A8.8.1: Layout of Compton Polarimetry

The system layout is illustrated in Fig. A8.8.1. In this setup, the electron bunches undergo nearly head-on collisions with a laser pulse. A dipole magnet, part of the CEPC Collider lattice, is used to bend and separate the scattered electrons from the primary electron beam and scattered photons. A free drift section, approximately 100 meters in length, is necessary to allow most of the scattered electrons to drift out of the beam pipe. Subsequently, the distance between the laser-electron interaction point (IP) and the detector spans approximately 157 meters. Position-sensitive detectors, situated perpendicular to the plane of beam propagation, are employed to capture and record the spatial distribution of the scattered electrons.

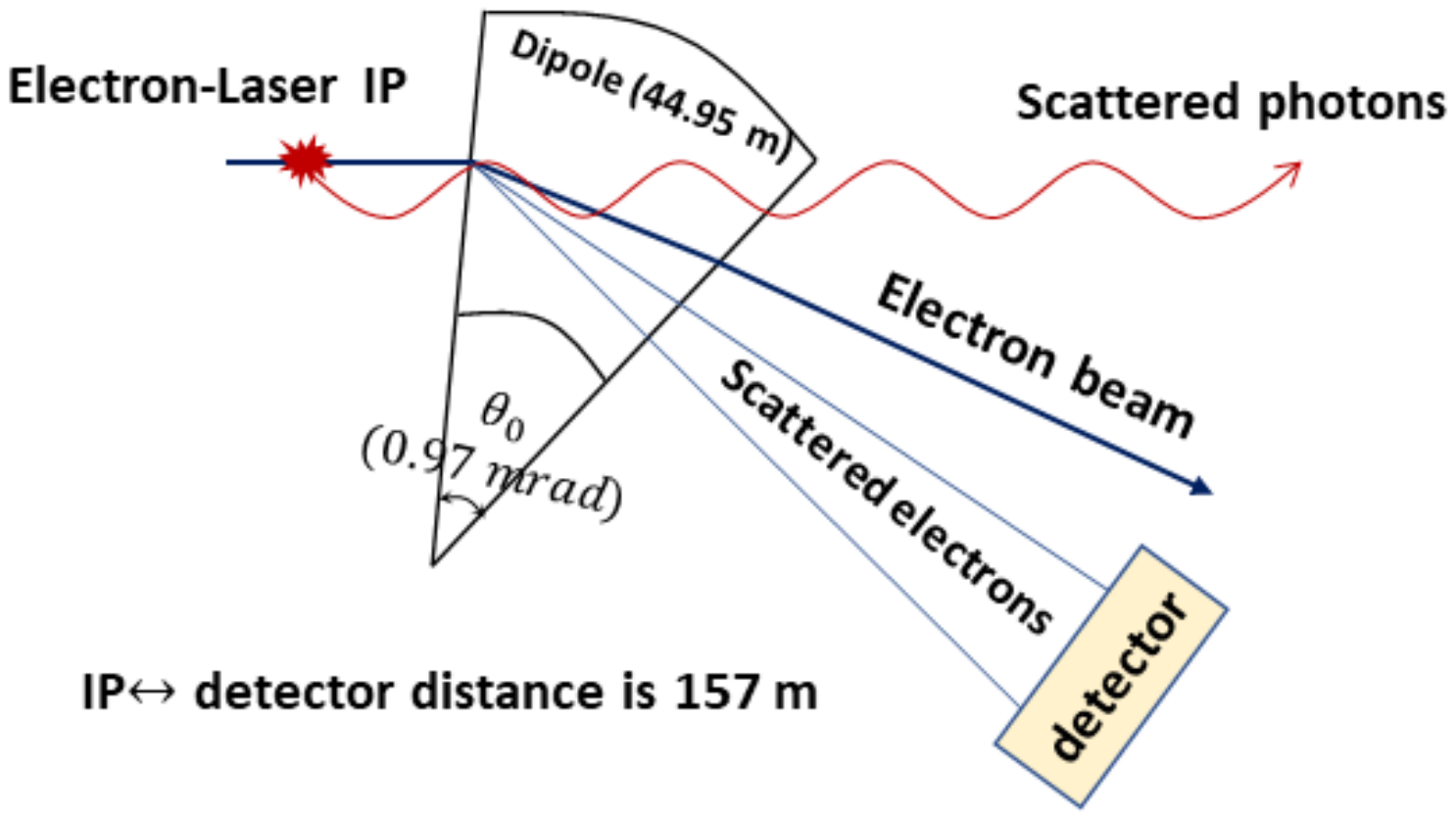

**Figure A8.8.1:** Schematic representation of the Compton polarimeter [44]: A transversely polarized electron bunch intersects with a circularly polarized 1064 nm laser pulse at the interaction point (IP). Following this collision, electrons are deflected away from the path of $\gamma$-rays by a dipole. The trajectory of the scattered $\gamma$-rays beam is illustrated by the red curve.

There are certain constraints related to the integration into the lattice. When employing a pulsed laser with a wavelength of 1064 nm, the minimum energy of scattered electrons is approximately 25.22 GeV. To ensure that scattered electrons within the energy interval of 25.11 GeV to 40 GeV can drift out of the beam tube, it is necessary for the drift distance ($L_2$) and the bending angle ($\theta_0$) of the bending magnet to satisfy the condition $L_2\theta_0 > 0.2255\, m$.

One of the candidate positions for the transverse polarimeter is located upstream, approximately 1 km before the electron-positron IP region. The selection of other candidate site locations is still under discussion. The implementation of polarization into the lattice will also necessitate some nontrivial local lattice adjustments, which will be addressed in the future.

### A8.8.2: Theoretical Basis

The principle of a Compton polarimeter is based on the Compton backscattering between a polarized electron beam and a circularly polarized laser. The Compton scattering process is described as: $\gamma(\omega_0, \mathbf{k}) + e^-(\varepsilon_0, \mathbf{p}) \to \gamma(\omega, \mathbf{k}') + e^-(\varepsilon, \mathbf{p}')$. The backscattered photons energy ($\omega$) is significantly larger than the laser photon energy ($\omega_0$):

$$\omega_{max} = \frac{\varepsilon_0^2}{\varepsilon_0 + m_e^2/4\omega_0} \qquad \text{(A8.8.1)}$$

where $\varepsilon_0$ and $\omega_0$ are, respectively, the initial energy of the electron beam and laser energy; $\omega$ is the energy of the scattered photons.

The spin-dependent differential Compton cross-section is given by [45]:

$$\frac{d\sigma}{d\omega d\varphi} = C[\frac{d\sigma_0}{d\omega d\varphi} + P_{\parallel}\frac{d\sigma_0}{d\omega d\varphi} + P_{\perp}\frac{d\sigma_{\perp}}{d\omega d\varphi}]$$

$$C = \frac{r_e^2(\varepsilon_0 + \omega_0 - \omega)}{\kappa^2(\varepsilon_0 + \omega_0)^2}$$

$$\frac{d\sigma_0}{d\omega} = \kappa\left(1 + \left(\frac{\varepsilon_0 + \omega_0}{\varepsilon_0 + \omega_0 - \omega}\right)^2\right) - 4\frac{(\varepsilon_0 + \omega_0)^2}{\kappa(\varepsilon_0 + \omega_0 - \omega)^2}(\kappa - \frac{\varepsilon_0 + \omega_0}{\varepsilon_0 + \omega_0 - \omega})$$

$$\frac{d\sigma_{\parallel}}{d\omega d\varphi} = \frac{\varepsilon_0 + \omega_0}{\varepsilon_0 + \omega_0 - \omega}(\frac{\varepsilon_0 + \omega_0}{\varepsilon_0 + \omega_0 - \omega} + 2)(\kappa - 2\frac{\varepsilon_0 + \omega_0}{\varepsilon_0 + \omega_0 - \omega})$$

$$\frac{d\sigma_{\perp}}{d\omega d\varphi} = 2\frac{\varepsilon_0 + \omega_0}{\varepsilon_0 + \omega_0 - \omega}\sqrt{\frac{\varepsilon_0 + \omega_0}{\varepsilon_0 + \omega_0 - \omega}(\kappa - \frac{\varepsilon_0 + \omega_0}{\varepsilon_0 + \omega_0 - \omega})}sin\varphi \qquad \text{(A8.8.2)}$$

where $d\sigma_0/d\omega d\varphi$ is the unpolarized differential cross-section, $d\sigma_{\parallel}/d\omega d\varphi$ represents the longitudinal differential cross-section, $d\sigma_{\perp}/d\omega d\varphi$ represents the part of transverse polarization. $r_e$ is the classical electron radius. The parameter $\kappa$ satisfies the relationship $\kappa = 4\varepsilon_0\omega_0/m_e^2$. $\varphi$ is the azimuthal angle. $P_{\parallel}$ is the laser and electron Stokes parameters, where laser is 100 % circularly polarized and electron is longitudinally polarized. $P_{\perp}$ denotes that laser is 100 % circularly polarized and electron is transversely polarized.

Longitudinal polarization primarily influences the energy distribution of scattered particles, as depicted in Fig. A8.8.2.

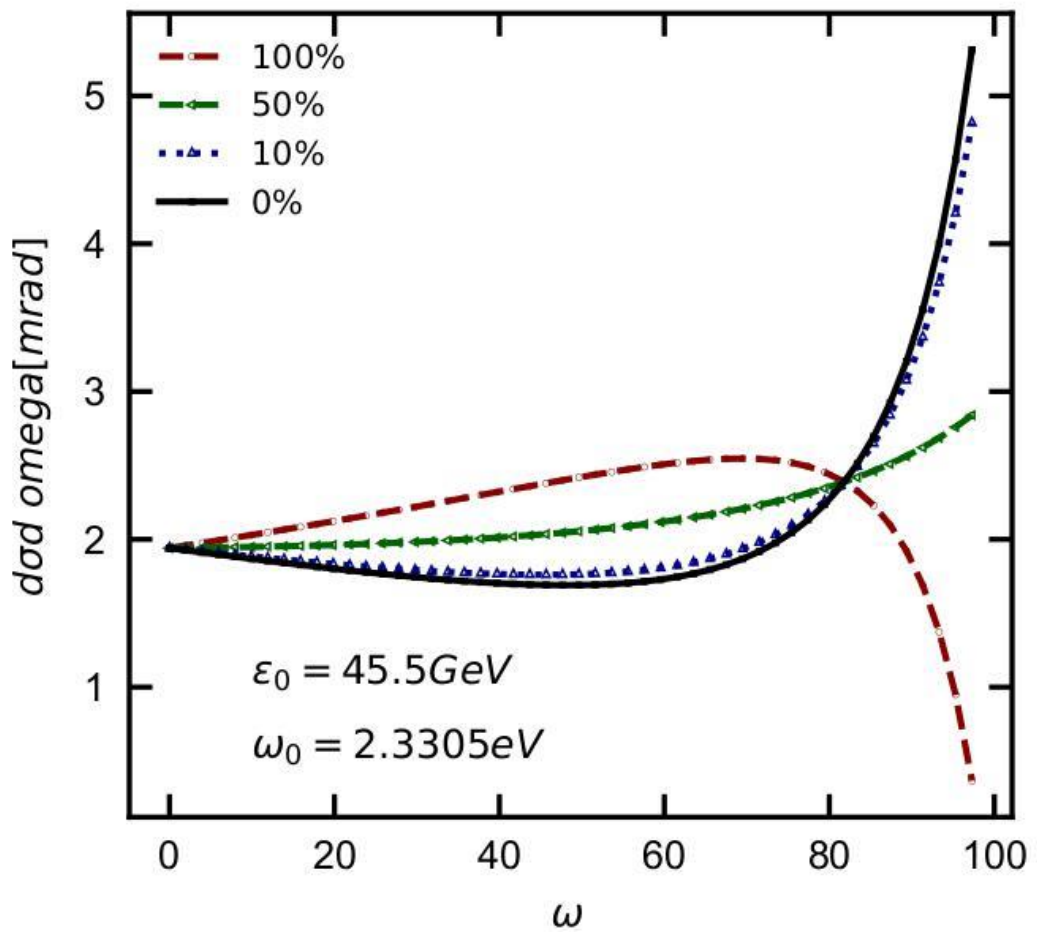

**Figure A8.8.3:** Diagram of the differential cross-section $d\sigma/d\omega$, with the energy of the scattered photon $\omega$ as the independent variable. Four distinct longitudinal polarization scenarios were examined individually: 100%, 50%, 10%, and 0%. The electron beam energy is 120 GeV, and the laser wavelength is 532 nm as an illustration.

A transversely polarized electron beam affects the spatial distribution of scattered particles in the angular plane. Gamma-ray photons generated by Compton scattering of relativistic electron beams and laser beams predominantly scatter within a narrow cone towards the electron beam direction. The photon spectrum exhibits spatial symmetry when the electron beam is unpolarized, as illustrated in Fig. A8.8.4 (a) and (b). Figures (c) through (f) demonstrate the impact of the initial polarization of the electron beam on the spatial position distribution of the scattered photons. In particular, (c) and (e) represent the disparity between the intensity spectral distribution of scattered photons containing 50% transverse polarization and the intensity spectral distribution of the unpolarized state. In this context, (c) pertains to a left-helicity laser, while (e) corresponds to a right-helicity laser. Additionally, (d) and (f) are presented as contour plots for a more detailed visualization.

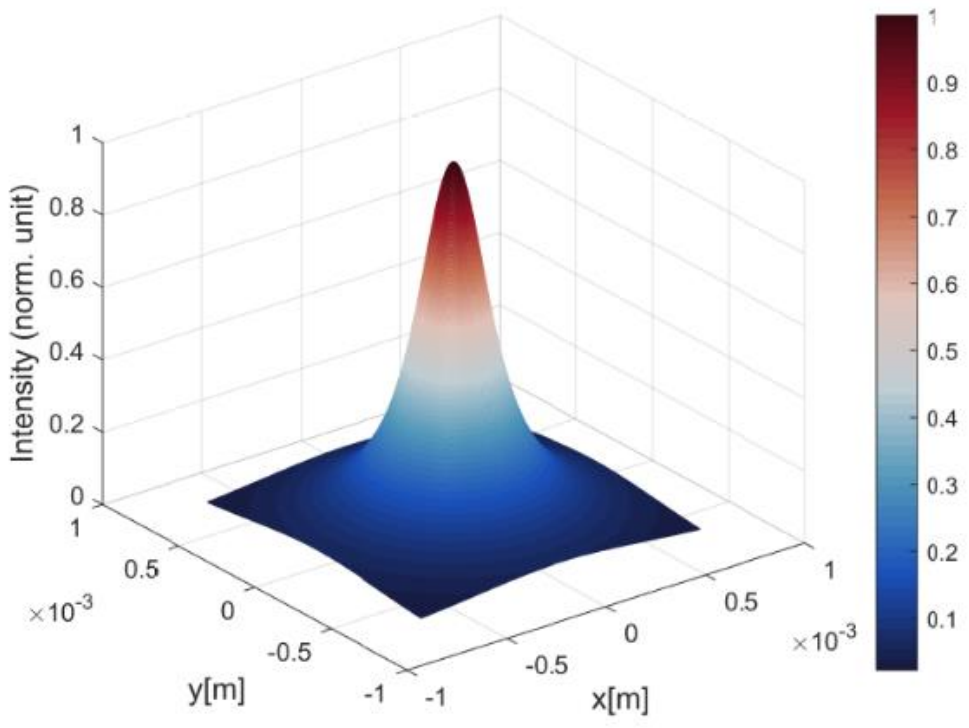

(a)

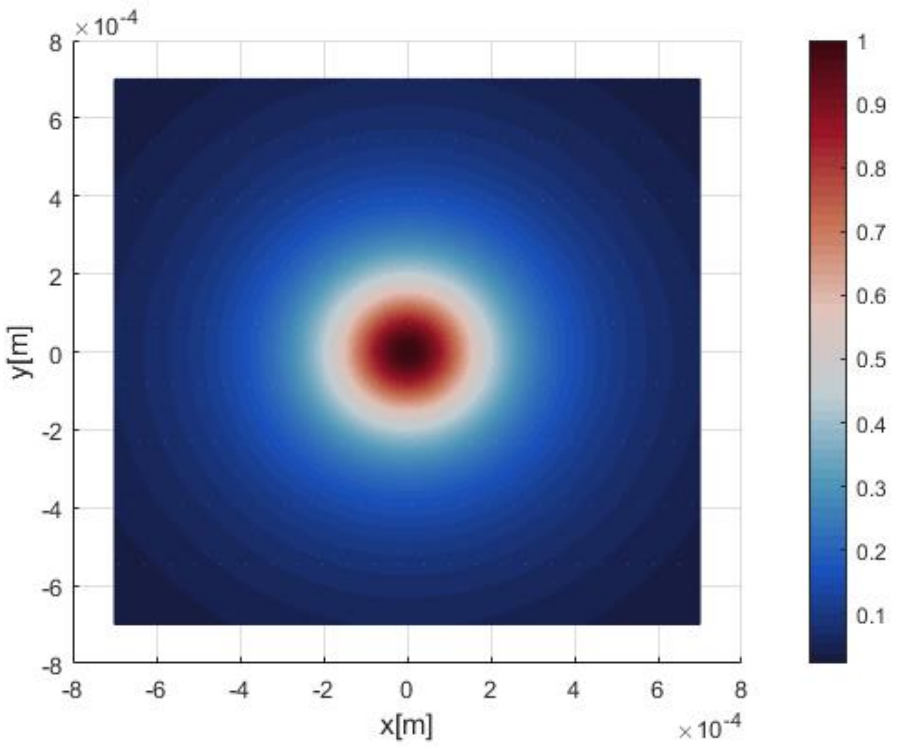

(b)

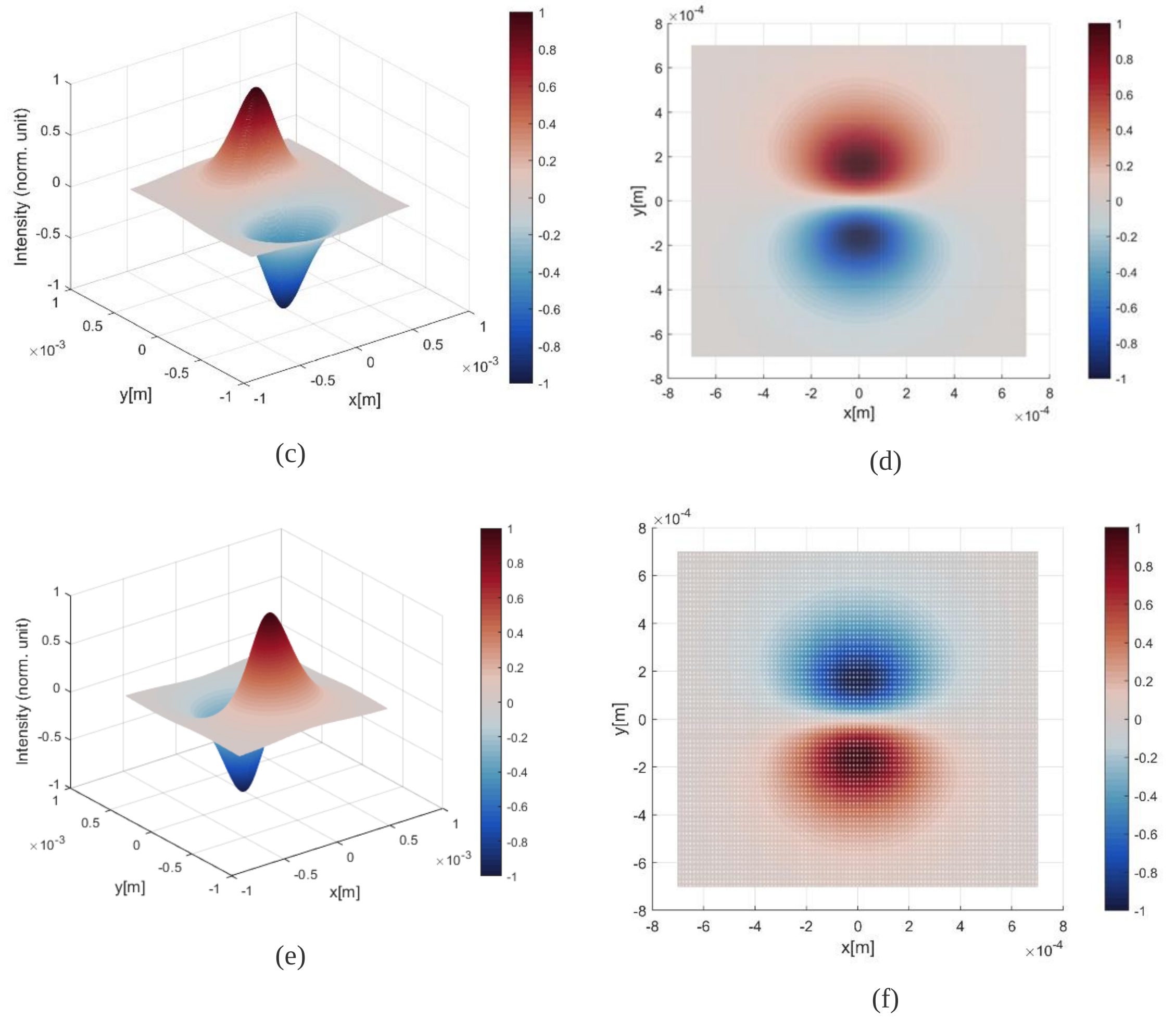

**Figure A8.8.4:** The computed spatial distribution of Compton photons [44]. A 120 GeV electron beam collides with a 532 nm laser, and the distribution is calculated 60 meters downstream from the collision point. (a) Intensity distribution with an unpolarized electron beam, (b) Contour plot of the gamma-beam intensity distribution with an unpolarized electron beam. (c) Intensity asymmetry due to a transversely polarized electron beam and a left-helicity laser. (d) Contour plot of the gamma-beam intensity distribution asymmetry with a transversely polarized electron beam and a left-helicity laser. (e) Intensity asymmetry distribution with a transversely polarized electron beam and a right-helicity laser. (f) Contour plot of the gamma-beam intensity distribution asymmetry with a transversely polarized electron beam and a right-helicity laser.

### A8.8.3: Monte-Carlo Simuation

In a Compton polarimeter, transverse beam polarization can be determined by measuring the experimental asymmetry of scattered electrons. Our data represent the Y-axis asymmetry in relation to the X-axis.

$$\mathrm{A}(X_e) = P_\perp \Pi(X_e) \tag{A8.8.3}$$

where A is the experimental asymmetry. Π is the theoretical asymmetry, which is called the analyzing power. The analyzing power can be obtained from the Compton scattering differential cross section [44]:

$$\Pi(X_e) = \frac{\int Y_e \frac{d\sigma}{dX_e dY_e} dY_e}{\int \frac{d\sigma}{dX_e dY_e} dY_e} \quad \text{(A8.8.4)}$$

Experimental asymmetry values are determined using scattered electrons detected by a position-sensitive detector, as illustrated in Fig. A8.8.5. The process involves several steps: firstly, obtaining the two-dimensional distribution of scattered electrons, denoted as $N(X_e, Y_e)$. Subsequently, calculating the average Y coordinate within each X-axis bin. Finally, determining the average Y coordinate of the Y coordinates projected onto the X-axis, denoted as ProfileX, $\overline{Y_e}_{|X_e}$.

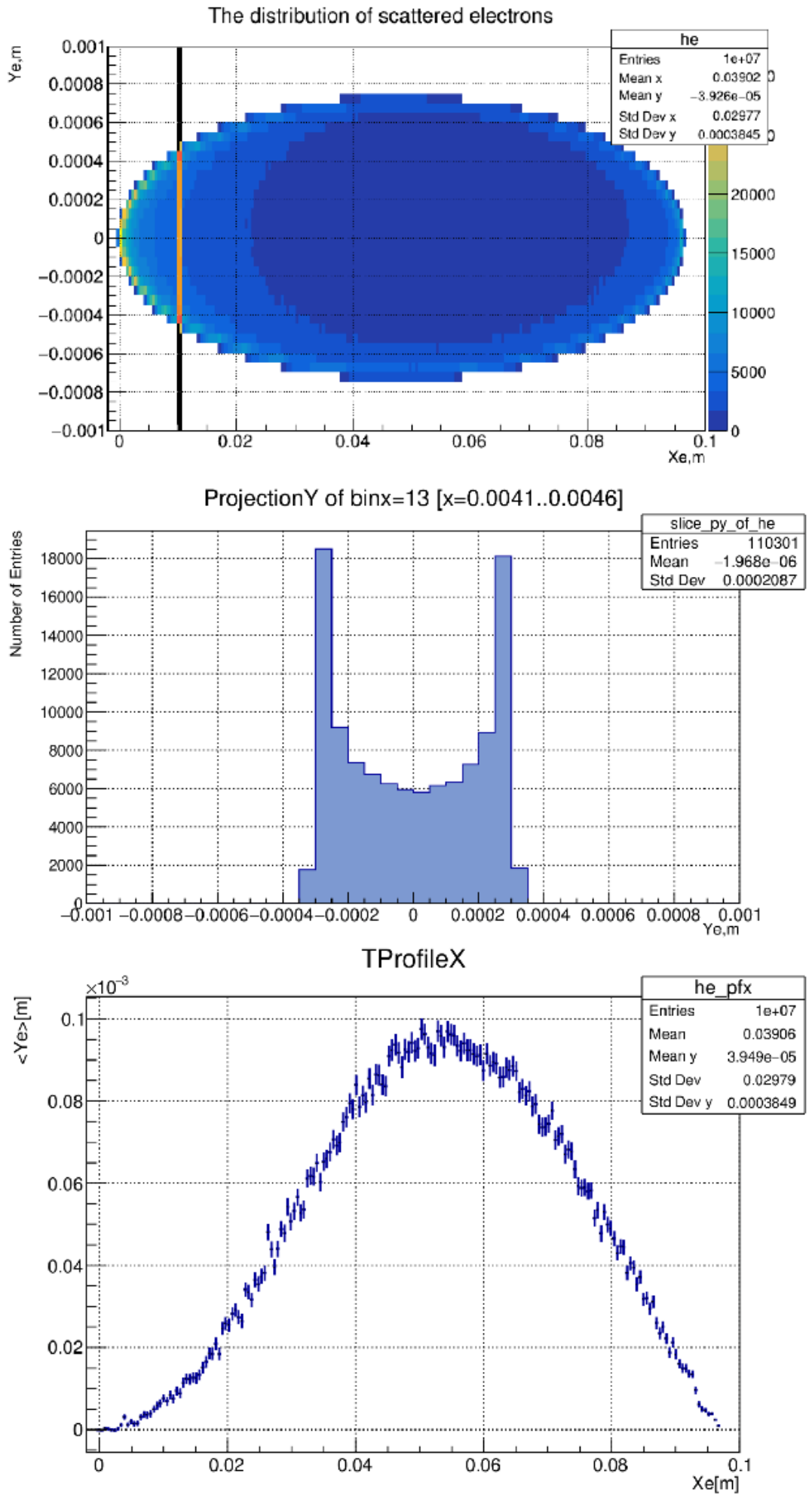

**Fig. A8.8.5:** Data postprocessing for extracting experimental asymmetry values [44].

Monte-Carlo simulations were employed to assess the polarimeter's performance. The simulations indicate that achieving a 1% statistical error is possible within a few minutes. These simulations utilize the analyzing power to fit the asymmetric positions of scattered electrons. The samples are derived from an electron bunch with an initial polarization of 10%. Scattered electrons falling outside the beampipe with energy range of 25 GeV to 39 GeV are detected. The fit result is presented in Fig. A8.8.6.

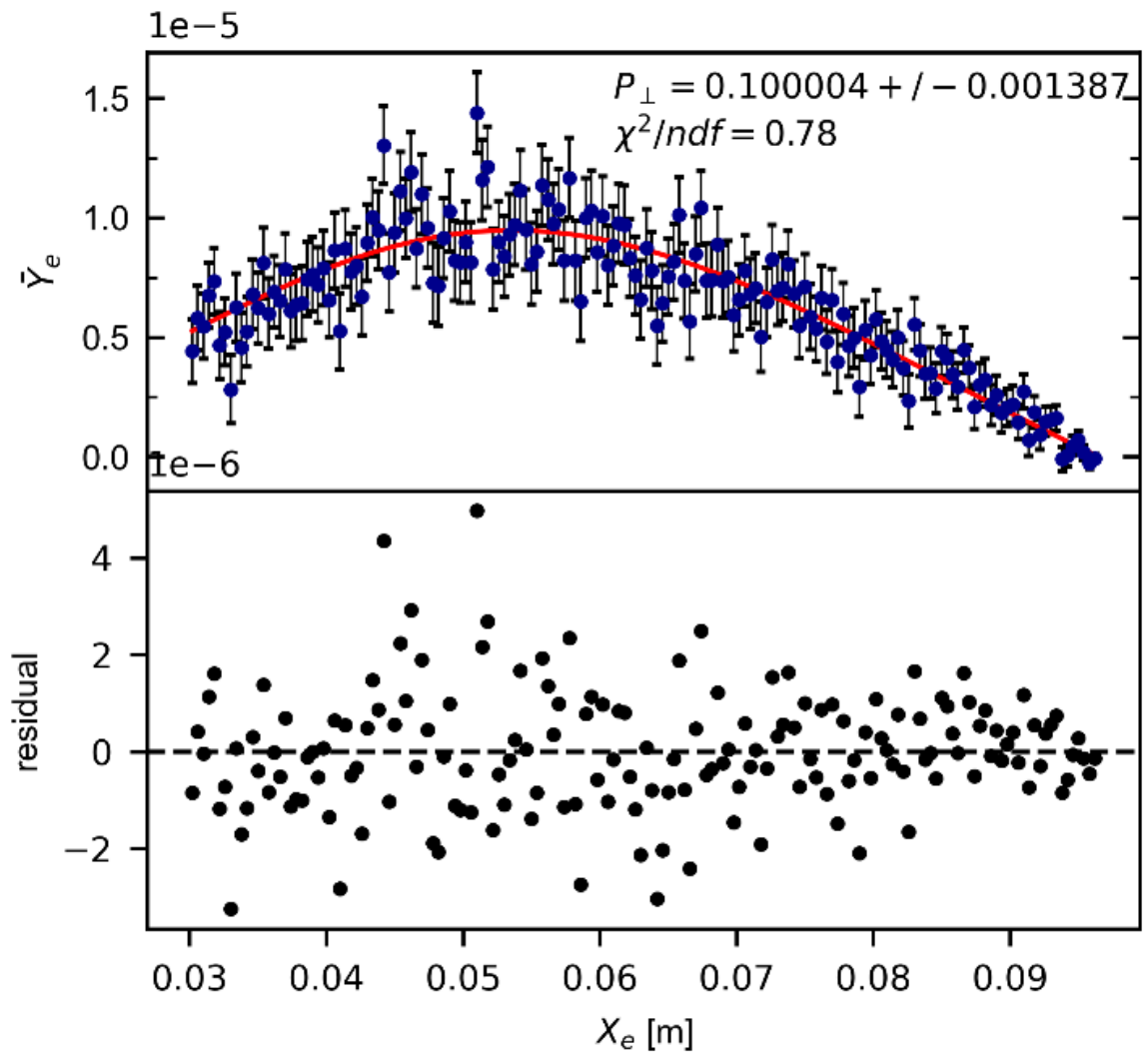

**Fig. A8.8.6:** Fitting results of the transverse polarization. The red line is the analyzing power. The blue circular point is the sample data. The residual of the sample and the fit function are also plotted.

### A8.8.4: Estimation of Errors

The total cross-section can be obtained by integrating the differential Compton cross-section:

$$\sigma_t = \pi r_e^2 \Gamma \left[\frac{2\Gamma^2 + 12\Gamma + 2}{(1-\Gamma)^2}\right] + \Gamma - 1 + \frac{6\Gamma^2 + 12\Gamma - 2}{(1-\Gamma)^3} ln\Gamma \qquad \text{(A8.8.5)}$$

where $\Gamma = 1/(1 + 4\varepsilon_0\omega_0/m_e^2)$; $\varepsilon_0, \omega_0$ is the initial energy of the electron beam and laser photons, respectively. For an electron beam interacting with a laser pulse at an angle of $\alpha$, the luminosity is given by:

$$L = N_e N_\gamma \frac{\cos(\alpha/2)}{2\pi} \frac{1}{\sqrt{\sigma_{e,y}^2 + \sigma_{\gamma,y}^2}\sqrt{(\sigma_{e,x}^2 + \sigma_{\gamma,x}^2)cos^2(\alpha/2)}\sqrt{(\sigma_{e,z}^2 + \sigma_{\gamma,z}^2)sin^2(\alpha/2)}}$$

(A8.8.6)

where $N_e$ is the electron beam population, $N_\gamma$ is the number of photons per laser pulse. The beam size parameter is $\sigma_{e,i}^2$, with $i = x, y, z$. The size parameter of the laser pulse spot is $\sigma_{\gamma,i}^2$, with $i = x, y, z$.

The electron beam parameters and various sets of laser parameters employed in the calculation are summarized in Table A8.8.1 and A8.8.2, respectively.

**Table A8.8.1:** Electron beam parameters

| Parameters | Symbol | Value |
|---|---|---|
| Mean energy | $E_b$ | 45.5 GeV |
| Bunch population | $N_b$ | $1.4 \times 10^{11}$ |
| Emittance parameters | $\epsilon_x$ | 0.27 nm |
| Emittance parameters | $\epsilon_y$ | 1.4 pm |
| Beta function | $\beta_x$ | 20 m |
| Beta function | $\beta_y$ | 40 m |
| Relative energy spread | $\sigma_E$ | 0.13% |
| Dispersion function | $D_x$ | 0.05 m |
| Beam size | $\sigma_x = \sqrt{\epsilon_x \cdot \beta_x + (\sigma_e \cdot D_x)^2}$ | 98 µm |
| Beam size | $\sigma_y = \sqrt{\epsilon_y \cdot \beta_y}$ | 7.5 µm |
| Bunch length | $\sigma_z$ | 8.7 mm |

**Table A8.8.2:** Laser parameters

| Parameters | Symbol | Version-1 | Version-2 | Version-3 |
|---|---|---|---|---|
| Each initial photon | $\omega_0$ | 1.165 eV | 1.165 eV | 2.3305 eV |
| Pulse energy | $E_L$ | 0.1 J | 0.1 J | 0.1 J |
| Laser power | $P_L$ | 5 W | 5 W | 5 W |
| Repetition frequency | $f_{rep}$ | 1 Hz | 10 Hz | 10 Hz |
| Pulse duration | $\tau_L$ | 5 ns | 5 ps | 5 ps |
| Laser spot | $\sigma_\gamma$ | 100 µm | 100 µm | 100 µm |

Taking into account the crossing angle between the electron beam and laser, which is α=2.35 mrad, the collision parameters for the three sets of laser parameters are detailed in Table A8.8.3.

**Table A8.8.3:** Colision parameters

| Parameter | Version-1 | Version-2 | Version-3 |
|---|---|---|---|
| Luminosity ($m^{-2}s^{-1}$) | $6.74 \times 10^{34}$ | $8.49 \times 10^{35}$ | $4.2 \times 10^{35}$ |
| Total cross section (mb) | 401 | 401 | 312 |
| Scattering rates ($s^{-1}$) | $2.7 \times 10^{6}$ | $3.4 \times 10^{8}$ | $1.3 \times 10^{8}$ |

Systematic uncertainties related to the polarimeter, stemming from factors such as dipole strength, drift distance, beam energy spread, and detector positioning, have been managed to be kept below 1%. These uncertainties are outlined in Table A8.8.4. Other corrections associated with the beam lattice are not addressed in this discussion.

**Table A8.8.4:** A list of polarimeter setup parameters expected to contribute to the systematic errors in the measurement of transverse polarization for a 10% polarization, with a cumulative impact of approximately 0.6%.

| **Source** | **Uncertainty** | $\Delta P/P$ |
|---|---|---|
| Dipole strength | $3.3 \times 10^{-7} T$ | 0.062% |
| L1 (IP-to-detector) | 1 cm | 0.007 % |
| L2 (Dipole-to-detector) | 1 cm | 0.051% |
| Beam energy | 100 keV | 0.0001% |
| Detector resolution | | 0.278% |
| Laser polarization | 0.2% | 0.2% |
| Total | | 0.6% |

## A8.8.5: Future Work

Establishing a Compton polarimeter represents a complex undertaking, combining the intricacies of physics and engineering. While we have introduced some fundamental considerations here, a multitude of engineering aspects still demand attention to ensure a coherent design. Once the Compton polarimeter is seamlessly integrated into the lattice, comprehensive assessments of beam-induced background effects on its performance must be conducted. This entails verifying the chosen location and enhancing our comprehension of systematic errors.

The successful operation of the laser systems and the laser-electron collision area is pivotal to maintaining stable beam performance. This encompasses the design of the laser beam transport system, dielectric mirrors, and the inclusion of a polarization analyzer to monitor laser polarization. Moreover, the incorporation of polarized pilot bunches for routine beam energy measurement by RD necessitates a decision on pulsed laser operation. Achieving optimal laser power and spot size is crucial, and simulations integrating the laser optics system with the desired polarization objectives are vital for achieving favorable polarization outcomes.

In our simulations, we employed silicon detectors for data collection. However, there are several alternatives such as diamond detectors, two-dimensional fiber array detectors, and silicon pixel detectors that can be considered for detecting the spatial distribution of scattered particles. Nevertheless, determining the detailed specifications and selecting the most suitable option among these candidates will require a comprehensive simulation and electronics design effort.

In principle, the Compton polarimeter can be designed to simultaneously measure various components of beam polarization, potentially accommodating the requirements of longitudinally polarized colliding beams. The number of polarimeters needed and their precision in measuring polarization components are fundamental questions to address in future studies.

## A8.9: Summary

This appendix outlines the possibilities for generating and utilizing beam polarization at the CEPC, primarily emphasizing Z energies. A comprehensive scheme for generating polarized beams, including their sourcing, transmission, and acceleration within the injector, leading to injection into the Collider, has undergone thorough examination. This scheme holds promise for facilitating both longitudinally polarized colliding beams and employing resonant depolarization for beam energy calibration. Furthermore, the modeling and simulations conducted to analyze beam polarization in these electron rings will lay the groundwork for exploring beam polarization generation and its applications at even higher beam energies.

These studies are currently reliant on the CEPC CDR parameters and lattice configurations, which have not been incorporated into the TDR baseline. In future iterations of the CEPC optics design, efforts will be directed toward investigating the integration of polarized beams. This may involve considerations like accommodating spin rotators and Compton polarimeters within the lattice. Furthermore, enhanced communication with the physics and detector community will be initiated to elucidate the significance and requirements of polarized beams for the scientific program, enabling the assignment of appropriate priority levels to different aspects.

## A8.10: References

# Appendix 9: Electronic Documentation System DeepC

## A9.1: Overview of DeepC

### A.9.1.1 What is DeepC

DeepC is a project management and technology asset platform designed specifically to handle the complex data management needs of CEPC's large scientific facilities. Utilizing the theory of multi-dimensional factor decomposition, this platform is capable of storing and managing engineering data related to CEPC projects, facilitating unified and efficient project and data management for CEPC's R&D and design teams.

In addition to serving as a management platform for CEPC's data and projects, DeepC can also be utilized by other large-scale projects or enterprises to solve their own data management challenges. As a technology spillover from CEPC projects, DeepC can serve as a valuable reference for the digital transformation of other industries.

DeepC has the ability to generate code-free data management systems for a wide range of business scenarios, creating a unified technical asset management platform (known as Data Middle Office) by encapsulating a set of open data management functional components. This platform enables integrated, standardized, structured, and ecological management of enterprise-level technical asset data, making it easier and more efficient for enterprises and users to manage various types of data generated by large and complex projects throughout their entire life cycle. The web service for DeepC has been developed and debugged, and a mobile app is currently under development.

### A.9.1.2 Data Management Features of the CEPC Project

The CEPC project encompasses diverse facets, including physical equipment engineering, civil construction, and general facility system support. In particular, there are 16 specialized areas within physics equipment engineering, such as power source, normal conducting and superconducting magnet, cryogenic system, vacuum system, microwave system, magnet system, magnet power supply system, mechanical system, radiation protection system, beam instrumentation, control system, superconducting radio frequency (RF) system, collimation system, silicon detector and vertex detector, calorimeter and more. Civil engineering and facility support involve creating a 100 km underground tunnel, surface buildings, and a science city. Given its considerable scale and the diversity of specialties involved, the project necessitates collaborative innovation through multi-organizational cooperation.

The CEPC project will generate massive amounts of data during its planning, design, construction, and operational phases. These data involve numerous specialties and organizations and have complex multidimensional attributes. Existing storage systems typically adopt a single directory structure, which is insufficient for managing multi-attribute project data. Additionally, the CEPC project possesses the dual attributes of complex scientific research and complex engineering, which are continuously evolving and updating based on existing complexities. Therefore, higher project management information system requirements are necessary.

## A9.2: DeepC-based CEPC Management Platform

### A.9.2.1 Architechture

(1) Define Data Management Types:

Using object-oriented element decomposition theory, we can further analyze and study the digital components of the complete CEPC project. This approach enables the rapid, code-free construction of CEPC project management organizational and knowledge management structures. In accordance with "The DAMA Guide to the Data Management Body of Knowledge (Second Edition)", critical technologies for managing multi-source data (i.e., structured and unstructured data) in CEPC have been completed, including file metadata management technology, multi-source unstructured data display, and structured and unstructured data fusion management technology. Our research focuses on object-oriented custom combinations of multi-dimensional elements, custom clustering of multi-source data, visualization technology, and the development of data management technology that enables the rapid construction of CEPC data without code or with low code on the web service platform.

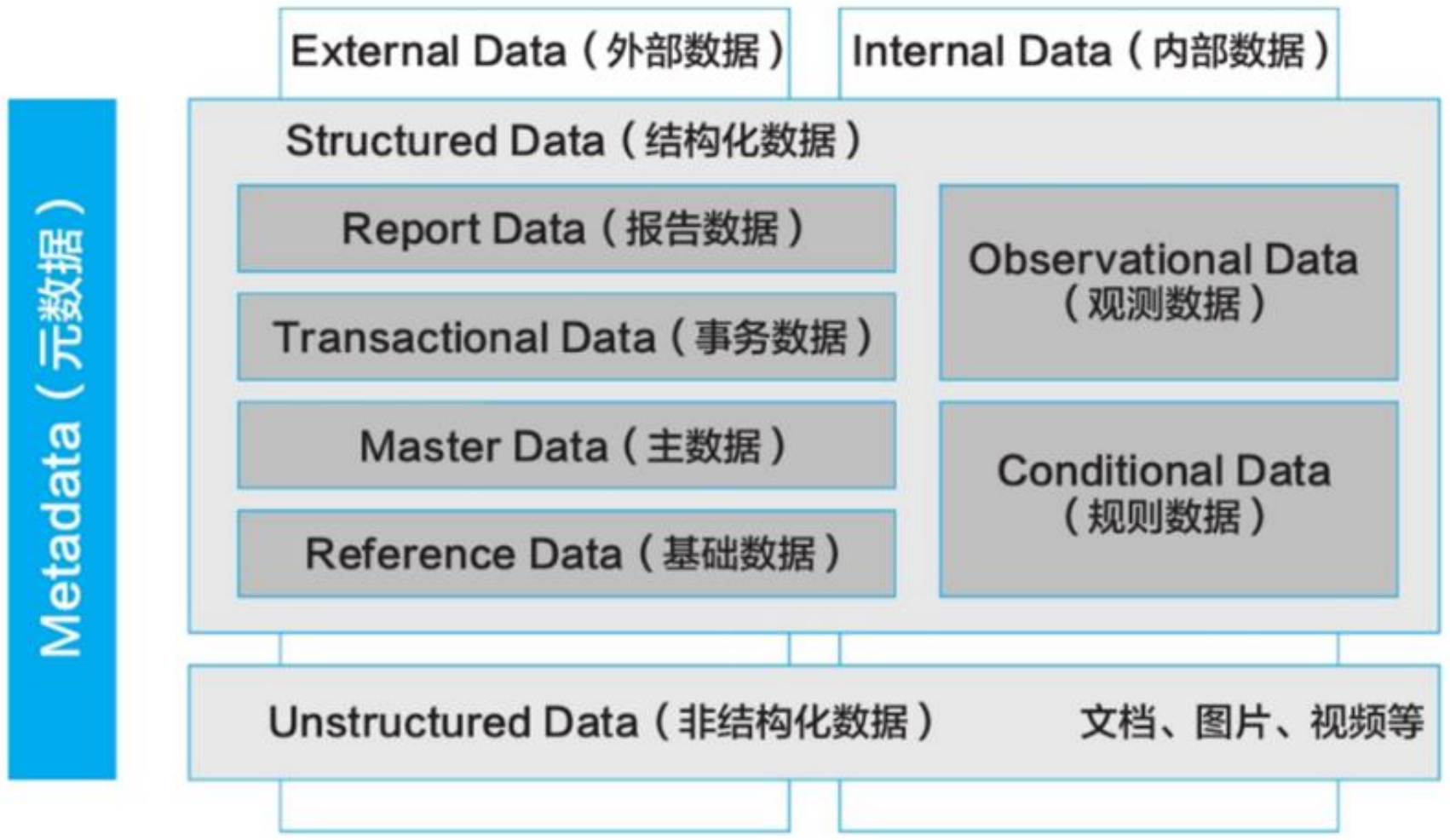

**Figure A9.1:** Schematic diagram of multi-source management data classification

(2) Establishment of CEPC Collaborative Data Management Foundations:

By collecting and studying project management experiences from major technological infrastructure projects, we will expand the project management methodology and tools to adapt to the characteristics of CEPC with its many scientific branches and iterative updates in the pre-study and design process. Furthermore, we will adapt to the characteristics of international cooperation in research and development. This will involve establishing a comprehensive project management system, methodology, and digital tools to support the planning and construction of major scientific and technological infrastructure.

(3) Composition of CEPC Management Data:

Based on the characteristics of CEPC, the DeepC system manages various types of data, such as conference data, report data (including TDR report and EDR report), test data, project progress data, policies and regulations, site selection reports, and R&D progress, among others. To effectively manage the large volume of data, it is necessary to investigate and study the CEPC data scale, temporal and spatial distribution, data type division, and data association framework. Additionally, a data management platform applicable to CEPC needs to be established, taking into consideration the computer implementation and deployment of data structure. Furthermore, a multi-terminal data query and display scheme should be proposed to facilitate the management of data.

In addition, considering the various data sources involved in CEPC and the need for data fusion, a multi-source data fusion and presentation method is proposed. This method will enable the integration of different types of data from multiple sources, allowing for more comprehensive and accurate data analysis. By applying this method, the DeepC system will be able to provide more comprehensive and accurate data management and analysis capabilities for the CEPC project.

### A.9.2.2 Multi-source Data Fusion and Presentation

Within the DeepC system, establish data boards specifically for various conferences, promotional presentations, CEPC-TDR reports, and knowledge bases for the CEPC-TDR. Conduct research on collaborative management solutions suitable for international multi-disciplinary cooperation to enhance the efficiency of cross-industry collaboration, and develop multi-source format data presentation technology for CEPC projects.

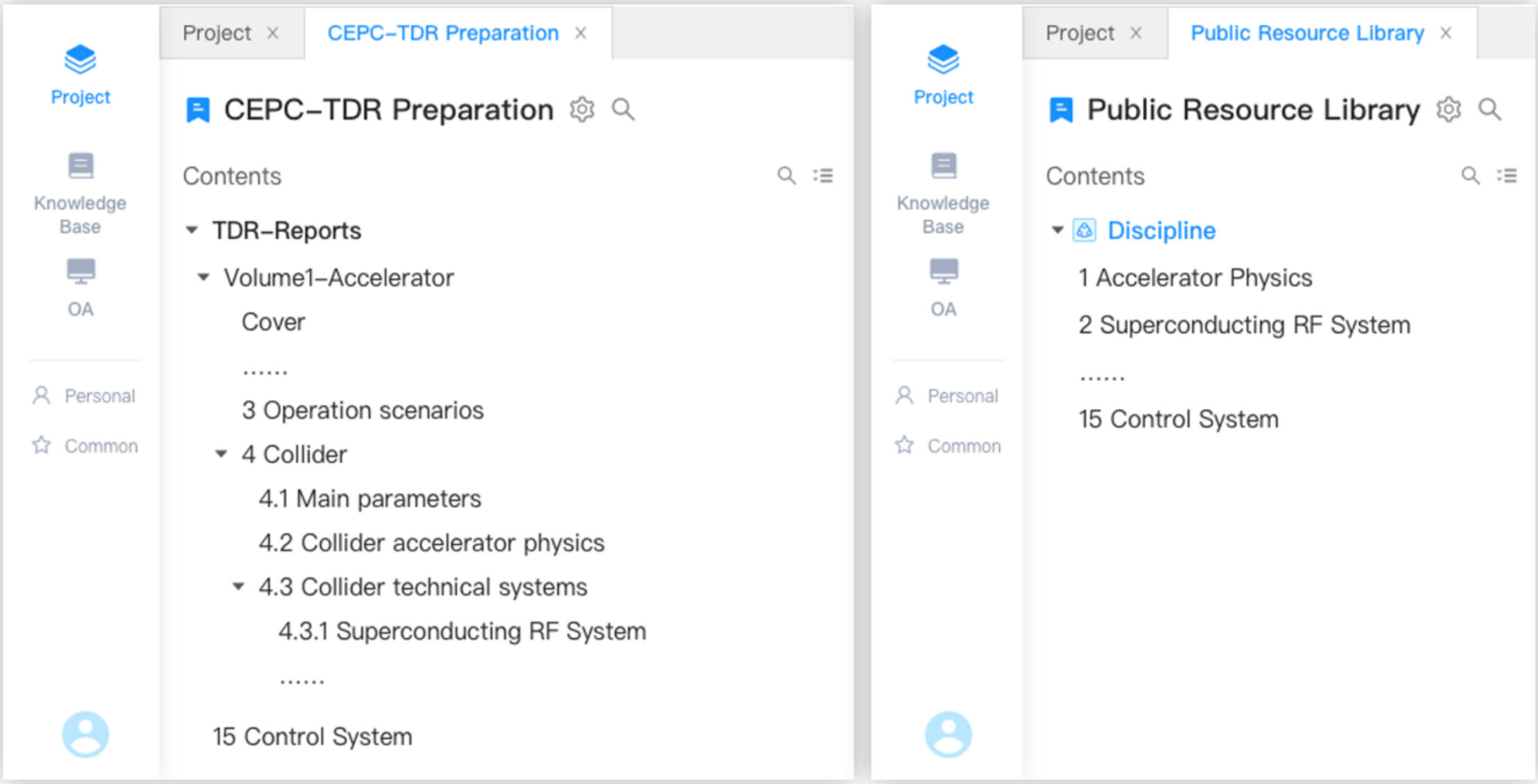

**Figure A9.2:** CEPC-TDR stage data board

### A.9.2.3 Construction of the CEPC Member Collaboration Management Platform

A new system framework is proposed for the digital collaborative management platform, taking into account the unique characteristics of the CEPC project and its multi-disciplinary requirements. The platform will be designed and developed through a combination of independent research and development, joint collaboration with other research institutions, and commissioned development. Additionally, the platform will consider the division of CEPC installation professional systems, the synergistic relationships between different majors, the nested relationships between various enterprises, international institutions, and CEPC majors, as well as the research and design depth of other majors at different stages.

As project management researchers, our goal is to streamline CEPC data management and reduce time costs without the need for extensive and complex project management knowledge. By leveraging the collaborative efforts of international multi-enterprises, we will establish a project data management platform in DeepC that can effectively manage and coordinate the complexities of the CEPC project.

# Appendix 10: Preliminary Construction Schedule

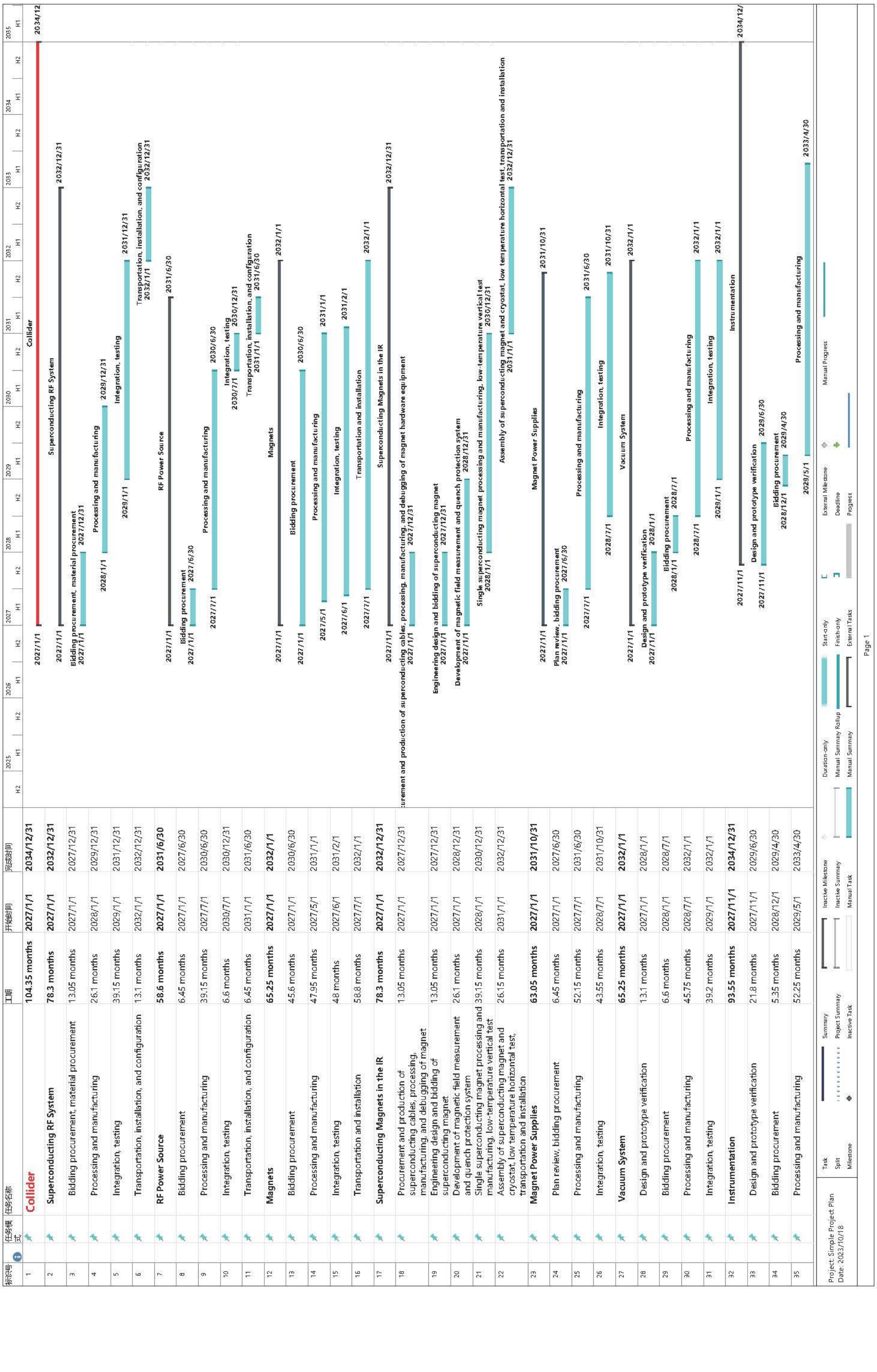

| 标识号 | 任务模式 | 任务名称 | 工期 | 开始时间 | 完成时间 |
|---|---|---|---|---|---|
| 1 | | **Collider** | **104.35 months** | **2027/1/1** | **2034/12/31** |
| 2 | | **Superconducting RF System** | **78.3 months** | **2027/1/1** | **2032/12/31** |
| 3 | | Bidding procurement, material procurement | 13.05 months | 2027/1/1 | 2027/12/31 |
| 4 | | Processing and manufacturing | 26.1 months | 2028/1/1 | 2029/12/31 |
| 5 | | Integration, testing | 39.15 months | 2029/1/1 | 2031/12/31 |
| 6 | | Transportation, installation, and configuration | 13.1 months | 2032/1/1 | 2032/12/31 |
| 7 | | **RF Power Source** | **58.6 months** | **2027/1/1** | **2031/6/30** |
| 8 | | Bidding procurement | 6.45 months | 2027/1/1 | 2027/6/30 |
| 9 | | Processing and manufacturing | 39.15 months | 2027/7/1 | 2030/6/30 |
| 10 | | Integration, testing | 6.6 months | 2030/7/1 | 2030/12/31 |
| 11 | | Transportation, installation, and configuration | 6.45 months | 2031/1/1 | 2031/6/30 |
| 12 | | **Magnets** | **65.25 months** | **2027/1/1** | **2032/1/1** |
| 13 | | Bidding procurement | 45.6 months | 2027/1/1 | 2030/6/30 |
| 14 | | Processing and manufacturing | 47.95 months | 2027/5/1 | 2031/1/1 |
| 15 | | Integration, testing | 48 months | 2027/6/1 | 2031/2/1 |
| 16 | | Transportation and installation | 58.8 months | 2027/7/1 | 2032/1/1 |
| 17 | | **Superconducting Magnets in the IR** | **78.3 months** | **2027/1/1** | **2032/12/31** |
| 18 | | Procurement and production of superconducting cables, processing, manufacturing, and debugging of magnet | 13.05 months | 2027/1/1 | 2027/12/31 |
| 19 | | Engineering design and bidding of superconducting magnet | 13.05 months | 2027/1/1 | 2027/12/31 |
| 20 | | Development of magnetic field measurement and quench protection system | 26.1 months | 2027/1/1 | 2028/12/31 |
| 21 | | Single superconducting magnet processing and manufacturing, low-temperature vertical test | 39.15 months | 2028/1/1 | 2030/12/31 |
| 22 | | Assembly of superconducting magnet and cryostat, low temperature horizontal test, transportation and installation | 26.15 months | 2031/1/1 | 2032/12/31 |
| 23 | | **Magnet Power Supplies** | **63.05 months** | **2027/1/1** | **2031/10/31** |
| 24 | | Plan review, bidding procurement | 6.45 months | 2027/1/1 | 2027/6/30 |
| 25 | | Processing and manufacturing | 52.15 months | 2027/7/1 | 2031/6/30 |
| 26 | | Integration, testing | 43.55 months | 2028/7/1 | 2031/10/31 |
| 27 | | **Vacuum System** | **65.25 months** | **2027/1/1** | **2032/1/1** |
| 28 | | Design and prototype verification | 13.1 months | 2027/1/1 | 2028/1/1 |
| 29 | | Bidding procurement | 6.6 months | 2028/1/1 | 2028/7/1 |
| 30 | | Processing and manufacturing | 45.75 months | 2028/7/1 | 2032/1/1 |
| 31 | | Integration, testing | 39.2 months | 2029/1/1 | 2032/1/1 |
| 32 | | **Instrumentation** | **93.55 months** | **2027/11/1** | **2034/12/31** |
| 33 | | Design and prototype verification | 21.8 months | 2027/11/1 | 2029/6/30 |
| 34 | | Bidding procurement | 5.35 months | 2028/12/1 | 2029/4/30 |
| 35 | | Processing and manufacturing | 52.25 months | 2029/5/1 | 2033/4/30 |

Project: Simple Project Plan
Date: 2023/10/18

Page 1

| 标识号 | 任务模式 | 任务名称 | 工期 | 开始时间 | 完成时间 |
|---|---|---|---|---|---|
| 36 | | Installation and offline debugging | 52.2 months | 2031/1/1 | 2034/12/31 |
| 37 | | **Inj. & Extr.** | **104.35 months** | **2027/1/1** | **2034/12/31** |
| 38 | | Design and prototype verification | 39.1 months | 2027/1/1 | 2029/12/31 |
| 39 | | Processing and manufacturing | 59.85 months | 2029/6/1 | 2033/12/31 |
| 40 | | Integration, testing | 39.2 months | 2031/1/1 | 2034/1/1 |
| 41 | | Installation and offline debugging | 13.1 months | 2034/1/1 | 2034/12/31 |
| 42 | | **Control System** | **65.25 months** | **2027/1/1** | **2032/1/1** |
| 43 | | Control scheme design and laboratory prototype testing | 13.1 months | 2027/1/1 | 2028/1/1 |
| 44 | | Equipment procurement | 19.6 months | 2028/1/1 | 2029/6/30 |
| 45 | | software development | 13.1 months | 2028/7/1 | 2029/6/30 |
| 46 | | Laboratory software and hardware testing | 8.65 months | 2028/12/1 | 2029/7/31 |
| 47 | | On site installation and debugging | 18.5 months | 2029/8/1 | 2030/12/31 |
| 48 | | Delivery, trial operation, joint debugging, etc | 13.1 months | 2031/1/1 | 2032/1/1 |
| 49 | | **Mechanical Systems** | **65.25 months** | **2027/1/1** | **2032/1/1** |
| 50 | | Bidding procurement | 4.3 months | 2027/1/1 | 2027/4/30 |
| 51 | | First machine processing, assembly, and acceptance | 43.6 months | 2027/5/1 | 2030/8/31 |
| 52 | | Batch processing and manufacturing | 56.6 months | 2027/7/1 | 2031/10/31 |
| 53 | | Testing, transportation, acceptance | 52.25 months | 2028/1/1 | 2032/1/1 |
| 54 | | **Booster** | **104.35 months** | **2027/1/1** | **2034/12/31** |
| 55 | | **Superconducting RF System** | 78.3 months | 2027/1/1 | 2032/12/31 |
| 56 | | Bidding procurement, material procurement | 13.05 months | 2027/1/1 | 2027/12/31 |
| 57 | | Processing and manufacturing | 26.1 months | 2028/1/1 | 2029/12/31 |
| 58 | | Integration, testing | 39.15 months | 2029/1/1 | 2031/12/31 |
| 59 | | Transportation, installation, and configuration | 13.1 months | 2032/1/1 | 2032/12/31 |
| 60 | | **RF Power Source** | 45.6 months | 2027/1/1 | 2030/6/30 |
| 61 | | Bidding procurement | 6.45 months | 2027/1/1 | 2027/6/30 |
| 62 | | Processing and manufacturing | 26.15 months | 2027/7/1 | 2029/6/30 |
| 63 | | Integration, testing | 6.6 months | 2029/7/1 | 2029/12/31 |
| 64 | | Transportation, installation, and configuration | 6.5 months | 2030/1/1 | 2030/6/30 |
| 65 | | **Magnets** | **84.75 months** | **2027/1/1** | **2033/6/30** |
| 66 | | Bidding procurement | 6.45 months | 2027/1/1 | 2027/6/30 |
| 67 | | Processing and manufacturing | 65.25 months | 2027/7/1 | 2032/6/30 |
| 68 | | Integration, testing | 65.25 months | 2028/7/1 | 2033/6/30 |
| 69 | | **Magnet Power Supplies** | **63.05 months** | **2027/1/1** | **2031/10/31** |
| 70 | | Plan review, bidding procurement | 6.45 months | 2027/1/1 | 2027/6/30 |
| 71 | | Processing and manufacturing | 52.15 months | 2027/7/1 | 2031/6/30 |

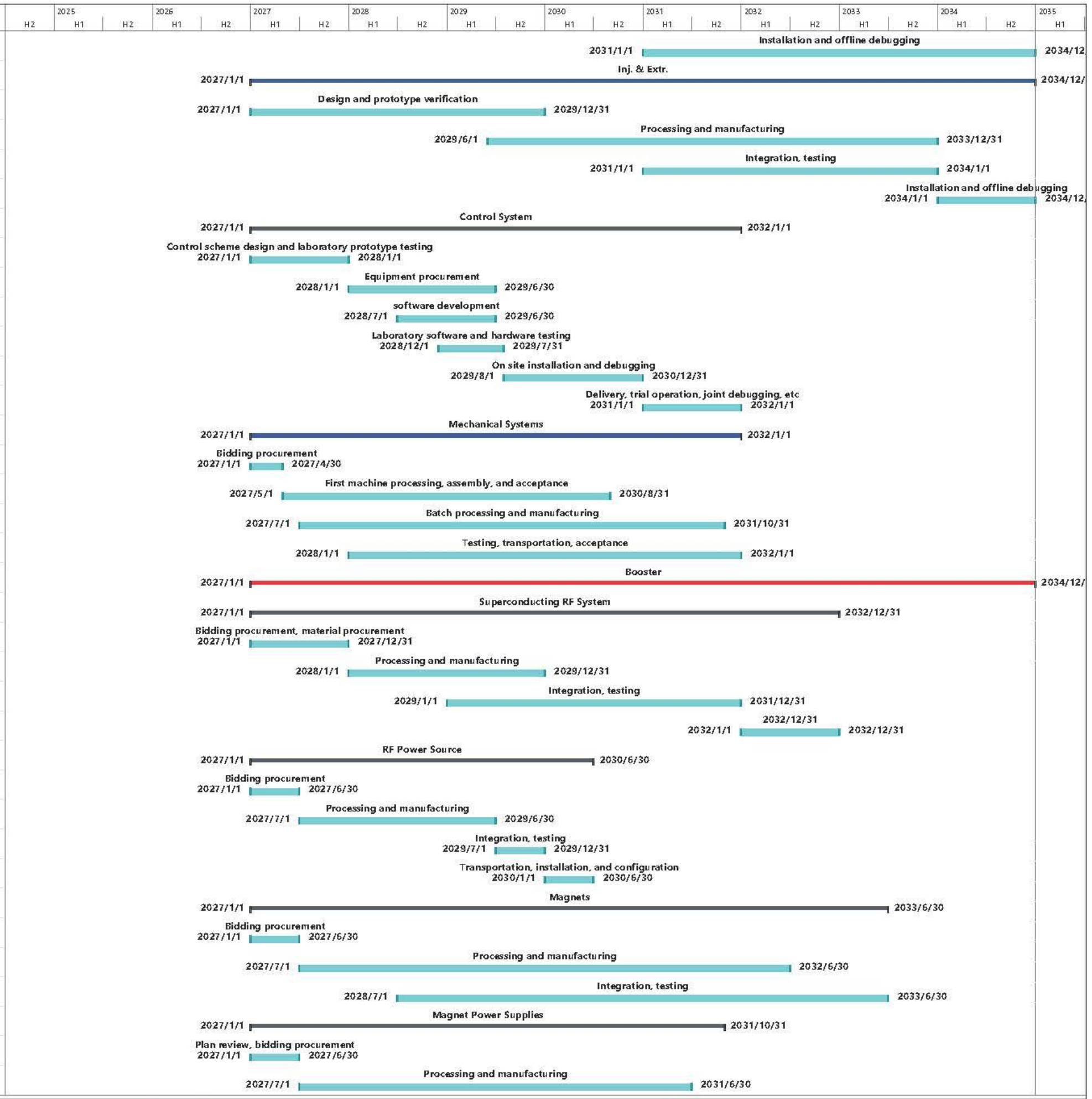

Project: Simple Project Plan
Date: 2023/10/18

| | | | | | | | |
|---|---|---|---|---|---|---|---|
| Task | Summary | Inactive Milestone | Duration-only | Start-only | External Milestone | Manual Progress |
| Split | Project Summary | Inactive Summary | Manual Summary Rollup | Finish-only | Deadline | |
| Milestone | Inactive Task | Manual Task | Manual Summary | External Tasks | Progress | |

| 标识号 | 任务模式 | 任务名称 | 工期 | 开始时间 | 完成时间 |
|---|---|---|---|---|---|
| 72 | | Integration, testing | 43.55 months | 2028/7/1 | 2031/10/31 |
| 73 | | **Vacuum System** | **65.25 months** | **2027/1/1** | **2032/1/1** |
| 74 | | Design and prototype verification | 13.1 months | 2027/1/1 | 2028/1/1 |
| 75 | | Bidding procurement | 6.6 months | 2028/1/1 | 2028/7/1 |
| 76 | | Processing and manufacturing | 45.75 months | 2028/7/1 | 2032/1/1 |
| 77 | | Integration, testing | 39.2 months | 2029/1/1 | 2032/1/1 |
| 78 | | **Instrumentation** | **93.55 months** | **2027/11/1** | **2034/12/31** |
| 79 | | Design and prototype verification | 21.8 months | 2027/11/1 | 2029/6/30 |
| 80 | | Bidding procurement | 5.35 months | 2028/12/1 | 2029/4/30 |
| 81 | | Processing and manufacturing | 52.25 months | 2029/5/1 | 2033/4/30 |
| 82 | | Installation and offline debugging | 52.2 months | 2031/1/1 | 2034/12/31 |
| 83 | | **Inj. & Extr.** | **104.35 months** | **2027/1/1** | **2034/12/31** |
| 84 | | Design and prototype verification | 39.1 months | 2027/1/1 | 2029/12/31 |
| 85 | | Processing and manufacturing | 59.85 months | 2029/6/1 | 2033/12/31 |
| 86 | | Integration, testing | 39.2 months | 2031/1/1 | 2034/1/1 |
| 87 | | Installation and offline debugging | 13.1 months | 2034/1/1 | 2034/12/31 |
| 88 | | **Control System** | **65.25 months** | **2027/1/1** | **2032/1/1** |
| 89 | | Control scheme design and laboratory prototype testing | 13.1 months | 2027/1/1 | 2028/1/1 |
| 90 | | Equipment procurement | 19.6 months | 2028/1/1 | 2029/6/30 |
| 91 | | software development | 13.1 months | 2028/7/1 | 2029/6/30 |
| 92 | | Laboratory software and hardware testing | 8.65 months | 2028/12/1 | 2029/7/31 |
| 93 | | On site installation and debugging | 18.5 months | 2029/8/1 | 2030/12/31 |
| 94 | | Delivery, trial operation, joint debugging, etc | 13.1 months | 2031/1/1 | 2032/1/1 |
| 95 | | **Mechanical Systems** | **65.25 months** | **2027/1/1** | **2032/1/1** |
| 96 | | Bidding procurement | 4.3 months | 2027/1/1 | 2027/4/30 |
| 97 | | First machine processing, assembly, and acceptance | 43.6 months | 2027/5/1 | 2030/8/31 |
| 98 | | Batch processing and manufacturing | 56.6 months | 2027/7/1 | 2031/10/31 |
| 99 | | Testing, transportation, acceptance | 52.25 months | 2028/1/1 | 2032/1/1 |
| 100 | | **Linac, Damping Ring and Sources** | **104.35 months** | **2027/1/1** | **2034/12/31** |
| 101 | | **e- source** | **52.15 months** | **2027/1/1** | **2030/12/31** |
| 102 | | Gun | 52.15 months | 2027/1/1 | 2030/12/31 |
| 103 | | Cathode-grid assembly | 52.15 months | 2027/1/1 | 2030/12/31 |
| 104 | | Pulser and High voltage system | 52.15 months | 2027/1/1 | 2030/12/31 |
| 105 | | **e+ source** | **52.15 months** | **2027/1/1** | **2030/12/31** |
| 106 | | Target | 52.15 months | 2027/1/1 | 2030/12/31 |
| 107 | | Positron converter device | 52.15 months | 2027/1/1 | 2030/12/31 |

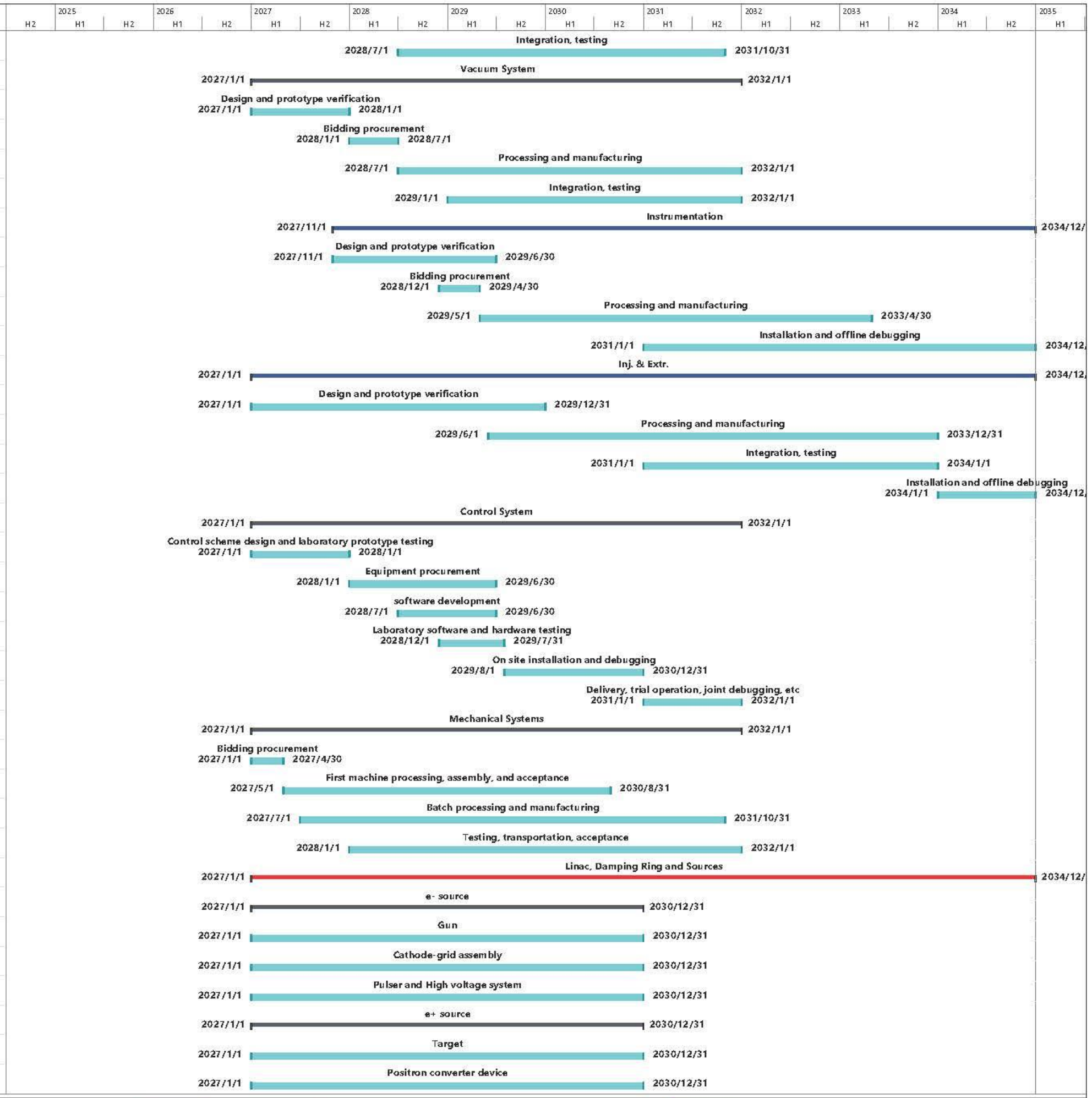

Project: Simple Project Plan
Date: 2023/10/18

Task | Split | Milestone | Summary | Project Summary | Inactive Task | Inactive Milestone | Inactive Summary | Manual Task | Duration-only | Manual Summary Rollup | Manual Summary | Start-only | Finish-only | External Tasks | External Milestone | Deadline | Progress | Manual Progress

| 标识号 | 任务模式 | 任务名称 | 工期 | 开始时间 | 完成时间 |
|---|---|---|---|---|---|
| 108 | | 15kA solid state modulator | 52.15 months | 2027/1/1 | 2030/12/31 |
| 109 | | **Linac** | **104.35 months** | **2027/1/1** | **2034/12/31** |
| 110 | | **RF System** | **52.15 months** | **2027/1/1** | **2030/12/31** |
| 111 | | Bunching cavities | 52.15 months | 2027/1/1 | 2030/12/31 |
| 112 | | Accelerating Structure | 52.15 months | 2027/1/1 | 2030/12/31 |
| 113 | | Pulse Compressor | 52.15 months | 2027/1/1 | 2030/12/31 |
| 114 | | RF Power Transfer Line | 52.15 months | 2027/1/1 | 2030/12/31 |
| 115 | | Low Level RF | 52.15 months | 2027/1/1 | 2030/12/31 |
| 116 | | **RF Power Source** | **45.6 months** | **2027/1/1** | **2030/6/30** |
| 117 | | Bidding procurement | 6.45 months | 2027/1/1 | 2027/6/30 |
| 118 | | Processing and manufacturing | 26.15 months | 2027/7/1 | 2029/6/30 |
| 119 | | Integration, testing | 6.6 months | 2029/7/1 | 2029/12/31 |
| 120 | | Transportation, installation, and configuration | 6.5 months | 2030/1/1 | 2030/6/30 |
| 121 | | **Magnets** | **44.65 months** | **2028/8/1** | **2032/1/1** |
| 122 | | Bidding procurement | 5.5 months | 2028/8/1 | 2028/12/31 |
| 123 | | Processing and manufacturing | 26.1 months | 2029/1/1 | 2030/12/31 |
| 124 | | Integration, testing | 26.15 months | 2030/1/1 | 2032/1/1 |
| 125 | | **Magnet Power Supplies** | **37 months** | **2029/1/1** | **2031/10/31** |
| 126 | | Plan review, bidding procurement | 6.55 months | 2029/1/1 | 2029/6/30 |
| 127 | | Processing and manufacturing | 26.1 months | 2029/7/1 | 2031/6/30 |
| 128 | | Integration, testing | 17.5 months | 2030/7/1 | 2031/10/31 |
| 129 | | **Vacuum System** | **65.25 months** | **2027/1/1** | **2032/1/1** |
| 130 | | Design and prototype verification | 13.1 months | 2027/1/1 | 2028/1/1 |
| 131 | | Bidding procurement | 6.6 months | 2028/1/1 | 2028/7/1 |
| 132 | | Processing and manufacturing | 45.75 months | 2028/7/1 | 2032/1/1 |
| 133 | | Integration, testing | 39.2 months | 2029/1/1 | 2032/1/1 |
| 134 | | **Instrumentation** | **93.55 months** | **2027/11/1** | **2034/12/31** |
| 135 | | Design and prototype verification | 14.2 months | 2027/11/1 | 2028/11/30 |
| 136 | | Bidding procurement | 5.35 months | 2028/12/1 | 2029/4/30 |
| 137 | | Processing and manufacturing | 57.6 months | 2028/12/1 | 2033/4/30 |
| 138 | | Installation and offline debugging | 52.2 months | 2031/1/1 | 2034/12/31 |
| 139 | | **Control System** | **45.6 months** | **2027/1/1** | **2030/6/30** |
| 140 | | Control scheme design and laboratory prototype testing | 6.45 months | 2027/1/1 | 2027/6/30 |
| 141 | | Equipment procurement | 17.5 months | 2027/7/1 | 2028/11/1 |
| 142 | | software development | 13.15 months | 2027/9/1 | 2028/9/1 |
| 143 | | Laboratory software and hardware testing | 5.5 months | 2028/9/1 | 2029/2/1 |

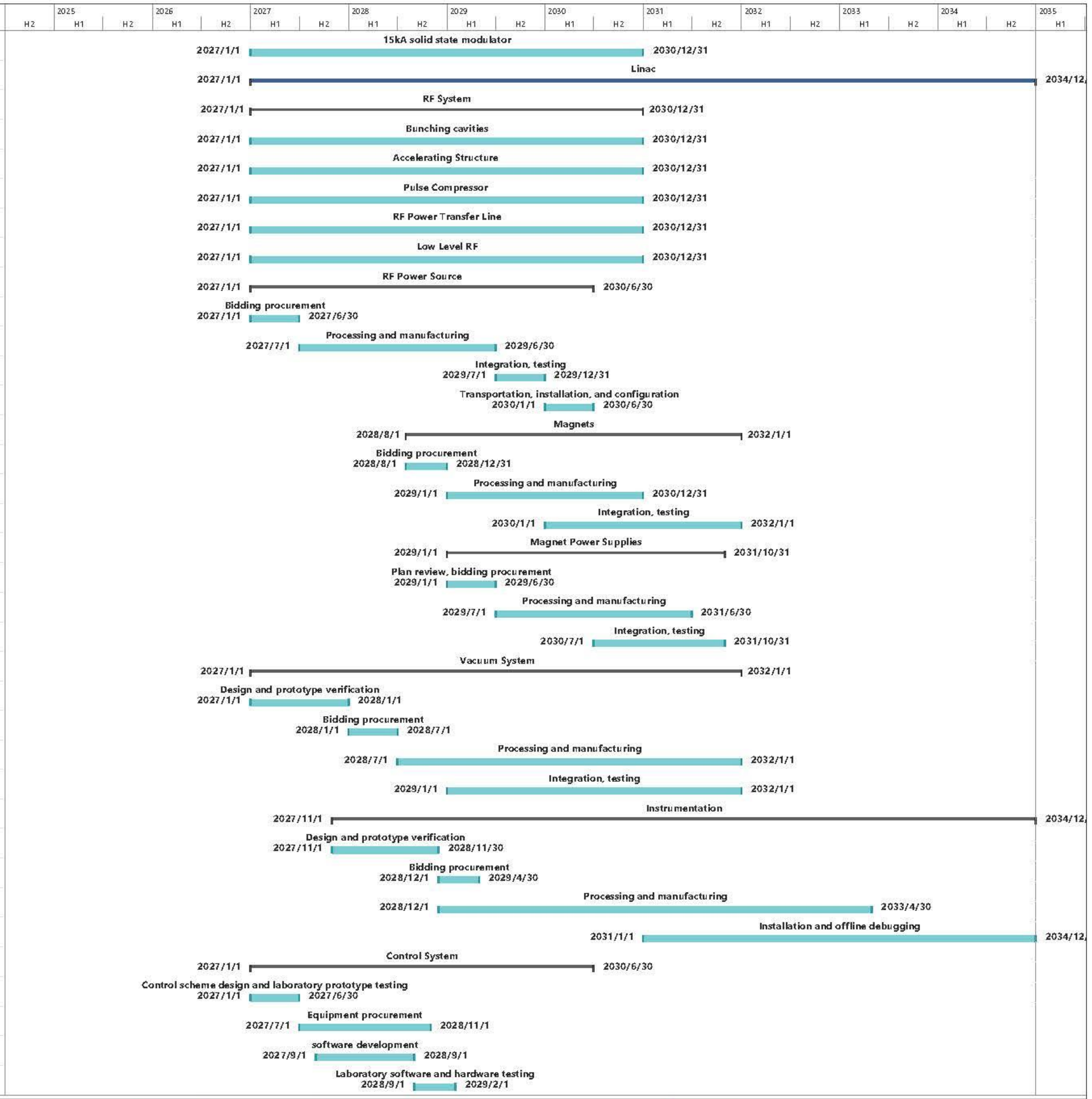

Project: Simple Project Plan
Date: 2023/10/18

| | | | | | |
|---|---|---|---|---|---|
| Task | Summary | Inactive Milestone | Duration-only | Start-only | External Milestone |
| Split | Project Summary | Inactive Summary | Manual Summary Rollup | Finish-only | Deadline |
| Milestone | Inactive Task | Manual Task | Manual Summary | External Tasks | Progress |
| Manual Progress | | | | | |

| 标识号 | 任务模式 | 任务名称 | 工期 | 开始时间 | 完成时间 |
|---|---|---|---|---|---|
| 144 | | On site installation and debugging | 8.65 months | 2029/2/1 | 2029/10/1 |
| 145 | | Delivery, trial operation, joint debugging, etc | 9.8 months | 2029/10/1 | 2030/6/30 |
| 146 | | **Mechanical Systems** | **32.75 months** | **2029/7/1** | **2032/1/1** |
| 147 | | Bidding procurement | 4.45 months | 2029/7/1 | 2029/10/31 |
| 148 | | First machine processing, assembly, and acceptance | 4.3 months | 2029/11/1 | 2030/2/28 |
| 149 | | Batch processing and manufacturing | 23.95 months | 2030/1/1 | 2031/10/31 |
| 150 | | Testing, transportation, acceptance | 2.25 months | 2031/11/1 | 2032/1/1 |
| 151 | | **Damping Ring** | **104.35 months** | **2027/1/1** | **2034/12/31** |
| 152 | | **RF System** | **52.15 months** | **2027/1/1** | **2030/12/31** |
| 153 | | Normal Conducting Cavity | 52.15 months | 2027/1/1 | 2030/12/31 |
| 154 | | RF Power Transfer Line | 52.15 months | 2027/1/1 | 2030/12/31 |
| 155 | | Low Level RF | 52.15 months | 2027/1/1 | 2030/12/31 |
| 156 | | **RF Power Source** | **32.6 months** | **2027/1/1** | **2029/6/30** |
| 157 | | Bidding procurement | 6.45 months | 2027/1/1 | 2027/6/30 |
| 158 | | Processing and manufacturing | 13.1 months | 2027/7/1 | 2028/6/30 |
| 159 | | Integration, testing | 6.6 months | 2028/7/1 | 2028/12/31 |
| 160 | | Transportation, installation, and configuration | 6.55 months | 2029/1/1 | 2029/6/30 |
| 161 | | **Magnets** | **44.65 months** | **2028/8/1** | **2032/1/1** |
| 162 | | Bidding procurement | 5.5 months | 2028/8/1 | 2028/12/31 |
| 163 | | Processing and manufacturing | 26.1 months | 2029/1/1 | 2030/12/31 |
| 164 | | Integration, testing | 26.15 months | 2030/1/1 | 2032/1/1 |
| 165 | | **Magnet Power Supplies** | **37 months** | **2029/1/1** | **2031/10/31** |
| 166 | | Plan review, bidding procurement | 6.55 months | 2029/1/1 | 2029/6/30 |
| 167 | | Processing and manufacturing | 26.1 months | 2029/7/1 | 2031/6/30 |
| 168 | | Integration, testing | 17.5 months | 2030/7/1 | 2031/10/31 |
| 169 | | **Vacuum System** | **65.25 months** | **2027/1/1** | **2032/1/1** |
| 170 | | Design and prototype verification | 13.1 months | 2027/1/1 | 2028/1/1 |
| 171 | | Bidding procurement | 6.6 months | 2028/1/1 | 2028/7/1 |
| 172 | | Processing and manufacturing | 45.75 months | 2028/7/1 | 2032/1/1 |
| 173 | | Integration, testing | 39.2 months | 2029/1/1 | 2032/1/1 |
| 174 | | **Instrumentation** | **93.55 months** | **2027/11/1** | **2034/12/31** |
| 175 | | Design and prototype verification | 14.2 months | 2027/11/1 | 2028/11/30 |
| 176 | | Bidding procurement | 5.35 months | 2028/12/1 | 2029/4/30 |
| 177 | | Processing and manufacturing | 52.25 months | 2029/5/1 | 2033/4/30 |
| 178 | | Installation and offline debugging | 52.2 months | 2031/1/1 | 2034/12/31 |
| 179 | | **Inj. & Extr.** | **91.35 months** | **2027/1/1** | **2033/12/31** |

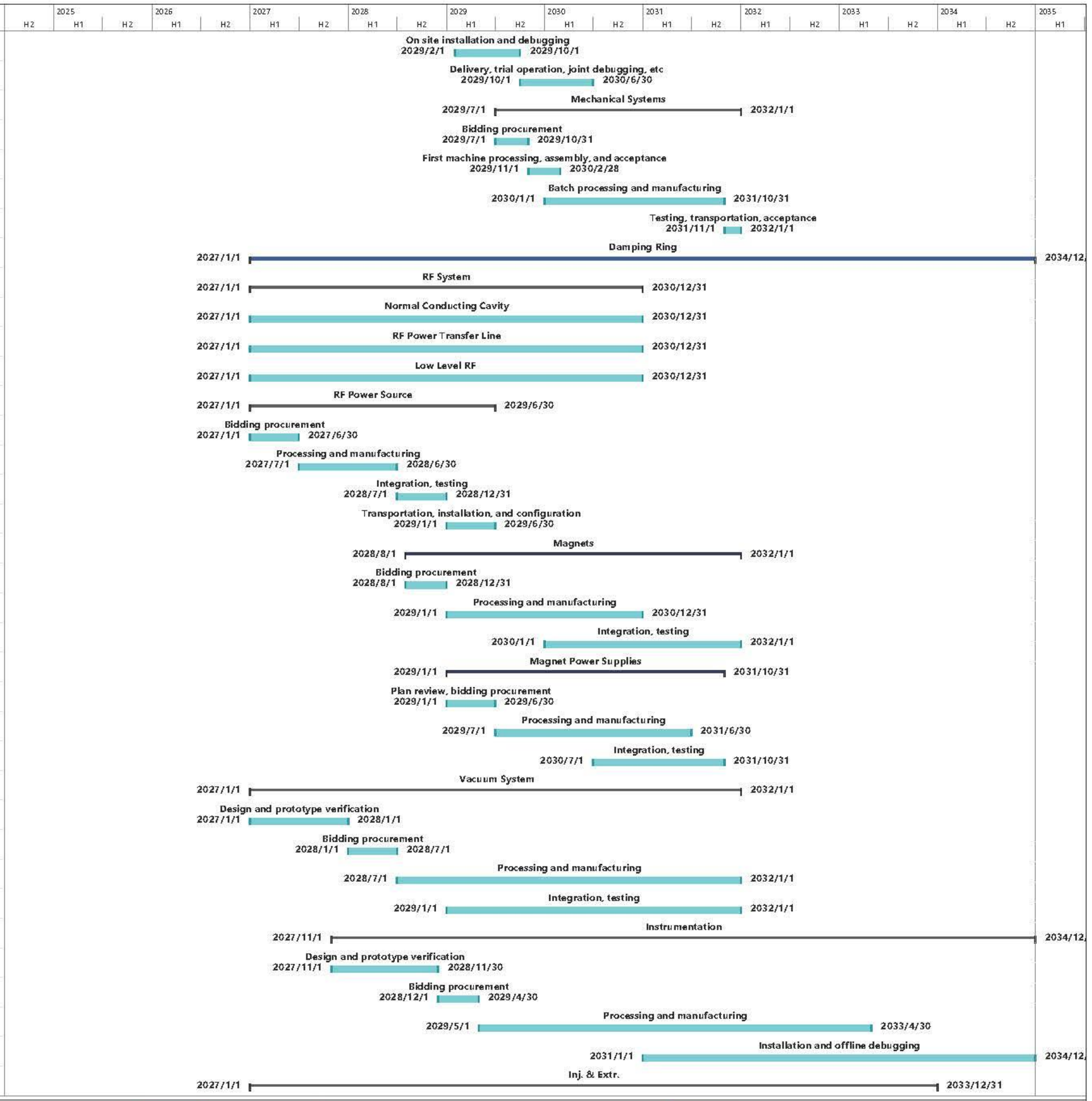

Project: Simple Project Plan
Date: 2023/10/18

Task | Summary | Inactive Milestone | Duration-only | Start-only | External Milestone | Manual Progress
Split | Project Summary | Inactive Summary | Manual Summary Rollup | Finish-only | Deadline
Milestone | Inactive Task | Manual Task | Manual Summary | External Tasks | Progress

| 标识号 | 任务模式 | 任务名称 | 工期 | 开始时间 | 完成时间 |
|---|---|---|---|---|---|
| 180 | | Design and prototype verification | 39.1 months | 2027/1/1 | 2029/12/31 |
| 181 | | Processing and manufacturing | 26.1 months | 2030/1/1 | 2031/12/31 |
| 182 | | Integration, testing | 13.1 months | 2032/1/1 | 2032/12/31 |
| 183 | | Installation and offline debugging | 13.1 months | 2033/1/1 | 2033/12/31 |
| 184 | | **Control System** | **39.15 months** | **2027/1/1** | **2030/1/1** |
| 185 | | Control scheme design and laboratoryprototype testing | 6.45 months | 2027/1/1 | 2027/6/30 |
| 186 | | Equipment procurement | 17.5 months | 2027/7/1 | 2028/11/1 |
| 187 | | software development | 13.15 months | 2027/9/1 | 2028/9/1 |
| 188 | | Laboratory software and hardware testing | 5.5 months | 2028/9/1 | 2029/2/1 |
| 189 | | On site installation and debugging | 6.5 months | 2029/2/1 | 2029/8/1 |
| 190 | | Delivery, trial operation, joint debugging, etc | 5.5 months | 2029/8/1 | 2030/1/1 |
| 191 | | **Mechanical Systems** | **32.75 months** | **2029/7/1** | **2032/1/1** |
| 192 | | Bidding procurement | 4.45 months | 2029/7/1 | 2029/10/31 |
| 193 | | First machine processing, assembly, and acceptance | 4.3 months | 2029/11/1 | 2030/2/28 |
| 194 | | Batch processing and manufacturing | 23.95 months | 2030/1/1 | 2031/10/31 |
| 195 | | Testing, transportation, acceptance | 2.25 months | 2031/11/1 | 2032/1/1 |
| 196 | | **Transport lines** | **104.35 months** | **2027/1/1** | **2034/12/31** |
| 197 | | **RF system** | **52.15 months** | **2027/1/1** | **2030/12/31** |
| 198 | | Accelering structure (1m) | 52.15 months | 2027/1/1 | 2030/12/31 |
| 199 | | pulse compressor | 52.15 months | 2027/1/1 | 2030/12/31 |
| 200 | | Waveguide system | 52.15 months | 2027/1/1 | 2030/12/31 |
| 201 | | load | 52.15 months | 2027/1/1 | 2030/12/31 |
| 202 | | Low leval RF&SSA | 52.15 months | 2027/1/1 | 2030/12/31 |
| 203 | | **RF Power Source** | **32.6 months** | **2027/1/1** | **2029/6/30** |
| 204 | | Bidding procurement | 6.45 months | 2027/1/1 | 2027/6/30 |
| 205 | | Processing and manufacturing | 13.05 months | 2027/7/1 | 2028/6/29 |
| 206 | | Integration, testing | 6.6 months | 2028/7/1 | 2028/12/31 |
| 207 | | Transportation, installation, and configuration | 6.55 months | 2029/1/1 | 2029/6/30 |
| 208 | | **Booster to collider e+ transport line** | **104.35 months** | **2027/1/1** | **2034/12/31** |
| 209 | | **Magnets** | **71.7 months** | **2027/1/1** | **2032/6/30** |
| 210 | | Bidding procurement | 6.45 months | 2027/1/1 | 2027/6/30 |
| 211 | | Processing and manufacturing | 65.25 months | 2027/7/1 | 2032/6/30 |
| 212 | | Integration, testing | 52.2 months | 2028/7/1 | 2032/6/30 |
| 213 | | **Magnet Power Supplies** | **37 months** | **2029/1/1** | **2031/10/31** |
| 214 | | Plan review, bidding procurement | 6.55 months | 2029/1/1 | 2029/6/30 |
| 215 | | Processing and manufacturing | 26.1 months | 2029/7/1 | 2031/6/30 |

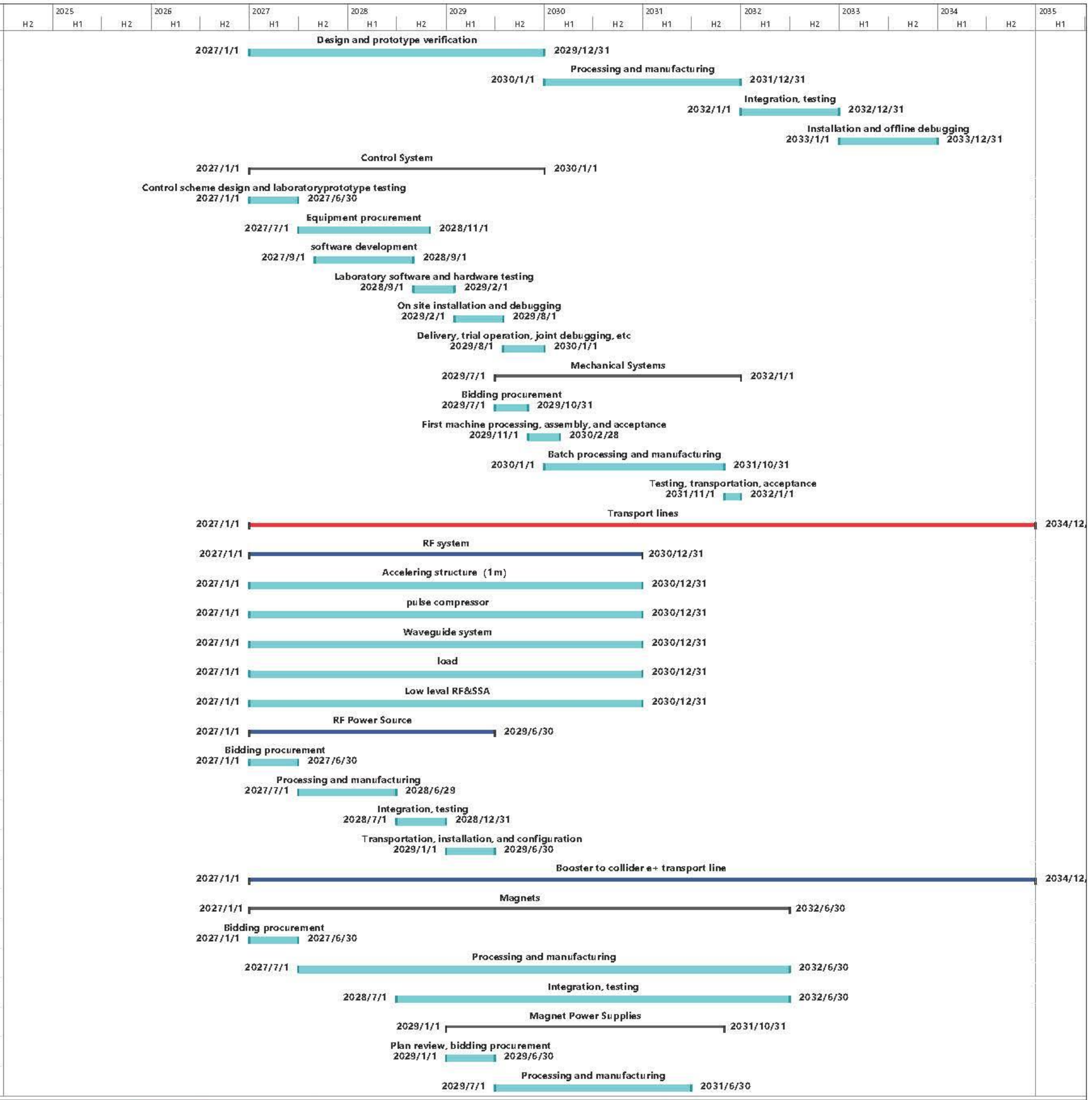

Project: Simple Project Plan
Date: 2023/10/18

Task | Split | Milestone | Summary | Project Summary | Inactive Task | Inactive Milestone | Inactive Summary | Manual Task | Duration-only | Manual Summary Rollup | Manual Summary | Start-only | Finish-only | External Tasks | External Milestone | Deadline | Progress | Manual Progress

| 标识号 | 任务模式 | 任务名称 | 工期 | 开始时间 | 完成时间 |
|---|---|---|---|---|---|
| 216 | | Integration, testing | 17.5 months | 2030/7/1 | 2031/10/31 |
| 217 | | **Vacuum System** | **65.25 months** | **2027/1/1** | **2032/1/1** |
| 218 | | Design and prototype verification | 13.1 months | 2027/1/1 | 2028/1/1 |
| 219 | | Bidding procurement | 6.6 months | 2028/1/1 | 2028/7/1 |
| 220 | | Processing and manufacturing | 45.75 months | 2028/7/1 | 2032/1/1 |
| 221 | | Integration, testing | 39.2 months | 2029/1/1 | 2032/1/1 |
| 222 | | **Instrumentation** | **93.55 months** | **2027/11/1** | **2034/12/31** |
| 223 | | Design and prototype verification | 14.2 months | 2027/11/1 | 2028/11/30 |
| 224 | | Bidding procurement | 5.35 months | 2028/12/1 | 2029/4/30 |
| 225 | | Processing and manufacturing | 52.25 months | 2029/5/1 | 2033/4/30 |
| 226 | | Installation and offline debugging | 52.2 months | 2031/1/1 | 2034/12/31 |
| 227 | | **Control System** | **65.25 months** | **2027/1/1** | **2032/1/1** |
| 228 | | Control scheme design and laboratory prototype testing | 13.1 months | 2027/1/1 | 2028/1/1 |
| 229 | | Equipment procurement | 19.6 months | 2028/1/2 | 2029/6/30 |
| 230 | | software development | 13.1 months | 2028/7/1 | 2029/6/30 |
| 231 | | Laboratory software and hardware testing | 8.65 months | 2028/12/2 | 2029/7/31 |
| 232 | | On site installation and debugging | 18.5 months | 2029/8/1 | 2030/12/31 |
| 233 | | Delivery, trial operation, joint debugging, etc | 13.1 months | 2031/1/1 | 2032/1/1 |
| 234 | | **Mechanical Systems** | **32.75 months** | **2029/7/1** | **2032/1/1** |
| 235 | | Bidding procurement | 4.45 months | 2029/7/1 | 2029/10/31 |
| 236 | | First machine processing, assembly, and acceptance | 4.3 months | 2029/11/1 | 2030/2/28 |
| 237 | | Batch processing and manufacturing | 23.95 months | 2030/1/1 | 2031/10/31 |
| 238 | | Testing, transportation, acceptance | 2.25 months | 2031/11/1 | 2032/1/1 |
| 239 | | **Booster to collider e- transport line** | **104.35 months** | **2027/1/1** | **2034/12/31** |
| 240 | | **Magnets** | **71.7 months** | **2027/1/1** | **2032/6/30** |
| 241 | | Bidding procurement | 6.45 months | 2027/1/1 | 2027/6/30 |
| 242 | | Processing and manufacturing | 65.25 months | 2027/7/1 | 2032/6/30 |
| 243 | | Integration, testing | 52.2 months | 2028/7/1 | 2032/6/30 |
| 244 | | **Magnet Power Supplies** | **37 months** | **2029/1/1** | **2031/10/31** |
| 245 | | Plan review, bidding procurement | 6.55 months | 2029/1/1 | 2029/6/30 |
| 246 | | Processing and manufacturing | 26.1 months | 2029/7/1 | 2031/6/30 |
| 247 | | Integration, testing | 17.5 months | 2030/7/1 | 2031/10/31 |
| 248 | | **Vacuum System** | **65.25 months** | **2027/1/1** | **2032/1/1** |
| 249 | | Design and prototype verification | 13.1 months | 2027/1/1 | 2028/1/1 |
| 250 | | Bidding procurement | 6.6 months | 2028/1/1 | 2028/7/1 |
| 251 | | Processing and manufacturing | 45.75 months | 2028/7/1 | 2032/1/1 |

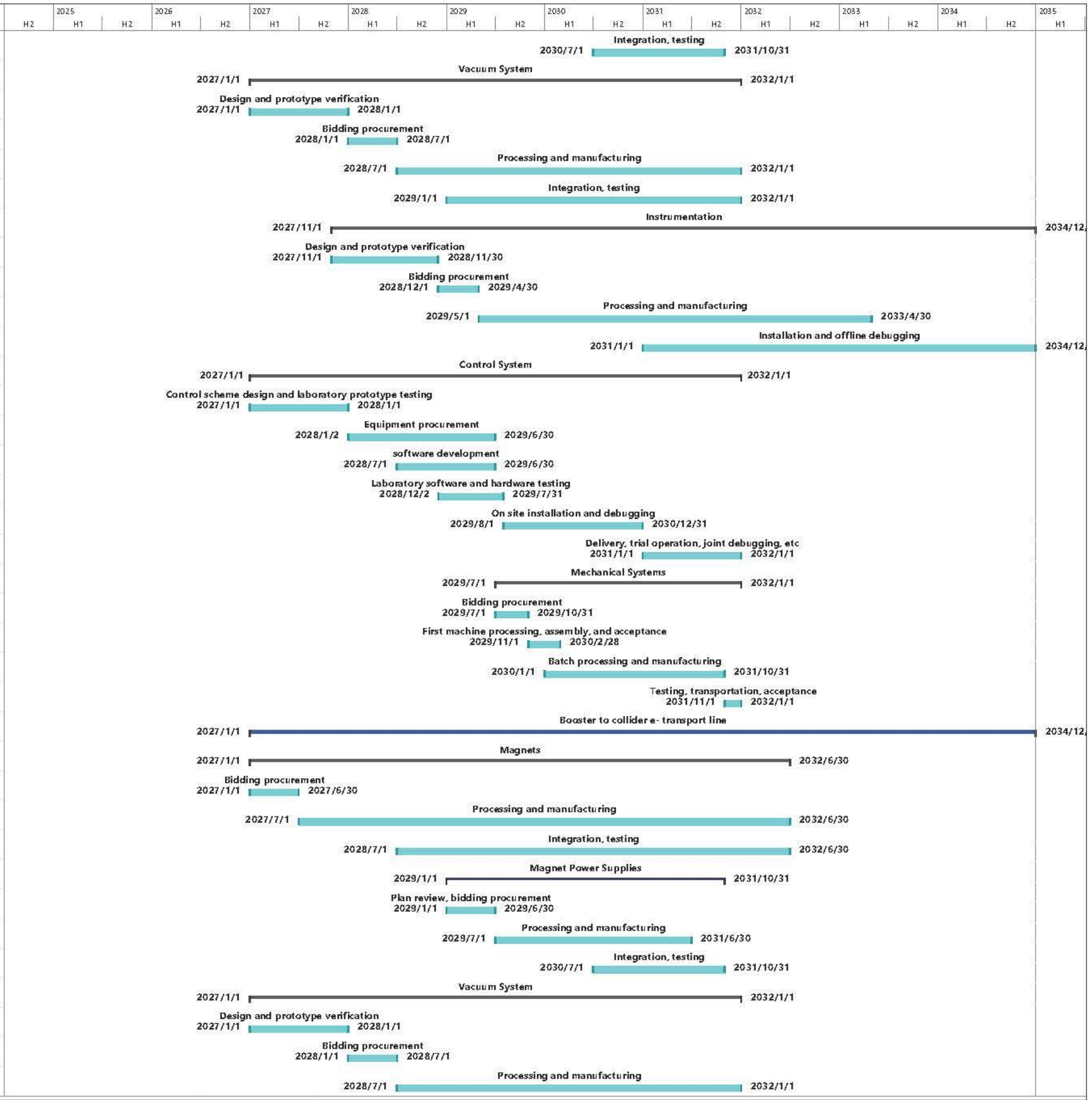

Project: Simple Project Plan
Date: 2023/10/18

Task, Split, Milestone, Summary, Project Summary, Inactive Task, Inactive Milestone, Inactive Summary, Manual Task, Duration-only, Manual Summary Rollup, Manual Summary, Start-only, Finish-only, External Tasks, External Milestone, Deadline, Progress, Manual Progress

| 标识号 | 任务模式 | 任务名称 | 工期 | 开始时间 | 完成时间 |
|---|---|---|---|---|---|
| 252 | | Integration, testing | 39.2 months | 2029/1/1 | 2032/1/1 |
| 253 | | **Instrumentation** | **93.55 months** | **2027/11/1** | **2034/12/31** |
| 254 | | Design and prototype verification | 14.2 months | 2027/11/1 | 2028/11/30 |
| 255 | | Bidding procurement | 5.35 months | 2028/12/1 | 2029/4/30 |
| 256 | | Processing and manufacturing | 52.25 months | 2029/5/1 | 2033/4/30 |
| 257 | | Installation and offline debugging | 52.2 months | 2031/1/1 | 2034/12/31 |
| 258 | | **Control System** | **65.25 months** | **2027/1/1** | **2032/1/1** |
| 259 | | Control scheme design and laboratory prototype testing | 13.1 months | 2027/1/1 | 2028/1/1 |
| 260 | | Equipment procurement | 19.6 months | 2028/1/2 | 2029/6/30 |
| 261 | | software development | 13.1 months | 2028/7/1 | 2029/6/30 |
| 262 | | Laboratory software and hardware testing | 8.65 months | 2028/12/2 | 2029/7/31 |
| 263 | | On site installation and debugging | 18.5 months | 2029/8/1 | 2030/12/31 |
| 264 | | Delivery, trial operation, joint debugging, etc | 13.1 months | 2031/1/1 | 2032/1/1 |
| 265 | | **Mechanical Systems** | **32.75 months** | **2029/7/1** | **2032/1/1** |
| 266 | | Bidding procurement | 4.45 months | 2029/7/1 | 2029/10/31 |
| 267 | | First machine processing, assembly, and acceptance | 4.3 months | 2029/11/1 | 2030/2/28 |
| 268 | | Batch processing and manufacturing | 23.95 months | 2030/1/1 | 2031/10/31 |
| 269 | | Testing, transportation, acceptance | 2.25 months | 2031/11/1 | 2032/1/1 |
| 270 | | **Linac to booster e+ transport line** | **104.35 months** | **2027/1/1** | **2034/12/31** |
| 271 | | **Magnets** | **71.7 months** | **2027/1/1** | **2032/6/30** |
| 272 | | Bidding procurement | 6.45 months | 2027/1/1 | 2027/6/30 |
| 273 | | Processing and manufacturing | 65.25 months | 2027/7/1 | 2032/6/30 |
| 274 | | Integration, testing | 52.2 months | 2028/7/1 | 2032/6/30 |
| 275 | | **Magnet Power Supplies** | **37 months** | **2029/1/1** | **2031/10/31** |
| 276 | | Plan review, bidding procurement | 6.55 months | 2029/1/1 | 2029/6/30 |
| 277 | | Processing and manufacturing | 26.1 months | 2029/7/1 | 2031/6/30 |
| 278 | | Integration, testing | 17.5 months | 2030/7/1 | 2031/10/31 |
| 279 | | **Vacuum System** | **65.25 months** | **2027/1/1** | **2032/1/1** |
| 280 | | Design and prototype verification | 13.1 months | 2027/1/1 | 2028/1/1 |
| 281 | | Bidding procurement | 6.6 months | 2028/1/1 | 2028/7/1 |
| 282 | | Processing and manufacturing | 45.75 months | 2028/7/1 | 2032/1/1 |
| 283 | | Integration, testing | 39.2 months | 2029/1/1 | 2032/1/1 |
| 284 | | **Instrumentation** | **93.55 months** | **2027/11/1** | **2034/12/31** |
| 285 | | Design and prototype verification | 14.2 months | 2027/11/1 | 2028/11/30 |
| 286 | | Bidding procurement | 5.35 months | 2028/12/1 | 2029/4/30 |
| 287 | | Processing and manufacturing | 52.25 months | 2029/5/1 | 2033/4/30 |

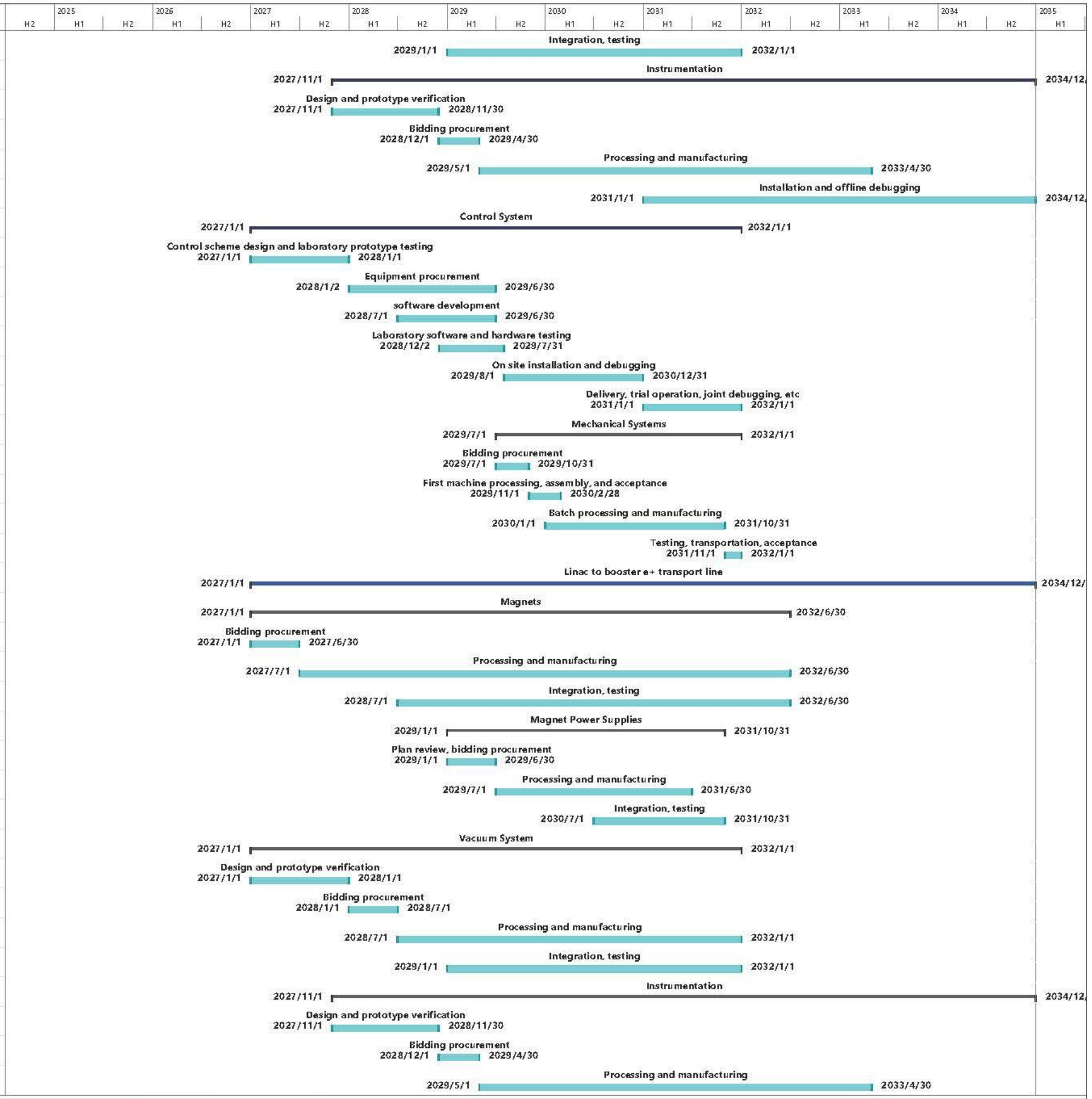

| 标识号 | 任务模式 | 任务名称 | 工期 | 开始时间 | 完成时间 |
|---|---|---|---|---|---|
| 288 | | Installation and offline debugging | 52.2 months | 2031/1/1 | 2034/12/31 |
| 289 | | **Control System** | **65.25 months** | **2027/1/1** | **2032/1/1** |
| 290 | | Control scheme design and laboratory prototype testing | 13.1 months | 2027/1/1 | 2028/1/1 |
| 291 | | Equipment procurement | 19.6 months | 2028/1/2 | 2029/6/30 |
| 292 | | software development | 13.1 months | 2028/7/1 | 2029/6/30 |
| 293 | | Laboratory software and hardware testing | 8.65 months | 2028/12/2 | 2029/7/31 |
| 294 | | On site installation and debugging | 18.5 months | 2029/8/1 | 2030/12/31 |
| 295 | | Delivery, trial operation, joint debugging, etc | 13.1 months | 2031/1/1 | 2032/1/1 |
| 296 | | **Mechanical Systems** | **32.75 months** | **2029/7/1** | **2032/1/1** |
| 297 | | Bidding procurement | 4.45 months | 2029/7/1 | 2029/10/31 |
| 298 | | First machine processing, assembly, and acceptance | 4.3 months | 2029/11/1 | 2030/2/28 |
| 299 | | Batch processing and manufacturing | 23.95 months | 2030/1/1 | 2031/10/31 |
| 300 | | Testing, transportation, acceptance | 2.25 months | 2031/11/1 | 2032/1/1 |
| 301 | | **Linac to booster e- transport line** | **104.35 months** | **2027/1/1** | **2034/12/31** |
| 302 | | **Magnets** | **71.7 months** | **2027/1/1** | **2032/6/30** |
| 303 | | Bidding procurement | 6.45 months | 2027/1/1 | 2027/6/30 |
| 304 | | Processing and manufacturing | 65.25 months | 2027/7/1 | 2032/6/30 |
| 305 | | Integration, testing | 52.2 months | 2028/7/1 | 2032/6/30 |
| 306 | | **Magnet Power Supplies** | **37 months** | **2029/1/1** | **2031/10/31** |
| 307 | | Plan review, bidding procurement | 6.55 months | 2029/1/1 | 2029/6/30 |
| 308 | | Processing and manufacturing | 26.1 months | 2029/7/1 | 2031/6/30 |
| 309 | | Integration, testing | 17.5 months | 2030/7/1 | 2031/10/31 |
| 310 | | **Vacuum System** | **65.25 months** | **2027/1/1** | **2032/1/1** |
| 311 | | Design and prototype verification | 13.1 months | 2027/1/1 | 2028/1/1 |
| 312 | | Bidding procurement | 6.6 months | 2028/1/1 | 2028/7/1 |
| 313 | | Processing and manufacturing | 45.75 months | 2028/7/1 | 2032/1/1 |
| 314 | | Integration, testing | 39.2 months | 2029/1/1 | 2032/1/1 |
| 315 | | **Instrumentation** | **93.55 months** | **2027/11/1** | **2034/12/31** |
| 316 | | Design and prototype verification | 14.2 months | 2027/11/1 | 2028/11/30 |
| 317 | | Bidding procurement | 5.35 months | 2028/12/1 | 2029/4/30 |
| 318 | | Processing and manufacturing | 52.25 months | 2029/5/1 | 2033/4/30 |
| 319 | | Installation and offline debugging | 52.2 months | 2031/1/1 | 2034/12/31 |
| 320 | | **Control System** | **65.25 months** | **2027/1/1** | **2032/1/1** |
| 321 | | Control scheme design and laboratory prototype testing | 13.1 months | 2027/1/1 | 2028/1/1 |
| 322 | | Equipment procurement | 19.6 months | 2028/1/2 | 2029/6/30 |
| 323 | | software development | 13.1 months | 2028/7/1 | 2029/6/30 |

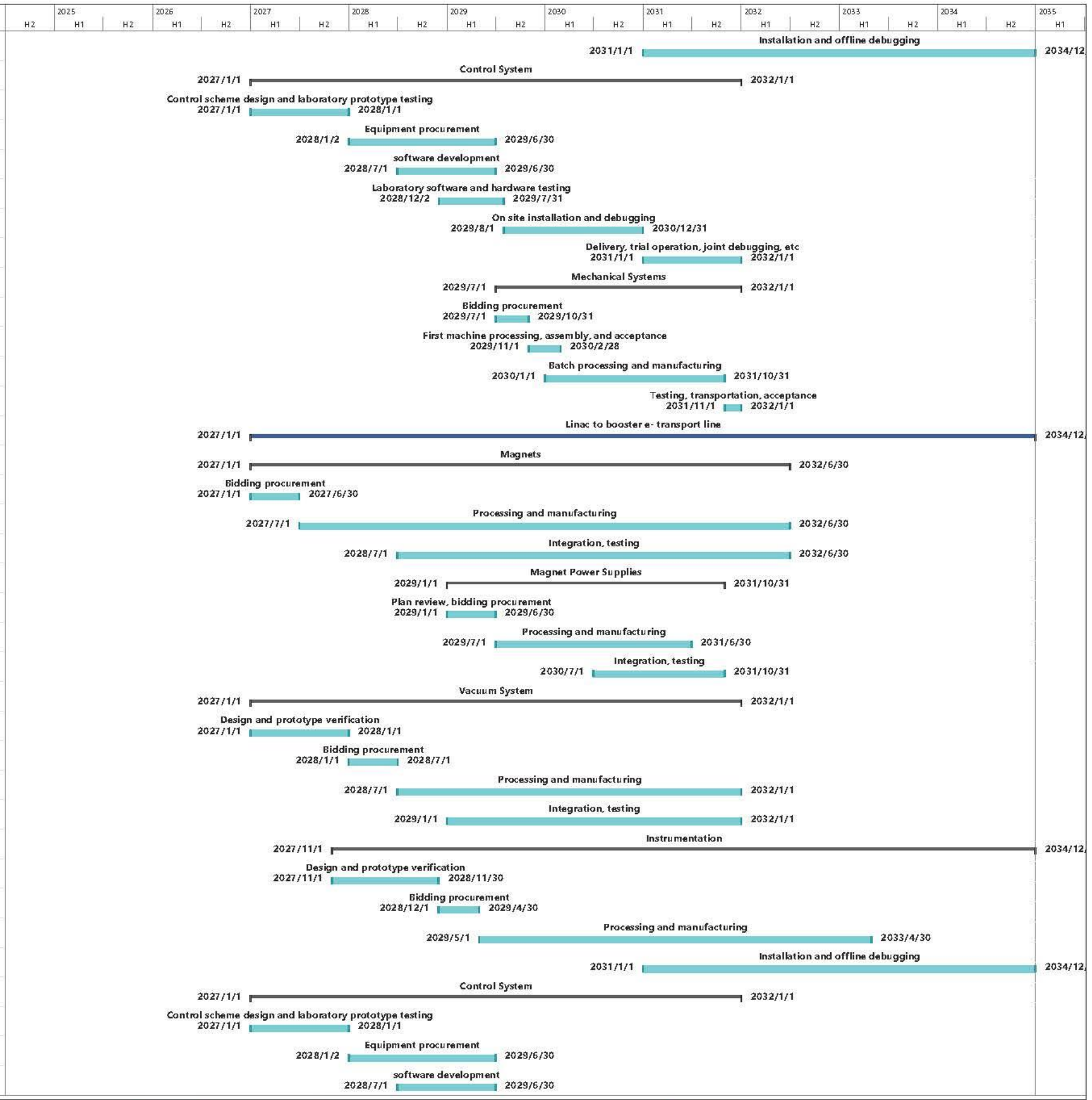

Project: Simple Project Plan
Date: 2023/10/18

Task | Split | Milestone | Summary | Project Summary | Inactive Task | Inactive Milestone | Inactive Summary | Manual Task | Duration-only | Manual Summary Rollup | Manual Summary | Start-only | Finish-only | External Tasks | External Milestone | Deadline | Progress | Manual Progress

| 标识号 | 任务模式 | 任务名称 | 工期 | 开始时间 | 完成时间 |
|---|---|---|---|---|---|
| 324 | | Laboratory software and hardware testing | 8.65 months | 2028/12/2 | 2029/7/31 |
| 325 | | On site installation and debugging | 18.5 months | 2029/8/1 | 2030/12/31 |
| 326 | | Delivery, trial operation, joint debugging, etc | 13.1 months | 2031/1/1 | 2032/1/1 |
| 327 | | **Mechanical Systems** | **32.75 months** | **2029/7/1** | **2032/1/1** |
| 328 | | Bidding procurement | 4.45 months | 2029/7/1 | 2029/10/31 |
| 329 | | First machine processing, assembly, and acceptance | 4.3 months | 2029/11/1 | 2030/2/28 |
| 330 | | Batch processing and manufacturing | 23.95 months | 2030/1/1 | 2031/10/31 |
| 331 | | Testing, transportation, acceptance | 2.25 months | 2031/11/1 | 2032/1/1 |
| 332 | | **Common systems** | **104.45 months** | **2025/1/1** | **2033/1/1** |
| 333 | | **Cryo** | **65.25 months** | **2027/1/1** | **2032/1/1** |
| 334 | | Process Design & Calculation | 7.6 months | 2027/1/1 | 2027/7/31 |
| 335 | | Bidding procurement (Cryoplant, tanks, et al) | 9.85 months | 2027/8/1 | 2028/5/1 |
| 336 | | Design work (unstandard equipment, purification and recling system design, et al) | 14.25 months | 2027/8/1 | 2028/8/31 |
| 337 | | manufacturing & assembling | 26.1 months | 2028/9/1 | 2030/8/31 |
| 338 | | Installation and offline debugging | 17.5 months | 2030/9/1 | 2032/1/1 |
| 339 | | **Survey & alignment** | **78.35 months** | **2027/1/1** | **2033/1/1** |
| 340 | | Bidding procurement | 13.1 months | 2027/1/1 | 2028/1/1 |
| 341 | | Processing and manufacturing | 26.15 months | 2028/1/1 | 2030/1/1 |
| 342 | | Integration, testing | 39.2 months | 2030/1/2 | 2033/1/1 |
| 343 | | **Radiation protection** | **91.35 months** | **2025/1/1** | **2032/1/1** |
| 344 | | Approval Documents Preparation | 26.15 months | 2025/1/1 | 2027/1/1 |
| 345 | | Radiation Shielding Design | 52.2 months | 2027/1/1 | 2031/1/1 |
| 346 | | Equipment procurement | 52.25 months | 2028/1/1 | 2032/1/1 |
| 347 | | software development | 26.15 months | 2028/1/1 | 2030/1/1 |
| 348 | | Laboratory software and hardware testing | 26.15 months | 2030/1/1 | 2032/1/1 |
| 349 | | **Conventional facilities** | **65.25 months** | **2027/1/1** | **2032/1/1** |
| 350 | | **Civil** | **58.65 months** | **2027/1/1** | **2031/7/1** |
| 351 | | **Preparation Engineering** | **8.7 months** | **2027/1/1** | **2027/9/1** |
| 352 | | External transportation road engineering | 5.4 months | 2027/1/1 | 2027/6/1 |
| 353 | | Site leveling | 3.35 months | 2027/6/1 | 2027/9/1 |
| 354 | | Construction power supply lines and substations | 8.7 months | 2027/1/1 | 2027/9/1 |
| 355 | | On site traffic road engineering | 3.35 months | 2027/6/1 | 2027/9/1 |
| 356 | | Temporary housing for construction | 4.35 months | 2027/1/1 | 2027/5/1 |
| 357 | | Construction of affiliated enterprises | 8.7 months | 2027/1/1 | 2027/9/1 |
| 358 | | **Main Ring Tunnel Engineering** | **46.75 months** | **2027/9/1** | **2031/4/1** |
| 359 | | Excavation and support of construction shaft | 3.3 months | 2027/9/1 | 2027/12/1 |

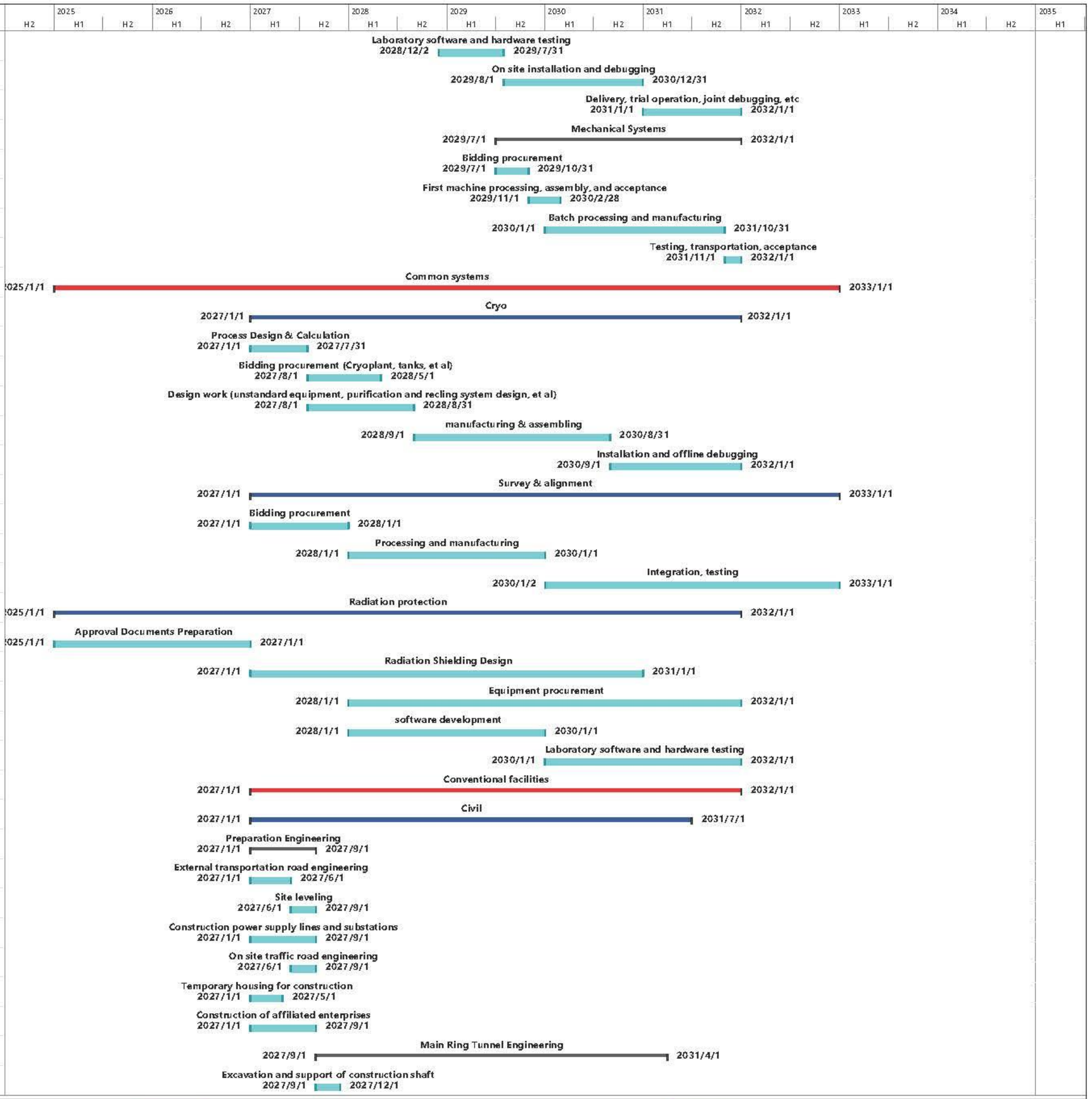

Project: Simple Project Plan
Date: 2023/10/18

Task | Split | Milestone | Summary | Project Summary | Inactive Task | Inactive Milestone | Inactive Summary | Manual Task | Duration-only | Manual Summary Rollup | Manual Summary | Start-only | Finish-only | External Tasks | External Milestone | Deadline | Progress | Manual Progress

| 标识号 | 任务模式 | 任务名称 | 工期 | 开始时间 | 完成时间 |
|---|---|---|---|---|---|
| 360 | | Concrete lining of construction shaft | 2.2 months | 2027/12/1 | 2028/1/31 |
| 361 | | Excavation and support of the main ring tunnel | 26.25 months | 2028/1/31 | 2030/2/1 |
| 362 | | Concrete lining of the main ring tunnel | 10.85 months | 2030/2/1 | 2030/12/1 |
| 363 | | Construction of waterproof layer for the main ring tunnel | 10.85 months | 2030/2/1 | 2030/12/1 |
| 364 | | Installation of vertical shaft ventilation facilities | 4.4 months | 2030/12/1 | 2031/4/1 |
| 365 | | **Experimental Hall Engineering** | **39.15 months** | **2027/9/1** | **2030/9/1** |
| 366 | | Construction of transportation channels in the experimental hall | 8.7 months | 2027/9/1 | 2028/5/1 |
| 367 | | Construction of pilot tunnel in a 15m equipment transportation shaft | 6.55 months | 2027/9/1 | 2028/3/1 |
| 368 | | Excavation and support of a 15m equipment transportation shaft | 6.65 months | 2028/3/1 | 2028/9/1 |
| 369 | | 15m equipment transportation shaft lining | 5.5 months | 2028/9/1 | 2029/2/1 |
| 370 | | Equipment transportation shaft opening equipment installation | 3.25 months | 2029/2/1 | 2029/5/2 |
| 371 | | Excavation and support of 10m traffic shaft | 8.7 months | 2027/9/1 | 2028/5/1 |
| 372 | | 10m traffic shaft lining | 10.95 months | 2028/5/1 | 2029/3/1 |
| 373 | | Excavation and support of 5m emergency escape shaft | 6.65 months | 2029/3/1 | 2029/9/1 |
| 374 | | 5m emergency escape shaft lining | 6.55 months | 2029/9/1 | 2030/3/1 |
| 375 | | Excavation and support of experimental hall | 13.1 months | 2028/5/1 | 2029/5/1 |
| 376 | | Lining of experimental hall | 8.8 months | 2029/7/1 | 2030/3/1 |
| 377 | | Excavation and lining of auxiliary tunnels | 6.65 months | 2029/3/1 | 2029/9/1 |
| 378 | | Construction of buildings in vertical shafts and installation of mechanical and electrical | 6.6 months | 2030/3/1 | 2030/9/1 |
| 379 | | **High frequency band engineering** | **18.6 months** | **2027/9/1** | **2029/2/1** |
| 380 | | Excavation and support of 10m traffic shaft | 3.85 months | 2027/9/1 | 2027/12/16 |
| 381 | | 10m traffic shaft lining | 4.9 months | 2027/12/16 | 2028/5/1 |
| 382 | | Excavation and lining of auxiliary tunnels | 6.65 months | 2028/5/1 | 2028/11/1 |
| 383 | | Installation of vertical shaft ventilation facilities | 3.35 months | 2028/11/1 | 2029/2/1 |
| 384 | | **Construction of ground ancillary buildings** | 10.85 months | 2030/2/1 | 2030/12/1 |
| 385 | | **Ventilation and electromechanical equipment installation** | 10.85 months | 2030/2/1 | 2030/12/1 |
| 386 | | **Project closure and completion** | 3.3 months | 2031/4/1 | 2031/7/1 |
| 387 | | **Mechanical and Electrical Systems** | 65.25 months | 2027/1/1 | 2032/1/1 |
| 388 | | **Mechanical Structure and Equipment** | 65.25 months | 2027/1/1 | 2032/1/1 |
| 389 | | **Temperary Construction** | 65.25 months | 2027/1/1 | 2032/1/1 |
| 390 | | **Independent Expenses** | 65.25 months | 2027/1/1 | 2032/1/1 |
| 391 | | **Other Expenses** | 65.25 months | 2027/1/1 | 2032/1/1 |

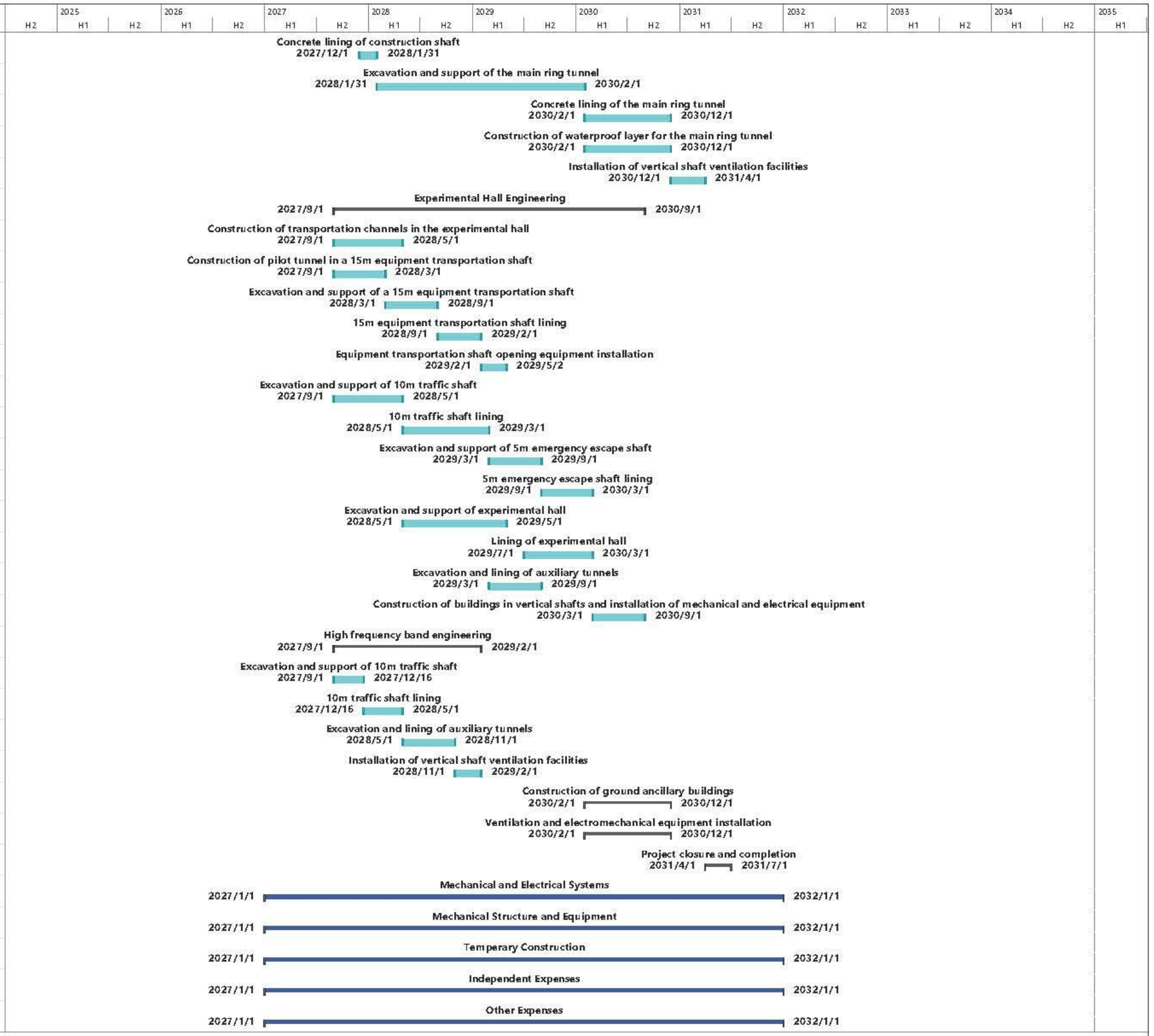

Project: Simple Project Plan
Date: 2023/10/18

| | | | | | | |
|---|---|---|---|---|---|---|
| Task | Summary | Inactive Milestone | Duration-only | Start-only | External Milestone | Manual Progress |
| Split | Project Summary | Inactive Summary | Manual Summary Rollup | Finish-only | Deadline | |
| Milestone | Inactive Task | Manual Task | Manual Summary | External Tasks | Progress | |

# Appendix 11: Advisory and Review Committees

Below is the list of members from an advisory committee (1) and several review committees (2-5). These committees have provided invaluable advice, comments, and recommendations during the development of this TDR document. Their contributions are greatly appreciated.

## 1. International Advisory Committee (IAC)

| | |
|---|---|
| Barry Barish | Caltech, Pasadena, CA |
| Maria Enrica Biagini | LNF-INFN, Frascati |
| Yuan-Hann Chang | Academia Sinica, Taipei |
| Andrew Cohen | HKUST, Hong Kong |
| Michael Davier | LAL, Center Scientifique d'Orsay, Orsay |
| Marcel Demarteau | ORNL, Oak Ridge, TN |
| Brian Foster | U. of Oxford, Oxford & DESY, Hamburg (Chair) |
| Rohini Godbole | CHEP, Indian Institute of Science, Bangalore |
| David Gross | UC Santa Barbara, CA |
| Beate Heinemann | DESY, Hamburg |
| Karl Jakobs | U. of Freiburg, Freiburg & CERN, Geneva |
| Eugene Levichev | BINP, Novosibirsk |
| Lucie Linssen | CERN, Geneva |
| Joe Lykken | Fermilab, Batavia, IL |
| Luciano Maiani | U. of Rome, Rome |
| Michelangelo Mangano | CERN, Geneva |
| Hitoshi Murayama | IPMU/UC Berkeley |
| Tatsuya Nakada | EPFL, Lausanne |
| Ian Shipsey | U. of Oxford, Oxford |
| Steinar Stapnes | CERN, Geneva |
| Geoffrey Taylor | U. of Melbourne, Parkville |
| Akira Yamamoto | KEK, Tsukuba |
| Hongwei Zhao | IMP, CAS, Lanzhou, Gansu |

## 2. International Technical Review Committee

| | |
|---|---|
| Philip Bambade | LAL, Center Scientifique d'Orsay, Orsay |
| Maria Enrica Biagini | LNF-INFN, Frascati |
| Manuela Boscolo | LNF-INFN, Frascati |
| Brian Foster | U. of Oxford & DESY, Hamburg (Observer) |
| Yoshihiro Funakoshi | KEK, Tsukuba |
| Kazuro Furukawa | KEK, Tsukuba |
| Roberto Kersevan | CERN, Geneva |
| Hélène Mainaud Durand | CERN, Geneva |
| Norihito Ohuchi | KEK, Tsukuba |
| Carlo Pagani | U. of Milano & INFN Milano, Segrate |
| Anatoly Sidorin | JINR, Dubna |
| Makoto Tobiyama | KEK, Tsukuba |
| Kay Wittenburg | DESY, Hamburg |
| Akira Yamamoto | KEK, Tsukuba |
| Zhentang Zhao | SINAP, CAS, Shanghai |
| Frank Zimmermann | CERN, Geneva (Chair) |

**3. International Cost Review Committee**

| | |
|---|---|
| Riccardo Bartolini | DESY, Hamburg |
| Brian Foster | U. of Oxford & DESY, Hamburg (Observer) |
| Philippe Lebrun | CERN, Geneva |
| Benno List | DESY, Hamburg |
| Vittorio Parma | CERN, Geneva |
| Lenny Rivkin | EPFL, Lausanne & PSI, Villigen (Chair) |
| Tetsuo Shidara | KEK, Tsukuba |
| Akira Yamamoto | KEK, Tsukuba & CERN, Geneva |

**4. Domestic Civil Engineering Cost Review Committee**

| | |
|---|---|
| Zongyin Guan | China Renewable Energy Engineering Institute, Beijing (Chair) |
| Yong Hu | Kunming Engineering Corporation, Ltd., Kunming, Yunnan |
| Jian Li | Sinohydro Bureau 14 Corporation, Ltd., Kunming, Yunnan |
| Pengxiong Li | Beijing Xiaguang Economic Technology Consulting Co., Ltd., Beijing |
| Yuanxin Luan | China Water Northeastern Investigation, Design and Research Co., Ltd., Changchun, Jilin |
| Shouzhe Zhou | Beijing Engineering Corporation, Ltd., Beijing |

**5. International Civil Engineering Cost Review Subpanel**

| | |
|---|---|
| Liam Bromiley | CERN, Geneva |
| Brian Foster | U. of Oxford & DESY, Hamburg (Observer) |
| John Osborne | CERN, Geneva (Chair) |
| Tetsuo Shidara | KEK, Tsukuba |
| Steinar Stapnes | CERN, Geneva |